%% file: main.tex
\documentclass[12pt,a4paper]{book}
\usepackage[utf8]{inputenc}
\usepackage[T1]{fontenc}
\usepackage{mathtools}
\usepackage{amsfonts}
\usepackage{fancyhdr}
\usepackage{amssymb}
\usepackage{xcolor}
\definecolor{Prune}{RGB}{99,0,60}
\usepackage{mdframed}
\usepackage{multirow}
\usepackage{multicol}
\usepackage{scrextend}
\usepackage{tikz}
\usepackage{graphicx}
\usepackage[absolute]{textpos} 
\usepackage{colortbl}
\usepackage{array}
\usepackage{geometry}
\usepackage{acronym}
\usepackage{longtable}
\usepackage{dcolumn}
\usepackage{bm}
\usepackage{feynmp-auto}
\usepackage{adjustbox}
\usepackage{footnote}
\usepackage[hyphens]{url}
\usepackage[breaklinks,colorlinks=true,urlcolor=blue,citecolor=blue]{hyperref}
\usepackage[hyphenbreaks]{breakurl}
\usepackage{float} 
\usepackage{epsfig}
\usepackage{rotating}
\usepackage{braket}
\usepackage{afterpage}
\usepackage{xspace}

\usepackage{caption}
\usepackage{subcaption}

\newcolumntype{M}[1]{>{\centering\arraybackslash}m{#1}}

\makeatletter
\newsavebox{\@brx}
\newcommand{\llangle}[1][]{\savebox{\@brx}{\(\m@th{#1\langle}\)}%
  \mathopen{\copy\@brx\mkern2mu\kern-0.9\wd\@brx\usebox{\@brx}}}
\newcommand{\rrangle}[1][]{\savebox{\@brx}{\(\m@th{#1\rangle}\)}%
  \mathclose{\copy\@brx\mkern2mu\kern-0.9\wd\@brx\usebox{\@brx}}}
\makeatother

\usepackage{empheq}

\usepackage{cancel}

\usepackage{lipsum}

\newcommand\Wider[2][3em]{%
\makebox[\linewidth][c]{%
  \begin{minipage}{\dimexpr\textwidth+#1\relax}
  \raggedright#2
  \end{minipage}%
  }%
}

\usepackage[toc]{glossaries}
\usepackage{minitoc}
\usepackage{pgfplots}
\usetikzlibrary{plotmarks}
\usetikzlibrary{shapes.geometric}
\pgfdeclareplotmark{mystar}{
    \node[star,star point ratio=2.25,minimum size=4pt,
          inner sep=0pt,draw=violet,solid,fill=violet] {};
}
\pgfdeclareplotmark{mystaro}{
    \node[star,star point ratio=2.25,minimum size=4pt,
          inner sep=0pt,draw=violet,solid,fill=white] {};
}

\newenvironment{customlegend}[1][]{%
    \begingroup
    \csname pgfplots@init@cleared@structures\endcsname
    \pgfplotsset{#1}%
}{%
    \csname pgfplots@createlegend\endcsname
    \endgroup
}%
\def\addlegendimage{\csname pgfplots@addlegendimage\endcsname}

\newcommand{\addlegendimageintext}[1]{%
    \tikz {
        \begin{customlegend}[
            legend entries={\empty},
            legend style={
                draw=none,
                inner sep=0pt,
                column sep=0pt,
                nodes={inner sep=0pt}}]
        \addlegendimage{#1}
        \end{customlegend}
    }%
}

\newcommand{\padeborel}{\!
  \addlegendimageintext{color = gray, mark = *, mark options = {scale=1, solid}, only marks, line width = 0.5pt}\!
}
\newcommand{\pbviolet}{\!
  \addlegendimageintext{dotted, color = violet, mark = *, mark options = {scale=1, solid}, line width = 1.0pt}\!
}
\newcommand{\pbgreen}{\!
  \addlegendimageintext{dashed,color = green!40!gray, mark = *, mark options = {scale=1, solid},  line width = 1.0pt}\!
}
\newcommand{\pborange}{\!
  \addlegendimageintext{dashdotted, color = orange, mark = *, mark options = {scale=1, solid, fill=white}, line width = 1.0pt}\!
}
\newcommand{\cmviolet}{\!
  \addlegendimageintext{dotted, color = violet, mark = mystar, mark options = {scale=1, solid}, line width = 1.0pt}\!
}
\newcommand{\cmvioleto}{\!
  \addlegendimageintext{dotted, color = violet, mark = mystaro, mark options = {scale=1, solid, fill=white}, line width = 1.0pt}\!
}

\newcommand{\mgviolet}{\!
  \addlegendimageintext{dotted, color = violet, mark = diamond*, mark options = {scale=1, solid, fill=violet}, line width = 1.0pt}\!
}
\newcommand{\mgvioleto}{\!
  \addlegendimageintext{dotted, color = violet, mark = diamond*, mark options = {scale=1, solid, fill=white}, line width = 1.0pt}\!
}

\usepackage{tabularx}
\makeglossaries

\fancypagestyle{plain}{}

\usepackage[style=numeric-comp,sorting=none,backend=bibtex]{biblatex}
\begin{document}

\unitlength=1mm


\pagenumbering{gobble}

\input{0FirstPage/FirstPage}

\cleardoublepage

\newgeometry{left=2.0cm,bottom=2.0cm, top=2.0cm, right=2.0cm}

\chapter*{Acknowledgements}
\input{0Acknowledgements/Acknowledgements}

\clearpage

\ifodd\value{page}\else
	\thispagestyle{empty}
\fi
\dominitoc

{\pagestyle{empty}
\tableofcontents}

\printglossary

\clearpage

\ifodd\value{page}\else
	\thispagestyle{empty}
\fi

\setcounter{chapter}{0}

\clearpage
\markboth{}{}

\include{Acronyms}

\clearpage

\ifodd\value{page}\else
	\thispagestyle{empty}
\fi

\pagenumbering{arabic}

\setcounter{page}{1}

\chapter{\label{chap:Intro1}Introduction}
\input{1ChapterIntroManyBodyProblem/IntroManyBodyProblem}

\clearpage

\ifodd\value{page}\else
	\thispagestyle{empty}
\fi

\chapter{\label{chap:Intro2}Setting the stage}
\input{2ChapterIntroPathIntegral/IntroPathIntegral}

\clearpage

\ifodd\value{page}\else
	\thispagestyle{empty}
\fi

\chapter{\label{chap:DiagTechniques}Diagrammatic techniques}
\input{4ChapterDiag/Diag}

\clearpage

\ifodd\value{page}\else
	\thispagestyle{empty}
\fi

\chapter{\label{chap:FRG}Functional renormalization group techniques}
\input{5ChapterFRG/FRG}

\clearpage

\ifodd\value{page}\else
	\thispagestyle{empty}
\fi

\chapter{\label{chap:Conclusion}Conclusion}
\input{6ChapterConclusion/Conclusion}

\clearpage

\ifodd\value{page}\else
	\thispagestyle{empty}
\fi

\appendix

\chapter{\label{ann:GaussianInt}Gaussian integration}
\input{7Appendix/GaussianIntegration}

\clearpage

\ifodd\value{page}\else
	\thispagestyle{empty}
\fi

\chapter{\label{ann:LargeN}$1/N$-expansion}
\input{7Appendix/LargeNExpansion}

\clearpage

\ifodd\value{page}\else
	\thispagestyle{empty}
\fi

\chapter{\label{ann:Diag}Diagrams and multiplicities}
\input{7Appendix/Diag}

\clearpage

\ifodd\value{page}\else
	\thispagestyle{empty}
\fi

\chapter{\label{ann:InversionMethod}Inversion method}
\input{7Appendix/IM}

\clearpage

\ifodd\value{page}\else
	\thispagestyle{empty}
\fi

\chapter{\label{ann:Derivations1PIFRG}1PI functional renormalization group}
\input{7Appendix/Derivations1PIFRG}

\clearpage

\ifodd\value{page}\else
	\thispagestyle{empty}
\fi

\chapter{\label{ann:Derivations2PIFRG}2PI functional renormalization group}
\input{7Appendix/Derivations2PIFRG}

\clearpage

\ifodd\value{page}\else
	\thispagestyle{empty}
\fi

\chapter{\label{ann:Derivations2PPIFRG}2PPI functional renormalization group}
\input{7Appendix/Derivations2PPIFRG}

\clearpage

\ifodd\value{page}\else
	\thispagestyle{empty}
\fi

\chapter{\label{ann:SummaryFrench}Summary in French - Résumé en français}
\input{7Appendix/SummaryFrench}

\clearpage

\ifodd\value{page}\else
	\thispagestyle{empty}
\fi

\sloppy

\markboth{}{}

\printbibliography

\clearpage

\ifodd\value{page}\else
	\thispagestyle{empty}
\fi

\input{8LastPage/LastPage}


\end{document}

%% file: 0FirstPage/FirstPage.tex
\begin{titlepage}

\newgeometry{left=7.5cm,bottom=2cm, top=1cm, right=1cm}

\tikz[remember picture,overlay] \node[opacity=1,inner sep=0pt] at (-28mm,-135mm){\includegraphics{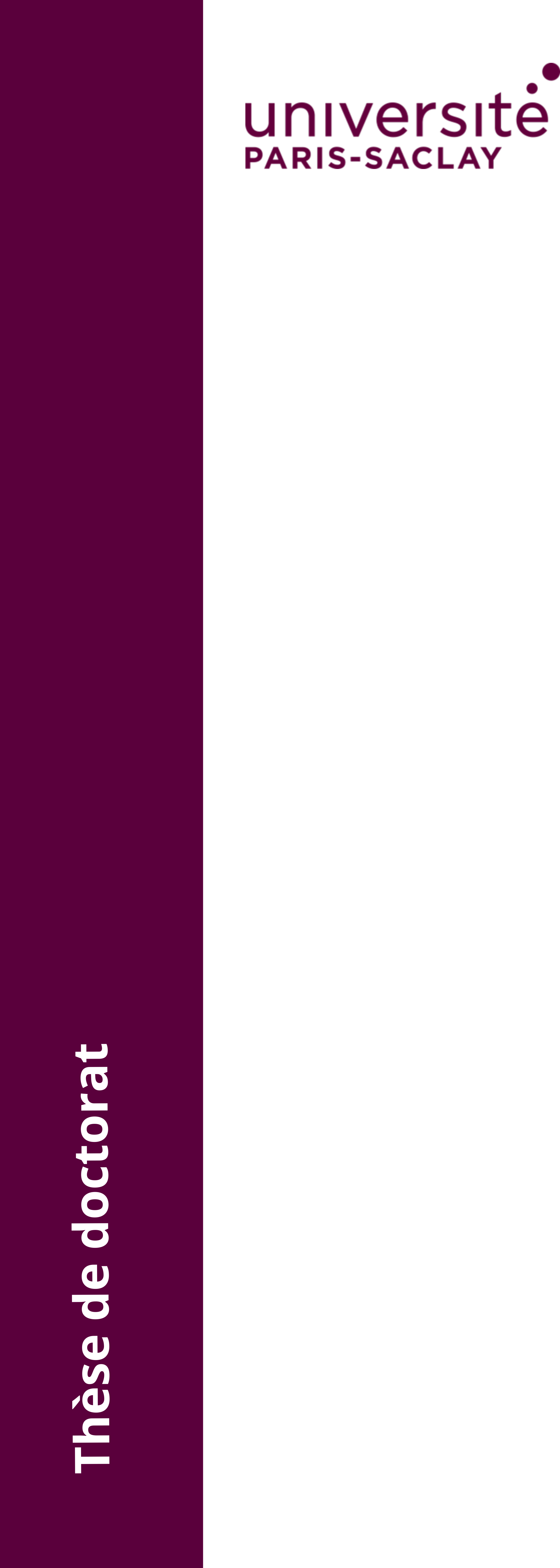}};

\fontfamily{fvs}\fontseries{m}\selectfont

\color{white}

\begin{picture}(0,0)

\put(-53,-258.25){\rotatebox{90}{NNT : 2021UPASP089}}

\end{picture}

\vspace{10mm}

\flushright
\vspace{10mm}
\color{Prune}
\fontfamily{fvs}\fontseries{m}\fontsize{22}{26}\selectfont
  Path-integral approaches to strongly-coupled quantum many-body systems\\
\bigskip
\color{black}
  \textit{Approches de type intégrale de chemin pour l'étude de systèmes quantiques à $N$ corps fortement corrélés}

\normalsize
\vspace{1.4cm}

\color{black}
\textbf{Thèse de doctorat de l'université Paris-Saclay}

\vspace{0.9cm}

École doctorale n$^{\circ}$ 576\\ Particules, Hadrons, Énergie et Noyau : Instrumentation, Imagerie, Cosmos et Simulation (PHENIICS)\\
\small Spécialité de doctorat : Structure et réactions nucléaires\\
\footnotesize Unité de recherche : Université Paris-Saclay, CEA, Laboratoire Matière sous conditions extrêmes, 91680, Bruyères-le-Châtel, France\\
\footnotesize Référent : Faculté des sciences d'Orsay

\vspace{0.9cm}

\textbf{Thèse présentée et soutenue à Orsay, le 23/09/2021, par}\\
\bigskip
\Large {\color{Prune} \textbf{Kilian FRABOULET}}

\vspace{1.4cm}

\vspace{\fill}

\Wider[9.4em]{
\flushleft \small \textbf{Composition du jury :}
\bigskip

\begingroup\scriptsize
\begin{tabular}{|p{11cm}l}
\arrayrulecolor{Prune}
\textbf{Nicolas DUPUIS} &  Président et Rapporteur \\
Directeur de recherche, \textit{LPTMC, Sorbonne Universit\'{e}, Paris, France}  &   \\ 
\textbf{Richard FURNSTAHL} &  Rapporteur et Examinateur \\ 
Professeur, \textit{Ohio State University, Colombus, USA}  &   \\ 
\textbf{Laura CLASSEN} &  Examinatrice \\ 
Chargée de recherche, \textit{MPI for Solid State Research, Stuttgart, Germany}   &   \\ 
\textbf{Jan PAWLOWSKI} &  Examinateur \\ 
Professeur, \textit{ITP, Heidelberg, Germany}   &   \\ 
\textbf{Janos POLONYI} &  Examinateur \\ 
Professeur, \textit{IPHC, Strasbourg, France}   &   \\ 

\end{tabular}
\endgroup

\bigskip

\flushleft \small \textbf{Direction de la thèse :}
\bigskip

\begingroup\scriptsize
\begin{tabular}{|p{11cm}l}\arrayrulecolor{Prune}
\textbf{Elias KHAN} &   Directeur de thèse\\ 
Professeur, IJCLab, Universit\'{e} Paris-Saclay, Orsay, France & \\
\textbf{Jean-Paul EBRAN} &   Co-encadrant de thèse\\ 
Ing\'{e}nieur-chercheur, CEA, DAM, DIF, Arpajon, France &   \\  

\end{tabular}
\endgroup}

\end{titlepage}

%% file: 0Acknowledgements/Acknowledgements.tex
Here we are, three years and a pandemic later.

\vspace{0.5cm}

First, I would like to thank some of my former professors from my master studies in Strasbourg, namely Janos Polonyi, Jérôme Baudot and Hervé Molique, who all led me to the present thesis in theoretical physics for different reasons.

\vspace{0.5cm}

In particular, I thank my thesis supervisors, Elias Khan and Jean-Paul Ebran, for giving me my chance. It turns out that the chosen directions have led me to work exclusively with Jean-Paul in the past three years. This thesis is a project that we have built together since day one, throughout many blackboard explanations, phone calls, email exchanges and Zoom meetings. It has been very stimulating working with Jean-Paul and I believe he taught me a lot, on theoretical physics itself of course but also on the way to handle a research project. Jean-Paul, thank you for everything!

\vspace{0.5cm}

I have definitely enjoyed a very rich scientific environment in Paris during the past three years, thus meeting many researchers from different labs. In that respect, I notably express my gratitude to those I have interacted with from IJCLab at Orsay and from CEA at Bruyères-le-Châtel or Saclay. In particular, I thank Guillaume Hupin for assuming the role of \textit{garant technique} on my behalf and Denis Lacroix for his advices in the preparation of the defense.

\vspace{0.5cm}

Of course, I can not avoid mentioning my fellow PhD students and postdocs: my office mate Yann Beaujeault-Taudière, Lysandra Batail, Ania Zdeb, Victor Tranchant, Mikael Frosini, Andrea Porro, Florian Mercier, Thomas Czuba, David Durel, Antoine Boulet and Julien Ripoche. I would like to thank them for our friendly scientific and non-scientific discussions. We have sometimes faced administrative issues together, which might bring back bad memories to Yann and Mikael. Congratulations also to both Ania and Mikael for their permanent positions! Finally, I want to thank Antoine and Florian for their $\mathtt{Mathematica}$ advices, as well as Victor who clearly spared me countless hours of bus with car sharing during the last months of my thesis.

\vspace{0.5cm}

Speaking of administrative issues, I want to thank Patricia B., the secretary of my group, for her kind help in facing them.

\vspace{0.5cm}

I also extend my gratitude to Nicolas Dupuis, Richard Furnstahl, Laura Classen, Jan Pawlowski and Janos Polonyi for accepting being part of my PhD jury. I am very glad to have exchanged with such experienced physicists from various fields on my work.

\vspace{0.5cm}

Finally, my warmest thanks go to my family to whom I owe so much.

%% file: Acronyms.tex
\chapter*{List of acronyms and abbreviations}

\begin{acronym}
\acro{BCS}{Bardeen-Cooper-Schrieffer}
\acro{Blaizot-M\'{e}ndez-Galain-Wschebor}{BMW}
\acro{BMBPT}{Bogoliubov many-body perturbation theory}
\acro{BRST}{Becchi-Rouet-Stora-Tyutin}
\acro{BVA}{bare vertex approximation}
\acro{CJT}{Cornwall-Jackiw-Tomboulis}
\acro{DE}{derivative expansion}
\acro{DFT}{density functional theory}
\acro{DMFT}{dynamical mean-field theory}
\acro{dof}{degree of freedom}
\acro{EA}{effective action}
\acro{EDF}{energy density functional}
\acro{EFT}{effective field theory}
\acro{FAC}{fastest apparent convergence}
\acro{fig.}{figure}
\acro{FRG}{functional renormalization group}
\acro{GCM}{generator coordinate method}
\acro{gs}{ground state}
\acro{HFB}{Hartree-Fock-Bogoliubov}
\acro{HST}{Hubbard-Stratonovich transformation}
\acro{IM}{inversion method}
\acro{IR}{infrared}
\acro{IS}{internal space}
\acro{KS-FRG}{Kohn-Sham functional renormalization group}
\acro{LDE}{linear delta-expansion}
\acro{LE}{loop expansion}
\acro{LHS}{left-hand side}
\acro{LOAF}{auxiliary field loop expansion}
\acro{LPA}{local-potential approximation}
\acro{MBPT}{many-body perturbation theory}
\acro{MFT}{mean-field theory}
\acro{MR}{multi-reference}
\acro{NLO}{next-to-leading order}
\acro{NN}{nucleon-nucleon}
\acro{$n$PI}{$n$-particle-irreducible}
\acro{$n$PI-FRG}{$n$-particle-irreducible functional renormalization group}
\acro{$n$PPI}{$n$-particle-point-irreducible}
\acro{$n$PPI-FRG}{$n$-particle-point-irreducible functional renormalization group}
\acro{$n$PR}{$n$-particle-reducible}
\acro{$n$VI}{$n$-vertex-irreducible}
\acro{$n$VR}{$n$-vertex-reducible}
\acro{OPT}{optimized perturbation theory}
\acro{ODM}{order-dependent mapping}
\acro{PGCM}{projected generator coordinate method}
\acro{PI}{path-integral}
\acro{PMS}{principle of minimal sensitivity}
\acro{PT}{perturbation theory}
\acro{QCD}{quantum chromodynamics}
\acro{QED}{quantum electrodynamics}
\acro{QFT}{quantum field theory}
\acro{RG}{renormalization group}
\acro{RHS}{right-hand side}
\acro{SCC}{self-consistent condition}
\acro{SCE}{self-consistent expansion}
\acro{SI2PI}{symmetry-improved 2-particle-irreducible}
\acro{SR}{single-reference}
\acro{SRG}{similarity renormalization group}
\acro{SSB}{spontaneous symmetry breaking}
\acro{tab.}{table}
\acro{TP}{turning point}
\acro{UV}{ultraviolet}
\acro{VPT}{variational perturbation theory}
\end{acronym}

%% file: 1ChapterIntroManyBodyProblem/IntroManyBodyProblem.tex
Path-integral (PI) approaches have been exploited since decades to treat strongly-coupled quantum many-body systems. Among the latter, we will pay particular attention throughout this thesis to those encountered in nuclear physics. The important diversity of phenomena occurring at the nuclear scale propelled developments of various approaches over the past decades, the most successful of which either focus on describing specific nuclear features or, on the contrary, assume the task of having to embrace all aspects of the nuclear phenomenology. On the one hand, we have for instance nuclear collective models~\cite{boh57,vil57,dav65,iac87}, which are based on global or bosonic degrees of freedom (dofs). Such models have addressed convincingly collective behaviors of vibrational and rotational nature. Other examples are the cluster models~\cite{wil58,hor87}, which involve cluster of nucleons as basal dofs and were developed to account for specific spectroscopic features of rather light nuclei. On the other hand, nuclear energy density functionals (EDFs)~\cite{ben03,sch19} were formulated with the ambition of ultimately describing a vast richness of nuclear phenomena, ranging from the structure and reactions in finite nuclei to the complex processes in neutron stars. These approaches describe the nucleus as a collection of dressed nucleons coupled through an effective interaction. The adjective ``effective'' is a key aspect here as it implies that part of the impact of the medium on the nucleon-nucleon (NN) interaction is taken into account implicitly via a fitting procedure, hence degrading the reliability of the EDF  method (as long as no hierarchization principles have been identified) while rendering its underlying numerical procedure less demanding. On the contrary, there are also nuclear approaches, coined as \textit{ab initio}, which are based on the NN interaction in free space and thus have the endeavor to describe nuclear phenomena from first principles. Such interactions explicitly take into account how correlations between nucleons are impacted by the presence of fellowmen and how the nucleus self-organizes in consequence. Early nuclear approaches used to employ phenomenological models of the NN force~\cite{mac01} (based on the seminal work of Yukawa~\cite{yuk35}) as input for ``traditional'' \textit{ab initio} schemes~\cite{glo71,pud97,suz98,nav98,kow04}.

\vspace{0.5cm}

Such a bushy proliferation of approaches may seem fundamentally flawed in the perspective of epistemological standards~\cite{bon20}, and may even seem to yield inconsistent viewpoints of the nucleus, e.g. the apparent conflicting pictures of a tightly bound liquid droplet on the one hand and of a delocalized shell-like structure on the other. Nonetheless, the philosophy underpinning the renormalization group (RG), namely the emergence and effectiveness concepts, promotes a description of complex systems in terms of a web of interlocking effective theories rather than based on a unique fundamental theory~\cite{bon20}. This thus supports the strategy instinctively adopted by nuclear physics since its infancy, although in an incomplete form due to the phenomenological construct of standard nuclear models. Even though the latter have given us access to a precious empirical knowledge about the emergent scales and associated dofs in nuclear physics, only a reformulation in the language of effective (field) theories (EFTs) can turn them into consistent and robust frameworks capable of reliable predictions~\cite{fur20}, i.e. with a domain of validity (e.g. in terms of energy scale) clearly identified, a systematic way of improving their results, a possibility to assess theoretical errors, ...

\vspace{0.5cm}

\begin{figure}[!hbt]
  \centering
  \includegraphics[clip, width=1.0\linewidth]{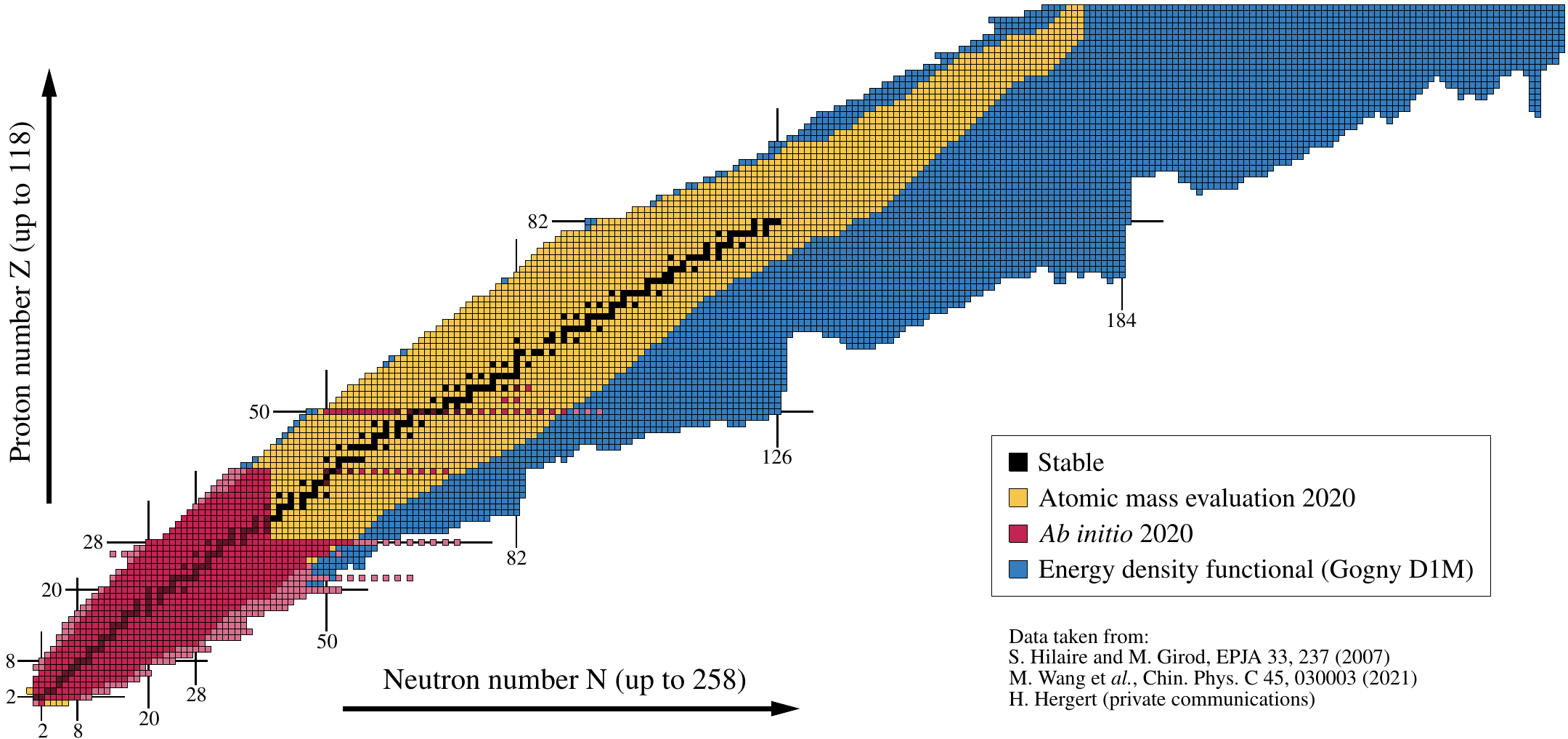}%
  \caption{Nuclear chart and range of applicability of the two main microscopic methods in nuclear theory (\textit{ab initio} and EDF), compared with the region of stable nuclei and that of nuclei that have been experimentally discovered until 2020 (Atomic mass evaluation 2020). Figure taken from ref.~\cite{bal21}.}
  \label{fig:NuclearChart}
\end{figure}

For the purpose of constructing such reliable approaches, a consistent description of inter-nucleon interactions (with respect to quantum chromodynamics (QCD)) has been achieved within chiral EFT~\cite{epe09,mac20,ham19}, which now constitutes a basis for various implementations of \textit{ab initio} approaches in nuclear physics~\cite{dic04bis,hag10,tsu11,som13,sig15,dug16,her16,tic18,dem20,tic20}. Likewise, halo/cluster features in light nuclei and collective rotational/vibrational behaviors in heavy systems can consistently be addressed within Halo/Cluster EFTs~\cite{ber02bis,bed03,ham19} and macroscopic EFTs~\cite{pap11,pap14,coe15,coe16,che18,che19} respectively. However, the framework of nuclear EDFs has yet to be reformulated in the language of EFT~\cite{fur20}. One can certainly argue that this is an appealing task as the EDF approach is currently providing the most complete and accurate (at least near the empirically known regions) description of ground state (gs) and excited state properties of atomic nuclei over the entire nuclear chart, as illustrated by fig.~\ref{fig:NuclearChart}. Traditional EDFs owe their success to several features, the first of which includes an efficient resummation of nucleon correlations at the level of the one-body reduced density matrix (hence the name of the approach) which captures quantitatively the bulk of properties deriving from the nuclear saturation phenomenon. In this way, the EDF method exhibits a strong resemblance to density functional theory (DFT)~\cite{hoh64,koh65,koh65bis}. However, another salient feature of EDFs, i.e. the optimal account of so-called non-dynamical correlations at the source of collective behaviors (deformation, superfluidity, clustering, ...) via the spontaneous breakdown and restoration of symmetries, distance them from DFT in the sense that they do not seem to align with the Kohn-Sham formulation of DFT, see e.g. refs.~\cite{mes11,les14} and references therein. Such a success has to be contrasted with the drawbacks related to the lack of rigorous foundations for the EDF method, e.g. the phenomenological character of the underlying NN interaction mentioned earlier (which translates into parametrization-dependent predictions away from known data) or the absence of a framework to design systematic improvements. Many paths towards a proper EFT formulation for nuclear EDFs have been envisioned to overcome these limitations~\cite{kai10,hol13,sap16,dyh17,nav18}, among which the functional integral or PI language provides a powerful frame to account for quantal fluctuations in a systematic way~\cite{ham00,fur01,fur02,pug03,bha05,bha05b,fur07,dru10,fur20}. The latter direction is further pursued in this thesis.

\vspace{0.5cm}

Let us analyze the key features of standard nuclear EDFs at the root of their success, i.e. their accuracy (near the empirically known regions) and their favorable scaling making them relevant for large-scale studies of nuclear systems (irrespective of the number of involved nucleons or their expected shell structure). Such features will guide the formulation of EDFs in the PI language. The EDF method typically unfolds two categories of expansions, neither of which are systematized\footnote{In other words, the corresponding results can not be improved in a systematic fashion, i.e. by going from one level of approximation to the next.}: a first expansion at the level of the effective vertex and a second in the form of a sequential integration of classes of correlations. We discuss below these two types of expansions in more detail:
\begin{itemize}
\item[1.] The first expansion involves the analytic form of the effective vertex at the heart of the EDF. More precisely, nuclear EDFs were originally built from an (ill-defined) effective vertex interpreted as an in-medium NN interaction~\cite{vau72,gog75,dec75,ser84,gam90}. A modern perspective seeks a general expression for the EDF without direct references to an effective interaction~\cite{car08,dug14,ebr19}, but requires the latter to derive from a pseudo Hamiltonian in order to avoid spurious self-interactions and self-pairing contributions~\cite{lac09,ben09bis}. In any case, an effective vertex is exploited as the generator of an EDF. Traditionally, its form follows from a heuristic argument of simplicity, i.e. the ability to reproduce some set of data with a minimal ``operatorial'' structure. Popular non-relativistic parametrizations, like the Skyrme~\cite{vau72} and Gogny~\cite{gog75,dec75} ones, involve central and spin-orbit contributions, while the covariant ones~\cite{ser84,gam90} use scalar-isoscalar, vector-isoscalar and vector-isovector channels (alongside with a Coulomb channel for both non-relativistic and covariant EDFs). In this way, the generalization of the effective vertex, e.g. the addition of a tensor term or more involved channels, is more a matter of art than dictated by some systematic arguments. Modern Skyrme-like parametrizations are built from a generic momentum expansion~\cite{car08}, preferably with a finite-range regulator and a three-nucleon channel~\cite{rai11,dob12,rai14,ben17,ben20}. If the resulting functional is shown to be systematically improvable, such an expansion is still not organized with respect to a genuine power counting. Another issue pertains to the presence of density-dependent terms, which on the one hand allow for a quantitative description of nuclear systems, while on the other hand are known to contaminate the EDF calculations with unphysical contributions~\cite{dug15}. This situation has triggered an effort to replace these density-dependent terms by well-defined operatorial forms, e.g. three-body terms~\cite{rai11,dob12,rai14,ben17,ben20} or a structure inspired by many-body perturbation theory (MBPT) beyond the first non-trivial order~\cite{dugebr15}. However, none of these approaches achieved the same accuracy and the same simplicity as density-dependent effective vertex. It is worth going back to the origin of such density-dependent terms to better understand how they efficiently capture the physics stemming from the saturation phenomenon. Early Hartree-Fock calculations in nuclear physics with a density-independent NN interaction adjusted to reproduce A-body observables (typically binding energies and radii) were unable to provide a simultaneous correct description of both binding energies and radii. The introduction of an explicit medium dependence, most generally in the form of a density dependence, magically resolved this issue. The so-called rearrangement term induced by the density-dependent interaction at the level of the equation of motion provides us with a more flexible relation between the binding energy of the system and the energies of the nucleon orbitals: it accommodates the computation of a correct binding energy with a sufficiently compressed single particle spectrum at the same time, thus yielding proper saturation properties. One of the questions explored in this thesis is related to the origin of such density-dependent terms. Namely, can we understand why they drastically improve the description of nuclear features? Can we find other objects than the density to grasp correlations as efficiently, but in a spuriosity-free fashion?

\item[2.] The second expansion takes the form of a sequential account of nucleonic correlations. So-called bulk correlations\footnote{The bulk correlations are by definition those whose contribution to the binding energy of nuclear systems varies continuously with the proton and/or neutron number(s). Such correlations encompass for instance nuclear saturation properties.} are efficiently grasped by a density-dependent vertex, from which one constructs an EDF. In a first step coined as single-reference (SR) EDF, the EDF is obtained as the expectation value of the effective Hamiltonian (containing the effective density-dependent vertex in addition to the kinetic energy operator) in a product state allowed to spontaneously break the symmetries of the original nuclear Hamiltonian (essentially the spatial translation group (nuclei are localized, self-bound systems), the rotational $SU(2)$ group (nuclei usually exhibit what is called nuclear deformation) and the $U(1)$ group associated with the conservation of nucleon number (most nuclei are superfluid)). The energy of the system then appears as a functional of its normal and anomalous one-body density matrices, found after solving the corresponding Hartree-Fock-Bogoliubov (HFB) equations~\cite{bog58,val58,deg66,rin80}, also referred to as Bogoliubov-de Gennes equations. These one-body density matrices can be parametrized by the (bosonic) order parameters associated with the aforementioned symmetries. It is through such a spontaneous symmetry breaking (SSB) procedure that the EDF is able to grasp non-trivial physics and more specifically non-dynamical correlations, which are responsible for deformation, clustering and superfluidity, at low cost (at the cost of a SR approach essentially). However, nuclei are finite-size (or mesoscopic) systems, so that they can not spontaneously break any symmetries. Quantum fluctuations of the order parameters around the values minimizing the SR EDF (or saddles) are not negligible and eventually preclude any SSBs by mixing the degenerate vacua defining the Goldstone manifold and yielding a unique gs with good symmetry properties. As such, the quantum fluctuations of the order parameters have to be accounted for in a further step, coined as multi-reference (MR) EDF. In MR EDF, the energy is computed as the expectation value of the effective Hamiltonian in a more general state, namely a non-orthogonal product of HFB states. The energy becomes a functional of transition density matrices connecting two HFB states. A full-fledged MR EDF description takes the form of the projected generator coordinate method (PGCM)~\cite{ben03}, where additional correlations stemming from the fluctuations of the order parameters are accounted for not only to describe the gs of nuclei, but also their (collective) excited spectra. In particular, one approximation of the PGCM, where one writes the expectation values of the effective Hamiltonian between two HFB states (i.e. the so-called energy kernels) as a Gaussian function (Gaussian Overlap Approximation), leads to a collective Hamiltonian whose dofs are (bosonic) collective coordinates (the order parameters discussed above). Such a collective Hamiltonian does not exhibit the spuriosities contaminating the full-fledged PGCM but its construction is not systematically improvable. One related question that the present thesis will try to address is whether one could arrive to a similar result, i.e. a framework where the original dofs have been integrated out of the theory in favor of collective coordinates, but in a rigorous and systematically improvable framework.

\end{itemize}

\vspace{0.3cm}

\begin{figure}[!hbt]
  \centering
    \hspace{-0cm}
  \includegraphics[clip, width=0.68\linewidth]{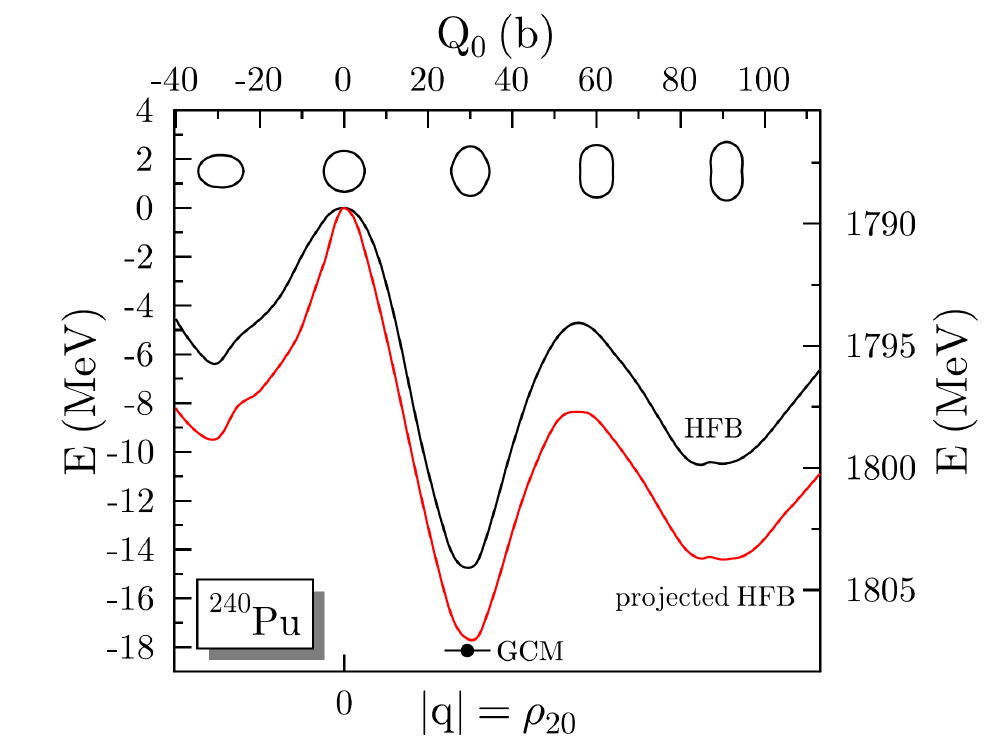}%
  \caption{Binding energy of the $^{240}$Pu calculated at the SR EDF level (HFB) and at the MR EDF one with projection (projected HFB) and configuration mixing (GCM). Figure taken from ref.~\cite{dug14}.}
  \label{fig:GCMProjHFB}
  \end{figure}

Such a sequential account of nucleonic correlations at the level of the SR, and then MR EDF, is relevant only if nucleonic correlations exhibit such a hierarchy. This is indeed what is empirically observed and illustrated in fig.~\ref{fig:GCMProjHFB}. The latter displays the $^{240}$Pu binding energy computed at both the SR and MR levels. The SR result that does not break any symmetry, corresponding to the black curve at $|q|=0$ ($q$ is the quadrupole moment of the density, which here plays the role of an order parameter for the rotational symmetry), grasps bulk correlations which constitute at least $98\%$ of the binding energy in this situation. Letting the reference state breaking down the rotational symmetry enables us to gain around 14 MeV of correlation energy, which represents around $2\%$ of the system's binding energy. Then, within the MR realization of the EDF method where only angular fluctuations of the order parameters are accounted for (with the so-called projection techniques), the restoration of the broken rotational (and particle-number) symmetry (symmetries) brings a further contribution to the binding energy, thus leading to the global minimum of the red curve. A full-fledged PGCM calculation further grasps correlations and yields the black dot dubbed ``GCM'', located about 0.5 MeV below the global minimum of the red curve. The behavior of $^{240}$Pu is representative of heavy doubly open-shell nuclei but remains one example among many. We can summarize the different categories of correlations that are treated within the framework of the EDF method, alongside with the corresponding energy scales, as done in tab.~\ref{tab2}.

\vspace{0.5cm}

\begin{table}[!htb]
\centering
\fontsize{8pt}{9pt}\selectfont
\caption{Categories of correlations treated by the EDF method.
$A_{\text{val}}$ and $G_{\text{deg}}$ denote respectively the number of valence nucleons and the degeneracy of the corresponding valence shell (i.e. the number of orbitals located in top of so-called magic configurations).}
\begin{tabularx}{\textwidth}{X X X X}
    \hline
   \textbf{Type of correlations}         & \textbf{Treatment}      & \textbf{Energy scale}  &    \textbf{Vary as} \\
    \hline
                                              &                                  &                               &                         \\ 
      bulk                           &  grasped in the functional &   $\sim$ 8 A MeV              &     $A$                   \\
                                              &                                  &                               &                         \\      
      collective static                   &  order parameter $|q|\neq$ 0 (SSB)& $\lesssim$ 25 MeV     &   $A_\text{val}$, $G_\text{deg}$\\                                                    
                                              &                                  &                               &                         \\      
      collective dynamical                  &  quantum fluctuations of $q$  & $\lesssim$ 5 MeV     &   $A_{\text{val}}$, $G_{\text{deg}}$\\                                                    
\hline
\end{tabularx}
\label{tab2}
\end{table}

\begin{figure}[!hbt]
  \centering
  \includegraphics[clip, width=1.05\linewidth]{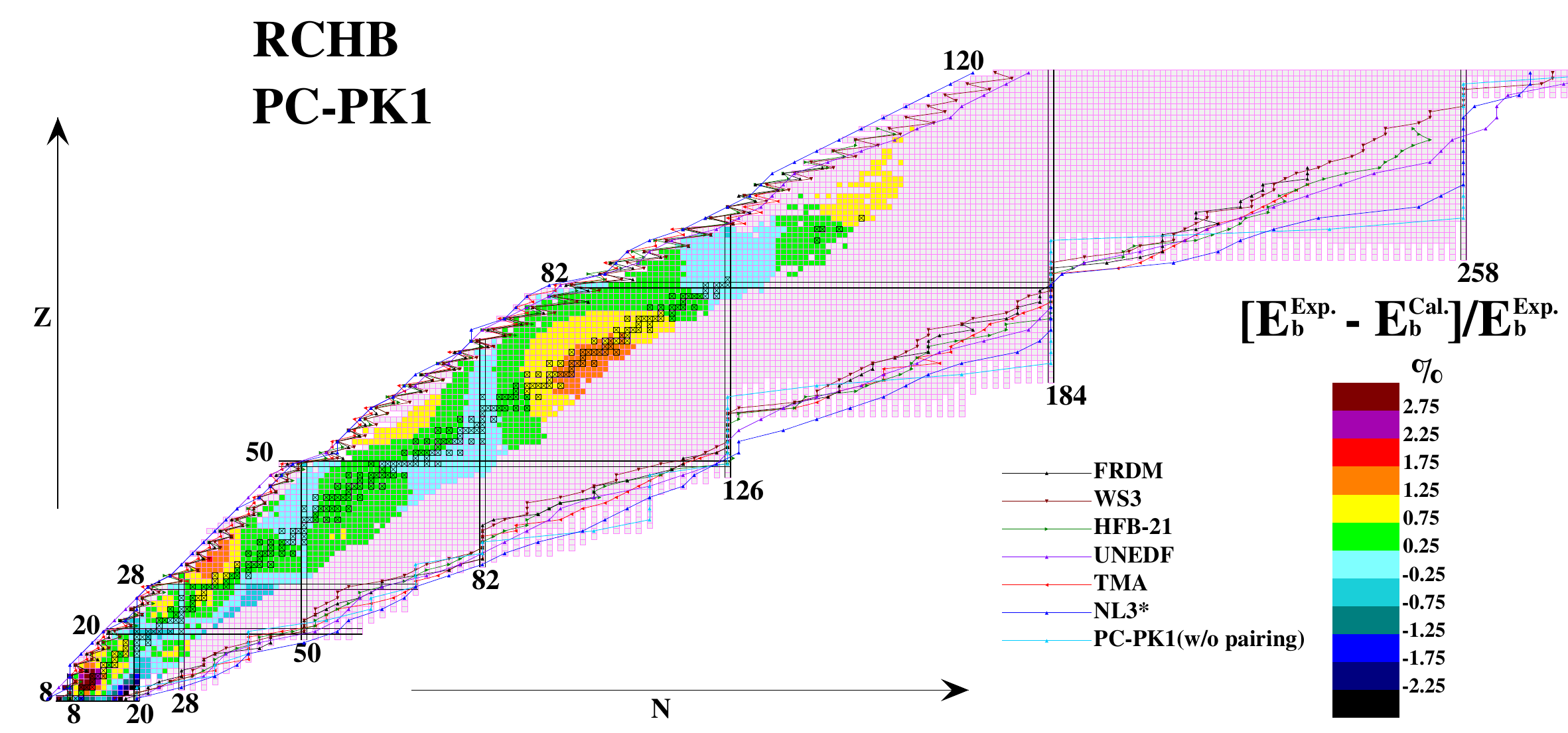}%
  \caption{Bound nuclei predicted via the PC-PK1 procedure used to parametrize a covariant EDF. Figure taken from ref.~\cite{xia18}.}
    \label{fig:drip}
\end{figure}

Besides the very good description of nuclei's gs and spectroscopic properties by the EDF method, the empirical nature of the latter approach, due to the parametrization of the underlying functionals, raises numerous sensible points: i) the connection between current functionals and elementary forces between nucleons is not explicit; ii) the predictive character of EDF predictions for experimentally undiscovered nuclei relies heavily on the area of the nuclear chart used to constrain the free parameters of the functionals. The second flaw in the predictive power of the EDF approach is illustrated for a covariant functional by fig.~\ref{fig:drip}. In the latter, one can see in particular the important variations of the driplines location\footnote{The driplines are the boundaries on the nuclear chart that separate the bound nuclei from the unbound ones.} with respect to the exploited parametrization procedure.
The extensive exploitation of traditional EDFs (Gogny, Skyrme or covariant) have shown numerous limitations of this framework related either to the parametrization procedure, the too simple nature of the analytical forms of the functionals or to the lack of firm theoretical foundations underlying their construction. Several approaches can be considered to overcome such limitations:
\begin{itemize}
\item[1.] Refine traditional empirical functionals either by improving the parametrization procedure to grasp more non-trivial physics or by using a richer analytical form for the functionals (finite-range features, addition of tensor terms, less drastic density dependences, ...). The latter strategy has notably led to the Gogny functional with various parametrizations such as D1N~\cite{d1n}, D1M~\cite{d1m09} and D2~\cite{chap15}.

\item[2.] Improve the MR EDF methods (PGCM, quasi-particle random phase approximation, ...) or construct new ones able to tackle excited states that are so far not reliably and/or not efficiently described by this framework.

\item[3.] Develop \textit{ab initio} approaches to extend their range of applicability and to enhance their accuracy.

\item[4.] One might also follow a parallel direction (with respect to 3.) with the aim of constructing less empirical EDFs, connected to the QCD vacuum (possibly at finite density). A relevant way to achieve this is to reformulate the EDF method as an EFT.
\end{itemize}

\vspace{0.3cm}

The reformulation of the EDF framework as an EFT can be done through numerous relevant perspectives~\cite{fur20}. It is possible for instance to employ a top-down approach in which one starts from \textit{ab initio} approaches to reach an EDF. The reverse can also be considered by following a bottom-up philosophy in which a general EDF is constructed from first principles, before being adjusted on experimental results or \textit{ab initio} calculations. This leads us to the goals of the present thesis work on PI approaches. Developing an EFT in nuclear physics amounts to constructing a Lagrangian or a classical action describing the bare NN interaction. The following step consisting in calculating nuclear observables from these functionals is usually referred to as the treatment of the (nuclear) many-body problem. The PI framework provides us with plenty of techniques to achieve the latter step. For instance, one can readily introduce in these approaches new (presumably collective) dofs, notably via exact identities called Hubbard-Stratonovich transformations (HSTs)~\cite{str57,hub59}. One can therefore naturally draw a parallel with the PGCM in that respect. Moreover, just like EFTs, most PI approaches are systematically improvable themselves, which enables us to control our approximations, another important feature lacking in current EDFs. One should also mention the natural connection that can be exhibited between DFT and the effective action (EA) formalism, which is a subpart of the PI framework. Along these lines, we can mention the work of Furnstahl and collaborators~\cite{ham00,fur01,fur02,pug03,bha05,bha05b,fur07,dru10,fur20}, and notably the work of Valiev and Fernando~\cite{val97} demonstrating that the 2-particle-point-irreducible (2PPI) EA is suited to formulate a Kohn-Sham DFT.

\vspace{0.5cm}

From these remarks, the PI framework seems indeed relevant to formulate reliable theoretical descriptions of nuclear systems. We will therefore aim at better understanding how this can be achieved throughout this thesis work. To that end, we will exploit a toy model as theoretical playground to perform a comparative study of state-of-the-art PI approaches. The chosen model is the (0+0)-D $O(N)$-symmetric $\varphi^4$-theory or, more simply, (0+0)-D $O(N)$ model\footnote{Throughout this thesis, we use ``$O(N)$ model'' as synonym for ``$O(N)$-symmetric $\varphi^4$-theory'', as is often the case.}, which will be presented in chapter~\ref{chap:Intro2} after giving a brief recall on the PI formalism. Considering the importance of $O(N)$ models~\cite{sta68,zin02,mos03} and PI techniques~\cite{neg98,zin02,kle06} in theoretical physics\footnote{To illustrate this, we point out that $O(N)$-symmetric $\varphi^4$-theories reduce to well-known models exploited in statistical physics, i.e. to the Ising model~\cite{isi25}, the XY model and the Heisenberg model at $N=1,2$ and $3$ respectively. At $N=4$, $O(N)$ models are also used to study the phase structure of QCD (see refs.~\cite{tol03,her13,dem14} as examples).} (especially due to the connection between these $O(N)$ models and universal properties of critical systems~\cite{wil71,wil71bis,wil72,wil74}), this comparative study is clearly fueled by other areas of physics, as pointed out in the forthcoming chapters and references therein, and this thesis thus certainly finds strong echoes in those areas as well. Besides this, a key feature of the toy model under consideration is the presence of the $O(N)$ symmetry, whose spontaneous breakdown will be studied with care in this comparative study. This will enable us in particular to draw connections with the SR and MR EDF schemes. We will also argue in chapter~\ref{chap:Intro2} that PI techniques can be split into two categories: functional renormalization group (FRG) approaches and the others, coined as diagrammatic techniques, which will be studied in chapters~\ref{chap:FRG} and~\ref{chap:DiagTechniques}, respectively. In these two chapters, the performances of these methods will be examined in the strongly-coupled regime of the model under consideration, in connection with the nuclear many-body problem. Different types of EAs and HSTs will also be exploited for many methods investigated in these two chapters, still with the aim of identifying the most relevant dofs to treat our problem, and we will also carefully discuss how the resulting conclusions can be extended to more realistic models, and notably to nuclear EDFs. Finally, chapter~\ref{chap:Conclusion} contains our concluding remarks and outlooks for this comparative study of PI techniques.

%% file: 2ChapterIntroPathIntegral/IntroPathIntegral.tex
\minitoc

\section{\label{sec:RelevantGenFunc}Relevant generating functionals and observables}

We aim at describing strongly-coupled quantum many-body systems hosting collective behaviors of bosonic nature. A possible angle of attack involves correlation functions which contain the complete physical information on the corresponding quantum many-body system\footnote{In an axiomatic approach to quantum field theory (QFT), the Wightman reconstruction theorem~\cite{wight56,swbook,ost73,ost75} states that a sequence of (tempered) $n$-point correlation functions completely determines the Hilbert space and algebra of fields (realized as operator-valued distributions), up to unitary equivalence.}.
In the canonical formulation of quantum mechanics, the constituents of the theory are represented by \textbf{operator-valued} distributions $\hat{\varphi}_{\alpha}$, with a generic index $\alpha \equiv (a,\boldsymbol{r},\tau,m_s,c) \equiv (a,x)$ collecting spacetime coordinates $(\boldsymbol{r},\tau)$ (with $\boldsymbol{r}$ and $\tau$ being respectively the $(D-1)$-dimensional space position vector and the imaginary time) and, if relevant, spin projection $m_s$, charge $c$, and internal $a$ labels, such that:
\begin{equation}
\hat{\varphi}_{\alpha}=\hat{\varphi}_{a,x}=\left\{
\begin{array}{lll}
        \displaystyle{\hat{\varphi}_{a,m_s}(\boldsymbol{r},\tau) \quad \mathrm{for}~c=-\;.} \\
        \\
        \displaystyle{\hat{\varphi}^{\dagger}_{a,m_s}(\boldsymbol{r},\tau) \quad \mathrm{for}~c=+\;.}
    \end{array}
\right.
\end{equation}
The $n$-point correlation function $G^{(n)}$ then stems from the expectation value, in the (interacting) many-body system gs $\ket{\mathrm{vac}}$, of a (time-ordered) product of $n$ field operators:
\begin{equation}
G^{(n)}_{\alpha_{1}\alpha_{2}\cdots\alpha_{n}} \equiv \braket{\mathrm{vac}|\mathrm{T}\hat{\varphi}_{\alpha_{1}}\hat{\varphi}_{\alpha_{2}}\cdots\hat{\varphi}_{\alpha_{n}}|\mathrm{vac}} \;,
\label{eq:CanonicalFormalismCorrelationFunction}
\end{equation}
where $\mathrm{T}$ stands for the time-ordered product.

\vspace{0.5cm}

Alternatively, correlation functions can be computed from standard generating functionals~\cite{sch51}, conveniently expressed as sum-over-histories in configuration space within Feynman's PI formulation of quantum mechanics~\cite{fey10,fri20}, where the dofs of the theory are now realized via \textbf{number-valued} fields\footnote{Throughout this thesis manuscript, fluctuating fields, i.e. fields displaying quantum fluctuations treated within a PI, are denoted with an upper tilde, like $\widetilde{\varphi}_{\alpha}$.}, such as $\widetilde{\varphi}_\alpha$\footnote{Although part of the derivations performed in this thesis are valid for both bosonic and fermionic field theories, we give in sections~\ref{sec:RelevantGenFunc} and~\ref{sec:WickTheorem} a brief introduction on the PI formalism assuming that $\widetilde{\varphi}_{\alpha}$ is a bosonic (i.e. non-Grassmann) field, as it is sufficient to present the main functionals exploited in the forthcoming chapters. More exhaustive introductions on the PI formalism for both bosonic and fermionic field theories can be found e.g. in refs.~\cite{neg98,kop10bis3}.}. Coupling the field $\widetilde{\varphi}_{\alpha}$ to a test (external) source $J_{\alpha}$ yields the action:
\begin{equation}
S_{J}[\widetilde{\varphi}] = S[\widetilde{\varphi}] - J_{\alpha} \widetilde{\varphi}_{\alpha} \;,
\label{eq:GeneralSJ}
\end{equation}
where $S[\widetilde{\varphi}]$ stands for the (Euclidean) classical action of the system and summation over repeated indices is assumed, i.e.:
\begin{equation}
\begin{split}
J_{\alpha} \widetilde{\varphi}_{\alpha} \equiv & \ J_{a,x} \widetilde{\varphi}_{a,x} \\ 
\equiv & \ \sum_{m_s,c,a}\int_{0}^{\hbar\beta} d\tau \int_{\mathbb{R}^{D-1}}d^{D-1}\boldsymbol{r} \ J_{a,m_s,c}(\boldsymbol{r},\tau) \widetilde{\varphi}_{a,m_s,-c}(\boldsymbol{r},\tau) \;, \\
\end{split}
\end{equation}
in a $D$-dimensional spacetime and with inverse temperature $\beta$. The (Euclidean) PI representation of the system's partition function in presence of the source (also called vacuum persistence amplitude) derives from these ingredients and is given by the functional integral (also called PI):
\begin{equation}
Z[J]=\mathcal{N}\int_\mathcal{C}\mathcal{D}\widetilde{\varphi} \ e^{-\frac{1}{\hbar}S_{J}[\widetilde{\varphi}]} \;,
\label{eq:GeneralGeneratingFunctional}
\end{equation}
where $\mathcal{N}$ is a normalization factor, $\mathcal{C}$ the space of configurations\footnote{The integration domain defining each functional integral will be left implicit in most cases.} for the fluctuating field $\widetilde{\varphi}_{\alpha}$ and $\mathcal{D}\widetilde{\varphi}$ the PI measure. The functional $Z[J]$ is the generating functional of correlation functions, namely:
\begin{equation}
\begin{split}
G^{(n)}_{\alpha_{1} \alpha_{2} \cdots \alpha_{n}} \equiv & \ \braket{\widetilde{\varphi}_{\alpha_{1}}\widetilde{\varphi}_{\alpha_{2}}\cdots\widetilde{\varphi}_{\alpha_{n}}}_{\mathrm{vac}} = \frac{\int \mathcal{D}\widetilde{\varphi} \ \widetilde{\varphi}_{\alpha_{1}}\widetilde{\varphi}_{\alpha_{2}}\cdots\widetilde{\varphi}_{\alpha_{n}} e^{-\frac{1}{\hbar} S[\widetilde{\varphi}]}}{\int \mathcal{D}\widetilde{\varphi} \ e^{-\frac{1}{\hbar} S[\widetilde{\varphi}]}} \\
= & \ \frac{\hbar^{n}}{Z[J=0]}\left.\frac{\delta^{n} Z[J]}{\delta J_{\alpha_{1}}\delta J_{\alpha_{2}}\cdots \delta J_{\alpha_{n}}}\right|_{J=0} \;.
\end{split}
\label{eq:Gn}
\end{equation}
A diagrammatic representation of $Z[J]$ consists of the sum of \textbf{all} vacuum diagrams, implying that the correlation functions~\eqref{eq:Gn} contain both connected and disconnected contributions. On the other hand, physically relevant observables often only involve the fully connected part of $Z[J]$, which can be summarized in terms of another generating functional called the Schwinger functional\footnote{In analogy with thermodynamics, $Z[J=0]$ is the partition function while $W[J=0]$ corresponds to (minus) the free energy (up to a constant proportional to the temperature).} $W[J]$ defined via:
\begin{equation}
Z[J] \equiv e^{\frac{1}{\hbar}W[J]} \;.
\end{equation}
The cumulants or \textbf{connected} correlation functions $G^{(n),\text{c}}$ then follow from the functional derivatives of $W[J]$:
\begin{equation}
G^{(n),\text{c}}_{\alpha_{1} \alpha_{2} \cdots \alpha_{n}} = \hbar^{n-1}\left.\frac{\delta^{n} W[J]}{\delta J_{\alpha_{1}}\delta J_{\alpha_{2}}\cdots \delta J_{\alpha_{n}}}\right|_{J=0} \;.
\end{equation}

\vspace{0.5cm}

An exact and compact representation of the generating functional $Z[J]$ can be achieved through the $n$-particle-irreducible ($n$PI) EA~\cite{jon64,ded64,cor74}. While the diagrammatic representation of $Z[J]$ consists of vacuum diagrams involving the \textbf{bare} propagator and vertices of the theory, the $n$PI EA provides a systematic method to perform non-perturbative resummations on the $m$-point correlation functions ($m\leq n$) of the theory, yielding a diagrammatic series in terms of the \textbf{dressed} propagator and $m$-point vertex functions with $m\leq n$. The diagrammatic series expressing a $n$PI EA involve $n$PI diagrams\footnote{A (connected) diagram is $n$PI if it remains connected after cutting $n$ non-equivalent propagator lines.} only, hence the name ``$n$PI'' EA. In that regard, we point out the works of Vasil'ev and collaborators~\cite{vas72,vas73,vas73bis,vas74,pis74,vas74bis}, which prove notably:
\begin{itemize}
\item The 1PI and 2PI natures of the diagrams expressing the 1PI and 2PI EAs, respectively~\cite{vas72}.
\item The close relation between the Schwinger-Dyson equations~\cite{dys49,sch51} and the gap equations of the $n$PI EA formalism~\cite{vas73}.
\item The convexity of $n$PI EAs with respect to each of their arguments~\cite{vas73bis}.
\item Expressions of $n$PI vertices in terms of 1PI vertices~\cite{vas74}.
\item The 3PI nature of the diagrams expressing the 3PI EA~\cite{pis74}.
\item The 4PI nature of the diagrams expressing the 4PI EA~\cite{vas74bis}.
\end{itemize}
As will be discussed throughout chapter~\ref{chap:FRG} in particular, the 2PI EA framework\footnote{See ref.~\cite{ber04} for a pedagogical introduction on the 2PI EA.} is a direct reformulation of the Green's function formalism based on Dyson equation and the Luttinger-Ward functional~\cite{dys49,lut60}. It was pioneered by the work of Lee, Yang, De Dominicis and others in statistical physics~\cite{lee60,lut60,ded62,ded64,ded64bis,bay62}, and subsequently extended by Cornwall, Jackiw and Tomboulis~\cite{cor74} to the framework of field theory discussed here, which is why the 2PI EA approach is also coined as CJT formalism.

\vspace{0.5cm}

Regarding the mathematical definitions of $n$PI EAs, one first introduces (external) sources $J_{\alpha}$, $K_{\alpha_1 \alpha_2}$, $L^{(3)}_{\alpha_1 \alpha_2 \alpha_3}$, ... , $L^{(n)}_{\alpha_1 \cdots \alpha_n}$ coupled to the local field $\widetilde{\varphi}_{\alpha}$ and the composite bilocal field $\widetilde{\varphi}_{\alpha_1}\widetilde{\varphi}_{\alpha_2}$, trilocal field $\widetilde{\varphi}_{\alpha_1}\widetilde{\varphi}_{\alpha_2}\widetilde{\varphi}_{\alpha_3}$, ..., $n$-local field $\widetilde{\varphi}_{\alpha_1}\cdots\widetilde{\varphi}_{\alpha_n}$, respectively:
\begin{equation}
\begin{split}
S_{JKL^{(3)}\cdots L^{(n)}}[\widetilde{\varphi}] \equiv & \ S\left[\widetilde{\varphi}\right] - J_{\alpha} \widetilde{\varphi}_{\alpha} - \frac{1}{2}  K_{\alpha_{1} \alpha_{2}} \widetilde{\varphi}_{\alpha_{1}} \widetilde{\varphi}_{\alpha_{2}} \\
& - \frac{1}{3!}  L^{(3)}_{\alpha_{1} \alpha_{2} \alpha_{3}} \widetilde{\varphi}_{\alpha_{1}} \widetilde{\varphi}_{\alpha_{2}}\widetilde{\varphi}_{\alpha_{3}}-\cdots - \frac{1}{n!}  L^{(n)}_{\alpha_1\cdots \alpha_n} \widetilde{\varphi}_{\alpha_1}\cdots \widetilde{\varphi}_{\alpha_n} \;,
\end{split}
\label{eq:GeneralSJKL}
\end{equation}
and
\begin{equation}
\begin{split}
Z\big[J,K,L^{(3)},\cdots\big] \equiv & \ e^{\frac{1}{\hbar}W\left[J,K,L^{(3)},\cdots\right]} \\
= & \ \mathcal{N}\int_\mathcal{C}\mathcal{D}\widetilde{\varphi} \ e^{-\frac{1}{\hbar}S_{JKL^{(3)}\cdots}[\widetilde{\varphi}]} \;.
\end{split}
\end{equation}
The $n$PI EA $\Gamma^{(n\mathrm{PI})}$ is then obtained after Legendre transforming the Schwinger functional with respect to the sources:
\begin{equation}
\begin{split}
\scalebox{0.99}{${\displaystyle \Gamma^{(n\mathrm{PI})} [\phi,G,V,\cdots] = }$} & \scalebox{0.99}{${\displaystyle -W\big[J,K,L^{(3)},\cdots\big] + J_{\alpha} \frac{\delta W\big[J,K,L^{(3)},\cdots\big]}{\delta J_{\alpha}} + K_{\alpha_{1}\alpha_{2}} \frac{\delta W\big[J,K,L^{(3)},\cdots\big]}{\delta K_{\alpha_{1}\alpha_{2}}} }$} \\
& \scalebox{0.99}{${\displaystyle + L^{(3)}_{\alpha_{1}\alpha_{2}\alpha_{3}} \frac{\delta W\big[J,K,L^{(3)},\cdots\big]}{\delta L^{(3)}_{\alpha_{1}\alpha_{2}\alpha_{3}}} + \cdots }$} \\
= & \scalebox{0.99}{${\displaystyle-W\big[J,K,L^{(3)},\cdots\big] + J_{\alpha} \phi_{\alpha} + \frac{1}{2} K_{\alpha_{1}\alpha_{2}} \left(\phi_{\alpha_{1}}\phi_{\alpha_{2}} + \hbar G_{\alpha_{1}\alpha_{2}}\right) }$} \\
& \scalebox{0.99}{${\displaystyle +\frac{1}{6}L^{(3)}_{\alpha_{1}\alpha_{2}\alpha_{3}}\left(\phi_{\alpha_{1}}\phi_{\alpha_{2}}\phi_{\alpha_{3}} +\hbar G_{\alpha_{1}\alpha_{2}}\phi_{\alpha_{3}} + \hbar G_{\alpha_{1}\alpha_{3}}\phi_{\alpha_{2}}+ \hbar G_{\alpha_{2}\alpha_{3}}\phi_{\alpha_{1}} + \hbar^2 V_{\alpha_{1}\alpha_{2}\alpha_{3}}\right) }$} \\
& \scalebox{0.99}{${\displaystyle + \cdots \;, }$}
\end{split}
\end{equation}
where the 1-point correlation function $\phi_{\alpha}$, the propagator $G_{\alpha_{1}\alpha_{2}}$, the 3-point vertex $V_{\alpha_{1}\alpha_{2}\alpha_{3}}$, ... satisfy:
\begin{equation}
\frac{\delta W\big[J,K,L^{(3)},\cdots\big]}{\delta J_{\alpha}} = \phi_{\alpha} \;,
\end{equation}
\begin{equation}
\frac{\delta W\big[J,K,L^{(3)},\cdots\big]}{\delta K_{\alpha_{1}\alpha_{2}}} = \frac{1}{2}\left[\phi_{\alpha_{1}} \phi_{\alpha_{2}}+\hbar G_{\alpha_{1}\alpha_{2}}\right] \;,
\end{equation}
\begin{equation}
\frac{\delta W\big[J,K,L^{(3)},\cdots\big]}{\delta L^{(3)}_{\alpha_{1}\alpha_{2}\alpha_{3}}} = \frac{1}{6}\left[\phi_{\alpha_{1}}\phi_{\alpha_{2}}\phi_{\alpha_{3}} + \hbar G_{\alpha_{1}\alpha_{2}}\phi_{\alpha_{3}} + \hbar G_{\alpha_{1}\alpha_{3}}\phi_{\alpha_{2}} + \hbar G_{\alpha_{2}\alpha_{3}}\phi_{\alpha_{1}} +\hbar^2 V_{\alpha_{1}\alpha_{2}\alpha_{3}}\right] \;,
\end{equation}
\begin{equation*}
\vdots
\end{equation*}
Within the $n$PI EA framework, the physical $m$-point functions (i.e. the $m$-point functions at vanishing sources) of the theory with $m\leq n$ are self-consistently dressed through a variational principle, i.e. by solving the gap equations:
\begin{equation}
\left.\frac{\delta\Gamma^{(n\mathrm{PI})}[\phi,G,V,\cdots]}{\delta\phi_{\alpha}}\right|_{\phi=\overline{\phi},G=\overline{G},V=\overline{V},\cdots} = 0 \mathrlap{\quad \forall \alpha \;,}
\label{eq:GapEquationphi}
\end{equation}
\begin{equation}
\left.\frac{\delta\Gamma^{(n\mathrm{PI})}[\phi,G,V,\cdots]}{\delta G_{\alpha_{1}\alpha_{2}}}\right|_{\phi=\overline{\phi},G=\overline{G},V=\overline{V},\cdots} = 0 \mathrlap{\quad \forall \alpha_{1},\alpha_{2} \;,}
\label{eq:GapEquationG}
\end{equation}
\begin{equation}
\left.\frac{\delta\Gamma^{(n\mathrm{PI})}[\phi,G,V,\cdots]}{\delta V_{\alpha_{1}\alpha_{2}\alpha_{3}}}\right|_{\phi=\overline{\phi},G=\overline{G},V=\overline{V},\cdots} = 0 \mathrlap{\quad \forall \alpha_{1},\alpha_{2},\alpha_{3} \;,}
\label{eq:GapEquationV}
\end{equation}
\begin{equation*}
\vdots
\end{equation*}
while the higher $m$-point functions (with $m>n$) coincide with the bare ones.

\vspace{0.5cm}

Basic properties about the system of interest can be obtained from the above generating functionals. Among these, we will focus throughout this thesis in particular on:
\begin{itemize}
\item The gs energy $E_{\mathrm{gs}}$ of the interacting system:
\begin{equation}
\begin{split}
E_{\mathrm{gs}} = & \ \underset{\beta\rightarrow\infty}{\lim} \left(-\frac{1}{\beta} \ln( Z[J=0,\cdots])\right) = \underset{\beta\rightarrow\infty}{\lim} \left(-\frac{1}{\hbar\beta} W[J=0,\cdots]\right) \\
= & \ \underset{\beta\rightarrow\infty}{\lim} \left(\frac{1}{\hbar\beta} \Gamma^{(n\mathrm{PI})}\big[\phi=\overline{\phi},\cdots\big]\right) \;.
\end{split}
\label{eq:Ener}
\end{equation}

\item The gs density $\rho_{\mathrm{gs}}$ of the interacting system:
\begin{equation}
\begin{split}
\rho_{\mathrm{gs}}(\boldsymbol{r},\tau) = & \ \braket{\widetilde{\varphi}_{\alpha}\widetilde{\varphi}_{\alpha}}_{\mathrm{vac}} \\
= & \ \frac{\hbar^2}{Z[J=0,K=0,\cdots]}\left.\frac{\delta^2 Z[J,K,\cdots]}{\delta J_{\alpha} \delta J_{\alpha}}\right|_{J=0,K=0,\cdots} \\
= & \ \hbar\left.\frac{\delta^2 W\left[J,K,\cdots\right]}{\delta J_{\alpha} \delta J_{\alpha}}\right|_{J=0,K=0,\cdots}+\overline{\phi}_{\alpha} \overline{\phi}_{\alpha} \\
= & \ 2\left.\frac{\delta W[J,K,\cdots]}{\delta K_{\alpha\alpha}}\right|_{J=0,K=0,\cdots} \;.
\end{split}
\label{eq:Dens}
\end{equation}

\item The effective potential $V_\text{eff}(\phi)$, which is determined from the 1PI EA evaluated at a uniform (i.e. spacetime-independent) field configuration $\phi_{\mathrm{u}}$ after factorizing the volume of Euclidean spacetime~\cite{col73}:
\begin{equation}
\Gamma^{(\mathrm{1PI})}[\phi_{\mathrm{u}}] = \int d\tau d^{D-1}\boldsymbol{r} \ V_{\mathrm{eff}}(\phi_{\mathrm{u}}) \;.
\label{eq:effpot}
\end{equation}
\end{itemize}

\section{\label{sec:WickTheorem}Wick's theorem and diagrammatic techniques}

In order to properly explain what we refer to as diagrammatic techniques in this thesis, we now illustrate how Wick's theorem~\cite{wic50} is implemented in the PI formalism. To that end, we first consider the non-interacting version of the generating functional $Z[J]$ defined by~\eqref{eq:GeneralSJ} and~\eqref{eq:GeneralGeneratingFunctional}:
\begin{equation}
\begin{split}
Z[J]= & \ \mathcal{N}\int\mathcal{D}\widetilde{\varphi} \ e^{\frac{1}{\hbar}\left(-S_{0}[\widetilde{\varphi}] + J_{\alpha} \widetilde{\varphi}_{\alpha}\right)} \\
= & \ \mathcal{N}\int\mathcal{D}\widetilde{\varphi} \ e^{\frac{1}{\hbar}\left(-\frac{1}{2}\widetilde{\varphi}_{\alpha_{1}} G^{-1}_{0,\alpha_{1}\alpha_{2}} \widetilde{\varphi}_{\alpha_{2}} + J_{\alpha_{1}} \widetilde{\varphi}_{\alpha_{1}}\right)} \;,
\end{split}
\label{eq:ZjKWick}
\end{equation}
where we just introduced the quadratic part of $S[\widetilde{\varphi}]$, i.e. the free classical action:
\begin{equation}
S_{0}[\widetilde{\varphi}] = \frac{1}{2}\widetilde{\varphi}_{\alpha_{1}} G^{-1}_{0,\alpha_{1}\alpha_{2}} \widetilde{\varphi}_{\alpha_{2}} \;,
\end{equation}
expressed here in terms of the free propagator $G_{0}$. We then exceptionally leave integration or summation over indices more implicit by using a compact vector notation (in which $X^{\mathrm{T}}$ denotes the transpose of $X$ notably) in order to calculate:
\begin{equation}
\begin{split}
-\frac{1}{2}\left(\widetilde{\varphi}-G_{0} J\right)^{\mathrm{T}}& G_{0}^{-1} \left(\widetilde{\varphi} - G_{0} J\right) + \frac{1}{2} J^{\mathrm{T}}G_{0}J \\
= \ & -\frac{1}{2} \widetilde{\varphi}^{\mathrm{T}} G_{0}^{-1} \widetilde{\varphi} + \frac{1}{2} \widetilde{\varphi}^{\mathrm{T}} G_{0}^{-1} G_{0} J + \frac{1}{2} J^{\mathrm{T}} G_{0}^{\mathrm{T}} G_{0}^{-1} \widetilde{\varphi} - \cancel{\frac{1}{2} J^{\mathrm{T}} G_{0}^{\mathrm{T}} G_{0}^{-1} G_{0} J} + \cancel{\frac{1}{2} J^{\mathrm{T}} G_{0} J} \\
= \ & -\frac{1}{2} \widetilde{\varphi}^{\mathrm{T}} G_{0}^{-1} \widetilde{\varphi} + J^{\mathrm{T}} \widetilde{\varphi} \\
= \ & -\frac{1}{2}\widetilde{\varphi}_{\alpha_{1}} G^{-1}_{0,\alpha_{1}\alpha_{2}} \widetilde{\varphi}_{\alpha_{2}} + J_{\alpha_{1}} \widetilde{\varphi}_{\alpha_{1}} \;,
\end{split}
\end{equation}
where we have notably exploited the symmetry of $G_{0}$ (i.e. $G_{0}^{\mathrm{T}}=G_{0}$) alongside with the relation $\widetilde{\varphi}^{\mathrm{T}} J = J^{\mathrm{T}} \widetilde{\varphi}$. Therefore,~\eqref{eq:ZjKWick} is equivalent to:
\begin{equation}
\begin{split}
Z[J] = & \ \mathcal{N}\int\mathcal{D}\widetilde{\varphi} \ e^{\frac{1}{\hbar}\left(-\frac{1}{2}\left(\widetilde{\varphi}-G_{0} J\right)^{\mathrm{T}} G_{0}^{-1} \left(\widetilde{\varphi} - G_{0} J\right) + \frac{1}{2} J^{\mathrm{T}}G_{0}J\right)} \\
= & \ \mathcal{N}\int\mathcal{D}\widetilde{\varphi}' \ e^{\frac{1}{\hbar}\left(-\frac{1}{2}\widetilde{\varphi}'_{\alpha_{1}} G_{0,\alpha_{1}\alpha_{2}}^{-1} \widetilde{\varphi}'_{\alpha_{2}} + \frac{1}{2} J_{\alpha_{1}}G_{0,\alpha_{1}\alpha_{2}}J_{\alpha_{2}}\right)} \;.
\end{split}
\label{eq:TrickWickTheorem}
\end{equation}
Besides introducing back $\alpha$-indices in the second line of~\eqref{eq:TrickWickTheorem}, we have also introduced $\widetilde{\varphi}'$ via the shift $\widetilde{\varphi} \rightarrow \widetilde{\varphi}' + G_{0} J$. From this, it directly follows that\footnote{The Jacobian underlying shifts like $\widetilde{\varphi} \rightarrow \widetilde{\varphi}' + G_{0} J$ is trivial such that $\mathcal{D}\widetilde{\varphi}'=\mathcal{D}\widetilde{\varphi}$ in the present case.}:
\begin{equation}
Z[J] = \mathcal{N} \ e^{\frac{1}{2\hbar} J_{\alpha_{1}}G_{0,\alpha_{1}\alpha_{2}}J_{\alpha_{2}}} \int\mathcal{D}\widetilde{\varphi} \ e^{-\frac{1}{2\hbar}\widetilde{\varphi}_{\alpha_{1}} G_{0,\alpha_{1}\alpha_{2}}^{-1} \widetilde{\varphi}_{\alpha_{2}}} \;.
\label{eq:TrickWickTheoremStep2}
\end{equation}
The integral in the right-hand side (RHS\footnote{In the same way, ``left-hand side'' will be denoted as LHS.}) of~\eqref{eq:TrickWickTheoremStep2} is just a constant, as can be shown from Gaussian integration (see appendix~\ref{ann:GaussianInt}). Replacing $Z[J]$ in~\eqref{eq:TrickWickTheoremStep2} with~\eqref{eq:ZjKWick} leads to:
\begin{equation}
\frac{\int\mathcal{D}\widetilde{\varphi} \ e^{\frac{1}{\hbar}\left(-\frac{1}{2} \widetilde{\varphi}_{\alpha_{1}} G_{0,\alpha_{1}\alpha_{2}}^{-1} \widetilde{\varphi}_{\alpha_{2}} + J_{\alpha_{1}}\widetilde{\varphi}_{\alpha_{1}}\right)}}{\int\mathcal{D}\widetilde{\varphi} \ e^{-\frac{1}{2\hbar} \widetilde{\varphi}_{\alpha_{1}} G_{0,\alpha_{1}\alpha_{2}}^{-1} \widetilde{\varphi}_{\alpha_{2}}}} = e^{\frac{1}{2\hbar} J_{\alpha_{1}}G_{0,\alpha_{1}\alpha_{2}}J_{\alpha_{2}}} \;.
\label{eq:TrickWickTheoremStep3}
\end{equation}
Finally, differentiating $n$ times ($n$ being even) both sides of~\eqref{eq:TrickWickTheoremStep3} with respect to $J$ before setting $J$ equal to zero gives us the following expression of the correlation functions of (even) order $n$ for the non-interacting theory specified by $S_{0}[\widetilde{\varphi}]$:
\begin{equation}
\scalebox{0.93}{${\displaystyle\left\langle \widetilde{\varphi}_{\alpha_{1}} \cdots \widetilde{\varphi}_{\alpha_{n}} \right\rangle_{0} \equiv \frac{\int\mathcal{D}\widetilde{\varphi} \ \widetilde{\varphi}_{\alpha_{1}} \cdots \widetilde{\varphi}_{\alpha_{n}} \ e^{\frac{1}{\hbar}\left(-\frac{1}{2} \widetilde{\varphi}_{\alpha_{1}} G_{0,\alpha_{1}\alpha_{2}}^{-1} \widetilde{\varphi}_{\alpha_{2}} + J_{\alpha_{1}}\widetilde{\varphi}_{\alpha_{1}}\right)}}{\int\mathcal{D}\widetilde{\varphi} \ e^{-\frac{1}{2\hbar} \widetilde{\varphi}_{\alpha_{1}} G_{0,\alpha_{1}\alpha_{2}}^{-1} \widetilde{\varphi}_{\alpha_{2}}}} = \hbar^{\frac{n}{2}}\sum_{P\in S_{n}} G_{0,\alpha_{P(1)}\alpha_{P(2)}}\cdots G_{0,\alpha_{P(n-1)}\alpha_{P(n)}} \;,}$}
\label{eq:TrickWickTheoremFinalStep}
\end{equation}
where the RHS involves a sum over all elements $P$ of the permutation group $S_{n}$ of even order $n$. Result~\eqref{eq:TrickWickTheoremFinalStep} formulates Wick's theorem in the PI formalism. All permutations between indices of propagators implied by the sum of the RHS of~\eqref{eq:TrickWickTheoremFinalStep} are most conveniently represented by diagrams weighted by numerical factors called multiplicities. In this thesis, we call diagrammatic techniques every approach relying on diagrammatic representations of the permutations set by Wick's theorem. This excludes notably FRG approaches, even though the latter are based on equations that can be represented diagrammatically as well. Note also that the quadratic part $S_{0}[\widetilde{\varphi}]$ retained in the generating functional $Z[J]$ to derive~\eqref{eq:TrickWickTheoremFinalStep} is not necessarily the free part of the classical action $S[\widetilde{\varphi}]$, such that $G_{0}$ is no longer a free propagator (but rather a more dressed object): this point is at the heart of chapter~\ref{chap:DiagTechniques}.

\section{\label{sec:KeyAspectsFRG}Selected topics on functional renormalization group}

FRG approaches rely on a master equation dictating the evolution of a functional of interest (typically a Schwinger functional or an EA) with respect to a parameter, called the flow parameter, which can either be a dimensionless or a dimensionful quantity representing typically the momentum scale of the theory under consideration. In practice, this master equation is treated via some expansion scheme (that we discuss in detail in chapter~\ref{chap:FRG}) to turn it into a set of coupled first-order integro-differential equations to solve, which contrasts with self-consistent equations such as~\eqref{eq:GapEquationphi} to~\eqref{eq:GapEquationV} encountered in diagrammatic EA approaches. All the technical details underpinning this procedure is outlined in chapter~\ref{chap:FRG} and corresponding appendices but we want to stress at this stage a few important aspects of current FRG studies of fermionic systems.

\vspace{0.5cm}

There are strong connections between several FRG methods and Wilson's RG~\cite{wil71,wil71bis,wil72} that turned out to be successful to describe critical phenomena\footnote{See ref.~\cite{del12} for a pedagogical introduction on Wilson's RG and the concept of coarse-graining.}. The idea underlying Wilson's approach, i.e. a step-by-step integration with respect to the momentum scale that we refer to as \textbf{Wilsonian momentum-shell integration}, was first implemented for an EA by Wetterich~\cite{wet93}, following up the previous work of Polchinski leading to the famous Polchinski or Wilson-Polchinski equation~\cite{pol84}. The flow parameter of Wetterich's FRG is thus the momentum scale $k$ (or another related parameter). Moreover, the corresponding master equation, coined as \textbf{Wetterich equation} (see section~\ref{sec:1PIFRGstateofplay} and appendix~\ref{sec:DerivMasterEq1PIFRG}), sets the scale dependence of the 1PI EA describing the studied quantum system, which is why we will use 1PI-FRG as synonym for Wetterich's FRG. This scale-dependent 1PI EA, also called average EA, describes a coarse-grained system in the spirit of Kadanoff and Wilson\footnote{Despite this, one should keep in mind that there are still fundamental differences between the 1PI-FRG and Wilson's RG: pedagogical explanations on this point can be found e.g. in ref.~\cite{del12}.}.

\vspace{0.5cm}

We put forward Wetterich's approach here as it is still clearly the most widespread FRG technique. This is notably due to a certain ease of implementation (e.g. as compared to FRG methods based on a $n$PI EA with $n > 1$), and also to the Wilsonian momentum-shell integration that makes it equipped to tackle critical phenomena and phase transitions. We also want to emphasize in the present section some important features of the implementation of Wetterich's FRG in the framework of fermionic systems. In particular, it has been proven very advantageous to introduce the so-called \textbf{scale-dependent or flowing bosonization} by combining this method with HSTs: either by exploiting a HST that is scale-dependent itself~\cite{flo09,flo10}, or rather by simply letting the expectation value of the Hubbard-Stratonovich field depend on the momentum scale~\cite{gie02,gie02bis,gie04}. The latter formulation has been generalized~\cite{paw07} such that only the fluctuating composite operators introduced via HSTs (and not their expectation values) are scale-dependent (see ref.~\cite{cyr18} for a concrete example of application in the context of QCD). A further generalization aiming at treating explicit symmetry breaking is developed in ref.~\cite{fu20}. We list below several reasons motivating the use of flowing bosonization:
\begin{itemize}
\item To \textbf{overcome the Fierz ambiguity}:\\
Some approximation schemes implemented in the framework of fermionic models can exhibit the so-called Fierz ambiguity which results from the possibility to apply Fierz transformations to multifermion interactions. This ambiguity translates notably into an unphysical dependence of the results obtained via mean-field theory (MFT) with respect to a mean-field parameter associated to those Fierz transformations\footnote{We refer to section~\ref{sec:1PIFRG0DON} for technical details on the implementation of MFT within the 1PI-FRG.}~\cite{jae03}. In the language of partial bosonization, MFT is implemented by neglecting quantum fluctuations of the Hubbard-Stratonovich field and the origin of the Fierz ambiguity can be traced back to the omission of those bosonic fluctuations. Hence, the incorporation of bosonic fluctuations on top of the MFT approximation allows for reducing the undesirable effects of the Fierz ambiguity, thus providing a fertile ground for quantitatively accurate results~\cite{jae03}. Furthermore, it has been shown that a proper treatment of bosonic fluctuations combined with a scale-dependent bosonization used to sweep away the problematic multifermion interactions at each momentum scale (which amounts to ``rebosonizing'' the multifermion interactions regenerated throughout the flow) can cure the Fierz ambiguity of MFT within the 1PI-FRG~\cite{jae03}.
\item To \textbf{bridge the gap between the ultraviolet (UV) and the infrared (IR) scales if the relevant dofs change during the flow}:\\
In practice, the differential equations underlying Wetterich's FRG are solved by letting the momentum scale flowing from an UV scale down to an IR scale at which the scale-dependent 1PI EA is supposedly close to the exact 1PI EA of the studied system. Furthermore, HSTs introduce bosonic fields in fermionic theories. This is interesting for example in FRG studies of QCD, in which the computations throughout the flow can be rendered more efficient by allowing scale-dependent quark and gluon couplings (more relevant at the UV scale) but also scale-dependent meson couplings (more relevant at the IR scale) in the scale-dependent 1PI EA~\cite{bra09}.
\item To \textbf{study SSBs and phase transitions}:\\
Phase transitions stemming from an onset of SSB might result in the divergence of some couplings (typically quartic couplings) involved in the ansatz of the scale-dependent 1PI EA. In this situation, it might be necessary to introduce further couplings in this ansatz (including for instance 8-fermion interactions in the flow) to have access to a proper order parameter for such transitions, which might render the implementation of the FRG procedure extremely difficult. If these divergences originate from quartic couplings $\lambda_{i}$, one can replace the latter by auxiliary fields with masses $m_{i} \sim 1/\lambda_{i}$ with the help of HSTs. In this way, the aforementioned divergences are therefore not present in the partially bosonized theory and there is hence no need to turn on further couplings during the flow to study phase transitions in this case. Note that, since the quartic couplings $\lambda_{i}$ depend on the momentum scale, the bosonization thus performed must also be scale-dependent to prevent the related divergences from pursuing the flow down to the IR scale.
\end{itemize}

\vspace{0.3cm}

The toy model studied in this thesis is not a relevant framework to investigate the technique of flowing bosonization owing to its simplicity (no fermionic fields in the original classical action, ...). Despite this, we believe that it is important to have the above three points in mind throughout our toy model study since they underlie many FRG applications to fermionic systems, whereas we will keep emphasizing the connections between the present work and more realistic (fermionic) applications in the forthcoming chapters.

\section{\label{sec:studiedtoymodel}Playground of this thesis: (0+0)-D $O(N)$-symmetric $\varphi^{4}$-theory}

We use the $O(N)$-symmetric $\varphi^{4}$-theory in arbitrary spacetime dimensions to investigate various expansion schemes expressed within the PI language, with a special emphasis on the broken-symmetry phase, and apply our results in the exactly solvable (0+0)-D case\footnote{By definition, a ($n$+$m$)-D theory lives in a spacetime with $n$ space and $m$ time dimensions, which means that there are neither space nor time dimensions in the case of the studied (0+0)-D toy model.}. QFTs formulated in zero dimension feature a base manifold $\mathcal{M}$ reducing exactly to one point\footnote{No notion of metric can be defined on the manifold $\mathcal{M}$, with the consequence that the Lorentz group and all of its representations are trivial, i.e. all fields living on $\mathcal{M}$ are point-like and must be scalars.}, i.e. $\mathcal{M} = \lbrace\bullet\rbrace$. All fields living on $\mathcal{M} = \lbrace\bullet\rbrace$ are completely specified by assigning a number (e.g. a real one) at this one point, such that the PI measure $\mathcal{D}\widetilde{\varphi}$ on $\mathcal{C}$ reduces to the standard Lebesgue measure $d\widetilde{\varphi}$ (e.g. on $\mathbb{R}$). The tremendous simplifications brought by the latter feature explain why (0+0)\nobreakdash-D QFTs serve as safe, more controllable and hence useful didactic playgrounds for exploring various aspects of more complicated QFTs, as they allow for explicit solutions that can not be obtained in higher dimensions~\cite{cai74,dos75,zin81,ben92,sch94,ban97,hoo99,ben00,mal01,kei12,kem13,bro15,ros16,lia18}. The dofs of the (0+0)-D $O(N)$ model are represented by real fluctuating bosonic fields $\widetilde{\varphi}_a: \lbrace\bullet\rbrace\rightarrow\mathbb{R}$ living on the base manifold $\mathcal{M} = \lbrace\bullet\rbrace$, i.e. real random variables, with $O(N)$-symmetric quartic self-interaction. We store them in the $O(N)$ scalar multiplet:
\begin{equation}
\vec{\widetilde{\varphi}} \equiv \begin{pmatrix}
\widetilde{\varphi}_{1} \\
\vdots \\
\widetilde{\varphi}_{N}
\end{pmatrix} \;,
\end{equation}
and consider the Lie group action $O(N) \curvearrowright \mathbb{R}^N$ defined by left multiplication, with the infinitesimal transformation given by:
\begin{equation}
\delta_\epsilon \widetilde{\varphi}_{a} = \sum_{b=1}^{N} \epsilon_{a b} \widetilde{\varphi}_{b} \;,
\end{equation}
characterized by real antisymmetric matrices (i.e. $\epsilon_{a b} = -\epsilon_{b a} \in\mathbb{R}$) and the indices $a$, $b$, ... (which label the $N$ orthogonal directions in the color space defined on $\mathbb{R}^N$) are referred to as color indices. The dynamics of the system is governed by the classical action\footnote{In a (0+0)-D spacetime, the classical action $S$ as well as all generating functionals characterizing the theory (such as Schwinger functionals or EAs) are functions rather than functionals.} $S:\mathbb{R}^N\rightarrow\mathbb{R}$, given by the expression:
\begin{equation}
S\Big(\vec{\widetilde{\varphi}}\Big)=\frac{m^{2}}{2}\vec{\widetilde{\varphi}}^{2}+\frac{\lambda}{4!}\left(\vec{\widetilde{\varphi}}^{2}\right)^{2} \;,
\label{eq:S0D}
\end{equation}
which is invariant under transformations of the $O(N)$ group, and where the 
real parameters $m^2$ and $\lambda$ stand for the bare squared mass and bare coupling constant, respectively. No derivative (i.e. kinetic) terms contribute to this action owing to the (0+0)-D nature of spacetime\footnote{This is why (0+0)-D QFTs are sometimes referred to as 
the static ultra-local limit of a QFT in finite $D$ dimensions.}. In the present work, we consider mostly two sources with $O(N)$ group structure, namely the local source $\vec{J}$ whose components $J_a$ are coupled to the fields $\widetilde{\varphi}_a$ and the bilocal source $\boldsymbol{K}$ whose elements $\boldsymbol{K}_{a b}$ are coupled to the composite fields $\widetilde{\varphi}_{a}\widetilde{\varphi}_{b}$. In this context, the Schwinger functional of the theory is given by:
\begin{equation}
Z\Big(\vec{J},\boldsymbol{K}\Big) = e^{\frac{1}{\hbar}W\big(\vec{J},\boldsymbol{K}\big)} = \int_{\mathbb{R}^N} d^N\vec{\widetilde{\varphi}} \ e^{-\frac{1}{\hbar}S_{JK}\big(\vec{\widetilde{\varphi}}\big)} \;,
\label{eq:ZJK}
\end{equation}
with
\begin{equation}
S_{JK}\Big(\vec{\widetilde{\varphi}}\Big)\equiv S\Big(\vec{\widetilde{\varphi}}\Big)-\vec{J}\cdot\vec{\widetilde{\varphi}}-\frac{1}{2}\vec{\widetilde{\varphi}}\cdot\left(\boldsymbol{K}\vec{\widetilde{\varphi}}\right) \;.
\label{eq:SJK0D}
\end{equation}
The symbol ``$\cdot$'' in~\eqref{eq:SJK0D} refers to the scalar product in color space defined as:
\begin{equation}
X \cdot Y \equiv \sum^{N}_{a=1} X_{a} Y_{a} \;,
\label{eq:ScalarProduct0DON}
\end{equation}
while $\vec{\widetilde{\varphi}}\cdot\left(\boldsymbol{K}\vec{\widetilde{\varphi}}\right)$ is short for:
\begin{equation}
\sum_{a,b=1}^{N} \widetilde{\varphi}_{a} \boldsymbol{K}_{a b} \widetilde{\varphi}_{b} \;.
\end{equation}  

\vspace{0.5cm}

\begin{figure}[!htb]
\captionsetup[subfigure]{labelformat=empty}
  \begin{center}
    \subfloat[]{
      \includegraphics[width=0.50\linewidth]{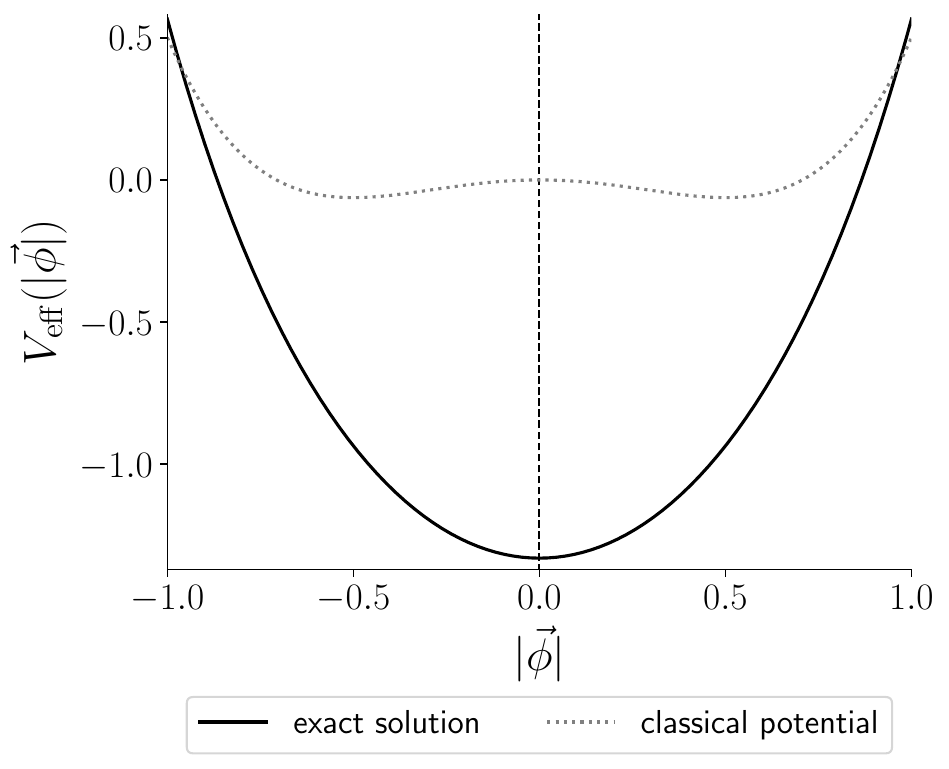}
                         }
    \subfloat[]{
      \includegraphics[width=0.50\linewidth]{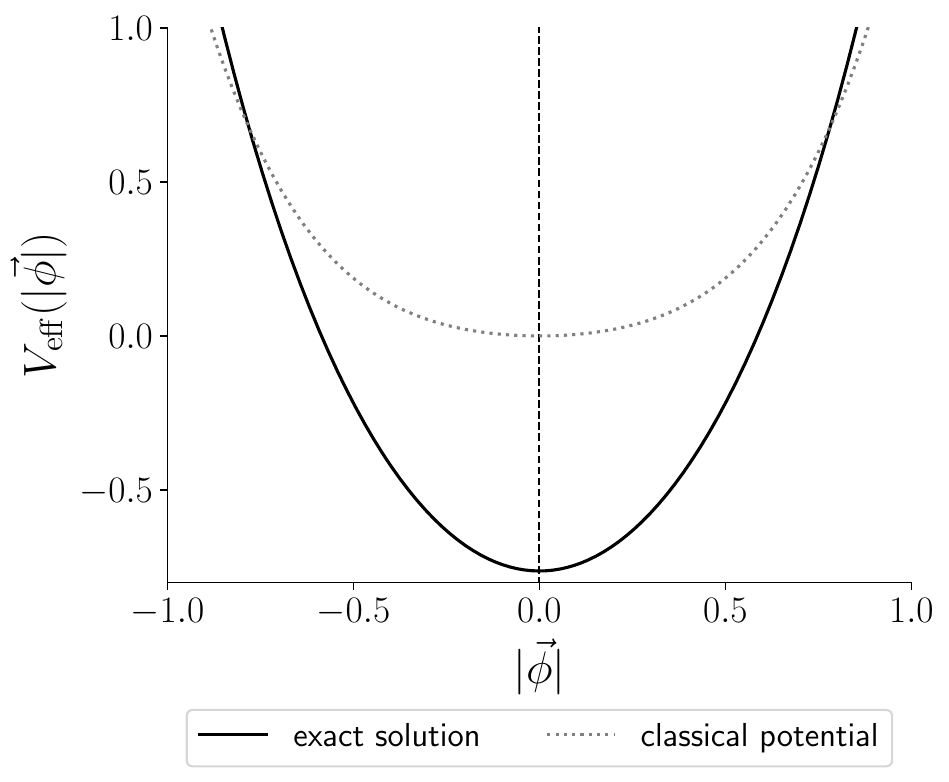}
                         }
    \caption{Classical and exact effective potentials as functions of the background constant field's modulus $\left|\vec{\phi}\right|$ at $N=2$, $\lambda/4! = 1$ and $m^2=-1$ (left) or $m^2=+1$ (right).}
    \label{fig:Exactpot}
  \end{center}
\end{figure}

The (0+0)-D $O(N)$ model with the generating functional~\eqref{eq:ZJK} takes the form of a probability theory for $N$ real stochastic variables $\widetilde{\varphi}_a$ whose probability distribution is given by $e^{-\frac{1}{\hbar}S_{JK}\left(\vec{\widetilde{\varphi}}\right)}$. The benefit of working in a (0+0)-D spacetime is already manifest from the fact that expression~\eqref{eq:ZJK} admits an analytical representation~\cite{kei12,ros16} in terms of the Kummer confluent hypergeometric function ${}_{1}F_{1}(a;b;z)$~\cite{abr65}. After rewriting the integral of~\eqref{eq:ZJK} in hyperspherical coordinates, the exact partition function reads:
\begin{equation}
Z^\text{exact}\Big(\vec{J}=\vec{0},\boldsymbol{K}=\boldsymbol{0}\Big) = e^{\frac{1}{\hbar}W^\text{exact}\big(\vec{J}=\vec{0},\boldsymbol{K}=\boldsymbol{0}\big)} = \Omega_{N}\mathcal{R}_{N-1} \;,
\label{eq:Z0DONexactsolution}
\end{equation}
with\footnote{$\Gamma$ exceptionally denotes Euler gamma function~\cite{abr65} and not an EA in~\eqref{eq:RN} and~\eqref{eq:SN}.}
\begin{equation}
\begin{split}
\mathcal{R}_N\big(\hbar ;m^2;\lambda \big) \equiv & \ \int_{0}^{\infty} d\widetilde{u} \ \widetilde{u}^{N} \ e^{-\frac{1}{\hbar}\left( \frac{m^{2}}{2}\widetilde{u}^{2}+\frac{\lambda}{4!}\widetilde{u}^{4}\right)} \\
= & \left\{
\begin{array}{lll}
        \displaystyle{2^\frac{N-1}{2} \left(\frac{\hbar}{m^2}\right)^\frac{N+1}{2}\Gamma\left(\frac{N+1}{2}\right) \quad \forall m^2 > 0 ~ \mathrm{and} ~ \lambda=0 \;,} \\
        \\
        \displaystyle{\infty \quad \forall m^2 \leq 0 ~ \mathrm{and} ~ \lambda=0 \;,} \\
        \\
        \displaystyle{2^{\frac{3N-5}{4}} 3^{\frac{N+1}{4}} \left(\frac{\lambda}{\hbar}\right)^{-\frac{N+3}{4}} \Bigg[\sqrt{\frac{\lambda}{\hbar}}\Gamma\bigg(\frac{N+1}{4}\bigg) {}_{1}F_{1}\bigg(\frac{N+1}{4};\frac{1}{2};\frac{3 m^{4}}{2\lambda\hbar}\bigg) } \\
        \hspace{3.92cm} \displaystyle{- \frac{m^{2}\sqrt{6}}{\hbar}\Gamma\bigg(\frac{N+3}{4}\bigg){}_{1}F_{1}\bigg(\frac{N+3}{4};\frac{3}{2};\frac{3 m^{4}}{2\lambda\hbar}\bigg)\Bigg] \quad \forall \lambda > 0 \;,}
    \end{array}
\right.
\end{split}
\label{eq:RN}
\end{equation}
for $N\in \mathbb{N}^*$. Note also that $\Omega_{N}$ denotes the surface area of the $N$-dimensional unit sphere:
\begin{equation}
\Omega_{N}=\frac{2\pi^{\frac{N}{2}}}{\Gamma\big(\frac{N}{2}\big)} \;.
\label{eq:SN}
\end{equation}
The gs energy\footnote{In a (0+0)-D spacetime, the extraction~\eqref{eq:Ener} of the gs energy from the partition function simplifies, after an arbitrary scaling that cancels  
the factor in front of the logarithm, to $E_{\mathrm{gs}} = -\ln\big(Z\big(\vec{J}=\vec{0},\boldsymbol{K}=\boldsymbol{0}\big)\big)$.} and density\footnote{Note that the denomination ``density'' is abusive in the case where $N=1$. Indeed, the $O(N)$ model does not exhibit any continuous symmetry in this situation, hence no conserved Noether current.} can be obtained from:
\begin{equation}
E_{\mathrm{gs}} = -\ln\Big( Z\Big(\vec{J}=\vec{0},\boldsymbol{K}=\boldsymbol{0}\Big)\Big) = -\frac{1}{\hbar}W\Big(\vec{J}=\vec{0},\boldsymbol{K}=\boldsymbol{0}\Big) \;,
\label{eq:DefEgsExactZexact0DON}
\end{equation}
\begin{equation}
\rho_{\mathrm{gs}} = \frac{1}{N}\left\langle\vec{\widetilde{\varphi}}^{2}\right\rangle = -\frac{2}{N} \left.\frac{\partial W\big(\vec{J},\boldsymbol{K}\big)}{\partial m^{2}}\right|_{\vec{J}=\vec{0}\atop\boldsymbol{K}=\boldsymbol{0}} \;,
\label{eq:DefrhogsExactwithExpectationValue0DON}
\end{equation}
with the expectation value defined as:
\begin{equation}
\big\langle \cdots \big\rangle \equiv \frac{1}{Z\big(\vec{J}=\vec{0},\boldsymbol{K}=\boldsymbol{0}\big)} \int d^{N}\vec{\widetilde{\varphi}} \ \cdots \ e^{-\frac{1}{\hbar}S\big(\vec{\widetilde{\varphi}}\big)}\;.
\label{eq:vacuumExpectationValue0DON}
\end{equation}
From these definitions, one can infer the exact solutions:
\begin{equation}
E^{\mathrm{exact}}_{\mathrm{gs}} = -\ln\big(\Omega_{N}\mathcal{R}_{N-1}\big) \;,
\end{equation}
\begin{equation}
\rho^{\mathrm{exact}}_{\mathrm{gs}} = \frac{\mathcal{R}_{N+1}}{N\mathcal{R}_{N-1}} \;.
\end{equation}
On the other hand, the computation of the 1-point correlation function:
\begin{equation}
\vec{\overline{\phi}} \equiv \left\langle\vec{\widetilde{\varphi}}\right\rangle = \left.\frac{\partial W\big(\vec{J},\boldsymbol{K}\big)}{\partial\vec{J}}\right|_{\vec{J}=\vec{0}\atop\boldsymbol{K}=\boldsymbol{0}} \;,
\end{equation}
or of the effective potential $V_{\mathrm{eff}}\big(\vec{\phi}\big)$ provides information on the occurrence of SSB~\cite{col73}. While the exact solution for the former reduces to $\vec{\overline{\phi}} \hspace{0.01cm} \rule{0cm}{0.4cm}^{\mathrm{exact}} = 0$ for all values of the coupling constant $\lambda$ and of the squared mass $m^2$, the latter derives from the 1PI EA according to\footnote{Relation~\eqref{eq:effpot} reduces to the leftmost equality of~\eqref{eq:DefEffectivePot1PIEA0DintroPI}, i.e. $V_{\mathrm{eff}}\big(\vec{\phi}\big) = \Gamma^{(\mathrm{1PI})}\big(\vec{\phi}\big)$, for our (0+0)-D $O(N)$ model.}:
\begin{equation}
V_{\mathrm{eff}}\Big(\vec{\phi}\Big) = \Gamma^{(\mathrm{1PI})}\Big(\vec{\phi}\Big) = -W\Big(\vec{J},\boldsymbol{K}=\boldsymbol{0}\Big) + \vec{J}\cdot\vec{\phi} \;,
\label{eq:DefEffectivePot1PIEA0DintroPI}
\end{equation}
with
\begin{equation}
\vec{\phi} = \left.\frac{\partial W\big(\vec{J},\boldsymbol{K}\big)}{\partial\vec{J}}\right|_{\boldsymbol{K}=\boldsymbol{0}} \;.
\label{eq:Defvecphi1PIEA0DintroPI}
\end{equation}
The exact effective potential $V_\text{eff}^\text{exact}\big(\vec{\phi}\big)$ is evaluated numerically from~\eqref{eq:DefEffectivePot1PIEA0DintroPI} combined with \eqref{eq:Defvecphi1PIEA0DintroPI}, and then plotted in fig.~\ref{fig:Exactpot} for $N=2$ together with the classical potential:
\begin{equation}
U\Big(\vec{\phi}\Big) = \frac{m^{2}}{2} \vec{\phi}^2 + \frac{\lambda}{4!}\left(\vec{\phi}^{2}\right)^{2} \;,
\label{eq:classpot}
\end{equation}
which coincides with the classical action~\eqref{eq:S0D} of the studied (0+0)-D model\footnote{At finite dimensions, the classical action $S$ differs from the corresponding classical potential $U$ since the latter does not contain any derivative (i.e. kinetic) terms.}.

\vspace{0.5cm}

The lowest energy states of the system at the classical (i.e. at the tree) level are given by the minima of the classical potential~\eqref{eq:classpot}. In the situation where the coupling constant $\lambda$ is real and zero or positive (which is a restriction followed to obtain all numerical results presented in this thesis), we can show by minimizing the classical potential~\eqref{eq:classpot} that these are:
\begin{itemize}
\item For $m^2>0$, a unique vacuum $\vec{\phi}=\vec{0}$ where the $O(N)$ symmetry is conserved (spontaneously as well as explicitly).
\item For $m^2<0$ and $\lambda\neq 0$, a manifold of degenerate vacua (satisfying $\vec{\phi}^2=-3! m^2/\lambda > 0$) where the original $O(N)$ symmetry is spontaneously (but not explicitly) broken down to $O(N-1)$.
\end{itemize}
In the full theory, the lowest energy states are found by minimizing the exact effective potential $V_\text{eff}^\text{exact}\big(\vec{\phi}\big)$, which yields a unique gs conserving the $O(N)$ symmetry (spontaneously and explicitly), irrespectively of the sign of $m^2$ as shown by fig.~\ref{fig:Exactpot}. In what follows, we will refer to the phase with $m^{2}<0$ ($m^{2}>0$) as broken-symmetry (unbroken-symmetry) regime or phase, even though one must keep in mind that the $O(N)$ symmetry is broken down only spontaneously and only at the classical level. This absence of broken symmetry in the exact solution of the toy model under consideration is consistent with the Mermin-Wagner theorem~\cite{mer66,hoh67,col73bis} and enables us to make an analogy with the study of finite-size systems (and nuclei notably) which do not exhibit any SSB, as we discussed in chapter~\ref{chap:Intro1}. We have also illustrated in this way with fig.~\ref{fig:Exactpot} a general result for spatial dimensions less than or equal to two, where the divergences of Goldstone propagators cause large quantum fluctuations that spread away any classically selected configuration, hence precluding the spontaneous breakdown of a continuous symmetry~\cite{col74}.

%% file: 4ChapterDiag/Diag.tex
\minitoc

\noindent
In this chapter, we investigate various diagrammatic PI techniques defined in section~\ref{sec:WickTheorem}. We indeed saw that the PI formalism offers a powerful language for a systematic, controlled treatment of quantum fluctuations by dressing an initial, reference configuration of a many-body system. It also provides us with convenient mathematical manipulations for reshuffling (bosonic) collective fluctuations so that it is efficiently captured from the leading order of the description, e.g. via a HST and/or by coupling sources to composite operators within the $n$PI EA framework. The goals we ultimately have in mind are: i) to find the theoretical construct underpinning the nuclear EDF method, for it to be rigorously formulated and turned into a systematically improvable approach; ii) to design other optimal strategies for catching efficiently the typical correlations at play in strongly-coupled many-body systems. To that end, we investigate several frameworks to incorporate non-perturbative collective features at the lowest non-trivial orders. To analyze these various diagrammatic expansions and their resummation properties, we work within the $O(N)$-symmetric $\varphi^4$-theory where the various strategies are first derived in arbitrary spacetime dimensions for both the unbroken- and broken-symmetry regimes. We then perform our numerical applications in the exactly solvable (0+0)-D situation presented in section~\ref{sec:studiedtoymodel}, where we have seen that radiative corrections restore the $O(N)$ symmetry spontaneously broken at the classical level in the phase with $m^{2}<0$.

\vspace{0.5cm}

More precisely, the chapter is organized as follows. First, a general presentation of the various studied strategies is given in section~\ref{sec:Intro}. In section~\ref{sec:PT}, we focus on perturbative schemes expanding around the non-interacting physics or the classical configuration. Because of the asymptotic nature of the perturbative series representation of our physical quantities, meaningful results can only be obtained after applying proper resummation techniques, which are examined in section~\ref{sec:Resum}. Then, we investigate the treatment of many-body systems within the optimized perturbation theory (OPT)~\cite{ste81,nev92,sis92a,sis92,sis93,kle98,kle11,kle18} in section~\ref{sec:OPT}. The latter provides a variational improvement of a perturbative expansion by adding and subtracting an arbitrary quadratic term in the classical action of the theory under consideration, then optimizing the perturbative expansion by imposing the resulting truncated series to be stationary with respect to this quadratic kernel \textbf{at the working order} rather than just at the leading order (as done for example in MBPT implemented on top of a self-consistent mean-field configuration such as the Hartree-Fock and HFB reference states). Finally, the $n$PI EA method is considered in section~\ref{sec:EA}. It offers a hybridation of variational and perturbative expansions, where a series of Feynman diagrams is involved, with 1- through $n$-point (connected) correlation functions obtained by self-consistently dressing the bare ones via variational equations of motion. In particular, for $n=2$, the so-called 2PPI reduction of the 2PI EA provides a firm theoretical ground to DFT~\cite{fuk94,fuk95,val97,sch04,fer08}, as will be explained in section~\ref{sec:2PPIEA}.

\section{Aim of the study}
\label{sec:Intro} 

We consider the finite-dimensional counterpart of the classical action~\eqref{eq:S0D}:
\begin{equation}
\begin{split}
S\Big[\vec{\widetilde{\varphi}}\Big] = & \ \int_x \left[ \frac{1}{2}\left(\nabla_x\widetilde{\varphi}_a(x)\right)\left(\nabla_x\widetilde{\varphi}^a(x)\right) +  \frac{m^2}{2}\widetilde{\varphi}_a(x)\widetilde{\varphi}^a(x)+\frac{\lambda}{4!}\left(\widetilde{\varphi}_a(x)\widetilde{\varphi}^a(x)\right)^2\right] \\
\equiv & \ \int_0^{\hbar\beta} d\tau \int_{\mathbb{R}^{D-1}}d^{D-1}\boldsymbol{r}\left[ \frac{1}{2}\left(\partial_\tau\vec{\widetilde{\varphi}}\cdot\partial_\tau\vec{\widetilde{\varphi}} + \nabla_r\vec{\widetilde{\varphi}}\cdot\nabla_r\vec{\widetilde{\varphi}} \right) +  \frac{m^2}{2}\vec{\widetilde{\varphi}}^2+\frac{\lambda}{4!}\left(\vec{\widetilde{\varphi}}^2\right)^2\right] \;,
\end{split}
\label{eq:SfiniteD}
\end{equation}
with implicit summation over repeated internal indices (also called color indices). Introducing a local source $\vec{J}(x)$ and a bilocal one $\boldsymbol{K}(x,y)$, a generating functional of the theory is given by:
\begin{equation}
Z\Big[\vec{J},\boldsymbol{K}\Big] = e^{\frac{1}{\hbar}W\big[\vec{J},\boldsymbol{K}\big]} = \int \mathcal{D}\vec{\widetilde{\varphi}} \ e^{-\frac{1}{\hbar}S_{JK}\big[\vec{\widetilde{\varphi}}\big]} \;,
\label{eq:ZJKfiniteD}
\end{equation}
with $\int \mathcal{D}\vec{\widetilde{\varphi}} = \int \mathcal{D}\widetilde{\varphi}_{1} \cdots \int \mathcal{D}\widetilde{\varphi}_{N}$ and
\begin{equation}
S_{JK}\Big[\vec{\widetilde{\varphi}}\Big] \equiv S\Big[\vec{\widetilde{\varphi}}\Big]-\int_x J_a(x)\widetilde{\varphi}^a(x)-\frac{1}{2}\int_{x,y} \widetilde{\varphi}^a(x)\boldsymbol{K}_{ab}(x,y) \widetilde{\varphi}^b(y) \;.
\label{eq:SJKfiniteD}
\end{equation}
Except for the zero-dimensional situation\footnote{To clarify, the zero-dimensional situation or zero-dimensional limit refers throughout this entire thesis to the (0+0)-D situation (and not to the (0+1)-D situation, as it is sometimes the case in the literature).}, the integral in~\eqref{eq:ZJKfiniteD} can not be computed exactly, and therefore needs to be treated approximatively, e.g. within the expansion schemes discussed subsequently.

\vspace{0.5cm}

The purpose of this work is to investigate various approaches for capturing efficiently, i.e. from the lowest non-trivial orders of the description, the correlated behavior of the system. In wavefunction-based theory, an expansion method is specified by a splitting of the many-body Hamiltonian $H$ into so-called unperturbed $H_0$ and residual $H_1$ parts. After solving (part of) the many-body problem for $H_0$, one goes from an eigenstate $\ket{\Theta^{(0)}}$ of $H_0$ to an exact eigenstate $\ket{\Psi}$ of $H$ by incorporating the physics encoded into $H_1$ within a given expansion method (of perturbative or non-perturbative nature). For the expansion method to converge efficiently (and even to start in open-shell systems), a careful partitioning between unperturbed and residual sectors must be employed. The unperturbed reference state $\ket{\Theta^{(0)}}$ can be chosen under the form of a product state, i.e. $H_0$ is quadratic in the fields (i.e. it is a 1-body operator). In finite-size systems featuring so-called static correlations, responsible for collective behaviors such as density oscillations or superfluidity, a more sophisticated choice for  $\ket{\Theta^{(0)}}$ is needed, e.g. under the form of a linear combination of non-orthogonal symmetry-breaking product states, each parametrized by a set of order parameters associated to the broken symmetries (i.e. a PGCM ansatz). Such a linear combination captures non-negligible fluctuations of the order parameters from the zeroth-order description, including the ones restoring the symmetries of the system.

\vspace{0.5cm}

Likewise, in the PI language, a given expansion method is based on a splitting of the classical action between an unperturbed and a residual part $S=S^0+S^1$. Such a partitioning is not unique, and it is the purpose of this study to explore various splittings and to analyze their effectiveness in catching correlations from the first non-trivial orders of the chosen approach. For instance, $S^0$ is usually obtained after reducing the interacting theory to a free theory, possibly with self-consistently determined parameters (e.g. in the self-consistent mean-field approach). Typically, the quartic self-interaction $\big(\vec{\widetilde{\varphi}}\cdot\vec{\widetilde{\varphi}}\big)^2$ would be replaced by a term proportional to $\big\langle\vec{\widetilde{\varphi}}\cdot\vec{\widetilde{\varphi}}\big\rangle_0\vec{\widetilde{\varphi}}\cdot\vec{\widetilde{\varphi}}$, where $\big\langle\vec{\widetilde{\varphi}}\cdot\vec{\widetilde{\varphi}}\big\rangle_0$ is the free field average. To systematically improve on such a $S^0$, one needs to identify some small expansion parameter, which is difficult in practice (especially in the treatment of strongly-coupled systems). In fact, the reduction yielding $S^0$ would be justified if, for some reasons, the (bosonic) fluctuations of the composite field $\vec{\widetilde{\varphi}}\cdot\vec{\widetilde{\varphi}}$ were negligible compared to the fluctuations of the original field $\vec{\widetilde{\varphi}}$ itself.

\vspace{0.5cm}

In contrast with the canonical formulation of quantum mechanics, the PI formalism makes it very convenient to include the physics associated to the composite field $\vec{\widetilde{\varphi}}\cdot\vec{\widetilde{\varphi}}$. In that respect, a possible option involves the HST, i.e. an exact manipulation by which the two-body interaction between the original dofs are decoupled at the price of introducing an extra, collective field (another way to achieve this being to work within the $n$PI EA framework). The HST is just based on standard Gaussian integral properties and, in the present situation, translates into the identity:
\begin{equation}
e^{-\frac{\lambda}{\hbar 4!}\int_x{\left(\vec{\widetilde{\varphi}}(x)\cdot\vec{\widetilde{\varphi}}(x) \right)^2}} = \sqrt{\frac{6\hbar}{\pi\lambda}}\int \mathcal{D}\widetilde{\sigma} \ e^{-\int_x\left[\frac{6\hbar}{\lambda}\widetilde{\sigma}^2(x)+i\widetilde{\sigma}(x)\vec{\widetilde{\varphi}}(x)\cdot\vec{\widetilde{\varphi}}(x)\right]} \;,
\end{equation}
where the Hubbard-Stratonovich (or auxiliary) field $\widetilde{\sigma}(x)$ is a collective quantum (i.e. fluctuating) field which is a scalar in color space. The original theory based on $\vec{\widetilde{\varphi}}(x)$ is then transformed into an equivalent one whose partition function reads, after the redefinition $\widetilde{\sigma}\rightarrow\sqrt{\frac{\lambda}{12\hbar^2}}\widetilde{\sigma}$, as follows:
\begin{equation}
Z_\text{mix} = \int\mathcal{D}\vec{\widetilde{\varphi}}\mathcal{D}\widetilde{\sigma} \ e^{-\frac{1}{\hbar}S_\text{mix}\big[\vec{\widetilde{\varphi}},\widetilde{\sigma}\big]} \propto Z\Big[\vec{J}=\vec{0},\boldsymbol{K}=\boldsymbol{0}\Big] \;,
\label{eq:Zmix}
\end{equation}
with
\begin{equation}
S_\text{mix}\Big[\vec{\widetilde{\varphi}},\widetilde{\sigma}\Big] =\frac{1}{2} \int_x \left[ \left(\nabla_x\vec{\widetilde{\varphi}}(x)\right)\cdot\left(\nabla_x\vec{\widetilde{\varphi}}(x)\right) +\left(m^2+i\sqrt{\frac{\lambda}{3}}\widetilde{\sigma}(x)\right)\vec{\widetilde{\varphi}}(x)\cdot\vec{\widetilde{\varphi}}(x)+\widetilde{\sigma}^2(x)\right] \;.
\label{eq:Smix}
\end{equation}

\vspace{0.5cm}

At this stage, one can work with the mixed system thus obtained, which involves both $\vec{\widetilde{\varphi}}$ and $\widetilde{\sigma}$ as dofs and where the original, cumbersome quartic interaction between the original dofs has been replaced by a Yukawa interaction between $\vec{\widetilde{\varphi}}$ and the collective field $\widetilde{\sigma}$. We can also exploit the fact that the mixed action~\eqref{eq:Smix} is now quadratic in the field $\vec{\widetilde{\varphi}}$, which can therefore be integrated out in the partition function~\eqref{eq:Zmix}, thus leading to:
\begin{equation}
Z_\text{col} = \int\mathcal{D}\widetilde{\sigma} \ e^{-\frac{1}{\hbar}S_\text{col}[\widetilde{\sigma}]} \propto Z\Big[\vec{J}=\vec{0},\boldsymbol{K}=\boldsymbol{0}\Big] \;,
\label{eq:Zcoll}
\end{equation}       
with
\begin{equation}
S_\text{col}[\widetilde{\sigma}]= \frac{1}{2}\int_x\widetilde{\sigma}^2(x) -\frac{1}{2} \mathrm{STr}\left[\ln(\boldsymbol{G}_{\widetilde{\sigma}})\right] \;,
\label{eq:Scoll}
\end{equation}
and 
\begin{equation}
\boldsymbol{G}^{-1}_{\widetilde{\sigma};ab}(x,y) = \left(-\nabla_x^2 + m^2 + i\sqrt{\frac{\lambda}{3}}\widetilde{\sigma}(x)\right)\delta_{ab} \delta(x-y) \;,
\label{eq:Gcoll}
\end{equation}
where the supertrace $\mathrm{STr}$ is by definition the trace taken with respect to both color and spacetime indices, i.e. $\mathrm{STr}\equiv\mathrm{Tr}_{a}\mathrm{Tr}_{x}$.

\vspace{0.5cm}

For each of these three representations, coined as original, mixed and collective hereafter, we will then proceed to a partitioning between unperturbed and residual sectors and account for the physics encoded in the residual part within a given (perturbative or non-perturbative) scheme. The performances of each strategy will be also tested in (0+0)-D by comparing the gs energy and density obtained at a given order of the studied expansions with their exact counterparts given respectively by $E^\text{exact}_\text{gs}$ and $\rho^\text{exact}_\text{gs}$ defined in section~\ref{sec:studiedtoymodel}. We also explained in the latter section that the phase of the $O(N)$ model under consideration with $m^2<0$ (and with $\lambda$ real and positive, which is a restriction often left implicit in our forthcoming discussions, although it only applies to our numerical applications and not to our formal derivations) exhibits a spontaneous breakdown of the $O(N)$ symmetry at the classical level whereas the field fluctuations lead to the restoration of this symmetry in the framework of the exact solution. The analysis of the gs symmetry properties at each order of the studied expansions will be carried out by systematically extending each expansion scheme to this phase.

\section{Loop expansions and perturbative treatment}
\label{sec:PT}

Among the various strategies one can elaborate to tackle a many-body problem, perturbation theory (PT) generally comes as a first attempt. In the PI formalism, PT is implemented by splitting the classical action (in presence of the sources) into a reference part $S_{JK}^0$ and a residual part $S_{JK}^1$, before perturbatively introducing the effects of the residual part in the $Z$ and $W$ functionals. A first (naive) approach is to organize the perturbative expansion by powers of the interaction after taking the action of the non-interacting system (i.e. at $\lambda=0$) as the reference configuration. This translates into:
\begin{equation}
S_{JK}\Big[\vec{\widetilde{\varphi}}\Big] = S^0_{JK}\Big[\vec{\widetilde{\varphi}}\Big]+S^1_{JK}\Big[\vec{\widetilde{\varphi}}\Big] \;,
\end{equation}
\begin{equation}
S^0_{JK}\Big[\vec{\widetilde{\varphi}}\Big] = \frac{1}{2} \int_{x,y} \widetilde{\varphi}^a(x)\boldsymbol{G}^{-1}_{0;K;ab}(x,y)\widetilde{\varphi}^b(y) -\int_x J_a(x)\widetilde{\varphi}^a(x) \;,
\end{equation}
\begin{equation}
S^1_{JK}\Big[\vec{\widetilde{\varphi}}\Big] = \frac{\lambda}{4!}\int_x\left(\widetilde{\varphi}_a(x)\widetilde{\varphi}^a(x)\right)^2 \;,
\end{equation}
with the free propagator of the field $\vec{\widetilde{\varphi}}$ in presence of the sources:
\begin{equation}
\boldsymbol{G}^{-1}_{0;K;ab}(x,y)=\left(-\nabla^2_x +m^2\right)\delta_{ab} \delta(x-y) - \boldsymbol{K}_{ab}(x,y) \;.
\label{eq:FreeG}
\end{equation}
In particular, such a $\lambda$-wise perturbative expansion breaks down in the broken-symmetry phase where the partition function of the non-interacting system diverges.

\vspace{0.5cm}

Let us briefly mention here another expansion method for many-body systems featuring some internal symmetry. When the dofs have $N$ components, scalar composite fields may exhibit small fluctuations in the large $N$ limit, thus providing us with a relevant leading order as a starting point for an efficient expansion (exploiting $1/N$ as expansion parameter). This yields the so-called large $N$ or $1/N$-expansions\footnote{See ref.~\cite{mos03} and references therein for a complete presentation of $1/N$-expansions.}~\cite{wil73}. The $1/N$-expansion of the zero-dimensional $O(N)$-symmetric $\varphi^4$ model is detailed in appendix~\ref{ann:LargeN}.

\vspace{0.5cm}

Another approach consists in implementing PT under the form of a loop expansion (LE), i.e. to organize the perturbative series by powers of the fluctuations of the field around its classical configuration. The main assumption behind the LE is that field configurations different from the field expectation value only give a small contribution to the functional integral. The LE will be carried out after considering three types of splitting of the classical action, involving the original, mixed and collective representations based on $S\big[\vec{\widetilde{\varphi}}\big]$, $S_{\text{mix}}\big[\vec{\widetilde{\varphi}},\widetilde{\sigma}\big]$ and $S_{\text{col}}[\widetilde{\sigma}]$ respectively, where correlations are differently shuffled between the unperturbed and residual parts.

\subsection{Splitting of the classical actions}

Within the LE framework, the splitting of each classical action $S\big[\vec{\widetilde{\varphi}}\big]$, $S_{\text{mix}}\big[\vec{\widetilde{\varphi}},\widetilde{\sigma}\big]$ and $S_{\text{col}}[\widetilde{\sigma}]$, added to the chosen source-dependent terms, between unperturbed and residual parts stems from a Taylor expansion around their saddle points.

\subsubsection{Original representation}

In the original theory with classical action in presence of the external sources~\eqref{eq:SJKfiniteD}, the saddle point $\vec{\varphi}_\text{cl}(x)$ satisfies:
\begin{equation}
\left.\frac{\delta S_{JK}\big[\vec{\widetilde{\varphi}}\big]}{\delta \vec{\widetilde{\varphi}}(x)}\right|_{\vec{\widetilde{\varphi}}=\vec{\varphi}_\text{cl}} = \left.\frac{\delta S\big[\vec{\widetilde{\varphi}}\big]}{\delta \vec{\widetilde{\varphi}}(x)}\right|_{\vec{\widetilde{\varphi}}=\vec{\varphi}_\text{cl}} - \vec{J}(x) - \int_y \boldsymbol{K}(x,y)\vec{\varphi}_\text{cl}(y) = \vec{0} \mathrlap{\quad \forall x \;.}
\label{eq:extrSJK}
\end{equation}
Setting\footnote{The $\sqrt{\hbar}$ factor is explicitly introduced such that $\vec{\widetilde{\chi}}\sim\mathcal{O}(1)$ and any dependence on the fluctuations around the saddle point translates into an appropriate power of the reduced Planck's constant $\hbar$.} $\vec{\widetilde{\varphi}} = \vec{\varphi}_\text{cl} + \sqrt{\hbar} \ \vec{\widetilde{\chi}}$, the Taylor expansion of $S_{JK}\big[\vec{\widetilde{\varphi}}\big]$ around $\vec{\widetilde{\varphi}}=\vec{\varphi}_\text{cl}$ leads to the following splitting:
\begin{equation}
S_{JK}\Big[\vec{\widetilde{\varphi}}=\vec{\varphi}_\text{cl} + \sqrt{\hbar} \ \vec{\widetilde{\chi}}\Big] = S_{\varphi_\text{cl};JK}^0\Big[\vec{\widetilde{\chi}}\Big] + S_{\varphi_\text{cl};JK}^1\Big[\vec{\widetilde{\chi}}\Big] \;,
\end{equation}
\begin{equation}
S_{\varphi_\text{cl};JK}^0\Big[\vec{\widetilde{\chi}}\Big] = S_{JK}[\vec{\varphi}_\text{cl}] + \frac{\hbar}{2}\int_{x,y}\widetilde{\chi}^a(x) \boldsymbol{G}^{-1}_{\varphi_\text{cl};JK;ab}(x,y) \widetilde{\chi}^b(y) \;,
\label{eq:LES1}
\end{equation}
\begin{equation}
S_{\varphi_\text{cl};JK}^1\Big[\vec{\widetilde{\chi}}\Big] = \frac{\hbar^{\frac{3}{2}}\lambda}{3!}\int_x \widetilde{\chi}^a(x)\widetilde{\chi}_a(x) \widetilde{\chi}^b(x)\varphi_{\text{cl};b}(x) + \frac{\hbar^{2}\lambda}{4!}\int_x \widetilde{\chi}^a(x)\widetilde{\chi}_a(x) \widetilde{\chi}^b(x)\widetilde{\chi}_b(x)\;,
\label{eq:LES2}
\end{equation}
with $\boldsymbol{G}_{\varphi_\text{cl};JK}^{-1}$ being the unperturbed inverse propagator in presence of the sources\footnote{The propagator $\boldsymbol{G}_{\varphi_\text{cl};JK}^{-1}$ implicitly depends on the source $\vec{J}$ through the saddle point $\vec{\varphi}_\text{cl}$.}:
\begin{equation}
\begin{split}
\boldsymbol{G}^{-1}_{\varphi_\text{cl};JK;ab}(x,y) \equiv & \ \left.\frac{\delta^{2} S_{JK}\big[\vec{\widetilde{\varphi}}\big]}{\delta \widetilde{\varphi}^{a}(x)\delta \widetilde{\varphi}^{b}(y)}\right|_{\vec{\widetilde{\varphi}}=\vec{\varphi}_\text{cl}} \\
= & \ \left(-\nabla_x^2 + m^2 + \frac{\lambda}{6} \varphi^c_\text{cl}(x)\varphi_{\text{cl};c}(x)\right)\delta_{ab} \delta(x-y)+ \frac{\lambda}{3}\varphi_{\text{cl};a}(x)\varphi_{\text{cl};b}(x)\delta(x-y) \\
& - \boldsymbol{K}_{ab}(x,y) \;.
\end{split}
\label{eq:LEG}
\end{equation}
The propagator~\eqref{eq:LEG} differs from that of~\eqref{eq:FreeG} involved  
in the $\lambda$-wise expansion as it is dressed by the classical solution $\vec{\varphi}_\text{cl}$. In the unbroken-symmetry phase, where $\vec{\varphi}_\text{cl} = \vec{0}$, the $\hbar$- and $\lambda$-expansions coincide. On the other hand, building the perturbative expansion on top of the symmetry-breaking saddle point allows for the exploration of the broken-symmetry phase within PT, contrary to the $\lambda$-expansion.

\subsubsection{Mixed representation}

The mixed representation can be based on generating functionals involving the local sources $\vec{J}(x)$, $j(x)$ as well as the bilocal ones $\boldsymbol{K}(x,y)$ and $k(x,y)$, associated with the original and collective fields respectively. This gives us e.g.:
\begin{equation}
Z_{\mathrm{mix}}\big[\mathcal{J},\mathcal{K}\big] = e^{\frac{1}{\hbar}W_{\mathrm{mix}}[\mathcal{J},\mathcal{K}]} = \int \mathcal{D}\vec{\widetilde{\varphi}} \mathcal{D}\widetilde{\sigma} \ e^{-\frac{1}{\hbar}S_{\mathrm{mix},\mathcal{J}\mathcal{K}}\big[\vec{\widetilde{\varphi}},\widetilde{\sigma}\big]}\;,
\label{eq:ZmixedmathcalJKLoopExpansion}
\end{equation}
where
\begin{equation}
S_{\text{mix},\mathcal{J}\mathcal{K}}\Big[\widetilde{\Psi}\Big] \equiv S_\text{mix}\Big[\widetilde{\Psi}\Big] - \int_{x} \mathcal{J}_{\alpha}(x) \widetilde{\Psi}^{\alpha}(x) - \frac{1}{2} \int_{x,y} \widetilde{\Psi}^{\alpha}(x) \mathcal{K}_{\alpha\beta}(x,y) \widetilde{\Psi}^{\beta}(y)\;,
\label{eq:SmixedJK}
\end{equation}
with $S_\text{mix}$ given by~\eqref{eq:Smix} and with superfields living in a ($N+1$)-dimensional extended color space aggregating the $N$ components of the original dofs and the collective one, i.e.:
\begin{equation}
\widetilde{\Psi}(x) \equiv \begin{pmatrix}
\vec{\widetilde{\varphi}}(x) \\
\widetilde{\sigma}(x)
\end{pmatrix} \;,
\label{eq:supernotationPsi}
\end{equation}
\begin{equation}
\mathcal{J}(x) \equiv \begin{pmatrix}
\vec{J}(x) \\
 j(x)
\end{pmatrix} \;,
\label{eq:supernotationJ}
\end{equation}
\begin{equation}
\mathcal{K}(x,y) \equiv \begin{pmatrix}
\boldsymbol{K}(x,y) & \vec{0} \\
\vec{0}^{\mathrm{T}} & k(x,y)
\end{pmatrix} \;.
\label{eq:supernotationK}
\end{equation}
In~\eqref{eq:SmixedJK} and hereafter, Greek indices run over the ($N+1$)-dimensional superspace while Latin indices run over the $N$-dimensional color space. The saddle point $\Psi_{\text{cl}}\equiv\begin{pmatrix}\vec{\varphi}_{\text{cl}} & \sigma_{\text{cl}}\end{pmatrix}^{\mathrm{T}}$ of $S_{\text{mix},\mathcal{J}\mathcal{K}}$ satisfies:
\begin{equation}
\left.\frac{\delta S_{\mathrm{mix},\mathcal{J}\mathcal{K}}\big[\widetilde{\Psi}\big]}{\delta \widetilde{\Psi}(x)}\right|_{\widetilde{\Psi}=\Psi_{\mathrm{cl}}} = 0 \mathrlap{\quad \forall x \;.}
\label{eq:minimizationSmixedKj}
\end{equation}
Setting $\widetilde{\Psi} = \Psi_\text{cl} + \sqrt{\hbar} \ \widetilde{\Xi}$ with $\widetilde{\Xi}\equiv\begin{pmatrix}\vec{\widetilde{\chi}} & \widetilde{\zeta}\end{pmatrix}^{\mathrm{T}}$, the Taylor expansion of $S_{\text{mix},\mathcal{J}\mathcal{K}}\big[\widetilde{\Psi}\big]$ around $\widetilde{\Psi}=\Psi_{\mathrm{cl}}$ leads to the splitting:
\begin{equation}
S_{\text{mix},\mathcal{J}\mathcal{K}}\Big[\widetilde{\Psi}= \Psi_\text{cl} + \sqrt{\hbar}\ \widetilde{\Xi}\Big] = S^0_{\mathrm{mix},\Psi_\text{cl};\mathcal{J}\mathcal{K}}\Big[\widetilde{\Xi}\Big]+S^1_{\mathrm{mix},\Psi_\text{cl};\mathcal{J}\mathcal{K}}\Big[\widetilde{\Xi}\Big] \;,
\end{equation}
\begin{equation}
S^0_{\mathrm{mix},\Psi_\text{cl};\mathcal{J}\mathcal{K}}\Big[\widetilde{\Xi}\Big] = S_{\mathrm{mix},\mathcal{J}\mathcal{K}}[\Psi_{\mathrm{cl}}]+ \frac{\hbar}{2}\int_{x,y}  \widetilde{\Xi}^\alpha (x)  \mathcal{G}^{-1}_{\Psi_\text{cl};\mathcal{JK};\alpha\beta}(x,y) \widetilde{\Xi}^\beta (y) \;,
\end{equation}
\begin{equation}
S^1_{\mathrm{mix},\Psi_\text{cl};\mathcal{J}\mathcal{K}}\left[\widetilde{\Xi}\right] = i\hbar^{\frac{3}{2}} \sqrt{\frac{\lambda}{12}} \int_{x} \widetilde{\zeta}(x) \widetilde{\chi}^a (x)\widetilde{\chi}_a (x) \;,
\end{equation}
where the unperturbed inverse propagator in presence of the sources reads:
\begin{equation}
\begin{split}
\mathcal{G}^{-1}_{\Psi_\text{cl};\mathcal{JK}}(x,y) \equiv & \ \left.\frac{\delta^2 S_{\text{mix},\mathcal{J}\mathcal{K}}\big[\widetilde{\Psi}\big]}{\delta \widetilde{\Psi}(x)\delta \widetilde{\Psi}(y)}\right|_{\widetilde{\Psi} = \Psi_\text{cl}} \\
= & \ \begin{pmatrix}
\left(-\nabla_x^2 + m^2 + i\sqrt{\frac{\lambda}{3}}\sigma_\text{cl}(x)\right)\mathbb{I}_{N} & i\sqrt{\frac{\lambda}{3}}\vec{\varphi}_\text{cl}(x) \\
i\sqrt{\frac{\lambda}{3}}\vec{\varphi}^\mathrm{T}_\text{cl}(x) & 1 \end{pmatrix}\delta(x-y)- \mathcal{K}(x,y) \;,
\end{split}
\label{eq:SuperSuperPropMixedLEmatrixmathcalK}
\end{equation}
with $\mathbb{I}_{D}$ being the $D$-dimensional identity matrix. As compared to the original representation, the classical configuration of the collective field $\sigma_\text{cl}$ dresses the propagator of the original dofs:
\begin{equation}
\boldsymbol{G}_{\sigma_\text{cl};\mathcal{JK};ab}^{-1}(x,y) =\left(-\nabla_x^2 + m^2 + i\sqrt{\frac{\lambda}{3}}\sigma_\text{cl}(x)\right)\delta_{ab}\delta(x-y)-\boldsymbol{K}_{ab}(x,y) \;.
\label{eq:mixLEG}
\end{equation}
From the equation of motion~\eqref{eq:minimizationSmixedKj} at vanishing sources, we show that the saddle points $\vec{\varphi}_\text{cl}$ and $\sigma_\text{cl}$ are related via:
\begin{equation}
\overline{\sigma}_\text{cl}(x) = -i\sqrt{\frac{\lambda}{12}}\vec{\overline{\varphi}}_\text{cl}(x)\cdot\vec{\overline{\varphi}}_\text{cl}(x) \;,
\label{eq:sigclmix}
\end{equation}
where $\vec{\overline{\varphi}}_\text{cl}$ and $\overline{\sigma}_\text{cl}$ denote respectively the configurations of $\vec{\varphi}_\text{cl}$ and $\sigma_\text{cl}$ when all sources are set equal to zero. One therefore recovers from~\eqref{eq:mixLEG} the inverse propagator in the original representation~\eqref{eq:LEG}, up to a term which is accounted for in off-diagonal parts of the superpropagator in the mixed representation. In other words, as long as one works within PT organized in powers of the \textbf{unperturbed} propagators, the same physics is encoded in the unperturbed sector of both the original and mixed representations, and it remains to be seen whether the incorporation of the residual physics is more efficient in one of these two representations.

\subsubsection{Collective representation}
\label{sec:DefCollectiveRepr}

Finally, the collective representation relies on the partition function~\eqref{eq:Zcoll} which becomes, after introducing the local source $\mathcal{J}(x)=\begin{pmatrix}
\vec{J}(x) & j(x)\end{pmatrix}^\mathrm{T}$:
\begin{equation}
Z_{\mathrm{col}}\big[\mathcal{J}\big] = e^{\frac{1}{\hbar}W_{\mathrm{col}}[\mathcal{J}]} = \int \mathcal{D}\widetilde{\sigma} \ e^{-\frac{1}{\hbar}S_{\mathrm{col},\mathcal{J}}[\widetilde{\sigma}]}\;,
\label{eq:ZbosonicKLoopExpansion}
\end{equation}
with
\begin{equation}
S_{\mathrm{col},\mathcal{J}}[\widetilde{\sigma}] = S_{\mathrm{col}}[\widetilde{\sigma}] - \int_{x} j(x) \widetilde{\sigma}(x)- \frac{1}{2} \int_{x,y} J^a(x) \boldsymbol{G}_{\widetilde{\sigma};ab}(x,y) J^b(y)\;,
\label{eq:SbosonicKLoopExpansion}
\end{equation}
where $S_{\mathrm{col}}$ and $\boldsymbol{G}_{\widetilde{\sigma}}$ are given by~\eqref{eq:Scoll} and~\eqref{eq:Gcoll} respectively. The saddle point $\sigma_\text{cl}$ of $S_{\mathrm{col},\mathcal{J}}$ satisfies:
\begin{equation}
\left.\frac{\delta S_{\mathrm{col},\mathcal{J}}[\widetilde{\sigma}]}{\delta \widetilde{\sigma}(x)}\right|_{\widetilde{\sigma}=\sigma_{\mathrm{cl}}} = 0 \mathrlap{\quad \forall x \;.}
\label{eq:minimizationSbosonicK}
\end{equation}
Setting $\widetilde{\sigma} = \sigma_\text{cl}+\sqrt{\hbar} \ \widetilde{\zeta}$, a Taylor expansion of $S_{\mathrm{col},\mathcal{J}}$ around $\widetilde{\sigma}=\sigma_\text{cl}$ leads to the splitting:
\begin{equation}
S_{\mathrm{col},\mathcal{J}}\Big[\widetilde{\sigma} = \sigma_\text{cl}+\sqrt{\hbar} \ \widetilde{\zeta}\Big] = S^0_{\mathrm{col},\sigma_\text{cl};\mathcal{J}}\Big[\widetilde{\zeta}\Big]+ S^1_{\mathrm{col},\sigma_\text{cl};\mathcal{J}}\Big[\widetilde{\zeta}\Big] \;,
\end{equation}
\begin{equation}
S^0_{\mathrm{col},\sigma_\text{cl};\mathcal{J}}\Big[\widetilde{\zeta}\Big] = S_{\mathrm{col},\mathcal{J}}[\sigma_{\mathrm{cl}}] + \frac{\hbar}{2} \int_{x,y} \widetilde{\zeta}(x) D^{-1}_{\sigma_\text{cl};\mathcal{J}}(x,y) \widetilde{\zeta}(y) \;,
\end{equation}
\begin{equation}
S^1_{\mathrm{col},\sigma_\text{cl};\mathcal{J}}\Big[\widetilde{\zeta}\Big] = \sum_{n=3}^{\infty} \frac{\hbar^{\frac{n}{2}}}{n!} \int_{x_{1},\cdots, x_{n}} S_{\mathrm{col}, \mathcal{J}}^{(n)}(x_{1}, \cdots ,x_{n}) \widetilde{\zeta}(x_{1}) \cdots \widetilde{\zeta}(x_{n}) \;,
\end{equation}
with
\begin{equation}
S_{\mathrm{col}, \mathcal{J}}^{(n)}(x_{1},\cdots,x_{n}) \equiv \left.\frac{\delta^n S_{\mathrm{col},\mathcal{J}}[\widetilde{\sigma}]}{\delta \widetilde{\sigma}(x_1)\cdots\delta \widetilde{\sigma}(x_n)}\right|_{\widetilde{\sigma}=\sigma_{\mathrm{cl}}} \;,
\end{equation}
and $D_{\sigma_\text{cl};\mathcal{J}}(x,y)$ being the propagator of the collective field in presence of the sources, i.e.:
\begin{equation}
D^{-1}_{\sigma_\text{cl};\mathcal{J}}(x,y) = S_{\mathrm{col}, \mathcal{J}}^{(2)}(x,y) \;.
\label{eq:DefDpropagCollectiveLE}
\end{equation}
The unperturbed and residual channels both involve the propagator of the original field $\vec{\widetilde{\varphi}}(x)$:
\begin{equation}
\boldsymbol{G}^{-1}_{\sigma_\text{cl};\mathcal{J};ab}(x,y) = \left(-\nabla_x^2 + m^2 + i\sqrt{\frac{\lambda}{3}}\sigma_\text{cl}(x)\right)\delta_{ab}\delta(x-y) \;,
\label{eq:DefGpropagCollectiveLE}
\end{equation} 
which coincides, in the limit where all sources vanish, with the mixed representation one~\eqref{eq:mixLEG}, with however a different configuration for the saddle point $\sigma_\text{cl}$ renormalizing the mass. We now proceed to the perturbative expansion based on each of the three aforementioned splittings.

\subsection{Loop expansions}

In each representation of the system, the LE is implemented after inserting the corresponding action split into unperturbed and residual parts in the partition function $Z$, and then Taylor expanding the exponential of $S^{1}$.

\subsubsection{Original loop expansion}
\label{sec:OriginalLE}

In the framework of the original theory, the $\hbar$-expansion for the partition function reads:
\begin{equation}
\begin{split}
\scalebox{0.867}{${\displaystyle Z^\text{LE;orig}\Big[\vec{J},\boldsymbol{K}\Big] = }$} & \ \scalebox{0.867}{${\displaystyle e^{-\frac{1}{\hbar}S_{JK}[\vec{\varphi}_\text{cl}]} \left(\int \mathcal{D}\vec{\widetilde{\chi}} \ e^{-\frac{1}{2}\int_{x,y} \vec{\widetilde{\chi}}(x) \cdot \left(\boldsymbol{G}^{-1}_{\varphi_\text{cl};JK}(x,y)\vec{\widetilde{\chi}}(y)\right)}\right) }$} \\ 
&\scalebox{0.867}{${\displaystyle\times\left[ 1 + \sum^{\infty}_{n=1} \frac{\left(-1\right)^n}{\left(3!\right)^n n!} \sum_{q=0}^n \begin{pmatrix}
n \\
q
\end{pmatrix} \frac{\hbar^\frac{n+q}{2}}{4^q} \left\langle \left(\lambda\int_x \vec{\widetilde{\chi}}^2(x)\vec{\widetilde{\chi}}(x)\cdot\vec{\varphi}_\text{cl}(x) \right)^{n-q}\left(\lambda \int_x \left(\vec{\widetilde{\chi}}^2(x) \right)^2 \right)^{q}\right\rangle_{0,JK}\right] \;,}$}
\end{split}
\label{eq:ZJKPT3}
\end{equation}
with the source-dependent expectation value defined as:
\begin{equation}
\big\langle\cdots\big\rangle_{0,JK} = \frac{1}{Z_{0}\big[\vec{J},\boldsymbol{K}\big]} \int \mathcal{D}\vec{\widetilde{\chi}} \ \cdots \ e^{-\frac{1}{\hbar}S_{\varphi_\text{cl};JK}^0\big[\vec{\widetilde{\chi}}\big]} \;,
\label{eq:SourceDepExpValueOriginalLE}
\end{equation}
and
\begin{equation}
Z_{0}\Big[\vec{J},\boldsymbol{K}\Big]=\int \mathcal{D}\vec{\widetilde{\chi}} \ e^{-\frac{1}{\hbar}S_{\varphi_\text{cl};JK}^0\big[\vec{\widetilde{\chi}}\big]} \;.
\label{eq:Z0JKoriginalLE}
\end{equation}
The $\hbar$-expansion of the Schwinger functional $W^\text{LE;orig}\big[\vec{J},\boldsymbol{K}\big] \equiv \hbar \ln\Big( Z^\text{LE;orig}\big[\vec{J},\boldsymbol{K}\big] \Big)$ derives from~\eqref{eq:ZJKPT3} together with the linked-cluster theorem~\cite{neg98}, by virtue of which one can substitute the correlation functions in~\eqref{eq:ZJKPT3} by their connected counterparts, thus leading to:
\begin{equation}
\begin{split}
\scalebox{0.905}{${\displaystyle W^\text{LE;orig}\Big[\vec{J},\boldsymbol{K}\Big] = }$} & \scalebox{0.905}{${\displaystyle -S_{JK}\big[\vec{\varphi}_\text{cl}\big] +\frac{\hbar}{2} \mathrm{STr}\left[\ln\big(\boldsymbol{G}_{\varphi_\text{cl};JK}\big) \right] }$} \\
& \scalebox{0.905}{${\displaystyle + \sum^{\infty}_{n=1} \frac{\left(-1\right)^n}{\left(3!\right)^n n!}\sum_{q=0}^n 
\begin{pmatrix}
n \\
q
\end{pmatrix} \frac{\hbar^\frac{n+q+2}{2}}{4^q}\left\langle \left(\lambda\int_x \vec{\widetilde{\chi}}^2(x)\vec{\widetilde{\chi}}(x)\cdot\vec{\varphi}_\text{cl}(x) \right)^{n-q}\left(\lambda \int_x \left( \vec{\widetilde{\chi}}^2(x) \right)^2 \right)^{q}\right\rangle^\text{c}_{0,JK} \;, }$}
\end{split}
\label{eq:WJKPT}
\end{equation}
where the term with the supertrace $\mathrm{STr}$ stems from Gaussian integration (see appendix~\ref{ann:GaussianInt}):
\begin{equation}
\int \mathcal{D}\vec{\widetilde{\chi}} \ e^{-\frac{1}{2}\int_{x,y} \vec{\widetilde{\chi}}(x) \cdot \big(\boldsymbol{G}^{-1}_{\varphi_\text{cl};JK}(x,y) \vec{\widetilde{\chi}}(y)\big)} = e^{\frac{1}{2}\mathrm{STr}\left[\ln(\boldsymbol{G}_{\varphi_\text{cl};JK})\right]} \;.
\label{eq:GaussianFormula1}
\end{equation}
Denoting the modulus of the classical solution $\vec{\varphi}_\text{cl}$ as $\varrho(x)\equiv\left|\vec{\varphi}_\text{cl}(x)\right|$, we can choose $a=N$ as the direction along which the SSB occurs in the broken-symmetry phase without any loss of generality, i.e.:
\begin{equation}
\vec{\varphi}_\text{cl}(x) = \varrho(x)\begin{pmatrix}
0 \\
\vdots\\
0 \\
1
\end{pmatrix} \;.
\label{eq:DefPhiclModulusRho}
\end{equation}
The propagator~\eqref{eq:LEG} can then be separated into the one associated to the $O(N-1)$ subspace (the Goldstone manifold in the broken-symmetry phase when $N\geq 2$), namely:
\begin{equation}
G^{-1}_{\varphi_\text{cl};JK;\mathfrak{g};ab}(x,y) = \left(-\nabla_x^2 + m^2 + \frac{\lambda}{6} \varrho^2(x)\right)\delta_{ab}\delta(x-y) - \boldsymbol{K}_{ab}(x,y) \quad \forall a,b\in [1,N-1] \;,  
\label{eq:LEGG}
\end{equation}
and the one of the remaining massive (or Higgs) mode for $a=N$, i.e.:
\begin{equation}
\boldsymbol{G}^{-1}_{\varphi_\text{cl};JK;NN}(x,y) = \left(-\nabla_x^2 + m^2 + \frac{\lambda}{2} \varrho^2(x)\right)\delta(x-y) - \boldsymbol{K}_{NN}(x,y) \;.
\label{eq:LEGN}
\end{equation}
The connected correlation functions in~\eqref{eq:WJKPT} can therefore be rewritten as:
\begin{equation}
\left\langle \left(\lambda\int_x \vec{\widetilde{\chi}}^2(x) \widetilde{\chi}_N(x) \varrho(x) \right)^{n-q}\left(\lambda \int_x \left(\vec{\widetilde{\chi}}^2(x) \right)^2 \right)^{q}\right\rangle^\text{c}_{0,JK} \;,
\end{equation}
and evaluated by means of Wick's theorem (as presented in section~\ref{sec:WickTheorem}), conveniently represented by a set of Feynman diagrams with the rules:
\begin{subequations}
\begin{align}
\begin{gathered}
\begin{fmffile}{Diagrams/LoopExpansion1_FeynRuleGbis}
\begin{fmfgraph*}(20,20)
\fmfleft{i0,i1,i2,i3}
\fmfright{o0,o1,o2,o3}
\fmflabel{$x, a$}{v1}
\fmflabel{$y, b$}{v2}
\fmf{phantom}{i1,v1}
\fmf{phantom}{i2,v1}
\fmf{plain,tension=0.6}{v1,v2}
\fmf{phantom}{v2,o1}
\fmf{phantom}{v2,o2}
\end{fmfgraph*}
\end{fmffile}
\end{gathered} \quad &\rightarrow \boldsymbol{G}_{\varphi_\text{cl};JK;ab}(x,y)\;,
\label{eq:FeynRulesLoopExpansionPropagator} \\
\begin{gathered}
\begin{fmffile}{Diagrams/LoopExpansion1_FeynRuleV3bis}
\begin{fmfgraph*}(20,20)
\fmfleft{i0,i1,i2,i3}
\fmfright{o0,o1,o2,o3}
\fmfv{decor.shape=cross,decor.angle=45,decor.size=3.5thick,foreground=(0,,0,,1)}{o2}
\fmf{phantom,tension=2.0}{i1,i1bis}
\fmf{plain,tension=2.0}{i1bis,v1}
\fmf{phantom,tension=2.0}{i2,i2bis}
\fmf{plain,tension=2.0}{i2bis,v1}
\fmf{dots,label=$x$,tension=0.6,foreground=(0,,0,,1)}{v1,v2}
\fmf{phantom,tension=2.0}{o1bis,o1}
\fmf{plain,tension=2.0}{v2,o1bis}
\fmf{phantom,tension=2.0}{o2bis,o2}
\fmf{phantom,tension=2.0}{v2,o2bis}
\fmf{dashes,tension=0.0,foreground=(0,,0,,1)}{v2,o2}
\fmflabel{$a$}{i1bis}
\fmflabel{$b$}{i2bis}
\fmflabel{$c$}{o1bis}
\fmflabel{$N$}{o2bis}
\end{fmfgraph*}
\end{fmffile}
\end{gathered} \quad &\rightarrow \lambda\varrho(x)\delta_{a b}\delta_{c N}\;,
\label{eq:FeynRulesLoopExpansion3legVertex} \\
\begin{gathered}
\begin{fmffile}{Diagrams/LoopExpansion1_FeynRuleV4bis}
\begin{fmfgraph*}(20,20)
\fmfleft{i0,i1,i2,i3}
\fmfright{o0,o1,o2,o3}
\fmf{phantom,tension=2.0}{i1,i1bis}
\fmf{plain,tension=2.0}{i1bis,v1}
\fmf{phantom,tension=2.0}{i2,i2bis}
\fmf{plain,tension=2.0}{i2bis,v1}
\fmf{zigzag,label=$x$,tension=0.6,foreground=(0,,0,,1)}{v1,v2}
\fmf{phantom,tension=2.0}{o1bis,o1}
\fmf{plain,tension=2.0}{v2,o1bis}
\fmf{phantom,tension=2.0}{o2bis,o2}
\fmf{plain,tension=2.0}{v2,o2bis}
\fmflabel{$a$}{i1bis}
\fmflabel{$b$}{i2bis}
\fmflabel{$c$}{o1bis}
\fmflabel{$d$}{o2bis}
\end{fmfgraph*}
\end{fmffile}
\end{gathered} \quad &\rightarrow \lambda\delta_{a b}\delta_{c d}\;.
\label{eq:FeynRulesLoopExpansion4legVertex}
\end{align}
\end{subequations}
Up to order $\mathcal{O}\big(\hbar^2\big)$\footnote{One might also refer to order $\mathcal{O}\big(\hbar^2\big)$ as the 2-loop order or 2-loop level. Indeed, for LEs of Schwinger functionals or diagrammatic expansions of EAs, the powers of $\hbar$ count the number of loops in the corresponding diagrams. In other words, $n$-loop order is synonymous with order $\mathcal{O}\big(\hbar^{n}\big)$ in the language used here.}, the perturbative series thus obtained for the Schwinger functional $W\big[\vec{J},\boldsymbol{K}\big]$ reads:
\begin{equation}
\begin{split}
W^\text{LE;orig}\Big[\vec{J},\boldsymbol{K}\Big] = & -S_{JK}\big[\vec{\varphi}_{\mathrm{cl}}\big] + \frac{\hbar}{2} \mathrm{STr}\left[\ln\big(\boldsymbol{G}_{\varphi_\text{cl};JK}\big)\right] \\
& + \hbar^{2} \left(\rule{0cm}{1.2cm}\right. -\frac{1}{24} \hspace{0.08cm} \begin{gathered}
\begin{fmffile}{Diagrams/LoopExpansion1_Hartree}
\begin{fmfgraph}(30,20)
\fmfleft{i}
\fmfright{o}
\fmf{phantom,tension=10}{i,i1}
\fmf{phantom,tension=10}{o,o1}
\fmf{plain,left,tension=0.5}{i1,v1,i1}
\fmf{plain,right,tension=0.5}{o1,v2,o1}
\fmf{zigzag,foreground=(0,,0,,1)}{v1,v2}
\end{fmfgraph}
\end{fmffile}
\end{gathered}
-\frac{1}{12}\begin{gathered}
\begin{fmffile}{Diagrams/LoopExpansion1_Fock}
\begin{fmfgraph}(15,15)
\fmfleft{i}
\fmfright{o}
\fmf{phantom,tension=11}{i,v1}
\fmf{phantom,tension=11}{v2,o}
\fmf{plain,left,tension=0.4}{v1,v2,v1}
\fmf{zigzag,foreground=(0,,0,,1)}{v1,v2}
\end{fmfgraph}
\end{fmffile}
\end{gathered} + \frac{1}{18} \ \ \begin{gathered}
\begin{fmffile}{Diagrams/LoopExpansion1_Diag1}
\begin{fmfgraph}(34,20)
\fmfleft{i}
\fmfright{o}
\fmfv{decor.shape=cross,decor.size=3.5thick,foreground=(0,,0,,1)}{i}
\fmfv{decor.shape=cross,decor.size=3.5thick,foreground=(0,,0,,1)}{o}
\fmf{dashes,tension=2.0,foreground=(0,,0,,1)}{i,v3}
\fmf{dashes,tension=2.0,foreground=(0,,0,,1)}{o,v4}
\fmf{plain,right,tension=0.7}{v2,v4}
\fmf{dots,left,tension=0.7,foreground=(0,,0,,1)}{v2,v4}
\fmf{plain,left,tension=0.7}{v1,v3}
\fmf{dots,right,tension=0.7,foreground=(0,,0,,1)}{v1,v3}
\fmf{plain,tension=1.5}{v1,v2}
\end{fmfgraph}
\end{fmffile}
\end{gathered} \\
& \hspace{1.35cm} + \frac{1}{18} \begin{gathered}
\begin{fmffile}{Diagrams/LoopExpansion1_Diag2}
\begin{fmfgraph}(27,15)
\fmfleft{i}
\fmfright{o}
\fmftop{vUp}
\fmfbottom{vDown}
\fmfv{decor.shape=cross,decor.size=3.5thick,foreground=(0,,0,,1)}{v1}
\fmfv{decor.shape=cross,decor.size=3.5thick,foreground=(0,,0,,1)}{v2}
\fmf{phantom,tension=10}{i,i1}
\fmf{phantom,tension=10}{o,o1}
\fmf{phantom,tension=2.2}{vUp,v5}
\fmf{phantom,tension=2.2}{vDown,v6}
\fmf{phantom,tension=0.5}{v3,v4}
\fmf{phantom,tension=10.0}{i1,v1}
\fmf{phantom,tension=10.0}{o1,v2}
\fmf{dashes,tension=2.0,foreground=(0,,0,,1)}{v1,v3}
\fmf{dots,left=0.4,tension=0.5,foreground=(0,,0,,1)}{v3,v5}
\fmf{plain,left=0.4,tension=0.5}{v5,v4}
\fmf{plain,right=0.4,tension=0.5}{v3,v6}
\fmf{dots,right=0.4,tension=0.5,foreground=(0,,0,,1)}{v6,v4}
\fmf{dashes,tension=2.0,foreground=(0,,0,,1)}{v4,v2}
\fmf{plain,tension=0}{v5,v6}
\end{fmfgraph}
\end{fmffile}
\end{gathered} + \frac{1}{36} \hspace{-0.15cm} \begin{gathered}
\begin{fmffile}{Diagrams/LoopExpansion1_Diag3}
\begin{fmfgraph}(25,20)
\fmfleft{i}
\fmfright{o}
\fmftop{vUp}
\fmfbottom{vDown}
\fmfv{decor.shape=cross,decor.angle=45,decor.size=3.5thick,foreground=(0,,0,,1)}{vUpbis}
\fmfv{decor.shape=cross,decor.angle=45,decor.size=3.5thick,foreground=(0,,0,,1)}{vDownbis}
\fmf{phantom,tension=0.8}{vUp,vUpbis}
\fmf{phantom,tension=0.8}{vDown,vDownbis}
\fmf{dashes,tension=0.5,foreground=(0,,0,,1)}{v3,vUpbis}
\fmf{phantom,tension=0.5}{v4,vUpbis}
\fmf{phantom,tension=0.5}{v3,vDownbis}
\fmf{dashes,tension=0.5,foreground=(0,,0,,1)}{v4,vDownbis}
\fmf{phantom,tension=11}{i,v1}
\fmf{phantom,tension=11}{v2,o}
\fmf{plain,left,tension=0.5}{v1,v2,v1}
\fmf{dots,tension=1.7,foreground=(0,,0,,1)}{v1,v3}
\fmf{plain}{v3,v4}
\fmf{dots,tension=1.7,foreground=(0,,0,,1)}{v4,v2}
\end{fmfgraph}
\end{fmffile}
\end{gathered} +\frac{1}{18} \ \begin{gathered}
\begin{fmffile}{Diagrams/LoopExpansion1_Diag4}
\begin{fmfgraph}(35,18)
\fmfleft{i}
\fmfright{o}
\fmftop{vUp}
\fmfbottom{vDown}
\fmfv{decor.shape=cross,decor.size=3.5thick,foreground=(0,,0,,1)}{v3bis}
\fmfv{decor.shape=cross,decor.size=3.5thick,foreground=(0,,0,,1)}{o}
\fmf{phantom,tension=10}{i,i1}
\fmf{dashes,tension=1.2,foreground=(0,,0,,1)}{o,v4}
\fmf{phantom,tension=0.5}{v3bis,i}
\fmf{phantom,tension=2.7}{v3bis,vUp}
\fmf{dashes,tension=0.9,foreground=(0,,0,,1)}{v3,v3bis}
\fmf{phantom,tension=0.5}{v4bis,i}
\fmf{phantom,tension=2.7}{v4bis,vDown}
\fmf{phantom,tension=0.9}{v3,v4bis}
\fmf{plain,left,tension=0.5}{i1,v1,i1}
\fmf{plain,right,tension=0.5}{v2,v4}
\fmf{dots,left,tension=0.5,foreground=(0,,0,,1)}{v2,v4}
\fmf{dots,foreground=(0,,0,,1)}{v1,v3}
\fmf{plain}{v3,v2}
\end{fmfgraph}
\end{fmffile}
\end{gathered} \\
& \hspace{1.35cm} +\frac{1}{72} \ \begin{gathered}
\begin{fmffile}{Diagrams/LoopExpansion1_Diag5}
\begin{fmfgraph}(40,18)
\fmfleft{i}
\fmfright{o}
\fmftop{vUp}
\fmfbottom{vDown}
\fmf{phantom,tension=1.0}{vUp,vUpbis}
\fmf{phantom,tension=1.0}{vDown,vDownbis}
\fmf{dashes,tension=0.5,foreground=(0,,0,,1)}{v3,vUpbis}
\fmf{phantom,tension=0.5}{v4,vUpbis}
\fmf{phantom,tension=0.5}{v3,vDownbis}
\fmf{dashes,tension=0.5,foreground=(0,,0,,1)}{v4,vDownbis}
\fmfv{decor.shape=cross,decor.angle=45,decor.size=3.5thick,foreground=(0,,0,,1)}{vUpbis}
\fmfv{decor.shape=cross,decor.angle=45,decor.size=3.5thick,foreground=(0,,0,,1)}{vDownbis}
\fmf{phantom,tension=10}{i,i1}
\fmf{phantom,tension=10}{o,o1}
\fmf{plain,left,tension=0.5}{i1,v1,i1}
\fmf{plain,right,tension=0.5}{o1,v2,o1}
\fmf{dots,tension=1.2,foreground=(0,,0,,1)}{v1,v3}
\fmf{plain,tension=0.6}{v3,v4}
\fmf{dots,tension=1.2,foreground=(0,,0,,1)}{v4,v2}
\end{fmfgraph}
\end{fmffile}
\end{gathered} \left.\rule{0cm}{1.2cm}\right) \\
& + \mathcal{O}\Big(\hbar^3\Big)\;.
\end{split}
\label{eq:WKjLoopExpansionStep3}
\end{equation}
Further details on the evaluation of the diagrams contributing to the latter diagrammatic series are given in appendix~\ref{sec:DiagLEO}.

\vspace{0.5cm}

In the (0+0)-D limit, the perturbative expansion of the Schwinger functional given by~\eqref{eq:WKjLoopExpansionStep3} reduces to:
\begin{equation}
\begin{split}
W^\text{LE;orig}\Big(\vec{J},\boldsymbol{K}\Big) = & -S_{JK}\big(\vec{\varphi}_{\mathrm{cl}}\big) + \frac{\hbar}{2}\left[(N-1)\ln\big(2\pi \mathfrak{G}_{\varphi_\text{cl};JK;\mathfrak{g}}\big)+\ln\big(2\pi \boldsymbol{G}_{\varphi_\text{cl};JK;NN}\big)\right] \\
& + \frac{\hbar^{2}\lambda}{72} \Big[ -3 \mathfrak{G}_{\varphi_\text{cl};JK;\mathfrak{g}}^2 \left(-1 + N^2\right) + 15 \boldsymbol{G}_{\varphi_\text{cl};JK;NN}^3 \lambda \varrho^2 \\
& \hspace{1.43cm} + \boldsymbol{G}_{\varphi_\text{cl};JK;NN}^2 \left(-9 + 6 \mathfrak{G}_{\varphi_\text{cl};JK;\mathfrak{g}} \left(-1 + N\right) \lambda \varrho^2\right) \\
& \hspace{1.43cm} + \boldsymbol{G}_{\varphi_\text{cl};JK;NN} \mathfrak{G}_{\varphi_\text{cl};JK;\mathfrak{g}} \left(-1 + N\right) \left(-6 + \mathfrak{G}_{\varphi_\text{cl};JK;\mathfrak{g}} \left(1 + N\right) \lambda \varrho^2\right) \Big] \\
&+ \mathcal{O}\big(\hbar^3\big)\;,
\end{split}
\label{eq:Worig}
\end{equation}
where we have assumed that $\boldsymbol{K}_{ab}=K\delta_{ab}$ so that~\eqref{eq:LEGG} becomes:
\begin{equation}
G^{-1}_{\varphi_\text{cl};JK;\mathfrak{g};ab} = \mathfrak{G}^{-1}_{\varphi_\text{cl};JK;\mathfrak{g}} \delta_{ab} \mathrlap{\quad \forall a,b\in [1,N-1] \;,}
\label{eq:origLEGoldstoneProp}
\end{equation}
with
\begin{equation}
\mathfrak{G}^{-1}_{\varphi_\text{cl};JK;\mathfrak{g}} = m^2 + \frac{\lambda}{6} \varrho^2 - K \mathrlap{\;.}
\label{eq:origLEGoldstonePropBis}
\end{equation}
When $\vec{J}=\vec{0}$ and $\boldsymbol{K}=\boldsymbol{0}$, the modulus $\varrho$ satisfies:
\begin{equation}
\overline{\varrho}^2 \equiv \varrho^2 \Big(\vec{J}=\vec{0},\boldsymbol{K}=\boldsymbol{0}\Big) = \begin{cases} \displaystyle{0} \quad \forall m^2\geq 0 \;, \\
\\
\displaystyle{-\frac{6m^2}{\lambda}} \quad \forall m^2< 0 ~ \text{and} ~ \lambda\neq 0 \;,  \end{cases}
\label{eq:SolutionsmodulusrhoLE}
\end{equation}
which yields the following expressions for $\mathfrak{G}_{\varphi_\text{cl};JK;\mathfrak{g}}$ and $\boldsymbol{G}_{\varphi_\text{cl};JK;NN}$ at vanishing sources:
\begin{equation}
\hspace{2.5cm} G_{\varphi_\text{cl};\mathfrak{g}} = \mathfrak{G}_{\varphi_\text{cl};\mathfrak{g}} \mathbb{I}_{N-1} = \begin{cases} \displaystyle{G_0 \ \mathbb{I}_{N-1} = \frac{1}{m^2} \ \mathbb{I}_{N-1} \quad \forall m^2> 0 \;,} \\
\\
\displaystyle{\infty \ \mathbb{I}_{N-1} \quad \forall m^2\leq 0 ~ \text{and} ~ \lambda\neq 0 \;,}  \end{cases} \qquad \forall N\geq 2 \;,
\end{equation}
\begin{equation}
\boldsymbol{G}_{\varphi_\text{cl},NN} = \begin{cases} \displaystyle{G_0 = \frac{1}{m^2} \quad \forall m^2\geq 0 \;,} \\
\\
\displaystyle{-\frac{1}{2m^2} \quad \forall m^2< 0 ~ \text{and} ~ \lambda\neq 0 \;,} \end{cases} \mathrlap{\qquad\forall N\geq 1 \;.}
\label{eq:origLEmassiveProp}
\end{equation}
From~\eqref{eq:DefEgsExactZexact0DON} and~\eqref{eq:DefrhogsExactwithExpectationValue0DON}, one can then obtain the corresponding series for the gs energy and density. Setting $x\equiv \hbar\lambda/m^4$ (thus showing that our loop-wise expansion amounts to an expansion in powers of the coupling constant $\lambda$), we get in the regime with $m^{2}>0$ and for $N\in\mathbb{N}^{*}$:
\begin{equation}
\begin{split}
E_\text{gs}^\text{LE;orig} = & -\frac{N}{2} \ln\bigg(\frac{2\pi}{m^{2}}\bigg) +  \frac{N\left(2+N\right)}{24}x- \frac{N \left(6 + 5 N + N^{2}\right)}{144}x^2 \\
& + \frac{N \left(120 + 128 N + 44 N^{2} + 5 N^{3}\right)}{2592}x^3 + \mathcal{O}\big(x^4\big)\;,
\end{split}
\label{eq:EgsPTpos}
\end{equation}
and
\begin{equation}
\rho_\mathrm{gs}^\text{LE;orig}=\frac{\hbar}{m^{2}}\left( 1 - \frac{2+N}{6}x+\frac{6 + 5 N + N^{2}}{18}x^2 - \frac{120 + 128 N + 44 N^{2} + 5 N^{3}}{216}x^3 + \mathcal{O}\big(x^4\big)\right)\;.
\end{equation}
In the regime with $m^{2}<0$, we have for $N=1$:
\begin{equation}
E^\text{LE;orig}_\mathrm{gs} = -\frac{3}{2 x} - \frac{1}{2} \ln\bigg(\frac{\pi}{-m^{2}}\bigg)-\frac{x}{8} - \frac{x^{2}}{12} - \frac{11 x^{3}}{96} + \mathcal{O}\big(x^{4}\big) \;,
\label{eq:ResultPTEgsm2neg}
\end{equation}
and
\begin{equation}
\rho^\text{LE;orig}_\mathrm{gs}= \frac{\hbar}{m^{2}}\left( -\frac{6}{x} + 1  + \frac{x}{2} + \frac{2 x^{2}}{3} + \frac{11 x^{3}}{8} + \mathcal{O}\big(x^{4}\big)\right)\;.
\label{eq:ResultPTrhogsm2neg}
\end{equation}
For $N>1$, no finite results can be obtained in the phase with $m^2<0$ for the gs energy and density, as long as we stick with the original dofs $\vec{\widetilde{\varphi}}$, because in this case the Goldstone propagator $G_{\varphi_\text{cl};\mathfrak{g}}$ exhibits (IR) divergences~\cite{col74} which, as discussed in section~\ref{sec:studiedtoymodel}, preclude the spontaneous breakdown of the continuous $O(N)$ symmetry.

\subsubsection{Mixed loop expansion}
\label{sec:MixedLoopExp}

Repeating the calculations outlined between~\eqref{eq:ZJKPT3} and~\eqref{eq:WJKPT} for the mixed case, the LE yields:
\begin{equation}
\begin{split}
W^\text{LE;mix}\big[\mathcal{J},\mathcal{K}\big] = & -S_{\text{mix},\mathcal{J}\mathcal{K}}[\Psi_{\mathrm{cl}}] + \frac{\hbar}{2} \mathcal{ST}r\left[\ln\big(\mathcal{G}_{\Psi_\text{cl};JK}\big)\right] \\
& + \sum^{\infty}_{n=1} \frac{(-1)^{n}}{(2n)!} \frac{\hbar^{n+1}\lambda^{n}}{12^{n}} \llangle[\Bigg]\left(\int_x  \widetilde{\zeta}(x)\vec{\widetilde{\chi}}(x)\cdot\vec{\widetilde{\chi}}(x)\right)^{2n}\rrangle[\Bigg]_{0,\mathcal{J}\mathcal{K}}^{\text{c}} \;,
\end{split}
\end{equation}
where the superpropagator $\mathcal{G}_{\Psi_\text{cl};\mathcal{JK}}(x,y)$ can be written in terms of the propagator $\boldsymbol{G}_{\sigma_\text{cl};\mathcal{JK}}(x,y)$ of the original dofs $\vec{\widetilde{\varphi}}$, the propagator $D_{\sigma_\text{cl};\mathcal{JK}}(x,y)$ of the collective dof $\widetilde{\sigma}$ and the mixed propagator $\vec{F}_{\varphi_\text{cl};\mathcal{JK}}(x,y)$ as follows:
\begin{equation}
\mathcal{G}_{\Psi_\text{cl};\mathcal{JK}} = \begin{pmatrix}
\boldsymbol{G}_{\sigma_\text{cl};\mathcal{JK}} & \vec{F}_{\varphi_\text{cl};\mathcal{JK}} \\
\vec{F}^{\mathrm{T}}_{\varphi_\text{cl};\mathcal{JK}} & D_{\sigma_\text{cl};\mathcal{JK}}
\end{pmatrix}\;,
\end{equation}
with the supertrace $\mathcal{ST}r$ satisfying $\mathcal{ST}r\equiv\mathrm{Tr}_{\alpha}\mathrm{Tr}_{x}$, while the source-dependent expectation value involves the following reference measure:
\begin{equation}
\llangle[\big] \cdots \rrangle[\big]_{0,\mathcal{J}\mathcal{K}} = \frac{1}{Z_{\mathrm{mix},0}\big[\mathcal{J},\mathcal{K}\big]} \int \mathcal{D}\widetilde{\Xi} \ \cdots \ e^{-\frac{1}{\hbar}S^0_{\mathrm{mix},\Psi_\text{cl};\mathcal{J}\mathcal{K}}\big[\widetilde{\Xi}\big]}\;,
\end{equation}
and
\begin{equation}
Z_{\mathrm{mix},0}\big[\mathcal{J},\mathcal{K}\big] =  \int \mathcal{D}\widetilde{\Xi}  \ e^{-\frac{1}{\hbar}S^0_{\mathrm{mix},\Psi_\text{cl};\mathcal{J}\mathcal{K}}\big[\widetilde{\Xi}\big]}\;.
\end{equation}
The connected correlation functions are once again conveniently represented by Feynman diagrams with the rules:
\begin{subequations}
\begin{align}
\left.
\begin{array}{ll}
\begin{gathered}
\begin{fmffile}{Diagrams/LoopExpansionMixedHS_FeynRuleVertexbis}
\begin{fmfgraph*}(4,4)
\fmfleft{i0,i1,i2,i3}
\fmfright{o0,o1,o2,o3}
\fmfv{label=$x$,label.angle=90,label.dist=4}{v1}
\fmfbottom{v2}
\fmf{phantom}{i1,v1}
\fmf{plain}{i2,v1}
\fmf{phantom}{v1,o1}
\fmf{plain}{v1,o2}
\fmf{wiggly,tension=0.6}{v1,v2}
\fmfv{decor.shape=circle,decor.size=2.0thick,foreground=(0,,0,,1)}{v1}
\fmflabel{$a$}{i2}
\fmflabel{$b$}{o2}
\end{fmfgraph*}
\end{fmffile}
\end{gathered} \\
\\
\begin{gathered}
\begin{fmffile}{Diagrams/LoopExpansionMixedHS_FeynRuleVertex2bis}
\begin{fmfgraph*}(4,4)
\fmfleft{i0,i1,i2,i3}
\fmfright{o0,o1,o2,o3}
\fmfv{label=$x$,label.angle=90,label.dist=4}{v1}
\fmfbottom{v2}
\fmf{phantom}{i1,v1}
\fmf{plain}{i2,v1}
\fmf{phantom}{v1,o1}
\fmf{plain}{v1,o2}
\fmf{dots,tension=0.6}{v1,v2}
\fmfv{decor.shape=circle,decor.size=2.0thick,foreground=(0,,0,,1)}{v1}
\fmflabel{$a$}{i2}
\fmflabel{$b$}{o2}
\end{fmfgraph*}
\end{fmffile}
\end{gathered} \\
\\
\begin{gathered}
\begin{fmffile}{Diagrams/LoopExpansionMixedHS_FeynRuleVertex3bis}
\begin{fmfgraph*}(4,4)
\fmfleft{i0,i1,i2,i3}
\fmfright{o0,o1,o2,o3}
\fmfv{label=$x$,label.angle=90,label.dist=4}{v1}
\fmfbottom{v2}
\fmf{phantom}{i1,v1}
\fmf{plain}{i2,v1}
\fmf{phantom}{v1,o1}
\fmf{dots}{v1,o2}
\fmf{dots,tension=0.6}{v1,v2}
\fmfv{decor.shape=circle,decor.size=2.0thick,foreground=(0,,0,,1)}{v1}
\fmflabel{$a$}{i2}
\fmflabel{$b$}{o2}
\end{fmfgraph*}
\end{fmffile}
\end{gathered}
\end{array}
\quad \right\rbrace &\rightarrow \sqrt{\lambda} \ \delta_{ab} \;, 
\label{eq:FeynRulesLoopExpansionMixedHSvertex} \\
\begin{gathered}
\begin{fmffile}{Diagrams/LoopExpansionMixedHS_FeynRuleGbis}
\begin{fmfgraph*}(20,20)
\fmfleft{i0,i1,i2,i3}
\fmfright{o0,o1,o2,o3}
\fmflabel{$x, a$}{v1}
\fmflabel{$y, b$}{v2}
\fmf{phantom}{i1,v1}
\fmf{phantom}{i2,v1}
\fmf{plain,tension=0.6}{v1,v2}
\fmf{phantom}{v2,o1}
\fmf{phantom}{v2,o2}
\end{fmfgraph*}
\end{fmffile}
\end{gathered} \quad &\rightarrow \boldsymbol{G}_{\sigma_\text{cl};\mathcal{JK};ab}(x,y)\;,
\label{eq:FeynRulesLoopExpansionMixedHSG} \\
\begin{gathered}
\begin{fmffile}{Diagrams/LoopExpansionMixedHS_FeynRuleDbis}
\begin{fmfgraph*}(20,20)
\fmfleft{i0,i1,i2,i3}
\fmfright{o0,o1,o2,o3}
\fmfv{label=$x$}{v1}
\fmfv{label=$y$}{v2}
\fmf{phantom}{i1,v1}
\fmf{phantom}{i2,v1}
\fmf{wiggly,tension=0.6}{v1,v2}
\fmf{phantom}{v2,o1}
\fmf{phantom}{v2,o2}
\end{fmfgraph*}
\end{fmffile}
\end{gathered} \quad &\rightarrow D_{\sigma_\text{cl};\mathcal{JK}}(x,y)\;,
\label{eq:FeynRulesLoopExpansionMixedHSD} \\
\begin{gathered}
\begin{fmffile}{Diagrams/LoopExpansionMixedHS_FeynRuleFbis}
\begin{fmfgraph*}(20,20)
\fmfleft{i0,i1,i2,i3}
\fmfright{o0,o1,o2,o3}
\fmflabel{$x, a$}{v1}
\fmfv{label=$y$}{v2}
\fmf{phantom}{i1,v1}
\fmf{phantom}{i2,v1}
\fmf{dashes,tension=0.6}{v1,v2}
\fmf{phantom}{v2,o1}
\fmf{phantom}{v2,o2}
\end{fmfgraph*}
\end{fmffile}
\end{gathered} \quad &\rightarrow F_{\varphi_\text{cl};\mathcal{JK};a}(x,y)\;.
\label{eq:FeynRulesLoopExpansionMixedHSF1}
\end{align}
\end{subequations}
Up to order $\mathcal{O}\big(\hbar^2\big)$, we derive in this way the following expression of the Schwinger functional:
\begin{equation}
\begin{split}
W^\text{LE;mix}\big[\mathcal{J},\mathcal{K}\big] = & -S_{\mathrm{mix},\mathcal{J}\mathcal{K}}[\Psi_{\mathrm{cl}}] + \frac{\hbar}{2} \mathcal{ST}r\left[\ln\big(\mathcal{G}_{\Psi_\text{cl};\mathcal{J}\mathcal{K}}\big)\right] \\
& - \hbar^{2} \left(\rule{0cm}{1.0cm}\right. \frac{1}{24}\begin{gathered}
\begin{fmffile}{Diagrams/LoopExpansionMixedHS_Hartree}
\begin{fmfgraph}(30,20)
\fmfleft{i}
\fmfright{o}
\fmfv{decor.shape=circle,decor.size=2.0thick,foreground=(0,,0,,1)}{v1}
\fmfv{decor.shape=circle,decor.size=2.0thick,foreground=(0,,0,,1)}{v2}
\fmf{phantom,tension=10}{i,i1}
\fmf{phantom,tension=10}{o,o1}
\fmf{plain,left,tension=0.5}{i1,v1,i1}
\fmf{plain,right,tension=0.5}{o1,v2,o1}
\fmf{wiggly}{v1,v2}
\end{fmfgraph}
\end{fmffile}
\end{gathered}
+\frac{1}{12}\begin{gathered}
\begin{fmffile}{Diagrams/LoopExpansionMixedHS_Fock}
\begin{fmfgraph}(15,15)
\fmfleft{i}
\fmfright{o}
\fmfv{decor.shape=circle,decor.size=2.0thick,foreground=(0,,0,,1)}{v1}
\fmfv{decor.shape=circle,decor.size=2.0thick,foreground=(0,,0,,1)}{v2}
\fmf{phantom,tension=11}{i,v1}
\fmf{phantom,tension=11}{v2,o}
\fmf{plain,left,tension=0.4}{v1,v2,v1}
\fmf{wiggly}{v1,v2}
\end{fmfgraph}
\end{fmffile}
\end{gathered} + \frac{1}{6}\begin{gathered}
\begin{fmffile}{Diagrams/LoopExpansionMixedHS_Diag1}
\begin{fmfgraph}(30,20)
\fmfleft{i}
\fmfright{o}
\fmfv{decor.shape=circle,decor.size=2.0thick,foreground=(0,,0,,1)}{v1}
\fmfv{decor.shape=circle,decor.size=2.0thick,foreground=(0,,0,,1)}{v2}
\fmf{phantom,tension=10}{i,i1}
\fmf{phantom,tension=10}{o,o1}
\fmf{dashes,left,tension=0.5}{i1,v1,i1}
\fmf{dashes,right,tension=0.5}{o1,v2,o1}
\fmf{plain}{v1,v2}
\end{fmfgraph}
\end{fmffile}
\end{gathered} \\
& \hspace{1.1cm} + \frac{1}{6}\begin{gathered}
\begin{fmffile}{Diagrams/LoopExpansionMixedHS_Diag2}
\begin{fmfgraph}(30,20)
\fmfleft{i}
\fmfright{o}
\fmfv{decor.shape=circle,decor.size=2.0thick,foreground=(0,,0,,1)}{v1}
\fmfv{decor.shape=circle,decor.size=2.0thick,foreground=(0,,0,,1)}{v2}
\fmf{phantom,tension=10}{i,i1}
\fmf{phantom,tension=10}{o,o1}
\fmf{plain,left,tension=0.5}{i1,v1,i1}
\fmf{dashes,right,tension=0.5}{o1,v2,o1}
\fmf{dashes}{v1,v2}
\end{fmfgraph}
\end{fmffile}
\end{gathered} + \frac{1}{6}\begin{gathered}
\begin{fmffile}{Diagrams/LoopExpansionMixedHS_Diag3}
\begin{fmfgraph}(15,15)
\fmfleft{i}
\fmfright{o}
\fmfv{decor.shape=circle,decor.size=2.0thick,foreground=(0,,0,,1)}{v1}
\fmfv{decor.shape=circle,decor.size=2.0thick,foreground=(0,,0,,1)}{v2}
\fmf{phantom,tension=11}{i,v1}
\fmf{phantom,tension=11}{v2,o}
\fmf{dashes,left,tension=0.4}{v1,v2,v1}
\fmf{plain}{v1,v2}
\end{fmfgraph}
\end{fmffile}
\end{gathered} \left.\rule{0cm}{1.0cm}\right) \\
& + \mathcal{O}\big(\hbar^{3}\big)\;,
\end{split}
\label{eq:WmixedKjLoopExpansionStep2}
\end{equation}
and we refer to appendix~\ref{sec:DiagLEM} for additional details on the determination of the diagrams.

\vspace{0.5cm}

In the (0+0)-D situation, the gs energy and density are obtained from the Schwinger functional according to the relations:
\begin{equation}
E^\text{LE;mix}_\text{gs} =  -\frac{1}{\hbar}W^\text{LE;mix}\big(\mathcal{J}=0,\mathcal{K}=0\big) \;,
\label{eq:AccessEgsFromWmix0DON}
\end{equation}
\begin{equation}
\rho^\text{LE;mix}_\text{gs} =\frac{2}{N}\mathrm{Tr}_{a}\hspace{-0.05cm}\left(\left.\frac{\partial W^\text{LE;mix}\big(\mathcal{J},\mathcal{K}\big)}{\partial \boldsymbol{K}}\right|_{\mathcal{J}=0 \atop \mathcal{K}=0}\right)=-\frac{2}{N}\frac{\partial W^\text{LE;mix}(\mathcal{J}=0,\mathcal{K}=0)}{\partial m^{2}} \;,
\label{eq:AccessRhogsFromWmix0DON}
\end{equation}
which are the counterparts of~\eqref{eq:DefEgsExactZexact0DON} and~\eqref{eq:DefrhogsExactwithExpectationValue0DON} in the mixed representation. When the sources $\mathcal{J}$ and $\mathcal{K}$ vanish, the saddle point $\Psi_\text{cl}$ reduces to:
\begin{equation}
\overline{\Psi}_\text{cl} \equiv \Psi_\text{cl}(\mathcal{J}=0,\mathcal{K}=0) = \begin{pmatrix}
\vec{\overline{\varphi}}_{\mathrm{cl}} \\
\overline{\sigma}_{\mathrm{cl}}
\end{pmatrix} = \begin{cases} \displaystyle{\begin{pmatrix}
\vec{0} \\
0
\end{pmatrix} \quad \forall m^2\geq 0\;,} \\
\\
\displaystyle{\begin{pmatrix}
\sqrt{\frac{-6m^2}{\lambda}}\vec{e} \\
im^2\sqrt{\frac{3}{\lambda}}
\end{pmatrix} \quad \forall m^2< 0 ~ \text{and} ~ \lambda\neq 0 \;,} \end{cases} 
\label{eq:SaddlePointPhiclSmix}
\end{equation}
with $\vec{e}$ a unit $N$-component vector. In the unbroken-symmetry regime and when all sources are set equal to zero, the inverse propagator $\mathcal{G}^{-1}_{\Psi_\text{cl}}=\mathcal{G}^{-1}_0$ is diagonal:
\begin{equation}
\mathcal{G}^{-1}_0 = \begin{pmatrix}
m^2 \mathbb{I}_{N} & \vec{0} \\
\vec{0}^\mathrm{T} & 1 \end{pmatrix} \;,
\end{equation}
so that all propagators of the theory reduce to the bare ones, i.e.:
\begin{equation}
\boldsymbol{G}_{\sigma_\text{cl}} = \boldsymbol{G}_0 = G_0 \ \mathbb{I}_{N} = \frac{1}{m^2} \ \mathbb{I}_{N} \;,
\end{equation}
\begin{equation}
D_{\sigma_\text{cl}} =  D_0 = 1 \;,
\end{equation}
\begin{equation}
\vec{F}_{\varphi_\text{cl}} = \vec{F}_0 = \vec{0} \;.
\end{equation}
In this case, the Schwinger functional~\eqref{eq:WmixedKjLoopExpansionStep2} reads up to order $\mathcal{O}\big(\hbar^{4}\big)$:
\begin{equation}
\begin{split}
W^\text{LE;mix}(\mathcal{J}=0,\mathcal{K}=0) = & \ \frac{\hbar}{2}\left[N\ln(2\pi G_0) + \ln(D_0)\right] - \frac{\hbar^{2}\lambda}{24} \left(2 D_0 G_0^2 N + D_0 G_0^2 N^2\right) \\
& + \frac{\hbar^{3}\lambda^{2}}{144} \left(6 D_0^2 G_0^4 N + 5 D_0^2 G_0^4 N^2 + D_0^2 G_0^4 N^3\right) \\
& - \frac{\hbar^{4}\lambda^{3}}{2592} \left(120 D_0^3 G_0^6 N + 128 D_0^3 G_0^6 N^2 + 44 D_0^3 G_0^6 N^3 + 5 D_0^3 G_0^6 N^4\right) \\
& + \mathcal{O}\big(\hbar^{5}\big)\;,
\end{split}
\label{eq:ResultLoopExpansionMixedAction0DON}
\end{equation}
which coincides with the Schwinger functional~\eqref{eq:Worig} of the original theory, still evaluated at $m^2 > 0$ and vanishing sources. Therefore, the same series representation of the gs energy and density as in the original theory are found in the mixed representation. Likewise, we get identical results in the regime with $m^2<0$. At $m^2<0$, $\lambda\neq 0$ and $N=1$ for instance, we can show from~\eqref{eq:SaddlePointPhiclSmix} that the inverse superpropagator at vanishing sources reads:
\begin{equation}
\mathcal{G}^{-1}_{\Psi_\text{cl}} = \begin{pmatrix}
 0  & -\sqrt{2m^{2}} \\
-\sqrt{2m^{2}} & 1 \end{pmatrix} \;,
\end{equation}
and thus
\begin{equation}
\mathcal{G}_{\Psi_\text{cl}} = -\begin{pmatrix}
\frac{1}{2 m^2}  & \frac{1}{\sqrt{2m^{2}}} \\
\frac{1}{\sqrt{2m^{2}}} & 0 \end{pmatrix} \;,
\end{equation}
leading to the same series as in the original representation in the broken-symmetry regime (and more specifically to the series~\eqref{eq:ResultPTEgsm2neg} and~\eqref{eq:ResultPTrhogsm2neg} for $E_{\mathrm{gs}}$ and $\rho_{\mathrm{gs}}$, respectively). For $N>1$, $\mathcal{G}^{-1}_{\Psi_\text{cl}}$ is a singular matrix and we face the same limitations as in the original representation. To conclude, the LE based on the mixed representation does not bring anything more compared to that of the original theory.

\subsubsection{Collective loop expansion}
\label{sec:CollLE}

We now focus on the LE in the collective representation, i.e. the collective LE. Such a technique is sometimes called the $\epsilon$-expansion, as e.g. in ref.~\cite{daw12}. This designation follows from early works with collective actions~\cite{ben77}, which name their expansion parameter $\epsilon$ instead of $\hbar$. For the sake of completeness, we also point out the works of refs.~\cite{mun80,kon81} which discuss the differences of renormalization issues between the collective case and the original one based on a $\varphi^{4}$ self-interaction for the original dofs. We do not address the matter of renormalization here as it is absent from our (0+0)-D applications. Furthermore, the collective LE is usually not exploited beyond its first non-trivial order. We will outline in this section how to construct the collective LE series up to their first non-trivial order for our $O(N)$ model at arbitrary dimensions but we will perform applications of this method up to its third non-trivial order and combine it with resummation procedures in (0+0)-D. To our knowledge, the collective LE has neither been pushed up to its third non-trivial order nor been combined with resummation theory so far, regardless of the model under consideration. Following the same steps as in the previous representations, the partition function of the theory based on~\eqref{eq:SbosonicKLoopExpansion} reads up to the first non-trivial order (i.e. up to order $\mathcal{O}(\hbar)$):
\begin{equation}
\begin{split}
\scalebox{0.99}{${\displaystyle Z^\text{LE;col}\big[\mathcal{J}\big] = }$} & \ \scalebox{0.99}{${\displaystyle e^{-\frac{1}{\hbar}S_{\mathrm{col},\mathcal{J}}[\sigma_{\mathrm{cl}}]} \left( \int\mathcal{D}\widetilde{\zeta} \ e^{-\frac{1}{2}\int_{x,y}\widetilde{\zeta}(x) D^{-1}_{\sigma_\text{cl};\mathcal{J}}(x,y) \widetilde{\zeta}(y)} \right) }$} \\
& \scalebox{0.99}{${\displaystyle \times \Bigg[1 -\frac{\hbar}{24} \int_{x,y,z,u} S_{\mathrm{col}, \mathcal{J}}^{(4)}(x,y,z,u) \ \left\langle \widetilde{\zeta}(x) \widetilde{\zeta}(y) \widetilde{\zeta}(z) \widetilde{\zeta}(u) \right\rangle_{0,\mathcal{J}} }$} \\
& \hspace{0.44cm} \scalebox{0.99}{${\displaystyle + \frac{\hbar}{72} \int_{{x_{\scalebox{0.4}{1}},y_{\scalebox{0.4}{1}},z_{\scalebox{0.4}{1}}}\atop{x_{\scalebox{0.4}{2}},y_{\scalebox{0.4}{2}},z_{\scalebox{0.4}{2}}}} \hspace{-0.2cm} S_{\mathrm{col}, \mathcal{J}}^{(3)}(x_{1},y_{1},z_{1}) S_{\mathrm{col}, \mathcal{J}}^{(3)}(x_{2},y_{2},z_{2}) \ \left\langle \widetilde{\zeta}(x_{1}) \widetilde{\zeta}(y_{1}) \widetilde{\zeta}(z_{1}) \widetilde{\zeta}(x_{2}) \widetilde{\zeta}(y_{2}) \widetilde{\zeta}(z_{2}) \right\rangle_{0,\mathcal{J}} }$} \\
& \hspace{0.44cm} \scalebox{0.99}{${\displaystyle + \mathcal{O}\big(\hbar^{2}\big)\Bigg]\;, }$}
\end{split}
\label{eq:Zlecol}
\end{equation}
where
\begin{equation}
\big\langle\cdots\big\rangle_{0,\mathcal{J}} = \frac{1}{Z_{\mathrm{col},0}\big[\mathcal{J}\big]} \int \mathcal{D}\widetilde{\zeta} \ \cdots \ e^{-\frac{1}{\hbar} S^0_{\mathrm{col},\sigma_\text{cl};\mathcal{J}}\big[\widetilde{\zeta}\big]}\;,
\end{equation}
and
\begin{equation}
Z_{\mathrm{col},0}\big[\mathcal{J}\big] = \int \mathcal{D}\widetilde{\zeta} \  e^{-\frac{1}{\hbar} S^0_{\mathrm{col},\sigma_\text{cl};\mathcal{J}}\big[\widetilde{\zeta}\big]}\;.
\end{equation}
Introducing the Feynman rules:
\begin{subequations}
\begin{align}
\begin{gathered}
\begin{fmffile}{Diagrams/LEcol-G}
\begin{fmfgraph*}(20,12)
\fmfleft{i0,i1,i2,i3}
\fmfright{o0,o1,o2,o3}
\fmflabel{$x, a$}{v1}
\fmflabel{$y, b$}{v2}
\fmf{phantom}{i1,v1}
\fmf{phantom}{i2,v1}
\fmf{plain,tension=0.6}{v1,v2}
\fmf{phantom}{v2,o1}
\fmf{phantom}{v2,o2}
\end{fmfgraph*}
\end{fmffile}
\end{gathered} \quad &\rightarrow \boldsymbol{G}_{\sigma_\text{cl};\mathcal{J};ab}(x,y)\;,
\label{eq:FeynRulesLoopExpansionBosonicHSG}\\
\begin{gathered}
\begin{fmffile}{Diagrams/LEcol-D}
\begin{fmfgraph*}(20,16)
\fmfleft{i0,i1,i2,i3}
\fmfright{o0,o1,o2,o3}
\fmfv{label=$x$}{v1}
\fmfv{label=$y$}{v2}
\fmf{phantom}{i1,v1}
\fmf{phantom}{i2,v1}
\fmf{wiggly,tension=0.6}{v1,v2}
\fmf{phantom}{v2,o1}
\fmf{phantom}{v2,o2}
\end{fmfgraph*}
\end{fmffile}
\end{gathered} \quad &\rightarrow D_{\sigma_\text{cl};\mathcal{J}}(x,y)\;,
\label{eq:FeynRulesLoopExpansionBosonicHSH}\\
\begin{gathered}
\begin{fmffile}{Diagrams/LEcol-V}
\begin{fmfgraph*}(5,5)
\fmfleft{i1}
\fmfright{o1}
\fmfv{label=$x$,label.angle=-90,label.dist=4,foreground=(0,,0,,1)}{v1}
\fmf{plain}{i1,v1}
\fmf{plain}{v1,o1}
\fmflabel{$a$}{i1}
\fmflabel{$b$}{o1}
\fmfdot{v1}
\end{fmfgraph*}
\end{fmffile}
\end{gathered}\qquad &\rightarrow i\sqrt{\frac{\lambda}{3}}\delta_{ab}\;,
\label{eq:FeynRulesLoopExpansionBosonicHSvertexDot} \\
\nonumber\\
\begin{gathered}
\begin{fmffile}{Diagrams/LEcol-J}
\begin{fmfgraph*}(10,5)
\fmfleft{i1}
\fmfright{o1}
\fmfv{decor.shape=circle,decor.filled=empty,decor.size=.26w,l=$\times$,label.dist=0}{v1}
\fmfv{label.angle=-90,label.dist=6}{v2}
\fmf{plain,tension=2.5}{i1,v1}
\fmf{phantom}{v1,o1}
\fmf{phantom,tension=2.5}{i1,v2}
\fmf{phantom}{v2,o1}
\fmflabel{$x, a$}{v2}
\end{fmfgraph*}
\end{fmffile}
\end{gathered} \hspace{-0.2cm} &\rightarrow J_{a}(x)\;,
\label{eq:FeynRulesLoopExpansionBosonicHSK} \\
\nonumber\\
\nonumber\\
\begin{gathered}
\begin{fmffile}{Diagrams/LoopExpansionBosonicHS_FeynRuleVertexS3}
\begin{fmfgraph*}(4,4)
\fmfleft{i0,i1,i2,i3}
\fmfright{o0,o1,o2,o3}
\fmfv{label=$x$,label.angle=-135,label.dist=11,decor.shape=triangle,decor.filled=shaded,decor.size=1.6w,foreground=(0,,0,,1)}{v1}
\fmfv{label=$y$,label.angle=-45,label.dist=11}{v2}
\fmfv{label=$z$,label.angle=90,label.dist=11}{v3}
\fmfbottom{v1bis}
\fmf{phantom}{i1,v1}
\fmf{phantom}{i2,v1}
\fmf{phantom}{v1,o1}
\fmf{phantom}{v1,o2}
\fmf{phantom}{i1,v2}
\fmf{phantom}{i2,v2}
\fmf{phantom}{v2,o1}
\fmf{phantom}{v2,o2}
\fmf{phantom}{i1,v3}
\fmf{phantom}{i2,v3}
\fmf{phantom}{v3,o1}
\fmf{phantom}{v3,o2}
\fmf{phantom,tension=0.6}{v1,v1bis}
\end{fmfgraph*}
\end{fmffile}
\end{gathered} \quad &\rightarrow S_{\mathrm{col},\mathcal{J}}^{(3)}(x,y,z)\;,
\label{eq:FeynRulesLoopExpansionBosonicHSvertexS3} \\
\nonumber \\
\nonumber\\
\begin{gathered}
\begin{fmffile}{Diagrams/LoopExpansionBosonicHS_FeynRuleVertexS4}
\begin{fmfgraph*}(4,4)
\fmfleft{i0,i1,i2,i3}
\fmfright{o0,o1,o2,o3}
\fmfv{label=$x$,label.angle=-135,label.dist=11,decor.shape=square,decor.filled=shaded,decor.size=1.2w,foreground=(0,,0,,1)}{v1}
\fmfv{label=$y$,label.angle=-45,label.dist=11}{v2}
\fmfv{label=$z$,label.angle=45,label.dist=11}{v3}
\fmfv{label=$u$,label.angle=135,label.dist=11}{v4}
\fmfbottom{v1bis}
\fmf{phantom}{i1,v1}
\fmf{phantom}{i2,v1}
\fmf{phantom}{v1,o1}
\fmf{phantom}{v1,o2}
\fmf{phantom}{i1,v2}
\fmf{phantom}{i2,v2}
\fmf{phantom}{v2,o1}
\fmf{phantom}{v2,o2}
\fmf{phantom}{i1,v3}
\fmf{phantom}{i2,v3}
\fmf{phantom}{v3,o1}
\fmf{phantom}{v3,o2}
\fmf{phantom}{i1,v4}
\fmf{phantom}{i2,v4}
\fmf{phantom}{v4,o1}
\fmf{phantom}{v4,o2}
\fmf{phantom,tension=0.6}{v1,v1bis}
\end{fmfgraph*}
\end{fmffile}
\end{gathered} \quad &\rightarrow S_{\mathrm{col},\mathcal{J}}^{(4)}(x,y,z,u)\;, \\
\nonumber
\label{eq:FeynRulesLoopExpansionBosonicHSvertexS4}
\end{align}
\end{subequations}
the terms involved in the brackets of the RHS of~\eqref{eq:Zlecol} read:
\begin{equation}
\int_{x,y,z,u} S_{\mathrm{col}, \mathcal{J}}^{(4)}(x,y,z,u) \left\langle \widetilde{\zeta}(x) \widetilde{\zeta}(y) \widetilde{\zeta}(z) \widetilde{\zeta}(u) \right\rangle_{0,\mathcal{J}} = \ 3 \ \ \begin{gathered}
\begin{fmffile}{Diagrams/LoopExpansionBosonicHS_ApplicationWickTheoremS4_Diag4}
\begin{fmfgraph}(20,20)
\fmfleft{i}
\fmfright{o}
\fmfv{decor.shape=square,decor.filled=shaded,decor.size=0.18w,foreground=(0,,0,,1)}{v1}
\fmf{phantom,tension=10}{i,i1}
\fmf{phantom,tension=10}{o,o1}
\fmf{wiggly,left,tension=0.4}{i1,v1,i1}
\fmf{wiggly,right,tension=0.4}{o1,v1,o1}
\end{fmfgraph}
\end{fmffile}
\end{gathered} \;,
\label{eq:ApplicationWickTheoremBosonicActionS4}
\end{equation}
and
\begin{equation}
\begin{split}
\scalebox{0.98}{${\displaystyle \int_{{x_{\scalebox{0.4}{1}},y_{\scalebox{0.4}{1}},z_{\scalebox{0.4}{1}}}\atop{x_{\scalebox{0.4}{2}},y_{\scalebox{0.4}{2}},z_{\scalebox{0.4}{2}}}} S_{\mathrm{col},\mathcal{J}}^{(3)}(x_{1},y_{1},z_{1}) S_{\mathrm{col}, \mathcal{J}}^{(3)}(x_{2},y_{2},z_{2}) \left\langle \widetilde{\zeta}(x_{1}) \widetilde{\zeta}(y_{1}) \widetilde{\zeta}(z_{1}) \widetilde{\zeta}(x_{2}) \widetilde{\zeta}(y_{2}) \widetilde{\zeta}(z_{2}) \right\rangle_{0,\mathcal{J}} = }$} & \ \scalebox{0.98}{${\displaystyle 9 \hspace{0.2cm} \begin{gathered}
\begin{fmffile}{Diagrams/LoopExpansionBosonicHS_ApplicationWickTheoremS4_Diag5}
\begin{fmfgraph}(28,20)
\fmfleft{i}
\fmfright{o}
\fmfv{decor.shape=triangle,decor.filled=shaded,decor.size=0.16w,foreground=(0,,0,,1)}{v1}
\fmfv{decor.shape=triangle,decor.filled=shaded,decor.size=0.16w,foreground=(0,,0,,1)}{v2}
\fmf{phantom,tension=25}{i,i1}
\fmf{phantom,tension=25}{o,o1}
\fmf{wiggly,left,tension=0.7}{i1,v1,i1}
\fmf{wiggly,right,tension=0.7}{o1,v2,o1}
\fmf{wiggly}{v1,v2}
\end{fmfgraph}
\end{fmffile}
\end{gathered} }$} \\
& \scalebox{0.98}{${\displaystyle + 6 \hspace{-0.4cm} \begin{gathered}
\begin{fmffile}{Diagrams/LoopExpansionBosonicHS_ApplicationWickTheoremS4_Diag6}
\begin{fmfgraph}(28,20)
\fmfleft{i}
\fmfright{o}
\fmfv{decor.shape=triangle,decor.filled=shaded,decor.size=0.16w,foreground=(0,,0,,1)}{v1}
\fmfv{decor.shape=triangle,decor.filled=shaded,decor.size=0.16w,foreground=(0,,0,,1)}{v2}
\fmf{phantom,tension=5}{i,v1}
\fmf{phantom,tension=5}{v2,o}
\fmf{wiggly,left,tension=0.8}{v1,v2,v1}
\fmf{wiggly}{v1,v2}
\end{fmfgraph}
\end{fmffile}
\end{gathered} \hspace{-0.5cm} \;.}$}
\end{split}
\label{eq:ApplicationWickTheoremBosonicActionS3}
\end{equation}
In these expressions, the propagator of the collective field as well as the vertex functions can be evaluated after exploiting the following expression for the derivative of the original field propagator $\boldsymbol{G}_{\widetilde{\sigma}}$ defined by~\eqref{eq:Gcoll}:
\begin{equation}
\begin{split}
\frac{\delta \boldsymbol{G}_{\widetilde{\sigma};ab}(x,y)}{\delta \widetilde{\sigma}(z)} = & \ \frac{\delta \left(\boldsymbol{G}^{-1}_{\widetilde{\sigma}}\right)_{ab}^{-1}(x,y)}{\delta \widetilde{\sigma}(z)} \\
= & - \int_{u,v} \left.\boldsymbol{G}_{\widetilde{\sigma};a}\right.^{c}(x,u) \frac{\delta \boldsymbol{G}^{-1}_{\widetilde{\sigma};cd}(u,v)}{\delta \widetilde{\sigma}(z)} \left.\left.\boldsymbol{G}_{\widetilde{\sigma};}\right.^{d}\right._{b}(v,y) \\
= & -i\sqrt{\frac{\lambda}{3}} \left.\boldsymbol{G}_{\widetilde{\sigma};a}\right.^c(x,z)\boldsymbol{G}_{\widetilde{\sigma};cb}(z,y) \;,
\end{split}
\label{eq:DiagramCalcCalculationGsigCollLE}
\end{equation}
where the derivative of the second line was evaluated with expression~\eqref{eq:Gcoll} of $\boldsymbol{G}^{-1}_{\widetilde{\sigma};ab}(x,y)$. Since the propagator $\boldsymbol{G}_{\sigma_{\mathrm{cl}};\mathcal{J}}$ (involved in~\eqref{eq:DefGpropagCollectiveLE} and~\eqref{eq:FeynRulesLoopExpansionBosonicHSG}) corresponds to $\boldsymbol{G}_{\widetilde{\sigma}}$ evaluated at $\widetilde{\sigma}=\sigma_{\mathrm{cl}}$, we show the following relations directly from~\eqref{eq:DiagramCalcCalculationGsigCollLE}:
\begin{equation}
D^{-1}_{\sigma_\text{cl};\mathcal{J}}(x,y) = - \hspace{0.2cm} \begin{gathered}
\begin{fmffile}{Diagrams/LoopExpansionBosonicHS_Vertex_Diag1}
\begin{fmfgraph*}(15,20)
\fmfleft{i1,i2}
\fmfright{o1,o2}
\fmfv{label=$x$,label.dist=4,foreground=(0,,0,,1)}{v1}
\fmfv{label=$y$,label.dist=4,foreground=(0,,0,,1)}{v2}
\fmfv{decor.shape=circle,decor.filled=empty,decor.size=.17w,l=$\times$,label.dist=0}{v3}
\fmfv{decor.shape=circle,decor.filled=empty,decor.size=.17w,l=$\times$,label.dist=0}{v4}
\fmf{phantom}{i1,v1}
\fmf{phantom}{i2,v3}
\fmf{phantom}{o1,v2}
\fmf{phantom}{o2,v4}
\fmf{plain,tension=0.2}{v1,v2}
\fmf{phantom,tension=0.2}{v3,v4}
\fmf{plain}{v1,v3}
\fmf{plain}{v2,v4}
\fmfdot{v1,v2}
\end{fmfgraph*}
\end{fmffile}
\end{gathered} \hspace{0.2cm} - \frac{1}{2} \hspace{0.3cm} \begin{gathered}
\begin{fmffile}{Diagrams/LoopExpansionBosonicHS_Vertex_Diag2}
\begin{fmfgraph*}(20,20)
\fmfleft{i0,i1,i2,i3}
\fmfright{o0,o1,o2,o3}
\fmfv{label=$x$,label.dist=4,foreground=(0,,0,,1)}{v1}
\fmfv{label=$y$,label.dist=4,foreground=(0,,0,,1)}{v2}
\fmf{phantom}{i1,v1}
\fmf{phantom}{i2,v1}
\fmf{plain,left,tension=0.4}{v1,v2,v1}
\fmf{phantom}{v2,o1}
\fmf{phantom}{v2,o2}
\fmfdot{v1,v2}
\end{fmfgraph*}
\end{fmffile}
\end{gathered} \hspace{0.3cm} + \delta(x-y)\;,
\label{eq:SbosonicKLoopExpansionH}
\end{equation}
\begin{equation}
\begin{gathered}
\begin{fmffile}{Diagrams/LoopExpansionBosonicHS_FeynRuleVertexS3}
\begin{fmfgraph*}(4,4)
\fmfleft{i0,i1,i2,i3}
\fmfright{o0,o1,o2,o3}
\fmfv{label=$x$,label.angle=-135,label.dist=11,decor.shape=triangle,decor.filled=shaded,decor.size=1.6w,foreground=(0,,0,,1)}{v1}
\fmfv{label=$y$,label.angle=-45,label.dist=11}{v2}
\fmfv{label=$z$,label.angle=90,label.dist=11}{v3}
\fmfbottom{v1bis}
\fmf{phantom}{i1,v1}
\fmf{phantom}{i2,v1}
\fmf{phantom}{v1,o1}
\fmf{phantom}{v1,o2}
\fmf{phantom}{i1,v2}
\fmf{phantom}{i2,v2}
\fmf{phantom}{v2,o1}
\fmf{phantom}{v2,o2}
\fmf{phantom}{i1,v3}
\fmf{phantom}{i2,v3}
\fmf{phantom}{v3,o1}
\fmf{phantom}{v3,o2}
\fmf{phantom,tension=0.6}{v1,v1bis}
\end{fmfgraph*}
\end{fmffile}
\end{gathered} \hspace{0.5cm} = \begin{gathered}
\begin{fmffile}{Diagrams/LoopExpansionBosonicHS_Vertex_Diag3}
\begin{fmfgraph*}(22,20)
\fmfleft{i1,i2}
\fmfright{o1,o2}
\fmfbottom{i0,o0}
\fmfv{label=$x$,label.dist=4,label.angle=-90,foreground=(0,,0,,1)}{v1}
\fmfv{label=$y$,label.dist=4,foreground=(0,,0,,1)}{v2}
\fmfv{label=$z$,label.dist=4,label.angle=180,foreground=(0,,0,,1)}{v3}
\fmfv{decor.shape=circle,decor.filled=empty,decor.size=.12w,l=$\times$,label.dist=0}{v1b}
\fmfv{decor.shape=circle,decor.filled=empty,decor.size=.12w,l=$\times$,label.dist=0}{v3b}
\fmf{phantom}{i1,v1}
\fmf{phantom}{i2,v3b}
\fmf{phantom}{o1,v2}
\fmf{phantom}{o2,v3b}
\fmf{phantom}{i0,v1b}
\fmf{phantom}{o0,v2b}
\fmf{plain,tension=0.5}{v1,v2}
\fmf{phantom}{v1,v3}
\fmf{plain}{v2,v3}
\fmf{plain,tension=0.2}{v1,v1b}
\fmf{phantom,tension=0.2}{v2,v2b}
\fmf{plain,tension=1.5}{v3,v3b}
\fmfdot{v1,v2,v3}
\end{fmfgraph*}  
\end{fmffile}
\end{gathered} + \begin{gathered}
\begin{fmffile}{Diagrams/LoopExpansionBosonicHS_Vertex_Diag4}
\begin{fmfgraph*}(22,20)
\fmfleft{i1,i2}
\fmfright{o1,o2}
\fmfbottom{i0,o0}
\fmfv{label=$x$,label.dist=4,foreground=(0,,0,,1)}{v1}
\fmfv{label=$y$,label.dist=4,label.angle=-90,foreground=(0,,0,,1)}{v2}
\fmfv{label=$z$,label.dist=4,label.angle=180,foreground=(0,,0,,1)}{v3}
\fmfv{decor.shape=circle,decor.filled=empty,decor.size=.12w,l=$\times$,label.dist=0}{v2b}
\fmfv{decor.shape=circle,decor.filled=empty,decor.size=.12w,l=$\times$,label.dist=0}{v3b}
\fmf{phantom}{i1,v1}
\fmf{phantom}{i2,v3b}
\fmf{phantom}{o1,v2}
\fmf{phantom}{o2,v3b}
\fmf{phantom}{i0,v1b}
\fmf{phantom}{o0,v2b}
\fmf{plain,tension=0.5}{v1,v2}
\fmf{plain}{v1,v3}
\fmf{phantom}{v2,v3}
\fmf{phantom,tension=0.2}{v1,v1b}
\fmf{plain,tension=0.2}{v2,v2b}
\fmf{plain,tension=1.5}{v3,v3b}
\fmfdot{v1,v2,v3}
\end{fmfgraph*}
\end{fmffile}
\end{gathered} + \begin{gathered}
\begin{fmffile}{Diagrams/LoopExpansionBosonicHS_Vertex_Diag5}
\begin{fmfgraph*}(22,20)
\fmfleft{i1,i2}
\fmfright{o1,o2}
\fmfbottom{i0,o0}
\fmfv{label=$x$,label.dist=4,label.angle=-90,foreground=(0,,0,,1)}{v1}
\fmfv{label=$y$,label.dist=4,label.angle=-90,foreground=(0,,0,,1)}{v2}
\fmfv{label=$z$,label.dist=4,label.angle=90,foreground=(0,,0,,1)}{v3}
\fmfv{decor.shape=circle,decor.filled=empty,decor.size=.12w,l=$\times$,label.dist=0}{v1b}
\fmfv{decor.shape=circle,decor.filled=empty,decor.size=.12w,l=$\times$,label.dist=0}{v2b}
\fmf{phantom}{i1,v1}
\fmf{phantom}{i2,v3b}
\fmf{phantom}{o1,v2}
\fmf{phantom}{o2,v3b}
\fmf{phantom}{i0,v1b}
\fmf{phantom}{o0,v2b}
\fmf{phantom,tension=0.5}{v1,v2}
\fmf{plain}{v1,v3}
\fmf{plain}{v2,v3}
\fmf{plain,tension=0.2}{v1,v1b}
\fmf{plain,tension=0.2}{v2,v2b}
\fmf{phantom,tension=1.5}{v3,v3b}
\fmfdot{v1,v2,v3}
\end{fmfgraph*}
\end{fmffile}
\end{gathered}+ \begin{gathered}
\begin{fmffile}{Diagrams/LoopExpansionBosonicHS_Vertex_Diag6}
\begin{fmfgraph*}(22,20)
\fmfleft{i1,i2}
\fmfright{o1,o2}
\fmfbottom{i0,o0}
\fmfv{label=$x$,label.dist=4,foreground=(0,,0,,1)}{v1}
\fmfv{label=$y$,label.dist=4,foreground=(0,,0,,1)}{v2}
\fmfv{label=$z$,label.dist=4,label.angle=90,foreground=(0,,0,,1)}{v3}
\fmf{phantom}{i1,v1}
\fmf{phantom}{i2,v3b}
\fmf{phantom}{o1,v2}
\fmf{phantom}{o2,v3b}
\fmf{phantom}{i0,v1b}
\fmf{phantom}{o0,v2b}
\fmf{plain,tension=0.5}{v1,v2}
\fmf{plain}{v1,v3}
\fmf{plain}{v2,v3}
\fmf{phantom,tension=0.2}{v1,v1b}
\fmf{phantom,tension=0.2}{v2,v2b}
\fmf{phantom,tension=1.5}{v3,v3b}
\fmfdot{v1,v2,v3}
\end{fmfgraph*}
\end{fmffile}
\end{gathered}\;,
\label{eq:SbosonicKLoopExpansionS3}
\end{equation}
and
\begin{equation}
\begin{split}
\scalebox{0.99}{${\displaystyle \begin{gathered}
\begin{fmffile}{Diagrams/LoopExpansionBosonicHS_FeynRuleVertexS4}
\begin{fmfgraph*}(4,4)
\fmfleft{i0,i1,i2,i3}
\fmfright{o0,o1,o2,o3}
\fmfv{label=$x$,label.angle=-135,label.dist=11,decor.shape=square,decor.filled=shaded,decor.size=1.2w,foreground=(0,,0,,1)}{v1}
\fmfv{label=$y$,label.angle=-45,label.dist=11}{v2}
\fmfv{label=$z$,label.angle=45,label.dist=11}{v3}
\fmfv{label=$u$,label.angle=135,label.dist=11}{v4}
\fmfbottom{v1bis}
\fmf{phantom}{i1,v1}
\fmf{phantom}{i2,v1}
\fmf{phantom}{v1,o1}
\fmf{phantom}{v1,o2}
\fmf{phantom}{i1,v2}
\fmf{phantom}{i2,v2}
\fmf{phantom}{v2,o1}
\fmf{phantom}{v2,o2}
\fmf{phantom}{i1,v3}
\fmf{phantom}{i2,v3}
\fmf{phantom}{v3,o1}
\fmf{phantom}{v3,o2}
\fmf{phantom}{i1,v4}
\fmf{phantom}{i2,v4}
\fmf{phantom}{v4,o1}
\fmf{phantom}{v4,o2}
\fmf{phantom,tension=0.6}{v1,v1bis}
\end{fmfgraph*}
\end{fmffile}
\end{gathered} \hspace{0.5cm} = }$} & \scalebox{0.99}{${\displaystyle -\left(\rule{0cm}{1.1cm}\right. \hspace{-0.31cm} \begin{gathered}
\begin{fmffile}{Diagrams/LoopExpansionBosonicHS_Vertex_Diag7}
\begin{fmfgraph*}(25,25)
\fmfleft{i1,i2}
\fmfright{o1,o2}
\fmfbottom{i0,o0}
\fmftop{i3,o3}
\fmfv{label=$x$,label.dist=4,label.angle=180,foreground=(0,,0,,1)}{v1}
\fmfv{label=$y$,label.dist=4,foreground=(0,,0,,1)}{v2}
\fmfv{label=$z$,label.dist=4,foreground=(0,,0,,1)}{v3}
\fmfv{label=$u$,label.dist=4,label.angle=180,foreground=(0,,0,,1)}{v4}
\fmfv{decor.shape=circle,decor.filled=empty,decor.size=.108w,l=$\times$,label.dist=0}{v1b}
\fmfv{decor.shape=circle,decor.filled=empty,decor.size=.108w,l=$\times$,label.dist=0}{v4b}
\fmf{phantom}{i1,v1}
\fmf{phantom}{i2,v4}
\fmf{phantom}{o1,v2}
\fmf{phantom}{o2,v3}
\fmf{phantom}{i3,v4b}
\fmf{phantom}{o3,v3b}
\fmf{phantom}{i0,v1b}
\fmf{phantom}{o0,v2b}
\fmf{plain,tension=1.6}{v1,v2}
\fmf{plain,tension=1.6}{v3,v4}
\fmf{phantom,tension=2.0}{v1,v4}
\fmf{plain,tension=2.0}{v2,v3}
\fmf{phantom,tension=0}{v1,v3}
\fmf{phantom,tension=0}{v2,v4}
\fmf{plain}{v1,v1b}
\fmf{phantom}{v2,v2b}
\fmf{phantom}{v3,v3b}
\fmf{plain}{v4,v4b}
\fmfdot{v1,v2,v3,v4}
\end{fmfgraph*}
\end{fmffile}
\end{gathered} + \begin{gathered}
\begin{fmffile}{Diagrams/LoopExpansionBosonicHS_Vertex_Diag8}
\begin{fmfgraph*}(25,25)
\fmfleft{i1,i2}
\fmfright{o1,o2}
\fmfbottom{i0,o0}
\fmftop{i3,o3}
\fmfv{label=$x$,label.dist=4,foreground=(0,,0,,1)}{v1}
\fmfv{label=$y$,label.dist=4,label.angle=0,foreground=(0,,0,,1)}{v2}
\fmfv{label=$z$,label.dist=4,foreground=(0,,0,,1)}{v3}
\fmfv{label=$u$,label.dist=4,label.angle=180,foreground=(0,,0,,1)}{v4}
\fmfv{decor.shape=circle,decor.filled=empty,decor.size=.108w,l=$\times$,label.dist=0}{v2b}
\fmfv{decor.shape=circle,decor.filled=empty,decor.size=.108w,l=$\times$,label.dist=0}{v4b}
\fmf{phantom}{i1,v1}
\fmf{phantom}{i2,v4}
\fmf{phantom}{o1,v2}
\fmf{phantom}{o2,v3}
\fmf{phantom}{i3,v4b}
\fmf{phantom}{o3,v3b}
\fmf{phantom}{i0,v1b}
\fmf{phantom}{o0,v2b}
\fmf{plain,tension=1.6}{v1,v2}
\fmf{plain,tension=1.6}{v3,v4}
\fmf{phantom,tension=2.0}{v1,v4}
\fmf{phantom,tension=2.0}{v2,v3}
\fmf{plain,tension=0}{v1,v3}
\fmf{phantom,tension=0}{v2,v4}
\fmf{phantom}{v1,v1b}
\fmf{plain}{v2,v2b}
\fmf{phantom}{v3,v3b}
\fmf{plain}{v4,v4b}
\fmfdot{v1,v2,v3,v4}
\end{fmfgraph*}
\end{fmffile}
\end{gathered} + \begin{gathered}
\begin{fmffile}{Diagrams/LoopExpansionBosonicHS_Vertex_Diag9}
\begin{fmfgraph*}(25,25)
\fmfleft{i1,i2}
\fmfright{o1,o2}
\fmfbottom{i0,o0}
\fmftop{i3,o3}
\fmfv{label=$x$,label.dist=4,foreground=(0,,0,,1)}{v1}
\fmfv{label=$y$,label.dist=4,label.angle=0,foreground=(0,,0,,1)}{v2}
\fmfv{label=$z$,label.dist=4,foreground=(0,,0,,1)}{v3}
\fmfv{label=$u$,label.dist=4,label.angle=180,foreground=(0,,0,,1)}{v4}
\fmfv{decor.shape=circle,decor.filled=empty,decor.size=.108w,l=$\times$,label.dist=0}{v2b}
\fmfv{decor.shape=circle,decor.filled=empty,decor.size=.108w,l=$\times$,label.dist=0}{v4b}
\fmf{phantom}{i1,v1}
\fmf{phantom}{i2,v4}
\fmf{phantom}{o1,v2}
\fmf{phantom}{o2,v3}
\fmf{phantom}{i3,v4b}
\fmf{phantom}{o3,v3b}
\fmf{phantom}{i0,v1b}
\fmf{phantom}{o0,v2b}
\fmf{phantom,tension=1.6}{v1,v2}
\fmf{phantom,tension=1.6}{v3,v4}
\fmf{plain,tension=2.0}{v1,v4}
\fmf{plain,tension=2.0}{v2,v3}
\fmf{plain,tension=0}{v1,v3}
\fmf{phantom,tension=0}{v2,v4}
\fmf{phantom}{v1,v1b}
\fmf{plain}{v2,v2b}
\fmf{phantom}{v3,v3b}
\fmf{plain}{v4,v4b}
\fmfdot{v1,v2,v3,v4}
\end{fmfgraph*}
\end{fmffile}
\end{gathered} + \begin{gathered}
\begin{fmffile}{Diagrams/LoopExpansionBosonicHS_Vertex_Diag10}
\begin{fmfgraph*}(25,25)
\fmfleft{i1,i2}
\fmfright{o1,o2}
\fmfbottom{i0,o0}
\fmftop{i3,o3}
\fmfv{label=$x$,label.dist=4,label.angle=180,foreground=(0,,0,,1)}{v1}
\fmfv{label=$y$,label.dist=4,label.angle=0,foreground=(0,,0,,1)}{v2}
\fmfv{label=$z$,label.dist=4,foreground=(0,,0,,1)}{v3}
\fmfv{label=$u$,label.dist=4,foreground=(0,,0,,1)}{v4}
\fmfv{decor.shape=circle,decor.filled=empty,decor.size=.108w,l=$\times$,label.dist=0}{v1b}
\fmfv{decor.shape=circle,decor.filled=empty,decor.size=.108w,l=$\times$,label.dist=0}{v2b}
\fmf{phantom}{i1,v1}
\fmf{phantom}{i2,v4}
\fmf{phantom}{o1,v2}
\fmf{phantom}{o2,v3}
\fmf{phantom}{i3,v4b}
\fmf{phantom}{o3,v3b}
\fmf{phantom}{i0,v1b}
\fmf{phantom}{o0,v2b}
\fmf{phantom,tension=1.6}{v1,v2}
\fmf{plain,tension=1.6}{v3,v4}
\fmf{plain,tension=2.0}{v1,v4}
\fmf{plain,tension=2.0}{v2,v3}
\fmf{phantom,tension=0}{v1,v3}
\fmf{phantom,tension=0}{v2,v4}
\fmf{plain}{v1,v1b}
\fmf{plain}{v2,v2b}
\fmf{phantom}{v3,v3b}
\fmf{phantom}{v4,v4b}
\fmfdot{v1,v2,v3,v4}
\end{fmfgraph*}
\end{fmffile}
\end{gathered} + \begin{gathered}
\begin{fmffile}{Diagrams/LoopExpansionBosonicHS_Vertex_Diag11}
\begin{fmfgraph*}(25,25)
\fmfleft{i1,i2}
\fmfright{o1,o2}
\fmfbottom{i0,o0}
\fmftop{i3,o3}
\fmfv{label=$x$,label.dist=4,label.angle=180,foreground=(0,,0,,1)}{v1}
\fmfv{label=$y$,label.dist=4,foreground=(0,,0,,1)}{v2}
\fmfv{label=$z$,label.dist=4,foreground=(0,,0,,1)}{v3}
\fmfv{label=$u$,label.dist=4,label.angle=180,foreground=(0,,0,,1)}{v4}
\fmfv{decor.shape=circle,decor.filled=empty,decor.size=.108w,l=$\times$,label.dist=0}{v1b}
\fmfv{decor.shape=circle,decor.filled=empty,decor.size=.108w,l=$\times$,label.dist=0}{v4b}
\fmf{phantom}{i1,v1}
\fmf{phantom}{i2,v4}
\fmf{phantom}{o1,v2}
\fmf{phantom}{o2,v3}
\fmf{phantom}{i3,v4b}
\fmf{phantom}{o3,v3b}
\fmf{phantom}{i0,v1b}
\fmf{phantom}{o0,v2b}
\fmf{phantom,tension=1.6}{v1,v2}
\fmf{phantom,tension=1.6}{v3,v4}
\fmf{phantom,tension=2.0}{v1,v4}
\fmf{plain,tension=2.0}{v2,v3}
\fmf{plain,tension=0}{v1,v3}
\fmf{plain,tension=0}{v2,v4}
\fmf{plain}{v1,v1b}
\fmf{phantom}{v2,v2b}
\fmf{phantom}{v3,v3b}
\fmf{plain}{v4,v4b}
\fmfdot{v1,v2,v3,v4}
\end{fmfgraph*}
\end{fmffile}
\end{gathered} }$} \\
& \hspace{0.5cm} \scalebox{0.99}{${\displaystyle + \begin{gathered}
\begin{fmffile}{Diagrams/LoopExpansionBosonicHS_Vertex_Diag12}
\begin{fmfgraph*}(25,25)
\fmfleft{i1,i2}
\fmfright{o1,o2}
\fmfbottom{i0,o0}
\fmftop{i3,o3}
\fmfv{label=$x$,label.dist=4,foreground=(0,,0,,1)}{v1}
\fmfv{label=$y$,label.dist=4,foreground=(0,,0,,1)}{v2}
\fmfv{label=$z$,label.dist=4,label.angle=0,foreground=(0,,0,,1)}{v3}
\fmfv{label=$u$,label.dist=4,label.angle=180,foreground=(0,,0,,1)}{v4}
\fmfv{decor.shape=circle,decor.filled=empty,decor.size=.108w,l=$\times$,label.dist=0}{v3b}
\fmfv{decor.shape=circle,decor.filled=empty,decor.size=.108w,l=$\times$,label.dist=0}{v4b}
\fmf{phantom}{i1,v1}
\fmf{phantom}{i2,v4}
\fmf{phantom}{o1,v2}
\fmf{phantom}{o2,v3}
\fmf{phantom}{i3,v4b}
\fmf{phantom}{o3,v3b}
\fmf{phantom}{i0,v1b}
\fmf{phantom}{o0,v2b}
\fmf{plain,tension=1.6}{v1,v2}
\fmf{phantom,tension=1.6}{v3,v4}
\fmf{phantom,tension=2.0}{v1,v4}
\fmf{phantom,tension=2.0}{v2,v3}
\fmf{plain,tension=0}{v1,v3}
\fmf{plain,tension=0}{v2,v4}
\fmf{phantom}{v1,v1b}
\fmf{phantom}{v2,v2b}
\fmf{plain}{v3,v3b}
\fmf{plain}{v4,v4b}
\fmfdot{v1,v2,v3,v4}
\end{fmfgraph*}
\end{fmffile}
\end{gathered} + \begin{gathered}
\begin{fmffile}{Diagrams/LoopExpansionBosonicHS_Vertex_Diag13}
\begin{fmfgraph*}(25,25)
\fmfleft{i1,i2}
\fmfright{o1,o2}
\fmfbottom{i0,o0}
\fmftop{i3,o3}
\fmfv{label=$x$,label.dist=4,foreground=(0,,0,,1)}{v1}
\fmfv{label=$y$,label.dist=4,foreground=(0,,0,,1)}{v2}
\fmfv{label=$z$,label.dist=4,label.angle=0,foreground=(0,,0,,1)}{v3}
\fmfv{label=$u$,label.dist=4,label.angle=180,foreground=(0,,0,,1)}{v4}
\fmfv{decor.shape=circle,decor.filled=empty,decor.size=.108w,l=$\times$,label.dist=0}{v3b}
\fmfv{decor.shape=circle,decor.filled=empty,decor.size=.108w,l=$\times$,label.dist=0}{v4b}
\fmf{phantom}{i1,v1}
\fmf{phantom}{i2,v4}
\fmf{phantom}{o1,v2}
\fmf{phantom}{o2,v3}
\fmf{phantom}{i3,v4b}
\fmf{phantom}{o3,v3b}
\fmf{phantom}{i0,v1b}
\fmf{phantom}{o0,v2b}
\fmf{plain,tension=1.6}{v1,v2}
\fmf{phantom,tension=1.6}{v3,v4}
\fmf{plain,tension=2.0}{v1,v4}
\fmf{plain,tension=2.0}{v2,v3}
\fmf{phantom,tension=0}{v1,v3}
\fmf{phantom,tension=0}{v2,v4}
\fmf{phantom}{v1,v1b}
\fmf{phantom}{v2,v2b}
\fmf{plain}{v3,v3b}
\fmf{plain}{v4,v4b}
\fmfdot{v1,v2,v3,v4}
\end{fmfgraph*}
\end{fmffile}
\end{gathered} + \begin{gathered}
\begin{fmffile}{Diagrams/LoopExpansionBosonicHS_Vertex_Diag14}
\begin{fmfgraph*}(25,25)
\fmfleft{i1,i2}
\fmfright{o1,o2}
\fmfbottom{i0,o0}
\fmftop{i3,o3}
\fmfv{label=$x$,label.dist=4,label.angle=180,foreground=(0,,0,,1)}{v1}
\fmfv{label=$y$,label.dist=4,foreground=(0,,0,,1)}{v2}
\fmfv{label=$z$,label.dist=4,label.angle=0,foreground=(0,,0,,1)}{v3}
\fmfv{label=$u$,label.dist=4,foreground=(0,,0,,1)}{v4}
\fmfv{decor.shape=circle,decor.filled=empty,decor.size=.108w,l=$\times$,label.dist=0}{v1b}
\fmfv{decor.shape=circle,decor.filled=empty,decor.size=.108w,l=$\times$,label.dist=0}{v3b}
\fmf{phantom}{i1,v1}
\fmf{phantom}{i2,v4}
\fmf{phantom}{o1,v2}
\fmf{phantom}{o2,v3}
\fmf{phantom}{i3,v4b}
\fmf{phantom}{o3,v3b}
\fmf{phantom}{i0,v1b}
\fmf{phantom}{o0,v2b}
\fmf{phantom,tension=1.6}{v1,v2}
\fmf{phantom,tension=1.6}{v3,v4}
\fmf{plain,tension=2.0}{v1,v4}
\fmf{plain,tension=2.0}{v2,v3}
\fmf{phantom,tension=0}{v1,v3}
\fmf{plain,tension=0}{v2,v4}
\fmf{plain}{v1,v1b}
\fmf{phantom}{v2,v2b}
\fmf{plain}{v3,v3b}
\fmf{phantom}{v4,v4b}
\fmfdot{v1,v2,v3,v4}
\end{fmfgraph*}
\end{fmffile}
\end{gathered} + \begin{gathered}
\begin{fmffile}{Diagrams/LoopExpansionBosonicHS_Vertex_Diag15}
\begin{fmfgraph*}(25,25)
\fmfleft{i1,i2}
\fmfright{o1,o2}
\fmfbottom{i0,o0}
\fmftop{i3,o3}
\fmfv{label=$x$,label.dist=4,label.angle=180,foreground=(0,,0,,1)}{v1}
\fmfv{label=$y$,label.dist=4,label.angle=0,foreground=(0,,0,,1)}{v2}
\fmfv{label=$z$,label.dist=4,foreground=(0,,0,,1)}{v3}
\fmfv{label=$u$,label.dist=4,foreground=(0,,0,,1)}{v4}
\fmfv{decor.shape=circle,decor.filled=empty,decor.size=.108w,l=$\times$,label.dist=0}{v1b}
\fmfv{decor.shape=circle,decor.filled=empty,decor.size=.108w,l=$\times$,label.dist=0}{v2b}
\fmf{phantom}{i1,v1}
\fmf{phantom}{i2,v4}
\fmf{phantom}{o1,v2}
\fmf{phantom}{o2,v3}
\fmf{phantom}{i3,v4b}
\fmf{phantom}{o3,v3b}
\fmf{phantom}{i0,v1b}
\fmf{phantom}{o0,v2b}
\fmf{phantom,tension=1.6}{v1,v2}
\fmf{plain,tension=1.6}{v3,v4}
\fmf{phantom,tension=2.0}{v1,v4}
\fmf{phantom,tension=2.0}{v2,v3}
\fmf{plain,tension=0}{v1,v3}
\fmf{plain,tension=0}{v2,v4}
\fmf{plain}{v1,v1b}
\fmf{plain}{v2,v2b}
\fmf{phantom}{v3,v3b}
\fmf{phantom}{v4,v4b}
\fmfdot{v1,v2,v3,v4}
\end{fmfgraph*}
\end{fmffile}
\end{gathered} + \begin{gathered}
\begin{fmffile}{Diagrams/LoopExpansionBosonicHS_Vertex_Diag16}
\begin{fmfgraph*}(25,25)
\fmfleft{i1,i2}
\fmfright{o1,o2}
\fmfbottom{i0,o0}
\fmftop{i3,o3}
\fmfv{label=$x$,label.dist=4,foreground=(0,,0,,1)}{v1}
\fmfv{label=$y$,label.dist=4,label.angle=0,foreground=(0,,0,,1)}{v2}
\fmfv{label=$z$,label.dist=4,label.angle=0,foreground=(0,,0,,1)}{v3}
\fmfv{label=$u$,label.dist=4,foreground=(0,,0,,1)}{v4}
\fmfv{decor.shape=circle,decor.filled=empty,decor.size=.108w,l=$\times$,label.dist=0}{v2b}
\fmfv{decor.shape=circle,decor.filled=empty,decor.size=.108w,l=$\times$,label.dist=0}{v3b}
\fmf{phantom}{i1,v1}
\fmf{phantom}{i2,v4}
\fmf{phantom}{o1,v2}
\fmf{phantom}{o2,v3}
\fmf{phantom}{i3,v4b}
\fmf{phantom}{o3,v3b}
\fmf{phantom}{i0,v1b}
\fmf{phantom}{o0,v2b}
\fmf{phantom,tension=1.6}{v1,v2}
\fmf{phantom,tension=1.6}{v3,v4}
\fmf{plain,tension=2.0}{v1,v4}
\fmf{phantom,tension=2.0}{v2,v3}
\fmf{plain,tension=0}{v1,v3}
\fmf{plain,tension=0}{v2,v4}
\fmf{phantom}{v1,v1b}
\fmf{plain}{v2,v2b}
\fmf{plain}{v3,v3b}
\fmf{phantom}{v4,v4b}
\fmfdot{v1,v2,v3,v4}
\end{fmfgraph*}
\end{fmffile}
\end{gathered} }$} \\
& \hspace{0.5cm} \scalebox{0.99}{${\displaystyle + \begin{gathered}
\begin{fmffile}{Diagrams/LoopExpansionBosonicHS_Vertex_Diag17}
\begin{fmfgraph*}(25,25)
\fmfleft{i1,i2}
\fmfright{o1,o2}
\fmfbottom{i0,o0}
\fmftop{i3,o3}
\fmfv{label=$x$,label.dist=4,foreground=(0,,0,,1)}{v1}
\fmfv{label=$y$,label.dist=4,label.angle=0,foreground=(0,,0,,1)}{v2}
\fmfv{label=$z$,label.dist=4,label.angle=0,foreground=(0,,0,,1)}{v3}
\fmfv{label=$u$,label.dist=4,foreground=(0,,0,,1)}{v4}
\fmfv{decor.shape=circle,decor.filled=empty,decor.size=.108w,l=$\times$,label.dist=0}{v2b}
\fmfv{decor.shape=circle,decor.filled=empty,decor.size=.108w,l=$\times$,label.dist=0}{v3b}
\fmf{phantom}{i1,v1}
\fmf{phantom}{i2,v4}
\fmf{phantom}{o1,v2}
\fmf{phantom}{o2,v3}
\fmf{phantom}{i3,v4b}
\fmf{phantom}{o3,v3b}
\fmf{phantom}{i0,v1b}
\fmf{phantom}{o0,v2b}
\fmf{plain,tension=1.6}{v1,v2}
\fmf{plain,tension=1.6}{v3,v4}
\fmf{plain,tension=2.0}{v1,v4}
\fmf{phantom,tension=2.0}{v2,v3}
\fmf{phantom,tension=0}{v1,v3}
\fmf{phantom,tension=0}{v2,v4}
\fmf{phantom}{v1,v1b}
\fmf{plain}{v2,v2b}
\fmf{plain}{v3,v3b}
\fmf{phantom}{v4,v4b}
\fmfdot{v1,v2,v3,v4}
\end{fmfgraph*}
\end{fmffile}
\end{gathered}+ \begin{gathered}
\begin{fmffile}{Diagrams/LoopExpansionBosonicHS_Vertex_Diag18}
\begin{fmfgraph*}(25,25)
\fmfleft{i1,i2}
\fmfright{o1,o2}
\fmfbottom{i0,o0}
\fmftop{i3,o3}
\fmfv{label=$x$,label.dist=4,label.angle=180,foreground=(0,,0,,1)}{v1}
\fmfv{label=$y$,label.dist=4,foreground=(0,,0,,1)}{v2}
\fmfv{label=$z$,label.dist=4,label.angle=0,foreground=(0,,0,,1)}{v3}
\fmfv{label=$u$,label.dist=4,foreground=(0,,0,,1)}{v4}
\fmfv{decor.shape=circle,decor.filled=empty,decor.size=.108w,l=$\times$,label.dist=0}{v1b}
\fmfv{decor.shape=circle,decor.filled=empty,decor.size=.108w,l=$\times$,label.dist=0}{v3b}
\fmf{phantom}{i1,v1}
\fmf{phantom}{i2,v4}
\fmf{phantom}{o1,v2}
\fmf{phantom}{o2,v3}
\fmf{phantom}{i3,v4b}
\fmf{phantom}{o3,v3b}
\fmf{phantom}{i0,v1b}
\fmf{phantom}{o0,v2b}
\fmf{plain,tension=1.6}{v1,v2}
\fmf{plain,tension=1.6}{v3,v4}
\fmf{phantom,tension=2.0}{v1,v4}
\fmf{phantom,tension=2.0}{v2,v3}
\fmf{phantom,tension=0}{v1,v3}
\fmf{plain,tension=0}{v2,v4}
\fmf{plain}{v1,v1b}
\fmf{phantom}{v2,v2b}
\fmf{plain}{v3,v3b}
\fmf{phantom}{v4,v4b}
\fmfdot{v1,v2,v3,v4}
\end{fmfgraph*}
\end{fmffile}
\end{gathered} + \begin{gathered}
\begin{fmffile}{Diagrams/LoopExpansionBosonicHS_Vertex_Diag19}
\begin{fmfgraph*}(25,25)
\fmfleft{i1,i2}
\fmfright{o1,o2}
\fmfbottom{i0,o0}
\fmftop{i3,o3}
\fmfv{label=$x$,label.dist=4,foreground=(0,,0,,1)}{v1}
\fmfv{label=$y$,label.dist=4,foreground=(0,,0,,1)}{v2}
\fmfv{label=$z$,label.dist=4,foreground=(0,,0,,1)}{v3}
\fmfv{label=$u$,label.dist=4,foreground=(0,,0,,1)}{v4}
\fmf{phantom}{i1,v1}
\fmf{phantom}{i2,v4}
\fmf{phantom}{o1,v2}
\fmf{phantom}{o2,v3}
\fmf{phantom}{i3,v4b}
\fmf{phantom}{o3,v3b}
\fmf{phantom}{i0,v1b}
\fmf{phantom}{o0,v2b}
\fmf{plain,tension=1.6}{v1,v2}
\fmf{plain,tension=1.6}{v3,v4}
\fmf{phantom,tension=2.0}{v1,v4}
\fmf{phantom,tension=2.0}{v2,v3}
\fmf{plain,tension=0}{v1,v3}
\fmf{plain,tension=0}{v2,v4}
\fmf{phantom}{v1,v1b}
\fmf{phantom}{v2,v2b}
\fmf{phantom}{v3,v3b}
\fmf{phantom}{v4,v4b}
\fmfdot{v1,v2,v3,v4}
\end{fmfgraph*}
\end{fmffile}
\end{gathered} + \begin{gathered}
\begin{fmffile}{Diagrams/LoopExpansionBosonicHS_Vertex_Diag20}
\begin{fmfgraph*}(25,25)
\fmfleft{i1,i2}
\fmfright{o1,o2}
\fmfbottom{i0,o0}
\fmftop{i3,o3}
\fmfv{label=$x$,label.dist=4,foreground=(0,,0,,1)}{v1}
\fmfv{label=$y$,label.dist=4,foreground=(0,,0,,1)}{v2}
\fmfv{label=$z$,label.dist=4,foreground=(0,,0,,1)}{v3}
\fmfv{label=$u$,label.dist=4,foreground=(0,,0,,1)}{v4}
\fmf{phantom}{i1,v1}
\fmf{phantom}{i2,v4}
\fmf{phantom}{o1,v2}
\fmf{phantom}{o2,v3}
\fmf{phantom}{i3,v4b}
\fmf{phantom}{o3,v3b}
\fmf{phantom}{i0,v1b}
\fmf{phantom}{o0,v2b}
\fmf{phantom,tension=1.6}{v1,v2}
\fmf{phantom,tension=1.6}{v3,v4}
\fmf{plain,tension=2.0}{v1,v4}
\fmf{plain,tension=2.0}{v2,v3}
\fmf{plain,tension=0}{v1,v3}
\fmf{plain,tension=0}{v2,v4}
\fmf{phantom}{v1,v1b}
\fmf{phantom}{v2,v2b}
\fmf{phantom}{v3,v3b}
\fmf{phantom}{v4,v4b}
\fmfdot{v1,v2,v3,v4}
\end{fmfgraph*}
\end{fmffile}
\end{gathered}+ \begin{gathered}
\begin{fmffile}{Diagrams/LoopExpansionBosonicHS_Vertex_Diag21}
\begin{fmfgraph*}(25,25)
\fmfleft{i1,i2}
\fmfright{o1,o2}
\fmfbottom{i0,o0}
\fmftop{i3,o3}
\fmfv{label=$x$,label.dist=4,foreground=(0,,0,,1)}{v1}
\fmfv{label=$y$,label.dist=4,foreground=(0,,0,,1)}{v2}
\fmfv{label=$z$,label.dist=4,foreground=(0,,0,,1)}{v3}
\fmfv{label=$u$,label.dist=4,foreground=(0,,0,,1)}{v4}
\fmf{phantom}{i1,v1}
\fmf{phantom}{i2,v4}
\fmf{phantom}{o1,v2}
\fmf{phantom}{o2,v3}
\fmf{phantom}{i3,v4b}
\fmf{phantom}{o3,v3b}
\fmf{phantom}{i0,v1b}
\fmf{phantom}{o0,v2b}
\fmf{plain,tension=1.6}{v1,v2}
\fmf{plain,tension=1.6}{v3,v4}
\fmf{plain,tension=2.0}{v1,v4}
\fmf{plain,tension=2.0}{v2,v3}
\fmf{phantom,tension=0}{v1,v3}
\fmf{phantom,tension=0}{v2,v4}
\fmf{phantom}{v1,v1b}
\fmf{phantom}{v2,v2b}
\fmf{phantom}{v3,v3b}
\fmf{phantom}{v4,v4b}
\fmfdot{v1,v2,v3,v4}
\end{fmfgraph*}
\end{fmffile}
\end{gathered} \hspace{-0.31cm} \left.\rule{0cm}{1.1cm}\right)\;. }$}
\end{split}
\label{eq:SbosonicKLoopExpansionS4}
\end{equation}
After plugging the vertex functions~\eqref{eq:SbosonicKLoopExpansionS3} and~\eqref{eq:SbosonicKLoopExpansionS4} into~\eqref{eq:ApplicationWickTheoremBosonicActionS4} and~\eqref{eq:ApplicationWickTheoremBosonicActionS3} combined with~\eqref{eq:Zlecol}, the Schwinger functional in the collective representation is expressed up to order $\mathcal{O}\big(\hbar^2\big)$ as follows:
\begin{equation}
\begin{split}
W^\text{LE;col}\big[\mathcal{J}\big] = & -S_{\mathrm{col},\mathcal{J}}[\sigma_\text{cl}]+\frac{\hbar}{2}\mathrm{Tr}\left[\ln\big(D_{\sigma_\text{cl};\mathcal{J}}\big)\right] \\
& + \hbar^{2} \left[\rule{0cm}{1.1cm}\right. \frac{1}{8} \left(\rule{0cm}{1.1cm}\right. 4 \hspace{-0.2cm} \begin{gathered}
\begin{fmffile}{Diagrams/LoopExpansionBosonicHS_W_Diag1}
\begin{fmfgraph*}(25,25)
\fmfleft{i1,i2}
\fmfright{o1,o2}
\fmfbottom{i0,o0}
\fmftop{i3,o3}
\fmfv{decor.shape=circle,decor.size=2.0thick,foreground=(0,,0,,1)}{v1}
\fmfv{decor.shape=circle,decor.size=2.0thick,foreground=(0,,0,,1)}{v2}
\fmfv{decor.shape=circle,decor.size=2.0thick,foreground=(0,,0,,1)}{v3}
\fmfv{decor.shape=circle,decor.size=2.0thick,foreground=(0,,0,,1)}{v4}
\fmfv{decor.shape=circle,decor.filled=empty,decor.size=.1w,l=$\times$,label.dist=0}{v3b}
\fmfv{decor.shape=circle,decor.filled=empty,decor.size=.1w,l=$\times$,label.dist=0}{v4b}
\fmf{phantom}{i1,v1}
\fmf{phantom}{i2,v4}
\fmf{phantom}{o1,v2}
\fmf{phantom}{o2,v3}
\fmf{phantom}{i3,v4b}
\fmf{phantom}{o3,v3b}
\fmf{phantom}{i0,v1b}
\fmf{phantom}{o0,v2b}
\fmf{plain,tension=1.6}{v1,v2}
\fmf{phantom,tension=1.6}{v3,v4}
\fmf{wiggly,tension=2.0}{v1,v4}
\fmf{wiggly,tension=2.0}{v2,v3}
\fmf{phantom,tension=0}{v1,v3}
\fmf{phantom,tension=0}{v2,v4}
\fmf{plain,right=0.8,tension=0}{v2,v3}
\fmf{plain,left=0.8,tension=0}{v1,v4}
\fmf{phantom}{v1,v1b}
\fmf{phantom}{v2,v2b}
\fmf{plain}{v3,v3b}
\fmf{plain}{v4,v4b}
\end{fmfgraph*}
\end{fmffile}
\end{gathered} \hspace{-0.2cm} + 4 \hspace{-0.2cm} \begin{gathered}
\begin{fmffile}{Diagrams/LoopExpansionBosonicHS_W_Diag2}
\begin{fmfgraph*}(25,25)
\fmfleft{i1,i2}
\fmfright{o1,o2}
\fmfbottom{i0,o0}
\fmftop{i3,o3}
\fmfv{decor.shape=circle,decor.size=2.0thick,foreground=(0,,0,,1)}{v1}
\fmfv{decor.shape=circle,decor.size=2.0thick,foreground=(0,,0,,1)}{v2}
\fmfv{decor.shape=circle,decor.size=2.0thick,foreground=(0,,0,,1)}{v3}
\fmfv{decor.shape=circle,decor.size=2.0thick,foreground=(0,,0,,1)}{v4}
\fmfv{decor.shape=circle,decor.filled=empty,decor.size=.1w,l=$\times$,label.dist=0}{v1b}
\fmfv{decor.shape=circle,decor.filled=empty,decor.size=.1w,l=$\times$,label.dist=0}{v4b}
\fmf{phantom}{i1,v1}
\fmf{phantom}{i2,v4}
\fmf{phantom}{o1,v2}
\fmf{phantom}{o2,v3}
\fmf{phantom}{i3,v4b}
\fmf{phantom}{o3,v3b}
\fmf{phantom}{i0,v1b}
\fmf{phantom}{o0,v2b}
\fmf{plain,tension=1.6}{v1,v2}
\fmf{plain,tension=1.6}{v3,v4}
\fmf{wiggly,tension=2.0}{v1,v4}
\fmf{wiggly,tension=2.0}{v2,v3}
\fmf{phantom,tension=0}{v1,v3}
\fmf{phantom,tension=0}{v2,v4}
\fmf{plain,right=0.8,tension=0}{v2,v3}
\fmf{phantom,left=0.8,tension=0}{v1,v4}
\fmf{plain}{v1,v1b}
\fmf{phantom}{v2,v2b}
\fmf{phantom}{v3,v3b}
\fmf{plain}{v4,v4b}
\end{fmfgraph*}
\end{fmffile}
\end{gathered} \hspace{-0.5cm} + 4 \hspace{-0.3cm} \begin{gathered}
\begin{fmffile}{Diagrams/LoopExpansionBosonicHS_W_Diag3}
\begin{fmfgraph*}(25,25)
\fmfleft{i1,i2}
\fmfright{o1,o2}
\fmfbottom{i0,o0}
\fmftop{i3,o3}
\fmfv{decor.shape=circle,decor.size=2.0thick,foreground=(0,,0,,1)}{v1}
\fmfv{decor.shape=circle,decor.size=2.0thick,foreground=(0,,0,,1)}{v2}
\fmfv{decor.shape=circle,decor.size=2.0thick,foreground=(0,,0,,1)}{v3}
\fmfv{decor.shape=circle,decor.size=2.0thick,foreground=(0,,0,,1)}{v4}
\fmfv{decor.shape=circle,decor.filled=empty,decor.size=.1w,l=$\times$,label.dist=0}{v3b}
\fmfv{decor.shape=circle,decor.filled=empty,decor.size=.1w,l=$\times$,label.dist=0}{v4b}
\fmf{phantom}{i1,v1}
\fmf{phantom}{i2,v4}
\fmf{phantom}{o1,v2}
\fmf{phantom}{o2,v3}
\fmf{phantom}{i3,v4b}
\fmf{phantom}{o3,v3b}
\fmf{phantom}{i0,v1b}
\fmf{phantom}{o0,v2b}
\fmf{plain,tension=1.6}{v1,v2}
\fmf{phantom,tension=1.6}{v3,v4}
\fmf{wiggly,tension=2.0}{v1,v4}
\fmf{wiggly,tension=2.0}{v2,v3}
\fmf{plain,tension=0}{v1,v3}
\fmf{plain,tension=0}{v2,v4}
\fmf{phantom,right=0.8,tension=0}{v2,v3}
\fmf{phantom,left=0.8,tension=0}{v1,v4}
\fmf{phantom}{v1,v1b}
\fmf{phantom}{v2,v2b}
\fmf{plain}{v3,v3b}
\fmf{plain}{v4,v4b}
\end{fmfgraph*}
\end{fmffile}
\end{gathered} \hspace{-0.4cm} + 2 \hspace{-0.4cm} \begin{gathered}
\begin{fmffile}{Diagrams/LoopExpansionBosonicHS_W_Diag5}
\begin{fmfgraph*}(25,25)
\fmfleft{i1,i2}
\fmfright{o1,o2}
\fmfbottom{i0,o0}
\fmftop{i3,o3}
\fmfv{decor.shape=circle,decor.size=2.0thick,foreground=(0,,0,,1)}{v1}
\fmfv{decor.shape=circle,decor.size=2.0thick,foreground=(0,,0,,1)}{v2}
\fmfv{decor.shape=circle,decor.size=2.0thick,foreground=(0,,0,,1)}{v3}
\fmfv{decor.shape=circle,decor.size=2.0thick,foreground=(0,,0,,1)}{v4}
\fmf{phantom}{i1,v1}
\fmf{phantom}{i2,v4}
\fmf{phantom}{o1,v2}
\fmf{phantom}{o2,v3}
\fmf{phantom}{i3,v4b}
\fmf{phantom}{o3,v3b}
\fmf{phantom}{i0,v1b}
\fmf{phantom}{o0,v2b}
\fmf{plain,tension=1.6}{v1,v2}
\fmf{plain,tension=1.6}{v3,v4}
\fmf{wiggly,tension=2.0}{v1,v4}
\fmf{wiggly,tension=2.0}{v2,v3}
\fmf{phantom,tension=0}{v1,v3}
\fmf{phantom,tension=0}{v2,v4}
\fmf{plain,right=0.8,tension=0}{v2,v3}
\fmf{plain,left=0.8,tension=0}{v1,v4}
\fmf{phantom}{v1,v1b}
\fmf{phantom}{v2,v2b}
\fmf{phantom}{v3,v3b}
\fmf{phantom}{v4,v4b}
\end{fmfgraph*}
\end{fmffile}
\end{gathered} \hspace{-0.5cm} + \hspace{-0.7cm} \begin{gathered}
\begin{fmffile}{Diagrams/LoopExpansionBosonicHS_W_Diag6}
\begin{fmfgraph*}(25,25)
\fmfleft{i1,i2}
\fmfright{o1,o2}
\fmfbottom{i0,o0}
\fmftop{i3,o3}
\fmfv{decor.shape=circle,decor.size=2.0thick,foreground=(0,,0,,1)}{v1}
\fmfv{decor.shape=circle,decor.size=2.0thick,foreground=(0,,0,,1)}{v2}
\fmfv{decor.shape=circle,decor.size=2.0thick,foreground=(0,,0,,1)}{v3}
\fmfv{decor.shape=circle,decor.size=2.0thick,foreground=(0,,0,,1)}{v4}
\fmf{phantom}{i1,v1}
\fmf{phantom}{i2,v4}
\fmf{phantom}{o1,v2}
\fmf{phantom}{o2,v3}
\fmf{phantom}{i3,v4b}
\fmf{phantom}{o3,v3b}
\fmf{phantom}{i0,v1b}
\fmf{phantom}{o0,v2b}
\fmf{plain,tension=1.6}{v1,v2}
\fmf{plain,tension=1.6}{v3,v4}
\fmf{wiggly,tension=2.0}{v1,v4}
\fmf{wiggly,tension=2.0}{v2,v3}
\fmf{plain,tension=0}{v1,v3}
\fmf{plain,tension=0}{v2,v4}
\fmf{phantom,right=0.8,tension=0}{v2,v3}
\fmf{phantom,left=0.8,tension=0}{v1,v4}
\fmf{phantom}{v1,v1b}
\fmf{phantom}{v2,v2b}
\fmf{phantom}{v3,v3b}
\fmf{phantom}{v4,v4b}
\end{fmfgraph*}
\end{fmffile}
\end{gathered} \hspace{-0.6cm} \left.\rule{0cm}{1.1cm}\right) \\
& \hspace{1.0cm} + \frac{1}{12} \left(\rule{0cm}{1.1cm}\right. 6 \begin{gathered}
\begin{fmffile}{Diagrams/LoopExpansionBosonicHS_W_Diag7}
\begin{fmfgraph*}(35,20)
\fmfleft{i1,i2}
\fmfright{o1,o2}
\fmfbottom{i0,o0}
\fmfbottom{b0}
\fmfbottom{b1}
\fmfbottom{b2}
\fmftop{i3,o3}
\fmfv{decor.shape=circle,decor.size=2.0thick,foreground=(0,,0,,1)}{v1}
\fmfv{decor.shape=circle,decor.size=2.0thick,foreground=(0,,0,,1)}{v2}
\fmfv{decor.shape=circle,decor.size=2.0thick,foreground=(0,,0,,1)}{v3}
\fmfv{decor.shape=circle,decor.size=2.0thick,foreground=(0,,0,,1)}{v4}
\fmfv{decor.shape=circle,decor.size=2.0thick,foreground=(0,,0,,1)}{v5}
\fmfv{decor.shape=circle,decor.size=2.0thick,foreground=(0,,0,,1)}{v6}
\fmfv{decor.shape=circle,decor.filled=empty,decor.size=.073w,l=$\times$,label.dist=0}{v1b}
\fmfv{decor.shape=circle,decor.filled=empty,decor.size=.073w,l=$\times$,label.dist=0}{v3b}
\fmfv{decor.shape=circle,decor.filled=empty,decor.size=.073w,l=$\times$,label.dist=0}{v4b}
\fmfv{decor.shape=circle,decor.filled=empty,decor.size=.073w,l=$\times$,label.dist=0}{v6b}
\fmf{phantom,tension=1.4}{i1,v1}
\fmf{phantom}{i2,v3b}
\fmf{phantom}{i0,v1b}
\fmf{phantom}{o2,v6b}
\fmf{phantom,tension=1.4}{o1,v5}
\fmf{phantom}{o0,v5b}
\fmf{phantom,tension=1.11}{v3b,v6b}
\fmf{phantom,tension=1.38}{i0,v2}
\fmf{phantom,tension=1.38}{o0,v2}
\fmf{phantom,tension=1.8}{i0,v2b}
\fmf{phantom,tension=1.2}{o0,v2b}
\fmf{phantom,tension=1.2}{b0,v2b}
\fmf{phantom,tension=1.2}{b1,v2b}
\fmf{phantom,tension=1.2}{b2,v2b}
\fmf{phantom,tension=1.38}{i0,v4}
\fmf{phantom,tension=1.38}{o0,v4}
\fmf{phantom,tension=1.2}{i0,v4b}
\fmf{phantom,tension=1.8}{o0,v4b}
\fmf{phantom,tension=1.2}{b0,v4b}
\fmf{phantom,tension=1.2}{b1,v4b}
\fmf{phantom,tension=1.2}{b2,v4b}
\fmf{phantom,tension=2}{i3,v3}
\fmf{phantom,tension=2}{o3,v6}
\fmf{phantom,tension=2}{i3,v3b}
\fmf{phantom,tension=0.8}{o3,v3b}
\fmf{phantom,tension=0.8}{i3,v6b}
\fmf{phantom,tension=2}{o3,v6b}
\fmf{plain,tension=1.4}{v1,v2}
\fmf{plain,tension=1.4}{v4,v5}
\fmf{phantom}{v1,v3}
\fmf{phantom,left=0.8,tension=0}{v1,v3}
\fmf{plain}{v5,v6}
\fmf{phantom,right=0.8,tension=0}{v5,v6}
\fmf{plain}{v2,v3}
\fmf{phantom}{v4,v6}
\fmf{wiggly,tension=0.5}{v2,v4}
\fmf{wiggly,tension=2}{v3,v6}
\fmf{wiggly,right=0.8,tension=0}{v1,v5}
\fmf{plain,tension=1}{v1,v1b}
\fmf{phantom,tension=0.2}{v2,v2b}
\fmf{plain,tension=1.5}{v3,v3b}
\fmf{plain,tension=0.2}{v4,v4b}
\fmf{phantom,tension=1}{v5,v5b}
\fmf{plain,tension=1.5}{v6,v6b}
\end{fmfgraph*}
\end{fmffile}
\end{gathered} \hspace{-0.5cm} + 3 \begin{gathered}
\begin{fmffile}{Diagrams/LoopExpansionBosonicHS_W_Diag8}
\begin{fmfgraph*}(35,20)
\fmfleft{i1,i2}
\fmfright{o1,o2}
\fmfbottom{i0,o0}
\fmfbottom{b0}
\fmfbottom{b1}
\fmfbottom{b2}
\fmftop{i3,o3}
\fmfv{decor.shape=circle,decor.size=2.0thick,foreground=(0,,0,,1)}{v1}
\fmfv{decor.shape=circle,decor.size=2.0thick,foreground=(0,,0,,1)}{v2}
\fmfv{decor.shape=circle,decor.size=2.0thick,foreground=(0,,0,,1)}{v3}
\fmfv{decor.shape=circle,decor.size=2.0thick,foreground=(0,,0,,1)}{v4}
\fmfv{decor.shape=circle,decor.size=2.0thick,foreground=(0,,0,,1)}{v5}
\fmfv{decor.shape=circle,decor.size=2.0thick,foreground=(0,,0,,1)}{v6}
\fmfv{decor.shape=circle,decor.filled=empty,decor.size=.073w,l=$\times$,label.dist=0}{v1b}
\fmfv{decor.shape=circle,decor.filled=empty,decor.size=.073w,l=$\times$,label.dist=0}{v3b}
\fmfv{decor.shape=circle,decor.filled=empty,decor.size=.073w,l=$\times$,label.dist=0}{v5b}
\fmfv{decor.shape=circle,decor.filled=empty,decor.size=.073w,l=$\times$,label.dist=0}{v6b}
\fmf{phantom,tension=1.4}{i1,v1}
\fmf{phantom}{i2,v3b}
\fmf{phantom}{i0,v1b}
\fmf{phantom}{o2,v6b}
\fmf{phantom,tension=1.4}{o1,v5}
\fmf{phantom}{o0,v5b}
\fmf{phantom,tension=1.11}{v3b,v6b}
\fmf{phantom,tension=1.38}{i0,v2}
\fmf{phantom,tension=1.38}{o0,v2}
\fmf{phantom,tension=1.8}{i0,v2b}
\fmf{phantom,tension=1.2}{o0,v2b}
\fmf{phantom,tension=1.2}{b0,v2b}
\fmf{phantom,tension=1.2}{b1,v2b}
\fmf{phantom,tension=1.2}{b2,v2b}
\fmf{phantom,tension=1.38}{i0,v4}
\fmf{phantom,tension=1.38}{o0,v4}
\fmf{phantom,tension=1.2}{i0,v4b}
\fmf{phantom,tension=1.8}{o0,v4b}
\fmf{phantom,tension=1.2}{b0,v4b}
\fmf{phantom,tension=1.2}{b1,v4b}
\fmf{phantom,tension=1.2}{b2,v4b}
\fmf{phantom,tension=2}{i3,v3}
\fmf{phantom,tension=2}{o3,v6}
\fmf{phantom,tension=2}{i3,v3b}
\fmf{phantom,tension=0.8}{o3,v3b}
\fmf{phantom,tension=0.8}{i3,v6b}
\fmf{phantom,tension=2}{o3,v6b}
\fmf{plain,tension=1.4}{v1,v2}
\fmf{plain,tension=1.4}{v4,v5}
\fmf{phantom}{v1,v3}
\fmf{phantom,left=0.8,tension=0}{v1,v3}
\fmf{phantom}{v5,v6}
\fmf{phantom,right=0.8,tension=0}{v5,v6}
\fmf{plain}{v2,v3}
\fmf{plain}{v4,v6}
\fmf{wiggly,tension=0.5}{v2,v4}
\fmf{wiggly,tension=2}{v3,v6}
\fmf{wiggly,right=0.8,tension=0}{v1,v5}
\fmf{plain,tension=1}{v1,v1b}
\fmf{phantom,tension=0.2}{v2,v2b}
\fmf{plain,tension=1.5}{v3,v3b}
\fmf{phantom,tension=0.2}{v4,v4b}
\fmf{plain,tension=1}{v5,v5b}
\fmf{plain,tension=1.5}{v6,v6b}
\end{fmfgraph*}
\end{fmffile}
\end{gathered} \hspace{-0.3cm} \\
& \hspace{2.2cm} + 6 \begin{gathered}
\begin{fmffile}{Diagrams/LoopExpansionBosonicHS_W_Diag9}
\begin{fmfgraph*}(35,20)
\fmfleft{i1,i2}
\fmfright{o1,o2}
\fmfbottom{i0,o0}
\fmfbottom{b0}
\fmfbottom{b1}
\fmfbottom{b2}
\fmftop{i3,o3}
\fmfv{decor.shape=circle,decor.size=2.0thick,foreground=(0,,0,,1)}{v1}
\fmfv{decor.shape=circle,decor.size=2.0thick,foreground=(0,,0,,1)}{v2}
\fmfv{decor.shape=circle,decor.size=2.0thick,foreground=(0,,0,,1)}{v3}
\fmfv{decor.shape=circle,decor.size=2.0thick,foreground=(0,,0,,1)}{v4}
\fmfv{decor.shape=circle,decor.size=2.0thick,foreground=(0,,0,,1)}{v5}
\fmfv{decor.shape=circle,decor.size=2.0thick,foreground=(0,,0,,1)}{v6}
\fmfv{decor.shape=circle,decor.filled=empty,decor.size=.073w,l=$\times$,label.dist=0}{v1b}
\fmfv{decor.shape=circle,decor.filled=empty,decor.size=.073w,l=$\times$,label.dist=0}{v3b}
\fmf{phantom,tension=1.4}{i1,v1}
\fmf{phantom}{i2,v3b}
\fmf{phantom}{i0,v1b}
\fmf{phantom}{o2,v6b}
\fmf{phantom,tension=1.4}{o1,v5}
\fmf{phantom}{o0,v5b}
\fmf{phantom,tension=1.11}{v3b,v6b}
\fmf{phantom,tension=1.38}{i0,v2}
\fmf{phantom,tension=1.38}{o0,v2}
\fmf{phantom,tension=1.8}{i0,v2b}
\fmf{phantom,tension=1.2}{o0,v2b}
\fmf{phantom,tension=1.2}{b0,v2b}
\fmf{phantom,tension=1.2}{b1,v2b}
\fmf{phantom,tension=1.2}{b2,v2b}
\fmf{phantom,tension=1.38}{i0,v4}
\fmf{phantom,tension=1.38}{o0,v4}
\fmf{phantom,tension=1.2}{i0,v4b}
\fmf{phantom,tension=1.8}{o0,v4b}
\fmf{phantom,tension=1.2}{b0,v4b}
\fmf{phantom,tension=1.2}{b1,v4b}
\fmf{phantom,tension=1.2}{b2,v4b}
\fmf{phantom,tension=2}{i3,v3}
\fmf{phantom,tension=2}{o3,v6}
\fmf{phantom,tension=2}{i3,v3b}
\fmf{phantom,tension=0.8}{o3,v3b}
\fmf{phantom,tension=0.8}{i3,v6b}
\fmf{phantom,tension=2}{o3,v6b}
\fmf{plain,tension=1.4}{v1,v2}
\fmf{plain,tension=1.4}{v4,v5}
\fmf{phantom}{v1,v3}
\fmf{phantom,left=0.8,tension=0}{v1,v3}
\fmf{plain}{v5,v6}
\fmf{phantom,right=0.8,tension=0}{v5,v6}
\fmf{plain}{v2,v3}
\fmf{plain}{v4,v6}
\fmf{wiggly,tension=0.5}{v2,v4}
\fmf{wiggly,tension=2}{v3,v6}
\fmf{wiggly,right=0.8,tension=0}{v1,v5}
\fmf{plain,tension=1}{v1,v1b}
\fmf{phantom,tension=0.2}{v2,v2b}
\fmf{plain,tension=1.5}{v3,v3b}
\fmf{phantom,tension=0.2}{v4,v4b}
\fmf{phantom,tension=1}{v5,v5b}
\fmf{phantom,tension=1.5}{v6,v6b}
\end{fmfgraph*}
\end{fmffile}
\end{gathered} \hspace{-0.5cm} + \hspace{-0.5cm} \begin{gathered}
\begin{fmffile}{Diagrams/LoopExpansionBosonicHS_W_Diag10}
\begin{fmfgraph*}(35,20)
\fmfleft{i1,i2}
\fmfright{o1,o2}
\fmfbottom{i0,o0}
\fmfbottom{b0}
\fmfbottom{b1}
\fmfbottom{b2}
\fmftop{i3,o3}
\fmfv{decor.shape=circle,decor.size=2.0thick,foreground=(0,,0,,1)}{v1}
\fmfv{decor.shape=circle,decor.size=2.0thick,foreground=(0,,0,,1)}{v2}
\fmfv{decor.shape=circle,decor.size=2.0thick,foreground=(0,,0,,1)}{v3}
\fmfv{decor.shape=circle,decor.size=2.0thick,foreground=(0,,0,,1)}{v4}
\fmfv{decor.shape=circle,decor.size=2.0thick,foreground=(0,,0,,1)}{v5}
\fmfv{decor.shape=circle,decor.size=2.0thick,foreground=(0,,0,,1)}{v6}
\fmf{phantom,tension=1.4}{i1,v1}
\fmf{phantom}{i2,v3b}
\fmf{phantom}{i0,v1b}
\fmf{phantom}{o2,v6b}
\fmf{phantom,tension=1.4}{o1,v5}
\fmf{phantom}{o0,v5b}
\fmf{phantom,tension=1.11}{v3b,v6b}
\fmf{phantom,tension=1.38}{i0,v2}
\fmf{phantom,tension=1.38}{o0,v2}
\fmf{phantom,tension=1.8}{i0,v2b}
\fmf{phantom,tension=1.2}{o0,v2b}
\fmf{phantom,tension=1.2}{b0,v2b}
\fmf{phantom,tension=1.2}{b1,v2b}
\fmf{phantom,tension=1.2}{b2,v2b}
\fmf{phantom,tension=1.38}{i0,v4}
\fmf{phantom,tension=1.38}{o0,v4}
\fmf{phantom,tension=1.2}{i0,v4b}
\fmf{phantom,tension=1.8}{o0,v4b}
\fmf{phantom,tension=1.2}{b0,v4b}
\fmf{phantom,tension=1.2}{b1,v4b}
\fmf{phantom,tension=1.2}{b2,v4b}
\fmf{phantom,tension=2}{i3,v3}
\fmf{phantom,tension=2}{o3,v6}
\fmf{phantom,tension=2}{i3,v3b}
\fmf{phantom,tension=0.8}{o3,v3b}
\fmf{phantom,tension=0.8}{i3,v6b}
\fmf{phantom,tension=2}{o3,v6b}
\fmf{plain,tension=1.4}{v1,v2}
\fmf{plain,tension=1.4}{v4,v5}
\fmf{plain}{v1,v3}
\fmf{phantom,left=0.8,tension=0}{v1,v3}
\fmf{plain}{v5,v6}
\fmf{phantom,right=0.8,tension=0}{v5,v6}
\fmf{plain}{v2,v3}
\fmf{plain}{v4,v6}
\fmf{wiggly,tension=0.5}{v2,v4}
\fmf{wiggly,tension=2}{v3,v6}
\fmf{wiggly,right=0.8,tension=0}{v1,v5}
\fmf{phantom,tension=1}{v1,v1b}
\fmf{phantom,tension=0.2}{v2,v2b}
\fmf{phantom,tension=1.5}{v3,v3b}
\fmf{phantom,tension=0.2}{v4,v4b}
\fmf{phantom,tension=1}{v5,v5b}
\fmf{phantom,tension=1.5}{v6,v6b}
\end{fmfgraph*}
\end{fmffile}
\end{gathered} \hspace{-0.6cm} \left.\rule{0cm}{1.1cm}\right) \\
& \hspace{1.0cm} + \frac{1}{8} \left(\rule{0cm}{1.1cm}\right. 4 \begin{gathered}
\begin{fmffile}{Diagrams/LoopExpansionBosonicHS_W_Diag11}
\begin{fmfgraph*}(35,20)
\fmfleft{i1,i2}
\fmfright{o1,o2}
\fmfbottom{i0,o0}
\fmfbottom{b0}
\fmfbottom{b1}
\fmfbottom{b2}
\fmftop{i3,o3}
\fmfv{decor.shape=circle,decor.size=2.0thick,foreground=(0,,0,,1)}{v1}
\fmfv{decor.shape=circle,decor.size=2.0thick,foreground=(0,,0,,1)}{v2}
\fmfv{decor.shape=circle,decor.size=2.0thick,foreground=(0,,0,,1)}{v3}
\fmfv{decor.shape=circle,decor.size=2.0thick,foreground=(0,,0,,1)}{v4}
\fmfv{decor.shape=circle,decor.size=2.0thick,foreground=(0,,0,,1)}{v5}
\fmfv{decor.shape=circle,decor.size=2.0thick,foreground=(0,,0,,1)}{v6}
\fmfv{decor.shape=circle,decor.filled=empty,decor.size=.073w,l=$\times$,label.dist=0}{v1b}
\fmfv{decor.shape=circle,decor.filled=empty,decor.size=.073w,l=$\times$,label.dist=0}{v3b}
\fmfv{decor.shape=circle,decor.filled=empty,decor.size=.073w,l=$\times$,label.dist=0}{v4b}
\fmfv{decor.shape=circle,decor.filled=empty,decor.size=.073w,l=$\times$,label.dist=0}{v6b}
\fmf{phantom,tension=1.4}{i1,v1}
\fmf{phantom}{i2,v3b}
\fmf{phantom}{i0,v1b}
\fmf{phantom}{o2,v6b}
\fmf{phantom,tension=1.4}{o1,v5}
\fmf{phantom}{o0,v5b}
\fmf{phantom,tension=1.11}{v3b,v6b}
\fmf{phantom,tension=1.38}{i0,v2}
\fmf{phantom,tension=1.38}{o0,v2}
\fmf{phantom,tension=1.8}{i0,v2b}
\fmf{phantom,tension=1.2}{o0,v2b}
\fmf{phantom,tension=1.2}{b0,v2b}
\fmf{phantom,tension=1.2}{b1,v2b}
\fmf{phantom,tension=1.2}{b2,v2b}
\fmf{phantom,tension=1.38}{i0,v4}
\fmf{phantom,tension=1.38}{o0,v4}
\fmf{phantom,tension=1.2}{i0,v4b}
\fmf{phantom,tension=1.8}{o0,v4b}
\fmf{phantom,tension=1.2}{b0,v4b}
\fmf{phantom,tension=1.2}{b1,v4b}
\fmf{phantom,tension=1.2}{b2,v4b}
\fmf{phantom,tension=2}{i3,v3}
\fmf{phantom,tension=2}{o3,v6}
\fmf{phantom,tension=2}{i3,v3b}
\fmf{phantom,tension=0.8}{o3,v3b}
\fmf{phantom,tension=0.8}{i3,v6b}
\fmf{phantom,tension=2}{o3,v6b}
\fmf{plain,tension=1.4}{v1,v2}
\fmf{plain,tension=1.4}{v4,v5}
\fmf{wiggly}{v1,v3}
\fmf{phantom,left=0.8,tension=0}{v1,v3}
\fmf{wiggly}{v5,v6}
\fmf{plain,right=0.8,tension=0}{v5,v6}
\fmf{plain}{v2,v3}
\fmf{phantom}{v4,v6}
\fmf{wiggly,tension=0.5}{v2,v4}
\fmf{phantom,tension=2}{v3,v6}
\fmf{phantom,right=0.8,tension=0}{v1,v5}
\fmf{plain,tension=1}{v1,v1b}
\fmf{phantom,tension=0.2}{v2,v2b}
\fmf{plain,tension=1.5}{v3,v3b}
\fmf{plain,tension=0.2}{v4,v4b}
\fmf{phantom,tension=1}{v5,v5b}
\fmf{plain,tension=1.5}{v6,v6b}
\end{fmfgraph*}
\end{fmffile}
\end{gathered} \hspace{-0.4cm} + 4 \begin{gathered}
\begin{fmffile}{Diagrams/LoopExpansionBosonicHS_W_Diag12}
\begin{fmfgraph*}(35,20)
\fmfleft{i1,i2}
\fmfright{o1,o2}
\fmfbottom{i0,o0}
\fmfbottom{b0}
\fmfbottom{b1}
\fmfbottom{b2}
\fmftop{i3,o3}
\fmfv{decor.shape=circle,decor.size=2.0thick,foreground=(0,,0,,1)}{v1}
\fmfv{decor.shape=circle,decor.size=2.0thick,foreground=(0,,0,,1)}{v2}
\fmfv{decor.shape=circle,decor.size=2.0thick,foreground=(0,,0,,1)}{v3}
\fmfv{decor.shape=circle,decor.size=2.0thick,foreground=(0,,0,,1)}{v4}
\fmfv{decor.shape=circle,decor.size=2.0thick,foreground=(0,,0,,1)}{v5}
\fmfv{decor.shape=circle,decor.size=2.0thick,foreground=(0,,0,,1)}{v6}
\fmfv{decor.shape=circle,decor.filled=empty,decor.size=.073w,l=$\times$,label.dist=0}{v2b}
\fmfv{decor.shape=circle,decor.filled=empty,decor.size=.073w,l=$\times$,label.dist=0}{v3b}
\fmfv{decor.shape=circle,decor.filled=empty,decor.size=.073w,l=$\times$,label.dist=0}{v4b}
\fmfv{decor.shape=circle,decor.filled=empty,decor.size=.073w,l=$\times$,label.dist=0}{v6b}
\fmf{phantom,tension=1.4}{i1,v1}
\fmf{phantom}{i2,v3b}
\fmf{phantom}{i0,v1b}
\fmf{phantom}{o2,v6b}
\fmf{phantom,tension=1.4}{o1,v5}
\fmf{phantom}{o0,v5b}
\fmf{phantom,tension=1.11}{v3b,v6b}
\fmf{phantom,tension=1.38}{i0,v2}
\fmf{phantom,tension=1.38}{o0,v2}
\fmf{phantom,tension=1.8}{i0,v2b}
\fmf{phantom,tension=1.2}{o0,v2b}
\fmf{phantom,tension=1.2}{b0,v2b}
\fmf{phantom,tension=1.2}{b1,v2b}
\fmf{phantom,tension=1.2}{b2,v2b}
\fmf{phantom,tension=1.38}{i0,v4}
\fmf{phantom,tension=1.38}{o0,v4}
\fmf{phantom,tension=1.2}{i0,v4b}
\fmf{phantom,tension=1.8}{o0,v4b}
\fmf{phantom,tension=1.2}{b0,v4b}
\fmf{phantom,tension=1.2}{b1,v4b}
\fmf{phantom,tension=1.2}{b2,v4b}
\fmf{phantom,tension=2}{i3,v3}
\fmf{phantom,tension=2}{o3,v6}
\fmf{phantom,tension=2}{i3,v3b}
\fmf{phantom,tension=0.8}{o3,v3b}
\fmf{phantom,tension=0.8}{i3,v6b}
\fmf{phantom,tension=2}{o3,v6b}
\fmf{plain,tension=1.4}{v1,v2}
\fmf{plain,tension=1.4}{v4,v5}
\fmf{wiggly}{v1,v3}
\fmf{plain,left=0.8,tension=0}{v1,v3}
\fmf{wiggly}{v5,v6}
\fmf{plain,right=0.8,tension=0}{v5,v6}
\fmf{phantom}{v2,v3}
\fmf{phantom}{v4,v6}
\fmf{wiggly,tension=0.5}{v2,v4}
\fmf{phantom,tension=2}{v3,v6}
\fmf{phantom,right=0.8,tension=0}{v1,v5}
\fmf{phantom,tension=1}{v1,v1b}
\fmf{plain,tension=0.2}{v2,v2b}
\fmf{plain,tension=1.5}{v3,v3b}
\fmf{plain,tension=0.2}{v4,v4b}
\fmf{phantom,tension=1}{v5,v5b}
\fmf{plain,tension=1.5}{v6,v6b}
\end{fmfgraph*}
\end{fmffile}
\end{gathered} \hspace{-0.4cm} + \hspace{-0.2cm} \begin{gathered}
\begin{fmffile}{Diagrams/LoopExpansionBosonicHS_W_Diag13}
\begin{fmfgraph*}(35,20)
\fmfleft{i1,i2}
\fmfright{o1,o2}
\fmfbottom{i0,o0}
\fmfbottom{b0}
\fmfbottom{b1}
\fmfbottom{b2}
\fmftop{i3,o3}
\fmfv{decor.shape=circle,decor.size=2.0thick,foreground=(0,,0,,1)}{v1}
\fmfv{decor.shape=circle,decor.size=2.0thick,foreground=(0,,0,,1)}{v2}
\fmfv{decor.shape=circle,decor.size=2.0thick,foreground=(0,,0,,1)}{v3}
\fmfv{decor.shape=circle,decor.size=2.0thick,foreground=(0,,0,,1)}{v4}
\fmfv{decor.shape=circle,decor.size=2.0thick,foreground=(0,,0,,1)}{v5}
\fmfv{decor.shape=circle,decor.size=2.0thick,foreground=(0,,0,,1)}{v6}
\fmfv{decor.shape=circle,decor.filled=empty,decor.size=.073w,l=$\times$,label.dist=0}{v1b}
\fmfv{decor.shape=circle,decor.filled=empty,decor.size=.073w,l=$\times$,label.dist=0}{v3b}
\fmfv{decor.shape=circle,decor.filled=empty,decor.size=.073w,l=$\times$,label.dist=0}{v5b}
\fmfv{decor.shape=circle,decor.filled=empty,decor.size=.073w,l=$\times$,label.dist=0}{v6b}
\fmf{phantom,tension=1.4}{i1,v1}
\fmf{phantom}{i2,v3b}
\fmf{phantom}{i0,v1b}
\fmf{phantom}{o2,v6b}
\fmf{phantom,tension=1.4}{o1,v5}
\fmf{phantom}{o0,v5b}
\fmf{phantom,tension=1.11}{v3b,v6b}
\fmf{phantom,tension=1.38}{i0,v2}
\fmf{phantom,tension=1.38}{o0,v2}
\fmf{phantom,tension=1.8}{i0,v2b}
\fmf{phantom,tension=1.2}{o0,v2b}
\fmf{phantom,tension=1.2}{b0,v2b}
\fmf{phantom,tension=1.2}{b1,v2b}
\fmf{phantom,tension=1.2}{b2,v2b}
\fmf{phantom,tension=1.38}{i0,v4}
\fmf{phantom,tension=1.38}{o0,v4}
\fmf{phantom,tension=1.2}{i0,v4b}
\fmf{phantom,tension=1.8}{o0,v4b}
\fmf{phantom,tension=1.2}{b0,v4b}
\fmf{phantom,tension=1.2}{b1,v4b}
\fmf{phantom,tension=1.2}{b2,v4b}
\fmf{phantom,tension=2}{i3,v3}
\fmf{phantom,tension=2}{o3,v6}
\fmf{phantom,tension=2}{i3,v3b}
\fmf{phantom,tension=0.8}{o3,v3b}
\fmf{phantom,tension=0.8}{i3,v6b}
\fmf{phantom,tension=2}{o3,v6b}
\fmf{plain,tension=1.4}{v1,v2}
\fmf{plain,tension=1.4}{v4,v5}
\fmf{wiggly}{v1,v3}
\fmf{phantom,left=0.8,tension=0}{v1,v3}
\fmf{wiggly}{v5,v6}
\fmf{phantom,right=0.8,tension=0}{v5,v6}
\fmf{plain}{v2,v3}
\fmf{plain}{v4,v6}
\fmf{wiggly,tension=0.5}{v2,v4}
\fmf{phantom,tension=2}{v3,v6}
\fmf{phantom,right=0.8,tension=0}{v1,v5}
\fmf{plain,tension=1}{v1,v1b}
\fmf{phantom,tension=0.2}{v2,v2b}
\fmf{plain,tension=1.5}{v3,v3b}
\fmf{phantom,tension=0.2}{v4,v4b}
\fmf{plain,tension=1}{v5,v5b}
\fmf{plain,tension=1.5}{v6,v6b}
\end{fmfgraph*}
\end{fmffile}
\end{gathered} \\
& \hspace{2.0cm} + 4 \hspace{-0.4cm} \begin{gathered}
\begin{fmffile}{Diagrams/LoopExpansionBosonicHS_W_Diag14}
\begin{fmfgraph*}(35,20)
\fmfleft{i1,i2}
\fmfright{o1,o2}
\fmfbottom{i0,o0}
\fmfbottom{b0}
\fmfbottom{b1}
\fmfbottom{b2}
\fmftop{i3,o3}
\fmfv{decor.shape=circle,decor.size=2.0thick,foreground=(0,,0,,1)}{v1}
\fmfv{decor.shape=circle,decor.size=2.0thick,foreground=(0,,0,,1)}{v2}
\fmfv{decor.shape=circle,decor.size=2.0thick,foreground=(0,,0,,1)}{v3}
\fmfv{decor.shape=circle,decor.size=2.0thick,foreground=(0,,0,,1)}{v4}
\fmfv{decor.shape=circle,decor.size=2.0thick,foreground=(0,,0,,1)}{v5}
\fmfv{decor.shape=circle,decor.size=2.0thick,foreground=(0,,0,,1)}{v6}
\fmfv{decor.shape=circle,decor.filled=empty,decor.size=.073w,l=$\times$,label.dist=0}{v2b}
\fmfv{decor.shape=circle,decor.filled=empty,decor.size=.073w,l=$\times$,label.dist=0}{v3b}
\fmf{phantom,tension=1.4}{i1,v1}
\fmf{phantom}{i2,v3b}
\fmf{phantom}{i0,v1b}
\fmf{phantom}{o2,v6b}
\fmf{phantom,tension=1.4}{o1,v5}
\fmf{phantom}{o0,v5b}
\fmf{phantom,tension=1.11}{v3b,v6b}
\fmf{phantom,tension=1.38}{i0,v2}
\fmf{phantom,tension=1.38}{o0,v2}
\fmf{phantom,tension=1.8}{i0,v2b}
\fmf{phantom,tension=1.2}{o0,v2b}
\fmf{phantom,tension=1.2}{b0,v2b}
\fmf{phantom,tension=1.2}{b1,v2b}
\fmf{phantom,tension=1.2}{b2,v2b}
\fmf{phantom,tension=1.38}{i0,v4}
\fmf{phantom,tension=1.38}{o0,v4}
\fmf{phantom,tension=1.2}{i0,v4b}
\fmf{phantom,tension=1.8}{o0,v4b}
\fmf{phantom,tension=1.2}{b0,v4b}
\fmf{phantom,tension=1.2}{b1,v4b}
\fmf{phantom,tension=1.2}{b2,v4b}
\fmf{phantom,tension=2}{i3,v3}
\fmf{phantom,tension=2}{o3,v6}
\fmf{phantom,tension=2}{i3,v3b}
\fmf{phantom,tension=0.8}{o3,v3b}
\fmf{phantom,tension=0.8}{i3,v6b}
\fmf{phantom,tension=2}{o3,v6b}
\fmf{plain,tension=1.4}{v1,v2}
\fmf{plain,tension=1.4}{v4,v5}
\fmf{wiggly}{v1,v3}
\fmf{plain,left=0.8,tension=0}{v1,v3}
\fmf{wiggly}{v5,v6}
\fmf{plain,right=0.8,tension=0}{v5,v6}
\fmf{phantom}{v2,v3}
\fmf{plain}{v4,v6}
\fmf{wiggly,tension=0.5}{v2,v4}
\fmf{phantom,tension=2}{v3,v6}
\fmf{phantom,right=0.8,tension=0}{v1,v5}
\fmf{phantom,tension=1}{v1,v1b}
\fmf{plain,tension=0.2}{v2,v2b}
\fmf{plain,tension=1.5}{v3,v3b}
\fmf{phantom,tension=0.2}{v4,v4b}
\fmf{phantom,tension=1}{v5,v5b}
\fmf{phantom,tension=1.5}{v6,v6b}
\end{fmfgraph*}
\end{fmffile}
\end{gathered} \hspace{-0.4cm} + 2 \begin{gathered}
\begin{fmffile}{Diagrams/LoopExpansionBosonicHS_W_Diag15}
\begin{fmfgraph*}(35,20)
\fmfleft{i1,i2}
\fmfright{o1,o2}
\fmfbottom{i0,o0}
\fmfbottom{b0}
\fmfbottom{b1}
\fmfbottom{b2}
\fmftop{i3,o3}
\fmfv{decor.shape=circle,decor.size=2.0thick,foreground=(0,,0,,1)}{v1}
\fmfv{decor.shape=circle,decor.size=2.0thick,foreground=(0,,0,,1)}{v2}
\fmfv{decor.shape=circle,decor.size=2.0thick,foreground=(0,,0,,1)}{v3}
\fmfv{decor.shape=circle,decor.size=2.0thick,foreground=(0,,0,,1)}{v4}
\fmfv{decor.shape=circle,decor.size=2.0thick,foreground=(0,,0,,1)}{v5}
\fmfv{decor.shape=circle,decor.size=2.0thick,foreground=(0,,0,,1)}{v6}
\fmfv{decor.shape=circle,decor.filled=empty,decor.size=.073w,l=$\times$,label.dist=0}{v1b}
\fmfv{decor.shape=circle,decor.filled=empty,decor.size=.073w,l=$\times$,label.dist=0}{v3b}
\fmf{phantom,tension=1.4}{i1,v1}
\fmf{phantom}{i2,v3b}
\fmf{phantom}{i0,v1b}
\fmf{phantom}{o2,v6b}
\fmf{phantom,tension=1.4}{o1,v5}
\fmf{phantom}{o0,v5b}
\fmf{phantom,tension=1.11}{v3b,v6b}
\fmf{phantom,tension=1.38}{i0,v2}
\fmf{phantom,tension=1.38}{o0,v2}
\fmf{phantom,tension=1.8}{i0,v2b}
\fmf{phantom,tension=1.2}{o0,v2b}
\fmf{phantom,tension=1.2}{b0,v2b}
\fmf{phantom,tension=1.2}{b1,v2b}
\fmf{phantom,tension=1.2}{b2,v2b}
\fmf{phantom,tension=1.38}{i0,v4}
\fmf{phantom,tension=1.38}{o0,v4}
\fmf{phantom,tension=1.2}{i0,v4b}
\fmf{phantom,tension=1.8}{o0,v4b}
\fmf{phantom,tension=1.2}{b0,v4b}
\fmf{phantom,tension=1.2}{b1,v4b}
\fmf{phantom,tension=1.2}{b2,v4b}
\fmf{phantom,tension=2}{i3,v3}
\fmf{phantom,tension=2}{o3,v6}
\fmf{phantom,tension=2}{i3,v3b}
\fmf{phantom,tension=0.8}{o3,v3b}
\fmf{phantom,tension=0.8}{i3,v6b}
\fmf{phantom,tension=2}{o3,v6b}
\fmf{plain,tension=1.4}{v1,v2}
\fmf{plain,tension=1.4}{v4,v5}
\fmf{wiggly}{v1,v3}
\fmf{phantom,left=0.8,tension=0}{v1,v3}
\fmf{wiggly}{v5,v6}
\fmf{plain,right=0.8,tension=0}{v5,v6}
\fmf{plain}{v2,v3}
\fmf{plain}{v4,v6}
\fmf{wiggly,tension=0.5}{v2,v4}
\fmf{phantom,tension=2}{v3,v6}
\fmf{phantom,right=0.8,tension=0}{v1,v5}
\fmf{plain,tension=1}{v1,v1b}
\fmf{phantom,tension=0.2}{v2,v2b}
\fmf{plain,tension=1.5}{v3,v3b}
\fmf{phantom,tension=0.2}{v4,v4b}
\fmf{phantom,tension=1}{v5,v5b}
\fmf{phantom,tension=1.5}{v6,v6b}
\end{fmfgraph*}
\end{fmffile}
\end{gathered} \hspace{-0.4cm} + \hspace{-0.4cm} \begin{gathered}
\begin{fmffile}{Diagrams/LoopExpansionBosonicHS_W_Diag16}
\begin{fmfgraph*}(35,20)
\fmfleft{i1,i2}
\fmfright{o1,o2}
\fmfbottom{i0,o0}
\fmfbottom{b0}
\fmfbottom{b1}
\fmfbottom{b2}
\fmftop{i3,o3}
\fmfv{decor.shape=circle,decor.size=2.0thick,foreground=(0,,0,,1)}{v1}
\fmfv{decor.shape=circle,decor.size=2.0thick,foreground=(0,,0,,1)}{v2}
\fmfv{decor.shape=circle,decor.size=2.0thick,foreground=(0,,0,,1)}{v3}
\fmfv{decor.shape=circle,decor.size=2.0thick,foreground=(0,,0,,1)}{v4}
\fmfv{decor.shape=circle,decor.size=2.0thick,foreground=(0,,0,,1)}{v5}
\fmfv{decor.shape=circle,decor.size=2.0thick,foreground=(0,,0,,1)}{v6}
\fmf{phantom,tension=1.4}{i1,v1}
\fmf{phantom}{i2,v3b}
\fmf{phantom}{i0,v1b}
\fmf{phantom}{o2,v6b}
\fmf{phantom,tension=1.4}{o1,v5}
\fmf{phantom}{o0,v5b}
\fmf{phantom,tension=1.11}{v3b,v6b}
\fmf{phantom,tension=1.38}{i0,v2}
\fmf{phantom,tension=1.38}{o0,v2}
\fmf{phantom,tension=1.8}{i0,v2b}
\fmf{phantom,tension=1.2}{o0,v2b}
\fmf{phantom,tension=1.2}{b0,v2b}
\fmf{phantom,tension=1.2}{b1,v2b}
\fmf{phantom,tension=1.2}{b2,v2b}
\fmf{phantom,tension=1.38}{i0,v4}
\fmf{phantom,tension=1.38}{o0,v4}
\fmf{phantom,tension=1.2}{i0,v4b}
\fmf{phantom,tension=1.8}{o0,v4b}
\fmf{phantom,tension=1.2}{b0,v4b}
\fmf{phantom,tension=1.2}{b1,v4b}
\fmf{phantom,tension=1.2}{b2,v4b}
\fmf{phantom,tension=2}{i3,v3}
\fmf{phantom,tension=2}{o3,v6}
\fmf{phantom,tension=2}{i3,v3b}
\fmf{phantom,tension=0.8}{o3,v3b}
\fmf{phantom,tension=0.8}{i3,v6b}
\fmf{phantom,tension=2}{o3,v6b}
\fmf{plain,tension=1.4}{v1,v2}
\fmf{plain,tension=1.4}{v4,v5}
\fmf{wiggly}{v1,v3}
\fmf{plain,left=0.8,tension=0}{v1,v3}
\fmf{wiggly}{v5,v6}
\fmf{plain,right=0.8,tension=0}{v5,v6}
\fmf{plain}{v2,v3}
\fmf{plain}{v4,v6}
\fmf{wiggly,tension=0.5}{v2,v4}
\fmf{phantom,tension=2}{v3,v6}
\fmf{phantom,right=0.8,tension=0}{v1,v5}
\fmf{phantom,tension=1}{v1,v1b}
\fmf{phantom,tension=0.2}{v2,v2b}
\fmf{phantom,tension=1.5}{v3,v3b}
\fmf{phantom,tension=0.2}{v4,v4b}
\fmf{phantom,tension=1}{v5,v5b}
\fmf{phantom,tension=1.5}{v6,v6b}
\end{fmfgraph*}
\end{fmffile}
\end{gathered} \hspace{-0.5cm} \left.\rule{0cm}{1.1cm}\right) \left.\rule{0cm}{1.2cm}\right] \\
& + \mathcal{O}\big(\hbar^{3}\big)\;.
\end{split}
\label{eq:SbosonicKLoopExpansionStep5}
\end{equation}

\vspace{0.5cm}

In the (0+0)-D limit, the Schwinger functional expanded up to the second non-trivial order (i.e. up to order $\mathcal{O}(\hbar^3)$) reads:
\begin{equation}
\begin{split}
W^\text{LE;col}(\mathcal{J})= & - S_{\mathrm{col},\mathcal{J}}(\sigma_{\mathrm{cl}})+{\frac{\hbar}{2}}\ln(D_{\sigma_\text{cl};\mathcal{J}}) \\
& -{\frac{\hbar^{2}}{648}} \ G_{\sigma_\text{cl};\mathcal{J}}^4 D_{\sigma_\text{cl};\mathcal{J}}^2 N \lambda^2 \Big[-27 - 108 G_{\sigma_\text{cl};\mathcal{J}} J^2 + 5 G_{\sigma_\text{cl};\mathcal{J}}^2 D_{\sigma_\text{cl};\mathcal{J}} N \lambda \\
& \hspace{4.47cm} + 30 G_{\sigma_\text{cl};\mathcal{J}}^3 D_{\sigma_\text{cl};\mathcal{J}} J^2 N \lambda + 45 G_{\sigma_\text{cl};\mathcal{J}}^4 D_{\sigma_\text{cl};\mathcal{J}} J^4 N \lambda\Big] \\
& + {\frac{\hbar^{3}}{11664}} \ G_{\sigma_\text{cl};\mathcal{J}}^6 D_{\sigma_\text{cl};\mathcal{J}}^3 N \lambda^3 \Big[-540 - 3240 G_{\sigma_\text{cl};\mathcal{J}} J^2 + 360 G_{\sigma_\text{cl};\mathcal{J}}^2 D_{\sigma_\text{cl};\mathcal{J}} N \lambda \\
& \hspace{4.87cm} + 2880 G_{\sigma_\text{cl};\mathcal{J}}^3 D_{\sigma_\text{cl};\mathcal{J}} J^2 N \lambda - 750 G_{\sigma_\text{cl};\mathcal{J}}^5 D_{\sigma_\text{cl};\mathcal{J}}^2 J^2 N^2 \lambda^2 \\
& \hspace{4.87cm} + 270 G_{\sigma_\text{cl};\mathcal{J}}^8 D_{\sigma_\text{cl};\mathcal{J}}^3 J^4 N^3 \lambda^3  + 540 G_{\sigma_\text{cl};\mathcal{J}}^9 D_{\sigma_\text{cl};\mathcal{J}}^3 J^6 N^3 \lambda^3 \\
& \hspace{4.87cm} + 405 G_{\sigma_\text{cl};\mathcal{J}}^{10} D_{\sigma_\text{cl};\mathcal{J}}^3 J^8 N^3 \lambda^3 + 3 G_{\sigma_\text{cl};\mathcal{J}}^4 D_{\sigma_\text{cl};\mathcal{J}} N \lambda \big(1836 J^4 \\
& \hspace{4.87cm} - 25 D_{\sigma_\text{cl};\mathcal{J}} N \lambda\big) + 5 G_{\sigma_\text{cl};\mathcal{J}}^6 D_{\sigma_\text{cl};\mathcal{J}}^2 N^2 \lambda^2 \big(-495 J^4 \\
& \hspace{4.87cm} + D_{\sigma_\text{cl};\mathcal{J}} N \lambda\big) + 60 G_{\sigma_\text{cl};\mathcal{J}}^7 D_{\sigma_\text{cl};\mathcal{J}}^2 J^2 N^2 \lambda^2 \big(-45 J^4 \\
& \hspace{4.87cm} + D_{\sigma_\text{cl};\mathcal{J}} N \lambda\big)\Big] \\
& + \mathcal{O}\big(\hbar^{4}\big) \;,
\end{split}
\label{eq:ResultWSbosonicKLoopExpansion0DON}
\end{equation}
where it follows from~\eqref{eq:SbosonicKLoopExpansionH} that:
\begin{equation}
D^{-1}_{\sigma_\text{cl};\mathcal{J}} = \frac{\lambda N}{6} \ G_{\sigma_\text{cl};\mathcal{J}}^{2} \left(2 G_{\sigma_\text{cl};\mathcal{J}} J^{2} + 1\right) + 1\;,
\label{eq:DsigmaJCollRepre}
\end{equation}
where we have used the relations $J_{a}=J$ $\forall a$ and $\boldsymbol{G}_{\sigma_\text{cl};\mathcal{J};ab}=G_{\sigma_\text{cl};\mathcal{J}}\delta_{ab}$ with $G_{\sigma_\text{cl};\mathcal{J}}$ given by:
\begin{equation}
G_{\sigma_\text{cl};\mathcal{J}}^{-1} = m^{2} + i\sqrt{\frac{\lambda}{3}}\sigma_{\mathrm{cl}}\;,
\label{eq:GsigmaCollRepreDiag}
\end{equation}
as can be deduced from~\eqref{eq:DefGpropagCollectiveLE}. In contrast with the original and mixed representations, the LE in the collective representation is not organized with respect to the coupling constant $\lambda$, thus making it non-perturbative. The corresponding expressions for the gs energy and density are deduced once again after imposing that all sources vanish. The saddle points of the collective classical action $S_{\mathrm{col}}$ (i.e. the solutions of~\eqref{eq:minimizationSbosonicK} at vanishing external sources and in (0+0)-D) read:
\begin{equation}
\overline{\sigma}_{\mathrm{cl}}\equiv\sigma_{\mathrm{cl}}[\mathcal{J}=0]=i\left(\frac{\sqrt{3}m^{2}\pm\sqrt{3m^{4}+2N\lambda}}{2\sqrt{\lambda}}\right)\;,
\label{eq:ClassicalSolutionBosonicAction}
\end{equation}
to be compared to the equivalent quantity~\eqref{eq:sigclmix} in the mixed representation. The propagator of the original field thus takes the same form $\boldsymbol{G}^{-1}_{\sigma_{\mathrm{cl}};ab} = G^{-1}_{\sigma_{\mathrm{cl}}} \delta_{ab} = \left(m^{2}+i\sqrt{\frac{\lambda}{3}}\overline{\sigma}_{\mathrm{cl}}\right)\delta_{ab}$ at vanishing sources in both the mixed and collective representations but, while the trivial expression for $\overline{\sigma}_{\mathrm{cl}}$ in the mixed representation leads to the same renormalization of the squared mass as in the original representation, the non-perturbative expression~\eqref{eq:ClassicalSolutionBosonicAction} for $\overline{\sigma}_{\mathrm{cl}}$ in the collective representation yields a non-trivial dressing of $\boldsymbol{G}_{\sigma_{\mathrm{cl}}}$, i.e.:
\begin{equation}
G^{-1}_{\sigma_{\mathrm{cl}}} =\frac{1}{2}\left(m^{2}\mp\sqrt{m^{4}+\frac{2}{3}\lambda N}\right) \;,
\label{eq:GsigmaBosonicActionClassicalSolution0DON}
\end{equation}
obtained after inserting~\eqref{eq:ClassicalSolutionBosonicAction} into~\eqref{eq:GsigmaCollRepreDiag} at vanishing $\mathcal{J}$ (i.e.~\eqref{eq:GsigmaBosonicActionClassicalSolution0DON} is the configuration of~\eqref{eq:GsigmaCollRepreDiag} when the source $\mathcal{J}$ vanishes). Similarly, in moving from the mixed to the collective representation, we go from a trivial collective field propagator to the non-perturbative expression:
\begin{equation}
D^{-1}_{\sigma_\text{cl}} = \frac{2 \lambda N}{3\left(m^{2} \mp \sqrt{m^{4}+ \frac{2}{3}\lambda N}\right)^{2}} + 1 \;,
\end{equation}
which results from the combination of~\eqref{eq:DsigmaJCollRepre} at vanishing $\mathcal{J}$ and~\eqref{eq:GsigmaBosonicActionClassicalSolution0DON}. We finally deduce the series expansions for the gs energy and density valid for all $N \geq 1$ and for both signs of $m^2$:
\begin{equation}
\begin{split}
E^\text{LE;col}_{\mathrm{gs}} = & -\frac{3 m^4 + N \lambda - m^{2} \sqrt{9 m^{4} + 6 N \lambda} + 2 N \lambda \ln\bigg(\frac{12 \pi}{3 m^{2} + \sqrt{9 m^{4} + 6 N \lambda}}\bigg)}{4 \hbar \lambda} \\
& + \frac{1}{2} \ln\Bigg(1 + \frac{6 N \lambda}{\left(3 m^{2} + \sqrt{9 m^{4} + 6 N \lambda}\right)^2}\Bigg) \\
& - \hbar \frac{N \lambda^2}{6 \left(3 m^{4} + 2 N \lambda + m^{2} \sqrt{9 m^{4} + 6 N \lambda}\right)^{3}}\Big[27 m^{4} + 8 N \lambda + 9 m^{2} \sqrt{9 m^{4} + 6 N \lambda}\Big] \\
& +\hbar^{2}\frac{10 m^{2} N \lambda^3}{\left(3 m^{4} + 2 N \lambda + m^{2} \sqrt{9 m^{4} + 6 N \lambda}\right)^{6}} \\
& \hspace{0.7cm} \times \Big[108 m^{10}+ 90 m^{6} N \lambda + 15 m^{2} N^2 \lambda^2 + \left( N^2 \lambda^2 + 18 m^{4} N \lambda + 36 m^{8}\right) \sqrt{9 m^{4} + 6 N \lambda} \Big] \\
& -\hbar^{3}\frac{2 N \lambda^{4}}{45 \left(3 m^{4} + 2 N \lambda + m^{2} \sqrt{9 m^{4} + 6 N \lambda}\right)^{9}} \\
& \hspace{0.7cm} \times \Big[9185400 m^{20} + 11136204 m^{16} N \lambda + 3610818 m^{12} N^{2} \lambda^{2} + 52650 m^{8} N^3 \lambda^3 \\
& \hspace{1.2cm} - 70605 m^{4} N^{4} \lambda^{4} - 1792 N^{5} \lambda^{5} + \big( - 7119 m^{2} N^{4} \lambda^{4} - 48474 m^{6} N^{3} \lambda^{3} \\
& \hspace{1.2cm} + 476550 m^{10} N^{2} \lambda^{2} + 2691468 m^{14} N \lambda + 3061800 m^{18} \big) \sqrt{9 m^{4} + 6 N \lambda} \Big] \\
& + \mathcal{O}\big(\hbar^{4}\big) \;,
\end{split}
\label{eq:ResultEgsBosonicAction0DON}
\end{equation}
and
\begin{equation}
\begin{split}
\rho^\text{LE;col}_{\mathrm{gs}} = & \ \frac{6}{3 m^{2} + \sqrt{9 m^{4} + 6 N \lambda}} \\
& + \frac{\sqrt{3} m^{2} + \sqrt{3 m^{4} + 2 N \lambda}}{\sqrt{3 m^{4} + 2 N \lambda}} \\
& \hspace{0.4cm} \times \Bigg\lbrace -\hbar\frac{6 \lambda }{ \left(3 m^{2} + \sqrt{9 m^{4} + 6 N \lambda}\right) \left(3 m^{4} + 2 N \lambda + m^{2} \sqrt{9 m^{4} + 6 N \lambda}\right)} \\
& \hspace{1.25cm} +\hbar^{2}\frac{2 \lambda^{2} \left(3 m^{2} + \sqrt{9 m^{4} + 6 N \lambda}\right)\left(9 m^{4} + N \lambda + 3 m^{2} \sqrt{9 m^{4} + 6 N \lambda}\right)}{\left(3 m^{4} + 2 N \lambda + m^{2} \sqrt{9 m^{4} + 6 N \lambda}\right)^{4}} \\
& \hspace{1.25cm} -\hbar^{3}\frac{10 \lambda^{3}\left(3 m^{2} + \sqrt{9 m^{4} + 6 N \lambda}\right)^{3}}{\left(3 m^{4} + 2 N \lambda + m^{2} \sqrt{9 m^{4} + 6 N \lambda}\right)^{7}} \\
& \hspace{1.95cm} \times \bigg[54 m^{8} + 15 m^{4} N \lambda - 2 N^{2} \lambda^{2} + \left(- m^{2} N \lambda + 18 m^{6}\right) \sqrt{9 m^{4} + 6 N \lambda}\bigg] \\
& \hspace{1.25cm} + \hbar^{4}\frac{2 \lambda^{4}  \left(3 m^{2} + \sqrt{9 m^{4} + 6 N \lambda}\right)^{5}}{9\left(3 m^{4} + 2 N \lambda + m^{2} \sqrt{9 m^{4} + 6 N \lambda}\right)^{10}} \\
& \hspace{1.95cm} \times \bigg[11340 m^{12} + 2538 m^{8} N \lambda - 1722 m^{4} N^{2} \lambda^2 - 21 N^{3} \lambda^{3} + \big(- 226 m^{2} N^{2} \lambda^2 \\
& \hspace{2.45cm} - 414 m^{6} N \lambda + 3780 m^{10}\big) \sqrt{9 m^{4} + 6 N \lambda} \bigg] + \mathcal{O}\big(\hbar^{5}\big) \Bigg\rbrace\;,
\end{split}
\label{eq:ResultrhogsBosonicAction0DON}
\end{equation}
where relations homologous to~\eqref{eq:AccessEgsFromWmix0DON} and~\eqref{eq:AccessRhogsFromWmix0DON} were exploited, i.e.:
\begin{equation}
E^\text{LE;col}_\text{gs} =  -\frac{1}{\hbar}W^\text{LE;col}\big(\mathcal{J}=0\big) \;,
\label{eq:GetEgscolFromWLEcolmathcalJ}
\end{equation}
\begin{equation}
\rho^\text{LE;col}_\text{gs} =-\frac{2}{N}\frac{\partial W^\text{LE;col}(\mathcal{J}=0)}{\partial m^{2}} \;.
\end{equation}

\subsection{Discussion}

\begin{figure}[!htb]
\captionsetup[subfigure]{labelformat=empty}
  \begin{center}
    \subfloat[]{
      \includegraphics[width=0.50\linewidth]{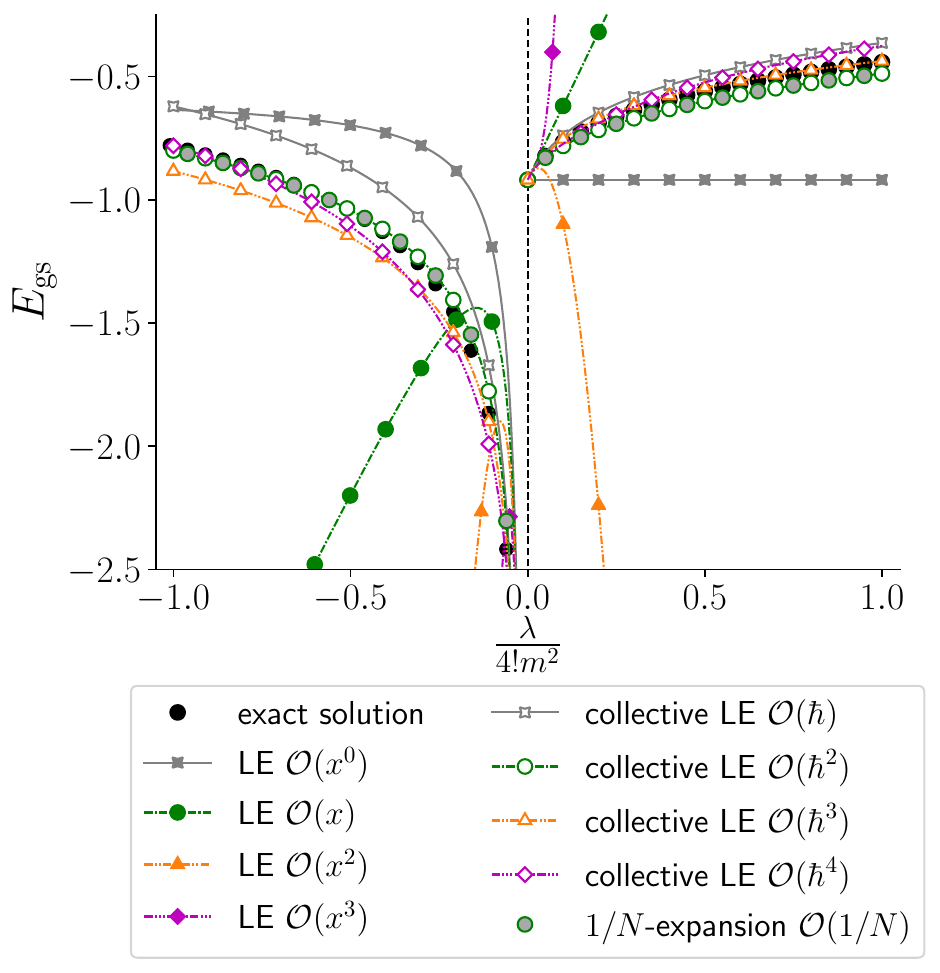}
                         }
    \subfloat[]{
      \includegraphics[width=0.50\linewidth]{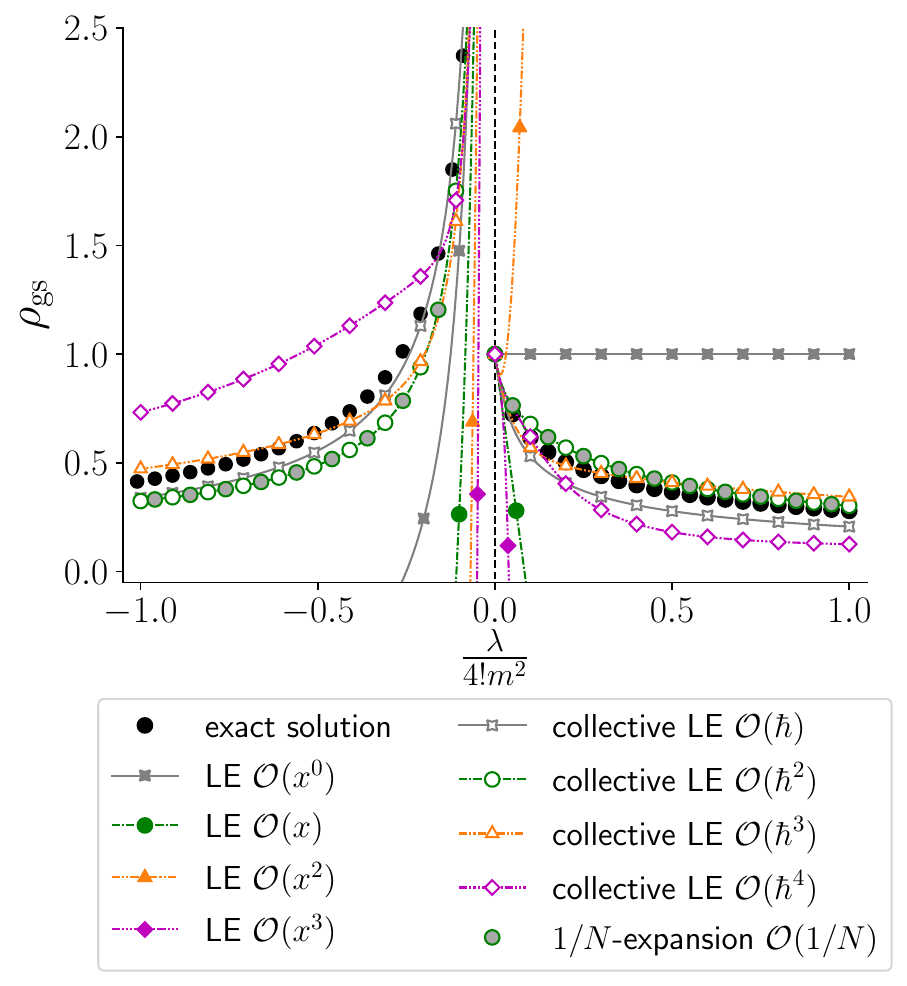}
                         }
    \caption{Gs energy $E_{\mathrm{gs}}$ (left) or density $\rho_{\mathrm{gs}}$ (right) calculated at $\hbar=1$, $m^{2}=\pm 1$ and $N=1$ ($\mathcal{R}e(\lambda)\geq 0$ and $\mathcal{I}m(\lambda)=0$), and compared with the corresponding exact solution (black dots). The indication ``$\mathcal{O}\big(\hbar^{n}\big)$'' for the collective LE results specifies that the series representing $W^\text{LE;col}$ has been exploited up to order $\mathcal{O}\big(\hbar^{n}\big)$ (which implies notably that the corresponding series for $E_{\mathrm{gs}}^\text{LE;col}$ is calculated up to order $\mathcal{O}(\hbar^{n-1})$ according to~\eqref{eq:GetEgscolFromWLEcolmathcalJ}). Recall also that the expansion parameter $x$ is defined as $x\equiv\hbar\lambda/m^4$.}
    \label{fig:O1PTcoll}
  \end{center}
\end{figure}

\begin{figure}[!htb]
\captionsetup[subfigure]{labelformat=empty}
  \begin{center}
    \subfloat[]{
      \includegraphics[width=0.50\linewidth]{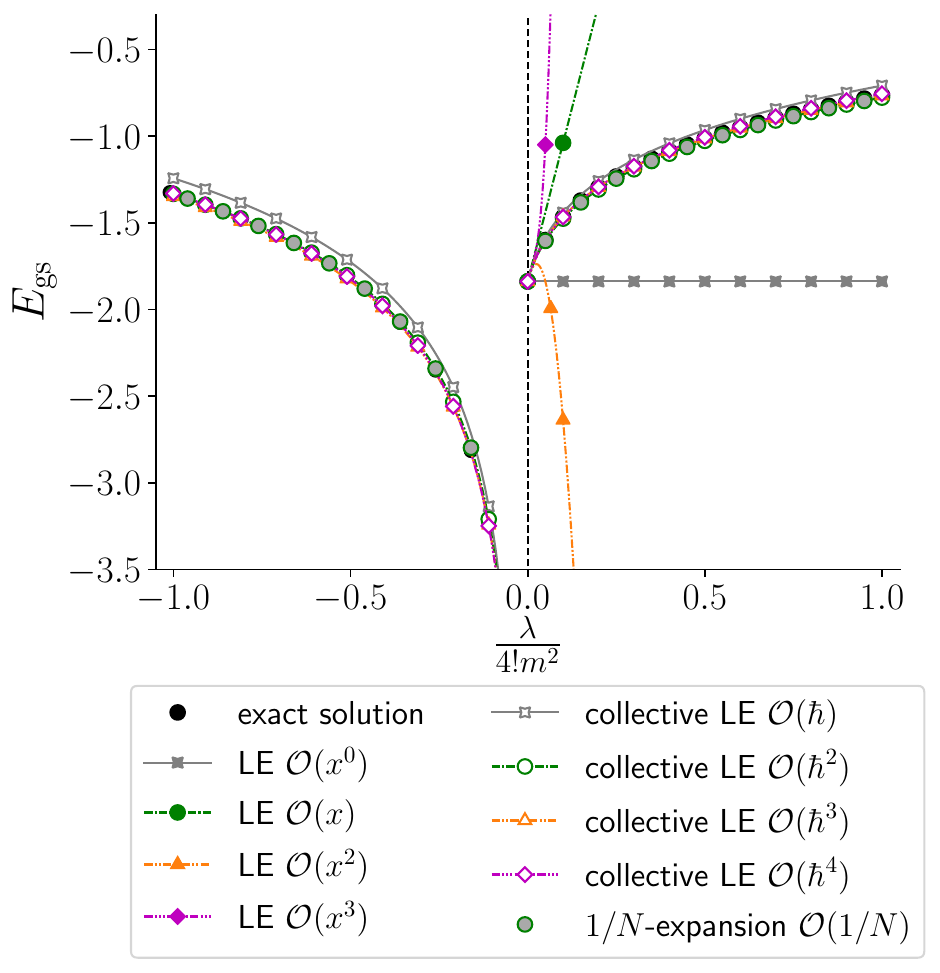}
                         }
    \subfloat[]{
      \includegraphics[width=0.50\linewidth]{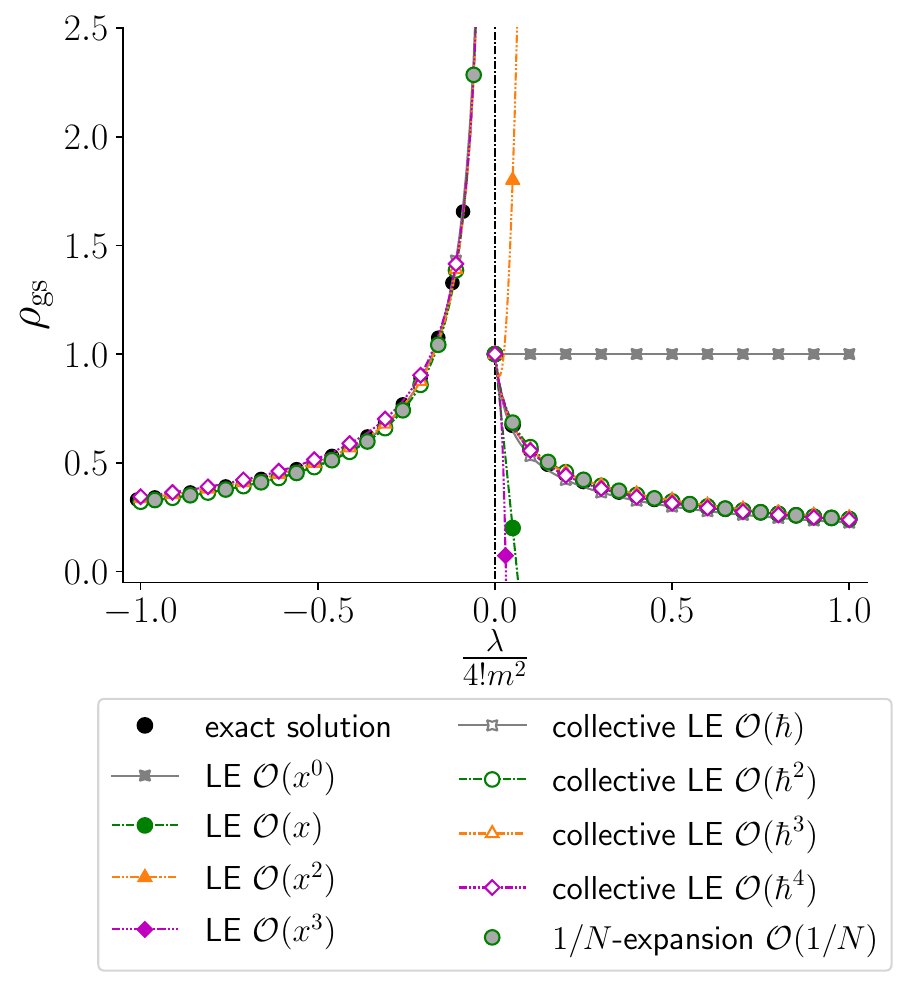}
                         }
    \caption{Same as fig.~\ref{fig:O1PTcoll} with $N=2$ instead. Note that no finite results can be obtained in the broken-symmetry phase (which corresponds to the left panel of both plots) from the LE in the original and mixed representations.}
    \label{fig:O2PTcoll}
  \end{center}
\end{figure}

\begin{figure}[!htb]
\vspace{0.26cm}
\begin{center}
      \includegraphics[width=0.6\linewidth]{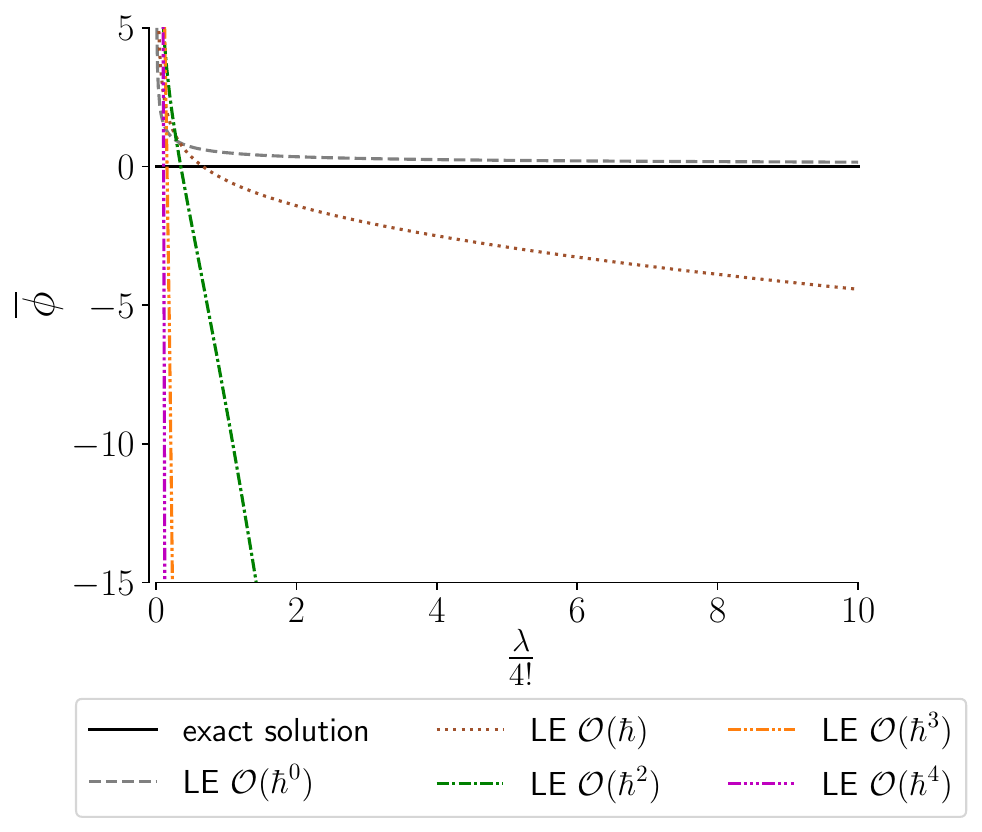}
      \end{center}
      \caption{1-point correlation function $\vec{\overline{\phi}}$ (defined from~\eqref{eq:DefvevphibarNumber1} and~\eqref{eq:DefvevphibarNumber2}) calculated at $\hbar=1$, $m^2=-1$ and $N=1$ ($\mathcal{R}e(\lambda) > 0$ and $\mathcal{I}m(\lambda)=0$) from the first orders of the original (and mixed) LE(s) as a function of the coupling constant $\lambda/4!$. Note that, at $N=1$, $\vec{\overline{\phi}}$ coincides with $\overline{\phi}\equiv\left\lvert\vec{\overline{\phi}}\right\rvert$.}
      \label{fig:vev}
\end{figure}

The comparison between the exact gs energy and density and the ones obtained within the LE based on the original (or mixed) and collective representations is performed in figs.~\ref{fig:O1PTcoll} and~\ref{fig:O2PTcoll} at $N=1$ and $2$, respectively. Moreover, we will also examine the vacuum expectation value of the original field, which signals the spontaneous breakdown of the $O(N)$ symmetry when becoming finite. It is defined as:
\begin{equation}
\overline{\phi}_{a}(x) \equiv \phi_{a}\Big[\vec{J}=\vec{0},\boldsymbol{K}=\boldsymbol{0};x\Big] \equiv \left.\rule{0cm}{0.6cm}\left\langle\widetilde{\varphi}_{a}(x)\right\rangle_{JK}\right|_{\vec{J}=\vec{0}\atop\boldsymbol{K}=\boldsymbol{0}} \;,
\label{eq:DefvevphibarNumber1}
\end{equation}
where
\begin{equation}
\big\langle\cdots\big\rangle_{JK} = \frac{1}{Z\big[\vec{J},\boldsymbol{K}\big]} \int \mathcal{D}\vec{\widetilde{\varphi}} \ \cdots \ e^{-\frac{1}{\hbar}S_{JK}\big[\vec{\widetilde{\varphi}}\big]} \;.
\label{eq:DefvevphibarNumber2}
\end{equation}
Note that $\vec{\overline{\phi}}$ can only be computed for $N=1$ in the original (or mixed) LE, but for all $N$ in the collective one. In the original and mixed representations, $\vec{\overline{\phi}}^{(n)}$, i.e. $\vec{\overline{\phi}}$ computed up to order $\mathcal{O}\big(\hbar^{n}\big)$ in the LE, stems from the derivative of the corresponding Schwinger functional $W$ with respect to the source $\vec{J}$, thus yielding the following relations from series~\eqref{eq:Worig} (at $\hbar = 1$, $m^2=-1$ and $N=1$):
\begin{equation}
\overline{\phi}^{(0)} =\sqrt{\frac{6}{\lambda}} = \overline{\varphi}_\text{cl} \;,
\end{equation}
\begin{equation}
\overline{\phi}^{(1)} =-\frac{\lambda-8}{4}\sqrt{\frac{3}{2\lambda}} \;,
\end{equation}
\begin{equation}
\overline{\phi}^{(2)} =-\frac{41\lambda^2+192\lambda-1536}{256\sqrt{6\lambda}} \;,
\end{equation}
\begin{equation}
\overline{\phi}^{(3)} =-\frac{321\lambda^3+82\lambda^2+384\lambda-3072}{512\sqrt{6\lambda}} \;,
\end{equation}
\begin{equation}
\overline{\phi}^{(4)} =-\frac{64573\lambda^4+30816\lambda^3+7872\lambda^2+36684\lambda-294912}{49152\sqrt{6\lambda}} \;,
\end{equation}
which are plotted in fig.~\ref{fig:vev}.

\vspace{0.5cm}


According to figs.~\ref{fig:O1PTcoll} and~\ref{fig:O2PTcoll} (and thus for $N=1$ and $2$), the original (as well as the mixed) LE(s) only yields (yield) a reasonable description of the gs energy and density for very small values of the coupling constant $\lambda\lesssim 0.2$ (hence $\lambda/4!\lesssim 8.10^{-3}$), as expected from the fact that PT based on the original (and mixed) dofs is, at best, an intrinsically weakly-interacting approach. The description quickly deteriorates for larger $\lambda$. Furthermore, fig.~\ref{fig:vev} shows that, at $N=1$, the expectation value of $\vec{\widetilde{\varphi}}$ is badly reproduced at each order of the original (and mixed) LE(s), with no sign of restoration of the (discrete) symmetry broken at the classical level. However, the first orders of the collective LE yield results close to the exact solutions, even in the broken-symmetry phase where no finite results can be obtained for $N\geq 2$ within the original (and mixed) LE(s). Regarding the expectation value of $\vec{\widetilde{\varphi}}$ in the framework of the collective LE, we deduce from the derivative of the generating functional $Z_{\mathrm{col}}$ (expressed by~\eqref{eq:ZbosonicKLoopExpansion}) with respect to the source $\vec{J}$ that:
\begin{equation}
\begin{split}
\vec{\overline{\phi}}(x) \propto & \ \left.\frac{\delta Z_{\mathrm{col}}\big[\vec{J},j\big]}{\delta\vec{J}(x)}\right|_{\vec{J}=\vec{0} \atop j=0} \\
\propto & \ \left[\int \mathcal{D}\widetilde{\sigma} \left(\int_{y}\boldsymbol{G}_{\widetilde{\sigma}}(x,y)\vec{J}(y)\right) e^{-\frac{1}{\hbar}S_{\text{col},\mathcal{J}}[\widetilde{\sigma}]} \right]_{\vec{J}=\vec{0} \atop j=0} \\
= & \ \vec{0} \;,
\end{split}
\end{equation}
to all orders of the collective LE, regardless of the dimension. In other words, the $O(N)$ symmetry, although possibly (spontaneously) broken down at the classical level, always gets exactly restored from the first non-trivial order of the collective LE.

\vspace{0.5cm}

In the original and mixed representations (as well as in the collective one, but to a lesser extent), the perturbative series derived so far show no signs of convergence: the results obtained for $E_{\mathrm{gs}}$ and $\rho_{\mathrm{gs}}$ worsen as the truncation order (with respect to $\hbar$) of these series increases, except for very small values of $\lambda$. This behavior signals the illegitimate application of PT to a system where the fundamental phenomena are non-perturbative in nature. Indeed, in quantum mechanics and QFT, PT typically produces asymptotic series with a zero radius of convergence, whose origin lies in instanton-like effects, i.e. an instability of the theory at some phase of the coupling (here for $\lambda<0$, where the potential becomes unbounded), which translates into a factorial growth of the number of Feynman diagrams with the order of the expansion~\cite{hur52,ben76}. Note furthermore that the $1/N$-expansion (which also typically produces asymptotic series) coincides with the collective LE at their first non-trivial orders according to figs.~\ref{fig:O1PTcoll} and~\ref{fig:O2PTcoll}. Although this equivalence breaks down at higher truncation orders, it illustrates the non-perturbative character of the collective LE. From this connection, we also expect the collective LE to be more and more performing as $N$ increases, which is in accordance with figs.~\ref{fig:O1PTcoll} and~\ref{fig:O2PTcoll}.

\vspace{0.5cm}

Asymptotic series however hide relevant information about the system, that needs to be deciphered through proper resummation techniques. Within this frame, PT is typically combined with a meticulously crafted analytic continuation function, yielding accurate results far beyond the weakly-interacting regime, and even allowing for the computation of genuinely non-perturbative features from low-order PT, as discussed in the next section.

\section{Resummation of the perturbative series}
\label{sec:Resum}
\subsection{Borel analysis}

The dominant method for giving a meaning to an asymptotic series relies on Borel analysis~\cite{bor28}. Let $P(x)$ be a physical quantity of interest, for which we only know a representation in terms of a divergent, asymptotic series expansion for small $x$:
\begin{equation}
P(x)\sim\sum_{n=0}^\infty p_n x^n \;,
\label{eq:P}
\end{equation}
with the generic large-order behavior:
\begin{equation}
p_n\underset{n\rightarrow\infty}{\sim}(-1)^n n! a^n n^b c \left(1+\mathcal{O}\left(\frac{1}{n}\right)\right) \;.
\label{eq:LOB}
\end{equation}   
Typically, the parameter $a$ only depends on the classical action of the system while the parameters $b$ and $c$ are specific to the quantity $P(x)$ under consideration. Borel analysis first consists in introducing the Borel transform of the asymptotic series under consideration, which translates for~\eqref{eq:P} into:
\begin{equation}
\mathcal{B}[P](\zeta) = \sum_{n=0}^\infty \frac{p_n}{\Gamma(n+1)}\zeta^n \mathrlap{\quad \forall\zeta\in\mathbb{C} \;,}
\label{eq:BtransFirst}
\end{equation}
which is a specific case of the Borel-Le Roy transform~\cite{ler1900}:
\begin{equation}
\mathcal{B}_s[P](\zeta) = \sum_{n=0}^\infty \frac{p_n}{\Gamma(n+s+1)}\zeta^n \mathrlap{\quad \forall s\in\mathbb{R}, \forall\zeta\in\mathbb{C} \;.}
\label{eq:Btrans}
\end{equation}
Both transforms~\eqref{eq:BtransFirst} and~\eqref{eq:Btrans} remove the factorial growth of the initial series coefficients $p_n$ and the $\zeta$-complex plane is referred to as Borel plane. The new series $\mathcal{B}_s[P](\zeta)$, also called Borel-Le Roy sum (or simply Borel sum at $s=0$), now has a finite non-zero radius of convergence and is analytic in a disk around the origin. The original function $P(x)$ is then recovered from $\mathcal{B}_s[P](\zeta)$ after taking the inverse Borel-Le Roy transform, i.e.:
\begin{equation}
P_{\mathcal{B}_s}(x) = \int_0^{\infty} d\zeta\ \zeta^s e^{-\zeta}\mathcal{B}_s[P](x\zeta) \;,
\label{eq:Bsum}
\end{equation}
which derives from the identity:
\begin{equation}
1=\frac{\int_0^\infty d\zeta \ \zeta^{n+s} e^{-\zeta}}{\Gamma(n+s+1)} \;.
\end{equation}
By construction, $P_{\mathcal{B}_s}(x)$ has the same asymptotic expansion as the function $P(x)$. It is actually an analytic continuation of $P(x)$ to a larger domain and may therefore provide sensible results out of the original asymptotic series. However, the Borel-Le Roy transform  $\mathcal{B}_s[P](\zeta)$ often exhibits poles and branch cuts along the integration path in~\eqref{eq:Bsum}, so that the integral in~\eqref{eq:Bsum} needs to be performed after deforming the integration contour in order to avoid the singularities in the Borel plane. $P(x)$ is said to be non Borel-summable when the result of the integration depends on the choice of contour, reflecting the fact that the half-line $[0,\infty)$ (where the perturbative expansion parameter takes values) is a so-called Stokes line. In this case, the singularities in the Borel plane induce a non-perturbative ambiguity: different integration paths yield functions with the same asymptotic behavior, but differing by exponentially suppressed terms which correspond to non-perturbative contributions. A unique well-defined resummation procedure can still be obtained after including the contributions from instanton-like configurations, thus resulting in a representation of the perturbative expansion under the more general form of a resurgent transseries. The latter can be derived for instance via the Picard-Lefschetz integration method, that we discuss next.

\subsection{Lefschetz thimbles decomposition}

Picard-Lefschetz theory provides an elegant framework for generating an ambiguous-free representation of a perturbative series~\cite{wit10,tan15}. For the sake of simplicity, we illustrate its application to the studied zero-dimensional model at $N=1$. Extensions to higher-dimensional PIs and systems invariant under continuous symmetries are detailed, e.g., in refs.~\cite{wit10,tan15}. Adding a multiplicative constant $1/\sqrt{\hbar}$ for later convenience, the partition function of the studied (0+0)-D $O(N)$ model at $N=1$ reads:
\begin{equation}
Z(m^2,\lambda,\hbar) = \frac{1}{\sqrt{\hbar}}\int_\mathbb{R} d\widetilde{\varphi} \ e^{-\frac{1}{\hbar} S(\widetilde{\varphi})} \;,
\end{equation}         
with classical action:
\begin{equation}
S(\widetilde{\varphi}) = \frac{m^2}{2}\widetilde{\varphi}^2 + \frac{\lambda}{4!}\widetilde{\varphi}^4 \;.
\end{equation}
The analysis can be straightforwardly extended to more general integrals of the form:
\begin{equation}
\int_\mathbb{R} d\widetilde{\varphi} \ e^{-\frac{1}{\hbar} S(\widetilde{\varphi})-J\widetilde{\varphi}-\frac{K}{2}\widetilde{\varphi}^2} \;,
\end{equation}
or
\begin{equation}
\int_\mathbb{R} d\widetilde{\varphi}\ p(\widetilde{\varphi})  e^{-\frac{1}{\hbar} S(\widetilde{\varphi})} \;,
\end{equation}
with $p(\widetilde{\varphi})$ a polynomial in the field $\widetilde{\varphi}$. Redefining the field via $\widetilde{\varphi} \rightarrow \widetilde{\varphi}/\sqrt{\lambda}$ yields:
\begin{equation}
Z(m^2,g) = \frac{1}{\sqrt{g}}\int_\mathbb{R} d\widetilde{\varphi}\ e^{-\frac{1}{g} V(\widetilde{\varphi})} \;,
\label{eq:oriint}
\end{equation}         
with $g\equiv\hbar\lambda$ and
\begin{equation}
V(\widetilde{\varphi})\equiv \left.S(\widetilde{\varphi})\right|_{\lambda=1} =  \frac{m^2}{2}\widetilde{\varphi}^2 + \frac{1}{4!}\widetilde{\varphi}^4 \;.
\end{equation}
The fundamental idea behind Picard-Lefshetz theory applied to the PI is appealingly summarized by Paul Painlev\'{e} in ref.~\cite{pai1900}: ``between two truths of the real domain, the easiest and shortest path quite often passes through the complex domain''. In the present situation, even if the integral $Z(m^2,g)$ is defined over real variables, the natural space in which the saddle (or critical) points of the action $V(\widetilde{\varphi})$ and their corresponding integration cycles live is the complexification of the original space. Understanding the behavior of $Z(m^2,g)$ for $g\in\mathbb{R}^+$ therefore passes by the study of its analytic continuation where $g=|g|e^{i\theta} \in\mathbb{C}$. The argument of the exponential in~\eqref{eq:oriint} becomes complex-valued, thus turning the partition function into a violently oscillating integral whose evaluation is difficult. One can significantly improve the properties of the integral by:
\begin{itemize}
\item Continuing the integrand into the complex plane, i.e. viewing the action $V(z)$ as a holomorphic function of the complex variable $z$, such that $Z$ now reads as an open contour integral:
\begin{equation}
Z(m^2,g) = \frac{1}{\sqrt{g}}\int_\mathcal{C} dz\ e^{-\frac{1}{g} V(z)} \;,
\label{eq:ACint}
\end{equation}
where $\mathcal{C}$ is a cycle with real dimension 1, coinciding with the real line when $g\in\mathbb{R}^+$.

\item Continuously deforming the integration domain as $g$ varies such that the integral~\eqref{eq:ACint} is convergent.

\end{itemize}

\vspace{0.3cm}

\begin{figure}[!htb]
\begin{center}
      \includegraphics[width=0.85\linewidth]{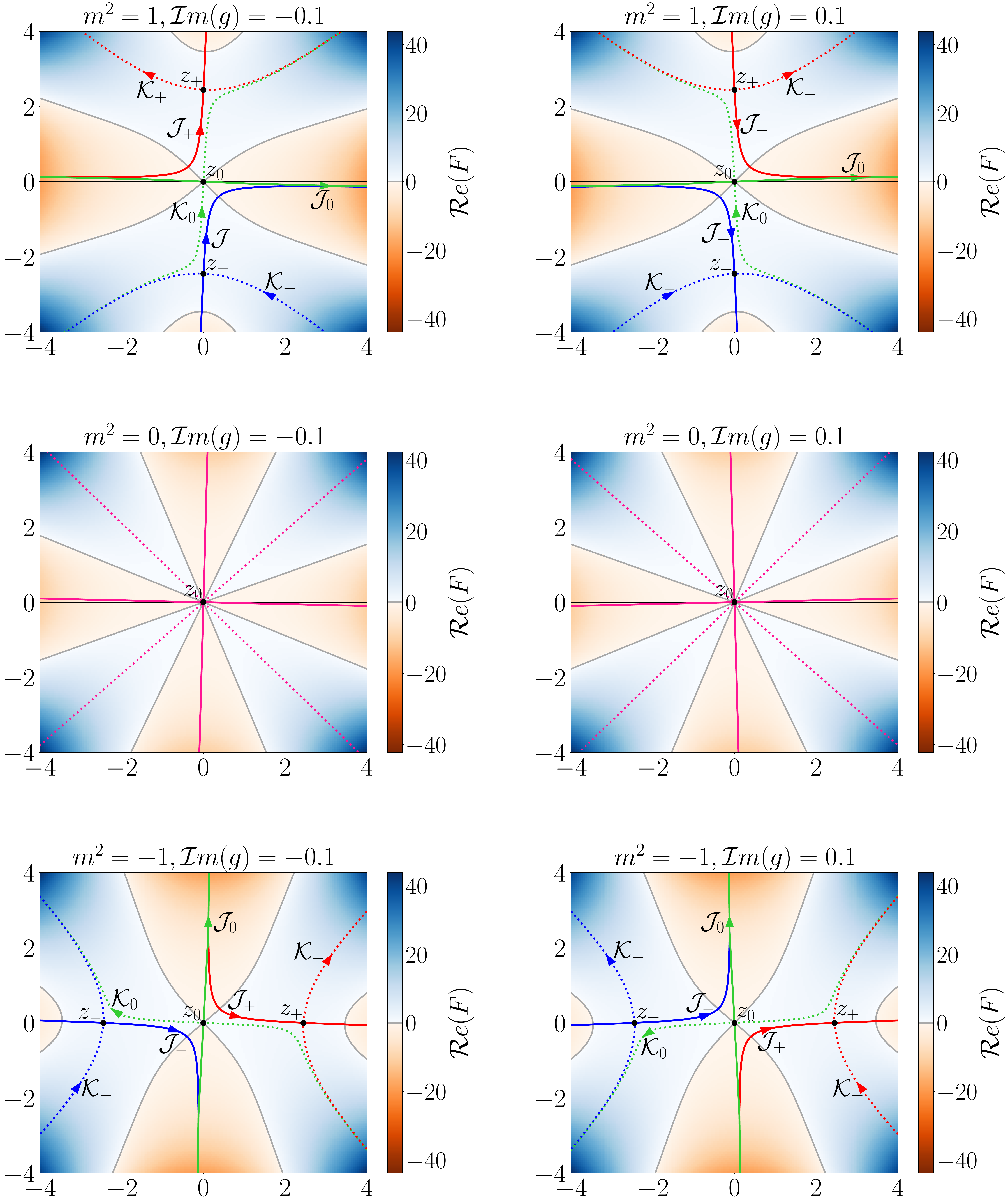}
      \end{center}
      \caption{Critical points (black dots) of $F(z)$ and their downward (solid lines) and  upward (dotted lines) flows  in  the $z$-complex plane, for $g= 1 - 0.1i$ (left column) and $g= 1+ 0.1i$ (right column),  and for  $m^2 = +1$ (upper panels), the degenerate case $m^2=0$ (middle panels) and  $m^2 = -1$ (lower panels). The value of $\mathcal{R}e(F)$ is given by the colormap.}
      \label{fig:PL}
\end{figure} 

Picard-Lefschetz integration method provides a decomposition of the integration cycle $\mathcal{C}$ into a linear combination $\mathcal{C} = \sum_i n_i \mathcal{J}_i$ (with $n_i\in \mathbb{Z}$) of nicer cycles (over which the integral is convergent) $\mathcal{J}_i$ attached to the saddle points $z^\star_i$ of $V(z)$\footnote{As a complex version of Morse theory~\cite{mor34}, this Picard-Lefschetz decomposition can only be performed for isolated critical points, i.e. for saddle points $z^\star_i$ such that $V''(z^\star_i)\neq 0$.}, and obtained after solving the gradient flow (or steepest descent) equations:
\begin{equation}
\begin{cases} \displaystyle{\frac{\partial z}{\partial\tau} = - \frac{\partial \bar{F}}{\partial \bar{z}} \;,} \\
\\
\displaystyle{\frac{\partial \bar{z}}{\partial\tau} = - \frac{\partial F}{\partial z} \;,} \end{cases}
\label{eq:PLd}
\end{equation}
where $F(z)\equiv -V(z)/g$, $\tau$ is the flow parameter and the upper bars denote the complex conjugation. The cycles $\mathcal{J}_i$ are called Lefschetz thimbles or simply thimbles. Along the flow, $\mathcal{R}e(F)$ is strictly decreasing (except for the trivial solution that sits at a saddle point $z^\star_i$ for all $\tau$) and $\mathcal{I}m(F)$ is constant and equal to $\mathcal{I}m(F(z^\star_i))$. In the studied case, the saddle points of $V(z)$ are $z^\star_0 = 0$ and $z^\star_\pm = \pm\sqrt{-6m^2}$, and sit on the imaginary (real) axis when $m^2\geq 0$ ($m^2\leq 0$). Saddle points and solutions of~\eqref{eq:PLd} are displayed in fig.~\ref{fig:PL} for different values of $m^2$ and $g = 1 \pm 0.1i$.

\vspace{0.5cm}

The integer coefficients $n_i$ are found after considering the upward flows $\mathcal{K}_i$ (which are called anti-thimbles), solutions of the converse (steepest ascent) equations:
\begin{equation}
\begin{cases} \displaystyle{\frac{\partial z}{\partial\tau} =+\frac{\partial \bar{F}}{\partial \bar{z}} \;,} \\
\\
\displaystyle{\frac{\partial \bar{z}}{\partial\tau} = +\frac{\partial F}{\partial z} \;,} \end{cases}
\label{eq:PLu}
\end{equation}
shown as dotted lines in fig.~\ref{fig:PL}. Along the anti-thimbles $\mathcal{K}_i$, $\mathcal{R}e(F)$ is monotonically increasing (making the integral divergent) and $\mathcal{I}m(F)$ is constant and equals to $\mathcal{I}m(F(z^\star_i))$. According to Picard-Lefschetz theory, $n_i$ corresponds to the intersection pairing of the original contour $\mathcal{C}$ and the upward flow $\mathcal{K}_i$. The partition function~\eqref{eq:ACint} can then be written as:
\begin{equation}
Z(m^2,g) = \sum_i n_i Z_i(m^2,g) \;,
\end{equation}
where 
\begin{equation}
Z_i(m^2,g) \equiv \frac{1}{\sqrt{g}}\int_{\mathcal{J}_i(\theta)} dz \ e^{-\frac{1}{g} V(z)} \;,
\end{equation}
admits an asymptotic power series expansion around the saddle point $z^\star_i$, which is Borel-summable to the exact result~\cite{ber91} (recall that $\theta \equiv \mathrm{arg}(g)$).

\vspace{0.5cm}

As can be seen in fig.~\ref{fig:PL}, the intersection numbers $(n_0=+1, n_\pm=0)$ of the upward flows $\mathcal{K}_i$ with the original integration cycle (i.e. the real axis) do not depend on the sign of $\mathcal{I}m(g)$ in the phase with $m^2>0$. In this case, the integration cycle coincides with a single thimble, which translates into:
\begin{equation}
Z(m^2>0,g) = Z_0(m^2>0,g)= \frac{1}{\sqrt{g}}\int_{\mathcal{J}_0(\theta)} dz\ e^{-\frac{1}{g} V(z)} \;.
\end{equation}
By expanding around the saddle point $z^\star_0=0$, the partition function of the theory can therefore be unambiguously represented by an asymptotic power series $ Z(m^2>0,g)= Z^{(0),m^2>0}(g)= \sum_n Z^{(0),m^2>0}_n g^n$  which is Borel-summable to the exact result, i.e. $Z(m^2>0,g) = Z^{(0),m^2>0}_{\mathcal{B}_s}(g)$ with:
\begin{equation}
Z^{(0),m^2>0}_{\mathcal{B}_s}(g) = \frac{1}{\sqrt{g}} e^{-\frac{1}{g}V(z^\star_0)} \int_0^{\infty}d\zeta\ \zeta^s e^{-\zeta}\mathcal{B}_s\Big[Z^{(0),m^2>0}\Big](g\zeta) \;,
\end{equation}
and
\begin{equation}
\mathcal{B}_s\Big[Z^{(0),m^2>0}\Big](\zeta) = \sum_{n=0}^{\infty} \frac{Z^{(0),m^2>0}_n}{\Gamma(n+s+1)}\zeta^n \;.  
\label{eq:BL}
\end{equation}

\vspace{0.5cm}

In the phase with $m^2<0$, the integral is on the Stokes line, which is reflected by the jump of the intersection numbers from $(n_0=+1, n_\pm=+1)$ for $\mathcal{I}m(g)<0$ to $(n_0=-1, n_\pm=+1)$ for $\mathcal{I}m(g)>0$. Since the integrals over the thimbles $\mathcal{J}_+$ and $\mathcal{J}_-$ yield the same result, the partition function of the theory can be written as follows:
\begin{equation}
\hspace{3.6cm} Z(m^2<0,g) = \pm Z_0(m^2<0,g) + 2 Z_+(m^2<0,g) \quad \forall \hspace{0.1cm} \mathcal{I}m(g)\lessgtr 0 \;,
\end{equation}
where each $Z_i$ can again be represented by an asymptotic series $Z^{(i),m^2<0}(g)= \sum_n Z^{(i),m^2<0}_n g^n$ after being expanded around the corresponding saddle point $z_i^\star$ and being Borel resummed, thus yielding the resurgent transseries:
\begin{equation}
\begin{split}
\hspace{2.0cm} Z(m^2<0,g) = & \ \frac{1}{\sqrt{g}}\Bigg\lbrace \pm e^{-\frac{1}{g}V(z^\star_0)}\int_0^{\infty}d\zeta\ \zeta^s e^{-\zeta}\mathcal{B}_s\Big[Z^{(0),m^2<0}\Big](g\zeta) \\
& \hspace{1.13cm} + 2e^{-\frac{1}{g}V(z^\star_+)}\int_0^{\infty}d\zeta\ \zeta^s e^{-\zeta}\mathcal{B}_s\Big[Z^{(+),m^2<0}\Big](g\zeta)\Bigg\rbrace \quad \forall \hspace{0.1cm} \mathcal{I}m(g)\lessgtr 0 \;.
\end{split}
\end{equation}

\vspace{0.5cm}

In practice, only the first terms of the asymptotic series $Z^{(i)}(g)$ are known, such that calculating the integral $\int_0^\infty d\zeta\ \zeta^s e^{-\zeta}\mathcal{B}_s\big[Z^{(i)}\big](g\zeta)$ only amounts to reinserting the $\Gamma(n+s+1)$ factors and leads back to the initial diverging series. Getting non-trivial results thus requires to make some assumptions about the unknown coefficients of the series, e.g. by re-expressing the Borel-Le Roy transform $\mathcal{B}_s\big[Z^{(i)}\big](\zeta)$ in terms of a non-polynomial function whose first Taylor coefficients match the known terms of the former. In what follows, we investigate three kinds of such functions, defining the so-called Pad\'e-Borel(-Le Roy), conformal mapping and Borel-hypergeometric resummations.

\subsection{\label{sec:PadBorResum}Pad\'e-Borel resummation}

We consider once again the generic physical quantity $P(x)$ represented by a factorially divergent asymptotic series~\eqref{eq:P}, which is transformed into~\eqref{eq:Btrans} and~\eqref{eq:Bsum} by means of Borel-Le Roy transforms. The idea behind Pad\'e-Borel-Le Roy resummation~\cite{pad1892,ler1900,bor28,ell96,ben99,kle01} is to rewrite the Borel-Le Roy transform $\mathcal{B}_s[P]$ as a Pad\'e approximant $\mathcal{P}^{U/V}\mathcal{B}_s[P]$, which is a rational function and can therefore develop a richer analytic behavior (with singularities in particular) as compared to the polynomial representing $\mathcal{B}_s[P]$ initially. The Pad\'e approximant $\mathcal{P}^{U/V}\mathcal{B}_s[P]$ is constructed from the knowledge of the Borel-Le Roy transform partial sum up to order $M$ as:
\begin{equation}
\mathcal{P}^{U/V}\mathcal{B}_s[P](\zeta) = \frac{\sum_{n=0}^U a_n \zeta^n}{1+\sum_{n=0}^V b_n \zeta^n} \;,
\label{eq:PBL}
\end{equation} 
with $U+V=M$. The coefficients $\{a_{n}\}$ and $\{b_{n}\}$ are fixed by equating order by order the Taylor series of~\eqref{eq:PBL} with the expansion~\eqref{eq:Btrans}, up to the desired order. The original function $P(x)$ is estimated after substituting the Borel-Le Roy sum $\mathcal{B}_s[P]$ by its Pad\'e approximant $\mathcal{P}^{U/V}\mathcal{B}_s[P]$ in the integral of~\eqref{eq:Bsum}. For our numerical applications, we will focus on the Pad\'e-Borel resummation, i.e. on the Pad\'e-Borel-Le Roy resummation with $s=0$, for which we define $\mathcal{P}\mathcal{B}_{P}[U/V]\equiv\mathcal{P}^{U/V}\mathcal{B}_{s=0}[P]$. We point out however that there exists recent studies, such as that of ref.~\cite{adz19}, discussing the determination of optimal values for the $s$ parameter within the framework of Pad\'e-Borel-Le Roy resummation, which is a task that we defer to future works for the $O(N)$ model considered in this thesis.

\vspace{0.5cm}

\begin{figure}[!htb]
\captionsetup[subfigure]{labelformat=empty}
  \begin{center}
    \subfloat[]{
      \includegraphics[width=0.50\linewidth]{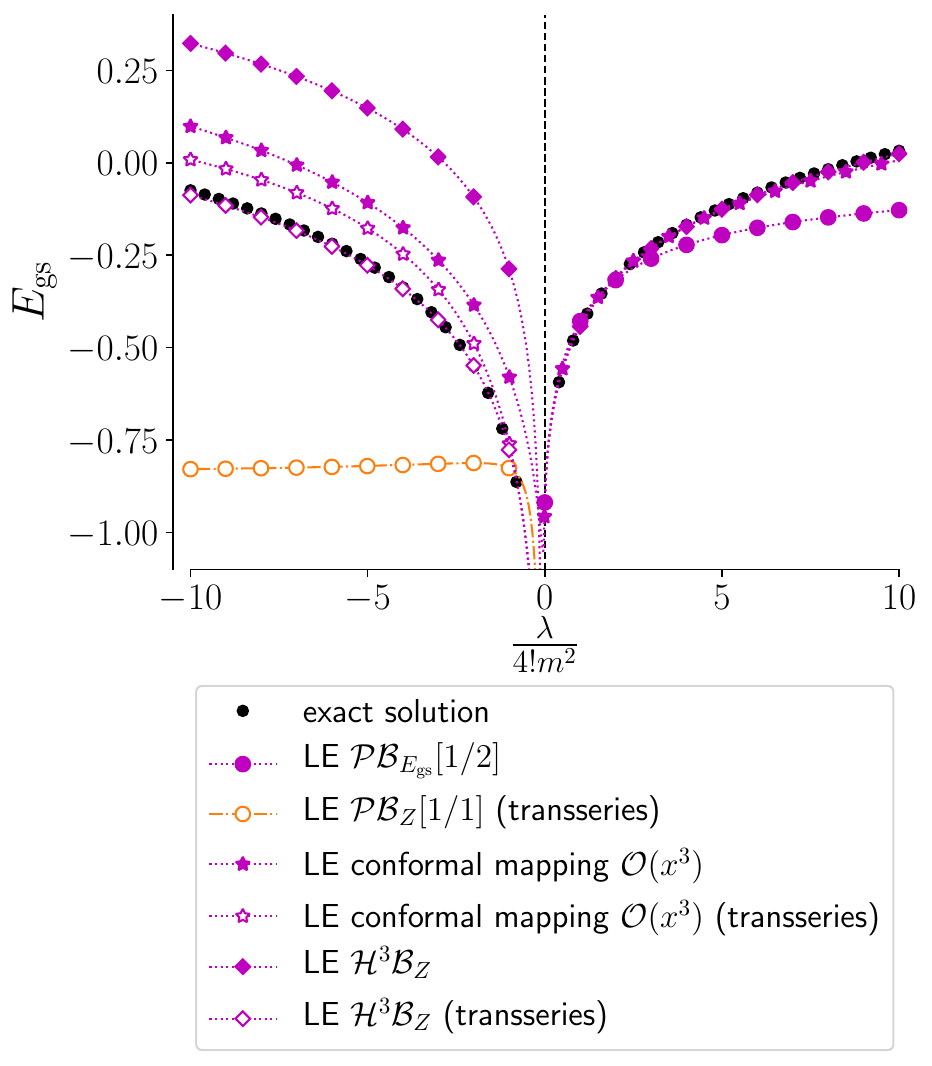}
                         }
    \subfloat[]{
      \includegraphics[width=0.50\linewidth]{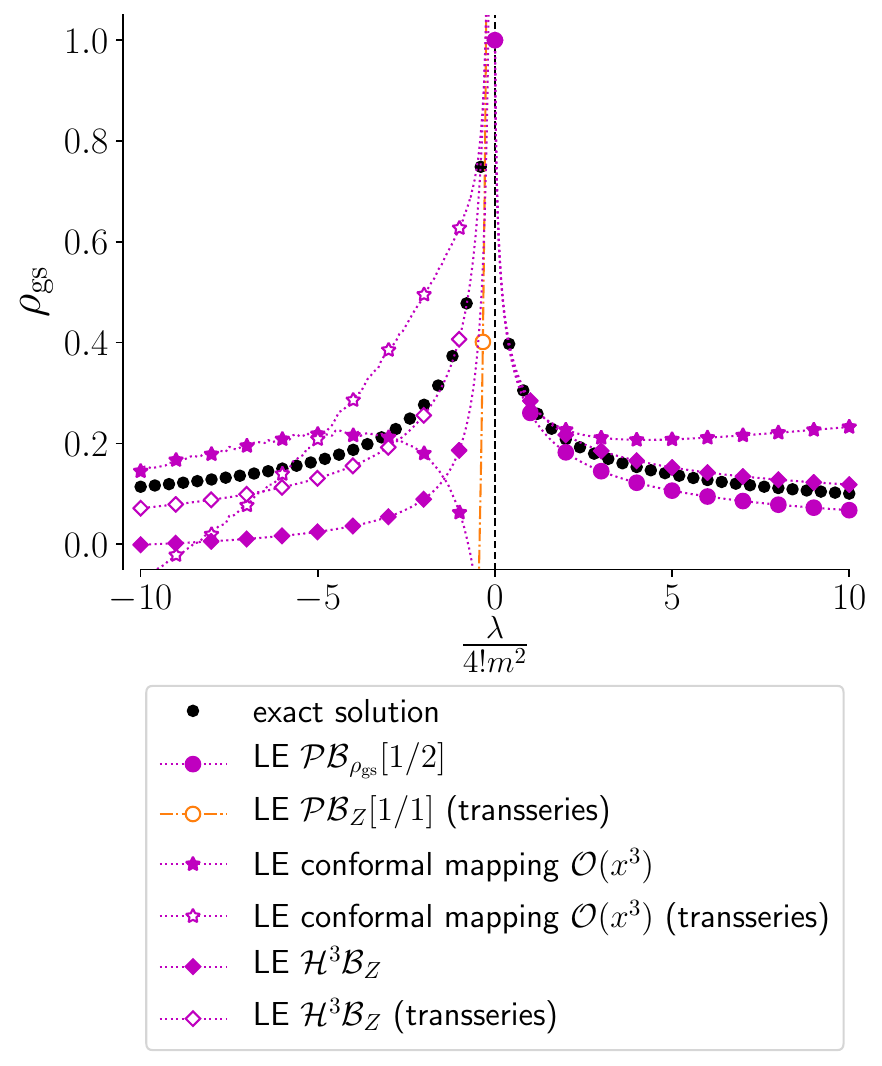}
                         }
    \caption{Gs energy $E_{\mathrm{gs}}$ (left) or density $\rho_{\mathrm{gs}}$ (right) calculated at $\hbar=1$, $m^{2}=\pm 1$ and $N=1$ ($\mathcal{R}e(\lambda)\geq 0$ and $\mathcal{I}m(\lambda)=0$), and compared with the corresponding exact solution (black dots). All presented results are obtained from series determined via the LE in the original representation.}
    \label{fig:O1ResumPTori}
  \end{center}
\end{figure}

\begin{figure}[!htb]
\captionsetup[subfigure]{labelformat=empty}
  \begin{center}
    \subfloat[]{
      \includegraphics[width=0.50\linewidth]{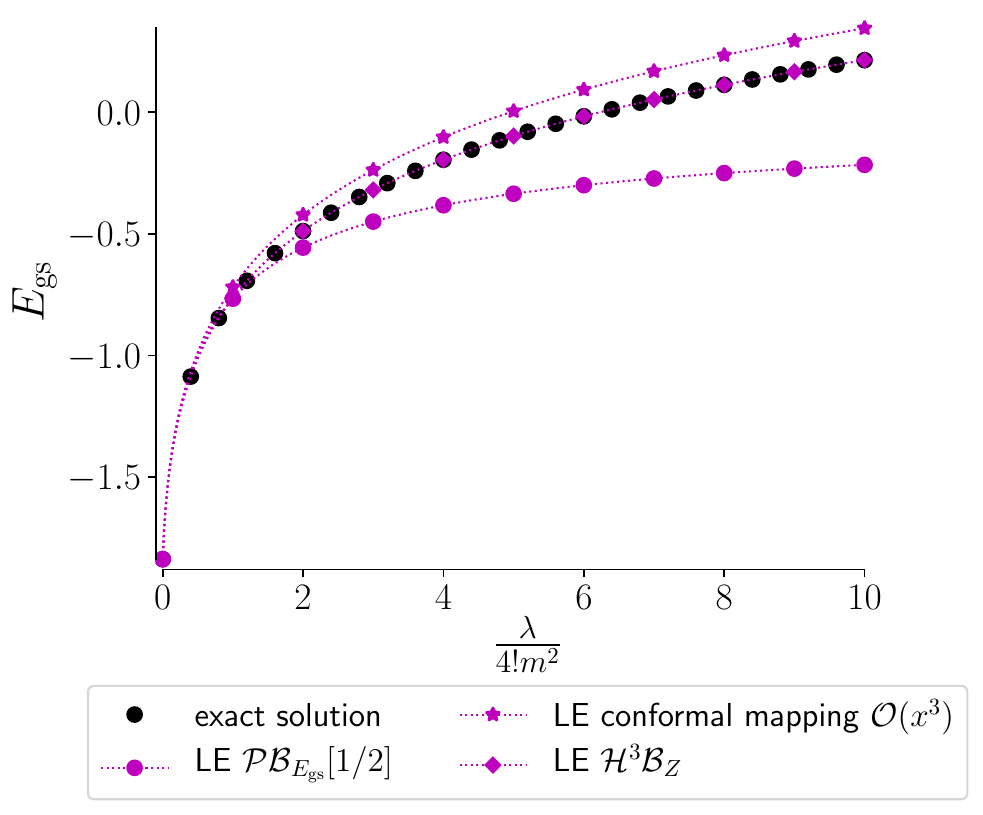}
                         }
    \subfloat[]{
      \includegraphics[width=0.50\linewidth]{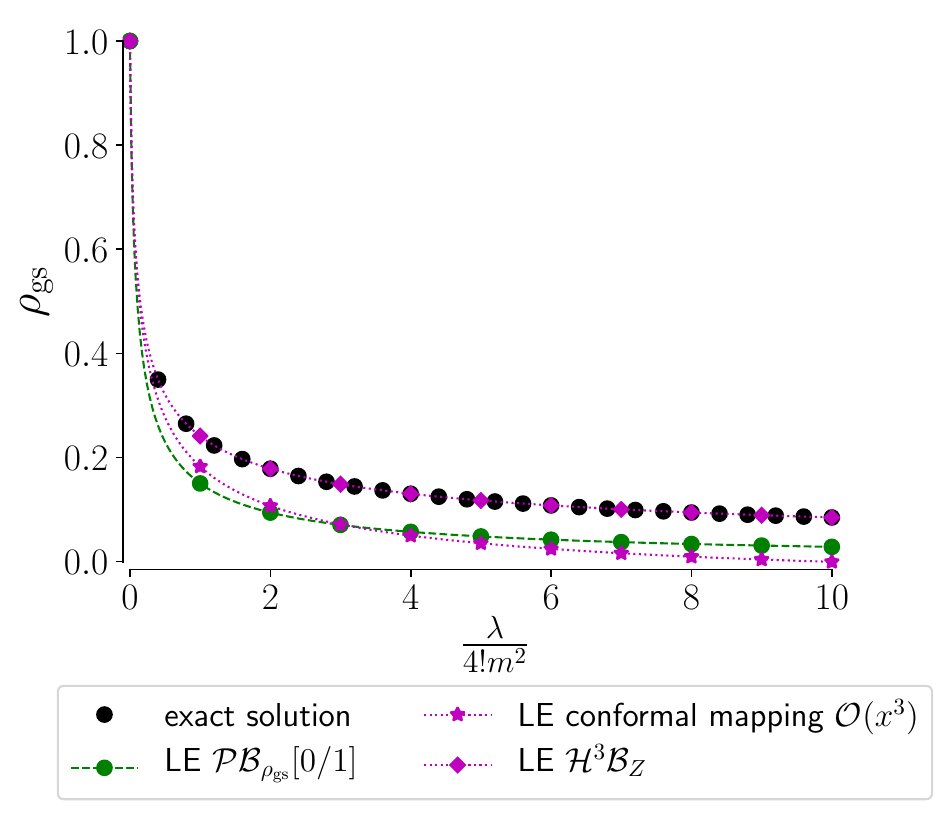}
                         }
    \caption{Same as fig.~\ref{fig:O1ResumPTori} with $m^{2}=+1$ and $N=2$ instead. As in fig.~\ref{fig:O2PTcoll}, no finite results can be obtained in the broken-symmetry phase from the LE in the original and mixed representations.}
    \label{fig:O2ResumPTori}
  \end{center}
\end{figure}

We compared the performances of the Pad\'e-Borel resummation procedure in reproducing the gs energy and density for the studied toy model under various settings, namely by following the step-by-step procedure: i) change the order $M$ of PT up to order $\mathcal{O}\big(x^3\big)$; ii) consider all the possible Pad\'e approximants\footnote{In this thesis, all Pad\'e approximants are determined with the $\mathtt{PadeApproximant}$ function of $\mathtt{Mathematica~12.1}$.} at a given order $M$; iii) either resum the power series or transseries representation of the partition function $Z$ and, from there, compute the gs energy and density (indicated by $\mathcal{PB}_{Z}$ in figs.~\ref{fig:O1ResumPTori} and~\ref{fig:O2ResumPTori}), or directly derive the perturbative expansion of the gs energy and density and then proceed with their resummation (indicated by $\mathcal{PB}_{E_{\mathrm{gs}}}$ and $\mathcal{PB}_{\rho_{\mathrm{gs}}}$ in figs.~\ref{fig:O1ResumPTori} and~\ref{fig:O2ResumPTori}). The best results are displayed as lines with a \padeborel symbol (the color indicates the truncation order in PT while the filled/open aspect refers to the power series/transseries representation of the resummed quantity) in figs.~\ref{fig:O1ResumPTori} and~\ref{fig:O2ResumPTori}, at $N=1$ and $2$ respectively. In the unbroken-symmetry regime, the best description of the gs energy on the one hand is obtained at $N=1$ and $2$ with the Pad\'e-Borel resummation of the perturbative series for $E_{\mathrm{gs}}$ pushed up to order $\mathcal{O}\big(x^3\big)$ (\pbviolet), with $[1/2]$ Pad\'e approximants. On the other hand, the best reproduction of the gs density in the unbroken-symmetry regime is achieved via Pad\'e-Borel resummation of the perturbative series for $\rho_{\mathrm{gs}}$ pushed up to order $\mathcal{O}\big(x^3\big)$ (\pbviolet) (up to order $\mathcal{O}(x)$ (\pbgreen)) with $[1/2]$ ($[0/1]$) Pad\'e approximants at $N=1$ ($N=2$). In the broken-symmetry regime at $N=1$, the best description of the gs energy and density is given by the Pad\'{e}-Borel-\'{E}calle resummation of the transseries representation of $Z$ (obtained from a slight modification of the integration path underlying the inverse Borel transforms in the corresponding Pad\'{e}-Borel resummation procedure~\cite{eca81}, thus avoiding singularities in the Borel plane) at order $\mathcal{O}\big(x^2\big)$ in PT (\pborange), with $[1/1]$ Pad\'e approximants. In all these cases, a major improvement over the bare PT results can be noticed, i.e. (except for $m^2<0$) the global behavior of the gs energy and density with respect to the coupling strength $\lambda/4!$ is now consistent with the exact trend over the whole range of tested values (i.e. for $\lambda/4!\in[0,10]$, which is wider than $[0,1]$ considered in figs.~\ref{fig:O1PTcoll} and~\ref{fig:O2PTcoll}), and even quantitatively reproduced up to $\lambda/4! \sim 2$.

\subsection{Conformal mapping}

While the Pad\'e-Borel(-Le Roy) resummation procedure only involves the knowledge about the first terms of the initial perturbative series~\eqref{eq:P}, the method of Borel(-Le Roy) transform with conformal mapping~\cite{leg80} aims at more reliable results by incorporating in addition the knowledge on the large-order behavior~\eqref{eq:LOB} of $p_n$. The parameters $a$, $b$ and $c$ in~\eqref{eq:LOB} can be computed via, e.g., the Lipatov method where the coefficients $p_n$ are represented by the contour integral $p_n=\frac{1}{2\pi i}\oint_{\mathcal{C}}d\zeta \ \frac{P(\zeta)}{\zeta^{n+1}}$ calculated for large $n$ through steepest descent. In particular, the coefficient $a$ determines the position of the singularity of the Borel-Le Roy transform~\eqref{eq:Btrans} which is the closest to the origin, i.e. $\mathcal{B}_s[P](\zeta)$ is analytical in a circle of radius $1/a$, with $a=2/3$ ($a=4/3$) at $N=1$ ($N=2$) for the studied (0+0)-D $O(N)$ model. One of the methods for continuing the Borel-Le Roy transform beyond its circle of convergence, and (to a certain extent) accelerating the convergence of $P_{\mathcal{B}_s}(x)$, relies on the conformal mapping of the Borel plane:
\begin{equation}
\begin{cases} \displaystyle{w(\zeta) = \frac{\sqrt{1+a\zeta}-1}{\sqrt{1+a\zeta}+1} \;.} \\
\\
\displaystyle{\zeta = \frac{4}{a}\frac{w}{(1-w)^2} \;.} \end{cases}
\end{equation}
Under this transformation, a point in the $\zeta$-complex plane is mapped within a disk of unit radius $|w|=1$. In particular, the origin is left invariant and the branch-cut singularity $\zeta\in]-\infty,-1/a]$ is mapped to the boundary of the $w$-unit disk, thus turning the Taylor expansion of the function $\widetilde{\mathcal{B}}_s[P](w)\equiv \mathcal{B}_s[P](\zeta(w))$ into a convergent one for $|w|<1$. The original quantity $P(x)$ is then determined after re-expanding $\mathcal{B}_s[P](x\zeta)$ in the new variable $w(x\zeta)$ in~\eqref{eq:Bsum}, i.e. after writing (assuming that only the first $M$ terms of the original series are known):
\begin{equation}
\mathcal{B}_s[P](\zeta) = \sum_{n=0}^M W_n \left(w(\zeta)\right)^n \;,
\end{equation}
with
\begin{equation}
W_n = \sum_{k=0}^n \frac{p_k}{\Gamma(k+s+1)} \left(\frac{4}{a}\right)^k \frac{(k+n-1)!}{(n-k)!(2k-1)!} \;.
\end{equation}
and we will also set $s=0$ in our numerical applications.

\vspace{0.5cm}

The gs energy and density obtained for our toy model after a conformal mapping resummation of the partition function $Z$ are displayed as \cmviolet (\cmvioleto) when $Z$ is represented by a power series (transseries) in figs.~\ref{fig:O1ResumPTori} and~\ref{fig:O2ResumPTori}, at $N=1$ and $2$ respectively. As far as the energy is concerned, the conformal mapping resummation applied to the partition function yields results in better agreement with the exact ones, especially in the strongly-interacting regime (for $\lambda/4!\gtrsim 2$ more specifically), in comparison with the estimates obtained via Pad\'e-Borel resummation of the energy perturbative series at the same order of PT. The results are however not as good for the gs density. In the phase with $m^2<0$ where the partition function is not Borel-summable, fig.~\ref{fig:O1ResumPTori} shows the gs energy and density of the system obtained both from $Z$ represented by an ambiguous power series and by a resurgent transseries. In the former case, the global behavior of the gs energy with the coupling strength is fairly well reproduced, but a quantitative reproduction of the exact result is not achieved, even in the weakly-interacting limit. The gs density, when deduced from a power series representation of $Z$, misses the monotonic decreasing displayed by the exact result when going from weak to strong couplings. Representing $Z$ by a transseries slightly improves the description of the gs energy and density of the system, with now a correct description of the range set by $\lambda/4!\lesssim 1$.

\subsection{Borel-hypergeometric resummation}

Borel-hypergeometric or Meijer-G resummation~\cite{mer15,mer16,ped16a,ped16b,san17,mer18,ant19} extends the idea behind Pad\'e approximants while trying to overcome the known issues of the latter\footnote{As rational functions, Pad\'e approximants built-in singularities are poles. Hence, since many poles are needed to mimic a branch cut, the Pad\'e-Borel resummation procedure converges slowly when the Borel transform to be approximated displays branch cuts.} by working with more sophisticated continuation functions, i.e. hypergeometric functions, which can notably mimic branch cuts in the complex plane, and whose (inverse) Borel transforms are known and conveniently represented by Meijer G-functions~\cite{mar83,bea13}. To our knowledge and unlike the other resummation methods discussed previously, the Borel-hypergeometric resummation has never been applied to any $O(N)$-symmetric theory, putting aside the (0+0)-D case at $N=1$~\cite{mer18}. We will push our investigations in (0+0)-D up to $N=4$ in the present study. Regardless of the model under consideration, the recipe underlying the Borel-hypergeometric resummation procedure can be presented as follows~\cite{mer18,ant19}:
\begin{enumerate}
\item[1.] As for the other resummation procedures, starting from the asymptotic series representing $P(x)$ truncated at an odd order $M$ (the case of even truncation orders will be discussed below), one starts by computing the coefficients $b_{n}\equiv p_{n}/n!=p_{n}/\Gamma(n+1)$ of the Borel transform $\mathcal{B}_{P}\equiv\mathcal{B}_{s=0}[P]$.

\item[2.] One then computes the $M$ ratios $b_1/b_0,\cdots,b_M/b_{M-1}$ of two consecutive coefficients of the Borel series and makes the ansatz that such ratios $b_{n+1}/b_n$ are rational functions of $n$, coined as $r_M(n)$ and defined as:
\begin{equation}
r_M(n)\equiv\frac{\sum_{k=0}^{l}u_k n^k}{1+\sum_{k=1}^{l}v_k n^k} \;,
\end{equation}
with $l=(M-1)/2$. The $\frac{M+1}{2}+\frac{M-1}{2}=M$ unknowns $u_k$ and $v_k$ are determined from the $M$ equations:
\begin{equation}
\frac{b_{n+1}}{b_n} = r_M(n) \;,
\end{equation} 
where $n$ runs from $0$ to $M-1$. 

\item[3.] Hypergeometric vectors $\bar{x}=(1,-x_1,\cdots,-x_l)$ and $\bar{y}=(-y_1,\cdots,-y_l)$ are then constructed via the equations:
\begin{equation}
\sum_{k=0}^l u_k x^k = 0 \;,
\end{equation}
\begin{equation}
1+\sum_{k=1}^l v_k y^k = 0 \;,
\end{equation}
and used to define the hypergeometric approximant of the Borel transform $\mathcal{B}_{P}$ in terms of the generalized hypergeometric function:
\begin{equation}
\mathcal{H}^M\mathcal{B}_{P}(\zeta)\equiv  \prescript{}{l+1}{F}_l\bigg(\bar{x},\bar{y},\frac{u_l}{v_l}\zeta\bigg) \;.
\end{equation}

\item[4.] One finally recovers the original function $P(x)$ through an inverse Borel transform, which can be represented in terms of a Meijer G-function $G^{m,n}_{p,q}
\Big(\begin{smallmatrix}
a_1,\cdots,a_p \\
b_1,\cdots,b_q
\end{smallmatrix}\Big| z \Big)$, i.e.:
\begin{equation}
\begin{split}
P_\mathcal{HB}(x) = & \ \int_0^{\infty} d\zeta \ e^{-\zeta}\mathcal{H}^M\mathcal{B}_{P}(x\zeta) = \frac{\prod_{k=1}^l\Gamma(-y_k)}{\prod_{k=1}^l\Gamma(-x_k)} G^{l+2,1}_{l+1,l+2}
\bigg(\begin{smallmatrix}
1,-y_1,\cdots,-y_l \\
1,1,-x_1,\cdots,-x_l
\end{smallmatrix}\bigg|
-\frac{v_l}{u_lx}\bigg) \;.
\end{split}
\end{equation}
\end{enumerate}

\vspace{0.3cm}

\noindent
For an even truncation order $M$, one first subtracts the constant zeroth-order term from the original series, then factors out the first-order term and finally follows the above recipe on the resulting series with an odd ($M-1$) truncation order. The final answer is obtained after re-multiplying the resummed series by the first-order term and re-adding the constant.

\vspace{0.5cm}

\begin{figure}[!htb]
\captionsetup[subfigure]{labelformat=empty}
  \begin{center}
    \subfloat[]{
      \includegraphics[width=0.50\linewidth]{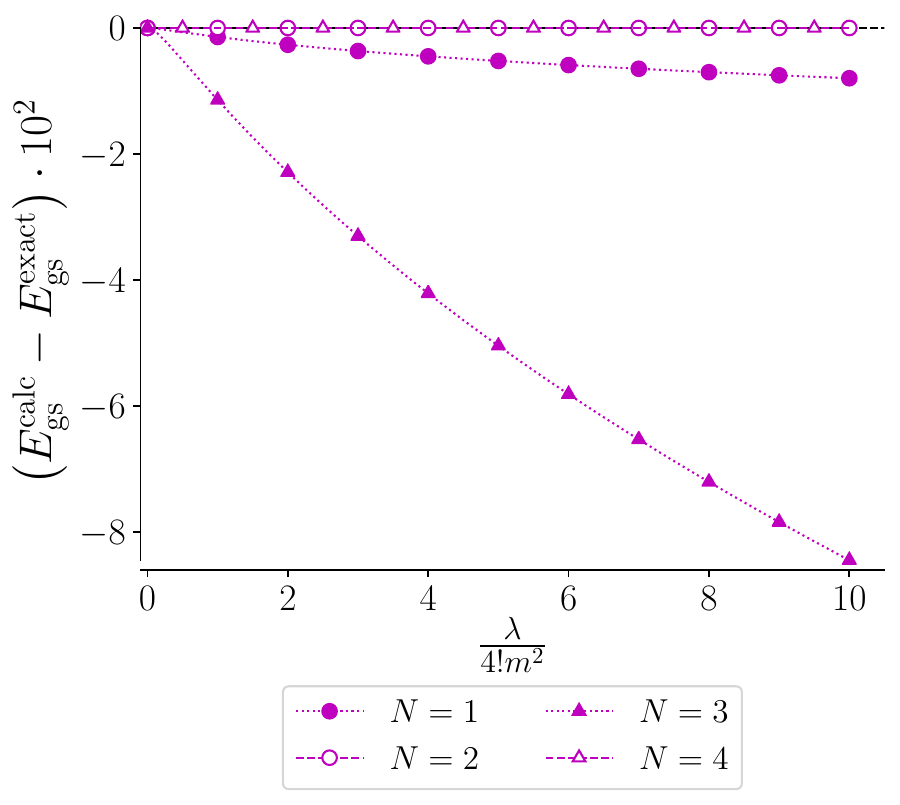}
                         }
    \subfloat[]{
      \includegraphics[width=0.50\linewidth]{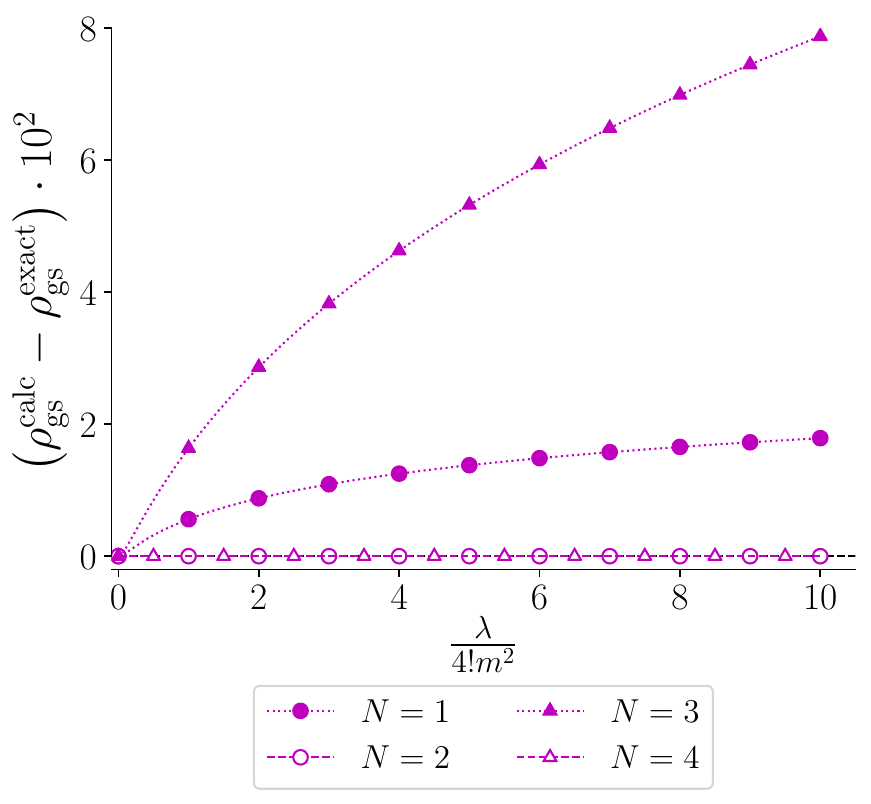}
                         }
    \caption{Gs energy $E_{\mathrm{gs}}^{\mathrm{calc}}$ (left) or density $\rho_{\mathrm{gs}}^{\mathrm{calc}}$ (right) calculated from Borel-hypergeometric resummation applied to the original (or mixed) LE series up to their third non-trivial order (notably labeled ``LE $\mathcal{H}^{3}\mathcal{B}_{Z}$'' in figs.~\ref{fig:O1ResumPTori} and~\ref{fig:O2ResumPTori}, which correspond to the present results (in the unbroken-symmetry regime) labeled ``$N=1$'' and ``$N=2$'', respectively). More specifically, we show here the difference between these results and the corresponding exact solution $E_{\mathrm{gs}}^{\mathrm{exact}}$ or $\rho_{\mathrm{gs}}^{\mathrm{exact}}$ at $\hbar=1$, $m^{2}=+1$ and $N=1,2,3$ and $4$ ($\mathcal{R}e(\lambda)\geq 0$ and $\mathcal{I}m(\lambda)=0$).}
    \label{fig:MeijerON}
  \end{center}
\end{figure}

The gs energy and density of the system obtained from the Borel-hypergeometric resummation of the partition function are reported in figs.~\ref{fig:O1ResumPTori} and~\ref{fig:O2ResumPTori} at $N=1$ and $2$ respectively. Results corresponding to the third non-trivial order of PT are displayed as \mgviolet (\mgvioleto) when $Z$ is represented by a power series (transseries). The Borel-hypergeometric resummation of the partition function yields the best results among all the resummation schemes, and even leads to an exact description of the partition function of the $O(2)$- and $O(4)$-symmetric theories from the third non-trivial order of the original LE, as shown by fig.~\ref{fig:MeijerON}.

\subsection{Conclusion}

Resummation techniques offer an impressive way of extracting sensible results from the very simple ordinary PT over a wide range of values for the coupling constant $\lambda/4!$, including the strongly-coupled regime. The various resummation techniques at our disposal actually render the LE (and all other techniques based on asymptotic series) rather versatile. The description of the gs energy and density (as well as the 1-point correlation function $\vec{\overline{\phi}}$ at $N=1$ in the phase with $m^2<0$) are significantly improved at trivial cost. However, a very accurate reproduction of the system's features requires to reach at least the third non-trivial order of PT, which can be difficult to determine in realistic cases. Besides, the theoretical foundation of the nuclear EDF can not be found in standard PT (even when completed by a resummation procedure), as no self-consistent dressing of, e.g., the field propagator is achieved and the energy is not obtained as a functional of the density in the spirit of DFT. We now turn to an optimized version of PT which bears stronger resemblance with EDFs via the use of self-consistent expansions.

\section{Optimized Perturbation Theory}
\label{sec:OPT}
\subsection{Spirit of the optimized perturbation theory}
\label{secSpiritOPT}

In standard PT formulated in the original representation (which was discussed in section~\ref{sec:PT}), the splitting of the classical action $S=S^0+S^1$ into an unperturbed reference part $S^0$ (which is supposedly simple enough to compute the corresponding energies and correlation functions without any approximation) and a residual part $S^1$ was performed within the LE, which amounts to a $\lambda$-wise expansion in the unbroken-symmetry phase of the studied $O(N)$ model. In this situation, $S^0$ coincides with the classical action of the non-interacting system and $S^1$ contains the interaction. OPT\footnote{We name OPT all strategies involving an optimized splitting into unperturbed and residual parts via some conditions, which includes more approaches than those exploited in refs.~\cite{ste81,kil81,oko87,oko87bis,oko88,dun88,oko89,dun89,jon89,jon90,coo91,kli91,buc92,buc92bis,sis91,sis92,sis92bis,kor93,kor93bis,sis93,sis94,sis94bis,kle98,kle11}, such as variational PT (VPT)~\cite{kle01,kle06,kle11bis}, linear delta-expansion (LDE)~\cite{jon90bis}, self-consistent expansion (SCE)~\cite{sch98,sch08}, self-similar PT~\cite{yuk99} or order-dependent mapping (ODM)~\cite{sez79,zin10}.} challenges such a splitting by optimizing the reference part $S^0$ around which one expands. Namely, there exists an infinite number of acceptable reference parts $S^0_\sigma$ depending on some parameter $\sigma(x)$ (not to be confused with the collective Hubbard-Stratonovich field $\widetilde{\sigma}(x)$), thus yielding the following splitting:
\begin{equation}
\begin{split}
S\Big[\vec{\widetilde{\varphi}}\Big] = & \ S^0_\sigma\Big[\vec{\widetilde{\varphi}}\Big]+ S^1_\sigma\Big[\vec{\widetilde{\varphi}}\Big] \\
\equiv & \ \left(S^0\Big[\vec{\widetilde{\varphi}}\Big]-\frac{1}{2}\int_x \sigma(x) \vec{\widetilde{\varphi}}(x)\cdot \vec{\widetilde{\varphi}}(x)\right) + \left(S^1\Big[\vec{\widetilde{\varphi}}\Big]+\frac{1}{2}\int_x \sigma(x) \vec{\widetilde{\varphi}}(x)\cdot \vec{\widetilde{\varphi}}(x)\right) \;,
\end{split}
\end{equation}
where the original splitting appears as the particular case where $\sigma(x) = 0$. Formally, one has done nothing but adding and subtracting an arbitrary quadratic term in the classical action. The idea behind OPT is then to exploit the introduction of such a Gaussian kernel to reorganize the partitioning into unperturbed and residual parts in a more flexible fashion, where non-perturbative correlations are shifted towards the easily solvable unperturbed channel. Indeed, when the perturbative expansion around $S^0_\sigma$ is truncated at some finite order, physical quantities exhibit an artificial dependence in the parameter $\sigma(x)$ that must be fixed. Relevant choices of $\sigma(x)$, such that $S^0_\sigma$ mimics as faithfully as possible the full action $S$, allow us to dress the propagator of the original field with non-trivial physics and turn the original divergent perturbative series into an exponentially-fast convergent one. At its first non-trivial order and depending on the condition chosen to fix $\sigma(x)$, OPT can lead to identical results as self-consistent mean-field approaches like the Hartree-Fock theory (which is different from standard PT in the original representation, presented in section~\ref{sec:PT}, whose propagator is not the one involved in Hartree-Fock theory but just corresponds to the bare propagator, possibly dressed by the vacuum expectation value of $\vec{\widetilde{\varphi}}(x)$ within the LE in the broken-symmetry regime), as will be illustrated later in section~\ref{sec:diagMixed2PIEA}. The key principle underlying OPT is that $\sigma(x)$ is optimized at the working order, i.e. is different at each truncation order, which is essential to obtain the aforementioned systematic improvement of our results and thus contrasts with e.g. the LE (combined with resummation theory).

\vspace{0.5cm}

Several optimization criteria can be devised to this end:
\begin{itemize}
\item In the ODM approach, $\sigma(x)$ is determined according to mathematical convergence properties of the series~\cite{zin10}.

\item Other classes of strategies are based on the fact that a given physical quantity $\mathcal{O}^{(k)}$ computed at $k$th order of the OPT expansion exhibits an artificial dependence with respect to $\sigma(x)$, so that one should try to make $\mathcal{O}^{(k)}$ minimally sensitive to it. This can be achieved via the so-called \textbf{principle of minimal sensitivity} (PMS)~\cite{ste81}, i.e.  a variational principle imposing:
\begin{equation}
\left.\frac{\delta\mathcal{O}^{(k)}}{\delta\sigma(x)}\right|_{\sigma=\sigma^{(k)}_{\text{PMS}}} = 0 \mathrlap{\quad \forall x \;,}
\end{equation}
or via the \textbf{turning point} (TP) method~\cite{kle05} where one rather looks for a plateau in the behavior of $\mathcal{O}^{(k)}$ with respect to $\sigma(x)$:
\begin{equation}
\left.\frac{\delta^2\mathcal{O}^{(k)}}{\delta\sigma(x)^2}\right|_{\sigma=\sigma^{(k)}_{\text{TP}}} = 0 \mathrlap{\quad \forall x \;,}
\end{equation}
or via the \textbf{fastest apparent convergence} (FAC)~\cite{ste81} where $\sigma(x)$ is fixed so that $\mathcal{O}$ calculated at two subsequent OPT orders yields the same result, i.e.:
\begin{equation}
\left[\mathcal{O}^{(k)}-\mathcal{O}^{(k-1)}\right]_{\sigma=\sigma^{(k)}_{\text{FAC}}} = 0 \;,
\end{equation}
which amounts to imposing that the $k$th coefficient in the OPT expansion is zero.

\item Another kind of optimization procedure involves a self-consistent condition (SCC) for $\sigma(x)$ where some physical features of the system are asked to be faithfully reproduced from the zeroth order of the description. In the spirit of Kohn-Sham DFT, one can ask that the 2-point correlation function of the system calculated at $k$th order of the OPT expansion $\left\langle\vec{\widetilde{\varphi}}(x)\cdot\vec{\widetilde{\varphi}}(y)\right\rangle^{(k)}$ (which reduces to the density at $y=x$) coincides with the zeroth-order one $\left\langle\vec{\widetilde{\varphi}}(x)\cdot\vec{\widetilde{\varphi}}(y)\right\rangle^{(0)}$:
\begin{equation}
\left[\left\langle\vec{\widetilde{\varphi}}(x)\cdot\vec{\widetilde{\varphi}}(y)\right\rangle^{(k)}-\left\langle\vec{\widetilde{\varphi}}(x)\cdot\vec{\widetilde{\varphi}}(y)\right\rangle^{(0)}\right]_{\sigma=\sigma^{(k)}_{\text{SCC}}} = 0 \mathrlap{\quad \forall x,y \;.}
\end{equation}
Such an optimization procedure, like the previous ones (PMS, TP, FAC), requires the calculation of a physical quantity at $k$th order of the OPT expansion, which is often difficult to achieve. We can use instead the following alternative implementation of the SCC:
\begin{equation}
\left[\left\langle\left(\vec{\widetilde{\varphi}}(x)\cdot\vec{\widetilde{\varphi}}(y)\right)^{m(k)}\right\rangle^{(1)}-\left\langle\left(\vec{\widetilde{\varphi}}(x)\cdot\vec{\widetilde{\varphi}}(y)\right)^{m(k)}\right\rangle^{(0)}\right]_{\sigma=\sigma^{(k)}_{\text{SCC}}} = 0 \mathrlap{\quad \forall x,y \;,}
\end{equation}
which is only a first-order relation, therefore easy to compute. The dependence in the working expansion order $k$ appears via the exponent of the correlation function $m(k)$. In particular, a dependence of the form $m(k)=k$ was studied in ref.~\cite{rem18} and shown to yield an exponentially-fast convergent series representation of physical quantities.  
One can understand the first-order nature of the last optimization procedure along the following lines: since the OPT expansion creates itself its partitioning between unperturbed and residual sectors such that the residual part is indeed small (in some sense) compared to the unperturbed one, we expect that the first-order correction will be the dominant one. In other words, the condition imposing a given correlation function computed at the first order of the OPT expansion to coincide with the zeroth-order one should not be very different from the same criteria for the correlation functions computed at the working order $k>1$ (instead of $k=1$), which is why the difference is expected to be captured by simple forms of $m(k)$ (such as $m(k)=k$).
\end{itemize}

\vspace{0.3cm}

OPT has been widely used for decades via the above optimization criteria~\cite{ste81,kil81,oko87,oko87bis,oko88,dun88,oko89,dun89,jon89,jon90,coo91,kli91,buc92,buc92bis,sis91,sis92,sis92bis,kor93,kor93bis,sis93,sis94,sis94bis,kle98,kle11,kle01,kle06,kle11bis,jon90bis,sch98,sch08,yuk99,sez79,zin10}. In particular, we can mention a previous study~\cite{ros16} of the unbroken-symmetry phase of the (0+0)-D $O(N)$ model considered in this thesis using OPT based on the PMS, the TP method and the FAC. However, this work exploits the (0+0)-D nature of the problem to directly expand the quantities of interest, thus bypassing notably the diagrammatic constructions underlying Wick's theorem that can hardly be avoided in finite dimensions. In what follows, we will not use such a shortcut and construct the diagrammatic series underlying OPT for our $O(N)$ model in arbitrary dimensions\footnote{Note that the construction of diagrammatic series in the framework of OPT has already been discussed in ref.~\cite{oko87} for a $\varphi^{4}$-theory, but not for the $O(N)$-symmetric case.} as usual. We will investigate the SCC as well and extend results of ref.~\cite{ros16} to the broken-symmetry regime of the studied (0+0)-D $O(N)$ model.

\subsection{Splitting of the classical action}
\label{sec:SplittingOPTfinitedimON}

We thus focus as a next step on the classical action of the studied (finite-dimensional) $O(N)$ model whose expression is recalled here for convenience:
\begin{equation}
S\Big[\vec{\widetilde{\varphi}}\Big] = \int_x \left[ \frac{1}{2}\left(\nabla_x\widetilde{\varphi}_a(x)\right)\left(\nabla_x\widetilde{\varphi}^a(x)\right) +  \frac{m^2}{2}\widetilde{\varphi}_a(x)\widetilde{\varphi}^a(x)+\frac{\lambda}{4!}\left(\widetilde{\varphi}_a(x)\widetilde{\varphi}^a(x)\right)^2\right] \;.
\end{equation}
The implementation of OPT first requires to introduce a set of non-fluctuating (i.e. classical) collective fields coupled to relevant bilinears in the original field(s). In the present case, the only relevant bilinear form in the field $\vec{\widetilde{\varphi}}(x)$ is $\vec{\widetilde{\varphi}}(x)\cdot \vec{\widetilde{\varphi}}(y)$, or its local counterpart $\vec{\widetilde{\varphi}}(x)\cdot \vec{\widetilde{\varphi}}(x)$. We therefore introduce a single collective field $\sigma(x)$\footnote{The absence of tilde on $\sigma$ highlights the fact that it is a non-fluctuating field.}, thus resulting in the following partitioning:
\begin{equation}
S\Big[\vec{\widetilde{\varphi}}\Big] = S^0_{\sigma}\Big[\vec{\widetilde{\varphi}}\Big] + S^1_{\sigma}\Big[\vec{\widetilde{\varphi}}\Big] \;,
\end{equation}
\begin{equation}
S^0_{\sigma}\Big[\vec{\widetilde{\varphi}}\Big] = \frac{1}{2}\int_{x,y}\widetilde{\varphi}^a(x) \boldsymbol{G}^{-1}_{\sigma;ab}(x,y)\widetilde{\varphi}^b(y) \;,
\end{equation}
\begin{equation}
S^1_{\sigma}\Big[\vec{\widetilde{\varphi}}\Big] = \int_x \left[ \frac{\lambda}{4!} \left(\vec{\widetilde{\varphi}}(x)\cdot\vec{\widetilde{\varphi}}(x)\right)^2+\frac{1}{2}\sigma(x)\vec{\widetilde{\varphi}}(x)\cdot\vec{\widetilde{\varphi}}(x)\right] \;,
\end{equation}
with the OPT propagator defined as follows:
\begin{equation}
\boldsymbol{G}^{-1}_{\sigma;ab}(x,y) = \left(-\nabla^2_x + m^2 - \sigma(x)\right)\delta_{ab}\delta(x-y) \;.
\label{eq:OPTpropagatorGsigma}
\end{equation}

\subsection{Perturbative expansion}

We proceed as before with the perturbative expansion of the partition function\footnote{We set $\hbar=1$ in this entire section on OPT.}:
\begin{equation}
Z=\int\mathcal{D}\vec{\widetilde{\varphi}} \ e^{-\left(S^0_{\sigma}\big[\vec{\widetilde{\varphi}}\big] + \delta S^1_{\sigma}\big[\vec{\widetilde{\varphi}}\big]\right)} \;,
\label{eq:PartitionFunctionOPT}
\end{equation}
where a fictitious factor $\delta$ has been introduced in order to keep track of the order for the OPT expansion ($\delta$ must therefore be set equal to 1 at the end of all calculations, which is the condition for~\eqref{eq:PartitionFunctionOPT} to reduce to the original partition function of the studied $O(N)$ model). Contrary to the mixed representation including the original field $\vec{\widetilde{\varphi}}(x)$ and the fluctuating collective Hubbard-Stratonovitch field $\widetilde{\sigma}(x)$, there is no path-integration over the configurations of $\sigma(x)$ in the present case, which is why we stressed above the non-fluctuating character of this field.

\vspace{0.5cm}

Taylor expanding the exponential of the residual action $S^1_\sigma$ in~\eqref{eq:PartitionFunctionOPT} yields the following expressions:
\begin{equation}
\begin{split}
Z^\text{OPT} = & \left(\int\mathcal{D}\vec{\widetilde{\varphi}} \ e^{-\int_{x,y}\vec{\widetilde{\varphi}}(x)\cdot\left(\boldsymbol{G}_\sigma^{-1}(x,y)\vec{\widetilde{\varphi}}(y)\right)}\right) \\
& \times \left[ 1 + \sum_{k=1}^{\infty}\frac{\left(-\delta\right)^k}{k!}\sum_{l=0}^k\begin{pmatrix}
k \\
l
\end{pmatrix} \left\langle \left(\frac{1}{2}\int_x\sigma(x)\vec{\widetilde{\varphi}}^{2}(x)\right)^{k-l} \left( \frac{\lambda}{4!} \int_x\left(\vec{\widetilde{\varphi}}^{2}(x)\right)^2\right)^l \right\rangle_{0,\sigma} \right] \;,
\end{split}
\end{equation}
and
\begin{equation}
W^\text{OPT}=\frac{1}{2}\mathrm{STr}\left[\ln (\boldsymbol{G}_\sigma)\right]+\sum_{k=1}^{\infty}\frac{\left(-\delta\right)^k}{k!}\sum_{l=0}^k\begin{pmatrix}
k \\
l
\end{pmatrix}\left\langle\left(\frac{1}{2}\int_x\sigma(x)\vec{\widetilde{\varphi}}^{2}(x)\right)^{k-l} \left( \frac{\lambda}{4!} \int_x\left(\vec{\widetilde{\varphi}}^{2}(x)\right)^2\right)^l\right\rangle_{0,\sigma}^{\text{c}} \;,
\label{eq:WOPTbeforeapplyingWick}
\end{equation}
where the $\sigma$-dependent expectation value is defined by:
\begin{equation}
\big\langle\cdots\big\rangle_{0,\sigma} \equiv \frac{1}{Z_{0,\sigma}}\int\mathcal{D}\vec{\widetilde{\varphi}} \ \cdots \ e^{-S^0_{\sigma}\big[\vec{\widetilde{\varphi}}\big]} \;,
\label{eq:OPTsigmaDepExpValueNumber1}
\end{equation}
with
\begin{equation}
Z_{0,\sigma} = \int\mathcal{D}\vec{\widetilde{\varphi}} \ e^{-S^0_{\sigma}\big[\vec{\widetilde{\varphi}}\big]} \;.
\label{eq:OPTsigmaDepExpValueNumber2}
\end{equation}
The connected correlation functions in~\eqref{eq:WOPTbeforeapplyingWick} are then rewritten with the help of Wick's theorem together with the Feynman rules:
\begin{subequations}
\begin{align}
\begin{gathered}
\begin{fmffile}{Diagrams/LoopExpansion1_FeynRuleGbis2}
\begin{fmfgraph*}(20,5)
\fmfleft{i0,i1,i2,i3}
\fmfright{o0,o1,o2,o3}
\fmflabel{$x, a$}{v1}
\fmflabel{$y, b$}{v2}
\fmf{phantom}{i1,v1}
\fmf{phantom}{i2,v1}
\fmf{plain,tension=0.6}{v1,v2}
\fmf{phantom}{v2,o1}
\fmf{phantom}{v2,o2}
\end{fmfgraph*}
\end{fmffile}
\end{gathered} \quad &\rightarrow \boldsymbol{G}_{\sigma;ab}(x,y)\;,
\label{eq:FeynRulesOPTPropagator} \\
\begin{gathered}
\begin{fmffile}{Diagrams/LoopExpansion1_FeynRuleV4bis}
\begin{fmfgraph*}(20,20)
\fmfleft{i0,i1,i2,i3}
\fmfright{o0,o1,o2,o3}
\fmf{phantom,tension=2.0}{i1,i1bis}
\fmf{plain,tension=2.0}{i1bis,v1}
\fmf{phantom,tension=2.0}{i2,i2bis}
\fmf{plain,tension=2.0}{i2bis,v1}
\fmf{zigzag,label=$x$,tension=0.6,foreground=(0,,0,,1)}{v1,v2}
\fmf{phantom,tension=2.0}{o1bis,o1}
\fmf{plain,tension=2.0}{v2,o1bis}
\fmf{phantom,tension=2.0}{o2bis,o2}
\fmf{plain,tension=2.0}{v2,o2bis}
\fmflabel{$a$}{i1bis}
\fmflabel{$b$}{i2bis}
\fmflabel{$c$}{o1bis}
\fmflabel{$d$}{o2bis}
\end{fmfgraph*}
\end{fmffile}
\end{gathered} \quad &\rightarrow \lambda\delta_{a b}\delta_{c d}\;,
\label{eq:FeynRuleOPT4legVertex} \\
\begin{gathered}
\begin{fmffile}{Diagrams/OPT_FeynRuleSig2}
\begin{fmfgraph*}(4,4)
\fmfleft{i1}
\fmfright{o1}
\fmfv{decor.shape=square,decor.size=2.5thick,label=$x$,label.angle=-90,foreground=(0,,0,,1)}{v3}
\fmflabel{$a$}{v1}
\fmflabel{$b$}{v2}
\fmf{plain}{i1,v1}
\fmf{plain}{v1,v3}
\fmf{plain}{v3,v2}
\fmf{plain}{v2,o1}
\end{fmfgraph*}
\end{fmffile}
\end{gathered} \hspace{0.5cm} &\rightarrow \sigma(x) \delta_{a b}\;.
\label{eq:FeynRulesOPTvertexSquare}
\end{align}
\end{subequations}
In this way, the Schwinger functional reads up to order $\mathcal{O}\big(\delta^2\big)$:
\begin{equation}
\begin{split}
W^\text{OPT} = & \ \frac{1}{2} \mathrm{STr}\left[\ln(\boldsymbol{G}_{\sigma})\right] \\
& - \delta \left(\rule{0cm}{1.2cm}\right. \frac{1}{2} \hspace{0.23cm} \begin{gathered}
\begin{fmffile}{Diagrams/OPT_Diag1}
\begin{fmfgraph}(10,10)
\fmfleft{i}
\fmfright{o}
\fmfv{decor.shape=square,decor.size=2.5thick,foreground=(0,,0,,1)}{v1}
\fmftop{v3}
\fmfbottom{v4}
\fmf{phantom,tension=30}{i,v1}
\fmf{phantom,tension=30}{v2,o}
\fmf{plain,left=0.43,tension=0.5}{v1,v3}
\fmf{plain,left=0.43,tension=0.5}{v3,v2}
\fmf{plain,left=0.43,tension=0.5}{v2,v4}
\fmf{plain,left=0.43,tension=0.5}{v4,v1}
\end{fmfgraph}
\end{fmffile}
\end{gathered} + \frac{1}{24} \hspace{0.08cm} \begin{gathered}
\begin{fmffile}{Diagrams/LoopExpansion1_Hartree}
\begin{fmfgraph}(30,20)
\fmfleft{i}
\fmfright{o}
\fmf{phantom,tension=10}{i,i1}
\fmf{phantom,tension=10}{o,o1}
\fmf{plain,left,tension=0.5}{i1,v1,i1}
\fmf{plain,right,tension=0.5}{o1,v2,o1}
\fmf{zigzag,foreground=(0,,0,,1)}{v1,v2}
\end{fmfgraph}
\end{fmffile}
\end{gathered}
+\frac{1}{12}\begin{gathered}
\begin{fmffile}{Diagrams/LoopExpansion1_Fock}
\begin{fmfgraph}(15,15)
\fmfleft{i}
\fmfright{o}
\fmf{phantom,tension=11}{i,v1}
\fmf{phantom,tension=11}{v2,o}
\fmf{plain,left,tension=0.4}{v1,v2,v1}
\fmf{zigzag,foreground=(0,,0,,1)}{v1,v2}
\end{fmfgraph}
\end{fmffile}
\end{gathered} \left.\rule{0cm}{1.2cm}\right) \\
& + \delta^{2} \left(\rule{0cm}{1.2cm}\right. \frac{1}{4} \hspace{0.23cm} \begin{gathered}
\begin{fmffile}{Diagrams/OPT_Diag2}
\begin{fmfgraph}(10,10)
\fmfleft{i}
\fmfright{o}
\fmfv{decor.shape=square,decor.size=2.5thick,foreground=(0,,0,,1)}{v1}
\fmfv{decor.shape=square,decor.size=2.5thick,foreground=(0,,0,,1)}{v2}
\fmftop{v3}
\fmfbottom{v4}
\fmf{phantom,tension=30}{i,v1}
\fmf{phantom,tension=30}{v2,o}
\fmf{plain,left=0.43,tension=0.5}{v1,v3}
\fmf{plain,left=0.43,tension=0.5}{v3,v2}
\fmf{plain,left=0.43,tension=0.5}{v2,v4}
\fmf{plain,left=0.43,tension=0.5}{v4,v1}
\end{fmfgraph}
\end{fmffile}
\end{gathered} \ + \frac{1}{12} \hspace{0.15cm} \begin{gathered}
\begin{fmffile}{Diagrams/OPT_Diag3}
\begin{fmfgraph}(32,20)
\fmfleft{i}
\fmfright{o}
\fmfv{decor.shape=square,decor.size=2.5thick,foreground=(0,,0,,1)}{v3}
\fmf{phantom,tension=10}{i,v3}
\fmf{phantom,tension=10}{o,v4}
\fmf{plain,left,tension=0.5}{i1,v1,i1}
\fmf{plain,right,tension=0.5}{v2,v4}
\fmf{plain,left,tension=0.5}{v2,v4}
\fmf{plain,left,tension=0.5}{v1,v3}
\fmf{plain,right,tension=0.5}{v1,v3}
\fmf{zigzag,foreground=(0,,0,,1)}{v1,v2}
\end{fmfgraph}
\end{fmffile}
\end{gathered} + \frac{1}{6} \hspace{0.15cm} \begin{gathered}
\begin{fmffile}{Diagrams/OPT_Diag4}
\begin{fmfgraph}(13,13)
\fmfleft{i}
\fmfright{o}
\fmfv{decor.shape=square,decor.size=2.5thick,foreground=(0,,0,,1)}{v3bis}
\fmftop{v3}
\fmfbottom{v4}
\fmf{phantom,tension=15.0}{v3,v3bis}
\fmf{phantom,tension=15.0}{v4,v4bis}
\fmf{phantom,tension=20}{i,v1}
\fmf{phantom,tension=20}{v2,o}
\fmf{phantom,left=0.43,tension=0.5}{v1,v3bis}
\fmf{phantom,left=0.43,tension=0.5}{v3bis,v2}
\fmf{phantom,left=0.43,tension=0.5}{v2,v4bis}
\fmf{phantom,left=0.43,tension=0.5}{v4bis,v1}
\fmf{plain,left,tension=0.0}{v1,v2,v1}
\fmf{zigzag,tension=0,foreground=(0,,0,,1)}{v1,v2}
\end{fmfgraph}
\end{fmffile}
\end{gathered} + \frac{1}{72} \hspace{0.38cm} \begin{gathered}
\begin{fmffile}{Diagrams/OPT_Diag5}
\begin{fmfgraph}(12,12)
\fmfleft{i0,i1}
\fmfright{o0,o1}
\fmftop{v1,vUp,v2}
\fmfbottom{v3,vDown,v4}
\fmf{phantom,tension=20}{i0,v1}
\fmf{phantom,tension=20}{i1,v3}
\fmf{phantom,tension=20}{o0,v2}
\fmf{phantom,tension=20}{o1,v4}
\fmf{plain,left=0.4,tension=0.5}{v3,v1}
\fmf{phantom,left=0.1,tension=0.5}{v1,vUp}
\fmf{phantom,left=0.1,tension=0.5}{vUp,v2}
\fmf{plain,left=0.4,tension=0.0}{v1,v2}
\fmf{plain,left=0.4,tension=0.5}{v2,v4}
\fmf{phantom,left=0.1,tension=0.5}{v4,vDown}
\fmf{phantom,left=0.1,tension=0.5}{vDown,v3}
\fmf{plain,left=0.4,tension=0.0}{v4,v3}
\fmf{zigzag,tension=0.5,foreground=(0,,0,,1)}{v1,v4}
\fmf{zigzag,tension=0.5,foreground=(0,,0,,1)}{v2,v3}
\end{fmfgraph}
\end{fmffile}
\end{gathered} \hspace{0.28cm} + \frac{1}{36} \hspace{0.38cm} \begin{gathered}
\begin{fmffile}{Diagrams/OPT_Diag6}
\begin{fmfgraph}(12,12)
\fmfleft{i0,i1}
\fmfright{o0,o1}
\fmftop{v1,vUp,v2}
\fmfbottom{v3,vDown,v4}
\fmf{phantom,tension=20}{i0,v1}
\fmf{phantom,tension=20}{i1,v3}
\fmf{phantom,tension=20}{o0,v2}
\fmf{phantom,tension=20}{o1,v4}
\fmf{plain,left=0.4,tension=0.5}{v3,v1}
\fmf{phantom,left=0.1,tension=0.5}{v1,vUp}
\fmf{phantom,left=0.1,tension=0.5}{vUp,v2}
\fmf{plain,left=0.4,tension=0.0}{v1,v2}
\fmf{plain,left=0.4,tension=0.5}{v2,v4}
\fmf{phantom,left=0.1,tension=0.5}{v4,vDown}
\fmf{phantom,left=0.1,tension=0.5}{vDown,v3}
\fmf{plain,left=0.4,tension=0.0}{v4,v3}
\fmf{zigzag,left=0.4,tension=0.5,foreground=(0,,0,,1)}{v1,v3}
\fmf{zigzag,right=0.4,tension=0.5,foreground=(0,,0,,1)}{v2,v4}
\end{fmfgraph}
\end{fmffile}
\end{gathered} \\
& \hspace{1.0cm} + \frac{1}{144} \hspace{0.38cm} \begin{gathered}
\begin{fmffile}{Diagrams/OPT_Diag7}
\begin{fmfgraph}(12,12)
\fmfleft{i0,i1}
\fmfright{o0,o1}
\fmftop{v1,vUp,v2}
\fmfbottom{v3,vDown,v4}
\fmf{phantom,tension=20}{i0,v1}
\fmf{phantom,tension=20}{i1,v3}
\fmf{phantom,tension=20}{o0,v2}
\fmf{phantom,tension=20}{o1,v4}
\fmf{plain,left=0.4,tension=0.5}{v3,v1}
\fmf{phantom,left=0.1,tension=0.5}{v1,vUp}
\fmf{phantom,left=0.1,tension=0.5}{vUp,v2}
\fmf{zigzag,left=0.4,tension=0.0,foreground=(0,,0,,1)}{v1,v2}
\fmf{plain,left=0.4,tension=0.5}{v2,v4}
\fmf{phantom,left=0.1,tension=0.5}{v4,vDown}
\fmf{phantom,left=0.1,tension=0.5}{vDown,v3}
\fmf{zigzag,left=0.4,tension=0.0,foreground=(0,,0,,1)}{v4,v3}
\fmf{plain,left=0.4,tension=0.5}{v1,v3}
\fmf{plain,right=0.4,tension=0.5}{v2,v4}
\end{fmfgraph}
\end{fmffile}
\end{gathered}\hspace{0.38cm}  + \frac{1}{36} \begin{gathered}
\begin{fmffile}{Diagrams/OPT_Diag8}
\begin{fmfgraph}(40,20)
\fmfleft{i}
\fmfright{o}
\fmftop{vUpLeft1,vUpLeft2,vUpLeft3,vUpRight1,vUpRight2,vUpRight3}
\fmfbottom{vDownLeft1,vDownLeft2,vDownLeft3,vDownRight1,vDownRight2,vDownRight3}
\fmf{phantom,tension=10}{i,v3}
\fmf{phantom,tension=10}{o,v4}
\fmf{plain,right=0.4,tension=0.5}{v1,vUpLeft}
\fmf{plain,right,tension=0.5}{vUpLeft,vDownLeft}
\fmf{plain,left=0.4,tension=0.5}{v1,vDownLeft}
\fmf{phantom,right=0.4,tension=0.5}{vUpRight,v2}
\fmf{phantom,left,tension=0.5}{vUpRight,vDownRight}
\fmf{phantom,left=0.4,tension=0.5}{vDownRight,v2}
\fmf{phantom,tension=0.3}{v2bis,o}
\fmf{plain,left,tension=0.1}{v2,v2bis,v2}
\fmf{zigzag,tension=2.7,foreground=(0,,0,,1)}{v1,v2}
\fmf{phantom,tension=2}{v1,v3}
\fmf{phantom,tension=2}{v2,v4}
\fmf{phantom,tension=2.4}{vUpLeft,vUpLeft2}
\fmf{phantom,tension=2.4}{vDownLeft,vDownLeft2}
\fmf{phantom,tension=2.4}{vUpRight,vUpRight2}
\fmf{phantom,tension=2.4}{vDownRight,vDownRight2}
\fmf{zigzag,tension=0,foreground=(0,,0,,1)}{vUpLeft,vDownLeft}
\end{fmfgraph}
\end{fmffile}
\end{gathered} \hspace{-0.4cm} + \frac{1}{144} \hspace{0.1cm} \begin{gathered}
\begin{fmffile}{Diagrams/OPT_Diag9}
\begin{fmfgraph}(45,18)
\fmfleft{i}
\fmfright{o}
\fmf{phantom,tension=10}{i,i1}
\fmf{phantom,tension=10}{o,o1}
\fmf{plain,left,tension=0.5}{i1,v1,i1}
\fmf{plain,right,tension=0.5}{o1,v2,o1}
\fmf{zigzag,foreground=(0,,0,,1)}{v1,v3}
\fmf{plain,left,tension=0.5}{v3,v4}
\fmf{plain,right,tension=0.5}{v3,v4}
\fmf{zigzag,foreground=(0,,0,,1)}{v4,v2}
\end{fmfgraph}
\end{fmffile}
\end{gathered} \left.\rule{0cm}{1.2cm}\right) \\
& + \mathcal{O}\Big(\delta^3\Big)\;,
\end{split}
\label{eq:WKjLoopExpansionStep3OPT}
\end{equation}
where further information on the determination of the diagrams can also be found in appendix~\ref{sec:DiagLEO}.

\vspace{0.5cm}

We deduce the expressions of the partition function and the Schwinger functional of our $O(N)$ model in the zero-dimensional limit (pushed up to order $\mathcal{O}(\delta^3)$ in the OPT expansion and setting $\delta=1$):
\begin{equation}
\begin{split}
Z^{\text{OPT};(3)} = & \ (2\pi G_\sigma)^\frac{N}{2}\Bigg[1-\frac{N\sigma G_\sigma}{2} + \frac{N(N+2)}{8} \left(\sigma^2-\frac{\lambda}{3}\right) G_\sigma^2+\frac{N(N+2)(N+4)}{48}\left(\lambda-\sigma^2\right)\sigma G_\sigma^3 \\
& \hspace{1.75cm} +\frac{N(N+2)(N+4)(N+6)}{192}\left(\frac{\lambda}{6}-\sigma^2\right)\lambda G_\sigma^4 \\
& \hspace{1.75cm} -\frac{N(N+2)(N+4)(N+6)(N+8)}{2304}\lambda^2\sigma G_\sigma^5 \\
& \hspace{1.75cm} -\frac{N(N+2)(N+4)(N+6)(N+8)(N+10)}{82944}\lambda^3G_\sigma^6\Bigg] \;,
\end{split}
\end{equation}
and
\begin{equation}
\begin{split}
W^{\text{OPT};(3)} = & \ \frac{N}{2}\ln(2\pi G_{\sigma}) - \frac{N\sigma G_{\sigma}}{2} - \frac{1}{24} N \left(\lambda\left(N+2\right)-6\sigma^2\right)G_{\sigma}^2 + \frac{1}{12} N \sigma \left(\lambda\left(N+2\right)-2\sigma^2\right)G_{\sigma}^3 \\
& +\frac{1}{144}\lambda N \left(N+2\right)\left(\lambda\left(N+3\right)-18\sigma^2\right)G_{\sigma}^4-\frac{1}{36}\lambda^2 N\sigma \left(N^2 + 5N +6\right)G_{\sigma}^5 \\
&-\frac{1}{2592}\lambda^3 N\left(5 N^3 + 44 N^2 + 128 N + 120\right)G_{\sigma}^6 \;,
\end{split}
\end{equation}
with the dressed propagator:
\begin{equation}
G_\sigma=\frac{1}{m^2-\sigma} \;,
\end{equation}
and $\sigma$ to be determined via one of the optimization conditions discussed previously. The gs energy and density of our (0+0)-D $O(N)$ model (for all $N\geq 1$ and for both the unbroken- and broken-symmetry phases) obtained from OPT thus read:
\begin{equation}
\begin{split}
E_{\text{gs}}^{\text{OPT};(3)} = & -\frac{N}{2}\ln\bigg(\frac{2\pi}{m^2-\sigma}\bigg) \\
& + \frac{N}{2592 (m^2 - \sigma)} \Bigg( 1296 \sigma + \frac{108 (\lambda (2 + N) - 6 \sigma^2)}{m^2 - \sigma} - \frac{216 \sigma (\lambda (2 + N) - 2 \sigma^2)}{(m^2 - \sigma)^2} \\
& \hspace{3.15cm} - \frac{18 \lambda (2 + N) (\lambda (3 + N) - 18 \sigma^2)}{(m^2 - \sigma)^3} + \frac{72 \lambda^2 (6 + 5 N + N^2) \sigma}{(m^2 - \sigma)^4} \\
& \hspace{3.15cm} + \frac{\lambda^3 (120 + 128 N + 44 N^2 + 5 N^3)}{(m^2 - \sigma)^5} \Bigg) \;,
\end{split}
\end{equation}
and
\begin{equation}
\begin{split}
\rho_{\text{gs}}^{\text{OPT};(3)} = & \ \frac{1}{m^2-\sigma} \\
& - \frac{1}{216 (m^2 - \sigma)^2} \Bigg( 216 \sigma + \frac{36 (\lambda (2 + N) - 6 \sigma^2)}{m^2 - \sigma} - \frac{108 (\lambda (2 + N) \sigma - 2 \sigma^3)}{(m^2 - \sigma)^2} \\
& \hspace{3.12cm} - \frac{12 \lambda (2 + N) (\lambda (3 + N) - 18 \sigma^2)}{(m^2 - \sigma)^3} + \frac{60 \lambda^2 (6 + 5 N + N^2) \sigma}{(m^2 - \sigma)^4} \\
& \hspace{3.12cm} + \frac{\lambda^3 (120 + 128 N + 44 N^2 + 5 N^3)}{(m^2 - \sigma)^5} \Bigg) \;.
\end{split}
\end{equation}
A definite estimate of the gs energy and density is only obtained after fixing a value for $\sigma$, as we discuss next.

\subsection{Optimization of $\sigma$}

What distinguishes OPT from many other approaches is the fact that the optimization of the field $\sigma$ depends on the working order $k$. We focus on three classes of optimization criteria for $\sigma$, namely the PMS, the TP method and the SCC\footnote{The study of ref.~\cite{ros16} has already shown that the FAC optimization procedure is less performing than the PMS approach when determining the gs energy, the self-energy and the fourth-order vertex function $\Gamma^{(\mathrm{1PI})(4)}\big(\vec{\phi}=\vec{0}\big)$ in the framework of the unbroken-symmetry phase of the toy model considered in this thesis.}.

\subsubsection{Principle of minimal sensitivity}

\begin{figure}[!htb]
\captionsetup[subfigure]{labelformat=empty}
  \begin{center}
    \subfloat[]{
      \includegraphics[width=0.50\linewidth]{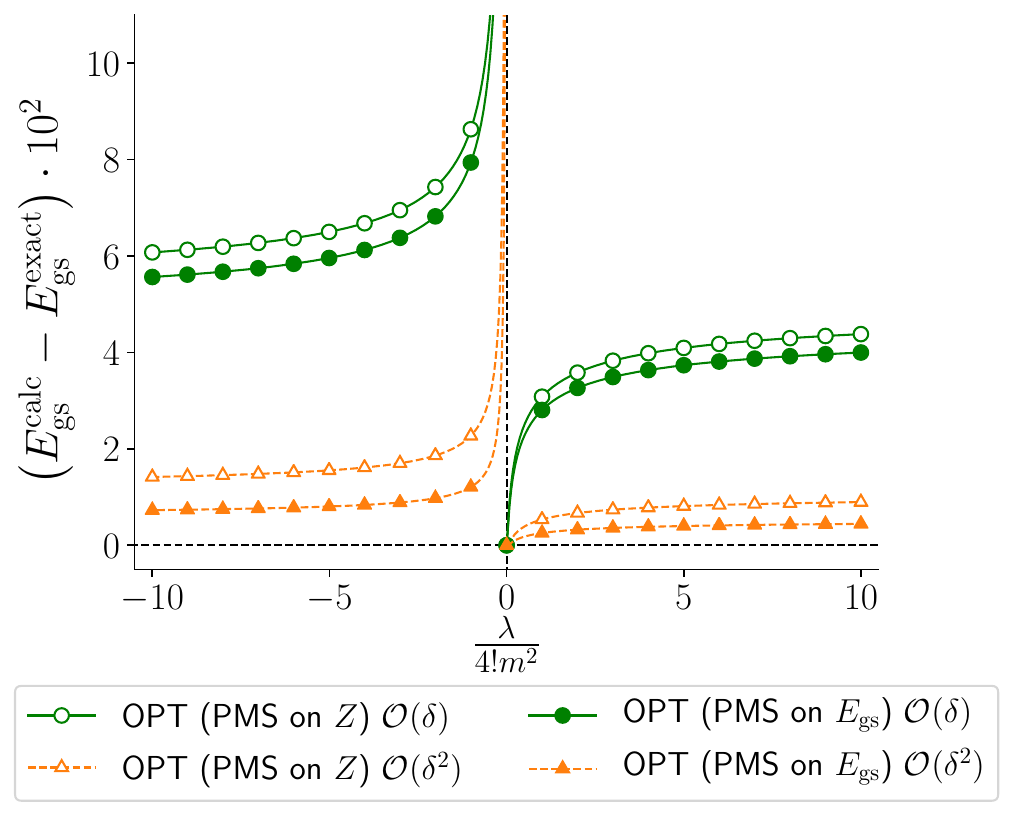}
                         }
    \subfloat[]{
      \includegraphics[width=0.50\linewidth]{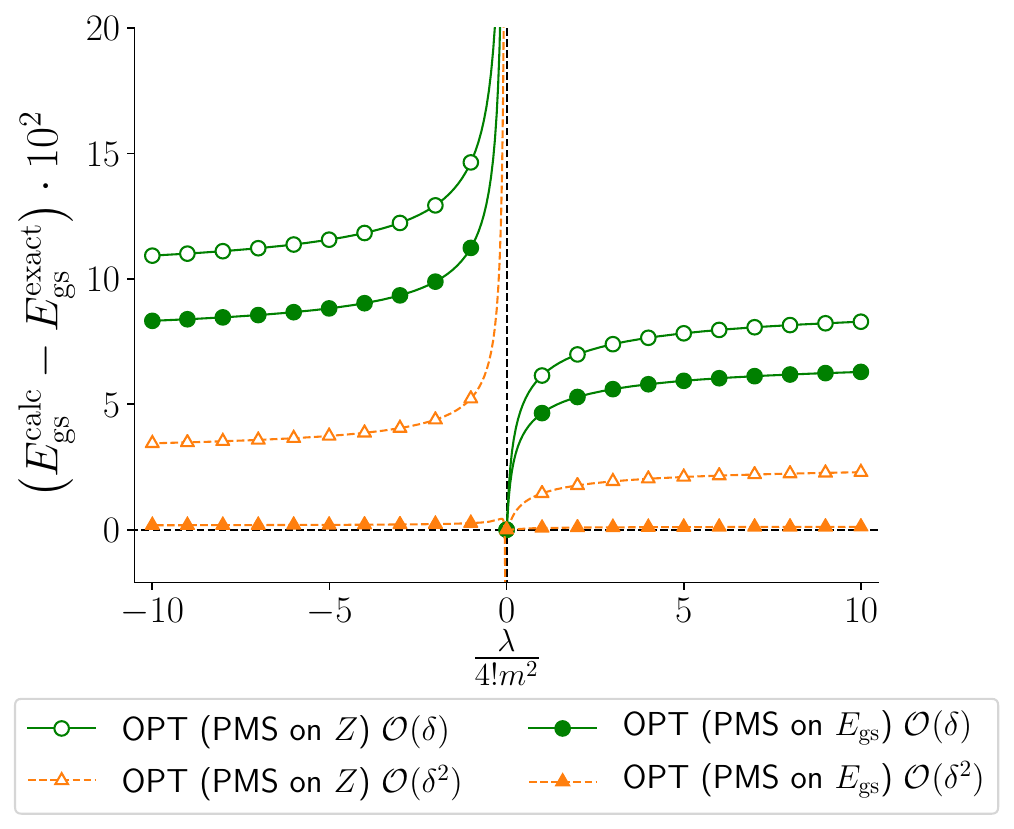}
                         }
    \caption{Difference between the calculated gs energy $E_{\mathrm{gs}}^{\mathrm{calc}}$ and the corresponding exact solution $E_{\mathrm{gs}}^{\mathrm{exact}}$ at $N=1$ (left) and $N=2$ (right), with $\hbar=1$ and $m^{2}=\pm 1$ ($\mathcal{R}e(\lambda)\geq 0$ and $\mathcal{I}m(\lambda)=0$).}
    \label{fig:OPTPMSonZorE}
  \end{center}
\end{figure}

We first consider the PMS, where we either ask the partition function $Z^\text{OPT;(k)}$, the gs energy $E^\text{OPT;(k)}_\text{gs}$ or the gs density $\rho^\text{OPT;(k)}_\text{gs}$, computed at order $k$ (i.e. up to order $\mathcal{O}(\delta^k)$) of the OPT expansion, to be extremal with respect to $\sigma$, i.e.:
\begin{equation}
\left.\frac{\partial Z^\text{OPT;(k)}}{\partial\sigma}\right|_{\sigma = \sigma^{(k)}_{\mathrm{PMS};Z}} = 0 \;,
\label{eq:PMSeqZ0DON}
\end{equation}
\begin{equation}
\left.\frac{\partial E^\text{OPT;(k)}_\text{gs}}{\partial\sigma}\right|_{\sigma = \sigma^{(k)}_{\mathrm{PMS};E}} = 0 \;,
\label{eq:PMSeqEgs0DON}
\end{equation}
\begin{equation}
\left.\frac{\partial\rho^\text{OPT;(k)}_\text{gs}}{\partial\sigma}\right|_{\sigma = \sigma^{(k)}_{\mathrm{PMS};\rho}} = 0 \;.
\label{eq:PMSeqrhogs0DON}
\end{equation}
For instance, at the first non-trivial order (i.e. up to order $\mathcal{O}(\delta)$) of the OPT expansion, the three above equations are polynomial and second-order with respect to $\sigma$ and their solutions read:
\begin{equation}
\sigma^{(1)}_{\mathrm{PMS};Z} = \frac{3 m^2\left(N+4\right) \pm \sqrt{3N\left(\lambda\left(N^2 +4N +4\right) + 3 N m^4\right)}}{6\left(N+2\right)} \;,
\end{equation}
\begin{equation}
\sigma^{(1)}_{\mathrm{PMS};E} = \frac{1}{2}\left(m^2 \pm \sqrt{m^{4} + \frac{2\lambda}{3}\left(N+2\right)}\right) \;,
\label{eq:solutionPMSEgs0DON}
\end{equation}
\begin{equation}
\sigma^{(1)}_{\mathrm{PMS};\rho} = \frac{1}{2}\left(m^2 \pm  \sqrt{m^4+\lambda\left(N+2\right)}\right) \;.
\label{eq:solutionPMSrhogs0DON}
\end{equation}
The complexity of the PMS equations~\eqref{eq:PMSeqZ0DON} to~\eqref{eq:PMSeqrhogs0DON} increases with the truncation order, i.e. with the working order $k$. We illustrate in fig.~\ref{fig:OPTPMSonZorE} that, for the purpose of determining $E_{\mathrm{gs}}$ or $\rho_{\mathrm{gs}}$, it is in general more efficient to apply the PMS directly on the OPT series representing $E_{\mathrm{gs}}$ or $\rho_{\mathrm{gs}}$ respectively (i.e. to exploit~\eqref{eq:PMSeqEgs0DON} and~\eqref{eq:PMSeqrhogs0DON}), rather than on $Z$ (i.e. rather than using~\eqref{eq:PMSeqZ0DON}). We can already appreciate in this figure the nice convergence properties of OPT at its first two non-trivial orders, which will be discussed further later in this section.

\subsubsection{Turning point}

The optimization of $\sigma$ via the TP method is based on the equations:
\begin{equation}
\left.\frac{\partial^2 E^\text{OPT;(k)}_\text{gs}}{\partial\sigma^2}\right|_{\sigma=\sigma^{(k)}_{\mathrm{TP};E}} = 0 \;,
\label{eq:solutionTPEgs0DON}
\end{equation}
\begin{equation}
\left.\frac{\partial^2\rho^\text{OPT;(k)}_\text{gs}}{\partial\sigma^2}\right|_{\sigma=\sigma^{(k)}_{\mathrm{TP};\rho}} = 0 \;,
\label{eq:solutionTPrhogs0DON}
\end{equation}
whose solutions are, at the first non-trivial order of the OPT expansion:
\begin{equation}
\sigma^{(1)}_{\mathrm{TP};E} = \pm \sqrt{m^4 + \frac{\lambda}{2}\left(N+2\right)} \;,
\end{equation}
\begin{equation}
\sigma^{(1)}_{\mathrm{TP};\rho} = \frac{1}{4}\left(m^2 \pm \sqrt{9 m^4 + 8 \lambda\left(N+2\right)}\right) \;.
\end{equation}
As for the PMS, the complexity of the TP optimization equations such as~\eqref{eq:solutionTPEgs0DON} and~\eqref{eq:solutionTPrhogs0DON} grows with the working order $k$.

\subsubsection{Self-consistent condition}

We implement the optimization of $\sigma$ via the SCC by solving the equation:
\begin{equation}
\left\langle\left(\vec{\widetilde{\varphi}}\cdot\vec{\widetilde{\varphi}}\right)^k\right\rangle_{0,\sigma=\sigma^{(k)}_\text{SCC}}^{(1)} = \left\langle\left(\vec{\widetilde{\varphi}}\cdot\vec{\widetilde{\varphi}}\right)^k\right\rangle_{0,\sigma=\sigma^{(k)}_\text{SCC}}^{(0)} \;,
\label{eq:SCCequation0DON}
\end{equation}
where the $\sigma$-dependent expectation value is already defined by~\eqref{eq:OPTsigmaDepExpValueNumber1} and~\eqref{eq:OPTsigmaDepExpValueNumber2} in arbitrary dimensions. Whatever the working order $k$ in the OPT expansion, the complexity of the SCC equation~\eqref{eq:SCCequation0DON} remains that of a second-order polynomial equation in $\sigma$. It is the order of the correlation functions involved in both the RHS and LHS of this SCC equation that changes with $k$. One can then write the solutions of~\eqref{eq:SCCequation0DON} for all $k$ as:
\begin{equation}
\sigma^{(k)}_\text{SCC} = \frac{1}{2}\left(m^2\pm\sqrt{m^4+\frac{2\lambda}{3}(N+k+1)}\right) \;.
\label{eq:SolutionSCC0DON}
\end{equation}
Hence, the solutions~\eqref{eq:solutionPMSEgs0DON} and~\eqref{eq:SolutionSCC0DON}, obtained respectively from the PMS and the SCC, coincide at $k=1$. In other words, we obtain the same solution for $\sigma$ either by imposing the gs energy calculated up to the first non-trivial order of the OPT expansion to be extremal with respect to $\sigma$ or by demanding that the density (or propagator) calculated at the first non-trivial order of the OPT expansion coincides with the one obtained from the zeroth order. We postpone to future works the investigation of the validity of this connection at higher truncation orders $k$, where the SCC is implemented in the form $\left\langle\vec{\widetilde{\varphi}}\cdot \vec{\widetilde{\varphi}}\right\rangle^{(k)}_{0,\sigma=\sigma^{(k)}_\text{SCC}} = \left\langle\vec{\widetilde{\varphi}}\cdot \vec{\widetilde{\varphi}}\right\rangle^{(0)}_{0,\sigma=\sigma^{(k)}_\text{SCC}}$.

\subsection{Discussion}

As explained in section~\ref{secSpiritOPT}, OPT optimizes the partitioning of the original classical action $S$ into an unperturbed reference part $S^0$ and a residual part $S^1$, which translates into a non-trivial dressing of the unperturbed field propagator $\boldsymbol{G}_{\sigma}$ (defined by~\eqref{eq:OPTpropagatorGsigma} for the studied $O(N)$ model in arbitrary dimensions) with non-perturbative correlations. It is instructive to compare how the propagator gets renormalized in the frameworks of both OPT and LEs, within the mixed and collective representations in particular. To that end, we write the dressed unperturbed propagators in the generic form:
\begin{equation}
\boldsymbol{G}^\star_{0;ab} = \frac{1}{m_\star^2} \delta_{ab} \;,
\end{equation}
with $m_\star$ being a renormalized mass. As discussed right below~\eqref{eq:ClassicalSolutionBosonicAction}, we have obtained for both the mixed and collective LEs:
\begin{equation}
m_{\star;\text{LE}}^2 = m^2+i\sqrt{\frac{\lambda}{3}}\overline{\sigma}_\text{cl} \;,
\label{eq:renormalizedmassmstart}
\end{equation}
with $\overline{\sigma}_\text{cl}$ being the saddle point of the mixed or collective classical action at vanishing sources, and where the collective Hubbard-Stratonovich dof $\widetilde{\sigma}$ is a fluctuating field participating to the PI measure. More specifically, we have found the following expressions for $\overline{\sigma}_\text{cl}$:
\begin{equation}
\overline{\sigma}_\text{cl;mix} = -i\sqrt{\frac{\lambda}{12}}\vec{\overline{\varphi}}_{\text{cl}}\cdot\vec{\overline{\varphi}}_{\text{cl}} =  \begin{cases} \displaystyle{ 0 \quad \forall m^2\geq 0 \;,} \\
\\
\displaystyle{ i\sqrt{\frac{3}{\lambda}}m^2 \quad \forall m^2< 0 ~ \text{and} ~ \lambda\neq 0 \;,} \end{cases}
\label{eq:sigmasaddleMixedLE}
\end{equation}
and\footnote{Only the minus sign solution (which is the physical solution) is taken in~\eqref{eq:ClassicalSolutionBosonicAction} to obtain~\eqref{eq:sigmasaddleCollectiveLE}.}
\begin{equation}
\overline{\sigma}_\text{cl;col} =\frac{i}{2}\sqrt{\frac{3}{\lambda}}\left(m^2-\sqrt{m^4+\frac{2\lambda}{3}N}\right) \;,
\label{eq:sigmasaddleCollectiveLE}
\end{equation}
according to~\eqref{eq:SaddlePointPhiclSmix} and~\eqref{eq:ClassicalSolutionBosonicAction} for the mixed and collective LEs, respectively. After inserting~\eqref{eq:sigmasaddleMixedLE} and~\eqref{eq:sigmasaddleCollectiveLE} into~\eqref{eq:renormalizedmassmstart}, we obtain the following unperturbed inverse propagators:
\begin{equation}
m_{\star;\text{LE;mix}}^2 = \begin{cases} \displaystyle{m^2  \quad \forall m^2\geq 0 \;,} \\
\\
\displaystyle{0 \quad \forall m^2< 0 ~ \text{and} ~ \lambda\neq 0 \;,}  \end{cases}
\label{eq:m2LEmix}
\end{equation}
and
\begin{equation}
m_{\star;\text{LE;col}}^2 =\frac{1}{2}\left(m^2 + \sqrt{m^4+\frac{2\lambda}{3}N}\right) \;.
\label{eq:m2LEcol}
\end{equation}
These (inverse) propagators are the same whatever the working order $k$ of the LE, i.e. regardless up to which order $\mathcal{O}\big(\hbar^k\big)$ the LE is carried out. In other words, only the series representing the quantities of interest (i.e. the gs energy and density in this study), which involve the dressed unperturbed (inverse) propagator $m_{\star}^2$, vary with the working order $k$ of the LE, whereas $m_{\star}^2$ remains independent of $k$ and is always given by~\eqref{eq:m2LEmix} or~\eqref{eq:m2LEcol} for the mixed and collective representations, respectively.

\vspace{0.5cm}

However, the dressed (inverse) propagator involved in OPT depends on the collective dof $\sigma$ (i.e. on a non-fluctuating field to be adjusted), which itself depends on the working order $k$ (i.e. it depends up to which order $\mathcal{O}(\delta^k)$ the OPT is carried out), namely $m_{\star;\text{OPT}}^2 = m_{\star;\text{OPT}}^2(\sigma(k))$. In other words, both the OPT series of the gs energy and density on the one hand and the dressed unperturbed propagator on the other hand change with the working order $k$. Focusing on the PMS and SCC optimization procedures only, we obtain\footnote{Similarly to~\eqref{eq:sigmasaddleCollectiveLE}, expressions~\eqref{eq:mstarOPTPMSE},~\eqref{eq:mstarOPTPMSrho} and~\eqref{eq:mstarOPTSCC} are all physical solutions. They are obtained by taking the minus sign solutions of~\eqref{eq:solutionPMSEgs0DON},~\eqref{eq:solutionPMSrhogs0DON} and~\eqref{eq:SolutionSCC0DON}, respectively. Their physical character can be seen from the fact that they all reduce to $m_{\star}^2=m^2$ at $\lambda=0$ and $m^2 \geq 0$.} at the first non-trivial order of the OPT expansion:
\begin{equation}
m_{\star;\text{OPT;PMS;$E$;(1)}}^2 = \frac{1}{2}\left(m^2 + \sqrt{m^{4} + \frac{2\lambda}{3}\left(N+2\right)}\right) \;,
\label{eq:mstarOPTPMSE}
\end{equation}
\begin{equation}
m_{\star;\text{OPT;PMS;$\rho$;(1)}}^2 = \frac{1}{2}\left(m^2 + \sqrt{m^4+\lambda\left(N+2\right)}\right) \;,
\label{eq:mstarOPTPMSrho}
\end{equation}
and, up to order $\mathcal{O}\big(\delta^k\big)$ of the OPT expansion,
\begin{equation}
m_{\star;\text{OPT;SCC;(k)}}^2 =\frac{1}{2}\left(m^2 + \sqrt{m^4+\frac{2\lambda}{3}(N+k+1)}\right) \;.
\label{eq:mstarOPTSCC}
\end{equation}
We recover of course the property $m_{\star;\text{OPT;SCC;(1)}}^2=m_{\star;\text{OPT;PMS;$E$;(1)}}^2$ discussed before right below~\eqref{eq:SolutionSCC0DON} but we can also see a significant resemblance between the renormalized masses \eqref{eq:mstarOPTPMSE} and~\eqref{eq:mstarOPTSCC} obtained from OPT via PMS and SCC on the one hand and, on the other hand, that given by~\eqref{eq:m2LEcol} for the collective LE.

\vspace{0.5cm}

\begin{figure}[!htb]
\captionsetup[subfigure]{labelformat=empty}
  \begin{center}
    \subfloat[]{
      \includegraphics[width=0.50\linewidth]{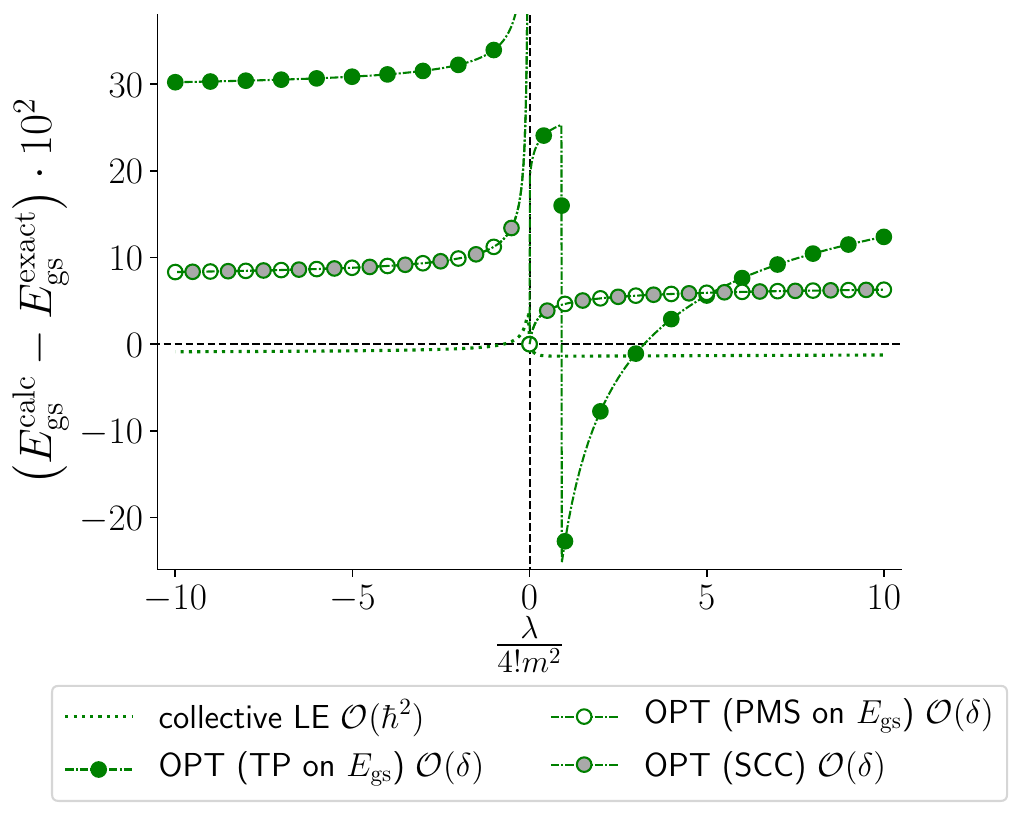}
                         }
    \subfloat[]{
      \includegraphics[width=0.50\linewidth]{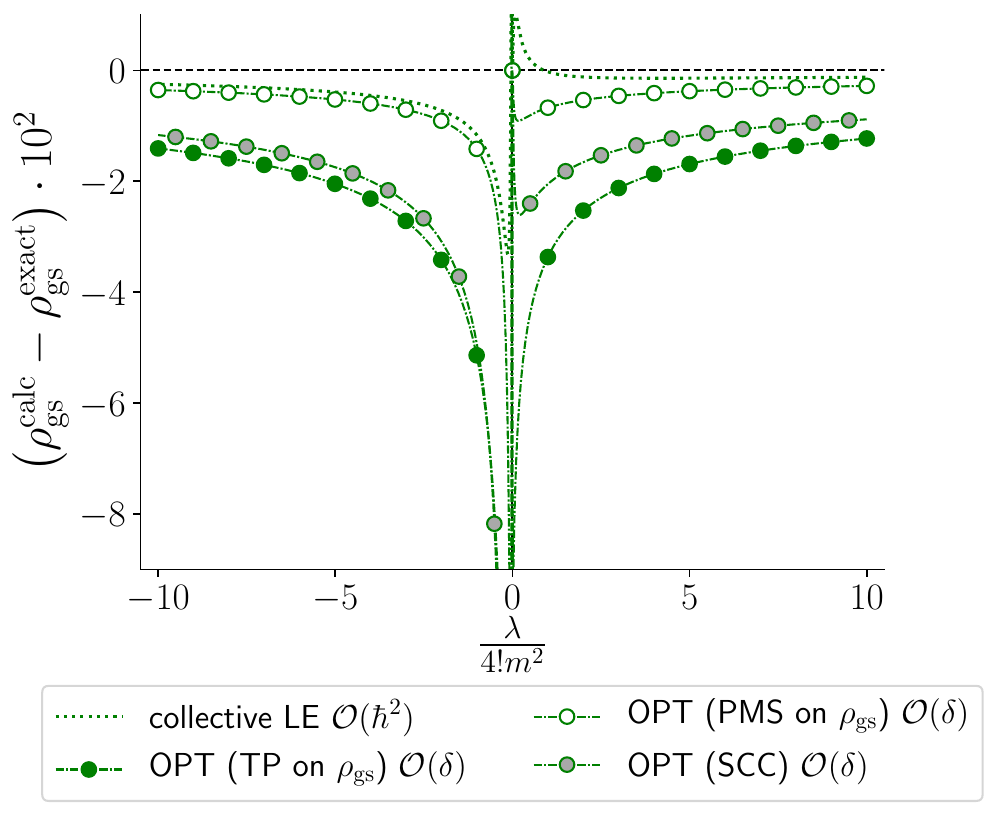}
                         }
    \caption{Difference between the calculated gs energy $E_{\mathrm{gs}}^{\mathrm{calc}}$ (left) or density $\rho_{\mathrm{gs}}^{\mathrm{calc}}$ (right) and the corresponding exact solution $E_{\mathrm{gs}}^{\mathrm{exact}}$ or $\rho_{\mathrm{gs}}^{\mathrm{exact}}$ at $\hbar=1$, $m^{2}=\pm 1$ and $N=2$ ($\mathcal{R}e(\lambda)\geq 0$ and $\mathcal{I}m(\lambda)=0$). The presented results are the first non-trivial orders of the collective LE and OPT with the three different tested optimization procedures. See also the caption of fig.~\ref{fig:O1PTcoll} for the meaning of the indication ``$\mathcal{O}\big(\hbar^{n}\big)$'' for the collective LE results.}
    \label{fig:O2OPTvsLE_1}
  \end{center}
\end{figure}

\begin{figure}[!htb]
\captionsetup[subfigure]{labelformat=empty}
  \begin{center}
    \subfloat[]{
      \includegraphics[width=0.50\linewidth]{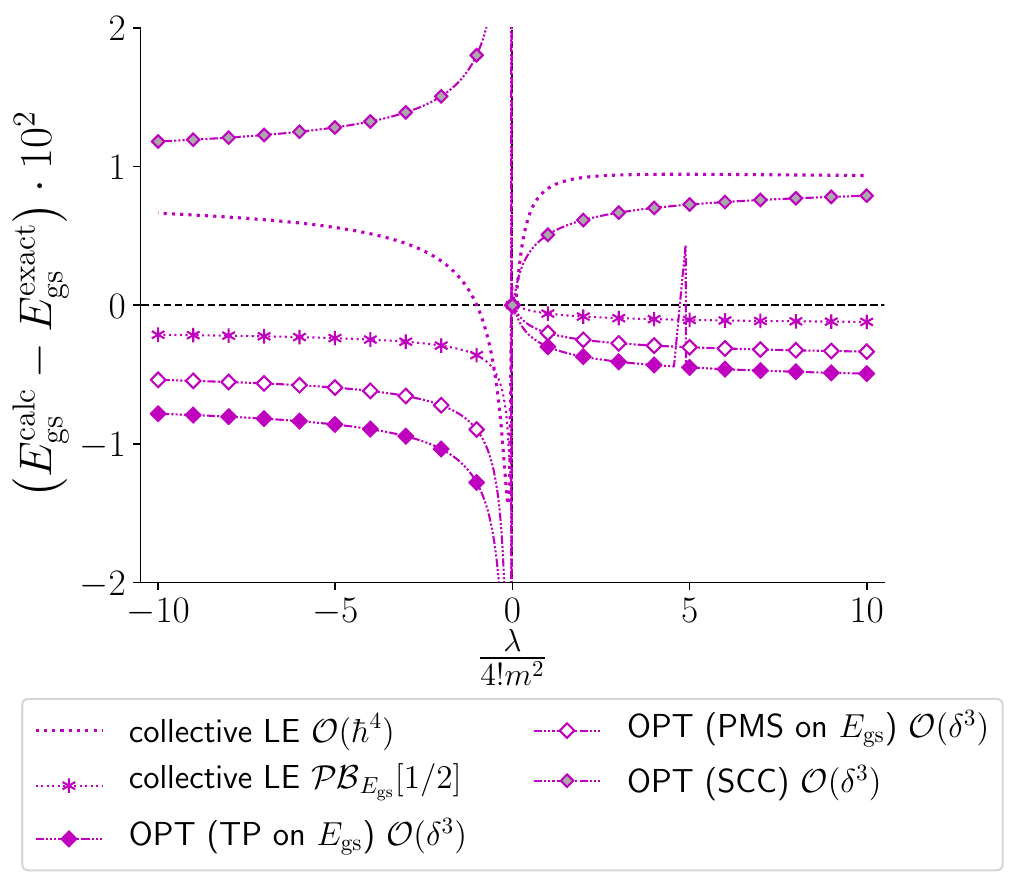}
                         }
    \subfloat[]{
      \includegraphics[width=0.50\linewidth]{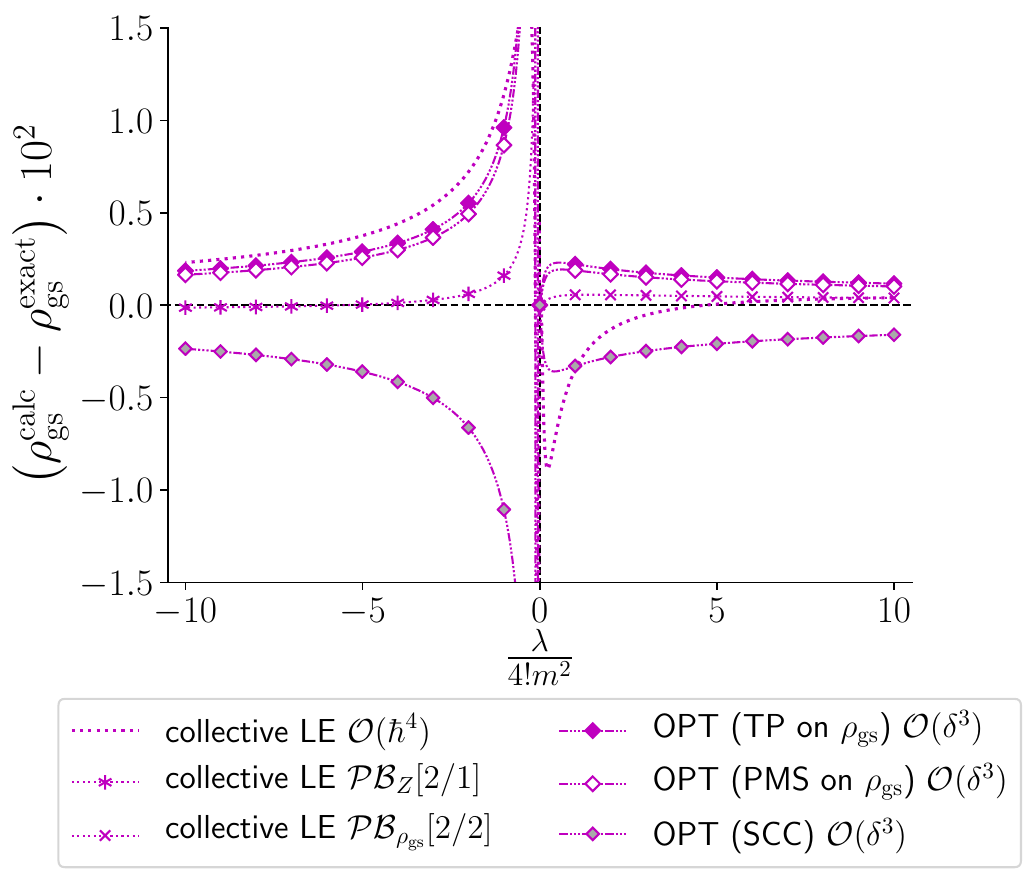}
                         }
    \caption{Same as fig.~\ref{fig:O2OPTvsLE_1} but for the third non-trivial orders of the collective LE (including the best tested resummation procedures among the Pad\'e-Borel and Borel-hypergeometric schemes) and of OPT.}
    \label{fig:O2OPTvsLE_3}
  \end{center}
\end{figure}

An excellent reproduction of the gs energy and density is achieved with OPT based on the PMS, which notably results in an accuracy around $0.5\%$ at the third non-trivial order in both the unbroken- and broken-symmetry regimes at $N=2$, as shown notably by fig.~\ref{fig:O2OPTvsLE_3}. At the first non-trivial order, the PMS and the SCC lead to identical results, as justified below~\eqref{eq:SolutionSCC0DON} and illustrated by fig.~\ref{fig:O2OPTvsLE_1} at $N=2$. However, fig.~\ref{fig:O2OPTvsLE_3} shows (still at $N=2$) that the PMS slightly outperforms the SCC: for example, the corresponding estimates for $E_{\mathrm{gs}}$ are respectively around $0.6\%$ and $1.1\%$ throughout most of the tested range of values for the coupling constant (i.e. for $\lambda/4! \in [0,10]$). This loss of accuracy of the SCC as compared to the PMS is rather small considering the simplicity of the underlying equations to solve: at the third non-trivial order (and regardless of the values of $N$ and $m^2$), the SCC still amounts to finding the roots of a quadratic polynomial whereas the PMS criterion is now a polynomial equation of order $6$. Note also that, according to figs.~\ref{fig:O2OPTvsLE_1} and~\ref{fig:O2OPTvsLE_3}, the TP method is clearly less performing than the SCC and the PMS at the first non-trivial order of the OPT expansion but becomes comparable to SCC at the third one, whereas the underpinning equations to solve are closer to those of the PMS in terms of complexity. The TP method is thus disappointing (as compared to both the PMS and the SCC) in that respect.

\vspace{0.5cm}

Comparing the two best approaches investigated so far, i.e. OPT and the collective LE, figs.~\ref{fig:O2OPTvsLE_1} and~\ref{fig:O2OPTvsLE_3} show that, after combination with resummation, the collective LE outperforms OPT at both the first and third non-trivial orders for both $E_{\mathrm{gs}}$ and $\rho_{\mathrm{gs}}$, at $N=2$. Furthermore, as discussed earlier from figs.~\ref{fig:O1PTcoll} and~\ref{fig:O2PTcoll}, the performances of the collective LE are expected to improve with $N$ due to its connection with the $1/N$-expansion whereas such an argument does not hold for OPT (for instance, the performances of OPT in fig.~\ref{fig:OPTPMSonZorE} do not differ significantly at $N=1$ and $2$). However, regarding the formalisms underpinning these two techniques, the diagrammatic representations of their respective expansions is much more demanding to determine on the side of the collective LE: whereas OPT diagrams are directly obtained by adding the square vertex~\eqref{eq:FeynRulesOPTvertexSquare} to the diagrams of the original LE at $\vec{\varphi}_{\mathrm{cl}}=\vec{0}$ (i.e. in the unbroken-symmetry regime) in all possible ways, the collective LE requires to construct the diagrammatic expressions of all vertex functions $S^{(n)}_{\mathrm{col},\mathcal{J}}$ for $n=2,\cdots,2k$ to determine the corresponding Schwinger functional up to order $\mathcal{O}\big(\hbar^{k}\big)$. As can be inferred from~\eqref{eq:SbosonicKLoopExpansionS3} and~\eqref{eq:SbosonicKLoopExpansionS4} expressing respectively $S^{(3)}_{\mathrm{col},\mathcal{J}}$ and $S^{(4)}_{\mathrm{col},\mathcal{J}}$, the determination of the diagrammatic expressions of $S^{(n)}_{\mathrm{col},\mathcal{J}}$ becomes quickly lengthy as $n$ increases. This cumbersomeness directly results from the logarithm structure of the collective classical action. For a complete comparison between the collective LE and OPT, one should also consider their ability to treat different (competing) channels in more realistic systems. This is indeed of crucial importance for the description of many realistic many-body systems, and of most atomic nuclei in particular. Although these two PI techniques were applied to a toy model involving a single channel, their implementation can in principle be extended to treat several channels in an equitable fashion. More precisely, for methods based on the mixed or the collective representation like the collective LE, one might use multi-channel HSTs, as was done e.g. in refs.~\cite{lan15,kug18}, and OPT can be implemented by introducing a classical field coupled to each relevant bilinear in the original field(s) of the theory under consideration~\cite{kle11}\footnote{OPT is sometimes put forward (notably in ref.~\cite{kle11}) as a technique designed to solve an old standing problem of HSTs leading to their inability to describe efficiently competing channels. One should stress however that the more recently introduced multi-channel HSTs also allow for a democratic treatment of several competing channels.}, which is a straightforward generalization of the recipe outlined in section~\ref{sec:SplittingOPTfinitedimON}.

\vspace{0.5cm}

In conclusion, OPT offers an interesting framework to describe strongly-coupled many-body systems at low cost. As opposed to LEs, OPT results are directly systematically improvable in the sense that they do not rely on resummation procedures: they take the form of (tremendously) fast convergent series (see notably refs.~\cite{buc93,dun93,ben94,arv95} for detailed studies on the convergence behavior of OPT series). However, the energy and the density are not tied in a functional as in the EDF approach. We now investigate the third and last family of approaches considered in the present chapter. These approaches are based on EAs from which one can work with functionals of the correlation functions of the theory, or their local versions that can coincide with e.g. the density of the system.

\section{Effective action}
\label{sec:EA}

The EA framework allows for representing the partition function of a given theory in an exact and compact fashion. As discussed in section~\ref{sec:RelevantGenFunc}, the $n$PI EA can be expressed in terms of diagrams with dressed propagators and dressed $k$-point vertex functions for $k\leq n$. We investigate the 1PI, 2P(P)I and 4PPI EAs either organized with respect to $\hbar$ or $\lambda$, for the original, mixed and collective representations of the studied $O(N)$ model, as summarized in tab.~\ref{tab:CJTEA0DON}. Within these various EA implementations, expressions will be worked out via the inversion method\footnote{The formalism of part of the 1PI and 2PI EA approaches discussed in this thesis (those expressed in the original representation and $\lambda$-wise organized) are discussed in ref.~\cite{hay91} for a $\varphi^{4}$-theory (but not for the $O(N)$-symmetric case) and QCD. As opposed to this work and for the sake of clarity, we explicitly construct here the 1PI EA via the IM instead of directly giving the 1PI diagrams contributing to it.} (IM) introduced by Fukuda and collaborators~\cite{fuk94,fuk95}, especially since it enables us to draw direct connections with Kohn-Sham DFT via the 2PPI EA~\cite{val97}. There are of course other methods to derive the diagrammatic expressions of EAs, like for instance a method developed by Carrington to exploit the 4PI EA~\cite{car04}. We choose to focus on the IM especially because of the links that it enlights between the EA formalism and DFT which is particularly precious to us in our aim to reformulate the nuclear EDF method. Hence, we will wait until section~\ref{sec:2PPIEA} on the 2PPI EA to present the general principles underlying the IM and its aforementioned connections with DFT but we stress that the derivations of all diagrammatic expressions of the EAs treated in this chapter are discussed in detail in appendix~\ref{ann:InversionMethod} within the IM framework.

\begin{table}[h]
\centering
\fontsize{9pt}{10pt}\selectfont
\caption{\label{tab:CJTEA0DON} All EA implementations investigated in the present study of the (0+0)-D $O(N)$-symmetric $\varphi^{4}$-theory. The designation ``no 1-pt'' indicates that all 1-point correlation functions (i.e. the 1-point correlation function of the original field and possibly that of the Hubbard-Stratonovich field) are imposed to vanish in the corresponding formalism. In addition, ``$\hbar$-expansion'' and ``$\lambda$-expansion'' refer to the parameter organizing the expansion (and therefore the truncation) of the EA, i.e. either the $\hbar$ constant or the coupling constant $\lambda$ of the studied $O(N)$ model.}
\begin{tabular}{M{0.13\textwidth}|M{0.13\textwidth}|M{0.105\textwidth}|M{0.105\textwidth}|M{0.105\textwidth}|M{0.105\textwidth}|M{0.105\textwidth}|M{0.105\textwidth}|}
\cline{3-8}
\multicolumn{2}{c|}{ } & \multicolumn{2}{|c|}{Original EA} & \multicolumn{2}{|c|}{Mixed EA} & \multicolumn{2}{|c|}{Collective EA} \\ \cline{3-8}
\multicolumn{2}{c|}{ } & $\hbar$-expansion & $\lambda$-expansion & $\hbar$-expansion & $\lambda$-expansion & $\hbar$-expansion & $\lambda$-expansion \\ \hline
\multicolumn{2}{|c|}{1PI EA} & \checkmark & \checkmark &  &  & \checkmark & \\ \hline
\multicolumn{2}{|c|}{2P(P)I} & \checkmark &  & \checkmark &  &  & \\ \hline
\multicolumn{2}{|c|}{2P(P)I EA (no 1-pt)} & \checkmark & \checkmark & \checkmark & \checkmark &  & \\
\multicolumn{2}{|c|}{4PPI EA (no 1-pt)} & \checkmark & \checkmark &  &  &  & \\ \hline
\end{tabular}
\end{table}

\subsection{\label{sec:1PIEA}1PI effective action}
\subsubsection{\label{sec:original1PIEA}Original effective action}
\paragraph{$\hbar$-expansion:}

We start our discussion on EA approaches with the original 1PI EA (i.e. the 1PI EA formulated in the original representation) organized with respect to $\hbar$.  Note that for any $n$PI EAs with a $\hbar$-expansion, we could actually directly express the EA by simply retaining all the $n$PI diagrams up to a certain power of $\hbar$ in the LE series of the corresponding Schwinger functional. We have however carried out the IM in such cases in order to illustrate how the IM formalism enables us to recover this diagrammatic property (see appendix~\ref{sec:1PIEAannIM}). The 1PI EA under consideration is defined by the following Legendre transform:
\begin{equation}
\begin{split}
\Gamma^{(\mathrm{1PI})}\Big[\vec{\phi}\Big] \equiv & -W\Big[\vec{J}\Big]+\int_{x}J^{a}(x) \frac{\delta W\big[\vec{J}\big]}{\delta J^{a}(x)} \\
= & -W\Big[\vec{J}\Big]+\int_{x}J^{a}(x) \phi_{a}(x) \;,
\end{split}
\label{eq:pure1PIEAdefinition0DONmain}
\end{equation}
with
\begin{equation}
\phi_{a}(x)=\frac{\delta W\big[\vec{J}\big]}{\delta J^{a}(x)} \;,
\label{eq:pure1PIEAdefinitionbis0DONmain}
\end{equation}
and $W\big[\vec{J}\big]$ corresponds to the original Schwinger functional determined in section~\ref{sec:PT} via the LE at $\boldsymbol{K}=\boldsymbol{0}$, i.e. $W\big[\vec{J}\big]\equiv W^{\text{LE};\text{orig}}\big[\vec{J},\boldsymbol{K}=\boldsymbol{0}\big]$. This EA can be expressed diagrammatically as (see appendix~\ref{sec:original1PIEAannIM}):
\begin{equation}
\begin{split}
\Gamma^{(\mathrm{1PI})}\Big[\vec{\phi}\Big] = & \ S\Big[\vec{\phi}\Big] -\frac{\hbar}{2}\mathrm{STr}\left[\ln\big(\boldsymbol{G}_{\phi}\big)\right] \\
& + \hbar^{2} \left(\rule{0cm}{1.2cm}\right. \frac{1}{24} \hspace{0.08cm} \begin{gathered}
\begin{fmffile}{Diagrams/1PIEA_Hartree}
\begin{fmfgraph}(30,20)
\fmfleft{i}
\fmfright{o}
\fmf{phantom,tension=10}{i,i1}
\fmf{phantom,tension=10}{o,o1}
\fmf{plain,left,tension=0.5,foreground=(1,,0,,0)}{i1,v1,i1}
\fmf{plain,right,tension=0.5,foreground=(1,,0,,0)}{o1,v2,o1}
\fmf{zigzag,foreground=(0,,0,,1)}{v1,v2}
\end{fmfgraph}
\end{fmffile}
\end{gathered}
+\frac{1}{12}\begin{gathered}
\begin{fmffile}{Diagrams/1PIEA_Fock}
\begin{fmfgraph}(15,15)
\fmfleft{i}
\fmfright{o}
\fmf{phantom,tension=11}{i,v1}
\fmf{phantom,tension=11}{v2,o}
\fmf{plain,left,tension=0.4,foreground=(1,,0,,0)}{v1,v2,v1}
\fmf{zigzag,foreground=(0,,0,,1)}{v1,v2}
\end{fmfgraph}
\end{fmffile}
\end{gathered}
- \frac{1}{18} \begin{gathered}
\begin{fmffile}{Diagrams/1PIEA_Diag1}
\begin{fmfgraph}(27,15)
\fmfleft{i}
\fmfright{o}
\fmftop{vUp}
\fmfbottom{vDown}
\fmfv{decor.shape=cross,decor.size=3.5thick,foreground=(1,,0,,0)}{v1}
\fmfv{decor.shape=cross,decor.size=3.5thick,foreground=(1,,0,,0)}{v2}
\fmf{phantom,tension=10}{i,i1}
\fmf{phantom,tension=10}{o,o1}
\fmf{phantom,tension=2.2}{vUp,v5}
\fmf{phantom,tension=2.2}{vDown,v6}
\fmf{phantom,tension=0.5}{v3,v4}
\fmf{phantom,tension=10.0}{i1,v1}
\fmf{phantom,tension=10.0}{o1,v2}
\fmf{dashes,tension=2.0,foreground=(0,,0,,1),foreground=(1,,0,,0)}{v1,v3}
\fmf{dots,left=0.4,tension=0.5,foreground=(0,,0,,1)}{v3,v5}
\fmf{plain,left=0.4,tension=0.5,foreground=(1,,0,,0)}{v5,v4}
\fmf{plain,right=0.4,tension=0.5,foreground=(1,,0,,0)}{v3,v6}
\fmf{dots,right=0.4,tension=0.5,foreground=(0,,0,,1)}{v6,v4}
\fmf{dashes,tension=2.0,foreground=(0,,0,,1),foreground=(1,,0,,0)}{v4,v2}
\fmf{plain,tension=0,foreground=(1,,0,,0)}{v5,v6}
\end{fmfgraph}
\end{fmffile}
\end{gathered} - \frac{1}{36} \hspace{-0.15cm} \begin{gathered}
\begin{fmffile}{Diagrams/1PIEA_Diag2}
\begin{fmfgraph}(25,20)
\fmfleft{i}
\fmfright{o}
\fmftop{vUp}
\fmfbottom{vDown}
\fmfv{decor.shape=cross,decor.angle=45,decor.size=3.5thick,foreground=(1,,0,,0)}{vUpbis}
\fmfv{decor.shape=cross,decor.angle=45,decor.size=3.5thick,foreground=(1,,0,,0)}{vDownbis}
\fmf{phantom,tension=0.8}{vUp,vUpbis}
\fmf{phantom,tension=0.8}{vDown,vDownbis}
\fmf{dashes,tension=0.5,foreground=(0,,0,,1),foreground=(1,,0,,0)}{v3,vUpbis}
\fmf{phantom,tension=0.5}{v4,vUpbis}
\fmf{phantom,tension=0.5}{v3,vDownbis}
\fmf{dashes,tension=0.5,foreground=(0,,0,,1),foreground=(1,,0,,0)}{v4,vDownbis}
\fmf{phantom,tension=11}{i,v1}
\fmf{phantom,tension=11}{v2,o}
\fmf{plain,left,tension=0.5,foreground=(1,,0,,0)}{v1,v2,v1}
\fmf{dots,tension=1.7,foreground=(0,,0,,1)}{v1,v3}
\fmf{plain,foreground=(1,,0,,0)}{v3,v4}
\fmf{dots,tension=1.7,foreground=(0,,0,,1)}{v4,v2}
\end{fmfgraph}
\end{fmffile}
\end{gathered} \hspace{-0.22cm} \left.\rule{0cm}{1.2cm}\right) \\
& + \mathcal{O}\big(\hbar^{3}\big)\;,
\end{split}
\label{eq:1PIEAfinalexpression}
\end{equation}
with the Feynman rules:
\begin{subequations}
\begin{align}
\begin{gathered}
\begin{fmffile}{Diagrams/1PIEA_G}
\begin{fmfgraph*}(20,20)
\fmfleft{i0,i1,i2,i3}
\fmfright{o0,o1,o2,o3}
\fmflabel{$x, a$}{v1}
\fmflabel{$y, b$}{v2}
\fmf{phantom}{i1,v1}
\fmf{phantom}{i2,v1}
\fmf{plain,tension=0.6,foreground=(1,,0,,0)}{v1,v2}
\fmf{phantom}{v2,o1}
\fmf{phantom}{v2,o2}
\end{fmfgraph*}
\end{fmffile}
\end{gathered} \quad &\rightarrow \boldsymbol{G}_{\phi;ab}(x,y)\;,
\label{eq:DefinitionG1PIEAhbarExpansion} \\
\begin{gathered}
\begin{fmffile}{Diagrams/1PIEA_V3}
\begin{fmfgraph*}(20,20)
\fmfleft{i0,i1,i2,i3}
\fmfright{o0,o1,o2,o3}
\fmfv{decor.shape=cross,decor.angle=45,decor.size=3.5thick,foreground=(1,,0,,0)}{o2}
\fmf{phantom,tension=2.0}{i1,i1bis}
\fmf{plain,tension=2.0,foreground=(1,,0,,0)}{i1bis,v1}
\fmf{phantom,tension=2.0}{i2,i2bis}
\fmf{plain,tension=2.0,foreground=(1,,0,,0)}{i2bis,v1}
\fmf{dots,label=$x$,tension=0.6,foreground=(0,,0,,1)}{v1,v2}
\fmf{phantom,tension=2.0}{o1bis,o1}
\fmf{plain,tension=2.0,foreground=(1,,0,,0)}{v2,o1bis}
\fmf{phantom,tension=2.0}{o2bis,o2}
\fmf{phantom,tension=2.0,foreground=(1,,0,,0)}{v2,o2bis}
\fmf{dashes,tension=0.0,foreground=(1,,0,,0)}{v2,o2}
\fmflabel{$a$}{i1bis}
\fmflabel{$b$}{i2bis}
\fmflabel{$c$}{o1bis}
\fmflabel{$N$}{o2bis}
\end{fmfgraph*}
\end{fmffile}
\end{gathered} \quad &\rightarrow \lambda\left|\vec{\phi}(x)\right|\delta_{a b}\delta_{c N}\;,
\label{eq:FeynRules1PIEA3legVertexSourceJ0main} \\
\begin{gathered}
\begin{fmffile}{Diagrams/1PIEA_V4}
\begin{fmfgraph*}(20,20)
\fmfleft{i0,i1,i2,i3}
\fmfright{o0,o1,o2,o3}
\fmf{phantom,tension=2.0}{i1,i1bis}
\fmf{plain,tension=2.0,foreground=(1,,0,,0)}{i1bis,v1}
\fmf{phantom,tension=2.0}{i2,i2bis}
\fmf{plain,tension=2.0,foreground=(1,,0,,0)}{i2bis,v1}
\fmf{zigzag,label=$x$,tension=0.6,foreground=(0,,0,,1)}{v1,v2}
\fmf{phantom,tension=2.0}{o1bis,o1}
\fmf{plain,tension=2.0,foreground=(1,,0,,0)}{v2,o1bis}
\fmf{phantom,tension=2.0}{o2bis,o2}
\fmf{plain,tension=2.0,foreground=(1,,0,,0)}{v2,o2bis}
\fmflabel{$a$}{i1bis}
\fmflabel{$b$}{i2bis}
\fmflabel{$c$}{o1bis}
\fmflabel{$d$}{o2bis}
\end{fmfgraph*}
\end{fmffile}
\end{gathered} \quad &\rightarrow \lambda\delta_{a b}\delta_{c d}\;,
\label{eq:FeynRules1PIEA4legVertexSourceJ0main}
\end{align}
\end{subequations}
where, as in our previous treatment of LEs in section~\ref{sec:PT}, we have fixed our coordinates in color space such that a spontaneous breakdown of the $O(N)$ symmetry can only occur in the direction set by $a=N$ (still without any loss of generality). Such a convention will be followed throughout the entire section~\ref{sec:EA}. This translates into:
\begin{equation}
\vec{\phi}(x) = \left|\vec{\phi}(x)\right| \begin{pmatrix}
0 \\
\vdots\\
0 \\
1
\end{pmatrix} \;.
\end{equation}
Moreover, the propagator $\boldsymbol{G}_{\phi}$ is not dressed by the classical configuration $\vec{\varphi}_\text{cl}$ of the original field as in~\eqref{eq:LEG} for the original LE, but by its 1-point correlation function $\vec{\phi}\equiv\left\langle\vec{\widetilde{\varphi}}\right\rangle_{J}$ (satisfying~\eqref{eq:pure1PIEAdefinitionbis0DONmain}), which contains quantal or radiative corrections. It is defined by:
\begin{equation}
\begin{split}
\boldsymbol{G}^{-1}_{\phi;ab}(x,y) \equiv & \ \left.\frac{\delta^{2} S[\vec{\widetilde{\varphi}}]}{\delta \widetilde{\varphi}^{a}(x)\delta \widetilde{\varphi}^{b}(y)}\right|_{\vec{\widetilde{\varphi}}=\vec{\phi}} \\
= & \left(-\nabla_x^2 + m^2 + \frac{\lambda}{6} \phi^c(x)\phi_{c}(x)\right)\delta_{ab}\delta(x-y)+ \frac{\lambda}{3}\phi_{a}(x)\phi_{b}(x)\delta(x-y) \;.
\end{split}
\label{eq:DefinitionG1PIEAhbarExpansionbis}
\end{equation}

\vspace{0.5cm}

Let us then evaluate $\Gamma^{(\mathrm{1PI})}$ in the (0+0)-D limit. As in section~\ref{sec:PT} (with~\eqref{eq:origLEGoldstoneProp} to~\eqref{eq:origLEmassiveProp} more specifically), we separate the (inverse) propagator $\boldsymbol{G}^{-1}_{\phi}$ into the Goldstone modes one $G^{-1}_{\phi;\mathfrak{g}} = \mathfrak{G}^{-1}_{\phi;\mathfrak{g}} \mathbb{I}_{N-1} = \left(m^{2} + \lambda \vec{\phi}^{2}/6\right) \mathbb{I}_{N-1}$ and that of the Higgs mode (associated to the direction $a=N$ in color space according to our choice of coordinates) $\boldsymbol{G}^{-1}_{\phi;NN}= m^{2} + \lambda \vec{\phi}^{2}/2$. From this, we evaluate the diagrams contributing to $\Gamma^{(\mathrm{1PI})}$ at order $\mathcal{O}\big(\hbar^2\big)$:
\begin{equation}
\begin{split}
\begin{gathered}
\begin{fmffile}{Diagrams/1PIEA_Hartree}
\begin{fmfgraph}(30,20)
\fmfleft{i}
\fmfright{o}
\fmf{phantom,tension=10}{i,i1}
\fmf{phantom,tension=10}{o,o1}
\fmf{plain,left,tension=0.5,foreground=(1,,0,,0)}{i1,v1,i1}
\fmf{plain,right,tension=0.5,foreground=(1,,0,,0)}{o1,v2,o1}
\fmf{zigzag,foreground=(0,,0,,1)}{v1,v2}
\end{fmfgraph}
\end{fmffile}
\end{gathered} = \lambda\left[\boldsymbol{G}_{\phi;NN} + \left(N-1\right)\mathfrak{G}_{\phi;\mathfrak{g}}\right]^{2} \;,
\end{split}
\label{eq:EvaluateDiag1PIEAnumber1}
\end{equation}

\vspace{-0.6cm}

\begin{equation}
\begin{split}
\begin{gathered}
\begin{fmffile}{Diagrams/1PIEA_Fock}
\begin{fmfgraph}(15,15)
\fmfleft{i}
\fmfright{o}
\fmf{phantom,tension=11}{i,v1}
\fmf{phantom,tension=11}{v2,o}
\fmf{plain,left,tension=0.4,foreground=(1,,0,,0)}{v1,v2,v1}
\fmf{zigzag,foreground=(0,,0,,1)}{v1,v2}
\end{fmfgraph}
\end{fmffile}
\end{gathered} = \lambda\left[\boldsymbol{G}_{\phi;NN}^{2} + \left(N-1\right)\mathfrak{G}_{\phi;\mathfrak{g}}^{2}\right] \;,
\end{split}
\end{equation}
\begin{equation}
\begin{split}
\begin{gathered}
\begin{fmffile}{Diagrams/1PIEA_Diag1}
\begin{fmfgraph}(27,15)
\fmfleft{i}
\fmfright{o}
\fmftop{vUp}
\fmfbottom{vDown}
\fmfv{decor.shape=cross,decor.size=3.5thick,foreground=(1,,0,,0)}{v1}
\fmfv{decor.shape=cross,decor.size=3.5thick,foreground=(1,,0,,0)}{v2}
\fmf{phantom,tension=10}{i,i1}
\fmf{phantom,tension=10}{o,o1}
\fmf{phantom,tension=2.2}{vUp,v5}
\fmf{phantom,tension=2.2}{vDown,v6}
\fmf{phantom,tension=0.5}{v3,v4}
\fmf{phantom,tension=10.0}{i1,v1}
\fmf{phantom,tension=10.0}{o1,v2}
\fmf{dashes,tension=2.0,foreground=(0,,0,,1),foreground=(1,,0,,0)}{v1,v3}
\fmf{dots,left=0.4,tension=0.5,foreground=(0,,0,,1)}{v3,v5}
\fmf{plain,left=0.4,tension=0.5,foreground=(1,,0,,0)}{v5,v4}
\fmf{plain,right=0.4,tension=0.5,foreground=(1,,0,,0)}{v3,v6}
\fmf{dots,right=0.4,tension=0.5,foreground=(0,,0,,1)}{v6,v4}
\fmf{dashes,tension=2.0,foreground=(0,,0,,1),foreground=(1,,0,,0)}{v4,v2}
\fmf{plain,tension=0,foreground=(1,,0,,0)}{v5,v6}
\end{fmfgraph}
\end{fmffile}
\end{gathered} = \lambda^{2} \phi_{N}^{2} \boldsymbol{G}_{\phi;NN}^{3} \;,
\end{split}
\end{equation}
\begin{equation}
\begin{split}
\begin{gathered}
\begin{fmffile}{Diagrams/1PIEA_Diag2}
\begin{fmfgraph}(25,20)
\fmfleft{i}
\fmfright{o}
\fmftop{vUp}
\fmfbottom{vDown}
\fmfv{decor.shape=cross,decor.angle=45,decor.size=3.5thick,foreground=(1,,0,,0)}{vUpbis}
\fmfv{decor.shape=cross,decor.angle=45,decor.size=3.5thick,foreground=(1,,0,,0)}{vDownbis}
\fmf{phantom,tension=0.8}{vUp,vUpbis}
\fmf{phantom,tension=0.8}{vDown,vDownbis}
\fmf{dashes,tension=0.5,foreground=(0,,0,,1),foreground=(1,,0,,0)}{v3,vUpbis}
\fmf{phantom,tension=0.5}{v4,vUpbis}
\fmf{phantom,tension=0.5}{v3,vDownbis}
\fmf{dashes,tension=0.5,foreground=(0,,0,,1),foreground=(1,,0,,0)}{v4,vDownbis}
\fmf{phantom,tension=11}{i,v1}
\fmf{phantom,tension=11}{v2,o}
\fmf{plain,left,tension=0.5,foreground=(1,,0,,0)}{v1,v2,v1}
\fmf{dots,tension=1.7,foreground=(0,,0,,1)}{v1,v3}
\fmf{plain,foreground=(1,,0,,0)}{v3,v4}
\fmf{dots,tension=1.7,foreground=(0,,0,,1)}{v4,v2}
\end{fmfgraph}
\end{fmffile}
\end{gathered} = \lambda^{2}\phi_{N}^{2}\boldsymbol{G}_{\phi;NN}\left[\boldsymbol{G}_{\phi;NN}^{2} + \left(N-1\right)\mathfrak{G}_{\phi;\mathfrak{g}}^{2} \right]\;.
\end{split}
\label{eq:EvaluateDiag1PIEAnumber4}
\end{equation}
According to~\eqref{eq:EvaluateDiag1PIEAnumber1} to~\eqref{eq:EvaluateDiag1PIEAnumber4}, it follows that~\eqref{eq:1PIEAfinalexpression} reduces in the zero-dimensional limit to:
\begin{equation}
\begin{split}
\Gamma^{(\mathrm{1PI})}\Big(\vec{\phi}\Big) = & \ S\Big(\vec{\phi}\Big) - \frac{\hbar}{2}\left[\left(N-1\right)\ln\big(2\pi \mathfrak{G}_{\phi;\mathfrak{g}}\big)+\ln\big(2\pi \boldsymbol{G}_{\phi;NN}\big)\right] \\
& + \hbar^{2}\Bigg[\frac{\lambda}{72}\Big(9\left(\boldsymbol{G}_{\phi;NN}\right)^{2}+3\left(\mathfrak{G}_{\phi;\mathfrak{g}}\right)^{2}\left(N^{2}-1\right)-6\left(\boldsymbol{G}_{\phi;NN}\right)^{3}\lambda\phi_{N}^{2} \\
& \hspace{0.9cm} -2\boldsymbol{G}_{\phi;NN}\mathfrak{G}_{\phi;\mathfrak{g}}\left(N-1\right)\left(-3+\mathfrak{G}_{\phi;\mathfrak{g}}\lambda\phi_{N}^{2}\right)\Big)\Bigg] \\
& + \mathcal{O}\big(\hbar^{3}\big) \;,
\end{split}
\label{eq:1PIEAfinalexpression0DON}
\end{equation}
with $S\big(\vec{\phi}\big)=m^{2}\phi^{2}_{N}/2+\lambda\phi^{4}_{N}/4!$ here. This expression of the 1PI EA is then exploited by fixing the configuration of the 1-point correlation function $\vec{\phi}$ and more specifically of its component $\phi_{N}$. This is achieved by solving the gap equation:
\begin{equation}
\begin{split}
0 = \left.\frac{\partial \Gamma^{(\mathrm{1PI})}\big(\vec{\phi}\big)}{\partial \phi_{N}}\right|_{\vec{\phi}=\vec{\overline{\phi}}} = & \ \left(\mathfrak{G}_{\overline{\phi};\mathfrak{g}}\right)^{-1}\overline{\phi}_{N} + \frac{\hbar}{2} \Bigg[\boldsymbol{G}_{\overline{\phi};NN} \lambda \overline{\phi}_{N} + \frac{1}{3} \mathfrak{G}_{\overline{\phi};\mathfrak{g}} \left(-1 + N\right) \lambda \overline{\phi}_{N}\Bigg] \\
& + \hbar^{2} \Bigg[ \frac{1}{108} \lambda^{2} \overline{\phi}_{N} \Bigg( -45 \left(\boldsymbol{G}_{\overline{\phi};NN}\right)^3 - 3 \left(\mathfrak{G}_{\overline{\phi};\mathfrak{g}}\right)^3 \left(-1 + N^2\right) \\
& \hspace{0.9cm} + 27 \left(\boldsymbol{G}_{\overline{\phi};NN}\right)^4 \lambda \overline{\phi}_{N}^2 + 3 \left(\boldsymbol{G}_{\overline{\phi};NN}\right)^2 \mathfrak{G}_{\overline{\phi};\mathfrak{g}} \left(-1 + N\right) \Big(-3 + \mathfrak{G}_{\overline{\phi};\mathfrak{g}} \lambda \overline{\phi}_{N}^2\Big) \\
& \hspace{0.9cm} + \boldsymbol{G}_{\overline{\phi};NN} \left(\mathfrak{G}_{\overline{\phi};\mathfrak{g}}\right)^2 \left(-1 + N\right) \Big(-9 + 2 \mathfrak{G}_{\overline{\phi};\mathfrak{g}} \lambda \overline{\phi}_{N}^2\Big) \Bigg) \Bigg] \\
& +\mathcal{O}\big(\hbar^{3}\big)\;,
\end{split}
\label{eq:1PIEAGapEquation0DON}
\end{equation}
with $\vec{\overline{\phi}}=\begin{pmatrix}
\overline{\phi}_{1} & \cdots & \overline{\phi}_{N-1} & \overline{\phi}_{N}
\end{pmatrix}^{\mathrm{T}}=\begin{pmatrix}
0 & \cdots & 0 & \overline{\phi}_{N}
\end{pmatrix}^{\mathrm{T}}$, $\mathfrak{G}^{-1}_{\overline{\phi};\mathfrak{g}} = m^{2} + \lambda \overline{\phi}_{N}^{2}/6$ and $\boldsymbol{G}^{-1}_{\overline{\phi};NN}= m^{2} + \lambda \overline{\phi}_{N}^{2}/2$ (the latter two quantities corresponding respectively to the configurations of $\mathfrak{G}^{-1}_{\phi;\mathfrak{g}}$ and $\boldsymbol{G}^{-1}_{\phi;NN}$ at $\vec{J}=\vec{0}$). The gs energy is subsequently inferred from the solution $\vec{\overline{\phi}}$ together with~\eqref{eq:1PIEAfinalexpression0DON} according to:
\begin{equation}
E^\text{1PI EA;orig}_{\mathrm{gs}} = \frac{1}{\hbar} \Gamma^{\mathrm{(1PI)}}\Big(\vec{\phi}=\vec{\overline{\phi}}\Big) \;.
\label{eq:DeduceEgs1PIEAdiag}
\end{equation}

\paragraph{$\lambda$-expansion:}
Exploiting instead $\lambda$ as expansion parameter\footnote{In this thesis, we always set $\hbar=1$ while treating $\lambda$-expansions of EAs.}, the 1PI EA (still defined by \eqref{eq:pure1PIEAdefinition0DONmain} and~\eqref{eq:pure1PIEAdefinitionbis0DONmain}) reads (see appendix~\ref{sec:original1PIEAannIM}):
\begin{equation}
\begin{split}
\Gamma^{(\mathrm{1PI})}\Big[\vec{\phi}\Big] = & \ S\Big[\vec{\phi}\Big] -\frac{1}{2}\mathrm{STr}\left[\ln\big(\boldsymbol{G}_{0}\big)\right] \\
& +\frac{1}{24} \hspace{0.08cm} \begin{gathered}
\begin{fmffile}{Diagrams/LoopExpansion1_Hartree}
\begin{fmfgraph}(30,20)
\fmfleft{i}
\fmfright{o}
\fmf{phantom,tension=10}{i,i1}
\fmf{phantom,tension=10}{o,o1}
\fmf{plain,left,tension=0.5}{i1,v1,i1}
\fmf{plain,right,tension=0.5}{o1,v2,o1}
\fmf{zigzag,foreground=(0,,0,,1)}{v1,v2}
\end{fmfgraph}
\end{fmffile}
\end{gathered}
+\frac{1}{12}\begin{gathered}
\begin{fmffile}{Diagrams/LoopExpansion1_Fock}
\begin{fmfgraph}(15,15)
\fmfleft{i}
\fmfright{o}
\fmf{phantom,tension=11}{i,v1}
\fmf{phantom,tension=11}{v2,o}
\fmf{plain,left,tension=0.4}{v1,v2,v1}
\fmf{zigzag,foreground=(0,,0,,1)}{v1,v2}
\end{fmfgraph}
\end{fmffile}
\end{gathered} + \frac{1}{12} \ \begin{gathered}
\begin{fmffile}{Diagrams/1PIEAlambda_Diag5bis2}
\begin{fmfgraph*}(25,13)
\fmfleft{i}
\fmfright{oDown,o,oUp}
\fmfv{decor.shape=cross,decor.angle=45,decor.size=3.5thick,foreground=(1,,0,,0)}{oUp}
\fmfv{decor.shape=cross,decor.angle=45,decor.size=3.5thick,foreground=(1,,0,,0)}{oDown}
\fmf{phantom,tension=10}{i,i1}
\fmf{phantom,tension=10}{o,o1}
\fmf{plain,left,tension=0.5}{i1,v1,i1}
\fmf{phantom,right,tension=0.5}{o1,v2,o1}
\fmf{dashes,tension=0,foreground=(1,,0,,0)}{v2,oUp}
\fmf{dashes,tension=0,foreground=(1,,0,,0)}{v2,oDown}
\fmf{zigzag,foreground=(0,,0,,1)}{v1,v2}
\end{fmfgraph*}
\end{fmffile}
\end{gathered} + \frac{1}{6} \ \begin{gathered}
\begin{fmffile}{Diagrams/1PIEAlambda_Diag6bis2}
\begin{fmfgraph*}(15,15)
\fmfleft{iDown,i,iUp}
\fmfright{oDown,o,oUp}
\fmfv{decor.shape=cross,decor.size=3.5thick,foreground=(1,,0,,0)}{iUp}
\fmfv{decor.shape=cross,decor.size=3.5thick,foreground=(1,,0,,0)}{oUp}
\fmf{dashes,tension=0,foreground=(1,,0,,0)}{v1,iUp}
\fmf{dashes,tension=0,foreground=(1,,0,,0)}{v2,oUp}
\fmf{phantom,tension=11}{i,v1}
\fmf{phantom,tension=11}{v2,o}
\fmf{plain,right,tension=0.4}{v1,v2}
\fmf{zigzag,foreground=(0,,0,,1)}{v1,v2}
\end{fmfgraph*}
\end{fmffile}
\end{gathered} \\
& + \mathcal{O}\big(\lambda^{2}\big)\;,
\end{split}
\label{eq:1PIEAlambdaEAStep30DON}
\end{equation}
with the Feynman rules:
\begin{subequations}
\begin{align}
\begin{gathered}
\begin{fmffile}{Diagrams/LoopExpansion1_FeynRuleGbis}
\begin{fmfgraph*}(20,20)
\fmfleft{i0,i1,i2,i3}
\fmfright{o0,o1,o2,o3}
\fmflabel{$x, a$}{v1}
\fmflabel{$y, b$}{v2}
\fmf{phantom}{i1,v1}
\fmf{phantom}{i2,v1}
\fmf{plain,tension=0.6}{v1,v2}
\fmf{phantom}{v2,o1}
\fmf{phantom}{v2,o2}
\end{fmfgraph*}
\end{fmffile}
\end{gathered} \quad &\rightarrow \boldsymbol{G}_{0;ab}(x,y)\;,\\
\begin{gathered}
\begin{fmffile}{Diagrams/1PIEAlambda_Phi2}
\begin{fmfgraph*}(20,20)
\fmfleft{i0,i1,i2,i3}
\fmfright{o0,o1,o2,o3}
\fmflabel{$x, a$}{v1}
\fmfv{decor.shape=cross,decor.size=3.5thick,foreground=(1,,0,,0)}{v2}
\fmf{phantom}{i1,v1}
\fmf{phantom}{i2,v1}
\fmf{dashes,tension=1.2,foreground=(1,,0,,0)}{v1,v2}
\fmf{phantom}{v2,o1}
\fmf{phantom}{v2,o2}
\end{fmfgraph*}
\end{fmffile}
\end{gathered} \hspace{-0.2cm} &\rightarrow \phi_{a}(x) \;,\\
\begin{gathered}
\begin{fmffile}{Diagrams/LoopExpansion1_FeynRuleV4bis}
\begin{fmfgraph*}(20,20)
\fmfleft{i0,i1,i2,i3}
\fmfright{o0,o1,o2,o3}
\fmf{phantom,tension=2.0}{i1,i1bis}
\fmf{plain,tension=2.0}{i1bis,v1}
\fmf{phantom,tension=2.0}{i2,i2bis}
\fmf{plain,tension=2.0}{i2bis,v1}
\fmf{zigzag,label=$x$,tension=0.6,foreground=(0,,0,,1)}{v1,v2}
\fmf{phantom,tension=2.0}{o1bis,o1}
\fmf{plain,tension=2.0}{v2,o1bis}
\fmf{phantom,tension=2.0}{o2bis,o2}
\fmf{plain,tension=2.0}{v2,o2bis}
\fmflabel{$a$}{i1bis}
\fmflabel{$b$}{i2bis}
\fmflabel{$c$}{o1bis}
\fmflabel{$d$}{o2bis}
\end{fmfgraph*}
\end{fmffile}
\end{gathered} \quad &\rightarrow \lambda\delta_{a b}\delta_{c d}\;.
\label{eq:FeynRulesLoopExpansion4legVertexbis}
\end{align}
\end{subequations}
We stress that, in the framework of the $\lambda$-expansion, the 1PI EA involves the bare propagator $\boldsymbol{G}_{0}$ which is no longer dressed by the 1-point correlation function $\vec{\phi}$ according to the relation:
\begin{equation}
\boldsymbol{G}^{-1}_{0;ab}(x,y) = \left(-\nabla_x^2 + m^2 \right)\delta_{ab}\delta(x-y) \;,
\label{eq:1PIEAlambdaExppDefG0}
\end{equation}
which is to be compared with~\eqref{eq:DefinitionG1PIEAhbarExpansionbis} for the $\hbar$-expansion.

\vspace{0.5cm}

As a next step, we take the zero-dimensional limit. The diagrams of~\eqref{eq:1PIEAlambdaEAStep30DON} are thus evaluated as follows:
\begin{equation}
\begin{gathered}
\begin{fmffile}{Diagrams/LoopExpansion1_Hartree}
\begin{fmfgraph}(30,20)
\fmfleft{i}
\fmfright{o}
\fmf{phantom,tension=10}{i,i1}
\fmf{phantom,tension=10}{o,o1}
\fmf{plain,left,tension=0.5}{i1,v1,i1}
\fmf{plain,right,tension=0.5}{o1,v2,o1}
\fmf{zigzag,foreground=(0,,0,,1)}{v1,v2}
\end{fmfgraph}
\end{fmffile}
\end{gathered} = \lambda N^{2} G^{2}_{0}\;,
\label{eq:1PIEAlambdaDiag1}
\end{equation}
\begin{equation}
\begin{gathered}
\begin{fmffile}{Diagrams/LoopExpansion1_Fock}
\begin{fmfgraph}(15,15)
\fmfleft{i}
\fmfright{o}
\fmf{phantom,tension=11}{i,v1}
\fmf{phantom,tension=11}{v2,o}
\fmf{plain,left,tension=0.4}{v1,v2,v1}
\fmf{zigzag,foreground=(0,,0,,1)}{v1,v2}
\end{fmfgraph}
\end{fmffile}
\end{gathered} = \lambda N G^{2}_{0} \;,
\end{equation}

\vspace{0.5cm}

\begin{equation}
\begin{gathered}
\begin{fmffile}{Diagrams/1PIEAlambda_Diag5bis2}
\begin{fmfgraph*}(25,13)
\fmfleft{i}
\fmfright{oDown,o,oUp}
\fmfv{decor.shape=cross,decor.angle=45,decor.size=3.5thick,foreground=(1,,0,,0)}{oUp}
\fmfv{decor.shape=cross,decor.angle=45,decor.size=3.5thick,foreground=(1,,0,,0)}{oDown}
\fmf{phantom,tension=10}{i,i1}
\fmf{phantom,tension=10}{o,o1}
\fmf{plain,left,tension=0.5}{i1,v1,i1}
\fmf{phantom,right,tension=0.5}{o1,v2,o1}
\fmf{dashes,tension=0,foreground=(1,,0,,0)}{v2,oUp}
\fmf{dashes,tension=0,foreground=(1,,0,,0)}{v2,oDown}
\fmf{zigzag,foreground=(0,,0,,1)}{v1,v2}
\end{fmfgraph*}
\end{fmffile}
\end{gathered} = \lambda N \phi^{2}_{N} G_{0} \;,
\end{equation}

\vspace{0.5cm}

\begin{equation}
\begin{gathered}
\begin{fmffile}{Diagrams/1PIEAlambda_Diag6bis2}
\begin{fmfgraph*}(15,15)
\fmfleft{iDown,i,iUp}
\fmfright{oDown,o,oUp}
\fmfv{decor.shape=cross,decor.size=3.5thick,foreground=(1,,0,,0)}{iUp}
\fmfv{decor.shape=cross,decor.size=3.5thick,foreground=(1,,0,,0)}{oUp}
\fmf{dashes,tension=0,foreground=(1,,0,,0)}{v1,iUp}
\fmf{dashes,tension=0,foreground=(1,,0,,0)}{v2,oUp}
\fmf{phantom,tension=11}{i,v1}
\fmf{phantom,tension=11}{v2,o}
\fmf{plain,right,tension=0.4}{v1,v2}
\fmf{zigzag,foreground=(0,,0,,1)}{v1,v2}
\end{fmfgraph*}
\end{fmffile}
\end{gathered} = \lambda \phi^{2}_{N} G_{0} \;,
\label{eq:1PIEAlambdaDiag4}
\end{equation}
with
\begin{equation}
\boldsymbol{G}^{-1}_{0;ab} = G^{-1}_{0}\delta_{ab} = m^{2}\delta_{ab} \;.
\end{equation}
After combining~\eqref{eq:1PIEAlambdaDiag1} to~\eqref{eq:1PIEAlambdaDiag4} with~\eqref{eq:1PIEAlambdaEAStep30DON}, we infer the following expression of $\Gamma^{(\mathrm{1PI})}$:
\begin{equation}
\Gamma^{(\mathrm{1PI})}\Big(\vec{\phi}\Big) = S\Big(\vec{\phi}\Big) - \frac{N}{2}\ln\big(2\pi G_{0}\big) +\lambda\left(\frac{N^{2}+2N}{24}G^{2}_{0} + \frac{N+2}{12}\phi^{2}_{N}G_{0}\right) + \mathcal{O}\big(\lambda^{2}\big)\;,
\label{eq:1PIEAlambdafinalexpression0DON}
\end{equation}
where we still have to specify a configuration for $\vec{\phi}$. This is done after solving the gap equation:
\begin{equation}
0 = \left.\frac{\partial \Gamma^{(\mathrm{1PI})}\big(\vec{\phi}\big)}{\partial \phi_{N}}\right|_{\vec{\phi}=\vec{\overline{\phi}}} = \overline{\phi}_{N} G_{0}^{-1} + \lambda\left(\frac{1}{6}\overline{\phi}_{N}^{3} + \frac{N+2}{6} \overline{\phi}_{N} G_{0}\right) +\mathcal{O}\big(\lambda^{2}\big) \;,
\label{eq:1PIEAlambdaGapEquation0DON}
\end{equation}
where $\vec{\overline{\phi}}$ is already defined right below~\eqref{eq:1PIEAGapEquation0DON}. The value of $\overline{\phi}_{N}$ thus obtained enables us to infer an estimate for the gs energy by using~\eqref{eq:DeduceEgs1PIEAdiag} as for the $\hbar$-expansion. It can also be noted that the results of the $\hbar$-expansion up to order $\mathcal{O}\big(\hbar^{2}\big)$ (given by~\eqref{eq:1PIEAfinalexpression0DON} and~\eqref{eq:1PIEAGapEquation0DON}) and those of the $\lambda$-expansion up to order $\mathcal{O}\big(\lambda\big)$ (given by~\eqref{eq:1PIEAlambdafinalexpression0DON} and~\eqref{eq:1PIEAlambdaGapEquation0DON}) coincide if $\vec{\phi}=\vec{\overline{\phi}}=\vec{0}$. This remark remains valid if we compare $\hbar$-expansion results up to order $\mathcal{O}\big(\hbar^{n+1}\big)$ and $\lambda$-expansion results up to order $\mathcal{O}\big(\lambda^{n}\big)$ for any $n\in\mathbb{N}$.

\subsubsection{\label{sec:1PIbosonicEA}Collective effective action}

Similarly to the analysis made in the section dedicated to the LE and for the same motivations (namely investigating whether the introduction of an auxiliary collective field helps in grasping more efficiently non-trivial correlations at low orders), we now focus on the collective 1PI EA, i.e. the 1PI EA implemented in the collective representation (defined in section~\ref{sec:DefCollectiveRepr}) where the original field $\vec{\widetilde{\varphi}}$ has been integrated out in favor of the collective Hubbard-Stratonovich field $\widetilde{\sigma}$. There are numerous works exploiting the collective 1PI EA. These applications take multiple forms for the following reasons:
\begin{itemize}
\item One can use the Schwinger-Dyson equations formalism, equivalent to that of the EA treated in this chapter~\cite{ben77,coo78,tam78,cam79,hay79,coo79,gol81,mun82,sar84,ben85,kul87,coo99,coo04,coo16}. The collective 1PI EA formalism in this form is referred to as mean-field PT, mean-field theory or self-consistent field approximation.
\item The EA can be expanded using different expansion parameters like $1/N$~\cite{can79,gol81,coo94,coo95,fei95,coo97,coo99,coo04,chi12} or $\hbar$~\cite{tam78bis,gur79,fur83,sac88,coo10,coo11,mih11bis,mih11,coo12,daw12,chi14,coo14,coo14bis,coo16}\footnote{Let us stress once again that the parameter that we refer to as $\hbar$ in the collective situation sometimes bear different names in the literature, like $\epsilon$~\cite{ben77} or $\theta$~\cite{sac88}.}.
\end{itemize}
We stress that all of these approaches are equivalent as long as the truncation of the EA is organized with respect to the same parameter, e.g. $1/N$ or $\hbar$ typically (see ref.~\cite{coo04} for an exhaustive discussion on truncation schemes of Schwinger-Dyson equations). If the parameter in question is $\hbar$, the resulting approach is sometimes called mean-field expansion~\cite{tam78bis,gur79} or, more recently, auxiliary field LE (LOAF)~\cite{coo10,coo11,mih11bis,mih11,coo12,daw12,chi14,coo14,coo14bis,coo16} (see ref.~\cite{coo16} for a detailed discussion on this technique). The LOAF should not be confused with the LOAF approximation which consists in keeping only the term of order $\mathcal{O}\big(\hbar^{0}\big)$ (i.e. the leading order) in the series representing the collective 1PI EA in the framework of the LOAF. For the purpose of determining the gs energy and density of the toy model under study, this amounts to considering the leading orders of the collective LE series~\eqref{eq:ResultEgsBosonicAction0DON} and~\eqref{eq:ResultrhogsBosonicAction0DON}\footnote{The leading order of the energy series expressed by~\eqref{eq:ResultEgsBosonicAction0DON} corresponds to a term of order $\mathcal{O}\big(\hbar^{-1}\big)$ and not $\mathcal{O}\big(\hbar^{0}\big)$.}. The latter remark is only true assuming that the physical configuration of the 1-point correlation function of the original field $\vec{\phi}$ determined from the gap equations in the LOAF approximation is zero (see appendix~\ref{sec:collective1PIEAannIM} for further details on that point).

\vspace{0.5cm}

The collective 1PI EA is also defined by Legendre transforming the corresponding Schwinger functional. This translates into:
\begin{equation}
\begin{split}
\Gamma_{\mathrm{col}}^{(\mathrm{1PI})}[\Phi] \equiv & -W_{\mathrm{col}}\big[\mathcal{J}\big] + \int_{x} \mathcal{J}^{\alpha}(x) \frac{\delta W_{\mathrm{col}}\big[\mathcal{J}\big]}{\delta\mathcal{J}^{\alpha}(x)} \\
= & -W_{\mathrm{col}}\big[\mathcal{J}\big]+\int_{\alpha}\mathcal{J}^{\alpha}(x)\Phi_{\alpha}(x)\;,
\end{split}
\label{eq:bosonic1PIEAdefinition0DONmain}
\end{equation}
with
\begin{equation}
\Phi_{\alpha}(x)=\frac{\delta W_{\mathrm{col}}\big[\mathcal{J}\big]}{\delta \mathcal{J}^{\alpha}(x)}\;,
\label{eq:bosonic1PIEAdefinitionbis0DONmain}
\end{equation}
or, in terms of the 1-point correlation functions of the original and Hubbard-Stratonovich fields (i.e. $\vec{\phi}(x) = \left\langle\vec{\widetilde{\varphi}}(x)\right\rangle$ and $\eta(x)=\left\langle\widetilde{\sigma}(x)\right\rangle$, respectively),
\begin{equation}
\Phi(x) = \begin{pmatrix}
\vec{\phi}(x) \\
\eta(x)
\end{pmatrix}\;.
\end{equation}
Note also that $W_{\mathrm{col}}\big[\mathcal{J}\big]$ is the collective Schwinger functional treated in section~\ref{sec:PT} with the LE, i.e. $W_{\mathrm{col}}\big[\mathcal{J}\big]\equiv W^\text{LE;col}\big[\mathcal{J}\big]$. The collective 1PI EA can be represented diagrammatically according to (see appendix~\ref{sec:collective1PIEAannIM}):
\begin{equation}
\begin{split}
\Gamma_{\mathrm{col}}^{(\mathrm{1PI})}[\Phi] = & \ S_{\mathrm{col}}[\eta] + \frac{1}{2} \int_{x,y}\phi^a(x) \boldsymbol{G}^{-1}_{\Phi;ab}(x,y) \phi^b(y) - \frac{\hbar}{2}\mathrm{Tr}\left[\ln\big(D_{\Phi}\big)\right] \\
& - \hbar^{2}  \left(\rule{0cm}{1.1cm}\right. \frac{1}{2} \hspace{-0.1cm} \begin{gathered}
\begin{fmffile}{Diagrams/bosonic1PIEA_Gamma2_Diag1bis}
\begin{fmfgraph*}(25,25)
\fmfleft{i1,i2}
\fmfright{o1,o2}
\fmfbottom{i0,o0}
\fmftop{i3,o3}
\fmfv{decor.shape=circle,decor.size=2.0thick,foreground=(0,,0,,1)}{v1}
\fmfv{decor.shape=circle,decor.size=2.0thick,foreground=(0,,0,,1)}{v2}
\fmfv{decor.shape=circle,decor.size=2.0thick,foreground=(0,,0,,1)}{v3}
\fmfv{decor.shape=circle,decor.size=2.0thick,foreground=(0,,0,,1)}{v4}
\fmfv{decor.shape=cross,decor.size=0.25cm,decor.angle=35,foreground=(1,,0,,0)}{v1b}
\fmfv{decor.shape=cross,decor.size=0.25cm,decor.angle=-35,foreground=(1,,0,,0)}{v4b}
\fmf{phantom}{i1,v1}
\fmf{phantom}{i2,v4}
\fmf{phantom}{o1,v2}
\fmf{phantom}{o2,v3}
\fmf{phantom}{i3,v4b}
\fmf{phantom}{o3,v3b}
\fmf{phantom}{i0,v1b}
\fmf{phantom}{o0,v2b}
\fmf{plain,tension=1.6,foreground=(1,,0,,0)}{v1,v2}
\fmf{plain,tension=1.6,foreground=(1,,0,,0)}{v3,v4}
\fmf{wiggly,tension=2.0,foreground=(1,,0,,0)}{v1,v4}
\fmf{wiggly,tension=2.0,foreground=(1,,0,,0)}{v2,v3}
\fmf{phantom,tension=0}{v1,v3}
\fmf{phantom,tension=0}{v2,v4}
\fmf{plain,right=0.8,tension=0,foreground=(1,,0,,0)}{v2,v3}
\fmf{phantom,left=0.8,tension=0}{v1,v4}
\fmf{dashes,foreground=(1,,0,,0)}{v1,v1b}
\fmf{phantom}{v2,v2b}
\fmf{phantom}{v3,v3b}
\fmf{dashes,foreground=(1,,0,,0)}{v4,v4b}
\end{fmfgraph*}
\end{fmffile}
\end{gathered} \hspace{-0.4cm} + \frac{1}{2} \hspace{-0.1cm} \begin{gathered}
\begin{fmffile}{Diagrams/bosonic1PIEA_Gamma2_Diag2bis}
\begin{fmfgraph*}(25,25)
\fmfleft{i1,i2}
\fmfright{o1,o2}
\fmfbottom{i0,o0}
\fmftop{i3,o3}
\fmfv{decor.shape=circle,decor.size=2.0thick,foreground=(0,,0,,1)}{v1}
\fmfv{decor.shape=circle,decor.size=2.0thick,foreground=(0,,0,,1)}{v2}
\fmfv{decor.shape=circle,decor.size=2.0thick,foreground=(0,,0,,1)}{v3}
\fmfv{decor.shape=circle,decor.size=2.0thick,foreground=(0,,0,,1)}{v4}
\fmfv{decor.shape=cross,decor.size=0.25cm,decor.angle=35,foreground=(1,,0,,0)}{v3b}
\fmfv{decor.shape=cross,decor.size=0.25cm,decor.angle=-35,foreground=(1,,0,,0)}{v4b}
\fmf{phantom}{i1,v1}
\fmf{phantom}{i2,v4}
\fmf{phantom}{o1,v2}
\fmf{phantom}{o2,v3}
\fmf{phantom}{i3,v4b}
\fmf{phantom}{o3,v3b}
\fmf{phantom}{i0,v1b}
\fmf{phantom}{o0,v2b}
\fmf{plain,tension=1.6,foreground=(1,,0,,0)}{v1,v2}
\fmf{phantom,tension=1.6}{v3,v4}
\fmf{wiggly,tension=2.0,foreground=(1,,0,,0)}{v1,v4}
\fmf{wiggly,tension=2.0,foreground=(1,,0,,0)}{v2,v3}
\fmf{plain,tension=0,foreground=(1,,0,,0)}{v1,v3}
\fmf{plain,tension=0,foreground=(1,,0,,0)}{v2,v4}
\fmf{phantom,right=0.8,tension=0}{v2,v3}
\fmf{phantom,left=0.8,tension=0}{v1,v4}
\fmf{phantom}{v1,v1b}
\fmf{phantom}{v2,v2b}
\fmf{dashes,foreground=(1,,0,,0)}{v3,v3b}
\fmf{dashes,foreground=(1,,0,,0)}{v4,v4b}
\end{fmfgraph*}
\end{fmffile}
\end{gathered} \hspace{-0.15cm} + \frac{1}{2} \hspace{-0.1cm} \begin{gathered}
\begin{fmffile}{Diagrams/bosonic1PIEA_Gamma2_Diag5bis}
\begin{fmfgraph*}(35,20)
\fmfleft{i1,i2}
\fmfright{o1,o2}
\fmfbottom{i0,o0}
\fmfbottom{b0}
\fmfbottom{b1}
\fmfbottom{b2}
\fmftop{i3,o3}
\fmfv{decor.shape=circle,decor.size=2.0thick,foreground=(0,,0,,1)}{v1}
\fmfv{decor.shape=circle,decor.size=2.0thick,foreground=(0,,0,,1)}{v2}
\fmfv{decor.shape=circle,decor.size=2.0thick,foreground=(0,,0,,1)}{v3}
\fmfv{decor.shape=circle,decor.size=2.0thick,foreground=(0,,0,,1)}{v4}
\fmfv{decor.shape=circle,decor.size=2.0thick,foreground=(0,,0,,1)}{v5}
\fmfv{decor.shape=circle,decor.size=2.0thick,foreground=(0,,0,,1)}{v6}
\fmfv{decor.shape=cross,decor.size=0.25cm,decor.angle=16,foreground=(1,,0,,0)}{v1b}
\fmfv{decor.shape=cross,decor.size=0.25cm,foreground=(1,,0,,0)}{v3b}
\fmfv{decor.shape=cross,decor.size=0.25cm,decor.angle=71,foreground=(1,,0,,0)}{v4b}
\fmfv{decor.shape=cross,decor.size=0.25cm,foreground=(1,,0,,0)}{v6b}
\fmf{phantom,tension=1.4}{i1,v1}
\fmf{phantom}{i2,v3b}
\fmf{phantom}{i0,v1b}
\fmf{phantom}{o2,v6b}
\fmf{phantom,tension=1.4}{o1,v5}
\fmf{phantom}{o0,v5b}
\fmf{phantom,tension=1.11}{v3b,v6b}
\fmf{phantom,tension=1.38}{i0,v2}
\fmf{phantom,tension=1.38}{o0,v2}
\fmf{phantom,tension=1.8}{i0,v2b}
\fmf{phantom,tension=1.2}{o0,v2b}
\fmf{phantom,tension=1.2}{b0,v2b}
\fmf{phantom,tension=1.2}{b1,v2b}
\fmf{phantom,tension=1.2}{b2,v2b}
\fmf{phantom,tension=1.38}{i0,v4}
\fmf{phantom,tension=1.38}{o0,v4}
\fmf{phantom,tension=1.2}{i0,v4b}
\fmf{phantom,tension=1.8}{o0,v4b}
\fmf{phantom,tension=1.2}{b0,v4b}
\fmf{phantom,tension=1.2}{b1,v4b}
\fmf{phantom,tension=1.2}{b2,v4b}
\fmf{phantom,tension=2}{i3,v3}
\fmf{phantom,tension=2}{o3,v6}
\fmf{phantom,tension=2}{i3,v3b}
\fmf{phantom,tension=0.8}{o3,v3b}
\fmf{phantom,tension=0.8}{i3,v6b}
\fmf{phantom,tension=2}{o3,v6b}
\fmf{plain,tension=1.4,foreground=(1,,0,,0)}{v1,v2}
\fmf{plain,tension=1.4,foreground=(1,,0,,0)}{v4,v5}
\fmf{phantom}{v1,v3}
\fmf{phantom,left=0.8,tension=0}{v1,v3}
\fmf{plain,foreground=(1,,0,,0)}{v5,v6}
\fmf{phantom,right=0.8,tension=0}{v5,v6}
\fmf{plain,foreground=(1,,0,,0)}{v2,v3}
\fmf{phantom}{v4,v6}
\fmf{wiggly,tension=0.5,foreground=(1,,0,,0)}{v2,v4}
\fmf{wiggly,tension=2,foreground=(1,,0,,0)}{v3,v6}
\fmf{wiggly,right=0.8,tension=0,foreground=(1,,0,,0)}{v1,v5}
\fmf{dashes,tension=1,foreground=(1,,0,,0)}{v1,v1b}
\fmf{phantom,tension=0.2}{v2,v2b}
\fmf{dashes,tension=1.5,foreground=(1,,0,,0)}{v3,v3b}
\fmf{dashes,tension=0.2,foreground=(1,,0,,0)}{v4,v4b}
\fmf{phantom,tension=1}{v5,v5b}
\fmf{dashes,tension=1.5,foreground=(1,,0,,0)}{v6,v6b}
\end{fmfgraph*}
\end{fmffile}
\end{gathered} \hspace{-0.5cm} +\frac{1}{4} \begin{gathered}
\begin{fmffile}{Diagrams/bosonic1PIEA_Gamma2_Diag6bis}
\begin{fmfgraph*}(35,20)
\fmfleft{i1,i2}
\fmfright{o1,o2}
\fmfbottom{i0,o0}
\fmfbottom{b0}
\fmfbottom{b1}
\fmfbottom{b2}
\fmftop{i3,o3}
\fmfv{decor.shape=circle,decor.size=2.0thick,foreground=(0,,0,,1)}{v1}
\fmfv{decor.shape=circle,decor.size=2.0thick,foreground=(0,,0,,1)}{v2}
\fmfv{decor.shape=circle,decor.size=2.0thick,foreground=(0,,0,,1)}{v3}
\fmfv{decor.shape=circle,decor.size=2.0thick,foreground=(0,,0,,1)}{v4}
\fmfv{decor.shape=circle,decor.size=2.0thick,foreground=(0,,0,,1)}{v5}
\fmfv{decor.shape=circle,decor.size=2.0thick,foreground=(0,,0,,1)}{v6}
\fmfv{decor.shape=cross,decor.size=0.25cm,decor.angle=16,foreground=(1,,0,,0)}{v1b}
\fmfv{decor.shape=cross,decor.size=0.25cm,foreground=(1,,0,,0)}{v3b}
\fmfv{decor.shape=cross,decor.size=0.25cm,decor.angle=-16,foreground=(1,,0,,0)}{v5b}
\fmfv{decor.shape=cross,decor.size=0.25cm,foreground=(1,,0,,0)}{v6b}
\fmf{phantom,tension=1.4}{i1,v1}
\fmf{phantom}{i2,v3b}
\fmf{phantom}{i0,v1b}
\fmf{phantom}{o2,v6b}
\fmf{phantom,tension=1.4}{o1,v5}
\fmf{phantom}{o0,v5b}
\fmf{phantom,tension=1.11}{v3b,v6b}
\fmf{phantom,tension=1.38}{i0,v2}
\fmf{phantom,tension=1.38}{o0,v2}
\fmf{phantom,tension=1.8}{i0,v2b}
\fmf{phantom,tension=1.2}{o0,v2b}
\fmf{phantom,tension=1.2}{b0,v2b}
\fmf{phantom,tension=1.2}{b1,v2b}
\fmf{phantom,tension=1.2}{b2,v2b}
\fmf{phantom,tension=1.38}{i0,v4}
\fmf{phantom,tension=1.38}{o0,v4}
\fmf{phantom,tension=1.2}{i0,v4b}
\fmf{phantom,tension=1.8}{o0,v4b}
\fmf{phantom,tension=1.2}{b0,v4b}
\fmf{phantom,tension=1.2}{b1,v4b}
\fmf{phantom,tension=1.2}{b2,v4b}
\fmf{phantom,tension=2}{i3,v3}
\fmf{phantom,tension=2}{o3,v6}
\fmf{phantom,tension=2}{i3,v3b}
\fmf{phantom,tension=0.8}{o3,v3b}
\fmf{phantom,tension=0.8}{i3,v6b}
\fmf{phantom,tension=2}{o3,v6b}
\fmf{plain,tension=1.4,foreground=(1,,0,,0)}{v1,v2}
\fmf{plain,tension=1.4,foreground=(1,,0,,0)}{v4,v5}
\fmf{phantom}{v1,v3}
\fmf{phantom,left=0.8,tension=0}{v1,v3}
\fmf{phantom}{v5,v6}
\fmf{phantom,right=0.8,tension=0}{v5,v6}
\fmf{plain,foreground=(1,,0,,0)}{v2,v3}
\fmf{plain,foreground=(1,,0,,0)}{v4,v6}
\fmf{wiggly,tension=0.5,foreground=(1,,0,,0)}{v2,v4}
\fmf{wiggly,tension=2,foreground=(1,,0,,0)}{v3,v6}
\fmf{wiggly,right=0.8,tension=0,foreground=(1,,0,,0)}{v1,v5}
\fmf{dashes,tension=1,foreground=(1,,0,,0)}{v1,v1b}
\fmf{phantom,tension=0.2}{v2,v2b}
\fmf{dashes,tension=1.5,foreground=(1,,0,,0)}{v3,v3b}
\fmf{phantom,tension=0.2}{v4,v4b}
\fmf{dashes,tension=1,foreground=(1,,0,,0)}{v5,v5b}
\fmf{dashes,tension=1.5,foreground=(1,,0,,0)}{v6,v6b}
\end{fmfgraph*}
\end{fmffile}
\end{gathered} \\
& \hspace{1.0cm} +\frac{1}{4} \hspace{-0.4cm} \begin{gathered}
\begin{fmffile}{Diagrams/bosonic1PIEA_Gamma2_Diag3bis}
\begin{fmfgraph*}(25,25)
\fmfleft{i1,i2}
\fmfright{o1,o2}
\fmfbottom{i0,o0}
\fmftop{i3,o3}
\fmfv{decor.shape=circle,decor.size=2.0thick,foreground=(0,,0,,1)}{v1}
\fmfv{decor.shape=circle,decor.size=2.0thick,foreground=(0,,0,,1)}{v2}
\fmfv{decor.shape=circle,decor.size=2.0thick,foreground=(0,,0,,1)}{v3}
\fmfv{decor.shape=circle,decor.size=2.0thick,foreground=(0,,0,,1)}{v4}
\fmf{phantom}{i1,v1}
\fmf{phantom}{i2,v4}
\fmf{phantom}{o1,v2}
\fmf{phantom}{o2,v3}
\fmf{phantom}{i3,v4b}
\fmf{phantom}{o3,v3b}
\fmf{phantom}{i0,v1b}
\fmf{phantom}{o0,v2b}
\fmf{plain,tension=1.6,foreground=(1,,0,,0)}{v1,v2}
\fmf{plain,tension=1.6,foreground=(1,,0,,0)}{v3,v4}
\fmf{wiggly,tension=2.0,foreground=(1,,0,,0)}{v1,v4}
\fmf{wiggly,tension=2.0,foreground=(1,,0,,0)}{v2,v3}
\fmf{phantom,tension=0}{v1,v3}
\fmf{phantom,tension=0}{v2,v4}
\fmf{plain,right=0.8,tension=0,foreground=(1,,0,,0)}{v2,v3}
\fmf{plain,left=0.8,tension=0,foreground=(1,,0,,0)}{v1,v4}
\fmf{phantom}{v1,v1b}
\fmf{phantom}{v2,v2b}
\fmf{phantom}{v3,v3b}
\fmf{phantom}{v4,v4b}
\end{fmfgraph*}
\end{fmffile}
\end{gathered} \hspace{-0.4cm} +\frac{1}{8} \hspace{-0.5cm} \begin{gathered}
\begin{fmffile}{Diagrams/bosonic1PIEA_Gamma2_Diag4bis}
\begin{fmfgraph*}(25,25)
\fmfleft{i1,i2}
\fmfright{o1,o2}
\fmfbottom{i0,o0}
\fmftop{i3,o3}
\fmfv{decor.shape=circle,decor.size=2.0thick,foreground=(0,,0,,1)}{v1}
\fmfv{decor.shape=circle,decor.size=2.0thick,foreground=(0,,0,,1)}{v2}
\fmfv{decor.shape=circle,decor.size=2.0thick,foreground=(0,,0,,1)}{v3}
\fmfv{decor.shape=circle,decor.size=2.0thick,foreground=(0,,0,,1)}{v4}
\fmf{phantom}{i1,v1}
\fmf{phantom}{i2,v4}
\fmf{phantom}{o1,v2}
\fmf{phantom}{o2,v3}
\fmf{phantom}{i3,v4b}
\fmf{phantom}{o3,v3b}
\fmf{phantom}{i0,v1b}
\fmf{phantom}{o0,v2b}
\fmf{plain,tension=1.6,foreground=(1,,0,,0)}{v1,v2}
\fmf{plain,tension=1.6,foreground=(1,,0,,0)}{v3,v4}
\fmf{wiggly,tension=2.0,foreground=(1,,0,,0)}{v1,v4}
\fmf{wiggly,tension=2.0,foreground=(1,,0,,0)}{v2,v3}
\fmf{plain,tension=0,foreground=(1,,0,,0)}{v1,v3}
\fmf{plain,tension=0,foreground=(1,,0,,0)}{v2,v4}
\fmf{phantom,right=0.8,tension=0}{v2,v3}
\fmf{phantom,left=0.8,tension=0}{v1,v4}
\fmf{phantom}{v1,v1b}
\fmf{phantom}{v2,v2b}
\fmf{phantom}{v3,v3b}
\fmf{phantom}{v4,v4b}
\end{fmfgraph*}
\end{fmffile}
\end{gathered} \hspace{-0.5cm} + \frac{1}{2} \begin{gathered}
\begin{fmffile}{Diagrams/bosonic1PIEA_Gamma2_Diag7bis}
\begin{fmfgraph*}(35,20)
\fmfleft{i1,i2}
\fmfright{o1,o2}
\fmfbottom{i0,o0}
\fmfbottom{b0}
\fmfbottom{b1}
\fmfbottom{b2}
\fmftop{i3,o3}
\fmfv{decor.shape=circle,decor.size=2.0thick,foreground=(0,,0,,1)}{v1}
\fmfv{decor.shape=circle,decor.size=2.0thick,foreground=(0,,0,,1)}{v2}
\fmfv{decor.shape=circle,decor.size=2.0thick,foreground=(0,,0,,1)}{v3}
\fmfv{decor.shape=circle,decor.size=2.0thick,foreground=(0,,0,,1)}{v4}
\fmfv{decor.shape=circle,decor.size=2.0thick,foreground=(0,,0,,1)}{v5}
\fmfv{decor.shape=circle,decor.size=2.0thick,foreground=(0,,0,,1)}{v6}
\fmfv{decor.shape=cross,decor.size=0.25cm,decor.angle=16,foreground=(1,,0,,0)}{v1b}
\fmfv{decor.shape=cross,decor.size=0.25cm,foreground=(1,,0,,0)}{v3b}
\fmf{phantom,tension=1.4}{i1,v1}
\fmf{phantom}{i2,v3b}
\fmf{phantom}{i0,v1b}
\fmf{phantom}{o2,v6b}
\fmf{phantom,tension=1.4}{o1,v5}
\fmf{phantom}{o0,v5b}
\fmf{phantom,tension=1.11}{v3b,v6b}
\fmf{phantom,tension=1.38}{i0,v2}
\fmf{phantom,tension=1.38}{o0,v2}
\fmf{phantom,tension=1.8}{i0,v2b}
\fmf{phantom,tension=1.2}{o0,v2b}
\fmf{phantom,tension=1.2}{b0,v2b}
\fmf{phantom,tension=1.2}{b1,v2b}
\fmf{phantom,tension=1.2}{b2,v2b}
\fmf{phantom,tension=1.38}{i0,v4}
\fmf{phantom,tension=1.38}{o0,v4}
\fmf{phantom,tension=1.2}{i0,v4b}
\fmf{phantom,tension=1.8}{o0,v4b}
\fmf{phantom,tension=1.2}{b0,v4b}
\fmf{phantom,tension=1.2}{b1,v4b}
\fmf{phantom,tension=1.2}{b2,v4b}
\fmf{phantom,tension=2}{i3,v3}
\fmf{phantom,tension=2}{o3,v6}
\fmf{phantom,tension=2}{i3,v3b}
\fmf{phantom,tension=0.8}{o3,v3b}
\fmf{phantom,tension=0.8}{i3,v6b}
\fmf{phantom,tension=2}{o3,v6b}
\fmf{plain,tension=1.4,foreground=(1,,0,,0)}{v1,v2}
\fmf{plain,tension=1.4,foreground=(1,,0,,0)}{v4,v5}
\fmf{phantom}{v1,v3}
\fmf{phantom,left=0.8,tension=0}{v1,v3}
\fmf{plain,foreground=(1,,0,,0)}{v5,v6}
\fmf{phantom,right=0.8,tension=0}{v5,v6}
\fmf{plain,foreground=(1,,0,,0)}{v2,v3}
\fmf{plain,foreground=(1,,0,,0)}{v4,v6}
\fmf{wiggly,tension=0.5,foreground=(1,,0,,0)}{v2,v4}
\fmf{wiggly,tension=2,foreground=(1,,0,,0)}{v3,v6}
\fmf{wiggly,right=0.8,tension=0,foreground=(1,,0,,0)}{v1,v5}
\fmf{dashes,tension=1,foreground=(1,,0,,0)}{v1,v1b}
\fmf{phantom,tension=0.2}{v2,v2b}
\fmf{dashes,tension=1.5,foreground=(1,,0,,0)}{v3,v3b}
\fmf{phantom,tension=0.2}{v4,v4b}
\fmf{phantom,tension=1}{v5,v5b}
\fmf{phantom,tension=1.5}{v6,v6b}
\end{fmfgraph*}
\end{fmffile}
\end{gathered} \hspace{-0.5cm} +\frac{1}{12} \hspace{-0.5cm} \begin{gathered}
\begin{fmffile}{Diagrams/bosonic1PIEA_Gamma2_Diag8bis}
\begin{fmfgraph*}(35,20)
\fmfleft{i1,i2}
\fmfright{o1,o2}
\fmfbottom{i0,o0}
\fmfbottom{b0}
\fmfbottom{b1}
\fmfbottom{b2}
\fmftop{i3,o3}
\fmfv{decor.shape=circle,decor.size=2.0thick,foreground=(0,,0,,1)}{v1}
\fmfv{decor.shape=circle,decor.size=2.0thick,foreground=(0,,0,,1)}{v2}
\fmfv{decor.shape=circle,decor.size=2.0thick,foreground=(0,,0,,1)}{v3}
\fmfv{decor.shape=circle,decor.size=2.0thick,foreground=(0,,0,,1)}{v4}
\fmfv{decor.shape=circle,decor.size=2.0thick,foreground=(0,,0,,1)}{v5}
\fmfv{decor.shape=circle,decor.size=2.0thick,foreground=(0,,0,,1)}{v6}
\fmf{phantom,tension=1.4}{i1,v1}
\fmf{phantom}{i2,v3b}
\fmf{phantom}{i0,v1b}
\fmf{phantom}{o2,v6b}
\fmf{phantom,tension=1.4}{o1,v5}
\fmf{phantom}{o0,v5b}
\fmf{phantom,tension=1.11}{v3b,v6b}
\fmf{phantom,tension=1.38}{i0,v2}
\fmf{phantom,tension=1.38}{o0,v2}
\fmf{phantom,tension=1.8}{i0,v2b}
\fmf{phantom,tension=1.2}{o0,v2b}
\fmf{phantom,tension=1.2}{b0,v2b}
\fmf{phantom,tension=1.2}{b1,v2b}
\fmf{phantom,tension=1.2}{b2,v2b}
\fmf{phantom,tension=1.38}{i0,v4}
\fmf{phantom,tension=1.38}{o0,v4}
\fmf{phantom,tension=1.2}{i0,v4b}
\fmf{phantom,tension=1.8}{o0,v4b}
\fmf{phantom,tension=1.2}{b0,v4b}
\fmf{phantom,tension=1.2}{b1,v4b}
\fmf{phantom,tension=1.2}{b2,v4b}
\fmf{phantom,tension=2}{i3,v3}
\fmf{phantom,tension=2}{o3,v6}
\fmf{phantom,tension=2}{i3,v3b}
\fmf{phantom,tension=0.8}{o3,v3b}
\fmf{phantom,tension=0.8}{i3,v6b}
\fmf{phantom,tension=2}{o3,v6b}
\fmf{plain,tension=1.4,foreground=(1,,0,,0)}{v1,v2}
\fmf{plain,tension=1.4,foreground=(1,,0,,0)}{v4,v5}
\fmf{plain,foreground=(1,,0,,0)}{v1,v3}
\fmf{phantom,left=0.8,tension=0}{v1,v3}
\fmf{plain,foreground=(1,,0,,0)}{v5,v6}
\fmf{phantom,right=0.8,tension=0}{v5,v6}
\fmf{plain,foreground=(1,,0,,0)}{v2,v3}
\fmf{plain,foreground=(1,,0,,0)}{v4,v6}
\fmf{wiggly,tension=0.5,foreground=(1,,0,,0)}{v2,v4}
\fmf{wiggly,tension=2,foreground=(1,,0,,0)}{v3,v6}
\fmf{wiggly,right=0.8,tension=0,foreground=(1,,0,,0)}{v1,v5}
\fmf{phantom,tension=1}{v1,v1b}
\fmf{phantom,tension=0.2}{v2,v2b}
\fmf{phantom,tension=1.5}{v3,v3b}
\fmf{phantom,tension=0.2}{v4,v4b}
\fmf{phantom,tension=1}{v5,v5b}
\fmf{phantom,tension=1.5}{v6,v6b}
\end{fmfgraph*}
\end{fmffile}
\end{gathered} \hspace{-0.6cm} \left.\rule{0cm}{1.1cm}\right) \\
& + \mathcal{O}\big(\hbar^{3}\big) \;,
\end{split}
\label{eq:bosonic1PIEAIMGamma0DON}
\end{equation}
where $\boldsymbol{G}_{\Phi}$ and $D_{\Phi}$ are the original and collective field propagators respectively, conveniently collected in the superpropagator $\mathcal{G}_{\Phi}$ as follows:
\begin{equation}
\mathcal{G}_{\Phi} = \begin{pmatrix}
\boldsymbol{G}_{\Phi} & \vec{0} \\
\vec{0}^{\mathrm{T}} & D_{\Phi}
\end{pmatrix}\;,
\end{equation}
\begin{equation}
\boldsymbol{G}^{-1}_{\Phi;ab}(x,y) = \left(-\nabla^2_x + m^{2} + i\sqrt{\frac{\lambda}{3}}\eta(x)\right) \delta_{ab}\delta(x-y) \;,
\label{eq:bosonic1PIEApropagatorGJ0main}
\end{equation}
\begin{equation}
D^{-1}_{\Phi}(x,y) = \left.\frac{\delta^{2} S_{\mathrm{col},\mathcal{J}}[\widetilde{\sigma}]}{\delta\widetilde{\sigma}(x) \delta\widetilde{\sigma}(y)}\right|_{\widetilde{\sigma}=\eta \atop \vec{J}=\vec{J}_{0}} \;,
\label{eq:bosonic1PIEApropagatorHJ0main}
\end{equation}
with $\vec{J}_{0}$ being a source coefficient introduced in the framework of the IM (see appendix~\ref{sec:collective1PIEAannIM}). Result~\eqref{eq:bosonic1PIEAIMGamma0DON} relies on the Feynman rules:
\begin{subequations}
\begin{align}
\begin{gathered}
\begin{fmffile}{Diagrams/1PIEAcol-G}
\begin{fmfgraph*}(20,16)
\fmfleft{i0,i1,i2,i3}
\fmfright{o0,o1,o2,o3}
\fmflabel{$x, a$}{v1}
\fmflabel{$y, b$}{v2}
\fmf{phantom}{i1,v1}
\fmf{phantom}{i2,v1}
\fmf{plain,tension=0.6,foreground=(1,,0,,0)}{v1,v2}
\fmf{phantom}{v2,o1}
\fmf{phantom}{v2,o2}
\end{fmfgraph*}
\end{fmffile}
\end{gathered} \quad &\rightarrow \boldsymbol{G}_{\Phi;ab}(x,y)\;, 
\label{eq:FeynRulesBosonic1PIEAJ0Gmain}\\
\begin{gathered}
\begin{fmffile}{Diagrams/1PIEAcol-D}
\begin{fmfgraph*}(20,20)
\fmfleft{i0,i1,i2,i3}
\fmfright{o0,o1,o2,o3}
\fmfv{label=$x$}{v1}
\fmfv{label=$y$}{v2}
\fmf{phantom}{i1,v1}
\fmf{phantom}{i2,v1}
\fmf{wiggly,tension=0.6,foreground=(1,,0,,0)}{v1,v2}
\fmf{phantom}{v2,o1}
\fmf{phantom}{v2,o2}
\end{fmfgraph*}
\end{fmffile}
\end{gathered} \quad &\rightarrow D_{\Phi}(x,y)\;,
\label{eq:FeynRulesBosonic1PIEAJ0Hmain} \\
\begin{gathered}
\begin{fmffile}{Diagrams/IPIEAcol-V}
\begin{fmfgraph*}(5,5)
\fmfleft{i1}
\fmfright{o1}
\fmfv{label=$x$,label.angle=-90,label.dist=4,foreground=(0,,0,,1)}{v1}
\fmf{plain,foreground=(1,,0,,0)}{i1,v1}
\fmf{plain,foreground=(1,,0,,0)}{v1,o1}
\fmflabel{$a$}{i1}
\fmflabel{$b$}{o1}
\fmfdot{v1}
\end{fmfgraph*}
\end{fmffile}
\end{gathered}\qquad &\rightarrow i\sqrt{\frac{\lambda}{3}}\delta_{ab}\;.
\label{eq:FeynRulesBosonic1PIEAJ0vertexDotmain}
\end{align}
\end{subequations}

\vspace{0.5cm}

As a next step, we study the collective 1PI EA in (0+0)-D. In this limit, the propagators~\eqref{eq:bosonic1PIEApropagatorGJ0main} and~\eqref{eq:bosonic1PIEApropagatorHJ0main} respectively satisfy:
\begin{equation}
\boldsymbol{G}_{\Phi;ab}^{-1} = G_{\Phi}^{-1}\delta_{ab}=\left(m^{2}+i\sqrt{\frac{\lambda}{3}} \eta\right)\delta_{ab} \;,
\label{eq:1PIEAcolGpropagator}
\end{equation}
\begin{equation}
D_{\Phi}^{-1} = \frac{\lambda}{3} G_{\Phi} \phi_{N}^{2}+\frac{\lambda}{6} N G_{\Phi}^{2} + 1 \;,
\label{eq:1PIEAcolDpropagator}
\end{equation}
and the diagrams involved in~\eqref{eq:bosonic1PIEAIMGamma0DON} read:
\begin{equation}
\begin{gathered}
\begin{fmffile}{Diagrams/bosonic1PIEA_Gamma2_Diag1bis}
\begin{fmfgraph*}(25,25)
\fmfleft{i1,i2}
\fmfright{o1,o2}
\fmfbottom{i0,o0}
\fmftop{i3,o3}
\fmfv{decor.shape=circle,decor.size=2.0thick,foreground=(0,,0,,1)}{v1}
\fmfv{decor.shape=circle,decor.size=2.0thick,foreground=(0,,0,,1)}{v2}
\fmfv{decor.shape=circle,decor.size=2.0thick,foreground=(0,,0,,1)}{v3}
\fmfv{decor.shape=circle,decor.size=2.0thick,foreground=(0,,0,,1)}{v4}
\fmfv{decor.shape=cross,decor.size=0.25cm,decor.angle=35,foreground=(1,,0,,0)}{v1b}
\fmfv{decor.shape=cross,decor.size=0.25cm,decor.angle=-35,foreground=(1,,0,,0)}{v4b}
\fmf{phantom}{i1,v1}
\fmf{phantom}{i2,v4}
\fmf{phantom}{o1,v2}
\fmf{phantom}{o2,v3}
\fmf{phantom}{i3,v4b}
\fmf{phantom}{o3,v3b}
\fmf{phantom}{i0,v1b}
\fmf{phantom}{o0,v2b}
\fmf{plain,tension=1.6,foreground=(1,,0,,0)}{v1,v2}
\fmf{plain,tension=1.6,foreground=(1,,0,,0)}{v3,v4}
\fmf{wiggly,tension=2.0,foreground=(1,,0,,0)}{v1,v4}
\fmf{wiggly,tension=2.0,foreground=(1,,0,,0)}{v2,v3}
\fmf{phantom,tension=0}{v1,v3}
\fmf{phantom,tension=0}{v2,v4}
\fmf{plain,right=0.8,tension=0,foreground=(1,,0,,0)}{v2,v3}
\fmf{phantom,left=0.8,tension=0}{v1,v4}
\fmf{dashes,foreground=(1,,0,,0)}{v1,v1b}
\fmf{phantom}{v2,v2b}
\fmf{phantom}{v3,v3b}
\fmf{dashes,foreground=(1,,0,,0)}{v4,v4b}
\end{fmfgraph*}
\end{fmffile}
\end{gathered} = \begin{gathered}
\begin{fmffile}{Diagrams/bosonic1PIEA_Gamma2_Diag2bis}
\begin{fmfgraph*}(25,25)
\fmfleft{i1,i2}
\fmfright{o1,o2}
\fmfbottom{i0,o0}
\fmftop{i3,o3}
\fmfv{decor.shape=circle,decor.size=2.0thick,foreground=(0,,0,,1)}{v1}
\fmfv{decor.shape=circle,decor.size=2.0thick,foreground=(0,,0,,1)}{v2}
\fmfv{decor.shape=circle,decor.size=2.0thick,foreground=(0,,0,,1)}{v3}
\fmfv{decor.shape=circle,decor.size=2.0thick,foreground=(0,,0,,1)}{v4}
\fmfv{decor.shape=cross,decor.size=0.25cm,decor.angle=35,foreground=(1,,0,,0)}{v3b}
\fmfv{decor.shape=cross,decor.size=0.25cm,decor.angle=-35,foreground=(1,,0,,0)}{v4b}
\fmf{phantom}{i1,v1}
\fmf{phantom}{i2,v4}
\fmf{phantom}{o1,v2}
\fmf{phantom}{o2,v3}
\fmf{phantom}{i3,v4b}
\fmf{phantom}{o3,v3b}
\fmf{phantom}{i0,v1b}
\fmf{phantom}{o0,v2b}
\fmf{plain,tension=1.6,foreground=(1,,0,,0)}{v1,v2}
\fmf{phantom,tension=1.6}{v3,v4}
\fmf{wiggly,tension=2.0,foreground=(1,,0,,0)}{v1,v4}
\fmf{wiggly,tension=2.0,foreground=(1,,0,,0)}{v2,v3}
\fmf{plain,tension=0,foreground=(1,,0,,0)}{v1,v3}
\fmf{plain,tension=0,foreground=(1,,0,,0)}{v2,v4}
\fmf{phantom,right=0.8,tension=0}{v2,v3}
\fmf{phantom,left=0.8,tension=0}{v1,v4}
\fmf{phantom}{v1,v1b}
\fmf{phantom}{v2,v2b}
\fmf{dashes,foreground=(1,,0,,0)}{v3,v3b}
\fmf{dashes,foreground=(1,,0,,0)}{v4,v4b}
\end{fmfgraph*}
\end{fmffile}
\end{gathered} = \left(i\sqrt{\frac{\lambda}{3}}\right)^{4} G_{\Phi}^{3} D_{\Phi}^{2} \phi_{N}^{2}\;,
\label{eq:col1PIDiagEvaluation1}
\end{equation}
\begin{equation}
\begin{gathered}
\begin{fmffile}{Diagrams/bosonic1PIEA_Gamma2_Diag3bis}
\begin{fmfgraph*}(25,25)
\fmfleft{i1,i2}
\fmfright{o1,o2}
\fmfbottom{i0,o0}
\fmftop{i3,o3}
\fmfv{decor.shape=circle,decor.size=2.0thick,foreground=(0,,0,,1)}{v1}
\fmfv{decor.shape=circle,decor.size=2.0thick,foreground=(0,,0,,1)}{v2}
\fmfv{decor.shape=circle,decor.size=2.0thick,foreground=(0,,0,,1)}{v3}
\fmfv{decor.shape=circle,decor.size=2.0thick,foreground=(0,,0,,1)}{v4}
\fmf{phantom}{i1,v1}
\fmf{phantom}{i2,v4}
\fmf{phantom}{o1,v2}
\fmf{phantom}{o2,v3}
\fmf{phantom}{i3,v4b}
\fmf{phantom}{o3,v3b}
\fmf{phantom}{i0,v1b}
\fmf{phantom}{o0,v2b}
\fmf{plain,tension=1.6,foreground=(1,,0,,0)}{v1,v2}
\fmf{plain,tension=1.6,foreground=(1,,0,,0)}{v3,v4}
\fmf{wiggly,tension=2.0,foreground=(1,,0,,0)}{v1,v4}
\fmf{wiggly,tension=2.0,foreground=(1,,0,,0)}{v2,v3}
\fmf{phantom,tension=0}{v1,v3}
\fmf{phantom,tension=0}{v2,v4}
\fmf{plain,right=0.8,tension=0,foreground=(1,,0,,0)}{v2,v3}
\fmf{plain,left=0.8,tension=0,foreground=(1,,0,,0)}{v1,v4}
\fmf{phantom}{v1,v1b}
\fmf{phantom}{v2,v2b}
\fmf{phantom}{v3,v3b}
\fmf{phantom}{v4,v4b}
\end{fmfgraph*}
\end{fmffile}
\end{gathered} = \begin{gathered}
\begin{fmffile}{Diagrams/bosonic1PIEA_Gamma2_Diag4bis}
\begin{fmfgraph*}(25,25)
\fmfleft{i1,i2}
\fmfright{o1,o2}
\fmfbottom{i0,o0}
\fmftop{i3,o3}
\fmfv{decor.shape=circle,decor.size=2.0thick,foreground=(0,,0,,1)}{v1}
\fmfv{decor.shape=circle,decor.size=2.0thick,foreground=(0,,0,,1)}{v2}
\fmfv{decor.shape=circle,decor.size=2.0thick,foreground=(0,,0,,1)}{v3}
\fmfv{decor.shape=circle,decor.size=2.0thick,foreground=(0,,0,,1)}{v4}
\fmf{phantom}{i1,v1}
\fmf{phantom}{i2,v4}
\fmf{phantom}{o1,v2}
\fmf{phantom}{o2,v3}
\fmf{phantom}{i3,v4b}
\fmf{phantom}{o3,v3b}
\fmf{phantom}{i0,v1b}
\fmf{phantom}{o0,v2b}
\fmf{plain,tension=1.6,foreground=(1,,0,,0)}{v1,v2}
\fmf{plain,tension=1.6,foreground=(1,,0,,0)}{v3,v4}
\fmf{wiggly,tension=2.0,foreground=(1,,0,,0)}{v1,v4}
\fmf{wiggly,tension=2.0,foreground=(1,,0,,0)}{v2,v3}
\fmf{plain,tension=0,foreground=(1,,0,,0)}{v1,v3}
\fmf{plain,tension=0,foreground=(1,,0,,0)}{v2,v4}
\fmf{phantom,right=0.8,tension=0}{v2,v3}
\fmf{phantom,left=0.8,tension=0}{v1,v4}
\fmf{phantom}{v1,v1b}
\fmf{phantom}{v2,v2b}
\fmf{phantom}{v3,v3b}
\fmf{phantom}{v4,v4b}
\end{fmfgraph*}
\end{fmffile}
\end{gathered} = \left(i\sqrt{\frac{\lambda}{3}}\right)^{4} N G_{\Phi}^{4} D_{\Phi}^{2}\;,
\end{equation}
\begin{equation}
\begin{gathered}
\begin{fmffile}{Diagrams/bosonic1PIEA_Gamma2_Diag5bis}
\begin{fmfgraph*}(35,20)
\fmfleft{i1,i2}
\fmfright{o1,o2}
\fmfbottom{i0,o0}
\fmfbottom{b0}
\fmfbottom{b1}
\fmfbottom{b2}
\fmftop{i3,o3}
\fmfv{decor.shape=circle,decor.size=2.0thick,foreground=(0,,0,,1)}{v1}
\fmfv{decor.shape=circle,decor.size=2.0thick,foreground=(0,,0,,1)}{v2}
\fmfv{decor.shape=circle,decor.size=2.0thick,foreground=(0,,0,,1)}{v3}
\fmfv{decor.shape=circle,decor.size=2.0thick,foreground=(0,,0,,1)}{v4}
\fmfv{decor.shape=circle,decor.size=2.0thick,foreground=(0,,0,,1)}{v5}
\fmfv{decor.shape=circle,decor.size=2.0thick,foreground=(0,,0,,1)}{v6}
\fmfv{decor.shape=cross,decor.size=0.25cm,decor.angle=16,foreground=(1,,0,,0)}{v1b}
\fmfv{decor.shape=cross,decor.size=0.25cm,foreground=(1,,0,,0)}{v3b}
\fmfv{decor.shape=cross,decor.size=0.25cm,decor.angle=71,foreground=(1,,0,,0)}{v4b}
\fmfv{decor.shape=cross,decor.size=0.25cm,foreground=(1,,0,,0)}{v6b}
\fmf{phantom,tension=1.4}{i1,v1}
\fmf{phantom}{i2,v3b}
\fmf{phantom}{i0,v1b}
\fmf{phantom}{o2,v6b}
\fmf{phantom,tension=1.4}{o1,v5}
\fmf{phantom}{o0,v5b}
\fmf{phantom,tension=1.11}{v3b,v6b}
\fmf{phantom,tension=1.38}{i0,v2}
\fmf{phantom,tension=1.38}{o0,v2}
\fmf{phantom,tension=1.8}{i0,v2b}
\fmf{phantom,tension=1.2}{o0,v2b}
\fmf{phantom,tension=1.2}{b0,v2b}
\fmf{phantom,tension=1.2}{b1,v2b}
\fmf{phantom,tension=1.2}{b2,v2b}
\fmf{phantom,tension=1.38}{i0,v4}
\fmf{phantom,tension=1.38}{o0,v4}
\fmf{phantom,tension=1.2}{i0,v4b}
\fmf{phantom,tension=1.8}{o0,v4b}
\fmf{phantom,tension=1.2}{b0,v4b}
\fmf{phantom,tension=1.2}{b1,v4b}
\fmf{phantom,tension=1.2}{b2,v4b}
\fmf{phantom,tension=2}{i3,v3}
\fmf{phantom,tension=2}{o3,v6}
\fmf{phantom,tension=2}{i3,v3b}
\fmf{phantom,tension=0.8}{o3,v3b}
\fmf{phantom,tension=0.8}{i3,v6b}
\fmf{phantom,tension=2}{o3,v6b}
\fmf{plain,tension=1.4,foreground=(1,,0,,0)}{v1,v2}
\fmf{plain,tension=1.4,foreground=(1,,0,,0)}{v4,v5}
\fmf{phantom}{v1,v3}
\fmf{phantom,left=0.8,tension=0}{v1,v3}
\fmf{plain,foreground=(1,,0,,0)}{v5,v6}
\fmf{phantom,right=0.8,tension=0}{v5,v6}
\fmf{plain,foreground=(1,,0,,0)}{v2,v3}
\fmf{phantom}{v4,v6}
\fmf{wiggly,tension=0.5,foreground=(1,,0,,0)}{v2,v4}
\fmf{wiggly,tension=2,foreground=(1,,0,,0)}{v3,v6}
\fmf{wiggly,right=0.8,tension=0,foreground=(1,,0,,0)}{v1,v5}
\fmf{dashes,tension=1,foreground=(1,,0,,0)}{v1,v1b}
\fmf{phantom,tension=0.2}{v2,v2b}
\fmf{dashes,tension=1.5,foreground=(1,,0,,0)}{v3,v3b}
\fmf{dashes,tension=0.2,foreground=(1,,0,,0)}{v4,v4b}
\fmf{phantom,tension=1}{v5,v5b}
\fmf{dashes,tension=1.5,foreground=(1,,0,,0)}{v6,v6b}
\end{fmfgraph*}
\end{fmffile}
\end{gathered} = \begin{gathered}
\begin{fmffile}{Diagrams/bosonic1PIEA_Gamma2_Diag6bis}
\begin{fmfgraph*}(35,20)
\fmfleft{i1,i2}
\fmfright{o1,o2}
\fmfbottom{i0,o0}
\fmfbottom{b0}
\fmfbottom{b1}
\fmfbottom{b2}
\fmftop{i3,o3}
\fmfv{decor.shape=circle,decor.size=2.0thick,foreground=(0,,0,,1)}{v1}
\fmfv{decor.shape=circle,decor.size=2.0thick,foreground=(0,,0,,1)}{v2}
\fmfv{decor.shape=circle,decor.size=2.0thick,foreground=(0,,0,,1)}{v3}
\fmfv{decor.shape=circle,decor.size=2.0thick,foreground=(0,,0,,1)}{v4}
\fmfv{decor.shape=circle,decor.size=2.0thick,foreground=(0,,0,,1)}{v5}
\fmfv{decor.shape=circle,decor.size=2.0thick,foreground=(0,,0,,1)}{v6}
\fmfv{decor.shape=cross,decor.size=0.25cm,decor.angle=16,foreground=(1,,0,,0)}{v1b}
\fmfv{decor.shape=cross,decor.size=0.25cm,foreground=(1,,0,,0)}{v3b}
\fmfv{decor.shape=cross,decor.size=0.25cm,decor.angle=-16,foreground=(1,,0,,0)}{v5b}
\fmfv{decor.shape=cross,decor.size=0.25cm,foreground=(1,,0,,0)}{v6b}
\fmf{phantom,tension=1.4}{i1,v1}
\fmf{phantom}{i2,v3b}
\fmf{phantom}{i0,v1b}
\fmf{phantom}{o2,v6b}
\fmf{phantom,tension=1.4}{o1,v5}
\fmf{phantom}{o0,v5b}
\fmf{phantom,tension=1.11}{v3b,v6b}
\fmf{phantom,tension=1.38}{i0,v2}
\fmf{phantom,tension=1.38}{o0,v2}
\fmf{phantom,tension=1.8}{i0,v2b}
\fmf{phantom,tension=1.2}{o0,v2b}
\fmf{phantom,tension=1.2}{b0,v2b}
\fmf{phantom,tension=1.2}{b1,v2b}
\fmf{phantom,tension=1.2}{b2,v2b}
\fmf{phantom,tension=1.38}{i0,v4}
\fmf{phantom,tension=1.38}{o0,v4}
\fmf{phantom,tension=1.2}{i0,v4b}
\fmf{phantom,tension=1.8}{o0,v4b}
\fmf{phantom,tension=1.2}{b0,v4b}
\fmf{phantom,tension=1.2}{b1,v4b}
\fmf{phantom,tension=1.2}{b2,v4b}
\fmf{phantom,tension=2}{i3,v3}
\fmf{phantom,tension=2}{o3,v6}
\fmf{phantom,tension=2}{i3,v3b}
\fmf{phantom,tension=0.8}{o3,v3b}
\fmf{phantom,tension=0.8}{i3,v6b}
\fmf{phantom,tension=2}{o3,v6b}
\fmf{plain,tension=1.4,foreground=(1,,0,,0)}{v1,v2}
\fmf{plain,tension=1.4,foreground=(1,,0,,0)}{v4,v5}
\fmf{phantom}{v1,v3}
\fmf{phantom,left=0.8,tension=0}{v1,v3}
\fmf{phantom}{v5,v6}
\fmf{phantom,right=0.8,tension=0}{v5,v6}
\fmf{plain,foreground=(1,,0,,0)}{v2,v3}
\fmf{plain,foreground=(1,,0,,0)}{v4,v6}
\fmf{wiggly,tension=0.5,foreground=(1,,0,,0)}{v2,v4}
\fmf{wiggly,tension=2,foreground=(1,,0,,0)}{v3,v6}
\fmf{wiggly,right=0.8,tension=0,foreground=(1,,0,,0)}{v1,v5}
\fmf{dashes,tension=1,foreground=(1,,0,,0)}{v1,v1b}
\fmf{phantom,tension=0.2}{v2,v2b}
\fmf{dashes,tension=1.5,foreground=(1,,0,,0)}{v3,v3b}
\fmf{phantom,tension=0.2}{v4,v4b}
\fmf{dashes,tension=1,foreground=(1,,0,,0)}{v5,v5b}
\fmf{dashes,tension=1.5,foreground=(1,,0,,0)}{v6,v6b}
\end{fmfgraph*}
\end{fmffile}
\end{gathered} = \left(i\sqrt{\frac{\lambda}{3}}\right)^{6} G_{\Phi}^{4} D_{\Phi}^{3} \phi_{N}^{4}\;,
\end{equation}
\begin{equation}
\begin{gathered}
\begin{fmffile}{Diagrams/bosonic1PIEA_Gamma2_Diag7bis}
\begin{fmfgraph*}(35,20)
\fmfleft{i1,i2}
\fmfright{o1,o2}
\fmfbottom{i0,o0}
\fmfbottom{b0}
\fmfbottom{b1}
\fmfbottom{b2}
\fmftop{i3,o3}
\fmfv{decor.shape=circle,decor.size=2.0thick,foreground=(0,,0,,1)}{v1}
\fmfv{decor.shape=circle,decor.size=2.0thick,foreground=(0,,0,,1)}{v2}
\fmfv{decor.shape=circle,decor.size=2.0thick,foreground=(0,,0,,1)}{v3}
\fmfv{decor.shape=circle,decor.size=2.0thick,foreground=(0,,0,,1)}{v4}
\fmfv{decor.shape=circle,decor.size=2.0thick,foreground=(0,,0,,1)}{v5}
\fmfv{decor.shape=circle,decor.size=2.0thick,foreground=(0,,0,,1)}{v6}
\fmfv{decor.shape=cross,decor.size=0.25cm,decor.angle=16,foreground=(1,,0,,0)}{v1b}
\fmfv{decor.shape=cross,decor.size=0.25cm,foreground=(1,,0,,0)}{v3b}
\fmf{phantom,tension=1.4}{i1,v1}
\fmf{phantom}{i2,v3b}
\fmf{phantom}{i0,v1b}
\fmf{phantom}{o2,v6b}
\fmf{phantom,tension=1.4}{o1,v5}
\fmf{phantom}{o0,v5b}
\fmf{phantom,tension=1.11}{v3b,v6b}
\fmf{phantom,tension=1.38}{i0,v2}
\fmf{phantom,tension=1.38}{o0,v2}
\fmf{phantom,tension=1.8}{i0,v2b}
\fmf{phantom,tension=1.2}{o0,v2b}
\fmf{phantom,tension=1.2}{b0,v2b}
\fmf{phantom,tension=1.2}{b1,v2b}
\fmf{phantom,tension=1.2}{b2,v2b}
\fmf{phantom,tension=1.38}{i0,v4}
\fmf{phantom,tension=1.38}{o0,v4}
\fmf{phantom,tension=1.2}{i0,v4b}
\fmf{phantom,tension=1.8}{o0,v4b}
\fmf{phantom,tension=1.2}{b0,v4b}
\fmf{phantom,tension=1.2}{b1,v4b}
\fmf{phantom,tension=1.2}{b2,v4b}
\fmf{phantom,tension=2}{i3,v3}
\fmf{phantom,tension=2}{o3,v6}
\fmf{phantom,tension=2}{i3,v3b}
\fmf{phantom,tension=0.8}{o3,v3b}
\fmf{phantom,tension=0.8}{i3,v6b}
\fmf{phantom,tension=2}{o3,v6b}
\fmf{plain,tension=1.4,foreground=(1,,0,,0)}{v1,v2}
\fmf{plain,tension=1.4,foreground=(1,,0,,0)}{v4,v5}
\fmf{phantom}{v1,v3}
\fmf{phantom,left=0.8,tension=0}{v1,v3}
\fmf{plain,foreground=(1,,0,,0)}{v5,v6}
\fmf{phantom,right=0.8,tension=0}{v5,v6}
\fmf{plain,foreground=(1,,0,,0)}{v2,v3}
\fmf{plain,foreground=(1,,0,,0)}{v4,v6}
\fmf{wiggly,tension=0.5,foreground=(1,,0,,0)}{v2,v4}
\fmf{wiggly,tension=2,foreground=(1,,0,,0)}{v3,v6}
\fmf{wiggly,right=0.8,tension=0,foreground=(1,,0,,0)}{v1,v5}
\fmf{dashes,tension=1,foreground=(1,,0,,0)}{v1,v1b}
\fmf{phantom,tension=0.2}{v2,v2b}
\fmf{dashes,tension=1.5,foreground=(1,,0,,0)}{v3,v3b}
\fmf{phantom,tension=0.2}{v4,v4b}
\fmf{phantom,tension=1}{v5,v5b}
\fmf{phantom,tension=1.5}{v6,v6b}
\end{fmfgraph*}
\end{fmffile}
\end{gathered} = \left(i\sqrt{\frac{\lambda}{3}}\right)^{6} N G_{\Phi}^{5} D_{\Phi}^{3} \phi_{N}^{2}\;,
\end{equation}
\begin{equation}
\begin{gathered}
\begin{fmffile}{Diagrams/bosonic1PIEA_Gamma2_Diag8bis}
\begin{fmfgraph*}(35,20)
\fmfleft{i1,i2}
\fmfright{o1,o2}
\fmfbottom{i0,o0}
\fmfbottom{b0}
\fmfbottom{b1}
\fmfbottom{b2}
\fmftop{i3,o3}
\fmfv{decor.shape=circle,decor.size=2.0thick,foreground=(0,,0,,1)}{v1}
\fmfv{decor.shape=circle,decor.size=2.0thick,foreground=(0,,0,,1)}{v2}
\fmfv{decor.shape=circle,decor.size=2.0thick,foreground=(0,,0,,1)}{v3}
\fmfv{decor.shape=circle,decor.size=2.0thick,foreground=(0,,0,,1)}{v4}
\fmfv{decor.shape=circle,decor.size=2.0thick,foreground=(0,,0,,1)}{v5}
\fmfv{decor.shape=circle,decor.size=2.0thick,foreground=(0,,0,,1)}{v6}
\fmf{phantom,tension=1.4}{i1,v1}
\fmf{phantom}{i2,v3b}
\fmf{phantom}{i0,v1b}
\fmf{phantom}{o2,v6b}
\fmf{phantom,tension=1.4}{o1,v5}
\fmf{phantom}{o0,v5b}
\fmf{phantom,tension=1.11}{v3b,v6b}
\fmf{phantom,tension=1.38}{i0,v2}
\fmf{phantom,tension=1.38}{o0,v2}
\fmf{phantom,tension=1.8}{i0,v2b}
\fmf{phantom,tension=1.2}{o0,v2b}
\fmf{phantom,tension=1.2}{b0,v2b}
\fmf{phantom,tension=1.2}{b1,v2b}
\fmf{phantom,tension=1.2}{b2,v2b}
\fmf{phantom,tension=1.38}{i0,v4}
\fmf{phantom,tension=1.38}{o0,v4}
\fmf{phantom,tension=1.2}{i0,v4b}
\fmf{phantom,tension=1.8}{o0,v4b}
\fmf{phantom,tension=1.2}{b0,v4b}
\fmf{phantom,tension=1.2}{b1,v4b}
\fmf{phantom,tension=1.2}{b2,v4b}
\fmf{phantom,tension=2}{i3,v3}
\fmf{phantom,tension=2}{o3,v6}
\fmf{phantom,tension=2}{i3,v3b}
\fmf{phantom,tension=0.8}{o3,v3b}
\fmf{phantom,tension=0.8}{i3,v6b}
\fmf{phantom,tension=2}{o3,v6b}
\fmf{plain,tension=1.4,foreground=(1,,0,,0)}{v1,v2}
\fmf{plain,tension=1.4,foreground=(1,,0,,0)}{v4,v5}
\fmf{plain,foreground=(1,,0,,0)}{v1,v3}
\fmf{phantom,left=0.8,tension=0}{v1,v3}
\fmf{plain,foreground=(1,,0,,0)}{v5,v6}
\fmf{phantom,right=0.8,tension=0}{v5,v6}
\fmf{plain,foreground=(1,,0,,0)}{v2,v3}
\fmf{plain,foreground=(1,,0,,0)}{v4,v6}
\fmf{wiggly,tension=0.5,foreground=(1,,0,,0)}{v2,v4}
\fmf{wiggly,tension=2,foreground=(1,,0,,0)}{v3,v6}
\fmf{wiggly,right=0.8,tension=0,foreground=(1,,0,,0)}{v1,v5}
\fmf{phantom,tension=1}{v1,v1b}
\fmf{phantom,tension=0.2}{v2,v2b}
\fmf{phantom,tension=1.5}{v3,v3b}
\fmf{phantom,tension=0.2}{v4,v4b}
\fmf{phantom,tension=1}{v5,v5b}
\fmf{phantom,tension=1.5}{v6,v6b}
\end{fmfgraph*}
\end{fmffile}
\end{gathered} = \left(i\sqrt{\frac{\lambda}{3}}\right)^{6} N^{2} G_{\Phi}^{6} D_{\Phi}^{3}\;.
\label{eq:col1PIDiagEvaluation5}
\end{equation}
Therefore, according to~\eqref{eq:bosonic1PIEAIMGamma0DON} as well as~\eqref{eq:col1PIDiagEvaluation1} to~\eqref{eq:col1PIDiagEvaluation5}, $\Gamma_{\mathrm{col}}^{(\mathrm{1PI})}$ becomes in (0+0)-D:
\begin{equation}
\begin{split}
\Gamma_{\mathrm{col}}^{(\mathrm{1PI})}\big(\Phi\big) = & \ S_{\mathrm{col}}(\eta) + \frac{1}{2}G_{\Phi}^{-1} \phi_{N}^{2} - \frac{\hbar}{2}\ln\big(D_{\Phi}\big) \\
& + \hbar^{2}\Bigg[-\frac{\lambda^{2}}{9}G_{\Phi}^{3}D_{\Phi}^{2}\phi_{N}^{2} -\frac{\lambda^{2}}{24} N G_{\Phi}^{4} D_{\Phi}^{2} + \frac{\lambda^{3}}{36}G_{\Phi}^{4} D_{\Phi}^{3} \phi_{N}^{4} + \frac{\lambda^{3}}{54} N G_{\Phi}^{5}D_\Phi^{3}\phi_{N}^{2} \\
& \hspace{1.15cm} + \frac{\lambda^{3}}{324} N^{2} G_{\Phi}^{6} D_{\Phi}^{3}\Bigg] \\
& + \mathcal{O}\big(\hbar^{3}\big)\;,
\end{split}
\label{eq:bosonic1PIEAfinalexpression0DON}
\end{equation}
with $S_{\mathrm{col}}(\eta)=\frac{1}{2}\eta^{2}-\frac{N}{2}\ln\big(2\pi G_{\Phi}\big)$. As before, the expression for the 1PI EA becomes exploitable after fixing the relevant 1-point correlation function(s) (i.e. the components of $\Phi$ here), which is now done via the gap equations:
\begin{equation}
\begin{split}
0 = \left.\frac{\partial \Gamma_{\mathrm{col}}^{(\mathrm{1PI})}\big(\Phi\big)}{\partial \phi_{N}}\right|_{\Phi=\overline{\Phi}} = & \ G_{\overline{\Phi}}^{-1}\overline{\phi}_{N} + \frac{\hbar}{3} \ G_{\overline{\Phi}} D_{\overline{\Phi}} \lambda \overline{\phi}_{N} \\
& + \hbar^{2} \Bigg[-\frac{1}{162} G_{\overline{\Phi}}^3 D_{\overline{\Phi}}^2 \lambda^2 \overline{\phi}_{N} \Big( 36 + G_{\overline{\Phi}}^4 D_{\overline{\Phi}}^2 N^2 \lambda^2 - 42 G_{\overline{\Phi}} D_{\overline{\Phi}} \lambda \overline{\phi}_{N}^2 \\
& \hspace{1.15cm} + 6 G_{\overline{\Phi}}^3 D_{\overline{\Phi}}^2 N \lambda^2 \overline{\phi}_{N}^2 + 3 G_{\overline{\Phi}}^2 D_{\overline{\Phi}} \lambda \left(-5 N + 3 D_{\overline{\Phi}} \lambda \overline{\phi}_{N}^4\right) \Big) \Bigg] \\
& +\mathcal{O}\big(\hbar^{3}\big)\;,
\end{split}
\label{eq:bosonic1PIEAGapEquationphiN0DON}
\end{equation}
and
\begin{equation}
\begin{split}
0 = \left.\frac{\partial \Gamma_{\mathrm{col}}^{(\mathrm{1PI})}\big(\Phi\big)}{\partial \eta}\right|_{\Phi=\overline{\Phi}} = & \ \overline{\eta} + \frac{i}{2}\sqrt{\frac{\lambda}{3}} \left(G_{\overline{\Phi}} N + \overline{\phi}_{N}^{2}\right) - \hbar \ \frac{i G_{\overline{\Phi}}^2 D_{\overline{\Phi}} \lambda^{\frac{3}{2}}}{6\sqrt{3}}\left(G_{\overline{\Phi}} N + \overline{\phi}_{N}^{2}\right) \\
& + \hbar^{2} \left[\rule{0cm}{0.8cm}\right. \frac{i G_{\overline{\Phi}}^4 D_{\overline{\Phi}}^{2} \lambda^{\frac{5}{2}}}{324\sqrt{3}} \Big( G_{\overline{\Phi}}^5 D_{\overline{\Phi}}^2 N^3 \lambda^2 + 108 \overline{\phi}_{N}^2 + 7 G_{\overline{\Phi}}^4 D_{\overline{\Phi}}^2 N^2 \lambda^2 \overline{\phi}_{N}^2 \\
& \hspace{1.0cm} + 6 G_{\overline{\Phi}} \Big(9 N - 10 D_{\overline{\Phi}} \lambda \overline{\phi}_{N}^4\Big) + 9 G_{\overline{\Phi}}^2 D_{\overline{\Phi}} \lambda \overline{\phi}_{N}^2 \left(-7 N + D_{\overline{\Phi}} \lambda \overline{\phi}_{N}^4\right) \\
& \hspace{1.0cm} + 15 G_{\overline{\Phi}}^3 D_{\overline{\Phi}} N \lambda \Big(-N + D_{\overline{\Phi}} \lambda \overline{\phi}_{N}^4\Big) \Big) \left.\rule{0cm}{0.8cm}\right] \\
& +\mathcal{O}\big(\hbar^{3}\big)\;,
\end{split}
\label{eq:bosonic1PIEAGapEquationeta0DON}
\end{equation}
with $\overline{\Phi}=\begin{pmatrix}
\vec{\overline{\phi}} & \overline{\eta} \end{pmatrix}^{\mathrm{T}} = \begin{pmatrix}
\overline{\phi}_{1} & \cdots & \overline{\phi}_{N-1} & \overline{\phi}_{N} & \overline{\eta}
\end{pmatrix}^{\mathrm{T}} = \begin{pmatrix}
0 & \cdots & 0 & \overline{\phi}_{N} & \overline{\eta}
\end{pmatrix}^{\mathrm{T}}$, $G_{\overline{\Phi}}^{-1} = m^{2}+i\sqrt{\frac{\lambda}{3}} \overline{\eta}$ and $D_{\overline{\Phi}}^{-1} = \frac{\lambda}{3} G_{\overline{\Phi}} \overline{\phi}_{N}^{2}+\frac{\lambda}{6} N G_{\overline{\Phi}}^{2} + 1$ (as follows respectively from~\eqref{eq:1PIEAcolGpropagator} and~\eqref{eq:1PIEAcolDpropagator} in the case where $\mathcal{J}$ vanishes). Finally, the gs energy and density are respectively deduced after plugging the solution $\overline{\Phi}$ into:
\begin{equation}
E^\text{1PI EA;col}_{\mathrm{gs}} = \frac{1}{\hbar} \Gamma_{\mathrm{col}}^{(\mathrm{1PI})}\big(\Phi=\overline{\Phi}\big) \;,
\label{eq:GetEgscol1PIEAfromGammacol1PIEA0DON}
\end{equation}
\begin{equation}
\rho^\text{1PI EA;col}_{\mathrm{gs}} = \frac{i}{N} \sqrt{\frac{12}{\lambda}} \overline{\eta} \;.
\label{eq:DeducerhogsbosonicGeneral1PIdiagEA0DON}
\end{equation}
The latter relation follows by considering the following classical equation of motion in the mixed representation:
\begin{equation}
\frac{\partial S_{\mathrm{mix}}\big(\vec{\widetilde{\varphi}},\widetilde{\sigma}\big)}{\partial\widetilde{\sigma}} = \widetilde{\sigma} + i\sqrt{\frac{\lambda}{12}} \ \vec{\widetilde{\varphi}}^{2} = 0 \;,
\label{eq:DerivSmixforrhogsinCollTheory}
\end{equation}
which can be inferred from~\eqref{eq:minimizationSmixedKj} in (0+0)-D with all external sources set equal to zero. In the spirit of the Schwinger-Dyson equations formalism, we take the expectation value of~\eqref{eq:DerivSmixforrhogsinCollTheory} to turn it into the equality\footnote{The definition of the gs density used in~\eqref{eq:DeducerhogsbosonicGeneral1PIFRG0DON} was introduced in~\eqref{eq:DefrhogsExactwithExpectationValue0DON} with an expectation value defined by~\eqref{eq:vacuumExpectationValue0DON}.}:
\begin{equation}
\rho_{\mathrm{gs}} \equiv \frac{1}{N}\left\langle \vec{\widetilde{\varphi}}^{2} \right\rangle = \frac{i}{N} \sqrt{\frac{12}{\lambda}} \ \overline{\eta} \;,
\label{eq:DeducerhogsbosonicGeneral1PIFRG0DON}
\end{equation}
in accordance with~\eqref{eq:DeducerhogsbosonicGeneral1PIdiagEA0DON}.

\vspace{0.5cm}

\begin{figure}[!htb]
\captionsetup[subfigure]{labelformat=empty}
  \begin{center}
    \subfloat[]{
      \includegraphics[width=0.50\linewidth]{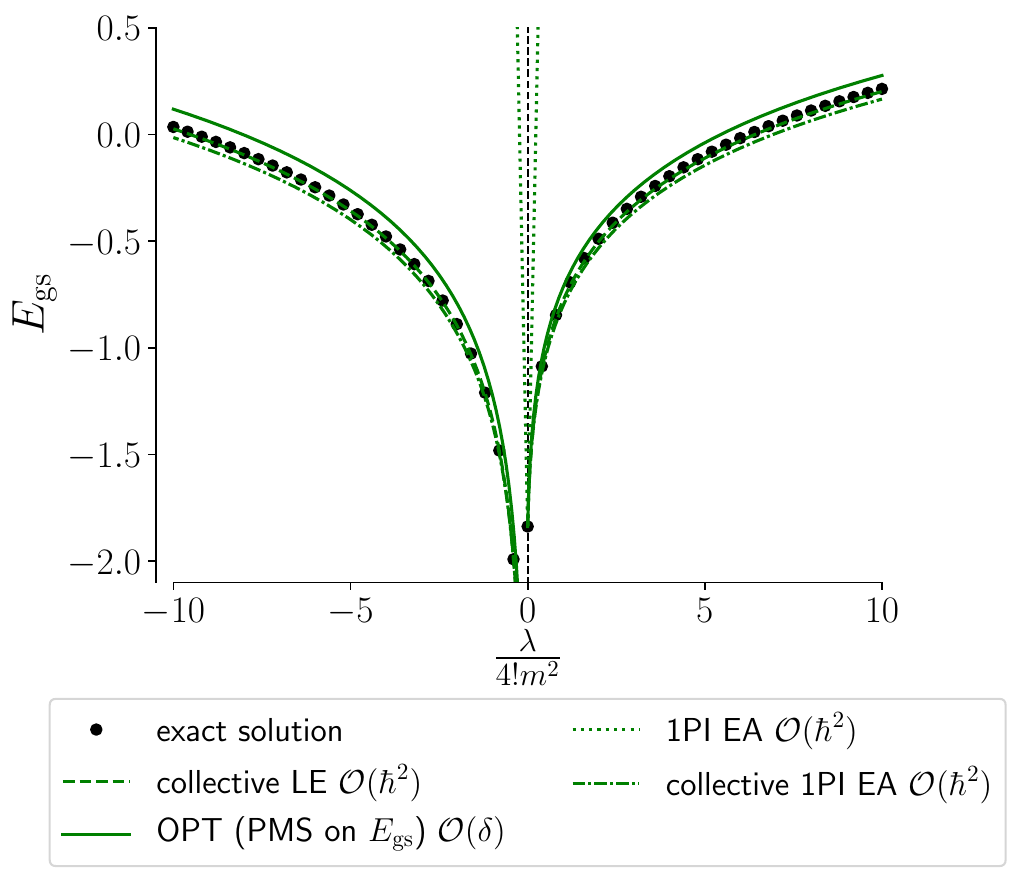}
                         }
    \subfloat[]{
      \includegraphics[width=0.50\linewidth]{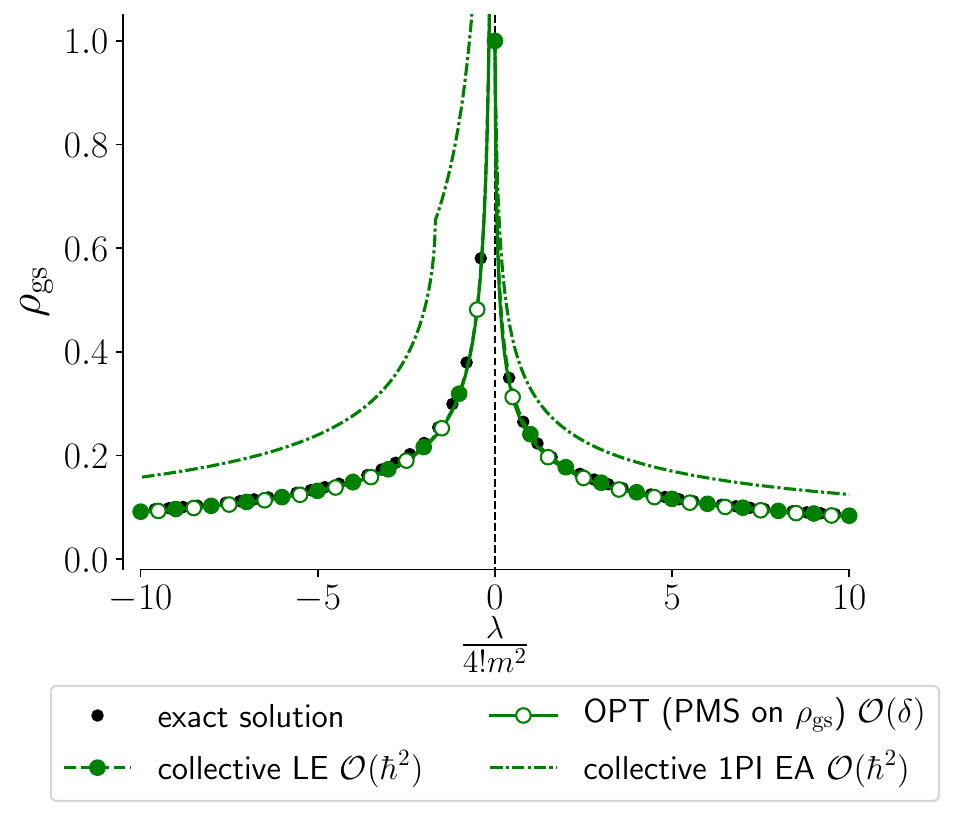}
                         }
    \caption{Gs energy $E_{\mathrm{gs}}$ or density $\rho_{\mathrm{gs}}$ calculated at $\hbar=1$, $m^{2}=\pm 1$ and $N=2$ ($\mathcal{R}e(\lambda)\geq 0$ and $\mathcal{I}m(\lambda)=0$). The indication ``$\mathcal{O}\big(\hbar^{n}\big)$'' for the results obtained from the $\hbar$-expanded 1PI EAs specifies that the series representing the EA in question has been exploited up to order $\mathcal{O}\big(\hbar^{n}\big)$ (which implies notably that the corresponding series for the gs energy is calculated up to order $\mathcal{O}(\hbar^{n-1})$ according to~\eqref{eq:DeduceEgs1PIEAdiag} and~\eqref{eq:GetEgscol1PIEAfromGammacol1PIEA0DON}).}
    \label{fig:1PIEA}
  \end{center}
\end{figure}

For both signs of $m^{2}$, the physical solutions found from the extremization of both the original 1PI EA (in the $\hbar$- and $\lambda$-expansions) and the collective 1PI EA at order $\mathcal{O}\big(\hbar^{2}\big)$ (i.e. from the resolution of the corresponding gap equations) possess a vanishing 1-point correlation function $\vec{\overline{\phi}}$ for the original field. As explained below~\eqref{eq:1PIEAlambdaGapEquation0DON}, this implies that the $\hbar$- and $\lambda$-expansions of the original 1PI EA are equivalent in this situation. Most importantly, this means that these 1PI EA approaches do not exhibit any spurious spontaneous breakdown of the $O(N)$ symmetry at their first non-trivial orders, which is in accordance with the minimum of the exact effective potential $V_\text{eff}^\text{exact}\big(\vec{\phi}\big)$ lying at $\vec{\phi}=\vec{0}$ regardless of the sign of $m^{2}$ (as discussed in section~\ref{sec:studiedtoymodel}). Even though this is a reassuring feature, it is also fatal for the original 1PI EA since $\vec{\overline{\phi}}$ is the only adjustable variable that can be used to grasp correlations in this framework. This is illustrated by fig.~\ref{fig:1PIEA} where the gs energy estimated from the original 1PI EA diverges very quickly as $\lambda$ increases. This approach is thus completely irrelevant to tackle the non-perturbative regime of our model, regardless of the calculated quantity (i.e. gs energy, gs density, ...).

\vspace{0.5cm}

While the constraint of the $O(N)$ symmetry is too strong for the original 1PI EA, the collective one manages to capture non-perturbative physics thanks to the 1-point correlation function $\eta$ of the Hubbard-Stratonovich field. This illustrates a key advantage of HSTs by which one introduces a new field in the arena that is not constrained by the symmetries of the model under consideration. The Hubbard-Stratonovich field being a scalar with respect to the $O(N)$ transformations in the present case, its expectation value can be finite without spoiling the $O(N)$ symmetry and can therefore dress the propagator $\boldsymbol{G}_{\Phi}$ with non-perturbative physics. However, we can question the efficiency of the collective 1PI EA from another angle as fig.~\ref{fig:1PIEA} shows for both $E_{\mathrm{gs}}$ and $\rho_{\mathrm{gs}}$ at $N=2$ that this EA approach is outperformed by the collective LE over the whole range of tested values for the coupling constant (i.e. for $\lambda/4!\in [0,10]$). Even though the determination of the diagrammatic representation of the 1PI EA is less demanding than that of the Schwinger functional (as the latter includes connected 1-particle-reducible (1PR) diagrams as opposed to the 1PI EA), the EA framework requires to solve also gap equations which are self-consistent for realistic models. This is a significant reason to favor the LE as compared to the EA method in this situation, especially considering the good performances of the collective LE beyond its first non-trivial order illustrated notably by figs.~\ref{fig:O2OPTvsLE_1} and~\ref{fig:O2OPTvsLE_3}. Besides this, it should also be noted that the diagrammatic constructions are significantly more demanding to develop in the framework of the collective representation, as compared to the original and mixed ones. Hence, we will investigate the 2PI EA in the next section for the original and mixed versions of the studied $O(N)$ model, still paying particular attention to the presence or absence of SSB in the solutions of the corresponding gap equations. The 2PI EA enables us to express the energy as a functional of the 2-point correlation function (or Green's function or full propagator) of the system, which gets dressed with non-perturbative physics via the resolution of the corresponding gap equation. We will also carefully illustrate how the mixed 2PI EA exploits the Hubbard-Stratonovich field to capture correlations.

\subsection{\label{sec:2PIEA}2PI effective action}
\subsubsection{Original effective action}

\paragraph{Full original 2PI EA:}

We first point out that the original 2PI EA has already been investigated for the $O(N)$-symmetric $\varphi^{4}$ model but many of these studies consider a $1/N$-expansion~\cite{baa03bis2,baa04,baa05}. We will focus in section~\ref{sec:2PIEA} on the $\hbar$- and $\lambda$-expansions exclusively and discuss the connections between these two expansion schemes as a next step.

\vspace{0.5cm}

As before, we start by giving the definition of the EA under consideration, i.e. the original 2PI EA here. It relies on the Legendre transform:
\begin{equation}
\begin{split}
\Gamma^{(\mathrm{2PI})}\Big[\vec{\phi},\boldsymbol{G}\Big] \equiv & -W\Big[\vec{J},\boldsymbol{K}\Big] + \int_{x}J^{a}(x) \frac{\delta W\big[\vec{J},\boldsymbol{K}\big]}{\delta J^{a}(x)} + \int_{x,y}\boldsymbol{K}^{ab}(x,y) \frac{\delta W\big[\vec{J},\boldsymbol{K}\big]}{\delta \boldsymbol{K}^{ba}(x,y)} \\
= & -W\Big[\vec{J},\boldsymbol{K}\Big] + \int_{x}J^{a}(x) \phi_{a}(x) + \frac{1}{2} \int_{x,y} \phi_{a}(x) \boldsymbol{K}^{ab}(x,y) \phi_{b}(y) \\
& + \frac{\hbar}{2} \int_{x,y} \boldsymbol{K}^{ab}(x,y) \boldsymbol{G}_{ba}(y,x) \;,
\end{split}
\label{eq:pure2PIEAdefinition0DONmain}
\end{equation}
with
\begin{equation}
\phi_{a}(x) = \frac{\delta W\big[\vec{J},\boldsymbol{K}\big]}{\delta J^{a}(x)} \;,
\label{eq:pure2PIEAdefinitionbis0DONmain}
\end{equation}
\begin{equation}
\boldsymbol{G}_{ab}(x,y) = \frac{\delta^{2} W\big[\vec{J},\boldsymbol{K}\big]}{\delta J^{a}(x)\delta J^{b}(y)} = \frac{2}{\hbar} \frac{\delta W\big[\vec{J},\boldsymbol{K}\big]}{\delta\boldsymbol{K}^{ab}(x,y)} - \frac{1}{\hbar} \phi_{a}(x) \phi_{b}(y) \;,
\label{eq:pure2PIEAdefinitionbis20DONmain}
\end{equation}
and $W\big[\vec{J},\boldsymbol{K}\big] \equiv W^{\text{LE};\text{orig}}\big[\vec{J},\boldsymbol{K}\big]$ has already been expressed diagrammatically via the LE in section~\ref{sec:PT}. From definition~\eqref{eq:pure2PIEAdefinition0DONmain}, it can be shown that the original 2PI EA can be expressed in terms of 2PI diagrams only, which translates for the studied $O(N)$ model into (see appendix~\ref{sec:original2PIEAannIM}):
\begin{equation}
\begin{split}
\scalebox{0.98}{${\displaystyle \Gamma^{(\mathrm{2PI})}\Big[\vec{\phi},\boldsymbol{G}\Big] = }$} & \ \scalebox{0.98}{${\displaystyle S\Big[\vec{\phi}\Big] -\frac{\hbar}{2}\mathrm{STr}\left[\ln\big(\boldsymbol{G}\big)\right] + \frac{\hbar}{2}\mathrm{STr}\left[\boldsymbol{G}^{-1}_\phi\boldsymbol{G}-\mathbb{I}\right] }$} \\
& \scalebox{0.98}{${\displaystyle + \hbar^{2} \left(\rule{0cm}{1.2cm}\right. \frac{1}{24} \hspace{0.08cm} \begin{gathered}
\begin{fmffile}{Diagrams/1PIEA_Hartree}
\begin{fmfgraph}(30,20)
\fmfleft{i}
\fmfright{o}
\fmf{phantom,tension=10}{i,i1}
\fmf{phantom,tension=10}{o,o1}
\fmf{plain,left,tension=0.5,foreground=(1,,0,,0)}{i1,v1,i1}
\fmf{plain,right,tension=0.5,foreground=(1,,0,,0)}{o1,v2,o1}
\fmf{zigzag,foreground=(0,,0,,1)}{v1,v2}
\end{fmfgraph}
\end{fmffile}
\end{gathered}
+\frac{1}{12}\begin{gathered}
\begin{fmffile}{Diagrams/1PIEA_Fock}
\begin{fmfgraph}(15,15)
\fmfleft{i}
\fmfright{o}
\fmf{phantom,tension=11}{i,v1}
\fmf{phantom,tension=11}{v2,o}
\fmf{plain,left,tension=0.4,foreground=(1,,0,,0)}{v1,v2,v1}
\fmf{zigzag,foreground=(0,,0,,1)}{v1,v2}
\end{fmfgraph}
\end{fmffile}
\end{gathered}
- \frac{1}{18} \begin{gathered}
\begin{fmffile}{Diagrams/1PIEA_Diag1}
\begin{fmfgraph}(27,15)
\fmfleft{i}
\fmfright{o}
\fmftop{vUp}
\fmfbottom{vDown}
\fmfv{decor.shape=cross,decor.size=3.5thick,foreground=(1,,0,,0)}{v1}
\fmfv{decor.shape=cross,decor.size=3.5thick,foreground=(1,,0,,0)}{v2}
\fmf{phantom,tension=10}{i,i1}
\fmf{phantom,tension=10}{o,o1}
\fmf{phantom,tension=2.2}{vUp,v5}
\fmf{phantom,tension=2.2}{vDown,v6}
\fmf{phantom,tension=0.5}{v3,v4}
\fmf{phantom,tension=10.0}{i1,v1}
\fmf{phantom,tension=10.0}{o1,v2}
\fmf{dashes,tension=2.0,foreground=(0,,0,,1),foreground=(1,,0,,0)}{v1,v3}
\fmf{dots,left=0.4,tension=0.5,foreground=(0,,0,,1)}{v3,v5}
\fmf{plain,left=0.4,tension=0.5,foreground=(1,,0,,0)}{v5,v4}
\fmf{plain,right=0.4,tension=0.5,foreground=(1,,0,,0)}{v3,v6}
\fmf{dots,right=0.4,tension=0.5,foreground=(0,,0,,1)}{v6,v4}
\fmf{dashes,tension=2.0,foreground=(0,,0,,1),foreground=(1,,0,,0)}{v4,v2}
\fmf{plain,tension=0,foreground=(1,,0,,0)}{v5,v6}
\end{fmfgraph}
\end{fmffile}
\end{gathered} - \frac{1}{36} \hspace{-0.15cm} \begin{gathered}
\begin{fmffile}{Diagrams/1PIEA_Diag2}
\begin{fmfgraph}(25,20)
\fmfleft{i}
\fmfright{o}
\fmftop{vUp}
\fmfbottom{vDown}
\fmfv{decor.shape=cross,decor.angle=45,decor.size=3.5thick,foreground=(1,,0,,0)}{vUpbis}
\fmfv{decor.shape=cross,decor.angle=45,decor.size=3.5thick,foreground=(1,,0,,0)}{vDownbis}
\fmf{phantom,tension=0.8}{vUp,vUpbis}
\fmf{phantom,tension=0.8}{vDown,vDownbis}
\fmf{dashes,tension=0.5,foreground=(0,,0,,1),foreground=(1,,0,,0)}{v3,vUpbis}
\fmf{phantom,tension=0.5}{v4,vUpbis}
\fmf{phantom,tension=0.5}{v3,vDownbis}
\fmf{dashes,tension=0.5,foreground=(0,,0,,1),foreground=(1,,0,,0)}{v4,vDownbis}
\fmf{phantom,tension=11}{i,v1}
\fmf{phantom,tension=11}{v2,o}
\fmf{plain,left,tension=0.5,foreground=(1,,0,,0)}{v1,v2,v1}
\fmf{dots,tension=1.7,foreground=(0,,0,,1)}{v1,v3}
\fmf{plain,foreground=(1,,0,,0)}{v3,v4}
\fmf{dots,tension=1.7,foreground=(0,,0,,1)}{v4,v2}
\end{fmfgraph}
\end{fmffile}
\end{gathered} \hspace{-0.2cm} \left.\rule{0cm}{1.2cm}\right) }$} \\
& \scalebox{0.98}{${\displaystyle + \mathcal{O}\big(\hbar^{3}\big)\;, }$}
\end{split}
\label{eq:2PIEAfinalexpression}
\end{equation}
with $\mathbb{I}$ being the identity with respect to both spacetime and color indices (i.e. $\mathbb{I}_{ab}(x,y)=\delta_{ab}\delta(x-y)$) and~\eqref{eq:2PIEAfinalexpression} is based on the Feynman rules:
\begin{subequations}
\begin{align}
\begin{gathered}
\begin{fmffile}{Diagrams/1PIEA_G}
\begin{fmfgraph*}(20,20)
\fmfleft{i0,i1,i2,i3}
\fmfright{o0,o1,o2,o3}
\fmflabel{$x, a$}{v1}
\fmflabel{$y, b$}{v2}
\fmf{phantom}{i1,v1}
\fmf{phantom}{i2,v1}
\fmf{plain,tension=0.6,foreground=(1,,0,,0)}{v1,v2}
\fmf{phantom}{v2,o1}
\fmf{phantom}{v2,o2}
\end{fmfgraph*}
\end{fmffile}
\end{gathered} \quad &\rightarrow \boldsymbol{G}_{ab}(x,y)\;,
\label{eq:FeynRuleorig2PIEAG} \\
\begin{gathered}
\begin{fmffile}{Diagrams/1PIEA_V3}
\begin{fmfgraph*}(20,20)
\fmfleft{i0,i1,i2,i3}
\fmfright{o0,o1,o2,o3}
\fmfv{decor.shape=cross,decor.angle=45,decor.size=3.5thick,foreground=(1,,0,,0)}{o2}
\fmf{phantom,tension=2.0}{i1,i1bis}
\fmf{plain,tension=2.0,foreground=(1,,0,,0)}{i1bis,v1}
\fmf{phantom,tension=2.0}{i2,i2bis}
\fmf{plain,tension=2.0,foreground=(1,,0,,0)}{i2bis,v1}
\fmf{dots,label=$x$,tension=0.6,foreground=(0,,0,,1)}{v1,v2}
\fmf{phantom,tension=2.0}{o1bis,o1}
\fmf{plain,tension=2.0,foreground=(1,,0,,0)}{v2,o1bis}
\fmf{phantom,tension=2.0}{o2bis,o2}
\fmf{phantom,tension=2.0,foreground=(1,,0,,0)}{v2,o2bis}
\fmf{dashes,tension=0.0,foreground=(1,,0,,0)}{v2,o2}
\fmflabel{$a$}{i1bis}
\fmflabel{$b$}{i2bis}
\fmflabel{$c$}{o1bis}
\fmflabel{$N$}{o2bis}
\end{fmfgraph*}
\end{fmffile}
\end{gathered} \quad &\rightarrow \lambda\left|\vec{\phi}(x)\right|\delta_{a b}\delta_{c N}\;,
\label{eq:FeynRules2PIEA3legVertexSourceJ0main} \\
\begin{gathered}
\begin{fmffile}{Diagrams/1PIEA_V4}
\begin{fmfgraph*}(20,20)
\fmfleft{i0,i1,i2,i3}
\fmfright{o0,o1,o2,o3}
\fmf{phantom,tension=2.0}{i1,i1bis}
\fmf{plain,tension=2.0,foreground=(1,,0,,0)}{i1bis,v1}
\fmf{phantom,tension=2.0}{i2,i2bis}
\fmf{plain,tension=2.0,foreground=(1,,0,,0)}{i2bis,v1}
\fmf{zigzag,label=$x$,tension=0.6,foreground=(0,,0,,1)}{v1,v2}
\fmf{phantom,tension=2.0}{o1bis,o1}
\fmf{plain,tension=2.0,foreground=(1,,0,,0)}{v2,o1bis}
\fmf{phantom,tension=2.0}{o2bis,o2}
\fmf{plain,tension=2.0,foreground=(1,,0,,0)}{v2,o2bis}
\fmflabel{$a$}{i1bis}
\fmflabel{$b$}{i2bis}
\fmflabel{$c$}{o1bis}
\fmflabel{$d$}{o2bis}
\end{fmfgraph*}
\end{fmffile}
\end{gathered} \quad &\rightarrow \lambda\delta_{a b}\delta_{c d}\;.
\label{eq:FeynRuleorig2PIEAV4}
\end{align}
\end{subequations}
Note that the propagator $\boldsymbol{G}$ is more general than the one involved in the original 1PI EA, i.e. $\boldsymbol{G}_{\phi}$. The former will be dressed by non-trivial correlations through the corresponding gap equations while the latter is the mere unperturbed propagator, however possibly dressed by the 1-point correlation function $\vec{\phi}$.

\vspace{0.5cm}

We then examine the original 2PI EA in the zero-dimensional limit. To that end, we evaluate the different contributions to the RHS of~\eqref{eq:2PIEAfinalexpression} in (0+0)-D:
\begin{equation}
\mathrm{STr}\left[\boldsymbol{G}^{-1}_\phi\boldsymbol{G}-\mathbb{I}\right] = \boldsymbol{G}^{-1}_{\phi;ab} \boldsymbol{G}^{ba} - \left.\delta_{a}\right.^a  = \left\{
\begin{array}{lll}
		\displaystyle{\boldsymbol{G}^{-1}_{\phi;11} \boldsymbol{G}_{11} - 1 \quad \mathrm{for} ~ N=1\;,} \\
		\\
		\displaystyle{\boldsymbol{G}^{-1}_{\phi;11} \boldsymbol{G}_{11} + \boldsymbol{G}^{-1}_{\phi;22} \boldsymbol{G}_{22} - 2 \quad \mathrm{for} ~ N=2\;,}
    \end{array}
\right.
\label{eq:Orig2PIEvaluateDiagBegin}
\end{equation}
\begin{equation}
\begin{gathered}
\begin{fmffile}{Diagrams/1PIEA_Hartree}
\begin{fmfgraph}(30,20)
\fmfleft{i}
\fmfright{o}
\fmf{phantom,tension=10}{i,i1}
\fmf{phantom,tension=10}{o,o1}
\fmf{plain,left,tension=0.5,foreground=(1,,0,,0)}{i1,v1,i1}
\fmf{plain,right,tension=0.5,foreground=(1,,0,,0)}{o1,v2,o1}
\fmf{zigzag,foreground=(0,,0,,1)}{v1,v2}
\end{fmfgraph}
\end{fmffile}
\end{gathered} = \lambda \left(\sum_{a=1}^{N} \boldsymbol{G}_{a a} \right)^{2} = \left\{
\begin{array}{lll}
		\displaystyle{\lambda \boldsymbol{G}_{11}^{2} \quad \mathrm{for} ~ N=1\;,} \\
		\\
		\displaystyle{\lambda \left(\boldsymbol{G}_{11}+\boldsymbol{G}_{22}\right)^{2} \quad \mathrm{for} ~ N=2\;,}
    \end{array}
\right.
\end{equation}
\begin{equation}
\begin{gathered}
\begin{fmffile}{Diagrams/1PIEA_Fock}
\begin{fmfgraph}(15,15)
\fmfleft{i}
\fmfright{o}
\fmf{phantom,tension=11}{i,v1}
\fmf{phantom,tension=11}{v2,o}
\fmf{plain,left,tension=0.4,foreground=(1,,0,,0)}{v1,v2,v1}
\fmf{zigzag,foreground=(0,,0,,1)}{v1,v2}
\end{fmfgraph}
\end{fmffile}
\end{gathered} = \lambda \sum_{a,b=1}^{N} \boldsymbol{G}^{2}_{a b} = \left\{
\begin{array}{lll}
		\displaystyle{\lambda \boldsymbol{G}_{11}^{2} \quad \mathrm{for} ~~ N=1\;,} \\
		\\
		\displaystyle{\lambda \left(\boldsymbol{G}_{11}^{2}+2\boldsymbol{G}^{2}_{12}+\boldsymbol{G}^{2}_{22}\right) \quad \mathrm{for} ~~ N=2\;,}
    \end{array}
\right.
\end{equation}
\begin{equation}
\begin{gathered}
\begin{fmffile}{Diagrams/1PIEA_Diag1}
\begin{fmfgraph}(27,15)
\fmfleft{i}
\fmfright{o}
\fmftop{vUp}
\fmfbottom{vDown}
\fmfv{decor.shape=cross,decor.size=3.5thick,foreground=(1,,0,,0)}{v1}
\fmfv{decor.shape=cross,decor.size=3.5thick,foreground=(1,,0,,0)}{v2}
\fmf{phantom,tension=10}{i,i1}
\fmf{phantom,tension=10}{o,o1}
\fmf{phantom,tension=2.2}{vUp,v5}
\fmf{phantom,tension=2.2}{vDown,v6}
\fmf{phantom,tension=0.5}{v3,v4}
\fmf{phantom,tension=10.0}{i1,v1}
\fmf{phantom,tension=10.0}{o1,v2}
\fmf{dashes,tension=2.0,foreground=(0,,0,,1),foreground=(1,,0,,0)}{v1,v3}
\fmf{dots,left=0.4,tension=0.5,foreground=(0,,0,,1)}{v3,v5}
\fmf{plain,left=0.4,tension=0.5,foreground=(1,,0,,0)}{v5,v4}
\fmf{plain,right=0.4,tension=0.5,foreground=(1,,0,,0)}{v3,v6}
\fmf{dots,right=0.4,tension=0.5,foreground=(0,,0,,1)}{v6,v4}
\fmf{dashes,tension=2.0,foreground=(0,,0,,1),foreground=(1,,0,,0)}{v4,v2}
\fmf{plain,tension=0,foreground=(1,,0,,0)}{v5,v6}
\end{fmfgraph}
\end{fmffile}
\end{gathered} = \lambda^{2}\phi^{2}_{N}\sum_{a,b=1}^{N} \boldsymbol{G}_{N a} \boldsymbol{G}_{a b} \boldsymbol{G}_{b N} = \left\{
\begin{array}{lll}
		\displaystyle{\lambda^{2}\phi^{2}_{1} \boldsymbol{G}_{11}^{3} \quad \mathrm{for} ~ N=1\;,} \\
		\\
		\displaystyle{\lambda^{2}\phi^{2}_{2} \left(\boldsymbol{G}^{2}_{12}\left(\boldsymbol{G}_{11}+2\boldsymbol{G}_{22}\right)+\boldsymbol{G}^{3}_{22}\right) \quad \mathrm{for} ~ N=2\;,}
    \end{array}
\right.
\end{equation}
\begin{equation}
\begin{gathered}
\begin{fmffile}{Diagrams/1PIEA_Diag2}
\begin{fmfgraph}(25,20)
\fmfleft{i}
\fmfright{o}
\fmftop{vUp}
\fmfbottom{vDown}
\fmfv{decor.shape=cross,decor.angle=45,decor.size=3.5thick,foreground=(1,,0,,0)}{vUpbis}
\fmfv{decor.shape=cross,decor.angle=45,decor.size=3.5thick,foreground=(1,,0,,0)}{vDownbis}
\fmf{phantom,tension=0.8}{vUp,vUpbis}
\fmf{phantom,tension=0.8}{vDown,vDownbis}
\fmf{dashes,tension=0.5,foreground=(0,,0,,1),foreground=(1,,0,,0)}{v3,vUpbis}
\fmf{phantom,tension=0.5}{v4,vUpbis}
\fmf{phantom,tension=0.5}{v3,vDownbis}
\fmf{dashes,tension=0.5,foreground=(0,,0,,1),foreground=(1,,0,,0)}{v4,vDownbis}
\fmf{phantom,tension=11}{i,v1}
\fmf{phantom,tension=11}{v2,o}
\fmf{plain,left,tension=0.5,foreground=(1,,0,,0)}{v1,v2,v1}
\fmf{dots,tension=1.7,foreground=(0,,0,,1)}{v1,v3}
\fmf{plain,foreground=(1,,0,,0)}{v3,v4}
\fmf{dots,tension=1.7,foreground=(0,,0,,1)}{v4,v2}
\end{fmfgraph}
\end{fmffile}
\end{gathered} = \lambda^{2}\phi^{2}_{N} \boldsymbol{G}_{NN} \sum_{a,b=1}^{N} \boldsymbol{G}^{2}_{a b} = \left\{
\begin{array}{lll}
		\displaystyle{\lambda^{2} \phi^{2}_{1} \boldsymbol{G}_{11}^{3} \quad \mathrm{for} ~ N=1\;,} \\
		\\
		\displaystyle{\lambda^{2} \phi^{2}_{2} \boldsymbol{G}_{22} \left(\boldsymbol{G}_{11}^{2} + 2 \boldsymbol{G}_{12}^{2} + \boldsymbol{G}_{22}^{2} \right) \quad \mathrm{for} ~ N=2\;,}
    \end{array}
\right.
\label{eq:Orig2PIEvaluateDiagEnd}
\end{equation}
where we have notably used the symmetry property of $\boldsymbol{G}$ (i.e. $\boldsymbol{G}_{ab}=\boldsymbol{G}_{ba}$ $\forall a,b$) to simplify our expressions for $N=2$. To further specify~\eqref{eq:Orig2PIEvaluateDiagBegin}, we also recall the expression of $\boldsymbol{G}^{-1}_{\phi}$ introduced for the 1PI EA in section~\ref{sec:original1PIEA}, based on a splitting between Goldstone and Higgs modes:
\begin{equation}
\boldsymbol{G}^{-1}_{\phi,ab} = \mathfrak{G}^{-1}_{\phi;\mathfrak{g}} \left(1 - \delta_{a N}\right)\left(1 - \delta_{b N}\right) + \boldsymbol{G}^{-1}_{\phi;N N} \delta_{a N} \delta_{b N} \;,
\end{equation}
with $\mathfrak{G}^{-1}_{\phi;\mathfrak{g}} = m^{2} + \lambda \vec{\phi}^{2}/6$ and $\boldsymbol{G}^{-1}_{\phi;NN}= m^{2} + \lambda \vec{\phi}^{2}/2$ (and $\vec{\phi}^{2}=\phi_{N}^{2}$ as usual). With the help of~\eqref{eq:Orig2PIEvaluateDiagBegin} to~\eqref{eq:Orig2PIEvaluateDiagEnd}, we show that expression~\eqref{eq:2PIEAfinalexpression} of the original 2PI EA becomes in (0+0)\nobreakdash-D:
\begin{itemize}
\item For $N=1$:
\begin{equation}
\begin{split}
\Gamma^{(\mathrm{2PI})}\Big(\vec{\phi},\boldsymbol{G}\Big) = & \ S\Big(\vec{\phi}\Big) + \hbar \Bigg[ -\frac{1}{2}\ln\big(2\pi\boldsymbol{G}_{11}\big) + \frac{1}{2} \left(m^{2}+\frac{\lambda}{2}\phi^{2}_{1}\right)\boldsymbol{G}_{11} - \frac{1}{2} \Bigg] \\
& + \hbar^{2}\Bigg[ \frac{\lambda}{8}\boldsymbol{G}^{2}_{11} - \frac{\lambda^{2}\phi^{2}_{1}}{12} \boldsymbol{G}^{3}_{11} \Bigg] \\
& + \mathcal{O}\big(\hbar^{3}\big)\;.
\end{split}
\label{eq:2PIEAfinalexpressionN10DON}
\end{equation}
\item For $N=2$:
\begin{equation}
\begin{split}
\Gamma^{(\mathrm{2PI})}\Big(\vec{\phi},\boldsymbol{G}\Big) = & \ S\Big(\vec{\phi}\Big) \\
& + \hbar \Bigg[ -\frac{1}{2}\left(\ln\big(2\pi\boldsymbol{G}_{11}\big) + \ln\big(2\pi\boldsymbol{G}_{22}\big) \right) + \frac{1}{2} \bigg(\left(m^{2}+\frac{\lambda}{6}\phi^{2}_{2}\right)\boldsymbol{G}_{11} \\
& \hspace{1.0cm} + \bigg(m^{2} + \frac{\lambda}{2}\phi^{2}_{2}\bigg)\boldsymbol{G}_{22}\bigg) - 1 \Bigg] \\
& + \hbar^{2}\Bigg[ \frac{\lambda}{72} \Big( \boldsymbol{G}_{11} \left(6\boldsymbol{G}_{22}-4\boldsymbol{G}^{2}_{12}\lambda\phi^{2}_{2}\right) + 3 \boldsymbol{G}^{2}_{22} \left(3-2\boldsymbol{G}_{22}\lambda\phi^{2}_{2}\right) \\
& \hspace{0.9cm} + \boldsymbol{G}^{2}_{11} \big(9 - 2\boldsymbol{G}_{22}\lambda\phi^{2}_{2}\big) - 12 \boldsymbol{G}^{2}_{12} \left( -1 + \boldsymbol{G}_{22} \lambda \phi^{2}_{2} \right) \Big) \Bigg] \\
& + \mathcal{O}\big(\hbar^{3}\big)\;.
\end{split}
\label{eq:2PIEAfinalexpressionN20DON}
\end{equation}
\end{itemize}
We stress that $\phi_{N}$ and $\boldsymbol{G}$ are independent. However, the components of $\boldsymbol{G}$ are constrained by its symmetry (i.e. $\boldsymbol{G}_{a b}=\boldsymbol{G}_{b a}$), so that the 2PI EA $\Gamma^{(\mathrm{2PI})}\big(\vec{\phi},\boldsymbol{G}\big)$ depends on $(N^{2}+N)/2+1$ independent variational variables forming the set $\left\lbrace \left. \phi_{N}, \boldsymbol{G}_{a b} \right| a, b \in \mathbb{N}^{*} , a \leq b \leq N \right\rbrace$. This enables us to derive the following gap equations for $\Gamma^{(\mathrm{2PI})}\big(\vec{\phi},\boldsymbol{G}\big)$:
\begin{itemize}
\item For $N=1$:
\begin{equation}
0 = \left.\frac{\partial \Gamma^{(\mathrm{2PI})}\big(\vec{\phi},\boldsymbol{G}\big)}{\partial \phi_{1}}\right|_{\vec{\phi}=\vec{\overline{\phi}} \atop \boldsymbol{G}=\overline{\boldsymbol{G}}} = m^{2} \overline{\phi}_{1} + \frac{\lambda}{6} \overline{\phi}^{3}_{1} + \hbar\left(\frac{1}{2} \lambda \overline{\phi}_{1} \overline{\boldsymbol{G}}_{11}\right) - \hbar^{2} \left(\frac{1}{6}\lambda^{2}\overline{\phi}_{1}\overline{\boldsymbol{G}}^{3}_{11}\right) + \mathcal{O}\big(\hbar^{3}\big)\;,
\label{eq:pure2PIEAGapEquationphiNN10DON}
\end{equation}
\begin{equation}
\begin{split}
0 = \left.\frac{\partial \Gamma^{(\mathrm{2PI})}\big(\vec{\phi},\boldsymbol{G}\big)}{\partial \boldsymbol{G}_{11}}\right|_{\vec{\phi}=\vec{\overline{\phi}} \atop \boldsymbol{G}=\overline{\boldsymbol{G}}} = & \ \frac{\hbar}{4} \overline{\boldsymbol{G}}^{-1}_{11} \left(-2 + 2 m^{2} \overline{\boldsymbol{G}}_{11} + \lambda \overline{\phi}^{2}_{1} \overline{\boldsymbol{G}}_{11}\right) + \frac{\hbar^{2}}{4} \left(\lambda \overline{\boldsymbol{G}}_{11}-\lambda^{2}\overline{\phi}^{2}_{1}\overline{\boldsymbol{G}}_{11}^{2}\right) \\
& + \mathcal{O}\big(\hbar^{3}\big) \;.
\end{split}
\label{eq:pure2PIEAGapEquationG11N10DON}
\end{equation}

\item For $N=2$:
\begin{equation}
\begin{split}
0 = \left.\frac{\partial \Gamma^{(\mathrm{2PI})}\big(\vec{\phi},\boldsymbol{G}\big)}{\partial \phi_{2}}\right|_{\vec{\phi}=\vec{\overline{\phi}} \atop \boldsymbol{G}=\overline{\boldsymbol{G}}} = & \ m^{2} \overline{\phi}_{2} + \frac{\lambda}{6} \overline{\phi}^{3}_{2} + \hbar\Bigg[ \frac{1}{6}\lambda\overline{\phi}_{2}\left(\overline{\boldsymbol{G}}_{11}+3\overline{\boldsymbol{G}}_{22}\right) \Bigg] \\
& - \hbar^{2} \Bigg[ \frac{1}{18} \lambda^{2} \overline{\phi}_{2} \left(2\overline{\boldsymbol{G}}_{11}\overline{\boldsymbol{G}}^{2}_{12}+\overline{\boldsymbol{G}}^{2}_{11}\overline{\boldsymbol{G}}_{22}+6\overline{\boldsymbol{G}}^{2}_{12}\overline{\boldsymbol{G}}_{22}+3\overline{\boldsymbol{G}}^{3}_{22}\right) \Bigg] \\
& +\mathcal{O}\big(\hbar^{3}\big)\;,
\end{split}
\label{eq:pure2PIEAGapEquationphiNN20DON}
\end{equation}
\begin{equation}
\begin{split}
0 = \left.\frac{\partial \Gamma^{(\mathrm{2PI})}\big(\vec{\phi},\boldsymbol{G}\big)}{\partial \boldsymbol{G}_{11}}\right|_{\vec{\phi}=\vec{\overline{\phi}} \atop \boldsymbol{G}=\overline{\boldsymbol{G}}} = & \ \hbar\Bigg[ -\frac{1}{2}\overline{\boldsymbol{G}}^{-1}_{11} + \frac{1}{2} \left(m^2 + \frac{\lambda}{6} \overline{\phi}_{2}^2\right) \Bigg] \\
& - \hbar^{2} \Bigg[ \frac{1}{36} \lambda \left(-3 \overline{\boldsymbol{G}}_{22} + 2 \overline{\boldsymbol{G}}_{12}^2 \lambda \overline{\phi}_{2}^2 + \overline{\boldsymbol{G}}_{11} \left(-9 + 2 \overline{\boldsymbol{G}}_{22} \lambda \overline{\phi}_{2}^2\right)\right) \Bigg] \\
& +\mathcal{O}\big(\hbar^{3}\big)\;,
\end{split}
\label{eq:pure2PIEAGapEquationG11N20DON}
\end{equation}
\begin{equation}
\begin{split}
0 = \left.\frac{\partial \Gamma^{(\mathrm{2PI})}\big(\vec{\phi},\boldsymbol{G}\big)}{\partial \boldsymbol{G}_{22}}\right|_{\vec{\phi}=\vec{\overline{\phi}} \atop \boldsymbol{G}=\overline{\boldsymbol{G}}} = & \ \hbar\Bigg[ -\frac{1}{2}\overline{\boldsymbol{G}}^{-1}_{22} + \frac{1}{2} \left(m^2 + \frac{\lambda}{2} \overline{\phi}_{2}^2\right) \Bigg] \\
& - \hbar^{2} \Bigg[ \frac{1}{36} \lambda \left(-3 \overline{\boldsymbol{G}}_{11} - 9 \overline{\boldsymbol{G}}_{22} + \overline{\boldsymbol{G}}_{11}^2 \lambda \overline{\phi}_{2}^2 +  6 \overline{\boldsymbol{G}}_{12}^2 \lambda \overline{\phi}_{2}^2 + 9 \overline{\boldsymbol{G}}_{22}^2 \lambda \overline{\phi}_{2}^2\right) \Bigg] \\
& +\mathcal{O}\big(\hbar^{3}\big)\;,
\end{split}
\label{eq:pure2PIEAGapEquationG22N20DON}
\end{equation}
\begin{equation}
0 = \left.\frac{\partial \Gamma^{(\mathrm{2PI})}\big(\vec{\phi},\boldsymbol{G}\big)}{\partial \boldsymbol{G}_{12}}\right|_{\vec{\phi}=\vec{\overline{\phi}} \atop \boldsymbol{G}=\overline{\boldsymbol{G}}} = - \hbar^{2} \Bigg[ \frac{1}{9} \overline{\boldsymbol{G}}_{12} \lambda \left(-3 + \overline{\boldsymbol{G}}_{11} \lambda \overline{\phi}_{2}^2 + 3 \overline{\boldsymbol{G}}_{22} \lambda \overline{\phi}_{2}^2\right) \Bigg] +\mathcal{O}\big(\hbar^{3}\big)\;.
\label{eq:pure2PIEAGapEquationG12N20DON}
\end{equation}

\end{itemize}
The gs energy and density can be obtained from the solutions of the latter gap equations via the relations:
\begin{equation}
E^\text{2PI EA;orig}_\text{gs} = \frac{1}{\hbar} \Gamma^{(\mathrm{2PI})}\Big(\vec{\phi}=\vec{\overline{\phi}},\boldsymbol{G}=\overline{\boldsymbol{G}}\Big) \;,
\label{eq:2PIorigE}
\end{equation}
\begin{equation}
\rho^\text{2PI EA;orig}_\text{gs} = \frac{1}{N} \left(\hbar\mathrm{Tr}_{a}\big(\overline{\boldsymbol{G}}\big) + \vec{\overline{\phi}}\cdot\vec{\overline{\phi}}\right) \;.
\label{eq:2PIorigRho}
\end{equation}

\begin{figure}[!htb]
\captionsetup[subfigure]{labelformat=empty}
  \begin{center}
    \subfloat[]{
      \includegraphics[width=0.85\linewidth]{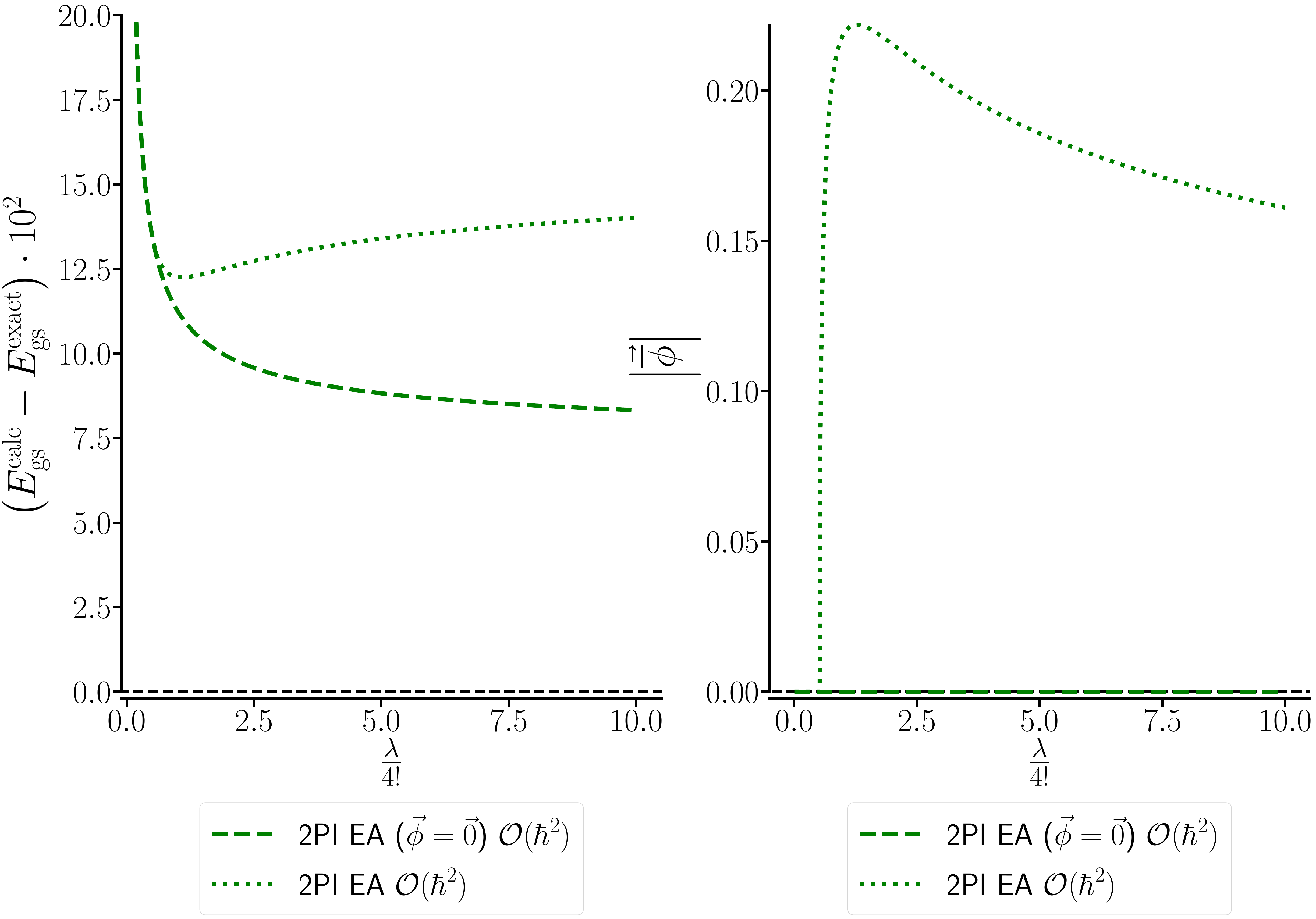}
                         }
    \caption{Two different solutions of the gap equations of the 2PI EA $\Gamma^{(\mathrm{2PI})}\big(\vec{\phi},\boldsymbol{G}\big)$ at its first non-trivial order for the gs energy $E_{\mathrm{gs}}$ or the 1-point correlation function $\vec{\overline{\phi}}$ at $\hbar=1$, $m^{2}=-1$ and $N=2$ ($\mathcal{R}e(\lambda)>0$ and $\mathcal{I}m(\lambda)=0$). More precisely, the left-hand plot shows the difference between the gs energy $E_{\mathrm{gs}}^{\mathrm{calc}}$ calculated from each of these solutions on the one hand and the corresponding exact solution $E_{\mathrm{gs}}^{\mathrm{exact}}$ on the other hand. See also the caption of fig.~\ref{fig:1PIEA} for the meaning of the indication ``$\mathcal{O}\big(\hbar^{n}\big)$'' for the results obtained from $\hbar$-expanded EAs.}
    \label{fig:2PIEAzeroVsNonzerovev}
  \end{center}
\end{figure}

Fig.~\ref{fig:2PIEAzeroVsNonzerovev} shows two different solutions obtained by solving the gap equations \eqref{eq:pure2PIEAGapEquationphiNN20DON} to \eqref{eq:pure2PIEAGapEquationG12N20DON} up to order $\mathcal{O}\big(\hbar^2\big)$: only one of these two solutions exhibits a spontaneous breakdown of the $O(N)$ symmetry, with a finite 1-point correlation function $\vec{\overline{\phi}}$ in the non-perturbative regime\footnote{It is acknowledged that solutions of gap equations for this 2PI EA have a tendency to violate Ward identities associated with global symmetries. This is synonymous to violations of Goldstone's theorem~\cite{nam60,gol61,gol62} and appearance of massive Goldstone bosons if SSB occurs. A recent extension of the 2PI EA formalism, known as symmetry-improved 2PI (SI2PI) EA or symmetry-improved CJT EA~\cite{pil13}, has been constructed to overcome the drawbacks of previous approaches that already address this issue~\cite{pet99,len00,nem00,van02,baa03,iva05,iva05bis,see12,mar13,gra13}. The SI2PI EA formalism consists in constraining the extremization of the standard 2PI EA by imposing the Ward identities of the model under consideration as a constraint via the method of Lagrange multipliers, thus resulting in a modified set of gap equations whose solutions can not violate Goldstone's theorem by construction. The SI2PI EA has already been successfully applied to several models~\cite{pil13,mao14,bro15bis,pil15,pil16}, especially to an $O(2)$-symmetric $\varphi^{4}$-theory~\cite{pil13,pil17}.}. It is however the solution without SSB (at least over the whole range of values for $\lambda/4!$ shown in fig.~\ref{fig:2PIEAzeroVsNonzerovev}) that minimizes the gs energy and can thus be coined as physical solution. This illustrates that, for the $O(N)$ model under consideration whose exact solution does not break the $O(N)$ symmetry, we can safely consider the 2PI EA at $\vec{\phi}=\vec{0}$ defined as $\Gamma^{(\mathrm{2PI})}[\boldsymbol{G}]\equiv\Gamma^{(\mathrm{2PI})}\big[\vec{\phi}=\vec{0},\boldsymbol{G}\big]$, without any loss of accuracy for a given truncation order. This could have been anticipated from our previous results of fig.~\ref{fig:1PIEA} which showed that the 1-point correlation function $\vec{\phi}$ is not capable of capturing non-trivial physics in the framework of the 1PI EA either.

\paragraph{Original 2PI EA with vanishing 1-point correlation function:}

Hence, we will now focus on the 2PI EA at $\vec{\phi}=\vec{0}$, i.e. $\Gamma^{(\mathrm{2PI})}[\boldsymbol{G}]$. This restriction imposes notably that all diagrams involving vertex~\eqref{eq:FeynRules2PIEA3legVertexSourceJ0main} vanish, thus rendering the enhanced complexity of the underpinning diagrammatic expansion with increasing truncation orders more tractable. We will thus push the present investigation up to its third non-trivial order (i.e. up to order $\mathcal{O}(\hbar^{4})$ for the EA). This will be done by determining every 2PI diagram contributing to $W^\text{LE;orig}\big[\vec{J}=\vec{0},\boldsymbol{K}\big]$ up to order $\mathcal{O}\big(\hbar^{4}\big)$ (the source $\vec{J}$ is no longer useful if we assume that $\vec{\phi}=\vec{0}$ since every non-vanishing correlation function can be expressed by differentiating the Schwinger functional with respect to $\boldsymbol{K}$ in this case). Note that all these diagrams and the corresponding multiplicities can be found in appendix~\ref{sec:DiagLEO}. This procedure leads to:
\begin{equation}
\begin{split}
\Gamma^{(\mathrm{2PI})}[\boldsymbol{G}] = & -\frac{\hbar}{2}\mathrm{STr}\left[\ln\big(\boldsymbol{G}\big)\right] + \frac{\hbar}{2}\mathrm{STr}\left[\boldsymbol{G}^{-1}_{0}\boldsymbol{G}-\mathbb{I}\right] \\
& + \hbar^{2} \left(\rule{0cm}{1.2cm}\right. \frac{1}{24} \hspace{0.08cm} \begin{gathered}
\begin{fmffile}{Diagrams/1PIEA_Hartree}
\begin{fmfgraph}(30,20)
\fmfleft{i}
\fmfright{o}
\fmf{phantom,tension=10}{i,i1}
\fmf{phantom,tension=10}{o,o1}
\fmf{plain,left,tension=0.5,foreground=(1,,0,,0)}{i1,v1,i1}
\fmf{plain,right,tension=0.5,foreground=(1,,0,,0)}{o1,v2,o1}
\fmf{zigzag,foreground=(0,,0,,1)}{v1,v2}
\end{fmfgraph}
\end{fmffile}
\end{gathered}
+\frac{1}{12}\begin{gathered}
\begin{fmffile}{Diagrams/1PIEA_Fock}
\begin{fmfgraph}(15,15)
\fmfleft{i}
\fmfright{o}
\fmf{phantom,tension=11}{i,v1}
\fmf{phantom,tension=11}{v2,o}
\fmf{plain,left,tension=0.4,foreground=(1,,0,,0)}{v1,v2,v1}
\fmf{zigzag,foreground=(0,,0,,1)}{v1,v2}
\end{fmfgraph}
\end{fmffile}
\end{gathered} \left.\rule{0cm}{1.2cm}\right) \\
& - \hbar^{3} \left(\rule{0cm}{1.2cm}\right. \frac{1}{72} \hspace{0.38cm} \begin{gathered}
\begin{fmffile}{Diagrams/2PIEAzerovev_Diag1}
\begin{fmfgraph}(12,12)
\fmfleft{i0,i1}
\fmfright{o0,o1}
\fmftop{v1,vUp,v2}
\fmfbottom{v3,vDown,v4}
\fmf{phantom,tension=20}{i0,v1}
\fmf{phantom,tension=20}{i1,v3}
\fmf{phantom,tension=20}{o0,v2}
\fmf{phantom,tension=20}{o1,v4}
\fmf{plain,left=0.4,tension=0.5,foreground=(1,,0,,0)}{v3,v1}
\fmf{phantom,left=0.1,tension=0.5}{v1,vUp}
\fmf{phantom,left=0.1,tension=0.5}{vUp,v2}
\fmf{plain,left=0.4,tension=0.0,foreground=(1,,0,,0)}{v1,v2}
\fmf{plain,left=0.4,tension=0.5,foreground=(1,,0,,0)}{v2,v4}
\fmf{phantom,left=0.1,tension=0.5}{v4,vDown}
\fmf{phantom,left=0.1,tension=0.5}{vDown,v3}
\fmf{plain,left=0.4,tension=0.0,foreground=(1,,0,,0)}{v4,v3}
\fmf{zigzag,tension=0.5,foreground=(0,,0,,1)}{v1,v4}
\fmf{zigzag,tension=0.5,foreground=(0,,0,,1)}{v2,v3}
\end{fmfgraph}
\end{fmffile}
\end{gathered} \hspace{0.3cm} + \frac{1}{144} \hspace{0.38cm} \begin{gathered}
\begin{fmffile}{Diagrams/2PIEAzerovev_Diag2}
\begin{fmfgraph}(12,12)
\fmfleft{i0,i1}
\fmfright{o0,o1}
\fmftop{v1,vUp,v2}
\fmfbottom{v3,vDown,v4}
\fmf{phantom,tension=20}{i0,v1}
\fmf{phantom,tension=20}{i1,v3}
\fmf{phantom,tension=20}{o0,v2}
\fmf{phantom,tension=20}{o1,v4}
\fmf{plain,left=0.4,tension=0.5,foreground=(1,,0,,0)}{v3,v1}
\fmf{phantom,left=0.1,tension=0.5}{v1,vUp}
\fmf{phantom,left=0.1,tension=0.5}{vUp,v2}
\fmf{zigzag,left=0.4,tension=0.0,foreground=(0,,0,,1)}{v1,v2}
\fmf{plain,left=0.4,tension=0.5,foreground=(1,,0,,0)}{v2,v4}
\fmf{phantom,left=0.1,tension=0.5}{v4,vDown}
\fmf{phantom,left=0.1,tension=0.5}{vDown,v3}
\fmf{zigzag,left=0.4,tension=0.0,foreground=(0,,0,,1)}{v4,v3}
\fmf{plain,left=0.4,tension=0.5,foreground=(1,,0,,0)}{v1,v3}
\fmf{plain,right=0.4,tension=0.5,foreground=(1,,0,,0)}{v2,v4}
\end{fmfgraph}
\end{fmffile}
\end{gathered} \hspace{0.35cm} \left.\rule{0cm}{1.2cm}\right) \\
& + \hbar^{4} \left(\rule{0cm}{1.2cm}\right. \frac{1}{324} \hspace{0.2cm} \begin{gathered}\begin{fmffile}{Diagrams/2PIEAzerovev_Diag3}
\begin{fmfgraph}(16,16)
\fmfleft{i}
\fmfright{o}
\fmftop{vUpLeft,vUp,vUpRight}
\fmfbottom{vDownLeft,vDown,vDownRight}
\fmf{phantom,tension=1}{i,v1}
\fmf{phantom,tension=1}{v2,o}
\fmf{phantom,tension=14.0}{v3,vUpLeft}
\fmf{phantom,tension=2.0}{v3,vUpRight}
\fmf{phantom,tension=4.0}{v3,i}
\fmf{phantom,tension=2.0}{v4,vUpLeft}
\fmf{phantom,tension=14.0}{v4,vUpRight}
\fmf{phantom,tension=4.0}{v4,o}
\fmf{phantom,tension=14.0}{v5,vDownLeft}
\fmf{phantom,tension=2.0}{v5,vDownRight}
\fmf{phantom,tension=4.0}{v5,i}
\fmf{phantom,tension=2.0}{v6,vDownLeft}
\fmf{phantom,tension=14.0}{v6,vDownRight}
\fmf{phantom,tension=4.0}{v6,o}
\fmf{zigzag,tension=0,foreground=(0,,0,,1)}{v1,v2}
\fmf{zigzag,tension=0.6,foreground=(0,,0,,1)}{v3,v6}
\fmf{zigzag,tension=0.6,foreground=(0,,0,,1)}{v5,v4}
\fmf{plain,left=0.18,tension=0,foreground=(1,,0,,0)}{v1,v3}
\fmf{plain,left=0.42,tension=0,foreground=(1,,0,,0)}{v3,v4}
\fmf{plain,left=0.18,tension=0,foreground=(1,,0,,0)}{v4,v2}
\fmf{plain,left=0.18,tension=0,foreground=(1,,0,,0)}{v2,v6}
\fmf{plain,left=0.42,tension=0,foreground=(1,,0,,0)}{v6,v5}
\fmf{plain,left=0.18,tension=0,foreground=(1,,0,,0)}{v5,v1}
\end{fmfgraph}
\end{fmffile}
\end{gathered} \hspace{0.15cm} + \frac{1}{108} \hspace{0.5cm} \begin{gathered}
\begin{fmffile}{Diagrams/2PIEAzerovev_Diag4}
\begin{fmfgraph}(12.5,12.5)
\fmfleft{i0,i1}
\fmfright{o0,o1}
\fmftop{v1,vUp,v2}
\fmfbottom{v3,vDown,v4}
\fmf{phantom,tension=20}{i0,v1}
\fmf{phantom,tension=20}{i1,v3}
\fmf{phantom,tension=20}{o0,v2}
\fmf{phantom,tension=20}{o1,v4}
\fmf{phantom,tension=0.005}{v5,v6}
\fmf{zigzag,left=0.4,tension=0,foreground=(0,,0,,1)}{v3,v1}
\fmf{phantom,left=0.1,tension=0}{v1,vUp}
\fmf{phantom,left=0.1,tension=0}{vUp,v2}
\fmf{plain,left=0.25,tension=0,foreground=(1,,0,,0)}{v1,v2}
\fmf{zigzag,left=0.4,tension=0,foreground=(0,,0,,1)}{v2,v4}
\fmf{phantom,left=0.1,tension=0}{v4,vDown}
\fmf{phantom,left=0.1,tension=0}{vDown,v3}
\fmf{plain,right=0.25,tension=0,foreground=(1,,0,,0)}{v3,v4}
\fmf{plain,left=0.2,tension=0.01,foreground=(1,,0,,0)}{v1,v5}
\fmf{plain,left=0.2,tension=0.01,foreground=(1,,0,,0)}{v5,v3}
\fmf{plain,right=0.2,tension=0.01,foreground=(1,,0,,0)}{v2,v6}
\fmf{plain,right=0.2,tension=0.01,foreground=(1,,0,,0)}{v6,v4}
\fmf{zigzag,tension=0,foreground=(0,,0,,1)}{v5,v6}
\end{fmfgraph}
\end{fmffile}
\end{gathered} \hspace{0.5cm} + \frac{1}{324} \hspace{0.4cm} \begin{gathered}
\begin{fmffile}{Diagrams/2PIEAzerovev_Diag5}
\begin{fmfgraph}(12.5,12.5)
\fmfleft{i0,i1}
\fmfright{o0,o1}
\fmftop{v1,vUp,v2}
\fmfbottom{v3,vDown,v4}
\fmf{phantom,tension=20}{i0,v1}
\fmf{phantom,tension=20}{i1,v3}
\fmf{phantom,tension=20}{o0,v2}
\fmf{phantom,tension=20}{o1,v4}
\fmf{phantom,tension=0.005}{v5,v6}
\fmf{plain,left=0.4,tension=0,foreground=(1,,0,,0)}{v3,v1}
\fmf{phantom,left=0.1,tension=0}{v1,vUp}
\fmf{phantom,left=0.1,tension=0}{vUp,v2}
\fmf{zigzag,left=0.25,tension=0,foreground=(0,,0,,1)}{v1,v2}
\fmf{plain,left=0.4,tension=0,foreground=(1,,0,,0)}{v2,v4}
\fmf{phantom,left=0.1,tension=0}{v4,vDown}
\fmf{phantom,left=0.1,tension=0}{vDown,v3}
\fmf{zigzag,right=0.25,tension=0,foreground=(0,,0,,1)}{v3,v4}
\fmf{plain,left=0.2,tension=0.01,foreground=(1,,0,,0)}{v1,v5}
\fmf{plain,left=0.2,tension=0.01,foreground=(1,,0,,0)}{v5,v3}
\fmf{plain,right=0.2,tension=0.01,foreground=(1,,0,,0)}{v2,v6}
\fmf{plain,right=0.2,tension=0.01,foreground=(1,,0,,0)}{v6,v4}
\fmf{zigzag,tension=0,foreground=(0,,0,,1)}{v5,v6}
\end{fmfgraph}
\end{fmffile}
\end{gathered} \hspace{0.35cm} + \frac{1}{216} \hspace{-0.35cm} \begin{gathered}
\begin{fmffile}{Diagrams/2PIEAzerovev_Diag6}
\begin{fmfgraph}(30,15)
\fmfleft{i0,i,i1}
\fmfright{o0,o,o1}
\fmftop{v1b,vUp,v2b}
\fmfbottom{v3b,vDown,v4b}
\fmf{phantom,tension=20}{i0,v1b}
\fmf{phantom,tension=20}{i1,v3b}
\fmf{phantom,tension=20}{o0,v2b}
\fmf{phantom,tension=20}{o1,v4b}
\fmf{phantom,tension=0.511}{i,v7}
\fmf{phantom,tension=0.11}{o,v7}
\fmf{phantom,tension=0.1}{v1,v1b}
\fmf{phantom,tension=0.1}{v2,v2b}
\fmf{phantom,tension=0.1}{v3,v3b}
\fmf{phantom,tension=0.1}{v4,v4b}
\fmf{phantom,tension=0.005}{v5,v6}
\fmf{phantom,left=0.1,tension=0.1}{v1,vUp}
\fmf{phantom,left=0.1,tension=0.1}{vUp,v2}
\fmf{zigzag,left=0.15,tension=0,foreground=(0,,0,,1)}{v1,v2}
\fmf{plain,left=0.4,tension=0,foreground=(1,,0,,0)}{v2,v4}
\fmf{plain,right=0.4,tension=0,foreground=(1,,0,,0)}{v2,v4}
\fmf{phantom,left=0.1,tension=0.1}{v4,vDown}
\fmf{phantom,left=0.1,tension=0.1}{vDown,v3}
\fmf{zigzag,left=0.15,tension=0,foreground=(0,,0,,1)}{v4,v3}
\fmf{plain,left=0.2,tension=0.01,foreground=(1,,0,,0)}{v1,v5}
\fmf{plain,left=0.2,tension=0.01,foreground=(1,,0,,0)}{v5,v3}
\fmf{plain,right=0.2,tension=0,foreground=(1,,0,,0)}{v1,v7}
\fmf{plain,right=0.2,tension=0,foreground=(1,,0,,0)}{v7,v3}
\fmf{phantom,right=0.2,tension=0.01}{v2,v6}
\fmf{phantom,right=0.2,tension=0.01}{v6,v4}
\fmf{zigzag,tension=0,foreground=(0,,0,,1)}{v5,v7}
\end{fmfgraph}
\end{fmffile}
\end{gathered} \\
& \hspace{1.0cm} + \frac{1}{1296} \hspace{-0.32cm} \begin{gathered}
\begin{fmffile}{Diagrams/2PIEAzerovev_Diag7}
\begin{fmfgraph}(30,14)
\fmfleft{i0,i,i1}
\fmfright{o0,o,o1}
\fmftop{v1b,vUp,v2b}
\fmfbottom{v3b,vDown,v4b}
\fmf{phantom,tension=5}{vUp,v5}
\fmf{phantom,tension=1}{v1b,v5}
\fmf{phantom,tension=5}{vUp,v6}
\fmf{phantom,tension=1}{v2b,v6}
\fmf{phantom,tension=20}{i0,v1b}
\fmf{phantom,tension=20}{i1,v3b}
\fmf{phantom,tension=20}{o0,v2b}
\fmf{phantom,tension=20}{o1,v4b}
\fmf{phantom,tension=0.1}{v1,v1b}
\fmf{phantom,tension=0.1}{v2,v2b}
\fmf{phantom,tension=0.1}{v3,v3b}
\fmf{phantom,tension=0.1}{v4,v4b}
\fmf{phantom,tension=0.005}{v5,v6}
\fmf{phantom,left=0.1,tension=0.1}{v1,vUp}
\fmf{phantom,left=0.1,tension=0.1}{vUp,v2}
\fmf{plain,left=0.4,tension=0.005,foreground=(1,,0,,0)}{v2,v4}
\fmf{plain,right=0.4,tension=0.005,foreground=(1,,0,,0)}{v2,v4}
\fmf{plain,left=0.4,tension=0.005,foreground=(1,,0,,0)}{v1,v3}
\fmf{plain,right=0.4,tension=0.005,foreground=(1,,0,,0)}{v1,v3}
\fmf{phantom,left=0.1,tension=0.1}{v4,vDown}
\fmf{phantom,left=0.1,tension=0.1}{vDown,v3}
\fmf{zigzag,left=0.05,tension=0,foreground=(0,,0,,1)}{v1,v5}
\fmf{plain,left,tension=0,foreground=(1,,0,,0)}{v5,v6}
\fmf{plain,right,tension=0,foreground=(1,,0,,0)}{v5,v6}
\fmf{zigzag,left=0.05,tension=0,foreground=(0,,0,,1)}{v6,v2}
\fmf{zigzag,left=0.15,tension=0,foreground=(0,,0,,1)}{v4,v3}
\end{fmfgraph}
\end{fmffile}
\end{gathered} \hspace{-0.35cm} \left.\rule{0cm}{1.2cm}\right) \\
& + \mathcal{O}\big(\hbar^{5}\big)\;,
\end{split}
\label{eq:2PIEAzerovevfinalexpression}
\end{equation}
where the Feynman rules are still given by~\eqref{eq:FeynRuleorig2PIEAG} and~\eqref{eq:FeynRuleorig2PIEAV4} and $\boldsymbol{G}_{0}$ is the bare propagator (defined by~\eqref{eq:1PIEAlambdaExppDefG0}) which coincides with $\boldsymbol{G}_{\phi}$ (defined by~\eqref{eq:DefinitionG1PIEAhbarExpansionbis}) at $\vec{\phi}=\vec{0}$.

\vspace{0.5cm}

As a next step, we consider the zero-dimensional situation. First of all, since SSB can not occur in the present approach because of the constraint $\vec{\phi}=\vec{0}$, $\boldsymbol{G}$ must also be invariant under $O(N)$ transformations, which means that its structure in color space must be trivial, i.e. $\boldsymbol{G}_{a b}= G \ \delta_{a b}$ $\forall a,b$. With this in mind, we evaluate in (0+0)-D the following supertrace:
\begin{equation}
\mathrm{STr}\left[\boldsymbol{G}^{-1}_{0}\boldsymbol{G}-\mathbb{I}\right] = N\left(m^{2}G - 1\right)\;,
\label{eq:pure2PIEAzeroVevSTraceTerm0DON}
\end{equation}
and the diagrams involved in~\eqref{eq:2PIEAzerovevfinalexpression}:
\begin{equation}
\begin{gathered}
\begin{fmffile}{Diagrams/1PIEA_Hartree}
\begin{fmfgraph}(30,20)
\fmfleft{i}
\fmfright{o}
\fmf{phantom,tension=10}{i,i1}
\fmf{phantom,tension=10}{o,o1}
\fmf{plain,left,tension=0.5,foreground=(1,,0,,0)}{i1,v1,i1}
\fmf{plain,right,tension=0.5,foreground=(1,,0,,0)}{o1,v2,o1}
\fmf{zigzag,foreground=(0,,0,,1)}{v1,v2}
\end{fmfgraph}
\end{fmffile}
\end{gathered} = \lambda N^{2} G^{2}\;,
\label{eq:pure2PIEAzeroVevDiag10DON}
\end{equation}

\vspace{-0.7cm}

\begin{equation}
\begin{gathered}
\begin{fmffile}{Diagrams/1PIEA_Fock}
\begin{fmfgraph}(15,15)
\fmfleft{i}
\fmfright{o}
\fmf{phantom,tension=11}{i,v1}
\fmf{phantom,tension=11}{v2,o}
\fmf{plain,left,tension=0.4,foreground=(1,,0,,0)}{v1,v2,v1}
\fmf{zigzag,foreground=(0,,0,,1)}{v1,v2}
\end{fmfgraph}
\end{fmffile}
\end{gathered} = \lambda N G^{2}\;,
\label{eq:pure2PIEAzeroVevDiag20DON}
\end{equation}
\begin{equation}
\begin{gathered}
\begin{fmffile}{Diagrams/2PIEAzerovev_Diag1}
\begin{fmfgraph}(12,12)
\fmfleft{i0,i1}
\fmfright{o0,o1}
\fmftop{v1,vUp,v2}
\fmfbottom{v3,vDown,v4}
\fmf{phantom,tension=20}{i0,v1}
\fmf{phantom,tension=20}{i1,v3}
\fmf{phantom,tension=20}{o0,v2}
\fmf{phantom,tension=20}{o1,v4}
\fmf{plain,left=0.4,tension=0.5,foreground=(1,,0,,0)}{v3,v1}
\fmf{phantom,left=0.1,tension=0.5}{v1,vUp}
\fmf{phantom,left=0.1,tension=0.5}{vUp,v2}
\fmf{plain,left=0.4,tension=0.0,foreground=(1,,0,,0)}{v1,v2}
\fmf{plain,left=0.4,tension=0.5,foreground=(1,,0,,0)}{v2,v4}
\fmf{phantom,left=0.1,tension=0.5}{v4,vDown}
\fmf{phantom,left=0.1,tension=0.5}{vDown,v3}
\fmf{plain,left=0.4,tension=0.0,foreground=(1,,0,,0)}{v4,v3}
\fmf{zigzag,tension=0.5,foreground=(0,,0,,1)}{v1,v4}
\fmf{zigzag,tension=0.5,foreground=(0,,0,,1)}{v2,v3}
\end{fmfgraph}
\end{fmffile}
\end{gathered} \hspace{0.3cm} = \lambda^{2} N G^{4}\;,
\label{eq:pure2PIEAzeroVevDiag30DON}
\end{equation}

\vspace{0.35cm}

\begin{equation}
\begin{gathered}
\begin{fmffile}{Diagrams/2PIEAzerovev_Diag2}
\begin{fmfgraph}(12,12)
\fmfleft{i0,i1}
\fmfright{o0,o1}
\fmftop{v1,vUp,v2}
\fmfbottom{v3,vDown,v4}
\fmf{phantom,tension=20}{i0,v1}
\fmf{phantom,tension=20}{i1,v3}
\fmf{phantom,tension=20}{o0,v2}
\fmf{phantom,tension=20}{o1,v4}
\fmf{plain,left=0.4,tension=0.5,foreground=(1,,0,,0)}{v3,v1}
\fmf{phantom,left=0.1,tension=0.5}{v1,vUp}
\fmf{phantom,left=0.1,tension=0.5}{vUp,v2}
\fmf{zigzag,left=0.4,tension=0.0,foreground=(0,,0,,1)}{v1,v2}
\fmf{plain,left=0.4,tension=0.5,foreground=(1,,0,,0)}{v2,v4}
\fmf{phantom,left=0.1,tension=0.5}{v4,vDown}
\fmf{phantom,left=0.1,tension=0.5}{vDown,v3}
\fmf{zigzag,left=0.4,tension=0.0,foreground=(0,,0,,1)}{v4,v3}
\fmf{plain,left=0.4,tension=0.5,foreground=(1,,0,,0)}{v1,v3}
\fmf{plain,right=0.4,tension=0.5,foreground=(1,,0,,0)}{v2,v4}
\end{fmfgraph}
\end{fmffile}
\end{gathered} \hspace{0.3cm} = \lambda^{2} N^{2} G^{4}\;,
\label{eq:pure2PIEAzeroVevDiag40DON}
\end{equation}
\begin{equation}
\begin{gathered}\begin{fmffile}{Diagrams/2PIEAzerovev_Diag3}
\begin{fmfgraph}(16,16)
\fmfleft{i}
\fmfright{o}
\fmftop{vUpLeft,vUp,vUpRight}
\fmfbottom{vDownLeft,vDown,vDownRight}
\fmf{phantom,tension=1}{i,v1}
\fmf{phantom,tension=1}{v2,o}
\fmf{phantom,tension=14.0}{v3,vUpLeft}
\fmf{phantom,tension=2.0}{v3,vUpRight}
\fmf{phantom,tension=4.0}{v3,i}
\fmf{phantom,tension=2.0}{v4,vUpLeft}
\fmf{phantom,tension=14.0}{v4,vUpRight}
\fmf{phantom,tension=4.0}{v4,o}
\fmf{phantom,tension=14.0}{v5,vDownLeft}
\fmf{phantom,tension=2.0}{v5,vDownRight}
\fmf{phantom,tension=4.0}{v5,i}
\fmf{phantom,tension=2.0}{v6,vDownLeft}
\fmf{phantom,tension=14.0}{v6,vDownRight}
\fmf{phantom,tension=4.0}{v6,o}
\fmf{zigzag,tension=0,foreground=(0,,0,,1)}{v1,v2}
\fmf{zigzag,tension=0.6,foreground=(0,,0,,1)}{v3,v6}
\fmf{zigzag,tension=0.6,foreground=(0,,0,,1)}{v5,v4}
\fmf{plain,left=0.18,tension=0,foreground=(1,,0,,0)}{v1,v3}
\fmf{plain,left=0.42,tension=0,foreground=(1,,0,,0)}{v3,v4}
\fmf{plain,left=0.18,tension=0,foreground=(1,,0,,0)}{v4,v2}
\fmf{plain,left=0.18,tension=0,foreground=(1,,0,,0)}{v2,v6}
\fmf{plain,left=0.42,tension=0,foreground=(1,,0,,0)}{v6,v5}
\fmf{plain,left=0.18,tension=0,foreground=(1,,0,,0)}{v5,v1}
\end{fmfgraph}
\end{fmffile}
\end{gathered} \hspace{0.15cm} = \hspace{0.5cm} \begin{gathered}
\begin{fmffile}{Diagrams/2PIEAzerovev_Diag4}
\begin{fmfgraph}(12.5,12.5)
\fmfleft{i0,i1}
\fmfright{o0,o1}
\fmftop{v1,vUp,v2}
\fmfbottom{v3,vDown,v4}
\fmf{phantom,tension=20}{i0,v1}
\fmf{phantom,tension=20}{i1,v3}
\fmf{phantom,tension=20}{o0,v2}
\fmf{phantom,tension=20}{o1,v4}
\fmf{phantom,tension=0.005}{v5,v6}
\fmf{zigzag,left=0.4,tension=0,foreground=(0,,0,,1)}{v3,v1}
\fmf{phantom,left=0.1,tension=0}{v1,vUp}
\fmf{phantom,left=0.1,tension=0}{vUp,v2}
\fmf{plain,left=0.25,tension=0,foreground=(1,,0,,0)}{v1,v2}
\fmf{zigzag,left=0.4,tension=0,foreground=(0,,0,,1)}{v2,v4}
\fmf{phantom,left=0.1,tension=0}{v4,vDown}
\fmf{phantom,left=0.1,tension=0}{vDown,v3}
\fmf{plain,right=0.25,tension=0,foreground=(1,,0,,0)}{v3,v4}
\fmf{plain,left=0.2,tension=0.01,foreground=(1,,0,,0)}{v1,v5}
\fmf{plain,left=0.2,tension=0.01,foreground=(1,,0,,0)}{v5,v3}
\fmf{plain,right=0.2,tension=0.01,foreground=(1,,0,,0)}{v2,v6}
\fmf{plain,right=0.2,tension=0.01,foreground=(1,,0,,0)}{v6,v4}
\fmf{zigzag,tension=0,foreground=(0,,0,,1)}{v5,v6}
\end{fmfgraph}
\end{fmffile}
\end{gathered} \hspace{0.5cm} = \lambda^{3} N G^{6}\;,
\label{eq:pure2PIEAzeroVevDiag50DON}
\end{equation}
\begin{equation}
\begin{gathered}
\begin{fmffile}{Diagrams/2PIEAzerovev_Diag5}
\begin{fmfgraph}(12.5,12.5)
\fmfleft{i0,i1}
\fmfright{o0,o1}
\fmftop{v1,vUp,v2}
\fmfbottom{v3,vDown,v4}
\fmf{phantom,tension=20}{i0,v1}
\fmf{phantom,tension=20}{i1,v3}
\fmf{phantom,tension=20}{o0,v2}
\fmf{phantom,tension=20}{o1,v4}
\fmf{phantom,tension=0.005}{v5,v6}
\fmf{plain,left=0.4,tension=0,foreground=(1,,0,,0)}{v3,v1}
\fmf{phantom,left=0.1,tension=0}{v1,vUp}
\fmf{phantom,left=0.1,tension=0}{vUp,v2}
\fmf{zigzag,left=0.25,tension=0,foreground=(0,,0,,1)}{v1,v2}
\fmf{plain,left=0.4,tension=0,foreground=(1,,0,,0)}{v2,v4}
\fmf{phantom,left=0.1,tension=0}{v4,vDown}
\fmf{phantom,left=0.1,tension=0}{vDown,v3}
\fmf{zigzag,right=0.25,tension=0,foreground=(0,,0,,1)}{v3,v4}
\fmf{plain,left=0.2,tension=0.01,foreground=(1,,0,,0)}{v1,v5}
\fmf{plain,left=0.2,tension=0.01,foreground=(1,,0,,0)}{v5,v3}
\fmf{plain,right=0.2,tension=0.01,foreground=(1,,0,,0)}{v2,v6}
\fmf{plain,right=0.2,tension=0.01,foreground=(1,,0,,0)}{v6,v4}
\fmf{zigzag,tension=0,foreground=(0,,0,,1)}{v5,v6}
\end{fmfgraph}
\end{fmffile}
\end{gathered} \hspace{0.35cm} = \hspace{-0.35cm} \begin{gathered}
\begin{fmffile}{Diagrams/2PIEAzerovev_Diag6}
\begin{fmfgraph}(30,15)
\fmfleft{i0,i,i1}
\fmfright{o0,o,o1}
\fmftop{v1b,vUp,v2b}
\fmfbottom{v3b,vDown,v4b}
\fmf{phantom,tension=20}{i0,v1b}
\fmf{phantom,tension=20}{i1,v3b}
\fmf{phantom,tension=20}{o0,v2b}
\fmf{phantom,tension=20}{o1,v4b}
\fmf{phantom,tension=0.511}{i,v7}
\fmf{phantom,tension=0.11}{o,v7}
\fmf{phantom,tension=0.1}{v1,v1b}
\fmf{phantom,tension=0.1}{v2,v2b}
\fmf{phantom,tension=0.1}{v3,v3b}
\fmf{phantom,tension=0.1}{v4,v4b}
\fmf{phantom,tension=0.005}{v5,v6}
\fmf{phantom,left=0.1,tension=0.1}{v1,vUp}
\fmf{phantom,left=0.1,tension=0.1}{vUp,v2}
\fmf{zigzag,left=0.15,tension=0,foreground=(0,,0,,1)}{v1,v2}
\fmf{plain,left=0.4,tension=0,foreground=(1,,0,,0)}{v2,v4}
\fmf{plain,right=0.4,tension=0,foreground=(1,,0,,0)}{v2,v4}
\fmf{phantom,left=0.1,tension=0.1}{v4,vDown}
\fmf{phantom,left=0.1,tension=0.1}{vDown,v3}
\fmf{zigzag,left=0.15,tension=0,foreground=(0,,0,,1)}{v4,v3}
\fmf{plain,left=0.2,tension=0.01,foreground=(1,,0,,0)}{v1,v5}
\fmf{plain,left=0.2,tension=0.01,foreground=(1,,0,,0)}{v5,v3}
\fmf{plain,right=0.2,tension=0,foreground=(1,,0,,0)}{v1,v7}
\fmf{plain,right=0.2,tension=0,foreground=(1,,0,,0)}{v7,v3}
\fmf{phantom,right=0.2,tension=0.01}{v2,v6}
\fmf{phantom,right=0.2,tension=0.01}{v6,v4}
\fmf{zigzag,tension=0,foreground=(0,,0,,1)}{v5,v7}
\end{fmfgraph}
\end{fmffile}
\end{gathered} \hspace{-0.35cm} = \lambda^{3} N^{2} G^{6}\;,
\label{eq:pure2PIEAzeroVevDiag60DON}
\end{equation}

\vspace{0.3cm}

\begin{equation}
\begin{gathered}
\begin{fmffile}{Diagrams/2PIEAzerovev_Diag7}
\begin{fmfgraph}(30,14)
\fmfleft{i0,i,i1}
\fmfright{o0,o,o1}
\fmftop{v1b,vUp,v2b}
\fmfbottom{v3b,vDown,v4b}
\fmf{phantom,tension=5}{vUp,v5}
\fmf{phantom,tension=1}{v1b,v5}
\fmf{phantom,tension=5}{vUp,v6}
\fmf{phantom,tension=1}{v2b,v6}
\fmf{phantom,tension=20}{i0,v1b}
\fmf{phantom,tension=20}{i1,v3b}
\fmf{phantom,tension=20}{o0,v2b}
\fmf{phantom,tension=20}{o1,v4b}
\fmf{phantom,tension=0.1}{v1,v1b}
\fmf{phantom,tension=0.1}{v2,v2b}
\fmf{phantom,tension=0.1}{v3,v3b}
\fmf{phantom,tension=0.1}{v4,v4b}
\fmf{phantom,tension=0.005}{v5,v6}
\fmf{phantom,left=0.1,tension=0.1}{v1,vUp}
\fmf{phantom,left=0.1,tension=0.1}{vUp,v2}
\fmf{plain,left=0.4,tension=0.005,foreground=(1,,0,,0)}{v2,v4}
\fmf{plain,right=0.4,tension=0.005,foreground=(1,,0,,0)}{v2,v4}
\fmf{plain,left=0.4,tension=0.005,foreground=(1,,0,,0)}{v1,v3}
\fmf{plain,right=0.4,tension=0.005,foreground=(1,,0,,0)}{v1,v3}
\fmf{phantom,left=0.1,tension=0.1}{v4,vDown}
\fmf{phantom,left=0.1,tension=0.1}{vDown,v3}
\fmf{zigzag,left=0.05,tension=0,foreground=(0,,0,,1)}{v1,v5}
\fmf{plain,left,tension=0,foreground=(1,,0,,0)}{v5,v6}
\fmf{plain,right,tension=0,foreground=(1,,0,,0)}{v5,v6}
\fmf{zigzag,left=0.05,tension=0,foreground=(0,,0,,1)}{v6,v2}
\fmf{zigzag,left=0.15,tension=0,foreground=(0,,0,,1)}{v4,v3}
\end{fmfgraph}
\end{fmffile}
\end{gathered} \hspace{-0.35cm} = \lambda^{3} N^{3} G^{6}\;.
\label{eq:pure2PIEAzeroVevDiag70DON}
\end{equation}
From~\eqref{eq:2PIEAzerovevfinalexpression} and~\eqref{eq:pure2PIEAzeroVevSTraceTerm0DON} to~\eqref{eq:pure2PIEAzeroVevDiag70DON}, it follows that:
\begin{equation}
\begin{split}
\Gamma^{(\mathrm{2PI})}(\boldsymbol{G}) = & \ \hbar\left(-\frac{N}{2}\ln(2\pi G) + \frac{N}{2}\left(m^{2} G - 1\right)\right) + \hbar^{2} \left(\frac{N^{2}+2N}{24}\lambda G^{2}\right) -\hbar^{3}\left(\frac{N^{2}+2N}{144}\lambda^{2}G^{4}\right) \\
& + \hbar^{4} \left(\frac{N^{3}+10N^{2}+16N}{1296}\lambda^{3}G^{6}\right) + \mathcal{O}\big(\hbar^{5}\big)\;.
\end{split}
\label{eq:2PIEAzerovevfinalexpression0DON}
\end{equation}
Finally, the differentiation of~\eqref{eq:2PIEAzerovevfinalexpression0DON} yields the following gap equation:
\begin{equation}
\begin{split}
0 = \left.\frac{\partial\Gamma^{(\mathrm{2PI})}(\boldsymbol{G})}{\partial G}\right|_{\boldsymbol{G}=\overline{\boldsymbol{G}}} = & \ \hbar\left(-\frac{N}{2}\overline{G}^{-1} + \frac{N}{2}m^{2}\right) + \hbar^{2} \left(\frac{N^{2}+2N}{12}\lambda \overline{G}\right) -\hbar^{3}\left(\frac{N^{2}+2N}{36}\lambda^{2}\overline{G}^{3}\right) \\
& + \hbar^{4} \left(\frac{N^{3}+10N^{2}+16N}{216}\lambda^{3}\overline{G}^{5}\right) + \mathcal{O}\big(\hbar^{5}\big)\;,
\end{split}
\label{eq:2PIEAzerovevgapequation0DON}
\end{equation}
with $\overline{\boldsymbol{G}}_{a b}=\overline{G} \ \delta_{a b}$ $\forall a,b$. The gs energy and density are obtained from the solution $\overline{G}$ as well as~\eqref{eq:2PIorigE} and~\eqref{eq:2PIorigRho} with $\vec{\overline{\phi}}=\vec{0}$. Besides, as was shown earlier for the 1PI EA at the first non-trivial orders, the $\hbar$- and $\lambda$-expansions for the 2PI EA coincide at $\vec{\phi}=\vec{0}$. It can indeed be seen in~\eqref{eq:2PIEAzerovevfinalexpression0DON} that the power series of $\Gamma^{(\mathrm{2PI})}(\boldsymbol{G})$ is equivalently organized with respect to $\hbar$ and $\lambda$. More specifically, $\hbar$-expansion results truncated at order $\mathcal{O}\big(\hbar^{n+1}\big)$ are equivalent to those of the $\lambda$-expansion truncated at order $\mathcal{O}\big(\lambda^{n}\big)$ for all $n\in\mathbb{N}$.

\vspace{0.5cm}

\begin{figure}[!htb]
\captionsetup[subfigure]{labelformat=empty}
  \begin{center}
    \subfloat[]{
      \includegraphics[width=0.50\linewidth]{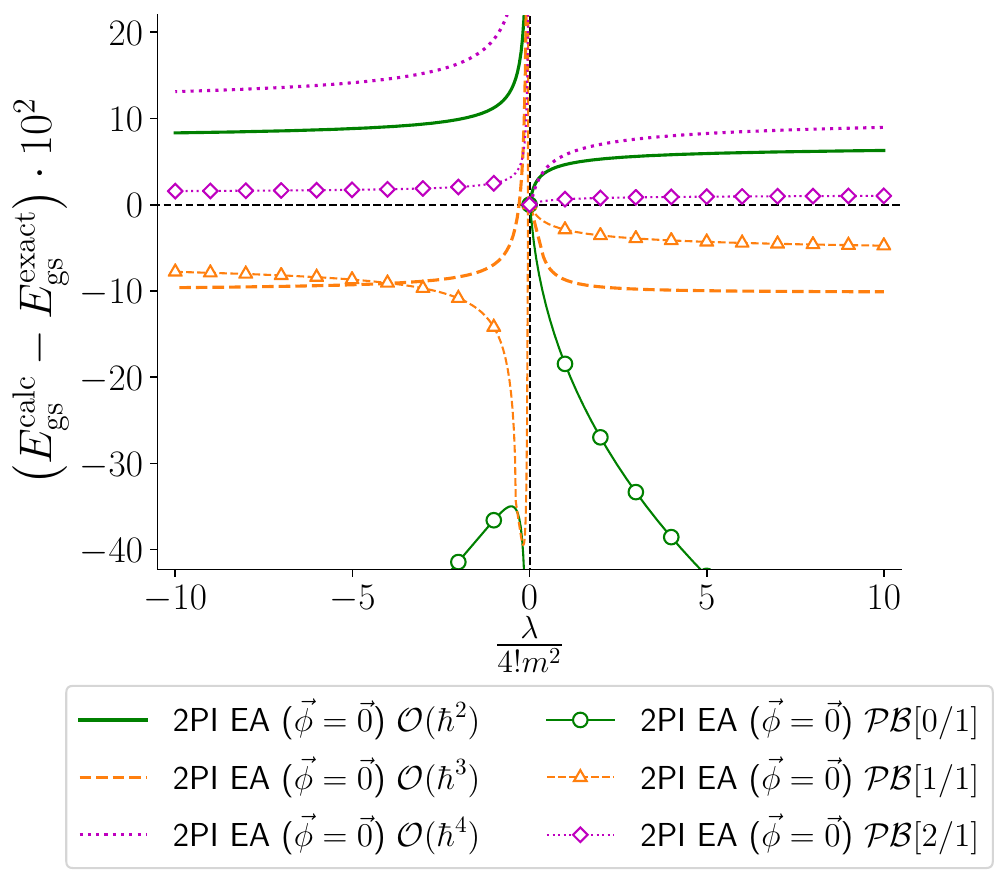}
                         }
    \subfloat[]{
      \includegraphics[width=0.50\linewidth]{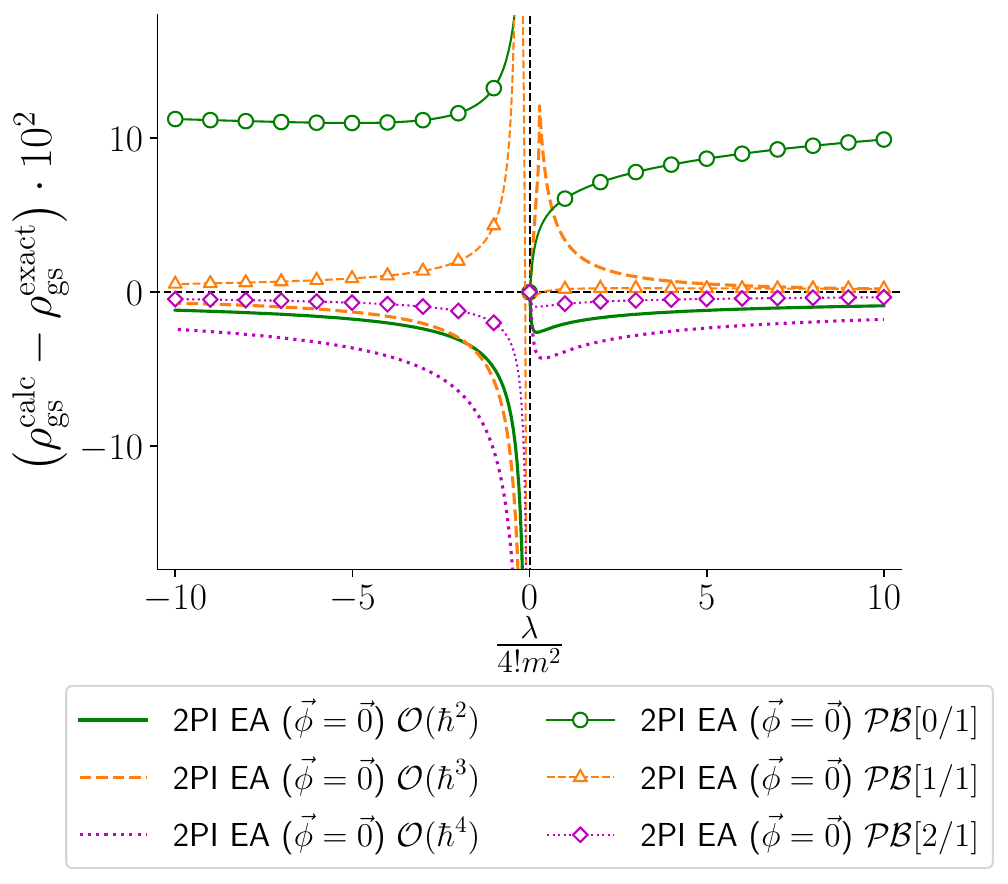}
                         }
    \caption{Difference between the calculated gs energy $E_{\mathrm{gs}}^{\mathrm{calc}}$ or density $\rho_{\mathrm{gs}}^{\mathrm{calc}}$ and the corresponding exact solution $E_{\mathrm{gs}}^{\mathrm{exact}}$ or $\rho_{\mathrm{gs}}^{\mathrm{exact}}$ at $\hbar=1$, $m^{2}=\pm 1$ and $N=2$ ($\mathcal{R}e(\lambda)\geq 0$ and $\mathcal{I}m(\lambda)=0$). See also the caption of fig.~\ref{fig:1PIEA} for the meaning of the indication ``$\mathcal{O}\big(\hbar^{n}\big)$'' for the results obtained from $\hbar$-expanded EAs.}
    \label{fig:2PIEAorigN2}
  \end{center}
\end{figure}

The results thus obtained from $\Gamma^{(\mathrm{2PI})}(\boldsymbol{G})$ are displayed in fig.~\ref{fig:2PIEAorigN2}. The first non-trivial order, which is implemented by truncating $\Gamma^{(\mathrm{2PI})}(\boldsymbol{G})$ right beyond order $\mathcal{O}\big(\hbar^{2}\big)$, coincides with the standard Hartree-Fock result as it can be shown that the gap equation~\eqref{eq:2PIEAzerovevgapequation0DON} is equivalent to a Dyson equation with Hartree-Fock self-energy if all terms of order $\mathcal{O}\big(\hbar^{3}\big)$ or higher are ignored. According to fig.~\ref{fig:2PIEAorigN2}, this truncation is barely affected as the coupling constant $\lambda$ increases in the regime set by $\lambda/4! \gtrsim 1$ (for both signs of $m^{2}$) and achieves an accuracy of about $10\%$ for both $E_{\mathrm{gs}}$ and $\rho_{\mathrm{gs}}$ at $N=2$ in this situation. We have illustrated in this way the non-perturbative character of the Hartree-Fock theory and we can thus see that the present 2PI EA approach is designed to improve this Hartree-Fock result in a systematic fashion. However, if we consider the solutions of the gap equation~\eqref{eq:2PIEAzerovevgapequation0DON} at the next two orders in $\hbar$, we actually observe the reverse: in almost the entire interval $\lambda/4! \in [0,10]$ for both signs of $m^{2}$, the resulting estimates of $E_{\mathrm{gs}}$ and $\rho_{\mathrm{gs}}$ worsen as the truncation order with respect to $\hbar$ increases. Note that, just like our perturbative series derived with the LE in section~\ref{sec:PT}, the series representing $\Gamma^{(\mathrm{2PI})}(\boldsymbol{G})$ (i.e.~\eqref{eq:2PIEAzerovevfinalexpression} or~\eqref{eq:2PIEAzerovevfinalexpression0DON} in (0+0)-D) is asymptotic and divergent, even after setting $\boldsymbol{G}=\overline{\boldsymbol{G}}$.

\vspace{0.5cm}

We therefore exploit a resummation procedure, and more specifically the Pad\'{e}-Borel resummation scheme, to illustrate the expected improvement with respect to the Hartree-Fock result. It amounts to modifying the underlying procedure as follows: the expression of $\Gamma^{(\mathrm{2PI})}(\boldsymbol{G})$ (given by~\eqref{eq:2PIEAzerovevfinalexpression0DON}) truncated at the chosen order with respect to $\hbar$ is replaced by a given Pad\'{e} approximant to subsequently derive the gap equations. In this way, the solution $\overline{\boldsymbol{G}}$ is systematically improved via resummation. As a next step, the gs energy and density are still inferred from~\eqref{eq:2PIorigE} and~\eqref{eq:2PIorigRho} at $\vec{\overline{\phi}}=\vec{0}$, with one additional peculiarity for $E_{\mathrm{gs}}$: $\Gamma^{(\mathrm{2PI})}\big(\vec{\phi}=\vec{0},\boldsymbol{G}=\overline{\boldsymbol{G}}\big)$ is rewritten in~\eqref{eq:2PIorigE} with the Pad\'{e}-Borel resummation procedure outlined in section~\ref{sec:PadBorResum}. We refer to this entire procedure as Pad\'{e}-Borel resummation of the EA, even though there is no Borel transform involved in the determination of $\rho_{\mathrm{gs}}$. The implementation of the Borel-hypergeometric resummation is not that straightforward for the EA formalism since we do not have analytical formulae to rewrite derivatives of Meijer G-functions with respect to each of their entries~\cite{fic80}. Considering the good performances of this resummation procedure at the level of the LE, we can definitely expect it to be relevant in the framework of EAs as well, but we postpone such an investigation to future works.

\vspace{0.5cm}

Regarding the numerical results thus obtained with the Pad\'{e}-Borel resummation, we can indeed see in fig.~\ref{fig:2PIEAorigN2}, which shows the results obtained from the best Pad\'{e} approximants at each of the three first non-trivial orders of $\Gamma^{(\mathrm{2PI})}(\boldsymbol{G})$, that a $[2/1]$ Pad\'{e} approximant reaches an accuracy of $1\%$ for both $E_{\mathrm{gs}}$ and $\rho_{\mathrm{gs}}$ for $\lambda/4!\in[0,10]$, which is to be compared with the $10\%$ of the Hartree-Fock result. However, the best Pad\'{e} approximants at the first two non-trivial orders, i.e. the $[0/1]$ and $[1/1]$ approximants, do not manage to clearly improve the corresponding bare results (labeled respectively ``2PI EA ($\vec{\phi}=\vec{0}$) $\mathcal{O}(\hbar^{2})$'' and ``2PI EA ($\vec{\phi}=\vec{0}$) $\mathcal{O}(\hbar^{3})$'' in fig.~\ref{fig:2PIEAorigN2}) for all values of the coupling constant $\lambda$ in both the unbroken- and broken-symmetry phases, which leads us to another important point on renormalization that we now discuss.

\pagebreak

The absence of integrals over spacetime indices in the studied (0+0)-D problem exempts us from renormalization issues. Until recently, there was no renormalization recipe to exploit reliably the EA approaches of this chapter beyond their lowest non-trivial orders (i.e. beyond the Hartree-Fock level). More specifically, whereas a procedure to renormalize 2PI EAs with counterterms has been put forward in the early 2000s~\cite{van02bis,bla04,ber05bis,rei10}, it is acknowledged that $n$PI EAs have a tendency to break gauge invariance at order $\mathcal{O}\big(\hbar^{m}\big)$ when $m>n$. In other words, $n$PI EAs are optimally exploited when they are considered at order $\mathcal{O}\big(\hbar^{n}\big)$~\cite{arr02,car05}, which implies that, to perform calculations beyond the Hartree-Fock level of 2PI EAs, one should handle 3PI or higher-order EAs, for which there was no renormalization recipe until recently. This limitation was overcome by a recent study~\cite{car19} developing a new renormalization scheme based on FRG that enables us to handle the divergences encountered in any $n$PI EA approaches, and thus safely exploit $n$PI EAs up to any order even for gauge theories. In addition to this longstanding lack of renormalization recipe for EA approaches beyond their lowest non-trivial orders, resummation procedures are in general inefficient at first non-trivial orders of diagrammatic expansions, as is illustrated with fig.~\ref{fig:2PIEAorigN2}. This explains that there are only very few studies~\cite{bro15} investigating the EA formalism in combination with resummation theory. However, the latter remains a key aspect of the present approach as it is the resummation that enables us to turn the EA techniques treated in this chapter into systematically improvable approaches. We will therefore not content ourselves with the present resummation analysis and perform similar applications to what will turn out to be the most performing EA method of this chapter, i.e. the mixed 2PI EA.

\subsubsection{\label{sec:diagMixed2PIEA}Mixed effective action}

\paragraph{$\hbar$-expansion for the full mixed 2PI EA:}

Although mixed EAs of $O(N)$ models were pioneered by the work of Coleman, Jackiw and Politzer~\cite{col74}, the developments of their 2PI versions were carried out by Cooper and collaborators~\cite{ben77,cho99,mih01,bla01,coo03,coo05} and later by Aarts \textit{et al.}~\cite{aar02} for instance. Note also some applications of this formalism to the study of chiral symmetry restoration~\cite{see12}. As for the collective LE discussed in section~\ref{sec:CollLE}, the present study is to our knowledge the first pushing this approach based on the mixed 2PI EA up to its third non-trivial order (i.e. up to order $\mathcal{O}(\hbar^{4})$ for the $\hbar$-expansion of this EA\footnote{To clarify, the present full mixed 2PI EA approach has to our knowledge never been pushed up to its third non-trivial order \textbf{regardless of the chosen expansion scheme}, even in the framework of the $1/N$-expansion which is often considered for 2PI EA studies notably~\cite{mih01,aar02,bla01,coo03}.}) and to combine it with a resummation procedure. Furthermore, a well-known implementation of the mixed 2PI EA is the bare vertex approximation (BVA)~\cite{mih01,bla01,coo03}, which is equivalent to the first non-trivial order of the $\hbar$-expansion (for which the EA is still considered up to order $\mathcal{O}(\hbar^{2})$). The BVA was notably shown to be successful in the framework of QFTs at finite temperature. For an $O(N)$-symmetric $\varphi^4$-theory in (1+1)-D for instance, the problematic absence of thermalization found in the framework of the Hartree approximation is cured by the BVA~\cite{coo03}.

\vspace{0.5cm}

The definition of the mixed 2PI EA can be inferred from that of the original 2PI EA (given by~\eqref{eq:pure2PIEAdefinition0DONmain} to~\eqref{eq:pure2PIEAdefinitionbis20DONmain}) by replacing correlation functions and sources by their supercounterparts, thus leading to:
\begin{equation}
\begin{split}
\Gamma_{\mathrm{mix}}^{(\mathrm{2PI})}\big[\Phi,\mathcal{G}\big] \equiv & -W_{\mathrm{mix}}\big[\mathcal{J},\mathcal{K}\big] + \int_{x}\mathcal{J}^{\alpha}(x) \frac{\delta W_{\mathrm{mix}}\big[\mathcal{J},\mathcal{K}\big]}{\delta \mathcal{J}^{\alpha}(x)} + \int_{x,y}\mathcal{K}^{\alpha\beta}(x,y) \frac{\delta W_{\mathrm{mix}}\big[\mathcal{J},\mathcal{K}\big]}{\delta \mathcal{K}^{\beta\alpha}(x,y)} \\
= & -W_{\mathrm{mix}}\big[\mathcal{J},\mathcal{K}\big] + \int_{x}\mathcal{J}^{\alpha}(x) \Phi_{\alpha}(x) + \frac{1}{2} \int_{x,y} \Phi_{\alpha}(x) \mathcal{K}^{\alpha\beta}(x,y) \Phi_{\beta}(y) \\
& + \frac{\hbar}{2} \int_{x,y} \mathcal{K}^{\alpha\beta}(x,y) \mathcal{G}_{\beta\alpha}(y,x) \;,
\end{split}
\label{eq:mixed2PIEAdefinition0DONmain}
\end{equation}
with
\begin{equation}
\Phi_{\alpha}(x) = \frac{\delta W_{\mathrm{mix}}\big[\mathcal{J},\mathcal{K}\big]}{\delta \mathcal{J}^{\alpha}(x)} \;,
\label{eq:mixed2PIEAdefinitionbis0DONmain}
\end{equation}
\begin{equation}
\mathcal{G}_{\alpha\beta}(x,y) = \frac{\delta^{2} W_{\mathrm{mix}}\big[\mathcal{J},\mathcal{K}\big]}{\delta \mathcal{J}^{\alpha}(x)\delta \mathcal{J}^{\beta}(y)} = \frac{2}{\hbar} \frac{\delta W_{\mathrm{mix}}\big[\mathcal{J},\mathcal{K}\big]}{\delta\mathcal{K}^{\alpha\beta}(x,y)} - \frac{1}{\hbar} \Phi_{\alpha}(x) \Phi_{\beta}(y) \;,
\label{eq:mixed2PIEAdefinitionbis20DONmain}
\end{equation}
or, to further specify our supernotations (still involving the 1-point correlation functions $\vec{\phi}(x) = \left\langle\vec{\widetilde{\varphi}}(x)\right\rangle$ and $\eta(x)=\left\langle\widetilde{\sigma}(x)\right\rangle$),
\begin{equation}
\Phi = \begin{pmatrix}
\vec{\phi} \\
\eta
\end{pmatrix} \;,
\end{equation}
\begin{equation}
\mathcal{G} = \begin{pmatrix}
\boldsymbol{G} & \vec{F} \\
\vec{F}^{\mathrm{T}} & D
\end{pmatrix} \;,
\label{eq:mixed2PIEADefSuperpropagator}
\end{equation}
and the Schwinger functional $W_{\mathrm{mix}}\big[\mathcal{J},\mathcal{K}\big] \equiv W^{\text{LE};\text{mix}}\big[\mathcal{J},\mathcal{K}\big]$ has already been introduced in section~\ref{sec:PT}. The mixed 2PI EA organized in powers of $\hbar$ can be written in terms of 2PI diagrams only, which yields for the $O(N)$ model under consideration (see appendix~\ref{sec:mixed2PIEAannIM}):
\begin{equation}
\begin{split}
\Gamma^{(\mathrm{2PI})}_{\mathrm{mix}}\big[\Phi,\mathcal{G}\big] = & \ S_{\mathrm{mix}}[\Phi] -\frac{\hbar}{2}\mathcal{ST}r\left[\ln\big(\mathcal{G}\big)\right] + \frac{\hbar}{2}\mathcal{ST}r\left[\mathcal{G}^{-1}_{\Phi}\mathcal{G}-\mathfrak{I}\right] \\
& + \hbar^{2} \left(\rule{0cm}{1.2cm}\right. \frac{1}{12} \hspace{0.1cm} \begin{gathered}
\begin{fmffile}{Diagrams/Mixed2PIEA_Fock}
\begin{fmfgraph}(15,15)
\fmfleft{i}
\fmfright{o}
\fmfv{decor.shape=circle,decor.size=2.0thick,foreground=(0,,0,,1)}{v1}
\fmfv{decor.shape=circle,decor.size=2.0thick,foreground=(0,,0,,1)}{v2}
\fmf{phantom,tension=11}{i,v1}
\fmf{phantom,tension=11}{v2,o}
\fmf{plain,left,tension=0.4,foreground=(1,,0,,0)}{v1,v2,v1}
\fmf{wiggly,foreground=(1,,0,,0)}{v1,v2}
\end{fmfgraph}
\end{fmffile}
\end{gathered} + \frac{1}{6} \hspace{0.1cm} \begin{gathered}
\begin{fmffile}{Diagrams/Mixed2PIEA_Diag1}
\begin{fmfgraph}(15,15)
\fmfleft{i}
\fmfright{o}
\fmfv{decor.shape=circle,decor.size=2.0thick,foreground=(0,,0,,1)}{v1}
\fmfv{decor.shape=circle,decor.size=2.0thick,foreground=(0,,0,,1)}{v2}
\fmf{phantom,tension=11}{i,v1}
\fmf{phantom,tension=11}{v2,o}
\fmf{dashes,left,tension=0.4,foreground=(1,,0,,0)}{v1,v2,v1}
\fmf{plain,foreground=(1,,0,,0)}{v1,v2}
\end{fmfgraph}
\end{fmffile}
\end{gathered} \left.\rule{0cm}{1.2cm}\right) \\
& - \hbar^{3} \left(\rule{0cm}{1.2cm}\right. \frac{1}{72} \hspace{0.35cm} \begin{gathered}
\begin{fmffile}{Diagrams/Mixed2PIEA_Diag2}
\begin{fmfgraph}(10,10)
\fmfleft{i0,i1}
\fmfright{o0,o1}
\fmftop{v1,vUp,v2}
\fmfbottom{v3,vDown,v4}
\fmfv{decor.shape=circle,decor.size=2.0thick,foreground=(0,,0,,1)}{v1}
\fmfv{decor.shape=circle,decor.size=2.0thick,foreground=(0,,0,,1)}{v2}
\fmfv{decor.shape=circle,decor.size=2.0thick,foreground=(0,,0,,1)}{v3}
\fmfv{decor.shape=circle,decor.size=2.0thick,foreground=(0,,0,,1)}{v4}
\fmf{phantom,tension=20}{i0,v1}
\fmf{phantom,tension=20}{i1,v3}
\fmf{phantom,tension=20}{o0,v2}
\fmf{phantom,tension=20}{o1,v4}
\fmf{plain,left=0.4,tension=0.5,foreground=(1,,0,,0)}{v3,v1}
\fmf{phantom,left=0.1,tension=0.5}{v1,vUp}
\fmf{phantom,left=0.1,tension=0.5}{vUp,v2}
\fmf{plain,left=0.4,tension=0.0,foreground=(1,,0,,0)}{v1,v2}
\fmf{plain,left=0.4,tension=0.5,foreground=(1,,0,,0)}{v2,v4}
\fmf{phantom,left=0.1,tension=0.5}{v4,vDown}
\fmf{phantom,left=0.1,tension=0.5}{vDown,v3}
\fmf{plain,left=0.4,tension=0.0,foreground=(1,,0,,0)}{v4,v3}
\fmf{wiggly,tension=0.5,foreground=(1,,0,,0)}{v1,v4}
\fmf{wiggly,tension=0.5,foreground=(1,,0,,0)}{v2,v3}
\end{fmfgraph}
\end{fmffile}
\end{gathered} \hspace{0.35cm} + \frac{1}{36} \hspace{0.35cm} \begin{gathered}
\begin{fmffile}{Diagrams/Mixed2PIEA_Diag3}
\begin{fmfgraph}(10,10)
\fmfleft{i0,i1}
\fmfright{o0,o1}
\fmftop{v1,vUp,v2}
\fmfbottom{v3,vDown,v4}
\fmfv{decor.shape=circle,decor.size=2.0thick,foreground=(0,,0,,1)}{v1}
\fmfv{decor.shape=circle,decor.size=2.0thick,foreground=(0,,0,,1)}{v2}
\fmfv{decor.shape=circle,decor.size=2.0thick,foreground=(0,,0,,1)}{v3}
\fmfv{decor.shape=circle,decor.size=2.0thick,foreground=(0,,0,,1)}{v4}
\fmf{phantom,tension=20}{i0,v1}
\fmf{phantom,tension=20}{i1,v3}
\fmf{phantom,tension=20}{o0,v2}
\fmf{phantom,tension=20}{o1,v4}
\fmf{dashes,left=0.4,tension=0.5,foreground=(1,,0,,0)}{v3,v1}
\fmf{phantom,left=0.1,tension=0.5}{v1,vUp}
\fmf{phantom,left=0.1,tension=0.5}{vUp,v2}
\fmf{dashes,left=0.4,tension=0.0,foreground=(1,,0,,0)}{v1,v2}
\fmf{dashes,left=0.4,tension=0.5,foreground=(1,,0,,0)}{v2,v4}
\fmf{phantom,left=0.1,tension=0.5}{v4,vDown}
\fmf{phantom,left=0.1,tension=0.5}{vDown,v3}
\fmf{dashes,left=0.4,tension=0.0,foreground=(1,,0,,0)}{v4,v3}
\fmf{plain,tension=0.5,foreground=(1,,0,,0)}{v1,v4}
\fmf{plain,tension=0.5,foreground=(1,,0,,0)}{v2,v3}
\end{fmfgraph}
\end{fmffile}
\end{gathered} \hspace{0.35cm} + \frac{1}{18} \hspace{0.35cm} \begin{gathered}
\begin{fmffile}{Diagrams/Mixed2PIEA_Diag4}
\begin{fmfgraph}(10,10)
\fmfleft{i0,i1}
\fmfright{o0,o1}
\fmftop{v1,vUp,v2}
\fmfbottom{v3,vDown,v4}
\fmfv{decor.shape=circle,decor.size=2.0thick,foreground=(0,,0,,1)}{v1}
\fmfv{decor.shape=circle,decor.size=2.0thick,foreground=(0,,0,,1)}{v2}
\fmfv{decor.shape=circle,decor.size=2.0thick,foreground=(0,,0,,1)}{v3}
\fmfv{decor.shape=circle,decor.size=2.0thick,foreground=(0,,0,,1)}{v4}
\fmf{phantom,tension=20}{i0,v1}
\fmf{phantom,tension=20}{i1,v3}
\fmf{phantom,tension=20}{o0,v2}
\fmf{phantom,tension=20}{o1,v4}
\fmf{plain,left=0.4,tension=0.5,foreground=(1,,0,,0)}{v3,v1}
\fmf{phantom,left=0.1,tension=0.5}{v1,vUp}
\fmf{phantom,left=0.1,tension=0.5}{vUp,v2}
\fmf{plain,left=0.4,tension=0.0,foreground=(1,,0,,0)}{v1,v2}
\fmf{dashes,left=0.4,tension=0.5,foreground=(1,,0,,0)}{v2,v4}
\fmf{phantom,left=0.1,tension=0.5}{v4,vDown}
\fmf{phantom,left=0.1,tension=0.5}{vDown,v3}
\fmf{dashes,left=0.4,tension=0.0,foreground=(1,,0,,0)}{v4,v3}
\fmf{wiggly,tension=0.5,foreground=(1,,0,,0)}{v1,v4}
\fmf{plain,tension=0.5,foreground=(1,,0,,0)}{v2,v3}
\end{fmfgraph}
\end{fmffile}
\end{gathered} \hspace{0.35cm} + \frac{1}{9} \hspace{0.35cm} \begin{gathered}
\begin{fmffile}{Diagrams/Mixed2PIEA_Diag5}
\begin{fmfgraph}(10,10)
\fmfleft{i0,i1}
\fmfright{o0,o1}
\fmftop{v1,vUp,v2}
\fmfbottom{v3,vDown,v4}
\fmfv{decor.shape=circle,decor.size=2.0thick,foreground=(0,,0,,1)}{v1}
\fmfv{decor.shape=circle,decor.size=2.0thick,foreground=(0,,0,,1)}{v2}
\fmfv{decor.shape=circle,decor.size=2.0thick,foreground=(0,,0,,1)}{v3}
\fmfv{decor.shape=circle,decor.size=2.0thick,foreground=(0,,0,,1)}{v4}
\fmf{phantom,tension=20}{i0,v1}
\fmf{phantom,tension=20}{i1,v3}
\fmf{phantom,tension=20}{o0,v2}
\fmf{phantom,tension=20}{o1,v4}
\fmf{dashes,left=0.4,tension=0.5,foreground=(1,,0,,0)}{v3,v1}
\fmf{phantom,left=0.1,tension=0.5}{v1,vUp}
\fmf{phantom,left=0.1,tension=0.5}{vUp,v2}
\fmf{plain,left=0.4,tension=0.0,foreground=(1,,0,,0)}{v1,v2}
\fmf{dashes,left=0.4,tension=0.5,foreground=(1,,0,,0)}{v2,v4}
\fmf{phantom,left=0.1,tension=0.5}{v4,vDown}
\fmf{phantom,left=0.1,tension=0.5}{vDown,v3}
\fmf{plain,left=0.4,tension=0.0,foreground=(1,,0,,0)}{v4,v3}
\fmf{wiggly,tension=0.5,foreground=(1,,0,,0)}{v1,v4}
\fmf{plain,tension=0.5,foreground=(1,,0,,0)}{v2,v3}
\end{fmfgraph}
\end{fmffile}
\end{gathered} \hspace{0.35cm} + \frac{1}{9} \hspace{0.35cm} \begin{gathered}
\begin{fmffile}{Diagrams/Mixed2PIEA_Diag6}
\begin{fmfgraph}(10,10)
\fmfleft{i0,i1}
\fmfright{o0,o1}
\fmftop{v1,vUp,v2}
\fmfbottom{v3,vDown,v4}
\fmfv{decor.shape=circle,decor.size=2.0thick,foreground=(0,,0,,1)}{v1}
\fmfv{decor.shape=circle,decor.size=2.0thick,foreground=(0,,0,,1)}{v2}
\fmfv{decor.shape=circle,decor.size=2.0thick,foreground=(0,,0,,1)}{v3}
\fmfv{decor.shape=circle,decor.size=2.0thick,foreground=(0,,0,,1)}{v4}
\fmf{phantom,tension=20}{i0,v1}
\fmf{phantom,tension=20}{i1,v3}
\fmf{phantom,tension=20}{o0,v2}
\fmf{phantom,tension=20}{o1,v4}
\fmf{plain,left=0.4,tension=0.5,foreground=(1,,0,,0)}{v3,v1}
\fmf{phantom,left=0.1,tension=0.5}{v1,vUp}
\fmf{phantom,left=0.1,tension=0.5}{vUp,v2}
\fmf{plain,left=0.4,tension=0.0,foreground=(1,,0,,0)}{v1,v2}
\fmf{dashes,left=0.4,tension=0.5,foreground=(1,,0,,0)}{v2,v4}
\fmf{phantom,left=0.1,tension=0.5}{v4,vDown}
\fmf{phantom,left=0.1,tension=0.5}{vDown,v3}
\fmf{dashes,left=0.4,tension=0.0,foreground=(1,,0,,0)}{v4,v3}
\fmf{dashes,tension=0.5,foreground=(1,,0,,0)}{v1,v4}
\fmf{dashes,tension=0.5,foreground=(1,,0,,0)}{v2,v3}
\end{fmfgraph}
\end{fmffile}
\end{gathered} \hspace{0.35cm} \left.\rule{0cm}{1.2cm}\right) \\
& + \hbar^{4} \left(\rule{0cm}{1.2cm}\right. \frac{1}{324} \hspace{0.3cm} \begin{gathered}
\begin{fmffile}{Diagrams/Mixed2PIEA_Diag7}
\begin{fmfgraph}(15,15)
\fmfleft{i}
\fmfright{o}
\fmftop{vUpLeft,vUp,vUpRight}
\fmfbottom{vDownLeft,vDown,vDownRight}
\fmfv{decor.shape=circle,decor.size=2.0thick,foreground=(0,,0,,1)}{v1}
\fmfv{decor.shape=circle,decor.size=2.0thick,foreground=(0,,0,,1)}{v2}
\fmfv{decor.shape=circle,decor.size=2.0thick,foreground=(0,,0,,1)}{v3}
\fmfv{decor.shape=circle,decor.size=2.0thick,foreground=(0,,0,,1)}{v4}
\fmfv{decor.shape=circle,decor.size=2.0thick,foreground=(0,,0,,1)}{v5}
\fmfv{decor.shape=circle,decor.size=2.0thick,foreground=(0,,0,,1)}{v6}
\fmf{phantom,tension=1}{i,v1}
\fmf{phantom,tension=1}{v2,o}
\fmf{phantom,tension=14.0}{v3,vUpLeft}
\fmf{phantom,tension=2.0}{v3,vUpRight}
\fmf{phantom,tension=4.0}{v3,i}
\fmf{phantom,tension=2.0}{v4,vUpLeft}
\fmf{phantom,tension=14.0}{v4,vUpRight}
\fmf{phantom,tension=4.0}{v4,o}
\fmf{phantom,tension=14.0}{v5,vDownLeft}
\fmf{phantom,tension=2.0}{v5,vDownRight}
\fmf{phantom,tension=4.0}{v5,i}
\fmf{phantom,tension=2.0}{v6,vDownLeft}
\fmf{phantom,tension=14.0}{v6,vDownRight}
\fmf{phantom,tension=4.0}{v6,o}
\fmf{wiggly,tension=0,foreground=(1,,0,,0)}{v1,v2}
\fmf{wiggly,tension=0.6,foreground=(1,,0,,0)}{v3,v6}
\fmf{wiggly,tension=0.6,foreground=(1,,0,,0)}{v5,v4}
\fmf{plain,left=0.18,tension=0,foreground=(1,,0,,0)}{v1,v3}
\fmf{plain,left=0.42,tension=0,foreground=(1,,0,,0)}{v3,v4}
\fmf{plain,left=0.18,tension=0,foreground=(1,,0,,0)}{v4,v2}
\fmf{plain,left=0.18,tension=0,foreground=(1,,0,,0)}{v2,v6}
\fmf{plain,left=0.42,tension=0,foreground=(1,,0,,0)}{v6,v5}
\fmf{plain,left=0.18,tension=0,foreground=(1,,0,,0)}{v5,v1}
\end{fmfgraph}
\end{fmffile}
\end{gathered} \hspace{0.3cm} + \frac{1}{108} \hspace{0.5cm} \begin{gathered}
\begin{fmffile}{Diagrams/Mixed2PIEA_Diag8}
\begin{fmfgraph}(12.5,12.5)
\fmfleft{i0,i1}
\fmfright{o0,o1}
\fmftop{v1,vUp,v2}
\fmfbottom{v3,vDown,v4}
\fmfv{decor.shape=circle,decor.size=2.0thick,foreground=(0,,0,,1)}{v1}
\fmfv{decor.shape=circle,decor.size=2.0thick,foreground=(0,,0,,1)}{v2}
\fmfv{decor.shape=circle,decor.size=2.0thick,foreground=(0,,0,,1)}{v3}
\fmfv{decor.shape=circle,decor.size=2.0thick,foreground=(0,,0,,1)}{v4}
\fmfv{decor.shape=circle,decor.size=2.0thick,foreground=(0,,0,,1)}{v5}
\fmfv{decor.shape=circle,decor.size=2.0thick,foreground=(0,,0,,1)}{v6}
\fmf{phantom,tension=20}{i0,v1}
\fmf{phantom,tension=20}{i1,v3}
\fmf{phantom,tension=20}{o0,v2}
\fmf{phantom,tension=20}{o1,v4}
\fmf{phantom,tension=0.005}{v5,v6}
\fmf{wiggly,left=0.4,tension=0,foreground=(1,,0,,0)}{v3,v1}
\fmf{phantom,left=0.1,tension=0}{v1,vUp}
\fmf{phantom,left=0.1,tension=0}{vUp,v2}
\fmf{plain,left=0.25,tension=0,foreground=(1,,0,,0)}{v1,v2}
\fmf{wiggly,left=0.4,tension=0,foreground=(1,,0,,0)}{v2,v4}
\fmf{phantom,left=0.1,tension=0}{v4,vDown}
\fmf{phantom,left=0.1,tension=0}{vDown,v3}
\fmf{plain,right=0.25,tension=0,foreground=(1,,0,,0)}{v3,v4}
\fmf{plain,left=0.2,tension=0.01,foreground=(1,,0,,0)}{v1,v5}
\fmf{plain,left=0.2,tension=0.01,foreground=(1,,0,,0)}{v5,v3}
\fmf{plain,right=0.2,tension=0.01,foreground=(1,,0,,0)}{v2,v6}
\fmf{plain,right=0.2,tension=0.01,foreground=(1,,0,,0)}{v6,v4}
\fmf{wiggly,tension=0,foreground=(1,,0,,0)}{v5,v6}
\end{fmfgraph}
\end{fmffile}
\end{gathered} \hspace{0.5cm} + \frac{1}{324} \hspace{0.4cm} \begin{gathered}
\begin{fmffile}{Diagrams/Mixed2PIEA_Diag9}
\begin{fmfgraph}(12.5,12.5)
\fmfleft{i0,i1}
\fmfright{o0,o1}
\fmftop{v1,vUp,v2}
\fmfbottom{v3,vDown,v4}
\fmfv{decor.shape=circle,decor.size=2.0thick,foreground=(0,,0,,1)}{v1}
\fmfv{decor.shape=circle,decor.size=2.0thick,foreground=(0,,0,,1)}{v2}
\fmfv{decor.shape=circle,decor.size=2.0thick,foreground=(0,,0,,1)}{v3}
\fmfv{decor.shape=circle,decor.size=2.0thick,foreground=(0,,0,,1)}{v4}
\fmfv{decor.shape=circle,decor.size=2.0thick,foreground=(0,,0,,1)}{v5}
\fmfv{decor.shape=circle,decor.size=2.0thick,foreground=(0,,0,,1)}{v6}
\fmf{phantom,tension=20}{i0,v1}
\fmf{phantom,tension=20}{i1,v3}
\fmf{phantom,tension=20}{o0,v2}
\fmf{phantom,tension=20}{o1,v4}
\fmf{phantom,tension=0.005}{v5,v6}
\fmf{plain,left=0.4,tension=0,foreground=(1,,0,,0)}{v3,v1}
\fmf{phantom,left=0.1,tension=0}{v1,vUp}
\fmf{phantom,left=0.1,tension=0}{vUp,v2}
\fmf{wiggly,left=0.25,tension=0,foreground=(1,,0,,0)}{v1,v2}
\fmf{plain,left=0.4,tension=0,foreground=(1,,0,,0)}{v2,v4}
\fmf{phantom,left=0.1,tension=0}{v4,vDown}
\fmf{phantom,left=0.1,tension=0}{vDown,v3}
\fmf{wiggly,right=0.25,tension=0,foreground=(1,,0,,0)}{v3,v4}
\fmf{plain,left=0.2,tension=0.01,foreground=(1,,0,,0)}{v1,v5}
\fmf{plain,left=0.2,tension=0.01,foreground=(1,,0,,0)}{v5,v3}
\fmf{plain,right=0.2,tension=0.01,foreground=(1,,0,,0)}{v2,v6}
\fmf{plain,right=0.2,tension=0.01,foreground=(1,,0,,0)}{v6,v4}
\fmf{wiggly,tension=0,foreground=(1,,0,,0)}{v5,v6}
\end{fmfgraph}
\end{fmffile}
\end{gathered} \hspace{0.4cm} + \mathcal{O}\Big(\vec{F}^{2}\Big) \left.\rule{0cm}{1.2cm}\right) \\
& + \mathcal{O}\big(\hbar^{5}\big)\;,
\end{split}
\label{eq:mixed2PIEAfinalexpression}
\end{equation}
with the superidentity $\mathfrak{I}_{\alpha\beta}(x,y)=\delta_{\alpha\beta}\delta(x-y)$ and the Feynman rules:\\
\begin{subequations}
\begin{align}
\left.
\begin{array}{ll}
\begin{gathered}
\begin{fmffile}{Diagrams/mixed2PIEA_FeynRuleVertexbis1}
\begin{fmfgraph*}(4,4)
\fmfleft{i0,i1,i2,i3}
\fmfright{o0,o1,o2,o3}
\fmfv{label=$x$,label.angle=90,label.dist=4}{v1}
\fmfbottom{v2}
\fmf{phantom}{i1,v1}
\fmf{plain,foreground=(1,,0,,0)}{i2,v1}
\fmf{phantom}{v1,o1}
\fmf{plain,foreground=(1,,0,,0)}{v1,o2}
\fmf{wiggly,tension=0.6,foreground=(1,,0,,0)}{v1,v2}
\fmfv{decor.shape=circle,decor.size=2.0thick,foreground=(0,,0,,1)}{v1}
\fmflabel{$a$}{i2}
\fmflabel{$b$}{o2}
\end{fmfgraph*}
\end{fmffile}
\end{gathered} \\
\\
\begin{gathered}
\begin{fmffile}{Diagrams/mixed2PIEA_FeynRuleVertexbis2}
\begin{fmfgraph*}(4,4)
\fmfleft{i0,i1,i2,i3}
\fmfright{o0,o1,o2,o3}
\fmfv{label=$x$,label.angle=90,label.dist=4}{v1}
\fmfbottom{v2}
\fmf{phantom}{i1,v1}
\fmf{plain,foreground=(1,,0,,0)}{i2,v1}
\fmf{phantom}{v1,o1}
\fmf{plain,foreground=(1,,0,,0)}{v1,o2}
\fmf{dots,tension=0.6,foreground=(1,,0,,0)}{v1,v2}
\fmfv{decor.shape=circle,decor.size=2.0thick,foreground=(0,,0,,1)}{v1}
\fmflabel{$a$}{i2}
\fmflabel{$b$}{o2}
\end{fmfgraph*}
\end{fmffile}
\end{gathered} \\
\\
\begin{gathered}
\begin{fmffile}{Diagrams/mixed2PIEA_FeynRuleVertexbis3}
\begin{fmfgraph*}(4,4)
\fmfleft{i0,i1,i2,i3}
\fmfright{o0,o1,o2,o3}
\fmfv{label=$x$,label.angle=90,label.dist=4}{v1}
\fmfbottom{v2}
\fmf{phantom}{i1,v1}
\fmf{plain,foreground=(1,,0,,0)}{i2,v1}
\fmf{phantom}{v1,o1}
\fmf{dots,foreground=(1,,0,,0)}{v1,o2}
\fmf{dots,tension=0.6,foreground=(1,,0,,0)}{v1,v2}
\fmfv{decor.shape=circle,decor.size=2.0thick,foreground=(0,,0,,1)}{v1}
\fmflabel{$a$}{i2}
\fmflabel{$b$}{o2}
\end{fmfgraph*}
\end{fmffile}
\end{gathered}
\end{array}
\quad \right\rbrace &\rightarrow \sqrt{\lambda} \ \delta_{ab} \;, 
\label{eq:mixed2PIEAvertex} \\
\begin{gathered}
\begin{fmffile}{Diagrams/mixed2PIEA_FeynRuleGbis}
\begin{fmfgraph*}(20,20)
\fmfleft{i0,i1,i2,i3}
\fmfright{o0,o1,o2,o3}
\fmflabel{$x, a$}{v1}
\fmflabel{$y, b$}{v2}
\fmf{phantom}{i1,v1}
\fmf{phantom}{i2,v1}
\fmf{plain,tension=0.6,foreground=(1,,0,,0)}{v1,v2}
\fmf{phantom}{v2,o1}
\fmf{phantom}{v2,o2}
\end{fmfgraph*}
\end{fmffile}
\end{gathered} \quad &\rightarrow \boldsymbol{G}_{ab}(x,y) \;,
\label{eq:mixed2PIEAFeynRuleG} \\
\begin{gathered}
\begin{fmffile}{Diagrams/mixed2PIEA_FeynRuleDbis}
\begin{fmfgraph*}(20,20)
\fmfleft{i0,i1,i2,i3}
\fmfright{o0,o1,o2,o3}
\fmfv{label=$x$}{v1}
\fmfv{label=$y$}{v2}
\fmf{phantom}{i1,v1}
\fmf{phantom}{i2,v1}
\fmf{wiggly,tension=0.6,foreground=(1,,0,,0)}{v1,v2}
\fmf{phantom}{v2,o1}
\fmf{phantom}{v2,o2}
\end{fmfgraph*}
\end{fmffile}
\end{gathered} \quad &\rightarrow D(x,y) \;,
\label{eq:mixed2PIEAFeynRuleD} \\
\begin{gathered}
\begin{fmffile}{Diagrams/mixed2PIEA_FeynRuleFbis}
\begin{fmfgraph*}(20,20)
\fmfleft{i0,i1,i2,i3}
\fmfright{o0,o1,o2,o3}
\fmflabel{$x, a$}{v1}
\fmfv{label=$y$}{v2}
\fmf{phantom}{i1,v1}
\fmf{phantom}{i2,v1}
\fmf{dashes,tension=0.6,foreground=(1,,0,,0)}{v1,v2}
\fmf{phantom}{v2,o1}
\fmf{phantom}{v2,o2}
\end{fmfgraph*}
\end{fmffile}
\end{gathered} \quad &\rightarrow F_{a}(x,y) \;,
\label{eq:mixed2PIEAFeynRuleF}
\end{align}
\end{subequations}
whereas the propagator $\mathcal{G}_{\Phi}$ satisfies:
\begin{equation}
\mathcal{G}^{-1}_{\Phi}(x,y) \equiv \left.\frac{\delta^2 S_{\text{mix}}\big[\widetilde{\Psi}\big]}{\delta \widetilde{\Psi}(x)\delta \widetilde{\Psi}(y)}\right|_{\widetilde{\Psi} = \Phi} = \begin{pmatrix}
\left(-\nabla_x^2 + m^2 + i\sqrt{\frac{\lambda}{3}}\eta(x)\right)\mathbb{I}_{N} & i\sqrt{\frac{\lambda}{3}}\vec{\phi}(x) \\
i\sqrt{\frac{\lambda}{3}}\vec{\phi}^\mathrm{T}(x) & 1 \end{pmatrix}\delta(x-y) \;.
\label{eq:mixed2PIEAMathcalGPhi}
\end{equation}
There is no direct analytical expression relating the components of $\mathcal{G}$ (i.e. $\boldsymbol{G}$, $D$ and $\vec{F}$) to the 1-point correlation function $\Phi$ (or to $\phi$ and $\eta$) because these propagators are introduced through~\eqref{eq:mixed2PIEADefSuperpropagator} and not as (inverses of) derivatives of the classical action $S_{\mathrm{mix}}$. This should be contrasted with~\eqref{eq:DefinitionG1PIEAhbarExpansionbis} (defining $\boldsymbol{G}_{\phi}$) as well as~\eqref{eq:bosonic1PIEApropagatorGJ0main} and~\eqref{eq:bosonic1PIEApropagatorHJ0main} (defining $\boldsymbol{G}_{\Phi}$ and $D_{\Phi}$), which constitute the propagator lines of the original and collective 1PI EAs in~\eqref{eq:1PIEAfinalexpression} and~\eqref{eq:bosonic1PIEAIMGamma0DON}, respectively. Hence, since propagator lines represent $\boldsymbol{G}$, $D$ and $\vec{F}$ in the present matrix implementation of mixed EAs and since such propagators are not easily tied to the corresponding 1-point correlation functions $\vec{\phi}$ and $\eta$, the mixed representation is not suited for a 1PI formulation, as opposed to the original and collective approaches discussed previously.

\vspace{0.5cm}

As a next step, we then turn to the (0+0)-D situation in which we evaluate the different contributions to~\eqref{eq:mixed2PIEAfinalexpression} as follows:
\begin{equation}
\begin{split}
\mathcal{ST}r\left[\mathcal{G}^{-1}_{\Phi}\mathcal{G}-\mathfrak{I}\right] = & \ \mathcal{G}^{-1}_{\Phi;\alpha\beta}\mathcal{G}^{\beta\alpha} - \left.\delta_{\alpha}\right.^{\alpha} \\
= & \ \left\{
\begin{array}{lll}
		\displaystyle{\mathcal{G}^{-1}_{\Phi;11} \mathcal{G}_{11} + \mathcal{G}^{-1}_{\Phi;12} \mathcal{G}_{21} + \mathcal{G}^{-1}_{\Phi;21} \mathcal{G}_{12} + \mathcal{G}^{-1}_{\Phi;22} \mathcal{G}_{22} - 2 \quad \mathrm{for} ~ N=1 \;,} \\
		\\
		\displaystyle{\mathcal{G}^{-1}_{\Phi;11} \mathcal{G}_{11} + \mathcal{G}^{-1}_{\Phi;22} \mathcal{G}_{22} + \mathcal{G}^{-1}_{\Phi;23} \mathcal{G}_{32} + \mathcal{G}^{-1}_{\Phi;32} \mathcal{G}_{23} + \mathcal{G}^{-1}_{\Phi;33} \mathcal{G}_{33} - 3 \quad \mathrm{for} ~ N=2\;,}
    \end{array}
\right.
\end{split}
\label{eq:mixed2PIEASTraceTerm0DON}
\end{equation}

\vspace{0.05cm}

\begin{equation}
\begin{split}
\begin{gathered}
\begin{fmffile}{Diagrams/Mixed2PIEA_Fock}
\begin{fmfgraph}(15,15)
\fmfleft{i}
\fmfright{o}
\fmfv{decor.shape=circle,decor.size=2.0thick,foreground=(0,,0,,1)}{v1}
\fmfv{decor.shape=circle,decor.size=2.0thick,foreground=(0,,0,,1)}{v2}
\fmf{phantom,tension=11}{i,v1}
\fmf{phantom,tension=11}{v2,o}
\fmf{plain,left,tension=0.4,foreground=(1,,0,,0)}{v1,v2,v1}
\fmf{wiggly,foreground=(1,,0,,0)}{v1,v2}
\end{fmfgraph}
\end{fmffile}
\end{gathered} = & \ \lambda D \sum_{a,b=1}^{N} \boldsymbol{G}^{2}_{a b} \\
= & \left\{
\begin{array}{lll}
		\displaystyle{\lambda D \boldsymbol{G}^{2}_{11} \quad \mathrm{for} ~ N=1\;,} \\
		\\
		\displaystyle{\lambda D \left(\boldsymbol{G}^{2}_{11}+2\boldsymbol{G}^{2}_{12}+\boldsymbol{G}^{2}_{22}\right) \quad \mathrm{for} ~ N=2\;,}
    \end{array}
\right.
\end{split}
\label{eq:mixed2PIEADiag10DON}
\end{equation}

\begin{equation}
\begin{split}
\begin{gathered}
\begin{fmffile}{Diagrams/Mixed2PIEA_Diag1}
\begin{fmfgraph}(15,15)
\fmfleft{i}
\fmfright{o}
\fmfv{decor.shape=circle,decor.size=2.0thick,foreground=(0,,0,,1)}{v1}
\fmfv{decor.shape=circle,decor.size=2.0thick,foreground=(0,,0,,1)}{v2}
\fmf{phantom,tension=11}{i,v1}
\fmf{phantom,tension=11}{v2,o}
\fmf{dashes,left,tension=0.4,foreground=(1,,0,,0)}{v1,v2,v1}
\fmf{plain,foreground=(1,,0,,0)}{v1,v2}
\end{fmfgraph}
\end{fmffile}
\end{gathered} = & \ \lambda \sum_{a,b=1}^{N} F_{a} \boldsymbol{G}_{a b} F_{b} \\
= & \left\{
\begin{array}{lll}
		\displaystyle{\lambda F_{1}^{2} \boldsymbol{G}_{11} \quad \mathrm{for} ~ N=1\;,} \\
		\\
		\displaystyle{\lambda \left(F_{1}^{2} \boldsymbol{G}_{11}+2F_{1}F_{2}\boldsymbol{G}_{12}+F_{2}^{2}\boldsymbol{G}_{22}\right) \quad \mathrm{for} ~ N=2\;,}
    \end{array}
\right.
\end{split}
\label{eq:mixed2PIEADiag20DON}
\end{equation}

\vspace{0.3cm}

\begin{equation}
\begin{split}
\begin{gathered}
\begin{fmffile}{Diagrams/Mixed2PIEA_Diag2}
\begin{fmfgraph}(10,10)
\fmfleft{i0,i1}
\fmfright{o0,o1}
\fmftop{v1,vUp,v2}
\fmfbottom{v3,vDown,v4}
\fmfv{decor.shape=circle,decor.size=2.0thick,foreground=(0,,0,,1)}{v1}
\fmfv{decor.shape=circle,decor.size=2.0thick,foreground=(0,,0,,1)}{v2}
\fmfv{decor.shape=circle,decor.size=2.0thick,foreground=(0,,0,,1)}{v3}
\fmfv{decor.shape=circle,decor.size=2.0thick,foreground=(0,,0,,1)}{v4}
\fmf{phantom,tension=20}{i0,v1}
\fmf{phantom,tension=20}{i1,v3}
\fmf{phantom,tension=20}{o0,v2}
\fmf{phantom,tension=20}{o1,v4}
\fmf{plain,left=0.4,tension=0.5,foreground=(1,,0,,0)}{v3,v1}
\fmf{phantom,left=0.1,tension=0.5}{v1,vUp}
\fmf{phantom,left=0.1,tension=0.5}{vUp,v2}
\fmf{plain,left=0.4,tension=0.0,foreground=(1,,0,,0)}{v1,v2}
\fmf{plain,left=0.4,tension=0.5,foreground=(1,,0,,0)}{v2,v4}
\fmf{phantom,left=0.1,tension=0.5}{v4,vDown}
\fmf{phantom,left=0.1,tension=0.5}{vDown,v3}
\fmf{plain,left=0.4,tension=0.0,foreground=(1,,0,,0)}{v4,v3}
\fmf{wiggly,tension=0.5,foreground=(1,,0,,0)}{v1,v4}
\fmf{wiggly,tension=0.5,foreground=(1,,0,,0)}{v2,v3}
\end{fmfgraph}
\end{fmffile}
\end{gathered} \hspace{0.35cm} = & \ \lambda^{2} D^{2} \sum_{a,b,c,d=1}^{N} \boldsymbol{G}_{a b} \boldsymbol{G}_{b c} \boldsymbol{G}_{c d} \boldsymbol{G}_{d a} \\
= & \left\{
\begin{array}{lll}
		\displaystyle{\lambda^{2} D^{2} \boldsymbol{G}_{11}^{4} \quad \mathrm{for} ~ N=1\;,} \\
		\\
		\displaystyle{\lambda^{2} D^{2} \left(\boldsymbol{G}_{11}^{4}+4\boldsymbol{G}_{11}^{2}\boldsymbol{G}_{12}^{2}+2\boldsymbol{G}_{12}^{4}+4\boldsymbol{G}_{11}\boldsymbol{G}_{12}^{2}\boldsymbol{G}_{22}+4\boldsymbol{G}_{12}^{2}\boldsymbol{G}_{22}^{2}+\boldsymbol{G}_{22}^{4}\right) \quad \mathrm{for} ~ N=2\;,}
    \end{array}
\right.
\end{split}
\label{eq:mixed2PIEADiag30DON}
\end{equation}

\vspace{0.6cm}

\begin{equation}
\begin{split}
\begin{gathered}
\begin{fmffile}{Diagrams/Mixed2PIEA_Diag3}
\begin{fmfgraph}(10,10)
\fmfleft{i0,i1}
\fmfright{o0,o1}
\fmftop{v1,vUp,v2}
\fmfbottom{v3,vDown,v4}
\fmfv{decor.shape=circle,decor.size=2.0thick,foreground=(0,,0,,1)}{v1}
\fmfv{decor.shape=circle,decor.size=2.0thick,foreground=(0,,0,,1)}{v2}
\fmfv{decor.shape=circle,decor.size=2.0thick,foreground=(0,,0,,1)}{v3}
\fmfv{decor.shape=circle,decor.size=2.0thick,foreground=(0,,0,,1)}{v4}
\fmf{phantom,tension=20}{i0,v1}
\fmf{phantom,tension=20}{i1,v3}
\fmf{phantom,tension=20}{o0,v2}
\fmf{phantom,tension=20}{o1,v4}
\fmf{dashes,left=0.4,tension=0.5,foreground=(1,,0,,0)}{v3,v1}
\fmf{phantom,left=0.1,tension=0.5}{v1,vUp}
\fmf{phantom,left=0.1,tension=0.5}{vUp,v2}
\fmf{dashes,left=0.4,tension=0.0,foreground=(1,,0,,0)}{v1,v2}
\fmf{dashes,left=0.4,tension=0.5,foreground=(1,,0,,0)}{v2,v4}
\fmf{phantom,left=0.1,tension=0.5}{v4,vDown}
\fmf{phantom,left=0.1,tension=0.5}{vDown,v3}
\fmf{dashes,left=0.4,tension=0.0,foreground=(1,,0,,0)}{v4,v3}
\fmf{plain,tension=0.5,foreground=(1,,0,,0)}{v1,v4}
\fmf{plain,tension=0.5,foreground=(1,,0,,0)}{v2,v3}
\end{fmfgraph}
\end{fmffile}
\end{gathered} \hspace{0.35cm} = & \ \lambda^{2} \left(\sum_{a,b=1}^{N}F_{a} \boldsymbol{G}_{ab} F_{b}\right)^{2} \\
= & \left\{
\begin{array}{lll}
		\displaystyle{\lambda^{2} F_{1} ^{4} \boldsymbol{G}_{11}^{2} \quad \mathrm{for} ~ N=1\;,} \\
		\\
		\displaystyle{\lambda^{2}\left(F_{1}^{2}\boldsymbol{G}_{11}+2F_{1}F_{2}\boldsymbol{G}_{12}+F_{2}^{2}\boldsymbol{G}_{22}\right)^{2} \quad \mathrm{for} ~ N=2\;,}
    \end{array}
\right.
\end{split}
\label{eq:mixed2PIEADiag40DON}
\end{equation}

\vspace{0.6cm}

\begin{equation}
\begin{split}
\begin{gathered}
\begin{fmffile}{Diagrams/Mixed2PIEA_Diag4}
\begin{fmfgraph}(10,10)
\fmfleft{i0,i1}
\fmfright{o0,o1}
\fmftop{v1,vUp,v2}
\fmfbottom{v3,vDown,v4}
\fmfv{decor.shape=circle,decor.size=2.0thick,foreground=(0,,0,,1)}{v1}
\fmfv{decor.shape=circle,decor.size=2.0thick,foreground=(0,,0,,1)}{v2}
\fmfv{decor.shape=circle,decor.size=2.0thick,foreground=(0,,0,,1)}{v3}
\fmfv{decor.shape=circle,decor.size=2.0thick,foreground=(0,,0,,1)}{v4}
\fmf{phantom,tension=20}{i0,v1}
\fmf{phantom,tension=20}{i1,v3}
\fmf{phantom,tension=20}{o0,v2}
\fmf{phantom,tension=20}{o1,v4}
\fmf{plain,left=0.4,tension=0.5,foreground=(1,,0,,0)}{v3,v1}
\fmf{phantom,left=0.1,tension=0.5}{v1,vUp}
\fmf{phantom,left=0.1,tension=0.5}{vUp,v2}
\fmf{plain,left=0.4,tension=0.0,foreground=(1,,0,,0)}{v1,v2}
\fmf{dashes,left=0.4,tension=0.5,foreground=(1,,0,,0)}{v2,v4}
\fmf{phantom,left=0.1,tension=0.5}{v4,vDown}
\fmf{phantom,left=0.1,tension=0.5}{vDown,v3}
\fmf{dashes,left=0.4,tension=0.0,foreground=(1,,0,,0)}{v4,v3}
\fmf{wiggly,tension=0.5,foreground=(1,,0,,0)}{v1,v4}
\fmf{plain,tension=0.5,foreground=(1,,0,,0)}{v2,v3}
\end{fmfgraph}
\end{fmffile}
\end{gathered} \hspace{0.35cm} = & \ \lambda^{2} D \left(\sum_{a=1}^{N}F_{a}^{2}\right)\left(\sum_{a,b,c=1}^{N} \boldsymbol{G}_{a b}\boldsymbol{G}_{b c}\boldsymbol{G}_{c a}\right) \\
= & \left\{
\begin{array}{lll}
		\displaystyle{\lambda^{2} D F_{1}^{2} \boldsymbol{G}_{11}^{3} \quad \mathrm{for} ~ N=1\;,} \\
		\\
		\displaystyle{\lambda^{2} D \left(F_{1}^{2}+F_{2}^{2}\right) \left(\boldsymbol{G}_{11}^{3}+3\boldsymbol{G}_{11}\boldsymbol{G}_{12}^{2}+3\boldsymbol{G}_{12}^{2}\boldsymbol{G}_{22}+\boldsymbol{G}_{22}^{3}\right) \quad \mathrm{for} ~ N=2\;,}
    \end{array}
\right.
\end{split}
\label{eq:mixed2PIEADiag50DON}
\end{equation}

\vspace{0.6cm}

\begin{equation}
\begin{split}
\begin{gathered}
\begin{fmffile}{Diagrams/Mixed2PIEA_Diag5}
\begin{fmfgraph}(10,10)
\fmfleft{i0,i1}
\fmfright{o0,o1}
\fmftop{v1,vUp,v2}
\fmfbottom{v3,vDown,v4}
\fmfv{decor.shape=circle,decor.size=2.0thick,foreground=(0,,0,,1)}{v1}
\fmfv{decor.shape=circle,decor.size=2.0thick,foreground=(0,,0,,1)}{v2}
\fmfv{decor.shape=circle,decor.size=2.0thick,foreground=(0,,0,,1)}{v3}
\fmfv{decor.shape=circle,decor.size=2.0thick,foreground=(0,,0,,1)}{v4}
\fmf{phantom,tension=20}{i0,v1}
\fmf{phantom,tension=20}{i1,v3}
\fmf{phantom,tension=20}{o0,v2}
\fmf{phantom,tension=20}{o1,v4}
\fmf{dashes,left=0.4,tension=0.5,foreground=(1,,0,,0)}{v3,v1}
\fmf{phantom,left=0.1,tension=0.5}{v1,vUp}
\fmf{phantom,left=0.1,tension=0.5}{vUp,v2}
\fmf{plain,left=0.4,tension=0.0,foreground=(1,,0,,0)}{v1,v2}
\fmf{dashes,left=0.4,tension=0.5,foreground=(1,,0,,0)}{v2,v4}
\fmf{phantom,left=0.1,tension=0.5}{v4,vDown}
\fmf{phantom,left=0.1,tension=0.5}{vDown,v3}
\fmf{plain,left=0.4,tension=0.0,foreground=(1,,0,,0)}{v4,v3}
\fmf{wiggly,tension=0.5,foreground=(1,,0,,0)}{v1,v4}
\fmf{plain,tension=0.5,foreground=(1,,0,,0)}{v2,v3}
\end{fmfgraph}
\end{fmffile}
\end{gathered} \hspace{0.35cm} = & \ \lambda^{2}D\sum_{a,b,c,d=1}^{N} F_{a} \boldsymbol{G}_{a b} \boldsymbol{G}_{b c} \boldsymbol{G}_{c d} F_{d} \\
= & \left\{
\begin{array}{lll}
		\displaystyle{\lambda^{2} D F_{1}^{2} \boldsymbol{G}_{11}^{3} \quad \mathrm{for} ~~ N=1\;,} \\
		\\
		\displaystyle{\lambda^{2} D \big(F_{1}^{2} \boldsymbol{G}_{11}^{3} + 2 F_{1}F_{2} \boldsymbol{G}_{11}^{2}\boldsymbol{G}_{12} + 2 F_{1}^{2}\boldsymbol{G}_{11}\boldsymbol{G}_{12}^{2} + F_{2}^{2} \boldsymbol{G}_{11} \boldsymbol{G}_{12}^{2} + 2 F_{1} F_{2} \boldsymbol{G}_{12}^{3}} \\
		\hspace{0.8cm} \displaystyle{+2F_{1}F_{2}\boldsymbol{G}_{11}\boldsymbol{G}_{12}\boldsymbol{G}_{22} + F_{1}^{2}\boldsymbol{G}_{12}^{2}\boldsymbol{G}_{22}+2 F_{2}^{2} \boldsymbol{G}_{12}^{2} \boldsymbol{G}_{22} + 2 F_{1} F_{2} \boldsymbol{G}_{12} \boldsymbol{G}_{22}^{2}} \\
		\hspace{0.8cm} \displaystyle{+ F_{2}^{2}\boldsymbol{G}_{22}^{3}\big) \quad \mathrm{for} ~ N=2\;,}
    \end{array}
\right.
\end{split}
\label{eq:mixed2PIEADiag60DON}
\end{equation}

\vspace{0.6cm}

\begin{equation}
\begin{split}
\begin{gathered}
\begin{fmffile}{Diagrams/Mixed2PIEA_Diag6}
\begin{fmfgraph}(10,10)
\fmfleft{i0,i1}
\fmfright{o0,o1}
\fmftop{v1,vUp,v2}
\fmfbottom{v3,vDown,v4}
\fmfv{decor.shape=circle,decor.size=2.0thick,foreground=(0,,0,,1)}{v1}
\fmfv{decor.shape=circle,decor.size=2.0thick,foreground=(0,,0,,1)}{v2}
\fmfv{decor.shape=circle,decor.size=2.0thick,foreground=(0,,0,,1)}{v3}
\fmfv{decor.shape=circle,decor.size=2.0thick,foreground=(0,,0,,1)}{v4}
\fmf{phantom,tension=20}{i0,v1}
\fmf{phantom,tension=20}{i1,v3}
\fmf{phantom,tension=20}{o0,v2}
\fmf{phantom,tension=20}{o1,v4}
\fmf{plain,left=0.4,tension=0.5,foreground=(1,,0,,0)}{v3,v1}
\fmf{phantom,left=0.1,tension=0.5}{v1,vUp}
\fmf{phantom,left=0.1,tension=0.5}{vUp,v2}
\fmf{plain,left=0.4,tension=0.0,foreground=(1,,0,,0)}{v1,v2}
\fmf{dashes,left=0.4,tension=0.5,foreground=(1,,0,,0)}{v2,v4}
\fmf{phantom,left=0.1,tension=0.5}{v4,vDown}
\fmf{phantom,left=0.1,tension=0.5}{vDown,v3}
\fmf{dashes,left=0.4,tension=0.0,foreground=(1,,0,,0)}{v4,v3}
\fmf{dashes,tension=0.5,foreground=(1,,0,,0)}{v1,v4}
\fmf{dashes,tension=0.5,foreground=(1,,0,,0)}{v2,v3}
\end{fmfgraph}
\end{fmffile}
\end{gathered} \hspace{0.35cm} = & \ \lambda^{2} \sum_{a,b,c,d=1}^{N} \boldsymbol{G}_{a b} \boldsymbol{G}_{b c} F_{a} F_{c} F_{d}^{2} \\
= & \left\{
\begin{array}{lll}
		\displaystyle{\lambda^{2} F_{1}^{4} \boldsymbol{G}_{11}^{2} \quad \mathrm{for} ~ N=1\;,} \\
		\\
		\displaystyle{\lambda^{2} \big( F_{1}^{4} \boldsymbol{G}_{11}^{2} + F_{1}^{2} F_{2}^{2} \boldsymbol{G}_{11}^{2} + 2 F_{1}^{3} F_{2} \boldsymbol{G}_{11} \boldsymbol{G}_{12} + 2 F_{1} F_{2}^{3} \boldsymbol{G}_{11} \boldsymbol{G}_{12} + F_{1}^{4} \boldsymbol{G}_{12}^{2} + 2 F_{1}^{2} F_{2}^{2} \boldsymbol{G}_{12}^{2}} \\
		\hspace{0.45cm} \displaystyle{+ F_{2}^{4}\boldsymbol{G}_{12}^{2} + 2 F_{1}^{3} F_{2} \boldsymbol{G}_{12} \boldsymbol{G}_{22} + 2 F_{1} F_{2}^{3} \boldsymbol{G}_{12} \boldsymbol{G}_{22} + F_{1}^{2} F_{2}^{2} \boldsymbol{G}_{22}^{2} + F_{2}^{4} \boldsymbol{G}_{22}^{2} \big) \quad \mathrm{for} ~ N=2\;,}
    \end{array}
\right.
\end{split}
\label{eq:mixed2PIEADiag70DON}
\end{equation}

\vspace{0.6cm}

\begin{equation}
\begin{split}
\begin{gathered}
\begin{fmffile}{Diagrams/Mixed2PIEA_Diag7}
\begin{fmfgraph}(15,15)
\fmfleft{i}
\fmfright{o}
\fmftop{vUpLeft,vUp,vUpRight}
\fmfbottom{vDownLeft,vDown,vDownRight}
\fmfv{decor.shape=circle,decor.size=2.0thick,foreground=(0,,0,,1)}{v1}
\fmfv{decor.shape=circle,decor.size=2.0thick,foreground=(0,,0,,1)}{v2}
\fmfv{decor.shape=circle,decor.size=2.0thick,foreground=(0,,0,,1)}{v3}
\fmfv{decor.shape=circle,decor.size=2.0thick,foreground=(0,,0,,1)}{v4}
\fmfv{decor.shape=circle,decor.size=2.0thick,foreground=(0,,0,,1)}{v5}
\fmfv{decor.shape=circle,decor.size=2.0thick,foreground=(0,,0,,1)}{v6}
\fmf{phantom,tension=1}{i,v1}
\fmf{phantom,tension=1}{v2,o}
\fmf{phantom,tension=14.0}{v3,vUpLeft}
\fmf{phantom,tension=2.0}{v3,vUpRight}
\fmf{phantom,tension=4.0}{v3,i}
\fmf{phantom,tension=2.0}{v4,vUpLeft}
\fmf{phantom,tension=14.0}{v4,vUpRight}
\fmf{phantom,tension=4.0}{v4,o}
\fmf{phantom,tension=14.0}{v5,vDownLeft}
\fmf{phantom,tension=2.0}{v5,vDownRight}
\fmf{phantom,tension=4.0}{v5,i}
\fmf{phantom,tension=2.0}{v6,vDownLeft}
\fmf{phantom,tension=14.0}{v6,vDownRight}
\fmf{phantom,tension=4.0}{v6,o}
\fmf{wiggly,tension=0,foreground=(1,,0,,0)}{v1,v2}
\fmf{wiggly,tension=0.6,foreground=(1,,0,,0)}{v3,v6}
\fmf{wiggly,tension=0.6,foreground=(1,,0,,0)}{v5,v4}
\fmf{plain,left=0.18,tension=0,foreground=(1,,0,,0)}{v1,v3}
\fmf{plain,left=0.42,tension=0,foreground=(1,,0,,0)}{v3,v4}
\fmf{plain,left=0.18,tension=0,foreground=(1,,0,,0)}{v4,v2}
\fmf{plain,left=0.18,tension=0,foreground=(1,,0,,0)}{v2,v6}
\fmf{plain,left=0.42,tension=0,foreground=(1,,0,,0)}{v6,v5}
\fmf{plain,left=0.18,tension=0,foreground=(1,,0,,0)}{v5,v1}
\end{fmfgraph}
\end{fmffile}
\end{gathered} \hspace{0.3cm} = \hspace{0.5cm} \begin{gathered}
\begin{fmffile}{Diagrams/Mixed2PIEA_Diag8}
\begin{fmfgraph}(12.5,12.5)
\fmfleft{i0,i1}
\fmfright{o0,o1}
\fmftop{v1,vUp,v2}
\fmfbottom{v3,vDown,v4}
\fmfv{decor.shape=circle,decor.size=2.0thick,foreground=(0,,0,,1)}{v1}
\fmfv{decor.shape=circle,decor.size=2.0thick,foreground=(0,,0,,1)}{v2}
\fmfv{decor.shape=circle,decor.size=2.0thick,foreground=(0,,0,,1)}{v3}
\fmfv{decor.shape=circle,decor.size=2.0thick,foreground=(0,,0,,1)}{v4}
\fmfv{decor.shape=circle,decor.size=2.0thick,foreground=(0,,0,,1)}{v5}
\fmfv{decor.shape=circle,decor.size=2.0thick,foreground=(0,,0,,1)}{v6}
\fmf{phantom,tension=20}{i0,v1}
\fmf{phantom,tension=20}{i1,v3}
\fmf{phantom,tension=20}{o0,v2}
\fmf{phantom,tension=20}{o1,v4}
\fmf{phantom,tension=0.005}{v5,v6}
\fmf{wiggly,left=0.4,tension=0,foreground=(1,,0,,0)}{v3,v1}
\fmf{phantom,left=0.1,tension=0}{v1,vUp}
\fmf{phantom,left=0.1,tension=0}{vUp,v2}
\fmf{plain,left=0.25,tension=0,foreground=(1,,0,,0)}{v1,v2}
\fmf{wiggly,left=0.4,tension=0,foreground=(1,,0,,0)}{v2,v4}
\fmf{phantom,left=0.1,tension=0}{v4,vDown}
\fmf{phantom,left=0.1,tension=0}{vDown,v3}
\fmf{plain,right=0.25,tension=0,foreground=(1,,0,,0)}{v3,v4}
\fmf{plain,left=0.2,tension=0.01,foreground=(1,,0,,0)}{v1,v5}
\fmf{plain,left=0.2,tension=0.01,foreground=(1,,0,,0)}{v5,v3}
\fmf{plain,right=0.2,tension=0.01,foreground=(1,,0,,0)}{v2,v6}
\fmf{plain,right=0.2,tension=0.01,foreground=(1,,0,,0)}{v6,v4}
\fmf{wiggly,tension=0,foreground=(1,,0,,0)}{v5,v6}
\end{fmfgraph}
\end{fmffile}
\end{gathered} \hspace{0.5cm} = & \ \lambda^{3} D^{3} \sum_{a,b,c,d,e,f=1}^{N} \boldsymbol{G}_{a b} \boldsymbol{G}_{b c} \boldsymbol{G}_{c d} \boldsymbol{G}_{d e} \boldsymbol{G}_{e f} \boldsymbol{G}_{f a} \\
= & \left\{
\begin{array}{lll}
		\displaystyle{\lambda^{3}D^{3}\boldsymbol{G}_{11}^{6} \quad \mathrm{for} ~ N=1\;,} \\
		\\
		\displaystyle{\lambda^{3}D^{3}\big(\boldsymbol{G}_{11}^{6}+6\boldsymbol{G}_{11}^{4}\boldsymbol{G}_{12}^{2}+9\boldsymbol{G}_{11}^{2}\boldsymbol{G}_{12}^{4}+2\boldsymbol{G}_{12}^{6}+6\boldsymbol{G}_{11}^{3}\boldsymbol{G}_{12}^{2}\boldsymbol{G}_{22}} \\
		\hspace{1.0cm} \displaystyle{+12\boldsymbol{G}_{11}\boldsymbol{G}_{12}^{4}\boldsymbol{G}_{22}+6\boldsymbol{G}_{11}^{2}\boldsymbol{G}_{12}^{2}\boldsymbol{G}_{22}^{2}+9\boldsymbol{G}_{12}^{4}\boldsymbol{G}_{22}^{2}} \\
		\hspace{1.0cm} \displaystyle{+6\boldsymbol{G}_{11}\boldsymbol{G}_{12}^{2}\boldsymbol{G}_{22}^{3}+6\boldsymbol{G}_{12}^{2}\boldsymbol{G}_{22}^{4}+\boldsymbol{G}_{22}^{6}\big) \quad \mathrm{for} ~ N=2\;,}
    \end{array}
\right.
\end{split}
\label{eq:mixed2PIEADiag80DON}
\end{equation}

\vspace{0.6cm}

\begin{equation}
\begin{split}
\begin{gathered}
\begin{fmffile}{Diagrams/Mixed2PIEA_Diag9}
\begin{fmfgraph}(12.5,12.5)
\fmfleft{i0,i1}
\fmfright{o0,o1}
\fmftop{v1,vUp,v2}
\fmfbottom{v3,vDown,v4}
\fmfv{decor.shape=circle,decor.size=2.0thick,foreground=(0,,0,,1)}{v1}
\fmfv{decor.shape=circle,decor.size=2.0thick,foreground=(0,,0,,1)}{v2}
\fmfv{decor.shape=circle,decor.size=2.0thick,foreground=(0,,0,,1)}{v3}
\fmfv{decor.shape=circle,decor.size=2.0thick,foreground=(0,,0,,1)}{v4}
\fmfv{decor.shape=circle,decor.size=2.0thick,foreground=(0,,0,,1)}{v5}
\fmfv{decor.shape=circle,decor.size=2.0thick,foreground=(0,,0,,1)}{v6}
\fmf{phantom,tension=20}{i0,v1}
\fmf{phantom,tension=20}{i1,v3}
\fmf{phantom,tension=20}{o0,v2}
\fmf{phantom,tension=20}{o1,v4}
\fmf{phantom,tension=0.005}{v5,v6}
\fmf{plain,left=0.4,tension=0,foreground=(1,,0,,0)}{v3,v1}
\fmf{phantom,left=0.1,tension=0}{v1,vUp}
\fmf{phantom,left=0.1,tension=0}{vUp,v2}
\fmf{wiggly,left=0.25,tension=0,foreground=(1,,0,,0)}{v1,v2}
\fmf{plain,left=0.4,tension=0,foreground=(1,,0,,0)}{v2,v4}
\fmf{phantom,left=0.1,tension=0}{v4,vDown}
\fmf{phantom,left=0.1,tension=0}{vDown,v3}
\fmf{wiggly,right=0.25,tension=0,foreground=(1,,0,,0)}{v3,v4}
\fmf{plain,left=0.2,tension=0.01,foreground=(1,,0,,0)}{v1,v5}
\fmf{plain,left=0.2,tension=0.01,foreground=(1,,0,,0)}{v5,v3}
\fmf{plain,right=0.2,tension=0.01,foreground=(1,,0,,0)}{v2,v6}
\fmf{plain,right=0.2,tension=0.01,foreground=(1,,0,,0)}{v6,v4}
\fmf{wiggly,tension=0,foreground=(1,,0,,0)}{v5,v6}
\end{fmfgraph}
\end{fmffile}
\end{gathered} \hspace{0.4cm} = & \ \lambda^{3} D^{3} \left(\sum_{a,b,c=1}^{N} \boldsymbol{G}_{a b}\boldsymbol{G}_{b c}\boldsymbol{G}_{c a}\right)^{2} \\
= & \left\{
\begin{array}{lll}
		\displaystyle{\lambda^{3}D^{3}\boldsymbol{G}_{11}^{6} \quad \mathrm{for} ~ N=1\;,} \\
		\\
		\displaystyle{\lambda^{3}D^{3}\left(\boldsymbol{G}_{11}^{3}+3\boldsymbol{G}_{11}\boldsymbol{G}_{12}^{2}+3\boldsymbol{G}_{12}^{2}\boldsymbol{G}_{22}+\boldsymbol{G}_{22}^{3}\right)^{2} \quad \mathrm{for} ~ N=2\;,}
    \end{array}
\right.
\end{split}
\label{eq:mixed2PIEADiag90DON}
\end{equation}
where we have exploited the symmetry property of $\boldsymbol{G}$ (i.e. $\boldsymbol{G}_{a b}=\boldsymbol{G}_{b a}$ $\forall a,b$) at $N=2$. Moreover, the mixed classical action reads $S_{\mathrm{mix}}(\Phi)=m^{2}\phi^{2}_{N}/2+\eta^{2}/2+i\sqrt{\lambda/12} \ \eta\phi^{2}_{N}$ in the present case and the components of $\mathcal{G}_{\Phi}^{-1}$ in~\eqref{eq:mixed2PIEASTraceTerm0DON} are found from the (0+0)-D version of~\eqref{eq:mixed2PIEAMathcalGPhi}, i.e.:
\begin{equation}
\begin{split}
\mathcal{G}^{-1}_{\Phi} = \begin{pmatrix}
\left(m^2 + i\sqrt{\frac{\lambda}{3}}\eta\right)\mathbb{I}_{N} & i\sqrt{\frac{\lambda}{3}}\vec{\phi} \\
i\sqrt{\frac{\lambda}{3}}\vec{\phi}^\mathrm{T} & 1 \end{pmatrix} \;.
\end{split}
\label{eq:mixed2PIEAMathcalGPhi0D}
\end{equation}
According to~\eqref{eq:mixed2PIEASTraceTerm0DON} to~\eqref{eq:mixed2PIEAMathcalGPhi0D},~\eqref{eq:mixed2PIEAfinalexpression} reduces in (0+0)-D to:
\begin{itemize}
\item For $N=1$:
\begin{equation}
\begin{split}
\scalebox{0.92}{${\displaystyle \Gamma_{\mathrm{mix}}^{(\mathrm{2PI})}\big(\Phi,\mathcal{G}\big) = }$} & \ \scalebox{0.92}{${\displaystyle S_{\mathrm{mix}}(\Phi) }$} \\
& \scalebox{0.92}{${\displaystyle + \hbar \Bigg[ -\frac{1}{2}\ln\big(2\pi\boldsymbol{G}_{11}\big) -\frac{1}{2}\ln(D) + \frac{1}{2} \Bigg(\left(m^{2}+i\sqrt{\frac{\lambda}{3}}\eta\right)\boldsymbol{G}_{11} + 2 i \sqrt{\frac{\lambda}{3}} \phi_{1} F_{1} + D \Bigg) - 1 \Bigg] }$} \\
& \scalebox{0.92}{${\displaystyle + \hbar^{2}\Bigg[ \frac{1}{12} \lambda \boldsymbol{G}_{11} \left(2 F_{1}^{2} + D \boldsymbol{G}_{11}\right) \Bigg] }$} \\
& \scalebox{0.92}{${\displaystyle + \hbar^{3}\Bigg[ \frac{1}{72} \lambda^{2} \boldsymbol{G}_{11}^2 \left(10 F_{1}^{4} + 12 D F_{1}^{2} \boldsymbol{G}_{11} + D^2 \boldsymbol{G}_{11}^2\right) \Bigg] }$} \\
& \scalebox{0.92}{${\displaystyle + \hbar^{4}\Bigg[ \frac{5}{324} \lambda^{3} D^3 \boldsymbol{G}_{11}^6 + \mathcal{O}\big(F^{2}\big) \Bigg] }$} \\
& \scalebox{0.92}{${\displaystyle + \mathcal{O}\big(\hbar^{5}\big)\;. }$}
\end{split}
\label{eq:mixed2PIEAfinalexpressionN10DON}
\end{equation}

\item For $N=2$:
\begin{equation}
\begin{split}
\scalebox{0.93}{${\displaystyle \Gamma_{\mathrm{mix}}^{(\mathrm{2PI})}\big(\Phi,\mathcal{G}\big) = }$} & \ \scalebox{0.93}{${\displaystyle S_{\mathrm{mix}}(\Phi) }$} \\
& \scalebox{0.93}{${\displaystyle + \hbar \Bigg[ -\frac{1}{2}\ln\big(2\pi\boldsymbol{G}_{11}\big) -\frac{1}{2}\ln\big(2\pi\boldsymbol{G}_{22}\big) -\frac{1}{2}\ln(D) + \frac{1}{2} \Bigg(\left(m^{2}+i\sqrt{\frac{\lambda}{3}}\eta\right)\left(\boldsymbol{G}_{11}+\boldsymbol{G}_{22}\right) }$} \\
& \hspace{0.8cm} \scalebox{0.93}{${\displaystyle + 2 i \sqrt{\frac{\lambda}{3}} \phi_{2} F_{2} + D \Bigg) - \frac{3}{2} \Bigg] }$} \\
& \scalebox{0.93}{${\displaystyle + \hbar^{2}\Bigg[ \frac{1}{12} \lambda \left(2 F_{1}^2 \boldsymbol{G}_{11} + 4 F_{1} F_{2} \boldsymbol{G}_{12} + 2 F_{2}^2 \boldsymbol{G}_{22} + D \left(\boldsymbol{G}_{11}^2 + 2 \boldsymbol{G}_{12}^2 + \boldsymbol{G}_{22}^2\right)\right) \Bigg] }$} \\
& \scalebox{0.93}{${\displaystyle + \hbar^{3}\Bigg[ \frac{1}{72} \lambda^{2} \Big(2 F_{1}^4 \left(5 \boldsymbol{G}_{11}^2 + 4 \boldsymbol{G}_{12}^2\right) + 8 F_{1}^3 F_{2} \boldsymbol{G}_{12} \left(3 \boldsymbol{G}_{11} + 2 \boldsymbol{G}_{22}\right) + 2 F_{2}^4 \left(4 \boldsymbol{G}_{12}^2 + 5 \boldsymbol{G}_{22}^2\right) }$} \\
& \hspace{0.8cm} \scalebox{0.93}{${\displaystyle + 4 D F_{2}^2 \left(\boldsymbol{G}_{11}^3 + 5 \boldsymbol{G}_{11} \boldsymbol{G}_{12}^2 + 7 \boldsymbol{G}_{12}^2 \boldsymbol{G}_{22} + 3 \boldsymbol{G}_{22}^3\right) + D^2 \big(\boldsymbol{G}_{11}^4 + 4 \boldsymbol{G}_{11}^2 \boldsymbol{G}_{12}^2 + 2 \boldsymbol{G}_{12}^4 }$} \\
& \hspace{0.8cm} \scalebox{0.93}{${\displaystyle + 4 \boldsymbol{G}_{11} \boldsymbol{G}_{12}^2 \boldsymbol{G}_{22} + 4 \boldsymbol{G}_{12}^2 \boldsymbol{G}_{22}^2 + \boldsymbol{G}_{22}^4\big) + 8 F_{1} F_{2} \boldsymbol{G}_{12} \big(F_{2}^2 \left(2 \boldsymbol{G}_{11} + 3 \boldsymbol{G}_{22}\right) + 2 D \big(\boldsymbol{G}_{11}^2 }$} \\
& \hspace{0.8cm} \scalebox{0.93}{${\displaystyle + \boldsymbol{G}_{12}^2 + \boldsymbol{G}_{11} \boldsymbol{G}_{22} + \boldsymbol{G}_{22}^2\big)\big) + 4 F_{1}^2 \big(F_{2}^2 \left(2 \boldsymbol{G}_{11}^2 + 6 \boldsymbol{G}_{12}^2 + \boldsymbol{G}_{11} \boldsymbol{G}_{22} + 2 \boldsymbol{G}_{22}^2\right) }$} \\
& \hspace{0.8cm} \scalebox{0.93}{${\displaystyle + D \big(3 \boldsymbol{G}_{11}^3 + 7 \boldsymbol{G}_{11} \boldsymbol{G}_{12}^2 + 5 \boldsymbol{G}_{12}^2 \boldsymbol{G}_{22} + \boldsymbol{G}_{22}^3\big)\big)\Big) \Bigg] }$} \\
& \scalebox{0.93}{${\displaystyle + \hbar^{4}\Bigg[ \frac{1}{324} \lambda^{3} D^3 \Big(5 \boldsymbol{G}_{11}^6 + 30 \boldsymbol{G}_{11}^4 \boldsymbol{G}_{12}^2 + 8 \boldsymbol{G}_{12}^6 + 45 \boldsymbol{G}_{12}^4 \boldsymbol{G}_{22}^2 + 30 \boldsymbol{G}_{12}^2 \boldsymbol{G}_{22}^4 + 5 \boldsymbol{G}_{22}^6 }$} \\
& \hspace{0.8cm} \scalebox{0.93}{${\displaystyle + 3 \boldsymbol{G}_{11}^2 \big(15 \boldsymbol{G}_{12}^4 + 8 \boldsymbol{G}_{12}^2 \boldsymbol{G}_{22}^2\big) + 2 \boldsymbol{G}_{11}^3 \left(15 \boldsymbol{G}_{12}^2 \boldsymbol{G}_{22} + \boldsymbol{G}_{22}^3\right) + 6 \boldsymbol{G}_{11} \big(11 \boldsymbol{G}_{12}^4 \boldsymbol{G}_{22} }$} \\
& \hspace{0.8cm} \scalebox{0.93}{${\displaystyle + 5 \boldsymbol{G}_{12}^2 \boldsymbol{G}_{22}^3\big)\Big) + \mathcal{O}\Big(\vec{F}^{2}\Big) \Bigg] }$} \\
& \scalebox{0.93}{${\displaystyle + \mathcal{O}\big(\hbar^{5}\big) \;.}$}
\end{split}
\label{eq:mixed2PIEAfinalexpressionN20DON}
\end{equation}
\end{itemize}
The gap equations associated to $\Gamma_{\mathrm{mix}}^{(\mathrm{2PI})}\big(\Phi,\mathcal{G}\big)$ are determined by differentiating~\eqref{eq:mixed2PIEAfinalexpressionN10DON} and \eqref{eq:mixed2PIEAfinalexpressionN20DON} (and by considering the symmetry of $\boldsymbol{G}$ discussed right below~\eqref{eq:2PIEAfinalexpressionN20DON}) as follows:
\begin{itemize}
\item For $N=1$:
\begin{equation}
0 = \left.\frac{\partial \Gamma_{\mathrm{mix}}^{(\mathrm{2PI})}\big(\Phi,\mathcal{G}\big)}{\partial \phi_{1}}\right|_{\Phi=\overline{\Phi} \atop \mathcal{G}=\overline{\mathcal{G}}} = m^{2} \overline{\phi}_{1} + i \sqrt{\frac{\lambda}{3}} \ \overline{\eta} \ \overline{\phi}_{1} + \hbar\left(i \sqrt{\frac{\lambda}{3}} \ \overline{F}_{1} \right) \;,
\label{eq:mixed2PIEAGapEquationphiNN10DON}
\end{equation}
\begin{equation}
0 = \left.\frac{\partial \Gamma_{\mathrm{mix}}^{(\mathrm{2PI})}\big(\Phi,\mathcal{G}\big)}{\partial \eta}\right|_{\Phi=\overline{\Phi} \atop \mathcal{G}=\overline{\mathcal{G}}} = \overline{\eta} + i \sqrt{\frac{\lambda}{12}} \ \overline{\phi}_{1}^{2} + \hbar\left(i\sqrt{\frac{\lambda}{12}} \ \overline{\boldsymbol{G}}_{11}\right)\;,
\label{eq:mixed2PIEAGapEquationetaN10DON}
\end{equation}
\begin{equation}
\begin{split}
0 = \left.\frac{\partial \Gamma_{\mathrm{mix}}^{(\mathrm{2PI})}\big(\Phi,\mathcal{G}\big)}{\partial \boldsymbol{G}_{11}}\right|_{\Phi=\overline{\Phi} \atop \mathcal{G}=\overline{\mathcal{G}}} = & \ \hbar \left[-\frac{1}{2}\overline{\boldsymbol{G}}_{11}^{-1} + \frac{1}{2} \left(m^{2} + i \sqrt{\frac{\lambda}{3}} \ \overline{\eta}\right)\right] + \hbar^{2} \left(\frac{1}{6} \overline{F}_{1}^{2}\lambda + \frac{1}{6} \overline{D} \ \overline{\boldsymbol{G}}_{11} \lambda\right) \\
& - \hbar^3 \left(\frac{5}{18} \overline{F}_{1}^4 \overline{\boldsymbol{G}}_{11} \lambda^2 + \frac{1}{2} \overline{D} \ \overline{F}_{1}^2 \overline{\boldsymbol{G}}_{11}^2 \lambda^2 + \frac{1}{18} \overline{D}^2 \overline{\boldsymbol{G}}_{11}^3 \lambda^2\right) \\
& + \hbar^{4}\left[ \frac{5}{54} \overline{D}^3 \overline{\boldsymbol{G}}_{11}^5 \lambda^3 + \mathcal{O}\Big(\overline{F}^{2}\Big) \right] \\
& + \mathcal{O}\big(\hbar^{5}\big)\;,
\end{split}
\label{eq:mixed2PIEAGapEquationG11N10DON}
\end{equation}
\begin{equation}
\begin{split}
0 = \left.\frac{\partial \Gamma_{\mathrm{mix}}^{(\mathrm{2PI})}\big(\Phi,\mathcal{G}\big)}{\partial D}\right|_{\Phi=\overline{\Phi} \atop \mathcal{G}=\overline{\mathcal{G}}} = & \ \hbar \left(- \frac{1}{2}\overline{D}^{-1} + \frac{1}{2}\right) + \hbar^{2} \left(\frac{1}{12} \overline{\boldsymbol{G}}_{11}^2 \lambda\right) \\
& - \hbar^3 \left(\frac{1}{6} \overline{F}_{1}^2 \overline{\boldsymbol{G}}_{11}^3 \lambda^2 + \frac{1}{36} \overline{D} \ \overline{\boldsymbol{G}}_{11}^4 \lambda^2 \right) \\
& + \hbar^4 \left[\frac{5}{108} \overline{D}^2 \overline{\boldsymbol{G}}_{11}^6 \lambda^3 + \mathcal{O}\Big(\overline{F}^{2}\Big)\right] \\
& + \mathcal{O}\big(\hbar^{5}\big)\;,
\end{split}
\label{eq:mixed2PIEAGapEquationDN10DON}
\end{equation}
\begin{equation}
\begin{split}
0 = \left.\frac{\partial \Gamma_{\mathrm{mix}}^{(\mathrm{2PI})}\big(\Phi,\mathcal{G}\big)}{\partial F_{1}}\right|_{\Phi=\overline{\Phi} \atop \mathcal{G}=\overline{\mathcal{G}}} = & \ \hbar\left(i\sqrt{\frac{\lambda}{3}} \ \overline{\phi}_{1}\right) + \hbar^2 \left(\frac{1}{3} \overline{F}_{1} \overline{\boldsymbol{G}}_{11} \lambda \right) \\
& - \hbar^3 \left(\frac{5}{9} \overline{F}_{1}^3 \overline{\boldsymbol{G}}_{11}^2 \lambda^2 + \frac{1}{3} \overline{D} \ \overline{F}_{1} \overline{\boldsymbol{G}}_{11}^3 \lambda^2\right) \\
& + \mathcal{O}\big(\hbar^{4}\big)\;.
\end{split}
\label{eq:mixed2PIEAGapEquationF1N10DON}
\end{equation}

\item For $N=2$:
\begin{equation}
0 = \left.\frac{\partial \Gamma_{\mathrm{mix}}^{(\mathrm{2PI})}\big(\Phi,\mathcal{G}\big)}{\partial \phi_{2}}\right|_{\Phi=\overline{\Phi} \atop \mathcal{G}=\overline{\mathcal{G}}} = m^{2} \overline{\phi}_{2} + i \sqrt{\frac{\lambda}{3}} \ \overline{\eta} \ \overline{\phi}_{2} + \hbar\left(i \sqrt{\frac{\lambda}{3}} \ \overline{F}_{2} \right)\;,
\label{eq:mixed2PIEAGapEquationphiNN20DON}
\end{equation}
\begin{equation}
0 = \left.\frac{\partial \Gamma_{\mathrm{mix}}^{(\mathrm{2PI})}\big(\Phi,\mathcal{G}\big)}{\partial \eta}\right|_{\Phi=\overline{\Phi} \atop \mathcal{G}=\overline{\mathcal{G}}} = \overline{\eta} + i \sqrt{\frac{\lambda}{12}} \ \overline{\phi}_{2}^{2} + \hbar\left[i\sqrt{\frac{\lambda}{12}}\left(\overline{\boldsymbol{G}}_{11}+\overline{\boldsymbol{G}}_{22}\right)\right] \;,
\label{eq:mixed2PIEAGapEquationetaN20DON}
\end{equation}
\begin{equation}
\begin{split}
0 = \left.\frac{\partial \Gamma_{\mathrm{mix}}^{(\mathrm{2PI})}\big(\Phi,\mathcal{G}\big)}{\partial \boldsymbol{G}_{11}}\right|_{\Phi=\overline{\Phi} \atop \mathcal{G}=\overline{\mathcal{G}}} = & \ \hbar\left[ -\frac{1}{2}\overline{\boldsymbol{G}}_{11}^{-1} + \frac{1}{2} \left(m^{2} + i\sqrt{\frac{\lambda}{3}} \ \overline{\eta}\right)\right] + \hbar^{2}\left[\frac{1}{6} \lambda \left( \overline{F}_{1}^{2} + \overline{D} \ \overline{\boldsymbol{G}}_{11}\right)\right] \\
& - \hbar^{3}\Bigg[ \frac{1}{18} \lambda^{2} \Big(5 \overline{F}_{1}^4 \overline{\boldsymbol{G}}_{11} + 6 \overline{F}_{1}^3 \overline{F}_{2} \overline{\boldsymbol{G}}_{12} + 4 \overline{F}_{1} \overline{F}_{2} \overline{\boldsymbol{G}}_{12} \Big(\overline{F}_{2}^2 + \overline{D} \big(2 \overline{\boldsymbol{G}}_{11} \\
& \hspace{0.9cm} + \overline{\boldsymbol{G}}_{22}\big)\Big) + \overline{F}_{1}^2 \Big(\overline{D} \Big(9 \overline{\boldsymbol{G}}_{11}^2 + 7 \overline{\boldsymbol{G}}_{12}^2\Big) + \overline{F}_{2}^2 \big(4 \overline{\boldsymbol{G}}_{11} + \overline{\boldsymbol{G}}_{22}\big)\Big) \\
& \hspace{0.9cm} + \overline{D} \Big(\overline{F}_{2}^2 \left(3 \overline{\boldsymbol{G}}_{11}^2 + 5 \overline{\boldsymbol{G}}_{12}^2\right) + \overline{D} \Big(\overline{\boldsymbol{G}}_{11}^3 + 2 \overline{\boldsymbol{G}}_{11} \overline{\boldsymbol{G}}_{12}^2 \\
& \hspace{0.9cm} + \overline{\boldsymbol{G}}_{12}^2 \overline{\boldsymbol{G}}_{22}\Big)\Big)\Big) \Bigg] \\
& + \hbar^{4} \Bigg[ \frac{1}{54} \lambda^{3} \overline{D}^3 \Big(5 \overline{\boldsymbol{G}}_{11}^5 + 20 \overline{\boldsymbol{G}}_{11}^3 \overline{\boldsymbol{G}}_{12}^2 + 11 \overline{\boldsymbol{G}}_{12}^4 \overline{\boldsymbol{G}}_{22} + 5 \overline{\boldsymbol{G}}_{12}^2 \overline{\boldsymbol{G}}_{22}^3 \\
& \hspace{0.9cm} + \overline{\boldsymbol{G}}_{11} \Big(15 \overline{\boldsymbol{G}}_{12}^4 + 8 \overline{\boldsymbol{G}}_{12}^2 \overline{\boldsymbol{G}}_{22}^2\Big) + \overline{\boldsymbol{G}}_{11}^2 \Big(15 \overline{\boldsymbol{G}}_{12}^2 \overline{\boldsymbol{G}}_{22} + \overline{\boldsymbol{G}}_{22}^3\Big)\Big) \\
& \hspace{0.9cm} + \mathcal{O}\hspace{-0.07cm}\left(\vec{\overline{F}}^{2}\right) \Bigg] \\
& + \mathcal{O}\big(\hbar^{5}\big) \;,
\end{split}
\label{eq:mixed2PIEAGapEquationG11N20DON}
\end{equation}
\begin{equation}
\begin{split}
0 = \left.\frac{\partial \Gamma_{\mathrm{mix}}^{(\mathrm{2PI})}\big(\Phi,\mathcal{G}\big)}{\partial \boldsymbol{G}_{22}}\right|_{\Phi=\overline{\Phi} \atop \mathcal{G}=\overline{\mathcal{G}}} = & \ \hbar\left[ -\frac{1}{2}\overline{\boldsymbol{G}}_{22}^{-1} + \frac{1}{2} \left(m^{2} + i\sqrt{\frac{\lambda}{3}} \ \overline{\eta}\right)\right] + \hbar^{2}\left[\frac{1}{6} \lambda \left( \overline{F}_{2}^{2} + \overline{D} \ \overline{\boldsymbol{G}}_{22}\right)\right] \\
& - \hbar^{3}\Bigg[ \frac{1}{18} \lambda^{2} \Big( 4 \overline{F}_{1}^3 \overline{F}_{2} \overline{\boldsymbol{G}}_{12} + 5 \overline{F}_{2}^4 \overline{\boldsymbol{G}}_{22} + \overline{D} \ \overline{F}_{2}^2 \left(7 \overline{\boldsymbol{G}}_{12}^2 + 9 \overline{\boldsymbol{G}}_{22}^2\right) \\
& \hspace{0.9cm} + \overline{D}^2 \Big(\overline{\boldsymbol{G}}_{11} \overline{\boldsymbol{G}}_{12}^2 + 2 \overline{\boldsymbol{G}}_{12}^2 \overline{\boldsymbol{G}}_{22} + \overline{\boldsymbol{G}}_{22}^3\Big) + \overline{F}_{1} \Big(6 \overline{F}_{2}^3 \overline{\boldsymbol{G}}_{12} \\
& \hspace{0.9cm} + 4 \overline{D} \ \overline{F}_{2} \overline{\boldsymbol{G}}_{12} \Big(\overline{\boldsymbol{G}}_{11} + 2 \overline{\boldsymbol{G}}_{22}\Big)\Big) + \overline{F}_{1}^2 \Big(\overline{F}_{2}^2 \left(\overline{\boldsymbol{G}}_{11} + 4 \overline{\boldsymbol{G}}_{22}\right) \\
& \hspace{0.9cm} + \overline{D} \Big(5 \overline{\boldsymbol{G}}_{12}^2 + 3 \overline{\boldsymbol{G}}_{22}^2\Big)\Big)\Big) \Bigg] \\
& + \hbar^{4}\Bigg[ \frac{1}{54} \lambda^{3} \overline{D}^3 \Big(8 \overline{\boldsymbol{G}}_{11}^2 \overline{\boldsymbol{G}}_{12}^2 \overline{\boldsymbol{G}}_{22} + \overline{\boldsymbol{G}}_{11}^3 \left(5 \overline{\boldsymbol{G}}_{12}^2 + \overline{\boldsymbol{G}}_{22}^2\right) \\
& \hspace{0.9cm} + \overline{\boldsymbol{G}}_{11} \Big(11 \overline{\boldsymbol{G}}_{12}^4 + 15 \overline{\boldsymbol{G}}_{12}^2 \overline{\boldsymbol{G}}_{22}^2\Big) + 5 \Big(3 \overline{\boldsymbol{G}}_{12}^4 \overline{\boldsymbol{G}}_{22} + 4 \overline{\boldsymbol{G}}_{12}^2 \overline{\boldsymbol{G}}_{22}^3 \\
& \hspace{0.9cm} + \overline{\boldsymbol{G}}_{22}^5\Big)\Big) + \mathcal{O}\hspace{-0.07cm}\left(\vec{\overline{F}}^{2}\right) \Bigg] \\
& + \mathcal{O}\big(\hbar^{5}\big)\;,
\end{split}
\label{eq:mixed2PIEAGapEquationG22N20DON}
\end{equation}
\begin{equation}
\begin{split}
0 = \left.\frac{\partial \Gamma_{\mathrm{mix}}^{(\mathrm{2PI})}\big(\Phi,\mathcal{G}\big)}{\partial \boldsymbol{G}_{12}}\right|_{\Phi=\overline{\Phi} \atop \mathcal{G}=\overline{\mathcal{G}}} = & \ \hbar^{2}\left[\frac{1}{3} \lambda \left( \overline{F}_{1} \overline{F}_{2} + \overline{D} \ \overline{\boldsymbol{G}}_{12} \right) \right] \\
& - \hbar^{3}\Bigg[ \frac{1}{9} \lambda^{2} \Big(2 \overline{F}_{1}^4 \overline{\boldsymbol{G}}_{12} + 2 \overline{F}_{2}^4 \overline{\boldsymbol{G}}_{12} + \overline{F}_{1}^3 \overline{F}_{2} \left(3 \overline{\boldsymbol{G}}_{11} + 2 \overline{\boldsymbol{G}}_{22}\right) \\
& \hspace{0.9cm} + \overline{F}_{1} \overline{F}_{2}^3 \left(2 \overline{\boldsymbol{G}}_{11} + 3 \overline{\boldsymbol{G}}_{22}\right) + \overline{D} \ \overline{F}_{2}^2 \overline{\boldsymbol{G}}_{12} \left(5 \overline{\boldsymbol{G}}_{11} + 7 \overline{\boldsymbol{G}}_{22}\right) \\
& \hspace{0.9cm} + \overline{F}_{1}^2 \overline{\boldsymbol{G}}_{12} \left(6 \overline{F}_{2}^2 + 7 \overline{D} \ \overline{\boldsymbol{G}}_{11} + 5 \overline{D} \ \overline{\boldsymbol{G}}_{22}\right) + \overline{D}^2 \overline{\boldsymbol{G}}_{12} \Big(\overline{\boldsymbol{G}}_{11}^2 \\
& \hspace{0.9cm} + \overline{\boldsymbol{G}}_{12}^2 + \overline{\boldsymbol{G}}_{11} \overline{\boldsymbol{G}}_{22} + \overline{\boldsymbol{G}}_{22}^2\Big) + 2 \overline{D} \ \overline{F}_{1} \overline{F}_{2} \Big(\overline{\boldsymbol{G}}_{11}^2 + 3 \overline{\boldsymbol{G}}_{12}^2 \\
& \hspace{0.9cm} + \overline{\boldsymbol{G}}_{11} \overline{\boldsymbol{G}}_{22} + \overline{\boldsymbol{G}}_{22}^2\Big)\Big) \Bigg] \\
& + \hbar^{4}\Bigg[ \frac{1}{27} \lambda^{3} \overline{D}^3 \overline{\boldsymbol{G}}_{12} \Big(5 \overline{\boldsymbol{G}}_{11}^4 + 4 \overline{\boldsymbol{G}}_{12}^4 + 5 \overline{\boldsymbol{G}}_{11}^3 \overline{\boldsymbol{G}}_{22} + 15 \overline{\boldsymbol{G}}_{12}^2 \overline{\boldsymbol{G}}_{22}^2 \\
& \hspace{0.9cm} + 5 \overline{\boldsymbol{G}}_{22}^4 + \overline{\boldsymbol{G}}_{11}^2 \Big(15 \overline{\boldsymbol{G}}_{12}^2 + 4 \overline{\boldsymbol{G}}_{22}^2\Big) + \overline{\boldsymbol{G}}_{11} \Big(22 \overline{\boldsymbol{G}}_{12}^2 \overline{\boldsymbol{G}}_{22} \\
& \hspace{0.9cm} + 5 \overline{\boldsymbol{G}}_{22}^3\Big)\Big) + \mathcal{O}\hspace{-0.07cm}\left(\vec{\overline{F}}^{2}\right) \Bigg] \\
& + \mathcal{O}\big(\hbar^{5}\big)\;,
\end{split}
\label{eq:mixed2PIEAGapEquationG12N20DON}
\end{equation}
\begin{equation}
\begin{split}
0 = \left.\frac{\partial \Gamma_{\mathrm{mix}}^{(\mathrm{2PI})}\big(\Phi,\mathcal{G}\big)}{\partial D}\right|_{\Phi=\overline{\Phi} \atop \mathcal{G}=\overline{\mathcal{G}}} = & \ \hbar \left(-\frac{1}{2}\overline{D}^{-1}+\frac{1}{2}\right) +\hbar^{2}\left[\frac{1}{12}\lambda \left(\overline{\boldsymbol{G}}_{11}^2 + 2 \overline{\boldsymbol{G}}_{12}^2 + \overline{\boldsymbol{G}}_{22}^2\right) \right] \\
& - \hbar^{3}\Bigg[ \frac{1}{36} \lambda^{2} \Big(8 \overline{F}_{1} \overline{F}_{2} \overline{\boldsymbol{G}}_{12} \left(\overline{\boldsymbol{G}}_{11}^2 + \overline{\boldsymbol{G}}_{12}^2 + \overline{\boldsymbol{G}}_{11} \overline{\boldsymbol{G}}_{22} + \overline{\boldsymbol{G}}_{22}^2\right) \\
& \hspace{0.9cm} + 2 \overline{F}_{1}^2 \Big(3 \overline{\boldsymbol{G}}_{11}^3 + 7 \overline{\boldsymbol{G}}_{11} \overline{\boldsymbol{G}}_{12}^2 + 5 \overline{\boldsymbol{G}}_{12}^2 \overline{\boldsymbol{G}}_{22} + \overline{\boldsymbol{G}}_{22}^3\Big) + 2 \overline{F}_{2}^2 \Big(\overline{\boldsymbol{G}}_{11}^3 \\
& \hspace{0.9cm} + 5 \overline{\boldsymbol{G}}_{11} \overline{\boldsymbol{G}}_{12}^2 + 7 \overline{\boldsymbol{G}}_{12}^2 \overline{\boldsymbol{G}}_{22} + 3 \overline{\boldsymbol{G}}_{22}^3\Big) + \overline{D} \Big(\overline{\boldsymbol{G}}_{11}^4 + 4 \overline{\boldsymbol{G}}_{11}^2 \overline{\boldsymbol{G}}_{12}^2 \\
& \hspace{0.9cm} + 2 \overline{\boldsymbol{G}}_{12}^4 + 4 \overline{\boldsymbol{G}}_{11} \overline{\boldsymbol{G}}_{12}^2 \overline{\boldsymbol{G}}_{22} + 4 \overline{\boldsymbol{G}}_{12}^2 \overline{\boldsymbol{G}}_{22}^2 + \overline{\boldsymbol{G}}_{22}^4\Big)\Big) \Bigg] \\
& + \hbar^{4}\Bigg[ \frac{1}{108} \lambda^{3} \overline{D}^2 \Big(5 \overline{\boldsymbol{G}}_{11}^6 + 30 \overline{\boldsymbol{G}}_{11}^4 \overline{\boldsymbol{G}}_{12}^2 + 8 \overline{\boldsymbol{G}}_{12}^6 + 45 \overline{\boldsymbol{G}}_{12}^4 \overline{\boldsymbol{G}}_{22}^2 \\
& \hspace{0.9cm} + 30 \overline{\boldsymbol{G}}_{12}^2 \overline{\boldsymbol{G}}_{22}^4 + 5 \overline{\boldsymbol{G}}_{22}^6 + 3 \overline{\boldsymbol{G}}_{11}^2 \left(15 \overline{\boldsymbol{G}}_{12}^4 + 8 \overline{\boldsymbol{G}}_{12}^2 \overline{\boldsymbol{G}}_{22}^2\right) \\
& \hspace{0.9cm} + 2 \overline{\boldsymbol{G}}_{11}^3 \left(15 \overline{\boldsymbol{G}}_{12}^2 \overline{\boldsymbol{G}}_{22} + \overline{\boldsymbol{G}}_{22}^3\right) + 6 \overline{\boldsymbol{G}}_{11} \Big(11 \overline{\boldsymbol{G}}_{12}^4 \overline{\boldsymbol{G}}_{22} \\
& \hspace{0.9cm} + 5 \overline{\boldsymbol{G}}_{12}^2 \overline{\boldsymbol{G}}_{22}^3\Big)\Big) + \mathcal{O}\hspace{-0.07cm}\left(\vec{\overline{F}}^{2}\right) \Bigg] \\
& + \mathcal{O}\big(\hbar^{5}\big)\;,
\end{split}
\label{eq:mixed2PIEAGapEquationDN20DON}
\end{equation}
\begin{equation}
\begin{split}
0 = \left.\frac{\partial \Gamma_{\mathrm{mix}}^{(\mathrm{2PI})}\big(\Phi,\mathcal{G}\big)}{\partial F_{1}}\right|_{\Phi=\overline{\Phi} \atop \mathcal{G}=\overline{\mathcal{G}}} = & \ \hbar^{2}\left[ \frac{1}{3} \lambda \left(\overline{F}_{1} \overline{\boldsymbol{G}}_{11} + \overline{F}_{2} \overline{\boldsymbol{G}}_{12}\right) \right] \\
& - \hbar^{3}\Bigg[ \frac{1}{9} \lambda^2 \Big(\overline{F}_{1}^3 \left(5 \overline{\boldsymbol{G}}_{11}^2 + 4 \overline{\boldsymbol{G}}_{12}^2\right) + 3 \overline{F}_{1}^2 \overline{F}_{2} \overline{\boldsymbol{G}}_{12} \left(3 \overline{\boldsymbol{G}}_{11} + 2 \overline{\boldsymbol{G}}_{22}\right) \\
& \hspace{0.9cm} + \overline{F}_{2}^3 \overline{\boldsymbol{G}}_{12} \left(2 \overline{\boldsymbol{G}}_{11} + 3 \overline{\boldsymbol{G}}_{22}\right) + 2 \overline{D} \ \overline{F}_{2} \overline{\boldsymbol{G}}_{12} \Big(\overline{\boldsymbol{G}}_{11}^2 + \overline{\boldsymbol{G}}_{12}^2 \\
& \hspace{0.9cm} + \overline{\boldsymbol{G}}_{11} \overline{\boldsymbol{G}}_{22} + \overline{\boldsymbol{G}}_{22}^2\Big) + \overline{F}_{1} \overline{F}_{2}^2 \Big(2 \overline{\boldsymbol{G}}_{11}^2 + 6 \overline{\boldsymbol{G}}_{12}^2 + \overline{\boldsymbol{G}}_{11} \overline{\boldsymbol{G}}_{22} \\
& \hspace{0.9cm} + 2 \overline{\boldsymbol{G}}_{22}^2\Big) + \overline{D} \ \overline{F}_{1} \Big(3 \overline{\boldsymbol{G}}_{11}^3 + 7 \overline{\boldsymbol{G}}_{11} \overline{\boldsymbol{G}}_{12}^2 + 5 \overline{\boldsymbol{G}}_{12}^2 \overline{\boldsymbol{G}}_{22} + \overline{\boldsymbol{G}}_{22}^3\Big) \Big) \Bigg] \\
& + \mathcal{O}\big(\hbar^{4}\big)\;,
\end{split}
\label{eq:mixed2PIEAGapEquationF1N20DON}
\end{equation}
\begin{equation}
\begin{split}
0 = \left.\frac{\partial \Gamma_{\mathrm{mix}}^{(\mathrm{2PI})}\big(\Phi,\mathcal{G}\big)}{\partial F_{2}}\right|_{\Phi=\overline{\Phi} \atop \mathcal{G}=\overline{\mathcal{G}}} = & \ \hbar\left(i\sqrt{\frac{\lambda}{3}} \ \overline{\phi}_{2}\right)  + \hbar^{2}\left[ \frac{1}{3} \lambda \left(\overline{F}_{1} \overline{\boldsymbol{G}}_{12} + \overline{F}_{2} \overline{\boldsymbol{G}}_{22}\right) \right] \\
& - \hbar^{3}\Bigg[ \frac{1}{9} \lambda^{2} \Big(\overline{F}_{1}^3 \overline{\boldsymbol{G}}_{12} \left(3 \overline{\boldsymbol{G}}_{11} + 2 \overline{\boldsymbol{G}}_{22}\right) + \overline{F}_{1}^2 \overline{F}_{2} \Big(2 \overline{\boldsymbol{G}}_{11}^2 + 6 \overline{\boldsymbol{G}}_{12}^2 \\
& \hspace{0.9cm} + \overline{\boldsymbol{G}}_{11} \overline{\boldsymbol{G}}_{22} + 2 \overline{\boldsymbol{G}}_{22}^2 \Big) + \overline{F}_{2}^3 \left(4 \overline{\boldsymbol{G}}_{12}^2 + 5 \overline{\boldsymbol{G}}_{22}^2\right) + \overline{D} \ \overline{F}_{2} \Big(\overline{\boldsymbol{G}}_{11}^3 \\
& \hspace{0.9cm} + 5 \overline{\boldsymbol{G}}_{11} \overline{\boldsymbol{G}}_{12}^2 + 7 \overline{\boldsymbol{G}}_{12}^2 \overline{\boldsymbol{G}}_{22} + 3 \overline{\boldsymbol{G}}_{22}^3\Big) + \overline{F}_{1} \overline{\boldsymbol{G}}_{12} \Big(\overline{F}_{2}^2 \big(6 \overline{\boldsymbol{G}}_{11} \\
& \hspace{0.9cm} + 9 \overline{\boldsymbol{G}}_{22}\big) + 2 \overline{D} \Big(\overline{\boldsymbol{G}}_{11}^2 + \overline{\boldsymbol{G}}_{12}^2 + \overline{\boldsymbol{G}}_{11} \overline{\boldsymbol{G}}_{22} + \overline{\boldsymbol{G}}_{22}^2\Big)\Big)\Big) \Bigg] \\
& + \mathcal{O}\big(\hbar^{4}\big)\;.
\end{split}
\label{eq:mixed2PIEAGapEquationF2N20DON}
\end{equation}

\end{itemize}
Note that the gap equations~\eqref{eq:mixed2PIEAGapEquationphiNN10DON} to~\eqref{eq:mixed2PIEAGapEquationF2N20DON} rely on the relations:
\begin{equation}
\overline{\Phi}=\begin{pmatrix}
\vec{\overline{\phi}} \\
\overline{\eta} \end{pmatrix} = \begin{pmatrix}
\overline{\phi}_{1} \\
\vdots \\
\overline{\phi}_{N-1} \\
\overline{\phi}_{N} \\
\overline{\eta}
\end{pmatrix} = \begin{pmatrix}
0 \\
\vdots \\
0 \\
\overline{\phi}_{N} \\
\overline{\eta}
\end{pmatrix} \;,
\end{equation}
and
\begin{equation}
\overline{\mathcal{G}} = \begin{pmatrix} \overline{\boldsymbol{G}} & \vec{\overline{F}} \\
 \vec{\overline{F}}^{\mathrm{T}} & \overline{D} \end{pmatrix} = \begin{pmatrix} \overline{\boldsymbol{G}}_{11} & \overline{\boldsymbol{G}}_{12} & \dotsb & \overline{\boldsymbol{G}}_{1N} & \overline{F}_{1} \\
 \overline{\boldsymbol{G}}_{21} & \overline{\boldsymbol{G}}_{22} & \dotsb & \overline{\boldsymbol{G}}_{2N} & \overline{F}_{2} \\
 \vdots & \vdots & \ddots & \vdots & \vdots \\
 \overline{\boldsymbol{G}}_{N1} & \overline{\boldsymbol{G}}_{N2} & \dotsb & \overline{\boldsymbol{G}}_{NN} & \overline{F}_{N} \\
 \overline{F}_{1} & \overline{F}_{2} & \dotsb & \overline{F}_{N} & \overline{D} \end{pmatrix}\;.
\label{eq:mixed2PIEAdefinitionGbar0DON}
\end{equation}
Similarly to~\eqref{eq:2PIorigE} and~\eqref{eq:2PIorigRho} for the original 2PI EA, the gs energy and density are now obtained from the solutions of the gap equations $\overline{\Phi}$ and $\overline{\mathcal{G}}$ alongside with the equalities:
\begin{equation}
E^\text{2PI EA;mix}_\text{gs} = \frac{1}{\hbar} \Gamma_{\mathrm{mix}}^{(\mathrm{2PI})}\big(\Phi=\overline{\Phi},\mathcal{G}=\overline{\mathcal{G}}\big) \;,
\label{eq:2PImixE}
\end{equation}
\begin{equation}
\rho^\text{2PI EA;mix}_\text{gs} = \frac{1}{N} \left(\hbar\mathrm{Tr}_{a}\big(\overline{\boldsymbol{G}}\big) + \vec{\overline{\phi}}\cdot\vec{\overline{\phi}}\right) \;.
\label{eq:2PImixRho}
\end{equation}
Furthermore, we assume no spontaneous breakdown of the $O(N)$ symmetry if the mixed 2PI EA is considered up to order $\mathcal{O}\big(\hbar^{4}\big)$, hence the term $\mathcal{O}\big(\vec{F}^{2}\big)$ in~\eqref{eq:mixed2PIEAfinalexpression}. This means that we can set $\vec{\overline{\phi}}=\vec{0}$, $\overline{\boldsymbol{G}}_{a b}=\overline{G} \ \delta_{a b}$ and $\vec{\overline{F}}=\vec{0}$ in the gap equations~\eqref{eq:mixed2PIEAGapEquationphiNN10DON} to~\eqref{eq:mixed2PIEAGapEquationF2N20DON} if the latter are exploited up to order $\mathcal{O}\big(\hbar^{4}\big)$.

\vspace{0.5cm}

\begin{figure}[!htb]
\captionsetup[subfigure]{labelformat=empty}
  \begin{center}
    \subfloat[]{
      \includegraphics[width=0.50\linewidth]{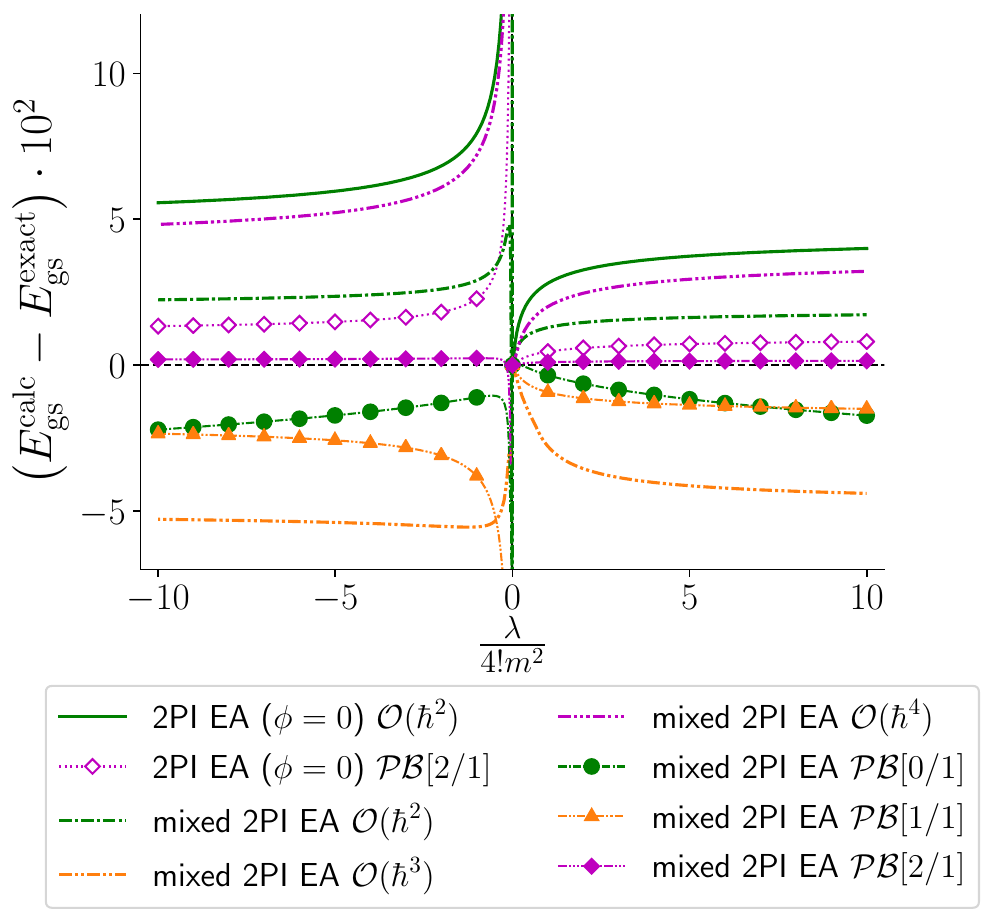}
                         }
    \subfloat[]{
      \includegraphics[width=0.50\linewidth]{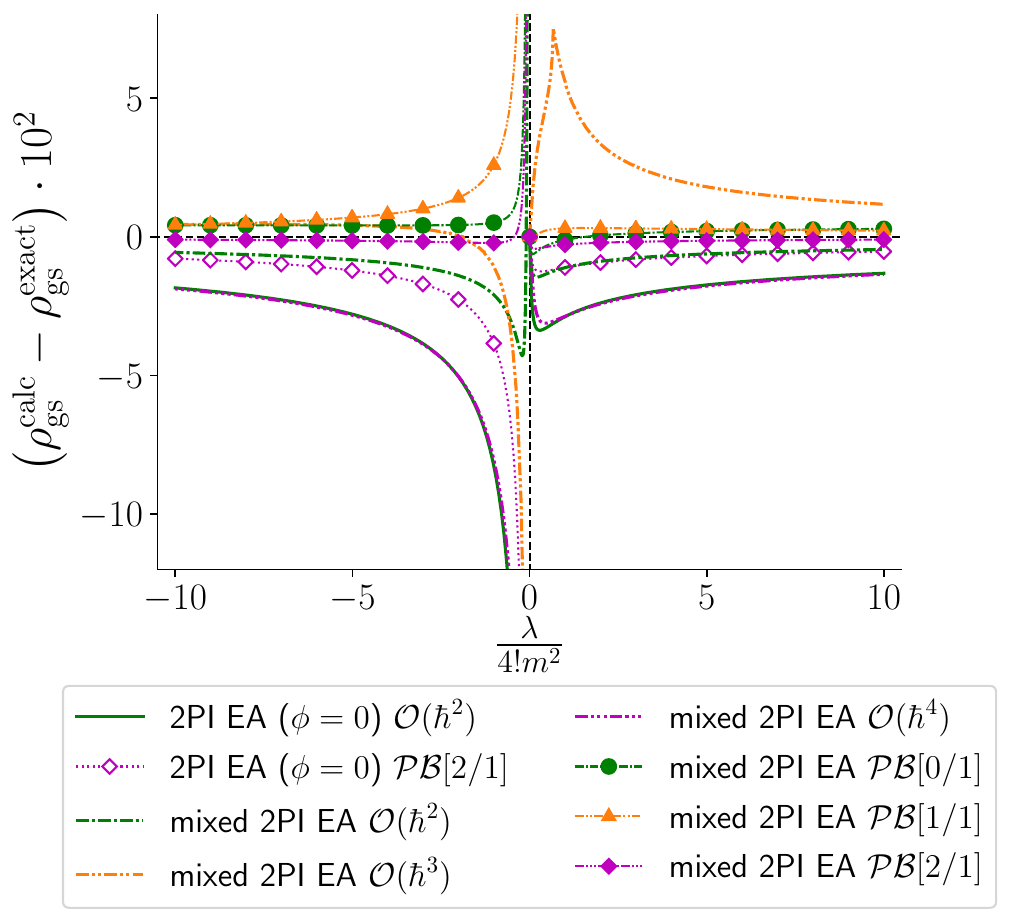}
                         }
    \caption{Difference between the calculated gs energy $E_{\mathrm{gs}}^{\mathrm{calc}}$ or density $\rho_{\mathrm{gs}}^{\mathrm{calc}}$ and the corresponding exact solution $E_{\mathrm{gs}}^{\mathrm{exact}}$ or $\rho_{\mathrm{gs}}^{\mathrm{exact}}$ at $\hbar=1$, $m^{2}=\pm 1$ and $N=1$ ($\mathcal{R}e(\lambda)\geq 0$ and $\mathcal{I}m(\lambda)=0$). See also the caption of fig.~\ref{fig:1PIEA} for the meaning of the indication ``$\mathcal{O}\big(\hbar^{n}\big)$'' for the results obtained from $\hbar$-expanded EAs.}
    \label{fig:2PIEAorigVsmixN1}
  \end{center}
\end{figure}

\begin{figure}[!htb]
\captionsetup[subfigure]{labelformat=empty}
  \begin{center}
    \subfloat[]{
      \includegraphics[width=0.50\linewidth]{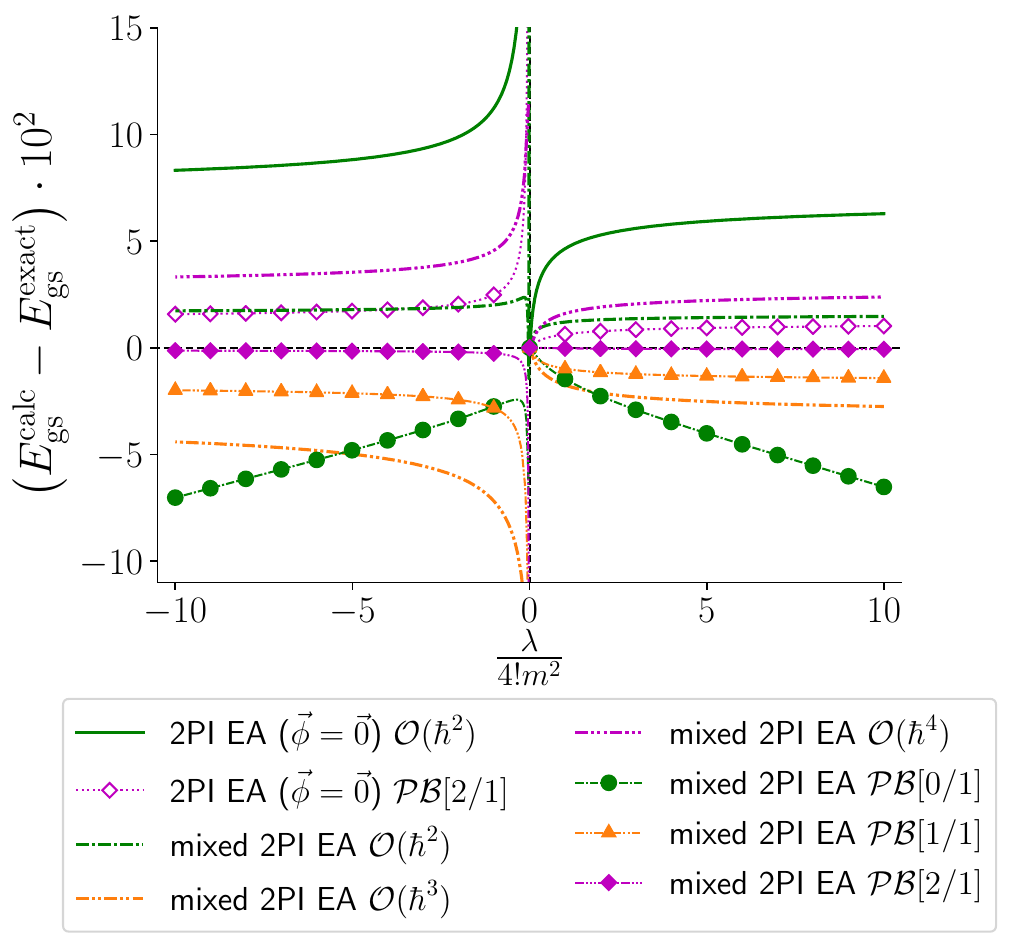}
                         }
    \subfloat[]{
      \includegraphics[width=0.50\linewidth]{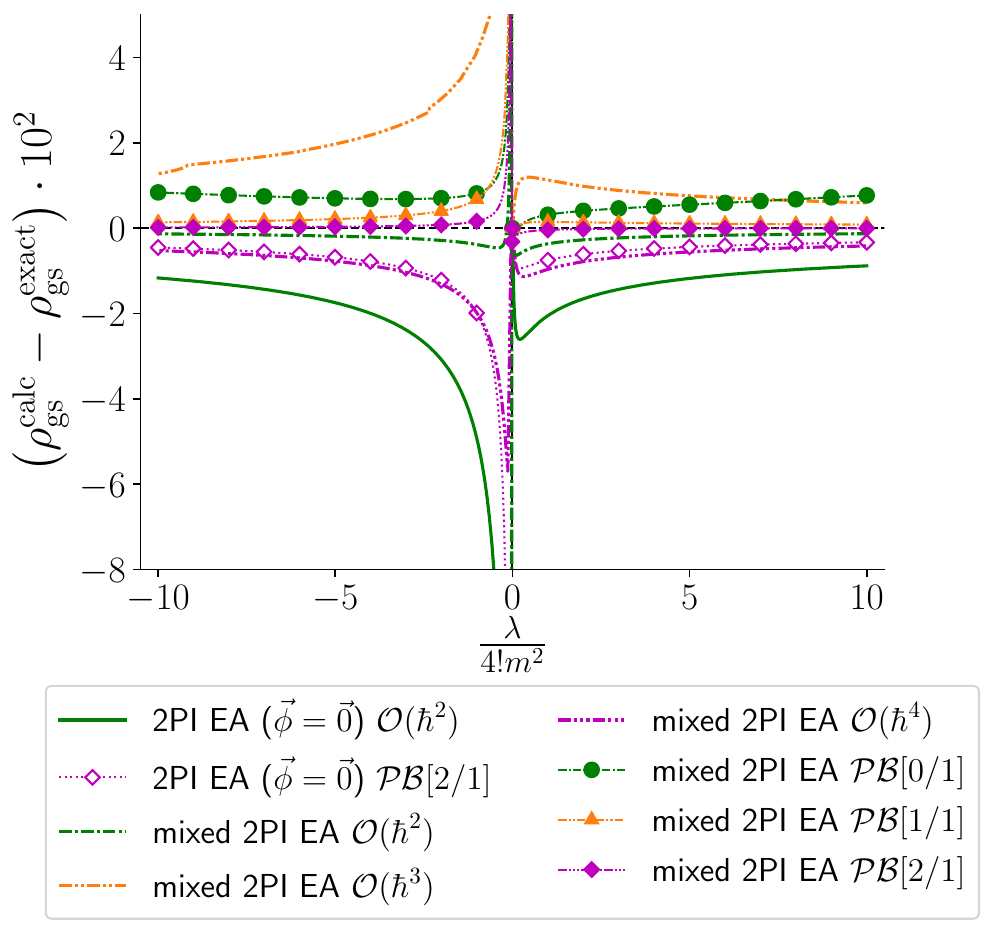}
                         }
    \caption{Same as fig.~\ref{fig:2PIEAorigVsmixN1} with $N=2$ instead.}
    \label{fig:2PIEAorigVsmixN2}
  \end{center}
\end{figure}

The estimations of the gs energy and density determined in this way from the mixed 2PI EA at $N=1$ and $N=2$ are presented in figs.~\ref{fig:2PIEAorigVsmixN1} and~\ref{fig:2PIEAorigVsmixN2}, respectively. With or without resummation, the mixed 2PI EA results outperform in general those of the original 2PI EA $\Gamma^{(\mathrm{2PI})}(\boldsymbol{G})$ for a given truncation with respect to $\hbar$. This is illustrated for both $E_{\mathrm{gs}}$ and $\rho_{\mathrm{gs}}$ in figs.~\ref{fig:2PIEAorigVsmixN1} and~\ref{fig:2PIEAorigVsmixN2} for the first non-trivial orders of these approaches and, after resummation, for their third non-trivial orders (with the $[2/1]$ Pad\'{e} approximants). In particular, these two figures show that, in the non-perturbative regime of the studied model at $N=1$ and $2$, the mixed 2PI EA achieves an accuracy of about $2\%$ for $E_{\mathrm{gs}}$ (to be compared with about $5\%$ to $8\%$ for the corresponding results of the original 2PI EA), and even less for $\rho_{\mathrm{gs}}$, already at its first non-trivial order, which corresponds to the BVA result mentioned previously. Furthermore, as the solutions of the gap equations leading to our mixed 2PI EA results never break the $O(N)$ symmetry (and notably always satisfy $\vec{\overline{F}}=\vec{0}$) at the first two non-trivial orders (i.e. when the mixed 2PI EA is considered up to orders $\mathcal{O}(\hbar^2)$ and $\mathcal{O}(\hbar^3)$), we can reasonably expect no SSB at the next non-trivial order, which motivates our previous assumption ignoring the contribution of $\vec{F}$-dependent diagrams at order $\mathcal{O}\big(\hbar^{4}\big)$ in~\eqref{eq:mixed2PIEAfinalexpression}. Interestingly, the absence of SSB at first non-trivial order (of the $\hbar$-expansion) for the mixed 2PI EA also implies that only the Fock diagram with a wiggly line (i.e. with a $D$ propagator) contributes to the BVA result, which means that the latter is equivalent to the result obtained from the same mixed 2PI EA at order $\mathcal{O}\big(1/N\big)$. Although we do not treat $1/N$-expansions of EAs in this study, it is interesting to know that its first non-trivial order in the expansion of the mixed 2PI EA coincides with the excellent BVA approximation for the studied model, which illustrates that $1/N$ can be a viable alternative to $\hbar$ or $\lambda$ as expansion parameter.

\vspace{0.5cm}

\begin{figure}[!htb]
\captionsetup[subfigure]{labelformat=empty}
  \begin{center}
    \subfloat[]{
      \includegraphics[width=0.50\linewidth]{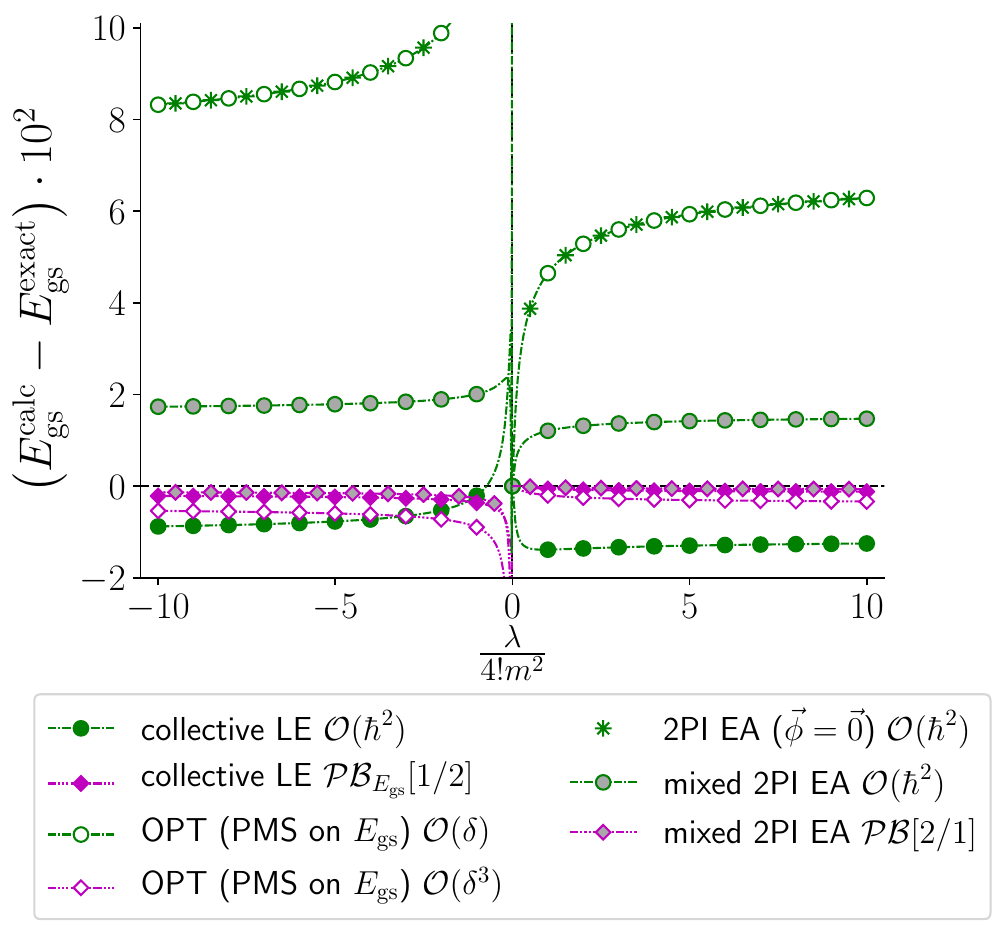}
                         }
    \subfloat[]{
      \includegraphics[width=0.50\linewidth]{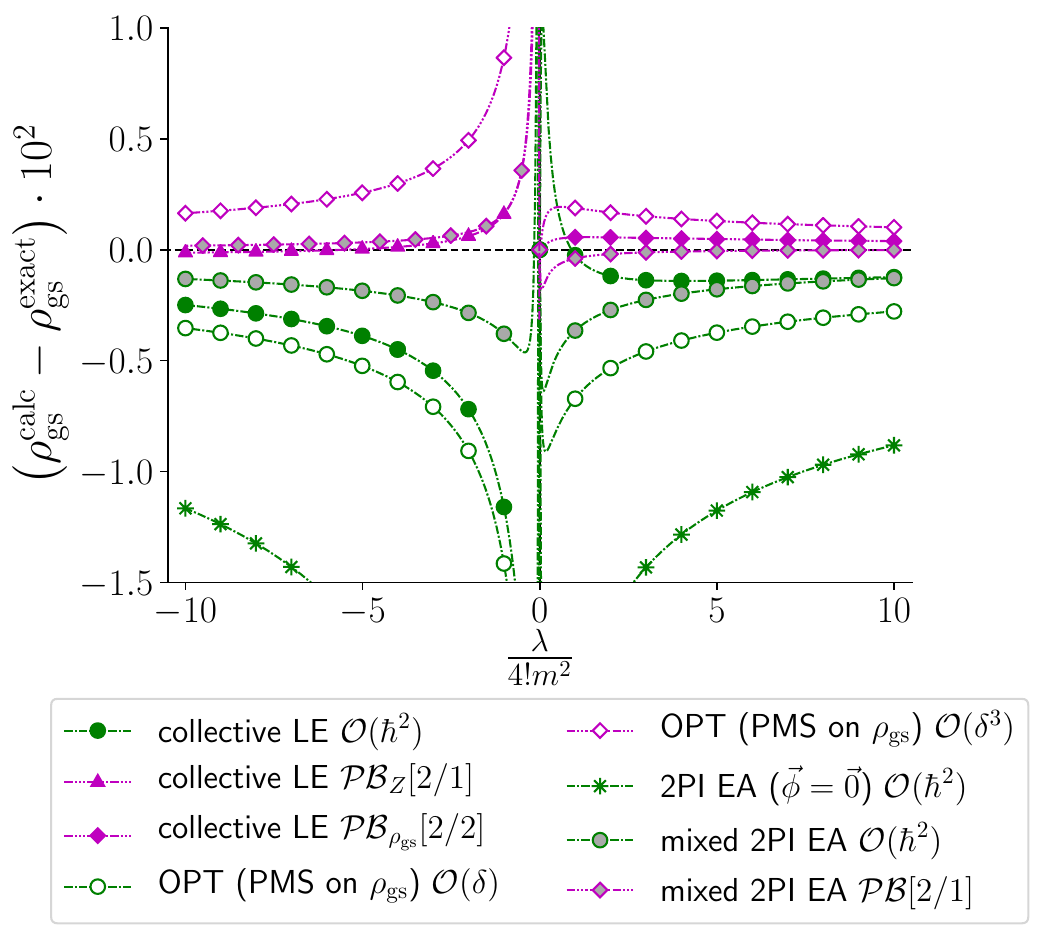}
                         }
    \caption{Difference between the calculated gs energy $E_{\mathrm{gs}}^{\mathrm{calc}}$ or density $\rho_{\mathrm{gs}}^{\mathrm{calc}}$ and the corresponding exact solution $E_{\mathrm{gs}}^{\mathrm{exact}}$ or $\rho_{\mathrm{gs}}^{\mathrm{exact}}$ at $\hbar=1$, $m^{2}=\pm 1$ and $N=2$ ($\mathcal{R}e(\lambda)\geq 0$ and $\mathcal{I}m(\lambda)=0$). See also the captions of figs.~\ref{fig:O1PTcoll} and~\ref{fig:1PIEA} for the meaning of the indication ``$\mathcal{O}\big(\hbar^{n}\big)$'' for the results obtained from the collective LE and $\hbar$-expanded EAs.}
    \label{fig:LEvsOPTvsmix2PIEAN2}
  \end{center}
\end{figure}

Furthermore, as the mixed 2PI EA method clearly stands out among the EA approaches treated so far, we compare it in fig.~\ref{fig:LEvsOPTvsmix2PIEAN2} with the best methods investigated in sections~\ref{sec:PT} and~\ref{sec:OPT} (i.e. the collective LE and OPT via PMS) for $N=2$, at first and third non-trivial orders. We can see on this figure that, after resummation if necessary, the excellent performances of the collective LE, OPT with PMS and the mixed 2PI EA are very close both at first and third non-trivial orders, with the exception of the first non-trivial order for OPT with PMS which is known to coincide with the Hartree-Fock result of the original 2PI EA $\Gamma^{(\mathrm{2PI})}(\boldsymbol{G})$ for $E_{\mathrm{gs}}$ (note however that the original 2PI EA as treated in this chapter relies on resummation procedures, whereas OPT does not), as can be seen from fig.~\ref{fig:LEvsOPTvsmix2PIEAN2}. These performances should of course be put in contrast with the ability to treat different (competing) channels in more realistic systems, in which case, as explained previously at the end of section~\ref{sec:OPT}, both the collective LE and the mixed 2PI EA approach might be reformulated in terms of multi-channel HSTs. We postpone such an application to more realistic systems with competing channels to future studies.

\vspace{0.5cm}

Although the BVA has already been applied to many QFTs, this is certainly not the case of higher truncation orders of the mixed 2PI EA (i.e. for truncations of this EA beyond order $\mathcal{O}(\hbar^2)$). In addition, we are performing to our knowledge the first applications of resummation theory to the mixed 2PI EA. Choosing once again the Pad\'{e}-Borel scheme as resummation procedure, we find that the best Pad\'{e} approximant obtained from the first non-trivial order of the mixed 2PI EA does not manage to clearly improve the corresponding bare results (i.e. the BVA results) in general, according to figs.~\ref{fig:2PIEAorigVsmixN1} and~\ref{fig:2PIEAorigVsmixN2}. Pad\'{e}-Borel resummation thus starts being efficient at order $\mathcal{O}\big(\hbar^{3}\big)$ and, at the third non-trivial order (and more specifically from $[2/1]$ Pad\'{e} approximants), yields excellent results which are barely distinguishable from the exact solution in both figs.~\ref{fig:2PIEAorigVsmixN1} and~\ref{fig:2PIEAorigVsmixN2}.

\vspace{0.5cm}

In conclusion, the significant improvement as compared to the original 2PI EA results can be attributed to both the 1-point correlation function $\eta$ and the propagator $D$ of the Hubbard-Stratonovich field since we always find $\vec{\overline{\phi}}=\vec{\overline{F}}=\vec{0}$ in the solutions of the mixed 2PI EA's gap equations leading to the results of figs.~\ref{fig:2PIEAorigVsmixN1} and~\ref{fig:2PIEAorigVsmixN2}, i.e. we do not find any spontaneous breakdown of the $O(N)$ symmetry in the framework of the mixed 2PI EA. We will thus determine as a next step how efficient the mixed 2PI EA formalism is if we impose the 1-point correlation function $\eta$ of the Hubbard-Stratonovich field to vanish, which will enable us to better understand the role of both $\eta$ and $D$ in the mixed 2PI EA approach.

\paragraph{$\hbar$-expansion for the mixed 2PI EA with vanishing 1-point correlation functions:}

We then study the mixed 2PI EA with both 1-point correlation functions $\vec{\phi}$ and $\eta$ set equal to zero, i.e. $\Gamma_{\mathrm{mix}}^{(\mathrm{2PI})}\big[\mathcal{G}\big]\equiv\Gamma_{\mathrm{mix}}^{(\mathrm{2PI})}\big[\Phi = 0,\mathcal{G}\big]$. This condition also imposes $\vec{F}=\vec{0}$, which enables us to discard all $\vec{F}$-dependent diagrams in~\eqref{eq:mixed2PIEAfinalexpression}:
\begin{equation}
\begin{split}
\Gamma^{(\mathrm{2PI})}_{\mathrm{mix}}\big[\mathcal{G}\big] = & -\frac{\hbar}{2}\mathcal{ST}r\left[\ln\big(\mathcal{G}\big)\right] + \frac{\hbar}{2}\mathcal{ST}r\left[\mathcal{G}^{-1}_{0}\mathcal{G}-\mathfrak{I}\right] \\
& + \frac{\hbar^{2}}{12} \hspace{0.1cm} \begin{gathered}
\begin{fmffile}{Diagrams/Mixed2PIEA_Fock}
\begin{fmfgraph}(15,15)
\fmfleft{i}
\fmfright{o}
\fmfv{decor.shape=circle,decor.size=2.0thick,foreground=(0,,0,,1)}{v1}
\fmfv{decor.shape=circle,decor.size=2.0thick,foreground=(0,,0,,1)}{v2}
\fmf{phantom,tension=11}{i,v1}
\fmf{phantom,tension=11}{v2,o}
\fmf{plain,left,tension=0.4,foreground=(1,,0,,0)}{v1,v2,v1}
\fmf{wiggly,foreground=(1,,0,,0)}{v1,v2}
\end{fmfgraph}
\end{fmffile}
\end{gathered} \hspace{0.15cm} - \frac{\hbar^{3}}{72} \hspace{0.35cm} \begin{gathered}
\begin{fmffile}{Diagrams/Mixed2PIEA_Diag2}
\begin{fmfgraph}(10,10)
\fmfleft{i0,i1}
\fmfright{o0,o1}
\fmftop{v1,vUp,v2}
\fmfbottom{v3,vDown,v4}
\fmfv{decor.shape=circle,decor.size=2.0thick,foreground=(0,,0,,1)}{v1}
\fmfv{decor.shape=circle,decor.size=2.0thick,foreground=(0,,0,,1)}{v2}
\fmfv{decor.shape=circle,decor.size=2.0thick,foreground=(0,,0,,1)}{v3}
\fmfv{decor.shape=circle,decor.size=2.0thick,foreground=(0,,0,,1)}{v4}
\fmf{phantom,tension=20}{i0,v1}
\fmf{phantom,tension=20}{i1,v3}
\fmf{phantom,tension=20}{o0,v2}
\fmf{phantom,tension=20}{o1,v4}
\fmf{plain,left=0.4,tension=0.5,foreground=(1,,0,,0)}{v3,v1}
\fmf{phantom,left=0.1,tension=0.5}{v1,vUp}
\fmf{phantom,left=0.1,tension=0.5}{vUp,v2}
\fmf{plain,left=0.4,tension=0.0,foreground=(1,,0,,0)}{v1,v2}
\fmf{plain,left=0.4,tension=0.5,foreground=(1,,0,,0)}{v2,v4}
\fmf{phantom,left=0.1,tension=0.5}{v4,vDown}
\fmf{phantom,left=0.1,tension=0.5}{vDown,v3}
\fmf{plain,left=0.4,tension=0.0,foreground=(1,,0,,0)}{v4,v3}
\fmf{wiggly,tension=0.5,foreground=(1,,0,,0)}{v1,v4}
\fmf{wiggly,tension=0.5,foreground=(1,,0,,0)}{v2,v3}
\end{fmfgraph}
\end{fmffile}
\end{gathered} \\
& + \hbar^{4} \left(\rule{0cm}{1.0cm}\right. \frac{1}{324} \hspace{0.3cm} \begin{gathered}
\begin{fmffile}{Diagrams/Mixed2PIEA_Diag7}
\begin{fmfgraph}(15,15)
\fmfleft{i}
\fmfright{o}
\fmftop{vUpLeft,vUp,vUpRight}
\fmfbottom{vDownLeft,vDown,vDownRight}
\fmfv{decor.shape=circle,decor.size=2.0thick,foreground=(0,,0,,1)}{v1}
\fmfv{decor.shape=circle,decor.size=2.0thick,foreground=(0,,0,,1)}{v2}
\fmfv{decor.shape=circle,decor.size=2.0thick,foreground=(0,,0,,1)}{v3}
\fmfv{decor.shape=circle,decor.size=2.0thick,foreground=(0,,0,,1)}{v4}
\fmfv{decor.shape=circle,decor.size=2.0thick,foreground=(0,,0,,1)}{v5}
\fmfv{decor.shape=circle,decor.size=2.0thick,foreground=(0,,0,,1)}{v6}
\fmf{phantom,tension=1}{i,v1}
\fmf{phantom,tension=1}{v2,o}
\fmf{phantom,tension=14.0}{v3,vUpLeft}
\fmf{phantom,tension=2.0}{v3,vUpRight}
\fmf{phantom,tension=4.0}{v3,i}
\fmf{phantom,tension=2.0}{v4,vUpLeft}
\fmf{phantom,tension=14.0}{v4,vUpRight}
\fmf{phantom,tension=4.0}{v4,o}
\fmf{phantom,tension=14.0}{v5,vDownLeft}
\fmf{phantom,tension=2.0}{v5,vDownRight}
\fmf{phantom,tension=4.0}{v5,i}
\fmf{phantom,tension=2.0}{v6,vDownLeft}
\fmf{phantom,tension=14.0}{v6,vDownRight}
\fmf{phantom,tension=4.0}{v6,o}
\fmf{wiggly,tension=0,foreground=(1,,0,,0)}{v1,v2}
\fmf{wiggly,tension=0.6,foreground=(1,,0,,0)}{v3,v6}
\fmf{wiggly,tension=0.6,foreground=(1,,0,,0)}{v5,v4}
\fmf{plain,left=0.18,tension=0,foreground=(1,,0,,0)}{v1,v3}
\fmf{plain,left=0.42,tension=0,foreground=(1,,0,,0)}{v3,v4}
\fmf{plain,left=0.18,tension=0,foreground=(1,,0,,0)}{v4,v2}
\fmf{plain,left=0.18,tension=0,foreground=(1,,0,,0)}{v2,v6}
\fmf{plain,left=0.42,tension=0,foreground=(1,,0,,0)}{v6,v5}
\fmf{plain,left=0.18,tension=0,foreground=(1,,0,,0)}{v5,v1}
\end{fmfgraph}
\end{fmffile}
\end{gathered} \hspace{0.3cm} + \frac{1}{108} \hspace{0.5cm} \begin{gathered}
\begin{fmffile}{Diagrams/Mixed2PIEA_Diag8}
\begin{fmfgraph}(12.5,12.5)
\fmfleft{i0,i1}
\fmfright{o0,o1}
\fmftop{v1,vUp,v2}
\fmfbottom{v3,vDown,v4}
\fmfv{decor.shape=circle,decor.size=2.0thick,foreground=(0,,0,,1)}{v1}
\fmfv{decor.shape=circle,decor.size=2.0thick,foreground=(0,,0,,1)}{v2}
\fmfv{decor.shape=circle,decor.size=2.0thick,foreground=(0,,0,,1)}{v3}
\fmfv{decor.shape=circle,decor.size=2.0thick,foreground=(0,,0,,1)}{v4}
\fmfv{decor.shape=circle,decor.size=2.0thick,foreground=(0,,0,,1)}{v5}
\fmfv{decor.shape=circle,decor.size=2.0thick,foreground=(0,,0,,1)}{v6}
\fmf{phantom,tension=20}{i0,v1}
\fmf{phantom,tension=20}{i1,v3}
\fmf{phantom,tension=20}{o0,v2}
\fmf{phantom,tension=20}{o1,v4}
\fmf{phantom,tension=0.005}{v5,v6}
\fmf{wiggly,left=0.4,tension=0,foreground=(1,,0,,0)}{v3,v1}
\fmf{phantom,left=0.1,tension=0}{v1,vUp}
\fmf{phantom,left=0.1,tension=0}{vUp,v2}
\fmf{plain,left=0.25,tension=0,foreground=(1,,0,,0)}{v1,v2}
\fmf{wiggly,left=0.4,tension=0,foreground=(1,,0,,0)}{v2,v4}
\fmf{phantom,left=0.1,tension=0}{v4,vDown}
\fmf{phantom,left=0.1,tension=0}{vDown,v3}
\fmf{plain,right=0.25,tension=0,foreground=(1,,0,,0)}{v3,v4}
\fmf{plain,left=0.2,tension=0.01,foreground=(1,,0,,0)}{v1,v5}
\fmf{plain,left=0.2,tension=0.01,foreground=(1,,0,,0)}{v5,v3}
\fmf{plain,right=0.2,tension=0.01,foreground=(1,,0,,0)}{v2,v6}
\fmf{plain,right=0.2,tension=0.01,foreground=(1,,0,,0)}{v6,v4}
\fmf{wiggly,tension=0,foreground=(1,,0,,0)}{v5,v6}
\end{fmfgraph}
\end{fmffile}
\end{gathered} \hspace{0.5cm} + \frac{1}{324} \hspace{0.4cm} \begin{gathered}
\begin{fmffile}{Diagrams/Mixed2PIEA_Diag9}
\begin{fmfgraph}(12.5,12.5)
\fmfleft{i0,i1}
\fmfright{o0,o1}
\fmftop{v1,vUp,v2}
\fmfbottom{v3,vDown,v4}
\fmfv{decor.shape=circle,decor.size=2.0thick,foreground=(0,,0,,1)}{v1}
\fmfv{decor.shape=circle,decor.size=2.0thick,foreground=(0,,0,,1)}{v2}
\fmfv{decor.shape=circle,decor.size=2.0thick,foreground=(0,,0,,1)}{v3}
\fmfv{decor.shape=circle,decor.size=2.0thick,foreground=(0,,0,,1)}{v4}
\fmfv{decor.shape=circle,decor.size=2.0thick,foreground=(0,,0,,1)}{v5}
\fmfv{decor.shape=circle,decor.size=2.0thick,foreground=(0,,0,,1)}{v6}
\fmf{phantom,tension=20}{i0,v1}
\fmf{phantom,tension=20}{i1,v3}
\fmf{phantom,tension=20}{o0,v2}
\fmf{phantom,tension=20}{o1,v4}
\fmf{phantom,tension=0.005}{v5,v6}
\fmf{plain,left=0.4,tension=0,foreground=(1,,0,,0)}{v3,v1}
\fmf{phantom,left=0.1,tension=0}{v1,vUp}
\fmf{phantom,left=0.1,tension=0}{vUp,v2}
\fmf{wiggly,left=0.25,tension=0,foreground=(1,,0,,0)}{v1,v2}
\fmf{plain,left=0.4,tension=0,foreground=(1,,0,,0)}{v2,v4}
\fmf{phantom,left=0.1,tension=0}{v4,vDown}
\fmf{phantom,left=0.1,tension=0}{vDown,v3}
\fmf{wiggly,right=0.25,tension=0,foreground=(1,,0,,0)}{v3,v4}
\fmf{plain,left=0.2,tension=0.01,foreground=(1,,0,,0)}{v1,v5}
\fmf{plain,left=0.2,tension=0.01,foreground=(1,,0,,0)}{v5,v3}
\fmf{plain,right=0.2,tension=0.01,foreground=(1,,0,,0)}{v2,v6}
\fmf{plain,right=0.2,tension=0.01,foreground=(1,,0,,0)}{v6,v4}
\fmf{wiggly,tension=0,foreground=(1,,0,,0)}{v5,v6}
\end{fmfgraph}
\end{fmffile}
\end{gathered} \hspace{0.4cm} \left.\rule{0cm}{1.0cm}\right) \\
& + \mathcal{O}\big(\hbar^{5}\big)\;,
\end{split}
\label{eq:mixed2PIEAZeroVevfinalexpression}
\end{equation}
with~\eqref{eq:mixed2PIEAvertex} to~\eqref{eq:mixed2PIEAFeynRuleD} as Feynman rules and the propagator $\mathcal{G}_{0}$ is given by:
\begin{equation}
\mathcal{G}^{-1}_{0}(x,y) = \begin{pmatrix}
\left(-\nabla_{x}^{2}+m^{2}\right)\mathbb{I}_{N} & \vec{0} \\
\vec{0}^{\mathrm{T}} & 1
\end{pmatrix} \delta(x-y) \;,
\end{equation}
which coincides with~\eqref{eq:mixed2PIEAMathcalGPhi} when $\Phi$ vanishes.

\vspace{0.5cm}

Let us then focus on the zero-dimensional situation by evaluating the different terms involved in the RHS of~\eqref{eq:mixed2PIEAZeroVevfinalexpression} in (0+0)-D. For that purpose, we first note that the constraint $\vec{\phi}=\vec{0}$ imposes that the $O(N)$ symmetry can not be broken down in the framework of the present approach and therefore $\boldsymbol{G}_{a b} = G \ \delta_{a b}$ $\forall a, b$, as discussed above~\eqref{eq:pure2PIEAzeroVevSTraceTerm0DON} for the original 2PI EA. This considerably simplifies~\eqref{eq:mixed2PIEAZeroVevfinalexpression} according to:
\begin{equation}
\mathcal{ST}r\left[\mathcal{G}^{-1}_{0}\mathcal{G}-\mathfrak{I}\right] = N m^{2} G + D - (N+1)\;,
\end{equation}
\begin{equation}
\begin{gathered}
\begin{fmffile}{Diagrams/Mixed2PIEA_Fock}
\begin{fmfgraph}(15,15)
\fmfleft{i}
\fmfright{o}
\fmfv{decor.shape=circle,decor.size=2.0thick,foreground=(0,,0,,1)}{v1}
\fmfv{decor.shape=circle,decor.size=2.0thick,foreground=(0,,0,,1)}{v2}
\fmf{phantom,tension=11}{i,v1}
\fmf{phantom,tension=11}{v2,o}
\fmf{plain,left,tension=0.4,foreground=(1,,0,,0)}{v1,v2,v1}
\fmf{wiggly,foreground=(1,,0,,0)}{v1,v2}
\end{fmfgraph}
\end{fmffile}
\end{gathered} = N \lambda D G^{2}\;,
\end{equation}
\begin{equation}
\begin{gathered}
\begin{fmffile}{Diagrams/Mixed2PIEA_Diag2}
\begin{fmfgraph}(10,10)
\fmfleft{i0,i1}
\fmfright{o0,o1}
\fmftop{v1,vUp,v2}
\fmfbottom{v3,vDown,v4}
\fmfv{decor.shape=circle,decor.size=2.0thick,foreground=(0,,0,,1)}{v1}
\fmfv{decor.shape=circle,decor.size=2.0thick,foreground=(0,,0,,1)}{v2}
\fmfv{decor.shape=circle,decor.size=2.0thick,foreground=(0,,0,,1)}{v3}
\fmfv{decor.shape=circle,decor.size=2.0thick,foreground=(0,,0,,1)}{v4}
\fmf{phantom,tension=20}{i0,v1}
\fmf{phantom,tension=20}{i1,v3}
\fmf{phantom,tension=20}{o0,v2}
\fmf{phantom,tension=20}{o1,v4}
\fmf{plain,left=0.4,tension=0.5,foreground=(1,,0,,0)}{v3,v1}
\fmf{phantom,left=0.1,tension=0.5}{v1,vUp}
\fmf{phantom,left=0.1,tension=0.5}{vUp,v2}
\fmf{plain,left=0.4,tension=0.0,foreground=(1,,0,,0)}{v1,v2}
\fmf{plain,left=0.4,tension=0.5,foreground=(1,,0,,0)}{v2,v4}
\fmf{phantom,left=0.1,tension=0.5}{v4,vDown}
\fmf{phantom,left=0.1,tension=0.5}{vDown,v3}
\fmf{plain,left=0.4,tension=0.0,foreground=(1,,0,,0)}{v4,v3}
\fmf{wiggly,tension=0.5,foreground=(1,,0,,0)}{v1,v4}
\fmf{wiggly,tension=0.5,foreground=(1,,0,,0)}{v2,v3}
\end{fmfgraph}
\end{fmffile}
\end{gathered} \hspace{0.35cm} = N \lambda^{2} D^{2} G^{4}\;,
\end{equation}
\begin{equation}
\begin{gathered}
\begin{fmffile}{Diagrams/Mixed2PIEA_Diag7}
\begin{fmfgraph}(15,15)
\fmfleft{i}
\fmfright{o}
\fmftop{vUpLeft,vUp,vUpRight}
\fmfbottom{vDownLeft,vDown,vDownRight}
\fmfv{decor.shape=circle,decor.size=2.0thick,foreground=(0,,0,,1)}{v1}
\fmfv{decor.shape=circle,decor.size=2.0thick,foreground=(0,,0,,1)}{v2}
\fmfv{decor.shape=circle,decor.size=2.0thick,foreground=(0,,0,,1)}{v3}
\fmfv{decor.shape=circle,decor.size=2.0thick,foreground=(0,,0,,1)}{v4}
\fmfv{decor.shape=circle,decor.size=2.0thick,foreground=(0,,0,,1)}{v5}
\fmfv{decor.shape=circle,decor.size=2.0thick,foreground=(0,,0,,1)}{v6}
\fmf{phantom,tension=1}{i,v1}
\fmf{phantom,tension=1}{v2,o}
\fmf{phantom,tension=14.0}{v3,vUpLeft}
\fmf{phantom,tension=2.0}{v3,vUpRight}
\fmf{phantom,tension=4.0}{v3,i}
\fmf{phantom,tension=2.0}{v4,vUpLeft}
\fmf{phantom,tension=14.0}{v4,vUpRight}
\fmf{phantom,tension=4.0}{v4,o}
\fmf{phantom,tension=14.0}{v5,vDownLeft}
\fmf{phantom,tension=2.0}{v5,vDownRight}
\fmf{phantom,tension=4.0}{v5,i}
\fmf{phantom,tension=2.0}{v6,vDownLeft}
\fmf{phantom,tension=14.0}{v6,vDownRight}
\fmf{phantom,tension=4.0}{v6,o}
\fmf{wiggly,tension=0,foreground=(1,,0,,0)}{v1,v2}
\fmf{wiggly,tension=0.6,foreground=(1,,0,,0)}{v3,v6}
\fmf{wiggly,tension=0.6,foreground=(1,,0,,0)}{v5,v4}
\fmf{plain,left=0.18,tension=0,foreground=(1,,0,,0)}{v1,v3}
\fmf{plain,left=0.42,tension=0,foreground=(1,,0,,0)}{v3,v4}
\fmf{plain,left=0.18,tension=0,foreground=(1,,0,,0)}{v4,v2}
\fmf{plain,left=0.18,tension=0,foreground=(1,,0,,0)}{v2,v6}
\fmf{plain,left=0.42,tension=0,foreground=(1,,0,,0)}{v6,v5}
\fmf{plain,left=0.18,tension=0,foreground=(1,,0,,0)}{v5,v1}
\end{fmfgraph}
\end{fmffile}
\end{gathered} \hspace{0.3cm} = \hspace{0.5cm} \begin{gathered}
\begin{fmffile}{Diagrams/Mixed2PIEA_Diag8}
\begin{fmfgraph}(12.5,12.5)
\fmfleft{i0,i1}
\fmfright{o0,o1}
\fmftop{v1,vUp,v2}
\fmfbottom{v3,vDown,v4}
\fmfv{decor.shape=circle,decor.size=2.0thick,foreground=(0,,0,,1)}{v1}
\fmfv{decor.shape=circle,decor.size=2.0thick,foreground=(0,,0,,1)}{v2}
\fmfv{decor.shape=circle,decor.size=2.0thick,foreground=(0,,0,,1)}{v3}
\fmfv{decor.shape=circle,decor.size=2.0thick,foreground=(0,,0,,1)}{v4}
\fmfv{decor.shape=circle,decor.size=2.0thick,foreground=(0,,0,,1)}{v5}
\fmfv{decor.shape=circle,decor.size=2.0thick,foreground=(0,,0,,1)}{v6}
\fmf{phantom,tension=20}{i0,v1}
\fmf{phantom,tension=20}{i1,v3}
\fmf{phantom,tension=20}{o0,v2}
\fmf{phantom,tension=20}{o1,v4}
\fmf{phantom,tension=0.005}{v5,v6}
\fmf{wiggly,left=0.4,tension=0,foreground=(1,,0,,0)}{v3,v1}
\fmf{phantom,left=0.1,tension=0}{v1,vUp}
\fmf{phantom,left=0.1,tension=0}{vUp,v2}
\fmf{plain,left=0.25,tension=0,foreground=(1,,0,,0)}{v1,v2}
\fmf{wiggly,left=0.4,tension=0,foreground=(1,,0,,0)}{v2,v4}
\fmf{phantom,left=0.1,tension=0}{v4,vDown}
\fmf{phantom,left=0.1,tension=0}{vDown,v3}
\fmf{plain,right=0.25,tension=0,foreground=(1,,0,,0)}{v3,v4}
\fmf{plain,left=0.2,tension=0.01,foreground=(1,,0,,0)}{v1,v5}
\fmf{plain,left=0.2,tension=0.01,foreground=(1,,0,,0)}{v5,v3}
\fmf{plain,right=0.2,tension=0.01,foreground=(1,,0,,0)}{v2,v6}
\fmf{plain,right=0.2,tension=0.01,foreground=(1,,0,,0)}{v6,v4}
\fmf{wiggly,tension=0,foreground=(1,,0,,0)}{v5,v6}
\end{fmfgraph}
\end{fmffile}
\end{gathered} \hspace{0.5cm} = N \lambda^{3} D^{3} G^{6}\;,
\end{equation}
\begin{equation}
\begin{gathered}
\begin{fmffile}{Diagrams/Mixed2PIEA_Diag9}
\begin{fmfgraph}(12.5,12.5)
\fmfleft{i0,i1}
\fmfright{o0,o1}
\fmftop{v1,vUp,v2}
\fmfbottom{v3,vDown,v4}
\fmfv{decor.shape=circle,decor.size=2.0thick,foreground=(0,,0,,1)}{v1}
\fmfv{decor.shape=circle,decor.size=2.0thick,foreground=(0,,0,,1)}{v2}
\fmfv{decor.shape=circle,decor.size=2.0thick,foreground=(0,,0,,1)}{v3}
\fmfv{decor.shape=circle,decor.size=2.0thick,foreground=(0,,0,,1)}{v4}
\fmfv{decor.shape=circle,decor.size=2.0thick,foreground=(0,,0,,1)}{v5}
\fmfv{decor.shape=circle,decor.size=2.0thick,foreground=(0,,0,,1)}{v6}
\fmf{phantom,tension=20}{i0,v1}
\fmf{phantom,tension=20}{i1,v3}
\fmf{phantom,tension=20}{o0,v2}
\fmf{phantom,tension=20}{o1,v4}
\fmf{phantom,tension=0.005}{v5,v6}
\fmf{plain,left=0.4,tension=0,foreground=(1,,0,,0)}{v3,v1}
\fmf{phantom,left=0.1,tension=0}{v1,vUp}
\fmf{phantom,left=0.1,tension=0}{vUp,v2}
\fmf{wiggly,left=0.25,tension=0,foreground=(1,,0,,0)}{v1,v2}
\fmf{plain,left=0.4,tension=0,foreground=(1,,0,,0)}{v2,v4}
\fmf{phantom,left=0.1,tension=0}{v4,vDown}
\fmf{phantom,left=0.1,tension=0}{vDown,v3}
\fmf{wiggly,right=0.25,tension=0,foreground=(1,,0,,0)}{v3,v4}
\fmf{plain,left=0.2,tension=0.01,foreground=(1,,0,,0)}{v1,v5}
\fmf{plain,left=0.2,tension=0.01,foreground=(1,,0,,0)}{v5,v3}
\fmf{plain,right=0.2,tension=0.01,foreground=(1,,0,,0)}{v2,v6}
\fmf{plain,right=0.2,tension=0.01,foreground=(1,,0,,0)}{v6,v4}
\fmf{wiggly,tension=0,foreground=(1,,0,,0)}{v5,v6}
\end{fmfgraph}
\end{fmffile}
\end{gathered} \hspace{0.4cm} = N^{2} \lambda^{3} D^{3} G^{6}\;.
\end{equation}
As a consequence, expression~\eqref{eq:mixed2PIEAZeroVevfinalexpression} of $\Gamma_{\mathrm{mix}}^{(\mathrm{2PI})}\big[\mathcal{G}\big]$ becomes in (0+0)-D:
\begin{equation}
\begin{split}
\Gamma_{\mathrm{mix}}^{(\mathrm{2PI})}\big(\mathcal{G}\big) = & \ \hbar \left(-\frac{N}{2}\ln\big(2\pi G\big) -\frac{1}{2}\ln(D) + \frac{1}{2}\left(N m^{2} G + D - N - 1\right) \right) \\
& + \hbar^{2}\left(\frac{N}{12}\lambda D G^{2} \right) - \hbar^{3}\left(\frac{N}{72}\lambda^{2} D^{2} G^{4} \right) + \hbar^{4}\left(\frac{N^{2}+4N}{324}\lambda^{3} D^{3} G^{6} \right) + \mathcal{O}\big(\hbar^{5}\big)\;,
\end{split}
\label{eq:mixed2PIEAzeroVevfinalexpression0DON}
\end{equation}
and the corresponding gap equations are:
\begin{equation}
\begin{split}
0 = \left.\frac{\partial \Gamma_{\mathrm{mix}}^{(\mathrm{2PI})}\big(\mathcal{G}\big)}{\partial G}\right|_{\mathcal{G}=\overline{\mathcal{G}}} = & \ \hbar\left(-\frac{N}{2}\overline{G}^{-1}+\frac{N}{2}m^{2}\right)+\hbar^{2}\left(\frac{N}{6}\lambda \overline{D} \ \overline{G}\right) -\hbar^{3}\left(\frac{N}{18}\lambda^{2} \overline{D}^{2} \overline{G}^{3}\right) \\
& +\hbar^{4}\left(\frac{N^{2}+4N}{54}\lambda^{3}\overline{D}^{3}\overline{G}^{5}\right) + \mathcal{O}\big(\hbar^{5}\big)\;,
\end{split}
\label{eq:mixed2PIEAzerovevGapEquationG0DON}
\end{equation}
and
\begin{equation}
\begin{split}
0 = \left.\frac{\partial \Gamma_{\mathrm{mix}}^{(\mathrm{2PI})}\big(\mathcal{G}\big)}{\partial D}\right|_{\mathcal{G}=\overline{\mathcal{G}}} = & \ \hbar\left(-\frac{1}{2}\overline{D}^{-1}+\frac{1}{2}\right)+\hbar^{2}\left(\frac{N}{12}\lambda \overline{G}^{2}\right)-\hbar^{3}\left(\frac{N}{36}\lambda^{2} \overline{D} \ \overline{G}^{4}\right) \\
& +\hbar^{4}\left(\frac{N^{2}+4N}{108}\lambda^{3}\overline{D}^{2}\overline{G}^{6}\right) + \mathcal{O}\big(\hbar^{5}\big)\;,
\end{split}
\label{eq:mixed2PIEAzerovevGapEquationD0DON}
\end{equation}
with
\begin{equation}
\overline{\mathcal{G}} = \begin{pmatrix} \overline{\boldsymbol{G}} & \vec{0} \\
 \vec{0}^{\mathrm{T}} & \overline{D} \end{pmatrix} = \begin{pmatrix} \overline{G} & 0 & \dotsb & 0 & 0 \\
 0 & \overline{G} & \dotsb & 0 & 0 \\
 \vdots & \vdots & \ddots & \vdots & \vdots \\
 0 & 0 & \dotsb & \overline{G} & 0 \\
 0 & 0 & \dotsb & 0 & \overline{D} \end{pmatrix}\;.
\label{eq:mixed2PIEAdzerovevdefinitionGbar0DON}
\end{equation}

\paragraph{$\lambda$-expansion for the mixed 2PI EA with vanishing 1-point correlation functions:}

In the framework of the mixed 2PI EA, $\hbar$- and $\lambda$-expansions are not equivalent even if all 1-point correlation functions are imposed to vanish. Hence, we now derive $\Gamma_{\mathrm{mix}}^{(\mathrm{2PI})}[\mathcal{G}]$ via the $\lambda$-expansion\footnote{We stress again that we set $\hbar=1$ while doing so.}. We thus obtain (see appendix~\ref{sec:mixed2PIEAannIM}):
\begin{equation}
\begin{split}
\Gamma_{\mathrm{mix}}^{(\mathrm{2PI})}[\mathcal{G}] = & -\frac{1}{2}\mathcal{ST}r\left[\ln\big(\mathcal{G}\big)\right] + \frac{1}{2}\mathcal{ST}r\left[\mathcal{G}^{-1}_{0}\mathcal{G}-\mathfrak{I}\right] \\
& + \left( \rule{0cm}{1.0cm} \right. \frac{1}{24}\begin{gathered}
\begin{fmffile}{Diagrams/Mixed2PIEAlambda_Hartree}
\begin{fmfgraph}(30,20)
\fmfleft{i}
\fmfright{o}
\fmfv{decor.shape=circle,decor.size=2.0thick,foreground=(0,,0,,1)}{v1}
\fmfv{decor.shape=circle,decor.size=2.0thick,foreground=(0,,0,,1)}{v2}
\fmf{phantom,tension=10}{i,i1}
\fmf{phantom,tension=10}{o,o1}
\fmf{plain,left,tension=0.5,foreground=(1,,0,,0)}{i1,v1,i1}
\fmf{plain,right,tension=0.5,foreground=(1,,0,,0)}{o1,v2,o1}
\fmf{wiggly,foreground=(1,,0,,0)}{v1,v2}
\end{fmfgraph}
\end{fmffile}
\end{gathered}
+\frac{1}{12}\begin{gathered}
\begin{fmffile}{Diagrams/Mixed2PIEAlambda_Fock}
\begin{fmfgraph}(15,15)
\fmfleft{i}
\fmfright{o}
\fmfv{decor.shape=circle,decor.size=2.0thick,foreground=(0,,0,,1)}{v1}
\fmfv{decor.shape=circle,decor.size=2.0thick,foreground=(0,,0,,1)}{v2}
\fmf{phantom,tension=11}{i,v1}
\fmf{phantom,tension=11}{v2,o}
\fmf{plain,left,tension=0.4,foreground=(1,,0,,0)}{v1,v2,v1}
\fmf{wiggly,foreground=(1,,0,,0)}{v1,v2}
\end{fmfgraph}
\end{fmffile}
\end{gathered} \left. \rule{0cm}{1.0cm} \right) \\
& + \left( \rule{0cm}{1.2cm} \right. -\frac{1}{72} \hspace{0.3cm} \begin{gathered}
\begin{fmffile}{Diagrams/Mixed2PIEAlambda_Diag1bis}
\begin{fmfgraph}(10,10)
\fmfleft{i0,i1}
\fmfright{o0,o1}
\fmftop{v1,vUp,v2}
\fmfbottom{v3,vDown,v4}
\fmfv{decor.shape=circle,decor.size=2.0thick,foreground=(0,,0,,1)}{v1}
\fmfv{decor.shape=circle,decor.size=2.0thick,foreground=(0,,0,,1)}{v2}
\fmfv{decor.shape=circle,decor.size=2.0thick,foreground=(0,,0,,1)}{v3}
\fmfv{decor.shape=circle,decor.size=2.0thick,foreground=(0,,0,,1)}{v4}
\fmf{phantom,tension=20}{i0,v1}
\fmf{phantom,tension=20}{i1,v3}
\fmf{phantom,tension=20}{o0,v2}
\fmf{phantom,tension=20}{o1,v4}
\fmf{plain,left=0.4,tension=0.5,foreground=(1,,0,,0)}{v3,v1}
\fmf{phantom,left=0.1,tension=0.5}{v1,vUp}
\fmf{phantom,left=0.1,tension=0.5}{vUp,v2}
\fmf{plain,left=0.4,tension=0.0,foreground=(1,,0,,0)}{v1,v2}
\fmf{plain,left=0.4,tension=0.5,foreground=(1,,0,,0)}{v2,v4}
\fmf{phantom,left=0.1,tension=0.5}{v4,vDown}
\fmf{phantom,left=0.1,tension=0.5}{vDown,v3}
\fmf{plain,left=0.4,tension=0.0,foreground=(1,,0,,0)}{v4,v3}
\fmf{wiggly,tension=0.5,foreground=(1,,0,,0)}{v1,v4}
\fmf{wiggly,tension=0.5,foreground=(1,,0,,0)}{v2,v3}
\end{fmfgraph}
\end{fmffile}
\end{gathered} \hspace{0.27cm} + \frac{1}{144} \hspace{0.1cm} \begin{gathered}
\begin{fmffile}{Diagrams/Mixed2PIEAlambda_Diag5bis}
\begin{fmfgraph}(30,15)
\fmfleft{i}
\fmfright{o}
\fmfv{decor.shape=circle,decor.size=2.0thick,foreground=(0,,0,,1)}{v1}
\fmfv{decor.shape=circle,decor.size=2.0thick,foreground=(0,,0,,1)}{v2}
\fmfv{decor.shape=circle,decor.size=2.0thick,foreground=(0,,0,,1)}{v3}
\fmfv{decor.shape=circle,decor.size=2.0thick,foreground=(0,,0,,1)}{v4}
\fmf{phantom,tension=10}{i,i1}
\fmf{phantom,tension=10}{o,o1}
\fmf{plain,left,tension=0.5,foreground=(1,,0,,0)}{i1,v1,i1}
\fmf{plain,right,tension=0.5,foreground=(1,,0,,0)}{o1,v2,o1}
\fmf{wiggly,foreground=(1,,0,,0)}{v1,v3}
\fmf{plain,left,tension=0.5,foreground=(1,,0,,0)}{v3,v4}
\fmf{plain,right,tension=0.5,foreground=(1,,0,,0)}{v3,v4}
\fmf{wiggly,foreground=(1,,0,,0)}{v4,v2}
\end{fmfgraph}
\end{fmffile}
\end{gathered} \hspace{0.1cm} + \frac{1}{576} \hspace{0.1cm} \begin{gathered}
\begin{fmffile}{Diagrams/Mixed2PIEAlambda_Diag6bis}
\begin{fmfgraph*}(25,12)
\fmfleft{i1,i2,i3,i4,i5,i6}
\fmfright{o1,o2,o3,o4,o5,o6}
\fmfv{decor.shape=circle,decor.size=2.0thick,foreground=(0,,0,,1)}{v1}
\fmfv{decor.shape=circle,decor.size=2.0thick,foreground=(0,,0,,1)}{v2}
\fmfv{decor.shape=circle,decor.size=2.0thick,foreground=(0,,0,,1)}{v3}
\fmfv{decor.shape=circle,decor.size=2.0thick,foreground=(0,,0,,1)}{v4}
\fmf{phantom,tension=3}{i2,v1}
\fmf{phantom,tension=1}{o2,v1}
\fmf{phantom,tension=3}{i5,v2}
\fmf{phantom,tension=1}{o5,v2}
\fmf{phantom,tension=1}{i2,v3}
\fmf{phantom,tension=3}{o2,v3}
\fmf{phantom,tension=1}{i5,v4}
\fmf{phantom,tension=3}{o5,v4}
\fmf{wiggly,tension=0,foreground=(1,,0,,0)}{v1,v3}
\fmf{wiggly,tension=0,foreground=(1,,0,,0)}{v2,v4}
\fmf{plain,left,tension=0,foreground=(1,,0,,0)}{v1,i2,v1}
\fmf{plain,left,tension=0,foreground=(1,,0,,0)}{v2,i5,v2}
\fmf{plain,left,tension=0,foreground=(1,,0,,0)}{v3,o2,v3}
\fmf{plain,left,tension=0,foreground=(1,,0,,0)}{v4,o5,v4}
\end{fmfgraph*}
\end{fmffile}
\end{gathered} \left. \rule{0cm}{1.2cm} \right) \\
& + \left( \rule{0cm}{1.3cm} \right. \frac{1}{324} \hspace{0.3cm} \begin{gathered}
\begin{fmffile}{Diagrams/Mixed2PIEA_Diag7}
\begin{fmfgraph}(15,15)
\fmfleft{i}
\fmfright{o}
\fmftop{vUpLeft,vUp,vUpRight}
\fmfbottom{vDownLeft,vDown,vDownRight}
\fmfv{decor.shape=circle,decor.size=2.0thick,foreground=(0,,0,,1)}{v1}
\fmfv{decor.shape=circle,decor.size=2.0thick,foreground=(0,,0,,1)}{v2}
\fmfv{decor.shape=circle,decor.size=2.0thick,foreground=(0,,0,,1)}{v3}
\fmfv{decor.shape=circle,decor.size=2.0thick,foreground=(0,,0,,1)}{v4}
\fmfv{decor.shape=circle,decor.size=2.0thick,foreground=(0,,0,,1)}{v5}
\fmfv{decor.shape=circle,decor.size=2.0thick,foreground=(0,,0,,1)}{v6}
\fmf{phantom,tension=1}{i,v1}
\fmf{phantom,tension=1}{v2,o}
\fmf{phantom,tension=14.0}{v3,vUpLeft}
\fmf{phantom,tension=2.0}{v3,vUpRight}
\fmf{phantom,tension=4.0}{v3,i}
\fmf{phantom,tension=2.0}{v4,vUpLeft}
\fmf{phantom,tension=14.0}{v4,vUpRight}
\fmf{phantom,tension=4.0}{v4,o}
\fmf{phantom,tension=14.0}{v5,vDownLeft}
\fmf{phantom,tension=2.0}{v5,vDownRight}
\fmf{phantom,tension=4.0}{v5,i}
\fmf{phantom,tension=2.0}{v6,vDownLeft}
\fmf{phantom,tension=14.0}{v6,vDownRight}
\fmf{phantom,tension=4.0}{v6,o}
\fmf{wiggly,tension=0,foreground=(1,,0,,0)}{v1,v2}
\fmf{wiggly,tension=0.6,foreground=(1,,0,,0)}{v3,v6}
\fmf{wiggly,tension=0.6,foreground=(1,,0,,0)}{v5,v4}
\fmf{plain,left=0.18,tension=0,foreground=(1,,0,,0)}{v1,v3}
\fmf{plain,left=0.42,tension=0,foreground=(1,,0,,0)}{v3,v4}
\fmf{plain,left=0.18,tension=0,foreground=(1,,0,,0)}{v4,v2}
\fmf{plain,left=0.18,tension=0,foreground=(1,,0,,0)}{v2,v6}
\fmf{plain,left=0.42,tension=0,foreground=(1,,0,,0)}{v6,v5}
\fmf{plain,left=0.18,tension=0,foreground=(1,,0,,0)}{v5,v1}
\end{fmfgraph}
\end{fmffile}
\end{gathered} \hspace{0.3cm} + \frac{1}{108} \hspace{0.5cm} \begin{gathered}
\begin{fmffile}{Diagrams/Mixed2PIEA_Diag8}
\begin{fmfgraph}(12.5,12.5)
\fmfleft{i0,i1}
\fmfright{o0,o1}
\fmftop{v1,vUp,v2}
\fmfbottom{v3,vDown,v4}
\fmfv{decor.shape=circle,decor.size=2.0thick,foreground=(0,,0,,1)}{v1}
\fmfv{decor.shape=circle,decor.size=2.0thick,foreground=(0,,0,,1)}{v2}
\fmfv{decor.shape=circle,decor.size=2.0thick,foreground=(0,,0,,1)}{v3}
\fmfv{decor.shape=circle,decor.size=2.0thick,foreground=(0,,0,,1)}{v4}
\fmfv{decor.shape=circle,decor.size=2.0thick,foreground=(0,,0,,1)}{v5}
\fmfv{decor.shape=circle,decor.size=2.0thick,foreground=(0,,0,,1)}{v6}
\fmf{phantom,tension=20}{i0,v1}
\fmf{phantom,tension=20}{i1,v3}
\fmf{phantom,tension=20}{o0,v2}
\fmf{phantom,tension=20}{o1,v4}
\fmf{phantom,tension=0.005}{v5,v6}
\fmf{wiggly,left=0.4,tension=0,foreground=(1,,0,,0)}{v3,v1}
\fmf{phantom,left=0.1,tension=0}{v1,vUp}
\fmf{phantom,left=0.1,tension=0}{vUp,v2}
\fmf{plain,left=0.25,tension=0,foreground=(1,,0,,0)}{v1,v2}
\fmf{wiggly,left=0.4,tension=0,foreground=(1,,0,,0)}{v2,v4}
\fmf{phantom,left=0.1,tension=0}{v4,vDown}
\fmf{phantom,left=0.1,tension=0}{vDown,v3}
\fmf{plain,right=0.25,tension=0,foreground=(1,,0,,0)}{v3,v4}
\fmf{plain,left=0.2,tension=0.01,foreground=(1,,0,,0)}{v1,v5}
\fmf{plain,left=0.2,tension=0.01,foreground=(1,,0,,0)}{v5,v3}
\fmf{plain,right=0.2,tension=0.01,foreground=(1,,0,,0)}{v2,v6}
\fmf{plain,right=0.2,tension=0.01,foreground=(1,,0,,0)}{v6,v4}
\fmf{wiggly,tension=0,foreground=(1,,0,,0)}{v5,v6}
\end{fmfgraph}
\end{fmffile}
\end{gathered} \hspace{0.5cm} + \frac{1}{324} \hspace{0.4cm} \begin{gathered}
\begin{fmffile}{Diagrams/Mixed2PIEA_Diag9}
\begin{fmfgraph}(12.5,12.5)
\fmfleft{i0,i1}
\fmfright{o0,o1}
\fmftop{v1,vUp,v2}
\fmfbottom{v3,vDown,v4}
\fmfv{decor.shape=circle,decor.size=2.0thick,foreground=(0,,0,,1)}{v1}
\fmfv{decor.shape=circle,decor.size=2.0thick,foreground=(0,,0,,1)}{v2}
\fmfv{decor.shape=circle,decor.size=2.0thick,foreground=(0,,0,,1)}{v3}
\fmfv{decor.shape=circle,decor.size=2.0thick,foreground=(0,,0,,1)}{v4}
\fmfv{decor.shape=circle,decor.size=2.0thick,foreground=(0,,0,,1)}{v5}
\fmfv{decor.shape=circle,decor.size=2.0thick,foreground=(0,,0,,1)}{v6}
\fmf{phantom,tension=20}{i0,v1}
\fmf{phantom,tension=20}{i1,v3}
\fmf{phantom,tension=20}{o0,v2}
\fmf{phantom,tension=20}{o1,v4}
\fmf{phantom,tension=0.005}{v5,v6}
\fmf{plain,left=0.4,tension=0,foreground=(1,,0,,0)}{v3,v1}
\fmf{phantom,left=0.1,tension=0}{v1,vUp}
\fmf{phantom,left=0.1,tension=0}{vUp,v2}
\fmf{wiggly,left=0.25,tension=0,foreground=(1,,0,,0)}{v1,v2}
\fmf{plain,left=0.4,tension=0,foreground=(1,,0,,0)}{v2,v4}
\fmf{phantom,left=0.1,tension=0}{v4,vDown}
\fmf{phantom,left=0.1,tension=0}{vDown,v3}
\fmf{wiggly,right=0.25,tension=0,foreground=(1,,0,,0)}{v3,v4}
\fmf{plain,left=0.2,tension=0.01,foreground=(1,,0,,0)}{v1,v5}
\fmf{plain,left=0.2,tension=0.01,foreground=(1,,0,,0)}{v5,v3}
\fmf{plain,right=0.2,tension=0.01,foreground=(1,,0,,0)}{v2,v6}
\fmf{plain,right=0.2,tension=0.01,foreground=(1,,0,,0)}{v6,v4}
\fmf{wiggly,tension=0,foreground=(1,,0,,0)}{v5,v6}
\end{fmfgraph}
\end{fmffile}
\end{gathered} \\
& \hspace{0.6cm} - \frac{1}{432} \hspace{0.1cm} \begin{gathered}
\begin{fmffile}{Diagrams/Mixed2PIEAlambda_Diag7}
\begin{fmfgraph}(36,12.5)
\fmfleft{i}
\fmfright{o}
\fmftop{vUp}
\fmfbottom{vDown}
\fmfv{decor.shape=circle,decor.size=2.0thick,foreground=(0,,0,,1)}{v1}
\fmfv{decor.shape=circle,decor.size=2.0thick,foreground=(0,,0,,1)}{v2}
\fmfv{decor.shape=circle,decor.size=2.0thick,foreground=(0,,0,,1)}{v3}
\fmfv{decor.shape=circle,decor.size=2.0thick,foreground=(0,,0,,1)}{v4}
\fmfv{decor.shape=circle,decor.size=2.0thick,foreground=(0,,0,,1)}{v5}
\fmfv{decor.shape=circle,decor.size=2.0thick,foreground=(0,,0,,1)}{v6}
\fmf{phantom,tension=10}{i,i1}
\fmf{phantom,tension=10}{o,o1}
\fmf{phantom,tension=2.2}{vUp,v5}
\fmf{phantom,tension=2.2}{vDown,v6}
\fmf{phantom,tension=0.5}{v3,v4}
\fmf{plain,left,tension=0.8,foreground=(1,,0,,0)}{i1,v1}
\fmf{plain,right,tension=0.8,foreground=(1,,0,,0)}{i1,v1}
\fmf{plain,right,tension=0.8,foreground=(1,,0,,0)}{o1,v2,o1}
\fmf{wiggly,tension=1.5,foreground=(1,,0,,0)}{v1,v3}
\fmf{plain,left=0.4,tension=0.5,foreground=(1,,0,,0)}{v3,v5}
\fmf{plain,left=0.4,tension=0.5,foreground=(1,,0,,0)}{v5,v4}
\fmf{plain,right=0.4,tension=0.5,foreground=(1,,0,,0)}{v3,v6}
\fmf{plain,right=0.4,tension=0.5,foreground=(1,,0,,0)}{v6,v4}
\fmf{wiggly,tension=1.5,foreground=(1,,0,,0)}{v4,v2}
\fmf{wiggly,tension=0,foreground=(1,,0,,0)}{v5,v6}
\end{fmfgraph}
\end{fmffile}
\end{gathered} + \frac{1}{864} \hspace{0.1cm} \begin{gathered}
\begin{fmffile}{Diagrams/Mixed2PIEAlambda_Diag8}
\begin{fmfgraph}(45,12.5)
\fmfleft{i}
\fmfright{o}
\fmfv{decor.shape=circle,decor.size=2.0thick,foreground=(0,,0,,1)}{v1}
\fmfv{decor.shape=circle,decor.size=2.0thick,foreground=(0,,0,,1)}{v2}
\fmfv{decor.shape=circle,decor.size=2.0thick,foreground=(0,,0,,1)}{v3}
\fmfv{decor.shape=circle,decor.size=2.0thick,foreground=(0,,0,,1)}{v4}
\fmfv{decor.shape=circle,decor.size=2.0thick,foreground=(0,,0,,1)}{v5}
\fmfv{decor.shape=circle,decor.size=2.0thick,foreground=(0,,0,,1)}{v6}
\fmf{phantom,tension=10}{i,i1}
\fmf{phantom,tension=10}{o,o1}
\fmf{plain,left,tension=0.5,foreground=(1,,0,,0)}{i1,v1,i1}
\fmf{plain,right,tension=0.5,foreground=(1,,0,,0)}{o1,v2,o1}
\fmf{wiggly,tension=1.0,foreground=(1,,0,,0)}{v1,v3}
\fmf{plain,left,tension=0.5,foreground=(1,,0,,0)}{v3,v4}
\fmf{plain,right,tension=0.5,foreground=(1,,0,,0)}{v3,v4}
\fmf{wiggly,tension=1.0,foreground=(1,,0,,0)}{v4,v5}
\fmf{plain,left,tension=0.5,foreground=(1,,0,,0)}{v5,v6}
\fmf{plain,right,tension=0.5,foreground=(1,,0,,0)}{v5,v6}
\fmf{wiggly,tension=1.0,foreground=(1,,0,,0)}{v6,v2}
\end{fmfgraph}
\end{fmffile}
\end{gathered} \\
& \hspace{0.6cm} + \frac{1}{864} \begin{gathered}
\begin{fmffile}{Diagrams/Mixed2PIEAlambda_Diag9}
\begin{fmfgraph}(36,15)
\fmfleft{i1,i,i4,i5}
\fmfright{o1,o,o4,o5}
\fmfv{decor.shape=circle,decor.size=2.0thick,foreground=(0,,0,,1)}{v1}
\fmfv{decor.shape=circle,decor.size=2.0thick,foreground=(0,,0,,1)}{v2}
\fmfv{decor.shape=circle,decor.size=2.0thick,foreground=(0,,0,,1)}{v3}
\fmfv{decor.shape=circle,decor.size=2.0thick,foreground=(0,,0,,1)}{v4}
\fmfv{decor.shape=circle,decor.size=2.0thick,foreground=(0,,0,,1)}{v5}
\fmfv{decor.shape=circle,decor.size=2.0thick,foreground=(0,,0,,1)}{v6}
\fmf{phantom,tension=10}{i,i1}
\fmf{phantom,tension=10}{o,o1}
\fmf{phantom,tension=4}{i4,i4bis}
\fmf{phantom,tension=1}{o4,i4bis}
\fmf{phantom,tension=1}{i4,o4bis}
\fmf{phantom,tension=4}{o4,o4bis}
\fmf{phantom,tension=2}{i4,v5}
\fmf{phantom,tension=1}{o4,v5}
\fmf{phantom,tension=1}{i4,v6}
\fmf{phantom,tension=2}{o4,v6}
\fmf{plain,left,tension=0.5,foreground=(1,,0,,0)}{i1,v1,i1}
\fmf{plain,right,tension=0.5,foreground=(1,,0,,0)}{o1,v2,o1}
\fmf{plain,left,tension=0,foreground=(1,,0,,0)}{i4bis,v5,i4bis}
\fmf{plain,right,tension=0,foreground=(1,,0,,0)}{o4bis,v6,o4bis}
\fmf{wiggly,foreground=(1,,0,,0)}{v1,v3}
\fmf{plain,left,tension=0.5,foreground=(1,,0,,0)}{v3,v4}
\fmf{plain,right,tension=0.5,foreground=(1,,0,,0)}{v3,v4}
\fmf{wiggly,foreground=(1,,0,,0)}{v4,v2}
\fmf{wiggly,tension=0.5,foreground=(1,,0,,0)}{v5,v6}
\end{fmfgraph}
\end{fmffile}
\end{gathered} + \frac{1}{5184} \begin{gathered}
\begin{fmffile}{Diagrams/Mixed2PIEAlambda_Diag10}
\begin{fmfgraph*}(25,17)
\fmfleft{i1,i2,i3,ibis1,iDown1,iDown2,ibis2,iUp1,iUp2,ibis3,i4,i5,i6}
\fmfright{o1,o2,o3,obis1,oDown1,oDown2,obis2,oUp1,oUp2,obis3,o4,o5,o6}
\fmfv{decor.shape=circle,decor.size=2.0thick,foreground=(0,,0,,1)}{v1}
\fmfv{decor.shape=circle,decor.size=2.0thick,foreground=(0,,0,,1)}{v2}
\fmfv{decor.shape=circle,decor.size=2.0thick,foreground=(0,,0,,1)}{v3}
\fmfv{decor.shape=circle,decor.size=2.0thick,foreground=(0,,0,,1)}{v4}
\fmfv{decor.shape=circle,decor.size=2.0thick,foreground=(0,,0,,1)}{v5}
\fmfv{decor.shape=circle,decor.size=2.0thick,foreground=(0,,0,,1)}{v6}
\fmf{phantom,tension=3}{i2,v1}
\fmf{phantom,tension=1}{o2,v1}
\fmf{phantom,tension=3}{i5,v2}
\fmf{phantom,tension=1}{o5,v2}
\fmf{phantom,tension=1}{i2,v3}
\fmf{phantom,tension=3}{o2,v3}
\fmf{phantom,tension=1}{i5,v4}
\fmf{phantom,tension=3}{o5,v4}
\fmf{phantom,tension=3}{ibis2,v5}
\fmf{phantom,tension=1}{obis2,v5}
\fmf{phantom,tension=1}{ibis2,v6}
\fmf{phantom,tension=3}{obis2,v6}
\fmf{phantom,tension=15}{ibis2,ibis2bis}
\fmf{phantom,tension=1}{obis2,ibis2bis}
\fmf{phantom,tension=1}{ibis2,obis2bis}
\fmf{phantom,tension=15}{obis2,obis2bis}
\fmf{wiggly,tension=0,foreground=(1,,0,,0)}{v1,v3}
\fmf{wiggly,tension=0,foreground=(1,,0,,0)}{v2,v4}
\fmf{wiggly,tension=0.52,foreground=(1,,0,,0)}{v5,v6}
\fmf{plain,left,tension=0,foreground=(1,,0,,0)}{v1,i2,v1}
\fmf{plain,left,tension=0,foreground=(1,,0,,0)}{v2,i5,v2}
\fmf{plain,left,tension=0,foreground=(1,,0,,0)}{v3,o2,v3}
\fmf{plain,left,tension=0,foreground=(1,,0,,0)}{v4,o5,v4}
\fmf{plain,left,tension=0.2,foreground=(1,,0,,0)}{v5,ibis2bis,v5}
\fmf{plain,left,tension=0.2,foreground=(1,,0,,0)}{v6,obis2bis,v6}
\end{fmfgraph*}
\end{fmffile}
\end{gathered} \left. \rule{0cm}{1.3cm} \right) \\
& + \mathcal{O}\big(\lambda^{4}\big) \;,
\end{split}
\label{eq:mixed2PIEAlambdafinalexpression}
\end{equation}
where we have used the Feynman rules~\eqref{eq:mixed2PIEAvertex} to~\eqref{eq:mixed2PIEAFeynRuleD} and the superpropagator $\mathcal{G}$ is now diagonal, i.e.:
\begin{equation}
\mathcal{G}=\begin{pmatrix}
\boldsymbol{G} & \vec{0} \\
\vec{0}^{\mathrm{T}} & D
\end{pmatrix} \;.
\end{equation}
The diagrams contributing to $\Gamma_{\mathrm{mix}}^{(\mathrm{2PI})}[\mathcal{G}]$ in~\eqref{eq:mixed2PIEAlambdafinalexpression} are not all 2PI (some are even disconnected). As opposed to the $\hbar$-expansion scheme, there is actually no guarantee that the diagrams generated by the Legendre transform of the 2PI EA cancel out with all 2PR graphs of the Schwinger functional in the framework of the $\lambda$-expansion (some disconnected diagrams are actually also generated by this Legendre transform), even though it turned out that the $\lambda$-expansion of the 1PI EA yielded $\Gamma^{(\mathrm{1PI})}$ in~\eqref{eq:1PIEAlambdaEAStep30DON} expressed in terms of 1PI diagrams only. Nevertheless, we keep referring to the functional $\Gamma_{\mathrm{mix}}^{(\mathrm{2PI})}[\mathcal{G}]$ as a 2PI EA in the present situation since the diagrams contributing to it are indeed all 2PI when expanded with respect to $\hbar$, as shown by~\eqref{eq:mixed2PIEAZeroVevfinalexpression}.

\vspace{0.5cm}

We then study~\eqref{eq:mixed2PIEAlambdafinalexpression} in the zero-dimensional situation. The absence of SSB still allows us to set $\boldsymbol{G}_{a b} = G_{a b} \ \delta_{a b}$. Hence, in (0+0)-D, the rightmost supertrace term and the diagrams of~\eqref{eq:mixed2PIEAlambdafinalexpression} reduce to:
\begin{equation}
\mathcal{ST}r\left[\mathcal{G}^{-1}_{0}\mathcal{G}-\mathfrak{I}\right] = N m^{2} G + D - (N+1) \;,
\end{equation}

\vspace{-0.4cm}

\begin{equation}
\begin{gathered}
\begin{fmffile}{Diagrams/Mixed2PIEAlambda_Hartree}
\begin{fmfgraph}(30,20)
\fmfleft{i}
\fmfright{o}
\fmfv{decor.shape=circle,decor.size=2.0thick,foreground=(0,,0,,1)}{v1}
\fmfv{decor.shape=circle,decor.size=2.0thick,foreground=(0,,0,,1)}{v2}
\fmf{phantom,tension=10}{i,i1}
\fmf{phantom,tension=10}{o,o1}
\fmf{plain,left,tension=0.5,foreground=(1,,0,,0)}{i1,v1,i1}
\fmf{plain,right,tension=0.5,foreground=(1,,0,,0)}{o1,v2,o1}
\fmf{wiggly,foreground=(1,,0,,0)}{v1,v2}
\end{fmfgraph}
\end{fmffile}
\end{gathered} = N^{2} \lambda D G^{2} \;,
\end{equation}

\vspace{-0.6cm}

\begin{equation}
\begin{gathered}
\begin{fmffile}{Diagrams/Mixed2PIEA_Fock}
\begin{fmfgraph}(15,15)
\fmfleft{i}
\fmfright{o}
\fmfv{decor.shape=circle,decor.size=2.0thick,foreground=(0,,0,,1)}{v1}
\fmfv{decor.shape=circle,decor.size=2.0thick,foreground=(0,,0,,1)}{v2}
\fmf{phantom,tension=11}{i,v1}
\fmf{phantom,tension=11}{v2,o}
\fmf{plain,left,tension=0.4,foreground=(1,,0,,0)}{v1,v2,v1}
\fmf{wiggly,foreground=(1,,0,,0)}{v1,v2}
\end{fmfgraph}
\end{fmffile}
\end{gathered} = N \lambda D G^{2} \;,
\end{equation}
\begin{equation}
\begin{gathered}
\begin{fmffile}{Diagrams/Mixed2PIEA_Diag2}
\begin{fmfgraph}(10,10)
\fmfleft{i0,i1}
\fmfright{o0,o1}
\fmftop{v1,vUp,v2}
\fmfbottom{v3,vDown,v4}
\fmfv{decor.shape=circle,decor.size=2.0thick,foreground=(0,,0,,1)}{v1}
\fmfv{decor.shape=circle,decor.size=2.0thick,foreground=(0,,0,,1)}{v2}
\fmfv{decor.shape=circle,decor.size=2.0thick,foreground=(0,,0,,1)}{v3}
\fmfv{decor.shape=circle,decor.size=2.0thick,foreground=(0,,0,,1)}{v4}
\fmf{phantom,tension=20}{i0,v1}
\fmf{phantom,tension=20}{i1,v3}
\fmf{phantom,tension=20}{o0,v2}
\fmf{phantom,tension=20}{o1,v4}
\fmf{plain,left=0.4,tension=0.5,foreground=(1,,0,,0)}{v3,v1}
\fmf{phantom,left=0.1,tension=0.5}{v1,vUp}
\fmf{phantom,left=0.1,tension=0.5}{vUp,v2}
\fmf{plain,left=0.4,tension=0.0,foreground=(1,,0,,0)}{v1,v2}
\fmf{plain,left=0.4,tension=0.5,foreground=(1,,0,,0)}{v2,v4}
\fmf{phantom,left=0.1,tension=0.5}{v4,vDown}
\fmf{phantom,left=0.1,tension=0.5}{vDown,v3}
\fmf{plain,left=0.4,tension=0.0,foreground=(1,,0,,0)}{v4,v3}
\fmf{wiggly,tension=0.5,foreground=(1,,0,,0)}{v1,v4}
\fmf{wiggly,tension=0.5,foreground=(1,,0,,0)}{v2,v3}
\end{fmfgraph}
\end{fmffile}
\end{gathered} \hspace{0.35cm} = N \lambda^{2} D^{2} G^{4} \;,
\end{equation}
\begin{equation}
\begin{gathered}
\begin{fmffile}{Diagrams/Mixed2PIEAlambda_Diag5bis}
\begin{fmfgraph}(30,15)
\fmfleft{i}
\fmfright{o}
\fmfv{decor.shape=circle,decor.size=2.0thick,foreground=(0,,0,,1)}{v1}
\fmfv{decor.shape=circle,decor.size=2.0thick,foreground=(0,,0,,1)}{v2}
\fmfv{decor.shape=circle,decor.size=2.0thick,foreground=(0,,0,,1)}{v3}
\fmfv{decor.shape=circle,decor.size=2.0thick,foreground=(0,,0,,1)}{v4}
\fmf{phantom,tension=10}{i,i1}
\fmf{phantom,tension=10}{o,o1}
\fmf{plain,left,tension=0.5,foreground=(1,,0,,0)}{i1,v1,i1}
\fmf{plain,right,tension=0.5,foreground=(1,,0,,0)}{o1,v2,o1}
\fmf{wiggly,foreground=(1,,0,,0)}{v1,v3}
\fmf{plain,left,tension=0.5,foreground=(1,,0,,0)}{v3,v4}
\fmf{plain,right,tension=0.5,foreground=(1,,0,,0)}{v3,v4}
\fmf{wiggly,foreground=(1,,0,,0)}{v4,v2}
\end{fmfgraph}
\end{fmffile}
\end{gathered} \hspace{0.1cm} = N^{3} \lambda^{2} D^{2} G^{4} \;,
\end{equation}
\begin{equation}
\begin{gathered}
\begin{fmffile}{Diagrams/Mixed2PIEAlambda_Diag6bis}
\begin{fmfgraph*}(25,12)
\fmfleft{i1,i2,i3,i4,i5,i6}
\fmfright{o1,o2,o3,o4,o5,o6}
\fmfv{decor.shape=circle,decor.size=2.0thick,foreground=(0,,0,,1)}{v1}
\fmfv{decor.shape=circle,decor.size=2.0thick,foreground=(0,,0,,1)}{v2}
\fmfv{decor.shape=circle,decor.size=2.0thick,foreground=(0,,0,,1)}{v3}
\fmfv{decor.shape=circle,decor.size=2.0thick,foreground=(0,,0,,1)}{v4}
\fmf{phantom,tension=3}{i2,v1}
\fmf{phantom,tension=1}{o2,v1}
\fmf{phantom,tension=3}{i5,v2}
\fmf{phantom,tension=1}{o5,v2}
\fmf{phantom,tension=1}{i2,v3}
\fmf{phantom,tension=3}{o2,v3}
\fmf{phantom,tension=1}{i5,v4}
\fmf{phantom,tension=3}{o5,v4}
\fmf{wiggly,tension=0,foreground=(1,,0,,0)}{v1,v3}
\fmf{wiggly,tension=0,foreground=(1,,0,,0)}{v2,v4}
\fmf{plain,left,tension=0,foreground=(1,,0,,0)}{v1,i2,v1}
\fmf{plain,left,tension=0,foreground=(1,,0,,0)}{v2,i5,v2}
\fmf{plain,left,tension=0,foreground=(1,,0,,0)}{v3,o2,v3}
\fmf{plain,left,tension=0,foreground=(1,,0,,0)}{v4,o5,v4}
\end{fmfgraph*}
\end{fmffile}
\end{gathered} = N^{4} \lambda^{2} D^{2} G^{4} \;,
\end{equation}
\begin{equation}
\begin{gathered}
\begin{fmffile}{Diagrams/Mixed2PIEA_Diag7}
\begin{fmfgraph}(15,15)
\fmfleft{i}
\fmfright{o}
\fmftop{vUpLeft,vUp,vUpRight}
\fmfbottom{vDownLeft,vDown,vDownRight}
\fmfv{decor.shape=circle,decor.size=2.0thick,foreground=(0,,0,,1)}{v1}
\fmfv{decor.shape=circle,decor.size=2.0thick,foreground=(0,,0,,1)}{v2}
\fmfv{decor.shape=circle,decor.size=2.0thick,foreground=(0,,0,,1)}{v3}
\fmfv{decor.shape=circle,decor.size=2.0thick,foreground=(0,,0,,1)}{v4}
\fmfv{decor.shape=circle,decor.size=2.0thick,foreground=(0,,0,,1)}{v5}
\fmfv{decor.shape=circle,decor.size=2.0thick,foreground=(0,,0,,1)}{v6}
\fmf{phantom,tension=1}{i,v1}
\fmf{phantom,tension=1}{v2,o}
\fmf{phantom,tension=14.0}{v3,vUpLeft}
\fmf{phantom,tension=2.0}{v3,vUpRight}
\fmf{phantom,tension=4.0}{v3,i}
\fmf{phantom,tension=2.0}{v4,vUpLeft}
\fmf{phantom,tension=14.0}{v4,vUpRight}
\fmf{phantom,tension=4.0}{v4,o}
\fmf{phantom,tension=14.0}{v5,vDownLeft}
\fmf{phantom,tension=2.0}{v5,vDownRight}
\fmf{phantom,tension=4.0}{v5,i}
\fmf{phantom,tension=2.0}{v6,vDownLeft}
\fmf{phantom,tension=14.0}{v6,vDownRight}
\fmf{phantom,tension=4.0}{v6,o}
\fmf{wiggly,tension=0,foreground=(1,,0,,0)}{v1,v2}
\fmf{wiggly,tension=0.6,foreground=(1,,0,,0)}{v3,v6}
\fmf{wiggly,tension=0.6,foreground=(1,,0,,0)}{v5,v4}
\fmf{plain,left=0.18,tension=0,foreground=(1,,0,,0)}{v1,v3}
\fmf{plain,left=0.42,tension=0,foreground=(1,,0,,0)}{v3,v4}
\fmf{plain,left=0.18,tension=0,foreground=(1,,0,,0)}{v4,v2}
\fmf{plain,left=0.18,tension=0,foreground=(1,,0,,0)}{v2,v6}
\fmf{plain,left=0.42,tension=0,foreground=(1,,0,,0)}{v6,v5}
\fmf{plain,left=0.18,tension=0,foreground=(1,,0,,0)}{v5,v1}
\end{fmfgraph}
\end{fmffile}
\end{gathered} \hspace{0.3cm} = \hspace{0.5cm} \begin{gathered}
\begin{fmffile}{Diagrams/Mixed2PIEA_Diag8}
\begin{fmfgraph}(12.5,12.5)
\fmfleft{i0,i1}
\fmfright{o0,o1}
\fmftop{v1,vUp,v2}
\fmfbottom{v3,vDown,v4}
\fmfv{decor.shape=circle,decor.size=2.0thick,foreground=(0,,0,,1)}{v1}
\fmfv{decor.shape=circle,decor.size=2.0thick,foreground=(0,,0,,1)}{v2}
\fmfv{decor.shape=circle,decor.size=2.0thick,foreground=(0,,0,,1)}{v3}
\fmfv{decor.shape=circle,decor.size=2.0thick,foreground=(0,,0,,1)}{v4}
\fmfv{decor.shape=circle,decor.size=2.0thick,foreground=(0,,0,,1)}{v5}
\fmfv{decor.shape=circle,decor.size=2.0thick,foreground=(0,,0,,1)}{v6}
\fmf{phantom,tension=20}{i0,v1}
\fmf{phantom,tension=20}{i1,v3}
\fmf{phantom,tension=20}{o0,v2}
\fmf{phantom,tension=20}{o1,v4}
\fmf{phantom,tension=0.005}{v5,v6}
\fmf{wiggly,left=0.4,tension=0,foreground=(1,,0,,0)}{v3,v1}
\fmf{phantom,left=0.1,tension=0}{v1,vUp}
\fmf{phantom,left=0.1,tension=0}{vUp,v2}
\fmf{plain,left=0.25,tension=0,foreground=(1,,0,,0)}{v1,v2}
\fmf{wiggly,left=0.4,tension=0,foreground=(1,,0,,0)}{v2,v4}
\fmf{phantom,left=0.1,tension=0}{v4,vDown}
\fmf{phantom,left=0.1,tension=0}{vDown,v3}
\fmf{plain,right=0.25,tension=0,foreground=(1,,0,,0)}{v3,v4}
\fmf{plain,left=0.2,tension=0.01,foreground=(1,,0,,0)}{v1,v5}
\fmf{plain,left=0.2,tension=0.01,foreground=(1,,0,,0)}{v5,v3}
\fmf{plain,right=0.2,tension=0.01,foreground=(1,,0,,0)}{v2,v6}
\fmf{plain,right=0.2,tension=0.01,foreground=(1,,0,,0)}{v6,v4}
\fmf{wiggly,tension=0,foreground=(1,,0,,0)}{v5,v6}
\end{fmfgraph}
\end{fmffile}
\end{gathered} \hspace{0.5cm} = N \lambda^{3} D^{3} G^{6} \;,
\end{equation}
\begin{equation}
\begin{gathered}
\begin{fmffile}{Diagrams/Mixed2PIEA_Diag9}
\begin{fmfgraph}(12.5,12.5)
\fmfleft{i0,i1}
\fmfright{o0,o1}
\fmftop{v1,vUp,v2}
\fmfbottom{v3,vDown,v4}
\fmfv{decor.shape=circle,decor.size=2.0thick,foreground=(0,,0,,1)}{v1}
\fmfv{decor.shape=circle,decor.size=2.0thick,foreground=(0,,0,,1)}{v2}
\fmfv{decor.shape=circle,decor.size=2.0thick,foreground=(0,,0,,1)}{v3}
\fmfv{decor.shape=circle,decor.size=2.0thick,foreground=(0,,0,,1)}{v4}
\fmfv{decor.shape=circle,decor.size=2.0thick,foreground=(0,,0,,1)}{v5}
\fmfv{decor.shape=circle,decor.size=2.0thick,foreground=(0,,0,,1)}{v6}
\fmf{phantom,tension=20}{i0,v1}
\fmf{phantom,tension=20}{i1,v3}
\fmf{phantom,tension=20}{o0,v2}
\fmf{phantom,tension=20}{o1,v4}
\fmf{phantom,tension=0.005}{v5,v6}
\fmf{plain,left=0.4,tension=0,foreground=(1,,0,,0)}{v3,v1}
\fmf{phantom,left=0.1,tension=0}{v1,vUp}
\fmf{phantom,left=0.1,tension=0}{vUp,v2}
\fmf{wiggly,left=0.25,tension=0,foreground=(1,,0,,0)}{v1,v2}
\fmf{plain,left=0.4,tension=0,foreground=(1,,0,,0)}{v2,v4}
\fmf{phantom,left=0.1,tension=0}{v4,vDown}
\fmf{phantom,left=0.1,tension=0}{vDown,v3}
\fmf{wiggly,right=0.25,tension=0,foreground=(1,,0,,0)}{v3,v4}
\fmf{plain,left=0.2,tension=0.01,foreground=(1,,0,,0)}{v1,v5}
\fmf{plain,left=0.2,tension=0.01,foreground=(1,,0,,0)}{v5,v3}
\fmf{plain,right=0.2,tension=0.01,foreground=(1,,0,,0)}{v2,v6}
\fmf{plain,right=0.2,tension=0.01,foreground=(1,,0,,0)}{v6,v4}
\fmf{wiggly,tension=0,foreground=(1,,0,,0)}{v5,v6}
\end{fmfgraph}
\end{fmffile}
\end{gathered} \hspace{0.4cm} = N^{2} \lambda^{3} D^{3} G^{6} \;,
\end{equation}
\begin{equation}
\begin{gathered}
\begin{fmffile}{Diagrams/Mixed2PIEAlambda_Diag7}
\begin{fmfgraph}(36,12.5)
\fmfleft{i}
\fmfright{o}
\fmftop{vUp}
\fmfbottom{vDown}
\fmfv{decor.shape=circle,decor.size=2.0thick,foreground=(0,,0,,1)}{v1}
\fmfv{decor.shape=circle,decor.size=2.0thick,foreground=(0,,0,,1)}{v2}
\fmfv{decor.shape=circle,decor.size=2.0thick,foreground=(0,,0,,1)}{v3}
\fmfv{decor.shape=circle,decor.size=2.0thick,foreground=(0,,0,,1)}{v4}
\fmfv{decor.shape=circle,decor.size=2.0thick,foreground=(0,,0,,1)}{v5}
\fmfv{decor.shape=circle,decor.size=2.0thick,foreground=(0,,0,,1)}{v6}
\fmf{phantom,tension=10}{i,i1}
\fmf{phantom,tension=10}{o,o1}
\fmf{phantom,tension=2.2}{vUp,v5}
\fmf{phantom,tension=2.2}{vDown,v6}
\fmf{phantom,tension=0.5}{v3,v4}
\fmf{plain,left,tension=0.8,foreground=(1,,0,,0)}{i1,v1}
\fmf{plain,right,tension=0.8,foreground=(1,,0,,0)}{i1,v1}
\fmf{plain,right,tension=0.8,foreground=(1,,0,,0)}{o1,v2,o1}
\fmf{wiggly,tension=1.5,foreground=(1,,0,,0)}{v1,v3}
\fmf{plain,left=0.4,tension=0.5,foreground=(1,,0,,0)}{v3,v5}
\fmf{plain,left=0.4,tension=0.5,foreground=(1,,0,,0)}{v5,v4}
\fmf{plain,right=0.4,tension=0.5,foreground=(1,,0,,0)}{v3,v6}
\fmf{plain,right=0.4,tension=0.5,foreground=(1,,0,,0)}{v6,v4}
\fmf{wiggly,tension=1.5,foreground=(1,,0,,0)}{v4,v2}
\fmf{wiggly,tension=0,foreground=(1,,0,,0)}{v5,v6}
\end{fmfgraph}
\end{fmffile}
\end{gathered} = N^{3} \lambda^{3} D^{3} G^{6} \;,
\end{equation}
\begin{equation}
\begin{gathered}
\begin{fmffile}{Diagrams/Mixed2PIEAlambda_Diag8}
\begin{fmfgraph}(45,12.5)
\fmfleft{i}
\fmfright{o}
\fmfv{decor.shape=circle,decor.size=2.0thick,foreground=(0,,0,,1)}{v1}
\fmfv{decor.shape=circle,decor.size=2.0thick,foreground=(0,,0,,1)}{v2}
\fmfv{decor.shape=circle,decor.size=2.0thick,foreground=(0,,0,,1)}{v3}
\fmfv{decor.shape=circle,decor.size=2.0thick,foreground=(0,,0,,1)}{v4}
\fmfv{decor.shape=circle,decor.size=2.0thick,foreground=(0,,0,,1)}{v5}
\fmfv{decor.shape=circle,decor.size=2.0thick,foreground=(0,,0,,1)}{v6}
\fmf{phantom,tension=10}{i,i1}
\fmf{phantom,tension=10}{o,o1}
\fmf{plain,left,tension=0.5,foreground=(1,,0,,0)}{i1,v1,i1}
\fmf{plain,right,tension=0.5,foreground=(1,,0,,0)}{o1,v2,o1}
\fmf{wiggly,tension=1.0,foreground=(1,,0,,0)}{v1,v3}
\fmf{plain,left,tension=0.5,foreground=(1,,0,,0)}{v3,v4}
\fmf{plain,right,tension=0.5,foreground=(1,,0,,0)}{v3,v4}
\fmf{wiggly,tension=1.0,foreground=(1,,0,,0)}{v4,v5}
\fmf{plain,left,tension=0.5,foreground=(1,,0,,0)}{v5,v6}
\fmf{plain,right,tension=0.5,foreground=(1,,0,,0)}{v5,v6}
\fmf{wiggly,tension=1.0,foreground=(1,,0,,0)}{v6,v2}
\end{fmfgraph}
\end{fmffile}
\end{gathered} = N^{4} \lambda^{3} D^{3} G^{6} \;,
\end{equation}
\begin{equation}
\begin{gathered}
\begin{fmffile}{Diagrams/Mixed2PIEAlambda_Diag9}
\begin{fmfgraph}(36,15)
\fmfleft{i1,i,i4,i5}
\fmfright{o1,o,o4,o5}
\fmfv{decor.shape=circle,decor.size=2.0thick,foreground=(0,,0,,1)}{v1}
\fmfv{decor.shape=circle,decor.size=2.0thick,foreground=(0,,0,,1)}{v2}
\fmfv{decor.shape=circle,decor.size=2.0thick,foreground=(0,,0,,1)}{v3}
\fmfv{decor.shape=circle,decor.size=2.0thick,foreground=(0,,0,,1)}{v4}
\fmfv{decor.shape=circle,decor.size=2.0thick,foreground=(0,,0,,1)}{v5}
\fmfv{decor.shape=circle,decor.size=2.0thick,foreground=(0,,0,,1)}{v6}
\fmf{phantom,tension=10}{i,i1}
\fmf{phantom,tension=10}{o,o1}
\fmf{phantom,tension=4}{i4,i4bis}
\fmf{phantom,tension=1}{o4,i4bis}
\fmf{phantom,tension=1}{i4,o4bis}
\fmf{phantom,tension=4}{o4,o4bis}
\fmf{phantom,tension=2}{i4,v5}
\fmf{phantom,tension=1}{o4,v5}
\fmf{phantom,tension=1}{i4,v6}
\fmf{phantom,tension=2}{o4,v6}
\fmf{plain,left,tension=0.5,foreground=(1,,0,,0)}{i1,v1,i1}
\fmf{plain,right,tension=0.5,foreground=(1,,0,,0)}{o1,v2,o1}
\fmf{plain,left,tension=0,foreground=(1,,0,,0)}{i4bis,v5,i4bis}
\fmf{plain,right,tension=0,foreground=(1,,0,,0)}{o4bis,v6,o4bis}
\fmf{wiggly,foreground=(1,,0,,0)}{v1,v3}
\fmf{plain,left,tension=0.5,foreground=(1,,0,,0)}{v3,v4}
\fmf{plain,right,tension=0.5,foreground=(1,,0,,0)}{v3,v4}
\fmf{wiggly,foreground=(1,,0,,0)}{v4,v2}
\fmf{wiggly,tension=0.5,foreground=(1,,0,,0)}{v5,v6}
\end{fmfgraph}
\end{fmffile}
\end{gathered} = N^{5} \lambda^{3} D^{3} G^{6} \;,
\end{equation}

\vspace{0.8cm}

\begin{equation}
\begin{gathered}
\begin{fmffile}{Diagrams/Mixed2PIEAlambda_Diag10}
\begin{fmfgraph*}(25,17)
\fmfleft{i1,i2,i3,ibis1,iDown1,iDown2,ibis2,iUp1,iUp2,ibis3,i4,i5,i6}
\fmfright{o1,o2,o3,obis1,oDown1,oDown2,obis2,oUp1,oUp2,obis3,o4,o5,o6}
\fmfv{decor.shape=circle,decor.size=2.0thick,foreground=(0,,0,,1)}{v1}
\fmfv{decor.shape=circle,decor.size=2.0thick,foreground=(0,,0,,1)}{v2}
\fmfv{decor.shape=circle,decor.size=2.0thick,foreground=(0,,0,,1)}{v3}
\fmfv{decor.shape=circle,decor.size=2.0thick,foreground=(0,,0,,1)}{v4}
\fmfv{decor.shape=circle,decor.size=2.0thick,foreground=(0,,0,,1)}{v5}
\fmfv{decor.shape=circle,decor.size=2.0thick,foreground=(0,,0,,1)}{v6}
\fmf{phantom,tension=3}{i2,v1}
\fmf{phantom,tension=1}{o2,v1}
\fmf{phantom,tension=3}{i5,v2}
\fmf{phantom,tension=1}{o5,v2}
\fmf{phantom,tension=1}{i2,v3}
\fmf{phantom,tension=3}{o2,v3}
\fmf{phantom,tension=1}{i5,v4}
\fmf{phantom,tension=3}{o5,v4}
\fmf{phantom,tension=3}{ibis2,v5}
\fmf{phantom,tension=1}{obis2,v5}
\fmf{phantom,tension=1}{ibis2,v6}
\fmf{phantom,tension=3}{obis2,v6}
\fmf{phantom,tension=15}{ibis2,ibis2bis}
\fmf{phantom,tension=1}{obis2,ibis2bis}
\fmf{phantom,tension=1}{ibis2,obis2bis}
\fmf{phantom,tension=15}{obis2,obis2bis}
\fmf{wiggly,tension=0,foreground=(1,,0,,0)}{v1,v3}
\fmf{wiggly,tension=0,foreground=(1,,0,,0)}{v2,v4}
\fmf{wiggly,tension=0.52,foreground=(1,,0,,0)}{v5,v6}
\fmf{plain,left,tension=0,foreground=(1,,0,,0)}{v1,i2,v1}
\fmf{plain,left,tension=0,foreground=(1,,0,,0)}{v2,i5,v2}
\fmf{plain,left,tension=0,foreground=(1,,0,,0)}{v3,o2,v3}
\fmf{plain,left,tension=0,foreground=(1,,0,,0)}{v4,o5,v4}
\fmf{plain,left,tension=0.2,foreground=(1,,0,,0)}{v5,ibis2bis,v5}
\fmf{plain,left,tension=0.2,foreground=(1,,0,,0)}{v6,obis2bis,v6}
\end{fmfgraph*}
\end{fmffile}
\end{gathered} = N^{6} \lambda^{3} D^{3} G^{6} \;.
\end{equation}

\begin{figure}[!htb]
\vspace{1.75cm}
\captionsetup[subfigure]{labelformat=empty}
  \begin{center}
    \subfloat[]{
      \includegraphics[width=0.70\linewidth]{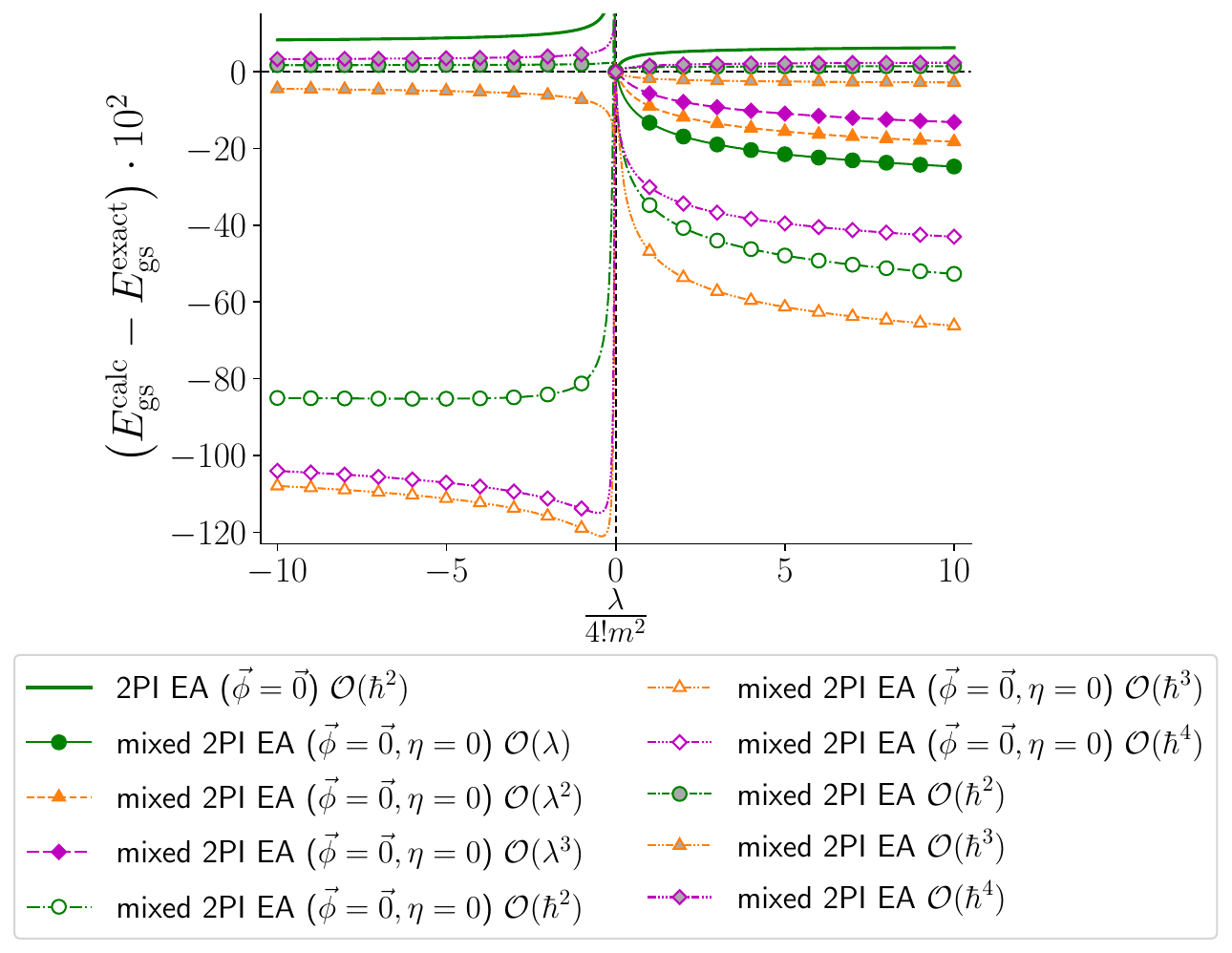}
                         }
   \\                     
    \subfloat[]{
      \includegraphics[width=0.70\linewidth]{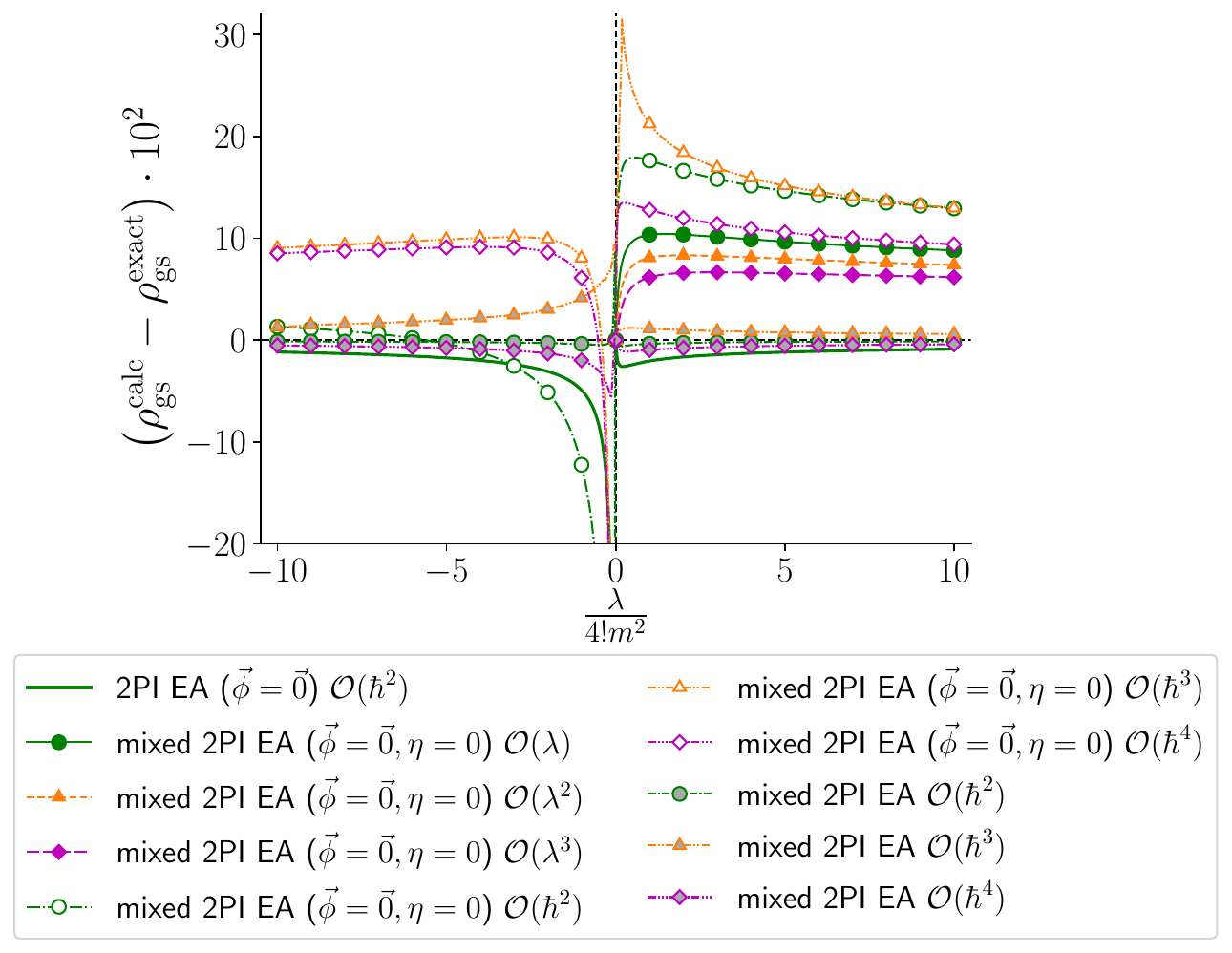}
                         }
\caption{Difference between the calculated gs energy $E_{\mathrm{gs}}^{\mathrm{calc}}$ or density $\rho_{\mathrm{gs}}^{\mathrm{calc}}$ and the corresponding exact solution $E_{\mathrm{gs}}^{\mathrm{exact}}$ or $\rho_{\mathrm{gs}}^{\mathrm{exact}}$ at $\hbar=1$, $m^{2}=\pm 1$ and $N=2$ ($\mathcal{R}e(\lambda)\geq 0$ and $\mathcal{I}m(\lambda)=0$). See notably the caption of fig.~\ref{fig:1PIEA} for the meaning of the indication ``$\mathcal{O}\big(\hbar^{n}\big)$'' for the results obtained from $\hbar$-expanded EAs. Note also that there are no results for the $\lambda$-expansion in the regime with $m^{2}<0$ as the corresponding approach is ill-defined in this case.}
\label{fig:2PIEAmixN2}
  \end{center}
\end{figure}

\noindent
From this,~\eqref{eq:mixed2PIEAlambdafinalexpression} becomes:
\begin{equation}
\begin{split}
\Gamma_{\mathrm{mix}}^{(\mathrm{2PI})}\big(\mathcal{G}\big) = & -\frac{N}{2}\ln(2\pi G) -\frac{1}{2}\ln(D) + \frac{1}{2}\left(N m^{2} G + D - N - 1\right) \\
& + \lambda \left(\frac{N^{2}+2N}{24} D G^{2} \right) + \lambda^{2} \left(\frac{N^{4}+4 N^{3}- 8N}{576} D^{2} G^{4} \right) \\
& + \lambda^{3} \left(\frac{N^{6} + 6 N^{5} + 6 N^{4} -12 N^{3} + 16 N^{2} + 64 N}{5184} D^{3} G^{6} \right) \\
& + \mathcal{O}\big(\lambda^{4}\big)\;,
\end{split}
\label{eq:mixed2PIEAlambdafinalexpression0DON}
\end{equation}
and the corresponding gap equations are given by:
\begin{equation}
\begin{split}
0 = \left.\frac{\partial \Gamma_{\mathrm{mix}}^{(\mathrm{2PI})}\big(\mathcal{G}\big)}{\partial G}\right|_{\mathcal{G}=\overline{\mathcal{G}}} = & -\frac{N}{2}\overline{G}^{-1}+\frac{N}{2}m^{2} +\lambda\left( \frac{N^{2}+2N}{12} \overline{D} \ \overline{G} \right)+\lambda^{2}\left( \frac{N^{4}+4 N^{3}- 8N}{144} \overline{D}^{2} \overline{G}^{3} \right) \\
& +\lambda^{3}\left( \frac{N^{6} + 6 N^{5} + 6 N^{4} -12 N^{3} + 16 N^{2} + 64 N}{864} \overline{D}^{3} \overline{G}^{5} \right) +\mathcal{O}\big(\lambda^{4}\big)\;,
\end{split}
\label{eq:mixed2PIEAlambdaGapEquationG0DON}
\end{equation}
\begin{equation}
\begin{split}
0 = \left.\frac{\partial \Gamma_{\mathrm{mix}}^{(\mathrm{2PI})}\big(\mathcal{G}\big)}{\partial D}\right|_{\mathcal{G}=\overline{\mathcal{G}}} = & -\frac{1}{2}\overline{D}^{-1}+\frac{1}{2} +\lambda\left( \frac{N^{2}+2N}{24} \overline{G}^{2} \right)+\lambda^{2}\left( \frac{N^{4}+4 N^{3}- 8N}{288} \overline{D} \ \overline{G}^{4} \right) \\
& +\lambda^{3}\left( \frac{N^{6} + 6 N^{5} + 6 N^{4} -12 N^{3} + 16 N^{2} + 64 N}{1728} \overline{D}^{2} \overline{G}^{6} \right) +\mathcal{O}\big(\lambda^{4}\big)\;,
\end{split}
\label{eq:mixed2PIEAlambdaGapEquationD0DON}
\end{equation}
with $\overline{G}$ and $\overline{D}$ defined via~\eqref{eq:mixed2PIEAdzerovevdefinitionGbar0DON}.

\vspace{0.5cm}

All results obtained from the mixed 2PI EA are compared in fig.~\ref{fig:2PIEAmixN2}, which logically illustrates that the full mixed 2PI EA method outperforms its homologous approaches imposing $(\vec{\phi},\eta)=(\vec{0},0)$. This clearly shows that the 1-point correlation function of the Hubbard-Stratonovich field, i.e. $\eta$, plays an essential role in the performances of the full mixed 2PI EA, both in the unbroken- and broken-symmetry regimes. We can also see in fig.~\ref{fig:2PIEAmixN2} that the mixed 2PI EA approach with $(\vec{\phi},\eta)=(\vec{0},0)$, based on the $\hbar$- or on the $\lambda$-expansion, does not outperform its original counterpart (which is based on the 2PI EA $\Gamma^{(\mathrm{2PI})}(\boldsymbol{G})$) at their first non-trivial orders. This section clearly puts forward the full mixed 2PI EA and the ability of the Hubbard-Stratonovich field to capture correlations. We will then investigate the 4PPI EA as another direction in the purpose of outperforming the original 2PI EA method via collective dofs. We will remain in the framework of the original representation to see how the addition of a new source (coupled to a quartic combination of the field) in the Schwinger functional could enable us to achieve this purpose. Before this, we will discuss the 2PPI EA formalism via the IM which is used thereafter to exploit the 4PPI EA.

\subsection{\label{sec:2PPIEA}2PPI effective action}

The IM was first developed in the framework of the 2PPI EA~\cite{fuk94,fuk95}, a few years after the introduction of the 2PPI EA itself~\cite{he90,ver92}. The original motivation for the IM was that, in finite dimensions, the Legendre transform defining the 2PPI EA (or any other $n$PPI EA) can not be done explicitly, as opposed to those of $n$PI EAs (the technical reason for this will be explained below in the present section). Hence, we were in need for a method that would be able to carry out the Legendre transform underlying the EAs indirectly or, more specifically, order by order with respect to a given expansion parameter, which is exactly what the IM does. This method has paved the way to a whole research effort which has notably led to refs.~\cite{yok95,val96,val97bis,ras98} for further applications of the 2PPI EA via the IM in atomic physics and especially ref.~\cite{Ina92} which is the first paper addressing superconductivity with the 2PPI EA. We will rather focus here, on the one hand, on the works of Fukuda \textit{et al.}~\cite{fuk94} and of Valiev and Fernando~\cite{val97}, who have shown that the 2PPI EA constitutes a means to implement Kohn-Sham DFT, and, on the other hand, on that of Okumura~\cite{oku96}, who studied exhaustively the diagrammatic properties of the 2PPI EA. Both refs.~\cite{val97} and~\cite{oku96} exploit the IM for a field theory involving a single field and impose the corresponding 1-point correlation function to vanish. We will therefore follow the same restriction here\footnote{Refs.~\cite{oku96,val97} also use a coupling constant instead of $\hbar$ as expansion parameter for the IM. We will stick to the $\hbar$-expansion scheme here to prepare the ground for the 4PPI EA treated in section~\ref{sec:4PPIEA} but one should keep in mind that, for the studied $O(N)$ model as well as for many other theories, $\hbar$- and $\lambda$-expansions of the 2PPI EA are equivalent if all 1-point correlation functions vanish, as was pointed out below~\eqref{eq:2PIEAzerovevgapequation0DON} for the original 2PI EA.}. The starting point of the IM consists in expanding the functionals of interest with respect to the chosen expansion parameter, i.e. $\hbar$ in the present case\footnote{In~\eqref{eq:pure2PPIEAGammaExpansion0DON} to~\eqref{eq:pure2PPIEArhoExpansion0DON}, we specify all variables on which $\Gamma^{(\mathrm{2PPI})}$, $W$, $K$ and $\rho$ depend but part of these dependences will be left implicit in the forthcoming derivations.}:
\begin{subequations}
\begin{empheq}[left=\empheqlbrace]{align}
& \hspace{0.1cm} \Gamma^{(\mathrm{2PPI})}[\rho;\hbar]=\sum_{n=0}^{\infty} \Gamma^{(\mathrm{2PPI})}_{n}[\rho]\hbar^{n}\;, \label{eq:pure2PPIEAGammaExpansion0DON}\\
\nonumber \\
& \hspace{0.1cm} W[K;\hbar]=\sum_{n=0}^{\infty} W_{n}[K]\hbar^{n}\;, \label{eq:pure2PPIEAWExpansion0DON} \\
\nonumber \\
& \hspace{0.1cm} K[\rho;\hbar]=\sum_{n=0}^{\infty} K_{n}[\rho]\hbar^{n}\;, \label{eq:pure2PPIEAKExpansion0DON}\\
\nonumber \\
& \hspace{0.1cm} \rho = \sum_{n=0}^{\infty} \rho_{n}[K]\hbar^{n}\;, \label{eq:pure2PPIEArhoExpansion0DON}
\end{empheq}
\end{subequations}
where the EA under consideration is now a functional of the density $\rho$:
\begin{equation}
\begin{split}
\Gamma^{(\mathrm{2PPI})}[\rho] \equiv & - W[K] + \int_{x} K^{a}[\rho;x] \frac{\delta W[K]}{\delta K^{a}(x)} \\
= & - W[K] + \frac{\hbar}{2} \int_{x} K^{a}[\rho;x] \rho_{a}(x) \;,
\end{split}
\label{eq:pure2PPIEAdefinition0DON}
\end{equation}
with
\begin{equation}
\rho_{a}(x) = \frac{2}{\hbar} \frac{\delta W[K]}{\delta K^{a}(x)} \;,
\label{eq:pure2PPIEAdefinitionbis20DON}
\end{equation}
and the Schwinger functional $W[K]$ defined by:
\begin{equation}
Z[K] = e^{\frac{1}{\hbar}W[K]} = \int \mathcal{D}\vec{\widetilde{\varphi}} \ e^{-\frac{1}{\hbar}S_{K}\big[\vec{\widetilde{\varphi}}\big]} \;,
\label{eq:ZJKfiniteD2PPIEA}
\end{equation}
\begin{equation}
S_{K}\Big[\vec{\widetilde{\varphi}}\Big] \equiv S\Big[\vec{\widetilde{\varphi}}\Big]-\frac{1}{2}\int_{x} K^{a}(x) \widetilde{\varphi}^{2}_{a}(x) \;.
\label{eq:SJKfiniteD2PPIEA}
\end{equation}
The $W_{n}$ coefficients are deduced from the LE result~\eqref{eq:WKjLoopExpansionStep3} for the studied model with $\vec{J}=\vec{0}$ and $\boldsymbol{K}_{a b}(x,y) = K_{a}(x)\delta_{a b}\delta(x-y)$. Note however that most of the present discussion on refs.~\cite{fuk94,val97,oku96} is model-independent.

\vspace{0.5cm}

Following the recipe of the IM, we combine the definition~\eqref{eq:pure2PPIEAdefinition0DON} with~\eqref{eq:pure2PPIEAGammaExpansion0DON},~\eqref{eq:pure2PPIEAWExpansion0DON} and~\eqref{eq:pure2PPIEAKExpansion0DON}:
\begin{equation}
\sum_{n=0}^{\infty} \Gamma^{(\mathrm{2PPI})}_{n}[\rho]\hbar^{n} = -\sum_{n=0}^{\infty} W_{n}\Bigg[\sum_{m=0}^{\infty} K_{m}[\rho]\hbar^{m}\Bigg]\hbar^{n}+\frac{\hbar}{2}\sum_{n=0}^{\infty} \int_{x} K^{a}_{n}[\rho;x]\rho_{a}(x)\hbar^{n}\;.
\label{eq:2PPIEAIMstep20DON}
\end{equation}
The Taylor expansion of the $W_{n}$ coefficients around $K=K_{0}$ in~\eqref{eq:2PPIEAIMstep20DON} yields\footnote{We exceptionally use $a_{1},a_{2},\cdots$ instead of $a,b,\cdots$ to denote color indices in~\eqref{eq:2PPIEAIMstep30DON}.}$^{,}$\footnote{Note that~\eqref{eq:2PPIEAIMstep30DON} involves the shorthand notation $\delta_{n \geq m} = \begin{cases} \displaystyle{1} \quad \mathrm{if} ~ n \geq q \;. \\
\displaystyle{0} \quad \mathrm{otherwise} \;. \end{cases}$}$^{,}$\footnote{The curly braces below discrete sums contain a condition that must be satisfied by each term of the sum in question.} (see appendix~\ref{sec:GammanCoeffIM4PPIEA} with $\zeta_{\alpha} = M_{\alpha}[\rho,\zeta] = M_{n,\alpha}[\rho,\zeta] = 0$ $\forall \alpha,n$):
\begin{equation}
\begin{split}
\scalebox{0.95}{${\displaystyle \Gamma^{(\mathrm{2PPI})}_{n}[\rho] = }$} & \scalebox{0.95}{${\displaystyle -W_{n}[K=K_{0}] -\sum_{m=1}^{n-1} \int_{x} \left.\frac{\delta W_{n-m}[K]}{\delta K_{a}(x)}\right|_{K=K_{0}} K_{m,a}[\rho;x] }$} \\
& \scalebox{0.95}{${\displaystyle -\sum_{m=2}^{n-1} \frac{1}{m!} \sum_{\underset{\lbrace n_{1} + \cdots + n_{m} \leq n\rbrace}{n_{1},\cdots,n_{m}=1}}^{n} \int_{x_{1},\cdots,x_{m}} \left.\frac{\delta^{m} W_{n-(n_{1}+\cdots+n_{m})}[K]}{\delta K_{a_{1}}(x_{1})\cdots\delta K_{a_{m}}(x_{m})}\right|_{K=K_{0}} K_{n_{1},a_{1}}[\rho;x_{1}]\cdots K_{n_{m},a_{m}}[\rho;x_{m}] }$} \\
& \scalebox{0.95}{${\displaystyle + \frac{1}{2}\int_{x} K^{a}_{n-1}[\rho;x] \rho_{a}(x) \delta_{n \geq 1} }$} \\
\scalebox{0.95}{${\displaystyle = }$} & \scalebox{0.95}{${\displaystyle -W_{n}[K=K_{0}] -\sum_{m=1}^{n-2} \int_{x} \left.\frac{\delta W_{n-m}[K]}{\delta K_{a}(x)}\right|_{K=K_{0}} K_{m,a}[\rho;x] }$} \\
& \scalebox{0.95}{${\displaystyle -\sum_{m=2}^{n-1} \frac{1}{m!} \sum_{\underset{\lbrace n_{1} + \cdots + n_{m} \leq n\rbrace}{n_{1},\cdots,n_{m}=1}}^{n} \int_{x_{1},\cdots,x_{m}} \left.\frac{\delta^{m} W_{n-(n_{1}+\cdots+n_{m})}[K]}{\delta K_{a_{1}}(x_{1})\cdots\delta K_{a_{m}}(x_{m})}\right|_{K=K_{0}} K_{n_{1},a_{1}}[\rho;x_{1}]\cdots K_{n_{m},a_{m}}[\rho;x_{m}] }$} \\
& \scalebox{0.95}{${\displaystyle + \frac{1}{2}\int_{x} K^{a}_{0}[\rho;x] \rho_{a}(x) \delta_{n 1}\;, }$}
\end{split}
\label{eq:2PPIEAIMstep30DON}
\end{equation}
where the second equality was obtained by using the relation:
\begin{equation}
\rho_{a}(x) = 2\left.\frac{\delta W_{1}[K]}{\delta K_{a}(x)}\right|_{K=K_{0}} \;.
\label{eq:2PPIEAIMdW1dK0DON}
\end{equation}
The latter equality results from the independence of $\rho$ with respect to $\hbar$, i.e. $\rho$ is a quantity of order $\mathcal{O}\big(\hbar^0\big)$. It might seem surprising at first for an object encoding quantum information like $\rho$ to be independent of $\hbar$. This striking feature is a direct consequence of the Legendre transform defining $\Gamma^{(\mathrm{2PPI})}$ in~\eqref{eq:pure2PPIEAdefinition0DON} (see~\eqref{eq:pure1PIEAIndependencephihbarstep20DON} in appendix~\ref{sec:1PIEAannIM} for a mathematical proof of this property). From this,~\eqref{eq:2PPIEAIMdW1dK0DON} directly follows after inserting~\eqref{eq:pure2PPIEAWExpansion0DON} into~\eqref{eq:pure2PPIEAdefinitionbis20DON} at $K=K_{0}$, i.e.:
\begin{equation}
\scalebox{0.99}{${\displaystyle \rho_{a}(x) = \underbrace{2 \left. \frac{\delta W_{0}[K]}{\delta K^{a}(x)} \right|_{K=K_{0}} \hbar^{-1}}_{0} + \underbrace{2 \left. \frac{\delta W_{1}[K]}{\delta K^{a}(x)} \right|_{K=K_{0}} }_{\rho_{a}(x)} + \underbrace{2 \left. \frac{\delta W_{2}[K]}{\delta K^{a}(x)} \right|_{K=K_{0}} \hbar + 2 \left. \frac{\delta W_{3}[K]}{\delta K^{a}(x)} \right|_{K=K_{0}} \hbar^{2} + \cdots}_{0} \;. }$}
\end{equation}

\vspace{0.5cm}

The zeroth-order coefficient of the 2PPI EA equals zero since the 1-point correlation function of $\vec{\widetilde{\varphi}}$ is imposed to vanish ($\Gamma^{(\mathrm{2PPI})}_{0}[\rho]=-W_{0}[K=K_{0}]=S\big(\vec{\widetilde{\varphi}}=\vec{0}\big)=0$) and we then discuss in further details the cases where $n=1$ and $2$ in~\eqref{eq:2PPIEAIMstep30DON}:
\begin{equation}
\Gamma^{(\mathrm{2PPI})}_{1}[\rho] = -W_{1}[K=K_{0}] + \frac{1}{2} \int_{x} K^{a}_{0}[\rho;x] \rho_{a}(x)\;,
\label{eq:2PPIEAIMGamma10DON}
\end{equation}
\begin{equation}
\Gamma^{(\mathrm{2PPI})}_{2}[\rho] = -W_{2}[K=K_{0}]\;.
\label{eq:2PPIEAIMGamma20DON}
\end{equation}
We then differentiate both sides of~\eqref{eq:2PPIEAIMGamma10DON} with respect to $\rho$:
\begin{equation}
\cancel{\frac{1}{2}K_{0,a}(x)} = -\int_{y} \frac{\delta W_{1}[K=K_{0}]}{\delta K_{0}^{b}(y)}\frac{\delta K_{0}^{b}(y)}{\delta \rho^{a}(x)} + \cancel{\frac{1}{2}K_{0,a}(x)} + \frac{1}{2}\int_{y} \frac{\delta K^{b}_{0}[\rho;y]}{\delta\rho^{a}(x)}\rho_{b}(y)\;,
\label{eq:2PPIEAIMK0cancelout0DON}
\end{equation}
where $K_{0}$ was introduced via:
\begin{equation}
\frac{\delta\Gamma^{(\mathrm{2PPI})}_{n}[\rho]}{\delta\rho^{a}(x)}=\frac{1}{2}K_{n-1,a}[\rho;x]\delta_{n \geq 1}\;,
\label{eq:2PPIEAdGamman0DON}
\end{equation}
which can be deduced from~\eqref{eq:pure2PPIEAdefinition0DON}. From~\eqref{eq:2PPIEAIMK0cancelout0DON}, it follows that:
\begin{equation}
0 = \int_{y} \left(-\frac{\delta W_{1}[K=K_{0}]}{\delta K_{0}^{b}(y)} + \frac{1}{2} \rho_{b}(y) \right)\frac{\delta K_{0}^{b}[\rho;y]}{\delta \rho_{a}(x)}\;.
\label{eq:2PPIEAIMK0canceloutbis0DON}
\end{equation}
We thus recover~\eqref{eq:2PPIEAIMdW1dK0DON} from~\eqref{eq:2PPIEAIMK0canceloutbis0DON} assuming that:
\begin{equation}
\frac{\delta W_{1}[K=K_{0}]}{\delta K_{0,a}(x)} = \left.\frac{\delta W_{1}[K]}{\delta K_{a}(x)}\right|_{K=K_{0}}\;,
\end{equation}
and
\begin{equation}
\frac{\delta K_{0}^{b}[\rho;y]}{\delta \rho_{a}(x)}\neq 0\;.
\label{eq:2PPIEAIMK0neq00DON}
\end{equation}
Valiev and Fernando have actually proven that~\eqref{eq:2PPIEAIMK0neq00DON} is a consequence of the strict concavity\footnote{In ref.~\cite{val97}, Valiev and Fernando use the convention $\Gamma^{(\mathrm{2PPI})}[\rho]=+W[K]-\int_{x}K^{a}[\rho;x]\frac{\delta W[K]}{\delta K^{a}(x)}$, as opposed to~\eqref{eq:pure2PPIEAdefinition0DON}, and therefore prove that $\Gamma^{(\mathrm{2PPI})}[\rho]$ is strictly convex instead of strictly concave.} of $\Gamma^{(\mathrm{2PPI})}[\rho]$ (or, equivalently, of $\Gamma_{1}^{(\mathrm{2PPI})}[\rho]$)~\cite{val97}, i.e.:
\begin{equation}
\frac{\delta K_{0}^{b}[\rho;y]}{\delta \rho_{a}(x)} = 2\frac{\delta^{2} \Gamma_{1}^{(\mathrm{2PPI})}[\rho]}{\delta \rho_{a}(x) \delta \rho_{b}(y)} < 0 \;,
\end{equation}
as follows from~\eqref{eq:2PPIEAdGamman0DON}. The strict concavity of the 2PPI EA results itself from that of the Schwinger functional~\cite{val97}:
\begin{equation}
W\big[\nu K+(1-\nu)K'\big] > \nu W[K] + (1-\nu) W\big[K'\big]\;,
\end{equation}
for $0 < \nu < 1$. From the latter result, it also follows that the mapping $\lbrace K \rbrace\rightarrow\lbrace \rho \rbrace$, between the set of allowable external sources $K$ and the corresponding densities $\rho$ generated by~\eqref{eq:pure2PPIEAdefinitionbis20DON}, is one-to-one. This is a very fundamental property for the IM applied to the 2PPI EA as it guarantees that~\eqref{eq:pure2PPIEAdefinitionbis20DON} can be inverted to obtain $K[\rho]$ without ambiguity. Besides the IM, it can also be viewed as an existence proof of the 2PPI EA, which plays the role of the density functional in the present analogy with DFT. In other words, the one-to-one nature of the mapping $\lbrace K \rbrace\rightarrow\lbrace \rho \rbrace$ implies that $\Gamma^{(\mathrm{2PPI})}[\rho]$, and therefore the corresponding gs energy, is a unique functional of the density $\rho$, which is none other than the first Hohenberg-Kohn theorem~\cite{hoh64}. This should be coupled with the proof that $\Gamma^{(\mathrm{2PPI})}[\rho]$ satisfies the second Hohenberg-Kohn theorem~\cite{hoh64}, which follows by shifting the external source by an arbitrary external potential $V_{\mathrm{ext}}$~\cite{fuk94}:
\begin{equation}
\begin{split}
\Gamma^{(\mathrm{2PPI})}[\rho]= & -W[K]+\frac{\hbar}{2}\int_{x} K^{a}[\rho;x] \rho_{a}(x) \\
= & \underbrace{-W_{V_{\mathrm{ext}}=0}[K-V_{\mathrm{ext}}]+ \frac{\hbar}{2} \int_{x} (K^{a}[\rho;x]-V^{a}_{\mathrm{ext}}(x)) \rho_{a}(x)}_{{\displaystyle\Gamma_{V_{\mathrm{ext}}=0}\left[\rho\right]}} + \frac{\hbar}{2} \int_{x} V^{a}_{\mathrm{ext}}(x) \rho_{a}(x) \;.
\end{split}
\label{eq:2PPIEAHohenbergKohnTheorem}
\end{equation}
Hence, the 2PPI EA satisfies the second Hohenberg-Kohn theorem simply because the Schwinger functional depends on $K$ and $V_{\mathrm{ext}}$ in the same fashion. These existence proofs shown in refs.~\cite{val97,fuk94} mean that the framework of the 2PPI EA contains all the ingredients of a DFT. Furthermore, as pointed out in ref.~\cite{val97} via the IM, the Kohn-Sham implementation~\cite{koh65,koh65bis} is also present in this framework. On the one hand, according to~\eqref{eq:2PPIEAIMdW1dK0DON} and the one-to-one nature of the mapping $\lbrace K \rbrace\rightarrow\lbrace \rho \rbrace$, the exact gs density of the interacting system can be obtained, for a single configuration $K_{0}$ of the source $K$, from the first-order coefficient $W_{1}[K]$ which is independent of the coupling constant $\lambda$\footnote{Recall that, since we have imposed the 1-point correlation function of $\vec{\widetilde{\varphi}}$ to vanish here, $\hbar$-expansion results truncated at order $\mathcal{O}\big(\hbar^{n+1}\big)$ coincide with those of the $\lambda$-expansion truncated at order $\mathcal{O}\big(\lambda^{n}\big)$ for all $n\in\mathbb{N}$. This implies that $W_{1}[K]$ does not depend on $\lambda$ in the $\hbar$-expansion scheme.}. On the other hand, the Kohn-Sham implementation states that there is a unique external potential $V_{\mathrm{ext},\mathrm{KS}}$ such that this exact gs density characterizing the interacting system coincides with that of an auxiliary non-interacting one submitted to $V_{\mathrm{ext},\mathrm{KS}}$. Therefore, the Kohn-Sham implementation acquires its meaning in the 2PPI EA formalism after noticing that $K_{0}$ plays the role of the Kohn-Sham potential. Moreover, we can also split the 2PPI EA according to the Kohn-Sham scheme: the kinetic part can be isolated through a derivative expansion of the EA and the exchange-correlation part can be approximated using usual methods (local density approximation, ...)~\cite{fuk94}. Thus, we can conclude that DFT is always present in the framework of the 2PPI EA according to the above proofs of the Hohenberg-Kohn theorems and exploiting the 2PPI EA via the IM amounts to implementing a Kohn-Sham DFT.

\vspace{0.5cm}

We will then explain how the Kohn-Sham self-consistent procedure works in the framework of the IM by differentiating expression~\eqref{eq:2PPIEAIMGamma20DON} of $\Gamma_{2}^{(\mathrm{2PPI})}$ with respect to $\rho$:
\begin{equation}
\begin{split}
\frac{\delta\Gamma^{(\mathrm{2PPI})}_{2}[\rho]}{\delta\rho^{a}(x)} = & -\frac{\delta W_{2}[K=K_{0}]}{\delta\rho^{a}(x)} \\
= & -\int_{y}\frac{\delta W_{2}[K=K_{0}]}{\delta K_{0,b}(y)}\frac{\delta K_{0,b}[\rho;y]}{\delta \rho^{a}(x)} \\
= & -\frac{1}{2}\int_{y}\frac{\delta W_{2}[K=K_{0}]}{\delta K_{0,b}(y)}\left(\frac{\delta^{2} W_{1}[K=K_{0}]}{\delta K_{0}(y) \delta K_{0}(x)}\right)_{ba}^{-1} \\
= & -\frac{1}{2}\int_{y} \boldsymbol{D}_{ab}^{-1}[K_{0};x,y] \frac{\delta W_{2}[K=K_{0}]}{\delta K_{0,b}(y)} \;,
\end{split}
\label{eq:2PPIEAK10DON}
\end{equation}
where the third line is obtained from~\eqref{eq:2PPIEAIMdW1dK0DON} and the following propagator was introduced:
\begin{equation}
\boldsymbol{D}_{a b}[K_{0};x,y] \equiv \frac{\delta^{2} W_{1}[K=K_{0}]}{\delta K^{a}_{0}(x) \delta K^{b}_{0}(y)}\;.
\label{eq:2PPIEADpropagator0DON}
\end{equation}
Hence, we have just used the chain rule:
\begin{equation}
\frac{\delta}{\delta\rho^{a}(x)} = \frac{1}{2}\int_{y} \boldsymbol{D}_{ab}^{-1}[K_{0};x,y] \frac{\delta}{\delta K_{0,b}(y)} \;.
\label{eq:2PPIEAchainrule0DON}
\end{equation}
As a next step, we infer from~\eqref{eq:2PPIEAdGamman0DON} and~\eqref{eq:2PPIEAK10DON}:
\begin{equation}
K_{1,a}[\rho;x] = -\int_{y} \boldsymbol{D}_{ab}^{-1}[K_{0};x,y] \frac{\delta W_{2}[K=K_{0}]}{\delta K_{0,b}(y)} \;.
\label{eq:2PPIEAK1bis0DON}
\end{equation}
This recipe can be applied at the next order, i.e. setting $n=3$ in~\eqref{eq:2PPIEAIMstep30DON} to obtain an expression for $\Gamma^{(\mathrm{2PPI})}_{3}[\rho]$ that we differentiate with respect to $\rho$ in order to deduce $K_{2}[\rho]$ according to~\eqref{eq:2PPIEAdGamman0DON}. We can proceed in this way up to any order, which would yield all $\Gamma^{(\mathrm{2PPI})}_{n}$ and $K_{n}$ coefficients thus derived expressed in terms of the $\boldsymbol{D}$ propagator and $W_{n}$ coefficients as well (recall that the $W_{n}$ coefficients are fully specified by the LE and are thus considered as inputs of the IM procedure). For a truncation order $N_{t}$, this procedure can be represented as follows:
\begin{equation}
\Gamma^{(\mathrm{2PPI})}_{1} \xrightarrow{\hspace*{0.6cm}} \Gamma^{(\mathrm{2PPI})}_{2} \underset{\scalebox{0.8}{${\displaystyle K_{1}[\rho]=2\frac{\delta \Gamma^{(\mathrm{2PPI})}_{2}[\rho]}{\delta\rho}}$}}{\xrightarrow{\hspace*{2.4cm}}} K_{1} \xrightarrow{\hspace*{0.6cm}} \Gamma^{(\mathrm{2PPI})}_{3} \underset{\scalebox{0.8}{${\displaystyle K_{2}[\rho]=2\frac{\delta \Gamma^{(\mathrm{2PPI})}_{3}[\rho]}{\delta\rho}}$}}{\xrightarrow{\hspace*{2.4cm}}} K_{2} \xrightarrow{\hspace*{0.6cm}} \cdots \xrightarrow{\hspace*{0.6cm}} \Gamma^{(\mathrm{2PPI})}_{N_{t}}
\label{eq:ProcedureDeterminationSourcesIM}
\end{equation}
As a result, the 2PPI EA is expressed in terms of the zeroth-order coefficient of the source (i.e. $K_{0}$ here) instead of the argument of the EA (i.e. $\rho$ here). In fact, the dependence of $\Gamma^{(\mathrm{2PPI})}[\rho]$ with respect to $\rho$ in its expression resulting from the IM is implicit through $K_{0}[\rho]$\footnote{Consequently, the $\boldsymbol{D}$ propagator is convenient here notably because it enables us through the chain rule~\eqref{eq:2PPIEAchainrule0DON} to turn derivatives of the $\Gamma^{(\mathrm{2PPI})}_{n}$ coefficients with respect to $\rho$ into derivatives with respect to $K_{0}$.}. The reason for this difference was given at the beginning of this section: for $n$PPI EAs and contrary to $n$PI EAs, the Legendre transform underlying the definition of the EA can not be done explicitly in finite dimensions. Technically, this translates into the fact that the relation~\eqref{eq:2PPIEAIMdW1dK0DON} can not be inverted to write $K_{0}$ \textbf{explicitly} in terms of $\rho$\footnote{On the contrary, we notably show in appendix~\ref{sec:1PIEAannIM} that the Legendre transform underlying the 1PI EA can be carried out explicitly. This can be seen in the framework of the IM from~\eqref{eq:pure1PIEAlambdaphi0DON} which can be inverted to express $\vec{J}_{0}\big[\vec{\phi}\big]$ explicitly in terms of the 1-point correlation function $\vec{\phi}$, as shown by~\eqref{eq:1PIEAJ0phi0DON}. In finite dimensions, there is no counterpart of this procedure for $K_{0}[\rho]$ in the present case.}. In (0+0)-D and after imposing that all 1-point correlation functions vanish (so that all bilocal sources involved in the 2PI EA formalism reduce to scalars in color space), this problem vanishes since the 2PPI EA reduces to the 2PI one in this situation, as discussed at the end of this section.

\vspace{0.5cm}

Hence, in the framework of the 2PPI EA, we must solve the gap equation for $K_{0}$ instead of $\rho$. This equation is therefore more conveniently rewritten as follows:
\begin{equation}
\scalebox{0.92}{${\displaystyle 0 = \left.\frac{\delta\Gamma^{(\mathrm{2PPI})}[\rho]}{\delta\rho_{a}(x)}\right|_{\rho=\overline{\rho}} = \underbrace{\left.\frac{\delta\Gamma_{0}^{(\mathrm{2PPI})}[\rho]}{\delta\rho_{a}(x)}\right|_{\rho=\overline{\rho}}}_{0} + \underbrace{\left.\frac{\delta\Gamma_{1}^{(\mathrm{2PPI})}[\rho]}{\delta\rho_{a}(x)}\right|_{\rho=\overline{\rho}}}_{\frac{1}{2}\overline{K}_{0,a}(x)} + \left.\frac{\delta\Gamma_{2}^{(\mathrm{2PPI})}[\rho]}{\delta\rho_{a}(x)}\right|_{\rho=\overline{\rho}} + \left.\frac{\delta\Gamma_{3}^{(\mathrm{2PPI})}[\rho]}{\delta\rho_{a}(x)}\right|_{\rho=\overline{\rho}} + \cdots \quad \forall a,x \;,}$}
\end{equation}
and, more conveniently:
\begin{equation}
\overline{K}_{0,a}(x) = - \sum_{n=2}^{\infty} \int_{y} \boldsymbol{D}_{ab}^{-1}\big[K_{0}=\overline{K}_{0};x,y\big] \left.\frac{\delta \Gamma^{(\mathrm{2PPI})}_{n}[\rho]}{\delta K_{0,b}(y)}\right|_{K_{0}=\overline{K}_{0}}\;,
\label{eq:2PPIEAgapequation0DON}
\end{equation}
where we have used the functional chain rule~\eqref{eq:2PPIEAchainrule0DON}. The self-consistent procedure aiming at extracting $\overline{K}_{0}$ from~\eqref{eq:2PPIEAgapequation0DON} is completely analogous to that treating the Kohn-Sham equations. It consists in the 5 following steps:
\begin{itemize}
\item[1)] Choose a truncation order $N_{t}$ with respect to $\hbar$.
\item[2)] Make an initial guess for $\overline{K}_{0}$ that we denote as $\overline{K}^{(\mathrm{old})}_{0}$.
\item[3)] Determine the derivatives $\left.\frac{\delta \Gamma^{(\mathrm{2PPI})}_{2}[\rho]}{\delta K_{0,a}(x)}\right|_{K_{0}=\overline{K}^{(\mathrm{old})}_{0}},\cdots,\left.\frac{\delta \Gamma^{(\mathrm{2PPI})}_{N_{t}}[\rho]}{\delta K_{0,a}(x)}\right|_{K_{0}=\overline{K}^{(\mathrm{old})}_{0}}$, where the coefficients $\Gamma^{(\mathrm{2PPI})}_{2}[\rho],\cdots,\Gamma^{(\mathrm{2PPI})}_{N_{t}}[\rho]$ are found from the IM.
\item[4)] Evaluate the new configuration $\overline{K}^{(\mathrm{new})}_{0}$ of the Kohn-Sham potential $K_{0}$ from~\eqref{eq:2PPIEAgapequation0DON} rewritten as:
\begin{equation}
\overline{K}^{(\mathrm{new})}_{0,a}(x) = - \sum_{n=2}^{N_{t}} \int_{y} \boldsymbol{D}_{ab}^{-1}\Big[K_{0}=\overline{K}^{(\mathrm{old})}_{0};x,y\Big] \left.\frac{\delta \Gamma^{(\mathrm{2PPI})}_{n}[\rho]}{\delta K_{0,b}(y)}\right|_{K_{0}=\overline{K}^{(\mathrm{old})}_{0}}\;,
\end{equation}
where the RHS is completely specified by steps 1, 2 and 3.
\item[5)] Replace $\overline{K}^{(\mathrm{old})}_{0}$ by $\overline{K}^{(\mathrm{new})}_{0}$ in step 2 and repeat steps 2 to 4 until the difference $\big|\overline{K}^{(\mathrm{new})}_{0}-\overline{K}^{(\mathrm{old})}_{0}\big|$ becomes reasonably small, i.e. until self-consistency is achieved.
\end{itemize}

\vspace{0.3cm}

\noindent
Such a self-consistent procedure is basically identical to that used to treat gap equations for other $n$PPI or $n$PI EAs, whether we solve these equations for zeroth-order coefficients of the sources or for the arguments of the EA. We just present this recipe in detail here to conclude our discussion on the connection between the 2PPI EA and Kohn-Sham DFT, which entirely follows from the fact that $K_{0}$ is a Kohn-Sham potential.

\vspace{0.5cm}

We then briefly discuss the diagrammatic properties of the 2PPI EA. The IM outlined above would yield an expression for the 2PPI EA in terms of the $K_{0}$ coefficient. Such an expression can be represented diagrammatically after constructing the Feynman rules for the $\boldsymbol{D}$ propagator as well as for other propagators and vertex functions involved in the LE of the Schwinger functional. It was shown in ref.~\cite{oku96} that the diagrams thus obtained are all 1-vertex-irreducible (1VI) beyond the Hartree-Fock level, i.e., for our $\hbar$-expansion, $\Delta\Gamma^{(\mathrm{2PPI})}[\rho] \equiv \Gamma^{(\mathrm{2PPI})}[\rho]-\Gamma_{0}^{(\mathrm{2PPI})}[\rho]-\Gamma_{1}^{(\mathrm{2PPI})}[\rho]-\Gamma_{2}^{(\mathrm{2PPI})}[\rho]$ is only given by 1VI diagrams\footnote{A diagram is 1-vertex-reducible (1VR) if at least one of its vertices can be split to render the diagram disconnected. If a diagram is neither a trivial skeleton (i.e. a diagram that does not contain any vertex, which usually represents $\mathrm{STr}\ln$ terms) nor 1VR, it is 1VI by definition. For example, in the case of the diagrams resulting from the LE in section~\ref{sec:OriginalLE}, splitting a zigzag vertex simply means cutting the corresponding zigzag line in half.}. More specifically, such a diagrammatic rule can be formulated as follows for our finite-dimensional $O(N)$-symmetric $\varphi^{4}$ model:
\begin{equation}
\Delta\Gamma^{(\mathrm{2PPI})}\sim\left.\frac{\int\mathcal{D}\vec{\widetilde{\varphi}}\mathcal{D}\vec{\widetilde{\Omega}} \ e^{-\frac{1}{\hbar}\breve{S}\big[\vec{\widetilde{\varphi}},\vec{\widetilde{\Omega}}\big]}}{\int\mathcal{D}\vec{\widetilde{\varphi}}\mathcal{D}\vec{\widetilde{\Omega}} \ e^{-\frac{1}{\hbar}\breve{S}_{0}\big[\vec{\widetilde{\varphi}},\vec{\widetilde{\Omega}}\big]}}\right|_{\mathrm{conn/tree/1VI/excl}}\;,
\label{eq:2PPIEAdiagrammaticrule0DON}
\end{equation}
with the classical action:
\begin{equation}
\begin{split}
\breve{S}\Big[\vec{\widetilde{\varphi}},\vec{\widetilde{\Omega}}\Big] = & \ \underbrace{\frac{1}{2}\int_{x,y}\widetilde{\varphi}^{a}(x)\boldsymbol{G}^{-1}_{ab}[K_{0};x,y]\widetilde{\varphi}^{b}(y)+\frac{1}{2}\int_{x,y}\widetilde{\Omega}^{a}(x) \boldsymbol{D}_{ab}[K_{0};x,y] \widetilde{\Omega}^{b}(y)}_{\breve{S}_{0}\big[\vec{\widetilde{\varphi}},\vec{\widetilde{\Omega}}\big]} \\
& + \frac{\lambda}{4!} \int_{x}\left(\widetilde{\varphi}_a(x)\widetilde{\varphi}^a(x)\right)^2 +\frac{1}{2}\int_{x}\widetilde{\Omega}^{a}(x)\widetilde{\varphi}^{2}_{a}(x) \;,
\end{split}
\end{equation}
where $\boldsymbol{G}[K_{0}] \equiv \boldsymbol{G}_{K}[K=K_{0}]$ is defined from the propagator $\boldsymbol{G}_{K}$ involved in the LE of $W[K]$, i.e.:
\begin{equation}
\boldsymbol{G}^{-1}_{K;ab}[K;x,y] = \left.\frac{\delta^{2}S\big[\vec{\widetilde{\varphi}}\big]}{\delta\widetilde{\varphi}^{a}(x) \delta\widetilde{\varphi}^{b}(y)}\right|_{\vec{\widetilde{\varphi}}=\vec{0}} - K_{a}(x) \delta_{ab} \delta(x-y) = \left(-\nabla_x^2 + m^2 - K_{a}(x) \right)\delta_{ab}\delta(x-y)\;.
\label{eq:2PPIEAdefinitionGKpropagator0DON}
\end{equation}
Owing to the absence of spontaneous breakdown of the $O(N)$ symmetry in the present framework, the source $K$ as well as all the $K_{n}$ coefficients are scalars in color space, i.e. $K_{a}(x)=K(x)$ and $K_{n,a}(x)=K_{n}(x)$ $\forall n,a,x$, which implies the same trivial structure for both $\boldsymbol{G}_{K}[K]$ and $\boldsymbol{G}[K_{0}]$, i.e. $\boldsymbol{G}_{K;ab}[K;x,y]=G_{K}[K;x,y]\delta_{ab}$ and $\boldsymbol{G}_{ab}[K_{0};x,y]=G[K_{0};x,y]\delta_{ab}$ $\forall a,b,x,y$. We also stress that $\vec{\widetilde{\Omega}}$ is a fake quantum field which has nothing to do with the auxiliary field introduced via HST or OPT. In other words, $\vec{\widetilde{\Omega}}$ is just a mathematical artifact introduced in~\eqref{eq:2PPIEAdiagrammaticrule0DON} after noticing that the diagrammatic expression of the 2PPI EA of a $\varphi^{4}$-theory resulting from the IM looks like that of a theory with both $\varphi^{4}$ and Yukawa interaction terms, in which the propagator associated to the ``meson'' field is $\boldsymbol{D}^{-1}[K_{0}]$\footnote{Note that our definition~\eqref{eq:2PPIEADpropagator0DON} of the $\boldsymbol{D}$ propagator is the inverse of that of Okumura in ref.~\cite{oku96} (see equation~(2.89) in ref.~\cite{oku96}).}. According to the rule~\eqref{eq:2PPIEAdiagrammaticrule0DON}, the diagrammatic expression resulting from the IM for the 2PPI EA with vanishing 1-point correlation function (for $\vec{\widetilde{\varphi}}$) can be equivalently obtained by performing a LE of the generating functional in the numerator of the RHS of~\eqref{eq:2PPIEAdiagrammaticrule0DON} around $\big(\vec{\widetilde{\varphi}},\vec{\widetilde{\Omega}}\big)=\big(\vec{0},\vec{0}\big)$. The indication ``$\mathrm{conn/tree/1VI/excl}$'' means that, among all diagrams generated by this LE, only those that are connected, 1VI and tree with respect to the $\boldsymbol{D}$ propagator\footnote{The expression ``tree with respect to the $\boldsymbol{D}$ propagator'' is synonymous to ``1PR with respect to each $\boldsymbol{D}$ propagator''. Namely, a diagram is tree with respect to the $\boldsymbol{D}$ propagator either if it does not contain any $\boldsymbol{D}$ propagator or if removing any of its $\boldsymbol{D}$ propagators renders this diagram disconnected.} are kept and we must also exclude every diagram that contains self connections, i.e.:
\begin{equation*}
\begin{gathered}
\begin{fmffile}{Diagrams/2PPIEA_SelfConnection}
\begin{fmfgraph*}(12,12)
\fmfleft{i}
\fmfright{o}
\fmf{phantom,tension=5.0}{i,vLeft}
\fmf{phantom,tension=0.5}{o,vLeft}
\fmf{phantom,tension=0.5}{i,vRight}
\fmf{phantom,tension=5.0}{o,vRight}
\fmf{plain,left,tension=0,foreground=(1,,0,,0)}{vLeft,vRight}
\fmf{plain,right,tension=0,foreground=(1,,0,,0)}{vLeft,vRight}
\fmf{plain,tension=0,foreground=(0,,0,,1)}{vLeft,i}
\end{fmfgraph*}
\end{fmffile}
\end{gathered}
\end{equation*}
and/or double connections, i.e.:
\begin{equation*}
\begin{gathered}
\begin{fmffile}{Diagrams/2PPIEA_DoubleConnection}
\begin{fmfgraph*}(12,12)
\fmfleft{i}
\fmfright{o}
\fmf{phantom,tension=5.0}{i,vLeft}
\fmf{phantom,tension=0.5}{o,vLeft}
\fmf{phantom,tension=0.5}{i,vRight}
\fmf{phantom,tension=5.0}{o,vRight}
\fmf{plain,left,tension=0,foreground=(1,,0,,0)}{vLeft,vRight}
\fmf{plain,right,tension=0,foreground=(1,,0,,0)}{vLeft,vRight}
\fmf{plain,tension=0,foreground=(0,,0,,1)}{vLeft,i}
\fmf{plain,tension=0,foreground=(0,,0,,1)}{vRight,o}
\end{fmfgraph*}
\end{fmffile}
\end{gathered}
\end{equation*}
with the Feynman rule:
\begin{equation}
\begin{gathered}
\begin{fmffile}{Diagrams/2PPIEA_FeynRuleGSourceGK0}
\begin{fmfgraph*}(20,20)
\fmfleft{i0,i1,i2,i3}
\fmfright{o0,o1,o2,o3}
\fmflabel{$x, a$}{v1}
\fmflabel{$y, b$}{v2}
\fmf{phantom}{i1,v1}
\fmf{phantom}{i2,v1}
\fmf{plain,tension=0.6,foreground=(1,,0,,0)}{v1,v2}
\fmf{phantom}{v2,o1}
\fmf{phantom}{v2,o2}
\end{fmfgraph*}
\end{fmffile}
\end{gathered} \hspace{0.5cm} \rightarrow \boldsymbol{G}_{a b}[K_{0};x,y] \;.
\label{eq:2PPIEAfeynRuleGK00DON}
\end{equation}
These restrictions are quite drastic so that, for the studied model, there are no diagrams involving $\boldsymbol{D}^{-1}[K_{0}]$ in the three first non-trivial orders of the 2PPI EA, i.e. in the $\Gamma_{n}^{(\mathrm{2PPI})}$ coefficients for $n\leq 4$. However, the following diagram contributes to $\Gamma^{(\mathrm{2PPI})}_{5}[\rho]$ for instance:
\begin{equation}
\begin{gathered}
\begin{fmffile}{Diagrams/2PPIEA_Gamma5Diag1}
\begin{fmfgraph}(35,17.5)
\fmfleft{iDown2,iDown1,i,iUp1,iUp2}
\fmfright{oDown2,oDown1,o,oUp1,oUp2}
\fmftop{vUpLeft,vUp,vUpRight}
\fmfbottom{vDownLeft,vDown,vDownRight}
\fmf{phantom,tension=1.6}{i,v1}
\fmf{phantom,tension=1.0}{o,v1}
\fmf{phantom,tension=1.0}{i,v2}
\fmf{phantom,tension=1.6}{o,v2}
\fmf{phantom,tension=1.3}{iUp1,v3}
\fmf{phantom,tension=1.6}{oUp1,v3}
\fmf{phantom,tension=1.6}{iDown1,v4}
\fmf{phantom,tension=1.3}{oDown1,v4}
\fmf{plain,tension=0,foreground=(1,,0,,0)}{v3,v4}
\fmfv{decor.shape=circle,decor.filled=empty,decor.size=1.5thick}{v1}
\fmfv{decor.shape=circle,decor.filled=empty,decor.size=1.5thick}{v2}
\fmf{plain,left=0.35,tension=0,foreground=(1,,0,,0)}{v1,v2}
\fmf{plain,right=0.35,tension=0,foreground=(1,,0,,0)}{v1,v2}
\fmf{plain,left=0.35,tension=0,foreground=(1,,0,,0)}{vUpLeft,vDownLeft}
\fmf{plain,right=0.35,tension=0,foreground=(1,,0,,0)}{vUpLeft,vDownLeft}
\fmf{plain,left=0.35,tension=0,foreground=(1,,0,,0)}{vUpRight,vDownRight}
\fmf{plain,right=0.35,tension=0,foreground=(1,,0,,0)}{vUpRight,vDownRight}
\fmf{plain,left=0.35,tension=0,foreground=(1,,0,,0)}{vUpL,vDownL}
\fmf{plain,right=0.35,tension=0,foreground=(1,,0,,0)}{vUpL,vDownL}
\fmf{plain,left=0.35,tension=0,foreground=(1,,0,,0)}{vUpR,vDownR}
\fmf{plain,right=0.35,tension=0,foreground=(1,,0,,0)}{vUpR,vDownR}
\fmf{zigzag,left=0.2,tension=2.2,foreground=(0,,0,,1)}{vUpLeft,vUpL}
\fmf{phantom,tension=1.0}{vUpRight,vUpL}
\fmf{phantom,tension=1.0}{vUpLeft,vUpR}
\fmf{zigzag,right=0.2,tension=2.2,foreground=(0,,0,,1)}{vUpRight,vUpR}
\fmf{zigzag,right=0.2,tension=2.2,foreground=(0,,0,,1)}{vDownLeft,vDownL}
\fmf{phantom,tension=1.0}{vDownRight,vDownL}
\fmf{phantom,tension=1.0}{vDownLeft,vDownR}
\fmf{zigzag,left=0.2,tension=2.2,foreground=(0,,0,,1)}{vDownRight,vDownR}
\end{fmfgraph}
\end{fmffile}
\end{gathered}\hspace{0.4cm}\;,
\label{eq:2PPIEAGamma5diagram0DON}
\end{equation}
where $\boldsymbol{D}^{-1}[K_{0}]$ is represented according to:
\begin{equation}
\begin{gathered}
\begin{fmffile}{Diagrams/2PPIEA_FeynRuleDminus1main}
\begin{fmfgraph*}(20,20)
\fmfleft{i0,i1,i2,i3}
\fmfright{o0,o1,o2,o3}
\fmftop{vUpL8,vUpL7,vUpL6,vUpL5,vUpL4,vUpL3,vUpL2,vUpL1,vUp,vUpR1,vUpR2,vUpR3,vUpR4,vUpR5,vUpR6,vUpR7,vUpR8}
\fmfbottom{vDownL8,vDownL7,vDownL6,vDownL5,vDownL4,vDownL3,vDownL2,vDownL1,vDown,vDownR1,vDownR2,vDownR3,vDownR4,vDownR5,vDownR6,vDownR7,vDownR8}
\fmf{phantom,tension=0.25}{vUpL3,vLeft}
\fmf{phantom,tension=5.5}{vDownL3,vLeft}
\fmf{phantom,tension=5.5}{vUpR3,vRight}
\fmf{phantom,tension=0.25}{vDownR3,vRight}
\fmflabel{$x, a$}{v1}
\fmflabel{$y, b$}{v2}
\fmf{phantom}{i1,v1}
\fmf{phantom}{i2,v1}
\fmf{phantom,tension=0.6,foreground=(1,,0,,0)}{v1,v2}
\fmf{plain,left=0.3,tension=0,foreground=(1,,0,,0)}{v1,v2}
\fmf{plain,right=0.3,tension=0,foreground=(1,,0,,0)}{v1,v2}
\fmf{phantom}{v2,o1}
\fmf{phantom}{v2,o2}
\fmf{plain,tension=0.8,foreground=(1,,0,,0)}{vLeft,vRight}
\end{fmfgraph*}
\end{fmffile}
\end{gathered} \hspace{0.5cm} \rightarrow \boldsymbol{D}^{-1}_{ab}[K_{0};x,y]\;,
\label{eq:2PPIEAfeynRuleDminus10DON}
\end{equation}
and the zigzag vertex is as usual given by:
\begin{equation}
\begin{gathered}
\begin{fmffile}{Diagrams/1PIEA_V4}
\begin{fmfgraph*}(20,20)
\fmfleft{i0,i1,i2,i3}
\fmfright{o0,o1,o2,o3}
\fmf{phantom,tension=2.0}{i1,i1bis}
\fmf{plain,tension=2.0,foreground=(1,,0,,0)}{i1bis,v1}
\fmf{phantom,tension=2.0}{i2,i2bis}
\fmf{plain,tension=2.0,foreground=(1,,0,,0)}{i2bis,v1}
\fmf{zigzag,label=$x$,tension=0.6,foreground=(0,,0,,1)}{v1,v2}
\fmf{phantom,tension=2.0}{o1bis,o1}
\fmf{plain,tension=2.0,foreground=(1,,0,,0)}{v2,o1bis}
\fmf{phantom,tension=2.0}{o2bis,o2}
\fmf{plain,tension=2.0,foreground=(1,,0,,0)}{v2,o2bis}
\fmflabel{$a$}{i1bis}
\fmflabel{$b$}{i2bis}
\fmflabel{$c$}{o1bis}
\fmflabel{$d$}{o2bis}
\end{fmfgraph*}
\end{fmffile}
\end{gathered} \quad \rightarrow \lambda\delta_{a b}\delta_{c d}\;.
\label{eq:2PPIEAfeynRuleZigzagVertex0DON}
\end{equation}
Therefore, as opposed to $n$PI EAs, some diagrams involved in the final expression of the 2PPI EA are not present at the level of the LE of the Schwinger functional in the framework of the $\hbar$-expansion. For example, the LE of $W[K]$ does not yield diagrams in terms of $\boldsymbol{D}^{-1}[K_{0}]$ such as~\eqref{eq:2PPIEAGamma5diagram0DON}. In conclusion, the diagrammatic construction of the 2PPI EA is based on more demanding rules than those of the 2PI EA but this must be contrasted with the numerical resolution of the gap equations which is more difficult in the situation of the 2PI EA since the latter is a functional of propagators (i.e. of bilocal objects) as opposed to the density functional $\Gamma^{(\mathrm{2PPI})}$.

\vspace{0.5cm}

Most of the present discussion on the IM readily extends to QFTs with several fields or to one field with additional external sources coupled with combinations of this field other than $\widetilde{\varphi}^{2}_{a}(x)$ encountered in~\eqref{eq:SJKfiniteD2PPIEA}. However, one has to keep in mind that this section does not discuss the most general implementation of the IM. If the Schwinger functional involves several sources (as e.g. for a 4PPI EA), the procedure~\eqref{eq:ProcedureDeterminationSourcesIM} is not convenient to determine the source coefficients so that one must rather follow the most general implementation of the IM~\cite{oku96,oko97} based on the series representing the arguments of the EA (such as~\eqref{eq:pure2PPIEArhoExpansion0DON} in the studied case) and outlined in appendix~\ref{ann:InversionMethod}. This remark applies in particular to the case of one field with a local source $\vec{J}$ added to~\eqref{eq:SJKfiniteD2PPIEA} such as:
\begin{equation}
S_{JK}\Big[\vec{\widetilde{\varphi}}\Big] \equiv S\Big[\vec{\widetilde{\varphi}}\Big] - \int_{x} J^{a}(x) \widetilde{\varphi}_{a}(x) -\frac{1}{2}\int_{x} K^{a}(x) \widetilde{\varphi}^{2}_{a}(x) \;,
\end{equation}
in order to include a finite 1-point correlation function in the 2PPI EA formalism. This would define a promising area to construct a DFT able to treat SSB, notably for $O(N)$-symmetric theories. There are already applications of the 2PPI EA to the $O(N)$-symmetric $\varphi^{4}$-theory which include a possible non-vanishing 1-point correlation function in the formalism~\cite{baa03,baa03bis,baa03bis2,baa04}. They have notably found an unphysical SSB at next-to-leading order (NLO) of a $1/N$-expansion in (1+1)-D at finite temperature. However, each approach aiming at avoiding possible violations of Ward identities in the framework of the 2PI EA, like the SI2PI EA~\cite{pil13}, the method of non-vanishing external sources of Garbrecht and Millington~\cite{gar16} or even the mixed 2PI EA~\cite{col74,ben77,cho99,mih01,bla01,coo05,aar02}, is straightforwardly adaptable to the 2PPI EA.

\vspace{0.5cm}

For the 2PPI EA with possible non-zero 1-point correlation function, we would find, as for the 2PI EA (see appendix~\ref{sec:original2PIEAannIM}), that the proliferation of diagrams is already cumbersome at the first non-trivial order in the framework of the IM. Unfortunately, as explained above, we can no longer simply pick the relevant diagrams among those of the LE of the Schwinger functional in the framework of the 2PPI EA. Nevertheless, one could tackle this issue by generalizing the rule~\eqref{eq:2PPIEAdiagrammaticrule0DON} and/or using the whole machinery of the IM presented above and in appendix~\ref{ann:InversionMethod} to write a code determining all the diagrams contributing to a 2PPI EA with finite 1-point correlation functions. A numerical tool constructing all the diagrams involved in Bogoliubov MBPT (BMBPT) up to any truncation order is discussed in refs.~\cite{art18,art19,art20} and could be used as a starting point. Such a code would be of great use for any $n$PPI or $n$PI EA, especially in the case of realistic models for which the diagrams generated by the LE might be very numerous at the first non-trivial orders as well.

\vspace{0.5cm}

Let us finish this general presentation of the 2PPI EA by discussing the work of Furnstahl and collaborators~\cite{pug03,bha05,fur07,dru10,fur12,fur20} who brought the 2PPI EA (still via the IM) in nuclear physics:
\begin{itemize}
\item Refs.~\cite{pug03,bha05} aim at turning the Hartree-Fock implementation of the Skyrme EDF into a systematically improvable approach by exploiting the 2PPI EA via the IM. In order to achieve this, the latter method is applied to a dilute fermion gas with short-range interaction. The authors use the product $a_{\mathrm{s}}k_{\mathrm{F}}$ instead of $\hbar$ as expansion parameter for the EA\footnote{The short-range feature of the interaction renders the product $a_{\mathrm{s}}k_{\mathrm{F}}$ small enough to be a relevant expansion parameter.}, with $a_{\mathrm{s}}$ and $k_{\mathrm{F}}$ being respectively the $s$-wave scattering length and the Fermi momentum of the gas. The diagrammatic underlying the expansion of the Schwinger functional used as input for the IM is that of an EFT combined with a renormalization scheme (based on dimensional regularization with minimal subtraction) worked out in ref.~\cite{ham00}.
\item The inclusion of pairing correlations in the framework of the 2PPI EA is worked out in ref.~\cite{fur07}. The latter shows notably how to recover results of the BCS theory with this PI technique.
\item Refs.~\cite{dru10,fur12,fur20} are reviews making a state of play of the methods relevant to turn the nuclear EDF approach into a DFT, like the 2PPI EA. In particular, these references discuss the techniques of quantization of gauge theories in the PI framework. Gauge symmetries and associated SSBs have proven to be powerful tools to describe open-shell nuclei, notably in the current EDF framework~\cite{ben03}. In technical terms, this requires the quantization of gauge theories. This can be achieved in the PI formalism by, e.g., introducing ghost fields via the Faddeev-Popov method~\cite{fad67}. Another possibility suggested in ref.~\cite{dru10,fur20} consists in exploiting the BRST symmetry~\cite{tyu75,bec76,iof76} to get rid of spurious dofs (and thus work with a unique gs)~\cite{bes90,bes16}. The latter direction might be less cumbersome than that of Faddeev-Popov, in particular in the case of non-Abelian gauge theories. However, both of these quantization techniques, and their combination with the PI techniques applied in the present comparative study, remain to be explored in the framework of nuclear physics.
\end{itemize}

\vspace{0.3cm}

Following the above IM procedure, the 2PPI EA of the studied $O(N)$ model with vanishing 1-point correlation function can be expressed as follows (see also appendix~\ref{sec:4PPIEAannIM} for a derivation of result~\eqref{eq:pure4PPIEAfinalexpressionmain} from which~\eqref{eq:pure2PPIEAfinalexpression} can be deduced by setting $M_{0,a}[\rho,\zeta;x]=0$ $\forall a,x$):
\begin{equation}
\begin{split}
\Gamma^{(\mathrm{2PPI})}[\rho] = & - \frac{\hbar}{2}\mathrm{STr}\left[\ln\big(\boldsymbol{G}[K_{0}]\big)\right] + \frac{\hbar}{2} \int_{x} K^{a}_{0}[\rho;x] \rho_{a}(x) \\
& + \hbar^{2} \left(\rule{0cm}{1.1cm}\right. \frac{1}{24} \hspace{0.08cm} \begin{gathered}
\begin{fmffile}{Diagrams/pure2PPIEA_Hartree}
\begin{fmfgraph}(30,20)
\fmfleft{i}
\fmfright{o}
\fmf{phantom,tension=10}{i,i1}
\fmf{phantom,tension=10}{o,o1}
\fmf{plain,left,tension=0.5,foreground=(1,,0,,0)}{i1,v1,i1}
\fmf{plain,right,tension=0.5,foreground=(1,,0,,0)}{o1,v2,o1}
\fmf{zigzag,foreground=(0,,0,,1)}{v1,v2}
\end{fmfgraph}
\end{fmffile}
\end{gathered}
+ \frac{1}{12}\begin{gathered}
\begin{fmffile}{Diagrams/pure2PPIEA_Fock}
\begin{fmfgraph}(15,15)
\fmfleft{i}
\fmfright{o}
\fmf{phantom,tension=11}{i,v1}
\fmf{phantom,tension=11}{v2,o}
\fmf{plain,left,tension=0.4,foreground=(1,,0,,0)}{v1,v2,v1}
\fmf{zigzag,foreground=(0,,0,,1)}{v1,v2}
\end{fmfgraph}
\end{fmffile}
\end{gathered} \left.\rule{0cm}{1.1cm}\right) \\
& - \hbar^{3} \left(\rule{0cm}{1.1cm}\right. \frac{1}{72} \hspace{0.38cm} \begin{gathered}
\begin{fmffile}{Diagrams/pure2PPIEA_Gamma3_Diag1}
\begin{fmfgraph}(12,12)
\fmfleft{i0,i1}
\fmfright{o0,o1}
\fmftop{v1,vUp,v2}
\fmfbottom{v3,vDown,v4}
\fmf{phantom,tension=20}{i0,v1}
\fmf{phantom,tension=20}{i1,v3}
\fmf{phantom,tension=20}{o0,v2}
\fmf{phantom,tension=20}{o1,v4}
\fmf{plain,left=0.4,tension=0.5,foreground=(1,,0,,0)}{v3,v1}
\fmf{phantom,left=0.1,tension=0.5}{v1,vUp}
\fmf{phantom,left=0.1,tension=0.5}{vUp,v2}
\fmf{plain,left=0.4,tension=0.0,foreground=(1,,0,,0)}{v1,v2}
\fmf{plain,left=0.4,tension=0.5,foreground=(1,,0,,0)}{v2,v4}
\fmf{phantom,left=0.1,tension=0.5}{v4,vDown}
\fmf{phantom,left=0.1,tension=0.5}{vDown,v3}
\fmf{plain,left=0.4,tension=0.0,foreground=(1,,0,,0)}{v4,v3}
\fmf{zigzag,tension=0.5,foreground=(0,,0,,1)}{v1,v4}
\fmf{zigzag,tension=0.5,foreground=(0,,0,,1)}{v2,v3}
\end{fmfgraph}
\end{fmffile}
\end{gathered} \hspace{0.28cm} + \frac{1}{144} \hspace{0.38cm} \begin{gathered}
\begin{fmffile}{Diagrams/pure2PPIEA_Gamma3_Diag2}
\begin{fmfgraph}(12,12)
\fmfleft{i0,i1}
\fmfright{o0,o1}
\fmftop{v1,vUp,v2}
\fmfbottom{v3,vDown,v4}
\fmf{phantom,tension=20}{i0,v1}
\fmf{phantom,tension=20}{i1,v3}
\fmf{phantom,tension=20}{o0,v2}
\fmf{phantom,tension=20}{o1,v4}
\fmf{plain,left=0.4,tension=0.5,foreground=(1,,0,,0)}{v3,v1}
\fmf{phantom,left=0.1,tension=0.5}{v1,vUp}
\fmf{phantom,left=0.1,tension=0.5}{vUp,v2}
\fmf{zigzag,left=0.4,tension=0.0,foreground=(0,,0,,1)}{v1,v2}
\fmf{plain,left=0.4,tension=0.5,foreground=(1,,0,,0)}{v2,v4}
\fmf{phantom,left=0.1,tension=0.5}{v4,vDown}
\fmf{phantom,left=0.1,tension=0.5}{vDown,v3}
\fmf{zigzag,left=0.4,tension=0.0,foreground=(0,,0,,1)}{v4,v3}
\fmf{plain,left=0.4,tension=0.5,foreground=(1,,0,,0)}{v1,v3}
\fmf{plain,right=0.4,tension=0.5,foreground=(1,,0,,0)}{v2,v4}
\end{fmfgraph}
\end{fmffile}
\end{gathered} \hspace{0.29cm} \left.\rule{0cm}{1.1cm}\right) \\
& + \hbar^{4} \left(\rule{0cm}{1.1cm}\right. \frac{1}{324} \hspace{0.2cm} \begin{gathered}
\begin{fmffile}{Diagrams/pure2PPIEA_Gamma4_Diag1}
\begin{fmfgraph}(16,16)
\fmfleft{i}
\fmfright{o}
\fmftop{vUpLeft,vUp,vUpRight}
\fmfbottom{vDownLeft,vDown,vDownRight}
\fmf{phantom,tension=1}{i,v1}
\fmf{phantom,tension=1}{v2,o}
\fmf{phantom,tension=14.0}{v3,vUpLeft}
\fmf{phantom,tension=2.0}{v3,vUpRight}
\fmf{phantom,tension=4.0}{v3,i}
\fmf{phantom,tension=2.0}{v4,vUpLeft}
\fmf{phantom,tension=14.0}{v4,vUpRight}
\fmf{phantom,tension=4.0}{v4,o}
\fmf{phantom,tension=14.0}{v5,vDownLeft}
\fmf{phantom,tension=2.0}{v5,vDownRight}
\fmf{phantom,tension=4.0}{v5,i}
\fmf{phantom,tension=2.0}{v6,vDownLeft}
\fmf{phantom,tension=14.0}{v6,vDownRight}
\fmf{phantom,tension=4.0}{v6,o}
\fmf{zigzag,tension=0,foreground=(0,,0,,1)}{v1,v2}
\fmf{zigzag,tension=0.6,foreground=(0,,0,,1)}{v3,v6}
\fmf{zigzag,tension=0.6,foreground=(0,,0,,1)}{v5,v4}
\fmf{plain,left=0.18,tension=0,foreground=(1,,0,,0)}{v1,v3}
\fmf{plain,left=0.42,tension=0,foreground=(1,,0,,0)}{v3,v4}
\fmf{plain,left=0.18,tension=0,foreground=(1,,0,,0)}{v4,v2}
\fmf{plain,left=0.18,tension=0,foreground=(1,,0,,0)}{v2,v6}
\fmf{plain,left=0.42,tension=0,foreground=(1,,0,,0)}{v6,v5}
\fmf{plain,left=0.18,tension=0,foreground=(1,,0,,0)}{v5,v1}
\end{fmfgraph}
\end{fmffile}
\end{gathered} \hspace{0.15cm} + \frac{1}{108} \hspace{0.5cm} \begin{gathered}
\begin{fmffile}{Diagrams/pure2PPIEA_Gamma4_Diag2}
\begin{fmfgraph}(12.5,12.5)
\fmfleft{i0,i1}
\fmfright{o0,o1}
\fmftop{v1,vUp,v2}
\fmfbottom{v3,vDown,v4}
\fmf{phantom,tension=20}{i0,v1}
\fmf{phantom,tension=20}{i1,v3}
\fmf{phantom,tension=20}{o0,v2}
\fmf{phantom,tension=20}{o1,v4}
\fmf{phantom,tension=0.005}{v5,v6}
\fmf{zigzag,left=0.4,tension=0,foreground=(0,,0,,1)}{v3,v1}
\fmf{phantom,left=0.1,tension=0}{v1,vUp}
\fmf{phantom,left=0.1,tension=0}{vUp,v2}
\fmf{plain,left=0.25,tension=0,foreground=(1,,0,,0)}{v1,v2}
\fmf{zigzag,left=0.4,tension=0,foreground=(0,,0,,1)}{v2,v4}
\fmf{phantom,left=0.1,tension=0}{v4,vDown}
\fmf{phantom,left=0.1,tension=0}{vDown,v3}
\fmf{plain,right=0.25,tension=0,foreground=(1,,0,,0)}{v3,v4}
\fmf{plain,left=0.2,tension=0.01,foreground=(1,,0,,0)}{v1,v5}
\fmf{plain,left=0.2,tension=0.01,foreground=(1,,0,,0)}{v5,v3}
\fmf{plain,right=0.2,tension=0.01,foreground=(1,,0,,0)}{v2,v6}
\fmf{plain,right=0.2,tension=0.01,foreground=(1,,0,,0)}{v6,v4}
\fmf{zigzag,tension=0,foreground=(0,,0,,1)}{v5,v6}
\end{fmfgraph}
\end{fmffile}
\end{gathered} \hspace{0.5cm} + \frac{1}{324} \hspace{0.4cm} \begin{gathered}
\begin{fmffile}{Diagrams/pure2PPIEA_Gamma4_Diag3}
\begin{fmfgraph}(12.5,12.5)
\fmfleft{i0,i1}
\fmfright{o0,o1}
\fmftop{v1,vUp,v2}
\fmfbottom{v3,vDown,v4}
\fmf{phantom,tension=20}{i0,v1}
\fmf{phantom,tension=20}{i1,v3}
\fmf{phantom,tension=20}{o0,v2}
\fmf{phantom,tension=20}{o1,v4}
\fmf{phantom,tension=0.005}{v5,v6}
\fmf{plain,left=0.4,tension=0,foreground=(1,,0,,0)}{v3,v1}
\fmf{phantom,left=0.1,tension=0}{v1,vUp}
\fmf{phantom,left=0.1,tension=0}{vUp,v2}
\fmf{zigzag,left=0.25,tension=0,foreground=(0,,0,,1)}{v1,v2}
\fmf{plain,left=0.4,tension=0,foreground=(1,,0,,0)}{v2,v4}
\fmf{phantom,left=0.1,tension=0}{v4,vDown}
\fmf{phantom,left=0.1,tension=0}{vDown,v3}
\fmf{zigzag,right=0.25,tension=0,foreground=(0,,0,,1)}{v3,v4}
\fmf{plain,left=0.2,tension=0.01,foreground=(1,,0,,0)}{v1,v5}
\fmf{plain,left=0.2,tension=0.01,foreground=(1,,0,,0)}{v5,v3}
\fmf{plain,right=0.2,tension=0.01,foreground=(1,,0,,0)}{v2,v6}
\fmf{plain,right=0.2,tension=0.01,foreground=(1,,0,,0)}{v6,v4}
\fmf{zigzag,tension=0,foreground=(0,,0,,1)}{v5,v6}
\end{fmfgraph}
\end{fmffile}
\end{gathered} \hspace{0.35cm} + \frac{1}{216} \hspace{-0.35cm} \begin{gathered}
\begin{fmffile}{Diagrams/pure2PPIEA_Gamma4_Diag4}
\begin{fmfgraph}(30,15)
\fmfleft{i0,i,i1}
\fmfright{o0,o,o1}
\fmftop{v1b,vUp,v2b}
\fmfbottom{v3b,vDown,v4b}
\fmf{phantom,tension=20}{i0,v1b}
\fmf{phantom,tension=20}{i1,v3b}
\fmf{phantom,tension=20}{o0,v2b}
\fmf{phantom,tension=20}{o1,v4b}
\fmf{phantom,tension=0.511}{i,v7}
\fmf{phantom,tension=0.11}{o,v7}
\fmf{phantom,tension=0.1}{v1,v1b}
\fmf{phantom,tension=0.1}{v2,v2b}
\fmf{phantom,tension=0.1}{v3,v3b}
\fmf{phantom,tension=0.1}{v4,v4b}
\fmf{phantom,tension=0.005}{v5,v6}
\fmf{phantom,left=0.1,tension=0.1}{v1,vUp}
\fmf{phantom,left=0.1,tension=0.1}{vUp,v2}
\fmf{zigzag,left=0.15,tension=0,foreground=(0,,0,,1)}{v1,v2}
\fmf{plain,left=0.4,tension=0,foreground=(1,,0,,0)}{v2,v4}
\fmf{plain,right=0.4,tension=0,foreground=(1,,0,,0)}{v2,v4}
\fmf{phantom,left=0.1,tension=0.1}{v4,vDown}
\fmf{phantom,left=0.1,tension=0.1}{vDown,v3}
\fmf{zigzag,left=0.15,tension=0,foreground=(0,,0,,1)}{v4,v3}
\fmf{plain,left=0.2,tension=0.01,foreground=(1,,0,,0)}{v1,v5}
\fmf{plain,left=0.2,tension=0.01,foreground=(1,,0,,0)}{v5,v3}
\fmf{plain,right=0.2,tension=0,foreground=(1,,0,,0)}{v1,v7}
\fmf{plain,right=0.2,tension=0,foreground=(1,,0,,0)}{v7,v3}
\fmf{phantom,right=0.2,tension=0.01}{v2,v6}
\fmf{phantom,right=0.2,tension=0.01}{v6,v4}
\fmf{zigzag,tension=0,foreground=(0,,0,,1)}{v5,v7}
\end{fmfgraph}
\end{fmffile}
\end{gathered} \\
& \hspace{1.1cm} + \frac{1}{1296} \hspace{-0.32cm} \begin{gathered}
\begin{fmffile}{Diagrams/pure2PPIEA_Gamma4_Diag5}
\begin{fmfgraph}(30,14)
\fmfleft{i0,i,i1}
\fmfright{o0,o,o1}
\fmftop{v1b,vUp,v2b}
\fmfbottom{v3b,vDown,v4b}
\fmf{phantom,tension=5}{vUp,v5}
\fmf{phantom,tension=1}{v1b,v5}
\fmf{phantom,tension=5}{vUp,v6}
\fmf{phantom,tension=1}{v2b,v6}
\fmf{phantom,tension=20}{i0,v1b}
\fmf{phantom,tension=20}{i1,v3b}
\fmf{phantom,tension=20}{o0,v2b}
\fmf{phantom,tension=20}{o1,v4b}
\fmf{phantom,tension=0.1}{v1,v1b}
\fmf{phantom,tension=0.1}{v2,v2b}
\fmf{phantom,tension=0.1}{v3,v3b}
\fmf{phantom,tension=0.1}{v4,v4b}
\fmf{phantom,tension=0.005}{v5,v6}
\fmf{phantom,left=0.1,tension=0.1}{v1,vUp}
\fmf{phantom,left=0.1,tension=0.1}{vUp,v2}
\fmf{plain,left=0.4,tension=0.005,foreground=(1,,0,,0)}{v2,v4}
\fmf{plain,right=0.4,tension=0.005,foreground=(1,,0,,0)}{v2,v4}
\fmf{plain,left=0.4,tension=0.005,foreground=(1,,0,,0)}{v1,v3}
\fmf{plain,right=0.4,tension=0.005,foreground=(1,,0,,0)}{v1,v3}
\fmf{phantom,left=0.1,tension=0.1}{v4,vDown}
\fmf{phantom,left=0.1,tension=0.1}{vDown,v3}
\fmf{zigzag,left=0.05,tension=0,foreground=(0,,0,,1)}{v1,v5}
\fmf{plain,left,tension=0,foreground=(1,,0,,0)}{v5,v6}
\fmf{plain,right,tension=0,foreground=(1,,0,,0)}{v5,v6}
\fmf{zigzag,left=0.05,tension=0,foreground=(0,,0,,1)}{v6,v2}
\fmf{zigzag,left=0.15,tension=0,foreground=(0,,0,,1)}{v4,v3}
\end{fmfgraph}
\end{fmffile}
\end{gathered} \hspace{-0.35cm} \left.\rule{0cm}{1.1cm}\right) \\
& + \mathcal{O}\big(\hbar^{5}\big)\;,
\end{split}
\label{eq:pure2PPIEAfinalexpression}
\end{equation}
where the Feynman rules are given by~\eqref{eq:2PPIEAfeynRuleGK00DON} and~\eqref{eq:2PPIEAfeynRuleZigzagVertex0DON}. The (0+0)-D version of~\eqref{eq:pure2PPIEAfinalexpression} (which can be directly deduced from a result of section~\ref{sec:4PPIEA} as well, i.e. by setting $M_{0}(\rho,\zeta)=0$ into~\eqref{eq:pure4PPIEAfinalexpression0DON}) satisfies:
\begin{equation}
\begin{split}
\Gamma^{(\mathrm{2PPI})}(\rho) = & \ \hbar\left[-\frac{N}{2}\ln(2\pi \rho(K_{0})) + \frac{N}{2}\left(m^{2} \rho(K_{0}) - 1\right)\right] + \hbar^{2} \left[\frac{1}{24}\lambda \left(N^{2}+2N\right) \left(\rho(K_{0})\right)^{2}\right] \\
& -\hbar^{3}\left[\frac{N^{2}+2N}{144}\lambda^{2} \left(\rho(K_{0})\right)^{4} \right] + \hbar^{4} \left[\frac{N^{3}+10N^{2}+16N}{1296}\lambda^{3} \left(\rho(K_{0})\right)^{6} \right] + \mathcal{O}\big(\hbar^{5}\big)\;,
\end{split}
\label{eq:pure2PPIEAfinalexpression0DON}
\end{equation}
with $\rho_{a}$ being also a scalar in color space, i.e. $\rho_{a} = \rho(K_{0}) = \left(m^{2}-K_{0}(\rho)\right)^{-1}$ $\forall a$. By comparing~\eqref{eq:2PIEAzerovevfinalexpression} with~\eqref{eq:pure2PPIEAfinalexpression}, we can see that the 2PI and 2PPI EAs with vanishing 1-point correlation function are expressed, up to order $\mathcal{O}\big(\hbar^{4}\big)$, in terms of the same diagrams which contribute to both EAs with the same numerical factors. The difference between these two EAs is hidden in the propagator which is dressed by the bilocal functional $\boldsymbol{K}_{0,ab}(x,y)$ for the 2PI EA on the one hand and by the local functional $K_{0,a}(x)$ for the 2PPI EA on the other hand. At order $\mathcal{O}\big(\hbar^{5}\big)$, the diagrams of the 2PPI and 2PI EAs no longer coincide due to the appearance of diagrams such as~\eqref{eq:2PPIEAGamma5diagram0DON} in the expression of the 2PPI EA. Nevertheless, in (0+0)-D, the expressions of the 2PI and 2PPI EAs coincide in the absence of SSB, i.e. results~\eqref{eq:2PIEAzerovevfinalexpression0DON} and~\eqref{eq:pure2PPIEAfinalexpression0DON} are equivalent, since the equality $\boldsymbol{G}_{a b}(x,y) = G \ \delta_{a b} = \rho(K_{0}) \ \delta_{a b}$ is valid in this case\footnote{The gap equation of the 2PPI EA expressed by~\eqref{eq:pure2PPIEAfinalexpression0DON} can therefore be directly deduced in (0+0)-D from~\eqref{eq:2PIEAzerovevgapequation0DON} after substituting $G$ by $\rho(K_{0})$.}. However, one should bear in mind that this equivalence \textit{a priori} does not hold (even in (0+0)-D) if the $O(N)$ symmetry is (spontaneously) broken down since there is no reason to enforce the relation $\boldsymbol{G}_{a b} = G \ \delta_{a b}$ for the 2PI EA in this situation.

\subsection{\label{sec:4PPIEA}4PPI effective action}

A first application of the 4PPI EA was carried out by Okopi\'{n}ska~\cite{oko97}. The 4PPI EA is deduced from the 4PI one by assuming that all sources are local. Hence, the 4PPI and 4PI EA formalisms are very similar and we can find most of the relevant information for the presentation of the 4PPI EA formalism in the literature related to the 4PI EA~\cite{vas74bis,car04,car13,car13bis}. As opposed to all EA approaches based on a $\hbar$-expansion that we have investigated in the rest of section~\ref{sec:EA}, no diagrammatic rule has been worked out so far for the 4PPI EA in order to determine general properties of the underpinning diagrams (as opposed to~\eqref{eq:2PPIEAdiagrammaticrule0DON} for the 2PPI EA with its 1VI diagrams notably). We will not develop such a diagrammatic rule below but construct instead the diagrammatic expression of the 4PPI EA via the IM for the $O(N)$ model under consideration (see appendix~\ref{sec:4PPIEAannIM}). Moreover, neither the 4PI nor the 4PPI EA has ever been exploited in the framework of an $O(N)$-symmetric theory (with $N>1$) to our knowledge. Since the 4PI and 4PPI EAs of our $O(N)$ model coincide in (0+0)-D and in the absence of spontaneous breakdown of the $O(N)$ symmetry\footnote{The 4PI and 4PPI EAs of the studied $O(N)$ model coincide without SSB and in (0+0)-D since the absence of SSB imposes that all non-local sources underlying the 4PI EA reduce to scalars in color space. Since spacetime indices vanish in (0+0)-D, this implies that all these sources become local, in which case the definitions of the 4PI and 4PPI EAs become indeed equivalent. For further details, one can compare the definition of the 4PI EA given in ref.~\cite{car04} with that of the 4PPI EA given by~\eqref{eq:pure4PPIEAdefinition0DON} to~\eqref{eq:SJKfiniteD4PPIEA}.}, the present study is the first to discuss the treatment of these EAs for such a model. More specifically, we now investigate the 4PPI EA with vanishing 1-point correlation function in the original representation, still in arbitrary dimensions as a first step. This functional is defined by the following Legendre transform:
\begin{equation}
\begin{split}
\scalebox{0.965}{${\displaystyle\Gamma^{(\mathrm{4PPI})}[\rho,\zeta] \equiv }$} & \scalebox{0.965}{${\displaystyle- W[K,M] + \int_{x} K^{a}[\rho,\zeta;x] \frac{\delta W[K,M]}{\delta K^{a}(x)} + \int_{x} M^{a}[\rho,\zeta;x] \frac{\delta W[K,M]}{\delta M^{a}(x)} }$} \\
\scalebox{0.965}{${\displaystyle = }$} & \scalebox{0.965}{${\displaystyle - W[K,M] + \frac{\hbar}{2} \int_{x} K^{a}[\rho,\zeta;x] \rho_{a}(x) + \frac{\hbar^{2}}{8} \int_{x} M^{a}[\rho,\zeta;x] \rho^{2}_{a}(x) + \frac{\hbar^{3}}{24} \int_{x} M^{a}[\rho,\zeta;x] \zeta_{a}(x) \;, }$}
\end{split}
\label{eq:pure4PPIEAdefinition0DON}
\end{equation}
with
\begin{equation}
\rho_{a}(x) = \frac{2}{\hbar} \frac{\delta W[K,M]}{\delta K^{a}(x)} \;,
\label{eq:pure4PPIEAdefinitionbis20DON}
\end{equation}
\begin{equation}
\zeta_{a}(x) = \frac{24}{\hbar^{3}} \frac{\delta W[K,M]}{\delta M^{a}(x)} - \frac{3}{\hbar} \rho^{2}_{a}(x) \;,
\label{eq:pure4PPIEAdefinitionbis30DON}
\end{equation}
and the Schwinger functional $W[K,M]$ satisfying:
\begin{equation}
Z[K,M] = e^{\frac{1}{\hbar}W[K,M]} = \int \mathcal{D}\vec{\widetilde{\varphi}} \ e^{-\frac{1}{\hbar}S_{KM}\big[\vec{\widetilde{\varphi}}\big]} \;,
\label{eq:ZJKfiniteD4PPIEA}
\end{equation}
\begin{equation}
S_{KM}\Big[\vec{\widetilde{\varphi}}\Big] \equiv S\Big[\vec{\widetilde{\varphi}}\Big]-\frac{1}{2}\int_{x} K^{a}(x) \widetilde{\varphi}^{2}_{a}(x)-\frac{1}{4!}\int_{x} M^{a}(x) \widetilde{\varphi}^{4}_{a}(x) \;.
\label{eq:SJKfiniteD4PPIEA}
\end{equation}
The IM is then implemented from the following power series:
\begin{subequations}
\begin{empheq}[left=\empheqlbrace]{align}
& \hspace{0.1cm} \Gamma^{(\mathrm{4PPI})}[\rho,\zeta;\hbar]=\sum_{n=0}^{\infty} \Gamma^{(\mathrm{4PPI})}_{n}[\rho,\zeta]\hbar^{n}\;. \label{eq:pure4PPIEAGammaExpansion0DONmain}\\
\nonumber \\
& \hspace{0.1cm} W[K,M;\hbar]=\sum_{n=0}^{\infty} W_{n}[K,M]\hbar^{n}\;. \label{eq:pure4PPIEAWExpansion0DONmain} \\
\nonumber \\
& \hspace{0.1cm} K[\rho,\zeta;\hbar]=\sum_{n=0}^{\infty} K_{n}[\rho,\zeta]\hbar^{n}\;. \label{eq:pure4PPIEAKExpansion0DONmain}\\
\nonumber \\
& \hspace{0.1cm} M[\rho,\zeta;\hbar]=\sum_{n=0}^{\infty} M_{n}[\rho,\zeta]\hbar^{n}\;. \label{eq:pure4PPIEAMExpansion0DONmain}\\
\nonumber \\
& \hspace{0.1cm} \rho = \sum_{n=0}^{\infty} \rho_{n}[K,M]\hbar^{n}\;. \label{eq:pure4PPIEArhoExpansion0DONmain} \\
\nonumber \\
& \hspace{0.1cm} \zeta = \sum_{n=0}^{\infty} \zeta_{n}[K,M]\hbar^{n}\;. \label{eq:pure4PPIEAzetaExpansion0DONmain}
\end{empheq}
\end{subequations}
From the IM, we show that the 4PPI EA~\eqref{eq:pure4PPIEAdefinition0DON} reads (see appendix~\ref{sec:4PPIEAannIM}):
\begin{equation}
\begin{split}
\Gamma^{(\mathrm{4PPI})}[\rho,\zeta] = & - \frac{\hbar}{2}\mathrm{STr}\left[\ln\big(\boldsymbol{G}[K_{0}]\big)\right] + \frac{\hbar}{2} \int_{x} K^{a}_{0}[\rho,\zeta;x] \rho_{a}(x) \\
& + \hbar^{2} \left(\rule{0cm}{1.1cm}\right. \frac{1}{24} \hspace{0.08cm} \begin{gathered}
\begin{fmffile}{Diagrams/pure4PPIEA_Hartree}
\begin{fmfgraph}(30,20)
\fmfleft{i}
\fmfright{o}
\fmf{phantom,tension=10}{i,i1}
\fmf{phantom,tension=10}{o,o1}
\fmf{plain,left,tension=0.5,foreground=(1,,0,,0)}{i1,v1,i1}
\fmf{plain,right,tension=0.5,foreground=(1,,0,,0)}{o1,v2,o1}
\fmf{zigzag,foreground=(1,,0,,0)}{v1,v2}
\end{fmfgraph}
\end{fmffile}
\end{gathered}
+ \frac{1}{12}\begin{gathered}
\begin{fmffile}{Diagrams/pure4PPIEA_Fock}
\begin{fmfgraph}(15,15)
\fmfleft{i}
\fmfright{o}
\fmf{phantom,tension=11}{i,v1}
\fmf{phantom,tension=11}{v2,o}
\fmf{plain,left,tension=0.4,foreground=(1,,0,,0)}{v1,v2,v1}
\fmf{zigzag,foreground=(1,,0,,0)}{v1,v2}
\end{fmfgraph}
\end{fmffile}
\end{gathered} + \frac{1}{8} \int_{x} M^{a}_{0}[\rho,\zeta;x] \rho_{a}^{2}(x) \left.\rule{0cm}{1.1cm}\right) \\
& + \hbar^{3} \left(\rule{0cm}{1.1cm}\right. -\frac{1}{72} \hspace{0.38cm} \begin{gathered}
\begin{fmffile}{Diagrams/pure4PPIEA_Gamma3_Diag4}
\begin{fmfgraph}(12,12)
\fmfleft{i0,i1}
\fmfright{o0,o1}
\fmftop{v1,vUp,v2}
\fmfbottom{v3,vDown,v4}
\fmf{phantom,tension=20}{i0,v1}
\fmf{phantom,tension=20}{i1,v3}
\fmf{phantom,tension=20}{o0,v2}
\fmf{phantom,tension=20}{o1,v4}
\fmf{plain,left=0.4,tension=0.5,foreground=(1,,0,,0)}{v3,v1}
\fmf{phantom,left=0.1,tension=0.5}{v1,vUp}
\fmf{phantom,left=0.1,tension=0.5}{vUp,v2}
\fmf{plain,left=0.4,tension=0.0,foreground=(1,,0,,0)}{v1,v2}
\fmf{plain,left=0.4,tension=0.5,foreground=(1,,0,,0)}{v2,v4}
\fmf{phantom,left=0.1,tension=0.5}{v4,vDown}
\fmf{phantom,left=0.1,tension=0.5}{vDown,v3}
\fmf{plain,left=0.4,tension=0.0,foreground=(1,,0,,0)}{v4,v3}
\fmf{zigzag,tension=0.5,foreground=(1,,0,,0)}{v1,v4}
\fmf{zigzag,tension=0.5,foreground=(1,,0,,0)}{v2,v3}
\end{fmfgraph}
\end{fmffile}
\end{gathered} \hspace{0.28cm} - \frac{1}{144} \hspace{0.38cm} \begin{gathered}
\begin{fmffile}{Diagrams/pure4PPIEA_Gamma3_Diag5}
\begin{fmfgraph}(12,12)
\fmfleft{i0,i1}
\fmfright{o0,o1}
\fmftop{v1,vUp,v2}
\fmfbottom{v3,vDown,v4}
\fmf{phantom,tension=20}{i0,v1}
\fmf{phantom,tension=20}{i1,v3}
\fmf{phantom,tension=20}{o0,v2}
\fmf{phantom,tension=20}{o1,v4}
\fmf{plain,left=0.4,tension=0.5,foreground=(1,,0,,0)}{v3,v1}
\fmf{phantom,left=0.1,tension=0.5}{v1,vUp}
\fmf{phantom,left=0.1,tension=0.5}{vUp,v2}
\fmf{zigzag,left=0.4,tension=0.0,foreground=(1,,0,,0)}{v1,v2}
\fmf{plain,left=0.4,tension=0.5,foreground=(1,,0,,0)}{v2,v4}
\fmf{phantom,left=0.1,tension=0.5}{v4,vDown}
\fmf{phantom,left=0.1,tension=0.5}{vDown,v3}
\fmf{zigzag,left=0.4,tension=0.0,foreground=(1,,0,,0)}{v4,v3}
\fmf{plain,left=0.4,tension=0.5,foreground=(1,,0,,0)}{v1,v3}
\fmf{plain,right=0.4,tension=0.5,foreground=(1,,0,,0)}{v2,v4}
\end{fmfgraph}
\end{fmffile}
\end{gathered} \hspace{0.29cm} + \frac{1}{24} \int_{x} M^{a}_{0}[\rho,\zeta;x] \zeta_{a}(x) \left.\rule{0cm}{1.1cm}\right) \\
& + \hbar^{4} \left(\rule{0cm}{1.1cm}\right. \frac{1}{324} \hspace{0.2cm} \begin{gathered}
\begin{fmffile}{Diagrams/pure4PPIEA_Gamma4_Diag1}
\begin{fmfgraph}(16,16)
\fmfleft{i}
\fmfright{o}
\fmftop{vUpLeft,vUp,vUpRight}
\fmfbottom{vDownLeft,vDown,vDownRight}
\fmf{phantom,tension=1}{i,v1}
\fmf{phantom,tension=1}{v2,o}
\fmf{phantom,tension=14.0}{v3,vUpLeft}
\fmf{phantom,tension=2.0}{v3,vUpRight}
\fmf{phantom,tension=4.0}{v3,i}
\fmf{phantom,tension=2.0}{v4,vUpLeft}
\fmf{phantom,tension=14.0}{v4,vUpRight}
\fmf{phantom,tension=4.0}{v4,o}
\fmf{phantom,tension=14.0}{v5,vDownLeft}
\fmf{phantom,tension=2.0}{v5,vDownRight}
\fmf{phantom,tension=4.0}{v5,i}
\fmf{phantom,tension=2.0}{v6,vDownLeft}
\fmf{phantom,tension=14.0}{v6,vDownRight}
\fmf{phantom,tension=4.0}{v6,o}
\fmf{zigzag,tension=0,foreground=(1,,0,,0)}{v1,v2}
\fmf{zigzag,tension=0.6,foreground=(1,,0,,0)}{v3,v6}
\fmf{zigzag,tension=0.6,foreground=(1,,0,,0)}{v5,v4}
\fmf{plain,left=0.18,tension=0,foreground=(1,,0,,0)}{v1,v3}
\fmf{plain,left=0.42,tension=0,foreground=(1,,0,,0)}{v3,v4}
\fmf{plain,left=0.18,tension=0,foreground=(1,,0,,0)}{v4,v2}
\fmf{plain,left=0.18,tension=0,foreground=(1,,0,,0)}{v2,v6}
\fmf{plain,left=0.42,tension=0,foreground=(1,,0,,0)}{v6,v5}
\fmf{plain,left=0.18,tension=0,foreground=(1,,0,,0)}{v5,v1}
\end{fmfgraph}
\end{fmffile}
\end{gathered} \hspace{0.15cm} + \frac{1}{108} \hspace{0.5cm} \begin{gathered}
\begin{fmffile}{Diagrams/pure4PPIEA_Gamma4_Diag2}
\begin{fmfgraph}(12.5,12.5)
\fmfleft{i0,i1}
\fmfright{o0,o1}
\fmftop{v1,vUp,v2}
\fmfbottom{v3,vDown,v4}
\fmf{phantom,tension=20}{i0,v1}
\fmf{phantom,tension=20}{i1,v3}
\fmf{phantom,tension=20}{o0,v2}
\fmf{phantom,tension=20}{o1,v4}
\fmf{phantom,tension=0.005}{v5,v6}
\fmf{zigzag,left=0.4,tension=0,foreground=(1,,0,,0)}{v3,v1}
\fmf{phantom,left=0.1,tension=0}{v1,vUp}
\fmf{phantom,left=0.1,tension=0}{vUp,v2}
\fmf{plain,left=0.25,tension=0,foreground=(1,,0,,0)}{v1,v2}
\fmf{zigzag,left=0.4,tension=0,foreground=(1,,0,,0)}{v2,v4}
\fmf{phantom,left=0.1,tension=0}{v4,vDown}
\fmf{phantom,left=0.1,tension=0}{vDown,v3}
\fmf{plain,right=0.25,tension=0,foreground=(1,,0,,0)}{v3,v4}
\fmf{plain,left=0.2,tension=0.01,foreground=(1,,0,,0)}{v1,v5}
\fmf{plain,left=0.2,tension=0.01,foreground=(1,,0,,0)}{v5,v3}
\fmf{plain,right=0.2,tension=0.01,foreground=(1,,0,,0)}{v2,v6}
\fmf{plain,right=0.2,tension=0.01,foreground=(1,,0,,0)}{v6,v4}
\fmf{zigzag,tension=0,foreground=(1,,0,,0)}{v5,v6}
\end{fmfgraph}
\end{fmffile}
\end{gathered} \hspace{0.5cm} + \frac{1}{324} \hspace{0.4cm} \begin{gathered}
\begin{fmffile}{Diagrams/pure4PPIEA_Gamma4_Diag3}
\begin{fmfgraph}(12.5,12.5)
\fmfleft{i0,i1}
\fmfright{o0,o1}
\fmftop{v1,vUp,v2}
\fmfbottom{v3,vDown,v4}
\fmf{phantom,tension=20}{i0,v1}
\fmf{phantom,tension=20}{i1,v3}
\fmf{phantom,tension=20}{o0,v2}
\fmf{phantom,tension=20}{o1,v4}
\fmf{phantom,tension=0.005}{v5,v6}
\fmf{plain,left=0.4,tension=0,foreground=(1,,0,,0)}{v3,v1}
\fmf{phantom,left=0.1,tension=0}{v1,vUp}
\fmf{phantom,left=0.1,tension=0}{vUp,v2}
\fmf{zigzag,left=0.25,tension=0,foreground=(1,,0,,0)}{v1,v2}
\fmf{plain,left=0.4,tension=0,foreground=(1,,0,,0)}{v2,v4}
\fmf{phantom,left=0.1,tension=0}{v4,vDown}
\fmf{phantom,left=0.1,tension=0}{vDown,v3}
\fmf{zigzag,right=0.25,tension=0,foreground=(1,,0,,0)}{v3,v4}
\fmf{plain,left=0.2,tension=0.01,foreground=(1,,0,,0)}{v1,v5}
\fmf{plain,left=0.2,tension=0.01,foreground=(1,,0,,0)}{v5,v3}
\fmf{plain,right=0.2,tension=0.01,foreground=(1,,0,,0)}{v2,v6}
\fmf{plain,right=0.2,tension=0.01,foreground=(1,,0,,0)}{v6,v4}
\fmf{zigzag,tension=0,foreground=(1,,0,,0)}{v5,v6}
\end{fmfgraph}
\end{fmffile}
\end{gathered} \hspace{0.35cm} + \frac{1}{216} \hspace{-0.35cm} \begin{gathered}
\begin{fmffile}{Diagrams/pure4PPIEA_Gamma4_Diag4}
\begin{fmfgraph}(30,15)
\fmfleft{i0,i,i1}
\fmfright{o0,o,o1}
\fmftop{v1b,vUp,v2b}
\fmfbottom{v3b,vDown,v4b}
\fmf{phantom,tension=20}{i0,v1b}
\fmf{phantom,tension=20}{i1,v3b}
\fmf{phantom,tension=20}{o0,v2b}
\fmf{phantom,tension=20}{o1,v4b}
\fmf{phantom,tension=0.511}{i,v7}
\fmf{phantom,tension=0.11}{o,v7}
\fmf{phantom,tension=0.1}{v1,v1b}
\fmf{phantom,tension=0.1}{v2,v2b}
\fmf{phantom,tension=0.1}{v3,v3b}
\fmf{phantom,tension=0.1}{v4,v4b}
\fmf{phantom,tension=0.005}{v5,v6}
\fmf{phantom,left=0.1,tension=0.1}{v1,vUp}
\fmf{phantom,left=0.1,tension=0.1}{vUp,v2}
\fmf{zigzag,left=0.15,tension=0,foreground=(1,,0,,0)}{v1,v2}
\fmf{plain,left=0.4,tension=0,foreground=(1,,0,,0)}{v2,v4}
\fmf{plain,right=0.4,tension=0,foreground=(1,,0,,0)}{v2,v4}
\fmf{phantom,left=0.1,tension=0.1}{v4,vDown}
\fmf{phantom,left=0.1,tension=0.1}{vDown,v3}
\fmf{zigzag,left=0.15,tension=0,foreground=(1,,0,,0)}{v4,v3}
\fmf{plain,left=0.2,tension=0.01,foreground=(1,,0,,0)}{v1,v5}
\fmf{plain,left=0.2,tension=0.01,foreground=(1,,0,,0)}{v5,v3}
\fmf{plain,right=0.2,tension=0,foreground=(1,,0,,0)}{v1,v7}
\fmf{plain,right=0.2,tension=0,foreground=(1,,0,,0)}{v7,v3}
\fmf{phantom,right=0.2,tension=0.01}{v2,v6}
\fmf{phantom,right=0.2,tension=0.01}{v6,v4}
\fmf{zigzag,tension=0,foreground=(1,,0,,0)}{v5,v7}
\end{fmfgraph}
\end{fmffile}
\end{gathered} \\
& \hspace{1.1cm} + \frac{1}{1296} \hspace{-0.32cm} \begin{gathered}
\begin{fmffile}{Diagrams/pure4PPIEA_Gamma4_Diag5}
\begin{fmfgraph}(30,14)
\fmfleft{i0,i,i1}
\fmfright{o0,o,o1}
\fmftop{v1b,vUp,v2b}
\fmfbottom{v3b,vDown,v4b}
\fmf{phantom,tension=5}{vUp,v5}
\fmf{phantom,tension=1}{v1b,v5}
\fmf{phantom,tension=5}{vUp,v6}
\fmf{phantom,tension=1}{v2b,v6}
\fmf{phantom,tension=20}{i0,v1b}
\fmf{phantom,tension=20}{i1,v3b}
\fmf{phantom,tension=20}{o0,v2b}
\fmf{phantom,tension=20}{o1,v4b}
\fmf{phantom,tension=0.1}{v1,v1b}
\fmf{phantom,tension=0.1}{v2,v2b}
\fmf{phantom,tension=0.1}{v3,v3b}
\fmf{phantom,tension=0.1}{v4,v4b}
\fmf{phantom,tension=0.005}{v5,v6}
\fmf{phantom,left=0.1,tension=0.1}{v1,vUp}
\fmf{phantom,left=0.1,tension=0.1}{vUp,v2}
\fmf{plain,left=0.4,tension=0.005,foreground=(1,,0,,0)}{v2,v4}
\fmf{plain,right=0.4,tension=0.005,foreground=(1,,0,,0)}{v2,v4}
\fmf{plain,left=0.4,tension=0.005,foreground=(1,,0,,0)}{v1,v3}
\fmf{plain,right=0.4,tension=0.005,foreground=(1,,0,,0)}{v1,v3}
\fmf{phantom,left=0.1,tension=0.1}{v4,vDown}
\fmf{phantom,left=0.1,tension=0.1}{vDown,v3}
\fmf{zigzag,left=0.05,tension=0,foreground=(1,,0,,0)}{v1,v5}
\fmf{plain,left,tension=0,foreground=(1,,0,,0)}{v5,v6}
\fmf{plain,right,tension=0,foreground=(1,,0,,0)}{v5,v6}
\fmf{zigzag,left=0.05,tension=0,foreground=(1,,0,,0)}{v6,v2}
\fmf{zigzag,left=0.15,tension=0,foreground=(1,,0,,0)}{v4,v3}
\end{fmfgraph}
\end{fmffile}
\end{gathered} \hspace{-0.35cm} \left.\rule{0cm}{1.1cm}\right) \\
& + \mathcal{O}\big(\hbar^{5}\big)\;,
\end{split}
\label{eq:pure4PPIEAfinalexpressionmain}
\end{equation}
and the corresponding Feynman rules are:
\begin{subequations}
\begin{align}
\begin{gathered}
\begin{fmffile}{Diagrams/2PPIEA_FeynRuleGSourceGK0}
\begin{fmfgraph*}(20,20)
\fmfleft{i0,i1,i2,i3}
\fmfright{o0,o1,o2,o3}
\fmflabel{$x, a$}{v1}
\fmflabel{$y, b$}{v2}
\fmf{phantom}{i1,v1}
\fmf{phantom}{i2,v1}
\fmf{plain,tension=0.6,foreground=(1,,0,,0)}{v1,v2}
\fmf{phantom}{v2,o1}
\fmf{phantom}{v2,o2}
\end{fmfgraph*}
\end{fmffile}
\end{gathered} \hspace{0.5cm} &\rightarrow \boldsymbol{G}_{a b}[K_{0};x,y] \;,
\label{eq:4PPIEAFeynRulesPropagatorK0M00DONmain} \\
\begin{gathered}
\begin{fmffile}{Diagrams/4PPIEA_FeynRuleV4}
\begin{fmfgraph*}(20,20)
\fmfleft{i0,i1,i2,i3}
\fmfright{o0,o1,o2,o3}
\fmf{phantom,tension=2.0}{i1,i1bis}
\fmf{plain,tension=2.0,foreground=(1,,0,,0)}{i1bis,v1}
\fmf{phantom,tension=2.0}{i2,i2bis}
\fmf{plain,tension=2.0,foreground=(1,,0,,0)}{i2bis,v1}
\fmf{zigzag,label=$x$,tension=0.6,foreground=(1,,0,,0)}{v1,v2}
\fmf{phantom,tension=2.0}{o1bis,o1}
\fmf{plain,tension=2.0,foreground=(1,,0,,0)}{v2,o1bis}
\fmf{phantom,tension=2.0}{o2bis,o2}
\fmf{plain,tension=2.0,foreground=(1,,0,,0)}{v2,o2bis}
\fmflabel{$a$}{i1bis}
\fmflabel{$b$}{i2bis}
\fmflabel{$c$}{o1bis}
\fmflabel{$d$}{o2bis}
\end{fmfgraph*}
\end{fmffile}
\end{gathered} \quad &\rightarrow \left(\lambda-M_{0,a}[\rho,\zeta;x]\delta_{a c}\right)\delta_{a b}\delta_{c d}\;.
\label{eq:4PPIEAFeynRules4legVertexK0M00DONmain}
\end{align}
\end{subequations}
where the propagator $\boldsymbol{G}[K_{0}] \equiv \boldsymbol{G}_{K}[K=K_{0}]$ is already defined by~\eqref{eq:2PPIEAdefinitionGKpropagator0DON} and remains a scalar in color space as for the 2PPI EA formalism discussed in section~\ref{sec:2PPIEA}. The coefficients $K_{0}$ and $M_{0}$, introduced via the power series~\eqref{eq:pure4PPIEAKExpansion0DONmain} and~\eqref{eq:pure4PPIEAMExpansion0DONmain}, also exhibit such a simple structure in color space and are the variational parameters of the present EA approach. Furthermore, we can see that~\eqref{eq:pure2PPIEAfinalexpression} and~\eqref{eq:pure4PPIEAfinalexpressionmain}, expressing respectively $\Gamma^{(\mathrm{2PPI})}[\rho]$ and $\Gamma^{(\mathrm{4PPI})}[\rho,\zeta]$, are very close: the diagrams have identical topologies and are weighted with the same factors up to order $\mathcal{O}\big(\hbar^{4}\big)$. However, the two expressions are of course not identical due to the presence of the $M_{0}$ coefficient in the case of the 4PPI EA (notably via the zigzag vertex~\eqref{eq:4PPIEAFeynRules4legVertexK0M00DONmain}).

\vspace{0.5cm}

We conclude this section by studying the zero-dimensional limit as usual. In order to evaluate the diagrams of~\eqref{eq:pure4PPIEAfinalexpressionmain} in (0+0)-D, we must have in mind that the zigzag vertex~\eqref{eq:4PPIEAFeynRules4legVertexK0M00DONmain} is dressed by $M_{0}[\rho,\zeta]$ if and only if the color indices at each four ends of this vertex are identical, as a result of the presence of $\delta_{ac}\delta_{ab}\delta_{cd}$ in~\eqref{eq:4PPIEAFeynRules4legVertexK0M00DONmain}. This yields:
\begin{equation}
\sum_{a=1}^{N} K_{0,a}(\rho,\zeta) \rho_{a} = \sum_{a=1}^{N} K_{0,a}(\rho,\zeta) \hspace{0.3cm} \begin{gathered}
\begin{fmffile}{Diagrams/pure4PPIEA_rho0D}
\begin{fmfgraph*}(12,12)
\fmfleft{i}
\fmfright{o}
\fmftop{vUp}
\fmfbottom{vDown}
\fmfv{decor.shape=circle,decor.filled=empty,decor.size=1.5thick,label.dist=0.15cm,label=$a$}{v1}
\fmf{phantom,tension=11}{i,v1}
\fmf{phantom,tension=11}{v2,o}
\fmf{plain,left,tension=0.4,foreground=(1,,0,,0)}{v1,v2,v1}
\fmf{phantom}{v1,v2}
\end{fmfgraph*}
\end{fmffile}
\end{gathered} = N K_{0}(\rho,\zeta) \rho(K_{0}) = N \left(m^{2} \rho(K_{0}) - 1\right) \;,
\label{eq:pure4PPIEAdiag1}
\end{equation}
\begin{equation}
\sum_{a=1}^{N} M_{0,a}(\rho,\zeta) \rho_{a}^{2} = \sum_{a=1}^{N} M_{0,a}(\rho,\zeta) \left(\rule{0cm}{0.6cm}\right. \hspace{0.3cm} \begin{gathered}
\begin{fmffile}{Diagrams/pure4PPIEA_rho0D}
\begin{fmfgraph*}(12,12)
\fmfleft{i}
\fmfright{o}
\fmftop{vUp}
\fmfbottom{vDown}
\fmfv{decor.shape=circle,decor.filled=empty,decor.size=1.5thick,label.dist=0.15cm,label=$a$}{v1}
\fmf{phantom,tension=11}{i,v1}
\fmf{phantom,tension=11}{v2,o}
\fmf{plain,left,tension=0.4,foreground=(1,,0,,0)}{v1,v2,v1}
\fmf{phantom}{v1,v2}
\end{fmfgraph*}
\end{fmffile}
\end{gathered} \left.\rule{0cm}{0.6cm}\right)^{2} = N M_{0}(\rho,\zeta) \left(\rho(K_{0})\right)^{2} \;,
\label{eq:pure4PPIEAdiag2}
\end{equation}
\begin{equation}
\sum_{a=1}^{N} M_{0,a}(\rho,\zeta) \zeta_{a} = - \sum_{a=1}^{N} M_{0,a}(\rho,\zeta) \hspace{-0.1cm} \begin{gathered}
\begin{fmffile}{Diagrams/pure4PPIEA_zeta0D}
\begin{fmfgraph*}(30,20)
\fmfleft{ibis,iUpbis}
\fmfright{obis,oUpbis}
\fmf{phantom,tension=3.0}{i,ibis}
\fmf{phantom,tension=3.0}{o,obis}
\fmf{phantom,tension=3.0}{iUp,iUpbis}
\fmf{phantom,tension=3.0}{oUp,oUpbis}
\fmfv{decor.shape=circle,decor.filled=empty,decor.size=1.5thick,label.angle=180,label.dist=0.1cm,label=$a$}{vIndex}
\fmf{plain,left,foreground=(1,,0,,0)}{i,vIndex,i}
\fmf{plain,left,foreground=(1,,0,,0)}{o,vIndex,o}
\fmf{zigzag,left=0.11,tension=1.5,foreground=(1,,0,,0)}{iUp,oUp}
\fmf{zigzag,left=1.0,tension=2.1,foreground=(1,,0,,0)}{i,iUp}
\fmf{zigzag,right=1.0,tension=2.1,foreground=(1,,0,,0)}{o,oUp}
\end{fmfgraph*}
\end{fmffile}
\end{gathered} \hspace{-0.15cm} = -N M_{0}(\rho,\zeta) \left(\lambda - M_{0}(\rho,\zeta)\right) \left(\rho(K_{0})\right)^{4} \;,
\label{eq:pure4PPIEAdiag3}
\end{equation}

\vspace{-0.5cm}

\begin{equation}
\begin{gathered}
\begin{fmffile}{Diagrams/pure4PPIEA_Hartree}
\begin{fmfgraph}(30,20)
\fmfleft{i}
\fmfright{o}
\fmf{phantom,tension=10}{i,i1}
\fmf{phantom,tension=10}{o,o1}
\fmf{plain,left,tension=0.5,foreground=(1,,0,,0)}{i1,v1,i1}
\fmf{plain,right,tension=0.5,foreground=(1,,0,,0)}{o1,v2,o1}
\fmf{zigzag,foreground=(1,,0,,0)}{v1,v2}
\end{fmfgraph}
\end{fmffile}
\end{gathered} = \left[ N \left(\lambda - M_{0}(\rho,\zeta)\right) + N \left(N-1\right) \lambda \right] \left(\rho(K_{0})\right)^{2} \;,
\label{eq:pure4PPIEAdiag4}
\end{equation}

\vspace{-0.5cm}

\begin{equation}
\begin{gathered}
\begin{fmffile}{Diagrams/pure4PPIEA_Fock}
\begin{fmfgraph}(15,15)
\fmfleft{i}
\fmfright{o}
\fmf{phantom,tension=11}{i,v1}
\fmf{phantom,tension=11}{v2,o}
\fmf{plain,left,tension=0.4,foreground=(1,,0,,0)}{v1,v2,v1}
\fmf{zigzag,foreground=(1,,0,,0)}{v1,v2}
\end{fmfgraph}
\end{fmffile}
\end{gathered} = N \left(\lambda - M_{0}(\rho,\zeta)\right) \left(\rho(K_{0})\right)^{2} \;,
\label{eq:pure4PPIEAdiag5}
\end{equation}
\begin{equation}
\begin{gathered}
\begin{fmffile}{Diagrams/pure4PPIEA_Gamma3_Diag4}
\begin{fmfgraph}(12,12)
\fmfleft{i0,i1}
\fmfright{o0,o1}
\fmftop{v1,vUp,v2}
\fmfbottom{v3,vDown,v4}
\fmf{phantom,tension=20}{i0,v1}
\fmf{phantom,tension=20}{i1,v3}
\fmf{phantom,tension=20}{o0,v2}
\fmf{phantom,tension=20}{o1,v4}
\fmf{plain,left=0.4,tension=0.5,foreground=(1,,0,,0)}{v3,v1}
\fmf{phantom,left=0.1,tension=0.5}{v1,vUp}
\fmf{phantom,left=0.1,tension=0.5}{vUp,v2}
\fmf{plain,left=0.4,tension=0.0,foreground=(1,,0,,0)}{v1,v2}
\fmf{plain,left=0.4,tension=0.5,foreground=(1,,0,,0)}{v2,v4}
\fmf{phantom,left=0.1,tension=0.5}{v4,vDown}
\fmf{phantom,left=0.1,tension=0.5}{vDown,v3}
\fmf{plain,left=0.4,tension=0.0,foreground=(1,,0,,0)}{v4,v3}
\fmf{zigzag,tension=0.5,foreground=(1,,0,,0)}{v1,v4}
\fmf{zigzag,tension=0.5,foreground=(1,,0,,0)}{v2,v3}
\end{fmfgraph}
\end{fmffile}
\end{gathered} \hspace{0.28cm} = N \left(\lambda - M_{0}(\rho,\zeta)\right)^{2} \left(\rho(K_{0})\right)^{4} \;,
\label{eq:pure4PPIEAdiag6}
\end{equation}

\vspace{0.2cm}

\begin{equation}
\begin{gathered}
\begin{fmffile}{Diagrams/pure4PPIEA_Gamma3_Diag5}
\begin{fmfgraph}(12,12)
\fmfleft{i0,i1}
\fmfright{o0,o1}
\fmftop{v1,vUp,v2}
\fmfbottom{v3,vDown,v4}
\fmf{phantom,tension=20}{i0,v1}
\fmf{phantom,tension=20}{i1,v3}
\fmf{phantom,tension=20}{o0,v2}
\fmf{phantom,tension=20}{o1,v4}
\fmf{plain,left=0.4,tension=0.5,foreground=(1,,0,,0)}{v3,v1}
\fmf{phantom,left=0.1,tension=0.5}{v1,vUp}
\fmf{phantom,left=0.1,tension=0.5}{vUp,v2}
\fmf{zigzag,left=0.4,tension=0.0,foreground=(1,,0,,0)}{v1,v2}
\fmf{plain,left=0.4,tension=0.5,foreground=(1,,0,,0)}{v2,v4}
\fmf{phantom,left=0.1,tension=0.5}{v4,vDown}
\fmf{phantom,left=0.1,tension=0.5}{vDown,v3}
\fmf{zigzag,left=0.4,tension=0.0,foreground=(1,,0,,0)}{v4,v3}
\fmf{plain,left=0.4,tension=0.5,foreground=(1,,0,,0)}{v1,v3}
\fmf{plain,right=0.4,tension=0.5,foreground=(1,,0,,0)}{v2,v4}
\end{fmfgraph}
\end{fmffile}
\end{gathered} \hspace{0.29cm} = \left[ N \left(\lambda - M_{0}(\rho,\zeta)\right)^{2} + N \left(N-1\right) \lambda^{2} \right] \left(\rho(K_{0})\right)^{4} \;,
\label{eq:pure4PPIEAdiag7}
\end{equation}

\vspace{0.2cm}

\begin{equation}
\begin{gathered}
\begin{fmffile}{Diagrams/pure4PPIEA_Gamma4_Diag1}
\begin{fmfgraph}(16,16)
\fmfleft{i}
\fmfright{o}
\fmftop{vUpLeft,vUp,vUpRight}
\fmfbottom{vDownLeft,vDown,vDownRight}
\fmf{phantom,tension=1}{i,v1}
\fmf{phantom,tension=1}{v2,o}
\fmf{phantom,tension=14.0}{v3,vUpLeft}
\fmf{phantom,tension=2.0}{v3,vUpRight}
\fmf{phantom,tension=4.0}{v3,i}
\fmf{phantom,tension=2.0}{v4,vUpLeft}
\fmf{phantom,tension=14.0}{v4,vUpRight}
\fmf{phantom,tension=4.0}{v4,o}
\fmf{phantom,tension=14.0}{v5,vDownLeft}
\fmf{phantom,tension=2.0}{v5,vDownRight}
\fmf{phantom,tension=4.0}{v5,i}
\fmf{phantom,tension=2.0}{v6,vDownLeft}
\fmf{phantom,tension=14.0}{v6,vDownRight}
\fmf{phantom,tension=4.0}{v6,o}
\fmf{zigzag,tension=0,foreground=(1,,0,,0)}{v1,v2}
\fmf{zigzag,tension=0.6,foreground=(1,,0,,0)}{v3,v6}
\fmf{zigzag,tension=0.6,foreground=(1,,0,,0)}{v5,v4}
\fmf{plain,left=0.18,tension=0,foreground=(1,,0,,0)}{v1,v3}
\fmf{plain,left=0.42,tension=0,foreground=(1,,0,,0)}{v3,v4}
\fmf{plain,left=0.18,tension=0,foreground=(1,,0,,0)}{v4,v2}
\fmf{plain,left=0.18,tension=0,foreground=(1,,0,,0)}{v2,v6}
\fmf{plain,left=0.42,tension=0,foreground=(1,,0,,0)}{v6,v5}
\fmf{plain,left=0.18,tension=0,foreground=(1,,0,,0)}{v5,v1}
\end{fmfgraph}
\end{fmffile}
\end{gathered} \hspace{0.15cm} = \hspace{0.5cm} \begin{gathered}
\begin{fmffile}{Diagrams/pure4PPIEA_Gamma4_Diag2}
\begin{fmfgraph}(12.5,12.5)
\fmfleft{i0,i1}
\fmfright{o0,o1}
\fmftop{v1,vUp,v2}
\fmfbottom{v3,vDown,v4}
\fmf{phantom,tension=20}{i0,v1}
\fmf{phantom,tension=20}{i1,v3}
\fmf{phantom,tension=20}{o0,v2}
\fmf{phantom,tension=20}{o1,v4}
\fmf{phantom,tension=0.005}{v5,v6}
\fmf{zigzag,left=0.4,tension=0,foreground=(1,,0,,0)}{v3,v1}
\fmf{phantom,left=0.1,tension=0}{v1,vUp}
\fmf{phantom,left=0.1,tension=0}{vUp,v2}
\fmf{plain,left=0.25,tension=0,foreground=(1,,0,,0)}{v1,v2}
\fmf{zigzag,left=0.4,tension=0,foreground=(1,,0,,0)}{v2,v4}
\fmf{phantom,left=0.1,tension=0}{v4,vDown}
\fmf{phantom,left=0.1,tension=0}{vDown,v3}
\fmf{plain,right=0.25,tension=0,foreground=(1,,0,,0)}{v3,v4}
\fmf{plain,left=0.2,tension=0.01,foreground=(1,,0,,0)}{v1,v5}
\fmf{plain,left=0.2,tension=0.01,foreground=(1,,0,,0)}{v5,v3}
\fmf{plain,right=0.2,tension=0.01,foreground=(1,,0,,0)}{v2,v6}
\fmf{plain,right=0.2,tension=0.01,foreground=(1,,0,,0)}{v6,v4}
\fmf{zigzag,tension=0,foreground=(1,,0,,0)}{v5,v6}
\end{fmfgraph}
\end{fmffile}
\end{gathered} \hspace{0.5cm} = N \left(\lambda - M_{0}(\rho,\zeta)\right)^{3} \left(\rho(K_{0})\right)^{6} \;,
\label{eq:pure4PPIEAdiag8}
\end{equation}

\vspace{0.2cm}

\begin{equation}
\begin{gathered}
\begin{fmffile}{Diagrams/pure4PPIEA_Gamma4_Diag3}
\begin{fmfgraph}(12.5,12.5)
\fmfleft{i0,i1}
\fmfright{o0,o1}
\fmftop{v1,vUp,v2}
\fmfbottom{v3,vDown,v4}
\fmf{phantom,tension=20}{i0,v1}
\fmf{phantom,tension=20}{i1,v3}
\fmf{phantom,tension=20}{o0,v2}
\fmf{phantom,tension=20}{o1,v4}
\fmf{phantom,tension=0.005}{v5,v6}
\fmf{plain,left=0.4,tension=0,foreground=(1,,0,,0)}{v3,v1}
\fmf{phantom,left=0.1,tension=0}{v1,vUp}
\fmf{phantom,left=0.1,tension=0}{vUp,v2}
\fmf{zigzag,left=0.25,tension=0,foreground=(1,,0,,0)}{v1,v2}
\fmf{plain,left=0.4,tension=0,foreground=(1,,0,,0)}{v2,v4}
\fmf{phantom,left=0.1,tension=0}{v4,vDown}
\fmf{phantom,left=0.1,tension=0}{vDown,v3}
\fmf{zigzag,right=0.25,tension=0,foreground=(1,,0,,0)}{v3,v4}
\fmf{plain,left=0.2,tension=0.01,foreground=(1,,0,,0)}{v1,v5}
\fmf{plain,left=0.2,tension=0.01,foreground=(1,,0,,0)}{v5,v3}
\fmf{plain,right=0.2,tension=0.01,foreground=(1,,0,,0)}{v2,v6}
\fmf{plain,right=0.2,tension=0.01,foreground=(1,,0,,0)}{v6,v4}
\fmf{zigzag,tension=0,foreground=(1,,0,,0)}{v5,v6}
\end{fmfgraph}
\end{fmffile}
\end{gathered} \hspace{0.35cm} = \left[ N \left(\lambda - M_{0}(\rho,\zeta)\right)^{3} + N \left(N-1\right) \lambda^{3} \right] \left(\rho(K_{0})\right)^{6} \;,
\label{eq:pure4PPIEAdiag9}
\end{equation}


\begin{equation}
\begin{gathered}
\begin{fmffile}{Diagrams/pure4PPIEA_Gamma4_Diag4}
\begin{fmfgraph}(30,15)
\fmfleft{i0,i,i1}
\fmfright{o0,o,o1}
\fmftop{v1b,vUp,v2b}
\fmfbottom{v3b,vDown,v4b}
\fmf{phantom,tension=20}{i0,v1b}
\fmf{phantom,tension=20}{i1,v3b}
\fmf{phantom,tension=20}{o0,v2b}
\fmf{phantom,tension=20}{o1,v4b}
\fmf{phantom,tension=0.511}{i,v7}
\fmf{phantom,tension=0.11}{o,v7}
\fmf{phantom,tension=0.1}{v1,v1b}
\fmf{phantom,tension=0.1}{v2,v2b}
\fmf{phantom,tension=0.1}{v3,v3b}
\fmf{phantom,tension=0.1}{v4,v4b}
\fmf{phantom,tension=0.005}{v5,v6}
\fmf{phantom,left=0.1,tension=0.1}{v1,vUp}
\fmf{phantom,left=0.1,tension=0.1}{vUp,v2}
\fmf{zigzag,left=0.15,tension=0,foreground=(1,,0,,0)}{v1,v2}
\fmf{plain,left=0.4,tension=0,foreground=(1,,0,,0)}{v2,v4}
\fmf{plain,right=0.4,tension=0,foreground=(1,,0,,0)}{v2,v4}
\fmf{phantom,left=0.1,tension=0.1}{v4,vDown}
\fmf{phantom,left=0.1,tension=0.1}{vDown,v3}
\fmf{zigzag,left=0.15,tension=0,foreground=(1,,0,,0)}{v4,v3}
\fmf{plain,left=0.2,tension=0.01,foreground=(1,,0,,0)}{v1,v5}
\fmf{plain,left=0.2,tension=0.01,foreground=(1,,0,,0)}{v5,v3}
\fmf{plain,right=0.2,tension=0,foreground=(1,,0,,0)}{v1,v7}
\fmf{plain,right=0.2,tension=0,foreground=(1,,0,,0)}{v7,v3}
\fmf{phantom,right=0.2,tension=0.01}{v2,v6}
\fmf{phantom,right=0.2,tension=0.01}{v6,v4}
\fmf{zigzag,tension=0,foreground=(1,,0,,0)}{v5,v7}
\end{fmfgraph}
\end{fmffile}
\end{gathered} \hspace{-0.32cm} = \left[ N \left(\lambda - M_{0}(\rho,\zeta)\right)^{2} + N \left(N-1\right) \lambda^{2} \right] \left(\lambda - M_{0}(\rho,\zeta)\right) \left(\rho(K_{0})\right)^{6} \;,
\label{eq:pure4PPIEAdiag10}
\end{equation}


\begin{equation}
\begin{split}
\begin{gathered}
\begin{fmffile}{Diagrams/pure4PPIEA_Gamma4_Diag5}
\begin{fmfgraph}(30,14)
\fmfleft{i0,i,i1}
\fmfright{o0,o,o1}
\fmftop{v1b,vUp,v2b}
\fmfbottom{v3b,vDown,v4b}
\fmf{phantom,tension=5}{vUp,v5}
\fmf{phantom,tension=1}{v1b,v5}
\fmf{phantom,tension=5}{vUp,v6}
\fmf{phantom,tension=1}{v2b,v6}
\fmf{phantom,tension=20}{i0,v1b}
\fmf{phantom,tension=20}{i1,v3b}
\fmf{phantom,tension=20}{o0,v2b}
\fmf{phantom,tension=20}{o1,v4b}
\fmf{phantom,tension=0.1}{v1,v1b}
\fmf{phantom,tension=0.1}{v2,v2b}
\fmf{phantom,tension=0.1}{v3,v3b}
\fmf{phantom,tension=0.1}{v4,v4b}
\fmf{phantom,tension=0.005}{v5,v6}
\fmf{phantom,left=0.1,tension=0.1}{v1,vUp}
\fmf{phantom,left=0.1,tension=0.1}{vUp,v2}
\fmf{plain,left=0.4,tension=0.005,foreground=(1,,0,,0)}{v2,v4}
\fmf{plain,right=0.4,tension=0.005,foreground=(1,,0,,0)}{v2,v4}
\fmf{plain,left=0.4,tension=0.005,foreground=(1,,0,,0)}{v1,v3}
\fmf{plain,right=0.4,tension=0.005,foreground=(1,,0,,0)}{v1,v3}
\fmf{phantom,left=0.1,tension=0.1}{v4,vDown}
\fmf{phantom,left=0.1,tension=0.1}{vDown,v3}
\fmf{zigzag,left=0.05,tension=0,foreground=(1,,0,,0)}{v1,v5}
\fmf{plain,left,tension=0,foreground=(1,,0,,0)}{v5,v6}
\fmf{plain,right,tension=0,foreground=(1,,0,,0)}{v5,v6}
\fmf{zigzag,left=0.05,tension=0,foreground=(1,,0,,0)}{v6,v2}
\fmf{zigzag,left=0.15,tension=0,foreground=(1,,0,,0)}{v4,v3}
\end{fmfgraph}
\end{fmffile}
\end{gathered} \hspace{-0.35cm} = & \ \Big[ N \left(\lambda - M_{0}(\rho,\zeta)\right)^{3} + 3 N \left(N-1\right) \lambda^{2} \left(\lambda - M_{0}(\rho,\zeta)\right) \\
& \hspace{0.1cm} + N \left(N-1\right) \left(N-2\right) \lambda^{3} \Big] \left(\rho(K_{0})\right)^{6} \;,
\end{split}
\label{eq:pure4PPIEAdiag11}
\end{equation}
where we have used relations derived in appendix~\ref{sec:4PPIEAannIM} (i.e.~\eqref{eq:pure4PPIIMrho0DON} and~\eqref{eq:pure4PPIIMzeta0DON}) in \eqref{eq:pure4PPIEAdiag1}, \eqref{eq:pure4PPIEAdiag2} and \eqref{eq:pure4PPIEAdiag3}. As mentioned earlier, the absence of spontaneous breakdown of the $O(N)$ symmetry in the present situation induces notably that $\boldsymbol{G}(K_{0})$, $K_{0}(\rho,\zeta)$ and $M_{0}(\rho,\zeta)$ all reduce to scalars in color space, which translates into the relations: $\boldsymbol{G}_{a b}(K_{0}) = \rho(K_{0}) \delta_{a b}$, $K_{0,a}(\rho,\zeta)=K_{0}(\rho,\zeta)$ $\forall a$ and $M_{0,a}(\rho,\zeta) = M_{0}(\rho,\zeta)$ $\forall a$. Moreover, by noticing that the contribution of each diagram must be of order $\mathcal{O}\big(N^{m}\big)$ when $M_{0}(\rho,\zeta)=0$, with $m$ being the number of independent propagator loops in the diagram under consideration, we can infer the following sum rules for the factors involved in the RHSs of~\eqref{eq:pure4PPIEAdiag4},~\eqref{eq:pure4PPIEAdiag7},~\eqref{eq:pure4PPIEAdiag9},~\eqref{eq:pure4PPIEAdiag10} and~\eqref{eq:pure4PPIEAdiag11}:
\begin{itemize}
\item For $m=2$ (i.e. for~\eqref{eq:pure4PPIEAdiag4},~\eqref{eq:pure4PPIEAdiag7},~\eqref{eq:pure4PPIEAdiag9} and~\eqref{eq:pure4PPIEAdiag10}):
\begin{equation}
N + N\left(N-1\right) = N^{2} \;.
\end{equation}
\item For $m=3$ (i.e. for~\eqref{eq:pure4PPIEAdiag11}):
\begin{equation}
N + 3 N \left(N-1\right) + N\left(N-1\right)\left(N-2\right) = N^{3} \;.
\end{equation}
\end{itemize}
According to~\eqref{eq:pure4PPIEAdiag1} to~\eqref{eq:pure4PPIEAdiag11},~\eqref{eq:pure4PPIEAfinalexpressionmain} reduces in (0+0)-D to:
\begin{equation}
\begin{split}
\Gamma^{(\mathrm{4PPI})}(\rho,\zeta) = & \ \hbar\bigg[-\frac{N}{2}\ln(2\pi \rho(K_{0})) + \frac{N}{2}\left(m^{2} \rho(K_{0}) - 1\right)\bigg] \\
& + \hbar^{2} \bigg[\frac{1}{24}\lambda \left(N^{2}+2N\right) \left(\rho(K_{0})\right)^{2}\bigg] \\
& - \hbar^{3}\bigg[ \frac{1}{144} N \left(-3 \left(M_{0}(\rho,\zeta)\right)^2 + \left(2 + N\right) \lambda^{2} \right) \left(\rho(K_{0})\right)^{4} \bigg] \\
& + \hbar^{4} \bigg[ \frac{1}{1296} N \Big( -27 \left(M_{0}(\rho,\zeta)\right)^{3} + 81 \left(M_{0}(\rho,\zeta)\right)^{2} \lambda - 9 M_{0}(\rho,\zeta) \left(8+N\right) \lambda^{2} \\
& \hspace{0.9cm} + \left(16 + 10 N + N^{2} \right) \lambda^{3} \big)\left(\rho(K_{0})\right)^{6} \bigg] \\
& + \mathcal{O}\big(\hbar^{5}\big)\;.
\end{split}
\label{eq:pure4PPIEAfinalexpression0DON}
\end{equation}
By exploiting the relations\footnote{The conservation of the $O(N)$ symmetry in the present framework also imposes that both arguments of the 4PPI EA, i.e. $\rho$ and $\zeta$, are scalars in color space.} $\rho_{a} = \rho(K_{0}) = \left(m^{2}-K_{0}(\rho,\zeta)\right)^{-1}$ $\forall a$ and $\zeta_{a} = \zeta(K_{0},M_{0}) = -\left(\lambda-M_{0}(\rho,\zeta)\right)\left(\rho(K_{0})\right)^{4}$ $\forall a$ (see appendix~\ref{sec:4PPIEAannIM} and more specifically~\eqref{eq:pure4PPIIMrho0DON} and~\eqref{eq:pure4PPIIMzeta0DON} for the derivation of these two relations in arbitrary dimensions), we then construct the following chain rules in order to derive the gap equations associated to $\Gamma^{(\mathrm{4PPI})}(\rho,\zeta)$:
\begin{equation}
\begin{split}
\frac{\partial}{\partial\rho} = & \ \frac{\partial K_{0}(\rho,\zeta)}{\partial \rho} \frac{\partial}{\partial K_{0}} + \underbrace{\frac{\partial M_{0}(\rho,\zeta)}{\partial \rho}}_{0} \frac{\partial}{\partial M_{0}} \\
= & \left(\frac{\partial \rho(K_{0})}{\partial K_{0}}\right)^{-1} \frac{\partial}{\partial K_{0}} \\
= & \left(\rho(K_{0})\right)^{-2} \frac{\partial}{\partial K_{0}} \;,
\end{split}
\label{eq:pure4PPIEAchainruleRho0DON}
\end{equation}
\begin{equation}
\begin{split}
\frac{\partial}{\partial\zeta} = & \ \frac{\partial K_{0}(\rho,\zeta)}{\partial \zeta} \frac{\partial}{\partial K_{0}} + \frac{\partial M_{0}(\rho,\zeta)}{\partial \zeta} \frac{\partial}{\partial M_{0}} \\
= & \left(\frac{\partial \zeta(K_{0},M_{0})}{\partial K_{0}}\right)^{-1} \frac{\partial}{\partial K_{0}} + \left(\frac{\partial \zeta(K_{0},M_{0})}{\partial M_{0}}\right)^{-1} \frac{\partial}{\partial M_{0}} \\
= & -\frac{1}{4}\left(\lambda-M_{0}(\rho,\zeta)\right)^{-1}\left(\rho(K_{0})\right)^{-5} \frac{\partial}{\partial K_{0}} + \left(\rho(K_{0})\right)^{-4} \frac{\partial}{\partial M_{0}}\;.
\end{split}
\label{eq:pure4PPIEAchainruleZeta0DON}
\end{equation}
The gap equations associated to $\Gamma^{(\mathrm{4PPI})}(\rho,\zeta)$ actually satisfy (see appendix~\ref{sec:4PPIEAannIM} and more specifically~\eqref{eq:pure4PPIEAdGammandRho0DON} and~\eqref{eq:pure4PPIEAdGammandZeta0DON} to justify the simplifications of the derivatives of $\Gamma^{(\mathrm{4PPI})}(\rho,\zeta)$ outlined in~\eqref{eq:pure4PPIEAgapequationRhostep10DON} and~\eqref{eq:pure4PPIEAgapequationZetastep10DON}):
\begin{equation}
\begin{split}
0 = \left.\frac{\partial\Gamma^{(\mathrm{4PPI})}(\rho,\zeta)}{\partial\rho}\right|_{\rho=\overline{\rho} \atop \zeta=\overline{\zeta}} = & \ \underbrace{\left.\frac{\partial\Gamma_{0}^{(\mathrm{4PPI})}(\rho,\zeta)}{\partial\rho}\right|_{\rho=\overline{\rho} \atop \zeta=\overline{\zeta}}}_{0} + \underbrace{\left.\frac{\partial\Gamma_{1}^{(\mathrm{4PPI})}(\rho,\zeta)}{\partial\rho}\right|_{\rho=\overline{\rho} \atop \zeta=\overline{\zeta}}}_{\frac{1}{2} \overline{K}_{0}}\hbar + \left.\frac{\partial\Gamma_{2}^{(\mathrm{4PPI})}(\rho,\zeta)}{\partial\rho}\right|_{\rho=\overline{\rho} \atop \zeta=\overline{\zeta}} \hbar^{2} \\
& + \left.\frac{\partial\Gamma_{3}^{(\mathrm{4PPI})}(\rho,\zeta)}{\partial\rho}\right|_{\rho=\overline{\rho} \atop \zeta=\overline{\zeta}} \hbar^{3} + \left.\frac{\partial\Gamma_{4}^{(\mathrm{4PPI})}(\rho,\zeta)}{\partial\rho}\right|_{\rho=\overline{\rho} \atop \zeta=\overline{\zeta}} \hbar^{4} + \mathcal{O}\big(\hbar^{5}\big)\;,
\end{split}
\label{eq:pure4PPIEAgapequationRhostep10DON}
\end{equation}
\begin{equation}
\begin{split}
0 = \left.\frac{\partial\Gamma^{(\mathrm{4PPI})}(\rho,\zeta)}{\partial\zeta}\right|_{\rho=\overline{\rho} \atop \zeta=\overline{\zeta}} = & \ \underbrace{\left.\frac{\partial\Gamma_{0}^{(\mathrm{4PPI})}(\rho,\zeta)}{\partial\zeta}\right|_{\rho=\overline{\rho} \atop \zeta=\overline{\zeta}}}_{0} + \underbrace{\left.\frac{\partial\Gamma_{1}^{(\mathrm{4PPI})}(\rho,\zeta)}{\partial\zeta}\right|_{\rho=\overline{\rho} \atop \zeta=\overline{\zeta}}}_{0}\hbar + \underbrace{\left.\frac{\partial\Gamma_{2}^{(\mathrm{4PPI})}(\rho,\zeta)}{\partial\zeta}\right|_{\rho=\overline{\rho} \atop \zeta=\overline{\zeta}}}_{0}\hbar^{2}\\
& + \underbrace{\left.\frac{\partial\Gamma_{3}^{(\mathrm{4PPI})}(\rho,\zeta)}{\partial\zeta}\right|_{\rho=\overline{\rho} \atop \zeta=\overline{\zeta}}}_{\frac{1}{24}\overline{M}_{0}}\hbar^{3} 
 + \left.\frac{\partial\Gamma_{4}^{(\mathrm{4PPI})}(\rho,\zeta)}{\partial\zeta}\right|_{\rho=\overline{\rho} \atop \zeta=\overline{\zeta}}\hbar^{4} + \mathcal{O}\big(\hbar^{5}\big)\;,
\end{split}
\label{eq:pure4PPIEAgapequationZetastep10DON}
\end{equation}
with $\overline{\rho}_{a} = \overline{\rho} = \left(m^{2}-\overline{K}_{0}\right)^{-1}$ $\forall a$ and $\overline{\zeta}_{a} = \overline{\zeta} = -\left(\lambda - \overline{M}_{0}\right)\overline{\rho}^{4}$ $\forall a$. Finally, after combining~\eqref{eq:pure4PPIEAfinalexpression0DON},~\eqref{eq:pure4PPIEAchainruleRho0DON} and~\eqref{eq:pure4PPIEAchainruleZeta0DON} with~\eqref{eq:pure4PPIEAgapequationRhostep10DON} and~\eqref{eq:pure4PPIEAgapequationZetastep10DON}, we obtain the two gap equations:
\begin{equation}
\begin{split}
0 = \left.\frac{\partial\Gamma^{(\mathrm{4PPI})}(\rho,\zeta)}{\partial\rho}\right|_{\rho=\overline{\rho} \atop \zeta=\overline{\zeta}} = & \ \frac{\hbar}{2} \overline{K}_{0} + \hbar^{2} \bigg[ \frac{1}{12} N \left(2 + N\right) \lambda \overline{\rho} \bigg] - \hbar^{3} \bigg[ \frac{1}{36} N \left(-3 \overline{M}_{0}^{2} + \left(2 + N\right) \lambda^2\right) \overline{\rho}^{3} \bigg] \\
& + \hbar^{4}\bigg[ \frac{1}{216} N \Big(-27 \overline{M}_{0}^{3} + 81 \overline{M}_{0}^{2} \lambda - 9 \overline{M}_{0} \left(8 + N\right) \lambda^2 \\
& \hspace{0.9cm} + \left(16 + 10 N + N^2\right) \lambda^3\Big) \overline{\rho}^{5} \bigg] \\
& + \mathcal{O}\big(\hbar^{5}\big)\;,
\end{split}
\label{eq:pure4PPIEAgapequationRhostep20DON}
\end{equation}
\begin{equation}
\begin{split}
0 = \left. \frac{\partial\Gamma^{(\mathrm{4PPI})}(\rho,\zeta)}{\partial\zeta} \right|_{\rho=\overline{\rho} \atop \zeta=\overline{\zeta}} = & \ \frac{\hbar^{3}}{24} \overline{M}_{0}\\
& - \hbar^{4} \bigg[ \frac{1}{864} \left(\lambda - \overline{M}_{0}\right)^{-1} N \Big(-81 \overline{M}_{0}^{3} + 243 \overline{M}_{0}^{2} \lambda - 3 \overline{M}_{0} \left(76 + 5 N\right) \lambda^{2} \\
& \hspace{0.9cm} + \left(8 + N \right)^{2} \lambda^{3} \Big) \overline{\rho}^{2} \bigg] \\
& + \mathcal{O}\big(\hbar^{5}\big)\;,
\end{split}
\label{eq:pure4PPIEAgapequationZetastep20DON}
\end{equation}
whereas the gs energy and density are deduced from the solutions of these two gap equations as follows:
\begin{equation}
E^\text{4PPI EA;orig}_\text{gs} = \frac{1}{\hbar} \Gamma^{(\mathrm{4PPI})}\big(\rho=\overline{\rho},\zeta=\overline{\zeta}\big) \;,
\end{equation}
\begin{equation}
\rho^\text{4PPI EA;orig}_\text{gs} = \hbar\overline{\rho} \;.
\end{equation}
One can directly infer from~\eqref{eq:pure4PPIEAgapequationZetastep10DON} that, if contributions to $\Gamma^{(\mathrm{4PPI})}$ of order $\mathcal{O}\big(\hbar^{4}\big)$ or higher are ignored, the solution of the gap equations for $M_{0}(\rho,\zeta)$ is inevitably trivial, i.e. $\overline{M}_{0}=0$. This implies that we need to push our investigations at least up to order $\mathcal{O}\big(\hbar^{4}\big)$ (i.e. up to the third non-trivial order) to find non-trivial solutions for $M_{0}(\rho,\zeta)$ so that the 4PPI EA formalism might improve the 2PPI one for a vanishing 1-point correlation function. This feature gets accentuated for higher $n$PPI EAs: the 4PPI and 6PPI EAs' results would not differ below the fifth non-trivial order (still for a vanishing 1-point correlation function) and so on\footnote{Although we do not show it here, we point out that the situation is slightly improved if the 1-point and 3-point correlation functions are allowed to take finite values. In this case, the 4PPI EA might improve the 2PPI EA results already at order $\mathcal{O}\big(\hbar^{3}\big)$ (i.e. at second non-trivial order) if the $n$-point correlation functions with $n$ odd are not constrained to vanish according to the symmetries of the problem. The latter condition does not hold in the case of the studied $O(N)$ model for which the 4PPI EA results are thus still expected to coincide with the 2PPI EA ones at second non-trivial order.}. We are just illustrating in this way a well-known property of EAs~\cite{ber04bis}.

\vspace{0.5cm}

All above remarks outlined below~\eqref{eq:pure4PPIEAgapequationZetastep20DON} are inherent to the present choice of expansion parameter, i.e. $\hbar$, and do not hold \textit{a priori} in the framework of another expansion scheme. Hence, at least in the framework of the $\hbar$-expansion, we conclude that, at lowest non-trivial orders, the treatment of $n$P(P)I EAs with $n>2$ presented here is not suited to improve the $2$P(P)I EA results obtained in section~\ref{sec:2PIEA} within the original representation of the studied model, which contrasts with the use of HST and the excellent results extracted previously from the mixed 2PI EA. Overall, we have illustrated in this chapter how HSTs and $n$P(P)I EAs with $n \geq 2$ can be used separately or together to efficiently introduce collective dofs in our description. Note that $n$PI EA formalisms are based in general on self-consistent gap equations involving bilocal, trilocal, ... and $n$-local objects as variational parameters, which might significantly burden the underpinning numerical procedure. We can circumvent this by dealing instead with $n$PPI EAs for which all variational parameters are local, but the price to pay is a significant complexification on the side of the formalism, as was illustrated in section~\ref{sec:2PPIEA} for the 2PPI EA (or in appendix~\ref{sec:4PPIEAannIM} for the 4PPI EA). Besides this, 2P(P)I EAs and higher-order EAs enable us to exploit densities as dofs (either through a propagator with e.g. the 2PI EA or directly with density functionals such as the 2PPI EA), which provides us with an interesting connection with the nuclear EDF formalism. We will remain in the EA framework in the next chapter on FRG techniques which rely on different expansion schemes for EAs (as compared to the $\hbar$-, $\lambda$- and $1/N$-expansions discussed in the present chapter), thus grasping non-perturbative physics in a very different manner.

%% file: 5ChapterFRG/FRG.tex
\minitoc

\noindent
We pursue our investigations on PI methods in this chapter, now focusing on FRG approaches. The questions underlying our study remain identical to those put forward in chapter~\ref{chap:DiagTechniques}: what are the most efficient methods in catching correlations at the non-perturbative level and what are the relevant dofs to achieve this? We will still exploit our (0+0)-D $O(N)$-symmetric $\varphi^{4}$-theory as a playground for our numerical applications and the latter question will be addressed by considering the mixed and collective representations of this model on various occasions. The link with the EDF formalism is less direct with FRG methods than with the EA approaches dealt with in chapter~\ref{chap:DiagTechniques}. All FRG techniques, which still belong to the EA formalism, rely on integro-differential equations whereas the EDF methods, as well as the diagrammatic EA techniques treated in chapter~\ref{chap:DiagTechniques} that we will now refer to as self-consistent PT for the sake of clarity, require to solve self-consistent equations for realistic models. In that respect, the FRG framework is a serious candidate to produce a new generation of approaches in nuclear theory. The study presented in this chapter aims at better understanding the ability of FRG techniques to achieve this and more generally to describe strongly-coupled quantum systems. To that end, we will discuss FRG techniques formulated from a 1PI EA, a 2PI EA and a 2PPI EA, coined respectively as 1PI-FRG~\cite{wet93,paw07}, 2PI-FRG~\cite{dup05,dup14} and 2PPI-FRG~\cite{pol02}, and we will emphasize in particular the connections between these different FRG approaches.

\vspace{0.5cm}

More specifically, the present chapter is split into three sections treating separately the 1PI-FRG, the 2PI-FRG and finally the 2PPI-FRG. Each of these sections contains two parts, the first one presenting the general formalism of the FRG approach under consideration at arbitrary dimensions and the second one specifying to our (0+0)-D $O(N)$ model. While there already exists plenty of applications of the 1PI-FRG to $O(N)$ models~\cite{ber96,tis00,ber02,tis02,tis03,can03,can03bis,jak14,def15,dup21}, we develop and apply the 2PI-FRG and the 2PPI-FRG to an $O(N)$ model for the first time to our knowledge. We will also illustrate the links between certain implementations of the 2PI-FRG and self-consistent PT. Some of these implementations are actually capable of taking the results of self-consistent PT as inputs. Since self-consistent PT is equivalent to Hartree-Fock(-Bogoliubov) theory at its first non-trivial order, this will enable us to draw an interesting parallel with the MR step of the nuclear EDF approach~\cite{ben03,sch19}.

\section{1PI functional renormalization group}
\label{sec:1PIFRG}
\subsection{State of play and general formalism}
\label{sec:1PIFRGstateofplay}

As we already discussed in chapter~\ref{chap:Intro2}, the most widespread FRG approach is the 1PI-FRG proposed by Wetterich in refs.~\cite{wet91,wet93,wet93bis,wet93bis2} alongside with others~\cite{rin90,bon93,ell93,ell94,mor94}. It remains an active area in numerous fields of physics, as e.g. in QCD~\cite{ber98,paw03,fis07,fis09,bra10,fis09bis,mit15,ren15bis,bra16,cyr16,cyr18,cyr18bis,fu20}, quantum gravity~\cite{gra98,reu02,lau02,lit04,man09,gro10,lit12bis,ben12bis,eic19,fal19}, condensed matter physics~\cite{bai04,bai05,kra07,dup08,bar09,kra09,kra09bis,fri11,dru12,del16bis,den20} or nuclear physics~\cite{ber03,dre16}. Note also some applications to out-of-equilibrium systems~\cite{sch00,jak07,gas08,pie08,ber09,ber12,sie13,chi17,tar17,tar18,tar19}. We will present the 1PI-FRG formalism for a general QFT involving a single fluctuating field $\widetilde{\varphi}_{\alpha}$ depending on a single index $\alpha\equiv(a,x)$, with $x\equiv (\boldsymbol{r},\tau)$ being the spacetime position ($\boldsymbol{r}$ and $\tau$ being respectively the space position and the imaginary time) and $a$ an internal quantum number that, if specified, coincides with the color index of an $O(N)$ model. The output of this method is the EA, as in all FRG approaches investigated in the present study. The 1PI-FRG relies on the scale-dependent generating functional\footnote{We set $\hbar=1$ throughout the entire chapter~\ref{chap:FRG} and corresponding appendices.}:
\begin{equation}
Z_{k}[J]=e^{W_{k}[J]}= \int \mathcal{D}\widetilde{\varphi} \ e^{-S[\widetilde{\varphi}]-\Delta S_{k}[\widetilde{\varphi}] +\int_{\alpha} J_{\alpha}\widetilde{\varphi}_{\alpha}}\;,
\label{eq:defZkWettFRG}
\end{equation}
where
\begin{equation}
\phi_{\alpha} \equiv \phi_{k,\alpha}[J] =\frac{\delta W_{k}[J]}{\delta J_{\alpha}}\;,
\label{eq:PhiEqualPhik1PIFRG}
\end{equation}
and
\begin{equation}
\Delta S_{k}\big[\widetilde{\varphi}\big]=\frac{1}{2}\int_{\alpha_{1},\alpha_{2}}\widetilde{\varphi}_{\alpha_{1}}R_{k,\alpha_{1} \alpha_{2}}\widetilde{\varphi}_{\alpha_{2}}\;,
\label{eq:DefDeltaS}
\end{equation}
using the shorthand notation for the integration:
\begin{equation}
\int_{\alpha}\equiv\sum_{a} \int_{x} \equiv\sum_{a} \int^{\beta}_{0} d\tau \int d^{D-1}\boldsymbol{r} \;,
\end{equation}
where $\beta$ still denotes the inverse temperature. Furthermore, $R_{k}$ is called cutoff function or regulator whereas $k$ denotes the momentum scale of the theory under consideration and will be referred to as flow parameter. The cutoff function $R_{k}$ plays a central role in the present approach, as in all FRG approaches. It must notably exhibit the same symmetry properties as the corresponding propagator. Therefore, as we are considering the case of a bosonic field, $R_{k}$ is symmetric, i.e.\footnote{In the present chapter and in the corresponding appendices, we will denote color indices by $a_{1}$, $a_{2}$, $a_{3}$, ... instead of $a$, $b$, $c$, ... as was done in the previous chapters.} $R_{k,\alpha_{1}\alpha_{2}} = R_{k,\alpha_{2}\alpha_{1}} = \delta_{a_{1}a_{2}} \mathfrak{R}_{k}(x_{1}-x_{2}) ~ \forall \alpha_{1},\alpha_{2}$ (for a fermionic field, $R_{k,\alpha_{1}\alpha_{2}} = -R_{k,\alpha_{2}\alpha_{1}} ~ \forall \alpha_{1},\alpha_{2}$). Note also that, for realistic applications, the final equations to solve are usually written in momentum space. It is indeed more natural to do so as the flow parameter corresponds to a momentum scale. Recalling that $\phi_{\alpha}\equiv\phi_{a}(x)$, this is achieved via the following Fourier transformations in $D$-dimensional Euclidean spacetime:
\begin{equation}
\breve{\phi}_{a}(p) = \int_{x} e^{-ipx} \phi_{a}(x) \;,
\end{equation}
\begin{equation}
\breve{\Gamma}_{a_{1},\cdots,a_{n}}^{(\mathrm{1PI})(n)}[\phi ; p_{1},\cdots,p_{n}] = (2\pi)^{-nD} \int_{x_{1},\cdots,x_{n}} e^{-i\sum_{m=1}^{n}p_{m}x_{m}} \Gamma_{a_{1},\cdots,a_{n}}^{(\mathrm{1PI})(n)}[\phi ; x_{1},\cdots,x_{n}]\;,
\end{equation}
where $\breve{\Gamma}_{a_{1} \cdots a_{n}}^{(\mathrm{1PI})(n)}[\phi ; p_{1},\cdots,p_{n}]\equiv\frac{\delta^{n}\Gamma^{(\mathrm{1PI})}[\phi]}{\delta \breve{\phi}_{a_{1}}(p_{1}) \cdots \delta \breve{\phi}_{a_{n}}(p_{n})}$, $\Gamma_{a_{1} \cdots a_{n}}^{(\mathrm{1PI})(n)}[\phi ; x_{1},\cdots,x_{n}]\equiv\frac{\delta^{n}\Gamma^{(\mathrm{1PI})}[\phi]}{\delta \phi_{a_{1}}(x_{1}) \cdots \delta \phi_{a_{n}}(x_{n})}$ and the momentum $p = (\boldsymbol{p},i\omega_{n})$ is such that $p x = \boldsymbol{p}\cdot \boldsymbol{r} - \omega_{n} \tau$, with $\omega_{n}$ being a Matsubara frequency. In this situation, we also Fourier transform the regulator $R_{k}$, thus leading to:
\begin{equation}
\Delta S_{k}\big[\phi\big] = \frac{1}{2} \sum_{a} \int \frac{d^{D}p}{(2\pi)^{D}} \breve{\phi}_{a}(p)\breve{\mathfrak{R}}_{k,a}(p)\breve{\phi}_{a}(-p)\;,
\end{equation}
with
\begin{equation}
\breve{R}_{k,a_{1} a_{2}}(p_{1},p_{2}) = (2\pi)^{D} \delta_{a_{1}a_{2}} \delta^{(D)}(p_{1}+p_{2}) \breve{\mathfrak{R}}_{k,a_{1}}(p_{1})\;,
\end{equation}
where $\delta^{(D)}$ is the $D$-dimensional Dirac delta function~\cite{dir30}. We will actually rather focus on the properties of $\breve{\mathfrak{R}}_{k}$ instead of those of $\mathfrak{R}_{k}$ in the discussions to come (we will switch to a more common notation thereafter, denoting $\breve{\mathfrak{R}}_{k}(p)$ rather as $R_{k}(p)$). However, we point out that, due to the absence of spacetime indices in the framework of our toy model study, we will not exploit Fourier transforms in the forthcoming applications of the 1PI-FRG, or of any other FRG approaches.

\vspace{0.5cm}

The flowing EA $\Gamma^{(\mathrm{1PI})}_{k}$ is defined through the modified Legendre transform:
\begin{equation}
\Gamma^{(\mathrm{1PI})}_{k}[\phi]=-W_{k}[J]+\int_{\alpha} J_{\alpha}\phi_{\alpha}-\Delta S_{k}[\phi]\;,
\label{eq:defGammakWettFRG}
\end{equation}
where the need for the rightmost term $-\Delta S_{k}[\phi]$ will be clarified below. After differentiating~\eqref{eq:defZkWettFRG} with respect to $k$, we end up with an exact flow equation for $W_{k}[J]$ which is equivalent to Polchinski's flow equation~\cite{pol84}. With the help of definition~\eqref{eq:defGammakWettFRG}, we can turn this into an exact flow equation for $\Gamma^{(\mathrm{1PI})}_{k}[\phi]$, which is nothing other than the Wetterich equation~\cite{wet93} (see appendix~\ref{sec:DerivMasterEq1PIFRG}):
\begin{equation}
\dot{\Gamma}^{(\mathrm{1PI})}_{k}[\phi] \equiv \partial_{k} \Gamma^{(\mathrm{1PI})}_{k}[\phi]=\frac{1}{2}\mathrm{STr}\left[\dot{R}_{k}\left(\Gamma^{(\mathrm{1PI})(2)}_{k}[\phi]+R_{k}\right)^{-1}\right] \;,
\label{eq:WetterichEq}
\end{equation}
where the matrix $\Gamma_{k}^{(\mathrm{1PI})(2)}[\phi]\equiv\frac{\delta^{2}\Gamma^{(\mathrm{1PI})}_{k}[\phi]}{\delta\phi\delta\phi}$ is the Hessian\footnote{The Hessian of $\Gamma^{(\mathrm{1PI})}_{k}$ is sometimes denoted as $\Gamma^{(1,1)}_{k}=\frac{\overrightarrow{\delta}}{\delta \Phi^{\mathrm{T}}}\Gamma_{k}\frac{\overleftarrow{\delta}}{\delta \Phi}$, where $\Phi=\frac{\delta W[J]}{\delta J^{\mathrm{T}}}$ is a column vector whose components are given by the (complex and real) fields on which the 1PI EA $\Gamma_{k}$ depends~\cite{jae03bis,bra06}.} of the flowing EA $\Gamma^{(\mathrm{1PI})}_{k}$ and $\mathrm{STr}$ denotes the supertrace as usual. We also stress that, in our notations, the dot will always denote derivatives with respect to the flow parameter, in the present as well as in the forthcoming sections treating other FRG approaches. The Wetterich equation is solved by evolving the flow parameter $k$ from the chosen UV cutoff $\Lambda$ to zero. Throughout this procedure, we actually probe the energy range [$0$,$\Lambda$] by incorporating quantum corrections to the classical action until it coincides with the full-fledged EA. Hence, the flowing EA $\Gamma^{(\mathrm{1PI})}_{k}$ must satisfy the following boundary conditions:
\begin{subequations}
\begin{empheq}[left=\empheqlbrace]{align}
& \Gamma^{(\mathrm{1PI})}_{k=\Lambda}[\phi] = S\big[\widetilde{\varphi}=\phi\big]\;. \label{eq:BoundaryGammaLambdaWettFRG}\\
\nonumber \\
& \Gamma^{(\mathrm{1PI})}_{k=0}[\phi] = \Gamma^{(\mathrm{1PI})}[\phi]\;. \label{eq:BoundaryGamma0WettFRG}
\end{empheq}
\end{subequations}
These boundary conditions are translated into constraints for the cutoff functions\footnote{According to~\eqref{eq:defZkWettFRG} and~\eqref{eq:DefDeltaS}, the introduction of the cutoff function $R_{k}$ dresses the (inverse) free propagator $C^{-1}$ as $C^{-1}\rightarrow C_{k}^{-1}=C^{-1} + R_{k}$. However, we can equivalently use a multiplicative cutoff function via $C^{-1}\rightarrow C_{k}^{-1}=R_{k}C^{-1}$, in which case~\eqref{eq:BoundaryR0WettFRG} is no longer valid. Hence, the constraints~\eqref{eq:BoundaryRLambdaWettFRG} and~\eqref{eq:BoundaryR0WettFRG} can be respectively rephrased as $C^{-1}_{k=\Lambda} = \infty$ and $C^{-1}_{k=0} = C^{-1}$ so that they become independent of the way $R_{k}$ is introduced.}:
\begin{subequations}
\begin{empheq}[left=\empheqlbrace]{align}
& R_{k=\Lambda} = \infty \;. \label{eq:BoundaryRLambdaWettFRG}\\
\nonumber \\
& R_{k=0} = 0 \;. \label{eq:BoundaryR0WettFRG}
\end{empheq}
\end{subequations}
The link between~\eqref{eq:BoundaryGamma0WettFRG} and~\eqref{eq:BoundaryR0WettFRG} can be directly established from the definition of $\Gamma^{(\mathrm{1PI})}_{k}$ given by~\eqref{eq:defGammakWettFRG}. It can be shown on the other hand that the constraint of~\eqref{eq:BoundaryRLambdaWettFRG} enables us to satisfy~\eqref{eq:BoundaryGammaLambdaWettFRG} by considering~\eqref{eq:defGammakWettFRG} in the form (see appendix~\ref{sec:DerivMasterEq1PIFRG} and more specifically~\eqref{eq:expressiondGammadphi1PIFRG} for a derivation of~\eqref{eq:DerivExpresionJalpha1for1PIFRG}):
\begin{equation}
J_{\alpha_{1}} = \frac{\delta\Gamma^{(\mathrm{1PI})}_{k}[\phi]}{\delta\phi_{\alpha_{1}}} + \int_{\alpha_{2}} R_{k,\alpha_{1}\alpha_{2}} \phi_{\alpha_{2}} \;.
\label{eq:DerivExpresionJalpha1for1PIFRG}
\end{equation}
By exploiting the latter equality at $k=\Lambda$, we can show that combining~\eqref{eq:defZkWettFRG} and~\eqref{eq:defGammakWettFRG} leads to:
\begin{equation}
\begin{split}
e^{-\Gamma^{(\mathrm{1PI})}_{k=\Lambda}[\phi]} = & \ e^{W_{k=\Lambda}[J]-\int_{\alpha} J_{\alpha} \phi_{\alpha} + \frac{1}{2} \int_{\alpha_{1},\alpha_{2}} \phi_{\alpha_{1}}R_{k=\Lambda,\alpha_{1} \alpha_{2}}\phi_{\alpha_{2}}} \\
=& \left(\int \mathcal{D}\widetilde{\varphi} \ e^{-S[\widetilde{\varphi}]-\frac{1}{2}\int_{\alpha_{1},\alpha_{2}}\widetilde{\varphi}_{\alpha_{1}}R_{k=\Lambda,\alpha_{1} \alpha_{2}}\widetilde{\varphi}_{\alpha_{2}} + \int_{\alpha} J_{\alpha} \widetilde{\varphi}_{\alpha}}\right) e^{-\int_{\alpha} J_{\alpha} \phi_{\alpha} + \frac{1}{2} \int_{\alpha_{1},\alpha_{2}} \phi_{\alpha_{1}}R_{k=\Lambda,\alpha_{1} \alpha_{2}}\phi_{\alpha_{2}}} \\
=& \left(\int \mathcal{D}\widetilde{\varphi} \ e^{-S[\widetilde{\varphi}]-\frac{1}{2}\int_{\alpha_{1},\alpha_{2}}\widetilde{\varphi}_{\alpha_{1}}R_{k=\Lambda,\alpha_{1} \alpha_{2}}\widetilde{\varphi}_{\alpha_{2}} + \int_{\alpha} \frac{\delta\Gamma^{(\mathrm{1PI})}_{k=\Lambda}[\phi]}{\delta\phi_{\alpha}} \widetilde{\varphi}_{\alpha} + \int_{\alpha_{1},\alpha_{2}} \phi_{\alpha_{1}} R_{k=\Lambda,\alpha_{1}\alpha_{2}} \widetilde{\varphi}_{\alpha_{2}}}\right) \\
& \times \ e^{-\int_{\alpha} \frac{\delta\Gamma^{(\mathrm{1PI})}_{k=\Lambda}[\phi]}{\delta\phi_{\alpha}} \phi_{\alpha} - \frac{1}{2} \int_{\alpha_{1},\alpha_{2}} \phi_{\alpha_{1}}R_{k=\Lambda,\alpha_{1} \alpha_{2}}\phi_{\alpha_{2}}} \\
=& \int \mathcal{D}\widetilde{\varphi} \ e^{-S[\widetilde{\varphi}]+\int_{\alpha} \frac{\delta \Gamma^{(\mathrm{1PI})}_{k=\Lambda}\left[\phi\right]}{\delta\phi_{\alpha}}\left(\widetilde{\varphi}_{\alpha}-\phi_{\alpha}\right)} \underbrace{e^{-\frac{1}{2}\int_{\alpha_{1},\alpha_{2}}\left(\widetilde{\varphi}_{\alpha_{1}}-\phi_{\alpha_{1}}\right)R_{k=\Lambda,\alpha_{1} \alpha_{2}}\left(\widetilde{\varphi}_{\alpha_{2}}-\phi_{\alpha_{2}}\right)}}_{\hspace{2.3cm} \sim \delta\left[\widetilde{\varphi}-\phi\right]~\mathrm{according~to~\eqref{eq:BoundaryRLambdaWettFRG}}}\;,
\end{split}
\end{equation}
which gives us:
\begin{equation}
\Gamma^{(\mathrm{1PI})}_{k=\Lambda}[\phi] \sim S[\phi]\;,
\label{eq:StartingPoint1PIFRG}
\end{equation}
as expected. The condition~\eqref{eq:StartingPoint1PIFRG} sets the starting point of the flow: needless to say that it is a very useful condition as the classical action is known in practice. Note however that~\eqref{eq:StartingPoint1PIFRG} would not be satisfied if we would not have modified the Legendre transform defining the flowing EA $\Gamma^{(\mathrm{1PI})}_{k}$, hence the relevance of the extra term $\Delta S_{k}$ in~\eqref{eq:defGammakWettFRG}.

\vspace{0.5cm}

\begin{figure}[H]
  \centering
  \includegraphics[width=.5\linewidth]{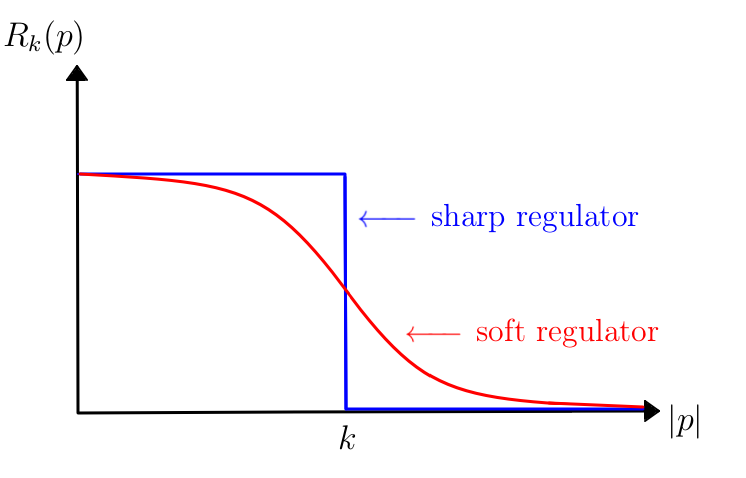}
  \caption{Typical shapes of soft and sharp cutoff functions.}
  \label{fig:FRGcutoff}
\end{figure}

We are now left with discussing the analytic form of the cutoff function $R_{k}$ for $0 < k < \Lambda$. In that respect, let us first point out that, except for simple toy models~\cite{kei12}, the Wetterich equation can not be directly integrated and must therefore be approximated in some way. Since approximations are almost always necessary to solve the Wetterich equation, the flow depends on the choice of $R_{k}$. This implies that physical results might themselves depend on the latter, even if the boundary conditions~\eqref{eq:BoundaryRLambdaWettFRG} and~\eqref{eq:BoundaryR0WettFRG} are fulfilled. The predictive power of the FRG approach can therefore be improved by implementing optimization procedures aiming at minimizing such an undesirable feature. Here are examples of optimization procedures developed so far: one based on the principle of minimal sensitivity~\cite{can03,can03bis,can05,mar14}, which determines the optimal values for the parameters of a given cutoff function, and the Litim-Pawlowski method~\cite{lit00,lit01,lit01bis,lit01bis2,lit02,paw07}, leading to the so-called theta or Litim regulator~\cite{lit00,lit01}:
\begin{equation}
R_{k}(p)=\mathcal{C}\left(k^{2}-p^{2}\right)\Theta\left(k^{2}-p^{2}\right) \;,
\label{eq:ThetaRegulator1PIFRG}
\end{equation}
where $\mathcal{C}$ is a constant of order unity and $\Theta$ is the Heaviside function~\cite{abr65}. The regulator~\eqref{eq:ThetaRegulator1PIFRG} is not suited to be combined with all approximations of the Wetterich equation (and in particular not with all orders of the derivative expansion discussed below) because of its non-smooth behavior (induced by the Heaviside function). However, in some situations, it has the advantage to allow for performing integrals analytically in the integro-differential equations deduced from the Wetterich equation. Another common choice of cutoff function is:
\begin{equation}
R_{k}(p) = \mathcal{C} \frac{p^2}{e^{p^2/k^2}-1} \;,
\label{eq:ExponentialRegulator1PIFRG}
\end{equation}
which is referred to as the exponential regulator. The cutoff functions~\eqref{eq:ThetaRegulator1PIFRG} and~\eqref{eq:ExponentialRegulator1PIFRG} are both considered as soft, in contrast to sharp regulators which are rather used in the framework of perturbative approaches (see fig.~\ref{fig:FRGcutoff}).

\vspace{0.5cm}

To summarize, there are essentially two features influencing the predictive power of the 1PI-FRG (and of all other FRG approaches): the choice of the cutoff function (alongside with the aforementioned optimization procedures) and the approximation used to solve the Wetterich equation (or other flow equations for other FRG approaches). The implementation of such an approximation is a two-step procedure: the Wetterich equation must be expanded in some way before being truncated. We will refer to the combination of these two steps as truncation scheme. For the 1PI-FRG, several truncation schemes have been developed and tested. As examples, we can mention:
\begin{itemize}
\item The \textbf{vertex expansion}~\cite{mor94,kop10bis}: it relies on a Taylor expansion of the flowing action $\Gamma^{(\mathrm{1PI})}_{k}\left[\phi\right]$ in powers of the field $\phi$ (and not of its spacetime derivatives):
\begin{equation}
\Gamma^{(\mathrm{1PI})}_{k}[\phi] = \overline{\Gamma}^{(\mathrm{1PI})}_{k} + \sum_{n=2}^{\infty}\frac{1}{n!}\int_{\alpha_{1},\cdots,\alpha_{n}} \overline{\Gamma}_{k,\alpha_{1} \cdots \alpha_{n}}^{(\mathrm{1PI})(n)} \left(\phi-\overline{\phi}_{k}\right)_{\alpha_{1}} \cdots \left(\phi-\overline{\phi}_{k}\right)_{\alpha_{n}} \;,
\label{eq:vertexexpansion1PIFRG}
\end{equation}
where $\overline{\phi}_{k,\alpha}=\left.\frac{\delta W_{k}[J]}{\delta J_{\alpha}}\right|_{J=0}$, $\overline{\Gamma}^{(\mathrm{1PI})}_{k}\equiv\Gamma^{(\mathrm{1PI})}_{k}\big[\phi=\overline{\phi}_{k}\big]$, $\overline{\Gamma}_{k,\alpha_{1} \cdots \alpha_{n}}^{(\mathrm{1PI})(n)}\equiv\left.\frac{\delta^{n}\Gamma_{k}^{(\mathrm{1PI})}[\phi]}{\delta\phi_{\alpha_{1}}\cdots\delta\phi_{\alpha_{n}}}\right|_{\phi=\overline{\phi}_{k}}$ and $\overline{\phi}_{k}$ must extremize the flowing EA, i.e.:
\begin{equation}
\left.\frac{\delta\Gamma^{(\mathrm{1PI})}_{k}[\phi]}{\delta\phi_{\alpha}}\right|_{\phi=\overline{\phi}_{k}} = 0 \mathrlap{\quad \forall \alpha, k\;.}
\label{eq:ExtremizeTruncationVertexExp1PIFRG}
\end{equation}
Inserting the expansion~\eqref{eq:vertexexpansion1PIFRG} into the Wetterich equation and identifying the terms with identical powers of $\phi-\overline{\phi}_{k}$ in the LHS and RHS turns it into an infinite tower of differential equations (see section~\ref{sec:1PIFRG0DON} for a concrete application). In order to deal with a closed finite set of equations, the simplest option consists in defining a truncation order $N_{\mathrm{max}}$ and imposing:
\begin{equation}
\hspace{5.2cm} \overline{\Gamma}_{k,\alpha_{1} \cdots \alpha_{n}}^{(\mathrm{1PI})(n)}=\overline{\Gamma}_{k=\Lambda,\alpha_{1} \cdots \alpha_{n}}^{(\mathrm{1PI})(n)} \quad \forall \alpha_{1},\cdots,\alpha_{n},k, ~ \forall n > N_{\mathrm{max}}\;.
\label{eq:1PIfrgSimplestTruncation}
\end{equation}

\vspace{0.5cm}

If the EA depends on several fields, it is convenient to rewrite first the Wetterich equation as:
\begin{equation}
\dot{\Gamma}^{(\mathrm{1PI})}_{k}=\frac{1}{2}\mathrm{STr}\left[\widetilde{\partial}_{k}\ln\left(\Gamma^{(\mathrm{1PI})(2)}_{k}+R_{k}\right)\right]\;,
\label{eq:1PIfrgWetterichEqNewDerivative}
\end{equation}
where the operator $\widetilde{\partial}_{k}$ is a derivative with respect to $k$ that only acts on the cutoff function $R_{k}$. The fluctuation matrix $\mathcal{F}_{k}$ is then introduced as:
\begin{equation}
\Gamma^{(\mathrm{1PI})(2)}_{k}+R_{k}=\mathcal{P}_{k}+\mathcal{F}_{k}\;,
\label{eq:1PIfrgFluctuatingMatrix}
\end{equation}
where $\mathcal{F}_{k}$ and $\mathcal{P}_{k}$ denote respectively the field-dependent and field-independent parts. Combining~\eqref{eq:1PIfrgWetterichEqNewDerivative} and~\eqref{eq:1PIfrgFluctuatingMatrix}, the vertex expansion can be implemented by expanding the Wetterich equation as:
\begin{equation}
\dot{\Gamma}^{(\mathrm{1PI})}_{k}=\frac{1}{2}\mathrm{STr}\left[\left(\widetilde{\partial}_{k}\mathcal{P}_{k}\right)\mathcal{P}_{k}^{-1}\right] + \frac{1}{2}\sum_{n=1}^{\infty}\frac{(-1)^{n-1}}{n}\mathrm{STr}\left[\widetilde{\partial}_{k}\left(\mathcal{P}_{k}^{-1}\mathcal{F}_{k}\right)^{n}\right]\;.
\label{eq:VertexExpSeveralField1PIFRG}
\end{equation}
The relevance of the vertex expansion relies on the validity of the Taylor expansion in~\eqref{eq:vertexexpansion1PIFRG} or~\eqref{eq:VertexExpSeveralField1PIFRG}. Put differently, the vertex expansion is useful as long as the momentum dependence of the vertex functions does not play a significant role in the physical process to be described.

\vspace{0.3cm}

\item The \textbf{derivative expansion} (DE): it mimics the expansion of the free energy in Landau theory for the flowing EA~\cite{mor94bis,ber96,can03,can03bis,kop10bis2,jak14}. The general idea of the DE is to expand the flowing EA with respect to the field and its spacetime derivatives. There are also different truncation orders for the DE, the simplest being the local-potential approximation (LPA). This scheme is most usually formulated for an $O(N)$ model in which case we rather deal with the $O(N)$-invariant $\rho\big[\vec{\phi}\big]\equiv\vec{\phi}^{2}/2$ or, more explicitly\footnote{The arrow still symbolizes the vector character in color space.}, $\rho\big[\vec{\phi}\big]\equiv\sum_{a=1}^{N}(\phi_{a}(x))^{2}/2$. The main reason for this is that the flowing classical action $S_{k}=S+\Delta S_{k}$ and the flowing EA $\Gamma^{(\mathrm{1PI})}_{k}$ exhibit the same linear symmetries (e.g. the $O(N)$ symmetry) and invariant of these symmetries (e.g. $\rho$ for the $O(N)$ symmetry) are natural choices to play the role of expansion parameter for $\Gamma^{(\mathrm{1PI})}_{k}$ in particular. The most widespread implementations of the DE are:
\begin{itemize}
\item LPA:
\begin{equation}
\Gamma^{(\mathrm{1PI})}_{k}\Big[\vec{\phi}\Big]=\int_{x}\left[\frac{1}{2}\left(\vec{\nabla}\vec{\phi}\right)^{2} + U_{k}[\rho]\right]\;,
\label{eq:LPA1PIFRG}
\end{equation}

\item LPA':
\begin{equation}
\Gamma^{(\mathrm{1PI})}_{k}\Big[\vec{\phi}\Big]=\int_{x}\left[\frac{Z_{k}}{2}\left(\vec{\nabla}\vec{\phi}\right)^{2} + U_{k}[\rho]\right]\;,
\label{eq:LPAp1PIFRG}
\end{equation}

\item DE$_{2}$:
\begin{equation}
\Gamma^{(\mathrm{1PI})}_{k}\Big[\vec{\phi}\Big]=\int_{x}\left[\frac{Z_{k}[\rho]}{2}\left(\vec{\nabla}\vec{\phi}\right)^{2} + \frac{1}{4} Y_{k}[\rho] \left(\vec{\nabla}\rho\right)^{2} + U_{k}[\rho]\right]\;,
\label{eq:DE21PIFRG}
\end{equation}
\end{itemize}
where the effective potential $U_{k}[\rho]$ only encompasses even powers of the field, as follows from the definition of $\rho$. The functional $U_{k}[\rho]$ gives us access to the most relevant information related to thermodynamics (equation of state, ...). The above ans$\ddot{\text{a}}$tze, and notably~\eqref{eq:DE21PIFRG}, have been proven to be very successful in treating $O(N)$~\cite{ber02} and Gross-Neveu~\cite{hoe02} models. However, for more complicated models, it turns out that additional approximations are almost always necessary, as illustrated by the work of Tissier \textit{et al.}~\cite{tis00,tis02,tis03}. In such situations, several invariants might come into play (instead of just $\rho$) and one usually needs new functions (other than $Z_{k}$ and $Y_{k}$) to achieve a reasonably good physical description, which renders the FRG procedure much more demanding numerically. To handle this, a convenient additional approximation is to expand $U_{k}$ and $Z_{k}$ with respect to all invariants and then truncate the power series thus obtained\footnote{See ref.~\cite{can03} for a concrete example.}.

\vspace{0.3cm}

Let us put aside these additional approximations for the rest of our discussion on the DE. In order to achieve a practical calculation based on the DE, we insert ans$\ddot{\text{a}}$tze~\eqref{eq:LPA1PIFRG}, \eqref{eq:LPAp1PIFRG} and~\eqref{eq:DE21PIFRG} into the Wetterich equation in order to extract from the latter an exact flow equation for the effective potential $U_{k}[\rho]$. For instance, after exploiting the $O(N)$ symmetry of the problem as well as the relation $\Gamma^{(\mathrm{1PI})}_{k}\big[\vec{\phi}=\vec{\phi}_{\mathrm{u}}\big]=\Omega U_{k}[\rho=\rho_{\mathrm{u}}]$ ($\Omega$ being the spacetime volume of the system and $\rho_{\mathrm{u}}\equiv\rho\big[\vec{\phi}=\vec{\phi}_{\mathrm{u}}\big]$ being evaluated by definition at a uniform field configuration $\vec{\phi}_{\mathrm{u}}$, uniform meaning $\vec{\phi}_{\mathrm{u}}(x) = \vec{\phi}_{\mathrm{u}} ~ \forall x$), we obtain for the LPA\footnote{We refer to refs.~\cite{kei12,del12} for more details on the derivation of~\eqref{eq:EquationULPAAPIFRG}.}$^{,}$\footnote{In a uniform field configuration, $\rho$ is no longer a function of $x$, which implies e.g. that $U_{k}$ is no longer a functional but a function, hence the notation $U_{k}(\rho)$ instead of $U_{k}[\rho]$.}:
\begin{equation}
\dot{U}_{k}[\rho] = \frac{1}{2} \int \frac{d^{D}p}{(2\pi)^{D}} \dot{R}_{k}(p) \left( G_{k,L}(\rho,p) + \left(N-1\right) G_{k,T}(\rho,p) \right) \;,
\label{eq:EquationULPAAPIFRG}
\end{equation}
where
\begin{subequations}
\begin{empheq}[left=\empheqlbrace]{align}
G_{k,\mathrm{L}}(\rho,p) & = \left( \Gamma_{k,\mathrm{L}}^{(\mathrm{1PI})(2)}(\rho,p) + R_{k}(p) \right)^{-1} \nonumber \\
& = \left( p^{2} + U^{(1)}_{k}(\rho) + 2 \rho U^{(2)}_{k}(\rho) + R_{k}(p) \right)^{-1} \;, \label{eq:GLPAAPIFRG}\\
\nonumber \\
G_{k,\mathrm{T}}(\rho,p) & = \left( \Gamma_{k,\mathrm{T}}^{(\mathrm{1PI})(2)}(\rho,p) + R_{k}(p) \right)^{-1} \nonumber \\
& = \left( p^{2} + U^{(1)}_{k}(\rho) + R_{k}(p) \right)^{-1} \;, \label{eq:GLTAAPIFRG}
\end{empheq}
\end{subequations}
with $U^{(n)}_{k}(\rho)$ denoting the $\mathrm{n^{th}}$-order derivative of $U_{k}$ with respect to $\rho$. The subscripts ``L'' and ``T'' respectively label longitudinal and transverse components with respect to the direction associated with SSB, e.g. $\Gamma_{k,\mathrm{L}}^{(\mathrm{1PI})(2)}$ denotes the second-order derivative of $\Gamma_{k}^{(\mathrm{1PI})}$ with respect to $\phi_{a=N}(x)$ if $\vec{\phi}^{2} = (\phi_{a=N}(x))^2$.

\vspace{0.3cm}

As its name suggests, DE$_{2}$ is referred to as the second order of the DE whereas the LPA and the LPA' can both be seen as its first order. The only difference between the LPA and the LPA' is the presence of the running field renormalization constant $Z_{k}$ in~\eqref{eq:LPAp1PIFRG}. Despite its simplicity, the LPA has notably proven successful in the study of SSBs in the framework of $O(N)$ models~\cite{def15}. Moreover, as shown in ref.~\cite{def15}, the LPA results satisfy the Mermin-Wagner theorem~\cite{mer66,hoh67,col73bis} assuming that the effective potential $U_{k}[\rho]$ in~\eqref{eq:LPA1PIFRG} is not further approximated. However, the LPA turns out to be disappointing in the determination of critical exponents, especially for the anomalous dimension $\eta$. More precisely, the LPA always yields $\eta=0$, as can be seen from the flowing propagator $G_{k}(\rho,p) = \left(\Gamma_{k}^{(\mathrm{1PI})(2)}(\rho,p)+R_{k}(p)\right)^{-1}$ which reduces to $G_{k=0}(\rho=0,p)=1/p^{2}$ at criticality which contrasts with the expected critical behavior $G_{k=0}(\rho=0,p) \sim 1/\lvert p \rvert^{2-\eta}$ with $\eta$ finite in less than 4 dimensions~\cite{lit98,bag01,pol03,ber02,ros12,del12,gie12,dup21}. The LPA' cures this problem thanks to $Z_{k}$: after choosing a relevant cutoff function~\cite{del16,dup21}, it enables us to define a flowing anomalous dimension $\eta_{k}=-k\partial_{k}\ln(Z_{k})$, so that $\eta=\underset{k \rightarrow 0}{\lim} \ \eta_{k}$ can be extracted at the end of the flow at criticality. Besides this improvement with respect to the LPA, the LPA' is often not sufficient to achieve a satisfactory quantitative accuracy in the determination of critical exponents. Nonetheless, let us point out that the successes of the LPA (and the LPA') suggest that the DE is performed around an efficiently chosen starting point, hence the motivation for exploiting the second order of the DE (in which $Z_{k}$ becomes a function of $\rho$ alongside with $Y_{k}$)~\cite{wet93bis,tet94,mor97,aok98,mor98,mor99,sei99,ger01,ber02}, or even higher orders such as the fourth~\cite{can03bis,lit12,dep20} and the sixth~\cite{bal19}.

\vspace{0.3cm}

As for the vertex expansion, the relevance of the ans$\ddot{\text{a}}$tze underlying the DE (i.e.~\eqref{eq:LPA1PIFRG}, \eqref{eq:LPAp1PIFRG} and~\eqref{eq:DE21PIFRG} for the first two orders) does not rely on the smallness of a given coupling constant so that the 1PI-FRG combined with the DE is still a non-perturbative method. However, the validity of these ans$\ddot{\text{a}}$tze is certainly not as clear as that of the vertex expansion. Firstly, we can argue that, thanks to the flow parameter $k$ acting as an IR regulator, the flowing EA $\Gamma^{(\mathrm{1PI})}_{k}$ does not suffer from the divergences characterizing critical behaviors\footnote{Technically, this translates into the fact that the derivative $\partial_{k}R_{k}(q)$ regularizes all momentum integrals involved in the set of integro-differential equations to solve.}. The latter only affect the physical EA, i.e. $\Gamma^{(\mathrm{1PI})}_{k=0}\big[\vec{\phi}\big]=\Gamma^{(\mathrm{1PI})}\big[\vec{\phi}\big]$. Therefore, the 1PI vertices $\Gamma_{k>0}^{(\mathrm{1PI})(n)}\big(\vec{\phi}_{\mathrm{u}},p_{1},\cdots,p_{n}\big)$ are smooth functions of the momenta $p_{i}$ so that $\Gamma^{(\mathrm{1PI})}_{k}$ can be expanded in terms of spacetime derivatives of the field $\vec{\phi}$ to capture the long-distance physics (typically the physics associated to a scale larger than $k^{-1}$ or $m^{-1}$, with $m$ being the mass of the field $\vec{\phi}$). More rigorously, the convergence of the DE relies on the fact that the expansion of $\Gamma^{(\mathrm{1PI})}_{k}$ with respect to $p^2/k^2$ has a finite radius of convergence $r_{\mathrm{conv}}$ on the one hand and, on the other hand, that the momentum cutoff $p_{\mathrm{cutoff}}$ is sufficiently small\footnote{Recall that the momentum $p_{\mathrm{cutoff}}$ is set by the derivative $\partial_{k}R_{k}(q)$ regularizing the momentum integrals so that only the momentum modes satisfying $\lvert p \rvert \lesssim p_{\mathrm{cutoff}}$ contribute to the flow, with $p_{\mathrm{cutoff}} \approx k$ as illustrated by fig.~\ref{fig:FRGcutoff}.} for $r_{\mathrm{conv}}$ to be significantly greater than $p_{\mathrm{cutoff}}^2/k^2$. Such conditions are usually satisfied in unitary theories for instance.

\vspace{0.3cm}

In conclusion, as opposed to the vertex expansion, the DE is particularly suited to catch long-distance physics and thus to study critical phenomena. However, such a truncation scheme does not retain the full momentum dependence of the vertex functions kept by the truncation. Although we discuss below other truncation schemes developed in order to cure this problem, we can also point out a slight modification of the LPA', called the LPA''~\cite{gue07,has12,ros18}, which enables us to capture the full momentum dependence of the propagator $G_{k}$, inducing notably better estimates of the critical exponents~\cite{has12,ros18}. The LPA'' has led to many successful applications, in equilibrium as well as in out-of-equilibrium physics~\cite{has11,can11,mat15,can16,fel18}.

\vspace{0.3cm}

\item The \textbf{BMW approximation} (named after its inventors Blaizot, M\'{e}ndez-Galain and Wschebor): it is inspired from previous studies of critical phenomena in the framework of liquid state theory~\cite{par84,par95} and relies on a set of coupled flow equations obtained by differentiating the Wetterich equation with respect to the arguments of the EA (hence $\phi$ for $\Gamma^{(\mathrm{1PI})}_{k}[\phi]$)~\cite{bla05,bla06,bla06bis,bla06bis2,bla07,ben08,ben09,ben12,ros15}\footnote{See also ref.~\cite{del12} for a pedagogical introduction on the BMW approximation.}. In this way, we obtain in momentum space a differential equation expressing e.g. $\partial_{k}\Gamma^{(\mathrm{1PI})(2)}_{k}(\phi_{\mathrm{u}},p)$ (evaluated in a uniform field configuration $\phi_{\mathrm{u}}$) in terms of $\Gamma^{(\mathrm{1PI})(3)}_{k}(\phi_{\mathrm{u}},p,-q,-(p+q))$ and $\Gamma^{(\mathrm{1PI})(4)}_{k}(\phi_{\mathrm{u}},p,-p,q,-q)$. Moreover, as the derivative $\partial_{k} R_{k}(q)$ regularizes the momentum integrals such that the terms with $\lvert q \rvert \gtrsim k$ vanish, we can consider the zero-momentum version of this equation (i.e. we set $q = 0$) as a first level of approximation. As $\Gamma^{(\mathrm{1PI})(3)}_{k}(\phi_{\mathrm{u}},p,-q,-(p+q))$ and $\Gamma^{(\mathrm{1PI})(4)}_{k}(\phi_{\mathrm{u}},p,-p,q,-q)$ are directly related to $\Gamma^{(\mathrm{1PI})(2)}_{k}(\phi_{\mathrm{u}},p)$ at $q = 0$, the equation thus obtained can be considered as a  closed set for $\Gamma^{(\mathrm{1PI})(2)}_{k}(\phi_{\mathrm{u}},p)$. One can also solve this integro-differential equation altogether with the exact equation~\eqref{eq:EquationULPAAPIFRG} for the effective potential $U_{k}$ derived in the framework of the LPA, which then gives us access to many physical quantities of interest. Hence, the implementation of the BMW approximation does not require to drop any vertices, as opposed to e.g. condition~\eqref{eq:1PIfrgSimplestTruncation} for the vertex expansion. Although the equations are solved in the zero-momentum sector (as we have set $q = 0$), the momentum dependence of the flowing vertices ($\Gamma^{(\mathrm{1PI})(2)}_{k}(\phi_{\mathrm{u}},p)$ in the above example) is fully taken into account whereas part of their field dependence is lost, which contrasts with the DE.

\end{itemize}

\vspace{0.3cm}

\noindent
The present discussion on the truncation schemes of the 1PI-FRG is not exhaustive. For instance, there are also truncation schemes for the 1PI-FRG which are inspired from the 2PI EA formalism, and more specifically from $\Phi$-derivable approximations\footnote{$\Phi$-derivable approximations are truncations of Luttinger-Ward functionals (and therefore of 2PI EAs) that satisfy specific conservation laws~\cite{bay61,bay62}. See also Refs.~\cite{van02,bla21} for more exhaustive discussions on $\Phi$-derivable approximations.}~\cite{bla11,bla21}, which has definitely shed some light on the renormalization and possible extensions of the latter type of approximations. For more details, we refer to the recent review~\cite{dup21} and references therein. This review notably compares the critical exponents (i.e. the correlation-length exponent $\nu$, the anomalous dimension $\eta$ and the correction-to-scaling exponent $\omega$) calculated with different orders of the DE and the BMW approximation.

\vspace{0.5cm}

\begin{figure}[t]
  \centering
  \includegraphics[width=.5\linewidth]{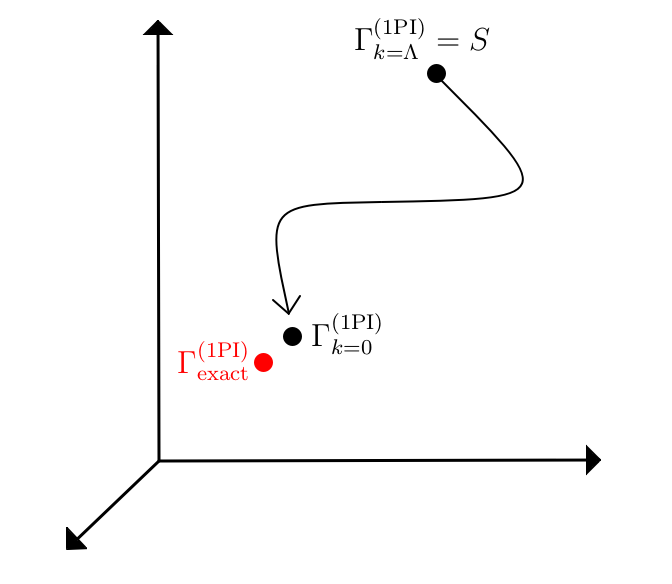}
  \caption{Schematic illustration of the 1PI-FRG flow.}
  \label{fig:1PIFRGtheoryspace}
\end{figure}

In conclusion, the FRG procedure based on the Wetterich equation starts from the classical theory (according to~\eqref{eq:StartingPoint1PIFRG}), incorporates progressively quantum correlations on top of it throughout the flow (i.e. by solving the set of integro-differential equations resulting from the vertex expansion, the DE, the BMW approximation or from any other truncation scheme of the Wetterich equation), so as to reach the corresponding quantum theory (or rather an approximated version of it in practice) at the end of the flow (see fig.~\ref{fig:1PIFRGtheoryspace}).

\subsection{Application to the (0+0)-D $O(N)$-symmetric $\varphi^4$-theory}
\label{sec:1PIFRG0DON}
\subsubsection{Original 1PI functional renormalization group}

Applications of the 1PI-FRG to the quantum anharmonic oscillator, i.e. to the (0+1)-D $\varphi^4$-theory, are presented in ref.~\cite{gie12}. We will deal here with the (0+0)-D situation via our $O(N)$ model introduced in section~\ref{sec:studiedtoymodel}. Note also that the 1PI-FRG has already been applied to this toy model using the vertex expansion, but only in its original representation and in its unbroken-symmetry phase~\cite{kei12}. However, it should be noted that, in (0+0)-D, the DE amounts to rewriting exactly the Wetterich equation at lowest order, i.e. the DE is no longer an approximate method in (0+0)\nobreakdash-D, as discussed in ref.~\cite{kei12}. This is simply due to the absence of spacetime indices in (0+0)-D. As a result, all derivative terms in the ans$\ddot{\text{a}}$tze~\eqref{eq:LPA1PIFRG} to~\eqref{eq:DE21PIFRG}, or in any other ans$\ddot{\text{a}}$tze underlying a DE, vanish, which implies that the effective potential $U_{k}$ coincides with the flowing 1PI EA $\Gamma^{(\mathrm{1PI})}_{k}$. Therefore, we will not consider the DE in the present toy model study. The BMW approximation will not be investigated in our analysis either and we will content ourselves with the vertex expansion. We will actually see later that we only exploit the vertex expansion to treat the exact flow equations of all FRG approaches tested in this study, although this may not appear clearly in all FRG formalisms (especially for the 2PI-FRG). Considering the aforementioned successes of both the DE and the BMW approximation for the 1PI-FRG, we will have to keep in mind, while drawing conclusions from the present (0+0)-D study and to prepare the ground for investigations based on more realistic models, that efficient truncation schemes other than the vertex expansion can be used in the framework of finite-dimensional models.

\vspace{0.5cm}

Exceptionally for the 1PI-FRG, we will directly take the zero-dimensional limit before carrying out the vertex expansion. The reason behind this is that we intend to reach quite high truncation orders for which the equations become already quite cumbersome even in (0+0)\nobreakdash-D. This is especially the case in the framework of the mixed theory that will be investigated to a lesser extent for other FRG approaches. In (0+0)-D, the Wetterich equation for an EA $\Gamma_{k}^{(\mathrm{1PI})}\big(\vec{\phi}\big)$ reads:
\begin{equation}
\dot{\Gamma}^{(\mathrm{1PI})}_{k}\Big(\vec{\phi}\Big) = \frac{1}{2}\mathrm{STr}\left[\dot{\boldsymbol{R}}_{k}\left(\Gamma^{(\mathrm{1PI})(2)}_{k}\Big(\vec{\phi}\Big)+\boldsymbol{R}_{k}\right)^{-1}\right] = \frac{1}{2}\sum_{a_{1},a_{2}=1}^{N} \dot{\boldsymbol{R}}_{k,a_{1}a_{2}} \boldsymbol{G}_{k,a_{2}a_{1}}\Big(\vec{\phi}\Big) \;,
\label{eq:WetterichEqpure1PIFRG0DON}
\end{equation}
with
\begin{equation}
\boldsymbol{G}^{-1}_{k,a_{1}a_{2}}\Big(\vec{\phi}\Big) \equiv \Gamma^{(\mathrm{1PI})(2)}_{k,a_{1}a_{2}}\Big(\vec{\phi}\Big)+\boldsymbol{R}_{k,a_{1}a_{2}} \;.
\label{eq:DefGkpure1PIFRG0DON}
\end{equation}
As can be deduced from the general formalism presented in section~\ref{sec:1PIFRGstateofplay}, the vertex expansion procedure applied to~\eqref{eq:WetterichEqpure1PIFRG0DON} starts from the Taylor expansion of $\Gamma^{(\mathrm{1PI})}_{k}$ around its flowing extremum at $\vec{\phi}=\vec{\overline{\phi}}_{k}$, which reduces in (0+0)-D to:
\begin{equation}
\Gamma^{(\mathrm{1PI})}_{k}\Big(\vec{\phi}\Big) = \overline{\Gamma}^{(\mathrm{1PI})}_{k} + \sum_{n=2}^{\infty} \frac{1}{n!} \sum_{a_{1},\cdots,a_{n}=1}^{N} \overline{\Gamma}_{k,a_{1} \cdots a_{n}}^{(\mathrm{1PI})(n)} \Big(\vec{\phi}-\vec{\overline{\phi}}_{k}\Big)_{a_{1}} \cdots \Big(\vec{\phi}-\vec{\overline{\phi}}_{k}\Big)_{a_{n}} \;.
\label{eq:VertexExppure1PIFRG0DON}
\end{equation}
We have used in~\eqref{eq:VertexExppure1PIFRG0DON} the definitions:
\begin{equation}
\overline{\Gamma}^{(\mathrm{1PI})}_{k} \equiv \Gamma^{(\mathrm{1PI})}_{k}\Big(\vec{\phi}=\vec{\overline{\phi}}_{k}\Big) \mathrlap{\quad \forall k \;,}
\label{eq:DefGammabarpure1PIFRG0DON}
\end{equation}
\begin{equation}
\overline{\Gamma}^{(\mathrm{1PI})(n)}_{k,a_{1} \cdots a_{n}} \equiv \left. \frac{\partial^{n} \Gamma^{(\mathrm{1PI})}_{k}\big(\vec{\phi}\big)}{\partial \phi_{a_{1}} \cdots \partial \phi_{a_{n}}} \right|_{\vec{\phi}=\vec{\overline{\phi}}_{k}} \mathrlap{\quad \forall a_{1},\cdots,a_{n},k \;,}
\label{eq:DefGammanbarpure1PIFRG0DON}
\end{equation}
and
\begin{equation}
\overline{\Gamma}_{k,a}^{(\mathrm{1PI})(1)} = 0 \mathrlap{\quad \forall a, k \;,}
\label{eq:DefGamma1barpure1PIFRG0DON}
\end{equation}
by construction. We will then distinguish two situations to pursue the vertex expansion further. On the one hand, in the unbroken-symmetry regime (i.e. in the phase with $m^{2}>0$) where $\vec{\overline{\phi}}_{k}=\vec{0}$ $\forall k$, the infinite set of differential equations resulting from the vertex expansion includes (see appendix~\ref{sec:VertexExpansionAppPure1PIFRG0DON}):
\begin{equation}
\dot{\overline{\Gamma}}^{(\mathrm{1PI})}_{k} = \frac{N}{2} \dot{R}_{k} \left( \overline{G}_{k} - \overline{G}^{(0)}_{k} \right) \;,
\label{eq:FlowEqGam0m2pos1PIFRG0DON}
\end{equation}
\begin{equation}
\dot{\overline{\Gamma}}^{(\mathrm{1PI})(2)}_{k} = -\frac{N+2}{6} \dot{R}_{k} \overline{G}^{2}_{k} \overline{\Gamma}^{(\mathrm{1PI})(4)}_{k}\;,
\label{eq:FlowEqGam2m2pos1PIFRG0DON}
\end{equation}
\begin{equation}
\dot{\overline{\Gamma}}^{(\mathrm{1PI})(4)}_{k} = \frac{N+8}{3}\dot{R}_{k} \overline{G}^{3}_{k}\left(\overline{\Gamma}^{(\mathrm{1PI})(4)}_{k}\right)^2 - \frac{N+4}{10}\dot{R}_{k} \overline{G}^{2}_{k} \overline{\Gamma}^{(\mathrm{1PI})(6)}_{k}\;,
\label{eq:FlowEqGam4m2pos1PIFRG0DON}
\end{equation}
\begin{equation}
\dot{\overline{\Gamma}}^{(\mathrm{1PI})(6)}_{k} = - \frac{5N+130}{3} \dot{R}_{k} \overline{G}^{4}_{k} \left(\overline{\Gamma}^{(\mathrm{1PI})(4)}_{k}\right)^{3} + \left(N+14\right) \dot{R}_{k} \overline{G}^{3}_{k} \overline{\Gamma}^{(\mathrm{1PI})(4)}_{k} \overline{\Gamma}^{(\mathrm{1PI})(6)}_{k} - \frac{N+6}{14}\dot{R}_{k} \overline{G}^{2}_{k} \overline{\Gamma}^{(\mathrm{1PI})(8)}_{k}\;,
\label{eq:FlowEqGam6m2pos1PIFRG0DON}
\end{equation}
where, as shown below by~\eqref{eq:propagatorm2pos1PIFRG0DON} and~\eqref{eq:G0m2pos1PIFRG0DON}, $\overline{G}_{k}$ and $\overline{G}^{(0)}_{k}$ are respectively the diagonal parts of the propagators $\overline{\boldsymbol{G}}_{k}$ and $\overline{\boldsymbol{G}}^{(0)}_{k}$ defined by:
\begin{equation}
\overline{\boldsymbol{G}}^{ \ -1}_{k,a_{1}a_{2}} \equiv \overline{\Gamma}^{(\mathrm{1PI})(2)}_{k,a_{1}a_{2}} + \boldsymbol{R}_{k,a_{1}a_{2}} \;,
\label{eq:propagator1PIFRG0DON}
\end{equation}
\begin{equation}
\left(\overline{\boldsymbol{G}}^{(0)}_{k}\right)_{a_{1}a_{2}}^{-1} \equiv \overline{\Gamma}^{(\mathrm{1PI})(2)}_{k=k_{\mathrm{i}},a_{1}a_{2}} + \boldsymbol{R}_{k,a_{1}a_{2}} \;,
\label{eq:G01PIFRG0DON}
\end{equation}
with
\begin{equation}
\boldsymbol{R}_{k,a_{1}a_{2}} = R_{k} \ \delta_{a_{1}a_{2}} \;,
\end{equation}
and $k_{\mathrm{i}}$ being the initial value of the flow parameter $k$. We have also used the following relations resulting from the $O(N)$ symmetry:
\begin{equation}
\overline{\Gamma}_{k,a_{1}a_{2}}^{(\mathrm{1PI})(2)}=\overline{\Gamma}_{k}^{(\mathrm{1PI})(2)} \ \delta_{a_{1}a_{2}} \mathrlap{\quad \forall a_{1},a_{2}\;,}
\label{eq:DefGamma2m2pos1PIFRG0DON}
\end{equation}
\begin{equation}
\hspace{2.75cm} \overline{\Gamma}_{k,a_{1}a_{2}a_{3}a_{4}}^{(\mathrm{1PI})(4)}=\frac{\overline{\Gamma}_{k}^{(\mathrm{1PI})(4)}}{3}\left(\delta_{a_{1}a_{2}}\delta_{a_{3}a_{4}}+\delta_{a_{1}a_{3}}\delta_{a_{2}a_{4}}+\delta_{a_{1}a_{4}}\delta_{a_{2}a_{3}}\right) \quad \forall a_{1},a_{2},a_{3},a_{4}\;,
\label{eq:DefGamma4m2pos1PIFRG0DON}
\end{equation}
\begin{equation}
\begin{split}
\overline{\Gamma}_{k,a_{1}a_{2}a_{3}a_{4}a_{5}a_{6}}^{(\mathrm{1PI})(6)} = \frac{\overline{\Gamma}_{k}^{(\mathrm{1PI})(6)}}{45} & \left(\delta_{a_{1}a_{2}}\delta_{a_{3}a_{4}}\delta_{a_{5}a_{6}}+\delta_{a_{1}a_{2}}\delta_{a_{3}a_{5}}\delta_{a_{4}a_{6}}+\delta_{a_{1}a_{2}}\delta_{a_{3}a_{6}}\delta_{a_{4}a_{5}}+\delta_{a_{1}a_{3}}\delta_{a_{2}a_{4}}\delta_{a_{5}a_{6}}\right. \\
& + \delta_{a_{1}a_{3}}\delta_{a_{2}a_{5}}\delta_{a_{4}a_{6}} + \delta_{a_{1}a_{3}}\delta_{a_{2}a_{6}}\delta_{a_{4}a_{5}} + \delta_{a_{1}a_{4}}\delta_{a_{2}a_{3}}\delta_{a_{5}a_{6}} + \delta_{a_{1}a_{4}}\delta_{a_{2}a_{5}}\delta_{a_{3}a_{6}} \\
& + \delta_{a_{1}a_{4}}\delta_{a_{2}a_{6}}\delta_{a_{3}a_{5}} + \delta_{a_{1}a_{5}}\delta_{a_{2}a_{3}}\delta_{a_{4}a_{6}} + \delta_{a_{1}a_{5}}\delta_{a_{2}a_{4}}\delta_{a_{3}a_{6}} + \delta_{a_{1}a_{5}}\delta_{a_{2}a_{6}}\delta_{a_{3}a_{4}} \\
& \left. + \delta_{a_{1}a_{6}}\delta_{a_{2}a_{3}}\delta_{a_{4}a_{5}} + \delta_{a_{1}a_{6}}\delta_{a_{2}a_{4}}\delta_{a_{3}a_{5}} + \delta_{a_{1}a_{6}}\delta_{a_{2}a_{5}}\delta_{a_{3}a_{4}} \right) \quad \forall a_{1},\cdots,a_{6}\;,
\label{eq:DefGamma6m2pos1PIFRG0DON}
\end{split}
\end{equation}
\begin{equation}
\overline{\Gamma}_{k,a_{1}\cdots a_{n}}^{(\mathrm{1PI})(n)} = 0 \mathrlap{\quad \forall a_{1},\cdots,a_{n},~\forall n ~ \mathrm{odd}\;.}
\label{eq:DefGammaoddm2pos1PIFRG0DON}
\end{equation}
As a consequence of~\eqref{eq:DefGamma2m2pos1PIFRG0DON}, the propagators $\overline{\boldsymbol{G}}_{k}$ and $\overline{\boldsymbol{G}}^{(0)}_{k}$ satisfy:
\begin{equation}
\overline{\boldsymbol{G}}_{k,a_{1}a_{2}} = \overline{G}_{k} \ \delta_{a_{1}a_{2}} = \left(\overline{\Gamma}^{(\mathrm{1PI})(2)}_{k} + R_{k}\right)^{-1} \delta_{a_{1}a_{2}} \;,
\label{eq:propagatorm2pos1PIFRG0DON}
\end{equation}
\begin{equation}
\overline{\boldsymbol{G}}^{(0)}_{k,a_{1}a_{2}} = \overline{G}^{(0)}_{k} \ \delta_{a_{1}a_{2}} = \left(\overline{\Gamma}^{(\mathrm{1PI})(2)}_{k=k_{\mathrm{i}}} + R_{k}\right)^{-1} \delta_{a_{1}a_{2}} \;.
\label{eq:G0m2pos1PIFRG0DON}
\end{equation}
Note that the differential equations~\eqref{eq:FlowEqGam0m2pos1PIFRG0DON} to~\eqref{eq:FlowEqGam6m2pos1PIFRG0DON} are already given in ref.~\cite{kei12}. On the other hand, in the broken-symmetry regime (i.e. in the phase with $m^{2}<0$), the homologous infinite set of differential equations contains at $N=1$ (see appendix~\ref{sec:VertexExpansionAppPure1PIFRG0DON} for the corresponding flow equations expressing the derivatives of the 1PI vertices of order 5 and 6 with respect to $k$):
\begin{equation}
\dot{\overline{\Gamma}}^{(\mathrm{1PI})}_{k} = \frac{1}{2} \dot{R}_{k} \left( \overline{G}_{k} - \overline{G}^{(0)}_{k} \right)\;,
\label{eq:FlowEqGam0m2neg1PIFRG0DON}
\end{equation}
\begin{equation}
\dot{\overline{\phi}}_{k}=\frac{1}{2\overline{\Gamma}^{(\mathrm{1PI})(2)}_{k}}\dot{R}_{k}\overline{G}_{k}^{2}\overline{\Gamma}^{(\mathrm{1PI})(3)}_{k} \;,
\label{eq:FlowEqPhim2neg1PIFRG0DON}
\end{equation}
\begin{equation}
\dot{\overline{\Gamma}}^{(\mathrm{1PI})(2)}_{k} = \dot{\overline{\phi}}_{k}\overline{\Gamma}^{(\mathrm{1PI})(3)}_{k} + \dot{R}_{k}\overline{G}_{k}^{3}\left(\overline{\Gamma}^{(\mathrm{1PI})(3)}_{k}\right)^{2}-\frac{1}{2}\dot{R}_{k}\overline{G}_{k}^{2}\overline{\Gamma}^{(\mathrm{1PI})(4)}_{k}\;,
\label{eq:FlowEqGam2m2neg1PIFRG0DON}
\end{equation}
\begin{equation}
\dot{\overline{\Gamma}}^{(\mathrm{1PI})(3)}_{k} = \dot{\overline{\phi}}_{k}\overline{\Gamma}^{(\mathrm{1PI})(4)}_{k}-3\dot{R}_{k}\overline{G}_{k}^{4}\left(\overline{\Gamma}^{(\mathrm{1PI})(3)}_{k}\right)^{3}+3\dot{R}_{k}\overline{G}_{k}^{3}\overline{\Gamma}^{(\mathrm{1PI})(3)}_{k}\overline{\Gamma}^{(\mathrm{1PI})(4)}_{k}-\frac{1}{2}\dot{R}_{k}\overline{G}_{k}^{2}\overline{\Gamma}^{(\mathrm{1PI})(5)}_{k}\;,
\label{eq:FlowEqGam3m2neg1PIFRG0DON}
\end{equation}
\begin{equation}
\begin{split}
\dot{\overline{\Gamma}}^{(\mathrm{1PI})(4)}_{k} = & \ \dot{\overline{\phi}}_{k}\overline{\Gamma}^{(\mathrm{1PI})(5)}_{k}+12\dot{R}_{k}\overline{G}_{k}^{5}\left(\overline{\Gamma}^{(\mathrm{1PI})(3)}_{k}\right)^{4}-18\dot{R}_{k}\overline{G}_{k}^{4}\left(\overline{\Gamma}^{(\mathrm{1PI})(3)}_{k}\right)^{2}\overline{\Gamma}^{(\mathrm{1PI})(4)}_{k} \\
& + 4\dot{R}_{k}\overline{G}^{3}_{k}\overline{\Gamma}^{(\mathrm{1PI})(3)}_{k}\overline{\Gamma}^{(\mathrm{1PI})(5)}_{k} + 3\dot{R}_{k}\overline{G}^{3}_{k}\left(\overline{\Gamma}^{(\mathrm{1PI})(4)}_{k}\right)^{2}-\frac{1}{2}\dot{R}_{k}\overline{G}^{2}_{k}\overline{\Gamma}^{(\mathrm{1PI})(6)}_{k}\;,
\label{eq:FlowEqGam4m2neg1PIFRG0DON}
\end{split}
\end{equation}
with $\overline{\Gamma}^{(\mathrm{1PI})(n)}_{k,1 \cdots 1} \equiv \overline{\Gamma}^{(\mathrm{1PI})(n)}_{k}$. Moreover, the propagators $\overline{G}_{k}\equiv\overline{\boldsymbol{G}}_{k,11}$ and $\overline{G}^{(0)}_{k}\equiv\overline{\boldsymbol{G}}^{(0)}_{k,11}$ are still given by~\eqref{eq:propagator1PIFRG0DON} and~\eqref{eq:G01PIFRG0DON}, respectively.

\vspace{0.5cm}

We also point out that the substitution $\overline{G}_{k}\rightarrow \overline{G}_{k}-\overline{G}^{(0)}_{k}$ was performed in the output of the vertex expansion procedure so as to obtain~\eqref{eq:FlowEqGam0m2pos1PIFRG0DON} (and~\eqref{eq:FlowEqGam0m2neg1PIFRG0DON} given afterwards), which is also the procedure followed in refs.~\cite{gie12,kei12}. Physical observables are not affected by this shift and the necessity of it can be seen by the fact that no quantum corrections must be added to $\overline{\Gamma}^{(\mathrm{1PI})}_{k}$ throughout the flow if $\lambda=0$, i.e. the relation $\overline{\Gamma}^{(\mathrm{1PI})}_{k=k_{\mathrm{f}}}=\overline{\Gamma}^{(\mathrm{1PI})}_{k=k_{\mathrm{i}}}$ must hold in the free case, which is indeed satisfied thanks to this operation. We note also that this induces a redefinition of the 1PI EA, which we nonetheless still denote as $\Gamma^{(\mathrm{1PI})}$ in what follows.

\vspace{0.5cm}

At the present stage, we still have not specified the model under consideration: all we know is that we are dealing with a 1PI EA depending on a single field $\vec{\phi}$ which is a vector in color space and lives in a zero-dimensional spacetime. In the framework of FRG approaches, the model (i.e. the classical action under consideration) is often only specified via the initial conditions used to solve the set of differential equations resulting from the truncation applied to the exact flow equation of the method (i.e. the Wetterich equation here). The initial conditions used to solve the above two sets of differential equations for our (0+0)-D $O(N)$ model can also be obtained by assuming that the $O(N)$ symmetry can only be spontaneously broken in the direction set by $a=N$ in color space, i.e. by assuming that $\vec{\phi}^2=\phi_{N}^2$. In this case, they are given by:
\begin{equation}
\overline{\phi}_{k=k_{\mathrm{i}},a} = \overline{\varphi}_{\mathrm{cl},a} = \left\{
\begin{array}{lll}
        \displaystyle{0  \quad \forall a, ~ \forall m^2 > 0\;,} \\
        \\
        \displaystyle{\pm \sqrt{-\frac{6 m^2}{\lambda}} \ \delta_{a N} \quad \forall a, ~ \forall m^2 < 0 ~ \mathrm{and} ~ \lambda\neq 0\;,}
    \end{array}
\right.
\label{eq:CIvev1PIFRG0DON}
\end{equation}
\begin{equation}
\overline{\Gamma}^{(\mathrm{1PI})}_{k=k_{\mathrm{i}}} = \left\{
\begin{array}{lll}
        \displaystyle{- \frac{N}{2}\ln\bigg(\frac{2\pi}{m^{2}}\bigg) + S\Big(\vec{\widetilde{\varphi}}=\vec{\overline{\phi}}_{k=k_{\mathrm{i}}}\Big) = - \frac{N}{2}\ln\bigg(\frac{2\pi}{m^{2}}\bigg) \quad \forall m^2 > 0\;,} \\
        \\
        \displaystyle{S\Big(\vec{\widetilde{\varphi}}=\vec{\overline{\phi}}_{k=k_{\mathrm{i}}}\Big) = \frac{m^{2}}{2} \vec{\overline{\phi}}_{k=k_{\mathrm{i}}}^{2} + \frac{\lambda}{4!}\left(\vec{\overline{\phi}}_{k=k_{\mathrm{i}}}^{2}\right)^{2} \quad \forall m^2 < 0\;,}
    \end{array}
\right.
\label{eq:CIGamma1PIFRG0DON}
\end{equation}
\begin{equation}
\begin{split}
\overline{\Gamma}^{(\mathrm{1PI})(2)}_{k=k_{\mathrm{i}},a_{1}a_{2}} = & \ \left.\frac{\partial^{2}S\big(\vec{\widetilde{\varphi}}\big)}{\partial\widetilde{\varphi}_{a_{1}}\partial\widetilde{\varphi}_{a_{2}}}\right|_{\vec{\widetilde{\varphi}}=\vec{\overline{\phi}}_{k=k_{\mathrm{i}}}} \\
= & \ \left(m^{2}+\frac{\lambda}{6}\overline{\phi}^{2}_{k=k_{\mathrm{i}},N}\right)\delta_{a_{1}a_{2}}+\frac{\lambda}{3}\overline{\phi}^{2}_{k=k_{\mathrm{i}},N}\delta_{a_{1}N}\delta_{a_{2}N} \quad \forall a_{1},a_{2} \;,
\end{split}
\label{eq:CIGamma21PIFRG0DON}
\end{equation}
\begin{equation}
\begin{split}
\overline{\Gamma}^{(\mathrm{1PI})(3)}_{k=k_{\mathrm{i}},a_{1}a_{2}a_{3}} = & \ \left.\frac{\partial^{3}S\big(\vec{\widetilde{\varphi}}\big)}{\partial\widetilde{\varphi}_{a_{1}}\partial\widetilde{\varphi}_{a_{2}}\partial\widetilde{\varphi}_{a_{3}}}\right|_{\vec{\widetilde{\varphi}}=\vec{\overline{\phi}}_{k=k_{\mathrm{i}}}} \\
= & \ \frac{\lambda}{3} \overline{\phi}_{k=k_{\mathrm{i}},N} \left(\delta_{a_{1}N}\delta_{a_{2}a_{3}}+\delta_{a_{2}N}\delta_{a_{1}a_{3}}+\delta_{a_{3}N}\delta_{a_{1}a_{2}}\right) \quad \forall a_{1}, a_{2}, a_{3} \;,
\end{split}
\label{eq:CIGamma31PIFRG0DON}
\end{equation}
\begin{equation}
\begin{split}
\overline{\Gamma}^{(\mathrm{1PI})(4)}_{k=k_{\mathrm{i}},a_{1}a_{2}a_{3}a_{4}} = & \ \left.\frac{\partial^{4}S\big(\vec{\widetilde{\varphi}}\big)}{\partial\widetilde{\varphi}_{a_{1}}\partial\widetilde{\varphi}_{a_{2}}\partial\widetilde{\varphi}_{a_{3}}\partial\widetilde{\varphi}_{a_{4}}}\right|_{\vec{\widetilde{\varphi}}=\vec{\overline{\phi}}_{k=k_{\mathrm{i}}}} \\
= & \ \frac{\lambda}{3} \left( \delta_{a_{1}a_{2}} \delta_{a_{3}a_{4}} + \delta_{a_{1}a_{3}} \delta_{a_{2}a_{4}} + \delta_{a_{1}a_{4}} \delta_{a_{2}a_{3}} \right) \quad \forall a_{1}, a_{2}, a_{3}, a_{4} \;,
\end{split}
\label{eq:CIGamma41PIFRG0DON}
\end{equation}
\begin{equation}
\begin{split}
\overline{\Gamma}^{(\mathrm{1PI})(n)}_{k=k_{\mathrm{i}},a_{1} \cdots a_{n}} = & \ \left.\frac{\partial^{n}S\big(\vec{\widetilde{\varphi}}\big)}{\partial\widetilde{\varphi}_{a_{1}}\cdots\partial\widetilde{\varphi}_{a_{n}}}\right|_{\vec{\widetilde{\varphi}}=\vec{\overline{\phi}}_{k=k_{\mathrm{i}}}} \\
= & \ 0 \quad \forall a_{1}, \cdots , a_{n}, ~ \forall n \geq 5 \;.
\end{split}
\label{eq:CIGamman1PIFRG0DON}
\end{equation}
From~\eqref{eq:CIGamma21PIFRG0DON},~\eqref{eq:CIGamma41PIFRG0DON} and~\eqref{eq:CIGamman1PIFRG0DON}, we readily deduce the initial conditions for the symmetric part $\Gamma_{k}^{(\mathrm{1PI})(n)}$ of the 1PI vertices of even order $n$ introduced via~\eqref{eq:DefGamma2m2pos1PIFRG0DON} to~\eqref{eq:DefGamma6m2pos1PIFRG0DON} for $m^2>0$:
\begin{equation}
\overline{\Gamma}^{(\mathrm{1PI})(2)}_{k=k_{\mathrm{i}}} = m^2 \mathrlap{\;,}
\label{eq:CIGamma2m2pos1PIFRG0DON}
\end{equation}
\begin{equation}
\overline{\Gamma}^{(\mathrm{1PI})(4)}_{k=k_{\mathrm{i}}} = \lambda \mathrlap{\;,}
\label{eq:CIGamma4m2pos1PIFRG0DON}
\end{equation}
\begin{equation}
\overline{\Gamma}^{(\mathrm{1PI})(n)}_{k=k_{\mathrm{i}}} = 0 \mathrlap{\quad \forall n \geq 6 \;.}
\label{eq:CIGammanm2pos1PIFRG0DON}
\end{equation}
In accordance with~\eqref{eq:1PIfrgSimplestTruncation}, the truncation of the infinite tower of differential equations containing either~\eqref{eq:FlowEqGam0m2pos1PIFRG0DON} to~\eqref{eq:FlowEqGam6m2pos1PIFRG0DON} (for all $N$ and $m^{2}>0$) or~\eqref{eq:FlowEqGam0m2neg1PIFRG0DON} to~\eqref{eq:FlowEqGam4m2neg1PIFRG0DON} (for $N=1$ and $m^{2}<0$) is implemented by the condition:
\begin{equation}
\overline{\Gamma}^{(\mathrm{1PI})(n)}_{k} = \overline{\Gamma}^{(\mathrm{1PI})(n)}_{k=k_{\mathrm{i}}} \mathrlap{\quad \forall k,~ \forall n > N_{\mathrm{max}} \;,}
\label{eq:pure1PIFRGtruncation0DON}
\end{equation}
where $\overline{\Gamma}_{k}^{(\mathrm{1PI})(n)}$ corresponds to: i) the symmetric part of the 1PI vertices of (even) order $n$ (as defined via~\eqref{eq:DefGamma2m2pos1PIFRG0DON} to~\eqref{eq:DefGamma6m2pos1PIFRG0DON} up to $n=6$) for all $N$ and $m^{2}>0$; ii) the 1PI vertices themselves for $N=1$ and $m^{2}<0$ according to the definition $\overline{\Gamma}_{k}^{(\mathrm{1PI})(n)}\equiv\overline{\Gamma}_{k,1 \cdots 1}^{(\mathrm{1PI})(n)}$. Note also that the logarithm term in~\eqref{eq:CIGamma1PIFRG0DON} was added to shift the calculated gs energy $E_{\mathrm{gs}}$ so that the latter coincides with the corresponding exact solution for $\lambda=0$ and $m^{2}>0$. We deduce indeed the gs energy from $\overline{\Gamma}^{(\mathrm{1PI})}_{k}$ at the end of the flow using the relation:
\begin{equation}
E^\text{1PI-FRG;orig}_{\mathrm{gs}} = \overline{\Gamma}^{(\mathrm{1PI})}_{k=k_{\mathrm{f}}} \;,
\label{eq:DeduceEgs1PIFRG0DON}
\end{equation}
which would hold exactly if the infinite tower of differential equations resulting from the vertex expansion was solved without approximation (such as~\eqref{eq:pure1PIFRGtruncation0DON}). Furthermore, the gs density $\rho_{\mathrm{gs}}$ is inferred at $m^{2}>0$ from~\eqref{eq:Dens} in the form:
\begin{equation}
\rho^\text{1PI-FRG;orig}_{\mathrm{gs}} = \frac{1}{N} \sum_{a=1}^{N} \left. \frac{\partial^{2} W_{k=k_{\mathrm{f}}}\big(\vec{J}\big)}{\partial J_{a}^{2}} \right|_{\vec{J}=\vec{0}} = \frac{1}{N} \sum_{a=1}^{N} \left(\overline{\Gamma}_{k=k_{\mathrm{f}}}^{(\mathrm{1PI})(2)}\right)_{aa}^{-1} = \left( \overline{\Gamma}_{k=k_{\mathrm{f}}}^{(\mathrm{1PI})(2)} \right)^{-1} \;,
\label{eq:Deducerhogs1PIFRG0DON}
\end{equation}
which results from~\eqref{eq:DefGamma2m2pos1PIFRG0DON}. We will actually neither calculate $E_{\mathrm{gs}}$ nor $\rho_{\mathrm{gs}}$ in the regime with $m^{2}<0$, as explained below. Furthermore, the chosen cutoff function for both $m^2<0$ and $m^2>0$ is:
\begin{equation}
\boldsymbol{R}_{k,a_{1}a_{2}} = R_{k} \ \delta_{a_{1}a_{2}} = \left(k^{-1} - 1\right) \delta_{a_{1}a_{2}} \mathrlap{\quad \forall a_{1}, a_{2}\;.}
\label{eq:choiceRkpure1PIFRG0DON}
\end{equation}
The cutoff function does not depend on position or momentum in (0+0)-D but its analytical form must still be consistent with conditions~\eqref{eq:BoundaryRLambdaWettFRG} and~\eqref{eq:BoundaryR0WettFRG}, that~\eqref{eq:choiceRkpure1PIFRG0DON} obviously satisfies if the flow parameter $k$ runs from $k_{\mathrm{i}}=0$ to $k_{\mathrm{f}}=1$ (which indeed implies that $R_{k=k_{\mathrm{i}}} = \infty$ and $R_{k=k_{\mathrm{f}}} = 0$).

\vspace{0.5cm}

\begin{figure}[!htb]
\captionsetup[subfigure]{labelformat=empty}
  \begin{center}
    \subfloat[]{
      \includegraphics[width=0.50\linewidth]{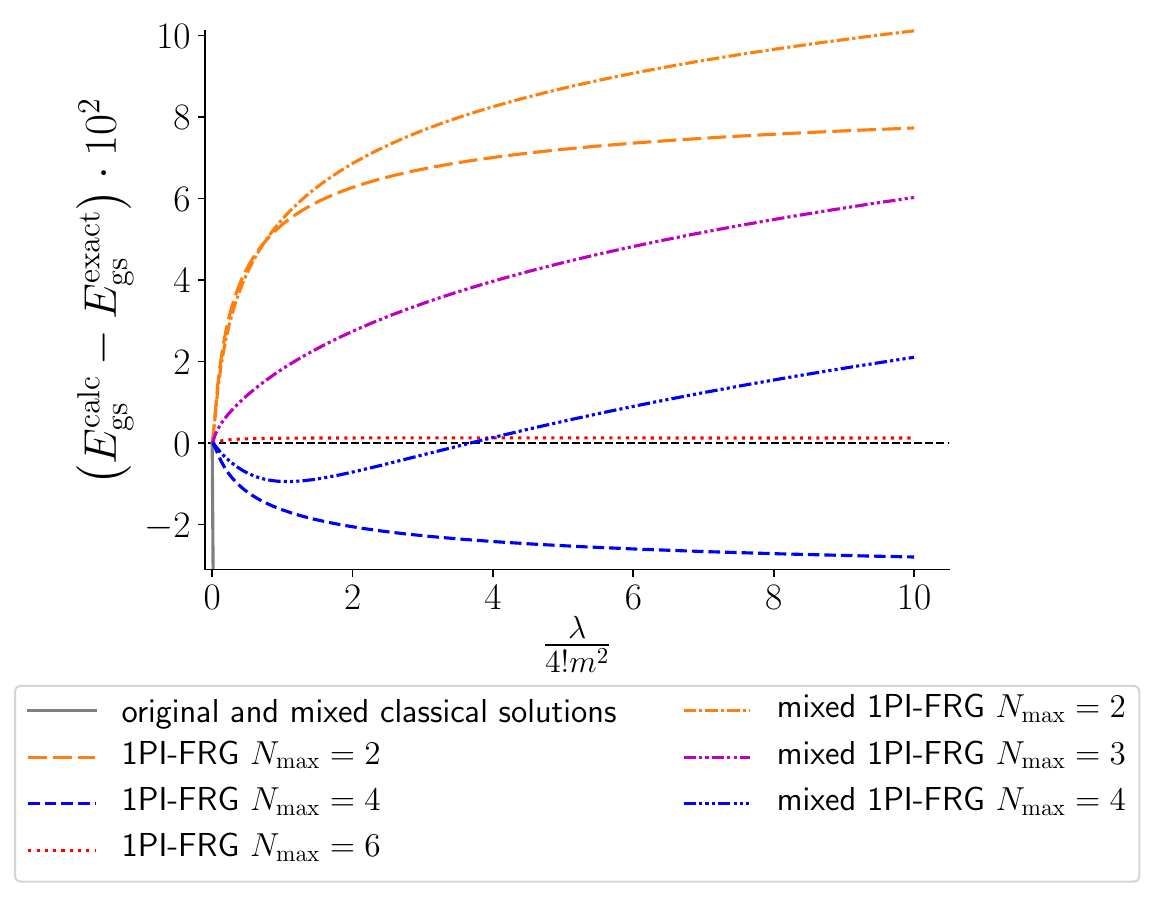}
                         }
    \subfloat[]{
      \includegraphics[width=0.50\linewidth]{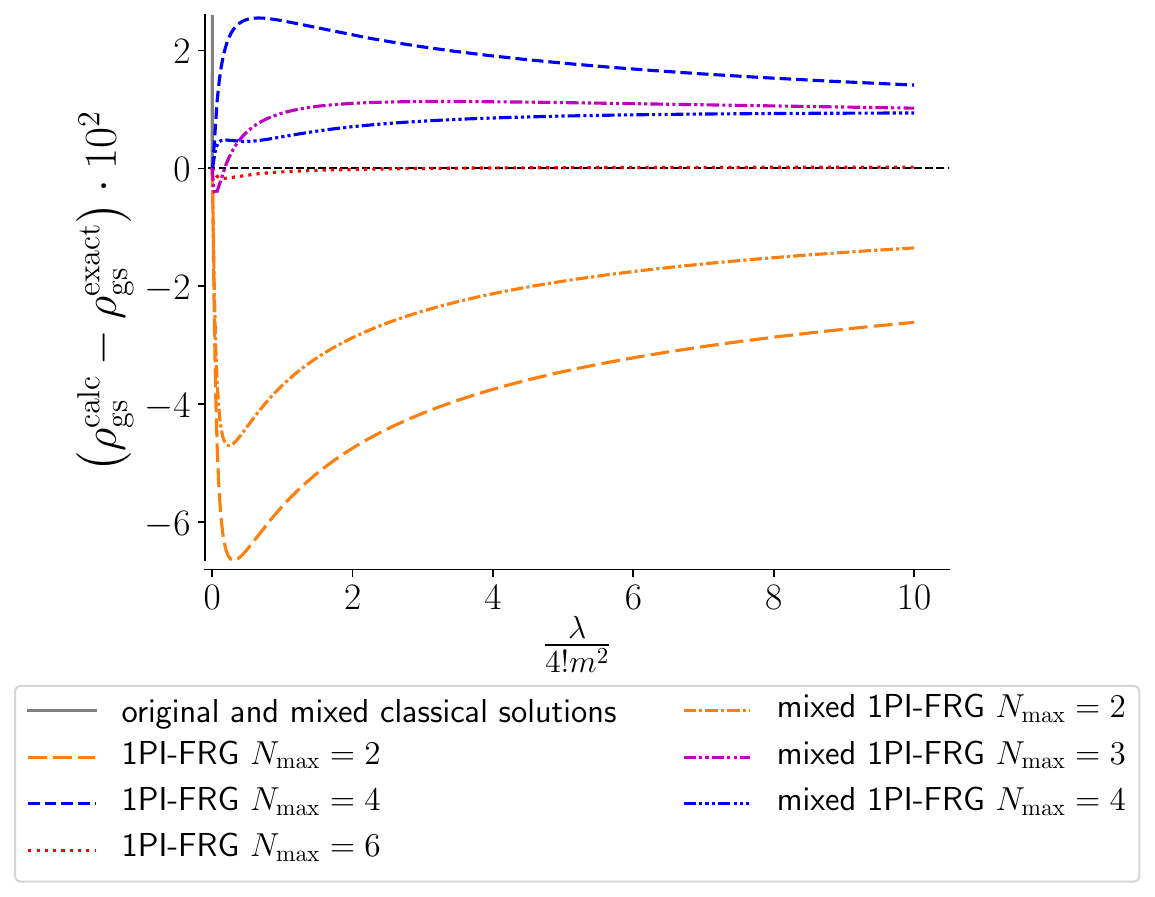}
                         }
\caption{Difference between the calculated gs energy $E_{\mathrm{gs}}^{\mathrm{calc}}$ (left) or density $\rho_{\mathrm{gs}}^{\mathrm{calc}}$ (right) and the corresponding exact solution $E_{\mathrm{gs}}^{\mathrm{exact}}$ or $\rho_{\mathrm{gs}}^{\mathrm{exact}}$ at $m^{2}=+1$ and $N=1$ ($\mathcal{R}e(\lambda)\geq 0$ and $\mathcal{I}m(\lambda)=0$).}
\label{fig:pureVsmixed1PIFRGlambdaN1}
  \end{center}
\end{figure}
\begin{figure}[!htb]
\captionsetup[subfigure]{labelformat=empty}
  \begin{center}
    \subfloat[]{
      \includegraphics[width=0.50\linewidth]{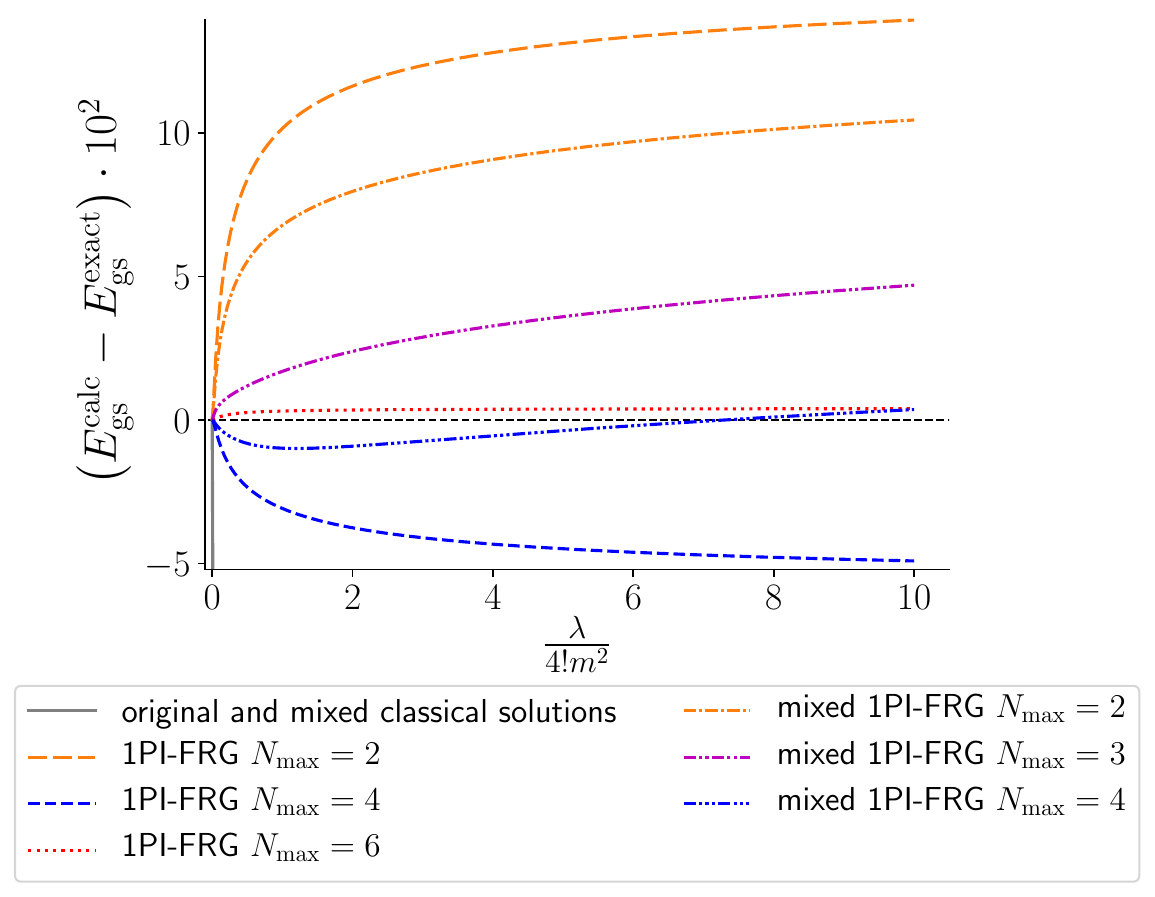}
                         }
    \subfloat[]{
      \includegraphics[width=0.50\linewidth]{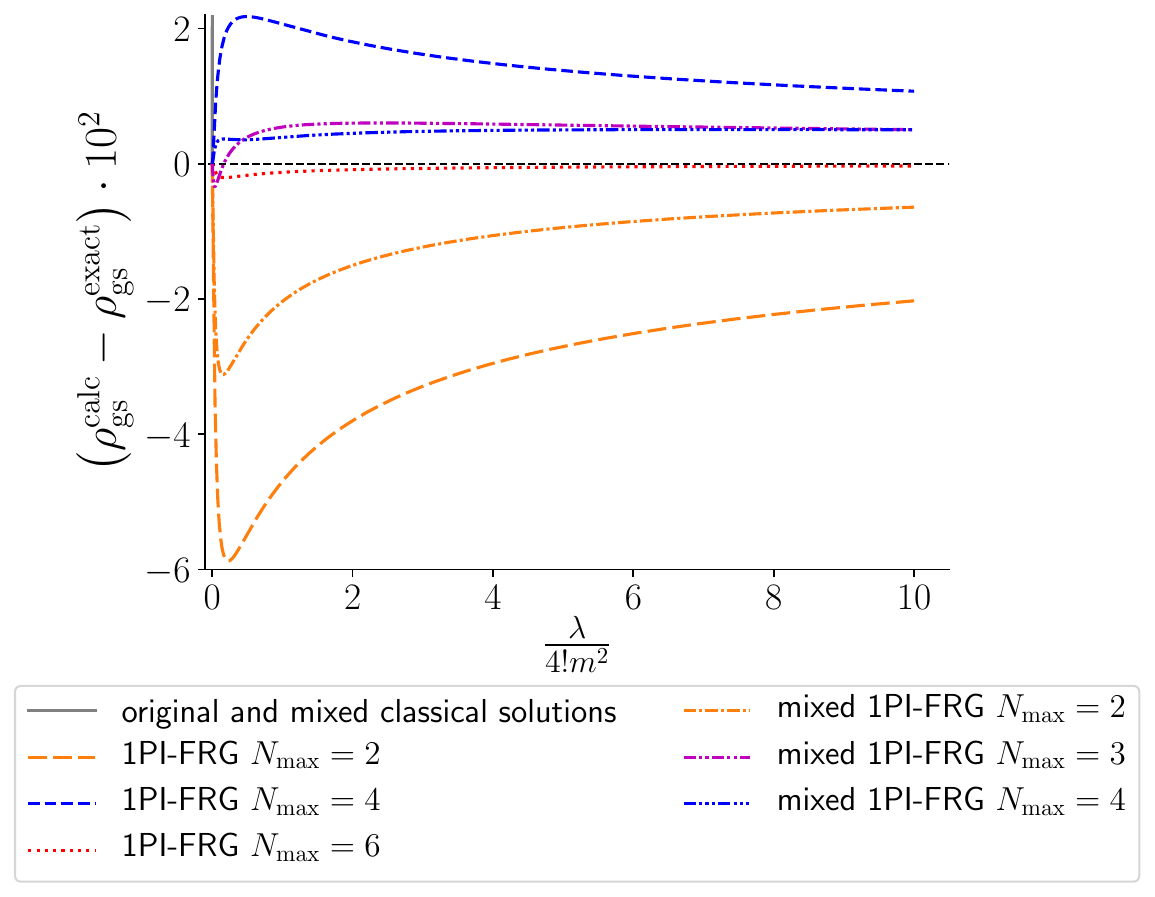}
                         }
\caption{Same as fig.~\ref{fig:pureVsmixed1PIFRGlambdaN1} with $N=2$ instead.}
\label{fig:pureVsmixed1PIFRGlambdaN2}
  \end{center}
\end{figure}

Regarding the regime with $m^{2}>0$, we solve the differential equations~\eqref{eq:FlowEqGam0m2pos1PIFRG0DON} to~\eqref{eq:FlowEqGam6m2pos1PIFRG0DON} up to $N_{\mathrm{max}}=6$, with the initial conditions~\eqref{eq:CIGamma2m2pos1PIFRG0DON} to~\eqref{eq:CIGammanm2pos1PIFRG0DON} and the cutoff function~\eqref{eq:choiceRkpure1PIFRG0DON}. Due to the symmetry constraint~\eqref{eq:DefGammaoddm2pos1PIFRG0DON}, this enables us to determine the first three non-trivial orders of the 1PI-FRG, set by the truncations $N_{\mathrm{max}}=2$, $4$ and $6$. The results thus obtained are displayed in figs.~\ref{fig:pureVsmixed1PIFRGlambdaN1} and~\ref{fig:pureVsmixed1PIFRGlambdaN2} for both $E_{\mathrm{gs}}$ and $\rho_{\mathrm{gs}}$ with $N=1$ or $2$. While the first two non-trivial order results all lie within a few percents away from the corresponding exact solution over the whole range of tested values for the coupling constant ($\lambda/4! \in [0,10]$ as usual), those obtained at $N_{\mathrm{max}}=6$ are hardly distinguishable from their exact solutions. The performances of this FRG approach are also barely affected as the coupling constant $\lambda/4!$ evolves, hence the non-perturbative character of this approach.

\vspace{0.5cm}

However, the resolution of the system given by~\eqref{eq:FlowEqGam0m2neg1PIFRG0DON} to~\eqref{eq:FlowEqGam4m2neg1PIFRG0DON} with $m^{2}<0$ (together with additional differential equations given in appendix~\ref{sec:VertexExpansionAppPure1PIFRG0DON} for $4 < N_{\mathrm{max}} \leq 6$), using the initial conditions~\eqref{eq:CIvev1PIFRG0DON} to~\eqref{eq:CIGamman1PIFRG0DON} and the cutoff function~\eqref{eq:choiceRkpure1PIFRG0DON}, is prevented, the system being too stiff (at least for the $\mathtt{NDSolve}$ function of $\mathtt{Mathematica~12.1}$). Hence, we only display 1PI-FRG results for the unbroken-symmetry phase in the framework of the original theory.

\vspace{0.5cm}

For purely fermionic systems nonetheless, the original field $\phi$ is of Grassmann nature. In this situation, the only truncation scheme at our disposal is a vertex expansion around the configuration where this field vanishes. This is because functionals of Grassmann variables can only be exploited via their Taylor expansions. Such an expansion actually yields a LE of the flow equations which implies that, after truncation, this FRG formulation is a perturbative approach. For fermionic models that can be reliably treated with dynamical MFT (DMFT)~\cite{met89,geo96}, such a limitation can be overcome with a recent extension of the 1PI-FRG called DMF$^2$RG~\cite{tar14,wen15,vil19}, in which the standard 1PI-FRG procedure relies on a cutoff function designed such that the flow exploits a correlated (non-perturbative) starting point in the form of DMFT results\footnote{DMFT is a non-perturbative method, and the same holds true for DMF$^2$RG as a consequence. DMFT consists in mapping a lattice model (i.e. a many-body system) onto an auxiliary system in the form of a quantum impurity model, like an Anderson impurity model~\cite{and61}, subject to a time-dependent mean-field adjusted self-consistently. Quantities characterizing the lattice model of interest (like susceptibilities) are then extracted from this optimized auxiliary system.}. HSTs also provide us with other means to tackle non-perturbative physics efficiently in the framework of the 1PI-FRG: this indeed enables us to introduce bosonic fields in the problem, which allows for expansions around non-trivial minima, and are thus more suited to grasp non-perturbative effects. This stresses the importance of the upcoming discussions on the mixed and collective representations. We will notably check if the stiffness of the equation systems to solve within the 1PI-FRG is sufficiently reduced in these situations to tackle the regime with $m^{2}<0$.

\subsubsection{Mixed 1PI functional renormalization group}

Let us now consider the Wetterich equation for our zero-dimensional $O(N)$ model after performing a HST, i.e. in the mixed representation. It reads:
\begin{equation}
\dot{\Gamma}^{(\mathrm{1PI})}_{\mathrm{mix},k}\Big(\vec{\phi},\eta\Big) = \frac{1}{2}\mathcal{ST}r\left[\dot{\mathcal{R}}_{k}\left(\Gamma^{(\mathrm{1PI})(2)}_{\mathrm{mix},k}\Big(\vec{\phi},\eta\Big)+\mathcal{R}_{k}\right)^{-1}\right] \;,
\label{eq:WetterichEqmixed1PIFRG0DON}
\end{equation}
where the Hessian of the flowing EA is now given by:
\begin{equation}
\Gamma^{(\mathrm{1PI})(2)}_{\mathrm{mix},k}\Big(\vec{\phi},\eta\Big) = \begin{pmatrix}
\frac{\partial^{2}\Gamma^{(\mathrm{1PI})}_{\mathrm{mix},k}}{\partial\vec{\phi}\partial\vec{\phi}} & \frac{\partial^{2}\Gamma^{(\mathrm{1PI})}_{\mathrm{mix},k}}{\partial\vec{\phi}\partial\eta} \\
\frac{\partial^{2}\Gamma^{(\mathrm{1PI})}_{\mathrm{mix},k}}{\partial\eta\partial\vec{\phi}} & \frac{\partial^{2}\Gamma^{(\mathrm{1PI})}_{\mathrm{mix},k}}{\partial\eta\partial\eta}
\end{pmatrix} \equiv \begin{pmatrix}
\Gamma^{(\mathrm{1PI})(2\phi)}_{\mathrm{mix},k} & \Gamma^{(\mathrm{1PI})(1\phi,1\eta)}_{\mathrm{mix},k} \\
\Gamma^{(\mathrm{1PI})(1\phi,1\eta)}_{\mathrm{mix},k} & \Gamma^{(\mathrm{1PI})(2\eta)}_{\mathrm{mix},k}
\end{pmatrix} \;,
\label{eq:Hessianmixed1PIFRG0DON}
\end{equation}
and the cutoff function $\mathcal{R}_{k}$ exhibits the following matrix structure in extended color space:
\begin{equation}
\mathcal{R}_{k} = \begin{pmatrix}
\boldsymbol{R}^{(\phi)}_{k} & \vec{0} \\
\vec{0}^{\mathrm{T}} & R^{(\eta)}_{k}
\end{pmatrix} = \begin{pmatrix}
R_{k} \mathbb{I}_{N} & \vec{0} \\
\vec{0}^{\mathrm{T}} & R_{k}
\end{pmatrix} = R_{k} \mathbb{I}_{N+1} \;.
\label{eq:mathcalRkmixed1PIFRG0DON}
\end{equation}
We will now apply the vertex expansion procedure to~\eqref{eq:WetterichEqmixed1PIFRG0DON}, starting from the Taylor series:
\begin{equation}
\scalebox{0.89}{${\displaystyle\Gamma^{(\mathrm{1PI})}_{\mathrm{mix},k}\Big(\vec{\phi},\eta\Big) = \overline{\Gamma}^{(\mathrm{1PI})}_{\mathrm{mix},k} + \sum_{n=2}^{\infty} \frac{1}{n!} \sum_{m=0}^{n} \begin{pmatrix}
n \\
m
\end{pmatrix} \sum_{a_{1},\cdots,a_{m}=1}^{N} \overline{\Gamma}_{\mathrm{mix},k,a_{1} \cdots a_{m}}^{(\mathrm{1PI})(m\phi,(n-m)\eta)} \Big(\vec{\phi}-\vec{\overline{\phi}}_{k}\Big)_{a_{1}} \cdots \Big(\vec{\phi}-\vec{\overline{\phi}}_{k}\Big)_{a_{m}} \left(\eta-\overline{\eta}_{k}\right)^{n-m} \;,}$}
\label{eq:TaylorExpmixed1PIFRG0DON}
\end{equation}
with
\begin{equation}
\overline{\Gamma}^{(\mathrm{1PI})}_{\mathrm{mix},k} \equiv \Gamma^{(\mathrm{1PI})}_{\mathrm{mix},k}\Big(\vec{\phi}=\vec{\overline{\phi}}_{k},\eta=\overline{\eta}_{k}\Big) \mathrlap{\quad \forall k \;,}
\label{eq:DefGammabarmixed1PIFRG0DON}
\end{equation}
\begin{equation}
\overline{\Gamma}^{(\mathrm{1PI})(n\phi,m\eta)}_{\mathrm{mix},k,a_{1} \cdots a_{n}} \equiv \left.\frac{\partial^{n+m}\Gamma^{(\mathrm{1PI})}_{\mathrm{mix},k}\big(\vec{\phi},\eta\big)}{\partial\phi_{a_{1}}\cdots\partial\phi_{a_{n}}\partial\eta^{m}}\right|_{\vec{\phi}=\vec{\overline{\phi}}_{k} \atop \eta=\overline{\eta}_{k}} \mathrlap{\quad \forall a_{1},\cdots,a_{n}, k \;,}
\label{eq:DefGammanmbarmixed1PIFRG0DON}
\end{equation}
and\footnote{The relations $\overline{\Gamma}^{(\mathrm{1PI})(n\phi)}_{\mathrm{mix},k}\equiv\overline{\Gamma}^{(\mathrm{1PI})(n\phi,0\eta)}_{\mathrm{mix},k}$ and $\overline{\Gamma}^{(\mathrm{1PI})(n\eta)}_{\mathrm{mix},k}\equiv\overline{\Gamma}^{(\mathrm{1PI})(0\phi,n\eta)}_{\mathrm{mix},k}$ are assumed for all $n$ in~\eqref{eq:DefGamma10and01barmixed1PIFRG0DON} as well as in subsequent equations.}
\begin{equation}
\overline{\Gamma}^{(\mathrm{1PI})(1\phi)}_{\mathrm{mix},k,a} = \overline{\Gamma}^{(\mathrm{1PI})(1\eta)}_{\mathrm{mix},k}=0 \mathrlap{\quad \forall a, k \;,}
\label{eq:DefGamma10and01barmixed1PIFRG0DON}
\end{equation}
since the flowing EA is now extremal at $\begin{pmatrix} \vec{\phi} & \eta \end{pmatrix} = \begin{pmatrix} \vec{\overline{\phi}}_{k} & \overline{\eta}_{k} \end{pmatrix}$. Furthermore, as the latter EA now depends on several fields, we follow the corresponding recipe outlined in section~\ref{sec:1PIFRGstateofplay} and perform the splitting put forward in~\eqref{eq:1PIfrgFluctuatingMatrix}:
\begin{equation}
\Gamma^{(\mathrm{1PI})(2)}_{\mathrm{mix},k}+\mathcal{R}_{k}=\mathcal{P}_{k}+\mathcal{F}_{k}\;,
\label{eq:SplittingPkFkmixed1PIFRG0DON}
\end{equation}
where the fluctuation matrix contains all the field dependence:
\begin{equation}
\mathcal{F}_{k} = \begin{pmatrix}
\Gamma^{(\mathrm{1PI})(2\phi)}_{\mathrm{mix},k}-\overline{\Gamma}^{(\mathrm{1PI})(2\phi)}_{\mathrm{mix},k} & \Gamma^{(\mathrm{1PI})(1\phi,1\eta)}_{\mathrm{mix},k}-\overline{\Gamma}^{(\mathrm{1PI})(1\phi,1\eta)}_{\mathrm{mix},k} \\
\Gamma^{(\mathrm{1PI})(1\phi,1\eta)}_{\mathrm{mix},k}-\overline{\Gamma}^{(\mathrm{1PI})(1\phi,1\eta)}_{\mathrm{mix},k} & \Gamma^{(\mathrm{1PI})(2\eta)}_{\mathrm{mix},k}-\overline{\Gamma}^{(\mathrm{1PI})(2\eta)}_{\mathrm{mix},k}
\end{pmatrix}\;,
\label{eq:Fkmixed1PIFRG0DON}
\end{equation}
which imposes that $\mathcal{P}_{k}$ satisfies:
\begin{equation}
\mathcal{P}_{k} = \begin{pmatrix}
\boldsymbol{R}^{(\phi)}_{k}+\overline{\Gamma}^{(\mathrm{1PI})(2\phi)}_{\mathrm{mix},k} & \overline{\Gamma}^{(\mathrm{1PI})(1\phi,1\eta)}_{\mathrm{mix},k} \\
\overline{\Gamma}^{(\mathrm{1PI})(1\phi,1\eta)}_{\mathrm{mix},k} & R^{(\eta)}_{k}+\overline{\Gamma}^{(\mathrm{1PI})(2\eta)}_{\mathrm{mix},k}
\end{pmatrix}\;,
\label{eq:Pkmixed1PIFRG0DON}
\end{equation}
to be consistent with~\eqref{eq:SplittingPkFkmixed1PIFRG0DON}. For $m^2>0$, we will use the counterparts of~\eqref{eq:DefGamma2m2pos1PIFRG0DON} to~\eqref{eq:DefGammaoddm2pos1PIFRG0DON} introducing the symmetric parts of 1PI vertices:
\begin{equation}
\overline{\Gamma}_{\mathrm{mix},k,a_{1}a_{2}}^{(\mathrm{1PI})(2\phi,n\eta)}=\overline{\Gamma}_{\mathrm{mix},k}^{(\mathrm{1PI})(2\phi,n\eta)} \ \delta_{a_{1}a_{2}} \mathrlap{\quad \forall a_{1},a_{2},n\;,}
\label{eq:DefGamma2m2posmixed1PIFRG0DON}
\end{equation}
\begin{equation}
\hspace{2.3cm} \overline{\Gamma}_{\mathrm{mix},k,a_{1}a_{2}a_{3}a_{4}}^{(\mathrm{1PI})(4\phi,n\eta)}=\overline{\Gamma}_{\mathrm{mix},k}^{(\mathrm{1PI})(4\phi,n\eta)}\left(\delta_{a_{1}a_{2}}\delta_{a_{3}a_{4}}+\delta_{a_{1}a_{3}}\delta_{a_{2}a_{4}}+\delta_{a_{1}a_{4}}\delta_{a_{2}a_{3}}\right) \quad \forall a_{1},a_{2},a_{3},a_{4},n\;,
\label{eq:DefGamma4m2posmixed1PIFRG0DON}
\end{equation}
\begin{equation}
\overline{\Gamma}_{\mathrm{mix},k,a_{1}\cdots a_{n}}^{(\mathrm{1PI})(n\phi,m\eta)} = 0 \mathrlap{\quad \forall a_{1},\cdots,a_{n},m, ~ \forall n ~ \mathrm{odd}\;.}
\label{eq:DefGammaoddm2posmixed1PIFRG0DON}
\end{equation}
As explained in section~\ref{sec:1PIFRGstateofplay}, the next step of the vertex expansion consists in carrying out matrix products between $\mathcal{P}_{k}^{-1}$ and $\mathcal{F}_{k}$. If $m^2<0$, we can not use the $O(N)$ symmetry to impose~\eqref{eq:DefGammaoddm2posmixed1PIFRG0DON}, which renders the resulting differential equations very cumbersome. For this reason, we will just present our analytical results of the mixed 1PI-FRG for the unbroken-symmetry phase (i.e. for $m^2>0$), in which $\vec{\overline{\phi}}_{k}=\vec{0}$ $\forall k$. In this situation, we show with the help of~\eqref{eq:Fkmixed1PIFRG0DON} and~\eqref{eq:Pkmixed1PIFRG0DON} as well as~\eqref{eq:DefGamma2m2posmixed1PIFRG0DON} to~\eqref{eq:DefGammaoddm2posmixed1PIFRG0DON} that the differential equations resulting from the vertex expansion procedure applied to~\eqref{eq:WetterichEqmixed1PIFRG0DON} are for example (see appendix~\ref{sec:VertexExpansionAppMixed1PIFRG0DON} for the corresponding flow equations expressing the derivatives of the 1PI vertices of order 3 and 4 with respect to $k$):
\begin{itemize}
\item For $N=1$:
\begin{equation}
\dot{\overline{\Gamma}}^{(\mathrm{1PI})}_{\mathrm{mix},k} = \ \frac{1}{2}\left(\overline{G}^{(\phi)}_{k}-\overline{G}^{(\phi)(0)}_{k}\right) + \frac{1}{2}\left(\overline{G}^{(\eta)}_{k}-\overline{G}^{(\eta)(0)}_{k}\right) \;,
\label{eq:DiffEqGam00N1mixed1PIFRG0DON}
\end{equation}
\begin{equation}
\dot{\overline{\eta}}_{k} = \frac{\dot{R}_{k}}{2\overline{\Gamma}^{(\mathrm{1PI})(2\eta)}_{\mathrm{mix},k}} \left(\overline{\Gamma}^{(\mathrm{1PI})(3\eta)}_{\mathrm{mix},k} \left(\overline{G}^{(\eta)}_{k}\right)^2 + \overline{\Gamma}^{(\mathrm{1PI})(2\phi,1\eta)}_{\mathrm{mix},k} \left(\overline{G}^{(\phi)}_{k}\right)^2\right) \;,
\label{eq:DiffEqetaN1mixed1PIFRG0DON}
\end{equation}
\begin{equation}
\begin{split}
\dot{\overline{\Gamma}}^{(\mathrm{1PI})(2\phi)}_{\mathrm{mix},k} = & \ \dot{\overline{\eta}}_{k}\overline{\Gamma}_{\mathrm{mix},k}^{(\mathrm{1PI})(2\phi,1\eta)} \\
& - \frac{1}{2} \dot{R}_{k} \bigg( \overline{\Gamma}_{\mathrm{mix},k}^{(\mathrm{1PI})(2\phi,2\eta)} \left(\overline{G}^{(\eta)}_{k}\right)^2 + \overline{G}^{(\phi)}_{k} \bigg(\overline{\Gamma}_{\mathrm{mix},k}^{(\mathrm{1PI})(4\phi)} \overline{G}^{(\phi)}_{k} \\
& \hspace{1.4cm} - 2 \left(\overline{\Gamma}_{\mathrm{mix},k}^{(\mathrm{1PI})(2\phi,1\eta)}\right)^2 \overline{G}^{(\eta)}_{k} \left(\overline{G}^{(\eta)}_{k} + \overline{G}^{(\phi)}_{k}\right)\bigg)\bigg) \;,
\end{split}
\label{eq:DiffEqGam20N1mixed1PIFRG0DON}
\end{equation}
\begin{equation}
\begin{split}
\dot{\overline{\Gamma}}^{(\mathrm{1PI})(2\eta)}_{\mathrm{mix},k} = & \ \dot{\overline{\eta}}_{k}\overline{\Gamma}_{\mathrm{mix},k}^{(\mathrm{1PI})(3\eta)} \\
& -\frac{1}{2} \dot{R}_{k} \bigg( \overline{\Gamma}_{\mathrm{mix},k}^{(\mathrm{1PI})(4\eta)} \left(\overline{G}^{(\eta)}_{k}\right)^2 - 2 \left(\overline{\Gamma}_{\mathrm{mix},k}^{(\mathrm{1PI})(3\eta)}\right)^2 \left(\overline{G}^{(\eta)}_{k}\right)^3 \\
& \hspace{1.4cm} + \left(\overline{G}^{(\phi)}_{k}\right)^2 \left(\overline{\Gamma}_{\mathrm{mix},k}^{(\mathrm{1PI})(2\phi,2\eta)} - 2 \left(\overline{\Gamma}_{\mathrm{mix},k}^{(\mathrm{1PI})(2\phi,1\eta)}\right)^2 \overline{G}^{(\phi)}_{k}\right) \bigg) \;,
\end{split}
\label{eq:DiffEqGam02N1mixed1PIFRG0DON}
\end{equation}

\item For $N=2$:
\begin{equation}
\dot{\overline{\Gamma}}^{(\mathrm{1PI})}_{\mathrm{mix},k} = \left(\overline{G}^{(\phi)}_{k}-\overline{G}^{(\phi)(0)}_{k}\right) + \frac{1}{2}\left(\overline{G}^{(\eta)}_{k}-\overline{G}^{(\eta)(0)}_{k}\right) \;,
\label{eq:DiffEqGam00N2mixed1PIFRG0DON}
\end{equation}
\begin{equation}
\dot{\overline{\eta}}_{k} = \frac{\dot{R}_{k}}{2\overline{\Gamma}^{(\mathrm{1PI})(2\eta)}_{\mathrm{mix},k}} \left(\overline{\Gamma}^{(\mathrm{1PI})(3\eta)}_{\mathrm{mix},k} \left(\overline{G}^{(\eta)}_{k}\right)^2 + 2 \overline{\Gamma}^{(\mathrm{1PI})(2\phi,1\eta)}_{\mathrm{mix},k} \left(\overline{G}^{(\phi)}_{k}\right)^2\right) \;,
\label{eq:DiffEqetaN2mixed1PIFRG0DON}
\end{equation}
\begin{equation}
\begin{split}
\dot{\overline{\Gamma}}^{(\mathrm{1PI})(2\phi)}_{\mathrm{mix},k} = & \ \dot{\overline{\eta}}_{k}\overline{\Gamma}_{\mathrm{mix},k}^{(\mathrm{1PI})(2\phi,1\eta)} \\
& -\frac{1}{6} \dot{R}_{k} \bigg(3 \overline{\Gamma}^{(\mathrm{1PI})(2\phi,2\eta)}_{\mathrm{mix},k} \left(\overline{G}^{(\eta)}_{k}\right)^2 + 4 \overline{\Gamma}_{\mathrm{mix},k}^{(\mathrm{1PI})(4\phi)} \left(\overline{G}^{(\phi)}_{k}\right)^2 \\
& \hspace{1.4cm} - 6 \left(\overline{\Gamma}_{\mathrm{mix},k}^{(\mathrm{1PI})(2\phi,1\eta)}\right)^2 \overline{G}^{(\eta)}_{k} \overline{G}^{(\phi)}_{k} \left(\overline{G}^{(\eta)}_{k} + \overline{G}^{(\phi)}_{k}\right)\bigg) \;,
\end{split}
\label{eq:DiffEqGam20N2mixed1PIFRG0DON}
\end{equation}
\begin{equation}
\begin{split}
\dot{\overline{\Gamma}}^{(\mathrm{1PI})(2\eta)}_{\mathrm{mix},k} = & \ \dot{\overline{\eta}}_{k}\overline{\Gamma}_{\mathrm{mix},k}^{(\mathrm{1PI})(3\eta)} \\
& + \dot{R}_{k} \bigg(-\frac{1}{2} \overline{\Gamma}^{(\mathrm{1PI})(4\eta)}_{\mathrm{mix},k} \left(\overline{G}^{(\eta)}_{k}\right)^2 + \left(\overline{\Gamma}^{(\mathrm{1PI})(3\eta)}_{\mathrm{mix},k}\right)^2 \left(\overline{G}^{(\eta)}_{k}\right)^3 - \overline{\Gamma}^{(\mathrm{1PI})(2\phi,2\eta)}_{\mathrm{mix},k} \left(\overline{G}^{(\phi)}_{k}\right)^2 \\
& \hspace{1.35cm} + 2 \left(\overline{\Gamma}^{(\mathrm{1PI})(2\phi,1\eta)}_{\mathrm{mix},k}\right)^2 \left(\overline{G}^{(\phi)}_{k}\right)^3 \bigg) \;,
\end{split}
\label{eq:DiffEqGam02N2mixed1PIFRG0DON}
\end{equation}

\end{itemize}
where we have introduced the propagators:
\begin{equation}
\left(\overline{\boldsymbol{G}}^{(\phi)}_{k}\right)^{-1}_{a_{1}a_{2}} = \left(\overline{G}^{(\phi)}_{k}\right)^{-1} \delta_{a_{1}a_{2}} = \left( \overline{\Gamma}^{(2\phi)}_{\mathrm{mix},k} + R_{k} \right) \delta_{a_{1}a_{2}} \mathrlap{\quad \forall a_{1}, a_{2} \;,}
\end{equation}
\begin{equation}
\left(\overline{G}^{(\eta)}_{k}\right)^{-1} = \overline{\Gamma}^{(2\eta)}_{\mathrm{mix},k} + R_{k} \mathrlap{\;,}
\end{equation}
and their classical counterparts introduced for the same reason as that mentioned below~\eqref{eq:FlowEqGam4m2neg1PIFRG0DON}:
\begin{equation}
\left(\overline{\boldsymbol{G}}^{(\phi)(0)}_{k}\right)^{-1}_{a_{1}a_{2}} = \left(\overline{G}^{(\phi)(0)}_{k}\right)^{-1} \delta_{a_{1}a_{2}} = \left( \overline{\Gamma}^{(2\phi)}_{\mathrm{mix},k=k_{\mathrm{i}}} + R_{k} \right) \delta_{a_{1}a_{2}} \mathrlap{\quad \forall a_{1}, a_{2} \;,}
\end{equation}
\begin{equation}
\left(\overline{G}^{(\eta)(0)}_{k}\right)^{-1} = \overline{\Gamma}^{(2\eta)}_{\mathrm{mix},k=k_{\mathrm{i}}} + R_{k} \mathrlap{\;.}
\end{equation}

\vspace{0.5cm}

Moreover, as a first level of approximation which is nothing other than the MFT already discussed in section~\ref{sec:KeyAspectsFRG}, we can set all bosonic entries of $\mathcal{P}_{k}^{-1}$ equal to zero~\cite{jae03}. For the toy model under consideration, this amounts to neglecting the bottom-right component of $\mathcal{P}_{k}^{-1}$, i.e. this amounts to setting $\mathcal{P}_{k,N+1 \hspace{0.04cm} N+1}^{-1}=\overline{G}^{(\eta)}_{k}=0$. The sets of differential equations to solve in the framework of MFT for $N_{\mathrm{max}} \leq 2$ can therefore be directly inferred from~\eqref{eq:DiffEqGam00N1mixed1PIFRG0DON} to~\eqref{eq:DiffEqGam02N1mixed1PIFRG0DON} for $N=1$ and from~\eqref{eq:DiffEqGam00N2mixed1PIFRG0DON} to~\eqref{eq:DiffEqGam02N2mixed1PIFRG0DON} for $N=2$ (and from the results of appendix~\ref{sec:VertexExpansionAppMixed1PIFRG0DON} for $N_{\mathrm{max}} \leq 4$ and $N=1~\mathrm{or}~2$) by setting $\overline{G}^{(\eta)}_{k}=0$.

\vspace{0.5cm}

Whether we restrict ourselves to MFT or not, the initial conditions used to solve the differential equations within the mixed 1PI-FRG, for all $N$ and for $m^{2}>0$, are inferred from the classical action $S_{\mathrm{mix}}$ (together with the definitions~\eqref{eq:DefGamma2m2posmixed1PIFRG0DON} to~\eqref{eq:DefGammaoddm2posmixed1PIFRG0DON} notably), which gives us:
\begin{equation}
\overline{\eta}_{k=k_{\mathrm{i}}} = \overline{\sigma}_{\mathrm{cl}} = 0 \mathrlap{\;,}
\label{eq:CIvevmix1PIFRG0DON}
\end{equation}
\begin{equation}
\overline{\Gamma}^{(\mathrm{1PI})}_{\mathrm{mix},k=k_{\mathrm{i}}} = - \frac{N}{2}\ln\bigg(\frac{2\pi}{m^{2}}\bigg) + S_{\mathrm{mix}}\Big(\vec{\widetilde{\varphi}}=\vec{\overline{\phi}}_{k=k_{\mathrm{i}}},\widetilde{\sigma}=\overline{\eta}_{k=k_{\mathrm{i}}}\Big) = - \frac{N}{2}\ln\bigg(\frac{2\pi}{m^{2}}\bigg) \mathrlap{\;,}
\label{eq:CIGammamix1PIFRG0DON}
\end{equation}
\begin{equation}
\overline{\Gamma}^{(\mathrm{1PI})(2\phi)}_{\mathrm{mix},k=k_{\mathrm{i}}} = m^2 \mathrlap{\;,}
\label{eq:CIGamma2phimix1PIFRG0DON}
\end{equation}
\begin{equation}
\overline{\Gamma}^{(\mathrm{1PI})(2\eta)}_{\mathrm{mix},k=k_{\mathrm{i}}} = 1 \mathrlap{\;,}
\label{eq:CIGamma2etamix1PIFRG0DON}
\end{equation}
\begin{equation}
\overline{\Gamma}^{(\mathrm{1PI})(2\phi,1\eta)}_{\mathrm{mix},k=k_{\mathrm{i}}} = i\sqrt{\frac{\lambda}{3}} \mathrlap{\;,}
\label{eq:CIGamma2phi1etamix1PIFRG0DON}
\end{equation}
\begin{equation}
\overline{\Gamma}^{(\mathrm{1PI})(3\eta)}_{\mathrm{mix},k=k_{\mathrm{i}}} = 0 \mathrlap{\;,}
\label{eq:CIGamma3etamix1PIFRG0DON}
\end{equation}
\begin{equation}
\overline{\Gamma}^{(\mathrm{1PI})(m\phi,n\eta)}_{\mathrm{mix},k=k_{\mathrm{i}}} = 0 \mathrlap{\quad \forall \ m+n \geq 4 \;.}
\label{eq:CIGammaOrder4ormoremix1PIFRG0DON}
\end{equation}
The truncation of the infinite set of differential equations containing e.g.~\eqref{eq:DiffEqGam00N1mixed1PIFRG0DON} to~\eqref{eq:DiffEqGam02N1mixed1PIFRG0DON} (at $N=1$ and $m^{2}>0$) or~\eqref{eq:DiffEqGam00N2mixed1PIFRG0DON} to~\eqref{eq:DiffEqGam02N2mixed1PIFRG0DON} (at $N=2$ and $m^{2}>0$) is now imposed by:
\begin{equation}
\overline{\Gamma}^{(\mathrm{1PI})(n\phi,m\eta)}_{\mathrm{mix},k} = \overline{\Gamma}^{(\mathrm{1PI})(n\phi,m\eta)}_{\mathrm{mix},k=k_{\mathrm{i}}} \quad \forall k,~ \forall \ n + m > N_{\mathrm{max}} \mathrlap{\;.}
\label{eq:mixed1PIFRGtruncation0DON}
\end{equation}
As explained below~\eqref{eq:pure1PIFRGtruncation0DON} for~\eqref{eq:CIGamma1PIFRG0DON}, the logarithm in~\eqref{eq:CIGammamix1PIFRG0DON} was only introduced to shift $E_{\mathrm{gs}}$, which is deduced in the present case from:
\begin{equation}
E^\text{1PI-FRG;mix}_{\mathrm{gs}} = \overline{\Gamma}^{(\mathrm{1PI})}_{\mathrm{mix},k=k_{\mathrm{f}}} \;,
\end{equation}
and the gs density follows from:
\begin{equation}
\rho^\text{1PI-FRG;mix}_{\mathrm{gs}} = \frac{1}{N} \sum_{a=1}^{N} \left(\overline{\Gamma}^{(\mathrm{1PI})(2\phi)}_{\mathrm{mix},k=k_{\mathrm{f}}}\right)_{aa}^{-1} = \left(\overline{\Gamma}^{(\mathrm{1PI})(2\phi)}_{\mathrm{mix},k=k_{\mathrm{f}}}\right)^{-1} \;,
\end{equation}
where $\overline{\Gamma}^{(\mathrm{1PI})(2\phi)}_{\mathrm{mix},k}$ is introduced in the RHS using~\eqref{eq:DefGamma2m2posmixed1PIFRG0DON} at $n=0$. We will also exploit the cutoff function~\eqref{eq:choiceRkpure1PIFRG0DON} for both the original and auxiliary field sectors:
\begin{equation}
R_{k} = k^{-1} - 1 \;.
\label{eq:choiceRkmix1PIFRG0DON}
\end{equation}

\vspace{0.5cm}

\begin{figure}[!htb]
\captionsetup[subfigure]{labelformat=empty}
  \begin{center}
    \subfloat[]{
      \includegraphics[width=0.50\linewidth]{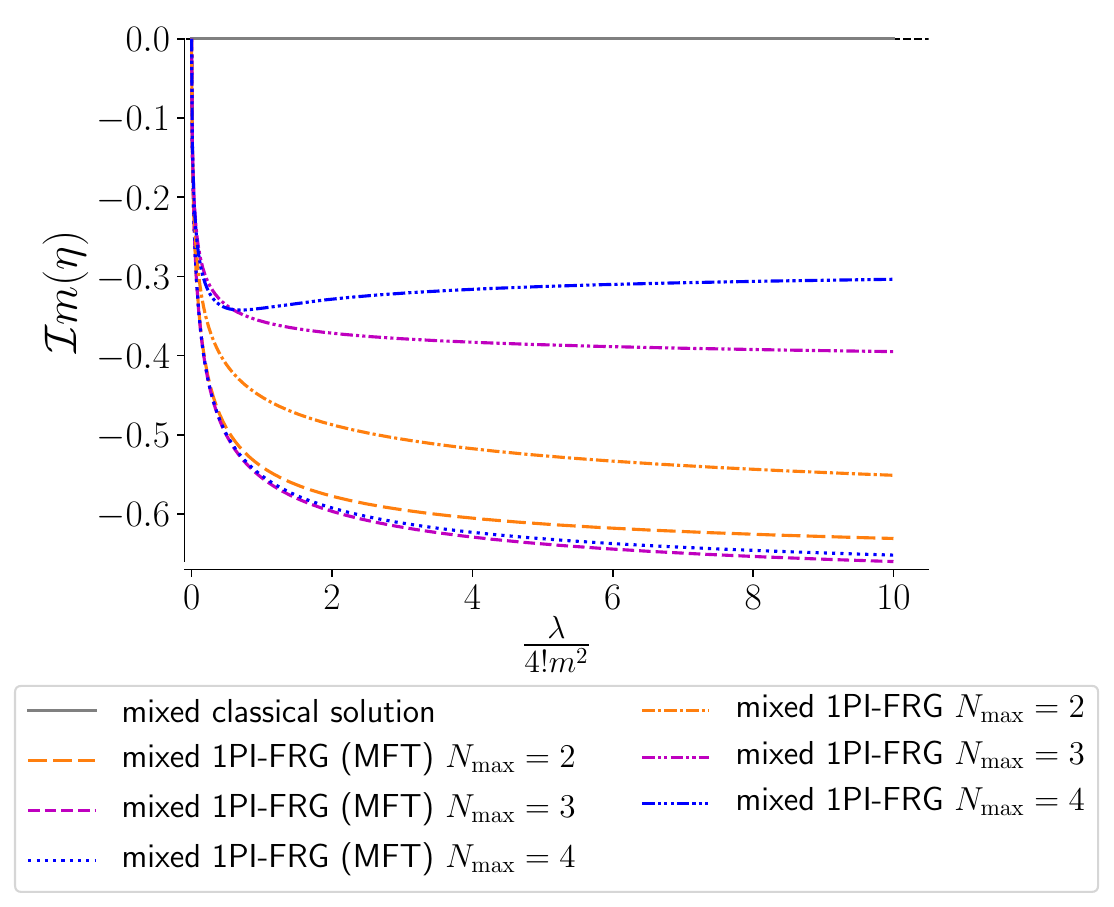}
                         }
    \subfloat[]{
      \includegraphics[width=0.50\linewidth]{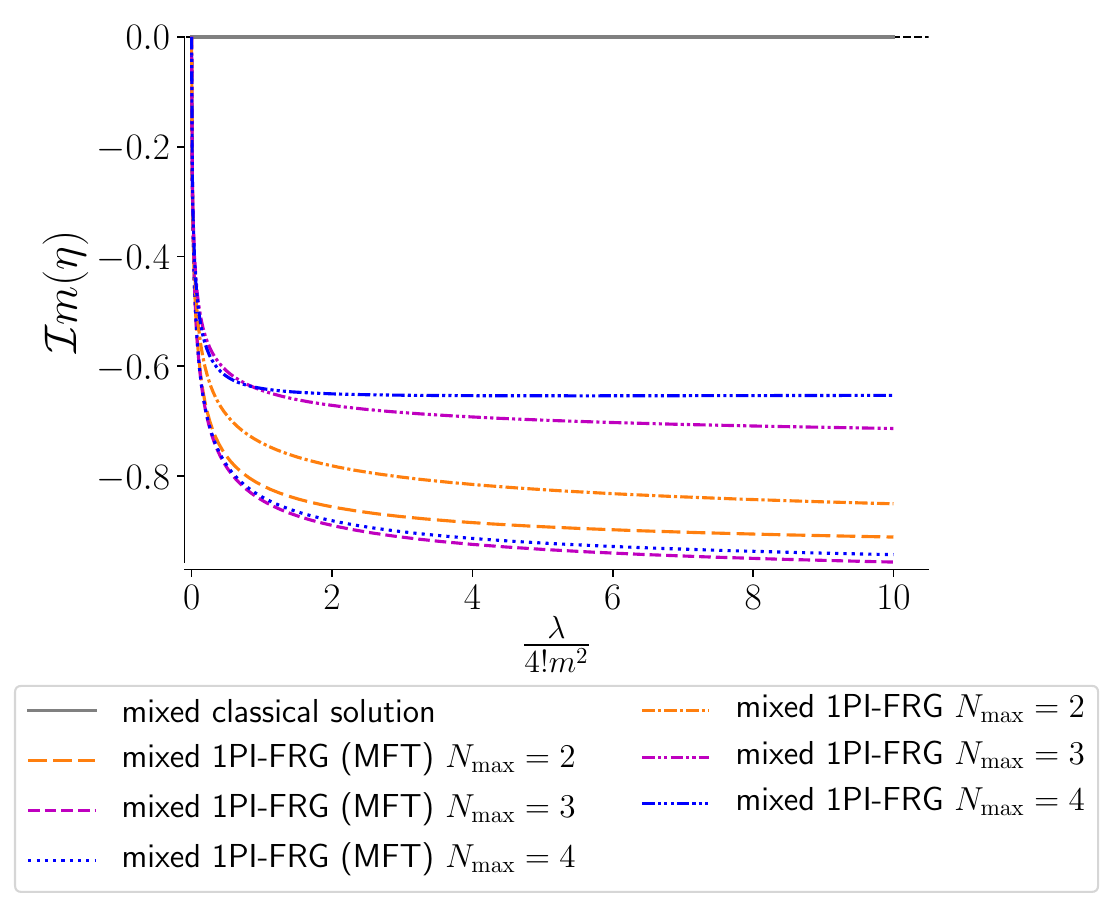}
                         }
\caption{Imaginary part of the 1-point correlation function of the auxiliary field in the framework of the mixed representation at $N=1$ (left) and $N=2$ (right) with $m^{2}=+1$ ($\mathcal{R}e(\lambda)\geq 0$ and $\mathcal{I}m(\lambda)=0$).}
\label{fig:Vevsmixed1PIFRGlambda}
  \end{center}
\end{figure}
\begin{figure}[!htb]
\captionsetup[subfigure]{labelformat=empty}
  \begin{center}
    \subfloat[]{
      \includegraphics[width=0.50\linewidth]{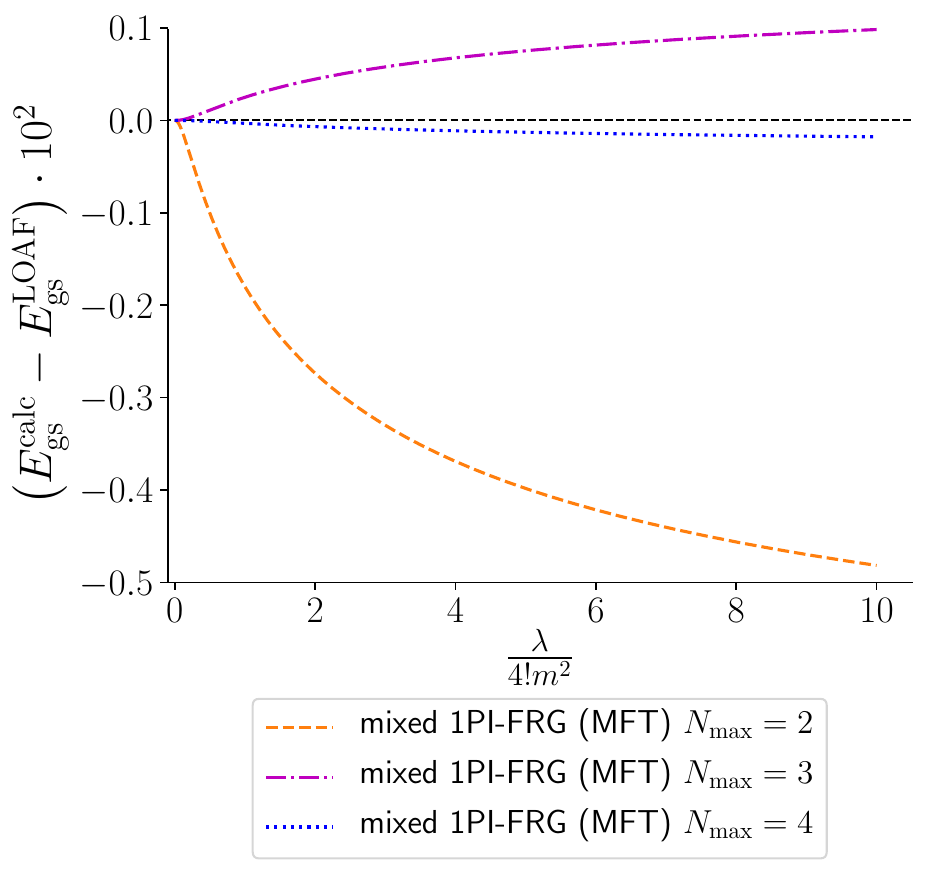}
                         }
    \subfloat[]{
      \includegraphics[width=0.50\linewidth]{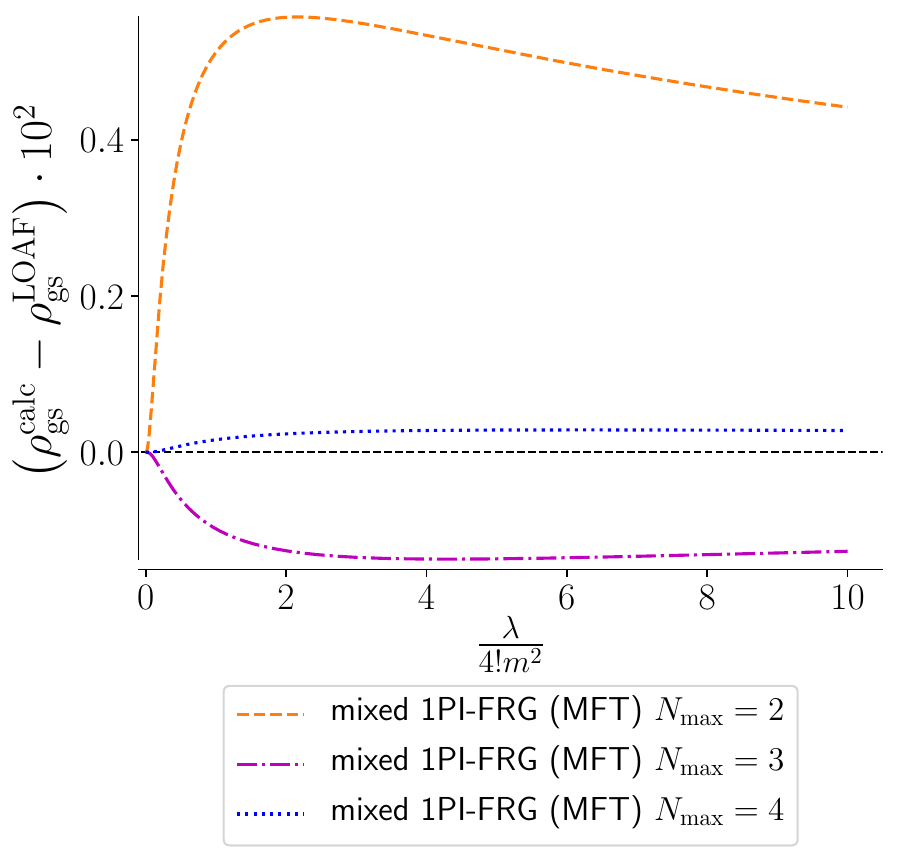}
                         }
\caption{Difference between the calculated gs energy $E_{\mathrm{gs}}^{\mathrm{calc}}$ (left) or density $\rho_{\mathrm{gs}}^{\mathrm{calc}}$ (right) and the corresponding LOAF approximation results $E_{\mathrm{gs}}^{\mathrm{LOAF}}$ or $\rho_{\mathrm{gs}}^{\mathrm{LOAF}}$ at $m^{2}=+1$ and $N=1$ ($\mathcal{R}e(\lambda)\geq 0$ and $\mathcal{I}m(\lambda)=0$).}
\label{fig:1PIFRGMFTvsLOAFlambdaN1}
  \end{center}
\end{figure}

Let us first concentrate our discussion on the regime with $m^{2}>0$. The corresponding mixed 1PI-FRG results are obtained by solving the equation system comprised of~\eqref{eq:DiffEqGam00N1mixed1PIFRG0DON} to~\eqref{eq:DiffEqGam02N1mixed1PIFRG0DON} for $N=1$ and of~\eqref{eq:DiffEqGam00N2mixed1PIFRG0DON} to~\eqref{eq:DiffEqGam02N2mixed1PIFRG0DON} for $N=2$ (together with additional differential equations given in appendix~\ref{sec:VertexExpansionAppMixed1PIFRG0DON} for $2 < N_{\mathrm{max}} \leq 4$), with the initial conditions~\eqref{eq:CIvevmix1PIFRG0DON} to~\eqref{eq:CIGammaOrder4ormoremix1PIFRG0DON} and the cutoff function~\eqref{eq:choiceRkmix1PIFRG0DON}. Without the MFT approximation, they exhibit a distinct convergence towards the exact solution, as can be seen in figs.~\ref{fig:pureVsmixed1PIFRGlambdaN1} and~\ref{fig:pureVsmixed1PIFRGlambdaN2} for both $E_{\mathrm{gs}}$ and $\rho_{\mathrm{gs}}$ with $N=1$ and $2$. More specifically, these figures show that the mixed 1PI-FRG outperforms the original 1PI-FRG at $N_{\mathrm{max}}=2$ and $4$ (we note however the specific case of the determination of $E_{\mathrm{gs}}$ at $N=1$ and $N_{\mathrm{max}}=2$ reported in fig.~\ref{fig:pureVsmixed1PIFRGlambdaN1}, where the original 1PI-FRG outperforms its mixed counterpart). The superiority of the mixed 1PI-FRG as compared to the original one for a given $N_{\mathrm{max}}$ can notably be attributed to the 1-point correlation function of the auxiliary field taking non-trivial values, as illustrated by fig.~\ref{fig:Vevsmixed1PIFRGlambda}. This echoes very clearly our comparison between the original and mixed 2PI EAs in chapter~\ref{chap:DiagTechniques} where the 1-point correlation function of the auxiliary field was also put forward to explain the difference between the BVA and the original Hartree-Fock result.

\vspace{0.5cm}

\begin{figure}[!htb]
\captionsetup[subfigure]{labelformat=empty}
  \begin{center}
    \subfloat[]{
      \includegraphics[width=0.50\linewidth]{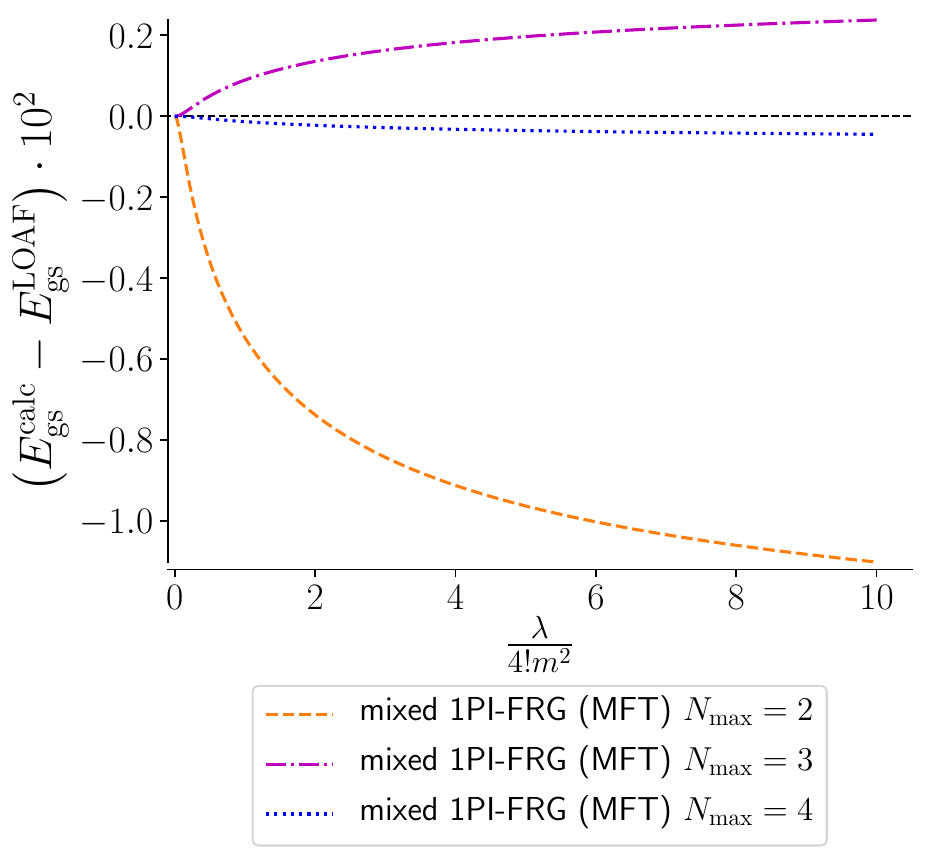}
                         }
    \subfloat[]{
      \includegraphics[width=0.50\linewidth]{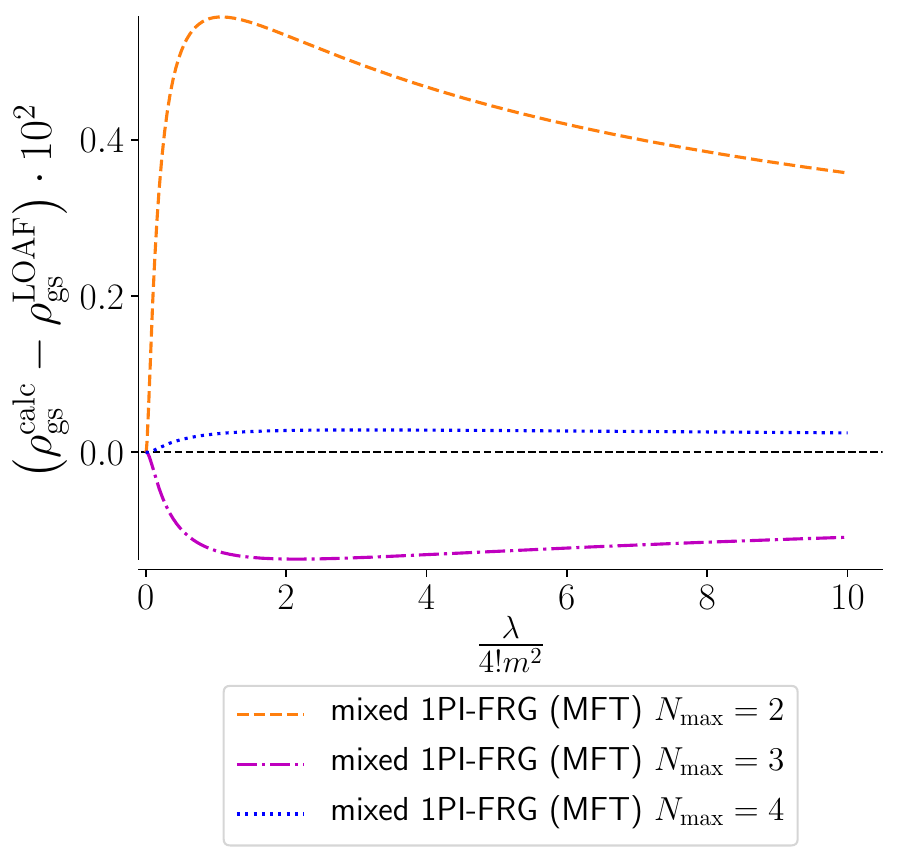}
                         }
\caption{Same as fig.~\ref{fig:1PIFRGMFTvsLOAFlambdaN1} with $N=2$ instead.}
\label{fig:1PIFRGMFTvsLOAFlambdaN2}
  \end{center}
\end{figure}
\begin{figure}[!htb]
\captionsetup[subfigure]{labelformat=empty}
  \begin{center}
    \subfloat[]{
      \includegraphics[width=0.50\linewidth]{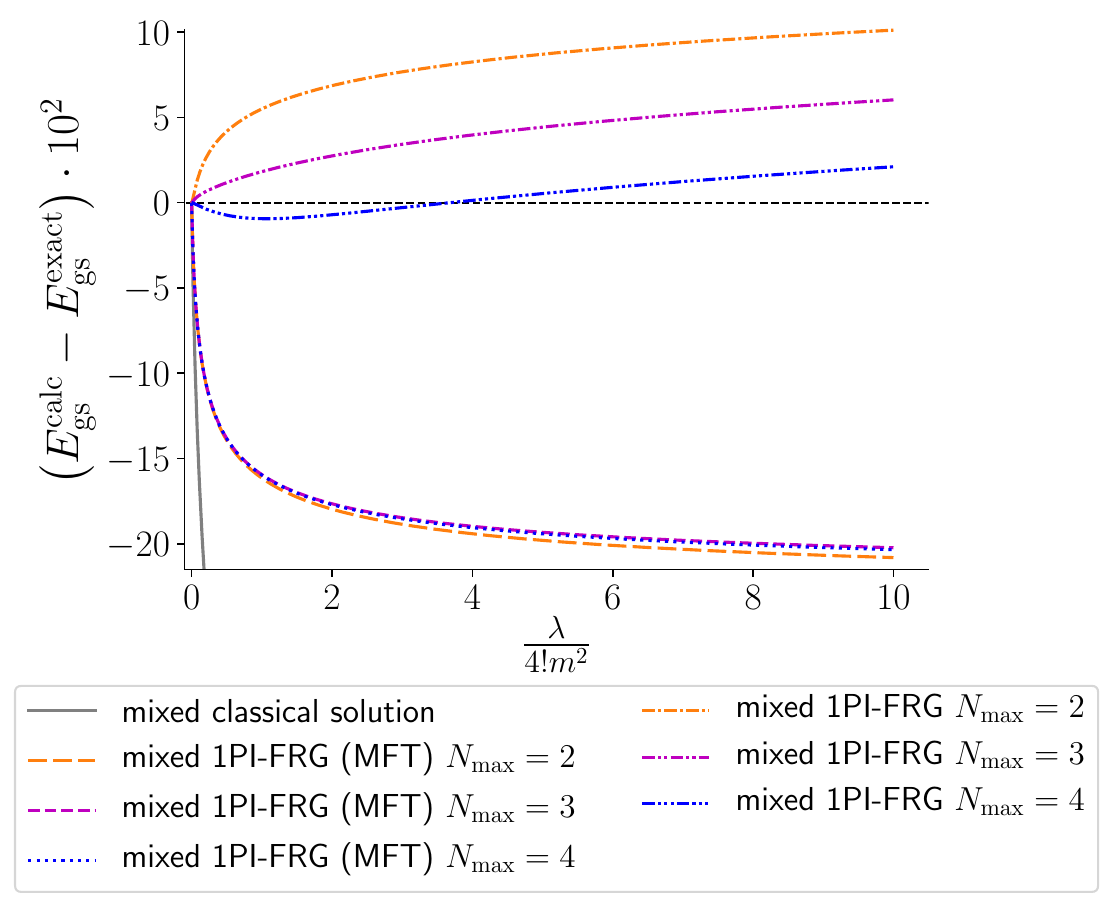}
                         }
    \subfloat[]{
      \includegraphics[width=0.50\linewidth]{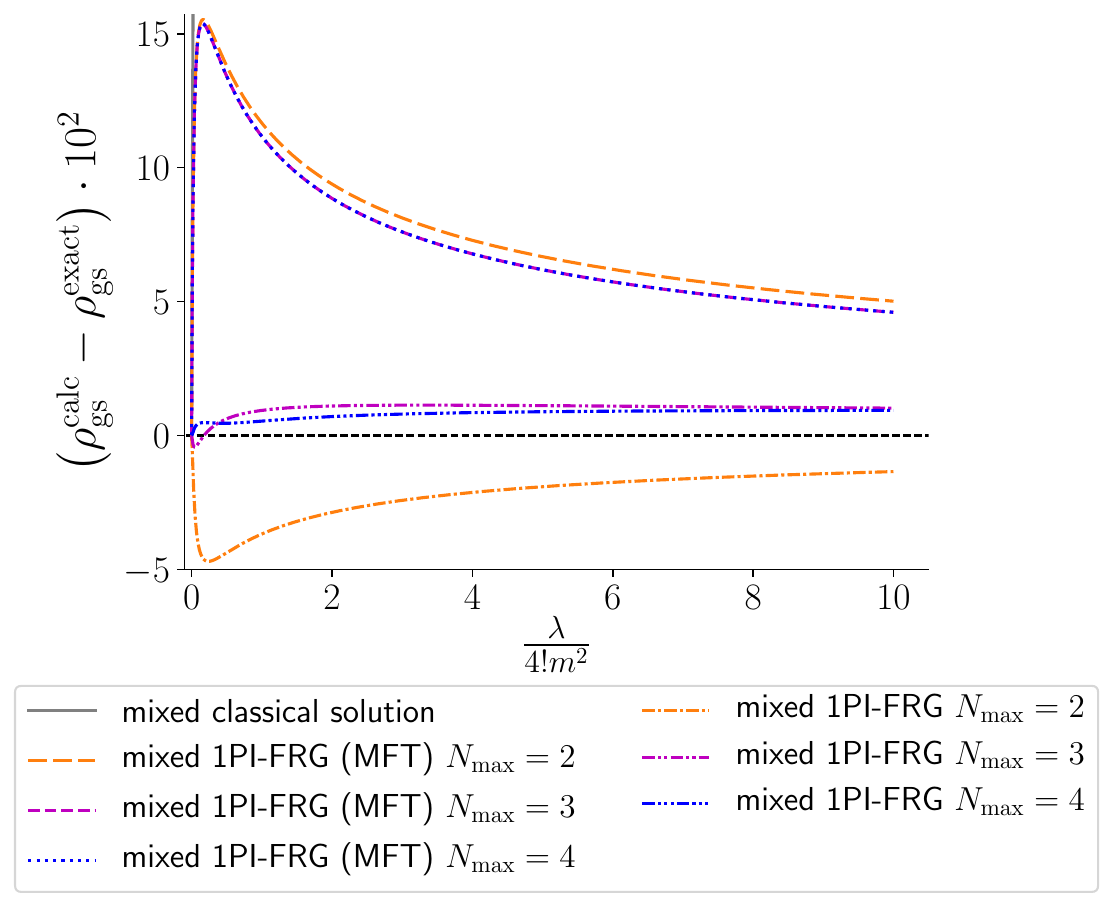}
                         }
\caption{Difference between the calculated gs energy $E_{\mathrm{gs}}^{\mathrm{calc}}$ (left) or density $\rho_{\mathrm{gs}}^{\mathrm{calc}}$ (right) and the corresponding exact solution $E_{\mathrm{gs}}^{\mathrm{exact}}$ or $\rho_{\mathrm{gs}}^{\mathrm{exact}}$ at $m^{2}=+1$ and $N=1$ ($\mathcal{R}e(\lambda)\geq 0$ and $\mathcal{I}m(\lambda)=0$).}
\label{fig:mixed1PIFRGlambdaN1}
  \end{center}
\end{figure}

We have also implemented the MFT by setting the propagator $\overline{G}^{(\eta)}_{k}$ equal to zero for all $k$. This amounts to setting the mass of the bosonic field $\widetilde{\sigma}$ (or equivalently the associated cutoff function $R^{(\eta)}_{k}$ introduced in~\eqref{eq:mathcalRkmixed1PIFRG0DON}) to infinity. This completely freezes the fluctuations of this field. In other words, the MFT can not capture radiative corrections associated with the auxiliary field. This notably excludes all contributions beyond the leading order of the collective LE. Therefore, the MFT can not, by construction, outperform the leading order of the collective LE, which coincides with the LOAF approximation as explained in chapter~\ref{chap:DiagTechniques}. This means that the MFT should tend to the LOAF approximation as the truncation order $N_{\mathrm{max}}$ increases, which is illustrated by figs.~\ref{fig:1PIFRGMFTvsLOAFlambdaN1} and~\ref{fig:1PIFRGMFTvsLOAFlambdaN2}.

\vspace{0.5cm}

\begin{figure}[!htb]
\captionsetup[subfigure]{labelformat=empty}
  \begin{center}
    \subfloat[]{
      \includegraphics[width=0.50\linewidth]{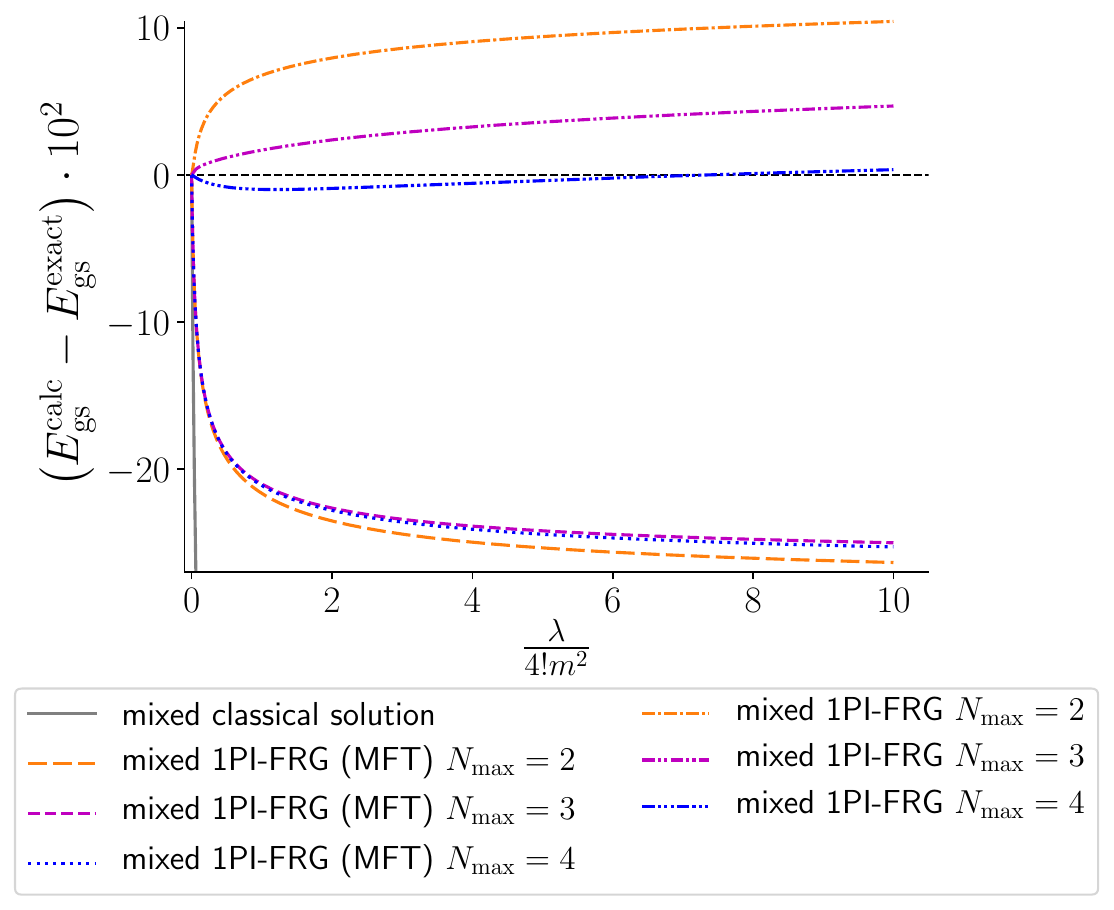}
                         }
    \subfloat[]{
      \includegraphics[width=0.50\linewidth]{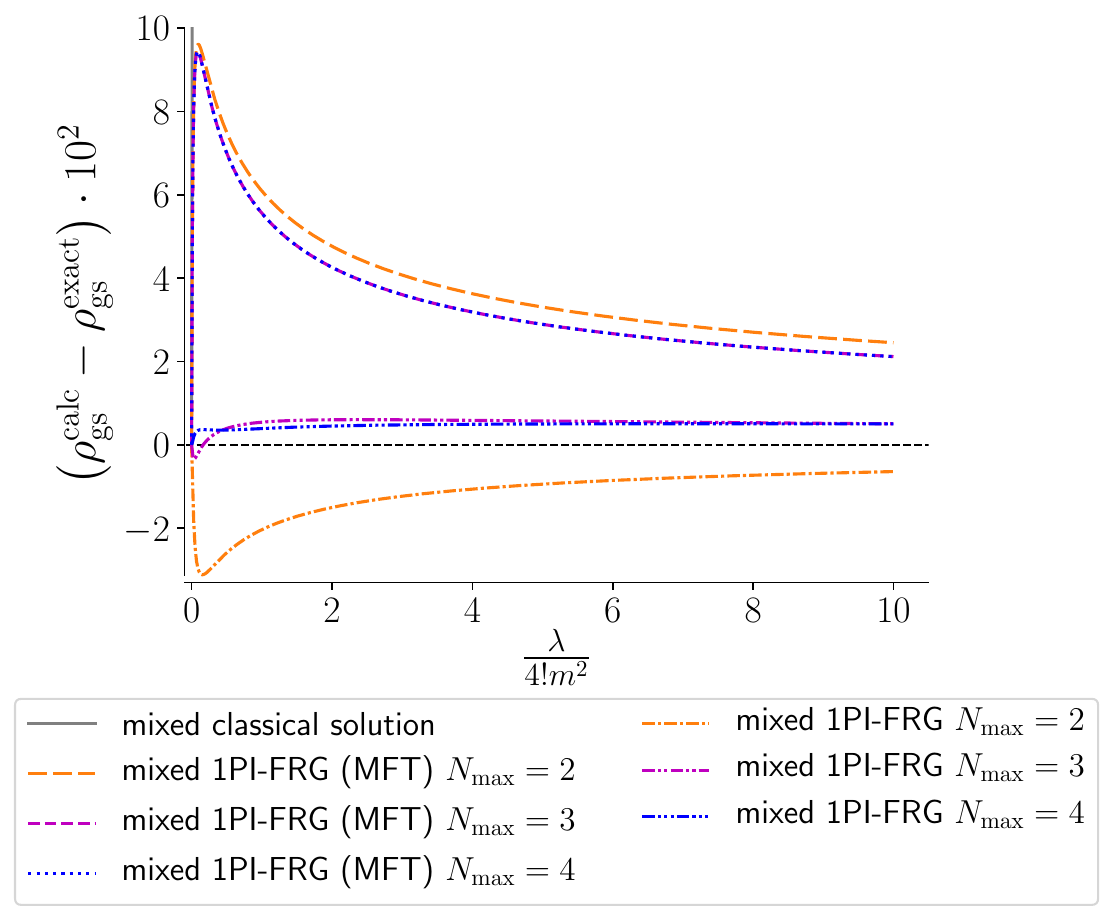}
                         }
\caption{Same as fig.~\ref{fig:mixed1PIFRGlambdaN1} with $N=2$ instead.}
\label{fig:mixed1PIFRGlambdaN2}
  \end{center}
\end{figure}

Finally, although figs.~\ref{fig:mixed1PIFRGlambdaN1} and~\ref{fig:mixed1PIFRGlambdaN2} show that the approximation underlying MFT induces a significant loss in the accuracy of mixed 1PI-FRG results, they also illustrate that its efficiency is not affected in the strongly-coupled regime: the MFT can therefore be considered as a first level of non-perturbative approximations.

\vspace{0.5cm}

Regarding the phase with $m^{2}<0$, we encounter the same limitation as in the original theory: the set of differential equations resulting from the vertex expansion procedure applied to~\eqref{eq:WetterichEqmixed1PIFRG0DON} is too stiff to be solved from $k_{\mathrm{i}}=0$ to $k_{\mathrm{f}}=1$ (still using the $\mathtt{NDSolve}$ function of $\mathtt{Mathematica~12.1}$). We will therefore turn to the collective representation as our last attempt to describe the broken-symmetry phase with the 1PI-FRG.

\subsubsection{Collective 1PI functional renormalization group}

In (0+0)-D and for the collective theory, the Wetterich equation reduces to:
\begin{equation}
\dot{\Gamma}^{(\mathrm{1PI})}_{\mathrm{col},k}\Big(\vec{\phi}\Big) = \frac{1}{2} \dot{R}_{k} D_{k}(\eta) \;,
\label{eq:WetterichEqbos1PIFRG0DON}
\end{equation}
with
\begin{equation}
D^{-1}_{k}(\eta) \equiv \Gamma^{(\mathrm{1PI})(2)}_{\mathrm{col},k}(\eta)+R_{k} \;.
\label{eq:DefGkbos1PIFRG0DON}
\end{equation}
Hence, the collective situation is very similar to the original one (based on~\eqref{eq:WetterichEqpure1PIFRG0DON}) with $N=1$. The output of the vertex expansion procedure applied to~\eqref{eq:WetterichEqbos1PIFRG0DON} can therefore be directly deduced from the set of differential equations presented in~\eqref{eq:FlowEqGam0m2neg1PIFRG0DON} to~\eqref{eq:FlowEqGam4m2neg1PIFRG0DON}. Up to $N_{\mathrm{max}}=4$, this gives us:
\begin{equation}
\dot{\overline{\Gamma}}^{(\mathrm{1PI})}_{\mathrm{col},k} = \frac{1}{2} \dot{R}_{k} \left( \overline{D}_{k} - \overline{D}^{(0)}_{k} \right)\;,
\label{eq:FlowEqGam0bos1PIFRG0DON}
\end{equation}
\begin{equation}
\dot{\overline{\eta}}_{k}=\frac{1}{2\overline{\Gamma}^{(\mathrm{1PI})(2)}_{\mathrm{col},k}}\dot{R}_{k}\overline{D}_{k}^{2}\overline{\Gamma}^{(\mathrm{1PI})(3)}_{\mathrm{col},k} \;,
\label{eq:FlowEqPhibos1PIFRG0DON}
\end{equation}
\begin{equation}
\dot{\overline{\Gamma}}^{(\mathrm{1PI})(2)}_{\mathrm{col},k}= \dot{\overline{\eta}}_{k}\overline{\Gamma}^{(\mathrm{1PI})(3)}_{\mathrm{col},k} + \dot{R}_{k}\overline{D}_{k}^{3}\left(\overline{\Gamma}^{(\mathrm{1PI})(3)}_{\mathrm{col},k}\right)^{2}-\frac{1}{2}\dot{R}_{k}\overline{D}_{k}^{2}\overline{\Gamma}^{(\mathrm{1PI})(4)}_{\mathrm{col},k}\;,
\label{eq:FlowEqGam2bos1PIFRG0DON}
\end{equation}
\begin{equation}
\dot{\overline{\Gamma}}^{(\mathrm{1PI})(3)}_{\mathrm{col},k}=\dot{\overline{\eta}}_{k}\overline{\Gamma}^{(\mathrm{1PI})(4)}_{\mathrm{col},k}-3\dot{R}_{k}\overline{D}_{k}^{4}\left(\overline{\Gamma}^{(\mathrm{1PI})(3)}_{\mathrm{col},k}\right)^{3}+3\dot{R}_{k}\overline{D}_{k}^{3}\overline{\Gamma}^{(\mathrm{1PI})(3)}_{\mathrm{col},k}\overline{\Gamma}^{(\mathrm{1PI})(4)}_{\mathrm{col},k}-\frac{1}{2}\dot{R}_{k}\overline{D}_{k}^{2}\overline{\Gamma}^{(\mathrm{1PI})(5)}_{\mathrm{col},k}\;,
\label{eq:FlowEqGam3bos1PIFRG0DON}
\end{equation}
\begin{equation}
\begin{split}
\dot{\overline{\Gamma}}^{(\mathrm{1PI})(4)}_{\mathrm{col},k} = & \ \dot{\overline{\eta}}_{k}\overline{\Gamma}^{(\mathrm{1PI})(5)}_{\mathrm{col},k}+12\dot{R}_{k}\overline{D}_{k}^{5}\left(\overline{\Gamma}^{(\mathrm{1PI})(3)}_{\mathrm{col},k}\right)^{4}-18\dot{R}_{k}\overline{D}_{k}^{4}\left(\overline{\Gamma}^{(\mathrm{1PI})(3)}_{\mathrm{col},k}\right)^{2}\overline{\Gamma}^{(\mathrm{1PI})(4)}_{\mathrm{col},k} \\
& + 4\dot{R}_{k}\overline{D}^{3}_{k}\overline{\Gamma}^{(\mathrm{1PI})(3)}_{\mathrm{col},k}\overline{\Gamma}^{(\mathrm{1PI})(5)}_{\mathrm{col},k} + 3\dot{R}_{k}\overline{D}^{3}_{k}\left(\overline{\Gamma}^{(\mathrm{1PI})(4)}_{\mathrm{col},k}\right)^{2}-\frac{1}{2}\dot{R}_{k}\overline{D}^{2}_{k}\overline{\Gamma}^{(\mathrm{1PI})(6)}_{\mathrm{col},k}\;,
\label{eq:FlowEqGam4bos1PIFRG0DON}
\end{split}
\end{equation}
with
\begin{equation}
\overline{\Gamma}^{(\mathrm{1PI})}_{\mathrm{col},k} \equiv \Gamma^{(\mathrm{1PI})}_{\mathrm{col},k}(\eta=\overline{\eta}_{k}) \mathrlap{\quad \forall k \;,}
\end{equation}
\begin{equation}
\overline{\Gamma}^{(\mathrm{1PI})(n)}_{\mathrm{col},k} \equiv \left.\frac{\partial^{n} \Gamma^{(\mathrm{1PI})}_{\mathrm{col},k}(\eta)}{\partial \eta^{n}} \right|_{\eta=\overline{\eta}_{k}} \mathrlap{\quad \forall k \;,}
\end{equation}
and
\begin{equation}
\overline{D}^{ \ -1}_{k} \equiv \overline{\Gamma}^{(\mathrm{1PI})(2)}_{\mathrm{col},k} + R_{k} \;,
\label{eq:propagatorbos1PIFRG0DON}
\end{equation}
\begin{equation}
\left(\overline{D}^{(0)}_{k}\right)^{-1} \equiv \overline{\Gamma}^{(\mathrm{1PI})(2)}_{\mathrm{col},k=k_{\mathrm{i}}} + R_{k} \;,
\label{eq:D0bos1PIFRG0DON}
\end{equation}
where $\overline{D}^{(0)}_{k}$ plays the same role as $\overline{G}^{(0)}_{k}$ in the original case (see the explanation below~\eqref{eq:FlowEqGam4m2neg1PIFRG0DON} for more details). The corresponding initial conditions are:
\begin{equation}
\overline{\eta}_{k=k_{\mathrm{i}}} = \overline{\sigma}_{\mathrm{cl}} = i\left(\frac{\sqrt{3}m^{2}-\sqrt{3m^{4}+2N\lambda}}{2\sqrt{\lambda}}\right) \mathrlap{\;,}
\label{eq:CIvevbos1PIFRG0DON}
\end{equation}
\begin{equation}
\overline{\Gamma}^{(\mathrm{1PI})}_{\mathrm{col},k=k_{\mathrm{i}}} = S_{\mathrm{col}}\big(\widetilde{\sigma}=\overline{\eta}_{k=k_{\mathrm{i}}}\big) = \frac{1}{2}\left(\overline{\eta}_{k=k_{\mathrm{i}}}\right)^{2} - \frac{N}{2} \ln\hspace{-0.05cm}\left(\frac{2\pi}{m^{2}+i\sqrt{\frac{\lambda}{3}}\overline{\eta}_{k=k_{\mathrm{i}}}}\right) \mathrlap{\;,}
\label{eq:CIGammabos1PIFRG0DON}
\end{equation}
\begin{equation}
\hspace{0.5cm} \overline{\Gamma}^{(\mathrm{1PI})(n)}_{\mathrm{col},k=k_{\mathrm{i}}} = \left.\frac{\partial^{n}S_{\mathrm{col}}\big(\widetilde{\sigma}\big)}{\partial\widetilde{\sigma}^{n}}\right|_{\widetilde{\sigma}=\overline{\eta}_{k=k_{\mathrm{i}}}} = \delta_{n 2} + (-1)^{n+1} \frac{N}{2} \left(n-1\right)! \left(\frac{i\sqrt{\frac{\lambda}{3}}}{m^{2}+i\sqrt{\frac{\lambda}{3}}\overline{\eta}_{k=k_{\mathrm{i}}}}\right)^{n} \quad \forall n \geq 2\;.
\label{eq:CIGammanbos1PIFRG0DON}
\end{equation}
In addition, the infinite tower of differential equations including~\eqref{eq:FlowEqGam0bos1PIFRG0DON} to~\eqref{eq:FlowEqGam4bos1PIFRG0DON} is truncated by imposing:
\begin{equation}
\overline{\Gamma}^{(\mathrm{1PI})(n)}_{\mathrm{col},k} = \overline{\Gamma}^{(\mathrm{1PI})(n)}_{\mathrm{col},k=k_{\mathrm{i}}} \mathrlap{\quad \forall k,~ \forall n > N_{\mathrm{max}} \;.}
\label{eq:bosonic1PIFRGtruncation0DON}
\end{equation}
Furthermore, the gs energy is deduced from:
\begin{equation}
E^\text{1PI-FRG;col}_{\mathrm{gs}} = \overline{\Gamma}^{(\mathrm{1PI})}_{\mathrm{col},k=k_{\mathrm{f}}} \;,
\label{eq:DeduceEgsbosonic1PIFRG0DON}
\end{equation}
as in~\eqref{eq:DeduceEgs1PIFRG0DON}, whereas the gs density is estimated by exploiting~\eqref{eq:DeducerhogsbosonicGeneral1PIFRG0DON} as follows:
\begin{equation}
\rho^\text{1PI-FRG;col}_{\mathrm{gs}} = \frac{i}{N} \sqrt{\frac{12}{\lambda}} \overline{\eta}_{k=k_{\mathrm{f}}} \;.
\label{eq:Deducerhogsbosonic1PIFRG0DON}
\end{equation}
Finally, we still use~\eqref{eq:choiceRkmix1PIFRG0DON} as cutoff function. Note that all analytical results given since equation~\eqref{eq:FlowEqGam0bos1PIFRG0DON} are valid for both $m^{2}<0$ and $m^{2}>0$. This follows from the fact that the $O(N)$ symmetry does not constrain the auxiliary field as it does for the original field via~\eqref{eq:DefGamma2m2pos1PIFRG0DON} to~\eqref{eq:DefGammaoddm2pos1PIFRG0DON}. Hence, as opposed to the original situation, there is no additional difficulty in treating the regime with $m^2<0$ instead of $m^2>0$ in the framework of the collective representation.

\vspace{0.5cm}

\begin{figure}[!htb]
\captionsetup[subfigure]{labelformat=empty}
  \begin{center}
    \subfloat[]{
      \includegraphics[width=0.50\linewidth]{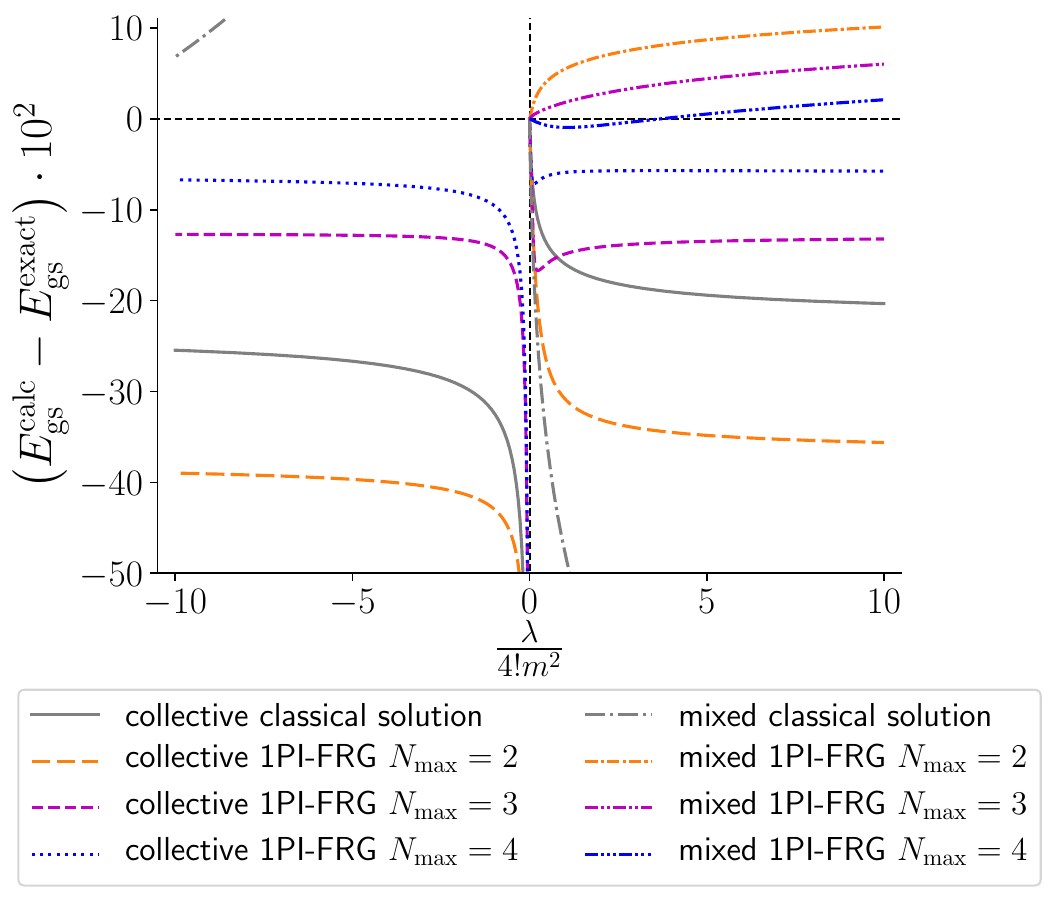}
                         }
    \subfloat[]{
      \includegraphics[width=0.50\linewidth]{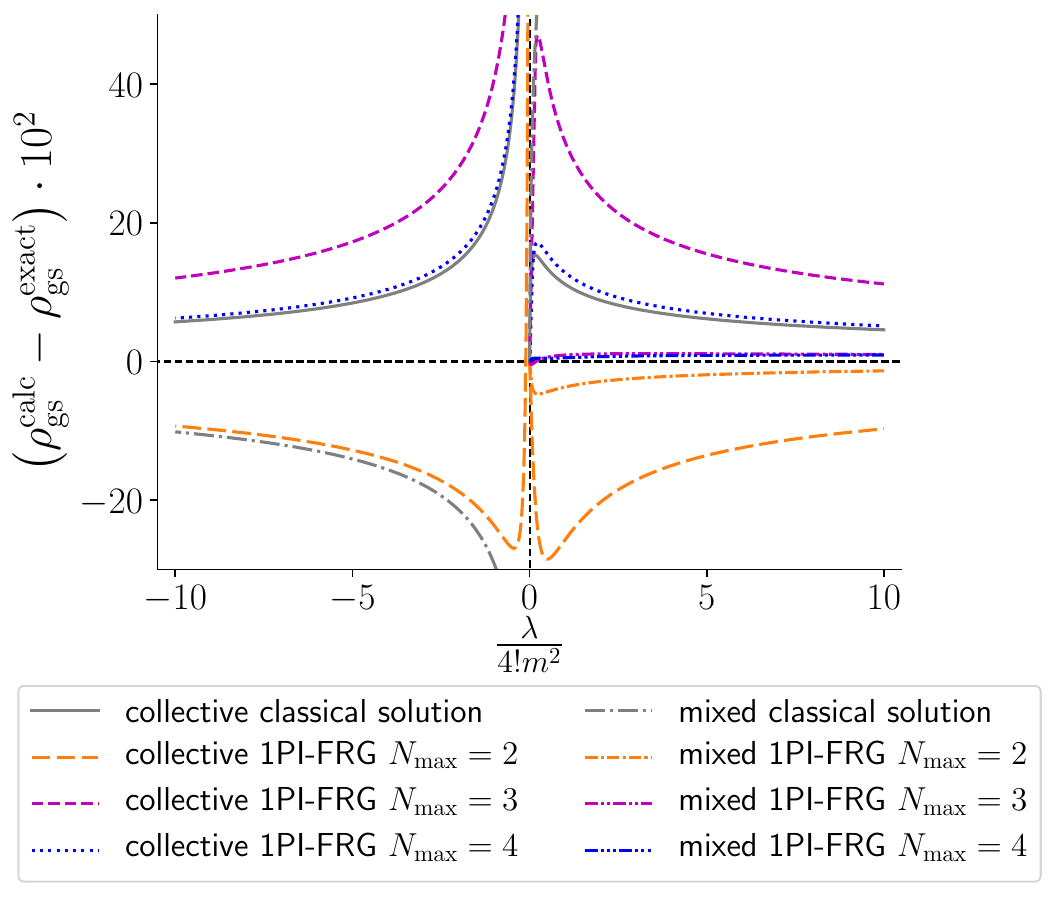}
                         }
\caption{Difference between the calculated gs energy $E_{\mathrm{gs}}^{\mathrm{calc}}$ (left) or density $\rho_{\mathrm{gs}}^{\mathrm{calc}}$ (right) and the corresponding exact solution $E_{\mathrm{gs}}^{\mathrm{exact}}$ or $\rho_{\mathrm{gs}}^{\mathrm{exact}}$ at $m^{2}=\pm 1$ and $N=1$ ($\mathcal{R}e(\lambda)\geq 0$ and $\mathcal{I}m(\lambda)=0$).}
\label{fig:bosonicVsmixed1PIFRGlambdaN1}
  \end{center}
\end{figure}
\begin{figure}[!htb]
\captionsetup[subfigure]{labelformat=empty}
  \begin{center}
    \subfloat[]{
      \includegraphics[width=0.50\linewidth]{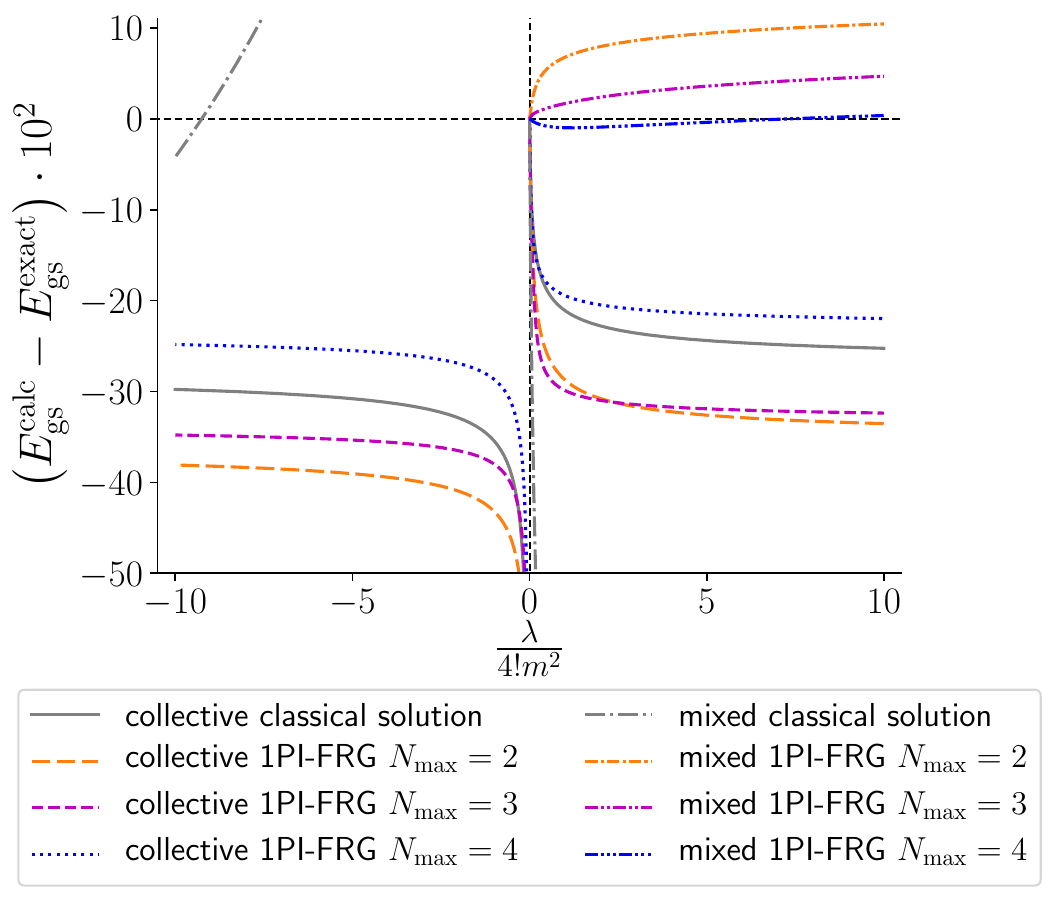}
                         }
    \subfloat[]{
      \includegraphics[width=0.50\linewidth]{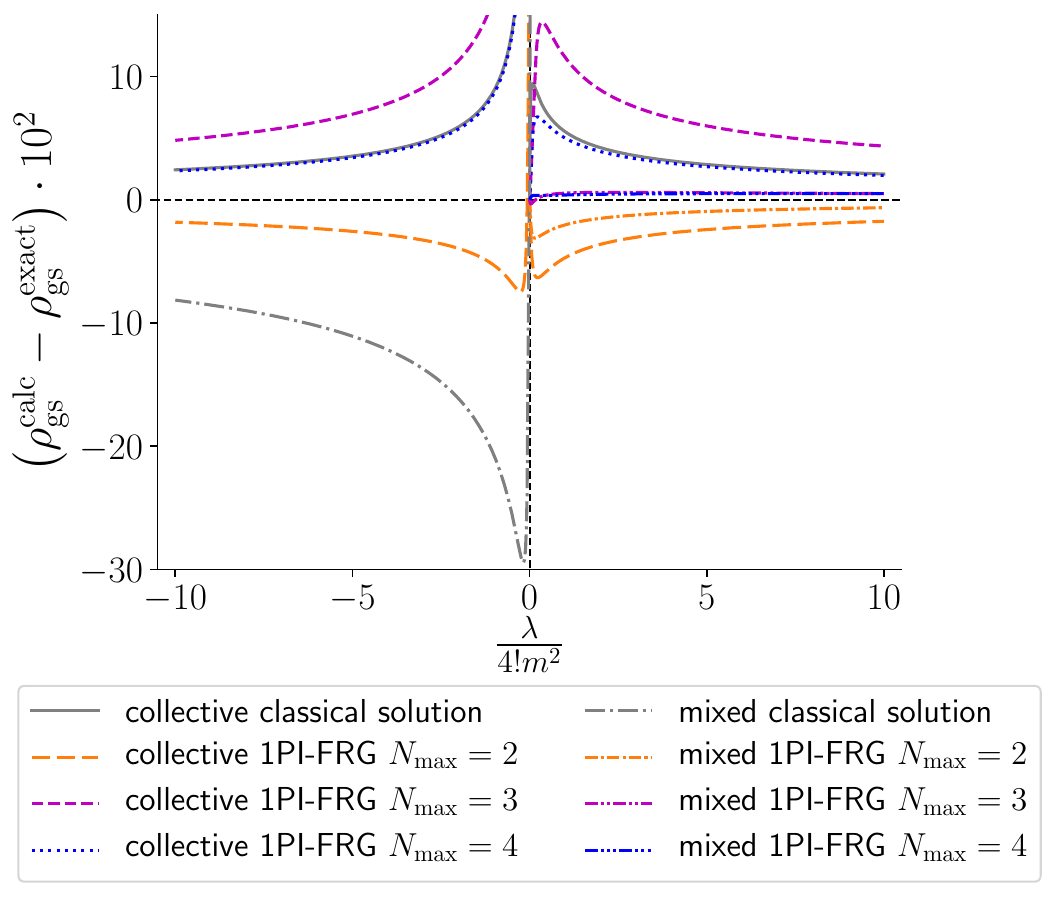}
                         }
\caption{Same as fig.~\ref{fig:bosonicVsmixed1PIFRGlambdaN1} with $N=2$ instead.}
\label{fig:bosonicVsmixed1PIFRGlambdaN2}
  \end{center}
\end{figure}

The collective 1PI-FRG procedure is carried out by solving the equation system given by~\eqref{eq:FlowEqGam0bos1PIFRG0DON} to~\eqref{eq:FlowEqGam4bos1PIFRG0DON}, with the initial conditions~\eqref{eq:CIvevbos1PIFRG0DON} to~\eqref{eq:CIGammanbos1PIFRG0DON} and the cutoff function \eqref{eq:choiceRkmix1PIFRG0DON}. From this, we are able to calculate $E_{\mathrm{gs}}$ and $\rho_{\mathrm{gs}}$ for all signs of $m^{2}$, which yields notably our first 1PI-FRG results for $m^{2}<0$. These results, shown in figs.~\ref{fig:bosonicVsmixed1PIFRGlambdaN1} and~\ref{fig:bosonicVsmixed1PIFRGlambdaN2}, are however disappointing in the sense that, for $m^{2}>0$, they are all outperformed by the original and mixed 1PI-FRG approaches at a given $N_{\mathrm{max}}$. Actually, for both $m^{2}<0$ and $m^{2}>0$, figs.~\ref{fig:bosonicVsmixed1PIFRGlambdaN1} and~\ref{fig:bosonicVsmixed1PIFRGlambdaN2} show that the collective 1PI-FRG must be pushed at least up to $N_{\mathrm{max}}=4$ to yield an accuracy below $10\%$, whereas this is already achieved by the mixed 1PI-FRG at $N_{\mathrm{max}}=2$ for $m^{2}>0$.

\vspace{0.5cm}

Note also that the connection between the collective 1PI-FRG and MFT is also clear. The starting point of the collective 1PI-FRG procedure coincides with the collective classical action, i.e. with the LOAF approximation towards which the MFT tends. Hence, the collective 1PI-FRG incorporates quantum corrections (which correspond to the bosonic fluctuations neglected by the MFT) on top of the LOAF approximation throughout the flow: it is therefore by construction more performing than the MFT version of the mixed 1PI-FRG.

\section{2PI functional renormalization group}
\label{sec:2PIFRG}
\subsection{State of play and general formalism}
\label{sec:2PIFRGstateofplay}

Formulations of FRG approaches for 2PI EAs have started since the 2000s~\cite{pol05,dup05,wet07,paw07}. For example, the detailed discussion of ref.~\cite{paw07} outlines the recipe to construct flow equations for any $n$PI EA (i.e. for $n$PI EAs with $n \geq 1$) by interpreting cutoff functions as shifts for the sources. Moreover, in ref.~\cite{pol05}, the ideas of the work of Alexandre, Polonyi and Sailer~\cite{ale01,ale02} deriving a generalization of the Callan-Symanzik equation~\cite{cal70,sym70,sym71} via the 1PI EA are exploited to determine a flow equation for the 2PI EA. The resulting approach, called internal space (IS) RG, has been compared with other RG methods, including the standard Callan-Symanzik RG, the Wegner-Houghton RG~\cite{weg73}, the LPA treatment of the 1PI-FRG, in the framework of a comparative study on a (0+1)-D $O(N)$-symmetric $\varphi^4$-theory~\cite{nag11}.

\vspace{0.5cm}

We will rather focus in this section on the more recent 2PI-FRG formalism put forward by Dupuis in refs.~\cite{dup05,dup14}, and more specifically on its different versions called C-flow~\cite{dup05}, U-flow~\cite{dup14} and CU-flow~\cite{dup14} that we will define further below. The U-flow and CU-flow can be both formulated via a modification of the Legendre transform defining the 2PI EA, in the same way as for the 1PI-FRG with the extra term $\Delta S_{k}[\phi]$ in~\eqref{eq:defGammakWettFRG}. In any case, the aim remains to obtain a starting point as convenient as possible for the flow: the presence of $\Delta S_{k}[\phi]$ in~\eqref{eq:defGammakWettFRG} enables us to start the 1PI-FRG procedure at the classical theory whereas the 2PI-FRG flow can begin at the result of self-consistent PT in this way\footnote{See more specifically the discussion below~\eqref{eq:2PIfrgmUflowExpressionBoldGDotNotApp} for a clarification on the link between modified Legendre transform and starting point of the 2PI-FRG flow.}. We save once again technical explanations for later discussions in this section but we just want to point out at this stage the connections between the 2PI-FRG \`{a} la Dupuis and the 1PI-FRG based on the Wetterich equation. In fact, Wetterich also developed a 2PI-FRG approach based on a modified Legendre transform as well~\cite{wet07}. However, as opposed to this work, Dupuis' 2PI-FRG is based on flow equations for the Luttinger-Ward functional and not for the 2PI EA itself, which significantly improves its convergence\footnote{This improvement will be thoroughly discussed below and illustrated via the numerical applications presented in section~\ref{sec:KSFRG0DON} on the 2PPI-FRG.}. Moreover, both Wetterich and Dupuis ignore the field dependence of the 2PI EA in their 2PI-FRG formulations, i.e. they consider $\Gamma^{(\mathrm{2PI})}[\phi=0,G]$, which is also referred to as bosonic EA~\cite{wet07}.

\vspace{0.5cm}

In particular, we can highlight the following appealing features of Dupuis' 2PI-FRG\footnote{In what follows, 2PI-FRG refers implicitly to the 2PI-FRG \`{a} la Dupuis developed in refs.~\cite{dup05,dup14}.}, thus stressing some advantages of this 2PI-FRG as compared to its 1PI counterpart:
\begin{itemize}
\item The simplified 2PI EA $\Gamma^{(\mathrm{2PI})}[\phi=0,G]$ does not depend on Grassmann variables. Hence, contrary to a 1PI EA depending only on Grassmann fields, expansions of $\Gamma^{(\mathrm{2PI})}[\phi=0,G]$ around \textbf{non-trivial minima} are possible since the propagator $G$ is a bosonic variable. This notably renders the 2PI-FRG well suited to grasp non-perturbative physics in fermionic systems, even without HSTs.
\item When the 2PI-FRG flow is designed to take results of self-consistent PT as inputs, the 2PI-FRG offers the possibility to \textbf{start the flow in a broken-symmetry phase}, which enables us to avoid phase transitions (and the associated problematic divergences encountered in the 1PI-FRG) during the flow.
\item Besides its convenient starting points, the 2PI-FRG is designed itself to avoid the undesirable divergences from which the 1PI-FRG suffers. This is simply because the quantities calculated during the flow are different: one calculates the 1PI vertices (i.e. derivatives of the 1PI EA) during the 1PI-FRG flow and the \textbf{2PI vertices} (i.e. derivatives of the Luttinger-Ward functional) during the 2PI-FRG flow. A concrete example is given in ref.~\cite{dup05} for an application of the C-flow implementation of the 2PI-FRG to the BCS theory: in this study, the entrance into the broken-symmetry phase during the flow is just signaled by a finite value of the anomalous self-energy and the divergences of certain response functions can only be noticed by solving \textit{a posteriori} the relevant Bethe-Salpeter equations with the calculated 2PI vertices.
\end{itemize}

\vspace{0.3cm}

In principle, scale-dependent bosonization could also be implemented in the framework of the 2PI-FRG, similarly to other 1PI-FRG studies discussed in section~\ref{sec:KeyAspectsFRG}. We can already point out an important reason to exploit such a bosonization procedure with the 2PI-FRG. It turns out that Bethe-Salpeter equation(s) must be solved at each step of the flow for most implementations of the U-flow and the CU-flow versions of the 2PI-FRG. The corresponding flow equations are thus quite demanding to solve numerically (and especially more demanding as compared to the corresponding 1PI-FRG equations). In that respect, it would be very interesting to freeze the evolution of the 2PI vertices that must be fed to this (these) Bethe-Salpeter equation(s) during the flow. In this situation, such an equation (such equations) must only be solved once and for all at the starting point of the flow, which would considerably lower the weight of the numerical procedure to implement, as we will discuss below.

\vspace{0.5cm}

It turns out that only a few applications of the 2PI-FRG have been carried out so far. Among these, we can mention the work of Rentrop, Jakobs and Meden on the (0+1)-D $\varphi^4$-theory~\cite{ren15} and on the (0+1)-D Anderson impurity model\footnote{Note that the work of refs.~\cite{ren15,ren16} echoes that of ref.~\cite{hed04} which treats the same models with the 1PI-FRG.}~\cite{ren16}. Note in addition the study of ref.~\cite{kug18} that discusses connections between the 2PI-FRG C-flow and the 1PI-FRG in the context of the Fermi-edge singularity, which is a model used to describe optical excitations in electronic systems. Notably, the 2PI-FRG has also been designed by Katanin so as to take the 2PI vertices calculated from DMFT as inputs~\cite{kat19}, thus developing a 2PI counterpart for the DMF$^2$RG discussed in section~\ref{sec:1PIFRG0DON}. Most importantly, the tower of differential equations resulting from this approach is tractable enough to tackle a (2+1)-D Hubbard model, as proven by the results presented in ref.~\cite{kat19}. Although the Hubbard model is most often taken as first playground to formulate 2PI-FRG approaches~\cite{wet07,dup14,kat19}, the work of ref.~\cite{kat19} is notably the first to present 2PI-FRG results for the self-energy of such a model, which are fairly close to the corresponding diagrammatic determinant Monte Carlo results~\cite{koz15}. Besides the 2PI-FRG study of the (3+1)\nobreakdash-D $\varphi^4$-theory reported in an article by Carrington and collaborators~\cite{car15}, ref.~\cite{kat19} shows to our knowledge the only 2PI-FRG application to a model with finite space dimensions, which illustrates that this promising 2PI approach has barely been beyond the stage of toy model applications.

\pagebreak

Let us then present the 2PI-FRG formalism. All implementations of this method investigated in the present study are based on the generating functional:
\begin{equation}
Z[K]=e^{W[K]}=\int\mathcal{D}\widetilde{\psi} \ e^{-S\big[\widetilde{\psi}\big] + \frac{1}{2}\int_{\alpha,\alpha'}\widetilde{\psi}_{\alpha}K_{\alpha\alpha'}\widetilde{\psi}_{\alpha'}} \;,
\label{eq:2PIFRGgeneratingFunc}
\end{equation}
where $\widetilde{\psi}$ is either a real bosonic field ($\zeta=+1$) or a real Grassmann field ($\zeta=-1$). The index $\alpha\equiv (a,x)$ used in~\eqref{eq:2PIFRGgeneratingFunc} combines this time an internal index $a$ (e.g. the color index for an $O(N)$ model) with $x\equiv (\boldsymbol{r}, \tau, m_{s}, c)$ including the space coordinate $\boldsymbol{r}$, the imaginary time $\tau$, the spin projection $m_{\sigma}$ and a charge index $c$ if necessary. Regarding the latter, note that, in the framework of the 2PI-FRG as presented here, $\widetilde{\psi}_{\alpha}$ is always mathematically treated as a real field with an extra index, instead of a complex one. This extra index is the charge index $c$ defined as follows\footnote{In other words, whether the relation $\widetilde{\psi}^{\dagger}=\widetilde{\psi}$ is satisfied or not, we will always use formulae pertaining to real fields in the framework of the 2PI-FRG, like Gaussian integral formulae (see appendix~\ref{ann:GaussianInt}).}:
\begin{equation}
\widetilde{\psi}_{\alpha}=\widetilde{\psi}_{a,x}=\left\{
\begin{array}{lll}
        \displaystyle{\widetilde{\psi}_{a,m_{s}}(\boldsymbol{r},\tau) \quad \mathrm{for}~c=-\;.} \\
        \\
        \displaystyle{\widetilde{\psi}^{\dagger}_{a,m_{s}}(\boldsymbol{r},\tau) \quad \mathrm{for}~c=+\;.}
    \end{array}
\right.
\label{eq:chargeIndexDefinition2PIFRG}
\end{equation}
In addition, we have exploited in~\eqref{eq:2PIFRGgeneratingFunc} the shorthand notation:
\begin{equation}
\int_{\alpha}\equiv\sum_{a} \int_{x} \equiv\sum_{a} \sum_{m_{s},c} \int^{\beta}_{0} d\tau \int d^{D-1}\boldsymbol{r}\;,
\label{eq:DefIntegralsFermionicIndices2PIFRG}
\end{equation}
assuming that the studied system lives in a $D$-dimensional spacetime. It will be also most convenient to group $\alpha$-indices by pairs via a bosonic index:
\begin{equation}
\gamma\equiv(\alpha,\alpha') \;.
\end{equation}
For instance, the connected correlation functions can be expressed in terms of such indices:
\begin{equation}
W^{(n)}_{\gamma_{1} \cdots \gamma_{n}}[K] \equiv \frac{\delta^{n}W[K]}{\delta K_{\gamma_{1}} \cdots \delta K_{\gamma_{n}}}=\frac{\delta^{n}W[K]}{\delta K_{\alpha_{1}\alpha'_{1}} \cdots \delta K_{\alpha_{n}\alpha'_{n}}} = \left\langle \widetilde{\psi}_{\alpha_{1}} \widetilde{\psi}_{\alpha'_{1}} \cdots \widetilde{\psi}_{\alpha_{n}} \widetilde{\psi}_{\alpha'_{n}} \right\rangle_{K} \;,
\label{eq:2PIFRGWnKbosonicIndices}
\end{equation}
which defines the connected propagator:
\begin{equation}
G_{\gamma} = W^{(1)}_{\gamma}[K] = \left\langle \widetilde{\psi}_{\alpha} \widetilde{\psi}_{\alpha'} \right\rangle_{K} \;,
\label{eq:definitionG2PIFRG}
\end{equation}
for $n=1$ (see appendix~\ref{ann:BosonicIndices2PI}), using an expectation value defined as:
\begin{equation}
\big\langle \cdots \big\rangle_{K} = \frac{1}{Z[K]} \int \mathcal{D}\widetilde{\psi} \ \cdots \ e^{-S\big[\widetilde{\psi}\big] + \frac{1}{2}\int_{\alpha,\alpha'}\widetilde{\psi}_{\alpha}K_{\alpha\alpha'}\widetilde{\psi}_{\alpha'}} \;.
\label{eq:2PIFRGKdependentExpectationValue}
\end{equation}
As the components of the source $K$ satisfy $K_{\alpha\alpha'} = \zeta K_{\alpha'\alpha}$, the correlation functions of~\eqref{eq:2PIFRGWnKbosonicIndices} possess the symmetry properties:
\begin{subequations}\label{eq:SymmetryW2PIFRG}
\begin{empheq}[left=\empheqlbrace]{align}
& W^{(n)}_{(\alpha_{1},\alpha_{1}')\cdots(\alpha_{i},\alpha_{i}')\cdots(\alpha_{n},\alpha_{n}')}[K]=\zeta W^{(n)}_{(\alpha_{1},\alpha_{1}')\cdots(\alpha_{i}',\alpha_{i})\cdots(\alpha_{n},\alpha_{n}')}[K] \;, \label{eq:SymmetryW2PIFRGUp}\\
\nonumber \\
& W^{(n)}_{\gamma_{1}\cdots\gamma_{n}}[K]=W^{(n)}_{\gamma_{P(1)}\cdots\gamma_{P(n)}}[K] \;, \label{eq:SymmetryW2PIFRGDown}
\end{empheq}
\end{subequations}
with $P$ denoting an arbitrary element of the permutation group of order $n$, and especially:
\begin{equation}
G_{\alpha\alpha'}=\zeta G_{\alpha'\alpha} \;,
\label{eq:SymmetryG2PIFRG}
\end{equation}
at $n=1$. With this bosonic index notation, the Legendre transform defining the 2PI EA under consideration reads:
\begin{equation}
\begin{split}
\Gamma^{(\mathrm{2PI})}[G] = & -W[K] + \frac{1}{2} \int_{\gamma} K_{\gamma} \frac{\delta W[K]}{\delta K_{\gamma}} \\
= & -W[K] + \mathrm{Tr}_{\gamma}(KG) \;,
\end{split}
\label{eq:LegendreTransform2PIeffActionWithTrace}
\end{equation}
where the second line was obtained using~\eqref{eq:definitionG2PIFRG} and the trace with respect to bosonic indices was introduced:
\begin{equation}
\mathrm{Tr}_{\gamma}(M) = \frac{1}{2} \int_{\gamma} M_{\gamma\gamma} = \frac{1}{2} \int_{\alpha,\alpha'} M_{(\alpha,\alpha')(\alpha,\alpha')}\;,
\label{eq:DefinitionTraceBosonicInd2PIFRG}
\end{equation}
with $M$ being an arbitrary bosonic matrix. In contrast, the trace (or supertrace) with respect to $\alpha$-indices will be denoted as $\mathrm{Tr}_{\alpha}$ in the whole section~\ref{sec:2PIFRG} (as well as in corresponding appendices) for the sake of clarity. Note also that, as in~\eqref{eq:DefinitionTraceBosonicInd2PIFRG}, an integration over bosonic indices amounts to integrating (or just summing) over all its constituent indices:
\begin{equation}
\int_{\gamma} \equiv \int_{\alpha,\alpha'} \;,
\end{equation}
where the integrals of the RHS are defined by~\eqref{eq:DefIntegralsFermionicIndices2PIFRG}. In what follows, we will use a DeWitt-like notation for the integration over bosonic indices. For $n$ arbitrary bosonic matrices $M_{m}$ (with $m=1,\dots,n$), it takes the following form:
\begin{equation}
M_{1,\gamma_{1}\hat{\gamma}_{1}} \cdots M_{n,\hat{\gamma}_{n-1}\gamma_{2}} = \frac{1}{2^{n-1}}\int_{\hat{\gamma}_{1},\cdots,\hat{\gamma}_{n-1}} M_{1,\gamma_{1}\hat{\gamma}_{1}} \cdots M_{n,\hat{\gamma}_{n-1}\gamma_{2}} \;,
\label{eq:DeWittNotation2PIFRG}
\end{equation}
where the hatted indices are all dummy and the non-hatted ones are all free by convention. The $1/2$ factors involved in~\eqref{eq:DefinitionTraceBosonicInd2PIFRG} and~\eqref{eq:DeWittNotation2PIFRG} are purely conventional but convenient as a result of the symmetry properties outlined in~\eqref{eq:SymmetryW2PIFRG} (see appendix~\ref{ann:BosonicIndices2PI}).

\vspace{0.5cm}

In the framework of the 2PI-FRG, it is also natural to consider the Luttinger-Ward functional $\Phi[G]$, which corresponds to the interaction part of the 2PI EA defined via~\eqref{eq:LegendreTransform2PIeffActionWithTrace}:
\begin{equation}
\Phi[G]\equiv\Gamma^{(\mathrm{2PI})}[G]-\Gamma^{(\mathrm{2PI})}_{0}[G] \;,
\label{eq:DefLWfunctional}
\end{equation}
where the free part of the 2PI EA is given by:
\begin{equation}
\Gamma_{0}^{(\mathrm{2PI})}[G] = -\frac{\zeta}{2} \mathrm{Tr}_{\alpha} \left[ \mathrm{ln}(G) \right] + \frac{\zeta}{2} \mathrm{Tr}_{\alpha}\left[GC^{-1}-\mathbb{I}\right]\;,
\label{eq:2PIFRGfreeGamma2PI}
\end{equation}
as can be derived via Gaussian functional integration (see appendix~\ref{ann:DysonEq}), with $\mathbb{I}$ denoting the identity with respect to $\alpha$-indices (i.e. $\mathbb{I}_{\alpha_{1}\alpha_{2}}=\delta_{\alpha_{1}\alpha_{2}}$) and $C$ being the free propagator, i.e.:
\begin{equation}
C_{\alpha\alpha'}^{-1} = \left. \frac{\delta^{2} S\big[\widetilde{\psi}\big]}{\delta\widetilde{\psi}_{\alpha}\delta\widetilde{\psi}_{\alpha'}} \right|_{\widetilde{\psi}=0} \;,
\end{equation}
in terms of the classical action $S$. Recall that the Luttinger-Ward functional is the sum of 2PI diagrams, with propagator lines corresponding to the full propagator $G$. The so-called 2PI vertices correspond to its derivatives, i.e. $\Phi_{\gamma_{1}\cdots\gamma_{n}}^{(n)}[G]\equiv\frac{\delta^{n}\Phi[G]}{\delta G_{\gamma_{1}} \cdots \delta G_{\gamma_{n}}}$. Note that, in order to determine gs energies, the thermodynamic potential:
\begin{equation}
\Omega[G]\equiv\frac{1}{\beta}\Gamma^{(\mathrm{2PI})}[G] \;,
\end{equation}
will also be considered.

\vspace{0.5cm}

In order to determine the physical configuration $\overline{G}_{\mathfrak{s}}$ of the propagator $G$ throughout the flow, we will consider Dyson equation. Note that the upper bars (as that in $\overline{G}_{\mathfrak{s}}$) label physical configurations as usual, i.e. a functional evaluated at vanishing source\footnote{For instance, the physical configurations of the 2PI EA under consideration and of the corresponding Luttinger-Ward functional respectively read $\overline{\Gamma}_{\mathfrak{s}}^{(\mathrm{2PI})}\equiv\Gamma^{(\mathrm{2PI})}\big[G=\overline{G}_{\mathfrak{s}}\big]$ and $\overline{\Phi}_{\mathfrak{s}}\equiv\Phi\big[G=\overline{G}_{\mathfrak{s}}\big]$.} (i.e. at $K_{\gamma}=0$ $\forall\gamma$ here). Furthermore, in the present framework, all these physical quantities are subject to evolve during the flow and they therefore all possess a subscript $\mathfrak{s}$ to stress that point, $\mathfrak{s}$ denoting the flow parameter for all 2PI-FRG approaches\footnote{We will see in our forthcoming discussions that the flow parameter $\mathfrak{s}$ coincides with a momentum scale in certain (but not all) versions of the 2PI-FRG implementing a Wilsonian momentum-shell approach.} as well as for all 2PPI-FRG implementations discussed in subsequent sections. Turning back to Dyson equation, we point out that the latter follows directly from the above definition of the Luttinger-Ward functional. In order to illustrate this, we can differentiate the definition of the 2PI EA given by~\eqref{eq:LegendreTransform2PIeffActionWithTrace} with respect to the source $K$, which leads to (see appendix~\ref{ann:BosonicIndices2PI}):
\begin{equation}
\frac{\delta\Gamma^{(\mathrm{2PI})}[G]}{\delta G_{\gamma}}=K_{\gamma} \;.
\label{eq:Extremize2PIEA2PIfrgNonZeroK}
\end{equation}
This equality can be shown to be equivalent to Dyson equation in the form (see appendix~\ref{ann:DysonEq}):
\begin{equation}
G^{-1}_{\gamma} = C^{-1}_{\gamma} - \Sigma_{\gamma}[G] - K_{\gamma} \;,
\label{eq:2PIfrgdysonEquation}
\end{equation}
by exploiting~\eqref{eq:DefLWfunctional} together with the following expression of the self-energy:
\begin{equation}
\Sigma_{\gamma}[G]\equiv-\frac{\delta\Phi[G]}{\delta G_{\gamma}} \;.
\label{eq:DefinitionSigForDysonEq}
\end{equation}
Between~\eqref{eq:2PIFRGgeneratingFunc} and~\eqref{eq:DefinitionSigForDysonEq}, we have only discussed various features of the 2PI EA formalism. Let us then specify to the 2PI-FRG formalism by introducing one or several cutoff functions into the generating functional~\eqref{eq:2PIFRGgeneratingFunc}. There are different ways to achieve this, which will be discussed thereafter, but our main point for the time being is that most quantities introduced since~\eqref{eq:2PIFRGgeneratingFunc} (e.g. $W[K]$, $\Gamma^{(\mathrm{2PI})}[G]$, $\Phi[G]$ and $\Sigma[G]$) now become dependent on the flow parameter $\mathfrak{s}$. In particular, by setting $K_{\gamma}=0$ $\forall\gamma$ in~\eqref{eq:Extremize2PIEA2PIfrgNonZeroK}, we then define $\overline{G}_{\mathfrak{s}}$ as the propagator configuration that extremizes the flow-dependent 2PI EA $\Gamma_{\mathfrak{s}}^{(\mathrm{2PI})}[G]$ at each step of the flow, i.e.:
\begin{equation}
\left. \frac{\delta\Gamma_{\mathfrak{s}}^{(\mathrm{2PI})}[G]}{\delta G_{\gamma}}\right|_{G=\overline{G}_{\mathfrak{s}}} = 0 \mathrlap{\quad \forall \gamma,\mathfrak{s} \;,}
\label{eq:2PIfrgExtremizeGamma2PI}
\end{equation}
which leads to~\eqref{eq:2PIfrgdysonEquation} with $K_{\gamma}=0$ $\forall\gamma$, i.e.:
\begin{equation}
\overline{G}_{\mathfrak{s},\gamma} = \left(C^{-1}-\overline{\Sigma}_{\mathfrak{s}}\right)_{\gamma}^{-1} \;.
\label{eq:2PIfrgDysonEqGbar}
\end{equation}
Differentiating both sides of this equation with respect to the flow parameter yields the first-order differential equation:
\begin{equation}
\dot{\overline{G}}_{\mathfrak{s},\alpha_{1}\alpha'_{1}} \equiv \partial_{\mathfrak{s}}\overline{G}_{\mathfrak{s},\alpha_{1}\alpha'_{1}} =-\int_{\alpha_{2},\alpha'_{2}}\overline{G}_{\mathfrak{s},\alpha_{1} \alpha_{2}}\left(\dot{C}^{-1}-\dot{\overline{\Sigma}}_{\mathfrak{s}}\right)_{\alpha_{2} \alpha'_{2}} \overline{G}_{\mathfrak{s},\alpha'_{2} \alpha'_{1}} \;,
\label{eq:2PIfrgFlowEqGbarviaDysonEq}
\end{equation}
where the dot still indicates a derivative with respect to the flow parameter, as will always be the case in what follows. Note also that, in the framework of the 2PI-FRG, the propagator $G$ is obtained from the flow-dependent Schwinger functional:
\begin{equation}
G_{\gamma} \equiv G_{\mathfrak{s},\gamma}[K] = W^{(1)}_{\mathfrak{s},\gamma}[K]\;,
\label{eq:GEqualGk2PIFRG}
\end{equation}
in parallel with~\eqref{eq:PhiEqualPhik1PIFRG} for the 1PI-FRG.

\vspace{0.5cm}

The general form of the equation system to solve in the framework of the 2PI-FRG corresponds to the following infinite tower of differential equations:
\begin{subequations}\label{eq:2PIfrgGeneralFlowEq}
\begin{empheq}[left=\empheqlbrace]{align}
& \dot{\overline{G}}_{\mathfrak{s},\alpha_{1}\alpha'_{1}} = -\int_{\alpha_{2},\alpha'_{2}}\overline{G}_{\mathfrak{s},\alpha_{1} \alpha_{2}}\left(\dot{C}^{-1}-\dot{\overline{\Sigma}}_{\mathfrak{s}}\right)_{\alpha_{2} \alpha'_{2}} \overline{G}_{\mathfrak{s},\alpha'_{2} \alpha'_{1}} \;, \label{eq:2PIfrgGeneralFlowEqG} \\
\nonumber \\
& \dot{\overline{\Omega}}_{\mathfrak{s}} = \mathfrak{F}\Big[C,\overline{G}_{\mathfrak{s}},\overline{\Phi}_{\mathfrak{s}},\overline{\Sigma}_{\mathfrak{s}},\left\lbrace\overline{\Phi}_{\mathfrak{s}}^{(n)}; n \geq 2\right\rbrace\Big] \;, \label{eq:2PIfrgGeneralFlowEqOmega}\\
\nonumber \\
& \dot{\overline{\Phi}}_{\mathfrak{s}} = \overline{\dot{\Phi}}_{\mathfrak{s}} + \dot{\overline{G}}_{\mathfrak{s},\hat{\gamma}} \overline{\Phi}^{(1)}_{\mathfrak{s},\hat{\gamma}} = \overline{\dot{\Phi}}_{\mathfrak{s}} - \dot{\overline{G}}_{\mathfrak{s},\hat{\gamma}} \overline{\Sigma}_{\mathfrak{s},\hat{\gamma}} \;, \label{eq:2PIfrgGeneralFlowEqPhi} \\
\nonumber \\
& \dot{\overline{\Sigma}}_{\mathfrak{s},\gamma} = \overline{\dot{\Sigma}}_{\mathfrak{s},\gamma} - \dot{\overline{G}}_{\mathfrak{s},\hat{\gamma}} \overline{\Phi}_{\mathfrak{s},\hat{\gamma}\gamma}^{(2)} \;, \label{eq:2PIfrgGeneralFlowEqSigma} \\
\nonumber \\
& \dot{\overline{\Phi}}_{\mathfrak{s},\gamma_{1}\cdots\gamma_{n}}^{(n)} = \overline{\dot{\Phi}}_{\mathfrak{s},\gamma_{1}\cdots\gamma_{n}}^{(n)} + \dot{\overline{G}}_{\mathfrak{s},\hat{\gamma}} \overline{\Phi}_{\mathfrak{s},\hat{\gamma}\gamma_{1}\cdots\gamma_{n}}^{(n+1)} \quad \forall n \geq 2 \;, \label{eq:2PIfrgGeneralFlowEqPhin}
\end{empheq}
\end{subequations}
where $\mathfrak{F}_{\mathfrak{s}}$ is a functional to be specified. The RHSs of~\eqref{eq:2PIfrgGeneralFlowEqPhi} to~\eqref{eq:2PIfrgGeneralFlowEqPhin} were obtained through the chain rule based on bosonic indices (see appendix~\ref{ann:BosonicIndices2PI}). Similarly to the 1PI-FRG procedure, the above flow equations are usually rewritten with the help of Fourier transformations. We refer to the works of Dupuis~\cite{dup14} and Katanin~\cite{kat19} for more details on these transformations and the associated conventions.

\vspace{0.5cm}

By introducing the Luttinger-Ward functional via the splitting set by~\eqref{eq:DefLWfunctional},~\eqref{eq:2PIfrgGeneralFlowEqG} to~\eqref{eq:2PIfrgGeneralFlowEqPhin} can equivalently be obtained by performing a vertex expansion of the flow-dependent 2PI EA $\Gamma^{(\mathrm{2PI})}_{\mathfrak{s}}[G]$ as follows:
\begin{equation}
\Gamma^{(\mathrm{2PI})}_{\mathfrak{s}}[G] = \overline{\Gamma}^{(\mathrm{2PI})}_{\mathfrak{s}} + \sum_{n=2}^{\infty}\frac{1}{n!}\int_{\gamma_{1},\cdots,\gamma_{n}} \overline{\Gamma}_{\mathfrak{s},\gamma_{1} \cdots \gamma_{n}}^{(\mathrm{2PI})(n)} \left(G-\overline{G}_{\mathfrak{s}}\right)_{\gamma_{1}} \cdots \left(G-\overline{G}_{\mathfrak{s}}\right)_{\gamma_{n}} \;,
\label{eq:VertexExpansion2PIFRG}
\end{equation}
with $\overline{\Gamma}^{(\mathrm{2PI})}_{\mathfrak{s}}\equiv\Gamma^{(\mathrm{2PI})}_{\mathfrak{s}}\big[G=\overline{G}_{\mathfrak{s}}\big]$, $\overline{\Gamma}_{\mathfrak{s},\gamma_{1} \cdots \gamma_{n}}^{(\mathrm{2PI})(n)}\equiv\left.\frac{\delta^{n}\Gamma_{\mathfrak{s}}^{(\mathrm{2PI})}[G]}{\delta G_{\gamma_{1}}\cdots\delta G_{\gamma_{n}}}\right|_{G=\overline{G}_{\mathfrak{s}}}$ and the first-order derivative of $\Gamma^{(\mathrm{2PI})}_{\mathfrak{s}}$ vanishes at $G=\overline{G}_{\mathfrak{s}}$ according to~\eqref{eq:2PIfrgExtremizeGamma2PI}, which is the counterpart of~\eqref{eq:ExtremizeTruncationVertexExp1PIFRG} underlying the vertex expansion for the 1PI-FRG. After plugging this expansion into the different master equations that can be derived for $\Gamma_{\mathfrak{s}}^{(\mathrm{2PI})}$ in the framework of the 2PI-FRG (see appendix~\ref{ann:2PIfrgFlowEquation}) and comparing the terms with identical powers of $G-\overline{G}_{\mathfrak{s}}$ in the LHS and RHS of the equations thus obtained, one should get back the differential equations expressed by~\eqref{eq:2PIfrgGeneralFlowEqG} to~\eqref{eq:2PIfrgGeneralFlowEqPhin}, thus specifying the analytical forms of the function $\mathfrak{F}$ in~\eqref{eq:2PIfrgGeneralFlowEqOmega} as well as those of the derivatives $\overline{\dot{\Phi}}_{\mathfrak{s}}$, $\overline{\dot{\Sigma}}_{\mathfrak{s}}$ and $\overline{\dot{\Phi}}_{\mathfrak{s}}^{(n)}$ (with $n \geq 2$)\footnote{We actually follow a different (although equivalent) path to derive the analytical forms of $\mathfrak{F}$, $\overline{\dot{\Phi}}_{\mathfrak{s}}$, $\overline{\dot{\Sigma}}_{\mathfrak{s}}$ and $\overline{\dot{\Phi}}_{\mathfrak{s}}^{(n)}$ (with $n \geq 2$) in appendix~\ref{ann:2PIfrgFlowEquation}.}.

\vspace{0.5cm}

\begin{figure}[t]
\begin{subfigure}{1.0\textwidth}
  \centering
  \includegraphics[width=0.45\linewidth]{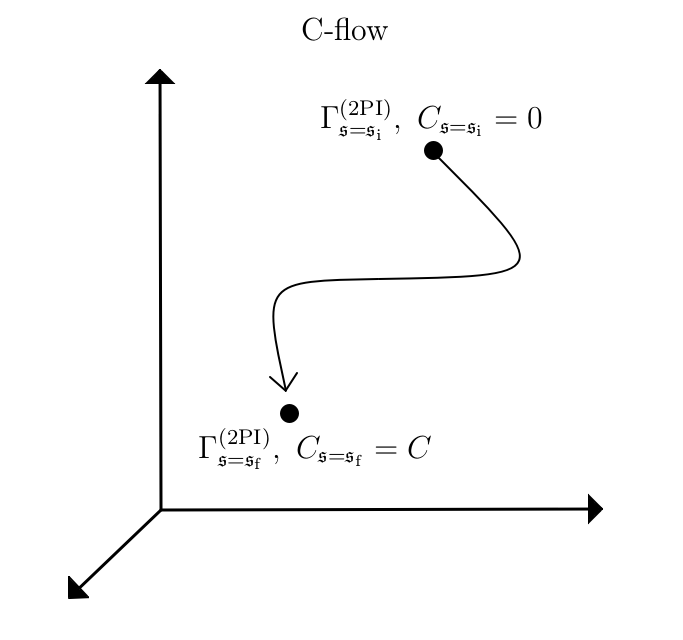}
  \includegraphics[width=0.45\linewidth]{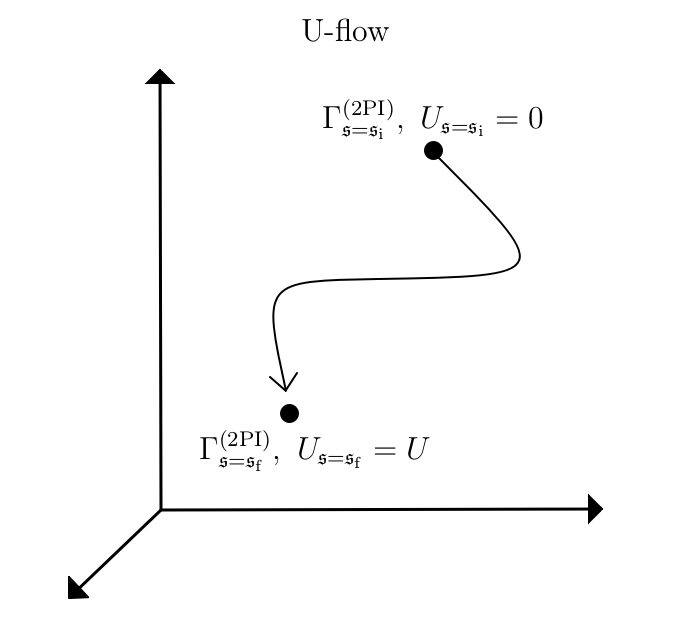}
\end{subfigure}

\vspace{0.1cm}\hfill

\begin{subfigure}{1.0\textwidth}
  \centering
  \includegraphics[width=.45\linewidth]{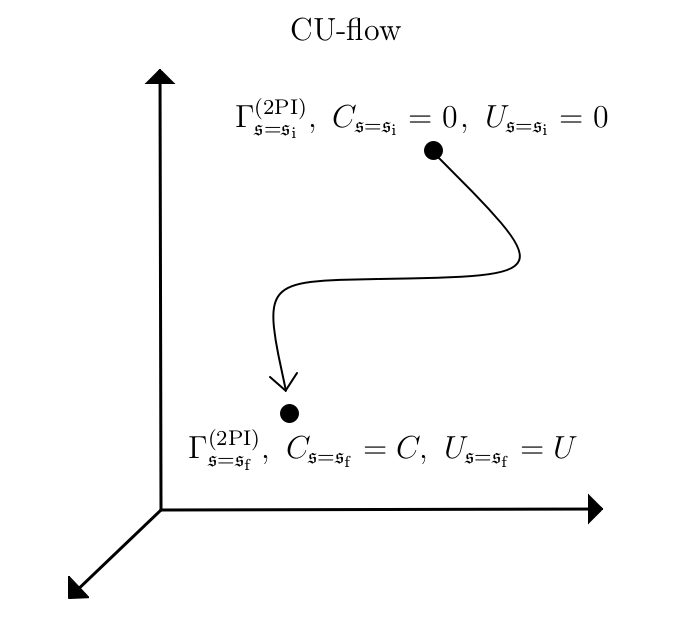}
\end{subfigure}
\caption{Schematic illustrations of the C-flow, U-flow and CU-flow for the 2PI-FRG. Recall that $C$ and $U$ are respectively the free propagator and the two-body interaction of the model under consideration (specified by the classical action expressed by~\eqref{eq:2PIFRGmostgeneralactionS}).}
\label{fig:2PIFRGtheoryspace}
\end{figure}

All 2PI-FRG approaches treated in this thesis can be applied to any system whose classical action can be put in the form:
\begin{equation}
S\Big[\widetilde{\psi}\Big] = S_{0}\Big[\widetilde{\psi}\Big] + S_{\mathrm{int}}\Big[\widetilde{\psi}\Big] = \frac{1}{2}\int_{\alpha_{1},\alpha_{2}}\widetilde{\psi}_{\alpha_{1}}C^{-1}_{\alpha_{1}\alpha_{2}}\widetilde{\psi}_{\alpha_{2}}+\frac{1}{4!}\int_{\alpha_{1},\alpha_{2},\alpha_{3},\alpha_{4}}U_{\alpha_{1}\alpha_{2}\alpha_{3}\alpha_{4}}\widetilde{\psi}_{\alpha_{1}}\widetilde{\psi}_{\alpha_{2}}\widetilde{\psi}_{\alpha_{3}}\widetilde{\psi}_{\alpha_{4}} \;,
\label{eq:2PIFRGmostgeneralactionS}
\end{equation}
with the two-body interaction $U$ satisfying the symmetry property:
\begin{equation}
U_{\alpha_{1}\alpha_{2}\alpha_{3}\alpha_{4}}=\zeta^{N(P)}U_{\alpha_{P(1)}\alpha_{P(2)}\alpha_{P(3)}\alpha_{P(4)}} \;,
\label{eq:SymmetryProperty2bodyInteraction2PIFRG}
\end{equation}
$N(P)$ being the number of inversions in the permutation $P$. Hence, in the present 2PI-FRG study, we are only treating systems with two-body interactions at most. However, it is straightforward to generalize the formalism presented in the whole section~\ref{sec:2PIFRG} to interactions which are three-body or more after including in $S_{\mathrm{int}}$ terms being sextic in the field (i.e. varying as $\sim\widetilde{\psi}^{6}$) or more. Following the derivations discussed in chapter~\ref{chap:DiagTechniques} for the 2PI EA, one can deduce from the classical action~\eqref{eq:2PIFRGmostgeneralactionS} the following expression of the Luttinger-Ward functional in terms of 2PI diagrams:
\begin{equation}
\begin{split}
\scalebox{0.92}{${\displaystyle \Phi[G] = }$} & \ \scalebox{0.92}{${\displaystyle \Phi_{\mathrm{SCPT}}[U,G] \equiv \frac{1}{8} \ \begin{gathered}
\begin{fmffile}{DiagramsFRG/2PIFRGphi_HartreeFock}
\begin{fmfgraph}(25,20)
\fmfleft{i}
\fmfright{o}
\fmfv{decor.shape=circle,decor.filled=full,decor.size=0.2cm}{v1}
\fmf{phantom,tension=10}{i,i1}
\fmf{phantom,tension=10}{o,o1}
\fmf{plain,left,tension=0.5,foreground=(1,,0,,0)}{i1,v1,i1}
\fmf{plain,right,tension=0.5,foreground=(1,,0,,0)}{o1,v1,o1}
\end{fmfgraph}
\end{fmffile}
\end{gathered} - \frac{1}{48} \hspace{0.25cm} \begin{gathered}
\begin{fmffile}{DiagramsFRG/2PIFRGphi_RPA}
\begin{fmfgraph}(20,20)
\fmftop{vUp}
\fmfbottom{vDown}
\fmfv{decor.shape=circle,decor.filled=full,decor.size=0.2cm}{vUp}
\fmfv{decor.shape=circle,decor.filled=full,decor.size=0.2cm}{vDown}
\fmf{plain,left,foreground=(1,,0,,0)}{vUp,vDown}
\fmf{plain,right,foreground=(1,,0,,0)}{vUp,vDown}
\fmf{plain,left=0.4,foreground=(1,,0,,0)}{vUp,vDown}
\fmf{plain,right=0.4,foreground=(1,,0,,0)}{vUp,vDown}
\end{fmfgraph}
\end{fmffile}
\end{gathered} \hspace{0.2cm} + \mathcal{O}\big(U^{3}\big) }$} \\
\scalebox{0.92}{${\displaystyle = }$} & \ \scalebox{0.92}{${\displaystyle \frac{1}{8}\int_{\gamma_{1},\gamma_{2}}U_{\gamma_{1}\gamma_{2}}G_{\gamma_{1}}G_{\gamma_{2}} - \frac{1}{48} \int_{\gamma_{1},\gamma_{2},\gamma_{3},\gamma_{4}} U_{\alpha_{1}\alpha_{2}\alpha_{3}\alpha_{4}} U_{\alpha'_{1}\alpha'_{2}\alpha'_{3}\alpha'_{4}} G_{\gamma_{1}}G_{\gamma_{2}}G_{\gamma_{3}}G_{\gamma_{4}} + \mathcal{O}\big(U^{3}\big) \;.}$}
\end{split}
\label{eq:2PIFRGperturbativeExpLWfunc}
\end{equation}
We have introduced in this way the functional $\Phi_{\mathrm{SCPT}}[U,G]$, which is identical to the Luttinger-Ward functional, although we stress its dependence with respect to the interaction $U$ (usually left implicit) and the subscript ``SCPT'' indicates that we consider its expression~\eqref{eq:2PIFRGperturbativeExpLWfunc} usually taken as input for the variational procedure underlying self-consistent PT. This concludes our introduction for the 2PI-FRG and we will then discuss in further details its specific implementations. The latter are coined as C-flow, U-flow or CU-flow, depending on the way the cutoff function $R_{\mathfrak{s}}$ is introduced in the classical action~\eqref{eq:2PIFRGmostgeneralactionS}: either in the free propagator (e.g. via $C^{-1} \rightarrow C_{\mathfrak{s}}^{-1}=C^{-1}+R_{\mathfrak{s}}$) for the C-flow, either in the two-body interaction (e.g. via $U \rightarrow U_{\mathfrak{s}} = U + R_{\mathfrak{s}}$) for the U-flow or in both for the CU-flow (see fig.~\ref{fig:2PIFRGtheoryspace}).

\subsubsection{C-flow}
\label{sec:Cflow2PIFRG}
\paragraph{Main features:}

The C-flow version of the 2PI-FRG was introduced in ref.~\cite{dup05}. The underlying idea remains to implement the momentum-shell integration \`{a} la Wilson, as in the 1PI-FRG based on Wetterich equation, except that we are now computing the 2PI EA $\Gamma^{(\mathrm{2PI})}[G]$ throughout the flow instead of the 1PI one. In particular, the flow parameter $\mathfrak{s}$ can be interpreted now as a momentum scale. By definition, the C-flow consists in considering a flow-dependent free propagator $C_{\mathfrak{s}}$. This amounts to inserting a cutoff function $R_{\mathfrak{s}}$ in the classical action~\eqref{eq:2PIFRGmostgeneralactionS} via the substitution $C^{-1} \rightarrow C^{-1}_{\mathfrak{s}} = R_{\mathfrak{s}} C^{-1}$ or equivalently $C^{-1} \rightarrow C^{-1}_{\mathfrak{s}} = C^{-1} + R_{\mathfrak{s}}$, which is exactly what is done for the 1PI-FRG by introducing the term $\Delta S_{k}$ in~\eqref{eq:defZkWettFRG}. The C-flow is therefore close in spirit to Wetterich's approach. Owing to such a connection, we can deduce the required values for $C_{\mathfrak{s}=\mathfrak{s}_{\mathrm{i}}}$ and $C_{\mathfrak{s}=\mathfrak{s}_{\mathrm{f}}}$ from the boundary conditions for $R_{k}$ set by~\eqref{eq:BoundaryRLambdaWettFRG} and~\eqref{eq:BoundaryR0WettFRG}:
\begin{subequations}
\begin{empheq}[left=\empheqlbrace]{align}
& C_{\mathfrak{s}=\mathfrak{s}_{\mathrm{i}},\gamma} = 0 \quad \forall \gamma \;, \label{eq:2PIfrgBoundaryCondforCkCflowUpper}\\
\nonumber \\
& C_{\mathfrak{s}=\mathfrak{s}_{\mathrm{f}}} = C \;, \label{eq:2PIfrgBoundaryCondforCkCflowBottom}
\end{empheq}
\end{subequations}
with $\mathfrak{s}_{\mathrm{i}}$ and $\mathfrak{s}_{\mathrm{f}}$ being respectively the values of the flow parameter at the beginning and at the end of the flow.

\vspace{0.5cm}

The Luttinger-Ward functional does not depend on the free propagator $C$, and therefore not on $\mathfrak{s}$ for the C-flow. Consequently, it is an invariant of the flow, which translates into\footnote{Besides the qualitative argument just given to justify~\eqref{eq:2PIfrgFlowEqCflowVersion1}, a mathematical proof of the latter is given in appendix~\ref{ann:2PIfrgFlowEquation}.}:
\begin{equation}
\dot{\Phi}_{\mathfrak{s}}[G]=0 \mathrlap{\quad \forall \mathfrak{s} \;.}
\label{eq:2PIfrgFlowEqCflowVersion1}
\end{equation}
In particular,~\eqref{eq:2PIfrgFlowEqCflowVersion1} implies that all components of $\overline{\dot{\Phi}}_{\mathfrak{s}}$, $\overline{\dot{\Sigma}}_{\mathfrak{s}}$ and $\overline{\dot{\Phi}}_{\mathfrak{s}}^{(n)}$ (with $n \geq 2$) vanish, which enables us to simplify the three lowest equalities in the set of~\eqref{eq:2PIfrgGeneralFlowEq} as:
\begin{subequations}\label{eq:2PIfrgFlowEqCflowVersion2}
\begin{empheq}[left=\empheqlbrace]{align}
& \dot{\overline{\Phi}}_{\mathfrak{s}} = - \dot{\overline{G}}_{\mathfrak{s},\hat{\gamma}} \overline{\Sigma}_{\mathfrak{s},\hat{\gamma}} \;. \label{eq:2PIfrgGeneralFlowEqPhiCflow} \\
\nonumber \\
& \dot{\overline{\Sigma}}_{\mathfrak{s},\gamma} = - \dot{\overline{G}}_{\mathfrak{s},\hat{\gamma}} \overline{\Phi}_{\mathfrak{s},\hat{\gamma}\gamma}^{(2)} \;. \label{eq:2PIfrgGeneralFlowEqSigmaCflow} \\
\nonumber \\
& \dot{\overline{\Phi}}_{\mathfrak{s},\gamma_{1}\cdots\gamma_{n}}^{(n)} = \dot{\overline{G}}_{\mathfrak{s},\hat{\gamma}} \overline{\Phi}_{\mathfrak{s},\hat{\gamma}\gamma_{1}\cdots\gamma_{n}}^{(n+1)} \quad \forall n \geq 2 \;. \label{eq:2PIfrgGeneralFlowEqPhinCflow}
\end{empheq}
\end{subequations}
The flow dependence of $\overline{G}_{\mathfrak{s}}$ involved in~\eqref{eq:2PIfrgGeneralFlowEqPhiCflow} to~\eqref{eq:2PIfrgGeneralFlowEqPhinCflow} follows from that of $C_{\mathfrak{s}}$ according to Dyson equation in the form of~\eqref{eq:2PIfrgDysonEqGbar}. Moreover, the flow equation~\eqref{eq:2PIfrgGeneralFlowEqOmega} expressing $\dot{\overline{\Omega}}_{\mathfrak{s}}=\frac{1}{\beta}\dot{\overline{\Gamma}}^{(\mathrm{2PI})}_{\mathfrak{s}}$ is basically determined by computing an expression for $\dot{\overline{\Gamma}}^{(\mathrm{2PI})}_{\mathfrak{s}}$ from the generating functional~\eqref{eq:2PIFRGgeneratingFunc} (see appendix~\ref{ann:2PIfrgFlowEquationCflow}). However, it is not tractable in this form because of the initial condition $\overline{G}_{\mathfrak{s}=\mathfrak{s}_{\mathrm{i}},\gamma}=0$ $\forall \gamma$ (see discussion around~\eqref{eq:2PIfrgInitialConditionsGki} for the justification of this initial condition) since it induces a divergence of $\frac{\zeta}{2} \mathrm{Tr}_{\alpha} \left[ \mathrm{ln}(\overline{G}_{\mathfrak{s}}) \right]$ in $\overline{\Gamma}^{(\mathrm{2PI})}_{0,\mathfrak{s}}$ (and thus in $\overline{\Omega}_{\mathfrak{s}}$) at the starting point of the flow, i.e. at $\mathfrak{s}=\mathfrak{s}_{\mathrm{i}}$. Therefore, we will calculate during the flow the following quantity:
\begin{equation}
\Delta \overline{\Omega}_{\mathfrak{s}} \equiv \frac{1}{\beta}\left(\overline{\Gamma}_{\mathfrak{s}}^{(\mathrm{2PI})} - \Gamma_{0,\mathfrak{s}}^{(\mathrm{2PI})}[G = C_{\mathfrak{s}}] \right) = \overline{\Omega}_{\mathfrak{s}} + \frac{\zeta}{2\beta} \mathrm{Tr}_{\alpha} \left[ \mathrm{ln}(C_{\mathfrak{s}}) \right] \;,
\label{eq:2PIfrgCflowDefDeltaOmega}
\end{equation}
instead of $\overline{\Omega}_{\mathfrak{s}}$. The extra term $\Gamma_{0,\mathfrak{s}}^{(\mathrm{2PI})}[G = C_{\mathfrak{s}}]$ eliminates the aforementioned divergence problem, as a result of~\eqref{eq:2PIfrgBoundaryCondforCkCflowUpper}.

\vspace{0.5cm}

In conclusion, the tower of differential equations underlying the C-flow version of the 2PI-FRG is thus given by (see appendix~\ref{ann:2PIfrgFlowEquationCflow}):
\begin{equation}
\dot{\overline{G}}_{\mathfrak{s},\alpha_{1} \alpha'_{1}}=-\int_{\alpha_{2},\alpha'_{2}}\overline{G}_{\mathfrak{s}, \alpha_{1} \alpha_{2}}\left(\dot{C}_{\mathfrak{s}}^{-1}-\dot{\overline{\Sigma}}_{\mathfrak{s}}\right)_{\alpha_{2} \alpha'_{2}} \overline{G}_{\mathfrak{s}, \alpha'_{2} \alpha'_{1}} \mathrlap{\;,}
\label{eq:2PIfrgFlowEquationsCflowG}
\end{equation}
\begin{equation}
\Delta \dot{\overline{\Omega}}_{\mathfrak{s}} = \frac{1}{\beta} \dot{C}_{\mathfrak{s},\hat{\gamma}}^{-1} \left(\overline{G}_{\mathfrak{s}}-C_{\mathfrak{s}}\right)_{\hat{\gamma}} \mathrlap{\;,}
\label{eq:2PIfrgFlowEquationsCflowDOmega}
\end{equation}
\begin{equation}
\dot{\overline{\Phi}}_{\mathfrak{s}} = - \dot{\overline{G}}_{\mathfrak{s},\hat{\gamma}} \overline{\Sigma}_{\mathfrak{s},\hat{\gamma}} \mathrlap{\;,}
\label{eq:2PIfrgFlowEquationsCflowPhi}
\end{equation}
\begin{equation}
\dot{\overline{\Sigma}}_{\mathfrak{s},\gamma} = - \dot{\overline{G}}_{\mathfrak{s},\hat{\gamma}} \overline{\Phi}_{\mathfrak{s},\hat{\gamma}\gamma}^{(2)} \mathrlap{\;,}
\label{eq:2PIfrgFlowEquationsCflowSigma}
\end{equation}
\begin{equation}
\dot{\overline{\Phi}}_{\mathfrak{s},\gamma_{1}\cdots\gamma_{n}}^{(n)} = \dot{\overline{G}}_{\mathfrak{s},\hat{\gamma}} \overline{\Phi}_{\mathfrak{s},\hat{\gamma}\gamma_{1}\cdots\gamma_{n}}^{(n+1)} \mathrlap{\quad \forall n \geq 2 \;.}
\label{eq:2PIfrgFlowEquationsCflowPhin}
\end{equation}
The desired value of the thermodynamic potential, i.e. $\overline{\Omega}_{\mathfrak{s}=\mathfrak{s}_{\mathrm{f}}}$, is readily obtained at the end of the flow of $\Delta\overline{\Omega}_{\mathfrak{s}}$ through~\eqref{eq:2PIfrgCflowDefDeltaOmega} in the form $\overline{\Omega}_{\mathfrak{s}=\mathfrak{s}_{\mathrm{f}}}=\Delta \overline{\Omega}_{\mathfrak{s}=\mathfrak{s}_{\mathrm{f}}}-\frac{\zeta}{2\beta} \mathrm{Tr}_{\alpha} \left[ \mathrm{ln}(C_{\mathfrak{s}=\mathfrak{s}_{\mathrm{f}}}) \right]$.

\paragraph{Initial conditions:}

From Dyson equation and more specifically from~\eqref{eq:2PIfrgDysonEqGbar}, it is clear that the initial condition for $C_{\mathfrak{s}}$ given by~\eqref{eq:2PIfrgBoundaryCondforCkCflowUpper} implies that:
\begin{equation}
\overline{G}_{\mathfrak{s}=\mathfrak{s}_{\mathrm{i}},\gamma}=0 \mathrlap{\quad \forall \gamma \;.}
\label{eq:2PIfrgInitialConditionsGki}
\end{equation}
This condition enables us to find the initial conditions for $\overline{\Phi}_{\mathfrak{s}}$ and the corresponding derivatives from the diagrammatic expansion of the Luttinger-Ward functional expressed by~\eqref{eq:2PIFRGperturbativeExpLWfunc} (see appendix~\ref{ann:2PIfrgCflowInitialConditionsGeneral} for the derivations of~\eqref{eq:2PIfrgCflowICPhi2} and~\eqref{eq:2PIfrgCflowICPhi4}):
\begin{equation}
\overline{\Phi}_{\mathfrak{s}=\mathfrak{s}_{\mathrm{i}}} = 0 \mathrlap{\;,}
\label{eq:2PIfrgCflowICPhis}
\end{equation}
\begin{equation}
\overline{\Sigma}_{\mathfrak{s}=\mathfrak{s}_{\mathrm{i}},\gamma} = 0 \mathrlap{\quad \forall \gamma \;,}
\label{eq:2PIfrgCflowICSigma}
\end{equation}
\begin{equation}
\overline{\Phi}_{\mathfrak{s} = \mathfrak{s}_{\mathrm{i}},\gamma_{1}\gamma_{2}}^{(2)} = U_{\gamma_{1}\gamma_{2}} \mathrlap{\;,}
\label{eq:2PIfrgCflowICPhi2}
\end{equation}
\begin{equation}
\scalebox{0.89}{${\displaystyle \overline{\Phi}_{\mathfrak{s} = \mathfrak{s}_{\mathrm{i}},\gamma_{1}\gamma_{2}\gamma_{3}\gamma_{4}}^{(4)} = -\frac{1}{2} \left[\left(\left\lbrace\left[ U_{\alpha_{1} \alpha_{2} \alpha_{3} \alpha_{4}} U_{\alpha'_{1} \alpha'_{2} \alpha'_{3} \alpha'_{4}} + \zeta \left(\alpha_{1} \leftrightarrow \alpha'_{1}\right) \right] + \zeta \left(\alpha_{2} \leftrightarrow \alpha'_{2}\right) \right\rbrace + \zeta \left(\alpha_{3} \leftrightarrow \alpha'_{3}\right) \right) + \zeta \left(\alpha_{4} \leftrightarrow \alpha'_{4} \right) \right] \;, }$}
\label{eq:2PIfrgCflowICPhi4}
\end{equation}
\begin{equation}
\overline{\Phi}_{\mathfrak{s} = \mathfrak{s}_{\mathrm{i}},\gamma_{1} \cdots \gamma_{n}}^{(n)}=0 \mathrlap{\quad \forall \gamma_{1}, \cdots, \gamma_{n} , ~ \forall n ~ \mathrm{odd} \;.}
\label{eq:2PIfrgCflowICPhiodd}
\end{equation}
Even though~\eqref{eq:2PIfrgCflowICPhis} to~\eqref{eq:2PIfrgCflowICPhiodd} are derived from a truncated result in practice, these equations are exact at $\mathfrak{s}=\mathfrak{s}_{\mathrm{i}}$ because of~\eqref{eq:2PIfrgInitialConditionsGki}. It remains to determine the initial condition for $\Delta\overline{\Omega}_{\mathfrak{s}}$. Combining~\eqref{eq:2PIfrgCflowDefDeltaOmega} with~\eqref{eq:DefLWfunctional} and~\eqref{eq:2PIFRGfreeGamma2PI} at $\mathfrak{s}=\mathfrak{s}_{\mathrm{i}}$ gives us:
\begin{equation}
\begin{split}
\Delta \overline{\Omega}_{\mathfrak{s}=\mathfrak{s}_{\mathrm{i}}} = & \ \frac{1}{\beta}\left(\overline{\Gamma}_{\mathfrak{s}=\mathfrak{s}_{\mathrm{i}}}^{(\mathrm{2PI})} - \Gamma_{0,\mathfrak{s}=\mathfrak{s}_{\mathrm{i}}}^{(\mathrm{2PI})}[G = C_{\mathfrak{s}=\mathfrak{s}_{\mathrm{i}}}] \right) \\
= & \ \frac{1}{\beta}\left(-\frac{\zeta}{2} \mathrm{Tr}_{\alpha} \left[ \mathrm{ln}\big(\overline{G}_{\mathfrak{s}=\mathfrak{s}_{\mathrm{i}}}\big) \right] + \frac{\zeta}{2} \mathrm{Tr}_{\alpha}\left[\overline{G}_{\mathfrak{s}=\mathfrak{s}_{\mathrm{i}}}C_{\mathfrak{s}=\mathfrak{s}_{\mathrm{i}}}^{-1}-\mathbb{I}\right] + \overline{\Phi}_{\mathfrak{s}=\mathfrak{s}_{\mathrm{i}}} + \frac{\zeta}{2} \mathrm{Tr}_{\alpha} \left[ \mathrm{ln}(C_{\mathfrak{s}=\mathfrak{s}_{\mathrm{i}}}) \right] \right) \;.
\end{split}
\end{equation}
According to~\eqref{eq:2PIfrgBoundaryCondforCkCflowUpper},~\eqref{eq:2PIfrgInitialConditionsGki} and~\eqref{eq:2PIfrgCflowICPhis}, this is equivalent to:
\begin{equation}
\Delta \overline{\Omega}_{\mathfrak{s}=\mathfrak{s}_{\mathrm{i}}} = 0 \;.
\label{eq:2PIfrgCflowICDeltaOmega}
\end{equation}

\paragraph{Truncations:}

\begin{itemize}
\item tC-flow:\\
The truncated C-flow (tC-flow) is a specific implementation of the C-flow in which the infinite tower of differential equations given by~\eqref{eq:2PIfrgFlowEquationsCflowG} to~\eqref{eq:2PIfrgFlowEquationsCflowPhin} is rendered finite according to the condition:
\begin{equation}
\overline{\Phi}_{\mathfrak{s}}^{(n)}=\overline{\Phi}_{\mathfrak{s}=\mathfrak{s}_{\mathrm{i}}}^{(n)} \mathrlap{\quad \forall \mathfrak{s}, ~ \forall n > N_{\mathrm{max}} \;.}
\label{eq:2PIfrgPhiBartCflow}
\end{equation}
In this way, the equation system made of~\eqref{eq:2PIfrgFlowEquationsCflowG} to~\eqref{eq:2PIfrgFlowEquationsCflowPhin} reduces to a set of $N_{\mathrm{max}}+2$ first-order differential equations. It is shown in ref.~\cite{ren15} that the tC-flow scheme with truncation order $N_{\mathrm{max}}=2 N_{\mathrm{SCPT}}-1$ or $2 N_{\mathrm{SCPT}}$ (with $N_{\mathrm{SCPT}} \in \mathbb{N}^{*}$) is equivalent to self-consistent PT up to order $\mathcal{O}\big(U^{N_{\mathrm{SCPT}}}\big)$, e.g. the tC-flow with $N_{\mathrm{max}}=1$ or $2$ is equivalent to Hartree-Fock theory.

\vspace{0.5cm}

Let us then prove the latter statement at $N_{\mathrm{max}}=1$ following the lines set out by ref.~\cite{ren15}. In this situation, we have notably $\overline{\Phi}_{\mathfrak{s}}^{(2)}=\overline{\Phi}_{\mathfrak{s}=\mathfrak{s}_{\mathrm{i}}}^{(2)}=U$ $\forall \mathfrak{s}$ according to~\eqref{eq:2PIfrgPhiBartCflow} alongside with the initial condition~\eqref{eq:2PIfrgCflowICPhi2}. In this way,~\eqref{eq:2PIfrgFlowEquationsCflowSigma} becomes:
\begin{equation}
\dot{\overline{\Sigma}}_{\mathfrak{s},\gamma} = - \dot{\overline{G}}_{\mathfrak{s},\hat{\gamma}} U_{\hat{\gamma}\gamma}\;,
\end{equation}
which, after integration with respect to $\mathfrak{s}$, yields\footnote{The integration constant in~\eqref{eq:2PIFRGHartreeFockSelfEnergy} equals zero according to the initial conditions~\eqref{eq:2PIfrgInitialConditionsGki} and~\eqref{eq:2PIfrgCflowICSigma}.}:
\begin{equation}
\overline{\Sigma}_{\mathfrak{s},\gamma} = - \overline{G}_{\mathfrak{s},\hat{\gamma}} U_{\hat{\gamma}\gamma}\;,
\label{eq:2PIFRGHartreeFockSelfEnergyFirst}
\end{equation}
and more specifically at $\mathfrak{s}=\mathfrak{s}_{\mathrm{f}}$:
\begin{equation}
\overline{\Sigma}_{\mathfrak{s}=\mathfrak{s}_{\mathrm{f}},\gamma} = - \overline{G}_{\mathfrak{s}=\mathfrak{s}_{\mathrm{f}},\hat{\gamma}} U_{\hat{\gamma}\gamma}\;,
\label{eq:2PIFRGHartreeFockSelfEnergy}
\end{equation}
which is nothing else than the Hartree-Fock self-energy. From~\eqref{eq:2PIFRGHartreeFockSelfEnergyFirst}, the flow equation~\eqref{eq:2PIfrgFlowEquationsCflowPhi} for the Luttinger-Ward functional is turned into the equality:
\begin{equation}
\dot{\overline{\Phi}}_{\mathfrak{s}} = \dot{\overline{G}}_{\mathfrak{s},\hat{\gamma}_{1}} U_{\hat{\gamma}_{1}\hat{\gamma}_{2}} \overline{G}_{\mathfrak{s},\hat{\gamma}_{2}} \;.
\end{equation}
After integration with respect to $\mathfrak{s}$, this gives us\footnote{The integration constant in~\eqref{eq:2PIFRGHartreeFockPhi} also vanishes, now according to the initial conditions~\eqref{eq:2PIfrgInitialConditionsGki} and~\eqref{eq:2PIfrgCflowICPhis}.}:
\begin{equation}
\overline{\Phi}_{\mathfrak{s}=\mathfrak{s}_{\mathrm{f}}} = \frac{1}{2} \overline{G}_{\mathfrak{s}=\mathfrak{s}_{\mathrm{f}},\hat{\gamma}_{1}} U_{\hat{\gamma}_{1}\hat{\gamma}_{2}} \overline{G}_{\mathfrak{s}=\mathfrak{s}_{\mathrm{f}},\hat{\gamma}_{2}} \;,
\label{eq:2PIFRGHartreeFockPhi}
\end{equation}
which now corresponds to the Hartree-Fock approximation of the Luttinger-Ward functional, as can be seen after comparison with~\eqref{eq:2PIFRGperturbativeExpLWfunc}. Therefore, according to~\eqref{eq:2PIFRGHartreeFockSelfEnergy} and~\eqref{eq:2PIFRGHartreeFockPhi}, all flowing quantities reduce to their Hartree-Fock results at the end of the flow. We have proven in this way that the tC-flow with truncation order $N_{\mathrm{max}}=1$ is equivalent to self-consistent PT at the Hartree-Fock level (i.e. at order $\mathcal{O}(U)$). This remark can be extended to $N_{\mathrm{max}}=2$ since:
\begin{equation}
\dot{\overline{\Phi}}_{\mathfrak{s},\gamma_{1}\gamma_{2}}^{(2)} = \dot{\overline{G}}_{\mathfrak{s},\hat{\gamma}} \overline{\Phi}_{\mathfrak{s},\hat{\gamma}\gamma_{1}\gamma_{2}}^{(3)} = 0 \mathrlap{\quad \forall \mathfrak{s} \;,}
\end{equation}
and thus $\overline{\Phi}_{\mathfrak{s}}^{(2)}=\overline{\Phi}_{\mathfrak{s}=\mathfrak{s}_{\mathrm{i}}}^{(2)}=U$ $\forall \mathfrak{s}$ in this case as well, as a consequence of~\eqref{eq:2PIfrgCflowICPhi2},~\eqref{eq:2PIfrgCflowICPhiodd} (for $n=3$) and~\eqref{eq:2PIfrgPhiBartCflow} (for $n=3$). Actually, we can show in the same manner that, if $N_{\mathrm{max}}$ is odd, the truncation orders $N_{\mathrm{max}}$ and $N_{\mathrm{max}}+1$ are equivalent in the framework of the tC-flow. In addition, the arguments leading to~\eqref{eq:2PIFRGHartreeFockSelfEnergy} and~\eqref{eq:2PIFRGHartreeFockPhi} can be straightforwardly extended to any truncation of the tC-flow (i.e. to any $N_{\mathrm{max}}\in\mathbb{N}^{*}$) in order to prove its connection with the corresponding order of self-consistent PT. In this fashion, we have shown that, for any non-zero positive integer $N_{\mathrm{SCPT}}$, the tC-flow with $N_{\mathrm{max}}=2 N_{\mathrm{SCPT}}-1$ or $2 N_{\mathrm{SCPT}}$ and $N_{\mathrm{SCPT}}^{\mathrm{th}}$-order self-consistent PT are equivalent, as stated above. This has the interesting consequence that identical results can be obtained by solving two different types of equations: first-order integro-differential equations for the tC-flow and self-consistent equations for self-consistent PT.

\item mC-flow:\\
The modified C-flow (mC-flow) implements a truncation of the C-flow's infinite tower via the following condition:
\begin{equation}
\overline{\Phi}_{\mathfrak{s}}^{(n)}=\left.\overline{\Phi}_{\mathrm{SCPT},N_{\mathrm{SCPT}},\mathfrak{s}}^{(n)}\right|_{U \rightarrow \overline{\Phi}_{\mathrm{sym},\mathfrak{s}}^{(2)}} \mathrlap{\quad \forall \mathfrak{s}, ~ \forall n > N_{\mathrm{max}} \;,}
\label{eq:2PIfrgTruncationmCflow}
\end{equation}
with
\begin{equation}
\overline{\Phi}_{\mathrm{sym},\mathfrak{s},\alpha_{1} \alpha_{2} \alpha_{3} \alpha_{4}}^{(2)} = \frac{1}{3}\left(\overline{\Phi}_{\mathfrak{s},\alpha_{1} \alpha_{2} \alpha_{3} \alpha_{4}}^{(2)}+\overline{\Phi}_{\mathfrak{s},\alpha_{2} \alpha_{3} \alpha_{1} \alpha_{4}}^{(2)}+\overline{\Phi}_{\mathfrak{s},\alpha_{3} \alpha_{1} \alpha_{2} \alpha_{4}}^{(2)}\right) \;.
\label{eq:DefPhi2sym2PIFRGmCflow}
\end{equation}
The truncation condition~\eqref{eq:2PIfrgTruncationmCflow} is an ansatz based on the perturbative expression of the Luttinger-Ward functional, i.e.~\eqref{eq:2PIFRGperturbativeExpLWfunc}. The functional $\Phi_{\mathrm{SCPT},N_{\mathrm{SCPT}}}[U,G]$ corresponds to the $N_{\mathrm{SCPT}}$ first terms of the perturbative series of the RHSs of~\eqref{eq:2PIFRGperturbativeExpLWfunc}. According to this definition, we have for example:
\begin{equation}
\left.\overline{\Phi}_{\mathrm{SCPT},N_{\mathrm{SCPT}}=1,\mathfrak{s}}\right|_{U\rightarrow\overline{\Phi}_{\mathrm{sym},\mathfrak{s}}^{(2)}}\equiv\frac{1}{8}\int_{\gamma_{1},\gamma_{2}}\overline{\Phi}_{\mathrm{sym},\mathfrak{s},\gamma_{1} \gamma_{2}}^{(2)}\overline{G}_{\mathfrak{s},\gamma_{1}}\overline{G}_{\mathfrak{s},\gamma_{2}} \;.
\end{equation}
Furthermore, the motivation for replacing $U$ by $\overline{\Phi}_{\mathfrak{s}}^{(2)}$ is set by the initial condition for $\overline{\Phi}_{\mathfrak{s}}^{(2)}$, i.e.~\eqref{eq:2PIfrgCflowICPhi2}. Generalizing the latter relation to all $\mathfrak{s}$ indeed suggests to substitute $U$ by $\overline{\Phi}_{\mathfrak{s}}^{(2)}$ in~\eqref{eq:2PIfrgTruncationmCflow} but the problem is that $U$ and $\overline{\Phi}_{\mathfrak{s}}^{(2)}$ have different symmetry properties. Indeed, the condition:
\begin{equation}
U_{\alpha_{1}\alpha_{2}\alpha_{3}\alpha_{4}}=\zeta^{N(P)}U_{\alpha_{P(1)}\alpha_{P(2)}\alpha_{P(3)}\alpha_{P(4)}} \;,
\end{equation}
imposes an invariance of $U$ (up to a sign) under $4!=24$ permutations, which is not equivalent to the conditions set by~\eqref{eq:SymmetryW2PIFRG} for $\overline{\Phi}^{(2)}_{\mathfrak{s}}$ involving also:
\begin{equation}
\overline{\Phi}^{(2)}_{\mathfrak{s},(\alpha_{1},\alpha'_{1})(\alpha_{2},\alpha'_{2})} = \zeta \overline{\Phi}^{(2)}_{\mathfrak{s},(\alpha'_{1},\alpha_{1})(\alpha_{2},\alpha'_{2})} \;,
\end{equation}
but not
\begin{equation}
\overline{\Phi}^{(2)}_{\mathfrak{s},(\alpha_{1},\alpha'_{1})(\alpha_{2},\alpha'_{2})}=\zeta\overline{\Phi}^{(2)}_{\mathfrak{s},(\alpha_{1},\alpha_{2})(\alpha'_{1},\alpha'_{2})} \;,
\end{equation}
for instance. This brings us to the relevance of $\overline{\Phi}_{\mathrm{sym},\mathfrak{s}}^{(2)}$ which is constructed so as to possess the same symmetry properties as $U$~\cite{ren15}:
\begin{equation}
\overline{\Phi}_{\mathrm{sym},\mathfrak{s},\alpha_{1}\alpha_{2}\alpha_{3}\alpha_{4}}^{(2)}=\zeta^{N(P)}\overline{\Phi}_{\mathrm{sym},\mathfrak{s},\alpha_{P(1)}\alpha_{P(2)}\alpha_{P(3)}\alpha_{P(4)}}^{(2)} \;,
\end{equation}
hence the substitution $U \rightarrow \overline{\Phi}_{\mathrm{sym},\mathfrak{s}}^{(2)}$ in~\eqref{eq:2PIfrgTruncationmCflow}.

\end{itemize}

\vspace{0.3cm}

In conclusion, the equation system to solve contains $N_{\mathrm{max}}+2$ differential equations for both the tC-flow and the mC-flow. Only the flow equations for the 2PI vertex of order $N_{\mathrm{max}}$ differ between these two versions of the C-flow:
\begin{subequations}\label{eq:ComparisontCflowmCflow2PIFRG}
\begin{empheq}[left=\empheqlbrace]{align}
& \dot{\overline{\Phi}}_{\mathfrak{s},\gamma_{1}\cdots\gamma_{N_{\mathrm{max}}}}^{(N_{\mathrm{max}})} = \dot{\overline{G}}_{\mathfrak{s},\hat{\gamma}} \overline{\Phi}_{\mathfrak{s}=\mathfrak{s}_{\mathrm{i}},\hat{\gamma}\gamma_{1}\cdots\gamma_{N_{\mathrm{max}}}}^{(N_{\mathrm{max}}+1)} \quad \text{for the tC-flow} \;, \label{eq:ComparisontCflowmCflow2PIFRGUp} \\
\nonumber \\
& \dot{\overline{\Phi}}_{\mathfrak{s},\gamma_{1}\cdots\gamma_{N_{\mathrm{max}}}}^{(N_{\mathrm{max}})} = \dot{\overline{G}}_{\mathfrak{s},\hat{\gamma}} \left(\left.\overline{\Phi}_{\mathrm{SCPT},N_{\mathrm{SCPT}},\mathfrak{s}}^{(N_{\mathrm{max}}+1)}\right|_{U \rightarrow \overline{\Phi}_{\mathrm{sym},\mathfrak{s}}^{(2)}}\right)_{\hat{\gamma}\gamma_{1}\cdots\gamma_{N_{\mathrm{max}}}} \quad \text{for the mC-flow} \;, \label{eq:ComparisontCflowmCflow2PIFRGDown}
\end{empheq}
\end{subequations}
where $N_{\mathrm{SCPT}}$ remains a positive integer to be specified. According to expression~\eqref{eq:2PIFRGperturbativeExpLWfunc} of the Luttinger-Ward functional, a choice $N_{\mathrm{SCPT}}\leq N_{\mathrm{max}}/2$ induces $\overline{\Phi}_{\mathrm{SCPT},N_{\mathrm{SCPT}},\mathfrak{s}}^{(N_{\mathrm{max}}+1)}[U,G] = 0$ $\forall\mathfrak{s}$, which implies that the mC-flow reduces to the tC-flow in this case.

\subsubsection{U-flow}
\label{sec:Uflow2PIFRG}
\paragraph{Main features:}

The U-flow scheme, which was put forward in refs.~\cite{dup14,ren15}, provides an alternative to the C-flow in which the cutoff function $R_{\mathfrak{s}}$ is inserted into the two-body interaction $U$ rather than in the free propagator $C$. In other words, it is based on the substitution $U \rightarrow U_{\mathfrak{s}} = R_{\mathfrak{s}} U$ or equivalently $U \rightarrow U_{\mathfrak{s}} = U + R_{\mathfrak{s}}$. Just like the C-flow, the U-flow implements in principle the Wilsonian momentum-shell integration, with $\mathfrak{s}$ being connected to the momentum scale. This is notably the case with the cutoff function $R_{\mathfrak{s}}$ chosen in ref.~\cite{dup14}\footnote{See notably section IV.C. of ref.~\cite{dup14}.} which plays the role of an IR regulator for (low-energy) collective fluctuations, thus preventing problematic divergences during the flow. However, as we will see in our zero-dimensional applications of section~\ref{sec:2PIFRGuflow0DON}, a perfectly valid choice (even in finite dimensions) for $R_{\mathfrak{s}}$ could be set by $U_{\mathfrak{s}} = R_{\mathfrak{s}} U = \mathfrak{s} U$, with $\mathfrak{s}$ a dimensionless parameter. This follows the philosophy of the 2PPI-FRG discussed in section~\ref{sec:2PPIFRG} and the resulting 2PI-FRG implementation does not carry out the momentum-shell integration \`{a} la Wilson. However, we will see that such a choice for $R_{\mathfrak{s}}$ does not necessarily diminish the power of the U-flow version of the 2PI-FRG (see notably the discussion below~\eqref{eq:TermModifLegendreTransf2PIFRGanySCPTstartingPoint} to conclude section~\ref{sec:Uflow2PIFRG}). The boundary conditions for $U_{\mathfrak{s}}$ are:
\begin{subequations}
\begin{empheq}[left=\empheqlbrace]{align}
& U_{\mathfrak{s}=\mathfrak{s}_{\mathrm{i}},\gamma_{1}\gamma_{2}} = 0 \quad \forall \gamma_{1}, \gamma_{2} \;. \label{eq:2PIfrgUflowBoundaryCondUpper}\\
\nonumber \\
& U_{\mathfrak{s}=\mathfrak{s}_{\mathrm{f}}} = U \;. \label{eq:2PIfrgUflowBoundaryCondBottom}
\end{empheq}
\end{subequations}
Hence, the starting point of the flow corresponds to the free theory (or to the results of self-consistent PT, as explained below in more detail) in the U-flow scheme and, just as for the C-flow, the fully interacting quantum theory is recovered at the end of the flow.

\vspace{0.5cm}

In the framework of the U-flow, we will frequently consider the pair propagator $\Pi[G]$ defined as:
\begin{equation}
\Pi_{\gamma_{1}\gamma_{2}}[G] \equiv W^{(2)}_{0,\gamma_{1}\gamma_{2}}[K(G)] \equiv \frac{\delta^{2}W_{0}[K(G)]}{\delta K_{\gamma_{1}} \delta K_{\gamma_{2}}} \;,
\label{eq:2PIfrgDefinitionPairPropagatorPi}
\end{equation}
from the free version of the generating functional given by~\eqref{eq:2PIFRGgeneratingFunc}, i.e.:
\begin{equation}
Z_{0}[K] = e^{W_{0}[K]}= \int\mathcal{D}\widetilde{\psi} \ e^{-S_{0}\big[\widetilde{\psi}\big] + \frac{1}{2}\int_{\alpha,\alpha'}\widetilde{\psi}_{\alpha}K_{\alpha\alpha'}\widetilde{\psi}_{\alpha'}} \;.
\end{equation}
It can be shown that definition~\eqref{eq:2PIfrgDefinitionPairPropagatorPi} is equivalent to (see appendix~\ref{ann:BetheSalpeterEq}):
\begin{equation}
\Pi_{\gamma_{1} \gamma_{2}}[G] = G_{\alpha_{1} \alpha'_{2}} G_{\alpha'_{1} \alpha_{2}} + \zeta G_{\alpha_{1} \alpha_{2}} G_{\alpha'_{1} \alpha'_{2}} \;.
\label{eq:2PIfrgExpressionPiandG}
\end{equation}
The inverse pair propagator is given by (see appendix~\ref{ann:BetheSalpeterEq}):
\begin{equation}
\Pi^{\mathrm{inv}}_{\gamma_{1} \gamma_{2}}[G] = G^{-1}_{\alpha_{1} \alpha'_{2}} G^{-1}_{\alpha'_{1} \alpha_{2}} + \zeta G^{-1}_{\alpha_{1} \alpha_{2}} G^{-1}_{\alpha'_{1} \alpha'_{2}} \;,
\label{eq:2PIFRGUflowDefInversePairPropagator}
\end{equation}
where the exponent ``$\mathrm{inv}$'' denotes an inverse with respect to bosonic indices, i.e.:
\begin{equation}
\mathcal{I}_{\gamma_{1}\gamma_{2}} = M_{\gamma_{1}\hat{\gamma}} M^{\mathrm{inv}}_{\hat{\gamma}\gamma_{2}} \;,
\label{eq:2PIfrgMinvBosonicIndices}
\end{equation}
for an arbitrary bosonic matrix $M$. Inverses with respect to $\alpha$-indices are still indicated via ``$-1$''  as an exponent. From~\eqref{eq:2PIfrgExpressionPiandG} and~\eqref{eq:2PIFRGUflowDefInversePairPropagator}, we infer the following symmetry properties of the pair propagator and its inverse:
\begin{subequations}\label{eq:2PIFRGUflowSymmetryPairPropagator}
\begin{empheq}[left=\empheqlbrace]{align}
& \Pi_{\gamma_{1}\gamma_{2}}^{(\mathrm{inv})}[G] = \zeta \Pi_{(\alpha'_{1},\alpha_{1})\gamma_{2}}^{(\mathrm{inv})}[G]= \zeta \Pi_{\gamma_{1}(\alpha'_{2},\alpha_{2})}^{(\mathrm{inv})}[G] = \Pi_{(\alpha'_{1},\alpha_{1})(\alpha'_{2},\alpha_{2})}^{(\mathrm{inv})}[G] \;, \label{eq:2PIFRGUflowSymmetryPairPropagatorUp}\\
\nonumber \\
& \Pi_{\gamma_{1}\gamma_{2}}^{(\mathrm{inv})}[G] = \Pi_{\gamma_{2}\gamma_{1}}^{(\mathrm{inv})}[G] \;, \label{eq:2PIFRGUflowSymmetryPairPropagatorDown}
\end{empheq}
\end{subequations}
which are similar to those of the connected correlation functions, i.e. to~\eqref{eq:SymmetryW2PIFRG}. It is also important to note that the pair propagator is related to the derivatives $W^{(2)}[K]$ and $\Phi^{(2)}[G]$ via the Bethe-Salpeter equation:
\begin{equation}
W^{(2)}[K] = \Pi[G] - \Pi[G] \Phi^{(2)}[G] W^{(2)}[K] \;.
\label{eq:2PIfrgBetheSalpeterEquation}
\end{equation}
This equation can be derived from the following equivalent relation (see appendix~\ref{ann:BetheSalpeterEq}):
\begin{equation}
W^{(2)}[K] = \left(\Gamma^{(\mathrm{2PI})(2)}[G]\right)^{\mathrm{inv}} = \left(\Pi^{\mathrm{inv}}[G] + \Phi^{(2)}[G]\right)^{\mathrm{inv}} \;.
\label{eq:2PIfrgUflowW2expression}
\end{equation}

\vspace{0.5cm}

Using the latter equalities (especially~\eqref{eq:2PIfrgExpressionPiandG} and~\eqref{eq:2PIfrgUflowW2expression}), the tower of differential equations associated with the U-flow implementation of the 2PI-FRG can be derived from the generating functional~\eqref{eq:2PIFRGgeneratingFunc} after introducing the cutoff function $R_{\mathfrak{s}}$ in the manner discussed previously. This leads to (see appendix~\ref{ann:2PIfrgFlowEquationUflow} for the corresponding flow equation expressing the derivative of the 2PI vertex of order 3 with respect to $\mathfrak{s}$):
\begin{equation}
\dot{\overline{G}}_{\mathfrak{s},\alpha_{1}\alpha'_{1}}=\int_{\alpha_{2},\alpha'_{2}}\overline{G}_{\mathfrak{s},\alpha_{1} \alpha_{2}} \dot{\overline{\Sigma}}_{\mathfrak{s},\alpha_{2} \alpha'_{2}} \overline{G}_{\mathfrak{s},\alpha'_{2} \alpha'_{1}} \;,
\label{eq:2PIfrgUflowEquationsGeneralFormG}
\end{equation}
\begin{equation}
\dot{\overline{\Omega}}_{\mathfrak{s}} = \frac{1}{6\beta} \dot{U}_{\mathfrak{s},\hat{\gamma}_{1} \hat{\gamma}_{2}} \left(\overline{W}_{\mathfrak{s}}^{(2)} + \frac{1}{2}\overline{\Pi}_{\mathfrak{s}}\right)_{\hat{\gamma}_{2}\hat{\gamma}_{1}} \;,
\label{eq:2PIfrgUflowEquationsGeneralFormOmega}
\end{equation}
\begin{equation}
\begin{split}
\dot{\overline{\Phi}}_{\mathfrak{s}} = & \ \frac{1}{6} \dot{U}_{\mathfrak{s},\hat{\gamma}_{1} \hat{\gamma}_{2}} \left(\overline{W}_{\mathfrak{s}}^{(2)} + \frac{1}{2}\overline{\Pi}_{\mathfrak{s}}\right)_{\hat{\gamma}_{2}\hat{\gamma}_{1}} \\
& +\frac{1}{6} \overline{\Sigma}_{\mathfrak{s},\hat{\gamma}_{1}} \overline{W}_{\mathfrak{s},\hat{\gamma}_{1} \hat{\gamma}_{2}}^{(2)} \dot{U}_{\mathfrak{s},\hat{\gamma}_{3} \hat{\gamma}_{4}} \left[ \overline{W}_{\mathfrak{s},\hat{\gamma}_{4} \hat{\gamma}_{5}}^{(2)} \left(\overline{\Pi}_{\mathfrak{s},\hat{\gamma}_{5}\hat{\gamma}_{6}}^{\mathrm{inv}} \frac{\delta \overline{\Pi}_{\mathfrak{s},\hat{\gamma}_{6}\hat{\gamma}_{7}}}{\delta \overline{G}_{\mathfrak{s},\hat{\gamma}_{2}}} \overline{\Pi}_{\mathfrak{s},\hat{\gamma}_{7}\hat{\gamma}_{8}}^{\mathrm{inv}} - \overline{\Phi}_{\mathfrak{s},\hat{\gamma}_{2} \hat{\gamma}_{5} \hat{\gamma}_{8}}^{(3)} \right) \overline{W}_{\mathfrak{s},\hat{\gamma}_{8} \hat{\gamma}_{3}}^{(2)} + \frac{1}{2} \frac{\delta \overline{\Pi}_{\mathfrak{s},\hat{\gamma}_{4} \hat{\gamma}_{3}}}{\delta \overline{G}_{\mathfrak{s},\hat{\gamma}_{2}}} \right] \;,
\end{split}
\label{eq:2PIfrgUflowEquationsGeneralFormPhi}
\end{equation}
\begin{equation}
\begin{split}
\dot{\overline{\Sigma}}_{\mathfrak{s},\gamma} = & -\frac{1}{3}\left(\mathcal{I} + \overline{\Pi}_{\mathfrak{s}} \overline{\Phi}_{\mathfrak{s}}^{(2)}\right)^{\mathrm{inv}}_{\gamma\hat{\gamma}_{1}} \left[ 2 \left(\mathcal{I} + \overline{\Pi}_{\mathfrak{s}} \overline{\Phi}_{\mathfrak{s}}^{(2)}\right)^{\mathrm{inv}} \dot{U}_{\mathfrak{s}} \left(\mathcal{I} + \overline{\Pi}_{\mathfrak{s}} \overline{\Phi}_{\mathfrak{s}}^{(2)}\right)^{\mathrm{inv}} + \dot{U}_{\mathfrak{s}} \right]_{\hat{\alpha}_{1} \hat{\alpha}_{2} \hat{\alpha}'_{2} \hat{\alpha}'_{1}} \overline{G}_{\mathfrak{s},\hat{\gamma}_{2}} \\
& +\frac{1}{6} \left(\mathcal{I} + \overline{\Pi}_{\mathfrak{s}} \overline{\Phi}_{\mathfrak{s}}^{(2)}\right)^{\mathrm{inv}}_{\gamma\hat{\gamma}_{1}} \dot{U}_{\mathfrak{s},\hat{\gamma}_{2} \hat{\gamma}_{3}} \overline{W}_{\mathfrak{s},\hat{\gamma}_{3} \hat{\gamma}_{4}}^{(2)} \overline{\Phi}_{\mathfrak{s},\hat{\gamma}_{1} \hat{\gamma}_{4} \hat{\gamma}_{5}}^{(3)} \overline{W}_{\mathfrak{s},\hat{\gamma}_{5} \hat{\gamma}_{2}}^{(2)} \;,
\end{split}
\label{eq:2PIfrgUflowEquationsGeneralFormSigma}
\end{equation}
\begin{equation}
\begin{split}
\dot{\overline{\Phi}}_{\mathfrak{s},\gamma_{1}\gamma_{2}}^{(2)} = \frac{1}{3} \dot{U}_{\mathfrak{s},\hat{\gamma}_{1}\hat{\gamma}_{2}} & \Bigg[\overline{W}_{\mathfrak{s},\hat{\gamma}_{2} \hat{\gamma}_{3}}^{(2)} \left(\overline{\Pi}_{\mathfrak{s},\hat{\gamma}_{3} \hat{\gamma}_{4}}^{\mathrm{inv}} \frac{\delta \overline{\Pi}_{\mathfrak{s},\hat{\gamma}_{4} \hat{\gamma}_{5}}}{\delta \overline{G}_{\mathfrak{s},\gamma_{1}}} \overline{\Pi}_{\mathfrak{s},\hat{\gamma}_{5} \hat{\gamma}_{6}}^{\mathrm{inv}} - \overline{\Phi}_{\mathfrak{s},\gamma_{1}\hat{\gamma}_{3}\hat{\gamma}_{6}}^{(3)}\right)\overline{W}_{\mathfrak{s},\hat{\gamma}_{6} \hat{\gamma}_{7}}^{(2)} \\
& \hspace{0.2cm} \times \left(\overline{\Pi}_{\mathfrak{s},\hat{\gamma}_{7} \hat{\gamma}_{8}}^{\mathrm{inv}} \frac{\delta \overline{\Pi}_{\mathfrak{s},\hat{\gamma}_{8} \hat{\gamma}_{9}}}{\delta \overline{G}_{\mathfrak{s},\gamma_{2}}} \overline{\Pi}_{\mathfrak{s},\hat{\gamma}_{9} \hat{\gamma}_{10}}^{\mathrm{inv}} - \overline{\Phi}_{\mathfrak{s},\gamma_{2} \hat{\gamma}_{7} \hat{\gamma}_{10}}^{(3)}\right) \overline{W}_{\mathfrak{s},\hat{\gamma}_{10} \hat{\gamma}_{1}}^{(2)} \\
& - \overline{W}_{\mathfrak{s},\hat{\gamma}_{2} \hat{\gamma}_{3}}^{(2)} \overline{\Pi}_{\mathfrak{s},\hat{\gamma}_{3} \hat{\gamma}_{4}}^{\mathrm{inv}} \frac{\delta \overline{\Pi}_{\mathfrak{s},\hat{\gamma}_{4} \hat{\gamma}_{5}}}{\delta \overline{G}_{\mathfrak{s},\gamma_{1}}} \overline{\Pi}_{\mathfrak{s},\hat{\gamma}_{5} \hat{\gamma}_{6}}^{\mathrm{inv}} \frac{\delta \overline{\Pi}_{\mathfrak{s},\hat{\gamma}_{6} \hat{\gamma}_{7}}}{\delta \overline{G}_{\mathfrak{s},\gamma_{2}}} \overline{\Pi}_{\mathfrak{s},\hat{\gamma}_{7} \hat{\gamma}_{8}}^{\mathrm{inv}} \overline{W}_{\mathfrak{s},\hat{\gamma}_{8} \hat{\gamma}_{1}}^{(2)} \\
& + \frac{1}{2} \overline{W}_{\mathfrak{s},\hat{\gamma}_{2} \hat{\gamma}_{3}}^{(2)} \left(\overline{\Pi}_{\mathfrak{s},\hat{\gamma}_{3} \hat{\gamma}_{4}}^{\mathrm{inv}} \frac{\delta^{2} \overline{\Pi}_{\mathfrak{s},\hat{\gamma}_{4} \hat{\gamma}_{5}}}{\delta \overline{G}_{\mathfrak{s},\gamma_{1}} \delta \overline{G}_{\mathfrak{s},\gamma_{2}}} \overline{\Pi}_{\mathfrak{s},\hat{\gamma}_{5} \hat{\gamma}_{6}}^{\mathrm{inv}} - \overline{\Phi}_{\mathfrak{s},\gamma_{1}\gamma_{2}\hat{\gamma}_{3}\hat{\gamma}_{6}}^{(4)}\right) \overline{W}_{\mathfrak{s},\hat{\gamma}_{6} \hat{\gamma}_{1}}^{(2)} \\
& + \frac{1}{4} \frac{\delta^{2}\overline{\Pi}_{\mathfrak{s},\hat{\gamma}_{2}\hat{\gamma}_{1}}}{\delta \overline{G}_{\mathfrak{s},\gamma_{1}} \delta \overline{G}_{\mathfrak{s},\gamma_{2}}}\Bigg] + \dot{\overline{G}}_{\mathfrak{s},\hat{\gamma}} \overline{\Phi}_{\mathfrak{s},\hat{\gamma}\gamma_{1}\gamma_{2}}^{(3)} \;,
\end{split}
\label{eq:2PIfrgUflowEquationsGeneralFormPhi2}
\end{equation}
where
\begin{equation}
\mathcal{I}_{\gamma_{1}\gamma_{2}} \equiv \frac{\delta G_{\gamma_{1}}}{\delta G_{\gamma_{2}}} = \delta_{\alpha_{1}\alpha_{2}} \delta_{\alpha'_{1}\alpha'_{2}} + \zeta \delta_{\alpha_{1}\alpha'_{2}} \delta_{\alpha'_{1}\alpha_{2}} \;,
\label{eq:IdentityBosonicMatrix2PIFRG}
\end{equation}
denotes the components of the identity matrix in the bosonic index formalism (see appendix~\ref{ann:BosonicIndices2PI}) and we have used the notation:
\begin{equation}
\frac{\delta^{n}\overline{\Pi}_{\mathfrak{s}}}{\delta \overline{G}_{\mathfrak{s},\gamma_{1}} \cdots \delta \overline{G}_{\mathfrak{s},\gamma_{n}}} \equiv \left.\frac{\delta^{n}\Pi[G]}{\delta G_{\gamma_{1}} \cdots \delta G_{\gamma_{n}}}\right|_{G=\overline{G}_{\mathfrak{s}}} \;.
\label{eq:ShorthandNotation2PIFRGUflow}
\end{equation}
The components of the bosonic matrix $\overline{W}_{\mathfrak{s}}^{(2)}$ involved in the differential equations~\eqref{eq:2PIfrgUflowEquationsGeneralFormOmega} to~\eqref{eq:2PIfrgUflowEquationsGeneralFormPhi2} must \textit{a priori} be determined by solving the Bethe-Salpeter equation (either in the form of~\eqref{eq:2PIfrgBetheSalpeterEquation} or~\eqref{eq:2PIfrgUflowW2expression}), unless we use a different (supposedly drastic) approximation for $\overline{W}_{\mathfrak{s}}^{(2)}$. This remark is not without consequences for the U-flow as this numerical resolution, which might be demanding for realistic models, must be repeated at each step of the flow.

\vspace{0.5cm}

We discuss two main implementations of the U-flow: the plain U-flow (pU-flow) and the modified U-flow (mU-flow). As opposed to the C-flow, such implementations do not only differ by the truncation of the hierarchy based on~\eqref{eq:2PIfrgUflowEquationsGeneralFormG} to~\eqref{eq:2PIfrgUflowEquationsGeneralFormPhi2}. We will see in particular that, as opposed to the pU-flow, the mU-flow does not directly rely on the latter equation system such that the starting point for these two versions of the U-flow are sharply different as well.

\paragraph{pU-flow:}

\begin{itemize}
\item Truncation:\\
The pU-flow consists in solving the tower of differential equations including~\eqref{eq:2PIfrgUflowEquationsGeneralFormG} to~\eqref{eq:2PIfrgUflowEquationsGeneralFormPhi2} with the truncation established by~\eqref{eq:2PIfrgPhiBartCflow}. In that respect, the pU-flow is the counterpart of the tC-flow for the U-flow.

\item Initial conditions:\\
As implied by~\eqref{eq:2PIfrgUflowBoundaryCondUpper}, the starting point of the pU-flow coincides with the free theory. In this situation, the Luttinger-Ward functional vanishes by definition, as well as the self-energy and all corresponding 2PI vertices, so that the initial conditions for the pU-flow read:
\begin{equation}
\overline{G}_{\mathfrak{s}=\mathfrak{s}_{\mathrm{i}}}=C \mathrlap{\;,}
\label{eq:InitialConditions2PIFRGpUflowG}
\end{equation}
\begin{equation}
\overline{\Omega}_{\mathfrak{s}=\mathfrak{s}_{\mathrm{i}}} = \frac{1}{\beta}\Gamma^{(\mathrm{2PI})}_{0}[G=C] = -\frac{\zeta}{2\beta} \mathrm{Tr}_{\alpha} \left[ \mathrm{ln}(C) \right] \mathrlap{\;,}
\label{eq:InitialConditions2PIFRGpUflowOmega}
\end{equation}
\begin{equation}
\overline{\Phi}_{\mathfrak{s}=\mathfrak{s}_{\mathrm{i}}}=0 \mathrlap{\;,}
\label{eq:InitialConditions2PIFRGpUflowPhi}
\end{equation}
\begin{equation}
\overline{\Sigma}_{\mathfrak{s}=\mathfrak{s}_{\mathrm{i}},\gamma}=0 \mathrlap{\quad \forall \gamma \;,}
\label{eq:InitialConditions2PIFRGpUflowSigma}
\end{equation}
\begin{equation}
\overline{\Phi}_{\mathfrak{s}=\mathfrak{s}_{\mathrm{i}},\gamma_{1},\cdots,\gamma_{n}}^{(n)}=0 \mathrlap{\quad \forall \gamma_{1}, \cdots, \gamma_{n} , ~ \forall n \geq 2 \;,}
\label{eq:InitialConditions2PIFRGpUflowPhin}
\end{equation}
where~\eqref{eq:InitialConditions2PIFRGpUflowOmega} is deduced from~\eqref{eq:2PIFRGfreeGamma2PI} with the condition $\overline{G}_{\mathfrak{s}=\mathfrak{s}_{\mathrm{i}}}=C$ which directly follows from Dyson equation~\eqref{eq:2PIfrgDysonEqGbar} combined with~\eqref{eq:InitialConditions2PIFRGpUflowSigma}.

\item tU-flow:\\
The truncated U-flow (tU-flow) is a computationally more affordable version of the pU-flow. In the framework of the tU-flow,~\eqref{eq:2PIfrgUflowW2expression} is truncated in a drastic fashion so as to obtain a simple expression of $\overline{W}_{\mathfrak{s}}^{(2)}$, thus bypassing the necessity to solve the Bethe-Salpeter equation. This approximation is applied to~\eqref{eq:2PIfrgUflowW2expression} as follows:
\begin{equation}
\overline{W}_{\mathfrak{s}}^{(2)}=\overline{\Pi}_{\mathfrak{s}}+\mathcal{O}\left(\overline{\Phi}_{\mathfrak{s}}^{(2)}\right) \simeq \overline{\Pi}_{\mathfrak{s}} \;.
\label{eq:2PIfrgExampleNmax2}
\end{equation}
Such a truncation indeed induces a cancellation of the inverse propagators involved e.g. in the flow equations~\eqref{eq:2PIfrgUflowEquationsGeneralFormPhi} and~\eqref{eq:2PIfrgUflowEquationsGeneralFormPhi2} via the relation:
\begin{equation}
\mathcal{I}_{\gamma_{1}\gamma_{2}} = \overline{\Pi}_{\mathfrak{s},\gamma_{1}\hat{\gamma}}\overline{\Pi}^{\mathrm{inv}}_{\mathfrak{s},\hat{\gamma}\gamma_{2}} \simeq \overline{W}^{(2)}_{\mathfrak{s},\gamma_{1}\hat{\gamma}}\overline{\Pi}^{\mathrm{inv}}_{\mathfrak{s},\hat{\gamma}\gamma_{2}} \;.
\label{eq:2PIfrgUflowUsefulIdentitytUflow}
\end{equation}
Finally, the flowing 2PI vertices in the tU-flow scheme are still selected according to the truncation condition~\eqref{eq:2PIfrgPhiBartCflow}. The tU-flow is actually particularly suited to deal with the truncation order $N_{\mathrm{max}}=2$. In this situation, the differential equation expressing $\dot{\overline{\Phi}}_{\mathfrak{s}}^{(2)}$ reduces to a form simple enough to be directly integrated so that the equation system to solve becomes (see appendix~\ref{ann:2PIfrgFlowEquationUflow}):
\begin{equation}
\dot{\overline{G}}_{\mathfrak{s},\alpha_{1}\alpha'_{1}}=\int_{\alpha_{2},\alpha'_{2}}\overline{G}_{\mathfrak{s},\alpha_{1} \alpha_{2}} \dot{\overline{\Sigma}}_{\mathfrak{s},\alpha_{2} \alpha'_{2}} \overline{G}_{\mathfrak{s},\alpha'_{2} \alpha'_{1}} \;,
\label{eq:2PIfrgtruncatedUflowExpressionGdot}
\end{equation}
\begin{equation}
\dot{\overline{\Omega}}_{\mathfrak{s}} = \frac{1}{6\beta} \dot{U}_{\mathfrak{s},\hat{\gamma}_{1}\hat{\gamma}_{2}} \left[\left(\overline{\Pi}^{\mathrm{inv}}_{\mathfrak{s}} + U_{\mathfrak{s}}\right)^{\mathrm{inv}} + \frac{1}{2}\overline{\Pi}_{\mathfrak{s}}\right]_{\hat{\gamma}_{2}\hat{\gamma}_{1}} \;,
\label{eq:2PIfrgtruncatedUflowExpressionOmegadot}
\end{equation}
\begin{equation}
\begin{split}
\scalebox{0.98}{${\displaystyle \dot{\overline{\Phi}}_{\mathfrak{s}} = }$} & \ \scalebox{0.98}{${\displaystyle \frac{1}{6} \dot{U}_{\mathfrak{s},\hat{\gamma}_{1}\hat{\gamma}_{2}} \left[\left(\overline{\Pi}^{\mathrm{inv}}_{\mathfrak{s}} + U_{\mathfrak{s}}\right)^{\mathrm{inv}} + \frac{1}{2}\overline{\Pi}_{\mathfrak{s}}\right]_{\hat{\gamma}_{2}\hat{\gamma}_{1}} }$} \\
& \scalebox{0.98}{${\displaystyle +\frac{1}{6} \overline{\Sigma}_{\mathfrak{s},\hat{\gamma}_{1}} \left(\overline{\Pi}^{\mathrm{inv}}_{\mathfrak{s}} + U_{\mathfrak{s}}\right)^{\mathrm{inv}}_{\hat{\gamma}_{1} \hat{\gamma}_{2}} \dot{U}_{\mathfrak{s},\hat{\gamma}_{3} \hat{\gamma}_{4}} \Bigg[\left(\mathcal{I} + \overline{\Pi}_{\mathfrak{s}} U_{\mathfrak{s}}\right)^{\mathrm{inv}}_{\hat{\gamma}_{4} \hat{\gamma}_{5}} \frac{\delta \overline{\Pi}_{\mathfrak{s},\hat{\gamma}_{5}\hat{\gamma}_{6}}}{\delta \overline{G}_{\mathfrak{s},\hat{\gamma}_{2}}} \left(\mathcal{I} + \overline{\Pi}_{\mathfrak{s}} U_{\mathfrak{s}}\right)^{\mathrm{inv}}_{\hat{\gamma}_{6} \hat{\gamma}_{3}} + \frac{1}{2} \frac{\delta \overline{\Pi}_{\mathfrak{s},\hat{\gamma}_{4} \hat{\gamma}_{3}}}{\delta \overline{G}_{\mathfrak{s},\hat{\gamma}_{2}}} \Bigg] \;, }$}
\end{split}
\label{eq:2PIfrgtruncatedUflowExpressionPhidot}
\end{equation}
\begin{equation}
\dot{\overline{\Sigma}}_{\mathfrak{s},\gamma} = -\frac{1}{6}\left(\mathcal{I} + \overline{\Pi}_{\mathfrak{s}} U_{\mathfrak{s}}\right)^{\mathrm{inv}}_{\gamma\hat{\gamma}_{1}} \dot{U}_{\mathfrak{s},\hat{\gamma}_{2} \hat{\gamma}_{3}} \Bigg[\left(\mathcal{I} + \overline{\Pi}_{\mathfrak{s}} U_{\mathfrak{s}}\right)_{\hat{\gamma}_{3}\hat{\gamma}_{4}}^{\mathrm{inv}} \frac{\delta \overline{\Pi}_{\mathfrak{s},\hat{\gamma}_{4}\hat{\gamma}_{5}}}{\delta \overline{G}_{\mathfrak{s},\hat{\gamma}_{1}}} \left(\mathcal{I} + \overline{\Pi}_{\mathfrak{s}} U_{\mathfrak{s}}\right)_{\hat{\gamma}_{5}\hat{\gamma}_{2}}^{\mathrm{inv}} + \frac{1}{2} \frac{\delta \overline{\Pi}_{\mathfrak{s},\hat{\gamma}_{3}\hat{\gamma}_{2}}}{\delta \overline{G}_{\mathfrak{s},\hat{\gamma}_{1}}} \Bigg] \;,
\label{eq:2PIfrgtruncatedUflowExpressionSigmadot}
\end{equation}
\begin{equation}
\overline{\Phi}_{\mathfrak{s},\gamma_{1}\gamma_{2}}^{(2)} = U_{\mathfrak{s},\gamma_{1}\gamma_{2}} \;.
\label{eq:2PIfrgtruncatedUflowExpressionPhi2dot}
\end{equation}
In summary,~\eqref{eq:2PIfrgtruncatedUflowExpressionGdot} to~\eqref{eq:2PIfrgtruncatedUflowExpressionPhi2dot} are respectively obtained from the pU-flow equations \eqref{eq:2PIfrgUflowEquationsGeneralFormG} to~\eqref{eq:2PIfrgUflowEquationsGeneralFormPhi2} by imposing that all components of $\overline{\Phi}_{\mathfrak{s}}^{(3)}$ and $\overline{\Phi}_{\mathfrak{s}}^{(4)}$ vanish (to enforce the truncation order $N_{\mathrm{max}}=2$) and by exploiting the approximation~\eqref{eq:2PIfrgExampleNmax2} in the form of~\eqref{eq:2PIfrgUflowUsefulIdentitytUflow} (to implement the tU-flow).

\end{itemize}

\paragraph{mU-flow:}

\begin{itemize}
\item Definition of the modified Luttinger-Ward functional:\\
The mU-flow is based on the following transformation of the Luttinger-Ward functional:
\begin{equation}
\boldsymbol{\Phi}_{\mathfrak{s}}[G] \equiv \Phi_{\mathfrak{s}}[G] + \Phi_{\mathrm{SCPT},N_{\mathrm{SCPT}}}[U,G] -\Phi_{\mathrm{SCPT},N_{\mathrm{SCPT}}}[U_{\mathfrak{s}},G] \;,
\label{eq:2PIfrgDefinitionmUflow}
\end{equation}
where $\boldsymbol{\Phi}_{\mathfrak{s}}[G]$ will be referred to as the modified Luttinger-Ward functional and the functional $\Phi_{\mathrm{SCPT},N_{\mathrm{SCPT}}}[U,G]$ was already introduced in~\eqref{eq:2PIfrgTruncationmCflow} to present the mC-flow. As a consequence of~\eqref{eq:2PIfrgDefinitionmUflow} combined with~\eqref{eq:DefLWfunctional}, we can also define a bold counterpart for the 2PI EA $\Gamma_{\mathfrak{s}}^{(\mathrm{2PI})}[G]$ as:
\begin{equation}
\boldsymbol{\Gamma}^{(\mathrm{2PI})}_{\mathfrak{s}}[G] \equiv \Gamma^{(\mathrm{2PI})}_{\mathfrak{s}}[G] + \Phi_{\mathrm{SCPT},N_{\mathrm{SCPT}}}[U,G]-\Phi_{\mathrm{SCPT},N_{\mathrm{SCPT}}}[U_{\mathfrak{s}},G] \;,
\label{eq:2PIfrgDefinitionmUflowEA}
\end{equation}
with the corresponding thermodynamic potential:
\begin{equation}
\boldsymbol{\Omega}_{\mathfrak{s}}[G]=\frac{1}{\beta}\boldsymbol{\Gamma}^{(\mathrm{2PI})}_{\mathfrak{s}}[G] \;,
\end{equation}
and the modified 2PI vertices $\boldsymbol{\Phi}^{(n)}_{\mathfrak{s},\gamma_{1}\cdots\gamma_{n}}[G] \equiv \frac{\delta^{n} \boldsymbol{\Phi}_{\mathfrak{s}}[G]}{\delta G_{\gamma_{1}} \cdots \delta G_{\gamma_{n}}}$ satisfy:
\begin{equation}
\hspace{1.8cm} \boldsymbol{\Phi}^{(n)}_{\mathfrak{s}}[G] \equiv \Phi^{(n)}_{\mathfrak{s}}[G] + \Phi^{(n)}_{\mathrm{SCPT},N_{\mathrm{SCPT}}}[U,G]-\Phi^{(n)}_{\mathrm{SCPT},N_{\mathrm{SCPT}}}[U_{\mathfrak{s}},G] \quad \forall n \in \mathbb{N}^{*} \;,
\label{eq:2PIfrgDefinitionmodified2PIverticesmUflow}
\end{equation}
which, at $n=1$, gives us the self-energy:
\begin{equation}
\boldsymbol{\Sigma}_{\mathfrak{s},\gamma}[G] \equiv - \frac{\delta \boldsymbol{\Phi}_{\mathfrak{s}}[G]}{\delta G_{\gamma}} \;.
\label{eq:boldselfenergy2PIFRGmUflow}
\end{equation}
We can also introduce the configuration $\overline{\boldsymbol{G}}_{\mathfrak{s}}$ of the propagator $G$ which extremizes the bold 2PI EA of~\eqref{eq:2PIfrgDefinitionmUflowEA} according to:
\begin{equation}
\left. \frac{\delta\boldsymbol{\Gamma}_{\mathfrak{s}}^{(\mathrm{2PI})}[G]}{\delta G_{\gamma}}\right|_{G=\overline{\boldsymbol{G}}_{\mathfrak{s}}} = 0 \mathrlap{\quad \forall \gamma,\mathfrak{s} \;,}
\label{eq:2PIfrgExtremizeboldGamma2PImUflow}
\end{equation}
which is the counterpart of~\eqref{eq:2PIfrgExtremizeGamma2PI}. Note that~\eqref{eq:2PIfrgExtremizeboldGamma2PImUflow} can also be rewritten in the form of a Dyson equation:
\begin{equation}
\overline{\boldsymbol{G}}_{\mathfrak{s},\gamma} = \left(C^{-1}-\overline{\boldsymbol{\Sigma}}_{\mathfrak{s}}\right)_{\gamma}^{-1} \;,
\label{eq:2PIfrgDysonEqGbarmUflow}
\end{equation}
using~\eqref{eq:boldselfenergy2PIFRGmUflow}. Although the relevance of the splitting of~\eqref{eq:2PIfrgDefinitionmUflow} will become clearer with the following discussion on the initial conditions for the present approach, we can already state at this stage that the general idea underlying the mU-flow is to calculate $\overline{\boldsymbol{\Phi}}_{\mathfrak{s}}\equiv\boldsymbol{\Phi}_{\mathfrak{s}}\big[G=\overline{\boldsymbol{G}}_{\mathfrak{s}}\big]$ and its derivatives:
\begin{equation}
\overline{\boldsymbol{\Phi}}^{(n)}_{\mathfrak{s},\gamma_{1} \cdots \gamma_{n}} \equiv \left.\frac{\delta^{n} \boldsymbol{\Phi}_{\mathfrak{s}}[G]}{\delta G_{\gamma_{1}} \cdots \delta G_{\gamma_{n}}}\right|_{G=\overline{\boldsymbol{G}}_{\mathfrak{s}}} \;,
\label{eq:2PIfrgDefinitionboldPhinmUflow}
\end{equation}
instead of $\Phi_{\mathfrak{s}}\big[G=\overline{G}_{\mathfrak{s}}\big]$ and the corresponding 2PI vertices during the flow. The mU-flow equations can therefore be obtained from the pU-flow ones (e.g.~\eqref{eq:2PIfrgUflowEquationsGeneralFormG} to~\eqref{eq:2PIfrgUflowEquationsGeneralFormPhi2}) by substituting the flowing quantities $\overline{G}_{\mathfrak{s}}$, $\overline{\Omega}_{\mathfrak{s}}$, $\overline{\Phi}_{\mathfrak{s}}$, $\overline{\Sigma}_{\mathfrak{s}}$ and $\overline{\Phi}^{(n)}_{\mathfrak{s}}$ (with $n \geq 2$) by their bold counterparts (i.e. $\overline{\boldsymbol{G}}_{\mathfrak{s}}$, $\overline{\boldsymbol{\Omega}}_{\mathfrak{s}}$, $\overline{\boldsymbol{\Phi}}_{\mathfrak{s}}$, $\overline{\boldsymbol{\Sigma}}_{\mathfrak{s}}$ and $\overline{\boldsymbol{\Phi}}^{(n)}_{\mathfrak{s}}$ with $n \geq 2$, respectively) according to the above definitions. That being so, we stress that, according to~\eqref{eq:2PIfrgUflowBoundaryCondBottom}, the functional $\boldsymbol{\Phi}_{\mathfrak{s}}[G]$ and all other bold entities introduced here are constructed such that they coincide with their original counterparts at the end of the flow, e.g. $\boldsymbol{\Phi}_{\mathfrak{s}=\mathfrak{s}_{\mathrm{f}}}[G]=\Phi_{\mathfrak{s}=\mathfrak{s}_{\mathrm{f}}}[G]$. Physical quantities are thus still recovered at $\mathfrak{s}=\mathfrak{s}_{\mathrm{f}}$. It is also important to keep in mind that, in the present discussion on the mU-flow and notably in~\eqref{eq:2PIfrgDysonEqGbarmUflow}, all upper bars label a functional evaluated at $G=\overline{\boldsymbol{G}}_{\mathfrak{s}}$ and not at $G=\overline{G}_{\mathfrak{s}}$ as opposed to all other 2PI-FRG implementations presented in this chapter.

\item Truncation:\\
The truncation is no longer implemented by the condition~\eqref{eq:2PIfrgPhiBartCflow} which is replaced by:
\begin{equation}
\overline{\boldsymbol{\Phi}}_{\mathfrak{s}}^{(n)}=\overline{\boldsymbol{\Phi}}_{\mathfrak{s}=\mathfrak{s}_{\mathrm{i}}}^{(n)} \mathrlap{\quad \forall \mathfrak{s}, ~ \forall n > N_{\mathrm{max}} \;.}
\label{eq:2PIfrgModifiedUflowTruncation}
\end{equation}
The truncation of the mU-flow is therefore more refined than that of the pU-flow. Indeed, whereas the initial conditions $\overline{\Phi}_{\mathfrak{s}=\mathfrak{s}_{\mathrm{i}}}^{(n)}$ are simply inferred from the free theory in~\eqref{eq:InitialConditions2PIFRGpUflowSigma} and~\eqref{eq:InitialConditions2PIFRGpUflowPhin}, the quantities $\overline{\boldsymbol{\Phi}}_{\mathfrak{s}=\mathfrak{s}_{\mathrm{i}}}^{(n)}$ contain non-perturbative information transcribing the correlations resummed via self-consistent PT according to~\eqref{eq:2PIfrgDefinitionmodified2PIverticesmUflow} and~\eqref{eq:2PIfrgDefinitionboldPhinmUflow} alongside with~\eqref{eq:2PIfrgUflowBoundaryCondUpper}, \eqref{eq:InitialConditions2PIFRGpUflowSigma} and \eqref{eq:InitialConditions2PIFRGpUflowPhin}.

\item Initial conditions:\\
Let us discuss the latter statement in more detail. According to the initial condition $U_{\mathfrak{s}=\mathfrak{s}_{\mathrm{i}}}=0$ set by~\eqref{eq:2PIfrgUflowBoundaryCondUpper},~\eqref{eq:2PIfrgDefinitionmUflow} and~\eqref{eq:2PIfrgDefinitionmodified2PIverticesmUflow} reduce at $\mathfrak{s}=\mathfrak{s}_{\mathrm{i}}$ to:
\begin{equation}
\boldsymbol{\Phi}_{\mathfrak{s}=\mathfrak{s}_{\mathrm{i}}}[G] = \Phi_{\mathrm{SCPT},N_{\mathrm{SCPT}}}[U,G] \mathrlap{\;,}
\label{eq:PhiboldInitialCondmUflow}
\end{equation}
\begin{equation}
\boldsymbol{\Sigma}_{\mathfrak{s}=\mathfrak{s}_{\mathrm{i}}}[G] = -\Phi^{(1)}_{\mathrm{SCPT},N_{\mathrm{SCPT}}}[U,G] \mathrlap{\;,}
\label{eq:SigmaboldInitialCondmUflow}
\end{equation}
\begin{equation}
\boldsymbol{\Phi}^{(n)}_{\mathfrak{s}=\mathfrak{s}_{\mathrm{i}}}[G] = \Phi^{(n)}_{\mathrm{SCPT},N_{\mathrm{SCPT}}}[U,G] \mathrlap{\quad \forall n \geq 2 \;.}
\label{eq:PhinboldInitialCondmUflow}
\end{equation}
Hence, at the starting point of the flow, $\boldsymbol{\Phi}_{\mathfrak{s}}[G]$, $\boldsymbol{\Sigma}_{\mathfrak{s}}[G]$ and $\boldsymbol{\Phi}^{(n)}_{\mathfrak{s}}[G]$ possess the analytical form respectively of $\Phi[G]$, $\Sigma[G]$ and $\Phi^{(n)}[G]$ in the framework of self-consistent PT up to order $\mathcal{O}\big(U^{N_{\mathrm{SCPT}}}\big)$. However, we are more specifically interested in these functionals evaluated at $G=\overline{\boldsymbol{G}}_{\mathfrak{s}}$, so we must now address the determination of $\overline{\boldsymbol{G}}_{\mathfrak{s}=\mathfrak{s}_{\mathrm{i}}}$. The latter configuration can be determined from Dyson equation~\eqref{eq:2PIfrgDysonEqGbarmUflow}, i.e. from:
\begin{equation}
\overline{\boldsymbol{G}}_{\mathfrak{s}=\mathfrak{s}_{\mathrm{i}},\gamma} = \left(C^{-1}-\overline{\boldsymbol{\Sigma}}_{\mathfrak{s}=\mathfrak{s}_{\mathrm{i}}}\right)_{\gamma}^{-1} \;.
\label{eq:2PIfrgDysonEqGbarsimUflow}
\end{equation}
Since, as we just discussed with~\eqref{eq:SigmaboldInitialCondmUflow}, $\boldsymbol{\Sigma}_{\mathfrak{s}=\mathfrak{s}_{\mathrm{i}}}[G]$ coincides with a truncated perturbative expression of the self-energy,~\eqref{eq:2PIfrgDysonEqGbarsimUflow} is nothing else than the usual gap equation for the propagator encountered in the CJT formalism. If we denote as $\overline{G}_{\mathrm{SCPT},N_{\mathrm{SCPT}}}$ a chosen solution\footnote{The self-consistent gap equation~\eqref{eq:2PIfrgDysonEqGbarsimUflow} possesses \textit{a priori} several solutions. In other words, there are several possible starting points of the mU-flow for a given approximation of $\Phi_{\mathrm{SCPT}}[U,G]$ (i.e. for a given $N_{\mathrm{SCPT}}$). Such a freedom can be seen as a considerable advantage of the mU-flow approach, as discussed below~\eqref{eq:TermModifLegendreTransf2PIFRGanySCPTstartingPoint}.} of this self-consistent equation, we have:
\begin{equation}
\overline{\boldsymbol{G}}_{\mathfrak{s}=\mathfrak{s}_{\mathrm{i}}} = \overline{G}_{\mathrm{SCPT},N_{\mathrm{SCPT}}} \;.
\label{eq:boldGinitialconditionmUflow}
\end{equation}
From~\eqref{eq:boldGinitialconditionmUflow} as well as~\eqref{eq:PhiboldInitialCondmUflow} to~\eqref{eq:PhinboldInitialCondmUflow}, we infer the rest of the relevant initial conditions for the mU-flow:
\begin{equation}
\begin{split}
\overline{\boldsymbol{\Omega}}_{\mathfrak{s}=\mathfrak{s}_{\mathrm{i}}} = & \ \frac{1}{\beta}\left(\Gamma^{(\mathrm{2PI})}_{0}\big[G=\overline{G}_{\mathrm{SCPT},N_{\mathrm{SCPT}}}\big] + \Phi_{\mathrm{SCPT},N_{\mathrm{SCPT}}}\big[U,G=\overline{G}_{\mathrm{SCPT},N_{\mathrm{SCPT}}}\big]\right) \\
= & \ \frac{1}{\beta}\bigg( -\frac{\zeta}{2} \mathrm{Tr}_{\alpha} \left[ \mathrm{ln}\big(\overline{G}_{\mathrm{SCPT},N_{\mathrm{SCPT}}}\big) \right] + \frac{\zeta}{2} \mathrm{Tr}_{\alpha}\left[\overline{G}_{\mathrm{SCPT},N_{\mathrm{SCPT}}}C^{-1}-\mathbb{I}\right] \\
& \hspace{0.8cm} + \Phi_{\mathrm{SCPT},N_{\mathrm{SCPT}}}\big[U,G=\overline{G}_{\mathrm{SCPT},N_{\mathrm{SCPT}}}\big]\bigg) \;,
\end{split}
\label{eq:InitialConditions2PIFRGmUflowOmega}
\end{equation}
\begin{equation}
\overline{\boldsymbol{\Phi}}_{\mathfrak{s}=\mathfrak{s}_{\mathrm{i}}} = \Phi_{\mathrm{SCPT},N_{\mathrm{SCPT}}}\big[U,G=\overline{G}_{\mathrm{SCPT},N_{\mathrm{SCPT}}}\big] \mathrlap{\;,}
\label{eq:InitialConditions2PIFRGmUflowPhi}
\end{equation}
\begin{equation}
\overline{\boldsymbol{\Sigma}}_{\mathfrak{s}=\mathfrak{s}_{\mathrm{i}}} = -\Phi^{(1)}_{\mathrm{SCPT},N_{\mathrm{SCPT}}}\big[U,G=\overline{G}_{\mathrm{SCPT},N_{\mathrm{SCPT}}}\big] \mathrlap{\;,}
\label{eq:InitialConditions2PIFRGmUflowSigma}
\end{equation}
\begin{equation}
\overline{\boldsymbol{\Phi}}_{\mathfrak{s}=\mathfrak{s}_{\mathrm{i}}}^{(n)}= \Phi^{(n)}_{\mathrm{SCPT},N_{\mathrm{SCPT}}}\big[U,G=\overline{G}_{\mathrm{SCPT},N_{\mathrm{SCPT}}}\big] \mathrlap{\quad \forall n \geq 2 \;.}
\label{eq:InitialConditions2PIFRGmUflowPhin}
\end{equation}
Therefore, as non-perturbative information has been incorporated into $\overline{G}_{\mathrm{SCPT},N_{\mathrm{SCPT}}}$ via the self-consistent procedure followed in the resolution of~\eqref{eq:2PIfrgDysonEqGbarsimUflow}, the same applies to $\overline{\boldsymbol{\Omega}}_{\mathfrak{s}=\mathfrak{s}_{\mathrm{i}}}$, $\overline{\boldsymbol{\Phi}}_{\mathfrak{s}=\mathfrak{s}_{\mathrm{i}}}$, $\overline{\boldsymbol{\Sigma}}_{\mathfrak{s}=\mathfrak{s}_{\mathrm{i}}}$ and $\overline{\boldsymbol{\Phi}}_{\mathfrak{s}=\mathfrak{s}_{\mathrm{i}}}^{(n)}$ (with $n \geq 2$). For example, we can already resum Hartree and Fock diagrams (e.g. for $N_{\mathrm{SCPT}}=1$) or even some ring diagrams (e.g. for $N_{\mathrm{SCPT}}=2$) before even starting to solve the flow equations, so we have now a very efficient starting point. In conclusion, the mU-flow is designed to take the results of self-consistent PT (with any approximation) as inputs for the FRG procedure\footnote{There is also an implementation of the 1PI-FRG for partially bosonized fermionic models (i.e. for fermionic models in their mixed representations) constructed to exploit the results of the self-consistent Hartree approximation (using the fermionic density as a variational parameter) as starting point of the flow~\cite{sch06}. The underlying formalism could be in principle adapted to treat the toy model considered in this thesis but we defer this to future works.}.

\item mU-flow equations with Hartree-Fock starting point:\\
Let us then set $N_{\mathrm{SCPT}}=1$ in order to specify to the situation where the starting point of the mU-flow coincides with the Hartree-Fock approximation. We can deduce from the perturbative series~\eqref{eq:2PIFRGperturbativeExpLWfunc} that:
\begin{equation}
\Phi_{\mathrm{SCPT},N_{\mathrm{SCPT}}=1}[U,G] = \frac{1}{8}\int_{\gamma_{1},\gamma_{2}}U_{\gamma_{1}\gamma_{2}}G_{\gamma_{1}}G_{\gamma_{2}} \;,
\end{equation}
which implies that the transformations underlying the mU-flow for $N_{\mathrm{SCPT}}=1$ read:
\begin{subequations}\label{eq:2PIfrgmUflowNpert1}
\begin{empheq}[left=\empheqlbrace]{align}
& \boldsymbol{\Omega}_{\mathfrak{s}}[G] = \Omega_{\mathfrak{s}}[G] + \frac{1}{2\beta}\left(U-U_{\mathfrak{s}}\right)_{\hat{\gamma}_{1}\hat{\gamma}_{2}}G_{\hat{\gamma}_{1}}G_{\hat{\gamma}_{2}} \;. \label{eq:2PIfrgmUflowNpert1Nb1} \\
\nonumber \\
& \boldsymbol{\Phi}_{\mathfrak{s}}[G] = \Phi_{\mathfrak{s}}[G] + \frac{1}{2}\left(U-U_{\mathfrak{s}}\right)_{\hat{\gamma}_{1}\hat{\gamma}_{2}}G_{\hat{\gamma}_{1}}G_{\hat{\gamma}_{2}} \;. \label{eq:2PIfrgmUflowNpert1Nb2}\\
\nonumber \\
& \boldsymbol{\Sigma}_{\mathfrak{s},\gamma}[G] = \Sigma_{\mathfrak{s},\gamma}[G] - \left(U-U_{\mathfrak{s}}\right)_{\gamma\hat{\gamma}}G_{\hat{\gamma}} \;. \label{eq:2PIfrgmUflowNpert1Nb3} \\
\nonumber \\
& \boldsymbol{\Phi}^{(2)}_{\mathfrak{s},\gamma_{1}\gamma_{2}}[G] = \Phi^{(2)}_{\mathfrak{s},\gamma_{1}\gamma_{2}}[G] + U_{\gamma_{1}\gamma_{2}} - U_{\mathfrak{s},\gamma_{1}\gamma_{2}} \;. \label{eq:2PIfrgmUflowNpert1Nb4} \\
\nonumber \\
& \boldsymbol{\Phi}^{(n)}_{\mathfrak{s},\gamma_{1}\cdots\gamma_{n}}[G] = \Phi^{(n)}_{\mathfrak{s},\gamma_{1}\cdots\gamma_{n}}[G] \quad \forall n \geq 3 \;. \label{eq:2PIfrgmUflowNpert1Nb5}
\end{empheq}
\end{subequations}
With the help of~\eqref{eq:2PIfrgmUflowNpert1Nb1} to~\eqref{eq:2PIfrgmUflowNpert1Nb5}, we can introduce $\boldsymbol{\Omega}_{\mathfrak{s}}[G]$, the modified Luttinger-Ward functional $\boldsymbol{\Phi}_{\mathfrak{s}}[G]$ and its derivatives into the pU-flow equations~\eqref{eq:2PIfrgUflowEquationsGeneralFormOmega} to~\eqref{eq:2PIfrgUflowEquationsGeneralFormPhi2}, thus obtaining the differential equations underlying the mU-flow with $N_{\mathrm{SCPT}}=1$. For instance,~\eqref{eq:2PIfrgUflowEquationsGeneralFormOmega} to~\eqref{eq:2PIfrgUflowEquationsGeneralFormSigma} become in this way (see appendix~\ref{ann:2PIfrgFlowEquationUflow} for the corresponding flow equation expressing the derivative of the modified 2PI vertex of order 2 with respect to $\mathfrak{s}$):
\begin{equation}
\dot{\overline{\boldsymbol{\Omega}}}_{\mathfrak{s}} = \frac{1}{6\beta} \dot{U}_{\mathfrak{s},\hat{\gamma}_{1} \hat{\gamma}_{2}} \left(\overline{W}_{\mathfrak{s}}^{(2)} - \overline{\Pi}_{\mathfrak{s}}\right)_{\hat{\gamma}_{2}\hat{\gamma}_{1}} \;,
\label{eq:2PIfrgmUflowExpressionBoldOmegaDotNotApp}
\end{equation}
\begin{equation}
\begin{split}
\scalebox{0.97}{${\displaystyle \dot{\overline{\boldsymbol{\Phi}}}_{\mathfrak{s}} = }$} & \ \scalebox{0.97}{${\displaystyle \frac{1}{6} \dot{U}_{\mathfrak{s},\hat{\gamma}_{1} \hat{\gamma}_{2}} \left(\overline{W}_{\mathfrak{s}}^{(2)} - \overline{\Pi}_{\mathfrak{s}}\right)_{\hat{\gamma}_{2}\hat{\gamma}_{1}} }$} \\
& \scalebox{0.97}{${\displaystyle +\frac{1}{6} \left[\overline{\boldsymbol{\Sigma}}_{\mathfrak{s},\gamma_{1}} + \left(U-U_{\mathfrak{s}}\right)_{\gamma_{1}\hat{\gamma}_{2}}\overline{\boldsymbol{G}}_{\mathfrak{s},\hat{\gamma}_{2}}\right] }$} \\
& \hspace{0.7cm} \scalebox{0.97}{${\displaystyle \times \overline{W}_{\mathfrak{s},\hat{\gamma}_{1} \hat{\gamma}_{3}}^{(2)} \dot{U}_{\mathfrak{s},\hat{\gamma}_{4} \hat{\gamma}_{5}} \left[ \overline{W}_{\mathfrak{s},\hat{\gamma}_{5} \hat{\gamma}_{6}}^{(2)} \left(\overline{\Pi}_{\mathfrak{s},\hat{\gamma}_{6}\hat{\gamma}_{7}}^{\mathrm{inv}} \frac{\delta \overline{\Pi}_{\mathfrak{s},\hat{\gamma}_{7}\hat{\gamma}_{8}}}{\delta \overline{\boldsymbol{G}}_{\mathfrak{s},\hat{\gamma}_{3}}} \overline{\Pi}_{\mathfrak{s},\hat{\gamma}_{8}\hat{\gamma}_{9}}^{\mathrm{inv}} - \overline{\boldsymbol{\Phi}}_{\mathfrak{s},\hat{\gamma}_{3} \hat{\gamma}_{6} \hat{\gamma}_{9}}^{(3)} \right) \overline{W}_{\mathfrak{s},\hat{\gamma}_{9} \hat{\gamma}_{4}}^{(2)} + \frac{1}{2} \frac{\delta \overline{\Pi}_{\mathfrak{s},\hat{\gamma}_{5} \hat{\gamma}_{4}}}{\delta \overline{\boldsymbol{G}}_{\mathfrak{s},\hat{\gamma}_{3}}} \right] \;, }$}
\end{split}
\label{eq:2PIfrgmUflowExpressionBoldPhiDotNotApp}
\end{equation}
\begin{equation}
\begin{split}
\scalebox{0.95}{${\displaystyle \dot{\overline{\boldsymbol{\Sigma}}}_{\mathfrak{s},\gamma} = }$} & \scalebox{0.95}{${\displaystyle - \frac{1}{3} \left(\mathcal{I} + \overline{\Pi}_{\mathfrak{s}} \overline{\boldsymbol{\Phi}}_{\mathfrak{s}}^{(2)}\right)^{\mathrm{inv}}_{\gamma\hat{\gamma}_{1}} }$} \\
& \scalebox{0.95}{${\displaystyle \times \Bigg( \left[ 2 \left(\mathcal{I} + \overline{\Pi}_{\mathfrak{s}} \left(\overline{\boldsymbol{\Phi}}_{\mathfrak{s}}^{(2)} - U + U_{\mathfrak{s}} \right)\right)^{\mathrm{inv}} \dot{U}_{\mathfrak{s}} \left(\mathcal{I} + \overline{\Pi}_{\mathfrak{s}} \left(\overline{\boldsymbol{\Phi}}_{\mathfrak{s}}^{(2)} - U + U_{\mathfrak{s}} \right)\right)^{\mathrm{inv}} + \dot{U}_{\mathfrak{s}} \right]_{\hat{\alpha}_{1} \hat{\alpha}_{2} \hat{\alpha}'_{2} \hat{\alpha}'_{1}} \hspace{-0.15cm} \overline{\boldsymbol{G}}_{\mathfrak{s},\hat{\gamma}_{2}} }$} \\
& \hspace{0.6cm} \scalebox{0.95}{${\displaystyle - \frac{1}{2} \dot{U}_{\mathfrak{s},\hat{\gamma}_{2} \hat{\gamma}_{3}} \overline{W}_{\mathfrak{s},\hat{\gamma}_{3} \hat{\gamma}_{4}}^{(2)} \overline{\boldsymbol{\Phi}}_{\mathfrak{s},\hat{\gamma}_{1} \hat{\gamma}_{4} \hat{\gamma}_{5}}^{(3)} \overline{W}_{\mathfrak{s},\hat{\gamma}_{5} \hat{\gamma}_{2}}^{(2)} - 3 \dot{U}_{\mathfrak{s},\hat{\gamma}_{1}\hat{\gamma}_{2}} \overline{\boldsymbol{G}}_{\mathfrak{s},\hat{\gamma}_{2}} \Bigg) \;, }$}
\end{split}
\label{eq:2PIfrgmUflowExpressionBoldSigmaDotNotApp}
\end{equation}
where $\overline{W}_{\mathfrak{s}}^{(2)}[K]$ is still given by the Bethe-Salpeter equation in the form:
\begin{equation}
W_{\mathfrak{s}}^{(2)}[K] = \left(\Pi^{\mathrm{inv}}[G] + \boldsymbol{\Phi}_{\mathfrak{s}}^{(2)}[G] - U + U_{\mathfrak{s}}\right)^{\mathrm{inv}} \;,
\end{equation}
as can be deduced from~\eqref{eq:2PIfrgUflowW2expression} and~\eqref{eq:2PIfrgmUflowNpert1Nb4}. However, the flow equation controlling the evolution of $\overline{\boldsymbol{G}}_{\mathfrak{s}}$ is not inferred from the pU-flow equations but simply by differentiating the Dyson equation~\eqref{eq:2PIfrgDysonEqGbarmUflow} with respect to the flow parameter\footnote{As opposed to the flow equations~\eqref{eq:2PIfrgmUflowExpressionBoldOmegaDotNotApp} to~\eqref{eq:2PIfrgmUflowExpressionBoldSigmaDotNotApp} which rely on the mU-flow transformations at $N_{\mathrm{SCPT}}=1$ given by~\eqref{eq:2PIfrgmUflowNpert1Nb1} to~\eqref{eq:2PIfrgmUflowNpert1Nb5},~\eqref{eq:2PIfrgmUflowExpressionBoldGDotNotApp} is valid regardless of the value of $N_{\mathrm{SCPT}}$.}:
\begin{equation}
\dot{\overline{\boldsymbol{G}}}_{\mathfrak{s},\alpha_{1}\alpha'_{1}}=\int_{\alpha_{2},\alpha'_{2}}\overline{\boldsymbol{G}}_{\mathfrak{s},\alpha_{1} \alpha_{2}} \dot{\overline{\boldsymbol{\Sigma}}}_{\mathfrak{s},\alpha_{2} \alpha'_{2}} \overline{\boldsymbol{G}}_{\mathfrak{s},\alpha'_{2} \alpha'_{1}} \;.
\label{eq:2PIfrgmUflowExpressionBoldGDotNotApp}
\end{equation}

\vspace{0.5cm}

Furthermore, the bold 2PI EA can be expressed by multiplying both sides of~\eqref{eq:2PIfrgmUflowNpert1Nb1} by~$\beta$:
\begin{equation}
\begin{split}
\boldsymbol{\Gamma}^{(\mathrm{2PI})}_{\mathfrak{s}}[G] = & \ \Gamma^{(\mathrm{2PI})}_{\mathfrak{s}}[G] + \frac{1}{2}\left(U-U_{\mathfrak{s}}\right)_{\hat{\gamma}_{1}\hat{\gamma}_{2}}G_{\hat{\gamma}_{1}}G_{\hat{\gamma}_{2}} \\
= & \ -W[K] + \mathrm{Tr}_{\gamma}(KG) + \frac{1}{2}\left(U-U_{\mathfrak{s}}\right)_{\hat{\gamma}_{1}\hat{\gamma}_{2}}G_{\hat{\gamma}_{1}}G_{\hat{\gamma}_{2}} \;.
\end{split}
\end{equation}
Therefore, the bold 2PI EA is defined via a Legendre transform modified by the term\footnote{We assume that $U_{\mathfrak{s}}=U+R_{\mathfrak{s}}$ in~\eqref{eq:TermModifLegendreTransf2PIFRGHFstartingPoint} only to clarify its link with~\eqref{eq:TermModifLegendreTransf1PIFRG} but we stress that all derivations presented so far are valid regardless of the analytical form of $U_{\mathfrak{s}}$.}:
\begin{equation}
\frac{1}{2}\left(U-U_{\mathfrak{s}}\right)_{\hat{\gamma}_{1}\hat{\gamma}_{2}}G_{\hat{\gamma}_{1}}G_{\hat{\gamma}_{2}} = \frac{1}{2}\left(U - (U + R_{\mathfrak{s}})\right)_{\hat{\gamma}_{1}\hat{\gamma}_{2}}G_{\hat{\gamma}_{1}}G_{\hat{\gamma}_{2}} = -\frac{1}{2} R_{\mathfrak{s},\hat{\gamma}_{1}\hat{\gamma}_{2}}G_{\hat{\gamma}_{1}}G_{\hat{\gamma}_{2}} \;,
\label{eq:TermModifLegendreTransf2PIFRGHFstartingPoint}
\end{equation}
for $N_{\mathrm{SCPT}}=1$, just like the Legendre transform underlying the 1PI EA is modified in~\eqref{eq:defGammakWettFRG} by the term:
\begin{equation}
-\Delta S_{k}[\phi] = -\frac{1}{2}\int_{\alpha_{1},\alpha_{2}}\phi_{\alpha_{1}}R_{k,\alpha_{1} \alpha_{2}}\phi_{\alpha_{2}} \;.
\label{eq:TermModifLegendreTransf1PIFRG}
\end{equation}
The same remark applies to any value of $N_{\mathrm{SCPT}}$ for the mU-flow, in which case the bold 2PI EA is defined by~\eqref{eq:2PIfrgDefinitionmUflowEA} and the term modifying the Legendre transform is:
\begin{equation}
\Phi_{\mathrm{SCPT},N_{\mathrm{SCPT}}}[U,G]-\Phi_{\mathrm{SCPT},N_{\mathrm{SCPT}}}[U_{\mathfrak{s}},G] \;.
\label{eq:TermModifLegendreTransf2PIFRGanySCPTstartingPoint}
\end{equation}
According to our previous explanations, the extra terms~\eqref{eq:TermModifLegendreTransf2PIFRGHFstartingPoint} to~\eqref{eq:TermModifLegendreTransf2PIFRGanySCPTstartingPoint} are just means to enforce a convenient starting point for the corresponding FRG procedure (the classical theory for the 1PI-FRG and self-consistent PT for the mU-flow version of the 2PI-FRG).

\vspace{0.5cm}

Finally, let us consider the situation where the choice for the scale-dependent interaction $U_{\mathfrak{s}}$ (or, equivalently, for the cutoff function $R_{\mathfrak{s}}$) is such that the Wilsonian momentum-shell integration is not implemented through the U-flow. At first sight, this could be taken as a severe limitation of this approach for the purpose of describing critical phenomena. However, this issue is remarkably circumvented in the framework of the mU-flow which can be designed to start in a broken-symmetry phase if necessary (or more generally in the phase that we aim to describe) so as to avoid undesirable phase transitions during the flow, depending on the solution $\overline{G}_{\mathrm{SCPT},N_{\mathrm{SCPT}}}$ chosen for the initial conditions. This remark only applies if the different solutions of the Dyson equation~\eqref{eq:2PIfrgDysonEqGbarsimUflow} give us access to each phase that we seek to describe, which works in principle for the $U(1)$ symmetry notably, and therefore for the description of superfluid systems, as the relevant order parameters can be identified as components of the propagator $G$ in such a situation\footnote{However, the mU-flow as presented here does not provide us with a similar freedom to tackle an $O(N)$ symmetry as neither $\Gamma^{(\mathrm{2PI})}[G]$ nor $\boldsymbol{\Gamma}^{(\mathrm{2PI})}[G]$ is capable of spontaneously breaking such a symmetry by construction (as they can not exhibit a non-zero 1-point correlation function for the field $\widetilde{\psi}$).}.

\end{itemize}

\subsubsection{CU-flow}
\label{sec:CUflow2PIFRG}
\paragraph{Main features:}

The CU-flow version of the 2PI-FRG was developed in ref.~\cite{dup14}. It essentially amounts to combining the C-flow and the U-flow, and more specifically the tC-flow and the pU-flow, together. A noticeable fact is that, as the C-flow and (if the corresponding cutoff function is chosen accordingly) the U-flow, such an approach also carries out the Wilsonian momentum-shell integration. Hence, the CU-flow relies on the transformation $C^{-1} \rightarrow C^{-1}_{\mathfrak{s}}=C^{-1} + R^{(C)}_{\mathfrak{s}}$ (or, equivalently, $C^{-1} \rightarrow C^{-1}_{\mathfrak{s}}=R^{(C)}_{\mathfrak{s}} C^{-1}$) combined with $U \rightarrow U_{\mathfrak{s}} = U + R^{(U)}_{\mathfrak{s}}$ (or, equivalently, $U \rightarrow U_{\mathfrak{s}} = R^{(U)}_{\mathfrak{s}} U$), with as before:
\begin{subequations}\label{eq:2PIfrgBoundaryCondforCUflow}
\begin{empheq}[left=\empheqlbrace]{align}
& C_{\mathfrak{s}=\mathfrak{s}_{\mathrm{i}},\gamma} = 0 \quad \forall \gamma \;. \label{eq:2PIfrgCUflowBoundaryCCondUpper}\\
\nonumber \\
& C_{\mathfrak{s}=\mathfrak{s}_{\mathrm{f}}} = C \;. \label{eq:2PIfrgCUflowBoundaryCCondBottom}\\
\nonumber \\
& U_{\mathfrak{s}=\mathfrak{s}_{\mathrm{i}},\gamma_{1}\gamma_{2}} = 0 \quad \forall \gamma_{1}, \gamma_{2} \;. \label{eq:2PIfrgCUflowBoundaryUCondUpper}\\
\nonumber \\
& U_{\mathfrak{s}=\mathfrak{s}_{\mathrm{f}}} = U \;. \label{eq:2PIfrgCUflowBoundaryUCondBottom}
\end{empheq}
\end{subequations}
Due to the initial condition~\eqref{eq:2PIfrgCUflowBoundaryCCondUpper}, the starting point of the CU-flow suffers from the same divergence problem as that of the C-flow, which is why we will consider the functional $\Delta\overline{\Omega}_{\mathfrak{s}}$ (defined by~\eqref{eq:2PIfrgCflowDefDeltaOmega}) in the present framework as well. Furthermore, we can expect the differential equations underlying the CU-flow to contain contributions from both the C-flow and the U-flow, besides the flow equation expressing $\dot{\overline{G}}_{\mathfrak{s}}$ which is given by~\eqref{eq:2PIfrgFlowEquationsCflowG} as in the C-flow. This can clearly be seen from the CU-flow equation expressing $\Delta\dot{\overline{\Omega}}_{\mathfrak{s}}$ (see appendix~\ref{ann:2PIfrgFlowEquationCUflow}):
\begin{equation}
\Delta \dot{\overline{\Omega}}_{\mathfrak{s}} = \underbrace{\frac{1}{\beta} \dot{C}_{\mathfrak{s},\hat{\gamma}}^{-1} \left(\overline{G}_{\mathfrak{s}}-C_{\mathfrak{s}}\right)_{\hat{\gamma}}}_{\text{C-flow contribution}} + \underbrace{\frac{1}{6\beta} \dot{U}_{\mathfrak{s},\hat{\gamma}_{1} \hat{\gamma}_{2}} \left(\overline{W}_{\mathfrak{s}}^{(2)} + \frac{1}{2}\overline{\Pi}_{\mathfrak{s}}\right)_{\hat{\gamma}_{2}\hat{\gamma}_{1}}}_{\text{U-flow contribution}} \;.
\label{eq:2PIFRGCUflowExpressionDeltaOmegadot}
\end{equation}
However, since the Luttinger-Ward functional is an invariant of the C-flow (according to \eqref{eq:2PIfrgFlowEqCflowVersion1}), the flow equations expressing $\dot{\overline{\Phi}}_{\mathfrak{s}}$, $\dot{\overline{\Sigma}}_{\mathfrak{s}}$ and $\dot{\overline{\Phi}}^{(n)}_{\mathfrak{s}}$ (with $n \geq 2$) coincide in principle with those of the pU-flow, and notably with~\eqref{eq:2PIfrgUflowEquationsGeneralFormPhi} for $\dot{\overline{\Phi}}_{\mathfrak{s}}$ and~\eqref{eq:2PIfrgUflowEquationsGeneralFormPhi2} for $\dot{\overline{\Phi}}^{(2)}_{\mathfrak{s}}$. There is however a subtlety that implies that~\eqref{eq:2PIfrgUflowEquationsGeneralFormSigma} (which expresses $\dot{\overline{\Sigma}}_{\mathfrak{s}}$ for the pU-flow) is not valid in the present situation. We have thus rewritten this equation in a form exploitable for the CU-flow (see appendix~\ref{ann:2PIfrgFlowEquationCUflow}):
\begin{equation}
\begin{split}
\dot{\overline{\Sigma}}_{\mathfrak{s},\gamma} = & -\frac{1}{3} \left[ 2 \left(\mathcal{I} + \overline{\Pi}_{\mathfrak{s}} \overline{\Phi}_{\mathfrak{s}}^{(2)}\right)^{\mathrm{inv}} \dot{U}_{\mathfrak{s}} \left(\mathcal{I} + \overline{\Pi}_{\mathfrak{s}} \overline{\Phi}_{\mathfrak{s}}^{(2)}\right)^{\mathrm{inv}} + \dot{U}_{\mathfrak{s}} \right]_{\alpha \hat{\alpha} \hat{\alpha}' \alpha'} \overline{G}_{\mathfrak{s},\hat{\gamma}} \\
& +\frac{1}{6} \dot{U}_{\mathfrak{s},\hat{\gamma}_{1} \hat{\gamma}_{2}} \overline{W}_{\mathfrak{s},\hat{\gamma}_{2} \hat{\gamma}_{3}}^{(2)} \overline{\Phi}_{\mathfrak{s},\gamma \hat{\gamma}_{3} \hat{\gamma}_{4}}^{(3)} \overline{W}_{\mathfrak{s},\hat{\gamma}_{4} \hat{\gamma}_{1}}^{(2)} - \dot{\overline{G}}_{\mathfrak{s},\hat{\gamma}} \overline{\Phi}^{(2)}_{\mathfrak{s},\hat{\gamma}\gamma} \;.
\end{split}
\label{eq:2PIFRGCUflowExpressionSigmadot}
\end{equation}

\paragraph{Truncations:}

The infinite tower of differential equations for the CU-flow is closed by enforcing the same truncation condition as in the tC-flow and the pU-flow (i.e.~\eqref{eq:2PIfrgPhiBartCflow}), which is why we mentioned above that the CU-flow is essentially a merger of the latter two approaches.

\paragraph{Initial conditions:}

On the one hand, as for the C-flow, the condition $C_{\mathfrak{s}=\mathfrak{s}_{\mathrm{i}},\gamma}=0$ $\forall \gamma$ (i.e.~\eqref{eq:2PIfrgCUflowBoundaryCCondUpper}) induces that:
\begin{equation}
\overline{G}_{\mathfrak{s}=\mathfrak{s}_{\mathrm{i}},\gamma}=0 \mathrlap{\quad \forall \gamma \;,}
\label{eq:2PIfrgCUflowInitialCondGs}
\end{equation}
\begin{equation}
\Delta\overline{\Omega}_{\mathfrak{s}=\mathfrak{s}_{\mathrm{i}}}=0 \mathrlap{\;.}
\label{eq:2PIfrgCUflowInitialCondOmegas}
\end{equation}
On the other hand, as for the pU-flow, the condition $U_{\mathfrak{s}=\mathfrak{s}_{\mathrm{i}},\gamma_{1}\gamma_{2}}=0$ $\forall\gamma_{1},\gamma_{2}$ (i.e.~\eqref{eq:2PIfrgCUflowBoundaryUCondUpper}) implies that:
\begin{equation}
\overline{\Phi}_{\mathfrak{s}=\mathfrak{s}_{\mathrm{i}}}=0 \mathrlap{\;,}
\label{eq:InitialConditions2PIFRGCUflowPhi}
\end{equation}
\begin{equation}
\overline{\Sigma}_{\mathfrak{s}=\mathfrak{s}_{\mathrm{i}},\gamma}=0 \mathrlap{\quad \forall \gamma \;,}
\label{eq:InitialConditions2PIFRGCUflowSigma}
\end{equation}
\begin{equation}
\overline{\Phi}_{\mathfrak{s}=\mathfrak{s}_{\mathrm{i}},\gamma_{1} \cdots \gamma_{n}}^{(n)}=0 \mathrlap{\quad \forall \gamma_{1},\cdots , \gamma_{n}, ~ \forall n \geq 2 \;.}
\label{eq:InitialConditions2PIFRGCUflowPhin}
\end{equation}

\subsection{Application to the (0+0)-D $O(N)$-symmetric $\varphi^4$-theory}
\label{sec:2PIFRG0DON}
\subsubsection{Symmetrization of the two-body interaction}
\label{sec:2PIFRG0DONSymmetrizationU}

We present in the whole section~\ref{sec:2PIFRG0DON} our applications of the 2PI-FRG to the studied zero-dimensional $O(N)$ model. The novel feature of this 2PI-FRG study is the treatment of the $O(N)$ symmetry (and especially of the broken-symmetry phase of an $O(N)$ model), which has never been tackled before with such a formalism to our knowledge. As a first step, we show that the 2PI-FRG is applicable to the $O(N)$ model under consideration. The C-flow, U-flow and CU-flow versions of the 2PI-FRG can indeed all be exploited to treat the original version of this model as its classical action is in accordance with the analytical form~\eqref{eq:2PIFRGmostgeneralactionS}, i.e.:
\begin{equation}
\begin{split}
S\Big(\vec{\widetilde{\varphi}}\Big) = & \ S_{0}\Big(\vec{\widetilde{\varphi}}\Big) + S_{\mathrm{int}}\Big(\vec{\widetilde{\varphi}}\Big) = \frac{1}{2}m^{2}\vec{\widetilde{\varphi}}^{2}+\frac{1}{4!}\lambda\left(\vec{\widetilde{\varphi}}^{2}\right)^{2} \\
= & \ \frac{1}{2}\sum_{a_{1},a_{2}=1}^{N}\widetilde{\varphi}_{a_{1}}C^{-1}_{a_{1}a_{2}}\widetilde{\varphi}_{a_{2}}+\frac{1}{4!}\sum_{a_{1},a_{2},a_{3},a_{4}=1}^{N} U_{a_{1}a_{2}a_{3}a_{4}}\widetilde{\varphi}_{a_{1}}\widetilde{\varphi}_{a_{2}}\widetilde{\varphi}_{a_{3}}\widetilde{\varphi}_{a_{4}} \\
= & \ \frac{1}{2}\int_{\alpha_{1},\alpha_{2}}\widetilde{\psi}_{\alpha_{1}}C^{-1}_{\alpha_{1}\alpha_{2}}\widetilde{\psi}_{\alpha_{2}}+\frac{1}{4!}\int_{\alpha_{1},\alpha_{2},\alpha_{3},\alpha_{4}}U_{\alpha_{1}\alpha_{2}\alpha_{3}\alpha_{4}}\widetilde{\psi}_{\alpha_{1}}\widetilde{\psi}_{\alpha_{2}}\widetilde{\psi}_{\alpha_{3}}\widetilde{\psi}_{\alpha_{4}} \;,
\end{split}
\label{eq:ClassicalAction2PIFRG0DON}
\end{equation}
where the fluctuating field $\widetilde{\psi}$ coincides with the bosonic field $\vec{\widetilde{\varphi}}$:
\begin{equation}
\widetilde{\psi}_{\alpha} = \widetilde{\varphi}_{a} \;,
\end{equation}
the integrals are now just discrete sums:
\begin{equation}
\int_{\gamma} = \sum_{a,a'=1}^{N} \;,
\label{eq:Summations2PIFRG0DON}
\end{equation}
as $\alpha$-indices reduce to color indices in the present situation:
\begin{equation}
\gamma \equiv (\alpha,\alpha') = (a,a') \;,
\label{eq:bosonicindices2PIFRG0DON}
\end{equation}
whereas the free propagator and the two-body interaction are respectively given by\footnote{As opposed to the conventions followed so far, matrices that live in color space are no longer written in bold characters in section~\ref{sec:2PIFRG0DON} (and corresponding appendices), e.g. the free propagator and the self-energy are denoted as $C$ and $\Sigma$ instead of $\boldsymbol{C}$ and $\boldsymbol{\Sigma}$, respectively. The reason behind this is to keep the bold characters for the modified 2PI vertices and other quantities introduced in the framework of the mU-flow, thus avoiding overlapping notations.}:
\begin{equation}
C^{-1}_{a_{1}a_{2}} = m^{2} \delta_{a_{1}a_{2}} \;,
\end{equation}
\begin{equation}
U_{a_{1}a_{2}a_{3}a_{4}} = \frac{\lambda}{3}\left(\delta_{a_{1}a_{2}}\delta_{a_{3}a_{4}}+\delta_{a_{1}a_{3}}\delta_{a_{2}a_{4}}+\delta_{a_{1}a_{4}}\delta_{a_{2}a_{3}}\right) \;.
\label{eq:2bodyinteraction2PIFRG0DON}
\end{equation}
For any application of the 2PI-FRG to a bosonic (fermionic) theory, the two-body interaction must be constructed so as to be symmetric (antisymmetric) under permutation of its indices, i.e. so as to satisfy~\eqref{eq:SymmetryProperty2bodyInteraction2PIFRG} which was widely exploited to derive the 2PI-FRG flow equations given in section~\ref{sec:2PIFRGstateofplay}. For our bosonic toy model, $U$ must therefore be symmetric, i.e.:
\begin{equation}
U_{a_{1}a_{2}a_{3}a_{4}}=U_{a_{P(1)}a_{P(2)}a_{P(3)}a_{P(4)}} \;,
\label{eq:SymmetrictwobodyInt2PIFRG0DON}
\end{equation}
under any permutation $P$, which is indeed satisfied by~\eqref{eq:2bodyinteraction2PIFRG0DON}. However, the classical actions underlying the mixed and collective representations (given in arbitrary dimensions by~\eqref{eq:Smix} and~\eqref{eq:Scoll} respectively) can not be written in the required form~\eqref{eq:2PIFRGmostgeneralactionS}. Therefore, the general 2PI-FRG formalism presented in section~\ref{sec:2PIFRGstateofplay} (and especially the U-flow and CU-flow implementations) can not be directly applied to these two versions of the toy model under consideration. We can nonetheless quite readily extend the C-flow formalism to these two situations by rederiving the initial conditions from the diagrammatic expressions of the corresponding Luttinger-Ward functionals (i.e. from the counterparts of~\eqref{eq:2PIFRGperturbativeExpLWfunc}) and by taking into account that a second fluctuating field enters the arena in the mixed situation. Formulations of the U-flow and CU-flow implementations of the 2PI-FRG in the framework of the mixed representation are however much less straightforward, as will be discussed in more detail in section~\ref{sec:2PIFRGuflow0DON}.

\subsubsection{Original 2PI functional renormalization group C-flow}

In almost all cases\footnote{There are only two exceptions: i) the application of the C-flow in the framework of the mixed representation for which we generalize in addition the C-flow formalism to a field theory involving two fields, as we just mentioned at the end of section~\ref{sec:2PIFRG0DONSymmetrizationU}; ii) the application of the mU-flow with $N=2$ and $N_{\mathrm{max}}=2$ or $3$ where we will exploit a trick (inherent to the (0+0)-D situation) in order to avoid evaluating the derivatives $\frac{\delta\overline{\Pi}_{\mathfrak{s}}}{\delta\overline{G}_{\mathfrak{s},\gamma}}$ and $\frac{\delta^{2}\overline{\Pi}_{\mathfrak{s}}}{\delta\overline{G}_{\mathfrak{s},\gamma_{1}}\delta\overline{G}_{\mathfrak{s},\gamma_{2}}}$ in the flow equations expressing the derivatives of the 2PI vertices of order 2 and 3 with respect to $\mathfrak{s}$ (see appendix~\ref{ann:2PIfrgmUflow0DON}).}, we will just rewrite the general 2PI-FRG flow equations presented in section~\ref{sec:2PIFRGstateofplay} for the studied toy model by simply replacing integrals over bosonic indices by summations over color indices, as follows from~\eqref{eq:Summations2PIFRG0DON}. We thus start by deducing in this way the tower of differential equations for the C-flow in the framework of the original theory from~\eqref{eq:2PIfrgFlowEquationsCflowG} to~\eqref{eq:2PIfrgFlowEquationsCflowPhin}:
\begin{equation}
\dot{\overline{G}}_{\mathfrak{s},a_{1} a'_{1}}=-\sum_{a_{2},a'_{2}=1}^{N} \overline{G}_{\mathfrak{s}, a_{1} a_{2}}\left(\dot{C}_{\mathfrak{s}}^{-1}-\dot{\overline{\Sigma}}_{\mathfrak{s}}\right)_{a_{2} a'_{2}} \overline{G}_{\mathfrak{s}, a'_{2} a'_{1}} \mathrlap{\;,}
\label{eq:2PIfrgFlowEquationsCflowG0DON}
\end{equation}
\begin{equation}
\Delta \dot{\overline{\Omega}}_{\mathfrak{s}} = \frac{1}{2} \sum_{a,a'=1}^{N} \dot{C}_{\mathfrak{s},a a'}^{-1} \left(\overline{G}_{\mathfrak{s}}-C_{\mathfrak{s}}\right)_{a a'} \mathrlap{\;,}
\label{eq:2PIfrgFlowEquationsCflowDOmega0DON}
\end{equation}
\begin{equation}
\dot{\overline{\Sigma}}_{\mathfrak{s},a_{1} a'_{1}} = - \frac{1}{2} \sum_{a_{2},a'_{2} = 1}^{N} \dot{\overline{G}}_{\mathfrak{s},a_{2} a'_{2}} \overline{\Phi}_{\mathfrak{s},(a_{2}, a'_{2})(a_{1}, a'_{1})}^{(2)} \mathrlap{\;,}
\label{eq:2PIfrgFlowEquationsCflowSigma0DON}
\end{equation}
\begin{equation}
\hspace{2.1cm} \dot{\overline{\Phi}}_{\mathfrak{s},(a_{1}, a'_{1})\cdots(a_{n}, a'_{n})}^{(n)} = \frac{1}{2} \sum_{a_{n+1},a'_{n+1} = 1}^{N} \dot{\overline{G}}_{\mathfrak{s},a_{n+1} a'_{n+1}} \overline{\Phi}_{\mathfrak{s},(a_{n+1}, a'_{n+1})(a_{1}, a'_{1})\cdots(a_{n}, a'_{n})}^{(n+1)} \quad \forall n \geq 2 \;,
\label{eq:2PIfrgFlowEquationsCflowPhin0DON}
\end{equation}
with $\Delta \overline{\Omega}_{\mathfrak{s}} = \overline{\Omega}_{\mathfrak{s}} + \frac{1}{2} \mathrm{Tr}_{a} \left[ \mathrm{ln}(2\pi C_{\mathfrak{s}})\right]$. We will exploit the flow equation expressing $\dot{\overline{\Phi}}_{\mathfrak{s}}$ (or $\dot{\overline{\boldsymbol{\Phi}}}_{\mathfrak{s}}$ for the mU-flow) in none of our 2PI-FRG applications. This stems from the fact that we are interested in the gs energy $E_{\mathrm{gs}}$ and gs density $\rho_{\mathrm{gs}}$, which are deduced respectively from $\overline{\Omega}_{\mathfrak{s}=\mathfrak{s}_{\mathrm{f}}}=\overline{\boldsymbol{\Omega}}_{\mathfrak{s}=\mathfrak{s}_{\mathrm{f}}}$ and $\overline{G}_{\mathfrak{s}=\mathfrak{s}_{\mathrm{f}}}=\overline{\boldsymbol{G}}_{\mathfrak{s}=\mathfrak{s}_{\mathrm{f}}}$ as follows:
\begin{equation}
E^\text{2PI-FRG}_{\mathrm{gs}} = \overline{\Omega}_{\mathfrak{s}=\mathfrak{s}_{\mathrm{f}}} = \overline{\boldsymbol{\Omega}}_{\mathfrak{s}=\mathfrak{s}_{\mathrm{f}}} \;,
\label{eq:Egs2PIFRG0DON}
\end{equation}
\begin{equation}
\rho^\text{2PI-FRG}_{\mathrm{gs}} = \frac{1}{N} \sum_{a=1}^{N} \overline{G}_{\mathfrak{s}=\mathfrak{s}_{\mathrm{f}},aa} = \frac{1}{N} \sum_{a=1}^{N} \overline{\boldsymbol{G}}_{\mathfrak{s}=\mathfrak{s}_{\mathrm{f}},aa} \;,
\label{eq:rhogs2PIFRG0DON}
\end{equation}
and the flow equations expressing $\dot{\overline{\Omega}}_{\mathfrak{s}}$ and $\dot{\overline{G}}_{\mathfrak{s}}$ (or $\dot{\overline{\boldsymbol{\Omega}}}_{\mathfrak{s}}$ and $\dot{\overline{\boldsymbol{G}}}_{\mathfrak{s}}$ for the mU-flow) never depend on $\overline{\Phi}_{\mathfrak{s}}$ (or $\overline{\boldsymbol{\Phi}}_{\mathfrak{s}}$ respectively), whether it is in the framework of the C-flow, the U-flow or the CU-flow. Relations~\eqref{eq:Egs2PIFRG0DON} and~\eqref{eq:rhogs2PIFRG0DON} will actually be used to estimate respectively $E_{\mathrm{gs}}$ and $\rho_{\mathrm{gs}}$ for all 2PI-FRG approaches (including the C-flow in the framework of the mixed representation) treated in this section~\ref{sec:2PIFRG0DON}. Furthermore, we have basically two symmetry arguments that allow us to simplify the flow equations~\eqref{eq:2PIfrgFlowEquationsCflowG0DON} to~\eqref{eq:2PIfrgFlowEquationsCflowPhin0DON}:
\begin{itemize}
\item General symmetry argument:\\
The symmetry properties of the correlation functions $W^{(n)}$ given by~\eqref{eq:SymmetryW2PIFRGUp} and~\eqref{eq:SymmetryW2PIFRGDown} are also exhibited by the propagator $\overline{G}_{\mathfrak{s}}$, the self-energy $\overline{\Sigma}_{\mathfrak{s}}$ and all other 2PI vertices $\overline{\Phi}_{\mathfrak{s}}^{(n)}$ (with $n \geq 2$). For the 2PI vertex $\overline{\Phi}^{(2)}_{\mathfrak{s}}$ at $N=2$ for instance, we have \textit{a priori} $2^{4}=16$ components $\overline{\Phi}^{(2)}_{\mathfrak{s},(a_{1},a'_{1})(a_{2},a'_{2})}$ to consider for the flow. However, since $\overline{\Phi}^{(2)}_{\mathfrak{s},(a_{1},a'_{1})(a_{2},a'_{2})}=\overline{\Phi}^{(2)}_{\mathfrak{s},(a'_{1},a_{1})(a_{2},a'_{2})}=\overline{\Phi}^{(2)}_{\mathfrak{s},(a_{1},a'_{1})(a'_{2},a_{2})}=\overline{\Phi}^{(2)}_{\mathfrak{s},(a'_{1},a_{1})(a'_{2},a_{2})}$ according to~\eqref{eq:SymmetryW2PIFRGUp} and $\overline{\Phi}^{(2)}_{\mathfrak{s},(a_{1},a'_{1})(a_{2},a'_{2})}=\overline{\Phi}^{(2)}_{\mathfrak{s},(a_{2},a'_{2})(a_{1},a'_{1})}$ according to~\eqref{eq:SymmetryW2PIFRGDown}, this set reduces to 6 flowing components, which are for instance: $\overline{\Phi}^{(2)}_{\mathfrak{s},(1,1)(1,1)}$, $\overline{\Phi}^{(2)}_{\mathfrak{s},(1,1)(1,2)}$, $\overline{\Phi}^{(2)}_{\mathfrak{s},(1,1)(2,2)}$, $\overline{\Phi}^{(2)}_{\mathfrak{s},(1,2)(1,2)}$, $\overline{\Phi}^{(2)}_{\mathfrak{s},(1,2)(2,2)}$ and $\overline{\Phi}^{(2)}_{\mathfrak{s},(2,2)(2,2)}$.

\item Symmetry argument inherent to the $O(N)$ symmetry:\\
Since the 2PI-FRG formalism was developed in a framework that can not exhibit any spontaneous breakdown of the $O(N)$ symmetry\footnote{We recall that the $O(N)$ symmetry can not be spontaneously broken down in the framework of the 2PI-FRG since its main functionals $\Gamma^{(\mathrm{2PI})}[G]\equiv\Gamma^{(\mathrm{2PI})}[\phi=0,G]$ and $\Phi[G]\equiv\Phi[\phi=0,G]$ are all defined in the configuration where the 1-point correlation function of the field $\widetilde{\psi}$ vanishes, i.e. where $\phi_{\alpha}=\left\langle\widetilde{\psi}_{\alpha}\right\rangle_{K}=0$ $\forall\alpha$.}, all matrices reduce to scalars in color space throughout the entire flow, i.e.:
\begin{equation}
C^{-1}_{\mathfrak{s},a a'} = C^{-1}_{\mathfrak{s}} \ \delta_{a a'} \mathrlap{\quad \forall\mathfrak{s} \;,}
\label{eq:diagonalC2PIFRGCflow0DON}
\end{equation}
\begin{equation}
\overline{G}_{\mathfrak{s},a a'} = \overline{G}_{\mathfrak{s}} \ \delta_{a a'} \mathrlap{\quad \forall\mathfrak{s} \;,}
\label{eq:diagonalG2PIFRGCflow0DON}
\end{equation}
\begin{equation}
\overline{\Sigma}_{\mathfrak{s},a a'} = \overline{\Sigma}_{\mathfrak{s}} \ \delta_{a a'} \mathrlap{\quad \forall\mathfrak{s} \;.}
\label{eq:diagonalSigma2PIFRGCflow0DON}
\end{equation}
The cutoff function $R_{\mathfrak{s}}$ must therefore be chosen such that condition~\eqref{eq:diagonalC2PIFRGCflow0DON} is fulfilled. After rewriting the flow equations~\eqref{eq:2PIfrgFlowEquationsCflowG0DON} to~\eqref{eq:2PIfrgFlowEquationsCflowPhin0DON} with~\eqref{eq:diagonalC2PIFRGCflow0DON} to~\eqref{eq:diagonalSigma2PIFRGCflow0DON}, we can see that the components of the 2PI vertices $\overline{\Phi}_{\mathfrak{s}}^{(n)}$ which have at least one bosonic index with distinct color indices (i.e. at least one index $\gamma=(a,a')$ with $a \neq a'$) are somehow cut out of the flow. In particular, this translates into the fact that they do not affect $\overline{\Omega}_{\mathfrak{s}=\mathfrak{s}_{\mathrm{f}}}$ and $\overline{G}_{\mathfrak{s}=\mathfrak{s}_{\mathrm{f}}}$ which are the quantities of interest for us in the present study. Getting back to our example on $\overline{\Phi}^{(2)}_{\mathfrak{s}}$, the number of corresponding components of interest for the flow is thus further reduced from 6 to 3: $\overline{\Phi}^{(2)}_{\mathfrak{s},(1,1)(1,1)}$, $\overline{\Phi}^{(2)}_{\mathfrak{s},(1,1)(2,2)}$ and $\overline{\Phi}^{(2)}_{\mathfrak{s},(2,2)(2,2)}$. Finally, since the color space is isotropic in the absence of SSB\footnote{For $\overline{\Phi}_{\mathfrak{s}}^{(3)}$, this implies $\overline{\Phi}^{(3)}_{\mathfrak{s},(2,2)(2,2)(2,2)}=\overline{\Phi}^{(3)}_{\mathfrak{s},(1,1)(1,1)(1,1)}$ $\forall\mathfrak{s}$, but also $\overline{\Phi}^{(3)}_{\mathfrak{s},(1,1)(2,2)(2,2)}=\overline{\Phi}^{(3)}_{\mathfrak{s},(1,1)(1,1)(2,2)}$ $\forall\mathfrak{s}$.}, we have $\overline{\Phi}^{(2)}_{\mathfrak{s},(2,2)(2,2)}=\overline{\Phi}^{(2)}_{\mathfrak{s},(1,1)(1,1)}$ $\forall\mathfrak{s}$, thus ending up with only 2 relevant components.

\end{itemize}

\vspace{0.3cm}

Such symmetry constraints significantly simplify the flow equations~\eqref{eq:2PIfrgFlowEquationsCflowG0DON} to~\eqref{eq:2PIfrgFlowEquationsCflowPhin0DON} for $N\geq 2$. We thus rewrite these differential equations explicitly for $N=1$ and $2$ by exploiting the above symmetry arguments in the latter case:
\begin{itemize}
\item For $N=1$ ($\forall N_{\mathrm{max}}$):
\begin{equation}
\dot{\overline{G}}_{\mathfrak{s}} =- \overline{G}_{\mathfrak{s}}^{2} \left(\dot{C}_{\mathfrak{s}}^{-1}-\dot{\overline{\Sigma}}_{\mathfrak{s}}\right) \mathrlap{\;,}
\label{eq:2PIfrgFlowEquationsCflowG0DONN1}
\end{equation}
\begin{equation}
\Delta \dot{\overline{\Omega}}_{\mathfrak{s}} = \frac{1}{2} \dot{C}_{\mathfrak{s}}^{-1} \left(\overline{G}_{\mathfrak{s}}-C_{\mathfrak{s}}\right) \mathrlap{\;,}
\label{eq:2PIfrgFlowEquationsCflowDOmega0DONN1}
\end{equation}
\begin{equation}
\dot{\overline{\Sigma}}_{\mathfrak{s}} = - \frac{1}{2} \dot{\overline{G}}_{\mathfrak{s}} \overline{\Phi}_{\mathfrak{s}}^{(2)} \mathrlap{\;,}
\label{eq:2PIfrgFlowEquationsCflowSigma0DONN1}
\end{equation}
\begin{equation}
\dot{\overline{\Phi}}_{\mathfrak{s}}^{(n)} = \frac{1}{2} \dot{\overline{G}}_{\mathfrak{s}} \overline{\Phi}_{\mathfrak{s}}^{(n+1)} \mathrlap{\quad \forall n \geq 2 \;.}
\label{eq:2PIfrgFlowEquationsCflowPhin0DONN1}
\end{equation}

\item For $N=2$ (up to $N_{\mathrm{max}}=3$):
\begin{equation}
\dot{\overline{G}}_{\mathfrak{s}} =- \overline{G}_{\mathfrak{s}}^{2} \left(\dot{C}_{\mathfrak{s}}^{-1}-\dot{\overline{\Sigma}}_{\mathfrak{s}}\right) \mathrlap{\;,}
\label{eq:2PIfrgFlowEquationsCflowG0DONN2}
\end{equation}
\begin{equation}
\Delta \dot{\overline{\Omega}}_{\mathfrak{s}} = \dot{C}_{\mathfrak{s}}^{-1} \left(\overline{G}_{\mathfrak{s}}-C_{\mathfrak{s}}\right) \mathrlap{\;,}
\label{eq:2PIfrgFlowEquationsCflowDOmega0DONN2}
\end{equation}
\begin{equation}
\dot{\overline{\Sigma}}_{\mathfrak{s}} = - \frac{1}{2} \dot{\overline{G}}_{\mathfrak{s}} \left(\overline{\Phi}_{\mathfrak{s},(1,1)(1,1)}^{(2)} + \overline{\Phi}_{\mathfrak{s},(1,1)(2,2)}^{(2)}\right) \mathrlap{\;,}
\label{eq:2PIfrgFlowEquationsCflowSigma0DONN2}
\end{equation}
\begin{equation}
\dot{\overline{\Phi}}_{\mathfrak{s},(1,1)(1,1)}^{(2)} = \frac{1}{2} \dot{\overline{G}}_{\mathfrak{s}} \left(\overline{\Phi}_{\mathfrak{s},(1,1)(1,1)(1,1)}^{(3)} + \overline{\Phi}_{\mathfrak{s},(1,1)(1,1)(2,2)}^{(3)}\right) \mathrlap{\;,}
\label{eq:2PIfrgFlowEquationsCflowPhi2s11110DONN2}
\end{equation}
\begin{equation}
\dot{\overline{\Phi}}_{\mathfrak{s},(1,1)(2,2)}^{(2)} = \frac{1}{2} \dot{\overline{G}}_{\mathfrak{s}} \left(\overline{\Phi}_{\mathfrak{s},(1,1)(1,1)(2,2)}^{(3)} + \overline{\Phi}_{\mathfrak{s},(1,1)(2,2)(2,2)}^{(3)}\right) \mathrlap{\;,}
\label{eq:2PIfrgFlowEquationsCflowPhi2s11220DONN2}
\end{equation}
\begin{equation}
\dot{\overline{\Phi}}_{\mathfrak{s},(1,1)(1,1)(1,1)}^{(3)} = \frac{1}{2} \dot{\overline{G}}_{\mathfrak{s}} \left(\overline{\Phi}_{\mathfrak{s},(1,1)(1,1)(1,1)(1,1)}^{(4)} + \overline{\Phi}_{\mathfrak{s},(1,1)(1,1)(1,1)(2,2)}^{(4)}\right) \mathrlap{\;,}
\label{eq:2PIfrgFlowEquationsCflowPhi3s1111110DONN2}
\end{equation}
\begin{equation}
\dot{\overline{\Phi}}_{\mathfrak{s},(1,1)(1,1)(2,2)}^{(3)} = \frac{1}{2} \dot{\overline{G}}_{\mathfrak{s}} \left(\overline{\Phi}_{\mathfrak{s},(1,1)(1,1)(1,1)(2,2)}^{(4)} + \overline{\Phi}_{\mathfrak{s},(1,1)(1,1)(2,2)(2,2)}^{(4)}\right) \mathrlap{\;.}
\label{eq:2PIfrgFlowEquationsCflowPhi3s1111220DONN2}
\end{equation}

\end{itemize}

\vspace{0.3cm}

We have introduced the shorthand notation $\overline{\Phi}_{\mathfrak{s}}^{(n)}\equiv\overline{\Phi}_{\mathfrak{s},(1,1)\cdots(1,1)}^{(n)}$ in~\eqref{eq:2PIfrgFlowEquationsCflowG0DONN1} to~\eqref{eq:2PIfrgFlowEquationsCflowPhin0DONN1}, which will be used again repeatedly in 2PI-FRG flow equations at $N=1$. The initial conditions required to solve the latter two sets of differential equations are directly deduced from those given by~\eqref{eq:2PIfrgInitialConditionsGki} to~\eqref{eq:2PIfrgCflowICDeltaOmega} in our previous general discussion on the C-flow. For all $N$, they are given by (see appendix~\ref{ann:2PIFRGCflowExpressionPhinsi0DONOrigRepr} for the expression of the components of $\overline{\Phi}_{\mathfrak{s}=\mathfrak{s}_{\mathrm{i}}}^{(4)}$):
\begin{equation}
\overline{G}_{\mathfrak{s}=\mathfrak{s}_{\mathrm{i}},a a'}=0 \mathrlap{\quad \forall a, a' \;,}
\label{eq:2PIfrgCflowICGki0DON}
\end{equation}
\begin{equation}
\Delta \overline{\Omega}_{\mathfrak{s}=\mathfrak{s}_{\mathrm{i}}} = 0 \mathrlap{\;,}
\label{eq:2PIfrgCflowICDeltaOmega0DON}
\end{equation}
\begin{equation}
\overline{\Sigma}_{\mathfrak{s}=\mathfrak{s}_{\mathrm{i}},a a'} = 0 \mathrlap{\quad \forall a, a' \;,}
\label{eq:2PIfrgCflowICSigma0DON}
\end{equation}
\begin{equation}
\overline{\Phi}_{\mathfrak{s} = \mathfrak{s}_{\mathrm{i}},(a_{1},a'_{1})(a_{2},a'_{2})}^{(2)} = U_{(a_{1},a'_{1})(a_{2},a'_{2})} = \frac{\lambda}{3}\left(\delta_{a_{1}a'_{1}}\delta_{a_{2}a'_{2}}+\delta_{a_{1}a_{2}}\delta_{a'_{1}a'_{2}}+\delta_{a_{1}a'_{2}}\delta_{a'_{1}a_{2}}\right) \mathrlap{\;,}
\label{eq:2PIfrgCflowICPhi20DON}
\end{equation}
\begin{equation}
\hspace{6.0cm} \overline{\Phi}_{\mathfrak{s} = \mathfrak{s}_{\mathrm{i}},(a_{1},a'_{1})\cdots(a_{n},a'_{n})}^{(n)} = 0 \quad \forall a_{1}, a'_{1},\cdots, a_{n}, a'_{n}, ~ \forall n ~ \mathrm{odd} \;,
\label{eq:2PIfrgCflowICPhiodd0DON}
\end{equation}
where~\eqref{eq:2bodyinteraction2PIFRG0DON} was used to express $U_{(a_{1},a'_{1})(a_{2},a'_{2})}$ in~\eqref{eq:2PIfrgCflowICPhi20DON}. We will implement the tC-flow up to $N_{\mathrm{max}}=10$ at $N=1$, which requires to determine all $\overline{\Phi}_{\mathfrak{s} = \mathfrak{s}_{\mathrm{i}}}^{(n)}$ up to $n=10$ (recall that the truncation orders $N_{\mathrm{max}}=9$ and $N_{\mathrm{max}}=10$ are equivalent for the tC-flow). This is achieved from the following perturbative expression of the Luttinger-Ward functional at $N=1$~\cite{bro15}:
\begin{equation}
\Phi_{\mathrm{SCPT}}(G) = \frac{1}{8} \lambda G^{2} - \frac{1}{48} \lambda^{2} G^{4} + \frac{1}{48} \lambda^{3} G^{6} - \frac{5}{128} \lambda^{4} G^{8} + \frac{101}{960} \lambda^{5} G^{10} + \mathcal{O}\big(\lambda^{6}\big) \;,
\label{eq:PertExpressionPhiN12PIFRGtCflow0DON}
\end{equation}
from which we infer:
\begin{equation}
\overline{\Phi}_{\mathfrak{s}=\mathfrak{s}_{\mathrm{i}}}^{(2)} = \lambda \;,
\label{eq:InitialCondPhi22PIFRGCflowN10DON}
\end{equation}
\begin{equation}
\overline{\Phi}_{\mathfrak{s}=\mathfrak{s}_{\mathrm{i}}}^{(4)} = - 8 \lambda^{2} \;,
\label{eq:InitialCondPhi42PIFRGCflowN10DON}
\end{equation}
\begin{equation}
\overline{\Phi}_{\mathfrak{s}=\mathfrak{s}_{\mathrm{i}}}^{(6)} = 960 \lambda^{3} \;,
\label{eq:InitialCondPhi62PIFRGCflowN10DON}
\end{equation}
\begin{equation}
\overline{\Phi}_{\mathfrak{s}=\mathfrak{s}_{\mathrm{i}}}^{(8)} = - 403200 \lambda^{4} \;,
\label{eq:InitialCondPhi82PIFRGCflowN10DON}
\end{equation}
\begin{equation}
\overline{\Phi}_{\mathfrak{s}=\mathfrak{s}_{\mathrm{i}}}^{(10)} = 390942720 \lambda^{5} \;,
\label{eq:InitialCondPhi102PIFRGCflowN10DON}
\end{equation}
where we have taken into account that the identity matrix of the bosonic index formalism (given by~\eqref{eq:IdentityBosonicMatrix2PIFRG}) reduces to 2 (and not 1) in the (0+0)-D limit at $N=1$, i.e.~\eqref{eq:IdentityBosonicMatrix2PIFRG} becomes in the (0+0)-D limit:
\begin{equation}
\mathcal{I}_{(a_{1},a'_{1})(a_{2},a'_{2})} \equiv \frac{\partial G_{a_{1}a'_{1}}}{\partial G_{a_{2}a'_{2}}} = \delta_{a_{1}a_{2}}\delta_{a'_{1}a'_{2}} + \delta_{a_{1}a'_{2}}\delta_{a'_{1}a_{2}} \;,
\label{eq:DefinitionBosonicIdentityMatrix0DON}
\end{equation}
which yields at $N=1$:
\begin{equation}
\mathcal{I} \equiv \frac{\partial G}{\partial G} = 2 \;.
\label{eq:OneEqualTwoN12PIFRG0DON}
\end{equation}
Hence, for the $O(N)$ model under consideration at $N=1$, the 2PI-FRG flow equations of our toy model can either be derived via standard derivation rules or by taking the (0+0)-D limit of their more general versions (written in terms of bosonic indices) using~\eqref{eq:OneEqualTwoN12PIFRG0DON}. We always follow the latter procedure in this study but solving the equations thus obtained in both situations leads in principle to identical results (see appendix~\ref{ann:ZeroDimLimitBosonicIndexFormalism}). Finally, the conditions implementing the truncations of the tC-flow and mC-flow schemes in the framework of the (0+0)-D $O(N)$-symmetric $\varphi^{4}$-theory are:
\begin{itemize}
\item For the tC-flow:
\begin{equation}
\overline{\Phi}_{\mathfrak{s}}^{(n)}=\overline{\Phi}_{\mathfrak{s}=\mathfrak{s}_{\mathrm{i}}}^{(n)} \mathrlap{\quad \forall \mathfrak{s}, ~ \forall n > N_{\mathrm{max}} \;.}
\label{eq:2PIfrgPhiBartCflow0DON}
\end{equation}

\pagebreak

\item For the mC-flow\footnote{Recall that, as discussed below~\eqref{eq:ComparisontCflowmCflow2PIFRGDown}, it is pointless to investigate the mC-flow with $N_{\mathrm{SCPT}} \leq N_{\mathrm{max}}/2$ as it reduces to the tC-flow in this situation.}$^{,}$\footnote{According to the definition of $\overline{\Phi}_{\mathrm{sym},\mathfrak{s}}^{(2)}$ given by~\eqref{eq:DefPhi2sym2PIFRGmCflow}, we have $\overline{\Phi}_{\mathrm{sym},\mathfrak{s}}^{(2)}=\overline{\Phi}_{\mathfrak{s}}^{(2)}$ for the studied toy model at $N=1$.} (at $N=1$):
\begin{itemize}
\item At $N_{\mathrm{max}}=2$:
\begin{itemize}
\item At $N_{\mathrm{SCPT}}=2$:
\begin{equation}
\overline{\Phi}^{(3)}_{\mathfrak{s}} = \left. \overline{\Phi}^{(3)}_{\mathrm{SCPT},N_{\mathrm{SCPT}}=2,\mathfrak{s}} \right|_{\lambda\rightarrow\overline{\Phi}_{\mathfrak{s}}^{(2)}} = - 4 \left(\overline{\Phi}_{\mathfrak{s}}^{(2)}\right)^{2} \overline{G}_{\mathfrak{s}} \;.
\label{eq:2PIFRGmCflowtruncationNmax2Nscpt20DON}
\end{equation}

\item At $N_{\mathrm{SCPT}}=3$:
\begin{equation}
\overline{\Phi}^{(3)}_{\mathfrak{s}} = \left. \overline{\Phi}^{(3)}_{\mathrm{SCPT},N_{\mathrm{SCPT}}=3,\mathfrak{s}} \right|_{\lambda\rightarrow\overline{\Phi}_{\mathfrak{s}}^{(2)}} = - 4 \left(\overline{\Phi}_{\mathfrak{s}}^{(2)}\right)^{2} \overline{G}_{\mathfrak{s}} + 20 \left(\overline{\Phi}_{\mathfrak{s}}^{(2)}\right)^{3} \overline{G}_{\mathfrak{s}}^{3} \;.
\label{eq:2PIFRGmCflowtruncationNmax2Nscpt30DON}
\end{equation}

\end{itemize}

\item At $N_{\mathrm{max}}=3$:
\begin{itemize}
\item At $N_{\mathrm{SCPT}}=2$:
\begin{equation}
\overline{\Phi}^{(4)}_{\mathfrak{s}} = \left. \overline{\Phi}^{(4)}_{\mathrm{SCPT},N_{\mathrm{SCPT}}=2,\mathfrak{s}} \right|_{\lambda\rightarrow\overline{\Phi}_{\mathfrak{s}}^{(2)}} = - 8 \left(\overline{\Phi}_{\mathfrak{s}}^{(2)}\right)^{2} \;.
\label{eq:2PIFRGmCflowtruncationNmax3Nscpt20DON}
\end{equation}

\item At $N_{\mathrm{SCPT}}=3$:
\begin{equation}
\overline{\Phi}^{(4)}_{\mathfrak{s}} = \left. \overline{\Phi}^{(4)}_{\mathrm{SCPT},N_{\mathrm{SCPT}}=3,\mathfrak{s}} \right|_{\lambda\rightarrow\overline{\Phi}_{\mathfrak{s}}^{(2)}} = - 8 \left(\overline{\Phi}_{\mathfrak{s}}^{(2)}\right)^{2} + 120 \left(\overline{\Phi}_{\mathfrak{s}}^{(2)}\right)^{3} \overline{G}_{\mathfrak{s}}^{2} \;.
\label{eq:2PIFRGmCflowtruncationNmax3Nscpt30DON}
\end{equation}

\end{itemize}

\item At $N_{\mathrm{max}}=4$:
\begin{itemize}
\item At $N_{\mathrm{SCPT}}=3$:
\begin{equation}
\overline{\Phi}^{(5)}_{\mathfrak{s}} = \left. \overline{\Phi}^{(5)}_{\mathrm{SCPT},N_{\mathrm{SCPT}}=3,\mathfrak{s}} \right|_{\lambda\rightarrow\overline{\Phi}_{\mathfrak{s}}^{(2)}} = 480 \left(\overline{\Phi}_{\mathfrak{s}}^{(2)}\right)^{3} \overline{G}_{\mathfrak{s}} \;.
\label{eq:2PIFRGmCflowtruncationNmax4Nscpt30DON}
\end{equation}

\end{itemize}

\end{itemize}

\end{itemize}

\vspace{0.3cm}

Finally, the cutoff function $R_{\mathfrak{s}}$ chosen for every application of the C-flow version of the 2PI-FRG in the framework of the original theory is identical to~\eqref{eq:choiceRkpure1PIFRG0DON} used for the 1PI-FRG, i.e.:
\begin{equation}
C_{\mathfrak{s},a_{1}a_{2}}^{-1} = C_{a_{1}a_{2}}^{-1} + R_{\mathfrak{s},a_{1}a_{2}} = \left(m^{2} + R_{\mathfrak{s}}\right) \delta_{a_{1}a_{2}} \mathrlap{\quad \forall a_{1}, a_{2} \;,}
\label{eq:2PIFRGCflowCutoff10DON}
\end{equation}
with
\begin{equation}
R_{\mathfrak{s}} = \mathfrak{s}^{-1} - 1 \;,
\label{eq:2PIFRGCflowCutoff20DON}
\end{equation}
which satisfies the required boundary conditions set by~\eqref{eq:2PIfrgBoundaryCondforCkCflowUpper} and~\eqref{eq:2PIfrgBoundaryCondforCkCflowBottom} as the flow parameter still runs from $\mathfrak{s}_{\mathrm{i}}=0$ to $\mathfrak{s}_{\mathrm{f}}=1$ during the flow. Note that, just like $k$ in our 1PI-FRG applications discussed in section~\ref{sec:1PIFRG0DON}, $\mathfrak{s}$ is also a dimensionless number here.

\vspace{0.5cm}

\begin{figure}[!t]
\captionsetup[subfigure]{labelformat=empty}
  \begin{center}
    \subfloat[]{
      \includegraphics[width=0.70\linewidth]{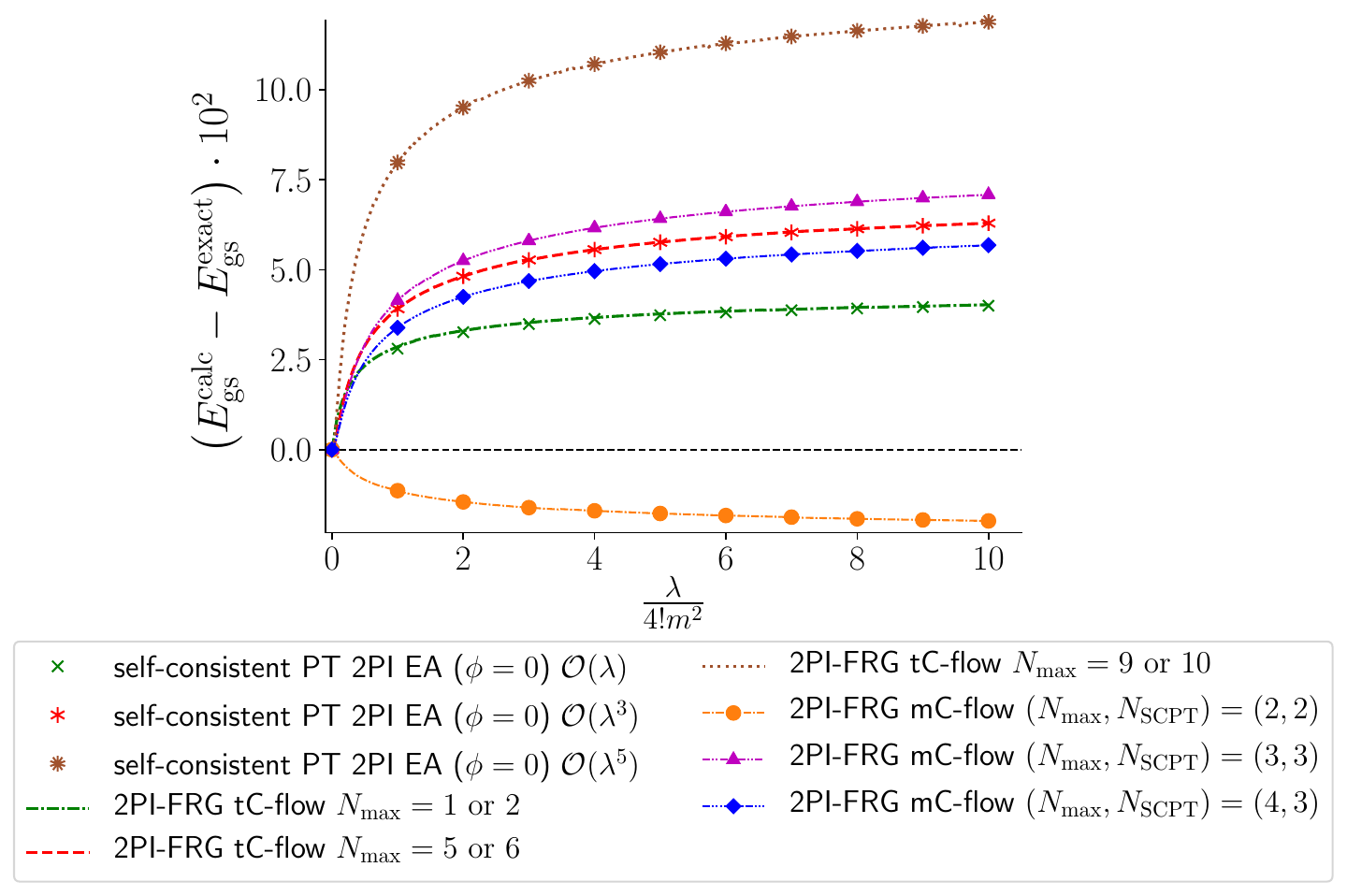}
                         }
   \\                     
    \subfloat[]{
      \includegraphics[width=0.70\linewidth]{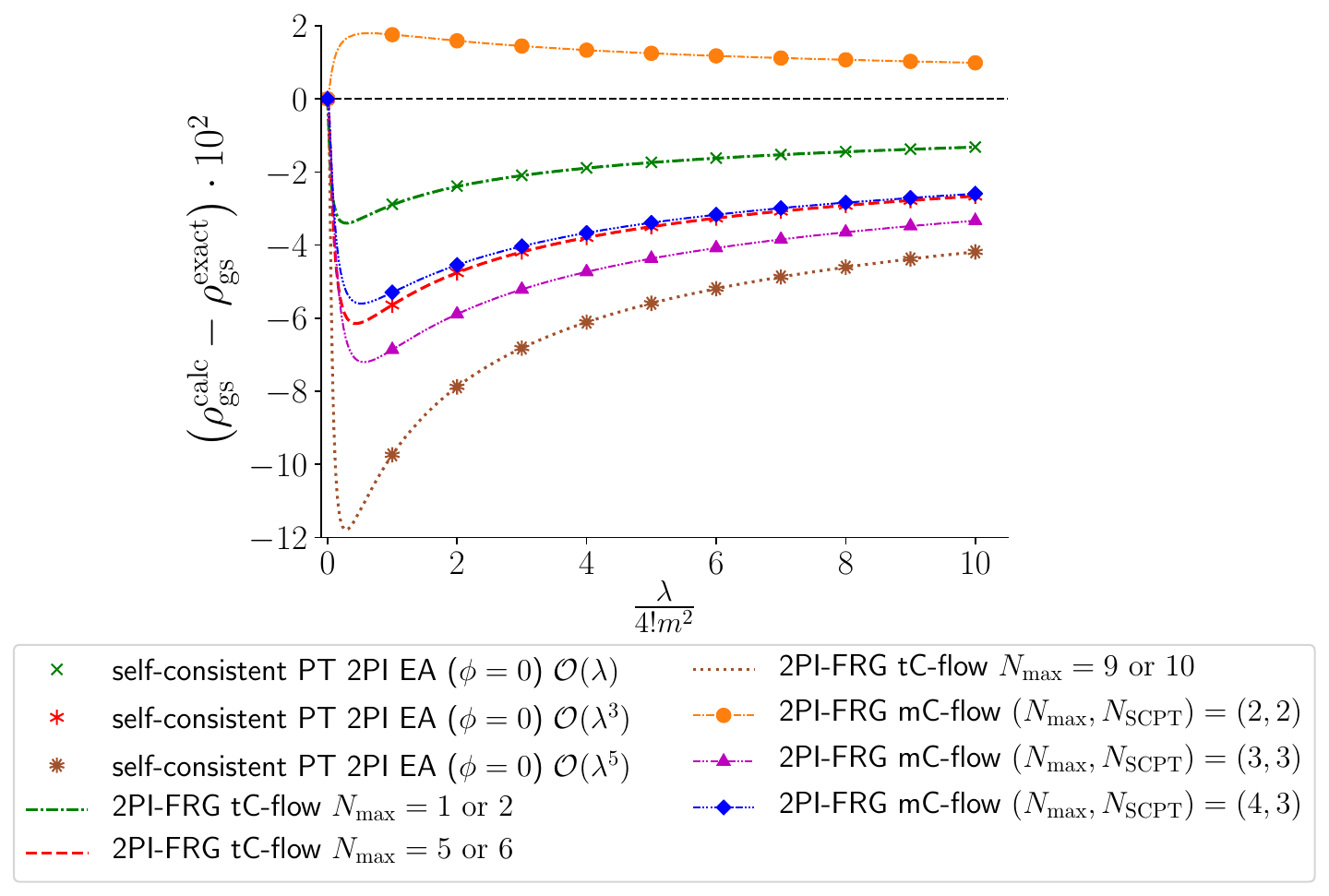}
                         }
\caption{Difference between the calculated gs energy $E_{\mathrm{gs}}^{\mathrm{calc}}$ or density $\rho_{\mathrm{gs}}^{\mathrm{calc}}$ and the corresponding exact solution $E_{\mathrm{gs}}^{\mathrm{exact}}$ or $\rho_{\mathrm{gs}}^{\mathrm{exact}}$ at $m^{2}=+1$ and $N=1$ ($\mathcal{R}e(\lambda)\geq 0$ and $\mathcal{I}m(\lambda)=0$).}
\label{fig:original2PIFRGCflowlambdaN1}
  \end{center}
\end{figure}

\begin{figure}[!htb]
\captionsetup[subfigure]{labelformat=empty}
  \begin{center}
    \subfloat[]{
      \includegraphics[width=0.90\linewidth]{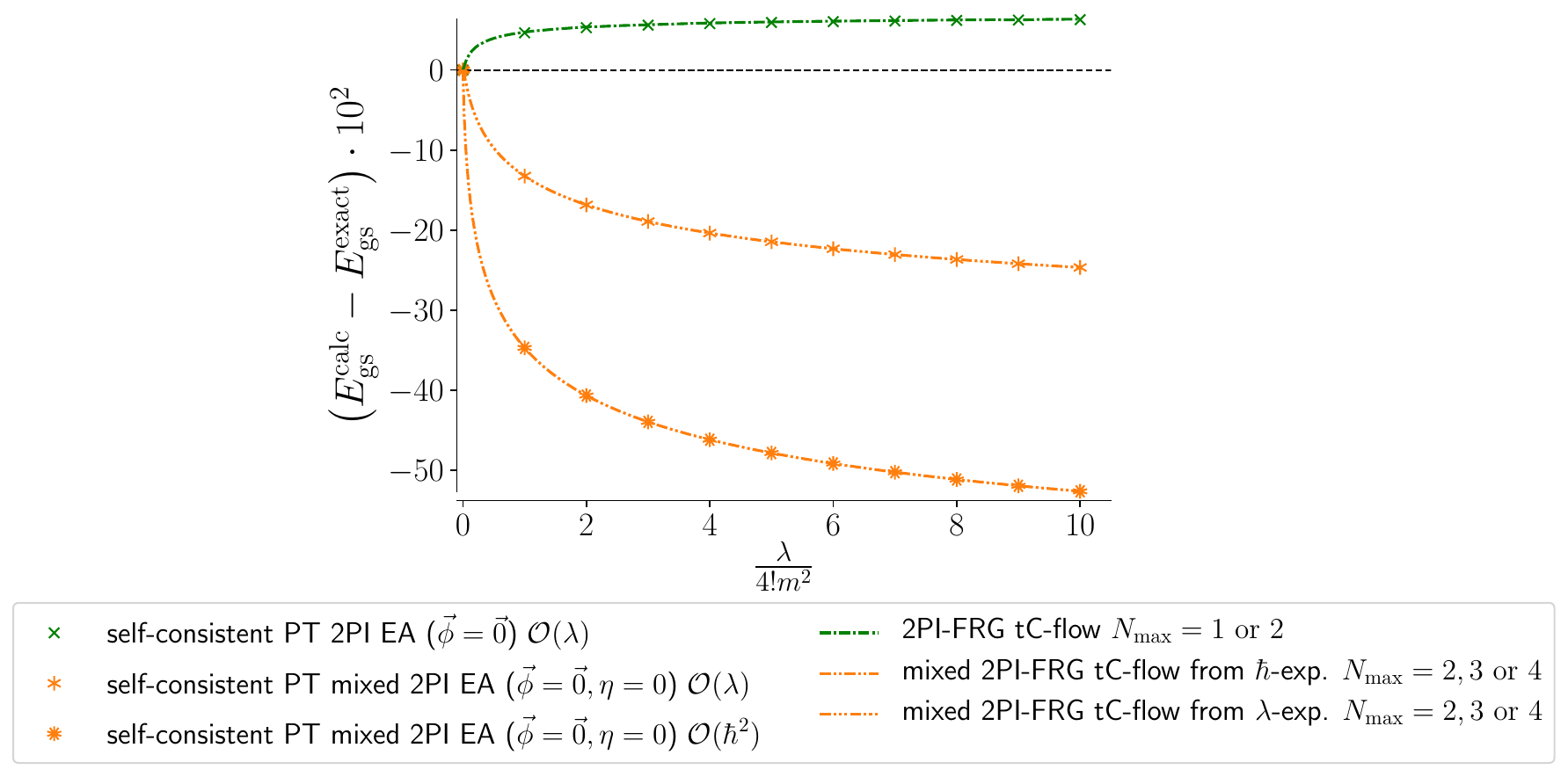}
                         }
   \\                     
    \subfloat[]{
      \includegraphics[width=0.90\linewidth]{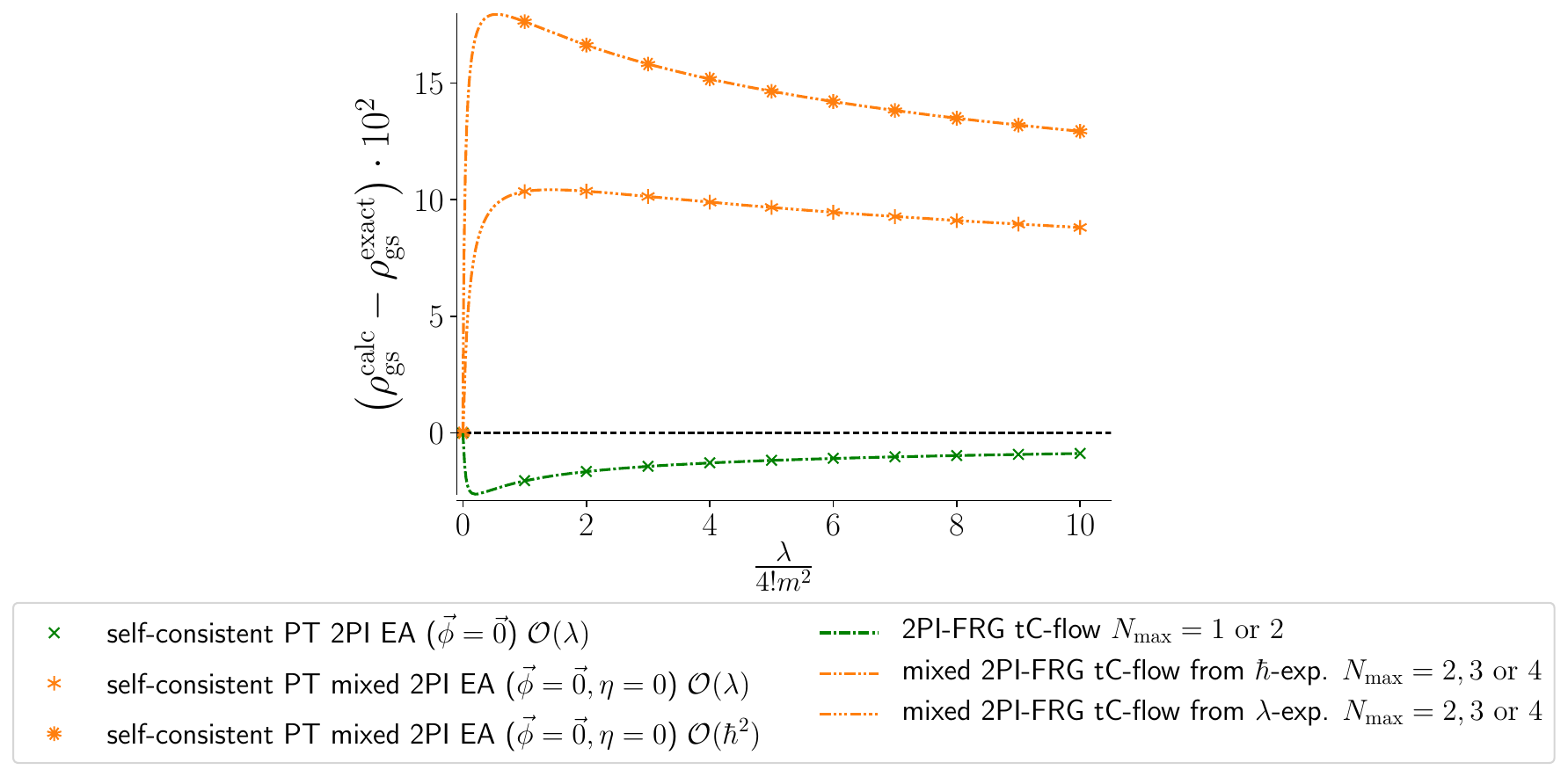}
                         }
\caption{Difference between the calculated gs energy $E_{\mathrm{gs}}^{\mathrm{calc}}$ or density $\rho_{\mathrm{gs}}^{\mathrm{calc}}$ and the corresponding exact solution $E_{\mathrm{gs}}^{\mathrm{exact}}$ or $\rho_{\mathrm{gs}}^{\mathrm{exact}}$ at $m^{2}=+1$ and $N=2$ ($\mathcal{R}e(\lambda)\geq 0$ and $\mathcal{I}m(\lambda)=0$). See notably the caption of fig.~\ref{fig:1PIEA} for the meaning of the indication ``$\mathcal{O}\big(\hbar^{n}\big)$'' for the results obtained from $\hbar$-expanded EAs within self-consitent PT.}
\label{fig:mixed2PIFRGCflowlambdaN2}
  \end{center}
\end{figure}

In conclusion, our C-flow results for the original theory are obtained by solving the differential equations~\eqref{eq:2PIfrgFlowEquationsCflowG0DONN1} to~\eqref{eq:2PIfrgFlowEquationsCflowPhin0DONN1} for $N=1$ (up to $N_{\mathrm{max}}=10$) and~\eqref{eq:2PIfrgFlowEquationsCflowG0DONN2} to~\eqref{eq:2PIfrgFlowEquationsCflowPhi3s1111220DONN2} for $N=2$ (up to $N_{\mathrm{max}}=4$), with initial conditions given by~\eqref{eq:2PIfrgCflowICGki0DON} to~\eqref{eq:2PIfrgCflowICPhiodd0DON} (along with~\eqref{eq:InitialCondPhi22PIFRGCflowN10DON} to~\eqref{eq:InitialCondPhi102PIFRGCflowN10DON} for $N=1$) and the cutoff function set by~\eqref{eq:2PIFRGCflowCutoff10DON} and~\eqref{eq:2PIFRGCflowCutoff20DON} for all $N$. Moreover, the truncations are imposed by~\eqref{eq:2PIfrgPhiBartCflow0DON} for all $N$ in the framework of the tC-flow and by~\eqref{eq:2PIFRGmCflowtruncationNmax2Nscpt20DON} to~\eqref{eq:2PIFRGmCflowtruncationNmax4Nscpt30DON} for $N=1$ in the framework of the mC-flow. This leads to all results presented in fig.~\ref{fig:original2PIFRGCflowlambdaN1} for $N=1$ and to the tC-flow result (obtained within the original representation) shown in fig.~\ref{fig:mixed2PIFRGCflowlambdaN2} for $N=2$. Focusing first on fig.~\ref{fig:original2PIFRGCflowlambdaN1}, we can see that, for both $E_{\mathrm{gs}}$ and $\rho_{\mathrm{gs}}$, the tC-flow curves are further and further away from the exact solution (except for $\lambda/4!\ll 1$) as the truncation order $N_{\mathrm{max}}$ increases. This behavior is consistent with the equivalence between the tC-flow and self-consistent PT discussed in section~\ref{sec:2PIFRGstateofplay}: such a worsening is thus a manifestation of the divergent character of the series underlying self-consistent PT, which is at the heart of chapter~\ref{chap:DiagTechniques}.

\vspace{0.5cm}


\begin{figure}[!htb]
  \centering
  \includegraphics[width=.7\linewidth]{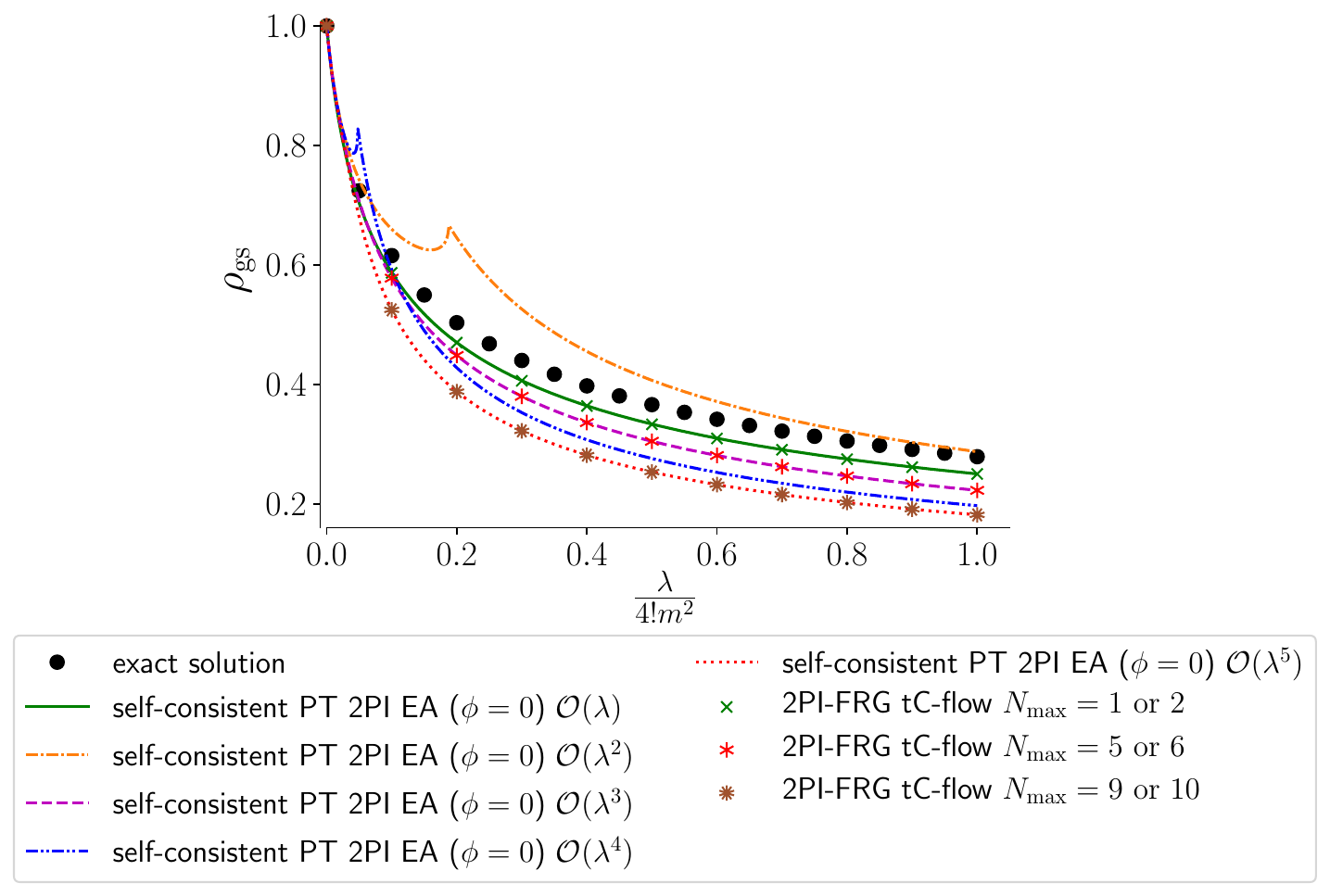}
  \caption{Gs density $\rho_{\mathrm{gs}}$ calculated at $m^{2}=+1$ and $N=1$ ($\mathcal{R}e(\lambda) \geq 0$ and $\mathcal{I}m(\lambda)=0$), and compared with the corresponding exact solution (black dots).}
  \label{fig:original2PIFRGCflowvsSCPTlambdaN1}
\end{figure}

We can also note that there are no tC-flow results for the truncation orders $N_{\mathrm{max}}=3 ~ \mathrm{or} ~ 4$ and $N_{\mathrm{max}}=7 ~ \mathrm{or} ~ 8$ in fig.~\ref{fig:original2PIFRGCflowlambdaN1} as we face the same stiffness issues (with the $\mathtt{NDSolve}$ function of $\mathtt{Mathematica~12.1}$) as those encountered in section~\ref{sec:1PIFRG0DON} for our 1PI-FRG applications in the broken-symmetry phase. We have also checked that the same problem manifests itself at $N=2$ with $N_{\mathrm{max}}=3 ~ \mathrm{or} ~ 4$: the only tC-flow curve for the original theory shown in fig.~\ref{fig:mixed2PIFRGCflowlambdaN2} is just obtained for $N_{\mathrm{max}}=1 ~ \mathrm{or} ~ 2$. However, all those tC-flow calculations (with or without stiffness issues) have been performed in the unbroken-symmetry phase, which suggests that the origins of these stiffness problems for the 1PI-FRG flow equations (only occurring in the broken-symmetry phase) on the one hand and for the 2PI-FRG tC-flow on the other hand are different. Despite such limitations, it is rather fruitful to further exploit the equivalence between the tC-flow implementation of the 2PI-FRG and self-consistent PT. To that end, we recall that, as we have done in chapter~\ref{chap:DiagTechniques}, self-consistent PT for $\Gamma^{(\mathrm{2PI})}(G)$ is carried out by solving the gap equations extremizing $\Gamma^{(\mathrm{2PI})}(G)$ with respect to $G$ and then picking up the physical solution\footnote{The physical solution $\overline{G}$ leading to all our results from self-consistent PT applied to $\Gamma^{(\mathrm{2PI})}(G)$ (shown notably in figs.~\ref{fig:2PIEAorigN2} and~\ref{fig:original2PIFRGCflowvsSCPTlambdaN1}) is defined here as the solution of the gap equation (for the propagator $G$) yielding the calculated complex gs density $\rho^{\mathrm{calc}}_{\mathrm{gs},\mathrm{comp}}$ that is closest to the corresponding exact solution $\rho_{\mathrm{gs}}^{\mathrm{exact}}$, i.e. that gives us the smallest norm $\big\lvert \rho^{\mathrm{calc}}_{\mathrm{gs},\mathrm{comp}}-\rho_{\mathrm{gs}}^{\mathrm{exact}} \big\rvert$. The term ``norm'' should be understood here as the norm of a complex number as $\rho^{\mathrm{calc}}_{\mathrm{gs},\mathrm{comp}}$ might have a non-zero imaginary part. Recall that $\rho_{\mathrm{gs}}^{\mathrm{calc}} = \mathcal{R}e\big(\rho_{\mathrm{gs},\mathrm{comp}}^{\mathrm{calc}}\big)$ in all our plots.} $\overline{G}$. Remarkably, the initial conditions of the C-flow are such that the tC-flow results coincide with those of our physical solutions for $N_{\mathrm{max}}=1 ~ \mathrm{or} ~ 2$, $N_{\mathrm{max}}=5 ~ \mathrm{or} ~ 6$ and $N_{\mathrm{max}}=9 ~ \mathrm{or} ~ 10$ at $N=1$ (according to fig.~\ref{fig:original2PIFRGCflowlambdaN1}) and for $N_{\mathrm{max}}=1 ~ \mathrm{or} ~ 2$ at $N=2$ (according to fig.~\ref{fig:mixed2PIFRGCflowlambdaN2}). Nevertheless, there is a change of physical solutions in the perturbative regime (i.e. for $\lambda/4!\ll 1$) for self-consistent PT applied up to order $\mathcal{O}(\lambda^{2})$ (or, equivalently, $\mathcal{O}(\hbar^{3})$) and $\mathcal{O}(\lambda^{4})$ (or, equivalently, $\mathcal{O}(\hbar^{5})$), as can be seen in fig.~\ref{fig:original2PIFRGCflowvsSCPTlambdaN1}. These correspond respectively to the tC-flow approach with truncation orders $N_{\mathrm{max}}=3 ~ \mathrm{or} ~ 4$ and $N_{\mathrm{max}}=7 ~ \mathrm{or} ~ 8$, which are precisely the $N_{\mathrm{max}}$ values where the stiffness problem arises. This illustrates that the tC-flow is not suited to fully reproduce self-consistent PT when there is a change of physical solutions involved in the latter framework for the chosen truncation of the EA. This limitation can be attributed to the fact that the initial conditions for the tC-flow are fixed once and for all (i.e. regardless of the values of coupling constants) from the perturbative expression of the Luttinger-Ward functional, which does not allow for reproducing the change of solutions observed in fig.~\ref{fig:original2PIFRGCflowvsSCPTlambdaN1}. This also implies that the stiffness problem arising in our tC-flow calculations is inherent to the C-flow formalism and not to the used numerical tools.

\vspace{0.5cm}

Regarding the mC-flow, we can see that the corresponding ansatz underlying the truncation manages to cure the aforementioned stiffness problem for $N_{\mathrm{max}}= 3 ~ \mathrm{or} ~ 4$, but not for all choices of $N_{\mathrm{SCPT}}$: the combination $(N_{\mathrm{max}},N_{\mathrm{SCPT}})=(3,2)$ set by~\eqref{eq:2PIFRGmCflowtruncationNmax3Nscpt20DON} still suffers from it for instance. Besides, it also introduces this issue at the truncation order $N_{\mathrm{max}}=2$ which is not affected in the framework of the tC-flow: this problem arises e.g. at the truncation $(N_{\mathrm{max}},N_{\mathrm{SCPT}})=(2,3)$ (set by~\eqref{eq:2PIFRGmCflowtruncationNmax2Nscpt30DON}), hence its absence from fig.~\ref{fig:original2PIFRGCflowlambdaN1}. Furthermore, we can also see that mC-flow results might deteriorate as $N_{\mathrm{max}}$ and/or $N_{\mathrm{SCPT}}$ increase(s), as can be seen by comparing the curves associated with $(N_{\mathrm{max}},N_{\mathrm{SCPT}})=(2,2)$ and $(N_{\mathrm{max}},N_{\mathrm{SCPT}})=(3,3)$ in fig.~\ref{fig:original2PIFRGCflowlambdaN1}. This is an important drawback as it shows that we lose accuracy while incorporating explicitly more information in our truncation. The ansatz underlying the mC-flow truncation is therefore not reliable. Moreover, although we have only investigated the mC-flow truncation scheme for $N=1$, the problematic features of the mC-flow that we have just put forward are expected to manifest themselves for any $N$. We thus consider the present discussion on fig.~\ref{fig:original2PIFRGCflowlambdaN1} to be sufficient to make our point for the mC-flow in the framework of the original representation.

\vspace{0.5cm}

All applications of the C-flow version of the 2PI-FRG have been discussed for $m^{2}>0$ so far. The corresponding formalism is actually not suited to treat the regime with $m^{2}<0$: if one might set $m^{2}$ equal to a negative value to solve the equation system made of~\eqref{eq:2PIfrgFlowEquationsCflowG0DON} to~\eqref{eq:2PIfrgFlowEquationsCflowPhin0DON}, the results thus obtained would be unphysical (with e.g. a negative estimate of the gs density $\rho_{\mathrm{gs}}$). Another way to tackle the broken-symmetry regime with a C-flow approach would be to add a linear source in the generating functional~\eqref{eq:2PIFRGgeneratingFunc} as follows:
\begin{equation}
Z[J,K]=e^{W[J,K]}=\int\mathcal{D}\widetilde{\psi} \ e^{-S\big[\widetilde{\psi}\big] + \int_{\alpha} J_{\alpha} \widetilde{\psi}_{\alpha} + \frac{1}{2}\int_{\alpha,\alpha'}\widetilde{\psi}_{\alpha}K_{\alpha\alpha'}\widetilde{\psi}_{\alpha'}} \;.
\label{eq:2PIFRGgeneratingFuncExtensionCflow}
\end{equation}
We would then reintroduce a cutoff function $R_{\mathfrak{s}}$ in the quadratic part of the classical action in order to develop an extension of the C-flow formalism able to treat regimes with a non-vanishing 1-point correlation function $\phi$ of the field $\widetilde{\psi}$. However, there are several reasons according to which such an extension is in general of little interest:
\begin{itemize}
\item The central object of the resulting approach would no longer be $\Gamma^{(\mathrm{2PI})}[G]\equiv\Gamma^{(\mathrm{2PI})}[\phi=0,G]$ but the full 2PI EA $\Gamma^{(\mathrm{2PI})}[\phi,G]$ instead, which would substancially complicate the underlying flow equations. Note also that, for purely fermionic systems, all the physical information is already accessible via $\Gamma^{(\mathrm{2PI})}[G]$ and there are no regimes with non-vanishing 1-point correlation functions. This makes the C-flow extension based on~\eqref{eq:2PIFRGgeneratingFuncExtensionCflow} particularly irrelevant for purely fermionic systems.
\item This extension would reduce to the usual C-flow approach developed for $\Gamma^{(\mathrm{2PI})}[G]$ in the regime with $m^{2}>0$. We have already shown that this C-flow approach possesses severe drawbacks (for its tC-flow as well as its mC-flow implementations), which would hold in principle in the regime with $m^{2}<0$ via such an extension.
\end{itemize}
Hence, we will stop our C-flow investigations at the present stage for the original theory and rather exploit other implementations of the 2PI-FRG to tackle the broken-symmetry regime.

\vspace{0.5cm}

In conclusion, we have studied the unbroken-symmetry regime of our $O(N)$ model by implementing higher truncation orders for the tC-flow and the mC-flow as compared to the applications presented in refs.~\cite{ren15,ren16} with the purpose of getting a clearer idea of the ability to control our approximations in these frameworks. The conclusion is rather negative as our results show that none of the two tested C-flow implementations are systematically improvable in a reliable fashion: i) the tC-flow worsens with increasing truncation orders and does not enable us to go reliably beyond its first non-trivial order (which coincides with the Hartree-Fock result); ii) the ansatz underlying the mC-flow does not seem reliable either as the corresponding results might also deteriorate as $N_{\mathrm{max}}$ and/or $N_{\mathrm{SCPT}}$ increase(s). Hence, we will then apply these methods to the mixed representation of the studied toy model (still in the regime with $m^{2}>0$) in order to check if these limitations can be lifted.

\subsubsection{Mixed 2PI functional renormalization group C-flow}

In the framework of the mixed theory, a second fluctuating field $\widetilde{\sigma}$ enters the arena and the Schwinger functional to consider involves the source-dependent terms $\frac{1}{2} \sum_{a_{1},a_{2}=1}^{N} \widetilde{\varphi}_{a_{1}} K_{a_{1}a_{2}}^{(\varphi)} \widetilde{\varphi}_{a_{2}}$ and $\frac{1}{2} K^{(\sigma)}\widetilde{\sigma}^{2}$ in that case, where $K^{(\varphi)}$ and $K^{(\sigma)}$ are both bilocal sources in finite dimensions. The corresponding Luttinger-Ward functional $\Phi_{\mathrm{mix}}(G,D)$ thus depends on two propagators, $D$ being the propagator associated with the Hubbard-Stratonovich field. Furthermore, the C-flow is now implemented by introducing two cutoff functions, $R_{\mathfrak{s}}^{(\varphi)}$ and $R_{\mathfrak{s}}^{(\sigma)}$, dressing respectively the free propagator of the original field (via $(C^{(\varphi)})^{-1}\rightarrow(C_{\mathfrak{s}}^{(\varphi)})^{-1}=(C^{(\varphi)})^{-1}+R_{\mathfrak{s}}^{(\varphi)}$ or $(C^{(\varphi)})^{-1}\rightarrow(C_{\mathfrak{s}}^{(\varphi)})^{-1}=R_{\mathfrak{s}}^{(\varphi)}(C^{(\varphi)})^{-1}$) and that of the Hubbard-Stratonovich field (via $(C^{(\sigma)})^{-1}\rightarrow(C_{\mathfrak{s}}^{(\sigma)})^{-1}=(C^{(\sigma)})^{-1}+R_{\mathfrak{s}}^{(\sigma)}$ or $(C^{(\sigma)})^{-1}\rightarrow(C_{\mathfrak{s}}^{(\sigma)})^{-1}=R_{\mathfrak{s}}^{(\sigma)}(C^{(\sigma)})^{-1}$). Hence, the contribution of the Hubbard-Stratonovich sector simply adds up to the flow equations of the original C-flow such that the tower of differential equations underlying the C-flow in the framework of the mixed representation of our (0+0)-D $O(N)$ model is directly deduced from~\eqref{eq:2PIfrgFlowEquationsCflowG0DON} to~\eqref{eq:2PIfrgFlowEquationsCflowPhin0DON}:
\begin{equation}
\dot{\overline{G}}_{\mathfrak{s},a_{1} a'_{1}}=-\sum_{a_{2},a'_{2}=1}^{N} \overline{G}_{\mathfrak{s}, a_{1} a_{2}}\left(\left(\dot{C}^{(\varphi)}_{\mathfrak{s}}\right)^{-1}-\dot{\overline{\Sigma}}^{(\varphi)}_{\mathfrak{s}}\right)_{a_{2} a'_{2}} \overline{G}_{\mathfrak{s}, a'_{2} a'_{1}}\;,
\label{eq:2PIfrgFlowEquationsmixedCflowG0DON}
\end{equation}
\begin{equation}
\dot{\overline{D}}_{\mathfrak{s}} = - \overline{D}_{\mathfrak{s}}^{2}\left(\left(\dot{C}^{(\sigma)}_{\mathfrak{s}}\right)^{-1}-\dot{\overline{\Sigma}}^{(\sigma)}_{\mathfrak{s}}\right)\;,
\label{eq:2PIfrgFlowEquationsmixedCflowD0DON}
\end{equation}
\begin{equation}
\Delta \dot{\overline{\Omega}}_{\mathfrak{s}} = \frac{1}{2} \sum_{a,a'=1}^{N} \left(\dot{C}^{(\varphi)}_{\mathfrak{s}}\right)_{a a'}^{-1} \left(\overline{G}_{\mathfrak{s}}-C^{(\varphi)}_{\mathfrak{s}}\right)_{a a'} + \frac{1}{2} \left(\dot{C}^{(\sigma)}_{\mathfrak{s}}\right)^{-1} \left(\overline{D}_{\mathfrak{s}}-C^{(\sigma)}_{\mathfrak{s}}\right)\;,
\label{eq:2PIfrgFlowEquationsmixedCflowDOmega0DON}
\end{equation}
\begin{equation}
\dot{\overline{\Sigma}}^{(\varphi)}_{\mathfrak{s},a_{1} a'_{1}} = - \frac{1}{2} \sum_{a_{2},a'_{2} = 1}^{N} \dot{\overline{G}}_{\mathfrak{s},a_{2} a'_{2}} \overline{\Phi}_{\mathrm{mix},\mathfrak{s},(a_{2}, a'_{2})(a_{1}, a'_{1})}^{(2G)} - \frac{1}{2} \dot{\overline{D}}_{\mathfrak{s}} \overline{\Phi}_{\mathrm{mix},\mathfrak{s},a_{1} a'_{1}}^{(1G,1D)} \;,
\label{eq:2PIfrgFlowEquationsmixedCflowSigmavarphi0DON}
\end{equation}
\begin{equation}
\dot{\overline{\Sigma}}^{(\sigma)}_{\mathfrak{s}} = - \frac{1}{2} \sum_{a_{1},a'_{1} = 1}^{N} \dot{\overline{G}}_{\mathfrak{s},a_{1} a'_{1}} \overline{\Phi}_{\mathrm{mix},\mathfrak{s},a_{1} a'_{1}}^{(1G,1D)} - \frac{1}{2} \dot{\overline{D}}_{\mathfrak{s}} \overline{\Phi}_{\mathrm{mix},\mathfrak{s}}^{(2D)} \;,
\label{eq:2PIfrgFlowEquationsmixedCflowSigmasigma0DON}
\end{equation}
\begin{equation}
\begin{split}
\dot{\overline{\Phi}}_{\mathrm{mix},\mathfrak{s},(a_{1}, a'_{1})\cdots(a_{n}, a'_{n})}^{(nG,mD)} = & \ \frac{1}{2} \sum_{a_{n+1},a'_{n+1} = 1}^{N} \dot{\overline{G}}_{\mathfrak{s},a_{n+1} a'_{n+1}} \overline{\Phi}_{\mathrm{mix},\mathfrak{s},(a_{n+1}, a'_{n+1})(a_{1}, a'_{1})\cdots(a_{n}, a'_{n})}^{((n+1)G,mD)} \\
& + \frac{1}{2} \dot{\overline{D}}_{\mathfrak{s}} \overline{\Phi}_{\mathrm{mix},\mathfrak{s},(a_{1}, a'_{1})\cdots(a_{n}, a'_{n})}^{(nG,(m+1)D)} \quad \forall (n,m) \setminus \lbrace(1,0),(0,1)\rbrace \;,
\end{split}
\label{eq:2PIfrgFlowEquationsmixedCflowPhin0DON}
\end{equation}
where $\Delta \overline{\Omega}_{\mathfrak{s}} = \overline{\Omega}_{\mathfrak{s}} + \frac{1}{2} \mathrm{Tr}_{a} \left[ \mathrm{ln}\Big(2\pi C^{(\varphi)}_{\mathfrak{s}}\Big) \right] + \frac{1}{2}\mathrm{ln}\Big(C^{(\sigma)}_{\mathfrak{s}}\Big)$ and the self-energies are defined as:
\begin{equation}
\Sigma_{a a'}^{(\varphi)}(G,D)=-\frac{\partial \Phi_{\mathrm{mix}}(G,D)}{\partial G_{a a'}} \;,
\label{eq:2PIfrgFlowEquationsmixedCflowDefSigmavarphi0DON}
\end{equation}
\begin{equation}
\Sigma^{(\sigma)}(G,D)=-\frac{\partial \Phi_{\mathrm{mix}}(G,D)}{\partial D} \;.
\label{eq:2PIfrgFlowEquationsmixedCflowDefSigmasigma0DON}
\end{equation}
Note also that we have used the shorthand notation $\overline{\Phi}^{(nG,mD)}_{\mathrm{mix},\mathfrak{s},(a_{1},a'_{1}) \cdots (a_{n},a'_{n})} \equiv \left.\frac{\partial^{n+m}\Phi_{\mathrm{mix},\mathfrak{s}}(G,D)}{\partial G_{a_{1}a'_{1}}\cdots\partial G_{a_{n}a'_{n}}\partial D^{m}}\right|_{G=\overline{G}_{\mathfrak{s}} \atop D=\overline{D}_{\mathfrak{s}}}$ together with $\overline{\Phi}^{(nG)}_{\mathrm{mix},\mathfrak{s}} \equiv \overline{\Phi}^{(nG,0D)}_{\mathrm{mix},\mathfrak{s}}$ and $\overline{\Phi}^{(nD)}_{\mathrm{mix},\mathfrak{s}} \equiv \overline{\Phi}^{(0G,nD)}_{\mathrm{mix},\mathfrak{s}}$ $\forall n$ in~\eqref{eq:2PIfrgFlowEquationsmixedCflowSigmavarphi0DON} and~\eqref{eq:2PIfrgFlowEquationsmixedCflowSigmasigma0DON}. The symmetry arguments put forward previously for the original C-flow also apply to the original sector in the present situation. This implies notably that:
\begin{equation}
\left(C_{\mathfrak{s}}^{(\varphi)}\right)^{-1}_{a a'} = \left(C^{(\varphi)}_{\mathfrak{s}}\right)^{-1} \delta_{a a'} \mathrlap{\quad \forall\mathfrak{s} \;,}
\label{eq:diagonalC2PIFRGmixedCflow0DON}
\end{equation}
\begin{equation}
\overline{G}_{\mathfrak{s},a a'} = \overline{G}_{\mathfrak{s}} \ \delta_{a a'} \mathrlap{\quad \forall\mathfrak{s} \;,}
\label{eq:diagonalG2PIFRGmixedCflow0DON}
\end{equation}
\begin{equation}
\overline{\Sigma}^{(\varphi)}_{\mathfrak{s},a a'} = \overline{\Sigma}^{(\varphi)}_{\mathfrak{s}} \ \delta_{a a'} \mathrlap{\quad \forall\mathfrak{s} \;.}
\label{eq:diagonalSigma2PIFRGmixedCflow0DON}
\end{equation}
Therefore, the sums over color indices in~\eqref{eq:2PIfrgFlowEquationsmixedCflowG0DON} to~\eqref{eq:2PIfrgFlowEquationsmixedCflowPhin0DON} can be dealt with in the same manner as in~\eqref{eq:2PIfrgFlowEquationsCflowG0DON} to~\eqref{eq:2PIfrgFlowEquationsCflowPhin0DON}. In this way,~\eqref{eq:2PIfrgFlowEquationsmixedCflowG0DON} to~\eqref{eq:2PIfrgFlowEquationsmixedCflowPhin0DON} become:
\begin{itemize}
\item For $N=1$ ($\forall N_{\mathrm{max}}$):
\begin{equation}
\dot{\overline{G}}_{\mathfrak{s}}=- \overline{G}_{\mathfrak{s}}^{2} \left(\left(\dot{C}^{(\varphi)}_{\mathfrak{s}}\right)^{-1}-\dot{\overline{\Sigma}}^{(\varphi)}_{\mathfrak{s}}\right) \mathrlap{\;,}
\label{eq:2PIfrgFlowEquationsmixedCflowG0DON1}
\end{equation}
\begin{equation}
\dot{\overline{D}}_{\mathfrak{s}} = - \overline{D}_{\mathfrak{s}}^{2}\left(\left(\dot{C}^{(\sigma)}_{\mathfrak{s}}\right)^{-1}-\dot{\overline{\Sigma}}^{(\sigma)}_{\mathfrak{s}}\right) \mathrlap{\;,}
\label{eq:2PIfrgFlowEquationsmixedCflowD0DON1}
\end{equation}
\begin{equation}
\Delta \dot{\overline{\Omega}}_{\mathfrak{s}} = \frac{1}{2} \left(\dot{C}^{(\varphi)}_{\mathfrak{s}}\right)^{-1} \left(\overline{G}_{\mathfrak{s}}-C^{(\varphi)}_{\mathfrak{s}}\right) + \frac{1}{2} \left(\dot{C}^{(\sigma)}_{\mathfrak{s}}\right)^{-1} \left(\overline{D}_{\mathfrak{s}}-C^{(\sigma)}_{\mathfrak{s}}\right) \mathrlap{\;,}
\label{eq:2PIfrgFlowEquationsmixedCflowDOmega0DON1}
\end{equation}
\begin{equation}
\dot{\overline{\Sigma}}^{(\varphi)}_{\mathfrak{s}} = - \frac{1}{2} \dot{\overline{G}}_{\mathfrak{s}} \overline{\Phi}_{\mathrm{mix},\mathfrak{s}}^{(2G)} - \frac{1}{2} \dot{\overline{D}}_{\mathfrak{s}} \overline{\Phi}_{\mathrm{mix},\mathfrak{s}}^{(1G,1D)} \mathrlap{\;,}
\label{eq:2PIfrgFlowEquationsmixedCflowSigmavarphi0DON1}
\end{equation}
\begin{equation}
\dot{\overline{\Sigma}}^{(\sigma)}_{\mathfrak{s}} = - \frac{1}{2} \dot{\overline{G}}_{\mathfrak{s}} \overline{\Phi}_{\mathrm{mix},\mathfrak{s}}^{(1G,1D)} - \frac{1}{2} \dot{\overline{D}}_{\mathfrak{s}} \overline{\Phi}_{\mathrm{mix},\mathfrak{s}}^{(2D)} \mathrlap{\;,}
\label{eq:2PIfrgFlowEquationsmixedCflowSigmasigma0DON1}
\end{equation}
\begin{equation}
\hspace{1.2cm} \dot{\overline{\Phi}}_{\mathrm{mix},\mathfrak{s}}^{(nG,mD)} = \frac{1}{2} \dot{\overline{G}}_{\mathfrak{s}} \overline{\Phi}_{\mathrm{mix},\mathfrak{s}}^{((n+1)G,mD)} + \frac{1}{2} \dot{\overline{D}}_{\mathfrak{s}} \overline{\Phi}_{\mathrm{mix},\mathfrak{s}}^{(nG,(m+1)D)} \quad \forall (n,m) \setminus \lbrace(1,0),(0,1)\rbrace \;.
\label{eq:2PIfrgFlowEquationsmixedCflowPhin0DON1}
\end{equation}

\item For $N=2$ (up to $N_{\mathrm{max}}=2$):
\begin{equation}
\dot{\overline{G}}_{\mathfrak{s}} =- \overline{G}_{\mathfrak{s}}^{2} \left(\left(\dot{C}^{(\varphi)}_{\mathfrak{s}}\right)^{-1}-\dot{\overline{\Sigma}}^{(\varphi)}_{\mathfrak{s}}\right) \mathrlap{\;,}
\label{eq:2PIfrgFlowEquationsmixedCflowG0DONN2}
\end{equation}
\begin{equation}
\dot{\overline{D}}_{\mathfrak{s}} =- \overline{D}_{\mathfrak{s}}^{2} \left(\left(\dot{C}^{(\sigma)}_{\mathfrak{s}}\right)^{-1}-\dot{\overline{\Sigma}}^{(\sigma)}_{\mathfrak{s}}\right) \mathrlap{\;,}
\label{eq:2PIfrgFlowEquationsmixedCflowD0DONN2}
\end{equation}
\begin{equation}
\Delta \dot{\overline{\Omega}}_{\mathfrak{s}} = \left(\dot{C}^{(\varphi)}_{\mathfrak{s}}\right)^{-1} \left(\overline{G}_{\mathfrak{s}}-C^{(\varphi)}_{\mathfrak{s}}\right) + \frac{1}{2} \left(\dot{C}^{(\sigma)}_{\mathfrak{s}}\right)^{-1} \left(\overline{D}_{\mathfrak{s}}-C^{(\sigma)}_{\mathfrak{s}}\right) \mathrlap{\;,}
\label{eq:2PIfrgFlowEquationsmixedCflowDOmega0DONN2}
\end{equation}
\begin{equation}
\dot{\overline{\Sigma}}^{(\varphi)}_{\mathfrak{s}} = - \frac{1}{2} \dot{\overline{G}}_{\mathfrak{s}} \left(\overline{\Phi}_{\mathrm{mix},\mathfrak{s},(1,1)(1,1)}^{(2G)} + \overline{\Phi}_{\mathrm{mix},\mathfrak{s},(1,1)(2,2)}^{(2G)}\right) - \frac{1}{2} \dot{\overline{D}}_{\mathfrak{s}} \overline{\Phi}^{(1G,1D)}_{\mathrm{mix},\mathfrak{s},(1,1)} \mathrlap{\;,}
\label{eq:2PIfrgFlowEquationsmixedCflowSigmavarphi0DONN2}
\end{equation}
\begin{equation}
\dot{\overline{\Sigma}}^{(\sigma)}_{\mathfrak{s}} = - \dot{\overline{G}}_{\mathfrak{s}} \overline{\Phi}_{\mathrm{mix},\mathfrak{s},(1,1)}^{(1G,1D)} - \frac{1}{2} \dot{\overline{D}}_{\mathfrak{s}} \overline{\Phi}^{(2D)}_{\mathrm{mix},\mathfrak{s}} \mathrlap{\;,}
\label{eq:2PIfrgFlowEquationsmixedCflowSigmasigma0DONN2}
\end{equation}
\begin{equation}
\dot{\overline{\Phi}}_{\mathrm{mix},\mathfrak{s},(1,1)(1,1)}^{(2G)} = \frac{1}{2} \dot{\overline{G}}_{\mathfrak{s}} \left(\overline{\Phi}_{\mathrm{mix},\mathfrak{s},(1,1)(1,1)(1,1)}^{(3G)} + \overline{\Phi}_{\mathrm{mix},\mathfrak{s},(1,1)(1,1)(2,2)}^{(3G)}\right) + \frac{1}{2} \dot{\overline{D}}_{\mathfrak{s}} \overline{\Phi}^{(2G,1D)}_{\mathrm{mix},\mathfrak{s},(1,1)(1,1)} \;,
\label{eq:2PIfrgFlowEquationsmixedCflowPhi2Gs11110DONN2}
\end{equation}
\begin{equation}
\dot{\overline{\Phi}}_{\mathrm{mix},\mathfrak{s},(1,1)(2,2)}^{(2G)} = \frac{1}{2} \dot{\overline{G}}_{\mathfrak{s}} \left(\overline{\Phi}_{\mathrm{mix},\mathfrak{s},(1,1)(1,1)(2,2)}^{(3G)} + \overline{\Phi}_{\mathrm{mix},\mathfrak{s},(1,1)(2,2)(2,2)}^{(3G)}\right) + \frac{1}{2} \dot{\overline{D}}_{\mathfrak{s}} \overline{\Phi}^{(2G,1D)}_{\mathrm{mix},\mathfrak{s},(1,1)(2,2)} \;,
\label{eq:2PIfrgFlowEquationsmixedCflowPhi2Gs11220DONN2}
\end{equation}
\begin{equation}
\dot{\overline{\Phi}}_{\mathrm{mix},\mathfrak{s}}^{(2D)} = \dot{\overline{G}}_{\mathfrak{s}} \overline{\Phi}_{\mathrm{mix},\mathfrak{s},(1,1)}^{(1G,2D)} + \frac{1}{2} \dot{\overline{D}}_{\mathfrak{s}} \overline{\Phi}_{\mathrm{mix},\mathfrak{s}}^{(3D)} \mathrlap{\;,}
\label{eq:2PIfrgFlowEquationsmixedCflowPhi2Ds0DONN2}
\end{equation}
\begin{equation}
\dot{\overline{\Phi}}_{\mathrm{mix},\mathfrak{s},(1,1)}^{(1G,1D)} = \frac{1}{2} \dot{\overline{G}}_{\mathfrak{s}} \left(\overline{\Phi}_{\mathrm{mix},\mathfrak{s},(1,1)(1,1)}^{(2G,1D)} + \overline{\Phi}_{\mathrm{mix},\mathfrak{s},(1,1)(2,2)}^{(2G,1D)} \right) + \frac{1}{2} \dot{\overline{D}}_{\mathfrak{s}} \overline{\Phi}_{\mathrm{mix},\mathfrak{s},(1,1)}^{(1G,2D)} \mathrlap{\;.}
\label{eq:2PIfrgFlowEquationsmixedCflowPhi1G1Ds110DONN2}
\end{equation}

\end{itemize}

\vspace{0.3cm}

The initial conditions required to solve the latter equation systems are determined in the same manner as for the original theory, with one additional subtlety: as was discussed in chapter~\ref{chap:DiagTechniques}, the series representing the Luttinger-Ward functional $\Phi_{\mathrm{mix}}(G,D)$ in the framework of self-consistent PT differ whether $\hbar$ or $\lambda$ is used as expansion parameter. Hence, we define these two distinct series from~\eqref{eq:mixed2PIEAZeroVevfinalexpression} and~\eqref{eq:mixed2PIEAlambdafinalexpression} as follows:
\begin{equation}
\begin{split}
\Phi_{\mathrm{mix},\mathrm{SCPT},\hbar\text{-exp}}(G,D) = & \ \frac{\hbar^{2}}{12}\begin{gathered}
\begin{fmffile}{DiagramsFRG/2PIFRGmixedCflow_PhiSCPT_Fock}
\begin{fmfgraph}(15,15)
\fmfleft{i}
\fmfright{o}
\fmfv{decor.shape=circle,decor.size=2.0thick,foreground=(0,,0,,1)}{v1}
\fmfv{decor.shape=circle,decor.size=2.0thick,foreground=(0,,0,,1)}{v2}
\fmf{phantom,tension=11}{i,v1}
\fmf{phantom,tension=11}{v2,o}
\fmf{plain,left,tension=0.4,foreground=(1,,0,,0)}{v1,v2,v1}
\fmf{wiggly,foreground=(1,,0,,0)}{v1,v2}
\end{fmfgraph}
\end{fmffile}
\end{gathered} + \mathcal{O}\big(\hbar^3\big) \\
= & \ \frac{1}{12} \hbar^{2} \lambda D \sum_{a_{1}, a_{2}= 1}^{N} G_{a_{1} a_{2}}^{2} + \mathcal{O}\big(\hbar^3\big) \;,
\end{split}
\label{eq:Cflow2PIFRGDefPhimixSCPThbarExp0DON}
\end{equation}
\begin{equation}
\begin{split}
\Phi_{\mathrm{mix},\mathrm{SCPT},\lambda\text{-exp}}(G,D) = & \ \frac{1}{24}\begin{gathered}
\begin{fmffile}{DiagramsFRG/2PIFRGmixedCflow_PhiSCPT_Hartree}
\begin{fmfgraph}(30,20)
\fmfleft{i}
\fmfright{o}
\fmfv{decor.shape=circle,decor.size=2.0thick,foreground=(0,,0,,1)}{v1}
\fmfv{decor.shape=circle,decor.size=2.0thick,foreground=(0,,0,,1)}{v2}
\fmf{phantom,tension=10}{i,i1}
\fmf{phantom,tension=10}{o,o1}
\fmf{plain,left,tension=0.5,foreground=(1,,0,,0)}{i1,v1,i1}
\fmf{plain,right,tension=0.5,foreground=(1,,0,,0)}{o1,v2,o1}
\fmf{wiggly,foreground=(1,,0,,0)}{v1,v2}
\end{fmfgraph}
\end{fmffile}
\end{gathered}
+\frac{1}{12}\begin{gathered}
\begin{fmffile}{DiagramsFRG/2PIFRGmixedCflow_PhiSCPT_Fock}
\begin{fmfgraph}(15,15)
\fmfleft{i}
\fmfright{o}
\fmfv{decor.shape=circle,decor.size=2.0thick,foreground=(0,,0,,1)}{v1}
\fmfv{decor.shape=circle,decor.size=2.0thick,foreground=(0,,0,,1)}{v2}
\fmf{phantom,tension=11}{i,v1}
\fmf{phantom,tension=11}{v2,o}
\fmf{plain,left,tension=0.4,foreground=(1,,0,,0)}{v1,v2,v1}
\fmf{wiggly,foreground=(1,,0,,0)}{v1,v2}
\end{fmfgraph}
\end{fmffile}
\end{gathered} + \mathcal{O}\big(\lambda^2\big) \\
= & \ \frac{1}{24} \lambda D \left(\sum_{a_{1}=1}^{N} G_{a_{1}a_{1}} \right)^{2} + \frac{1}{12} \lambda D \sum_{a_{1}, a_{2} = 1}^{N} G_{a_{1} a_{2}}^{2} + \mathcal{O}\big(\lambda^3\big) \;.
\end{split}
\label{eq:Cflow2PIFRGDefPhimixSCPTlambdaExp0DON}
\end{equation}
Relations~\eqref{eq:Cflow2PIFRGDefPhimixSCPThbarExp0DON} and~\eqref{eq:Cflow2PIFRGDefPhimixSCPTlambdaExp0DON} are thus exploited to determine the initial conditions $\overline{\Phi}_{\mathfrak{s}=\mathfrak{s}_{\mathrm{i}}}^{(nG,mD)}$ which are identical whether we consider~\eqref{eq:Cflow2PIFRGDefPhimixSCPThbarExp0DON} or~\eqref{eq:Cflow2PIFRGDefPhimixSCPTlambdaExp0DON} for all combinations of $m$ and $n$ except for $m=2n$, as a consequence of $\overline{G}_{\mathfrak{s}=\mathfrak{s}_{\mathrm{i}},a a'}=\overline{D}_{\mathfrak{s}=\mathfrak{s}_{\mathrm{i}}}=0$ $\forall a, a'$. In conclusion, we obtain the following initial conditions for all $N$ (see appendix~\ref{ann:2PIFRGCflowExpressionPhinsi0DONMixedRepr}):
\begin{equation}
\overline{G}_{\mathfrak{s}=\mathfrak{s}_{\mathrm{i}},a a'}=0 \mathrlap{\quad \forall a, a' \;,}
\label{eq:2PIfrgmixedCflowICGki0DON}
\end{equation}
\begin{equation}
\overline{D}_{\mathfrak{s}=\mathfrak{s}_{\mathrm{i}}}=0 \mathrlap{\;,}
\label{eq:2PIfrgmixedCflowICDki0DON}
\end{equation}
\begin{equation}
\Delta \overline{\Omega}_{\mathfrak{s}=\mathfrak{s}_{\mathrm{i}}} = 0 \mathrlap{\;,}
\label{eq:2PIfrgmixedCflowICDeltaOmega0DON}
\end{equation}
\begin{equation}
\overline{\Sigma}^{(\varphi)}_{\mathfrak{s}=\mathfrak{s}_{\mathrm{i}},a a'} = 0 \mathrlap{\quad \forall a, a' \;,}
\label{eq:2PIfrgmixedCflowICSigmavarphi0DON}
\end{equation}
\begin{equation}
\overline{\Sigma}^{(\sigma)}_{\mathfrak{s}=\mathfrak{s}_{\mathrm{i}}} = 0 \mathrlap{\;,}
\label{eq:2PIfrgmixedCflowICSigmasigma0DON}
\end{equation}
\begin{equation}
\scalebox{0.877}{${\displaystyle\overline{\Phi}_{\mathrm{mix},\mathfrak{s} = \mathfrak{s}_{\mathrm{i}},(a_{1},a'_{1})(a_{2},a'_{2})}^{(2G,1D)} = \left\{
\begin{array}{lll}
        \displaystyle{\frac{2\lambda}{3}\left(\delta_{a_{1}a_{2}}\delta_{a'_{1}a'_{2}}+\delta_{a_{1}a'_{2}}\delta_{a'_{1}a_{2}}\right) \quad \text{from the $\hbar$-expansion (i.e. from~\eqref{eq:Cflow2PIFRGDefPhimixSCPThbarExp0DON})} \;,} \\
        \\
        \displaystyle{\frac{2\lambda}{3}\left(\delta_{a_{1}a'_{1}}\delta_{a_{2}a'_{2}}+\delta_{a_{1}a_{2}}\delta_{a'_{1}a'_{2}}+\delta_{a_{1}a'_{2}}\delta_{a'_{1}a_{2}}\right) \quad \text{from the $\lambda$-expansion (i.e. from~\eqref{eq:Cflow2PIFRGDefPhimixSCPTlambdaExp0DON})} \;,}
    \end{array}
\right.}$}
\label{eq:2PIfrgmixedCflowICPhi2G1D0DON}
\end{equation}
\begin{equation}
\hspace{5.55cm} \overline{\Phi}_{\mathrm{mix},\mathfrak{s} = \mathfrak{s}_{\mathrm{i}},(a_{1},a'_{1})\cdots(a_{n},a'_{n})}^{(nG,mD)} = 0 \quad \forall a_{1}, a'_{1},\cdots, a_{n}, a'_{n}, ~ \forall n \neq 2m \;.
\label{eq:2PIfrgmixedCflowICPhinGmD0DON}
\end{equation}
Finally, we address the truncation of the present C-flow approach. For the mC-flow, we will exploit the three first non-trivial orders of both $\Phi_{\mathrm{mix},\mathrm{SCPT},\hbar\text{-exp}}(G,D)$ and $\Phi_{\mathrm{mix},\mathrm{SCPT},\lambda\text{-exp}}(G,D)$ (deduced respectively from~\eqref{eq:mixed2PIEAzeroVevfinalexpression0DON} and~\eqref{eq:mixed2PIEAlambdafinalexpression0DON}) at $N=1$:
\begin{equation}
\Phi_{\mathrm{mix},\mathrm{SCPT},\hbar\text{-exp}}(G,D) = \frac{1}{12} \hbar^{2} \lambda D G^{2} - \frac{1}{72} \hbar^{3} \lambda^{2} D^{2} G^{4} + \frac{5}{324} \hbar^{4} \lambda^{3} D^{3} G^{6} + \mathcal{O}\big(\hbar^{5}\big) \;,
\label{eq:Cflow2PIFRGDefPhimixSCPThbarExp0DONN1}
\end{equation}
\begin{equation}
\Phi_{\mathrm{mix},\mathrm{SCPT},\lambda\text{-exp}}(G,D) = \frac{1}{8} \lambda D G^{2} - \frac{1}{192} \lambda^{2} D^{2} G^{4} + \frac{1}{64} \lambda^{3} D^{3} G^{6} + \mathcal{O}\big(\lambda^{4}\big) \;.
\label{eq:Cflow2PIFRGDefPhimixSCPTlambdaExp0DONN1}
\end{equation}
Hence, the truncation of the tower of differential equations for the C-flow in the framework of the mixed representation is set by (see appendix~\ref{ann:2PIfrgmCflowTruncationConditions0DON} for the mC-flow's truncation conditions up to $N_{\mathrm{SCPT}}=3$ with $N_{\mathrm{max}}=1$):
\begin{itemize}
\item For the tC-flow:
\begin{equation}
\overline{\Phi}_{\mathfrak{s}}^{(nG,mD)}=\overline{\Phi}_{\mathfrak{s}=\mathfrak{s}_{\mathrm{i}}}^{(nG,mD)} \mathrlap{\quad \forall \mathfrak{s}, ~ \forall \ n+m > N_{\mathrm{max}} \;.}
\label{eq:2PIfrgPhiBarmodifiedtCflow0DON}
\end{equation}

\pagebreak

\item For the mC-flow\footnote{The initial condition~\eqref{eq:2PIfrgmixedCflowICPhi2G1D0DON} induces $\overline{\Phi}_{\mathfrak{s}=\mathfrak{s}_{\mathrm{i}}}^{(2G,1D)}=4\lambda/3$ and $\overline{\Phi}_{\mathfrak{s}=\mathfrak{s}_{\mathrm{i}}}^{(2G,1D)}=2\lambda$ at $N=1$, respectively from the $\hbar$-expansion and from the $\lambda$-expansion, which explains the substitutions $\lambda \rightarrow 3\overline{\Phi}_{\mathfrak{s}}^{(2G,1D)}/4$ and $\lambda \rightarrow \overline{\Phi}_{\mathfrak{s}}^{(2G,1D)}/2$ in the corresponding implementations of the mC-flow.} (at $N=1$ and $N_{\mathrm{max}}=N_{\mathrm{SCPT}}=1$):
\begin{itemize}
\item From the $\hbar$-expansion at $\hbar=1$ (i.e. from~\eqref{eq:Cflow2PIFRGDefPhimixSCPThbarExp0DONN1}):
\begin{equation}
\overline{\Phi}^{(2G)}_{\mathfrak{s}} = \left. \overline{\Phi}^{(2G)}_{\mathrm{mix},\mathrm{SCPT},\hbar\text{-exp},N_{\mathrm{SCPT}}=1,\mathfrak{s}} \right|_{\lambda \rightarrow \frac{3}{4}\overline{\Phi}^{(2G,1D)}_{\mathfrak{s}}} = \frac{1}{2} \overline{\Phi}^{(2G,1D)}_{\mathfrak{s}} \overline{D}_{\mathfrak{s}} \;,
\label{eq:2PIFRGmixedmCflowPhi2GhbarNSCPT20DON}
\end{equation}
\begin{equation}
\overline{\Phi}^{(2D)}_{\mathfrak{s}} = \left. \overline{\Phi}^{(2D)}_{\mathrm{mix},\mathrm{SCPT},\hbar\text{-exp},N_{\mathrm{SCPT}}=1,\mathfrak{s}} \right|_{\lambda \rightarrow \frac{3}{4}\overline{\Phi}^{(2G,1D)}_{\mathfrak{s}}} = 0 \;,
\label{eq:2PIFRGmixedmCflowPhi2DhbarNSCPT20DON}
\end{equation}
\begin{equation}
\overline{\Phi}^{(1G,1D)}_{\mathfrak{s}} = \left. \overline{\Phi}^{(1G,1D)}_{\mathrm{mix},\mathrm{SCPT},\hbar\text{-exp},N_{\mathrm{SCPT}}=1,\mathfrak{s}} \right|_{\lambda \rightarrow \frac{3}{4}\overline{\Phi}^{(2G,1D)}_{\mathfrak{s}}} = \frac{1}{2} \overline{\Phi}^{(2G,1D)}_{\mathfrak{s}} \overline{G}_{\mathfrak{s}} \;.
\label{eq:2PIFRGmixedmCflowPhi1G1DhbarNSCPT20DON}
\end{equation}

\item From the $\lambda$-expansion (i.e. from~\eqref{eq:Cflow2PIFRGDefPhimixSCPTlambdaExp0DONN1}):
\begin{equation}
\overline{\Phi}^{(2G)}_{\mathfrak{s}} = \left. \overline{\Phi}^{(2G)}_{\mathrm{mix},\mathrm{SCPT},\lambda\text{-exp},N_{\mathrm{SCPT}}=1,\mathfrak{s}} \right|_{\lambda \rightarrow \frac{1}{2}\overline{\Phi}^{(2G,1D)}_{\mathfrak{s}}} = \frac{1}{2} \overline{\Phi}^{(2G,1D)}_{\mathfrak{s}} \overline{D}_{\mathfrak{s}} \;,
\label{eq:2PIFRGmixedmCflowPhi2GlambdaNSCPT20DON}
\end{equation}
\begin{equation}
\overline{\Phi}^{(2D)}_{\mathfrak{s}} = \left. \overline{\Phi}^{(2D)}_{\mathrm{mix},\mathrm{SCPT},\lambda\text{-exp},N_{\mathrm{SCPT}}=1,\mathfrak{s}} \right|_{\lambda \rightarrow \frac{1}{2}\overline{\Phi}^{(2G,1D)}_{\mathfrak{s}}} = 0 \;,
\label{eq:2PIFRGmixedmCflowPhi2DlambdaNSCPT20DON}
\end{equation}
\begin{equation}
\overline{\Phi}^{(1G,1D)}_{\mathfrak{s}} = \left. \overline{\Phi}^{(1G,1D)}_{\mathrm{mix},\mathrm{SCPT},\lambda\text{-exp},N_{\mathrm{SCPT}}=1,\mathfrak{s}} \right|_{\lambda \rightarrow \frac{1}{2}\overline{\Phi}^{(2G,1D)}_{\mathfrak{s}}} = \frac{1}{2} \overline{\Phi}^{(2G,1D)}_{\mathfrak{s}} \overline{G}_{\mathfrak{s}} \;.
\label{eq:2PIFRGmixedmCflowPhi1G1DlambdaNSCPT20DON}
\end{equation}

\end{itemize}

\end{itemize}

\vspace{0.3cm}

Note that the integer $N_{\mathrm{SCPT}}$ still indicates that the $N_{\mathrm{SCPT}}$ first terms of the chosen perturbative expression of the Luttinger-Ward functional (i.e. either $\Phi_{\mathrm{mix},\mathrm{SCPT},\hbar\text{-exp}}(G,D)$ in~\eqref{eq:Cflow2PIFRGDefPhimixSCPThbarExp0DONN1} up to order $\mathcal{O}(\hbar^{N_{\mathrm{SCPT}}+1})$ or $\Phi_{\mathrm{mix},\mathrm{SCPT},\lambda\text{-exp}}(G,D)$ in~\eqref{eq:Cflow2PIFRGDefPhimixSCPTlambdaExp0DONN1} up to order $\mathcal{O}(\lambda^{N_{\mathrm{SCPT}}})$) have been considered to establish the truncation conditions of the mC-flow. The last ingredient required to perform our calculations are the analytic forms of the cutoff functions $R^{(\varphi)}_{\mathfrak{s}}$ and $R^{(\sigma)}_{\mathfrak{s}}$. We choose:
\begin{equation}
\left(C^{(\varphi)}_{\mathfrak{s}}\right)_{a_{1}a_{2}}^{-1} = \left(C^{(\varphi)}\right)_{a_{1}a_{2}}^{-1} + R^{(\varphi)}_{\mathfrak{s},a_{1}a_{2}} = \left(m^{2} + R_{\mathfrak{s}}\right) \delta_{a_{1}a_{2}} \mathrlap{\quad \forall a_{1}, a_{2} \;,}
\label{eq:2PIFRGCflowCutoffRsvarphi0DON}
\end{equation}
\begin{equation}
\left(C^{(\sigma)}_{\mathfrak{s}}\right)^{-1} = \left(C^{(\sigma)}\right)^{-1} + R^{(\sigma)}_{\mathfrak{s}} = 1 + R_{\mathfrak{s}} \mathrlap{\;,}
\label{eq:2PIFRGCflowCutoffRssigma0DON}
\end{equation}
and
\begin{equation}
R_{\mathfrak{s}} = \mathfrak{s}^{-1} - 1 \;,
\label{eq:2PIFRGmixedCflowCutoffRk0DON}
\end{equation}
with $\mathfrak{s}_{\mathrm{i}}=0$ and $\mathfrak{s}_{\mathrm{f}}=1$, similarly to~\eqref{eq:2PIFRGCflowCutoff20DON} for the original representation.

\vspace{0.5cm}

\begin{figure}[!t]
\captionsetup[subfigure]{labelformat=empty}
  \begin{center}
    \subfloat[]{
      \includegraphics[width=0.9\linewidth]{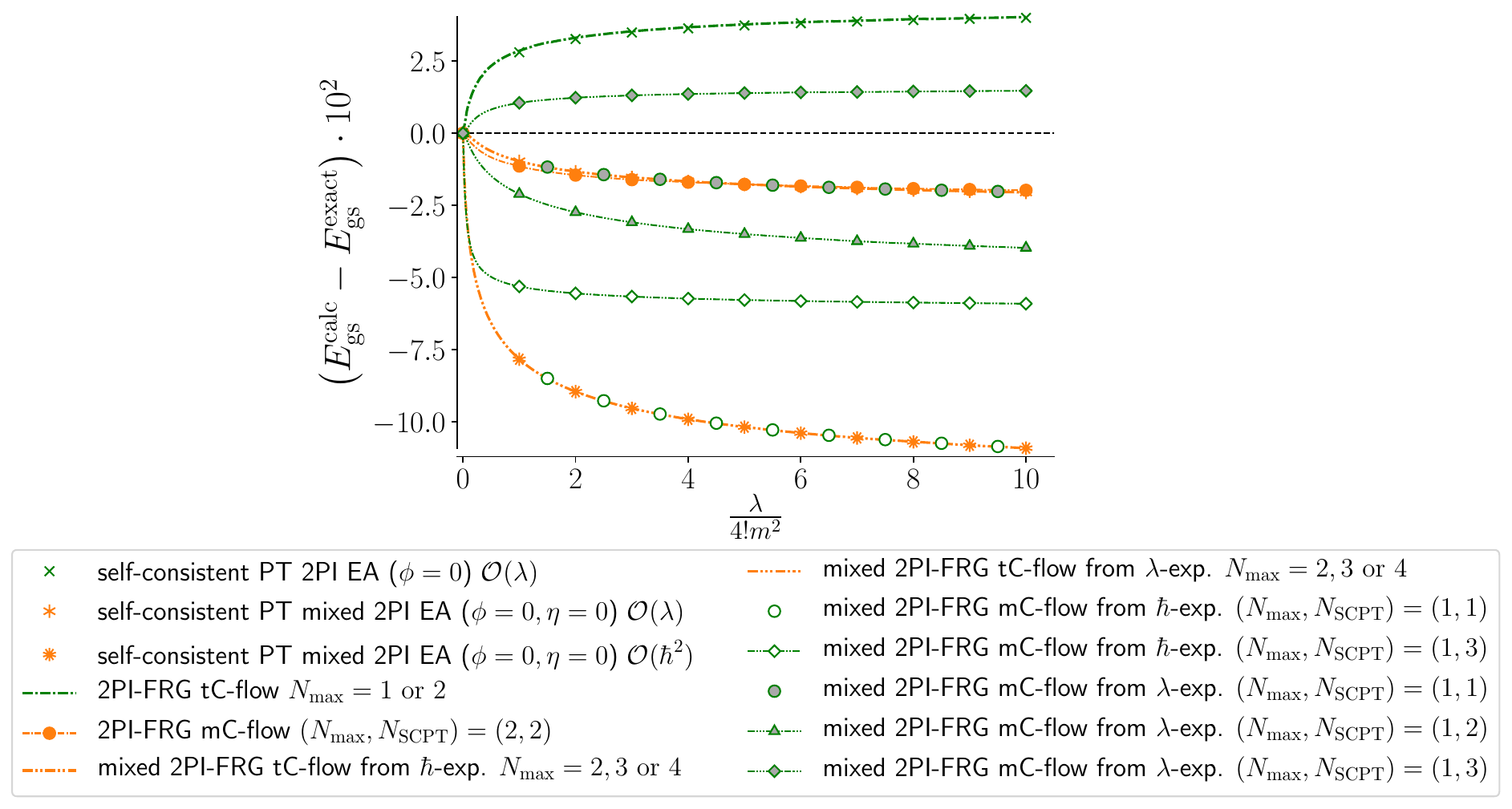}
                         }
   \\                     
    \subfloat[]{
      \includegraphics[width=0.9\linewidth]{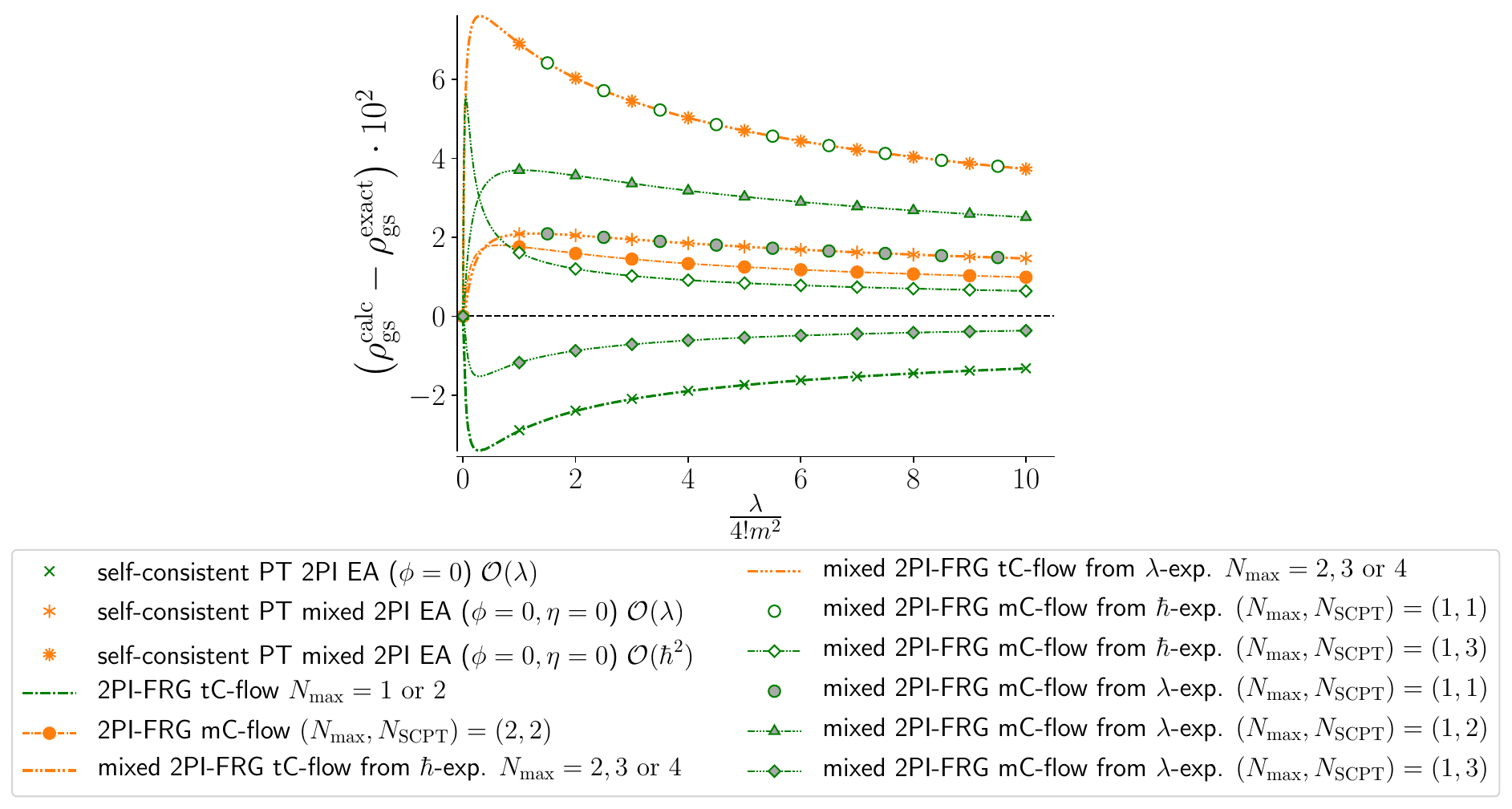}
                         }
\caption{Same as fig.~\ref{fig:mixed2PIFRGCflowlambdaN2} with $N=1$ instead.}
\label{fig:mixed2PIFRGCflowlambdaN1}
  \end{center}
\end{figure}

To summarize, our numerical results for the C-flow in the framework of the mixed theory are determined up to $N_{\mathrm{max}}=4$ by solving the differential equation system made of~\eqref{eq:2PIfrgFlowEquationsmixedCflowG0DON1} to~\eqref{eq:2PIfrgFlowEquationsmixedCflowPhin0DON1} for $N=1$ and of~\eqref{eq:2PIfrgFlowEquationsmixedCflowG0DONN2} to~\eqref{eq:2PIfrgFlowEquationsmixedCflowPhi1G1Ds110DONN2} for $N=2$, with initial conditions set by~\eqref{eq:2PIfrgmixedCflowICGki0DON} to~\eqref{eq:2PIfrgmixedCflowICPhinGmD0DON}, truncation conditions given by~\eqref{eq:2PIfrgPhiBarmodifiedtCflow0DON} for the tC-flow and by~\eqref{eq:2PIFRGmixedmCflowPhi2GhbarNSCPT20DON} to~\eqref{eq:2PIFRGmixedmCflowPhi1G1DlambdaNSCPT20DON} for the mC-flow at $N=1$ and $N_{\mathrm{max}}=N_{\mathrm{SCPT}}=1$ (see appendix~\ref{ann:2PIfrgmCflowTruncationConditions0DON} for $N_{\mathrm{SCPT}}=1,2~\mathrm{or}~3$ with $N_{\mathrm{max}}=1$). Finally,~\eqref{eq:2PIFRGCflowCutoffRsvarphi0DON} to~\eqref{eq:2PIFRGmixedCflowCutoffRk0DON} set the chosen cutoff functions. The gs energy and density thus calculated are presented by figs.~\ref{fig:mixed2PIFRGCflowlambdaN1} for $N=1$ and~\ref{fig:mixed2PIFRGCflowlambdaN2} for $N=2$. First of all, we can see in both of these figures, and thus for $N=1$ and $N=2$, that the first non-trivial order of self-consistent PT is well reproduced by the tC-flow, in the framework of the $\hbar$-expansion as well as the $\lambda$-expansion. The formal proof of the equivalence between the tC-flow and self-consistent PT, given in the discussion of section~\ref{sec:2PIFRGstateofplay} on the C-flow, is indeed straightforwardly generalizable to the mixed situation. We thus conclude that, in the framework of the original and of the mixed representations, the tC-flow approach is not systematically improvable in itself. One might further investigate the properties of the underlying series~\eqref{eq:VertexExpansion2PIFRG} representing the 2PI EA to see whether a combination of the tC-flow with resummation theory might be relevant to overcome this limitation.

\vspace{0.5cm}

Regarding the mC-flow results of fig.~\ref{fig:mixed2PIFRGCflowlambdaN1}, our qualitative conclusions do not change either as compared to the original theory. For $E_{\mathrm{gs}}$ and $\rho_{\mathrm{gs}}$, the tC-flow at $N_{\mathrm{max}}=2$ and mC-flow at $N_{\mathrm{max}}=N_{\mathrm{SCPT}}=1$ coincide for both the $\hbar$-expansion and the $\lambda$-expansion. This could have been directly deduced from the truncation conditions~\eqref{eq:2PIFRGmixedmCflowPhi2GhbarNSCPT20DON} to~\eqref{eq:2PIFRGmixedmCflowPhi1G1DlambdaNSCPT20DON} which are equivalent to the differential equations expressing $\dot{\overline{\Phi}}^{(2G)}_{\mathfrak{s}}$, $\dot{\overline{\Phi}}^{(2D)}_{\mathfrak{s}}$ and $\dot{\overline{\Phi}}^{(1G,1D)}_{\mathfrak{s}}$ in the framework of the tC-flow at $N_{\mathrm{max}}=2$. Furthermore, as for the tC-flow and self-consistent PT, the mC-flow results determined from the $\lambda$-expansion always outperform in fig.~\ref{fig:mixed2PIFRGCflowlambdaN1} those obtained from the $\hbar$-expansion for a given choice of $N_{\mathrm{max}}$ and $N_{\mathrm{SCPT}}$ (note that the mC-flow with $(N_{\mathrm{max}},N_{\mathrm{SCPT}})=(1,2)$ and truncation deduced from the $\hbar$-expansion of the Luttinger-Ward functional is absent from fig.~\ref{fig:mixed2PIFRGCflowlambdaN1} due to the stiffness problem discussed earlier, slightly after~\eqref{eq:2PIFRGCflowCutoff20DON}). However, at $N_{\mathrm{max}}=1$, the mC-flow results associated with the $\lambda$-expansion clearly worsen from $N_{\mathrm{SCPT}}=1$ to $N_{\mathrm{SCPT}}=2$ (at least for $\lambda/4! \in [0,10]$) and, in particular for $E_{\mathrm{gs}}$, the corresponding curve at $N_{\mathrm{SCPT}}=3$ is barely closer to the exact solution than that at $N_{\mathrm{SCPT}}=1$. Furthermore, the mixed mC-flow results of fig.~\ref{fig:mixed2PIFRGCflowlambdaN1}, which are all determined for $N_{\mathrm{max}}=1$ (but up to $N_{\mathrm{SCPT}}=3$, which can turn out to be a very demanding truncation to reach for a realistic theory), are either worse or only slightly better than the best mC-flow estimates for $E_{\mathrm{gs}}$ and $\rho_{\mathrm{gs}}$ obtained in the original theory (with $(N_{\mathrm{max}},N_{\mathrm{SCPT}})=(2,2)$) in fig.~\ref{fig:original2PIFRGCflowlambdaN1}. Hence, the present mC-flow approach does not appear to be very efficient at exploiting the Hubbard-Stratonovich field to grasp further correlations but, besides the performance, our main point is that the mC-flow truncation is not more reliable in the mixed representation than it is in the original one.

\vspace{0.5cm}

In conclusion, we have contented ourselves with the original and mixed situations for the C-flow (and only at $N=1$ for the mC-flow) as we believe that such applications are sufficient to make our point: the C-flow is rather disappointing and not reliable to design well-controlled systematically improvable approximation schemes. However, we have illustrated the interesting equivalence between the tC-flow and self-consistent PT (with a restriction to odd truncation orders $N_{\mathrm{SCPT}}$ for self-consistent PT in the framework of the original theory). There is no reason that such observations change for the collective representation (for which the determination of the initial conditions would actually be significantly more cumbersome due to the more involved nature of the diagrammatic representation of the Luttinger-Ward functional in this situation). Unless some extensions of these methods are designed to change these qualitative features, these approaches are therefore of little interest for us in our aim to construct reliable approaches to study strongly-coupled quantum many-body systems. Thus, we now turn to the U-flow in order to see how this can be achieved via the 2PI-FRG.

\subsubsection{2PI functional renormalization group U-flow}
\label{sec:2PIFRGuflow0DON}
\paragraph{Plain U-flow:}

Let us first stress that the symmetry arguments put forward for the C-flow also hold for the U-flow as none of the 2PI-FRG approaches treated in this study are able to exhibit a spontaneous breakdown of the $O(N)$ symmetry since they are all based on the 2PI EA $\Gamma^{(\mathrm{2PI})}[G]\equiv\Gamma^{(\mathrm{2PI})}[\phi=0,G]$ with vanishing 1-point correlation function $\phi$. In particular, the propagator $\overline{G}_{\mathfrak{s}}$ and the self-energy $\overline{\Sigma}_{\mathfrak{s}}$ still satisfy respectively~\eqref{eq:diagonalG2PIFRGCflow0DON} and~\eqref{eq:diagonalSigma2PIFRGCflow0DON} for all 2PI-FRG applications to the $O(N)$ model under consideration. This implies that the corresponding pair propagator, defined previously by~\eqref{eq:2PIfrgExpressionPiandG}, reduces to:
\begin{equation}
\overline{\Pi}_{\mathfrak{s},(a_{1},a'_{1}) (a_{2},a'_{2})} = \overline{G}_{\mathfrak{s},a_{1} a'_{2}} \overline{G}_{\mathfrak{s},a'_{1} a_{2}} + \overline{G}_{\mathfrak{s},a_{1} a_{2}} \overline{G}_{\mathfrak{s},a'_{1} a'_{2}} = \overline{G}_{\mathfrak{s}}^{2} \left(\delta_{a_{1} a'_{2}} \delta_{a'_{1} a_{2}} + \delta_{a_{1} a_{2}} \delta_{a'_{1} a'_{2}}\right) \;,
\label{eq:2PIfrgExpressionPiandG0DON}
\end{equation}
and, according to~\eqref{eq:2PIfrgUflowW2expression} and~\eqref{eq:IdentityBosonicMatrix2PIFRG} expressing respectively the derivative $W^{(2)}$ and the bosonic identity matrix, this also leads to:
\begin{equation}
\begin{split}
\overline{W}_{\mathfrak{s},(a_{1},a'_{1})(a_{2},a'_{2})}^{(2)} = & \left(\overline{\Pi}_{\mathfrak{s}}^{\mathrm{inv}} + \overline{\Phi}_{\mathfrak{s}}^{(2)}\right)_{(a_{1},a'_{1})(a_{2},a'_{2})}^{\mathrm{inv}} \\
= & \ \frac{1}{2} \sum_{a_{3},a'_{3}=1}^{N} \overline{\Pi}_{\mathfrak{s},(a_{1},a'_{1})(a_{3},a'_{3})} \left(\mathcal{I} + \overline{\Pi}_{\mathfrak{s}} \overline{\Phi}_{\mathfrak{s}}^{(2)}\right)_{(a_{3},a'_{3})(a_{2},a'_{2})}^{\mathrm{inv}} \\
= & \ \overline{G}_{\mathfrak{s}}^{2} \left(\mathcal{I} + \overline{\Pi}_{\mathfrak{s}} \overline{\Phi}_{\mathfrak{s}}^{(2)}\right)_{(a_{1},a'_{1})(a_{2},a'_{2})}^{\mathrm{inv}} \;,
\end{split}
\label{eq:2PIFRGUflowExpressionW2forallN0DON}
\end{equation}
with
\begin{equation}
\begin{split}
\left(\mathcal{I}+\overline{\Pi}_{\mathfrak{s}}\overline{\Phi}_{\mathfrak{s}}^{(2)}\right)_{(a_{1},a'_{1})(a_{2},a'_{2})} = & \ \mathcal{I}_{(a_{1},a'_{1})(a_{2},a'_{2})} + \frac{1}{2} \sum_{a_{3},a'_{3}=1}^{N} \overline{\Pi}_{\mathfrak{s},(a_{1},a'_{1})(a_{3},a'_{3})} \overline{\Phi}_{\mathfrak{s},(a_{3},a'_{3})(a_{2},a'_{2})}^{(2)} \\
= & \ \delta_{a_{1}a_{2}} \delta_{a'_{1}a'_{2}} + \delta_{a_{1}a'_{2}} \delta_{a'_{1}a_{2}} + \overline{G}_{\mathfrak{s}}^{2} \overline{\Phi}^{(2)}_{\mathfrak{s},(a_{1},a'_{1})(a_{2},a'_{2})} \;,
\end{split}
\label{eq:2PIFRGUflowMatrixToInverse0DON}
\end{equation}
where~\eqref{eq:DefinitionBosonicIdentityMatrix0DON} was used to replace $\mathcal{I}_{(a_{1},a'_{1})(a_{2},a'_{2})}$. Furthermore, we already specify the chosen cutoff function for $U_{\mathfrak{s}}$:
\begin{equation}
U_{\mathfrak{s},a_{1}a_{2}a_{3}a_{4}} = R_{\mathfrak{s}} U_{a_{1}a_{2}a_{3}a_{4}} = \mathfrak{s} U_{a_{1}a_{2}a_{3}a_{4}} = \frac{\mathfrak{s}\lambda}{3}\left(\delta_{a_{1}a_{2}}\delta_{a_{3}a_{4}}+\delta_{a_{1}a_{3}}\delta_{a_{2}a_{4}}+\delta_{a_{1}a_{4}}\delta_{a_{2}a_{3}}\right) \quad \forall a_{1},a_{2},a_{3},a_{4} \;,
\label{eq:choiceCutoffUs2PIFRGUflow0DON}
\end{equation}
as this will enable us to simplify the final forms of our flow equations thanks to Kronecker deltas brought in by the interaction $U$ (expressed by~\eqref{eq:2bodyinteraction2PIFRG0DON}). Note that the flow parameter $\mathfrak{s}$ still runs from $\mathfrak{s}_{\mathrm{i}}=0$ to $\mathfrak{s}_{\mathrm{f}}=1$ in the present situation, thus implying that~\eqref{eq:choiceCutoffUs2PIFRGUflow0DON} satisfies the required boundary conditions~\eqref{eq:2PIfrgUflowBoundaryCondUpper} and~\eqref{eq:2PIfrgUflowBoundaryCondBottom}. With all of this in mind, we rewrite the general pU-flow equations given by~\eqref{eq:2PIfrgUflowEquationsGeneralFormG},~\eqref{eq:2PIfrgUflowEquationsGeneralFormOmega} and~\eqref{eq:2PIfrgUflowEquationsGeneralFormSigma} in the framework of the studied toy model:
\begin{equation}
\dot{\overline{G}}_{\mathfrak{s}} = \overline{G}^{2}_{\mathfrak{s}} \dot{\overline{\Sigma}}_{\mathfrak{s}} \;,
\label{eq:2PIFRGpUflowEquationG0DON}
\end{equation}
\begin{equation}
\scalebox{0.99}{${\displaystyle\dot{\overline{\Omega}}_{\mathfrak{s}} = \frac{\lambda}{72} \overline{G}_{\mathfrak{s}}^{2} \left(\sum_{a_{1},a_{2}=1}^{N}\left(\mathcal{I}+\overline{\Pi}_{\mathfrak{s}}\overline{\Phi}_{\mathfrak{s}}^{(2)}\right)_{(a_{1},a_{1})(a_{2},a_{2})}^{\mathrm{inv}} + 2\sum_{a_{1},a'_{1}=1}^{N}\left(\mathcal{I}+\overline{\Pi}_{\mathfrak{s}}\overline{\Phi}_{\mathfrak{s}}^{(2)}\right)_{(a_{1},a'_{1})(a_{1},a'_{1})}^{\mathrm{inv}} + N\left(2+N\right) \right) \;,}$}
\label{eq:2PIFRGpUflowEquationOmega0DON}
\end{equation}
\begin{equation}
\begin{split}
\dot{\overline{\Sigma}}_{\mathfrak{s}} = & - \frac{\lambda}{72} \overline{G}_{\mathfrak{s}} \sum_{a_{1},a'_{1},a_{2}=1}^{N} \left(\mathcal{I} + \overline{\Pi}_{\mathfrak{s}}\overline{\Phi}_{\mathfrak{s}}^{(2)}\right)_{(1,1)(a_{1},a'_{1})}^{\mathrm{inv}} \\
& \hspace{2.8cm} \times \Bigg( \sum_{a_{3},a_{4}=1}^{N} \left(\mathcal{I}+\overline{\Pi}_{\mathfrak{s}}\overline{\Phi}_{\mathfrak{s}}^{(2)}\right)_{(a_{1},a_{2})(a_{3},a_{3})}^{\mathrm{inv}} \left(\mathcal{I}+\overline{\Pi}_{\mathfrak{s}}\overline{\Phi}_{\mathfrak{s}}^{(2)}\right)_{(a_{4},a_{4})(a_{2},a'_{1})}^{\mathrm{inv}}  \\
& \hspace{3.4cm} + 2 \sum_{a_{3},a'_{3}=1}^{N} \left(\mathcal{I}+\overline{\Pi}_{\mathfrak{s}}\overline{\Phi}_{\mathfrak{s}}^{(2)}\right)_{(a_{1},a_{2})(a_{3},a'_{3})}^{\mathrm{inv}} \left(\mathcal{I}+\overline{\Pi}_{\mathfrak{s}}\overline{\Phi}_{\mathfrak{s}}^{(2)}\right)_{(a_{3},a'_{3})(a_{2},a'_{1})}^{\mathrm{inv}} \Bigg) \\
& - \frac{\lambda}{36} \overline{G}_{\mathfrak{s}} \left(N+2\right) \sum_{a_{1}=1}^{N} \left(\mathcal{I}+\overline{\Pi}_{\mathfrak{s}}\overline{\Phi}_{\mathfrak{s}}^{(2)}\right)_{(1,1)(a_{1},a_{1})}^{\mathrm{inv}} \\
& + \frac{\lambda}{576} \overline{G}_{\mathfrak{s}}^{4} \sum_{a_{1},a'_{1},a_{2},a_{3},a_{4},a'_{4},a_{5},a'_{5}=1}^{N} \left(\mathcal{I}+\overline{\Pi}_{\mathfrak{s}}\overline{\Phi}_{\mathfrak{s}}^{(2)}\right)_{(1,1)(a_{1},a'_{1})}^{\mathrm{inv}} \left(\mathcal{I}+\overline{\Pi}_{\mathfrak{s}}\overline{\Phi}_{\mathfrak{s}}^{(2)}\right)_{(a_{2},a_{2})(a_{4},a'_{4})}^{\mathrm{inv}} \\
& \hspace{5.0cm} \times \overline{\Phi}_{\mathfrak{s},(a_{1},a'_{1})(a_{4},a'_{4})(a_{5},a'_{5})}^{(3)} \left(\mathcal{I}+\overline{\Pi}_{\mathfrak{s}}\overline{\Phi}_{\mathfrak{s}}^{(2)}\right)_{(a_{5},a'_{5})(a_{3},a_{3})}^{\mathrm{inv}} \\
& + \frac{\lambda}{288} \overline{G}_{\mathfrak{s}}^{4} \sum_{a_{1},a'_{1},a_{2},a'_{2},a_{3},a'_{3},a_{4},a'_{4}=1}^{N} \left(\mathcal{I}+\overline{\Pi}_{\mathfrak{s}}\overline{\Phi}_{\mathfrak{s}}^{(2)}\right)_{(1,1)(a_{1},a'_{1})}^{\mathrm{inv}} \left(\mathcal{I}+\overline{\Pi}_{\mathfrak{s}}\overline{\Phi}_{\mathfrak{s}}^{(2)}\right)_{(a_{2},a'_{2})(a_{3},a'_{3})}^{\mathrm{inv}} \\
& \hspace{5.0cm} \times \overline{\Phi}_{\mathfrak{s},(a_{1},a'_{1})(a_{3},a'_{3})(a_{4},a'_{4})}^{(3)} \left(\mathcal{I}+\overline{\Pi}_{\mathfrak{s}}\overline{\Phi}_{\mathfrak{s}}^{(2)}\right)_{(a_{4},a'_{4})(a_{2},a'_{2})}^{\mathrm{inv}} \;,
\end{split}
\label{eq:2PIFRGpUflowEquationSigma0DON}
\end{equation}
where the color indices set equal to 1 in the RHS of~\eqref{eq:2PIFRGpUflowEquationSigma0DON} just result from our convention $\overline{\Sigma}_{\mathfrak{s}} \equiv \overline{\Sigma}_{\mathfrak{s},11}$ (which is arbitrary in the sense that $\overline{\Sigma}_{\mathfrak{s},11} = \overline{\Sigma}_{\mathfrak{s},a a}$ $\forall a$ owing to the conservation of the $O(N)$ symmetry). Hence, the resolution of the Bethe-Salpeter equation during the flow now simply amounts to inverting $\left(\mathcal{I}+\overline{\Pi}_{\mathfrak{s}}\overline{\Phi}_{\mathfrak{s}}^{(2)}\right)$, i.e. to solve:
\begin{equation}
\mathcal{I}_{(a_{1},a'_{1})(a_{2},a'_{2})} = \frac{1}{2} \sum_{a_{3},a'_{3}=1}^{N} \left(\mathcal{I}+\overline{\Pi}_{\mathfrak{s}}\overline{\Phi}_{\mathfrak{s}}^{(2)}\right)_{(a_{1},a'_{1})(a_{3},a'_{3})} \left(\mathcal{I}+\overline{\Pi}_{\mathfrak{s}}\overline{\Phi}_{\mathfrak{s}}^{(2)}\right)_{(a_{3},a'_{3})(a_{2},a'_{2})}^{\mathrm{inv}} \;,
\label{eq:MatrixToInvertpUflow0DON}
\end{equation}
which is nothing other than a set of $N^4$ coupled algebraic equations.

\vspace{0.5cm}

Furthermore, as opposed to our presentation of the C-flow equations, we will not evaluate explicitly the sums over color indices for $N=2$ in the U-flow equations as the relations thus obtained (in particular from~\eqref{eq:2PIFRGpUflowEquationSigma0DON}) would be extremely cumbersome and certainly pointless for our discussion. We simply carry out these summations numerically instead. However, we will still pay particular attention to the case with $N=1$ on which we now focus. In this situation, the definitions given in our general presentation of the U-flow in section~\ref{sec:2PIFRGstateofplay} take a very simple form. For example, the definition~\eqref{eq:2PIfrgMinvBosonicIndices} of the inverse of a given bosonic matrix $M$ becomes:
\begin{equation}
M^{\mathrm{inv}} = \frac{4}{M} \;,
\label{eq:2PIFRGuflowInvBosonicMatrixODON1}
\end{equation}
as a result of~\eqref{eq:OneEqualTwoN12PIFRG0DON}, whereas the pair propagator and its inverse read:
\begin{equation}
\Pi(G) = 2 G^{2} \;,
\end{equation}
\begin{equation}
\Pi^{\mathrm{inv}}(G) = \frac{2}{G^{2}} \;,
\end{equation}
from which we can deduce the following expression of $\overline{W}_{\mathfrak{s}}^{(2)}$:
\begin{equation}
\overline{W}_{\mathfrak{s}}^{(2)} = \frac{4}{2 \overline{G}_{\mathfrak{s}}^{-2} +\overline{\Phi}_{\mathfrak{s}}^{(2)}} \;,
\end{equation}
according to~\eqref{eq:2PIfrgUflowW2expression} (or~\eqref{eq:2PIFRGUflowExpressionW2forallN0DON}) and the derivatives:
\begin{equation}
\frac{\partial \Pi(G)}{\partial G} = 8 G \;,
\end{equation}
\begin{equation}
\frac{\partial^{2} \Pi(G)}{\partial G^{2}} = 16 \;,
\label{eq:2PIFRGuflowD2PiG20DON1}
\end{equation}
still according to~\eqref{eq:OneEqualTwoN12PIFRG0DON}. From~\eqref{eq:2PIfrgUflowEquationsGeneralFormG},~\eqref{eq:2PIfrgUflowEquationsGeneralFormOmega},~\eqref{eq:2PIfrgUflowEquationsGeneralFormSigma} and~\eqref{eq:2PIfrgUflowEquationsGeneralFormPhi2} as well as~\eqref{eq:choiceCutoffUs2PIFRGUflow0DON} specifying the chosen cutoff function for $U_{\mathfrak{s}}$, we infer the following pU-flow equations for the studied $O(N)$ model at $N=1$ with the help of~\eqref{eq:2PIFRGuflowInvBosonicMatrixODON1} to~\eqref{eq:2PIFRGuflowD2PiG20DON1} (see appendix~\ref{ann:2PIfrgpUandmUflow0DON1} for the corresponding flow equation expressing the derivative of the 2PI vertex of order 3 with respect to $\mathfrak{s}$):
\begin{equation}
\dot{\overline{G}}_{\mathfrak{s}} = \overline{G}_{\mathfrak{s}}^{2} \dot{\overline{\Sigma}}_{\mathfrak{s}} \;,
\label{eq:2PIFRGpuflowFlowEqG0DON1}
\end{equation}
\begin{equation}
\dot{\overline{\Omega}}_{\mathfrak{s}} = \frac{\lambda}{24} \left( 4\left( 2\overline{G}_{\mathfrak{s}}^{-2} + \overline{\Phi}_{\mathfrak{s}}^{(2)} \right)^{-1} + \overline{G}_{\mathfrak{s}}^{2} \right) \;,
\label{eq:2PIFRGpuflowFlowEqOmega0DON1}
\end{equation}
\begin{equation}
\dot{\overline{\Sigma}}_{\mathfrak{s}} = -\frac{\lambda}{3} \left( 2 + \overline{G}_{\mathfrak{s}}^{2} \overline{\Phi}_{\mathfrak{s}}^{(2)} \right)^{-1} \left( \left(2\overline{G}_{\mathfrak{s}}^{-2} + \overline{\Phi}_{\mathfrak{s}}^{(2)}\right)^{-2} \left(8\overline{G}_{\mathfrak{s}}^{-3} - \overline{\Phi}_{\mathfrak{s}}^{(3)}\right) + \overline{G}_{\mathfrak{s}} \right) \;,
\label{eq:2PIFRGpuflowFlowEqSigma0DON1}
\end{equation}
\begin{equation}
\begin{split}
\dot{\overline{\Phi}}_{\mathfrak{s}}^{(2)} = & \ \frac{\lambda}{6} \bigg( 2\left( 2\overline{G}_{\mathfrak{s}}^{-2} + \overline{\Phi}_{\mathfrak{s}}^{(2)} \right)^{-3} \left( 8 \overline{G}_{\mathfrak{s}}^{-3} - \overline{\Phi}_{\mathfrak{s}}^{(3)} \right)^{2} - 64 \left( 2\overline{G}_{\mathfrak{s}}^{-2} + \overline{\Phi}_{\mathfrak{s}}^{(2)} \right)^{-2} \overline{G}_{\mathfrak{s}}^{-4} \\
& \hspace{0.5cm} + \left( 2\overline{G}_{\mathfrak{s}}^{-2} + \overline{\Phi}_{\mathfrak{s}}^{(2)} \right)^{-2} \left(16\overline{G}_{\mathfrak{s}}^{-4} - \overline{\Phi}_{\mathfrak{s}}^{(4)} \right) + 2 \bigg) + \frac{1}{2} \dot{\overline{G}}_{\mathfrak{s}} \overline{\Phi}_{\mathfrak{s}}^{(3)} \;,
\end{split}
\label{eq:2PIFRGpuflowFlowEqPhi20DON1}
\end{equation}
where~\eqref{eq:2PIFRGpuflowFlowEqG0DON1},~\eqref{eq:2PIFRGpuflowFlowEqOmega0DON1} and~\eqref{eq:2PIFRGpuflowFlowEqSigma0DON1} are respectively equivalent to~\eqref{eq:2PIFRGpUflowEquationG0DON},~\eqref{eq:2PIFRGpUflowEquationOmega0DON} and~\eqref{eq:2PIFRGpUflowEquationSigma0DON} at $N=1$. We deduce in the same way the tU-flow equations at $N_{\mathrm{max}}=2$ and $N=1$ from~\eqref{eq:2PIfrgtruncatedUflowExpressionGdot},~\eqref{eq:2PIfrgtruncatedUflowExpressionOmegadot},~\eqref{eq:2PIfrgtruncatedUflowExpressionSigmadot} and~\eqref{eq:2PIfrgtruncatedUflowExpressionPhi2dot}:
\begin{equation}
\dot{\overline{G}}_{\mathfrak{s}} = \overline{G}_{\mathfrak{s}}^{2} \dot{\overline{\Sigma}}_{\mathfrak{s}} \;,
\label{eq:2PIFRGtuflowFlowEqG0DON1}
\end{equation}
\begin{equation}
\dot{\overline{\Omega}}_{\mathfrak{s}} = \frac{\lambda}{24} \left( 4\left( 2\overline{G}_{\mathfrak{s}}^{-2} + \mathfrak{s} \lambda \right)^{-1} + \overline{G}_{\mathfrak{s}}^{2} \right) \;,
\label{eq:2PIFRGtuflowFlowEqOmega0DON1}
\end{equation}
\begin{equation}
\dot{\overline{\Sigma}}_{\mathfrak{s}} = -\frac{\lambda}{3} \overline{G}_{\mathfrak{s}} \left( 2 + \mathfrak{s} \lambda \overline{G}_{\mathfrak{s}}^{2} \right)^{-1} \left( 8 \left(2 + \mathfrak{s} \lambda \overline{G}_{\mathfrak{s}}^{2} \right)^{-2} + 1 \right) \;,
\label{eq:2PIFRGtuflowFlowEqSigma0DON1}
\end{equation}
\begin{equation}
\overline{\Phi}_{\mathfrak{s}}^{(2)} = \mathfrak{s} \lambda \;.
\label{eq:2PIFRGtuflowFlowEqPhi20DON1}
\end{equation}
The initial conditions required to solve the pU-flow equations (including the tU-flow ones) in the framework of the toy model under consideration are:
\begin{equation}
\overline{G}_{\mathfrak{s}=\mathfrak{s}_{\mathrm{i}},a a'} = C_{a a'} = \frac{1}{m^{2}} \delta_{a a'} \mathrlap{\;,}
\label{eq:2PIFRGpuflowICG0DON}
\end{equation}
\begin{equation}
\overline{\Omega}_{\mathfrak{s}=\mathfrak{s}_{\mathrm{i}}} = -\frac{N}{2} \ln\bigg(\frac{2\pi}{m^{2}}\bigg) \mathrlap{\;,}
\label{eq:2PIFRGpuflowICOmega0DON}
\end{equation}
\begin{equation}
\overline{\Sigma}_{\mathfrak{s}=\mathfrak{s}_{\mathrm{i}},a a'} = 0 \mathrlap{\quad \forall a, a' \;,}
\label{eq:2PIFRGpuflowICSigma0DON}
\end{equation}
\begin{equation}
\hspace{6.0cm} \overline{\Phi}_{\mathfrak{s} = \mathfrak{s}_{\mathrm{i}},(a_{1},a'_{1})\cdots(a_{n},a'_{n})}^{(n)} = 0 \quad \forall a_{1}, a'_{1},\cdots, a_{n}, a'_{n}, ~ \forall n \geq 2 \;,
\label{eq:2PIFRGpuflowICPhin0DON}
\end{equation}
and the truncation of the corresponding tower of differential equations is set by~\eqref{eq:2PIfrgPhiBartCflow0DON}. We stress that condition~\eqref{eq:2PIfrgPhiBartCflow0DON} is already implemented in the tU-flow equations~\eqref{eq:2PIFRGtuflowFlowEqG0DON1} to~\eqref{eq:2PIFRGtuflowFlowEqPhi20DON1} for the truncation at $N_{\mathrm{max}}=2$.

\vspace{0.5cm}

\begin{figure}[!htb]
\captionsetup[subfigure]{labelformat=empty}
  \begin{center}
    \subfloat[]{
      \includegraphics[width=0.50\linewidth]{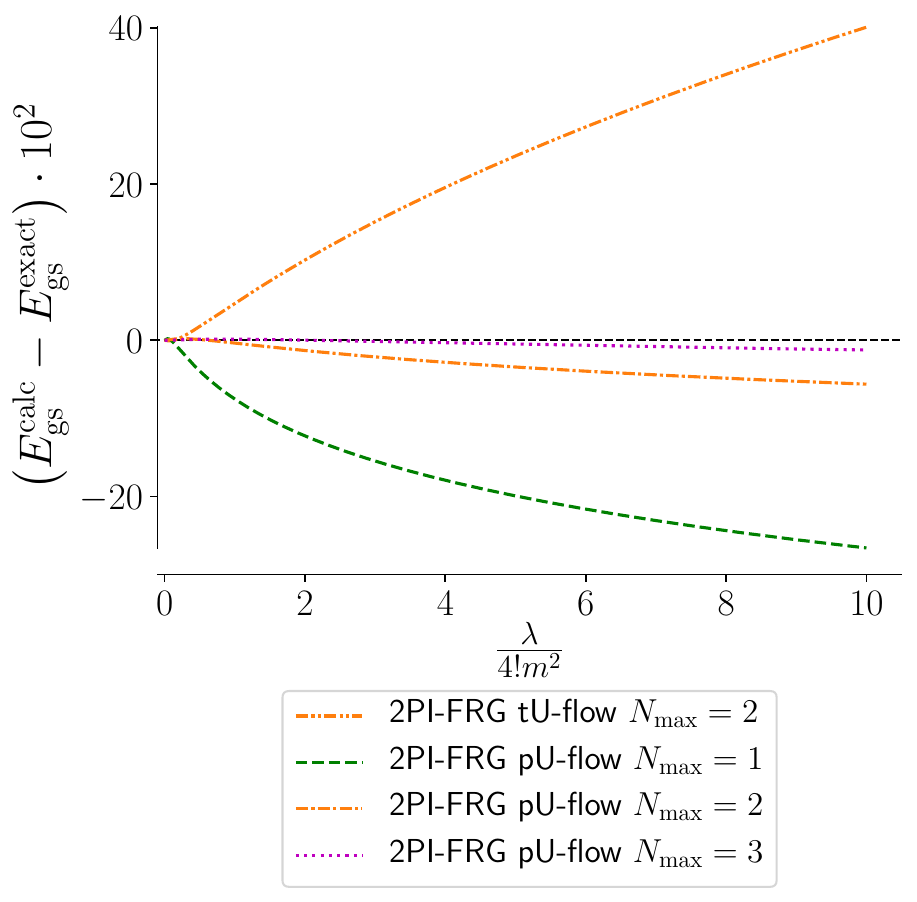}
                         }
    \subfloat[]{
      \includegraphics[width=0.50\linewidth]{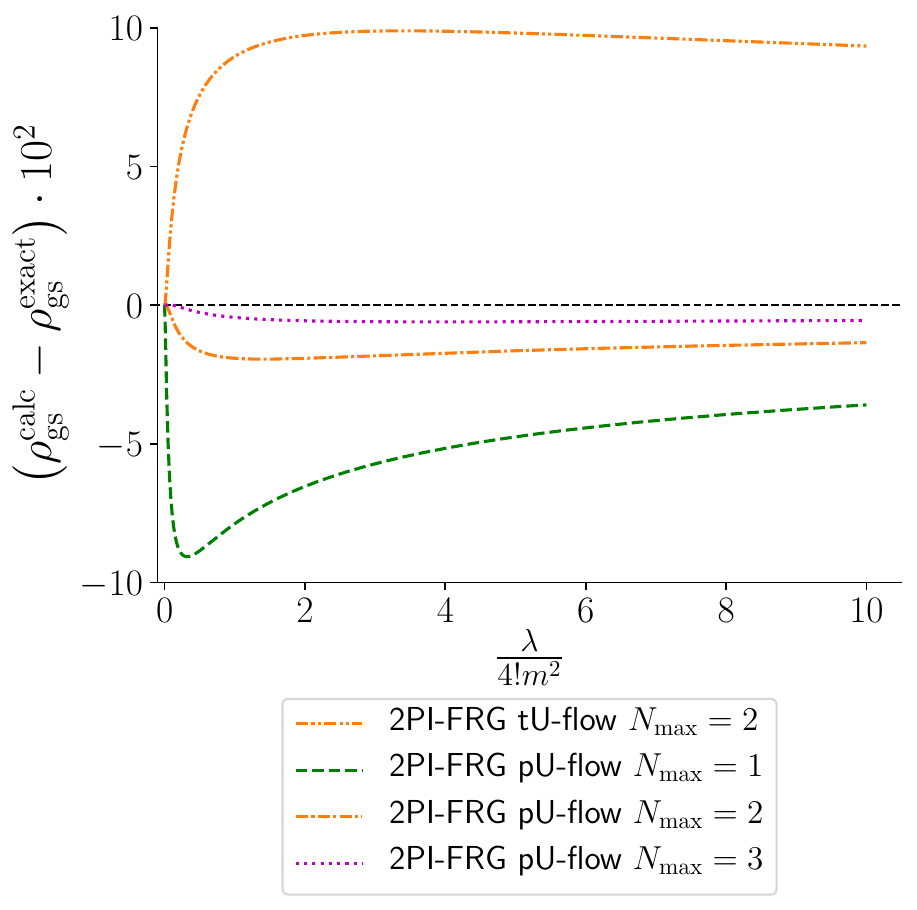}
                         }
\caption{Difference between the calculated gs energy $E_{\mathrm{gs}}^{\mathrm{calc}}$ or density $\rho_{\mathrm{gs}}^{\mathrm{calc}}$ and the corresponding exact solution $E_{\mathrm{gs}}^{\mathrm{exact}}$ or $\rho_{\mathrm{gs}}^{\mathrm{exact}}$ at $m^{2}=+1$ and $N=1$ ($\mathcal{R}e(\lambda)\geq 0$ and $\mathcal{I}m(\lambda)=0$).}
\label{fig:2PIFRGUflowtUflowVspUflowlambdaN1}
  \end{center}
\end{figure}

Thus, our pU-flow calculations are performed by solving the equation system made of~\eqref{eq:2PIFRGpUflowEquationG0DON} to~\eqref{eq:2PIFRGpUflowEquationSigma0DON} at $N_{\mathrm{max}}=1$ for all $N$ (for $N=2$ especially),~\eqref{eq:2PIFRGpuflowFlowEqG0DON1} to~\eqref{eq:2PIFRGpuflowFlowEqPhi20DON1} up to $N_{\mathrm{max}}=2$ at $N=1$ or~\eqref{eq:2PIFRGtuflowFlowEqG0DON1} to~\eqref{eq:2PIFRGtuflowFlowEqPhi20DON1} for the tU-flow at $N_{\mathrm{max}}=2$ and $N=1$. The initial conditions used to solve these differential equations are given by~\eqref{eq:2PIFRGpuflowICG0DON} to~\eqref{eq:2PIFRGpuflowICPhin0DON}, the associated truncation condition is expressed by~\eqref{eq:2PIfrgPhiBartCflow0DON} whereas~\eqref{eq:choiceCutoffUs2PIFRGUflow0DON} sets the chosen cutoff function for $U_{\mathfrak{s}}$. The pU-flow results thus obtained for $N=1$ are notably presented in fig.~\ref{fig:2PIFRGUflowtUflowVspUflowlambdaN1}. The latter shows that the pU-flow exhibits a clear convergence from $N_{\mathrm{max}}=1$ to $N_{\mathrm{max}}=3$ towards the exact solution for the gs energy $E_{\mathrm{gs}}$ and density $\rho_{\mathrm{gs}}$, thus achieving an accuracy of about $1\%$ or less at $\lambda/4!=10$ and $N_{\mathrm{max}}=3$ for both $E_{\mathrm{gs}}$ and $\rho_{\mathrm{gs}}$. As could have been expected from the drastic character of the approximation underpinning the tU-flow, the latter approach is significantly less performing than the standard implementation of the pU-flow, as can be seen in fig.~\ref{fig:2PIFRGUflowtUflowVspUflowlambdaN1} at $N_{\mathrm{max}}=2$. It shows actually that the tU-flow at $N_{\mathrm{max}}=2$ is even less efficient than the standard pU-flow at $N_{\mathrm{max}}=1$ in almost the entire interval $\lambda/4! \in [0,10]$ for both $E_{\mathrm{gs}}$ and $\rho_{\mathrm{gs}}$. We recall here that the idea motivating the introduction of the tU-flow was to avoid solving the Bethe-Salpeter equation repeatedly throughout the flow. However, the standard implementation of the pU-flow only requires to solve this equation at $N_{\mathrm{max}}\geq 2$. At $N_{\mathrm{max}}=1$, we have indeed $\overline{\Phi}_{\mathfrak{s}}^{(2)}=\overline{\Phi}_{\mathfrak{s}=\mathfrak{s}_{\mathrm{i}}}^{(2)}=0$ $\forall \mathfrak{s}$ according to the truncation condition~\eqref{eq:2PIfrgPhiBartCflow0DON}, which implies that $\mathcal{I}+\overline{\Pi}_{\mathfrak{s}}\overline{\Phi}^{(2)}_{\mathfrak{s}}=\mathcal{I}$ $\forall\mathfrak{s}$ and the Bethe-Salpeter equation in the form of~\eqref{eq:MatrixToInvertpUflow0DON} becomes immediately trivial. Therefore, since the pU-flow at $N_{\mathrm{max}}=1$ outperforms in general its tU-flow simplification at $N_{\mathrm{max}}=2$, the latter is of very little relevance, at least at $N_{\mathrm{max}}=2$. As truncation orders $N_{\mathrm{max}}$ larger than 2 are already quite demanding to reach for realistic theories, we thus conclude that we must rely on other approximations to circumvent the numerical weight underlying the implementation of the pU-flow (or of the mU-flow) version of the 2PI-FRG. Note also that our pU-flow results at $N=2$ and $N_{\mathrm{max}}=1$ are shown in fig.~\ref{fig:2PIFRGCflowVsUflowVsCUflowlambdaN2} for $E_{\mathrm{gs}}$ and $\rho_{\mathrm{gs}}$ in comparison with other approaches like the mU-flow on which we then focus.

\paragraph{Modified U-flow:}

We will start by giving the mU-flow equations expressing $\dot{\overline{\boldsymbol{G}}}_{\mathfrak{s}}$, $\dot{\overline{\boldsymbol{\Omega}}}_{\mathfrak{s}}$ and $\dot{\overline{\boldsymbol{\Sigma}}}_{\mathfrak{s}}$ with $N_{\mathrm{SCPT}}=1$ in the framework of our (0+0)-D $O(N)$ model for all $N$. There are essentially two manners to achieve this at this stage: either we deduce such equations from our general results of section~\ref{sec:2PIFRGstateofplay} (i.e. from~\eqref{eq:2PIfrgmUflowExpressionBoldOmegaDotNotApp},~\eqref{eq:2PIfrgmUflowExpressionBoldSigmaDotNotApp} and~\eqref{eq:2PIfrgmUflowExpressionBoldGDotNotApp}) by exploiting the $O(N)$ symmetry in the same way as for the pU-flow or directly by rewriting the pU-flow equations~\eqref{eq:2PIFRGpUflowEquationG0DON} to~\eqref{eq:2PIFRGpUflowEquationSigma0DON} in terms of the bold quantities underlying the mU-flow, which involves notably the modified Luttinger-Ward functional defined as:
\begin{equation}
\begin{split}
\boldsymbol{\Phi}_{\mathfrak{s}}(G) \equiv & \ \Phi_{\mathfrak{s}}(G) + \Phi_{\mathrm{SCPT},N_{\mathrm{SCPT}}=1}(U,G) - \Phi_{\mathrm{SCPT},N_{\mathrm{SCPT}}=1}(U_{\mathfrak{s}},G) \\
= & \ \Phi_{\mathfrak{s}}(G) + \lambda \left(1-\mathfrak{s}\right) \left( \frac{1}{24}\left(\sum_{a_{1}=1}^{N} G_{a_{1}a_{1}} \right)^{2} + \frac{1}{12} \sum_{a_{1},a_{2}=1}^{N} G_{a_{1}a_{2}}^{2} \right) \;,
\end{split}
\label{eq:DefinitionModifiedLWFunctionalNSCPT10DON}
\end{equation}
with
\begin{equation}
\begin{split}
\Phi_{\mathrm{SCPT},N_{\mathrm{SCPT}}=1}(U,G) = & \ \frac{1}{24} \hspace{0.08cm} \begin{gathered}
\begin{fmffile}{DiagramsFRG/2PIEAzerovev_Hartree}
\begin{fmfgraph}(30,20)
\fmfleft{i}
\fmfright{o}
\fmf{phantom,tension=10}{i,i1}
\fmf{phantom,tension=10}{o,o1}
\fmf{plain,left,tension=0.5,foreground=(1,,0,,0)}{i1,v1,i1}
\fmf{plain,right,tension=0.5,foreground=(1,,0,,0)}{o1,v2,o1}
\fmf{zigzag,foreground=(0,,0,,1)}{v1,v2}
\end{fmfgraph}
\end{fmffile}
\end{gathered}
+\frac{1}{12}\begin{gathered}
\begin{fmffile}{DiagramsFRG/2PIEAzerovev_Fock}
\begin{fmfgraph}(15,15)
\fmfleft{i}
\fmfright{o}
\fmf{phantom,tension=11}{i,v1}
\fmf{phantom,tension=11}{v2,o}
\fmf{plain,left,tension=0.4,foreground=(1,,0,,0)}{v1,v2,v1}
\fmf{zigzag,foreground=(0,,0,,1)}{v1,v2}
\end{fmfgraph}
\end{fmffile}
\end{gathered} \\
= & \ \frac{\lambda}{24}\left(\sum_{a_{1}=1}^{N} G_{a_{1}a_{1}} \right)^{2} + \frac{\lambda}{12} \sum_{a_{1},a_{2}=1}^{N} G_{a_{1}a_{2}}^{2} \;.
\end{split}
\label{eq:DefinitionmUflowPhiboldNscpt10DON}
\end{equation}
In any case, the $O(N)$ symmetry implies:
\begin{equation}
\overline{\boldsymbol{G}}_{\mathfrak{s},a a'} = \overline{\boldsymbol{G}}_{\mathfrak{s}} \ \delta_{a a'} \mathrlap{\quad \forall\mathfrak{s} \;,}
\label{eq:diagonalboldG2PIFRGmUflow0DON}
\end{equation}
\begin{equation}
\overline{\boldsymbol{\Sigma}}_{\mathfrak{s},a a'} = \overline{\boldsymbol{\Sigma}}_{\mathfrak{s}} \ \delta_{a a'} \mathrlap{\quad \forall\mathfrak{s} \;,}
\label{eq:diagonalboldSigma2PIFRGmUflow0DON}
\end{equation}
\begin{equation}
\overline{\Pi}_{\mathfrak{s},(a_{1},a'_{1}) (a_{2},a'_{2})} = \overline{\boldsymbol{G}}_{\mathfrak{s}}^{2} \left(\delta_{a_{1} a'_{2}} \delta_{a'_{1} a_{2}} + \delta_{a_{1} a_{2}} \delta_{a'_{1} a'_{2}}\right) \mathrlap{\;,}
\label{eq:2PIfrgmUflowboldPiandG0DON}
\end{equation}
and the expression of $\overline{W}_{\mathfrak{s}}^{(2)}$ to consider now can be directly inferred from~\eqref{eq:2PIFRGUflowExpressionW2forallN0DON} (alongside with~\eqref{eq:DefinitionmUflowPhiboldNscpt10DON} and~\eqref{eq:2PIfrgmUflowboldPiandG0DON}):
\begin{equation}
\begin{split}
\overline{W}_{\mathfrak{s},(a_{1},a'_{1})(a_{2},a'_{2})}^{(2)} = & \left(\overline{\Pi}_{\mathfrak{s}}^{\mathrm{inv}} + \overline{\boldsymbol{\Phi}}_{\mathfrak{s}}^{(2)} + \overline{\Phi}^{(2)}_{\mathrm{SCPT},N_{\mathrm{SCPT}}=1,\mathfrak{s}}(U_{\mathfrak{s}}) - \overline{\Phi}^{(2)}_{\mathrm{SCPT},N_{\mathrm{SCPT}}=1,\mathfrak{s}}(U) \right)_{(a_{1},a'_{1})(a_{2},a'_{2})}^{\mathrm{inv}} \\
= & \ \frac{1}{2} \sum_{a_{3},a'_{3}=1}^{N} \overline{\Pi}_{\mathfrak{s},(a_{1},a'_{1})(a_{3},a'_{3})} \overline{\Upsilon}_{\mathfrak{s},(a_{3},a'_{3})(a_{2},a'_{2})}^{\mathrm{inv}} \\
= & \ \overline{G}_{\mathfrak{s}}^{2} \overline{\Upsilon}_{\mathfrak{s},(a_{1},a'_{1})(a_{2},a'_{2})}^{\mathrm{inv}} \;,
\end{split}
\label{eq:2PIFRGmUflowExpressionW2forallN0DON}
\end{equation}
with
\begin{equation}
\begin{split}
\overline{\Upsilon}_{\mathfrak{s},(a_{1},a'_{1})(a_{2},a'_{2})} \equiv & \ \left(\mathcal{I} + \overline{\Pi}_{\mathfrak{s}} \left( \overline{\boldsymbol{\Phi}}_{\mathfrak{s}}^{(2)} + \overline{\Phi}^{(2)}_{\mathrm{SCPT},N_{\mathrm{SCPT}}=1,\mathfrak{s}}(U_{\mathfrak{s}}) - \overline{\Phi}^{(2)}_{\mathrm{SCPT},N_{\mathrm{SCPT}}=1,\mathfrak{s}}(U) \right)\right)_{(a_{1},a'_{1})(a_{2},a'_{2})} \\
= & \ \delta_{a_{1}a_{2}} \delta_{a'_{1}a'_{2}} + \delta_{a_{1}a'_{2}} \delta_{a'_{1}a_{2}} + \overline{\boldsymbol{G}}_{\mathfrak{s}}^{2} \overline{\boldsymbol{\Phi}}^{(2)}_{\mathfrak{s},(a_{1},a'_{1})(a_{2},a'_{2})} \\
& - \frac{\lambda}{3}(1-\mathfrak{s}) \overline{\boldsymbol{G}}^{2}_{\mathfrak{s}} \left( \delta_{a_{1} a'_{1}} \delta_{a_{2} a'_{2}} + \delta_{a_{1} a_{2}} \delta_{a'_{1} a'_{2}} + \delta_{a_{1} a'_{2}} \delta_{a'_{1} a_{2}} \right) \;,
\end{split}
\label{eq:2PIFRGmUflowDefinitionUpsilon0DON}
\end{equation}
where the last line simply follows from the definition~\eqref{eq:DefinitionBosonicIdentityMatrix0DON} of the bosonic identity matrix and:
\begin{equation}
\scalebox{0.90}{${\displaystyle \left(\overline{\Phi}^{(2)}_{\mathrm{SCPT},N_{\mathrm{SCPT}}=1,\mathfrak{s}}(U_{\mathfrak{s}}) - \overline{\Phi}^{(2)}_{\mathrm{SCPT},N_{\mathrm{SCPT}}=1,\mathfrak{s}}(U)\right)_{(a_{1},a'_{1})(a_{2},a'_{2})} = -\frac{\lambda}{3}(1-\mathfrak{s}) \left( \delta_{a_{1} a'_{1}} \delta_{a_{2} a'_{2}} + \delta_{a_{1} a_{2}} \delta_{a'_{1} a'_{2}} + \delta_{a_{1} a'_{2}} \delta_{a'_{1} a_{2}} \right) \;,}$}
\end{equation}
as a result of~\eqref{eq:DefinitionmUflowPhiboldNscpt10DON}. We also recall once again that the bar labels functions evaluated at $G=\overline{\boldsymbol{G}}_{\mathfrak{s}}$ instead of $G=\overline{G}_{\mathfrak{s}}$ in the framework of the mU-flow. The three mU-flow equations thus obtained are:
\begin{equation}
\dot{\overline{\boldsymbol{G}}}_{\mathfrak{s}} = \overline{\boldsymbol{G}}_{\mathfrak{s}}^{2} \dot{\overline{\boldsymbol{\Sigma}}}_{\mathfrak{s}} \;,
\label{eq:2PIFRGmUflowNSCPT1EquationG0DON}
\end{equation}
\begin{equation}
\dot{\overline{\boldsymbol{\Omega}}}_{\mathfrak{s}} = \frac{\lambda}{72} \overline{\boldsymbol{G}}_{\mathfrak{s}}^{2} \left( \sum_{a_{1},a_{2}=1}^{N} \overline{\Upsilon}^{\mathrm{inv}}_{\mathfrak{s},(a_{1},a_{1})(a_{2},a_{2})} + 2 \sum_{a_{1},a'_{1}=1}^{N} \overline{\Upsilon}^{\mathrm{inv}}_{\mathfrak{s},(a_{1},a'_{1})(a_{1},a'_{1})} - 2 N \left( N + 2 \right) \right) \;,
\label{eq:2PIFRGmUflowNSCPT1EquationOmega0DON}
\end{equation}
\begin{equation}
\begin{split}
\scalebox{0.91}{${\displaystyle \dot{\overline{\boldsymbol{\Sigma}}}_{\mathfrak{s}} =}$} & \scalebox{0.91}{${\displaystyle -\frac{\lambda}{72} \overline{\boldsymbol{G}}_{\mathfrak{s}} \sum_{a_{1},a'_{1},a_{2}=1}^{N} \left(\mathcal{I} + \overline{\Pi}_{\mathfrak{s}}\overline{\boldsymbol{\Phi}}_{\mathfrak{s}}^{(2)}\right)_{(1,1)(a_{1},a'_{1})}^{\mathrm{inv}} }$} \\
& \hspace{2.55cm} \scalebox{0.91}{${\displaystyle \times \Bigg( \sum_{a_{3},a_{4}=1}^{N} \overline{\Upsilon}_{\mathfrak{s},(a_{1},a_{2})(a_{3},a_{3})}^{\mathrm{inv}} \overline{\Upsilon}_{\mathfrak{s},(a_{4},a_{4})(a_{2},a'_{1})}^{\mathrm{inv}} + 2 \sum_{a_{3},a'_{3}=1}^{N} \overline{\Upsilon}_{\mathfrak{s},(a_{1},a_{2})(a_{3},a'_{3})}^{\mathrm{inv}} \overline{\Upsilon}_{\mathfrak{s},(a_{3},a'_{3})(a_{2},a'_{1})}^{\mathrm{inv}} \Bigg) }$} \\
& \scalebox{0.91}{${\displaystyle + \frac{\lambda}{18} \overline{\boldsymbol{G}}_{\mathfrak{s}} \left(N+2\right) \sum_{a_{1}=1}^{N} \left(\mathcal{I}+\overline{\Pi}_{\mathfrak{s}}\overline{\boldsymbol{\Phi}}_{\mathfrak{s}}^{(2)}\right)_{(1,1)(a_{1},a_{1})}^{\mathrm{inv}} }$} \\
& \scalebox{0.91}{${\displaystyle + \frac{\lambda}{576} \overline{\boldsymbol{G}}_{\mathfrak{s}}^{4} \sum_{a_{1},a'_{1},a_{2},a_{3},a_{4},a'_{4},a_{5},a'_{5}=1}^{N} \left(\mathcal{I}+\overline{\Pi}_{\mathfrak{s}}\overline{\boldsymbol{\Phi}}_{\mathfrak{s}}^{(2)}\right)_{(1,1)(a_{1},a'_{1})}^{\mathrm{inv}} \overline{\Upsilon}_{\mathfrak{s},(a_{2},a_{2})(a_{4},a'_{4})}^{\mathrm{inv}} \overline{\boldsymbol{\Phi}}_{\mathfrak{s},(a_{1},a'_{1})(a_{4},a'_{4})(a_{5},a'_{5})}^{(3)} \overline{\Upsilon}_{\mathfrak{s},(a_{5},a'_{5})(a_{3},a_{3})}^{\mathrm{inv}} }$} \\
& \scalebox{0.91}{${\displaystyle + \frac{\lambda}{288} \overline{\boldsymbol{G}}_{\mathfrak{s}}^{4} \sum_{a_{1},a'_{1},a_{2},a'_{2},a_{3},a'_{3},a_{4},a'_{4}=1}^{N} \left(\mathcal{I}+\overline{\Pi}_{\mathfrak{s}}\overline{\boldsymbol{\Phi}}_{\mathfrak{s}}^{(2)}\right)_{(1,1)(a_{1},a'_{1})}^{\mathrm{inv}} \overline{\Upsilon}_{\mathfrak{s},(a_{2},a'_{2})(a_{3},a'_{3})}^{\mathrm{inv}} \overline{\boldsymbol{\Phi}}_{\mathfrak{s},(a_{1},a'_{1})(a_{3},a'_{3})(a_{4},a'_{4})}^{(3)} \overline{\Upsilon}_{\mathfrak{s},(a_{4},a'_{4})(a_{2},a'_{2})}^{\mathrm{inv}} \;,}$}
\end{split}
\label{eq:2PIFRGmUflowNSCPT1EquationSigma0DON}
\end{equation}
where the color indices equal to 1 in the RHS of~\eqref{eq:2PIFRGmUflowNSCPT1EquationSigma0DON} follow from the convention $\overline{\boldsymbol{\Sigma}}_{\mathfrak{s}} \equiv \overline{\boldsymbol{\Sigma}}_{\mathfrak{s},11} = \overline{\boldsymbol{\Sigma}}_{\mathfrak{s},a a}$ $\forall a$, similarly to~\eqref{eq:2PIFRGpUflowEquationSigma0DON}. We have pushed our mU-flow derivations so as to be able to perform calculations up to $N_{\mathrm{max}}=3$ for all $N$. However, instead of calculating the differential equations involving $\dot{\overline{\boldsymbol{\Phi}}}_{\mathfrak{s}}^{(2)}$ and $\dot{\overline{\boldsymbol{\Phi}}}_{\mathfrak{s}}^{(3)}$ to achieve this, we have exploited the fact that the 2PI and 2PPI EAs of the studied (0+0)-D model coincide in the absence of SSB to develop a mU-flow formulation of the 2PPI-FRG treated in section~\ref{sec:2PPIFRG} (see appendix~\ref{ann:2PIfrgmUflow0DON}). The derivation of the underpinning differential equations is less demanding than for the present 2PI-FRG approach. In particular, it does not require to evaluate derivatives such as $\frac{\delta\overline{\Pi}_{\mathfrak{s}}}{\delta\overline{G}_{\mathfrak{s},\gamma}}$ and $\frac{\delta^{2}\overline{\Pi}_{\mathfrak{s}}}{\delta\overline{G}_{\mathfrak{s},\gamma_{1}}\delta\overline{G}_{\mathfrak{s},\gamma_{2}}}$. For instance, the evaluation of $\frac{\delta\overline{\Pi}_{\mathfrak{s}}}{\delta\overline{G}_{\mathfrak{s},\gamma}}$ led to a consequent step in the derivation of~\eqref{eq:2PIFRGmUflowNSCPT1EquationSigma0DON}\footnote{For the same reasons, we could develop a similar 2PPI-FRG approach to push our pU-flow and CU-flow calculations to higher truncation orders for $N \geq 2$ but we rather focus on the mU-flow which will turn out to be the most performing 2PI-FRG method tested in this study.} (see notably~\eqref{eq:DetailedCalculation1Sigma2PIFRGUflowAppendix} and~\eqref{eq:DetailedCalculation2Sigma2PIFRGUflowAppendix} in appendix~\ref{ann:2PIfrgFlowEquationUflow}). This 2PPI-FRG approach is not suited to treat finite-dimensional systems (as we explain technically in appendix~\ref{ann:2PIfrgmUflow0DON}) but the existence of such a shortcut illustrates an important advantage of the zero-dimensional toy model chosen for this comparative study.

\vspace{0.5cm}

Let us nevertheless focus on the quantities involved in~\eqref{eq:2PIFRGmUflowNSCPT1EquationG0DON} to~\eqref{eq:2PIFRGmUflowNSCPT1EquationSigma0DON}, which will enable us to further clarify the general features of the implementation of the mU-flow. The bosonic matrix $\mathcal{I}+\overline{\Pi}_{\mathfrak{s}}\overline{\boldsymbol{\Phi}}_{\mathfrak{s}}^{(2)}$ satisfies the equality:
\begin{equation}
\left(\mathcal{I}+\overline{\Pi}_{\mathfrak{s}}\overline{\boldsymbol{\Phi}}_{\mathfrak{s}}^{(2)}\right)_{(a_{1},a'_{1})(a_{2},a'_{2})} = \delta_{a_{1}a_{2}} \delta_{a'_{1}a'_{2}} + \delta_{a_{1}a'_{2}} \delta_{a'_{1}a_{2}} + \overline{\boldsymbol{G}}_{\mathfrak{s}}^{2} \overline{\boldsymbol{\Phi}}^{(2)}_{\mathfrak{s},(a_{1},a'_{1})(a_{2},a'_{2})} \;,
\label{eq:2PIFRGIplusPiPhi2mUflow0DON}
\end{equation}
which can be derived in the same way as in~\eqref{eq:2PIFRGUflowMatrixToInverse0DON} with $\overline{G}_{\mathfrak{s}}$ and $\overline{\Phi}_{\mathfrak{s}}^{(2)}$ respectively replaced by $\overline{\boldsymbol{G}}_{\mathfrak{s}}$ and $\overline{\boldsymbol{\Phi}}_{\mathfrak{s}}^{(2)}$. Hence, as can be seen from~\eqref{eq:2PIFRGmUflowNSCPT1EquationOmega0DON} and~\eqref{eq:2PIFRGmUflowNSCPT1EquationSigma0DON}, solving the Bethe-Salpeter equation now translates into inverting the bosonic matrices $\mathcal{I}+\overline{\Pi}_{\mathfrak{s}}\overline{\boldsymbol{\Phi}}_{\mathfrak{s}}^{(2)}$ and $\overline{\Upsilon}_{\mathfrak{s}}$. Actually, by comparing~\eqref{eq:2PIFRGmUflowDefinitionUpsilon0DON} with~\eqref{eq:2PIFRGIplusPiPhi2mUflow0DON}, we can notice that $\overline{\Upsilon}_{\mathfrak{s}}$ reduces to $\mathcal{I}+\overline{\Pi}_{\mathfrak{s}}\overline{\boldsymbol{\Phi}}_{\mathfrak{s}}^{(2)}$ by setting $\mathfrak{s}=1$ in the very last line of~\eqref{eq:2PIFRGmUflowDefinitionUpsilon0DON}. Hence, the components of $\left(\mathcal{I}+\overline{\Pi}_{\mathfrak{s}}\overline{\boldsymbol{\Phi}}_{\mathfrak{s}}^{(2)}\right)^{\mathrm{inv}}$ can be directly deduced from those of $\overline{\Upsilon}_{\mathfrak{s}}^{\mathrm{inv}}$ as well and we can content ourselves to invert $\overline{\Upsilon}_{\mathfrak{s}}$, i.e. to solve:
\begin{equation}
\mathcal{I}_{(a_{1},a'_{1})(a_{2},a'_{2})} = \frac{1}{2} \sum_{a_{3},a'_{3}=1}^{N} \overline{\Upsilon}_{\mathfrak{s},(a_{1},a'_{1})(a_{3},a'_{3})} \overline{\Upsilon}_{\mathfrak{s},(a_{3},a'_{3})(a_{2},a'_{2})}^{\mathrm{inv}} \;.
\label{eq:MatrixToInvertmUflowNSCPT10DON}
\end{equation}
The inversion of $\overline{\Upsilon}_{\mathfrak{s}}$ is not trivial at $N_{\mathrm{max}}=1$ and $N \geq 2$, as opposed to $\mathcal{I}+\overline{\Pi}_{\mathfrak{s}}\overline{\Phi}_{\mathfrak{s}}^{(2)}$ for the pU-flow\footnote{As we explained when treating the pU-flow, the triviality of the inversion of $\mathcal{I}+\overline{\Pi}_{\mathfrak{s}}\overline{\Phi}_{\mathfrak{s}}^{(2)}$ at $N_{\mathrm{max}}=1$ $\forall N$ follows from the truncation condition $\overline{\Phi}_{\mathfrak{s}}^{(2)} = \overline{\Phi}_{\mathfrak{s}=\mathfrak{s}_{\mathrm{i}}}^{(2)} = 0$, thus implying that $\mathcal{I}+\overline{\Pi}_{\mathfrak{s}}\overline{\Phi}_{\mathfrak{s}}^{(2)}$ reduces to the identity $\mathcal{I}$. We clearly do not have such a condition at our disposal here to simplify $\overline{\Upsilon}_{\mathfrak{s}}$ that drastically, even at $N_{\mathrm{max}}=1$.}. At $N=1$ however, $\overline{\Upsilon}_{\mathfrak{s}}^{\mathrm{inv}}$ is readily found according to~\eqref{eq:2PIFRGuflowInvBosonicMatrixODON1}:
\begin{equation}
\overline{\Upsilon}_{\mathfrak{s}}^{\mathrm{inv}} = \frac{4}{\overline{\Upsilon}_{\mathfrak{s}}} \;,
\end{equation}
with $\overline{\Upsilon}_{\mathfrak{s}} \equiv \overline{\Upsilon}_{\mathfrak{s},(1,1)(1,1)}$ whereas, at $N=2$, we solve $2^{4}=16$ coupled algebraic equations deduced from~\eqref{eq:MatrixToInvertmUflowNSCPT10DON} combined with~\eqref{eq:2PIFRGmUflowDefinitionUpsilon0DON}. Among the 16 components of $\overline{\Upsilon}_{\mathfrak{s}}^{\mathrm{inv}}$ at $N=2$, only 8 differ from zero and are given by:
\begin{equation}
\overline{\Upsilon}_{\mathfrak{s},(1,1)(1,1)}^{\mathrm{inv}} = \overline{\Upsilon}_{\mathfrak{s},(2,2)(2,2)}^{\mathrm{inv}} = \frac{9\left(2+\mathfrak{s}\lambda\overline{\boldsymbol{G}}_{\mathfrak{s}}^{2}\right)}{9+9\mathfrak{s}\lambda\overline{\boldsymbol{G}}_{\mathfrak{s}}^{2}+2\mathfrak{s}^{2}\lambda^{2}\overline{\boldsymbol{G}}_{\mathfrak{s}}^{4}} \;,
\end{equation}
\begin{equation}
\overline{\Upsilon}_{\mathfrak{s},(1,1)(2,2)}^{\mathrm{inv}} = \overline{\Upsilon}_{\mathfrak{s},(2,2)(1,1)}^{\mathrm{inv}} = -\frac{3\mathfrak{s}\lambda\overline{\boldsymbol{G}}_{\mathfrak{s}}^{2}}{9+9\mathfrak{s}\lambda\overline{\boldsymbol{G}}_{\mathfrak{s}}^{2}+2\mathfrak{s}^{2}\lambda^{2}\overline{\boldsymbol{G}}_{\mathfrak{s}}^{4}} \;,
\end{equation}
\begin{equation}
\overline{\Upsilon}_{\mathfrak{s},(1,2)(1,2)}^{\mathrm{inv}} = \overline{\Upsilon}_{\mathfrak{s},(1,2)(2,1)}^{\mathrm{inv}} = \overline{\Upsilon}_{\mathfrak{s},(2,1)(1,2)}^{\mathrm{inv}} = \overline{\Upsilon}_{\mathfrak{s},(2,1)(2,1)}^{\mathrm{inv}} = \frac{3}{3+\mathfrak{s}\lambda\overline{\boldsymbol{G}}_{\mathfrak{s}}^{2}} \;.
\end{equation}
Following our previous explanation right below~\eqref{eq:2PIFRGIplusPiPhi2mUflow0DON}, we directly infer from this the 8 non-vanishing components of $\left(\mathcal{I}+\overline{\Pi}_{\mathfrak{s}}\overline{\boldsymbol{\Phi}}_{\mathfrak{s}}^{(2)}\right)^{\mathrm{inv}}$:
\begin{equation}
\left(\mathcal{I}+\overline{\Pi}_{\mathfrak{s}}\overline{\boldsymbol{\Phi}}_{\mathfrak{s}}^{(2)}\right)_{(1,1)(1,1)}^{\mathrm{inv}} = \left(\mathcal{I}+\overline{\Pi}_{\mathfrak{s}}\overline{\boldsymbol{\Phi}}_{\mathfrak{s}}^{(2)}\right)_{(2,2)(2,2)}^{\mathrm{inv}} = \frac{9\left(2+\lambda\overline{\boldsymbol{G}}_{\mathfrak{s}}^{2}\right)}{9+9\lambda\overline{\boldsymbol{G}}_{\mathfrak{s}}^{2}+2\lambda^{2}\overline{\boldsymbol{G}}_{\mathfrak{s}}^{4}} \;,
\end{equation}
\begin{equation}
\left(\mathcal{I}+\overline{\Pi}_{\mathfrak{s}}\overline{\boldsymbol{\Phi}}_{\mathfrak{s}}^{(2)}\right)_{(1,1)(2,2)}^{\mathrm{inv}} = \left(\mathcal{I}+\overline{\Pi}_{\mathfrak{s}}\overline{\boldsymbol{\Phi}}_{\mathfrak{s}}^{(2)}\right)_{(2,2)(1,1)}^{\mathrm{inv}} = -\frac{3\lambda\overline{\boldsymbol{G}}_{\mathfrak{s}}^{2}}{9+9\lambda\overline{\boldsymbol{G}}_{\mathfrak{s}}^{2}+2\lambda^{2}\overline{\boldsymbol{G}}_{\mathfrak{s}}^{4}} \;,
\end{equation}
\begin{equation}
\scalebox{0.9}{${\displaystyle \left(\mathcal{I}+\overline{\Pi}_{\mathfrak{s}}\overline{\boldsymbol{\Phi}}_{\mathfrak{s}}^{(2)}\right)_{(1,2)(1,2)}^{\mathrm{inv}} = \left(\mathcal{I}+\overline{\Pi}_{\mathfrak{s}}\overline{\boldsymbol{\Phi}}_{\mathfrak{s}}^{(2)}\right)_{(1,2)(2,1)}^{\mathrm{inv}} = \left(\mathcal{I}+\overline{\Pi}_{\mathfrak{s}}\overline{\boldsymbol{\Phi}}_{\mathfrak{s}}^{(2)}\right)_{(2,1)(1,2)}^{\mathrm{inv}} = \left(\mathcal{I}+\overline{\Pi}_{\mathfrak{s}}\overline{\boldsymbol{\Phi}}_{\mathfrak{s}}^{(2)}\right)_{(2,1)(2,1)}^{\mathrm{inv}} = \frac{3}{3+\lambda\overline{\boldsymbol{G}}_{\mathfrak{s}}^{2}} \;. }$}
\end{equation}
We will also treat the mU-flow with $N_{\mathrm{SCPT}}$ up to 3 at $N=1$. Hence, we investigate the 2PI-FRG with starting point coinciding with each of the three first non-trivial orders of self-consistent PT. Note that the mU-flow has only been tested at $N_{\mathrm{SCPT}}=1$ in the comparative studies of 2PI-FRG approaches mentioned earlier (i.e. in refs.~\cite{ren15,ren16}). The expression of the modified Luttinger-Ward $\boldsymbol{\Phi}_{\mathfrak{s}}(G)$ at $N_{\mathrm{SCPT}}=3$ can be deduced from the perturbative expression~\eqref{eq:PertExpressionPhiN12PIFRGtCflow0DON}, which leads to:
\begin{equation}
\begin{split}
\boldsymbol{\Phi}_{\mathfrak{s}}(G) = & \ \Phi_{\mathfrak{s}}(G) + \Phi_{\mathrm{SCPT},N_{\mathrm{SCPT}}=3}(U,G) - \Phi_{\mathrm{SCPT},N_{\mathrm{SCPT}}=3}(U_{\mathfrak{s}},G) \\
= & \ \Phi_{\mathfrak{s}}(G) + \frac{1}{8} \lambda G^{2} \left(1-\mathfrak{s}\right) - \frac{1}{48} \lambda^{2} G^{4} \left(1-\mathfrak{s}^{2}\right) + \frac{1}{48} \lambda^{3} G^{6} \left(1-\mathfrak{s}^{3}\right) \;.
\end{split}
\end{equation}
From this relation, we can determine the differential equations underlying the mU-flow up to $N_{\mathrm{SCPT}}=3$ from the pU-flow equations~\eqref{eq:2PIFRGpuflowFlowEqG0DON1} to~\eqref{eq:2PIFRGpuflowFlowEqSigma0DON1}. We obtain in this way (see appendix~\ref{ann:2PIfrgpUandmUflow0DON1} for the corresponding flow equations expressing the derivative of the modified 2PI vertices of order 2 and 3 with respect to $\mathfrak{s}$):
\begin{itemize}
\item For $N_{\mathrm{SCPT}}=1$:
\begin{equation}
\dot{\overline{\boldsymbol{G}}}_{\mathfrak{s}} = \overline{\boldsymbol{G}}_{\mathfrak{s}}^{2} \dot{\overline{\boldsymbol{\Sigma}}}_{\mathfrak{s}} \;,
\label{eq:2PIFRGmuflowNSCPT1FlowEqG0DON1}
\end{equation}
\begin{equation}
\dot{\overline{\boldsymbol{\Omega}}}_{\mathfrak{s}} = \frac{\lambda}{24} \left( 4\left( 2\overline{\boldsymbol{G}}_{\mathfrak{s}}^{-2} + \overline{\boldsymbol{\Phi}}_{\mathfrak{s}}^{(2)} - \lambda\left(1-\mathfrak{s}\right) \right)^{-1} + \overline{\boldsymbol{G}}_{\mathfrak{s}}^{2} \right) - \frac{1}{8} \lambda \overline{\boldsymbol{G}}_{\mathfrak{s}}^{2} \;,
\label{eq:2PIFRGmuflowNSCPT1FlowEqOmega0DON1}
\end{equation}
\begin{equation}
\dot{\overline{\boldsymbol{\Sigma}}}_{\mathfrak{s}} = -\frac{\lambda}{3} \left( 2 + \overline{\boldsymbol{G}}_{\mathfrak{s}}^{2} \overline{\boldsymbol{\Phi}}_{\mathfrak{s}}^{(2)} \right)^{-1} \bigg( \left(2\overline{\boldsymbol{G}}_{\mathfrak{s}}^{-2} + \overline{\boldsymbol{\Phi}}_{\mathfrak{s}}^{(2)} - \lambda\left(1-\mathfrak{s}\right) \right)^{-2} \left(8\overline{\boldsymbol{G}}_{\mathfrak{s}}^{-3} - \overline{\boldsymbol{\Phi}}_{\mathfrak{s}}^{(3)} \right) - 2\overline{\boldsymbol{G}}_{\mathfrak{s}} \bigg) \;.
\label{eq:2PIFRGmuflowNSCPT1FlowEqSigma0DON1}
\end{equation}

\item For $N_{\mathrm{SCPT}}=2$:
\begin{equation}
\dot{\overline{\boldsymbol{G}}}_{\mathfrak{s}} = \overline{\boldsymbol{G}}_{\mathfrak{s}}^{2} \dot{\overline{\boldsymbol{\Sigma}}}_{\mathfrak{s}} \;,
\label{eq:2PIFRGmuflowNSCPT2FlowEqG0DON1}
\end{equation}
\begin{equation}
\dot{\overline{\boldsymbol{\Omega}}}_{\mathfrak{s}} = \frac{\lambda}{24} \left( 4\left( 2\overline{\boldsymbol{G}}_{\mathfrak{s}}^{-2} + \overline{\boldsymbol{\Phi}}_{\mathfrak{s}}^{(2)} - \lambda\left(1-\mathfrak{s}\right) + \lambda^{2}\overline{\boldsymbol{G}}_{\mathfrak{s}}^{2}\left(1-\mathfrak{s}^{2}\right) \right)^{-1} + \overline{\boldsymbol{G}}_{\mathfrak{s}}^{2} \right) - \frac{1}{8} \lambda \overline{\boldsymbol{G}}_{\mathfrak{s}}^{2} + \frac{1}{24} \mathfrak{s} \lambda^{2} \overline{\boldsymbol{G}}_{\mathfrak{s}}^{4} \;,
\label{eq:2PIFRGmuflowNSCPT2FlowEqOmega0DON1}
\end{equation}
\begin{equation}
\begin{split}
\dot{\overline{\boldsymbol{\Sigma}}}_{\mathfrak{s}} = & -\frac{\lambda}{3} \left( 2 + \overline{\boldsymbol{G}}_{\mathfrak{s}}^{2} \overline{\boldsymbol{\Phi}}_{\mathfrak{s}}^{(2)} \right)^{-1} \bigg( \left(2\overline{\boldsymbol{G}}_{\mathfrak{s}}^{-2} + \overline{\boldsymbol{\Phi}}_{\mathfrak{s}}^{(2)} - \lambda\left(1-\mathfrak{s}\right) + \lambda^{2}\overline{\boldsymbol{G}}_{\mathfrak{s}}^{2}\left(1-\mathfrak{s}^{2}\right) \right)^{-2} \\
& \hspace{4.25cm} \times\left(8\overline{\boldsymbol{G}}_{\mathfrak{s}}^{-3} - \overline{\boldsymbol{\Phi}}_{\mathfrak{s}}^{(3)} - 4 \lambda^{2} \overline{\boldsymbol{G}}_{\mathfrak{s}} \left(1-\mathfrak{s}^{2}\right) \right) - 2\overline{\boldsymbol{G}}_{\mathfrak{s}} + 2\mathfrak{s}\lambda\overline{\boldsymbol{G}}_{\mathfrak{s}}^{3} \bigg) \;.
\end{split}
\label{eq:2PIFRGmuflowNSCPT2FlowEqSigma0DON1}
\end{equation}

\item For $N_{\mathrm{SCPT}}=3$:
\begin{equation}
\dot{\overline{\boldsymbol{G}}}_{\mathfrak{s}} = \overline{\boldsymbol{G}}_{\mathfrak{s}}^{2} \dot{\overline{\boldsymbol{\Sigma}}}_{\mathfrak{s}} \;,
\label{eq:2PIFRGmuflowNSCPT3FlowEqG0DON1}
\end{equation}
\begin{equation}
\begin{split}
\dot{\overline{\boldsymbol{\Omega}}}_{\mathfrak{s}} = & \ \frac{\lambda}{24} \left( 4\left( 2\overline{\boldsymbol{G}}_{\mathfrak{s}}^{-2} + \overline{\boldsymbol{\Phi}}_{\mathfrak{s}}^{(2)} - \lambda\left(1-\mathfrak{s}\right) + \lambda^{2}\overline{\boldsymbol{G}}_{\mathfrak{s}}^{2}\left(1-\mathfrak{s}^{2}\right) - \frac{5}{2}\lambda^{3}\overline{\boldsymbol{G}}_{\mathfrak{s}}^{4}\left(1-\mathfrak{s}^{3}\right) \right)^{-1} + \overline{\boldsymbol{G}}_{\mathfrak{s}}^{2} \right) \\
& - \frac{1}{8} \lambda \overline{\boldsymbol{G}}_{\mathfrak{s}}^{2} + \frac{1}{24} \mathfrak{s} \lambda^{2} \overline{\boldsymbol{G}}_{\mathfrak{s}}^{4} - \frac{1}{16} \mathfrak{s}^{2} \lambda^{3} \overline{\boldsymbol{G}}_{\mathfrak{s}}^{6} \;,
\end{split}
\label{eq:2PIFRGmuflowNSCPT3FlowEqOmega0DON1}
\end{equation}
\begin{equation}
\begin{split}
\scalebox{0.96}{${\displaystyle \dot{\overline{\boldsymbol{\Sigma}}}_{\mathfrak{s}} = }$} & \scalebox{0.96}{${\displaystyle -\frac{\lambda}{3} \left( 2 + \overline{\boldsymbol{G}}_{\mathfrak{s}}^{2} \overline{\boldsymbol{\Phi}}_{\mathfrak{s}}^{(2)} \right)^{-1} \bigg( \left(2\overline{\boldsymbol{G}}_{\mathfrak{s}}^{-2} + \overline{\boldsymbol{\Phi}}_{\mathfrak{s}}^{(2)} - \lambda\left(1-\mathfrak{s}\right) + \lambda^{2}\overline{\boldsymbol{G}}_{\mathfrak{s}}^{2}\left(1-\mathfrak{s}^{2}\right) - \frac{5}{2}\lambda^{3}\overline{\boldsymbol{G}}_{\mathfrak{s}}^{4}\left(1-\mathfrak{s}^{3}\right) \right)^{-2} }$} \\
& \hspace{4.0cm} \scalebox{0.96}{${\displaystyle \times\left(8\overline{\boldsymbol{G}}_{\mathfrak{s}}^{-3} - \overline{\boldsymbol{\Phi}}_{\mathfrak{s}}^{(3)} - 4 \lambda^{2} \overline{\boldsymbol{G}}_{\mathfrak{s}} \left(1-\mathfrak{s}^{2}\right) + 20 \lambda^{3} \overline{\boldsymbol{G}}_{\mathfrak{s}}^{3} \left(1-\mathfrak{s}^{3}\right) \right) }$} \\
& \hspace{3.7cm} \scalebox{0.96}{${\displaystyle - 2\overline{\boldsymbol{G}}_{\mathfrak{s}} + 2\mathfrak{s}\lambda\overline{\boldsymbol{G}}_{\mathfrak{s}}^{3} - \frac{9}{2}\mathfrak{s}^{2}\lambda^{2}\overline{\boldsymbol{G}}_{\mathfrak{s}}^{5} \bigg) \;.}$}
\end{split}
\label{eq:2PIFRGmuflowNSCPT3FlowEqSigma0DON1}
\end{equation}

\end{itemize}

\vspace{0.3cm}

As already stated by~\eqref{eq:InitialConditions2PIFRGmUflowOmega} to~\eqref{eq:InitialConditions2PIFRGmUflowPhin} as well as~\eqref{eq:boldGinitialconditionmUflow}, the initial conditions of the mU-flow are directly inferred from the self-consistent PT results up to the chosen order $\mathcal{O}\big(U^{N_{\mathrm{SCPT}}}\big)$. For the present application to the (0+0)-D $O(N)$-symmetric $\varphi^{4}$-theory, they read:
\begin{equation}
\overline{\boldsymbol{G}}_{\mathfrak{s}=\mathfrak{s}_{\mathrm{i}}} = \overline{G}_{\mathrm{SCPT},N_{\mathrm{SCPT}}} \mathrlap{\;,}
\label{eq:boldGinitialconditionmUflow0DON}
\end{equation}
\begin{equation}
\overline{\boldsymbol{\Omega}}_{\mathfrak{s}=\mathfrak{s}_{\mathrm{i}}} = E_{\mathrm{gs},\mathrm{SCPT},N_{\mathrm{SCPT}}} \mathrlap{\;,}
\label{eq:InitialConditions2PIFRGmUflowOmega0DON}
\end{equation}
\begin{equation}
\overline{\boldsymbol{\Sigma}}_{\mathfrak{s}=\mathfrak{s}_{\mathrm{i}}} = -\Phi^{(1)}_{\mathrm{SCPT},N_{\mathrm{SCPT}}}\big(U,G=\overline{G}_{\mathrm{SCPT},N_{\mathrm{SCPT}}}\big) \mathrlap{\;,}
\label{eq:InitialConditions2PIFRGmUflowSigma0DON}
\end{equation}
\begin{equation}
\overline{\boldsymbol{\Phi}}_{\mathfrak{s}=\mathfrak{s}_{\mathrm{i}}}^{(n)}= \Phi^{(n)}_{\mathrm{SCPT},N_{\mathrm{SCPT}}}\big(U,G=\overline{G}_{\mathrm{SCPT},N_{\mathrm{SCPT}}}\big) \mathrlap{\quad \forall n \geq 2 \;.}
\label{eq:InitialConditions2PIFRGmUflowPhin0DON}
\end{equation}
We thus point out a peculiarity of (0+0)-D applications: $\overline{\boldsymbol{\Omega}}_{\mathfrak{s}=\mathfrak{s}_{\mathrm{i}}}$ directly coincides with the gs energy calculated from self-consistent PT up to order $\mathcal{O}\big(U^{N_{\mathrm{SCPT}}}\big)$, i.e. $E_{\mathrm{gs},\mathrm{SCPT},N_{\mathrm{SCPT}}}$, since the zero-temperature limit defining the gs energy vanishes in (0+0)-D (as was already discussed when introducing~\eqref{eq:DefEgsExactZexact0DON}). At $N_{\mathrm{SCPT}}=1$, the initial conditions~\eqref{eq:boldGinitialconditionmUflow0DON} to~\eqref{eq:InitialConditions2PIFRGmUflowPhin0DON} reduce to:
\begin{equation}
\overline{\boldsymbol{G}}_{\mathfrak{s}=\mathfrak{s}_{\mathrm{i}}} = \overline{G}_{\mathrm{SCPT},N_{\mathrm{SCPT}}=1} \mathrlap{\;,}
\label{eq:boldGinitialconditionmUflow0DON2}
\end{equation}
\begin{equation}
\overline{\boldsymbol{\Omega}}_{\mathfrak{s}=\mathfrak{s}_{\mathrm{i}}} = E_{\mathrm{gs},\mathrm{SCPT},N_{\mathrm{SCPT}}=1} \mathrlap{\;,}
\label{eq:InitialConditions2PIFRGmUflowOmega0DON2}
\end{equation}
\begin{equation}
\overline{\boldsymbol{\Sigma}}_{\mathfrak{s}=\mathfrak{s}_{\mathrm{i}},(a_{1},a'_{1})} = -\frac{\lambda}{6} \delta_{a_{1} a'_{1}} \sum_{a_{2}=1}^{N} \overline{G}_{\mathrm{SCPT},N_{\mathrm{SCPT}}=1, a_{2} a_{2}} - \frac{\lambda}{3} \overline{G}_{\mathrm{SCPT},N_{\mathrm{SCPT}}=1, a_{1} a'_{1}} \mathrlap{\;,}
\label{eq:InitialConditions2PIFRGmUflowSigma0DON2}
\end{equation}
\begin{equation}
\overline{\boldsymbol{\Phi}}^{(2)}_{\mathfrak{s}=\mathfrak{s}_{\mathrm{i}},(a_{1},a'_{1})(a_{2},a'_{2})} = U_{(a_{1},a'_{1})(a_{2},a'_{2})} = \frac{\lambda}{3}\left(\delta_{a_{1}a'_{1}}\delta_{a_{2}a'_{2}}+\delta_{a_{1}a_{2}}\delta_{a'_{1}a'_{2}}+\delta_{a_{1}a'_{2}}\delta_{a'_{1}a_{2}}\right) \mathrlap{\;,}
\label{eq:InitialConditions2PIFRGmUflowPhi20DON2}
\end{equation}
\begin{equation}
\hspace{5.9cm} \overline{\boldsymbol{\Phi}}_{\mathfrak{s}=\mathfrak{s}_{\mathrm{i}},(a_{1},a'_{1}) \cdots (a_{n},a'_{n})}^{(n)}= 0 \quad \forall a_{1}, a'_{1},\cdots, a_{n}, a'_{n}, ~ \forall n \geq 3 \;,
\label{eq:InitialConditions2PIFRGmUflowPhin0DON2}
\end{equation}
as can be deduced from~\eqref{eq:DefinitionmUflowPhiboldNscpt10DON} together with identity~\eqref{eq:DefinitionBosonicIdentityMatrix0DON}. Finally, we recall that, regardless of the chosen value for $N_{\mathrm{SCPT}}$, the infinite tower of differential equations underlying the mU-flow is truncated with the condition:
\begin{equation}
\overline{\boldsymbol{\Phi}}_{\mathfrak{s}}^{(n)}=\overline{\boldsymbol{\Phi}}_{\mathfrak{s}=\mathfrak{s}_{\mathrm{i}}}^{(n)} \mathrlap{\quad \forall \mathfrak{s}, ~ \forall n > N_{\mathrm{max}} \;.}
\label{eq:2PIfrgModifiedUflowTruncation0DON}
\end{equation}

\vspace{0.5cm}

\begin{figure}[!htb]
\captionsetup[subfigure]{labelformat=empty}
  \begin{center}
    \subfloat[]{
      \includegraphics[width=0.50\linewidth]{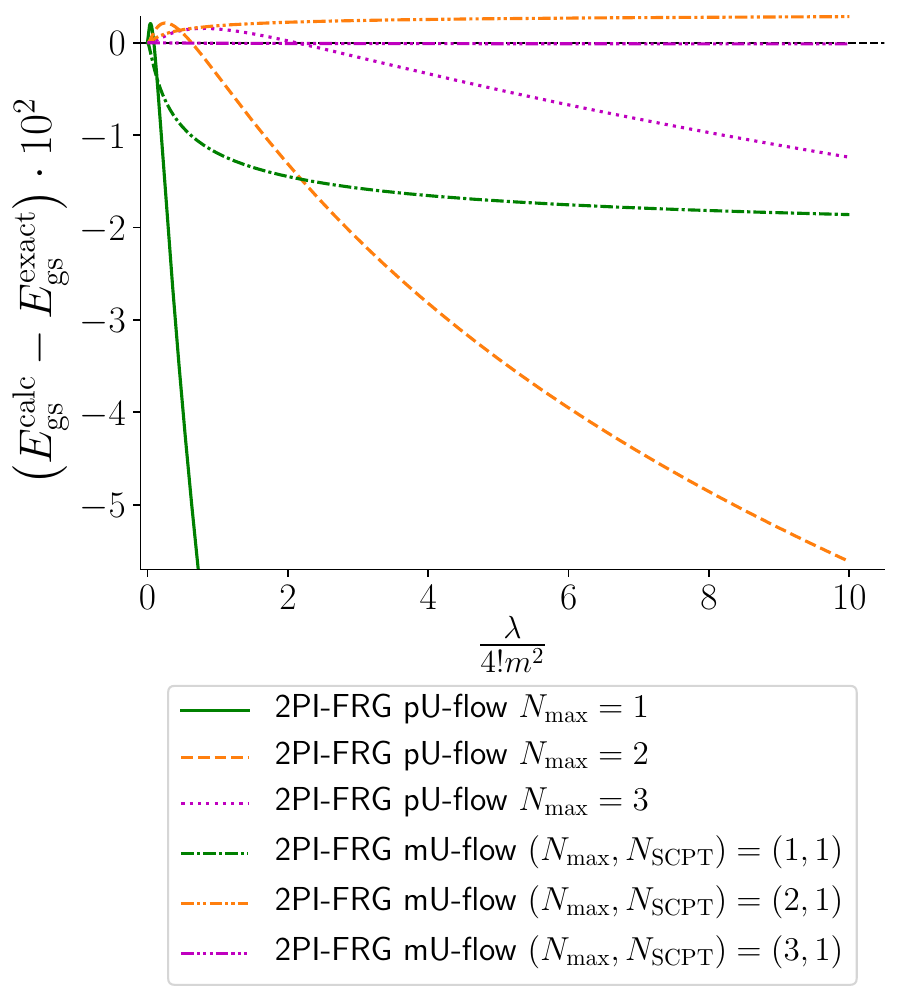}
                         }
    \subfloat[]{
      \includegraphics[width=0.50\linewidth]{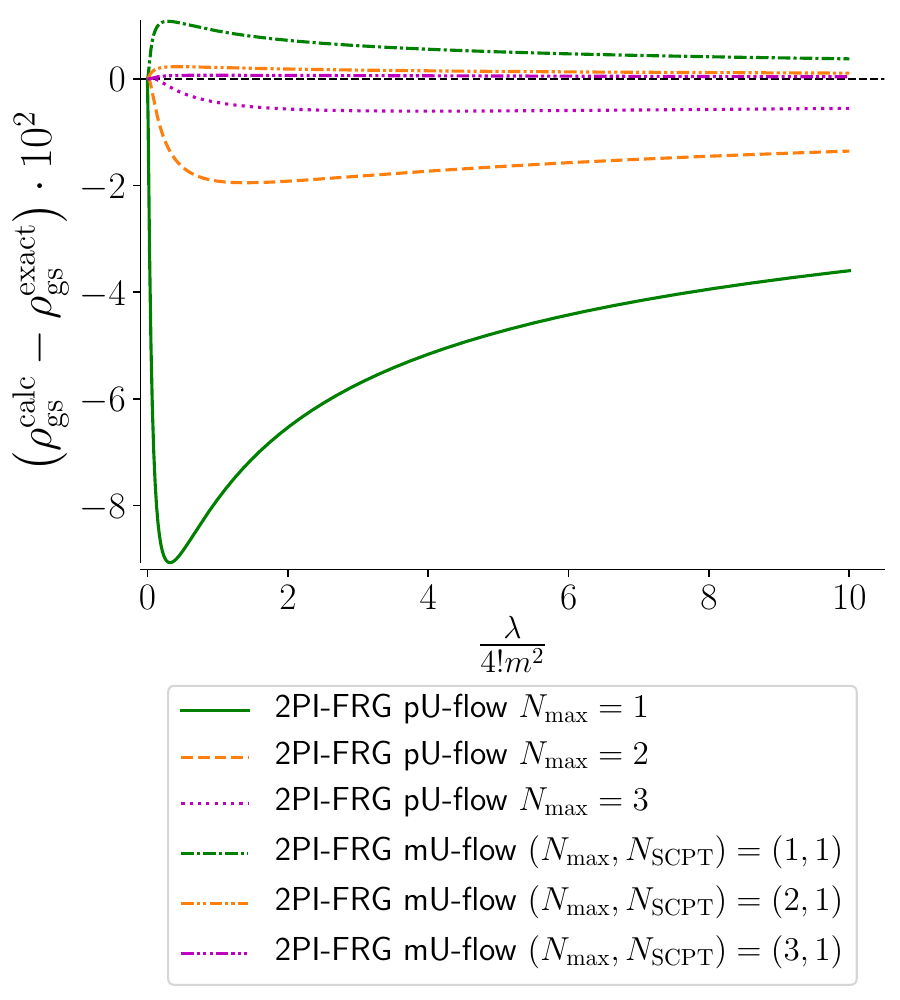}
                         }
\caption{Difference between the calculated gs energy $E_{\mathrm{gs}}^{\mathrm{calc}}$ or density $\rho_{\mathrm{gs}}^{\mathrm{calc}}$ and the corresponding exact solution $E_{\mathrm{gs}}^{\mathrm{exact}}$ or $\rho_{\mathrm{gs}}^{\mathrm{exact}}$ at $m^{2}=+1$ and $N=1$ ($\mathcal{R}e(\lambda)\geq 0$ and $\mathcal{I}m(\lambda)=0$).}
\label{fig:2PIFRGUflowpUflowVsmUflowlambdaN1}
  \end{center}
\end{figure}
\begin{figure}[!htb]
\captionsetup[subfigure]{labelformat=empty}
  \begin{center}
    \subfloat[]{
      \includegraphics[width=0.80\linewidth]{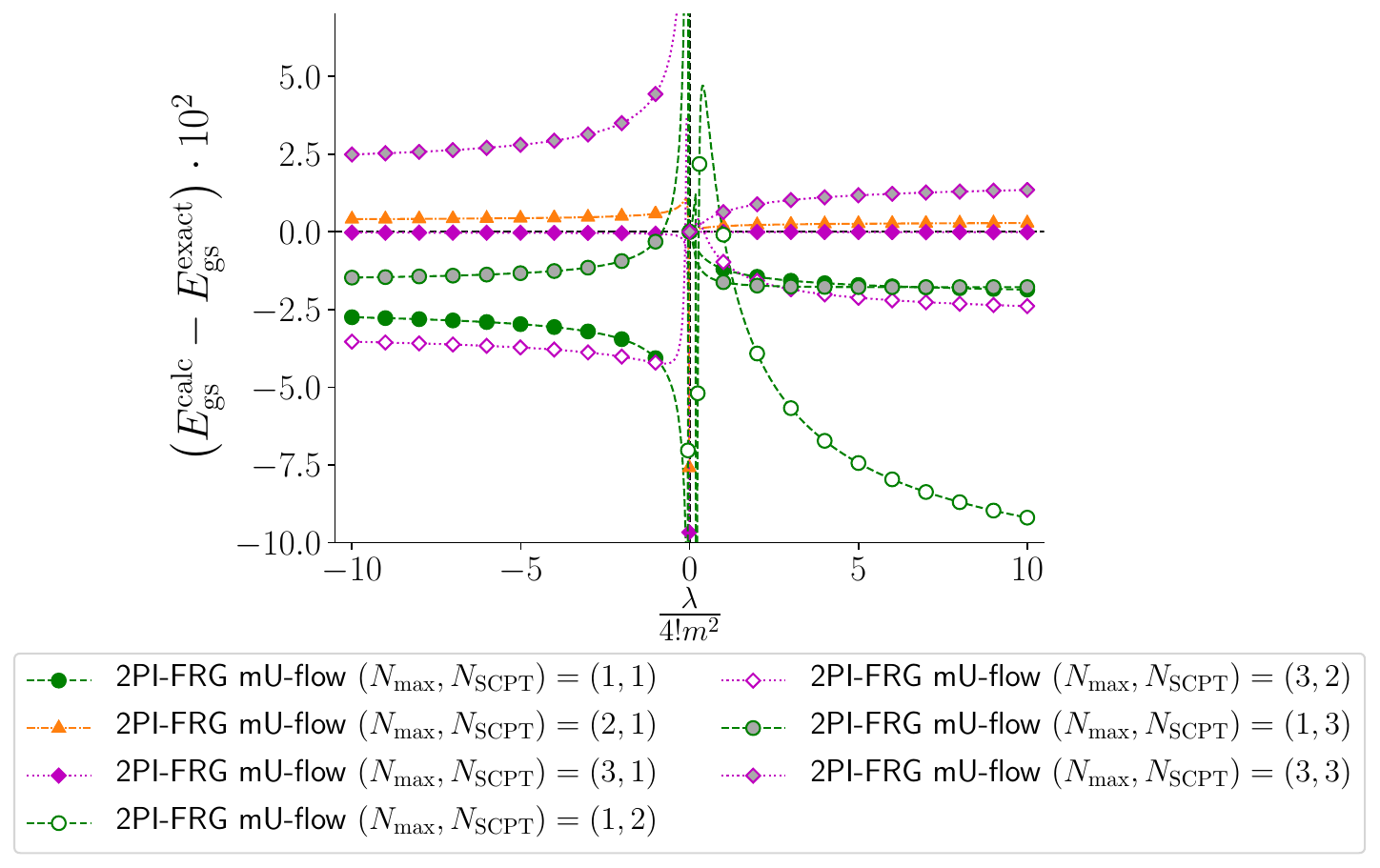}
                         }
   \\                     
    \subfloat[]{
      \includegraphics[width=0.80\linewidth]{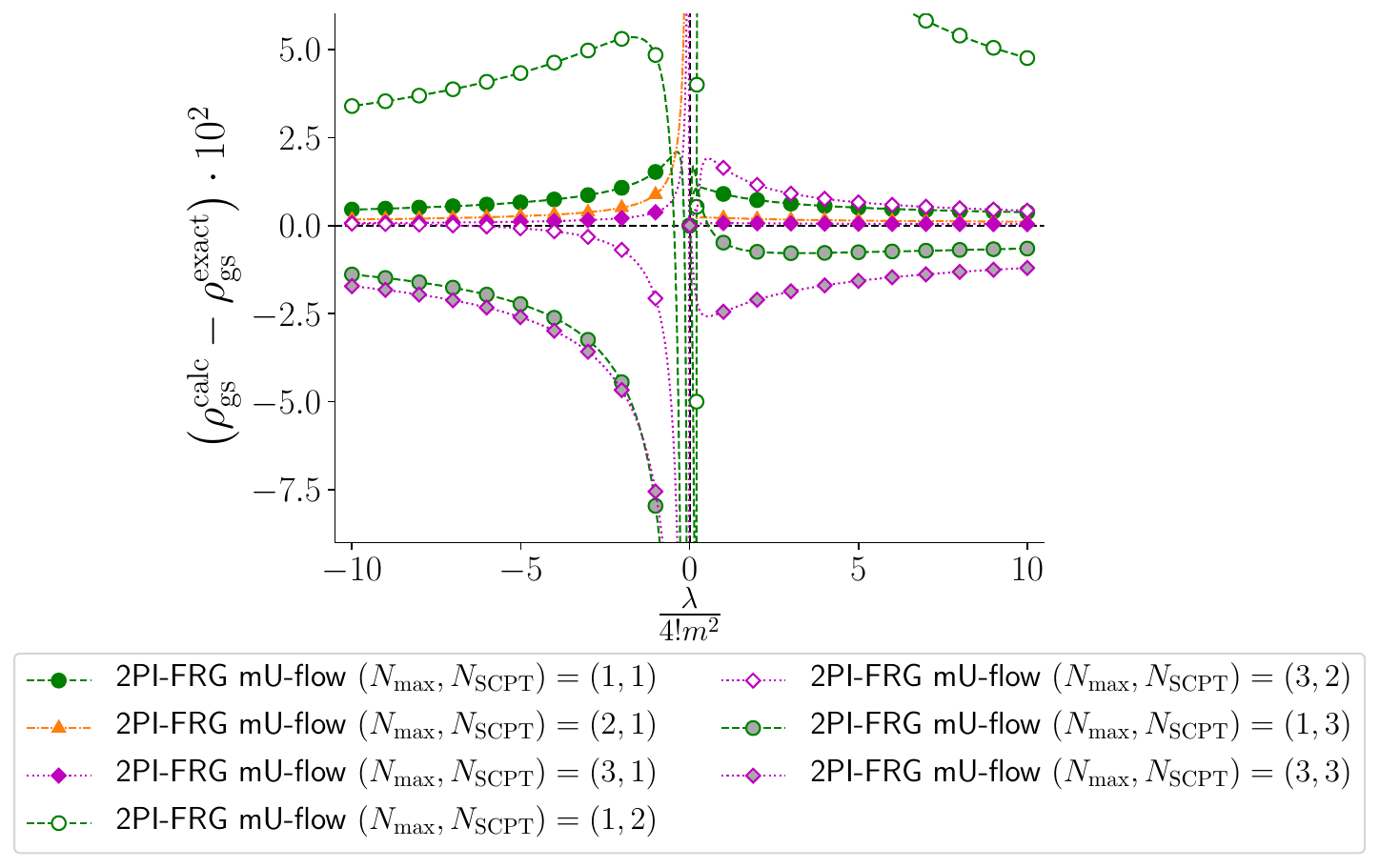}
                         }
\caption{Difference between the calculated gs energy $E_{\mathrm{gs}}^{\mathrm{calc}}$ or density $\rho_{\mathrm{gs}}^{\mathrm{calc}}$ and the corresponding exact solution $E_{\mathrm{gs}}^{\mathrm{exact}}$ or $\rho_{\mathrm{gs}}^{\mathrm{exact}}$ at $m^{2}=\pm 1$ and $N=1$ ($\mathcal{R}e(\lambda)\geq 0$ and $\mathcal{I}m(\lambda)=0$).}
\label{fig:2PIFRGUflowmUflowDifferentNSCPTlambdaN1}
  \end{center}
\end{figure}

In summary, our mU-flow results at $N_{\mathrm{max}}=1$ are determined by solving the set of differential equations~\eqref{eq:2PIFRGmUflowNSCPT1EquationG0DON} to~\eqref{eq:2PIFRGmUflowNSCPT1EquationSigma0DON} for all $N$ (for $N=2$ especially) and at $N_{\mathrm{SCPT}}=1$ as well as~\eqref{eq:2PIFRGmuflowNSCPT1FlowEqG0DON1} to~\eqref{eq:2PIFRGmuflowNSCPT3FlowEqSigma0DON1} at $N=1$ for more involved starting points (i.e. for $N_{\mathrm{SCPT}}=1,2~\mathrm{or}~3$). The corresponding initial conditions are inferred from the perturbative expression of the Luttinger-Ward functional (given e.g. by~\eqref{eq:DefinitionmUflowPhiboldNscpt10DON} for all $N$ and by~\eqref{eq:PertExpressionPhiN12PIFRGtCflow0DON} at $N=1$), as shown by~\eqref{eq:boldGinitialconditionmUflow0DON} to~\eqref{eq:InitialConditions2PIFRGmUflowPhin0DON} which reduce to~\eqref{eq:boldGinitialconditionmUflow0DON2} to~\eqref{eq:InitialConditions2PIFRGmUflowPhin0DON2} at $N_{\mathrm{SCPT}}=1$ for all $N$. Finally, the truncation condition for the mU-flow is given by~\eqref{eq:2PIfrgModifiedUflowTruncation0DON} and the cutoff function for the two-body interaction is still set by~\eqref{eq:choiceCutoffUs2PIFRGUflow0DON}. Comparing first the pU-flow and the mU-flow with the simplest starting point (i.e. with a Hartree-Fock starting point corresponding to $N_{\mathrm{SCPT}}=1$) in fig.~\ref{fig:2PIFRGUflowpUflowVsmUflowlambdaN1}, we can see that the convergence is significantly faster for the latter approach: at the first non-trivial order for instance, the mU-flow results of fig.~\ref{fig:2PIFRGUflowpUflowVsmUflowlambdaN1} exhibit an accuracy below $2\%$ for $E_{\mathrm{gs}}$, and even less for $\rho_{\mathrm{gs}}$, over the whole range $\lambda/4! \in [0,10]$ with $m^{2} > 0$. This could have been expected from the quality of the starting points of the two methods thus compared: the Hartree-Fock starting point of the mU-flow already incorporates a consequent part of the correlations within the studied systems (even at the non-perturbative level), as opposed to the free theory for the pU-flow.

\vspace{0.5cm}

\begin{figure}[!t]
\captionsetup[subfigure]{labelformat=empty}
  \begin{center}
    \subfloat[]{
      \includegraphics[width=0.50\linewidth]{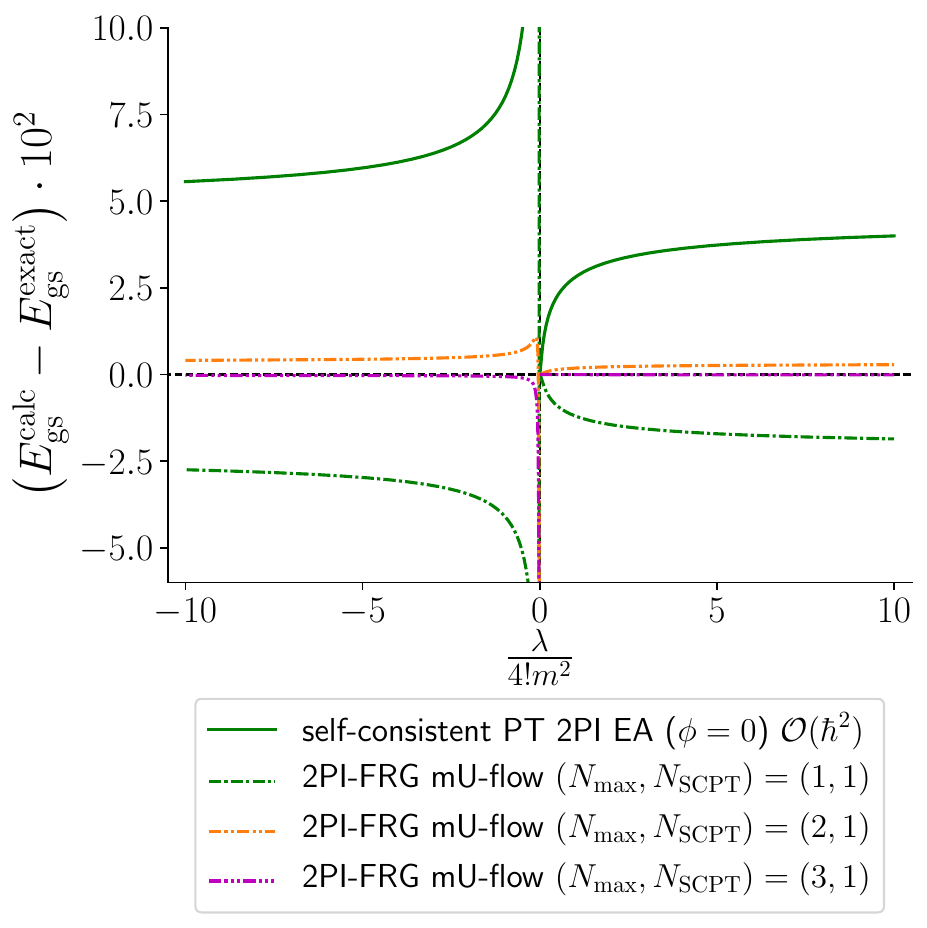}
                         }
    \subfloat[]{
      \includegraphics[width=0.50\linewidth]{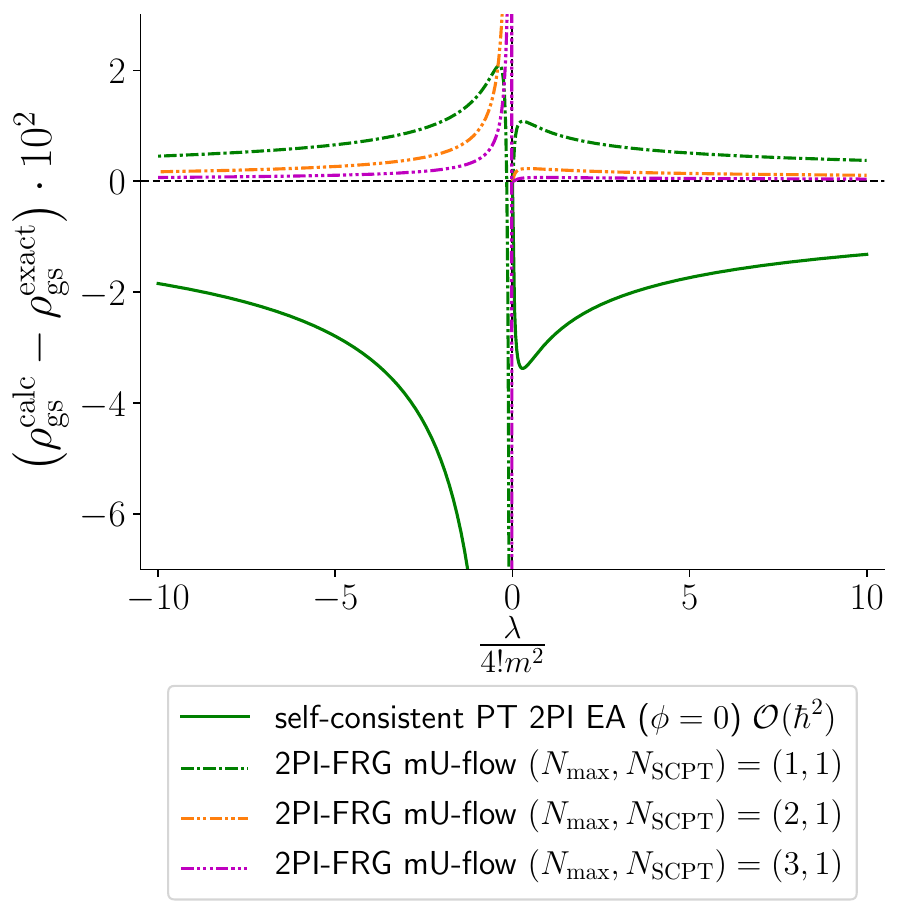}
                         }
\caption{Difference between the calculated gs energy $E_{\mathrm{gs}}^{\mathrm{calc}}$ or density $\rho_{\mathrm{gs}}^{\mathrm{calc}}$ and the corresponding exact solution $E_{\mathrm{gs}}^{\mathrm{exact}}$ or $\rho_{\mathrm{gs}}^{\mathrm{exact}}$ at $m^{2}=\pm 1$ and $N=1$ ($\mathcal{R}e(\lambda)\geq 0$ and $\mathcal{I}m(\lambda)=0$). See also the caption of fig.~\ref{fig:1PIEA} for the meaning of the indication ``$\mathcal{O}\big(\hbar^{n}\big)$'' for the results obtained from $\hbar$-expanded EAs within self-consitent PT.}
\label{fig:2PIFRGUflowmUflowHFstartingpointlambdaN1}
  \end{center}
\end{figure}
\begin{figure}[!htb]
\captionsetup[subfigure]{labelformat=empty}
  \begin{center}
    \subfloat[]{
      \includegraphics[width=0.50\linewidth]{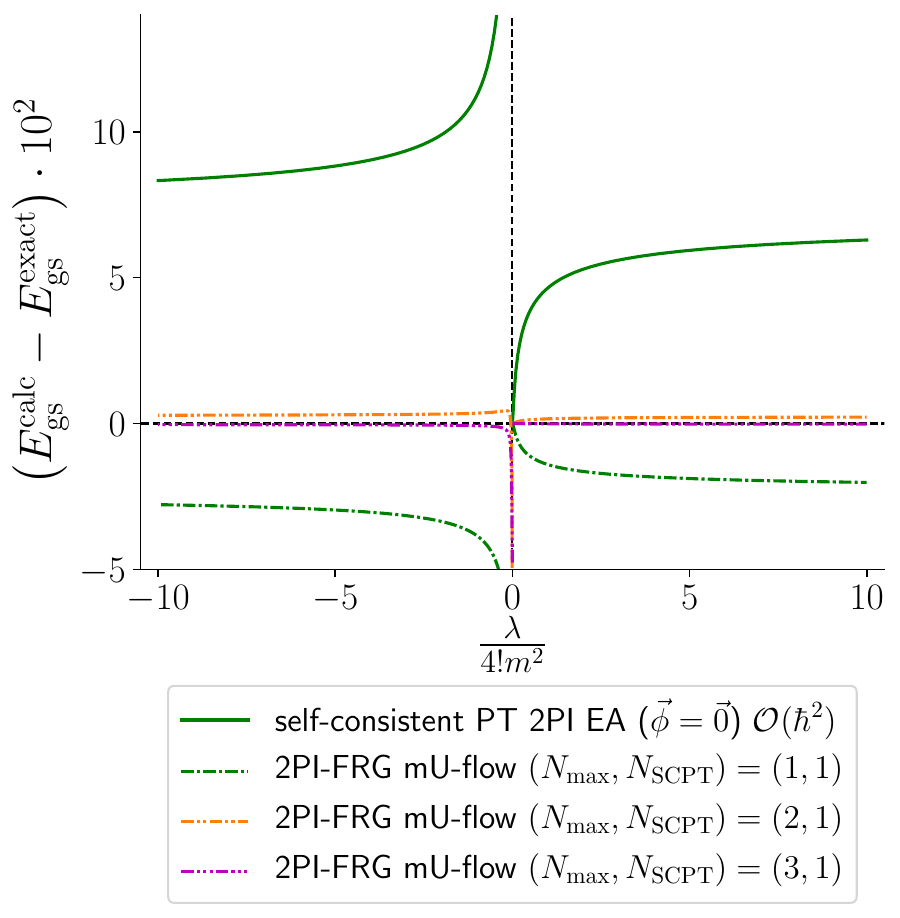}
                         }
    \subfloat[]{
      \includegraphics[width=0.50\linewidth]{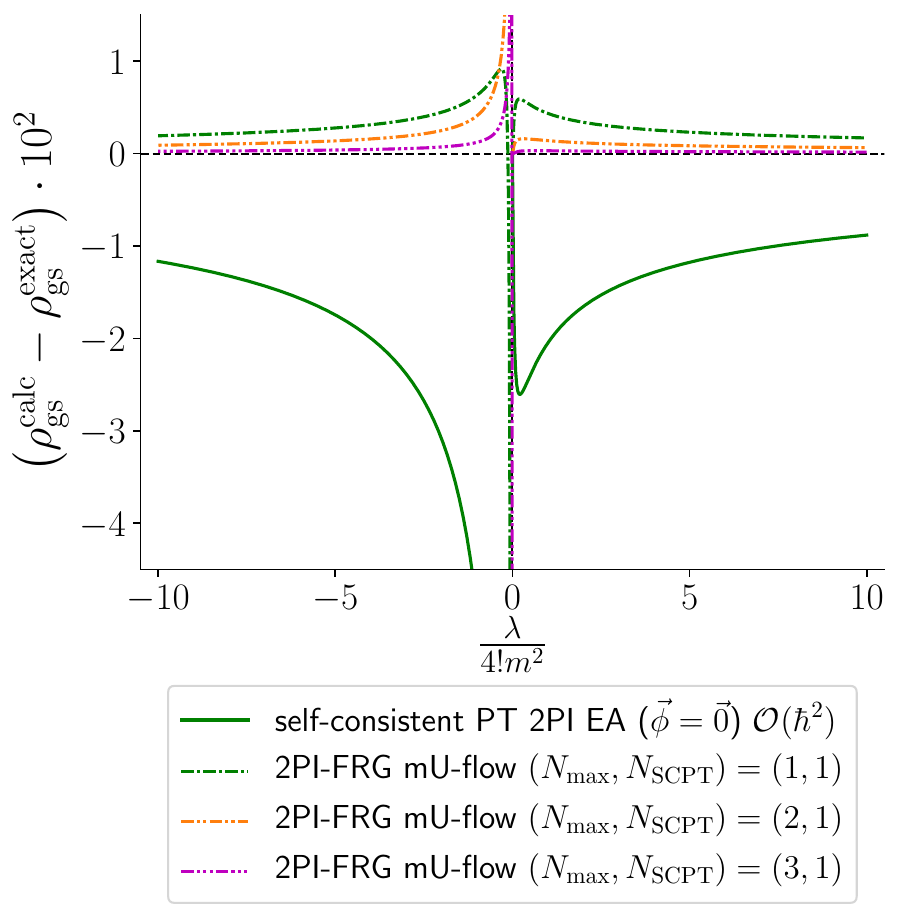}
                         }
\caption{Same as fig.~\ref{fig:2PIFRGUflowmUflowHFstartingpointlambdaN1} with $N=2$ instead.}
\label{fig:2PIFRGUflowmUflowHFstartingpointlambdaN2}
  \end{center}
\end{figure}

It is then quite natural to test the mU-flow for more involved starting points, i.e. for $N_{\mathrm{SCPT}}>1$, which has never been done so far to our knowledge. Hence, fig.~\ref{fig:2PIFRGUflowmUflowDifferentNSCPTlambdaN1} shows mU-flow results up to $N_{\mathrm{max}}=3$ with $N_{\mathrm{SCPT}}=1,2$ and $3$ for both $E_{\mathrm{gs}}$ and $\rho_{\mathrm{gs}}$ in the unbroken- and broken-symmetry regimes of our toy model. We point out first of all the appearance of the same stiffness issues as before (still with the $\mathtt{NDSolve}$ function of $\mathtt{Mathematica~12.1}$) for the mU-flow at $N_{\mathrm{SCPT}}=2$ and $3$ with $N_{\mathrm{max}}=2$, hence explaining the absence of the corresponding curves in fig.~\ref{fig:2PIFRGUflowmUflowDifferentNSCPTlambdaN1}. This does not prevent us from noticing in this figure that, besides a few exceptions, the mU-flow at $N_{\mathrm{SCPT}}=1$ is more performing than at $N_{\mathrm{SCPT}}=2$ or $3$ for a given truncation order $N_{\mathrm{max}}$, for $E_{\mathrm{gs}}$ as well as for $\rho_{\mathrm{gs}}$. Although the starting point contains more and more information about the system to describe as $N_{\mathrm{SCPT}}$ increases, we recall that it is the bare self-consistent PT results (i.e. without resummation) that the mU-flow procedure takes as inputs. Such results take the form of diverging asymptotic series for $\Gamma^{(\mathrm{2PI})}(G)$ (as we have shown in chapter~\ref{chap:DiagTechniques}) and the corresponding estimates for $E_{\mathrm{gs}}$ and $\rho_{\mathrm{gs}}$ all worsen as the truncation order $N_{\mathrm{SCPT}}$ increases. It was only after applying a resummation procedure that we managed to turn self-consistent PT into a systematically improvable technique in chapter~\ref{chap:DiagTechniques}. In the light of the latter comments, we can conclude that, unlike resummation procedures, the mU-flow approach is not suited to extract the full information from the asymptotic series representing the 2PI EA (and the corresponding 2PI vertices) taken as input(s). The mU-flow is thus most efficient at $N_{\mathrm{SCPT}}=1$, which is why we have tested this approach in the unbroken- and broken-symmetry regimes for $N=1$ and $2$, as shown by figs.~\ref{fig:2PIFRGUflowmUflowHFstartingpointlambdaN1} and~\ref{fig:2PIFRGUflowmUflowHFstartingpointlambdaN2}. Our qualitative conclusions on the two latter figures are identical: for $E_{\mathrm{gs}}$ and $\rho_{\mathrm{gs}}$ and for both signs of $m^{2}$, we see that the mU-flow procedure at $N_{\mathrm{max}}=1$ clearly improves the Hartree-Fock curve representing its starting point, and this mU-flow result is itself improved by increasing the truncation order $N_{\mathrm{max}}$ until the curve corresponding to $N_{\mathrm{max}}=3$ becomes barely distinguishable from the exact solution.

\vspace{0.5cm}

\begin{figure}[!t]
\captionsetup[subfigure]{labelformat=empty}
  \begin{center}
    \subfloat[]{
      \includegraphics[width=0.50\linewidth]{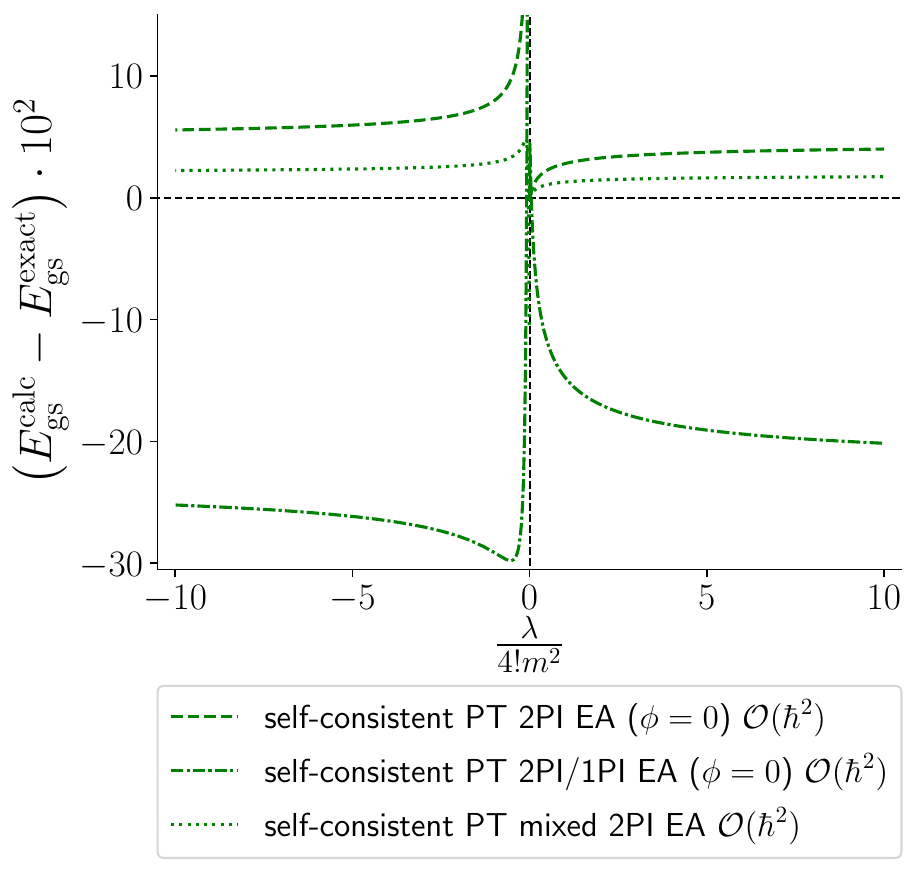}
                         }
    \subfloat[]{
      \includegraphics[width=0.50\linewidth]{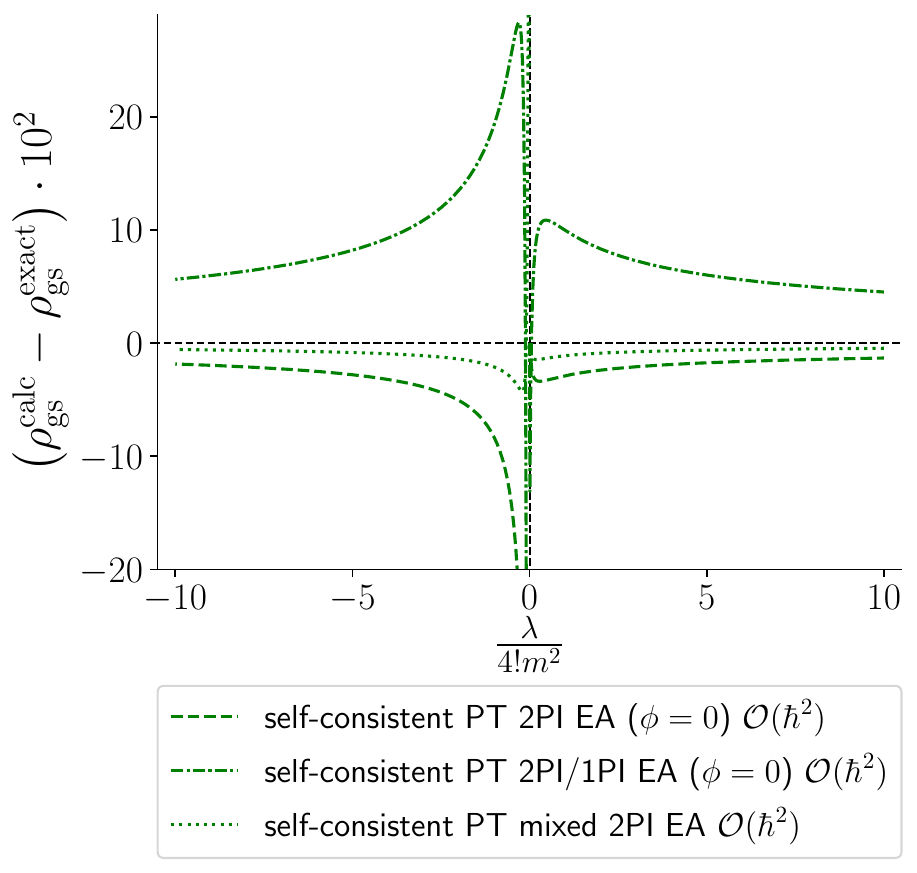}
                         }
\caption{Difference between the calculated gs energy $E_{\mathrm{gs}}^{\mathrm{calc}}$ or density $\rho_{\mathrm{gs}}^{\mathrm{calc}}$ and the corresponding exact solution $E_{\mathrm{gs}}^{\mathrm{exact}}$ or $\rho_{\mathrm{gs}}^{\mathrm{exact}}$ at $m^{2}=\pm 1$ and $N=1$ ($\mathcal{R}e(\lambda)\geq 0$ and $\mathcal{I}m(\lambda)=0$). See also the caption of fig.~\ref{fig:1PIEA} for the meaning of the indication ``$\mathcal{O}\big(\hbar^{n}\big)$'' for the results obtained from $\hbar$-expanded EAs within self-consitent PT.}
\label{fig:mixed2PIFRGmUflowHFstartingpointlambdaN1}
  \end{center}
\end{figure}

Besides these appealing performances of the mU-flow, one might also address the numerical weight of the underpinning numerical procedure which is significantly increased by the Bethe-Salpeter equation to solve throughout the flow. A possibility to diminish the weight of this numerical procedure would be to freeze the evolution of $\overline{\boldsymbol{\Phi}}_{\mathfrak{s}}^{(2)}$ (or $\overline{\Phi}_{\mathfrak{s}}^{(2)}$ for the pU-flow) in the spirit of the scale-dependent HSTs discussed in section~\ref{sec:KeyAspectsFRG} for the 1PI-FRG. To achieve this, we would need to develop a U-flow version of the 2PI-FRG in the mixed representation, i.e. for a theory based on a Yukawa interaction. Such approaches would be based on generating functionals like:
\begin{equation}
Z_{\mathrm{mix}}[j,K] = e^{W_{\mathrm{mix}}[j,K]}=\int\mathcal{D}\widetilde{\psi}\mathcal{D}\widetilde{\sigma} \ e^{-S_{\mathrm{mix},0}\big[\widetilde{\psi},\widetilde{\sigma}\big] - \frac{1}{2}\int_{x_{1},\gamma_{2}} U_{x_{1}\gamma_{2}} \widetilde{\sigma}_{x_{1}} \widetilde{\psi}_{\alpha_{2}} \widetilde{\psi}_{\alpha'_{2}} + \int_{\alpha} j_{\alpha} \widetilde{\sigma}_{\alpha} + \frac{1}{2}\int_{\alpha,\alpha'}\widetilde{\psi}_{\alpha}K_{\alpha\alpha'}\widetilde{\psi}_{\alpha'}} \;,
\label{eq:mixed2PIFRGgeneratingFunc1}
\end{equation}
or
\begin{equation}
Z_{\mathrm{mix}}[\mathcal{J},\mathcal{K}] = e^{W_{\mathrm{mix}}[\mathcal{J},\mathcal{K}]}=\int\mathcal{D}\widetilde{\Psi} \ e^{-S_{\mathrm{mix},0}\big[\widetilde{\Psi}\big] - \frac{1}{2}\int_{\beta_{1},\gamma_{2}} \mathcal{U}_{\beta_{1}\gamma_{2}} \widetilde{\Psi}_{\beta_{1}} \widetilde{\Psi}_{\beta_{2}} \widetilde{\Psi}_{\beta'_{2}} + \int_{\beta} \mathcal{J}_{\beta} \widetilde{\Psi}_{\beta} + \frac{1}{2}\int_{\beta,\beta'}\widetilde{\Psi}_{\beta}\mathcal{K}_{\beta\beta'}\widetilde{\Psi}_{\beta'}} \;,
\label{eq:mixed2PIFRGgeneratingFunc2}
\end{equation}
instead of~\eqref{eq:2PIFRGgeneratingFunc}. Note that $S_{\mathrm{mix},0}$ denotes the free part of the classical action $S_{\mathrm{mix}}$, the supernotations used in~\eqref{eq:mixed2PIFRGgeneratingFunc2} are introduced in chapter~\ref{chap:DiagTechniques} via~\eqref{eq:supernotationPsi} to~\eqref{eq:supernotationK} and the super Yukawa interaction $\mathcal{U}$ is such that:
\begin{equation}
\int_{\beta_{1},\gamma_{2}} \mathcal{U}_{\beta_{1}\gamma_{2}} \widetilde{\Psi}_{\beta_{1}} \widetilde{\Psi}_{\beta_{2}} \widetilde{\Psi}_{\beta'_{2}} = \int_{x_{1},\gamma_{2}} U_{x_{1}\gamma_{2}} \widetilde{\sigma}_{x_{1}} \widetilde{\psi}_{\alpha_{2}} \widetilde{\psi}_{\alpha'_{2}} \;.
\end{equation}
According to the excellent performances of the 1PI-FRG combined with HSTs, one might assume that a linear source directly coupled to the bosonic field $\widetilde{\sigma}$ would be sufficient to add so as to extend the original U-flow into an efficient approach in the mixed representation, as was done in~\eqref{eq:mixed2PIFRGgeneratingFunc1}. In order to develop a mU-flow approach, it should however be noted that the corresponding self-consistent PT results (referred to as ``self-consistent PT 2PI/1PI EA'' in fig.~\ref{fig:mixed2PIFRGmUflowHFstartingpointlambdaN1}) are however worse than those of the original 2PI EA $\Gamma^{(\mathrm{2PI})}(G)$ for the studied toy model, as can be seen from fig.~\ref{fig:mixed2PIFRGmUflowHFstartingpointlambdaN1}. According to the latter, it is clearly appealing to design a mU-flow approach starting from the self-consistent PT results of the full mixed 2PI EA\footnote{Recall that self-consistent PT based on the full mixed 2PI EA is presented in section~\ref{sec:diagMixed2PIEA} and definitely stood out among the diagrammatic techniques tested in chapter~\ref{chap:DiagTechniques}.}, although it would inevitably be more demanding to implement than the method based on~\eqref{eq:mixed2PIFRGgeneratingFunc1}. This can be done by considering the generating functional~\eqref{eq:mixed2PIFRGgeneratingFunc2} instead of~\eqref{eq:mixed2PIFRGgeneratingFunc1}. Therefore, extending the U-flow to the mixed representation could not only allow for monitoring the numerical weight of the 2PI-FRG procedure but also to improve the performances of the FRG approach itself by exploiting the Hubbard-Stratonovich field to grasp correlations before even starting the flow (via self-consistent PT) as well as throughout the flow. However, because of the index structure of the Yukawa interaction $U$ (in~\eqref{eq:mixed2PIFRGgeneratingFunc1}) or $\mathcal{U}$ (in~\eqref{eq:mixed2PIFRGgeneratingFunc2}), the derivation of the corresponding flow equations (following the lines set out in appendix~\ref{ann:2PIfrgFlowEquationUflow} for the original U-flow version of the 2PI-FRG) requires to invert matrices with components of the form:
\begin{equation}
\mathcal{M}_{\alpha_{1}\gamma_{2}} \;,
\end{equation}
for~\eqref{eq:mixed2PIFRGgeneratingFunc1}, with the bosonic index defined as usual as $\gamma\equiv(\alpha,\alpha')$, or:
\begin{equation}
\mathcal{M}_{\beta_{1}\gamma_{2}} \;,
\end{equation}
for~\eqref{eq:mixed2PIFRGgeneratingFunc2}, with $\gamma\equiv(\beta,\beta')$. The definitions of the corresponding inverse(s) do not straightforwardly follow from the bosonic index formalism as exploited so far. In the case of~\eqref{eq:mixed2PIFRGgeneratingFunc1} for instance, the definitions set by:
\begin{equation}
\int_{\alpha_{3}} \mathcal{M}_{\gamma_{1}\alpha_{3}} \mathcal{M}^{\mathrm{inv}}_{\alpha_{3}\gamma_{2}} = \mathcal{I}_{\gamma_{1}\gamma_{2}} \;,
\end{equation}
and
\begin{equation}
\frac{1}{2} \int_{\gamma_{3}} \mathcal{M}_{\alpha_{1}\gamma_{3}} \mathcal{M}^{\mathrm{inv}}_{\gamma_{3}\alpha_{2}} = \delta_{\alpha_{1}\alpha_{2}} \;,
\end{equation}
do not provide the right number of conditions to fix the components of $\mathcal{M}^{\mathrm{inv}}$ in an unambiguous manner. Hence, extending the mU-flow implementation of the 2PI-FRG in the framework of the mixed theory still requires a consequent work on the side of the formalism. The present discussion has put forward several appealing features of such a direction that we postpone to subsequent projects.

\subsubsection{2PI functional renormalization group CU-flow}
\label{sec:2PIFRGcuflow0DON}

As for the mU-flow, we have two options at our disposal to derive the CU-flow equations expressing $\dot{\overline{G}}_{\mathfrak{s}}$, $\Delta\dot{\overline{\Omega}}_{\mathfrak{s}}$ and $\dot{\overline{\Sigma}}_{\mathfrak{s}}$ for the studied $O(N)$ model: either we combine the corresponding C-flow and pU-flow equations already formulated for our (0+0)-D model in the present section~\ref{sec:2PIFRG0DON} by following the general recipe set out in the CU-flow discussion of section~\ref{sec:2PIFRGstateofplay} or we start our derivations from the results of the latter section (i.e. from~\eqref{eq:2PIfrgFlowEquationsCflowG},~\eqref{eq:2PIFRGCUflowExpressionDeltaOmegadot} and~\eqref{eq:2PIFRGCUflowExpressionSigmadot}) and simplify them by exploiting the $O(N)$ symmetry as we already did for the C-flow and U-flow approaches. In all cases, the obtained differential equations can be put in the form:
\begin{equation}
\dot{\overline{G}}_{\mathfrak{s}} = -\overline{G}_{\mathfrak{s}}^{2} \left( \dot{C}_{\mathfrak{s}}^{-1} - \dot{\overline{\Sigma}}_{\mathfrak{s}} \right) \;,
\label{eq:2PIFRGCUflowExpressionGdot0DON}
\end{equation}
\begin{equation}
\begin{split}
\scalebox{0.96}{${\displaystyle \Delta\dot{\overline{\Omega}}_{\mathfrak{s}} = }$} & \ \scalebox{0.96}{${\displaystyle \frac{N}{2} \dot{C}_{\mathfrak{s}}^{-1} \left( \overline{G}_{\mathfrak{s}} - C_{\mathfrak{s}} \right) }$} \\
& \scalebox{0.96}{${\displaystyle + \frac{\lambda}{72} \overline{G}_{\mathfrak{s}}^{2} \left(\sum_{a_{1},a_{2}=1}^{N}\left(\mathcal{I}+\overline{\Pi}_{\mathfrak{s}}\overline{\Phi}_{\mathfrak{s}}^{(2)}\right)_{(a_{1},a_{1})(a_{2},a_{2})}^{\mathrm{inv}} + 2\sum_{a_{1},a'_{1}=1}^{N}\left(\mathcal{I}+\overline{\Pi}_{\mathfrak{s}}\overline{\Phi}_{\mathfrak{s}}^{(2)}\right)_{(a_{1},a'_{1})(a_{1},a'_{1})}^{\mathrm{inv}} + N\left(2+N\right) \right) \;, }$}
\end{split}
\label{eq:2PIFRGCUflowExpressionDeltaOmegadot0DON}
\end{equation}
\begin{equation}
\begin{split}
\scalebox{0.999}{${\displaystyle \dot{\overline{\Sigma}}_{\mathfrak{s}} = }$} & \scalebox{0.999}{${\displaystyle - \frac{\lambda}{36} \overline{G}_{\mathfrak{s}} \sum_{a_{1},a_{2},a_{3}=1}^{N} \left(\mathcal{I}+\overline{\Pi}_{\mathfrak{s}}\overline{\Phi}_{\mathfrak{s}}^{(2)}\right)^{\mathrm{inv}}_{(1,a_{1})(a_{2},a_{2})} \left(\mathcal{I}+\overline{\Pi}_{\mathfrak{s}}\overline{\Phi}_{\mathfrak{s}}^{(2)}\right)^{\mathrm{inv}}_{(a_{3},a_{3})(a_{1},1)} }$} \\
& \scalebox{0.999}{${\displaystyle - \frac{\lambda}{18} \overline{G}_{\mathfrak{s}} \sum_{a_{1},a_{2},a'_{2}=1}^{N} \left(\mathcal{I}+\overline{\Pi}_{\mathfrak{s}}\overline{\Phi}_{\mathfrak{s}}^{(2)}\right)^{\mathrm{inv}}_{(1,a_{1})(a_{2},a'_{2})} \left(\mathcal{I}+\overline{\Pi}_{\mathfrak{s}}\overline{\Phi}_{\mathfrak{s}}^{(2)}\right)^{\mathrm{inv}}_{(a_{2},a'_{2})(a_{1},1)} }$} \\
& \scalebox{0.999}{${\displaystyle + \frac{\lambda}{288} \overline{G}_{\mathfrak{s}}^{4} \sum_{a_{1},a_{2},a_{3},a'_{3},a_{4},a'_{4}=1}^{N} \left(\mathcal{I}+\overline{\Pi}_{\mathfrak{s}}\overline{\Phi}_{\mathfrak{s}}^{(2)}\right)^{\mathrm{inv}}_{(a_{1},a_{1})(a_{3},a'_{3})} \overline{\Phi}^{(3)}_{\mathfrak{s},(1,1)(a_{3},a'_{3})(a_{4},a'_{4})} \left(\mathcal{I}+\overline{\Pi}_{\mathfrak{s}}\overline{\Phi}_{\mathfrak{s}}^{(2)}\right)^{\mathrm{inv}}_{(a_{4},a'_{4})(a_{2},a_{2})} }$} \\
& \scalebox{0.999}{${\displaystyle + \frac{\lambda}{144} \overline{G}_{\mathfrak{s}}^{4} \sum_{a_{1},a'_{1},a_{2},a'_{2},a_{3},a'_{3}=1}^{N} \left(\mathcal{I}+\overline{\Pi}_{\mathfrak{s}}\overline{\Phi}_{\mathfrak{s}}^{(2)}\right)^{\mathrm{inv}}_{(a_{1},a'_{1})(a_{2},a'_{2})} \overline{\Phi}^{(3)}_{\mathfrak{s},(1,1)(a_{2},a'_{2})(a_{3},a'_{3})} \left(\mathcal{I}+\overline{\Pi}_{\mathfrak{s}}\overline{\Phi}_{\mathfrak{s}}^{(2)}\right)^{\mathrm{inv}}_{(a_{3},a'_{3})(a_{1},a'_{1})} }$} \\
& \scalebox{0.999}{${\displaystyle -\frac{\lambda}{18} \overline{G}_{\mathfrak{s}} \left(N+2\right) -\frac{1}{2} \dot{\overline{G}}_{\mathfrak{s}} \sum_{a_{1}=1}^{N} \overline{\Phi}_{\mathfrak{s},(a_{1},a_{1})(1,1)}^{(2)} \;, }$}
\end{split}
\label{eq:2PIFRGCUflowExpressionSigmadot0DON}
\end{equation}
where the definition $\overline{\Sigma}_{\mathfrak{s}}\equiv\overline{\Sigma}_{\mathfrak{s},11}$ was also used in~\eqref{eq:2PIFRGCUflowExpressionSigmadot0DON} to set color indices equal to 1. In particular,~\eqref{eq:2PIFRGCUflowExpressionGdot0DON} to~\eqref{eq:2PIFRGCUflowExpressionSigmadot0DON} were derived from relations~\eqref{eq:diagonalC2PIFRGCflow0DON} to~\eqref{eq:diagonalSigma2PIFRGCflow0DON} implementing the conservation of the $O(N)$ symmetry during the flow. We also choose the same cutoff functions as those used previously for the C-flow and the U-flow, which are specified by~\eqref{eq:2PIFRGCflowCutoff20DON} (along with~\eqref{eq:2PIFRGCflowCutoff10DON}) and~\eqref{eq:choiceCutoffUs2PIFRGUflow0DON}. The CU-flow will notably be investigated up to the truncation order $N_{\mathrm{max}}=3$ at $N=1$, in which case~\eqref{eq:2PIFRGCUflowExpressionGdot0DON} to~\eqref{eq:2PIFRGCUflowExpressionSigmadot0DON} reduce to (see appendix~\ref{ann:2PIfrgpUandmUflow0DON1} for the corresponding flow equations expressing the derivative of the 2PI vertices of order 2 and 3 with respect to $\mathfrak{s}$):
\begin{equation}
\dot{\overline{G}}_{\mathfrak{s}} = -\overline{G}_{\mathfrak{s}}^{2} \left( \dot{C}_{\mathfrak{s}}^{-1} - \dot{\overline{\Sigma}}_{\mathfrak{s}} \right) \;,
\label{eq:2PIFRGCUflowExpressionGdot0DON1}
\end{equation}
\begin{equation}
\Delta\dot{\overline{\Omega}}_{\mathfrak{s}} = \frac{1}{2} \dot{C}_{\mathfrak{s}}^{-1} \left( \overline{G}_{\mathfrak{s}} - C_{\mathfrak{s}} \right) + \frac{\lambda}{24} \left( 4\left(2\overline{G}_{\mathfrak{s}}^{-2} + \overline{\Phi}_{\mathfrak{s}}^{(2)} \right)^{-1} + \overline{G}_{\mathfrak{s}}^{2} \right) \;,
\label{eq:2PIFRGCUflowExpressionDeltaOmegadot0DON1}
\end{equation}
\begin{equation}
\dot{\overline{\Sigma}}_{\mathfrak{s}} = \frac{\lambda}{3} \overline{G}_{\mathfrak{s}} \left( 2 + \overline{G}_{\mathfrak{s}}^{2} \overline{\Phi}_{\mathfrak{s}}^{(2)} \right)^{-2} \left( \frac{1}{2} \overline{G}_{\mathfrak{s}}^{3} \overline{\Phi}_{\mathfrak{s}}^{(3)} - 4 \right) - \frac{\lambda}{6} \overline{G}_{\mathfrak{s}} - \frac{1}{2} \dot{\overline{G}}_{\mathfrak{s}} \overline{\Phi}_{\mathfrak{s}}^{(2)} \;.
\label{eq:2PIFRGCUflowExpressionSigmadot0DON1}
\end{equation}
Finally, we recall that, in the framework of the CU-flow, all quantities calculated throughout the flow vanish at the starting point, i.e.:
\begin{equation}
\overline{G}_{\mathfrak{s}=\mathfrak{s}_{\mathrm{i}},a a'}=0 \mathrlap{\quad \forall a, a' \;,}
\label{eq:2PIfrgCUflowInitialCondGs0DON}
\end{equation}
\begin{equation}
\Delta \overline{\Omega}_{\mathfrak{s}=\mathfrak{s}_{\mathrm{i}}} = 0 \mathrlap{\;,}
\label{eq:2PIfrgCUflowInitialCondOmegas0DON}
\end{equation}
\begin{equation}
\overline{\Sigma}_{\mathfrak{s}=\mathfrak{s}_{\mathrm{i}},a a'} = 0 \mathrlap{\quad \forall a, a' \;,}
\label{eq:InitialConditions2PIFRGCUflowSigma0DON}
\end{equation}
\begin{equation}
\hspace{6.0cm} \overline{\Phi}_{\mathfrak{s} = \mathfrak{s}_{\mathrm{i}},(a_{1},a'_{1})\cdots(a_{n},a'_{n})}^{(n)} = 0 \quad \forall a_{1}, a'_{1},\cdots, a_{n}, a'_{n}, ~ \forall n \geq 2 \;,
\label{eq:InitialConditions2PIFRGCUflowPhin0DON}
\end{equation}
as was already indicated by~\eqref{eq:2PIfrgCUflowInitialCondGs} to~\eqref{eq:InitialConditions2PIFRGCUflowPhin}, and the associated infinite tower of differential equations is truncated by imposing~\eqref{eq:2PIfrgPhiBartCflow0DON}.

\vspace{0.5cm}

\begin{figure}[!htb]
\captionsetup[subfigure]{labelformat=empty}
  \begin{center}
    \subfloat[]{
      \includegraphics[width=0.7\linewidth]{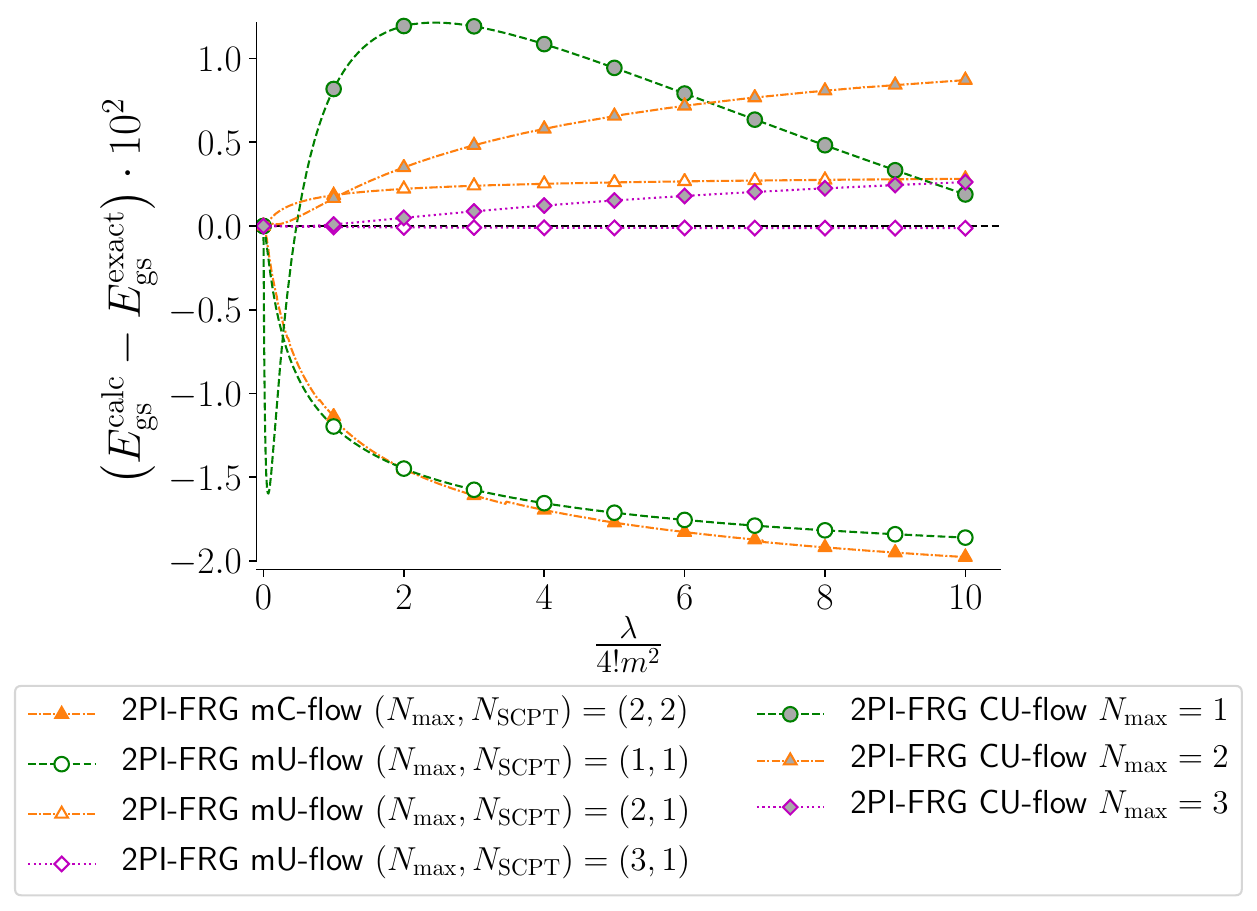}
                         }
   \\                     
    \subfloat[]{
      \includegraphics[width=0.7\linewidth]{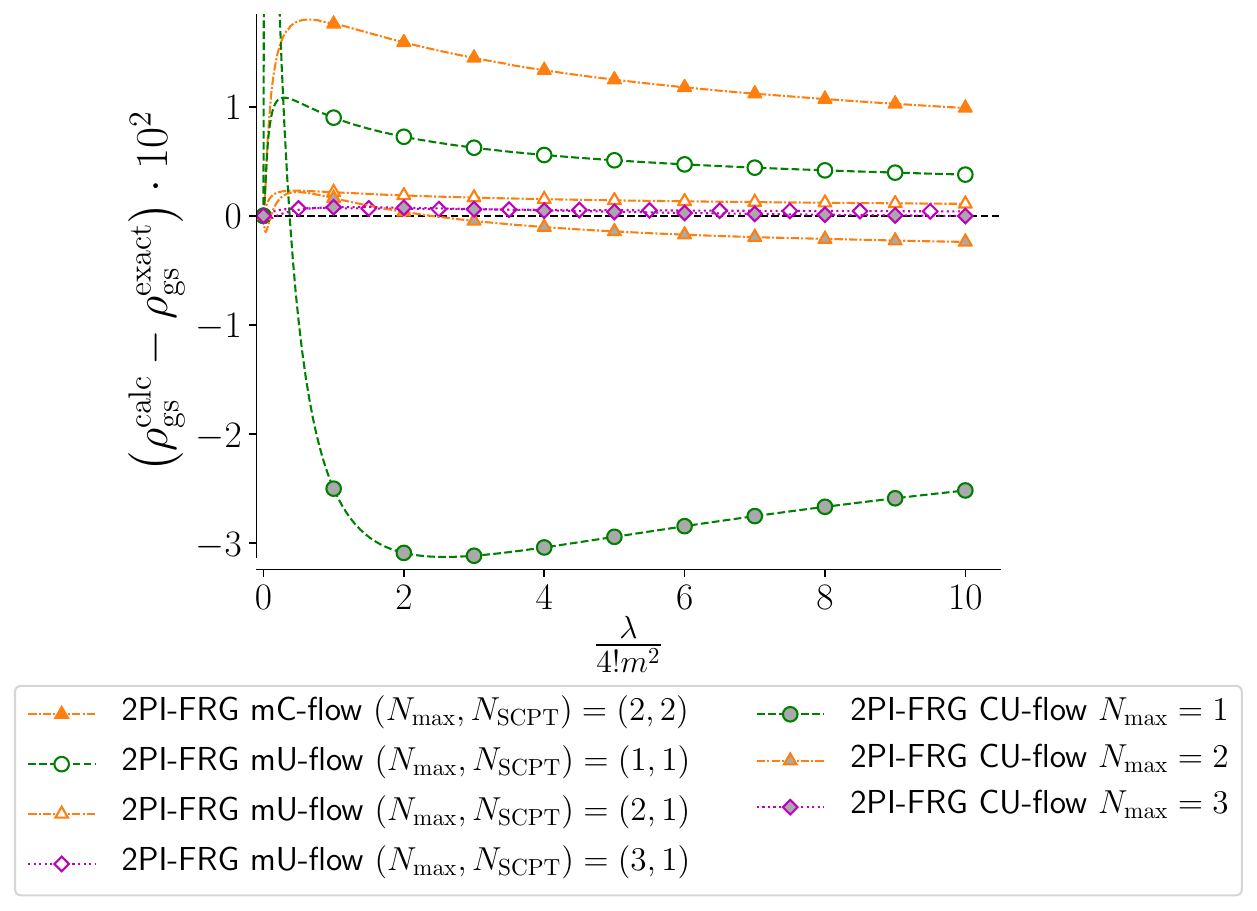}
                         }
\caption{Difference between the calculated gs energy $E_{\mathrm{gs}}^{\mathrm{calc}}$ or density $\rho_{\mathrm{gs}}^{\mathrm{calc}}$ and the corresponding exact solution $E_{\mathrm{gs}}^{\mathrm{exact}}$ or $\rho_{\mathrm{gs}}^{\mathrm{exact}}$ at $m^{2}=+1$ and $N=1$ ($\mathcal{R}e(\lambda)\geq 0$ and $\mathcal{I}m(\lambda)=0$). The mC-flow curve represents the best C-flow result up to $N_{\mathrm{max}}=3$ whereas the green, orange and purple mU-flow curves correspond to the best U-flow results obtained at $N_{\mathrm{max}}=1,2~\mathrm{and}~3$, respectively.}
\label{fig:2PIFRGCflowVsUflowVsCUflowlambdaN1}
  \end{center}
\end{figure}
\begin{figure}[!htb]
\captionsetup[subfigure]{labelformat=empty}
  \begin{center}
    \subfloat[]{
      \includegraphics[width=0.50\linewidth]{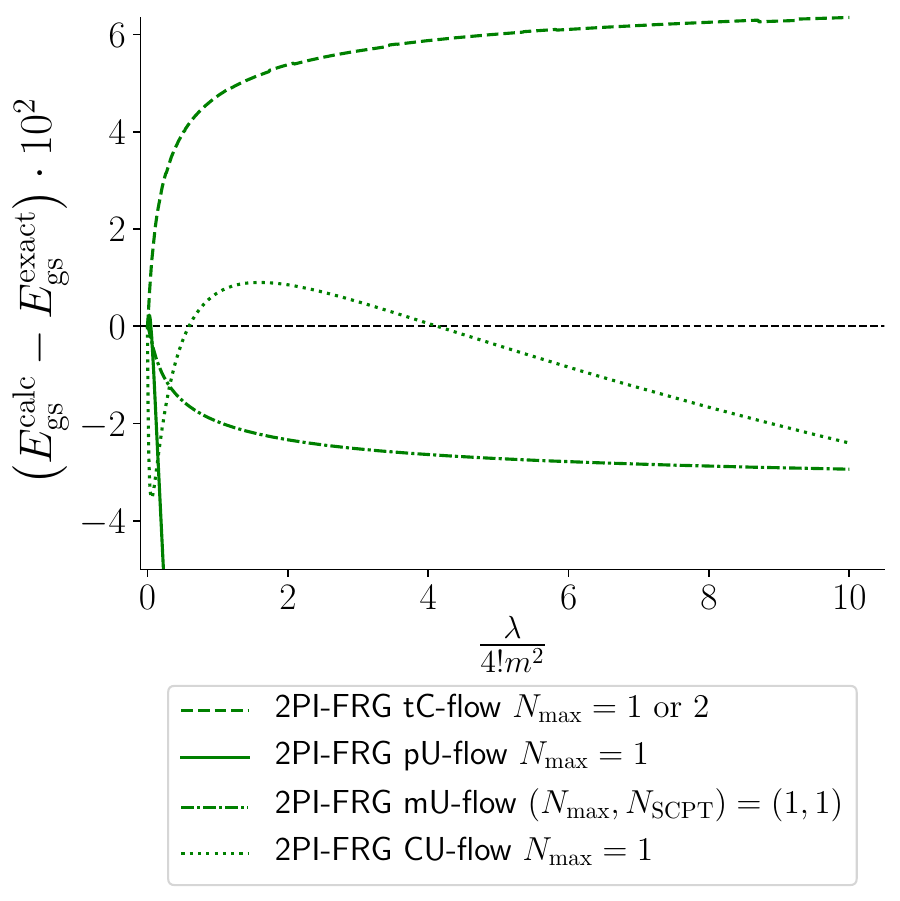}
                         }
    \subfloat[]{
      \includegraphics[width=0.50\linewidth]{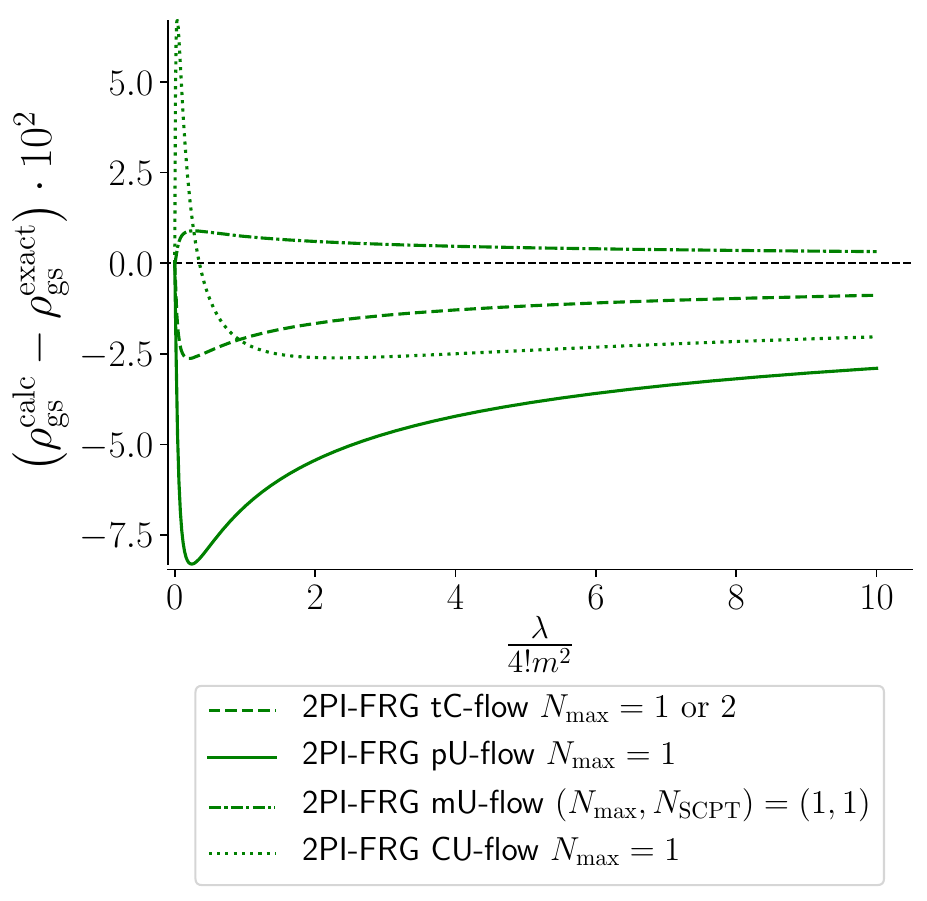}
                         }
\caption{Difference between the calculated gs energy $E_{\mathrm{gs}}^{\mathrm{calc}}$ or density $\rho_{\mathrm{gs}}^{\mathrm{calc}}$ and the corresponding exact solution $E_{\mathrm{gs}}^{\mathrm{exact}}$ or $\rho_{\mathrm{gs}}^{\mathrm{exact}}$ at $m^{2}=+1$ and $N=2$ ($\mathcal{R}e(\lambda)\geq 0$ and $\mathcal{I}m(\lambda)=0$). The tC-flow and mU-flow curves correspond respectively to the best C-flow and U-flow results obtained at $N_{\mathrm{max}}=1$.}
\label{fig:2PIFRGCflowVsUflowVsCUflowlambdaN2}
  \end{center}
\end{figure}

Hence, our CU-flow results at $N_{\mathrm{max}}=1$ are obtained by solving~\eqref{eq:2PIFRGCUflowExpressionGdot0DON} to~\eqref{eq:2PIFRGCUflowExpressionSigmadot0DON} (which reduce respectively to~\eqref{eq:2PIFRGCUflowExpressionGdot0DON1} to~\eqref{eq:2PIFRGCUflowExpressionSigmadot0DON1} at $N=1$), with the initial conditions~\eqref{eq:2PIfrgCUflowInitialCondGs0DON} to~\eqref{eq:InitialConditions2PIFRGCUflowPhin0DON} and the truncation condition~\eqref{eq:2PIfrgPhiBartCflow0DON}. The corresponding cutoff functions are still set by~\eqref{eq:2PIFRGCflowCutoff20DON} (with~\eqref{eq:2PIFRGCflowCutoff10DON}) for $C_{\mathfrak{s}}^{-1}$ and by~\eqref{eq:choiceCutoffUs2PIFRGUflow0DON} for $U_{\mathfrak{s}}$. In fig.~\ref{fig:2PIFRGCflowVsUflowVsCUflowlambdaN1} which shows the results thus obtained at $N=1$, the CU-flow estimates for $E_{\mathrm{gs}}$ and $\rho_{\mathrm{gs}}$ clearly outperform at $N_{\mathrm{max}}=2$ the mC-flow curve which can be considered as our best C-flow result (including those obtained in the mixed representation), as can be seen from fig.~\ref{fig:mixed2PIFRGCflowlambdaN1}. Furthermore, fig.~\ref{fig:2PIFRGCflowVsUflowVsCUflowlambdaN1} also shows that the CU-flow results are comparable to the best mU-flow ones (i.e. to the mU-flow results at $N_{\mathrm{SCPT}}=1$) up to $N_{\mathrm{max}}=3$ for both $E_{\mathrm{gs}}$ and $\rho_{\mathrm{gs}}$. These remarks also apply to the case where $N=2$ at $N_{\mathrm{max}}=1$, as can be checked from fig.~\ref{fig:2PIFRGCflowVsUflowVsCUflowlambdaN2}. Note however that, for the same reasons as the C-flow, the CU-flow is not suited to treat the regime with $m^{2}<0$.

\section{2PPI functional renormalization group}
\label{sec:2PPIFRG}
\subsection{State of play and general formalism}
\label{sec:2PPIFRGstateofplay}

The 2PPI-FRG was first developed in the early 2000s by Polonyi and Sailer~\cite{pol02} and discussed in the context of quantum electrodynamics (QED). Links between this approach and DFT were also emphasized in ref.~\cite{pol02}, as well as in ref.~\cite{sch04} which puts forward the 2PPI-FRG as a means for calculating properties of nuclear systems in a systematic manner. The first numerical application of this method came out almost a decade later with the work of Kemler and Braun~\cite{kem13}, who took as theoretical laboratories the (0+0)-D $\varphi^4$-theory in its unbroken-symmetry regime (i.e. the studied toy model with $N=1$ and $m^2>0$) and the (0+1)-D $\varphi^4$-theory, still in the phase without SSB. Corrections of the application to the latter toy model were pointed out subsequently by Rentrop and collaborators in ref.~\cite{ren15}. A few years later, an extension of the 2PPI-FRG formalism, coined as Kohn-Sham FRG (KS-FRG) due to its connection with the Kohn-Sham scheme, was developed by Liang, Niu and Hatsuda~\cite{lia18}. To our knowledge, the KS-FRG has only been applied to the toy model considered in this thesis with $N=1$ and in its unbroken-symmetry regime.

\pagebreak

The 2PPI-FRG practitioners have also managed to treat a (1+1)-D model~\cite{kem16,kem16bis,yok19,yok19bis,yok19bis3}, called the Alexandrou-Negele nuclei as a consequence of an earlier work of Alexandrou, Myczkowski and Negele on this model using a Monte Carlo approach~\cite{ale89}. Such a model reproduces some basic properties of the nuclear force (short-range repulsive and long-range attractive). As any other toy model, the Alexandrou-Negele nuclei have been used to benchmark different theoretical approaches (see ref.~\cite{jur09} for the similarity RG (SRG)) but they have also been exploited to describe real physical systems such as ultracold fermionic atoms interacting via a dipolar interaction~\cite{deu13}. For that reason in addition to the technical difficulties related to the inclusion of a space dimension, the work of Kemler and Braun presented in ref.~\cite{kem16} can be considered as a pioneering work for the 2PPI-FRG community. More specifically, it presents results obtained for the gs energies (in comparison with Monte Carlo results~\cite{ale89}), intrinsic densities and density correlation functions. Other subsequent applications were carried out for an infinite number of particles by Yokota and collaborators in order to study spinless nuclear matter with this model: this led to the determination of the nuclear saturation curve and other gs properties on the one hand~\cite{yok19} and to the calculation of spectral functions for the study of excited states on the other hand~\cite{yok19bis}. Note also that other 2PPI-FRG results for a (1+1)\nobreakdash-D fermionic system are reported in ref.~\cite{ram17}.

\vspace{0.5cm}

Then, applications to higher-dimensional systems were performed recently by Yokota and collaborators on a (2+1)-D homogeneous electron gas~\cite{yok19bis2,yok19bis3} and on a (3+1)-D homogeneous electron gas~\cite{yok21}, thus achieving the first two-dimensional and three-dimensional applications of the 2PPI-FRG. Note also the work of ref.~\cite{yok21bis} which designs a 2PPI-FRG approach to describe classical liquids. This paper shows through an application to a (1+0)-D toy model that this novel approach compares favorably with more conventional methods based on integral equations. Finally, the 2PPI-FRG formalism has also been generalized to treat superfluid systems~\cite{yok20}, thus marking a significant step towards the description of systems with competing instabilities. The resulting approach can actually be considered as a DFT for systems with pairing correlations and echoes the work of Furnstahl and collaborators with the 2PPI EA and the IM~\cite{fur07}.

\vspace{0.5cm}

As a first step, we now outline the basic ingredients of the 2PPI-FRG formalism in order to prepare the ground for our toy model applications. The generating functional underlying this FRG approach is given by\footnote{\label{footnote:notationZKWK2PPIFRG}Although the notations for the generating functionals $Z[K]$ and $W[K]$ are identical to those of~\eqref{eq:2PIFRGgeneratingFunc} related to the 2PI-FRG, we stress that $Z[K]$ and $W[K]$ are always defined by~\eqref{eq:GeneratingFunctional2PPIFRG} in the whole section~\ref{sec:2PPIFRG} and in corresponding appendices.}:
\begin{equation}
Z[K] = e^{W[K]} = \int \mathcal{D}\widetilde{\psi}^{\dagger}\mathcal{D}\widetilde{\psi} \ e^{-S\big[\widetilde{\psi}^{\dagger},\widetilde{\psi}\big] + \int_{\alpha} K_{\alpha}\widetilde{\psi}_{\alpha}^{\dagger}\widetilde{\psi}_{\alpha}} \;,
\label{eq:GeneratingFunctional2PPIFRG}
\end{equation}
where we now consider a complex field $\widetilde{\psi}$ which is either bosonic or fermionic. The equations underlying the 2PPI-FRG are barely modified if $\widetilde{\psi}$ is a real field (which is the case of interest for our toy model study), besides a few numerical factors. We choose to keep the present discussion to the framework of a complex field $\widetilde{\psi}$ and postpone the discussion of such details to section~\ref{sec:2PPIFRG0DON}. The different configurations of $\widetilde{\psi}$ are now specified by an $\alpha$-index which is essentially the same as that used in section~\ref{sec:2PIFRGstateofplay} in our presentation of the 2PI-FRG, with the exception of the charge index $c$. The latter is indeed no longer necessary as we distinguish explicitly $\widetilde{\psi}$ from its complex conjugate $\widetilde{\psi}^{\dagger}$ in the present formalism. Therefore, the following conventions are used in~\eqref{eq:GeneratingFunctional2PPIFRG} and will hold in the rest of section~\ref{sec:2PPIFRGstateofplay}:
\begin{equation}
\int_{\alpha}\equiv\sum_{a} \int_{x} \equiv\sum_{a} \sum_{m_{s}} \int^{\beta}_{0} d\tau \int d^{D-1}\boldsymbol{r} \;,
\end{equation}
with $\alpha\equiv (a,x)$ and $x\equiv (\boldsymbol{r}, \tau, m_{s})$ ($a$, $\boldsymbol{r}$, $\tau$ and $m_{s}$ being defined between~\eqref{eq:2PIFRGgeneratingFunc} and~\eqref{eq:chargeIndexDefinition2PIFRG}). The connected correlation functions deduced from~\eqref{eq:GeneratingFunctional2PPIFRG} are:
\begin{equation}
W_{\alpha_{1}\cdots\alpha_{n}}^{(n)}[K] \equiv \frac{\delta^{n} W[K]}{\delta K_{\alpha_{1}} \cdots \delta K_{\alpha_{n}}} = \left\langle \widetilde{\psi}^{\dagger}_{\alpha_{1}} \widetilde{\psi}_{\alpha_{1}} \cdots \widetilde{\psi}^{\dagger}_{\alpha_{n}} \widetilde{\psi}_{\alpha_{n}} \right\rangle_{K} \;,
\label{eq:DefinitionWn2PPIFRG}
\end{equation}
which yields the density:
\begin{equation}
\rho_{\alpha} = W_{\alpha}^{(1)}[K] = \left\langle \widetilde{\psi}^{\dagger}_{\alpha} \widetilde{\psi}_{\alpha} \right\rangle_{K} \;,
\label{eq:rho2PPIFRG}
\end{equation}
at $n=1$, where we have just introduced the expectation value\footnote{Following up the remark of footnote~\ref{footnote:notationZKWK2PPIFRG}, the expectation value $\big\langle \cdots \big\rangle_{K}$ is given by~\eqref{eq:2PPIFRGKdependentExpectationValue} and not by~\eqref{eq:2PIFRGKdependentExpectationValue} in all 2PPI-FRG discussions.}:
\begin{equation}
\big\langle \cdots \big\rangle_{K} = \frac{1}{Z[K]} \int \mathcal{D}\widetilde{\psi}^{\dagger} \mathcal{D}\widetilde{\psi} \ \cdots \ e^{-S\big[\widetilde{\psi}^{\dagger},\widetilde{\psi}\big] + \int_{\alpha} K_{\alpha}\widetilde{\psi}_{\alpha}^{\dagger}\widetilde{\psi}_{\alpha}} \;.
\label{eq:2PPIFRGKdependentExpectationValue}
\end{equation}
As opposed to the 2PI-FRG, the source $K$ is now local and does not exhibit antisymmetric properties, even if $\widetilde{\psi}$ is a Grassmann field. This also translates into simpler symmetry properties for the underlying correlation functions, with e.g. for those of~\eqref{eq:DefinitionWn2PPIFRG}:
\begin{equation}
W^{(n)}_{\alpha_{1}\cdots\alpha_{n}}[K]=W^{(n)}_{\alpha_{P(1)}\cdots\alpha_{P(n)}}[K] \;,
\end{equation}
to be compared with~\eqref{eq:SymmetryW2PIFRGUp} and~\eqref{eq:SymmetryW2PIFRGDown} for the 2PI-FRG formalism. The following 2PPI EA is at the heart of the present approach:
\begin{equation}
\begin{split}
\Gamma^{(\mathrm{2PPI})}[\rho] = & -W[K] + \int_{\alpha} K_{\alpha} \frac{\delta W[K]}{\delta K_{\alpha}} \\
= & -W[K] + \int_{\alpha} K_{\alpha} \rho_{\alpha} \;,
\end{split}
\label{eq:LegendreTransform2PPIeffActionFor2PPIFRG}
\end{equation}
where the second line follows from~\eqref{eq:rho2PPIFRG}.

\vspace{0.5cm}

All implementations of the 2PPI-FRG treated below are applicable to any system whose classical action can be written as:
\begin{equation}
S\Big[\widetilde{\psi}^{\dagger},\widetilde{\psi}\Big] = S_{0}\Big[\widetilde{\psi}^{\dagger},\widetilde{\psi}\Big] + S_{\mathrm{int}}\Big[\widetilde{\psi}^{\dagger},\widetilde{\psi}\Big] = \int_{\alpha} \widetilde{\psi}_{\alpha}^{\dagger}\left(\hat{O}_{\mathrm{kin},\alpha} + V_{\alpha} - \mu\right)\widetilde{\psi}_{\alpha} + \frac{1}{2}\int_{\alpha_{1},\alpha_{2}} \widetilde{\psi}_{\alpha_{1}}^{\dagger}\widetilde{\psi}_{\alpha_{2}}^{\dagger} U_{\alpha_{1}\alpha_{2}} \widetilde{\psi}_{\alpha_{2}}\widetilde{\psi}_{\alpha_{1}} \;,
\label{eq:ClassicalAction2PPIFRG}
\end{equation}
where $\hat{\mathcal{O}}_{\mathrm{kin}}$ contains the kinetic operator\footnote{For instance, $\hat{\mathcal{O}}_{\mathrm{kin},\alpha}=\partial_{\tau}-\frac{\boldsymbol{\nabla}^{2}}{2m}$ for a non-relativistic system of mass $m$.} and we place ourselves in the grand canonical ensemble here by using a chemical potential $\mu$ to monitor the particle number\footnote{Alternatively, one can also simply impose a given particle number at the initial conditions of the 2PPI-FRG procedure since the particle number is conserved during the flow, as shown in ref.~\cite{kem16}.}. In the framework of the 2PPI-FRG, it is typically the one-body potential $V$ and the two-body interaction $U$ which are rendered flow-dependent by introducing cutoff functions, i.e. the Schwinger functional $W[K]$ and the 2PPI EA $\Gamma^{(\mathrm{2PPI})}[\rho]$ become dependent on $\mathfrak{s}$ via the substitution $U \rightarrow U_{\mathfrak{s}}$ sometimes combined with $V \rightarrow V_{\mathfrak{s}}$. After doing so, we exploit in particular the following convention:
\begin{equation}
\rho_{\alpha} \equiv \rho_{\mathfrak{s},\alpha}[K] = W^{(1)}_{\mathfrak{s},\alpha}[K] \;,
\label{eq:rhoEqualrhok2PPIFRG}
\end{equation}
which is the counterpart of~\eqref{eq:PhiEqualPhik1PIFRG} and~\eqref{eq:GEqualGk2PIFRG} used respectively for the 1PI-FRG and the 2PI-FRG. Moreover, although the classical action~\eqref{eq:ClassicalAction2PPIFRG} only contains a two-body interaction, the present FRG method can be straightforwardly extended to treat three-body and even higher-body interactions by including higher powers of $\widetilde{\psi}^{\dagger}\widetilde{\psi}$ into~\eqref{eq:ClassicalAction2PPIFRG}, similarly to~\eqref{eq:2PIFRGmostgeneralactionS} for the 2PI-FRG.

\vspace{0.5cm}

As a side comment, we want to point out that any theoretical approach based on the generating functional~\eqref{eq:GeneratingFunctional2PPIFRG} is not relevant to describe superfluid systems, as opposed to~\eqref{eq:2PIFRGgeneratingFunc} for the 2PI-FRG. In that respect, we would indeed need to calculate the expectation values of $\widetilde{\psi}\widetilde{\psi}$ and $\widetilde{\psi}^{\dagger}\widetilde{\psi}^{\dagger}$, and the generating functional~\eqref{eq:GeneratingFunctional2PPIFRG} is not suited to achieve this. Following the steps of ref.~\cite{yok20}, we can generalize~\eqref{eq:GeneratingFunctional2PPIFRG} as follows in that purpose:
\begin{equation}
\scalebox{0.89}{${\displaystyle Z\Big[K^{(\rho)},K^{(\kappa)},\left(K^{(\kappa)}\right)^{\dagger}\Big] = e^{W\big[K^{(\rho)},K^{(\kappa)},\left(K^{(\kappa)}\right)^{\dagger}\big]} = \int \mathcal{D}\widetilde{\psi}^{\dagger}\mathcal{D}\widetilde{\psi} \ e^{-S\big[\widetilde{\psi}^{\dagger},\widetilde{\psi}\big] + \int_{\alpha} K^{(\rho)}_{\alpha}\widetilde{\psi}_{\alpha}^{\dagger}\widetilde{\psi}_{\alpha} + \int_{\alpha} K^{(\kappa)}_{\alpha}\widetilde{\psi}_{\alpha}\widetilde{\psi}_{\alpha} + \int_{\alpha} \left(K^{(\kappa)}\right)^{\dagger}_{\alpha}\widetilde{\psi}_{\alpha}^{\dagger}\widetilde{\psi}_{\alpha}^{\dagger}} \;,}$}
\label{eq:GeneratingFunctional2PPIFRGpairing}
\end{equation}
where the anomalous densities can now be accessed via:
\begin{equation}
\frac{\delta W[K]}{\delta K^{(\kappa)}_{\alpha}} = \left\langle \widetilde{\psi}_{\alpha}\widetilde{\psi}_{\alpha} \right\rangle_{K} \;,
\end{equation}
\begin{equation}
\frac{\delta W[K]}{\delta \left(K^{(\kappa)}\right)^{\dagger}_{\alpha}} = \left\langle \widetilde{\psi}^{\dagger}_{\alpha}\widetilde{\psi}^{\dagger}_{\alpha} \right\rangle_{K} \;.
\end{equation}
Such an extension is not necessary for the 2PI-FRG as presented previously since some components of the propagator $G_{\alpha\alpha'}$ already coincide with expectation values of $\widetilde{\psi}\widetilde{\psi}$ and $\widetilde{\psi}^{\dagger}\widetilde{\psi}^{\dagger}$ (for charge indices satisfying $c=c'=+$ or $c=c'=-$ according to~\eqref{eq:definitionG2PIFRG}). There are also some subtleties underpinning the implementation of a 2PPI-FRG procedure from the generating functional~\eqref{eq:GeneratingFunctional2PPIFRGpairing}. This is beyond the scope of the present discussion but we refer to the work of Yokota and collaborators~\cite{yok20} for more details on this topic. Although this formulation of the 2PPI-FRG is suited to describe SSBs (related to superfluid systems at least), it should be noted that it does not carry out the momentum-shell integration \`{a} la Wilson, and neither do all of the 2PPI-FRG approaches treated thereafter\footnote{To be more specific, none of the 2PPI-FRG implementations exploited so far, i.e. the approach of Polonyi and Sailer~\cite{pol02} and the KS-FRG of Liang and collaborators~\cite{lia18}, implements the Wilsonian momentum-shell integration (since they rely on a dimensionless flow parameter), although this is in principle possible via suitable choices of cutoff functions, as for the 2PI-FRG.}. Our main point here is that, unlike the 2PI-FRG considered in section~\ref{sec:2PIFRG}, the 2PPI-FRG formalism exploited in this thesis is not adapted to tackle superfluid systems, although the necessary extensions to achieve this (which are irrelevant for our toy model study) are well established~\cite{yok20}.

\subsubsection{Standard 2PPI functional renormalization group}
\label{sec:standard2PPIFRG}
\paragraph{Main features:}

We start by discussing the standard version of the 2PPI-FRG, as proposed by Polonyi and Sailer~\cite{pol02} and then exploited e.g. by Kemler and Braun~\cite{kem13,kem16}. After performing the substitutions $V \rightarrow V_{\mathfrak{s}}$ and $U \rightarrow U_{\mathfrak{s}}$ into the classical action~\eqref{eq:ClassicalAction2PPIFRG}, we can show that the corresponding flow-dependent 2PPI EA $\Gamma_{\mathfrak{s}}^{(\mathrm{2PPI})}[\rho]$, originally expressed via~\eqref{eq:LegendreTransform2PPIeffActionFor2PPIFRG}, satisfies the master equation (see appendix~\ref{sec:DerivMasterEq2PPIFRG}):
\begin{equation}
\dot{\Gamma}^{(\mathrm{2PPI})}_{\mathfrak{s}}[\rho] = \int_{\alpha} \dot{V}_{\mathfrak{s},\alpha} \rho_{\alpha} + \frac{1}{2} \mathrm{STr}\left[\dot{U}_{\mathfrak{s}} \left(\Gamma_{\mathfrak{s}}^{(\mathrm{2PPI})(2)}[\rho]\right)^{-1} \right] + \frac{1}{2} \int_{\alpha_{1},\alpha_{2}} \dot{U}_{\mathfrak{s},\alpha_{1}\alpha_{2}} \rho_{\alpha_{1}} \rho_{\alpha_{2}} \;,
\label{eq:2PPIFRGflowEquationGamma}
\end{equation}
where we just introduced back the notation $\mathrm{STr}$ to indicate the trace with respect to $\alpha$-indices\footnote{We recall that $\mathrm{Tr}_{\alpha}$ was introduced in section~\ref{sec:2PIFRG} on the 2PI-FRG to denote the trace with respect to $\alpha$-indices. The underlying reason was specific to the 2PI-FRG formalism as we stressed in this way the difference with the trace taken with respect to bosonic indices (i.e. $\mathrm{Tr}_{\gamma}$).} and the 2PPI vertices satisfy:
\begin{equation}
\Gamma_{\mathfrak{s},\alpha_{1}\cdots\alpha_{n}}^{(\mathrm{2PPI})(n)}[\rho] \equiv \frac{\delta^{n}\Gamma_{\mathfrak{s}}^{(\mathrm{2PPI})}[\rho]}{\delta\rho_{\alpha_{1}} \cdots \delta\rho_{\alpha_{n}}} \;.
\end{equation}
Moreover, it can also be rather advantageous to consider the flow equation for the Schwinger functional instead of that of the corresponding EA (see appendix~\ref{sec:DerivMasterEq2PPIFRG}):
\begin{equation}
\dot{W}_{\mathfrak{s}}[K] = -\int_{\alpha} \dot{V}_{\mathfrak{s},\alpha} W^{(1)}_{\mathfrak{s},\alpha}[K] - \frac{1}{2} \int_{\alpha_{1},\alpha_{2}} \dot{U}_{\mathfrak{s},\alpha_{1}\alpha_{2}} W_{\mathfrak{s},\alpha_{1} \alpha_{2}}^{(2)}[K] - \frac{1}{2} \int_{\alpha_{1},\alpha_{2}} \dot{U}_{\mathfrak{s},\alpha_{1}\alpha_{2}} W^{(1)}_{\mathfrak{s},\alpha_{1}}[K] W^{(1)}_{\mathfrak{s},\alpha_{2}}[K] \;,
\label{eq:FlowequationWs2PPIFRG}
\end{equation}
as was done in ref.~\cite{kem16}. Although~\eqref{eq:2PPIFRGflowEquationGamma} and~\eqref{eq:FlowequationWs2PPIFRG} are fully equivalent (i.e. they can be deduced from each other without approximation), the initial conditions might be much less cumbersome to determine in the latter case, as explained in more detail below.

\vspace{0.5cm}

In practice, the exact flow equation~\eqref{eq:2PPIFRGflowEquationGamma} is turned into an infinite tower of differential equations with a vertex expansion of $\Gamma_{\mathfrak{s}}^{(\mathrm{2PPI})}[\rho]$:
\begin{equation}
\Gamma^{(\mathrm{2PPI})}_{\mathfrak{s}}[\rho] = \overline{\Gamma}^{(\mathrm{2PPI})}_{\mathfrak{s}} + \sum_{n=2}^{\infty} \frac{1}{n!} \int_{\alpha_{1},\cdots,\alpha_{n}} \overline{\Gamma}_{\mathfrak{s},\alpha_{1}\cdots\alpha_{n}}^{(\mathrm{2PPI})(n)} \left(\rho-\overline{\rho}_{\mathfrak{s}}\right)_{\alpha_{1}} \cdots \left(\rho-\overline{\rho}_{\mathfrak{s}}\right)_{\alpha_{n}} \;,
\label{eq:vertexExpansion2PPIFRG}
\end{equation}
where $\overline{\Gamma}^{(\mathrm{2PPI})}_{\mathfrak{s}} \equiv \Gamma^{(\mathrm{2PPI})}_{\mathfrak{s}}[\rho=\overline{\rho}_{\mathfrak{s}}]$, $\overline{\Gamma}^{(\mathrm{2PPI})(n)}_{\mathfrak{s},\alpha_{1}\cdots\alpha_{n}} \equiv \left.\frac{\delta^{n}\Gamma_{\mathfrak{s}}^{(\mathrm{2PPI})}[\rho]}{\delta \rho_{\alpha_{1}}\cdots\delta \rho_{\alpha_{n}}}\right|_{\rho=\overline{\rho}_{\mathfrak{s}}}$ and the flowing density $\overline{\rho}_{\mathfrak{s}}$ extremizes the flowing 2PPI EA:
\begin{equation}
\left. \frac{\delta\Gamma_{\mathfrak{s}}^{(\mathrm{2PPI})}[\rho]}{\delta \rho_{\alpha}}\right|_{\rho=\overline{\rho}_{\mathfrak{s}}} = 0 \mathrlap{\quad \forall \alpha,\mathfrak{s} \;,}
\label{eq:2PPIfrgExtremizeGamma2PPI}
\end{equation}
in the same manner as~\eqref{eq:ExtremizeTruncationVertexExp1PIFRG} for the 1PI-FRG and~\eqref{eq:2PIfrgExtremizeGamma2PI} for the 2PI-FRG. The flow equation~\eqref{eq:FlowequationWs2PPIFRG} is treated in the same spirit by Taylor expanding $W_{\mathfrak{s}}[K]$ around the configuration where the source $K$ vanishes, i.e.:
\begin{equation}
W_{\mathfrak{s}}[K] = \overline{W}_{\mathfrak{s}} + \sum_{n=1}^{\infty} \int_{\alpha_{1},\cdots,\alpha_{n}} \overline{W}_{\mathfrak{s},\alpha_{1}\cdots\alpha_{n}}^{(n)} K_{\alpha_{1}} \cdots K_{\alpha_{n}} \;,
\label{eq:WExpansion2PPIFRG}
\end{equation}
with
\begin{equation}
\overline{W}_{\mathfrak{s},\alpha}^{(1)} = \overline{\rho}_{\mathfrak{s},\alpha} \;,
\end{equation}
according to~\eqref{eq:rhoEqualrhok2PPIFRG}. Finally, the expansion of the 2PPI EA given by~\eqref{eq:vertexExpansion2PPIFRG} is inserted into the corresponding master equation and the terms of identical powers of $\rho-\overline{\rho}_{\mathfrak{s}}$ are then identified to deduce an infinite set of differential equations for $\overline{\rho}_{\mathfrak{s}}$ and the 2PPI vertices $\overline{\Gamma}^{(\mathrm{2PPI})(n)}_{\mathfrak{s}}$. The same is done for the expansion~\eqref{eq:WExpansion2PPIFRG} of $W_{\mathfrak{s}}[K]$ except that the identification is performed with respect to the powers of the source $K$.

\vspace{0.5cm}

Regarding the boundary conditions, $U_{\mathfrak{s}}$ must satisfy the same relations as those encountered in the U-flow implementation of the 2PI-FRG, i.e.:
\begin{subequations}
\begin{empheq}[left=\empheqlbrace]{align}
& U_{\mathfrak{s}=\mathfrak{s}_{\mathrm{i}},\alpha_{1} \alpha_{2}} = 0 \quad \forall \alpha_{1}, \alpha_{2} \;. \label{eq:2PPIfrgUflowBoundaryCondUpper}\\
\nonumber \\
& U_{\mathfrak{s}=\mathfrak{s}_{\mathrm{f}}} = U \;. \label{eq:2PPIfrgUflowBoundaryCondBottom}
\end{empheq}
\end{subequations}
However, the initial condition~\eqref{eq:2PPIfrgUflowBoundaryCondUpper} does not always imply that the starting point of the flow coincides with the free theory. This actually depends on the analytical form chosen for the one-body potential $V_{\mathfrak{s}}$, and more specifically on the initial condition $V_{\mathfrak{s}=\mathfrak{s}_{\mathrm{i}}}$ ($V_{\mathfrak{s}=\mathfrak{s}_{\mathrm{f}}}=V$ is always imposed to recover~\eqref{eq:ClassicalAction2PPIFRG} at the end of the flow). In that respect, we can mainly distinguish two situations:
\begin{itemize}
\item $V_{\mathfrak{s}}=V$ $\forall \mathfrak{s}$:\\
A possibility consists in choosing $V_{\mathfrak{s}}$ independent from $\mathfrak{s}$, i.e. $V_{\mathfrak{s}}=V ~ \forall \mathfrak{s}$. As a result, the two-body interaction $U_{\mathfrak{s}}$ is gradually turned on during the flow while $V_{\mathfrak{s}}$ is kept equal to its initial value $V_{\mathfrak{s}=\mathfrak{s}_{\mathrm{i}}}=V$. Hence, this approach can be qualified as a \textbf{U-flow} scheme. However, it should be noted that the 2PPI-FRG based on classical action~\eqref{eq:ClassicalAction2PPIFRG} also relies in principle on a flow equation for the chemical potential $\mu$ (which will be discarded in our forthcoming discussions since it will not be used for our (0+0)\nobreakdash-D applications) and the flow of the chemical potential is essentially a C-flow contribution as $\mu$ dresses the (inverse) free propagator of the theory (as can be seen from~\eqref{eq:ClassicalAction2PPIFRG} or~\eqref{eq:GKminus1OkinVmu2PPIFRG}). In other words, even when one imposes that $V_{\mathfrak{s}}=V$ $\forall \mathfrak{s}$, the evolution of the chemical potential throughout the flow can also be considered as a flow-dependent shift to the one-body potential $V$, and therefore as a renormalization of $V$.

\item $V_{\mathfrak{s}}=(1-\mathfrak{s})V_{\mathrm{KS}}$ $\forall \mathfrak{s}$ (with $\mathfrak{s}_{\mathfrak{i}}=0$ and $\mathfrak{s}_{\mathfrak{f}}=1$):\\
Self-bound systems such as nuclei are by definition not subjected to an external potential such that the condition $V_{\mathfrak{s}=\mathfrak{s}_{\mathrm{f}}=1,\alpha}=0$ $\forall \alpha$ must be fulfilled. In this case, a natural option consists in starting the flow at the Kohn-Sham system with the choice $V_{\mathfrak{s}}=(1-\mathfrak{s})V_{\mathrm{KS}}$ such that $V_{\mathfrak{s}=\mathfrak{s}_{\mathrm{i}}=0}=V_{\mathrm{KS}}$, with $V_{\mathrm{KS}}$ being the solution of the Kohn-Sham equation for the optimal one-body potential~\cite{sch04}. This induces a particularly elegant structure of the flow: the two-body interaction $U_{\mathfrak{s}}$ is gradually turned on and the mean-field $V_{\mathfrak{s}}$ gradually switched off during the flow until the Kohn-Sham system at $\mathfrak{s}=\mathfrak{s}_{\mathrm{i}}=0$ is transformed into the fully self-bound system at $\mathfrak{s}=\mathfrak{s}_{\mathrm{f}}=1$. We implement in this way a \textbf{CU-flow}. It should also be pointed out that, even in the situation where $V_{\mathfrak{s}}=V$ $\forall \mathfrak{s}$, it is possible to incorporate information on $V_{\mathrm{KS}}$ already at the starting point of the 2PPI-FRG flow by adjusting the chemical potential so that it coincides with that of the Kohn-Sham system at $\mathfrak{s}=\mathfrak{s}_{\mathrm{i}}$, as is done e.g. in ref.~\cite{yok19bis2}.

\end{itemize}

\vspace{0.3cm}

In what follows, we will focus on the U-flow implementation of the standard 2PPI-FRG, which is ubiquitous in the 2PPI-FRG studies mentioned previously. However, the forthcoming discussion remains valid regardless of the analytical form chosen for $U_{\mathfrak{s}}$. As for the 2PI-FRG, the cutoff function $R_{\mathfrak{s}}$ can be introduced equivalently via $U_{\mathfrak{s}}=R_{\mathfrak{s}}U$ or $U_{\mathfrak{s}}=U+R_{\mathfrak{s}}$, although the former remains the most common choice especially with $U_{\mathfrak{s}}=\mathfrak{s}U$. Hence, in the context of the U-flow (i.e. if $\dot{V}_{\mathfrak{s}}=0$ $\forall \mathfrak{s}$), the master equation for $\Gamma^{(\mathrm{2PPI})}_{\mathfrak{s}}[\rho]$ given by~\eqref{eq:2PPIFRGflowEquationGamma} reduces to:
\begin{equation}
\dot{\Gamma}^{(\mathrm{2PPI})}_{\mathfrak{s}}[\rho] = \frac{1}{2} \mathrm{STr}\left[\dot{U}_{\mathfrak{s}} \left(\Gamma_{\mathfrak{s}}^{(\mathrm{2PPI})(2)}[\rho]\right)^{-1} \right] + \frac{1}{2} \int_{\alpha_{1},\alpha_{2}} \dot{U}_{\mathfrak{s},\alpha_{1}\alpha_{2}} \rho_{\alpha_{1}} \rho_{\alpha_{2}} \;,
\label{eq:2PPIFRGflowEquationGammaUflow}
\end{equation}
and the starting point of the flow coincides with the free theory\footnote{To be precise, the starting point of what we define here as the U-flow (i.e. the flow scheme with $V_{\mathfrak{s}}=V$ $\forall \mathfrak{s}$) is not necessarily fully uncorrelated. Indeed, as mentioned earlier, the chemical potential can still be chosen to coincide with that of the Kohn-Sham system at $\mathfrak{s}=\mathfrak{s}_{\mathrm{i}}$, thus including information about the interaction already at the starting point of the flow. However, we stress again that we discard the flow of the chemical potential in the present discussion.}, as we now discuss in further details.

\paragraph{Initial conditions:}

The free theory is specified by retaining only the quadratic part of the classical action $S$, i.e. $S_{0}$, in the generating functional~\eqref{eq:GeneratingFunctional2PPIFRG}, thus obtaining\footnote{To clarify, we have used the relation $S_{0}\Big[\widetilde{\psi}^{\dagger},\widetilde{\psi}\Big] - \int_{\alpha} K_{\alpha} \widetilde{\psi}^{\dagger}_{\alpha} \widetilde{\psi}_{\alpha}=\int_{\alpha} \widetilde{\psi}_{\alpha}^{\dagger}\left(\hat{O}_{\mathrm{kin},\alpha} + V_{\alpha} - \mu\right)\widetilde{\psi}_{\alpha} - \int_{\alpha} K_{\alpha} \widetilde{\psi}^{\dagger}_{\alpha} \widetilde{\psi}_{\alpha} = \int_{\alpha_{1},\alpha_{2}} \widetilde{\psi}_{\alpha_{1}} G_{K,\alpha_{1}\alpha_{2}}^{-1} \widetilde{\psi}_{\alpha_{2}}$ in order to carry out Gaussian integration.}:
\begin{equation}
\begin{split}
Z_{0}[K] = e^{W_{0}[K]} = \int \mathcal{D}\widetilde{\psi}^{\dagger}\mathcal{D}\widetilde{\psi} \ e^{-S_{0}\big[\widetilde{\psi}^{\dagger},\widetilde{\psi}\big] + \int_{\alpha} K_{\alpha}\widetilde{\psi}_{\alpha}^{\dagger}\widetilde{\psi}_{\alpha}} = e^{\zeta \mathrm{STr}[\ln(G_{K})]} \;,
\end{split}
\label{eq:FreeGeneratingFunctional2PPIFRG}
\end{equation}
with
\begin{equation}
G^{-1}_{K,\alpha_{1}\alpha_{2}} = \left(\hat{O}_{\mathrm{kin},\alpha_{1}} + V_{\alpha_{1}} - \mu - K_{\alpha_{1}}\right) \delta_{\alpha_{1}\alpha_{2}}\;,
\label{eq:GKminus1OkinVmu2PPIFRG}
\end{equation}
and $\zeta=\pm 1$ to distinguish the cases where $\widetilde{\psi}$ is a bosonic or a Grassmann field, as usual. Hence, the initial conditions for the derivatives of the Schwinger functional are readily obtained by differentiating the $\mathrm{STr}\ln$ term expressing $W_{0}[K]$ in~\eqref{eq:FreeGeneratingFunctional2PPIFRG} with respect to $K$:
\begin{equation}
\overline{W}^{(n)}_{\mathfrak{s}=\mathfrak{s}_{\mathrm{i}},\alpha_{1}\cdots\alpha_{n}}[K] = \overline{W}^{(n)}_{0,\alpha_{1}\cdots\alpha_{n}} \;,
\end{equation}
which constitutes all initial conditions required to treat the flow equation for $W_{\mathfrak{s}}[K]$, i.e.~\eqref{eq:FlowequationWs2PPIFRG} (with $\dot{V}_{\mathfrak{s},\alpha}=0$ $\forall \alpha,\mathfrak{s}$ in the present case). Regarding the flow equation for the EA, we can not simply express the free 2PPI EA $\Gamma_{0}^{(\mathrm{2PPI})}[\rho]$ explicitly in terms of $\rho$ as we did for the free 2PI EA in terms of $G$ with~\eqref{eq:2PIFRGfreeGamma2PI}. This stems from the fact that the Legendre transform can not be carried out explicitly for the 2PPI EA, as we explained in detail in section~\ref{sec:2PPIEA}. Instead, the 2PPI vertices can be determined from the derivatives of the Schwinger functional via the relation (see appendix~\ref{sec:DerivMasterEq2PPIFRG}):
\begin{equation}
\Gamma_{\mathfrak{s},\alpha_{1}\alpha_{2}}^{(\mathrm{2PPI})(2)}[\rho] = \left(W^{(2)}_{\mathfrak{s}}[K]\right)_{\alpha_{1}\alpha_{2}}^{-1} \;.
\label{eq:Gamma2W22PPIFRG}
\end{equation}
This is done with the help of the chain rule:
\begin{equation}
\frac{\delta}{\delta\rho_{\alpha_{1}}} = \int_{\alpha_{2}} \frac{\delta K_{\alpha_{2}}}{\delta \rho_{\alpha_{1}}} \frac{\delta}{\delta K_{\alpha_{2}}} = \int_{\alpha_{2}} \left(W^{(2)}_{\mathfrak{s}}[K]\right)_{\alpha_{1}\alpha_{2}}^{-1} \frac{\delta}{\delta K_{\alpha_{2}}} \;,
\end{equation}
based on~\eqref{eq:rhoEqualrhok2PPIFRG}. For example, $\Gamma_{\mathfrak{s},\alpha_{1}\alpha_{2}}^{(\mathrm{2PPI})(3)}[\rho]$ is deduced from~\eqref{eq:Gamma2W22PPIFRG} as follows:
\begin{equation}
\begin{split}
\Gamma_{\mathfrak{s},\alpha_{1}\alpha_{2}\alpha_{3}}^{(\mathrm{2PPI})(3)}[\rho] = & \int_{\alpha_{4}} \left(W^{(2)}_{\mathfrak{s}}[K]\right)_{\alpha_{3}\alpha_{4}}^{-1} \frac{\delta}{\delta K_{\alpha_{4}}} \left(W^{(2)}_{\mathfrak{s}}[K]\right)_{\alpha_{1}\alpha_{2}}^{-1} \\
= & -\int_{\alpha_{4},\alpha_{5},\alpha_{6}} \left(W^{(2)}_{\mathfrak{s}}[K]\right)_{\alpha_{3}\alpha_{4}}^{-1} \left(W^{(2)}_{\mathfrak{s}}[K]\right)_{\alpha_{1}\alpha_{5}}^{-1} W^{(3)}_{\mathfrak{s},\alpha_{4}\alpha_{5}\alpha_{6}}[K] \left(W^{(2)}_{\mathfrak{s}}[K]\right)_{\alpha_{6}\alpha_{2}}^{-1} \;,
\end{split}
\label{eq:Gamma3W2W32PPIFRG}
\end{equation}
and we can further differentiate this expression of $\Gamma_{\mathfrak{s},\alpha_{1}\alpha_{2}\alpha_{3}}^{(\mathrm{2PPI})(3)}[\rho]$ to obtain a similar result for $\Gamma_{\mathfrak{s},\alpha_{1}\alpha_{2}\alpha_{3}\alpha_{4}}^{(\mathrm{2PPI})(4)}[\rho]$ and so on. What is of interest for us is the free version of such relations for the purpose of determining the initial conditions for the 2PPI vertices, which are thus given by:
\begin{equation}
\overline{\Gamma}_{\mathfrak{s}=\mathfrak{s}_{\mathrm{i}},\alpha_{1}\alpha_{2}}^{(\mathrm{2PPI})(2)} = \left(\overline{W}^{(2)}_{0}\right)_{\alpha_{1}\alpha_{2}}^{-1} \;,
\end{equation}
\begin{equation}
\overline{\Gamma}_{\mathfrak{s}=\mathfrak{s}_{\mathrm{i}},\alpha_{1}\alpha_{2}\alpha_{3}}^{(\mathrm{2PPI})(3)} = -\int_{\alpha_{4},\alpha_{5},\alpha_{6}} \left(\overline{W}^{(2)}_{0}\right)_{\alpha_{3}\alpha_{4}}^{-1} \left(\overline{W}^{(2)}_{0}\right)_{\alpha_{1}\alpha_{5}}^{-1} \overline{W}^{(3)}_{0,\alpha_{4}\alpha_{5}\alpha_{6}} \left(\overline{W}^{(2)}_{0}\right)_{\alpha_{6}\alpha_{2}}^{-1} \;,
\end{equation}
and so on, or more generally for $n \geq 3$:
\begin{equation}
\scalebox{0.89}{${\displaystyle\overline{\Gamma}_{\mathfrak{s}=\mathfrak{s}_{\mathrm{i}},\alpha_{1}\alpha_{2}\alpha_{3} \cdots \alpha_{n}}^{(\mathrm{2PPI})(n)} = \left. \int_{\alpha_{2n-2}} \left(W^{(2)}_{0}[K]\right)_{\alpha_{n}\alpha_{2n-2}}^{-1}  \frac{\delta}{\delta K_{\alpha_{2n-2}}} \cdots \int_{\alpha_{n+1}} \left(W^{(2)}_{0}[K]\right)_{\alpha_{3}\alpha_{n+1}}^{-1} \frac{\delta}{\delta K_{\alpha_{n+1}}} \left(W^{(2)}_{0}[K]\right)_{\alpha_{1}\alpha_{2}}^{-1} \right|_{K=0} \;,}$}
\label{eq:GeneralFormulaDeterminationIC2PPIFRG}
\end{equation}
which is to be combined with the initial condition for the flowing density:
\begin{equation}
\overline{\rho}_{\mathfrak{s}=\mathfrak{s}_{\mathrm{i}},\alpha} = \overline{W}_{\mathfrak{s}=\mathfrak{s}_{\mathrm{i}},\alpha}^{(1)} \;,
\end{equation}
resulting from~\eqref{eq:rhoEqualrhok2PPIFRG}. Besides being generally more tedious than the derivation of $\overline{W}^{(n)}_{\mathfrak{s}=\mathfrak{s}_{\mathrm{i}}}$ (with $n \geq 1$), the calculation of the derivatives $\overline{\Gamma}_{\mathfrak{s}=\mathfrak{s}_{\mathrm{i}}}^{(\mathrm{2PPI})(n)}$ (with $n \geq 2$) also requires the inversion of $\overline{W}^{(2)}_{0}$, which might be very involved for theories that are not invariant under time and space translations. This clarifies why it might be more efficient to exploit the master equation for the Schwinger functional (i.e.~\eqref{eq:FlowequationWs2PPIFRG}) instead of that for the 2PPI EA (i.e.~\eqref{eq:2PPIFRGflowEquationGamma}) in some situations. In any case, the 2PPI vertices can be deduced at the end of the flow from the derivatives of the Schwinger functional (and vice versa) via relations such as~\eqref{eq:Gamma2W22PPIFRG} and~\eqref{eq:Gamma3W2W32PPIFRG} if necessary.

\paragraph{Truncations:}

The infinite tower of differential equations resulting from the vertex expansion~\eqref{eq:vertexExpansion2PPIFRG} combined with the master equation for $\Gamma_{\mathfrak{s}}^{(\mathrm{2PPI})}[\rho]$ is truncated once again by ignoring the flow of the vertices of order larger than a given integer $N_{\mathrm{max}}$. This is achieved by imposing different conditions on these vertices, which define different implementations of the standard 2PPI-FRG:
\begin{itemize}
\item For the sU-flow:
\begin{equation}
\overline{\Gamma}_{\mathfrak{s}}^{(\mathrm{2PPI})(n)} = 0 \mathrlap{\quad \forall \mathfrak{s}, ~ \forall n > N_{\mathrm{max}} \;.}
\label{eq:DefinitionsUflow2PPIFRG}
\end{equation}

\item For the pU-flow:
\begin{equation}
\overline{\Gamma}_{\mathfrak{s}}^{(\mathrm{2PPI})(n)} = \overline{\Gamma}_{\mathfrak{s}=\mathfrak{s}_{\mathrm{i}}}^{(\mathrm{2PPI})(n)} \mathrlap{\quad \forall \mathfrak{s}, ~ \forall n > N_{\mathrm{max}} \;.}
\label{eq:DefinitionpUflow2PPIFRG}
\end{equation}

\item For the iU-flow:
\begin{equation}
\overline{\Gamma}_{\mathfrak{s}}^{(\mathrm{2PPI})(n)} = \left.\overline{\Gamma}_{\mathfrak{s}=\mathfrak{s}_{\mathrm{i}}}^{(\mathrm{2PPI})(n)}\right|_{m^{2}[\overline{\rho}_{0}] \rightarrow m^{2}[\overline{\rho}_{\mathfrak{s}}]} \mathrlap{\quad \forall \mathfrak{s}, ~ \forall n > N_{\mathrm{max}} \;.}
\label{eq:DefinitioniUflow2PPIFRG}
\end{equation}

\end{itemize}

Although the standard 2PPI-FRG has already been applied to the (0+0)-D model considered in this thesis (only in its unbroken-symmetry regime with $N=1$)~\cite{kem13}, it should be stressed at the present stage that only the simplest truncation scheme, i.e. the sU-flow, was exploited in this case. Indeed, the sU-flow truncation is clearly the most drastic whereas the iU-flow one is \textit{a priori} the most refined. The condition underpinning the sU-flow enables us to content ourselves with the determination of the initial conditions $\overline{\Gamma}_{\mathfrak{s}=\mathfrak{s}_{\mathrm{i}}}^{(\mathrm{2PPI})(n)}$ up to $n=N_{\mathrm{max}}$. However, the pU-flow requires us to pursue this procedure up to $n=N_{\mathrm{max}}+2$ since the differential equations (resulting from the vertex expansion) expressing $\dot{\overline{\Gamma}}_{\mathfrak{s}}^{(\mathrm{2PPI})(n)}$ (with $2\leq n \leq N_{\mathrm{max}}$) depend on other 2PPI vertices $\overline{\Gamma}_{\mathfrak{s}}^{(\mathrm{2PPI})(m)}$ of order up to $m=n+2$. Finally, the iU-flow was first introduced in ref.~\cite{kem13} under the name ``RG improvement'' or ``RGi''. The associated truncation is implemented by expressing the squared mass of the studied system (which coincides with the oscillator frequency $\omega$ for the toy models studied in ref.~\cite{kem13}) in terms of its free density $\overline{\rho}_{0}$ and then replace it by the flowing density $\overline{\rho}_{\mathfrak{s}}$ in the expression of $\overline{\Gamma}_{\mathfrak{s}=\mathfrak{s}_{\mathrm{i}}}^{(\mathrm{2PPI})(n)}$ for $n > N_{\mathrm{max}}$, as stated by~\eqref{eq:DefinitioniUflow2PPIFRG}. This assumption is thus quite close in spirit to the truncation condition used in the mC-flow implementation of the 2PI-FRG.

\subsubsection{Kohn-Sham functional renormalization group}
\paragraph{Main features:}

As mentioned at the beginning of section~\ref{sec:2PPIFRG}, the KS-FRG was introduced by Liang and collaborators~\cite{lia18}. This approach was put forward as a ``novel optimization theory of FRG with faster convergence'', as compared to other 2PPI-FRG approaches (especially the sU-flow and pU-flow of the standard 2PPI-FRG) treated previously. The underlying idea is to ``split the total EA into the mean-field part $\Gamma_{\mathrm{KS},\mathfrak{s}}$ and the correlation part $\gamma_{\mathfrak{s}}$'':
\begin{equation}
\Gamma^{(\mathrm{2PPI})}_{\mathfrak{s}}[\rho] = \Gamma_{\mathrm{KS},\mathfrak{s}}[\rho] + \gamma_{\mathfrak{s}}[\rho] \;.
\label{eq:splittingKSFRG}
\end{equation}
Keeping in mind that the flow dependence of $\Gamma^{(\mathrm{2PPI})}_{\mathfrak{s}}[\rho]$ results in principle from that of the one-body potential $V_{\mathfrak{s}}$ (unless we specify to the U-flow scheme) and of the two-body interaction $U_{\mathfrak{s}}$, we can define the flowing 2PPI EA as:
\begin{equation}
\Gamma^{(\mathrm{2PPI})}_{\mathfrak{s}}[\rho] \equiv \Gamma^{(\mathrm{2PPI})}[\rho;V_{\mathfrak{s}},U_{\mathfrak{s}}] \;,
\end{equation}
which implies that its flowing mean-field part satisfies:
\begin{equation}
\Gamma_{\mathrm{KS},\mathfrak{s}}[\rho] \equiv \Gamma^{(\mathrm{2PPI})}[\rho;V_{\mathfrak{s}}=V_{\mathrm{KS},\mathfrak{s}},U_{\mathfrak{s}}=0] \;,
\label{eq:LinkGammaKSVKS}
\end{equation}
where $V_{\mathrm{KS},\mathfrak{s}}$ is the Kohn-Sham potential for the studied system at the stage of the flow where the flow parameter has value $\mathfrak{s}$. The flowing density $\overline{\rho}_{\mathfrak{s}}$ must now extremize both $\Gamma^{(\mathrm{2PPI})}_{\mathfrak{s}}[\rho]$ and $\Gamma_{\mathrm{KS},\mathfrak{s}}[\rho]$ according to:
\begin{equation}
\left. \frac{\delta\Gamma_{\mathrm{KS},\mathfrak{s}}[\rho]}{\delta \rho_{\alpha}}\right|_{\rho=\overline{\rho}_{\mathfrak{s}}} = 0 \mathrlap{\quad \forall \alpha,\mathfrak{s} \;,}
\label{eq:2PPIfrgExtremizeGammaKS2PPI}
\end{equation}
to be combined with~\eqref{eq:2PPIfrgExtremizeGamma2PPI}. As a consequence of definition~\eqref{eq:splittingKSFRG}, this implies that the configuration $\overline{\rho}_{\mathfrak{s}}$ of the density also corresponds to an extremum of the correlation part $\gamma_{\mathfrak{s}}[\rho]$:
\begin{equation}
\left. \frac{\delta\gamma_{\mathfrak{s}}[\rho]}{\delta \rho_{\alpha}}\right|_{\rho=\overline{\rho}_{\mathfrak{s}}} = 0 \mathrlap{\quad \forall \alpha,\mathfrak{s} \;.}
\label{eq:2PPIfrgExtremizeCorrgamma2PPI}
\end{equation}
Relation~\eqref{eq:2PPIfrgExtremizeGammaKS2PPI} is actually equivalent to the Kohn-Sham equation~\cite{koh65,koh65bis}, hence~\eqref{eq:2PPIfrgExtremizeGammaKS2PPI} is not at all a new artificial condition. We are just exploiting in this way a well-known equation of quantum many-body physics to improve the 2PPI-FRG approach. This also explains the origin of the name KS-FRG: the mean-field part $\Gamma_{\mathrm{KS},\mathfrak{s}}[\rho]$ of the 2PPI EA is determined by solving the Kohn-Sham equation in the form of~\eqref{eq:2PPIfrgExtremizeGammaKS2PPI} at each step of the flow\footnote{At each step of the flow, it is more precisely the flow-dependent Kohn-Sham potential $V_{\mathrm{KS},\mathfrak{s}}$ that is determined from~\eqref{eq:2PPIfrgExtremizeGammaKS2PPI} and $\Gamma_{\mathrm{KS},\mathfrak{s}}[\rho]$ is then deduced from $V_{\mathrm{KS},\mathfrak{s}}$, as the relation between $\Gamma_{\mathrm{KS},\mathfrak{s}}[\rho]$ and $V_{\mathrm{KS},\mathfrak{s}}$ can be inferred from Gaussian integration, as will be illustrated in section~\ref{sec:KSFRG0DON}.} and its correlation part $\gamma_{\mathfrak{s}}[\rho]$ is calculated by solving a set of integro-differential equations, like those encountered in any FRG procedure. To obtain such an equation system, we must rewrite the exact flow equation for the 2PPI EA, i.e.~\eqref{eq:2PPIFRGflowEquationGamma}, as a (still exact) flow equation for $\gamma_{\mathfrak{s}}[\rho]$. This is achieved by plugging the splitting~\eqref{eq:splittingKSFRG} into~\eqref{eq:2PPIFRGflowEquationGamma} as a first step:
\begin{equation}
\dot{\gamma}_{\mathfrak{s}}[\rho] = - \dot{\Gamma}_{\mathrm{KS},\mathfrak{s}}[\rho] + \int_{\alpha} \dot{V}_{\mathfrak{s},\alpha} \rho_{\alpha} + \frac{1}{2} \mathrm{STr}\left[\dot{U}_{\mathfrak{s}} \left( \Gamma^{(2)}_{\mathrm{KS},\mathfrak{s}}[\rho] + \gamma^{(2)}_{\mathfrak{s}}[\rho] \right)^{-1} \right] + \frac{1}{2} \int_{\alpha_{1},\alpha_{2}} \dot{U}_{\mathfrak{s},\alpha_{1}\alpha_{2}} \rho_{\alpha_{1}} \rho_{\alpha_{2}} \;,
\label{eq:MasterEquationKSFRGversion1}
\end{equation}
where $\Gamma^{(n)}_{\mathrm{KS},\mathfrak{s},\alpha_{1}\cdots\alpha_{n}}\equiv\frac{\delta^{n} \Gamma_{\mathrm{KS},\mathfrak{s}}[\rho]}{\delta\rho_{\alpha_{1}}\cdots\delta\rho_{\alpha_{n}}}$ and $\gamma_{\mathfrak{s},\alpha_{1}\cdots\alpha_{n}}^{(n)}\equiv\frac{\delta^{n}\gamma_{\mathfrak{s}}[\rho]}{\delta\rho_{\alpha_{1}}\cdots\delta\rho_{\alpha_{n}}}$. We then rewrite the derivative $\dot{\Gamma}_{\mathrm{KS},\mathfrak{s}}[\rho]$ with the help of the chain rule:
\begin{equation}
\dot{\Gamma}_{\mathrm{KS},\mathfrak{s}}[\rho] = \int_{\alpha_{1},\alpha_{2}} \frac{\delta \Gamma_{\mathrm{KS},\mathfrak{s}}[\rho]}{\delta V_{\mathrm{KS},\mathfrak{s},\alpha_{1}}} \frac{\delta V_{\mathrm{KS},\mathfrak{s},\alpha_{1}}}{\delta \overline{\rho}_{\mathfrak{s},\alpha_{2}}} \dot{\overline{\rho}}_{\mathfrak{s},\alpha_{2}} \;.
\label{eq:ChainRuleVersion1KSFRG}
\end{equation}
The whole dependence of $\Gamma_{\mathrm{KS},\mathfrak{s}}[\rho]$ with respect to the flow parameter $\mathfrak{s}$ stems from that of $V_{\mathrm{KS},\mathfrak{s}}$ according to its definition given by~\eqref{eq:LinkGammaKSVKS}. We also consider the definition of $\Gamma_{\mathrm{KS},\mathfrak{s}}[\rho]$ in terms of the corresponding Schwinger functional $W_{\mathrm{KS},\mathfrak{s}}[K]$:
\begin{equation}
\begin{split}
\Gamma_{\textcolor{blue}{\mathrm{KS}},\mathfrak{s}}[\rho] = & -W_{\textcolor{blue}{\mathrm{KS}},\mathfrak{s}}[K] + \int_{\alpha} K_{\alpha} \underbrace{\frac{\delta W_{\textcolor{blue}{\mathrm{KS}},\mathfrak{s}}[K]}{\delta K_{\alpha}}}_{\rho_{\alpha}} \\
= & -\ln\bigg(\int\mathcal{D}\widetilde{\psi}^{\dagger}\mathcal{D}\widetilde{\psi} \ e^{-S_{\textcolor{blue}{\mathrm{KS}},\mathfrak{s}}\big[\widetilde{\psi}^{\dagger},\widetilde{\psi}\big]+\int_{\alpha} K_{\alpha} \widetilde{\psi}_{\alpha}^{\dagger}\widetilde{\psi}_{\alpha}} \bigg) + \int_{\alpha} K_{\alpha} \rho_{\alpha} \\
= & -\ln\bigg(\int\mathcal{D}\widetilde{\psi}^{\dagger}\mathcal{D}\widetilde{\psi} \ e^{-\int_{\alpha} \widetilde{\psi}_{\alpha}^{\dagger}\left(\hat{O}_{\mathrm{kin},\alpha} + V_{\textcolor{blue}{\mathrm{KS}},\alpha} - \mu\right)\widetilde{\psi}_{\alpha}+\int_{\alpha} K_{\alpha} \widetilde{\psi}_{\alpha}^{\dagger}\widetilde{\psi}_{\alpha}} \bigg) + \int_{\alpha} K_{\alpha} \rho_{\alpha} \;,
\end{split}
\label{eq:GammaKSLegendreTransformKSFRG}
\end{equation}
which, after differentiation with respect to $V_{\textcolor{blue}{\mathrm{KS}},\alpha}$, gives us:
\begin{equation}
\begin{split}
\frac{\delta \Gamma_{\textcolor{blue}{\mathrm{KS}},\mathfrak{s}}[\rho]}{\delta V_{\mathrm{KS},\alpha}} = & \ \frac{1}{Z_{\textcolor{blue}{\mathrm{KS}},\mathfrak{s}}[K]} \int\mathcal{D}\widetilde{\psi}^{\dagger}\mathcal{D}\widetilde{\psi} \ \widetilde{\psi}_{\alpha}^{\dagger}\widetilde{\psi}_{\alpha} \ e^{-\int_{\alpha} \widetilde{\psi}_{\alpha}^{\dagger}\left(\hat{O}_{\mathrm{kin},\alpha} + V_{\textcolor{blue}{\mathrm{KS}},\alpha} - \mu\right)\widetilde{\psi}_{\alpha}+\int_{\alpha} K_{\alpha} \widetilde{\psi}_{\alpha}^{\dagger}\widetilde{\psi}_{\alpha}} \\
= & \ \frac{\delta W_{\textcolor{blue}{\mathrm{KS}},\mathfrak{s}}[K]}{\delta K_{\alpha}} \\
= & \ \rho_{\alpha} \;.
\end{split}
\label{eq:DerivativeGammaKSVKSKSFRG}
\end{equation}
Note that we have just used in both~\eqref{eq:GammaKSLegendreTransformKSFRG} and~\eqref{eq:DerivativeGammaKSVKSKSFRG} the relation:
\begin{equation}
\rho_{\alpha} = \frac{\delta W_{\textcolor{blue}{\mathrm{KS}},\mathfrak{s}}[K]}{\delta K_{\alpha}} \;,
\end{equation}
which asserts that the exact source-dependent density is obtained from the generating functionals of the Kohn-Sham system. Such a relation has already been discussed in section~\ref{sec:2PPIEA} via the IM (see~\eqref{eq:2PPIEAIMdW1dK0DON}). At $K_{\alpha}=0$ $\forall\alpha$, this coincides with the Kohn-Sham scheme~\cite{koh65,koh65bis} stating that, for any interacting system, there is a unique non-interacting system (i.e. a unique system whose classical action is quadratic) with the same gs density, the one-body potential of this auxiliary non-interacting system being the Kohn-Sham potential. Furthermore, we also know, notably from our previous discussion on DFT in section~\ref{sec:2PPIEA} as well, that the dependence of the density functional $\Gamma_{\mathrm{KS},\mathfrak{s}}[\rho]$ with respect to the corresponding one-body potential has the form of a convolution with the density $\rho_{\alpha}$. This leads to:
\begin{equation}
\frac{\delta\overline{\Gamma}_{\mathrm{KS},\mathfrak{s}}}{\delta\overline{\rho}_{\mathfrak{s},\alpha_{1}}} = - \frac{\delta}{\delta\overline{\rho}_{\mathfrak{s},\alpha_{1}}} \int_{\alpha_{2}} V_{\mathrm{KS},\mathfrak{s},\alpha_{2}} \overline{\rho}_{\mathfrak{s},\alpha_{2}} = - V_{\mathrm{KS},\mathfrak{s},\alpha_{1}} \;.
\label{eq:DerivativeGammaKSrhosKSFRG}
\end{equation}
Using~\eqref{eq:DerivativeGammaKSVKSKSFRG} and~\eqref{eq:DerivativeGammaKSrhosKSFRG}, the chain rule~\eqref{eq:ChainRuleVersion1KSFRG} is rewritten as follows:
\begin{equation}
\dot{\Gamma}_{\mathrm{KS},\mathfrak{s}}[\rho] = \int_{\alpha_{1},\alpha_{2}} \underbrace{\frac{\delta \Gamma_{\mathrm{KS},\mathfrak{s}}[\rho]}{\delta V_{\mathrm{KS},\mathfrak{s},\alpha_{1}}}}_{\rho_{\alpha_{1}}} \underbrace{\frac{\delta V_{\mathrm{KS},\mathfrak{s},\alpha_{1}}}{\delta \overline{\rho}_{\mathfrak{s},\alpha_{2}}}}_{-\frac{\delta^{2} \overline{\Gamma}_{\mathrm{KS},\mathfrak{s}}}{\delta\overline{\rho}_{\mathfrak{s},\alpha_{2}}\delta\overline{\rho}_{\mathfrak{s},\alpha_{1}}}} \dot{\overline{\rho}}_{\mathfrak{s},\alpha_{2}} = - \int_{\alpha_{1},\alpha_{2}} \rho_{\alpha_{1}} \overline{\Gamma}^{(2)}_{\mathrm{KS},\mathfrak{s},\alpha_{1}\alpha_{2}} \dot{\overline{\rho}}_{\mathfrak{s},\alpha_{2}} \;,
\label{eq:ChainRuleVersion2KSFRG}
\end{equation}
assuming that $\frac{\delta^{n} \overline{\Gamma}_{\mathrm{KS},\mathfrak{s}}}{\delta\overline{\rho}_{\mathfrak{s},\alpha_{1}}\cdots\delta\overline{\rho}_{\mathfrak{s},\alpha_{n}}} = \left.\frac{\delta^{n} \Gamma_{\mathrm{KS},\mathfrak{s}}[\rho]}{\delta\rho_{\alpha_{1}}\cdots\delta\rho_{\alpha_{n}}}\right|_{\rho=\overline{\rho}_{\mathfrak{s}}} \equiv \overline{\Gamma}^{(n)}_{\mathrm{KS},\mathfrak{s},\alpha_{1}\cdots\alpha_{n}}$ $\forall \alpha_{1},\cdots,\alpha_{n},\mathfrak{s}$. After combining \eqref{eq:ChainRuleVersion2KSFRG} with~\eqref{eq:MasterEquationKSFRGversion1}, we obtain the following master equation for $\gamma_{\mathfrak{s}}[\rho]$:
\begin{equation}
\scalebox{0.92}{${\displaystyle\dot{\gamma}_{\mathfrak{s}}[\rho] = \int_{\alpha_{1}} \rho_{\alpha_{1}} \left(\dot{V}_{\mathfrak{s},\alpha_{1}} + \int_{\alpha_{2}} \overline{\Gamma}^{(2)}_{\mathrm{KS},\mathfrak{s},\alpha_{1}\alpha_{2}} \dot{\overline{\rho}}_{\mathfrak{s},\alpha_{2}} \right) + \frac{1}{2} \mathrm{STr}\left[\dot{U}_{\mathfrak{s}} \left( \Gamma^{(2)}_{\mathrm{KS},\mathfrak{s}}[\rho] + \gamma^{(2)}_{\mathfrak{s}}[\rho] \right)^{-1} \right] + \frac{1}{2} \int_{\alpha_{1},\alpha_{2}} \dot{U}_{\mathfrak{s},\alpha_{1}\alpha_{2}} \rho_{\alpha_{1}} \rho_{\alpha_{2}} \;. }$}
\label{eq:MasterEquationKSFRGfinalversion}
\end{equation}
This master equation is turned into an infinite tower of differential equations for $\overline{\gamma}_{\mathfrak{s}}\equiv\gamma_{\mathfrak{s}}[\rho=\overline{\rho}_{\mathfrak{s}}]$ and the derivatives $\overline{\gamma}_{\mathfrak{s},\alpha_{1}\cdots\alpha_{n}}^{(n)}\equiv\left.\frac{\delta^{n}\gamma_{\mathfrak{s}}[\rho]}{\delta\rho_{\alpha_{1}}\cdots\delta\rho_{\alpha_{n}}}\right|_{\rho=\overline{\rho}_{\mathfrak{s}}}$ from the expansion:
\begin{equation}
\gamma_{\mathfrak{s}}[\rho] = \overline{\gamma}_{\mathfrak{s}} + \sum_{n=2}^{\infty} \int_{\alpha_{1},\cdots,\alpha_{n}} \overline{\gamma}^{(n)}_{\mathfrak{s},\alpha_{1}\cdots\alpha_{n}} \left(\rho-\overline{\rho}_{\mathfrak{s}}\right)_{\alpha_{1}} \cdots \left(\rho-\overline{\rho}_{\mathfrak{s}}\right)_{\alpha_{n}} \;,
\label{eq:ExpansionCorrgammaKSFRG}
\end{equation}
which is simplified according to~\eqref{eq:2PPIfrgExtremizeCorrgamma2PPI} imposing that $\overline{\gamma}_{\mathfrak{s}}^{(1)}$ vanishes. It is finally by inserting this expansion into~\eqref{eq:MasterEquationKSFRGfinalversion} and identifying the terms with identical powers of $\rho-\overline{\rho}_{\mathfrak{s}}$ on both sides of the equation thus derived that we obtain the equation system to solve for the KS-FRG. Note once again that this equation system includes in addition the Kohn-Sham equation in the form~\eqref{eq:2PPIfrgExtremizeGammaKS2PPI} in order to determine the Kohn-Sham potential $V_{\mathrm{KS},\mathfrak{s}}$, and therefore the mean-field part $\Gamma_{\mathrm{KS},\mathfrak{s}}[\rho]$, at each step of the flow. It should also be stressed that, as for the standard 2PPI-FRG, we can split the KS-FRG into a U-flow and a CU-flow scheme, whether we choose $V_{\mathfrak{s}}$ independent or dependent from $\mathfrak{s}$, respectively. However, in both situations, the Kohn-Sham potential $V_{\mathrm{KS},\mathfrak{s}}$ is subject to evolve during the flow. Furthermore, we can already see at this stage a strong connection between the U-flow implementation of the KS-FRG and the pU-flow version of the 2PI-FRG for the following reasons:
\begin{itemize}
\item They all rely on a \textbf{splitting} of the EA under consideration introducing its correlation part ($\gamma_{\mathfrak{s}}[\rho]$ for the KS-FRG and the Luttinger-Ward functional $\Phi_{\mathfrak{s}}[G]$ for the 2PI-FRG), although this analogy must be taken with care: $\Phi_{\mathfrak{s}}[G]$ encompasses the entire information from the interaction in the case of the 2PI-FRG whereas some of it is recast into the mean-field part $\Gamma_{\mathrm{KS},\mathfrak{s}}$ via $V_{\mathrm{KS},\mathfrak{s}}$ for the 2PPI-FRG.
\item They can be both formulated as a \textbf{vertex expansion} of the EA under consideration. This was already stated by~\eqref{eq:VertexExpansion2PIFRG} for the 2PI-FRG, whereas the tower of differential equations underlying the KS-FRG can be obtained equivalently by directly Taylor expanding $\Gamma_{\mathfrak{s}}^{(\mathrm{2PPI})}[\rho]$ and performing the splitting $\Gamma^{(\mathrm{2PPI})}_{\mathfrak{s}}[\rho] = \Gamma_{\mathrm{KS},\mathfrak{s}}[\rho] + \gamma_{\mathfrak{s}}[\rho]$ \textit{a posteriori} (instead of applying the latter splitting first and expanding the correlation part via~\eqref{eq:ExpansionCorrgammaKSFRG} afterwards).
\item The correlation part of the EA is \textbf{truncated} in the same manner, as can be seen by comparing~\eqref{eq:2PIfrgPhiBartCflow} with~\eqref{eq:KSFRGtruncationScheme} below.
\end{itemize}
The link between these two FRG approaches will be further discussed from our toy model applications presented in section~\ref{sec:2PPIFRG0DON}.

\vspace{0.5cm}

In conclusion, besides the direct connection with DFT provided by the 2PPI EA formalism, the KS-FRG enables us to exploit the Kohn-Sham scheme to improve the convergence of the method, as compared to other standard 2PPI-FRG approaches. Technically, this amounts to adding the Kohn-Sham equation~\eqref{eq:2PPIfrgExtremizeGammaKS2PPI} to the set of integro-differential equations to solve. The KS-FRG has also been put forward with a method to estimate theoretical errors~\cite{lia18}, that we will not discuss further besides pointing out that the approach thus obtained is equivalent to a Kohn-Sham DFT with uncertainty quantification of the results. Finally, another interesting feature of the KS-FRG is the determination of the corresponding initial conditions which is much less cumbersome as compared to that of the standard 2PPI-FRG, as we explain below.

\paragraph{Initial conditions:}

Whether we consider the U-flow or CU-flow implementation of the KS-FRG, the initial conditions for $\overline{\gamma}_{\mathfrak{s}}$ and the corresponding derivatives are:
\begin{equation}
\overline{\gamma}_{\mathfrak{s}=\mathfrak{s}_{\mathrm{i}}} = 0 \mathrlap{\;,}
\label{eq:InitialCondCorrgammaKSFRG}
\end{equation}
\begin{equation}
\overline{\gamma}^{(n)}_{\mathfrak{s}=\mathfrak{s}_{\mathrm{i}},\alpha_{1}, \cdots, \alpha_{n}} = 0 \mathrlap{\quad \forall \alpha_{1}, \cdots, \alpha_{n}, ~ \forall n \geq 2 \;.}
\label{eq:InitialCondCorrgammanKSFRG}
\end{equation}
In the case of the U-flow, the solution of $V_{\mathrm{KS},\mathfrak{s}}$ found from~\eqref{eq:2PPIfrgExtremizeGammaKS2PPI} is trivial at the starting point of the flow such that $\Gamma_{\mathrm{KS},\mathfrak{s}=\mathfrak{s}_{\mathrm{i}}}[\rho]$ coincides with the free part of $\Gamma^{(\mathrm{2PPI})}[\rho]$. For the CU-flow, such a solution is non-trivial at $\mathfrak{s}=\mathfrak{s}_{\mathrm{i}}$ and $\Gamma_{\mathrm{KS},\mathfrak{s}=\mathfrak{s}_{\mathrm{i}}}[\rho]$ thus already incorporates information about the interaction. In both cases, the exactly solvable system used as starting point of the flow is fully specified by the mean-field part of the EA, hence the initial condition $\gamma_{\mathfrak{s}=\mathfrak{s}_{\mathrm{i}}}[\rho]=0$ which translates into~\eqref{eq:InitialCondCorrgammaKSFRG} and~\eqref{eq:InitialCondCorrgammanKSFRG} according to~\eqref{eq:ExpansionCorrgammaKSFRG}.

\paragraph{Truncation:}

Finally, the truncation of the infinite set of differential equations extracted from the master equation~\eqref{eq:MasterEquationKSFRGfinalversion} with~\eqref{eq:ExpansionCorrgammaKSFRG} is simply implemented by imposing:
\begin{equation}
\overline{\gamma}_{\mathfrak{s}}^{(n)} = \overline{\gamma}_{\mathfrak{s}=\mathfrak{s}_{\mathrm{i}}}^{(n)} \mathrlap{\quad \forall \mathfrak{s}, ~ \forall n > N_{\mathrm{max}} \;,}
\label{eq:KSFRGtruncationScheme}
\end{equation}
for a given truncation order $N_{\mathrm{max}}$.

\subsection{Application to the (0+0)-D $O(N)$-symmetric $\varphi^4$-theory}
\label{sec:2PPIFRG0DON}
\subsubsection{Standard 2PPI functional renormalization group}
\label{sec:standard2PPIFRG0DON}

We now turn back to the studied (0+0)-D $O(N)$ model for which the $\alpha$-indices all reduce to color ones. In particular, the two-body interaction $U$ can be directly expressed in terms of the coupling constant $\lambda$ as follows:
\begin{equation}
U_{\alpha_{1}\alpha_{2}} = U_{a_{1}a_{2}} = \frac{\lambda}{12} \mathrlap{\quad \forall a_{1},a_{2} \;.}
\label{eq:2PPIFRGTwoInteraction0DON}
\end{equation}
One can indeed check that this definition is consistent with:
\begin{equation}
S_{\mathrm{int}}\Big[\widetilde{\psi}^{\dagger},\widetilde{\psi}\Big] = \frac{1}{2}\int_{\alpha_{1},\alpha_{2}} \widetilde{\psi}_{\alpha_{1}}^{\dagger}\widetilde{\psi}_{\alpha_{2}}^{\dagger} U_{\alpha_{1}\alpha_{2}} \widetilde{\psi}_{\alpha_{2}}\widetilde{\psi}_{\alpha_{1}} = \frac{\lambda}{4!} \sum_{a_{1},a_{2}=1}^{N} \widetilde{\varphi}_{a_{1}}^{2} \widetilde{\varphi}_{a_{2}}^{2} = \frac{\lambda}{4!} \vec{\widetilde{\varphi}}^{2} \;,
\end{equation}
where
\begin{equation}
\widetilde{\psi}_{\alpha} = \widetilde{\psi}^{\dagger}_{\alpha} = \widetilde{\varphi}_{a} \;,
\end{equation}
as required by comparing~\eqref{eq:ClassicalAction2PPIFRG} with the original classical action $S\big(\vec{\widetilde{\varphi}}\big)$ of our toy model. Moreover, we also choose the same cutoff function $R_{\mathfrak{s}}$ as for the U-flow and CU-flow implementations of the 2PI-FRG treated in section~\ref{sec:2PIFRG0DON}, i.e.:
\begin{equation}
U_{\mathfrak{s},a_{1}a_{2}} = R_{\mathfrak{s}} U_{a_{1}a_{2}} = \mathfrak{s} U_{a_{1}a_{2}} = \mathfrak{s} \frac{\lambda}{12} \mathrlap{\quad \forall a_{1}, a_{2} \;,}
\label{eq:choiceCutoffUs2PPIFRG0DON}
\end{equation}
in accordance with~\eqref{eq:choiceCutoffUs2PIFRGUflow0DON}. We also use $\mathfrak{s}_{\mathrm{i}}=0$ and $\mathfrak{s}_{\mathrm{f}}=1$ as for the 2PI-FRG. From~\eqref{eq:2PPIFRGTwoInteraction0DON} to~\eqref{eq:choiceCutoffUs2PPIFRG0DON}, we infer that the master equation~\eqref{eq:2PPIFRGflowEquationGamma} reduces in the present (0+0)-D framework to:
\begin{equation}
\dot{\Gamma}^{(\mathrm{2PPI})}_{\mathfrak{s}}(\rho) = \frac{\lambda}{24} \left( \sum_{a_{1},a_{2}=1}^{N} \boldsymbol{G}_{\mathfrak{s},a_{1}a_{2}}(\rho) + \sum_{a_{1},a_{2}=1}^{N} \rho_{a_{1}} \rho_{a_{2}} \right) \;,
\label{eq:2PPIFRGmasterEquation0DON}
\end{equation}
with
\begin{equation}
\boldsymbol{G}^{-1}_{\mathfrak{s},a_{1}a_{2}}(\rho) \equiv \Gamma_{\mathfrak{s},a_{1}a_{2}}^{(\mathrm{2PPI})(2)}(\rho) \;.
\label{eq:standard2PPIFRGvertExpPropagator}
\end{equation}

\vspace{0.5cm}

Similarly to the previous FRG approaches, we treat~\eqref{eq:2PPIFRGmasterEquation0DON} with a vertex expansion procedure based on the Taylor expansion of $\Gamma_{\mathfrak{s}}^{(\mathrm{2PPI})}(\rho)$ around its extremum at $\rho=\overline{\rho}_{\mathfrak{s}}$, i.e.:
\begin{equation}
\Gamma_{\mathfrak{s}}^{(\mathrm{2PPI})}(\rho) = \overline{\Gamma}_{\mathfrak{s}}^{(\mathrm{2PPI})} + \sum_{n=2}^{\infty} \frac{1}{n!} \sum_{a_{1},\cdots,a_{n}=1}^{N} \overline{\Gamma}_{\mathfrak{s},a_{1} \cdots a_{n}}^{(\mathrm{2PPI})(n)} \left(\rho-\overline{\rho}_{\mathfrak{s}}\right)_{a_{1}} \cdots \left(\rho-\overline{\rho}_{\mathfrak{s}}\right)_{a_{n}} \;,
\label{eq:Standard2PPIFRGVertexExpansion}
\end{equation}
where
\begin{equation}
\overline{\Gamma}^{(\mathrm{2PPI})}_{\mathfrak{s}} \equiv \Gamma^{(\mathrm{2PPI})}_{\mathfrak{s}}\big(\rho=\overline{\rho}_{\mathfrak{s}}\big) \mathrlap{\quad \forall \mathfrak{s} \;,}
\label{eq:DefGammabarpure2PPIFRG0DON}
\end{equation}
\begin{equation}
\overline{\Gamma}^{(\mathrm{2PPI})(n)}_{\mathfrak{s},a_{1} \cdots a_{n}} \equiv \left. \frac{\partial^{n} \Gamma^{(\mathrm{2PPI})}_{\mathfrak{s}}(\rho)}{\partial \rho_{a_{1}} \cdots \partial \rho_{a_{n}}} \right|_{\rho=\overline{\rho}_{\mathfrak{s}}} \mathrlap{\quad \forall a_{1},\cdots,a_{n},\mathfrak{s} \;,}
\label{eq:DefGammanbarpure2PPIFRG0DON}
\end{equation}
and, in particular,
\begin{equation}
\overline{\Gamma}_{\mathfrak{s},a}^{(\mathrm{2PPI})(1)} = 0 \mathrlap{\quad \forall a, \mathfrak{s} \;.}
\label{eq:DefGamma1barpure2PPIFRG0DON}
\end{equation}
The infinite tower of differential equations resulting from this vertex expansion procedure includes notably (see appendix~\ref{sec:standard2PPIFRGVertexExpansionAppendix} for the corresponding flow equations expressing the derivatives of the 2PPI vertices of order 3 and 4 with respect to $\mathfrak{s}$):
\begin{equation}
\dot{\overline{\Gamma}}_{\mathfrak{s}}^{(\mathrm{2PPI})} = \frac{\lambda}{24} \sum_{a_{1},a_{2}=1}^{N} \left( \overline{\boldsymbol{G}}_{\mathfrak{s},a_{1}a_{2}} + \overline{\rho}_{\mathfrak{s},a_{1}} \overline{\rho}_{\mathfrak{s},a_{2}} \right) \;,
\label{eq:standard2PPIFRGflowEqExpressionGammadot}
\end{equation}
\begin{equation}
\dot{\overline{\rho}}_{\mathfrak{s},a_{1}} = \frac{\lambda}{24} \sum_{a_{2}=1}^{N} \overline{\boldsymbol{G}}_{\mathfrak{s},a_{1}a_{2}} \left( \sum_{a_{3},a_{4},a_{5},a_{6}=1}^{N} \overline{\boldsymbol{G}}_{\mathfrak{s},a_{3}a_{5}} \overline{\Gamma}_{\mathfrak{s},a_{2}a_{5}a_{6}}^{(\mathrm{2PPI})(3)} \overline{\boldsymbol{G}}_{\mathfrak{s},a_{6}a_{4}} - 2 \sum_{a_{3}=1}^{N} \overline{\rho}_{\mathfrak{s},a_{3}} \right) \;,
\label{eq:standard2PPIFRGflowEqExpressionrhodot}
\end{equation}
\begin{equation}
\begin{split}
\dot{\overline{\Gamma}}_{\mathfrak{s},a_{1}a_{2}}^{(\mathrm{2PPI})(2)} = & \sum_{a_{3}=1}^{N} \dot{\overline{\rho}}_{\mathfrak{s},a_{3}} \overline{\Gamma}^{(\mathrm{2PPI})(3)}_{\mathfrak{s},a_{3}a_{1}a_{2}} \\
& + \frac{\lambda}{24} \Bigg( 2 + 2 \sum_{a_{3},a_{4},a_{5},a_{6},a_{7},a_{8}=1}^{N} \overline{\boldsymbol{G}}_{\mathfrak{s},a_{3}a_{5}} \overline{\Gamma}_{\mathfrak{s},a_{1}a_{5}a_{6}}^{(\mathrm{2PPI})(3)} \overline{\boldsymbol{G}}_{\mathfrak{s},a_{6}a_{7}} \overline{\Gamma}_{\mathfrak{s},a_{2}a_{7}a_{8}}^{(\mathrm{2PPI})(3)} \overline{\boldsymbol{G}}_{\mathfrak{s},a_{8}a_{4}} \\
& \ \hspace{1.05cm} - \sum_{a_{3},a_{4},a_{5},a_{6}=1}^{N} \overline{\boldsymbol{G}}_{\mathfrak{s},a_{3}a_{5}} \overline{\Gamma}_{\mathfrak{s},a_{1}a_{2}a_{5}a_{6}}^{(\mathrm{2PPI})(4)} \overline{\boldsymbol{G}}_{\mathfrak{s},a_{6}a_{4}} \Bigg) \;.
\end{split}
\label{eq:standard2PPIFRGflowEqExpressionGamma2dot}
\end{equation}
Regarding the initial conditions, we can actually bypass the lengthy procedure outlined in section~\ref{sec:standard2PPIFRG} by exploiting the fact that the 2PI EA coincides with the 2PPI one in the present framework, i.e. by exploiting the relation $\Gamma^{(\mathrm{2PPI})}(\rho)=\Gamma^{(\mathrm{2PI})}(\boldsymbol{G})$ (which results itself from both the (0+0)-D character of the studied model and from the fact that the 1-point correlation function of $\vec{\widetilde{\varphi}}$ is imposed to be zero in the 2PPI-FRG framework). This enables us to find the initial conditions for the 2PPI vertices by differentiating the free 2PI EA, given by~\eqref{eq:2PIFRGfreeGamma2PI} that reduces in (0+0)-D to:
\begin{equation}
\Gamma_{0}^{(\mathrm{2PI})}(\boldsymbol{G}) = -\frac{1}{2}\mathrm{Tr}_{a}\big[\ln(2\pi\boldsymbol{G})\big] + \frac{1}{2}\mathrm{Tr}_{a}\big(C^{-1}\boldsymbol{G}-\mathbb{I}_{N}\big) \;,
\label{eq:free2PIEAforCI2PPIFRG0DON}
\end{equation}
and then replacing $\boldsymbol{G}$ by $\rho$ via $\boldsymbol{G}_{a_{1}a_{2}} = \rho_{a_{1}} \delta_{a_{1}a_{2}}$. This leads to:
\begin{equation}
\overline{\rho}_{\mathfrak{s}=\mathfrak{s}_{\mathrm{i}},a} = \frac{1}{m^{2}} \mathrlap{\quad \forall a \;,}
\label{eq:standard2PPIFRGICExpressionrho00DON}
\end{equation}
\begin{equation}
\overline{\Gamma}_{\mathfrak{s}=\mathfrak{s}_{\mathrm{i}},a_{1}a_{2}}^{(\mathrm{2PPI})} = - \frac{N}{2} \ln\bigg(\frac{2\pi}{m^{2}}\bigg) \mathrlap{\;,}
\label{eq:standard2PPIFRGICExpressionGamma00DON}
\end{equation}
\begin{equation}
\overline{\Gamma}_{\mathfrak{s}=\mathfrak{s}_{\mathrm{i}},a_{1}a_{2}}^{(\mathrm{2PPI})(2)} = \frac{1}{2} \overline{\rho}_{\mathfrak{s}=\mathfrak{s}_{\mathrm{i}},a_{1}}^{-2} \delta_{a_{1}a_{2}} \mathrlap{\quad \forall a_{1},a_{2} \;,}
\label{eq:standard2PPIFRGICExpressionGamma020DON}
\end{equation}
\begin{equation}
\overline{\Gamma}_{\mathfrak{s}=\mathfrak{s}_{\mathrm{i}},a_{1}a_{2}a_{3}}^{(\mathrm{2PPI})(3)} = - \overline{\rho}_{\mathfrak{s}=\mathfrak{s}_{\mathrm{i}},a_{1}}^{-3} \delta_{a_{1}a_{2}} \delta_{a_{1}a_{3}} \mathrlap{\quad \forall a_{1},a_{2},a_{3} \;,}
\label{eq:standard2PPIFRGICExpressionGamma030DON}
\end{equation}
\begin{equation}
\overline{\Gamma}_{\mathfrak{s}=\mathfrak{s}_{\mathrm{i}},a_{1}a_{2}a_{3}a_{4}}^{(\mathrm{2PPI})(4)} = 3 \overline{\rho}_{\mathfrak{s}=\mathfrak{s}_{\mathrm{i}},a_{1}}^{-4} \delta_{a_{1}a_{2}} \delta_{a_{1}a_{3}} \delta_{a_{1}a_{4}} \mathrlap{\quad \forall a_{1},a_{2},a_{3},a_{4} \;,}
\label{eq:standard2PPIFRGICExpressionGamma040DON}
\end{equation}
\begin{equation}
\hspace{3.6cm} \overline{\Gamma}_{\mathfrak{s}=\mathfrak{s}_{\mathrm{i}},a_{1}a_{2}a_{3}a_{4}a_{5}}^{(\mathrm{2PPI})(5)} = - 12 \overline{\rho}_{\mathfrak{s}=\mathfrak{s}_{\mathrm{i}},a_{1}}^{-5} \delta_{a_{1}a_{2}} \delta_{a_{1}a_{3}} \delta_{a_{1}a_{4}} \delta_{a_{1}a_{5}} \quad \forall a_{1},a_{2},a_{3},a_{4},a_{5} \;,
\label{eq:standard2PPIFRGICExpressionGamma050DON}
\end{equation}
\begin{equation}
\hspace{3.7cm} \overline{\Gamma}_{\mathfrak{s}=\mathfrak{s}_{\mathrm{i}},a_{1}a_{2}a_{3}a_{4}a_{5}a_{6}}^{(\mathrm{2PPI})(6)} = 120 \overline{\rho}_{\mathfrak{s}=\mathfrak{s}_{\mathrm{i}},a_{1}}^{-6} \delta_{a_{1}a_{2}} \delta_{a_{1}a_{3}} \delta_{a_{1}a_{4}} \delta_{a_{1}a_{5}} \delta_{a_{1}a_{6}} \quad \forall a_{1},a_{2},a_{3},a_{4},a_{5},a_{6} \;.
\label{eq:standard2PPIFRGICExpressionGamma060DON}
\end{equation}
The infinite tower of differential equations containing~\eqref{eq:standard2PPIFRGflowEqExpressionGammadot} to~\eqref{eq:standard2PPIFRGflowEqExpressionGamma2dot} is truncated according to the three following conditions (given by~\eqref{eq:DefinitionsUflow2PPIFRG} to~\eqref{eq:DefinitioniUflow2PPIFRG} in our general presentation of the standard 2PPI-FRG):
\begin{itemize}
\item For the sU-flow:
\begin{equation}
\overline{\Gamma}_{\mathfrak{s}}^{(\mathrm{2PPI})(n)} = 0 \mathrlap{\quad \forall \mathfrak{s}, ~ \forall n > N_{\mathrm{max}} \;.}
\label{eq:DefinitionsUflow2PPIFRG0DON}
\end{equation}

\item For the pU-flow:
\begin{equation}
\overline{\Gamma}_{\mathfrak{s}}^{(\mathrm{2PPI})(n)} = \overline{\Gamma}_{\mathfrak{s}=\mathfrak{s}_{\mathrm{i}}}^{(\mathrm{2PPI})(n)} \mathrlap{\quad \forall \mathfrak{s}, ~ \forall n > N_{\mathrm{max}} \;.}
\label{eq:DefinitionpUflow2PPIFRG0DON}
\end{equation}

\item For the iU-flow:
\begin{equation}
\overline{\Gamma}_{\mathfrak{s}}^{(\mathrm{2PPI})(n)} = \left.\overline{\Gamma}_{\mathfrak{s}=\mathfrak{s}_{\mathrm{i}}}^{(\mathrm{2PPI})(n)}\right|_{\overline{\rho}_{\mathfrak{s}=\mathfrak{s}_{\mathrm{i}}}\rightarrow\overline{\rho}_{\mathfrak{s}}} \mathrlap{\quad \forall \mathfrak{s}, ~ \forall n > N_{\mathrm{max}} \;.}
\label{eq:DefinitioniUflow2PPIFRG0DON}
\end{equation}

\end{itemize}

\vspace{0.3cm}

In particular, the substitution $m^{2}[\overline{\rho}_{0}] \rightarrow m^{2}[\overline{\rho}_{\mathfrak{s}}]$ in~\eqref{eq:DefinitioniUflow2PPIFRG} has been replaced by $\overline{\rho}_{\mathfrak{s}=\mathfrak{s}_{\mathrm{i}}}\rightarrow\overline{\rho}_{\mathfrak{s}}$ in its zero-dimensional counterpart given by~\eqref{eq:DefinitioniUflow2PPIFRG0DON}. This can be explained as follows: for the studied (0+0)-D model, $m^{2}(\overline{\rho}_{0})=1/\overline{\rho}_{0} = (1/N) \sum_{a=1}^{N} 1/\overline{\rho}_{\mathfrak{s}=\mathfrak{s}_{\mathrm{i}},a}$ according to~\eqref{eq:standard2PPIFRGICExpressionrho00DON}, which implies that the substitution $m^{2}(\overline{\rho}_{0}) \rightarrow m^{2}(\overline{\rho}_{\mathfrak{s}})= (1/N) \sum_{a=1}^{N} 1/\overline{\rho}_{\mathfrak{s},a}$ is equivalent to $\overline{\rho}_{\mathfrak{s}=\mathfrak{s}_{\mathrm{i}}}\rightarrow\overline{\rho}_{\mathfrak{s}}$. Note that we could not directly replace $\overline{\rho}_{\mathfrak{s}=\mathfrak{s}_{\mathrm{i}}}$ for finite-dimensional systems since $\overline{\Gamma}_{\mathfrak{s}=\mathfrak{s}_{\mathrm{i}}}^{(\mathrm{2PPI})(n)}$ can not be written explicitly in terms of $\overline{\rho}_{\mathfrak{s}=\mathfrak{s}_{\mathrm{i}}}$ in such cases. Finally, let us point out that the gs energy and density are both directly obtained at the end of the flow via the relations:
\begin{equation}
E^\text{s2PPI-FRG}_{\mathrm{gs}} = \overline{\Gamma}_{\mathfrak{s}=\mathfrak{s}_{\mathrm{f}}}^{(\mathrm{2PPI})} \;,
\label{eq:deduceEgsstandard2PPIFRG}
\end{equation}
\begin{equation}
\rho^\text{s2PPI-FRG}_{\mathrm{gs}} = \frac{1}{N} \sum_{a=1}^{N} \overline{\rho}_{\mathfrak{s}=\mathfrak{s}_{\mathrm{f}},a} \;.
\label{eq:deducerhogsstandard2PPIFRG}
\end{equation}
The application of the standard 2PPI-FRG designed since the beginning of section~\ref{sec:standard2PPIFRG0DON} is only able to treat the unbroken-symmetry regime of our toy model. It can actually be readily extended to tackle the broken-symmetry one by using a CU-flow implementation, as will be illustrated later with the most performing 2PPI-FRG approach tested in this study, i.e. the KS-FRG.

\vspace{0.5cm}

\begin{figure}[!t]
\captionsetup[subfigure]{labelformat=empty}
  \begin{center}
    \subfloat[]{
      \includegraphics[width=0.50\linewidth]{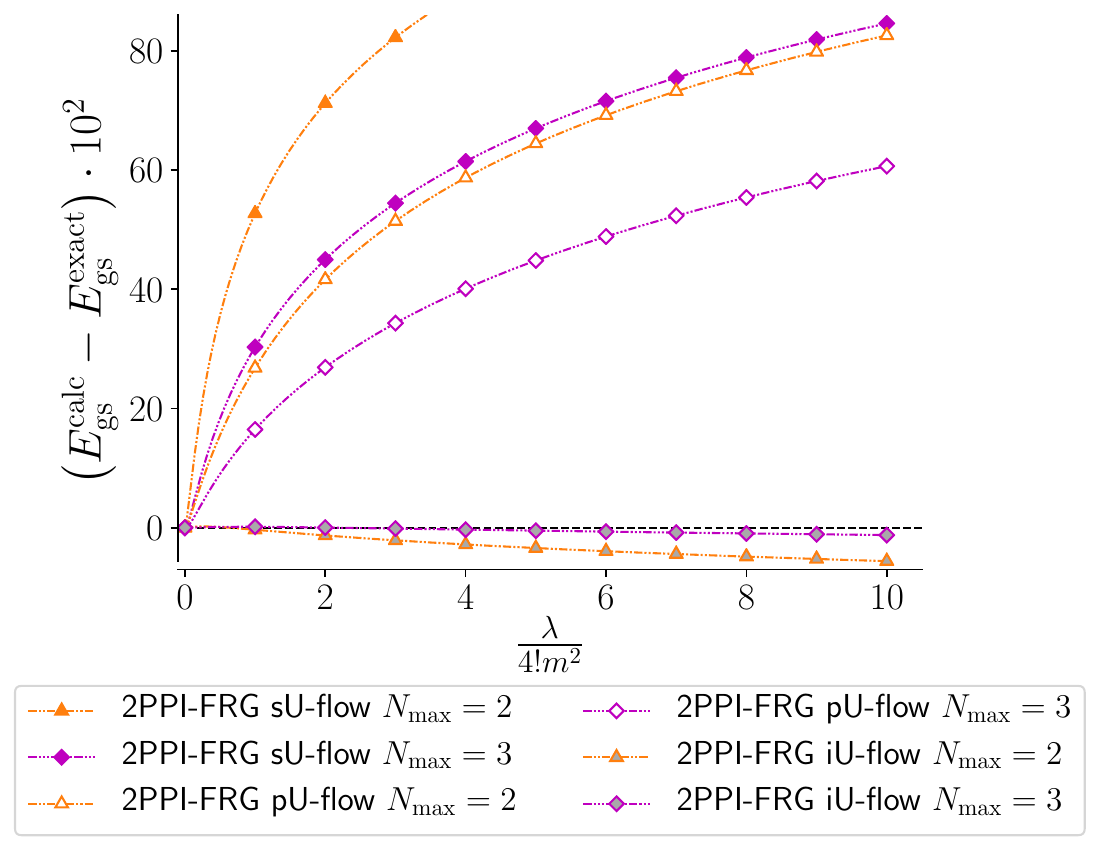}
                         }
    \subfloat[]{
      \includegraphics[width=0.50\linewidth]{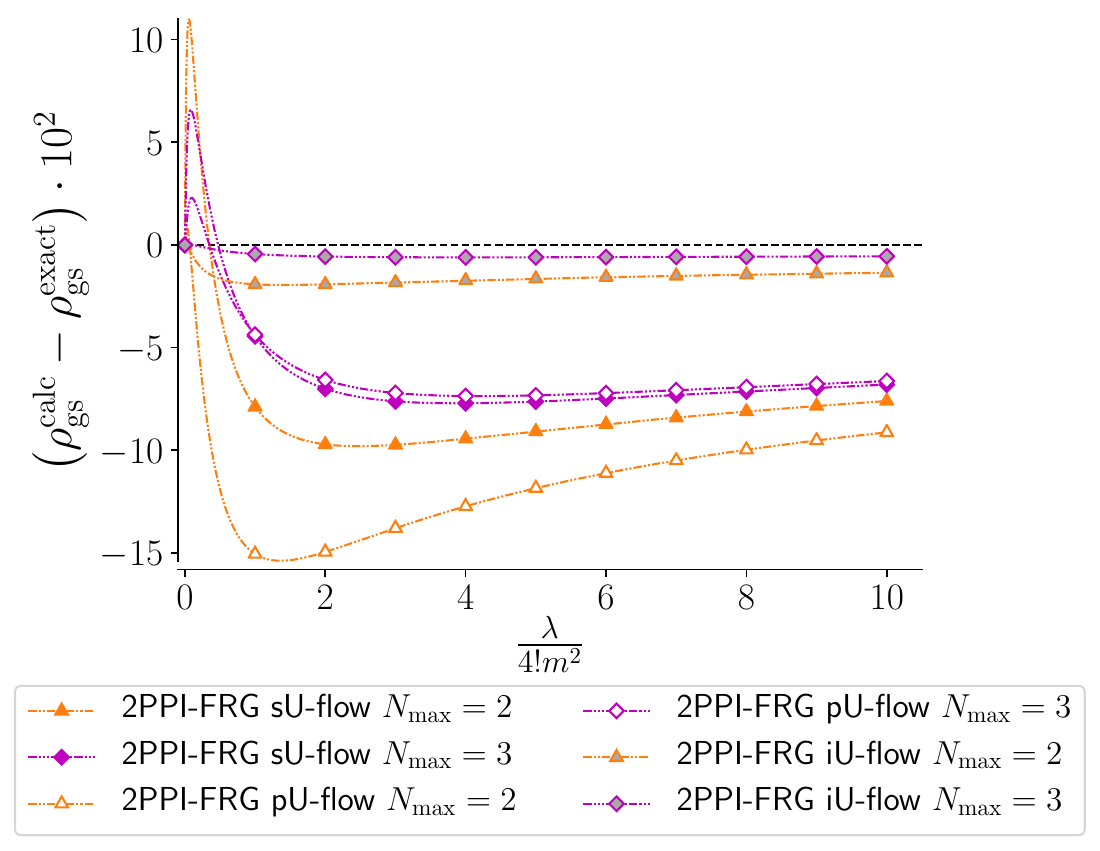}
                         }
\caption{Difference between the calculated gs energy $E_{\mathrm{gs}}^{\mathrm{calc}}$ or density $\rho_{\mathrm{gs}}^{\mathrm{calc}}$ and the corresponding exact solution $E_{\mathrm{gs}}^{\mathrm{exact}}$ or $\rho_{\mathrm{gs}}^{\mathrm{exact}}$ at $m^{2}=+1$ and $N=1$ ($\mathcal{R}e(\lambda)\geq 0$ and $\mathcal{I}m(\lambda)=0$).}
\label{fig:standard2PPIFRGallVersionslambdaN1}
  \end{center}
\end{figure}
\begin{figure}[!t]
\captionsetup[subfigure]{labelformat=empty}
  \begin{center}
    \subfloat[]{
      \includegraphics[width=0.50\linewidth]{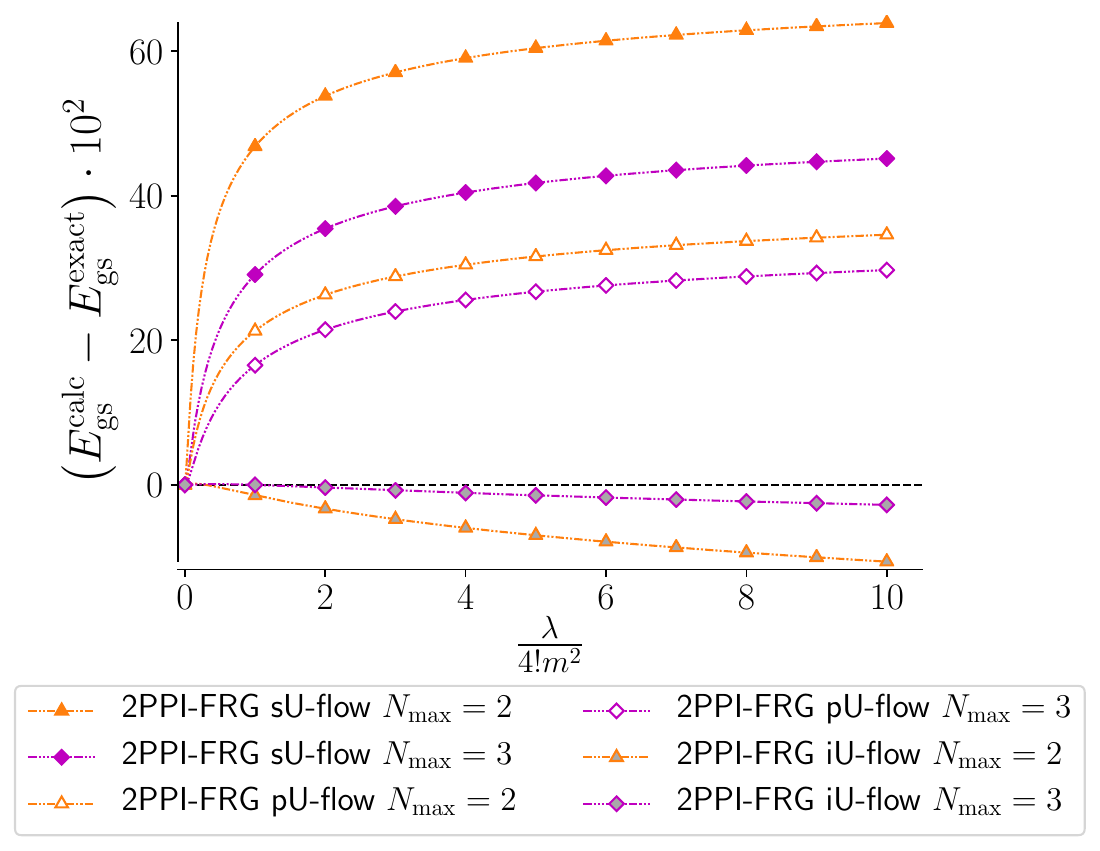}
                         }
    \subfloat[]{
      \includegraphics[width=0.50\linewidth]{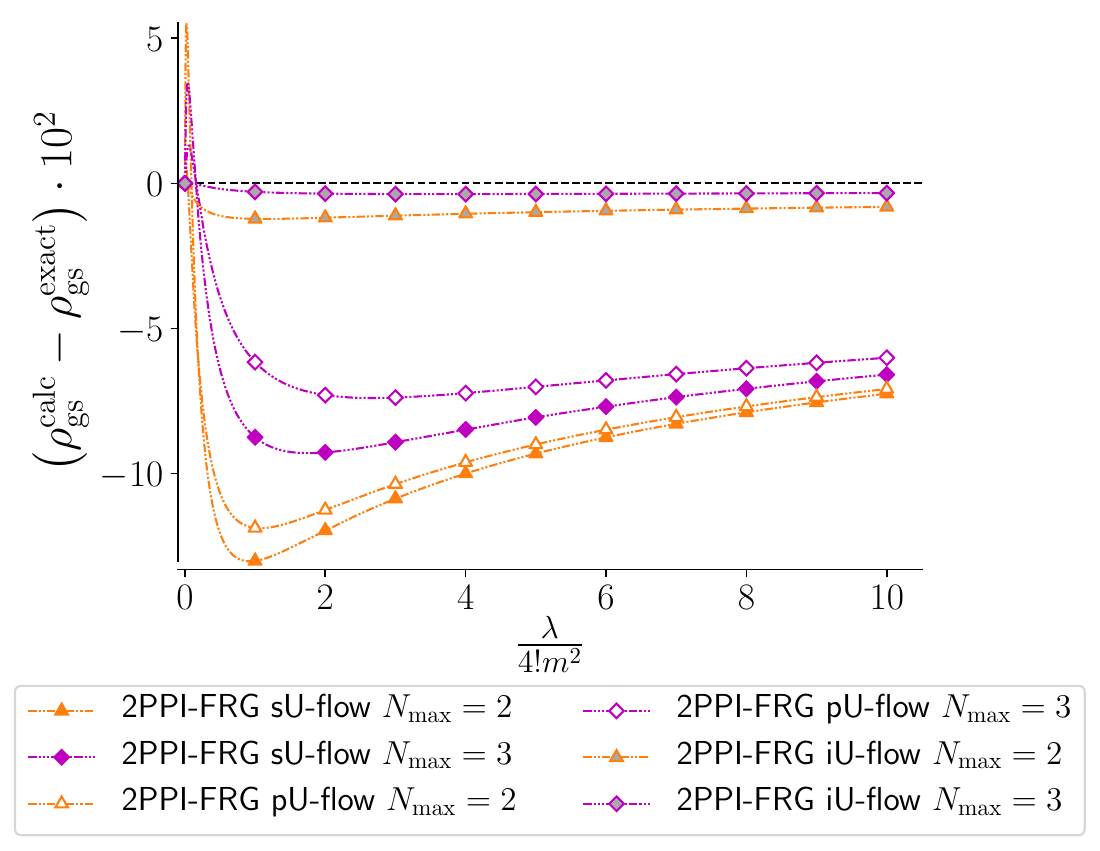}
                         }
\caption{Same as fig.~\ref{fig:standard2PPIFRGallVersionslambdaN1} with $N=2$ instead.}
\label{fig:standard2PPIFRGallVersionslambdaN2}
  \end{center}
\end{figure}

To summarize, our calculations for the standard 2PPI-FRG up to $N_{\mathrm{max}}=2$ are carried out by solving the differential equations~\eqref{eq:standard2PPIFRGflowEqExpressionGammadot} to~\eqref{eq:standard2PPIFRGflowEqExpressionGamma2dot}, with initial conditions given by~\eqref{eq:standard2PPIFRGICExpressionrho00DON} to~\eqref{eq:standard2PPIFRGICExpressionGamma040DON}. The associated truncation conditions are set by~\eqref{eq:DefinitionsUflow2PPIFRG0DON},~\eqref{eq:DefinitionpUflow2PPIFRG0DON} and~\eqref{eq:DefinitioniUflow2PPIFRG0DON} for the sU-flow, the pU-flow and the iU-flow respectively. Note also that the chosen cutoff function for the two-body interaction is expressed by~\eqref{eq:choiceCutoffUs2PPIFRG0DON}. The sU-flow can not be implemented at $N_{\mathrm{max}}=1$ since~\eqref{eq:DefinitionsUflow2PPIFRG0DON} implies that $\overline{\boldsymbol{G}}^{-1}_{\mathfrak{s},a_{1}a_{2}}=\overline{\Gamma}_{\mathfrak{s},a_{1}a_{2}}^{(\mathrm{2PPI})(2)}=0$ $\forall a_{1},a_{2},\mathfrak{s}$ at this truncation order, which renders the associated flow equations (i.e.~\eqref{eq:standard2PPIFRGflowEqExpressionGammadot} and~\eqref{eq:standard2PPIFRGflowEqExpressionrhodot}) ill-defined. We thus compare the three aforementioned truncation schemes of the standard 2PPI-FRG at $N=1$ and $N=2$, in figs.~\ref{fig:standard2PPIFRGallVersionslambdaN1} and~\ref{fig:standard2PPIFRGallVersionslambdaN2} respectively. From these two figures and for both $E_{\mathrm{gs}}$ and $\rho_{\mathrm{gs}}$, it can clearly be seen that the sU-flow is outperformed by the pU-flow which is itself less performing than the iU-flow. This was expected from definitions~\eqref{eq:DefinitionsUflow2PPIFRG0DON} to~\eqref{eq:DefinitioniUflow2PPIFRG0DON} that put forward the iU-flow as the most refined truncation scheme among the three. As can be seen from figs.~\ref{fig:standard2PPIFRGallVersionslambdaN1} and~\ref{fig:standard2PPIFRGallVersionslambdaN2} as well as from figs.~\ref{fig:ComparisonKSFRGpUflowiUflowlambdaN1} and~\ref{fig:ComparisonKSFRGpUflowiUflowlambdaN2}, the iU-flow results for the gs energy and density are systematically improved from $N_{\mathrm{max}}=1$ to $N_{\mathrm{max}}=3$. This could be regarded as a surprising feature considering the similarity between the truncation conditions~\eqref{eq:DefinitioniUflow2PPIFRG} for the iU-flow and~\eqref{eq:2PIfrgTruncationmCflow} (with~\eqref{eq:DefPhi2sym2PIFRGmCflow}) for the mC-flow implementation of the 2PI-FRG which did not prove to be that reliable.

\subsubsection{Kohn-Sham functional renormalization group}
\label{sec:KSFRG0DON}

From~\eqref{eq:2PPIFRGTwoInteraction0DON} to~\eqref{eq:choiceCutoffUs2PPIFRG0DON}, we can show that the master equation of the KS-FRG expressed by~\eqref{eq:MasterEquationKSFRGfinalversion} reduces for our zero-dimensional $O(N)$ model to:
\begin{equation}
\dot{\gamma}_{\mathfrak{s}}(\rho) = \frac{1}{2}\dot{V}_{\mathfrak{s}} \sum_{a_{1}=1}^{N} \rho_{a_{1}} + \sum_{a_{1},a_{2}=1}^{N} \rho_{a_{1}} \overline{\Gamma}^{(2)}_{\mathrm{KS},\mathfrak{s},a_{1}a_{2}} \dot{\overline{\rho}}_{\mathfrak{s},a_{2}} + \frac{\lambda}{24} \left( \sum_{a_{1},a_{2}=1}^{N} \boldsymbol{G}_{\mathfrak{s},a_{1}a_{2}}(\rho) + \sum_{a_{1},a_{2}=1}^{N} \rho_{a_{1}} \rho_{a_{2}} \right) \;,
\label{eq:KSFRGmasterEquation0DON}
\end{equation}
where
\begin{equation}
\overline{\Gamma}_{\mathrm{KS},\mathfrak{s}} \equiv \Gamma_{\mathrm{KS},\mathfrak{s}}\big(\rho=\overline{\rho}_{\mathfrak{s}}\big) \mathrlap{\quad \forall \mathfrak{s} \;,}
\label{eq:DefGammaKSbarpureKSFRG0DON}
\end{equation}
\begin{equation}
\overline{\Gamma}^{(n)}_{\mathrm{KS},\mathfrak{s},a_{1} \cdots a_{n}} \equiv \left. \frac{\partial^{n} \Gamma_{\mathrm{KS},\mathfrak{s}}(\rho)}{\partial \rho_{a_{1}} \cdots \partial \rho_{a_{n}}} \right|_{\rho=\overline{\rho}_{\mathfrak{s}}} \mathrlap{\quad \forall a_{1},\cdots,a_{n},\mathfrak{s} \;,}
\label{eq:DefGammaKSnbarpureKSFRG0DON}
\end{equation}
and the propagator $\boldsymbol{G}_{\mathfrak{s}}(\rho)$ is already defined by~\eqref{eq:standard2PPIFRGvertExpPropagator} which can be put in the form:
\begin{equation}
\boldsymbol{G}^{-1}_{\mathfrak{s},a_{1}a_{2}}(\rho) \equiv \Gamma_{\mathfrak{s},a_{1}a_{2}}^{(\mathrm{2PPI})(2)}(\rho) = \Gamma_{\mathrm{KS},\mathfrak{s},a_{1}a_{2}}^{(2)}(\rho) + \gamma^{(2)}_{\mathfrak{s},a_{1}a_{2}}(\rho) \;.
\label{eq:KSFRGvertExpPropagator}
\end{equation}
As opposed to our previous discussion on the standard 2PPI-FRG, we assume that the one-body potential $V_{\mathfrak{s}}$ can vary throughout the flow to treat the regime with $m^{2}<0$, hence exploiting a CU-flow approach. More specifically, the free part of the classical action~\eqref{eq:ClassicalAction2PPIFRG} now satisfies:
\begin{equation}
S_{0}\Big[\widetilde{\psi}^{\dagger},\widetilde{\psi}\Big] = \int_{\alpha} \widetilde{\psi}_{\alpha}^{\dagger}\left(\hat{O}_{\mathrm{kin},\alpha} + V_{\alpha} - \mu\right)\widetilde{\psi}_{\alpha} = \frac{1}{2} V \sum_{a=1}^{N} \widetilde{\varphi}_{a}^{2} = \frac{1}{2} m^{2} \vec{\widetilde{\varphi}}^{2} \;,
\end{equation}
with $V = m^{2}$, and we choose the following flow-dependent one-body potential:
\begin{equation}
V_{\mathfrak{s}} = \left\{
\begin{array}{lll}
        \displaystyle{m^{2} \quad \forall m^{2} > 0 \;,} \\
        \\
        \displaystyle{\left(2\mathfrak{s} - 1\right)m^{2} \quad \forall m^{2} < 0 \;,}
    \end{array}
\right.
\label{eq:choiceVsKSFRG0DON}
\end{equation}
alongside with~\eqref{eq:choiceCutoffUs2PPIFRG0DON} for the flow-dependent two-body interaction $U_{\mathfrak{s}}$, still using $\mathfrak{s}_{\mathrm{i}}=0$ and $\mathfrak{s}_{\mathrm{f}}=1$ as boundary values for $\mathfrak{s}$. Hence, the derivative $\dot{V}_{\mathfrak{s}}$ vanishes in the regime with $m^{2}>0$ in which case we thus recover a U-flow formulation. Actually, our choice~\eqref{eq:choiceVsKSFRG0DON} for $m^{2}<0$ has nothing to do with a CU-flow implementation designed so that the starting point of the flow coincides with the Kohn-Sham system (as described right above~\eqref{eq:2PPIFRGflowEquationGammaUflow}). It is simply a trick to tackle the regime with $m^{2}<0$ for the studied (0+0)-D model by using the free theory with squared mass $-m^{2}>0$ as starting point for the flow ($V_{\mathfrak{s}=\mathfrak{s}_{\mathrm{i}}}=-m^{2}$ at $\mathfrak{s}_{\mathrm{i}} = 0$ for $m^{2}<0$ according to~\eqref{eq:choiceVsKSFRG0DON}). We thus circumvent in this way the problem of divergence of the corresponding partition function or Schwinger functional at $\lambda=0$ and $m^{2}<0$, which restricts the U-flow implementation of all 2PPI-FRG approaches treated here to the unbroken-symmetry phase. Note that this trick can also be exploited to tackle the latter phase with the pU-flow implementation of the 2PI-FRG.

\vspace{0.5cm}

Furthermore, the vertex expansion recipe is applied to~\eqref{eq:KSFRGmasterEquation0DON} using notably the Taylor expansion of $\gamma_{\mathfrak{s}}(\rho)$ around its extremum at $\rho=\overline{\rho}_{\mathfrak{s}}$, i.e.:
\begin{equation}
\gamma_{\mathfrak{s}}(\rho) = \overline{\gamma}_{\mathfrak{s}} + \sum_{n=2}^{\infty} \frac{1}{n!} \sum_{a_{1},\cdots,a_{n}=1}^{N} \overline{\gamma}_{\mathfrak{s},a_{1} \cdots a_{n}}^{(n)} \left(\rho-\overline{\rho}_{\mathfrak{s}}\right)_{a_{1}} \cdots \left(\rho-\overline{\rho}_{\mathfrak{s}}\right)_{a_{n}} \;,
\label{eq:KSFRGVertexExpansion}
\end{equation}
with
\begin{equation}
\overline{\gamma}_{\mathfrak{s}} \equiv \gamma_{\mathfrak{s}}\big(\rho=\overline{\rho}_{\mathfrak{s}}\big) \mathrlap{\quad \forall \mathfrak{s} \;,}
\label{eq:DefgammabarpureKSFRG0DON}
\end{equation}
\begin{equation}
\overline{\gamma}^{(n)}_{\mathfrak{s},a_{1} \cdots a_{n}} \equiv \left. \frac{\partial^{n} \gamma_{\mathfrak{s}}(\rho)}{\partial \rho_{a_{1}} \cdots \partial \rho_{a_{n}}} \right|_{\rho=\overline{\rho}_{\mathfrak{s}}} \mathrlap{\quad \forall a_{1},\cdots,a_{n},\mathfrak{s} \;,}
\label{eq:DefgammanbarpureKSFRG0DON}
\end{equation}
and
\begin{equation}
\overline{\gamma}_{\mathfrak{s},a}^{(1)} = 0 \mathrlap{\quad \forall a, \mathfrak{s} \;.}
\label{eq:Defgamma1barpureKSFRG0DON}
\end{equation}
This vertex expansion leads as usual to an infinite tower of differential equations which contains for instance (see appendix~\ref{sec:KSFRGVertexExpansionAppendix} for the corresponding flow equations expressing the derivatives $\dot{\overline{\gamma}}_{\mathfrak{s}}^{(3)}$ and $\dot{\overline{\gamma}}_{\mathfrak{s}}^{(4)}$):
\begin{equation}
\dot{\overline{\gamma}}_{\mathfrak{s}} = \frac{1}{2}\dot{V}_{\mathfrak{s}} \sum_{a_{1}=1}^{N} \overline{\rho}_{\mathfrak{s},a_{1}} + \frac{1}{2} \sum_{a_{1}=1}^{N} \overline{\rho}^{-1}_{\mathfrak{s},a_{1}} \dot{\overline{\rho}}_{\mathfrak{s},a_{1}} + \frac{\lambda}{24} \sum_{a_{1},a_{2}=1}^{N} \left( \overline{\boldsymbol{G}}_{\mathfrak{s},a_{1}a_{2}} + \overline{\rho}_{\mathfrak{s},a_{1}} \overline{\rho}_{\mathfrak{s},a_{2}} \right) \;,
\label{eq:KSFRGEqExpressionGammadot}
\end{equation}
\begin{equation}
\begin{split}
\dot{\overline{\rho}}_{\mathfrak{s},a_{1}} = & - \frac{1}{2}\dot{V}_{\mathfrak{s}} \sum_{a_{2}=1}^{N} \overline{\boldsymbol{G}}_{\mathfrak{s},a_{1}a_{2}} \\
& + \frac{\lambda}{24} \sum_{a_{2}=1}^{N} \overline{\boldsymbol{G}}_{\mathfrak{s},a_{1}a_{2}} \left( \sum_{a_{3},a_{4},a_{5},a_{6}=1}^{N} \overline{\boldsymbol{G}}_{\mathfrak{s},a_{3}a_{5}} \left(\overline{\Gamma}_{\mathrm{KS},\mathfrak{s}}^{(3)}+\overline{\gamma}_{\mathfrak{s}}^{(3)}\right)_{a_{2}a_{5}a_{6}} \overline{\boldsymbol{G}}_{\mathfrak{s},a_{6}a_{4}} - 2 \sum_{a_{3}=1}^{N} \overline{\rho}_{\mathfrak{s},a_{3}} \right) \;,
\end{split}
\label{eq:KSFRGEqExpressionrhodot}
\end{equation}
\begin{equation}
\begin{split}
\scalebox{0.99}{${\displaystyle \dot{\overline{\gamma}}_{\mathfrak{s},a_{1}a_{2}}^{(2)} = }$} & \ \scalebox{0.99}{${\displaystyle \sum_{a_{3}=1}^{N} \dot{\overline{\rho}}_{\mathfrak{s},a_{3}} \overline{\gamma}^{(3)}_{\mathfrak{s},a_{3}a_{1}a_{2}} }$} \\
& \scalebox{0.99}{${\displaystyle + \frac{\lambda}{24} \Bigg( 2 + 2 \sum_{a_{3},a_{4},a_{5},a_{6},a_{7},a_{8}=1}^{N} \overline{\boldsymbol{G}}_{\mathfrak{s},a_{3}a_{5}} \left(\overline{\Gamma}_{\mathrm{KS},\mathfrak{s}}^{(3)} + \overline{\gamma}^{(3)}_{\mathfrak{s}}\right)_{a_{1}a_{5}a_{6}} \overline{\boldsymbol{G}}_{\mathfrak{s},a_{6}a_{7}} \left(\overline{\Gamma}_{\mathrm{KS},\mathfrak{s}}^{(3)} + \overline{\gamma}^{(3)}_{\mathfrak{s}}\right)_{a_{2}a_{7}a_{8}} \overline{\boldsymbol{G}}_{\mathfrak{s},a_{8}a_{4}} }$} \\
& \ \hspace{0.85cm} \scalebox{0.99}{${\displaystyle - \sum_{a_{3},a_{4},a_{5},a_{6}=1}^{N} \overline{\boldsymbol{G}}_{\mathfrak{s},a_{3}a_{5}} \left(\overline{\Gamma}_{\mathrm{KS},\mathfrak{s}}^{(4)} + \overline{\gamma}^{(4)}_{\mathfrak{s}}\right)_{a_{1}a_{2}a_{5}a_{6}} \overline{\boldsymbol{G}}_{\mathfrak{s},a_{6}a_{4}} \Bigg) \;. }$}
\end{split}
\label{eq:KSFRGEqExpressiongamma2dot}
\end{equation}
The initial conditions used to solve the resulting equation systems are:
\begin{equation}
\overline{\gamma}_{\mathfrak{s}=\mathfrak{s}_{\mathrm{i}}} = 0 \mathrlap{\;,}
\label{eq:InitialCondCorrgammaKSFRG0DON}
\end{equation}
\begin{equation}
\overline{\gamma}^{(n)}_{\mathfrak{s}=\mathfrak{s}_{\mathrm{i}},a_{1}, \cdots, a_{n}} = 0 \mathrlap{\quad \forall a_{1}, \cdots, a_{n}, ~ \forall n \geq 2 \;,}
\label{eq:InitialCondCorrgammanKSFRG0DON}
\end{equation}
and these equation systems are closed with the condition:
\begin{equation}
\overline{\gamma}_{\mathfrak{s}}^{(n)} = \overline{\gamma}_{\mathfrak{s}=\mathfrak{s}_{\mathrm{i}}}^{(n)} \mathrlap{\quad \forall \mathfrak{s}, ~ \forall n > N_{\mathrm{max}} \;,}
\label{eq:KSFRGtruncationScheme0DON}
\end{equation}
after choosing a truncation order $N_{\mathrm{max}}$. Finally, an expression for the mean-field part $\Gamma_{\mathrm{KS},\mathfrak{s}}(\rho)$ can be directly inferred from the free 2PI EA expressed by~\eqref{eq:free2PIEAforCI2PPIFRG0DON}, after performing the substitutions $\boldsymbol{G}_{a_{1}a_{2}}\rightarrow\rho_{a_{1}}\delta_{a_{1}a_{2}}$ and $C^{-1}_{a_{1}a_{2}} \rightarrow V_{\mathrm{KS},\mathfrak{s},a_{1}}\delta_{a_{1}a_{2}}$ (thus introducing the Kohn-Sham potential at the expense of the free propagator). This leads to:
\begin{equation}
\Gamma_{\mathrm{KS},\mathfrak{s}}(\rho) = - \frac{1}{2} \sum_{a=1}^{N} \ln(2\pi\rho_{a}) + \frac{1}{2} \sum_{a=1}^{N} V_{\mathrm{KS},\mathfrak{s},a} \rho_{a} - \frac{N}{2} \;.
\label{eq:GammaKSsexpressionKSFRG0DON}
\end{equation}
We directly deduce from~\eqref{eq:GammaKSsexpressionKSFRG0DON} the following derivatives:
\begin{equation}
\overline{\Gamma}_{\mathrm{KS},\mathfrak{s},a}^{(1)} = -\frac{1}{2} \overline{\rho}_{\mathfrak{s},a}^{-1} + \frac{1}{2} V_{\mathrm{KS},\mathfrak{s},a} \mathrlap{\quad \forall a,\mathfrak{s} \;,}
\label{eq:KSFRGExpressionGammaKS10DON}
\end{equation}
\begin{equation}
\overline{\Gamma}_{\mathrm{KS},\mathfrak{s},a_{1}a_{2}}^{(2)} = \frac{1}{2} \overline{\rho}_{\mathfrak{s},a_{1}}^{-2} \delta_{a_{1}a_{2}} \mathrlap{\quad \forall a_{1},a_{2},\mathfrak{s} \;,}
\label{eq:KSFRGExpressionGammaKS20DON}
\end{equation}
\begin{equation}
\overline{\Gamma}_{\mathrm{KS},\mathfrak{s},a_{1}a_{2}a_{3}}^{(3)} = - \overline{\rho}_{\mathfrak{s},a_{1}}^{-3} \delta_{a_{1}a_{2}} \delta_{a_{1}a_{3}} \mathrlap{\quad \forall a_{1},a_{2},a_{3},\mathfrak{s} \;,}
\label{eq:KSFRGExpressionGammaKS30DON}
\end{equation}
\begin{equation}
\overline{\Gamma}_{\mathrm{KS},\mathfrak{s},a_{1}a_{2}a_{3}a_{4}}^{(4)} = 3 \overline{\rho}_{\mathfrak{s},a_{1}}^{-4} \delta_{a_{1}a_{2}} \delta_{a_{1}a_{3}} \delta_{a_{1}a_{4}} \mathrlap{\quad \forall a_{1},a_{2},a_{3},a_{4},\mathfrak{s} \;,}
\label{eq:KSFRGExpressionGammaKS40DON}
\end{equation}
\begin{equation}
\hspace{3.7cm} \overline{\Gamma}_{\mathrm{KS},\mathfrak{s},a_{1}a_{2}a_{3}a_{4}a_{5}}^{(5)} = - 12 \overline{\rho}_{\mathfrak{s},a_{1}}^{-5} \delta_{a_{1}a_{2}} \delta_{a_{1}a_{3}} \delta_{a_{1}a_{4}} \delta_{a_{1}a_{5}} \quad \forall a_{1},a_{2},a_{3},a_{4},a_{5},\mathfrak{s} \;,
\label{eq:KSFRGExpressionGammaKS50DON}
\end{equation}
\begin{equation}
\hspace{3.7cm} \overline{\Gamma}_{\mathrm{KS},\mathfrak{s},a_{1}a_{2}a_{3}a_{4}a_{5}a_{6}}^{(6)} = 120 \overline{\rho}_{\mathfrak{s},a_{1}}^{-6} \delta_{a_{1}a_{2}} \delta_{a_{1}a_{3}} \delta_{a_{1}a_{4}} \delta_{a_{1}a_{5}} \delta_{a_{1}a_{6}} \quad \forall a_{1},a_{2},a_{3},a_{4},a_{5},a_{6},\mathfrak{s} \;,
\label{eq:KSFRGExpressionGammaKS60DON}
\end{equation}
which coincide with the corresponding derivatives of the free version of the 2PPI EA $\Gamma^{(\mathrm{2PPI})}_{\mathfrak{s}}(\rho)$ at $n \geq 2$ (as can be seen up to $n=6$ by comparing~\eqref{eq:KSFRGExpressionGammaKS20DON} to~\eqref{eq:KSFRGExpressionGammaKS60DON} with~\eqref{eq:standard2PPIFRGICExpressionGamma020DON} to~\eqref{eq:standard2PPIFRGICExpressionGamma060DON}). We then consider~\eqref{eq:KSFRGExpressionGammaKS10DON} together with the condition of extremization of $\Gamma_{\mathrm{KS},\mathfrak{s}}(\rho)$:
\begin{equation}
\overline{\Gamma}_{\mathrm{KS},\mathfrak{s},a}^{(1)} = 0 \mathrlap{\quad \forall a, \mathfrak{s} \;,}
\end{equation}
which enables us to find the following simple expression for the flow-dependent Kohn-Sham potential:
\begin{equation}
V_{\mathrm{KS},\mathfrak{s},a} = \overline{\rho}_{\mathfrak{s},a}^{-1} \mathrlap{\quad \forall a, \mathfrak{s} \;.}
\label{eq:KSpotentialKSFRG0DON}
\end{equation}
Plugging~\eqref{eq:KSpotentialKSFRG0DON} into~\eqref{eq:GammaKSsexpressionKSFRG0DON} leads to:
\begin{equation}
\overline{\Gamma}_{\mathrm{KS},\mathfrak{s}} = - \frac{1}{2} \sum_{a=1}^{N} \ln\big(2\pi\overline{\rho}_{\mathfrak{s},a}\big) \;,
\end{equation}
and we can thus deduce the gs energy at the end of the flow by using the relation:
\begin{equation}
E^\text{KS-FRG}_{\mathrm{gs}} = \overline{\Gamma}_{\mathfrak{s}=\mathfrak{s}_{\mathrm{f}}}^{(\mathrm{2PPI})} = \overline{\Gamma}_{\mathrm{KS},\mathfrak{s}=\mathfrak{s}_{\mathrm{f}}} + \overline{\gamma}_{\mathfrak{s}=\mathfrak{s}_{\mathrm{f}}} = - \frac{1}{2} \sum_{a=1}^{N} \ln\big(2\pi\overline{\rho}_{\mathfrak{s}=\mathfrak{s}_{\mathrm{f}},a}\big) + \overline{\gamma}_{\mathfrak{s}=\mathfrak{s}_{\mathrm{f}}} \;,
\end{equation}
whereas the gs density is obtained from:
\begin{equation}
\rho^\text{KS-FRG}_{\mathrm{gs}} = \frac{1}{N} \sum_{a=1}^{N} \overline{\rho}_{\mathfrak{s}=\mathfrak{s}_{\mathrm{f}},a} \;.
\end{equation}

\pagebreak

\begin{figure}[!t]
\captionsetup[subfigure]{labelformat=empty}
  \begin{center}
    \subfloat[]{
      \includegraphics[width=0.50\linewidth]{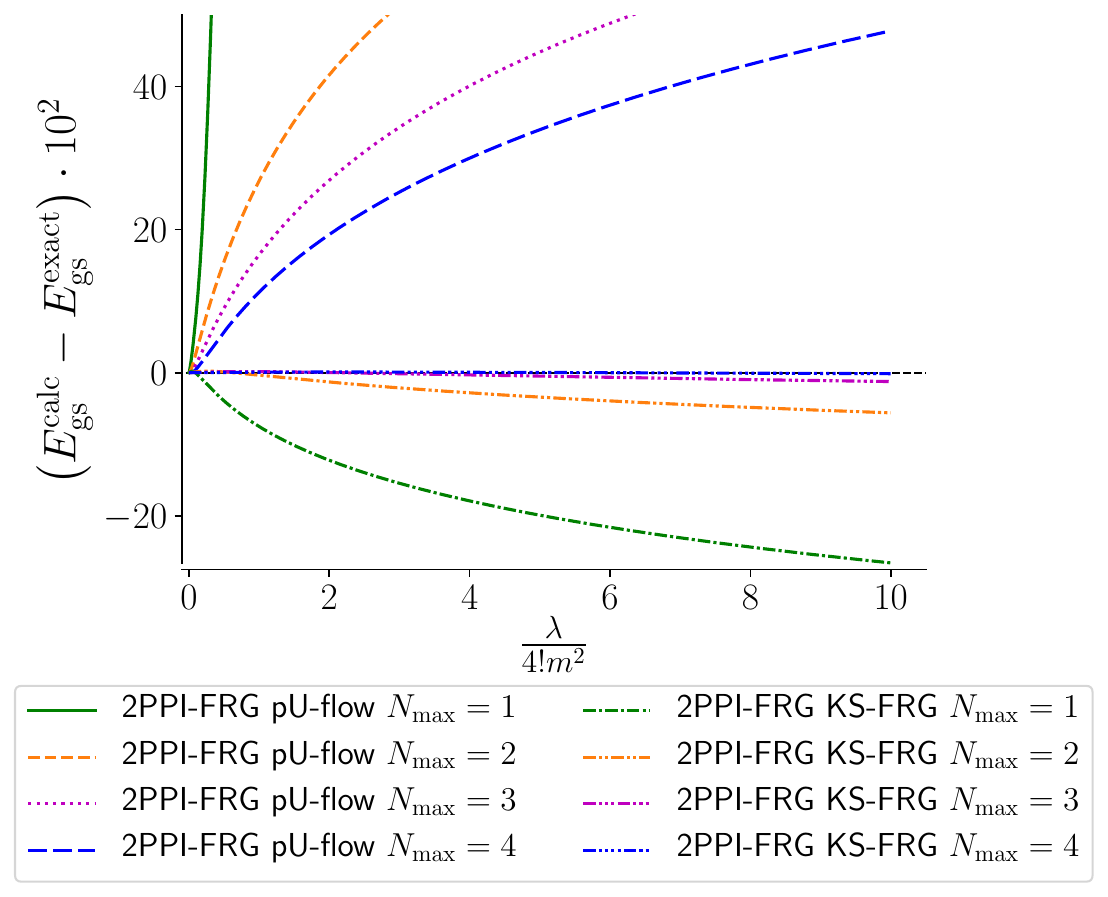}
                         }
    \subfloat[]{
      \includegraphics[width=0.50\linewidth]{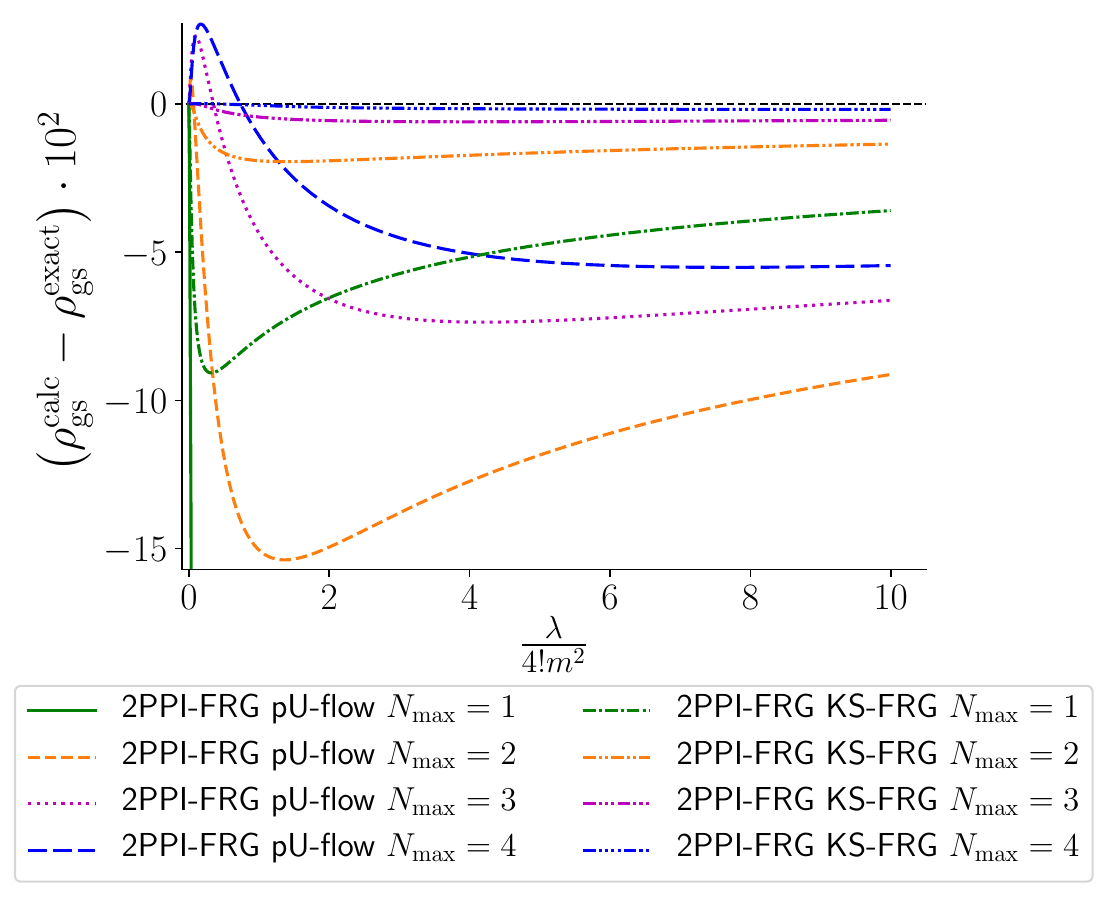}
                         }
\caption{Difference between the calculated gs energy $E_{\mathrm{gs}}^{\mathrm{calc}}$ or density $\rho_{\mathrm{gs}}^{\mathrm{calc}}$ and the corresponding exact solution $E_{\mathrm{gs}}^{\mathrm{exact}}$ or $\rho_{\mathrm{gs}}^{\mathrm{exact}}$ at $m^{2}=+1$ and $N=1$ ($\mathcal{R}e(\lambda)\geq 0$ and $\mathcal{I}m(\lambda)=0$).}
\label{fig:KSFRGvs2PPIFRGpUflowlambdaN1}
  \end{center}
\end{figure}
\begin{figure}[!t]
\captionsetup[subfigure]{labelformat=empty}
  \begin{center}
    \subfloat[]{
      \includegraphics[width=0.50\linewidth]{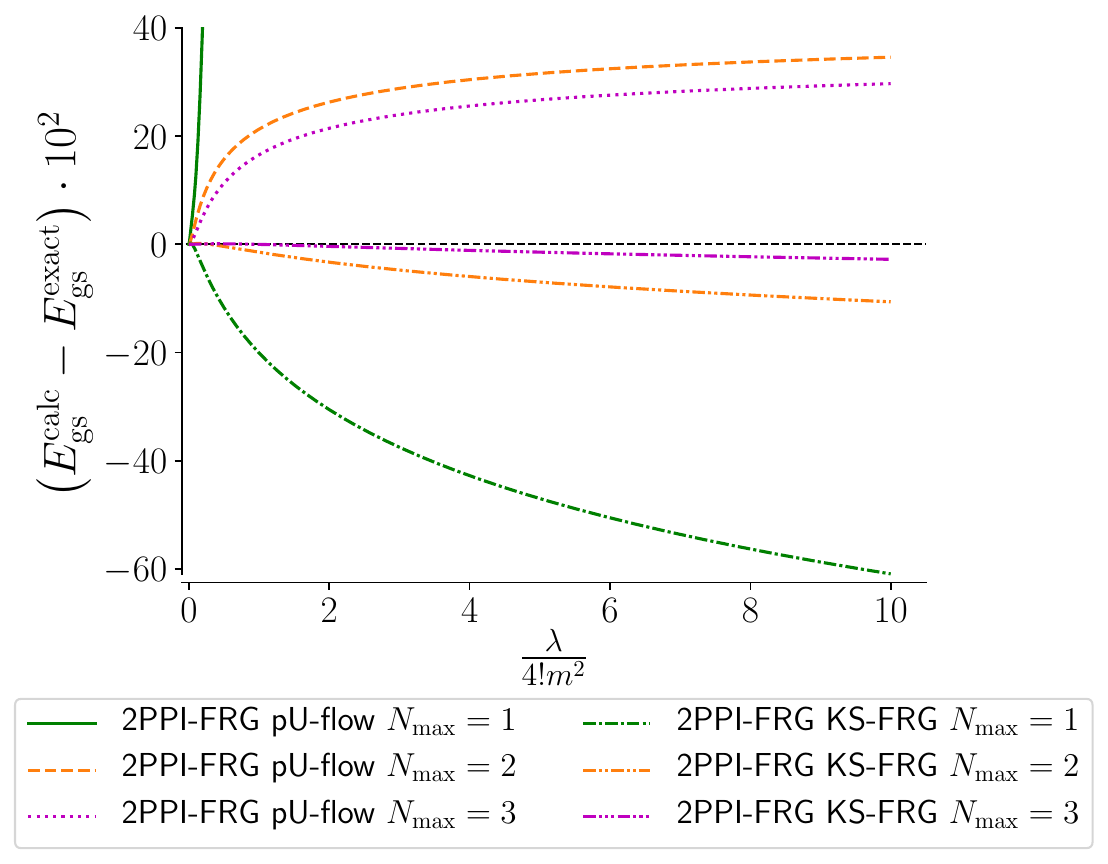}
                         }
    \subfloat[]{
      \includegraphics[width=0.50\linewidth]{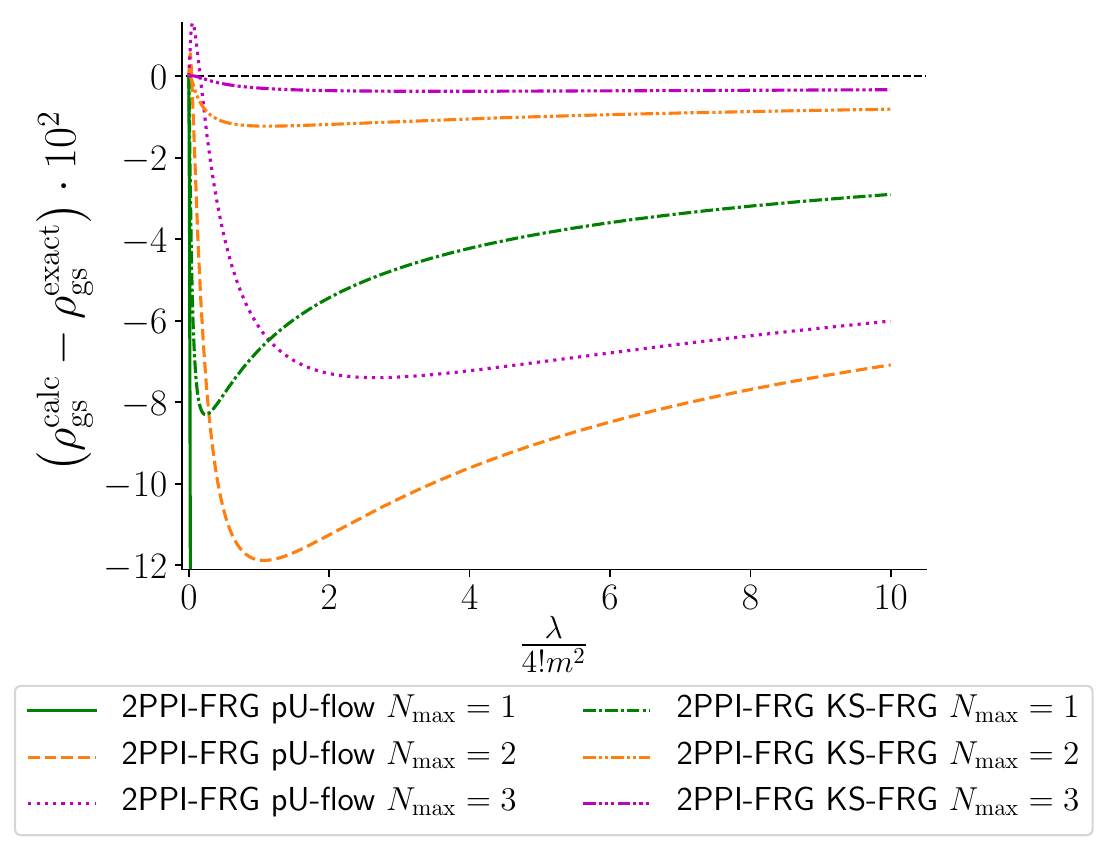}
                         }
\caption{Same as fig.~\ref{fig:KSFRGvs2PPIFRGpUflowlambdaN1} with $N=2$ instead.}
\label{fig:KSFRGvs2PPIFRGpUflowlambdaN2}
  \end{center}
\end{figure}
\begin{figure}[!t]
\captionsetup[subfigure]{labelformat=empty}
  \begin{center}
    \subfloat[]{
      \includegraphics[width=0.50\linewidth]{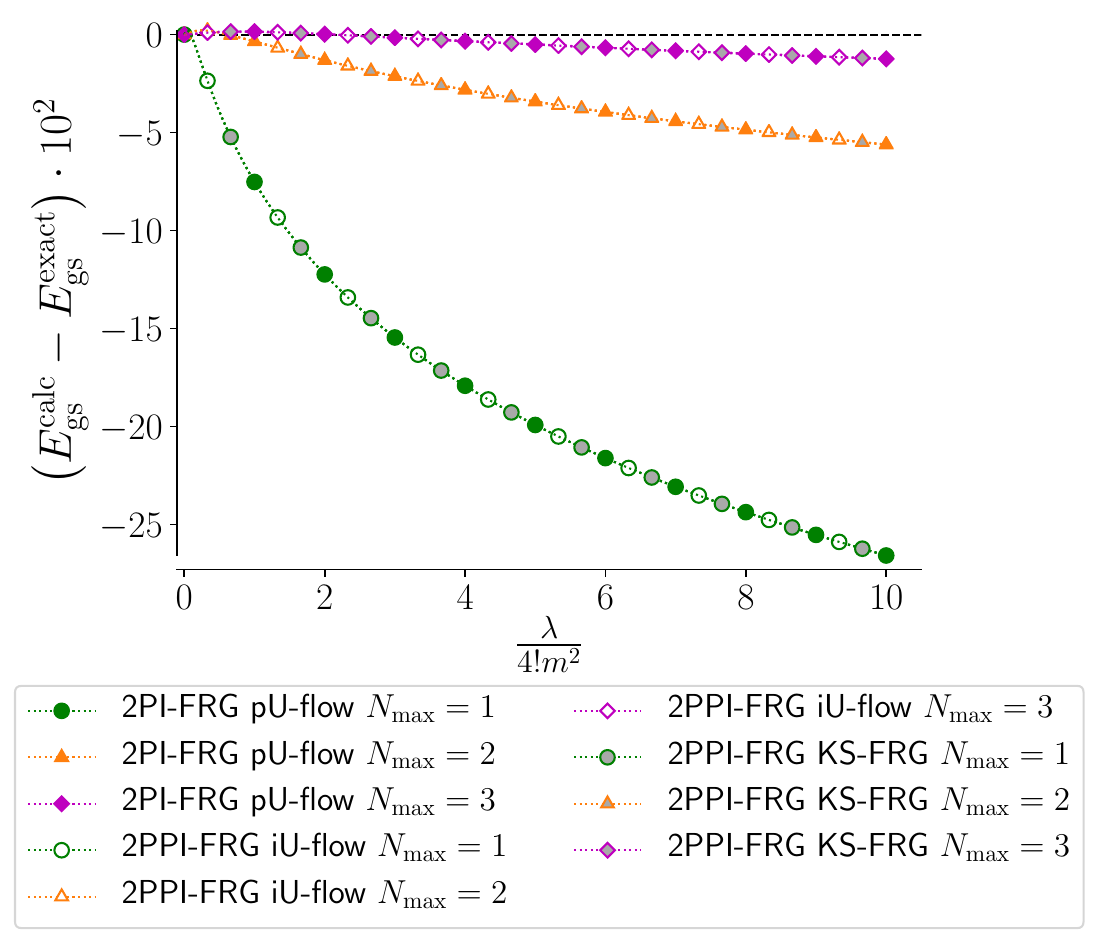}
                         }
    \subfloat[]{
      \includegraphics[width=0.50\linewidth]{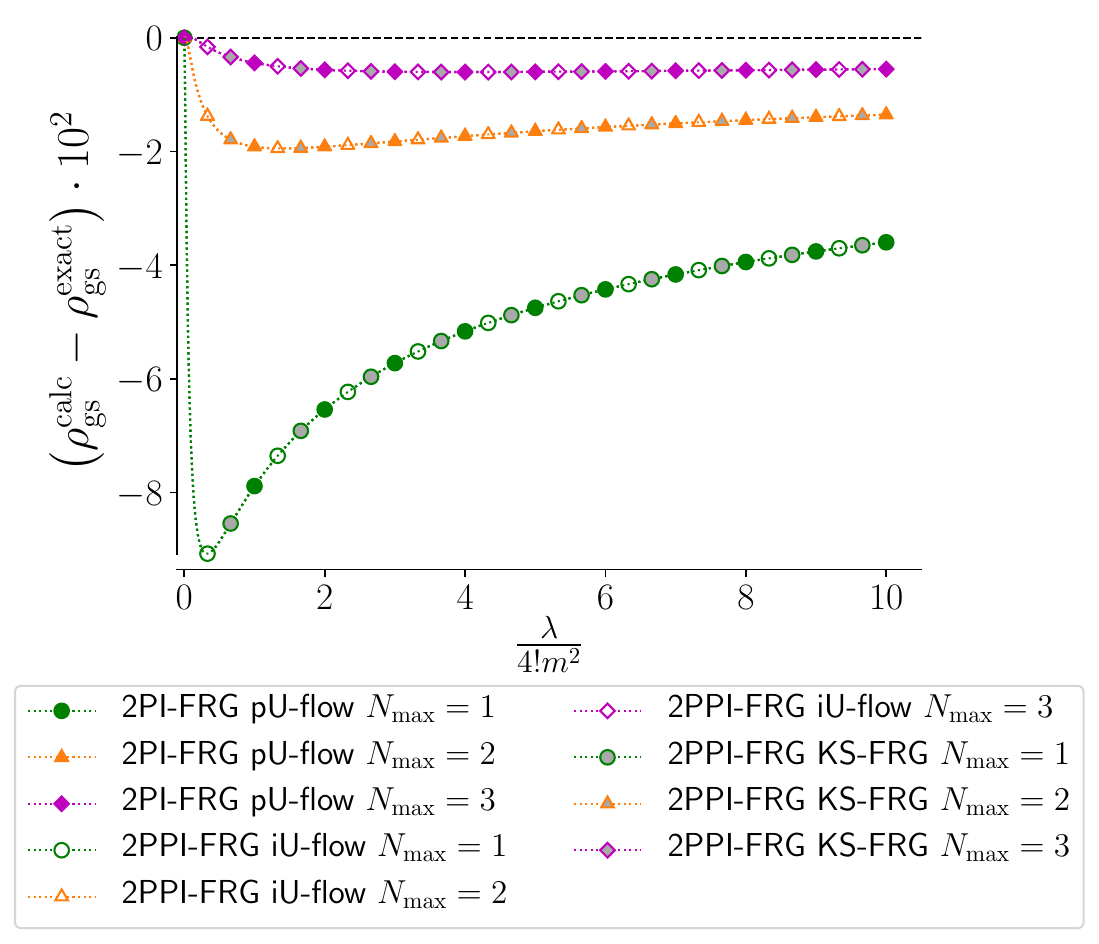}
                         }
\caption{Difference between the calculated gs energy $E_{\mathrm{gs}}^{\mathrm{calc}}$ or density $\rho_{\mathrm{gs}}^{\mathrm{calc}}$ and the corresponding exact solution $E_{\mathrm{gs}}^{\mathrm{exact}}$ or $\rho_{\mathrm{gs}}^{\mathrm{exact}}$ at $m^{2}=+1$ and $N=1$ ($\mathcal{R}e(\lambda)\geq 0$ and $\mathcal{I}m(\lambda)=0$).}
\label{fig:ComparisonKSFRGpUflowiUflowlambdaN1}
  \end{center}
\end{figure}
\begin{figure}[!t]
\captionsetup[subfigure]{labelformat=empty}
  \begin{center}
    \subfloat[]{
      \includegraphics[width=0.50\linewidth]{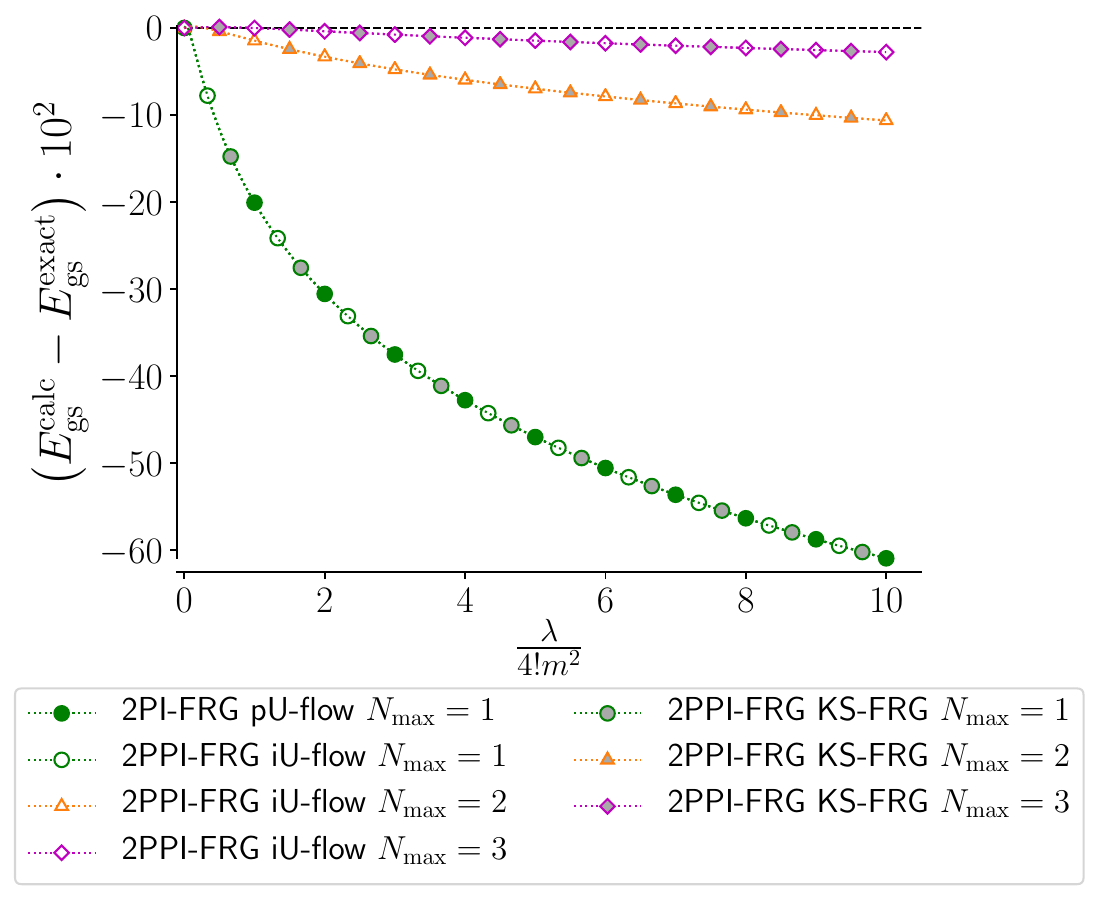}
                         }
    \subfloat[]{
      \includegraphics[width=0.50\linewidth]{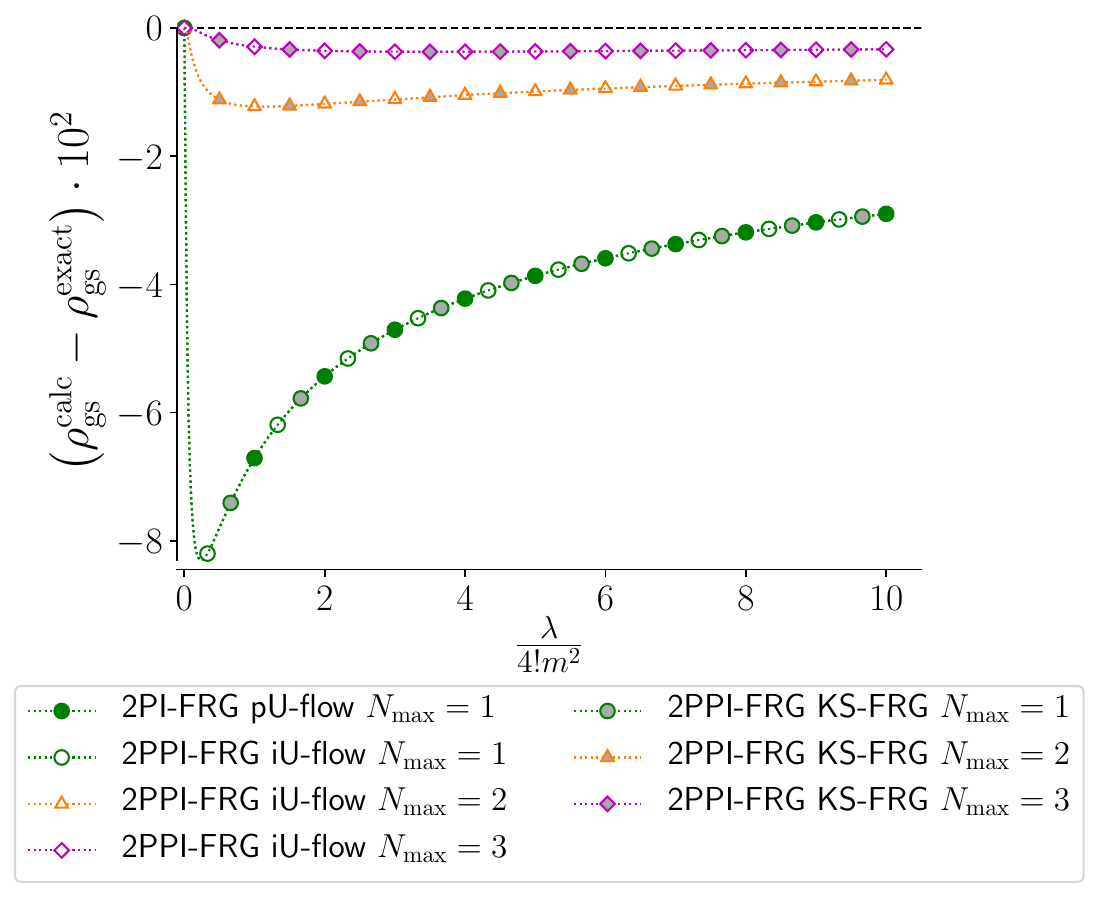}
                         }
\caption{Same as fig.~\ref{fig:ComparisonKSFRGpUflowiUflowlambdaN1} with $N=2$ instead.}
\label{fig:ComparisonKSFRGpUflowiUflowlambdaN2}
  \end{center}
\end{figure}

Hence, our KS-FRG results up to $N_{\mathrm{max}}=2$ are obtained by solving the equation system containing~\eqref{eq:KSFRGEqExpressionGammadot} to~\eqref{eq:KSFRGEqExpressiongamma2dot} (combined with~\eqref{eq:KSFRGExpressionGammaKS20DON} to~\eqref{eq:KSFRGExpressionGammaKS40DON}), with initial conditions set by~\eqref{eq:InitialCondCorrgammaKSFRG0DON} and~\eqref{eq:InitialCondCorrgammanKSFRG0DON} and truncation condition imposed by~\eqref{eq:KSFRGtruncationScheme0DON}. The flow-dependent one-body potential $V_{\mathfrak{s}}$ and Kohn-Sham potential $V_{\mathrm{KS},\mathfrak{s}}$ are respectively given by~\eqref{eq:choiceVsKSFRG0DON} and~\eqref{eq:KSpotentialKSFRG0DON} whereas the cutoff function for the two-body interaction $U_{\mathfrak{s}}$ is still expressed by~\eqref{eq:choiceCutoffUs2PPIFRG0DON}. The KS-FRG was introduced to improve the convergence of the standard 2PPI-FRG in its most basic implementations, i.e. with the sU-flow or pU-flow as truncation schemes. This improvement is illustrated by figs.~\ref{fig:KSFRGvs2PPIFRGpUflowlambdaN1} and~\ref{fig:KSFRGvs2PPIFRGpUflowlambdaN2}, which clearly show that the KS-FRG outperforms the pU-flow version of the standard 2PPI-FRG\footnote{Recall that the pU-flow outperforms itself the sU-flow implementation of the standard 2PPI-FRG, as was illustrated by figs.~\ref{fig:standard2PPIFRGallVersionslambdaN1} and~\ref{fig:standard2PPIFRGallVersionslambdaN2}.} over the whole range of tested values for the coupling constant (i.e. for $\lambda/4! \in [0,10]$) for both $E_{\mathrm{gs}}$ and $\rho_{\mathrm{gs}}$ up to $N_{\mathrm{max}}=4$ at $N=1$ and up to $N_{\mathrm{max}}=3$ at $N=2$. We also point out interesting connections between different FRG approaches tested so far: figs.~\ref{fig:ComparisonKSFRGpUflowiUflowlambdaN1} and~\ref{fig:ComparisonKSFRGpUflowiUflowlambdaN2} illustrate that the pU-flow version of the 2PI-FRG, the iU-flow version of the 2PPI-FRG and the KS-FRG lead to identical results for the studied toy model. The equivalence between the pU-flow of the 2PI-FRG on the one hand and the latter two 2PPI-FRG approaches on the other hand would no longer be valid at finite dimensions since the 2PI EA $\Gamma^{(\mathrm{2PI})}(G)$ and the 2PPI EA $\Gamma^{(\mathrm{2PPI})}(\rho)$ only coincide in (0+0)-D. However, the connection between the iU-flow and the KS-FRG remains unaffected as dimension increases since both of these approaches are based on the 2PPI EA $\Gamma^{(\mathrm{2PPI})}(\rho)$. The latter remark enables us to characterize the KS-FRG as a more easily implementable version of the iU-flow (which clearly stands out among the standard 2PPI-FRG techniques) for the two following reasons: i)~the substitution $m^{2}[\overline{\rho}_{0}] \rightarrow m^{2}[\overline{\rho}_{\mathfrak{s}}]$ (introduced in~\eqref{eq:DefinitioniUflow2PPIFRG}) underpinning the iU-flow is not readily generalizable to all theories and the KS-FRG does not rely on it; ii)~the KS-FRG completely bypasses the lengthy determination of the initial conditions $\overline{\Gamma}_{\mathfrak{s}=\mathfrak{s}_{\mathrm{i}}}^{(\mathrm{2PPI})(n)}$ (outlined from~\eqref{eq:Gamma2W22PPIFRG} to~\eqref{eq:GeneralFormulaDeterminationIC2PPIFRG}) that must be performed up to $n=N_{\mathrm{max}}+2$ within the iU-flow.

\vspace{0.5cm}

\begin{figure}[!t]
\captionsetup[subfigure]{labelformat=empty}
  \begin{center}
    \subfloat[]{
      \includegraphics[width=0.8\linewidth]{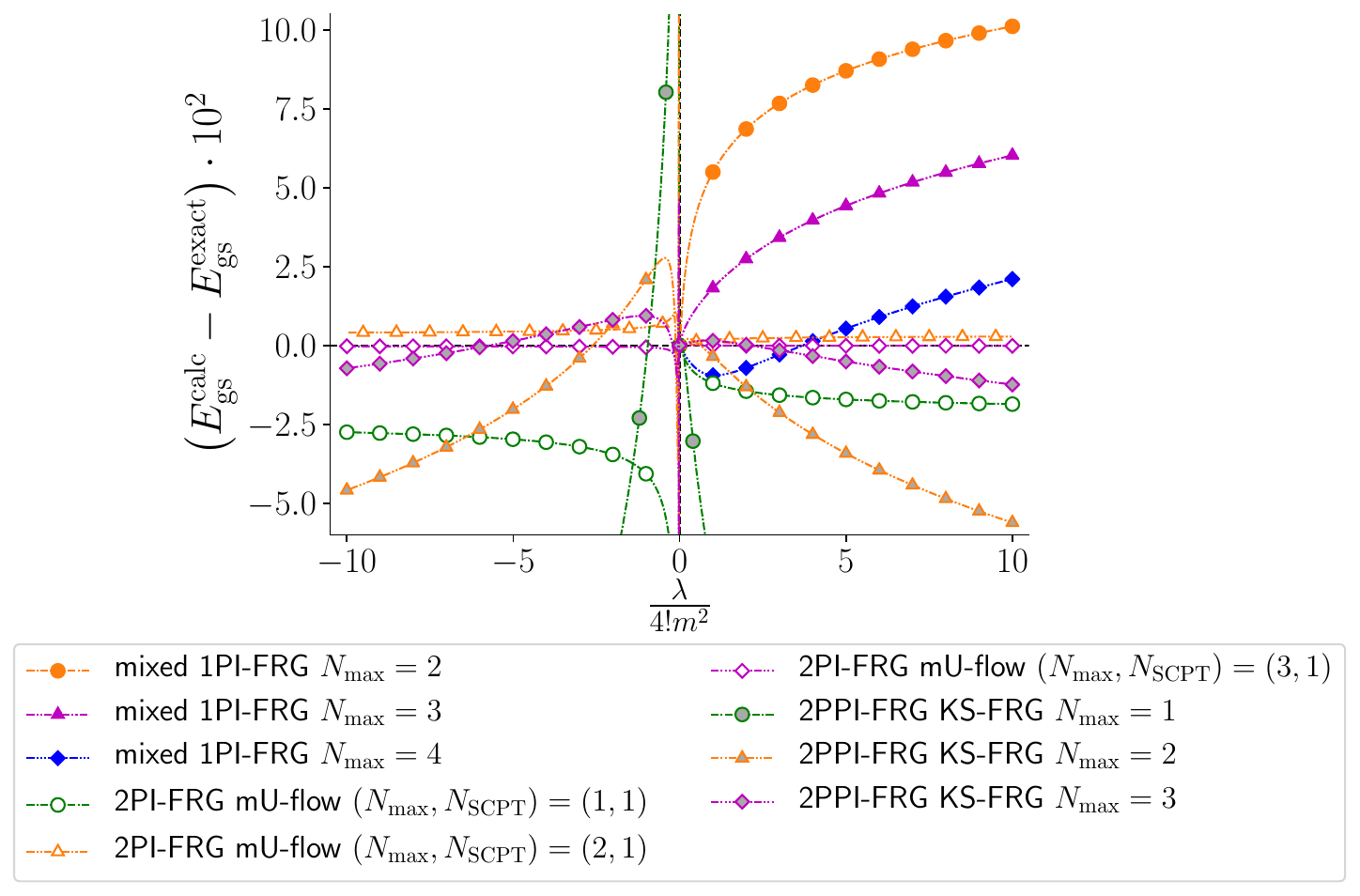}
                         }
   \\                     
    \subfloat[]{
      \includegraphics[width=0.8\linewidth]{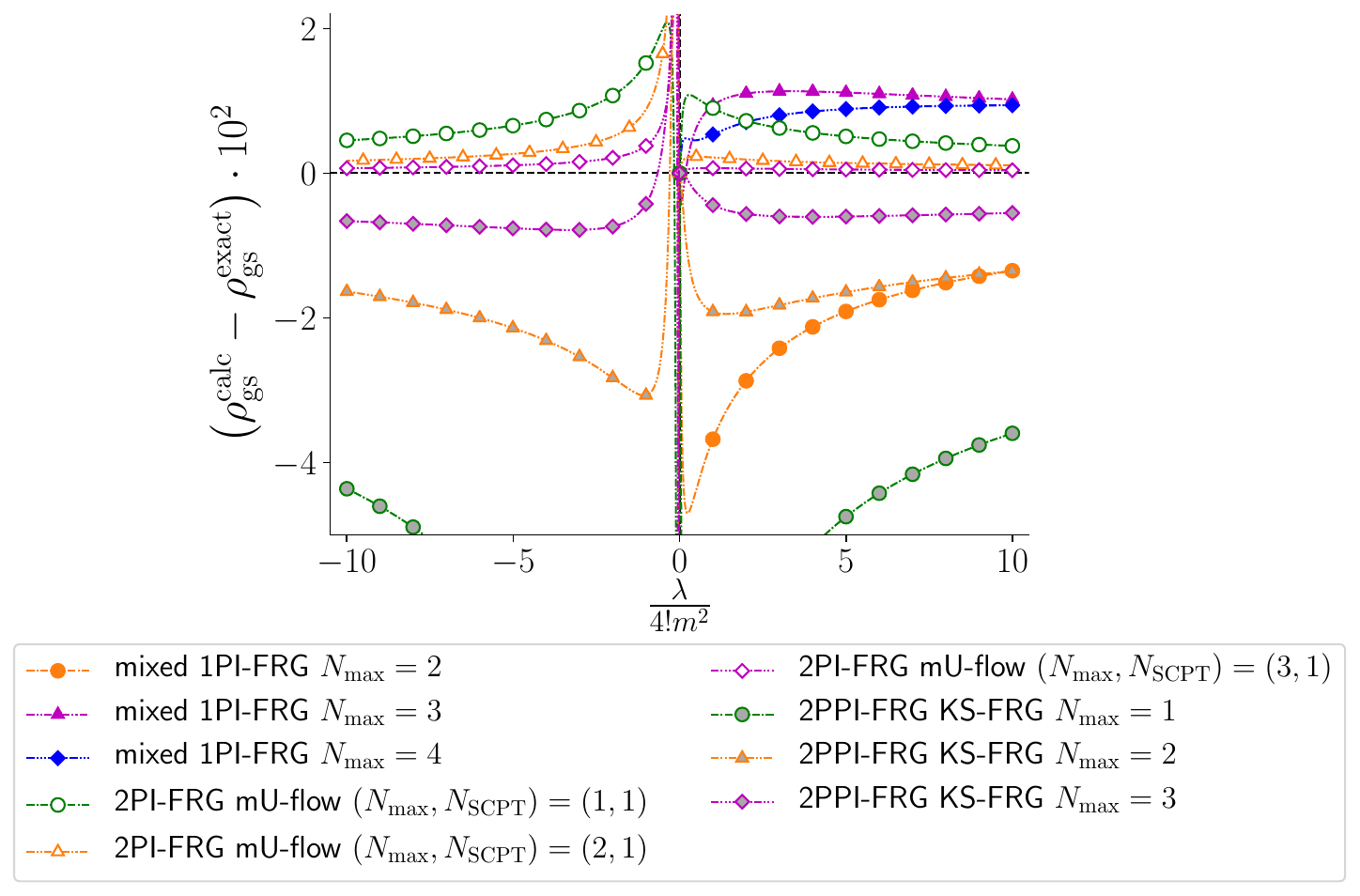}
                         }
\caption{Difference between the calculated gs energy $E_{\mathrm{gs}}^{\mathrm{calc}}$ or density $\rho_{\mathrm{gs}}^{\mathrm{calc}}$ and the corresponding exact solution $E_{\mathrm{gs}}^{\mathrm{exact}}$ or $\rho_{\mathrm{gs}}^{\mathrm{exact}}$ at $m^{2}=\pm 1$ and $N=1$ ($\mathcal{R}e(\lambda)\geq 0$ and $\mathcal{I}m(\lambda)=0$).}
\label{fig:KSFRGvs2PIFRGmUflowlambdaN1}
  \end{center}
\end{figure}
\begin{figure}[!t]
\captionsetup[subfigure]{labelformat=empty}
  \begin{center}
    \subfloat[]{
      \includegraphics[width=0.8\linewidth]{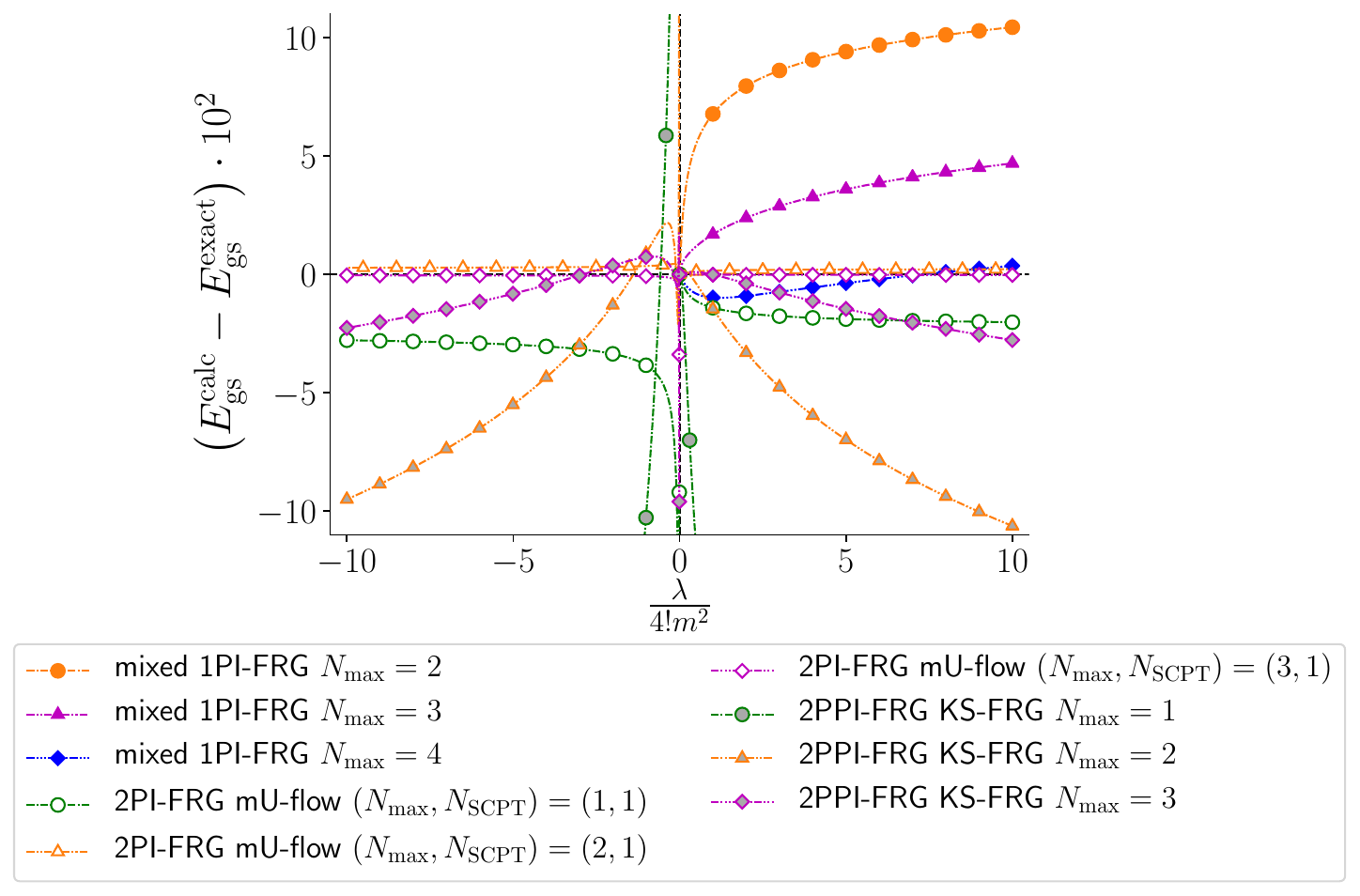}
                         }
   \\                     
    \subfloat[]{
      \includegraphics[width=0.8\linewidth]{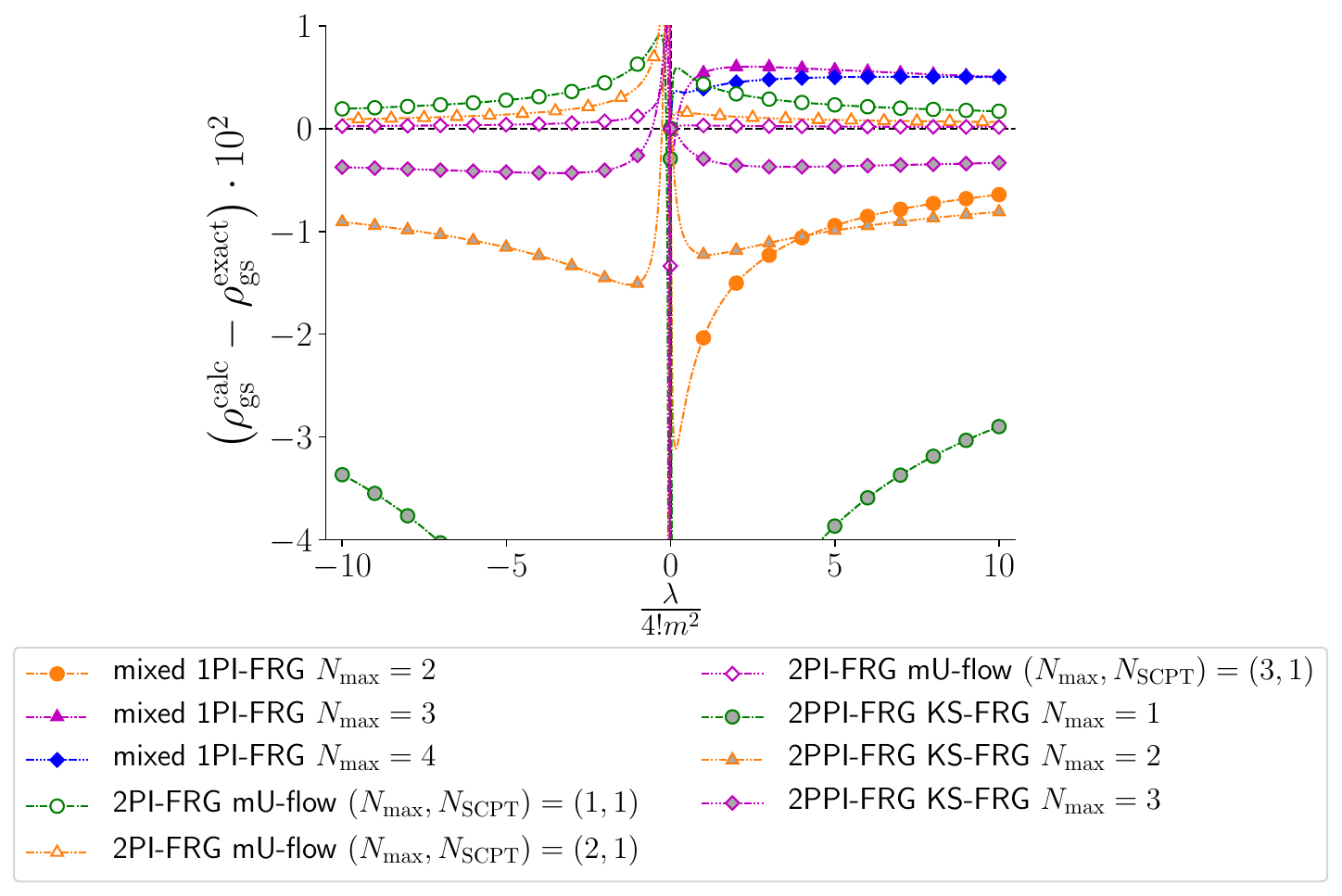}
                         }
\caption{Same as fig.~\ref{fig:KSFRGvs2PIFRGmUflowlambdaN1} with $N=2$ instead.}
\label{fig:KSFRGvs2PIFRGmUflowlambdaN2}
  \end{center}
\end{figure}

Finally, as the KS-FRG stands out as the most efficient 2PPI-FRG approach tested in the present study, we compare this method in figs.~\ref{fig:KSFRGvs2PIFRGmUflowlambdaN1} and~\ref{fig:KSFRGvs2PIFRGmUflowlambdaN2} with the most performing 1PI-FRG and 2PI-FRG techniques tested in previous sections, i.e. the mixed 1PI-FRG and the mU-flow version of the 2PI-FRG at $N_{\mathrm{SCPT}}=1$. Except for the mixed 1PI-FRG, this comparison is extended to the regime with $m^{2}<0$ thanks to the choice of flowing one-body potential $V_{\mathfrak{s}}$ put forward in~\eqref{eq:choiceVsKSFRG0DON}, which leads to KS-FRG results of similar accuracy for both signs of $m^{2}$ at a given truncation order $N_{\mathrm{max}}$. Since the pU-flow version of the 2PI-FRG is equivalent to the KS-FRG for the toy model under consideration, figs.~\ref{fig:KSFRGvs2PIFRGmUflowlambdaN1} and~\ref{fig:KSFRGvs2PIFRGmUflowlambdaN2} basically compare the mU-flow and pU-flow implementations of the 2PI-FRG in the unbroken- and broken-symmetry regimes. As was pointed out in section~\ref{sec:2PIFRG0DON}, the better performances of the mU-flow can be attributed to the quality of its starting point: the Hartree-Fock theory for the mU-flow and the free theory (at squared mass $|m^{2}|$) for the pU-flow or the KS-FRG. Furthermore, in figs.~\ref{fig:KSFRGvs2PIFRGmUflowlambdaN1} and~\ref{fig:KSFRGvs2PIFRGmUflowlambdaN2} and in the regime with $m^{2}>0$, the mU-flow of the 2PI-FRG clearly outperforms the mixed 1PI-FRG whereas the KS-FRG yields a similar or a slightly better accuracy than the latter. In addition to the absence of mixed 1PI-FRG results in the broken-symmetry phase, it can be said that our best 1PI-FRG results are rather disappointing as compared to the 2PI-FRG and 2PPI-FRG ones. However, it should be stressed that many state-of-the-art implementations of the 1PI-FRG discussed in section~\ref{sec:1PIFRGstateofplay} have not been tested in this study, either because their implementation in the framework of our zero-dimensional toy model was not relevant\footnote{We have notably mentioned alternatives to the vertex expansion, such as the DE which is an exact method for the (0+0)\nobreakdash-D model under consideration.} or simply for the sake of conciseness. The DMF$^2$RG implementations of the 1PI-FRG and the 2PI-FRG have not been tested in this comparative study either. It would actually be enlightening to compare the ability of 1PI-FRG and 2PI-FRG approaches to describe the momentum dependence of 1PI vertex functions as they are extracted from the flow in very different fashions within these two schemes: in the framework of the 2PI-FRG and as opposed to the fermionic 1PI-FRG, the momentum dependence of the flowing 1PI vertex $\overline{\Gamma}_{\mathfrak{s}}^{(\mathrm{1PI})(4)}$ or $\overline{\Gamma}_{k}^{(\mathrm{1PI})(4)}$ is determined from $\overline{\Phi}_{\mathfrak{s}}^{(2)}$ or $\overline{\boldsymbol{\Phi}}_{\mathfrak{s}}^{(2)}$ by solving a Bethe-Salpeter equation. Such a comparison could be done using a (1+1)\nobreakdash-D or a (2+1)\nobreakdash-D Hubbard model for instance. Note that there is also an implementation of the KS-FRG left to test, i.e. its CU-flow implementation designed such that the KS-FRG's starting point coincides with the Kohn-Sham system (i.e. using the flowing one-body potential $V_{\mathfrak{s}}=(1-\mathfrak{s})V_{\mathrm{KS}}$ $\forall \mathfrak{s}$ introduced right above~\eqref{eq:2PPIFRGflowEquationGammaUflow}). It would also be interesting to compare its performances with those of the mU-flow of the 2PI-FRG, which would amount to comparing the relevance of the Hartree-Fock and the Kohn-Sham systems as starting points of the flow, but we defer this to future investigations.

\vspace{0.5cm}

Getting back to our main point, we finally stress that, considering the limitations of the present study and especially the (0+0)-D character of the toy model under consideration, it can not be concluded from our results that the 1PI-FRG is less efficient than the 2PI-FRG and 2PPI-FRG schemes, especially considering the many successful applications to strongly-coupled systems outlined in section~\ref{sec:1PIFRGstateofplay}. The efficiency of the 1PI-FRG, 2PI-FRG and 2PPI-FRG would be compared in a more reliable manner by further discussing the numerical implementations of these three FRG schemes in the framework of finite-dimensional models. However, we do expect the excellent performances of the mU-flow and CU-flow versions of the 2PI-FRG to hold in the framework of more realistic (fermionic) systems, notably because they are designed to tackle both particle-hole and particle-particle channels on an equal footing.

%% file: 6ChapterConclusion/Conclusion.tex
Let us first summarize the results obtained from the (0+0)-D $O(N)$ model studied in this thesis.

\vspace{0.5cm}

\begin{figure}[!htb]
\captionsetup[subfigure]{labelformat=empty}
  \begin{center}
    \subfloat[]{
      \includegraphics[width=0.50\linewidth]{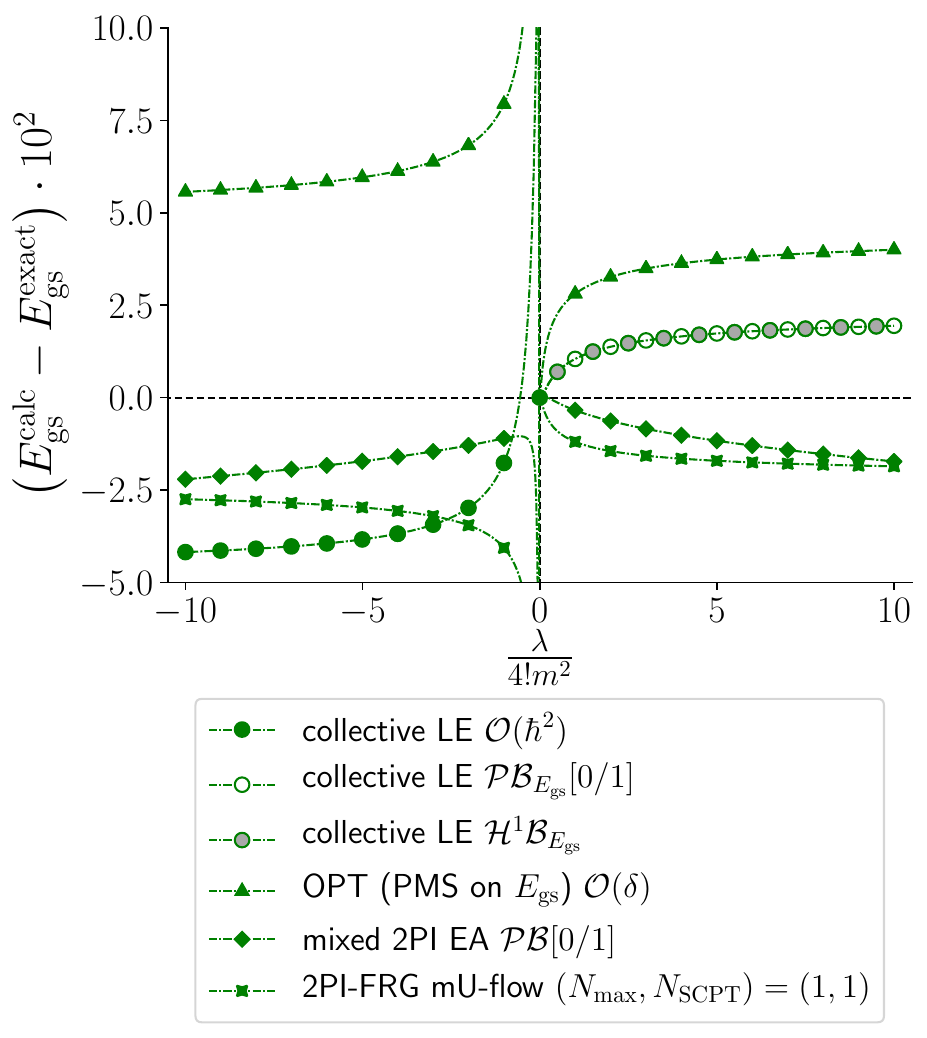}
                         }
    \subfloat[]{
      \includegraphics[width=0.50\linewidth]{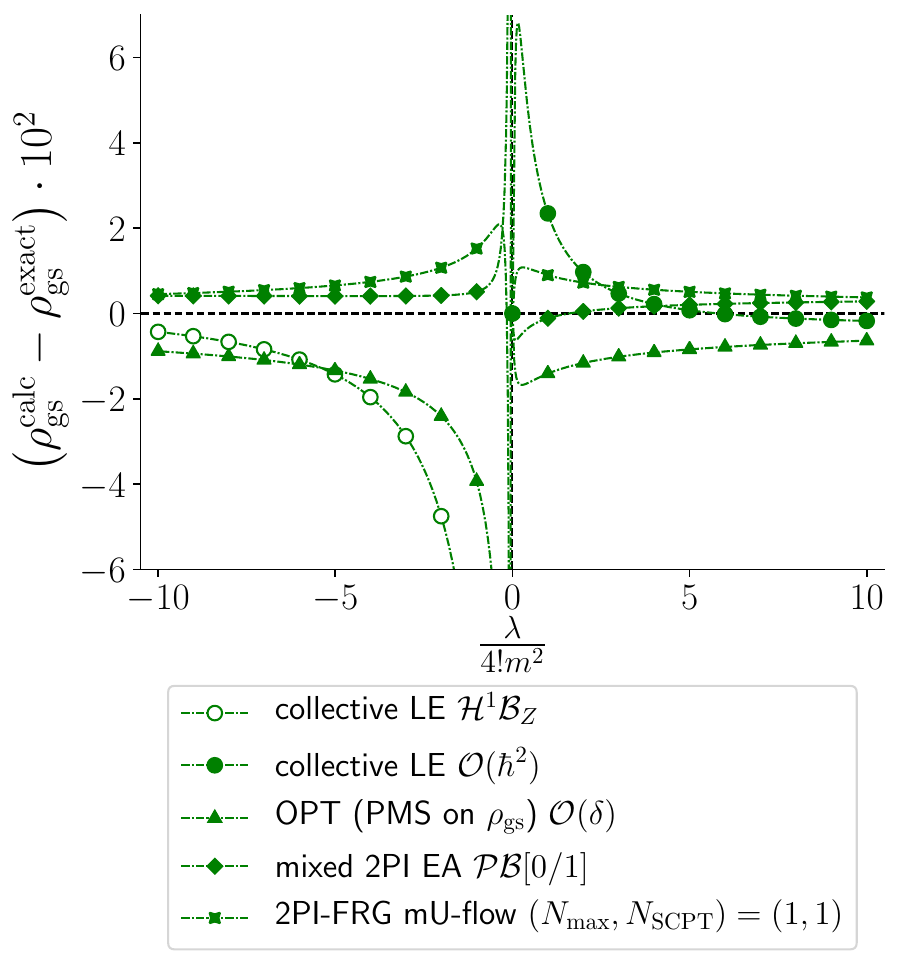}
                         }
\caption{Difference between the calculated gs energy $E_{\mathrm{gs}}^{\mathrm{calc}}$ or density $\rho_{\mathrm{gs}}^{\mathrm{calc}}$ and the corresponding exact solution $E_{\mathrm{gs}}^{\mathrm{exact}}$ or $\rho_{\mathrm{gs}}^{\mathrm{exact}}$ at $\hbar=1$, $m^{2}=\pm 1$ and $N=1$ ($\mathcal{R}e(\lambda)\geq 0$ and $\mathcal{I}m(\lambda)=0$).}
\label{fig:ConcluOrder1N1}
  \end{center}
\end{figure}

\begin{figure}[!htb]
\captionsetup[subfigure]{labelformat=empty}
  \begin{center}
    \subfloat[]{
      \includegraphics[width=0.50\linewidth]{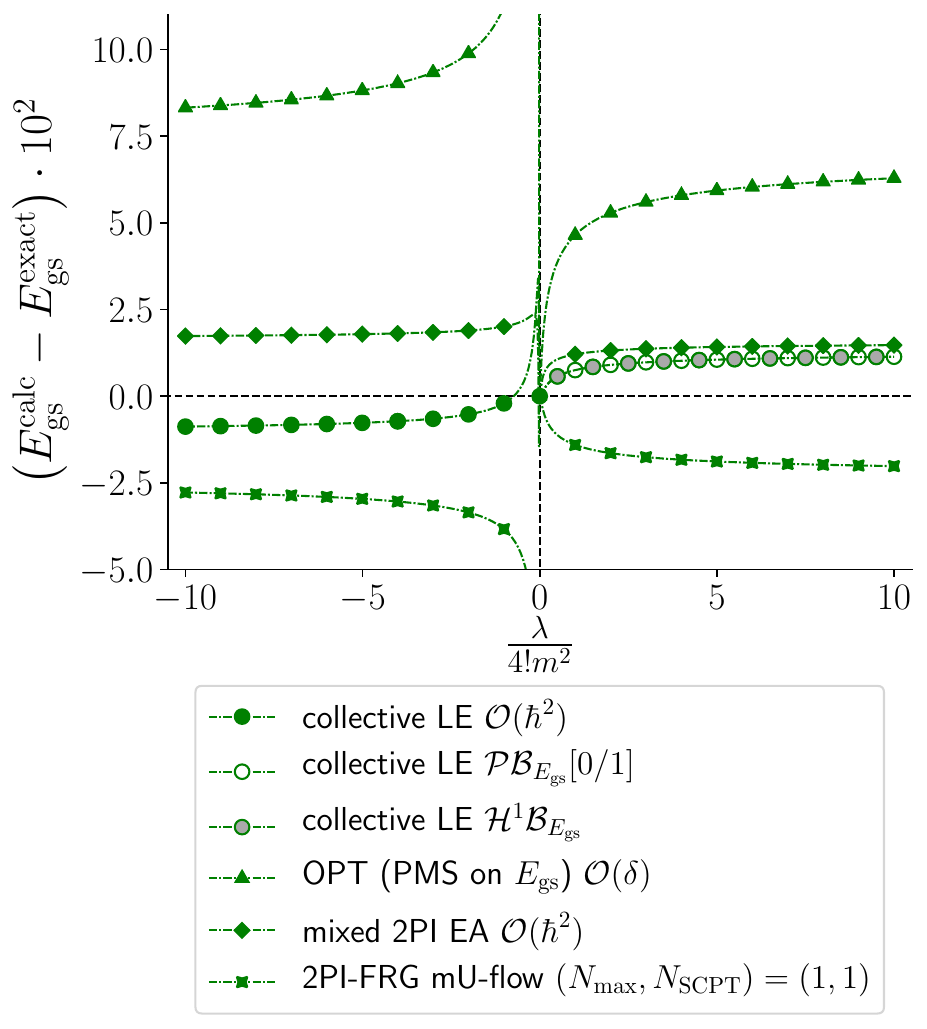}
                         }
    \subfloat[]{
      \includegraphics[width=0.50\linewidth]{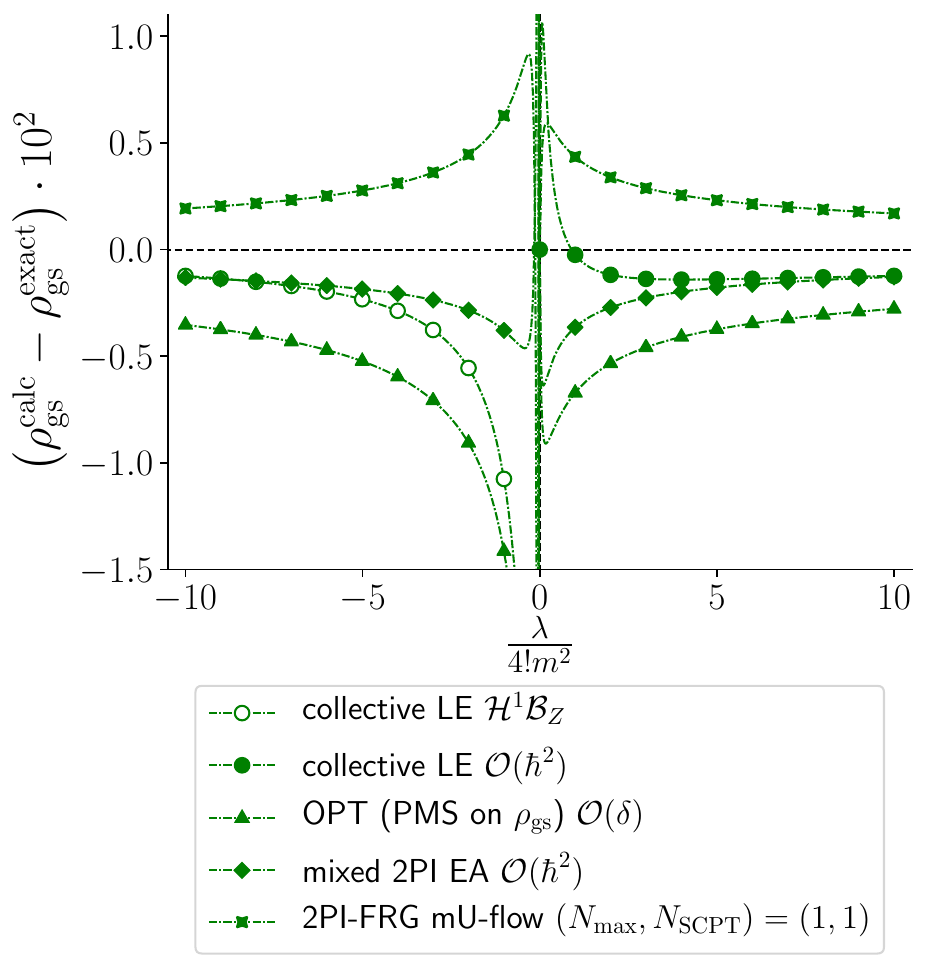}
                         }
\caption{Same as fig.~\ref{fig:ConcluOrder1N1} with $N=2$ instead.}
\label{fig:ConcluOrder1N2}
  \end{center}
\end{figure}

\begin{figure}[!htb]
\captionsetup[subfigure]{labelformat=empty}
  \begin{center}
    \subfloat[]{
      \includegraphics[width=0.50\linewidth]{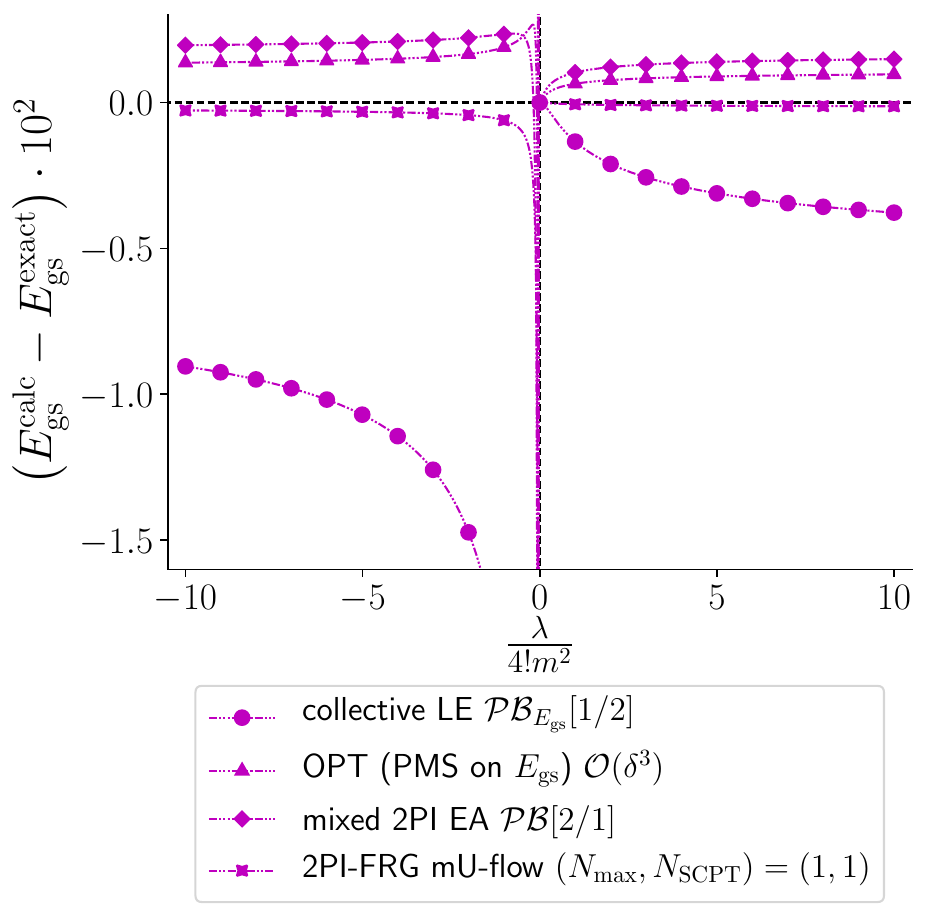}
                         }
    \subfloat[]{
      \includegraphics[width=0.50\linewidth]{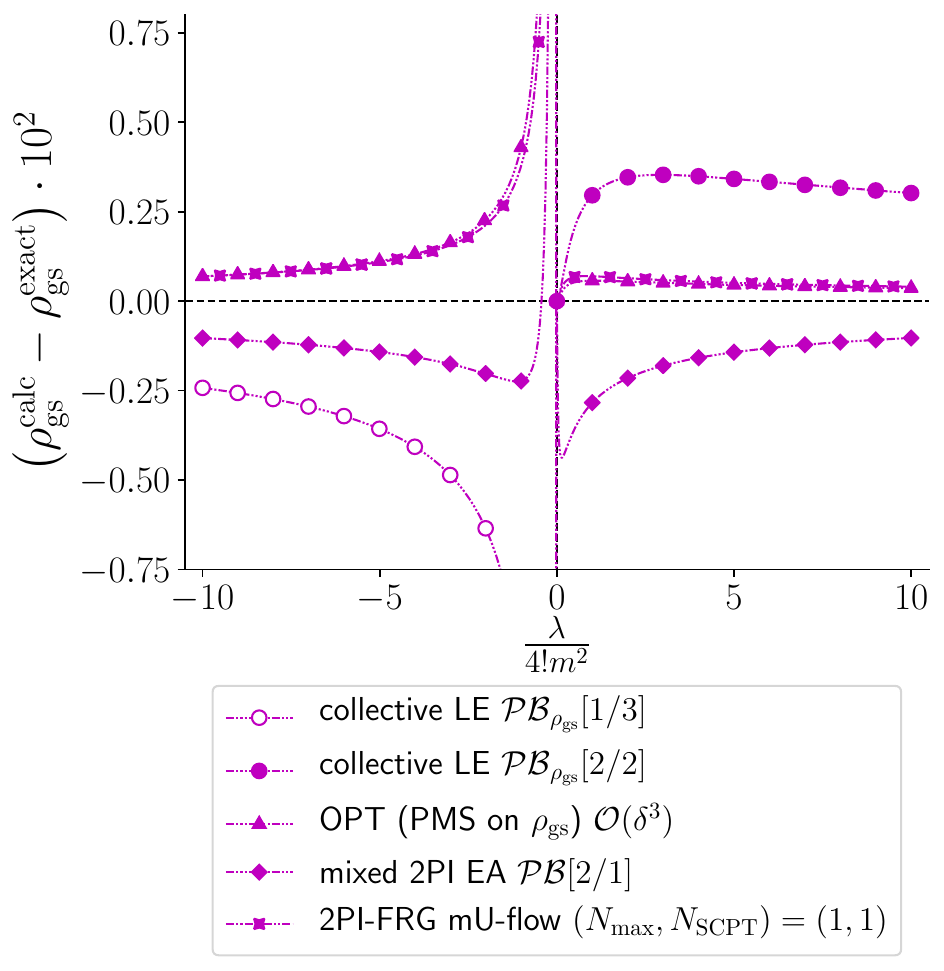}
                         }
\caption{Difference between the calculated gs energy $E_{\mathrm{gs}}^{\mathrm{calc}}$ or density $\rho_{\mathrm{gs}}^{\mathrm{calc}}$ and the corresponding exact solution $E_{\mathrm{gs}}^{\mathrm{exact}}$ or $\rho_{\mathrm{gs}}^{\mathrm{exact}}$ at $\hbar=1$, $m^{2}=\pm 1$ and $N=1$ ($\mathcal{R}e(\lambda)\geq 0$ and $\mathcal{I}m(\lambda)=0$).}
\label{fig:ConcluOrder3N1}
  \end{center}
\end{figure}

\begin{figure}[!htb]
\captionsetup[subfigure]{labelformat=empty}
  \begin{center}
    \subfloat[]{
      \includegraphics[width=0.50\linewidth]{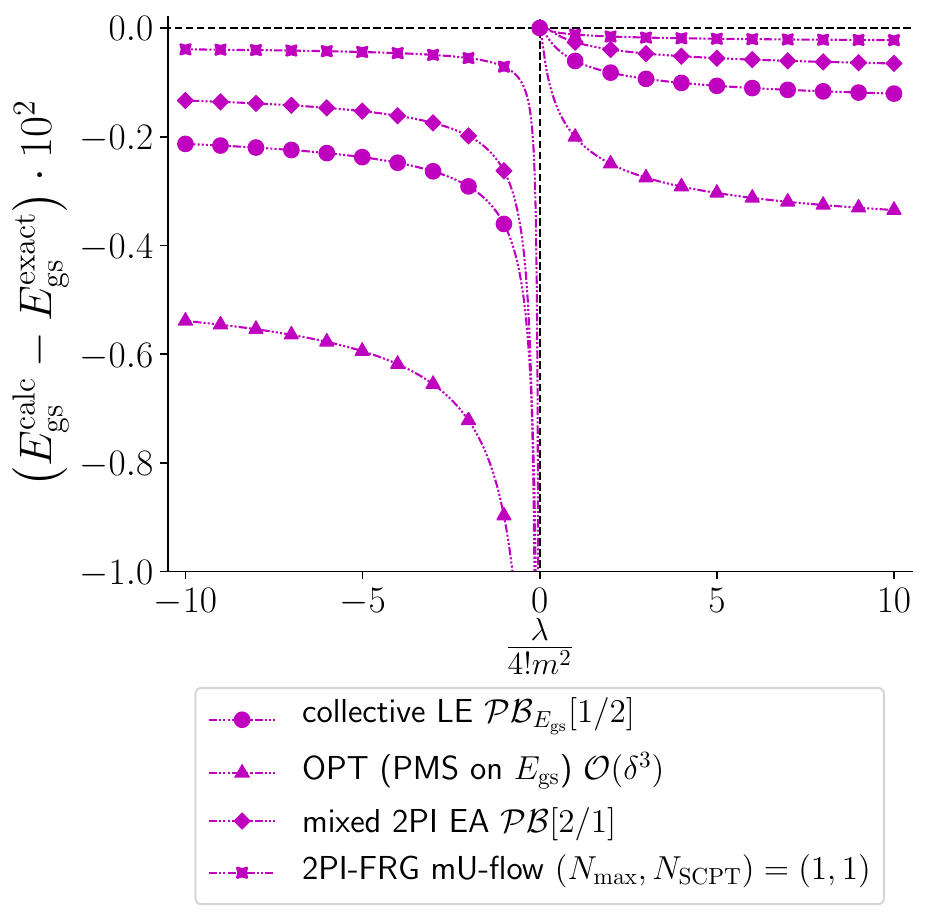}
                         }
    \subfloat[]{
      \includegraphics[width=0.50\linewidth]{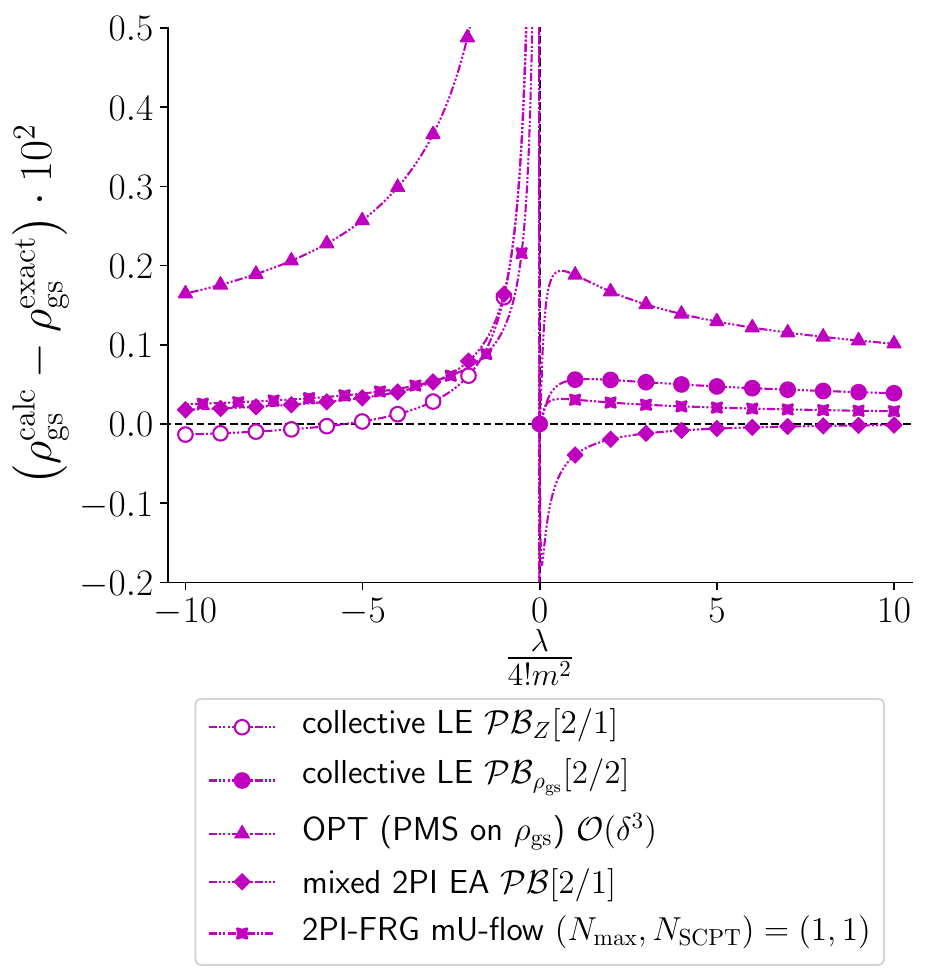}
                         }
\caption{Same as fig.~\ref{fig:ConcluOrder3N1} with $N=2$ instead.}
\label{fig:ConcluOrder3N2}
  \end{center}
\end{figure}

Concerning the \textbf{diagrammatic techniques} discussed in chapter~\ref{chap:DiagTechniques}, the simplest approach to implement is certainly the standard LE for the original theory. Combining the latter with resummation theory (Pad\'{e}-Borel resummation~\cite{pad1892,bor28}, conformal mapping~\cite{leg80} or Borel-hypergeometric resummation~\cite{mer15,mer18}) leads to decent results only at the third non-trivial order in the unbroken-symmetry regime. This is not even the case in the broken-symmetry phase: although a transseries representation combined with Borel-hypergeometric resummation yields satisfactory results at the third non-trivial order for $N=1$, no (finite) result can be extracted from the LE (even with a transseries representation) for $N \geq 2$ due to divergences inherent to Golstone's theorem~\cite{gol62}. We have also applied a HST to the studied toy model, with or without integrating out the original field, thus leading to its collective and mixed representations, respectively. The latter problem related to Golstone's theorem is also present in the mixed case but not in the collective one. The diagrammatic series is technically more demanding to derive in the framework of the collective representation but it turns out to be very successful even at the first non-trivial order (with or without resummation). The present study is to our knowledge the first to combine this collective LE with resummation theory (see e.g. results of figs.~\ref{fig:ConcluOrder1N1} to~\ref{fig:ConcluOrder3N2}) and to push it up to its third non-trivial order (see e.g. results of figs.~\ref{fig:ConcluOrder3N1} and~\ref{fig:ConcluOrder3N2}) where we find, as expected, a significant improvement of our results for all calculated quantities, i.e. for the gs energy $E_{\mathrm{gs}}$ and density $\rho_{\mathrm{gs}}$, with the help of resummation. We have also exhibited an interesting link between the collective LE and the $1/N$-expansion~\cite{wil73,mos03}, which coincide at their first non-trivial orders.

\vspace{0.5cm}

We have also exploited a modified version of PT called OPT~\cite{ste81}. As opposed to the LE, OPT does not suffer from Goldstone's theorem in the sense that it can tackle the broken-symmetry phase of our model at $N \geq 2$ within its original representation. The simplicity of OPT is also appealing: it simply amounts to adding and subtracting a quadratic term in the classical action of the theory under consideration, thus introducing a classical field (which contrasts with the quantum field introduced via HST) which is then adjusted in order to optimize the truncation of the perturbative series. Several optimization procedures have been investigated, including the SCC which is close in spirit to Kohn-Sham DFT, and the more demanding PMS, which gave us our best OPT results in both the unbroken- and broken-symmetry regimes.

\vspace{0.5cm}

Finally, among the self-consistent PT or diagrammatic EA approaches, the best results are obtained from the 2PI EA in the framework of the mixed representation, whereas other more involved techniques such as the 4PPI EA method~\cite{oko97} turned out to be less efficient. More specifically, this mixed 2PI EA approach, whose first non-trivial order in the $\hbar$-expansion is known as the BVA~\cite{mih01}, has been combined with a resummation procedure (see e.g. results of figs.~\ref{fig:ConcluOrder1N1},~\ref{fig:ConcluOrder3N1} and~\ref{fig:ConcluOrder3N2}) and pushed up to its third non-trivial order (see e.g. results of figs.~\ref{fig:ConcluOrder3N1} and~\ref{fig:ConcluOrder3N2}) for the first time in this thesis. It clearly stands out from our study thanks to the expectation value of the auxiliary field which has proven to be very effective in capturing quantum correlations. We have also carefully checked that the physical solution of the gap equations corresponding to this mixed 2PI EA never exhibits a spurious breakdown of the $O(N)$ symmetry at the first non-trivial order or beyond, in accordance with the exact solution of our toy model discussed in chapter~\ref{chap:Intro2}.

\vspace{0.5cm}

In conclusion, the most performing diagrammatic techniques (in the determination of $E_{\mathrm{gs}}$ and $\rho_{\mathrm{gs}}$) of this study are the collective LE, OPT with PMS and the mixed 2PI EA (or, more accurately, self-consistent PT based on the mixed 2PI EA). We have notably detailed in appendix~\ref{ann:Diag} the determination of the diagrams (including their multiplicities) underlying the latter two approaches up to their third non-trivial orders. The performances of these three techniques are compared at their first non-trivial orders in figs.~\ref{fig:ConcluOrder1N1} and~\ref{fig:ConcluOrder1N2} and at their third non-trivial orders in figs.~\ref{fig:ConcluOrder3N1} and~\ref{fig:ConcluOrder3N2}, still for the gs energy and density at $N=1$ and $2$. For the collective LE and the mixed 2PI EA approach, the resummation procedure might slightly improve the first non-trivial order obtained from the bare series, which is why figs.~\ref{fig:ConcluOrder1N1} and~\ref{fig:ConcluOrder1N2} sometimes display results from Pad\'{e}-Borel or Borel-hypergeometric resummation (which might actually be identical for $E_{\mathrm{gs}}$ in the unbroken-symmetry phase). To fully appreciate the merits of the resummation, we need to focus on higher truncation orders like those leading to figs.~\ref{fig:ConcluOrder3N1} and~\ref{fig:ConcluOrder3N2}. The results shown in figs.~\ref{fig:ConcluOrder1N1} to~\ref{fig:ConcluOrder3N2} clearly show that the collective LE, OPT with PMS and the mixed 2PI EA all yield similar performances in both the unbroken- and broken-symmetry regimes of our toy model. It can be noted however that OPT with PMS gives a slightly worse estimate of the gs energy than the two other methods at first non-trivial order, in which case it simply coincides with the Hartree-Fock result of the original 2PI EA. However, this worse estimate should be put in contrast with a certain ease of implementation underlying the OPT approach. Indeed, its variational parameter (the auxiliary classical field) is in principle local at finite dimensions, which contrasts with the bilocal objects (the propagators) involved in the 2PI EA formalism. Moreover, according to the aforementioned connection between the $1/N$-expansion and the collective LE, the latter is expected to be more efficient as $N$ increases, which is in accordance with the results of figs.~\ref{fig:ConcluOrder1N1} to~\ref{fig:ConcluOrder3N2}. Finally, we stress again that some of the most performing tested techniques, i.e. the collective LE and the mixed 2PI EA, have been pushed beyond their first non-trivial orders and combined with resummation procedures for the first time in this work. This is of particular importance considering that resummation is an essential ingredient to turn these approaches into reliable and systematically improvable techniques.

\vspace{0.5cm}

Then, regarding the study of chapter~\ref{chap:FRG} on \textbf{FRG techniques}, various implementations of the 1PI-FRG~\cite{wet93}, the 2PPI-FRG~\cite{pol02} and the 2PI-FRG~\cite{dup05,dup14} were investigated. The 1PI-FRG approach put forward by Wetterich exhibits nice convergence properties for both $E_{\mathrm{gs}}$ and $\rho_{\mathrm{gs}}$ in the framework of the original, mixed and collective representations, although the equation systems underlying the 1PI-FRG in the first two representations turned out to be too stiff (at least for the numerical tools used to perform this study, i.e. the $\mathtt{NDSolve}$ function of $\mathtt{Mathematica~12.1}$ in particular) to tackle the broken-symmetry regime. We have also shown how to recover the leading order of the collective LE, referred to as the LOAF approximation~\cite{coo10}, via the 1PI-FRG in the mixed representation, i.e. via the mixed 1PI-FRG, combined with MFT~\cite{jae03}. Furthermore, it should be stressed that our 1PI-FRG study was restricted to the vertex expansion as a means to treat the Wetterich equation. There are however alternative approaches to this vertex expansion, e.g. the DE and the BMW approximation, that are either not relevant to test in (0+0)-D or were just discarded for the sake of conciseness, but have already proven very efficient in the framework of finite-dimensional models.

\vspace{0.5cm}

There are also various implementations of the 2PPI-FRG and of the 2PI-FRG. To our knowledge, none of these FRG approaches have ever been applied to an $O(N)$ model (with $N>1$), which is why we have discussed the treatment of the $O(N)$ symmetry in their respective formalisms in an exhaustive fashion. As can be expected from ref.~\cite{lia18}, our calculations show that the most efficient implementation of the 2PPI-FRG is the KS-FRG that relates the 2PPI-FRG to the Kohn-Sham scheme~\cite{koh65,koh65bis} of DFT. We have also shown that the KS-FRG results can be recovered from a truncation scheme of the standard 2PPI-FRG, coined as iU-flow. We have thus clarified in this way the utility of the KS-FRG scheme, whose implementation is more readily adaptable to higher-dimensional theories than that of the iU-flow. In addition, a cutoff function has been designed for the KS-FRG to treat the broken-symmetry phase of the studied model as efficiently as the unbroken-symmetry one, for both $E_{\mathrm{gs}}$ and $\rho_{\mathrm{gs}}$.

\vspace{0.5cm}

Concerning the 2PI-FRG, there are mainly two implementations, i.e. the C-flow and the U-flow, which can be combined into a third, i.e. the CU-flow. For the C-flow, the cutoff function is introduced in the quadratic part of the classical action, as for the 1PI-FRG of Wetterich, whereas the U-flow is closer in spirit to the 2PPI-FRG as the cutoff function is inserted into the interaction part of the classical action. We have established an interesting connection with the 2PPI-FRG, namely that the pU-flow results of the 2PI-FRG are identical to those obtained from the KS-FRG for the studied (0+0)-D toy model. Such a link does not hold for finite-dimensional theories but gives us a clearer idea on the closeness between the 2PPI-FRG and the 2PI-FRG formalisms. Moreover, the performances of the C-flow of the 2PI-FRG are rather disappointing, for the original as well as for the mixed theories. We have also encountered some stiffness problems in our C-flow calculations, whose origins have been clarified and certainly differ from those leading to the stiffness issues in our 1PI-FRG applications. The results obtained from both the U-flow and the CU-flow in the original representation are much more satisfactory. We have also discussed extensions of these two methods to the mixed theory but deferred their implementations to future investigations. Within the U-flow approaches, our study clearly puts forward the mU-flow designed such that the starting point of the flow coincides with the Hartree-Fock result of the original 2PI EA treated via self-consistent PT. This mU-flow implementation of the 2PI-FRG clearly stands out among the FRG results presented in chapter~\ref{chap:FRG}, in both the unbroken- and broken-symmetry phases of our toy model. We can actually see in figs.~\ref{fig:ConcluOrder1N1} to~\ref{fig:ConcluOrder3N2} that this approach reproduces the gs energy and density of our model with an accuracy which is comparable to that of the best diagrammatic techniques of chapter~\ref{chap:DiagTechniques}.

\vspace{0.5cm}

Therefore, we have applied and illustrated the efficiency of many non-perturbative PI methods throughout the comparative study of this thesis. However, in order to fully exploit the benefits of this toy model study, we must also be aware of its limitations. The latter are essentially due to two features of the chosen toy model: i) the absence of competing channels; ii) the (0+0)-D character. Concerning the first point, one should stress that the PI techniques based on the mixed or the collective representation (i.e. the PI techniques relying on a HST) can be reformulated in terms of multi-channel HSTs in the context of competing channels. However, such bosonization schemes might be severely plagued by the Fierz ambiguity and a thorough investigation of this point would be required to fully understand the reliability of the method designed in this way. Note that such a study was performed in ref.~\cite{jae03} for the mixed 1PI-FRG, thus highlighting the merit of flowing bosonization, as was discussed in chapter~\ref{chap:Intro2}. Regarding the second source of limitation, i.e. the (0+0)-D character of our toy model, it prevented us from properly apprehending the numerical weight of the tested methods\footnote{It should be noted that the (0+0)-D nature of our $O(N)$ model also led to reasonable computation times for all tested techniques and was therefore an advantage in itself as well.}, although we have partially compensated this point with the inclusion of the $O(N)$ symmetry. It should be stressed that, at finite dimensions, the gap equations to solve within diagrammatic EA approaches would be in principle self-consistent whereas the FRG flow equations would become integro-differential. In particular, for the purpose of describing more realistic strongly-coupled systems, further studies of the mU-flow with models including at least one spatial dimension should be performed to better illustrate how demanding this implementation of the 2PI-FRG (which relies on the resolution of the Bethe-Salpeter equation at each step of the flow) is, as compared to 2PPI-FRG approaches especially. A relevant arena to achieve this is the Alexandrou-Negele nuclei's model studied in refs.~\cite{kem16,yok19,yok19bis} with the 2PPI-FRG.

\vspace{0.5cm}

In order to better apprehend the power and limitations of the aforementioned PI techniques for the treatment of (realistic) strongly-coupled fermionic systems, we thus suggest other comparative studies of finite-dimensional fermionic models with competing instabilities as outlook of the present work. Some possible playgrounds to achieve this are the (2+1)-D Hubbard model of ref.~\cite{dup14} or the uniform fermionic system with short-range interaction treated in ref.~\cite{fur07} constructing a DFT for superfluid systems from the 2PPI EA and the IM. Even though the obtained results might be disappointing, such comparative studies might trigger some modifications or extensions of the methods under consideration that would make them more suitable to achieve our ultimate goal, i.e. to achieve an accurate and computationally affordable description of strongly-coupled fermionic systems. An example of such extensions of the 2PI-FRG is the DMF$^2$RG implemention put forward by Katanin in ref.~\cite{kat19} presenting a successful application of this formalism to the (2+1)-D Hubbard model. Nonetheless, sticking to the mU-flow version of the 2PI-FRG provides us with an interesting connection with the nuclear EDF approach: while the MR EDF procedures all take as inputs the one-body density matrices calculated from the HFB equations, the mU-flow can start from the propagator (from which we can infer the densities of the studied system) determined from self-consistent PT via the Hartree-Fock approximation. It is tempting to push this analogy with the MR EDF scheme further by saying that the mU-flow also has the power to describe collective phenomena by restoring the symmetries broken down by its starting point, but this remains to be proven.

\newpage


%% file: 7Appendix/GaussianIntegration.tex
In this appendix, we derive Gaussian integral formulae for real bosonic and fermionic (i.e. Grassmann) fields:
\begin{itemize}
\item Bosonic scalar field:\\
For a real non-Grassmann variable $y$, we have the following Gaussian integral formula:
\begin{equation}
\int_{-\infty}^{\infty} \frac{dy}{\sqrt{2\pi}} \ e^{-\frac{1}{2}a y^{2}}=\frac{1}{\sqrt{a}}\;,
\label{eq:GaussianIntcomplexNumb}
\end{equation}
with $a > 0$. As a next step, we generalize this result to $n$ dimensions by introducing a $n$-component vector $\boldsymbol{y}=\left( \begin{matrix} y_{1} & \dotsb & y_{n} \end{matrix} \right)^{\mathrm{T}}$ and a $n \times n$ positive definite\footnote{The positive definiteness of $A$ means that this matrix satisfies the relation $\boldsymbol{y}^{\mathrm{T}} A \boldsymbol{y} > 0$ for any non-zero $n$-component column vector $\boldsymbol{y}$.} real symmetric matrix $A$. Assuming that $\boldsymbol{y}$ is expressed in a basis where $A$ is diagonal, we obtain:
\begin{equation}
\begin{split}
\int_{\mathbb{R}^{\overset{n}{}}} \frac{d^{n}\boldsymbol{y}}{(2\pi)^{n/2}} \ e^{-\frac{1}{2}\boldsymbol{y}^{\mathrm{T}} A \boldsymbol{y}}=& \int_{\mathbb{R}^{\overset{n}{}}} \left(\prod_{i=1}^{n} \frac{dy_{i}}{\sqrt{2\pi}}\right) e^{-\frac{1}{2}\sum_{i=1}^{n} y_{i} A_{ii} y_{i}} \\
=& \prod_{i=1}^{n} \underbrace{\int_{-\infty}^{\infty} \frac{dy_{i}}{\sqrt{2\pi}} \ e^{-\frac{1}{2} y_{i}A_{ii}y_{i}}}_{\frac{1}{\sqrt{A_{ii}}}} \\
=& \left(\prod_{i=1}^{n} A_{ii}\right)^{-1/2} \\
=& \left[\mathrm{Det}(A)\right]^{-1/2}\;.
\end{split}
\label{eq:GaussianNComponentVecNG}
\end{equation}
The determinant being a basis-independent quantity, the latter result holds in any basis. The Gaussian functional integral formula for a non-Grassmann scalar field is given by the limit $n\rightarrow\infty$ of result~\eqref{eq:GaussianNComponentVecNG}:
\begin{equation}
\int\mathcal{D}\widetilde{\psi} \ e^{-\frac{1}{2}\int_{\alpha_{1},\alpha_{2}} \widetilde{\psi}_{\alpha_{1}} A_{\alpha_{1}\alpha_{2}} \widetilde{\psi}_{\alpha_{2}}} = \left[\mathrm{Det}(A)\right]^{\textcolor{blue}{-1/2}}\;,
\label{eq:GaussianComplexNGfield}
\end{equation}
where $\widetilde{\psi}_{\alpha}$ is a real bosonic field, the $\alpha$-indices gather all relevant indices labeling the field $\widetilde{\psi}$ (spacetime indices, spin projections, ...) and $A$ is henceforth a functional matrix.

\pagebreak

\item Fermionic scalar field:\\
We consider a complex Grassmann variable $\eta$. Owing to the identity $\eta^{2}=0$, any element $f$ of a Grassmann algebra with $m$ generators $\eta_{1}$, ..., $\eta_{m}$ can be expressed as:
\begin{equation}
f(\eta_{1},...,\eta_{m})=1 + \sum_{k=1}^{m} \sum_{l_{1},...,l_{k}=1}^{m} c_{l_{1}...l_{k}} \eta_{l_{1}} ... \eta_{l_{k}}\;.
\end{equation}
In particular, the function $f(\eta,\eta^{*})=e^{-\eta^{*}a\eta}$ can be considered as an element of a Grassmann algebra with 2 generators $\eta$ and $\eta^{*}$ so that:
\begin{equation}
e^{-\eta^{*}a\eta}=1+c_{1}\eta+c_{2}\eta^{*}+c_{12}\eta\eta^{*}=1-\eta^{*}a\eta\;,
\end{equation}
where the coefficients $c_{1}=c_{2}=0$ and $c_{12}=-a$ are found by Taylor expanding the exponential function of the LHS. This enables us to infer the following Gaussian integral formula for complex Grassmann variables\footnote{Recall that, for Grassmann variables, differentiation and integration are identical operations, i.e. $\frac{\partial}{\partial\eta}\eta=\int d\eta \ \eta = 1$, $\frac{\partial}{\partial\eta^{*}}\eta^{*}=\int d\eta^{*} \ \eta^{*} = 1$ and $\frac{\partial^{2}}{\partial\eta^{*}\partial\eta}\eta^{*}\eta=\int d\eta^{*}d\eta \ \eta^{*}\eta = -1$. The minus sign in the latter relation results from the anticommuting property of Grassmann variables, i.e. $\eta\eta^{*}=-\eta^{*}\eta$.}:
\begin{equation}
\int d\eta^{*}d\eta \ e^{-\eta^{*}a\eta} = \int d\eta^{*}d\eta \left( 1-\eta^{*}a\eta \right) = + a \;.
\label{eq:GaussianFormulaComplexGrassmannVar}
\end{equation}
Let $A$ and $\boldsymbol{\eta}=\left( \begin{matrix} \eta_{1} & \dotsb & \eta_{n} \end{matrix} \right)^{\mathrm{T}}$ be respectively an arbitrary complex matrix and a $n$-component vector. We assume once again that $A$ is diagonal in the chosen basis in the first instance so that the generalization of result~\eqref{eq:GaussianFormulaComplexGrassmannVar} to $n$ dimensions reads:
\begin{equation}
\begin{split}
\int d^{n}\boldsymbol{\eta}^{\dagger} d^{n}\boldsymbol{\eta} \ e^{-\boldsymbol{\eta}^{\dagger} A \boldsymbol{\eta}}=& \int \left(\prod_{i=1}^{n} d\eta_{i}^{*} d\eta_{i}\right) \ e^{-\sum_{i=1}^{n} \eta^{*}_{i} A_{ii} \eta_{i}} \\
=& \prod_{i=1}^{n} \underbrace{\int d\eta_{i}^{*} d\eta_{i} \ e^{-\eta^{*}_{i}A_{ii}\eta_{i}}}_{A_{ii}} \\
=& \prod_{i=1}^{n} A_{ii} \\
=& \mathrm{Det}(A) \;.
\end{split}
\label{eq:GaussianNComponentVecGrass}
\end{equation}
The properties of the determinant can once again be invoked to justify that the latter result does not depend on the chosen basis. Let us now focus on the real case by introducing a vector $\boldsymbol{\theta}=\left( \begin{matrix} \theta_{1} & \dotsb & \theta_{n} \end{matrix} \right)^{\mathrm{T}}$ with $n$ real Grassmann scalar variables as components. The corresponding Gaussian integral reads:
\begin{equation}
\int d^{n}\boldsymbol{\theta} \ e^{-\frac{1}{2}\boldsymbol{\theta}^{\mathrm{T}} A \boldsymbol{\theta}}\;,
\end{equation}
where the integer $n$ must now be even. Taylor expanding the exponential function leads to\footnote{Due to the anticommuting property of Grassmann variables, we have $\sum_{i,j=1}^{n} \theta_{i} A_{ij} \theta_{j}=-\sum_{i,j=1}^{n} \theta_{j} A_{ij} \theta_{i}=-\sum_{i,j=1}^{n} \theta_{i} A_{ji} \theta_{j}$. Therefore, only the antisymmetric part of $A$ contributes to the sums involved in~\eqref{eq:CalculationGaussianRealGrassmann}.}:
\begin{equation}
\begin{split}
\scalebox{0.97}{${\displaystyle\int d^{n}\boldsymbol{\theta} \ e^{-\frac{1}{2}\boldsymbol{\theta}^{\mathrm{T}} A \boldsymbol{\theta}}=}$} & \scalebox{0.97}{${\displaystyle\int \left(\prod_{i=1}^{n} d\theta_{i}\right) \ e^{-\frac{1}{2}\sum_{i,j=1}^{n} \theta_{i} A_{ij} \theta_{j}}}$} \\
\scalebox{0.97}{${\displaystyle =}$} & \scalebox{0.97}{${\displaystyle\sum_{m=0}^{\infty} \frac{\left(-1\right)^{m}}{2^{m} m!} \int \left(\prod_{i=1}^{n} d\theta_{i}\right) \left(\sum_{i,j=1}^{n} \theta_{i} A_{ij} \theta_{j}\right)^{m}}$} \\
\scalebox{0.97}{${\displaystyle =}$} & \scalebox{0.97}{${\displaystyle\frac{\left(-1\right)^{n/2}}{2^{n/2}\left(\frac{n}{2}\right)!} \int \left(\prod_{i=1}^{n} d\theta_{i}\right) \left(\sum_{i,j=1}^{n} \theta_{i} A_{ij} \theta_{j}\right)^{n/2}}$} \\
\scalebox{0.97}{${\displaystyle =}$} & \scalebox{0.97}{${\displaystyle\frac{\left(-1\right)^{n/2}}{2^{n/2}\left(\frac{n}{2}\right)!} \sum_{P\in S_{n}} \int \left(\prod_{i=1}^{n} d\theta_{i}\right) \theta_{P(1)} A_{P(1)P(2)} \theta_{P(2)} \dotsb \theta_{P(n-1)} A_{P(n-1)P(n)} \theta_{P(n)}}$} \\
\scalebox{0.97}{${\displaystyle =}$} & \scalebox{0.97}{${\displaystyle\frac{\left(-1\right)^{n/2}}{2^{n/2}\left(\frac{n}{2}\right)!} \sum_{P\in S_{n}} \mathrm{sgn}(P) \underbrace{\left(\int d\theta_{1} \ \theta_{1}\right)}_{1} \dotsb \underbrace{\left(\int d\theta_{n} \ \theta_{n}\right)}_{1} A_{P(1)P(2)} \dotsb A_{P(n-1)P(n)}}$} \\
\scalebox{0.97}{${\displaystyle =}$} & \scalebox{0.97}{${\displaystyle\frac{\left(-1\right)^{n/2}}{2^{n/2}\left(\frac{n}{2}\right)!} \sum_{P\in S_{n}} \mathrm{sgn}(P) \ A_{P(1)P(2)} \dotsb A_{P(n-1)P(n)} \;,}$}
\end{split}
\label{eq:CalculationGaussianRealGrassmann}
\end{equation}
where $\mathrm{sgn}(P)$ is the signature of permutation $P$. The only non-vanishing terms in the sum $\sum_{m=0}^{\infty}$ of the second line must contain the product $\theta_{1}\dotsb\theta_{n}$ up to a permutation, i.e. they must satisfy $m=n/2$: those with $m<n/2$ all vanish due to the relation $\int d\theta \ 1 = 0$ and those with $m>n/2$ all equal zero because of the anticommuting property of Grassmann variables in the form $\theta_{i}^{2}=0$ $\forall i$. The last line of~\eqref{eq:CalculationGaussianRealGrassmann} involves the Pfaffian of matrix $A$ so that we can write:
\begin{equation}
\int \frac{d^{n}\boldsymbol{\theta}}{\left(-1\right)^{n/2}} \ e^{-\frac{1}{2}\boldsymbol{\theta}^{\mathrm{T}} A \boldsymbol{\theta}}=\mathrm{Pf}(A)\;.
\label{eq:PfaffianGaussianRealGrassmann}
\end{equation}
This Pfaffian is related to the determinant of $A$. By taking the square of both sides of~\eqref{eq:PfaffianGaussianRealGrassmann}, it can be shown that:
\begin{equation}
\int d^{n}\boldsymbol{\theta}d^{n}\boldsymbol{\theta}' \ e^{-\frac{1}{2}\boldsymbol{\theta}^{\mathrm{T}} A \boldsymbol{\theta}-\frac{1}{2}\boldsymbol{\theta}'^{\mathrm{T}} A \boldsymbol{\theta}'}=\left[\mathrm{Pf}(A)\right]^{2}\;.
\label{eq:ProofPfaffianGaussianRealGrassmann}
\end{equation}
We then perform the change of variables:
\begin{equation}
\left\{
\begin{array}{lll}
        \displaystyle{\boldsymbol{\eta}=\frac{1}{\sqrt{2}}\left(\boldsymbol{\theta}+i\boldsymbol{\theta}'\right)\;,} \\
        \\
        \displaystyle{\boldsymbol{\eta}^{*}=\frac{1}{\sqrt{2}}\left(\boldsymbol{\theta}-i\boldsymbol{\theta}'\right)\;,}
    \end{array}
\right.
\label{eq:ChangeVariableGaussianRealGrassmann}
\end{equation}
with
\begin{equation}
\begin{split}
d^{n}\boldsymbol{\theta}d^{n}\boldsymbol{\theta}' = & \ d\theta_{1} \dotsb d\theta_{n} d\theta'_{1} \dotsb d\theta'_{n} \quad \quad \\
= & \left(-1\right)^{n/2} d\theta_{1} d\theta'_{1} \dotsb d\theta_{n} d\theta'_{n} \\
= & \left(-1\right)^{n/2} \left| \begin{matrix} \frac{\partial \theta_{1}}{\partial \eta^{*}_{1}} & \frac{\partial \theta_{1}}{\partial \eta_{1}} \\
\frac{\partial \theta'_{1}}{\partial \eta^{*}_{1}} & \frac{\partial \theta'_{1}}{\partial \eta_{1}}
\end{matrix} \right| d\eta^{*}_{1} d\eta_{1} \dotsb \left| \begin{matrix} \frac{\partial \theta_{n}}{\partial \eta^{*}_{n}} & \frac{\partial \theta_{n}}{\partial \eta_{n}} \\
\frac{\partial \theta'_{n}}{\partial \eta^{*}_{n}} & \frac{\partial \theta'_{n}}{\partial \eta_{n}}
\end{matrix} \right| d\eta^{*}_{n} d\eta_{n} \\
= & \left(-1\right)^{n/2} \left| \begin{matrix} \frac{1}{\sqrt{2}} & \frac{1}{\sqrt{2}} \\
\frac{i}{\sqrt{2}} & -\frac{i}{\sqrt{2}}
\end{matrix} \right| d\eta^{*}_{1} d\eta_{1} \dotsb \left| \begin{matrix} \frac{1}{\sqrt{2}} & \frac{1}{\sqrt{2}} \\
\frac{i}{\sqrt{2}} & -\frac{i}{\sqrt{2}}
\end{matrix} \right| d\eta^{*}_{n} d\eta_{n} \\
= & \left(-1\right)^{n/2} \underbrace{\left(-1\right)^{n}}_{+1} \underbrace{i^{n}}_{(-1)^{n/2}} \underbrace{d\eta^{*}_{1} d\eta_{1} \dotsb d\eta^{*}_{n} d\eta_{n}}_{d^{n}\boldsymbol{\eta}^{\dagger} d^{n}\boldsymbol{\eta}} \\
= & \ d^{n}\boldsymbol{\eta}^{\dagger} d^{n}\boldsymbol{\eta} \;,
\end{split}
\label{eq:ChangeVariableGaussianRealGrassmannIntMeasure}
\end{equation}
where the Jacobians are determined from the inverse relations of~\eqref{eq:ChangeVariableGaussianRealGrassmann}, i.e.:
\begin{equation}
\left\{
\begin{array}{lll}
        \displaystyle{\boldsymbol{\theta}=\frac{1}{\sqrt{2}}\left(\boldsymbol{\eta}+\boldsymbol{\eta}^{*}\right)\;.} \\
        \\
        \displaystyle{\boldsymbol{\theta}'=-\frac{i}{\sqrt{2}}\left(\boldsymbol{\eta}-\boldsymbol{\eta}^{*}\right)\;.}
    \end{array}
\right.
\end{equation}
These two relations also allow for calculating:
\begin{equation}
\begin{split}
-\frac{1}{2}\boldsymbol{\theta}^{\mathrm{T}} A \boldsymbol{\theta} - \frac{1}{2}\boldsymbol{\theta}'^{\mathrm{T}} A \boldsymbol{\theta}' = & -\frac{1}{4}\left(\boldsymbol{\eta}+\boldsymbol{\eta}^{*}\right)^{\mathrm{T}} A \left(\boldsymbol{\eta}+\boldsymbol{\eta}^{*}\right) + \frac{1}{4}\left(\boldsymbol{\eta}-\boldsymbol{\eta}^{*}\right)^{\mathrm{T}} A \left(\boldsymbol{\eta}-\boldsymbol{\eta}^{*}\right) \\
= & -\frac{1}{2} \boldsymbol{\eta}^{\mathrm{T}} A \boldsymbol{\eta}^{*} -\frac{1}{2} \boldsymbol{\eta}^{*\mathrm{T}} A \boldsymbol{\eta} \\
= & -\boldsymbol{\eta}^{\dagger} A \boldsymbol{\eta} \;.
\end{split}
\label{eq:ChangeVariableGaussianRealGrassmannExponent}
\end{equation}
Combining~\eqref{eq:ChangeVariableGaussianRealGrassmannIntMeasure} and~\eqref{eq:ChangeVariableGaussianRealGrassmannExponent} with~\eqref{eq:ProofPfaffianGaussianRealGrassmann}, we end up with:
\begin{equation}
\int d^{n}\boldsymbol{\eta}^{\dagger} d^{n}\boldsymbol{\eta} \ e^{-\boldsymbol{\eta}^{\dagger} A \boldsymbol{\eta}}=\left[\mathrm{Pf}(A)\right]^{2}\;.
\label{eq:NewExpressionForPfaffianSquared}
\end{equation}
According to result~\eqref{eq:GaussianNComponentVecGrass}, the LHS of~\eqref{eq:NewExpressionForPfaffianSquared} can be identified as the determinant of $A$ so that:
\begin{equation}
\mathrm{Det}(A)=\left[\mathrm{Pf}(A)\right]^{2}\;.
\end{equation}
From this, we infer that~\eqref{eq:PfaffianGaussianRealGrassmann} is equivalent to:
\begin{equation}
\int \frac{d^{n}\boldsymbol{\theta}}{\pm\left(-1\right)^{n/2}} \ e^{-\frac{1}{2}\boldsymbol{\theta}^{\mathrm{T}} A \boldsymbol{\theta}}=\left[\mathrm{Det}(A)\right]^{+1/2}\;.
\label{eq:GaussianRealGrassmannNcomponents}
\end{equation}
The homologous relation for fields is obtained in the continuum limit $n\rightarrow\infty$:
\begin{equation}
\int\mathcal{D}\widetilde{\psi} \ e^{-\frac{1}{2}\int_{\alpha_{1},\alpha_{2}} \widetilde{\psi}_{\alpha_{1}} A_{\alpha_{1}\alpha_{2}} \widetilde{\psi}_{\alpha_{2}}} = \left[\mathrm{Det}(A)\right]^{\textcolor{blue}{+1/2}}\;,
\label{eq:GaussianComplexGrassmannfield}
\end{equation}
with $\widetilde{\psi}_{\alpha}$ being now a real Grassmann field and the $\alpha$-indices play the same role as in~\eqref{eq:GaussianComplexNGfield}.

\end{itemize}

\vspace{0.3cm}

According to~\eqref{eq:GaussianComplexNGfield} and~\eqref{eq:GaussianComplexGrassmannfield}, the Gaussian functional integral formula for an arbitrary real field is\footnote{The integration measures involved in~\eqref{eq:GaussianNComponentVecNG} and~\eqref{eq:GaussianRealGrassmannNcomponents} set the convention for the PI measure $\mathcal{D}\widetilde{\psi}$, which hence absorbs a factor $\underset{n\rightarrow\infty}{\mathrm{lim}}(2\pi)^{-n/2}$ for a bosonic field and $\underset{n\rightarrow\infty}{\mathrm{lim}}\pm(-1)^{n/2}$ for a Grassmann one, respectively.}:
\begin{equation}
\int\mathcal{D}\widetilde{\psi} \ e^{-\frac{1}{2}\int_{\alpha_{1},\alpha_{2}} \widetilde{\psi}_{\alpha_{1}} A_{\alpha_{1}\alpha_{2}} \widetilde{\psi}_{\alpha_{2}}} = \left[\mathrm{Det}(A)\right]^{\textcolor{blue}{-\zeta/2}} = e^{\textcolor{blue}{-\frac{\zeta}{2}}\mathrm{STr}[\ln(A)]}\;,
\label{eq:GaussianArbitraryComplexfield}
\end{equation}
with $\zeta$ equaling \textcolor{blue}{+1} if $\widetilde{\psi}$ is a bosonic field and \textcolor{blue}{-1} if $\widetilde{\psi}$ is a Grassmann field. The supertrace $\mathrm{STr}$ introduced in~\eqref{eq:GaussianArbitraryComplexfield} is by definition the trace taken with respect to $\alpha$-indices.

%% file: 7Appendix/LargeNExpansion.tex
We summarize below the derivation of the $1/N$-expansion for the (0+0)-D $O(N)$-symmetric $\varphi^{4}$-theory given by Keitel and Bartosch in ref.~\cite{kei12} which is also exploited by Rosa \textit{et al.} in ref.~\cite{ros16}. We will pursue the derivations up to higher orders for the gs density and also explain how to extend their results in the broken-symmetry regime. Let us start by considering the partition function of the studied toy model\footnote{We set $\hbar=1$ in this entire appendix on the $1/N$-expansion.}:
\begin{equation}
Z = \int_{\mathbb{R}^N} d^{N}\vec{\widetilde{\varphi}} \ e^{-S\big(\vec{\widetilde{\varphi}}\big)} = \int_{\mathbb{R}^N} d^{N}\vec{\widetilde{\varphi}} \ e^{-\frac{m^{2}}{2}\vec{\widetilde{\varphi}}^{2}-\frac{\lambda}{4!}\left(\vec{\widetilde{\varphi}}^{2}\right)^{2}} \;.
\label{eq:1overNpartitionfunction}
\end{equation}
In the limit $N\rightarrow\infty$, our zero-dimensional toy model possesses an infinite number of dofs and fluctuations are therefore suppressed in this case. This translates into the constraint that the integrand vanishes in~\eqref{eq:1overNpartitionfunction} in this limit, which can be achieved by imposing that the classical action $S\big(\vec{\widetilde{\varphi}}\big)$ is of order $\mathcal{O}(N)$. In this way, we deduce that $\vec{\widetilde{\varphi}}^{2}=\mathcal{O}(N)$ since the mass $m$ does not depend on $N$. In order to ensure dimensional consistency, we infer from this that $\lambda=\mathcal{O}\big(N^{-1}\big)$. This dimensional analysis suggests to introduce the $N$-independent quantities $\widetilde{y}\equiv N^{-1}\vec{\widetilde{\varphi}}^{2}$ and $\breve{\lambda}\equiv N\lambda$. After defining the norm $\widetilde{u}\equiv \Big|\vec{\widetilde{\varphi}}\Big|=\sqrt{N\widetilde{y}}$, we rewrite~\eqref{eq:1overNpartitionfunction} in hyperspherical coordinates as follows\footnote{Since the scalar product $\vec{\widetilde{\varphi}}^{2}$ and therefore the integrand in~\eqref{eq:1overNpartitionfunction} are isotropic in color space, we can carry out integration over angular variables in the entire color space, thus introducing the surface area of the $N$-dimensional unit sphere $\Omega_{N}=2\pi^{N/2}/\Gamma(N/2)$ (with $\Gamma$ being Euler gamma function~\cite{abr65}) into~\eqref{eq:1overNintroducefy}.}:
\begin{equation}
\begin{split}
Z^{\text{$1/N$-exp}} = & \ \Omega_{N} \int_{0}^{\infty} d\widetilde{u} \ \widetilde{u}^{N-1} e^{-\frac{m^{2}}{2}\widetilde{u}^{2}-\frac{\lambda}{4!}\widetilde{u}^{4}} \\
= & \ \Omega_{N} N^{\frac{N-1}{2}} \int_{0}^{\infty} \left(\frac{1}{2}\sqrt{\frac{N}{\widetilde{y}}} \ d\widetilde{y}\right) \widetilde{y}^{\frac{N-1}{2}} e^{-N\left(\frac{m^{2}}{2}\widetilde{y}+\frac{\breve{\lambda}}{4!}\widetilde{y}^{2}\right)} \\
= & \ \frac{1}{2} \Omega_{N} N^{\frac{N}{2}} \int_{0}^{\infty} \frac{d\widetilde{y}}{\widetilde{y}} \ e^{-N f(\widetilde{y})} \;,
\end{split}
\label{eq:1overNintroducefy}
\end{equation}
where we have notably used $d\widetilde{u}=\sqrt{N/\widetilde{y}} \ d\widetilde{y}/2$ to obtain the second line as well as the following definition in the third line:
\begin{equation}
f\big(\widetilde{y}\big) \equiv \frac{m^{2}}{2}\widetilde{y}+\frac{\breve{\lambda}}{4!}\widetilde{y}^{2} -\frac{1}{2}\ln(\widetilde{y})\;.
\label{eq:1overNdefinitiony}
\end{equation}
At this stage, we are in a situation very similar to that of the starting point of the original LE in~\eqref{eq:ZJKfiniteD}, except that the role of the expansion parameter is now played by $1/N$ instead of $\hbar$. Hence, we will carry out a saddle point approximation by solving:
\begin{equation}
\left.\frac{\partial f\big(\widetilde{y}\big)}{\partial\widetilde{y}}\right|_{\widetilde{y}=\overline{y}} = 0\;,
\end{equation}
which has two solutions, whose physical relevance depends on the value of $m^{2}$, as in~\eqref{eq:SolutionsmodulusrhoLE} (combined with~\eqref{eq:DefPhiclModulusRho}). We find for $\lambda \neq 0$:
\begin{equation}
\overline{y} = \left\{
\begin{array}{lll}
		\displaystyle{\frac{3 m^{2}}{\breve{\lambda}}\left(\sqrt{1+\frac{2\breve{\lambda}}{3 m^{4}}} - 1\right) \quad \forall m^2 > 0\;.} \\
		\\
		\displaystyle{\frac{3 m^{2}}{\breve{\lambda}}\left(-\sqrt{1+\frac{2\breve{\lambda}}{3 m^{4}}} - 1\right) \quad \forall m^2 < 0\;.}
    \end{array}
\right.
\label{eq:1overNsaddlepointapprox0DON}
\end{equation}
Only the upper solution was exploited in refs.~\cite{kei12} and~\cite{ros16} as these works focus on the unbroken-symmetry phase in their applications of the $1/N$-expansion. Then, the rest of the procedure is also quite similar to that of a LE. We Taylor expand $f\big(\widetilde{y}\big)$ as well as $1/\widetilde{y}$ around $\widetilde{y}=\overline{y}$ up to a chosen order in $\widetilde{y}-\overline{y}$ and finally expand the partition function thus obtained. If $f\big(\widetilde{y}\big)$ and $1/\widetilde{y}$ are respectively expanded up to fourth and second orders in $\widetilde{y}-\overline{y}$, this leads to:
\begin{equation}
Z^{\text{$1/N$-exp}} = \Omega_{N}N^{\frac{N}{2}}\sqrt{\frac{2\pi}{4\overline{y}^{2}\overline{f}^{(2)}}} \ e^{-N \overline{f}} \left[1+\frac{12 m^{4} \overline{y}^{2}-27m^{2}\overline{y}+16}{6N\left(2-m^{2}\overline{y}\right)^{3}}\right]\left[1+\mathcal{O}\hspace{-0.07cm}\left(\frac{1}{N}\right)\right]\;,
\label{eq:1overNZseries0DON}
\end{equation}
where the function $f$ and its second-order derivative can be rewritten as follows when evaluated at $\widetilde{y}=\overline{y}$:
\begin{equation}
\overline{f} \equiv f\big(\widetilde{y}=\overline{y}\big) = \frac{m^{2}}{4}\overline{y}+\frac{1}{4}-\frac{1}{2}\ln(\overline{y})\;,
\end{equation}
\begin{equation}
\overline{f}^{(2)} \equiv \left.\frac{\partial^{2} f\big(\widetilde{y}\big)}{\partial\widetilde{y}^{2}}\right|_{\widetilde{y}=\overline{y}} = \frac{1}{\overline{y}^{2}}-\frac{m^{2}}{2\overline{y}}\;.
\end{equation}
The gs energy $E^{\text{$1/N$-exp}}_{\mathrm{gs}}=-\ln\big(Z^{\text{$1/N$-exp}}\big)$ can be expressed by expanding $\ln\big(Z^{\text{$1/N$-exp}}\big)$, with $Z^{\text{$1/N$-exp}}$ given by~\eqref{eq:1overNZseries0DON}\footnote{Note that only the leftmost logarithm term in our expression~\eqref{eq:1overNEgsseries0DON} for the gs energy differs from that of the interaction-induced shift of the free energy $\Gamma^{(0)}$ given by equation~(A.5) in ref.~\cite{kei12} since $\Gamma^{(0)}=-\ln(Z)+\ln(Z_{0})=E_{\mathrm{gs}}+N\ln\hspace{-0.07cm}\big(2\pi/m^{2}\big)/2$.}:
\begin{equation}
E^{\text{$1/N$-exp}}_{\mathrm{gs}} = N\Bigg[\frac{m^{2}}{4}\overline{y}-\frac{1}{4}-\frac{1}{2}\ln(2\pi\overline{y})\Bigg]+\frac{1}{2}\ln\hspace{-0.07cm}\Big(2-m^{2}\overline{y}\Big)-\frac{1}{N}\Bigg[\frac{\left(8+m^{2}\overline{y}\right)\left(m^{2}\overline{y}-1\right)^{2}}{6\left(2-m^{2}\overline{y}\right)^{3}}\Bigg]+\mathcal{O}\hspace{-0.07cm}\left(\frac{1}{N^{2}}\right)\;,
\label{eq:1overNEgsseries0DON}
\end{equation}
and the gs density is determined by differentiating~\eqref{eq:1overNEgsseries0DON} with respect to $m^{2}$ according to the relation\footnote{The dependence of $\overline{y}$ with respect to $m^{2}$, which is set by~\eqref{eq:1overNsaddlepointapprox0DON}, must be taken into account when differentiating $E^{\text{$1/N$-exp}}_{\mathrm{gs}}$ to deduce $\rho^{\text{$1/N$-exp}}_{\mathrm{gs}}$.} $\rho^{\text{$1/N$-exp}}_{\mathrm{gs}}=\frac{2}{N}\frac{\partial E^{\text{$1/N$-exp}}_{\mathrm{gs}}}{\partial m^{2}}$ (which follows from~\eqref{eq:DefEgsExactZexact0DON} to~\eqref{eq:vacuumExpectationValue0DON}):
\begin{equation}
\scalebox{0.94}{${\displaystyle \rho^{\text{$1/N$-exp}}_{\mathrm{gs}} = \overline{y} +\frac{1}{N}\left[\frac{\frac{2}{\frac{\breve{\lambda}}{3}\overline{y}+m^{2}}-2\overline{y}}{2-m^{2}\overline{y}}\right] +\frac{1}{N^{2}}\left[\frac{4\left(1-m^{2}\overline{y}\right)\left(1+2m^{2}\overline{y}\right)\left(-\sqrt{3}+m^{2}\overline{y}+\frac{\breve{\lambda}}{3}\overline{y}^{2}\right)}{\left(\frac{\breve{\lambda}}{3}\overline{y}+m^{2}\right)\left(-2+m^{2}\overline{y}\right)^{4}}\right] + \mathcal{O}\hspace{-0.07cm}\left(\frac{1}{N^{3}}\right)\;. }$}
\label{eq:1overNrhogsseries0DON}
\end{equation}
For $m^{2}>0$, the first-order coefficient in~\eqref{eq:1overNrhogsseries0DON} reduces to the free gs density (i.e. the exact gs density at $\lambda=0$) in the limit of vanishing coupling constant, i.e.:
\begin{equation}
\underset{\lambda \rightarrow 0}{\lim} ~ \overline{y} = \frac{1}{m^{2}} \mathrlap{\quad \forall m^{2}>0\;,}
\end{equation}
as expected. Note that an alternative derivation of result~\eqref{eq:1overNEgsseries0DON} was developed by Schelstraete and Verschelde in ref.~\cite{sch94}.

%% file: 7Appendix/Diag.tex
\section{Original loop expansion and optimized perturbation theory}
\label{sec:DiagLEO}

In the framework of the original LE, each diagram contributing to the Schwinger functional $W^\text{LE;orig}\big[\vec{J},\boldsymbol{K}\big]$ (expressed by~\eqref{eq:WKjLoopExpansionStep3}) comes with a multiplicity\footnote{The multiplicity corresponds to the number of Wick's contractions represented by a given diagram, i.e. the number of different ways of pairing the indices in the RHS of~\eqref{eq:TrickWickTheoremFinalStep}.} given by:
\begin{equation}
\mathcal{M}_{\mathrm{LE},\mathrm{orig}}=\frac{(2p)!!4^{p}(2q)!!}{(2!)^{S+D} N_{\mathrm{V}}} \;,
\label{eq:MultiplicityDiagLoopExpansion}
\end{equation}
which can be deduced from the work of ref.~\cite{kle00}. The integers $S$ and $D$ denote respectively the number of self and double connections with the propagator lines~\eqref{eq:FeynRulesLoopExpansionPropagator} (representing $\boldsymbol{G}_{\varphi_{\mathrm{cl}};JK;ab}(x,y)$) whereas $q$ and $p$ are the number of vertices~\eqref{eq:FeynRulesLoopExpansion3legVertex} and~\eqref{eq:FeynRulesLoopExpansion4legVertex} involved in the diagram under consideration. Finally, $N_{\mathrm{V}}$ corresponds to the number of vertex permutations that leave the diagram unchanged. These numbers are given in tab.~\ref{tab:MultiplicityOPTdiagramsON} for the diagrams of~\eqref{eq:WKjLoopExpansionStep3}. Formula~\eqref{eq:MultiplicityDiagLoopExpansion} can be generalized for the OPT expansion by taking into account that the underlying diagrams might possess 2-leg vertices as well:
\begin{equation}
\mathcal{M}_{\mathrm{OPT}}=\frac{(2!)^{r}(r!)(2p)!!4^{p}(2q)!!}{(2!)^{S+D} N_{\mathrm{V}}}\;,
\label{eq:MultiplicityDiagOPT}
\end{equation}
where $r$ denotes the number of 2-leg vertices~\eqref{eq:FeynRulesOPTvertexSquare}, whereas $S$ and $D$ are now respectively the number of self and double connections with the propagator lines~\eqref{eq:FeynRulesOPTPropagator} (representing $\boldsymbol{G}_{\sigma;ab}(x,y)$). Every diagram contributing to $W^\text{OPT}$ (expressed by~\eqref{eq:WKjLoopExpansionStep3OPT}) up to order $\mathcal{O}\big(\delta^{3}\big)$ is given in tab.~\ref{tab:MultiplicityOPTdiagramsON} together with the corresponding $S$, $D$ and $N_{\mathrm{V}}$ factors. Note that we always have $q=0$ in the framework of OPT as treated in section~\ref{sec:OPT} since the OPT expansion was performed around a trivial saddle point (i.e. for $\vec{\varphi}_{\mathrm{cl}}=\vec{0}$) in this section. It is also possible to design an OPT expansion around a non-trivial (i.e. finite) saddle point, just like the original LE for the studied toy model in the broken-symmetry regime (for which $\vec{\varphi}^{2}_{\mathrm{cl}}=-6m^{2}/\lambda$). Such an expansion is of little relevance for the zero-dimensional $O(N)$ model under consideration since the minimum of its exact effective potential $V_\text{eff}^\text{exact}\big(\vec{\phi}\big)$ always lies at $\vec{\phi}=\vec{0}$, as was discussed in section~\ref{sec:studiedtoymodel}. However, the additional OPT diagrams appearing when $\vec{\varphi}_{\mathrm{cl}} \neq \vec{0}$ are given in tab.~\ref{tab:MultiplicityOPTdiagramsON} as well.

\vspace{0.5cm}

Summations over internal and spacetime indices are also implicitly accounted for by all diagrammatic representations used in chapter~\ref{chap:DiagTechniques} (and especially in~\eqref{eq:WKjLoopExpansionStep3} and~\eqref{eq:WKjLoopExpansionStep3OPT} for the original LE and OPT expansions). In order to explain how to evaluate the sums over color (i.e. internal) indices, let us assume exceptionally that the latter sums are not taken into account by the diagrams. To that end, let us focus temporarily on the original LE (based on the expectation value $\big\langle\cdots\big\rangle_{0,JK}$ defined by~\eqref{eq:SourceDepExpValueOriginalLE} and~\eqref{eq:Z0JKoriginalLE}) and we make use of the multinomial theorem in order to write:
\begin{equation}
\begin{split}
\scalebox{0.96}{${\displaystyle \lambda \left\langle \left(\vec{\widetilde{\chi}}^{2}\right)^{2} \right\rangle_{0,JK} = }$} & \ \scalebox{0.96}{${\displaystyle \lambda \left\langle \left(\int_{x}\widetilde{\chi}_{1,x}^{2} + \int_{x}\widetilde{\chi}_{2,x}^{2} + \cdots + \int_{x}\widetilde{\chi}_{N,x}^{2}\right)^{2} \right\rangle_{0,JK} }$} \\
\scalebox{0.96}{${\displaystyle = }$} & \ \scalebox{0.96}{${\displaystyle \lambda \sum_{\underset{\{i_{1} + i_{2} + \cdots + i_{N}=2\}}{i_{1},i_{2},\cdots,i_{N}=0}} \begin{pmatrix}
2 \\
i_{1},i_{2},\cdots,i_{N}
\end{pmatrix} }$} \\
& \hspace{2.2cm} \scalebox{0.96}{${\displaystyle \times \int_{x^{(1)}_{1},\cdots,x^{(1)}_{i_{1}},\cdots,x^{(N)}_{1},\cdots,x^{(N)}_{i_{N}}} \left\langle \widetilde{\chi}_{1,x^{(1)}_{1}}^{2} \cdots \widetilde{\chi}_{1,x^{(1)}_{i_{1}}}^{2} \cdots \widetilde{\chi}_{N,x^{(N)}_{1}}^{2} \cdots \widetilde{\chi}_{N,x^{(N)}_{i_{N}}}^{2} \right\rangle_{0,JK} }$} \\
\scalebox{0.96}{${\displaystyle = }$} & \ \scalebox{0.96}{${\displaystyle \lambda \sum_{a=1}^{N} \int_{x_{1}^{(a)},x_{2}^{(a)}} \left\langle \widetilde{\chi}_{a,x^{(a)}_{1}}^{2} \widetilde{\chi}_{a,x^{(a)}_{2}}^{2} \right\rangle_{0,JK} + 2\lambda \sum_{\underset{\{a > b\}}{a,b=1}}^{N} \int_{x_{1}^{(a)},x_{1}^{(b)}} \left\langle \widetilde{\chi}_{a,x^{(a)}_{1}}^{2} \widetilde{\chi}_{b,x^{(b)}_{1}}^{2} \right\rangle_{0,JK} }$} \\
\scalebox{0.96}{${\displaystyle = }$} & \ \scalebox{0.96}{${\displaystyle \lambda \sum_{a=1}^{N} \int_{x_{1}^{(a)},x_{2}^{(a)}} \left\langle \widetilde{\chi}_{a,x^{(a)}_{1}}^{2} \widetilde{\chi}_{a,x^{(a)}_{2}}^{2} \right\rangle_{0,JK} + \lambda \sum_{\underset{\{a \neq b\}}{a,b=1}}^{N} \int_{x_{1}^{(a)},x_{1}^{(b)}} \left\langle \widetilde{\chi}_{a,x^{(a)}_{1}}^{2} \widetilde{\chi}_{b,x^{(b)}_{1}}^{2} \right\rangle_{0,JK} }$} \\
\scalebox{0.96}{${\displaystyle = }$} & \ \scalebox{0.96}{${\displaystyle \lambda N \int_{x_{1}^{(1)},x_{2}^{(1)}} \left\langle \widetilde{\chi}_{1,x^{(1)}_{1}}^{2} \widetilde{\chi}_{1,x^{(1)}_{2}}^{2} \right\rangle_{0,JK} + \lambda N(N-1) \int_{x_{1}^{(1)},x_{1}^{(2)}} \left\langle \widetilde{\chi}_{1,x^{(1)}_{1}}^{2} \widetilde{\chi}_{2,x^{(2)}_{1}}^{2} \right\rangle_{0,JK} }$} \\
\scalebox{0.96}{${\displaystyle = }$} & \ \scalebox{0.96}{${\displaystyle N \left(\rule{0cm}{1.0cm}\right. \begin{gathered}
\begin{fmffile}{Diagrams/LoopExpansion1_Hartree}
\begin{fmfgraph}(30,20)
\fmfleft{i}
\fmfright{o}
\fmf{phantom,tension=10}{i,i1}
\fmf{phantom,tension=10}{o,o1}
\fmf{plain,left,tension=0.5}{i1,v1,i1}
\fmf{plain,right,tension=0.5}{o1,v2,o1}
\fmf{zigzag,foreground=(0,,0,,1)}{v1,v2}
\end{fmfgraph}
\end{fmffile}
\end{gathered}
+ 2 \begin{gathered}
\begin{fmffile}{Diagrams/LoopExpansion1_Fock}
\begin{fmfgraph}(15,15)
\fmfleft{i}
\fmfright{o}
\fmf{phantom,tension=11}{i,v1}
\fmf{phantom,tension=11}{v2,o}
\fmf{plain,left,tension=0.4}{v1,v2,v1}
\fmf{zigzag,foreground=(0,,0,,1)}{v1,v2}
\end{fmfgraph}
\end{fmffile}
\end{gathered} \left.\rule{0cm}{1.0cm}\right) + N(N-1)\begin{gathered}
\begin{fmffile}{Diagrams/LoopExpansion1_Hartree}
\begin{fmfgraph}(30,20)
\fmfleft{i}
\fmfright{o}
\fmf{phantom,tension=10}{i,i1}
\fmf{phantom,tension=10}{o,o1}
\fmf{plain,left,tension=0.5}{i1,v1,i1}
\fmf{plain,right,tension=0.5}{o1,v2,o1}
\fmf{zigzag,foreground=(0,,0,,1)}{v1,v2}
\end{fmfgraph}
\end{fmffile}
\end{gathered} }$} \\
\scalebox{0.96}{${\displaystyle = }$} & \ \scalebox{0.96}{${\displaystyle N^{2} \begin{gathered}
\begin{fmffile}{Diagrams/LoopExpansion1_Hartree}
\begin{fmfgraph}(30,20)
\fmfleft{i}
\fmfright{o}
\fmf{phantom,tension=10}{i,i1}
\fmf{phantom,tension=10}{o,o1}
\fmf{plain,left,tension=0.5}{i1,v1,i1}
\fmf{plain,right,tension=0.5}{o1,v2,o1}
\fmf{zigzag,foreground=(0,,0,,1)}{v1,v2}
\end{fmfgraph}
\end{fmffile}
\end{gathered} + 2 N \begin{gathered}
\begin{fmffile}{Diagrams/LoopExpansion1_Fock}
\begin{fmfgraph}(15,15)
\fmfleft{i}
\fmfright{o}
\fmf{phantom,tension=11}{i,v1}
\fmf{phantom,tension=11}{v2,o}
\fmf{plain,left,tension=0.4}{v1,v2,v1}
\fmf{zigzag,foreground=(0,,0,,1)}{v1,v2}
\end{fmfgraph}
\end{fmffile}
\end{gathered}\;,}$}
\end{split}
\label{eq:NfactorPropagatorLoop0DON}
\end{equation}
where $x_{n}$, $x^{(a)}_{n}$ and $x^{(b)}_{n}$ denote spacetime indices for all $n$. The curly braces in the discrete sums of the second, third and fourth equalities of~\eqref{eq:NfactorPropagatorLoop0DON} contain conditions that must be satisfied by all terms generated by the corresponding sum. We have also exploited the properties $\big\langle \widetilde{\chi}_{a,x_{1}}^{2} \widetilde{\chi}_{a,x_{2}}^{2} \big\rangle_{0,JK} = \big\langle \widetilde{\chi}_{1,x_{1}}^{2} \widetilde{\chi}_{1,x_{2}}^{2} \big\rangle_{0,JK}$ $\forall a,x_{1},x_{2}$ and $\big\langle \widetilde{\chi}_{a,x_{1}}^{2} \widetilde{\chi}_{b,x_{2}}^{2} \big\rangle_{0,JK} = \big\langle \widetilde{\chi}_{1,x_{1}}^{2} \widetilde{\chi}_{2,x_{2}}^{2} \big\rangle_{0,JK}$ $\forall x_{1}, x_{2}$ with $a \neq b$ to obtain the fifth equality, as well as $\boldsymbol{G}_{\varphi_{\mathrm{cl}};JK,(a,x_{1})(b,x_{2})} = 0$ $\forall x_{1}, x_{2}$ with $a \neq b$\footnote{We stress that, although this is not directly apparent from definition~\eqref{eq:LEG}, the condition $\boldsymbol{G}_{\varphi_{\mathrm{cl}};JK,(a,x_{1})(b,x_{2})} = 0$ $\forall x_{1}, x_{2}$ with $a \neq b$ also holds in the broken-symmetry phase of the studied $O(N)$ model in the framework of our convention~\eqref{eq:DefPhiclModulusRho} which does not induce any loss of generality in our derivations.} to get the sixth equality. Hence, summations over color indices yield a factor $N^2$ and $N$ for a Hartree and a Fock diagram respectively. This illustrates a more general rule, namely that the contribution of each diagram with $m$ propagator loops is proportional to $N^{m}$. Note that the loop in question must be exclusively made of propagator lines, e.g. the contribution of
\begin{equation*}
\begin{gathered}
\begin{fmffile}{Diagrams/LoopExpansion1_Diag2}
\begin{fmfgraph}(27,15)
\fmfleft{i}
\fmfright{o}
\fmftop{vUp}
\fmfbottom{vDown}
\fmfv{decor.shape=cross,decor.size=3.5thick,foreground=(0,,0,,1)}{v1}
\fmfv{decor.shape=cross,decor.size=3.5thick,foreground=(0,,0,,1)}{v2}
\fmf{phantom,tension=10}{i,i1}
\fmf{phantom,tension=10}{o,o1}
\fmf{phantom,tension=2.2}{vUp,v5}
\fmf{phantom,tension=2.2}{vDown,v6}
\fmf{phantom,tension=0.5}{v3,v4}
\fmf{phantom,tension=10.0}{i1,v1}
\fmf{phantom,tension=10.0}{o1,v2}
\fmf{dashes,tension=2.0,foreground=(0,,0,,1)}{v1,v3}
\fmf{dots,left=0.4,tension=0.5,foreground=(0,,0,,1)}{v3,v5}
\fmf{plain,left=0.4,tension=0.5}{v5,v4}
\fmf{plain,right=0.4,tension=0.5}{v3,v6}
\fmf{dots,right=0.4,tension=0.5,foreground=(0,,0,,1)}{v6,v4}
\fmf{dashes,tension=2.0,foreground=(0,,0,,1)}{v4,v2}
\fmf{plain,tension=0}{v5,v6}
\end{fmfgraph}
\end{fmffile}
\end{gathered}
\end{equation*}
is independent of $N$ whereas
\begin{equation*}
\begin{gathered}
\begin{fmffile}{Diagrams/LoopExpansion1_Diag4}
\begin{fmfgraph}(35,18)
\fmfleft{i}
\fmfright{o}
\fmftop{vUp}
\fmfbottom{vDown}
\fmfv{decor.shape=cross,decor.size=3.5thick,foreground=(0,,0,,1)}{v3bis}
\fmfv{decor.shape=cross,decor.size=3.5thick,foreground=(0,,0,,1)}{o}
\fmf{phantom,tension=10}{i,i1}
\fmf{dashes,tension=1.2,foreground=(0,,0,,1)}{o,v4}
\fmf{phantom,tension=0.5}{v3bis,i}
\fmf{phantom,tension=2.7}{v3bis,vUp}
\fmf{dashes,tension=0.9,foreground=(0,,0,,1)}{v3,v3bis}
\fmf{phantom,tension=0.5}{v4bis,i}
\fmf{phantom,tension=2.7}{v4bis,vDown}
\fmf{phantom,tension=0.9}{v3,v4bis}
\fmf{plain,left,tension=0.5}{i1,v1,i1}
\fmf{plain,right,tension=0.5}{v2,v4}
\fmf{dots,left,tension=0.5,foreground=(0,,0,,1)}{v2,v4}
\fmf{dots,foreground=(0,,0,,1)}{v1,v3}
\fmf{plain}{v3,v2}
\end{fmfgraph}
\end{fmffile}
\end{gathered}
\end{equation*}
yields a contribution of order $\mathcal{O}(N)$.

\vspace{0.5cm}

Finally, we illustrate an advantage of the studied toy model resulting from its (0+0)-D nature: in (0+0)-D and as opposed to finite-dimensional problems, partition functions or Schwinger functionals can be expressed in terms of a Lebesgue integral which can be directly expanded with respect to $\hbar$ or another chosen parameter after a relevant change of coordinates, thus bypassing the increasing complexity of the diagrammatic with the truncation order. For the original LE, the change of coordinates we are referring to consists in rewriting in hyperspherical coordinates the zero-dimensional counterpart of the generating functional~\eqref{eq:ZJKfiniteD} after the saddle point approximation introducing $\vec{\widetilde{\chi}}$ via $\vec{\widetilde{\varphi}}=\vec{\varphi}_{\mathrm{cl}}+\sqrt{\hbar} \ \vec{\widetilde{\chi}}$. This generating functional can be put in the form:
\begin{equation}
\begin{split}
\scalebox{0.95}{${\displaystyle Z\Big(\vec{J},\boldsymbol{K}\Big) = }$} & \ \scalebox{0.95}{${\displaystyle e^{-\frac{1}{\hbar}S_{JK}(\vec{\varphi}_{\mathrm{cl}})} \int_{\mathbb{R}^N} d^{N}\vec{\widetilde{\chi}} \ e^{-\frac{1}{2} \vec{\widetilde{\chi}} \cdot \big(\boldsymbol{G}^{-1}_{\varphi_{\text{cl}};JK}\vec{\widetilde{\chi}}\big)-\frac{\hbar^{\frac{1}{2}}\lambda}{3!} \vec{\widetilde{\chi}}^{2} \left(\vec{\varphi}_{\mathrm{cl}}\cdot\vec{\widetilde{\chi}}\right) - \frac{\hbar\lambda}{4!}\left(\vec{\widetilde{\chi}}^{2}\right)^{2}} }$} \\
\scalebox{0.95}{${\displaystyle = }$} & \ \scalebox{0.95}{${\displaystyle e^{-\frac{1}{\hbar}S_{JK}(\vec{\varphi}_{\mathrm{cl}})} \int_{\mathbb{R}^N} d^{N}\vec{\widetilde{\chi}} \ e^{-\frac{1}{2} \mathfrak{G}^{-1}_{\varphi_\text{cl};JK;\mathfrak{g}} \vec{\widetilde{A}}^{2} - \frac{1}{2} \boldsymbol{G}^{-1}_{\varphi_\text{cl};JK;NN} \widetilde{\chi}^{2}_{N} -\frac{\hbar^{\frac{1}{2}}\lambda}{3!}\varrho\widetilde{\chi}^{}_{N}\left(\vec{\widetilde{A}}^{2}+\widetilde{\chi}^{2}_{N}\right)-\frac{\hbar\lambda}{4!} \left(\left(\vec{\widetilde{A}}^{2}\right)^{2}+2\vec{\widetilde{A}}^{2}\widetilde{\chi}^{2}_{N}+\widetilde{\chi}^{4}_{N}\right)} \;,}$}
\end{split}
\label{eq:ZKjHypersphericalCoordinatesStep1}
\end{equation}
where $\vec{\widetilde{A}}$ satisfies:
\begin{equation}
\vec{\widetilde{\chi}} \equiv \begin{pmatrix}
\widetilde{\chi}^{}_{1} \\
\vdots \\
\widetilde{\chi}^{}_{N-1} \\
\widetilde{\chi}^{}_{N}
\end{pmatrix} \equiv \begin{pmatrix}
\vec{\widetilde{A}} \\
\widetilde{\chi}^{}_{N}
\end{pmatrix}\;,
\label{eq:AvectorLoopExpansion0DON}
\end{equation}
and $\mathfrak{G}^{-1}_{\varphi_\text{cl};JK;\mathfrak{g}} = m^2 + \frac{\lambda}{6} \varrho^2 - K$ is the diagonal part of the inverse Goldstone propagator defined by~\eqref{eq:origLEGoldstoneProp} and~\eqref{eq:origLEGoldstonePropBis} whereas $\boldsymbol{G}^{-1}_{\varphi_\text{cl};JK;NN}=m^2 + \frac{\lambda}{2} \varrho^2 - K$ is associated to the Higgs mode (and expressed by~\eqref{eq:LEGN} in arbitrary dimensions), still assuming that the source $\boldsymbol{K}$ is a scalar in color space (i.e. $\boldsymbol{K}_{ab}=K\delta_{ab}$). Note also that $\varrho$ is the modulus introduced via the choice of coordinates~\eqref{eq:DefPhiclModulusRho} and expressed by~\eqref{eq:SolutionsmodulusrhoLE} at vanishing sources. Taking into account that the $O(N)$ symmetry of our toy model can be spontaneously broken (in the direction set by $a=N$ in color space according to~\eqref{eq:DefPhiclModulusRho}), isotropy in color space is only exhibited in the subspace of dimension $N-1$ in which $\vec{\widetilde{A}}$ lives. Performing the aforementioned change of coordinates within this subspace and carrying out integration over angular variables lead to:
\begin{equation}
Z\Big(\vec{J},\boldsymbol{K}\Big) = \Omega_{N-1} \ e^{-\frac{1}{\hbar}S_{JK}(\vec{\varphi}_{\mathrm{cl}})} \int_{-\infty}^{\infty} d\widetilde{\chi}^{}_{N} \ \mathcal{P}_{N-2}(\widetilde{\chi}^{}_{N},\varrho) \ e^{-\frac{1}{2} \boldsymbol{G}^{-1}_{\varphi_\text{cl};JK;NN} \widetilde{\chi}^{2}_{N}-\frac{\hbar^{\frac{1}{2}}\lambda}{3!}\varrho\widetilde{\chi}^{3}_{N}-\frac{\hbar\lambda}{4!}\widetilde{\chi}^{4}_{N}}\;,
\label{eq:ZKjHypersphericalCoordinatesStep2}
\end{equation}
where $\Omega_{N}=2\pi^{N/2}/\Gamma(N/2)$ (with $\Gamma$ denoting Euler gamma function~\cite{abr65}) is the surface area of the $N$-dimensional unit sphere and we have also:
\begin{equation}
\mathcal{P}_{N}(\widetilde{\chi}^{}_{N},\varrho) = \int_{0}^{\infty} d\widetilde{a} \ \widetilde{a}^{N} e^{-\frac{1}{2} \mathfrak{G}^{-1}_{\varphi_\text{cl};JK;\mathfrak{g}} \widetilde{a}^{2}-\frac{\hbar^{\frac{1}{2}}\lambda}{3!}\varrho\widetilde{\chi}^{}_{N} \widetilde{a}^{2}-\frac{\hbar\lambda}{12}\widetilde{\chi}^{2}_{N} \widetilde{a}^{2}-\frac{\hbar\lambda}{4!} \widetilde{a}^{4}}\;,
\label{eq:ZKjHypersphericalCoordinatesPN}
\end{equation}
with $\widetilde{a}$ being the norm of $\vec{\widetilde{A}}$, i.e. $\widetilde{a}\equiv\Big|\vec{\widetilde{A}}\Big|$. Expanding the RHS of~\eqref{eq:ZKjHypersphericalCoordinatesStep2} with respect to $\hbar$ yields:
\begin{equation}
\begin{split}
\scalebox{0.76}{${\displaystyle Z^\text{LE;orig}\Big(\vec{J},\boldsymbol{K}\Big) = }$} & \scalebox{0.76}{${\displaystyle \ e^{-\frac{1}{\hbar}S_{JK}(\vec{\varphi}_{\mathrm{cl}})} (2\pi)^{\frac{N}{2}} \mathfrak{G}_{\varphi_\text{cl};JK;\mathfrak{g}}^{\frac{N-1}{2}} \sqrt{\boldsymbol{G}_{\varphi_\text{cl};JK;NN}} \ \Bigg\{ 1 }$} \\
& \scalebox{0.76}{${\displaystyle + \frac{\hbar\lambda}{72} \Big[-3 \mathfrak{G}_{\varphi_\text{cl};JK;\mathfrak{g}}^2 \left(-1 + N^2\right) + 15 \boldsymbol{G}_{\varphi_\text{cl};JK;NN}^3 \lambda \varrho^{2} + \boldsymbol{G}_{\varphi_\text{cl};JK;NN}^2 \left(-9 + 6 \mathfrak{G}_{\varphi_\text{cl};JK;\mathfrak{g}} \left(-1 + N\right) \lambda \varrho^{2}\right) }$} \\
& \hspace{0.885cm} \scalebox{0.76}{${\displaystyle + \boldsymbol{G}_{\varphi_\text{cl};JK;NN} \mathfrak{G}_{\varphi_\text{cl};JK;\mathfrak{g}} \left(-1 + N\right) \left(-6 + \mathfrak{G}_{\varphi_\text{cl};JK;\mathfrak{g}} \left(1 + N\right) \lambda \varrho^{2}\right) \Big] }$} \\
& \scalebox{0.76}{${\displaystyle + \frac{\hbar^{2}\lambda^{2}}{10368} \Big[ 9 \mathfrak{G}_{\varphi_\text{cl};JK;\mathfrak{g}}^4 \left(-15 - 8 N + 14 N^2 + 8 N^3 + N^4\right) + 3465 \boldsymbol{G}_{\varphi_\text{cl};JK;NN}^6 \lambda^2 \varrho^{4} }$} \\
& \hspace{1.15cm} \scalebox{0.76}{${\displaystyle + 630 \boldsymbol{G}_{\varphi_\text{cl};JK;NN}^5 \lambda \varrho^{2} \left(-9 + 2 \mathfrak{G}_{\varphi_\text{cl};JK;\mathfrak{g}} \left(-1 + N\right) \lambda \varrho^{2}\right) }$} \\
& \hspace{1.15cm} \scalebox{0.76}{${\displaystyle - 6 \boldsymbol{G}_{\varphi_\text{cl};JK;NN} \mathfrak{G}_{\varphi_\text{cl};JK;\mathfrak{g}}^3 \left(-3 - N + 3 N^2 + N^3\right) \left(-6 + \mathfrak{G}_{\varphi_\text{cl};JK;\mathfrak{g}} \left(5 + N\right) \lambda \varrho^{2}\right) }$} \\
& \hspace{1.15cm} \scalebox{0.76}{${\displaystyle + 105 \boldsymbol{G}_{\varphi_\text{cl};JK;NN}^4 \left(9 - 24 \mathfrak{G}_{\varphi_\text{cl};JK;\mathfrak{g}} \left(-1 + N\right) \lambda \varrho^{2} + 2 \mathfrak{G}_{\varphi_\text{cl};JK;\mathfrak{g}}^2 \left(-1 + N^2\right) \lambda^2 \varrho^{4} \right) }$} \\
& \hspace{1.15cm} \scalebox{0.76}{${\displaystyle + 20 \boldsymbol{G}_{\varphi_\text{cl};JK;NN}^3 \mathfrak{G}_{\varphi_\text{cl};JK;\mathfrak{g}} \left(-1 + N\right) \left(27 - 27 \mathfrak{G}_{\varphi_\text{cl};JK;\mathfrak{g}} \left(1 + N\right) \lambda \varrho^{2} + \mathfrak{G}_{\varphi_\text{cl};JK;\mathfrak{g}}^2 \left(3 + 4 N + N^2\right) \lambda^2 \varrho^{4} \right) }$} \\
& \hspace{1.15cm} \scalebox{0.76}{${\displaystyle + \boldsymbol{G}_{\varphi_\text{cl};JK;NN}^2 \mathfrak{G}_{\varphi_\text{cl};JK;\mathfrak{g}}^2 \left(-1 + N^2\right) \left(162 - 72 \mathfrak{G}_{\varphi_\text{cl};JK;\mathfrak{g}} \left(3 + N\right) \lambda \varrho^{2} + \mathfrak{G}_{\varphi_\text{cl};JK;\mathfrak{g}}^2 \left(15 + 8 N + N^2\right) \lambda^2 \varrho^{4} \right) \Big] }$} \\
& \scalebox{0.76}{${\displaystyle + \frac{\hbar^{3}\lambda^{3}}{2239488} \Big[ -27 \mathfrak{G}_{\varphi_\text{cl};JK;\mathfrak{g}}^6 \left(-945 - 744 N + 739 N^2 + 720 N^3 + 205 N^4 + 24 N^5 + N^6\right) }$} \\
& \hspace{1.631cm} \scalebox{0.76}{${\displaystyle + 2297295 \boldsymbol{G}_{\varphi_\text{cl};JK;NN}^9 \lambda^3 \varrho^{6} + 405405 \boldsymbol{G}_{\varphi_\text{cl};JK;NN}^8 \lambda^2 \varrho^{4} \left(-15 + 2 \mathfrak{G}_{\varphi_\text{cl};JK;\mathfrak{g}} \left(-1 + N\right) \lambda \varrho^{2}\right) }$} \\
& \hspace{1.631cm} \scalebox{0.76}{${\displaystyle + 27 \boldsymbol{G}_{\varphi_\text{cl};JK;NN} \mathfrak{G}_{\varphi_\text{cl};JK;\mathfrak{g}}^5 \left(-1 + N\right) \left(105 + 176 N + 86 N^2 + 16 N^3 + N^4\right) \left(-6 + \mathfrak{G}_{\varphi_\text{cl};JK;\mathfrak{g}} \left(9 + N\right) \lambda \varrho^{2}\right) }$} \\
& \hspace{1.631cm} \scalebox{0.76}{${\displaystyle + 135135 \boldsymbol{G}_{\varphi_\text{cl};JK;NN}^7 \lambda \varrho^{2} \left(27 - 18 \mathfrak{G}_{\varphi_\text{cl};JK;\mathfrak{g}} \left(-1 + N\right) \lambda \varrho^{2} + \mathfrak{G}_{\varphi_\text{cl};JK;\mathfrak{g}}^2 \left(-1 + N^2\right) \lambda^2 \varrho^{4} \right) }$} \\
& \hspace{1.631cm} \scalebox{0.76}{${\displaystyle - 9 \boldsymbol{G}_{\varphi_\text{cl};JK;NN}^2 \mathfrak{G}_{\varphi_\text{cl};JK;\mathfrak{g}}^4 \left(-15 - 8 N + 14 N^2 + 8 N^3 + N^4\right) \Big(135 - 54 \mathfrak{G}_{\varphi_\text{cl};JK;\mathfrak{g}} \left(7 + N\right) \lambda \varrho^{2} }$} \\
& \hspace{1.631cm} \scalebox{0.76}{${\displaystyle + \mathfrak{G}_{\varphi_\text{cl};JK;\mathfrak{g}}^2 \left(63 + 16 N + N^2\right) \lambda^2 \varrho^{4} \Big) + 3465 \boldsymbol{G}_{\varphi_\text{cl};JK;NN}^6 \Big(-81 + 486 \mathfrak{G}_{\varphi_\text{cl};JK;\mathfrak{g}} \left(-1 + N\right) \lambda \varrho^{2} }$} \\
& \hspace{1.631cm} \scalebox{0.76}{${\displaystyle - 135 \mathfrak{G}_{\varphi_\text{cl};JK;\mathfrak{g}}^2 \left(-1 + N^2\right) \lambda^2 \varrho^{4} + 4 \mathfrak{G}_{\varphi_\text{cl};JK;\mathfrak{g}}^3 \left(-3 - N + 3 N^2 + N^3\right) \lambda^3 \varrho^{6}\Big) }$} \\
& \hspace{1.631cm} \scalebox{0.76}{${\displaystyle + \boldsymbol{G}_{\varphi_\text{cl};JK;NN}^3 \mathfrak{G}_{\varphi_\text{cl};JK;\mathfrak{g}}^3 \left(-3 - N + 3 N^2 + N^3\right) \Big(-8100 + 6075 \mathfrak{G}_{\varphi_\text{cl};JK;\mathfrak{g}} \left(5 + N\right) \lambda \varrho^{2} }$} \\
& \hspace{1.631cm} \scalebox{0.76}{${\displaystyle - 270 \mathfrak{G}_{\varphi_\text{cl};JK;\mathfrak{g}}^2 \left(35 + 12 N + N^2\right) \lambda^2 \varrho^{4} + \mathfrak{G}_{\varphi_\text{cl};JK;\mathfrak{g}}^3 \left(315 + 143 N + 21 N^2 + N^3\right) \lambda^3 \varrho^{6} \Big) }$} \\
& \hspace{1.631cm} \scalebox{0.76}{${\displaystyle + 945 \boldsymbol{G}_{\varphi_\text{cl};JK;NN}^5 \mathfrak{G}_{\varphi_\text{cl};JK;\mathfrak{g}} \Big(-162 \left(-1 + N\right) + 405 \mathfrak{G}_{\varphi_\text{cl};JK;\mathfrak{g}} \left(-1 + N^2\right) \lambda \varrho^{2} }$} \\
& \hspace{1.631cm} \scalebox{0.76}{${\displaystyle - 60 \mathfrak{G}_{\varphi_\text{cl};JK;\mathfrak{g}}^2 \left(-3 - N + 3 N^2 + N^3\right) \lambda^2 \varrho^{4} + \mathfrak{G}_{\varphi_\text{cl};JK;\mathfrak{g}}^3 \left(-15 - 8 N + 14 N^2 + 8 N^3 + N^4\right) \lambda^3 \varrho^{6} \Big) }$} \\
& \hspace{1.631cm} \scalebox{0.76}{${\displaystyle + 21 \boldsymbol{G}_{\varphi_\text{cl};JK;NN}^4 \mathfrak{G}_{\varphi_\text{cl};JK;\mathfrak{g}}^2 \Big(-2025 \left(-1 + N^2\right) + 2700 \mathfrak{G}_{\varphi_\text{cl};JK;\mathfrak{g}} \left(-3 - N + 3 N^2 + N^3\right) \lambda \varrho^{2} }$} \\
& \hspace{1.631cm} \scalebox{0.76}{${\displaystyle - 225 \mathfrak{G}_{\varphi_\text{cl};JK;\mathfrak{g}}^2 \left(-15 - 8 N + 14 N^2 + 8 N^3 + N^4\right) \lambda^2 \varrho^{4} }$} \\
& \hspace{1.631cm} \scalebox{0.76}{${\displaystyle + 2 \mathfrak{G}_{\varphi_\text{cl};JK;\mathfrak{g}}^3 \left(-105 - 71 N + 90 N^2 + 70 N^3 + 15 N^4 + N^5\right) \lambda^3 \varrho^{6} \Big) \Big] }$} \\
& \scalebox{0.76}{${\displaystyle + \mathcal{O}\big(\hbar^{4}\big) \Bigg\}\;,}$}
\end{split}
\label{eq:ResultZLoopExpansion0DONAppendix}
\end{equation}
or, considering $W^\text{LE;orig}\big(\vec{J},\boldsymbol{K}\big)\equiv\hbar\ln\Big(Z^\text{LE;orig}\big(\vec{J},\boldsymbol{K}\big)\Big)$ instead of $Z^\text{LE;orig}\big(\vec{J},\boldsymbol{K}\big)$:
\begin{equation*}
\begin{split}
\scalebox{0.74}{${\displaystyle W^\text{LE;orig}\Big(\vec{J},\boldsymbol{K}\Big) = }$} & \scalebox{0.74}{${\displaystyle -S_{JK}(\vec{\varphi}_{\mathrm{cl}}) + \frac{\hbar}{2} \left[(N-1)\ln\big(2\pi \mathfrak{G}_{\varphi_\text{cl};JK;\mathfrak{g}}\big)+\ln\big(2\pi \boldsymbol{G}_{\varphi_\text{cl};JK;NN}\big)\right] }$} \\
& \scalebox{0.74}{${\displaystyle + \frac{\hbar^{2}\lambda}{72} \Big[ -3 \mathfrak{G}_{\varphi_\text{cl};JK;\mathfrak{g}}^2 \left(-1 + N^2\right) + 15 \boldsymbol{G}_{\varphi_\text{cl};JK;NN}^3 \lambda \varrho^2 + \boldsymbol{G}_{\varphi_\text{cl};JK;NN}^2 \left(-9 + 6 \mathfrak{G}_{\varphi_\text{cl};JK;\mathfrak{g}} \left(-1 + N\right) \lambda \varrho^2\right) }$} \\
& \hspace{0.99cm} \scalebox{0.74}{${\displaystyle + \boldsymbol{G}_{\varphi_\text{cl};JK;NN} \mathfrak{G}_{\varphi_\text{cl};JK;\mathfrak{g}} \left(-1 + N\right) \left(-6 + \mathfrak{G}_{\varphi_\text{cl};JK;\mathfrak{g}} \left(1 + N\right) \lambda \varrho^2\right) \Big] }$} \\
& \scalebox{0.74}{${\displaystyle + \frac{\hbar^{3}\lambda^{2}}{1296} \Big[ 9 \mathfrak{G}_{\varphi_\text{cl};JK;\mathfrak{g}}^4 \left(-2 - N + 2 N^2 + N^3\right) + 405 \boldsymbol{G}_{\varphi_\text{cl};JK;NN}^6 \lambda^2 \varrho^{4} + 135 \boldsymbol{G}_{\varphi_\text{cl};JK;NN}^5 \lambda \varrho^2 \left(-5 + \mathfrak{G}_{\varphi_\text{cl};JK;\mathfrak{g}} \left(-1 + N\right) \lambda \varrho^2\right) }$} \\
& \hspace{1.0cm} \scalebox{0.74}{${\displaystyle - 6 \boldsymbol{G}_{\varphi_\text{cl};JK;NN} \mathfrak{G}_{\varphi_\text{cl};JK;\mathfrak{g}}^3 \left(-1 + N^2\right) \left(-3 + \mathfrak{G}_{\varphi_\text{cl};JK;\mathfrak{g}} \left(2 + N\right) \lambda \varrho^2\right) }$} \\
& \hspace{1.0cm} \scalebox{0.74}{${\displaystyle + \boldsymbol{G}_{\varphi_\text{cl};JK;NN}^3 \mathfrak{G}_{\varphi_\text{cl};JK;\mathfrak{g}} \left(-1 + N\right) \left(54 - 9 \mathfrak{G}_{\varphi_\text{cl};JK;\mathfrak{g}} \left(7 + 5 N\right) \lambda \varrho^2 + \mathfrak{G}_{\varphi_\text{cl};JK;\mathfrak{g}}^2 \left(9 + 10 N + N^2\right) \lambda^2 \varrho^{4}\right) }$} \\
& \hspace{1.0cm} \scalebox{0.74}{${\displaystyle + 9 \boldsymbol{G}_{\varphi_\text{cl};JK;NN}^4 \left(12 - 31 \mathfrak{G}_{\varphi_\text{cl};JK;\mathfrak{g}} \left(-1 + N \right) \lambda \varrho^2 + \mathfrak{G}_{\varphi_\text{cl};JK;\mathfrak{g}}^2 \left(-3 + N + 2 N^2\right) \lambda^2 \varrho^{4}\right) }$} \\
& \hspace{1.0cm} \scalebox{0.74}{${\displaystyle + \boldsymbol{G}_{\varphi_\text{cl};JK;NN}^2 \mathfrak{G}_{\varphi_\text{cl};JK;\mathfrak{g}}^2 \left(-1 + N\right) \Big( 18 - 33 \mathfrak{G}_{\varphi_\text{cl};JK;\mathfrak{g}} \lambda \varrho^2 + 2 \mathfrak{G}_{\varphi_\text{cl};JK;\mathfrak{g}}^2 \lambda^2 \varrho^{4} + \mathfrak{G}_{\varphi_\text{cl};JK;\mathfrak{g}} N^2 \lambda \varrho^2 \left(-3 + \mathfrak{G}_{\varphi_\text{cl};JK;\mathfrak{g}} \lambda \varrho^2\right) }$}
\end{split}
\end{equation*}
\begin{equation}
\begin{split}
\hspace{2.4cm} & \hspace{1.0cm} \scalebox{0.74}{${\displaystyle + 3 N \left(3 - 12 \mathfrak{G}_{\varphi_\text{cl};JK;\mathfrak{g}} \lambda \varrho^2 + \mathfrak{G}_{\varphi_\text{cl};JK;\mathfrak{g}}^2 \lambda^2 \varrho^{4}\right) \Big) \Big] }$} \\
& \scalebox{0.74}{${\displaystyle + \frac{\hbar^{4}\lambda^{3}}{279936} \Big[ -108 \mathfrak{G}_{\varphi_\text{cl};JK;\mathfrak{g}}^6 \left(-31 - 24 N + 26 N^2 + 24 N^3 + 5 N^4\right) + 268515 \boldsymbol{G}_{\varphi_\text{cl};JK;NN}^9 \lambda^3 \varrho^{6} }$} \\
& \hspace{1.427cm} \scalebox{0.74}{${\displaystyle + 3645 \boldsymbol{G}_{\varphi_\text{cl};JK;NN}^8 \lambda^2 \varrho^{4} \left(-197 + 24 \mathfrak{G}_{\varphi_\text{cl};JK;\mathfrak{g}} \left(-1 + N\right) \lambda \varrho^2\right) }$} \\
& \hspace{1.427cm} \scalebox{0.74}{${\displaystyle + 2430 \boldsymbol{G}_{\varphi_\text{cl};JK;NN}^7 \lambda \varrho^2 \left(178 - 110 \mathfrak{G}_{\varphi_\text{cl};JK;\mathfrak{g}} \left(-1 + N\right) \lambda \varrho^2 + \mathfrak{G}_{\varphi_\text{cl};JK;\mathfrak{g}}^2 \left(-7 + 2 N + 5 N^2\right) \lambda^2 \varrho^{4}\right) }$} \\
& \hspace{1.427cm} \scalebox{0.74}{${\displaystyle + 27 \boldsymbol{G}_{\varphi_\text{cl};JK;NN}^5 \mathfrak{G}_{\varphi_\text{cl};JK;\mathfrak{g}} \left(-1 + N\right) \Big(-576 + 12 \mathfrak{G}_{\varphi_\text{cl};JK;\mathfrak{g}} \left(139 + 102 N\right) \lambda \varrho^2 }$} \\
& \hspace{1.427cm} \scalebox{0.74}{${\displaystyle - 4 \mathfrak{G}_{\varphi_\text{cl};JK;\mathfrak{g}}^2 \left(217 + 242 N + 31 N^2\right) \lambda^2 \varrho^{4} + \mathfrak{G}_{\varphi_\text{cl};JK;\mathfrak{g}}^3 \left(71 + 99 N + 29 N^2 + N^3\right) \lambda^3 \varrho^{6}\Big) }$} \\
& \hspace{1.427cm} \scalebox{0.74}{${\displaystyle + 54 \boldsymbol{G}_{\varphi_\text{cl};JK;NN}^6 \Big(-594 + 3438 \mathfrak{G}_{\varphi_\text{cl};JK;\mathfrak{g}} \left(-1 + N\right) \lambda \varrho^2 - 3 \mathfrak{G}_{\varphi_\text{cl};JK;\mathfrak{g}}^2 \left(-351 + 92 N + 259 N^2\right) \lambda^2 \varrho^{4} }$} \\
& \hspace{1.427cm} \scalebox{0.74}{${\displaystyle + 2 \mathfrak{G}_{\varphi_\text{cl};JK;\mathfrak{g}}^3 \left(-53 - 6 N + 51 N^2 + 8 N^3\right) \lambda^3 \varrho^{6}\Big) }$} \\
& \hspace{1.427cm} \scalebox{0.74}{${\displaystyle + 108 \boldsymbol{G}_{\varphi_\text{cl};JK;NN} \mathfrak{G}_{\varphi_\text{cl};JK;\mathfrak{g}}^5 \left(-1 + N^2\right) \Big(-24 + 31 \mathfrak{G}_{\varphi_\text{cl};JK;\mathfrak{g}} \lambda \varrho^2 + 5 \mathfrak{G}_{\varphi_\text{cl};JK;\mathfrak{g}} N^2 \lambda \varrho^2 }$} \\
& \hspace{1.427cm} \scalebox{0.74}{${\displaystyle + 12 N \left(-1 + 2 \mathfrak{G}_{\varphi_\text{cl};JK;\mathfrak{g}} \lambda \varrho^2\right)\Big) }$} \\
& \hspace{1.427cm} \scalebox{0.74}{${\displaystyle -36 \boldsymbol{G}_{\varphi_\text{cl};JK;NN}^2 \mathfrak{G}_{\varphi_\text{cl};JK;\mathfrak{g}}^4 \left(-1 + N^2\right) \Big( 63 - 201 \mathfrak{G}_{\varphi_\text{cl};JK;\mathfrak{g}} \lambda \varrho^2 + 31 \mathfrak{G}_{\varphi_\text{cl};JK;\mathfrak{g}}^2 \lambda^2 \varrho^{4} }$} \\
& \hspace{1.427cm} \scalebox{0.74}{${\displaystyle + \mathfrak{G}_{\varphi_\text{cl};JK;\mathfrak{g}} N^2 \lambda \varrho^2 \left(-9 + 5 \mathfrak{G}_{\varphi_\text{cl};JK;\mathfrak{g}} \lambda \varrho^2\right) + 6 N \left(3 - 19 \mathfrak{G}_{\varphi_\text{cl};JK;\mathfrak{g}} \lambda \varrho^2 + 4 \mathfrak{G}_{\varphi_\text{cl};JK;\mathfrak{g}}^2 \lambda^2 \varrho^{4}\right)\Big) }$} \\
& \hspace{1.427cm} \scalebox{0.74}{${\displaystyle + 4 \boldsymbol{G}_{\varphi_\text{cl};JK;NN}^3 \mathfrak{G}_{\varphi_\text{cl};JK;\mathfrak{g}}^3 \left(-1 + N\right) \Big( -864 + 2943 \mathfrak{G}_{\varphi_\text{cl};JK;\mathfrak{g}} \lambda \varrho^2 - 990 \mathfrak{G}_{\varphi_\text{cl};JK;\mathfrak{g}}^2 \lambda^2 \varrho^{4} + 31 \mathfrak{G}_{\varphi_\text{cl};JK;\mathfrak{g}}^3 \lambda^3 \varrho^{6} }$} \\
& \hspace{1.427cm} \scalebox{0.74}{${\displaystyle + \mathfrak{G}_{\varphi_\text{cl};JK;\mathfrak{g}} N^3 \lambda \varrho^2 \left(27 - 54 \mathfrak{G}_{\varphi_\text{cl};JK;\mathfrak{g}} \lambda \varrho^2 + 5 \mathfrak{G}_{\varphi_\text{cl};JK;\mathfrak{g}}^2 \lambda^2 \varrho^{4}\right) }$} \\
& \hspace{1.427cm} \scalebox{0.74}{${\displaystyle + N^2 \left(-54 + 1188 \mathfrak{G}_{\varphi_\text{cl};JK;\mathfrak{g}} \lambda \varrho^2 - 630 \mathfrak{G}_{\varphi_\text{cl};JK;\mathfrak{g}}^2 \lambda^2 \varrho^{4} + 29 \mathfrak{G}_{\varphi_\text{cl};JK;\mathfrak{g}}^3 \lambda^3 \varrho^{6}\right) }$} \\
& \hspace{1.427cm} \scalebox{0.74}{${\displaystyle + N \left(-864 + 4104 \mathfrak{G}_{\varphi_\text{cl};JK;\mathfrak{g}} \lambda \varrho^2 - 1566 \mathfrak{G}_{\varphi_\text{cl};JK;\mathfrak{g}}^2 \lambda^2 \varrho^{4} + 55 \mathfrak{G}_{\varphi_\text{cl};JK;\mathfrak{g}}^3 \lambda^3 \varrho^{6}\right) \Big) }$} \\
& \hspace{1.427cm} \scalebox{0.74}{${\displaystyle + 9 \boldsymbol{G}_{\varphi_\text{cl};JK;NN}^4 \mathfrak{G}_{\varphi_\text{cl};JK;\mathfrak{g}}^2 \left(-1 + N\right) \Big( -540 + 2664 \mathfrak{G}_{\varphi_\text{cl};JK;\mathfrak{g}} \lambda \varrho^2 - 1047 \mathfrak{G}_{\varphi_\text{cl};JK;\mathfrak{g}}^2 \lambda^2 \varrho^{4} + 68 \mathfrak{G}_{\varphi_\text{cl};JK;\mathfrak{g}}^3 \lambda^3 \varrho^{6} }$} \\
& \hspace{1.427cm} \scalebox{0.74}{${\displaystyle + \mathfrak{G}_{\varphi_\text{cl};JK;\mathfrak{g}}^2 N^3 \lambda^2 \varrho^{4} \left(-13 + 4 \mathfrak{G}_{\varphi_\text{cl};JK;\mathfrak{g}} \lambda \varrho^2\right) + \mathfrak{G}_{\varphi_\text{cl};JK;\mathfrak{g}} N^2 \lambda \varrho^2 \left(324 - 429 \mathfrak{G}_{\varphi_\text{cl};JK;\mathfrak{g}} \lambda \varrho^2 + 44 \mathfrak{G}_{\varphi_\text{cl};JK;\mathfrak{g}}^2 \lambda^2 \varrho^{4}\right) }$} \\
& \hspace{1.427cm} \scalebox{0.74}{${\displaystyle + N \left(-324 + 2916 \mathfrak{G}_{\varphi_\text{cl};JK;\mathfrak{g}} \lambda \varrho^2 - 1463 \mathfrak{G}_{\varphi_\text{cl};JK;\mathfrak{g}}^2 \lambda^2 \varrho^{4} + 108 \mathfrak{G}_{\varphi_\text{cl};JK;\mathfrak{g}}^3 \lambda^3 \varrho^{6}\right) \Big) \Big] }$} \\
& \scalebox{0.74}{${\displaystyle + \mathcal{O}\big(\hbar^{5}\big)\;.}$}
\end{split}
\label{eq:ResultWLoopExpansion0DONAppendix}
\end{equation}
It is then straightforward to deduce an expression for the gs energy and density from~\eqref{eq:ResultZLoopExpansion0DONAppendix} or~\eqref{eq:ResultWLoopExpansion0DONAppendix} combined with the relations:
\begin{equation}
E^\text{LE;orig}_{\mathrm{gs}} = -\ln\left( Z^\text{LE;orig}\Big(\vec{J}=\vec{0},\boldsymbol{K}=\boldsymbol{0}\Big)\right) = -\frac{1}{\hbar}W^\text{LE;orig}\Big(\vec{J}=\vec{0},\boldsymbol{K}=\boldsymbol{0}\Big) \;,
\label{eq:DefEgsExactZexact0DONAppendix}
\end{equation}
\begin{equation}
\rho^\text{LE;orig}_{\mathrm{gs}} = \frac{2\hbar}{N} \frac{\partial E^\text{LE;orig}_{\mathrm{gs}}}{\partial m^{2}} \;,
\label{eq:DefrhogsExactwithExpectationValue0DONAppendix}
\end{equation}
which follow respectively from~\eqref{eq:DefEgsExactZexact0DON} and~\eqref{eq:DefrhogsExactwithExpectationValue0DON}. Finally, we stress that the rule of evaluation of sums over color indices illustrated with~\eqref{eq:NfactorPropagatorLoop0DON} (i.e. that the contribution of each diagram with $m$ propagator loops\footnote{The components of the propagators in question must depend on two color indices, which is e.g. not the case of $D_{\sigma_{\mathrm{cl}};\mathcal{J}\mathcal{K}}$ and $\vec{F}_{\varphi_{\mathrm{cl}};\mathcal{J}\mathcal{K}}$ in the mixed LE.} is proportional to $N^{m}$) applies to all diagrammatic techniques investigated in chapter~\ref{chap:DiagTechniques}. Furthermore, the direct expansions of the generating functionals $Z$ or $W$ enable us to obtain directly the series representations of $E_{\mathrm{gs}}$ and $\rho_{\mathrm{gs}}$ in the framework of LEs and OPT. The underlying recipe is always essentially the same as that outlined between~\eqref{eq:ZKjHypersphericalCoordinatesStep1} and~\eqref{eq:DefrhogsExactwithExpectationValue0DONAppendix}, although some steps can sometimes be skipped: i) for the OPT expansion performed around a trivial saddle point (as discussed in section~\ref{sec:OPT}), the $O(N)$ symmetry can not be spontaneously broken, which implies that the angular integration can be carried out over the whole $N$-dimensional color space (instead of its ($N-1$)-dimensional subspace) at the stage of~\eqref{eq:ZKjHypersphericalCoordinatesStep2}; ii) There is no need of hyperspherical coordinates at all in the framework of the collective LE since the $N$-component original field is integrated out in this situation.


\newpage
\begin{tiny}
\begin{longtable}{|M{0.14\linewidth}|M{0.14\linewidth}M{0.14\linewidth}M{0.14\linewidth}M{0.14\linewidth}M{0.14\linewidth}|}
\caption{\label{tab:MultiplicityOPTdiagramsON} Diagrams involved in the expansion of $W^\text{OPT}$ (expressed by~\eqref{eq:WKjLoopExpansionStep3OPT}) up to order $\mathcal{O}\big(\delta^{3}\big)$. The diagrams are sorted according to the connected correlation functions that generate them in the OPT expansion (the corresponding expectation value $\big\langle\cdots\big\rangle_{0,\sigma}$ is defined by~\eqref{eq:OPTsigmaDepExpValueNumber1} and~\eqref{eq:OPTsigmaDepExpValueNumber2}), as indicated by the left-hand column. The multiplicity $\mathcal{M}_{\mathrm{OPT}}$ of a given diagram can be deduced from~\eqref{eq:MultiplicityDiagOPT} together with the corresponding vector $(S,D;N_{\mathrm{V}})$ given below (see text below~\eqref{eq:MultiplicityDiagLoopExpansion} and~\eqref{eq:MultiplicityDiagOPT} for the definitions of $S$, $D$ and $N_{\mathrm{V}}$).} \\ \hline
  & & & & & \\
 \scalebox{1.7}{\hspace{-0.028cm}Contribution} & & & \scalebox{1.7}{Diagrams} & & \\
  & & & & & \\ \hline
  & & & & & \\
 \scalebox{1.1}{$\sigma\left\langle\vec{\widetilde{\chi}}^{2}\right\rangle_{0,\sigma}^{\text{c}}$} & & & \begin{fmffile}{Diagrams/OPTtable_Diag1}
\begin{fmfgraph}(10,10)
\fmfleft{i}
\fmfright{o}
\fmfv{decor.shape=square,decor.size=2.5thick,foreground=(0,,0,,1)}{v1}
\fmftop{v3}
\fmfbottom{v4}
\fmf{phantom,tension=30}{i,v1}
\fmf{phantom,tension=30}{v2,o}
\fmf{plain,left=0.43,tension=0.5}{v1,v3}
\fmf{plain,left=0.43,tension=0.5}{v3,v2}
\fmf{plain,left=0.43,tension=0.5}{v2,v4}
\fmf{plain,left=0.43,tension=0.5}{v4,v1}
\end{fmfgraph}
\end{fmffile} & & \\
  & & & \vspace{-0.18cm}\scalebox{1.3}{\hspace{0.0365cm}$(1,0;1)$} & & \\
  & & & & & \\
\hline
\end{longtable}
\end{tiny}

\section{Mixed loop expansion}
\label{sec:DiagLEM}

Similarly to the path followed in section~\ref{sec:DiagLEO}, we work out a formula expressing the multiplicities of all diagrams resulting from the LE in the mixed representation. In this way, we obtain:
\begin{equation}
\mathcal{M}_{\mathrm{LE},\mathrm{mix}}=\frac{(2p)!4^{p} N_{\mathrm{F}}}{(2!)^{S+D} N_{\mathrm{V}}}\;,
\label{eq:MultiplicityDiagLoopExpansionMixed}
\end{equation}
where $S$ and $D$ are now respectively the number of self and double connections with the propagator lines~\eqref{eq:FeynRulesLoopExpansionMixedHSG} and~\eqref{eq:FeynRulesLoopExpansionMixedHSF1} representing respectively $\boldsymbol{G}_{\sigma_{\mathrm{cl}};\mathcal{J}\mathcal{K}}$ and $\vec{F}_{\varphi_{\mathrm{cl}};\mathcal{J}\mathcal{K}}$ (self and double connections made of $D_{\sigma_{\mathrm{cl}};\mathcal{J}\mathcal{K}}$ propagators, i.e. made of wiggly lines according to~\eqref{eq:FeynRulesLoopExpansionMixedHSD}, are not possible). Note that $N_{\mathrm{V}}$ still denotes the number of vertex permutations that leave the diagram unchanged, $N_{\mathrm{F}}$ is to be specified below and $p$ equals half the number of vertices~\eqref{eq:FeynRulesLoopExpansionMixedHSvertex} involved in the diagram under consideration. Result~\eqref{eq:MultiplicityDiagLoopExpansionMixed} can actually be determined through slight modifications of~\eqref{eq:MultiplicityDiagLoopExpansion}. The $\boldsymbol{G}_{\sigma_{\mathrm{cl}};\mathcal{J}\mathcal{K}}$ and $D_{\sigma_{\mathrm{cl}};\mathcal{J}\mathcal{K}}$ propagators (i.e.~\eqref{eq:FeynRulesLoopExpansionMixedHSG} and~\eqref{eq:FeynRulesLoopExpansionMixedHSD}) contribute to $\mathcal{M}_{\mathrm{LE},\mathrm{mix}}$ respectively in the same way as the $\boldsymbol{G}_{\varphi_{\mathrm{cl}};JK}$ propagator (i.e.~\eqref{eq:FeynRulesLoopExpansionPropagator}) and the zigzag vertex (i.e.~\eqref{eq:FeynRulesLoopExpansion4legVertex}) to $\mathcal{M}_{\mathrm{LE},\mathrm{orig}}$. Due to the HST performed in the mixed representation, we have swapped the two interaction terms (with associated vertices~\eqref{eq:FeynRulesLoopExpansion3legVertex} and~\eqref{eq:FeynRulesLoopExpansion4legVertex}) of the original $\varphi^{4}$-theory for a Yukawa interaction (corresponding to~\eqref{eq:FeynRulesLoopExpansionMixedHSvertex}), thus inducing that the factor $(2p)!!(2q)!!$ in~\eqref{eq:MultiplicityDiagLoopExpansion} is replaced by $(2p)!$ in~\eqref{eq:MultiplicityDiagLoopExpansionMixed}. The other and last difference between the diagrammatic of the original and mixed LEs is the presence in the mixed case of the $\vec{F}_{\varphi_{\mathrm{cl}};\mathcal{J}\mathcal{K}}$ propagator which has no counterpart in the original LE. As opposed to all other propagators introduced so far, we must account for the possibility to exchange the extremities of the $\vec{F}_{\varphi_{\mathrm{cl}};\mathcal{J}\mathcal{K}}$ propagator (even though $\vec{F}^{\mathrm{T}}_{\varphi_{\mathrm{cl}};\mathcal{J}\mathcal{K};a}(x,y) \equiv \vec{F}_{\varphi_{\mathrm{cl}};\mathcal{J}\mathcal{K};a}(y,x) = \vec{F}_{\varphi_{\mathrm{cl}};\mathcal{J}\mathcal{K};a}(x,y)$ to ensure that $\mathcal{G}_{\Phi_{\mathrm{cl}};\mathcal{J}\mathcal{K}}^{\mathrm{T}}=\mathcal{G}_{\Phi_{\mathrm{cl}};\mathcal{J}\mathcal{K}}$) since only one of its two extremities is associated to a color index, as can be seen from~\eqref{eq:FeynRulesLoopExpansionMixedHSF1}. More specifically, for every loop exclusively made of $\vec{F}_{\varphi_{\mathrm{cl}};\mathcal{J}\mathcal{K}}$ propagators, we can switch the extremities of all these propagators at once without affecting the nature of the studied diagram. Such a property contributes a factor 2 to the multiplicity of the latter. This explains the dependence of $\mathcal{M}_{\mathrm{LE},\mathrm{mix}}$ with respect to the number $N_{\mathrm{F}}$ of loops exclusively made of $\vec{F}_{\varphi_{\mathrm{cl}};\mathcal{J}\mathcal{K}}$ propagators. The factors $S$, $D$, $N_{\mathrm{V}}$ and $N_{\mathrm{F}}$ associated to 2PI diagrams expressing the mixed 2PI EA $\Gamma_{\mathrm{mix}}^{(\mathrm{2PI})}\big[\Phi,\mathcal{G}\big]$ introduced in section~\ref{sec:diagMixed2PIEA} are given by tab.~\ref{tab:MultiplicityMixedEAdiagramsON} and their multiplicities can also be inferred from~\eqref{eq:MultiplicityDiagLoopExpansionMixed} (2PI diagrams contributing to $W^\text{LE;mix}[\mathcal{J},\mathcal{K}]$ determined from the mixed LE in section~\ref{sec:MixedLoopExp} are also essentially given by tab.~\ref{tab:MultiplicityMixedEAdiagramsON}, except that all of their propagator lines are black and not red).


\begin{longtable}{|M{0.14\textwidth}|M{0.14\textwidth}M{0.14\textwidth}M{0.14\textwidth}M{0.14\textwidth}M{0.14\textwidth}|}
\caption{\label{tab:MultiplicityMixedEAdiagramsON} Diagrams contributing to the 2PI EA $\Gamma_{\mathrm{mix}}^{(\mathrm{2PI})}\big[\Phi,\mathcal{G}\big]$ up to order $\mathcal{O}\big(\hbar^{4}\big)$ (with $\vec{F}=\vec{0}$ at order $\mathcal{O}(\hbar^{4})$). The left-hand column indicates the order at which the diagrams contribute to the expansion of $\Gamma_{\mathrm{mix}}^{(\mathrm{2PI})}\big[\Phi,\mathcal{G}\big]$. The multiplicity $\mathcal{M}_{\mathrm{LE},\mathrm{mix}}$ of a given diagram can be deduced from~\eqref{eq:MultiplicityDiagLoopExpansionMixed} together with the corresponding vector $(S,D;N_{\mathrm{V}},N_{\mathrm{F}})$ given below (see text below~\eqref{eq:MultiplicityDiagLoopExpansionMixed} for the definitions of $S$, $D$, $N_{\mathrm{V}}$ and $N_{\mathrm{F}}$).} \\ \hline
 \scalebox{1.0}{Order} & & & \scalebox{1.0}{Diagrams} & & \\ \hline
  & & & & & \\
 \scalebox{1.0}{$\mathcal{O}\big(\hbar^{2}\big)$} & \multicolumn{5}{l|}{\hspace{3.325cm}${\displaystyle\begin{gathered}
\begin{fmffile}{Diagrams/MixedEAtable_Fock}
\begin{fmfgraph}(12.5,12.5)
\fmfleft{i}
\fmfright{o}
\fmfv{decor.shape=circle,decor.size=2.0thick,foreground=(0,,0,,1)}{v1}
\fmfv{decor.shape=circle,decor.size=2.0thick,foreground=(0,,0,,1)}{v2}
\fmf{phantom,tension=11}{i,v1}
\fmf{phantom,tension=11}{v2,o}
\fmf{plain,left,tension=0.4,foreground=(1,,0,,0)}{v1,v2,v1}
\fmf{wiggly,foreground=(1,,0,,0)}{v1,v2}
\end{fmfgraph}
\end{fmffile}
\end{gathered} \hspace{4.35cm} \begin{gathered}
\begin{fmffile}{Diagrams/MixedEAtable_Diag1}
\begin{fmfgraph}(12.5,12.5)
\fmfleft{i}
\fmfright{o}
\fmfv{decor.shape=circle,decor.size=2.0thick,foreground=(0,,0,,1)}{v1}
\fmfv{decor.shape=circle,decor.size=2.0thick,foreground=(0,,0,,1)}{v2}
\fmf{phantom,tension=11}{i,v1}
\fmf{phantom,tension=11}{v2,o}
\fmf{dashes,left,tension=0.4,foreground=(1,,0,,0)}{v1,v2,v1}
\fmf{plain,foreground=(1,,0,,0)}{v1,v2}
\end{fmfgraph}
\end{fmffile}
\end{gathered} }$} \\
  & \multicolumn{5}{l|}{\hspace{3.275cm}\scalebox{0.8}{$(0,1;2,1)$} \hspace{4.105cm}\scalebox{0.8}{$(0,1;2,2)$} } \\
  & & & & & \\ \hline
  & & & & & \\
 \scalebox{1.0}{$\mathcal{O}\big(\hbar^{3}\big)$} & \multicolumn{5}{l|}{\hspace{0.64cm}${\displaystyle\begin{gathered}
\begin{fmffile}{Diagrams/MixedEAtable_Diag2}
\begin{fmfgraph}(10,10)
\fmfleft{i0,i1}
\fmfright{o0,o1}
\fmftop{v1,vUp,v2}
\fmfbottom{v3,vDown,v4}
\fmfv{decor.shape=circle,decor.size=2.0thick,foreground=(0,,0,,1)}{v1}
\fmfv{decor.shape=circle,decor.size=2.0thick,foreground=(0,,0,,1)}{v2}
\fmfv{decor.shape=circle,decor.size=2.0thick,foreground=(0,,0,,1)}{v3}
\fmfv{decor.shape=circle,decor.size=2.0thick,foreground=(0,,0,,1)}{v4}
\fmf{phantom,tension=20}{i0,v1}
\fmf{phantom,tension=20}{i1,v3}
\fmf{phantom,tension=20}{o0,v2}
\fmf{phantom,tension=20}{o1,v4}
\fmf{plain,left=0.4,tension=0.5,foreground=(1,,0,,0)}{v3,v1}
\fmf{phantom,left=0.1,tension=0.5}{v1,vUp}
\fmf{phantom,left=0.1,tension=0.5}{vUp,v2}
\fmf{plain,left=0.4,tension=0.0,foreground=(1,,0,,0)}{v1,v2}
\fmf{plain,left=0.4,tension=0.5,foreground=(1,,0,,0)}{v2,v4}
\fmf{phantom,left=0.1,tension=0.5}{v4,vDown}
\fmf{phantom,left=0.1,tension=0.5}{vDown,v3}
\fmf{plain,left=0.4,tension=0.0,foreground=(1,,0,,0)}{v4,v3}
\fmf{wiggly,tension=0.5,foreground=(1,,0,,0)}{v1,v4}
\fmf{wiggly,tension=0.5,foreground=(1,,0,,0)}{v2,v3}
\end{fmfgraph}
\end{fmffile}
\end{gathered} \hspace{1.8cm} \begin{gathered}
\begin{fmffile}{Diagrams/MixedEAtable_Diag3}
\begin{fmfgraph}(10,10)
\fmfleft{i0,i1}
\fmfright{o0,o1}
\fmftop{v1,vUp,v2}
\fmfbottom{v3,vDown,v4}
\fmfv{decor.shape=circle,decor.size=2.0thick,foreground=(0,,0,,1)}{v1}
\fmfv{decor.shape=circle,decor.size=2.0thick,foreground=(0,,0,,1)}{v2}
\fmfv{decor.shape=circle,decor.size=2.0thick,foreground=(0,,0,,1)}{v3}
\fmfv{decor.shape=circle,decor.size=2.0thick,foreground=(0,,0,,1)}{v4}
\fmf{phantom,tension=20}{i0,v1}
\fmf{phantom,tension=20}{i1,v3}
\fmf{phantom,tension=20}{o0,v2}
\fmf{phantom,tension=20}{o1,v4}
\fmf{dashes,left=0.4,tension=0.5,foreground=(1,,0,,0)}{v3,v1}
\fmf{phantom,left=0.1,tension=0.5}{v1,vUp}
\fmf{phantom,left=0.1,tension=0.5}{vUp,v2}
\fmf{dashes,left=0.4,tension=0.0,foreground=(1,,0,,0)}{v1,v2}
\fmf{dashes,left=0.4,tension=0.5,foreground=(1,,0,,0)}{v2,v4}
\fmf{phantom,left=0.1,tension=0.5}{v4,vDown}
\fmf{phantom,left=0.1,tension=0.5}{vDown,v3}
\fmf{dashes,left=0.4,tension=0.0,foreground=(1,,0,,0)}{v4,v3}
\fmf{plain,tension=0.5,foreground=(1,,0,,0)}{v1,v4}
\fmf{plain,tension=0.5,foreground=(1,,0,,0)}{v2,v3}
\end{fmfgraph}
\end{fmffile}
\end{gathered} \hspace{1.8cm} \begin{gathered}
\begin{fmffile}{Diagrams/MixedEAtable_Diag4}
\begin{fmfgraph}(10,10)
\fmfleft{i0,i1}
\fmfright{o0,o1}
\fmftop{v1,vUp,v2}
\fmfbottom{v3,vDown,v4}
\fmfv{decor.shape=circle,decor.size=2.0thick,foreground=(0,,0,,1)}{v1}
\fmfv{decor.shape=circle,decor.size=2.0thick,foreground=(0,,0,,1)}{v2}
\fmfv{decor.shape=circle,decor.size=2.0thick,foreground=(0,,0,,1)}{v3}
\fmfv{decor.shape=circle,decor.size=2.0thick,foreground=(0,,0,,1)}{v4}
\fmf{phantom,tension=20}{i0,v1}
\fmf{phantom,tension=20}{i1,v3}
\fmf{phantom,tension=20}{o0,v2}
\fmf{phantom,tension=20}{o1,v4}
\fmf{plain,left=0.4,tension=0.5,foreground=(1,,0,,0)}{v3,v1}
\fmf{phantom,left=0.1,tension=0.5}{v1,vUp}
\fmf{phantom,left=0.1,tension=0.5}{vUp,v2}
\fmf{plain,left=0.4,tension=0.0,foreground=(1,,0,,0)}{v1,v2}
\fmf{dashes,left=0.4,tension=0.5,foreground=(1,,0,,0)}{v2,v4}
\fmf{phantom,left=0.1,tension=0.5}{v4,vDown}
\fmf{phantom,left=0.1,tension=0.5}{vDown,v3}
\fmf{dashes,left=0.4,tension=0.0,foreground=(1,,0,,0)}{v4,v3}
\fmf{wiggly,tension=0.5,foreground=(1,,0,,0)}{v1,v4}
\fmf{plain,tension=0.5,foreground=(1,,0,,0)}{v2,v3}
\end{fmfgraph}
\end{fmffile}
\end{gathered} \hspace{1.8cm} \begin{gathered}
\begin{fmffile}{Diagrams/MixedEAtable_Diag5}
\begin{fmfgraph}(10,10)
\fmfleft{i0,i1}
\fmfright{o0,o1}
\fmftop{v1,vUp,v2}
\fmfbottom{v3,vDown,v4}
\fmfv{decor.shape=circle,decor.size=2.0thick,foreground=(0,,0,,1)}{v1}
\fmfv{decor.shape=circle,decor.size=2.0thick,foreground=(0,,0,,1)}{v2}
\fmfv{decor.shape=circle,decor.size=2.0thick,foreground=(0,,0,,1)}{v3}
\fmfv{decor.shape=circle,decor.size=2.0thick,foreground=(0,,0,,1)}{v4}
\fmf{phantom,tension=20}{i0,v1}
\fmf{phantom,tension=20}{i1,v3}
\fmf{phantom,tension=20}{o0,v2}
\fmf{phantom,tension=20}{o1,v4}
\fmf{dashes,left=0.4,tension=0.5,foreground=(1,,0,,0)}{v3,v1}
\fmf{phantom,left=0.1,tension=0.5}{v1,vUp}
\fmf{phantom,left=0.1,tension=0.5}{vUp,v2}
\fmf{plain,left=0.4,tension=0.0,foreground=(1,,0,,0)}{v1,v2}
\fmf{dashes,left=0.4,tension=0.5,foreground=(1,,0,,0)}{v2,v4}
\fmf{phantom,left=0.1,tension=0.5}{v4,vDown}
\fmf{phantom,left=0.1,tension=0.5}{vDown,v3}
\fmf{plain,left=0.4,tension=0.0,foreground=(1,,0,,0)}{v4,v3}
\fmf{wiggly,tension=0.5,foreground=(1,,0,,0)}{v1,v4}
\fmf{plain,tension=0.5,foreground=(1,,0,,0)}{v2,v3}
\end{fmfgraph}
\end{fmffile}
\end{gathered} \hspace{1.8cm} \begin{gathered}
\begin{fmffile}{Diagrams/MixedEAtable_Diag6}
\begin{fmfgraph}(10,10)
\fmfleft{i0,i1}
\fmfright{o0,o1}
\fmftop{v1,vUp,v2}
\fmfbottom{v3,vDown,v4}
\fmfv{decor.shape=circle,decor.size=2.0thick,foreground=(0,,0,,1)}{v1}
\fmfv{decor.shape=circle,decor.size=2.0thick,foreground=(0,,0,,1)}{v2}
\fmfv{decor.shape=circle,decor.size=2.0thick,foreground=(0,,0,,1)}{v3}
\fmfv{decor.shape=circle,decor.size=2.0thick,foreground=(0,,0,,1)}{v4}
\fmf{phantom,tension=20}{i0,v1}
\fmf{phantom,tension=20}{i1,v3}
\fmf{phantom,tension=20}{o0,v2}
\fmf{phantom,tension=20}{o1,v4}
\fmf{plain,left=0.4,tension=0.5,foreground=(1,,0,,0)}{v3,v1}
\fmf{phantom,left=0.1,tension=0.5}{v1,vUp}
\fmf{phantom,left=0.1,tension=0.5}{vUp,v2}
\fmf{plain,left=0.4,tension=0.0,foreground=(1,,0,,0)}{v1,v2}
\fmf{dashes,left=0.4,tension=0.5,foreground=(1,,0,,0)}{v2,v4}
\fmf{phantom,left=0.1,tension=0.5}{v4,vDown}
\fmf{phantom,left=0.1,tension=0.5}{vDown,v3}
\fmf{dashes,left=0.4,tension=0.0,foreground=(1,,0,,0)}{v4,v3}
\fmf{dashes,tension=0.5,foreground=(1,,0,,0)}{v1,v4}
\fmf{dashes,tension=0.5,foreground=(1,,0,,0)}{v2,v3}
\end{fmfgraph}
\end{fmffile}
\end{gathered} \hspace{-1.5cm} }$} \\
  & \multicolumn{5}{l|}{\hspace{0.46cm}\scalebox{0.8}{$(0,0;8,1)$} \hspace{1.33cm}\scalebox{0.8}{$(0,0;8,2)$} \hspace{1.32cm}\scalebox{0.8}{$(0,0;2,1)$} \hspace{1.32cm}\scalebox{0.8}{$(0,0;1,1)$} \hspace{1.33cm}\scalebox{0.8}{$(0,0;2,2)$}} \\
  & & & & & \\ \hline
  & & & & & \\
 \scalebox{1.0}{$\mathcal{O}\big(\hbar^{4}\big)$} & \multicolumn{5}{l|}{\hspace{1.67cm}${\displaystyle\begin{gathered}
 \begin{fmffile}{Diagrams/MixedEAtable_Diag7}
\begin{fmfgraph}(15,15)
\fmfleft{i}
\fmfright{o}
\fmftop{vUpLeft,vUp,vUpRight}
\fmfbottom{vDownLeft,vDown,vDownRight}
\fmfv{decor.shape=circle,decor.size=2.0thick,foreground=(0,,0,,1)}{v1}
\fmfv{decor.shape=circle,decor.size=2.0thick,foreground=(0,,0,,1)}{v2}
\fmfv{decor.shape=circle,decor.size=2.0thick,foreground=(0,,0,,1)}{v3}
\fmfv{decor.shape=circle,decor.size=2.0thick,foreground=(0,,0,,1)}{v4}
\fmfv{decor.shape=circle,decor.size=2.0thick,foreground=(0,,0,,1)}{v5}
\fmfv{decor.shape=circle,decor.size=2.0thick,foreground=(0,,0,,1)}{v6}
\fmf{phantom,tension=1}{i,v1}
\fmf{phantom,tension=1}{v2,o}
\fmf{phantom,tension=14.0}{v3,vUpLeft}
\fmf{phantom,tension=2.0}{v3,vUpRight}
\fmf{phantom,tension=4.0}{v3,i}
\fmf{phantom,tension=2.0}{v4,vUpLeft}
\fmf{phantom,tension=14.0}{v4,vUpRight}
\fmf{phantom,tension=4.0}{v4,o}
\fmf{phantom,tension=14.0}{v5,vDownLeft}
\fmf{phantom,tension=2.0}{v5,vDownRight}
\fmf{phantom,tension=4.0}{v5,i}
\fmf{phantom,tension=2.0}{v6,vDownLeft}
\fmf{phantom,tension=14.0}{v6,vDownRight}
\fmf{phantom,tension=4.0}{v6,o}
\fmf{wiggly,tension=0,foreground=(1,,0,,0)}{v1,v2}
\fmf{wiggly,tension=0.6,foreground=(1,,0,,0)}{v3,v6}
\fmf{wiggly,tension=0.6,foreground=(1,,0,,0)}{v5,v4}
\fmf{plain,left=0.18,tension=0,foreground=(1,,0,,0)}{v1,v3}
\fmf{plain,left=0.42,tension=0,foreground=(1,,0,,0)}{v3,v4}
\fmf{plain,left=0.18,tension=0,foreground=(1,,0,,0)}{v4,v2}
\fmf{plain,left=0.18,tension=0,foreground=(1,,0,,0)}{v2,v6}
\fmf{plain,left=0.42,tension=0,foreground=(1,,0,,0)}{v6,v5}
\fmf{plain,left=0.18,tension=0,foreground=(1,,0,,0)}{v5,v1}
\end{fmfgraph}
\end{fmffile}
\end{gathered} \hspace{3.0cm} \begin{gathered}
\begin{fmffile}{Diagrams/MixedEAtable_Diag8}
\begin{fmfgraph}(12.5,12.5)
\fmfleft{i0,i1}
\fmfright{o0,o1}
\fmftop{v1,vUp,v2}
\fmfbottom{v3,vDown,v4}
\fmfv{decor.shape=circle,decor.size=2.0thick,foreground=(0,,0,,1)}{v1}
\fmfv{decor.shape=circle,decor.size=2.0thick,foreground=(0,,0,,1)}{v2}
\fmfv{decor.shape=circle,decor.size=2.0thick,foreground=(0,,0,,1)}{v3}
\fmfv{decor.shape=circle,decor.size=2.0thick,foreground=(0,,0,,1)}{v4}
\fmfv{decor.shape=circle,decor.size=2.0thick,foreground=(0,,0,,1)}{v5}
\fmfv{decor.shape=circle,decor.size=2.0thick,foreground=(0,,0,,1)}{v6}
\fmf{phantom,tension=20}{i0,v1}
\fmf{phantom,tension=20}{i1,v3}
\fmf{phantom,tension=20}{o0,v2}
\fmf{phantom,tension=20}{o1,v4}
\fmf{phantom,tension=0.005}{v5,v6}
\fmf{wiggly,left=0.4,tension=0,foreground=(1,,0,,0)}{v3,v1}
\fmf{phantom,left=0.1,tension=0}{v1,vUp}
\fmf{phantom,left=0.1,tension=0}{vUp,v2}
\fmf{plain,left=0.25,tension=0,foreground=(1,,0,,0)}{v1,v2}
\fmf{wiggly,left=0.4,tension=0,foreground=(1,,0,,0)}{v2,v4}
\fmf{phantom,left=0.1,tension=0}{v4,vDown}
\fmf{phantom,left=0.1,tension=0}{vDown,v3}
\fmf{plain,right=0.25,tension=0,foreground=(1,,0,,0)}{v3,v4}
\fmf{plain,left=0.2,tension=0.01,foreground=(1,,0,,0)}{v1,v5}
\fmf{plain,left=0.2,tension=0.01,foreground=(1,,0,,0)}{v5,v3}
\fmf{plain,right=0.2,tension=0.01,foreground=(1,,0,,0)}{v2,v6}
\fmf{plain,right=0.2,tension=0.01,foreground=(1,,0,,0)}{v6,v4}
\fmf{wiggly,tension=0,foreground=(1,,0,,0)}{v5,v6}
\end{fmfgraph}
\end{fmffile}
\end{gathered} \hspace{3.0cm} \begin{gathered}
\begin{fmffile}{Diagrams/MixedEAtable_Diag9}
\begin{fmfgraph}(12.5,12.5)
\fmfleft{i0,i1}
\fmfright{o0,o1}
\fmftop{v1,vUp,v2}
\fmfbottom{v3,vDown,v4}
\fmfv{decor.shape=circle,decor.size=2.0thick,foreground=(0,,0,,1)}{v1}
\fmfv{decor.shape=circle,decor.size=2.0thick,foreground=(0,,0,,1)}{v2}
\fmfv{decor.shape=circle,decor.size=2.0thick,foreground=(0,,0,,1)}{v3}
\fmfv{decor.shape=circle,decor.size=2.0thick,foreground=(0,,0,,1)}{v4}
\fmfv{decor.shape=circle,decor.size=2.0thick,foreground=(0,,0,,1)}{v5}
\fmfv{decor.shape=circle,decor.size=2.0thick,foreground=(0,,0,,1)}{v6}
\fmf{phantom,tension=20}{i0,v1}
\fmf{phantom,tension=20}{i1,v3}
\fmf{phantom,tension=20}{o0,v2}
\fmf{phantom,tension=20}{o1,v4}
\fmf{phantom,tension=0.005}{v5,v6}
\fmf{plain,left=0.4,tension=0,foreground=(1,,0,,0)}{v3,v1}
\fmf{phantom,left=0.1,tension=0}{v1,vUp}
\fmf{phantom,left=0.1,tension=0}{vUp,v2}
\fmf{wiggly,left=0.25,tension=0,foreground=(1,,0,,0)}{v1,v2}
\fmf{plain,left=0.4,tension=0,foreground=(1,,0,,0)}{v2,v4}
\fmf{phantom,left=0.1,tension=0}{v4,vDown}
\fmf{phantom,left=0.1,tension=0}{vDown,v3}
\fmf{wiggly,right=0.25,tension=0,foreground=(1,,0,,0)}{v3,v4}
\fmf{plain,left=0.2,tension=0.01,foreground=(1,,0,,0)}{v1,v5}
\fmf{plain,left=0.2,tension=0.01,foreground=(1,,0,,0)}{v5,v3}
\fmf{plain,right=0.2,tension=0.01,foreground=(1,,0,,0)}{v2,v6}
\fmf{plain,right=0.2,tension=0.01,foreground=(1,,0,,0)}{v6,v4}
\fmf{wiggly,tension=0,foreground=(1,,0,,0)}{v5,v6}
\end{fmfgraph}
\end{fmffile}
\end{gathered} }$} \\
 \scalebox{0.9}{(with $\vec{F}=\vec{0}$)} & \multicolumn{5}{l|}{\hspace{1.652cm}\scalebox{0.8}{$(0,0;12,1)$} \hspace{2.823cm}\scalebox{0.8}{$(0,0;4,1)$} \hspace{2.6825cm}\scalebox{0.8}{$(0,0;12,1)$} } \\
  & & & & & \\ \hline
\end{longtable}

%% file: 7Appendix/IM.tex
\section{\label{sec:ForewordNotationsIM}Foreword and notations}

As explained in chapter~\ref{chap:DiagTechniques}, the IM is a technique designed to derive diagrammatic expressions of EAs. We will present in this appendix the most general implementation of the IM, which is in particular more general than the implementation discussed in section~\ref{sec:2PPIEA} (and based notably on~\eqref{eq:ProcedureDeterminationSourcesIM}) in the sense that it is more suited to treat EAs based on Schwinger functionals depending on several sources (such as 2P(P)I EAs with non-vanishing 1-point correlation function(s) and 4P(P)I EAs) and/or for theories involving several quantum fields.

\vspace{0.5cm}

Some of the underlying equations being significantly cluttered, we will use some notations which are slightly more compact than those exploited in chapter~\ref{chap:DiagTechniques}. In particular, color (spacetime) indices will be denoted as $a_{1}$, $a_{2}$, ... instead of $a$, $b$, ... ($x_{1}$, $x_{2}$, ... instead of $x$, $y$, ...). Furthermore, color and spacetime indices will be collected into $\alpha$-indices defined as $\alpha\equiv (a,x)$. Summation over color indices and integration over spacetime positions will thus be encompassed through the following shorthand notation:
\begin{equation}
\int_{\alpha}\equiv\sum_{a=1}^{N} \int_{x} \equiv\sum_{a=1}^{N} \int^{\hbar/T}_{0} d\tau \int d^{D-1}\boldsymbol{r} \;,
\end{equation}
at temperature $T$. Hence, the notations used in this appendix~\ref{ann:InversionMethod} coincide with those of section~\ref{sec:1PIFRGstateofplay} of the FRG chapter. Furthermore, the superindices introduced in the framework of the mixed representation are denoted by $b_{1}$, $b_{2}$, ..., and gathered with spacetime indices into $\beta$-indices as $\beta\equiv (b,x)$. In this case, the shorthand notation for sums and integrals reads:
\begin{equation}
\int_{\beta}\equiv\sum_{b=1}^{N+1} \int_{x} \equiv\sum_{b=1}^{N+1} \int^{\hbar/T}_{0} d\tau \int d^{D-1}\boldsymbol{r} \;.
\end{equation}

\vspace{0.5cm}

We will thus carry out the IM in sections~\ref{sec:1PIEAannIM},~\ref{sec:2PIEAannIM} and~\ref{sec:4PPIEAannIM} for 1PI, 2PI and 4PPI EAs, respectively. The purpose of section~\ref{sec:1PIEAannIM}, which tackles both the original and collective 1PI EAs, is mostly pedagogical since it shows how to recover a well-known diagrammatic rule via the IM, i.e. that 1PI EAs can be expressed in terms of 1PI graphs only. In this way, we prepare the ground for the subsequent sections~\ref{sec:2PIEAannIM} and~\ref{sec:4PPIEAannIM}, and more specifically for the treatments of the $\lambda$-expanded mixed 2PI EA and of the original 4PPI EA, for which there is no such a simple diagrammatic rule. Note also that one of the purposes of section~\ref{sec:2PIEAannIM} is to illustrate the cumbersomeness of the IM for the full 2PI EAs (i.e. for 2PI EAs with non-vanishing 1-point correlation function(s)). Finally, let us point out that the IM for a $m$P(P)I EA relies on the power series:
\begin{equation}
\Gamma^{(m\mathrm{P(P)I})} = \sum_{n=0}^{\infty} \Gamma_{n}^{(m\mathrm{P(P)I})} g^{n} \;,
\label{eq:GammanPPIcoeffAppendix}
\end{equation}
where $g$ is the chosen expansion parameter (i.e. $\hbar$ or $\lambda$ in the studied situations). The IM procedure requires to express the $\Gamma_{n}^{(m\mathrm{P(P)I})}$ coefficients in terms of the corresponding Schwinger functional. The derivations of such expressions being quite technical, we will disentangle them from the core of the IM procedure and postpone them to the end of this appendix in section~\ref{sec:DerivEAcoeff}.

\section{\label{sec:1PIEAannIM}1PI effective action}
\subsection{\label{sec:original1PIEAannIM}Original effective action}
\paragraph{$\hbar$-expansion:}

The starting point of the IM consists in expanding the EA under consideration, its arguments, the corresponding Schwinger functional and the associated sources with respect to the chosen expansion parameter. For the 1PI EA in the framework of the $\hbar$-expansion, this translates into:
\begin{subequations}
\begin{empheq}[left=\empheqlbrace]{align}
& \hspace{0.1cm} \Gamma^{(\mathrm{1PI})}\Big[\vec{\phi};\hbar\Big]=\sum_{n=0}^{\infty} \Gamma^{(\mathrm{1PI})}_{n}\Big[\vec{\phi}\Big]\hbar^{n}\;, \label{eq:pure1PIEAGammaExpansion0DONAppendix}\\
\nonumber \\
& \hspace{0.1cm} W\Big[\vec{J};\hbar\Big]=\sum_{n=0}^{\infty} W_{n}\Big[\vec{J}\Big]\hbar^{n}\;, \label{eq:pure1PIEAWExpansion0DONAppendix} \\
\nonumber \\
& \hspace{0.1cm} \vec{J}\Big[\vec{\phi};\hbar\Big]=\sum_{n=0}^{\infty} \vec{J}_{n}\Big[\vec{\phi}\Big]\hbar^{n}\;, \label{eq:pure1PIEAJExpansion0DONAppendix}\\
\nonumber \\
& \hspace{0.1cm} \vec{\phi}=\sum_{n=0}^{\infty} \vec{\phi}_{n}\Big[\vec{J}\Big]\hbar^{n}\;, \label{eq:pure1PIEAphiExpansion0DONAppendix}
\end{empheq}
\end{subequations}
and we recall the definition of $\Gamma^{(\mathrm{1PI})}$:
\begin{equation}
\begin{split}
\Gamma^{(\mathrm{1PI})}\Big[\vec{\phi}\Big] \equiv & -W\Big[\vec{J}\Big]+\int_{\alpha}J_{\alpha}\Big[\vec{\phi}\Big] \frac{\delta W\big[\vec{J}\big]}{\delta J_{\alpha}} \\
= & -W\Big[\vec{J}\Big]+\int_{\alpha}J_{\alpha}\Big[\vec{\phi}\Big]\phi_{\alpha}\;,
\end{split}
\label{eq:pure1PIEAdefinition0DONAppendix}
\end{equation}
with
\begin{equation}
\phi_{\alpha} = \frac{\delta W\big[\vec{J}\big]}{\delta J_{\alpha}}\;.
\label{eq:pure1PIEAdefinitionbis0DONAppendix}
\end{equation}
Note also the following subtlety about the 1-point correlation function in~\eqref{eq:pure1PIEAphiExpansion0DONAppendix}: $\vec{\phi}$ is independent of $\vec{J}$ (as explained from~\eqref{eq:pure1PIEAphi0DON} below~\eqref{eq:pure1PIEAGJ0minus1expressArbitaryDimON}) and of $\hbar$ (as proven by~\eqref{eq:pure1PIEAIndependencephihbarstep20DON}) but this does not prevent the $\vec{\phi}_{n}$ coefficients from depending on the source $\vec{J}$. For the sake of brevity, we will not indicate in what follows $\hbar$ as argument of the EA, the Schwinger functional or the sources, except for a few cases in which it clarifies the discussion. Determining a diagrammatic expression for the EA $\Gamma^{(\mathrm{1PI})}$ up to order $\mathcal{O}\big(\hbar^{m}\big)$ amounts to specifying all the $\Gamma^{(\mathrm{1PI})}_{n}$ coefficients up to $n=m$: this is the purpose of the IM. It will become clear below why we also need to determine the $W_{n}$, $\vec{J}_{n}$ and $\vec{\phi}_{n}$ coefficients to achieve this. We can already point out at this stage that combining~\eqref{eq:pure1PIEAWExpansion0DONAppendix} and~\eqref{eq:pure1PIEAphiExpansion0DONAppendix} with~\eqref{eq:pure1PIEAdefinitionbis0DONAppendix} leads to:
\begin{equation}
\phi_{n,\alpha}\Big[\vec{J}\Big]=\frac{\delta W_{n}\big[\vec{J}\big]}{\delta J_{\alpha}}\;.
\label{eq:pure1PIEAphinCoeff0DON}
\end{equation}
Diagrammatic expressions of the $W_{n}$ coefficients can be directly deduced from the LE of $W\big[\vec{J},\boldsymbol{K}\big]$ carried out in section~\ref{sec:OriginalLE}, and more specifically from result~\eqref{eq:WKjLoopExpansionStep3} expressing $W^\text{LE;orig}\big[\vec{J},\boldsymbol{K}\big]$, after setting $\boldsymbol{K}=\boldsymbol{0}$. For $n=0,1~\mathrm{and}~2$, we have:
\begin{equation}
W_{0}\Big[\vec{J}\Big] = -S\big[\vec{\varphi}_{\mathrm{cl}}\big] +\int_{\alpha}J_{\alpha}\Big[\vec{\phi}\Big]\varphi_{\mathrm{cl},\alpha}\Big[\vec{J}\Big]\;,
\label{eq:pure1PIEAIMW00DON}
\end{equation}
\begin{equation}
W_{1}\Big[\vec{J}\Big] = \frac{1}{2} \mathrm{STr}\left[\ln\Big(\boldsymbol{G}_{\varphi_\text{cl};J}\Big[\vec{J}\Big]\Big)\right]\;,
\label{eq:pure1PIEAIMW10DON}
\end{equation}
\begin{equation}
\begin{split}
W_{2}\Big[\vec{J}\Big] = & -\frac{1}{24} \hspace{0.08cm} \begin{gathered}
\begin{fmffile}{Diagrams/LoopExpansion1_Hartree}
\begin{fmfgraph}(30,20)
\fmfleft{i}
\fmfright{o}
\fmf{phantom,tension=10}{i,i1}
\fmf{phantom,tension=10}{o,o1}
\fmf{plain,left,tension=0.5}{i1,v1,i1}
\fmf{plain,right,tension=0.5}{o1,v2,o1}
\fmf{zigzag,foreground=(0,,0,,1)}{v1,v2}
\end{fmfgraph}
\end{fmffile}
\end{gathered}
-\frac{1}{12}\begin{gathered}
\begin{fmffile}{Diagrams/LoopExpansion1_Fock}
\begin{fmfgraph}(15,15)
\fmfleft{i}
\fmfright{o}
\fmf{phantom,tension=11}{i,v1}
\fmf{phantom,tension=11}{v2,o}
\fmf{plain,left,tension=0.4}{v1,v2,v1}
\fmf{zigzag,foreground=(0,,0,,1)}{v1,v2}
\end{fmfgraph}
\end{fmffile}
\end{gathered}
+ \frac{1}{18} \ \ \begin{gathered}
\begin{fmffile}{Diagrams/LoopExpansion1_Diag1}
\begin{fmfgraph}(34,20)
\fmfleft{i}
\fmfright{o}
\fmfv{decor.shape=cross,decor.size=3.5thick,foreground=(0,,0,,1)}{i}
\fmfv{decor.shape=cross,decor.size=3.5thick,foreground=(0,,0,,1)}{o}
\fmf{dashes,tension=2.0,foreground=(0,,0,,1)}{i,v3}
\fmf{dashes,tension=2.0,foreground=(0,,0,,1)}{o,v4}
\fmf{plain,right,tension=0.7}{v2,v4}
\fmf{dots,left,tension=0.7,foreground=(0,,0,,1)}{v2,v4}
\fmf{plain,left,tension=0.7}{v1,v3}
\fmf{dots,right,tension=0.7,foreground=(0,,0,,1)}{v1,v3}
\fmf{plain,tension=1.5}{v1,v2}
\end{fmfgraph}
\end{fmffile}
\end{gathered}
\ + \frac{1}{18} \begin{gathered}
\begin{fmffile}{Diagrams/LoopExpansion1_Diag2}
\begin{fmfgraph}(27,15)
\fmfleft{i}
\fmfright{o}
\fmftop{vUp}
\fmfbottom{vDown}
\fmfv{decor.shape=cross,decor.size=3.5thick,foreground=(0,,0,,1)}{v1}
\fmfv{decor.shape=cross,decor.size=3.5thick,foreground=(0,,0,,1)}{v2}
\fmf{phantom,tension=10}{i,i1}
\fmf{phantom,tension=10}{o,o1}
\fmf{phantom,tension=2.2}{vUp,v5}
\fmf{phantom,tension=2.2}{vDown,v6}
\fmf{phantom,tension=0.5}{v3,v4}
\fmf{phantom,tension=10.0}{i1,v1}
\fmf{phantom,tension=10.0}{o1,v2}
\fmf{dashes,tension=2.0,foreground=(0,,0,,1)}{v1,v3}
\fmf{dots,left=0.4,tension=0.5,foreground=(0,,0,,1)}{v3,v5}
\fmf{plain,left=0.4,tension=0.5}{v5,v4}
\fmf{plain,right=0.4,tension=0.5}{v3,v6}
\fmf{dots,right=0.4,tension=0.5,foreground=(0,,0,,1)}{v6,v4}
\fmf{dashes,tension=2.0,foreground=(0,,0,,1)}{v4,v2}
\fmf{plain,tension=0}{v5,v6}
\end{fmfgraph}
\end{fmffile}
\end{gathered} \\
& + \frac{1}{36} \hspace{-0.15cm} \begin{gathered}
\begin{fmffile}{Diagrams/LoopExpansion1_Diag3}
\begin{fmfgraph}(25,20)
\fmfleft{i}
\fmfright{o}
\fmftop{vUp}
\fmfbottom{vDown}
\fmfv{decor.shape=cross,decor.angle=45,decor.size=3.5thick,foreground=(0,,0,,1)}{vUpbis}
\fmfv{decor.shape=cross,decor.angle=45,decor.size=3.5thick,foreground=(0,,0,,1)}{vDownbis}
\fmf{phantom,tension=0.8}{vUp,vUpbis}
\fmf{phantom,tension=0.8}{vDown,vDownbis}
\fmf{dashes,tension=0.5,foreground=(0,,0,,1)}{v3,vUpbis}
\fmf{phantom,tension=0.5}{v4,vUpbis}
\fmf{phantom,tension=0.5}{v3,vDownbis}
\fmf{dashes,tension=0.5,foreground=(0,,0,,1)}{v4,vDownbis}
\fmf{phantom,tension=11}{i,v1}
\fmf{phantom,tension=11}{v2,o}
\fmf{plain,left,tension=0.5}{v1,v2,v1}
\fmf{dots,tension=1.7,foreground=(0,,0,,1)}{v1,v3}
\fmf{plain}{v3,v4}
\fmf{dots,tension=1.7,foreground=(0,,0,,1)}{v4,v2}
\end{fmfgraph}
\end{fmffile}
\end{gathered}
+\frac{1}{18} \ \begin{gathered}
\begin{fmffile}{Diagrams/LoopExpansion1_Diag4}
\begin{fmfgraph}(35,18)
\fmfleft{i}
\fmfright{o}
\fmftop{vUp}
\fmfbottom{vDown}
\fmfv{decor.shape=cross,decor.size=3.5thick,foreground=(0,,0,,1)}{v3bis}
\fmfv{decor.shape=cross,decor.size=3.5thick,foreground=(0,,0,,1)}{o}
\fmf{phantom,tension=10}{i,i1}
\fmf{dashes,tension=1.2,foreground=(0,,0,,1)}{o,v4}
\fmf{phantom,tension=0.5}{v3bis,i}
\fmf{phantom,tension=2.7}{v3bis,vUp}
\fmf{dashes,tension=0.9,foreground=(0,,0,,1)}{v3,v3bis}
\fmf{phantom,tension=0.5}{v4bis,i}
\fmf{phantom,tension=2.7}{v4bis,vDown}
\fmf{phantom,tension=0.9}{v3,v4bis}
\fmf{plain,left,tension=0.5}{i1,v1,i1}
\fmf{plain,right,tension=0.5}{v2,v4}
\fmf{dots,left,tension=0.5,foreground=(0,,0,,1)}{v2,v4}
\fmf{dots,foreground=(0,,0,,1)}{v1,v3}
\fmf{plain}{v3,v2}
\end{fmfgraph}
\end{fmffile}
\end{gathered}
\hspace{0.2cm} + \frac{1}{72} \ \begin{gathered}
\begin{fmffile}{Diagrams/LoopExpansion1_Diag5}
\begin{fmfgraph}(40,18)
\fmfleft{i}
\fmfright{o}
\fmftop{vUp}
\fmfbottom{vDown}
\fmf{phantom,tension=1.0}{vUp,vUpbis}
\fmf{phantom,tension=1.0}{vDown,vDownbis}
\fmf{dashes,tension=0.5,foreground=(0,,0,,1)}{v3,vUpbis}
\fmf{phantom,tension=0.5}{v4,vUpbis}
\fmf{phantom,tension=0.5}{v3,vDownbis}
\fmf{dashes,tension=0.5,foreground=(0,,0,,1)}{v4,vDownbis}
\fmfv{decor.shape=cross,decor.angle=45,decor.size=3.5thick,foreground=(0,,0,,1)}{vUpbis}
\fmfv{decor.shape=cross,decor.angle=45,decor.size=3.5thick,foreground=(0,,0,,1)}{vDownbis}
\fmf{phantom,tension=10}{i,i1}
\fmf{phantom,tension=10}{o,o1}
\fmf{plain,left,tension=0.5}{i1,v1,i1}
\fmf{plain,right,tension=0.5}{o1,v2,o1}
\fmf{dots,tension=1.2,foreground=(0,,0,,1)}{v1,v3}
\fmf{plain,tension=0.6}{v3,v4}
\fmf{dots,tension=1.2,foreground=(0,,0,,1)}{v4,v2}
\end{fmfgraph}
\end{fmffile}
\end{gathered}\;.
\end{split}
\label{eq:pure1PIEAIMW20DON}
\end{equation}
Such relations involve the classical 1-point correlation function $\vec{\varphi}_{\mathrm{cl}}\big[\vec{J}\big]$ and the propagator $\boldsymbol{G}_{\varphi_\text{cl};J}\big[\vec{J}\big] = \boldsymbol{G}_{\varphi_\text{cl};JK}\big[\vec{J},\boldsymbol{K}=\boldsymbol{0}\big]$ which are both functions of an arbitrary external source $\vec{J}$. They satisfy:
\begin{subequations}\label{eq:pure1PIEAphiJsubeq}
\begin{empheq}[left=\empheqlbrace]{align}
& \hspace{0.1cm} \varphi_{\mathrm{cl},\alpha}\Big[\vec{J}\Big] = \phi_{0,\alpha}\Big[\vec{J}\Big] = \frac{\delta W_{0}\big[\vec{J}\big]}{\delta J_{\alpha}}\;, \label{eq:pure1PIEAphicl0DON}\\
\nonumber \\
& \hspace{0.1cm} \boldsymbol{G}_{\varphi_\text{cl};J,\alpha_{1}\alpha_{2}}\Big[\vec{J}\Big] \equiv \left(\left.\frac{\delta^{2} S\big[\vec{\widetilde{\varphi}}\big]}{\delta\vec{\widetilde{\varphi}}\delta\vec{\widetilde{\varphi}}}\right|_{\vec{\widetilde{\varphi}}=\vec{\varphi}_{\mathrm{cl}}}\right)_{\alpha_{1}\alpha_{2}}^{-1} = \frac{\delta^{2} W_{0}\big[\vec{J}\big]}{\delta J_{\alpha_{1}} \delta J_{\alpha_{2}}} \;, \label{eq:pure1PIEAGJ0DON}
\end{empheq}
\end{subequations}
where
\begin{equation}
\boldsymbol{G}^{-1}_{\varphi_\text{cl};J,\alpha_{1}\alpha_{2}}\Big[\vec{J}\Big] = \left(-\nabla_{x_{1}}^{2} + m^2 + \frac{\lambda}{6} \vec{\varphi}_{\mathrm{cl},x_{1}}^{2}\Big[\vec{J}\Big]\right) \delta_{\alpha_{1}\alpha_{2}} + \frac{\lambda}{3}\varphi_{\mathrm{cl},\alpha_{1}}\Big[\vec{J}\Big] \varphi_{\mathrm{cl},\alpha_{2}}\Big[\vec{J}\Big] \delta_{x_{1}x_{2}} \;,
\label{eq:pure1PIEAGJminus1expressArbitaryDimON}
\end{equation}
and~\eqref{eq:pure1PIEAphicl0DON} is equivalent to~\eqref{eq:pure1PIEAphinCoeff0DON} at $n=0$. Note however that the argument of $\Gamma^{(\mathrm{1PI})}$ is the quantum 1-point correlation function $\vec{\phi}=\vec{\phi}_{0}\big[\vec{J}=\vec{J}_{0}\big]$ and not the classical one $\vec{\varphi}_{\mathrm{cl}}\big[\vec{J}\big]$. Setting $\vec{J}=\vec{J}_{0}$ into~\eqref{eq:pure1PIEAphicl0DON} and~\eqref{eq:pure1PIEAGJ0DON}, we obtain:
\begin{subequations}\label{eq:pure1PIEAphiJ0subeq}
\begin{empheq}[left=\empheqlbrace]{align}
& \hspace{0.1cm} \phi_{\alpha} = \varphi_{\mathrm{cl},\alpha}\Big[\vec{J}=\vec{J}_{0}\Big] = \phi_{0,\alpha}\Big[\vec{J}=\vec{J}_{0}\Big] = \left.\frac{\delta W_{0}\big[\vec{J}\big]}{\delta J_{\alpha}}\right|_{\vec{J}=\vec{J}_{0}}\;, \label{eq:pure1PIEAphi0DON}\\
\nonumber \\
& \hspace{0.1cm} \boldsymbol{G}_{\phi,\alpha_{1}\alpha_{2}}\Big[\vec{\phi}\Big] \equiv \boldsymbol{G}_{\varphi_{\mathrm{cl}};J,\alpha_{1}\alpha_{2}}\Big[\vec{J}=\vec{J}_{0}\Big] = \left(\left.\frac{\delta^{2} S\big[\vec{\widetilde{\varphi}}\big]}{\delta\vec{\widetilde{\varphi}}\delta\vec{\widetilde{\varphi}}}\right|_{\vec{\widetilde{\varphi}}=\vec{\phi}}\right)_{\alpha_{1}\alpha_{2}}^{-1} = \left.\frac{\delta^{2} W_{0}\big[\vec{J}\big]}{\delta J_{\alpha_{1}} \delta J_{\alpha_{2}}}\right|_{\vec{J}=\vec{J}_{0}} \;, \label{eq:pure1PIEAGJ00DON}
\end{empheq}
\end{subequations}
with
\begin{equation}
\boldsymbol{G}^{-1}_{\phi,\alpha_{1}\alpha_{2}}\Big[\vec{\phi}\Big] = \left(-\nabla_{x_{1}}^{2} + m^2 + \frac{\lambda}{6} \vec{\phi}^{2}_{x_{1}}\right) \delta_{\alpha_{1}\alpha_{2}} + \frac{\lambda}{3}\phi_{\alpha_{1}} \phi_{\alpha_{2}} \delta_{x_{1}x_{2}} \;.
\label{eq:pure1PIEAGJ0minus1expressArbitaryDimON}
\end{equation}
Since $\vec{\phi}$ equals $\vec{\varphi}_{\mathrm{cl}}$ evaluated at $\vec{J}=\vec{J}_{0}$ according to~\eqref{eq:pure1PIEAphi0DON}, it is clear that $\vec{\phi}$ does not depend on the external source $\vec{J}$. Relations~\eqref{eq:pure1PIEAphi0DON} and~\eqref{eq:pure1PIEAGJ00DON} emphasize the key role of the zeroth-order coefficient of the source(s) (i.e. $\vec{J}_{0}$ here) in the IM. We then make two comments about~\eqref{eq:pure1PIEAphiJsubeq} and~\eqref{eq:pure1PIEAphiJ0subeq}:
\begin{itemize}
\item The relation $\vec{\phi}=\vec{\phi}_{0}\big[\vec{J}=\vec{J}_{0}\big]$ imposes that $\vec{\phi}$ is a quantity of order $\mathcal{O}\big(\hbar^{0}\big)$, i.e. that it is independent of $\hbar$. This property directly results from the Legendre transform defining $\Gamma^{(\mathrm{1PI})}$. Considering $\hbar$ and $\vec{\phi}$ as independent variables, taking the differential of both sides of~\eqref{eq:pure1PIEAdefinition0DONAppendix} yields:
\begin{equation}
\begin{split}
\delta\Gamma^{(\mathrm{1PI})}\Big[\vec{\phi},\hbar\Big] = & -\cancel{\int_{\alpha}\underbrace{\frac{\delta W\big[\vec{J},\hbar\big]}{\delta J_{\alpha}}}_{\phi_{\alpha}} \delta J_{\alpha}\Big[\vec{\phi}\Big]} -\frac{\partial W\big[\vec{J},\hbar\big]}{\partial\hbar} d\hbar + \cancel{\int_{\alpha} \delta J_{\alpha}\Big[\vec{\phi}\Big] \phi_{\alpha}} + \int_{\alpha}J_{\alpha}\Big[\vec{\phi}\Big] \delta\phi_{\alpha} \\
= & -\frac{\partial W\big[\vec{J},\hbar\big]}{\partial\hbar} d\hbar + \int_{\alpha}J_{\alpha}\Big[\vec{\phi}\Big] \delta\phi_{\alpha} \;.
\end{split}
\label{eq:pure1PIEAIndependencephihbarstep20DON}
\end{equation}
The derivation of~\eqref{eq:pure1PIEAIndependencephihbarstep20DON} indeed shows that $\Gamma^{(\mathrm{1PI})}$ is independent of the external source $\vec{J}$ (as should be expected from the Legendre transform) if and only if $\vec{\phi}$ and $\hbar$ are independent variables. Hence, the independence of $\vec{\phi}$ and $\hbar$ is an essential feature for the IM. Note also that the reasoning based on~\eqref{eq:pure1PIEAIndependencephihbarstep20DON} can be adapted to any EA in order to show the independence of its argument(s) with respect to $\hbar$.

\item We will distinguish the Feynman rules of the propagators and vertices evaluated at arbitrary external source $\vec{J}$ or at $\vec{J}=\vec{J}_{0}$. More specifically, for the original 1PI EA, we choose the following Feynman rules at arbitrary external source $\vec{J}$:
\begin{subequations}
\begin{align}
\begin{gathered}
\begin{fmffile}{Diagrams/LoopExpansion1_FeynRuleGbis_Appendix}
\begin{fmfgraph*}(20,20)
\fmfleft{i0,i1,i2,i3}
\fmfright{o0,o1,o2,o3}
\fmflabel{$\alpha_{1}$}{v1}
\fmflabel{$\alpha_{2}$}{v2}
\fmf{phantom}{i1,v1}
\fmf{phantom}{i2,v1}
\fmf{plain,tension=0.6}{v1,v2}
\fmf{phantom}{v2,o1}
\fmf{phantom}{v2,o2}
\end{fmfgraph*}
\end{fmffile}
\end{gathered} \quad &\rightarrow \boldsymbol{G}_{\varphi_\text{cl};J,\alpha_{1}\alpha_{2}}\Big[\vec{J}\Big]\;,
\label{eq:FeynRulesLoopExpansionPropagatorAppendix} \\
\begin{gathered}
\begin{fmffile}{Diagrams/LoopExpansion1_FeynRuleV3bis_Appendix}
\begin{fmfgraph*}(20,20)
\fmfleft{i0,i1,i2,i3}
\fmfright{o0,o1,o2,o3}
\fmfv{decor.shape=cross,decor.angle=45,decor.size=3.5thick,foreground=(0,,0,,1)}{o2}
\fmf{phantom,tension=2.0}{i1,i1bis}
\fmf{plain,tension=2.0}{i1bis,v1}
\fmf{phantom,tension=2.0}{i2,i2bis}
\fmf{plain,tension=2.0}{i2bis,v1}
\fmf{dots,label=$x$,tension=0.6,foreground=(0,,0,,1)}{v1,v2}
\fmf{phantom,tension=2.0}{o1bis,o1}
\fmf{plain,tension=2.0}{v2,o1bis}
\fmf{phantom,tension=2.0}{o2bis,o2}
\fmf{phantom,tension=2.0}{v2,o2bis}
\fmf{dashes,tension=0.0,foreground=(0,,0,,1)}{v2,o2}
\fmflabel{$a_{1}$}{i1bis}
\fmflabel{$a_{2}$}{i2bis}
\fmflabel{$a_{3}$}{o1bis}
\fmflabel{$N$}{o2bis}
\end{fmfgraph*}
\end{fmffile}
\end{gathered} \quad &\rightarrow \lambda\left|\vec{\varphi}_{\mathrm{cl}}\Big[\vec{J}\Big]\right|\delta_{a_{1} a_{2}}\delta_{a_{3} N} \;,
\label{eq:FeynRulesLoopExpansion3legVertexAppendix} \\
\begin{gathered}
\begin{fmffile}{Diagrams/LoopExpansion1_FeynRuleV4bis_Appendix}
\begin{fmfgraph*}(20,20)
\fmfleft{i0,i1,i2,i3}
\fmfright{o0,o1,o2,o3}
\fmf{phantom,tension=2.0}{i1,i1bis}
\fmf{plain,tension=2.0}{i1bis,v1}
\fmf{phantom,tension=2.0}{i2,i2bis}
\fmf{plain,tension=2.0}{i2bis,v1}
\fmf{zigzag,label=$x$,tension=0.6,foreground=(0,,0,,1)}{v1,v2}
\fmf{phantom,tension=2.0}{o1bis,o1}
\fmf{plain,tension=2.0}{v2,o1bis}
\fmf{phantom,tension=2.0}{o2bis,o2}
\fmf{plain,tension=2.0}{v2,o2bis}
\fmflabel{$a_{1}$}{i1bis}
\fmflabel{$a_{2}$}{i2bis}
\fmflabel{$a_{3}$}{o1bis}
\fmflabel{$a_{4}$}{o2bis}
\end{fmfgraph*}
\end{fmffile}
\end{gathered} \quad &\rightarrow \lambda\delta_{a_{1} a_{2}}\delta_{a_{3} a_{4}}\;,
\label{eq:FeynRulesLoopExpansion4legVertexAppendix}
\end{align}
\end{subequations}
which respectively correspond to~\eqref{eq:FeynRulesLoopExpansionPropagator},~\eqref{eq:FeynRulesLoopExpansion3legVertex} and~\eqref{eq:FeynRulesLoopExpansion4legVertex} evaluated at $\boldsymbol{K}=\boldsymbol{0}$, and $\left|\vec{\varphi}_{\mathrm{cl}}\Big[\vec{J}\Big]\right|=\left.\varrho\right|_{\boldsymbol{K}=\boldsymbol{0}}$ is also defined by~\eqref{eq:DefPhiclModulusRho}. On the other hand, at $\vec{J}=\vec{J}_{0}$, we have:
\begin{subequations}
\begin{align}
\begin{gathered}
\begin{fmffile}{Diagrams/1PIEA_G_Appendix}
\begin{fmfgraph*}(20,20)
\fmfleft{i0,i1,i2,i3}
\fmfright{o0,o1,o2,o3}
\fmflabel{$\alpha_{1}$}{v1}
\fmflabel{$\alpha_{2}$}{v2}
\fmf{phantom}{i1,v1}
\fmf{phantom}{i2,v1}
\fmf{plain,tension=0.6,foreground=(1,,0,,0)}{v1,v2}
\fmf{phantom}{v2,o1}
\fmf{phantom}{v2,o2}
\end{fmfgraph*}
\end{fmffile}
\end{gathered} \quad &\rightarrow \boldsymbol{G}_{\phi,\alpha_{1}\alpha_{2}}\Big[\vec{\phi}\Big]\;,
\label{eq:DefinitionG1PIEAhbarExpansionAppendix} \\
\begin{gathered}
\begin{fmffile}{Diagrams/1PIEA_V3_Appendix}
\begin{fmfgraph*}(20,20)
\fmfleft{i0,i1,i2,i3}
\fmfright{o0,o1,o2,o3}
\fmfv{decor.shape=cross,decor.angle=45,decor.size=3.5thick,foreground=(1,,0,,0)}{o2}
\fmf{phantom,tension=2.0}{i1,i1bis}
\fmf{plain,tension=2.0,foreground=(1,,0,,0)}{i1bis,v1}
\fmf{phantom,tension=2.0}{i2,i2bis}
\fmf{plain,tension=2.0,foreground=(1,,0,,0)}{i2bis,v1}
\fmf{dots,label=$x$,tension=0.6,foreground=(0,,0,,1)}{v1,v2}
\fmf{phantom,tension=2.0}{o1bis,o1}
\fmf{plain,tension=2.0,foreground=(1,,0,,0)}{v2,o1bis}
\fmf{phantom,tension=2.0}{o2bis,o2}
\fmf{phantom,tension=2.0,foreground=(1,,0,,0)}{v2,o2bis}
\fmf{dashes,tension=0.0,foreground=(1,,0,,0)}{v2,o2}
\fmflabel{$a_{1}$}{i1bis}
\fmflabel{$a_{2}$}{i2bis}
\fmflabel{$a_{3}$}{o1bis}
\fmflabel{$N$}{o2bis}
\end{fmfgraph*}
\end{fmffile}
\end{gathered} \quad &\rightarrow \lambda\left|\vec{\phi}\right|\delta_{a_{1} a_{2}}\delta_{a_{3} N}\;,
\label{eq:FeynRules1PIEA3legVertexSourceJ0mainAppendix} \\
\begin{gathered}
\begin{fmffile}{Diagrams/1PIEA_V4_Appendix}
\begin{fmfgraph*}(20,20)
\fmfleft{i0,i1,i2,i3}
\fmfright{o0,o1,o2,o3}
\fmf{phantom,tension=2.0}{i1,i1bis}
\fmf{plain,tension=2.0,foreground=(1,,0,,0)}{i1bis,v1}
\fmf{phantom,tension=2.0}{i2,i2bis}
\fmf{plain,tension=2.0,foreground=(1,,0,,0)}{i2bis,v1}
\fmf{zigzag,label=$x$,tension=0.6,foreground=(0,,0,,1)}{v1,v2}
\fmf{phantom,tension=2.0}{o1bis,o1}
\fmf{plain,tension=2.0,foreground=(1,,0,,0)}{v2,o1bis}
\fmf{phantom,tension=2.0}{o2bis,o2}
\fmf{plain,tension=2.0,foreground=(1,,0,,0)}{v2,o2bis}
\fmflabel{$a_{1}$}{i1bis}
\fmflabel{$a_{2}$}{i2bis}
\fmflabel{$a_{3}$}{o1bis}
\fmflabel{$a_{4}$}{o2bis}
\end{fmfgraph*}
\end{fmffile}
\end{gathered} \quad &\rightarrow \lambda\delta_{a_{1} a_{2}}\delta_{a_{3} a_{4}}\;,
\label{eq:FeynRules1PIEA4legVertexSourceJ0mainAppendix}
\end{align}
\end{subequations}
which are equivalent to~\eqref{eq:DefinitionG1PIEAhbarExpansion},~\eqref{eq:FeynRules1PIEA3legVertexSourceJ0main} and~\eqref{eq:FeynRules1PIEA4legVertexSourceJ0main}, respectively.

\end{itemize}

\vspace{0.3cm}

We are now equipped to pursue the IM. The next step consists in inserting the power series~\eqref{eq:pure1PIEAGammaExpansion0DONAppendix},~\eqref{eq:pure1PIEAWExpansion0DONAppendix} and~\eqref{eq:pure1PIEAJExpansion0DONAppendix} into the Legendre transform definition of the EA given by~\eqref{eq:pure1PIEAdefinition0DONAppendix}, thus leading to:
\begin{equation}
\sum_{n=0}^{\infty} \Gamma^{(\mathrm{1PI})}_{n}\Big[\vec{\phi}\Big]\hbar^{n} = -\sum_{n=0}^{\infty} W_{n}\Bigg[\sum_{m=0}^{\infty} \vec{J}_{m}\Big[\vec{\phi}\Big]\hbar^{m}\Bigg]\hbar^{n}+\sum_{n=0}^{\infty} \int_{\alpha} J_{n,\alpha}\Big[\vec{\phi}\Big] \phi_{\alpha} \hbar^{n}\;.
\label{eq:pure1PIEAIMstep20DON}
\end{equation}
By Taylor expanding the $W_{n}$ coefficients in the RHS of~\eqref{eq:pure1PIEAIMstep20DON} around $\vec{J}=\vec{J}_{0}$, one obtains (see section~\ref{sec:GammanCoeffIM1PIEA}):
\begin{equation}
\begin{split}
\Gamma^{(\mathrm{1PI})}_{n}\Big[\vec{\phi}\Big] = & -W_{n}\Big[\vec{J}=\vec{J}_{0}\Big] -\sum_{m=1}^{n} \int_{\alpha} \left.\frac{\delta W_{n-m}\big[\vec{J}\big]}{\delta J_{\alpha}}\right|_{\vec{J}=\vec{J}_{0}} J_{m,\alpha}\Big[\vec{\phi}\Big] \\
& -\sum_{m=2}^{n} \frac{1}{m!} \sum_{\underset{\lbrace n_{1} + \cdots + n_{m} \leq n\rbrace}{n_{1},\cdots,n_{m}=1}}^{n} \int_{\alpha_{1},\cdots,\alpha_{m}} \left.\frac{\delta^{m} W_{n-(n_{1}+\cdots+n_{m})}\big[\vec{J}\big]}{\delta J_{\alpha_{1}}\cdots\delta J_{\alpha_{m}}}\right|_{\vec{J}=\vec{J}_{0}} J_{n_{1},\alpha_{1}}\Big[\vec{\phi}\Big]\cdots J_{n_{m},\alpha_{m}}\Big[\vec{\phi}\Big] \\
& + \int_{\alpha} J_{n,\alpha}\Big[\vec{\phi}\Big] \phi_{\alpha}\;.
\end{split}
\label{eq:pure1PIEAIMstep50DON}
\end{equation}
Let us recall here that curly braces below discrete sums contain a condition that must satisfied by each term of the sum in question, as already used in~\eqref{eq:NfactorPropagatorLoop0DON}. We also point out that a discrete sum vanishes if the upper boundary of the running index is less than its starting value, e.g. the second line of~\eqref{eq:pure1PIEAIMstep50DON} vanishes if $n<2$ due to $\sum_{m=2}^{n}$. We can further simplify~\eqref{eq:pure1PIEAIMstep50DON} by noticing that~\eqref{eq:pure1PIEAphi0DON} enables us to write:
\begin{equation}
\begin{split}
-\sum_{m=1}^{\textcolor{red}{n}} \int_{\alpha} \left.\frac{\delta W_{n-m}\big[\vec{J}\big]}{\delta J_{\alpha}}\right|_{\vec{J}=\vec{J}_{0}} J_{m,\alpha}\Big[\vec{\phi}\Big]+\int_{\alpha} J_{n,\alpha}\Big[\vec{\phi}\Big] \phi_{\alpha} = & -\sum_{m=1}^{\textcolor{red}{n-1}} \int_{\alpha} \left.\frac{\delta W_{n-m}\big[\vec{J}\big]}{\delta J_{\alpha}}\right|_{\vec{J}=\vec{J}_{0}} J_{m,\alpha}\Big[\vec{\phi}\Big] \\
& +\int_{\alpha} J_{0,\alpha}\Big[\vec{\phi}\Big] \phi_{\alpha} \delta_{n 0}\;.
\end{split}
\label{eq:pure1PIEAIMSimplifystep50DON}
\end{equation}
After combining~\eqref{eq:pure1PIEAIMSimplifystep50DON} with~\eqref{eq:pure1PIEAIMstep50DON}, we have:
\begin{equation}
\begin{split}
\Gamma^{(\mathrm{1PI})}_{n}\Big[\vec{\phi}\Big] = & -W_{n}\Big[\vec{J}=\vec{J}_{0}\Big] -\sum_{m=1}^{n-1} \int_{\alpha} \left.\frac{\delta W_{n-m}\big[\vec{J}\big]}{\delta J_{\alpha}}\right|_{\vec{J}=\vec{J}_{0}} J_{m,\alpha}\Big[\vec{\phi}\Big] \\
& -\sum_{m=2}^{n} \frac{1}{m!} \sum_{\underset{\lbrace n_{1} + \cdots + n_{m} \leq n\rbrace}{n_{1},\cdots,n_{m}=1}}^{n} \int_{\alpha_{1},\cdots,\alpha_{m}} \left.\frac{\delta^{m} W_{n-(n_{1}+\cdots+n_{m})}\big[\vec{J}\big]}{\delta J_{\alpha_{1}}\cdots\delta J_{\alpha_{m}}}\right|_{\vec{J}=\vec{J}_{0}} J_{n_{1},\alpha_{1}}\Big[\vec{\phi}\Big]\cdots J_{n_{m},\alpha_{m}}\Big[\vec{\phi}\Big] \\
& + \int_{\alpha} J_{0,\alpha}\Big[\vec{\phi}\Big] \phi_{\alpha} \delta_{n 0}\;.
\end{split}
\label{eq:pure1PIEAIMstep60DON}
\end{equation}
We will derive in the present section the first non-trivial order of the 1PI EA via the IM, which amounts to determining $\Gamma^{(\mathrm{1PI})}$ up to order $\mathcal{O}\big(\hbar^{2}\big)$. Therefore, we will consider~\eqref{eq:pure1PIEAIMstep60DON} at $n=0,1~\mathrm{and}~2$:
\begin{equation}
\Gamma^{(\mathrm{1PI})}_{0}\Big[\vec{\phi}\Big] = -W_{0}\Big[\vec{J}=\vec{J}_{0}\Big] + \int_{\alpha} J_{0,\alpha}\Big[\vec{\phi}\Big] \phi_{\alpha}\;,
\label{eq:pure1PIEAIMGamma00DON}
\end{equation}
\begin{equation}
\Gamma^{(\mathrm{1PI})}_{1}\Big[\vec{\phi}\Big] = -W_{1}\Big[\vec{J}=\vec{J}_{0}\Big]\;,
\label{eq:pure1PIEAIMGamma10DON}
\end{equation}
\begin{equation}
\scalebox{0.98}{${\displaystyle\Gamma^{(\mathrm{1PI})}_{2}\Big[\vec{\phi}\Big] = -W_{2}\Big[\vec{J}=\vec{J}_{0}\Big] -\int_{\alpha} \left.\frac{\delta W_{1}\big[\vec{J}\big]}{\delta J_{\alpha}}\right|_{\vec{J}=\vec{J}_{0}} J_{1,\alpha}\Big[\vec{\phi}\Big] -\frac{1}{2} \int_{\alpha_{1},\alpha_{2}} \left.\frac{\delta^{2} W_{0}\big[\vec{J}\big]}{\delta J_{\alpha_{1}} \delta J_{\alpha_{2}}}\right|_{\vec{J}=\vec{J}_{0}} J_{1,\alpha_{1}}\Big[\vec{\phi}\Big] J_{1,\alpha_{2}}\Big[\vec{\phi}\Big]\;.}$}
\label{eq:pure1PIEAIMGamma20DON}
\end{equation}
From~\eqref{eq:pure1PIEAIMstep60DON} (as well as~\eqref{eq:pure1PIEAIMGamma00DON} to~\eqref{eq:pure1PIEAIMGamma20DON}) together with~\eqref{eq:pure1PIEAGammaExpansion0DONAppendix}, it is clear that $\Gamma^{(\mathrm{1PI})}$ is completely specified by the $W_{n}$ and $\vec{J}_{n}$ coefficients. In particular, we must evaluate derivatives of the $W_{n}$ coefficients. According to~\eqref{eq:pure1PIEAIMGamma00DON} to~\eqref{eq:pure1PIEAIMGamma20DON}, we only need two of them to determine the EA up to order $\mathcal{O}\big(\hbar^{2}\big)$. One of these two derivatives is already given by~\eqref{eq:pure1PIEAGJ00DON}:
\begin{equation}
\left.\frac{\delta^{2} W_{0}\big[\vec{J}\big]}{\delta J_{\alpha_{1}} \delta J_{\alpha_{2}}}\right|_{\vec{J}=\vec{J}_{0}} = \hspace{0.3cm} \begin{gathered}
\begin{fmffile}{Diagrams/1PIEA_DerivW0J0J0}
\begin{fmfgraph*}(20,20)
\fmfleft{i0,i1,i2,i3}
\fmfright{o0,o1,o2,o3}
\fmfv{decor.shape=circle,decor.filled=empty,decor.size=1.5thick,label=$\alpha_{1}$}{v1}
\fmfv{decor.shape=circle,decor.filled=empty,decor.size=1.5thick,label=$\alpha_{2}$}{v2}
\fmf{phantom}{i1,v1}
\fmf{phantom}{i2,v1}
\fmf{plain,tension=0.6,foreground=(1,,0,,0)}{v1,v2}
\fmf{phantom}{v2,o1}
\fmf{phantom}{v2,o2}
\end{fmfgraph*}
\end{fmffile}
\end{gathered}\hspace{0.3cm}\;.
\label{eq:pure1PIEADerivW0j0j00DON}
\end{equation}
Note that, in~\eqref{eq:pure1PIEADerivW0j0j00DON} and in all forthcoming diagrams in appendix~\ref{ann:InversionMethod}, external points are represented by an empty dot with its corresponding index (except for Feynman rules). If the index is not indicated, then integration is implicitly carried out over this index (see~\eqref{eq:pure1PIEADeterminationJ1step30DON} for an application of this convention). The other derivative that we need is calculated using~\eqref{eq:pure1PIEAIMW10DON}:
\begin{equation}
\begin{split}
\frac{\delta W_{1}\big[\vec{J}\big]}{\delta J_{\alpha_{1}}} = & \ \frac{\delta}{\delta J_{\alpha_{1}}}\left(\frac{1}{2} \mathrm{STr}\left[\ln\Big(\boldsymbol{G}_{\varphi_{\mathrm{cl}};J}\Big[\vec{J}\Big]\Big)\right]\right) \\
= & \ \frac{1}{2} \frac{\delta}{\delta J_{\alpha_{1}}} \int_{\alpha_{2}}\ln\Big(\boldsymbol{G}_{\varphi_{\mathrm{cl}};J,\alpha_{2}\alpha_{2}}\Big[\vec{J}\Big]\Big) \\
= & \ \frac{1}{2} \int_{\alpha_{2},\alpha_{3}} \boldsymbol{G}^{-1}_{\varphi_{\mathrm{cl}};J,\alpha_{2}\alpha_{3}}\Big[\vec{J}\Big] \frac{\delta \boldsymbol{G}_{\varphi_{\mathrm{cl}};J,\alpha_{3}\alpha_{2}}\big[\vec{J}\big]}{\delta J_{\alpha_{1}}}\;.
\end{split}
\label{eq:pure1PIEADerivW1j0step10DON}
\end{equation}
In order to evaluate the derivative of the last line, we will make use of the three following relations:
\begin{equation}
\frac{\delta}{\delta J_{\alpha_{1}}}=\int_{\alpha_{2}} \frac{\delta \varphi_{\mathrm{cl},\alpha_{2}}\big[\vec{J}\big]}{\delta J_{\alpha_{1}}} \frac{\delta}{\delta \varphi_{\mathrm{cl},\alpha_{2}}}\;,
\label{eq:pure1PIEAchainRule0DON}
\end{equation}
\begin{equation}
\frac{\delta\varphi_{\mathrm{cl},\alpha_{1}}\big[\vec{J}\big]}{\delta J_{\alpha_{2}}} = \frac{\delta^{2} W_{0}\big[\vec{J}\big]}{\delta J_{\alpha_{2}}\delta J_{\alpha_{1}}} = \boldsymbol{G}_{\varphi_{\mathrm{cl}};J,\alpha_{2}\alpha_{1}}\Big[\vec{J}\Big]\;,
\label{eq:pure1PIEADerivphiclJ0DON}
\end{equation}
\begin{equation}
\begin{split}
\frac{\delta \boldsymbol{G}^{-1}_{\varphi_{\mathrm{cl}};J,\alpha_{1}\alpha_{2}}\big[\vec{J}\big]}{\delta\varphi_{\mathrm{cl},\alpha_{3}}} = & \ \frac{\delta}{\delta\varphi_{\mathrm{cl},\alpha_{3}}} \left(\left(-\nabla^{2}_{x_{1}} + m^{2} + \frac{\lambda}{6}\vec{\varphi}^{2}_{\mathrm{cl},x_{1}}\Big[\vec{J}\Big]\right)\delta_{\alpha_{1}\alpha_{2}} +\frac{\lambda}{3}\varphi_{\mathrm{cl},\alpha_{1}}\Big[\vec{J}\Big]\varphi_{\mathrm{cl},\alpha_{2}}\Big[\vec{J}\Big]\delta_{x_{1}x_{2}}\right) \\
= & \ \frac{\lambda}{6} \underbrace{\frac{\delta\vec{\varphi}^{2}_{\mathrm{cl},x_{1}}\big[\vec{J}\big]}{\delta\varphi_{\mathrm{cl},\alpha_{3}}}}_{2\varphi_{\mathrm{cl},\alpha_{3}}\delta_{x_{1}x_{3}}} \delta_{\alpha_{1}\alpha_{2}} +\frac{\lambda}{3}\delta_{\alpha_{1}\alpha_{3}}\varphi_{\mathrm{cl},\alpha_{2}}\Big[\vec{J}\Big] \delta_{x_{1}x_{2}} +\frac{\lambda}{3}\varphi_{\mathrm{cl},\alpha_{1}}\Big[\vec{J}\Big] \delta_{\alpha_{2}\alpha_{3}} \delta_{x_{1}x_{2}} \\
= & \ \frac{\lambda}{3}\left(\varphi_{\mathrm{cl},\alpha_{3}}\Big[\vec{J}\Big]\delta_{\alpha_{1}\alpha_{2}}\delta_{x_{1}x_{3}}+\varphi_{\mathrm{cl},\alpha_{2}}\Big[\vec{J}\Big] \delta_{\alpha_{1}\alpha_{3}} \delta_{x_{1}x_{2}} + \varphi_{\mathrm{cl},\alpha_{1}}\Big[\vec{J}\Big] \delta_{\alpha_{2}\alpha_{3}} \delta_{x_{1}x_{2}} \right)\;,
\end{split}
\label{eq:pure1PIEADerivGminus1phicl0DON}
\end{equation}
where we have combined~\eqref{eq:pure1PIEAphicl0DON} with~\eqref{eq:pure1PIEAGJ0DON} to obtain~\eqref{eq:pure1PIEADerivphiclJ0DON} and used~\eqref{eq:pure1PIEAGJminus1expressArbitaryDimON} as expression of $\boldsymbol{G}^{-1}_{\varphi_{\mathrm{cl}};J,\alpha_{1}\alpha_{2}}\big[\vec{J}\big]$ to start the derivation of~\eqref{eq:pure1PIEADerivGminus1phicl0DON}. By exploiting~\eqref{eq:pure1PIEAchainRule0DON},~\eqref{eq:pure1PIEADerivphiclJ0DON} and~\eqref{eq:pure1PIEADerivGminus1phicl0DON}, we calculate:
\begin{equation}
\begin{split}
\frac{\delta \boldsymbol{G}_{\varphi_{\mathrm{cl}};J,\alpha_{3}\alpha_{2}}\big[\vec{J}\big]}{\delta J_{\alpha_{1}}} = & - \int_{\alpha_{4},\alpha_{5}} \boldsymbol{G}_{\varphi_{\mathrm{cl}};J,\alpha_{3}\alpha_{4}}\Big[\vec{J}\Big] \frac{\delta \boldsymbol{G}^{-1}_{\varphi_{\mathrm{cl}};J,\alpha_{4}\alpha_{5}}\big[\vec{J}\big]}{\delta J_{\alpha_{1}}} \boldsymbol{G}_{\varphi_{\mathrm{cl}};J,\alpha_{5}\alpha_{2}}\Big[\vec{J}\Big] \\
= & - \int_{\alpha_{4},\alpha_{5},\alpha_{6}} \boldsymbol{G}_{\varphi_{\mathrm{cl}};J,\alpha_{3}\alpha_{4}}\Big[\vec{J}\Big] \frac{\delta \varphi_{\mathrm{cl},\alpha_{6}}\big[\vec{J}\big]}{\delta J_{\alpha_{1}}} \frac{\delta \boldsymbol{G}^{-1}_{\varphi_{\mathrm{cl}};J,\alpha_{4}\alpha_{5}}\big[\vec{J}\big]}{\delta \varphi_{\mathrm{cl},\alpha_{6}}} \boldsymbol{G}_{\varphi_{\mathrm{cl}};J,\alpha_{5}\alpha_{2}}\Big[\vec{J}\Big] \\
= & - \frac{\lambda}{3} \int_{\alpha_{4},\alpha_{5},\alpha_{6}} \boldsymbol{G}_{\varphi_{\mathrm{cl}};J,\alpha_{3}\alpha_{4}}\Big[\vec{J}\Big] \boldsymbol{G}_{\varphi_{\mathrm{cl}};J,\alpha_{1}\alpha_{6}}\Big[\vec{J}\Big] \\
& \hspace{1.95cm} \times \left(\varphi_{\mathrm{cl},\alpha_{6}}\Big[\vec{J}\Big]\delta_{\alpha_{4}\alpha_{5}}\delta_{x_{4}x_{6}}+\varphi_{\mathrm{cl},\alpha_{5}}\Big[\vec{J}\Big]\delta_{\alpha_{4}\alpha_{6}}\delta_{x_{4}x_{5}}+\varphi_{\mathrm{cl},\alpha_{4}}\Big[\vec{J}\Big]\delta_{\alpha_{5}\alpha_{6}}\delta_{x_{4}x_{5}}\right) \\
& \hspace{1.95cm} \times \boldsymbol{G}_{\varphi_{\mathrm{cl}};J,\alpha_{5}\alpha_{2}}\Big[\vec{J}\Big] \\
= & - \frac{\lambda}{3} \Bigg(\int_{\alpha_{4},\alpha_{6}} \varphi_{\mathrm{cl},\alpha_{6}}\Big[\vec{J}\Big] \boldsymbol{G}_{\varphi_{\mathrm{cl}};J,\alpha_{3}\alpha_{4}}\Big[\vec{J}\Big] \boldsymbol{G}_{\varphi_{\mathrm{cl}};J,\alpha_{1}\alpha_{6}}\Big[\vec{J}\Big] \boldsymbol{G}_{\varphi_{\mathrm{cl}};J,\alpha_{4}\alpha_{2}}\Big[\vec{J}\Big] \delta_{x_{4}x_{6}} \\
& \hspace{0.75cm} + \int_{\alpha_{4},\alpha_{5}} \varphi_{\mathrm{cl},\alpha_{5}}\Big[\vec{J}\Big] \boldsymbol{G}_{\varphi_{\mathrm{cl}};J,\alpha_{3}\alpha_{4}}\Big[\vec{J}\Big] \boldsymbol{G}_{\varphi_{\mathrm{cl}};J,\alpha_{1}\alpha_{4}}\Big[\vec{J}\Big] \boldsymbol{G}_{\varphi_{\mathrm{cl}};J,\alpha_{5}\alpha_{2}}\Big[\vec{J}\Big]\delta_{x_{4}x_{5}} \\
& \hspace{0.75cm} + \int_{\alpha_{4},\alpha_{5}} \varphi_{\mathrm{cl},\alpha_{4}}\Big[\vec{J}\Big] \boldsymbol{G}_{\varphi_{\mathrm{cl}};J,\alpha_{3}\alpha_{4}}\Big[\vec{J}\Big] \boldsymbol{G}_{\varphi_{\mathrm{cl}};J,\alpha_{1}\alpha_{5}}\Big[\vec{J}\Big] \boldsymbol{G}_{\varphi_{\mathrm{cl}};J,\alpha_{5}\alpha_{2}}\Big[\vec{J}\Big]\delta_{x_{4}x_{5}}\Bigg) \\
= & - \frac{1}{3} \left(\rule{0cm}{0.8cm}\right. \hspace{0.5cm} \begin{gathered}
\begin{fmffile}{Diagrams/1PIEA_DerivGJ1}
\begin{fmfgraph*}(20,20)
\fmfleft{i0,i1,i2,i3}
\fmfright{o0,o1,o2,o3}
\fmfv{decor.shape=cross,decor.angle=45,decor.size=3.5thick,foreground=(0,,0,,1)}{o2}
\fmfv{decor.shape=circle,decor.filled=empty,decor.size=1.5thick,label=$\alpha_{1}$}{o1}
\fmfv{decor.shape=circle,decor.filled=empty,decor.size=1.5thick,label=$\alpha_{2}$}{i1}
\fmfv{decor.shape=circle,decor.filled=empty,decor.size=1.5thick,label=$\alpha_{3}$}{i2}
\fmf{phantom,tension=2.0}{i1,i1bis}
\fmf{phantom,tension=2.0}{i1bis,v1}
\fmf{phantom,tension=2.0}{i2,i2bis}
\fmf{phantom,tension=2.0}{i2bis,v1}
\fmf{dots,tension=0.6,foreground=(0,,0,,1)}{v1,v2}
\fmf{phantom,tension=2.0}{o1bis,o1}
\fmf{phantom,tension=2.0}{v2,o1bis}
\fmf{phantom,tension=2.0}{o2bis,o2}
\fmf{phantom,tension=2.0}{v2,o2bis}
\fmf{plain,tension=0.0}{i2,v1}
\fmf{plain,tension=0.0}{i1,v1}
\fmf{dashes,tension=0.0,foreground=(0,,0,,1)}{v2,o2}
\fmf{plain,tension=0.0}{v2,o1}
\end{fmfgraph*}
\end{fmffile}
\end{gathered} \hspace{0.5cm} + \hspace{0.6cm} \begin{gathered}
\begin{fmffile}{Diagrams/1PIEA_DerivGJ2}
\begin{fmfgraph*}(20,20)
\fmfleft{i0,i1,i2,i3}
\fmfright{o0,o1,o2,o3}
\fmfv{decor.shape=cross,decor.angle=45,decor.size=3.5thick,foreground=(0,,0,,1)}{o2}
\fmfv{decor.shape=circle,decor.filled=empty,decor.size=1.5thick,label=$\alpha_{2}$}{o1}
\fmfv{decor.shape=circle,decor.filled=empty,decor.size=1.5thick,label=$\alpha_{1}$}{i1}
\fmfv{decor.shape=circle,decor.filled=empty,decor.size=1.5thick,label=$\alpha_{3}$}{i2}
\fmf{phantom,tension=2.0}{i1,i1bis}
\fmf{phantom,tension=2.0}{i1bis,v1}
\fmf{phantom,tension=2.0}{i2,i2bis}
\fmf{phantom,tension=2.0}{i2bis,v1}
\fmf{dots,tension=0.6,foreground=(0,,0,,1)}{v1,v2}
\fmf{phantom,tension=2.0}{o1bis,o1}
\fmf{phantom,tension=2.0}{v2,o1bis}
\fmf{phantom,tension=2.0}{o2bis,o2}
\fmf{phantom,tension=2.0}{v2,o2bis}
\fmf{plain,tension=0.0}{i2,v1}
\fmf{plain,tension=0.0}{i1,v1}
\fmf{dashes,tension=0.0,foreground=(0,,0,,1)}{v2,o2}
\fmf{plain,tension=0.0}{v2,o1}
\end{fmfgraph*}
\end{fmffile}
\end{gathered} \hspace{0.5cm} + \hspace{0.6cm} \begin{gathered}
\begin{fmffile}{Diagrams/1PIEA_DerivGJ3}
\begin{fmfgraph*}(20,20)
\fmfleft{i0,i1,i2,i3}
\fmfright{o0,o1,o2,o3}
\fmfv{decor.shape=cross,decor.angle=45,decor.size=3.5thick,foreground=(0,,0,,1)}{o2}
\fmfv{decor.shape=circle,decor.filled=empty,decor.size=1.5thick,label=$\alpha_{3}$}{o1}
\fmfv{decor.shape=circle,decor.filled=empty,decor.size=1.5thick,label=$\alpha_{1}$}{i1}
\fmfv{decor.shape=circle,decor.filled=empty,decor.size=1.5thick,label=$\alpha_{2}$}{i2}
\fmf{phantom,tension=2.0}{i1,i1bis}
\fmf{phantom,tension=2.0}{i1bis,v1}
\fmf{phantom,tension=2.0}{i2,i2bis}
\fmf{phantom,tension=2.0}{i2bis,v1}
\fmf{dots,tension=0.6,foreground=(0,,0,,1)}{v1,v2}
\fmf{phantom,tension=2.0}{o1bis,o1}
\fmf{phantom,tension=2.0}{v2,o1bis}
\fmf{phantom,tension=2.0}{o2bis,o2}
\fmf{phantom,tension=2.0}{v2,o2bis}
\fmf{plain,tension=0.0}{i2,v1}
\fmf{plain,tension=0.0}{i1,v1}
\fmf{dashes,tension=0.0,foreground=(0,,0,,1)}{v2,o2}
\fmf{plain,tension=0.0}{v2,o1}
\end{fmfgraph*}
\end{fmffile}
\end{gathered} \hspace{0.6cm} \left.\rule{0cm}{0.8cm}\right)\;.
\end{split}
\label{eq:pure1PIEADerivGJ0DON}
\end{equation}
We then insert~\eqref{eq:pure1PIEADerivGJ0DON} into~\eqref{eq:pure1PIEADerivW1j0step10DON} and set $\vec{J}=\vec{J}_{0}$:
\begin{equation}
\left.\frac{\delta W_{1}\big[\vec{J}\big]}{\delta J_{\alpha}}\right|_{\vec{J}=\vec{J}_{0}} = - \frac{1}{6} \left(\rule{0cm}{1.2cm}\right. \hspace{0.08cm} \begin{gathered}
\begin{fmffile}{Diagrams/1PIEA_DerivW1J1}
\begin{fmfgraph*}(25,12)
\fmfleft{i0,i,i1}
\fmfright{o0,o,o1}
\fmfv{decor.shape=cross,decor.angle=45,decor.size=3.5thick,foreground=(1,,0,,0)}{o1}
\fmfv{decor.shape=circle,decor.filled=empty,decor.size=1.5thick,label.dist=0.15cm,label=$\alpha$}{o0}
\fmf{phantom,tension=10}{i,i1bis}
\fmf{phantom,tension=10}{o,o1bis}
\fmf{plain,left,tension=0.5,foreground=(1,,0,,0)}{i1bis,v1,i1bis}
\fmf{phantom,right,tension=0.5}{o1bis,v2,o1bis}
\fmf{dashes,tension=0,foreground=(1,,0,,0)}{v2,o1}
\fmf{plain,tension=0,foreground=(1,,0,,0)}{v2,o0}
\fmf{dots,foreground=(0,,0,,1)}{v1,v2}
\end{fmfgraph*}
\end{fmffile}
\end{gathered}
+ 2 \hspace{0.2cm}\begin{gathered}
\begin{fmffile}{Diagrams/1PIEA_DerivW1J2}
\begin{fmfgraph*}(22,12)
\fmfleft{i0,i,i1}
\fmfright{o0,o,o1}
\fmfv{decor.shape=cross,decor.angle=65,decor.size=3.5thick,foreground=(1,,0,,0)}{o1}
\fmfv{decor.shape=circle,decor.filled=empty,decor.size=1.5thick,label.dist=0.15cm,label=$\alpha$}{i1}
\fmf{phantom,tension=11}{i,v1}
\fmf{phantom,tension=11}{v2,o}
\fmf{plain,tension=0,foreground=(1,,0,,0)}{v1,i1}
\fmf{dashes,tension=0,foreground=(1,,0,,0)}{v2,o1}
\fmf{plain,right,tension=0.4,foreground=(1,,0,,0)}{v1,v2}
\fmf{dots,tension=4.0,foreground=(0,,0,,1)}{v1,v2}
\end{fmfgraph*}
\end{fmffile}
\end{gathered} \left.\rule{0cm}{1.2cm}\right)\;.
\label{eq:pure1PIEADerivW1j0step20DON}
\end{equation}
The only quantity left to determine in~\eqref{eq:pure1PIEAIMGamma00DON} to~\eqref{eq:pure1PIEAIMGamma20DON} is the source coefficient $\vec{J}_{1}$. To that end, we will make use of the power series of $\vec{\phi}$, i.e.~\eqref{eq:pure1PIEAphiExpansion0DONAppendix}. By Taylor expanding the $\vec{\phi}_{n}$ coefficients around $\vec{J}=\vec{J}_{0}$, we obtain:
\begin{equation}
\begin{split}
\phi_{\alpha_{1}} = & \sum_{n=0}^{\infty} \phi_{n,\alpha_{1}}\Big[\vec{J}\Big]\hbar^{n} \\
= & \ \phi_{0,\alpha_{1}}\Big[\vec{J}\Big] + \phi_{1,\alpha_{1}}\Big[\vec{J}\Big]\hbar + \mathcal{O}\big(\hbar^{2}\big) \\
= & \ \phi_{0,\alpha_{1}}\Big[\vec{J}=\vec{J}_{0}\Big] + \int_{\alpha_{2}} \left.\frac{\delta \phi_{0,\alpha_{1}}\big[\vec{J}\big]}{\delta J_{\alpha_{2}}}\right|_{\vec{J}=\vec{J}_{0}} \left(J_{\alpha_{2}}\Big[\vec{\phi}\Big]-J_{0,\alpha_{2}}\Big[\vec{\phi}\Big]\right) + \mathcal{O}\Bigg(\left(\vec{J}-\vec{J_{0}}\right)^{2}\Bigg) \\
& + \left(\phi_{1,\alpha_{1}}\Big[\vec{J}=\vec{J}_{0}\Big] + \mathcal{O}\Big(\vec{J}-\vec{J_{0}}\Big)\right)\hbar + \mathcal{O}\big(\hbar^{2}\big) \\
= & \ \phi_{0,\alpha_{1}}\Big[\vec{J}=\vec{J}_{0}\Big] + \hbar\left(\int_{\alpha_{2}} \left.\frac{\delta \phi_{0,\alpha_{1}}\big[\vec{J}\big]}{\delta J_{\alpha_{2}}}\right|_{\vec{J}=\vec{J}_{0}} J_{1,\alpha_{2}}\Big[\vec{\phi}\Big] + \phi_{1,\alpha_{1}}\Big[\vec{J}=\vec{J}_{0}\Big]\right) + \mathcal{O}\big(\hbar^{2}\big)\;,
\end{split}
\label{eq:pure1PIEAphiExpansionAroundJ00DON}
\end{equation}
where the last line was obtained by using the equality $\vec{J}\big[\vec{\phi}\big]-\vec{J}_{0}\big[\vec{\phi}\big]=\vec{J}_{1}\big[\vec{\phi}\big] \hbar + \mathcal{O}\big(\hbar^{2}\big)$ resulting directly from~\eqref{eq:pure1PIEAJExpansion0DONAppendix}. By setting $\vec{J}=\vec{J}_{0}$ in~\eqref{eq:pure1PIEAphiExpansionAroundJ00DON}, we turn the latter result into an infinite tower of coupled integro-algebraic equations as a consequence of~\eqref{eq:pure1PIEAphi0DON} (i.e. as a consequence of the fact that $\vec{\phi}$ is of order $\mathcal{O}(\hbar^{0})$):
\begin{subequations}
\begin{empheq}[left=\empheqlbrace]{align}
& \hspace{0.1cm} \mathrm{Order}~\mathcal{O}\big(\hbar^{0}\big):~\phi_{\alpha} = \phi_{0,\alpha}\Big[\vec{J}=\vec{J}_{0}\Big] \quad \forall \alpha\;, \label{eq:pure1PIEATowerEquationJn10DON}\\
\nonumber \\
& \hspace{0.1cm} \mathrm{Order}~\mathcal{O}(\hbar):~0 = \int_{\alpha_{2}} \left.\frac{\delta \phi_{0,\alpha_{1}}\big[\vec{J}\big]}{\delta J_{\alpha_{2}}}\right|_{\vec{J}=\vec{J}_{0}} J_{1,\alpha_{2}}\Big[\vec{\phi}\Big] + \phi_{1,\alpha_{1}}\Big[\vec{J}=\vec{J}_{0}\Big] \quad \forall \alpha_{1}\;, \label{eq:pure1PIEATowerEquationJn20DON}\\
\nonumber \\
& \hspace{6.5cm} \vdots \nonumber
\end{empheq}
\end{subequations}
where~\eqref{eq:pure1PIEATowerEquationJn10DON} is equivalent to~\eqref{eq:pure1PIEAphi0DON}. We can then exploit~\eqref{eq:pure1PIEATowerEquationJn20DON} to determine $\vec{J}_{1}$. In order to achieve this, we must find diagrammatic expressions for the other quantities involved in this equation. To that end, we can use~\eqref{eq:pure1PIEAphinCoeff0DON} to directly deduce these quantities from~\eqref{eq:pure1PIEADerivW0j0j00DON} and~\eqref{eq:pure1PIEADerivW1j0step20DON}, i.e.:
\begin{equation}
\left.\frac{\delta \phi_{0,\alpha_{1}}\big[\vec{J}\big]}{\delta J_{\alpha_{2}}}\right|_{\vec{J}=\vec{J}_{0}}=\left.\frac{\delta^{2} W_{0}\big[\vec{J}\big]}{\delta J_{\alpha_{2}} \delta J_{\alpha_{1}}}\right|_{\vec{J}=\vec{J}_{0}} = \hspace{0.3cm} \begin{gathered}
\begin{fmffile}{Diagrams/1PIEA_DerivW0J0J0}
\begin{fmfgraph*}(20,20)
\fmfleft{i0,i1,i2,i3}
\fmfright{o0,o1,o2,o3}
\fmfv{decor.shape=circle,decor.filled=empty,decor.size=1.5thick,label=$\alpha_{1}$}{v1}
\fmfv{decor.shape=circle,decor.filled=empty,decor.size=1.5thick,label=$\alpha_{2}$}{v2}
\fmf{phantom}{i1,v1}
\fmf{phantom}{i2,v1}
\fmf{plain,tension=0.6,foreground=(1,,0,,0)}{v1,v2}
\fmf{phantom}{v2,o1}
\fmf{phantom}{v2,o2}
\end{fmfgraph*}
\end{fmffile}
\end{gathered}\hspace{0.3cm}\;,
\label{eq:pure1PIEADerivphi0J0DON}
\end{equation}
\begin{equation}
\phi_{1,\alpha}\Big[\vec{J}=\vec{J}_{0}\Big]=\left.\frac{\delta W_{1}\big[\vec{J}\big]}{\delta J_{\alpha}}\right|_{\vec{J}=\vec{J}_{0}} = - \frac{1}{6} \left(\rule{0cm}{1.2cm}\right. \hspace{0.08cm} \begin{gathered}
\begin{fmffile}{Diagrams/1PIEA_phi11}
\begin{fmfgraph*}(25,12)
\fmfleft{i0,i,i1}
\fmfright{o0,o,o1}
\fmfv{decor.shape=cross,decor.angle=45,decor.size=3.5thick,foreground=(1,,0,,0)}{o1}
\fmfv{decor.shape=circle,decor.filled=empty,decor.size=1.5thick,label.dist=0.15cm,label=$\alpha$}{o0}
\fmf{phantom,tension=10}{i,i1bis}
\fmf{phantom,tension=10}{o,o1bis}
\fmf{plain,left,tension=0.5,foreground=(1,,0,,0)}{i1bis,v1,i1bis}
\fmf{phantom,right,tension=0.5}{o1bis,v2,o1bis}
\fmf{dashes,tension=0,foreground=(1,,0,,0)}{v2,o1}
\fmf{plain,tension=0,foreground=(1,,0,,0)}{v2,o0}
\fmf{dots,foreground=(0,,0,,1)}{v1,v2}
\end{fmfgraph*}
\end{fmffile}
\end{gathered}
+ 2 \hspace{0.2cm}\begin{gathered}
\begin{fmffile}{Diagrams/1PIEA_phi12}
\begin{fmfgraph*}(22,12)
\fmfleft{i0,i,i1}
\fmfright{o0,o,o1}
\fmfv{decor.shape=cross,decor.angle=65,decor.size=3.5thick,foreground=(1,,0,,0)}{o1}
\fmfv{decor.shape=circle,decor.filled=empty,decor.size=1.5thick,label.dist=0.15cm,label=$\alpha$}{i1}
\fmf{phantom,tension=11}{i,v1}
\fmf{phantom,tension=11}{v2,o}
\fmf{plain,tension=0,foreground=(1,,0,,0)}{v1,i1}
\fmf{dashes,tension=0,foreground=(1,,0,,0)}{v2,o1}
\fmf{plain,right,tension=0.4,foreground=(1,,0,,0)}{v1,v2}
\fmf{dots,tension=4.0,foreground=(0,,0,,1)}{v1,v2}
\end{fmfgraph*}
\end{fmffile}
\end{gathered} \left.\rule{0cm}{1.2cm}\right)\;.
\label{eq:pure1PIEAphi10DON}
\end{equation}
After inserting~\eqref{eq:pure1PIEADerivphi0J0DON} and~\eqref{eq:pure1PIEAphi10DON} into~\eqref{eq:pure1PIEATowerEquationJn20DON}, we obtain:
\begin{equation}
0 = \hspace{0.3cm} \begin{gathered}
\begin{fmffile}{Diagrams/1PIEA_DeterminationJ11}
\begin{fmfgraph*}(20,20)
\fmfleft{i0,i1,i2,i3}
\fmfright{o0,o1,o2,o3}
\fmfv{decor.shape=circle,decor.filled=empty,decor.size=1.5thick,label=$\alpha$}{v1}
\fmfv{decor.shape=circle,decor.filled=empty,decor.size=0.4cm,label=$1$,label.dist=0}{v2}
\fmf{phantom}{i1,v1}
\fmf{phantom}{i2,v1}
\fmf{plain,tension=0.6,foreground=(1,,0,,0)}{v1,v2}
\fmf{phantom}{v2,o1}
\fmf{phantom}{v2,o2}
\end{fmfgraph*}
\end{fmffile}
\end{gathered} - \frac{1}{6} \left(\rule{0cm}{1.2cm}\right. \hspace{0.08cm} \begin{gathered}
\begin{fmffile}{Diagrams/1PIEA_phi11}
\begin{fmfgraph*}(25,12)
\fmfleft{i0,i,i1}
\fmfright{o0,o,o1}
\fmfv{decor.shape=cross,decor.angle=45,decor.size=3.5thick,foreground=(1,,0,,0)}{o1}
\fmfv{decor.shape=circle,decor.filled=empty,decor.size=1.5thick,label.dist=0.15cm,label=$\alpha$}{o0}
\fmf{phantom,tension=10}{i,i1bis}
\fmf{phantom,tension=10}{o,o1bis}
\fmf{plain,left,tension=0.5,foreground=(1,,0,,0)}{i1bis,v1,i1bis}
\fmf{phantom,right,tension=0.5}{o1bis,v2,o1bis}
\fmf{dashes,tension=0,foreground=(1,,0,,0)}{v2,o1}
\fmf{plain,tension=0,foreground=(1,,0,,0)}{v2,o0}
\fmf{dots,foreground=(0,,0,,1)}{v1,v2}
\end{fmfgraph*}
\end{fmffile}
\end{gathered}
+ 2 \hspace{0.2cm}\begin{gathered}
\begin{fmffile}{Diagrams/1PIEA_phi12}
\begin{fmfgraph*}(22,12)
\fmfleft{i0,i,i1}
\fmfright{o0,o,o1}
\fmfv{decor.shape=cross,decor.angle=65,decor.size=3.5thick,foreground=(1,,0,,0)}{o1}
\fmfv{decor.shape=circle,decor.filled=empty,decor.size=1.5thick,label.dist=0.15cm,label=$\alpha$}{i1}
\fmf{phantom,tension=11}{i,v1}
\fmf{phantom,tension=11}{v2,o}
\fmf{plain,tension=0,foreground=(1,,0,,0)}{v1,i1}
\fmf{dashes,tension=0,foreground=(1,,0,,0)}{v2,o1}
\fmf{plain,right,tension=0.4,foreground=(1,,0,,0)}{v1,v2}
\fmf{dots,tension=4.0,foreground=(0,,0,,1)}{v1,v2}
\end{fmfgraph*}
\end{fmffile}
\end{gathered} \left.\rule{0cm}{1.2cm}\right)\;,
\label{eq:pure1PIEADeterminationJ1step20DON}
\end{equation}
with
\begin{equation}
\begin{gathered}
\begin{fmffile}{Diagrams/1PIEA_FeynRuleJn}
\begin{fmfgraph*}(6,4)
\fmfleft{i1}
\fmfright{o1}
\fmfv{decor.shape=circle,decor.filled=empty,decor.size=0.4cm,label=$n$,label.dist=0}{v1}
\fmfv{label=$\alpha$,label.angle=-90,label.dist=7}{v2}
\fmf{plain,tension=0.5,foreground=(1,,0,,0)}{i1,v1}
\fmf{phantom}{v1,o1}
\fmf{phantom,tension=0.5}{i1,v2}
\fmf{phantom}{v2,o1}
\end{fmfgraph*}
\end{fmffile}
\end{gathered} \hspace{0.1cm} \rightarrow J_{n,\alpha}\Big[\vec{\phi}\Big]\;.
\label{eq:pure1PIEAfeynRuleJn0DON}
\end{equation}
By introducing the inverse propagator:
\begin{equation}
\begin{gathered}
\begin{fmffile}{Diagrams/1PIEA_FeynRuleGJ0minus1}
\begin{fmfgraph*}(20,20)
\fmfleft{i0,i1,i2,i3}
\fmfright{o0,o1,o2,o3}
\fmftop{vUpL8,vUpL7,vUpL6,vUpL5,vUpL4,vUpL3,vUpL2,vUpL1,vUp,vUpR1,vUpR2,vUpR3,vUpR4,vUpR5,vUpR6,vUpR7,vUpR8}
\fmfbottom{vDownL8,vDownL7,vDownL6,vDownL5,vDownL4,vDownL3,vDownL2,vDownL1,vDown,vDownR1,vDownR2,vDownR3,vDownR4,vDownR5,vDownR6,vDownR7,vDownR8}
\fmf{phantom,tension=1.0}{vUpL1,vLeft}
\fmf{phantom,tension=1.5}{vDownL1,vLeft}
\fmf{phantom,tension=1.5}{vUpR1,vRight}
\fmf{phantom,tension=1.0}{vDownR1,vRight}
\fmfv{label=$\alpha_{1}$}{v1}
\fmfv{label=$\alpha_{2}$}{v2}
\fmf{phantom}{i1,v1}
\fmf{phantom}{i2,v1}
\fmf{plain,tension=0.6,foreground=(1,,0,,0)}{v1,v2}
\fmf{phantom}{v2,o1}
\fmf{phantom}{v2,o2}
\fmf{plain,tension=0,foreground=(1,,0,,0)}{vLeft,vRight}
\end{fmfgraph*}
\end{fmffile}
\end{gathered} \quad \rightarrow \boldsymbol{G}^{-1}_{\phi,\alpha_{1}\alpha_{2}}\Big[\vec{\phi}\Big]\;,
\label{eq:pure1PIEAinversepropagator0DON}
\end{equation}
we can isolate $\vec{J}_{1}$ in~\eqref{eq:pure1PIEADeterminationJ1step20DON} by multiplying the latter equation by $\boldsymbol{G}_{\phi}^{-1}$ and integrating over the relevant index. Indeed, the term involving $\vec{J}_{1}$ becomes in this way:
\begin{equation}
\begin{gathered}
\begin{fmffile}{Diagrams/1PIEA_DeterminationJ12}
\begin{fmfgraph*}(25,15)
\fmfleft{i0,i1,i2,i3}
\fmfright{o0,o1,o2,o3}
\fmftop{vUp}
\fmfbottom{vDown}
\fmfv{decor.shape=circle,decor.filled=empty,decor.size=1.5thick,label=$\alpha_{1}$}{v1}
\fmfv{decor.shape=circle,decor.filled=empty,decor.size=1.5thick}{vMiddle}
\fmfv{decor.shape=circle,decor.filled=empty,decor.size=0.4cm,label=$1$,label.dist=0}{v2}
\fmf{phantom,tension=3.4}{i1,vSlashDown}
\fmf{phantom,tension=1.0}{o1,vSlashDown}
\fmf{phantom,tension=2.0}{i2,vSlashUp}
\fmf{phantom,tension=1.0}{o2,vSlashUp}
\fmf{plain,tension=0,foreground=(1,,0,,0)}{vSlashDown,vSlashUp}
\fmf{phantom}{vUp,vMiddle}
\fmf{phantom}{vDown,vMiddle}
\fmf{phantom}{i1,v1}
\fmf{phantom}{i2,v1}
\fmf{plain,tension=0.1,foreground=(1,,0,,0)}{v1,vMiddle}
\fmf{plain,tension=0.1,foreground=(1,,0,,0)}{vMiddle,v2}
\fmf{phantom}{v2,o1}
\fmf{phantom}{v2,o2}
\end{fmfgraph*}
\end{fmffile}
\end{gathered} \hspace{0.2cm} =\int_{\alpha_{2},\alpha_{3}} \boldsymbol{G}^{-1}_{\phi,\alpha_{1}\alpha_{2}}\Big[\vec{\phi}\Big] \boldsymbol{G}_{\phi,\alpha_{2}\alpha_{3}}\Big[\vec{\phi}\Big] J_{1,\alpha_{3}}\Big[\vec{\phi}\Big]=\int_{\alpha_{3}} \delta_{\alpha_{1}\alpha_{3}} J_{1,\alpha_{3}}\Big[\vec{\phi}\Big] = J_{1,\alpha_{1}}\Big[\vec{\phi}\Big]\;.
\label{eq:pure1PIEADeterminationJ1step30DON}
\end{equation}
Therefore,~\eqref{eq:pure1PIEADeterminationJ1step20DON} is rewritten as:
\begin{equation}
J_{1,\alpha}\Big[\vec{\phi}\Big] = J_{1,a,x}\Big[\vec{\phi}\Big] = \frac{1}{6} \hspace{0.08cm} \begin{gathered}
\begin{fmffile}{Diagrams/1PIEA_J11}
\begin{fmfgraph*}(25,12)
\fmfleft{i0,i,i1}
\fmfright{o0,o,o1}
\fmftop{vUp}
\fmfbottom{vDown}
\fmfv{decor.shape=cross,decor.angle=45,decor.size=3.5thick,foreground=(1,,0,,0)}{o1}
\fmfv{decor.shape=circle,decor.filled=empty,decor.size=1.5thick,label.dist=0.15cm,label=$a$}{v2}
\fmf{phantom,tension=1.0}{i,vBis}
\fmf{phantom,tension=2.3}{vBis,o}
\fmf{phantom,tension=0.4}{vDown,vBis}
\fmf{phantom,tension=10}{i,i1bis}
\fmf{phantom,tension=10}{o,o1bis}
\fmf{plain,left,tension=0.5,foreground=(1,,0,,0)}{i1bis,v1,i1bis}
\fmf{phantom,right,tension=0.5}{o1bis,v2,o1bis}
\fmf{dashes,tension=0,foreground=(1,,0,,0)}{v2,o1}
\fmf{dots,label=$x$,label.dist=0.1cm,foreground=(0,,0,,1)}{v1,v2}
\end{fmfgraph*}
\end{fmffile}
\end{gathered}
+ \frac{1}{3} \begin{gathered}
\begin{fmffile}{Diagrams/1PIEA_J12}
\begin{fmfgraph*}(22,12)
\fmfleft{i0,i,i1}
\fmfright{o0,o,o1}
\fmftop{vUp}
\fmfbottom{vDown}
\fmfv{decor.shape=cross,decor.angle=65,decor.size=3.5thick,foreground=(1,,0,,0)}{o1}
\fmfv{decor.shape=circle,decor.filled=empty,decor.size=1.5thick,label.dist=0.15cm,label=$a$}{v1}
\fmf{phantom,tension=4.5}{i,vBis}
\fmf{phantom,tension=1.0}{vBis,o}
\fmf{phantom,tension=0.8}{vUp,vBis}
\fmf{phantom,tension=11}{i,v1}
\fmf{phantom,tension=11}{v2,o}
\fmf{dashes,tension=0,foreground=(1,,0,,0)}{v2,o1}
\fmf{plain,right,tension=0.4,foreground=(1,,0,,0)}{v1,v2}
\fmf{dots,label=$x$,label.dist=0.1cm,tension=4.0,foreground=(0,,0,,1)}{v1,v2}
\end{fmfgraph*}
\end{fmffile}
\end{gathered} \;.
\label{eq:pure1PIEADeterminationJ1step40DON}
\end{equation}
We are now able to determine a diagrammatic expression for $\Gamma_{2}^{(\mathrm{1PI})}$ from~\eqref{eq:pure1PIEAIMGamma20DON}. For that purpose, let us focus on each term of~\eqref{eq:pure1PIEAIMGamma20DON} separately. From~\eqref{eq:pure1PIEADerivW0j0j00DON},~\eqref{eq:pure1PIEADerivW1j0step20DON} and~\eqref{eq:pure1PIEADeterminationJ1step40DON}, it follows that:
\begin{equation}
\begin{split}
\int_{\alpha} \left.\frac{\delta W_{1}\big[\vec{J}\big]}{\delta J_{\alpha}}\right|_{\vec{J}=\vec{J}_{0}} J_{1,\alpha}\Big[\vec{\phi}\Big] = & - \frac{1}{9} \ \ \begin{gathered}
\begin{fmffile}{Diagrams/1PIEA_1PRDiag1}
\begin{fmfgraph}(34,20)
\fmfleft{i}
\fmfright{o}
\fmfv{decor.shape=cross,decor.size=3.5thick,foreground=(1,,0,,0)}{i}
\fmfv{decor.shape=cross,decor.size=3.5thick,foreground=(1,,0,,0)}{o}
\fmf{dashes,tension=2.0,foreground=(1,,0,,0)}{i,v3}
\fmf{dashes,tension=2.0,foreground=(1,,0,,0)}{o,v4}
\fmf{plain,right,tension=0.7,foreground=(1,,0,,0)}{v2,v4}
\fmf{dots,left,tension=0.7,foreground=(0,,0,,1)}{v2,v4}
\fmf{plain,left,tension=0.7,foreground=(1,,0,,0)}{v1,v3}
\fmf{dots,right,tension=0.7,foreground=(0,,0,,1)}{v1,v3}
\fmf{plain,tension=1.5,foreground=(1,,0,,0)}{v1,v2}
\end{fmfgraph}
\end{fmffile}
\end{gathered}
\ -\frac{1}{9} \ \begin{gathered}
\begin{fmffile}{Diagrams/1PIEA_1PRDiag2}
\begin{fmfgraph}(35,18)
\fmfleft{i}
\fmfright{o}
\fmftop{vUp}
\fmfbottom{vDown}
\fmfv{decor.shape=cross,decor.size=3.5thick,foreground=(1,,0,,0)}{v3bis}
\fmfv{decor.shape=cross,decor.size=3.5thick,foreground=(1,,0,,0)}{o}
\fmf{phantom,tension=10}{i,i1}
\fmf{dashes,tension=1.2,foreground=(1,,0,,0)}{o,v4}
\fmf{phantom,tension=0.5}{v3bis,i}
\fmf{phantom,tension=2.7}{v3bis,vUp}
\fmf{dashes,tension=0.9,foreground=(1,,0,,0)}{v3,v3bis}
\fmf{phantom,tension=0.5}{v4bis,i}
\fmf{phantom,tension=2.7}{v4bis,vDown}
\fmf{phantom,tension=0.9}{v3,v4bis}
\fmf{plain,left,tension=0.5,foreground=(1,,0,,0)}{i1,v1,i1}
\fmf{plain,right,tension=0.5,foreground=(1,,0,,0)}{v2,v4}
\fmf{dots,left,tension=0.5,foreground=(0,,0,,1)}{v2,v4}
\fmf{dots,foreground=(0,,0,,1)}{v1,v3}
\fmf{plain,foreground=(1,,0,,0)}{v3,v2}
\end{fmfgraph}
\end{fmffile}
\end{gathered} \\
& - \frac{1}{36} \ \begin{gathered}
\begin{fmffile}{Diagrams/1PIEA_1PRDiag3}
\begin{fmfgraph}(40,18)
\fmfleft{i}
\fmfright{o}
\fmftop{vUp}
\fmfbottom{vDown}
\fmf{phantom,tension=1.0}{vUp,vUpbis}
\fmf{phantom,tension=1.0}{vDown,vDownbis}
\fmf{dashes,tension=0.5,foreground=(1,,0,,0)}{v3,vUpbis}
\fmf{phantom,tension=0.5}{v4,vUpbis}
\fmf{phantom,tension=0.5}{v3,vDownbis}
\fmf{dashes,tension=0.5,foreground=(1,,0,,0)}{v4,vDownbis}
\fmfv{decor.shape=cross,decor.angle=45,decor.size=3.5thick,foreground=(1,,0,,0)}{vUpbis}
\fmfv{decor.shape=cross,decor.angle=45,decor.size=3.5thick,foreground=(1,,0,,0)}{vDownbis}
\fmf{phantom,tension=10}{i,i1}
\fmf{phantom,tension=10}{o,o1}
\fmf{plain,left,tension=0.5,foreground=(1,,0,,0)}{i1,v1,i1}
\fmf{plain,right,tension=0.5,foreground=(1,,0,,0)}{o1,v2,o1}
\fmf{dots,tension=1.2,foreground=(0,,0,,1)}{v1,v3}
\fmf{plain,tension=0.6,foreground=(1,,0,,0)}{v3,v4}
\fmf{dots,tension=1.2,foreground=(0,,0,,1)}{v4,v2}
\end{fmfgraph}
\end{fmffile}
\end{gathered}\;,
\end{split}
\label{eq:pure1PIEAGamma2coeffstep10DON}
\end{equation}
\begin{equation}
\begin{split}
\frac{1}{2} \int_{\alpha_{1},\alpha_{2}} \left.\frac{\delta^{2} W_{0}\big[\vec{J}\big]}{\delta J_{\alpha_{1}} \delta J_{\alpha_{2}}}\right|_{\vec{J}=\vec{J}_{0}} J_{1,\alpha_{1}}\Big[\vec{\phi}\Big] J_{1,\alpha_{2}}\Big[\vec{\phi}\Big] = & \ \frac{1}{18} \ \ \begin{gathered}
\begin{fmffile}{Diagrams/1PIEA_1PRDiag1}
\begin{fmfgraph}(34,20)
\fmfleft{i}
\fmfright{o}
\fmfv{decor.shape=cross,decor.size=3.5thick,foreground=(1,,0,,0)}{i}
\fmfv{decor.shape=cross,decor.size=3.5thick,foreground=(1,,0,,0)}{o}
\fmf{dashes,tension=2.0,foreground=(1,,0,,0)}{i,v3}
\fmf{dashes,tension=2.0,foreground=(1,,0,,0)}{o,v4}
\fmf{plain,right,tension=0.7,foreground=(1,,0,,0)}{v2,v4}
\fmf{dots,left,tension=0.7,foreground=(0,,0,,1)}{v2,v4}
\fmf{plain,left,tension=0.7,foreground=(1,,0,,0)}{v1,v3}
\fmf{dots,right,tension=0.7,foreground=(0,,0,,1)}{v1,v3}
\fmf{plain,tension=1.5,foreground=(1,,0,,0)}{v1,v2}
\end{fmfgraph}
\end{fmffile}
\end{gathered}
\ +\frac{1}{18} \ \begin{gathered}
\begin{fmffile}{Diagrams/1PIEA_1PRDiag2}
\begin{fmfgraph}(35,18)
\fmfleft{i}
\fmfright{o}
\fmftop{vUp}
\fmfbottom{vDown}
\fmfv{decor.shape=cross,decor.size=3.5thick,foreground=(1,,0,,0)}{v3bis}
\fmfv{decor.shape=cross,decor.size=3.5thick,foreground=(1,,0,,0)}{o}
\fmf{phantom,tension=10}{i,i1}
\fmf{dashes,tension=1.2,foreground=(1,,0,,0)}{o,v4}
\fmf{phantom,tension=0.5}{v3bis,i}
\fmf{phantom,tension=2.7}{v3bis,vUp}
\fmf{dashes,tension=0.9,foreground=(1,,0,,0)}{v3,v3bis}
\fmf{phantom,tension=0.5}{v4bis,i}
\fmf{phantom,tension=2.7}{v4bis,vDown}
\fmf{phantom,tension=0.9}{v3,v4bis}
\fmf{plain,left,tension=0.5,foreground=(1,,0,,0)}{i1,v1,i1}
\fmf{plain,right,tension=0.5,foreground=(1,,0,,0)}{v2,v4}
\fmf{dots,left,tension=0.5,foreground=(0,,0,,1)}{v2,v4}
\fmf{dots,foreground=(0,,0,,1)}{v1,v3}
\fmf{plain,foreground=(1,,0,,0)}{v3,v2}
\end{fmfgraph}
\end{fmffile}
\end{gathered} \\
& + \frac{1}{72} \ \begin{gathered}
\begin{fmffile}{Diagrams/1PIEA_1PRDiag3}
\begin{fmfgraph}(40,18)
\fmfleft{i}
\fmfright{o}
\fmftop{vUp}
\fmfbottom{vDown}
\fmf{phantom,tension=1.0}{vUp,vUpbis}
\fmf{phantom,tension=1.0}{vDown,vDownbis}
\fmf{dashes,tension=0.5,foreground=(1,,0,,0)}{v3,vUpbis}
\fmf{phantom,tension=0.5}{v4,vUpbis}
\fmf{phantom,tension=0.5}{v3,vDownbis}
\fmf{dashes,tension=0.5,foreground=(1,,0,,0)}{v4,vDownbis}
\fmfv{decor.shape=cross,decor.angle=45,decor.size=3.5thick,foreground=(1,,0,,0)}{vUpbis}
\fmfv{decor.shape=cross,decor.angle=45,decor.size=3.5thick,foreground=(1,,0,,0)}{vDownbis}
\fmf{phantom,tension=10}{i,i1}
\fmf{phantom,tension=10}{o,o1}
\fmf{plain,left,tension=0.5,foreground=(1,,0,,0)}{i1,v1,i1}
\fmf{plain,right,tension=0.5,foreground=(1,,0,,0)}{o1,v2,o1}
\fmf{dots,tension=1.2,foreground=(0,,0,,1)}{v1,v3}
\fmf{plain,tension=0.6,foreground=(1,,0,,0)}{v3,v4}
\fmf{dots,tension=1.2,foreground=(0,,0,,1)}{v4,v2}
\end{fmfgraph}
\end{fmffile}
\end{gathered}\;.
\end{split}
\label{eq:pure1PIEAGamma2coeffstep20DON}
\end{equation}
By collecting~\eqref{eq:pure1PIEAGamma2coeffstep10DON},~\eqref{eq:pure1PIEAGamma2coeffstep20DON} as well as~\eqref{eq:pure1PIEAIMW00DON} to~\eqref{eq:pure1PIEAIMW20DON}, we can turn~\eqref{eq:pure1PIEAIMGamma00DON} to~\eqref{eq:pure1PIEAIMGamma20DON} into our final expression for $\Gamma^{(\mathrm{1PI})}$ up to order $\mathcal{O}\big(\hbar^{2}\big)$:
\begin{equation}
\begin{split}
\Gamma^{(\mathrm{1PI})}\Big[\vec{\phi}\Big] = & \ S\Big[\vec{\phi}\Big] -\frac{\hbar}{2}\mathrm{STr}\left[\ln\Big(\boldsymbol{G}_{\phi}\Big[\vec{\phi}\Big]\Big)\right] \\
& + \hbar^{2} \left(\rule{0cm}{1.2cm}\right. \frac{1}{24} \hspace{0.08cm} \begin{gathered}
\begin{fmffile}{Diagrams/1PIEA_Hartree}
\begin{fmfgraph}(30,20)
\fmfleft{i}
\fmfright{o}
\fmf{phantom,tension=10}{i,i1}
\fmf{phantom,tension=10}{o,o1}
\fmf{plain,left,tension=0.5,foreground=(1,,0,,0)}{i1,v1,i1}
\fmf{plain,right,tension=0.5,foreground=(1,,0,,0)}{o1,v2,o1}
\fmf{zigzag,foreground=(0,,0,,1)}{v1,v2}
\end{fmfgraph}
\end{fmffile}
\end{gathered}
+\frac{1}{12}\begin{gathered}
\begin{fmffile}{Diagrams/1PIEA_Fock}
\begin{fmfgraph}(15,15)
\fmfleft{i}
\fmfright{o}
\fmf{phantom,tension=11}{i,v1}
\fmf{phantom,tension=11}{v2,o}
\fmf{plain,left,tension=0.4,foreground=(1,,0,,0)}{v1,v2,v1}
\fmf{zigzag,foreground=(0,,0,,1)}{v1,v2}
\end{fmfgraph}
\end{fmffile}
\end{gathered}
- \frac{1}{18} \begin{gathered}
\begin{fmffile}{Diagrams/1PIEA_Diag1}
\begin{fmfgraph}(27,15)
\fmfleft{i}
\fmfright{o}
\fmftop{vUp}
\fmfbottom{vDown}
\fmfv{decor.shape=cross,decor.size=3.5thick,foreground=(1,,0,,0)}{v1}
\fmfv{decor.shape=cross,decor.size=3.5thick,foreground=(1,,0,,0)}{v2}
\fmf{phantom,tension=10}{i,i1}
\fmf{phantom,tension=10}{o,o1}
\fmf{phantom,tension=2.2}{vUp,v5}
\fmf{phantom,tension=2.2}{vDown,v6}
\fmf{phantom,tension=0.5}{v3,v4}
\fmf{phantom,tension=10.0}{i1,v1}
\fmf{phantom,tension=10.0}{o1,v2}
\fmf{dashes,tension=2.0,foreground=(0,,0,,1),foreground=(1,,0,,0)}{v1,v3}
\fmf{dots,left=0.4,tension=0.5,foreground=(0,,0,,1)}{v3,v5}
\fmf{plain,left=0.4,tension=0.5,foreground=(1,,0,,0)}{v5,v4}
\fmf{plain,right=0.4,tension=0.5,foreground=(1,,0,,0)}{v3,v6}
\fmf{dots,right=0.4,tension=0.5,foreground=(0,,0,,1)}{v6,v4}
\fmf{dashes,tension=2.0,foreground=(0,,0,,1),foreground=(1,,0,,0)}{v4,v2}
\fmf{plain,tension=0,foreground=(1,,0,,0)}{v5,v6}
\end{fmfgraph}
\end{fmffile}
\end{gathered} - \frac{1}{36} \hspace{-0.15cm} \begin{gathered}
\begin{fmffile}{Diagrams/1PIEA_Diag2}
\begin{fmfgraph}(25,20)
\fmfleft{i}
\fmfright{o}
\fmftop{vUp}
\fmfbottom{vDown}
\fmfv{decor.shape=cross,decor.angle=45,decor.size=3.5thick,foreground=(1,,0,,0)}{vUpbis}
\fmfv{decor.shape=cross,decor.angle=45,decor.size=3.5thick,foreground=(1,,0,,0)}{vDownbis}
\fmf{phantom,tension=0.8}{vUp,vUpbis}
\fmf{phantom,tension=0.8}{vDown,vDownbis}
\fmf{dashes,tension=0.5,foreground=(0,,0,,1),foreground=(1,,0,,0)}{v3,vUpbis}
\fmf{phantom,tension=0.5}{v4,vUpbis}
\fmf{phantom,tension=0.5}{v3,vDownbis}
\fmf{dashes,tension=0.5,foreground=(0,,0,,1),foreground=(1,,0,,0)}{v4,vDownbis}
\fmf{phantom,tension=11}{i,v1}
\fmf{phantom,tension=11}{v2,o}
\fmf{plain,left,tension=0.5,foreground=(1,,0,,0)}{v1,v2,v1}
\fmf{dots,tension=1.7,foreground=(0,,0,,1)}{v1,v3}
\fmf{plain,foreground=(1,,0,,0)}{v3,v4}
\fmf{dots,tension=1.7,foreground=(0,,0,,1)}{v4,v2}
\end{fmfgraph}
\end{fmffile}
\end{gathered} \hspace{-0.22cm} \left.\rule{0cm}{1.2cm}\right) \\
& + \mathcal{O}\big(\hbar^{3}\big)\;.
\end{split}
\label{eq:1PIEAfinalexpressionAppendix}
\end{equation}
We first notice that all 1PR graphs of $W_{2}\big[\vec{J}=\vec{J}_{0}\big]$ are canceled out by those of~\eqref{eq:pure1PIEAGamma2coeffstep10DON} and~\eqref{eq:pure1PIEAGamma2coeffstep20DON} in~\eqref{eq:pure1PIEAIMGamma20DON}, thus leaving~$\Gamma^{(\mathrm{1PI})}_{2}$ (and therefore $\Gamma^{(\mathrm{1PI})}$ up to order $\mathcal{O}(\hbar^{2})$) expressed in terms of 1PI diagrams only, as expected. The recipe of the IM outlined above to derive~\eqref{eq:1PIEAfinalexpressionAppendix} can be straightforwardly generalized to derive $\Gamma^{(\mathrm{1PI})}$ up to any desired order in $\hbar$: we have determined $\Gamma^{(\mathrm{1PI})}$ up to order $\mathcal{O}\big(\hbar^{2}\big)$ by calculating $\vec{\phi}_{1}$ and $\vec{J}_{1}$, which translates in general into $\Gamma^{(\mathrm{1PI})}$ is determined up to order $\mathcal{O}\big(\hbar^{n}\big)$ (with $n \geq 2$) via the IM from $\vec{\phi}_{1},\cdots,\vec{\phi}_{n-1}$ and $\vec{J}_{1},\cdots,\vec{J}_{n-1}$. In summary, the IM consists in ``inverting'' the relation $\phi_{\alpha}=\frac{\delta W[\vec{J}]}{\delta J_{\alpha}}$ to determine $\vec{J}\big[\vec{\phi}\big]$ and, as a result, the 1PI EA through~\eqref{eq:pure1PIEAdefinition0DONAppendix}.

\paragraph{$\lambda$-expansion:}

We then determine the 1PI EA via the IM with the coupling constant $\lambda$ instead of $\hbar$ as expansion parameter\footnote{We therefore set $\hbar=1$ while developing the formalism.}. This version of the IM has already been developed by Okumura for a $\varphi^{4}$-theory which does not exhibit any $O(N)$ symmetry~\cite{oku96}. As opposed to the $\hbar$-expansion treated earlier, the coefficients of the Schwinger functional $W\big[\vec{J}\big]$ are not already at our disposal thanks to the LE or other methods treated in previous sections. Therefore, we start by expanding $W\big[\vec{J}\big]$ for our $O(N)$-symmetric $\varphi^{4}$-theory in arbitrary dimensions as follows:
\begin{equation}
\begin{split}
e^{W\big[\vec{J}\big]} = & \int \mathcal{D}\vec{\widetilde{\varphi}} \ e^{-S\big[\vec{\widetilde{\varphi}}\big]+\int_{\alpha} J_{\alpha}\widetilde{\varphi}_{\alpha}} \\
= & \int \mathcal{D}\vec{\widetilde{\varphi}} \ e^{-\frac{1}{2}\int_{\alpha_{\scalebox{0.4}{1}},\alpha_{\scalebox{0.4}{2}}}\widetilde{\varphi}_{\alpha_{\scalebox{0.4}{1}}}\boldsymbol{G}^{-1}_{0,\alpha_{\scalebox{0.4}{1}}\alpha_{\scalebox{0.4}{2}}}\widetilde{\varphi}_{\alpha_{\scalebox{0.4}{2}}} -\frac{\lambda}{4!}\sum_{a_{\scalebox{0.4}{1}},a_{\scalebox{0.4}{2}}=1}^{N}\int_{x} \varphi^{2}_{a_{\scalebox{0.4}{1}},x} \varphi^{2}_{a_{\scalebox{0.4}{2}},x} +\int_{\alpha} J_{\alpha}\widetilde{\varphi}_{\alpha}} \\
= & \ e^{-\frac{\lambda}{4!}\sum_{a_{\scalebox{0.4}{1}},a_{\scalebox{0.4}{2}}=1}^{N}\int_{x} \left(\frac{\delta}{\delta J_{a_{\scalebox{0.4}{1}},x}}\right)^{2}\left(\frac{\delta}{\delta J_{a_{\scalebox{0.4}{2}},x}}\right)^{2}} \int \mathcal{D}\vec{\widetilde{\varphi}} \ e^{-\frac{1}{2}\int_{\alpha_{\scalebox{0.4}{1}},\alpha_{\scalebox{0.4}{2}}}\widetilde{\varphi}_{\alpha_{\scalebox{0.4}{1}}}\boldsymbol{G}^{-1}_{0,\alpha_{\scalebox{0.4}{1}}\alpha_{\scalebox{0.4}{2}}}\widetilde{\varphi}_{\alpha_{\scalebox{0.4}{2}}} +\int_{\alpha}J_{\alpha}\widetilde{\varphi}_{\alpha}}\;,
\end{split}
\label{eq:1PIEAlambdaWexpansionStep10DON}
\end{equation}
with the free propagator $\boldsymbol{G}_{0}$ defined by~\eqref{eq:1PIEAlambdaExppDefG0} recalled below:
\begin{equation}
\boldsymbol{G}^{-1}_{0,\alpha_{1}\alpha_{2}} = \left(-\nabla^{2}_{x_{1}} + m^{2}\right) \delta_{\alpha_{1}\alpha_{2}}\;.
\label{eq:G0propagator0Nmodel}
\end{equation}
We then complete the square in the last line of~\eqref{eq:1PIEAlambdaWexpansionStep10DON} by calculating:
\begin{equation}
\begin{split}
-\frac{1}{2}\int_{\alpha_{1},\alpha_{2}} & \left(\vec{\widetilde{\varphi}}-\boldsymbol{G}_{0}\vec{J}\right)_{\alpha_{1}}\boldsymbol{G}^{-1}_{0,\alpha_{1}\alpha_{2}}\left(\vec{\widetilde{\varphi}}-\boldsymbol{G}_{0}\vec{J}\right)_{\alpha_{2}} + \frac{1}{2} \int_{\alpha_{1},\alpha_{2}} J_{\alpha_{1}} \boldsymbol{G}_{0,\alpha_{1}\alpha_{2}} J_{\alpha_{2}} \\
= & -\int_{\alpha_{1},\alpha_{2}}\frac{1}{2}\widetilde{\varphi}_{\alpha_{1}} \boldsymbol{G}^{-1}_{0,\alpha_{1}\alpha_{2}} \widetilde{\varphi}_{\alpha_{2}} + \frac{1}{2} \int_{\alpha_{1},\alpha_{2},\alpha_{3}} J_{\alpha_{1}} \boldsymbol{G}_{0,\alpha_{1}\alpha_{2}} \boldsymbol{G}^{-1}_{0,\alpha_{2}\alpha_{3}} \widetilde{\varphi}_{\alpha_{3}} \\
& + \frac{1}{2} \int_{\alpha_{1},\alpha_{2},\alpha_{3}} \widetilde{\varphi}_{\alpha_{1}} \boldsymbol{G}^{-1}_{0,\alpha_{1}\alpha_{2}} \boldsymbol{G}_{0,\alpha_{2}\alpha_{3}} J_{\alpha_{3}} - \frac{1}{2} \int_{\alpha_{1},\alpha_{2},\alpha_{3},\alpha_{4}} J_{\alpha_{1}} \boldsymbol{G}_{0,\alpha_{1}\alpha_{2}} \boldsymbol{G}^{-1}_{0,\alpha_{2}\alpha_{3}} \boldsymbol{G}_{0,\alpha_{3}\alpha_{4}} J_{\alpha_{4}} \\
& + \frac{1}{2} \int_{\alpha_{1},\alpha_{2}} J_{\alpha_{1}} \boldsymbol{G}_{0,\alpha_{1}\alpha_{2}} J_{\alpha_{2}} \\
= & -\int_{\alpha_{1},\alpha_{2}}\frac{1}{2}\widetilde{\varphi}_{\alpha_{1}} \boldsymbol{G}^{-1}_{0,\alpha_{1}\alpha_{2}} \widetilde{\varphi}_{\alpha_{2}} + \frac{1}{2} \int_{\alpha_{1},\alpha_{3}} J_{\alpha_{1}} \underbrace{\int_{\alpha_{2}} \boldsymbol{G}_{0,\alpha_{1}\alpha_{2}} \boldsymbol{G}^{-1}_{0,\alpha_{2}\alpha_{3}}}_{\delta_{\alpha_{1}\alpha_{3}}} \widetilde{\varphi}_{\alpha_{3}} \\
& + \frac{1}{2} \int_{\alpha_{1},\alpha_{3}} \widetilde{\varphi}_{\alpha_{1}} \underbrace{\int_{\alpha_{2}} \boldsymbol{G}^{-1}_{0,\alpha_{1}\alpha_{2}} \boldsymbol{G}_{0,\alpha_{2}\alpha_{3}}}_{\delta_{\alpha_{1}\alpha_{3}}} J_{\alpha_{3}} - \frac{1}{2} \int_{\alpha_{1},\alpha_{4}} J_{\alpha_{1}} \underbrace{\int_{\alpha_{2},\alpha_{3}} \boldsymbol{G}_{0,\alpha_{1}\alpha_{2}} \boldsymbol{G}^{-1}_{0,\alpha_{2}\alpha_{3}} \boldsymbol{G}_{0,\alpha_{3}\alpha_{4}}}_{\boldsymbol{G}_{0,\alpha_{1}\alpha_{4}}} J_{\alpha_{4}} \\
& + \frac{1}{2} \int_{\alpha_{1},\alpha_{2}} J_{\alpha_{1}} \boldsymbol{G}_{0,\alpha_{1}\alpha_{2}} J_{\alpha_{2}} \\
= & -\frac{1}{2}\int_{\alpha_{1},\alpha_{2}}\widetilde{\varphi}_{\alpha_{1}}\boldsymbol{G}^{-1}_{0,\alpha_{1}\alpha_{2}}\widetilde{\varphi}_{\alpha_{2}} +\int_{\alpha}J_{\alpha}\widetilde{\varphi}_{\alpha}\;.
\end{split}
\label{eq:1PIEAlambdaSquareCompletion0DON}
\end{equation}
With the help of the latter result,~\eqref{eq:1PIEAlambdaWexpansionStep10DON} is rewritten as:
\begin{equation}
\begin{split}
\scalebox{0.98}{${\displaystyle e^{W\big[\vec{J}\big]} =}$} & \ \scalebox{0.98}{${\displaystyle e^{-\frac{\lambda}{4!}\sum_{a_{\scalebox{0.4}{1}},a_{\scalebox{0.4}{2}}=1}^{N}\int_{x} \left(\frac{\delta}{\delta J_{a_{\scalebox{0.4}{1}},x}}\right)^{2}\left(\frac{\delta}{\delta J_{a_{\scalebox{0.4}{2}},x}}\right)^{2}} \int \mathcal{D}\vec{\widetilde{\varphi}} \ e^{-\frac{1}{2}\int_{\alpha_{\scalebox{0.4}{1}},\alpha_{\scalebox{0.4}{2}}} \left(\vec{\widetilde{\varphi}}-\boldsymbol{G}_{0}\vec{J}\right)_{\alpha_{\scalebox{0.4}{1}}}\boldsymbol{G}^{-1}_{0,\alpha_{\scalebox{0.4}{1}}\alpha_{\scalebox{0.4}{2}}}\left(\vec{\widetilde{\varphi}}-\boldsymbol{G}_{0}\vec{J}\right)_{\alpha_{\scalebox{0.4}{2}}} + \frac{1}{2} \int_{\alpha_{\scalebox{0.4}{1}},\alpha_{\scalebox{0.4}{2}}} J_{\alpha_{\scalebox{0.4}{1}}} \boldsymbol{G}_{0,\alpha_{\scalebox{0.4}{1}}\alpha_{\scalebox{0.4}{2}}} J_{\alpha_{\scalebox{0.4}{2}}}} }$} \\
\scalebox{0.98}{${\displaystyle = }$} & \ \scalebox{0.98}{${\displaystyle e^{-\frac{\lambda}{4!}\sum_{a_{\scalebox{0.4}{1}},a_{\scalebox{0.4}{2}}=1}^{N}\int_{x} \left(\frac{\delta}{\delta J_{a_{\scalebox{0.4}{1}},x}}\right)^{2}\left(\frac{\delta}{\delta J_{a_{\scalebox{0.4}{2}},x}}\right)^{2}} e^{\frac{1}{2} \int_{\alpha_{\scalebox{0.4}{1}},\alpha_{\scalebox{0.4}{2}}} J_{\alpha_{\scalebox{0.4}{1}}} \boldsymbol{G}_{0,\alpha_{\scalebox{0.4}{1}}\alpha_{\scalebox{0.4}{2}}} J_{\alpha_{\scalebox{0.4}{2}}}} \int \mathcal{D}\vec{\widetilde{\varphi}} \ e^{-\frac{1}{2}\int_{\alpha_{\scalebox{0.4}{1}},\alpha_{\scalebox{0.4}{2}}} \left(\vec{\widetilde{\varphi}}-\boldsymbol{G}_{0}\vec{J}\right)_{\alpha_{\scalebox{0.4}{1}}}\boldsymbol{G}^{-1}_{0,\alpha_{\scalebox{0.4}{1}}\alpha_{\scalebox{0.4}{2}}}\left(\vec{\widetilde{\varphi}}-\boldsymbol{G}_{0}\vec{J}\right)_{\alpha_{\scalebox{0.4}{2}}}}\;. }$}
\end{split}
\label{eq:1PIEAlambdaWexpansionStep20DON}
\end{equation}
After the substitution $\vec{\widetilde{\varphi}} \rightarrow \vec{\widetilde{\varphi}}'=\vec{\widetilde{\varphi}}-\boldsymbol{G}_{0}\vec{J}$ (whose Jacobian is trivial, i.e. $\mathcal{D}\vec{\widetilde{\varphi}}'=\mathcal{D}\vec{\widetilde{\varphi}}$),~\eqref{eq:1PIEAlambdaWexpansionStep20DON} becomes:
\begin{equation}
\begin{split}
\scalebox{0.95}{${\displaystyle e^{W\big[\vec{J}\big]} =}$} & \ \scalebox{0.95}{${\displaystyle e^{-\frac{\lambda}{4!}\sum_{a_{\scalebox{0.4}{1}},a_{\scalebox{0.4}{2}}=1}^{N}\int_{x} \left(\frac{\delta}{\delta J_{a_{\scalebox{0.4}{1}},x}}\right)^{2}\left(\frac{\delta}{\delta J_{a_{\scalebox{0.4}{2}},x}}\right)^{2}} e^{\frac{1}{2} \int_{\alpha_{\scalebox{0.4}{1}},\alpha_{\scalebox{0.4}{2}}} J_{\alpha_{\scalebox{0.4}{1}}} \boldsymbol{G}_{0,\alpha_{\scalebox{0.4}{1}}\alpha_{\scalebox{0.4}{2}}} J_{\alpha_{\scalebox{0.4}{2}}}} \underbrace{\int \mathcal{D}\vec{\widetilde{\varphi}}' \ e^{-\frac{1}{2}\int_{\alpha_{\scalebox{0.4}{1}},\alpha_{\scalebox{0.4}{2}}} \widetilde{\varphi}'_{\alpha_{\scalebox{0.4}{1}}} \boldsymbol{G}^{-1}_{0,\alpha_{\scalebox{0.4}{1}}\alpha_{\scalebox{0.4}{2}}} \widetilde{\varphi}'_{\alpha_{\scalebox{0.4}{2}}}}}_{e^{\frac{1}{2}\mathrm{STr}[\ln(\boldsymbol{G}_{0})]}} }$} \\
\scalebox{0.95}{${\displaystyle = }$} & \ \scalebox{0.95}{${\displaystyle e^{\frac{1}{2}\mathrm{STr}[\ln(\boldsymbol{G}_{0})]} e^{-\frac{\lambda}{4!}\sum_{a_{\scalebox{0.4}{1}},a_{\scalebox{0.4}{2}}=1}^{N}\int_{x} \left(\frac{\delta}{\delta J_{a_{\scalebox{0.4}{1}},x}}\right)^{2}\left(\frac{\delta}{\delta J_{a_{\scalebox{0.4}{2}},x}}\right)^{2}} e^{\frac{1}{2} \int_{\alpha,\alpha'} J_{\alpha} \boldsymbol{G}_{0,\alpha\alpha'} J_{\alpha'}} }$} \\
\scalebox{0.95}{${\displaystyle = }$} & \ \scalebox{0.95}{${\displaystyle e^{\frac{1}{2}\mathrm{STr}[\ln(\boldsymbol{G}_{0})]} \sum^{\infty}_{n=0} \frac{(-1)^{n}}{n!} \left(\frac{\lambda}{4!}\right)^{n} }$} \\
& \scalebox{0.95}{${\displaystyle \times \sum_{a_{1},\cdots,a_{2n}=1}^{N} \int_{x_{1},\cdots,x_{n}} \left(\frac{\delta}{\delta J_{a_{1},x_{1}}}\right)^{2} \left(\frac{\delta}{\delta J_{a_{2},x_{1}}}\right)^{2} \cdots \left(\frac{\delta}{\delta J_{a_{2n-1},x_{n}}}\right)^{2} \left(\frac{\delta}{\delta J_{a_{2n},x_{n}}}\right)^{2} e^{\frac{1}{2} \int_{\alpha,\alpha'} J_{\alpha} \boldsymbol{G}_{0,\alpha\alpha'} J_{\alpha'}}\;. }$}
\end{split}
\label{eq:1PIEAlambdaWexpansionStep30DON}
\end{equation}
A diagrammatic representation of $W\big[\vec{J}\big]$ follows by carrying out the functional derivatives with respect to $\vec{J}$ in the last line. In this way, we find:
\begin{equation}
\begin{split}
\scalebox{0.93}{${\displaystyle \lambda\left(\frac{\delta}{\delta J_{a_{1},x}}\right)^{2} \left(\frac{\delta}{\delta J_{a_{2},x}}\right)^{2} e^{\frac{1}{2} \int_{\alpha,\alpha'} J_{\alpha} \boldsymbol{G}_{0,\alpha\alpha'} J_{\alpha'}} = }$} & \ \scalebox{0.93}{${\displaystyle \left(\rule{0cm}{1.2cm}\right.  \hspace{0.08cm} \begin{gathered}
\begin{fmffile}{Diagrams/1PIEAlambda_Hartree1}
\begin{fmfgraph*}(30,20)
\fmfleft{i}
\fmfright{o}
\fmfv{decor.shape=circle,decor.filled=empty,decor.size=1.5thick,label.dist=0.1cm,label=$a_{1}$}{v1}
\fmfv{decor.shape=circle,decor.filled=empty,decor.size=1.5thick,label.dist=0.15cm,label=$a_{2}$}{v2}
\fmf{phantom,tension=10}{i,i1}
\fmf{phantom,tension=10}{o,o1}
\fmf{plain,left,tension=0.5}{i1,v1,i1}
\fmf{plain,right,tension=0.5}{o1,v2,o1}
\fmf{zigzag,label=$x$,foreground=(0,,0,,1)}{v1,v2}
\end{fmfgraph*}
\end{fmffile}
\end{gathered}
+ 2 \hspace{0.5cm} \begin{gathered}
\begin{fmffile}{Diagrams/1PIEAlambda_Fock1}
\begin{fmfgraph*}(15,15)
\fmfleft{i}
\fmfright{o}
\fmfv{decor.shape=circle,decor.filled=empty,decor.size=1.5thick,label.dist=0.1cm,label=$a_{1}$}{v1}
\fmfv{decor.shape=circle,decor.filled=empty,decor.size=1.5thick,label.dist=0.15cm,label=$a_{2}$}{v2}
\fmf{phantom,tension=11}{i,v1}
\fmf{phantom,tension=11}{v2,o}
\fmf{plain,left,tension=0.4}{v1,v2,v1}
\fmf{zigzag,label=$x$,foreground=(0,,0,,1)}{v1,v2}
\end{fmfgraph*}
\end{fmffile}
\end{gathered}
\hspace{0.4cm} + \begin{gathered}
\begin{fmffile}{Diagrams/1PIEAlambda_Diag1}
\begin{fmfgraph*}(30,15)
\fmfleft{i}
\fmfright{oDown,o,oUp}
\fmfv{decor.shape=circle,decor.filled=empty,decor.size=1.5thick,label.dist=0.1cm,label=$a_{1}$}{v1}
\fmfv{decor.shape=circle,decor.filled=empty,decor.size=1.5thick,label.dist=0.15cm,label=$a_{2}$}{v2}
\fmfv{decor.shape=circle,decor.filled=empty,decor.size=0.25cm,l=$\times$,label.dist=0}{oUp}
\fmfv{decor.shape=circle,decor.filled=empty,decor.size=0.25cm,l=$\times$,label.dist=0}{oDown}
\fmf{phantom,tension=10}{i,i1}
\fmf{phantom,tension=10}{o,o1}
\fmf{plain,left,tension=0.5}{i1,v1,i1}
\fmf{phantom,right,tension=0.5}{o1,v2,o1}
\fmf{plain,tension=0}{v2,oUp}
\fmf{plain,tension=0}{v2,oDown}
\fmf{zigzag,label=$x$,foreground=(0,,0,,1)}{v1,v2}
\end{fmfgraph*}
\end{fmffile}
\end{gathered} }$} \\
& \hspace{0.4cm} \scalebox{0.93}{${\displaystyle + \begin{gathered}
\begin{fmffile}{Diagrams/1PIEAlambda_Diag2}
\begin{fmfgraph*}(30,15)
\fmfleft{iDown,i,iUp}
\fmfright{o}
\fmfv{decor.shape=circle,decor.filled=empty,decor.size=1.5thick,label.dist=0.15cm,label=$a_{1}$}{v1}
\fmfv{decor.shape=circle,decor.filled=empty,decor.size=1.5thick,label.dist=0.15cm,label=$a_{2}$}{v2}
\fmfv{decor.shape=circle,decor.filled=empty,decor.size=0.25cm,l=$\times$,label.dist=0}{iUp}
\fmfv{decor.shape=circle,decor.filled=empty,decor.size=0.25cm,l=$\times$,label.dist=0}{iDown}
\fmf{phantom,tension=10}{i,i1}
\fmf{phantom,tension=10}{o,o1}
\fmf{phantom,left,tension=0.5}{i1,v1,i1}
\fmf{plain,right,tension=0.5}{o1,v2,o1}
\fmf{plain,tension=0}{v1,iUp}
\fmf{plain,tension=0}{v1,iDown}
\fmf{zigzag,label=$x$,foreground=(0,,0,,1)}{v1,v2}
\end{fmfgraph*}
\end{fmffile}
\end{gathered} + 4 \hspace{0.5cm} \begin{gathered}
\begin{fmffile}{Diagrams/1PIEAlambda_Diag3}
\begin{fmfgraph*}(15,15)
\fmfleft{iDown,i,iUp}
\fmfright{oDown,o,oUp}
\fmfv{decor.shape=circle,decor.filled=empty,decor.size=1.5thick,label.dist=0.1cm,label=$a_{1}$}{v1}
\fmfv{decor.shape=circle,decor.filled=empty,decor.size=1.5thick,label.dist=0.15cm,label=$a_{2}$}{v2}
\fmfv{decor.shape=circle,decor.filled=empty,decor.size=0.25cm,l=$\times$,label.dist=0}{iUp}
\fmfv{decor.shape=circle,decor.filled=empty,decor.size=0.25cm,l=$\times$,label.dist=0}{oUp}
\fmf{plain,tension=0}{v1,iUp}
\fmf{plain,tension=0}{v2,oUp}
\fmf{phantom,tension=11}{i,v1}
\fmf{phantom,tension=11}{v2,o}
\fmf{plain,right,tension=0.4}{v1,v2}
\fmf{zigzag,label=$x$,foreground=(0,,0,,1)}{v1,v2}
\end{fmfgraph*}
\end{fmffile}
\end{gathered} \hspace{0.4cm} + \begin{gathered}
\begin{fmffile}{Diagrams/1PIEAlambda_Diag4}
\begin{fmfgraph*}(30,15)
\fmfleft{iDown,i,iUp}
\fmfright{oDown,o,oUp}
\fmfv{decor.shape=circle,decor.filled=empty,decor.size=1.5thick,label.dist=0.15cm,label=$a_{1}$}{v1}
\fmfv{decor.shape=circle,decor.filled=empty,decor.size=1.5thick,label.dist=0.15cm,label=$a_{2}$}{v2}
\fmfv{decor.shape=circle,decor.filled=empty,decor.size=0.25cm,l=$\times$,label.dist=0}{iUp}
\fmfv{decor.shape=circle,decor.filled=empty,decor.size=0.25cm,l=$\times$,label.dist=0}{iDown}
\fmfv{decor.shape=circle,decor.filled=empty,decor.size=0.25cm,l=$\times$,label.dist=0}{oUp}
\fmfv{decor.shape=circle,decor.filled=empty,decor.size=0.25cm,l=$\times$,label.dist=0}{oDown}
\fmf{phantom,tension=10}{i,i1}
\fmf{phantom,tension=10}{o,o1}
\fmf{phantom,left,tension=0.5}{i1,v1,i1}
\fmf{phantom,right,tension=0.5}{o1,v2,o1}
\fmf{plain,tension=0}{v1,iUp}
\fmf{plain,tension=0}{v1,iDown}
\fmf{plain,tension=0}{v2,oUp}
\fmf{plain,tension=0}{v2,oDown}
\fmf{zigzag,label=$x$,foreground=(0,,0,,1)}{v1,v2}
\end{fmfgraph*}
\end{fmffile}
\end{gathered} \left.\rule{0cm}{1.2cm}\right) }$} \\
& \hspace{0.2cm} \scalebox{0.93}{${\displaystyle \times e^{\frac{1}{2} \int_{\alpha,\alpha'} J_{\alpha} \boldsymbol{G}_{0,\alpha\alpha'} J_{\alpha'}}\;,}$}
\end{split}
\label{eq:1PIEAlambdaWexpansionStep40DON}
\end{equation}
with the Feynman rules:
\begin{subequations}
\begin{align}
\begin{gathered}
\begin{fmffile}{Diagrams/LoopExpansion1_FeynRuleGbis_Appendix}
\begin{fmfgraph*}(20,20)
\fmfleft{i0,i1,i2,i3}
\fmfright{o0,o1,o2,o3}
\fmflabel{$\alpha_{1}$}{v1}
\fmflabel{$\alpha_{2}$}{v2}
\fmf{phantom}{i1,v1}
\fmf{phantom}{i2,v1}
\fmf{plain,tension=0.6}{v1,v2}
\fmf{phantom}{v2,o1}
\fmf{phantom}{v2,o2}
\end{fmfgraph*}
\end{fmffile}
\end{gathered} \quad &\rightarrow \quad \boldsymbol{G}_{0,\alpha_{1}\alpha_{2}}\;,
\label{eq:1PIEAlambdapropagator0DON} \\
\begin{gathered}
\begin{fmffile}{Diagrams/LoopExpansionBosonicHS_FeynRuleK}
\begin{fmfgraph*}(10,5)
\fmfleft{i1}
\fmfright{o1}
\fmfv{decor.shape=circle,decor.filled=empty,decor.size=.26w,l=$\times$,label.dist=0}{v1}
\fmfv{label=$\alpha$,label.angle=-90,label.dist=6}{v2}
\fmf{plain,tension=2.5}{i1,v1}
\fmf{phantom}{v1,o1}
\fmf{phantom,tension=2.5}{i1,v2}
\fmf{phantom}{v2,o1}
\end{fmfgraph*}
\end{fmffile}
\end{gathered} \quad &\rightarrow \quad J_{\alpha} \;,
\label{eq:1PIEAlambdaSource0DON} \\
\begin{gathered}
\begin{fmffile}{Diagrams/LoopExpansion1_FeynRuleV4bis_Appendix}
\begin{fmfgraph*}(20,20)
\fmfleft{i0,i1,i2,i3}
\fmfright{o0,o1,o2,o3}
\fmf{phantom,tension=2.0}{i1,i1bis}
\fmf{plain,tension=2.0}{i1bis,v1}
\fmf{phantom,tension=2.0}{i2,i2bis}
\fmf{plain,tension=2.0}{i2bis,v1}
\fmf{zigzag,label=$x$,tension=0.6,foreground=(0,,0,,1)}{v1,v2}
\fmf{phantom,tension=2.0}{o1bis,o1}
\fmf{plain,tension=2.0}{v2,o1bis}
\fmf{phantom,tension=2.0}{o2bis,o2}
\fmf{plain,tension=2.0}{v2,o2bis}
\fmflabel{$a_{1}$}{i1bis}
\fmflabel{$a_{2}$}{i2bis}
\fmflabel{$a_{3}$}{o1bis}
\fmflabel{$a_{4}$}{o2bis}
\end{fmfgraph*}
\end{fmffile}
\end{gathered} \quad &\rightarrow \quad \lambda\delta_{a_{1} a_{2}}\delta_{a_{3} a_{4}}\;,
\label{eq:1PIEAlambda4legvertex0DON}
\end{align}
\end{subequations}
and external points are still indicated by empty dots. If we combine~\eqref{eq:1PIEAlambdaWexpansionStep40DON} with~\eqref{eq:1PIEAlambdaWexpansionStep30DON}, we obtain the expansion of $W\big[\vec{J}\big]$ up to order $\mathcal{O}(\lambda)$:
\begin{equation}
\begin{split}
e^{W\big[\vec{J}\big]} = & \ e^{\frac{1}{2}\mathrm{STr}[\ln(\boldsymbol{G}_{0})]+\frac{1}{2}\int_{\alpha_{\scalebox{0.4}{1}},\alpha_{\scalebox{0.4}{2}}}J_{\alpha_{\scalebox{0.4}{1}}}\boldsymbol{G}_{0,\alpha_{\scalebox{0.4}{1}}\alpha_{\scalebox{0.4}{2}}}J_{\alpha_{\scalebox{0.4}{2}}}} \\
& \times \left[\rule{0cm}{1.2cm}\right. 1 - \left(\rule{0cm}{1.2cm}\right. \frac{1}{24} \hspace{0.08cm} \begin{gathered}
\begin{fmffile}{Diagrams/LoopExpansion1_Hartree}
\begin{fmfgraph}(30,20)
\fmfleft{i}
\fmfright{o}
\fmf{phantom,tension=10}{i,i1}
\fmf{phantom,tension=10}{o,o1}
\fmf{plain,left,tension=0.5}{i1,v1,i1}
\fmf{plain,right,tension=0.5}{o1,v2,o1}
\fmf{zigzag,foreground=(0,,0,,1)}{v1,v2}
\end{fmfgraph}
\end{fmffile}
\end{gathered}
+\frac{1}{12}\begin{gathered}
\begin{fmffile}{Diagrams/LoopExpansion1_Fock}
\begin{fmfgraph}(15,15)
\fmfleft{i}
\fmfright{o}
\fmf{phantom,tension=11}{i,v1}
\fmf{phantom,tension=11}{v2,o}
\fmf{plain,left,tension=0.4}{v1,v2,v1}
\fmf{zigzag,foreground=(0,,0,,1)}{v1,v2}
\end{fmfgraph}
\end{fmffile}
\end{gathered} + \frac{1}{12} \ \begin{gathered}
\begin{fmffile}{Diagrams/1PIEAlambda_Diag5}
\begin{fmfgraph*}(30,15)
\fmfleft{i}
\fmfright{oDown,o,oUp}
\fmfv{decor.shape=circle,decor.filled=empty,decor.size=0.25cm,l=$\times$,label.dist=0}{oUp}
\fmfv{decor.shape=circle,decor.filled=empty,decor.size=0.25cm,l=$\times$,label.dist=0}{oDown}
\fmf{phantom,tension=10}{i,i1}
\fmf{phantom,tension=10}{o,o1}
\fmf{plain,left,tension=0.5}{i1,v1,i1}
\fmf{phantom,right,tension=0.5}{o1,v2,o1}
\fmf{plain,tension=0}{v2,oUp}
\fmf{plain,tension=0}{v2,oDown}
\fmf{zigzag,foreground=(0,,0,,1)}{v1,v2}
\end{fmfgraph*}
\end{fmffile}
\end{gathered} + \frac{1}{6} \ \begin{gathered}
\begin{fmffile}{Diagrams/1PIEAlambda_Diag6}
\begin{fmfgraph*}(15,15)
\fmfleft{iDown,i,iUp}
\fmfright{oDown,o,oUp}
\fmfv{decor.shape=circle,decor.filled=empty,decor.size=0.25cm,l=$\times$,label.dist=0}{iUp}
\fmfv{decor.shape=circle,decor.filled=empty,decor.size=0.25cm,l=$\times$,label.dist=0}{oUp}
\fmf{plain,tension=0}{v1,iUp}
\fmf{plain,tension=0}{v2,oUp}
\fmf{phantom,tension=11}{i,v1}
\fmf{phantom,tension=11}{v2,o}
\fmf{plain,right,tension=0.4}{v1,v2}
\fmf{zigzag,foreground=(0,,0,,1)}{v1,v2}
\end{fmfgraph*}
\end{fmffile}
\end{gathered} \\
& \hspace{1.8cm} + \frac{1}{24} \ \begin{gathered}
\begin{fmffile}{Diagrams/1PIEAlambda_Diag7}
\begin{fmfgraph*}(30,15)
\fmfleft{iDown,i,iUp}
\fmfright{oDown,o,oUp}
\fmfv{decor.shape=circle,decor.filled=empty,decor.size=0.25cm,l=$\times$,label.dist=0}{iUp}
\fmfv{decor.shape=circle,decor.filled=empty,decor.size=0.25cm,l=$\times$,label.dist=0}{iDown}
\fmfv{decor.shape=circle,decor.filled=empty,decor.size=0.25cm,l=$\times$,label.dist=0}{oUp}
\fmfv{decor.shape=circle,decor.filled=empty,decor.size=0.25cm,l=$\times$,label.dist=0}{oDown}
\fmf{phantom,tension=10}{i,i1}
\fmf{phantom,tension=10}{o,o1}
\fmf{phantom,left,tension=0.5}{i1,v1,i1}
\fmf{phantom,right,tension=0.5}{o1,v2,o1}
\fmf{plain,tension=0}{v1,iUp}
\fmf{plain,tension=0}{v1,iDown}
\fmf{plain,tension=0}{v2,oUp}
\fmf{plain,tension=0}{v2,oDown}
\fmf{zigzag,foreground=(0,,0,,1)}{v1,v2}
\end{fmfgraph*}
\end{fmffile}
\end{gathered} + \mathcal{O}\big(\lambda^{2}\big) \left.\rule{0cm}{1.2cm}\right) \left.\rule{0cm}{1.2cm}\right]\;,
\end{split}
\label{eq:1PIEAlambdaWexpansionStep50DON}
\end{equation}
which, after taking the logarithm, gives us:
\begin{equation}
\begin{split}
W\Big[\vec{J}\Big] = & \ \frac{1}{2}\mathrm{STr}[\ln(\boldsymbol{G}_{0})]+\frac{1}{2} \begin{gathered}
\begin{fmffile}{Diagrams/1PIEAlambda_Diag0}
\begin{fmfgraph*}(20,20)
\fmfleft{i0,i1,i2,i3}
\fmfright{o0,o1,o2,o3}
\fmfv{decor.shape=circle,decor.filled=empty,decor.size=0.25cm,l=$\times$,label.dist=0}{v1}
\fmfv{decor.shape=circle,decor.filled=empty,decor.size=0.25cm,l=$\times$,label.dist=0}{v2}
\fmf{phantom}{i1,v1}
\fmf{phantom}{i2,v1}
\fmf{plain,tension=0.6}{v1,v2}
\fmf{phantom}{v2,o1}
\fmf{phantom}{v2,o2}
\end{fmfgraph*}
\end{fmffile}
\end{gathered} \\
& -\frac{1}{24} \hspace{0.08cm} \begin{gathered}
\begin{fmffile}{Diagrams/LoopExpansion1_Hartree}
\begin{fmfgraph}(30,20)
\fmfleft{i}
\fmfright{o}
\fmf{phantom,tension=10}{i,i1}
\fmf{phantom,tension=10}{o,o1}
\fmf{plain,left,tension=0.5}{i1,v1,i1}
\fmf{plain,right,tension=0.5}{o1,v2,o1}
\fmf{zigzag,foreground=(0,,0,,1)}{v1,v2}
\end{fmfgraph}
\end{fmffile}
\end{gathered}
-\frac{1}{12}\begin{gathered}
\begin{fmffile}{Diagrams/LoopExpansion1_Fock}
\begin{fmfgraph}(15,15)
\fmfleft{i}
\fmfright{o}
\fmf{phantom,tension=11}{i,v1}
\fmf{phantom,tension=11}{v2,o}
\fmf{plain,left,tension=0.4}{v1,v2,v1}
\fmf{zigzag,foreground=(0,,0,,1)}{v1,v2}
\end{fmfgraph}
\end{fmffile}
\end{gathered} - \frac{1}{12} \ \begin{gathered}
\begin{fmffile}{Diagrams/1PIEAlambda_Diag5}
\begin{fmfgraph*}(30,15)
\fmfleft{i}
\fmfright{oDown,o,oUp}
\fmfv{decor.shape=circle,decor.filled=empty,decor.size=0.25cm,l=$\times$,label.dist=0}{oUp}
\fmfv{decor.shape=circle,decor.filled=empty,decor.size=0.25cm,l=$\times$,label.dist=0}{oDown}
\fmf{phantom,tension=10}{i,i1}
\fmf{phantom,tension=10}{o,o1}
\fmf{plain,left,tension=0.5}{i1,v1,i1}
\fmf{phantom,right,tension=0.5}{o1,v2,o1}
\fmf{plain,tension=0}{v2,oUp}
\fmf{plain,tension=0}{v2,oDown}
\fmf{zigzag,foreground=(0,,0,,1)}{v1,v2}
\end{fmfgraph*}
\end{fmffile}
\end{gathered} - \frac{1}{6} \ \begin{gathered}
\begin{fmffile}{Diagrams/1PIEAlambda_Diag6}
\begin{fmfgraph*}(15,15)
\fmfleft{iDown,i,iUp}
\fmfright{oDown,o,oUp}
\fmfv{decor.shape=circle,decor.filled=empty,decor.size=0.25cm,l=$\times$,label.dist=0}{iUp}
\fmfv{decor.shape=circle,decor.filled=empty,decor.size=0.25cm,l=$\times$,label.dist=0}{oUp}
\fmf{plain,tension=0}{v1,iUp}
\fmf{plain,tension=0}{v2,oUp}
\fmf{phantom,tension=11}{i,v1}
\fmf{phantom,tension=11}{v2,o}
\fmf{plain,right,tension=0.4}{v1,v2}
\fmf{zigzag,foreground=(0,,0,,1)}{v1,v2}
\end{fmfgraph*}
\end{fmffile}
\end{gathered} \\
& -\frac{1}{24} \ \begin{gathered}
\begin{fmffile}{Diagrams/1PIEAlambda_Diag7}
\begin{fmfgraph*}(30,15)
\fmfleft{iDown,i,iUp}
\fmfright{oDown,o,oUp}
\fmfv{decor.shape=circle,decor.filled=empty,decor.size=0.25cm,l=$\times$,label.dist=0}{iUp}
\fmfv{decor.shape=circle,decor.filled=empty,decor.size=0.25cm,l=$\times$,label.dist=0}{iDown}
\fmfv{decor.shape=circle,decor.filled=empty,decor.size=0.25cm,l=$\times$,label.dist=0}{oUp}
\fmfv{decor.shape=circle,decor.filled=empty,decor.size=0.25cm,l=$\times$,label.dist=0}{oDown}
\fmf{phantom,tension=10}{i,i1}
\fmf{phantom,tension=10}{o,o1}
\fmf{phantom,left,tension=0.5}{i1,v1,i1}
\fmf{phantom,right,tension=0.5}{o1,v2,o1}
\fmf{plain,tension=0}{v1,iUp}
\fmf{plain,tension=0}{v1,iDown}
\fmf{plain,tension=0}{v2,oUp}
\fmf{plain,tension=0}{v2,oDown}
\fmf{zigzag,foreground=(0,,0,,1)}{v1,v2}
\end{fmfgraph*}
\end{fmffile}
\end{gathered} \\
\\
& + \mathcal{O}\big(\lambda^{2}\big)\;.
\end{split}
\label{eq:1PIEAlambdaWexpansionStep60DON}
\end{equation}

\vspace{0.5cm}

We then turn to the IM by giving the following series:
\begin{subequations}
\begin{empheq}[left=\empheqlbrace]{align}
& \hspace{0.1cm} \Gamma^{(\mathrm{1PI})}\Big[\vec{\phi};\lambda\Big]=\sum_{n=0}^{\infty} \Gamma^{(\mathrm{1PI})}_{n}\Big[\vec{\phi};\lambda\Big]\;, \label{eq:1PIEAlambdaGammaExpansion0DON}\\
\nonumber \\
& \hspace{0.1cm} W\Big[\vec{J};\lambda\Big]=\sum_{n=0}^{\infty} W_{n}\Big[\vec{J};\lambda\Big]\;, \label{eq:1PIEAlambdaWExpansion0DON} \\
\nonumber \\
& \hspace{0.1cm} \vec{J}\Big[\vec{\phi};\lambda\Big]=\sum_{n=0}^{\infty} \vec{J}_{n}\Big[\vec{\phi};\lambda\Big]\;, \label{eq:1PIEAlambdaJExpansion0DON}\\
\nonumber \\
& \hspace{0.1cm} \vec{\phi}=\sum_{n=0}^{\infty} \vec{\phi}_{n}\Big[\vec{J};\lambda\Big]\;, \label{eq:1PIEAlambdaphiExpansion0DON}
\end{empheq}
\end{subequations}
where $\Gamma^{(\mathrm{1PI})}$ is still defined by~\eqref{eq:pure1PIEAdefinition0DONAppendix}. Note that the $\lambda$-dependence of the functionals involved in~\eqref{eq:1PIEAlambdaGammaExpansion0DON} to~\eqref{eq:1PIEAlambdaphiExpansion0DON} will be left implicit below, as was done above for $\hbar$. As the coupling constant $\lambda$ is incorporated into the vertex function~\eqref{eq:1PIEAlambda4legvertex0DON} underlying our diagrammatic representation, it will be more convenient to deal with $\lambda$-dependent $\Gamma^{(\mathrm{1PI})}_{n}$, $W_{n}$, $\vec{J}_{n}$ and $\vec{\phi}_{n}$ coefficients in order to carry out the IM. Hence, it should be understood that the functionals $\Gamma^{(\mathrm{1PI})}_{n}$, $W_{n}$, $\vec{J}_{n}$ and $\vec{\phi}_{n}$ are of order $\mathcal{O}\big(\lambda^{n}\big)$ by definition and are \textit{a priori} different from their eponymous counterparts of~\eqref{eq:pure1PIEAGammaExpansion0DONAppendix} to~\eqref{eq:pure1PIEAphiExpansion0DONAppendix}. The general recipe of the IM outlined earlier for the $\hbar$-expansion remains identical, as it would be for any other expansion parameter. In particular, $\vec{\phi}$ remains independent of $\vec{J}$ and of the expansion parameter. Moreover, similarly to~\eqref{eq:pure1PIEAphicl0DON} and~\eqref{eq:pure1PIEAphi0DON}, the classical and quantum 1-point correlation functions satisfy:
\begin{equation}
\varphi_{\mathrm{cl},\alpha}\Big[\vec{J}\Big] = \phi_{0,\alpha}\Big[\vec{J}\Big] = \frac{\delta W_{0}\big[\vec{J}\big]}{\delta J_{\alpha}} = \frac{\delta}{\delta J_{\alpha}}\Bigg(\frac{1}{2}\mathrm{STr}[\ln(\boldsymbol{G}_{0})]+\frac{1}{2} \begin{gathered}
\begin{fmffile}{Diagrams/1PIEAlambda_Diag0}
\begin{fmfgraph*}(20,20)
\fmfleft{i0,i1,i2,i3}
\fmfright{o0,o1,o2,o3}
\fmfv{decor.shape=circle,decor.filled=empty,decor.size=0.25cm,l=$\times$,label.dist=0}{v1}
\fmfv{decor.shape=circle,decor.filled=empty,decor.size=0.25cm,l=$\times$,label.dist=0}{v2}
\fmf{phantom}{i1,v1}
\fmf{phantom}{i2,v1}
\fmf{plain,tension=0.6}{v1,v2}
\fmf{phantom}{v2,o1}
\fmf{phantom}{v2,o2}
\end{fmfgraph*}
\end{fmffile}
\end{gathered}\Bigg) = \hspace{0.2cm} \begin{gathered}
\begin{fmffile}{Diagrams/1PIEAlambda_phicl}
\begin{fmfgraph*}(20,20)
\fmfleft{i0,i1,i2,i3}
\fmfright{o0,o1,o2,o3}
\fmfv{decor.shape=circle,decor.filled=empty,decor.size=1.5thick,label=$\alpha$}{v1}
\fmfv{decor.shape=circle,decor.filled=empty,decor.size=0.25cm,l=$\times$,label.dist=0}{v2}
\fmf{phantom}{i1,v1}
\fmf{phantom}{i2,v1}
\fmf{plain,tension=0.6}{v1,v2}
\fmf{phantom}{v2,o1}
\fmf{phantom}{v2,o2}
\end{fmfgraph*}
\end{fmffile}
\end{gathered}\;,
\label{eq:pure1PIEAlambdaphicl0DON}
\end{equation}
and
\begin{equation}
\phi_{\alpha} = \varphi_{\mathrm{cl},\alpha}\Big[\vec{J}=\vec{J}_{0}\Big] = \phi_{0,\alpha}\Big[\vec{J}=\vec{J}_{0}\Big] = \left.\frac{\delta W_{0}\big[\vec{J}\big]}{\delta J_{\alpha}}\right|_{\vec{J}=\vec{J}_{0}} = \hspace{0.2cm} \begin{gathered}
\begin{fmffile}{Diagrams/1PIEAlambda_phi}
\begin{fmfgraph*}(20,20)
\fmfleft{i0,i1,i2,i3}
\fmfright{o0,o1,o2,o3}
\fmfv{decor.shape=circle,decor.filled=empty,decor.size=1.5thick,label=$\alpha$}{v1}
\fmfv{decor.shape=circle,decor.filled=empty,decor.size=0.28cm,label=$\mathrm{o}$,label.dist=0}{v2}
\fmf{phantom}{i1,v1}
\fmf{phantom}{i2,v1}
\fmf{plain,tension=0.6}{v1,v2}
\fmf{phantom}{v2,o1}
\fmf{phantom}{v2,o2}
\end{fmfgraph*}
\end{fmffile}
\end{gathered}\;,
\label{eq:pure1PIEAlambdaphi0DON}
\end{equation}
where we have notably used:
\begin{equation}
\begin{gathered}
\begin{fmffile}{Diagrams/1PIEAlambda_FeynRuleJn}
\begin{fmfgraph*}(6,4)
\fmfleft{i1}
\fmfright{o1}
\fmfv{decor.shape=circle,decor.filled=empty,decor.size=0.4cm,label=$n$,label.dist=0}{v1}
\fmfv{label=$\alpha$,label.angle=-90,label.dist=7}{v2}
\fmf{plain,tension=0.5}{i1,v1}
\fmf{phantom}{v1,o1}
\fmf{phantom,tension=0.5}{i1,v2}
\fmf{phantom}{v2,o1}
\end{fmfgraph*}
\end{fmffile}
\end{gathered} \hspace{0.1cm} \rightarrow J_{n,\alpha}\Big[\vec{\phi}\Big]\;,
\label{eq:1PIEAlambdafeynRuleJn0DON}
\end{equation}
as in~\eqref{eq:pure1PIEAfeynRuleJn0DON}. We can also straightforwardly derive the homologous result of~\eqref{eq:pure1PIEAIMstep60DON} for the $\lambda$-expansion, that we directly give below:
\begin{equation}
\begin{split}
\Gamma^{(\mathrm{1PI})}_{n}\Big[\vec{\phi}\Big] = & -W_{n}\Big[\vec{J}=\vec{J}_{0}\Big] -\sum_{m=1}^{n-1} \int_{\alpha} \left.\frac{\delta W_{n-m}\big[\vec{J}\big]}{\delta J_{\alpha}}\right|_{\vec{J}=\vec{J}_{0}} J_{m,\alpha}\Big[\vec{\phi}\Big] \\
& -\sum_{m=2}^{n} \frac{1}{m!} \sum_{\underset{\lbrace n_{1} + \cdots + n_{m} \leq n\rbrace}{n_{1},\cdots,n_{m}=1}}^{n} \int_{\alpha_{1},\cdots,\alpha_{m}} \left.\frac{\delta^{m} W_{n-(n_{1}+\cdots+n_{m})}\big[\vec{J}\big]}{\delta J_{\alpha_{1}}\cdots\delta J_{\alpha_{m}}}\right|_{\vec{J}=\vec{J}_{0}} J_{n_{1},\alpha_{1}}\Big[\vec{\phi}\Big]\cdots J_{n_{m},\alpha_{m}}\Big[\vec{\phi}\Big] \\
& + \int_{\alpha} J_{0,\alpha}\Big[\vec{\phi}\Big] \phi_{\alpha} \delta_{n 0}\;.
\end{split}
\label{eq:1PIEAlambdaGammanCoeff0DON}
\end{equation}
In the present situation, the first non-trivial order corresponds to $n=1$ (i.e. to order $\mathcal{O}(\lambda)$) instead of $n=2$ (i.e. instead of order $\mathcal{O}(\hbar^{2})$) in the framework of the $\hbar$-expansion. Hence, contenting ourselves with this order, we are interested in~\eqref{eq:1PIEAlambdaGammanCoeff0DON} evaluated at $n=0~\mathrm{and}~1$:
\begin{equation}
\Gamma^{(\mathrm{1PI})}_{0}\Big[\vec{\phi}\Big] = -W_{0}\Big[\vec{J}=\vec{J}_{0}\Big] + \int_{\alpha} J_{0,\alpha}\Big[\vec{\phi}\Big] \phi_{\alpha}\;,
\label{eq:pure1PIEAlambdaIMGamma00DON}
\end{equation}
\begin{equation}
\Gamma^{(\mathrm{1PI})}_{1}\Big[\vec{\phi}\Big] = -W_{1}\Big[\vec{J}=\vec{J}_{0}\Big]\;.
\label{eq:pure1PIEAlambdaIMGamma10DON}
\end{equation}
Hence, we no longer need to determine the $\vec{J}_{1}$ and $\vec{\phi}_{1}$ coefficients in order to calculate the first non-trivial order. From~\eqref{eq:pure1PIEAlambdaIMGamma00DON} and~\eqref{eq:pure1PIEAlambdaIMGamma10DON} as well as~\eqref{eq:1PIEAlambdaWexpansionStep60DON}, we directly infer:
\begin{equation}
\begin{split}
\Gamma^{(\mathrm{1PI})}\Big[\vec{\phi}\Big] = & \ -\frac{1}{2}\mathrm{STr}[\ln(\boldsymbol{G}_{0})]-\frac{1}{2} \begin{gathered}
\begin{fmffile}{Diagrams/1PIEAlambda_Diag0bis}
\begin{fmfgraph*}(20,20)
\fmfleft{i0,i1,i2,i3}
\fmfright{o0,o1,o2,o3}
\fmfv{decor.shape=circle,decor.filled=empty,decor.size=0.28cm,label=$\mathrm{o}$,label.dist=0}{v1}
\fmfv{decor.shape=circle,decor.filled=empty,decor.size=0.28cm,label=$\mathrm{o}$,label.dist=0}{v2}
\fmf{phantom}{i1,v1}
\fmf{phantom}{i2,v1}
\fmf{plain,tension=0.6}{v1,v2}
\fmf{phantom}{v2,o1}
\fmf{phantom}{v2,o2}
\end{fmfgraph*}
\end{fmffile}
\end{gathered} + \int_{\alpha} J_{0,\alpha}\Big[\vec{\phi}\Big] \phi_{\alpha} \\
& +\frac{1}{24} \hspace{0.08cm} \begin{gathered}
\begin{fmffile}{Diagrams/LoopExpansion1_Hartree}
\begin{fmfgraph}(30,20)
\fmfleft{i}
\fmfright{o}
\fmf{phantom,tension=10}{i,i1}
\fmf{phantom,tension=10}{o,o1}
\fmf{plain,left,tension=0.5}{i1,v1,i1}
\fmf{plain,right,tension=0.5}{o1,v2,o1}
\fmf{zigzag,foreground=(0,,0,,1)}{v1,v2}
\end{fmfgraph}
\end{fmffile}
\end{gathered}
+\frac{1}{12}\begin{gathered}
\begin{fmffile}{Diagrams/LoopExpansion1_Fock}
\begin{fmfgraph}(15,15)
\fmfleft{i}
\fmfright{o}
\fmf{phantom,tension=11}{i,v1}
\fmf{phantom,tension=11}{v2,o}
\fmf{plain,left,tension=0.4}{v1,v2,v1}
\fmf{zigzag,foreground=(0,,0,,1)}{v1,v2}
\end{fmfgraph}
\end{fmffile}
\end{gathered} + \frac{1}{12} \ \begin{gathered}
\begin{fmffile}{Diagrams/1PIEAlambda_Diag5bis}
\begin{fmfgraph*}(30,15)
\fmfleft{i}
\fmfright{oDown,o,oUp}
\fmfv{decor.shape=circle,decor.filled=empty,decor.size=0.28cm,label=$\mathrm{o}$,label.dist=0}{oUp}
\fmfv{decor.shape=circle,decor.filled=empty,decor.size=0.28cm,label=$\mathrm{o}$,label.dist=0}{oDown}
\fmf{phantom,tension=10}{i,i1}
\fmf{phantom,tension=10}{o,o1}
\fmf{plain,left,tension=0.5}{i1,v1,i1}
\fmf{phantom,right,tension=0.5}{o1,v2,o1}
\fmf{plain,tension=0}{v2,oUp}
\fmf{plain,tension=0}{v2,oDown}
\fmf{zigzag,foreground=(0,,0,,1)}{v1,v2}
\end{fmfgraph*}
\end{fmffile}
\end{gathered} + \frac{1}{6} \ \begin{gathered}
\begin{fmffile}{Diagrams/1PIEAlambda_Diag6bis}
\begin{fmfgraph*}(15,15)
\fmfleft{iDown,i,iUp}
\fmfright{oDown,o,oUp}
\fmfv{decor.shape=circle,decor.filled=empty,decor.size=0.28cm,label=$\mathrm{o}$,label.dist=0}{iUp}
\fmfv{decor.shape=circle,decor.filled=empty,decor.size=0.28cm,label=$\mathrm{o}$,label.dist=0}{oUp}
\fmf{plain,tension=0}{v1,iUp}
\fmf{plain,tension=0}{v2,oUp}
\fmf{phantom,tension=11}{i,v1}
\fmf{phantom,tension=11}{v2,o}
\fmf{plain,right,tension=0.4}{v1,v2}
\fmf{zigzag,foreground=(0,,0,,1)}{v1,v2}
\end{fmfgraph*}
\end{fmffile}
\end{gathered} \\
& +\frac{1}{24} \ \begin{gathered}
\begin{fmffile}{Diagrams/1PIEAlambda_Diag7bis}
\begin{fmfgraph*}(30,15)
\fmfleft{iDown,i,iUp}
\fmfright{oDown,o,oUp}
\fmfv{decor.shape=circle,decor.filled=empty,decor.size=0.28cm,label=$\mathrm{o}$,label.dist=0}{iUp}
\fmfv{decor.shape=circle,decor.filled=empty,decor.size=0.28cm,label=$\mathrm{o}$,label.dist=0}{iDown}
\fmfv{decor.shape=circle,decor.filled=empty,decor.size=0.28cm,label=$\mathrm{o}$,label.dist=0}{oUp}
\fmfv{decor.shape=circle,decor.filled=empty,decor.size=0.28cm,label=$\mathrm{o}$,label.dist=0}{oDown}
\fmf{phantom,tension=10}{i,i1}
\fmf{phantom,tension=10}{o,o1}
\fmf{phantom,left,tension=0.5}{i1,v1,i1}
\fmf{phantom,right,tension=0.5}{o1,v2,o1}
\fmf{plain,tension=0}{v1,iUp}
\fmf{plain,tension=0}{v1,iDown}
\fmf{plain,tension=0}{v2,oUp}
\fmf{plain,tension=0}{v2,oDown}
\fmf{zigzag,foreground=(0,,0,,1)}{v1,v2}
\end{fmfgraph*}
\end{fmffile}
\end{gathered} \\
\\
& + \mathcal{O}\big(\lambda^{2}\big)\;.
\end{split}
\label{eq:1PIEAlambdaEAStep10DON}
\end{equation}
According to~\eqref{eq:pure1PIEAlambdaphi0DON}, we can show that:
\begin{equation}
\begin{split}
-\frac{1}{2} \begin{gathered}
\begin{fmffile}{Diagrams/1PIEAlambda_Diag0bis}
\begin{fmfgraph*}(20,20)
\fmfleft{i0,i1,i2,i3}
\fmfright{o0,o1,o2,o3}
\fmfv{decor.shape=circle,decor.filled=empty,decor.size=0.28cm,label=$\mathrm{o}$,label.dist=0}{v1}
\fmfv{decor.shape=circle,decor.filled=empty,decor.size=0.28cm,label=$\mathrm{o}$,label.dist=0}{v2}
\fmf{phantom}{i1,v1}
\fmf{phantom}{i2,v1}
\fmf{plain,tension=0.6}{v1,v2}
\fmf{phantom}{v2,o1}
\fmf{phantom}{v2,o2}
\end{fmfgraph*}
\end{fmffile}
\end{gathered} + \int_{\alpha} J_{0,\alpha}\Big[\vec{\phi}\Big] \phi_{\alpha} + \frac{1}{24} \ \begin{gathered}
\begin{fmffile}{Diagrams/1PIEAlambda_Diag7bis}
\begin{fmfgraph*}(30,15)
\fmfleft{iDown,i,iUp}
\fmfright{oDown,o,oUp}
\fmfv{decor.shape=circle,decor.filled=empty,decor.size=0.28cm,label=$\mathrm{o}$,label.dist=0}{iUp}
\fmfv{decor.shape=circle,decor.filled=empty,decor.size=0.28cm,label=$\mathrm{o}$,label.dist=0}{iDown}
\fmfv{decor.shape=circle,decor.filled=empty,decor.size=0.28cm,label=$\mathrm{o}$,label.dist=0}{oUp}
\fmfv{decor.shape=circle,decor.filled=empty,decor.size=0.28cm,label=$\mathrm{o}$,label.dist=0}{oDown}
\fmf{phantom,tension=10}{i,i1}
\fmf{phantom,tension=10}{o,o1}
\fmf{phantom,left,tension=0.5}{i1,v1,i1}
\fmf{phantom,right,tension=0.5}{o1,v2,o1}
\fmf{plain,tension=0}{v1,iUp}
\fmf{plain,tension=0}{v1,iDown}
\fmf{plain,tension=0}{v2,oUp}
\fmf{plain,tension=0}{v2,oDown}
\fmf{zigzag,foreground=(0,,0,,1)}{v1,v2}
\end{fmfgraph*}
\end{fmffile}
\end{gathered} = & \ \frac{1}{2} \begin{gathered}
\begin{fmffile}{Diagrams/1PIEAlambda_Diag0bis}
\begin{fmfgraph*}(20,20)
\fmfleft{i0,i1,i2,i3}
\fmfright{o0,o1,o2,o3}
\fmfv{decor.shape=circle,decor.filled=empty,decor.size=0.28cm,label=$\mathrm{o}$,label.dist=0}{v1}
\fmfv{decor.shape=circle,decor.filled=empty,decor.size=0.28cm,label=$\mathrm{o}$,label.dist=0}{v2}
\fmf{phantom}{i1,v1}
\fmf{phantom}{i2,v1}
\fmf{plain,tension=0.6}{v1,v2}
\fmf{phantom}{v2,o1}
\fmf{phantom}{v2,o2}
\end{fmfgraph*}
\end{fmffile}
\end{gathered} + \frac{1}{24} \ \begin{gathered}
\begin{fmffile}{Diagrams/1PIEAlambda_Diag7bis2}
\begin{fmfgraph*}(25,13)
\fmfleft{iDown,i,iUp}
\fmfright{oDown,o,oUp}
\fmfv{decor.shape=cross,decor.angle=45,decor.size=3.5thick,foreground=(1,,0,,0)}{iUp}
\fmfv{decor.shape=cross,decor.angle=45,decor.size=3.5thick,foreground=(1,,0,,0)}{iDown}
\fmfv{decor.shape=cross,decor.angle=45,decor.size=3.5thick,foreground=(1,,0,,0)}{oUp}
\fmfv{decor.shape=cross,decor.angle=45,decor.size=3.5thick,foreground=(1,,0,,0)}{oDown}
\fmf{phantom,tension=10}{i,i1}
\fmf{phantom,tension=10}{o,o1}
\fmf{phantom,left,tension=0.5}{i1,v1,i1}
\fmf{phantom,right,tension=0.5}{o1,v2,o1}
\fmf{dashes,tension=0.2,foreground=(1,,0,,0)}{v1,iUp}
\fmf{dashes,tension=0.2,foreground=(1,,0,,0)}{v1,iDown}
\fmf{dashes,tension=0.2,foreground=(1,,0,,0)}{v2,oUp}
\fmf{dashes,tension=0.2,foreground=(1,,0,,0)}{v2,oDown}
\fmf{zigzag,foreground=(0,,0,,1)}{v1,v2}
\end{fmfgraph*}
\end{fmffile}
\end{gathered} \\
= & \ \frac{1}{2}\int_{\alpha_{1},\alpha_{2}} J_{0,\alpha_{1}} \boldsymbol{G}_{0,\alpha_{1}\alpha_{2}} J_{0,\alpha_{2}} \\
& + \frac{\lambda}{4!} \sum_{a_{1},a_{2}=1}^{N} \int_{x} \phi_{a_{1},x}^{2} \phi_{a_{2},x}^{2} \\
= & \ \frac{1}{2}\int_{\alpha_{1},\alpha_{2}} \phi_{\alpha_{1}} \boldsymbol{G}^{-1}_{0,\alpha_{1}\alpha_{2}} \phi_{\alpha_{2}} \\
& + \frac{\lambda}{4!} \sum_{a_{1},a_{2}=1}^{N} \int_{x} \phi_{a_{1},x}^{2} \phi_{a_{2},x}^{2} \\
= & \ S\Big[\vec{\phi}\Big] \;,
\label{eq:1PIEAlambdaEAStep20DON}
\end{split}
\end{equation}
with
\begin{equation}
\begin{gathered}
\begin{fmffile}{Diagrams/1PIEAlambda_FeynRulephi}
\begin{fmfgraph*}(20,20)
\fmfleft{i0,i1,i2,i3}
\fmfright{o0,o1,o2,o3}
\fmfv{label.dist=0.15cm,label=$\alpha$}{v1}
\fmfv{decor.shape=cross,decor.size=3.5thick,foreground=(1,,0,,0)}{v2}
\fmf{phantom}{i1,v1}
\fmf{phantom}{i2,v1}
\fmf{dashes,tension=1.2,foreground=(1,,0,,0)}{v1,v2}
\fmf{phantom}{v2,o1}
\fmf{phantom}{v2,o2}
\end{fmfgraph*}
\end{fmffile}
\end{gathered} \hspace{-0.3cm} \rightarrow \phi_{\alpha} \;.
\end{equation}
With the help of~\eqref{eq:pure1PIEAlambdaphi0DON} and~\eqref{eq:1PIEAlambdaEAStep20DON},~\eqref{eq:1PIEAlambdaEAStep10DON} can be simplified as follows:
\begin{equation}
\begin{split}
\Gamma^{(\mathrm{1PI})}\Big[\vec{\phi}\Big] = & \ S\Big[\vec{\phi}\Big] -\frac{1}{2}\mathrm{STr}[\ln(\boldsymbol{G}_{0})] \\
& +\frac{1}{24} \hspace{0.08cm} \begin{gathered}
\begin{fmffile}{Diagrams/LoopExpansion1_Hartree}
\begin{fmfgraph}(30,20)
\fmfleft{i}
\fmfright{o}
\fmf{phantom,tension=10}{i,i1}
\fmf{phantom,tension=10}{o,o1}
\fmf{plain,left,tension=0.5}{i1,v1,i1}
\fmf{plain,right,tension=0.5}{o1,v2,o1}
\fmf{zigzag,foreground=(0,,0,,1)}{v1,v2}
\end{fmfgraph}
\end{fmffile}
\end{gathered}
+\frac{1}{12}\begin{gathered}
\begin{fmffile}{Diagrams/LoopExpansion1_Fock}
\begin{fmfgraph}(15,15)
\fmfleft{i}
\fmfright{o}
\fmf{phantom,tension=11}{i,v1}
\fmf{phantom,tension=11}{v2,o}
\fmf{plain,left,tension=0.4}{v1,v2,v1}
\fmf{zigzag,foreground=(0,,0,,1)}{v1,v2}
\end{fmfgraph}
\end{fmffile}
\end{gathered} + \frac{1}{12} \ \begin{gathered}
\begin{fmffile}{Diagrams/1PIEAlambda_Diag5bis2}
\begin{fmfgraph*}(25,13)
\fmfleft{i}
\fmfright{oDown,o,oUp}
\fmfv{decor.shape=cross,decor.angle=45,decor.size=3.5thick,foreground=(1,,0,,0)}{oUp}
\fmfv{decor.shape=cross,decor.angle=45,decor.size=3.5thick,foreground=(1,,0,,0)}{oDown}
\fmf{phantom,tension=10}{i,i1}
\fmf{phantom,tension=10}{o,o1}
\fmf{plain,left,tension=0.5}{i1,v1,i1}
\fmf{phantom,right,tension=0.5}{o1,v2,o1}
\fmf{dashes,tension=0,foreground=(1,,0,,0)}{v2,oUp}
\fmf{dashes,tension=0,foreground=(1,,0,,0)}{v2,oDown}
\fmf{zigzag,foreground=(0,,0,,1)}{v1,v2}
\end{fmfgraph*}
\end{fmffile}
\end{gathered} + \frac{1}{6} \ \begin{gathered}
\begin{fmffile}{Diagrams/1PIEAlambda_Diag6bis2}
\begin{fmfgraph*}(15,15)
\fmfleft{iDown,i,iUp}
\fmfright{oDown,o,oUp}
\fmfv{decor.shape=cross,decor.size=3.5thick,foreground=(1,,0,,0)}{iUp}
\fmfv{decor.shape=cross,decor.size=3.5thick,foreground=(1,,0,,0)}{oUp}
\fmf{dashes,tension=0,foreground=(1,,0,,0)}{v1,iUp}
\fmf{dashes,tension=0,foreground=(1,,0,,0)}{v2,oUp}
\fmf{phantom,tension=11}{i,v1}
\fmf{phantom,tension=11}{v2,o}
\fmf{plain,right,tension=0.4}{v1,v2}
\fmf{zigzag,foreground=(0,,0,,1)}{v1,v2}
\end{fmfgraph*}
\end{fmffile}
\end{gathered} \\
& + \mathcal{O}\big(\lambda^{2}\big)\;.
\end{split}
\label{eq:1PIEAlambdaEAStep30DONAppendix}
\end{equation}
In the previous work of Okumura~\cite{oku96}, it is also pointed out that the diagrammatic expression of the 1PI EA can be determined by integrating the equation:
\begin{equation}
\frac{\delta \Gamma^{(\mathrm{1PI})}_{n}\big[\vec{\phi}\big]}{\delta \phi_{\alpha}}=J_{n,\alpha}\Big[\vec{\phi}\Big]\;,
\end{equation}
which directly follows from definition~\eqref{eq:pure1PIEAdefinition0DONAppendix} together with the power series \eqref{eq:1PIEAlambdaGammaExpansion0DON} and \eqref{eq:1PIEAlambdaJExpansion0DON}. Following the lines set out by ref.~\cite{oku96}, $\vec{J}_{0}$ is determined by inverting~\eqref{eq:pure1PIEAlambdaphi0DON}, thus yielding:
\begin{equation}
J_{0,\alpha_{1}}\Big[\vec{\phi}\Big] = \int_{\alpha_{2}} \boldsymbol{G}^{-1}_{0,\alpha_{1}\alpha_{2}} \phi_{\alpha_{2}}\;,
\label{eq:1PIEAJ0phi0DON}
\end{equation}
and the $\vec{J}_{n}$ coefficients (with $n \geq 1$) are calculated in the same way as $\vec{J}_{1}$ via~\eqref{eq:pure1PIEATowerEquationJn20DON} above. In summary, comparing the additional method proposed by Okumura and the IM formulation presented here, the $\vec{J}_{n}$ coefficients are determined in the same manner and the two approaches lead to identical expressions of the 1PI EA (i.e.~\eqref{eq:1PIEAlambdaEAStep30DONAppendix}) up to $\vec{\phi}$-independent (or, equivalently, $\vec{J}_{0}$-independent) terms.

\subsection{\label{sec:collective1PIEAannIM}Collective effective action}

Let us then turn to a more involved situation by carrying out the IM in order to determine the collective 1PI EA $\Gamma_{\mathrm{col}}^{(\mathrm{1PI})}$ at the lowest non-trivial order in the framework of the $\hbar$-expansion, i.e. up to order $\mathcal{O}\big(\hbar^{2}\big)$. Most of the derivations are very similar to the IM applied to $\Gamma^{(\mathrm{1PI})}$, the only difference being that the functionals involved in the calculations are labeled by the superindices defined in section~\ref{sec:ForewordNotationsIM}, i.e. $\beta\equiv(b,x)$ (instead of $\alpha\equiv(a,x)$). A few additional complexities arise at some stages in which we will decouple in our equations the auxiliary field sector (i.e. $b=N+1$) to that of the original field (i.e. $b=1,\cdots,N$). The power series underpinning the first step of the IM are now given by:
\begin{subequations}
\begin{empheq}[left=\empheqlbrace]{align}
& \hspace{0.1cm} \Gamma_{\mathrm{col}}^{(\mathrm{1PI})}\big[\Phi;\hbar\big]=\sum_{n=0}^{\infty} \Gamma^{(\mathrm{1PI})}_{\mathrm{col},n}[\Phi]\hbar^{n}\;, \label{eq:bosonic1PIEAGammaExpansion0DONAppendix}\\
\nonumber \\
& \hspace{0.1cm} W_{\mathrm{col}}\big[\mathcal{J};\hbar\big]=\sum_{n=0}^{\infty} W_{\mathrm{col},n}\big[\mathcal{J}\big]\hbar^{n}\;, \label{eq:bosonic1PIEAWExpansion0DONAppendix} \\
\nonumber \\
& \hspace{0.1cm} \mathcal{J}\big[\Phi;\hbar\big]=\sum_{n=0}^{\infty} \mathcal{J}_{n}[\Phi]\hbar^{n}\;, \label{eq:bosonic1PIEAJExpansion0DONAppendix}\\
\nonumber \\
& \hspace{0.1cm} \Phi = \sum_{n=0}^{\infty} \Phi_{n}\big[\mathcal{J}\big]\hbar^{n}\;, \label{eq:bosonic1PIEAphiExpansion0DONAppendix}
\end{empheq}
\end{subequations}
and the EA $\Gamma_{\mathrm{col}}^{(\mathrm{1PI})}$ is defined from the corresponding Schwinger functional $W_{\mathrm{col}}\big[\mathcal{J}\big]$ by~\eqref{eq:bosonic1PIEAdefinition0DONmain} and~\eqref{eq:bosonic1PIEAdefinitionbis0DONmain}, i.e.:
\begin{equation}
\begin{split}
\Gamma_{\mathrm{col}}^{(\mathrm{1PI})}[\Phi] \equiv & -W_{\mathrm{col}}\big[\mathcal{J}\big] + \int_{\beta}\mathcal{J}_{\beta}[\Phi] \frac{\delta W_{\mathrm{col}}\big[\mathcal{J}\big]}{\delta\mathcal{J}_{\beta}} \\
= & -W_{\mathrm{col}}\big[\mathcal{J}\big]+\int_{\beta}\mathcal{J}_{\beta}[\Phi] \Phi_{\beta}\;,
\end{split}
\label{eq:bosonic1PIEAdefinition0DON}
\end{equation}
with
\begin{equation}
\Phi_{\beta}=\frac{\delta W_{\mathrm{col}}\big[\mathcal{J}\big]}{\delta \mathcal{J}_{\beta}}\;.
\label{eq:bosonic1PIEAdefinitionbis0DON}
\end{equation}
Combining~\eqref{eq:bosonic1PIEAdefinitionbis0DON} with~\eqref{eq:bosonic1PIEAWExpansion0DONAppendix} and~\eqref{eq:bosonic1PIEAphiExpansion0DONAppendix}, the $\Phi_{n}$ coefficients can be expressed as follows:
\begin{equation}
\Phi_{n,\beta}\big[\mathcal{J}\big]=\frac{\delta W_{\mathrm{col},n}\big[\mathcal{J}\big]}{\delta \mathcal{J}_{\beta}}\;,
\label{eq:bosonic1PIEAphinCoeff0DON}
\end{equation}
where the $W_{\mathrm{col},n}$ coefficients are deduced from the collective LE result~\eqref{eq:SbosonicKLoopExpansionStep5}:
\begin{equation}
W_{\mathrm{col},0}\big[\mathcal{J}\big]=-S_{\mathrm{col}}[\sigma_{\mathrm{cl}}] + \int_{x} j_{x}[\Phi] \sigma_{\mathrm{cl},x}\big[\mathcal{J}\big] + \frac{1}{2} \int_{\alpha_{1},\alpha_{2}} J_{\alpha_{1}}[\Phi] \boldsymbol{G}_{\sigma_\text{cl};\mathcal{J},\alpha_{1}\alpha_{2}}\big[\mathcal{J}\big] J_{\alpha_{2}}[\Phi]\;,
\label{eq:bosonic1PIEAIMW00DON}
\end{equation}
\begin{equation}
W_{\mathrm{col},1}\big[\mathcal{J}\big]=\frac{1}{2}\mathrm{Tr}\left[\ln\big(D_{\sigma_{\text{cl}};\mathcal{J}}\big[\mathcal{J}\big]\big)\right]\;,
\label{eq:bosonic1PIEAIMW10DON}
\end{equation}
\begin{equation}
\begin{split}
W_{\mathrm{col},2}\big[\mathcal{J}\big] = & \ \frac{1}{8} \left(\rule{0cm}{1.1cm}\right. 4 \hspace{-0.2cm} \begin{gathered}
\begin{fmffile}{Diagrams/LoopExpansionBosonicHS_W_Diag1}
\begin{fmfgraph*}(25,25)
\fmfleft{i1,i2}
\fmfright{o1,o2}
\fmfbottom{i0,o0}
\fmftop{i3,o3}
\fmfv{decor.shape=circle,decor.size=2.0thick,foreground=(0,,0,,1)}{v1}
\fmfv{decor.shape=circle,decor.size=2.0thick,foreground=(0,,0,,1)}{v2}
\fmfv{decor.shape=circle,decor.size=2.0thick,foreground=(0,,0,,1)}{v3}
\fmfv{decor.shape=circle,decor.size=2.0thick,foreground=(0,,0,,1)}{v4}
\fmfv{decor.shape=circle,decor.filled=empty,decor.size=.1w,l=$\times$,label.dist=0}{v3b}
\fmfv{decor.shape=circle,decor.filled=empty,decor.size=.1w,l=$\times$,label.dist=0}{v4b}
\fmf{phantom}{i1,v1}
\fmf{phantom}{i2,v4}
\fmf{phantom}{o1,v2}
\fmf{phantom}{o2,v3}
\fmf{phantom}{i3,v4b}
\fmf{phantom}{o3,v3b}
\fmf{phantom}{i0,v1b}
\fmf{phantom}{o0,v2b}
\fmf{plain,tension=1.6}{v1,v2}
\fmf{phantom,tension=1.6}{v3,v4}
\fmf{wiggly,tension=2.0}{v1,v4}
\fmf{wiggly,tension=2.0}{v2,v3}
\fmf{phantom,tension=0}{v1,v3}
\fmf{phantom,tension=0}{v2,v4}
\fmf{plain,right=0.8,tension=0}{v2,v3}
\fmf{plain,left=0.8,tension=0}{v1,v4}
\fmf{phantom}{v1,v1b}
\fmf{phantom}{v2,v2b}
\fmf{plain}{v3,v3b}
\fmf{plain}{v4,v4b}
\end{fmfgraph*}
\end{fmffile}
\end{gathered} \hspace{-0.2cm} + 4 \hspace{-0.2cm} \begin{gathered}
\begin{fmffile}{Diagrams/LoopExpansionBosonicHS_W_Diag2}
\begin{fmfgraph*}(25,25)
\fmfleft{i1,i2}
\fmfright{o1,o2}
\fmfbottom{i0,o0}
\fmftop{i3,o3}
\fmfv{decor.shape=circle,decor.size=2.0thick,foreground=(0,,0,,1)}{v1}
\fmfv{decor.shape=circle,decor.size=2.0thick,foreground=(0,,0,,1)}{v2}
\fmfv{decor.shape=circle,decor.size=2.0thick,foreground=(0,,0,,1)}{v3}
\fmfv{decor.shape=circle,decor.size=2.0thick,foreground=(0,,0,,1)}{v4}
\fmfv{decor.shape=circle,decor.filled=empty,decor.size=.1w,l=$\times$,label.dist=0}{v1b}
\fmfv{decor.shape=circle,decor.filled=empty,decor.size=.1w,l=$\times$,label.dist=0}{v4b}
\fmf{phantom}{i1,v1}
\fmf{phantom}{i2,v4}
\fmf{phantom}{o1,v2}
\fmf{phantom}{o2,v3}
\fmf{phantom}{i3,v4b}
\fmf{phantom}{o3,v3b}
\fmf{phantom}{i0,v1b}
\fmf{phantom}{o0,v2b}
\fmf{plain,tension=1.6}{v1,v2}
\fmf{plain,tension=1.6}{v3,v4}
\fmf{wiggly,tension=2.0}{v1,v4}
\fmf{wiggly,tension=2.0}{v2,v3}
\fmf{phantom,tension=0}{v1,v3}
\fmf{phantom,tension=0}{v2,v4}
\fmf{plain,right=0.8,tension=0}{v2,v3}
\fmf{phantom,left=0.8,tension=0}{v1,v4}
\fmf{plain}{v1,v1b}
\fmf{phantom}{v2,v2b}
\fmf{phantom}{v3,v3b}
\fmf{plain}{v4,v4b}
\end{fmfgraph*}
\end{fmffile}
\end{gathered} \hspace{-0.5cm} + 4 \hspace{-0.3cm} \begin{gathered}
\begin{fmffile}{Diagrams/LoopExpansionBosonicHS_W_Diag3}
\begin{fmfgraph*}(25,25)
\fmfleft{i1,i2}
\fmfright{o1,o2}
\fmfbottom{i0,o0}
\fmftop{i3,o3}
\fmfv{decor.shape=circle,decor.size=2.0thick,foreground=(0,,0,,1)}{v1}
\fmfv{decor.shape=circle,decor.size=2.0thick,foreground=(0,,0,,1)}{v2}
\fmfv{decor.shape=circle,decor.size=2.0thick,foreground=(0,,0,,1)}{v3}
\fmfv{decor.shape=circle,decor.size=2.0thick,foreground=(0,,0,,1)}{v4}
\fmfv{decor.shape=circle,decor.filled=empty,decor.size=.1w,l=$\times$,label.dist=0}{v3b}
\fmfv{decor.shape=circle,decor.filled=empty,decor.size=.1w,l=$\times$,label.dist=0}{v4b}
\fmf{phantom}{i1,v1}
\fmf{phantom}{i2,v4}
\fmf{phantom}{o1,v2}
\fmf{phantom}{o2,v3}
\fmf{phantom}{i3,v4b}
\fmf{phantom}{o3,v3b}
\fmf{phantom}{i0,v1b}
\fmf{phantom}{o0,v2b}
\fmf{plain,tension=1.6}{v1,v2}
\fmf{phantom,tension=1.6}{v3,v4}
\fmf{wiggly,tension=2.0}{v1,v4}
\fmf{wiggly,tension=2.0}{v2,v3}
\fmf{plain,tension=0}{v1,v3}
\fmf{plain,tension=0}{v2,v4}
\fmf{phantom,right=0.8,tension=0}{v2,v3}
\fmf{phantom,left=0.8,tension=0}{v1,v4}
\fmf{phantom}{v1,v1b}
\fmf{phantom}{v2,v2b}
\fmf{plain}{v3,v3b}
\fmf{plain}{v4,v4b}
\end{fmfgraph*}
\end{fmffile}
\end{gathered} \hspace{-0.4cm} + 2 \hspace{-0.4cm} \begin{gathered}
\begin{fmffile}{Diagrams/LoopExpansionBosonicHS_W_Diag5}
\begin{fmfgraph*}(25,25)
\fmfleft{i1,i2}
\fmfright{o1,o2}
\fmfbottom{i0,o0}
\fmftop{i3,o3}
\fmfv{decor.shape=circle,decor.size=2.0thick,foreground=(0,,0,,1)}{v1}
\fmfv{decor.shape=circle,decor.size=2.0thick,foreground=(0,,0,,1)}{v2}
\fmfv{decor.shape=circle,decor.size=2.0thick,foreground=(0,,0,,1)}{v3}
\fmfv{decor.shape=circle,decor.size=2.0thick,foreground=(0,,0,,1)}{v4}
\fmf{phantom}{i1,v1}
\fmf{phantom}{i2,v4}
\fmf{phantom}{o1,v2}
\fmf{phantom}{o2,v3}
\fmf{phantom}{i3,v4b}
\fmf{phantom}{o3,v3b}
\fmf{phantom}{i0,v1b}
\fmf{phantom}{o0,v2b}
\fmf{plain,tension=1.6}{v1,v2}
\fmf{plain,tension=1.6}{v3,v4}
\fmf{wiggly,tension=2.0}{v1,v4}
\fmf{wiggly,tension=2.0}{v2,v3}
\fmf{phantom,tension=0}{v1,v3}
\fmf{phantom,tension=0}{v2,v4}
\fmf{plain,right=0.8,tension=0}{v2,v3}
\fmf{plain,left=0.8,tension=0}{v1,v4}
\fmf{phantom}{v1,v1b}
\fmf{phantom}{v2,v2b}
\fmf{phantom}{v3,v3b}
\fmf{phantom}{v4,v4b}
\end{fmfgraph*}
\end{fmffile}
\end{gathered} \hspace{-0.5cm} + \hspace{-0.7cm} \begin{gathered}
\begin{fmffile}{Diagrams/LoopExpansionBosonicHS_W_Diag6}
\begin{fmfgraph*}(25,25)
\fmfleft{i1,i2}
\fmfright{o1,o2}
\fmfbottom{i0,o0}
\fmftop{i3,o3}
\fmfv{decor.shape=circle,decor.size=2.0thick,foreground=(0,,0,,1)}{v1}
\fmfv{decor.shape=circle,decor.size=2.0thick,foreground=(0,,0,,1)}{v2}
\fmfv{decor.shape=circle,decor.size=2.0thick,foreground=(0,,0,,1)}{v3}
\fmfv{decor.shape=circle,decor.size=2.0thick,foreground=(0,,0,,1)}{v4}
\fmf{phantom}{i1,v1}
\fmf{phantom}{i2,v4}
\fmf{phantom}{o1,v2}
\fmf{phantom}{o2,v3}
\fmf{phantom}{i3,v4b}
\fmf{phantom}{o3,v3b}
\fmf{phantom}{i0,v1b}
\fmf{phantom}{o0,v2b}
\fmf{plain,tension=1.6}{v1,v2}
\fmf{plain,tension=1.6}{v3,v4}
\fmf{wiggly,tension=2.0}{v1,v4}
\fmf{wiggly,tension=2.0}{v2,v3}
\fmf{plain,tension=0}{v1,v3}
\fmf{plain,tension=0}{v2,v4}
\fmf{phantom,right=0.8,tension=0}{v2,v3}
\fmf{phantom,left=0.8,tension=0}{v1,v4}
\fmf{phantom}{v1,v1b}
\fmf{phantom}{v2,v2b}
\fmf{phantom}{v3,v3b}
\fmf{phantom}{v4,v4b}
\end{fmfgraph*}
\end{fmffile}
\end{gathered} \hspace{-0.6cm} \left.\rule{0cm}{1.1cm}\right) \\
& + \frac{1}{12} \left(\rule{0cm}{1.1cm}\right. 6 \begin{gathered}
\begin{fmffile}{Diagrams/LoopExpansionBosonicHS_W_Diag7}
\begin{fmfgraph*}(35,20)
\fmfleft{i1,i2}
\fmfright{o1,o2}
\fmfbottom{i0,o0}
\fmfbottom{b0}
\fmfbottom{b1}
\fmfbottom{b2}
\fmftop{i3,o3}
\fmfv{decor.shape=circle,decor.size=2.0thick,foreground=(0,,0,,1)}{v1}
\fmfv{decor.shape=circle,decor.size=2.0thick,foreground=(0,,0,,1)}{v2}
\fmfv{decor.shape=circle,decor.size=2.0thick,foreground=(0,,0,,1)}{v3}
\fmfv{decor.shape=circle,decor.size=2.0thick,foreground=(0,,0,,1)}{v4}
\fmfv{decor.shape=circle,decor.size=2.0thick,foreground=(0,,0,,1)}{v5}
\fmfv{decor.shape=circle,decor.size=2.0thick,foreground=(0,,0,,1)}{v6}
\fmfv{decor.shape=circle,decor.filled=empty,decor.size=.073w,l=$\times$,label.dist=0}{v1b}
\fmfv{decor.shape=circle,decor.filled=empty,decor.size=.073w,l=$\times$,label.dist=0}{v3b}
\fmfv{decor.shape=circle,decor.filled=empty,decor.size=.073w,l=$\times$,label.dist=0}{v4b}
\fmfv{decor.shape=circle,decor.filled=empty,decor.size=.073w,l=$\times$,label.dist=0}{v6b}
\fmf{phantom,tension=1.4}{i1,v1}
\fmf{phantom}{i2,v3b}
\fmf{phantom}{i0,v1b}
\fmf{phantom}{o2,v6b}
\fmf{phantom,tension=1.4}{o1,v5}
\fmf{phantom}{o0,v5b}
\fmf{phantom,tension=1.11}{v3b,v6b}
\fmf{phantom,tension=1.38}{i0,v2}
\fmf{phantom,tension=1.38}{o0,v2}
\fmf{phantom,tension=1.8}{i0,v2b}
\fmf{phantom,tension=1.2}{o0,v2b}
\fmf{phantom,tension=1.2}{b0,v2b}
\fmf{phantom,tension=1.2}{b1,v2b}
\fmf{phantom,tension=1.2}{b2,v2b}
\fmf{phantom,tension=1.38}{i0,v4}
\fmf{phantom,tension=1.38}{o0,v4}
\fmf{phantom,tension=1.2}{i0,v4b}
\fmf{phantom,tension=1.8}{o0,v4b}
\fmf{phantom,tension=1.2}{b0,v4b}
\fmf{phantom,tension=1.2}{b1,v4b}
\fmf{phantom,tension=1.2}{b2,v4b}
\fmf{phantom,tension=2}{i3,v3}
\fmf{phantom,tension=2}{o3,v6}
\fmf{phantom,tension=2}{i3,v3b}
\fmf{phantom,tension=0.8}{o3,v3b}
\fmf{phantom,tension=0.8}{i3,v6b}
\fmf{phantom,tension=2}{o3,v6b}
\fmf{plain,tension=1.4}{v1,v2}
\fmf{plain,tension=1.4}{v4,v5}
\fmf{phantom}{v1,v3}
\fmf{phantom,left=0.8,tension=0}{v1,v3}
\fmf{plain}{v5,v6}
\fmf{phantom,right=0.8,tension=0}{v5,v6}
\fmf{plain}{v2,v3}
\fmf{phantom}{v4,v6}
\fmf{wiggly,tension=0.5}{v2,v4}
\fmf{wiggly,tension=2}{v3,v6}
\fmf{wiggly,right=0.8,tension=0}{v1,v5}
\fmf{plain,tension=1}{v1,v1b}
\fmf{phantom,tension=0.2}{v2,v2b}
\fmf{plain,tension=1.5}{v3,v3b}
\fmf{plain,tension=0.2}{v4,v4b}
\fmf{phantom,tension=1}{v5,v5b}
\fmf{plain,tension=1.5}{v6,v6b}
\end{fmfgraph*}
\end{fmffile}
\end{gathered} \hspace{-0.5cm} + 3 \begin{gathered}
\begin{fmffile}{Diagrams/LoopExpansionBosonicHS_W_Diag8}
\begin{fmfgraph*}(35,20)
\fmfleft{i1,i2}
\fmfright{o1,o2}
\fmfbottom{i0,o0}
\fmfbottom{b0}
\fmfbottom{b1}
\fmfbottom{b2}
\fmftop{i3,o3}
\fmfv{decor.shape=circle,decor.size=2.0thick,foreground=(0,,0,,1)}{v1}
\fmfv{decor.shape=circle,decor.size=2.0thick,foreground=(0,,0,,1)}{v2}
\fmfv{decor.shape=circle,decor.size=2.0thick,foreground=(0,,0,,1)}{v3}
\fmfv{decor.shape=circle,decor.size=2.0thick,foreground=(0,,0,,1)}{v4}
\fmfv{decor.shape=circle,decor.size=2.0thick,foreground=(0,,0,,1)}{v5}
\fmfv{decor.shape=circle,decor.size=2.0thick,foreground=(0,,0,,1)}{v6}
\fmfv{decor.shape=circle,decor.filled=empty,decor.size=.073w,l=$\times$,label.dist=0}{v1b}
\fmfv{decor.shape=circle,decor.filled=empty,decor.size=.073w,l=$\times$,label.dist=0}{v3b}
\fmfv{decor.shape=circle,decor.filled=empty,decor.size=.073w,l=$\times$,label.dist=0}{v5b}
\fmfv{decor.shape=circle,decor.filled=empty,decor.size=.073w,l=$\times$,label.dist=0}{v6b}
\fmf{phantom,tension=1.4}{i1,v1}
\fmf{phantom}{i2,v3b}
\fmf{phantom}{i0,v1b}
\fmf{phantom}{o2,v6b}
\fmf{phantom,tension=1.4}{o1,v5}
\fmf{phantom}{o0,v5b}
\fmf{phantom,tension=1.11}{v3b,v6b}
\fmf{phantom,tension=1.38}{i0,v2}
\fmf{phantom,tension=1.38}{o0,v2}
\fmf{phantom,tension=1.8}{i0,v2b}
\fmf{phantom,tension=1.2}{o0,v2b}
\fmf{phantom,tension=1.2}{b0,v2b}
\fmf{phantom,tension=1.2}{b1,v2b}
\fmf{phantom,tension=1.2}{b2,v2b}
\fmf{phantom,tension=1.38}{i0,v4}
\fmf{phantom,tension=1.38}{o0,v4}
\fmf{phantom,tension=1.2}{i0,v4b}
\fmf{phantom,tension=1.8}{o0,v4b}
\fmf{phantom,tension=1.2}{b0,v4b}
\fmf{phantom,tension=1.2}{b1,v4b}
\fmf{phantom,tension=1.2}{b2,v4b}
\fmf{phantom,tension=2}{i3,v3}
\fmf{phantom,tension=2}{o3,v6}
\fmf{phantom,tension=2}{i3,v3b}
\fmf{phantom,tension=0.8}{o3,v3b}
\fmf{phantom,tension=0.8}{i3,v6b}
\fmf{phantom,tension=2}{o3,v6b}
\fmf{plain,tension=1.4}{v1,v2}
\fmf{plain,tension=1.4}{v4,v5}
\fmf{phantom}{v1,v3}
\fmf{phantom,left=0.8,tension=0}{v1,v3}
\fmf{phantom}{v5,v6}
\fmf{phantom,right=0.8,tension=0}{v5,v6}
\fmf{plain}{v2,v3}
\fmf{plain}{v4,v6}
\fmf{wiggly,tension=0.5}{v2,v4}
\fmf{wiggly,tension=2}{v3,v6}
\fmf{wiggly,right=0.8,tension=0}{v1,v5}
\fmf{plain,tension=1}{v1,v1b}
\fmf{phantom,tension=0.2}{v2,v2b}
\fmf{plain,tension=1.5}{v3,v3b}
\fmf{phantom,tension=0.2}{v4,v4b}
\fmf{plain,tension=1}{v5,v5b}
\fmf{plain,tension=1.5}{v6,v6b}
\end{fmfgraph*}
\end{fmffile}
\end{gathered} \hspace{-0.3cm} \\
& \hspace{1.2cm} + 6 \begin{gathered}
\begin{fmffile}{Diagrams/LoopExpansionBosonicHS_W_Diag9}
\begin{fmfgraph*}(35,20)
\fmfleft{i1,i2}
\fmfright{o1,o2}
\fmfbottom{i0,o0}
\fmfbottom{b0}
\fmfbottom{b1}
\fmfbottom{b2}
\fmftop{i3,o3}
\fmfv{decor.shape=circle,decor.size=2.0thick,foreground=(0,,0,,1)}{v1}
\fmfv{decor.shape=circle,decor.size=2.0thick,foreground=(0,,0,,1)}{v2}
\fmfv{decor.shape=circle,decor.size=2.0thick,foreground=(0,,0,,1)}{v3}
\fmfv{decor.shape=circle,decor.size=2.0thick,foreground=(0,,0,,1)}{v4}
\fmfv{decor.shape=circle,decor.size=2.0thick,foreground=(0,,0,,1)}{v5}
\fmfv{decor.shape=circle,decor.size=2.0thick,foreground=(0,,0,,1)}{v6}
\fmfv{decor.shape=circle,decor.filled=empty,decor.size=.073w,l=$\times$,label.dist=0}{v1b}
\fmfv{decor.shape=circle,decor.filled=empty,decor.size=.073w,l=$\times$,label.dist=0}{v3b}
\fmf{phantom,tension=1.4}{i1,v1}
\fmf{phantom}{i2,v3b}
\fmf{phantom}{i0,v1b}
\fmf{phantom}{o2,v6b}
\fmf{phantom,tension=1.4}{o1,v5}
\fmf{phantom}{o0,v5b}
\fmf{phantom,tension=1.11}{v3b,v6b}
\fmf{phantom,tension=1.38}{i0,v2}
\fmf{phantom,tension=1.38}{o0,v2}
\fmf{phantom,tension=1.8}{i0,v2b}
\fmf{phantom,tension=1.2}{o0,v2b}
\fmf{phantom,tension=1.2}{b0,v2b}
\fmf{phantom,tension=1.2}{b1,v2b}
\fmf{phantom,tension=1.2}{b2,v2b}
\fmf{phantom,tension=1.38}{i0,v4}
\fmf{phantom,tension=1.38}{o0,v4}
\fmf{phantom,tension=1.2}{i0,v4b}
\fmf{phantom,tension=1.8}{o0,v4b}
\fmf{phantom,tension=1.2}{b0,v4b}
\fmf{phantom,tension=1.2}{b1,v4b}
\fmf{phantom,tension=1.2}{b2,v4b}
\fmf{phantom,tension=2}{i3,v3}
\fmf{phantom,tension=2}{o3,v6}
\fmf{phantom,tension=2}{i3,v3b}
\fmf{phantom,tension=0.8}{o3,v3b}
\fmf{phantom,tension=0.8}{i3,v6b}
\fmf{phantom,tension=2}{o3,v6b}
\fmf{plain,tension=1.4}{v1,v2}
\fmf{plain,tension=1.4}{v4,v5}
\fmf{phantom}{v1,v3}
\fmf{phantom,left=0.8,tension=0}{v1,v3}
\fmf{plain}{v5,v6}
\fmf{phantom,right=0.8,tension=0}{v5,v6}
\fmf{plain}{v2,v3}
\fmf{plain}{v4,v6}
\fmf{wiggly,tension=0.5}{v2,v4}
\fmf{wiggly,tension=2}{v3,v6}
\fmf{wiggly,right=0.8,tension=0}{v1,v5}
\fmf{plain,tension=1}{v1,v1b}
\fmf{phantom,tension=0.2}{v2,v2b}
\fmf{plain,tension=1.5}{v3,v3b}
\fmf{phantom,tension=0.2}{v4,v4b}
\fmf{phantom,tension=1}{v5,v5b}
\fmf{phantom,tension=1.5}{v6,v6b}
\end{fmfgraph*}
\end{fmffile}
\end{gathered} \hspace{-0.5cm} + \hspace{-0.5cm} \begin{gathered}
\begin{fmffile}{Diagrams/LoopExpansionBosonicHS_W_Diag10}
\begin{fmfgraph*}(35,20)
\fmfleft{i1,i2}
\fmfright{o1,o2}
\fmfbottom{i0,o0}
\fmfbottom{b0}
\fmfbottom{b1}
\fmfbottom{b2}
\fmftop{i3,o3}
\fmfv{decor.shape=circle,decor.size=2.0thick,foreground=(0,,0,,1)}{v1}
\fmfv{decor.shape=circle,decor.size=2.0thick,foreground=(0,,0,,1)}{v2}
\fmfv{decor.shape=circle,decor.size=2.0thick,foreground=(0,,0,,1)}{v3}
\fmfv{decor.shape=circle,decor.size=2.0thick,foreground=(0,,0,,1)}{v4}
\fmfv{decor.shape=circle,decor.size=2.0thick,foreground=(0,,0,,1)}{v5}
\fmfv{decor.shape=circle,decor.size=2.0thick,foreground=(0,,0,,1)}{v6}
\fmf{phantom,tension=1.4}{i1,v1}
\fmf{phantom}{i2,v3b}
\fmf{phantom}{i0,v1b}
\fmf{phantom}{o2,v6b}
\fmf{phantom,tension=1.4}{o1,v5}
\fmf{phantom}{o0,v5b}
\fmf{phantom,tension=1.11}{v3b,v6b}
\fmf{phantom,tension=1.38}{i0,v2}
\fmf{phantom,tension=1.38}{o0,v2}
\fmf{phantom,tension=1.8}{i0,v2b}
\fmf{phantom,tension=1.2}{o0,v2b}
\fmf{phantom,tension=1.2}{b0,v2b}
\fmf{phantom,tension=1.2}{b1,v2b}
\fmf{phantom,tension=1.2}{b2,v2b}
\fmf{phantom,tension=1.38}{i0,v4}
\fmf{phantom,tension=1.38}{o0,v4}
\fmf{phantom,tension=1.2}{i0,v4b}
\fmf{phantom,tension=1.8}{o0,v4b}
\fmf{phantom,tension=1.2}{b0,v4b}
\fmf{phantom,tension=1.2}{b1,v4b}
\fmf{phantom,tension=1.2}{b2,v4b}
\fmf{phantom,tension=2}{i3,v3}
\fmf{phantom,tension=2}{o3,v6}
\fmf{phantom,tension=2}{i3,v3b}
\fmf{phantom,tension=0.8}{o3,v3b}
\fmf{phantom,tension=0.8}{i3,v6b}
\fmf{phantom,tension=2}{o3,v6b}
\fmf{plain,tension=1.4}{v1,v2}
\fmf{plain,tension=1.4}{v4,v5}
\fmf{plain}{v1,v3}
\fmf{phantom,left=0.8,tension=0}{v1,v3}
\fmf{plain}{v5,v6}
\fmf{phantom,right=0.8,tension=0}{v5,v6}
\fmf{plain}{v2,v3}
\fmf{plain}{v4,v6}
\fmf{wiggly,tension=0.5}{v2,v4}
\fmf{wiggly,tension=2}{v3,v6}
\fmf{wiggly,right=0.8,tension=0}{v1,v5}
\fmf{phantom,tension=1}{v1,v1b}
\fmf{phantom,tension=0.2}{v2,v2b}
\fmf{phantom,tension=1.5}{v3,v3b}
\fmf{phantom,tension=0.2}{v4,v4b}
\fmf{phantom,tension=1}{v5,v5b}
\fmf{phantom,tension=1.5}{v6,v6b}
\end{fmfgraph*}
\end{fmffile}
\end{gathered} \hspace{-0.6cm} \left.\rule{0cm}{1.1cm}\right) \\
& + \frac{1}{8} \left(\rule{0cm}{1.1cm}\right. 4 \begin{gathered}
\begin{fmffile}{Diagrams/LoopExpansionBosonicHS_W_Diag11}
\begin{fmfgraph*}(35,20)
\fmfleft{i1,i2}
\fmfright{o1,o2}
\fmfbottom{i0,o0}
\fmfbottom{b0}
\fmfbottom{b1}
\fmfbottom{b2}
\fmftop{i3,o3}
\fmfv{decor.shape=circle,decor.size=2.0thick,foreground=(0,,0,,1)}{v1}
\fmfv{decor.shape=circle,decor.size=2.0thick,foreground=(0,,0,,1)}{v2}
\fmfv{decor.shape=circle,decor.size=2.0thick,foreground=(0,,0,,1)}{v3}
\fmfv{decor.shape=circle,decor.size=2.0thick,foreground=(0,,0,,1)}{v4}
\fmfv{decor.shape=circle,decor.size=2.0thick,foreground=(0,,0,,1)}{v5}
\fmfv{decor.shape=circle,decor.size=2.0thick,foreground=(0,,0,,1)}{v6}
\fmfv{decor.shape=circle,decor.filled=empty,decor.size=.073w,l=$\times$,label.dist=0}{v1b}
\fmfv{decor.shape=circle,decor.filled=empty,decor.size=.073w,l=$\times$,label.dist=0}{v3b}
\fmfv{decor.shape=circle,decor.filled=empty,decor.size=.073w,l=$\times$,label.dist=0}{v4b}
\fmfv{decor.shape=circle,decor.filled=empty,decor.size=.073w,l=$\times$,label.dist=0}{v6b}
\fmf{phantom,tension=1.4}{i1,v1}
\fmf{phantom}{i2,v3b}
\fmf{phantom}{i0,v1b}
\fmf{phantom}{o2,v6b}
\fmf{phantom,tension=1.4}{o1,v5}
\fmf{phantom}{o0,v5b}
\fmf{phantom,tension=1.11}{v3b,v6b}
\fmf{phantom,tension=1.38}{i0,v2}
\fmf{phantom,tension=1.38}{o0,v2}
\fmf{phantom,tension=1.8}{i0,v2b}
\fmf{phantom,tension=1.2}{o0,v2b}
\fmf{phantom,tension=1.2}{b0,v2b}
\fmf{phantom,tension=1.2}{b1,v2b}
\fmf{phantom,tension=1.2}{b2,v2b}
\fmf{phantom,tension=1.38}{i0,v4}
\fmf{phantom,tension=1.38}{o0,v4}
\fmf{phantom,tension=1.2}{i0,v4b}
\fmf{phantom,tension=1.8}{o0,v4b}
\fmf{phantom,tension=1.2}{b0,v4b}
\fmf{phantom,tension=1.2}{b1,v4b}
\fmf{phantom,tension=1.2}{b2,v4b}
\fmf{phantom,tension=2}{i3,v3}
\fmf{phantom,tension=2}{o3,v6}
\fmf{phantom,tension=2}{i3,v3b}
\fmf{phantom,tension=0.8}{o3,v3b}
\fmf{phantom,tension=0.8}{i3,v6b}
\fmf{phantom,tension=2}{o3,v6b}
\fmf{plain,tension=1.4}{v1,v2}
\fmf{plain,tension=1.4}{v4,v5}
\fmf{wiggly}{v1,v3}
\fmf{phantom,left=0.8,tension=0}{v1,v3}
\fmf{wiggly}{v5,v6}
\fmf{plain,right=0.8,tension=0}{v5,v6}
\fmf{plain}{v2,v3}
\fmf{phantom}{v4,v6}
\fmf{wiggly,tension=0.5}{v2,v4}
\fmf{phantom,tension=2}{v3,v6}
\fmf{phantom,right=0.8,tension=0}{v1,v5}
\fmf{plain,tension=1}{v1,v1b}
\fmf{phantom,tension=0.2}{v2,v2b}
\fmf{plain,tension=1.5}{v3,v3b}
\fmf{plain,tension=0.2}{v4,v4b}
\fmf{phantom,tension=1}{v5,v5b}
\fmf{plain,tension=1.5}{v6,v6b}
\end{fmfgraph*}
\end{fmffile}
\end{gathered} \hspace{-0.4cm} + 4 \begin{gathered}
\begin{fmffile}{Diagrams/LoopExpansionBosonicHS_W_Diag12}
\begin{fmfgraph*}(35,20)
\fmfleft{i1,i2}
\fmfright{o1,o2}
\fmfbottom{i0,o0}
\fmfbottom{b0}
\fmfbottom{b1}
\fmfbottom{b2}
\fmftop{i3,o3}
\fmfv{decor.shape=circle,decor.size=2.0thick,foreground=(0,,0,,1)}{v1}
\fmfv{decor.shape=circle,decor.size=2.0thick,foreground=(0,,0,,1)}{v2}
\fmfv{decor.shape=circle,decor.size=2.0thick,foreground=(0,,0,,1)}{v3}
\fmfv{decor.shape=circle,decor.size=2.0thick,foreground=(0,,0,,1)}{v4}
\fmfv{decor.shape=circle,decor.size=2.0thick,foreground=(0,,0,,1)}{v5}
\fmfv{decor.shape=circle,decor.size=2.0thick,foreground=(0,,0,,1)}{v6}
\fmfv{decor.shape=circle,decor.filled=empty,decor.size=.073w,l=$\times$,label.dist=0}{v2b}
\fmfv{decor.shape=circle,decor.filled=empty,decor.size=.073w,l=$\times$,label.dist=0}{v3b}
\fmfv{decor.shape=circle,decor.filled=empty,decor.size=.073w,l=$\times$,label.dist=0}{v4b}
\fmfv{decor.shape=circle,decor.filled=empty,decor.size=.073w,l=$\times$,label.dist=0}{v6b}
\fmf{phantom,tension=1.4}{i1,v1}
\fmf{phantom}{i2,v3b}
\fmf{phantom}{i0,v1b}
\fmf{phantom}{o2,v6b}
\fmf{phantom,tension=1.4}{o1,v5}
\fmf{phantom}{o0,v5b}
\fmf{phantom,tension=1.11}{v3b,v6b}
\fmf{phantom,tension=1.38}{i0,v2}
\fmf{phantom,tension=1.38}{o0,v2}
\fmf{phantom,tension=1.8}{i0,v2b}
\fmf{phantom,tension=1.2}{o0,v2b}
\fmf{phantom,tension=1.2}{b0,v2b}
\fmf{phantom,tension=1.2}{b1,v2b}
\fmf{phantom,tension=1.2}{b2,v2b}
\fmf{phantom,tension=1.38}{i0,v4}
\fmf{phantom,tension=1.38}{o0,v4}
\fmf{phantom,tension=1.2}{i0,v4b}
\fmf{phantom,tension=1.8}{o0,v4b}
\fmf{phantom,tension=1.2}{b0,v4b}
\fmf{phantom,tension=1.2}{b1,v4b}
\fmf{phantom,tension=1.2}{b2,v4b}
\fmf{phantom,tension=2}{i3,v3}
\fmf{phantom,tension=2}{o3,v6}
\fmf{phantom,tension=2}{i3,v3b}
\fmf{phantom,tension=0.8}{o3,v3b}
\fmf{phantom,tension=0.8}{i3,v6b}
\fmf{phantom,tension=2}{o3,v6b}
\fmf{plain,tension=1.4}{v1,v2}
\fmf{plain,tension=1.4}{v4,v5}
\fmf{wiggly}{v1,v3}
\fmf{plain,left=0.8,tension=0}{v1,v3}
\fmf{wiggly}{v5,v6}
\fmf{plain,right=0.8,tension=0}{v5,v6}
\fmf{phantom}{v2,v3}
\fmf{phantom}{v4,v6}
\fmf{wiggly,tension=0.5}{v2,v4}
\fmf{phantom,tension=2}{v3,v6}
\fmf{phantom,right=0.8,tension=0}{v1,v5}
\fmf{phantom,tension=1}{v1,v1b}
\fmf{plain,tension=0.2}{v2,v2b}
\fmf{plain,tension=1.5}{v3,v3b}
\fmf{plain,tension=0.2}{v4,v4b}
\fmf{phantom,tension=1}{v5,v5b}
\fmf{plain,tension=1.5}{v6,v6b}
\end{fmfgraph*}
\end{fmffile}
\end{gathered} \hspace{-0.4cm} + \hspace{-0.2cm} \begin{gathered}
\begin{fmffile}{Diagrams/LoopExpansionBosonicHS_W_Diag13}
\begin{fmfgraph*}(35,20)
\fmfleft{i1,i2}
\fmfright{o1,o2}
\fmfbottom{i0,o0}
\fmfbottom{b0}
\fmfbottom{b1}
\fmfbottom{b2}
\fmftop{i3,o3}
\fmfv{decor.shape=circle,decor.size=2.0thick,foreground=(0,,0,,1)}{v1}
\fmfv{decor.shape=circle,decor.size=2.0thick,foreground=(0,,0,,1)}{v2}
\fmfv{decor.shape=circle,decor.size=2.0thick,foreground=(0,,0,,1)}{v3}
\fmfv{decor.shape=circle,decor.size=2.0thick,foreground=(0,,0,,1)}{v4}
\fmfv{decor.shape=circle,decor.size=2.0thick,foreground=(0,,0,,1)}{v5}
\fmfv{decor.shape=circle,decor.size=2.0thick,foreground=(0,,0,,1)}{v6}
\fmfv{decor.shape=circle,decor.filled=empty,decor.size=.073w,l=$\times$,label.dist=0}{v1b}
\fmfv{decor.shape=circle,decor.filled=empty,decor.size=.073w,l=$\times$,label.dist=0}{v3b}
\fmfv{decor.shape=circle,decor.filled=empty,decor.size=.073w,l=$\times$,label.dist=0}{v5b}
\fmfv{decor.shape=circle,decor.filled=empty,decor.size=.073w,l=$\times$,label.dist=0}{v6b}
\fmf{phantom,tension=1.4}{i1,v1}
\fmf{phantom}{i2,v3b}
\fmf{phantom}{i0,v1b}
\fmf{phantom}{o2,v6b}
\fmf{phantom,tension=1.4}{o1,v5}
\fmf{phantom}{o0,v5b}
\fmf{phantom,tension=1.11}{v3b,v6b}
\fmf{phantom,tension=1.38}{i0,v2}
\fmf{phantom,tension=1.38}{o0,v2}
\fmf{phantom,tension=1.8}{i0,v2b}
\fmf{phantom,tension=1.2}{o0,v2b}
\fmf{phantom,tension=1.2}{b0,v2b}
\fmf{phantom,tension=1.2}{b1,v2b}
\fmf{phantom,tension=1.2}{b2,v2b}
\fmf{phantom,tension=1.38}{i0,v4}
\fmf{phantom,tension=1.38}{o0,v4}
\fmf{phantom,tension=1.2}{i0,v4b}
\fmf{phantom,tension=1.8}{o0,v4b}
\fmf{phantom,tension=1.2}{b0,v4b}
\fmf{phantom,tension=1.2}{b1,v4b}
\fmf{phantom,tension=1.2}{b2,v4b}
\fmf{phantom,tension=2}{i3,v3}
\fmf{phantom,tension=2}{o3,v6}
\fmf{phantom,tension=2}{i3,v3b}
\fmf{phantom,tension=0.8}{o3,v3b}
\fmf{phantom,tension=0.8}{i3,v6b}
\fmf{phantom,tension=2}{o3,v6b}
\fmf{plain,tension=1.4}{v1,v2}
\fmf{plain,tension=1.4}{v4,v5}
\fmf{wiggly}{v1,v3}
\fmf{phantom,left=0.8,tension=0}{v1,v3}
\fmf{wiggly}{v5,v6}
\fmf{phantom,right=0.8,tension=0}{v5,v6}
\fmf{plain}{v2,v3}
\fmf{plain}{v4,v6}
\fmf{wiggly,tension=0.5}{v2,v4}
\fmf{phantom,tension=2}{v3,v6}
\fmf{phantom,right=0.8,tension=0}{v1,v5}
\fmf{plain,tension=1}{v1,v1b}
\fmf{phantom,tension=0.2}{v2,v2b}
\fmf{plain,tension=1.5}{v3,v3b}
\fmf{phantom,tension=0.2}{v4,v4b}
\fmf{plain,tension=1}{v5,v5b}
\fmf{plain,tension=1.5}{v6,v6b}
\end{fmfgraph*}
\end{fmffile}
\end{gathered} \\
& \hspace{1.0cm} + 4 \hspace{-0.4cm} \begin{gathered}
\begin{fmffile}{Diagrams/LoopExpansionBosonicHS_W_Diag14}
\begin{fmfgraph*}(35,20)
\fmfleft{i1,i2}
\fmfright{o1,o2}
\fmfbottom{i0,o0}
\fmfbottom{b0}
\fmfbottom{b1}
\fmfbottom{b2}
\fmftop{i3,o3}
\fmfv{decor.shape=circle,decor.size=2.0thick,foreground=(0,,0,,1)}{v1}
\fmfv{decor.shape=circle,decor.size=2.0thick,foreground=(0,,0,,1)}{v2}
\fmfv{decor.shape=circle,decor.size=2.0thick,foreground=(0,,0,,1)}{v3}
\fmfv{decor.shape=circle,decor.size=2.0thick,foreground=(0,,0,,1)}{v4}
\fmfv{decor.shape=circle,decor.size=2.0thick,foreground=(0,,0,,1)}{v5}
\fmfv{decor.shape=circle,decor.size=2.0thick,foreground=(0,,0,,1)}{v6}
\fmfv{decor.shape=circle,decor.filled=empty,decor.size=.073w,l=$\times$,label.dist=0}{v2b}
\fmfv{decor.shape=circle,decor.filled=empty,decor.size=.073w,l=$\times$,label.dist=0}{v3b}
\fmf{phantom,tension=1.4}{i1,v1}
\fmf{phantom}{i2,v3b}
\fmf{phantom}{i0,v1b}
\fmf{phantom}{o2,v6b}
\fmf{phantom,tension=1.4}{o1,v5}
\fmf{phantom}{o0,v5b}
\fmf{phantom,tension=1.11}{v3b,v6b}
\fmf{phantom,tension=1.38}{i0,v2}
\fmf{phantom,tension=1.38}{o0,v2}
\fmf{phantom,tension=1.8}{i0,v2b}
\fmf{phantom,tension=1.2}{o0,v2b}
\fmf{phantom,tension=1.2}{b0,v2b}
\fmf{phantom,tension=1.2}{b1,v2b}
\fmf{phantom,tension=1.2}{b2,v2b}
\fmf{phantom,tension=1.38}{i0,v4}
\fmf{phantom,tension=1.38}{o0,v4}
\fmf{phantom,tension=1.2}{i0,v4b}
\fmf{phantom,tension=1.8}{o0,v4b}
\fmf{phantom,tension=1.2}{b0,v4b}
\fmf{phantom,tension=1.2}{b1,v4b}
\fmf{phantom,tension=1.2}{b2,v4b}
\fmf{phantom,tension=2}{i3,v3}
\fmf{phantom,tension=2}{o3,v6}
\fmf{phantom,tension=2}{i3,v3b}
\fmf{phantom,tension=0.8}{o3,v3b}
\fmf{phantom,tension=0.8}{i3,v6b}
\fmf{phantom,tension=2}{o3,v6b}
\fmf{plain,tension=1.4}{v1,v2}
\fmf{plain,tension=1.4}{v4,v5}
\fmf{wiggly}{v1,v3}
\fmf{plain,left=0.8,tension=0}{v1,v3}
\fmf{wiggly}{v5,v6}
\fmf{plain,right=0.8,tension=0}{v5,v6}
\fmf{phantom}{v2,v3}
\fmf{plain}{v4,v6}
\fmf{wiggly,tension=0.5}{v2,v4}
\fmf{phantom,tension=2}{v3,v6}
\fmf{phantom,right=0.8,tension=0}{v1,v5}
\fmf{phantom,tension=1}{v1,v1b}
\fmf{plain,tension=0.2}{v2,v2b}
\fmf{plain,tension=1.5}{v3,v3b}
\fmf{phantom,tension=0.2}{v4,v4b}
\fmf{phantom,tension=1}{v5,v5b}
\fmf{phantom,tension=1.5}{v6,v6b}
\end{fmfgraph*}
\end{fmffile}
\end{gathered} \hspace{-0.4cm} + 2 \begin{gathered}
\begin{fmffile}{Diagrams/LoopExpansionBosonicHS_W_Diag15}
\begin{fmfgraph*}(35,20)
\fmfleft{i1,i2}
\fmfright{o1,o2}
\fmfbottom{i0,o0}
\fmfbottom{b0}
\fmfbottom{b1}
\fmfbottom{b2}
\fmftop{i3,o3}
\fmfv{decor.shape=circle,decor.size=2.0thick,foreground=(0,,0,,1)}{v1}
\fmfv{decor.shape=circle,decor.size=2.0thick,foreground=(0,,0,,1)}{v2}
\fmfv{decor.shape=circle,decor.size=2.0thick,foreground=(0,,0,,1)}{v3}
\fmfv{decor.shape=circle,decor.size=2.0thick,foreground=(0,,0,,1)}{v4}
\fmfv{decor.shape=circle,decor.size=2.0thick,foreground=(0,,0,,1)}{v5}
\fmfv{decor.shape=circle,decor.size=2.0thick,foreground=(0,,0,,1)}{v6}
\fmfv{decor.shape=circle,decor.filled=empty,decor.size=.073w,l=$\times$,label.dist=0}{v1b}
\fmfv{decor.shape=circle,decor.filled=empty,decor.size=.073w,l=$\times$,label.dist=0}{v3b}
\fmf{phantom,tension=1.4}{i1,v1}
\fmf{phantom}{i2,v3b}
\fmf{phantom}{i0,v1b}
\fmf{phantom}{o2,v6b}
\fmf{phantom,tension=1.4}{o1,v5}
\fmf{phantom}{o0,v5b}
\fmf{phantom,tension=1.11}{v3b,v6b}
\fmf{phantom,tension=1.38}{i0,v2}
\fmf{phantom,tension=1.38}{o0,v2}
\fmf{phantom,tension=1.8}{i0,v2b}
\fmf{phantom,tension=1.2}{o0,v2b}
\fmf{phantom,tension=1.2}{b0,v2b}
\fmf{phantom,tension=1.2}{b1,v2b}
\fmf{phantom,tension=1.2}{b2,v2b}
\fmf{phantom,tension=1.38}{i0,v4}
\fmf{phantom,tension=1.38}{o0,v4}
\fmf{phantom,tension=1.2}{i0,v4b}
\fmf{phantom,tension=1.8}{o0,v4b}
\fmf{phantom,tension=1.2}{b0,v4b}
\fmf{phantom,tension=1.2}{b1,v4b}
\fmf{phantom,tension=1.2}{b2,v4b}
\fmf{phantom,tension=2}{i3,v3}
\fmf{phantom,tension=2}{o3,v6}
\fmf{phantom,tension=2}{i3,v3b}
\fmf{phantom,tension=0.8}{o3,v3b}
\fmf{phantom,tension=0.8}{i3,v6b}
\fmf{phantom,tension=2}{o3,v6b}
\fmf{plain,tension=1.4}{v1,v2}
\fmf{plain,tension=1.4}{v4,v5}
\fmf{wiggly}{v1,v3}
\fmf{phantom,left=0.8,tension=0}{v1,v3}
\fmf{wiggly}{v5,v6}
\fmf{plain,right=0.8,tension=0}{v5,v6}
\fmf{plain}{v2,v3}
\fmf{plain}{v4,v6}
\fmf{wiggly,tension=0.5}{v2,v4}
\fmf{phantom,tension=2}{v3,v6}
\fmf{phantom,right=0.8,tension=0}{v1,v5}
\fmf{plain,tension=1}{v1,v1b}
\fmf{phantom,tension=0.2}{v2,v2b}
\fmf{plain,tension=1.5}{v3,v3b}
\fmf{phantom,tension=0.2}{v4,v4b}
\fmf{phantom,tension=1}{v5,v5b}
\fmf{phantom,tension=1.5}{v6,v6b}
\end{fmfgraph*}
\end{fmffile}
\end{gathered} \hspace{-0.4cm} + \hspace{-0.4cm} \begin{gathered}
\begin{fmffile}{Diagrams/LoopExpansionBosonicHS_W_Diag16}
\begin{fmfgraph*}(35,20)
\fmfleft{i1,i2}
\fmfright{o1,o2}
\fmfbottom{i0,o0}
\fmfbottom{b0}
\fmfbottom{b1}
\fmfbottom{b2}
\fmftop{i3,o3}
\fmfv{decor.shape=circle,decor.size=2.0thick,foreground=(0,,0,,1)}{v1}
\fmfv{decor.shape=circle,decor.size=2.0thick,foreground=(0,,0,,1)}{v2}
\fmfv{decor.shape=circle,decor.size=2.0thick,foreground=(0,,0,,1)}{v3}
\fmfv{decor.shape=circle,decor.size=2.0thick,foreground=(0,,0,,1)}{v4}
\fmfv{decor.shape=circle,decor.size=2.0thick,foreground=(0,,0,,1)}{v5}
\fmfv{decor.shape=circle,decor.size=2.0thick,foreground=(0,,0,,1)}{v6}
\fmf{phantom,tension=1.4}{i1,v1}
\fmf{phantom}{i2,v3b}
\fmf{phantom}{i0,v1b}
\fmf{phantom}{o2,v6b}
\fmf{phantom,tension=1.4}{o1,v5}
\fmf{phantom}{o0,v5b}
\fmf{phantom,tension=1.11}{v3b,v6b}
\fmf{phantom,tension=1.38}{i0,v2}
\fmf{phantom,tension=1.38}{o0,v2}
\fmf{phantom,tension=1.8}{i0,v2b}
\fmf{phantom,tension=1.2}{o0,v2b}
\fmf{phantom,tension=1.2}{b0,v2b}
\fmf{phantom,tension=1.2}{b1,v2b}
\fmf{phantom,tension=1.2}{b2,v2b}
\fmf{phantom,tension=1.38}{i0,v4}
\fmf{phantom,tension=1.38}{o0,v4}
\fmf{phantom,tension=1.2}{i0,v4b}
\fmf{phantom,tension=1.8}{o0,v4b}
\fmf{phantom,tension=1.2}{b0,v4b}
\fmf{phantom,tension=1.2}{b1,v4b}
\fmf{phantom,tension=1.2}{b2,v4b}
\fmf{phantom,tension=2}{i3,v3}
\fmf{phantom,tension=2}{o3,v6}
\fmf{phantom,tension=2}{i3,v3b}
\fmf{phantom,tension=0.8}{o3,v3b}
\fmf{phantom,tension=0.8}{i3,v6b}
\fmf{phantom,tension=2}{o3,v6b}
\fmf{plain,tension=1.4}{v1,v2}
\fmf{plain,tension=1.4}{v4,v5}
\fmf{wiggly}{v1,v3}
\fmf{plain,left=0.8,tension=0}{v1,v3}
\fmf{wiggly}{v5,v6}
\fmf{plain,right=0.8,tension=0}{v5,v6}
\fmf{plain}{v2,v3}
\fmf{plain}{v4,v6}
\fmf{wiggly,tension=0.5}{v2,v4}
\fmf{phantom,tension=2}{v3,v6}
\fmf{phantom,right=0.8,tension=0}{v1,v5}
\fmf{phantom,tension=1}{v1,v1b}
\fmf{phantom,tension=0.2}{v2,v2b}
\fmf{phantom,tension=1.5}{v3,v3b}
\fmf{phantom,tension=0.2}{v4,v4b}
\fmf{phantom,tension=1}{v5,v5b}
\fmf{phantom,tension=1.5}{v6,v6b}
\end{fmfgraph*}
\end{fmffile}
\end{gathered} \hspace{-0.5cm} \left.\rule{0cm}{1.1cm}\right)\;.
\end{split}
\label{eq:bosonic1PIEAIMW20DON}
\end{equation}
The classical 1-point correlation function $\Psi_{\mathrm{cl}}\big[\mathcal{J}\big]$ and the propagator $\mathcal{G}_{\Psi_{\mathrm{cl}};\mathcal{J}}\big[\mathcal{J}\big]$ are given at arbitrary external source $\mathcal{J}$ by:
\begin{subequations}
\begin{empheq}[left=\empheqlbrace]{align}
& \hspace{0.1cm} \Psi_{\mathrm{cl},\beta}\big[\mathcal{J}\big] = \Phi_{0,\beta}\big[\mathcal{J}\big] = \frac{\delta W_{\mathrm{col},0}\big[\mathcal{J}\big]}{\delta \mathcal{J}_{\beta}} = \left(1-\delta_{b \hspace{0.04cm} N+1}\right) \varphi_{\mathrm{cl},\alpha}\big[\mathcal{J}\big] + \delta_{b \hspace{0.04cm} N+1} \sigma_{\mathrm{cl},x}\big[\mathcal{J}\big] \;, \label{eq:bosonic1PIEAphicl0DON}\\
\nonumber \\
& \hspace{0.1cm} \mathcal{G}_{\Psi_{\mathrm{cl}};\mathcal{J},\beta_{1}\beta_{2}}\big[\mathcal{J}\big] = \frac{\delta^{2} W_{\mathrm{col},0}\big[\mathcal{J}\big]}{\delta \mathcal{J}_{\beta_{1}} \delta \mathcal{J}_{\beta_{2}}} = \left(1-\delta_{b_{1} N+1}\right)\left(1-\delta_{b_{2} N+1}\right) \boldsymbol{G}_{\sigma_\text{cl};\mathcal{J},\alpha_{1}\alpha_{2}}\big[\mathcal{J}\big] \nonumber \\
& \hspace{0.1cm} \hspace{5.5cm} + \delta_{b_{1} N+1}\delta_{b_{2} N+1} D_{\sigma_\text{cl};\mathcal{J},x_{1}x_{2}}\big[\mathcal{J}\big] \;, \label{eq:bosonic1PIEAGJ0DON}
\end{empheq}
\end{subequations}
or, in matrix form:
\begin{subequations}
\begin{empheq}[left=\empheqlbrace]{align}
& \hspace{0.1cm} \Psi_{\mathrm{cl}}\big[\mathcal{J}\big]=\begin{pmatrix}
\vec{\varphi}_{\mathrm{cl}}\big[\mathcal{J}\big] & \sigma_{\mathrm{cl}}\big[\mathcal{J}\big]
\end{pmatrix}^{\mathrm{T}}\;, \\
\nonumber \\
& \hspace{0.1cm} \mathcal{G}_{\Psi_{\mathrm{cl}};\mathcal{J}}\big[\mathcal{J}\big] = \begin{pmatrix}
\boldsymbol{G}_{\sigma_{\mathrm{cl}};\mathcal{J}}\big[\mathcal{J}\big] & \vec{0} \\
\vec{0}^{\mathrm{T}} & D_{\sigma_{\mathrm{cl}};\mathcal{J}}\big[\mathcal{J}\big]
\end{pmatrix}\;,
\end{empheq}
\end{subequations}
where $\boldsymbol{G}_{\sigma_{\mathrm{cl}};\mathcal{J}}$ and $D_{\sigma_{\mathrm{cl}};\mathcal{J}}$ are respectively defined by~\eqref{eq:DefGpropagCollectiveLE} and~\eqref{eq:DefDpropagCollectiveLE}, i.e.:
\begin{equation}
\boldsymbol{G}^{-1}_{\sigma_{\mathrm{cl}};\mathcal{J},\alpha_{1}\alpha_{2}}\big[\mathcal{J}\big] \equiv \left(-\nabla^{2}_{x_{1}} + m^{2} + i\sqrt{\frac{\lambda}{3}}\sigma_{\mathrm{cl},x_{1}}\big[\mathcal{J}\big]\right) \delta_{\alpha_{1}\alpha_{2}}\;,
\label{eq:bosonic1PIEApropagatorGJ}
\end{equation}
and
\begin{equation}
D^{-1}_{\sigma_{\mathrm{cl}};\mathcal{J},x_{1}x_{2}}\big[\mathcal{J}\big] \equiv \left.\frac{\delta^{2} S_{\mathrm{col},\mathcal{J}}[\widetilde{\sigma}]}{\delta\widetilde{\sigma}_{x_{1}} \delta\widetilde{\sigma}_{x_{2}}}\right|_{\widetilde{\sigma}=\sigma_{\mathrm{cl}}}\;.
\label{eq:bosonic1PIEApropagatorHJ}
\end{equation}
If~\eqref{eq:bosonic1PIEAphicl0DON} and~\eqref{eq:bosonic1PIEAGJ0DON} are evaluated at $\mathcal{J}=\mathcal{J}_{0}$, we have:
\begin{subequations}
\begin{empheq}[left=\empheqlbrace]{align}
& \hspace{0.1cm} \Phi_{\beta} = \Psi_{\mathrm{cl},\beta}\big[\mathcal{J}=\mathcal{J}_{0}\big] = \Phi_{0,\beta}\big[\mathcal{J}=\mathcal{J}_{0}\big] = \left.\frac{\delta W_{\mathrm{col},0}\big[\mathcal{J}\big]}{\delta \mathcal{J}_{\beta}}\right|_{\mathcal{J}=\mathcal{J}_{0}} \nonumber \\
& \hspace{0.1cm} \hspace{0.5cm} = \left(1-\delta_{b \hspace{0.04cm} N+1}\right) \phi_{\alpha} + \delta_{b \hspace{0.04cm} N+1} \eta_{x}\;, \label{eq:bosonic1PIEAphi0DON}\\
\nonumber \\
& \hspace{0.1cm} \mathcal{G}_{\beta_{1}\beta_{2}}[\Phi] = \mathcal{G}_{\Psi_{\mathrm{cl}};\mathcal{J},\beta_{1}\beta_{2}}\big[\mathcal{J}=\mathcal{J}_{0}\big] = \left.\frac{\delta^{2} W_{\mathrm{col},0}\big[\mathcal{J}\big]}{\delta \mathcal{J}_{\beta_{1}} \delta \mathcal{J}_{\beta_{2}}}\right|_{\mathcal{J}=\mathcal{J}_{0}} \nonumber \\
& \hspace{0.1cm} \hspace{1.421cm} = \left(1-\delta_{b_{1} N+1}\right)\left(1-\delta_{b_{2} N+1}\right) \boldsymbol{G}_{\Phi,\alpha_{1}\alpha_{2}}[\Phi] + \delta_{b_{1} N+1}\delta_{b_{2} N+1} D_{\Phi, x_{1}x_{2}}[\Phi]\;, \label{eq:bosonic1PIEAGJ00DON}
\end{empheq}
\end{subequations}
and, in matrix form:
\begin{subequations}
\begin{empheq}[left=\empheqlbrace]{align}
& \hspace{0.1cm} \Phi = \begin{pmatrix}
\vec{\phi} & \eta
\end{pmatrix}^{\mathrm{T}}\;, \\
\nonumber \\
& \hspace{0.1cm} \mathcal{G}[\Phi] = \begin{pmatrix}
\boldsymbol{G}_{\Phi}[\Phi] & \vec{0} \\
\vec{0}^{\mathrm{T}} & D_{\Phi}[\Phi]
\end{pmatrix}\;,
\end{empheq}
\end{subequations}
where, according to~\eqref{eq:bosonic1PIEApropagatorGJ0main} and~\eqref{eq:bosonic1PIEApropagatorHJ0main}, we have:
\begin{equation}
\boldsymbol{G}^{-1}_{\Phi,\alpha_{1}\alpha_{2}}[\Phi] \equiv \left(-\nabla^{2}_{x_{1}} + m^{2} + i\sqrt{\frac{\lambda}{3}}\eta_{x_{1}}\right) \delta_{\alpha_{1}\alpha_{2}}\;,
\label{eq:bosonic1PIEApropagatorGJ0}
\end{equation}
and
\begin{equation}
D^{-1}_{\Phi,x_{1}x_{2}}[\Phi] \equiv \left.\frac{\delta^{2} S_{\mathrm{col},\mathcal{J}}[\widetilde{\sigma}]}{\delta\widetilde{\sigma}_{x_{1}} \delta\widetilde{\sigma}_{x_{2}}}\right|_{\widetilde{\sigma}=\eta \atop \vec{J}=\vec{J}_{0}}\;.
\label{eq:bosonic1PIEApropagatorHJ0}
\end{equation}
By following the reasoning of~\eqref{eq:pure1PIEAIndependencephihbarstep20DON}, it can be proven from the definition of $\Gamma_{\mathrm{col}}^{(\mathrm{1PI})}$ (i.e.~\eqref{eq:bosonic1PIEAdefinition0DON}) that $\Phi$ is independent of $\hbar$. From the latter definitions, we then work out all the Feynman rules that we will use below to determine $\Gamma_{\mathrm{col}}^{(\mathrm{1PI})}$ via the IM. At arbitrary external source $\mathcal{J}$, these Feynman rules are identical to those of the collective LE (i.e. identical to~\eqref{eq:FeynRulesLoopExpansionBosonicHSG},~\eqref{eq:FeynRulesLoopExpansionBosonicHSH},~\eqref{eq:FeynRulesLoopExpansionBosonicHSvertexDot} and~\eqref{eq:FeynRulesLoopExpansionBosonicHSK}), which are:
\begin{subequations}
\begin{align}
\begin{gathered}
\begin{fmffile}{Diagrams/LEcol-G_Appendix}
\begin{fmfgraph*}(20,12)
\fmfleft{i0,i1,i2,i3}
\fmfright{o0,o1,o2,o3}
\fmflabel{$\alpha_{1}$}{v1}
\fmflabel{$\alpha_{2}$}{v2}
\fmf{phantom}{i1,v1}
\fmf{phantom}{i2,v1}
\fmf{plain,tension=0.6}{v1,v2}
\fmf{phantom}{v2,o1}
\fmf{phantom}{v2,o2}
\end{fmfgraph*}
\end{fmffile}
\end{gathered} \quad &\rightarrow \boldsymbol{G}_{\sigma_\text{cl};\mathcal{J},\alpha_{1}\alpha_{2}}\big[\mathcal{J}\big]\;,
\label{eq:FeynRulesLoopExpansionBosonicHSGAppendix}\\
\begin{gathered}
\begin{fmffile}{Diagrams/LEcol-D_Appendix}
\begin{fmfgraph*}(20,16)
\fmfleft{i0,i1,i2,i3}
\fmfright{o0,o1,o2,o3}
\fmfv{label=$x_{1}$}{v1}
\fmfv{label=$x_{2}$}{v2}
\fmf{phantom}{i1,v1}
\fmf{phantom}{i2,v1}
\fmf{wiggly,tension=0.6}{v1,v2}
\fmf{phantom}{v2,o1}
\fmf{phantom}{v2,o2}
\end{fmfgraph*}
\end{fmffile}
\end{gathered} \quad &\rightarrow D_{\sigma_\text{cl};\mathcal{J},x_{1}x_{2}}\big[\mathcal{J}\big]\;,
\label{eq:FeynRulesLoopExpansionBosonicHSHAppendix}\\
\begin{gathered}
\begin{fmffile}{Diagrams/LEcol-V_Appendix}
\begin{fmfgraph*}(5,5)
\fmfleft{i1}
\fmfright{o1}
\fmfv{label=$x$,label.angle=-90,label.dist=4,foreground=(0,,0,,1)}{v1}
\fmf{plain}{i1,v1}
\fmf{plain}{v1,o1}
\fmflabel{$a_{1}$}{i1}
\fmflabel{$a_{2}$}{o1}
\fmfdot{v1}
\end{fmfgraph*}
\end{fmffile}
\end{gathered}\qquad &\rightarrow i\sqrt{\frac{\lambda}{3}}\delta_{a_{1}a_{2}}\;,
\label{eq:FeynRulesLoopExpansionBosonicHSvertexDotAppendix} \\
\nonumber\\
\begin{gathered}
\begin{fmffile}{Diagrams/LEcol-J_Appendix}
\begin{fmfgraph*}(10,5)
\fmfleft{i1}
\fmfright{o1}
\fmfv{decor.shape=circle,decor.filled=empty,decor.size=.26w,l=$\times$,label.dist=0}{v1}
\fmfv{label.angle=-90,label.dist=6}{v2}
\fmf{plain,tension=2.5}{i1,v1}
\fmf{phantom}{v1,o1}
\fmf{phantom,tension=2.5}{i1,v2}
\fmf{phantom}{v2,o1}
\fmflabel{$\alpha$}{v2}
\end{fmfgraph*}
\end{fmffile}
\end{gathered} \hspace{-0.2cm} &\rightarrow J_{\alpha}\;,
\label{eq:FeynRulesLoopExpansionBosonicHSKAppendix}
\end{align}
\end{subequations}
whereas, at $\mathcal{J}=\mathcal{J}_{0}$, we will rather use:
\begin{subequations}
\begin{align}
\begin{gathered}
\begin{fmffile}{Diagrams/1PIEAcol-G_Appendix}
\begin{fmfgraph*}(20,16)
\fmfleft{i0,i1,i2,i3}
\fmfright{o0,o1,o2,o3}
\fmflabel{$\alpha_{1}$}{v1}
\fmflabel{$\alpha_{2}$}{v2}
\fmf{phantom}{i1,v1}
\fmf{phantom}{i2,v1}
\fmf{plain,tension=0.6,foreground=(1,,0,,0)}{v1,v2}
\fmf{phantom}{v2,o1}
\fmf{phantom}{v2,o2}
\end{fmfgraph*}
\end{fmffile}
\end{gathered} \quad &\rightarrow \boldsymbol{G}_{\Phi,\alpha_{1}\alpha_{2}}[\Phi]\;, 
\label{eq:FeynRulesBosonic1PIEAJ0GAppendix}\\
\begin{gathered}
\begin{fmffile}{Diagrams/1PIEAcol-D_Appendix}
\begin{fmfgraph*}(20,20)
\fmfleft{i0,i1,i2,i3}
\fmfright{o0,o1,o2,o3}
\fmfv{label=$x_{1}$}{v1}
\fmfv{label=$x_{2}$}{v2}
\fmf{phantom}{i1,v1}
\fmf{phantom}{i2,v1}
\fmf{wiggly,tension=0.6,foreground=(1,,0,,0)}{v1,v2}
\fmf{phantom}{v2,o1}
\fmf{phantom}{v2,o2}
\end{fmfgraph*}
\end{fmffile}
\end{gathered} \quad &\rightarrow D_{\Phi,x_{1}x_{2}}[\Phi]\;,
\label{eq:FeynRulesBosonic1PIEAJ0HAppendix} \\
\begin{gathered}
\begin{fmffile}{Diagrams/IPIEAcol-V_Appendix}
\begin{fmfgraph*}(5,5)
\fmfleft{i1}
\fmfright{o1}
\fmfv{label=$x$,label.angle=-90,label.dist=4,foreground=(0,,0,,1)}{v1}
\fmf{plain,foreground=(1,,0,,0)}{i1,v1}
\fmf{plain,foreground=(1,,0,,0)}{v1,o1}
\fmflabel{$a_{1}$}{i1}
\fmflabel{$a_{2}$}{o1}
\fmfdot{v1}
\end{fmfgraph*}
\end{fmffile}
\end{gathered}\qquad &\rightarrow i\sqrt{\frac{\lambda}{3}}\delta_{a_{1}a_{2}}\;,
\label{eq:FeynRulesBosonic1PIEAJ0vertexDotAppendix} \\
\nonumber \\
\begin{gathered}
\begin{fmffile}{Diagrams/bosonic1PIEA_FeynRuleK}
\begin{fmfgraph*}(6,8)
\fmfleft{i1}
\fmfright{o1}
\fmfv{decor.shape=circle,decor.filled=empty,decor.size=0.4cm,label=$n$,label.dist=0}{v1}
\fmfv{label=$\alpha$,label.angle=-90,label.dist=7}{v2}
\fmf{plain,tension=0.5,foreground=(1,,0,,0)}{i1,v1}
\fmf{phantom}{v1,o1}
\fmf{phantom,tension=0.5}{i1,v2}
\fmf{phantom}{v2,o1}
\end{fmfgraph*}
\end{fmffile}
\end{gathered} \quad &\rightarrow J_{n,\alpha}[\Phi]\;,
\label{eq:FeynRulesBosonic1PIEAJ0K} \\
\begin{gathered}
\begin{fmffile}{Diagrams/bosonic1PIEA_FeynRulemathcalG}
\begin{fmfgraph*}(20,20)
\fmfleft{i0,i1,i2,i3}
\fmfright{o0,o1,o2,o3}
\fmfv{label=$\beta_{1}$}{v1}
\fmfv{label=$\beta_{2}$}{v2}
\fmf{phantom}{i1,v1}
\fmf{phantom}{i2,v1}
\fmf{plain,tension=0,foreground=(1,,0,,0)}{v1,v2}
\fmf{wiggly,tension=0.6,foreground=(1,,0,,0)}{v1,v2}
\fmf{phantom}{v2,o1}
\fmf{phantom}{v2,o2}
\end{fmfgraph*}
\end{fmffile}
\end{gathered} \quad &\rightarrow \mathcal{G}_{\beta_{1}\beta_{2}}[\Phi]\;,
\label{eq:FeynRulesBosonic1PIEAJ0mathcalG}
\end{align}
\end{subequations}
where~\eqref{eq:FeynRulesBosonic1PIEAJ0GAppendix},~\eqref{eq:FeynRulesBosonic1PIEAJ0HAppendix} and~\eqref{eq:FeynRulesBosonic1PIEAJ0vertexDotAppendix} are equivalent to~\eqref{eq:FeynRulesBosonic1PIEAJ0Gmain},~\eqref{eq:FeynRulesBosonic1PIEAJ0Hmain} and~\eqref{eq:FeynRulesBosonic1PIEAJ0vertexDotmain}, respectively. As a next step, we combine the series~\eqref{eq:bosonic1PIEAGammaExpansion0DONAppendix},~\eqref{eq:bosonic1PIEAWExpansion0DONAppendix} and~\eqref{eq:bosonic1PIEAJExpansion0DONAppendix} with~\eqref{eq:bosonic1PIEAdefinition0DON} (i.e. with the Legendre transform definition of $\Gamma_{\mathrm{col}}^{(\mathrm{1PI})}$):
\begin{equation}
\sum_{n=0}^{\infty} \Gamma_{\mathrm{col},n}^{(\mathrm{1PI})}[\Phi]\hbar^{n} = -\sum_{n=0}^{\infty} W_{\mathrm{col},n}\Bigg[\sum_{m=0}^{\infty} \mathcal{J}_{m}[\Phi]\hbar^{m}\Bigg]\hbar^{n}+\sum_{n=0}^{\infty} \int_{\beta}\mathcal{J}_{n,\beta}[\Phi]\Phi_{\beta} \ \hbar^{n}\;.
\end{equation}
From this, we derive in the same way as for~\eqref{eq:pure1PIEAIMstep60DON}:
\begin{equation}
\begin{split}
\Gamma_{\mathrm{col},n}^{(\mathrm{1PI})}[\Phi] = & -W_{\mathrm{col},n}\big[\mathcal{J}=\mathcal{J}_{0}\big] -\sum_{m=1}^{n-1} \int_{\beta} \left.\frac{\delta W_{\mathrm{col},n-m}\big[\mathcal{J}\big]}{\delta J_{\beta}}\right|_{\mathcal{J}=\mathcal{J}_{0}} \mathcal{J}_{m,\beta}[\Phi] \\
& -\sum_{m=2}^{n} \frac{1}{m!} \sum_{\underset{\lbrace n_{1} + \cdots + n_{m} \leq n\rbrace}{n_{1},\cdots,n_{m}=1}}^{n} \int_{\beta_{1},\cdots,\beta_{m}} \left.\frac{\delta^{m} W_{\mathrm{col},n-(n_{1}+\cdots+n_{m})}\big[\mathcal{J}\big]}{\delta \mathcal{J}_{\beta_{1}}\cdots\delta \mathcal{J}_{\beta_{m}}}\right|_{\mathcal{J}=\mathcal{J}_{0}} \mathcal{J}_{n_{1},\beta_{1}}[\Phi]\cdots \mathcal{J}_{n_{m},\beta_{m}}[\Phi] \\
& + \int_{\beta} \mathcal{J}_{0,\beta}[\Phi] \Phi_{\beta} \delta_{n 0}\;.
\end{split}
\label{eq:bosonic1PIEAIMstep20DON}
\end{equation}
As we are interested in the lowest non-trivial order, we will determine $\Gamma_{\mathrm{col},0}^{(\mathrm{1PI})}$, $\Gamma_{\mathrm{col},1}^{(\mathrm{1PI})}$ and $\Gamma_{\mathrm{col},2}^{(\mathrm{1PI})}$ which, according to~\eqref{eq:bosonic1PIEAIMstep20DON}, read:
\begin{equation}
\Gamma_{\mathrm{col},0}^{(\mathrm{1PI})}[\Phi] = -W_{0}\big[\mathcal{J}=\mathcal{J}_{0}\big] + \int_{\beta} \mathcal{J}_{0,\beta}[\Phi] \Phi_{\beta}\;,
\label{eq:bosonic1PIEAIMGamma00DON}
\end{equation}
\begin{equation}
\Gamma_{\mathrm{col},1}^{(\mathrm{1PI})}[\Phi] = -W_{1}\big[\mathcal{J}=\mathcal{J}_{0}\big]\;,
\label{eq:bosonic1PIEAIMGamma10DON}
\end{equation}
\begin{equation}
\scalebox{0.95}{${\displaystyle\Gamma_{\mathrm{col},2}^{(\mathrm{1PI})}[\Phi] = -W_{2}\big[\mathcal{J}=\mathcal{J}_{0}\big] -\int_{\beta} \left.\frac{\delta W_{1}\big[\mathcal{J}\big]}{\delta \mathcal{J}_{\beta}}\right|_{\mathcal{J}=\mathcal{J}_{0}} \mathcal{J}_{1,\beta}[\Phi] -\frac{1}{2} \int_{\beta_{1},\beta_{2}} \left.\frac{\delta^{2} W_{0}\big[\mathcal{J}\big]}{\delta \mathcal{J}_{\beta_{1}} \delta \mathcal{J}_{\beta_{2}}}\right|_{\mathcal{J}=\mathcal{J}_{0}} \mathcal{J}_{1,\beta_{1}}[\Phi] \mathcal{J}_{1,\beta_{2}}[\Phi]\;.}$}
\label{eq:bosonic1PIEAIMGamma20DON}
\end{equation}
We can further specify $\Gamma_{\mathrm{col},0}^{(\mathrm{1PI})}$ by deriving from~\eqref{eq:bosonic1PIEAIMW00DON} and~\eqref{eq:bosonic1PIEAphicl0DON} some relations analogous to~\eqref{eq:pure1PIEAlambdaphicl0DON} and~\eqref{eq:pure1PIEAlambdaphi0DON} exploited above in the framework of the $\lambda$-expansion of $\Gamma^{(\mathrm{1PI})}$:
\begin{equation}
\begin{split}
\varphi_{\mathrm{cl},\alpha_{1}}\big[\mathcal{J}\big] = & \ \frac{\delta W_{\mathrm{col},0}\big[\mathcal{J}\big]}{\delta J_{\alpha_{1}}} \\
= & \ \frac{\delta}{\delta J_{\alpha_{1}}}\left(-S_{\mathrm{col}}[\sigma_{\mathrm{cl}}] + \int_{x_{2}} j_{x_{2}}[\Phi] \sigma_{\mathrm{cl},x_{2}}\big[\mathcal{J}\big] + \frac{1}{2} \int_{\alpha_{2},\alpha_{3}} J_{\alpha_{2}}[\Phi] \boldsymbol{G}_{\sigma_\text{cl};\mathcal{J},\alpha_{2}\alpha_{3}}\big[\mathcal{J}\big] J_{\alpha_{3}}[\Phi]\right) \\
= & \int_{\alpha_{2}} \boldsymbol{G}_{\sigma_\text{cl};\mathcal{J},\alpha_{1}\alpha_{2}}\big[\mathcal{J}\big] J_{\alpha_{2}}[\Phi]\;,
\end{split}
\end{equation}
and, at $\mathcal{J}=\mathcal{J}_{0}$:
\begin{equation}
\phi_{\alpha_{1}} = \int_{\alpha_{2}} \boldsymbol{G}_{\Phi,\alpha_{1}\alpha_{2}}[\Phi] J_{0,\alpha_{2}}[\Phi]\;,
\label{eq:bosonic1PIEAphiGJ00DON}
\end{equation}
or, equivalently:
\begin{equation}
\begin{gathered}
\begin{fmffile}{Diagrams/bosonic1PIEA_phiDiag1}
\begin{fmfgraph*}(20,20)
\fmfleft{i0,i1,i2,i3}
\fmfright{o0,o1,o2,o3}
\fmfv{decor.shape=circle,decor.filled=empty,decor.size=1.5thick,label=$\alpha$}{v1}
\fmfv{decor.shape=cross,decor.size=3.5thick,foreground=(1,,0,,0)}{v2}
\fmf{phantom}{i1,v1}
\fmf{phantom}{i2,v1}
\fmf{dashes,tension=1.2,foreground=(1,,0,,0)}{v1,v2}
\fmf{phantom}{v2,o1}
\fmf{phantom}{v2,o2}
\end{fmfgraph*}
\end{fmffile}
\end{gathered} \hspace{-0.1cm} = \hspace{0.15cm} \begin{gathered}
\begin{fmffile}{Diagrams/bosonic1PIEA_phiDiag2}
\begin{fmfgraph*}(20,20)
\fmfleft{i0,i1,i2,i3}
\fmfright{o0,o1,o2,o3}
\fmfv{decor.shape=circle,decor.filled=empty,decor.size=1.5thick,label=$\alpha$}{v1}
\fmfv{decor.shape=circle,decor.filled=empty,decor.size=0.28cm,label=$\mathrm{o}$,label.dist=0}{v2}
\fmf{phantom}{i1,v1}
\fmf{phantom}{i2,v1}
\fmf{plain,tension=1,foreground=(1,,0,,0)}{v1,v2}
\fmf{phantom}{v2,o1}
\fmf{phantom}{v2,o2}
\end{fmfgraph*}
\end{fmffile}
\end{gathered}\;.
\label{eq:bosonic1PIEAphiGJ0diagrams0DON}
\end{equation}
From~\eqref{eq:bosonic1PIEAIMW00DON},~\eqref{eq:bosonic1PIEAIMGamma00DON} and~\eqref{eq:bosonic1PIEAphiGJ00DON}, we show that:
\begin{equation}
\begin{split}
\Gamma_{\mathrm{col},0}^{(\mathrm{1PI})}[\Phi] = & -W_{0}[\mathcal{J}=\mathcal{J}_{0}] + \int_{\beta} \mathcal{J}_{0,\beta}[\Phi] \Phi_{\beta} \\
= & -\Bigg(-S_{\mathrm{col}}[\eta] + \int_{x} j_{0,x}[\Phi] \eta_{x} + \frac{1}{2} \int_{\alpha_{1},\alpha_{2}} J_{0,\alpha_{1}}[\Phi] \boldsymbol{G}_{\alpha_{1}\alpha_{2}}[\Phi] J_{0,\alpha_{2}}[\Phi]\Bigg) + \int_{\beta} \mathcal{J}_{0,\beta}[\Phi] \Phi_{\beta} \\
= & \ S_{\mathrm{col}}[\eta] \underbrace{- \int_{x} j_{0,x}[\Phi] \eta_{x} - \int_{\alpha_{1},\alpha_{2}} J_{0,\alpha_{1}}[\Phi] \boldsymbol{G}_{\alpha_{1}\alpha_{2}}[\Phi] J_{0,\alpha_{2}}[\Phi]}_{-\int_{\beta} \mathcal{J}_{0,\beta}[\Phi] \Phi_{\beta}} \\
& + \frac{1}{2} \underbrace{\int_{\alpha_{1},\alpha_{2}} J_{0,\alpha_{1}}[\Phi] \boldsymbol{G}_{\alpha_{1}\alpha_{2}}[\Phi] J_{0,\alpha_{2}}[\Phi]}_{\int_{\alpha_{\scalebox{0.4}{1}},\alpha_{\scalebox{0.4}{2}}}\phi_{\alpha_{\scalebox{0.4}{1}}} \boldsymbol{G}^{-1}_{\alpha_{\scalebox{0.4}{1}}\alpha_{\scalebox{0.4}{2}}}[\Phi] \phi_{\alpha_{\scalebox{0.4}{2}}}} + \int_{\beta} \mathcal{J}_{0,\beta}[\Phi] \Phi_{\beta} \\
& = \ S_{\mathrm{col}}[\eta] + \frac{1}{2} \int_{\alpha_{1},\alpha_{1}}\phi_{\alpha_{1}} \boldsymbol{G}^{-1}_{\alpha_{1}\alpha_{2}}[\Phi] \phi_{\alpha_{2}} \;.
\end{split}
\label{eq:bosonic1PIEAIMGamma0step20DON}
\end{equation}
Even though the original field has been integrated out in the collective representation, it can still be seen at the level of the LE that this field still plays a role through its source $\vec{J}$, which gives us access to the corresponding correlation functions (and to the gs density $\rho_{\mathrm{gs}}$ in particular). The presence of the original field is perhaps more apparent at the level of the EA since we can introduce back the 1-point correlation function $\vec{\phi}$ through~\eqref{eq:bosonic1PIEAphiGJ00DON}, which notably yields a kinetic term\footnote{This kinetic term is the reason why the LOAF approximation coincides with the collective LE at its leading order if and only if the gap equations of $\Gamma_{\mathrm{col}}^{(\mathrm{1PI})}$ yield $\vec{\overline{\phi}}=\vec{0}$, as stated at the beginning of section~\ref{sec:1PIbosonicEA}.} for $\vec{\phi}$ in $\Gamma_{\mathrm{col},0}^{(\mathrm{1PI})}$ according to~\eqref{eq:bosonic1PIEAIMGamma0step20DON}.

\vspace{0.5cm}

Let us then focus on $\Gamma_{\mathrm{col},2}^{(\mathrm{1PI})}$ by determining the two derivatives involved in~\eqref{eq:bosonic1PIEAIMGamma20DON}. One of these two derivatives directly follows from~\eqref{eq:bosonic1PIEAGJ00DON}:
\begin{equation}
\left.\frac{\delta^{2} W_{\mathrm{col},0}\big[\mathcal{J}\big]}{\delta \mathcal{J}_{\beta_{1}} \delta \mathcal{J}_{\beta_{2}}}\right|_{\mathcal{J}=\mathcal{J}_{0}} = \hspace{0.3cm} \begin{gathered}
\begin{fmffile}{Diagrams/bosonic1PIEA_DerivW0J0J0}
\begin{fmfgraph*}(20,20)
\fmfleft{i0,i1,i2,i3}
\fmfright{o0,o1,o2,o3}
\fmfv{decor.shape=circle,decor.filled=empty,decor.size=1.5thick,label=$\beta_{1}$}{v1}
\fmfv{decor.shape=circle,decor.filled=empty,decor.size=1.5thick,label=$\beta_{2}$}{v2}
\fmf{phantom}{i1,v1}
\fmf{phantom}{i2,v1}
\fmf{plain,tension=0.6,foreground=(1,,0,,0)}{v1,v2}
\fmf{wiggly,tension=0,foreground=(1,,0,,0)}{v1,v2}
\fmf{phantom}{v2,o1}
\fmf{phantom}{v2,o2}
\end{fmfgraph*}
\end{fmffile}
\end{gathered}\hspace{0.3cm}\;,
\label{eq:bosonic1PIEADerivW0j0j00DON}
\end{equation}
and the other derivative that we seek is calculated from~\eqref{eq:bosonic1PIEAIMW10DON} as follows:
\begin{equation}
\begin{split}
\frac{\delta W_{\mathrm{col},1}\big[\mathcal{J}\big]}{\delta \mathcal{J}_{\beta_{1}}} = & \ \frac{\delta}{\delta \mathcal{J}_{\beta_{1}}}\left(\frac{1}{2}\mathrm{Tr}\left[\ln\big(D_{\sigma_\text{cl};\mathcal{J}}\big[\mathcal{J}\big]\big)\right]\right) \\
= & \ \frac{1}{2} \frac{\delta}{\delta \mathcal{J}_{\beta_{1}}} \int_{x_{2}} \ln\big(D_{\sigma_\text{cl};\mathcal{J},x_{2}x_{2}}\big[\mathcal{J}\big]\big) \\
= & \ \frac{1}{2} \int_{x_{2},x_{3}} D^{-1}_{\sigma_\text{cl};\mathcal{J},x_{2}x_{3}}\big[\mathcal{J}\big] \frac{\delta D_{\sigma_\text{cl};\mathcal{J},x_{3}x_{2}}\big[\mathcal{J}\big]}{\delta \mathcal{J}_{\beta_{1}}} \\
= & -\frac{1}{2} \int_{x_{2},x_{3},x_{4},x_{5}} D^{-1}_{\sigma_\text{cl};\mathcal{J},x_{2}x_{3}}\big[\mathcal{J}\big] D_{\sigma_\text{cl};\mathcal{J},x_{3}x_{4}}\big[\mathcal{J}\big] \frac{\delta D^{-1}_{\sigma_\text{cl};\mathcal{J},x_{4}x_{5}}\big[\mathcal{J}\big]}{\delta \mathcal{J}_{\beta_{1}}} D_{\sigma_\text{cl};\mathcal{J},x_{5}x_{2}}\big[\mathcal{J}\big] \\
= & -\frac{1}{2} \int_{x_{2},x_{3}} \frac{\delta D^{-1}_{\sigma_\text{cl};\mathcal{J},x_{2}x_{3}}\big[\mathcal{J}\big]}{\delta \mathcal{J}_{\beta_{1}}} D_{\sigma_\text{cl};\mathcal{J},x_{3}x_{2}}\big[\mathcal{J}\big] \;.
\end{split}
\label{eq:bosonic1PIEADerivW1j00DON}
\end{equation}
The derivative of $D^{-1}_{\sigma_\text{cl};\mathcal{J}}$ with respect to $\mathcal{J}$ in the last line of~\eqref{eq:bosonic1PIEADerivW1j00DON} will be evaluated by making use of:
\begin{equation}
\frac{\delta J_{\alpha_{2}}}{\delta \mathcal{J}_{\beta_{1}}} = \delta_{\alpha_{1}\alpha_{2}} \left(1-\delta_{b_{1} N+1}\right)\;,
\label{eq:bosonic1PIEADerivHminus1J1}
\end{equation}
and
\begin{equation}
\begin{split}
\scalebox{0.93}{${\displaystyle\frac{\delta\boldsymbol{G}_{\sigma_\text{cl};\mathcal{J},\alpha_{2}\alpha_{3}}\big[\mathcal{J}\big]}{\delta\mathcal{J}_{\beta_{1}}} = }$} & \scalebox{0.93}{${\displaystyle - \int_{\alpha_{4},\alpha_{5}} \boldsymbol{G}_{\sigma_\text{cl};\mathcal{J},\alpha_{2}\alpha_{4}}\big[\mathcal{J}\big] \frac{\delta\boldsymbol{G}^{-1}_{\sigma_\text{cl};\mathcal{J},\alpha_{4}\alpha_{5}}\big[\mathcal{J}\big]}{\delta\mathcal{J}_{\beta_{1}}} \boldsymbol{G}_{\sigma_\text{cl};\mathcal{J},\alpha_{5}\alpha_{3}}\big[\mathcal{J}\big] }$} \\
\scalebox{0.93}{${\displaystyle = }$} & \scalebox{0.93}{${\displaystyle - \int_{\alpha_{4},\alpha_{5}} \boldsymbol{G}_{\sigma_\text{cl};\mathcal{J},\alpha_{2}\alpha_{4}}\big[\mathcal{J}\big] \frac{\delta}{\delta\mathcal{J}_{\beta_{1}}}\left(\left(-\nabla^{2}_{x_{4}} + m^{2} + i\sqrt{\frac{\lambda}{3}}\sigma_{\mathrm{cl},x_{4}}\big[\mathcal{J}\big]\right) \delta_{\alpha_{4}\alpha_{5}}\right) \boldsymbol{G}_{\sigma_\text{cl};\mathcal{J},\alpha_{5}\alpha_{3}}\big[\mathcal{J}\big] }$} \\
\scalebox{0.93}{${\displaystyle = }$} & \scalebox{0.93}{${\displaystyle - \int_{\alpha_{4}} \boldsymbol{G}_{\sigma_\text{cl};\mathcal{J},\alpha_{2}\alpha_{4}}\big[\mathcal{J}\big] \Bigg(i\sqrt{\frac{\lambda}{3}}\underbrace{\frac{\delta\sigma_{\mathrm{cl},x_{4}}\big[\mathcal{J}\big]}{\delta\mathcal{J}_{\beta_{1}}}}_{\frac{\delta\sigma_{\mathrm{cl},x_{\scalebox{0.4}{4}}}[\mathcal{J}]}{\delta j_{x_{\scalebox{0.4}{1}}}}\delta_{b_{\scalebox{0.4}{1}} N+1}}\Bigg) \boldsymbol{G}_{\sigma_\text{cl};\mathcal{J},\alpha_{4}\alpha_{3}}\big[\mathcal{J}\big] }$} \\
\scalebox{0.93}{${\displaystyle = }$} & \scalebox{0.93}{${\displaystyle - i\sqrt{\frac{\lambda}{3}} \ \delta_{b_{1} N+1} \int_{\alpha_{4}} \boldsymbol{G}_{\sigma_\text{cl};\mathcal{J},\alpha_{2}\alpha_{4}}\big[\mathcal{J}\big] \underbrace{\frac{\delta\sigma_{\mathrm{cl},x_{4}}\big[\mathcal{J}\big]}{\delta j_{x_{1}}}}_{D_{\sigma_\text{cl};\mathcal{J},x_{\scalebox{0.4}{1}}x_{\scalebox{0.4}{4}}}[\mathcal{J}]} \boldsymbol{G}_{\sigma_\text{cl};\mathcal{J},\alpha_{4}\alpha_{3}}\big[\mathcal{J}\big] }$} \\
\scalebox{0.93}{${\displaystyle = }$} & \scalebox{0.93}{${\displaystyle - i\sqrt{\frac{\lambda}{3}} \ \delta_{b_{1} N+1} \int_{\alpha_{4}} \boldsymbol{G}_{\sigma_\text{cl};\mathcal{J},\alpha_{2}\alpha_{4}}\big[\mathcal{J}\big] D_{\sigma_\text{cl};\mathcal{J},x_{1}x_{4}}\big[\mathcal{J}\big] \boldsymbol{G}_{\sigma_\text{cl};\mathcal{J},\alpha_{4}\alpha_{3}}\big[\mathcal{J}\big] \;, }$}
\end{split}
\label{eq:bosonic1PIEADerivHminus1J2}
\end{equation}
which follows from~\eqref{eq:bosonic1PIEApropagatorGJ} and where the relation:
\begin{equation}
D_{\sigma_\text{cl};\mathcal{J},x_{1}x_{2}}\big[\mathcal{J}\big] = \frac{\delta^{2} W_{\mathrm{col},0}\big[\mathcal{J}\big]}{\delta j_{x_{1}} \delta j_{x_{2}}} = \frac{\delta\eta_{\mathrm{cl},x_{2}}}{\delta j_{x_{1}}}\;,
\end{equation}
is equivalent to~\eqref{eq:bosonic1PIEAGJ0DON} (combined with~\eqref{eq:bosonic1PIEAphicl0DON}) at $b_{1}=b_{2}=N+1$. From~\eqref{eq:bosonic1PIEADerivHminus1J1} and~\eqref{eq:bosonic1PIEADerivHminus1J2} as well as~\eqref{eq:SbosonicKLoopExpansionH}, we calculate:
\begin{equation}
\begin{split}
\scalebox{0.91}{${\displaystyle \frac{\delta D^{-1}_{\sigma_\text{cl};\mathcal{J},x_{2}x_{3}}\big[\mathcal{J}\big]}{\delta\mathcal{J}_{\beta_{1}}} = }$} & \ \scalebox{0.91}{${\displaystyle \frac{\delta}{\delta\mathcal{J}_{\beta_{1}}} \Bigg( \frac{\lambda}{3} \sum^{N}_{a_{6},a_{7}=1} \int_{\alpha_{4},\alpha_{5}} J_{\alpha_{4}} \boldsymbol{G}_{\sigma_\text{cl};\mathcal{J},\alpha_{4}(a_{6},x_{2})}\big[\mathcal{J}\big] \boldsymbol{G}_{\sigma_\text{cl};\mathcal{J},(a_{6},x_{2})(a_{7},x_{3})}\big[\mathcal{J}\big] \boldsymbol{G}_{\sigma_\text{cl};\mathcal{J},(a_{7},x_{3})\alpha_{5}}\big[\mathcal{J}\big] J_{\alpha_{5}} }$} \\
& \hspace{1.1cm} \scalebox{0.91}{${\displaystyle + \frac{\lambda}{6} \sum^{N}_{a_{4},a_{5}=1} \boldsymbol{G}_{\sigma_\text{cl};\mathcal{J},(a_{4},x_{2})(a_{5},x_{3})}\big[\mathcal{J}\big] \boldsymbol{G}_{\sigma_\text{cl};\mathcal{J},(a_{5},x_{3})(a_{4},x_{2})}\big[\mathcal{J}\big] + \delta_{x_{2}x_{3}} \Bigg) }$} \\
\scalebox{0.91}{${\displaystyle = }$} & \scalebox{0.91}{${\displaystyle -\left(1-\delta_{b_{1} N+1}\right) \left(\rule{0cm}{1.0cm}\right. \begin{gathered}
\begin{fmffile}{Diagrams/bosonic1PIEA_DerivHminus1Diag1}
\begin{fmfgraph*}(25,25)
\fmfleft{i1,i2}
\fmfright{o1,o2}
\fmfbottom{i0,o0}
\fmftop{i3,o3}
\fmfv{decor.shape=circle,decor.filled=empty,decor.size=1.5thick,label.dist=4,label=$\alpha_{1}$}{v4}
\fmfv{label=$x_{2}$,label.dist=4,foreground=(0,,0,,1)}{v1}
\fmfv{label=$x_{3}$,label.dist=4,foreground=(0,,0,,1)}{v2}
\fmfv{decor.shape=circle,decor.filled=empty,decor.size=.1w,l=$\times$,label.dist=0}{v3}
\fmf{phantom}{i1,v1}
\fmf{phantom}{i2,v4}
\fmf{phantom}{o1,v2}
\fmf{phantom}{o2,v3}
\fmf{phantom}{i3,v4b}
\fmf{phantom}{o3,v3b}
\fmf{phantom}{i0,v1b}
\fmf{phantom}{o0,v2b}
\fmf{plain,tension=1.6}{v1,v2}
\fmf{phantom,tension=1.6}{v3,v4}
\fmf{plain,tension=2.0}{v1,v4}
\fmf{plain,tension=2.0}{v2,v3}
\fmf{phantom,tension=0}{v1,v3}
\fmf{phantom,tension=0}{v2,v4}
\fmf{phantom}{v1,v1b}
\fmf{phantom}{v2,v2b}
\fmf{phantom}{v3,v3b}
\fmf{phantom}{v4,v4b}
\fmfdot{v1,v2}
\end{fmfgraph*}
\end{fmffile}
\end{gathered} + \begin{gathered}
\begin{fmffile}{Diagrams/bosonic1PIEA_DerivHminus1Diag2}
\begin{fmfgraph*}(25,25)
\fmfleft{i1,i2}
\fmfright{o1,o2}
\fmfbottom{i0,o0}
\fmftop{i3,o3}
\fmfv{decor.shape=circle,decor.filled=empty,decor.size=1.5thick,label.dist=4,label=$\alpha_{1}$}{v3}
\fmfv{label=$x_{2}$,label.dist=4,foreground=(0,,0,,1)}{v1}
\fmfv{label=$x_{3}$,label.dist=4,foreground=(0,,0,,1)}{v2}
\fmfv{decor.shape=circle,decor.filled=empty,decor.size=.1w,l=$\times$,label.dist=0}{v4}
\fmf{phantom}{i1,v1}
\fmf{phantom}{i2,v4}
\fmf{phantom}{o1,v2}
\fmf{phantom}{o2,v3}
\fmf{phantom}{i3,v4b}
\fmf{phantom}{o3,v3b}
\fmf{phantom}{i0,v1b}
\fmf{phantom}{o0,v2b}
\fmf{plain,tension=1.6}{v1,v2}
\fmf{phantom,tension=1.6}{v3,v4}
\fmf{plain,tension=2.0}{v1,v4}
\fmf{plain,tension=2.0}{v2,v3}
\fmf{phantom,tension=0}{v1,v3}
\fmf{phantom,tension=0}{v2,v4}
\fmf{phantom}{v1,v1b}
\fmf{phantom}{v2,v2b}
\fmf{phantom}{v3,v3b}
\fmf{phantom}{v4,v4b}
\fmfdot{v1,v2}
\end{fmfgraph*}
\end{fmffile}
\end{gathered} \left.\rule{0cm}{1.0cm}\right) }$} \\
& \scalebox{0.91}{${\displaystyle + \delta_{b_{1} N+1} \left(\rule{0cm}{1.5cm}\right. \begin{gathered}
\begin{fmffile}{Diagrams/bosonic1PIEA_DerivHminus1Diag3}
\begin{fmfgraph*}(22,18)
\fmftop{vUpL2,vUpL1,vUp,vUpR1,vUpR2}
\fmfleft{i1,i2}
\fmfright{o1,o2}
\fmfbottom{i0,o0}
\fmfv{decor.shape=circle,decor.filled=empty,decor.size=1.5thick,label.dist=4,label=$x_{1}$}{vUpL1}
\fmfv{decor.shape=circle,decor.filled=empty,decor.size=.115w,l=$\times$,label.dist=0}{v2b}
\fmfv{decor.shape=circle,decor.filled=empty,decor.size=.115w,l=$\times$,label.dist=0}{vUpR1}
\fmfv{label=$x_{2}$,label.dist=4,foreground=(0,,0,,1)}{v1}
\fmfv{label=$x_{3}$,label.angle=-90,label.dist=4,foreground=(0,,0,,1)}{v2}
\fmfv{foreground=(0,,0,,1)}{v3}
\fmf{wiggly,tension=0.4}{v3,vUpL1}
\fmf{plain,tension=0.4}{v3,vUpR1}
\fmf{phantom}{i1,v1}
\fmf{phantom}{i2,v3b}
\fmf{phantom}{o1,v2}
\fmf{phantom}{o2,v3b}
\fmf{phantom}{i0,v1b}
\fmf{phantom}{o0,v2b}
\fmf{plain,tension=0.5}{v1,v2}
\fmf{plain}{v1,v3}
\fmf{phantom}{v2,v3}
\fmf{phantom,tension=0.4}{v1,v1b}
\fmf{plain,tension=0.4}{v2,v2b}
\fmf{phantom,tension=1.5}{v3,v3b}
\fmfdot{v1,v2,v3}
\end{fmfgraph*}
\end{fmffile}
\end{gathered} + \begin{gathered}
\begin{fmffile}{Diagrams/bosonic1PIEA_DerivHminus1Diag4}
\begin{fmfgraph*}(22,18)
\fmftop{vUpL2,vUpL1,vUp,vUpR1,vUpR2}
\fmfleft{i1,i2}
\fmfright{o1,o2}
\fmfbottom{i0,o0}
\fmfv{decor.shape=circle,decor.filled=empty,decor.size=1.5thick,label.dist=4,label=$x_{1}$}{vUp}
\fmfv{decor.shape=circle,decor.filled=empty,decor.size=.115w,l=$\times$,label.dist=0}{v1b}
\fmfv{decor.shape=circle,decor.filled=empty,decor.size=.115w,l=$\times$,label.dist=0}{v2b}
\fmfv{label=$x_{2}$,label.angle=-90,label.dist=4,foreground=(0,,0,,1)}{v1}
\fmfv{label=$x_{3}$,label.angle=-90,label.dist=4,foreground=(0,,0,,1)}{v2}
\fmfv{foreground=(0,,0,,1)}{v3}
\fmf{wiggly,tension=0}{v3,vUp}
\fmf{phantom,tension=0.4}{v3,vUpL1}
\fmf{phantom,tension=0.4}{v3,vUpR1}
\fmf{phantom}{i1,v1}
\fmf{phantom}{i2,v3b}
\fmf{phantom}{o1,v2}
\fmf{phantom}{o2,v3b}
\fmf{phantom}{i0,v1b}
\fmf{phantom}{o0,v2b}
\fmf{phantom,tension=0.5}{v1,v2}
\fmf{plain}{v1,v3}
\fmf{plain}{v2,v3}
\fmf{plain,tension=0.4}{v1,v1b}
\fmf{plain,tension=0.4}{v2,v2b}
\fmf{phantom,tension=1.5}{v3,v3b}
\fmfdot{v1,v2,v3}
\end{fmfgraph*}
\end{fmffile}
\end{gathered} + \begin{gathered}
\begin{fmffile}{Diagrams/bosonic1PIEA_DerivHminus1Diag5}
\begin{fmfgraph*}(22,18)
\fmftop{vUpL2,vUpL1,vUp,vUpR1,vUpR2}
\fmfleft{i1,i2}
\fmfright{o1,o2}
\fmfbottom{i0,o0}
\fmfv{decor.shape=circle,decor.filled=empty,decor.size=1.5thick,label.dist=4,label=$x_{1}$}{vUpL1}
\fmfv{decor.shape=circle,decor.filled=empty,decor.size=.115w,l=$\times$,label.dist=0}{v1b}
\fmfv{decor.shape=circle,decor.filled=empty,decor.size=.115w,l=$\times$,label.dist=0}{vUpR1}
\fmfv{label=$x_{2}$,label.angle=-90,label.dist=4,foreground=(0,,0,,1)}{v1}
\fmfv{label=$x_{3}$,label.dist=4,foreground=(0,,0,,1)}{v2}
\fmfv{foreground=(0,,0,,1)}{v3}
\fmf{wiggly,tension=0.4}{v3,vUpL1}
\fmf{plain,tension=0.4}{v3,vUpR1}
\fmf{phantom}{i1,v1}
\fmf{phantom}{i2,v3b}
\fmf{phantom}{o1,v2}
\fmf{phantom}{o2,v3b}
\fmf{phantom}{i0,v1b}
\fmf{phantom}{o0,v2b}
\fmf{plain,tension=0.5}{v1,v2}
\fmf{phantom}{v1,v3}
\fmf{plain}{v2,v3}
\fmf{plain,tension=0.4}{v1,v1b}
\fmf{phantom,tension=0.4}{v2,v2b}
\fmf{phantom,tension=1.5}{v3,v3b}
\fmfdot{v1,v2,v3}
\end{fmfgraph*}
\end{fmffile}
\end{gathered} + \begin{gathered}
\begin{fmffile}{Diagrams/bosonic1PIEA_DerivHminus1Diag6}
\begin{fmfgraph*}(22,18)
\fmftop{vUpL2,vUpL1,vUp,vUpR1,vUpR2}
\fmfleft{i1,i2}
\fmfright{o1,o2}
\fmfbottom{i0,o0}
\fmfv{decor.shape=circle,decor.filled=empty,decor.size=1.5thick,label.dist=4,label=$x_{1}$}{vUp}
\fmfv{label=$x_{2}$,label.dist=4,foreground=(0,,0,,1)}{v1}
\fmfv{label=$x_{3}$,label.dist=4,foreground=(0,,0,,1)}{v2}
\fmfv{foreground=(0,,0,,1)}{v3}
\fmf{wiggly,tension=0}{v3,vUp}
\fmf{phantom,tension=0.4}{v3,vUpL1}
\fmf{phantom,tension=0.4}{v3,vUpR1}
\fmf{phantom}{i1,v1}
\fmf{phantom}{i2,v3b}
\fmf{phantom}{o1,v2}
\fmf{phantom}{o2,v3b}
\fmf{phantom}{i0,v1b}
\fmf{phantom}{o0,v2b}
\fmf{plain,tension=0.5}{v1,v2}
\fmf{plain}{v1,v3}
\fmf{plain}{v2,v3}
\fmf{phantom,tension=0.4}{v1,v1b}
\fmf{phantom,tension=0.4}{v2,v2b}
\fmf{phantom,tension=1.5}{v3,v3b}
\fmfdot{v1,v2,v3}
\end{fmfgraph*}
\end{fmffile}
\end{gathered} \left.\rule{0cm}{1.5cm}\right)\;. }$}
\end{split}
\label{eq:bosonic1PIEADerivHminus1J}
\end{equation}
The derivative of $W_{\mathrm{col},1}\big[\mathcal{J}\big]$ involved in~\eqref{eq:bosonic1PIEAIMGamma20DON} is obtained by inserting~\eqref{eq:bosonic1PIEADerivHminus1J} into~\eqref{eq:bosonic1PIEADerivW1j00DON} before setting $\mathcal{J}=\mathcal{J}_{0}$:
\begin{equation}
\begin{split}
\left.\frac{\delta W_{\mathrm{col},1}\big[\mathcal{J}\big]}{\delta \mathcal{J}_{\beta_{1}}}\right|_{\mathcal{J}=\mathcal{J}_{0}} = & \ \left(1-\delta_{b_{1} N+1}\right) \hspace{-0.3cm} \begin{gathered}
\begin{fmffile}{Diagrams/bosonic1PIEA_DerivW1J0Diag1}
\begin{fmfgraph*}(25,25)
\fmfleft{i1,i2}
\fmfright{o1,o2}
\fmfbottom{i0,o0}
\fmftop{i3,o3}
\fmfv{decor.shape=circle,decor.filled=empty,decor.size=1.5thick,label.dist=4,label=$\alpha_{1}$}{v4}
\fmfv{foreground=(0,,0,,1)}{v1}
\fmfv{foreground=(0,,0,,1)}{v2}
\fmfv{decor.shape=circle,decor.filled=empty,decor.size=0.28cm,label=$\mathrm{o}$,label.dist=0}{v3}
\fmf{phantom}{i1,v1}
\fmf{phantom}{i2,v4}
\fmf{phantom}{o1,v2}
\fmf{phantom}{o2,v3}
\fmf{phantom}{i3,v4b}
\fmf{phantom}{o3,v3b}
\fmf{phantom}{i0,v1b}
\fmf{phantom}{o0,v2b}
\fmf{wiggly,tension=1.6,foreground=(1,,0,,0)}{v1,v2}
\fmf{plain,left=0.6,tension=0,foreground=(1,,0,,0)}{v1,v2}
\fmf{phantom,tension=1.6}{v3,v4}
\fmf{plain,tension=2.0,foreground=(1,,0,,0)}{v1,v4}
\fmf{plain,tension=2.0,foreground=(1,,0,,0)}{v2,v3}
\fmf{phantom,tension=0}{v1,v3}
\fmf{phantom,tension=0}{v2,v4}
\fmf{phantom}{v1,v1b}
\fmf{phantom}{v2,v2b}
\fmf{phantom}{v3,v3b}
\fmf{phantom}{v4,v4b}
\fmfdot{v1,v2}
\end{fmfgraph*}
\end{fmffile}
\end{gathered} \\
& - \frac{1}{2} \ \delta_{b_{1} N+1} \left(\rule{0cm}{1.5cm}\right. 2 \begin{gathered}
\begin{fmffile}{Diagrams/bosonic1PIEA_DerivW1J0Diag2}
\begin{fmfgraph*}(22,18)
\fmftop{vUpL2,vUpL1,vUp,vUpR1,vUpR2}
\fmfleft{i1,i2}
\fmfright{o1,o2}
\fmfbottom{i0,o0}
\fmfv{decor.shape=circle,decor.filled=empty,decor.size=1.5thick,label.dist=4,label=$x_{1}$}{vUpL1}
\fmfv{decor.shape=circle,decor.filled=empty,decor.size=0.28cm,label=$\mathrm{o}$,label.dist=0}{v2b}
\fmfv{decor.shape=circle,decor.filled=empty,decor.size=0.28cm,label=$\mathrm{o}$,label.dist=0}{vUpR1}
\fmfv{foreground=(0,,0,,1)}{v1}
\fmfv{foreground=(0,,0,,1)}{v2}
\fmfv{foreground=(0,,0,,1)}{v3}
\fmf{wiggly,tension=0.4,foreground=(1,,0,,0)}{v3,vUpL1}
\fmf{plain,tension=0.4,foreground=(1,,0,,0)}{v3,vUpR1}
\fmf{phantom}{i1,v1}
\fmf{phantom}{i2,v3b}
\fmf{phantom}{o1,v2}
\fmf{phantom}{o2,v3b}
\fmf{phantom}{i0,v1b}
\fmf{phantom}{o0,v2b}
\fmf{wiggly,tension=0.5,foreground=(1,,0,,0)}{v1,v2}
\fmf{plain,right=0.6,tension=0,foreground=(1,,0,,0)}{v1,v2}
\fmf{plain,foreground=(1,,0,,0)}{v1,v3}
\fmf{phantom}{v2,v3}
\fmf{phantom,tension=0.4}{v1,v1b}
\fmf{plain,tension=0.4,foreground=(1,,0,,0)}{v2,v2b}
\fmf{phantom,tension=1.5}{v3,v3b}
\fmfdot{v1,v2,v3}
\end{fmfgraph*}
\end{fmffile}
\end{gathered} + \begin{gathered}
\begin{fmffile}{Diagrams/bosonic1PIEA_DerivW1J0Diag3}
\begin{fmfgraph*}(22,18)
\fmftop{vUpL2,vUpL1,vUp,vUpR1,vUpR2}
\fmfleft{i1,i2}
\fmfright{o1,o2}
\fmfbottom{i0,o0}
\fmfv{decor.shape=circle,decor.filled=empty,decor.size=1.5thick,label.dist=4,label=$x_{1}$}{vUp}
\fmfv{decor.shape=circle,decor.filled=empty,decor.size=0.28cm,label=$\mathrm{o}$,label.dist=0}{v1b}
\fmfv{decor.shape=circle,decor.filled=empty,decor.size=0.28cm,label=$\mathrm{o}$,label.dist=0}{v2b}
\fmfv{foreground=(0,,0,,1)}{v1}
\fmfv{foreground=(0,,0,,1)}{v2}
\fmfv{foreground=(0,,0,,1)}{v3}
\fmf{wiggly,tension=0,foreground=(1,,0,,0)}{v3,vUp}
\fmf{phantom,tension=0.4}{v3,vUpL1}
\fmf{phantom,tension=0.4}{v3,vUpR1}
\fmf{phantom}{i1,v1}
\fmf{phantom}{i2,v3b}
\fmf{phantom}{o1,v2}
\fmf{phantom}{o2,v3b}
\fmf{phantom}{i0,v1b}
\fmf{phantom}{o0,v2b}
\fmf{wiggly,tension=0.5,foreground=(1,,0,,0)}{v1,v2}
\fmf{plain,foreground=(1,,0,,0)}{v1,v3}
\fmf{plain,foreground=(1,,0,,0)}{v2,v3}
\fmf{plain,tension=0.4,foreground=(1,,0,,0)}{v1,v1b}
\fmf{plain,tension=0.4,foreground=(1,,0,,0)}{v2,v2b}
\fmf{phantom,tension=1.5}{v3,v3b}
\fmfdot{v1,v2,v3}
\end{fmfgraph*}
\end{fmffile}
\end{gathered} + \begin{gathered}
\begin{fmffile}{Diagrams/bosonic1PIEA_DerivW1J0Diag4}
\begin{fmfgraph*}(22,18)
\fmftop{vUpL2,vUpL1,vUp,vUpR1,vUpR2}
\fmfleft{i1,i2}
\fmfright{o1,o2}
\fmfbottom{i0,o0}
\fmfv{decor.shape=circle,decor.filled=empty,decor.size=1.5thick,label.dist=4,label=$x_{1}$}{vUp}
\fmfv{foreground=(0,,0,,1)}{v1}
\fmfv{foreground=(0,,0,,1)}{v2}
\fmfv{foreground=(0,,0,,1)}{v3}
\fmf{wiggly,tension=0,foreground=(1,,0,,0)}{v3,vUp}
\fmf{phantom,tension=0.4}{v3,vUpL1}
\fmf{phantom,tension=0.4}{v3,vUpR1}
\fmf{phantom}{i1,v1}
\fmf{phantom}{i2,v3b}
\fmf{phantom}{o1,v2}
\fmf{phantom}{o2,v3b}
\fmf{phantom}{i0,v1b}
\fmf{phantom}{o0,v2b}
\fmf{wiggly,tension=0.5,foreground=(1,,0,,0)}{v1,v2}
\fmf{plain,right=0.6,tension=0,foreground=(1,,0,,0)}{v1,v2}
\fmf{plain,foreground=(1,,0,,0)}{v1,v3}
\fmf{plain,foreground=(1,,0,,0)}{v2,v3}
\fmf{phantom,tension=0.4}{v1,v1b}
\fmf{phantom,tension=0.4}{v2,v2b}
\fmf{phantom,tension=1.5}{v3,v3b}
\fmfdot{v1,v2,v3}
\end{fmfgraph*}
\end{fmffile}
\end{gathered} \left.\rule{0cm}{1.5cm}\right)\;.
\end{split}
\label{eq:bosonic1PIEADerivW1j0step20DON}
\end{equation}
As a next step, we determine the $\mathcal{J}_{1}$ coefficient from the power series of $\Phi$. The procedure is analogous to that leading to $\vec{J}_{1}$ for $\Gamma^{(\mathrm{1PI})}$: we Taylor expand the $\Phi_{n}$ coefficients around $\mathcal{J}=\mathcal{J}_{0}$ in~\eqref{eq:bosonic1PIEAphiExpansion0DONAppendix} and identify the following relations by exploiting the fact that $\Phi$ is of order $\mathcal{O}\big(\hbar^{0}\big)$:
\begin{subequations}
\begin{empheq}[left=\empheqlbrace]{align}
& \hspace{0.1cm} \mathrm{Order}~\mathcal{O}\big(\hbar^{0}\big):~\Phi_{\beta} = \Phi_{0,\beta}\big[\mathcal{J}=\mathcal{J}_{0}\big]\;, \label{eq:bosonic1PIEATowerEquationJn10DON}\\
\nonumber \\
& \hspace{0.1cm} \mathrm{Order}~\mathcal{O}(\hbar):~0 = \int_{\beta_{2}} \left.\frac{\delta \Phi_{0,\beta_{1}}\big[\mathcal{J}\big]}{\delta \mathcal{J}_{\beta_{2}}}\right|_{\mathcal{J}=\mathcal{J}_{0}} \mathcal{J}_{1,\beta_{2}}[\Phi] + \Phi_{1,\beta_{1}}\big[\mathcal{J}=\mathcal{J}_{0}\big] \quad \forall \beta_{1}\;, \label{eq:bosonic1PIEATowerEquationJn20DON}\\
\nonumber \\
& \hspace{6.5cm} \vdots \nonumber
\end{empheq}
\end{subequations}
where it follows from~\eqref{eq:bosonic1PIEAphinCoeff0DON} that:
\begin{equation}
\frac{\delta \Phi_{0,\beta_{1}}\big[\mathcal{J}\big]}{\delta \mathcal{J}_{\beta_{2}}} = \frac{\delta^{2} W_{\mathrm{col},0}\big[\mathcal{J}\big]}{\delta \mathcal{J}_{\beta_{2}} \delta \mathcal{J}_{\beta_{1}}}\;,
\label{eq:bosonic1PIEADerivPhi0J0DON}
\end{equation}
\begin{equation}
\Phi_{1,\beta}\big[\mathcal{J}\big] = \frac{\delta W_{\mathrm{col},1}\big[\mathcal{J}\big]}{\delta \mathcal{J}_{\beta}}\;.
\label{eq:bosonic1PIEAPhi1J0DON}
\end{equation}
According to~\eqref{eq:bosonic1PIEADerivPhi0J0DON} and~\eqref{eq:bosonic1PIEAPhi1J0DON},~\eqref{eq:bosonic1PIEATowerEquationJn20DON} can be rewritten with the help of~\eqref{eq:bosonic1PIEADerivW0j0j00DON} and~\eqref{eq:bosonic1PIEADerivW1j0step20DON} as:
\begin{equation}
\begin{split}
0 = & \hspace{0.3cm} \begin{gathered}
\begin{fmffile}{Diagrams/bosonic1PIEA_DeterminationJ11}
\begin{fmfgraph*}(20,20)
\fmfleft{i0,i1,i2,i3}
\fmfright{o0,o1,o2,o3}
\fmfv{decor.shape=circle,decor.filled=empty,decor.size=1.5thick,label=$\beta$}{v1}
\fmfv{decor.shape=circle,decor.filled=empty,decor.size=0.4cm,label=$1$,label.dist=0}{v2}
\fmf{phantom}{i1,v1}
\fmf{phantom}{i2,v1}
\fmf{plain,tension=0.6,foreground=(1,,0,,0)}{v1,v2}
\fmf{wiggly,tension=0,foreground=(1,,0,,0)}{v1,v2}
\fmf{phantom}{v2,o1}
\fmf{phantom}{v2,o2}
\end{fmfgraph*}
\end{fmffile}
\end{gathered} + \left(1-\delta_{b \hspace{0.04cm} N+1}\right) \hspace{-0.3cm} \begin{gathered}
\begin{fmffile}{Diagrams/bosonic1PIEA_DeterminationJ12}
\begin{fmfgraph*}(25,25)
\fmfleft{i1,i2}
\fmfright{o1,o2}
\fmfbottom{i0,o0}
\fmftop{i3,o3}
\fmfv{decor.shape=circle,decor.filled=empty,decor.size=1.5thick,label.dist=4,label=$\alpha$}{v4}
\fmfv{foreground=(0,,0,,1)}{v1}
\fmfv{foreground=(0,,0,,1)}{v2}
\fmfv{decor.shape=circle,decor.filled=empty,decor.size=0.28cm,label=$\mathrm{o}$,label.dist=0}{v3}
\fmf{phantom}{i1,v1}
\fmf{phantom}{i2,v4}
\fmf{phantom}{o1,v2}
\fmf{phantom}{o2,v3}
\fmf{phantom}{i3,v4b}
\fmf{phantom}{o3,v3b}
\fmf{phantom}{i0,v1b}
\fmf{phantom}{o0,v2b}
\fmf{wiggly,tension=1.6,foreground=(1,,0,,0)}{v1,v2}
\fmf{plain,left=0.6,tension=0,foreground=(1,,0,,0)}{v1,v2}
\fmf{phantom,tension=1.6}{v3,v4}
\fmf{plain,tension=2.0,foreground=(1,,0,,0)}{v1,v4}
\fmf{plain,tension=2.0,foreground=(1,,0,,0)}{v2,v3}
\fmf{phantom,tension=0}{v1,v3}
\fmf{phantom,tension=0}{v2,v4}
\fmf{phantom}{v1,v1b}
\fmf{phantom}{v2,v2b}
\fmf{phantom}{v3,v3b}
\fmf{phantom}{v4,v4b}
\fmfdot{v1,v2}
\end{fmfgraph*}
\end{fmffile}
\end{gathered} \\
& - \frac{1}{2} \ \delta_{b \hspace{0.04cm} N+1} \left(\rule{0cm}{1.5cm}\right. 2 \begin{gathered}
\begin{fmffile}{Diagrams/bosonic1PIEA_DeterminationJ13}
\begin{fmfgraph*}(22,18)
\fmftop{vUpL2,vUpL1,vUp,vUpR1,vUpR2}
\fmfleft{i1,i2}
\fmfright{o1,o2}
\fmfbottom{i0,o0}
\fmfv{decor.shape=circle,decor.filled=empty,decor.size=1.5thick,label.dist=4,label=$x$}{vUpL1}
\fmfv{decor.shape=circle,decor.filled=empty,decor.size=0.28cm,label=$\mathrm{o}$,label.dist=0}{v2b}
\fmfv{decor.shape=circle,decor.filled=empty,decor.size=0.28cm,label=$\mathrm{o}$,label.dist=0}{vUpR1}
\fmfv{foreground=(0,,0,,1)}{v1}
\fmfv{foreground=(0,,0,,1)}{v2}
\fmfv{foreground=(0,,0,,1)}{v3}
\fmf{wiggly,tension=0.4,foreground=(1,,0,,0)}{v3,vUpL1}
\fmf{plain,tension=0.4,foreground=(1,,0,,0)}{v3,vUpR1}
\fmf{phantom}{i1,v1}
\fmf{phantom}{i2,v3b}
\fmf{phantom}{o1,v2}
\fmf{phantom}{o2,v3b}
\fmf{phantom}{i0,v1b}
\fmf{phantom}{o0,v2b}
\fmf{wiggly,tension=0.5,foreground=(1,,0,,0)}{v1,v2}
\fmf{plain,right=0.6,tension=0,foreground=(1,,0,,0)}{v1,v2}
\fmf{plain,foreground=(1,,0,,0)}{v1,v3}
\fmf{phantom}{v2,v3}
\fmf{phantom,tension=0.4}{v1,v1b}
\fmf{plain,tension=0.4,foreground=(1,,0,,0)}{v2,v2b}
\fmf{phantom,tension=1.5}{v3,v3b}
\fmfdot{v1,v2,v3}
\end{fmfgraph*}
\end{fmffile}
\end{gathered} + \begin{gathered}
\begin{fmffile}{Diagrams/bosonic1PIEA_DeterminationJ14}
\begin{fmfgraph*}(22,18)
\fmftop{vUpL2,vUpL1,vUp,vUpR1,vUpR2}
\fmfleft{i1,i2}
\fmfright{o1,o2}
\fmfbottom{i0,o0}
\fmfv{decor.shape=circle,decor.filled=empty,decor.size=1.5thick,label.dist=4,label=$x$}{vUp}
\fmfv{decor.shape=circle,decor.filled=empty,decor.size=0.28cm,label=$\mathrm{o}$,label.dist=0}{v1b}
\fmfv{decor.shape=circle,decor.filled=empty,decor.size=0.28cm,label=$\mathrm{o}$,label.dist=0}{v2b}
\fmfv{foreground=(0,,0,,1)}{v1}
\fmfv{foreground=(0,,0,,1)}{v2}
\fmfv{foreground=(0,,0,,1)}{v3}
\fmf{wiggly,tension=0,foreground=(1,,0,,0)}{v3,vUp}
\fmf{phantom,tension=0.4}{v3,vUpL1}
\fmf{phantom,tension=0.4}{v3,vUpR1}
\fmf{phantom}{i1,v1}
\fmf{phantom}{i2,v3b}
\fmf{phantom}{o1,v2}
\fmf{phantom}{o2,v3b}
\fmf{phantom}{i0,v1b}
\fmf{phantom}{o0,v2b}
\fmf{wiggly,tension=0.5,foreground=(1,,0,,0)}{v1,v2}
\fmf{plain,foreground=(1,,0,,0)}{v1,v3}
\fmf{plain,foreground=(1,,0,,0)}{v2,v3}
\fmf{plain,tension=0.4,foreground=(1,,0,,0)}{v1,v1b}
\fmf{plain,tension=0.4,foreground=(1,,0,,0)}{v2,v2b}
\fmf{phantom,tension=1.5}{v3,v3b}
\fmfdot{v1,v2,v3}
\end{fmfgraph*}
\end{fmffile}
\end{gathered} + \begin{gathered}
\begin{fmffile}{Diagrams/bosonic1PIEA_DeterminationJ15}
\begin{fmfgraph*}(22,18)
\fmftop{vUpL2,vUpL1,vUp,vUpR1,vUpR2}
\fmfleft{i1,i2}
\fmfright{o1,o2}
\fmfbottom{i0,o0}
\fmfv{decor.shape=circle,decor.filled=empty,decor.size=1.5thick,label.dist=4,label=$x$}{vUp}
\fmfv{foreground=(0,,0,,1)}{v1}
\fmfv{foreground=(0,,0,,1)}{v2}
\fmfv{foreground=(0,,0,,1)}{v3}
\fmf{wiggly,tension=0,foreground=(1,,0,,0)}{v3,vUp}
\fmf{phantom,tension=0.4}{v3,vUpL1}
\fmf{phantom,tension=0.4}{v3,vUpR1}
\fmf{phantom}{i1,v1}
\fmf{phantom}{i2,v3b}
\fmf{phantom}{o1,v2}
\fmf{phantom}{o2,v3b}
\fmf{phantom}{i0,v1b}
\fmf{phantom}{o0,v2b}
\fmf{wiggly,tension=0.5,foreground=(1,,0,,0)}{v1,v2}
\fmf{plain,right=0.6,tension=0,foreground=(1,,0,,0)}{v1,v2}
\fmf{plain,foreground=(1,,0,,0)}{v1,v3}
\fmf{plain,foreground=(1,,0,,0)}{v2,v3}
\fmf{phantom,tension=0.4}{v1,v1b}
\fmf{phantom,tension=0.4}{v2,v2b}
\fmf{phantom,tension=1.5}{v3,v3b}
\fmfdot{v1,v2,v3}
\end{fmfgraph*}
\end{fmffile}
\end{gathered} \left.\rule{0cm}{1.5cm}\right)\;.
\end{split}
\label{eq:bosonic1PIEADeterminationJ10DON}
\end{equation}
With the inverse propagator:
\begin{equation}
\begin{gathered}
\begin{fmffile}{Diagrams/bosonic1PIEA_FeynRuleGJ0minus1}
\begin{fmfgraph*}(20,20)
\fmfleft{i0,i1,i2,i3}
\fmfright{o0,o1,o2,o3}
\fmftop{vUpL8,vUpL7,vUpL6,vUpL5,vUpL4,vUpL3,vUpL2,vUpL1,vUp,vUpR1,vUpR2,vUpR3,vUpR4,vUpR5,vUpR6,vUpR7,vUpR8}
\fmfbottom{vDownL8,vDownL7,vDownL6,vDownL5,vDownL4,vDownL3,vDownL2,vDownL1,vDown,vDownR1,vDownR2,vDownR3,vDownR4,vDownR5,vDownR6,vDownR7,vDownR8}
\fmf{phantom,tension=1.0}{vUpL1,vLeft}
\fmf{phantom,tension=1.5}{vDownL1,vLeft}
\fmf{phantom,tension=1.5}{vUpR1,vRight}
\fmf{phantom,tension=1.0}{vDownR1,vRight}
\fmfv{label=$\beta_{1}$}{v1}
\fmfv{label=$\beta_{2}$}{v2}
\fmf{phantom}{i1,v1}
\fmf{phantom}{i2,v1}
\fmf{plain,tension=0.6,foreground=(1,,0,,0)}{v1,v2}
\fmf{wiggly,tension=0,foreground=(1,,0,,0)}{v1,v2}
\fmf{phantom}{v2,o1}
\fmf{phantom}{v2,o2}
\fmf{plain,tension=0,foreground=(1,,0,,0)}{vLeft,vRight}
\end{fmfgraph*}
\end{fmffile}
\end{gathered} \quad \rightarrow \mathcal{G}^{-1}_{\beta_{1}\beta_{2}}[\Phi]\;,
\end{equation}
we isolate $\mathcal{J}_{1}$ in~\eqref{eq:bosonic1PIEADeterminationJ10DON}:
\begin{equation}
\begin{split}
\mathcal{J}_{1,\beta}[\Phi] = & - i\sqrt{\frac{\lambda}{3}}\left(1-\delta_{b \hspace{0.04cm} N+1}\right) \hspace{-0.3cm} \begin{gathered}
\begin{fmffile}{Diagrams/bosonic1PIEA_DeterminationJ12bis}
\begin{fmfgraph*}(25,25)
\fmfleft{i1,i2}
\fmfright{o1,o2}
\fmfbottom{i0,o0}
\fmftop{i3,o3}
\fmfv{decor.shape=circle,decor.filled=empty,decor.size=1.5thick,label.angle=180,label.dist=4,label=$\alpha$}{v1}
\fmfv{foreground=(0,,0,,1)}{v2}
\fmfv{decor.shape=circle,decor.filled=empty,decor.size=0.28cm,label=$\mathrm{o}$,label.dist=0}{v3}
\fmf{phantom}{i1,v1}
\fmf{phantom}{i2,v4}
\fmf{phantom}{o1,v2}
\fmf{phantom}{o2,v3}
\fmf{phantom}{i3,v4b}
\fmf{phantom}{o3,v3b}
\fmf{phantom}{i0,v1b}
\fmf{phantom}{o0,v2b}
\fmf{wiggly,tension=1.6,foreground=(1,,0,,0)}{v1,v2}
\fmf{plain,left=0.6,tension=0,foreground=(1,,0,,0)}{v1,v2}
\fmf{phantom,tension=1.6}{v3,v4}
\fmf{phantom,tension=2.0,foreground=(1,,0,,0)}{v1,v4}
\fmf{plain,tension=2.0,foreground=(1,,0,,0)}{v2,v3}
\fmf{phantom,tension=0}{v1,v3}
\fmf{phantom,tension=0}{v2,v4}
\fmf{phantom}{v1,v1b}
\fmf{phantom}{v2,v2b}
\fmf{phantom}{v3,v3b}
\fmf{phantom}{v4,v4b}
\fmfdot{v2}
\end{fmfgraph*}
\end{fmffile}
\end{gathered} \\
& + \frac{i}{2}\sqrt{\frac{\lambda}{3}} \ \delta_{b \hspace{0.04cm} N+1} \left(\rule{0cm}{1.4cm}\right. 2 \hspace{-0.3cm} \begin{gathered}
\begin{fmffile}{Diagrams/bosonic1PIEA_DeterminationJ13bis}
\begin{fmfgraph*}(22,18)
\fmftop{vUpL2,vUpL1,vUp,vUpR1,vUpR2}
\fmfleft{i1,i2}
\fmfright{o1,o2}
\fmfbottom{i0,o0}
\fmfv{decor.shape=circle,decor.filled=empty,decor.size=1.5thick,label.angle=0,label.dist=4,label=$x$}{v3}
\fmfv{decor.shape=circle,decor.filled=empty,decor.size=0.28cm,label=$\mathrm{o}$,label.dist=0}{v2b}
\fmfv{decor.shape=circle,decor.filled=empty,decor.size=0.28cm,label=$\mathrm{o}$,label.dist=0}{vUp}
\fmfv{foreground=(0,,0,,1)}{v1}
\fmfv{foreground=(0,,0,,1)}{v2}
\fmf{phantom,tension=0.4,foreground=(1,,0,,0)}{v3,vUpL1}
\fmf{phantom,tension=0.4,foreground=(1,,0,,0)}{v3,vUpR1}
\fmf{plain,tension=0,foreground=(1,,0,,0)}{v3,vUp}
\fmf{phantom}{i1,v1}
\fmf{phantom}{i2,v3b}
\fmf{phantom}{o1,v2}
\fmf{phantom}{o2,v3b}
\fmf{phantom}{i0,v1b}
\fmf{phantom}{o0,v2b}
\fmf{wiggly,tension=0.5,foreground=(1,,0,,0)}{v1,v2}
\fmf{plain,right=0.6,tension=0,foreground=(1,,0,,0)}{v1,v2}
\fmf{plain,foreground=(1,,0,,0)}{v1,v3}
\fmf{phantom}{v2,v3}
\fmf{phantom,tension=0.4}{v1,v1b}
\fmf{plain,tension=0.4,foreground=(1,,0,,0)}{v2,v2b}
\fmf{phantom,tension=1.5}{v3,v3b}
\fmfdot{v1,v2}
\end{fmfgraph*}
\end{fmffile}
\end{gathered} + \begin{gathered}
\begin{fmffile}{Diagrams/bosonic1PIEA_DeterminationJ14bis}
\begin{fmfgraph*}(22,18)
\fmftop{vUpL2,vUpL1,vUp,vUpR1,vUpR2}
\fmfleft{i1,i2}
\fmfright{o1,o2}
\fmfbottom{i0,o0}
\fmfv{decor.shape=circle,decor.filled=empty,decor.size=1.5thick,label.dist=4,label=$x$}{v3}
\fmfv{decor.shape=circle,decor.filled=empty,decor.size=0.28cm,label=$\mathrm{o}$,label.dist=0}{v1b}
\fmfv{decor.shape=circle,decor.filled=empty,decor.size=0.28cm,label=$\mathrm{o}$,label.dist=0}{v2b}
\fmfv{foreground=(0,,0,,1)}{v1}
\fmfv{foreground=(0,,0,,1)}{v2}
\fmf{phantom,tension=0.4}{v3,vUpL1}
\fmf{phantom,tension=0.4}{v3,vUpR1}
\fmf{phantom}{i1,v1}
\fmf{phantom}{i2,v3b}
\fmf{phantom}{o1,v2}
\fmf{phantom}{o2,v3b}
\fmf{phantom}{i0,v1b}
\fmf{phantom}{o0,v2b}
\fmf{wiggly,tension=0.5,foreground=(1,,0,,0)}{v1,v2}
\fmf{plain,foreground=(1,,0,,0)}{v1,v3}
\fmf{plain,foreground=(1,,0,,0)}{v2,v3}
\fmf{plain,tension=0.4,foreground=(1,,0,,0)}{v1,v1b}
\fmf{plain,tension=0.4,foreground=(1,,0,,0)}{v2,v2b}
\fmf{phantom,tension=1.5}{v3,v3b}
\fmfdot{v1,v2}
\end{fmfgraph*}
\end{fmffile}
\end{gathered} + \begin{gathered}
\begin{fmffile}{Diagrams/bosonic1PIEA_DeterminationJ15bis}
\begin{fmfgraph*}(22,18)
\fmftop{vUpL2,vUpL1,vUp,vUpR1,vUpR2}
\fmfleft{i1,i2}
\fmfright{o1,o2}
\fmfbottom{i0,o0}
\fmfv{decor.shape=circle,decor.filled=empty,decor.size=1.5thick,label.dist=4,label=$x$}{v3}
\fmfv{foreground=(0,,0,,1)}{v1}
\fmfv{foreground=(0,,0,,1)}{v2}
\fmf{phantom,tension=0.4}{v3,vUpL1}
\fmf{phantom,tension=0.4}{v3,vUpR1}
\fmf{phantom}{i1,v1}
\fmf{phantom}{i2,v3b}
\fmf{phantom}{o1,v2}
\fmf{phantom}{o2,v3b}
\fmf{phantom}{i0,v1b}
\fmf{phantom}{o0,v2b}
\fmf{wiggly,tension=0.5,foreground=(1,,0,,0)}{v1,v2}
\fmf{plain,right=0.6,tension=0,foreground=(1,,0,,0)}{v1,v2}
\fmf{plain,foreground=(1,,0,,0)}{v1,v3}
\fmf{plain,foreground=(1,,0,,0)}{v2,v3}
\fmf{phantom,tension=0.4}{v1,v1b}
\fmf{phantom,tension=0.4}{v2,v2b}
\fmf{phantom,tension=1.5}{v3,v3b}
\fmfdot{v1,v2}
\end{fmfgraph*}
\end{fmffile}
\end{gathered} \left.\rule{0cm}{1.4cm}\right)\;,
\end{split}
\label{eq:bosonic1PIEADeterminationJ1step20DON}
\end{equation}
where we have used:
\begin{equation}
\int_{\beta_{3}} \left(1-\delta_{b_{3} N+1}\right) \hspace{0.2cm} \begin{gathered}
\begin{fmffile}{Diagrams/bosonic1PIEA_IsolateJ11}
\begin{fmfgraph*}(25,15)
\fmfleft{i0,i1,i2,i3}
\fmfright{o0,o1,o2,o3}
\fmftop{vUp}
\fmfbottom{vDown}
\fmfv{decor.shape=circle,decor.filled=empty,decor.size=1.5thick,label.angle=-90,label=$\beta_{1}$}{v1}
\fmfv{decor.shape=circle,decor.filled=empty,decor.size=1.5thick,label.angle=-90,label=$\beta_{3}$}{vMiddle}
\fmfv{decor.shape=circle,decor.filled=empty,decor.size=1.5thick,label.angle=-90,label=$\alpha_{2}$}{v2}
\fmf{phantom,tension=3.4}{i1,vSlashDown}
\fmf{phantom,tension=1.0}{o1,vSlashDown}
\fmf{phantom,tension=2.0}{i2,vSlashUp}
\fmf{phantom,tension=1.0}{o2,vSlashUp}
\fmf{plain,tension=0,foreground=(1,,0,,0)}{vSlashDown,vSlashUp}
\fmf{phantom}{vUp,vMiddle}
\fmf{phantom}{vDown,vMiddle}
\fmf{phantom}{i1,v1}
\fmf{phantom}{i2,v1}
\fmf{plain,tension=0.1,foreground=(1,,0,,0)}{v1,vMiddle}
\fmf{wiggly,tension=0,foreground=(1,,0,,0)}{v1,vMiddle}
\fmf{plain,tension=0.1,foreground=(1,,0,,0)}{vMiddle,v2}
\fmf{phantom}{v2,o1}
\fmf{phantom}{v2,o2}
\end{fmfgraph*}
\end{fmffile}
\end{gathered} \hspace{0.2cm} = \ \left(1-\delta_{b_{1} N+1}\right) \delta_{\alpha_{1}\alpha_{2}}\;,
\end{equation}
\begin{equation}
\int_{\beta_{3}} \delta_{b_{3} N+1} \hspace{0.2cm} \begin{gathered}
\begin{fmffile}{Diagrams/bosonic1PIEA_IsolateJ12}
\begin{fmfgraph*}(25,15)
\fmfleft{i0,i1,i2,i3}
\fmfright{o0,o1,o2,o3}
\fmftop{vUp}
\fmfbottom{vDown}
\fmfv{decor.shape=circle,decor.filled=empty,decor.size=1.5thick,label.angle=-90,label=$\beta_{1}$}{v1}
\fmfv{decor.shape=circle,decor.filled=empty,decor.size=1.5thick,label.angle=-90,label=$\beta_{3}$}{vMiddle}
\fmfv{decor.shape=circle,decor.filled=empty,decor.size=1.5thick,label.angle=-90,label=$x_{2}$}{v2}
\fmf{phantom,tension=3.4}{i1,vSlashDown}
\fmf{phantom,tension=1.0}{o1,vSlashDown}
\fmf{phantom,tension=2.0}{i2,vSlashUp}
\fmf{phantom,tension=1.0}{o2,vSlashUp}
\fmf{plain,tension=0,foreground=(1,,0,,0)}{vSlashDown,vSlashUp}
\fmf{phantom}{vUp,vMiddle}
\fmf{phantom}{vDown,vMiddle}
\fmf{phantom}{i1,v1}
\fmf{phantom}{i2,v1}
\fmf{plain,tension=0.1,foreground=(1,,0,,0)}{v1,vMiddle}
\fmf{wiggly,tension=0,foreground=(1,,0,,0)}{v1,vMiddle}
\fmf{wiggly,tension=0.1,foreground=(1,,0,,0)}{vMiddle,v2}
\fmf{phantom}{v2,o1}
\fmf{phantom}{v2,o2}
\end{fmfgraph*}
\end{fmffile}
\end{gathered} \hspace{0.2cm} = \ \delta_{b_{1} N+1} \delta_{x_{1}x_{2}}\;.
\end{equation}
From derivatives~\eqref{eq:bosonic1PIEADerivW0j0j00DON} and~\eqref{eq:bosonic1PIEADerivW1j0step20DON} as well as expression~\eqref{eq:bosonic1PIEADeterminationJ1step20DON} of $\mathcal{J}_{1}$, we calculate:
\begin{equation}
\begin{split}
\int_{\beta} \left.\frac{\delta W_{1}\big[\mathcal{J}\big]}{\delta \mathcal{J}_{\beta}}\right|_{\mathcal{J}=\mathcal{J}_{0}} \mathcal{J}_{1,\beta}[\Phi] = & -\begin{gathered}
\begin{fmffile}{Diagrams/bosonic1PIEA_DerivW1J0J1_Diag1}
\begin{fmfgraph*}(25,25)
\fmfleft{i1,i2}
\fmfright{o1,o2}
\fmfbottom{i0,o0}
\fmftop{i3,o3}
\fmfv{decor.shape=circle,decor.size=2.0thick,foreground=(0,,0,,1)}{v1}
\fmfv{decor.shape=circle,decor.size=2.0thick,foreground=(0,,0,,1)}{v2}
\fmfv{decor.shape=circle,decor.size=2.0thick,foreground=(0,,0,,1)}{v3}
\fmfv{decor.shape=circle,decor.size=2.0thick,foreground=(0,,0,,1)}{v4}
\fmfv{decor.shape=circle,decor.filled=empty,decor.size=0.28cm,label=$\mathrm{o}$,label.dist=0}{v3b}
\fmfv{decor.shape=circle,decor.filled=empty,decor.size=0.28cm,label=$\mathrm{o}$,label.dist=0}{v4b}
\fmf{phantom}{i1,v1}
\fmf{phantom}{i2,v4}
\fmf{phantom}{o1,v2}
\fmf{phantom}{o2,v3}
\fmf{phantom}{i3,v4b}
\fmf{phantom}{o3,v3b}
\fmf{phantom}{i0,v1b}
\fmf{phantom}{o0,v2b}
\fmf{plain,tension=1.6,foreground=(1,,0,,0)}{v1,v2}
\fmf{phantom,tension=1.6}{v3,v4}
\fmf{wiggly,tension=2.0,foreground=(1,,0,,0)}{v1,v4}
\fmf{wiggly,tension=2.0,foreground=(1,,0,,0)}{v2,v3}
\fmf{phantom,tension=0}{v1,v3}
\fmf{phantom,tension=0}{v2,v4}
\fmf{plain,right=0.8,tension=0,foreground=(1,,0,,0)}{v2,v3}
\fmf{plain,left=0.8,tension=0,foreground=(1,,0,,0)}{v1,v4}
\fmf{phantom}{v1,v1b}
\fmf{phantom}{v2,v2b}
\fmf{plain,foreground=(1,,0,,0)}{v3,v3b}
\fmf{plain,foreground=(1,,0,,0)}{v4,v4b}
\end{fmfgraph*}
\end{fmffile}
\end{gathered} - \begin{gathered}
\begin{fmffile}{Diagrams/bosonic1PIEA_DerivW1J0J1_Diag2}
\begin{fmfgraph*}(35,20)
\fmfleft{i1,i2}
\fmfright{o1,o2}
\fmfbottom{i0,o0}
\fmfbottom{b0}
\fmfbottom{b1}
\fmfbottom{b2}
\fmftop{i3,o3}
\fmfv{decor.shape=circle,decor.size=2.0thick,foreground=(0,,0,,1)}{v1}
\fmfv{decor.shape=circle,decor.size=2.0thick,foreground=(0,,0,,1)}{v2}
\fmfv{decor.shape=circle,decor.size=2.0thick,foreground=(0,,0,,1)}{v3}
\fmfv{decor.shape=circle,decor.size=2.0thick,foreground=(0,,0,,1)}{v4}
\fmfv{decor.shape=circle,decor.size=2.0thick,foreground=(0,,0,,1)}{v5}
\fmfv{decor.shape=circle,decor.size=2.0thick,foreground=(0,,0,,1)}{v6}
\fmfv{decor.shape=circle,decor.filled=empty,decor.size=0.28cm,label=$\mathrm{o}$,label.dist=0}{v2b}
\fmfv{decor.shape=circle,decor.filled=empty,decor.size=0.28cm,label=$\mathrm{o}$,label.dist=0}{v3b}
\fmfv{decor.shape=circle,decor.filled=empty,decor.size=0.28cm,label=$\mathrm{o}$,label.dist=0}{v4b}
\fmfv{decor.shape=circle,decor.filled=empty,decor.size=0.28cm,label=$\mathrm{o}$,label.dist=0}{v6b}
\fmf{phantom,tension=1.4}{i1,v1}
\fmf{phantom}{i2,v3b}
\fmf{phantom}{i0,v1b}
\fmf{phantom}{o2,v6b}
\fmf{phantom,tension=1.4}{o1,v5}
\fmf{phantom}{o0,v5b}
\fmf{phantom,tension=1.11}{v3b,v6b}
\fmf{phantom,tension=1.38}{i0,v2}
\fmf{phantom,tension=1.38}{o0,v2}
\fmf{phantom,tension=1.8}{i0,v2b}
\fmf{phantom,tension=1.2}{o0,v2b}
\fmf{phantom,tension=1.2}{b0,v2b}
\fmf{phantom,tension=1.2}{b1,v2b}
\fmf{phantom,tension=1.2}{b2,v2b}
\fmf{phantom,tension=1.38}{i0,v4}
\fmf{phantom,tension=1.38}{o0,v4}
\fmf{phantom,tension=1.2}{i0,v4b}
\fmf{phantom,tension=1.8}{o0,v4b}
\fmf{phantom,tension=1.2}{b0,v4b}
\fmf{phantom,tension=1.2}{b1,v4b}
\fmf{phantom,tension=1.2}{b2,v4b}
\fmf{phantom,tension=2}{i3,v3}
\fmf{phantom,tension=2}{o3,v6}
\fmf{phantom,tension=2}{i3,v3b}
\fmf{phantom,tension=0.8}{o3,v3b}
\fmf{phantom,tension=0.8}{i3,v6b}
\fmf{phantom,tension=2}{o3,v6b}
\fmf{plain,tension=1.4,foreground=(1,,0,,0)}{v1,v2}
\fmf{plain,tension=1.4,foreground=(1,,0,,0)}{v4,v5}
\fmf{wiggly,foreground=(1,,0,,0)}{v1,v3}
\fmf{plain,left=0.8,tension=0,foreground=(1,,0,,0)}{v1,v3}
\fmf{wiggly,foreground=(1,,0,,0)}{v5,v6}
\fmf{plain,right=0.8,tension=0,foreground=(1,,0,,0)}{v5,v6}
\fmf{phantom}{v2,v3}
\fmf{phantom}{v4,v6}
\fmf{wiggly,tension=0.5,foreground=(1,,0,,0)}{v2,v4}
\fmf{phantom,tension=2}{v3,v6}
\fmf{phantom,right=0.8,tension=0}{v1,v5}
\fmf{phantom,tension=1}{v1,v1b}
\fmf{plain,tension=0.2,foreground=(1,,0,,0)}{v2,v2b}
\fmf{plain,tension=1.5,foreground=(1,,0,,0)}{v3,v3b}
\fmf{plain,tension=0.2,foreground=(1,,0,,0)}{v4,v4b}
\fmf{phantom,tension=1}{v5,v5b}
\fmf{plain,tension=1.5,foreground=(1,,0,,0)}{v6,v6b}
\end{fmfgraph*}
\end{fmffile}
\end{gathered} - \begin{gathered}
\begin{fmffile}{Diagrams/bosonic1PIEA_DerivW1J0J1_Diag3}
\begin{fmfgraph*}(35,20)
\fmfleft{i1,i2}
\fmfright{o1,o2}
\fmfbottom{i0,o0}
\fmfbottom{b0}
\fmfbottom{b1}
\fmfbottom{b2}
\fmftop{i3,o3}
\fmfv{decor.shape=circle,decor.size=2.0thick,foreground=(0,,0,,1)}{v1}
\fmfv{decor.shape=circle,decor.size=2.0thick,foreground=(0,,0,,1)}{v2}
\fmfv{decor.shape=circle,decor.size=2.0thick,foreground=(0,,0,,1)}{v3}
\fmfv{decor.shape=circle,decor.size=2.0thick,foreground=(0,,0,,1)}{v4}
\fmfv{decor.shape=circle,decor.size=2.0thick,foreground=(0,,0,,1)}{v5}
\fmfv{decor.shape=circle,decor.size=2.0thick,foreground=(0,,0,,1)}{v6}
\fmfv{decor.shape=circle,decor.filled=empty,decor.size=0.28cm,label=$\mathrm{o}$,label.dist=0}{v1b}
\fmfv{decor.shape=circle,decor.filled=empty,decor.size=0.28cm,label=$\mathrm{o}$,label.dist=0}{v3b}
\fmfv{decor.shape=circle,decor.filled=empty,decor.size=0.28cm,label=$\mathrm{o}$,label.dist=0}{v4b}
\fmfv{decor.shape=circle,decor.filled=empty,decor.size=0.28cm,label=$\mathrm{o}$,label.dist=0}{v6b}
\fmf{phantom,tension=1.4}{i1,v1}
\fmf{phantom}{i2,v3b}
\fmf{phantom}{i0,v1b}
\fmf{phantom}{o2,v6b}
\fmf{phantom,tension=1.4}{o1,v5}
\fmf{phantom}{o0,v5b}
\fmf{phantom,tension=1.11}{v3b,v6b}
\fmf{phantom,tension=1.38}{i0,v2}
\fmf{phantom,tension=1.38}{o0,v2}
\fmf{phantom,tension=1.8}{i0,v2b}
\fmf{phantom,tension=1.2}{o0,v2b}
\fmf{phantom,tension=1.2}{b0,v2b}
\fmf{phantom,tension=1.2}{b1,v2b}
\fmf{phantom,tension=1.2}{b2,v2b}
\fmf{phantom,tension=1.38}{i0,v4}
\fmf{phantom,tension=1.38}{o0,v4}
\fmf{phantom,tension=1.2}{i0,v4b}
\fmf{phantom,tension=1.8}{o0,v4b}
\fmf{phantom,tension=1.2}{b0,v4b}
\fmf{phantom,tension=1.2}{b1,v4b}
\fmf{phantom,tension=1.2}{b2,v4b}
\fmf{phantom,tension=2}{i3,v3}
\fmf{phantom,tension=2}{o3,v6}
\fmf{phantom,tension=2}{i3,v3b}
\fmf{phantom,tension=0.8}{o3,v3b}
\fmf{phantom,tension=0.8}{i3,v6b}
\fmf{phantom,tension=2}{o3,v6b}
\fmf{plain,tension=1.4,foreground=(1,,0,,0)}{v1,v2}
\fmf{plain,tension=1.4,foreground=(1,,0,,0)}{v4,v5}
\fmf{wiggly,foreground=(1,,0,,0)}{v1,v3}
\fmf{phantom,left=0.8,tension=0}{v1,v3}
\fmf{wiggly,foreground=(1,,0,,0)}{v5,v6}
\fmf{plain,right=0.8,tension=0,foreground=(1,,0,,0)}{v5,v6}
\fmf{plain,foreground=(1,,0,,0)}{v2,v3}
\fmf{phantom}{v4,v6}
\fmf{wiggly,tension=0.5,foreground=(1,,0,,0)}{v2,v4}
\fmf{phantom,tension=2}{v3,v6}
\fmf{phantom,right=0.8,tension=0}{v1,v5}
\fmf{plain,tension=1,foreground=(1,,0,,0)}{v1,v1b}
\fmf{phantom,tension=0.2}{v2,v2b}
\fmf{plain,tension=1.5,foreground=(1,,0,,0)}{v3,v3b}
\fmf{plain,tension=0.2,foreground=(1,,0,,0)}{v4,v4b}
\fmf{phantom,tension=1}{v5,v5b}
\fmf{plain,tension=1.5,foreground=(1,,0,,0)}{v6,v6b}
\end{fmfgraph*}
\end{fmffile}
\end{gathered} \\
& -\frac{1}{4} \begin{gathered}
\begin{fmffile}{Diagrams/bosonic1PIEA_DerivW1J0J1_Diag4}
\begin{fmfgraph*}(35,20)
\fmfleft{i1,i2}
\fmfright{o1,o2}
\fmfbottom{i0,o0}
\fmfbottom{b0}
\fmfbottom{b1}
\fmfbottom{b2}
\fmftop{i3,o3}
\fmfv{decor.shape=circle,decor.size=2.0thick,foreground=(0,,0,,1)}{v1}
\fmfv{decor.shape=circle,decor.size=2.0thick,foreground=(0,,0,,1)}{v2}
\fmfv{decor.shape=circle,decor.size=2.0thick,foreground=(0,,0,,1)}{v3}
\fmfv{decor.shape=circle,decor.size=2.0thick,foreground=(0,,0,,1)}{v4}
\fmfv{decor.shape=circle,decor.size=2.0thick,foreground=(0,,0,,1)}{v5}
\fmfv{decor.shape=circle,decor.size=2.0thick,foreground=(0,,0,,1)}{v6}
\fmfv{decor.shape=circle,decor.filled=empty,decor.size=0.28cm,label=$\mathrm{o}$,label.dist=0}{v1b}
\fmfv{decor.shape=circle,decor.filled=empty,decor.size=0.28cm,label=$\mathrm{o}$,label.dist=0}{v3b}
\fmfv{decor.shape=circle,decor.filled=empty,decor.size=0.28cm,label=$\mathrm{o}$,label.dist=0}{v5b}
\fmfv{decor.shape=circle,decor.filled=empty,decor.size=0.28cm,label=$\mathrm{o}$,label.dist=0}{v6b}
\fmf{phantom,tension=1.4}{i1,v1}
\fmf{phantom}{i2,v3b}
\fmf{phantom}{i0,v1b}
\fmf{phantom}{o2,v6b}
\fmf{phantom,tension=1.4}{o1,v5}
\fmf{phantom}{o0,v5b}
\fmf{phantom,tension=1.11}{v3b,v6b}
\fmf{phantom,tension=1.38}{i0,v2}
\fmf{phantom,tension=1.38}{o0,v2}
\fmf{phantom,tension=1.8}{i0,v2b}
\fmf{phantom,tension=1.2}{o0,v2b}
\fmf{phantom,tension=1.2}{b0,v2b}
\fmf{phantom,tension=1.2}{b1,v2b}
\fmf{phantom,tension=1.2}{b2,v2b}
\fmf{phantom,tension=1.38}{i0,v4}
\fmf{phantom,tension=1.38}{o0,v4}
\fmf{phantom,tension=1.2}{i0,v4b}
\fmf{phantom,tension=1.8}{o0,v4b}
\fmf{phantom,tension=1.2}{b0,v4b}
\fmf{phantom,tension=1.2}{b1,v4b}
\fmf{phantom,tension=1.2}{b2,v4b}
\fmf{phantom,tension=2}{i3,v3}
\fmf{phantom,tension=2}{o3,v6}
\fmf{phantom,tension=2}{i3,v3b}
\fmf{phantom,tension=0.8}{o3,v3b}
\fmf{phantom,tension=0.8}{i3,v6b}
\fmf{phantom,tension=2}{o3,v6b}
\fmf{plain,tension=1.4,foreground=(1,,0,,0)}{v1,v2}
\fmf{plain,tension=1.4,foreground=(1,,0,,0)}{v4,v5}
\fmf{wiggly,foreground=(1,,0,,0)}{v1,v3}
\fmf{phantom,left=0.8,tension=0,foreground=(1,,0,,0)}{v1,v3}
\fmf{wiggly,foreground=(1,,0,,0)}{v5,v6}
\fmf{phantom,right=0.8,tension=0}{v5,v6}
\fmf{plain,foreground=(1,,0,,0)}{v2,v3}
\fmf{plain,foreground=(1,,0,,0)}{v4,v6}
\fmf{wiggly,tension=0.5,foreground=(1,,0,,0)}{v2,v4}
\fmf{phantom,tension=2}{v3,v6}
\fmf{phantom,right=0.8,tension=0}{v1,v5}
\fmf{plain,tension=1,foreground=(1,,0,,0)}{v1,v1b}
\fmf{phantom,tension=0.2}{v2,v2b}
\fmf{plain,tension=1.5,foreground=(1,,0,,0)}{v3,v3b}
\fmf{phantom,tension=0.2}{v4,v4b}
\fmf{plain,tension=1,foreground=(1,,0,,0)}{v5,v5b}
\fmf{plain,tension=1.5,foreground=(1,,0,,0)}{v6,v6b}
\end{fmfgraph*}
\end{fmffile}
\end{gathered} - \begin{gathered}
\begin{fmffile}{Diagrams/bosonic1PIEA_DerivW1J0J1_Diag5}
\begin{fmfgraph*}(35,20)
\fmfleft{i1,i2}
\fmfright{o1,o2}
\fmfbottom{i0,o0}
\fmfbottom{b0}
\fmfbottom{b1}
\fmfbottom{b2}
\fmftop{i3,o3}
\fmfv{decor.shape=circle,decor.size=2.0thick,foreground=(0,,0,,1)}{v1}
\fmfv{decor.shape=circle,decor.size=2.0thick,foreground=(0,,0,,1)}{v2}
\fmfv{decor.shape=circle,decor.size=2.0thick,foreground=(0,,0,,1)}{v3}
\fmfv{decor.shape=circle,decor.size=2.0thick,foreground=(0,,0,,1)}{v4}
\fmfv{decor.shape=circle,decor.size=2.0thick,foreground=(0,,0,,1)}{v5}
\fmfv{decor.shape=circle,decor.size=2.0thick,foreground=(0,,0,,1)}{v6}
\fmfv{decor.shape=circle,decor.filled=empty,decor.size=0.28cm,label=$\mathrm{o}$,label.dist=0}{v2b}
\fmfv{decor.shape=circle,decor.filled=empty,decor.size=0.28cm,label=$\mathrm{o}$,label.dist=0}{v3b}
\fmf{phantom,tension=1.4}{i1,v1}
\fmf{phantom}{i2,v3b}
\fmf{phantom}{i0,v1b}
\fmf{phantom}{o2,v6b}
\fmf{phantom,tension=1.4}{o1,v5}
\fmf{phantom}{o0,v5b}
\fmf{phantom,tension=1.11}{v3b,v6b}
\fmf{phantom,tension=1.38}{i0,v2}
\fmf{phantom,tension=1.38}{o0,v2}
\fmf{phantom,tension=1.8}{i0,v2b}
\fmf{phantom,tension=1.2}{o0,v2b}
\fmf{phantom,tension=1.2}{b0,v2b}
\fmf{phantom,tension=1.2}{b1,v2b}
\fmf{phantom,tension=1.2}{b2,v2b}
\fmf{phantom,tension=1.38}{i0,v4}
\fmf{phantom,tension=1.38}{o0,v4}
\fmf{phantom,tension=1.2}{i0,v4b}
\fmf{phantom,tension=1.8}{o0,v4b}
\fmf{phantom,tension=1.2}{b0,v4b}
\fmf{phantom,tension=1.2}{b1,v4b}
\fmf{phantom,tension=1.2}{b2,v4b}
\fmf{phantom,tension=2}{i3,v3}
\fmf{phantom,tension=2}{o3,v6}
\fmf{phantom,tension=2}{i3,v3b}
\fmf{phantom,tension=0.8}{o3,v3b}
\fmf{phantom,tension=0.8}{i3,v6b}
\fmf{phantom,tension=2}{o3,v6b}
\fmf{plain,tension=1.4,foreground=(1,,0,,0)}{v1,v2}
\fmf{plain,tension=1.4,foreground=(1,,0,,0)}{v4,v5}
\fmf{wiggly,foreground=(1,,0,,0)}{v1,v3}
\fmf{plain,left=0.8,tension=0,foreground=(1,,0,,0)}{v1,v3}
\fmf{wiggly,foreground=(1,,0,,0)}{v5,v6}
\fmf{plain,right=0.8,tension=0,foreground=(1,,0,,0)}{v5,v6}
\fmf{phantom}{v2,v3}
\fmf{plain,foreground=(1,,0,,0)}{v4,v6}
\fmf{wiggly,tension=0.5,foreground=(1,,0,,0)}{v2,v4}
\fmf{phantom,tension=2}{v3,v6}
\fmf{phantom,right=0.8,tension=0}{v1,v5}
\fmf{phantom,tension=1}{v1,v1b}
\fmf{plain,tension=0.2,foreground=(1,,0,,0)}{v2,v2b}
\fmf{plain,tension=1.5,foreground=(1,,0,,0)}{v3,v3b}
\fmf{phantom,tension=0.2}{v4,v4b}
\fmf{phantom,tension=1}{v5,v5b}
\fmf{phantom,tension=1.5}{v6,v6b}
\end{fmfgraph*}
\end{fmffile}
\end{gathered} - \frac{1}{2} \begin{gathered}
\begin{fmffile}{Diagrams/bosonic1PIEA_DerivW1J0J1_Diag6}
\begin{fmfgraph*}(35,20)
\fmfleft{i1,i2}
\fmfright{o1,o2}
\fmfbottom{i0,o0}
\fmfbottom{b0}
\fmfbottom{b1}
\fmfbottom{b2}
\fmftop{i3,o3}
\fmfv{decor.shape=circle,decor.size=2.0thick,foreground=(0,,0,,1)}{v1}
\fmfv{decor.shape=circle,decor.size=2.0thick,foreground=(0,,0,,1)}{v2}
\fmfv{decor.shape=circle,decor.size=2.0thick,foreground=(0,,0,,1)}{v3}
\fmfv{decor.shape=circle,decor.size=2.0thick,foreground=(0,,0,,1)}{v4}
\fmfv{decor.shape=circle,decor.size=2.0thick,foreground=(0,,0,,1)}{v5}
\fmfv{decor.shape=circle,decor.size=2.0thick,foreground=(0,,0,,1)}{v6}
\fmfv{decor.shape=circle,decor.filled=empty,decor.size=0.28cm,label=$\mathrm{o}$,label.dist=0}{v1b}
\fmfv{decor.shape=circle,decor.filled=empty,decor.size=0.28cm,label=$\mathrm{o}$,label.dist=0}{v3b}
\fmf{phantom,tension=1.4}{i1,v1}
\fmf{phantom}{i2,v3b}
\fmf{phantom}{i0,v1b}
\fmf{phantom}{o2,v6b}
\fmf{phantom,tension=1.4}{o1,v5}
\fmf{phantom}{o0,v5b}
\fmf{phantom,tension=1.11}{v3b,v6b}
\fmf{phantom,tension=1.38}{i0,v2}
\fmf{phantom,tension=1.38}{o0,v2}
\fmf{phantom,tension=1.8}{i0,v2b}
\fmf{phantom,tension=1.2}{o0,v2b}
\fmf{phantom,tension=1.2}{b0,v2b}
\fmf{phantom,tension=1.2}{b1,v2b}
\fmf{phantom,tension=1.2}{b2,v2b}
\fmf{phantom,tension=1.38}{i0,v4}
\fmf{phantom,tension=1.38}{o0,v4}
\fmf{phantom,tension=1.2}{i0,v4b}
\fmf{phantom,tension=1.8}{o0,v4b}
\fmf{phantom,tension=1.2}{b0,v4b}
\fmf{phantom,tension=1.2}{b1,v4b}
\fmf{phantom,tension=1.2}{b2,v4b}
\fmf{phantom,tension=2}{i3,v3}
\fmf{phantom,tension=2}{o3,v6}
\fmf{phantom,tension=2}{i3,v3b}
\fmf{phantom,tension=0.8}{o3,v3b}
\fmf{phantom,tension=0.8}{i3,v6b}
\fmf{phantom,tension=2}{o3,v6b}
\fmf{plain,tension=1.4,foreground=(1,,0,,0)}{v1,v2}
\fmf{plain,tension=1.4,foreground=(1,,0,,0)}{v4,v5}
\fmf{wiggly,foreground=(1,,0,,0)}{v1,v3}
\fmf{phantom,left=0.8,tension=0}{v1,v3}
\fmf{wiggly,foreground=(1,,0,,0)}{v5,v6}
\fmf{plain,right=0.8,tension=0,foreground=(1,,0,,0)}{v5,v6}
\fmf{plain,foreground=(1,,0,,0)}{v2,v3}
\fmf{plain,foreground=(1,,0,,0)}{v4,v6}
\fmf{wiggly,tension=0.5,foreground=(1,,0,,0)}{v2,v4}
\fmf{phantom,tension=2}{v3,v6}
\fmf{phantom,right=0.8,tension=0}{v1,v5}
\fmf{plain,tension=1,foreground=(1,,0,,0)}{v1,v1b}
\fmf{phantom,tension=0.2}{v2,v2b}
\fmf{plain,tension=1.5,foreground=(1,,0,,0)}{v3,v3b}
\fmf{phantom,tension=0.2}{v4,v4b}
\fmf{phantom,tension=1}{v5,v5b}
\fmf{phantom,tension=1.5}{v6,v6b}
\end{fmfgraph*}
\end{fmffile}
\end{gathered} \\
& -\frac{1}{4} \begin{gathered}
\begin{fmffile}{Diagrams/bosonic1PIEA_DerivW1J0J1_Diag7}
\begin{fmfgraph*}(35,20)
\fmfleft{i1,i2}
\fmfright{o1,o2}
\fmfbottom{i0,o0}
\fmfbottom{b0}
\fmfbottom{b1}
\fmfbottom{b2}
\fmftop{i3,o3}
\fmfv{decor.shape=circle,decor.size=2.0thick,foreground=(0,,0,,1)}{v1}
\fmfv{decor.shape=circle,decor.size=2.0thick,foreground=(0,,0,,1)}{v2}
\fmfv{decor.shape=circle,decor.size=2.0thick,foreground=(0,,0,,1)}{v3}
\fmfv{decor.shape=circle,decor.size=2.0thick,foreground=(0,,0,,1)}{v4}
\fmfv{decor.shape=circle,decor.size=2.0thick,foreground=(0,,0,,1)}{v5}
\fmfv{decor.shape=circle,decor.size=2.0thick,foreground=(0,,0,,1)}{v6}
\fmf{phantom,tension=1.4}{i1,v1}
\fmf{phantom}{i2,v3b}
\fmf{phantom}{i0,v1b}
\fmf{phantom}{o2,v6b}
\fmf{phantom,tension=1.4}{o1,v5}
\fmf{phantom}{o0,v5b}
\fmf{phantom,tension=1.11}{v3b,v6b}
\fmf{phantom,tension=1.38}{i0,v2}
\fmf{phantom,tension=1.38}{o0,v2}
\fmf{phantom,tension=1.8}{i0,v2b}
\fmf{phantom,tension=1.2}{o0,v2b}
\fmf{phantom,tension=1.2}{b0,v2b}
\fmf{phantom,tension=1.2}{b1,v2b}
\fmf{phantom,tension=1.2}{b2,v2b}
\fmf{phantom,tension=1.38}{i0,v4}
\fmf{phantom,tension=1.38}{o0,v4}
\fmf{phantom,tension=1.2}{i0,v4b}
\fmf{phantom,tension=1.8}{o0,v4b}
\fmf{phantom,tension=1.2}{b0,v4b}
\fmf{phantom,tension=1.2}{b1,v4b}
\fmf{phantom,tension=1.2}{b2,v4b}
\fmf{phantom,tension=2}{i3,v3}
\fmf{phantom,tension=2}{o3,v6}
\fmf{phantom,tension=2}{i3,v3b}
\fmf{phantom,tension=0.8}{o3,v3b}
\fmf{phantom,tension=0.8}{i3,v6b}
\fmf{phantom,tension=2}{o3,v6b}
\fmf{plain,tension=1.4,foreground=(1,,0,,0)}{v1,v2}
\fmf{plain,tension=1.4,foreground=(1,,0,,0)}{v4,v5}
\fmf{wiggly,foreground=(1,,0,,0)}{v1,v3}
\fmf{plain,left=0.8,tension=0,foreground=(1,,0,,0)}{v1,v3}
\fmf{wiggly,foreground=(1,,0,,0)}{v5,v6}
\fmf{plain,right=0.8,tension=0,foreground=(1,,0,,0)}{v5,v6}
\fmf{plain,foreground=(1,,0,,0)}{v2,v3}
\fmf{plain,foreground=(1,,0,,0)}{v4,v6}
\fmf{wiggly,tension=0.5,foreground=(1,,0,,0)}{v2,v4}
\fmf{phantom,tension=2}{v3,v6}
\fmf{phantom,right=0.8,tension=0}{v1,v5}
\fmf{phantom,tension=1}{v1,v1b}
\fmf{phantom,tension=0.2}{v2,v2b}
\fmf{phantom,tension=1.5}{v3,v3b}
\fmf{phantom,tension=0.2}{v4,v4b}
\fmf{phantom,tension=1}{v5,v5b}
\fmf{phantom,tension=1.5}{v6,v6b}
\end{fmfgraph*}
\end{fmffile}
\end{gathered} \hspace{-0.4cm} \;,
\end{split}
\label{eq:bosonic1PIEADerivW1J0J10DON}
\end{equation}
\begin{equation}
\begin{split}
\frac{1}{2} \int_{\beta_{1},\beta_{2}} \left.\frac{\delta^{2} W_{0}\big[\mathcal{J}\big]}{\delta \mathcal{J}_{\beta_{1}} \delta \mathcal{J}_{\beta_{2}}}\right|_{\mathcal{J}=\mathcal{J}_{0}} \mathcal{J}_{1,\beta_{1}}[\Phi] \mathcal{J}_{1,\beta_{2}}[\Phi] = & \ \frac{1}{2}\begin{gathered}
\begin{fmffile}{Diagrams/bosonic1PIEA_DerivW1J0J1_Diag1}
\begin{fmfgraph*}(25,25)
\fmfleft{i1,i2}
\fmfright{o1,o2}
\fmfbottom{i0,o0}
\fmftop{i3,o3}
\fmfv{decor.shape=circle,decor.size=2.0thick,foreground=(0,,0,,1)}{v1}
\fmfv{decor.shape=circle,decor.size=2.0thick,foreground=(0,,0,,1)}{v2}
\fmfv{decor.shape=circle,decor.size=2.0thick,foreground=(0,,0,,1)}{v3}
\fmfv{decor.shape=circle,decor.size=2.0thick,foreground=(0,,0,,1)}{v4}
\fmfv{decor.shape=circle,decor.filled=empty,decor.size=0.28cm,label=$\mathrm{o}$,label.dist=0}{v3b}
\fmfv{decor.shape=circle,decor.filled=empty,decor.size=0.28cm,label=$\mathrm{o}$,label.dist=0}{v4b}
\fmf{phantom}{i1,v1}
\fmf{phantom}{i2,v4}
\fmf{phantom}{o1,v2}
\fmf{phantom}{o2,v3}
\fmf{phantom}{i3,v4b}
\fmf{phantom}{o3,v3b}
\fmf{phantom}{i0,v1b}
\fmf{phantom}{o0,v2b}
\fmf{plain,tension=1.6,foreground=(1,,0,,0)}{v1,v2}
\fmf{phantom,tension=1.6}{v3,v4}
\fmf{wiggly,tension=2.0,foreground=(1,,0,,0)}{v1,v4}
\fmf{wiggly,tension=2.0,foreground=(1,,0,,0)}{v2,v3}
\fmf{phantom,tension=0}{v1,v3}
\fmf{phantom,tension=0}{v2,v4}
\fmf{plain,right=0.8,tension=0,foreground=(1,,0,,0)}{v2,v3}
\fmf{plain,left=0.8,tension=0,foreground=(1,,0,,0)}{v1,v4}
\fmf{phantom}{v1,v1b}
\fmf{phantom}{v2,v2b}
\fmf{plain,foreground=(1,,0,,0)}{v3,v3b}
\fmf{plain,foreground=(1,,0,,0)}{v4,v4b}
\end{fmfgraph*}
\end{fmffile}
\end{gathered} + \frac{1}{2} \begin{gathered}
\begin{fmffile}{Diagrams/bosonic1PIEA_DerivW1J0J1_Diag2}
\begin{fmfgraph*}(35,20)
\fmfleft{i1,i2}
\fmfright{o1,o2}
\fmfbottom{i0,o0}
\fmfbottom{b0}
\fmfbottom{b1}
\fmfbottom{b2}
\fmftop{i3,o3}
\fmfv{decor.shape=circle,decor.size=2.0thick,foreground=(0,,0,,1)}{v1}
\fmfv{decor.shape=circle,decor.size=2.0thick,foreground=(0,,0,,1)}{v2}
\fmfv{decor.shape=circle,decor.size=2.0thick,foreground=(0,,0,,1)}{v3}
\fmfv{decor.shape=circle,decor.size=2.0thick,foreground=(0,,0,,1)}{v4}
\fmfv{decor.shape=circle,decor.size=2.0thick,foreground=(0,,0,,1)}{v5}
\fmfv{decor.shape=circle,decor.size=2.0thick,foreground=(0,,0,,1)}{v6}
\fmfv{decor.shape=circle,decor.filled=empty,decor.size=0.28cm,label=$\mathrm{o}$,label.dist=0}{v2b}
\fmfv{decor.shape=circle,decor.filled=empty,decor.size=0.28cm,label=$\mathrm{o}$,label.dist=0}{v3b}
\fmfv{decor.shape=circle,decor.filled=empty,decor.size=0.28cm,label=$\mathrm{o}$,label.dist=0}{v4b}
\fmfv{decor.shape=circle,decor.filled=empty,decor.size=0.28cm,label=$\mathrm{o}$,label.dist=0}{v6b}
\fmf{phantom,tension=1.4}{i1,v1}
\fmf{phantom}{i2,v3b}
\fmf{phantom}{i0,v1b}
\fmf{phantom}{o2,v6b}
\fmf{phantom,tension=1.4}{o1,v5}
\fmf{phantom}{o0,v5b}
\fmf{phantom,tension=1.11}{v3b,v6b}
\fmf{phantom,tension=1.38}{i0,v2}
\fmf{phantom,tension=1.38}{o0,v2}
\fmf{phantom,tension=1.8}{i0,v2b}
\fmf{phantom,tension=1.2}{o0,v2b}
\fmf{phantom,tension=1.2}{b0,v2b}
\fmf{phantom,tension=1.2}{b1,v2b}
\fmf{phantom,tension=1.2}{b2,v2b}
\fmf{phantom,tension=1.38}{i0,v4}
\fmf{phantom,tension=1.38}{o0,v4}
\fmf{phantom,tension=1.2}{i0,v4b}
\fmf{phantom,tension=1.8}{o0,v4b}
\fmf{phantom,tension=1.2}{b0,v4b}
\fmf{phantom,tension=1.2}{b1,v4b}
\fmf{phantom,tension=1.2}{b2,v4b}
\fmf{phantom,tension=2}{i3,v3}
\fmf{phantom,tension=2}{o3,v6}
\fmf{phantom,tension=2}{i3,v3b}
\fmf{phantom,tension=0.8}{o3,v3b}
\fmf{phantom,tension=0.8}{i3,v6b}
\fmf{phantom,tension=2}{o3,v6b}
\fmf{plain,tension=1.4,foreground=(1,,0,,0)}{v1,v2}
\fmf{plain,tension=1.4,foreground=(1,,0,,0)}{v4,v5}
\fmf{wiggly,foreground=(1,,0,,0)}{v1,v3}
\fmf{plain,left=0.8,tension=0,foreground=(1,,0,,0)}{v1,v3}
\fmf{wiggly,foreground=(1,,0,,0)}{v5,v6}
\fmf{plain,right=0.8,tension=0,foreground=(1,,0,,0)}{v5,v6}
\fmf{phantom}{v2,v3}
\fmf{phantom}{v4,v6}
\fmf{wiggly,tension=0.5,foreground=(1,,0,,0)}{v2,v4}
\fmf{phantom,tension=2}{v3,v6}
\fmf{phantom,right=0.8,tension=0}{v1,v5}
\fmf{phantom,tension=1}{v1,v1b}
\fmf{plain,tension=0.2,foreground=(1,,0,,0)}{v2,v2b}
\fmf{plain,tension=1.5,foreground=(1,,0,,0)}{v3,v3b}
\fmf{plain,tension=0.2,foreground=(1,,0,,0)}{v4,v4b}
\fmf{phantom,tension=1}{v5,v5b}
\fmf{plain,tension=1.5,foreground=(1,,0,,0)}{v6,v6b}
\end{fmfgraph*}
\end{fmffile}
\end{gathered} \\
& + \frac{1}{2} \begin{gathered}
\begin{fmffile}{Diagrams/bosonic1PIEA_DerivW1J0J1_Diag3}
\begin{fmfgraph*}(35,20)
\fmfleft{i1,i2}
\fmfright{o1,o2}
\fmfbottom{i0,o0}
\fmfbottom{b0}
\fmfbottom{b1}
\fmfbottom{b2}
\fmftop{i3,o3}
\fmfv{decor.shape=circle,decor.size=2.0thick,foreground=(0,,0,,1)}{v1}
\fmfv{decor.shape=circle,decor.size=2.0thick,foreground=(0,,0,,1)}{v2}
\fmfv{decor.shape=circle,decor.size=2.0thick,foreground=(0,,0,,1)}{v3}
\fmfv{decor.shape=circle,decor.size=2.0thick,foreground=(0,,0,,1)}{v4}
\fmfv{decor.shape=circle,decor.size=2.0thick,foreground=(0,,0,,1)}{v5}
\fmfv{decor.shape=circle,decor.size=2.0thick,foreground=(0,,0,,1)}{v6}
\fmfv{decor.shape=circle,decor.filled=empty,decor.size=0.28cm,label=$\mathrm{o}$,label.dist=0}{v1b}
\fmfv{decor.shape=circle,decor.filled=empty,decor.size=0.28cm,label=$\mathrm{o}$,label.dist=0}{v3b}
\fmfv{decor.shape=circle,decor.filled=empty,decor.size=0.28cm,label=$\mathrm{o}$,label.dist=0}{v4b}
\fmfv{decor.shape=circle,decor.filled=empty,decor.size=0.28cm,label=$\mathrm{o}$,label.dist=0}{v6b}
\fmf{phantom,tension=1.4}{i1,v1}
\fmf{phantom}{i2,v3b}
\fmf{phantom}{i0,v1b}
\fmf{phantom}{o2,v6b}
\fmf{phantom,tension=1.4}{o1,v5}
\fmf{phantom}{o0,v5b}
\fmf{phantom,tension=1.11}{v3b,v6b}
\fmf{phantom,tension=1.38}{i0,v2}
\fmf{phantom,tension=1.38}{o0,v2}
\fmf{phantom,tension=1.8}{i0,v2b}
\fmf{phantom,tension=1.2}{o0,v2b}
\fmf{phantom,tension=1.2}{b0,v2b}
\fmf{phantom,tension=1.2}{b1,v2b}
\fmf{phantom,tension=1.2}{b2,v2b}
\fmf{phantom,tension=1.38}{i0,v4}
\fmf{phantom,tension=1.38}{o0,v4}
\fmf{phantom,tension=1.2}{i0,v4b}
\fmf{phantom,tension=1.8}{o0,v4b}
\fmf{phantom,tension=1.2}{b0,v4b}
\fmf{phantom,tension=1.2}{b1,v4b}
\fmf{phantom,tension=1.2}{b2,v4b}
\fmf{phantom,tension=2}{i3,v3}
\fmf{phantom,tension=2}{o3,v6}
\fmf{phantom,tension=2}{i3,v3b}
\fmf{phantom,tension=0.8}{o3,v3b}
\fmf{phantom,tension=0.8}{i3,v6b}
\fmf{phantom,tension=2}{o3,v6b}
\fmf{plain,tension=1.4,foreground=(1,,0,,0)}{v1,v2}
\fmf{plain,tension=1.4,foreground=(1,,0,,0)}{v4,v5}
\fmf{wiggly,foreground=(1,,0,,0)}{v1,v3}
\fmf{phantom,left=0.8,tension=0}{v1,v3}
\fmf{wiggly,foreground=(1,,0,,0)}{v5,v6}
\fmf{plain,right=0.8,tension=0,foreground=(1,,0,,0)}{v5,v6}
\fmf{plain,foreground=(1,,0,,0)}{v2,v3}
\fmf{phantom}{v4,v6}
\fmf{wiggly,tension=0.5,foreground=(1,,0,,0)}{v2,v4}
\fmf{phantom,tension=2}{v3,v6}
\fmf{phantom,right=0.8,tension=0}{v1,v5}
\fmf{plain,tension=1,foreground=(1,,0,,0)}{v1,v1b}
\fmf{phantom,tension=0.2}{v2,v2b}
\fmf{plain,tension=1.5,foreground=(1,,0,,0)}{v3,v3b}
\fmf{plain,tension=0.2,foreground=(1,,0,,0)}{v4,v4b}
\fmf{phantom,tension=1}{v5,v5b}
\fmf{plain,tension=1.5,foreground=(1,,0,,0)}{v6,v6b}
\end{fmfgraph*}
\end{fmffile}
\end{gathered} +\frac{1}{8} \begin{gathered}
\begin{fmffile}{Diagrams/bosonic1PIEA_DerivW1J0J1_Diag4}
\begin{fmfgraph*}(35,20)
\fmfleft{i1,i2}
\fmfright{o1,o2}
\fmfbottom{i0,o0}
\fmfbottom{b0}
\fmfbottom{b1}
\fmfbottom{b2}
\fmftop{i3,o3}
\fmfv{decor.shape=circle,decor.size=2.0thick,foreground=(0,,0,,1)}{v1}
\fmfv{decor.shape=circle,decor.size=2.0thick,foreground=(0,,0,,1)}{v2}
\fmfv{decor.shape=circle,decor.size=2.0thick,foreground=(0,,0,,1)}{v3}
\fmfv{decor.shape=circle,decor.size=2.0thick,foreground=(0,,0,,1)}{v4}
\fmfv{decor.shape=circle,decor.size=2.0thick,foreground=(0,,0,,1)}{v5}
\fmfv{decor.shape=circle,decor.size=2.0thick,foreground=(0,,0,,1)}{v6}
\fmfv{decor.shape=circle,decor.filled=empty,decor.size=0.28cm,label=$\mathrm{o}$,label.dist=0}{v1b}
\fmfv{decor.shape=circle,decor.filled=empty,decor.size=0.28cm,label=$\mathrm{o}$,label.dist=0}{v3b}
\fmfv{decor.shape=circle,decor.filled=empty,decor.size=0.28cm,label=$\mathrm{o}$,label.dist=0}{v5b}
\fmfv{decor.shape=circle,decor.filled=empty,decor.size=0.28cm,label=$\mathrm{o}$,label.dist=0}{v6b}
\fmf{phantom,tension=1.4}{i1,v1}
\fmf{phantom}{i2,v3b}
\fmf{phantom}{i0,v1b}
\fmf{phantom}{o2,v6b}
\fmf{phantom,tension=1.4}{o1,v5}
\fmf{phantom}{o0,v5b}
\fmf{phantom,tension=1.11}{v3b,v6b}
\fmf{phantom,tension=1.38}{i0,v2}
\fmf{phantom,tension=1.38}{o0,v2}
\fmf{phantom,tension=1.8}{i0,v2b}
\fmf{phantom,tension=1.2}{o0,v2b}
\fmf{phantom,tension=1.2}{b0,v2b}
\fmf{phantom,tension=1.2}{b1,v2b}
\fmf{phantom,tension=1.2}{b2,v2b}
\fmf{phantom,tension=1.38}{i0,v4}
\fmf{phantom,tension=1.38}{o0,v4}
\fmf{phantom,tension=1.2}{i0,v4b}
\fmf{phantom,tension=1.8}{o0,v4b}
\fmf{phantom,tension=1.2}{b0,v4b}
\fmf{phantom,tension=1.2}{b1,v4b}
\fmf{phantom,tension=1.2}{b2,v4b}
\fmf{phantom,tension=2}{i3,v3}
\fmf{phantom,tension=2}{o3,v6}
\fmf{phantom,tension=2}{i3,v3b}
\fmf{phantom,tension=0.8}{o3,v3b}
\fmf{phantom,tension=0.8}{i3,v6b}
\fmf{phantom,tension=2}{o3,v6b}
\fmf{plain,tension=1.4,foreground=(1,,0,,0)}{v1,v2}
\fmf{plain,tension=1.4,foreground=(1,,0,,0)}{v4,v5}
\fmf{wiggly,foreground=(1,,0,,0)}{v1,v3}
\fmf{phantom,left=0.8,tension=0,foreground=(1,,0,,0)}{v1,v3}
\fmf{wiggly,foreground=(1,,0,,0)}{v5,v6}
\fmf{phantom,right=0.8,tension=0}{v5,v6}
\fmf{plain,foreground=(1,,0,,0)}{v2,v3}
\fmf{plain,foreground=(1,,0,,0)}{v4,v6}
\fmf{wiggly,tension=0.5,foreground=(1,,0,,0)}{v2,v4}
\fmf{phantom,tension=2}{v3,v6}
\fmf{phantom,right=0.8,tension=0}{v1,v5}
\fmf{plain,tension=1,foreground=(1,,0,,0)}{v1,v1b}
\fmf{phantom,tension=0.2}{v2,v2b}
\fmf{plain,tension=1.5,foreground=(1,,0,,0)}{v3,v3b}
\fmf{phantom,tension=0.2}{v4,v4b}
\fmf{plain,tension=1,foreground=(1,,0,,0)}{v5,v5b}
\fmf{plain,tension=1.5,foreground=(1,,0,,0)}{v6,v6b}
\end{fmfgraph*}
\end{fmffile}
\end{gathered} \\
& + \frac{1}{2} \begin{gathered}
\begin{fmffile}{Diagrams/bosonic1PIEA_DerivW1J0J1_Diag5}
\begin{fmfgraph*}(35,20)
\fmfleft{i1,i2}
\fmfright{o1,o2}
\fmfbottom{i0,o0}
\fmfbottom{b0}
\fmfbottom{b1}
\fmfbottom{b2}
\fmftop{i3,o3}
\fmfv{decor.shape=circle,decor.size=2.0thick,foreground=(0,,0,,1)}{v1}
\fmfv{decor.shape=circle,decor.size=2.0thick,foreground=(0,,0,,1)}{v2}
\fmfv{decor.shape=circle,decor.size=2.0thick,foreground=(0,,0,,1)}{v3}
\fmfv{decor.shape=circle,decor.size=2.0thick,foreground=(0,,0,,1)}{v4}
\fmfv{decor.shape=circle,decor.size=2.0thick,foreground=(0,,0,,1)}{v5}
\fmfv{decor.shape=circle,decor.size=2.0thick,foreground=(0,,0,,1)}{v6}
\fmfv{decor.shape=circle,decor.filled=empty,decor.size=0.28cm,label=$\mathrm{o}$,label.dist=0}{v2b}
\fmfv{decor.shape=circle,decor.filled=empty,decor.size=0.28cm,label=$\mathrm{o}$,label.dist=0}{v3b}
\fmf{phantom,tension=1.4}{i1,v1}
\fmf{phantom}{i2,v3b}
\fmf{phantom}{i0,v1b}
\fmf{phantom}{o2,v6b}
\fmf{phantom,tension=1.4}{o1,v5}
\fmf{phantom}{o0,v5b}
\fmf{phantom,tension=1.11}{v3b,v6b}
\fmf{phantom,tension=1.38}{i0,v2}
\fmf{phantom,tension=1.38}{o0,v2}
\fmf{phantom,tension=1.8}{i0,v2b}
\fmf{phantom,tension=1.2}{o0,v2b}
\fmf{phantom,tension=1.2}{b0,v2b}
\fmf{phantom,tension=1.2}{b1,v2b}
\fmf{phantom,tension=1.2}{b2,v2b}
\fmf{phantom,tension=1.38}{i0,v4}
\fmf{phantom,tension=1.38}{o0,v4}
\fmf{phantom,tension=1.2}{i0,v4b}
\fmf{phantom,tension=1.8}{o0,v4b}
\fmf{phantom,tension=1.2}{b0,v4b}
\fmf{phantom,tension=1.2}{b1,v4b}
\fmf{phantom,tension=1.2}{b2,v4b}
\fmf{phantom,tension=2}{i3,v3}
\fmf{phantom,tension=2}{o3,v6}
\fmf{phantom,tension=2}{i3,v3b}
\fmf{phantom,tension=0.8}{o3,v3b}
\fmf{phantom,tension=0.8}{i3,v6b}
\fmf{phantom,tension=2}{o3,v6b}
\fmf{plain,tension=1.4,foreground=(1,,0,,0)}{v1,v2}
\fmf{plain,tension=1.4,foreground=(1,,0,,0)}{v4,v5}
\fmf{wiggly,foreground=(1,,0,,0)}{v1,v3}
\fmf{plain,left=0.8,tension=0,foreground=(1,,0,,0)}{v1,v3}
\fmf{wiggly,foreground=(1,,0,,0)}{v5,v6}
\fmf{plain,right=0.8,tension=0,foreground=(1,,0,,0)}{v5,v6}
\fmf{phantom}{v2,v3}
\fmf{plain,foreground=(1,,0,,0)}{v4,v6}
\fmf{wiggly,tension=0.5,foreground=(1,,0,,0)}{v2,v4}
\fmf{phantom,tension=2}{v3,v6}
\fmf{phantom,right=0.8,tension=0}{v1,v5}
\fmf{phantom,tension=1}{v1,v1b}
\fmf{plain,tension=0.2,foreground=(1,,0,,0)}{v2,v2b}
\fmf{plain,tension=1.5,foreground=(1,,0,,0)}{v3,v3b}
\fmf{phantom,tension=0.2}{v4,v4b}
\fmf{phantom,tension=1}{v5,v5b}
\fmf{phantom,tension=1.5}{v6,v6b}
\end{fmfgraph*}
\end{fmffile}
\end{gathered} + \frac{1}{4} \begin{gathered}
\begin{fmffile}{Diagrams/bosonic1PIEA_DerivW1J0J1_Diag6}
\begin{fmfgraph*}(35,20)
\fmfleft{i1,i2}
\fmfright{o1,o2}
\fmfbottom{i0,o0}
\fmfbottom{b0}
\fmfbottom{b1}
\fmfbottom{b2}
\fmftop{i3,o3}
\fmfv{decor.shape=circle,decor.size=2.0thick,foreground=(0,,0,,1)}{v1}
\fmfv{decor.shape=circle,decor.size=2.0thick,foreground=(0,,0,,1)}{v2}
\fmfv{decor.shape=circle,decor.size=2.0thick,foreground=(0,,0,,1)}{v3}
\fmfv{decor.shape=circle,decor.size=2.0thick,foreground=(0,,0,,1)}{v4}
\fmfv{decor.shape=circle,decor.size=2.0thick,foreground=(0,,0,,1)}{v5}
\fmfv{decor.shape=circle,decor.size=2.0thick,foreground=(0,,0,,1)}{v6}
\fmfv{decor.shape=circle,decor.filled=empty,decor.size=0.28cm,label=$\mathrm{o}$,label.dist=0}{v1b}
\fmfv{decor.shape=circle,decor.filled=empty,decor.size=0.28cm,label=$\mathrm{o}$,label.dist=0}{v3b}
\fmf{phantom,tension=1.4}{i1,v1}
\fmf{phantom}{i2,v3b}
\fmf{phantom}{i0,v1b}
\fmf{phantom}{o2,v6b}
\fmf{phantom,tension=1.4}{o1,v5}
\fmf{phantom}{o0,v5b}
\fmf{phantom,tension=1.11}{v3b,v6b}
\fmf{phantom,tension=1.38}{i0,v2}
\fmf{phantom,tension=1.38}{o0,v2}
\fmf{phantom,tension=1.8}{i0,v2b}
\fmf{phantom,tension=1.2}{o0,v2b}
\fmf{phantom,tension=1.2}{b0,v2b}
\fmf{phantom,tension=1.2}{b1,v2b}
\fmf{phantom,tension=1.2}{b2,v2b}
\fmf{phantom,tension=1.38}{i0,v4}
\fmf{phantom,tension=1.38}{o0,v4}
\fmf{phantom,tension=1.2}{i0,v4b}
\fmf{phantom,tension=1.8}{o0,v4b}
\fmf{phantom,tension=1.2}{b0,v4b}
\fmf{phantom,tension=1.2}{b1,v4b}
\fmf{phantom,tension=1.2}{b2,v4b}
\fmf{phantom,tension=2}{i3,v3}
\fmf{phantom,tension=2}{o3,v6}
\fmf{phantom,tension=2}{i3,v3b}
\fmf{phantom,tension=0.8}{o3,v3b}
\fmf{phantom,tension=0.8}{i3,v6b}
\fmf{phantom,tension=2}{o3,v6b}
\fmf{plain,tension=1.4,foreground=(1,,0,,0)}{v1,v2}
\fmf{plain,tension=1.4,foreground=(1,,0,,0)}{v4,v5}
\fmf{wiggly,foreground=(1,,0,,0)}{v1,v3}
\fmf{phantom,left=0.8,tension=0}{v1,v3}
\fmf{wiggly,foreground=(1,,0,,0)}{v5,v6}
\fmf{plain,right=0.8,tension=0,foreground=(1,,0,,0)}{v5,v6}
\fmf{plain,foreground=(1,,0,,0)}{v2,v3}
\fmf{plain,foreground=(1,,0,,0)}{v4,v6}
\fmf{wiggly,tension=0.5,foreground=(1,,0,,0)}{v2,v4}
\fmf{phantom,tension=2}{v3,v6}
\fmf{phantom,right=0.8,tension=0}{v1,v5}
\fmf{plain,tension=1,foreground=(1,,0,,0)}{v1,v1b}
\fmf{phantom,tension=0.2}{v2,v2b}
\fmf{plain,tension=1.5,foreground=(1,,0,,0)}{v3,v3b}
\fmf{phantom,tension=0.2}{v4,v4b}
\fmf{phantom,tension=1}{v5,v5b}
\fmf{phantom,tension=1.5}{v6,v6b}
\end{fmfgraph*}
\end{fmffile}
\end{gathered} \\
& +\frac{1}{8} \begin{gathered}
\begin{fmffile}{Diagrams/bosonic1PIEA_DerivW1J0J1_Diag7}
\begin{fmfgraph*}(35,20)
\fmfleft{i1,i2}
\fmfright{o1,o2}
\fmfbottom{i0,o0}
\fmfbottom{b0}
\fmfbottom{b1}
\fmfbottom{b2}
\fmftop{i3,o3}
\fmfv{decor.shape=circle,decor.size=2.0thick,foreground=(0,,0,,1)}{v1}
\fmfv{decor.shape=circle,decor.size=2.0thick,foreground=(0,,0,,1)}{v2}
\fmfv{decor.shape=circle,decor.size=2.0thick,foreground=(0,,0,,1)}{v3}
\fmfv{decor.shape=circle,decor.size=2.0thick,foreground=(0,,0,,1)}{v4}
\fmfv{decor.shape=circle,decor.size=2.0thick,foreground=(0,,0,,1)}{v5}
\fmfv{decor.shape=circle,decor.size=2.0thick,foreground=(0,,0,,1)}{v6}
\fmf{phantom,tension=1.4}{i1,v1}
\fmf{phantom}{i2,v3b}
\fmf{phantom}{i0,v1b}
\fmf{phantom}{o2,v6b}
\fmf{phantom,tension=1.4}{o1,v5}
\fmf{phantom}{o0,v5b}
\fmf{phantom,tension=1.11}{v3b,v6b}
\fmf{phantom,tension=1.38}{i0,v2}
\fmf{phantom,tension=1.38}{o0,v2}
\fmf{phantom,tension=1.8}{i0,v2b}
\fmf{phantom,tension=1.2}{o0,v2b}
\fmf{phantom,tension=1.2}{b0,v2b}
\fmf{phantom,tension=1.2}{b1,v2b}
\fmf{phantom,tension=1.2}{b2,v2b}
\fmf{phantom,tension=1.38}{i0,v4}
\fmf{phantom,tension=1.38}{o0,v4}
\fmf{phantom,tension=1.2}{i0,v4b}
\fmf{phantom,tension=1.8}{o0,v4b}
\fmf{phantom,tension=1.2}{b0,v4b}
\fmf{phantom,tension=1.2}{b1,v4b}
\fmf{phantom,tension=1.2}{b2,v4b}
\fmf{phantom,tension=2}{i3,v3}
\fmf{phantom,tension=2}{o3,v6}
\fmf{phantom,tension=2}{i3,v3b}
\fmf{phantom,tension=0.8}{o3,v3b}
\fmf{phantom,tension=0.8}{i3,v6b}
\fmf{phantom,tension=2}{o3,v6b}
\fmf{plain,tension=1.4,foreground=(1,,0,,0)}{v1,v2}
\fmf{plain,tension=1.4,foreground=(1,,0,,0)}{v4,v5}
\fmf{wiggly,foreground=(1,,0,,0)}{v1,v3}
\fmf{plain,left=0.8,tension=0,foreground=(1,,0,,0)}{v1,v3}
\fmf{wiggly,foreground=(1,,0,,0)}{v5,v6}
\fmf{plain,right=0.8,tension=0,foreground=(1,,0,,0)}{v5,v6}
\fmf{plain,foreground=(1,,0,,0)}{v2,v3}
\fmf{plain,foreground=(1,,0,,0)}{v4,v6}
\fmf{wiggly,tension=0.5,foreground=(1,,0,,0)}{v2,v4}
\fmf{phantom,tension=2}{v3,v6}
\fmf{phantom,right=0.8,tension=0}{v1,v5}
\fmf{phantom,tension=1}{v1,v1b}
\fmf{phantom,tension=0.2}{v2,v2b}
\fmf{phantom,tension=1.5}{v3,v3b}
\fmf{phantom,tension=0.2}{v4,v4b}
\fmf{phantom,tension=1}{v5,v5b}
\fmf{phantom,tension=1.5}{v6,v6b}
\end{fmfgraph*}
\end{fmffile}
\end{gathered} \hspace{-0.4cm} \;.
\end{split}
\label{eq:bosonic1PIEADerivW0J0J0J1J10DON}
\end{equation}
After inserting~\eqref{eq:bosonic1PIEAIMW20DON},~\eqref{eq:bosonic1PIEADerivW1J0J10DON} and~\eqref{eq:bosonic1PIEADerivW0J0J0J1J10DON} into~\eqref{eq:bosonic1PIEAIMGamma20DON}, $\Gamma_{\mathrm{col},2}^{(\mathrm{1PI})}$ becomes:
\begin{equation}
\begin{split}
\Gamma_{\mathrm{col},2}^{(\mathrm{1PI})}[\Phi] = & -\frac{1}{2} \hspace{-0.1cm} \begin{gathered}
\begin{fmffile}{Diagrams/bosonic1PIEA_Gamma2_Diag1}
\begin{fmfgraph*}(25,25)
\fmfleft{i1,i2}
\fmfright{o1,o2}
\fmfbottom{i0,o0}
\fmftop{i3,o3}
\fmfv{decor.shape=circle,decor.size=2.0thick,foreground=(0,,0,,1)}{v1}
\fmfv{decor.shape=circle,decor.size=2.0thick,foreground=(0,,0,,1)}{v2}
\fmfv{decor.shape=circle,decor.size=2.0thick,foreground=(0,,0,,1)}{v3}
\fmfv{decor.shape=circle,decor.size=2.0thick,foreground=(0,,0,,1)}{v4}
\fmfv{decor.shape=circle,decor.filled=empty,decor.size=0.28cm,label=$\mathrm{o}$,label.dist=0}{v1b}
\fmfv{decor.shape=circle,decor.filled=empty,decor.size=0.28cm,label=$\mathrm{o}$,label.dist=0}{v4b}
\fmf{phantom}{i1,v1}
\fmf{phantom}{i2,v4}
\fmf{phantom}{o1,v2}
\fmf{phantom}{o2,v3}
\fmf{phantom}{i3,v4b}
\fmf{phantom}{o3,v3b}
\fmf{phantom}{i0,v1b}
\fmf{phantom}{o0,v2b}
\fmf{plain,tension=1.6,foreground=(1,,0,,0)}{v1,v2}
\fmf{plain,tension=1.6,foreground=(1,,0,,0)}{v3,v4}
\fmf{wiggly,tension=2.0,foreground=(1,,0,,0)}{v1,v4}
\fmf{wiggly,tension=2.0,foreground=(1,,0,,0)}{v2,v3}
\fmf{phantom,tension=0}{v1,v3}
\fmf{phantom,tension=0}{v2,v4}
\fmf{plain,right=0.8,tension=0,foreground=(1,,0,,0)}{v2,v3}
\fmf{phantom,left=0.8,tension=0}{v1,v4}
\fmf{plain,foreground=(1,,0,,0)}{v1,v1b}
\fmf{phantom}{v2,v2b}
\fmf{phantom}{v3,v3b}
\fmf{plain,foreground=(1,,0,,0)}{v4,v4b}
\end{fmfgraph*}
\end{fmffile}
\end{gathered} \hspace{-0.4cm} -\frac{1}{2} \hspace{-0.1cm} \begin{gathered}
\begin{fmffile}{Diagrams/bosonic1PIEA_Gamma2_Diag2}
\begin{fmfgraph*}(25,25)
\fmfleft{i1,i2}
\fmfright{o1,o2}
\fmfbottom{i0,o0}
\fmftop{i3,o3}
\fmfv{decor.shape=circle,decor.size=2.0thick,foreground=(0,,0,,1)}{v1}
\fmfv{decor.shape=circle,decor.size=2.0thick,foreground=(0,,0,,1)}{v2}
\fmfv{decor.shape=circle,decor.size=2.0thick,foreground=(0,,0,,1)}{v3}
\fmfv{decor.shape=circle,decor.size=2.0thick,foreground=(0,,0,,1)}{v4}
\fmfv{decor.shape=circle,decor.filled=empty,decor.size=0.28cm,label=$\mathrm{o}$,label.dist=0}{v3b}
\fmfv{decor.shape=circle,decor.filled=empty,decor.size=0.28cm,label=$\mathrm{o}$,label.dist=0}{v4b}
\fmf{phantom}{i1,v1}
\fmf{phantom}{i2,v4}
\fmf{phantom}{o1,v2}
\fmf{phantom}{o2,v3}
\fmf{phantom}{i3,v4b}
\fmf{phantom}{o3,v3b}
\fmf{phantom}{i0,v1b}
\fmf{phantom}{o0,v2b}
\fmf{plain,tension=1.6,foreground=(1,,0,,0)}{v1,v2}
\fmf{phantom,tension=1.6}{v3,v4}
\fmf{wiggly,tension=2.0,foreground=(1,,0,,0)}{v1,v4}
\fmf{wiggly,tension=2.0,foreground=(1,,0,,0)}{v2,v3}
\fmf{plain,tension=0,foreground=(1,,0,,0)}{v1,v3}
\fmf{plain,tension=0,foreground=(1,,0,,0)}{v2,v4}
\fmf{phantom,right=0.8,tension=0}{v2,v3}
\fmf{phantom,left=0.8,tension=0}{v1,v4}
\fmf{phantom}{v1,v1b}
\fmf{phantom}{v2,v2b}
\fmf{plain,foreground=(1,,0,,0)}{v3,v3b}
\fmf{plain,foreground=(1,,0,,0)}{v4,v4b}
\end{fmfgraph*}
\end{fmffile}
\end{gathered} \hspace{-0.1cm} -\frac{1}{2} \hspace{-0.1cm} \begin{gathered}
\begin{fmffile}{Diagrams/bosonic1PIEA_Gamma2_Diag5}
\begin{fmfgraph*}(35,20)
\fmfleft{i1,i2}
\fmfright{o1,o2}
\fmfbottom{i0,o0}
\fmfbottom{b0}
\fmfbottom{b1}
\fmfbottom{b2}
\fmftop{i3,o3}
\fmfv{decor.shape=circle,decor.size=2.0thick,foreground=(0,,0,,1)}{v1}
\fmfv{decor.shape=circle,decor.size=2.0thick,foreground=(0,,0,,1)}{v2}
\fmfv{decor.shape=circle,decor.size=2.0thick,foreground=(0,,0,,1)}{v3}
\fmfv{decor.shape=circle,decor.size=2.0thick,foreground=(0,,0,,1)}{v4}
\fmfv{decor.shape=circle,decor.size=2.0thick,foreground=(0,,0,,1)}{v5}
\fmfv{decor.shape=circle,decor.size=2.0thick,foreground=(0,,0,,1)}{v6}
\fmfv{decor.shape=circle,decor.filled=empty,decor.size=0.28cm,label=$\mathrm{o}$,label.dist=0}{v1b}
\fmfv{decor.shape=circle,decor.filled=empty,decor.size=0.28cm,label=$\mathrm{o}$,label.dist=0}{v3b}
\fmfv{decor.shape=circle,decor.filled=empty,decor.size=0.28cm,label=$\mathrm{o}$,label.dist=0}{v4b}
\fmfv{decor.shape=circle,decor.filled=empty,decor.size=0.28cm,label=$\mathrm{o}$,label.dist=0}{v6b}
\fmf{phantom,tension=1.4}{i1,v1}
\fmf{phantom}{i2,v3b}
\fmf{phantom}{i0,v1b}
\fmf{phantom}{o2,v6b}
\fmf{phantom,tension=1.4}{o1,v5}
\fmf{phantom}{o0,v5b}
\fmf{phantom,tension=1.11}{v3b,v6b}
\fmf{phantom,tension=1.38}{i0,v2}
\fmf{phantom,tension=1.38}{o0,v2}
\fmf{phantom,tension=1.8}{i0,v2b}
\fmf{phantom,tension=1.2}{o0,v2b}
\fmf{phantom,tension=1.2}{b0,v2b}
\fmf{phantom,tension=1.2}{b1,v2b}
\fmf{phantom,tension=1.2}{b2,v2b}
\fmf{phantom,tension=1.38}{i0,v4}
\fmf{phantom,tension=1.38}{o0,v4}
\fmf{phantom,tension=1.2}{i0,v4b}
\fmf{phantom,tension=1.8}{o0,v4b}
\fmf{phantom,tension=1.2}{b0,v4b}
\fmf{phantom,tension=1.2}{b1,v4b}
\fmf{phantom,tension=1.2}{b2,v4b}
\fmf{phantom,tension=2}{i3,v3}
\fmf{phantom,tension=2}{o3,v6}
\fmf{phantom,tension=2}{i3,v3b}
\fmf{phantom,tension=0.8}{o3,v3b}
\fmf{phantom,tension=0.8}{i3,v6b}
\fmf{phantom,tension=2}{o3,v6b}
\fmf{plain,tension=1.4,foreground=(1,,0,,0)}{v1,v2}
\fmf{plain,tension=1.4,foreground=(1,,0,,0)}{v4,v5}
\fmf{phantom}{v1,v3}
\fmf{phantom,left=0.8,tension=0}{v1,v3}
\fmf{plain,foreground=(1,,0,,0)}{v5,v6}
\fmf{phantom,right=0.8,tension=0}{v5,v6}
\fmf{plain,foreground=(1,,0,,0)}{v2,v3}
\fmf{phantom}{v4,v6}
\fmf{wiggly,tension=0.5,foreground=(1,,0,,0)}{v2,v4}
\fmf{wiggly,tension=2,foreground=(1,,0,,0)}{v3,v6}
\fmf{wiggly,right=0.8,tension=0,foreground=(1,,0,,0)}{v1,v5}
\fmf{plain,tension=1,foreground=(1,,0,,0)}{v1,v1b}
\fmf{phantom,tension=0.2}{v2,v2b}
\fmf{plain,tension=1.5,foreground=(1,,0,,0)}{v3,v3b}
\fmf{plain,tension=0.2,foreground=(1,,0,,0)}{v4,v4b}
\fmf{phantom,tension=1}{v5,v5b}
\fmf{plain,tension=1.5,foreground=(1,,0,,0)}{v6,v6b}
\end{fmfgraph*}
\end{fmffile}
\end{gathered} \hspace{-0.5cm} -\frac{1}{4} \begin{gathered}
\begin{fmffile}{Diagrams/bosonic1PIEA_Gamma2_Diag6}
\begin{fmfgraph*}(35,20)
\fmfleft{i1,i2}
\fmfright{o1,o2}
\fmfbottom{i0,o0}
\fmfbottom{b0}
\fmfbottom{b1}
\fmfbottom{b2}
\fmftop{i3,o3}
\fmfv{decor.shape=circle,decor.size=2.0thick,foreground=(0,,0,,1)}{v1}
\fmfv{decor.shape=circle,decor.size=2.0thick,foreground=(0,,0,,1)}{v2}
\fmfv{decor.shape=circle,decor.size=2.0thick,foreground=(0,,0,,1)}{v3}
\fmfv{decor.shape=circle,decor.size=2.0thick,foreground=(0,,0,,1)}{v4}
\fmfv{decor.shape=circle,decor.size=2.0thick,foreground=(0,,0,,1)}{v5}
\fmfv{decor.shape=circle,decor.size=2.0thick,foreground=(0,,0,,1)}{v6}
\fmfv{decor.shape=circle,decor.filled=empty,decor.size=0.28cm,label=$\mathrm{o}$,label.dist=0}{v1b}
\fmfv{decor.shape=circle,decor.filled=empty,decor.size=0.28cm,label=$\mathrm{o}$,label.dist=0}{v3b}
\fmfv{decor.shape=circle,decor.filled=empty,decor.size=0.28cm,label=$\mathrm{o}$,label.dist=0}{v5b}
\fmfv{decor.shape=circle,decor.filled=empty,decor.size=0.28cm,label=$\mathrm{o}$,label.dist=0}{v6b}
\fmf{phantom,tension=1.4}{i1,v1}
\fmf{phantom}{i2,v3b}
\fmf{phantom}{i0,v1b}
\fmf{phantom}{o2,v6b}
\fmf{phantom,tension=1.4}{o1,v5}
\fmf{phantom}{o0,v5b}
\fmf{phantom,tension=1.11}{v3b,v6b}
\fmf{phantom,tension=1.38}{i0,v2}
\fmf{phantom,tension=1.38}{o0,v2}
\fmf{phantom,tension=1.8}{i0,v2b}
\fmf{phantom,tension=1.2}{o0,v2b}
\fmf{phantom,tension=1.2}{b0,v2b}
\fmf{phantom,tension=1.2}{b1,v2b}
\fmf{phantom,tension=1.2}{b2,v2b}
\fmf{phantom,tension=1.38}{i0,v4}
\fmf{phantom,tension=1.38}{o0,v4}
\fmf{phantom,tension=1.2}{i0,v4b}
\fmf{phantom,tension=1.8}{o0,v4b}
\fmf{phantom,tension=1.2}{b0,v4b}
\fmf{phantom,tension=1.2}{b1,v4b}
\fmf{phantom,tension=1.2}{b2,v4b}
\fmf{phantom,tension=2}{i3,v3}
\fmf{phantom,tension=2}{o3,v6}
\fmf{phantom,tension=2}{i3,v3b}
\fmf{phantom,tension=0.8}{o3,v3b}
\fmf{phantom,tension=0.8}{i3,v6b}
\fmf{phantom,tension=2}{o3,v6b}
\fmf{plain,tension=1.4,foreground=(1,,0,,0)}{v1,v2}
\fmf{plain,tension=1.4,foreground=(1,,0,,0)}{v4,v5}
\fmf{phantom}{v1,v3}
\fmf{phantom,left=0.8,tension=0}{v1,v3}
\fmf{phantom}{v5,v6}
\fmf{phantom,right=0.8,tension=0}{v5,v6}
\fmf{plain,foreground=(1,,0,,0)}{v2,v3}
\fmf{plain,foreground=(1,,0,,0)}{v4,v6}
\fmf{wiggly,tension=0.5,foreground=(1,,0,,0)}{v2,v4}
\fmf{wiggly,tension=2,foreground=(1,,0,,0)}{v3,v6}
\fmf{wiggly,right=0.8,tension=0,foreground=(1,,0,,0)}{v1,v5}
\fmf{plain,tension=1,foreground=(1,,0,,0)}{v1,v1b}
\fmf{phantom,tension=0.2}{v2,v2b}
\fmf{plain,tension=1.5,foreground=(1,,0,,0)}{v3,v3b}
\fmf{phantom,tension=0.2}{v4,v4b}
\fmf{plain,tension=1,foreground=(1,,0,,0)}{v5,v5b}
\fmf{plain,tension=1.5,foreground=(1,,0,,0)}{v6,v6b}
\end{fmfgraph*}
\end{fmffile}
\end{gathered} \\
& -\frac{1}{4} \hspace{-0.4cm} \begin{gathered}
\begin{fmffile}{Diagrams/bosonic1PIEA_Gamma2_Diag3}
\begin{fmfgraph*}(25,25)
\fmfleft{i1,i2}
\fmfright{o1,o2}
\fmfbottom{i0,o0}
\fmftop{i3,o3}
\fmfv{decor.shape=circle,decor.size=2.0thick,foreground=(0,,0,,1)}{v1}
\fmfv{decor.shape=circle,decor.size=2.0thick,foreground=(0,,0,,1)}{v2}
\fmfv{decor.shape=circle,decor.size=2.0thick,foreground=(0,,0,,1)}{v3}
\fmfv{decor.shape=circle,decor.size=2.0thick,foreground=(0,,0,,1)}{v4}
\fmf{phantom}{i1,v1}
\fmf{phantom}{i2,v4}
\fmf{phantom}{o1,v2}
\fmf{phantom}{o2,v3}
\fmf{phantom}{i3,v4b}
\fmf{phantom}{o3,v3b}
\fmf{phantom}{i0,v1b}
\fmf{phantom}{o0,v2b}
\fmf{plain,tension=1.6,foreground=(1,,0,,0)}{v1,v2}
\fmf{plain,tension=1.6,foreground=(1,,0,,0)}{v3,v4}
\fmf{wiggly,tension=2.0,foreground=(1,,0,,0)}{v1,v4}
\fmf{wiggly,tension=2.0,foreground=(1,,0,,0)}{v2,v3}
\fmf{phantom,tension=0}{v1,v3}
\fmf{phantom,tension=0}{v2,v4}
\fmf{plain,right=0.8,tension=0,foreground=(1,,0,,0)}{v2,v3}
\fmf{plain,left=0.8,tension=0,foreground=(1,,0,,0)}{v1,v4}
\fmf{phantom}{v1,v1b}
\fmf{phantom}{v2,v2b}
\fmf{phantom}{v3,v3b}
\fmf{phantom}{v4,v4b}
\end{fmfgraph*}
\end{fmffile}
\end{gathered} \hspace{-0.4cm} -\frac{1}{8} \hspace{-0.5cm} \begin{gathered}
\begin{fmffile}{Diagrams/bosonic1PIEA_Gamma2_Diag4}
\begin{fmfgraph*}(25,25)
\fmfleft{i1,i2}
\fmfright{o1,o2}
\fmfbottom{i0,o0}
\fmftop{i3,o3}
\fmfv{decor.shape=circle,decor.size=2.0thick,foreground=(0,,0,,1)}{v1}
\fmfv{decor.shape=circle,decor.size=2.0thick,foreground=(0,,0,,1)}{v2}
\fmfv{decor.shape=circle,decor.size=2.0thick,foreground=(0,,0,,1)}{v3}
\fmfv{decor.shape=circle,decor.size=2.0thick,foreground=(0,,0,,1)}{v4}
\fmf{phantom}{i1,v1}
\fmf{phantom}{i2,v4}
\fmf{phantom}{o1,v2}
\fmf{phantom}{o2,v3}
\fmf{phantom}{i3,v4b}
\fmf{phantom}{o3,v3b}
\fmf{phantom}{i0,v1b}
\fmf{phantom}{o0,v2b}
\fmf{plain,tension=1.6,foreground=(1,,0,,0)}{v1,v2}
\fmf{plain,tension=1.6,foreground=(1,,0,,0)}{v3,v4}
\fmf{wiggly,tension=2.0,foreground=(1,,0,,0)}{v1,v4}
\fmf{wiggly,tension=2.0,foreground=(1,,0,,0)}{v2,v3}
\fmf{plain,tension=0,foreground=(1,,0,,0)}{v1,v3}
\fmf{plain,tension=0,foreground=(1,,0,,0)}{v2,v4}
\fmf{phantom,right=0.8,tension=0}{v2,v3}
\fmf{phantom,left=0.8,tension=0}{v1,v4}
\fmf{phantom}{v1,v1b}
\fmf{phantom}{v2,v2b}
\fmf{phantom}{v3,v3b}
\fmf{phantom}{v4,v4b}
\end{fmfgraph*}
\end{fmffile}
\end{gathered} \hspace{-0.5cm} -\frac{1}{2} \begin{gathered}
\begin{fmffile}{Diagrams/bosonic1PIEA_Gamma2_Diag7}
\begin{fmfgraph*}(35,20)
\fmfleft{i1,i2}
\fmfright{o1,o2}
\fmfbottom{i0,o0}
\fmfbottom{b0}
\fmfbottom{b1}
\fmfbottom{b2}
\fmftop{i3,o3}
\fmfv{decor.shape=circle,decor.size=2.0thick,foreground=(0,,0,,1)}{v1}
\fmfv{decor.shape=circle,decor.size=2.0thick,foreground=(0,,0,,1)}{v2}
\fmfv{decor.shape=circle,decor.size=2.0thick,foreground=(0,,0,,1)}{v3}
\fmfv{decor.shape=circle,decor.size=2.0thick,foreground=(0,,0,,1)}{v4}
\fmfv{decor.shape=circle,decor.size=2.0thick,foreground=(0,,0,,1)}{v5}
\fmfv{decor.shape=circle,decor.size=2.0thick,foreground=(0,,0,,1)}{v6}
\fmfv{decor.shape=circle,decor.filled=empty,decor.size=0.28cm,label=$\mathrm{o}$,label.dist=0}{v1b}
\fmfv{decor.shape=circle,decor.filled=empty,decor.size=0.28cm,label=$\mathrm{o}$,label.dist=0}{v3b}
\fmf{phantom,tension=1.4}{i1,v1}
\fmf{phantom}{i2,v3b}
\fmf{phantom}{i0,v1b}
\fmf{phantom}{o2,v6b}
\fmf{phantom,tension=1.4}{o1,v5}
\fmf{phantom}{o0,v5b}
\fmf{phantom,tension=1.11}{v3b,v6b}
\fmf{phantom,tension=1.38}{i0,v2}
\fmf{phantom,tension=1.38}{o0,v2}
\fmf{phantom,tension=1.8}{i0,v2b}
\fmf{phantom,tension=1.2}{o0,v2b}
\fmf{phantom,tension=1.2}{b0,v2b}
\fmf{phantom,tension=1.2}{b1,v2b}
\fmf{phantom,tension=1.2}{b2,v2b}
\fmf{phantom,tension=1.38}{i0,v4}
\fmf{phantom,tension=1.38}{o0,v4}
\fmf{phantom,tension=1.2}{i0,v4b}
\fmf{phantom,tension=1.8}{o0,v4b}
\fmf{phantom,tension=1.2}{b0,v4b}
\fmf{phantom,tension=1.2}{b1,v4b}
\fmf{phantom,tension=1.2}{b2,v4b}
\fmf{phantom,tension=2}{i3,v3}
\fmf{phantom,tension=2}{o3,v6}
\fmf{phantom,tension=2}{i3,v3b}
\fmf{phantom,tension=0.8}{o3,v3b}
\fmf{phantom,tension=0.8}{i3,v6b}
\fmf{phantom,tension=2}{o3,v6b}
\fmf{plain,tension=1.4,foreground=(1,,0,,0)}{v1,v2}
\fmf{plain,tension=1.4,foreground=(1,,0,,0)}{v4,v5}
\fmf{phantom}{v1,v3}
\fmf{phantom,left=0.8,tension=0}{v1,v3}
\fmf{plain,foreground=(1,,0,,0)}{v5,v6}
\fmf{phantom,right=0.8,tension=0}{v5,v6}
\fmf{plain,foreground=(1,,0,,0)}{v2,v3}
\fmf{plain,foreground=(1,,0,,0)}{v4,v6}
\fmf{wiggly,tension=0.5,foreground=(1,,0,,0)}{v2,v4}
\fmf{wiggly,tension=2,foreground=(1,,0,,0)}{v3,v6}
\fmf{wiggly,right=0.8,tension=0,foreground=(1,,0,,0)}{v1,v5}
\fmf{plain,tension=1,foreground=(1,,0,,0)}{v1,v1b}
\fmf{phantom,tension=0.2}{v2,v2b}
\fmf{plain,tension=1.5,foreground=(1,,0,,0)}{v3,v3b}
\fmf{phantom,tension=0.2}{v4,v4b}
\fmf{phantom,tension=1}{v5,v5b}
\fmf{phantom,tension=1.5}{v6,v6b}
\end{fmfgraph*}
\end{fmffile}
\end{gathered} \hspace{-0.5cm} -\frac{1}{12} \hspace{-0.5cm} \begin{gathered}
\begin{fmffile}{Diagrams/bosonic1PIEA_Gamma2_Diag8}
\begin{fmfgraph*}(35,20)
\fmfleft{i1,i2}
\fmfright{o1,o2}
\fmfbottom{i0,o0}
\fmfbottom{b0}
\fmfbottom{b1}
\fmfbottom{b2}
\fmftop{i3,o3}
\fmfv{decor.shape=circle,decor.size=2.0thick,foreground=(0,,0,,1)}{v1}
\fmfv{decor.shape=circle,decor.size=2.0thick,foreground=(0,,0,,1)}{v2}
\fmfv{decor.shape=circle,decor.size=2.0thick,foreground=(0,,0,,1)}{v3}
\fmfv{decor.shape=circle,decor.size=2.0thick,foreground=(0,,0,,1)}{v4}
\fmfv{decor.shape=circle,decor.size=2.0thick,foreground=(0,,0,,1)}{v5}
\fmfv{decor.shape=circle,decor.size=2.0thick,foreground=(0,,0,,1)}{v6}
\fmf{phantom,tension=1.4}{i1,v1}
\fmf{phantom}{i2,v3b}
\fmf{phantom}{i0,v1b}
\fmf{phantom}{o2,v6b}
\fmf{phantom,tension=1.4}{o1,v5}
\fmf{phantom}{o0,v5b}
\fmf{phantom,tension=1.11}{v3b,v6b}
\fmf{phantom,tension=1.38}{i0,v2}
\fmf{phantom,tension=1.38}{o0,v2}
\fmf{phantom,tension=1.8}{i0,v2b}
\fmf{phantom,tension=1.2}{o0,v2b}
\fmf{phantom,tension=1.2}{b0,v2b}
\fmf{phantom,tension=1.2}{b1,v2b}
\fmf{phantom,tension=1.2}{b2,v2b}
\fmf{phantom,tension=1.38}{i0,v4}
\fmf{phantom,tension=1.38}{o0,v4}
\fmf{phantom,tension=1.2}{i0,v4b}
\fmf{phantom,tension=1.8}{o0,v4b}
\fmf{phantom,tension=1.2}{b0,v4b}
\fmf{phantom,tension=1.2}{b1,v4b}
\fmf{phantom,tension=1.2}{b2,v4b}
\fmf{phantom,tension=2}{i3,v3}
\fmf{phantom,tension=2}{o3,v6}
\fmf{phantom,tension=2}{i3,v3b}
\fmf{phantom,tension=0.8}{o3,v3b}
\fmf{phantom,tension=0.8}{i3,v6b}
\fmf{phantom,tension=2}{o3,v6b}
\fmf{plain,tension=1.4,foreground=(1,,0,,0)}{v1,v2}
\fmf{plain,tension=1.4,foreground=(1,,0,,0)}{v4,v5}
\fmf{plain,foreground=(1,,0,,0)}{v1,v3}
\fmf{phantom,left=0.8,tension=0}{v1,v3}
\fmf{plain,foreground=(1,,0,,0)}{v5,v6}
\fmf{phantom,right=0.8,tension=0}{v5,v6}
\fmf{plain,foreground=(1,,0,,0)}{v2,v3}
\fmf{plain,foreground=(1,,0,,0)}{v4,v6}
\fmf{wiggly,tension=0.5,foreground=(1,,0,,0)}{v2,v4}
\fmf{wiggly,tension=2,foreground=(1,,0,,0)}{v3,v6}
\fmf{wiggly,right=0.8,tension=0,foreground=(1,,0,,0)}{v1,v5}
\fmf{phantom,tension=1}{v1,v1b}
\fmf{phantom,tension=0.2}{v2,v2b}
\fmf{phantom,tension=1.5}{v3,v3b}
\fmf{phantom,tension=0.2}{v4,v4b}
\fmf{phantom,tension=1}{v5,v5b}
\fmf{phantom,tension=1.5}{v6,v6b}
\end{fmfgraph*}
\end{fmffile}
\end{gathered} \hspace{-0.5cm} \;.
\end{split}
\label{eq:bosonic1PIEAIMGamma2step20DON}
\end{equation}
Finally, we conclude the above derivations by giving our final expression for the collective 1PI EA which results from~\eqref{eq:bosonic1PIEAIMGamma0step20DON},~\eqref{eq:bosonic1PIEAIMGamma10DON} (combined with~\eqref{eq:bosonic1PIEAIMW10DON}) and~\eqref{eq:bosonic1PIEAIMGamma2step20DON} (combined with~\eqref{eq:bosonic1PIEAphiGJ0diagrams0DON}):
\begin{equation}
\begin{split}
\Gamma_{\mathrm{col}}^{(\mathrm{1PI})}[\Phi] = & \ S_{\mathrm{col}}[\eta] + \frac{1}{2} \int_{\alpha_{1},\alpha_{2}} \phi_{\alpha_{1}} \boldsymbol{G}^{-1}_{\Phi;\alpha_{1}\alpha_{2}}[\Phi] \phi_{\alpha_{2}} - \frac{\hbar}{2}\mathrm{Tr}\left[\ln\big(D_{\Phi}[\Phi]\big)\right] \\
& - \hbar^{2} \left(\rule{0cm}{1.1cm}\right. \frac{1}{2} \hspace{-0.1cm} \begin{gathered}
\begin{fmffile}{Diagrams/bosonic1PIEA_Gamma2_Diag1bis}
\begin{fmfgraph*}(25,25)
\fmfleft{i1,i2}
\fmfright{o1,o2}
\fmfbottom{i0,o0}
\fmftop{i3,o3}
\fmfv{decor.shape=circle,decor.size=2.0thick,foreground=(0,,0,,1)}{v1}
\fmfv{decor.shape=circle,decor.size=2.0thick,foreground=(0,,0,,1)}{v2}
\fmfv{decor.shape=circle,decor.size=2.0thick,foreground=(0,,0,,1)}{v3}
\fmfv{decor.shape=circle,decor.size=2.0thick,foreground=(0,,0,,1)}{v4}
\fmfv{decor.shape=cross,decor.size=0.25cm,decor.angle=35,foreground=(1,,0,,0)}{v1b}
\fmfv{decor.shape=cross,decor.size=0.25cm,decor.angle=-35,foreground=(1,,0,,0)}{v4b}
\fmf{phantom}{i1,v1}
\fmf{phantom}{i2,v4}
\fmf{phantom}{o1,v2}
\fmf{phantom}{o2,v3}
\fmf{phantom}{i3,v4b}
\fmf{phantom}{o3,v3b}
\fmf{phantom}{i0,v1b}
\fmf{phantom}{o0,v2b}
\fmf{plain,tension=1.6,foreground=(1,,0,,0)}{v1,v2}
\fmf{plain,tension=1.6,foreground=(1,,0,,0)}{v3,v4}
\fmf{wiggly,tension=2.0,foreground=(1,,0,,0)}{v1,v4}
\fmf{wiggly,tension=2.0,foreground=(1,,0,,0)}{v2,v3}
\fmf{phantom,tension=0}{v1,v3}
\fmf{phantom,tension=0}{v2,v4}
\fmf{plain,right=0.8,tension=0,foreground=(1,,0,,0)}{v2,v3}
\fmf{phantom,left=0.8,tension=0}{v1,v4}
\fmf{dashes,foreground=(1,,0,,0)}{v1,v1b}
\fmf{phantom}{v2,v2b}
\fmf{phantom}{v3,v3b}
\fmf{dashes,foreground=(1,,0,,0)}{v4,v4b}
\end{fmfgraph*}
\end{fmffile}
\end{gathered} \hspace{-0.4cm} + \frac{1}{2} \hspace{-0.1cm} \begin{gathered}
\begin{fmffile}{Diagrams/bosonic1PIEA_Gamma2_Diag2bis}
\begin{fmfgraph*}(25,25)
\fmfleft{i1,i2}
\fmfright{o1,o2}
\fmfbottom{i0,o0}
\fmftop{i3,o3}
\fmfv{decor.shape=circle,decor.size=2.0thick,foreground=(0,,0,,1)}{v1}
\fmfv{decor.shape=circle,decor.size=2.0thick,foreground=(0,,0,,1)}{v2}
\fmfv{decor.shape=circle,decor.size=2.0thick,foreground=(0,,0,,1)}{v3}
\fmfv{decor.shape=circle,decor.size=2.0thick,foreground=(0,,0,,1)}{v4}
\fmfv{decor.shape=cross,decor.size=0.25cm,decor.angle=35,foreground=(1,,0,,0)}{v3b}
\fmfv{decor.shape=cross,decor.size=0.25cm,decor.angle=-35,foreground=(1,,0,,0)}{v4b}
\fmf{phantom}{i1,v1}
\fmf{phantom}{i2,v4}
\fmf{phantom}{o1,v2}
\fmf{phantom}{o2,v3}
\fmf{phantom}{i3,v4b}
\fmf{phantom}{o3,v3b}
\fmf{phantom}{i0,v1b}
\fmf{phantom}{o0,v2b}
\fmf{plain,tension=1.6,foreground=(1,,0,,0)}{v1,v2}
\fmf{phantom,tension=1.6}{v3,v4}
\fmf{wiggly,tension=2.0,foreground=(1,,0,,0)}{v1,v4}
\fmf{wiggly,tension=2.0,foreground=(1,,0,,0)}{v2,v3}
\fmf{plain,tension=0,foreground=(1,,0,,0)}{v1,v3}
\fmf{plain,tension=0,foreground=(1,,0,,0)}{v2,v4}
\fmf{phantom,right=0.8,tension=0}{v2,v3}
\fmf{phantom,left=0.8,tension=0}{v1,v4}
\fmf{phantom}{v1,v1b}
\fmf{phantom}{v2,v2b}
\fmf{dashes,foreground=(1,,0,,0)}{v3,v3b}
\fmf{dashes,foreground=(1,,0,,0)}{v4,v4b}
\end{fmfgraph*}
\end{fmffile}
\end{gathered} \hspace{-0.15cm} + \frac{1}{2} \hspace{-0.1cm} \begin{gathered}
\begin{fmffile}{Diagrams/bosonic1PIEA_Gamma2_Diag5bis}
\begin{fmfgraph*}(35,20)
\fmfleft{i1,i2}
\fmfright{o1,o2}
\fmfbottom{i0,o0}
\fmfbottom{b0}
\fmfbottom{b1}
\fmfbottom{b2}
\fmftop{i3,o3}
\fmfv{decor.shape=circle,decor.size=2.0thick,foreground=(0,,0,,1)}{v1}
\fmfv{decor.shape=circle,decor.size=2.0thick,foreground=(0,,0,,1)}{v2}
\fmfv{decor.shape=circle,decor.size=2.0thick,foreground=(0,,0,,1)}{v3}
\fmfv{decor.shape=circle,decor.size=2.0thick,foreground=(0,,0,,1)}{v4}
\fmfv{decor.shape=circle,decor.size=2.0thick,foreground=(0,,0,,1)}{v5}
\fmfv{decor.shape=circle,decor.size=2.0thick,foreground=(0,,0,,1)}{v6}
\fmfv{decor.shape=cross,decor.size=0.25cm,decor.angle=16,foreground=(1,,0,,0)}{v1b}
\fmfv{decor.shape=cross,decor.size=0.25cm,foreground=(1,,0,,0)}{v3b}
\fmfv{decor.shape=cross,decor.size=0.25cm,decor.angle=71,foreground=(1,,0,,0)}{v4b}
\fmfv{decor.shape=cross,decor.size=0.25cm,foreground=(1,,0,,0)}{v6b}
\fmf{phantom,tension=1.4}{i1,v1}
\fmf{phantom}{i2,v3b}
\fmf{phantom}{i0,v1b}
\fmf{phantom}{o2,v6b}
\fmf{phantom,tension=1.4}{o1,v5}
\fmf{phantom}{o0,v5b}
\fmf{phantom,tension=1.11}{v3b,v6b}
\fmf{phantom,tension=1.38}{i0,v2}
\fmf{phantom,tension=1.38}{o0,v2}
\fmf{phantom,tension=1.8}{i0,v2b}
\fmf{phantom,tension=1.2}{o0,v2b}
\fmf{phantom,tension=1.2}{b0,v2b}
\fmf{phantom,tension=1.2}{b1,v2b}
\fmf{phantom,tension=1.2}{b2,v2b}
\fmf{phantom,tension=1.38}{i0,v4}
\fmf{phantom,tension=1.38}{o0,v4}
\fmf{phantom,tension=1.2}{i0,v4b}
\fmf{phantom,tension=1.8}{o0,v4b}
\fmf{phantom,tension=1.2}{b0,v4b}
\fmf{phantom,tension=1.2}{b1,v4b}
\fmf{phantom,tension=1.2}{b2,v4b}
\fmf{phantom,tension=2}{i3,v3}
\fmf{phantom,tension=2}{o3,v6}
\fmf{phantom,tension=2}{i3,v3b}
\fmf{phantom,tension=0.8}{o3,v3b}
\fmf{phantom,tension=0.8}{i3,v6b}
\fmf{phantom,tension=2}{o3,v6b}
\fmf{plain,tension=1.4,foreground=(1,,0,,0)}{v1,v2}
\fmf{plain,tension=1.4,foreground=(1,,0,,0)}{v4,v5}
\fmf{phantom}{v1,v3}
\fmf{phantom,left=0.8,tension=0}{v1,v3}
\fmf{plain,foreground=(1,,0,,0)}{v5,v6}
\fmf{phantom,right=0.8,tension=0}{v5,v6}
\fmf{plain,foreground=(1,,0,,0)}{v2,v3}
\fmf{phantom}{v4,v6}
\fmf{wiggly,tension=0.5,foreground=(1,,0,,0)}{v2,v4}
\fmf{wiggly,tension=2,foreground=(1,,0,,0)}{v3,v6}
\fmf{wiggly,right=0.8,tension=0,foreground=(1,,0,,0)}{v1,v5}
\fmf{dashes,tension=1,foreground=(1,,0,,0)}{v1,v1b}
\fmf{phantom,tension=0.2}{v2,v2b}
\fmf{dashes,tension=1.5,foreground=(1,,0,,0)}{v3,v3b}
\fmf{dashes,tension=0.2,foreground=(1,,0,,0)}{v4,v4b}
\fmf{phantom,tension=1}{v5,v5b}
\fmf{dashes,tension=1.5,foreground=(1,,0,,0)}{v6,v6b}
\end{fmfgraph*}
\end{fmffile}
\end{gathered} \hspace{-0.5cm} +\frac{1}{4} \begin{gathered}
\begin{fmffile}{Diagrams/bosonic1PIEA_Gamma2_Diag6bis}
\begin{fmfgraph*}(35,20)
\fmfleft{i1,i2}
\fmfright{o1,o2}
\fmfbottom{i0,o0}
\fmfbottom{b0}
\fmfbottom{b1}
\fmfbottom{b2}
\fmftop{i3,o3}
\fmfv{decor.shape=circle,decor.size=2.0thick,foreground=(0,,0,,1)}{v1}
\fmfv{decor.shape=circle,decor.size=2.0thick,foreground=(0,,0,,1)}{v2}
\fmfv{decor.shape=circle,decor.size=2.0thick,foreground=(0,,0,,1)}{v3}
\fmfv{decor.shape=circle,decor.size=2.0thick,foreground=(0,,0,,1)}{v4}
\fmfv{decor.shape=circle,decor.size=2.0thick,foreground=(0,,0,,1)}{v5}
\fmfv{decor.shape=circle,decor.size=2.0thick,foreground=(0,,0,,1)}{v6}
\fmfv{decor.shape=cross,decor.size=0.25cm,decor.angle=16,foreground=(1,,0,,0)}{v1b}
\fmfv{decor.shape=cross,decor.size=0.25cm,foreground=(1,,0,,0)}{v3b}
\fmfv{decor.shape=cross,decor.size=0.25cm,decor.angle=-16,foreground=(1,,0,,0)}{v5b}
\fmfv{decor.shape=cross,decor.size=0.25cm,foreground=(1,,0,,0)}{v6b}
\fmf{phantom,tension=1.4}{i1,v1}
\fmf{phantom}{i2,v3b}
\fmf{phantom}{i0,v1b}
\fmf{phantom}{o2,v6b}
\fmf{phantom,tension=1.4}{o1,v5}
\fmf{phantom}{o0,v5b}
\fmf{phantom,tension=1.11}{v3b,v6b}
\fmf{phantom,tension=1.38}{i0,v2}
\fmf{phantom,tension=1.38}{o0,v2}
\fmf{phantom,tension=1.8}{i0,v2b}
\fmf{phantom,tension=1.2}{o0,v2b}
\fmf{phantom,tension=1.2}{b0,v2b}
\fmf{phantom,tension=1.2}{b1,v2b}
\fmf{phantom,tension=1.2}{b2,v2b}
\fmf{phantom,tension=1.38}{i0,v4}
\fmf{phantom,tension=1.38}{o0,v4}
\fmf{phantom,tension=1.2}{i0,v4b}
\fmf{phantom,tension=1.8}{o0,v4b}
\fmf{phantom,tension=1.2}{b0,v4b}
\fmf{phantom,tension=1.2}{b1,v4b}
\fmf{phantom,tension=1.2}{b2,v4b}
\fmf{phantom,tension=2}{i3,v3}
\fmf{phantom,tension=2}{o3,v6}
\fmf{phantom,tension=2}{i3,v3b}
\fmf{phantom,tension=0.8}{o3,v3b}
\fmf{phantom,tension=0.8}{i3,v6b}
\fmf{phantom,tension=2}{o3,v6b}
\fmf{plain,tension=1.4,foreground=(1,,0,,0)}{v1,v2}
\fmf{plain,tension=1.4,foreground=(1,,0,,0)}{v4,v5}
\fmf{phantom}{v1,v3}
\fmf{phantom,left=0.8,tension=0}{v1,v3}
\fmf{phantom}{v5,v6}
\fmf{phantom,right=0.8,tension=0}{v5,v6}
\fmf{plain,foreground=(1,,0,,0)}{v2,v3}
\fmf{plain,foreground=(1,,0,,0)}{v4,v6}
\fmf{wiggly,tension=0.5,foreground=(1,,0,,0)}{v2,v4}
\fmf{wiggly,tension=2,foreground=(1,,0,,0)}{v3,v6}
\fmf{wiggly,right=0.8,tension=0,foreground=(1,,0,,0)}{v1,v5}
\fmf{dashes,tension=1,foreground=(1,,0,,0)}{v1,v1b}
\fmf{phantom,tension=0.2}{v2,v2b}
\fmf{dashes,tension=1.5,foreground=(1,,0,,0)}{v3,v3b}
\fmf{phantom,tension=0.2}{v4,v4b}
\fmf{dashes,tension=1,foreground=(1,,0,,0)}{v5,v5b}
\fmf{dashes,tension=1.5,foreground=(1,,0,,0)}{v6,v6b}
\end{fmfgraph*}
\end{fmffile}
\end{gathered} \\
& \hspace{1.0cm} +\frac{1}{4} \hspace{-0.4cm} \begin{gathered}
\begin{fmffile}{Diagrams/bosonic1PIEA_Gamma2_Diag3bis}
\begin{fmfgraph*}(25,25)
\fmfleft{i1,i2}
\fmfright{o1,o2}
\fmfbottom{i0,o0}
\fmftop{i3,o3}
\fmfv{decor.shape=circle,decor.size=2.0thick,foreground=(0,,0,,1)}{v1}
\fmfv{decor.shape=circle,decor.size=2.0thick,foreground=(0,,0,,1)}{v2}
\fmfv{decor.shape=circle,decor.size=2.0thick,foreground=(0,,0,,1)}{v3}
\fmfv{decor.shape=circle,decor.size=2.0thick,foreground=(0,,0,,1)}{v4}
\fmf{phantom}{i1,v1}
\fmf{phantom}{i2,v4}
\fmf{phantom}{o1,v2}
\fmf{phantom}{o2,v3}
\fmf{phantom}{i3,v4b}
\fmf{phantom}{o3,v3b}
\fmf{phantom}{i0,v1b}
\fmf{phantom}{o0,v2b}
\fmf{plain,tension=1.6,foreground=(1,,0,,0)}{v1,v2}
\fmf{plain,tension=1.6,foreground=(1,,0,,0)}{v3,v4}
\fmf{wiggly,tension=2.0,foreground=(1,,0,,0)}{v1,v4}
\fmf{wiggly,tension=2.0,foreground=(1,,0,,0)}{v2,v3}
\fmf{phantom,tension=0}{v1,v3}
\fmf{phantom,tension=0}{v2,v4}
\fmf{plain,right=0.8,tension=0,foreground=(1,,0,,0)}{v2,v3}
\fmf{plain,left=0.8,tension=0,foreground=(1,,0,,0)}{v1,v4}
\fmf{phantom}{v1,v1b}
\fmf{phantom}{v2,v2b}
\fmf{phantom}{v3,v3b}
\fmf{phantom}{v4,v4b}
\end{fmfgraph*}
\end{fmffile}
\end{gathered} \hspace{-0.4cm} +\frac{1}{8} \hspace{-0.5cm} \begin{gathered}
\begin{fmffile}{Diagrams/bosonic1PIEA_Gamma2_Diag4bis}
\begin{fmfgraph*}(25,25)
\fmfleft{i1,i2}
\fmfright{o1,o2}
\fmfbottom{i0,o0}
\fmftop{i3,o3}
\fmfv{decor.shape=circle,decor.size=2.0thick,foreground=(0,,0,,1)}{v1}
\fmfv{decor.shape=circle,decor.size=2.0thick,foreground=(0,,0,,1)}{v2}
\fmfv{decor.shape=circle,decor.size=2.0thick,foreground=(0,,0,,1)}{v3}
\fmfv{decor.shape=circle,decor.size=2.0thick,foreground=(0,,0,,1)}{v4}
\fmf{phantom}{i1,v1}
\fmf{phantom}{i2,v4}
\fmf{phantom}{o1,v2}
\fmf{phantom}{o2,v3}
\fmf{phantom}{i3,v4b}
\fmf{phantom}{o3,v3b}
\fmf{phantom}{i0,v1b}
\fmf{phantom}{o0,v2b}
\fmf{plain,tension=1.6,foreground=(1,,0,,0)}{v1,v2}
\fmf{plain,tension=1.6,foreground=(1,,0,,0)}{v3,v4}
\fmf{wiggly,tension=2.0,foreground=(1,,0,,0)}{v1,v4}
\fmf{wiggly,tension=2.0,foreground=(1,,0,,0)}{v2,v3}
\fmf{plain,tension=0,foreground=(1,,0,,0)}{v1,v3}
\fmf{plain,tension=0,foreground=(1,,0,,0)}{v2,v4}
\fmf{phantom,right=0.8,tension=0}{v2,v3}
\fmf{phantom,left=0.8,tension=0}{v1,v4}
\fmf{phantom}{v1,v1b}
\fmf{phantom}{v2,v2b}
\fmf{phantom}{v3,v3b}
\fmf{phantom}{v4,v4b}
\end{fmfgraph*}
\end{fmffile}
\end{gathered} \hspace{-0.5cm} + \frac{1}{2} \begin{gathered}
\begin{fmffile}{Diagrams/bosonic1PIEA_Gamma2_Diag7bis}
\begin{fmfgraph*}(35,20)
\fmfleft{i1,i2}
\fmfright{o1,o2}
\fmfbottom{i0,o0}
\fmfbottom{b0}
\fmfbottom{b1}
\fmfbottom{b2}
\fmftop{i3,o3}
\fmfv{decor.shape=circle,decor.size=2.0thick,foreground=(0,,0,,1)}{v1}
\fmfv{decor.shape=circle,decor.size=2.0thick,foreground=(0,,0,,1)}{v2}
\fmfv{decor.shape=circle,decor.size=2.0thick,foreground=(0,,0,,1)}{v3}
\fmfv{decor.shape=circle,decor.size=2.0thick,foreground=(0,,0,,1)}{v4}
\fmfv{decor.shape=circle,decor.size=2.0thick,foreground=(0,,0,,1)}{v5}
\fmfv{decor.shape=circle,decor.size=2.0thick,foreground=(0,,0,,1)}{v6}
\fmfv{decor.shape=cross,decor.size=0.25cm,decor.angle=16,foreground=(1,,0,,0)}{v1b}
\fmfv{decor.shape=cross,decor.size=0.25cm,foreground=(1,,0,,0)}{v3b}
\fmf{phantom,tension=1.4}{i1,v1}
\fmf{phantom}{i2,v3b}
\fmf{phantom}{i0,v1b}
\fmf{phantom}{o2,v6b}
\fmf{phantom,tension=1.4}{o1,v5}
\fmf{phantom}{o0,v5b}
\fmf{phantom,tension=1.11}{v3b,v6b}
\fmf{phantom,tension=1.38}{i0,v2}
\fmf{phantom,tension=1.38}{o0,v2}
\fmf{phantom,tension=1.8}{i0,v2b}
\fmf{phantom,tension=1.2}{o0,v2b}
\fmf{phantom,tension=1.2}{b0,v2b}
\fmf{phantom,tension=1.2}{b1,v2b}
\fmf{phantom,tension=1.2}{b2,v2b}
\fmf{phantom,tension=1.38}{i0,v4}
\fmf{phantom,tension=1.38}{o0,v4}
\fmf{phantom,tension=1.2}{i0,v4b}
\fmf{phantom,tension=1.8}{o0,v4b}
\fmf{phantom,tension=1.2}{b0,v4b}
\fmf{phantom,tension=1.2}{b1,v4b}
\fmf{phantom,tension=1.2}{b2,v4b}
\fmf{phantom,tension=2}{i3,v3}
\fmf{phantom,tension=2}{o3,v6}
\fmf{phantom,tension=2}{i3,v3b}
\fmf{phantom,tension=0.8}{o3,v3b}
\fmf{phantom,tension=0.8}{i3,v6b}
\fmf{phantom,tension=2}{o3,v6b}
\fmf{plain,tension=1.4,foreground=(1,,0,,0)}{v1,v2}
\fmf{plain,tension=1.4,foreground=(1,,0,,0)}{v4,v5}
\fmf{phantom}{v1,v3}
\fmf{phantom,left=0.8,tension=0}{v1,v3}
\fmf{plain,foreground=(1,,0,,0)}{v5,v6}
\fmf{phantom,right=0.8,tension=0}{v5,v6}
\fmf{plain,foreground=(1,,0,,0)}{v2,v3}
\fmf{plain,foreground=(1,,0,,0)}{v4,v6}
\fmf{wiggly,tension=0.5,foreground=(1,,0,,0)}{v2,v4}
\fmf{wiggly,tension=2,foreground=(1,,0,,0)}{v3,v6}
\fmf{wiggly,right=0.8,tension=0,foreground=(1,,0,,0)}{v1,v5}
\fmf{dashes,tension=1,foreground=(1,,0,,0)}{v1,v1b}
\fmf{phantom,tension=0.2}{v2,v2b}
\fmf{dashes,tension=1.5,foreground=(1,,0,,0)}{v3,v3b}
\fmf{phantom,tension=0.2}{v4,v4b}
\fmf{phantom,tension=1}{v5,v5b}
\fmf{phantom,tension=1.5}{v6,v6b}
\end{fmfgraph*}
\end{fmffile}
\end{gathered} \hspace{-0.5cm} +\frac{1}{12} \hspace{-0.5cm} \begin{gathered}
\begin{fmffile}{Diagrams/bosonic1PIEA_Gamma2_Diag8bis}
\begin{fmfgraph*}(35,20)
\fmfleft{i1,i2}
\fmfright{o1,o2}
\fmfbottom{i0,o0}
\fmfbottom{b0}
\fmfbottom{b1}
\fmfbottom{b2}
\fmftop{i3,o3}
\fmfv{decor.shape=circle,decor.size=2.0thick,foreground=(0,,0,,1)}{v1}
\fmfv{decor.shape=circle,decor.size=2.0thick,foreground=(0,,0,,1)}{v2}
\fmfv{decor.shape=circle,decor.size=2.0thick,foreground=(0,,0,,1)}{v3}
\fmfv{decor.shape=circle,decor.size=2.0thick,foreground=(0,,0,,1)}{v4}
\fmfv{decor.shape=circle,decor.size=2.0thick,foreground=(0,,0,,1)}{v5}
\fmfv{decor.shape=circle,decor.size=2.0thick,foreground=(0,,0,,1)}{v6}
\fmf{phantom,tension=1.4}{i1,v1}
\fmf{phantom}{i2,v3b}
\fmf{phantom}{i0,v1b}
\fmf{phantom}{o2,v6b}
\fmf{phantom,tension=1.4}{o1,v5}
\fmf{phantom}{o0,v5b}
\fmf{phantom,tension=1.11}{v3b,v6b}
\fmf{phantom,tension=1.38}{i0,v2}
\fmf{phantom,tension=1.38}{o0,v2}
\fmf{phantom,tension=1.8}{i0,v2b}
\fmf{phantom,tension=1.2}{o0,v2b}
\fmf{phantom,tension=1.2}{b0,v2b}
\fmf{phantom,tension=1.2}{b1,v2b}
\fmf{phantom,tension=1.2}{b2,v2b}
\fmf{phantom,tension=1.38}{i0,v4}
\fmf{phantom,tension=1.38}{o0,v4}
\fmf{phantom,tension=1.2}{i0,v4b}
\fmf{phantom,tension=1.8}{o0,v4b}
\fmf{phantom,tension=1.2}{b0,v4b}
\fmf{phantom,tension=1.2}{b1,v4b}
\fmf{phantom,tension=1.2}{b2,v4b}
\fmf{phantom,tension=2}{i3,v3}
\fmf{phantom,tension=2}{o3,v6}
\fmf{phantom,tension=2}{i3,v3b}
\fmf{phantom,tension=0.8}{o3,v3b}
\fmf{phantom,tension=0.8}{i3,v6b}
\fmf{phantom,tension=2}{o3,v6b}
\fmf{plain,tension=1.4,foreground=(1,,0,,0)}{v1,v2}
\fmf{plain,tension=1.4,foreground=(1,,0,,0)}{v4,v5}
\fmf{plain,foreground=(1,,0,,0)}{v1,v3}
\fmf{phantom,left=0.8,tension=0}{v1,v3}
\fmf{plain,foreground=(1,,0,,0)}{v5,v6}
\fmf{phantom,right=0.8,tension=0}{v5,v6}
\fmf{plain,foreground=(1,,0,,0)}{v2,v3}
\fmf{plain,foreground=(1,,0,,0)}{v4,v6}
\fmf{wiggly,tension=0.5,foreground=(1,,0,,0)}{v2,v4}
\fmf{wiggly,tension=2,foreground=(1,,0,,0)}{v3,v6}
\fmf{wiggly,right=0.8,tension=0,foreground=(1,,0,,0)}{v1,v5}
\fmf{phantom,tension=1}{v1,v1b}
\fmf{phantom,tension=0.2}{v2,v2b}
\fmf{phantom,tension=1.5}{v3,v3b}
\fmf{phantom,tension=0.2}{v4,v4b}
\fmf{phantom,tension=1}{v5,v5b}
\fmf{phantom,tension=1.5}{v6,v6b}
\end{fmfgraph*}
\end{fmffile}
\end{gathered} \hspace{-0.6cm} \left.\rule{0cm}{1.1cm}\right) \\
& + \mathcal{O}\big(\hbar^{3}\big) \;,
\end{split}
\label{eq:bosonic1PIEAIMGamma0DONAppendix}
\end{equation}
where all diagrams are 1PI (with respect to both $\boldsymbol{G}_{\Phi}$ and $D_{\Phi}$), as expected.

\section{\label{sec:2PIEAannIM}2PI effective action}
\subsection{\label{sec:original2PIEAannIM}Original effective action}

With $\hbar$ as expansion parameter, the investigation of the 2PI EA via the IM starts with the following power series:
\begin{subequations}
\begin{empheq}[left=\empheqlbrace]{align}
& \hspace{0.1cm} \Gamma^{(\mathrm{2PI})}\Big[\vec{\phi},\boldsymbol{G};\hbar\Big]=\sum_{n=0}^{\infty} \Gamma^{(\mathrm{2PI})}_{n}\Big[\vec{\phi},\boldsymbol{G}\Big]\hbar^{n}\;, \label{eq:pure2PIEAGammaExpansion0DON}\\
\nonumber \\
& \hspace{0.1cm} W\Big[\vec{J},\boldsymbol{K};\hbar\Big]=\sum_{n=0}^{\infty} W_{n}\Big[\vec{J},\boldsymbol{K}\Big]\hbar^{n}\;, \label{eq:pure2PIEAWExpansion0DON} \\
\nonumber \\
& \hspace{0.1cm} \vec{J}\Big[\vec{\phi},\boldsymbol{G};\hbar\Big]=\sum_{n=0}^{\infty} \vec{J}_{n}\Big[\vec{\phi},\boldsymbol{G}\Big]\hbar^{n}\;, \label{eq:pure2PIEAJExpansion0DON}\\
\nonumber \\
& \hspace{0.1cm} \boldsymbol{K}\Big[\vec{\phi},\boldsymbol{G};\hbar\Big]=\sum_{n=0}^{\infty} \boldsymbol{K}_{n}\Big[\vec{\phi},\boldsymbol{G}\Big]\hbar^{n}\;, \label{eq:pure2PIEAKExpansion0DON}\\
\nonumber \\
& \hspace{0.1cm} \vec{\phi}=\sum_{n=0}^{\infty} \vec{\phi}_{n}\Big[\vec{J},\boldsymbol{K}\Big]\hbar^{n}\;, \label{eq:pure2PIEAphiExpansion0DON} \\
\nonumber \\
& \hspace{0.1cm} \boldsymbol{G}=\sum_{n=0}^{\infty} \boldsymbol{G}_{n}\Big[\vec{J},\boldsymbol{K}\Big]\hbar^{n}\;, \label{eq:pure2PIEAGExpansion0DON}
\end{empheq}
\end{subequations}
with the definition:
\begin{equation}
\begin{split}
\Gamma^{(\mathrm{2PI})}\Big[\vec{\phi},\boldsymbol{G}\Big] \equiv & - W\Big[\vec{J},\boldsymbol{K}\Big] + \int_{\alpha} J_{\alpha}\Big[\vec{\phi},\boldsymbol{G}\Big] \frac{\delta W\big[\vec{J},\boldsymbol{K}\big]}{\delta J_{\alpha}} + \int_{\alpha_{1},\alpha_{2}} \boldsymbol{K}_{\alpha_{1}\alpha_{2}}\Big[\vec{\phi},\boldsymbol{G}\Big] \frac{\delta W\big[\vec{J},\boldsymbol{K}\big]}{\delta \boldsymbol{K}_{\alpha_{1}\alpha_{2}}} \\
= & - W\Big[\vec{J},\boldsymbol{K}\Big] + \int_{\alpha} J_{\alpha}\Big[\vec{\phi},\boldsymbol{G}\Big] \phi_{\alpha} + \frac{1}{2} \int_{\alpha_{1},\alpha_{2}} \phi_{\alpha_{1}} \boldsymbol{K}_{\alpha_{1}\alpha_{2}}\Big[\vec{\phi},\boldsymbol{G}\Big] \phi_{\alpha_{2}} \\
& + \frac{\hbar}{2} \int_{\alpha_{1},\alpha_{2}}\boldsymbol{K}_{\alpha_{1}\alpha_{2}}\Big[\vec{\phi},\boldsymbol{G}\Big]\boldsymbol{G}_{\alpha_{1}\alpha_{2}} \;,
\end{split}
\label{eq:pure2PIEAdefinition0DON}
\end{equation}
where
\begin{equation}
\phi_{\alpha}=\frac{\delta W\big[\vec{J},\boldsymbol{K}\big]}{\delta J_{\alpha}}\;,
\label{eq:pure2PIEAdefinitionbis0DON}
\end{equation}
\begin{equation}
\boldsymbol{G}_{\alpha_{1}\alpha_{2}} = \frac{\delta^{2} W\big[\vec{J},\boldsymbol{K}\big]}{\delta J_{\alpha_{1}}\delta J_{\alpha_{2}}} = \frac{2}{\hbar} \frac{\delta W\big[\vec{J},\boldsymbol{K}\big]}{\delta \boldsymbol{K}_{\alpha_{1}\alpha_{2}}} - \frac{1}{\hbar} \phi_{\alpha_{1}} \phi_{\alpha_{2}}\;.
\label{eq:pure2PIEAdefinitionbis20DON}
\end{equation}
The $W_{n}$ coefficients are directly inferred from the LE result~\eqref{eq:WKjLoopExpansionStep3} at $n=0,1~\mathrm{and}~2$. At arbitrary external sources $\vec{J}$ and $\boldsymbol{K}$, the 1-point correlation function and propagator of interest are:
\begin{subequations}
\begin{empheq}[left=\empheqlbrace]{align}
& \hspace{0.1cm} \scalebox{0.91}{${\displaystyle \varphi_{\mathrm{cl},\alpha}\Big[\vec{J},\boldsymbol{K}\Big] = \phi_{0,\alpha}\Big[\vec{J},\boldsymbol{K}\Big] = \frac{\delta W_{0}\big[\vec{J},\boldsymbol{K}\big]}{\delta J_{\alpha}}\;, }$} \label{eq:pure2PIEAphicl0DON}\\
\nonumber \\
& \hspace{0.1cm} \scalebox{0.91}{${\displaystyle \boldsymbol{G}_{\varphi_{\mathrm{cl}};JK,\alpha_{1}\alpha_{2}}\Big[\vec{J},\boldsymbol{K}\Big] = \boldsymbol{G}_{0,\alpha_{1}\alpha_{2}}\Big[\vec{J},\boldsymbol{K}\Big] }$} \nonumber \\
& \hspace{0.1cm} \scalebox{0.91}{${\displaystyle \hspace{3.48cm} = \left(\left.\frac{\delta^{2} S\big[\vec{\widetilde{\varphi}}\big]}{\delta\vec{\widetilde{\varphi}}\delta\vec{\widetilde{\varphi}}}\right|_{\vec{\widetilde{\varphi}}=\vec{\varphi}_{\mathrm{cl}}}-\boldsymbol{K}\Big[\vec{\phi},\boldsymbol{G}\Big]\right)_{\alpha_{1}\alpha_{2}}^{-1} = \frac{\delta^{2} W_{0}\big[\vec{J},\boldsymbol{K}\big]}{\delta J_{\alpha_{1}} \delta J_{\alpha_{2}}} }$} \nonumber \\
& \hspace{0.1cm} \scalebox{0.91}{${\displaystyle \hspace{3.48cm} = 2 \frac{\delta W_{1}\big[\vec{J},\boldsymbol{K}\big]}{\delta \boldsymbol{K}_{\alpha_{1}\alpha_{2}}} - \frac{\delta W_{1}\big[\vec{J},\boldsymbol{K}\big]}{\delta J_{\alpha_{1}}} \frac{\delta W_{0}\big[\vec{J},\boldsymbol{K}\big]}{\delta J_{\alpha_{2}}} - \frac{\delta W_{0}\big[\vec{J},\boldsymbol{K}\big]}{\delta J_{\alpha_{1}}} \frac{\delta W_{1}\big[\vec{J},\boldsymbol{K}\big]}{\delta J_{\alpha_{2}}}\;, }$} \label{eq:pure2PIEAGJK0DON}
\end{empheq}
\end{subequations}
where the penultimate line results from~\eqref{eq:pure2PIEAdefinitionbis20DON} combined with~\eqref{eq:pure2PIEAWExpansion0DON} and~\eqref{eq:pure2PIEAdefinitionbis0DON}. At $\big(\vec{J},\boldsymbol{K}\big)=\big(\vec{J}_{0},\boldsymbol{K}_{0}\big)$, we have:
\begin{subequations}
\begin{empheq}[left=\empheqlbrace]{align}
& \hspace{0.1cm} \scalebox{0.91}{${\displaystyle \vec{\phi}=\varphi_{\mathrm{cl},\alpha}\Big[\vec{J}=\vec{J}_{0},\boldsymbol{K}=\boldsymbol{K}_{0}\Big] = \phi_{0,\alpha}\Big[\vec{J}=\vec{J}_{0},\boldsymbol{K}=\boldsymbol{K}_{0}\Big] = \left.\frac{\delta W_{0}\big[\vec{J},\boldsymbol{K}\big]}{\delta J_{\alpha}}\right|_{\vec{J}=\vec{J}_{0}\atop\boldsymbol{K}=\boldsymbol{K}_{0}} \;. }$} \label{eq:pure2PIEAphi0DON}\\
\nonumber \\
& \hspace{0.1cm} \scalebox{0.91}{${\displaystyle \boldsymbol{G}_{\alpha_{1}\alpha_{2}} = \boldsymbol{G}_{\varphi_{\mathrm{cl}};JK,\alpha_{1}\alpha_{2}}\Big[\vec{J}=\vec{J}_{0},\boldsymbol{K}=\boldsymbol{K}_{0}\Big] = \boldsymbol{G}_{0,\alpha_{1}\alpha_{2}}\Big[\vec{J}=\vec{J}_{0},\boldsymbol{K}=\boldsymbol{K}_{0}\Big] }$} \nonumber \\
& \hspace{0.1cm} \scalebox{0.91}{${\displaystyle \hspace{1.18cm} = \left(\left.\frac{\delta^{2} S\big[\vec{\widetilde{\varphi}}\big]}{\delta\vec{\widetilde{\varphi}}\delta\vec{\widetilde{\varphi}}}\right|_{\vec{\widetilde{\varphi}}=\vec{\phi}}-\boldsymbol{K}_{0}\Big[\vec{\phi},\boldsymbol{G}\Big]\right)_{\alpha_{1}\alpha_{2}}^{-1} = \left.\frac{\delta^{2} W_{0}\big[\vec{J},\boldsymbol{K}\big]}{\delta J_{\alpha_{1}} \delta J_{\alpha_{2}}}\right|_{\vec{J}=\vec{J}_{0}\atop\boldsymbol{K}=\boldsymbol{K}_{0}} }$} \nonumber \\
& \hspace{0.1cm} \scalebox{0.91}{${\displaystyle \hspace{1.18cm} = 2 \left.\frac{\delta W_{1}\big[\vec{J},\boldsymbol{K}\big]}{\delta \boldsymbol{K}_{\alpha_{1}\alpha_{2}}}\right|_{\vec{J}=\vec{J}_{0}\atop\boldsymbol{K}=\boldsymbol{K}_{0}} - \left.\frac{\delta W_{1}\big[\vec{J},\boldsymbol{K}\big]}{\delta J_{\alpha_{1}}}\right|_{\vec{J}=\vec{J}_{0}\atop\boldsymbol{K}=\boldsymbol{K}_{0}} \left.\frac{\delta W_{0}\big[\vec{J},\boldsymbol{K}\big]}{\delta J_{\alpha_{2}}}\right|_{\vec{J}=\vec{J}_{0}\atop\boldsymbol{K}=\boldsymbol{K}_{0}} }$} \nonumber \\
& \hspace{0.1cm} \scalebox{0.91}{${\displaystyle \hspace{1.58cm} - \left.\frac{\delta W_{0}\big[\vec{J},\boldsymbol{K}\big]}{\delta J_{\alpha_{1}}}\right|_{\vec{J}=\vec{J}_{0}\atop\boldsymbol{K}=\boldsymbol{K}_{0}}\left.\frac{\delta W_{1}\big[\vec{J},\boldsymbol{K}\big]}{\delta J_{\alpha_{2}}}\right|_{\vec{J}=\vec{J}_{0}\atop\boldsymbol{K}=\boldsymbol{K}_{0}} \;. }$} \label{eq:pure2PIEAG0DON}
\end{empheq}
\end{subequations}
As was proven for $\vec{\phi}$ in the framework of the 1PI EA, we can prove similarly from~\eqref{eq:pure2PIEAdefinition0DON} that both $\vec{\phi}$ and $\boldsymbol{G}$ are independent of $\hbar$ in the present case. Note also that, thanks to the presence of $\boldsymbol{K}_{0}$ in~\eqref{eq:pure2PIEAG0DON}, $\vec{\phi}$ and $\boldsymbol{G}$ can be considered as independent as well: this essential property, which marks an important difference between the 1PI EA and 2PI EA formalisms, is used in the derivation of the gap equations associated to $\Gamma^{(\mathrm{2PI})}$ in section~\ref{sec:2PIEA}. As a next step, we give the relevant Feynman rules to develop the IM for $\Gamma^{(\mathrm{2PI})}$. At arbitrary external sources $\vec{J}$ and $\boldsymbol{K}$, we have the Feynman rules~\eqref{eq:FeynRulesLoopExpansionPropagator} to~\eqref{eq:FeynRulesLoopExpansion4legVertex} underpinning the LE of section~\ref{sec:OriginalLE} recalled below:
\begin{subequations}
\begin{align}
\begin{gathered}
\begin{fmffile}{Diagrams/LoopExpansion1_FeynRuleGbis_Appendix2PI}
\begin{fmfgraph*}(20,20)
\fmfleft{i0,i1,i2,i3}
\fmfright{o0,o1,o2,o3}
\fmflabel{$\alpha_{1}$}{v1}
\fmflabel{$\alpha_{2}$}{v2}
\fmf{phantom}{i1,v1}
\fmf{phantom}{i2,v1}
\fmf{plain,tension=0.6}{v1,v2}
\fmf{phantom}{v2,o1}
\fmf{phantom}{v2,o2}
\end{fmfgraph*}
\end{fmffile}
\end{gathered} \quad &\rightarrow \boldsymbol{G}_{\varphi_\text{cl};JK,\alpha_{1}\alpha_{2}}\Big[\vec{J},\boldsymbol{K}\Big]\;,
\label{eq:FeynRules2PIEAPropagatorSourceJ} \\
\begin{gathered}
\begin{fmffile}{Diagrams/LoopExpansion1_FeynRuleV3bis_Appendix2PI}
\begin{fmfgraph*}(20,20)
\fmfleft{i0,i1,i2,i3}
\fmfright{o0,o1,o2,o3}
\fmfv{decor.shape=cross,decor.angle=45,decor.size=3.5thick,foreground=(0,,0,,1)}{o2}
\fmf{phantom,tension=2.0}{i1,i1bis}
\fmf{plain,tension=2.0}{i1bis,v1}
\fmf{phantom,tension=2.0}{i2,i2bis}
\fmf{plain,tension=2.0}{i2bis,v1}
\fmf{dots,label=$x$,tension=0.6,foreground=(0,,0,,1)}{v1,v2}
\fmf{phantom,tension=2.0}{o1bis,o1}
\fmf{plain,tension=2.0}{v2,o1bis}
\fmf{phantom,tension=2.0}{o2bis,o2}
\fmf{phantom,tension=2.0}{v2,o2bis}
\fmf{dashes,tension=0.0,foreground=(0,,0,,1)}{v2,o2}
\fmflabel{$a_{1}$}{i1bis}
\fmflabel{$a_{2}$}{i2bis}
\fmflabel{$a_{3}$}{o1bis}
\fmflabel{$N$}{o2bis}
\end{fmfgraph*}
\end{fmffile}
\end{gathered} \quad &\rightarrow \lambda \left|\vec{\varphi}_{\mathrm{cl},\alpha}\Big[\vec{J},\boldsymbol{K}\Big]\right|\delta_{a_{1} a_{2}}\delta_{a_{3} N}\;,
\label{eq:FeynRules2PIEA3legVertexSourceJ} \\
\begin{gathered}
\begin{fmffile}{Diagrams/LoopExpansion1_FeynRuleV4bis_Appendix2PI}
\begin{fmfgraph*}(20,20)
\fmfleft{i0,i1,i2,i3}
\fmfright{o0,o1,o2,o3}
\fmf{phantom,tension=2.0}{i1,i1bis}
\fmf{plain,tension=2.0}{i1bis,v1}
\fmf{phantom,tension=2.0}{i2,i2bis}
\fmf{plain,tension=2.0}{i2bis,v1}
\fmf{zigzag,label=$x$,tension=0.6,foreground=(0,,0,,1)}{v1,v2}
\fmf{phantom,tension=2.0}{o1bis,o1}
\fmf{plain,tension=2.0}{v2,o1bis}
\fmf{phantom,tension=2.0}{o2bis,o2}
\fmf{plain,tension=2.0}{v2,o2bis}
\fmflabel{$a_{1}$}{i1bis}
\fmflabel{$a_{2}$}{i2bis}
\fmflabel{$a_{3}$}{o1bis}
\fmflabel{$a_{4}$}{o2bis}
\end{fmfgraph*}
\end{fmffile}
\end{gathered} \quad &\rightarrow \lambda\delta_{a_{1} a_{2}}\delta_{a_{3} a_{4}}\;,
\label{eq:FeynRules2PIEA4legVertexSourceJ}
\end{align}
\end{subequations}
and, at $\big(\vec{J},\boldsymbol{K}\big)=\big(\vec{J}_{0},\boldsymbol{K}_{0}\big)$,~\eqref{eq:FeynRules2PIEAPropagatorSourceJ} to~\eqref{eq:FeynRules2PIEA4legVertexSourceJ} coincide with~\eqref{eq:FeynRuleorig2PIEAG} to~\eqref{eq:FeynRuleorig2PIEAV4}, i.e.:
\begin{subequations}
\begin{align}
\begin{gathered}
\begin{fmffile}{Diagrams/1PIEA_G_Appendix2PI}
\begin{fmfgraph*}(20,20)
\fmfleft{i0,i1,i2,i3}
\fmfright{o0,o1,o2,o3}
\fmflabel{$\alpha_{1}$}{v1}
\fmflabel{$\alpha_{2}$}{v2}
\fmf{phantom}{i1,v1}
\fmf{phantom}{i2,v1}
\fmf{plain,tension=0.6,foreground=(1,,0,,0)}{v1,v2}
\fmf{phantom}{v2,o1}
\fmf{phantom}{v2,o2}
\end{fmfgraph*}
\end{fmffile}
\end{gathered} \quad &\rightarrow \boldsymbol{G}_{\alpha_{1}\alpha_{2}} \;,
\label{eq:FeynRules2PIEAPropagatorSourceJ0} \\
\begin{gathered}
\begin{fmffile}{Diagrams/1PIEA_V3_Appendix2PI}
\begin{fmfgraph*}(20,20)
\fmfleft{i0,i1,i2,i3}
\fmfright{o0,o1,o2,o3}
\fmfv{decor.shape=cross,decor.angle=45,decor.size=3.5thick,foreground=(1,,0,,0)}{o2}
\fmf{phantom,tension=2.0}{i1,i1bis}
\fmf{plain,tension=2.0,foreground=(1,,0,,0)}{i1bis,v1}
\fmf{phantom,tension=2.0}{i2,i2bis}
\fmf{plain,tension=2.0,foreground=(1,,0,,0)}{i2bis,v1}
\fmf{dots,label=$x$,tension=0.6,foreground=(0,,0,,1)}{v1,v2}
\fmf{phantom,tension=2.0}{o1bis,o1}
\fmf{plain,tension=2.0,foreground=(1,,0,,0)}{v2,o1bis}
\fmf{phantom,tension=2.0}{o2bis,o2}
\fmf{phantom,tension=2.0,foreground=(1,,0,,0)}{v2,o2bis}
\fmf{dashes,tension=0.0,foreground=(1,,0,,0)}{v2,o2}
\fmflabel{$a_{1}$}{i1bis}
\fmflabel{$a_{2}$}{i2bis}
\fmflabel{$a_{3}$}{o1bis}
\fmflabel{$N$}{o2bis}
\end{fmfgraph*}
\end{fmffile}
\end{gathered} \quad &\rightarrow \lambda\left|\vec{\phi}\right|\delta_{a_{1} a_{2}}\delta_{a_{3} N}\;,
\label{eq:FeynRules2PIEA3legVertexSourceJ0} \\
\begin{gathered}
\begin{fmffile}{Diagrams/1PIEA_V4_Appendix2PI}
\begin{fmfgraph*}(20,20)
\fmfleft{i0,i1,i2,i3}
\fmfright{o0,o1,o2,o3}
\fmf{phantom,tension=2.0}{i1,i1bis}
\fmf{plain,tension=2.0,foreground=(1,,0,,0)}{i1bis,v1}
\fmf{phantom,tension=2.0}{i2,i2bis}
\fmf{plain,tension=2.0,foreground=(1,,0,,0)}{i2bis,v1}
\fmf{zigzag,label=$x$,tension=0.6,foreground=(0,,0,,1)}{v1,v2}
\fmf{phantom,tension=2.0}{o1bis,o1}
\fmf{plain,tension=2.0,foreground=(1,,0,,0)}{v2,o1bis}
\fmf{phantom,tension=2.0}{o2bis,o2}
\fmf{plain,tension=2.0,foreground=(1,,0,,0)}{v2,o2bis}
\fmflabel{$a_{1}$}{i1bis}
\fmflabel{$a_{2}$}{i2bis}
\fmflabel{$a_{3}$}{o1bis}
\fmflabel{$a_{4}$}{o2bis}
\end{fmfgraph*}
\end{fmffile}
\end{gathered} \quad &\rightarrow \lambda\delta_{a_{1} a_{2}}\delta_{a_{3} a_{4}}\;.
\label{eq:FeynRules2PIEA4legVertexSourceJ0}
\end{align}
\end{subequations}
As a next step, we determine an expression for the $\Gamma^{(\mathrm{2PI})}_{n}$ coefficients by combining~\eqref{eq:pure2PIEAdefinition0DON} with~\eqref{eq:pure2PIEAGammaExpansion0DON},~\eqref{eq:pure2PIEAWExpansion0DON},~\eqref{eq:pure2PIEAJExpansion0DON} and~\eqref{eq:pure2PIEAKExpansion0DON}:
\begin{equation}
\begin{split}
\scalebox{0.96}{${\displaystyle \sum_{n=0}^{\infty}\Gamma_{n}^{(\mathrm{2PI})}\Big[\vec{\phi},\boldsymbol{G}\Big]\hbar^{n}= }$} & \scalebox{0.96}{${\displaystyle - \sum_{n=0}^{\infty} W_{n}\Bigg[\sum_{m=0}^{\infty} \vec{J}_{m}\Big[\vec{\phi},\boldsymbol{G}\Big]\hbar^{m},\sum_{m=0}^{\infty} \boldsymbol{K}_{m}\Big[\vec{\phi},\boldsymbol{G}\Big]\hbar^{m}\Bigg] \hbar^{n} + \sum_{n=0}^{\infty}\int_{\alpha} J_{n,\alpha}\Big[\vec{\phi},\boldsymbol{G}\Big] \phi_{\alpha} \hbar^{n} }$} \\
& \scalebox{0.96}{${\displaystyle + \frac{1}{2} \sum_{n=0}^{\infty} \int_{\alpha_{1},\alpha_{2}} \phi_{\alpha_{1}} \boldsymbol{K}_{n,\alpha_{1}\alpha_{2}}\Big[\vec{\phi},\boldsymbol{G}\Big]\phi_{\alpha_{2}} \hbar^{n} + \frac{1}{2} \sum_{n=0}^{\infty} \int_{\alpha_{1},\alpha_{2}} \boldsymbol{K}_{n,\alpha_{1}\alpha_{2}}\Big[\vec{\phi},\boldsymbol{G}\Big]\boldsymbol{G}_{\alpha_{1}\alpha_{2}} \hbar^{n+1}\;. }$}
\end{split}
\label{eq:pure2PIEAIMstep20DON}
\end{equation}
This leads to (see section~\ref{sec:GammanCoeffIM2PIEA}):
\begin{equation}
\begin{split}
\scalebox{0.99}{${\displaystyle \Gamma_{n}^{(\mathrm{2PI})}\Big[\vec{\phi},\boldsymbol{G}\Big] = }$} & \scalebox{0.99}{${\displaystyle - W_{n}\Big[\vec{J}=\vec{J}_{0},\boldsymbol{K}=\boldsymbol{K}_{0}\Big] - \sum_{m=1}^{n} \int_{\alpha} \left.\frac{\delta W_{n-m}\big[\vec{J},\boldsymbol{K}\big]}{\delta J_{\alpha}}\right|_{\vec{J}=\vec{J}_{0} \atop \boldsymbol{K}=\boldsymbol{K}_{0}} J_{m,\alpha}\Big[\vec{\phi},\boldsymbol{K}\Big] }$} \\
& \scalebox{0.99}{${\displaystyle - \sum_{m=1}^{n} \int_{\alpha_{1},\alpha_{2}} \left.\frac{\delta W_{n-m}\big[\vec{J},\boldsymbol{K}\big]}{\delta \boldsymbol{K}_{\alpha_{1}\alpha_{2}}}\right|_{\vec{J}=\vec{J}_{0} \atop \boldsymbol{K}=\boldsymbol{K}_{0}} \boldsymbol{K}_{m,\alpha_{1}\alpha_{2}}\Big[\vec{\phi},\boldsymbol{G}\Big] }$} \\
& \scalebox{0.99}{${\displaystyle - \sum_{m=2}^{n} \frac{1}{m!} \sum_{\underset{\lbrace l+l'=m \rbrace}{l,l'=1}}^{m} \sum^{n}_{\underset{\lbrace n_{1} + \cdots + n_{l} + \hat{n}_{1} + \cdots + \hat{n}_{l'} \leq n\rbrace}{n_{1},\cdots,n_{l},\hat{n}_{1},\cdots,\hat{n}_{l'}=1}} \begin{pmatrix}
m \\
l
\end{pmatrix} }$} \\
& \hspace{0.85cm} \scalebox{0.99}{${\displaystyle \times \int_{\alpha_{1},\cdots,\alpha_{l} \atop \hat{\alpha}_{1},\cdots,\hat{\alpha}_{2l'}} \left.\frac{\delta^{m} W_{n-(n_{1}+\cdots+n_{l}+\hat{n}_{1}+\cdots+\hat{n}_{l'})}\big[\vec{J},\boldsymbol{K}\big]}{\delta J_{\alpha_{1}}\cdots\delta J_{\alpha_{l}} \delta \boldsymbol{K}_{\hat{\alpha}_{1}\hat{\alpha}_{2}}\cdots\delta \boldsymbol{K}_{\hat{\alpha}_{2 l'-1}\hat{\alpha}_{2 l'}}}\right|_{\vec{J}=\vec{J}_{0} \atop \boldsymbol{K}=\boldsymbol{K}_{0}} }$} \\
& \hspace{2.55cm} \scalebox{0.99}{${\displaystyle \times J_{n_{1},\alpha_{1}}\Big[\vec{\phi},\boldsymbol{G}\Big] \cdots J_{n_{l},\alpha_{l}}\Big[\vec{\phi},\boldsymbol{G}\Big] \boldsymbol{K}_{\hat{n}_{1},\hat{\alpha}_{1}\hat{\alpha}_{2}}\Big[\vec{\phi},\boldsymbol{G}\Big] \cdots \boldsymbol{K}_{\hat{n}_{l'},\hat{\alpha}_{2 l' -1}\hat{\alpha}_{2 l'}}\Big[\vec{\phi},\boldsymbol{G}\Big] }$} \\
& \scalebox{0.99}{${\displaystyle + \int_{\alpha} J_{n,\alpha}\Big[\vec{\phi},\boldsymbol{G}\Big] \phi_{\alpha} + \frac{1}{2} \int_{\alpha_{1},\alpha_{2}} \phi_{\alpha_{1}} \boldsymbol{K}_{n,\alpha_{1}\alpha_{2}}\Big[\vec{\phi},\boldsymbol{G}\Big]\phi_{\alpha_{2}} }$} \\
& \scalebox{0.99}{${\displaystyle + \frac{1}{2} \int_{\alpha_{1},\alpha_{2}} \boldsymbol{K}_{n-1,\alpha_{1}\alpha_{2}}\Big[\vec{\phi},\boldsymbol{G}\Big] \boldsymbol{G}_{\alpha_{1}\alpha_{2}} \delta_{n \geq 1} \;. }$}
\end{split}
\label{eq:pure2PIEAIMstep50DON}
\end{equation}
We then rewrite~\eqref{eq:pure2PIEAIMstep50DON} by considering~\eqref{eq:pure2PIEAphi0DON} as well as the relation:
\begin{equation}
\left.\frac{\delta W_{0}\big[\vec{J},\boldsymbol{K}\big]}{\delta\boldsymbol{K}_{\alpha_{1}\alpha_{2}}}\right|_{\vec{J}=\vec{J}_{0} \atop \boldsymbol{K}=\boldsymbol{K}_{0}} = \frac{1}{2}\phi_{\alpha_{1}}\phi_{\alpha_{2}}\;,
\end{equation}
which can be inferred from the equality $W_{0}\big[\vec{J},\boldsymbol{K}\big]=-S_{JK}\big[\vec{\varphi}_{\mathrm{cl}}\big]$ that follows from~\eqref{eq:WKjLoopExpansionStep3}. Similarly to the derivation of~\eqref{eq:pure1PIEAIMSimplifystep50DON}, these relations enable us to show that:
\begin{equation}
\begin{split}
& -\sum_{m=1}^{\textcolor{red}{n}} \int_{\alpha} \left.\frac{\delta W_{n-m}\big[\vec{J},\boldsymbol{K}\big]}{\delta J_{\alpha}}\right|_{\vec{J}=\vec{J}_{0} \atop \boldsymbol{K}=\boldsymbol{K}_{0}} J_{m,\alpha}\Big[\vec{\phi},\boldsymbol{G}\Big]+\int_{\alpha} J_{n,\alpha}\Big[\vec{\phi},\boldsymbol{G}\Big] \phi_{\alpha} \\
& \hspace{0.8cm} =-\sum_{m=1}^{\textcolor{red}{n-1}} \int_{\alpha} \left.\frac{\delta W_{n-m}\big[\vec{J},\boldsymbol{K}\big]}{\delta J_{\alpha}}\right|_{\vec{J}=\vec{J}_{0} \atop \boldsymbol{K}=\boldsymbol{K}_{0}} J_{m,\alpha}\Big[\vec{\phi},\boldsymbol{G}\Big] +\int_{\alpha} J_{0,\alpha}\Big[\vec{\phi},\boldsymbol{G}\Big] \phi_{\alpha} \delta_{n 0}\;,
\end{split}
\label{eq:pure2PIEAIMSimplifystep5a0DON}
\end{equation}
and
\begin{equation}
\begin{split}
& - \sum_{m=1}^{\textcolor{red}{n}} \int_{\alpha_{1},\alpha_{2}} \left.\frac{\delta W_{n-m}\big[\vec{J},\boldsymbol{K}\big]}{\delta \boldsymbol{K}_{\alpha_{1}\alpha_{2}}}\right|_{\vec{J}=\vec{J}_{0} \atop \boldsymbol{K}=\boldsymbol{K}_{0}} \boldsymbol{K}_{m,\alpha_{1}\alpha_{2}}\Big[\vec{\phi},\boldsymbol{G}\Big]+ \frac{1}{2} \int_{\alpha_{1},\alpha_{2}} \phi_{\alpha_{1}} \boldsymbol{K}_{n,\alpha_{1}\alpha_{2}}\Big[\vec{\phi},\boldsymbol{G}\Big]\phi_{\alpha_{2}} \\
& \hspace{0.8cm} = -\sum_{m=1}^{\textcolor{red}{n-1}} \int_{\alpha_{1},\alpha_{2}} \left.\frac{\delta W_{n-m}\big[\vec{J},\boldsymbol{K}\big]}{\delta \boldsymbol{K}_{\alpha_{1}\alpha_{2}}}\right|_{\vec{J}=\vec{J}_{0} \atop \boldsymbol{K}=\boldsymbol{K}_{0}} \boldsymbol{K}_{m,\alpha_{1}\alpha_{2}}\Big[\vec{\phi},\boldsymbol{G}\Big] + \frac{1}{2} \int_{\alpha_{1},\alpha_{2}} \phi_{\alpha_{1}} \boldsymbol{K}_{0,\alpha_{1}\alpha_{2}}\Big[\vec{\phi},\boldsymbol{G}\Big]\phi_{\alpha_{2}} \delta_{n 0}\;,
\end{split}
\label{eq:pure2PIEAIMSimplifystep5b0DON}
\end{equation}
which implies that~\eqref{eq:pure2PIEAIMstep50DON} is equivalent to:
\begin{equation}
\begin{split}
\scalebox{0.99}{${\displaystyle \Gamma_{n}^{(\mathrm{2PI})}\Big[\vec{\phi},\boldsymbol{G}\Big] = }$} & \scalebox{0.99}{${\displaystyle - W_{n}\Big[\vec{J}=\vec{J}_{0},\boldsymbol{K}=\boldsymbol{K}_{0}\Big] - \sum_{m=1}^{n-1} \int_{\alpha} \left.\frac{\delta W_{n-m}\big[\vec{J},\boldsymbol{K}\big]}{\delta J_{\alpha}}\right|_{\vec{J}=\vec{J}_{0} \atop \boldsymbol{K}=\boldsymbol{K}_{0}} J_{m,\alpha}\Big[\vec{\phi},\boldsymbol{K}\Big] }$} \\
& \scalebox{0.99}{${\displaystyle - \sum_{m=1}^{n-1} \int_{\alpha_{1},\alpha_{2}} \left.\frac{\delta W_{n-m}\big[\vec{J},\boldsymbol{K}\big]}{\delta \boldsymbol{K}_{\alpha_{1}\alpha_{2}}}\right|_{\vec{J}=\vec{J}_{0} \atop \boldsymbol{K}=\boldsymbol{K}_{0}} \boldsymbol{K}_{m,\alpha_{1}\alpha_{2}}\Big[\vec{\phi},\boldsymbol{G}\Big] }$} \\
& \scalebox{0.99}{${\displaystyle - \sum_{m=2}^{n} \frac{1}{m!} \sum_{\underset{\lbrace l+l'=m \rbrace}{l,l'=1}}^{m} \sum^{n}_{\underset{\lbrace n_{1} + \cdots + n_{l} + \hat{n}_{1} + \cdots + \hat{n}_{l'} \leq n\rbrace}{n_{1},\cdots,n_{l},\hat{n}_{1},\cdots,\hat{n}_{l'}=1}} \begin{pmatrix}
m \\
l
\end{pmatrix} }$} \\
& \hspace{0.85cm} \scalebox{0.99}{${\displaystyle \times \int_{\alpha_{1},\cdots,\alpha_{l} \atop \hat{\alpha}_{1},\cdots,\hat{\alpha}_{2l'}} \left.\frac{\delta^{m} W_{n-(n_{1}+\cdots+n_{l}+\hat{n}_{1}+\cdots+\hat{n}_{l'})}\big[\vec{J},\boldsymbol{K}\big]}{\delta J_{\alpha_{1}}\cdots\delta J_{\alpha_{l}} \delta \boldsymbol{K}_{\hat{\alpha}_{1}\hat{\alpha}_{2}}\cdots\delta \boldsymbol{K}_{\hat{\alpha}_{2 l'-1}\hat{\alpha}_{2 l'}}}\right|_{\vec{J}=\vec{J}_{0} \atop \boldsymbol{K}=\boldsymbol{K}_{0}} }$} \\
& \hspace{2.55cm} \scalebox{0.99}{${\displaystyle \times J_{n_{1},\alpha_{1}}\Big[\vec{\phi},\boldsymbol{G}\Big] \cdots J_{n_{l},\alpha_{l}}\Big[\vec{\phi},\boldsymbol{G}\Big] \boldsymbol{K}_{\hat{n}_{1},\hat{\alpha}_{1}\hat{\alpha}_{2}}\Big[\vec{\phi},\boldsymbol{G}\Big] \cdots \boldsymbol{K}_{\hat{n}_{l'},\hat{\alpha}_{2 l' -1}\hat{\alpha}_{2 l'}}\Big[\vec{\phi},\boldsymbol{G}\Big] }$} \\
& \scalebox{0.99}{${\displaystyle + \int_{\alpha} J_{0,\alpha}\Big[\vec{\phi},\boldsymbol{G}\Big] \phi_{\alpha} \delta_{n 0} + \frac{1}{2} \int_{\alpha_{1},\alpha_{2}} \phi_{\alpha_{1}} \boldsymbol{K}_{0,\alpha_{1}\alpha_{2}}\Big[\vec{\phi},\boldsymbol{G}\Big]\phi_{\alpha_{2}} \delta_{n 0} }$} \\
& \scalebox{0.99}{${\displaystyle + \frac{1}{2} \int_{\alpha_{1},\alpha_{2}} \boldsymbol{K}_{n-1,\alpha_{1}\alpha_{2}}\Big[\vec{\phi},\boldsymbol{G}\Big] \boldsymbol{G}_{\alpha_{1}\alpha_{2}} \delta_{n \geq 1}\;. }$}
\end{split}
\label{eq:pure2PIEAIMstep60DON}
\end{equation}
Still focusing on the first non-trivial order, we will exploit~\eqref{eq:pure2PIEAIMstep60DON} evaluated at $n=0,1~\mathrm{and}~2$, i.e.:
\begin{equation}
\Gamma^{(\mathrm{2PI})}_{0}\Big[\vec{\phi},\boldsymbol{G}\Big] = -W_{0}\Big[\vec{J}=\vec{J}_{0},\boldsymbol{K}=\boldsymbol{K}_{0}\Big] + \int_{\alpha} J_{0,\alpha}\Big[\vec{\phi},\boldsymbol{G}\Big] \phi_{\alpha} + \frac{1}{2}\int_{\alpha_{1},\alpha_{2}}\phi_{\alpha_{1}} \boldsymbol{K}_{0,\alpha_{1}\alpha_{2}}\Big[\vec{\phi},\boldsymbol{G}\Big] \phi_{\alpha_{2}}\;,
\label{eq:pure2PIEAIMGamma00DON}
\end{equation}
\begin{equation}
\Gamma^{(\mathrm{2PI})}_{1}\Big[\vec{\phi},\boldsymbol{G}\Big] = -W_{1}\Big[\vec{J}=\vec{J}_{0},\boldsymbol{K}=\boldsymbol{K}_{0}\Big]+\frac{1}{2}\int_{\alpha_{1},\alpha_{2}} \boldsymbol{K}_{0,\alpha_{1}\alpha_{2}}\Big[\vec{\phi},\boldsymbol{G}\Big] \boldsymbol{G}_{\alpha_{1}\alpha_{2}}\;,
\label{eq:pure2PIEAIMGamma10DON}
\end{equation}
\begin{equation}
\begin{split}
\Gamma^{(\mathrm{2PI})}_{2}\Big[\vec{\phi},\boldsymbol{G}\Big] = & -W_{2}\Big[\vec{J}=\vec{J}_{0},\boldsymbol{K}=\boldsymbol{K}_{0}\Big] -\int_{\alpha} \left.\frac{\delta W_{1}\big[\vec{J},\boldsymbol{K}\big]}{\delta J_{\alpha}}\right|_{\vec{J}=\vec{J}_{0} \atop \boldsymbol{K}=\boldsymbol{K}_{0}} J_{1,\alpha}\Big[\vec{\phi},\boldsymbol{G}\Big] \\
& -\int_{\alpha_{1},\alpha_{2}} \left.\frac{\delta W_{1}\big[\vec{J},\boldsymbol{K}\big]}{\delta \boldsymbol{K}_{\alpha_{1}\alpha_{2}}}\right|_{\vec{J}=\vec{J}_{0} \atop \boldsymbol{K}=\boldsymbol{K}_{0}} \boldsymbol{K}_{1,\alpha_{1}\alpha_{2}}\Big[\vec{\phi},\boldsymbol{G}\Big] \\
& -\frac{1}{2} \int_{\alpha_{1},\alpha_{2}} \left.\frac{\delta^{2} W_{0}\big[\vec{J},\boldsymbol{K}\big]}{\delta J_{\alpha_{1}} \delta J_{\alpha_{2}}}\right|_{\vec{J}=\vec{J}_{0} \atop \boldsymbol{K}=\boldsymbol{K}_{0}} J_{1,\alpha_{1}}\Big[\vec{\phi},\boldsymbol{G}\Big] J_{1,\alpha_{2}}\Big[\vec{\phi},\boldsymbol{G}\Big] \\
& -\frac{1}{2} \int_{\alpha_{1},\alpha_{2},\alpha_{3},\alpha_{4}} \left.\frac{\delta^{2} W_{0}\big[\vec{J},\boldsymbol{K}\big]}{\delta \boldsymbol{K}_{\alpha_{1}\alpha_{2}} \delta \boldsymbol{K}_{\alpha_{3}\alpha_{4}}}\right|_{\vec{J}=\vec{J}_{0} \atop \boldsymbol{K}=\boldsymbol{K}_{0}} \boldsymbol{K}_{1,\alpha_{1}\alpha_{2}}\Big[\vec{\phi},\boldsymbol{G}\Big] \boldsymbol{K}_{1,\alpha_{3}\alpha_{4}}\Big[\vec{\phi},\boldsymbol{G}\Big] \\
& - \int_{\alpha_{1},\alpha_{2},\alpha_{3}} \left.\frac{\delta^{2} W_{0}\big[\vec{J},\boldsymbol{K}\big]}{\delta J_{\alpha_{1}} \delta \boldsymbol{K}_{\alpha_{2}\alpha_{3}}}\right|_{\vec{J}=\vec{J}_{0} \atop \boldsymbol{K}=\boldsymbol{K}_{0}} J_{1,\alpha_{1}}\Big[\vec{\phi},\boldsymbol{G}\Big] \boldsymbol{K}_{1,\alpha_{2}\alpha_{3}}\Big[\vec{\phi},\boldsymbol{G}\Big] \\
& + \frac{1}{2} \int_{\alpha_{1},\alpha_{2}} \boldsymbol{K}_{1,\alpha_{1}\alpha_{2}}\Big[\vec{\phi},\boldsymbol{G}\Big] \boldsymbol{G}_{\alpha_{1}\alpha_{2}} \;.
\end{split}
\label{eq:pure2PIEAIMGamma20DON}
\end{equation}
It is customary to rewrite the rightmost term in~\eqref{eq:pure2PIEAIMGamma10DON} using the relation:
\begin{equation}
\boldsymbol{K}_{0}\Big[\vec{\phi},\boldsymbol{G}\Big] = \boldsymbol{G}^{-1}_{\phi}\Big[\vec{\phi}\Big] - \boldsymbol{G}^{-1}\;,
\end{equation}
with
\begin{equation}
\boldsymbol{G}^{-1}_{\phi,\alpha_{1}\alpha_{2}}\Big[\vec{\phi}\Big] \equiv \left.\frac{\delta^{2} S\big[\vec{\widetilde{\varphi}}\big]}{\delta\widetilde{\varphi}_{\alpha_{1}}\delta\widetilde{\varphi}_{\alpha_{2}}}\right|_{\vec{\widetilde{\varphi}}=\vec{\phi}} = \bigg(-\nabla_{x_{1}}^{2} + m^2 + \frac{\lambda}{6} \vec{\phi}^{2}_{x_{1}}\bigg) \delta_{\alpha_{1}\alpha_{2}} + \frac{\lambda}{3}\phi_{\alpha_{1}} \phi_{\alpha_{2}}\delta_{x_{1}x_{2}}\;,
\label{eq:pure2PIEADefinitionG00DON}
\end{equation}
as follows from~\eqref{eq:pure2PIEAG0DON}. In this way,~\eqref{eq:pure2PIEAIMGamma10DON} becomes:
\begin{equation}
\Gamma^{(\mathrm{2PI})}_{1}\Big[\vec{\phi},\boldsymbol{G}\Big] = -W_{1}\Big[\vec{J}=\vec{J}_{0},\boldsymbol{K}=\boldsymbol{K}_{0}\Big]+\frac{1}{2}\mathrm{STr}\left[\boldsymbol{G}^{-1}_{\phi}\Big[\vec{\phi}\Big]\boldsymbol{G}-\mathbb{I}\right]\;,
\label{eq:pure2PIEAIMGamma1step20DON}
\end{equation}
where $\mathbb{I}_{\alpha_{1}\alpha_{2}}=\delta_{\alpha_{1}\alpha_{2}}=\delta_{a_{1}a_{2}}\delta_{x_{1}x_{2}}$. We then focus on $\Gamma_{2}^{(\mathrm{2PI})}$ and~\eqref{eq:pure2PIEAIMGamma20DON}. As for the 1PI EA with result~\eqref{eq:pure1PIEAphi10DON}, we derive:
\begin{equation}
\phi_{1,\alpha}\Big[\vec{J},\boldsymbol{K}\Big]=\frac{\delta W_{1}\big[\vec{J},\boldsymbol{G}\big]}{\delta J_{\alpha}} = - \frac{1}{6} \hspace{0.08cm} \begin{gathered}
\begin{fmffile}{Diagrams/2PIEA_phi11}
\begin{fmfgraph*}(25,12)
\fmfleft{i0,i,i1}
\fmfright{o0,o,o1}
\fmfv{decor.shape=cross,decor.angle=45,decor.size=3.5thick,foreground=(0,,0,,1)}{o1}
\fmfv{decor.shape=circle,decor.filled=empty,decor.size=1.5thick,label.dist=0.15cm,label=$\alpha$}{o0}
\fmf{phantom,tension=10}{i,i1bis}
\fmf{phantom,tension=10}{o,o1bis}
\fmf{plain,left,tension=0.5}{i1bis,v1,i1bis}
\fmf{phantom,right,tension=0.5}{o1bis,v2,o1bis}
\fmf{dashes,tension=0,foreground=(0,,0,,1)}{v2,o1}
\fmf{plain,tension=0}{v2,o0}
\fmf{dots,foreground=(0,,0,,1)}{v1,v2}
\end{fmfgraph*}
\end{fmffile}
\end{gathered}
-\frac{1}{3} \hspace{0.2cm}\begin{gathered}
\begin{fmffile}{Diagrams/2PIEA_phi12}
\begin{fmfgraph*}(22,12)
\fmfleft{i0,i,i1}
\fmfright{o0,o,o1}
\fmfv{decor.shape=cross,decor.angle=65,decor.size=3.5thick,foreground=(0,,0,,1)}{o1}
\fmfv{decor.shape=circle,decor.filled=empty,decor.size=1.5thick,label.dist=0.15cm,label=$\alpha$}{i1}
\fmf{phantom,tension=11}{i,v1}
\fmf{phantom,tension=11}{v2,o}
\fmf{plain,tension=0}{v1,i1}
\fmf{dashes,tension=0,foreground=(0,,0,,1)}{v2,o1}
\fmf{plain,right,tension=0.4}{v1,v2}
\fmf{dots,tension=4.0,foreground=(0,,0,,1)}{v1,v2}
\end{fmfgraph*}
\end{fmffile}
\end{gathered}\;.
\label{eq:pure2PIEAphi10DON}
\end{equation}
In order to determine the source coefficients $\vec{J}_{1}$ and $\boldsymbol{K}_{1}$, we need to find a diagrammatic expression for $\boldsymbol{G}_{1}$ as well. This can be achieved by further differentiating~\eqref{eq:pure2PIEAphi10DON}, i.e.:
\begin{equation}
\boldsymbol{G}_{1}\Big[\vec{J},\boldsymbol{K}\Big] = \frac{\delta^{2} W_{1}\big[\vec{J},\boldsymbol{G}\big]}{\delta J_{\alpha_{1}} \delta J_{\alpha_{2}}} = \frac{\delta}{\delta J_{\alpha_{1}}}\left(\rule{0cm}{1.2cm}\right. - \frac{1}{6} \hspace{0.08cm} \begin{gathered}
\begin{fmffile}{Diagrams/2PIEA_phi11bis}
\begin{fmfgraph*}(25,12)
\fmfleft{i0,i,i1}
\fmfright{o0,o,o1}
\fmfv{decor.shape=cross,decor.angle=45,decor.size=3.5thick,foreground=(0,,0,,1)}{o1}
\fmfv{decor.shape=circle,decor.filled=empty,decor.size=1.5thick,label.dist=0.15cm,label=$\alpha_{2}$}{o0}
\fmf{phantom,tension=10}{i,i1bis}
\fmf{phantom,tension=10}{o,o1bis}
\fmf{plain,left,tension=0.5}{i1bis,v1,i1bis}
\fmf{phantom,right,tension=0.5}{o1bis,v2,o1bis}
\fmf{dashes,tension=0,foreground=(0,,0,,1)}{v2,o1}
\fmf{plain,tension=0}{v2,o0}
\fmf{dots,foreground=(0,,0,,1)}{v1,v2}
\end{fmfgraph*}
\end{fmffile}
\end{gathered}
\hspace{0.1cm} -\frac{1}{3} \hspace{0.3cm}\begin{gathered}
\begin{fmffile}{Diagrams/2PIEA_phi12bis}
\begin{fmfgraph*}(22,12)
\fmfleft{i0,i,i1}
\fmfright{o0,o,o1}
\fmfv{decor.shape=cross,decor.angle=65,decor.size=3.5thick,foreground=(0,,0,,1)}{o1}
\fmfv{decor.shape=circle,decor.filled=empty,decor.size=1.5thick,label.dist=0.15cm,label=$\alpha_{2}$}{i1}
\fmf{phantom,tension=11}{i,v1}
\fmf{phantom,tension=11}{v2,o}
\fmf{plain,tension=0}{v1,i1}
\fmf{dashes,tension=0,foreground=(0,,0,,1)}{v2,o1}
\fmf{plain,right,tension=0.4}{v1,v2}
\fmf{dots,tension=4.0,foreground=(0,,0,,1)}{v1,v2}
\end{fmfgraph*}
\end{fmffile}
\end{gathered} \left.\rule{0cm}{1.2cm}\right)\;.
\end{equation}
The derivative in the rightmost side can be evaluated with the help of the two equalities given below:
\begin{equation}
\frac{\delta \varphi_{\mathrm{cl},\alpha_{2}}\big[\vec{J},\boldsymbol{K}\big]}{\delta J_{\alpha_{1}}} = \frac{\delta^{2} W_{0}\big[\vec{J}\big]}{\delta J_{\alpha_{1}} \delta J_{\alpha_{2}}} = \boldsymbol{G}_{\varphi_{\mathrm{cl}};JK,\alpha_{1}\alpha_{2}}\Big[\vec{J},\boldsymbol{K}\Big] = \hspace{0.3cm} \begin{gathered}
\begin{fmffile}{Diagrams/2PIEA_DerivW0J0J0}
\begin{fmfgraph*}(20,20)
\fmfleft{i0,i1,i2,i3}
\fmfright{o0,o1,o2,o3}
\fmfv{decor.shape=circle,decor.filled=empty,decor.size=1.5thick,label=$\alpha_{1}$}{v1}
\fmfv{decor.shape=circle,decor.filled=empty,decor.size=1.5thick,label=$\alpha_{2}$}{v2}
\fmf{phantom}{i1,v1}
\fmf{phantom}{i2,v1}
\fmf{plain,tension=0.6}{v1,v2}
\fmf{phantom}{v2,o1}
\fmf{phantom}{v2,o2}
\end{fmfgraph*}
\end{fmffile}
\end{gathered}\hspace{0.3cm}\;,
\label{eq:pure2PIEADerivW0j0j00DON}
\end{equation}
\begin{equation}
\frac{\delta \boldsymbol{G}_{\varphi_{\mathrm{cl}};JK,\alpha_{2}\alpha_{3}}\big[\vec{J},\boldsymbol{K}\big]}{\delta J_{\alpha_{1}}} = - \frac{1}{3} \left(\rule{0cm}{0.8cm}\right. \hspace{0.5cm} \begin{gathered}
\begin{fmffile}{Diagrams/1PIEA_DerivGJ1}
\begin{fmfgraph*}(20,20)
\fmfleft{i0,i1,i2,i3}
\fmfright{o0,o1,o2,o3}
\fmfv{decor.shape=cross,decor.angle=45,decor.size=3.5thick,foreground=(0,,0,,1)}{o2}
\fmfv{decor.shape=circle,decor.filled=empty,decor.size=1.5thick,label=$\alpha_{1}$}{o1}
\fmfv{decor.shape=circle,decor.filled=empty,decor.size=1.5thick,label=$\alpha_{2}$}{i1}
\fmfv{decor.shape=circle,decor.filled=empty,decor.size=1.5thick,label=$\alpha_{3}$}{i2}
\fmf{phantom,tension=2.0}{i1,i1bis}
\fmf{phantom,tension=2.0}{i1bis,v1}
\fmf{phantom,tension=2.0}{i2,i2bis}
\fmf{phantom,tension=2.0}{i2bis,v1}
\fmf{dots,tension=0.6,foreground=(0,,0,,1)}{v1,v2}
\fmf{phantom,tension=2.0}{o1bis,o1}
\fmf{phantom,tension=2.0}{v2,o1bis}
\fmf{phantom,tension=2.0}{o2bis,o2}
\fmf{phantom,tension=2.0}{v2,o2bis}
\fmf{plain,tension=0.0}{i2,v1}
\fmf{plain,tension=0.0}{i1,v1}
\fmf{dashes,tension=0.0,foreground=(0,,0,,1)}{v2,o2}
\fmf{plain,tension=0.0}{v2,o1}
\end{fmfgraph*}
\end{fmffile}
\end{gathered} \hspace{0.5cm} + \hspace{0.6cm} \begin{gathered}
\begin{fmffile}{Diagrams/1PIEA_DerivGJ2}
\begin{fmfgraph*}(20,20)
\fmfleft{i0,i1,i2,i3}
\fmfright{o0,o1,o2,o3}
\fmfv{decor.shape=cross,decor.angle=45,decor.size=3.5thick,foreground=(0,,0,,1)}{o2}
\fmfv{decor.shape=circle,decor.filled=empty,decor.size=1.5thick,label=$\alpha_{2}$}{o1}
\fmfv{decor.shape=circle,decor.filled=empty,decor.size=1.5thick,label=$\alpha_{1}$}{i1}
\fmfv{decor.shape=circle,decor.filled=empty,decor.size=1.5thick,label=$\alpha_{3}$}{i2}
\fmf{phantom,tension=2.0}{i1,i1bis}
\fmf{phantom,tension=2.0}{i1bis,v1}
\fmf{phantom,tension=2.0}{i2,i2bis}
\fmf{phantom,tension=2.0}{i2bis,v1}
\fmf{dots,tension=0.6,foreground=(0,,0,,1)}{v1,v2}
\fmf{phantom,tension=2.0}{o1bis,o1}
\fmf{phantom,tension=2.0}{v2,o1bis}
\fmf{phantom,tension=2.0}{o2bis,o2}
\fmf{phantom,tension=2.0}{v2,o2bis}
\fmf{plain,tension=0.0}{i2,v1}
\fmf{plain,tension=0.0}{i1,v1}
\fmf{dashes,tension=0.0,foreground=(0,,0,,1)}{v2,o2}
\fmf{plain,tension=0.0}{v2,o1}
\end{fmfgraph*}
\end{fmffile}
\end{gathered} \hspace{0.5cm} + \hspace{0.6cm} \begin{gathered}
\begin{fmffile}{Diagrams/1PIEA_DerivGJ3}
\begin{fmfgraph*}(20,20)
\fmfleft{i0,i1,i2,i3}
\fmfright{o0,o1,o2,o3}
\fmfv{decor.shape=cross,decor.angle=45,decor.size=3.5thick,foreground=(0,,0,,1)}{o2}
\fmfv{decor.shape=circle,decor.filled=empty,decor.size=1.5thick,label=$\alpha_{3}$}{o1}
\fmfv{decor.shape=circle,decor.filled=empty,decor.size=1.5thick,label=$\alpha_{1}$}{i1}
\fmfv{decor.shape=circle,decor.filled=empty,decor.size=1.5thick,label=$\alpha_{2}$}{i2}
\fmf{phantom,tension=2.0}{i1,i1bis}
\fmf{phantom,tension=2.0}{i1bis,v1}
\fmf{phantom,tension=2.0}{i2,i2bis}
\fmf{phantom,tension=2.0}{i2bis,v1}
\fmf{dots,tension=0.6,foreground=(0,,0,,1)}{v1,v2}
\fmf{phantom,tension=2.0}{o1bis,o1}
\fmf{phantom,tension=2.0}{v2,o1bis}
\fmf{phantom,tension=2.0}{o2bis,o2}
\fmf{phantom,tension=2.0}{v2,o2bis}
\fmf{plain,tension=0.0}{i2,v1}
\fmf{plain,tension=0.0}{i1,v1}
\fmf{dashes,tension=0.0,foreground=(0,,0,,1)}{v2,o2}
\fmf{plain,tension=0.0}{v2,o1}
\end{fmfgraph*}
\end{fmffile}
\end{gathered} \hspace{0.6cm} \left.\rule{0cm}{0.8cm}\right)\;,
\label{eq:pure2PIEADerivGJ0DON}
\end{equation}
which are the counterparts of~\eqref{eq:pure1PIEADerivW0j0j00DON} and~\eqref{eq:pure1PIEADerivGJ0DON} for the 2PI EA, respectively. After following this recipe, we find an expression of $\boldsymbol{G}_{1}$ involving $13$ distinct diagrams. Let us bear this number in mind and derive the equations to solve for $\vec{J}_{1}$ and $\boldsymbol{K}_{1}$. For that purpose, we Taylor expand the $\vec{\phi}_{n}$ and $\boldsymbol{G}_{n}$ coefficients around $\big(\vec{J},\boldsymbol{K}\big)=\big(\vec{J}_{0},\boldsymbol{K}_{0}\big)$ in the power series~\eqref{eq:pure2PIEAphiExpansion0DON} and~\eqref{eq:pure2PIEAGExpansion0DON}:
\begin{equation}
\begin{split}
\phi_{\alpha_{1}} = & \sum_{n=0}^{\infty} \phi_{n,\alpha_{1}}\Big[\vec{J},\boldsymbol{K}\Big]\hbar^{n} \\
= & \ \phi_{0,\alpha_{1}}\Big[\vec{J},\boldsymbol{K}\Big] + \phi_{1,\alpha_{1}}\Big[\vec{J},\boldsymbol{K}\Big]\hbar + \mathcal{O}\big(\hbar^{2}\big) \\
= & \ \phi_{0,\alpha_{1}}\Big[\vec{J}=\vec{J}_{0},\boldsymbol{K}=\boldsymbol{K}_{0}\Big] + \int_{\alpha_{2}} \left.\frac{\delta \phi_{0,\alpha_{1}}\big[\vec{J},\boldsymbol{K}\big]}{\delta J_{\alpha_{2}}}\right|_{\vec{J}=\vec{J}_{0} \atop \boldsymbol{K}=\boldsymbol{K}_{0}} \left(J_{\alpha_{2}}\Big[\vec{\phi},\boldsymbol{K}\Big]-J_{0,\alpha_{2}}\Big[\vec{\phi},\boldsymbol{K}\Big]\right) \\
& + \int_{\alpha_{2},\alpha_{3}} \left.\frac{\delta \phi_{0,\alpha_{1}}\big[\vec{J},\boldsymbol{K}\big]}{\delta \boldsymbol{K}_{\alpha_{2}\alpha_{3}}}\right|_{\vec{J}=\vec{J}_{0} \atop \boldsymbol{K}=\boldsymbol{K}_{0}} \left(\boldsymbol{K}_{\alpha_{2}\alpha_{3}}\Big[\vec{\phi},\boldsymbol{K}\Big]-\boldsymbol{K}_{0,\alpha_{2}\alpha_{3}}\Big[\vec{\phi},\boldsymbol{K}\Big]\right) \\
& + \phi_{1,\alpha_{1}}\Big[\vec{J}=\vec{J}_{0},\boldsymbol{K}=\boldsymbol{K}_{0}\Big] \hbar + \mathcal{O}\big(\hbar^{2}\big) \\
= & \ \phi_{0,\alpha_{1}}\Big[\vec{J}=\vec{J}_{0},\boldsymbol{K}=\boldsymbol{K}_{0}\Big] \\
& + \hbar \left(\rule{0cm}{0.6cm}\right. \int_{\alpha_{2}} \left.\frac{\delta \phi_{0,\alpha_{1}}\big[\vec{J},\boldsymbol{K}\big]}{\delta J_{\alpha_{2}}}\right|_{\vec{J}=\vec{J}_{0}\atop \boldsymbol{K}=\boldsymbol{K}_{0}} J_{1,\alpha_{2}}\Big[\vec{\phi},\boldsymbol{K}\Big] + \int_{\alpha_{2},\alpha_{3}} \left.\frac{\delta \phi_{0,\alpha_{1}}\big[\vec{J},\boldsymbol{K}\big]}{\delta \boldsymbol{K}_{\alpha_{2}\alpha_{3}}}\right|_{\vec{J}=\vec{J}_{0} \atop \boldsymbol{K}=\boldsymbol{K}_{0}} \boldsymbol{K}_{1,\alpha_{2}\alpha_{3}}\Big[\vec{\phi},\boldsymbol{K}\Big] \\
& \hspace{1.0cm} + \phi_{1,\alpha_{1}}\Big[\vec{J}=\vec{J}_{0},\boldsymbol{K}=\boldsymbol{K}_{0}\Big]\left.\rule{0cm}{0.6cm}\right) \\
& + \mathcal{O}\big(\hbar^{2}\big)\;,
\end{split}
\label{eq:pure2PIEAphiExpansionAroundJ00DON}
\end{equation}
\begin{equation}
\begin{split}
\boldsymbol{G}_{\alpha_{1}\alpha_{2}} = & \sum_{n=0}^{\infty} \boldsymbol{G}_{n,\alpha_{1}\alpha_{2}}\Big[\vec{J},\boldsymbol{K}\Big]\hbar^{n} \\
= & \ \boldsymbol{G}_{0,\alpha_{1}\alpha_{2}}\Big[\vec{J},\boldsymbol{K}\Big] + \boldsymbol{G}_{1,\alpha_{1}\alpha_{2}}\Big[\vec{J},\boldsymbol{K}\Big]\hbar + \mathcal{O}\big(\hbar^{2}\big) \\
= & \ \boldsymbol{G}_{0,\alpha_{1}\alpha_{2}}\Big[\vec{J}=\vec{J}_{0},\boldsymbol{K}=\boldsymbol{K}_{0}\Big] + \int_{\alpha_{3}} \left.\frac{\delta \boldsymbol{G}_{0,\alpha_{1}\alpha_{2}}\big[\vec{J},\boldsymbol{K}\big]}{\delta J_{\alpha_{3}}}\right|_{\vec{J}=\vec{J}_{0} \atop \boldsymbol{K}=\boldsymbol{K}_{0}} \left(J_{\alpha_{3}}\Big[\vec{\phi},\boldsymbol{K}\Big]-J_{0,\alpha_{3}}\Big[\vec{\phi},\boldsymbol{K}\Big]\right) \\
& + \int_{\alpha_{3},\alpha_{4}} \left.\frac{\delta \boldsymbol{G}_{0,\alpha_{1}\alpha_{2}}\big[\vec{J},\boldsymbol{K}\big]}{\delta \boldsymbol{K}_{\alpha_{3}\alpha_{4}}}\right|_{\vec{J}=\vec{J}_{0} \atop \boldsymbol{K}=\boldsymbol{K}_{0}} \left(\boldsymbol{K}_{\alpha_{3}\alpha_{4}}\Big[\vec{\phi},\boldsymbol{K}\Big]-\boldsymbol{K}_{0,\alpha_{3}\alpha_{4}}\Big[\vec{\phi},\boldsymbol{K}\Big]\right) \\
& + \boldsymbol{G}_{1,\alpha_{1}\alpha_{2}}\Big[\vec{J}=\vec{J}_{0},\boldsymbol{K}=\boldsymbol{K}_{0}\Big] \hbar + \mathcal{O}\big(\hbar^{2}\big) \\
= & \ \boldsymbol{G}_{0,\alpha_{1}\alpha_{2}}\Big[\vec{J}=\vec{J}_{0},\boldsymbol{K}=\boldsymbol{K}_{0}\Big] \\
& + \hbar \left(\rule{0cm}{0.6cm}\right. \int_{\alpha_{3}} \left.\frac{\delta \boldsymbol{G}_{0,\alpha_{1}\alpha_{2}}\big[\vec{J},\boldsymbol{K}\big]}{\delta J_{\alpha_{3}}}\right|_{\vec{J}=\vec{J}_{0} \atop \boldsymbol{K}=\boldsymbol{K}_{0}} J_{1,\alpha_{3}}\Big[\vec{\phi},\boldsymbol{K}\Big] \\
& \hspace{1.0cm} + \int_{\alpha_{3},\alpha_{4}} \left.\frac{\delta \boldsymbol{G}_{0,\alpha_{1}\alpha_{2}}\big[\vec{J},\boldsymbol{K}\big]}{\delta \boldsymbol{K}_{\alpha_{3}\alpha_{4}}}\right|_{\vec{J}=\vec{J}_{0} \atop \boldsymbol{K}=\boldsymbol{K}_{0}} \boldsymbol{K}_{1,\alpha_{3}\alpha_{4}}\Big[\vec{\phi},\boldsymbol{K}\Big] + \boldsymbol{G}_{1,\alpha_{1}\alpha_{2}}\Big[\vec{J}=\vec{J}_{0},\boldsymbol{K}=\boldsymbol{K}_{0}\Big]\left.\rule{0cm}{0.6cm}\right) \\
& + \mathcal{O}\big(\hbar^{2}\big)\;.
\end{split}
\label{eq:pure2PIEAGExpansionAroundJ00DON}
\end{equation}
Exploiting the fact that $\vec{\phi}$ and $\boldsymbol{G}$ are independent of $\hbar$, we extract the following equations from~\eqref{eq:pure2PIEAphiExpansionAroundJ00DON} and~\eqref{eq:pure2PIEAGExpansionAroundJ00DON}:
\begin{subequations}
\begin{empheq}[left=\empheqlbrace]{align}
& \hspace{0.1cm} \scalebox{0.92}{${\displaystyle 0 = \int_{\alpha_{2}} \left.\frac{\delta \phi_{0,\alpha_{1}}\big[\vec{J},\boldsymbol{K}\big]}{\delta J_{\alpha_{2}}}\right|_{\vec{J}=\vec{J}_{0}\atop \boldsymbol{K}=\boldsymbol{K}_{0}} J_{1,\alpha_{2}}\Big[\vec{\phi},\boldsymbol{K}\Big] + \int_{\alpha_{2},\alpha_{3}} \left.\frac{\delta \phi_{0,\alpha_{1}}\big[\vec{J},\boldsymbol{K}\big]}{\delta \boldsymbol{K}_{\alpha_{2}\alpha_{3}}}\right|_{\vec{J}=\vec{J}_{0} \atop \boldsymbol{K}=\boldsymbol{K}_{0}} \boldsymbol{K}_{1,\alpha_{2}\alpha_{3}}\Big[\vec{\phi},\boldsymbol{K}\Big] }$} \nonumber \\
& \hspace{0.1cm} \scalebox{0.92}{${\displaystyle \hspace{0.72cm} + \phi_{1,\alpha_{1}}\Big[\vec{J}=\vec{J}_{0},\boldsymbol{K}=\boldsymbol{K}_{0}\Big] \;. }$} \label{eq:pure2PIEATowerEquationJ1K1n10DON}\\
\nonumber \\
& \hspace{0.1cm} \scalebox{0.92}{${\displaystyle 0 = \int_{\alpha_{3}} \left.\frac{\delta \boldsymbol{G}_{0,\alpha_{1}\alpha_{2}}\big[\vec{J},\boldsymbol{K}\big]}{\delta J_{\alpha_{3}}}\right|_{\vec{J}=\vec{J}_{0} \atop \boldsymbol{K}=\boldsymbol{K}_{0}} J_{1,\alpha_{3}}\Big[\vec{\phi},\boldsymbol{K}\Big] + \int_{\alpha_{3},\alpha_{4}} \left.\frac{\delta \boldsymbol{G}_{0,\alpha_{1}\alpha_{2}}\big[\vec{J},\boldsymbol{K}\big]}{\delta \boldsymbol{K}_{\alpha_{3}\alpha_{4}}}\right|_{\vec{J}=\vec{J}_{0} \atop \boldsymbol{K}=\boldsymbol{K}_{0}} \boldsymbol{K}_{1,\alpha_{3}\alpha_{4}}\Big[\vec{\phi},\boldsymbol{K}\Big] }$} \nonumber \\
& \hspace{0.1cm} \scalebox{0.92}{${\displaystyle \hspace{0.72cm} + \boldsymbol{G}_{1,\alpha_{1}\alpha_{2}}\Big[\vec{J}=\vec{J}_{0},\boldsymbol{K}=\boldsymbol{K}_{0}\Big] \;.}$} \label{eq:pure2PIEATowerEquationJ1K1n20DON}
\end{empheq}
\end{subequations}
It is clear at this stage that, since $\boldsymbol{G}_{1}$ brings up a dozen diagrams in the latter equation system, both $\vec{J}_{1}$ and $\boldsymbol{K}_{1}$ will be given by more than $10$ diagrams, regardless of how we isolate them in~\eqref{eq:pure2PIEATowerEquationJ1K1n10DON} and~\eqref{eq:pure2PIEATowerEquationJ1K1n20DON}. Therefore, some terms in the RHS of~\eqref{eq:pure2PIEAIMGamma20DON} will yield more than a hundred terms to the expression of $\Gamma^{(\mathrm{2PI})}_{2}$, which renders the IM already very cumbersome to carry out at the first non-trivial order for the 2PI EA. However, we point out that the formalism developed so far already gives us the essential ingredients to determine $n$PI EAs with (mixed and collective representations) and without HST (original representation). In particular, the separation between the original field and auxiliary field sectors with Kronecker deltas carried out in section~\ref{sec:collective1PIEAannIM} can be used for any mixed or collective $n$P(P)I EAs. Luckily, in the present case, we know that all 2PR diagrams should be canceled out from the expressions of the $\Gamma^{(\mathrm{2PI})}_{n}$ coefficients (with $n \geq 2$), notably from the work of Vasil'ev and collaborators~\cite{vas72}. We therefore exploit this property to express the full original 2PI EA up to order $\mathcal{O}\big(\hbar^{2}\big)$ directly from~\eqref{eq:pure2PIEAIMGamma00DON} and~\eqref{eq:pure2PIEAIMGamma1step20DON} alongside with the LE result~\eqref{eq:WKjLoopExpansionStep3}:
\begin{equation}
\begin{split}
\scalebox{0.98}{${\displaystyle \Gamma^{(\mathrm{2PI})}\Big[\vec{\phi},\boldsymbol{G}\Big] = }$} & \ \scalebox{0.98}{${\displaystyle S\Big[\vec{\phi}\Big] -\frac{\hbar}{2}\mathrm{STr}\left[\ln\big(\boldsymbol{G}\big)\right] + \frac{\hbar}{2}\mathrm{STr}\left[\boldsymbol{G}^{-1}_{\phi}\Big[\vec{\phi}\Big]\boldsymbol{G}-\mathbb{I}\right] }$} \\
& \scalebox{0.98}{${\displaystyle + \hbar^{2} \left(\rule{0cm}{1.2cm}\right. \frac{1}{24} \hspace{0.08cm} \begin{gathered}
\begin{fmffile}{Diagrams/1PIEA_Hartree}
\begin{fmfgraph}(30,20)
\fmfleft{i}
\fmfright{o}
\fmf{phantom,tension=10}{i,i1}
\fmf{phantom,tension=10}{o,o1}
\fmf{plain,left,tension=0.5,foreground=(1,,0,,0)}{i1,v1,i1}
\fmf{plain,right,tension=0.5,foreground=(1,,0,,0)}{o1,v2,o1}
\fmf{zigzag,foreground=(0,,0,,1)}{v1,v2}
\end{fmfgraph}
\end{fmffile}
\end{gathered}
+\frac{1}{12}\begin{gathered}
\begin{fmffile}{Diagrams/1PIEA_Fock}
\begin{fmfgraph}(15,15)
\fmfleft{i}
\fmfright{o}
\fmf{phantom,tension=11}{i,v1}
\fmf{phantom,tension=11}{v2,o}
\fmf{plain,left,tension=0.4,foreground=(1,,0,,0)}{v1,v2,v1}
\fmf{zigzag,foreground=(0,,0,,1)}{v1,v2}
\end{fmfgraph}
\end{fmffile}
\end{gathered}
- \frac{1}{18} \begin{gathered}
\begin{fmffile}{Diagrams/1PIEA_Diag1}
\begin{fmfgraph}(27,15)
\fmfleft{i}
\fmfright{o}
\fmftop{vUp}
\fmfbottom{vDown}
\fmfv{decor.shape=cross,decor.size=3.5thick,foreground=(1,,0,,0)}{v1}
\fmfv{decor.shape=cross,decor.size=3.5thick,foreground=(1,,0,,0)}{v2}
\fmf{phantom,tension=10}{i,i1}
\fmf{phantom,tension=10}{o,o1}
\fmf{phantom,tension=2.2}{vUp,v5}
\fmf{phantom,tension=2.2}{vDown,v6}
\fmf{phantom,tension=0.5}{v3,v4}
\fmf{phantom,tension=10.0}{i1,v1}
\fmf{phantom,tension=10.0}{o1,v2}
\fmf{dashes,tension=2.0,foreground=(0,,0,,1),foreground=(1,,0,,0)}{v1,v3}
\fmf{dots,left=0.4,tension=0.5,foreground=(0,,0,,1)}{v3,v5}
\fmf{plain,left=0.4,tension=0.5,foreground=(1,,0,,0)}{v5,v4}
\fmf{plain,right=0.4,tension=0.5,foreground=(1,,0,,0)}{v3,v6}
\fmf{dots,right=0.4,tension=0.5,foreground=(0,,0,,1)}{v6,v4}
\fmf{dashes,tension=2.0,foreground=(0,,0,,1),foreground=(1,,0,,0)}{v4,v2}
\fmf{plain,tension=0,foreground=(1,,0,,0)}{v5,v6}
\end{fmfgraph}
\end{fmffile}
\end{gathered} - \frac{1}{36} \hspace{-0.15cm} \begin{gathered}
\begin{fmffile}{Diagrams/1PIEA_Diag2}
\begin{fmfgraph}(25,20)
\fmfleft{i}
\fmfright{o}
\fmftop{vUp}
\fmfbottom{vDown}
\fmfv{decor.shape=cross,decor.angle=45,decor.size=3.5thick,foreground=(1,,0,,0)}{vUpbis}
\fmfv{decor.shape=cross,decor.angle=45,decor.size=3.5thick,foreground=(1,,0,,0)}{vDownbis}
\fmf{phantom,tension=0.8}{vUp,vUpbis}
\fmf{phantom,tension=0.8}{vDown,vDownbis}
\fmf{dashes,tension=0.5,foreground=(0,,0,,1),foreground=(1,,0,,0)}{v3,vUpbis}
\fmf{phantom,tension=0.5}{v4,vUpbis}
\fmf{phantom,tension=0.5}{v3,vDownbis}
\fmf{dashes,tension=0.5,foreground=(0,,0,,1),foreground=(1,,0,,0)}{v4,vDownbis}
\fmf{phantom,tension=11}{i,v1}
\fmf{phantom,tension=11}{v2,o}
\fmf{plain,left,tension=0.5,foreground=(1,,0,,0)}{v1,v2,v1}
\fmf{dots,tension=1.7,foreground=(0,,0,,1)}{v1,v3}
\fmf{plain,foreground=(1,,0,,0)}{v3,v4}
\fmf{dots,tension=1.7,foreground=(0,,0,,1)}{v4,v2}
\end{fmfgraph}
\end{fmffile}
\end{gathered} \hspace{-0.2cm} \left.\rule{0cm}{1.2cm}\right) }$} \\
& \scalebox{0.98}{${\displaystyle + \mathcal{O}\big(\hbar^{3}\big)\;. }$}
\end{split}
\label{eq:2PIEAfinalexpressionAppendix}
\end{equation}
Since all 2-particle-reducible (2PR) diagrams are also 1PR at order $\mathcal{O}\big(\hbar^{2}\big)$ of the original LE series of the Schwinger functional (i.e.~\eqref{eq:WKjLoopExpansionStep3}), the diagrams contributing to $\Gamma^{(\mathrm{2PI})}$ are identical to those of $\Gamma^{(\mathrm{1PI})}$ at this order (although the plain red lines do not represent the same propagator, i.e.~\eqref{eq:DefinitionG1PIEAhbarExpansionAppendix} for the 1PI EA and~\eqref{eq:FeynRules2PIEAPropagatorSourceJ0} for the 2PI EA). This is no longer the case at the next order, i.e. at order $\mathcal{O}\big(\hbar^{3}\big)$.

\subsection{\label{sec:mixed2PIEAannIM}Mixed effective action}
\paragraph{$\hbar$-expansion for the full mixed 2PI EA:}

As before with the IM, we start by presenting this formalism with some definitions and relevant Feynman rules. To begin with, we have the following power series:
\begin{subequations}
\begin{empheq}[left=\empheqlbrace]{align}
& \hspace{0.1cm} \Gamma_{\mathrm{mix}}^{(\mathrm{2PI})}\big[\Phi,\mathcal{G};\hbar\big]=\sum_{n=0}^{\infty} \Gamma^{(\mathrm{2PI})}_{n}\big[\Phi,\mathcal{G}\big]\hbar^{n}\;, \label{eq:mixed2PIEAGammaExpansion0DON}\\
\nonumber \\
& \hspace{0.1cm} W_{\mathrm{mix}}\big[\mathcal{J},\mathcal{K};\hbar\big]=\sum_{n=0}^{\infty} W_{\mathrm{mix},n}\big[\mathcal{J},\mathcal{K}\big]\hbar^{n}\;, \label{eq:mixed2PIEAWExpansion0DON} \\
\nonumber \\
& \hspace{0.1cm} \mathcal{J}\big[\Phi,\mathcal{G};\hbar\big]=\sum_{n=0}^{\infty} \mathcal{J}_{n}\big[\Phi,\mathcal{G}\big]\hbar^{n}\;, \label{eq:mixed2PIEAJExpansion0DON}\\
\nonumber \\
& \hspace{0.1cm} \mathcal{K}\big[\Phi,\mathcal{G};\hbar\big]=\sum_{n=0}^{\infty} \mathcal{K}_{n}\big[\Phi,\mathcal{G}\big]\hbar^{n}\;, \label{eq:mixed2PIEAKExpansion0DON}\\
\nonumber \\
& \hspace{0.1cm} \Phi=\sum_{n=0}^{\infty} \Phi_{n}\big[\mathcal{J},\mathcal{K}\big]\hbar^{n}\;, \label{eq:mixed2PIEAphiExpansion0DON} \\
\nonumber \\
& \hspace{0.1cm} \mathcal{G}=\sum_{n=0}^{\infty} \mathcal{G}_{n}\big[\mathcal{J},\mathcal{K}\big]\hbar^{n}\;, \label{eq:mixed2PIEAGExpansion0DON}
\end{empheq}
\end{subequations}
where
\begin{equation}
\begin{split}
\Gamma_{\mathrm{mix}}^{(\mathrm{2PI})}\big[\Phi,\mathcal{G}\big] \equiv & -W_{\mathrm{mix}}\big[\mathcal{J},\mathcal{K}\big] + \int_{\beta} \mathcal{J}_{\beta}\big[\Phi,\mathcal{G}\big] \frac{\delta W_{\mathrm{mix}}\big[\mathcal{J},\mathcal{K}\big]}{\delta \mathcal{J}_{\beta}} + \int_{\beta_{1},\beta_{2}} \mathcal{K}_{\beta_{1}\beta_{2}}\big[\Phi,\mathcal{G}\big] \frac{\delta W_{\mathrm{mix}}\big[\mathcal{J},\mathcal{K}\big]}{\delta\mathcal{K}_{\beta_{1}\beta_{2}}} \\
= & -W_{\mathrm{mix}}\big[\mathcal{J},\mathcal{K}\big] + \int_{\beta} \mathcal{J}_{\beta}\big[\Phi,\mathcal{G}\big] \Phi_{\beta} + \frac{1}{2} \int_{\beta_{1},\beta_{2}} \Phi_{\beta_{1}} \mathcal{K}_{\beta_{1}\beta_{2}}\big[\Phi,\mathcal{G}\big] \Phi_{\beta_{2}} \\
& + \frac{\hbar}{2}\int_{\beta_{1},\beta_{2}} \mathcal{K}_{\beta_{1}\beta_{2}}\big[\Phi,\mathcal{G}\big] \mathcal{G}_{\beta_{1}\beta_{2}}\;,
\end{split}
\label{eq:mixed2PIEAdefinition0DON}
\end{equation}
and
\begin{equation}
\Phi_{\beta}=\frac{\delta W_{\mathrm{mix}}\big[\mathcal{J},\mathcal{K}\big]}{\delta \mathcal{J}_{\beta}}\;,
\label{eq:mixed2PIEAdefinitionbis0DON}
\end{equation}
\begin{equation}
\mathcal{G}_{\beta_{1}\beta_{2}} = \frac{\delta^{2} W_{\mathrm{mix}}\big[\mathcal{J},\mathcal{G}\big]}{\delta \mathcal{J}_{\beta_{1}} \delta \mathcal{J}_{\beta_{2}}} = \frac{2}{\hbar} \frac{\delta W_{\mathrm{mix}}\big[\mathcal{J},\mathcal{K}\big]}{\delta \mathcal{K}_{\beta_{1}\beta_{2}}} - \frac{1}{\hbar} \Phi_{\beta_{1}} \Phi_{\beta_{2}}\;.
\label{eq:mixed2PIEAdefinitionbis20DON}
\end{equation}
The $W_{\mathrm{mix},n}$ coefficients are now given by the mixed LE result~\eqref{eq:WmixedKjLoopExpansionStep2} for $n=0,1~\mathrm{and}~2$. On the one hand, at arbitrary external sources $\mathcal{J}$ and $\mathcal{K}$, the 1-point correlation function and propagator to consider for the mixed 2PI EA formalism are:
\begin{subequations}
\begin{empheq}[left=\empheqlbrace]{align}
& \hspace{0.1cm} \Psi_{\mathrm{cl},\beta}\big[\mathcal{J},\mathcal{K}\big] = \Phi_{0,\beta}\big[\mathcal{J},\mathcal{K}\big] = \frac{\delta W_{\mathrm{mix},0}\big[\mathcal{J},\mathcal{K}\big]}{\delta \mathcal{J}_{\beta}} \\
& \hspace{0.1cm} \hspace{2.045cm} = \left(1-\delta_{b \hspace{0.04cm} N+1}\right) \varphi_{\mathrm{cl},\alpha}\big[\mathcal{J},\mathcal{K}\big] + \delta_{b \hspace{0.04cm} N+1} \sigma_{\mathrm{cl},x}\big[\mathcal{J},\mathcal{K}\big] \;, \label{eq:mixed2PIEAphicl0DON}\\
\nonumber \\
& \hspace{0.1cm} \mathcal{G}_{\Psi_{\mathrm{cl}};\mathcal{J}\mathcal{K},\beta_{1}\beta_{2}}\big[\mathcal{J},\mathcal{K}\big] = \mathcal{G}_{0,\beta_{1}\beta_{2}}\big[\mathcal{J},\mathcal{K}\big] = \frac{\delta^{2} W_{\mathrm{mix},0}\big[\mathcal{J},\mathcal{K}\big]}{\delta \mathcal{J}_{\beta_{1}} \delta \mathcal{J}_{\beta_{2}}} \nonumber \\
& \hspace{0.1cm} \hspace{3.21cm} = \left(\left.\frac{\delta^{2}S_{\mathrm{mix}}\big[\widetilde{\Psi}\big]}{\delta\widetilde{\Psi}\delta\widetilde{\Psi}}\right|_{\widetilde{\Psi}=\Psi_{\mathrm{cl}}} - \mathcal{K}\big[\Phi,\mathcal{G}\big]\right)^{-1}_{\beta_{1}\beta_{2}} \nonumber \\
& \hspace{0.1cm} \hspace{3.21cm} = \left(1-\delta_{b_{1} N+1}\right)\left(1-\delta_{b_{2} N+1}\right) \boldsymbol{G}_{\sigma_{\mathrm{cl}};\mathcal{J}\mathcal{K},\alpha_{1}\alpha_{2}}\big[\mathcal{J},\mathcal{K}\big] \nonumber \\
& \hspace{0.1cm} \hspace{3.61cm} + \delta_{b_{1} N+1}\delta_{b_{2} N+1} D_{\sigma_{\mathrm{cl}};\mathcal{J}\mathcal{K},x_{1}x_{2}}\big[\mathcal{J},\mathcal{K}\big] \nonumber \\
& \hspace{0.1cm} \hspace{3.61cm} + \left(1-\delta_{b_{1} N+1}\right)\delta_{b_{2} N+1} F_{\varphi_{\mathrm{cl}};\mathcal{J}\mathcal{K},\alpha_{1}x_{2}}\big[\mathcal{J},\mathcal{K}\big] \nonumber \\
& \hspace{0.1cm} \hspace{3.61cm} + \delta_{b_{1} N+1}\left(1-\delta_{b_{2} N+1}\right) F_{\varphi_{\mathrm{cl}};\mathcal{J}\mathcal{K},x_{1}\alpha_{2}}\big[\mathcal{J},\mathcal{K}\big]\;, \label{eq:mixed2PIEAGJK0DON}
\end{empheq}
\end{subequations}
and, in matrix form:
\begin{subequations}
\begin{empheq}[left=\empheqlbrace]{align}
& \hspace{0.1cm} \Psi_{\mathrm{cl}}\big[\mathcal{J},\mathcal{K}\big]=\begin{pmatrix}
\vec{\varphi}_{\mathrm{cl}}\big[\mathcal{J},\mathcal{K}\big] & \sigma_{\mathrm{cl}}\big[\mathcal{J},\mathcal{K}\big]
\end{pmatrix}^{\mathrm{T}}\;. \\
\nonumber \\
& \hspace{0.1cm} \mathcal{G}_{\Psi_{\mathrm{cl}};\mathcal{J}\mathcal{K}}\big[\mathcal{J},\mathcal{K}\big] = \begin{pmatrix}
\boldsymbol{G}_{\sigma_{\mathrm{cl}};\mathcal{J}\mathcal{K}}\big[\mathcal{J},\mathcal{K}\big] & \vec{F}_{\varphi_{\mathrm{cl}};\mathcal{J}\mathcal{K}}\big[\mathcal{J},\mathcal{K}\big] \\
\vec{F}^{\mathrm{T}}_{\varphi_{\mathrm{cl}};\mathcal{J}\mathcal{K}}\big[\mathcal{J},\mathcal{K}\big] & D_{\sigma_{\mathrm{cl}};\mathcal{J}\mathcal{K}}\big[\mathcal{J},\mathcal{K}\big] \label{eq:mixed2PIEAGJKmatrix0DON}
\end{pmatrix}\;.
\end{empheq}
\end{subequations}
On the other hand, we have at $\big(\mathcal{J},\mathcal{K}\big)=\big(\mathcal{J}_{0},\mathcal{K}_{0}\big)$:
\begin{subequations}
\begin{empheq}[left=\empheqlbrace]{align}
& \hspace{0.1cm} \Phi_{\beta} = \Psi_{\mathrm{cl},\beta}\big[\mathcal{J}=\mathcal{J}_{0},\mathcal{K}=\mathcal{K}_{0}\big] = \Phi_{0,\beta}\big[\mathcal{J}=\mathcal{J}_{0},\mathcal{K}=\mathcal{K}_{0}\big] \\
& \hspace{0.1cm} \hspace{0.5cm} = \left.\frac{\delta W_{\mathrm{mix},0}\big[\mathcal{J},\mathcal{K}\big]}{\delta \mathcal{J}_{\beta}}\right|_{\mathcal{J}=\mathcal{J}_{0} \atop \mathcal{K}=\mathcal{K}_{0}} \nonumber \\
& \hspace{0.1cm} \hspace{0.5cm} = \left(1-\delta_{b \hspace{0.04cm} N+1}\right) \phi_{\alpha} + \delta_{b \hspace{0.04cm} N+1} \eta_{x} \;, \label{eq:mixed2PIEAphi0DON}\\
\nonumber \\
& \hspace{0.1cm} \mathcal{G}_{\beta_{1}\beta_{2}} = \mathcal{G}_{\Psi_{\mathrm{cl}};\mathcal{J}\mathcal{K},\beta_{1}\beta_{2}}\big[\mathcal{J}=\mathcal{J}_{0},\mathcal{K}=\mathcal{K}_{0}\big] = \mathcal{G}_{0,\beta_{1}\beta_{2}}\big[\mathcal{J}=\mathcal{J}_{0},\mathcal{K}=\mathcal{K}_{0}\big] \nonumber \\
& \hspace{0.1cm} \hspace{0.9cm} = \left.\frac{\delta^{2} W_{\mathrm{mix},0}\big[\mathcal{J},\mathcal{K}\big]}{\delta \mathcal{J}_{\beta_{1}} \delta \mathcal{J}_{\beta_{2}}}\right|_{\mathcal{J}=\mathcal{J}_{0} \atop \mathcal{K}=\mathcal{K}_{0}} \nonumber \\
& \hspace{0.1cm} \hspace{0.9cm} = \left(\left.\frac{\delta^{2}S_{\mathrm{mix}}\big[\widetilde{\Psi}\big]}{\delta\widetilde{\Psi}\delta\widetilde{\Psi}}\right|_{\widetilde{\Psi}=\Phi} - \mathcal{K}_{0}\big[\Phi,\mathcal{G}\big]\right)^{-1}_{\beta_{1}\beta_{2}} \nonumber \\
& \hspace{0.1cm} \hspace{0.9cm} = \left(1-\delta_{b_{1} N+1}\right)\left(1-\delta_{b_{2} N+1}\right) \boldsymbol{G}_{\alpha_{1}\alpha_{2}} + \delta_{b_{1} N+1}\delta_{b_{2} N+1} D_{x_{1}x_{2}} \nonumber \\
& \hspace{0.1cm} \hspace{1.3cm} + \left(1-\delta_{b_{1} N+1}\right)\delta_{b_{2} N+1} F_{\alpha_{1}x_{2}} + \delta_{b_{1} N+1}\left(1-\delta_{b_{2} N+1}\right) F_{x_{1}\alpha_{2}}\;, \label{eq:mixed2PIEAGJ0K00DON}
\end{empheq}
\end{subequations}
or, in matrix form:
\begin{subequations}
\begin{empheq}[left=\empheqlbrace]{align}
& \hspace{0.1cm} \Phi = \begin{pmatrix}
\vec{\phi} & \eta
\end{pmatrix}^{\mathrm{T}}\;. \\
\nonumber \\
& \hspace{0.1cm} \mathcal{G} = \begin{pmatrix}
\boldsymbol{G} & \vec{F} \\
\vec{F}^{\mathrm{T}} & D
\end{pmatrix} \label{eq:mixed2PIEAGJ0K0matrix0DON}\;.
\end{empheq}
\end{subequations}
The independence of (each component of) $\mathcal{G}$ and $\Phi$ with respect to $\hbar$ follows from the Legendre transform in~\eqref{eq:mixed2PIEAdefinition0DON}, as for the 1PI and 2PI EAs treated previously. We then recall the Feynman rules (already given by~\eqref{eq:mixed2PIEAvertex} to~\eqref{eq:mixed2PIEAFeynRuleF}) that we use to express the mixed 2PI EA:
\begin{subequations}
\begin{align}
\left.
\begin{array}{ll}
\begin{gathered}
\begin{fmffile}{Diagrams/mixed2PIEA_FeynRuleVertexbis1_AppendixFull2PI}
\begin{fmfgraph*}(4,4)
\fmfleft{i0,i1,i2,i3}
\fmfright{o0,o1,o2,o3}
\fmfv{label=$x$,label.angle=90,label.dist=4}{v1}
\fmfbottom{v2}
\fmf{phantom}{i1,v1}
\fmf{plain,foreground=(1,,0,,0)}{i2,v1}
\fmf{phantom}{v1,o1}
\fmf{plain,foreground=(1,,0,,0)}{v1,o2}
\fmf{wiggly,tension=0.6,foreground=(1,,0,,0)}{v1,v2}
\fmfv{decor.shape=circle,decor.size=2.0thick,foreground=(0,,0,,1)}{v1}
\fmflabel{$a_{1}$}{i2}
\fmflabel{$a_{2}$}{o2}
\end{fmfgraph*}
\end{fmffile}
\end{gathered} \\
\\
\begin{gathered}
\begin{fmffile}{Diagrams/mixed2PIEA_FeynRuleVertexbis2_AppendixFull2PI}
\begin{fmfgraph*}(4,4)
\fmfleft{i0,i1,i2,i3}
\fmfright{o0,o1,o2,o3}
\fmfv{label=$x$,label.angle=90,label.dist=4}{v1}
\fmfbottom{v2}
\fmf{phantom}{i1,v1}
\fmf{plain,foreground=(1,,0,,0)}{i2,v1}
\fmf{phantom}{v1,o1}
\fmf{plain,foreground=(1,,0,,0)}{v1,o2}
\fmf{dots,tension=0.6,foreground=(1,,0,,0)}{v1,v2}
\fmfv{decor.shape=circle,decor.size=2.0thick,foreground=(0,,0,,1)}{v1}
\fmflabel{$a_{1}$}{i2}
\fmflabel{$a_{2}$}{o2}
\end{fmfgraph*}
\end{fmffile}
\end{gathered} \\
\\
\begin{gathered}
\begin{fmffile}{Diagrams/mixed2PIEA_FeynRuleVertexbis3_AppendixFull2PI}
\begin{fmfgraph*}(4,4)
\fmfleft{i0,i1,i2,i3}
\fmfright{o0,o1,o2,o3}
\fmfv{label=$x$,label.angle=90,label.dist=4}{v1}
\fmfbottom{v2}
\fmf{phantom}{i1,v1}
\fmf{plain,foreground=(1,,0,,0)}{i2,v1}
\fmf{phantom}{v1,o1}
\fmf{dots,foreground=(1,,0,,0)}{v1,o2}
\fmf{dots,tension=0.6,foreground=(1,,0,,0)}{v1,v2}
\fmfv{decor.shape=circle,decor.size=2.0thick,foreground=(0,,0,,1)}{v1}
\fmflabel{$a_{1}$}{i2}
\fmflabel{$a_{2}$}{o2}
\end{fmfgraph*}
\end{fmffile}
\end{gathered}
\end{array}
\quad \right\rbrace &\rightarrow \sqrt{\lambda} \ \delta_{a_{1}a_{2}} \;, 
\label{eq:Mixed2PIEAFeynRulesvertex} \\
\begin{gathered}
\begin{fmffile}{Diagrams/mixed2PIEA_FeynRuleGbis_AppendixFull2PI}
\begin{fmfgraph*}(20,20)
\fmfleft{i0,i1,i2,i3}
\fmfright{o0,o1,o2,o3}
\fmflabel{$\alpha_{1}$}{v1}
\fmflabel{$\alpha_{2}$}{v2}
\fmf{phantom}{i1,v1}
\fmf{phantom}{i2,v1}
\fmf{plain,tension=0.6,foreground=(1,,0,,0)}{v1,v2}
\fmf{phantom}{v2,o1}
\fmf{phantom}{v2,o2}
\end{fmfgraph*}
\end{fmffile}
\end{gathered} \quad &\rightarrow \boldsymbol{G}_{\alpha_{1}\alpha_{2}} \;,
\label{eq:Mixed2PIEAFeynRulesHSG} \\
\begin{gathered}
\begin{fmffile}{Diagrams/mixed2PIEA_FeynRuleDbis_AppendixFull2PI}
\begin{fmfgraph*}(20,20)
\fmfleft{i0,i1,i2,i3}
\fmfright{o0,o1,o2,o3}
\fmfv{label=$x_{1}$}{v1}
\fmfv{label=$x_{2}$}{v2}
\fmf{phantom}{i1,v1}
\fmf{phantom}{i2,v1}
\fmf{wiggly,tension=0.6,foreground=(1,,0,,0)}{v1,v2}
\fmf{phantom}{v2,o1}
\fmf{phantom}{v2,o2}
\end{fmfgraph*}
\end{fmffile}
\end{gathered} \quad &\rightarrow D_{x_{1}x_{2}} \;,
\label{eq:Mixed2PIEAFeynRulesHSD} \\
\begin{gathered}
\begin{fmffile}{Diagrams/mixed2PIEA_FeynRuleFbis_AppendixFull2PI}
\begin{fmfgraph*}(20,20)
\fmfleft{i0,i1,i2,i3}
\fmfright{o0,o1,o2,o3}
\fmflabel{$\alpha_{1}$}{v1}
\fmfv{label=$x_{2}$}{v2}
\fmf{phantom}{i1,v1}
\fmf{phantom}{i2,v1}
\fmf{dashes,tension=0.6,foreground=(1,,0,,0)}{v1,v2}
\fmf{phantom}{v2,o1}
\fmf{phantom}{v2,o2}
\end{fmfgraph*}
\end{fmffile}
\end{gathered} \quad &\rightarrow F_{\alpha_{1} x_{2}} = F_{x_{2} \alpha_{1}} \;.
\label{eq:Mixed2PIEAFeynRulesHSF}
\end{align}
\end{subequations}
From analogous derivations as those leading to~\eqref{eq:pure2PIEAIMGamma00DON} and~\eqref{eq:pure2PIEAIMGamma10DON} (combined with~\eqref{eq:pure2PIEAIMGamma1step20DON}) for the original 2PI EA, we prove the relations:
\begin{equation}
\Gamma^{(\mathrm{2PI})}_{\mathrm{mix},0}\big[\Phi,\mathcal{G}\big] = -W_{\mathrm{mix},0}\big[\mathcal{J}=\mathcal{J}_{0},\mathcal{K}=\mathcal{K}_{0}\big] + \int_{\beta} \mathcal{J}_{0,\beta}\big[\Phi,\mathcal{G}\big] \Phi_{\beta} + \frac{1}{2}\int_{\beta_{1},\beta_{2}}\Phi_{\beta_{1}} \mathcal{K}_{0,\beta_{1}\beta_{2}}\big[\Phi,\mathcal{G}\big] \Phi_{\beta_{2}}\;,
\label{eq:mixed2PIEAIMGamma00DON}
\end{equation}
\begin{equation}
\begin{split}
\Gamma^{(\mathrm{2PI})}_{\mathrm{mix},1}\big[\Phi,\mathcal{G}\big] = & -W_{\mathrm{mix},1}\big[\mathcal{J}=\mathcal{J}_{0},\mathcal{K}=\mathcal{K}_{0}\big]+\frac{1}{2}\int_{\beta_{1},\beta_{2}} \mathcal{K}_{0,\beta_{1}\beta_{2}}\big[\Phi,\mathcal{G}\big] \mathcal{G}_{\beta_{1}\beta_{2}} \\
= & -W_{\mathrm{mix},1}\big[\mathcal{J}=\mathcal{J}_{0},\mathcal{K}=\mathcal{K}_{0}\big]+\frac{1}{2}\mathcal{ST}r\left[\mathcal{G}^{-1}_{\Phi}[\Phi]\mathcal{G}-\mathfrak{I}\right]\;,
\label{eq:mixed2PIEAIMGamma10DON}
\end{split}
\end{equation}
where $\mathfrak{I}=\delta_{\beta_{1}\beta_{2}}=\delta_{b_{1}b_{2}}\delta_{x_{1}x_{2}}$, and:
\begin{equation}
\begin{split}
\mathcal{G}^{-1}_{\Phi,\beta_{1}\beta_{2}}[\Phi] \equiv \left.\frac{\delta^{2} S_{\mathrm{mix}}\big[\widetilde{\Psi}\big]}{\delta\widetilde{\Psi}_{\beta_{1}}\delta\widetilde{\Psi}_{\beta_{2}}}\right|_{\widetilde{\Psi}=\Phi} = & \ \left(-\nabla_{x_{1}}^{2}+m^{2}+i\sqrt{\frac{\lambda}{3}}\eta_{x_{1}}\right)\delta_{\alpha_{1}\alpha_{2}}\left(1-\delta_{b_{1} N+1}\right)\left(1-\delta_{b_{2} N+1}\right) \\
& + \delta_{x_{1}x_{2}}\delta_{b_{1} N+1}\delta_{b_{2} N+1} \\
& + i\sqrt{\frac{\lambda}{3}}\phi_{\alpha_{1}}\delta_{x_{1}x_{2}}\left(1-\delta_{b_{1} N+1}\right)\delta_{b_{2} N+1} \\
& + i\sqrt{\frac{\lambda}{3}}\phi_{\alpha_{2}}\delta_{x_{1}x_{2}}\delta_{b_{1} N+1}\left(1-\delta_{b_{2} N+1}\right)\;,
\end{split}
\label{eq:mixed2PIEADefinitionG00DON}
\end{equation}
or, in matrix form:
\begin{equation}
\mathcal{G}^{-1}_{\Phi,x_{1}x_{2}}[\Phi] = \begin{pmatrix}
\left(-\nabla_{x_{1}}^{2}+m^{2}+i\sqrt{\frac{\lambda}{3}}\eta_{x_{1}}\right)\mathbb{I}_{N} & i\sqrt{\frac{\lambda}{3}}\vec{\phi}_{x_{1}} \\
i\sqrt{\frac{\lambda}{3}}\vec{\phi}_{x_{1}}^{\mathrm{T}} & 1
\end{pmatrix} \delta_{x_{1}x_{2}}\;.
\end{equation}
Finally, we construct a diagrammatic expression for $\Gamma^{(\mathrm{2PI})}_{\mathrm{mix}}\big[\Phi,\mathcal{G}\big]$ up to the third non-trivial order, i.e. up to order $\mathcal{O}\big(\hbar^{4}\big)$. To that end, we first determine all the 2PI diagrams contributing to this EA (or to the LE of $W_{\mathrm{mix}}\big[\mathcal{J},\mathcal{K}\big]$) up to order $\mathcal{O}\big(\hbar^{3}\big)$. At order $\mathcal{O}\big(\hbar^{4}\big)$, the number of relevant 2PI diagrams is already considerable (roughly a hundred). However, since we always find in our zero-dimensional applications (with or without Pad\'{e}-Borel resummation) $\vec{\overline{\phi}}=\vec{\overline{F}}=\vec{0}$ as optimal solution of our gap equations at the first two non-trivial orders (i.e. up to order $\mathcal{O}(\hbar^{2})$ and $\mathcal{O}(\hbar^{3})$) both in the unbroken- and broken-symmetry regimes, we can reasonably expect that the $O(N)$ symmetry is also fully conserved at the third non-trivial order as well (i.e. up to order $\mathcal{O}(\hbar^{4})$) and therefore safely ignore $\vec{F}$-dependent diagrams. In this way, using in addition~\eqref{eq:mixed2PIEAIMGamma00DON} and~\eqref{eq:mixed2PIEAIMGamma10DON} combined with the LE result~\eqref{eq:WmixedKjLoopExpansionStep2}, we obtain:
\begin{equation}
\begin{split}
\Gamma^{(\mathrm{2PI})}_{\mathrm{mix}}\big[\Phi,\mathcal{G}\big] = & \ S_{\mathrm{mix}}[\Phi] -\frac{\hbar}{2}\mathcal{ST}r\left[\ln\big(\mathcal{G}\big)\right] + \frac{\hbar}{2}\mathcal{ST}r\left[\mathcal{G}^{-1}_{\Phi}[\Phi]\mathcal{G}-\mathfrak{I}\right] \\
& + \hbar^{2} \left(\rule{0cm}{1.2cm}\right. \frac{1}{12} \hspace{0.1cm} \begin{gathered}
\begin{fmffile}{Diagrams/Mixed2PIEA_Fock}
\begin{fmfgraph}(15,15)
\fmfleft{i}
\fmfright{o}
\fmfv{decor.shape=circle,decor.size=2.0thick,foreground=(0,,0,,1)}{v1}
\fmfv{decor.shape=circle,decor.size=2.0thick,foreground=(0,,0,,1)}{v2}
\fmf{phantom,tension=11}{i,v1}
\fmf{phantom,tension=11}{v2,o}
\fmf{plain,left,tension=0.4,foreground=(1,,0,,0)}{v1,v2,v1}
\fmf{wiggly,foreground=(1,,0,,0)}{v1,v2}
\end{fmfgraph}
\end{fmffile}
\end{gathered} + \frac{1}{6} \hspace{0.1cm} \begin{gathered}
\begin{fmffile}{Diagrams/Mixed2PIEA_Diag1}
\begin{fmfgraph}(15,15)
\fmfleft{i}
\fmfright{o}
\fmfv{decor.shape=circle,decor.size=2.0thick,foreground=(0,,0,,1)}{v1}
\fmfv{decor.shape=circle,decor.size=2.0thick,foreground=(0,,0,,1)}{v2}
\fmf{phantom,tension=11}{i,v1}
\fmf{phantom,tension=11}{v2,o}
\fmf{dashes,left,tension=0.4,foreground=(1,,0,,0)}{v1,v2,v1}
\fmf{plain,foreground=(1,,0,,0)}{v1,v2}
\end{fmfgraph}
\end{fmffile}
\end{gathered} \left.\rule{0cm}{1.2cm}\right) \\
& - \hbar^{3} \left(\rule{0cm}{1.2cm}\right. \frac{1}{72} \hspace{0.35cm} \begin{gathered}
\begin{fmffile}{Diagrams/Mixed2PIEA_Diag2}
\begin{fmfgraph}(10,10)
\fmfleft{i0,i1}
\fmfright{o0,o1}
\fmftop{v1,vUp,v2}
\fmfbottom{v3,vDown,v4}
\fmfv{decor.shape=circle,decor.size=2.0thick,foreground=(0,,0,,1)}{v1}
\fmfv{decor.shape=circle,decor.size=2.0thick,foreground=(0,,0,,1)}{v2}
\fmfv{decor.shape=circle,decor.size=2.0thick,foreground=(0,,0,,1)}{v3}
\fmfv{decor.shape=circle,decor.size=2.0thick,foreground=(0,,0,,1)}{v4}
\fmf{phantom,tension=20}{i0,v1}
\fmf{phantom,tension=20}{i1,v3}
\fmf{phantom,tension=20}{o0,v2}
\fmf{phantom,tension=20}{o1,v4}
\fmf{plain,left=0.4,tension=0.5,foreground=(1,,0,,0)}{v3,v1}
\fmf{phantom,left=0.1,tension=0.5}{v1,vUp}
\fmf{phantom,left=0.1,tension=0.5}{vUp,v2}
\fmf{plain,left=0.4,tension=0.0,foreground=(1,,0,,0)}{v1,v2}
\fmf{plain,left=0.4,tension=0.5,foreground=(1,,0,,0)}{v2,v4}
\fmf{phantom,left=0.1,tension=0.5}{v4,vDown}
\fmf{phantom,left=0.1,tension=0.5}{vDown,v3}
\fmf{plain,left=0.4,tension=0.0,foreground=(1,,0,,0)}{v4,v3}
\fmf{wiggly,tension=0.5,foreground=(1,,0,,0)}{v1,v4}
\fmf{wiggly,tension=0.5,foreground=(1,,0,,0)}{v2,v3}
\end{fmfgraph}
\end{fmffile}
\end{gathered} \hspace{0.35cm} + \frac{1}{36} \hspace{0.35cm} \begin{gathered}
\begin{fmffile}{Diagrams/Mixed2PIEA_Diag3}
\begin{fmfgraph}(10,10)
\fmfleft{i0,i1}
\fmfright{o0,o1}
\fmftop{v1,vUp,v2}
\fmfbottom{v3,vDown,v4}
\fmfv{decor.shape=circle,decor.size=2.0thick,foreground=(0,,0,,1)}{v1}
\fmfv{decor.shape=circle,decor.size=2.0thick,foreground=(0,,0,,1)}{v2}
\fmfv{decor.shape=circle,decor.size=2.0thick,foreground=(0,,0,,1)}{v3}
\fmfv{decor.shape=circle,decor.size=2.0thick,foreground=(0,,0,,1)}{v4}
\fmf{phantom,tension=20}{i0,v1}
\fmf{phantom,tension=20}{i1,v3}
\fmf{phantom,tension=20}{o0,v2}
\fmf{phantom,tension=20}{o1,v4}
\fmf{dashes,left=0.4,tension=0.5,foreground=(1,,0,,0)}{v3,v1}
\fmf{phantom,left=0.1,tension=0.5}{v1,vUp}
\fmf{phantom,left=0.1,tension=0.5}{vUp,v2}
\fmf{dashes,left=0.4,tension=0.0,foreground=(1,,0,,0)}{v1,v2}
\fmf{dashes,left=0.4,tension=0.5,foreground=(1,,0,,0)}{v2,v4}
\fmf{phantom,left=0.1,tension=0.5}{v4,vDown}
\fmf{phantom,left=0.1,tension=0.5}{vDown,v3}
\fmf{dashes,left=0.4,tension=0.0,foreground=(1,,0,,0)}{v4,v3}
\fmf{plain,tension=0.5,foreground=(1,,0,,0)}{v1,v4}
\fmf{plain,tension=0.5,foreground=(1,,0,,0)}{v2,v3}
\end{fmfgraph}
\end{fmffile}
\end{gathered} \hspace{0.35cm} + \frac{1}{18} \hspace{0.35cm} \begin{gathered}
\begin{fmffile}{Diagrams/Mixed2PIEA_Diag4}
\begin{fmfgraph}(10,10)
\fmfleft{i0,i1}
\fmfright{o0,o1}
\fmftop{v1,vUp,v2}
\fmfbottom{v3,vDown,v4}
\fmfv{decor.shape=circle,decor.size=2.0thick,foreground=(0,,0,,1)}{v1}
\fmfv{decor.shape=circle,decor.size=2.0thick,foreground=(0,,0,,1)}{v2}
\fmfv{decor.shape=circle,decor.size=2.0thick,foreground=(0,,0,,1)}{v3}
\fmfv{decor.shape=circle,decor.size=2.0thick,foreground=(0,,0,,1)}{v4}
\fmf{phantom,tension=20}{i0,v1}
\fmf{phantom,tension=20}{i1,v3}
\fmf{phantom,tension=20}{o0,v2}
\fmf{phantom,tension=20}{o1,v4}
\fmf{plain,left=0.4,tension=0.5,foreground=(1,,0,,0)}{v3,v1}
\fmf{phantom,left=0.1,tension=0.5}{v1,vUp}
\fmf{phantom,left=0.1,tension=0.5}{vUp,v2}
\fmf{plain,left=0.4,tension=0.0,foreground=(1,,0,,0)}{v1,v2}
\fmf{dashes,left=0.4,tension=0.5,foreground=(1,,0,,0)}{v2,v4}
\fmf{phantom,left=0.1,tension=0.5}{v4,vDown}
\fmf{phantom,left=0.1,tension=0.5}{vDown,v3}
\fmf{dashes,left=0.4,tension=0.0,foreground=(1,,0,,0)}{v4,v3}
\fmf{wiggly,tension=0.5,foreground=(1,,0,,0)}{v1,v4}
\fmf{plain,tension=0.5,foreground=(1,,0,,0)}{v2,v3}
\end{fmfgraph}
\end{fmffile}
\end{gathered} \hspace{0.35cm} + \frac{1}{9} \hspace{0.35cm} \begin{gathered}
\begin{fmffile}{Diagrams/Mixed2PIEA_Diag5}
\begin{fmfgraph}(10,10)
\fmfleft{i0,i1}
\fmfright{o0,o1}
\fmftop{v1,vUp,v2}
\fmfbottom{v3,vDown,v4}
\fmfv{decor.shape=circle,decor.size=2.0thick,foreground=(0,,0,,1)}{v1}
\fmfv{decor.shape=circle,decor.size=2.0thick,foreground=(0,,0,,1)}{v2}
\fmfv{decor.shape=circle,decor.size=2.0thick,foreground=(0,,0,,1)}{v3}
\fmfv{decor.shape=circle,decor.size=2.0thick,foreground=(0,,0,,1)}{v4}
\fmf{phantom,tension=20}{i0,v1}
\fmf{phantom,tension=20}{i1,v3}
\fmf{phantom,tension=20}{o0,v2}
\fmf{phantom,tension=20}{o1,v4}
\fmf{dashes,left=0.4,tension=0.5,foreground=(1,,0,,0)}{v3,v1}
\fmf{phantom,left=0.1,tension=0.5}{v1,vUp}
\fmf{phantom,left=0.1,tension=0.5}{vUp,v2}
\fmf{plain,left=0.4,tension=0.0,foreground=(1,,0,,0)}{v1,v2}
\fmf{dashes,left=0.4,tension=0.5,foreground=(1,,0,,0)}{v2,v4}
\fmf{phantom,left=0.1,tension=0.5}{v4,vDown}
\fmf{phantom,left=0.1,tension=0.5}{vDown,v3}
\fmf{plain,left=0.4,tension=0.0,foreground=(1,,0,,0)}{v4,v3}
\fmf{wiggly,tension=0.5,foreground=(1,,0,,0)}{v1,v4}
\fmf{plain,tension=0.5,foreground=(1,,0,,0)}{v2,v3}
\end{fmfgraph}
\end{fmffile}
\end{gathered} \hspace{0.35cm} + \frac{1}{9} \hspace{0.35cm} \begin{gathered}
\begin{fmffile}{Diagrams/Mixed2PIEA_Diag6}
\begin{fmfgraph}(10,10)
\fmfleft{i0,i1}
\fmfright{o0,o1}
\fmftop{v1,vUp,v2}
\fmfbottom{v3,vDown,v4}
\fmfv{decor.shape=circle,decor.size=2.0thick,foreground=(0,,0,,1)}{v1}
\fmfv{decor.shape=circle,decor.size=2.0thick,foreground=(0,,0,,1)}{v2}
\fmfv{decor.shape=circle,decor.size=2.0thick,foreground=(0,,0,,1)}{v3}
\fmfv{decor.shape=circle,decor.size=2.0thick,foreground=(0,,0,,1)}{v4}
\fmf{phantom,tension=20}{i0,v1}
\fmf{phantom,tension=20}{i1,v3}
\fmf{phantom,tension=20}{o0,v2}
\fmf{phantom,tension=20}{o1,v4}
\fmf{plain,left=0.4,tension=0.5,foreground=(1,,0,,0)}{v3,v1}
\fmf{phantom,left=0.1,tension=0.5}{v1,vUp}
\fmf{phantom,left=0.1,tension=0.5}{vUp,v2}
\fmf{plain,left=0.4,tension=0.0,foreground=(1,,0,,0)}{v1,v2}
\fmf{dashes,left=0.4,tension=0.5,foreground=(1,,0,,0)}{v2,v4}
\fmf{phantom,left=0.1,tension=0.5}{v4,vDown}
\fmf{phantom,left=0.1,tension=0.5}{vDown,v3}
\fmf{dashes,left=0.4,tension=0.0,foreground=(1,,0,,0)}{v4,v3}
\fmf{dashes,tension=0.5,foreground=(1,,0,,0)}{v1,v4}
\fmf{dashes,tension=0.5,foreground=(1,,0,,0)}{v2,v3}
\end{fmfgraph}
\end{fmffile}
\end{gathered} \hspace{0.35cm} \left.\rule{0cm}{1.2cm}\right) \\
& + \hbar^{4} \left(\rule{0cm}{1.2cm}\right. \frac{1}{324} \hspace{0.3cm} \begin{gathered}
\begin{fmffile}{Diagrams/Mixed2PIEA_Diag7}
\begin{fmfgraph}(15,15)
\fmfleft{i}
\fmfright{o}
\fmftop{vUpLeft,vUp,vUpRight}
\fmfbottom{vDownLeft,vDown,vDownRight}
\fmfv{decor.shape=circle,decor.size=2.0thick,foreground=(0,,0,,1)}{v1}
\fmfv{decor.shape=circle,decor.size=2.0thick,foreground=(0,,0,,1)}{v2}
\fmfv{decor.shape=circle,decor.size=2.0thick,foreground=(0,,0,,1)}{v3}
\fmfv{decor.shape=circle,decor.size=2.0thick,foreground=(0,,0,,1)}{v4}
\fmfv{decor.shape=circle,decor.size=2.0thick,foreground=(0,,0,,1)}{v5}
\fmfv{decor.shape=circle,decor.size=2.0thick,foreground=(0,,0,,1)}{v6}
\fmf{phantom,tension=1}{i,v1}
\fmf{phantom,tension=1}{v2,o}
\fmf{phantom,tension=14.0}{v3,vUpLeft}
\fmf{phantom,tension=2.0}{v3,vUpRight}
\fmf{phantom,tension=4.0}{v3,i}
\fmf{phantom,tension=2.0}{v4,vUpLeft}
\fmf{phantom,tension=14.0}{v4,vUpRight}
\fmf{phantom,tension=4.0}{v4,o}
\fmf{phantom,tension=14.0}{v5,vDownLeft}
\fmf{phantom,tension=2.0}{v5,vDownRight}
\fmf{phantom,tension=4.0}{v5,i}
\fmf{phantom,tension=2.0}{v6,vDownLeft}
\fmf{phantom,tension=14.0}{v6,vDownRight}
\fmf{phantom,tension=4.0}{v6,o}
\fmf{wiggly,tension=0,foreground=(1,,0,,0)}{v1,v2}
\fmf{wiggly,tension=0.6,foreground=(1,,0,,0)}{v3,v6}
\fmf{wiggly,tension=0.6,foreground=(1,,0,,0)}{v5,v4}
\fmf{plain,left=0.18,tension=0,foreground=(1,,0,,0)}{v1,v3}
\fmf{plain,left=0.42,tension=0,foreground=(1,,0,,0)}{v3,v4}
\fmf{plain,left=0.18,tension=0,foreground=(1,,0,,0)}{v4,v2}
\fmf{plain,left=0.18,tension=0,foreground=(1,,0,,0)}{v2,v6}
\fmf{plain,left=0.42,tension=0,foreground=(1,,0,,0)}{v6,v5}
\fmf{plain,left=0.18,tension=0,foreground=(1,,0,,0)}{v5,v1}
\end{fmfgraph}
\end{fmffile}
\end{gathered} \hspace{0.3cm} + \frac{1}{108} \hspace{0.5cm} \begin{gathered}
\begin{fmffile}{Diagrams/Mixed2PIEA_Diag8}
\begin{fmfgraph}(12.5,12.5)
\fmfleft{i0,i1}
\fmfright{o0,o1}
\fmftop{v1,vUp,v2}
\fmfbottom{v3,vDown,v4}
\fmfv{decor.shape=circle,decor.size=2.0thick,foreground=(0,,0,,1)}{v1}
\fmfv{decor.shape=circle,decor.size=2.0thick,foreground=(0,,0,,1)}{v2}
\fmfv{decor.shape=circle,decor.size=2.0thick,foreground=(0,,0,,1)}{v3}
\fmfv{decor.shape=circle,decor.size=2.0thick,foreground=(0,,0,,1)}{v4}
\fmfv{decor.shape=circle,decor.size=2.0thick,foreground=(0,,0,,1)}{v5}
\fmfv{decor.shape=circle,decor.size=2.0thick,foreground=(0,,0,,1)}{v6}
\fmf{phantom,tension=20}{i0,v1}
\fmf{phantom,tension=20}{i1,v3}
\fmf{phantom,tension=20}{o0,v2}
\fmf{phantom,tension=20}{o1,v4}
\fmf{phantom,tension=0.005}{v5,v6}
\fmf{wiggly,left=0.4,tension=0,foreground=(1,,0,,0)}{v3,v1}
\fmf{phantom,left=0.1,tension=0}{v1,vUp}
\fmf{phantom,left=0.1,tension=0}{vUp,v2}
\fmf{plain,left=0.25,tension=0,foreground=(1,,0,,0)}{v1,v2}
\fmf{wiggly,left=0.4,tension=0,foreground=(1,,0,,0)}{v2,v4}
\fmf{phantom,left=0.1,tension=0}{v4,vDown}
\fmf{phantom,left=0.1,tension=0}{vDown,v3}
\fmf{plain,right=0.25,tension=0,foreground=(1,,0,,0)}{v3,v4}
\fmf{plain,left=0.2,tension=0.01,foreground=(1,,0,,0)}{v1,v5}
\fmf{plain,left=0.2,tension=0.01,foreground=(1,,0,,0)}{v5,v3}
\fmf{plain,right=0.2,tension=0.01,foreground=(1,,0,,0)}{v2,v6}
\fmf{plain,right=0.2,tension=0.01,foreground=(1,,0,,0)}{v6,v4}
\fmf{wiggly,tension=0,foreground=(1,,0,,0)}{v5,v6}
\end{fmfgraph}
\end{fmffile}
\end{gathered} \hspace{0.5cm} + \frac{1}{324} \hspace{0.4cm} \begin{gathered}
\begin{fmffile}{Diagrams/Mixed2PIEA_Diag9}
\begin{fmfgraph}(12.5,12.5)
\fmfleft{i0,i1}
\fmfright{o0,o1}
\fmftop{v1,vUp,v2}
\fmfbottom{v3,vDown,v4}
\fmfv{decor.shape=circle,decor.size=2.0thick,foreground=(0,,0,,1)}{v1}
\fmfv{decor.shape=circle,decor.size=2.0thick,foreground=(0,,0,,1)}{v2}
\fmfv{decor.shape=circle,decor.size=2.0thick,foreground=(0,,0,,1)}{v3}
\fmfv{decor.shape=circle,decor.size=2.0thick,foreground=(0,,0,,1)}{v4}
\fmfv{decor.shape=circle,decor.size=2.0thick,foreground=(0,,0,,1)}{v5}
\fmfv{decor.shape=circle,decor.size=2.0thick,foreground=(0,,0,,1)}{v6}
\fmf{phantom,tension=20}{i0,v1}
\fmf{phantom,tension=20}{i1,v3}
\fmf{phantom,tension=20}{o0,v2}
\fmf{phantom,tension=20}{o1,v4}
\fmf{phantom,tension=0.005}{v5,v6}
\fmf{plain,left=0.4,tension=0,foreground=(1,,0,,0)}{v3,v1}
\fmf{phantom,left=0.1,tension=0}{v1,vUp}
\fmf{phantom,left=0.1,tension=0}{vUp,v2}
\fmf{wiggly,left=0.25,tension=0,foreground=(1,,0,,0)}{v1,v2}
\fmf{plain,left=0.4,tension=0,foreground=(1,,0,,0)}{v2,v4}
\fmf{phantom,left=0.1,tension=0}{v4,vDown}
\fmf{phantom,left=0.1,tension=0}{vDown,v3}
\fmf{wiggly,right=0.25,tension=0,foreground=(1,,0,,0)}{v3,v4}
\fmf{plain,left=0.2,tension=0.01,foreground=(1,,0,,0)}{v1,v5}
\fmf{plain,left=0.2,tension=0.01,foreground=(1,,0,,0)}{v5,v3}
\fmf{plain,right=0.2,tension=0.01,foreground=(1,,0,,0)}{v2,v6}
\fmf{plain,right=0.2,tension=0.01,foreground=(1,,0,,0)}{v6,v4}
\fmf{wiggly,tension=0,foreground=(1,,0,,0)}{v5,v6}
\end{fmfgraph}
\end{fmffile}
\end{gathered} \hspace{0.4cm} + \mathcal{O}\Big(\vec{F}^{2}\Big) \left.\rule{0cm}{1.2cm}\right) \\
& + \mathcal{O}\big(\hbar^{5}\big)\;.
\end{split}
\label{eq:mixed2PIEAfinalexpressionAppendix}
\end{equation}

\paragraph{$\lambda$-expansion for the mixed 2PI EA with vanishing 1-point correlation functions:}

We start by expanding the Schwinger functional of the studied $O(N)$ model in arbitrary dimensions and involving the source-dependent action $S_{\text{mix},\mathcal{K}}$ (which coincides with $S_{\text{mix},\mathcal{J}\mathcal{K}}$ defined by~\eqref{eq:SmixedJK} when $\mathcal{J}$ vanishes), i.e.\footnote{We set once again $\hbar=1$ while $\lambda$ is the expansion parameter under consideration.}:
\begin{equation}
\begin{split}
\scalebox{0.96}{${\displaystyle e^{W_{\mathrm{mix}}[\mathcal{K}]} =}$} & \scalebox{0.96}{${\displaystyle \int \mathcal{D}\widetilde{\Psi} \ e^{-S_{\mathrm{mix},\mathcal{K}}\big[\widetilde{\Psi}\big]} }$} \\
\scalebox{0.96}{${\displaystyle = }$} & \scalebox{0.96}{${\displaystyle \int \mathcal{D}\widetilde{\Psi} \ e^{-S_{\mathrm{mix}}\big[\widetilde{\Psi}\big] + \frac{1}{2} \int_{\beta_{\scalebox{0.4}{1}},\beta_{\scalebox{0.4}{2}}} \widetilde{\Psi}_{\beta_{\scalebox{0.4}{1}}} \mathcal{K}_{\beta_{\scalebox{0.4}{1}}\beta_{\scalebox{0.4}{2}}} \widetilde{\Psi}_{\beta_{\scalebox{0.4}{2}}}} }$} \\
\scalebox{0.96}{${\displaystyle = }$} & \scalebox{0.96}{${\displaystyle \int \mathcal{D}\vec{\widetilde{\varphi}} \mathcal{D}\widetilde{\sigma} \ e^{-\frac{1}{2}\int_{\alpha_{1},\alpha_{2}}\widetilde{\varphi}_{\alpha_{1}}\left[\left(-\nabla_{x_{1}}^{2} + m^{2}\right)\delta_{\alpha_{1}\alpha_{2}}-\boldsymbol{K}_{\alpha_{1}\alpha_{2}}\right]\widetilde{\varphi}_{\alpha_{2}} - \frac{1}{2}\int_{x_{\scalebox{0.4}{1}},x_{\scalebox{0.4}{2}}} \widetilde{\sigma}_{x_{\scalebox{0.4}{1}}}\left(\delta_{x_{\scalebox{0.4}{1}}x_{\scalebox{0.4}{2}}}-k_{x_{\scalebox{0.4}{1}}x_{\scalebox{0.4}{2}}}\right)\widetilde{\sigma}_{x_{\scalebox{0.4}{2}}} - i\sqrt{\frac{\lambda}{12}} \sum_{a=1}^{N} \int_{x} \widetilde{\sigma}_{x} \widetilde{\varphi}^{2}_{a,x}}\;, }$}
\end{split}
\label{eq:mixed2PIEAlambdaWExpansionstep10DON}
\end{equation}
where we have exploited the diagonality of the external source $\mathcal{K}$:
\begin{equation}
\mathcal{K}=\begin{pmatrix}
\boldsymbol{K} & \vec{0} \\
\vec{0}^{\mathrm{T}} & k
\end{pmatrix}\;.
\end{equation}
Using the definitions:
\begin{equation}
\boldsymbol{G}^{-1}_{\mathcal{K},\alpha_{1}\alpha_{2}}[\mathcal{K}] \equiv \left.\frac{\delta^{2}S_{\text{mix},\mathcal{K}}\big[\widetilde{\Psi}\big]}{\delta\widetilde{\varphi}_{\alpha_{1}}\delta\widetilde{\varphi}_{\alpha_{2}}}\right|_{\widetilde{\Psi}=0} = \left(-\nabla_{x_{1}}^{2} + m^{2}\right)\delta_{\alpha_{1}\alpha_{2}}-\boldsymbol{K}_{\alpha_{1}\alpha_{2}}\;,
\end{equation}
\begin{equation}
D^{-1}_{\mathcal{K},x_{1}x_{2}}[\mathcal{K}] \equiv \left.\frac{\delta^{2}S_{\text{mix},\mathcal{K}}\big[\widetilde{\Psi}\big]}{\delta\widetilde{\sigma}_{x_{1}}\delta\widetilde{\sigma}_{x_{2}}}\right|_{\widetilde{\Psi}=0} = \delta_{x_{1}x_{2}}-k_{x_{1}x_{2}}\;,
\end{equation}
as well as the source-dependent expectation value:
\begin{equation}
\llangle[\big] \cdots \rrangle[\big]_{0,\mathcal{K}} = \frac{1}{Z_{\mathrm{mix},0}[\mathcal{K}]} \int \mathcal{D}\widetilde{\Psi} \ \cdots \ e^{S_{\mathrm{mix},\mathcal{K}}^{0}\big[\widetilde{\Psi}\big]} \;,
\end{equation}
with
\begin{equation}
Z_{\mathrm{mix},0}[\mathcal{K}] = \int \mathcal{D}\widetilde{\Psi} \ \cdots \ e^{S_{\mathrm{mix},\mathcal{K}}^{0}\big[\widetilde{\Psi}\big]} \;,
\end{equation}
and
\begin{equation}
S_{\mathrm{mix},\mathcal{K}}^{0}\Big[\widetilde{\Psi}\Big] = \frac{1}{2}\int_{\alpha_{1},\alpha_{2}}\widetilde{\varphi}_{\alpha_{1}}\left[\left(-\nabla^{2}_{x_{1}} + m^{2}\right)\delta_{\alpha_{1}\alpha_{2}}-\boldsymbol{K}_{\alpha_{1}\alpha_{2}}\right]\widetilde{\varphi}_{\alpha_{2}} + \frac{1}{2}\int_{x_{1},x_{2}} \widetilde{\sigma}_{x_{1}}\left(\delta_{x_{1}x_{2}}-k_{x_{1}x_{2}}\right)\widetilde{\sigma}_{x_{2}} \;,
\end{equation}
we can rewrite~\eqref{eq:mixed2PIEAlambdaWExpansionstep10DON} as follows:
\begin{equation}
\begin{split}
e^{W_{\mathrm{mix}}[\mathcal{K}]} = & \int \mathcal{D}\vec{\widetilde{\varphi}} \mathcal{D}\widetilde{\sigma} \ e^{-\frac{1}{2}\int_{\alpha_{\scalebox{0.4}{1}},\alpha_{\scalebox{0.4}{2}}}\widetilde{\varphi}_{\alpha_{\scalebox{0.4}{1}}}\boldsymbol{G}^{-1}_{\mathcal{K},\alpha_{\scalebox{0.4}{1}}\alpha_{\scalebox{0.4}{2}}}[\mathcal{K}]\widetilde{\varphi}_{\alpha_{\scalebox{0.4}{2}}} - \frac{1}{2}\int_{x_{\scalebox{0.4}{1}},x_{\scalebox{0.4}{2}}} \widetilde{\sigma}_{x_{\scalebox{0.4}{1}}}D^{-1}_{\mathcal{K},x_{\scalebox{0.4}{1}}x_{\scalebox{0.4}{2}}}[\mathcal{K}]\widetilde{\sigma}_{x_{\scalebox{0.4}{2}}} - i\sqrt{\frac{\lambda}{12}} \sum_{a=1}^{N} \int_{x} \widetilde{\sigma}_{x} \widetilde{\varphi}^{2}_{a,x}} \\
= & \ Z_{\mathrm{mix},0}[\mathcal{K}] \llangle[\bigg] e^{- i\sqrt{\frac{\lambda}{12}} \sum_{a=1}^{N} \int_{x} \widetilde{\sigma}_{x} \widetilde{\varphi}^{2}_{a,x}} \rrangle[\bigg]_{0,\mathcal{K}} \\
= & \ e^{\frac{1}{2}\mathrm{STr}[\ln(\boldsymbol{G}_{\mathcal{K}}[\mathcal{K}])]+\frac{1}{2}\mathrm{Tr}[\ln(D_{\mathcal{K}}[\mathcal{K}])]} \\
& \times \sum_{n=0}^{\infty} \frac{\left(-i\right)^{n}}{n!} \left(\frac{\lambda}{12}\right)^{\frac{n}{2}} \sum_{a_{1},\cdots,a_{n}=1}^{N} \int_{x_{1},\cdots,x_{n}} \llangle[\Big] \widetilde{\sigma}_{x_{1}} \widetilde{\varphi}^{2}_{a_{1},x_{1}} \cdots \widetilde{\sigma}_{x_{n}} \widetilde{\varphi}^{2}_{a_{n},x_{n}} \rrangle[\Big]_{0,\mathcal{K}} \\
= & \ e^{\frac{1}{2}\mathrm{STr}[\ln(\boldsymbol{G}_{\mathcal{K}}[\mathcal{K}])]+\frac{1}{2}\mathrm{Tr}[\ln(D_{\mathcal{K}}[\mathcal{K}])]} \\
& \times \sum_{n=0}^{\infty} \frac{\left(-1\right)^{n}}{(2n)!} \left(\frac{\lambda}{12}\right)^{n} \sum_{a_{1},\cdots,a_{2n}=1}^{N} \int_{x_{1},\cdots,x_{2n}} \llangle[\Big] \widetilde{\sigma}_{x_{1}} \widetilde{\varphi}^{2}_{a_{1},x_{1}} \cdots \widetilde{\sigma}_{x_{2n}} \widetilde{\varphi}^{2}_{a_{2n},x_{2n}} \rrangle[\Big]_{0,\mathcal{K}} \;.
\end{split}
\label{eq:mixed2PIEAlambdaWExpansionstep20DON}
\end{equation}
The last line of~\eqref{eq:mixed2PIEAlambdaWExpansionstep20DON} was deduced from the previous one by exploiting the fact that all correlation functions containing an odd number of $\widetilde{\sigma}$ fields vanish in the present approach, as a consequence of our condition to impose each 1-point correlation function to vanish. Taking the logarithm in~\eqref{eq:mixed2PIEAlambdaWExpansionstep20DON} enables us to express the Schwinger functional in terms of connected correlation functions only, as usual. This translates into:
\begin{equation*}
\begin{split}
W_{\mathrm{mix}}[\mathcal{K}] = & \ \frac{1}{2}\mathrm{STr}\left[\ln\big(\boldsymbol{G}_{\mathcal{K}}[\mathcal{K}]\big)\right]+\frac{1}{2}\mathrm{Tr}\left[\ln\big(D_{\mathcal{K}}[\mathcal{K}]\big)\right] \\
& + \sum_{n=1}^{\infty} \frac{\left(-1\right)^{n}}{(2n)!} \left(\frac{\lambda}{12}\right)^{n} \sum_{a_{1},\cdots,a_{2n}=1}^{N} \int_{x_{1},\cdots,x_{2n}} \llangle[\Big] \widetilde{\sigma}_{x_{1}} \widetilde{\varphi}^{2}_{a_{1},x_{1}} \cdots \widetilde{\sigma}_{x_{2n}} \widetilde{\varphi}^{2}_{a_{2n},x_{2n}} \rrangle[\Big]^{\text{c}}_{0,\mathcal{K}} \\
= & \ \frac{1}{2}\mathrm{STr}\left[\ln\big(\boldsymbol{G}_{\mathcal{K}}[\mathcal{K}]\big)\right]+\frac{1}{2}\mathrm{Tr}\left[\ln\big(D_{\mathcal{K}}[\mathcal{K}]\big)\right] \\
& - \left(\rule{0cm}{1.0cm}\right. \frac{1}{24}\begin{gathered}
\begin{fmffile}{Diagrams/LoopExpansionMixedHS_Hartree}
\begin{fmfgraph}(30,20)
\fmfleft{i}
\fmfright{o}
\fmfv{decor.shape=circle,decor.size=2.0thick,foreground=(0,,0,,1)}{v1}
\fmfv{decor.shape=circle,decor.size=2.0thick,foreground=(0,,0,,1)}{v2}
\fmf{phantom,tension=10}{i,i1}
\fmf{phantom,tension=10}{o,o1}
\fmf{plain,left,tension=0.5}{i1,v1,i1}
\fmf{plain,right,tension=0.5}{o1,v2,o1}
\fmf{wiggly}{v1,v2}
\end{fmfgraph}
\end{fmffile}
\end{gathered}
+\frac{1}{12}\begin{gathered}
\begin{fmffile}{Diagrams/LoopExpansionMixedHS_Fock}
\begin{fmfgraph}(15,15)
\fmfleft{i}
\fmfright{o}
\fmfv{decor.shape=circle,decor.size=2.0thick,foreground=(0,,0,,1)}{v1}
\fmfv{decor.shape=circle,decor.size=2.0thick,foreground=(0,,0,,1)}{v2}
\fmf{phantom,tension=11}{i,v1}
\fmf{phantom,tension=11}{v2,o}
\fmf{plain,left,tension=0.4}{v1,v2,v1}
\fmf{wiggly}{v1,v2}
\end{fmfgraph}
\end{fmffile}
\end{gathered} \left.\rule{0cm}{1.0cm}\right)
\end{split}
\end{equation*}
\begin{equation}
\begin{split}
\hspace{3.02cm} & + \left(\rule{0cm}{1.0cm}\right. \frac{1}{72} \hspace{0.3cm} \begin{gathered}
\begin{fmffile}{Diagrams/Mixed2PIEAlambda_Diag1}
\begin{fmfgraph}(10,10)
\fmfleft{i0,i1}
\fmfright{o0,o1}
\fmftop{v1,vUp,v2}
\fmfbottom{v3,vDown,v4}
\fmfv{decor.shape=circle,decor.size=2.0thick,foreground=(0,,0,,1)}{v1}
\fmfv{decor.shape=circle,decor.size=2.0thick,foreground=(0,,0,,1)}{v2}
\fmfv{decor.shape=circle,decor.size=2.0thick,foreground=(0,,0,,1)}{v3}
\fmfv{decor.shape=circle,decor.size=2.0thick,foreground=(0,,0,,1)}{v4}
\fmf{phantom,tension=20}{i0,v1}
\fmf{phantom,tension=20}{i1,v3}
\fmf{phantom,tension=20}{o0,v2}
\fmf{phantom,tension=20}{o1,v4}
\fmf{plain,left=0.4,tension=0.5}{v3,v1}
\fmf{phantom,left=0.1,tension=0.5}{v1,vUp}
\fmf{phantom,left=0.1,tension=0.5}{vUp,v2}
\fmf{plain,left=0.4,tension=0.0}{v1,v2}
\fmf{plain,left=0.4,tension=0.5}{v2,v4}
\fmf{phantom,left=0.1,tension=0.5}{v4,vDown}
\fmf{phantom,left=0.1,tension=0.5}{vDown,v3}
\fmf{plain,left=0.4,tension=0.0}{v4,v3}
\fmf{wiggly,tension=0.5}{v1,v4}
\fmf{wiggly,tension=0.5}{v2,v3}
\end{fmfgraph}
\end{fmffile}
\end{gathered} \hspace{0.27cm} + \frac{1}{36} \hspace{0.3cm} \begin{gathered}
\begin{fmffile}{Diagrams/Mixed2PIEAlambda_Diag2}
\begin{fmfgraph}(10,10)
\fmfleft{i0,i1}
\fmfright{o0,o1}
\fmftop{v1,vUp,v2}
\fmfbottom{v3,vDown,v4}
\fmfv{decor.shape=circle,decor.size=2.0thick,foreground=(0,,0,,1)}{v1}
\fmfv{decor.shape=circle,decor.size=2.0thick,foreground=(0,,0,,1)}{v2}
\fmfv{decor.shape=circle,decor.size=2.0thick,foreground=(0,,0,,1)}{v3}
\fmfv{decor.shape=circle,decor.size=2.0thick,foreground=(0,,0,,1)}{v4}
\fmf{phantom,tension=20}{i0,v1}
\fmf{phantom,tension=20}{i1,v3}
\fmf{phantom,tension=20}{o0,v2}
\fmf{phantom,tension=20}{o1,v4}
\fmf{plain,left=0.4,tension=0.5}{v3,v1}
\fmf{phantom,left=0.1,tension=0.5}{v1,vUp}
\fmf{phantom,left=0.1,tension=0.5}{vUp,v2}
\fmf{plain,left=0.4,tension=0.0}{v1,v2}
\fmf{plain,left=0.4,tension=0.5}{v2,v4}
\fmf{phantom,left=0.1,tension=0.5}{v4,vDown}
\fmf{phantom,left=0.1,tension=0.5}{vDown,v3}
\fmf{plain,left=0.4,tension=0.0}{v4,v3}
\fmf{wiggly,left=0.4,tension=0.5}{v1,v3}
\fmf{wiggly,right=0.4,tension=0.5}{v2,v4}
\end{fmfgraph}
\end{fmffile}
\end{gathered} \hspace{0.27cm} + \frac{1}{36} \hspace{-0.15cm} \begin{gathered}
\begin{fmffile}{Diagrams/Mixed2PIEAlambda_Diag3}
\begin{fmfgraph}(30,15)
\fmfleft{i}
\fmfright{o}
\fmftop{vUpLeft1,vUpLeft2,vUpLeft3,vUpRight1,vUpRight2,vUpRight3}
\fmfbottom{vDownLeft1,vDownLeft2,vDownLeft3,vDownRight1,vDownRight2,vDownRight3}
\fmfv{decor.shape=circle,decor.size=2.0thick,foreground=(0,,0,,1)}{v1}
\fmfv{decor.shape=circle,decor.size=2.0thick,foreground=(0,,0,,1)}{v2}
\fmfv{decor.shape=circle,decor.size=2.0thick,foreground=(0,,0,,1)}{vUpLeft}
\fmfv{decor.shape=circle,decor.size=2.0thick,foreground=(0,,0,,1)}{vDownLeft}
\fmf{phantom,tension=10}{i,v3}
\fmf{phantom,tension=10}{o,v4}
\fmf{plain,right=0.4,tension=0.5}{v1,vUpLeft}
\fmf{plain,right,tension=0.5}{vUpLeft,vDownLeft}
\fmf{plain,left=0.4,tension=0.5}{v1,vDownLeft}
\fmf{phantom,right=0.4,tension=0.5}{vUpRight,v2}
\fmf{phantom,left,tension=0.5}{vUpRight,vDownRight}
\fmf{phantom,left=0.4,tension=0.5}{vDownRight,v2}
\fmf{phantom,tension=0.3}{v2bis,o}
\fmf{plain,left,tension=0.1}{v2,v2bis,v2}
\fmf{wiggly,tension=2.7}{v1,v2}
\fmf{phantom,tension=2}{v1,v3}
\fmf{phantom,tension=2}{v2,v4}
\fmf{phantom,tension=2.4}{vUpLeft,vUpLeft2}
\fmf{phantom,tension=2.4}{vDownLeft,vDownLeft2}
\fmf{phantom,tension=2.4}{vUpRight,vUpRight2}
\fmf{phantom,tension=2.4}{vDownRight,vDownRight2}
\fmf{wiggly,tension=0}{vUpLeft,vDownLeft}
\end{fmfgraph}
\end{fmffile}
\end{gathered} \hspace{-0.35cm} + \frac{1}{144} \hspace{0.35cm} \begin{gathered}
\begin{fmffile}{Diagrams/Mixed2PIEAlambda_Diag4}
\begin{fmfgraph}(12,12)
\fmfleft{i0,i1}
\fmfright{o0,o1}
\fmftop{v1,vUp,v2}
\fmfbottom{v3,vDown,v4}
\fmfv{decor.shape=circle,decor.size=2.0thick,foreground=(0,,0,,1)}{v1}
\fmfv{decor.shape=circle,decor.size=2.0thick,foreground=(0,,0,,1)}{v2}
\fmfv{decor.shape=circle,decor.size=2.0thick,foreground=(0,,0,,1)}{v3}
\fmfv{decor.shape=circle,decor.size=2.0thick,foreground=(0,,0,,1)}{v4}
\fmf{phantom,tension=20}{i0,v1}
\fmf{phantom,tension=20}{i1,v3}
\fmf{phantom,tension=20}{o0,v2}
\fmf{phantom,tension=20}{o1,v4}
\fmf{plain,left=0.4,tension=0.5}{v3,v1}
\fmf{phantom,left=0.1,tension=0.5}{v1,vUp}
\fmf{phantom,left=0.1,tension=0.5}{vUp,v2}
\fmf{wiggly,tension=0.0}{v1,v2}
\fmf{plain,left=0.4,tension=0.5}{v2,v4}
\fmf{phantom,left=0.1,tension=0.5}{v4,vDown}
\fmf{phantom,left=0.1,tension=0.5}{vDown,v3}
\fmf{wiggly,tension=0.0}{v4,v3}
\fmf{plain,left=0.4,tension=0.5}{v1,v3}
\fmf{plain,right=0.4,tension=0.5}{v2,v4}
\end{fmfgraph}
\end{fmffile}
\end{gathered} \\
& \hspace{0.6cm} + \frac{1}{144} \hspace{0.1cm} \begin{gathered}
\begin{fmffile}{Diagrams/Mixed2PIEAlambda_Diag5}
\begin{fmfgraph}(30,15)
\fmfleft{i}
\fmfright{o}
\fmfv{decor.shape=circle,decor.size=2.0thick,foreground=(0,,0,,1)}{v1}
\fmfv{decor.shape=circle,decor.size=2.0thick,foreground=(0,,0,,1)}{v2}
\fmfv{decor.shape=circle,decor.size=2.0thick,foreground=(0,,0,,1)}{v3}
\fmfv{decor.shape=circle,decor.size=2.0thick,foreground=(0,,0,,1)}{v4}
\fmf{phantom,tension=10}{i,i1}
\fmf{phantom,tension=10}{o,o1}
\fmf{plain,left,tension=0.5}{i1,v1,i1}
\fmf{plain,right,tension=0.5}{o1,v2,o1}
\fmf{wiggly}{v1,v3}
\fmf{plain,left,tension=0.5}{v3,v4}
\fmf{plain,right,tension=0.5}{v3,v4}
\fmf{wiggly}{v4,v2}
\end{fmfgraph}
\end{fmffile}
\end{gathered} \left.\rule{0cm}{1.0cm}\right) \\
& + \mathcal{O}\Big(\lambda^{3}\Big)\;,
\end{split}
\label{eq:mixed2PIEAlambdaWExpansionstep30DON}
\end{equation}
with the Feynman rules:
\begin{subequations}
\begin{align}
\left.
\begin{array}{ll}
\begin{gathered}
\begin{fmffile}{Diagrams/LoopExpansionMixedHS_FeynRuleVertexbis_Appendix2PIlambda}
\begin{fmfgraph*}(4,4)
\fmfleft{i0,i1,i2,i3}
\fmfright{o0,o1,o2,o3}
\fmfv{label=$x$,label.angle=90,label.dist=4}{v1}
\fmfbottom{v2}
\fmf{phantom}{i1,v1}
\fmf{plain}{i2,v1}
\fmf{phantom}{v1,o1}
\fmf{plain}{v1,o2}
\fmf{wiggly,tension=0.6}{v1,v2}
\fmfv{decor.shape=circle,decor.size=2.0thick,foreground=(0,,0,,1)}{v1}
\fmflabel{$a_{1}$}{i2}
\fmflabel{$a_{2}$}{o2}
\end{fmfgraph*}
\end{fmffile}
\end{gathered} \\
\\
\begin{gathered}
\begin{fmffile}{Diagrams/LoopExpansionMixedHS_FeynRuleVertex2bis_Appendix2PIlambda}
\begin{fmfgraph*}(4,4)
\fmfleft{i0,i1,i2,i3}
\fmfright{o0,o1,o2,o3}
\fmfv{label=$x$,label.angle=90,label.dist=4}{v1}
\fmfbottom{v2}
\fmf{phantom}{i1,v1}
\fmf{plain}{i2,v1}
\fmf{phantom}{v1,o1}
\fmf{plain}{v1,o2}
\fmf{dots,tension=0.6}{v1,v2}
\fmfv{decor.shape=circle,decor.size=2.0thick,foreground=(0,,0,,1)}{v1}
\fmflabel{$a_{1}$}{i2}
\fmflabel{$a_{2}$}{o2}
\end{fmfgraph*}
\end{fmffile}
\end{gathered} \\
\\
\begin{gathered}
\begin{fmffile}{Diagrams/LoopExpansionMixedHS_FeynRuleVertex3bis_Appendix2PIlambda}
\begin{fmfgraph*}(4,4)
\fmfleft{i0,i1,i2,i3}
\fmfright{o0,o1,o2,o3}
\fmfv{label=$x$,label.angle=90,label.dist=4}{v1}
\fmfbottom{v2}
\fmf{phantom}{i1,v1}
\fmf{plain}{i2,v1}
\fmf{phantom}{v1,o1}
\fmf{dots}{v1,o2}
\fmf{dots,tension=0.6}{v1,v2}
\fmfv{decor.shape=circle,decor.size=2.0thick,foreground=(0,,0,,1)}{v1}
\fmflabel{$a_{1}$}{i2}
\fmflabel{$a_{2}$}{o2}
\end{fmfgraph*}
\end{fmffile}
\end{gathered}
\end{array}
\quad \right\rbrace &\rightarrow \sqrt{\lambda} \ \delta_{a_{1}a_{2}} \;, 
\label{eq:mixed2PIEAlambdaFeynRulesvertexAppendix} \\
\begin{gathered}
\begin{fmffile}{Diagrams/LoopExpansionMixedHS_FeynRuleGbis_Appendix2PIlambda}
\begin{fmfgraph*}(20,20)
\fmfleft{i0,i1,i2,i3}
\fmfright{o0,o1,o2,o3}
\fmflabel{$\alpha_{1}$}{v1}
\fmflabel{$\alpha_{2}$}{v2}
\fmf{phantom}{i1,v1}
\fmf{phantom}{i2,v1}
\fmf{plain,tension=0.6}{v1,v2}
\fmf{phantom}{v2,o1}
\fmf{phantom}{v2,o2}
\end{fmfgraph*}
\end{fmffile}
\end{gathered} \quad &\rightarrow \boldsymbol{G}_{\mathcal{K},\alpha_{1}\alpha_{2}}[\mathcal{K}] \;,
\label{eq:mixed2PIEAlambdaFeynRulesGAppendix} \\
\begin{gathered}
\begin{fmffile}{Diagrams/LoopExpansionMixedHS_FeynRuleDbis_Appendix2PIlambda}
\begin{fmfgraph*}(20,20)
\fmfleft{i0,i1,i2,i3}
\fmfright{o0,o1,o2,o3}
\fmfv{label=$x_{1}$}{v1}
\fmfv{label=$x_{2}$}{v2}
\fmf{phantom}{i1,v1}
\fmf{phantom}{i2,v1}
\fmf{wiggly,tension=0.6}{v1,v2}
\fmf{phantom}{v2,o1}
\fmf{phantom}{v2,o2}
\end{fmfgraph*}
\end{fmffile}
\end{gathered} \quad &\rightarrow D_{\mathcal{K},x_{1} x_{2}}[\mathcal{K}] \;.
\label{eq:mixed2PIEAlambdaFeynRulesDAppendix}
\end{align}
\end{subequations}
Note that the multiplicities of the diagrams involved in~\eqref{eq:mixed2PIEAlambdaWExpansionstep30DON} are still deduced from~\eqref{eq:MultiplicityDiagLoopExpansionMixed} and the diagrams contributing at order $\mathcal{O}\big(\lambda^{3}\big)$ in this expansion can also be inferred from the graphs expressing $\lambda^{3}\left\langle\left(\vec{\widetilde{\chi}}^{2}\right)^{6}\right\rangle_{0,\sigma}^{\text{c}}$ in tab.~\ref{tab:MultiplicityOPTdiagramsON} by replacing each zigzag vertex~\eqref{eq:FeynRuleOPT4legVertex} by a wiggly line~\eqref{eq:mixed2PIEAlambdaFeynRulesDAppendix}.

\vspace{0.5cm}

Let us then turn to the IM, starting with the power series:
\begin{subequations}
\begin{empheq}[left=\empheqlbrace]{align}
& \hspace{0.1cm} \Gamma_{\mathrm{mix}}^{(\mathrm{2PI})}\big[\mathcal{G};\lambda\big]=\sum_{n=0}^{\infty} \Gamma_{\mathrm{mix},n}^{(\mathrm{2PI})}\big[\mathcal{G};\lambda\big]\;, \label{eq:2PIEAlambdaGammaExpansion0DON}\\
\nonumber \\
& \hspace{0.1cm} W_{\mathrm{mix}}\big[\mathcal{K};\lambda\big]=\sum_{n=0}^{\infty} W_{\mathrm{mix},n}\big[\mathcal{K};\lambda\big]\;, \label{eq:2PIEAlambdaWExpansion0DON} \\
\nonumber \\
& \hspace{0.1cm} \mathcal{K}\big[\mathcal{G};\lambda\big]=\sum_{n=0}^{\infty} \mathcal{K}_{n}\big[\mathcal{K};\lambda\big]\;, \label{eq:2PIEAlambdaKExpansion0DON}\\
\nonumber \\
& \hspace{0.1cm} \mathcal{G}=\sum_{n=0}^{\infty} \mathcal{G}_{n}\big[\mathcal{K};\lambda\big]\;, \label{eq:2PIEAlambdaGExpansion0DON}
\end{empheq}
\end{subequations}
with $\Gamma_{\mathrm{mix}}^{(\mathrm{2PI})}[\mathcal{G}]\equiv\Gamma_{\mathrm{mix}}^{(\mathrm{2PI})}\big[\Phi=0,\mathcal{G}\big]$ satisfying~\eqref{eq:mixed2PIEAdefinition0DON} (with $\mathcal{J}$ and $\Phi$ equaling zero), from which we can infer the relation:
\begin{equation}
\frac{\delta \Gamma_{\mathrm{mix},n}^{(\mathrm{2PI})}[\mathcal{G}]}{\delta\mathcal{G}_{\beta_{1}\beta_{2}}} = \frac{1}{2}\mathcal{K}_{n,\beta_{1}\beta_{2}}[\mathcal{G}]\;,
\label{eq:mixed2PIEAlambdadGammanG}
\end{equation}
that we will exploit later. Here, $\Gamma_{\mathrm{mix},n}^{(\mathrm{2PI})}$, $W_{\mathrm{mix},n}$, $\mathcal{K}_{n}$ and $\mathcal{G}_{n}$ are all functionals of order $\mathcal{O}\big(\lambda^{n}\big)$ by definition. On the one hand, we have the propagator $\mathcal{G}_{\mathcal{K}}[\mathcal{K}]$ involved in the $\lambda$-expansion of $W_{\mathrm{mix}}$ and evaluated at arbitrary external source $\mathcal{K}$:
\begin{equation}
\mathcal{G}_{\mathcal{K},\beta_{1}\beta_{2}}[\mathcal{K}] = \left.\frac{\delta^{2} S_{\mathrm{mix}}\big[\widetilde{\Psi}\big]}{\delta\widetilde{\Psi}_{\beta_{1}}\delta\widetilde{\Psi}_{\beta_{2}}}\right|_{\widetilde{\Psi}=0} - \mathcal{K}_{\beta_{1}\beta_{2}}[\mathcal{G}] = 2\frac{\delta W_{\mathrm{mix},0}[\mathcal{K}]}{\delta\mathcal{K}_{\beta_{1}\beta_{2}}}\;,
\label{eq:mixed2PIlambdaGK}
\end{equation}
or, in matrix form:
\begin{equation}
\mathcal{G}_{\mathcal{K}}[\mathcal{K}]=\begin{pmatrix}
\boldsymbol{G}_{\mathcal{K}}[\mathcal{K}] & \vec{0} \\
\vec{0}^{\mathrm{T}} & D_{\mathcal{K}}[\mathcal{K}]
\end{pmatrix}\;,
\label{eq:mixed2PIlambdaGKmatrix}
\end{equation}
and, on the other hand, the argument of $\Gamma_{\mathrm{mix}}^{(\mathrm{2PI})}[\mathcal{G}]$, i.e. the propagator $\mathcal{G}_{\mathcal{K}}[\mathcal{K}]$ evaluated at $\mathcal{K}=\mathcal{K}_{0}$:
\begin{equation}
\mathcal{G}_{\beta_{1}\beta_{2}}=\mathcal{G}_{\mathcal{K},\beta_{1}\beta_{2}}[\mathcal{K}=\mathcal{K}_{0}]= \left.\frac{\delta^{2} S_{\mathrm{mix}}\big[\widetilde{\Psi}\big]}{\delta\widetilde{\Psi}_{\beta_{1}}\delta\widetilde{\Psi}_{\beta_{2}}}\right|_{\widetilde{\Psi}=0} - \mathcal{K}_{0,\beta_{1}\beta_{2}}[\mathcal{G}] = 2\left.\frac{\delta W_{\mathrm{mix},0}[\mathcal{K}]}{\delta\mathcal{K}_{\beta_{1}\beta_{2}}}\right|_{\mathcal{K}=\mathcal{K}_{0}} = \hspace{0.3cm} \begin{gathered}
\begin{fmffile}{Diagrams/bosonic1PIEA_DerivW0J0J0}
\begin{fmfgraph*}(20,20)
\fmfleft{i0,i1,i2,i3}
\fmfright{o0,o1,o2,o3}
\fmfv{decor.shape=circle,decor.filled=empty,decor.size=1.5thick,label=$\beta_{1}$}{v1}
\fmfv{decor.shape=circle,decor.filled=empty,decor.size=1.5thick,label=$\beta_{2}$}{v2}
\fmf{phantom}{i1,v1}
\fmf{phantom}{i2,v1}
\fmf{plain,tension=0.6,foreground=(1,,0,,0)}{v1,v2}
\fmf{wiggly,tension=0,foreground=(1,,0,,0)}{v1,v2}
\fmf{phantom}{v2,o1}
\fmf{phantom}{v2,o2}
\end{fmfgraph*}
\end{fmffile}
\end{gathered} \hspace{0.2cm} \;,
\label{eq:mixed2PIlambdaGK0}
\end{equation}
and, in matrix form:
\begin{equation}
\mathcal{G}=\begin{pmatrix}
\boldsymbol{G} & \vec{0} \\
\vec{0}^{\mathrm{T}} & D
\end{pmatrix}\;.
\label{eq:mixed2PIlambdaGK0matrix}
\end{equation}
In order to develop the IM for the $\lambda$-expansion of $\Gamma_{\mathrm{mix}}^{(\mathrm{2PI})}$, we will exploit the following Feynman rules:
\begin{subequations}
\begin{align}
\left.
\begin{array}{ll}
\begin{gathered}
\begin{fmffile}{Diagrams/mixed2PIEA_FeynRuleVertexbis1_AppendixFull2PI}
\begin{fmfgraph*}(4,4)
\fmfleft{i0,i1,i2,i3}
\fmfright{o0,o1,o2,o3}
\fmfv{label=$x$,label.angle=90,label.dist=4}{v1}
\fmfbottom{v2}
\fmf{phantom}{i1,v1}
\fmf{plain,foreground=(1,,0,,0)}{i2,v1}
\fmf{phantom}{v1,o1}
\fmf{plain,foreground=(1,,0,,0)}{v1,o2}
\fmf{wiggly,tension=0.6,foreground=(1,,0,,0)}{v1,v2}
\fmfv{decor.shape=circle,decor.size=2.0thick,foreground=(0,,0,,1)}{v1}
\fmflabel{$a_{1}$}{i2}
\fmflabel{$a_{2}$}{o2}
\end{fmfgraph*}
\end{fmffile}
\end{gathered} \\
\\
\begin{gathered}
\begin{fmffile}{Diagrams/mixed2PIEA_FeynRuleVertexbis2_AppendixFull2PI}
\begin{fmfgraph*}(4,4)
\fmfleft{i0,i1,i2,i3}
\fmfright{o0,o1,o2,o3}
\fmfv{label=$x$,label.angle=90,label.dist=4}{v1}
\fmfbottom{v2}
\fmf{phantom}{i1,v1}
\fmf{plain,foreground=(1,,0,,0)}{i2,v1}
\fmf{phantom}{v1,o1}
\fmf{plain,foreground=(1,,0,,0)}{v1,o2}
\fmf{dots,tension=0.6,foreground=(1,,0,,0)}{v1,v2}
\fmfv{decor.shape=circle,decor.size=2.0thick,foreground=(0,,0,,1)}{v1}
\fmflabel{$a_{1}$}{i2}
\fmflabel{$a_{2}$}{o2}
\end{fmfgraph*}
\end{fmffile}
\end{gathered} \\
\\
\begin{gathered}
\begin{fmffile}{Diagrams/mixed2PIEA_FeynRuleVertexbis3_AppendixFull2PI}
\begin{fmfgraph*}(4,4)
\fmfleft{i0,i1,i2,i3}
\fmfright{o0,o1,o2,o3}
\fmfv{label=$x$,label.angle=90,label.dist=4}{v1}
\fmfbottom{v2}
\fmf{phantom}{i1,v1}
\fmf{plain,foreground=(1,,0,,0)}{i2,v1}
\fmf{phantom}{v1,o1}
\fmf{dots,foreground=(1,,0,,0)}{v1,o2}
\fmf{dots,tension=0.6,foreground=(1,,0,,0)}{v1,v2}
\fmfv{decor.shape=circle,decor.size=2.0thick,foreground=(0,,0,,1)}{v1}
\fmflabel{$a_{1}$}{i2}
\fmflabel{$a_{2}$}{o2}
\end{fmfgraph*}
\end{fmffile}
\end{gathered}
\end{array}
\quad \right\rbrace &\rightarrow \sqrt{\lambda} \ \delta_{a_{1}a_{2}} \;, 
\label{eq:Mixed2PIEAlambdaFeynRulesvertexK0} \\
\begin{gathered}
\begin{fmffile}{Diagrams/mixed2PIEA_FeynRuleGbis_AppendixFull2PI}
\begin{fmfgraph*}(20,20)
\fmfleft{i0,i1,i2,i3}
\fmfright{o0,o1,o2,o3}
\fmflabel{$\alpha_{1}$}{v1}
\fmflabel{$\alpha_{2}$}{v2}
\fmf{phantom}{i1,v1}
\fmf{phantom}{i2,v1}
\fmf{plain,tension=0.6,foreground=(1,,0,,0)}{v1,v2}
\fmf{phantom}{v2,o1}
\fmf{phantom}{v2,o2}
\end{fmfgraph*}
\end{fmffile}
\end{gathered} \quad &\rightarrow \boldsymbol{G}_{\alpha_{1}\alpha_{2}} \;,
\label{eq:Mixed2PIEAlambdaFeynRulesGK0} \\
\begin{gathered}
\begin{fmffile}{Diagrams/mixed2PIEA_FeynRuleDbis_AppendixFull2PI}
\begin{fmfgraph*}(20,20)
\fmfleft{i0,i1,i2,i3}
\fmfright{o0,o1,o2,o3}
\fmfv{label=$x_{1}$}{v1}
\fmfv{label=$x_{2}$}{v2}
\fmf{phantom}{i1,v1}
\fmf{phantom}{i2,v1}
\fmf{wiggly,tension=0.6,foreground=(1,,0,,0)}{v1,v2}
\fmf{phantom}{v2,o1}
\fmf{phantom}{v2,o2}
\end{fmfgraph*}
\end{fmffile}
\end{gathered} \quad &\rightarrow D_{x_{1}x_{2}} \;,
\label{eq:Mixed2PIEAlambdaFeynRulesDK0} \\
\begin{gathered}
\begin{fmffile}{Diagrams/bosonic1PIEA_FeynRulemathcalG}
\begin{fmfgraph*}(20,20)
\fmfleft{i0,i1,i2,i3}
\fmfright{o0,o1,o2,o3}
\fmfv{label=$\beta_{1}$}{v1}
\fmfv{label=$\beta_{2}$}{v2}
\fmf{phantom}{i1,v1}
\fmf{phantom}{i2,v1}
\fmf{plain,tension=0,foreground=(1,,0,,0)}{v1,v2}
\fmf{wiggly,tension=0.6,foreground=(1,,0,,0)}{v1,v2}
\fmf{phantom}{v2,o1}
\fmf{phantom}{v2,o2}
\end{fmfgraph*}
\end{fmffile}
\end{gathered} \quad &\rightarrow \mathcal{G}_{\beta_{1}\beta_{2}}\;.
\label{eq:Mixed2PIEAlambdaFeynRulesmathcalG}
\end{align}
\end{subequations}
We then combine the series~\eqref{eq:2PIEAlambdaGammaExpansion0DON},~\eqref{eq:2PIEAlambdaWExpansion0DON} and~\eqref{eq:2PIEAlambdaKExpansion0DON} with the definition of the mixed 2PI EA $\Gamma_{\mathrm{mix}}^{(\mathrm{2PI})}[\mathcal{G}]$ in terms of Legendre transform of $W_{\mathrm{mix}}[\mathcal{K}]$:
\begin{equation}
\sum_{n=0}^{\infty} \Gamma_{\mathrm{mix},n}^{(\mathrm{2PI})}[\mathcal{G}] = -\sum_{n=0}^{\infty} W_{\mathrm{mix},n}\Bigg[\sum_{m=0}^{\infty} \mathcal{K}_{m}[\mathcal{G}]\Bigg]+\frac{1}{2}\sum_{n=0}^{\infty} \int_{\beta_{1},\beta_{2}}\mathcal{K}_{n,\beta_{1}\beta_{2}}[\mathcal{G}]\mathcal{G}_{\beta_{1}\beta_{2}}\;.
\end{equation}
After Taylor expanding the $W_{\mathrm{mix},n}$ coefficients of the RHS around $\mathcal{K}=\mathcal{K}_{0}$, we obtain the following expression:
\begin{equation}
\begin{split}
\scalebox{0.94}{${\displaystyle \Gamma_{\mathrm{mix},n}^{(\mathrm{2PI})}[\mathcal{G}] = }$} & \scalebox{0.94}{${\displaystyle -W_{\mathrm{mix},n}[\mathcal{K}=\mathcal{K}_{0}] -\sum_{m=1}^{n} \int_{\beta_{1},\beta_{2}} \left.\frac{\delta W_{\mathrm{mix},n-m}[\mathcal{K}]}{\delta \mathcal{K}_{\beta_{1}\beta_{2}}}\right|_{\mathcal{K}=\mathcal{K}_{0}} \mathcal{K}_{m,\beta_{1}\beta_{2}}[\mathcal{G}] }$} \\
& \scalebox{0.94}{${\displaystyle -\sum_{m=2}^{n} \frac{1}{m!} \sum_{\underset{\lbrace n_{1} + \cdots + n_{m} \leq n\rbrace}{n_{1},\cdots,n_{m}=1}}^{n} \int_{\beta_{1},\cdots,\beta_{2m}} \left.\frac{\delta^{m} W_{\mathrm{mix},n-(n_{1}+\cdots+n_{m})}[\mathcal{K}]}{\delta \mathcal{K}_{\beta_{1}\beta_{2}}\cdots\delta \mathcal{K}_{\beta_{2m-1}\beta_{2m}}}\right|_{\mathcal{K}=\mathcal{K}_{0}} \mathcal{K}_{n_{1},\beta_{1}\beta_{2}}[\mathcal{G}]\cdots \mathcal{K}_{n_{m},\beta_{2m-1}\beta_{2m}}[\mathcal{G}] }$} \\
& \scalebox{0.94}{${\displaystyle + \frac{1}{2}\int_{\beta_{1},\beta_{2}} \mathcal{K}_{n,\beta_{1}\beta_{2}}[\mathcal{G}] \mathcal{G}_{\beta_{1}\beta_{2}}\;, }$}
\end{split}
\label{eq:mixed2PIEAlambdaIMstep10DON}
\end{equation}
which can also be directly deduced from~\eqref{eq:pure2PIEAIMstep50DON} after replacing $\boldsymbol{K}$ by $\mathcal{K}$ and discarding the contributions of both $\vec{J}$ and $\vec{\phi}$. Furthermore, according to~\eqref{eq:mixed2PIlambdaGK0}, we have the equality:
\begin{equation}
\begin{split}
& -\sum_{m=1}^{\textcolor{red}{n}} \int_{\beta_{1},\beta_{2}} \left.\frac{\delta W_{\mathrm{mix},n-m}[\mathcal{K}]}{\delta \mathcal{K}_{\beta_{1}\beta_{2}}}\right|_{\mathcal{K}=\mathcal{K}_{0}} \mathcal{K}_{m,\beta_{1}\beta_{2}}[\mathcal{G}] + \frac{1}{2}\int_{\beta_{1},\beta_{2}} \mathcal{K}_{n,\beta_{1}\beta_{2}}[\mathcal{G}] \mathcal{G}_{\beta_{1}\beta_{2}} \\
& \hspace{0.8cm} = -\sum_{m=1}^{\textcolor{red}{n-1}} \int_{\beta_{1},\beta_{2}} \left.\frac{\delta W_{\mathrm{mix},n-m}[\mathcal{K}]}{\delta \mathcal{K}_{\beta_{1}\beta_{2}}}\right|_{\mathcal{K}=\mathcal{K}_{0}} \mathcal{K}_{m,\beta_{1}\beta_{2}}[\mathcal{G}] + \frac{1}{2}\int_{\beta_{1},\beta_{2}} \mathcal{K}_{0,\beta_{1}\beta_{2}}[\mathcal{G}] \mathcal{G}_{\beta_{1}\beta_{2}} \delta_{n 0}\;,
\end{split}
\end{equation}
which enables us to simplify~\eqref{eq:mixed2PIEAlambdaIMstep10DON} as follows:
\begin{equation}
\begin{split}
\scalebox{0.94}{${\displaystyle \Gamma_{\mathrm{mix},n}^{(\mathrm{2PI})}[\mathcal{G}] = }$} & \scalebox{0.94}{${\displaystyle -W_{\mathrm{mix},n}[\mathcal{K}=\mathcal{K}_{0}] -\sum_{m=1}^{n-1} \int_{\beta_{1},\beta_{2}} \left.\frac{\delta W_{\mathrm{mix},n-m}[\mathcal{K}]}{\delta \mathcal{K}_{\beta_{1}\beta_{2}}}\right|_{\mathcal{K}=\mathcal{K}_{0}} \mathcal{K}_{m,\beta_{1}\beta_{2}}[\mathcal{G}] }$} \\
& \scalebox{0.94}{${\displaystyle -\sum_{m=2}^{n} \frac{1}{m!} \sum_{\underset{\lbrace n_{1} + \cdots + n_{m} \leq n\rbrace}{n_{1},\cdots,n_{m}=1}}^{n} \int_{\beta_{1},\cdots,\beta_{2m}} \left.\frac{\delta^{m} W_{\mathrm{mix},n-(n_{1}+\cdots+n_{m})}[\mathcal{K}]}{\delta \mathcal{K}_{\beta_{1}\beta_{2}}\cdots\delta \mathcal{K}_{\beta_{2m-1}\beta_{2m}}}\right|_{\mathcal{K}=\mathcal{K}_{0}} \mathcal{K}_{n_{1},\beta_{1}\beta_{2}}[\mathcal{G}]\cdots \mathcal{K}_{n_{m},\beta_{2m-1}\beta_{2m}}[\mathcal{G}] }$} \\
& \scalebox{0.94}{${\displaystyle + \frac{1}{2}\int_{\beta_{1},\beta_{2}} \mathcal{K}_{0,\beta_{1}\beta_{2}}[\mathcal{G}] \mathcal{G}_{\beta_{1}\beta_{2}} \delta_{n 0}\;. }$}
\end{split}
\label{eq:mixed2PIEAlambdaIMstep20DON}
\end{equation}
At $n=0,1,2~\mathrm{and}~3$,~\eqref{eq:mixed2PIEAlambdaIMstep20DON} becomes:
\begin{equation}
\begin{split}
\Gamma_{\mathrm{mix},0}^{(\mathrm{2PI})}[\mathcal{G}] = & \ -W_{\mathrm{mix},0}[\mathcal{K}=\mathcal{K}_{0}]+\frac{1}{2}\int_{\beta_{1},\beta_{2}} \mathcal{K}_{0,\beta_{1}\beta_{2}}[\mathcal{G}] \mathcal{G}_{\beta_{1}\beta_{2}} \\
= & \ -W_{\mathrm{mix},0}[\mathcal{K}=\mathcal{K}_{0}] + \frac{1}{2}\mathcal{ST}r\left[\mathcal{G}^{-1}_{0}\mathcal{G}-\mathfrak{I}\right] \;,
\end{split}
\label{eq:mixed2PIEAlambdaGamma00DON}
\end{equation}
\begin{equation}
\Gamma_{\mathrm{mix},1}^{(\mathrm{2PI})}[\mathcal{G}] = -W_{\mathrm{mix},1}[\mathcal{K}=\mathcal{K}_{0}]\;,
\label{eq:mixed2PIEAlambdaGamma10DON}
\end{equation}
\begin{equation}
\begin{split}
\Gamma_{\mathrm{mix},2}^{(\mathrm{2PI})}[\mathcal{G}] = & -W_{\mathrm{mix},2}[\mathcal{K}=\mathcal{K}_{0}] - \int_{\beta_{1},\beta_{2}} \left.\frac{\delta W_{\mathrm{mix},1}[\mathcal{K}]}{\delta \mathcal{K}_{\beta_{1}\beta_{2}}}\right|_{\mathcal{K}=\mathcal{K}_{0}}\mathcal{K}_{1,\beta_{1}\beta_{2}}[\mathcal{G}] \\
& -\frac{1}{2}\int_{\beta_{1},\beta_{2},\beta_{3},\beta_{4}} \left.\frac{\delta^{2} W_{\mathrm{mix},0}[\mathcal{K}]}{\delta \mathcal{K}_{\beta_{1}\beta_{2}}\delta \mathcal{K}_{\beta_{3}\beta_{4}}}\right|_{\mathcal{K}=\mathcal{K}_{0}}\mathcal{K}_{1,\beta_{1}\beta_{2}}[\mathcal{G}]\mathcal{K}_{1,\beta_{3}\beta_{4}}[\mathcal{G}]\;,
\end{split}
\label{eq:mixed2PIEAlambdaGamma20DON}
\end{equation}
\begin{equation}
\begin{split}
\Gamma_{\mathrm{mix},3}^{(\mathrm{2PI})}[\mathcal{G}] = & -W_{\mathrm{mix},3}[\mathcal{K}=\mathcal{K}_{0}] - \int_{\beta_{1},\beta_{2}} \left.\frac{\delta W_{\mathrm{mix},2}[\mathcal{K}]}{\delta \mathcal{K}_{\beta_{1}\beta_{2}}}\right|_{\mathcal{K}=\mathcal{K}_{0}}\mathcal{K}_{1,\beta_{1}\beta_{2}}[\mathcal{G}] \\
& - \int_{\beta_{1},\beta_{2}} \left.\frac{\delta W_{\mathrm{mix},1}[\mathcal{K}]}{\delta \mathcal{K}_{\beta_{1}\beta_{2}}}\right|_{\mathcal{K}=\mathcal{K}_{0}}\mathcal{K}_{2,\beta_{1}\beta_{2}}[\mathcal{G}] \\
& -\frac{1}{2}\int_{\beta_{1},\beta_{2},\beta_{3},\beta_{4}} \left.\frac{\delta^{2} W_{\mathrm{mix},1}[\mathcal{K}]}{\delta \mathcal{K}_{\beta_{1}\beta_{2}}\delta \mathcal{K}_{\beta_{3}\beta_{4}}}\right|_{\mathcal{K}=\mathcal{K}_{0}}\mathcal{K}_{1,\beta_{1}\beta_{2}}[\mathcal{G}]\mathcal{K}_{1,\beta_{3}\beta_{4}}[\mathcal{G}] \\
& -\int_{\beta_{1},\beta_{2},\beta_{3},\beta_{4}} \left.\frac{\delta^{2} W_{\mathrm{mix},0}[\mathcal{K}]}{\delta \mathcal{K}_{\beta_{1}\beta_{2}}\delta \mathcal{K}_{\beta_{3}\beta_{4}}}\right|_{\mathcal{K}=\mathcal{K}_{0}}\mathcal{K}_{1,\beta_{1}\beta_{2}}[\mathcal{G}]\mathcal{K}_{2,\beta_{3}\beta_{4}}[\mathcal{G}] \\
& - \frac{1}{6} \int_{\beta_{1},\beta_{2},\beta_{3},\beta_{4},\beta_{5},\beta_{6}} \left.\frac{\delta^{3} W_{\mathrm{mix},0}[\mathcal{K}]}{\delta \mathcal{K}_{\beta_{1}\beta_{2}}\delta \mathcal{K}_{\beta_{3}\beta_{4}}\delta \mathcal{K}_{\beta_{5}\beta_{6}}}\right|_{\mathcal{K}=\mathcal{K}_{0}}\mathcal{K}_{1,\beta_{1}\beta_{2}}[\mathcal{G}]\mathcal{K}_{1,\beta_{3}\beta_{4}}[\mathcal{G}]\mathcal{K}_{1,\beta_{5}\beta_{6}}[\mathcal{G}]\;,
\end{split}
\label{eq:mixed2PIEAlambdaGamma30DON}
\end{equation}
with
\begin{equation}
\mathcal{G}^{-1}_{0,x_{1}x_{2}} = \begin{pmatrix}
\left(-\nabla_{x_{1}}^{2}+m^{2}\right)\mathbb{I}_{N} & \vec{0} \\
\vec{0}^{\mathrm{T}} & 1
\end{pmatrix} \delta_{x_{1}x_{2}}\;,
\end{equation}
and the supertrace term in~\eqref{eq:mixed2PIEAlambdaGamma00DON} is obtained in the same way as in~\eqref{eq:pure2PIEAIMGamma1step20DON} and~\eqref{eq:mixed2PIEAIMGamma10DON}. From~\eqref{eq:mixed2PIEAlambdaGamma00DON} and~\eqref{eq:mixed2PIEAlambdaGamma10DON} combined with~\eqref{eq:mixed2PIEAlambdaWExpansionstep30DON}, it directly follows that:
\begin{equation}
\Gamma_{\mathrm{mix},0}^{(\mathrm{2PI})}[\mathcal{G}] = -\frac{1}{2}\mathcal{ST}r\left[\ln\big(\mathcal{G}\big)\right] + \frac{1}{2}\mathcal{ST}r\left[\mathcal{G}^{-1}_{0}\mathcal{G}-\mathfrak{I}\right]\;,
\label{eq:mixed2PIEAlambdaGamma0bis0DON}
\end{equation}
\begin{equation}
\Gamma_{\mathrm{mix},1}^{(\mathrm{2PI})}[\mathcal{G}] = \frac{1}{24}\begin{gathered}
\begin{fmffile}{Diagrams/Mixed2PIEAlambda_Hartree}
\begin{fmfgraph}(30,20)
\fmfleft{i}
\fmfright{o}
\fmfv{decor.shape=circle,decor.size=2.0thick,foreground=(0,,0,,1)}{v1}
\fmfv{decor.shape=circle,decor.size=2.0thick,foreground=(0,,0,,1)}{v2}
\fmf{phantom,tension=10}{i,i1}
\fmf{phantom,tension=10}{o,o1}
\fmf{plain,left,tension=0.5,foreground=(1,,0,,0)}{i1,v1,i1}
\fmf{plain,right,tension=0.5,foreground=(1,,0,,0)}{o1,v2,o1}
\fmf{wiggly,foreground=(1,,0,,0)}{v1,v2}
\end{fmfgraph}
\end{fmffile}
\end{gathered}
+\frac{1}{12}\begin{gathered}
\begin{fmffile}{Diagrams/Mixed2PIEAlambda_Fock}
\begin{fmfgraph}(15,15)
\fmfleft{i}
\fmfright{o}
\fmfv{decor.shape=circle,decor.size=2.0thick,foreground=(0,,0,,1)}{v1}
\fmfv{decor.shape=circle,decor.size=2.0thick,foreground=(0,,0,,1)}{v2}
\fmf{phantom,tension=11}{i,v1}
\fmf{phantom,tension=11}{v2,o}
\fmf{plain,left,tension=0.4,foreground=(1,,0,,0)}{v1,v2,v1}
\fmf{wiggly,foreground=(1,,0,,0)}{v1,v2}
\end{fmfgraph}
\end{fmffile}
\end{gathered}\;.
\label{eq:mixed2PIEAlambdaGamma1bis0DON}
\end{equation}
In order to determine a diagrammatic expression of $\Gamma_{\mathrm{mix},2}^{(\mathrm{2PI})}$ from~\eqref{eq:mixed2PIEAlambdaGamma20DON}, we point out that differentiating propagator lines with respect to $\mathcal{K}$ amounts to cutting these lines in half, i.e.:
\begin{equation}
\left.\frac{\delta \mathcal{G}_{\mathcal{K},\beta_{1}\beta_{2}}[\mathcal{K}]}{\delta\mathcal{K}_{\beta_{3}\beta_{4}}}\right|_{\mathcal{K}=\mathcal{K}_{0}} = \hspace{0.6cm} \begin{gathered}
\begin{fmffile}{Diagrams/Mixed2PIEAlambda_DerivGK}
\begin{fmfgraph*}(15,15)
\fmfleft{i,i0,i1,i2,i3,ibis}
\fmfright{o,o0,o1,o2,o3,obis}
\fmfv{decor.shape=circle,decor.filled=empty,decor.size=1.5thick,label.angle=180,label=$\beta_{4}$}{i0}
\fmfv{decor.shape=circle,decor.filled=empty,decor.size=1.5thick,label.angle=0,label=$\beta_{2}$}{o0}
\fmfv{decor.shape=circle,decor.filled=empty,decor.size=1.5thick,label.angle=180,label=$\beta_{1}$}{i3}
\fmfv{decor.shape=circle,decor.filled=empty,decor.size=1.5thick,label.angle=0,label=$\beta_{3}$}{o3}
\fmf{plain,tension=1.0,foreground=(1,,0,,0)}{i0,o0}
\fmf{wiggly,tension=1.0,foreground=(1,,0,,0)}{i0,o0}
\fmf{plain,tension=1.0,foreground=(1,,0,,0)}{i3,o3}
\fmf{wiggly,tension=1.0,foreground=(1,,0,,0)}{i3,o3}
\end{fmfgraph*}
\end{fmffile}
\end{gathered} \hspace{0.6cm}\;,
\label{eq:mixed2PIEAlambdadmathcalGdK0DON}
\end{equation}
and, for the components of $\mathcal{G}_{\mathcal{K}}$:
\begin{equation}
\left.\frac{\delta \boldsymbol{G}_{\mathcal{K},\alpha_{1}\alpha_{2}}[\mathcal{K}]}{\delta\mathcal{K}_{\beta_{3}\beta_{4}}}\right|_{\mathcal{K}=\mathcal{K}_{0}} = \left(1-\delta_{b_{3} N+1}\right) \left(1-\delta_{b_{4} N+1}\right) \hspace{0.6cm} \begin{gathered}
\begin{fmffile}{Diagrams/Mixed2PIEAlambda_DerivmathbfGK}
\begin{fmfgraph*}(15,15)
\fmfleft{i,i0,i1,i2,i3,ibis}
\fmfright{o,o0,o1,o2,o3,obis}
\fmfv{decor.shape=circle,decor.filled=empty,decor.size=1.5thick,label.angle=180,label=$\alpha_{4}$}{i0}
\fmfv{decor.shape=circle,decor.filled=empty,decor.size=1.5thick,label.angle=0,label=$\alpha_{2}$}{o0}
\fmfv{decor.shape=circle,decor.filled=empty,decor.size=1.5thick,label.angle=180,label=$\alpha_{1}$}{i3}
\fmfv{decor.shape=circle,decor.filled=empty,decor.size=1.5thick,label.angle=0,label=$\alpha_{3}$}{o3}
\fmf{plain,tension=1.0,foreground=(1,,0,,0)}{i0,o0}
\fmf{plain,tension=1.0,foreground=(1,,0,,0)}{i3,o3}
\end{fmfgraph*}
\end{fmffile}
\end{gathered} \hspace{0.6cm}\;,
\label{eq:mixed2PIEAlambdadmathbfGdK0DON}
\end{equation}
\begin{equation}
\left.\frac{\delta D_{\mathcal{K},x_{1}x_{2}}[\mathcal{K}]}{\delta\mathcal{K}_{\beta_{3}\beta_{4}}}\right|_{\mathcal{K}=\mathcal{K}_{0}} = \delta_{b_{3} N+1} \delta_{b_{4} N+1} \hspace{0.6cm} \begin{gathered}
\begin{fmffile}{Diagrams/Mixed2PIEAlambda_DerivDK}
\begin{fmfgraph*}(15,15)
\fmfleft{i,i0,i1,i2,i3,ibis}
\fmfright{o,o0,o1,o2,o3,obis}
\fmfv{decor.shape=circle,decor.filled=empty,decor.size=1.5thick,label.angle=180,label=$x_{4}$}{i0}
\fmfv{decor.shape=circle,decor.filled=empty,decor.size=1.5thick,label.angle=0,label=$x_{2}$}{o0}
\fmfv{decor.shape=circle,decor.filled=empty,decor.size=1.5thick,label.angle=180,label=$x_{1}$}{i3}
\fmfv{decor.shape=circle,decor.filled=empty,decor.size=1.5thick,label.angle=0,label=$x_{3}$}{o3}
\fmf{wiggly,tension=1.0,foreground=(1,,0,,0)}{i0,o0}
\fmf{wiggly,tension=1.0,foreground=(1,,0,,0)}{i3,o3}
\end{fmfgraph*}
\end{fmffile}
\end{gathered} \hspace{0.6cm}\;.
\label{eq:mixed2PIEAlambdadDK0DON}
\end{equation}
From~\eqref{eq:mixed2PIEAlambdadmathcalGdK0DON} as well as~\eqref{eq:mixed2PIlambdaGK}, we can evaluate the second-order derivative in~\eqref{eq:mixed2PIEAlambdaGamma20DON}:
\begin{equation}
\left.\frac{\delta^{2} W_{\mathrm{mix},0}[\mathcal{K}]}{\delta\mathcal{K}_{\beta_{1}\beta_{2}}\delta\mathcal{K}_{\beta_{3}\beta_{4}}}\right|_{\mathcal{K}=\mathcal{K}_{0}} = \frac{1}{2}\left.\frac{\delta \mathcal{G}_{\mathcal{K},\beta_{3}\beta_{4}}[\mathcal{K}]}{\delta\mathcal{K}_{\beta_{1}\beta_{2}}}\right|_{\mathcal{K}=\mathcal{K}_{0}} = \frac{1}{2} \hspace{0.6cm} \begin{gathered}
\begin{fmffile}{Diagrams/Mixed2PIEAlambda_DerivW0KK}
\begin{fmfgraph*}(15,15)
\fmfleft{i,i0,i1,i2,i3,ibis}
\fmfright{o,o0,o1,o2,o3,obis}
\fmfv{decor.shape=circle,decor.filled=empty,decor.size=1.5thick,label.angle=180,label=$\beta_{2}$}{i0}
\fmfv{decor.shape=circle,decor.filled=empty,decor.size=1.5thick,label.angle=0,label=$\beta_{4}$}{o0}
\fmfv{decor.shape=circle,decor.filled=empty,decor.size=1.5thick,label.angle=180,label=$\beta_{3}$}{i3}
\fmfv{decor.shape=circle,decor.filled=empty,decor.size=1.5thick,label.angle=0,label=$\beta_{1}$}{o3}
\fmf{plain,tension=1.0,foreground=(1,,0,,0)}{i0,o0}
\fmf{wiggly,tension=1.0,foreground=(1,,0,,0)}{i0,o0}
\fmf{plain,tension=1.0,foreground=(1,,0,,0)}{i3,o3}
\fmf{wiggly,tension=1.0,foreground=(1,,0,,0)}{i3,o3}
\end{fmfgraph*}
\end{fmffile}
\end{gathered} \hspace{0.6cm}\;.
\label{eq:mixed2PIEAlambdadW0dKdK0DON}
\end{equation}
The first-order derivative in~\eqref{eq:mixed2PIEAlambdaGamma20DON} can be determined from~\eqref{eq:mixed2PIEAlambdadmathbfGdK0DON} and~\eqref{eq:mixed2PIEAlambdadDK0DON}:
\begin{equation}
\begin{split}
\left.\frac{\delta W_{\mathrm{mix},1}[\mathcal{K}]}{\delta\mathcal{K}_{\beta_{1}\beta_{2}}}\right|_{\mathcal{K}=\mathcal{K}_{0}} = & -\frac{1}{12}\left(1-\delta_{b_{1} N+1}\right)\left(1-\delta_{b_{2} N+1}\right)\left(\rule{0cm}{1.2cm}\right. \hspace{0.2cm} \begin{gathered}
\begin{fmffile}{Diagrams/Mixed2PIEAlambda_dW1dK_Diag1}
\begin{fmfgraph*}(25,12)
\fmfleft{i0,i,i1}
\fmfright{o0,o,o1}
\fmfv{decor.shape=circle,decor.size=2.0thick,foreground=(0,,0,,1)}{v1}
\fmfv{decor.shape=circle,decor.size=2.0thick,foreground=(0,,0,,1)}{v2}
\fmfv{decor.shape=circle,decor.filled=empty,decor.size=1.5thick,label.dist=0.15cm,label=$\alpha_{1}$}{i1}
\fmfv{decor.shape=circle,decor.filled=empty,decor.size=1.5thick,label.dist=0.15cm,label=$\alpha_{2}$}{i0}
\fmf{phantom,tension=10}{i,i1bis}
\fmf{phantom,tension=10}{o,o1bis}
\fmf{phantom,left,tension=0.5}{i1bis,v1,i1bis}
\fmf{plain,right,tension=0.5,foreground=(1,,0,,0)}{o1bis,v2,o1bis}
\fmf{plain,tension=0,foreground=(1,,0,,0)}{v1,i1}
\fmf{plain,tension=0,foreground=(1,,0,,0)}{v1,i0}
\fmf{wiggly,foreground=(1,,0,,0)}{v1,v2}
\end{fmfgraph*}
\end{fmffile}
\end{gathered}
\hspace{0.1cm} + 2 \hspace{0.3cm}\begin{gathered}
\begin{fmffile}{Diagrams/Mixed2PIEAlambda_dW1dK_Diag2}
\begin{fmfgraph*}(22,12)
\fmfleft{i0,i,i1}
\fmfright{o0,o,o1}
\fmfv{decor.shape=circle,decor.size=2.0thick,foreground=(0,,0,,1)}{v1}
\fmfv{decor.shape=circle,decor.size=2.0thick,foreground=(0,,0,,1)}{v2}
\fmfv{decor.shape=circle,decor.filled=empty,decor.size=1.5thick,label.dist=0.15cm,label=$\alpha_{1}$}{i1}
\fmfv{decor.shape=circle,decor.filled=empty,decor.size=1.5thick,label.dist=0.15cm,label=$\alpha_{2}$}{o1}
\fmf{phantom,tension=11}{i,v1}
\fmf{phantom,tension=11}{v2,o}
\fmf{plain,tension=0,foreground=(1,,0,,0)}{v1,i1}
\fmf{plain,tension=0,foreground=(1,,0,,0)}{v2,o1}
\fmf{plain,right,tension=0.4,foreground=(1,,0,,0)}{v1,v2}
\fmf{wiggly,tension=4.0,foreground=(1,,0,,0)}{v1,v2}
\end{fmfgraph*}
\end{fmffile}
\end{gathered} \hspace{0.2cm} \left.\rule{0cm}{1.2cm} \right) \\
& -\frac{1}{24}\delta_{b_{1} N+1}\delta_{b_{2} N+1}\left(\rule{0cm}{1.2cm} \right. \begin{gathered}
\begin{fmffile}{Diagrams/Mixed2PIEAlambda_dW1dK_Diag3}
\begin{fmfgraph*}(25,12)
\fmfleft{i1,i2,i3,i4,i5,i6}
\fmfright{o1,o2,o3,o4,o5,o6}
\fmfv{decor.shape=circle,decor.size=2.0thick,foreground=(0,,0,,1)}{v1}
\fmfv{decor.shape=circle,decor.size=2.0thick,foreground=(0,,0,,1)}{v2}
\fmfv{decor.shape=circle,decor.filled=empty,decor.size=1.5thick,label.angle=0,label.dist=0.15cm,label=$x_{1}$}{v3}
\fmfv{decor.shape=circle,decor.filled=empty,decor.size=1.5thick,label.angle=0,label.dist=0.15cm,label=$x_{2}$}{v4}
\fmf{phantom,tension=3}{i2,v1}
\fmf{phantom,tension=1}{o2,v1}
\fmf{phantom,tension=3}{i5,v2}
\fmf{phantom,tension=1}{o5,v2}
\fmf{phantom,tension=1}{i2,v3}
\fmf{phantom,tension=3}{o2,v3}
\fmf{phantom,tension=1}{i5,v4}
\fmf{phantom,tension=3}{o5,v4}
\fmf{wiggly,tension=0,foreground=(1,,0,,0)}{v1,v3}
\fmf{wiggly,tension=0,foreground=(1,,0,,0)}{v2,v4}
\fmf{plain,left,tension=0,foreground=(1,,0,,0)}{v1,i2,v1}
\fmf{plain,left,tension=0,foreground=(1,,0,,0)}{v2,i5,v2}
\end{fmfgraph*}
\end{fmffile}
\end{gathered} + 2 \hspace{0.7cm} \begin{gathered}
\begin{fmffile}{Diagrams/Mixed2PIEAlambda_dW1dK_Diag4}
\begin{fmfgraph*}(25,12)
\fmfleft{i}
\fmfright{o}
\fmfv{decor.shape=circle,decor.size=2.0thick,foreground=(0,,0,,1)}{v1}
\fmfv{decor.shape=circle,decor.size=2.0thick,foreground=(0,,0,,1)}{v2}
\fmfv{decor.shape=circle,decor.filled=empty,decor.size=1.5thick,label.angle=180,label.dist=0.15cm,label=$x_{1}$}{i}
\fmfv{decor.shape=circle,decor.filled=empty,decor.size=1.5thick,label.angle=0,label.dist=0.15cm,label=$x_{2}$}{o}
\fmf{phantom,tension=2}{i,v1}
\fmf{phantom,tension=1}{o,v1}
\fmf{phantom,tension=1}{i,v2}
\fmf{phantom,tension=2}{o,v2}
\fmf{plain,left,tension=0,foreground=(1,,0,,0)}{v1,v2,v1}
\fmf{wiggly,tension=0,foreground=(1,,0,,0)}{i,v1}
\fmf{wiggly,tension=0,foreground=(1,,0,,0)}{v2,o}
\end{fmfgraph*}
\end{fmffile}
\end{gathered} \hspace{0.4cm} \left.\rule{0cm}{1.2cm} \right)\;,
\end{split}
\label{eq:mixed2PIEAlambdadW1dK0DON}
\end{equation}
where the $W_{\mathrm{mix},1}$ coefficient is given by~\eqref{eq:mixed2PIEAlambdaWExpansionstep30DON}. Only the $\mathcal{K}_{1}$ coefficient is left to determine in expression~\eqref{eq:mixed2PIEAlambdaGamma20DON} of $\Gamma^{(\mathrm{2PI})}_{\mathrm{mix},2}$. As we are dealing with a single external source in the present situation, the source coefficients can all be determined via~\eqref{eq:mixed2PIEAlambdadGammanG} (thus following the simplified implementation of the IM presented in section~\ref{sec:2PPIEA} via~\eqref{eq:ProcedureDeterminationSourcesIM}) from which we can infer:
\begin{equation}
\begin{split}
\mathcal{K}_{n,\beta_{1}\beta_{2}}[\mathcal{G}] = & \ 2 \frac{\delta \Gamma_{\mathrm{mix},n}^{(\mathrm{2PI})}[\mathcal{G}]}{\delta\mathcal{G}_{\beta_{1}\beta_{2}}} \\
= & \ 2 \int_{\beta_{3},\beta_{4}} \frac{\delta \mathcal{K}_{0,\beta_{3}\beta_{4}}[\mathcal{G}]}{\delta\mathcal{G}_{\beta_{1}\beta_{2}}} \frac{\delta \Gamma_{\mathrm{mix},n}^{(\mathrm{2PI})}[\mathcal{G}]}{\delta\mathcal{K}_{0,\beta_{3}\beta_{4}}} \\
= & \ 2 \int_{\beta_{3},\beta_{4}} \left(\frac{\delta \mathcal{G}_{\beta_{1}\beta_{2}}[\mathcal{K}_{0}]}{\delta \mathcal{K}_{0,\beta_{3}\beta_{4}}}\right)^{-1} \frac{\delta \Gamma_{\mathrm{mix},n}^{(\mathrm{2PI})}[\mathcal{G}]}{\delta\mathcal{K}_{0,\beta_{3}\beta_{4}}} \\
= & \ 2 \int_{\beta_{3},\beta_{4}} \Bigg(-\int_{\beta_{5},\beta_{6}} \mathcal{G}_{\beta_{1}\beta_{5}}[\mathcal{K}_{0}]\underbrace{\frac{\delta \mathcal{G}^{-1}_{\beta_{5}\beta_{6}}[\mathcal{K}_{0}]}{\delta \mathcal{K}_{0,\beta_{3}\beta_{4}}}}_{-\delta_{\beta_{5}\beta_{3}}\delta_{\beta_{6}\beta_{4}}} \mathcal{G}_{\beta_{6}\beta_{2}}[\mathcal{K}_{0}]\Bigg)^{-1} \frac{\delta \Gamma_{\mathrm{mix},n}^{(\mathrm{2PI})}[\mathcal{G}]}{\delta\mathcal{K}_{0,\beta_{3}\beta_{4}}} \\
= & \ 2 \int_{\beta_{3},\beta_{4}} \left( \mathcal{G}_{\beta_{1}\beta_{3}}[\mathcal{K}_{0}] \mathcal{G}_{\beta_{4}\beta_{2}}[\mathcal{K}_{0}] \right)^{-1} \frac{\delta \Gamma_{\mathrm{mix},n}^{(\mathrm{2PI})}[\mathcal{G}]}{\delta\mathcal{K}_{0,\beta_{3}\beta_{4}}}\;,
\end{split}
\label{eq:mixed2PIEAlambdaKn}
\end{equation}
where we have used~\eqref{eq:mixed2PIlambdaGK0} to introduce the Kronecker deltas in the fourth line. It should be stressed that, in the present situation, this procedure is strictly equivalent to the most general implementation of the IM (outlined e.g. in section~\ref{sec:1PIEAannIM} for the 1PI EA with~\eqref{eq:pure1PIEAphiExpansionAroundJ00DON} notably), which would amount here to Taylor expanding the $\mathcal{G}_{n}$ coefficients in the power series~\eqref{eq:2PIEAlambdaGExpansion0DON} around $\mathcal{K}=\mathcal{K}_{0}$. However, the trick of~\eqref{eq:mixed2PIEAlambdaKn} would not enable us to isolate all the source coefficients if several sources are involved in the formalism (like $\vec{J}$ and $\boldsymbol{K}$ for $\Gamma^{(\mathrm{2PI})}\big[\vec{\phi},\boldsymbol{G}\big]$), in which case the most general implementation of the IM is relevant. In order to determine $\mathcal{K}_{1}$, we can combine~\eqref{eq:mixed2PIEAlambdaGamma10DON} with~\eqref{eq:mixed2PIEAlambdaKn} at $n=1$, thus leading to:
\begin{equation}
\begin{split}
\mathcal{K}_{1,\beta_{1}\beta_{2}}[\mathcal{G}] = & \ 2 \int_{\beta_{3},\beta_{4}} \left( \mathcal{G}_{\beta_{1}\beta_{3}}[\mathcal{K}_{0}] \mathcal{G}_{\beta_{4}\beta_{2}}[\mathcal{K}_{0}] \right)^{-1} \frac{\delta \Gamma_{\mathrm{mix},1}^{(\mathrm{2PI})}[\mathcal{G}]}{\delta\mathcal{K}_{0,\beta_{3}\beta_{4}}} \\
= & - 2 \int_{\beta_{3},\beta_{4}} \left( \mathcal{G}_{\beta_{1}\beta_{3}}[\mathcal{K}_{0}] \mathcal{G}_{\beta_{4}\beta_{2}}[\mathcal{K}_{0}] \right)^{-1} \frac{\delta W_{\mathrm{mix},1}[\mathcal{K}=\mathcal{K}_{0}]}{\delta\mathcal{K}_{0,\beta_{3}\beta_{4}}}\;.
\end{split}
\label{eq:mixed2PIEAlambdaK1}
\end{equation}
Then, assuming that:
\begin{equation}
\frac{\delta W_{\mathrm{mix},1}[\mathcal{K}=\mathcal{K}_{0}]}{\delta\mathcal{K}_{0,\beta_{3}\beta_{4}}}=\left.\frac{\delta W_{\mathrm{mix},1}[\mathcal{K}]}{\delta\mathcal{K}_{\beta_{3}\beta_{4}}}\right|_{\mathcal{K}=\mathcal{K}_{0}}\;,
\end{equation}
we can insert~\eqref{eq:mixed2PIEAlambdadW1dK0DON} into~\eqref{eq:mixed2PIEAlambdaK1} to express the $\mathcal{K}_{1}$ coefficient diagrammatically. By doing so, the factor $\left( \mathcal{G}_{\beta_{1}\beta_{3}}[\mathcal{K}_{0}] \mathcal{G}_{\beta_{4}\beta_{2}}[\mathcal{K}_{0}] \right)^{-1}$ in~\eqref{eq:mixed2PIEAlambdaK1} will cancel out with all propagator lines connected to an external point in~\eqref{eq:mixed2PIEAlambdadW1dK0DON}, thus yielding:
\begin{equation}
\begin{split}
\mathcal{K}_{1,\beta_{1}\beta_{2}}[\mathcal{G}] = & \ \left(1-\delta_{b_{1} N+1}\right)\left(1-\delta_{b_{2} N+1}\right)\left(\rule{0cm}{1.0cm}\right. \frac{1}{6} \hspace{-0.15cm} \begin{gathered}
\begin{fmffile}{Diagrams/Mixed2PIEAlambda_K1_Diag1}
\begin{fmfgraph*}(25,12)
\fmfleft{i0,i,i1}
\fmfright{o0,o,o1}
\fmfv{decor.shape=circle,decor.size=2.0thick,foreground=(0,,0,,1)}{v1}
\fmfv{decor.shape=circle,decor.size=2.0thick,foreground=(0,,0,,1)}{v2}
\fmfv{decor.shape=circle,decor.filled=empty,decor.size=1.5thick,label.angle=135,label.dist=0.1cm,label=$\alpha_{1}$}{i1bis2}
\fmfv{decor.shape=circle,decor.filled=empty,decor.size=1.5thick,label.angle=-135,label.dist=0.1cm,label=$\alpha_{2}$}{i0bis2}
\fmf{phantom,tension=2.5}{i,i1bis2}
\fmf{phantom,tension=1.27}{o,i1bis2}
\fmf{phantom,tension=0.7}{i1,i1bis2}
\fmf{phantom,tension=2.5}{i,i0bis2}
\fmf{phantom,tension=1.27}{o,i0bis2}
\fmf{phantom,tension=0.7}{i0,i0bis2}
\fmf{phantom,tension=10}{i,i1bis}
\fmf{phantom,tension=10}{o,o1bis}
\fmf{phantom,left,tension=0.5}{i1bis,v1,i1bis}
\fmf{plain,right,tension=0.5,foreground=(1,,0,,0)}{o1bis,v2,o1bis}
\fmf{phantom,tension=0}{v1,i1}
\fmf{phantom,tension=0}{v1,i0}
\fmf{wiggly,foreground=(1,,0,,0)}{v1,v2}
\end{fmfgraph*}
\end{fmffile}
\end{gathered}
\hspace{0.1cm} + \frac{1}{3} \hspace{0.1cm}\begin{gathered}
\begin{fmffile}{Diagrams/Mixed2PIEAlambda_K1_Diag2}
\begin{fmfgraph*}(22,12)
\fmfleft{i0,i,i1}
\fmfright{o0,o,o1}
\fmfv{decor.shape=circle,decor.size=2.0thick,foreground=(0,,0,,1)}{v1}
\fmfv{decor.shape=circle,decor.size=2.0thick,foreground=(0,,0,,1)}{v2}
\fmfv{decor.shape=circle,decor.filled=empty,decor.size=1.5thick,label.angle=135,label.dist=0.1cm,label=$\alpha_{1}$}{i1bis2}
\fmfv{decor.shape=circle,decor.filled=empty,decor.size=1.5thick,label.angle=45,label.dist=0.1cm,label=$\alpha_{2}$}{o1bis2}
\fmf{phantom,tension=3.15}{i,i1bis2}
\fmf{phantom,tension=1.1}{o,i1bis2}
\fmf{phantom,tension=1.3}{i1,i1bis2}
\fmf{phantom,tension=1.1}{i,o1bis2}
\fmf{phantom,tension=3.15}{o,o1bis2}
\fmf{phantom,tension=1.3}{o1,o1bis2}
\fmf{phantom,tension=11}{i,v1}
\fmf{phantom,tension=11}{v2,o}
\fmf{phantom,tension=0,foreground=(1,,0,,0)}{v1,i1}
\fmf{phantom,tension=0,foreground=(1,,0,,0)}{v2,o1}
\fmf{plain,right,tension=0.4,foreground=(1,,0,,0)}{v1,v2}
\fmf{wiggly,tension=4.0,foreground=(1,,0,,0)}{v1,v2}
\end{fmfgraph*}
\end{fmffile}
\end{gathered} \left.\rule{0cm}{1.0cm} \right) \\
& + \delta_{b_{1} N+1}\delta_{b_{2} N+1}\left(\rule{0cm}{1.0cm} \right. \frac{1}{12} \hspace{0.15cm} \begin{gathered}
\begin{fmffile}{Diagrams/Mixed2PIEAlambda_K1_Diag3}
\begin{fmfgraph*}(25,12)
\fmfleft{i1,i2,i3,i4,i5,i6}
\fmfright{o1,o2,o3,o4,o5,o6}
\fmfv{decor.shape=circle,decor.size=2.0thick,foreground=(0,,0,,1)}{v1}
\fmfv{decor.shape=circle,decor.size=2.0thick,foreground=(0,,0,,1)}{v2}
\fmfv{decor.shape=circle,decor.filled=empty,decor.size=1.5thick,label.angle=0,label.dist=0.15cm,label=$x_{1}$}{v3}
\fmfv{decor.shape=circle,decor.filled=empty,decor.size=1.5thick,label.angle=0,label.dist=0.15cm,label=$x_{2}$}{v4}
\fmf{phantom,tension=3}{i2,v1}
\fmf{phantom,tension=1}{o2,v1}
\fmf{phantom,tension=3}{i5,v2}
\fmf{phantom,tension=1}{o5,v2}
\fmf{phantom,tension=3}{i2,v3}
\fmf{phantom,tension=1.35}{o2,v3}
\fmf{phantom,tension=3}{i5,v4}
\fmf{phantom,tension=1.35}{o5,v4}
\fmf{plain,left,tension=0,foreground=(1,,0,,0)}{v1,i2,v1}
\fmf{plain,left,tension=0,foreground=(1,,0,,0)}{v2,i5,v2}
\end{fmfgraph*}
\end{fmffile}
\end{gathered} \hspace{-1.1cm} + \frac{1}{6} \begin{gathered}
\begin{fmffile}{Diagrams/Mixed2PIEAlambda_K1_Diag4}
\begin{fmfgraph*}(25,12)
\fmfleft{i}
\fmfright{o}
\fmfv{decor.shape=circle,decor.size=2.0thick,foreground=(0,,0,,1)}{v1}
\fmfv{decor.shape=circle,decor.size=2.0thick,foreground=(0,,0,,1)}{v2}
\fmfv{decor.shape=circle,decor.filled=empty,decor.size=1.5thick,label.angle=180,label.dist=0.15cm,label=$x_{1}$}{ibis}
\fmfv{decor.shape=circle,decor.filled=empty,decor.size=1.5thick,label.angle=0,label.dist=0.15cm,label=$x_{2}$}{obis}
\fmf{phantom,tension=2.6}{i,ibis}
\fmf{phantom,tension=1}{o,ibis}
\fmf{phantom,tension=1}{i,obis}
\fmf{phantom,tension=2.6}{o,obis}
\fmf{phantom,tension=2}{i,v1}
\fmf{phantom,tension=1}{o,v1}
\fmf{phantom,tension=1}{i,v2}
\fmf{phantom,tension=2}{o,v2}
\fmf{plain,left,tension=0,foreground=(1,,0,,0)}{v1,v2,v1}
\end{fmfgraph*}
\end{fmffile}
\end{gathered} \hspace{-0.3cm} \left.\rule{0cm}{1.0cm} \right)\;.
\end{split}
\label{eq:mixed2PIEAlambdaK1bis}
\end{equation}
As a next step, we obtain from~\eqref{eq:mixed2PIEAlambdadW0dKdK0DON},~\eqref{eq:mixed2PIEAlambdadW1dK0DON} and~\eqref{eq:mixed2PIEAlambdaK1bis}:
\begin{equation}
\begin{split}
\int_{\beta_{1},\beta_{2}} \left.\frac{\delta W_{\mathrm{mix},1}[\mathcal{K}]}{\delta \mathcal{K}_{\beta_{1}\beta_{2}}}\right|_{\mathcal{K}=\mathcal{K}_{0}} \mathcal{K}_{1,\beta_{1}\beta_{2}}[\mathcal{G}] = & - \frac{1}{18} \hspace{0.3cm} \begin{gathered}
\begin{fmffile}{Diagrams/Mixed2PIEAlambda_Diag2bis}
\begin{fmfgraph}(10,10)
\fmfleft{i0,i1}
\fmfright{o0,o1}
\fmftop{v1,vUp,v2}
\fmfbottom{v3,vDown,v4}
\fmfv{decor.shape=circle,decor.size=2.0thick,foreground=(0,,0,,1)}{v1}
\fmfv{decor.shape=circle,decor.size=2.0thick,foreground=(0,,0,,1)}{v2}
\fmfv{decor.shape=circle,decor.size=2.0thick,foreground=(0,,0,,1)}{v3}
\fmfv{decor.shape=circle,decor.size=2.0thick,foreground=(0,,0,,1)}{v4}
\fmf{phantom,tension=20}{i0,v1}
\fmf{phantom,tension=20}{i1,v3}
\fmf{phantom,tension=20}{o0,v2}
\fmf{phantom,tension=20}{o1,v4}
\fmf{plain,left=0.4,tension=0.5,foreground=(1,,0,,0)}{v3,v1}
\fmf{phantom,left=0.1,tension=0.5}{v1,vUp}
\fmf{phantom,left=0.1,tension=0.5}{vUp,v2}
\fmf{plain,left=0.4,tension=0.0,foreground=(1,,0,,0)}{v1,v2}
\fmf{plain,left=0.4,tension=0.5,foreground=(1,,0,,0)}{v2,v4}
\fmf{phantom,left=0.1,tension=0.5}{v4,vDown}
\fmf{phantom,left=0.1,tension=0.5}{vDown,v3}
\fmf{plain,left=0.4,tension=0.0,foreground=(1,,0,,0)}{v4,v3}
\fmf{wiggly,left=0.4,tension=0.5,foreground=(1,,0,,0)}{v1,v3}
\fmf{wiggly,right=0.4,tension=0.5,foreground=(1,,0,,0)}{v2,v4}
\end{fmfgraph}
\end{fmffile}
\end{gathered} \hspace{0.27cm} - \frac{1}{18} \hspace{-0.15cm} \begin{gathered}
\begin{fmffile}{Diagrams/Mixed2PIEAlambda_Diag3bis}
\begin{fmfgraph}(30,15)
\fmfleft{i}
\fmfright{o}
\fmftop{vUpLeft1,vUpLeft2,vUpLeft3,vUpRight1,vUpRight2,vUpRight3}
\fmfbottom{vDownLeft1,vDownLeft2,vDownLeft3,vDownRight1,vDownRight2,vDownRight3}
\fmfv{decor.shape=circle,decor.size=2.0thick,foreground=(0,,0,,1)}{v1}
\fmfv{decor.shape=circle,decor.size=2.0thick,foreground=(0,,0,,1)}{v2}
\fmfv{decor.shape=circle,decor.size=2.0thick,foreground=(0,,0,,1)}{vUpLeft}
\fmfv{decor.shape=circle,decor.size=2.0thick,foreground=(0,,0,,1)}{vDownLeft}
\fmf{phantom,tension=10}{i,v3}
\fmf{phantom,tension=10}{o,v4}
\fmf{plain,right=0.4,tension=0.5,foreground=(1,,0,,0)}{v1,vUpLeft}
\fmf{plain,right,tension=0.5,foreground=(1,,0,,0)}{vUpLeft,vDownLeft}
\fmf{plain,left=0.4,tension=0.5,foreground=(1,,0,,0)}{v1,vDownLeft}
\fmf{phantom,right=0.4,tension=0.5}{vUpRight,v2}
\fmf{phantom,left,tension=0.5}{vUpRight,vDownRight}
\fmf{phantom,left=0.4,tension=0.5}{vDownRight,v2}
\fmf{phantom,tension=0.3}{v2bis,o}
\fmf{plain,left,tension=0.1,foreground=(1,,0,,0)}{v2,v2bis,v2}
\fmf{wiggly,tension=2.7,foreground=(1,,0,,0)}{v1,v2}
\fmf{phantom,tension=2}{v1,v3}
\fmf{phantom,tension=2}{v2,v4}
\fmf{phantom,tension=2.4}{vUpLeft,vUpLeft2}
\fmf{phantom,tension=2.4}{vDownLeft,vDownLeft2}
\fmf{phantom,tension=2.4}{vUpRight,vUpRight2}
\fmf{phantom,tension=2.4}{vDownRight,vDownRight2}
\fmf{wiggly,tension=0,foreground=(1,,0,,0)}{vUpLeft,vDownLeft}
\end{fmfgraph}
\end{fmffile}
\end{gathered} \hspace{-0.35cm} - \frac{1}{72} \hspace{0.35cm} \begin{gathered}
\begin{fmffile}{Diagrams/Mixed2PIEAlambda_Diag4bis}
\begin{fmfgraph}(12,12)
\fmfleft{i0,i1}
\fmfright{o0,o1}
\fmftop{v1,vUp,v2}
\fmfbottom{v3,vDown,v4}
\fmfv{decor.shape=circle,decor.size=2.0thick,foreground=(0,,0,,1)}{v1}
\fmfv{decor.shape=circle,decor.size=2.0thick,foreground=(0,,0,,1)}{v2}
\fmfv{decor.shape=circle,decor.size=2.0thick,foreground=(0,,0,,1)}{v3}
\fmfv{decor.shape=circle,decor.size=2.0thick,foreground=(0,,0,,1)}{v4}
\fmf{phantom,tension=20}{i0,v1}
\fmf{phantom,tension=20}{i1,v3}
\fmf{phantom,tension=20}{o0,v2}
\fmf{phantom,tension=20}{o1,v4}
\fmf{plain,left=0.4,tension=0.5,foreground=(1,,0,,0)}{v3,v1}
\fmf{phantom,left=0.1,tension=0.5}{v1,vUp}
\fmf{phantom,left=0.1,tension=0.5}{vUp,v2}
\fmf{wiggly,tension=0.0,foreground=(1,,0,,0)}{v1,v2}
\fmf{plain,left=0.4,tension=0.5,foreground=(1,,0,,0)}{v2,v4}
\fmf{phantom,left=0.1,tension=0.5}{v4,vDown}
\fmf{phantom,left=0.1,tension=0.5}{vDown,v3}
\fmf{wiggly,tension=0.0,foreground=(1,,0,,0)}{v4,v3}
\fmf{plain,left=0.4,tension=0.5,foreground=(1,,0,,0)}{v1,v3}
\fmf{plain,right=0.4,tension=0.5,foreground=(1,,0,,0)}{v2,v4}
\end{fmfgraph}
\end{fmffile}
\end{gathered} \\
& - \frac{1}{36} \hspace{0.1cm} \begin{gathered}
\begin{fmffile}{Diagrams/Mixed2PIEAlambda_Diag5bis}
\begin{fmfgraph}(30,15)
\fmfleft{i}
\fmfright{o}
\fmfv{decor.shape=circle,decor.size=2.0thick,foreground=(0,,0,,1)}{v1}
\fmfv{decor.shape=circle,decor.size=2.0thick,foreground=(0,,0,,1)}{v2}
\fmfv{decor.shape=circle,decor.size=2.0thick,foreground=(0,,0,,1)}{v3}
\fmfv{decor.shape=circle,decor.size=2.0thick,foreground=(0,,0,,1)}{v4}
\fmf{phantom,tension=10}{i,i1}
\fmf{phantom,tension=10}{o,o1}
\fmf{plain,left,tension=0.5,foreground=(1,,0,,0)}{i1,v1,i1}
\fmf{plain,right,tension=0.5,foreground=(1,,0,,0)}{o1,v2,o1}
\fmf{wiggly,foreground=(1,,0,,0)}{v1,v3}
\fmf{plain,left,tension=0.5,foreground=(1,,0,,0)}{v3,v4}
\fmf{plain,right,tension=0.5,foreground=(1,,0,,0)}{v3,v4}
\fmf{wiggly,foreground=(1,,0,,0)}{v4,v2}
\end{fmfgraph}
\end{fmffile}
\end{gathered} \hspace{0.1cm} -\frac{1}{288} \hspace{0.1cm} \begin{gathered}
\begin{fmffile}{Diagrams/Mixed2PIEAlambda_Diag6bis}
\begin{fmfgraph*}(25,12)
\fmfleft{i1,i2,i3,i4,i5,i6}
\fmfright{o1,o2,o3,o4,o5,o6}
\fmfv{decor.shape=circle,decor.size=2.0thick,foreground=(0,,0,,1)}{v1}
\fmfv{decor.shape=circle,decor.size=2.0thick,foreground=(0,,0,,1)}{v2}
\fmfv{decor.shape=circle,decor.size=2.0thick,foreground=(0,,0,,1)}{v3}
\fmfv{decor.shape=circle,decor.size=2.0thick,foreground=(0,,0,,1)}{v4}
\fmf{phantom,tension=3}{i2,v1}
\fmf{phantom,tension=1}{o2,v1}
\fmf{phantom,tension=3}{i5,v2}
\fmf{phantom,tension=1}{o5,v2}
\fmf{phantom,tension=1}{i2,v3}
\fmf{phantom,tension=3}{o2,v3}
\fmf{phantom,tension=1}{i5,v4}
\fmf{phantom,tension=3}{o5,v4}
\fmf{wiggly,tension=0,foreground=(1,,0,,0)}{v1,v3}
\fmf{wiggly,tension=0,foreground=(1,,0,,0)}{v2,v4}
\fmf{plain,left,tension=0,foreground=(1,,0,,0)}{v1,i2,v1}
\fmf{plain,left,tension=0,foreground=(1,,0,,0)}{v2,i5,v2}
\fmf{plain,left,tension=0,foreground=(1,,0,,0)}{v3,o2,v3}
\fmf{plain,left,tension=0,foreground=(1,,0,,0)}{v4,o5,v4}
\end{fmfgraph*}
\end{fmffile}
\end{gathered} \;,
\end{split}
\label{eq:mixed2PIlambdadW1dK0K1}
\end{equation}
\begin{equation}
\begin{split}
\frac{1}{2}\int_{\beta_{1},\beta_{2},\beta_{3},\beta_{4}} \left.\frac{\delta^{2} W_{\mathrm{mix},0}[\mathcal{K}]}{\delta \mathcal{K}_{\beta_{1}\beta_{2}}\delta \mathcal{K}_{\beta_{3}\beta_{4}}}\right|_{\mathcal{K}=\mathcal{K}_{0}}\mathcal{K}_{1,\beta_{1}\beta_{2}}[\mathcal{G}]\mathcal{K}_{1,\beta_{3}\beta_{4}}[\mathcal{G}] = & \ \frac{1}{36} \hspace{0.3cm} \begin{gathered}
\begin{fmffile}{Diagrams/Mixed2PIEAlambda_Diag2bis}
\begin{fmfgraph}(10,10)
\fmfleft{i0,i1}
\fmfright{o0,o1}
\fmftop{v1,vUp,v2}
\fmfbottom{v3,vDown,v4}
\fmfv{decor.shape=circle,decor.size=2.0thick,foreground=(0,,0,,1)}{v1}
\fmfv{decor.shape=circle,decor.size=2.0thick,foreground=(0,,0,,1)}{v2}
\fmfv{decor.shape=circle,decor.size=2.0thick,foreground=(0,,0,,1)}{v3}
\fmfv{decor.shape=circle,decor.size=2.0thick,foreground=(0,,0,,1)}{v4}
\fmf{phantom,tension=20}{i0,v1}
\fmf{phantom,tension=20}{i1,v3}
\fmf{phantom,tension=20}{o0,v2}
\fmf{phantom,tension=20}{o1,v4}
\fmf{plain,left=0.4,tension=0.5,foreground=(1,,0,,0)}{v3,v1}
\fmf{phantom,left=0.1,tension=0.5}{v1,vUp}
\fmf{phantom,left=0.1,tension=0.5}{vUp,v2}
\fmf{plain,left=0.4,tension=0.0,foreground=(1,,0,,0)}{v1,v2}
\fmf{plain,left=0.4,tension=0.5,foreground=(1,,0,,0)}{v2,v4}
\fmf{phantom,left=0.1,tension=0.5}{v4,vDown}
\fmf{phantom,left=0.1,tension=0.5}{vDown,v3}
\fmf{plain,left=0.4,tension=0.0,foreground=(1,,0,,0)}{v4,v3}
\fmf{wiggly,left=0.4,tension=0.5,foreground=(1,,0,,0)}{v1,v3}
\fmf{wiggly,right=0.4,tension=0.5,foreground=(1,,0,,0)}{v2,v4}
\end{fmfgraph}
\end{fmffile}
\end{gathered} \hspace{0.27cm} + \frac{1}{36} \hspace{-0.15cm} \begin{gathered}
\begin{fmffile}{Diagrams/Mixed2PIEAlambda_Diag3bis}
\begin{fmfgraph}(30,15)
\fmfleft{i}
\fmfright{o}
\fmftop{vUpLeft1,vUpLeft2,vUpLeft3,vUpRight1,vUpRight2,vUpRight3}
\fmfbottom{vDownLeft1,vDownLeft2,vDownLeft3,vDownRight1,vDownRight2,vDownRight3}
\fmfv{decor.shape=circle,decor.size=2.0thick,foreground=(0,,0,,1)}{v1}
\fmfv{decor.shape=circle,decor.size=2.0thick,foreground=(0,,0,,1)}{v2}
\fmfv{decor.shape=circle,decor.size=2.0thick,foreground=(0,,0,,1)}{vUpLeft}
\fmfv{decor.shape=circle,decor.size=2.0thick,foreground=(0,,0,,1)}{vDownLeft}
\fmf{phantom,tension=10}{i,v3}
\fmf{phantom,tension=10}{o,v4}
\fmf{plain,right=0.4,tension=0.5,foreground=(1,,0,,0)}{v1,vUpLeft}
\fmf{plain,right,tension=0.5,foreground=(1,,0,,0)}{vUpLeft,vDownLeft}
\fmf{plain,left=0.4,tension=0.5,foreground=(1,,0,,0)}{v1,vDownLeft}
\fmf{phantom,right=0.4,tension=0.5}{vUpRight,v2}
\fmf{phantom,left,tension=0.5}{vUpRight,vDownRight}
\fmf{phantom,left=0.4,tension=0.5}{vDownRight,v2}
\fmf{phantom,tension=0.3}{v2bis,o}
\fmf{plain,left,tension=0.1,foreground=(1,,0,,0)}{v2,v2bis,v2}
\fmf{wiggly,tension=2.7,foreground=(1,,0,,0)}{v1,v2}
\fmf{phantom,tension=2}{v1,v3}
\fmf{phantom,tension=2}{v2,v4}
\fmf{phantom,tension=2.4}{vUpLeft,vUpLeft2}
\fmf{phantom,tension=2.4}{vDownLeft,vDownLeft2}
\fmf{phantom,tension=2.4}{vUpRight,vUpRight2}
\fmf{phantom,tension=2.4}{vDownRight,vDownRight2}
\fmf{wiggly,tension=0,foreground=(1,,0,,0)}{vUpLeft,vDownLeft}
\end{fmfgraph}
\end{fmffile}
\end{gathered} \\
& + \frac{1}{144} \hspace{0.35cm} \begin{gathered}
\begin{fmffile}{Diagrams/Mixed2PIEAlambda_Diag4bis}
\begin{fmfgraph}(12,12)
\fmfleft{i0,i1}
\fmfright{o0,o1}
\fmftop{v1,vUp,v2}
\fmfbottom{v3,vDown,v4}
\fmfv{decor.shape=circle,decor.size=2.0thick,foreground=(0,,0,,1)}{v1}
\fmfv{decor.shape=circle,decor.size=2.0thick,foreground=(0,,0,,1)}{v2}
\fmfv{decor.shape=circle,decor.size=2.0thick,foreground=(0,,0,,1)}{v3}
\fmfv{decor.shape=circle,decor.size=2.0thick,foreground=(0,,0,,1)}{v4}
\fmf{phantom,tension=20}{i0,v1}
\fmf{phantom,tension=20}{i1,v3}
\fmf{phantom,tension=20}{o0,v2}
\fmf{phantom,tension=20}{o1,v4}
\fmf{plain,left=0.4,tension=0.5,foreground=(1,,0,,0)}{v3,v1}
\fmf{phantom,left=0.1,tension=0.5}{v1,vUp}
\fmf{phantom,left=0.1,tension=0.5}{vUp,v2}
\fmf{wiggly,tension=0.0,foreground=(1,,0,,0)}{v1,v2}
\fmf{plain,left=0.4,tension=0.5,foreground=(1,,0,,0)}{v2,v4}
\fmf{phantom,left=0.1,tension=0.5}{v4,vDown}
\fmf{phantom,left=0.1,tension=0.5}{vDown,v3}
\fmf{wiggly,tension=0.0,foreground=(1,,0,,0)}{v4,v3}
\fmf{plain,left=0.4,tension=0.5,foreground=(1,,0,,0)}{v1,v3}
\fmf{plain,right=0.4,tension=0.5,foreground=(1,,0,,0)}{v2,v4}
\end{fmfgraph}
\end{fmffile}
\end{gathered} \hspace{0.25cm} + \frac{1}{72} \hspace{0.1cm} \begin{gathered}
\begin{fmffile}{Diagrams/Mixed2PIEAlambda_Diag5bis}
\begin{fmfgraph}(30,15)
\fmfleft{i}
\fmfright{o}
\fmfv{decor.shape=circle,decor.size=2.0thick,foreground=(0,,0,,1)}{v1}
\fmfv{decor.shape=circle,decor.size=2.0thick,foreground=(0,,0,,1)}{v2}
\fmfv{decor.shape=circle,decor.size=2.0thick,foreground=(0,,0,,1)}{v3}
\fmfv{decor.shape=circle,decor.size=2.0thick,foreground=(0,,0,,1)}{v4}
\fmf{phantom,tension=10}{i,i1}
\fmf{phantom,tension=10}{o,o1}
\fmf{plain,left,tension=0.5,foreground=(1,,0,,0)}{i1,v1,i1}
\fmf{plain,right,tension=0.5,foreground=(1,,0,,0)}{o1,v2,o1}
\fmf{wiggly,foreground=(1,,0,,0)}{v1,v3}
\fmf{plain,left,tension=0.5,foreground=(1,,0,,0)}{v3,v4}
\fmf{plain,right,tension=0.5,foreground=(1,,0,,0)}{v3,v4}
\fmf{wiggly,foreground=(1,,0,,0)}{v4,v2}
\end{fmfgraph}
\end{fmffile}
\end{gathered} \\
& + \frac{1}{576} \hspace{0.1cm} \begin{gathered}
\begin{fmffile}{Diagrams/Mixed2PIEAlambda_Diag6bis}
\begin{fmfgraph*}(25,12)
\fmfleft{i1,i2,i3,i4,i5,i6}
\fmfright{o1,o2,o3,o4,o5,o6}
\fmfv{decor.shape=circle,decor.size=2.0thick,foreground=(0,,0,,1)}{v1}
\fmfv{decor.shape=circle,decor.size=2.0thick,foreground=(0,,0,,1)}{v2}
\fmfv{decor.shape=circle,decor.size=2.0thick,foreground=(0,,0,,1)}{v3}
\fmfv{decor.shape=circle,decor.size=2.0thick,foreground=(0,,0,,1)}{v4}
\fmf{phantom,tension=3}{i2,v1}
\fmf{phantom,tension=1}{o2,v1}
\fmf{phantom,tension=3}{i5,v2}
\fmf{phantom,tension=1}{o5,v2}
\fmf{phantom,tension=1}{i2,v3}
\fmf{phantom,tension=3}{o2,v3}
\fmf{phantom,tension=1}{i5,v4}
\fmf{phantom,tension=3}{o5,v4}
\fmf{wiggly,tension=0,foreground=(1,,0,,0)}{v1,v3}
\fmf{wiggly,tension=0,foreground=(1,,0,,0)}{v2,v4}
\fmf{plain,left,tension=0,foreground=(1,,0,,0)}{v1,i2,v1}
\fmf{plain,left,tension=0,foreground=(1,,0,,0)}{v2,i5,v2}
\fmf{plain,left,tension=0,foreground=(1,,0,,0)}{v3,o2,v3}
\fmf{plain,left,tension=0,foreground=(1,,0,,0)}{v4,o5,v4}
\end{fmfgraph*}
\end{fmffile}
\end{gathered} \;.
\end{split}
\label{eq:mixed2PIlambdadW0dK0dK0K1K1}
\end{equation}
According to~\eqref{eq:mixed2PIlambdadW1dK0K1},~\eqref{eq:mixed2PIlambdadW0dK0dK0K1K1} and the expression of $W_{\mathrm{mix},2}$ given by~\eqref{eq:mixed2PIEAlambdaWExpansionstep30DON},~\eqref{eq:mixed2PIEAlambdaGamma20DON} can be rewritten as follows:
\begin{equation}
\Gamma_{\mathrm{mix},2}^{(\mathrm{2PI})}[\mathcal{G}] = -\frac{1}{72} \hspace{0.3cm} \begin{gathered}
\begin{fmffile}{Diagrams/Mixed2PIEAlambda_Diag1bis}
\begin{fmfgraph}(10,10)
\fmfleft{i0,i1}
\fmfright{o0,o1}
\fmftop{v1,vUp,v2}
\fmfbottom{v3,vDown,v4}
\fmfv{decor.shape=circle,decor.size=2.0thick,foreground=(0,,0,,1)}{v1}
\fmfv{decor.shape=circle,decor.size=2.0thick,foreground=(0,,0,,1)}{v2}
\fmfv{decor.shape=circle,decor.size=2.0thick,foreground=(0,,0,,1)}{v3}
\fmfv{decor.shape=circle,decor.size=2.0thick,foreground=(0,,0,,1)}{v4}
\fmf{phantom,tension=20}{i0,v1}
\fmf{phantom,tension=20}{i1,v3}
\fmf{phantom,tension=20}{o0,v2}
\fmf{phantom,tension=20}{o1,v4}
\fmf{plain,left=0.4,tension=0.5,foreground=(1,,0,,0)}{v3,v1}
\fmf{phantom,left=0.1,tension=0.5}{v1,vUp}
\fmf{phantom,left=0.1,tension=0.5}{vUp,v2}
\fmf{plain,left=0.4,tension=0.0,foreground=(1,,0,,0)}{v1,v2}
\fmf{plain,left=0.4,tension=0.5,foreground=(1,,0,,0)}{v2,v4}
\fmf{phantom,left=0.1,tension=0.5}{v4,vDown}
\fmf{phantom,left=0.1,tension=0.5}{vDown,v3}
\fmf{plain,left=0.4,tension=0.0,foreground=(1,,0,,0)}{v4,v3}
\fmf{wiggly,tension=0.5,foreground=(1,,0,,0)}{v1,v4}
\fmf{wiggly,tension=0.5,foreground=(1,,0,,0)}{v2,v3}
\end{fmfgraph}
\end{fmffile}
\end{gathered} \hspace{0.27cm} + \frac{1}{144} \hspace{0.1cm} \begin{gathered}
\begin{fmffile}{Diagrams/Mixed2PIEAlambda_Diag5bis}
\begin{fmfgraph}(30,15)
\fmfleft{i}
\fmfright{o}
\fmfv{decor.shape=circle,decor.size=2.0thick,foreground=(0,,0,,1)}{v1}
\fmfv{decor.shape=circle,decor.size=2.0thick,foreground=(0,,0,,1)}{v2}
\fmfv{decor.shape=circle,decor.size=2.0thick,foreground=(0,,0,,1)}{v3}
\fmfv{decor.shape=circle,decor.size=2.0thick,foreground=(0,,0,,1)}{v4}
\fmf{phantom,tension=10}{i,i1}
\fmf{phantom,tension=10}{o,o1}
\fmf{plain,left,tension=0.5,foreground=(1,,0,,0)}{i1,v1,i1}
\fmf{plain,right,tension=0.5,foreground=(1,,0,,0)}{o1,v2,o1}
\fmf{wiggly,foreground=(1,,0,,0)}{v1,v3}
\fmf{plain,left,tension=0.5,foreground=(1,,0,,0)}{v3,v4}
\fmf{plain,right,tension=0.5,foreground=(1,,0,,0)}{v3,v4}
\fmf{wiggly,foreground=(1,,0,,0)}{v4,v2}
\end{fmfgraph}
\end{fmffile}
\end{gathered} \hspace{0.1cm} + \frac{1}{576} \hspace{0.1cm} \begin{gathered}
\begin{fmffile}{Diagrams/Mixed2PIEAlambda_Diag6bis}
\begin{fmfgraph*}(25,12)
\fmfleft{i1,i2,i3,i4,i5,i6}
\fmfright{o1,o2,o3,o4,o5,o6}
\fmfv{decor.shape=circle,decor.size=2.0thick,foreground=(0,,0,,1)}{v1}
\fmfv{decor.shape=circle,decor.size=2.0thick,foreground=(0,,0,,1)}{v2}
\fmfv{decor.shape=circle,decor.size=2.0thick,foreground=(0,,0,,1)}{v3}
\fmfv{decor.shape=circle,decor.size=2.0thick,foreground=(0,,0,,1)}{v4}
\fmf{phantom,tension=3}{i2,v1}
\fmf{phantom,tension=1}{o2,v1}
\fmf{phantom,tension=3}{i5,v2}
\fmf{phantom,tension=1}{o5,v2}
\fmf{phantom,tension=1}{i2,v3}
\fmf{phantom,tension=3}{o2,v3}
\fmf{phantom,tension=1}{i5,v4}
\fmf{phantom,tension=3}{o5,v4}
\fmf{wiggly,tension=0,foreground=(1,,0,,0)}{v1,v3}
\fmf{wiggly,tension=0,foreground=(1,,0,,0)}{v2,v4}
\fmf{plain,left,tension=0,foreground=(1,,0,,0)}{v1,i2,v1}
\fmf{plain,left,tension=0,foreground=(1,,0,,0)}{v2,i5,v2}
\fmf{plain,left,tension=0,foreground=(1,,0,,0)}{v3,o2,v3}
\fmf{plain,left,tension=0,foreground=(1,,0,,0)}{v4,o5,v4}
\end{fmfgraph*}
\end{fmffile}
\end{gathered} \;.
\label{eq:mixed2PIEAlambdaGamma2bis0DON}
\end{equation}
Then, in order to find a homologous expression for $\Gamma_{\mathrm{mix},3}^{(\mathrm{2PI})}$ from~\eqref{eq:mixed2PIEAlambdaGamma30DON}, we determine notably the $\mathcal{K}_{2}$ coefficient from~\eqref{eq:mixed2PIEAlambdaKn}:
\begin{equation}
\begin{split}
\mathcal{K}_{2,\beta_{1}\beta_{2}}[\mathcal{G}] = & \ \left(1-\delta_{b_{1} N+1}\right)\left(1-\delta_{b_{2} N+1}\right)\left(\rule{0cm}{1.0cm}\right. -\frac{1}{9} \hspace{0.7cm} \begin{gathered}
\begin{fmffile}{Diagrams/Mixed2PIEAlambda_K2_Diag1}
\begin{fmfgraph*}(13,10)
\fmfleft{i0,i1}
\fmfright{o0,o1}
\fmftop{v1UpL,v1,v2UpL,v3UpL,vUp,v3UpR,v2UpR,v2,v1UpR}
\fmfbottom{v1DownL,v3,v2DownL,v3DownL,vDown,v3DownR,v2DownR,v4,v1DownR}
\fmfv{decor.shape=circle,decor.filled=empty,decor.size=1.5thick,label.angle=180,label.dist=0.1cm,label=$\alpha_{1}$}{v1bis}
\fmfv{decor.shape=circle,decor.filled=empty,decor.size=1.5thick,label.angle=180,label.dist=0.1cm,label=$\alpha_{2}$}{v3bis}
\fmf{phantom,tension=10}{v1UpL,v1bis}
\fmf{phantom,tension=1.2}{v1,v1bis}
\fmf{phantom,tension=10}{v1DownL,v3bis}
\fmf{phantom,tension=1.2}{v3,v3bis}
\fmfv{decor.shape=circle,decor.size=2.0thick,foreground=(0,,0,,1)}{v1}
\fmfv{decor.shape=circle,decor.size=2.0thick,foreground=(0,,0,,1)}{v2}
\fmfv{decor.shape=circle,decor.size=2.0thick,foreground=(0,,0,,1)}{v3}
\fmfv{decor.shape=circle,decor.size=2.0thick,foreground=(0,,0,,1)}{v4}
\fmf{phantom,tension=20}{i0,v1}
\fmf{phantom,tension=20}{i1,v3}
\fmf{phantom,tension=20}{o0,v2}
\fmf{phantom,tension=20}{o1,v4}
\fmf{phantom,left=0.4,tension=0.5,foreground=(1,,0,,0)}{v3,v1}
\fmf{phantom,left=0.1,tension=0.5}{v1,vUp}
\fmf{phantom,left=0.1,tension=0.5}{vUp,v2}
\fmf{plain,left=0.4,tension=0.0,foreground=(1,,0,,0)}{v1,v2}
\fmf{plain,left=0.4,tension=0.5,foreground=(1,,0,,0)}{v2,v4}
\fmf{phantom,left=0.1,tension=0.5}{v4,vDown}
\fmf{phantom,left=0.1,tension=0.5}{vDown,v3}
\fmf{plain,left=0.4,tension=0.0,foreground=(1,,0,,0)}{v4,v3}
\fmf{wiggly,tension=0.5,foreground=(1,,0,,0)}{v1,v4}
\fmf{wiggly,tension=0.5,foreground=(1,,0,,0)}{v2,v3}
\end{fmfgraph*}
\end{fmffile}
\end{gathered} \hspace{0.27cm} + \frac{1}{36} \hspace{0.1cm} \begin{gathered}
\begin{fmffile}{Diagrams/Mixed2PIEAlambda_K2_Diag2}
\begin{fmfgraph*}(30,15)
\fmfleft{i}
\fmfright{o}
\fmftop{vUp}
\fmfbottom{vDown}
\fmfv{decor.shape=circle,decor.filled=empty,decor.size=1.5thick,label.angle=90,label.dist=0.1cm,label=$\alpha_{1}$}{v1bis}
\fmfv{decor.shape=circle,decor.filled=empty,decor.size=1.5thick,label.angle=90,label.dist=0.1cm,label=$\alpha_{2}$}{v2bis}
\fmf{phantom,tension=2.0}{i,v1bis}
\fmf{phantom,tension=1.25}{o,v1bis}
\fmf{phantom,tension=0.8}{vUp,v1bis}
\fmf{phantom,tension=1.25}{i,v2bis}
\fmf{phantom,tension=2.0}{o,v2bis}
\fmf{phantom,tension=0.8}{vUp,v2bis}
\fmfv{decor.shape=circle,decor.size=2.0thick,foreground=(0,,0,,1)}{v1}
\fmfv{decor.shape=circle,decor.size=2.0thick,foreground=(0,,0,,1)}{v2}
\fmfv{decor.shape=circle,decor.size=2.0thick,foreground=(0,,0,,1)}{v3}
\fmfv{decor.shape=circle,decor.size=2.0thick,foreground=(0,,0,,1)}{v4}
\fmf{phantom,tension=10}{i,i1}
\fmf{phantom,tension=10}{o,o1}
\fmf{plain,left,tension=0.5,foreground=(1,,0,,0)}{i1,v1,i1}
\fmf{plain,right,tension=0.5,foreground=(1,,0,,0)}{o1,v2,o1}
\fmf{wiggly,foreground=(1,,0,,0)}{v1,v3}
\fmf{phantom,left,tension=0.5,foreground=(1,,0,,0)}{v3,v4}
\fmf{plain,right,tension=0.5,foreground=(1,,0,,0)}{v3,v4}
\fmf{wiggly,foreground=(1,,0,,0)}{v4,v2}
\end{fmfgraph*}
\end{fmffile}
\end{gathered} \\
& \hspace{5.09cm} + \frac{1}{36} \hspace{0.1cm} \begin{gathered}
\begin{fmffile}{Diagrams/Mixed2PIEAlambda_K2_Diag3}
\begin{fmfgraph*}(30,15)
\fmfleft{iDown,i,iUp}
\fmfright{oDown,o,oUp}
\fmfv{decor.shape=circle,decor.filled=empty,decor.size=1.5thick,label.angle=135,label.dist=0.1cm,label=$\alpha_{1}$}{v1Up}
\fmfv{decor.shape=circle,decor.filled=empty,decor.size=1.5thick,label.angle=-135,label.dist=0.1cm,label=$\alpha_{2}$}{v1Down}
\fmf{phantom,tension=5}{i,v1Up}
\fmf{phantom,tension=1.2}{o,v1Up}
\fmf{phantom,tension=1.1}{iUp,v1Up}
\fmf{phantom,tension=5}{i,v1Down}
\fmf{phantom,tension=1.2}{o,v1Down}
\fmf{phantom,tension=1.1}{iDown,v1Down}
\fmfv{decor.shape=circle,decor.size=2.0thick,foreground=(0,,0,,1)}{v1}
\fmfv{decor.shape=circle,decor.size=2.0thick,foreground=(0,,0,,1)}{v2}
\fmfv{decor.shape=circle,decor.size=2.0thick,foreground=(0,,0,,1)}{v3}
\fmfv{decor.shape=circle,decor.size=2.0thick,foreground=(0,,0,,1)}{v4}
\fmf{phantom,tension=10}{i,i1}
\fmf{phantom,tension=10}{o,o1}
\fmf{phantom,left,tension=0.5,foreground=(1,,0,,0)}{i1,v1,i1}
\fmf{plain,right,tension=0.5,foreground=(1,,0,,0)}{o1,v2,o1}
\fmf{wiggly,foreground=(1,,0,,0)}{v1,v3}
\fmf{plain,left,tension=0.5,foreground=(1,,0,,0)}{v3,v4}
\fmf{plain,right,tension=0.5,foreground=(1,,0,,0)}{v3,v4}
\fmf{wiggly,foreground=(1,,0,,0)}{v4,v2}
\end{fmfgraph*}
\end{fmffile}
\end{gathered} \hspace{0.1cm} + \frac{1}{72} \hspace{0.1cm} \begin{gathered}
\begin{fmffile}{Diagrams/Mixed2PIEAlambda_K2_Diag4}
\begin{fmfgraph*}(25,12)
\fmfleft{i1,i2,i3,i4,i5,i6}
\fmfright{o1,o2,o3,o4,o5,o6}
\fmftop{vUp}
\fmfbottom{vDown}
\fmfv{decor.shape=circle,decor.filled=empty,decor.size=1.5thick,label.angle=135,label.dist=0.1cm,label=$\alpha_{1}$}{v2Up}
\fmfv{decor.shape=circle,decor.filled=empty,decor.size=1.5thick,label.angle=-135,label.dist=0.1cm,label=$\alpha_{2}$}{v2Down}
\fmf{phantom,tension=3.2}{i6,v2Up}
\fmf{phantom,tension=11.0}{i5,v2Up}
\fmf{phantom,tension=0.685}{o5,v2Up}
\fmf{phantom,tension=7.0}{vUp,v2Up}
\fmf{phantom,tension=6.8}{i5,v2Down}
\fmf{phantom,tension=1.4}{o5,v2Down}
\fmf{phantom,tension=0.8}{vDown,v2Down}
\fmfv{decor.shape=circle,decor.size=2.0thick,foreground=(0,,0,,1)}{v1}
\fmfv{decor.shape=circle,decor.size=2.0thick,foreground=(0,,0,,1)}{v2}
\fmfv{decor.shape=circle,decor.size=2.0thick,foreground=(0,,0,,1)}{v3}
\fmfv{decor.shape=circle,decor.size=2.0thick,foreground=(0,,0,,1)}{v4}
\fmf{phantom,tension=3}{i2,v1}
\fmf{phantom,tension=1}{o2,v1}
\fmf{phantom,tension=3}{i5,v2}
\fmf{phantom,tension=1}{o5,v2}
\fmf{phantom,tension=1}{i2,v3}
\fmf{phantom,tension=3}{o2,v3}
\fmf{phantom,tension=1}{i5,v4}
\fmf{phantom,tension=3}{o5,v4}
\fmf{wiggly,tension=0,foreground=(1,,0,,0)}{v1,v3}
\fmf{wiggly,tension=0,foreground=(1,,0,,0)}{v2,v4}
\fmf{plain,left,tension=0,foreground=(1,,0,,0)}{v1,i2,v1}
\fmf{phantom,left,tension=0,foreground=(1,,0,,0)}{v2,i5,v2}
\fmf{plain,left,tension=0,foreground=(1,,0,,0)}{v3,o2,v3}
\fmf{plain,left,tension=0,foreground=(1,,0,,0)}{v4,o5,v4}
\end{fmfgraph*}
\end{fmffile}
\end{gathered} \left.\rule{0cm}{1.0cm} \right) \\
& + \delta_{b_{1} N+1}\delta_{b_{2} N+1}\left(\rule{0cm}{1.3cm} \right. - \frac{1}{18} \begin{gathered}
\begin{fmffile}{Diagrams/Mixed2PIEAlambda_K2_Diag5}
\begin{fmfgraph*}(25,12)
\fmfleft{i}
\fmfright{o}
\fmftop{vUp}
\fmfbottom{vDown}
\fmfv{decor.shape=circle,decor.size=2.0thick,foreground=(0,,0,,1)}{v1}
\fmfv{decor.shape=circle,decor.size=2.0thick,foreground=(0,,0,,1)}{v2}
\fmfv{decor.shape=circle,decor.size=2.0thick,foreground=(0,,0,,1)}{v3}
\fmfv{decor.shape=circle,decor.size=2.0thick,foreground=(0,,0,,1)}{v4}
\fmfv{decor.shape=circle,decor.filled=empty,decor.size=1.5thick,label.angle=180,label.dist=0.15cm,label=$x_{1}$}{ibis}
\fmfv{decor.shape=circle,decor.filled=empty,decor.size=1.5thick,label.angle=0,label.dist=0.15cm,label=$x_{2}$}{obis}
\fmf{phantom,tension=4.5}{vUp,v3}
\fmf{phantom,tension=0.85}{vDown,v3}
\fmf{phantom,tension=0.85}{vUp,v4}
\fmf{phantom,tension=4.5}{vDown,v4}
\fmf{phantom,tension=2.6}{i,ibis}
\fmf{phantom,tension=1}{o,ibis}
\fmf{phantom,tension=1}{i,obis}
\fmf{phantom,tension=2.6}{o,obis}
\fmf{phantom,tension=2}{i,v1}
\fmf{phantom,tension=1}{o,v1}
\fmf{phantom,tension=1}{i,v2}
\fmf{phantom,tension=2}{o,v2}
\fmf{wiggly,tension=0,foreground=(1,,0,,0)}{v3,v4}
\fmf{plain,left,tension=0,foreground=(1,,0,,0)}{v1,v2,v1}
\end{fmfgraph*}
\end{fmffile}
\end{gathered} + \frac{1}{72} \hspace{0.1cm} \begin{gathered}
\begin{fmffile}{Diagrams/Mixed2PIEAlambda_K2_Diag6}
\begin{fmfgraph*}(30,15)
\fmfleft{iDownbis,i,iUp1,iUp2,iUpbis}
\fmfright{oDownbis,o,oUp1,oUp2,oUpbis}
\fmfv{decor.shape=circle,decor.filled=empty,decor.size=1.5thick,label.angle=0,label.dist=0.15cm,label=$x_{1}$}{vbis2}
\fmfv{decor.shape=circle,decor.filled=empty,decor.size=1.5thick,label.angle=0,label.dist=0.15cm,label=$x_{2}$}{vbisbis2}
\fmf{phantom,tension=1}{oUp2,vbis}
\fmf{phantom,tension=3}{iUp2,vbis}
\fmf{phantom,tension=1.01}{oUp2,vbis2}
\fmf{phantom,tension=3}{iUp2,vbis2}
\fmf{phantom,tension=2.025}{o,vbisbis2}
\fmf{phantom,tension=1.1}{i,vbisbis2}
\fmf{plain,left,tension=0.5,foreground=(1,,0,,0)}{iUp2,vbis,iUp2}
\fmfv{decor.shape=circle,decor.size=2.0thick,foreground=(0,,0,,1)}{vbis}
\fmfv{decor.shape=circle,decor.size=2.0thick,foreground=(0,,0,,1)}{v1}
\fmfv{decor.shape=circle,decor.size=2.0thick,foreground=(0,,0,,1)}{v3}
\fmfv{decor.shape=circle,decor.size=2.0thick,foreground=(0,,0,,1)}{v4}
\fmf{phantom,tension=10}{i,i1}
\fmf{phantom,tension=10}{o,o1}
\fmf{plain,left,tension=0.5,foreground=(1,,0,,0)}{i1,v1,i1}
\fmf{phantom,right,tension=0.5,foreground=(1,,0,,0)}{o1,v2,o1}
\fmf{wiggly,foreground=(1,,0,,0)}{v1,v3}
\fmf{plain,left,tension=0.5,foreground=(1,,0,,0)}{v3,v4}
\fmf{plain,right,tension=0.5,foreground=(1,,0,,0)}{v3,v4}
\fmf{phantom,foreground=(1,,0,,0)}{v4,v2}
\end{fmfgraph*}
\end{fmffile}
\end{gathered} \\
& \hspace{3.23cm} + \frac{1}{72} \hspace{0.1cm} \begin{gathered}
\begin{fmffile}{Diagrams/Mixed2PIEAlambda_K2_Diag7}
\begin{fmfgraph*}(30,15)
\fmfleft{iDownbis,i,iUp1,iUp2,iUpbis}
\fmfright{oDownbis,o,oUp1,oUp2,oUpbis}
\fmfv{decor.shape=circle,decor.filled=empty,decor.size=1.5thick,label.angle=0,label.dist=0.15cm,label=$x_{2}$}{vbis2}
\fmfv{decor.shape=circle,decor.filled=empty,decor.size=1.5thick,label.angle=0,label.dist=0.15cm,label=$x_{1}$}{vbisbis2}
\fmf{phantom,tension=1}{oUp2,vbis}
\fmf{phantom,tension=3}{iUp2,vbis}
\fmf{phantom,tension=1.01}{oUp2,vbis2}
\fmf{phantom,tension=3}{iUp2,vbis2}
\fmf{phantom,tension=2.025}{o,vbisbis2}
\fmf{phantom,tension=1.1}{i,vbisbis2}
\fmf{plain,left,tension=0.5,foreground=(1,,0,,0)}{iUp2,vbis,iUp2}
\fmfv{decor.shape=circle,decor.size=2.0thick,foreground=(0,,0,,1)}{vbis}
\fmfv{decor.shape=circle,decor.size=2.0thick,foreground=(0,,0,,1)}{v1}
\fmfv{decor.shape=circle,decor.size=2.0thick,foreground=(0,,0,,1)}{v3}
\fmfv{decor.shape=circle,decor.size=2.0thick,foreground=(0,,0,,1)}{v4}
\fmf{phantom,tension=10}{i,i1}
\fmf{phantom,tension=10}{o,o1}
\fmf{plain,left,tension=0.5,foreground=(1,,0,,0)}{i1,v1,i1}
\fmf{phantom,right,tension=0.5,foreground=(1,,0,,0)}{o1,v2,o1}
\fmf{wiggly,foreground=(1,,0,,0)}{v1,v3}
\fmf{plain,left,tension=0.5,foreground=(1,,0,,0)}{v3,v4}
\fmf{plain,right,tension=0.5,foreground=(1,,0,,0)}{v3,v4}
\fmf{phantom,foreground=(1,,0,,0)}{v4,v2}
\end{fmfgraph*}
\end{fmffile}
\end{gathered} \hspace{-0.5cm} + \frac{1}{144} \begin{gathered}
\begin{fmffile}{Diagrams/Mixed2PIEAlambda_K2_Diag8}
\begin{fmfgraph*}(25,17)
\fmfleft{i1,i2,i3,ibis1,iDown1,iDown2,ibis2,iUp1,iUp2,ibis3,i4,i5,i6}
\fmfright{o1,o2,o3,obis1,oDown1,oDown2,obis2,oUp1,oUp2,obis3,o4,o5,o6}
\fmfv{decor.shape=circle,decor.filled=empty,decor.size=1.5thick,label.angle=0,label.dist=0.15cm,label=$x_{2}$}{vA}
\fmfv{decor.shape=circle,decor.filled=empty,decor.size=1.5thick,label.angle=0,label.dist=0.15cm,label=$x_{1}$}{vB}
\fmfv{decor.shape=circle,decor.size=2.0thick,foreground=(0,,0,,1)}{v1}
\fmfv{decor.shape=circle,decor.size=2.0thick,foreground=(0,,0,,1)}{v2}
\fmfv{decor.shape=circle,decor.size=2.0thick,foreground=(0,,0,,1)}{v3}
\fmfv{decor.shape=circle,decor.size=2.0thick,foreground=(0,,0,,1)}{v5}
\fmf{phantom,tension=3}{i2,v1}
\fmf{phantom,tension=1}{o2,v1}
\fmf{phantom,tension=3}{i5,v2}
\fmf{phantom,tension=1}{o5,v2}
\fmf{phantom,tension=1}{i2,v3}
\fmf{phantom,tension=3}{o2,v3}
\fmf{phantom,tension=1}{i5,v4}
\fmf{phantom,tension=3}{o5,v4}
\fmf{phantom,tension=3}{i5,vB}
\fmf{phantom,tension=1.405}{o5,vB}
\fmf{phantom,tension=3}{ibis2,v5}
\fmf{phantom,tension=1}{obis2,v5}
\fmf{phantom,tension=1}{ibis2,v6}
\fmf{phantom,tension=3}{obis2,v6}
\fmf{phantom,tension=3}{ibis2,vA}
\fmf{phantom,tension=1.57}{obis2,vA}
\fmf{phantom,tension=15}{ibis2,ibis2bis}
\fmf{phantom,tension=1}{obis2,ibis2bis}
\fmf{phantom,tension=1}{ibis2,obis2bis}
\fmf{phantom,tension=15}{obis2,obis2bis}
\fmf{wiggly,tension=0,foreground=(1,,0,,0)}{v1,v3}
\fmf{phantom,tension=0,foreground=(1,,0,,0)}{v2,v4}
\fmf{phantom,tension=0.52,foreground=(1,,0,,0)}{v5,v6}
\fmf{plain,left,tension=0,foreground=(1,,0,,0)}{v1,i2,v1}
\fmf{plain,left,tension=0,foreground=(1,,0,,0)}{v2,i5,v2}
\fmf{plain,left,tension=0,foreground=(1,,0,,0)}{v3,o2,v3}
\fmf{phantom,left,tension=0,foreground=(1,,0,,0)}{v4,o5,v4}
\fmf{plain,left,tension=0.2,foreground=(1,,0,,0)}{v5,ibis2bis,v5}
\fmf{phantom,left,tension=0.2,foreground=(1,,0,,0)}{v6,obis2bis,v6}
\end{fmfgraph*}
\end{fmffile}
\end{gathered} \left.\rule{0cm}{1.3cm} \right)\;.
\end{split}
\label{eq:mixed2PIEAlambdaK2}
\end{equation}
From this, we determine a diagrammatic expression for $\Gamma_{\mathrm{mix},3}^{(\mathrm{2PI})}$ and combine it with~\eqref{eq:mixed2PIEAlambdaGamma0bis0DON}, \eqref{eq:mixed2PIEAlambdaGamma1bis0DON} and~\eqref{eq:mixed2PIEAlambdaGamma2bis0DON}, which gives us:
\begin{equation}
\begin{split}
\Gamma_{\mathrm{mix}}^{(\mathrm{2PI})}[\mathcal{G}] = & -\frac{1}{2}\mathcal{ST}r\left[\ln\big(\mathcal{G}\big)\right] + \frac{1}{2}\mathcal{ST}r\left[\mathcal{G}^{-1}_{0}\mathcal{G}-\mathfrak{I}\right] \\
& + \left( \rule{0cm}{1.0cm} \right. \frac{1}{24}\begin{gathered}
\begin{fmffile}{Diagrams/Mixed2PIEAlambda_Hartree}
\begin{fmfgraph}(30,20)
\fmfleft{i}
\fmfright{o}
\fmfv{decor.shape=circle,decor.size=2.0thick,foreground=(0,,0,,1)}{v1}
\fmfv{decor.shape=circle,decor.size=2.0thick,foreground=(0,,0,,1)}{v2}
\fmf{phantom,tension=10}{i,i1}
\fmf{phantom,tension=10}{o,o1}
\fmf{plain,left,tension=0.5,foreground=(1,,0,,0)}{i1,v1,i1}
\fmf{plain,right,tension=0.5,foreground=(1,,0,,0)}{o1,v2,o1}
\fmf{wiggly,foreground=(1,,0,,0)}{v1,v2}
\end{fmfgraph}
\end{fmffile}
\end{gathered}
+\frac{1}{12}\begin{gathered}
\begin{fmffile}{Diagrams/Mixed2PIEAlambda_Fock}
\begin{fmfgraph}(15,15)
\fmfleft{i}
\fmfright{o}
\fmfv{decor.shape=circle,decor.size=2.0thick,foreground=(0,,0,,1)}{v1}
\fmfv{decor.shape=circle,decor.size=2.0thick,foreground=(0,,0,,1)}{v2}
\fmf{phantom,tension=11}{i,v1}
\fmf{phantom,tension=11}{v2,o}
\fmf{plain,left,tension=0.4,foreground=(1,,0,,0)}{v1,v2,v1}
\fmf{wiggly,foreground=(1,,0,,0)}{v1,v2}
\end{fmfgraph}
\end{fmffile}
\end{gathered} \left. \rule{0cm}{1.0cm} \right) \\
& + \left( \rule{0cm}{1.2cm} \right. -\frac{1}{72} \hspace{0.3cm} \begin{gathered}
\begin{fmffile}{Diagrams/Mixed2PIEAlambda_Diag1bis}
\begin{fmfgraph}(10,10)
\fmfleft{i0,i1}
\fmfright{o0,o1}
\fmftop{v1,vUp,v2}
\fmfbottom{v3,vDown,v4}
\fmfv{decor.shape=circle,decor.size=2.0thick,foreground=(0,,0,,1)}{v1}
\fmfv{decor.shape=circle,decor.size=2.0thick,foreground=(0,,0,,1)}{v2}
\fmfv{decor.shape=circle,decor.size=2.0thick,foreground=(0,,0,,1)}{v3}
\fmfv{decor.shape=circle,decor.size=2.0thick,foreground=(0,,0,,1)}{v4}
\fmf{phantom,tension=20}{i0,v1}
\fmf{phantom,tension=20}{i1,v3}
\fmf{phantom,tension=20}{o0,v2}
\fmf{phantom,tension=20}{o1,v4}
\fmf{plain,left=0.4,tension=0.5,foreground=(1,,0,,0)}{v3,v1}
\fmf{phantom,left=0.1,tension=0.5}{v1,vUp}
\fmf{phantom,left=0.1,tension=0.5}{vUp,v2}
\fmf{plain,left=0.4,tension=0.0,foreground=(1,,0,,0)}{v1,v2}
\fmf{plain,left=0.4,tension=0.5,foreground=(1,,0,,0)}{v2,v4}
\fmf{phantom,left=0.1,tension=0.5}{v4,vDown}
\fmf{phantom,left=0.1,tension=0.5}{vDown,v3}
\fmf{plain,left=0.4,tension=0.0,foreground=(1,,0,,0)}{v4,v3}
\fmf{wiggly,tension=0.5,foreground=(1,,0,,0)}{v1,v4}
\fmf{wiggly,tension=0.5,foreground=(1,,0,,0)}{v2,v3}
\end{fmfgraph}
\end{fmffile}
\end{gathered} \hspace{0.27cm} + \frac{1}{144} \hspace{0.1cm} \begin{gathered}
\begin{fmffile}{Diagrams/Mixed2PIEAlambda_Diag5bis}
\begin{fmfgraph}(30,15)
\fmfleft{i}
\fmfright{o}
\fmfv{decor.shape=circle,decor.size=2.0thick,foreground=(0,,0,,1)}{v1}
\fmfv{decor.shape=circle,decor.size=2.0thick,foreground=(0,,0,,1)}{v2}
\fmfv{decor.shape=circle,decor.size=2.0thick,foreground=(0,,0,,1)}{v3}
\fmfv{decor.shape=circle,decor.size=2.0thick,foreground=(0,,0,,1)}{v4}
\fmf{phantom,tension=10}{i,i1}
\fmf{phantom,tension=10}{o,o1}
\fmf{plain,left,tension=0.5,foreground=(1,,0,,0)}{i1,v1,i1}
\fmf{plain,right,tension=0.5,foreground=(1,,0,,0)}{o1,v2,o1}
\fmf{wiggly,foreground=(1,,0,,0)}{v1,v3}
\fmf{plain,left,tension=0.5,foreground=(1,,0,,0)}{v3,v4}
\fmf{plain,right,tension=0.5,foreground=(1,,0,,0)}{v3,v4}
\fmf{wiggly,foreground=(1,,0,,0)}{v4,v2}
\end{fmfgraph}
\end{fmffile}
\end{gathered} \hspace{0.1cm} + \frac{1}{576} \hspace{0.1cm} \begin{gathered}
\begin{fmffile}{Diagrams/Mixed2PIEAlambda_Diag6bis}
\begin{fmfgraph*}(25,12)
\fmfleft{i1,i2,i3,i4,i5,i6}
\fmfright{o1,o2,o3,o4,o5,o6}
\fmfv{decor.shape=circle,decor.size=2.0thick,foreground=(0,,0,,1)}{v1}
\fmfv{decor.shape=circle,decor.size=2.0thick,foreground=(0,,0,,1)}{v2}
\fmfv{decor.shape=circle,decor.size=2.0thick,foreground=(0,,0,,1)}{v3}
\fmfv{decor.shape=circle,decor.size=2.0thick,foreground=(0,,0,,1)}{v4}
\fmf{phantom,tension=3}{i2,v1}
\fmf{phantom,tension=1}{o2,v1}
\fmf{phantom,tension=3}{i5,v2}
\fmf{phantom,tension=1}{o5,v2}
\fmf{phantom,tension=1}{i2,v3}
\fmf{phantom,tension=3}{o2,v3}
\fmf{phantom,tension=1}{i5,v4}
\fmf{phantom,tension=3}{o5,v4}
\fmf{wiggly,tension=0,foreground=(1,,0,,0)}{v1,v3}
\fmf{wiggly,tension=0,foreground=(1,,0,,0)}{v2,v4}
\fmf{plain,left,tension=0,foreground=(1,,0,,0)}{v1,i2,v1}
\fmf{plain,left,tension=0,foreground=(1,,0,,0)}{v2,i5,v2}
\fmf{plain,left,tension=0,foreground=(1,,0,,0)}{v3,o2,v3}
\fmf{plain,left,tension=0,foreground=(1,,0,,0)}{v4,o5,v4}
\end{fmfgraph*}
\end{fmffile}
\end{gathered} \left. \rule{0cm}{1.2cm} \right) \\
& + \left( \rule{0cm}{1.3cm} \right. \frac{1}{324} \hspace{0.3cm} \begin{gathered}
\begin{fmffile}{Diagrams/Mixed2PIEA_Diag7}
\begin{fmfgraph}(15,15)
\fmfleft{i}
\fmfright{o}
\fmftop{vUpLeft,vUp,vUpRight}
\fmfbottom{vDownLeft,vDown,vDownRight}
\fmfv{decor.shape=circle,decor.size=2.0thick,foreground=(0,,0,,1)}{v1}
\fmfv{decor.shape=circle,decor.size=2.0thick,foreground=(0,,0,,1)}{v2}
\fmfv{decor.shape=circle,decor.size=2.0thick,foreground=(0,,0,,1)}{v3}
\fmfv{decor.shape=circle,decor.size=2.0thick,foreground=(0,,0,,1)}{v4}
\fmfv{decor.shape=circle,decor.size=2.0thick,foreground=(0,,0,,1)}{v5}
\fmfv{decor.shape=circle,decor.size=2.0thick,foreground=(0,,0,,1)}{v6}
\fmf{phantom,tension=1}{i,v1}
\fmf{phantom,tension=1}{v2,o}
\fmf{phantom,tension=14.0}{v3,vUpLeft}
\fmf{phantom,tension=2.0}{v3,vUpRight}
\fmf{phantom,tension=4.0}{v3,i}
\fmf{phantom,tension=2.0}{v4,vUpLeft}
\fmf{phantom,tension=14.0}{v4,vUpRight}
\fmf{phantom,tension=4.0}{v4,o}
\fmf{phantom,tension=14.0}{v5,vDownLeft}
\fmf{phantom,tension=2.0}{v5,vDownRight}
\fmf{phantom,tension=4.0}{v5,i}
\fmf{phantom,tension=2.0}{v6,vDownLeft}
\fmf{phantom,tension=14.0}{v6,vDownRight}
\fmf{phantom,tension=4.0}{v6,o}
\fmf{wiggly,tension=0,foreground=(1,,0,,0)}{v1,v2}
\fmf{wiggly,tension=0.6,foreground=(1,,0,,0)}{v3,v6}
\fmf{wiggly,tension=0.6,foreground=(1,,0,,0)}{v5,v4}
\fmf{plain,left=0.18,tension=0,foreground=(1,,0,,0)}{v1,v3}
\fmf{plain,left=0.42,tension=0,foreground=(1,,0,,0)}{v3,v4}
\fmf{plain,left=0.18,tension=0,foreground=(1,,0,,0)}{v4,v2}
\fmf{plain,left=0.18,tension=0,foreground=(1,,0,,0)}{v2,v6}
\fmf{plain,left=0.42,tension=0,foreground=(1,,0,,0)}{v6,v5}
\fmf{plain,left=0.18,tension=0,foreground=(1,,0,,0)}{v5,v1}
\end{fmfgraph}
\end{fmffile}
\end{gathered} \hspace{0.3cm} + \frac{1}{108} \hspace{0.5cm} \begin{gathered}
\begin{fmffile}{Diagrams/Mixed2PIEA_Diag8}
\begin{fmfgraph}(12.5,12.5)
\fmfleft{i0,i1}
\fmfright{o0,o1}
\fmftop{v1,vUp,v2}
\fmfbottom{v3,vDown,v4}
\fmfv{decor.shape=circle,decor.size=2.0thick,foreground=(0,,0,,1)}{v1}
\fmfv{decor.shape=circle,decor.size=2.0thick,foreground=(0,,0,,1)}{v2}
\fmfv{decor.shape=circle,decor.size=2.0thick,foreground=(0,,0,,1)}{v3}
\fmfv{decor.shape=circle,decor.size=2.0thick,foreground=(0,,0,,1)}{v4}
\fmfv{decor.shape=circle,decor.size=2.0thick,foreground=(0,,0,,1)}{v5}
\fmfv{decor.shape=circle,decor.size=2.0thick,foreground=(0,,0,,1)}{v6}
\fmf{phantom,tension=20}{i0,v1}
\fmf{phantom,tension=20}{i1,v3}
\fmf{phantom,tension=20}{o0,v2}
\fmf{phantom,tension=20}{o1,v4}
\fmf{phantom,tension=0.005}{v5,v6}
\fmf{wiggly,left=0.4,tension=0,foreground=(1,,0,,0)}{v3,v1}
\fmf{phantom,left=0.1,tension=0}{v1,vUp}
\fmf{phantom,left=0.1,tension=0}{vUp,v2}
\fmf{plain,left=0.25,tension=0,foreground=(1,,0,,0)}{v1,v2}
\fmf{wiggly,left=0.4,tension=0,foreground=(1,,0,,0)}{v2,v4}
\fmf{phantom,left=0.1,tension=0}{v4,vDown}
\fmf{phantom,left=0.1,tension=0}{vDown,v3}
\fmf{plain,right=0.25,tension=0,foreground=(1,,0,,0)}{v3,v4}
\fmf{plain,left=0.2,tension=0.01,foreground=(1,,0,,0)}{v1,v5}
\fmf{plain,left=0.2,tension=0.01,foreground=(1,,0,,0)}{v5,v3}
\fmf{plain,right=0.2,tension=0.01,foreground=(1,,0,,0)}{v2,v6}
\fmf{plain,right=0.2,tension=0.01,foreground=(1,,0,,0)}{v6,v4}
\fmf{wiggly,tension=0,foreground=(1,,0,,0)}{v5,v6}
\end{fmfgraph}
\end{fmffile}
\end{gathered} \hspace{0.5cm} + \frac{1}{324} \hspace{0.4cm} \begin{gathered}
\begin{fmffile}{Diagrams/Mixed2PIEA_Diag9}
\begin{fmfgraph}(12.5,12.5)
\fmfleft{i0,i1}
\fmfright{o0,o1}
\fmftop{v1,vUp,v2}
\fmfbottom{v3,vDown,v4}
\fmfv{decor.shape=circle,decor.size=2.0thick,foreground=(0,,0,,1)}{v1}
\fmfv{decor.shape=circle,decor.size=2.0thick,foreground=(0,,0,,1)}{v2}
\fmfv{decor.shape=circle,decor.size=2.0thick,foreground=(0,,0,,1)}{v3}
\fmfv{decor.shape=circle,decor.size=2.0thick,foreground=(0,,0,,1)}{v4}
\fmfv{decor.shape=circle,decor.size=2.0thick,foreground=(0,,0,,1)}{v5}
\fmfv{decor.shape=circle,decor.size=2.0thick,foreground=(0,,0,,1)}{v6}
\fmf{phantom,tension=20}{i0,v1}
\fmf{phantom,tension=20}{i1,v3}
\fmf{phantom,tension=20}{o0,v2}
\fmf{phantom,tension=20}{o1,v4}
\fmf{phantom,tension=0.005}{v5,v6}
\fmf{plain,left=0.4,tension=0,foreground=(1,,0,,0)}{v3,v1}
\fmf{phantom,left=0.1,tension=0}{v1,vUp}
\fmf{phantom,left=0.1,tension=0}{vUp,v2}
\fmf{wiggly,left=0.25,tension=0,foreground=(1,,0,,0)}{v1,v2}
\fmf{plain,left=0.4,tension=0,foreground=(1,,0,,0)}{v2,v4}
\fmf{phantom,left=0.1,tension=0}{v4,vDown}
\fmf{phantom,left=0.1,tension=0}{vDown,v3}
\fmf{wiggly,right=0.25,tension=0,foreground=(1,,0,,0)}{v3,v4}
\fmf{plain,left=0.2,tension=0.01,foreground=(1,,0,,0)}{v1,v5}
\fmf{plain,left=0.2,tension=0.01,foreground=(1,,0,,0)}{v5,v3}
\fmf{plain,right=0.2,tension=0.01,foreground=(1,,0,,0)}{v2,v6}
\fmf{plain,right=0.2,tension=0.01,foreground=(1,,0,,0)}{v6,v4}
\fmf{wiggly,tension=0,foreground=(1,,0,,0)}{v5,v6}
\end{fmfgraph}
\end{fmffile}
\end{gathered} \\
& \hspace{0.6cm} - \frac{1}{432} \hspace{0.1cm} \begin{gathered}
\begin{fmffile}{Diagrams/Mixed2PIEAlambda_Diag7}
\begin{fmfgraph}(36,12.5)
\fmfleft{i}
\fmfright{o}
\fmftop{vUp}
\fmfbottom{vDown}
\fmfv{decor.shape=circle,decor.size=2.0thick,foreground=(0,,0,,1)}{v1}
\fmfv{decor.shape=circle,decor.size=2.0thick,foreground=(0,,0,,1)}{v2}
\fmfv{decor.shape=circle,decor.size=2.0thick,foreground=(0,,0,,1)}{v3}
\fmfv{decor.shape=circle,decor.size=2.0thick,foreground=(0,,0,,1)}{v4}
\fmfv{decor.shape=circle,decor.size=2.0thick,foreground=(0,,0,,1)}{v5}
\fmfv{decor.shape=circle,decor.size=2.0thick,foreground=(0,,0,,1)}{v6}
\fmf{phantom,tension=10}{i,i1}
\fmf{phantom,tension=10}{o,o1}
\fmf{phantom,tension=2.2}{vUp,v5}
\fmf{phantom,tension=2.2}{vDown,v6}
\fmf{phantom,tension=0.5}{v3,v4}
\fmf{plain,left,tension=0.8,foreground=(1,,0,,0)}{i1,v1}
\fmf{plain,right,tension=0.8,foreground=(1,,0,,0)}{i1,v1}
\fmf{plain,right,tension=0.8,foreground=(1,,0,,0)}{o1,v2,o1}
\fmf{wiggly,tension=1.5,foreground=(1,,0,,0)}{v1,v3}
\fmf{plain,left=0.4,tension=0.5,foreground=(1,,0,,0)}{v3,v5}
\fmf{plain,left=0.4,tension=0.5,foreground=(1,,0,,0)}{v5,v4}
\fmf{plain,right=0.4,tension=0.5,foreground=(1,,0,,0)}{v3,v6}
\fmf{plain,right=0.4,tension=0.5,foreground=(1,,0,,0)}{v6,v4}
\fmf{wiggly,tension=1.5,foreground=(1,,0,,0)}{v4,v2}
\fmf{wiggly,tension=0,foreground=(1,,0,,0)}{v5,v6}
\end{fmfgraph}
\end{fmffile}
\end{gathered} + \frac{1}{864} \hspace{0.1cm} \begin{gathered}
\begin{fmffile}{Diagrams/Mixed2PIEAlambda_Diag8}
\begin{fmfgraph}(45,12.5)
\fmfleft{i}
\fmfright{o}
\fmfv{decor.shape=circle,decor.size=2.0thick,foreground=(0,,0,,1)}{v1}
\fmfv{decor.shape=circle,decor.size=2.0thick,foreground=(0,,0,,1)}{v2}
\fmfv{decor.shape=circle,decor.size=2.0thick,foreground=(0,,0,,1)}{v3}
\fmfv{decor.shape=circle,decor.size=2.0thick,foreground=(0,,0,,1)}{v4}
\fmfv{decor.shape=circle,decor.size=2.0thick,foreground=(0,,0,,1)}{v5}
\fmfv{decor.shape=circle,decor.size=2.0thick,foreground=(0,,0,,1)}{v6}
\fmf{phantom,tension=10}{i,i1}
\fmf{phantom,tension=10}{o,o1}
\fmf{plain,left,tension=0.5,foreground=(1,,0,,0)}{i1,v1,i1}
\fmf{plain,right,tension=0.5,foreground=(1,,0,,0)}{o1,v2,o1}
\fmf{wiggly,tension=1.0,foreground=(1,,0,,0)}{v1,v3}
\fmf{plain,left,tension=0.5,foreground=(1,,0,,0)}{v3,v4}
\fmf{plain,right,tension=0.5,foreground=(1,,0,,0)}{v3,v4}
\fmf{wiggly,tension=1.0,foreground=(1,,0,,0)}{v4,v5}
\fmf{plain,left,tension=0.5,foreground=(1,,0,,0)}{v5,v6}
\fmf{plain,right,tension=0.5,foreground=(1,,0,,0)}{v5,v6}
\fmf{wiggly,tension=1.0,foreground=(1,,0,,0)}{v6,v2}
\end{fmfgraph}
\end{fmffile}
\end{gathered} \\
& \hspace{0.6cm} + \frac{1}{864} \begin{gathered}
\begin{fmffile}{Diagrams/Mixed2PIEAlambda_Diag9}
\begin{fmfgraph}(36,15)
\fmfleft{i1,i,i4,i5}
\fmfright{o1,o,o4,o5}
\fmfv{decor.shape=circle,decor.size=2.0thick,foreground=(0,,0,,1)}{v1}
\fmfv{decor.shape=circle,decor.size=2.0thick,foreground=(0,,0,,1)}{v2}
\fmfv{decor.shape=circle,decor.size=2.0thick,foreground=(0,,0,,1)}{v3}
\fmfv{decor.shape=circle,decor.size=2.0thick,foreground=(0,,0,,1)}{v4}
\fmfv{decor.shape=circle,decor.size=2.0thick,foreground=(0,,0,,1)}{v5}
\fmfv{decor.shape=circle,decor.size=2.0thick,foreground=(0,,0,,1)}{v6}
\fmf{phantom,tension=10}{i,i1}
\fmf{phantom,tension=10}{o,o1}
\fmf{phantom,tension=4}{i4,i4bis}
\fmf{phantom,tension=1}{o4,i4bis}
\fmf{phantom,tension=1}{i4,o4bis}
\fmf{phantom,tension=4}{o4,o4bis}
\fmf{phantom,tension=2}{i4,v5}
\fmf{phantom,tension=1}{o4,v5}
\fmf{phantom,tension=1}{i4,v6}
\fmf{phantom,tension=2}{o4,v6}
\fmf{plain,left,tension=0.5,foreground=(1,,0,,0)}{i1,v1,i1}
\fmf{plain,right,tension=0.5,foreground=(1,,0,,0)}{o1,v2,o1}
\fmf{plain,left,tension=0,foreground=(1,,0,,0)}{i4bis,v5,i4bis}
\fmf{plain,right,tension=0,foreground=(1,,0,,0)}{o4bis,v6,o4bis}
\fmf{wiggly,foreground=(1,,0,,0)}{v1,v3}
\fmf{plain,left,tension=0.5,foreground=(1,,0,,0)}{v3,v4}
\fmf{plain,right,tension=0.5,foreground=(1,,0,,0)}{v3,v4}
\fmf{wiggly,foreground=(1,,0,,0)}{v4,v2}
\fmf{wiggly,tension=0.5,foreground=(1,,0,,0)}{v5,v6}
\end{fmfgraph}
\end{fmffile}
\end{gathered} + \frac{1}{5184} \begin{gathered}
\begin{fmffile}{Diagrams/Mixed2PIEAlambda_Diag10}
\begin{fmfgraph*}(25,17)
\fmfleft{i1,i2,i3,ibis1,iDown1,iDown2,ibis2,iUp1,iUp2,ibis3,i4,i5,i6}
\fmfright{o1,o2,o3,obis1,oDown1,oDown2,obis2,oUp1,oUp2,obis3,o4,o5,o6}
\fmfv{decor.shape=circle,decor.size=2.0thick,foreground=(0,,0,,1)}{v1}
\fmfv{decor.shape=circle,decor.size=2.0thick,foreground=(0,,0,,1)}{v2}
\fmfv{decor.shape=circle,decor.size=2.0thick,foreground=(0,,0,,1)}{v3}
\fmfv{decor.shape=circle,decor.size=2.0thick,foreground=(0,,0,,1)}{v4}
\fmfv{decor.shape=circle,decor.size=2.0thick,foreground=(0,,0,,1)}{v5}
\fmfv{decor.shape=circle,decor.size=2.0thick,foreground=(0,,0,,1)}{v6}
\fmf{phantom,tension=3}{i2,v1}
\fmf{phantom,tension=1}{o2,v1}
\fmf{phantom,tension=3}{i5,v2}
\fmf{phantom,tension=1}{o5,v2}
\fmf{phantom,tension=1}{i2,v3}
\fmf{phantom,tension=3}{o2,v3}
\fmf{phantom,tension=1}{i5,v4}
\fmf{phantom,tension=3}{o5,v4}
\fmf{phantom,tension=3}{ibis2,v5}
\fmf{phantom,tension=1}{obis2,v5}
\fmf{phantom,tension=1}{ibis2,v6}
\fmf{phantom,tension=3}{obis2,v6}
\fmf{phantom,tension=15}{ibis2,ibis2bis}
\fmf{phantom,tension=1}{obis2,ibis2bis}
\fmf{phantom,tension=1}{ibis2,obis2bis}
\fmf{phantom,tension=15}{obis2,obis2bis}
\fmf{wiggly,tension=0,foreground=(1,,0,,0)}{v1,v3}
\fmf{wiggly,tension=0,foreground=(1,,0,,0)}{v2,v4}
\fmf{wiggly,tension=0.52,foreground=(1,,0,,0)}{v5,v6}
\fmf{plain,left,tension=0,foreground=(1,,0,,0)}{v1,i2,v1}
\fmf{plain,left,tension=0,foreground=(1,,0,,0)}{v2,i5,v2}
\fmf{plain,left,tension=0,foreground=(1,,0,,0)}{v3,o2,v3}
\fmf{plain,left,tension=0,foreground=(1,,0,,0)}{v4,o5,v4}
\fmf{plain,left,tension=0.2,foreground=(1,,0,,0)}{v5,ibis2bis,v5}
\fmf{plain,left,tension=0.2,foreground=(1,,0,,0)}{v6,obis2bis,v6}
\end{fmfgraph*}
\end{fmffile}
\end{gathered} \left. \rule{0cm}{1.3cm} \right) \\
& + \mathcal{O}\big(\lambda^{4}\big) \;.
\end{split}
\label{eq:mixed2PIEAlambdafinalexpressionAppendix}
\end{equation}

\section{\label{sec:4PPIEAannIM}4PPI effective action}

For the original 4PPI EA with vanishing 1-point correlation function, the IM relies on the following power series:
\begin{subequations}
\begin{empheq}[left=\empheqlbrace]{align}
& \hspace{0.1cm} \Gamma^{(\mathrm{4PPI})}[\rho,\zeta;\hbar]=\sum_{n=0}^{\infty} \Gamma^{(\mathrm{4PPI})}_{n}[\rho,\zeta]\hbar^{n}\;, \label{eq:pure4PPIEAGammaExpansion0DON}\\
\nonumber \\
& \hspace{0.1cm} W[K,M;\hbar]=\sum_{n=0}^{\infty} W_{n}[K,M]\hbar^{n}\;, \label{eq:pure4PPIEAWExpansion0DON} \\
\nonumber \\
& \hspace{0.1cm} K[\rho,\zeta;\hbar]=\sum_{n=0}^{\infty} K_{n}[\rho,\zeta]\hbar^{n}\;, \label{eq:pure4PPIEAKExpansion0DON}\\
\nonumber \\
& \hspace{0.1cm} M[\rho,\zeta;\hbar]=\sum_{n=0}^{\infty} M_{n}[\rho,\zeta]\hbar^{n}\;, \label{eq:pure4PPIEAMExpansion0DON}\\
\nonumber \\
& \hspace{0.1cm} \rho = \sum_{n=0}^{\infty} \rho_{n}[K,M]\hbar^{n}\;, \label{eq:pure4PPIEArhoExpansion0DON} \\
\nonumber \\
& \hspace{0.1cm} \zeta = \sum_{n=0}^{\infty} \zeta_{n}[K,M]\hbar^{n}\;, \label{eq:pure4PPIEAzetaExpansion0DON}
\end{empheq}
\end{subequations}
and the definition of the 4PPI EA with vanishing 1-point correlation function is:
\begin{equation}
\begin{split}
\Gamma^{(\mathrm{4PPI})}[\rho,\zeta] \equiv & - W[K,M] + \int_{\alpha} K_{\alpha}[\rho,\zeta] \frac{\delta W[K,M]}{\delta K_{\alpha}} + \int_{\alpha} M_{\alpha}[\rho,\zeta] \frac{\delta W[K,M]}{\delta M_{\alpha}} \\
= & - W[K,M] + \frac{\hbar}{2} \int_{\alpha} K_{\alpha}[\rho,\zeta] \rho_{\alpha} + \frac{\hbar^{2}}{8} \int_{\alpha} M_{\alpha}[\rho,\zeta] \rho_{\alpha}^{2} + \frac{\hbar^{3}}{24} \int_{\alpha} M_{\alpha}[\rho,\zeta] \zeta_{\alpha} \;,
\end{split}
\label{eq:pure4PPIEAdefinition0DONAppendix}
\end{equation}
where
\begin{equation}
\rho_{\alpha} = \frac{2}{\hbar} \frac{\delta W[K,M]}{\delta K_{\alpha}}\;,
\label{eq:pure4PPIEAdefinitionrho0DON}
\end{equation}
\begin{equation}
\zeta_{\alpha} = \frac{24}{\hbar^{3}} \frac{\delta W[K,M]}{\delta M_{\alpha}} - \frac{3}{\hbar} \rho_{\alpha}^{2}\;.
\label{eq:pure4PPIEAdefinitionzeta0DON}
\end{equation}
If we differentiate the second line of~\eqref{eq:pure4PPIEAdefinition0DONAppendix}, we obtain:
\begin{subequations}
\begin{empheq}[left=\empheqlbrace]{align}
& \hspace{0.1cm} \frac{\delta\Gamma^{(\mathrm{4PPI})}[\rho,\zeta]}{\delta\rho_{\alpha}} = \frac{\hbar}{2}K_{\alpha}[\rho,\zeta] + \frac{\hbar^{2}}{4}M_{\alpha}[\rho,\zeta] \rho_{\alpha}\;. \label{eq:pure4PPIEAdGammadRho0DON} \\
\nonumber \\
& \hspace{0.1cm} \frac{\delta\Gamma^{(\mathrm{4PPI})}[\rho,\zeta]}{\delta\zeta_{\alpha}} = \frac{\hbar^{3}}{24} M_{\alpha}[\rho,\zeta] \;. \label{eq:pure4PPIEAdGammadZeta0DON}
\end{empheq}
\end{subequations}
If we insert as a next step the power series~\eqref{eq:pure4PPIEAGammaExpansion0DON},~\eqref{eq:pure4PPIEAKExpansion0DON} and~\eqref{eq:pure4PPIEAMExpansion0DON} into~\eqref{eq:pure4PPIEAdGammadRho0DON} and~\eqref{eq:pure4PPIEAdGammadZeta0DON}, the independence of both $\rho$ and $\zeta$ with respect to $\hbar$ enables us to identify the terms of order $\mathcal{O}\big(\hbar^{n}\big)$ in the relations thus derived to obtain:
\begin{subequations}
\begin{empheq}[left=\empheqlbrace]{align}
& \hspace{0.1cm} \frac{\delta\Gamma_{n}^{(\mathrm{4PPI})}[\rho,\zeta]}{\delta\rho_{\alpha}} = \frac{1}{2} K_{n-1,\alpha}[\rho,\zeta] \delta_{n \geq 1} + \frac{1}{4} M_{n-2,\alpha}[\rho,\zeta] \rho_{\alpha} \delta_{n \geq 2}\;. \label{eq:pure4PPIEAdGammandRho0DON} \\
\nonumber \\
& \hspace{0.1cm} \frac{\delta\Gamma_{n}^{(\mathrm{4PPI})}[\rho,\zeta]}{\delta\zeta_{\alpha}} = \frac{1}{24} M_{n-3,\alpha}[\rho,\zeta] \delta_{n \geq 3} \;. \label{eq:pure4PPIEAdGammandZeta0DON}
\end{empheq}
\end{subequations}
Note that, as usual, the independence of the arguments of the EA, i.e. $\rho$ and $\zeta$ here, with respect to $\hbar$ can be proven using the reasoning of~\eqref{eq:pure1PIEAIndependencephihbarstep20DON} showing that the 1-point correlation function $\vec{\phi}$ is of order $\mathcal{O}\big(\hbar^{0}\big)$ in the framework of the 1PI EA. Furthermore, the diagrammatic expressions of the $W_{n}$ coefficients can still be inferred from the original LE discussed in section~\ref{sec:OriginalLE} but one must take into account that both $\vec{\varphi}_{\text{cl}}$ and $\vec{J}$ vanish in the present case (as all $n$-point correlation functions with $n$ odd vanish in the present situation) and that the quartic vertex function represented by a zigzag line (according to rule~\eqref{eq:FeynRulesLoopExpansion4legVertex}) is now dressed by the external source $M$\footnote{This is actually the advantage of exploiting the 4PI (or the 4PPI) EA at the expense of the 2PI (or the 2PPI) one, namely using extra sources to dress vertex functions in addition to propagators.}. More accurately, for $n=0,1,2~\mathrm{and}~3$, we have:
\begin{equation}
W_{0}[K,M] = 0 \;,
\label{eq:pure4PPIEAIMW00DON}
\end{equation}
\begin{equation}
W_{1}[K,M] = W_{1}[K] = \frac{1}{2}\mathrm{STr}\left[\ln\big(\boldsymbol{G}_{K}[K]\big)\right] \;,
\label{eq:pure4PPIEAIMW10DON}
\end{equation}
\begin{equation}
\begin{split}
W_{2}[K,M] = & -\frac{1}{24} \hspace{0.08cm} \begin{gathered}
\begin{fmffile}{Diagrams/LoopExpansion1_Hartree}
\begin{fmfgraph}(30,20)
\fmfleft{i}
\fmfright{o}
\fmf{phantom,tension=10}{i,i1}
\fmf{phantom,tension=10}{o,o1}
\fmf{plain,left,tension=0.5}{i1,v1,i1}
\fmf{plain,right,tension=0.5}{o1,v2,o1}
\fmf{zigzag,foreground=(0,,0,,1)}{v1,v2}
\end{fmfgraph}
\end{fmffile}
\end{gathered}
-\frac{1}{12}\begin{gathered}
\begin{fmffile}{Diagrams/LoopExpansion1_Fock}
\begin{fmfgraph}(15,15)
\fmfleft{i}
\fmfright{o}
\fmf{phantom,tension=11}{i,v1}
\fmf{phantom,tension=11}{v2,o}
\fmf{plain,left,tension=0.4}{v1,v2,v1}
\fmf{zigzag,foreground=(0,,0,,1)}{v1,v2}
\end{fmfgraph}
\end{fmffile}
\end{gathered} \;,
\end{split}
\label{eq:pure4PPIEAIMW20DON}
\end{equation}
\begin{equation}
\begin{split}
W_{3}[K,M] = & \ \frac{1}{72} \hspace{0.38cm} \begin{gathered}
\begin{fmffile}{Diagrams/OPT_Diag5}
\begin{fmfgraph}(12,12)
\fmfleft{i0,i1}
\fmfright{o0,o1}
\fmftop{v1,vUp,v2}
\fmfbottom{v3,vDown,v4}
\fmf{phantom,tension=20}{i0,v1}
\fmf{phantom,tension=20}{i1,v3}
\fmf{phantom,tension=20}{o0,v2}
\fmf{phantom,tension=20}{o1,v4}
\fmf{plain,left=0.4,tension=0.5}{v3,v1}
\fmf{phantom,left=0.1,tension=0.5}{v1,vUp}
\fmf{phantom,left=0.1,tension=0.5}{vUp,v2}
\fmf{plain,left=0.4,tension=0.0}{v1,v2}
\fmf{plain,left=0.4,tension=0.5}{v2,v4}
\fmf{phantom,left=0.1,tension=0.5}{v4,vDown}
\fmf{phantom,left=0.1,tension=0.5}{vDown,v3}
\fmf{plain,left=0.4,tension=0.0}{v4,v3}
\fmf{zigzag,tension=0.5,foreground=(0,,0,,1)}{v1,v4}
\fmf{zigzag,tension=0.5,foreground=(0,,0,,1)}{v2,v3}
\end{fmfgraph}
\end{fmffile}
\end{gathered} \hspace{0.28cm} + \frac{1}{36} \hspace{0.38cm} \begin{gathered}
\begin{fmffile}{Diagrams/OPT_Diag6}
\begin{fmfgraph}(12,12)
\fmfleft{i0,i1}
\fmfright{o0,o1}
\fmftop{v1,vUp,v2}
\fmfbottom{v3,vDown,v4}
\fmf{phantom,tension=20}{i0,v1}
\fmf{phantom,tension=20}{i1,v3}
\fmf{phantom,tension=20}{o0,v2}
\fmf{phantom,tension=20}{o1,v4}
\fmf{plain,left=0.4,tension=0.5}{v3,v1}
\fmf{phantom,left=0.1,tension=0.5}{v1,vUp}
\fmf{phantom,left=0.1,tension=0.5}{vUp,v2}
\fmf{plain,left=0.4,tension=0.0}{v1,v2}
\fmf{plain,left=0.4,tension=0.5}{v2,v4}
\fmf{phantom,left=0.1,tension=0.5}{v4,vDown}
\fmf{phantom,left=0.1,tension=0.5}{vDown,v3}
\fmf{plain,left=0.4,tension=0.0}{v4,v3}
\fmf{zigzag,left=0.4,tension=0.5,foreground=(0,,0,,1)}{v1,v3}
\fmf{zigzag,right=0.4,tension=0.5,foreground=(0,,0,,1)}{v2,v4}
\end{fmfgraph}
\end{fmffile}
\end{gathered} \hspace{0.28cm} + \frac{1}{144} \hspace{0.38cm} \begin{gathered}
\begin{fmffile}{Diagrams/OPT_Diag7}
\begin{fmfgraph}(12,12)
\fmfleft{i0,i1}
\fmfright{o0,o1}
\fmftop{v1,vUp,v2}
\fmfbottom{v3,vDown,v4}
\fmf{phantom,tension=20}{i0,v1}
\fmf{phantom,tension=20}{i1,v3}
\fmf{phantom,tension=20}{o0,v2}
\fmf{phantom,tension=20}{o1,v4}
\fmf{plain,left=0.4,tension=0.5}{v3,v1}
\fmf{phantom,left=0.1,tension=0.5}{v1,vUp}
\fmf{phantom,left=0.1,tension=0.5}{vUp,v2}
\fmf{zigzag,left=0.4,tension=0.0,foreground=(0,,0,,1)}{v1,v2}
\fmf{plain,left=0.4,tension=0.5}{v2,v4}
\fmf{phantom,left=0.1,tension=0.5}{v4,vDown}
\fmf{phantom,left=0.1,tension=0.5}{vDown,v3}
\fmf{zigzag,left=0.4,tension=0.0,foreground=(0,,0,,1)}{v4,v3}
\fmf{plain,left=0.4,tension=0.5}{v1,v3}
\fmf{plain,right=0.4,tension=0.5}{v2,v4}
\end{fmfgraph}
\end{fmffile}
\end{gathered} \hspace{0.29cm} + \frac{1}{36} \hspace{-0.22cm} \begin{gathered}
\begin{fmffile}{Diagrams/OPT_Diag8}
\begin{fmfgraph}(40,20)
\fmfleft{i}
\fmfright{o}
\fmftop{vUpLeft1,vUpLeft2,vUpLeft3,vUpRight1,vUpRight2,vUpRight3}
\fmfbottom{vDownLeft1,vDownLeft2,vDownLeft3,vDownRight1,vDownRight2,vDownRight3}
\fmf{phantom,tension=10}{i,v3}
\fmf{phantom,tension=10}{o,v4}
\fmf{plain,right=0.4,tension=0.5}{v1,vUpLeft}
\fmf{plain,right,tension=0.5}{vUpLeft,vDownLeft}
\fmf{plain,left=0.4,tension=0.5}{v1,vDownLeft}
\fmf{phantom,right=0.4,tension=0.5}{vUpRight,v2}
\fmf{phantom,left,tension=0.5}{vUpRight,vDownRight}
\fmf{phantom,left=0.4,tension=0.5}{vDownRight,v2}
\fmf{phantom,tension=0.3}{v2bis,o}
\fmf{plain,left,tension=0.1}{v2,v2bis,v2}
\fmf{zigzag,tension=2.7,foreground=(0,,0,,1)}{v1,v2}
\fmf{phantom,tension=2}{v1,v3}
\fmf{phantom,tension=2}{v2,v4}
\fmf{phantom,tension=2.4}{vUpLeft,vUpLeft2}
\fmf{phantom,tension=2.4}{vDownLeft,vDownLeft2}
\fmf{phantom,tension=2.4}{vUpRight,vUpRight2}
\fmf{phantom,tension=2.4}{vDownRight,vDownRight2}
\fmf{zigzag,tension=0,foreground=(0,,0,,1)}{vUpLeft,vDownLeft}
\end{fmfgraph}
\end{fmffile}
\end{gathered} \\
& + \frac{1}{144} \hspace{0.1cm} \begin{gathered}
\begin{fmffile}{Diagrams/OPT_Diag9}
\begin{fmfgraph}(45,18)
\fmfleft{i}
\fmfright{o}
\fmf{phantom,tension=10}{i,i1}
\fmf{phantom,tension=10}{o,o1}
\fmf{plain,left,tension=0.5}{i1,v1,i1}
\fmf{plain,right,tension=0.5}{o1,v2,o1}
\fmf{zigzag,foreground=(0,,0,,1)}{v1,v3}
\fmf{plain,left,tension=0.5}{v3,v4}
\fmf{plain,right,tension=0.5}{v3,v4}
\fmf{zigzag,foreground=(0,,0,,1)}{v4,v2}
\end{fmfgraph}
\end{fmffile}
\end{gathered} \;,
\end{split}
\label{eq:pure4PPIEAIMW30DON}
\end{equation}
where the propagator involved in the LE of $W[K,M]$ is given by:
\begin{equation}
\begin{split}
\boldsymbol{G}^{-1}_{K,\alpha_{1}\alpha_{2}}[K] = & \ \left.\frac{\delta^{2}S\big[\vec{\widetilde{\varphi}}\big]}{\delta\widetilde{\varphi}_{\alpha_{1}}\delta\widetilde{\varphi}_{\alpha_{2}}}\right|_{\vec{\widetilde{\varphi}}=0} - K_{\alpha_{1}}[\rho,\zeta] \delta_{\alpha_{1}\alpha_{2}} \\
= & \left(-\nabla_{x_{1}}^{2} + m^2 - K_{\alpha_{1}}[\rho,\zeta] \right) \delta_{\alpha_{1}\alpha_{2}} \;,
\end{split}
\label{eq:pure4PPIEAGKLE0DON}
\end{equation}
and we have used the following Feynman rules at arbitrary external sources $K$ and $M$:
\begin{subequations}
\begin{align}
\begin{gathered}
\begin{fmffile}{Diagrams/4PPIEA_FeynRuleGSourceGK_AppendixAtK}
\begin{fmfgraph*}(20,20)
\fmfleft{i0,i1,i2,i3}
\fmfright{o0,o1,o2,o3}
\fmflabel{$\alpha_{1}$}{v1}
\fmflabel{$\alpha_{2}$}{v2}
\fmf{phantom}{i1,v1}
\fmf{phantom}{i2,v1}
\fmf{plain,tension=0.6}{v1,v2}
\fmf{phantom}{v2,o1}
\fmf{phantom}{v2,o2}
\end{fmfgraph*}
\end{fmffile}
\end{gathered} \hspace{0.5cm} &\rightarrow \boldsymbol{G}_{K,\alpha_{1}\alpha_{2}}[K] \;,
\label{eq:4PPIEAFeynRulesPropagator0DON} \\
\begin{gathered}
\begin{fmffile}{Diagrams/4PPIEA_FeynRuleV4_AppendixAtK}
\begin{fmfgraph*}(20,20)
\fmfleft{i0,i1,i2,i3}
\fmfright{o0,o1,o2,o3}
\fmf{phantom,tension=2.0}{i1,i1bis}
\fmf{plain,tension=2.0}{i1bis,v1}
\fmf{phantom,tension=2.0}{i2,i2bis}
\fmf{plain,tension=2.0}{i2bis,v1}
\fmf{zigzag,label=$x$,tension=0.6,foreground=(0,,0,,1)}{v1,v2}
\fmf{phantom,tension=2.0}{o1bis,o1}
\fmf{plain,tension=2.0}{v2,o1bis}
\fmf{phantom,tension=2.0}{o2bis,o2}
\fmf{plain,tension=2.0}{v2,o2bis}
\fmflabel{$a_{1}$}{i1bis}
\fmflabel{$a_{2}$}{i2bis}
\fmflabel{$a_{3}$}{o1bis}
\fmflabel{$a_{4}$}{o2bis}
\end{fmfgraph*}
\end{fmffile}
\end{gathered} \quad &\rightarrow \left(\lambda-M_{a_{1},x}[\rho,\zeta]\delta_{a_{1} a_{3}}\right)\delta_{a_{1} a_{2}}\delta_{a_{3} a_{4}}\;.
\label{eq:4PPIEAFeynRules4legVertex0DON}
\end{align}
\end{subequations}
The diagrams contributing to $W_{4}[K,M]$ are identical to those representing $\lambda^{3}\left\langle\left(\vec{\widetilde{\chi}}^{2}\right)^{6}\right\rangle_{0,\sigma}^{\text{c}}$ in tab.~\ref{tab:MultiplicityOPTdiagramsON} and their multiplicities are also expressed by~\eqref{eq:MultiplicityDiagLoopExpansion}. We then express the following non-fluctuating fields evaluated at arbitrary external sources $K$ and $M$:
\begin{subequations}
\begin{empheq}[left=\empheqlbrace]{align}
& \hspace{0.1cm} \rho_{K,\alpha}[K] = \boldsymbol{G}_{K,\alpha\alpha}[K] = \rho_{0,\alpha}[K,M] = \rho_{0,\alpha}[K] = 2 \frac{\delta W_{1}[K]}{\delta K_{\alpha}}\;. \label{eq:pure4PPIEAGK0DON} \\
\nonumber \\
& \hspace{0.1cm} \zeta_{KM,\alpha}[K,M] = \zeta_{0,\alpha}[K,M] = 24 \left(\frac{\delta W_{3}[K,M]}{\delta M_{\alpha}} - \frac{\delta W_{1}[K]}{\delta K_{\alpha}}\frac{\delta W_{2}[K,M]}{\delta K_{\alpha}}\right)\;. \label{eq:pure4PPIEAzetaK0DON}
\end{empheq}
\end{subequations}
At $(K,M)=(K_{0},M_{0})$, these two fields coincide with the arguments of the 4PPI EA with vanishing 1-point correlation function:
\begin{subequations}
\begin{empheq}[left=\empheqlbrace]{align}
& \hspace{0.1cm} \rho_{\alpha} = \boldsymbol{G}_{\alpha\alpha}[K_{0}] = \boldsymbol{G}_{K,\alpha\alpha}[K=K_{0}] = \rho_{0,\alpha}[K=K_{0}] = 2 \left.\frac{\delta W_{1}[K]}{\delta K_{\alpha}}\right|_{K=K_{0}}\;. \label{eq:pure4PPIEAGK00DON} \\
\nonumber \\
& \hspace{0.1cm} \zeta_{\alpha} = \zeta_{KM,\alpha}[K=K_{0},M=M_{0}] = \zeta_{0,\alpha}[K=K_{0},M=M_{0}] \nonumber \\
& \hspace{0.1cm} \hspace{0.383cm} = 24 \left(\left.\frac{\delta W_{3}[K,M]}{\delta M_{\alpha}}\right|_{K=K_{0} \atop M=M_{0}} - \left.\frac{\delta W_{1}[K]}{\delta K_{\alpha}}\right|_{K = K_{0}} \left.\frac{\delta W_{2}[K,M]}{\delta K_{\alpha}}\right|_{K=K_{0} \atop M=M_{0}}\right) \;. \label{eq:pure4PPIEAzetaK00DON}
\end{empheq}
\end{subequations}
The above expressions of $\rho_{\alpha}$ and $\zeta_{\alpha}$ in terms of derivatives of $W_{n}$ coefficients result from~\eqref{eq:pure4PPIEAdefinitionrho0DON} and~\eqref{eq:pure4PPIEAdefinitionzeta0DON}, as well as from the power series~\eqref{eq:pure4PPIEArhoExpansion0DON} and~\eqref{eq:pure4PPIEAzetaExpansion0DON}, combined with their independence with respect to $\hbar$. Moreover, we point out that $\rho_{K,\alpha}[K]$ introduced in~\eqref{eq:pure4PPIEAGK0DON} is just the diagonal part of the propagator $\boldsymbol{G}_{K}[K]$ (defined by~\eqref{eq:pure4PPIEAGKLE0DON}) which underlies our diagrammatic representation for the LE of the Schwinger functional $W[K,M]$. In the same way, $\rho_{\alpha}$ is the diagonal part of the propagator $\boldsymbol{G}[K_{0}]$ involved in the diagrammatic expression of the 4PPI EA that we seek. The associated Feynman rules correspond to those of~\eqref{eq:4PPIEAFeynRulesPropagator0DON} and~\eqref{eq:4PPIEAFeynRules4legVertex0DON} evaluated at $(K,M)=(K_{0},M_{0})$:
\begin{subequations}
\begin{align}
\begin{gathered}
\begin{fmffile}{Diagrams/4PPIEA_FeynRuleGSourceGK_AppendixAtK0}
\begin{fmfgraph*}(20,20)
\fmfleft{i0,i1,i2,i3}
\fmfright{o0,o1,o2,o3}
\fmflabel{$\alpha_{1}$}{v1}
\fmflabel{$\alpha_{2}$}{v2}
\fmf{phantom}{i1,v1}
\fmf{phantom}{i2,v1}
\fmf{plain,tension=0.6,foreground=(1,,0,,0)}{v1,v2}
\fmf{phantom}{v2,o1}
\fmf{phantom}{v2,o2}
\end{fmfgraph*}
\end{fmffile}
\end{gathered} \hspace{0.5cm} &\rightarrow \boldsymbol{G}_{\alpha_{1}\alpha_{2}}[K_{0}] \;,
\label{eq:4PPIEAFeynRulesPropagatorK0M00DON} \\
\begin{gathered}
\begin{fmffile}{Diagrams/4PPIEA_FeynRuleV4_AppendixAtK0}
\begin{fmfgraph*}(20,20)
\fmfleft{i0,i1,i2,i3}
\fmfright{o0,o1,o2,o3}
\fmf{phantom,tension=2.0}{i1,i1bis}
\fmf{plain,tension=2.0,foreground=(1,,0,,0)}{i1bis,v1}
\fmf{phantom,tension=2.0}{i2,i2bis}
\fmf{plain,tension=2.0,foreground=(1,,0,,0)}{i2bis,v1}
\fmf{zigzag,label=$x$,tension=0.6,foreground=(1,,0,,0)}{v1,v2}
\fmf{phantom,tension=2.0}{o1bis,o1}
\fmf{plain,tension=2.0,foreground=(1,,0,,0)}{v2,o1bis}
\fmf{phantom,tension=2.0}{o2bis,o2}
\fmf{plain,tension=2.0,foreground=(1,,0,,0)}{v2,o2bis}
\fmflabel{$a_{1}$}{i1bis}
\fmflabel{$a_{2}$}{i2bis}
\fmflabel{$a_{3}$}{o1bis}
\fmflabel{$a_{4}$}{o2bis}
\end{fmfgraph*}
\end{fmffile}
\end{gathered} \quad &\rightarrow \left(\lambda-M_{0,a_{1},x}[\rho,\zeta]\delta_{a_{1} a_{3}}\right)\delta_{a_{1} a_{2}}\delta_{a_{3} a_{4}}\;.
\label{eq:4PPIEAFeynRules4legVertexK0M00DON}
\end{align}
\end{subequations}
Then, we insert the power series~\eqref{eq:pure4PPIEAGammaExpansion0DON},~\eqref{eq:pure4PPIEAWExpansion0DON},~\eqref{eq:pure4PPIEAKExpansion0DON} and~\eqref{eq:pure4PPIEAMExpansion0DON} into definition \eqref{eq:pure4PPIEAdefinition0DONAppendix} of our 4PPI EA:
\begin{equation}
\begin{split}
\sum_{n=0}^{\infty} \Gamma^{(\mathrm{4PPI})}_{n}[\rho,\zeta]\hbar^{n} = & - \sum_{n=0}^{\infty} W_{n}\Bigg[\sum_{m=0}^{\infty} K_{m}[\rho,\zeta]\hbar^{m},\sum_{m=0}^{\infty} M_{m}[\rho,\zeta]\hbar^{m}\Bigg]\hbar^{n} + \frac{1}{2} \sum_{n=0}^{\infty} \int_{\alpha} K_{n,\alpha}[\rho,\zeta] \rho_{\alpha} \hbar^{n+1} \\
& + \frac{1}{8} \sum_{n=0}^{\infty} \int_{\alpha} M_{n,\alpha}[\rho,\zeta] \rho_{\alpha}^{2} \hbar^{n+2} + \frac{1}{24} \sum_{n=0}^{\infty} \int_{\alpha} M_{n,\alpha}[\rho,\zeta] \zeta_{\alpha} \hbar^{n+3} \;.
\end{split}
\label{eq:pure4PPIEAIMstep10DON}
\end{equation}
The Taylor expansion of the $W_{n}$ coefficients in the RHS of~\eqref{eq:pure4PPIEAIMstep10DON} around $(K,M)=(K_{0},M_{0})$ yields (see section~\ref{sec:GammanCoeffIM4PPIEA}):
\begin{equation}
\begin{split}
\Gamma_{n}^{(\mathrm{4PPI})}[\rho,\zeta] = & - W_{n}[K=K_{0},M=M_{0}] - \sum_{m=1}^{n} \int_{\alpha} \left.\frac{\delta W_{n-m}[K,M]}{\delta K_{\alpha}}\right|_{K=K_{0} \atop M=M_{0}} K_{m,\alpha}[\rho,\zeta] \\
& - \sum_{m=1}^{n} \int_{\alpha} \left.\frac{\delta W_{n-m}[K,M]}{\delta M_{\alpha}}\right|_{K=K_{0} \atop M=M_{0}} M_{m,\alpha}[\rho,\zeta] \\
& - \sum_{m=2}^{n} \frac{1}{m!} \sum_{\underset{\lbrace l+l'=m \rbrace}{l,l'=1}}^{m} \sum^{n}_{\underset{\lbrace n_{1} + \cdots + n_{l} + \hat{n}_{1} + \cdots + \hat{n}_{l'} \leq n\rbrace}{n_{1},\cdots,n_{l},\hat{n}_{1},\cdots,\hat{n}_{l'}=1}} \begin{pmatrix}
m \\
l
\end{pmatrix} \\
& \hspace{0.8cm} \times \int_{\alpha_{1},\cdots,\alpha_{l} \atop \hat{\alpha}_{1},\cdots,\hat{\alpha}_{l'}} \left.\frac{\delta^{m} W_{n-(n_{1}+\cdots+n_{l}+\hat{n}_{1}+\cdots+\hat{n}_{l'})}[K,M]}{\delta K_{\alpha_{1}}\cdots\delta K_{\alpha_{l}} \delta M_{\hat{\alpha}_{1}}\cdots\delta M_{\hat{\alpha}_{l'}}}\right|_{K=K_{0} \atop M=M_{0}} \\
& \hspace{2.5cm} \times K_{n_{1},\alpha_{1}}[\rho,\zeta] \cdots K_{n_{l},\alpha_{l}}[\rho,\zeta] M_{\hat{n}_{1},\hat{\alpha}_{1}}[\rho,\zeta] \cdots M_{\hat{n}_{l'},\hat{\alpha}_{l'}}[\rho,\zeta] \\
& + \frac{1}{2} \int_{\alpha} K_{n-1,\alpha}[\rho,\zeta] \rho_{\alpha} \delta_{n\geq 1} + \frac{1}{8} \int_{\alpha} M_{n-2,\alpha}[\rho,\zeta] \rho_{\alpha}^{2} \delta_{n\geq 2} + \frac{1}{24} \int_{\alpha} M_{n-3,\alpha}[\rho,\zeta] \zeta_{\alpha} \delta_{n\geq 3} \\
= & - W_{n}[K=K_{0},M=M_{0}] - \sum_{m=1}^{n-2} \int_{\alpha} \left.\frac{\delta W_{n-m}[K,M]}{\delta K_{\alpha}}\right|_{K=K_{0} \atop M=M_{0}} K_{m,\alpha}[\rho,\zeta] \\
& - \sum_{m=1}^{n-3} \int_{\alpha} \left.\frac{\delta W_{n-m}[K,M]}{\delta M_{\alpha}}\right|_{K=K_{0} \atop M=M_{0}} M_{m,\alpha}[\rho,\zeta] \\
& - \sum_{m=2}^{n-1} \frac{1}{m!} \sum_{\underset{\lbrace l+l'=m \rbrace}{l,l'=1}}^{m} \sum^{n}_{\underset{\lbrace n_{1} + \cdots + n_{l} + \hat{n}_{1} + \cdots + \hat{n}_{l'} \leq n\rbrace}{n_{1},\cdots,n_{l},\hat{n}_{1},\cdots,\hat{n}_{l'}=1}} \begin{pmatrix}
m \\
l
\end{pmatrix} \\
& \hspace{0.8cm} \times \int_{\alpha_{1},\cdots,\alpha_{l} \atop \hat{\alpha}_{1},\cdots,\hat{\alpha}_{l'}} \left.\frac{\delta^{m} W_{n-(n_{1}+\cdots+n_{l}+\hat{n}_{1}+\cdots+\hat{n}_{l'})}[K,M]}{\delta K_{\alpha_{1}}\cdots\delta K_{\alpha_{l}} \delta M_{\hat{\alpha}_{1}}\cdots\delta M_{\hat{\alpha}_{l'}}}\right|_{K=K_{0} \atop M=M_{0}} \\
& \hspace{2.5cm} \times K_{n_{1},\alpha_{1}}[\rho,\zeta] \cdots K_{n_{l},\alpha_{l}}[\rho,\zeta] M_{\hat{n}_{1},\hat{\alpha}_{1}}[\rho,\zeta] \cdots M_{\hat{n}_{l'},\hat{\alpha}_{l'}}[\rho,\zeta] \\
& + \frac{1}{2} \int_{\alpha} K_{0,\alpha}[\rho,\zeta] \rho_{\alpha} \delta_{n 1} + \frac{1}{8} \int_{\alpha} M_{0,\alpha}[\rho,\zeta] \rho_{\alpha}^{2} \delta_{n 2} + \frac{1}{24} \int_{\alpha} M_{n-3,\alpha}[\rho,\zeta] \zeta_{\alpha} \delta_{n\geq 3} \;,
\end{split}
\label{eq:pure4PPIEAIMstep40DON}
\end{equation}
where the latter equality was obtained by exploiting~\eqref{eq:pure4PPIEAIMW00DON}, expression~\eqref{eq:pure4PPIEAIMW10DON} of $W_{1}[K,M]$ in the form:
\begin{equation}
\frac{\delta W_{1}[K,M]}{\delta M_{\alpha}} = 0 \mathrlap{\quad \forall \alpha \;,}
\end{equation}
as well as~\eqref{eq:pure4PPIEAGK00DON} via the relation:
\begin{equation}
\begin{split}
& - \sum_{m=1}^{\textcolor{red}{n-1}} \int_{\alpha} \left.\frac{\delta W_{n-m}[K,M]}{\delta K_{\alpha}}\right|_{K=K_{0} \atop M=M_{0}} K_{m,\alpha}[\rho,\zeta] + \frac{1}{2} \int_{\alpha} K_{n-1,\alpha}[\rho,\zeta] \rho_{\alpha} \delta_{n\geq 1} \\
& \hspace{0.8cm} = - \sum_{m=1}^{\textcolor{red}{n-2}} \int_{\alpha} \left.\frac{\delta W_{n-m}[K,M]}{\delta K_{\alpha}}\right|_{K=K_{0} \atop M=M_{0}} K_{m,\alpha}[\rho,\zeta] + \frac{1}{2} \int_{\alpha} K_{0,\alpha}[\rho,\zeta] \rho_{\alpha} \delta_{n 1} \;,
\end{split}
\end{equation}
and expression~\eqref{eq:pure4PPIEAIMW20DON} of $W_{2}[K,M]$ in the form:
\begin{equation}
\begin{split}
& - \sum_{m=1}^{\textcolor{red}{n-2}} \int_{\alpha} \left.\frac{\delta W_{n-m}[K,M]}{\delta M_{\alpha}}\right|_{K=K_{0} \atop M=M_{0}} M_{m,\alpha}[\rho,\zeta] + \frac{1}{8} \int_{\alpha} M_{n-2,\alpha}[\rho,\zeta] \rho_{\alpha}^{2} \delta_{n\geq 2} \\
& \hspace{0.8cm} = - \sum_{m=1}^{\textcolor{red}{n-3}} \int_{\alpha} \left.\frac{\delta W_{n-m}[K,M]}{\delta M_{\alpha}}\right|_{K=K_{0} \atop M=M_{0}} M_{m,\alpha}[\rho,\zeta] + \frac{1}{8} \int_{\alpha} M_{0,\alpha}[\rho,\zeta] \rho_{\alpha}^{2} \delta_{n 2} \;.
\end{split}
\end{equation}
For $n=0,1,2,3~\mathrm{and}~4$,~\eqref{eq:pure4PPIEAIMstep40DON} translates into:
\begin{equation}
\Gamma_{0}^{(\mathrm{4PPI})}[\rho,\zeta] = - W_{0}[K=K_{0},M=M_{0}] \;,
\label{eq:pure4PPIEAIMGamma00DON}
\end{equation}
\begin{equation}
\Gamma_{1}^{(\mathrm{4PPI})}[\rho,\zeta] = - W_{1}[K=K_{0},M=M_{0}] + \frac{1}{2} \int_{\alpha} K_{0,\alpha}[\rho,\zeta] \rho_{\alpha} \;,
\label{eq:pure4PPIEAIMGamma10DON}
\end{equation}
\begin{equation}
\Gamma_{2}^{(\mathrm{4PPI})}[\rho,\zeta] = - W_{2}[K=K_{0},M=M_{0}] + \frac{1}{8} \int_{\alpha} M_{0,\alpha}[\rho,\zeta] \rho_{\alpha}^{2} \;,
\label{eq:pure4PPIEAIMGamma20DON}
\end{equation}
\begin{equation}
\begin{split}
\Gamma_{3}^{(\mathrm{4PPI})}[\rho,\zeta] = & - W_{3}[K=K_{0},M=M_{0}] - \int_{\alpha} \left.\frac{\delta W_{2}[K,M]}{\delta K_{\alpha}}\right|_{K=K_{0} \atop M=M_{0}} K_{1,\alpha}[\rho,\zeta] \\
& - \frac{1}{2} \int_{\alpha_{1},\alpha_{2}} \left.\frac{\delta^{2}W_{1}[K,M]}{\delta K_{\alpha_{1}} \delta K_{\alpha_{2}}} \right|_{K=K_{0} \atop M=M_{0}} K_{1,\alpha_{1}}[\rho,\zeta] K_{1,\alpha_{2}}[\rho,\zeta] + \frac{1}{24} \int_{\alpha} M_{0,\alpha}[\rho,\zeta] \zeta_{\alpha} \;,
\end{split}
\label{eq:pure4PPIEAIMGamma30DON}
\end{equation}
\begin{equation}
\begin{split}
\Gamma_{4}^{(\mathrm{4PPI})}[\rho,\zeta] = & - W_{4}[K=K_{0},M=M_{0}] - \int_{\alpha} \left.\frac{\delta W_{3}[K,M]}{\delta K_{\alpha}}\right|_{K=K_{0} \atop M=M_{0}} K_{1,\alpha}[\rho,\zeta] \\
& - \int_{\alpha} \left.\frac{\delta W_{2}[K,M]}{\delta K_{\alpha}}\right|_{K=K_{0} \atop M=M_{0}} K_{2,\alpha}[\rho,\zeta] - \int_{\alpha} \left.\frac{\delta W_{3}[K,M]}{\delta M_{\alpha}}\right|_{K=K_{0} \atop M=M_{0}} M_{1,\alpha}[\rho,\zeta] \\
& - \frac{1}{2} \int_{\alpha_{1},\alpha_{2}} \left.\frac{\delta^{2}W_{2}[K,M]}{\delta K_{\alpha_{1}} \delta K_{\alpha_{2}}} \right|_{K=K_{0} \atop M=M_{0}} K_{1,\alpha_{1}}[\rho,\zeta] K_{1,\alpha_{2}}[\rho,\zeta] \\
& - \frac{1}{2} \int_{\alpha_{1},\alpha_{2}} \left.\frac{\delta^{2}W_{2}[K,M]}{\delta M_{\alpha_{1}} \delta M_{\alpha_{2}}} \right|_{K=K_{0} \atop M=M_{0}} M_{1,\alpha_{1}}[\rho,\zeta] M_{1,\alpha_{2}}[\rho,\zeta] \\
& - \int_{\alpha_{1},\alpha_{2}} \left.\frac{\delta^{2}W_{2}[K,M]}{\delta K_{\alpha_{1}} \delta M_{\alpha_{2}}} \right|_{K=K_{0} \atop M=M_{0}} K_{1,\alpha_{1}}[\rho,\zeta] M_{1,\alpha_{2}}[\rho,\zeta] \\
& - \int_{\alpha_{1},\alpha_{2}} \left.\frac{\delta^{2}W_{1}[K,M]}{\delta K_{\alpha_{1}} \delta K_{\alpha_{2}}} \right|_{K=K_{0} \atop M=M_{0}} K_{1,\alpha_{1}}[\rho,\zeta] K_{2,\alpha_{2}}[\rho,\zeta] \\
& - \frac{1}{6} \int_{\alpha_{1},\alpha_{2},\alpha_{3}} \left.\frac{\delta^{3}W_{1}[K,M]}{\delta K_{\alpha_{1}} \delta K_{\alpha_{2}} \delta K_{\alpha_{3}}} \right|_{K=K_{0} \atop M=M_{0}} K_{1,\alpha_{1}}[\rho,\zeta] K_{1,\alpha_{2}}[\rho,\zeta] K_{1,\alpha_{3}}[\rho,\zeta] \\
& + \frac{1}{24} \int_{\alpha} M_{1,\alpha}[\rho,\zeta] \zeta_{\alpha} \;.
\end{split}
\label{eq:pure4PPIEAIMGamma40DON}
\end{equation}
From expressions~\eqref{eq:pure4PPIEAIMW00DON} to~\eqref{eq:pure4PPIEAIMW20DON} of the $W_{n}$ coefficients, we can directly rewrite~\eqref{eq:pure4PPIEAIMGamma00DON} to~\eqref{eq:pure4PPIEAIMGamma20DON} as:
\begin{equation}
\Gamma_{0}^{(\mathrm{4PPI})}[\rho,\zeta] = 0 \;,
\label{eq:pure4PPIEAIMGamma0bis0DON}
\end{equation}
\begin{equation}
\Gamma_{1}^{(\mathrm{4PPI})}[\rho,\zeta] = - \frac{1}{2}\mathrm{STr}\left[\ln\big(\boldsymbol{G}[K_{0}]\big)\right] + \frac{1}{2} \int_{\alpha} K_{0,\alpha}[\rho,\zeta] \rho_{\alpha} \;,
\label{eq:pure4PPIEAIMGamma1bis0DON}
\end{equation}
\begin{equation}
\begin{split}
\Gamma_{2}^{(\mathrm{4PPI})}[\rho,\zeta] = & \ \frac{1}{24} \hspace{0.08cm} \begin{gathered}
\begin{fmffile}{Diagrams/pure4PPIEA_Hartree}
\begin{fmfgraph}(30,20)
\fmfleft{i}
\fmfright{o}
\fmf{phantom,tension=10}{i,i1}
\fmf{phantom,tension=10}{o,o1}
\fmf{plain,left,tension=0.5,foreground=(1,,0,,0)}{i1,v1,i1}
\fmf{plain,right,tension=0.5,foreground=(1,,0,,0)}{o1,v2,o1}
\fmf{zigzag,foreground=(1,,0,,0)}{v1,v2}
\end{fmfgraph}
\end{fmffile}
\end{gathered}
+ \frac{1}{12}\begin{gathered}
\begin{fmffile}{Diagrams/pure4PPIEA_Fock}
\begin{fmfgraph}(15,15)
\fmfleft{i}
\fmfright{o}
\fmf{phantom,tension=11}{i,v1}
\fmf{phantom,tension=11}{v2,o}
\fmf{plain,left,tension=0.4,foreground=(1,,0,,0)}{v1,v2,v1}
\fmf{zigzag,foreground=(1,,0,,0)}{v1,v2}
\end{fmfgraph}
\end{fmffile}
\end{gathered} + \frac{1}{8} \int_{\alpha} M_{0,\alpha}[\rho,\zeta] \rho_{\alpha}^{2} \\
= & \ \frac{1}{24} \sum_{a_{1},a_{2}=1}^{N} \int_{x} \boldsymbol{G}_{(a_{1},x)(a_{1},x)}[K_{0}] \boldsymbol{G}_{(a_{2},x)(a_{2},x)}[K_{0}] \left(\lambda - \cancel{M_{0,a_{1},x}[\rho,\zeta] \delta_{a_{1}a_{2}}}\right) \\
& + \frac{1}{12} \sum_{a_{1},a_{2}=1}^{N} \int_{x} \left(\boldsymbol{G}_{(a_{1},x)(a_{2},x)}[K_{0}]\right)^{2} \left(\lambda - \cancel{M_{0,a_{1},x}[\rho,\zeta] \delta_{a_{1}a_{2}}}\right) \\
& + \cancel{\frac{1}{8} \int_{\alpha} M_{0,\alpha}[\rho,\zeta] \rho_{\alpha}^{2}} \\
= & \ \frac{\lambda}{24} \sum_{a_{1},a_{2}=1}^{N} \int_{x} \boldsymbol{G}_{(a_{1},x)(a_{1},x)}[K_{0}] \boldsymbol{G}_{(a_{2},x)(a_{2},x)}[K_{0}] + \frac{\lambda}{12} \sum_{a_{1},a_{2}=1}^{N} \int_{x} \left(\boldsymbol{G}_{(a_{1},x)(a_{2},x)}[K_{0}]\right)^{2} \\
= & \ \frac{1}{24} \hspace{0.08cm} \begin{gathered}
\begin{fmffile}{Diagrams/1PIEA_Hartree}
\begin{fmfgraph}(30,20)
\fmfleft{i}
\fmfright{o}
\fmf{phantom,tension=10}{i,i1}
\fmf{phantom,tension=10}{o,o1}
\fmf{plain,left,tension=0.5,foreground=(1,,0,,0)}{i1,v1,i1}
\fmf{plain,right,tension=0.5,foreground=(1,,0,,0)}{o1,v2,o1}
\fmf{zigzag,foreground=(0,,0,,1)}{v1,v2}
\end{fmfgraph}
\end{fmffile}
\end{gathered}
+\frac{1}{12}\begin{gathered}
\begin{fmffile}{Diagrams/1PIEA_Fock}
\begin{fmfgraph}(15,15)
\fmfleft{i}
\fmfright{o}
\fmf{phantom,tension=11}{i,v1}
\fmf{phantom,tension=11}{v2,o}
\fmf{plain,left,tension=0.4,foreground=(1,,0,,0)}{v1,v2,v1}
\fmf{zigzag,foreground=(0,,0,,1)}{v1,v2}
\end{fmfgraph}
\end{fmffile}
\end{gathered} \;.
\end{split}
\label{eq:pure4PPIEAIMGamma2bis0DON}
\end{equation}
According to~\eqref{eq:pure4PPIEAIMGamma0bis0DON},~\eqref{eq:pure4PPIEAIMGamma1bis0DON} and~\eqref{eq:pure4PPIEAIMGamma2bis0DON}, $\Gamma^{(\mathrm{4PPI})}$ does not depend on $M_{0}$ if truncated at the lowest non-trivial order (i.e. at order $\mathcal{O}(\hbar^{2})$), which means that we need to calculate at least the second non-trivial order (i.e. order $\mathcal{O}(\hbar^{3})$) here to appreciate a possible improvement of the 2PPI EA of section~\ref{sec:2PPIEA} with the present 4PPI EA formalism. We recall that we have actually shown in section~\ref{sec:4PPIEA} with~\eqref{eq:pure4PPIEAgapequationRhostep10DON} and~\eqref{eq:pure4PPIEAgapequationZetastep10DON} that results obtained from $\Gamma^{(\mathrm{2PPI})}[\rho]$ and $\Gamma^{(\mathrm{4PPI})}[\rho,\zeta]$ coincide at second non-trivial order as well. We then evaluate the derivatives involved in the RHSs of~\eqref{eq:pure4PPIEAIMGamma30DON} and~\eqref{eq:pure4PPIEAIMGamma40DON} in order to rewrite $\Gamma^{(\mathrm{4PPI})}_{3}[\rho,\zeta]$ and $\Gamma^{(\mathrm{4PPI})}_{4}[\rho,\zeta]$. To that end, let us first point out that, when applied to diagrammatic expressions of the $W_{n}$ coefficients, derivatives with respect to $K_{\alpha}$ only act on propagator lines as:
\begin{equation}
\begin{split}
\frac{\delta}{\delta K_{\alpha_{1}}} \hspace{0.5cm} \begin{gathered}
\begin{fmffile}{Diagrams/pure4PPIEA_DerivK0_Diag1}
\begin{fmfgraph*}(20,20)
\fmfleft{i0,i1,i2,i3}
\fmfright{o0,o1,o2,o3}
\fmfv{decor.shape=circle,decor.filled=empty,decor.size=1.5thick,label=$\alpha_{2}$}{v1}
\fmfv{decor.shape=circle,decor.filled=empty,decor.size=1.5thick,label=$\alpha_{3}$}{v2}
\fmf{phantom}{i1,v1}
\fmf{phantom}{i2,v1}
\fmf{plain,tension=0.6}{v1,v2}
\fmf{phantom}{v2,o1}
\fmf{phantom}{v2,o2}
\end{fmfgraph*}
\end{fmffile}
\end{gathered} \hspace{0.4cm} = & -\int_{\alpha_{4},\alpha_{5}} \boldsymbol{G}_{K,\alpha_{2}\alpha_{4}}[K] \frac{\delta \boldsymbol{G}^{-1}_{K,\alpha_{4}\alpha_{5}}[K]}{\delta K_{\alpha_{1}}} \boldsymbol{G}_{K,\alpha_{5}\alpha_{3}}[K] \\
= & \int_{\alpha_{4},\alpha_{5}} \boldsymbol{G}_{K,\alpha_{2}\alpha_{4}}[K] \underbrace{\frac{\delta K_{\alpha_{4}}[\rho,\zeta]}{\delta K_{\alpha_{1}}}}_{\delta_{\alpha_{4}\alpha_{1}}} \delta_{\alpha_{4}\alpha_{5}} \boldsymbol{G}_{K,\alpha_{5}\alpha_{3}}[K] \\
= & \ \boldsymbol{G}_{K,\alpha_{2}\alpha_{1}}[K] \boldsymbol{G}_{K,\alpha_{1}\alpha_{3}}[K] \\
= & \hspace{-0.1cm} \begin{gathered}
\begin{fmffile}{Diagrams/pure4PPIEA_DerivK0_Diag2}
\begin{fmfgraph*}(28,20)
\fmfleft{i0,i1,i2,i3}
\fmfright{o0,o1,o2,o3}
\fmfv{decor.shape=circle,decor.filled=empty,decor.size=1.5thick,label.angle=-90,label.dist=0.15cm,label=$\alpha_{2}$}{v1}
\fmfv{decor.shape=circle,decor.filled=empty,decor.size=1.5thick,label.angle=-90,label.dist=0.15cm,label=$\alpha_{3}$}{v2}
\fmfv{decor.shape=circle,decor.filled=empty,decor.size=1.5thick,label.angle=-90,label.dist=0.15cm,label=$\alpha_{1}$}{v3}
\fmf{phantom}{i1,v3}
\fmf{phantom}{i2,v3}
\fmf{phantom}{o1,v3}
\fmf{phantom}{o2,v3}
\fmf{phantom}{i1,v1}
\fmf{phantom}{i2,v1}
\fmf{plain,tension=0.6}{v1,v2}
\fmf{phantom}{v2,o1}
\fmf{phantom}{v2,o2}
\end{fmfgraph*}
\end{fmffile}
\end{gathered} \;,
\end{split}
\label{eq:pure4PPIEAderivWithRespectToK0DON}
\end{equation}
whereas those with respect to $M_{\alpha}$ only affect vertices as:
\begin{equation}
\frac{\delta}{\delta M_{\alpha_{1}}} \hspace{0.8cm} \begin{gathered}
\begin{fmffile}{Diagrams/pure4PPIEA_DerivM0_Diag1}
\begin{fmfgraph*}(20,20)
\fmfleft{i0,i1,i2,i3}
\fmfright{o0,o1,o2,o3}
\fmfv{decor.shape=circle,decor.filled=empty,decor.size=1.5thick,label=$\alpha_{4}$}{o2}
\fmfv{decor.shape=circle,decor.filled=empty,decor.size=1.5thick,label=$\alpha_{5}$}{o1}
\fmfv{decor.shape=circle,decor.filled=empty,decor.size=1.5thick,label=$\alpha_{3}$}{i1}
\fmfv{decor.shape=circle,decor.filled=empty,decor.size=1.5thick,label=$\alpha_{2}$}{i2}
\fmf{phantom,tension=2.0}{i1,i1bis}
\fmf{phantom,tension=2.0}{i1bis,v1}
\fmf{phantom,tension=2.0}{i2,i2bis}
\fmf{phantom,tension=2.0}{i2bis,v1}
\fmf{zigzag,tension=0.6,foreground=(0,,0,,1)}{v1,v2}
\fmf{phantom,tension=2.0}{o1bis,o1}
\fmf{phantom,tension=2.0}{v2,o1bis}
\fmf{phantom,tension=2.0}{o2bis,o2}
\fmf{phantom,tension=2.0}{v2,o2bis}
\fmf{plain,tension=0.0}{i2,v1}
\fmf{plain,tension=0.0}{i1,v1}
\fmf{plain,tension=0.0}{v2,o2}
\fmf{plain,tension=0.0}{v2,o1}
\end{fmfgraph*}
\end{fmffile}
\end{gathered} \hspace{0.7cm} = - \hspace{0.8cm} \begin{gathered}
\begin{fmffile}{Diagrams/pure4PPIEA_DerivM0_Diag2}
\begin{fmfgraph*}(15,20)
\fmfleft{i0,i1,i2,i3}
\fmfright{o0,o1,o2,o3}
\fmfv{decor.shape=circle,decor.filled=empty,decor.size=1.5thick,label=$\alpha_{4}$}{o2}
\fmfv{decor.shape=circle,decor.filled=empty,decor.size=1.5thick,label=$\alpha_{5}$}{o1}
\fmfv{decor.shape=circle,decor.filled=empty,decor.size=1.5thick,label=$\alpha_{3}$}{i1}
\fmfv{decor.shape=circle,decor.filled=empty,decor.size=1.5thick,label=$\alpha_{2}$}{i2}
\fmfv{decor.shape=circle,decor.filled=empty,decor.size=1.5thick,label=$\alpha_{1}$}{vCenter}
\fmf{phantom,tension=2.0}{i1,i1bis}
\fmf{phantom,tension=2.0}{i1bis,v1}
\fmf{phantom,tension=2.0}{i2,i2bis}
\fmf{phantom,tension=2.0}{i2bis,v1}
\fmf{phantom,tension=0.6,foreground=(0,,0,,1)}{v1,v2}
\fmf{phantom,tension=2.0}{o1bis,o1}
\fmf{phantom,tension=2.0}{v2,o1bis}
\fmf{phantom,tension=2.0}{o2bis,o2}
\fmf{phantom,tension=2.0}{v2,o2bis}
\fmf{plain,tension=0.5}{i2,vCenter}
\fmf{plain,tension=0.5}{i1,vCenter}
\fmf{plain,tension=0.5}{vCenter,o2}
\fmf{plain,tension=0.5}{vCenter,o1}
\end{fmfgraph*}
\end{fmffile}
\end{gathered} \hspace{0.7cm} \;.
\label{eq:pure4PPIEAderivWithRespectToM0DON}
\end{equation}
By comparing~\eqref{eq:pure4PPIEAderivWithRespectToK0DON} with, e.g.,~\eqref{eq:mixed2PIEAlambdadmathcalGdK0DON}, one can see that derivatives with respect to the external source dressing the propagator (i.e. $K_{\alpha}[\rho,\zeta]$ here) now amounts to inserting external points into propagator lines instead of cutting them as in the 2PI EA formalism. This results from the definitions of $n$PPI EAs, and more specifically from the restriction that $K$ is a local source in the present situation. Using~\eqref{eq:pure4PPIEAderivWithRespectToK0DON} and~\eqref{eq:pure4PPIEAderivWithRespectToM0DON} together with~\eqref{eq:pure4PPIEAIMW10DON} to~\eqref{eq:pure4PPIEAIMW30DON}, we are equipped to evaluate the following derivatives of the $W_{n}$ coefficients:
\begin{equation}
\left.\frac{\delta W_{2}[K,M]}{\delta K_{\alpha}}\right|_{K=K_{0} \atop M=M_{0}} = -\frac{1}{12} \hspace{0.45cm} \begin{gathered}
\begin{fmffile}{Diagrams/pure4PPIEA_dW2dK_Diag1}
\begin{fmfgraph*}(30,20)
\fmfleft{i}
\fmfright{o}
\fmfv{decor.shape=circle,decor.filled=empty,decor.size=1.5thick,label.dist=0.15cm,label=$\alpha$}{i1}
\fmf{phantom,tension=10}{i,i1}
\fmf{phantom,tension=10}{o,o1}
\fmf{plain,left,tension=0.5,foreground=(1,,0,,0)}{i1,v1,i1}
\fmf{plain,right,tension=0.5,foreground=(1,,0,,0)}{o1,v2,o1}
\fmf{zigzag,foreground=(1,,0,,0)}{v1,v2}
\end{fmfgraph*}
\end{fmffile}
\end{gathered}
\hspace{0.1cm} - \frac{1}{6}\begin{gathered}
\begin{fmffile}{Diagrams/pure4PPIEA_dW2dK_Diag2}
\begin{fmfgraph*}(15,15)
\fmfleft{i}
\fmfright{o}
\fmftop{vUp}
\fmfbottom{vDown}
\fmfv{decor.shape=circle,decor.filled=empty,decor.size=1.5thick,label.dist=0.15cm,label=$\alpha$}{vBis}
\fmf{phantom,tension=7}{vUp,vBis}
\fmf{phantom,tension=1}{vDown,vBis}
\fmf{phantom,tension=11}{i,v1}
\fmf{phantom,tension=11}{v2,o}
\fmf{plain,left,tension=0.4,foreground=(1,,0,,0)}{v1,v2,v1}
\fmf{zigzag,foreground=(1,,0,,0)}{v1,v2}
\end{fmfgraph*}
\end{fmffile}
\end{gathered} \;,
\label{eq:pure4PPIEAdW2dK0DON}
\end{equation}
\begin{equation}
\left.\frac{\delta^{2} W_{1}[K,M]}{\delta K_{\alpha_{1}} \delta K_{\alpha_{2}}}\right|_{K=K_{0} \atop M=M_{0}} = \frac{1}{2} \hspace{0.5cm} \begin{gathered}
\begin{fmffile}{Diagrams/pure4PPIEA_dW1dKdK_Diag1}
\begin{fmfgraph*}(15,15)
\fmfleft{i}
\fmfright{o}
\fmftop{vUp}
\fmfbottom{vDown}
\fmfv{decor.shape=circle,decor.filled=empty,decor.size=1.5thick,label.dist=0.15cm,label=$\alpha_{1}$}{v1}
\fmfv{decor.shape=circle,decor.filled=empty,decor.size=1.5thick,label.dist=0.15cm,label=$\alpha_{2}$}{v2}
\fmf{phantom,tension=11}{i,v1}
\fmf{phantom,tension=11}{v2,o}
\fmf{plain,left,tension=0.4,foreground=(1,,0,,0)}{v1,v2,v1}
\fmf{phantom}{v1,v2}
\end{fmfgraph*}
\end{fmffile}
\end{gathered} \hspace{0.4cm} \;,
\label{eq:pure4PPIEAdW1dKdK0DON}
\end{equation}
\begin{equation}
\begin{split}
\scalebox{0.97}{${\displaystyle \left.\frac{\delta W_{3}[K,M]}{\delta K_{\alpha}}\right|_{K=K_{0} \atop M=M_{0}} = }$} & \ \scalebox{0.97}{${\displaystyle \frac{1}{18} \hspace{0.55cm} \begin{gathered}
\begin{fmffile}{Diagrams/pure4PPIEA_dW3dK_Diag1}
\begin{fmfgraph*}(15,15)
\fmfleft{i0,i1}
\fmfright{o0,o1}
\fmftop{v1bis,vUp,v2bis}
\fmfbottom{v3bis,vDown,v4bis}
\fmfleft{vLeft}
\fmfright{vRight}
\fmf{phantom,tension=3.2}{v1,v1bis}
\fmf{phantom,tension=3.2}{v2,v2bis}
\fmf{phantom,tension=3.2}{v3,v3bis}
\fmf{phantom,tension=3.2}{v4,v4bis}
\fmfv{decor.shape=circle,decor.filled=empty,decor.size=1.5thick,label.dist=0.15cm,label=$\alpha$}{vIndex}
\fmf{phantom,tension=-30}{vLeft,vIndex}
\fmf{phantom,tension=0.5}{vRight,vIndex}
\fmf{phantom,tension=20}{i0,v1}
\fmf{phantom,tension=20}{i1,v3}
\fmf{phantom,tension=20}{o0,v2}
\fmf{phantom,tension=20}{o1,v4}
\fmf{plain,right=0.4,tension=0.5,foreground=(1,,0,,0)}{v3,v1}
\fmf{phantom,left=0.1,tension=0.5}{v1,vUp}
\fmf{phantom,left=0.1,tension=0.5}{vUp,v2}
\fmf{plain,right=0.4,tension=0.0,foreground=(1,,0,,0)}{v1,v2}
\fmf{plain,right=0.4,tension=0.5,foreground=(1,,0,,0)}{v2,v4}
\fmf{phantom,left=0.1,tension=0.5}{v4,vDown}
\fmf{phantom,left=0.1,tension=0.5}{vDown,v3}
\fmf{plain,right=0.4,tension=0.0,foreground=(1,,0,,0)}{v4,v3}
\fmf{zigzag,tension=0.5,foreground=(1,,0,,0)}{v1,v4}
\fmf{zigzag,tension=0.5,foreground=(1,,0,,0)}{v2,v3}
\end{fmfgraph*}
\end{fmffile}
\end{gathered} \hspace{0.2cm} + \frac{1}{18} \hspace{0.55cm} \begin{gathered}
\begin{fmffile}{Diagrams/pure4PPIEA_dW3dK_Diag2}
\begin{fmfgraph*}(15,15)
\fmfleft{i0,i1}
\fmfright{o0,o1}
\fmftop{v1bis,vUp,v2bis}
\fmfbottom{v3bis,vDown,v4bis}
\fmfleft{vLeft}
\fmfright{vRight}
\fmf{phantom,tension=3.2}{v1,v1bis}
\fmf{phantom,tension=3.2}{v2,v2bis}
\fmf{phantom,tension=3.2}{v3,v3bis}
\fmf{phantom,tension=3.2}{v4,v4bis}
\fmfv{decor.shape=circle,decor.filled=empty,decor.size=1.5thick,label.dist=0.15cm,label=$\alpha$}{vIndex}
\fmf{phantom,tension=-30}{vLeft,vIndex}
\fmf{phantom,tension=0.5}{vRight,vIndex}
\fmf{phantom,tension=20}{i0,v1}
\fmf{phantom,tension=20}{i1,v3}
\fmf{phantom,tension=20}{o0,v2}
\fmf{phantom,tension=20}{o1,v4}
\fmf{plain,right=0.4,tension=0.5,foreground=(1,,0,,0)}{v3,v1}
\fmf{phantom,left=0.1,tension=0.5}{v1,vUp}
\fmf{phantom,left=0.1,tension=0.5}{vUp,v2}
\fmf{plain,right=0.4,tension=0.0,foreground=(1,,0,,0)}{v1,v2}
\fmf{plain,right=0.4,tension=0.5,foreground=(1,,0,,0)}{v2,v4}
\fmf{phantom,left=0.1,tension=0.5}{v4,vDown}
\fmf{phantom,left=0.1,tension=0.5}{vDown,v3}
\fmf{plain,right=0.4,tension=0.0,foreground=(1,,0,,0)}{v4,v3}
\fmf{phantom,tension=0.5}{v1,v4}
\fmf{phantom,tension=0.5}{v2,v3}
\fmf{zigzag,right=0.4,tension=0.5,foreground=(1,,0,,0)}{v1,v3}
\fmf{zigzag,left=0.4,tension=0.5,foreground=(1,,0,,0)}{v2,v4}
\end{fmfgraph*}
\end{fmffile}
\end{gathered} \hspace{0.2cm} + \frac{1}{18} \hspace{0.25cm} \begin{gathered}
\begin{fmffile}{Diagrams/pure4PPIEA_dW3dK_Diag3}
\begin{fmfgraph*}(15,15)
\fmfleft{i0,i1}
\fmfright{o0,o1}
\fmftop{v1bis,vUp,v2bis}
\fmfbottom{v3bis,vDown,v4bis}
\fmf{phantom,tension=3.2}{v1,v1bis}
\fmf{phantom,tension=3.2}{v2,v2bis}
\fmf{phantom,tension=3.2}{v3,v3bis}
\fmf{phantom,tension=3.2}{v4,v4bis}
\fmfv{decor.shape=circle,decor.filled=empty,decor.size=1.5thick,label.dist=0.15cm,label=$\alpha$}{vIndex}
\fmf{phantom,tension=-10}{vUp,vIndex}
\fmf{phantom,tension=-0.25}{vDown,vIndex}
\fmf{phantom,tension=20}{i0,v1}
\fmf{phantom,tension=20}{i1,v3}
\fmf{phantom,tension=20}{o0,v2}
\fmf{phantom,tension=20}{o1,v4}
\fmf{plain,right=0.4,tension=0.5,foreground=(1,,0,,0)}{v3,v1}
\fmf{phantom,left=0.1,tension=0.5}{v1,vUp}
\fmf{phantom,left=0.1,tension=0.5}{vUp,v2}
\fmf{plain,right=0.4,tension=0.0,foreground=(1,,0,,0)}{v1,v2}
\fmf{plain,right=0.4,tension=0.5,foreground=(1,,0,,0)}{v2,v4}
\fmf{phantom,left=0.1,tension=0.5}{v4,vDown}
\fmf{phantom,left=0.1,tension=0.5}{vDown,v3}
\fmf{plain,right=0.4,tension=0.0,foreground=(1,,0,,0)}{v4,v3}
\fmf{phantom,tension=0.5}{v1,v4}
\fmf{phantom,tension=0.5}{v2,v3}
\fmf{zigzag,right=0.4,tension=0.5,foreground=(1,,0,,0)}{v1,v3}
\fmf{zigzag,left=0.4,tension=0.5,foreground=(1,,0,,0)}{v2,v4}
\end{fmfgraph*}
\end{fmffile}
\end{gathered} \hspace{0.2cm} + \frac{1}{36} \hspace{0.55cm} \begin{gathered}
\begin{fmffile}{Diagrams/pure4PPIEA_dW3dK_Diag4}
\begin{fmfgraph*}(15,15)
\fmfleft{i0,i1}
\fmfright{o0,o1}
\fmftop{v1bis,vUp,v2bis}
\fmfbottom{v3bis,vDown,v4bis}
\fmfleft{vLeft}
\fmfright{vRight}
\fmf{phantom,tension=3.2}{v1,v1bis}
\fmf{phantom,tension=3.2}{v2,v2bis}
\fmf{phantom,tension=3.2}{v3,v3bis}
\fmf{phantom,tension=3.2}{v4,v4bis}
\fmfv{decor.shape=circle,decor.filled=empty,decor.size=1.5thick,label.dist=0.15cm,label=$\alpha$}{vIndex}
\fmf{phantom,tension=-30}{vLeft,vIndex}
\fmf{phantom,tension=0.5}{vRight,vIndex}
\fmf{phantom,tension=20}{i0,v1}
\fmf{phantom,tension=20}{i1,v3}
\fmf{phantom,tension=20}{o0,v2}
\fmf{phantom,tension=20}{o1,v4}
\fmf{plain,right=0.4,tension=0.5,foreground=(1,,0,,0)}{v3,v1}
\fmf{phantom,left=0.1,tension=0.5}{v1,vUp}
\fmf{phantom,left=0.1,tension=0.5}{vUp,v2}
\fmf{zigzag,right=0.4,tension=0.0,foreground=(1,,0,,0)}{v1,v2}
\fmf{plain,right=0.4,tension=0.5,foreground=(1,,0,,0)}{v2,v4}
\fmf{phantom,left=0.1,tension=0.5}{v4,vDown}
\fmf{phantom,left=0.1,tension=0.5}{vDown,v3}
\fmf{zigzag,right=0.4,tension=0.0,foreground=(1,,0,,0)}{v4,v3}
\fmf{phantom,tension=0.5}{v1,v4}
\fmf{phantom,tension=0.5}{v2,v3}
\fmf{plain,right=0.4,tension=0.5,foreground=(1,,0,,0)}{v1,v3}
\fmf{plain,left=0.4,tension=0.5,foreground=(1,,0,,0)}{v2,v4}
\end{fmfgraph*}
\end{fmffile}
\end{gathered} }$} \\
& \scalebox{0.97}{${\displaystyle + \frac{1}{36} \hspace{0.2cm} \begin{gathered}
\begin{fmffile}{Diagrams/pure4PPIEA_dW3dK_Diag5}
\begin{fmfgraph*}(40,20)
\fmfleft{i}
\fmfright{o}
\fmftop{vUpLeft1,vUpLeft2,vUpLeft3,vUpRight1,vUpRight2,vUpRight3}
\fmfbottom{vDownLeft1,vDownLeft2,vDownLeft3,vDownRight1,vDownRight2,vDownRight3}
\fmfv{decor.shape=circle,decor.filled=empty,decor.size=1.5thick,label.dist=0.15cm,label=$\alpha$}{vIndex}
\fmf{phantom,tension=10.5}{i,vIndex}
\fmf{phantom,tension=1}{o,vIndex}
\fmf{phantom,tension=10}{i,v3}
\fmf{phantom,tension=10}{o,v4}
\fmf{plain,right=0.4,tension=0.5,foreground=(1,,0,,0)}{v1,vUpLeft}
\fmf{plain,right,tension=0.5,foreground=(1,,0,,0)}{vUpLeft,vDownLeft}
\fmf{plain,left=0.4,tension=0.5,foreground=(1,,0,,0)}{v1,vDownLeft}
\fmf{phantom,right=0.4,tension=0.5}{vUpRight,v2}
\fmf{phantom,left,tension=0.5}{vUpRight,vDownRight}
\fmf{phantom,left=0.4,tension=0.5}{vDownRight,v2}
\fmf{phantom,tension=0.3}{v2bis,o}
\fmf{plain,left,tension=0.1,foreground=(1,,0,,0)}{v2,v2bis,v2}
\fmf{zigzag,tension=2.7,foreground=(1,,0,,0)}{v1,v2}
\fmf{phantom,tension=2}{v1,v3}
\fmf{phantom,tension=2}{v2,v4}
\fmf{phantom,tension=2.4}{vUpLeft,vUpLeft2}
\fmf{phantom,tension=2.4}{vDownLeft,vDownLeft2}
\fmf{phantom,tension=2.4}{vUpRight,vUpRight2}
\fmf{phantom,tension=2.4}{vDownRight,vDownRight2}
\fmf{zigzag,tension=0,foreground=(1,,0,,0)}{vUpLeft,vDownLeft}
\end{fmfgraph*}
\end{fmffile}
\end{gathered} \hspace{-0.4cm} + \frac{1}{18} \hspace{-0.22cm} \begin{gathered}
\begin{fmffile}{Diagrams/pure4PPIEA_dW3dK_Diag6}
\begin{fmfgraph*}(40,20)
\fmfleft{i}
\fmfright{o}
\fmftop{vUpLeft1,vUpLeft2,vUpLeft3,vUpRight1,vUpRight2,vUpRight3}
\fmfbottom{vDownLeft1,vDownLeft2,vDownLeft3,vDownRight1,vDownRight2,vDownRight3}
\fmfv{decor.shape=circle,decor.filled=empty,decor.size=1.5thick,label.angle=45,label.dist=0.15cm,label=$\alpha$}{vIndex}
\fmf{phantom,tension=3.5}{vUpLeft3,vIndex}
\fmf{phantom,tension=3.6}{i,vIndex}
\fmf{phantom,tension=1.4}{o,vIndex}
\fmf{phantom,tension=10}{i,v3}
\fmf{phantom,tension=10}{o,v4}
\fmf{plain,right=0.4,tension=0.5,foreground=(1,,0,,0)}{v1,vUpLeft}
\fmf{plain,right,tension=0.5,foreground=(1,,0,,0)}{vUpLeft,vDownLeft}
\fmf{plain,left=0.4,tension=0.5,foreground=(1,,0,,0)}{v1,vDownLeft}
\fmf{phantom,right=0.4,tension=0.5}{vUpRight,v2}
\fmf{phantom,left,tension=0.5}{vUpRight,vDownRight}
\fmf{phantom,left=0.4,tension=0.5}{vDownRight,v2}
\fmf{phantom,tension=0.3}{v2bis,o}
\fmf{plain,left,tension=0.1,foreground=(1,,0,,0)}{v2,v2bis,v2}
\fmf{zigzag,tension=2.7,foreground=(1,,0,,0)}{v1,v2}
\fmf{phantom,tension=2}{v1,v3}
\fmf{phantom,tension=2}{v2,v4}
\fmf{phantom,tension=2.4}{vUpLeft,vUpLeft2}
\fmf{phantom,tension=2.4}{vDownLeft,vDownLeft2}
\fmf{phantom,tension=2.4}{vUpRight,vUpRight2}
\fmf{phantom,tension=2.4}{vDownRight,vDownRight2}
\fmf{zigzag,tension=0,foreground=(1,,0,,0)}{vUpLeft,vDownLeft}
\end{fmfgraph*}
\end{fmffile}
\end{gathered} \hspace{-0.4cm} + \frac{1}{36} \hspace{-0.22cm} \begin{gathered}
\begin{fmffile}{Diagrams/pure4PPIEA_dW3dK_Diag7}
\begin{fmfgraph*}(40,20)
\fmfleft{i}
\fmfright{o}
\fmftop{vUpLeft1,vUpLeft2,vUpLeft3,vUpRight1,vUpRight2,vUpRight3}
\fmfbottom{vDownLeft1,vDownLeft2,vDownLeft3,vDownRight1,vDownRight2,vDownRight3}
\fmfv{decor.shape=circle,decor.filled=empty,decor.size=1.5thick,label.dist=0.15cm,label=$\alpha$}{v2bis}
\fmf{phantom,tension=10}{i,v3}
\fmf{phantom,tension=10}{o,v4}
\fmf{plain,right=0.4,tension=0.5,foreground=(1,,0,,0)}{v1,vUpLeft}
\fmf{plain,right,tension=0.5,foreground=(1,,0,,0)}{vUpLeft,vDownLeft}
\fmf{plain,left=0.4,tension=0.5,foreground=(1,,0,,0)}{v1,vDownLeft}
\fmf{phantom,right=0.4,tension=0.5}{vUpRight,v2}
\fmf{phantom,left,tension=0.5}{vUpRight,vDownRight}
\fmf{phantom,left=0.4,tension=0.5}{vDownRight,v2}
\fmf{phantom,tension=0.3}{v2bis,o}
\fmf{plain,left,tension=0.1,foreground=(1,,0,,0)}{v2,v2bis,v2}
\fmf{zigzag,tension=2.7,foreground=(1,,0,,0)}{v1,v2}
\fmf{phantom,tension=2}{v1,v3}
\fmf{phantom,tension=2}{v2,v4}
\fmf{phantom,tension=2.4}{vUpLeft,vUpLeft2}
\fmf{phantom,tension=2.4}{vDownLeft,vDownLeft2}
\fmf{phantom,tension=2.4}{vUpRight,vUpRight2}
\fmf{phantom,tension=2.4}{vDownRight,vDownRight2}
\fmf{zigzag,tension=0,foreground=(1,,0,,0)}{vUpLeft,vDownLeft}
\end{fmfgraph*}
\end{fmffile}
\end{gathered} }$} \\
& \scalebox{0.97}{${\displaystyle + \frac{1}{72} \hspace{0.45cm} \begin{gathered}
\begin{fmffile}{Diagrams/pure4PPIEA_dW3dK_Diag8}
\begin{fmfgraph*}(45,18)
\fmfleft{i}
\fmfright{o}
\fmfv{decor.shape=circle,decor.filled=empty,decor.size=1.5thick,label.dist=0.15cm,label=$\alpha$}{i1}
\fmf{phantom,tension=10}{i,i1}
\fmf{phantom,tension=10}{o,o1}
\fmf{plain,left,tension=0.5,foreground=(1,,0,,0)}{i1,v1,i1}
\fmf{plain,right,tension=0.5,foreground=(1,,0,,0)}{o1,v2,o1}
\fmf{zigzag,foreground=(1,,0,,0)}{v1,v3}
\fmf{plain,left,tension=0.5,foreground=(1,,0,,0)}{v3,v4}
\fmf{plain,right,tension=0.5,foreground=(1,,0,,0)}{v3,v4}
\fmf{zigzag,foreground=(1,,0,,0)}{v4,v2}
\end{fmfgraph*}
\end{fmffile}
\end{gathered} + \frac{1}{72} \hspace{0.1cm} \begin{gathered}
\begin{fmffile}{Diagrams/pure4PPIEA_dW3dK_Diag9}
\begin{fmfgraph*}(45,18)
\fmfleft{i}
\fmfright{o}
\fmftop{vUp}
\fmfbottom{vDown}
\fmfv{decor.shape=circle,decor.filled=empty,decor.size=1.5thick,label.dist=0.15cm,label=$\alpha$}{vIndex}
\fmf{phantom,tension=2.8}{vUp,vIndex}
\fmf{phantom,tension=1}{vDown,vIndex}
\fmf{phantom,tension=10}{i,i1}
\fmf{phantom,tension=10}{o,o1}
\fmf{plain,left,tension=0.5,foreground=(1,,0,,0)}{i1,v1,i1}
\fmf{plain,right,tension=0.5,foreground=(1,,0,,0)}{o1,v2,o1}
\fmf{zigzag,foreground=(1,,0,,0)}{v1,v3}
\fmf{plain,left,tension=0.5,foreground=(1,,0,,0)}{v3,v4}
\fmf{plain,right,tension=0.5,foreground=(1,,0,,0)}{v3,v4}
\fmf{zigzag,foreground=(1,,0,,0)}{v4,v2}
\end{fmfgraph*}
\end{fmffile}
\end{gathered} \;,}$}
\end{split}
\label{eq:pure4PPIEAdW3dK0DON}
\end{equation}
\begin{equation}
\left.\frac{\delta W_{3}[K,M]}{\delta M_{\alpha}}\right|_{K=K_{0} \atop M=M_{0}} = - \frac{1}{24} \hspace{-0.1cm} \begin{gathered}
\begin{fmffile}{Diagrams/pure4PPIEA_dW3dM_Diag1}
\begin{fmfgraph*}(30,20)
\fmfleft{ibis,iUpbis}
\fmfright{obis,oUpbis}
\fmf{phantom,tension=3.0}{i,ibis}
\fmf{phantom,tension=3.0}{o,obis}
\fmf{phantom,tension=3.0}{iUp,iUpbis}
\fmf{phantom,tension=3.0}{oUp,oUpbis}
\fmfv{decor.shape=circle,decor.filled=empty,decor.size=1.5thick,label.angle=180,label.dist=0.1cm,label=$\alpha$}{vIndex}
\fmf{plain,left,foreground=(1,,0,,0)}{i,vIndex,i}
\fmf{plain,left,foreground=(1,,0,,0)}{o,vIndex,o}
\fmf{zigzag,left=0.11,tension=1.5,foreground=(1,,0,,0)}{iUp,oUp}
\fmf{zigzag,left=1.0,tension=2.1,foreground=(1,,0,,0)}{i,iUp}
\fmf{zigzag,right=1.0,tension=2.1,foreground=(1,,0,,0)}{o,oUp}
\end{fmfgraph*}
\end{fmffile}
\end{gathered} \hspace{-0.15cm} - \frac{1}{12} \hspace{-0.2cm} \begin{gathered}
\begin{fmffile}{Diagrams/pure4PPIEA_dW3dM_Diag2}
\begin{fmfgraph*}(38,18)
\fmfleft{i}
\fmfright{o}
\fmftop{vUpLeft1,vUpLeft2,vUpLeft3,vUpRight1,vUpRight2,vUpRight3}
\fmfbottom{vDownLeft1,vDownLeft2,vDownLeft3,vDownRight1,vDownRight2,vDownRight3}
\fmfv{decor.shape=circle,decor.filled=empty,decor.size=1.5thick,label.angle=0,label.dist=0.15cm,label=$\alpha$}{v1}
\fmf{phantom,tension=1.1}{i,vbis}
\fmf{phantom,tension=2}{o,vbis}
\fmf{plain,left,tension=0,foreground=(1,,0,,0)}{vbis,v1,vbis}
\fmf{phantom,tension=10}{i,v3}
\fmf{phantom,tension=10}{o,v4}
\fmf{plain,right=0.4,tension=0.5,foreground=(1,,0,,0)}{v1,vUpLeft}
\fmf{plain,right,tension=0.5,foreground=(1,,0,,0)}{vUpLeft,vDownLeft}
\fmf{plain,left=0.4,tension=0.5,foreground=(1,,0,,0)}{v1,vDownLeft}
\fmf{phantom,right=0.4,tension=0.5}{vUpRight,v2}
\fmf{phantom,left,tension=0.5}{vUpRight,vDownRight}
\fmf{phantom,left=0.4,tension=0.5}{vDownRight,v2}
\fmf{phantom,tension=0.3}{v2bis,o}
\fmf{phantom,left,tension=0.1,foreground=(1,,0,,0)}{v2,v2bis,v2}
\fmf{phantom,tension=2.7,foreground=(1,,0,,0)}{v1,v2}
\fmf{phantom,tension=2}{v1,v3}
\fmf{phantom,tension=2}{v2,v4}
\fmf{phantom,tension=2.4}{vUpLeft,vUpLeft2}
\fmf{phantom,tension=2.4}{vDownLeft,vDownLeft2}
\fmf{phantom,tension=2.4}{vUpRight,vUpRight2}
\fmf{phantom,tension=2.4}{vDownRight,vDownRight2}
\fmf{zigzag,tension=0,foreground=(1,,0,,0)}{vUpLeft,vDownLeft}
\end{fmfgraph*}
\end{fmffile}
\end{gathered} \hspace{-1.2cm} - \frac{1}{24} \hspace{0.1cm} \begin{gathered}
\begin{fmffile}{Diagrams/pure4PPIEA_dW3dM_Diag3}
\begin{fmfgraph*}(35,18)
\fmfleft{i}
\fmfright{o}
\fmfv{decor.shape=circle,decor.filled=empty,decor.size=1.5thick,label.dist=0.15cm,label=$\alpha$}{v3}
\fmf{phantom,tension=10}{i,i1}
\fmf{phantom,tension=10}{o,o1}
\fmf{plain,left,tension=0.5,foreground=(1,,0,,0)}{i1,v3,i1}
\fmf{plain,right,tension=0.5,foreground=(1,,0,,0)}{o1,v2,o1}
\fmf{plain,left,tension=0.5,foreground=(1,,0,,0)}{v3,v4}
\fmf{plain,right,tension=0.5,foreground=(1,,0,,0)}{v3,v4}
\fmf{zigzag,foreground=(1,,0,,0)}{v4,v2}
\end{fmfgraph*}
\end{fmffile}
\end{gathered} \;,
\label{eq:pure4PPIEAdW3dM0DON}
\end{equation}
\begin{equation}
\scalebox{0.96}{${\displaystyle \left.\frac{\delta^{2} W_{2}[K,M]}{\delta K_{\alpha_{1}} \delta K_{\alpha_{2}}}\right|_{K=K_{0} \atop M=M_{0}} = -\frac{1}{6} \hspace{0.1cm} \begin{gathered}
\begin{fmffile}{Diagrams/pure4PPIEA_dW2dKdK_Diag1}
\begin{fmfgraph*}(30,20)
\fmfleft{i}
\fmfright{o}
\fmftop{vUpL,vUp,vUpR}
\fmfbottom{vDownL,vDown,vDownR}
\fmfv{decor.shape=circle,decor.filled=empty,decor.size=1.5thick,label.angle=90,label.dist=0.15cm,label=$\alpha_{1}$}{vIndex1}
\fmfv{decor.shape=circle,decor.filled=empty,decor.size=1.5thick,label.angle=-90,label.dist=0.15cm,label=$\alpha_{2}$}{vIndex2}
\fmf{phantom,tension=12.0}{vUpL,vIndex1}
\fmf{phantom,tension=4.9}{i,vIndex1}
\fmf{phantom,tension=4}{o,vIndex1}
\fmf{phantom,tension=12.0}{vDownL,vIndex2}
\fmf{phantom,tension=4.9}{i,vIndex2}
\fmf{phantom,tension=4}{o,vIndex2}
\fmf{phantom,tension=10}{i,i1}
\fmf{phantom,tension=10}{o,o1}
\fmf{plain,left,tension=0.5,foreground=(1,,0,,0)}{i1,v1,i1}
\fmf{plain,right,tension=0.5,foreground=(1,,0,,0)}{o1,v2,o1}
\fmf{zigzag,foreground=(1,,0,,0)}{v1,v2}
\end{fmfgraph*}
\end{fmffile}
\end{gathered}
\hspace{0.1cm} -\frac{1}{12} \hspace{0.6cm} \begin{gathered}
\begin{fmffile}{Diagrams/pure4PPIEA_dW2dKdK_Diag2}
\begin{fmfgraph*}(30,20)
\fmfleft{i}
\fmfright{o}
\fmfv{decor.shape=circle,decor.filled=empty,decor.size=1.5thick,label.dist=0.15cm,label=$\alpha_{1}$}{i1}
\fmfv{decor.shape=circle,decor.filled=empty,decor.size=1.5thick,label.dist=0.15cm,label=$\alpha_{2}$}{o1}
\fmf{phantom,tension=10}{i,i1}
\fmf{phantom,tension=10}{o,o1}
\fmf{plain,left,tension=0.5,foreground=(1,,0,,0)}{i1,v1,i1}
\fmf{plain,right,tension=0.5,foreground=(1,,0,,0)}{o1,v2,o1}
\fmf{zigzag,foreground=(1,,0,,0)}{v1,v2}
\end{fmfgraph*}
\end{fmffile}
\end{gathered}
\hspace{0.5cm} - \frac{1}{3}\begin{gathered}
\begin{fmffile}{Diagrams/pure4PPIEA_dW2dKdK_Diag3}
\begin{fmfgraph*}(15,15)
\fmfleft{i}
\fmfright{o}
\fmftop{vUp}
\fmfbottom{vDown}
\fmfv{decor.shape=circle,decor.filled=empty,decor.size=1.5thick,label.angle=135,label.dist=0.1cm,label=$\alpha_{1}$}{vBis}
\fmfv{decor.shape=circle,decor.filled=empty,decor.size=1.5thick,label.angle=45,label.dist=0.1cm,label=$\alpha_{2}$}{vBis2}
\fmf{phantom,tension=5}{o,vBis2}
\fmf{phantom,tension=8}{vUp,vBis2}
\fmf{phantom,tension=-0.2}{vDown,vBis2}
\fmf{phantom,tension=5}{i,vBis}
\fmf{phantom,tension=8}{vUp,vBis}
\fmf{phantom,tension=-0.2}{vDown,vBis}
\fmf{phantom,tension=11}{i,v1}
\fmf{phantom,tension=11}{v2,o}
\fmf{plain,left,tension=0.4,foreground=(1,,0,,0)}{v1,v2,v1}
\fmf{zigzag,foreground=(1,,0,,0)}{v1,v2}
\end{fmfgraph*}
\end{fmffile}
\end{gathered} - \frac{1}{6}\begin{gathered}
\begin{fmffile}{Diagrams/pure4PPIEA_dW2dKdK_Diag4}
\begin{fmfgraph*}(15,15)
\fmfleft{i}
\fmfright{o}
\fmftop{vUp}
\fmfbottom{vDown}
\fmfv{decor.shape=circle,decor.filled=empty,decor.size=1.5thick,label.dist=0.15cm,label=$\alpha_{1}$}{vBis}
\fmfv{decor.shape=circle,decor.filled=empty,decor.size=1.5thick,label.dist=0.15cm,label=$\alpha_{2}$}{vBis2}
\fmf{phantom,tension=7}{vUp,vBis}
\fmf{phantom,tension=1}{vDown,vBis}
\fmf{phantom,tension=1}{vUp,vBis2}
\fmf{phantom,tension=7}{vDown,vBis2}
\fmf{phantom,tension=11}{i,v1}
\fmf{phantom,tension=11}{v2,o}
\fmf{plain,left,tension=0.4,foreground=(1,,0,,0)}{v1,v2,v1}
\fmf{zigzag,foreground=(1,,0,,0)}{v1,v2}
\end{fmfgraph*}
\end{fmffile}
\end{gathered} \hspace{-0.05cm} \;,}$}
\label{eq:pure4PPIEAdW2dKdK0DON}
\end{equation}
\begin{equation}
\left.\frac{\delta^{2} W_{2}[K,M]}{\delta M_{\alpha_{1}} \delta M_{\alpha_{2}}}\right|_{K=K_{0} \atop M=M_{0}} = 0 \;,
\label{eq:pure4PPIEAdW2dMdM0DON}
\end{equation}
\begin{equation}
\left.\frac{\delta^{2} W_{2}[K,M]}{\delta K_{\alpha_{1}} \delta M_{\alpha_{2}}}\right|_{K=K_{0} \atop M=M_{0}} = \frac{1}{4} \hspace{0.7cm} \begin{gathered}
\begin{fmffile}{Diagrams/pure4PPIEA_dW2dKdM_Diag1}
\begin{fmfgraph*}(20,10)
\fmfleft{i}
\fmfright{o}
\fmfv{decor.shape=circle,decor.filled=empty,decor.size=1.5thick,label.angle=180,label.dist=0.15cm,label=$\alpha_{1}$}{i}
\fmfv{decor.shape=circle,decor.filled=empty,decor.size=1.5thick,label.angle=180,label.dist=0.15cm,label=$\alpha_{2}$}{v1}
\fmf{plain,left,tension=0.1,foreground=(1,,0,,0)}{i,v1,i}
\fmf{plain,left,tension=0.1,foreground=(1,,0,,0)}{o,v1,o}
\end{fmfgraph*}
\end{fmffile}
\end{gathered} \hspace{0.1cm} \;,
\label{eq:pure4PPIEAdW2dMdK0DON}
\end{equation}
\begin{equation}
\left.\frac{\delta^{3} W_{1}[K,M]}{\delta K_{\alpha_{1}} \delta K_{\alpha_{2}} \delta K_{\alpha_{3}}}\right|_{K=K_{0} \atop M=M_{0}} = \hspace{0.4cm} \begin{gathered}
\begin{fmffile}{Diagrams/pure4PPIEA_dW1dKdKdK_Diag1}
\begin{fmfgraph*}(15,15)
\fmfleft{i}
\fmfright{o}
\fmftop{vUp}
\fmfbottom{vDown}
\fmfv{decor.shape=circle,decor.filled=empty,decor.size=1.5thick,label.dist=0.15cm,label=$\alpha_{1}$}{v1}
\fmfv{decor.shape=circle,decor.filled=empty,decor.size=1.5thick,label.dist=0.15cm,label=$\alpha_{2}$}{v2}
\fmfv{decor.shape=circle,decor.filled=empty,decor.size=1.5thick,label.dist=0.15cm,label=$\alpha_{3}$}{v3}
\fmf{phantom,tension=7.0}{vUp,v3}
\fmf{phantom,tension=1}{vDown,v3}
\fmf{phantom,tension=11}{i,v1}
\fmf{phantom,tension=11}{v2,o}
\fmf{plain,left,tension=0.4,foreground=(1,,0,,0)}{v1,v2,v1}
\fmf{phantom}{v1,v2}
\end{fmfgraph*}
\end{fmffile}
\end{gathered} \hspace{0.4cm} \;.
\label{eq:pure4PPIEAdW1dKdKdK0DON}
\end{equation}
We are left to determine the source coefficients $K_{1}[\rho,\zeta]$, $K_{2}[\rho,\zeta]$ and $M_{1}[\rho,\zeta]$ in order to rewrite expressions~\eqref{eq:pure4PPIEAIMGamma30DON} and~\eqref{eq:pure4PPIEAIMGamma40DON} of $\Gamma^{(\mathrm{4PPI})}_{3}[\rho,\zeta]$ and $\Gamma^{(\mathrm{4PPI})}_{4}[\rho,\zeta]$. As the present formalism involves more than one external source, we will do so by following the recipe of section~\ref{sec:1PIEAannIM} and Taylor expand the coefficients of the power series~\eqref{eq:pure4PPIEArhoExpansion0DON} and~\eqref{eq:pure4PPIEAzetaExpansion0DON} around $(K,M)=(K_{0},M_{0})$:
\begin{equation}
\begin{split}
\rho_{\alpha_{1}} = & \sum_{n=0}^{\infty} \rho_{n,\alpha_{1}}[K,M] \hbar^{n} \\
= & \ \rho_{0,\alpha_{1}}[K] + \rho_{1,\alpha_{1}}[K,M]\hbar + \rho_{2,\alpha_{1}}[K,M]\hbar^{2} + \mathcal{O}\big(\hbar^{3}\big) \\
= & \ \rho_{0,\alpha_{1}}[K=K_{0}] + \int_{\alpha_{2}} \left.\frac{\delta \rho_{0,\alpha_{1}}[K]}{\delta K_{\alpha_{2}}}\right|_{K=K_{0}} \left(K_{1,\alpha_{2}}[\rho,\zeta] \hbar + K_{2,\alpha_{2}}[\rho,\zeta] \hbar^{2}\right) \\
& + \frac{1}{2} \int_{\alpha_{2},\alpha_{3}} \left.\frac{\delta^{2} \rho_{0,\alpha_{1}}[K]}{\delta K_{\alpha_{2}}\delta K_{\alpha_{3}}}\right|_{K=K_{0}} \left(K_{1,\alpha_{2}}[\rho,\zeta] \hbar\right) \left(K_{1,\alpha_{3}}[\rho,\zeta] \hbar\right) \\
& + \Bigg[ \rho_{1,\alpha_{1}}[K=K_{0},M=M_{0}] + \int_{\alpha_{2}} \left.\frac{\delta \rho_{1,\alpha_{1}}[K,M]}{\delta K_{\alpha_{2}}}\right|_{K=K_{0} \atop M=M_{0}} \left(K_{1,\alpha_{2}}[\rho,\zeta] \hbar\right) \\
& \hspace{0.58cm} + \int_{\alpha_{2}} \left.\frac{\delta \rho_{1,\alpha_{1}}[K,M]}{\delta M_{\alpha_{2}}}\right|_{K=K_{0} \atop M=M_{0}} \left(M_{1,\alpha_{2}}[\rho,\zeta] \hbar\right) \Bigg] \hbar \\
& + \rho_{2,\alpha_{1}}[K=K_{0},M=M_{0}] \hbar^{2} + \mathcal{O}\big(\hbar^{3}\big) \\
& \\
= & \ \rho_{0,\alpha_{1}}[K=K_{0}] \\
& + \hbar \bigg(\int_{\alpha_{2}} \left.\frac{\delta \rho_{0,\alpha_{1}}[K]}{\delta K_{\alpha_{2}}}\right|_{K=K_{0}} K_{1,\alpha_{2}}[\rho,\zeta] + \rho_{1,\alpha_{1}}[K=K_{0},M=M_{0}] \bigg) \\
& + \hbar^{2} \Bigg( \int_{\alpha_{2}} \left.\frac{\delta \rho_{0,\alpha_{1}}[K]}{\delta K_{\alpha_{2}}}\right|_{K=K_{0}} K_{2,\alpha_{2}}[\rho,\zeta] + \frac{1}{2} \int_{\alpha_{2},\alpha_{3}} \left.\frac{\delta^{2} \rho_{0,\alpha_{1}}[K]}{\delta K_{\alpha_{2}}\delta K_{\alpha_{3}}}\right|_{K=K_{0}} K_{1,\alpha_{2}}[\rho,\zeta] K_{1,\alpha_{3}}[\rho,\zeta] \\
& \hspace{1.07cm} + \int_{\alpha_{2}} \left.\frac{\delta \rho_{1,\alpha_{1}}[K,M]}{\delta K_{\alpha_{2}}}\right|_{K=K_{0} \atop M=M_{0}} K_{1,\alpha_{2}}[\rho,\zeta] + \int_{\alpha_{2}} \left.\frac{\delta \rho_{1,\alpha_{1}}[K,M]}{\delta M_{\alpha_{2}}}\right|_{K=K_{0} \atop M=M_{0}} M_{1,\alpha_{2}}[\rho,\zeta] \\
& \hspace{1.07cm} + \rho_{2,\alpha_{1}}[K=K_{0},M=M_{0}] \Bigg) \\
& + \mathcal{O}\big(\hbar^{3}\big) \;,
\end{split}
\label{eq:pure4PPIEArhoExpansionAroundK0M00DON}
\end{equation}
\begin{equation*}
\begin{split}
\zeta_{\alpha_{1}} = & \sum_{n=0}^{\infty} \zeta_{n,\alpha_{1}}[K,M] \hbar^{n} \\
= & \ \zeta_{0,\alpha_{1}}[K,M] + \zeta_{1,\alpha_{1}}[K,M]\hbar + \zeta_{2,\alpha_{1}}[K,M]\hbar^{2} + \mathcal{O}\big(\hbar^{3}\big) \\
= & \ \zeta_{0,\alpha_{1}}[K=K_{0},M=M_{0}] + \int_{\alpha_{2}} \left.\frac{\delta \zeta_{0,\alpha_{1}}[K,M]}{\delta K_{\alpha_{2}}}\right|_{K=K_{0} \atop M=M_{0}} \left(K_{1,\alpha_{2}}[\rho,\zeta] \hbar + K_{2,\alpha_{2}}[\rho,\zeta] \hbar^{2}\right) \\
& + \int_{\alpha_{2}} \left.\frac{\delta \zeta_{0,\alpha_{1}}[K,M]}{\delta M_{\alpha_{2}}}\right|_{K=K_{0} \atop M=M_{0}} \left(M_{1,\alpha_{2}}[\rho,\zeta] \hbar + M_{2,\alpha_{2}}[\rho,\zeta] \hbar^{2}\right)
\end{split}
\end{equation*}
\begin{equation}
\begin{split}
\hspace{1.2cm} & + \Bigg[ \zeta_{1,\alpha_{1}}[K=K_{0},M=M_{0}] + \int_{\alpha_{2}} \left.\frac{\delta \zeta_{1,\alpha_{1}}[K,M]}{\delta K_{\alpha_{2}}}\right|_{K=K_{0} \atop M=M_{0}} \left(K_{1,\alpha_{2}}[\rho,\zeta] \hbar\right) \\
& \hspace{0.58cm} + \int_{\alpha_{2}} \left.\frac{\delta \zeta_{1,\alpha_{1}}[K,M]}{\delta M_{\alpha_{2}}}\right|_{K=K_{0} \atop M=M_{0}} \left(M_{1,\alpha_{2}}[\rho,\zeta] \hbar\right) \Bigg] \hbar \\
& + \zeta_{2,\alpha_{1}}[K=K_{0},M=M_{0}] \hbar^{2} + \mathcal{O}\big(\hbar^{3}\big) \\
& \\
= & \ \zeta_{0,\alpha_{1}}[K=K_{0},M=M_{0}] \\
& + \hbar \Bigg(\int_{\alpha_{2}} \left.\frac{\delta \zeta_{0,\alpha_{1}}[K,M]}{\delta K_{\alpha_{2}}}\right|_{K=K_{0} \atop M=M_{0}} K_{1,\alpha_{2}}[\rho,\zeta] + \int_{\alpha_{2}} \left.\frac{\delta \zeta_{0,\alpha_{1}}[K,M]}{\delta M_{\alpha_{2}}}\right|_{K=K_{0} \atop M=M_{0}} M_{1,\alpha_{2}}[\rho,\zeta] \\
& \hspace{0.9cm} + \zeta_{1,\alpha_{1}}[K=K_{0},M=M_{0}] \Bigg) \\
& + \hbar^{2} \Bigg( \int_{\alpha_{2}} \left.\frac{\delta \zeta_{0,\alpha_{1}}[K,M]}{\delta K_{\alpha_{2}}}\right|_{K=K_{0} \atop M=M_{0}} K_{2,\alpha_{2}}[\rho,\zeta] + \int_{\alpha_{2}} \left.\frac{\delta \zeta_{0,\alpha_{1}}[K,M]}{\delta M_{\alpha_{2}}}\right|_{K=K_{0} \atop M=M_{0}} M_{2,\alpha_{2}}[\rho,\zeta] \\
& \hspace{1.07cm} + \int_{\alpha_{2}} \left.\frac{\delta \zeta_{1,\alpha_{1}}[K,M]}{\delta K_{\alpha_{2}}}\right|_{K=K_{0} \atop M=M_{0}} K_{1,\alpha_{2}}[\rho,\zeta] + \int_{\alpha_{2}} \left.\frac{\delta \zeta_{1,\alpha_{1}}[K,M]}{\delta M_{\alpha_{2}}}\right|_{K=K_{0} \atop M=M_{0}} M_{1,\alpha_{2}}[\rho,\zeta] \\
& \hspace{1.07cm} + \zeta_{2,\alpha_{1}}[K=K_{0},M=M_{0}] \Bigg) \\
& + \mathcal{O}\big(\hbar^{3}\big) \;.
\end{split}
\label{eq:pure4PPIEAzetaExpansionAroundK0M00DON}
\end{equation}
We have used in~\eqref{eq:pure4PPIEArhoExpansionAroundK0M00DON} and~\eqref{eq:pure4PPIEAzetaExpansionAroundK0M00DON} the relations $\frac{\delta \rho_{0,\alpha_{1}}[K,M]}{\delta M_{\alpha_{2}}} = \frac{\delta \rho_{0,\alpha_{1}}[K]}{\delta M_{\alpha_{2}}} = 0$ $\forall \alpha_{1},\alpha_{2}$ and $\frac{\delta^{2} \zeta_{0,\alpha_{1}}[K,M]}{\delta M_{\alpha_{2}}\delta M_{\alpha_{3}}} = 0$ $\forall \alpha_{1},\alpha_{2},\alpha_{3}$, which both follow from expressions~\eqref{eq:pure4PPIEAGK00DON} and~\eqref{eq:pure4PPIEAzetaK00DON} of $\rho_{0,\alpha}$ and $\zeta_{0,\alpha}$ after evaluating the derivatives of the $W_{n}$ coefficients. At $(K,M)=(K_{0},M_{0})$, the latter expressions read:
\begin{equation}
\begin{split}
\rho_{\alpha} = \rho_{0,\alpha}[K=K_{0}] = & \ 2 \left.\frac{\delta W_{1}[K]}{\delta K_{\alpha}}\right|_{K=K_{0}} \\
= & \ 2 \left.\frac{\delta}{\delta K_{\alpha}} \left(\frac{1}{2}\mathrm{STr}\left[\ln\big(\boldsymbol{G}_{K}[K]\big)\right]\right) \right|_{K=K_{0}} \\
= & \ \frac{\delta}{\delta K_{\alpha}} \begin{gathered}
\begin{fmffile}{Diagrams/pure4PPIEA_rho_Diag1}
\begin{fmfgraph*}(15,15)
\fmfleft{i}
\fmfright{o}
\fmftop{vUp}
\fmfbottom{vDown}
\fmf{phantom,tension=11}{i,v1}
\fmf{phantom,tension=11}{v2,o}
\fmf{plain,left,tension=0.4}{v1,v2,v1}
\fmf{phantom}{v1,v2}
\end{fmfgraph*}
\end{fmffile}
\end{gathered}  \left. \rule{0cm}{0.65cm} \right|_{K=K_{0}} \\
= & \hspace{0.4cm} \begin{gathered}
\begin{fmffile}{Diagrams/pure4PPIEA_rho_Diag2}
\begin{fmfgraph*}(15,15)
\fmfleft{i}
\fmfright{o}
\fmftop{vUp}
\fmfbottom{vDown}
\fmfv{decor.shape=circle,decor.filled=empty,decor.size=1.5thick,label.dist=0.15cm,label=$\alpha$}{v1}
\fmf{phantom,tension=11}{i,v1}
\fmf{phantom,tension=11}{v2,o}
\fmf{plain,left,tension=0.4,foreground=(1,,0,,0)}{v1,v2,v1}
\fmf{phantom}{v1,v2}
\end{fmfgraph*}
\end{fmffile}
\end{gathered} \;,
\end{split}
\label{eq:pure4PPIIMrho0DON}
\end{equation}
\begin{equation}
\begin{split}
\zeta_{\alpha} = \zeta_{0,\alpha}[K=K_{0},M=M_{0}] = & \ 24 \left(\left.\frac{\delta W_{3}[K,M]}{\delta M_{\alpha}}\right|_{K=K_{0} \atop M=M_{0}} - \left.\frac{\delta W_{1}[K]}{\delta K_{\alpha}}\right|_{K = K_{0}} \left.\frac{\delta W_{2}[K,M]}{\delta K_{\alpha}}\right|_{K=K_{0} \atop M=M_{0}}\right) \\
= & - \hspace{-0.1cm} \begin{gathered}
\begin{fmffile}{Diagrams/pure4PPIEA_dW3dM_Diag1}
\begin{fmfgraph*}(30,20)
\fmfleft{ibis,iUpbis}
\fmfright{obis,oUpbis}
\fmf{phantom,tension=3.0}{i,ibis}
\fmf{phantom,tension=3.0}{o,obis}
\fmf{phantom,tension=3.0}{iUp,iUpbis}
\fmf{phantom,tension=3.0}{oUp,oUpbis}
\fmfv{decor.shape=circle,decor.filled=empty,decor.size=1.5thick,label.angle=180,label.dist=0.1cm,label=$\alpha$}{vIndex}
\fmf{plain,left,foreground=(1,,0,,0)}{i,vIndex,i}
\fmf{plain,left,foreground=(1,,0,,0)}{o,vIndex,o}
\fmf{zigzag,left=0.11,tension=1.5,foreground=(1,,0,,0)}{iUp,oUp}
\fmf{zigzag,left=1.0,tension=2.1,foreground=(1,,0,,0)}{i,iUp}
\fmf{zigzag,right=1.0,tension=2.1,foreground=(1,,0,,0)}{o,oUp}
\end{fmfgraph*}
\end{fmffile}
\end{gathered} \hspace{-0.15cm} \;,
\end{split}
\label{eq:pure4PPIIMzeta0DON}
\end{equation}
where the derivatives of $W_{2}$ and $W_{3}$ are given by~\eqref{eq:pure4PPIEAdW2dK0DON} and~\eqref{eq:pure4PPIEAdW3dM0DON}. As a next step, we set $K=K_{0}$ and $M=M_{0}$ in~\eqref{eq:pure4PPIEArhoExpansionAroundK0M00DON} and~\eqref{eq:pure4PPIEAzetaExpansionAroundK0M00DON}. By exploiting the fact that both $\rho$ and $\zeta$ are quantities of order $\mathcal{O}\big(\hbar^{0}\big)$, we turn the relations thus obtained into the following system of coupled equations:
\begin{subequations}
\begin{empheq}[left=\empheqlbrace]{align}
& \hspace{0.1cm} \mathrm{Order}~\mathcal{O}\big(\hbar^{0}\big):~\rho_{\alpha} = \rho_{0,\alpha}[K=K_{0}] \quad \forall \alpha \;, \label{eq:pure4PPIEATowerEquation10DON}\\
& \hspace{0.1cm} \hspace{2.835cm} \zeta_{\alpha} = \zeta_{0,\alpha}[K=K_{0},M=M_{0}] \quad \forall \alpha\;, \label{eq:pure4PPIEATowerEquation1bis0DON}\\
\nonumber \\
& \hspace{0.1cm} \mathrm{Order}~\mathcal{O}(\hbar):~0 = \int_{\alpha_{2}} \left.\frac{\delta \rho_{0,\alpha_{1}}[K]}{\delta K_{\alpha_{2}}}\right|_{K=K_{0}} K_{1,\alpha_{2}}[\rho,\zeta] \nonumber \\
& \hspace{0.1cm} \hspace{3.22cm} + \rho_{1,\alpha_{1}}[K=K_{0},M=M_{0}] \quad \forall \alpha_{1}\;, \label{eq:pure4PPIEATowerEquation20DON}\\
& \hspace{0.1cm} \hspace{2.57cm} 0 = \int_{\alpha_{2}} \left.\frac{\delta \zeta_{0,\alpha_{1}}[K,M]}{\delta K_{\alpha_{2}}}\right|_{K=K_{0} \atop M=M_{0}} K_{1,\alpha_{2}}[\rho,\zeta] \nonumber \\
& \hspace{0.1cm} \hspace{3.22cm} + \int_{\alpha_{2}} \left.\frac{\delta \zeta_{0,\alpha_{1}}[K,M]}{\delta M_{\alpha_{2}}}\right|_{K=K_{0} \atop M=M_{0}} M_{1,\alpha_{2}}[\rho,\zeta] \nonumber \\
& \hspace{0.1cm} \hspace{3.22cm} + \zeta_{1,\alpha_{1}}[K=K_{0},M=M_{0}] \quad \forall \alpha_{1} \;, \label{eq:pure4PPIEATowerEquation2bis0DON} \\
\nonumber \\
& \hspace{0.1cm} \mathrm{Order}~\mathcal{O}\big(\hbar^{2}\big):~0 = \int_{\alpha_{2}} \left.\frac{\delta \rho_{0,\alpha_{1}}[K]}{\delta K_{\alpha_{2}}}\right|_{K=K_{0}} K_{2,\alpha_{2}}[\rho,\zeta] \nonumber \\
& \hspace{0.1cm} \hspace{3.453cm} + \frac{1}{2} \int_{\alpha_{2},\alpha_{3}} \left.\frac{\delta^{2} \rho_{0,\alpha_{1}}[K]}{\delta K_{\alpha_{2}}\delta K_{\alpha_{3}}}\right|_{K=K_{0}} K_{1,\alpha_{2}}[\rho,\zeta] K_{1,\alpha_{3}}[\rho,\zeta] \nonumber \\
& \hspace{0.1cm} \hspace{3.453cm} + \int_{\alpha_{2}} \left.\frac{\delta \rho_{1,\alpha_{1}}[K,M]}{\delta K_{\alpha_{2}}}\right|_{K=K_{0} \atop M=M_{0}} K_{1,\alpha_{2}}[\rho,\zeta] \nonumber \\
& \hspace{0.1cm} \hspace{3.453cm} + \int_{\alpha_{2}} \left.\frac{\delta \rho_{1,\alpha_{1}}[K,M]}{\delta M_{\alpha_{2}}}\right|_{K=K_{0} \atop M=M_{0}} M_{1,\alpha_{2}}[\rho,\zeta] \nonumber \\
& \hspace{0.1cm} \hspace{3.453cm} + \rho_{2,\alpha_{1}}[K=K_{0},M=M_{0}] \quad \forall \alpha_{1} \;, \label{eq:pure4PPIEATowerEquation30DON} \\
& \hspace{0.1cm} \hspace{2.803cm} 0 = \int_{\alpha_{2}} \left.\frac{\delta \zeta_{0,\alpha_{1}}[K,M]}{\delta K_{\alpha_{2}}}\right|_{K=K_{0} \atop M=M_{0}} K_{2,\alpha_{2}}[\rho,\zeta] \nonumber \\
& \hspace{0.1cm} \hspace{3.453cm} + \int_{\alpha_{2}} \left.\frac{\delta \zeta_{0,\alpha_{1}}[K,M]}{\delta M_{\alpha_{2}}}\right|_{K=K_{0} \atop M=M_{0}} M_{2,\alpha_{2}}[\rho,\zeta] \nonumber \\
& \hspace{0.1cm} \hspace{3.453cm} + \int_{\alpha_{2}} \left.\frac{\delta \zeta_{1,\alpha_{1}}[K,M]}{\delta K_{\alpha_{2}}}\right|_{K=K_{0} \atop M=M_{0}} K_{1,\alpha_{2}}[\rho,\zeta] \nonumber \\
& \hspace{0.1cm} \hspace{3.453cm} + \int_{\alpha_{2}} \left.\frac{\delta \zeta_{1,\alpha_{1}}[K,M]}{\delta M_{\alpha_{2}}}\right|_{K=K_{0} \atop M=M_{0}} M_{1,\alpha_{2}}[\rho,\zeta] \nonumber \\
& \hspace{0.1cm} \hspace{3.453cm} + \zeta_{2,\alpha_{1}}[K=K_{0},M=M_{0}] \quad \forall \alpha_{1} \;, \label{eq:pure4PPIEATowerEquation3bis0DON} \\
\nonumber \\
& \hspace{7.0cm} \vdots \nonumber
\end{empheq}
\end{subequations}
where~\eqref{eq:pure4PPIEATowerEquation10DON} and~\eqref{eq:pure4PPIEATowerEquation1bis0DON} are already imposed by~\eqref{eq:pure4PPIEAGK00DON} and~\eqref{eq:pure4PPIEAzetaK00DON} respectively. Since
\begin{equation}
\left.\frac{\delta \rho_{0,\alpha_{1}}[K]}{\delta K_{\alpha_{2}}}\right|_{K=K_{0}} = 2 \left.\frac{\delta^{2} W_{1}[K]}{\delta K_{\alpha_{2}} \delta K_{\alpha_{1}}} \right|_{K=K_{0}} = \hspace{0.5cm} \begin{gathered}
\begin{fmffile}{Diagrams/pure4PPIEA_dW1dKdK_Diag1}
\begin{fmfgraph*}(15,15)
\fmfleft{i}
\fmfright{o}
\fmftop{vUp}
\fmfbottom{vDown}
\fmfv{decor.shape=circle,decor.filled=empty,decor.size=1.5thick,label.dist=0.15cm,label=$\alpha_{1}$}{v1}
\fmfv{decor.shape=circle,decor.filled=empty,decor.size=1.5thick,label.dist=0.15cm,label=$\alpha_{2}$}{v2}
\fmf{phantom,tension=11}{i,v1}
\fmf{phantom,tension=11}{v2,o}
\fmf{plain,left,tension=0.4,foreground=(1,,0,,0)}{v1,v2,v1}
\fmf{phantom}{v1,v2}
\end{fmfgraph*}
\end{fmffile}
\end{gathered} \hspace{0.4cm} \;,
\label{eq:pure4PPIEAdrho0dK0DON}
\end{equation}
according to~\eqref{eq:pure4PPIEAdW1dKdK0DON}, we can determine the $K_{1}$ coefficient from~\eqref{eq:pure4PPIEATowerEquation20DON} by introducing the 2-particle inverse propagator:
\begin{equation}
\begin{gathered}
\begin{fmffile}{Diagrams/2PPIEA_FeynRuleDminus1}
\begin{fmfgraph*}(20,20)
\fmfleft{i0,i1,i2,i3}
\fmfright{o0,o1,o2,o3}
\fmftop{vUpL8,vUpL7,vUpL6,vUpL5,vUpL4,vUpL3,vUpL2,vUpL1,vUp,vUpR1,vUpR2,vUpR3,vUpR4,vUpR5,vUpR6,vUpR7,vUpR8}
\fmfbottom{vDownL8,vDownL7,vDownL6,vDownL5,vDownL4,vDownL3,vDownL2,vDownL1,vDown,vDownR1,vDownR2,vDownR3,vDownR4,vDownR5,vDownR6,vDownR7,vDownR8}
\fmf{phantom,tension=1.0}{vUpL1,vLeft}
\fmf{phantom,tension=2.5}{vDownL1,vLeft}
\fmf{phantom,tension=2.5}{vUpR1,vRight}
\fmf{phantom,tension=1.0}{vDownR1,vRight}
\fmfv{label=$\alpha_{1}$}{v1}
\fmfv{label=$\alpha_{2}$}{v2}
\fmf{phantom}{i1,v1}
\fmf{phantom}{i2,v1}
\fmf{phantom,tension=0.6,foreground=(1,,0,,0)}{v1,v2}
\fmf{plain,left=0.3,tension=0,foreground=(1,,0,,0)}{v1,v2}
\fmf{plain,right=0.3,tension=0,foreground=(1,,0,,0)}{v1,v2}
\fmf{phantom}{v2,o1}
\fmf{phantom}{v2,o2}
\fmf{plain,tension=0,foreground=(1,,0,,0)}{vLeft,vRight}
\end{fmfgraph*}
\end{fmffile}
\end{gathered} \quad \rightarrow \left(\boldsymbol{G}^{2}[K_{0}]\right)_{\alpha_{1}\alpha_{2}}^{-1} \;.
\label{eq:4PPIEAfeynRuleInverse2particleG0DON}
\end{equation}
Practically, we multiply both sides of~\eqref{eq:pure4PPIEATowerEquation20DON} by this inverse propagator and integrate over the relevant index to isolate $K_{1}$, which leads to:
\begin{equation}
0 = \hspace{0.1cm} \underbrace{\hspace{0.3cm} \begin{gathered}
\begin{fmffile}{Diagrams/pure4PPIEA_DeterminationK1_Diag1}
\begin{fmfgraph*}(25,10)
\fmfleft{i0,i1,i,i2,i3}
\fmfright{o0,o1,o,o2,o3}
\fmftop{vUpL,vUp,vUpR}
\fmfbottom{vDownL,vDown,vDownR}
\fmfv{decor.shape=circle,decor.filled=empty,label.dist=0.15cm,decor.size=1.5thick,label=$\alpha$}{v1}
\fmfv{decor.shape=circle,decor.filled=empty,decor.size=1.5thick}{vMiddle}
\fmfv{decor.shape=square,decor.filled=empty,decor.size=0.40cm,label=$1$,label.dist=0}{o}
\fmf{phantom,tension=0.5}{vUpL,vSlashUp}
\fmf{phantom,tension=1.0}{vUp,vSlashUp}
\fmf{phantom,tension=1.0}{vDownL,vSlashDown}
\fmf{phantom,tension=0.5}{vDown,vSlashDown}
\fmf{phantom,tension=3.6}{i1,vSlashDown}
\fmf{phantom,tension=1.0}{o1,vSlashDown}
\fmf{phantom,tension=2.2}{i2,vSlashUp}
\fmf{phantom,tension=1.0}{o2,vSlashUp}
\fmf{plain,tension=0,foreground=(1,,0,,0)}{vSlashDown,vSlashUp}
\fmf{phantom}{vUp,vMiddle}
\fmf{phantom}{vDown,vMiddle}
\fmf{phantom}{i1,v1}
\fmf{phantom}{i2,v1}
\fmf{plain,left=0.3,tension=0.05,foreground=(1,,0,,0)}{v1,vMiddle}
\fmf{plain,right=0.3,tension=0.05,foreground=(1,,0,,0)}{v1,vMiddle}
\fmf{plain,left=0.3,tension=0.05,foreground=(1,,0,,0)}{vMiddle,v2}
\fmf{plain,right=0.3,tension=0.05,foreground=(1,,0,,0)}{vMiddle,v2}
\fmf{phantom}{v2,o1}
\fmf{phantom}{v2,o2}
\end{fmfgraph*}
\end{fmffile}
\end{gathered} \hspace{0.2cm}}_{K_{1,\alpha}[\rho,\zeta]} \hspace{0.2cm} - \frac{1}{6} \hspace{0.6cm} \begin{gathered}
\begin{fmffile}{Diagrams/pure4PPIEA_DeterminationK1_Diag2}
\begin{fmfgraph*}(35,20)
\fmfleft{iDown2,iDown1,iDown,i,iUp,iUp1,iUp2}
\fmfright{oDown2,oDown1,oDown,o,oUp,oUp1,oUp2}
\fmfv{decor.shape=circle,decor.filled=empty,decor.size=1.5thick,label.dist=0.15cm,label=$\alpha$}{i}
\fmf{phantom,tension=6.2}{iUp,vSlashUp}
\fmf{phantom,tension=1.2}{oUp,vSlashUp}
\fmf{phantom,tension=10.5}{iDown,vSlashDown}
\fmf{phantom,tension=1.0}{oDown,vSlashDown}
\fmf{plain,tension=0,foreground=(1,,0,,0)}{vSlashDown,vSlashUp}
\fmfv{decor.shape=circle,decor.filled=empty,decor.size=1.5thick}{i1}
\fmf{plain,left=0.3,tension=0.8,foreground=(1,,0,,0)}{i,i1}
\fmf{plain,right=0.3,tension=0.8,foreground=(1,,0,,0)}{i,i1}
\fmf{phantom,tension=10}{o,o1}
\fmf{plain,left,tension=1.0,foreground=(1,,0,,0)}{i1,v1,i1}
\fmf{plain,right,tension=1.0,foreground=(1,,0,,0)}{o1,v2,o1}
\fmf{zigzag,tension=1.5,foreground=(1,,0,,0)}{v1,v2}
\end{fmfgraph*}
\end{fmffile}
\end{gathered}
\hspace{0.05cm} - \frac{1}{3} \hspace{0.35cm} \begin{gathered}
\begin{fmffile}{Diagrams/pure4PPIEA_DeterminationK1_Diag3}
\begin{fmfgraph*}(20,13)
\fmfleft{iDown4,iDown3,iDown2,iDown1,i,iUp1,iUp2,iUp3,iUp4}
\fmfright{oDown4,oDown3,oDown2,oDown1,o,oUp1,oUp2,oUp3,oUp4}
\fmftop{vUpL3,vUpL2,vUpL1,vUp,vUpR1,vUpR2,vUpR3}
\fmfbottom{vDownL3,vDownL2,vDownL1,vDown,vDownR1,vDownR2,vDownR3}
\fmfv{decor.shape=circle,decor.filled=empty,decor.size=1.5thick,label.dist=0.15cm,label=$\alpha$}{ibis}
\fmfv{decor.shape=circle,decor.filled=empty,decor.size=1.5thick}{ibis2}
\fmf{plain,left=0.3,tension=0,foreground=(1,,0,,0)}{ibis,ibis2}
\fmf{plain,right=0.3,tension=0,foreground=(1,,0,,0)}{ibis,ibis2}
\fmf{phantom,tension=7.0}{ibis,i}
\fmf{phantom,tension=1.0}{ibis,o}
\fmf{phantom,tension=1.0}{ibis2,i}
\fmf{phantom,tension=1.16}{ibis2,o}
\fmf{phantom,tension=3.2}{iUp2,vSlashUp}
\fmf{phantom,tension=1.9}{oUp2,vSlashUp}
\fmf{phantom,tension=3.5}{iDown2,vSlashDown}
\fmf{phantom,tension=1.3}{oDown2,vSlashDown}
\fmf{plain,tension=0,foreground=(1,,0,,0)}{vSlashDown,vSlashUp}
\fmf{plain,left,tension=1.0,foreground=(1,,0,,0)}{vUpR2,vDownR2,vUpR2}
\fmf{zigzag,tension=1.0,foreground=(1,,0,,0)}{vUpR2,vDownR2}
\end{fmfgraph*}
\end{fmffile}
\end{gathered} \hspace{0.4cm} \;,
\label{eq:4PPIEADeterminationK1step10DON}
\end{equation}
where we have introduced the Feynman rule:
\begin{equation}
\begin{gathered}
\begin{fmffile}{Diagrams/pure4PPIEA_FeynRuleKn}
\begin{fmfgraph*}(6,4)
\fmfleft{i1}
\fmfright{o1}
\fmfv{decor.shape=square,decor.filled=empty,decor.size=0.4cm,label=$n$,label.dist=0}{v1}
\fmfv{label=$\alpha$,label.angle=-90,label.dist=7}{v2}
\fmf{plain,tension=0.5,foreground=(1,,0,,0)}{i1,v1}
\fmf{phantom}{v1,o1}
\fmf{phantom,tension=0.5}{i1,v2}
\fmf{phantom}{v2,o1}
\end{fmfgraph*}
\end{fmffile}
\end{gathered} \hspace{0.1cm} \rightarrow K_{n,\alpha}[\rho,\zeta]\;,
\label{eq:4PPIEAlambdafeynRuleKn0DON}
\end{equation}
and exploited the following expression of $\rho_{1}[K=K_{0},M=M_{0}]$:
\begin{equation}
\rho_{1}[K=K_{0},M=M_{0}] = 2 \left.\frac{\delta W_{2}[K,M]}{\delta K_{\alpha}}\right|_{K=K_{0} \atop M=M_{0}} = -\frac{1}{6} \hspace{0.45cm} \begin{gathered}
\begin{fmffile}{Diagrams/pure4PPIEA_dW2dK_Diag1}
\begin{fmfgraph*}(30,20)
\fmfleft{i}
\fmfright{o}
\fmfv{decor.shape=circle,decor.filled=empty,decor.size=1.5thick,label.dist=0.15cm,label=$\alpha$}{i1}
\fmf{phantom,tension=10}{i,i1}
\fmf{phantom,tension=10}{o,o1}
\fmf{plain,left,tension=0.5,foreground=(1,,0,,0)}{i1,v1,i1}
\fmf{plain,right,tension=0.5,foreground=(1,,0,,0)}{o1,v2,o1}
\fmf{zigzag,foreground=(1,,0,,0)}{v1,v2}
\end{fmfgraph*}
\end{fmffile}
\end{gathered}
\hspace{0.1cm} - \frac{1}{3}\begin{gathered}
\begin{fmffile}{Diagrams/pure4PPIEA_dW2dK_Diag2}
\begin{fmfgraph*}(15,15)
\fmfleft{i}
\fmfright{o}
\fmftop{vUp}
\fmfbottom{vDown}
\fmfv{decor.shape=circle,decor.filled=empty,decor.size=1.5thick,label.dist=0.15cm,label=$\alpha$}{vBis}
\fmf{phantom,tension=7}{vUp,vBis}
\fmf{phantom,tension=1}{vDown,vBis}
\fmf{phantom,tension=11}{i,v1}
\fmf{phantom,tension=11}{v2,o}
\fmf{plain,left,tension=0.4,foreground=(1,,0,,0)}{v1,v2,v1}
\fmf{zigzag,foreground=(1,,0,,0)}{v1,v2}
\end{fmfgraph*}
\end{fmffile}
\end{gathered} \;,
\label{eq:pure4PPIEArho10DON}
\end{equation}
which is deduced from~\eqref{eq:pure4PPIEAdW2dK0DON}. Then, we rewrite~\eqref{eq:4PPIEADeterminationK1step10DON} as:
\begin{equation}
\begin{split}
K_{1,\alpha_{1}}[\rho,\zeta] = & \ \frac{1}{6} \hspace{0.7cm} \begin{gathered}
\begin{fmffile}{Diagrams/pure4PPIEA_DeterminationK1_Diag4}
\begin{fmfgraph*}(35,20)
\fmfleft{iDown2,iDown1,iDown,i,iUp,iUp1,iUp2}
\fmfright{oDown2,oDown1,oDown,o,oUp,oUp1,oUp2}
\fmfv{decor.shape=circle,decor.filled=empty,decor.size=1.5thick,label.dist=0.15cm,label=$\alpha_{1}$}{i}
\fmf{phantom,tension=6.2}{iUp,vSlashUp}
\fmf{phantom,tension=1.2}{oUp,vSlashUp}
\fmf{phantom,tension=10.5}{iDown,vSlashDown}
\fmf{phantom,tension=1.0}{oDown,vSlashDown}
\fmf{plain,tension=0,foreground=(1,,0,,0)}{vSlashDown,vSlashUp}
\fmfv{decor.shape=circle,decor.filled=empty,decor.size=1.5thick}{i1}
\fmf{plain,left=0.3,tension=0.8,foreground=(1,,0,,0)}{i,i1}
\fmf{plain,right=0.3,tension=0.8,foreground=(1,,0,,0)}{i,i1}
\fmf{phantom,tension=10}{o,o1}
\fmf{plain,left,tension=1.0,foreground=(1,,0,,0)}{i1,v1,i1}
\fmf{plain,right,tension=1.0,foreground=(1,,0,,0)}{o1,v2,o1}
\fmf{zigzag,tension=1.5,foreground=(1,,0,,0)}{v1,v2}
\end{fmfgraph*}
\end{fmffile}
\end{gathered}
\hspace{0.05cm} + \frac{1}{3} \hspace{0.45cm} \begin{gathered}
\begin{fmffile}{Diagrams/pure4PPIEA_DeterminationK1_Diag5}
\begin{fmfgraph*}(20,13)
\fmfleft{iDown4,iDown3,iDown2,iDown1,i,iUp1,iUp2,iUp3,iUp4}
\fmfright{oDown4,oDown3,oDown2,oDown1,o,oUp1,oUp2,oUp3,oUp4}
\fmftop{vUpL3,vUpL2,vUpL1,vUp,vUpR1,vUpR2,vUpR3}
\fmfbottom{vDownL3,vDownL2,vDownL1,vDown,vDownR1,vDownR2,vDownR3}
\fmfv{decor.shape=circle,decor.filled=empty,decor.size=1.5thick,label.dist=0.15cm,label=$\alpha_{1}$}{ibis}
\fmfv{decor.shape=circle,decor.filled=empty,decor.size=1.5thick}{ibis2}
\fmf{plain,left=0.3,tension=0,foreground=(1,,0,,0)}{ibis,ibis2}
\fmf{plain,right=0.3,tension=0,foreground=(1,,0,,0)}{ibis,ibis2}
\fmf{phantom,tension=7.0}{ibis,i}
\fmf{phantom,tension=1.0}{ibis,o}
\fmf{phantom,tension=1.0}{ibis2,i}
\fmf{phantom,tension=1.16}{ibis2,o}
\fmf{phantom,tension=3.2}{iUp2,vSlashUp}
\fmf{phantom,tension=1.9}{oUp2,vSlashUp}
\fmf{phantom,tension=3.5}{iDown2,vSlashDown}
\fmf{phantom,tension=1.3}{oDown2,vSlashDown}
\fmf{plain,tension=0,foreground=(1,,0,,0)}{vSlashDown,vSlashUp}
\fmf{plain,left,tension=1.0,foreground=(1,,0,,0)}{vUpR2,vDownR2,vUpR2}
\fmf{zigzag,tension=1.0,foreground=(1,,0,,0)}{vUpR2,vDownR2}
\end{fmfgraph*}
\end{fmffile}
\end{gathered} \\
= & \ \frac{1}{6} \sum_{a_{2},a_{3},a_{4}=1}^{N} \int_{x_{2},x_{3}} \left(\boldsymbol{G}^{2}[K_{0}]\right)_{\alpha_{1}(a_{2},x_{2})}^{-1} \boldsymbol{G}^{2}_{(a_{2},x_{2})(a_{3},x_{3})}[K_{0}] \boldsymbol{G}_{(a_{4},x_{3})(a_{4},x_{3})}[K_{0}] \\
& \hspace{2.8cm} \times \left(\lambda - M_{0,a_{3},x_{3}}[\rho,\zeta]\delta_{a_{3}a_{4}}\right) \\
& + \frac{1}{3} \hspace{0.45cm} \begin{gathered}
\begin{fmffile}{Diagrams/pure4PPIEA_DeterminationK1_Diag5}
\begin{fmfgraph*}(20,13)
\fmfleft{iDown4,iDown3,iDown2,iDown1,i,iUp1,iUp2,iUp3,iUp4}
\fmfright{oDown4,oDown3,oDown2,oDown1,o,oUp1,oUp2,oUp3,oUp4}
\fmftop{vUpL3,vUpL2,vUpL1,vUp,vUpR1,vUpR2,vUpR3}
\fmfbottom{vDownL3,vDownL2,vDownL1,vDown,vDownR1,vDownR2,vDownR3}
\fmfv{decor.shape=circle,decor.filled=empty,decor.size=1.5thick,label.dist=0.15cm,label=$\alpha_{1}$}{ibis}
\fmfv{decor.shape=circle,decor.filled=empty,decor.size=1.5thick}{ibis2}
\fmf{plain,left=0.3,tension=0,foreground=(1,,0,,0)}{ibis,ibis2}
\fmf{plain,right=0.3,tension=0,foreground=(1,,0,,0)}{ibis,ibis2}
\fmf{phantom,tension=7.0}{ibis,i}
\fmf{phantom,tension=1.0}{ibis,o}
\fmf{phantom,tension=1.0}{ibis2,i}
\fmf{phantom,tension=1.16}{ibis2,o}
\fmf{phantom,tension=3.2}{iUp2,vSlashUp}
\fmf{phantom,tension=1.9}{oUp2,vSlashUp}
\fmf{phantom,tension=3.5}{iDown2,vSlashDown}
\fmf{phantom,tension=1.3}{oDown2,vSlashDown}
\fmf{plain,tension=0,foreground=(1,,0,,0)}{vSlashDown,vSlashUp}
\fmf{plain,left,tension=1.0,foreground=(1,,0,,0)}{vUpR2,vDownR2,vUpR2}
\fmf{zigzag,tension=1.0,foreground=(1,,0,,0)}{vUpR2,vDownR2}
\end{fmfgraph*}
\end{fmffile}
\end{gathered} \\
= & \ \frac{1}{6} \sum_{a_{4}=1}^{N} \boldsymbol{G}_{(a_{4},x_{1})(a_{4},x_{1})}[K_{0}] \left(\lambda - M_{0,\alpha_{1}}[\rho,\zeta]\delta_{a_{1}a_{4}}\right) + \frac{1}{3} \hspace{0.45cm} \begin{gathered}
\begin{fmffile}{Diagrams/pure4PPIEA_DeterminationK1_Diag5}
\begin{fmfgraph*}(20,13)
\fmfleft{iDown4,iDown3,iDown2,iDown1,i,iUp1,iUp2,iUp3,iUp4}
\fmfright{oDown4,oDown3,oDown2,oDown1,o,oUp1,oUp2,oUp3,oUp4}
\fmftop{vUpL3,vUpL2,vUpL1,vUp,vUpR1,vUpR2,vUpR3}
\fmfbottom{vDownL3,vDownL2,vDownL1,vDown,vDownR1,vDownR2,vDownR3}
\fmfv{decor.shape=circle,decor.filled=empty,decor.size=1.5thick,label.dist=0.15cm,label=$\alpha_{1}$}{ibis}
\fmfv{decor.shape=circle,decor.filled=empty,decor.size=1.5thick}{ibis2}
\fmf{plain,left=0.3,tension=0,foreground=(1,,0,,0)}{ibis,ibis2}
\fmf{plain,right=0.3,tension=0,foreground=(1,,0,,0)}{ibis,ibis2}
\fmf{phantom,tension=7.0}{ibis,i}
\fmf{phantom,tension=1.0}{ibis,o}
\fmf{phantom,tension=1.0}{ibis2,i}
\fmf{phantom,tension=1.16}{ibis2,o}
\fmf{phantom,tension=3.2}{iUp2,vSlashUp}
\fmf{phantom,tension=1.9}{oUp2,vSlashUp}
\fmf{phantom,tension=3.5}{iDown2,vSlashDown}
\fmf{phantom,tension=1.3}{oDown2,vSlashDown}
\fmf{plain,tension=0,foreground=(1,,0,,0)}{vSlashDown,vSlashUp}
\fmf{plain,left,tension=1.0,foreground=(1,,0,,0)}{vUpR2,vDownR2,vUpR2}
\fmf{zigzag,tension=1.0,foreground=(1,,0,,0)}{vUpR2,vDownR2}
\end{fmfgraph*}
\end{fmffile}
\end{gathered} \\
= & \ \frac{1}{6} \hspace{-0.35cm} \begin{gathered}
\begin{fmffile}{Diagrams/pure4PPIEA_DeterminationK1_Diag6}
\begin{fmfgraph*}(30,20)
\fmfleft{i}
\fmfright{o}
\fmfv{decor.shape=circle,decor.filled=empty,decor.size=1.5thick,label.dist=0.15cm,label=$a_{1}$}{v1}
\fmf{phantom,tension=10}{i,i1}
\fmf{phantom,tension=10}{o,o1}
\fmf{phantom,left,tension=0.5}{i1,v1,i1}
\fmf{plain,right,tension=0.5,foreground=(1,,0,,0)}{o1,v2,o1}
\fmf{zigzag,label=$x_{1}$,foreground=(1,,0,,0)}{v1,v2}
\end{fmfgraph*}
\end{fmffile}
\end{gathered} \hspace{0.05cm} + \frac{1}{3} \hspace{0.45cm} \begin{gathered}
\begin{fmffile}{Diagrams/pure4PPIEA_DeterminationK1_Diag5}
\begin{fmfgraph*}(20,13)
\fmfleft{iDown4,iDown3,iDown2,iDown1,i,iUp1,iUp2,iUp3,iUp4}
\fmfright{oDown4,oDown3,oDown2,oDown1,o,oUp1,oUp2,oUp3,oUp4}
\fmftop{vUpL3,vUpL2,vUpL1,vUp,vUpR1,vUpR2,vUpR3}
\fmfbottom{vDownL3,vDownL2,vDownL1,vDown,vDownR1,vDownR2,vDownR3}
\fmfv{decor.shape=circle,decor.filled=empty,decor.size=1.5thick,label.dist=0.15cm,label=$\alpha_{1}$}{ibis}
\fmfv{decor.shape=circle,decor.filled=empty,decor.size=1.5thick}{ibis2}
\fmf{plain,left=0.3,tension=0,foreground=(1,,0,,0)}{ibis,ibis2}
\fmf{plain,right=0.3,tension=0,foreground=(1,,0,,0)}{ibis,ibis2}
\fmf{phantom,tension=7.0}{ibis,i}
\fmf{phantom,tension=1.0}{ibis,o}
\fmf{phantom,tension=1.0}{ibis2,i}
\fmf{phantom,tension=1.16}{ibis2,o}
\fmf{phantom,tension=3.2}{iUp2,vSlashUp}
\fmf{phantom,tension=1.9}{oUp2,vSlashUp}
\fmf{phantom,tension=3.5}{iDown2,vSlashDown}
\fmf{phantom,tension=1.3}{oDown2,vSlashDown}
\fmf{plain,tension=0,foreground=(1,,0,,0)}{vSlashDown,vSlashUp}
\fmf{plain,left,tension=1.0,foreground=(1,,0,,0)}{vUpR2,vDownR2,vUpR2}
\fmf{zigzag,tension=1.0,foreground=(1,,0,,0)}{vUpR2,vDownR2}
\end{fmfgraph*}
\end{fmffile}
\end{gathered} \hspace{0.4cm} \;.
\end{split}
\label{eq:4PPIEADeterminationK1step20DON}
\end{equation}
From~\eqref{eq:4PPIEADeterminationK1step20DON} as well as derivatives~\eqref{eq:pure4PPIEAdW2dK0DON} and~\eqref{eq:pure4PPIEAdW1dKdK0DON}, we infer that:
\begin{equation}
\begin{split}
\int_{\alpha} \left.\frac{\delta W_{2}[K,M]}{\delta K_{\alpha}}\right|_{K=K_{0} \atop M=M_{0}} K_{1,\alpha}[\rho,\zeta] = & - \frac{1}{18} \hspace{0.38cm} \begin{gathered}
\begin{fmffile}{Diagrams/pure4PPIEA_Gamma3_Diag1}
\begin{fmfgraph}(12,12)
\fmfleft{i0,i1}
\fmfright{o0,o1}
\fmftop{v1,vUp,v2}
\fmfbottom{v3,vDown,v4}
\fmf{phantom,tension=20}{i0,v1}
\fmf{phantom,tension=20}{i1,v3}
\fmf{phantom,tension=20}{o0,v2}
\fmf{phantom,tension=20}{o1,v4}
\fmf{plain,left=0.4,tension=0.5,foreground=(1,,0,,0)}{v3,v1}
\fmf{phantom,left=0.1,tension=0.5}{v1,vUp}
\fmf{phantom,left=0.1,tension=0.5}{vUp,v2}
\fmf{plain,left=0.4,tension=0.0,foreground=(1,,0,,0)}{v1,v2}
\fmf{plain,left=0.4,tension=0.5,foreground=(1,,0,,0)}{v2,v4}
\fmf{phantom,left=0.1,tension=0.5}{v4,vDown}
\fmf{phantom,left=0.1,tension=0.5}{vDown,v3}
\fmf{plain,left=0.4,tension=0.0,foreground=(1,,0,,0)}{v4,v3}
\fmf{zigzag,left=0.4,tension=0.5,foreground=(1,,0,,0)}{v1,v3}
\fmf{zigzag,right=0.4,tension=0.5,foreground=(1,,0,,0)}{v2,v4}
\end{fmfgraph}
\end{fmffile}
\end{gathered} \hspace{0.28cm} - \frac{1}{18} \hspace{-0.22cm} \begin{gathered}
\begin{fmffile}{Diagrams/pure4PPIEA_Gamma3_Diag2}
\begin{fmfgraph}(40,20)
\fmfleft{i}
\fmfright{o}
\fmftop{vUpLeft1,vUpLeft2,vUpLeft3,vUpRight1,vUpRight2,vUpRight3}
\fmfbottom{vDownLeft1,vDownLeft2,vDownLeft3,vDownRight1,vDownRight2,vDownRight3}
\fmf{phantom,tension=10}{i,v3}
\fmf{phantom,tension=10}{o,v4}
\fmf{plain,right=0.4,tension=0.5,foreground=(1,,0,,0)}{v1,vUpLeft}
\fmf{plain,right,tension=0.5,foreground=(1,,0,,0)}{vUpLeft,vDownLeft}
\fmf{plain,left=0.4,tension=0.5,foreground=(1,,0,,0)}{v1,vDownLeft}
\fmf{phantom,right=0.4,tension=0.5}{vUpRight,v2}
\fmf{phantom,left,tension=0.5}{vUpRight,vDownRight}
\fmf{phantom,left=0.4,tension=0.5}{vDownRight,v2}
\fmf{phantom,tension=0.3}{v2bis,o}
\fmf{plain,left,tension=0.1,foreground=(1,,0,,0)}{v2,v2bis,v2}
\fmf{zigzag,tension=2.7,foreground=(1,,0,,0)}{v1,v2}
\fmf{phantom,tension=2}{v1,v3}
\fmf{phantom,tension=2}{v2,v4}
\fmf{phantom,tension=2.4}{vUpLeft,vUpLeft2}
\fmf{phantom,tension=2.4}{vDownLeft,vDownLeft2}
\fmf{phantom,tension=2.4}{vUpRight,vUpRight2}
\fmf{phantom,tension=2.4}{vDownRight,vDownRight2}
\fmf{zigzag,tension=0,foreground=(1,,0,,0)}{vUpLeft,vDownLeft}
\end{fmfgraph}
\end{fmffile}
\end{gathered} \\
& - \frac{1}{72} \hspace{0.1cm} \begin{gathered}
\begin{fmffile}{Diagrams/pure4PPIEA_Gamma3_Diag3}
\begin{fmfgraph}(45,18)
\fmfleft{i}
\fmfright{o}
\fmf{phantom,tension=10}{i,i1}
\fmf{phantom,tension=10}{o,o1}
\fmf{plain,left,tension=0.5,foreground=(1,,0,,0)}{i1,v1,i1}
\fmf{plain,right,tension=0.5,foreground=(1,,0,,0)}{o1,v2,o1}
\fmf{zigzag,foreground=(1,,0,,0)}{v1,v3}
\fmf{plain,left,tension=0.5,foreground=(1,,0,,0)}{v3,v4}
\fmf{plain,right,tension=0.5,foreground=(1,,0,,0)}{v3,v4}
\fmf{zigzag,foreground=(1,,0,,0)}{v4,v2}
\end{fmfgraph}
\end{fmffile}
\end{gathered} \;,
\end{split}
\label{eq:pure4PPIEAIMGamma3dW2dKK10DON}
\end{equation}
\begin{equation}
\begin{split}
\frac{1}{2} \int_{\alpha_{1},\alpha_{2}} \left.\frac{\delta^{2}W_{1}[K,M]}{\delta K_{\alpha_{1}} \delta K_{\alpha_{2}}} \right|_{K=K_{0} \atop M=M_{0}} K_{1,\alpha_{1}}[\rho,\zeta] K_{1,\alpha_{2}}[\rho,\zeta] = & \ \frac{1}{36} \hspace{0.38cm} \begin{gathered}
\begin{fmffile}{Diagrams/pure4PPIEA_Gamma3_Diag1}
\begin{fmfgraph}(12,12)
\fmfleft{i0,i1}
\fmfright{o0,o1}
\fmftop{v1,vUp,v2}
\fmfbottom{v3,vDown,v4}
\fmf{phantom,tension=20}{i0,v1}
\fmf{phantom,tension=20}{i1,v3}
\fmf{phantom,tension=20}{o0,v2}
\fmf{phantom,tension=20}{o1,v4}
\fmf{plain,left=0.4,tension=0.5,foreground=(1,,0,,0)}{v3,v1}
\fmf{phantom,left=0.1,tension=0.5}{v1,vUp}
\fmf{phantom,left=0.1,tension=0.5}{vUp,v2}
\fmf{plain,left=0.4,tension=0.0,foreground=(1,,0,,0)}{v1,v2}
\fmf{plain,left=0.4,tension=0.5,foreground=(1,,0,,0)}{v2,v4}
\fmf{phantom,left=0.1,tension=0.5}{v4,vDown}
\fmf{phantom,left=0.1,tension=0.5}{vDown,v3}
\fmf{plain,left=0.4,tension=0.0,foreground=(1,,0,,0)}{v4,v3}
\fmf{zigzag,left=0.4,tension=0.5,foreground=(1,,0,,0)}{v1,v3}
\fmf{zigzag,right=0.4,tension=0.5,foreground=(1,,0,,0)}{v2,v4}
\end{fmfgraph}
\end{fmffile}
\end{gathered} \hspace{0.28cm} + \frac{1}{36} \hspace{-0.22cm} \begin{gathered}
\begin{fmffile}{Diagrams/pure4PPIEA_Gamma3_Diag2}
\begin{fmfgraph}(40,20)
\fmfleft{i}
\fmfright{o}
\fmftop{vUpLeft1,vUpLeft2,vUpLeft3,vUpRight1,vUpRight2,vUpRight3}
\fmfbottom{vDownLeft1,vDownLeft2,vDownLeft3,vDownRight1,vDownRight2,vDownRight3}
\fmf{phantom,tension=10}{i,v3}
\fmf{phantom,tension=10}{o,v4}
\fmf{plain,right=0.4,tension=0.5,foreground=(1,,0,,0)}{v1,vUpLeft}
\fmf{plain,right,tension=0.5,foreground=(1,,0,,0)}{vUpLeft,vDownLeft}
\fmf{plain,left=0.4,tension=0.5,foreground=(1,,0,,0)}{v1,vDownLeft}
\fmf{phantom,right=0.4,tension=0.5}{vUpRight,v2}
\fmf{phantom,left,tension=0.5}{vUpRight,vDownRight}
\fmf{phantom,left=0.4,tension=0.5}{vDownRight,v2}
\fmf{phantom,tension=0.3}{v2bis,o}
\fmf{plain,left,tension=0.1,foreground=(1,,0,,0)}{v2,v2bis,v2}
\fmf{zigzag,tension=2.7,foreground=(1,,0,,0)}{v1,v2}
\fmf{phantom,tension=2}{v1,v3}
\fmf{phantom,tension=2}{v2,v4}
\fmf{phantom,tension=2.4}{vUpLeft,vUpLeft2}
\fmf{phantom,tension=2.4}{vDownLeft,vDownLeft2}
\fmf{phantom,tension=2.4}{vUpRight,vUpRight2}
\fmf{phantom,tension=2.4}{vDownRight,vDownRight2}
\fmf{zigzag,tension=0,foreground=(1,,0,,0)}{vUpLeft,vDownLeft}
\end{fmfgraph}
\end{fmffile}
\end{gathered} \\
& + \frac{1}{144} \hspace{0.1cm} \begin{gathered}
\begin{fmffile}{Diagrams/pure4PPIEA_Gamma3_Diag3}
\begin{fmfgraph}(45,18)
\fmfleft{i}
\fmfright{o}
\fmf{phantom,tension=10}{i,i1}
\fmf{phantom,tension=10}{o,o1}
\fmf{plain,left,tension=0.5,foreground=(1,,0,,0)}{i1,v1,i1}
\fmf{plain,right,tension=0.5,foreground=(1,,0,,0)}{o1,v2,o1}
\fmf{zigzag,foreground=(1,,0,,0)}{v1,v3}
\fmf{plain,left,tension=0.5,foreground=(1,,0,,0)}{v3,v4}
\fmf{plain,right,tension=0.5,foreground=(1,,0,,0)}{v3,v4}
\fmf{zigzag,foreground=(1,,0,,0)}{v4,v2}
\end{fmfgraph}
\end{fmffile}
\end{gathered} \;,
\end{split}
\label{eq:pure4PPIEAIMGamma3dW1dKdKK1K10DON}
\end{equation}
where we have used equalities like:
\begin{equation}
\begin{gathered}
\begin{fmffile}{Diagrams/pure4PPIEA_NontrivialProduct_DiagToCancel2}
\begin{fmfgraph}(37.5,16.5)
\fmfleft{i}
\fmfright{o}
\fmfleft{vAddLeft1,vAddLeft2,vAddLeft3,vAddLeft4,vAddLeft5,vAddLeft6}
\fmfright{vAddRight1,vAddRight2,vAddRight3,vAddRight4,vAddRight5,vAddRight6}
\fmftop{vUpLeft1,vUpLeft2,vUpLeft3,vUpRight1,vUpRight2,vUpRight3}
\fmfbottom{vDownLeft1,vDownLeft2,vDownLeft3,vDownRight1,vDownRight2,vDownRight3}
\fmf{phantom,tension=20}{vUpLeft2,vUpLeft}
\fmf{phantom,tension=20}{vUpRight2,vUpRight}
\fmf{phantom,tension=20}{vDownLeft2,vDownLeft}
\fmf{phantom,tension=20}{vDownRight2,vDownRight}
\fmfv{decor.shape=circle,decor.filled=empty,decor.size=1.5thick}{v1}
\fmfv{decor.shape=circle,decor.filled=empty,decor.size=1.5thick}{v2}
\fmf{phantom,tension=0.81}{i,v1}
\fmf{phantom,tension=0.81}{o,v2}
\fmf{plain,left,tension=1.0,foreground=(1,,0,,0)}{vUpLeft,vDownLeft}
\fmf{plain,right,tension=1.0,foreground=(1,,0,,0)}{vUpLeft,vDownLeft}
\fmf{zigzag,tension=1.0,foreground=(1,,0,,0)}{vUpLeft,vDownLeft}
\fmf{plain,left,tension=1.0,foreground=(1,,0,,0)}{vUpRight,vDownRight}
\fmf{plain,right,tension=1.0,foreground=(1,,0,,0)}{vUpRight,vDownRight}
\fmf{zigzag,tension=1.0,foreground=(1,,0,,0)}{vUpRight,vDownRight}
\fmf{phantom,tension=1.0,foreground=(1,,0,,0)}{v1,v2}
\fmf{plain,tension=0,left=0.3,foreground=(1,,0,,0)}{v1,v2}
\fmf{plain,tension=0,right=0.3,foreground=(1,,0,,0)}{v1,v2}
\fmf{phantom,tension=0.82}{vAddUp,vAddLeft5}
\fmf{phantom,tension=1.0}{vAddUp,vAddRight5}
\fmf{phantom,tension=1.0}{vAddDown,vAddLeft2}
\fmf{phantom,tension=0.82}{vAddDown,vAddRight2}
\fmf{plain,tension=0.4,foreground=(1,,0,,0)}{vAddUp,vAddDown}
\end{fmfgraph}
\end{fmffile}
\end{gathered} = \hspace{0.38cm} \begin{gathered}
\begin{fmffile}{Diagrams/pure4PPIEA_Gamma3_Diag1}
\begin{fmfgraph}(12,12)
\fmfleft{i0,i1}
\fmfright{o0,o1}
\fmftop{v1,vUp,v2}
\fmfbottom{v3,vDown,v4}
\fmf{phantom,tension=20}{i0,v1}
\fmf{phantom,tension=20}{i1,v3}
\fmf{phantom,tension=20}{o0,v2}
\fmf{phantom,tension=20}{o1,v4}
\fmf{plain,left=0.4,tension=0.5,foreground=(1,,0,,0)}{v3,v1}
\fmf{phantom,left=0.1,tension=0.5}{v1,vUp}
\fmf{phantom,left=0.1,tension=0.5}{vUp,v2}
\fmf{plain,left=0.4,tension=0.0,foreground=(1,,0,,0)}{v1,v2}
\fmf{plain,left=0.4,tension=0.5,foreground=(1,,0,,0)}{v2,v4}
\fmf{phantom,left=0.1,tension=0.5}{v4,vDown}
\fmf{phantom,left=0.1,tension=0.5}{vDown,v3}
\fmf{plain,left=0.4,tension=0.0,foreground=(1,,0,,0)}{v4,v3}
\fmf{zigzag,left=0.4,tension=0.5,foreground=(1,,0,,0)}{v1,v3}
\fmf{zigzag,right=0.4,tension=0.5,foreground=(1,,0,,0)}{v2,v4}
\end{fmfgraph}
\end{fmffile}
\end{gathered} \hspace{0.28cm} \;,
\label{eq:NonTrivialRelation4PPIEAIMGdiagonal}
\end{equation}
which holds in particular thanks to the trivial structure of the propagator $\boldsymbol{G}[K_{0}]$ in color space, i.e. $\boldsymbol{G}_{(a_{1},x_{1})(a_{2},x_{2})}[K_{0}] = G_{x_{1}x_{2}}[K_{0}]\delta_{a_{1}a_{2}}$ as follows from the absence of spontaneous breakdown of the $O(N)$ symmetry in the present framework (see also the remark right below~\eqref{eq:2PPIEAdefinitionGKpropagator0DON}). According to~\eqref{eq:pure4PPIEAIMGamma3dW2dKK10DON} and~\eqref{eq:pure4PPIEAIMGamma3dW1dKdKK1K10DON} as well as expression~\eqref{eq:pure4PPIEAIMW30DON} of the $W_{3}$ coefficient,~\eqref{eq:pure4PPIEAIMGamma30DON} can be rewritten as follows:\\
\begin{equation}
\Gamma_{3}^{(\mathrm{4PPI})}[\rho,\zeta] = - \frac{1}{72} \hspace{0.38cm} \begin{gathered}
\begin{fmffile}{Diagrams/pure4PPIEA_Gamma3_Diag4}
\begin{fmfgraph}(12,12)
\fmfleft{i0,i1}
\fmfright{o0,o1}
\fmftop{v1,vUp,v2}
\fmfbottom{v3,vDown,v4}
\fmf{phantom,tension=20}{i0,v1}
\fmf{phantom,tension=20}{i1,v3}
\fmf{phantom,tension=20}{o0,v2}
\fmf{phantom,tension=20}{o1,v4}
\fmf{plain,left=0.4,tension=0.5,foreground=(1,,0,,0)}{v3,v1}
\fmf{phantom,left=0.1,tension=0.5}{v1,vUp}
\fmf{phantom,left=0.1,tension=0.5}{vUp,v2}
\fmf{plain,left=0.4,tension=0.0,foreground=(1,,0,,0)}{v1,v2}
\fmf{plain,left=0.4,tension=0.5,foreground=(1,,0,,0)}{v2,v4}
\fmf{phantom,left=0.1,tension=0.5}{v4,vDown}
\fmf{phantom,left=0.1,tension=0.5}{vDown,v3}
\fmf{plain,left=0.4,tension=0.0,foreground=(1,,0,,0)}{v4,v3}
\fmf{zigzag,tension=0.5,foreground=(1,,0,,0)}{v1,v4}
\fmf{zigzag,tension=0.5,foreground=(1,,0,,0)}{v2,v3}
\end{fmfgraph}
\end{fmffile}
\end{gathered} \hspace{0.28cm} - \frac{1}{144} \hspace{0.38cm} \begin{gathered}
\begin{fmffile}{Diagrams/pure4PPIEA_Gamma3_Diag5}
\begin{fmfgraph}(12,12)
\fmfleft{i0,i1}
\fmfright{o0,o1}
\fmftop{v1,vUp,v2}
\fmfbottom{v3,vDown,v4}
\fmf{phantom,tension=20}{i0,v1}
\fmf{phantom,tension=20}{i1,v3}
\fmf{phantom,tension=20}{o0,v2}
\fmf{phantom,tension=20}{o1,v4}
\fmf{plain,left=0.4,tension=0.5,foreground=(1,,0,,0)}{v3,v1}
\fmf{phantom,left=0.1,tension=0.5}{v1,vUp}
\fmf{phantom,left=0.1,tension=0.5}{vUp,v2}
\fmf{zigzag,left=0.4,tension=0.0,foreground=(1,,0,,0)}{v1,v2}
\fmf{plain,left=0.4,tension=0.5,foreground=(1,,0,,0)}{v2,v4}
\fmf{phantom,left=0.1,tension=0.5}{v4,vDown}
\fmf{phantom,left=0.1,tension=0.5}{vDown,v3}
\fmf{zigzag,left=0.4,tension=0.0,foreground=(1,,0,,0)}{v4,v3}
\fmf{plain,left=0.4,tension=0.5,foreground=(1,,0,,0)}{v1,v3}
\fmf{plain,right=0.4,tension=0.5,foreground=(1,,0,,0)}{v2,v4}
\end{fmfgraph}
\end{fmffile}
\end{gathered} \hspace{0.29cm} + \frac{1}{24} \int_{\alpha} M_{0,\alpha}[\rho,\zeta] \zeta_{\alpha} \;.
\label{eq:pure4PPIEAIMGamma3bis0DON}
\end{equation}\\
Diagrammatic expressions for the source coefficients $K_{2}$ and $M_{1}$ are then determined in order to find from~\eqref{eq:pure4PPIEAIMGamma40DON} the counterpart of result~\eqref{eq:pure4PPIEAIMGamma3bis0DON} for $\Gamma_{4}^{(\mathrm{4PPI})}[\rho,\zeta]$. This can be achieved from the coupled equations~\eqref{eq:pure4PPIEATowerEquation2bis0DON} and~\eqref{eq:pure4PPIEATowerEquation30DON}, whose derivatives of the $\rho_{n}$ and $\zeta_{n}$ coefficients are evaluated below:
\begin{equation}
\begin{split}
\left.\frac{\delta \zeta_{0,\alpha_{1}}[K,M]}{\delta K_{\alpha_{2}}}\right|_{K=K_{0} \atop M=M_{0}} = & \ 24 \Bigg(\left.\frac{\delta^{2} W_{3}[K,M]}{\delta K_{\alpha_{2}} \delta M_{\alpha_{1}}}\right|_{K=K_{0} \atop M=M_{0}} - \left.\frac{\delta^{2} W_{1}[K]}{\delta K_{\alpha_{2}} \delta K_{\alpha_{1}}}\right|_{K = K_{0}} \left.\frac{\delta W_{2}[K,M]}{\delta K_{\alpha_{1}}}\right|_{K=K_{0} \atop M=M_{0}} \\
& \hspace{0.7cm} - \left.\frac{\delta W_{1}[K]}{\delta K_{\alpha_{1}}}\right|_{K = K_{0}} \left.\frac{\delta^{2} W_{2}[K,M]}{\delta K_{\alpha_{2}} \delta K_{\alpha_{1}}}\right|_{K=K_{0} \atop M=M_{0}}\Bigg) \\
= & \ - 4 \hspace{-0.1cm} \begin{gathered}
\begin{fmffile}{Diagrams/pure4PPIEA_dzeta0dK_Diag1}
\begin{fmfgraph*}(30,20)
\fmfleft{ibis,iUpbis}
\fmfright{obis,oUpbis}
\fmftop{vUpL,vUp,vUpR}
\fmfbottom{vDownL,vDown,vDownR}
\fmf{phantom,tension=3.0}{i,ibis}
\fmf{phantom,tension=3.0}{o,obis}
\fmf{phantom,tension=3.0}{iUp,iUpbis}
\fmf{phantom,tension=3.0}{oUp,oUpbis}
\fmfv{decor.shape=circle,decor.filled=empty,decor.size=1.5thick,label.angle=180,label.dist=0.1cm,label=$\alpha_{1}$}{vIndex}
\fmfv{decor.shape=circle,decor.filled=empty,decor.size=1.5thick,label.angle=-90,label.dist=0.15cm,label=$\alpha_{2}$}{vIndex2}
\fmf{phantom,tension=1.45}{vDownL,vIndex2}
\fmf{phantom,tension=4.3}{vDown,vIndex2}
\fmf{phantom,tension=0.57}{vUp,vIndex2}
\fmf{plain,left,foreground=(1,,0,,0)}{i,vIndex,i}
\fmf{plain,left,foreground=(1,,0,,0)}{o,vIndex,o}
\fmf{zigzag,left=0.11,tension=1.5,foreground=(1,,0,,0)}{iUp,oUp}
\fmf{zigzag,left=1.0,tension=2.1,foreground=(1,,0,,0)}{i,iUp}
\fmf{zigzag,right=1.0,tension=2.1,foreground=(1,,0,,0)}{o,oUp}
\end{fmfgraph*}
\end{fmffile}
\end{gathered} \;,
\end{split}
\label{eq:pure4PPIEAIMdzeta0dK0DON}
\end{equation}
\begin{equation}
\begin{split}
\left.\frac{\delta \zeta_{0,\alpha_{1}}[K,M]}{\delta M_{\alpha_{2}}}\right|_{K=K_{0} \atop M=M_{0}} = & \ 24 \Bigg(\left.\frac{\delta^{2} W_{3}[K,M]}{\delta M_{\alpha_{2}} \delta M_{\alpha_{1}}}\right|_{K=K_{0} \atop M=M_{0}} - \left.\frac{\delta^{2} W_{1}[K]}{\delta M_{\alpha_{2}} \delta K_{\alpha_{1}}}\right|_{K = K_{0}} \left.\frac{\delta W_{2}[K,M]}{\delta K_{\alpha_{1}}}\right|_{K=K_{0} \atop M=M_{0}} \\
& \hspace{0.7cm} - \left.\frac{\delta W_{1}[K]}{\delta K_{\alpha_{1}}}\right|_{K = K_{0}} \left.\frac{\delta^{2} W_{2}[K,M]}{\delta M_{\alpha_{2}} \delta K_{\alpha_{1}}}\right|_{K=K_{0} \atop M=M_{0}}\Bigg) \\
= & \hspace{0.1cm} \begin{gathered}
\begin{fmffile}{Diagrams/pure4PPIEA_dzeta0dM_Diag1}
\begin{fmfgraph*}(16,16)
\fmftop{vUpBis}
\fmfbottom{vDownBis}
\fmf{phantom,tension=28.0}{vUp,vUpBis}
\fmf{phantom,tension=0.5}{vDown,vUpBis}
\fmf{phantom,tension=0.5}{vUp,vDownBis}
\fmf{phantom,tension=28.0}{vDown,vDownBis}
\fmfv{decor.shape=circle,decor.filled=empty,decor.size=1.5thick,label.angle=90,label.dist=0.15cm,label=$\alpha_{1}$}{vUp}
\fmfv{decor.shape=circle,decor.filled=empty,decor.size=1.5thick,label.angle=-90,label.dist=0.15cm,label=$\alpha_{2}$}{vDown}
\fmf{plain,left,foreground=(1,,0,,0)}{vUp,vDown}
\fmf{plain,left=0.3,foreground=(1,,0,,0)}{vUp,vDown}
\fmf{plain,right,foreground=(1,,0,,0)}{vUp,vDown}
\fmf{plain,right=0.3,foreground=(1,,0,,0)}{vUp,vDown}
\end{fmfgraph*}
\end{fmffile}
\end{gathered} \;,
\end{split}
\label{eq:pure4PPIEAIMdzeta0dM0DON}
\end{equation}
\begin{equation}
\left.\frac{\delta^{2} \rho_{0,\alpha_{1}}[K]}{\delta K_{\alpha_{2}} \delta K_{\alpha_{3}}}\right|_{K=K_{0}} = 2 \left. \frac{\delta^{3} W_{1}[K]}{\delta K_{\alpha_{2}} \delta K_{\alpha_{3}} \delta K_{\alpha_{1}}} \right|_{K=K_{0}} = 2 \hspace{0.55cm} \begin{gathered}
\begin{fmffile}{Diagrams/pure4PPIEA_dW1dKdKdK_Diag1}
\begin{fmfgraph*}(15,15)
\fmfleft{i}
\fmfright{o}
\fmftop{vUp}
\fmfbottom{vDown}
\fmfv{decor.shape=circle,decor.filled=empty,decor.size=1.5thick,label.dist=0.15cm,label=$\alpha_{1}$}{v1}
\fmfv{decor.shape=circle,decor.filled=empty,decor.size=1.5thick,label.dist=0.15cm,label=$\alpha_{2}$}{v2}
\fmfv{decor.shape=circle,decor.filled=empty,decor.size=1.5thick,label.dist=0.15cm,label=$\alpha_{3}$}{v3}
\fmf{phantom,tension=7.0}{vUp,v3}
\fmf{phantom,tension=1}{vDown,v3}
\fmf{phantom,tension=11}{i,v1}
\fmf{phantom,tension=11}{v2,o}
\fmf{plain,left,tension=0.4,foreground=(1,,0,,0)}{v1,v2,v1}
\fmf{phantom}{v1,v2}
\end{fmfgraph*}
\end{fmffile}
\end{gathered} \hspace{0.4cm} \;,
\label{eq:pure4PPIEAIMdrho0dKdK0DON}
\end{equation}
\begin{equation}
\begin{split}
\scalebox{0.97}{${\displaystyle \left.\frac{\delta \rho_{1,\alpha_{1}}[K,M]}{\delta K_{\alpha_{2}}}\right|_{K=K_{0} \atop M=M_{0}} = }$} & \ \scalebox{0.97}{${\displaystyle 2 \left. \frac{\delta^{2} W_{2}[K,M]}{\delta K_{\alpha_{2}} \delta K_{\alpha_{1}}} \right|_{K=K_{0} \atop M=M_{0}} }$} \\
\scalebox{0.97}{${\displaystyle = }$} & \scalebox{0.97}{${\displaystyle -\frac{1}{3} \hspace{0.1cm} \begin{gathered}
\begin{fmffile}{Diagrams/pure4PPIEA_dW2dKdK_Diag1}
\begin{fmfgraph*}(30,20)
\fmfleft{i}
\fmfright{o}
\fmftop{vUpL,vUp,vUpR}
\fmfbottom{vDownL,vDown,vDownR}
\fmfv{decor.shape=circle,decor.filled=empty,decor.size=1.5thick,label.angle=90,label.dist=0.15cm,label=$\alpha_{1}$}{vIndex1}
\fmfv{decor.shape=circle,decor.filled=empty,decor.size=1.5thick,label.angle=-90,label.dist=0.15cm,label=$\alpha_{2}$}{vIndex2}
\fmf{phantom,tension=12.0}{vUpL,vIndex1}
\fmf{phantom,tension=4.9}{i,vIndex1}
\fmf{phantom,tension=4}{o,vIndex1}
\fmf{phantom,tension=12.0}{vDownL,vIndex2}
\fmf{phantom,tension=4.9}{i,vIndex2}
\fmf{phantom,tension=4}{o,vIndex2}
\fmf{phantom,tension=10}{i,i1}
\fmf{phantom,tension=10}{o,o1}
\fmf{plain,left,tension=0.5,foreground=(1,,0,,0)}{i1,v1,i1}
\fmf{plain,right,tension=0.5,foreground=(1,,0,,0)}{o1,v2,o1}
\fmf{zigzag,foreground=(1,,0,,0)}{v1,v2}
\end{fmfgraph*}
\end{fmffile}
\end{gathered}
\hspace{0.1cm} -\frac{1}{6} \hspace{0.6cm} \begin{gathered}
\begin{fmffile}{Diagrams/pure4PPIEA_dW2dKdK_Diag2}
\begin{fmfgraph*}(30,20)
\fmfleft{i}
\fmfright{o}
\fmfv{decor.shape=circle,decor.filled=empty,decor.size=1.5thick,label.dist=0.15cm,label=$\alpha_{1}$}{i1}
\fmfv{decor.shape=circle,decor.filled=empty,decor.size=1.5thick,label.dist=0.15cm,label=$\alpha_{2}$}{o1}
\fmf{phantom,tension=10}{i,i1}
\fmf{phantom,tension=10}{o,o1}
\fmf{plain,left,tension=0.5,foreground=(1,,0,,0)}{i1,v1,i1}
\fmf{plain,right,tension=0.5,foreground=(1,,0,,0)}{o1,v2,o1}
\fmf{zigzag,foreground=(1,,0,,0)}{v1,v2}
\end{fmfgraph*}
\end{fmffile}
\end{gathered}
\hspace{0.5cm} - \frac{2}{3}\begin{gathered}
\begin{fmffile}{Diagrams/pure4PPIEA_dW2dKdK_Diag3}
\begin{fmfgraph*}(15,15)
\fmfleft{i}
\fmfright{o}
\fmftop{vUp}
\fmfbottom{vDown}
\fmfv{decor.shape=circle,decor.filled=empty,decor.size=1.5thick,label.angle=135,label.dist=0.1cm,label=$\alpha_{1}$}{vBis}
\fmfv{decor.shape=circle,decor.filled=empty,decor.size=1.5thick,label.angle=45,label.dist=0.1cm,label=$\alpha_{2}$}{vBis2}
\fmf{phantom,tension=5}{o,vBis2}
\fmf{phantom,tension=8}{vUp,vBis2}
\fmf{phantom,tension=-0.2}{vDown,vBis2}
\fmf{phantom,tension=5}{i,vBis}
\fmf{phantom,tension=8}{vUp,vBis}
\fmf{phantom,tension=-0.2}{vDown,vBis}
\fmf{phantom,tension=11}{i,v1}
\fmf{phantom,tension=11}{v2,o}
\fmf{plain,left,tension=0.4,foreground=(1,,0,,0)}{v1,v2,v1}
\fmf{zigzag,foreground=(1,,0,,0)}{v1,v2}
\end{fmfgraph*}
\end{fmffile}
\end{gathered} - \frac{1}{3}\begin{gathered}
\begin{fmffile}{Diagrams/pure4PPIEA_dW2dKdK_Diag4}
\begin{fmfgraph*}(15,15)
\fmfleft{i}
\fmfright{o}
\fmftop{vUp}
\fmfbottom{vDown}
\fmfv{decor.shape=circle,decor.filled=empty,decor.size=1.5thick,label.dist=0.15cm,label=$\alpha_{1}$}{vBis}
\fmfv{decor.shape=circle,decor.filled=empty,decor.size=1.5thick,label.dist=0.15cm,label=$\alpha_{2}$}{vBis2}
\fmf{phantom,tension=7}{vUp,vBis}
\fmf{phantom,tension=1}{vDown,vBis}
\fmf{phantom,tension=1}{vUp,vBis2}
\fmf{phantom,tension=7}{vDown,vBis2}
\fmf{phantom,tension=11}{i,v1}
\fmf{phantom,tension=11}{v2,o}
\fmf{plain,left,tension=0.4,foreground=(1,,0,,0)}{v1,v2,v1}
\fmf{zigzag,foreground=(1,,0,,0)}{v1,v2}
\end{fmfgraph*}
\end{fmffile}
\end{gathered} \;,}$}
\end{split}
\label{eq:pure4PPIEAIMdrho1dK0DON}
\end{equation}
\begin{equation}
\left.\frac{\delta \rho_{1,\alpha_{1}}[K,M]}{\delta M_{\alpha_{2}}}\right|_{K=K_{0} \atop M=M_{0}} = 2 \left. \frac{\delta^{2} W_{2}[K,M]}{\delta M_{\alpha_{2}} \delta K_{\alpha_{1}}} \right|_{K=K_{0} \atop M=M_{0}} = \frac{1}{2} \hspace{0.7cm} \begin{gathered}
\begin{fmffile}{Diagrams/pure4PPIEA_dW2dKdM_Diag1}
\begin{fmfgraph*}(20,10)
\fmfleft{i}
\fmfright{o}
\fmfv{decor.shape=circle,decor.filled=empty,decor.size=1.5thick,label.angle=180,label.dist=0.15cm,label=$\alpha_{1}$}{i}
\fmfv{decor.shape=circle,decor.filled=empty,decor.size=1.5thick,label.angle=180,label.dist=0.15cm,label=$\alpha_{2}$}{v1}
\fmf{plain,left,tension=0.1,foreground=(1,,0,,0)}{i,v1,i}
\fmf{plain,left,tension=0.1,foreground=(1,,0,,0)}{o,v1,o}
\end{fmfgraph*}
\end{fmffile}
\end{gathered} \hspace{0.1cm} \;,
\label{eq:pure4PPIEAIMdrho1dM0DON}
\end{equation}
and $\left.\frac{\delta \rho_{0,\alpha_{1}}[K]}{\delta K_{\alpha_{2}}}\right|_{K=K_{0}}$ is already given by~\eqref{eq:pure4PPIEAdrho0dK0DON}. Note also that~\eqref{eq:pure4PPIEAIMdrho0dKdK0DON},~\eqref{eq:pure4PPIEAIMdrho1dK0DON} and~\eqref{eq:pure4PPIEAIMdrho1dM0DON} correspond respectively to~\eqref{eq:pure4PPIEAdW1dKdKdK0DON},~\eqref{eq:pure4PPIEAdW2dKdK0DON} and~\eqref{eq:pure4PPIEAdW2dMdK0DON} up to a factor $2$. According to~\eqref{eq:pure4PPIEAIMdzeta0dM0DON}, $M_{1}$ can be isolated in~\eqref{eq:pure4PPIEATowerEquation2bis0DON} by introducing the 4-particle inverse propagator:
\begin{equation}
\begin{gathered}
\begin{fmffile}{Diagrams/pure4PPIEA_FeynRuleGminus4}
\begin{fmfgraph*}(20,20)
\fmfleft{i0,i1,i2,i3}
\fmfright{o0,o1,o2,o3}
\fmftop{vUpL8,vUpL7,vUpL6,vUpL5,vUpL4,vUpL3,vUpL2,vUpL1,vUp,vUpR1,vUpR2,vUpR3,vUpR4,vUpR5,vUpR6,vUpR7,vUpR8}
\fmfbottom{vDownL8,vDownL7,vDownL6,vDownL5,vDownL4,vDownL3,vDownL2,vDownL1,vDown,vDownR1,vDownR2,vDownR3,vDownR4,vDownR5,vDownR6,vDownR7,vDownR8}
\fmf{phantom,tension=1.0}{vUpL1,vLeft}
\fmf{phantom,tension=25.0}{vDownL2,vLeft}
\fmf{phantom,tension=25.0}{vUpR2,vRight}
\fmf{phantom,tension=1.0}{vDownR1,vRight}
\fmfv{label=$\alpha_{1}$}{v1}
\fmfv{label=$\alpha_{2}$}{v2}
\fmf{phantom}{i1,v1}
\fmf{phantom}{i2,v1}
\fmf{phantom,tension=0.6,foreground=(1,,0,,0)}{v1,v2}
\fmf{plain,left=0.3,tension=0,foreground=(1,,0,,0)}{v1,v2}
\fmf{plain,right=0.3,tension=0,foreground=(1,,0,,0)}{v1,v2}
\fmf{plain,left,tension=0,foreground=(1,,0,,0)}{v1,v2}
\fmf{plain,right,tension=0,foreground=(1,,0,,0)}{v1,v2}
\fmf{phantom}{v2,o1}
\fmf{phantom}{v2,o2}
\fmf{plain,tension=0,foreground=(1,,0,,0)}{vLeft,vRight}
\end{fmfgraph*}
\end{fmffile}
\end{gathered} \quad \rightarrow \left(\boldsymbol{G}^{4}[K_{0}]\right)_{\alpha_{1}\alpha_{2}}^{-1} \;.
\label{eq:4PPIEAfeynRuleInverse4particleG0DON}
\end{equation}
After combining~\eqref{eq:4PPIEADeterminationK1step20DON},~\eqref{eq:pure4PPIEAIMdzeta0dK0DON} and~\eqref{eq:pure4PPIEAIMdzeta0dM0DON} with~\eqref{eq:pure4PPIEATowerEquation2bis0DON}, this procedure leads to:
\begin{equation}
M_{1,\alpha}[\rho,\zeta] = -\frac{2}{3} \hspace{-0.7cm} \begin{gathered}
\begin{fmffile}{Diagrams/pure4PPIEA_M1_Diag1}
\begin{fmfgraph*}(35,20)
\fmfleft{iDown3,iDown2,iDown1,i,iUp1,iUp2,iUp3}
\fmfright{oDown3,oDown2,oDown1,o,oUp1,oUp2,oUp3}
\fmftop{vUpL3,vUpL2,vUpL1bis2,vUpL1bis,vUpL1,vUp,vUpR1,vUpR1bis,vUpR1bis2,vUpR2,vUpR3}
\fmfbottom{vDownL3,vDownL2,vDownL1bis2,vDownL1bis,vDownL1,vDown,vDownR1,vDownR1bis,vDownR1bis2,vDownR2,vDownR3}
\fmfv{decor.shape=circle,decor.filled=empty,decor.size=1.5thick}{vC1}
\fmfv{decor.shape=circle,decor.filled=empty,decor.size=1.5thick,label.angle=90,label.dist=0.1cm,label=$\alpha$}{vC2}
\fmf{phantom,tension=4.0}{iDown1,vC1}
\fmf{phantom,tension=4.0}{oDown1,vC1}
\fmf{phantom,tension=4.0}{iUp1,vC2}
\fmf{phantom,tension=4.0}{oUp1,vC2}
\fmf{phantom,tension=5.7}{vSlashUp,vUpL1bis}
\fmf{phantom,tension=3.7}{vSlashUp,vDownL1bis}
\fmf{phantom,tension=2.9}{vSlashDown,vUpR1bis}
\fmf{phantom,tension=3.7}{vSlashDown,vDownR1bis}
\fmf{plain,foreground=(1,,0,,0)}{vSlashUp,vSlashDown}
\fmf{plain,right=0.2,tension=0,foreground=(1,,0,,0)}{vC1,iDown2bis}
\fmf{plain,left=0.2,tension=0,foreground=(1,,0,,0)}{vC1,oDown2bis}
\fmf{plain,tension=0,foreground=(1,,0,,0)}{vC1,vDownL1}
\fmf{plain,tension=0,foreground=(1,,0,,0)}{vC1,vDownR1}
\fmf{phantom,tension=3.1}{iDown2bis,iDown2}
\fmf{phantom,tension=1.0}{iDown2bis,oDown2}
\fmf{phantom,tension=3.1}{oDown2bis,oDown2}
\fmf{phantom,tension=1.0}{oDown2bis,iDown2}
\fmf{plain,right,tension=0,foreground=(1,,0,,0)}{iDown2bis,vDownL1}
\fmf{plain,left,tension=0,foreground=(1,,0,,0)}{oDown2bis,vDownR1}
\fmf{phantom,left,tension=0.8,foreground=(1,,0,,0)}{vC2,vUp,vC2}
\fmf{plain,left,tension=0.2,foreground=(1,,0,,0)}{vC1,vC2}
\fmf{plain,left=0.3,tension=0.2,foreground=(1,,0,,0)}{vC1,vC2}
\fmf{plain,right,tension=0.2,foreground=(1,,0,,0)}{vC1,vC2}
\fmf{plain,right=0.3,tension=0.2,foreground=(1,,0,,0)}{vC1,vC2}
\fmf{phantom,tension=1.29}{iDown2bis2,iDown2}
\fmf{phantom,tension=1.0}{iDown2bis2,oDown2}
\fmf{phantom,tension=1.29}{oDown2bis2,oDown2}
\fmf{phantom,tension=1.0}{oDown2bis2,iDown2}
\fmf{zigzag,right=0.5,tension=0,foreground=(1,,0,,0)}{vDownL1,vDownR1}
\fmf{zigzag,right=0.5,tension=0,foreground=(1,,0,,0)}{iDown2bis2,oDown2bis2}
\end{fmfgraph*}
\end{fmffile}
\end{gathered} \hspace{-0.7cm} -\frac{2}{3} \hspace{-0.7cm} \begin{gathered}
\begin{fmffile}{Diagrams/pure4PPIEA_M1_Diag2}
\begin{fmfgraph*}(35,20)
\fmfleft{iDown3,iDown2,iDown1,i,iUp1,iUp2,iUp3}
\fmfright{oDown3,oDown2,oDown1,o,oUp1,oUp2,oUp3}
\fmftop{vUpL3,vUpL2,vUpL1bis2,vUpL1bis,vUpL1,vUp,vUpR1,vUpR1bis,vUpR1bis2,vUpR2,vUpR3}
\fmfbottom{vDownL3,vDownL2,vDownL1bis2,vDownL1bis,vDownL1,vDown,vDownR1,vDownR1bis,vDownR1bis2,vDownR2,vDownR3}
\fmfv{decor.shape=circle,decor.filled=empty,decor.size=1.5thick}{vC1}
\fmfv{decor.shape=circle,decor.filled=empty,decor.size=1.5thick,label.angle=90,label.dist=0.1cm,label=$\alpha$}{vC2}
\fmf{phantom,tension=4.0}{iDown1,vC1}
\fmf{phantom,tension=4.0}{oDown1,vC1}
\fmf{phantom,tension=4.0}{iUp1,vC2}
\fmf{phantom,tension=4.0}{oUp1,vC2}
\fmf{phantom,tension=5.7}{vSlashUp,vUpL1bis}
\fmf{phantom,tension=3.7}{vSlashUp,vDownL1bis}
\fmf{phantom,tension=2.9}{vSlashDown,vUpR1bis}
\fmf{phantom,tension=3.7}{vSlashDown,vDownR1bis}
\fmf{plain,foreground=(1,,0,,0)}{vSlashUp,vSlashDown}
\fmf{plain,right=0.2,tension=0,foreground=(1,,0,,0)}{vC1,iDown2bis}
\fmf{plain,left=0.2,tension=0,foreground=(1,,0,,0)}{vC1,oDown2bis}
\fmf{plain,tension=0,foreground=(1,,0,,0)}{vC1,vDownL1}
\fmf{plain,tension=0,foreground=(1,,0,,0)}{vC1,vDownR1}
\fmf{phantom,tension=3.1}{iDown2bis,iDown2}
\fmf{phantom,tension=1.0}{iDown2bis,oDown2}
\fmf{phantom,tension=3.1}{oDown2bis,oDown2}
\fmf{phantom,tension=1.0}{oDown2bis,iDown2}
\fmf{plain,right,tension=0,foreground=(1,,0,,0)}{iDown2bis,vDownL1}
\fmf{plain,left,tension=0,foreground=(1,,0,,0)}{oDown2bis,vDownR1}
\fmf{phantom,left,tension=0.8,foreground=(1,,0,,0)}{vC2,vUp,vC2}
\fmf{plain,left,tension=0.2,foreground=(1,,0,,0)}{vC1,vC2}
\fmf{plain,left=0.3,tension=0.2,foreground=(1,,0,,0)}{vC1,vC2}
\fmf{plain,right,tension=0.2,foreground=(1,,0,,0)}{vC1,vC2}
\fmf{plain,right=0.3,tension=0.2,foreground=(1,,0,,0)}{vC1,vC2}
\fmf{phantom,tension=1.29}{iDown2bis2,iDown2}
\fmf{phantom,tension=1.0}{iDown2bis2,oDown2}
\fmf{phantom,tension=1.29}{oDown2bis2,oDown2}
\fmf{phantom,tension=1.0}{oDown2bis2,iDown2}
\fmf{zigzag,tension=0,foreground=(1,,0,,0)}{iDown2bis,vDownL1}
\fmf{zigzag,right=0.5,tension=0,foreground=(1,,0,,0)}{iDown2bis2,oDown2bis2}
\end{fmfgraph*}
\end{fmffile}
\end{gathered} \hspace{-0.7cm} - \frac{1}{6} \hspace{-0.55cm} \begin{gathered}
\begin{fmffile}{Diagrams/pure4PPIEA_M1_Diag3}
\begin{fmfgraph*}(35,20)
\fmfleft{iDown3,iDown2,iDown1,i,iUp1,iUp2,iUp3}
\fmfright{oDown3,oDown2,oDown1,o,oUp1,oUp2,oUp3}
\fmftop{vUpL3,vUpL2,vUpL1bis2,vUpL1bis,vUpL1,vUp,vUpR1,vUpR1bis,vUpR1bis2,vUpR2,vUpR3}
\fmfbottom{vDownL3,vDownL2,vDownL1bis2,vDownL1bis,vDownL1,vDown,vDownR1,vDownR1bis,vDownR1bis2,vDownR2,vDownR3}
\fmfv{decor.shape=circle,decor.filled=empty,decor.size=1.5thick}{vC1}
\fmfv{decor.shape=circle,decor.filled=empty,decor.size=1.5thick,label.angle=90,label.dist=0.1cm,label=$\alpha$}{vC2}
\fmf{phantom,tension=4.0}{iDown1,vC1}
\fmf{phantom,tension=4.0}{oDown1,vC1}
\fmf{phantom,tension=4.0}{iUp1,vC2}
\fmf{phantom,tension=4.0}{oUp1,vC2}
\fmf{phantom,tension=5.7}{vSlashUp,vUpL1bis}
\fmf{phantom,tension=3.7}{vSlashUp,vDownL1bis}
\fmf{phantom,tension=2.9}{vSlashDown,vUpR1bis}
\fmf{phantom,tension=3.7}{vSlashDown,vDownR1bis}
\fmf{plain,foreground=(1,,0,,0)}{vSlashUp,vSlashDown}
\fmf{plain,right=0.2,tension=0,foreground=(1,,0,,0)}{vC1,iDown2bis}
\fmf{plain,left=0.2,tension=0,foreground=(1,,0,,0)}{vC1,oDown2bis}
\fmf{plain,tension=0,foreground=(1,,0,,0)}{vC1,vDownL1bis}
\fmf{plain,tension=0,foreground=(1,,0,,0)}{vC1,vDownR1bis}
\fmf{phantom,tension=4.0}{iDown2bis,iDown2}
\fmf{phantom,tension=1.0}{iDown2bis,oDown2}
\fmf{phantom,tension=4.0}{oDown2bis,oDown2}
\fmf{phantom,tension=1.0}{oDown2bis,iDown2}
\fmf{plain,right,tension=0,foreground=(1,,0,,0)}{iDown2bis,vDownL1bis}
\fmf{plain,left,tension=0,foreground=(1,,0,,0)}{oDown2bis,vDownR1bis}
\fmf{phantom,left,tension=0.8,foreground=(1,,0,,0)}{vC2,vUp,vC2}
\fmf{plain,left,tension=0.2,foreground=(1,,0,,0)}{vC1,vC2}
\fmf{plain,left=0.3,tension=0.2,foreground=(1,,0,,0)}{vC1,vC2}
\fmf{plain,right,tension=0.2,foreground=(1,,0,,0)}{vC1,vC2}
\fmf{plain,right=0.3,tension=0.2,foreground=(1,,0,,0)}{vC1,vC2}
\fmf{phantom,tension=1.29}{iDown2bis2,iDown2}
\fmf{phantom,tension=1.0}{iDown2bis2,oDown2}
\fmf{phantom,tension=1.29}{oDown2bis2,oDown2}
\fmf{phantom,tension=1.0}{oDown2bis2,iDown2}
\fmf{phantom,tension=1.25}{iDown3,vBisL}
\fmf{phantom,tension=1.0}{oDown3,vBisL}
\fmf{phantom,tension=1.0}{iDown3,vBisR}
\fmf{phantom,tension=1.25}{oDown3,vBisR}
\fmf{plain,left,tension=0,foreground=(1,,0,,0)}{vBisL,vBisR,vBisL}
\fmf{zigzag,left=0.2,tension=0,foreground=(1,,0,,0)}{vBisL,vDownL1bis}
\fmf{zigzag,right=0.2,tension=0,foreground=(1,,0,,0)}{vBisR,vDownR1bis}
\end{fmfgraph*}
\end{fmffile}
\end{gathered} \hspace{-0.55cm} \;.
\label{eq:4PPIEADeterminationM10DON}
\end{equation}
Thanks to the latter result, we have everything to determine a homologous expression of $K_{2}$ from~\eqref{eq:pure4PPIEATowerEquation30DON}. This is accomplished by inserting the expression of $\rho_{2}[K=K_{0},M=M_{0}]$ deduced from~\eqref{eq:pure4PPIEAdW3dK0DON}, i.e.:
\begin{equation}
\begin{split}
\scalebox{0.95}{${\displaystyle \rho_{2}[K=K_{0},M=M_{0}] = }$} & \ \scalebox{0.95}{${\displaystyle 2 \left.\frac{\delta W_{3}[K,M]}{\delta K_{\alpha}}\right|_{K=K_{0} \atop M=M_{0}} }$} \\
\scalebox{0.95}{${\displaystyle = }$} & \ \scalebox{0.95}{${\displaystyle \frac{1}{9} \hspace{0.55cm} \begin{gathered}
\begin{fmffile}{Diagrams/pure4PPIEA_dW3dK_Diag1}
\begin{fmfgraph*}(15,15)
\fmfleft{i0,i1}
\fmfright{o0,o1}
\fmftop{v1bis,vUp,v2bis}
\fmfbottom{v3bis,vDown,v4bis}
\fmfleft{vLeft}
\fmfright{vRight}
\fmf{phantom,tension=3.2}{v1,v1bis}
\fmf{phantom,tension=3.2}{v2,v2bis}
\fmf{phantom,tension=3.2}{v3,v3bis}
\fmf{phantom,tension=3.2}{v4,v4bis}
\fmfv{decor.shape=circle,decor.filled=empty,decor.size=1.5thick,label.dist=0.15cm,label=$\alpha$}{vIndex}
\fmf{phantom,tension=-30}{vLeft,vIndex}
\fmf{phantom,tension=0.5}{vRight,vIndex}
\fmf{phantom,tension=20}{i0,v1}
\fmf{phantom,tension=20}{i1,v3}
\fmf{phantom,tension=20}{o0,v2}
\fmf{phantom,tension=20}{o1,v4}
\fmf{plain,right=0.4,tension=0.5,foreground=(1,,0,,0)}{v3,v1}
\fmf{phantom,left=0.1,tension=0.5}{v1,vUp}
\fmf{phantom,left=0.1,tension=0.5}{vUp,v2}
\fmf{plain,right=0.4,tension=0.0,foreground=(1,,0,,0)}{v1,v2}
\fmf{plain,right=0.4,tension=0.5,foreground=(1,,0,,0)}{v2,v4}
\fmf{phantom,left=0.1,tension=0.5}{v4,vDown}
\fmf{phantom,left=0.1,tension=0.5}{vDown,v3}
\fmf{plain,right=0.4,tension=0.0,foreground=(1,,0,,0)}{v4,v3}
\fmf{zigzag,tension=0.5,foreground=(1,,0,,0)}{v1,v4}
\fmf{zigzag,tension=0.5,foreground=(1,,0,,0)}{v2,v3}
\end{fmfgraph*}
\end{fmffile}
\end{gathered} \hspace{0.2cm} + \frac{1}{9} \hspace{0.55cm} \begin{gathered}
\begin{fmffile}{Diagrams/pure4PPIEA_dW3dK_Diag2}
\begin{fmfgraph*}(15,15)
\fmfleft{i0,i1}
\fmfright{o0,o1}
\fmftop{v1bis,vUp,v2bis}
\fmfbottom{v3bis,vDown,v4bis}
\fmfleft{vLeft}
\fmfright{vRight}
\fmf{phantom,tension=3.2}{v1,v1bis}
\fmf{phantom,tension=3.2}{v2,v2bis}
\fmf{phantom,tension=3.2}{v3,v3bis}
\fmf{phantom,tension=3.2}{v4,v4bis}
\fmfv{decor.shape=circle,decor.filled=empty,decor.size=1.5thick,label.dist=0.15cm,label=$\alpha$}{vIndex}
\fmf{phantom,tension=-30}{vLeft,vIndex}
\fmf{phantom,tension=0.5}{vRight,vIndex}
\fmf{phantom,tension=20}{i0,v1}
\fmf{phantom,tension=20}{i1,v3}
\fmf{phantom,tension=20}{o0,v2}
\fmf{phantom,tension=20}{o1,v4}
\fmf{plain,right=0.4,tension=0.5,foreground=(1,,0,,0)}{v3,v1}
\fmf{phantom,left=0.1,tension=0.5}{v1,vUp}
\fmf{phantom,left=0.1,tension=0.5}{vUp,v2}
\fmf{plain,right=0.4,tension=0.0,foreground=(1,,0,,0)}{v1,v2}
\fmf{plain,right=0.4,tension=0.5,foreground=(1,,0,,0)}{v2,v4}
\fmf{phantom,left=0.1,tension=0.5}{v4,vDown}
\fmf{phantom,left=0.1,tension=0.5}{vDown,v3}
\fmf{plain,right=0.4,tension=0.0,foreground=(1,,0,,0)}{v4,v3}
\fmf{phantom,tension=0.5}{v1,v4}
\fmf{phantom,tension=0.5}{v2,v3}
\fmf{zigzag,right=0.4,tension=0.5,foreground=(1,,0,,0)}{v1,v3}
\fmf{zigzag,left=0.4,tension=0.5,foreground=(1,,0,,0)}{v2,v4}
\end{fmfgraph*}
\end{fmffile}
\end{gathered} \hspace{0.2cm} + \frac{1}{9} \hspace{0.25cm} \begin{gathered}
\begin{fmffile}{Diagrams/pure4PPIEA_dW3dK_Diag3}
\begin{fmfgraph*}(15,15)
\fmfleft{i0,i1}
\fmfright{o0,o1}
\fmftop{v1bis,vUp,v2bis}
\fmfbottom{v3bis,vDown,v4bis}
\fmf{phantom,tension=3.2}{v1,v1bis}
\fmf{phantom,tension=3.2}{v2,v2bis}
\fmf{phantom,tension=3.2}{v3,v3bis}
\fmf{phantom,tension=3.2}{v4,v4bis}
\fmfv{decor.shape=circle,decor.filled=empty,decor.size=1.5thick,label.dist=0.15cm,label=$\alpha$}{vIndex}
\fmf{phantom,tension=-10}{vUp,vIndex}
\fmf{phantom,tension=-0.25}{vDown,vIndex}
\fmf{phantom,tension=20}{i0,v1}
\fmf{phantom,tension=20}{i1,v3}
\fmf{phantom,tension=20}{o0,v2}
\fmf{phantom,tension=20}{o1,v4}
\fmf{plain,right=0.4,tension=0.5,foreground=(1,,0,,0)}{v3,v1}
\fmf{phantom,left=0.1,tension=0.5}{v1,vUp}
\fmf{phantom,left=0.1,tension=0.5}{vUp,v2}
\fmf{plain,right=0.4,tension=0.0,foreground=(1,,0,,0)}{v1,v2}
\fmf{plain,right=0.4,tension=0.5,foreground=(1,,0,,0)}{v2,v4}
\fmf{phantom,left=0.1,tension=0.5}{v4,vDown}
\fmf{phantom,left=0.1,tension=0.5}{vDown,v3}
\fmf{plain,right=0.4,tension=0.0,foreground=(1,,0,,0)}{v4,v3}
\fmf{phantom,tension=0.5}{v1,v4}
\fmf{phantom,tension=0.5}{v2,v3}
\fmf{zigzag,right=0.4,tension=0.5,foreground=(1,,0,,0)}{v1,v3}
\fmf{zigzag,left=0.4,tension=0.5,foreground=(1,,0,,0)}{v2,v4}
\end{fmfgraph*}
\end{fmffile}
\end{gathered} \hspace{0.2cm} + \frac{1}{18} \hspace{0.55cm} \begin{gathered}
\begin{fmffile}{Diagrams/pure4PPIEA_dW3dK_Diag4}
\begin{fmfgraph*}(15,15)
\fmfleft{i0,i1}
\fmfright{o0,o1}
\fmftop{v1bis,vUp,v2bis}
\fmfbottom{v3bis,vDown,v4bis}
\fmfleft{vLeft}
\fmfright{vRight}
\fmf{phantom,tension=3.2}{v1,v1bis}
\fmf{phantom,tension=3.2}{v2,v2bis}
\fmf{phantom,tension=3.2}{v3,v3bis}
\fmf{phantom,tension=3.2}{v4,v4bis}
\fmfv{decor.shape=circle,decor.filled=empty,decor.size=1.5thick,label.dist=0.15cm,label=$\alpha$}{vIndex}
\fmf{phantom,tension=-30}{vLeft,vIndex}
\fmf{phantom,tension=0.5}{vRight,vIndex}
\fmf{phantom,tension=20}{i0,v1}
\fmf{phantom,tension=20}{i1,v3}
\fmf{phantom,tension=20}{o0,v2}
\fmf{phantom,tension=20}{o1,v4}
\fmf{plain,right=0.4,tension=0.5,foreground=(1,,0,,0)}{v3,v1}
\fmf{phantom,left=0.1,tension=0.5}{v1,vUp}
\fmf{phantom,left=0.1,tension=0.5}{vUp,v2}
\fmf{zigzag,right=0.4,tension=0.0,foreground=(1,,0,,0)}{v1,v2}
\fmf{plain,right=0.4,tension=0.5,foreground=(1,,0,,0)}{v2,v4}
\fmf{phantom,left=0.1,tension=0.5}{v4,vDown}
\fmf{phantom,left=0.1,tension=0.5}{vDown,v3}
\fmf{zigzag,right=0.4,tension=0.0,foreground=(1,,0,,0)}{v4,v3}
\fmf{phantom,tension=0.5}{v1,v4}
\fmf{phantom,tension=0.5}{v2,v3}
\fmf{plain,right=0.4,tension=0.5,foreground=(1,,0,,0)}{v1,v3}
\fmf{plain,left=0.4,tension=0.5,foreground=(1,,0,,0)}{v2,v4}
\end{fmfgraph*}
\end{fmffile}
\end{gathered} }$} \\
& \scalebox{0.95}{${\displaystyle + \frac{1}{18} \hspace{0.2cm} \begin{gathered}
\begin{fmffile}{Diagrams/pure4PPIEA_dW3dK_Diag5}
\begin{fmfgraph*}(40,20)
\fmfleft{i}
\fmfright{o}
\fmftop{vUpLeft1,vUpLeft2,vUpLeft3,vUpRight1,vUpRight2,vUpRight3}
\fmfbottom{vDownLeft1,vDownLeft2,vDownLeft3,vDownRight1,vDownRight2,vDownRight3}
\fmfv{decor.shape=circle,decor.filled=empty,decor.size=1.5thick,label.dist=0.15cm,label=$\alpha$}{vIndex}
\fmf{phantom,tension=10.5}{i,vIndex}
\fmf{phantom,tension=1}{o,vIndex}
\fmf{phantom,tension=10}{i,v3}
\fmf{phantom,tension=10}{o,v4}
\fmf{plain,right=0.4,tension=0.5,foreground=(1,,0,,0)}{v1,vUpLeft}
\fmf{plain,right,tension=0.5,foreground=(1,,0,,0)}{vUpLeft,vDownLeft}
\fmf{plain,left=0.4,tension=0.5,foreground=(1,,0,,0)}{v1,vDownLeft}
\fmf{phantom,right=0.4,tension=0.5}{vUpRight,v2}
\fmf{phantom,left,tension=0.5}{vUpRight,vDownRight}
\fmf{phantom,left=0.4,tension=0.5}{vDownRight,v2}
\fmf{phantom,tension=0.3}{v2bis,o}
\fmf{plain,left,tension=0.1,foreground=(1,,0,,0)}{v2,v2bis,v2}
\fmf{zigzag,tension=2.7,foreground=(1,,0,,0)}{v1,v2}
\fmf{phantom,tension=2}{v1,v3}
\fmf{phantom,tension=2}{v2,v4}
\fmf{phantom,tension=2.4}{vUpLeft,vUpLeft2}
\fmf{phantom,tension=2.4}{vDownLeft,vDownLeft2}
\fmf{phantom,tension=2.4}{vUpRight,vUpRight2}
\fmf{phantom,tension=2.4}{vDownRight,vDownRight2}
\fmf{zigzag,tension=0,foreground=(1,,0,,0)}{vUpLeft,vDownLeft}
\end{fmfgraph*}
\end{fmffile}
\end{gathered} \hspace{-0.4cm} + \frac{1}{9} \hspace{-0.22cm} \begin{gathered}
\begin{fmffile}{Diagrams/pure4PPIEA_dW3dK_Diag6}
\begin{fmfgraph*}(40,20)
\fmfleft{i}
\fmfright{o}
\fmftop{vUpLeft1,vUpLeft2,vUpLeft3,vUpRight1,vUpRight2,vUpRight3}
\fmfbottom{vDownLeft1,vDownLeft2,vDownLeft3,vDownRight1,vDownRight2,vDownRight3}
\fmfv{decor.shape=circle,decor.filled=empty,decor.size=1.5thick,label.angle=45,label.dist=0.15cm,label=$\alpha$}{vIndex}
\fmf{phantom,tension=3.5}{vUpLeft3,vIndex}
\fmf{phantom,tension=3.6}{i,vIndex}
\fmf{phantom,tension=1.4}{o,vIndex}
\fmf{phantom,tension=10}{i,v3}
\fmf{phantom,tension=10}{o,v4}
\fmf{plain,right=0.4,tension=0.5,foreground=(1,,0,,0)}{v1,vUpLeft}
\fmf{plain,right,tension=0.5,foreground=(1,,0,,0)}{vUpLeft,vDownLeft}
\fmf{plain,left=0.4,tension=0.5,foreground=(1,,0,,0)}{v1,vDownLeft}
\fmf{phantom,right=0.4,tension=0.5}{vUpRight,v2}
\fmf{phantom,left,tension=0.5}{vUpRight,vDownRight}
\fmf{phantom,left=0.4,tension=0.5}{vDownRight,v2}
\fmf{phantom,tension=0.3}{v2bis,o}
\fmf{plain,left,tension=0.1,foreground=(1,,0,,0)}{v2,v2bis,v2}
\fmf{zigzag,tension=2.7,foreground=(1,,0,,0)}{v1,v2}
\fmf{phantom,tension=2}{v1,v3}
\fmf{phantom,tension=2}{v2,v4}
\fmf{phantom,tension=2.4}{vUpLeft,vUpLeft2}
\fmf{phantom,tension=2.4}{vDownLeft,vDownLeft2}
\fmf{phantom,tension=2.4}{vUpRight,vUpRight2}
\fmf{phantom,tension=2.4}{vDownRight,vDownRight2}
\fmf{zigzag,tension=0,foreground=(1,,0,,0)}{vUpLeft,vDownLeft}
\end{fmfgraph*}
\end{fmffile}
\end{gathered} \hspace{-0.4cm} + \frac{1}{18} \hspace{-0.22cm} \begin{gathered}
\begin{fmffile}{Diagrams/pure4PPIEA_dW3dK_Diag7}
\begin{fmfgraph*}(40,20)
\fmfleft{i}
\fmfright{o}
\fmftop{vUpLeft1,vUpLeft2,vUpLeft3,vUpRight1,vUpRight2,vUpRight3}
\fmfbottom{vDownLeft1,vDownLeft2,vDownLeft3,vDownRight1,vDownRight2,vDownRight3}
\fmfv{decor.shape=circle,decor.filled=empty,decor.size=1.5thick,label.dist=0.15cm,label=$\alpha$}{v2bis}
\fmf{phantom,tension=10}{i,v3}
\fmf{phantom,tension=10}{o,v4}
\fmf{plain,right=0.4,tension=0.5,foreground=(1,,0,,0)}{v1,vUpLeft}
\fmf{plain,right,tension=0.5,foreground=(1,,0,,0)}{vUpLeft,vDownLeft}
\fmf{plain,left=0.4,tension=0.5,foreground=(1,,0,,0)}{v1,vDownLeft}
\fmf{phantom,right=0.4,tension=0.5}{vUpRight,v2}
\fmf{phantom,left,tension=0.5}{vUpRight,vDownRight}
\fmf{phantom,left=0.4,tension=0.5}{vDownRight,v2}
\fmf{phantom,tension=0.3}{v2bis,o}
\fmf{plain,left,tension=0.1,foreground=(1,,0,,0)}{v2,v2bis,v2}
\fmf{zigzag,tension=2.7,foreground=(1,,0,,0)}{v1,v2}
\fmf{phantom,tension=2}{v1,v3}
\fmf{phantom,tension=2}{v2,v4}
\fmf{phantom,tension=2.4}{vUpLeft,vUpLeft2}
\fmf{phantom,tension=2.4}{vDownLeft,vDownLeft2}
\fmf{phantom,tension=2.4}{vUpRight,vUpRight2}
\fmf{phantom,tension=2.4}{vDownRight,vDownRight2}
\fmf{zigzag,tension=0,foreground=(1,,0,,0)}{vUpLeft,vDownLeft}
\end{fmfgraph*}
\end{fmffile}
\end{gathered} }$} \\
& \scalebox{0.95}{${\displaystyle + \frac{1}{36} \hspace{0.45cm} \begin{gathered}
\begin{fmffile}{Diagrams/pure4PPIEA_dW3dK_Diag8}
\begin{fmfgraph*}(45,18)
\fmfleft{i}
\fmfright{o}
\fmfv{decor.shape=circle,decor.filled=empty,decor.size=1.5thick,label.dist=0.15cm,label=$\alpha$}{i1}
\fmf{phantom,tension=10}{i,i1}
\fmf{phantom,tension=10}{o,o1}
\fmf{plain,left,tension=0.5,foreground=(1,,0,,0)}{i1,v1,i1}
\fmf{plain,right,tension=0.5,foreground=(1,,0,,0)}{o1,v2,o1}
\fmf{zigzag,foreground=(1,,0,,0)}{v1,v3}
\fmf{plain,left,tension=0.5,foreground=(1,,0,,0)}{v3,v4}
\fmf{plain,right,tension=0.5,foreground=(1,,0,,0)}{v3,v4}
\fmf{zigzag,foreground=(1,,0,,0)}{v4,v2}
\end{fmfgraph*}
\end{fmffile}
\end{gathered} + \frac{1}{36} \hspace{0.1cm} \begin{gathered}
\begin{fmffile}{Diagrams/pure4PPIEA_dW3dK_Diag9}
\begin{fmfgraph*}(45,18)
\fmfleft{i}
\fmfright{o}
\fmftop{vUp}
\fmfbottom{vDown}
\fmfv{decor.shape=circle,decor.filled=empty,decor.size=1.5thick,label.dist=0.15cm,label=$\alpha$}{vIndex}
\fmf{phantom,tension=2.8}{vUp,vIndex}
\fmf{phantom,tension=1}{vDown,vIndex}
\fmf{phantom,tension=10}{i,i1}
\fmf{phantom,tension=10}{o,o1}
\fmf{plain,left,tension=0.5,foreground=(1,,0,,0)}{i1,v1,i1}
\fmf{plain,right,tension=0.5,foreground=(1,,0,,0)}{o1,v2,o1}
\fmf{zigzag,foreground=(1,,0,,0)}{v1,v3}
\fmf{plain,left,tension=0.5,foreground=(1,,0,,0)}{v3,v4}
\fmf{plain,right,tension=0.5,foreground=(1,,0,,0)}{v3,v4}
\fmf{zigzag,foreground=(1,,0,,0)}{v4,v2}
\end{fmfgraph*}
\end{fmffile}
\end{gathered} \;,}$}
\end{split}
\label{eq:pure4PPIEArho20DON}
\end{equation}
as well as~\eqref{eq:pure4PPIEAdrho0dK0DON}~\eqref{eq:4PPIEADeterminationK1step20DON},~\eqref{eq:pure4PPIEAIMdrho0dKdK0DON},~\eqref{eq:pure4PPIEAIMdrho1dK0DON},~\eqref{eq:pure4PPIEAIMdrho1dM0DON} and~\eqref{eq:4PPIEADeterminationM10DON} into~\eqref{eq:pure4PPIEATowerEquation30DON} before isolating $K_{2}$ with the help of the 2-particle inverse propagator~\eqref{eq:4PPIEAfeynRuleInverse2particleG0DON}. In this way, we obtain:
\begin{equation}
\begin{split}
K_{2,\alpha}[\rho,\zeta] = & -\frac{4}{9} \hspace{-0.7cm} \begin{gathered}
\begin{fmffile}{Diagrams/pure4PPIEA_K2_Diag1}
\begin{fmfgraph*}(35,20)
\fmfleft{iDown3,iDown2,iDown1,i,iUp1,iUp2,iUp3}
\fmfright{oDown3,oDown2,oDown1,o,oUp1,oUp2,oUp3}
\fmftop{vUpL3,vUpL2,vUpL1bis2,vUpL1bis,vUpL1,vUp,vUpR1,vUpR1bis,vUpR1bis2,vUpR2,vUpR3}
\fmfbottom{vDownL3,vDownL2,vDownL1bis2,vDownL1bis,vDownL1,vDown,vDownR1,vDownR1bis,vDownR1bis2,vDownR2,vDownR3}
\fmfv{decor.shape=circle,decor.filled=empty,decor.size=1.5thick}{vC1}
\fmfv{decor.shape=circle,decor.filled=empty,decor.size=1.5thick,label.angle=90,label.dist=0.1cm,label=$\alpha$}{vC2}
\fmf{phantom,tension=4.0}{iDown1,vC1}
\fmf{phantom,tension=4.0}{oDown1,vC1}
\fmf{phantom,tension=4.0}{iUp1,vC2}
\fmf{phantom,tension=4.0}{oUp1,vC2}
\fmf{phantom,tension=5.7}{vSlashUp,vUpL1bis}
\fmf{phantom,tension=3.7}{vSlashUp,vDownL1bis}
\fmf{phantom,tension=2.9}{vSlashDown,vUpR1bis}
\fmf{phantom,tension=3.7}{vSlashDown,vDownR1bis}
\fmf{plain,foreground=(1,,0,,0)}{vSlashUp,vSlashDown}
\fmf{plain,right=0.2,tension=0,foreground=(1,,0,,0)}{vC1,iDown2bis}
\fmf{plain,left=0.2,tension=0,foreground=(1,,0,,0)}{vC1,oDown2bis}
\fmf{plain,tension=0,foreground=(1,,0,,0)}{vC1,vDownL1}
\fmf{plain,tension=0,foreground=(1,,0,,0)}{vC1,vDownR1}
\fmf{phantom,tension=3.1}{iDown2bis,iDown2}
\fmf{phantom,tension=1.0}{iDown2bis,oDown2}
\fmf{phantom,tension=3.1}{oDown2bis,oDown2}
\fmf{phantom,tension=1.0}{oDown2bis,iDown2}
\fmf{plain,right,tension=0,foreground=(1,,0,,0)}{iDown2bis,vDownL1}
\fmf{plain,left,tension=0,foreground=(1,,0,,0)}{oDown2bis,vDownR1}
\fmf{plain,left,tension=0.8,foreground=(1,,0,,0)}{vC2,vUp,vC2}
\fmf{plain,left,tension=0.2,foreground=(1,,0,,0)}{vC1,vC2}
\fmf{plain,left=0.3,tension=0.2,foreground=(1,,0,,0)}{vC1,vC2}
\fmf{plain,right,tension=0.2,foreground=(1,,0,,0)}{vC1,vC2}
\fmf{plain,right=0.3,tension=0.2,foreground=(1,,0,,0)}{vC1,vC2}
\fmf{phantom,tension=1.29}{iDown2bis2,iDown2}
\fmf{phantom,tension=1.0}{iDown2bis2,oDown2}
\fmf{phantom,tension=1.29}{oDown2bis2,oDown2}
\fmf{phantom,tension=1.0}{oDown2bis2,iDown2}
\fmf{zigzag,right=0.5,tension=0,foreground=(1,,0,,0)}{vDownL1,vDownR1}
\fmf{zigzag,right=0.5,tension=0,foreground=(1,,0,,0)}{iDown2bis2,oDown2bis2}
\end{fmfgraph*}
\end{fmffile}
\end{gathered} \hspace{-0.7cm} -\frac{2}{3} \hspace{-0.7cm} \begin{gathered}
\begin{fmffile}{Diagrams/pure4PPIEA_K2_Diag2}
\begin{fmfgraph*}(35,20)
\fmfleft{iDown3,iDown2,iDown1,i,iUp1,iUp2,iUp3}
\fmfright{oDown3,oDown2,oDown1,o,oUp1,oUp2,oUp3}
\fmftop{vUpL3,vUpL2,vUpL1bis2,vUpL1bis,vUpL1,vUp,vUpR1,vUpR1bis,vUpR1bis2,vUpR2,vUpR3}
\fmfbottom{vDownL3,vDownL2,vDownL1bis2,vDownL1bis,vDownL1,vDown,vDownR1,vDownR1bis,vDownR1bis2,vDownR2,vDownR3}
\fmfv{decor.shape=circle,decor.filled=empty,decor.size=1.5thick}{vC1}
\fmfv{decor.shape=circle,decor.filled=empty,decor.size=1.5thick,label.angle=90,label.dist=0.1cm,label=$\alpha$}{vC2}
\fmf{phantom,tension=4.0}{iDown1,vC1}
\fmf{phantom,tension=4.0}{oDown1,vC1}
\fmf{phantom,tension=4.0}{iUp1,vC2}
\fmf{phantom,tension=4.0}{oUp1,vC2}
\fmf{phantom,tension=5.7}{vSlashUp,vUpL1bis}
\fmf{phantom,tension=3.7}{vSlashUp,vDownL1bis}
\fmf{phantom,tension=2.9}{vSlashDown,vUpR1bis}
\fmf{phantom,tension=3.7}{vSlashDown,vDownR1bis}
\fmf{plain,foreground=(1,,0,,0)}{vSlashUp,vSlashDown}
\fmf{plain,right=0.2,tension=0,foreground=(1,,0,,0)}{vC1,iDown2bis}
\fmf{plain,left=0.2,tension=0,foreground=(1,,0,,0)}{vC1,oDown2bis}
\fmf{plain,tension=0,foreground=(1,,0,,0)}{vC1,vDownL1}
\fmf{plain,tension=0,foreground=(1,,0,,0)}{vC1,vDownR1}
\fmf{phantom,tension=3.1}{iDown2bis,iDown2}
\fmf{phantom,tension=1.0}{iDown2bis,oDown2}
\fmf{phantom,tension=3.1}{oDown2bis,oDown2}
\fmf{phantom,tension=1.0}{oDown2bis,iDown2}
\fmf{plain,right,tension=0,foreground=(1,,0,,0)}{iDown2bis,vDownL1}
\fmf{plain,left,tension=0,foreground=(1,,0,,0)}{oDown2bis,vDownR1}
\fmf{plain,left,tension=0.8,foreground=(1,,0,,0)}{vC2,vUp,vC2}
\fmf{plain,left,tension=0.2,foreground=(1,,0,,0)}{vC1,vC2}
\fmf{plain,left=0.3,tension=0.2,foreground=(1,,0,,0)}{vC1,vC2}
\fmf{plain,right,tension=0.2,foreground=(1,,0,,0)}{vC1,vC2}
\fmf{plain,right=0.3,tension=0.2,foreground=(1,,0,,0)}{vC1,vC2}
\fmf{phantom,tension=1.29}{iDown2bis2,iDown2}
\fmf{phantom,tension=1.0}{iDown2bis2,oDown2}
\fmf{phantom,tension=1.29}{oDown2bis2,oDown2}
\fmf{phantom,tension=1.0}{oDown2bis2,iDown2}
\fmf{zigzag,tension=0,foreground=(1,,0,,0)}{iDown2bis,vDownL1}
\fmf{zigzag,right=0.5,tension=0,foreground=(1,,0,,0)}{iDown2bis2,oDown2bis2}
\end{fmfgraph*}
\end{fmffile}
\end{gathered} \hspace{-0.7cm} - \frac{1}{6} \hspace{-0.55cm} \begin{gathered}
\begin{fmffile}{Diagrams/pure4PPIEA_K2_Diag3}
\begin{fmfgraph*}(35,20)
\fmfleft{iDown3,iDown2,iDown1,i,iUp1,iUp2,iUp3}
\fmfright{oDown3,oDown2,oDown1,o,oUp1,oUp2,oUp3}
\fmftop{vUpL3,vUpL2,vUpL1bis2,vUpL1bis,vUpL1,vUp,vUpR1,vUpR1bis,vUpR1bis2,vUpR2,vUpR3}
\fmfbottom{vDownL3,vDownL2,vDownL1bis2,vDownL1bis,vDownL1,vDown,vDownR1,vDownR1bis,vDownR1bis2,vDownR2,vDownR3}
\fmfv{decor.shape=circle,decor.filled=empty,decor.size=1.5thick}{vC1}
\fmfv{decor.shape=circle,decor.filled=empty,decor.size=1.5thick,label.angle=90,label.dist=0.1cm,label=$\alpha$}{vC2}
\fmf{phantom,tension=4.0}{iDown1,vC1}
\fmf{phantom,tension=4.0}{oDown1,vC1}
\fmf{phantom,tension=4.0}{iUp1,vC2}
\fmf{phantom,tension=4.0}{oUp1,vC2}
\fmf{phantom,tension=5.7}{vSlashUp,vUpL1bis}
\fmf{phantom,tension=3.7}{vSlashUp,vDownL1bis}
\fmf{phantom,tension=2.9}{vSlashDown,vUpR1bis}
\fmf{phantom,tension=3.7}{vSlashDown,vDownR1bis}
\fmf{plain,foreground=(1,,0,,0)}{vSlashUp,vSlashDown}
\fmf{plain,right=0.2,tension=0,foreground=(1,,0,,0)}{vC1,iDown2bis}
\fmf{plain,left=0.2,tension=0,foreground=(1,,0,,0)}{vC1,oDown2bis}
\fmf{plain,tension=0,foreground=(1,,0,,0)}{vC1,vDownL1bis}
\fmf{plain,tension=0,foreground=(1,,0,,0)}{vC1,vDownR1bis}
\fmf{phantom,tension=4.0}{iDown2bis,iDown2}
\fmf{phantom,tension=1.0}{iDown2bis,oDown2}
\fmf{phantom,tension=4.0}{oDown2bis,oDown2}
\fmf{phantom,tension=1.0}{oDown2bis,iDown2}
\fmf{plain,right,tension=0,foreground=(1,,0,,0)}{iDown2bis,vDownL1bis}
\fmf{plain,left,tension=0,foreground=(1,,0,,0)}{oDown2bis,vDownR1bis}
\fmf{plain,left,tension=0.8,foreground=(1,,0,,0)}{vC2,vUp,vC2}
\fmf{plain,left,tension=0.2,foreground=(1,,0,,0)}{vC1,vC2}
\fmf{plain,left=0.3,tension=0.2,foreground=(1,,0,,0)}{vC1,vC2}
\fmf{plain,right,tension=0.2,foreground=(1,,0,,0)}{vC1,vC2}
\fmf{plain,right=0.3,tension=0.2,foreground=(1,,0,,0)}{vC1,vC2}
\fmf{phantom,tension=1.29}{iDown2bis2,iDown2}
\fmf{phantom,tension=1.0}{iDown2bis2,oDown2}
\fmf{phantom,tension=1.29}{oDown2bis2,oDown2}
\fmf{phantom,tension=1.0}{oDown2bis2,iDown2}
\fmf{phantom,tension=1.25}{iDown3,vBisL}
\fmf{phantom,tension=1.0}{oDown3,vBisL}
\fmf{phantom,tension=1.0}{iDown3,vBisR}
\fmf{phantom,tension=1.25}{oDown3,vBisR}
\fmf{plain,left,tension=0,foreground=(1,,0,,0)}{vBisL,vBisR,vBisL}
\fmf{zigzag,left=0.2,tension=0,foreground=(1,,0,,0)}{vBisL,vDownL1bis}
\fmf{zigzag,right=0.2,tension=0,foreground=(1,,0,,0)}{vBisR,vDownR1bis}
\end{fmfgraph*}
\end{fmffile}
\end{gathered} \hspace{-0.55cm} + \frac{1}{9} \hspace{0.2cm} \begin{gathered}
\begin{fmffile}{Diagrams/pure4PPIEA_K2_Diag4}
\begin{fmfgraph*}(30,15)
\fmfleft{i}
\fmfright{o}
\fmftop{vUpLeft1,vUpLeft2,vUpLeft3,vUpRight1,vUpRight2,vUpRight3}
\fmfbottom{vDownLeft1,vDownLeft2,vDownLeft3,vDownRight1,vDownRight2,vDownRight3}
\fmfv{decor.shape=circle,decor.filled=empty,decor.size=1.5thick}{v1}
\fmfv{decor.shape=circle,decor.filled=empty,decor.size=1.5thick,label.angle=0,label.dist=0.15cm,label=$\alpha$}{v2}
\fmf{phantom,tension=3}{vUpRight2,vSlashUp}
\fmf{phantom,tension=3.5}{vUpLeft3,vSlashUp}
\fmf{phantom,tension=2.5}{vDownRight1,vSlashUp}
\fmf{phantom,tension=3.0}{vDownRight1,vSlashDown}
\fmf{phantom,tension=2.6}{vDownLeft3,vSlashDown}
\fmf{phantom,tension=2.1}{vUpLeft3,vSlashDown}
\fmf{plain,tension=0,foreground=(1,,0,,0)}{vSlashUp,vSlashDown}
\fmf{phantom,tension=1.235}{i,v1}
\fmf{phantom,tension=6.0}{o,v2right}
\fmf{phantom,tension=20}{vUpLeft,vUpLeft2}
\fmf{phantom,tension=20}{vDownLeft,vDownLeft2}
\fmf{phantom,tension=2.0}{v1,v2}
\fmf{plain,left=0.3,tension=0,foreground=(1,,0,,0)}{v1,v2}
\fmf{plain,right=0.3,tension=0,foreground=(1,,0,,0)}{v1,v2}
\fmf{plain,right,tension=0.5,foreground=(1,,0,,0)}{vUpLeft,vDownLeft}
\fmf{plain,left,tension=0.5,foreground=(1,,0,,0)}{vUpLeft,vDownLeft}
\fmf{phantom,left,tension=1.0}{v2,v2right,v2}
\fmf{phantom,tension=1.0}{i,v3}
\fmf{phantom,tension=1.4}{vUpLeft1,v3}
\fmf{phantom,tension=1.0}{vUpLeft2,v3}
\fmf{phantom,tension=12.0}{i,v4}
\fmf{phantom,tension=10.0}{vUpLeft2,v4}
\fmf{phantom,tension=12.0}{vUpRight2,v4}
\fmf{phantom,tension=1.0}{i,v5}
\fmf{phantom,tension=1.4}{vDownLeft1,v5}
\fmf{phantom,tension=1.0}{vDownLeft2,v5}
\fmf{phantom,tension=12.0}{i,v6}
\fmf{phantom,tension=10.0}{vDownLeft2,v6}
\fmf{phantom,tension=12.0}{vDownRight2,v6}
\fmf{zigzag,tension=0,foreground=(1,,0,,0)}{v3,v6}
\fmf{zigzag,tension=0,foreground=(1,,0,,0)}{v4,v5}
\end{fmfgraph*}
\end{fmffile}
\end{gathered} \\
\\
& + \frac{1}{18} \hspace{0.35cm} \begin{gathered}
\begin{fmffile}{Diagrams/pure4PPIEA_K2_Diag5}
\begin{fmfgraph*}(35,13)
\fmfleft{iDown2,iDown1,i,iUp1,iUp2}
\fmfright{oDown2,oDown1,o,oUp1,oUp2}
\fmftop{vUpLeft,vUp,vUpRight}
\fmfbottom{vDownLeft,vDown,vDownRight}
\fmfv{decor.shape=circle,decor.filled=empty,decor.size=1.5thick}{v1}
\fmfv{decor.shape=circle,decor.filled=empty,decor.size=1.5thick,label.angle=0,label.dist=0.15cm,label=$\alpha$}{v2}
\fmf{phantom,tension=1.74}{i,v1}
\fmf{phantom,tension=1.0}{o,v1}
\fmf{phantom,tension=1.0}{i,v2}
\fmf{phantom,tension=1.6}{o,v2}
\fmf{phantom,tension=1.35}{iUp1,v3}
\fmf{phantom,tension=1.6}{oUp1,v3}
\fmf{phantom,tension=1.68}{iDown1,v4}
\fmf{phantom,tension=1.3}{oDown1,v4}
\fmf{plain,tension=0,foreground=(1,,0,,0)}{v3,v4}
\fmf{plain,left=0.3,tension=0,foreground=(1,,0,,0)}{v1,v2}
\fmf{plain,right=0.3,tension=0,foreground=(1,,0,,0)}{v1,v2}
\fmf{plain,left=0.35,tension=0,foreground=(1,,0,,0)}{vUpLeft,vDownLeft}
\fmf{plain,right=0.35,tension=0,foreground=(1,,0,,0)}{vUpLeft,vDownLeft}
\fmf{plain,left=0.35,tension=0,foreground=(1,,0,,0)}{vUpL,vDownL}
\fmf{plain,right=0.35,tension=0,foreground=(1,,0,,0)}{vUpL,vDownL}
\fmf{zigzag,left=0.2,tension=2.2,foreground=(1,,0,,0)}{vUpLeft,vUpL}
\fmf{phantom,tension=1.0}{vUpRight,vUpL}
\fmf{phantom,tension=1.0}{vUpLeft,vUpR}
\fmf{phantom,right=0.2,tension=2.2}{vUpRight,vUpR}
\fmf{zigzag,right=0.2,tension=2.2,foreground=(1,,0,,0)}{vDownLeft,vDownL}
\fmf{phantom,tension=1.0}{vDownRight,vDownL}
\fmf{phantom,tension=1.0}{vDownLeft,vDownR}
\fmf{phantom,left=0.2,tension=2.2}{vDownRight,vDownR}
\end{fmfgraph*}
\end{fmffile}
\end{gathered} \hspace{-0.85cm} \;.
\end{split}
\label{eq:4PPIEADeterminationK20DON}
\end{equation}
From the determination of the $K_{n}$ and $M_{n}$ coefficients presented in this section, we can infer the general procedure to calculate any source coefficient of any $n$PPI EA: the coefficients of a source coupled to a local product of $m$ fields in the source-dependent partition function $Z$ are isolated in the equations homologous to~\eqref{eq:pure4PPIEATowerEquation10DON} to~\eqref{eq:pure4PPIEATowerEquation3bis0DON} by introducing a $m$-particle inverse propagator, such as~\eqref{eq:4PPIEAfeynRuleInverse2particleG0DON} (used to determine all $K_{n}$ coefficients here) at $m=2$ and~\eqref{eq:4PPIEAfeynRuleInverse4particleG0DON} (used to determine all $M_{n}$ coefficients here) at $m=4$. The latter statement is also valid if $p$-point correlation functions with $p$ odd are accessible in the formalism, in which case we would consider additional local products of $m$ fields with $m$ odd: at $m=1$, we would make use of the 1-particle inverse propagator like~\eqref{eq:pure1PIEAinversepropagator0DON} used for the 1PI EA, etc. Getting back to the present derivation, we have specified each entity of expression~\eqref{eq:pure4PPIEAIMGamma40DON} of $\Gamma_{4}^{(\mathrm{4PPI})}[\rho,\zeta]$ according to results~\eqref{eq:4PPIEADeterminationK1step20DON},~\eqref{eq:4PPIEADeterminationM10DON} and~\eqref{eq:4PPIEADeterminationK20DON} for the source coefficients as well as derivatives~\eqref{eq:pure4PPIEAdW2dK0DON} to~\eqref{eq:pure4PPIEAdW1dKdKdK0DON}. From this, we determine all diagrams contributing to $\Gamma_{4}^{(\mathrm{4PPI})}[\rho,\zeta]$. In doing so, we notably exploit relations similar to~\eqref{eq:NonTrivialRelation4PPIEAIMGdiagonal} such as:
\begin{equation}
\begin{gathered}
\begin{fmffile}{Diagrams/pure4PPIEA_NontrivialProduct_DiagToCancel}
\begin{fmfgraph}(35,17.5)
\fmfleft{iDown2,iDown1,i,iUp1,iUp2}
\fmfright{oDown2,oDown1,o,oUp1,oUp2}
\fmftop{vUpLeft,vUp,vUpRight}
\fmfbottom{vDownLeft,vDown,vDownRight}
\fmf{phantom,tension=3.4}{vUpLeft,vaddUp1}
\fmf{phantom,tension=1.0}{vUpRight,vaddUp1}
\fmf{phantom,tension=3.4}{vDownLeft,vaddDown1}
\fmf{phantom,tension=1.0}{vDownRight,vaddDown1}
\fmf{zigzag,tension=0.2,foreground=(1,,0,,0)}{vaddUp1,vaddDown1}
\fmf{plain,left,tension=0.2,foreground=(1,,0,,0)}{vaddUp1,vaddDown1}
\fmf{plain,right,tension=0.2,foreground=(1,,0,,0)}{vaddUp1,vaddDown1}
\fmf{phantom,tension=1.6}{i,v1}
\fmf{phantom,tension=1.0}{o,v1}
\fmf{phantom,tension=1.0}{i,v2}
\fmf{phantom,tension=1.6}{o,v2}
\fmf{phantom,tension=1.3}{iUp1,v3}
\fmf{phantom,tension=1.6}{oUp1,v3}
\fmf{phantom,tension=1.6}{iDown1,v4}
\fmf{phantom,tension=1.3}{oDown1,v4}
\fmf{plain,tension=0,foreground=(1,,0,,0)}{v3,v4}
\fmfv{decor.shape=circle,decor.filled=empty,decor.size=1.5thick}{v1}
\fmfv{decor.shape=circle,decor.filled=empty,decor.size=1.5thick}{v2}
\fmf{plain,left=0.35,tension=0,foreground=(1,,0,,0)}{v1,v2}
\fmf{plain,right=0.35,tension=0,foreground=(1,,0,,0)}{v1,v2}
\fmf{phantom,left=0.35,tension=0,foreground=(1,,0,,0)}{vUpLeft,vDownLeft}
\fmf{phantom,right=0.35,tension=0,foreground=(1,,0,,0)}{vUpLeft,vDownLeft}
\fmf{plain,left=0.35,tension=0,foreground=(1,,0,,0)}{vUpRight,vDownRight}
\fmf{plain,right=0.35,tension=0,foreground=(1,,0,,0)}{vUpRight,vDownRight}
\fmf{phantom,left=0.35,tension=0,foreground=(1,,0,,0)}{vUpL,vDownL}
\fmf{phantom,right=0.35,tension=0,foreground=(1,,0,,0)}{vUpL,vDownL}
\fmf{plain,left=0.35,tension=0,foreground=(1,,0,,0)}{vUpR,vDownR}
\fmf{plain,right=0.35,tension=0,foreground=(1,,0,,0)}{vUpR,vDownR}
\fmf{phantom,left=0.2,tension=2.2,foreground=(0,,0,,1)}{vUpLeft,vUpL}
\fmf{phantom,tension=1.0}{vUpRight,vUpL}
\fmf{phantom,tension=1.0}{vUpLeft,vUpR}
\fmf{zigzag,right=0.2,tension=2.2,foreground=(1,,0,,0)}{vUpRight,vUpR}
\fmf{phantom,right=0.2,tension=2.2,foreground=(0,,0,,1)}{vDownLeft,vDownL}
\fmf{phantom,tension=1.0}{vDownRight,vDownL}
\fmf{phantom,tension=1.0}{vDownLeft,vDownR}
\fmf{zigzag,left=0.2,tension=2.2,foreground=(1,,0,,0)}{vDownRight,vDownR}
\end{fmfgraph}
\end{fmffile}
\end{gathered} \hspace{0.4cm} = \hspace{-0.15cm} \begin{gathered}
\begin{fmffile}{Diagrams/pure4PPIEA_NontrivialProduct_Diag4}
\begin{fmfgraph}(30,15)
\fmfleft{i0,i,i1}
\fmfright{o0,o,o1}
\fmftop{v1b,vUp,v2b}
\fmfbottom{v3b,vDown,v4b}
\fmf{phantom,tension=20}{i0,v1b}
\fmf{phantom,tension=20}{i1,v3b}
\fmf{phantom,tension=20}{o0,v2b}
\fmf{phantom,tension=20}{o1,v4b}
\fmf{phantom,tension=8.0}{i,v5}
\fmf{phantom,tension=4.0}{vUp,v5}
\fmf{phantom,tension=0.1}{vDown,v5}
\fmf{phantom,tension=8.0}{i,v6}
\fmf{phantom,tension=4.0}{vDown,v6}
\fmf{phantom,tension=0.1}{vUp,v6}
\fmf{phantom,tension=0.1}{v1,v1b}
\fmf{phantom,tension=0.1}{v2,v2b}
\fmf{phantom,tension=0.1}{v3,v3b}
\fmf{phantom,tension=0.1}{v4,v4b}
\fmf{phantom,left=0.1,tension=0.1}{v1,vUp}
\fmf{phantom,left=0.1,tension=0.1}{vUp,v2}
\fmf{zigzag,left=0.15,tension=0,foreground=(1,,0,,0)}{v1,v2}
\fmf{plain,left=0.4,tension=0,foreground=(1,,0,,0)}{v1,v3}
\fmf{plain,right=0.1,tension=0,foreground=(1,,0,,0)}{v1,v5}
\fmf{plain,right,tension=0,foreground=(1,,0,,0)}{v5,v6}
\fmf{zigzag,left,tension=0,foreground=(1,,0,,0)}{v5,v6}
\fmf{plain,right=0.1,tension=0,foreground=(1,,0,,0)}{v6,v3}
\fmf{plain,left=0.4,tension=0,foreground=(1,,0,,0)}{v2,v4}
\fmf{plain,right=0.4,tension=0,foreground=(1,,0,,0)}{v2,v4}
\fmf{phantom,left=0.1,tension=0.1}{v4,vDown}
\fmf{phantom,left=0.1,tension=0.1}{vDown,v3}
\fmf{zigzag,left=0.15,tension=0,foreground=(1,,0,,0)}{v4,v3}
\end{fmfgraph}
\end{fmffile}
\end{gathered} \hspace{-0.3cm} \;.
\end{equation}
By combining the expression of $\Gamma_{4}^{(\mathrm{4PPI})}[\rho,\zeta]$ thus obtained with the other $\Gamma_{n}^{(\mathrm{4PPI})}$ coefficients given by~\eqref{eq:pure4PPIEAIMGamma0bis0DON},~\eqref{eq:pure4PPIEAIMGamma1bis0DON},~\eqref{eq:pure4PPIEAIMGamma2bis0DON} and~\eqref{eq:pure4PPIEAIMGamma3bis0DON}, we derive our final result for the 4PPI EA $\Gamma^{(\mathrm{4PPI})}[\rho,\zeta]$ of the studied $O(N)$ model up to the third non-trivial order:
\begin{equation}
\begin{split}
\Gamma^{(\mathrm{4PPI})}[\rho,\zeta] = & - \frac{\hbar}{2}\mathrm{STr}\left[\ln\big(\boldsymbol{G}[K_{0}]\big)\right] + \frac{\hbar}{2} \int_{\alpha} K_{0,\alpha}[\rho,\zeta] \rho_{\alpha} \\
& + \hbar^{2} \left(\rule{0cm}{1.1cm}\right. \frac{1}{24} \hspace{0.08cm} \begin{gathered}
\begin{fmffile}{Diagrams/pure4PPIEA_Hartree}
\begin{fmfgraph}(30,20)
\fmfleft{i}
\fmfright{o}
\fmf{phantom,tension=10}{i,i1}
\fmf{phantom,tension=10}{o,o1}
\fmf{plain,left,tension=0.5,foreground=(1,,0,,0)}{i1,v1,i1}
\fmf{plain,right,tension=0.5,foreground=(1,,0,,0)}{o1,v2,o1}
\fmf{zigzag,foreground=(1,,0,,0)}{v1,v2}
\end{fmfgraph}
\end{fmffile}
\end{gathered}
+ \frac{1}{12}\begin{gathered}
\begin{fmffile}{Diagrams/pure4PPIEA_Fock}
\begin{fmfgraph}(15,15)
\fmfleft{i}
\fmfright{o}
\fmf{phantom,tension=11}{i,v1}
\fmf{phantom,tension=11}{v2,o}
\fmf{plain,left,tension=0.4,foreground=(1,,0,,0)}{v1,v2,v1}
\fmf{zigzag,foreground=(1,,0,,0)}{v1,v2}
\end{fmfgraph}
\end{fmffile}
\end{gathered} + \frac{1}{8} \int_{\alpha} M_{0,\alpha}[\rho,\zeta] \rho_{\alpha}^{2} \left.\rule{0cm}{1.1cm}\right) \\
& + \hbar^{3} \left(\rule{0cm}{1.1cm}\right. -\frac{1}{72} \hspace{0.38cm} \begin{gathered}
\begin{fmffile}{Diagrams/pure4PPIEA_Gamma3_Diag4}
\begin{fmfgraph}(12,12)
\fmfleft{i0,i1}
\fmfright{o0,o1}
\fmftop{v1,vUp,v2}
\fmfbottom{v3,vDown,v4}
\fmf{phantom,tension=20}{i0,v1}
\fmf{phantom,tension=20}{i1,v3}
\fmf{phantom,tension=20}{o0,v2}
\fmf{phantom,tension=20}{o1,v4}
\fmf{plain,left=0.4,tension=0.5,foreground=(1,,0,,0)}{v3,v1}
\fmf{phantom,left=0.1,tension=0.5}{v1,vUp}
\fmf{phantom,left=0.1,tension=0.5}{vUp,v2}
\fmf{plain,left=0.4,tension=0.0,foreground=(1,,0,,0)}{v1,v2}
\fmf{plain,left=0.4,tension=0.5,foreground=(1,,0,,0)}{v2,v4}
\fmf{phantom,left=0.1,tension=0.5}{v4,vDown}
\fmf{phantom,left=0.1,tension=0.5}{vDown,v3}
\fmf{plain,left=0.4,tension=0.0,foreground=(1,,0,,0)}{v4,v3}
\fmf{zigzag,tension=0.5,foreground=(1,,0,,0)}{v1,v4}
\fmf{zigzag,tension=0.5,foreground=(1,,0,,0)}{v2,v3}
\end{fmfgraph}
\end{fmffile}
\end{gathered} \hspace{0.28cm} - \frac{1}{144} \hspace{0.38cm} \begin{gathered}
\begin{fmffile}{Diagrams/pure4PPIEA_Gamma3_Diag5}
\begin{fmfgraph}(12,12)
\fmfleft{i0,i1}
\fmfright{o0,o1}
\fmftop{v1,vUp,v2}
\fmfbottom{v3,vDown,v4}
\fmf{phantom,tension=20}{i0,v1}
\fmf{phantom,tension=20}{i1,v3}
\fmf{phantom,tension=20}{o0,v2}
\fmf{phantom,tension=20}{o1,v4}
\fmf{plain,left=0.4,tension=0.5,foreground=(1,,0,,0)}{v3,v1}
\fmf{phantom,left=0.1,tension=0.5}{v1,vUp}
\fmf{phantom,left=0.1,tension=0.5}{vUp,v2}
\fmf{zigzag,left=0.4,tension=0.0,foreground=(1,,0,,0)}{v1,v2}
\fmf{plain,left=0.4,tension=0.5,foreground=(1,,0,,0)}{v2,v4}
\fmf{phantom,left=0.1,tension=0.5}{v4,vDown}
\fmf{phantom,left=0.1,tension=0.5}{vDown,v3}
\fmf{zigzag,left=0.4,tension=0.0,foreground=(1,,0,,0)}{v4,v3}
\fmf{plain,left=0.4,tension=0.5,foreground=(1,,0,,0)}{v1,v3}
\fmf{plain,right=0.4,tension=0.5,foreground=(1,,0,,0)}{v2,v4}
\end{fmfgraph}
\end{fmffile}
\end{gathered} \hspace{0.29cm} + \frac{1}{24} \int_{\alpha} M_{0,\alpha}[\rho,\zeta] \zeta_{\alpha} \left.\rule{0cm}{1.1cm}\right) \\
& + \hbar^{4} \left(\rule{0cm}{1.1cm}\right. \frac{1}{324} \hspace{0.2cm} \begin{gathered}
\begin{fmffile}{Diagrams/pure4PPIEA_Gamma4_Diag1}
\begin{fmfgraph}(16,16)
\fmfleft{i}
\fmfright{o}
\fmftop{vUpLeft,vUp,vUpRight}
\fmfbottom{vDownLeft,vDown,vDownRight}
\fmf{phantom,tension=1}{i,v1}
\fmf{phantom,tension=1}{v2,o}
\fmf{phantom,tension=14.0}{v3,vUpLeft}
\fmf{phantom,tension=2.0}{v3,vUpRight}
\fmf{phantom,tension=4.0}{v3,i}
\fmf{phantom,tension=2.0}{v4,vUpLeft}
\fmf{phantom,tension=14.0}{v4,vUpRight}
\fmf{phantom,tension=4.0}{v4,o}
\fmf{phantom,tension=14.0}{v5,vDownLeft}
\fmf{phantom,tension=2.0}{v5,vDownRight}
\fmf{phantom,tension=4.0}{v5,i}
\fmf{phantom,tension=2.0}{v6,vDownLeft}
\fmf{phantom,tension=14.0}{v6,vDownRight}
\fmf{phantom,tension=4.0}{v6,o}
\fmf{zigzag,tension=0,foreground=(1,,0,,0)}{v1,v2}
\fmf{zigzag,tension=0.6,foreground=(1,,0,,0)}{v3,v6}
\fmf{zigzag,tension=0.6,foreground=(1,,0,,0)}{v5,v4}
\fmf{plain,left=0.18,tension=0,foreground=(1,,0,,0)}{v1,v3}
\fmf{plain,left=0.42,tension=0,foreground=(1,,0,,0)}{v3,v4}
\fmf{plain,left=0.18,tension=0,foreground=(1,,0,,0)}{v4,v2}
\fmf{plain,left=0.18,tension=0,foreground=(1,,0,,0)}{v2,v6}
\fmf{plain,left=0.42,tension=0,foreground=(1,,0,,0)}{v6,v5}
\fmf{plain,left=0.18,tension=0,foreground=(1,,0,,0)}{v5,v1}
\end{fmfgraph}
\end{fmffile}
\end{gathered} \hspace{0.15cm} + \frac{1}{108} \hspace{0.5cm} \begin{gathered}
\begin{fmffile}{Diagrams/pure4PPIEA_Gamma4_Diag2}
\begin{fmfgraph}(12.5,12.5)
\fmfleft{i0,i1}
\fmfright{o0,o1}
\fmftop{v1,vUp,v2}
\fmfbottom{v3,vDown,v4}
\fmf{phantom,tension=20}{i0,v1}
\fmf{phantom,tension=20}{i1,v3}
\fmf{phantom,tension=20}{o0,v2}
\fmf{phantom,tension=20}{o1,v4}
\fmf{phantom,tension=0.005}{v5,v6}
\fmf{zigzag,left=0.4,tension=0,foreground=(1,,0,,0)}{v3,v1}
\fmf{phantom,left=0.1,tension=0}{v1,vUp}
\fmf{phantom,left=0.1,tension=0}{vUp,v2}
\fmf{plain,left=0.25,tension=0,foreground=(1,,0,,0)}{v1,v2}
\fmf{zigzag,left=0.4,tension=0,foreground=(1,,0,,0)}{v2,v4}
\fmf{phantom,left=0.1,tension=0}{v4,vDown}
\fmf{phantom,left=0.1,tension=0}{vDown,v3}
\fmf{plain,right=0.25,tension=0,foreground=(1,,0,,0)}{v3,v4}
\fmf{plain,left=0.2,tension=0.01,foreground=(1,,0,,0)}{v1,v5}
\fmf{plain,left=0.2,tension=0.01,foreground=(1,,0,,0)}{v5,v3}
\fmf{plain,right=0.2,tension=0.01,foreground=(1,,0,,0)}{v2,v6}
\fmf{plain,right=0.2,tension=0.01,foreground=(1,,0,,0)}{v6,v4}
\fmf{zigzag,tension=0,foreground=(1,,0,,0)}{v5,v6}
\end{fmfgraph}
\end{fmffile}
\end{gathered} \hspace{0.5cm} + \frac{1}{324} \hspace{0.4cm} \begin{gathered}
\begin{fmffile}{Diagrams/pure4PPIEA_Gamma4_Diag3}
\begin{fmfgraph}(12.5,12.5)
\fmfleft{i0,i1}
\fmfright{o0,o1}
\fmftop{v1,vUp,v2}
\fmfbottom{v3,vDown,v4}
\fmf{phantom,tension=20}{i0,v1}
\fmf{phantom,tension=20}{i1,v3}
\fmf{phantom,tension=20}{o0,v2}
\fmf{phantom,tension=20}{o1,v4}
\fmf{phantom,tension=0.005}{v5,v6}
\fmf{plain,left=0.4,tension=0,foreground=(1,,0,,0)}{v3,v1}
\fmf{phantom,left=0.1,tension=0}{v1,vUp}
\fmf{phantom,left=0.1,tension=0}{vUp,v2}
\fmf{zigzag,left=0.25,tension=0,foreground=(1,,0,,0)}{v1,v2}
\fmf{plain,left=0.4,tension=0,foreground=(1,,0,,0)}{v2,v4}
\fmf{phantom,left=0.1,tension=0}{v4,vDown}
\fmf{phantom,left=0.1,tension=0}{vDown,v3}
\fmf{zigzag,right=0.25,tension=0,foreground=(1,,0,,0)}{v3,v4}
\fmf{plain,left=0.2,tension=0.01,foreground=(1,,0,,0)}{v1,v5}
\fmf{plain,left=0.2,tension=0.01,foreground=(1,,0,,0)}{v5,v3}
\fmf{plain,right=0.2,tension=0.01,foreground=(1,,0,,0)}{v2,v6}
\fmf{plain,right=0.2,tension=0.01,foreground=(1,,0,,0)}{v6,v4}
\fmf{zigzag,tension=0,foreground=(1,,0,,0)}{v5,v6}
\end{fmfgraph}
\end{fmffile}
\end{gathered} \hspace{0.35cm} + \frac{1}{216} \hspace{-0.35cm} \begin{gathered}
\begin{fmffile}{Diagrams/pure4PPIEA_Gamma4_Diag4}
\begin{fmfgraph}(30,15)
\fmfleft{i0,i,i1}
\fmfright{o0,o,o1}
\fmftop{v1b,vUp,v2b}
\fmfbottom{v3b,vDown,v4b}
\fmf{phantom,tension=20}{i0,v1b}
\fmf{phantom,tension=20}{i1,v3b}
\fmf{phantom,tension=20}{o0,v2b}
\fmf{phantom,tension=20}{o1,v4b}
\fmf{phantom,tension=0.511}{i,v7}
\fmf{phantom,tension=0.11}{o,v7}
\fmf{phantom,tension=0.1}{v1,v1b}
\fmf{phantom,tension=0.1}{v2,v2b}
\fmf{phantom,tension=0.1}{v3,v3b}
\fmf{phantom,tension=0.1}{v4,v4b}
\fmf{phantom,tension=0.005}{v5,v6}
\fmf{phantom,left=0.1,tension=0.1}{v1,vUp}
\fmf{phantom,left=0.1,tension=0.1}{vUp,v2}
\fmf{zigzag,left=0.15,tension=0,foreground=(1,,0,,0)}{v1,v2}
\fmf{plain,left=0.4,tension=0,foreground=(1,,0,,0)}{v2,v4}
\fmf{plain,right=0.4,tension=0,foreground=(1,,0,,0)}{v2,v4}
\fmf{phantom,left=0.1,tension=0.1}{v4,vDown}
\fmf{phantom,left=0.1,tension=0.1}{vDown,v3}
\fmf{zigzag,left=0.15,tension=0,foreground=(1,,0,,0)}{v4,v3}
\fmf{plain,left=0.2,tension=0.01,foreground=(1,,0,,0)}{v1,v5}
\fmf{plain,left=0.2,tension=0.01,foreground=(1,,0,,0)}{v5,v3}
\fmf{plain,right=0.2,tension=0,foreground=(1,,0,,0)}{v1,v7}
\fmf{plain,right=0.2,tension=0,foreground=(1,,0,,0)}{v7,v3}
\fmf{phantom,right=0.2,tension=0.01}{v2,v6}
\fmf{phantom,right=0.2,tension=0.01}{v6,v4}
\fmf{zigzag,tension=0,foreground=(1,,0,,0)}{v5,v7}
\end{fmfgraph}
\end{fmffile}
\end{gathered} \\
& \hspace{1.1cm} + \frac{1}{1296} \hspace{-0.32cm} \begin{gathered}
\begin{fmffile}{Diagrams/pure4PPIEA_Gamma4_Diag5}
\begin{fmfgraph}(30,14)
\fmfleft{i0,i,i1}
\fmfright{o0,o,o1}
\fmftop{v1b,vUp,v2b}
\fmfbottom{v3b,vDown,v4b}
\fmf{phantom,tension=5}{vUp,v5}
\fmf{phantom,tension=1}{v1b,v5}
\fmf{phantom,tension=5}{vUp,v6}
\fmf{phantom,tension=1}{v2b,v6}
\fmf{phantom,tension=20}{i0,v1b}
\fmf{phantom,tension=20}{i1,v3b}
\fmf{phantom,tension=20}{o0,v2b}
\fmf{phantom,tension=20}{o1,v4b}
\fmf{phantom,tension=0.1}{v1,v1b}
\fmf{phantom,tension=0.1}{v2,v2b}
\fmf{phantom,tension=0.1}{v3,v3b}
\fmf{phantom,tension=0.1}{v4,v4b}
\fmf{phantom,tension=0.005}{v5,v6}
\fmf{phantom,left=0.1,tension=0.1}{v1,vUp}
\fmf{phantom,left=0.1,tension=0.1}{vUp,v2}
\fmf{plain,left=0.4,tension=0.005,foreground=(1,,0,,0)}{v2,v4}
\fmf{plain,right=0.4,tension=0.005,foreground=(1,,0,,0)}{v2,v4}
\fmf{plain,left=0.4,tension=0.005,foreground=(1,,0,,0)}{v1,v3}
\fmf{plain,right=0.4,tension=0.005,foreground=(1,,0,,0)}{v1,v3}
\fmf{phantom,left=0.1,tension=0.1}{v4,vDown}
\fmf{phantom,left=0.1,tension=0.1}{vDown,v3}
\fmf{zigzag,left=0.05,tension=0,foreground=(1,,0,,0)}{v1,v5}
\fmf{plain,left,tension=0,foreground=(1,,0,,0)}{v5,v6}
\fmf{plain,right,tension=0,foreground=(1,,0,,0)}{v5,v6}
\fmf{zigzag,left=0.05,tension=0,foreground=(1,,0,,0)}{v6,v2}
\fmf{zigzag,left=0.15,tension=0,foreground=(1,,0,,0)}{v4,v3}
\end{fmfgraph}
\end{fmffile}
\end{gathered} \hspace{-0.35cm} \left.\rule{0cm}{1.1cm}\right) \\
& + \mathcal{O}\big(\hbar^{5}\big)\;.
\end{split}
\label{eq:pure4PPIEAfinalexpressionAppendix}
\end{equation}
Result~\eqref{eq:pure4PPIEAfinalexpressionAppendix} shows that all inverse propagators involved in expressions~\eqref{eq:4PPIEADeterminationM10DON} and~\eqref{eq:4PPIEADeterminationK20DON} of $M_{1}$ and $K_{2}$ have been canceled out in $\Gamma_{4}^{(\mathrm{4PPI})}[\rho,\zeta]$. However, as discussed for the 2PPI EA with~\eqref{eq:2PPIEAGamma5diagram0DON}, these inverse propagators will appear in diagrams contributing to the 4PPI EA at order $\mathcal{O}\big(\hbar^{5}\big)$.

\section{\label{sec:DerivEAcoeff}Derivation of effective action coefficients}
\subsection{\label{sec:GammanCoeffIM1PIEA}1PI effective action}

Finally, we derive general expressions for $\Gamma_{n}^{(m\mathrm{P(P)I})}$ coefficients used in the previous sections, starting with the $\Gamma^{(\mathrm{1PI})}_{n}$ coefficients introduced in~\eqref{eq:pure1PIEAGammaExpansion0DONAppendix}. To that end, we start by considering~\eqref{eq:pure1PIEAIMstep20DON} recalled below:
\begin{equation}
\sum_{n=0}^{\infty} \Gamma^{(\mathrm{1PI})}_{n}\Big[\vec{\phi}\Big]\hbar^{n} = -\sum_{n=0}^{\infty} W_{n}\Bigg[\sum_{m=0}^{\infty} \vec{J}_{m}\Big[\vec{\phi}\Big]\hbar^{m}\Bigg]\hbar^{n}+\sum_{n=0}^{\infty} \int_{\alpha} J_{n,\alpha}\Big[\vec{\phi}\Big] \phi_{\alpha} \hbar^{n}\;.
\label{eq:pure1PIEAIMstep2bis0DON}
\end{equation}
The $W_{n}$ coefficients are then Taylor expanded around $\vec{J}=\vec{J}_{0}$:
\begin{equation}
\begin{split}
W_{n}\Bigg[\sum_{m=0}^{\infty} \vec{J}_{m}\Big[\vec{\phi}\Big]\hbar^{m}\Bigg] = \ & W_{n}\Bigg[\vec{J}_{0}\Big[\vec{\phi}\Big] + \sum_{m=1}^{\infty} \vec{J}_{m}\Big[\vec{\phi}\Big]\hbar^{m}\Bigg] \\
= \ & W_{n}\Big[\vec{J}=\vec{J}_{0}\Big] + \int_{\alpha} \left.\frac{\delta W_{n}\big[\vec{J}\big]}{\delta J_{\alpha}}\right|_{\vec{J}=\vec{J}_{0}} \left(\sum_{m=1}^{\infty} J_{m,\alpha}\Big[\vec{\phi}\Big]\hbar^{m}\right) \\
& + \sum_{m=2}^{\infty} \frac{1}{m!} \int_{\alpha_{1},\cdots,\alpha_{m}} \left.\frac{\delta^{m} W_{n}\big[\vec{J}\big]}{\delta J_{\alpha_{1}}\cdots\delta J_{\alpha_{m}}}\right|_{\vec{J}=\vec{J}_{0}} \\
& \hspace{3.1cm} \times \left(\sum_{n_{1}=1}^{\infty} J_{n_{1},\alpha_{1}}\Big[\vec{\phi}\Big]\hbar^{n_{1}}\right) \cdots \left(\sum_{n_{m}=1}^{\infty} J_{n_{m},\alpha_{m}}\Big[\vec{\phi}\Big]\hbar^{n_{m}}\right) \\
= \ & W_{n}\Big[\vec{J}=\vec{J}_{0}\Big] + \sum_{m=1}^{\infty} \int_{\alpha} \left.\frac{\delta W_{n}\big[\vec{J}\big]}{\delta J_{\alpha}}\right|_{\vec{J}=\vec{J}_{0}} J_{m,\alpha}\Big[\vec{\phi}\Big]\hbar^{m} \\
& + \sum_{m=2}^{\infty} \frac{1}{m!} \sum_{n_{1},\cdots,n_{m}=1}^{\infty} \int_{\alpha_{1},\cdots,\alpha_{m}} \left.\frac{\delta^{m} W_{n}\big[\vec{J}\big]}{\delta J_{\alpha_{1}}\cdots\delta J_{\alpha_{m}}}\right|_{\vec{J}=\vec{J}_{0}} \\
& \hspace{4.8cm} \times J_{n_{1},\alpha_{1}}\Big[\vec{\phi}\Big]\cdots J_{n_{m},\alpha_{m}}\Big[\vec{\phi}\Big]\hbar^{n_{1}+\cdots+n_{m}}\;.
\end{split}
\label{eq:1PIEAWnTaylorExpansion0DON}
\end{equation}
The latter relation is just the result of a functional generalization of the Taylor expansion for a $p$-variable function $f(x_{1},\cdots,x_{p})$ around the point $(x_{1},\cdots,x_{p})=(\overline{x}_{1},\cdots,\overline{x}_{p})$ given by\footnote{In the example of~\eqref{eq:TaylorExpansionpvariablefunction}, the functional generalization amounts to taking the limit $p\rightarrow\infty$.}:
\begin{equation}
\begin{split}
f(x_{1},\cdots,x_{p})= \ & f(\overline{x}_{1},\cdots,\overline{x}_{p})+\sum_{l=1}^{p}\left.\frac{\partial f(x_{1},\cdots,x_{p})}{\partial x_{l}}\right|_{x_{1}=\overline{x}_{1},\cdots,x_{p}=\overline{x}_{p}}\left(x_{l}-\overline{x}_{l}\right) \\
& + \sum_{m=2}^{\infty} \frac{1}{m!} \sum_{\underset{\{k_{1}+\cdots+k_{p}=m\}}{k_{1},\cdots,k_{p} \geq 0}} \left.\frac{\partial^{m} f(x_{1},...,x_{p})}{\partial x_{1}^{k_{1}}...\partial x_{p}^{k_{p}}}\right|_{x_{1}=\overline{x}_{1},\cdots,x_{p}=\overline{x}_{p}}\left(x_{1}-\overline{x}_{1}\right)^{k_{1}}\cdots\left(x_{p}-\overline{x}_{p}\right)^{k_{p}}\;.
\end{split}
\label{eq:TaylorExpansionpvariablefunction}
\end{equation}
Inserting~\eqref{eq:1PIEAWnTaylorExpansion0DON} into~\eqref{eq:pure1PIEAIMstep2bis0DON} leads to:
\begin{equation}
\begin{split}
\sum_{n=0}^{\infty} \Gamma^{(\mathrm{1PI})}_{n}\Big[\vec{\phi}\Big]\hbar^{n} = & -\sum_{n=0}^{\infty} W_{n}\Big[\vec{J}=\vec{J}_{0}\Big] \hbar^{n} - \underbrace{\sum_{n=0}^{\infty}\sum_{m=1}^{\infty} \int_{\alpha} \left.\frac{\delta W_{n}\big[\vec{J}\big]}{\delta J_{\alpha}}\right|_{\vec{J}=\vec{J}_{0}} J_{m,\alpha}\Big[\vec{\phi}\Big]\hbar^{m+n}}_{A_{1}} \\
& - \underbrace{\sum_{n=0}^{\infty}\sum_{m=2}^{\infty} \frac{1}{m!} \sum_{n_{1},\cdots,n_{m}=1}^{\infty} \int_{\alpha_{1},\cdots,\alpha_{m}} \left.\frac{\delta^{m} W_{n}\big[\vec{J}\big]}{\delta J_{\alpha_{1}}\cdots\delta J_{\alpha_{m}}}\right|_{\vec{J}=\vec{J}_{0}} J_{n_{1},\alpha_{1}}\Big[\vec{\phi}\Big]\cdots J_{n_{m},\alpha_{m}}\Big[\vec{\phi}\Big]\hbar^{n_{1}+\cdots+n_{m}+n}}_{A_{2}} \\
& +\sum_{n=0}^{\infty} \int_{\alpha} J_{n,\alpha}\Big[\vec{\phi}\Big] \phi_{\alpha} \ \hbar^{n}\;.
\end{split}
\label{eq:pure1PIEAIMstep30DON}
\end{equation}
The powers of $\hbar$ involved in the latter equation suggest to perform the substitutions:
\begin{itemize}
\item For $A_{1}$:
\begin{equation}
n \rightarrow n-m\;.
\end{equation}
\item For $A_{2}$:
\begin{equation}
n \rightarrow n-\left(n_{1}+\cdots+n_{m}\right)\;.
\end{equation}
\end{itemize}
This implies that the discrete sums of terms $A_{1}$ and $A_{2}$ change as follows:
\begin{itemize}
\item For $A_{1}$:
\begin{equation}
\sum_{n=0}^{\infty}\sum_{m=1}^{\infty} \rightarrow \sum_{n=1}^{\infty}\sum_{m=1}^{n}\;.
\end{equation}
\item For $A_{2}$:
\begin{equation}
\sum_{n=0}^{\infty}\sum_{m=2}^{\infty}\sum_{n_{1},\cdots,n_{m}=1}^{\infty} \rightarrow \sum_{n=1}^{\infty}\sum_{m=2}^{n}\sum_{\underset{\lbrace n_{1}+\cdots+n_{m} \leq n\rbrace}{n_{1},\cdots,n_{m}=1}}^{n}\;.
\end{equation}
\end{itemize}
From this,~\eqref{eq:pure1PIEAIMstep30DON} becomes:
\begin{equation}
\begin{split}
\sum_{n=0}^{\infty} \Gamma^{(\mathrm{1PI})}_{n}\Big[\vec{\phi}\Big]\hbar^{n} = & -\sum_{n=0}^{\infty} W_{n}\Big[\vec{J}=\vec{J}_{0}\Big] \hbar^{n} - \sum_{n=1}^{\infty}\sum_{m=1}^{n} \int_{\alpha} \left.\frac{\delta W_{n-m}\big[\vec{J}\big]}{\delta J_{\alpha}}\right|_{\vec{J}=\vec{J}_{0}} J_{m,\alpha}\Big[\vec{\phi}\Big]\hbar^{n} \\
& - \sum_{n=1}^{\infty}\sum_{m=2}^{n} \frac{1}{m!} \sum_{\underset{\lbrace n_{1} + \cdots + n_{m} \leq n\rbrace}{n_{1},\cdots,n_{m}=1}}^{n} \int_{\alpha_{1},\cdots,\alpha_{m}} \left.\frac{\delta^{m} W_{n-(n_{1}+\cdots+n_{m})}\big[\vec{J}\big]}{\delta J_{\alpha_{1}}\cdots\delta J_{\alpha_{m}}}\right|_{\vec{J}=\vec{J}_{0}} \\
& \hspace{5.8cm} \times J_{n_{1},\alpha_{1}}\Big[\vec{\phi}\Big]\cdots J_{n_{m},\alpha_{m}}\Big[\vec{\phi}\Big]\hbar^{n} \\
& +\sum_{n=0}^{\infty} \int_{\alpha} J_{n,\alpha}\Big[\vec{\phi}\Big] \phi_{\alpha} \ \hbar^{n}\;.
\end{split}
\label{eq:pure1PIEAIMstep40DON}
\end{equation}
Since $\vec{\phi}$ and $\hbar$ are independent, we can identify the terms of order $\mathcal{O}\big(\hbar^{n}\big)$ in both sides of~\eqref{eq:pure1PIEAIMstep40DON}, thus leading to:
\begin{equation}
\begin{split}
\Gamma^{(\mathrm{1PI})}_{n}\Big[\vec{\phi}\Big] = & -W_{n}\Big[\vec{J}=\vec{J}_{0}\Big] -\sum_{m=1}^{n} \int_{\alpha} \left.\frac{\delta W_{n-m}\big[\vec{J}\big]}{\delta J_{\alpha}}\right|_{\vec{J}=\vec{J}_{0}} J_{m,\alpha}\Big[\vec{\phi}\Big] \\
& -\sum_{m=2}^{n} \frac{1}{m!} \sum_{\underset{\lbrace n_{1} + \cdots + n_{m} \leq n\rbrace}{n_{1},\cdots,n_{m}=1}}^{n} \int_{\alpha_{1},\cdots,\alpha_{m}} \left.\frac{\delta^{m} W_{n-(n_{1}+\cdots+n_{m})}\big[\vec{J}\big]}{\delta J_{\alpha_{1}}\cdots\delta J_{\alpha_{m}}}\right|_{\vec{J}=\vec{J}_{0}} J_{n_{1},\alpha_{1}}\Big[\vec{\phi}\Big]\cdots J_{n_{m},\alpha_{m}}\Big[\vec{\phi}\Big] \\
& + \int_{\alpha} J_{n,\alpha}\Big[\vec{\phi}\Big] \phi_{\alpha}\;.
\end{split}
\label{eq:pure1PIEAIMstep5bis0DON}
\end{equation}

\subsection{\label{sec:GammanCoeffIM2PIEA}2PI effective action}

We then derive an expression for the $\Gamma_{n}^{(\mathrm{2PI})}$ coefficients introduced in~\eqref{eq:pure2PIEAGammaExpansion0DON}. Our derivations start from~\eqref{eq:pure2PIEAIMstep20DON}, i.e.:
\begin{equation}
\begin{split}
\scalebox{0.96}{${\displaystyle \sum_{n=0}^{\infty}\Gamma_{n}^{(\mathrm{2PI})}\Big[\vec{\phi},\boldsymbol{G}\Big]\hbar^{n}= }$} & \scalebox{0.96}{${\displaystyle - \sum_{n=0}^{\infty} W_{n}\Bigg[\sum_{m=0}^{\infty} \vec{J}_{m}\Big[\vec{\phi},\boldsymbol{G}\Big]\hbar^{m},\sum_{m=0}^{\infty} \boldsymbol{K}_{m}\Big[\vec{\phi},\boldsymbol{G}\Big]\hbar^{m}\Bigg] \hbar^{n} + \sum_{n=0}^{\infty} \int_{\alpha} J_{n,\alpha}\Big[\vec{\phi},\boldsymbol{G}\Big] \phi_{\alpha} \hbar^{n} }$} \\
& \scalebox{0.96}{${\displaystyle + \frac{1}{2} \sum_{n=0}^{\infty} \int_{\alpha_{1},\alpha_{2}} \phi_{\alpha_{1}} \boldsymbol{K}_{n,\alpha_{1}\alpha_{2}}\Big[\vec{\phi},\boldsymbol{G}\Big]\phi_{\alpha_{2}} \hbar^{n} + \frac{1}{2} \sum_{n=0}^{\infty} \int_{\alpha_{1},\alpha_{2}} \boldsymbol{K}_{n,\alpha_{1}\alpha_{2}}\Big[\vec{\phi},\boldsymbol{G}\Big]\boldsymbol{G}_{\alpha_{1}\alpha_{2}} \hbar^{n+1}\;. }$}
\end{split}
\label{eq:pure2PIEAIMstep2bis0DON}
\end{equation}
The recipe is very similar to that leading to expression~\eqref{eq:pure1PIEAIMstep5bis0DON} of the $\Gamma_{n}^{(\mathrm{1PI})}$ coefficients. Hence, we start by Taylor expanding $W_{n}\big[\vec{J},\boldsymbol{K}\big]$ around $\big(\vec{J},\boldsymbol{K}\big)=\big(\vec{J}_{0},\boldsymbol{K}_{0}\big)$:
\begin{equation}
\begin{split}
& W_{n} \Bigg[\sum_{m=0}^{\infty} \vec{J}_{m}\Big[\vec{\phi},\boldsymbol{G}\Big]\hbar^{m},\sum_{m=0}^{\infty} \boldsymbol{K}_{m}\Big[\vec{\phi},\boldsymbol{G}\Big]\hbar^{m}\Bigg] \\
& \hspace{0.8cm} = W_{n}\Bigg[\vec{J}_{0}\Big[\vec{\phi},\boldsymbol{G}\Big]+\sum_{m=1}^{\infty} \vec{J}_{m}\Big[\vec{\phi},\boldsymbol{G}\Big]\hbar^{m},\boldsymbol{K}_{0}\Big[\vec{\phi},\boldsymbol{G}\Big]+\sum_{m=0}^{\infty} \boldsymbol{K}_{m}\Big[\vec{\phi},\boldsymbol{G}\Big]\hbar^{m}\Bigg] \\
& \hspace{0.8cm} = W_{n}\Big[\vec{J}=\vec{J}_{0},\boldsymbol{K}=\boldsymbol{K}_{0}\Big] + \int_{\alpha} \left.\frac{\delta W_{n}\big[\vec{J},\boldsymbol{K}\big]}{\delta J_{\alpha}}\right|_{\vec{J}=\vec{J}_{0} \atop \boldsymbol{K}=\boldsymbol{K}_{0}} \left(\sum_{m=1}^{\infty} J_{m,\alpha}\Big[\vec{\phi},\boldsymbol{G}\Big]\hbar^{m}\right) \\
& \hspace{0.8cm} \hspace{0.4cm} + \int_{\alpha_{1},\alpha_{2}} \left.\frac{\delta W_{n}\big[\vec{J},\boldsymbol{K}\big]}{\delta \boldsymbol{K}_{\alpha_{1}\alpha_{2}}}\right|_{\vec{J}=\vec{J}_{0} \atop \boldsymbol{K}=\boldsymbol{K}_{0}} \left(\sum_{m=1}^{\infty} \boldsymbol{K}_{m,\alpha_{1}\alpha_{2}}\Big[\vec{\phi},\boldsymbol{G}\Big]\hbar^{m}\right) \\
& \hspace{0.8cm} \hspace{0.4cm} + \sum_{m=2}^{\infty} \frac{1}{m!} \sum_{\underset{\lbrace l+l'=m \rbrace}{l,l'=1}}^{m} \begin{pmatrix}
m \\
l
\end{pmatrix} \\
& \hspace{0.8cm} \hspace{0.4cm} \hspace{0.8cm} \times \int_{\alpha_{1},\cdots,\alpha_{l} \atop \hat{\alpha}_{1},\cdots,\hat{\alpha}_{2l'}} \left.\frac{\delta^{m} W_{n}\big[\vec{J},\boldsymbol{K}\big]}{\delta J_{\alpha_{1}}\cdots\delta J_{\alpha_{l}} \delta \boldsymbol{K}_{\hat{\alpha}_{1}\hat{\alpha}_{2}}\cdots\delta \boldsymbol{K}_{\hat{\alpha}_{2 l'-1}\hat{\alpha}_{2 l'}}}\right|_{\vec{J}=\vec{J}_{0} \atop \boldsymbol{K}=\boldsymbol{K}_{0}} \\
& \hspace{0.8cm} \hspace{0.4cm} \hspace{2.5cm} \times \left(\sum_{n_{1}=1}^{\infty} J_{n_{1},\alpha_{1}}\Big[\vec{\phi},\boldsymbol{G}\Big]\hbar^{n_{1}}\right)\cdots\left(\sum_{n_{l}=1}^{\infty} J_{n_{l},\alpha_{l}}\Big[\vec{\phi},\boldsymbol{G}\Big]\hbar^{n_{l}}\right) \\
& \hspace{0.8cm} \hspace{0.4cm} \hspace{2.5cm} \times \left(\sum_{\hat{n}_{1}=1}^{\infty} \boldsymbol{K}_{\hat{n}_{1},\hat{\alpha}_{1}\hat{\alpha}_{2}}\Big[\vec{\phi},\boldsymbol{G}\Big]\hbar^{\hat{n}_{1}}\right) \cdots \left(\sum_{\hat{n}_{l'}=1}^{\infty} \boldsymbol{K}_{\hat{n}_{l'},\hat{\alpha}_{2 l' -1}\hat{\alpha}_{2 l'}}\Big[\vec{\phi},\boldsymbol{G}\Big]\hbar^{\hat{n}_{l'}}\right) \\
& \hspace{0.8cm} = W_{n}\Big[\vec{J}=\vec{J}_{0},\boldsymbol{K}=\boldsymbol{K}_{0}\Big] + \sum_{m=1}^{\infty} \int_{\alpha} \left.\frac{\delta W_{n}\big[\vec{J},\boldsymbol{K}\big]}{\delta J_{\alpha}}\right|_{\vec{J}=\vec{J}_{0} \atop \boldsymbol{K}=\boldsymbol{K}_{0}} J_{m,\alpha}\Big[\vec{\phi},\boldsymbol{K}\Big]\hbar^{m} \\
& \hspace{0.8cm} \hspace{0.4cm} + \sum_{m=1}^{\infty} \int_{\alpha_{1},\alpha_{2}} \left.\frac{\delta W_{n}\big[\vec{J},\boldsymbol{K}\big]}{\delta \boldsymbol{K}_{\alpha_{1}\alpha_{2}}}\right|_{\vec{J}=\vec{J}_{0} \atop \boldsymbol{K}=\boldsymbol{K}_{0}} \boldsymbol{K}_{m,\alpha_{1}\alpha_{2}}\Big[\vec{\phi},\boldsymbol{G}\Big]\hbar^{m} \\
& \hspace{0.8cm} \hspace{0.4cm} + \sum_{m=2}^{\infty} \frac{1}{m!} \sum_{\underset{\lbrace l+l'=m \rbrace}{l,l'=1}}^{m} \sum^{\infty}_{n_{1},\cdots,n_{l},\hat{n}_{1},\cdots,\hat{n}_{l'}=1} \begin{pmatrix}
m \\
l
\end{pmatrix} \\
& \hspace{0.8cm} \hspace{0.4cm} \hspace{0.8cm} \times \int_{\alpha_{1},\cdots,\alpha_{l} \atop \hat{\alpha}_{1},\cdots,\hat{\alpha}_{2l'}} \left.\frac{\delta^{m} W_{n}\big[\vec{J},\boldsymbol{K}\big]}{\delta J_{\alpha_{1}}\cdots\delta J_{\alpha_{l}} \delta \boldsymbol{K}_{\hat{\alpha}_{1}\hat{\alpha}_{2}}\cdots\delta \boldsymbol{K}_{\hat{\alpha}_{2 l'-1}\hat{\alpha}_{2 l'}}}\right|_{\vec{J}=\vec{J}_{0} \atop \boldsymbol{K}=\boldsymbol{K}_{0}} \\
& \hspace{0.8cm} \hspace{0.4cm} \hspace{2.5cm} \times J_{n_{1},\alpha_{1}}\Big[\vec{\phi},\boldsymbol{G}\Big] \cdots J_{n_{l},\alpha_{l}}\Big[\vec{\phi},\boldsymbol{G}\Big] \boldsymbol{K}_{\hat{n}_{1},\hat{\alpha}_{1}\hat{\alpha}_{2}}\Big[\vec{\phi},\boldsymbol{G}\Big] \cdots \boldsymbol{K}_{\hat{n}_{l'},\hat{\alpha}_{2 l'-1}\hat{\alpha}_{2 l'}}\Big[\vec{\phi},\boldsymbol{G}\Big] \\
& \hspace{0.8cm} \hspace{0.4cm} \hspace{2.5cm} \times \hbar^{n_{1}+\cdots+n_{l}+\hat{n}_{1}+\cdots+\hat{n}_{l'}}\;.
\end{split}
\label{eq:2PIEAWnTaylorExpansion0DON}
\end{equation}
We combine~\eqref{eq:2PIEAWnTaylorExpansion0DON} with~\eqref{eq:pure2PIEAIMstep2bis0DON}:
\begin{equation}
\begin{split}
\scalebox{0.93}{${\displaystyle \sum_{n=0}^{\infty}\Gamma_{n}^{(\mathrm{2PI})}\Big[\vec{\phi},\boldsymbol{G}\Big]\hbar^{n} = }$} & \scalebox{0.93}{${\displaystyle - \sum_{n=0}^{\infty} W_{n}\Big[\vec{J}=\vec{J}_{0},\boldsymbol{K}=\boldsymbol{K}_{0}\Big] \hbar^{n} - \underbrace{\sum_{n=0}^{\infty} \sum_{m=1}^{\infty} \int_{\alpha} \left.\frac{\delta W_{n}\big[\vec{J},\boldsymbol{K}\big]}{\delta J_{\alpha}}\right|_{\vec{J}=\vec{J}_{0} \atop \boldsymbol{K}=\boldsymbol{K}_{0}} J_{m,\alpha}\Big[\vec{\phi},\boldsymbol{K}\Big]\hbar^{m+n}}_{B_{1}} }$} \\
& \scalebox{0.93}{${\displaystyle - \underbrace{\sum_{n=0}^{\infty} \sum_{m=1}^{\infty} \int_{\alpha_{1},\alpha_{2}} \left.\frac{\delta W_{n}\big[\vec{J},\boldsymbol{K}\big]}{\delta \boldsymbol{K}_{\alpha_{1}\alpha_{2}}}\right|_{\vec{J}=\vec{J}_{0} \atop \boldsymbol{K}=\boldsymbol{K}_{0}} \boldsymbol{K}_{m,\alpha_{1}\alpha_{2}}\Big[\vec{\phi},\boldsymbol{G}\Big]\hbar^{m+n}}_{B_{2}} }$} \\
& \scalebox{0.93}{${\displaystyle \underbrace{\begin{aligned} & - \sum_{n=0}^{\infty} \sum_{m=2}^{\infty} \frac{1}{m!} \sum_{\underset{\lbrace l+l'=m \rbrace}{l,l'=1}}^{m} \sum^{\infty}_{n_{1},\cdots,n_{l},\hat{n}_{1},\cdots,\hat{n}_{l'}=1} \begin{pmatrix}
m \\
l
\end{pmatrix} \int_{\alpha_{1},\cdots,\alpha_{l} \atop \hat{\alpha}_{1},\cdots,\hat{\alpha}_{2l'}} \left.\frac{\delta^{m} W_{n}\big[\vec{J},\boldsymbol{K}\big]}{\delta J_{\alpha_{1}}\cdots\delta J_{\alpha_{l}} \delta \boldsymbol{K}_{\hat{\alpha}_{1}\hat{\alpha}_{2}}\cdots\delta \boldsymbol{K}_{\hat{\alpha}_{2 l'-1}\hat{\alpha}_{2 l'}}}\right|_{\vec{J}=\vec{J}_{0} \atop \boldsymbol{K}=\boldsymbol{K}_{0}} \\ & \hspace{0.5cm} \times J_{n_{1},\alpha_{1}}\Big[\vec{\phi},\boldsymbol{G}\Big] \cdots J_{n_{l},\alpha_{l}}\Big[\vec{\phi},\boldsymbol{G}\Big] \boldsymbol{K}_{\hat{n}_{1},\hat{\alpha}_{1}\hat{\alpha}_{2}}\Big[\vec{\phi},\boldsymbol{G}\Big] \cdots \boldsymbol{K}_{\hat{n}_{l'},\hat{\alpha}_{2 l' -1}\hat{\alpha}_{2 l'}}\Big[\vec{\phi},\boldsymbol{G}\Big] \hbar^{n_{1}+\cdots+n_{l}+\hat{n}_{1}+\cdots+\hat{n}_{l'}+n} \end{aligned}}_{B_{3}} }$} \\
& \scalebox{0.93}{${\displaystyle + \sum_{n=0}^{\infty} \int_{\alpha} J_{n,\alpha}\Big[\vec{\phi},\boldsymbol{G}\Big] \phi_{\alpha} \hbar^{n} + \frac{1}{2} \sum_{n=0}^{\infty} \int_{\alpha_{1},\alpha_{2}} \phi_{\alpha_{1}} \boldsymbol{K}_{n,\alpha_{1}\alpha_{2}}\Big[\vec{\phi},\boldsymbol{G}\Big]\phi_{\alpha_{2}} \hbar^{n} }$} \\
& \scalebox{0.93}{${\displaystyle + \frac{1}{2} \sum_{n=0}^{\infty} \int_{\alpha_{1},\alpha_{2}} \boldsymbol{K}_{n,\alpha_{1}\alpha_{2}}\Big[\vec{\phi},\boldsymbol{G}\Big] \boldsymbol{G}_{\alpha_{1}\alpha_{2}} \hbar^{n+1} \;. }$}
\end{split}
\label{eq:pure2PIEAIMstep30DON}
\end{equation}
We then perform the following substitutions in~\eqref{eq:pure2PIEAIMstep30DON}:
\begin{itemize}
\item For $B_{1}$ and $B_{2}$:
\begin{equation}
n \rightarrow n-m\;.
\label{eq:pure2PIEAchangeIndexB1B20DON}
\end{equation}
\item For $B_{3}$:
\begin{equation}
n \rightarrow n-\left(n_{1}+\cdots+n_{l}+\hat{n}_{1}+\cdots+\hat{n}_{l'}\right)\;.
\label{eq:pure2PIEAchangeIndexB30DON}
\end{equation}
\end{itemize}
As a consequence, the discrete sums of $B_{1}$, $B_{2}$ and $B_{3}$ are affected as follows:
\begin{itemize}
\item For $B_{1}$ and $B_{2}$:
\begin{equation}
\sum_{n=0}^{\infty}\sum_{m=1}^{\infty} \rightarrow \sum_{n=1}^{\infty}\sum_{m=1}^{n}\;.
\label{eq:pure2PIEAsumIndexB1B20DON}
\end{equation}
\item For $B_{3}$:
\begin{equation}
\sum_{n=0}^{\infty}\sum_{m=2}^{\infty}\sum^{\infty}_{n_{1},\cdots,n_{l},\hat{n}_{1},\cdots,\hat{n}_{l'}=1} \rightarrow \sum_{n=1}^{\infty}\sum_{m=2}^{n}\sum^{n}_{\underset{\lbrace n_{1} + \cdots + n_{l} + \hat{n}_{1} + \cdots + \hat{n}_{l'} \leq n\rbrace}{n_{1},\cdots,n_{l},\hat{n}_{1},\cdots,\hat{n}_{l'}=1}} \;.
\label{eq:pure2PIEAchangesumB30DON}
\end{equation}
\end{itemize}
According to this,~\eqref{eq:pure2PIEAIMstep30DON} becomes:
\begin{equation}
\begin{split}
\sum_{n=0}^{\infty}\Gamma_{n}^{(\mathrm{2PI})}\Big[\vec{\phi},\boldsymbol{G}\Big]\hbar^{n}= & - \sum_{n=0}^{\infty} W_{n}\Big[\vec{J}=\vec{J}_{0},\boldsymbol{K}=\boldsymbol{K}_{0}\Big] \hbar^{n} \\
& - \sum_{n=1}^{\infty}\sum_{m=1}^{n} \int_{\alpha} \left.\frac{\delta W_{n-m}\big[\vec{J},\boldsymbol{K}\big]}{\delta J_{\alpha}}\right|_{\vec{J}=\vec{J}_{0} \atop \boldsymbol{K}=\boldsymbol{K}_{0}} J_{m,\alpha}\Big[\vec{\phi},\boldsymbol{K}\Big]\hbar^{n} \\
& - \sum_{n=1}^{\infty}\sum_{m=1}^{n} \int_{\alpha_{1},\alpha_{2}} \left.\frac{\delta W_{n-m}\big[\vec{J},\boldsymbol{K}\big]}{\delta \boldsymbol{K}_{\alpha_{1}\alpha_{2}}}\right|_{\vec{J}=\vec{J}_{0} \atop \boldsymbol{K}=\boldsymbol{K}_{0}} \boldsymbol{K}_{m,\alpha_{1}\alpha_{2}}\Big[\vec{\phi},\boldsymbol{G}\Big]\hbar^{n} \\
& - \sum_{n=1}^{\infty}\sum_{m=2}^{n} \frac{1}{m!} \sum_{\underset{\lbrace l+l'=m \rbrace}{l,l'=1}}^{m} \sum^{n}_{\underset{\lbrace n_{1} + \cdots + n_{l} + \hat{n}_{1} + \cdots + \hat{n}_{l'} \leq n\rbrace}{n_{1},\cdots,n_{l},\hat{n}_{1},\cdots,\hat{n}_{l'}=1}} \begin{pmatrix}
m \\
l
\end{pmatrix} \\
& \hspace{0.8cm} \times \int_{\alpha_{1},\cdots,\alpha_{l} \atop \hat{\alpha}_{1},\cdots,\hat{\alpha}_{2l'}} \left.\frac{\delta^{m} W_{n-(n_{1}+\cdots+n_{l}+\hat{n}_{1}+\cdots+\hat{n}_{l'})}\big[\vec{J},\boldsymbol{K}\big]}{\delta J_{\alpha_{1}}\cdots\delta J_{\alpha_{l}} \delta \boldsymbol{K}_{\hat{\alpha}_{1}\hat{\alpha}_{2}}\cdots\delta \boldsymbol{K}_{\hat{\alpha}_{2 l'-1}\hat{\alpha}_{2 l'}}}\right|_{\vec{J}=\vec{J}_{0} \atop \boldsymbol{K}=\boldsymbol{K}_{0}} \\
& \hspace{2.5cm} \times J_{n_{1},\alpha_{1}}\Big[\vec{\phi},\boldsymbol{G}\Big] \cdots J_{n_{l},\alpha_{l}}\Big[\vec{\phi},\boldsymbol{G}\Big] \\
& \hspace{2.5cm} \times \boldsymbol{K}_{\hat{n}_{1},\hat{\alpha}_{1}\hat{\alpha}_{2}}\Big[\vec{\phi},\boldsymbol{G}\Big] \cdots \boldsymbol{K}_{\hat{n}_{l'},\hat{\alpha}_{2 l' -1}\hat{\alpha}_{2 l'}}\Big[\vec{\phi},\boldsymbol{G}\Big] \hbar^{n} \\
& + \sum_{n=0}^{\infty} \int_{\alpha} J_{n,\alpha}\Big[\vec{\phi},\boldsymbol{G}\Big] \phi_{\alpha} \hbar^{n} + \frac{1}{2} \sum_{n=0}^{\infty} \int_{\alpha_{1},\alpha_{2}} \phi_{\alpha_{1}} \boldsymbol{K}_{n,\alpha_{1}\alpha_{2}}\Big[\vec{\phi},\boldsymbol{G}\Big]\phi_{\alpha_{2}} \hbar^{n} \\
& + \frac{1}{2} \sum_{n=0}^{\infty} \int_{\alpha_{1},\alpha_{2}} \boldsymbol{K}_{n,\alpha_{1}\alpha_{2}}\Big[\vec{\phi},\boldsymbol{G}\Big] \boldsymbol{G}_{\alpha_{1}\alpha_{2}} \hbar^{n+1}\;.
\end{split}
\label{eq:pure2PIEAIMstep40DON}
\end{equation}
Finally, we exploit the fact that both $\vec{\phi}$ and $\boldsymbol{G}$ are independent of $\hbar$ to identify the terms of order $\mathcal{O}\big(\hbar^{n}\big)$ in~\eqref{eq:pure2PIEAIMstep40DON}:
\begin{equation}
\begin{split}
\scalebox{0.91}{${\displaystyle \Gamma_{n}^{(\mathrm{2PI})}\Big[\vec{\phi},\boldsymbol{G}\Big] = }$} & \scalebox{0.91}{${\displaystyle - W_{n}\Big[\vec{J}=\vec{J}_{0},\boldsymbol{K}=\boldsymbol{K}_{0}\Big] - \sum_{m=1}^{n} \int_{\alpha} \left.\frac{\delta W_{n-m}\big[\vec{J},\boldsymbol{K}\big]}{\delta J_{\alpha}}\right|_{\vec{J}=\vec{J}_{0} \atop \boldsymbol{K}=\boldsymbol{K}_{0}} J_{m,\alpha}\Big[\vec{\phi},\boldsymbol{K}\Big] }$} \\
& \scalebox{0.91}{${\displaystyle - \sum_{m=1}^{n} \int_{\alpha_{1},\alpha_{2}} \left.\frac{\delta W_{n-m}\big[\vec{J},\boldsymbol{K}\big]}{\delta \boldsymbol{K}_{\alpha_{1}\alpha_{2}}}\right|_{\vec{J}=\vec{J}_{0} \atop \boldsymbol{K}=\boldsymbol{K}_{0}} \boldsymbol{K}_{m,\alpha_{1}\alpha_{2}}\Big[\vec{\phi},\boldsymbol{G}\Big] }$} \\
& \scalebox{0.91}{${\displaystyle - \sum_{m=2}^{n} \frac{1}{m!} \sum_{\underset{\lbrace l+l'=m \rbrace}{l,l'=1}}^{m} \sum^{n}_{\underset{\lbrace n_{1} + \cdots + n_{l} + \hat{n}_{1} + \cdots + \hat{n}_{l'} \leq n\rbrace}{n_{1},\cdots,n_{l},\hat{n}_{1},\cdots,\hat{n}_{l'}=1}} \begin{pmatrix}
m \\
l
\end{pmatrix} }$} \\
& \hspace{0.8cm} \scalebox{0.91}{${\displaystyle \times \int_{\alpha_{1},\cdots,\alpha_{l} \atop \hat{\alpha}_{1},\cdots,\hat{\alpha}_{2l'}} \left.\frac{\delta^{m} W_{n-(n_{1}+\cdots+n_{l}+\hat{n}_{1}+\cdots+\hat{n}_{l'})}\big[\vec{J},\boldsymbol{K}\big]}{\delta J_{\alpha_{1}}\cdots\delta J_{\alpha_{l}} \delta \boldsymbol{K}_{\hat{\alpha}_{1}\hat{\alpha}_{2}}\cdots\delta \boldsymbol{K}_{\hat{\alpha}_{2 l'-1}\hat{\alpha}_{2 l'}}}\right|_{\vec{J}=\vec{J}_{0} \atop \boldsymbol{K}=\boldsymbol{K}_{0}} }$} \\
& \hspace{2.35cm} \scalebox{0.91}{${\displaystyle \times J_{n_{1},\alpha_{1}}\Big[\vec{\phi},\boldsymbol{G}\Big] \cdots J_{n_{l},\alpha_{l}}\Big[\vec{\phi},\boldsymbol{G}\Big] \boldsymbol{K}_{\hat{n}_{1},\hat{\alpha}_{1}\hat{\alpha}_{2}}\Big[\vec{\phi},\boldsymbol{G}\Big] \cdots \boldsymbol{K}_{\hat{n}_{l'},\hat{\alpha}_{2 l' -1}\hat{\alpha}_{2 l'}}\Big[\vec{\phi},\boldsymbol{G}\Big] }$} \\
& \scalebox{0.91}{${\displaystyle + \int_{\alpha} J_{n,\alpha}\Big[\vec{\phi},\boldsymbol{G}\Big] \phi_{\alpha} + \frac{1}{2} \int_{\alpha_{1},\alpha_{2}} \phi_{\alpha_{1}} \boldsymbol{K}_{n,\alpha_{1}\alpha_{2}}\Big[\vec{\phi},\boldsymbol{G}\Big]\phi_{\alpha_{2}} + \frac{1}{2} \int_{\alpha_{1},\alpha_{2}} \boldsymbol{K}_{n-1,\alpha_{1}\alpha_{2}}\Big[\vec{\phi},\boldsymbol{G}\Big] \boldsymbol{G}_{\alpha_{1}\alpha_{2}} \delta_{n \geq 1} \;. }$}
\end{split}
\label{eq:pure2PIEAIMstep5bis0DON}
\end{equation}

\subsection{\label{sec:GammanCoeffIM4PPIEA}4PPI effective action}

Finally, we determine an expression for the $\Gamma_{n}^{(\mathrm{4PPI})}$ coefficients introduced in the power series~\eqref{eq:pure4PPIEAGammaExpansion0DON} for the original 4PPI EA with vanishing 1-point correlation function. The derivations start from~\eqref{eq:pure4PPIEAIMstep10DON} recalled below:
\begin{equation}
\begin{split}
\sum_{n=0}^{\infty} \Gamma^{(\mathrm{4PPI})}_{n}[\rho,\zeta]\hbar^{n} = & - \sum_{n=0}^{\infty} W_{n}\Bigg[\sum_{m=0}^{\infty} K_{m}[\rho,\zeta]\hbar^{m},\sum_{m=0}^{\infty} M_{m}[\rho,\zeta]\hbar^{m}\Bigg]\hbar^{n} + \frac{1}{2} \sum_{n=0}^{\infty} \int_{\alpha} K_{n,\alpha}[\rho,\zeta] \rho_{\alpha} \hbar^{n+1} \\
& + \frac{1}{8} \sum_{n=0}^{\infty} \int_{\alpha} M_{n,\alpha}[\rho,\zeta] \rho_{\alpha}^{2} \hbar^{n+2} + \frac{1}{24} \sum_{n=0}^{\infty} \int_{\alpha} M_{n,\alpha}[\rho,\zeta] \zeta_{\alpha} \hbar^{n+3} \;.
\end{split}
\label{eq:pure4PPIEAIMstep1bis0DON}
\end{equation}
Once again, the general calculation steps are essentially the same as those of sections~\ref{sec:GammanCoeffIM1PIEA} and~\ref{sec:GammanCoeffIM2PIEA} treating the 1PI and 2PI EAs, as they would be for any other $n$PI or $n$PPI EAs. We thus Taylor expand the $W_{n}$ coefficients in~\eqref{eq:pure4PPIEAIMstep1bis0DON} around $(K,M)=(K_{0},M_{0})$:
\begin{equation}
\begin{split}
& W_{n}\Bigg[\sum_{m=0}^{\infty} K_{m}[\rho,\zeta]\hbar^{m},\sum_{m=0}^{\infty} M_{m}[\rho,\zeta]\hbar^{m}\Bigg] \\
& \hspace{0.8cm} = W_{n}\Bigg[K_{0}[\rho,\zeta] + \sum_{m=1}^{\infty} K_{m}[\rho,\zeta]\hbar^{m},M_{0}[\rho,\zeta] + \sum_{m=1}^{\infty} M_{m}[\rho,\zeta]\hbar^{m}\Bigg] \\
& \hspace{0.8cm} = W_{n}[K = K_{0},M = M_{0}] + \int_{\alpha} \left.\frac{\delta W_{n}[K,M]}{\delta K_{\alpha}}\right|_{K = K_{0} \atop M = M_{0}} \left(\sum_{m=1}^{\infty} K_{m,\alpha}[\rho,\zeta]\hbar^{m}\right) \\
& \hspace{0.8cm} \hspace{0.4cm} + \int_{\alpha} \left.\frac{\delta W_{n}[K,M]}{\delta M_{\alpha}}\right|_{K = K_{0} \atop M = M_{0}} \left(\sum_{m=1}^{\infty} M_{m,\alpha}[\rho,\zeta]\hbar^{m}\right) \\
& \hspace{0.8cm} \hspace{0.4cm} + \sum_{m=2}^{\infty} \frac{1}{m!} \sum_{\underset{\lbrace l+l'=m \rbrace}{l,l'=1}}^{m} \begin{pmatrix}
m \\
l
\end{pmatrix} \\
& \hspace{0.8cm} \hspace{0.4cm} \hspace{0.8cm} \times\int_{\alpha_{1},\cdots,\alpha_{l} \atop \hat{\alpha}_{1},\cdots,\hat{\alpha}_{l'}} \left.\frac{\delta^{m} W_{n}[K,M]}{\delta K_{\alpha_{1}}\cdots\delta K_{\alpha_{l}} \delta M_{\hat{\alpha}_{1}}\cdots\delta M_{\hat{\alpha}_{l'}}}\right|_{K=K_{0} \atop M=M_{0}} \\
& \hspace{0.8cm} \hspace{0.4cm} \hspace{2.5cm} \times \left(\sum_{n_{1}=1}^{\infty} K_{n_{1},\alpha_{1}}[\rho,\zeta]\hbar^{n_{1}}\right)\cdots\left(\sum_{n_{l}=1}^{\infty} K_{n_{l},\alpha_{l}}[\rho,\zeta]\hbar^{n_{l}}\right) \\
& \hspace{0.8cm} \hspace{0.4cm} \hspace{2.5cm} \times \left(\sum_{\hat{n}_{1}=1}^{\infty} M_{\hat{n}_{1},\hat{\alpha}_{1}}[\rho,\zeta]\hbar^{\hat{n}_{1}}\right) \cdots \left(\sum_{\hat{n}_{l'}=1}^{\infty} M_{\hat{n}_{l'},\hat{\alpha}_{l'}}[\rho,\zeta]\hbar^{\hat{n}_{l'}}\right) \\
& \hspace{0.8cm} = W_{n}[K=K_{0},M=M_{0}] + \sum_{m=1}^{\infty} \int_{\alpha} \left.\frac{\delta W_{n}[K,M]}{\delta K_{\alpha}}\right|_{K = K_{0} \atop M = M_{0}} K_{m,\alpha}[\rho,\zeta]\hbar^{m} \\
& \hspace{0.8cm} \hspace{0.4cm} + \sum_{m=1}^{\infty} \int_{\alpha} \left.\frac{\delta W_{n}[K,M]}{\delta M_{\alpha}}\right|_{K = K_{0} \atop M = M_{0}} M_{m,\alpha}[\rho,\zeta]\hbar^{m} \\
& \hspace{0.8cm} \hspace{0.4cm} + \sum_{m=2}^{\infty} \frac{1}{m!} \sum_{\underset{\lbrace l+l'=m \rbrace}{l,l'=1}}^{m} \sum^{\infty}_{n_{1},\cdots,n_{l},\hat{n}_{1},\cdots,\hat{n}_{l'}=1} \begin{pmatrix}
m \\
l
\end{pmatrix} \\
& \hspace{0.8cm} \hspace{0.4cm} \hspace{0.8cm} \times \int_{\alpha_{1},\cdots,\alpha_{l} \atop \hat{\alpha}_{1},\cdots,\hat{\alpha}_{l'}} \left.\frac{\delta^{m} W_{n}[K,M]}{\delta K_{\alpha_{1}}\cdots\delta K_{\alpha_{l}} \delta M_{\hat{\alpha}_{1}}\cdots\delta M_{\hat{\alpha}_{l'}}}\right|_{K = K_{0} \atop M = M_{0}} \\
& \hspace{0.8cm} \hspace{0.4cm} \hspace{2.5cm} \times K_{n_{1},\alpha_{1}}[\rho,\zeta] \cdots K_{n_{l},\alpha_{l}}[\rho,\zeta] M_{\hat{n}_{1},\hat{\alpha}_{1}}[\rho,\zeta] \cdots M_{\hat{n}_{l'},\hat{\alpha}_{l'}}[\rho,\zeta] \hbar^{n_{1}+\cdots+n_{l}+\hat{n}_{1}+\cdots+\hat{n}_{l'}}\;.
\end{split}
\label{eq:4PPIEAWnTaylorExpansion0DON}
\end{equation}
As a next step, we insert~\eqref{eq:4PPIEAWnTaylorExpansion0DON} into~\eqref{eq:pure4PPIEAIMstep1bis0DON}:
\begin{equation}
\begin{split}
\scalebox{0.99}{${\displaystyle \sum_{n=0}^{\infty}\Gamma_{n}^{(\mathrm{4PPI})}[\rho,\zeta]\hbar^{n} = }$} & \scalebox{0.99}{${\displaystyle - \sum_{n=0}^{\infty} W_{n}[K=K_{0},M=M_{0}] \hbar^{n} - \underbrace{\sum_{n=0}^{\infty} \sum_{m=1}^{\infty} \int_{\alpha} \left.\frac{\delta W_{n}[K,M]}{\delta K_{\alpha}}\right|_{K=K_{0} \atop M=M_{0}} K_{m,\alpha}[\rho,\zeta]\hbar^{m+n}}_{C_{1}} }$} \\
& \scalebox{0.99}{${\displaystyle - \underbrace{\sum_{n=0}^{\infty} \sum_{m=1}^{\infty} \int_{\alpha} \left.\frac{\delta W_{n}[K,M]}{\delta M_{\alpha}}\right|_{K=K_{0} \atop M=M_{0}} M_{m,\alpha}[\rho,\zeta]\hbar^{m+n}}_{C_{2}} }$} \\
& \scalebox{0.99}{${\displaystyle \underbrace{\begin{aligned} & - \sum_{n=0}^{\infty} \sum_{m=2}^{\infty} \frac{1}{m!} \sum_{\underset{\lbrace l+l'=m \rbrace}{l,l'=1}}^{m} \sum^{\infty}_{n_{1},\cdots,n_{l},\hat{n}_{1},\cdots,\hat{n}_{l'}=1} \begin{pmatrix}
m \\
l
\end{pmatrix} \int_{\alpha_{1},\cdots,\alpha_{l} \atop \hat{\alpha}_{1},\cdots,\hat{\alpha}_{l'}} \left.\frac{\delta^{m} W_{n}[K,M]}{\delta K_{\alpha_{1}}\cdots\delta K_{\alpha_{l}} \delta M_{\hat{\alpha}_{1}}\cdots\delta M_{\hat{\alpha}_{l'}}}\right|_{K=K_{0} \atop M=M_{0}} \\ & \hspace{1.9cm} \times K_{n_{1},\alpha_{1}}[\rho,\zeta] \cdots K_{n_{l},\alpha_{l}}[\rho,\zeta] M_{\hat{n}_{1},\hat{\alpha}_{1}}[\rho,\zeta] \cdots M_{\hat{n}_{l'},\hat{\alpha}_{l'}}[\rho,\zeta] \hbar^{n_{1}+\cdots+n_{l}+\hat{n}_{1}+\cdots+\hat{n}_{l'}+n} \end{aligned}}_{C_{3}} }$} \\
& \scalebox{0.99}{${\displaystyle + \frac{1}{2} \sum_{n=0}^{\infty} \int_{\alpha} K_{n,\alpha}[\rho,\zeta] \rho_{\alpha} \hbar^{n+1} + \frac{1}{8} \sum_{n=0}^{\infty} \int_{\alpha} M_{n,\alpha}[\rho,\zeta] \rho_{\alpha}^{2} \hbar^{n+2} + \frac{1}{24} \sum_{n=0}^{\infty} \int_{\alpha} M_{n,\alpha}[\rho,\zeta] \zeta_{\alpha} \hbar^{n+3}\;.}$}
\end{split}
\label{eq:pure4PPIEAIMstep20DON}
\end{equation}
We then rename indices in $C_{1}$, $C_{2}$ and $C_{3}$ respectively as in $B_{1}$, $B_{2}$ and $B_{3}$ in section~\ref{sec:GammanCoeffIM2PIEA}. According to~\eqref{eq:pure2PIEAchangeIndexB1B20DON} to~\eqref{eq:pure2PIEAchangesumB30DON},~\eqref{eq:pure4PPIEAIMstep20DON} is equivalent to:
\begin{equation}
\begin{split}
\scalebox{0.97}{${\displaystyle \sum_{n=0}^{\infty}\Gamma_{n}^{(\mathrm{4PPI})}[\rho,\zeta]\hbar^{n}= }$} & \scalebox{0.97}{${\displaystyle - \sum_{n=0}^{\infty} W_{n}[K=K_{0},M=M_{0}] \hbar^{n} - \sum_{n=1}^{\infty}\sum_{m=1}^{n} \int_{\alpha} \left.\frac{\delta W_{n-m}[K,M]}{\delta K_{\alpha}}\right|_{K=K_{0} \atop M=M_{0}} K_{m,\alpha}[\rho,\zeta]\hbar^{n} }$} \\
& \scalebox{0.97}{${\displaystyle - \sum_{n=1}^{\infty}\sum_{m=1}^{n} \int_{\alpha} \left.\frac{\delta W_{n-m}[K,M]}{\delta M_{\alpha}}\right|_{K=K_{0} \atop M=M_{0}} M_{m,\alpha}[\rho,\zeta]\hbar^{n} }$} \\
& \scalebox{0.97}{${\displaystyle - \sum_{n=1}^{\infty}\sum_{m=2}^{n} \frac{1}{m!} \sum_{\underset{\lbrace l+l'=m \rbrace}{l,l'=1}}^{m} \sum^{n}_{\underset{\lbrace n_{1} + \cdots + n_{l} + \hat{n}_{1} + \cdots + \hat{n}_{l'} \leq n\rbrace}{n_{1},\cdots,n_{l},\hat{n}_{1},\cdots,\hat{n}_{l'}=1}} \begin{pmatrix}
m \\
l
\end{pmatrix} }$} \\
& \scalebox{0.97}{${\displaystyle \hspace{0.8cm} \times \int_{\alpha_{1},\cdots,\alpha_{l} \atop \hat{\alpha}_{1},\cdots,\hat{\alpha}_{l'}} \left.\frac{\delta^{m} W_{n-(n_{1}+\cdots+n_{l}+\hat{n}_{1}+\cdots+\hat{n}_{l'})}[K,M]}{\delta K_{\alpha_{1}}\cdots\delta K_{\alpha_{l}} \delta M_{\hat{\alpha}_{1}}\cdots\delta M_{\hat{\alpha}_{l'}}}\right|_{K=K_{0} \atop M=M_{0}} }$} \\
& \scalebox{0.97}{${\displaystyle \hspace{2.5cm} \times K_{n_{1},\alpha_{1}}[\rho,\zeta] \cdots K_{n_{l},\alpha_{l}}[\rho,\zeta] M_{\hat{n}_{1},\hat{\alpha}_{1}}[\rho,\zeta] \cdots M_{\hat{n}_{l'},\hat{\alpha}_{l'}}[\rho,\zeta] \hbar^{n} }$} \\
& \scalebox{0.97}{${\displaystyle + \frac{1}{2} \sum_{n=0}^{\infty} \int_{\alpha} K_{n,\alpha}[\rho,\zeta] \rho_{\alpha} \hbar^{n+1} + \frac{1}{8} \sum_{n=0}^{\infty} \int_{\alpha} M_{n,\alpha}[\rho,\zeta] \rho_{\alpha}^{2} \hbar^{n+2} }$} \\
& \scalebox{0.97}{${\displaystyle + \frac{1}{24} \sum_{n=0}^{\infty} \int_{\alpha} M_{n,\alpha}[\rho,\zeta] \zeta_{\alpha} \hbar^{n+3} \;.}$}
\end{split}
\label{eq:pure4PPIEAIMstep30DON}
\end{equation}
The independence of both $\rho$ and $\zeta$ with respect to $\hbar$ enables us to identify the terms of order $\mathcal{O}\big(\hbar^{n}\big)$ in~\eqref{eq:pure4PPIEAIMstep30DON} to obtain:
\begin{equation}
\begin{split}
\Gamma_{n}^{(\mathrm{4PPI})}[\rho,\zeta] = & - W_{n}[K=K_{0},M=M_{0}] - \sum_{m=1}^{n} \int_{\alpha} \left.\frac{\delta W_{n-m}[K,M]}{\delta K_{\alpha}}\right|_{K=K_{0} \atop M=M_{0}} K_{m,\alpha}[\rho,\zeta] \\
& - \sum_{m=1}^{n} \int_{\alpha} \left.\frac{\delta W_{n-m}[K,M]}{\delta M_{\alpha}}\right|_{K=K_{0} \atop M=M_{0}} M_{m,\alpha}[\rho,\zeta] \\
& - \sum_{m=2}^{n} \frac{1}{m!} \sum_{\underset{\lbrace l+l'=m \rbrace}{l,l'=1}}^{m} \sum^{n}_{\underset{\lbrace n_{1} + \cdots + n_{l} + \hat{n}_{1} + \cdots + \hat{n}_{l'} \leq n\rbrace}{n_{1},\cdots,n_{l},\hat{n}_{1},\cdots,\hat{n}_{l'}=1}} \begin{pmatrix}
m \\
l
\end{pmatrix} \\
& \hspace{0.8cm} \times \int_{\alpha_{1},\cdots,\alpha_{l} \atop \hat{\alpha}_{1},\cdots,\hat{\alpha}_{l'}} \left.\frac{\delta^{m} W_{n-(n_{1}+\cdots+n_{l}+\hat{n}_{1}+\cdots+\hat{n}_{l'})}[K,M]}{\delta K_{\alpha_{1}}\cdots\delta K_{\alpha_{l}} \delta M_{\hat{\alpha}_{1}}\cdots\delta M_{\hat{\alpha}_{l'}}}\right|_{K=K_{0} \atop M=M_{0}} \\
& \hspace{2.5cm} \times K_{n_{1},\alpha_{1}}[\rho,\zeta] \cdots K_{n_{l},\alpha_{l}}[\rho,\zeta] M_{\hat{n}_{1},\hat{\alpha}_{1}}[\rho,\zeta] \cdots M_{\hat{n}_{l'},\hat{\alpha}_{l'}}[\rho,\zeta] \\
& + \frac{1}{2} \int_{\alpha} K_{n-1,\alpha}[\rho,\zeta] \rho_{\alpha} \delta_{n\geq 1} + \frac{1}{8} \int_{\alpha} M_{n-2,\alpha}[\rho,\zeta] \rho_{\alpha}^{2} \delta_{n\geq 2} + \frac{1}{24} \int_{\alpha} M_{n-3,\alpha}[\rho,\zeta] \zeta_{\alpha} \delta_{n\geq 3} \;.
\end{split}
\label{eq:pure4PPIEAIMstep4bis0DON}
\end{equation}

%% file: 7Appendix/Derivations1PIFRG.tex
\section{\label{sec:DerivMasterEq1PIFRG}Master equation (Wetterich equation)}

We will derive in this appendix all general equations that led to our results for the 1PI-FRG. We will start by deriving the master equation of this approach, i.e. the Wetterich equation~\cite{wet93}. To that end, let us first introduce the flow-dependent expectation value:
\begin{equation}
\big\langle \cdots \big\rangle_{J,k} = \frac{1}{Z_{k}[J]} \int \mathcal{D}\widetilde{\varphi} \ \cdots \ e^{-S[\widetilde{\varphi}]-\Delta S_{k}[\widetilde{\varphi}] +\int_{\alpha} J_{\alpha}\widetilde{\varphi}_{\alpha}}\;,
\end{equation}
and then differentiate~\eqref{eq:defZkWettFRG} with respect to $k$ at a fixed configuration of the source $J$:
\begin{equation}
\begin{split}
\left. \dot{W}_{k}[J] \right|_{J} = - \frac{1}{2} \int_{\alpha_{1},\alpha_{2}} \dot{R}_{k,\alpha_{1}\alpha_{2}} \left\langle \widetilde{\varphi}_{\alpha_{1}} \widetilde{\varphi}_{\alpha_{2}} \right\rangle_{J,k}\;.
\end{split}
\label{eq:Wdot1PIFRG}
\end{equation}
We also calculate:
\begin{equation}
\begin{split}
W^{(2)}_{k,\alpha_{1}\alpha_{2}}[J] \equiv \frac{\delta^{2} W_{k}[J]}{\delta J_{\alpha_{1}} \delta J_{\alpha_{2}}} = & \ \frac{\delta}{\delta J_{\alpha_{1}}} \left\langle \widetilde{\varphi}_{\alpha_{2}} \right\rangle_{J,k} \\
= & \left\langle \widetilde{\varphi}_{\alpha_{1}} \widetilde{\varphi}_{\alpha_{2}} \right\rangle_{J,k} - \left\langle \widetilde{\varphi}_{\alpha_{1}} \right\rangle_{J,k} \left\langle \widetilde{\varphi}_{\alpha_{2}} \right\rangle_{J,k} \\
= & \left\langle \widetilde{\varphi}_{\alpha_{1}} \widetilde{\varphi}_{\alpha_{2}} \right\rangle_{J,k} - \phi_{\alpha_{1}} \phi_{\alpha_{2}} \;,
\end{split}
\end{equation}
which gives us:
\begin{equation}
\left\langle \widetilde{\varphi}_{\alpha_{1}} \widetilde{\varphi}_{\alpha_{2}} \right\rangle_{J,k} = W^{(2)}_{k,\alpha_{1}\alpha_{2}}[J] + \phi_{\alpha_{1}} \phi_{\alpha_{2}} \;.
\label{eq:ExpValpsi21PIFRG}
\end{equation}
After combining~\eqref{eq:ExpValpsi21PIFRG} with~\eqref{eq:Wdot1PIFRG}, we obtain the relation:
\begin{equation}
\left. \dot{W}_{k}[J] \right|_{J} = - \frac{1}{2} \int_{\alpha_{1},\alpha_{2}} \dot{R}_{k,\alpha_{1}\alpha_{2}} \left( W^{(2)}_{k,\alpha_{1}\alpha_{2}}[J] + \phi_{\alpha_{1}} \phi_{\alpha_{2}} \right) \;,
\label{eq:PolchinskiEquation}
\end{equation}
which is fully equivalent to the Polchinski equation~\cite{pol84}. We will then turn this into a flow equation for $\Gamma^{(\mathrm{1PI})}_{k}[\phi]$ by differentiating~\eqref{eq:defGammakWettFRG} with respect to $k$ while keeping the source $J$ fixed:
\begin{equation}
\begin{split}
\left.\dot{\Gamma}^{(\mathrm{1PI})}_{k}[\phi]\right|_{J} = & -\left.\dot{W}_{k}[J]\right|_{J} + \int_{\alpha} J_{\alpha}\left.\dot{\phi}_{\alpha}\right|_{J} - \frac{1}{2}\int_{\alpha_{1},\alpha_{2}} \dot{R}_{k,\alpha_{1}\alpha_{2}} \phi_{\alpha_{1}} \phi_{\alpha_{2}} - \int_{\alpha_{1},\alpha_{2}} R_{k,\alpha_{1}\alpha_{2}} \left(\left.\dot{\phi}_{\alpha_{1}}\right|_{J}\right) \phi_{\alpha_{2}} \\
= & -\left.\dot{W}_{k}[J]\right|_{J} + \int_{\alpha} \frac{\delta\Gamma^{(\mathrm{1PI})}_{k}[\phi]}{\delta\phi_{\alpha}} \left.\dot{\phi}_{\alpha}\right|_{J} - \frac{1}{2}\int_{\alpha_{1},\alpha_{2}} \dot{R}_{k,\alpha_{1}\alpha_{2}} \phi_{\alpha_{1}} \phi_{\alpha_{2}} \;,
\end{split}
\label{eq:expressionGammadot1PIFRG}
\end{equation}
where the last line follows from:
\begin{equation}
\begin{split}
\frac{\delta\Gamma^{(\mathrm{1PI})}_{k}[\phi]}{\delta\phi_{\alpha_{1}}} = & -\int_{\alpha_{2}} \frac{\delta J_{\alpha_{2}}}{\delta \phi_{\alpha_{1}}} \underbrace{\frac{\delta W_{k}[J]}{\delta J_{\alpha_{2}}}}_{\phi_{\alpha_{2}}} + \int_{\alpha_{2}} \frac{\delta J_{\alpha_{2}}}{\delta \phi_{\alpha_{1}}} \phi_{\alpha_{2}} + J_{\alpha_{1}} - \int_{\alpha_{2}} R_{k,\alpha_{1}\alpha_{2}} \phi_{\alpha_{2}} \\
= & \ J_{\alpha_{1}} - \int_{\alpha_{2}} R_{k,\alpha_{1}\alpha_{2}} \phi_{\alpha_{2}} \;,
\end{split}
\label{eq:expressiondGammadphi1PIFRG}
\end{equation}
deduced from~\eqref{eq:defGammakWettFRG} as well. As a next step, we use the following chain rule:
\begin{equation}
\left.\frac{\partial}{\partial k}\right|_{J}=\left.\frac{\partial}{\partial k}\right|_{\phi}+\int_{\alpha}\left.\dot{\phi}_{\alpha}\right|_{J}\frac{\delta}{\delta\phi_{\alpha}}\;.
\label{eq:ChainRule1PIFRG}
\end{equation}
Note that such a chain rule trick will be used throughout all derivations of FRG flow equations in this thesis. Reorganizing the terms in~\eqref{eq:expressionGammadot1PIFRG}, we obtain the equality:
\begin{equation}
\left.\dot{\Gamma}^{(\mathrm{1PI})}_{k}[\phi]\right|_{J} - \int_{\alpha} \frac{\delta\Gamma^{(\mathrm{1PI})}_{k}[\phi]}{\delta\phi_{\alpha}} \left.\dot{\phi}_{\alpha}\right|_{J} = -\left.\dot{W}_{k}[J]\right|_{J} - \frac{1}{2}\int_{\alpha_{1},\alpha_{2}} \dot{R}_{k,\alpha_{1}\alpha_{2}} \phi_{\alpha_{1}} \phi_{\alpha_{2}}\;,
\end{equation}
which, according to~\eqref{eq:PolchinskiEquation} and~\eqref{eq:ChainRule1PIFRG}, is equivalent to:
\begin{equation}
\left.\dot{\Gamma}^{(\mathrm{1PI})}_{k}[\phi]\right|_{\phi} = \frac{1}{2} \int_{\alpha_{1},\alpha_{2}} \dot{R}_{k,\alpha_{1}\alpha_{2}} W^{(2)}_{k,\alpha_{1}\alpha_{2}}[J] \;.
\label{eq:expression2Gammadot1PIFRG}
\end{equation}
We then need to rewrite the derivative of $W_{k}[J]$ in the RHS of~\eqref{eq:expression2Gammadot1PIFRG} in terms of $\Gamma^{(\mathrm{1PI})}_{k}$. This can be done by differentiating~\eqref{eq:expressiondGammadphi1PIFRG} with respect to $\phi_{\alpha}$, thus leading to:
\begin{equation}
\begin{split}
\Gamma^{(\mathrm{1PI})(2)}_{k,\alpha_{1}\alpha_{2}}[\phi] \equiv \frac{\delta^{2}\Gamma^{(\mathrm{1PI})}_{k}[\phi]}{\delta\phi_{\alpha_{1}}\delta\phi_{\alpha_{2}}} = & \ \frac{\delta J_{\alpha_{2}}}{\delta \phi_{\alpha_{1}}} - R_{k,\alpha_{1}\alpha_{2}} \\
= & \left(W^{(2)}_{k}[J]\right)_{\alpha_{1}\alpha_{2}}^{-1} - R_{k,\alpha_{1}\alpha_{2}} \;,
\end{split}
\end{equation}
or, equivalently,
\begin{equation}
W^{(2)}_{k,\alpha_{1}\alpha_{2}}[J] = \left( \Gamma^{(\mathrm{1PI})(2)}_{k}[\phi] + R_{k} \right)_{\alpha_{1}\alpha_{2}}^{-1} \;.
\label{eq:InvGamma2W21PIFRG}
\end{equation}
The Wetterich equation~\cite{wet93} directly follows after inserting~\eqref{eq:InvGamma2W21PIFRG} into~\eqref{eq:expression2Gammadot1PIFRG}:
\begin{equation}
\dot{\Gamma}^{(\mathrm{1PI})}_{k}[\phi] \equiv \left.\dot{\Gamma}^{(\mathrm{1PI})}_{k}[\phi]\right|_{\phi} = \frac{1}{2} \int_{\alpha_{1},\alpha_{2}} \dot{R}_{k,\alpha_{1}\alpha_{2}} \left( \Gamma^{(\mathrm{1PI})(2)}_{k}[\phi] + R_{k} \right)_{\alpha_{1}\alpha_{2}}^{-1} \;,
\label{eq:WetterichEqGammaDotAppendix}
\end{equation}
or, in a more common form,
\begin{equation}
\dot{\Gamma}^{(\mathrm{1PI})}_{k}[\phi]=\frac{1}{2}\mathrm{STr}\left[\dot{R}_{k}\left(\Gamma^{(\mathrm{1PI})(2)}_{k}[\phi]+R_{k}\right)^{-1}\right] \;.
\end{equation}

\section{\label{sec:VertexExpansionApp1PIFRG}Vertex expansion}
\subsection{\label{sec:VertexExpansionAppPure1PIFRG0DON}Original 1PI functional renormalization group}

As a next step, we carry out a vertex expansion of the original 1PI EA of the studied zero-dimensional $O(N)$ model in order to turn the corresponding Wetterich equation into a tower of coupled differential equations. This Wetterich equation is given by~\eqref{eq:WetterichEqpure1PIFRG0DON} recalled below:
\begin{equation}
\dot{\Gamma}^{(\mathrm{1PI})}_{k}\Big(\vec{\phi}\Big) = \frac{1}{2}\mathrm{STr}\left[\dot{\boldsymbol{R}}_{k}\left(\Gamma^{(\mathrm{1PI})(2)}_{k}\Big(\vec{\phi}\Big)+\boldsymbol{R}_{k}\right)^{-1}\right] = \frac{1}{2}\sum_{a_{1},a_{2}=1}^{N} \dot{\boldsymbol{R}}_{k,a_{1}a_{2}} \boldsymbol{G}_{k,a_{2}a_{1}}\Big(\vec{\phi}\Big) \;.
\label{eq:WetterichEqpure1PIFRG0DONappendix}
\end{equation}
On the one hand, the Taylor expansion of $\Gamma^{(\mathrm{1PI})}_{k}$ performed around an extremum ($\overline{\Gamma}^{(\mathrm{1PI})(1)}_{k,a}=0$ $\forall a,k$), given by:
\begin{equation}
\Gamma^{(\mathrm{1PI})}_{k}\Big(\vec{\phi}\Big) = \overline{\Gamma}^{(\mathrm{1PI})}_{k} + \sum^{\infty}_{n=2} \frac{1}{n!} \sum_{a_{1},\cdots,a_{n}=1}^{N} \overline{\Gamma}_{k,a_{1} \cdots a_{n}}^{(\mathrm{1PI})(n)} \Big(\vec{\phi}-\vec{\overline{\phi}}_{k}\Big)_{a_{1}} \cdots \Big(\vec{\phi}-\vec{\overline{\phi}}_{k}\Big)_{a_{n}} \;,
\end{equation}
is used to rewrite the LHS of~\eqref{eq:WetterichEqpure1PIFRG0DONappendix} in the following form:
\begin{equation}
\begin{split}
\scalebox{0.98}{${\displaystyle\dot{\Gamma}^{(\mathrm{1PI})}_{k}\Big(\vec{\phi}\Big) =}$} & \ \scalebox{0.98}{${\displaystyle\dot{\overline{\Gamma}}^{(\mathrm{1PI})}_{k} - \sum_{a_{1},a_{2}=1}^{N} \dot{\overline{\phi}}_{k,a_{2}} \overline{\Gamma}_{k,a_{2} a_{1}}^{(\mathrm{1PI})(2)} \Big(\vec{\phi}-\vec{\overline{\phi}}_{k}\Big)_{a_{1}} }$} \\
& \scalebox{0.98}{${\displaystyle + \sum^{\infty}_{n=2} \frac{1}{n!} \sum_{a_{1},\cdots,a_{n}=1}^{N} \left(\dot{\overline{\Gamma}}_{k,a_{1} \cdots a_{n}}^{(\mathrm{1PI})(n)} - \sum_{a_{n+1}=1}^{N} \dot{\overline{\phi}}_{k,a_{n+1}} \overline{\Gamma}_{k,a_{n+1} a_{1} \cdots a_{n}}^{(\mathrm{1PI})(n+1)} \right) \Big(\vec{\phi}-\vec{\overline{\phi}}_{k}\Big)_{a_{1}} \cdots \Big(\vec{\phi}-\vec{\overline{\phi}}_{k}\Big)_{a_{n}} \;,}$}
\end{split}
\label{eq:WetterichEqLHSpure1PIFRG0DONappendix}
\end{equation}
which is to be associated with definitions~\eqref{eq:DefGammabarpure1PIFRG0DON},~\eqref{eq:DefGammanbarpure1PIFRG0DON} and~\eqref{eq:DefGamma1barpure1PIFRG0DON}. On the other hand, the RHS of~\eqref{eq:WetterichEqpure1PIFRG0DONappendix} is rewritten by Taylor expanding the propagator $\boldsymbol{G}_{k}\big(\vec{\phi}\big)$ around $\vec{\phi}=\vec{\overline{\phi}}_{k}$:
\begin{equation}
\boldsymbol{G}_{k,a_{1}a_{2}}\Big(\vec{\phi}\Big) = \overline{\boldsymbol{G}}_{k,a_{1}a_{2}} + \sum_{n=1}^{\infty} \frac{1}{n!} \sum_{a_{3}, \cdots, a_{n+2}=1}^{N} \left.\frac{\partial^{n} \boldsymbol{G}_{k,a_{1}a_{2}}\big(\vec{\phi}\big)}{\partial\phi_{a_{3}}\cdots\partial\phi_{a_{n+2}}}\right|_{\vec{\phi}=\vec{\overline{\phi}}_{k}} \left(\vec{\phi}-\vec{\overline{\phi}}_{k}\right)_{a_{3}} \cdots \left(\vec{\phi}-\vec{\overline{\phi}}_{k}\right)_{a_{n+2}}\;,
\label{eq:GkExpansionpure1PIFRG0DON}
\end{equation}
where $\overline{\boldsymbol{G}}_{k}\equiv\boldsymbol{G}_{k}\big(\vec{\phi}=\vec{\overline{\phi}}_{k}\big)$. With the help of~\eqref{eq:DefGkpure1PIFRG0DON}, the derivatives of $\boldsymbol{G}_{k,a_{1}a_{2}}\big(\vec{\phi}\big)$ in~\eqref{eq:GkExpansionpure1PIFRG0DON} can be evaluated to obtain:
\begin{equation}
\begin{split}
\boldsymbol{G}_{k,a_{1}a_{2}}\Big(\vec{\phi}\Big) = & \ \overline{\boldsymbol{G}}_{k,a_{1}a_{2}} - \sum_{a_{3},a_{4},a_{5}=1}^{N} \overline{\boldsymbol{G}}_{k,a_{1}a_{4}} \overline{\Gamma}^{(\mathrm{1PI})(3)}_{k,a_{3}a_{4}a_{5}} \overline{\boldsymbol{G}}_{k,a_{5}a_{2}} \left(\vec{\phi}-\vec{\overline{\phi}}_{k}\right)_{a_{3}} \\
& + \frac{1}{2} \sum_{a_{3},a_{4}=1}^{N} \Bigg[2\sum_{a_{5},a_{6},a_{7},a_{8}=1}^{N} \overline{\boldsymbol{G}}_{k,a_{1}a_{5}} \overline{\Gamma}^{(\mathrm{1PI})(3)}_{k,a_{3}a_{5}a_{6}} \overline{\boldsymbol{G}}_{k,a_{6}a_{7}} \overline{\Gamma}^{(\mathrm{1PI})(3)}_{k,a_{4}a_{7}a_{8}} \overline{\boldsymbol{G}}_{k,a_{8}a_{2}} \\
& \hspace{2.1cm} - \sum_{a_{5},a_{6}=1}^{N} \overline{\boldsymbol{G}}_{k,a_{1}a_{5}} \overline{\Gamma}^{(\mathrm{1PI})(4)}_{k,a_{3}a_{4}a_{5}a_{6}} \overline{\boldsymbol{G}}_{k,a_{6}a_{2}} \Bigg] \left(\vec{\phi}-\vec{\overline{\phi}}_{k}\right)_{a_{3}} \left(\vec{\phi}-\vec{\overline{\phi}}_{k}\right)_{a_{4}} \\
& + \mathcal{O}\bigg(\left|\vec{\phi}-\vec{\overline{\phi}}_{k}\right|^3\bigg) \;.
\end{split}
\label{eq:GkExpansionBispure1PIFRG0DON}
\end{equation}
We can already see at this stage that the expansion of $\boldsymbol{G}_{k}$ can be conveniently represented in a diagrammatic fashion. Following this direction, we can more readily push the expansion in~\eqref{eq:GkExpansionBispure1PIFRG0DON} further, thus leading to:
\begin{equation}
\begin{split}
\boldsymbol{G}_{k,a_{1}a_{2}}\Big(\vec{\phi}\Big) = & \hspace{0.5cm} \begin{gathered}
\begin{fmffile}{DiagramsFRG/pure1PIFRG_Gexpansion_Diag1}
\begin{fmfgraph*}(20,20)
\fmfleft{i0,i1,i2,i3}
\fmfright{o0,o1,o2,o3}
\fmfv{decor.shape=circle,decor.filled=empty,decor.size=1.5thick,label=$a_{1}$}{v1}
\fmfv{decor.shape=circle,decor.filled=empty,decor.size=1.5thick,label=$a_{2}$}{v2}
\fmf{phantom}{i1,v1}
\fmf{phantom}{i2,v1}
\fmf{plain,tension=0.6}{v1,v2}
\fmf{phantom}{v2,o1}
\fmf{phantom}{v2,o2}
\end{fmfgraph*}
\end{fmffile}
\end{gathered} \hspace{0.4cm} - \hspace{0.5cm} \begin{gathered}
\begin{fmffile}{DiagramsFRG/pure1PIFRG_Gexpansion_Diag2}
\begin{fmfgraph*}(20,20)
\fmftop{vtopLeft,vtopRight}
\fmfbottom{vbottomLeft,vbottomRight}
\fmfv{decor.shape=circle,decor.filled=empty,decor.size=1.5thick,label.dist=0.15cm,label=$a_{1}$}{vtopLeft}
\fmfv{decor.shape=circle,decor.filled=empty,decor.size=1.5thick,label.dist=0.15cm,label=$a_{2}$}{vtopRight}
\fmfv{decor.shape=circle,decor.filled=empty,decor.size=0.5cm,label=$3$,label.dist=0}{v1}
\fmfv{decor.shape=cross,decor.size=3.5thick}{v2}
\fmf{plain}{v1,vtopLeft}
\fmf{plain}{v1,vtopRight}
\fmf{phantom}{v1,vbottomLeft}
\fmf{phantom}{v1,vbottomRight}
\fmf{phantom}{v2,vbottomLeft}
\fmf{phantom}{v2,vbottomRight}
\fmf{dashes,tension=0}{v1,v2}
\end{fmfgraph*}
\end{fmffile}
\end{gathered} \hspace{0.5cm} \\
& + \frac{1}{2} \left(\rule{0cm}{1.5cm}\right. 2 \hspace{0.4cm} \begin{gathered}
\begin{fmffile}{DiagramsFRG/pure1PIFRG_Gexpansion_Diag3}
\begin{fmfgraph*}(30,17)
\fmfleft{iDown,iMid,iUp}
\fmfright{oDown,oMid,oUp}
\fmfv{decor.shape=circle,decor.filled=empty,decor.size=1.5thick,label.dist=0.15cm,label=$a_{1}$}{iUp}
\fmfv{decor.shape=circle,decor.filled=empty,decor.size=1.5thick,label.dist=0.15cm,label=$a_{2}$}{oUp}
\fmfv{decor.shape=circle,decor.filled=empty,decor.size=0.5cm,label=$3$,label.dist=0}{v1}
\fmfv{decor.shape=circle,decor.filled=empty,decor.size=0.5cm,label=$3$,label.dist=0}{v2}
\fmfv{decor.shape=cross,decor.size=3.5thick,decor.angle=50}{iDown}
\fmfv{decor.shape=cross,decor.size=3.5thick,decor.angle=-50}{oDown}
\fmf{plain,tension=1.0}{v1,iUp}
\fmf{dashes,tension=1.0}{v1,iDown}
\fmf{plain,tension=1.0}{v2,oUp}
\fmf{dashes,tension=1.0}{v2,oDown}
\fmf{plain,tension=1.0}{v1,v2}
\end{fmfgraph*}
\end{fmffile}
\end{gathered} \hspace{0.4cm} - \hspace{0.5cm} \begin{gathered}
\begin{fmffile}{DiagramsFRG/pure1PIFRG_Gexpansion_Diag4}
\begin{fmfgraph*}(20,20)
\fmftop{vtopLeft,vtopRight}
\fmfbottom{vbottomLeft,vbottomRight}
\fmfv{decor.shape=circle,decor.filled=empty,decor.size=1.5thick,label.dist=0.15cm,label=$a_{1}$}{vtopLeft}
\fmfv{decor.shape=circle,decor.filled=empty,decor.size=1.5thick,label.dist=0.15cm,label=$a_{2}$}{vtopRight}
\fmfv{decor.shape=circle,decor.filled=empty,decor.size=0.5cm,label=$4$,label.dist=0}{v1}
\fmfv{decor.shape=cross,decor.size=3.5thick,decor.angle=40}{vbottomLeft}
\fmfv{decor.shape=cross,decor.size=3.5thick,decor.angle=-40}{vbottomRight}
\fmf{plain}{v1,vtopLeft}
\fmf{plain}{v1,vtopRight}
\fmf{dashes}{v1,vbottomLeft}
\fmf{dashes}{v1,vbottomRight}
\end{fmfgraph*}
\end{fmffile}
\end{gathered} \hspace{0.5cm} \left.\rule{0cm}{1.5cm}\right) \\
& + \frac{1}{6} \left(\rule{0cm}{1.5cm}\right. - 6 \hspace{0.4cm} \begin{gathered}
\begin{fmffile}{DiagramsFRG/pure1PIFRG_Gexpansion_Diag5}
\begin{fmfgraph*}(40,17)
\fmfleft{iDown,iMid,iUp}
\fmfright{oDown,oMid,oUp}
\fmfbottom{vDown}
\fmfv{decor.shape=circle,decor.filled=empty,decor.size=1.5thick,label.dist=0.15cm,label=$a_{1}$}{iUp}
\fmfv{decor.shape=circle,decor.filled=empty,decor.size=1.5thick,label.dist=0.15cm,label=$a_{2}$}{oUp}
\fmfv{decor.shape=circle,decor.filled=empty,decor.size=0.5cm,label=$3$,label.dist=0}{v1}
\fmfv{decor.shape=circle,decor.filled=empty,decor.size=0.5cm,label=$3$,label.dist=0}{v2}
\fmfv{decor.shape=circle,decor.filled=empty,decor.size=0.5cm,label=$3$,label.dist=0}{v3}
\fmfv{decor.shape=cross,decor.size=3.5thick,decor.angle=55}{iDown}
\fmfv{decor.shape=cross,decor.size=3.5thick,decor.angle=-55}{oDown}
\fmfv{decor.shape=cross,decor.size=3.5thick}{vDown}
\fmf{plain,tension=1.0}{v1,iUp}
\fmf{dashes,tension=1.0}{v1,iDown}
\fmf{plain,tension=1.0}{v2,oUp}
\fmf{dashes,tension=1.0}{v2,oDown}
\fmf{plain,tension=1.0}{v1,v3}
\fmf{plain,tension=1.0}{v3,v2}
\fmf{dashes,tension=0}{v3,vDown}
\end{fmfgraph*}
\end{fmffile}
\end{gathered} \hspace{0.4cm} + 3 \hspace{0.4cm} \begin{gathered}
\begin{fmffile}{DiagramsFRG/pure1PIFRG_Gexpansion_Diag6}
\begin{fmfgraph*}(30,17)
\fmfleft{iDown,iMid,iUp}
\fmfright{oDown,oMid,oUp}
\fmfbottom{vDown}
\fmfv{decor.shape=circle,decor.filled=empty,decor.size=1.5thick,label.dist=0.15cm,label=$a_{1}$}{iUp}
\fmfv{decor.shape=circle,decor.filled=empty,decor.size=1.5thick,label.dist=0.15cm,label=$a_{2}$}{oUp}
\fmfv{decor.shape=circle,decor.filled=empty,decor.size=0.5cm,label=$4$,label.dist=0}{v1}
\fmfv{decor.shape=circle,decor.filled=empty,decor.size=0.5cm,label=$3$,label.dist=0}{v2}
\fmfv{decor.shape=cross,decor.size=3.5thick,decor.angle=50}{iDown}
\fmfv{decor.shape=cross,decor.size=3.5thick,decor.angle=-50}{oDown}
\fmfv{decor.shape=cross,decor.size=3.5thick,decor.angle=-50}{vDown}
\fmf{plain,tension=1.0}{v1,iUp}
\fmf{dashes,tension=1.0}{v1,iDown}
\fmf{plain,tension=1.0}{v2,oUp}
\fmf{dashes,tension=1.0}{v2,oDown}
\fmf{plain,tension=1.0}{v1,v2}
\fmf{dashes,tension=0}{v1,vDown}
\end{fmfgraph*}
\end{fmffile}
\end{gathered} \\
& \hspace{1.3cm} + 3 \hspace{0.4cm} \begin{gathered}
\begin{fmffile}{DiagramsFRG/pure1PIFRG_Gexpansion_Diag7}
\begin{fmfgraph*}(30,17)
\fmfleft{iDown,iMid,iUp}
\fmfright{oDown,oMid,oUp}
\fmfbottom{vDown}
\fmfv{decor.shape=circle,decor.filled=empty,decor.size=1.5thick,label.dist=0.15cm,label=$a_{1}$}{iUp}
\fmfv{decor.shape=circle,decor.filled=empty,decor.size=1.5thick,label.dist=0.15cm,label=$a_{2}$}{oUp}
\fmfv{decor.shape=circle,decor.filled=empty,decor.size=0.5cm,label=$3$,label.dist=0}{v1}
\fmfv{decor.shape=circle,decor.filled=empty,decor.size=0.5cm,label=$4$,label.dist=0}{v2}
\fmfv{decor.shape=cross,decor.size=3.5thick,decor.angle=50}{iDown}
\fmfv{decor.shape=cross,decor.size=3.5thick,decor.angle=-50}{oDown}
\fmfv{decor.shape=cross,decor.size=3.5thick,decor.angle=50}{vDown}
\fmf{plain,tension=1.0}{v1,iUp}
\fmf{dashes,tension=1.0}{v1,iDown}
\fmf{plain,tension=1.0}{v2,oUp}
\fmf{dashes,tension=1.0}{v2,oDown}
\fmf{plain,tension=1.0}{v1,v2}
\fmf{dashes,tension=0}{v2,vDown}
\end{fmfgraph*}
\end{fmffile}
\end{gathered} \hspace{0.4cm} - \hspace{0.5cm} \begin{gathered}
\begin{fmffile}{DiagramsFRG/pure1PIFRG_Gexpansion_Diag8}
\begin{fmfgraph*}(20,20)
\fmftop{vtopLeft,vtopRight}
\fmfbottom{vbottomLeft,vbottomRight}
\fmfbottom{vDown}
\fmfv{decor.shape=circle,decor.filled=empty,decor.size=1.5thick,label.dist=0.15cm,label=$a_{1}$}{vtopLeft}
\fmfv{decor.shape=circle,decor.filled=empty,decor.size=1.5thick,label.dist=0.15cm,label=$a_{2}$}{vtopRight}
\fmfv{decor.shape=circle,decor.filled=empty,decor.size=0.5cm,label=$5$,label.dist=0}{v1}
\fmfv{decor.shape=cross,decor.size=3.5thick,decor.angle=40}{vbottomLeft}
\fmfv{decor.shape=cross,decor.size=3.5thick,decor.angle=-40}{vbottomRight}
\fmfv{decor.shape=cross,decor.size=3.5thick}{vDown}
\fmf{plain}{v1,vtopLeft}
\fmf{plain}{v1,vtopRight}
\fmf{dashes}{v1,vbottomLeft}
\fmf{dashes}{v1,vbottomRight}
\fmf{dashes,tension=0}{v1,vDown}
\end{fmfgraph*}
\end{fmffile}
\end{gathered} \hspace{0.5cm} \left.\rule{0cm}{1.5cm}\right) \\
& + \frac{1}{24} \left(\rule{0cm}{1.5cm}\right. 24 \hspace{0.4cm} \begin{gathered}
\begin{fmffile}{DiagramsFRG/pure1PIFRG_Gexpansion_Diag9}
\begin{fmfgraph*}(50,17)
\fmfleft{iDown,iMid,iUp}
\fmfright{oDown,oMid,oUp}
\fmfbottom{vDown1,vDown1bis,vDown2bis,vDown2}
\fmfv{decor.shape=circle,decor.filled=empty,decor.size=1.5thick,label.dist=0.15cm,label=$a_{1}$}{iUp}
\fmfv{decor.shape=circle,decor.filled=empty,decor.size=1.5thick,label.dist=0.15cm,label=$a_{2}$}{oUp}
\fmfv{decor.shape=circle,decor.filled=empty,decor.size=0.5cm,label=$3$,label.dist=0}{v1}
\fmfv{decor.shape=circle,decor.filled=empty,decor.size=0.5cm,label=$3$,label.dist=0}{v2}
\fmfv{decor.shape=circle,decor.filled=empty,decor.size=0.5cm,label=$3$,label.dist=0}{v3}
\fmfv{decor.shape=circle,decor.filled=empty,decor.size=0.5cm,label=$3$,label.dist=0}{v4}
\fmfv{decor.shape=cross,decor.size=3.5thick,decor.angle=55}{iDown}
\fmfv{decor.shape=cross,decor.size=3.5thick,decor.angle=-55}{oDown}
\fmfv{decor.shape=cross,decor.size=3.5thick}{v3bis}
\fmfv{decor.shape=cross,decor.size=3.5thick}{v4bis}
\fmf{plain,tension=1.0}{v1,iUp}
\fmf{dashes,tension=1.0}{v1,iDown}
\fmf{plain,tension=1.0}{v2,oUp}
\fmf{dashes,tension=1.0}{v2,oDown}
\fmf{plain,tension=1.0}{v1,v3}
\fmf{plain,tension=1.0}{v3,v4}
\fmf{plain,tension=1.0}{v4,v2}
\fmf{dashes,tension=0}{v3,v3bis}
\fmf{dashes,tension=0}{v4,v4bis}
\fmf{phantom,tension=4.05}{vDown1bis,v3bis}
\fmf{phantom,tension=1.0}{vDown2bis,v3bis}
\fmf{phantom,tension=1.0}{vDown1bis,v4bis}
\fmf{phantom,tension=4.05}{vDown2bis,v4bis}
\end{fmfgraph*}
\end{fmffile}
\end{gathered} \hspace{0.4cm} - 12 \hspace{0.4cm} \begin{gathered}
\begin{fmffile}{DiagramsFRG/pure1PIFRG_Gexpansion_Diag10}
\begin{fmfgraph*}(40,17)
\fmfleft{iDown,iMid,iUp}
\fmfright{oDown,oMid,oUp}
\fmfbottom{vDown}
\fmfv{decor.shape=circle,decor.filled=empty,decor.size=1.5thick,label.dist=0.15cm,label=$a_{1}$}{iUp}
\fmfv{decor.shape=circle,decor.filled=empty,decor.size=1.5thick,label.dist=0.15cm,label=$a_{2}$}{oUp}
\fmfv{decor.shape=circle,decor.filled=empty,decor.size=0.5cm,label=$4$,label.dist=0}{v1}
\fmfv{decor.shape=circle,decor.filled=empty,decor.size=0.5cm,label=$3$,label.dist=0}{v2}
\fmfv{decor.shape=circle,decor.filled=empty,decor.size=0.5cm,label=$3$,label.dist=0}{v3}
\fmfv{decor.shape=cross,decor.size=3.5thick,decor.angle=55}{iDown}
\fmfv{decor.shape=cross,decor.size=3.5thick,decor.angle=-55}{oDown}
\fmfv{decor.shape=cross,decor.size=3.5thick}{vDown}
\fmfv{decor.shape=cross,decor.size=3.5thick,decor.angle=-55}{vNew}
\fmf{plain,tension=1.0}{v1,iUp}
\fmf{dashes,tension=1.0}{v1,iDown}
\fmf{plain,tension=1.0}{v2,oUp}
\fmf{dashes,tension=1.0}{v2,oDown}
\fmf{plain,tension=1.0}{v1,v3}
\fmf{plain,tension=1.0}{v3,v2}
\fmf{dashes,tension=0}{v3,vDown}
\fmf{phantom,tension=1.5}{vDown,vNew}
\fmf{phantom,tension=1.0}{iDown,vNew}
\fmf{dashes,tension=0}{v1,vNew}
\end{fmfgraph*}
\end{fmffile}
\end{gathered} \\
\\
& \hspace{1.3cm} - 12 \hspace{0.4cm} \begin{gathered}
\begin{fmffile}{DiagramsFRG/pure1PIFRG_Gexpansion_Diag11}
\begin{fmfgraph*}(40,17)
\fmfleft{iDown,iMid,iUp}
\fmfright{oDown,oMid,oUp}
\fmfbottom{vDown}
\fmfv{decor.shape=circle,decor.filled=empty,decor.size=1.5thick,label.dist=0.15cm,label=$a_{1}$}{iUp}
\fmfv{decor.shape=circle,decor.filled=empty,decor.size=1.5thick,label.dist=0.15cm,label=$a_{2}$}{oUp}
\fmfv{decor.shape=circle,decor.filled=empty,decor.size=0.5cm,label=$3$,label.dist=0}{v1}
\fmfv{decor.shape=circle,decor.filled=empty,decor.size=0.5cm,label=$3$,label.dist=0}{v2}
\fmfv{decor.shape=circle,decor.filled=empty,decor.size=0.5cm,label=$4$,label.dist=0}{v3}
\fmfv{decor.shape=cross,decor.size=3.5thick,decor.angle=55}{iDown}
\fmfv{decor.shape=cross,decor.size=3.5thick,decor.angle=-55}{oDown}
\fmfv{decor.shape=cross,decor.size=3.5thick,decor.angle=55}{vDownL}
\fmfv{decor.shape=cross,decor.size=3.5thick,decor.angle=-55}{vDownR}
\fmf{plain,tension=1.0}{v1,iUp}
\fmf{dashes,tension=1.0}{v1,iDown}
\fmf{plain,tension=1.0}{v2,oUp}
\fmf{dashes,tension=1.0}{v2,oDown}
\fmf{plain,tension=1.0}{v1,v3}
\fmf{plain,tension=1.0}{v3,v2}
\fmf{dashes,tension=0}{v3,vDownL}
\fmf{dashes,tension=0}{v3,vDownR}
\fmf{phantom,tension=1.0}{vDownL,iDown}
\fmf{phantom,tension=2.0}{vDownL,vDown}
\fmf{phantom,tension=1.0}{vDownR,oDown}
\fmf{phantom,tension=2.0}{vDownR,vDown}
\end{fmfgraph*}
\end{fmffile}
\end{gathered} \hspace{0.4cm} - 12 \hspace{0.4cm} \begin{gathered}
\begin{fmffile}{DiagramsFRG/pure1PIFRG_Gexpansion_Diag12}
\begin{fmfgraph*}(40,17)
\fmfleft{iDown,iMid,iUp}
\fmfright{oDown,oMid,oUp}
\fmfbottom{vDown}
\fmfv{decor.shape=circle,decor.filled=empty,decor.size=1.5thick,label.dist=0.15cm,label=$a_{1}$}{iUp}
\fmfv{decor.shape=circle,decor.filled=empty,decor.size=1.5thick,label.dist=0.15cm,label=$a_{2}$}{oUp}
\fmfv{decor.shape=circle,decor.filled=empty,decor.size=0.5cm,label=$3$,label.dist=0}{v1}
\fmfv{decor.shape=circle,decor.filled=empty,decor.size=0.5cm,label=$4$,label.dist=0}{v2}
\fmfv{decor.shape=circle,decor.filled=empty,decor.size=0.5cm,label=$3$,label.dist=0}{v3}
\fmfv{decor.shape=cross,decor.size=3.5thick,decor.angle=55}{iDown}
\fmfv{decor.shape=cross,decor.size=3.5thick,decor.angle=-55}{oDown}
\fmfv{decor.shape=cross,decor.size=3.5thick}{vDown}
\fmfv{decor.shape=cross,decor.size=3.5thick,decor.angle=55}{vNew}
\fmf{plain,tension=1.0}{v1,iUp}
\fmf{dashes,tension=1.0}{v1,iDown}
\fmf{plain,tension=1.0}{v2,oUp}
\fmf{dashes,tension=1.0}{v2,oDown}
\fmf{plain,tension=1.0}{v1,v3}
\fmf{plain,tension=1.0}{v3,v2}
\fmf{dashes,tension=0}{v3,vDown}
\fmf{phantom,tension=1.5}{vDown,vNew}
\fmf{phantom,tension=1.0}{oDown,vNew}
\fmf{dashes,tension=0}{v2,vNew}
\end{fmfgraph*}
\end{fmffile}
\end{gathered} \\
\\
\\
& \hspace{1.3cm} + 6 \hspace{0.4cm} \begin{gathered}
\begin{fmffile}{DiagramsFRG/pure1PIFRG_Gexpansion_Diag13}
\begin{fmfgraph*}(40,17)
\fmfleft{iDown,iMid,iUp}
\fmfright{oDown,oMid,oUp}
\fmfbottom{vDown}
\fmfv{decor.shape=circle,decor.filled=empty,decor.size=1.5thick,label.dist=0.15cm,label=$a_{1}$}{iUp}
\fmfv{decor.shape=circle,decor.filled=empty,decor.size=1.5thick,label.dist=0.15cm,label=$a_{2}$}{oUp}
\fmfv{decor.shape=circle,decor.filled=empty,decor.size=0.5cm,label=$4$,label.dist=0}{v1}
\fmfv{decor.shape=circle,decor.filled=empty,decor.size=0.5cm,label=$4$,label.dist=0}{v2}
\fmfv{decor.shape=cross,decor.size=3.5thick,decor.angle=50}{iDown}
\fmfv{decor.shape=cross,decor.size=3.5thick,decor.angle=-50}{oDown}
\fmfv{decor.shape=cross,decor.size=3.5thick,decor.angle=-50}{vDownL}
\fmfv{decor.shape=cross,decor.size=3.5thick,decor.angle=50}{vDownR}
\fmf{plain,tension=1.0}{v1,iUp}
\fmf{dashes,tension=1.0}{v1,iDown}
\fmf{plain,tension=1.0}{v2,oUp}
\fmf{dashes,tension=1.0}{v2,oDown}
\fmf{plain,tension=0.7}{v1,v2}
\fmf{phantom,tension=1.0}{vDownL,iDown}
\fmf{phantom,tension=4.0}{vDownL,vDown}
\fmf{phantom,tension=1.0}{vDownR,oDown}
\fmf{phantom,tension=4.0}{vDownR,vDown}
\fmf{dashes,tension=0}{v1,vDownL}
\fmf{dashes,tension=0}{v2,vDownR}
\end{fmfgraph*}
\end{fmffile}
\end{gathered} \hspace{0.4cm} + 4 \hspace{0.4cm} \begin{gathered}
\begin{fmffile}{DiagramsFRG/pure1PIFRG_Gexpansion_Diag14}
\begin{fmfgraph*}(35,17)
\fmfleft{iDown,iMid,iUp}
\fmfright{oDown,oMid,oUp}
\fmfbottom{vDown}
\fmfv{decor.shape=circle,decor.filled=empty,decor.size=1.5thick,label.dist=0.15cm,label=$a_{1}$}{iUp}
\fmfv{decor.shape=circle,decor.filled=empty,decor.size=1.5thick,label.dist=0.15cm,label=$a_{2}$}{oUp}
\fmfv{decor.shape=circle,decor.filled=empty,decor.size=0.5cm,label=$5$,label.dist=0}{v1}
\fmfv{decor.shape=circle,decor.filled=empty,decor.size=0.5cm,label=$3$,label.dist=0}{v2}
\fmfv{decor.shape=cross,decor.size=3.5thick,decor.angle=50}{iDown}
\fmfv{decor.shape=cross,decor.size=3.5thick,decor.angle=-50}{oDown}
\fmfv{decor.shape=cross,decor.size=3.5thick,decor.angle=-50}{vDown}
\fmfv{decor.shape=cross,decor.size=3.5thick}{vNew}
\fmf{plain,tension=1.0}{v1,iUp}
\fmf{dashes,tension=1.0}{v1,iDown}
\fmf{plain,tension=1.0}{v2,oUp}
\fmf{dashes,tension=1.0}{v2,oDown}
\fmf{plain,tension=1.0}{v1,v2}
\fmf{dashes,tension=0}{v1,vDown}
\fmf{phantom,tension=3.0}{vNew,iDown}
\fmf{phantom,tension=1.0}{vNew,oDown}
\fmf{dashes,tension=0}{v1,vNew}
\end{fmfgraph*}
\end{fmffile}
\end{gathered} \\
\\
& \hspace{1.3cm} + 4 \hspace{0.4cm} \begin{gathered}
\begin{fmffile}{DiagramsFRG/pure1PIFRG_Gexpansion_Diag15}
\begin{fmfgraph*}(35,17)
\fmfleft{iDown,iMid,iUp}
\fmfright{oDown,oMid,oUp}
\fmfbottom{vDown}
\fmfv{decor.shape=circle,decor.filled=empty,decor.size=1.5thick,label.dist=0.15cm,label=$a_{1}$}{iUp}
\fmfv{decor.shape=circle,decor.filled=empty,decor.size=1.5thick,label.dist=0.15cm,label=$a_{2}$}{oUp}
\fmfv{decor.shape=circle,decor.filled=empty,decor.size=0.5cm,label=$3$,label.dist=0}{v1}
\fmfv{decor.shape=circle,decor.filled=empty,decor.size=0.5cm,label=$5$,label.dist=0}{v2}
\fmfv{decor.shape=cross,decor.size=3.5thick,decor.angle=50}{iDown}
\fmfv{decor.shape=cross,decor.size=3.5thick,decor.angle=-50}{oDown}
\fmfv{decor.shape=cross,decor.size=3.5thick,decor.angle=50}{vDown}
\fmfv{decor.shape=cross,decor.size=3.5thick}{vNew}
\fmf{plain,tension=1.0}{v1,iUp}
\fmf{dashes,tension=1.0}{v1,iDown}
\fmf{plain,tension=1.0}{v2,oUp}
\fmf{dashes,tension=1.0}{v2,oDown}
\fmf{plain,tension=1.0}{v1,v2}
\fmf{dashes,tension=0}{v2,vDown}
\fmf{phantom,tension=1.0}{vNew,iDown}
\fmf{phantom,tension=3.0}{vNew,oDown}
\fmf{dashes,tension=0}{v2,vNew}
\end{fmfgraph*}
\end{fmffile}
\end{gathered} \hspace{0.4cm} - \hspace{0.5cm} \begin{gathered}
\begin{fmffile}{DiagramsFRG/pure1PIFRG_Gexpansion_Diag16}
\begin{fmfgraph*}(20,20)
\fmftop{vtopLeft,vtopRight}
\fmfbottom{vbottomLeft,vbottomRight}
\fmfbottom{vDown}
\fmfv{decor.shape=circle,decor.filled=empty,decor.size=1.5thick,label.dist=0.15cm,label=$a_{1}$}{vtopLeft}
\fmfv{decor.shape=circle,decor.filled=empty,decor.size=1.5thick,label.dist=0.15cm,label=$a_{2}$}{vtopRight}
\fmfv{decor.shape=circle,decor.filled=empty,decor.size=0.5cm,label=$6$,label.dist=0}{v1}
\fmfv{decor.shape=cross,decor.size=3.5thick,decor.angle=40}{vbottomLeft}
\fmfv{decor.shape=cross,decor.size=3.5thick,decor.angle=-40}{vbottomRight}
\fmfv{decor.shape=cross,decor.size=3.5thick,decor.angle=-15}{vDownL}
\fmfv{decor.shape=cross,decor.size=3.5thick,decor.angle=15}{vDownR}
\fmf{plain}{v1,vtopLeft}
\fmf{plain}{v1,vtopRight}
\fmf{dashes}{v1,vbottomLeft}
\fmf{dashes}{v1,vbottomRight}
\fmf{dashes,tension=0}{v1,vDownL}
\fmf{dashes,tension=0}{v1,vDownR}
\fmf{phantom,tension=3.0}{vDownL,vDown}
\fmf{phantom,tension=1.0}{vDownL,vbottomLeft}
\fmf{phantom,tension=3.0}{vDownR,vDown}
\fmf{phantom,tension=1.0}{vDownR,vbottomRight}
\end{fmfgraph*}
\end{fmffile}
\end{gathered} \hspace{0.5cm} \left.\rule{0cm}{1.5cm}\right) \\
& + \mathcal{O}\bigg(\left|\vec{\phi}-\vec{\overline{\phi}}_{k}\right|^5\bigg) \;,
\end{split}
\label{eq:DiagrammaticExpressionGk0DON1PIFRG}
\end{equation}
with the diagrammatic rules:
\begin{subequations}
\begin{align}
\begin{gathered}
\begin{fmffile}{DiagramsFRG/pure1PIFRG_Gexpansion_FeynRuleCross}
\begin{fmfgraph*}(20,20)
\fmfleft{i0,i1,i2,i3}
\fmfright{o0,o1,o2,o3}
\fmfv{label=$a$,decor.shape=cross,decor.size=3.5thick}{v2}
\fmf{phantom}{i1,v1}
\fmf{phantom}{i2,v1}
\fmf{dashes,tension=0.6}{v1,v2}
\fmf{phantom}{v2,o1}
\fmf{phantom}{v2,o2}
\end{fmfgraph*}
\end{fmffile}
\end{gathered} \quad &\rightarrow \quad \left(\vec{\phi} - \vec{\overline{\phi}}_{k}\right)_{a} \;,
\label{eq:DiagramsGkFeynRulesphik0DON1PIFRG} \\
\begin{gathered}
\begin{fmffile}{DiagramsFRG/pure1PIFRG_Gexpansion_FeynRule1PIvertex}
\begin{fmfgraph*}(8,6)
\fmfleft{iDown,i1,iUp}
\fmfright{oDown,o1,oUp}
\fmftop{vUpL,vUp,vUpR}
\fmfbottom{vDownL,vDown,vDownR}
\fmfv{label=$a_{1}$,label.angle=180,label.dist=0.06cm}{i1}
\fmfv{label=$a_{n}$,label.angle=135,label.dist=0.1cm}{iUp}
\fmfv{label=$a_{2}$,label.angle=-135,label.dist=0.06cm}{iDown}
\fmfv{label=$.$,label.dist=0cm}{o1}
\fmfv{label=$.$,label.angle=-135,label.dist=-0.03cm}{oUp}
\fmfv{label=$.$,label.angle=135,label.dist=-0.03cm}{oDown}
\fmfv{label=$.$,label.angle=-90,label.dist=0.07cm}{vDown}
\fmfv{label=$.$,label.angle=90,label.dist=0.07cm}{vUp}
\fmfv{decor.shape=circle,decor.filled=empty,decor.size=0.5cm,label=$n$,label.dist=0}{v1}
\fmf{plain,tension=0.5}{i1,v1}
\fmf{plain,tension=0.5}{iDown,v1}
\fmf{plain,tension=0.5}{iUp,v1}
\fmf{phantom,tension=0.5}{vDown,v1}
\fmf{phantom,tension=0.5}{v1,o1}
\fmf{phantom,tension=0.5}{v1,oUp}
\fmf{phantom,tension=0.5}{v1,oDown}
\fmf{phantom,tension=0.5}{v1,vUp}
\end{fmfgraph*}
\end{fmffile}
\end{gathered} \quad &\rightarrow \quad \overline{\Gamma}_{k,a_{1} \cdots a_{n}}^{(\mathrm{1PI})(n)} \;,
\label{eq:DiagramsGkFeynRulesGamman0DON1PIFRG} \\
\begin{gathered}
\begin{fmffile}{DiagramsFRG/pure1PIFRG_Gexpansion_FeynRuleG}
\begin{fmfgraph*}(20,20)
\fmfleft{i0,i1,i2,i3}
\fmfright{o0,o1,o2,o3}
\fmfv{label=$a_{1}$}{v1}
\fmfv{label=$a_{2}$}{v2}
\fmf{phantom}{i1,v1}
\fmf{phantom}{i2,v1}
\fmf{plain,tension=0.6}{v1,v2}
\fmf{phantom}{v2,o1}
\fmf{phantom}{v2,o2}
\end{fmfgraph*}
\end{fmffile}
\end{gathered} \quad &\rightarrow \quad \overline{\boldsymbol{G}}_{k,a_{1}a_{2}} \;,
\label{eq:DiagramsGkFeynRulesGk0DON1PIFRG}
\end{align}
\end{subequations}
and the empty dots still indicate external points. We then specify to the unbroken-symmetry regime, in which $\vec{\overline{\phi}}_{k}$ and all 1PI vertices $\overline{\Gamma}_{k}^{(\mathrm{1PI})(n)}$ vanish for $n$ odd and for all $k$ to ensure invariance under $O(N)$ rotations. In this situation,~\eqref{eq:DiagrammaticExpressionGk0DON1PIFRG} reduces to:\\

\vspace{-1.3cm}

\begin{equation}
\begin{split}
\boldsymbol{G}_{k,a_{1}a_{2}}\Big(\vec{\phi}\Big) = & \hspace{0.5cm} \begin{gathered}
\begin{fmffile}{DiagramsFRG/pure1PIFRG_Gexpansion_Diag1}
\begin{fmfgraph*}(20,20)
\fmfleft{i0,i1,i2,i3}
\fmfright{o0,o1,o2,o3}
\fmfv{decor.shape=circle,decor.filled=empty,decor.size=1.5thick,label=$a_{1}$}{v1}
\fmfv{decor.shape=circle,decor.filled=empty,decor.size=1.5thick,label=$a_{2}$}{v2}
\fmf{phantom}{i1,v1}
\fmf{phantom}{i2,v1}
\fmf{plain,tension=0.6}{v1,v2}
\fmf{phantom}{v2,o1}
\fmf{phantom}{v2,o2}
\end{fmfgraph*}
\end{fmffile}
\end{gathered} \\
& - \frac{1}{2} \hspace{0.7cm} \begin{gathered}
\begin{fmffile}{DiagramsFRG/pure1PIFRG_Gexpansion_Diag4}
\begin{fmfgraph*}(20,20)
\fmftop{vtopLeft,vtopRight}
\fmfbottom{vbottomLeft,vbottomRight}
\fmfv{decor.shape=circle,decor.filled=empty,decor.size=1.5thick,label.dist=0.15cm,label=$a_{1}$}{vtopLeft}
\fmfv{decor.shape=circle,decor.filled=empty,decor.size=1.5thick,label.dist=0.15cm,label=$a_{2}$}{vtopRight}
\fmfv{decor.shape=circle,decor.filled=empty,decor.size=0.5cm,label=$4$,label.dist=0}{v1}
\fmfv{decor.shape=cross,decor.size=3.5thick,decor.angle=40}{vbottomLeft}
\fmfv{decor.shape=cross,decor.size=3.5thick,decor.angle=-40}{vbottomRight}
\fmf{plain}{v1,vtopLeft}
\fmf{plain}{v1,vtopRight}
\fmf{dashes}{v1,vbottomLeft}
\fmf{dashes}{v1,vbottomRight}
\end{fmfgraph*}
\end{fmffile}
\end{gathered} \\
\\
& + \frac{1}{4} \hspace{0.4cm} \begin{gathered}
\begin{fmffile}{DiagramsFRG/pure1PIFRG_Gexpansion_Diag13}
\begin{fmfgraph*}(40,17)
\fmfleft{iDown,iMid,iUp}
\fmfright{oDown,oMid,oUp}
\fmfbottom{vDown}
\fmfv{decor.shape=circle,decor.filled=empty,decor.size=1.5thick,label.dist=0.15cm,label=$a_{1}$}{iUp}
\fmfv{decor.shape=circle,decor.filled=empty,decor.size=1.5thick,label.dist=0.15cm,label=$a_{2}$}{oUp}
\fmfv{decor.shape=circle,decor.filled=empty,decor.size=0.5cm,label=$4$,label.dist=0}{v1}
\fmfv{decor.shape=circle,decor.filled=empty,decor.size=0.5cm,label=$4$,label.dist=0}{v2}
\fmfv{decor.shape=cross,decor.size=3.5thick,decor.angle=50}{iDown}
\fmfv{decor.shape=cross,decor.size=3.5thick,decor.angle=-50}{oDown}
\fmfv{decor.shape=cross,decor.size=3.5thick,decor.angle=-50}{vDownL}
\fmfv{decor.shape=cross,decor.size=3.5thick,decor.angle=50}{vDownR}
\fmf{plain,tension=1.0}{v1,iUp}
\fmf{dashes,tension=1.0}{v1,iDown}
\fmf{plain,tension=1.0}{v2,oUp}
\fmf{dashes,tension=1.0}{v2,oDown}
\fmf{plain,tension=0.7}{v1,v2}
\fmf{phantom,tension=1.0}{vDownL,iDown}
\fmf{phantom,tension=4.0}{vDownL,vDown}
\fmf{phantom,tension=1.0}{vDownR,oDown}
\fmf{phantom,tension=4.0}{vDownR,vDown}
\fmf{dashes,tension=0}{v1,vDownL}
\fmf{dashes,tension=0}{v2,vDownR}
\end{fmfgraph*}
\end{fmffile}
\end{gathered} \hspace{0.4cm} - \frac{1}{24} \hspace{0.7cm} \begin{gathered}
\begin{fmffile}{DiagramsFRG/pure1PIFRG_Gexpansion_Diag16}
\begin{fmfgraph*}(20,20)
\fmftop{vtopLeft,vtopRight}
\fmfbottom{vbottomLeft,vbottomRight}
\fmfbottom{vDown}
\fmfv{decor.shape=circle,decor.filled=empty,decor.size=1.5thick,label.dist=0.15cm,label=$a_{1}$}{vtopLeft}
\fmfv{decor.shape=circle,decor.filled=empty,decor.size=1.5thick,label.dist=0.15cm,label=$a_{2}$}{vtopRight}
\fmfv{decor.shape=circle,decor.filled=empty,decor.size=0.5cm,label=$6$,label.dist=0}{v1}
\fmfv{decor.shape=cross,decor.size=3.5thick,decor.angle=40}{vbottomLeft}
\fmfv{decor.shape=cross,decor.size=3.5thick,decor.angle=-40}{vbottomRight}
\fmfv{decor.shape=cross,decor.size=3.5thick,decor.angle=-15}{vDownL}
\fmfv{decor.shape=cross,decor.size=3.5thick,decor.angle=15}{vDownR}
\fmf{plain}{v1,vtopLeft}
\fmf{plain}{v1,vtopRight}
\fmf{dashes}{v1,vbottomLeft}
\fmf{dashes}{v1,vbottomRight}
\fmf{dashes,tension=0}{v1,vDownL}
\fmf{dashes,tension=0}{v1,vDownR}
\fmf{phantom,tension=3.0}{vDownL,vDown}
\fmf{phantom,tension=1.0}{vDownL,vbottomLeft}
\fmf{phantom,tension=3.0}{vDownR,vDown}
\fmf{phantom,tension=1.0}{vDownR,vbottomRight}
\end{fmfgraph*}
\end{fmffile}
\end{gathered} \\
\\
& - \frac{1}{8} \hspace{0.4cm} \begin{gathered}
\begin{fmffile}{DiagramsFRG/pure1PIFRG_Gexpansion_Diag17}
\begin{fmfgraph*}(40,17)
\fmfleft{iDown,iMid,iUp}
\fmfright{oDown,oMid,oUp}
\fmftop{vUp}
\fmfbottom{vDown}
\fmfv{decor.shape=circle,decor.filled=empty,decor.size=1.5thick,label.dist=0.15cm,label=$a_{1}$}{iUp}
\fmfv{decor.shape=circle,decor.filled=empty,decor.size=1.5thick,label.dist=0.15cm,label=$a_{2}$}{oUp}
\fmfv{decor.shape=circle,decor.filled=empty,decor.size=0.5cm,label=$4$,label.dist=0}{v1}
\fmfv{decor.shape=circle,decor.filled=empty,decor.size=0.5cm,label=$4$,label.dist=0}{v2}
\fmfv{decor.shape=circle,decor.filled=empty,decor.size=0.5cm,label=$4$,label.dist=0}{v3}
\fmfv{decor.shape=cross,decor.size=3.5thick,decor.angle=50}{iDown}
\fmfv{decor.shape=cross,decor.size=3.5thick,decor.angle=-50}{oDown}
\fmfv{decor.shape=cross,decor.size=3.5thick,decor.angle=-50}{vDownL}
\fmfv{decor.shape=cross,decor.size=3.5thick,decor.angle=50}{vDownR}
\fmfv{decor.shape=cross,decor.size=3.5thick}{vUp}
\fmfv{decor.shape=cross,decor.size=3.5thick}{vDown}
\fmf{plain,tension=1.0}{v1,iUp}
\fmf{dashes,tension=1.0}{v1,iDown}
\fmf{plain,tension=1.0}{v2,oUp}
\fmf{dashes,tension=1.0}{v2,oDown}
\fmf{plain,tension=1.0}{v1,v3}
\fmf{plain,tension=1.0}{v3,v2}
\fmf{phantom,tension=1.0}{vDownL,iDown}
\fmf{phantom,tension=2.0}{vDownL,vDown}
\fmf{phantom,tension=1.0}{vDownR,oDown}
\fmf{phantom,tension=2.0}{vDownR,vDown}
\fmf{dashes,tension=0}{v1,vDownL}
\fmf{dashes,tension=0}{v2,vDownR}
\fmf{dashes,tension=0}{v3,vUp}
\fmf{dashes,tension=0}{v3,vDown}
\end{fmfgraph*}
\end{fmffile}
\end{gathered} \hspace{0.4cm} + \frac{1}{48} \hspace{0.4cm} \begin{gathered}
\begin{fmffile}{DiagramsFRG/pure1PIFRG_Gexpansion_Diag18}
\begin{fmfgraph*}(30,17)
\fmfleft{iDown,iMid,iUp}
\fmfright{oDown,oMid,oUp}
\fmftop{vUp}
\fmfbottom{vDown}
\fmfv{decor.shape=circle,decor.filled=empty,decor.size=1.5thick,label.dist=0.15cm,label=$a_{1}$}{iUp}
\fmfv{decor.shape=circle,decor.filled=empty,decor.size=1.5thick,label.dist=0.15cm,label=$a_{2}$}{oUp}
\fmfv{decor.shape=circle,decor.filled=empty,decor.size=0.5cm,label=$6$,label.dist=0}{v1}
\fmfv{decor.shape=circle,decor.filled=empty,decor.size=0.5cm,label=$4$,label.dist=0}{v2}
\fmfv{decor.shape=cross,decor.size=3.5thick,decor.angle=60}{iDown}
\fmfv{decor.shape=cross,decor.size=3.5thick,decor.angle=-60}{oDown}
\fmfv{decor.shape=cross,decor.size=3.5thick,decor.angle=-60}{vDownL}
\fmfv{decor.shape=cross,decor.size=3.5thick,decor.angle=60}{vDownR}
\fmfv{decor.shape=cross,decor.size=3.5thick}{iMid}
\fmfv{decor.shape=cross,decor.size=3.5thick,decor.angle=60}{vUpL}
\fmf{plain,tension=1.0}{v1,iUp}
\fmf{dashes,tension=1.0}{v1,iDown}
\fmf{plain,tension=1.0}{v2,oUp}
\fmf{dashes,tension=1.0}{v2,oDown}
\fmf{plain,tension=0.5}{v1,v2}
\fmf{phantom,tension=1.0}{vDownL,iDown}
\fmf{phantom,tension=2.0}{vDownL,vDown}
\fmf{phantom,tension=1.0}{vDownR,oDown}
\fmf{phantom,tension=2.0}{vDownR,vDown}
\fmf{dashes,tension=0}{v1,vDownL}
\fmf{dashes,tension=0}{v2,vDownR}
\fmf{dashes,tension=0}{iMid,v1}
\fmf{phantom,tension=1.0}{vUpL,iUp}
\fmf{phantom,tension=2.0}{vUpL,vUp}
\fmf{dashes,tension=0}{v1,vUpL}
\end{fmfgraph*}
\end{fmffile}
\end{gathered} \\
\\
& + \frac{1}{48} \hspace{0.4cm} \begin{gathered}
\begin{fmffile}{DiagramsFRG/pure1PIFRG_Gexpansion_Diag20}
\begin{fmfgraph*}(30,17)
\fmfleft{iDown,iMid,iUp}
\fmfright{oDown,oMid,oUp}
\fmftop{vUp}
\fmfbottom{vDown}
\fmfv{decor.shape=circle,decor.filled=empty,decor.size=1.5thick,label.dist=0.15cm,label=$a_{1}$}{iUp}
\fmfv{decor.shape=circle,decor.filled=empty,decor.size=1.5thick,label.dist=0.15cm,label=$a_{2}$}{oUp}
\fmfv{decor.shape=circle,decor.filled=empty,decor.size=0.5cm,label=$4$,label.dist=0}{v1}
\fmfv{decor.shape=circle,decor.filled=empty,decor.size=0.5cm,label=$6$,label.dist=0}{v2}
\fmfv{decor.shape=cross,decor.size=3.5thick,decor.angle=60}{iDown}
\fmfv{decor.shape=cross,decor.size=3.5thick,decor.angle=-60}{oDown}
\fmfv{decor.shape=cross,decor.size=3.5thick,decor.angle=-60}{vDownL}
\fmfv{decor.shape=cross,decor.size=3.5thick,decor.angle=60}{vDownR}
\fmfv{decor.shape=cross,decor.size=3.5thick}{oMid}
\fmfv{decor.shape=cross,decor.size=3.5thick,decor.angle=-60}{vUpR}
\fmf{plain,tension=1.0}{v1,iUp}
\fmf{dashes,tension=1.0}{v1,iDown}
\fmf{plain,tension=1.0}{v2,oUp}
\fmf{dashes,tension=1.0}{v2,oDown}
\fmf{plain,tension=0.5}{v1,v2}
\fmf{phantom,tension=1.0}{vDownL,iDown}
\fmf{phantom,tension=2.0}{vDownL,vDown}
\fmf{phantom,tension=1.0}{vDownR,oDown}
\fmf{phantom,tension=2.0}{vDownR,vDown}
\fmf{dashes,tension=0}{v1,vDownL}
\fmf{dashes,tension=0}{v2,vDownR}
\fmf{dashes,tension=0}{oMid,v2}
\fmf{phantom,tension=1.0}{vUpR,oUp}
\fmf{phantom,tension=2.0}{vUpR,vUp}
\fmf{dashes,tension=0}{v2,vUpR}
\end{fmfgraph*}
\end{fmffile}
\end{gathered} \hspace{0.5cm} - \frac{1}{720} \hspace{0.7cm} \begin{gathered}
\begin{fmffile}{DiagramsFRG/pure1PIFRG_Gexpansion_Diag19}
\begin{fmfgraph*}(20,20)
\fmftop{vtopLeft,vtopRight}
\fmfbottom{vbottomLeft,vbottomRight}
\fmfbottom{vDown}
\fmfleft{i1}
\fmfright{o1}
\fmfv{decor.shape=circle,decor.filled=empty,decor.size=1.5thick,label.dist=0.15cm,label=$a_{1}$}{vtopLeft}
\fmfv{decor.shape=circle,decor.filled=empty,decor.size=1.5thick,label.dist=0.15cm,label=$a_{2}$}{vtopRight}
\fmfv{decor.shape=circle,decor.filled=empty,decor.size=0.5cm,label=$8$,label.dist=0}{v1}
\fmfv{decor.shape=cross,decor.size=3.5thick,decor.angle=40}{vbottomLeft}
\fmfv{decor.shape=cross,decor.size=3.5thick,decor.angle=-40}{vbottomRight}
\fmfv{decor.shape=cross,decor.size=3.5thick,decor.angle=-15}{vDownL}
\fmfv{decor.shape=cross,decor.size=3.5thick,decor.angle=15}{vDownR}
\fmfv{decor.shape=cross,decor.size=3.5thick}{o1}
\fmfv{decor.shape=cross,decor.size=3.5thick}{i1}
\fmf{plain}{v1,vtopLeft}
\fmf{plain}{v1,vtopRight}
\fmf{dashes}{v1,vbottomLeft}
\fmf{dashes}{v1,vbottomRight}
\fmf{dashes,tension=0}{v1,vDownL}
\fmf{dashes,tension=0}{v1,vDownR}
\fmf{phantom,tension=3.0}{vDownL,vDown}
\fmf{phantom,tension=1.0}{vDownL,vbottomLeft}
\fmf{phantom,tension=3.0}{vDownR,vDown}
\fmf{phantom,tension=1.0}{vDownR,vbottomRight}
\fmf{dashes,tension=0}{v1,o1}
\fmf{dashes,tension=0}{v1,i1}
\end{fmfgraph*}
\end{fmffile}
\end{gathered} \\
\\
& + \mathcal{O}\bigg(\left|\vec{\phi}\right|^8\bigg) \;.
\end{split}
\label{eq:DiagrammaticExpressionGknoSSB0DON1PIFRG}
\end{equation}
As a next step, we insert this diagrammatic expression into the RHS of~\eqref{eq:WetterichEqpure1PIFRG0DONappendix}. As a result of the presence of the supertrace operator in the latter relation, this basically amounts to joining the external points in~\eqref{eq:DiagrammaticExpressionGknoSSB0DON1PIFRG} as follows:
\begin{equation}
\begin{gathered}
\begin{fmffile}{DiagramsFRG/pure1PIFRG_ConsequenceSTrOperator_Diag1}
\begin{fmfgraph*}(20,20)
\fmfleft{i0,i1,i2,i3}
\fmfright{o0,o1,o2,o3}
\fmfv{decor.shape=circle,decor.filled=empty,decor.size=1.5thick,label=$a_{1}$}{v1}
\fmfv{decor.shape=circle,decor.filled=empty,decor.size=1.5thick,label=$a_{2}$}{v2}
\fmfv{decor.shape=circle,decor.filled=shaded,decor.size=8.0thick}{v3}
\fmf{phantom}{i1,v1}
\fmf{phantom}{i2,v1}
\fmf{plain,tension=0.6}{v1,v3}
\fmf{plain,tension=0.6}{v3,v2}
\fmf{phantom}{v2,o1}
\fmf{phantom}{v2,o2}
\end{fmfgraph*}
\end{fmffile}
\end{gathered} \hspace{0.8cm} \xrightarrow{\hspace*{0.6cm}} \hspace{-0.1cm} \begin{gathered}
\begin{fmffile}{DiagramsFRG/pure1PIFRG_ConsequenceSTrOperator_Diag2}
\begin{fmfgraph*}(20,20)
\fmfleft{i0,i1,i2,i3}
\fmfright{o0,o1,o2,o3}
\fmftop{vUp}
\fmfv{decor.shape=circle,decor.filled=shaded,decor.size=8.0thick}{v3}
\fmfv{decor.shape=cross,decor.size=6.0thick}{vUp}
\fmf{phantom}{i1,v1}
\fmf{phantom}{i2,v1}
\fmf{phantom,tension=0.6}{v1,v3}
\fmf{phantom,tension=0.6}{v3,v2}
\fmf{phantom}{v2,o1}
\fmf{phantom}{v2,o2}
\fmf{plain,left,tension=0}{v3,vUp}
\fmf{plain,right,tension=0}{v3,vUp}
\end{fmfgraph*}
\end{fmffile}
\end{gathered} \hspace{-0.2cm} \;,
\end{equation}
where the cross indicates an insertion of $\dot{\boldsymbol{R}}_{k}$ according to the rule:
\begin{equation}
\begin{gathered}
\begin{fmffile}{DiagramsFRG/pure1PIFRG_RHSexpansion_FeynRuleRkdot}
\begin{fmfgraph*}(8,4)
\fmfleft{i1}
\fmfright{o1}
\fmfv{decor.shape=cross,decor.size=6.0thick,label=$a_{1}$,label.angle=0,label.dist=16}{v1}
\fmfv{label=$a_{2}$,label.angle=180,label.dist=16}{v2}
\fmf{plain}{i1,v1}
\fmf{plain}{v1,o1}
\fmf{plain}{i1,v2}
\fmf{plain}{v2,o1}
\end{fmfgraph*}
\end{fmffile}
\end{gathered} \quad \quad \rightarrow \quad \dot{\boldsymbol{R}}_{k,a_{1}a_{2}} = \dot{R}_{k} \delta_{a_{1}a_{2}}\;,
\label{eq:DiagramsRHSFeynRulesRkdot0DON1PIFRG}
\end{equation}
and the shaded blobs encompass any combinations of 1PI vertices and field insertions~\eqref{eq:DiagramsGkFeynRulesphik0DON1PIFRG} that can be found in~\eqref{eq:DiagrammaticExpressionGknoSSB0DON1PIFRG}. Hence, the RHS of~\eqref{eq:WetterichEqpure1PIFRG0DONappendix} becomes in this way:
\begin{equation}
\begin{split}
\frac{1}{2}\sum_{a_{1},a_{2}=1}^{N} \dot{\boldsymbol{R}}_{k,a_{1}a_{2}} \boldsymbol{G}_{k,a_{2}a_{1}}\Big(\vec{\phi}\Big) = & \ \frac{1}{2} \hspace{0.2cm} \begin{gathered}
\begin{fmffile}{DiagramsFRG/pure1PIFRG_RHSexpansion_Diag1}
\begin{fmfgraph*}(10,10)
\fmftop{vUp}
\fmfbottom{vDown}
\fmfv{decor.shape=cross,decor.size=6.0thick}{vUp}
\fmf{plain,left}{vUp,vDown}
\fmf{plain,right}{vUp,vDown}
\end{fmfgraph*}
\end{fmffile}
\end{gathered} \\
& - \frac{1}{4} \hspace{0.4cm} \begin{gathered}
\begin{fmffile}{DiagramsFRG/pure1PIFRG_RHSexpansion_Diag2}
\begin{fmfgraph*}(20,20)
\fmftop{vUp}
\fmftop{vtopLeft,vtopRight}
\fmfbottom{vbottomLeft,vbottomRight}
\fmfv{decor.shape=cross,decor.size=6.0thick}{vUp}
\fmfv{decor.shape=circle,decor.filled=empty,decor.size=0.5cm,label=$4$,label.dist=0}{v1}
\fmfv{decor.shape=cross,decor.size=3.5thick,decor.angle=40}{vbottomLeft}
\fmfv{decor.shape=cross,decor.size=3.5thick,decor.angle=-40}{vbottomRight}
\fmf{phantom}{v1,vtopLeft}
\fmf{phantom}{v1,vtopRight}
\fmf{dashes}{v1,vbottomLeft}
\fmf{dashes}{v1,vbottomRight}
\fmf{plain,left,tension=0}{v1,vUp}
\fmf{plain,right,tension=0}{v1,vUp}
\end{fmfgraph*}
\end{fmffile}
\end{gathered} \\
& + \frac{1}{8} \hspace{0.1cm} \begin{gathered}
\begin{fmffile}{DiagramsFRG/pure1PIFRG_RHSexpansion_Diag3}
\begin{fmfgraph*}(35,15)
\fmfleft{iDown,iMid,iUp}
\fmfright{oDown,oMid,oUp}
\fmftop{vUp}
\fmfbottom{vDown}
\fmfv{decor.shape=circle,decor.filled=empty,decor.size=0.5cm,label=$4$,label.dist=0}{v1}
\fmfv{decor.shape=circle,decor.filled=empty,decor.size=0.5cm,label=$4$,label.dist=0}{v2}
\fmfv{decor.shape=cross,decor.size=3.5thick,decor.angle=45}{iDown}
\fmfv{decor.shape=cross,decor.size=3.5thick,decor.angle=-45}{oDown}
\fmfv{decor.shape=cross,decor.size=3.5thick,decor.angle=-45}{iUp}
\fmfv{decor.shape=cross,decor.size=3.5thick,decor.angle=45}{oUp}
\fmfv{decor.shape=cross,decor.size=6.0thick}{vCross}
\fmf{dashes,tension=1.0}{v1,iUp}
\fmf{dashes,tension=1.0}{v1,iDown}
\fmf{dashes,tension=1.0}{v2,oUp}
\fmf{dashes,tension=1.0}{v2,oDown}
\fmf{plain,right,tension=0.7}{v1,v2}
\fmf{plain,left,tension=0.7}{v1,v2}
\fmf{phantom,tension=1.0}{vDownL,iDown}
\fmf{phantom,tension=4.0}{vDownL,vDown}
\fmf{phantom,tension=1.0}{vDownR,oDown}
\fmf{phantom,tension=4.0}{vDownR,vDown}
\fmf{phantom,tension=8.0}{vCross,vUp}
\fmf{phantom,tension=1.0}{vCross,vDown}
\end{fmfgraph*}
\end{fmffile}
\end{gathered} \hspace{0.1cm} - \frac{1}{48} \hspace{0.4cm} \begin{gathered}
\begin{fmffile}{DiagramsFRG/pure1PIFRG_RHSexpansion_Diag4}
\begin{fmfgraph*}(20,20)
\fmftop{vtopLeft,vtopRight}
\fmfbottom{vbottomLeft,vbottomRight}
\fmfbottom{vDown}
\fmftop{vUp}
\fmfv{decor.shape=circle,decor.filled=empty,decor.size=0.5cm,label=$6$,label.dist=0}{v1}
\fmfv{decor.shape=cross,decor.size=3.5thick,decor.angle=40}{vbottomLeft}
\fmfv{decor.shape=cross,decor.size=3.5thick,decor.angle=-40}{vbottomRight}
\fmfv{decor.shape=cross,decor.size=3.5thick,decor.angle=-15}{vDownL}
\fmfv{decor.shape=cross,decor.size=3.5thick,decor.angle=15}{vDownR}
\fmfv{decor.shape=cross,decor.size=6.0thick}{vUp}
\fmf{phantom}{v1,vtopLeft}
\fmf{phantom}{v1,vtopRight}
\fmf{dashes}{v1,vbottomLeft}
\fmf{dashes}{v1,vbottomRight}
\fmf{dashes,tension=0}{v1,vDownL}
\fmf{dashes,tension=0}{v1,vDownR}
\fmf{phantom,tension=3.0}{vDownL,vDown}
\fmf{phantom,tension=1.0}{vDownL,vbottomLeft}
\fmf{phantom,tension=3.0}{vDownR,vDown}
\fmf{phantom,tension=1.0}{vDownR,vbottomRight}
\fmf{plain,left,tension=0}{v1,vUp}
\fmf{plain,right,tension=0}{v1,vUp}
\end{fmfgraph*}
\end{fmffile}
\end{gathered} \\
\\
& - \frac{1}{16} \begin{gathered}
\begin{fmffile}{DiagramsFRG/pure1PIFRG_RHSexpansion_Diag5}
\begin{fmfgraph*}(40,17)
\fmfleft{iDown,iMid,iUp}
\fmfright{oDown,oMid,oUp}
\fmftop{vUp}
\fmfbottom{vDown}
\fmfv{decor.shape=circle,decor.filled=empty,decor.size=1.5thick}{iUp}
\fmfv{decor.shape=circle,decor.filled=empty,decor.size=1.5thick}{oUp}
\fmfv{decor.shape=circle,decor.filled=empty,decor.size=0.5cm,label=$4$,label.dist=0}{v1}
\fmfv{decor.shape=circle,decor.filled=empty,decor.size=0.5cm,label=$4$,label.dist=0}{v2}
\fmfv{decor.shape=circle,decor.filled=empty,decor.size=0.5cm,label=$4$,label.dist=0}{v3}
\fmfv{decor.shape=cross,decor.size=3.5thick,decor.angle=58}{iDown}
\fmfv{decor.shape=cross,decor.size=3.5thick,decor.angle=-58}{oDown}
\fmfv{decor.shape=cross,decor.size=3.5thick,decor.angle=56}{vDownL}
\fmfv{decor.shape=cross,decor.size=3.5thick,decor.angle=-56}{vDownR}
\fmfv{decor.shape=cross,decor.size=3.5thick,decor.angle=-58}{iUp}
\fmfv{decor.shape=cross,decor.size=3.5thick,decor.angle=58}{oUp}
\fmfv{decor.shape=cross,decor.size=6.0thick}{vUp}
\fmf{dashes,tension=1.0}{v1,iUp}
\fmf{dashes,tension=1.0}{v1,iDown}
\fmf{dashes,tension=1.0}{v2,oUp}
\fmf{dashes,tension=1.0}{v2,oDown}
\fmf{plain,tension=1.0}{v1,v3}
\fmf{plain,tension=1.0}{v3,v2}
\fmf{phantom,tension=1.0}{vDownL,iDown}
\fmf{phantom,tension=2.0}{vDownL,vDown}
\fmf{phantom,tension=1.0}{vDownR,oDown}
\fmf{phantom,tension=2.0}{vDownR,vDown}
\fmf{dashes,tension=0}{v3,vDownL}
\fmf{dashes,tension=0}{v3,vDownR}
\fmf{phantom,tension=1.0}{v3,vUp}
\fmf{phantom,tension=1.0}{v3,vDown}
\fmf{plain,left=0.35,tension=0}{v1,vUp}
\fmf{plain,right=0.35,tension=0}{v2,vUp}
\end{fmfgraph*}
\end{fmffile}
\end{gathered} \hspace{0.1cm} + \frac{1}{48} \hspace{0.4cm} \begin{gathered}
\begin{fmffile}{DiagramsFRG/pure1PIFRG_RHSexpansion_Diag6}
\begin{fmfgraph*}(30,17)
\fmfleft{iDown,iMid,iUp}
\fmfright{oDown,oMid,oUp}
\fmftop{vUp}
\fmfbottom{vDown}
\fmfv{decor.shape=circle,decor.filled=empty,decor.size=0.5cm,label=$6$,label.dist=0}{v1}
\fmfv{decor.shape=circle,decor.filled=empty,decor.size=0.5cm,label=$4$,label.dist=0}{v2}
\fmfv{decor.shape=cross,decor.size=3.5thick,decor.angle=60}{iDown}
\fmfv{decor.shape=cross,decor.size=3.5thick,decor.angle=-60}{oDown}
\fmfv{decor.shape=cross,decor.size=3.5thick,decor.angle=-60}{vDownL}
\fmfv{decor.shape=cross,decor.size=3.5thick,decor.angle=60}{vDownR}
\fmfv{decor.shape=cross,decor.size=3.5thick}{iMid}
\fmfv{decor.shape=cross,decor.size=3.5thick,decor.angle=-60}{iUp}
\fmfv{decor.shape=cross,decor.size=6.0thick}{vUp}
\fmf{dashes,tension=1.0}{v1,iUp}
\fmf{dashes,tension=1.0}{v1,iDown}
\fmf{phantom,tension=1.0}{v2,oUp}
\fmf{dashes,tension=1.0}{v2,oDown}
\fmf{plain,tension=0.5}{v1,v2}
\fmf{phantom,tension=1.0}{vDownL,iDown}
\fmf{phantom,tension=2.0}{vDownL,vDown}
\fmf{phantom,tension=1.0}{vDownR,oDown}
\fmf{phantom,tension=2.0}{vDownR,vDown}
\fmf{dashes,tension=0}{v1,vDownL}
\fmf{dashes,tension=0}{v2,vDownR}
\fmf{dashes,tension=0}{iMid,v1}
\fmf{phantom,tension=1.0}{vUpL,iUp}
\fmf{phantom,tension=2.0}{vUpL,vUp}
\fmf{plain,left=0.45,tension=0}{v1,vUp}
\fmf{plain,right=0.45,tension=0}{v2,vUp}
\end{fmfgraph*}
\end{fmffile}
\end{gathered} \\
\\
& - \frac{1}{1440} \hspace{0.4cm} \begin{gathered}
\begin{fmffile}{DiagramsFRG/pure1PIFRG_RHSexpansion_Diag7}
\begin{fmfgraph*}(20,20)
\fmftop{vtopLeft,vtopRight}
\fmfbottom{vbottomLeft,vbottomRight}
\fmfbottom{vDown}
\fmftop{vUp}
\fmfleft{i1}
\fmfright{o1}
\fmfv{decor.shape=circle,decor.filled=empty,decor.size=0.5cm,label=$8$,label.dist=0}{v1}
\fmfv{decor.shape=cross,decor.size=3.5thick,decor.angle=40}{vbottomLeft}
\fmfv{decor.shape=cross,decor.size=3.5thick,decor.angle=-40}{vbottomRight}
\fmfv{decor.shape=cross,decor.size=3.5thick,decor.angle=-15}{vDownL}
\fmfv{decor.shape=cross,decor.size=3.5thick,decor.angle=15}{vDownR}
\fmfv{decor.shape=cross,decor.size=3.5thick}{o1}
\fmfv{decor.shape=cross,decor.size=3.5thick}{i1}
\fmfv{decor.shape=cross,decor.size=6.0thick}{vUp}
\fmf{phantom}{v1,vtopLeft}
\fmf{phantom}{v1,vtopRight}
\fmf{dashes}{v1,vbottomLeft}
\fmf{dashes}{v1,vbottomRight}
\fmf{dashes,tension=0}{v1,vDownL}
\fmf{dashes,tension=0}{v1,vDownR}
\fmf{phantom,tension=3.0}{vDownL,vDown}
\fmf{phantom,tension=1.0}{vDownL,vbottomLeft}
\fmf{phantom,tension=3.0}{vDownR,vDown}
\fmf{phantom,tension=1.0}{vDownR,vbottomRight}
\fmf{dashes,tension=0}{v1,o1}
\fmf{dashes,tension=0}{v1,i1}
\fmf{plain,left,tension=0}{v1,vUp}
\fmf{plain,right,tension=0}{v1,vUp}
\end{fmfgraph*}
\end{fmffile}
\end{gathered} \\
\\
& + \mathcal{O}\bigg(\left|\vec{\phi}\right|^8\bigg) \;.
\end{split}
\label{eq:DiagrammaticExpressionRHSnoSSB0DON1PIFRG}
\end{equation}
Moreover, in the framework of the unbroken-symmetry regime,~\eqref{eq:WetterichEqLHSpure1PIFRG0DONappendix} can be rewritten as:
\begin{equation}
\dot{\Gamma}^{(\mathrm{1PI})}_{k}\Big(\vec{\phi}\Big) = \dot{\overline{\Gamma}}^{(\mathrm{1PI})}_{k} + \sum_{\underset{\lbrace \text{n even}\rbrace}{n=2}}^{\infty}\frac{1}{n!}\sum_{a_{1},\cdots,a_{n}=1}^{N} \dot{\overline{\Gamma}}_{k,a_{1} \cdots a_{n}}^{(\mathrm{1PI})(n)} \phi_{a_{1}} \cdots \phi_{a_{n}} \;.
\label{eq:WetterichEqLHSnoSSBpure1PIFRG0DONappendix}
\end{equation}
At this stage, we have expanded the RHS and the LHS of~\eqref{eq:WetterichEqpure1PIFRG0DONappendix}, via~\eqref{eq:DiagrammaticExpressionRHSnoSSB0DON1PIFRG} and~\eqref{eq:WetterichEqLHSnoSSBpure1PIFRG0DONappendix} respectively. The next step consists in identifying the terms with identical powers of the field $\vec{\phi}$ in~\eqref{eq:DiagrammaticExpressionRHSnoSSB0DON1PIFRG} and~\eqref{eq:WetterichEqLHSnoSSBpure1PIFRG0DONappendix}, which yields:
\begin{equation}
\dot{\overline{\Gamma}}^{(\mathrm{1PI})}_{k} = \frac{1}{2} \hspace{0.2cm} \begin{gathered}
\begin{fmffile}{DiagramsFRG/pure1PIFRG_RHSexpansion_Diag1}
\begin{fmfgraph*}(10,10)
\fmftop{vUp}
\fmfbottom{vDown}
\fmfv{decor.shape=cross,decor.size=6.0thick}{vUp}
\fmf{plain,left}{vUp,vDown}
\fmf{plain,right}{vUp,vDown}
\end{fmfgraph*}
\end{fmffile}
\end{gathered} \hspace{0.2cm} \;,
\label{eq:Gamma0IdentificnoSSBpure1PIFRG0DONappendix}
\end{equation}
\begin{equation}
\sum_{a_{1},a_{2}=1}^{N} \dot{\overline{\Gamma}}_{k,a_{1} a_{2}}^{(\mathrm{1PI})(2)} \phi_{a_{1}} \phi_{a_{2}} = - \frac{1}{2} \hspace{0.4cm} \begin{gathered}
\begin{fmffile}{DiagramsFRG/pure1PIFRG_RHSexpansion_Diag2}
\begin{fmfgraph*}(20,20)
\fmftop{vUp}
\fmftop{vtopLeft,vtopRight}
\fmfbottom{vbottomLeft,vbottomRight}
\fmfv{decor.shape=cross,decor.size=6.0thick}{vUp}
\fmfv{decor.shape=circle,decor.filled=empty,decor.size=0.5cm,label=$4$,label.dist=0}{v1}
\fmfv{decor.shape=cross,decor.size=3.5thick,decor.angle=40}{vbottomLeft}
\fmfv{decor.shape=cross,decor.size=3.5thick,decor.angle=-40}{vbottomRight}
\fmf{phantom}{v1,vtopLeft}
\fmf{phantom}{v1,vtopRight}
\fmf{dashes}{v1,vbottomLeft}
\fmf{dashes}{v1,vbottomRight}
\fmf{plain,left,tension=0}{v1,vUp}
\fmf{plain,right,tension=0}{v1,vUp}
\end{fmfgraph*}
\end{fmffile}
\end{gathered} \hspace{0.2cm} \;,
\label{eq:Gamma2IdentificnoSSBpure1PIFRG0DONappendix}
\end{equation}

\vspace{0.4cm}

\begin{equation}
\sum_{a_{1},a_{2},a_{3},a_{4}=1}^{N} \dot{\overline{\Gamma}}_{k,a_{1} a_{2} a_{3} a_{4}}^{(\mathrm{1PI})(4)} \phi_{a_{1}} \phi_{a_{2}} \phi_{a_{3}} \phi_{a_{4}} = 3 \hspace{0.1cm} \begin{gathered}
\begin{fmffile}{DiagramsFRG/pure1PIFRG_RHSexpansion_Diag3}
\begin{fmfgraph*}(35,15)
\fmfleft{iDown,iMid,iUp}
\fmfright{oDown,oMid,oUp}
\fmftop{vUp}
\fmfbottom{vDown}
\fmfv{decor.shape=circle,decor.filled=empty,decor.size=0.5cm,label=$4$,label.dist=0}{v1}
\fmfv{decor.shape=circle,decor.filled=empty,decor.size=0.5cm,label=$4$,label.dist=0}{v2}
\fmfv{decor.shape=cross,decor.size=3.5thick,decor.angle=45}{iDown}
\fmfv{decor.shape=cross,decor.size=3.5thick,decor.angle=-45}{oDown}
\fmfv{decor.shape=cross,decor.size=3.5thick,decor.angle=-45}{iUp}
\fmfv{decor.shape=cross,decor.size=3.5thick,decor.angle=45}{oUp}
\fmfv{decor.shape=cross,decor.size=6.0thick}{vCross}
\fmf{dashes,tension=1.0}{v1,iUp}
\fmf{dashes,tension=1.0}{v1,iDown}
\fmf{dashes,tension=1.0}{v2,oUp}
\fmf{dashes,tension=1.0}{v2,oDown}
\fmf{plain,right,tension=0.7}{v1,v2}
\fmf{plain,left,tension=0.7}{v1,v2}
\fmf{phantom,tension=1.0}{vDownL,iDown}
\fmf{phantom,tension=4.0}{vDownL,vDown}
\fmf{phantom,tension=1.0}{vDownR,oDown}
\fmf{phantom,tension=4.0}{vDownR,vDown}
\fmf{phantom,tension=8.0}{vCross,vUp}
\fmf{phantom,tension=1.0}{vCross,vDown}
\end{fmfgraph*}
\end{fmffile}
\end{gathered} \hspace{0.1cm} - \frac{1}{2} \hspace{0.4cm} \begin{gathered}
\begin{fmffile}{DiagramsFRG/pure1PIFRG_RHSexpansion_Diag4}
\begin{fmfgraph*}(20,20)
\fmftop{vtopLeft,vtopRight}
\fmfbottom{vbottomLeft,vbottomRight}
\fmfbottom{vDown}
\fmftop{vUp}
\fmfv{decor.shape=circle,decor.filled=empty,decor.size=0.5cm,label=$6$,label.dist=0}{v1}
\fmfv{decor.shape=cross,decor.size=3.5thick,decor.angle=40}{vbottomLeft}
\fmfv{decor.shape=cross,decor.size=3.5thick,decor.angle=-40}{vbottomRight}
\fmfv{decor.shape=cross,decor.size=3.5thick,decor.angle=-15}{vDownL}
\fmfv{decor.shape=cross,decor.size=3.5thick,decor.angle=15}{vDownR}
\fmfv{decor.shape=cross,decor.size=6.0thick}{vUp}
\fmf{phantom}{v1,vtopLeft}
\fmf{phantom}{v1,vtopRight}
\fmf{dashes}{v1,vbottomLeft}
\fmf{dashes}{v1,vbottomRight}
\fmf{dashes,tension=0}{v1,vDownL}
\fmf{dashes,tension=0}{v1,vDownR}
\fmf{phantom,tension=3.0}{vDownL,vDown}
\fmf{phantom,tension=1.0}{vDownL,vbottomLeft}
\fmf{phantom,tension=3.0}{vDownR,vDown}
\fmf{phantom,tension=1.0}{vDownR,vbottomRight}
\fmf{plain,left,tension=0}{v1,vUp}
\fmf{plain,right,tension=0}{v1,vUp}
\end{fmfgraph*}
\end{fmffile}
\end{gathered} \hspace{0.2cm} \;,
\label{eq:Gamma4IdentificnoSSBpure1PIFRG0DONappendix}
\end{equation}

\vspace{0.7cm}

\begin{equation}
\begin{split}
\sum_{a_{1},a_{2},a_{3},a_{4},a_{5},a_{6}=1}^{N} \dot{\overline{\Gamma}}_{k,a_{1} a_{2} a_{3} a_{4} a_{5} a_{6}}^{(\mathrm{1PI})(6)} \phi_{a_{1}} \phi_{a_{2}} \phi_{a_{3}} \phi_{a_{4}} \phi_{a_{5}} \phi_{a_{6}} = & - 45 \begin{gathered}
\begin{fmffile}{DiagramsFRG/pure1PIFRG_RHSexpansion_Diag5}
\begin{fmfgraph*}(40,17)
\fmfleft{iDown,iMid,iUp}
\fmfright{oDown,oMid,oUp}
\fmftop{vUp}
\fmfbottom{vDown}
\fmfv{decor.shape=circle,decor.filled=empty,decor.size=1.5thick}{iUp}
\fmfv{decor.shape=circle,decor.filled=empty,decor.size=1.5thick}{oUp}
\fmfv{decor.shape=circle,decor.filled=empty,decor.size=0.5cm,label=$4$,label.dist=0}{v1}
\fmfv{decor.shape=circle,decor.filled=empty,decor.size=0.5cm,label=$4$,label.dist=0}{v2}
\fmfv{decor.shape=circle,decor.filled=empty,decor.size=0.5cm,label=$4$,label.dist=0}{v3}
\fmfv{decor.shape=cross,decor.size=3.5thick,decor.angle=58}{iDown}
\fmfv{decor.shape=cross,decor.size=3.5thick,decor.angle=-58}{oDown}
\fmfv{decor.shape=cross,decor.size=3.5thick,decor.angle=56}{vDownL}
\fmfv{decor.shape=cross,decor.size=3.5thick,decor.angle=-56}{vDownR}
\fmfv{decor.shape=cross,decor.size=3.5thick,decor.angle=-58}{iUp}
\fmfv{decor.shape=cross,decor.size=3.5thick,decor.angle=58}{oUp}
\fmfv{decor.shape=cross,decor.size=6.0thick}{vUp}
\fmf{dashes,tension=1.0}{v1,iUp}
\fmf{dashes,tension=1.0}{v1,iDown}
\fmf{dashes,tension=1.0}{v2,oUp}
\fmf{dashes,tension=1.0}{v2,oDown}
\fmf{plain,tension=1.0}{v1,v3}
\fmf{plain,tension=1.0}{v3,v2}
\fmf{phantom,tension=1.0}{vDownL,iDown}
\fmf{phantom,tension=2.0}{vDownL,vDown}
\fmf{phantom,tension=1.0}{vDownR,oDown}
\fmf{phantom,tension=2.0}{vDownR,vDown}
\fmf{dashes,tension=0}{v3,vDownL}
\fmf{dashes,tension=0}{v3,vDownR}
\fmf{phantom,tension=1.0}{v3,vUp}
\fmf{phantom,tension=1.0}{v3,vDown}
\fmf{plain,left=0.35,tension=0}{v1,vUp}
\fmf{plain,right=0.35,tension=0}{v2,vUp}
\end{fmfgraph*}
\end{fmffile}
\end{gathered} \\
\\
& + 15 \hspace{0.4cm} \begin{gathered}
\begin{fmffile}{DiagramsFRG/pure1PIFRG_RHSexpansion_Diag6}
\begin{fmfgraph*}(30,17)
\fmfleft{iDown,iMid,iUp}
\fmfright{oDown,oMid,oUp}
\fmftop{vUp}
\fmfbottom{vDown}
\fmfv{decor.shape=circle,decor.filled=empty,decor.size=0.5cm,label=$6$,label.dist=0}{v1}
\fmfv{decor.shape=circle,decor.filled=empty,decor.size=0.5cm,label=$4$,label.dist=0}{v2}
\fmfv{decor.shape=cross,decor.size=3.5thick,decor.angle=60}{iDown}
\fmfv{decor.shape=cross,decor.size=3.5thick,decor.angle=-60}{oDown}
\fmfv{decor.shape=cross,decor.size=3.5thick,decor.angle=-60}{vDownL}
\fmfv{decor.shape=cross,decor.size=3.5thick,decor.angle=60}{vDownR}
\fmfv{decor.shape=cross,decor.size=3.5thick}{iMid}
\fmfv{decor.shape=cross,decor.size=3.5thick,decor.angle=-60}{iUp}
\fmfv{decor.shape=cross,decor.size=6.0thick}{vUp}
\fmf{dashes,tension=1.0}{v1,iUp}
\fmf{dashes,tension=1.0}{v1,iDown}
\fmf{phantom,tension=1.0}{v2,oUp}
\fmf{dashes,tension=1.0}{v2,oDown}
\fmf{plain,tension=0.5}{v1,v2}
\fmf{phantom,tension=1.0}{vDownL,iDown}
\fmf{phantom,tension=2.0}{vDownL,vDown}
\fmf{phantom,tension=1.0}{vDownR,oDown}
\fmf{phantom,tension=2.0}{vDownR,vDown}
\fmf{dashes,tension=0}{v1,vDownL}
\fmf{dashes,tension=0}{v2,vDownR}
\fmf{dashes,tension=0}{iMid,v1}
\fmf{phantom,tension=1.0}{vUpL,iUp}
\fmf{phantom,tension=2.0}{vUpL,vUp}
\fmf{plain,left=0.45,tension=0}{v1,vUp}
\fmf{plain,right=0.45,tension=0}{v2,vUp}
\end{fmfgraph*}
\end{fmffile}
\end{gathered} \\
\\
& - \frac{1}{2} \hspace{0.4cm} \begin{gathered}
\begin{fmffile}{DiagramsFRG/pure1PIFRG_RHSexpansion_Diag7}
\begin{fmfgraph*}(20,20)
\fmftop{vtopLeft,vtopRight}
\fmfbottom{vbottomLeft,vbottomRight}
\fmfbottom{vDown}
\fmftop{vUp}
\fmfleft{i1}
\fmfright{o1}
\fmfv{decor.shape=circle,decor.filled=empty,decor.size=0.5cm,label=$8$,label.dist=0}{v1}
\fmfv{decor.shape=cross,decor.size=3.5thick,decor.angle=40}{vbottomLeft}
\fmfv{decor.shape=cross,decor.size=3.5thick,decor.angle=-40}{vbottomRight}
\fmfv{decor.shape=cross,decor.size=3.5thick,decor.angle=-15}{vDownL}
\fmfv{decor.shape=cross,decor.size=3.5thick,decor.angle=15}{vDownR}
\fmfv{decor.shape=cross,decor.size=3.5thick}{o1}
\fmfv{decor.shape=cross,decor.size=3.5thick}{i1}
\fmfv{decor.shape=cross,decor.size=6.0thick}{vUp}
\fmf{phantom}{v1,vtopLeft}
\fmf{phantom}{v1,vtopRight}
\fmf{dashes}{v1,vbottomLeft}
\fmf{dashes}{v1,vbottomRight}
\fmf{dashes,tension=0}{v1,vDownL}
\fmf{dashes,tension=0}{v1,vDownR}
\fmf{phantom,tension=3.0}{vDownL,vDown}
\fmf{phantom,tension=1.0}{vDownL,vbottomLeft}
\fmf{phantom,tension=3.0}{vDownR,vDown}
\fmf{phantom,tension=1.0}{vDownR,vbottomRight}
\fmf{dashes,tension=0}{v1,o1}
\fmf{dashes,tension=0}{v1,i1}
\fmf{plain,left,tension=0}{v1,vUp}
\fmf{plain,right,tension=0}{v1,vUp}
\end{fmfgraph*}
\end{fmffile}
\end{gathered} \hspace{0.4cm} \;.
\label{eq:Gamma6IdentificnoSSBpure1PIFRG0DONappendix}
\end{split}
\end{equation}
In order to further simplify~\eqref{eq:Gamma0IdentificnoSSBpure1PIFRG0DONappendix} to~\eqref{eq:Gamma6IdentificnoSSBpure1PIFRG0DONappendix}, we introduce the symmetric parts of the propagator $\overline{G}_{k}$ and of the 1PI vertices of even order:
\begin{equation}
\overline{\boldsymbol{G}}_{k,a_{1}a_{2}} = \overline{G}_{k} \ \delta_{a_{1}a_{2}} \mathrlap{\;,}
\label{eq:DefGkm2pos1PIFRG0DONappendix}
\end{equation}
\begin{equation}
\overline{\Gamma}_{k,a_{1}a_{2}}^{(\mathrm{1PI})(2)}=\overline{\Gamma}_{k}^{(\mathrm{1PI})(2)} \ \delta_{a_{1}a_{2}} \mathrlap{\quad \forall a_{1},a_{2}\;,}
\label{eq:DefGamma2m2pos1PIFRG0DONappendix}
\end{equation}
\begin{equation}
\hspace{2.9cm} \overline{\Gamma}_{k,a_{1}a_{2}a_{3}a_{4}}^{(\mathrm{1PI})(4)}=\frac{\overline{\Gamma}_{k}^{(\mathrm{1PI})(4)}}{3}\left(\delta_{a_{1}a_{2}}\delta_{a_{3}a_{4}}+\delta_{a_{1}a_{3}}\delta_{a_{2}a_{4}}+\delta_{a_{1}a_{4}}\delta_{a_{2}a_{3}}\right) \quad \forall a_{1},a_{2},a_{3},a_{4}\;,
\label{eq:DefGamma4m2pos1PIFRG0DONappendix}
\end{equation}
\begin{equation}
\begin{split}
\overline{\Gamma}_{k,a_{1}a_{2}a_{3}a_{4}a_{5}a_{6}}^{(\mathrm{1PI})(6)} = \frac{\overline{\Gamma}_{k}^{(\mathrm{1PI})(6)}}{45} & \left(\delta_{a_{1}a_{2}}\delta_{a_{3}a_{4}}\delta_{a_{5}a_{6}}+\delta_{a_{1}a_{2}}\delta_{a_{3}a_{5}}\delta_{a_{4}a_{6}}+\delta_{a_{1}a_{2}}\delta_{a_{3}a_{6}}\delta_{a_{4}a_{5}}+\delta_{a_{1}a_{3}}\delta_{a_{2}a_{4}}\delta_{a_{5}a_{6}}\right. \\
& + \delta_{a_{1}a_{3}}\delta_{a_{2}a_{5}}\delta_{a_{4}a_{6}} + \delta_{a_{1}a_{3}}\delta_{a_{2}a_{6}}\delta_{a_{4}a_{5}} + \delta_{a_{1}a_{4}}\delta_{a_{2}a_{3}}\delta_{a_{5}a_{6}} + \delta_{a_{1}a_{4}}\delta_{a_{2}a_{5}}\delta_{a_{3}a_{6}} \\
& + \delta_{a_{1}a_{4}}\delta_{a_{2}a_{6}}\delta_{a_{3}a_{5}} + \delta_{a_{1}a_{5}}\delta_{a_{2}a_{3}}\delta_{a_{4}a_{6}} + \delta_{a_{1}a_{5}}\delta_{a_{2}a_{4}}\delta_{a_{3}a_{6}} + \delta_{a_{1}a_{5}}\delta_{a_{2}a_{6}}\delta_{a_{3}a_{4}} \\
& \left. + \delta_{a_{1}a_{6}}\delta_{a_{2}a_{3}}\delta_{a_{4}a_{5}} + \delta_{a_{1}a_{6}}\delta_{a_{2}a_{4}}\delta_{a_{3}a_{5}} + \delta_{a_{1}a_{6}}\delta_{a_{2}a_{5}}\delta_{a_{3}a_{4}} \right) \quad \forall a_{1},\cdots,a_{6}\;,
\label{eq:DefGamma6m2pos1PIFRG0DONappendix}
\end{split}
\end{equation}
consistently with the absence of spontaneous breakdown of the $O(N)$ symmetry, and the symmetric part of the cutoff function:
\begin{equation}
\boldsymbol{R}_{k,a_{1}a_{2}} = R_{k} \ \delta_{a_{1}a_{2}} \;.
\label{eq:DefRkm2pos1PIFRG0DONappendix}
\end{equation}
More specifically, we aim at turning~\eqref{eq:Gamma0IdentificnoSSBpure1PIFRG0DONappendix} to~\eqref{eq:Gamma6IdentificnoSSBpure1PIFRG0DONappendix} into expressions for the $\beta$-functions $\dot{\overline{\Gamma}}_{k}^{(\mathrm{1PI})(n)}$. In the case of~\eqref{eq:Gamma0IdentificnoSSBpure1PIFRG0DONappendix}, it directly follows from~\eqref{eq:DefGkm2pos1PIFRG0DONappendix} and~\eqref{eq:DefRkm2pos1PIFRG0DONappendix} that:
\begin{equation}
\dot{\overline{\Gamma}}^{(\mathrm{1PI})}_{k} = \frac{N}{2} \overline{G}_{k} \;,
\end{equation}
since the sums over color indices yield a factor $N$ for any closed propagator loop, as discussed in appendix~\ref{sec:DiagLEO}. In order to evaluate such sums in~\eqref{eq:Gamma2IdentificnoSSBpure1PIFRG0DONappendix} to~\eqref{eq:Gamma6IdentificnoSSBpure1PIFRG0DONappendix}, we must open up the 1PI vertices according to~\eqref{eq:DefGamma4m2pos1PIFRG0DONappendix} and~\eqref{eq:DefGamma6m2pos1PIFRG0DONappendix}. For $\overline{\Gamma}_{k}^{(\mathrm{1PI})(4)}$, this amounts to writing:
\begin{equation}
\begin{gathered}
\begin{fmffile}{DiagramsFRG/pure1PIFRG_OpenGamma4_Diag1}
\begin{fmfgraph*}(15,15)
\fmftop{vtopLeft,vtopRight}
\fmfbottom{vbottomLeft,vbottomRight}
\fmfv{decor.shape=circle,decor.filled=empty,decor.size=1.5thick,label.dist=0.15cm,label=$a_{1}$}{vtopLeft}
\fmfv{decor.shape=circle,decor.filled=empty,decor.size=1.5thick,label.dist=0.15cm,label=$a_{2}$}{vbottomLeft}
\fmfv{decor.shape=circle,decor.filled=empty,decor.size=1.5thick,label.dist=0.15cm,label=$a_{3}$}{vtopRight}
\fmfv{decor.shape=circle,decor.filled=empty,decor.size=1.5thick,label.dist=0.15cm,label=$a_{4}$}{vbottomRight}
\fmfv{decor.shape=circle,decor.filled=empty,decor.size=0.5cm,label=$4$,label.dist=0}{v1}
\fmf{plain}{v1,vtopLeft}
\fmf{plain}{v1,vtopRight}
\fmf{plain}{v1,vbottomLeft}
\fmf{plain}{v1,vbottomRight}
\end{fmfgraph*}
\end{fmffile}
\end{gathered} \hspace{0.5cm} = \frac{1}{3} \left(\rule{0cm}{1.7cm}\right. \hspace{0.5cm} \begin{gathered}
\begin{fmffile}{DiagramsFRG/pure1PIFRG_OpenGamma4_Diag2}
\begin{fmfgraph*}(20,25)
\fmftop{vtopLeft,vtopRight}
\fmfbottom{vbottomLeft,vbottomRight}
\fmfv{decor.shape=circle,decor.filled=empty,decor.size=1.5thick,label.dist=0.15cm,label=$a_{1}$}{vtopLeft}
\fmfv{decor.shape=circle,decor.filled=empty,decor.size=1.5thick,label.dist=0.15cm,label=$a_{2}$}{vbottomLeft}
\fmfv{decor.shape=circle,decor.filled=empty,decor.size=1.5thick,label.dist=0.15cm,label=$a_{3}$}{vtopRight}
\fmfv{decor.shape=circle,decor.filled=empty,decor.size=1.5thick,label.dist=0.15cm,label=$a_{4}$}{vbottomRight}
\fmfv{decor.shape=circle,decor.filled=empty,decor.size=0.5cm,label=$4$,label.dist=0}{v2}
\fmf{plain}{v1,vtopLeft}
\fmf{plain}{v1,vbottomLeft}
\fmf{dots,tension=1.0}{v1,v2}
\fmf{dots,tension=1.0}{v2,v3}
\fmf{plain}{v3,vtopRight}
\fmf{plain}{v3,vbottomRight}
\end{fmfgraph*}
\end{fmffile}
\end{gathered} \hspace{0.5cm} + \hspace{0.5cm} \begin{gathered}
\begin{fmffile}{DiagramsFRG/pure1PIFRG_OpenGamma4_Diag3}
\begin{fmfgraph*}(20,25)
\fmftop{vtopLeft,vtopRight}
\fmfbottom{vbottomLeft,vbottomRight}
\fmfv{decor.shape=circle,decor.filled=empty,decor.size=1.5thick,label.dist=0.15cm,label=$a_{1}$}{vtopLeft}
\fmfv{decor.shape=circle,decor.filled=empty,decor.size=1.5thick,label.dist=0.15cm,label=$a_{2}$}{vbottomLeft}
\fmfv{decor.shape=circle,decor.filled=empty,decor.size=1.5thick,label.dist=0.15cm,label=$a_{3}$}{vtopRight}
\fmfv{decor.shape=circle,decor.filled=empty,decor.size=1.5thick,label.dist=0.15cm,label=$a_{4}$}{vbottomRight}
\fmfv{decor.shape=circle,decor.filled=empty,decor.size=0.5cm,label=$4$,label.dist=0}{v2}
\fmf{phantom}{v1,vtopLeft}
\fmf{plain,tension=0}{v1,vtopRight}
\fmf{plain}{v1,vbottomLeft}
\fmf{dots,tension=1.0}{v1,v2}
\fmf{dots,tension=1.0}{v2,v3}
\fmf{phantom}{v3,vtopRight}
\fmf{plain,tension=0}{v3,vtopLeft}
\fmf{plain}{v3,vbottomRight}
\end{fmfgraph*}
\end{fmffile}
\end{gathered} \hspace{0.5cm} + \hspace{0.5cm} \begin{gathered}
\begin{fmffile}{DiagramsFRG/pure1PIFRG_OpenGamma4_Diag4}
\begin{fmfgraph*}(20,25)
\fmftop{vtopLeft,vtopRight}
\fmfbottom{vbottomLeft,vbottomRight}
\fmfv{decor.shape=circle,decor.filled=empty,decor.size=1.5thick,label.dist=0.15cm,label=$a_{1}$}{vtopLeft}
\fmfv{decor.shape=circle,decor.filled=empty,decor.size=1.5thick,label.dist=0.15cm,label=$a_{2}$}{vbottomLeft}
\fmfv{decor.shape=circle,decor.filled=empty,decor.size=1.5thick,label.dist=0.15cm,label=$a_{3}$}{vtopRight}
\fmfv{decor.shape=circle,decor.filled=empty,decor.size=1.5thick,label.dist=0.15cm,label=$a_{4}$}{vbottomRight}
\fmfv{decor.shape=circle,decor.filled=empty,decor.size=0.5cm,label=$4$,label.dist=0}{v2}
\fmf{phantom}{v1,vtopLeft}
\fmf{plain}{v1,vbottomLeft}
\fmf{plain,tension=0}{v1,vbottomRight}
\fmf{dots,tension=1.0}{v1,v2}
\fmf{dots,tension=1.0}{v2,v3}
\fmf{plain}{v3,vtopRight}
\fmf{phantom}{v3,vbottomRight}
\fmf{plain,tension=0}{v3,vtopLeft}
\end{fmfgraph*}
\end{fmffile}
\end{gathered} \hspace{0.5cm} \left.\rule{0cm}{1.7cm}\right) \;,
\label{eq:OpenGamma4pure1PIFRG0DONappendix}
\end{equation}
with the following rule:
\begin{equation}
\begin{gathered}
\begin{fmffile}{DiagramsFRG/pure1PIFRG_OpenGamman_FeynRule}
\begin{fmfgraph*}(15,10)
\fmfleft{iDown,i1,iUp}
\fmfright{oDown,o1,oUp}
\fmftop{vUpL,vUp,vUpR}
\fmfbottom{vDownL,vDown,vDownR}
\fmfv{label=$.$,label.dist=-0.4cm}{o1}
\fmfv{label=$.$,label.angle=-135,label.dist=0.325cm}{oUp}
\fmfv{label=$.$,label.angle=135,label.dist=0.325cm}{oDown}
\fmfv{label=$.$,label.angle=-90,label.dist=-0.14cm}{vDown}
\fmfv{label=$.$,label.angle=90,label.dist=-0.14cm}{vUp}
\fmfv{decor.shape=circle,decor.filled=empty,decor.size=0.5cm,label=$n$,label.dist=0}{v1}
\fmf{dots,tension=0.5}{i1,v1}
\fmf{dots,tension=0.5}{iDown,v1}
\fmf{dots,tension=0.5}{iUp,v1}
\fmf{phantom,tension=0.5}{vDown,v1}
\fmf{phantom,tension=0.5}{v1,o1}
\fmf{phantom,tension=0.5}{v1,oUp}
\fmf{phantom,tension=0.5}{v1,oDown}
\fmf{phantom,tension=0.5}{v1,vUp}
\end{fmfgraph*}
\end{fmffile}
\end{gathered} \hspace{0.1cm} \rightarrow \quad \overline{\Gamma}^{(\mathrm{1PI})(n)}_{k} \;,
\label{eq:OpenGamma4FeynRulepure1PIFRG0DONappendix}
\end{equation}
where there are $n/2$ dotted lines that leave the circle containing the integer $n$ (which is always even here). Rule~\eqref{eq:OpenGamma4FeynRulepure1PIFRG0DONappendix} is to be distinguished from~\eqref{eq:DiagramsGkFeynRulesGamman0DON1PIFRG} according to~\eqref{eq:DefGamma4m2pos1PIFRG0DONappendix} and~\eqref{eq:DefGamma6m2pos1PIFRG0DONappendix} notably. After inserting~\eqref{eq:DefGamma2m2pos1PIFRG0DONappendix} and~\eqref{eq:OpenGamma4pure1PIFRG0DONappendix} into~\eqref{eq:Gamma2IdentificnoSSBpure1PIFRG0DONappendix}, we obtain:
\begin{equation}
\begin{split}
\dot{\overline{\Gamma}}_{k}^{(\mathrm{1PI})(2)} \delta_{a_{1} a_{2}} = & - \frac{1}{6} \left(\rule{0cm}{1.7cm}\right. \hspace{0.5cm} \begin{gathered}
\begin{fmffile}{DiagramsFRG/pure1PIFRG_OpenGamma4_Diag5}
\begin{fmfgraph*}(25,25)
\fmftop{vtopLeft,vtopRight}
\fmfbottom{vbottomLeft,vbottomRight}
\fmfright{o1}
\fmfv{decor.shape=circle,decor.filled=empty,decor.size=1.5thick,label.dist=0.15cm,label=$a_{1}$}{vtopLeft}
\fmfv{decor.shape=circle,decor.filled=empty,decor.size=1.5thick,label.dist=0.15cm,label=$a_{2}$}{vbottomLeft}
\fmfv{decor.shape=circle,decor.filled=empty,decor.size=0.5cm,label=$4$,label.dist=0}{v2}
\fmfv{decor.shape=cross,decor.size=6.0thick}{o1}
\fmf{plain}{v1,vtopLeft}
\fmf{plain}{v1,vbottomLeft}
\fmf{dots,tension=1.0}{v1,v2}
\fmf{dots,tension=1.0}{v2,v3}
\fmf{phantom,tension=0.5}{v3,vtopRight}
\fmf{phantom,tension=0.5}{v3,vbottomRight}
\fmf{plain,left,tension=0}{v3,o1}
\fmf{plain,right,tension=0}{v3,o1}
\end{fmfgraph*}
\end{fmffile}
\end{gathered} \hspace{0.5cm} + 2 \hspace{0.5cm} \begin{gathered}
\begin{fmffile}{DiagramsFRG/pure1PIFRG_OpenGamma4_Diag6}
\begin{fmfgraph*}(20,25)
\fmftop{vUp}
\fmfbottom{vDown}
\fmftop{vtopLeft,vtopRight}
\fmfbottom{vbottomLeft,vbottomRight}
\fmfv{decor.shape=circle,decor.filled=empty,decor.size=1.5thick,label.dist=0.15cm,label=$a_{1}$}{vbottomLeft}
\fmfv{decor.shape=circle,decor.filled=empty,decor.size=1.5thick,label.dist=0.15cm,label=$a_{2}$}{vbottomRight}
\fmfv{decor.shape=circle,decor.filled=empty,decor.size=0.5cm,label=$4$,label.dist=0}{v2}
\fmfv{decor.shape=cross,decor.size=6.0thick}{vUpbis}
\fmf{phantom}{v1,vtopLeft}
\fmf{plain}{v1,vbottomLeft}
\fmf{dots,tension=1.0}{v1,v2}
\fmf{dots,tension=1.0}{v2,v3}
\fmf{phantom}{v3,vtopRight}
\fmf{plain}{v3,vbottomRight}
\fmf{phantom,tension=4.5}{vUp,vUpbis}
\fmf{phantom,tension=1.0}{vDown,vUpbis}
\fmf{plain,left=0.45,tension=0}{v1,vUpbis}
\fmf{plain,right=0.45,tension=0}{v3,vUpbis}
\end{fmfgraph*}
\end{fmffile}
\end{gathered} \hspace{0.5cm} \left.\rule{0cm}{1.7cm}\right) \\
\\
= & -\frac{N+2}{6} \dot{R}_{k} \overline{G}^{2}_{k} \overline{\Gamma}^{(\mathrm{1PI})(4)}_{k} \delta_{a_{1}a_{2}} \;,
\end{split}
\label{eq:Gamma2dotnoSSBpure1PIFRG0DONappendix}
\end{equation}
which, at $a_{1}=a_{2}$, corresponds to our final expression for $\dot{\overline{\Gamma}}_{k}^{(\mathrm{1PI})(2)}$:
\begin{equation}
\dot{\overline{\Gamma}}_{k}^{(\mathrm{1PI})(2)} = -\frac{N+2}{6} \dot{R}_{k} \overline{G}^{2}_{k} \overline{\Gamma}^{(\mathrm{1PI})(4)}_{k} \;.
\label{eq:Gamma2dotbisnoSSBpure1PIFRG0DONappendix}
\end{equation}
In order to rewrite~\eqref{eq:Gamma4IdentificnoSSBpure1PIFRG0DONappendix} in the same fashion, we need to open up $\overline{\Gamma}_{k}^{(\mathrm{1PI})(6)}$ as well. According to~\eqref{eq:DefGamma6m2pos1PIFRG0DONappendix}, there are 15 combinations to consider:
\begin{equation}
\begin{gathered}
\begin{fmffile}{DiagramsFRG/pure1PIFRG_OpenGamma6_Diag1}
\begin{fmfgraph*}(15,15)
\fmftop{vtopLeft,vUp,vtopRight}
\fmfbottom{vbottomLeft,vDown,vbottomRight}
\fmfv{decor.shape=circle,decor.filled=empty,decor.size=1.5thick,label.dist=0.15cm,label=$a_{1}$}{vtopLeft}
\fmfv{decor.shape=circle,decor.filled=empty,decor.size=1.5thick,label.dist=0.15cm,label=$a_{2}$}{vUp}
\fmfv{decor.shape=circle,decor.filled=empty,decor.size=1.5thick,label.dist=0.15cm,label=$a_{3}$}{vtopRight}
\fmfv{decor.shape=circle,decor.filled=empty,decor.size=1.5thick,label.dist=0.15cm,label=$a_{4}$}{vbottomLeft}
\fmfv{decor.shape=circle,decor.filled=empty,decor.size=1.5thick,label.dist=0.15cm,label=$a_{5}$}{vDown}
\fmfv{decor.shape=circle,decor.filled=empty,decor.size=1.5thick,label.dist=0.15cm,label=$a_{6}$}{vbottomRight}
\fmfv{decor.shape=circle,decor.filled=empty,decor.size=0.5cm,label=$6$,label.dist=0}{v1}
\fmf{plain}{v1,vtopLeft}
\fmf{plain}{v1,vtopRight}
\fmf{plain}{v1,vbottomLeft}
\fmf{plain}{v1,vbottomRight}
\fmf{plain}{v1,vUp}
\fmf{plain}{v1,vDown}
\end{fmfgraph*}
\end{fmffile}
\end{gathered} \hspace{0.5cm} = \frac{1}{45} \left(\rule{0cm}{1.8cm}\right. \hspace{0.5cm} \begin{gathered}
\begin{fmffile}{DiagramsFRG/pure1PIFRG_OpenGamma6_Diag2}
\begin{fmfgraph*}(26,24)
\fmftop{vUpL0,vUpL1,vUpL2,vUpL3,vUpL4,vUp,vUpR1,vUpR2,vUpR3,vUpR4,vUpR5}
\fmfbottom{vDownL1,vDownL2,vDownL3,vDown,vDownR1,vDownR2,vDownR3}
\fmfleft{i1,i2,i3,i4,i5,i6,i7}
\fmfright{o1,o2,o3,o4,o5,o6,o7}
\fmfv{decor.shape=circle,decor.filled=empty,decor.size=1.5thick,label.dist=0.15cm,label=$a_{1}$}{i5}
\fmfv{decor.shape=circle,decor.filled=empty,decor.size=1.5thick,label.dist=0.15cm,label=$a_{2}$}{vUpL2}
\fmfv{decor.shape=circle,decor.filled=empty,decor.size=1.5thick,label.dist=0.15cm,label=$a_{3}$}{vUpR3}
\fmfv{decor.shape=circle,decor.filled=empty,decor.size=1.5thick,label.dist=0.15cm,label=$a_{4}$}{o5}
\fmfv{decor.shape=circle,decor.filled=empty,decor.size=1.5thick,label.dist=0.15cm,label=$a_{5}$}{vDownL3}
\fmfv{decor.shape=circle,decor.filled=empty,decor.size=1.5thick,label.dist=0.15cm,label=$a_{6}$}{vDownR1}
\fmfv{decor.shape=circle,decor.filled=empty,decor.size=0.5cm,label=$6$,label.dist=0}{v1}
\fmf{phantom}{v1,vUp}
\fmf{phantom}{v1,vDown}
\fmf{plain,tension=1.0}{i5,v2}
\fmf{plain,tension=1.0}{vUpL2,v2}
\fmf{dots,tension=1.0}{v1,v2}
\fmf{plain,tension=1.0}{o5,v3}
\fmf{plain,tension=1.0}{vUpR3,v3}
\fmf{dots,tension=1.0}{v1,v3}
\fmf{plain,tension=1.0}{vDownL3,v4}
\fmf{plain,tension=1.0}{vDownR1,v4}
\fmf{dots,tension=1.0}{v1,v4}
\end{fmfgraph*}
\end{fmffile}
\end{gathered} \hspace{0.5cm} + \hspace{0.5cm} \begin{gathered}
\begin{fmffile}{DiagramsFRG/pure1PIFRG_OpenGamma6_Diag3}
\begin{fmfgraph*}(26,24)
\fmftop{vUpL0,vUpL1,vUpL2,vUpL3,vUpL4,vUp,vUpR1,vUpR2,vUpR3,vUpR4,vUpR5}
\fmfbottom{vDownL1,vDownL2,vDownL3,vDown,vDownR1,vDownR2,vDownR3}
\fmfleft{i1,i2,i3,i4,i5,i6,i7}
\fmfright{o1,o2,o3,o4,o5,o6,o7}
\fmfv{decor.shape=circle,decor.filled=empty,decor.size=1.5thick,label.dist=0.15cm,label=$a_{1}$}{i5}
\fmfv{decor.shape=circle,decor.filled=empty,decor.size=1.5thick,label.dist=0.15cm,label=$a_{2}$}{vUpL2}
\fmfv{decor.shape=circle,decor.filled=empty,decor.size=1.5thick,label.dist=0.15cm,label=$a_{3}$}{vUpR3}
\fmfv{decor.shape=circle,decor.filled=empty,decor.size=1.5thick,label.dist=0.15cm,label=$a_{4}$}{o5}
\fmfv{decor.shape=circle,decor.filled=empty,decor.size=1.5thick,label.dist=0.15cm,label=$a_{5}$}{vDownL3}
\fmfv{decor.shape=circle,decor.filled=empty,decor.size=1.5thick,label.dist=0.15cm,label=$a_{6}$}{vDownR1}
\fmfv{decor.shape=circle,decor.filled=empty,decor.size=0.5cm,label=$6$,label.dist=0}{v1}
\fmf{phantom}{v1,vUp}
\fmf{phantom}{v1,vDown}
\fmf{plain,tension=1.0}{i5,v2}
\fmf{plain,tension=1.0}{vUpL2,v2}
\fmf{dots,tension=1.0}{v1,v2}
\fmf{phantom,tension=1.0}{o5,v3}
\fmf{plain,tension=1.0}{vUpR3,v3}
\fmf{dots,tension=1.0}{v1,v3}
\fmf{plain,tension=1.0}{vDownL3,v4}
\fmf{phantom,tension=1.0}{vDownR1,v4}
\fmf{dots,tension=1.0}{v1,v4}
\fmf{plain,tension=0}{o5,v4}
\fmf{plain,tension=0}{vDownR1,v3}
\end{fmfgraph*}
\end{fmffile}
\end{gathered} \hspace{0.5cm} + \text{13 other diagrams}  \left.\rule{0cm}{1.8cm}\right) \;.
\label{eq:OpenGamma6pure1PIFRG0DONappendix}
\end{equation}
Following the lines set out by~\eqref{eq:Gamma2dotnoSSBpure1PIFRG0DONappendix}, we can now exploit~\eqref{eq:OpenGamma4pure1PIFRG0DONappendix} and~\eqref{eq:OpenGamma6pure1PIFRG0DONappendix} as well as~\eqref{eq:DefGamma4m2pos1PIFRG0DONappendix} and~\eqref{eq:DefGamma6m2pos1PIFRG0DONappendix} to simplify~\eqref{eq:Gamma4IdentificnoSSBpure1PIFRG0DONappendix} as follows:
\begin{equation}
\begin{split}
\scalebox{0.975}{${\displaystyle\frac{1}{3}\dot{\overline{\Gamma}}_{k}^{(\mathrm{1PI})(4)}\left(\delta_{a_{1}a_{2}}\delta_{a_{3}a_{4}}+\delta_{a_{1}a_{3}}\delta_{a_{2}a_{4}}+\delta_{a_{1}a_{4}}\delta_{a_{2}a_{3}}\right) =}$} & \ \scalebox{0.975}{${\displaystyle\left(\frac{N+8}{9}\dot{R}_{k} \overline{G}^{3}_{k}\left(\overline{\Gamma}^{(\mathrm{1PI})(4)}_{k}\right)^2 - \frac{N+4}{30}\dot{R}_{k} \overline{G}^{2}_{k} \overline{\Gamma}^{(\mathrm{1PI})(6)}_{k}\right)}$} \\
& \scalebox{0.975}{${\displaystyle \times \left(\delta_{a_{1}a_{2}}\delta_{a_{3}a_{4}}+\delta_{a_{1}a_{3}}\delta_{a_{2}a_{4}}+\delta_{a_{1}a_{4}}\delta_{a_{2}a_{3}}\right) \;,}$}
\end{split}
\label{eq:Gamma4dotnoSSBpure1PIFRG0DONappendix}
\end{equation}
which directly yields:
\begin{equation}
\dot{\overline{\Gamma}}^{(\mathrm{1PI})(4)}_{k} = \frac{N+8}{3}\dot{R}_{k} \overline{G}^{3}_{k}\left(\overline{\Gamma}^{(\mathrm{1PI})(4)}_{k}\right)^2 - \frac{N+4}{10}\dot{R}_{k} \overline{G}^{2}_{k} \overline{\Gamma}^{(\mathrm{1PI})(6)}_{k} \;.
\end{equation}
Finally, it remains us to open up $\overline{\Gamma}^{(\mathrm{1PI})(8)}_{k}$ in order to rewrite~\eqref{eq:Gamma6IdentificnoSSBpure1PIFRG0DONappendix} via the same procedure. Gathering up all differential equations obtained in this way, we have:
\begin{equation}
\dot{\overline{\Gamma}}_{k} = \frac{N}{2} \dot{R}_{k} \left( \overline{G}_{k} - \overline{G}^{(0)}_{k} \right) \;,
\label{eq:FlowEqGam0m2pos1PIFRG0DONappendix}
\end{equation}
\begin{equation}
\dot{\overline{\Gamma}}^{(\mathrm{1PI})(2)}_{k} = -\frac{N+2}{6} \dot{R}_{k} \overline{G}^{2}_{k} \overline{\Gamma}^{(\mathrm{1PI})(4)}_{k}\;,
\label{eq:FlowEqGam2m2pos1PIFRG0DONappendix}
\end{equation}
\begin{equation}
\dot{\overline{\Gamma}}^{(\mathrm{1PI})(4)}_{k} = \frac{N+8}{3}\dot{R}_{k} \overline{G}^{3}_{k}\left(\overline{\Gamma}^{(\mathrm{1PI})(4)}_{k}\right)^2 - \frac{N+4}{10}\dot{R}_{k} \overline{G}^{2}_{k} \overline{\Gamma}^{(\mathrm{1PI})(6)}_{k}\;,
\label{eq:FlowEqGam4m2pos1PIFRG0DONappendix}
\end{equation}
\begin{equation}
\dot{\overline{\Gamma}}^{(\mathrm{1PI})(6)}_{k} = - \frac{5N+130}{3} \dot{R}_{k} \overline{G}^{4}_{k} \left(\overline{\Gamma}^{(\mathrm{1PI})(4)}_{k}\right)^{3} + \left(N+14\right) \dot{R}_{k} \overline{G}^{3}_{k} \overline{\Gamma}^{(\mathrm{1PI})(4)}_{k} \overline{\Gamma}^{(\mathrm{1PI})(6)}_{k} - \frac{N+6}{14}\dot{R}_{k} \overline{G}^{2}_{k} \overline{\Gamma}^{(\mathrm{1PI})(8)}_{k}\;,
\label{eq:FlowEqGam6m2pos1PIFRG0DONappendix}
\end{equation}
where the introduction of $\overline{G}^{(0)}_{k}$ is justified below~\eqref{eq:FlowEqGam4m2neg1PIFRG0DON}.

\vspace{0.5cm}

In order to treat the regime with $m^{2}<0$, let us then assume that the $O(N)$ symmetry can be spontaneously broken. We restrict ourselves to the situation where $N=1$, which implies that the vector $\vec{\phi}$ has a single component $\phi$. We will outline the main steps of the vertex expansion as well. To begin with, we consider the Wetterich equation~\eqref{eq:WetterichEqpure1PIFRG0DONappendix} at $N=1$:
\begin{equation}
\dot{\Gamma}^{(\mathrm{1PI})}_{k}(\phi) = \frac{1}{2}\dot{R}_{k}\left(\Gamma^{(\mathrm{1PI})(2)}_{k}(\phi)+R_{k}\right)^{-1} = \frac{1}{2} \dot{R}_{k} G_{k}(\phi) \;.
\label{eq:WetterichEqN1withSSBpure1PIFRG0DONappendix}
\end{equation}
As mentioned in section~\ref{sec:1PIFRG0DON}, this situation is equivalent to that of the collective representation. Hence, the present discussion can also be considered as the recipe to treat the Wetterich equation of the collective theory with the vertex expansion. The vertex expansion procedure is simply carried out by Taylor expanding the LHS of~\eqref{eq:WetterichEqN1withSSBpure1PIFRG0DONappendix} as:
\begin{equation}
\dot{\Gamma}^{(\mathrm{1PI})}_{k}(\phi) = \dot{\overline{\Gamma}}^{(\mathrm{1PI})}_{k} - \dot{\overline{\phi}}_{k} \overline{\Gamma}_{k}^{(\mathrm{1PI})(2)} \left(\phi-\overline{\phi}_{k}\right) + \sum^{\infty}_{n=2} \frac{1}{n!} \left(\dot{\overline{\Gamma}}_{k}^{(\mathrm{1PI})(n)} - \dot{\overline{\phi}}_{k} \overline{\Gamma}_{k}^{(\mathrm{1PI})(n+1)} \right) \left(\phi-\overline{\phi}_{k}\right)^{n} \;,
\label{eq:WetterichEqN1withSSBLHSpure1PIFRG0DONappendix}
\end{equation}
(as follows from~\eqref{eq:WetterichEqLHSpure1PIFRG0DONappendix} with $N=1$) and its RHS as:
\begin{equation}
\begin{split}
\frac{1}{2} \dot{\boldsymbol{R}}_{k} G_{k}(\phi) = & \ \frac{1}{2} \dot{R}_{k} \overline{G}_{k} \\
& - \frac{1}{2}\dot{R}_{k}\overline{G}_{k}^{2}\overline{\Gamma}^{(\mathrm{1PI})(3)}_{k} \left(\phi-\overline{\phi}_{k}\right) \\
& + \frac{1}{2} \left(\dot{R}_{k}\overline{G}_{k}^{3}\left(\overline{\Gamma}^{(\mathrm{1PI})(3)}_{k}\right)^{2}-\frac{1}{2}\dot{R}_{k}\overline{G}_{k}^{2}\overline{\Gamma}^{(\mathrm{1PI})(4)}_{k}\right) \left(\phi-\overline{\phi}_{k}\right)^2 \\
& + \frac{1}{6} \left( -3\dot{R}_{k}\overline{G}_{k}^{4}\left(\overline{\Gamma}^{(\mathrm{1PI})(3)}_{k}\right)^{3}+3\dot{R}_{k}\overline{G}_{k}^{3}\overline{\Gamma}^{(\mathrm{1PI})(3)}_{k}\overline{\Gamma}^{(\mathrm{1PI})(4)}_{k}-\frac{1}{2}\dot{R}_{k}\overline{G}_{k}^{2}\overline{\Gamma}^{(\mathrm{1PI})(5)}_{k} \right) \left(\phi-\overline{\phi}_{k}\right)^3 \\
& + \frac{1}{24} \bigg( 12\dot{R}_{k}\overline{G}_{k}^{5}\left(\overline{\Gamma}^{(\mathrm{1PI})(3)}_{k}\right)^{4}-18\dot{R}_{k}\overline{G}_{k}^{4}\left(\overline{\Gamma}^{(\mathrm{1PI})(3)}_{k}\right)^{2}\overline{\Gamma}^{(\mathrm{1PI})(4)}_{k}+4\dot{R}_{k}\overline{G}^{3}_{k}\overline{\Gamma}^{(\mathrm{1PI})(3)}_{k}\overline{\Gamma}^{(\mathrm{1PI})(5)}_{k} \\
& \hspace{1.1cm} + 3\dot{R}_{k}\overline{G}^{3}_{k}\left(\overline{\Gamma}^{(\mathrm{1PI})(4)}_{k}\right)^{2}-\frac{1}{2}\dot{R}_{k}\overline{G}^{2}_{k}\overline{\Gamma}^{(\mathrm{1PI})(6)}_{k} \bigg) \left(\phi-\overline{\phi}_{k}\right)^4 \\
& + \frac{1}{120} \bigg( - 60\dot{R}_{k}\overline{G}_{k}^{6}\left(\overline{\Gamma}^{(\mathrm{1PI})(3)}_{k}\right)^{5} + 130\dot{R}_{k}\overline{G}_{k}^{5}\left(\overline{\Gamma}^{(\mathrm{1PI})(3)}_{k}\right)^{3}\overline{\Gamma}^{(\mathrm{1PI})(4)}_{k} \\
& \hspace{1.5cm} - 41\dot{R}_{k}\overline{G}_{k}^{4}\overline{\Gamma}^{(\mathrm{1PI})(3)}_{k}\left(\overline{\Gamma}^{(\mathrm{1PI})(4)}_{k}\right)^{2} - 30\dot{R}_{k}\overline{G}_{k}^{4}\left(\overline{\Gamma}^{(\mathrm{1PI})(3)}_{k}\right)^{2}\overline{\Gamma}^{(\mathrm{1PI})(5)}_{k} \\
& \hspace{1.5cm} + 10\dot{R}_{k}\overline{G}_{k}^{3}\overline{\Gamma}^{(\mathrm{1PI})(4)}_{k}\overline{\Gamma}^{(\mathrm{1PI})(5)}_{k} + 5\dot{R}_{k}\overline{G}_{k}^{3}\overline{\Gamma}^{(\mathrm{1PI})(3)}_{k}\overline{\Gamma}^{(\mathrm{1PI})(6)}_{k} \\
& \hspace{1.5cm} - \frac{1}{2} \dot{R}_{k} \overline{G}_{k}^{2} \overline{\Gamma}^{(\mathrm{1PI})(7)}_{k} \bigg) \left(\phi-\overline{\phi}_{k}\right)^5 \\
& + \frac{1}{720} \bigg( + 360\dot{R}_{k}\overline{G}_{k}^{7}\left(\overline{\Gamma}^{(\mathrm{1PI})(3)}_{k}\right)^{6}-950\dot{R}_{k}\overline{G}_{k}^{6}\left(\overline{\Gamma}^{(\mathrm{1PI})(3)}_{k}\right)^{4}\overline{\Gamma}^{(\mathrm{1PI})(4)}_{k} \\
& \hspace{1.5cm} +554\dot{R}_{k}\overline{G}_{k}^{5}\left(\overline{\Gamma}^{(\mathrm{1PI})(3)}_{k}\right)^{2}\left(\overline{\Gamma}^{(\mathrm{1PI})(4)}_{k}\right)^{2}+250\dot{R}_{k}\overline{G}_{k}^{5}\left(\overline{\Gamma}^{(\mathrm{1PI})(3)}_{k}\right)^{3}\overline{\Gamma}^{(\mathrm{1PI})(5)}_{k} \\
& \hspace{1.5cm} -41\dot{R}_{k}\overline{G}_{k}^{4}\left(\overline{\Gamma}^{(\mathrm{1PI})(4)}_{k}\right)^{3} - 172\dot{R}_{k}\overline{G}_{k}^{4} \overline{\Gamma}^{(\mathrm{1PI})(3)}_{k} \overline{\Gamma}^{(\mathrm{1PI})(4)}_{k} \overline{\Gamma}^{(\mathrm{1PI})(5)}_{k} \\
& \hspace{1.5cm} + 10 \dot{R}_{k}\overline{G}_{k}^{3}\left(\overline{\Gamma}^{(\mathrm{1PI})(5)}_{k}\right)^{2} - 45 \dot{R}_{k}\overline{G}_{k}^{4}\left(\overline{\Gamma}^{(\mathrm{1PI})(3)}_{k}\right)^{2} \overline{\Gamma}^{(\mathrm{1PI})(6)}_{k} \\
& \hspace{1.5cm} + 15 \dot{R}_{k}\overline{G}_{k}^{3} \overline{\Gamma}^{(\mathrm{1PI})(4)}_{k} \overline{\Gamma}^{(\mathrm{1PI})(6)}_{k} + 6 \dot{R}_{k}\overline{G}_{k}^{3} \overline{\Gamma}^{(\mathrm{1PI})(3)}_{k} \overline{\Gamma}^{(\mathrm{1PI})(7)}_{k} \\
& \hspace{1.5cm} + \frac{1}{2} \dot{R}_{k}\overline{G}_{k}^{2} \overline{\Gamma}^{(\mathrm{1PI})(8)}_{k} \bigg) \left(\phi-\overline{\phi}_{k}\right)^6 \\
& + \mathcal{O}\big((\phi-\overline{\phi}_{k})^7\big) \;.
\end{split}
\label{eq:WetterichEqN1withSSBRHSpure1PIFRG0DONappendix}
\end{equation}
Finally, identifying the terms with the same powers of $\phi-\overline{\phi}_{k}$ leads to the following set of differential equations:
\begin{equation}
\dot{\overline{\Gamma}}^{(\mathrm{1PI})}_{k} = \frac{1}{2} \dot{R}_{k} \left( \overline{G}_{k} - \overline{G}^{(0)}_{k} \right)\;,
\label{eq:FlowEqGam0m2neg1PIFRG0DONappendix}
\end{equation}
\begin{equation}
\dot{\overline{\phi}}_{k}=\frac{1}{2\overline{\Gamma}^{(\mathrm{1PI})(2)}_{k}}\dot{R}_{k}\overline{G}_{k}^{2}\overline{\Gamma}^{(\mathrm{1PI})(3)}_{k} \;,
\label{eq:FlowEqPhim2neg1PIFRG0DONappendix}
\end{equation}
\begin{equation}
\dot{\overline{\Gamma}}^{(\mathrm{1PI})(2)}_{k}= \dot{\overline{\phi}}_{k}\overline{\Gamma}^{(\mathrm{1PI})(3)}_{k} + \dot{R}_{k}\overline{G}_{k}^{3}\left(\overline{\Gamma}^{(\mathrm{1PI})(3)}_{k}\right)^{2}-\frac{1}{2}\dot{R}_{k}\overline{G}_{k}^{2}\overline{\Gamma}^{(\mathrm{1PI})(4)}_{k}\;,
\label{eq:FlowEqGam2m2neg1PIFRG0DONappendix}
\end{equation}
\begin{equation}
\dot{\overline{\Gamma}}^{(\mathrm{1PI})(3)}_{k}= \dot{\overline{\phi}}_{k}\overline{\Gamma}^{(\mathrm{1PI})(4)}_{k}-3\dot{R}_{k}\overline{G}_{k}^{4}\left(\overline{\Gamma}^{(\mathrm{1PI})(3)}_{k}\right)^{3}+3\dot{R}_{k}\overline{G}_{k}^{3}\overline{\Gamma}^{(\mathrm{1PI})(3)}_{k}\overline{\Gamma}^{(\mathrm{1PI})(4)}_{k}-\frac{1}{2}\dot{R}_{k}\overline{G}_{k}^{2}\overline{\Gamma}^{(\mathrm{1PI})(5)}_{k}\;,
\label{eq:FlowEqGam3m2neg1PIFRG0DONappendix}
\end{equation}
\begin{equation}
\begin{split}
\dot{\overline{\Gamma}}^{(\mathrm{1PI})(4)}_{k} = & \ \dot{\overline{\phi}}_{k}\overline{\Gamma}^{(\mathrm{1PI})(5)}_{k}+12\dot{R}_{k}\overline{G}_{k}^{5}\left(\overline{\Gamma}^{(\mathrm{1PI})(3)}_{k}\right)^{4}-18\dot{R}_{k}\overline{G}_{k}^{4}\left(\overline{\Gamma}^{(\mathrm{1PI})(3)}_{k}\right)^{2}\overline{\Gamma}^{(\mathrm{1PI})(4)}_{k} \\
& + 4\dot{R}_{k}\overline{G}^{3}_{k}\overline{\Gamma}^{(\mathrm{1PI})(3)}_{k}\overline{\Gamma}^{(\mathrm{1PI})(5)}_{k} + 3\dot{R}_{k}\overline{G}^{3}_{k}\left(\overline{\Gamma}^{(\mathrm{1PI})(4)}_{k}\right)^{2}-\frac{1}{2}\dot{R}_{k}\overline{G}^{2}_{k}\overline{\Gamma}^{(\mathrm{1PI})(6)}_{k}\;,
\label{eq:FlowEqGam4m2neg1PIFRG0DONappendix}
\end{split}
\end{equation}
\begin{equation}
\begin{split}
\dot{\overline{\Gamma}}^{(\mathrm{1PI})(5)}_{k} = & \ \dot{\overline{\phi}}_{k}\overline{\Gamma}^{(\mathrm{1PI})(6)}_{k} - 60\dot{R}_{k}\overline{G}_{k}^{6}\left(\overline{\Gamma}^{(\mathrm{1PI})(3)}_{k}\right)^{5} + 130\dot{R}_{k}\overline{G}_{k}^{5}\left(\overline{\Gamma}^{(\mathrm{1PI})(3)}_{k}\right)^{3}\overline{\Gamma}^{(\mathrm{1PI})(4)}_{k} \\
& - 41\dot{R}_{k}\overline{G}_{k}^{4}\overline{\Gamma}^{(\mathrm{1PI})(3)}_{k}\left(\overline{\Gamma}^{(\mathrm{1PI})(4)}_{k}\right)^{2} - 30\dot{R}_{k}\overline{G}_{k}^{4}\left(\overline{\Gamma}^{(\mathrm{1PI})(3)}_{k}\right)^{2}\overline{\Gamma}^{(\mathrm{1PI})(5)}_{k}+10\dot{R}_{k}\overline{G}_{k}^{3}\overline{\Gamma}^{(\mathrm{1PI})(4)}_{k}\overline{\Gamma}^{(\mathrm{1PI})(5)}_{k} \\
& + 5\dot{R}_{k}\overline{G}_{k}^{3}\overline{\Gamma}^{(\mathrm{1PI})(3)}_{k}\overline{\Gamma}^{(\mathrm{1PI})(6)}_{k} - \frac{1}{2} \dot{R}_{k} \overline{G}_{k}^{2} \overline{\Gamma}^{(\mathrm{1PI})(7)}_{k} \;,
\end{split}
\label{eq:FlowEqGam5m2neg1PIFRG0DONappendix}
\end{equation}
\begin{equation}
\begin{split}
\dot{\overline{\Gamma}}^{(\mathrm{1PI})(6)}_{k} = & \ \dot{\overline{\phi}}_{k}\overline{\Gamma}^{(\mathrm{1PI})(7)}_{k} + 360\dot{R}_{k}\overline{G}_{k}^{7}\left(\overline{\Gamma}^{(\mathrm{1PI})(3)}_{k}\right)^{6}-950\dot{R}_{k}\overline{G}_{k}^{6}\left(\overline{\Gamma}^{(\mathrm{1PI})(3)}_{k}\right)^{4}\overline{\Gamma}^{(\mathrm{1PI})(4)}_{k} \\
& +554\dot{R}_{k}\overline{G}_{k}^{5}\left(\overline{\Gamma}^{(\mathrm{1PI})(3)}_{k}\right)^{2}\left(\overline{\Gamma}^{(\mathrm{1PI})(4)}_{k}\right)^{2}+250\dot{R}_{k}\overline{G}_{k}^{5}\left(\overline{\Gamma}^{(\mathrm{1PI})(3)}_{k}\right)^{3}\overline{\Gamma}^{(\mathrm{1PI})(5)}_{k} \\
& -41\dot{R}_{k}\overline{G}_{k}^{4}\left(\overline{\Gamma}^{(\mathrm{1PI})(4)}_{k}\right)^{3} - 172\dot{R}_{k}\overline{G}_{k}^{4} \overline{\Gamma}^{(\mathrm{1PI})(3)}_{k} \overline{\Gamma}^{(\mathrm{1PI})(4)}_{k} \overline{\Gamma}^{(\mathrm{1PI})(5)}_{k} + 10 \dot{R}_{k}\overline{G}_{k}^{3}\left(\overline{\Gamma}^{(\mathrm{1PI})(5)}_{k}\right)^{2} \\
& - 45 \dot{R}_{k}\overline{G}_{k}^{4}\left(\overline{\Gamma}^{(\mathrm{1PI})(3)}_{k}\right)^{2} \overline{\Gamma}^{(\mathrm{1PI})(6)}_{k} + 15 \dot{R}_{k}\overline{G}_{k}^{3} \overline{\Gamma}^{(\mathrm{1PI})(4)}_{k} \overline{\Gamma}^{(\mathrm{1PI})(6)}_{k} + 6 \dot{R}_{k}\overline{G}_{k}^{3} \overline{\Gamma}^{(\mathrm{1PI})(3)}_{k} \overline{\Gamma}^{(\mathrm{1PI})(7)}_{k} \\
& + \frac{1}{2} \dot{R}_{k}\overline{G}_{k}^{2} \overline{\Gamma}^{(\mathrm{1PI})(8)}_{k} \;,
\end{split}
\label{eq:FlowEqGam6m2neg1PIFRG0DONappendix}
\end{equation}
where the reason behind the introduction of $\overline{G}^{(0)}_{k}$ in~\eqref{eq:FlowEqGam0m2neg1PIFRG0DONappendix} is once again discussed below~\eqref{eq:FlowEqGam4m2neg1PIFRG0DON}.

\subsection{\label{sec:VertexExpansionAppMixed1PIFRG0DON}Mixed 1PI functional renormalization group}

We then outline the main steps of the vertex expansion procedure treating the Wetterich equation for the toy model under consideration in the framework of the mixed representation. The latter equation is already given by~\eqref{eq:WetterichEqmixed1PIFRG0DON} in the form:
\begin{equation}
\dot{\Gamma}^{(\mathrm{1PI})}_{\mathrm{mix},k}\Big(\vec{\phi},\eta\Big) = \frac{1}{2}\mathcal{ST}r\left[\dot{\mathcal{R}}_{k}\left(\Gamma^{(\mathrm{1PI})(2)}_{\mathrm{mix},k}\Big(\vec{\phi},\eta\Big)+\mathcal{R}_{k}\right)^{-1}\right] \;,
\label{eq:WetterichEqmixed1PIFRG0DONappendix}
\end{equation}
with
\begin{equation}
\mathcal{R}_{k} = \begin{pmatrix}
\boldsymbol{R}^{(\phi)}_{k} & \vec{0} \\
\vec{0}^{\mathrm{T}} & R^{(\eta)}_{k}
\end{pmatrix} = \begin{pmatrix}
R_{k} \mathbb{I}_{N} & \vec{0} \\
\vec{0}^{\mathrm{T}} & R_{k}
\end{pmatrix} = R_{k} \mathbb{I}_{N+1} \;.
\label{eq:mathcalRkmixed1PIFRG0DONAppendix}
\end{equation}
As a first step, we expand the mixed 1PI EA around an extremum ($\overline{\Gamma}_{\mathrm{mix},k,a}^{(\mathrm{1PI})(1\phi)}=\overline{\Gamma}_{\mathrm{mix},k}^{(\mathrm{1PI})(1\eta)}=0$ $\forall a,k$) via the relation:
\begin{equation}
\scalebox{0.89}{${\displaystyle\Gamma^{(\mathrm{1PI})}_{\mathrm{mix},k}\Big(\vec{\phi},\eta\Big) = \overline{\Gamma}^{(\mathrm{1PI})}_{\mathrm{mix},k} + \sum^{\infty}_{n=2} \frac{1}{n!} \sum_{m=0}^{n} \begin{pmatrix}
n \\
m
\end{pmatrix} \sum_{a_{1},\cdots,a_{m}=1}^{N} \overline{\Gamma}_{\mathrm{mix},k,a_{1} \cdots a_{m}}^{(\mathrm{1PI})(m\phi,(n-m)\eta)} \Big(\vec{\phi}-\vec{\overline{\phi}}_{k}\Big)_{a_{1}} \cdots \Big(\vec{\phi}-\vec{\overline{\phi}}_{k}\Big)_{a_{m}} \left(\eta-\overline{\eta}_{k}\right)^{n-m} \;,}$}
\label{eq:TaylorExpmixed1PIFRG0DONappendix}
\end{equation}
from which we deduce the following expression of the LHS of~\eqref{eq:WetterichEqmixed1PIFRG0DONappendix}:
\begin{equation}
\begin{split}
\scalebox{0.79}{${\displaystyle \dot{\Gamma}^{(\mathrm{1PI})}_{\mathrm{mix},k}\Big(\vec{\phi},\eta\Big) =}$} & \ \scalebox{0.79}{${\displaystyle\dot{\overline{\Gamma}}^{(\mathrm{1PI})}_{\mathrm{mix},k} - \sum_{a_{1}=1}^{N} \left( \sum_{a_{2}=1}^{N} \dot{\overline{\phi}}_{k,a_{2}} \overline{\Gamma}_{\mathrm{mix},k,a_{2} a_{1}}^{(\mathrm{1PI})(2\phi)} + \dot{\overline{\eta}}_{k} \overline{\Gamma}_{\mathrm{mix},k,a_{1}}^{(\mathrm{1PI})(1\phi ,1\eta)} \right) \Big(\vec{\phi}-\vec{\overline{\phi}}_{k}\Big)_{a_{1}} }$} \\
& \scalebox{0.79}{${\displaystyle - \left( \sum_{a_{1}=1}^{N} \dot{\overline{\phi}}_{k,a_{1}} \overline{\Gamma}_{\mathrm{mix},k,a_{1}}^{(\mathrm{1PI})(1\phi ,1\eta)} + \dot{\overline{\eta}}_{k} \overline{\Gamma}_{\mathrm{mix},k}^{(\mathrm{1PI})(2\eta)} \right) \left(\eta-\overline{\eta}_{k}\right) }$} \\
& \scalebox{0.79}{${\displaystyle + \sum^{\infty}_{n=2} \frac{1}{n!} \sum_{m=0}^{n} \begin{pmatrix}
n \\
m
\end{pmatrix} \sum_{a_{1},\cdots,a_{m}=1}^{N} \left(\dot{\overline{\Gamma}}_{\mathrm{mix},k,a_{1} \cdots a_{m}}^{(\mathrm{1PI})(m\phi,(n-m)\eta)} - \sum_{a_{m+1}=1}^{N} \dot{\overline{\phi}}_{k,a_{m+1}} \overline{\Gamma}_{\mathrm{mix},k,a_{m+1} a_{1} \cdots a_{m}}^{(\mathrm{1PI})((m+1)\phi,(n-m)\eta)} - \dot{\overline{\eta}}_{k} \overline{\Gamma}_{\mathrm{mix},k,a_{1} \cdots a_{m}}^{(\mathrm{1PI})(m\phi,(n-m+1)\eta)} \right)}$} \\
& \hspace{3.8cm} \scalebox{0.79}{${\displaystyle \times \Big(\vec{\phi}-\vec{\overline{\phi}}_{k}\Big)_{a_{1}} \cdots \Big(\vec{\phi}-\vec{\overline{\phi}}_{k}\Big)_{a_{m}} \left(\eta-\overline{\eta}_{k}\right)^{n-m} \;.}$}
\end{split}
\label{eq:TaylorExpBismixed1PIFRG0DONappendix}
\end{equation}
Then, the RHS of~\eqref{eq:WetterichEqmixed1PIFRG0DONappendix} is expanded by introducing the matrices $\mathcal{P}_{k}$ and $\mathcal{F}_{k}$ as:
\begin{equation}
\Gamma^{(\mathrm{1PI})(2)}_{\mathrm{mix},k}+\mathcal{R}_{k}=\mathcal{P}_{k}+\mathcal{F}_{k}\;.
\label{eq:SplittingPkFkmixed1PIFRG0DONappendix}
\end{equation}
As $\mathcal{F}_{k}$ must encompass the whole field dependence, we must have:
\begin{equation}
\mathcal{F}_{k} = \begin{pmatrix}
\Gamma^{(\mathrm{1PI})(2\phi)}_{\mathrm{mix},k}-\overline{\Gamma}^{(\mathrm{1PI})(2\phi)}_{\mathrm{mix},k} & \Gamma^{(\mathrm{1PI})(1\phi,1\eta)}_{\mathrm{mix},k}-\overline{\Gamma}^{(\mathrm{1PI})(1\phi,1\eta)}_{\mathrm{mix},k} \\
\Gamma^{(\mathrm{1PI})(1\phi,1\eta)}_{\mathrm{mix},k}-\overline{\Gamma}^{(\mathrm{1PI})(1\phi,1\eta)}_{\mathrm{mix},k} & \Gamma^{(\mathrm{1PI})(2\eta)}_{\mathrm{mix},k}-\overline{\Gamma}^{(\mathrm{1PI})(2\eta)}_{\mathrm{mix},k}
\end{pmatrix}\;,
\label{eq:Fkmixed1PIFRG0DONappendix}
\end{equation}
and
\begin{equation}
\mathcal{P}_{k} = \begin{pmatrix}
\boldsymbol{R}^{(\phi)}_{k}+\overline{\Gamma}^{(\mathrm{1PI})(2\phi)}_{\mathrm{mix},k} & \overline{\Gamma}^{(\mathrm{1PI})(1\phi,1\eta)}_{\mathrm{mix},k} \\
\overline{\Gamma}^{(\mathrm{1PI})(1\phi,1\eta)}_{\mathrm{mix},k} & R^{(\eta)}_{k}+\overline{\Gamma}^{(\mathrm{1PI})(2\eta)}_{\mathrm{mix},k}
\end{pmatrix}\;.
\label{eq:Pkmixed1PIFRG0DONappendix}
\end{equation}
Furthermore, each component of $\mathcal{F}_{k}$ can be expanded by differentiating~\eqref{eq:TaylorExpmixed1PIFRG0DONappendix}, thus leading to:
\begin{equation}
\begin{split}
\scalebox{0.96}{${\displaystyle \mathcal{F}_{k,a_{1}a_{2}} \equiv }$} & \ \scalebox{0.96}{${\displaystyle\Gamma^{(\mathrm{1PI})(2\phi)}_{\mathrm{mix},k,a_{1}a_{2}}-\overline{\Gamma}^{(\mathrm{1PI})(2\phi)}_{\mathrm{mix},k,a_{1}a_{2}} }$} \\
\scalebox{0.96}{${\displaystyle = }$} & \scalebox{0.96}{${\displaystyle \sum^{\infty}_{n=1} \frac{1}{n!} \sum_{m=0}^{n} \begin{pmatrix}
n \\
m
\end{pmatrix} \sum_{a_{3},\cdots,a_{m+2}=1}^{N} \overline{\Gamma}_{\mathrm{mix},k,a_{1} \cdots a_{m+2}}^{(\mathrm{1PI})((m+2)\phi,(n-m)\eta)} \Big(\vec{\phi}-\vec{\overline{\phi}}_{k}\Big)_{a_{3}} \cdots \Big(\vec{\phi}-\vec{\overline{\phi}}_{k}\Big)_{a_{m+2}} \left(\eta-\overline{\eta}_{k}\right)^{n-m} \;,}$}
\end{split}
\label{eq:Faamixed1PIFRGappendix}
\end{equation}
\begin{equation}
\begin{split}
\scalebox{0.92}{${\displaystyle\mathcal{F}_{k,a_{1} \hspace{0.04cm} N+1} =}$} & \ \scalebox{0.92}{${\displaystyle \mathcal{F}_{k,N+1 \hspace{0.04cm} a_{1}} \equiv \Gamma^{(\mathrm{1PI})(1\phi,1\eta)}_{\mathrm{mix},k,a_{1}}-\overline{\Gamma}^{(\mathrm{1PI})(1\phi,1\eta)}_{\mathrm{mix},k,a_{1}} }$} \\
\scalebox{0.92}{${\displaystyle = }$} & \scalebox{0.92}{${\displaystyle\sum^{\infty}_{n=1} \frac{1}{n!} \sum_{m=0}^{n} \begin{pmatrix}
n \\
m
\end{pmatrix} \sum_{a_{2},\cdots,a_{m+1}=1}^{N} \overline{\Gamma}_{\mathrm{mix},k,a_{1} \cdots a_{m+1}}^{(\mathrm{1PI})((m+1)\phi,(n-m+1)\eta)} \Big(\vec{\phi}-\vec{\overline{\phi}}_{k}\Big)_{a_{2}} \cdots \Big(\vec{\phi}-\vec{\overline{\phi}}_{k}\Big)_{a_{m+1}} \left(\eta-\overline{\eta}_{k}\right)^{n-m} \;,}$}
\end{split}
\label{eq:FaNp1mixed1PIFRGappendix}
\end{equation}
\begin{equation}
\begin{split}
\scalebox{0.97}{${\displaystyle \mathcal{F}_{k,N+1 \hspace{0.04cm} N+1} \equiv }$} & \ \scalebox{0.97}{${\displaystyle \Gamma^{(\mathrm{1PI})(2\eta)}_{\mathrm{mix},k}-\overline{\Gamma}^{(\mathrm{1PI})(2\eta)}_{\mathrm{mix},k} }$} \\
\scalebox{0.97}{${\displaystyle = }$} & \scalebox{0.97}{${\displaystyle \sum^{\infty}_{n=1} \frac{1}{n!} \sum_{m=0}^{n} \begin{pmatrix}
n \\
m
\end{pmatrix} \sum_{a_{1},\cdots,a_{m}=1}^{N} \overline{\Gamma}_{\mathrm{mix},k,a_{1} \cdots a_{m}}^{(\mathrm{1PI})(m\phi,(n-m+2)\eta)} \Big(\vec{\phi}-\vec{\overline{\phi}}_{k}\Big)_{a_{1}} \cdots \Big(\vec{\phi}-\vec{\overline{\phi}}_{k}\Big)_{a_{m}} \left(\eta-\overline{\eta}_{k}\right)^{n-m} \;.}$}
\end{split}
\label{eq:FNp1Np1mixed1PIFRGappendix}
\end{equation}
The expansion of the RHS is then carried out by matrix multiplications between $\mathcal{P}_{k}^{-1}$ and $\mathcal{F}_{k}$ according to:
\begin{equation}
\begin{split}
\frac{1}{2}\mathcal{ST}r\left[\dot{\mathcal{R}}_{k}\left(\Gamma^{(\mathrm{1PI})(2)}_{\mathrm{mix},k}\Big(\vec{\phi},\eta\Big)+\mathcal{R}_{k}\right)^{-1}\right] = & \ \frac{1}{2}\mathcal{ST}r\left[\dot{\mathcal{R}}_{k}\left(\mathcal{P}_{k}+\mathcal{F}_{k}\right)^{-1}\right] \\
= & \ \frac{1}{2}\mathcal{ST}r\left[\dot{\mathcal{R}}_{k}\mathcal{P}_{k}^{-1}\left(\mathbb{I}_{N+1}+\mathcal{P}_{k}^{-1}\mathcal{F}_{k}\right)^{-1}\right] \\
= & \ \frac{1}{2}\mathcal{ST}r\left[\dot{\mathcal{R}}_{k}\mathcal{P}_{k}^{-1}\left(\mathbb{I}_{N+1}+\sum_{n=1}^{\infty}(-1)^{n}\left(\mathcal{P}_{k}^{-1}\mathcal{F}_{k}\right)^{n}\right)\right] \\
= & \ \frac{1}{2}\mathcal{ST}r\left[\dot{\mathcal{R}}_{k}\mathcal{P}_{k}^{-1}\right] \\
& + \frac{1}{2}\sum_{n=1}^{\infty}(-1)^{n}\mathcal{ST}r\left[\dot{\mathcal{R}}_{k}\mathcal{P}_{k}^{-1}\left(\mathcal{P}_{k}^{-1}\mathcal{F}_{k}\right)^{n}\right] \;,
\end{split}
\label{eq:expansionRHSmixed1PIFRGappendix}
\end{equation}
where the third line was obtained using the matrix generalization of the Taylor series:
\begin{equation}
\frac{1}{1+x} = \sum_{n=0}^{\infty} (-1)^{n} x^{n}\;.
\end{equation}
We also point out that the expansion procedure outlined in~\eqref{eq:expansionRHSmixed1PIFRGappendix} is strictly equivalent to that of~\eqref{eq:VertexExpSeveralField1PIFRG}.

\vspace{0.5cm}

We will focus on the unbroken-symmetry regime in order to go further in the vertex expansion procedure. In this situation, we can exploit the following relations:
\begin{equation}
\overline{\Gamma}_{\mathrm{mix},k,a_{1}a_{2}}^{(\mathrm{1PI})(2\phi,n\eta)}=\overline{\Gamma}_{\mathrm{mix},k}^{(\mathrm{1PI})(2\phi,n\eta)} \ \delta_{a_{1}a_{2}} \mathrlap{\quad \forall a_{1},a_{2},n\;,}
\label{eq:DefGamma2m2posmixed1PIFRG0DONappendix}
\end{equation}
\begin{equation}
\hspace{2.3cm} \overline{\Gamma}_{\mathrm{mix},k,a_{1}a_{2}a_{3}a_{4}}^{(\mathrm{1PI})(4\phi,n\eta)}=\overline{\Gamma}_{\mathrm{mix},k}^{(\mathrm{1PI})(4\phi,n\eta)}\left(\delta_{a_{1}a_{2}}\delta_{a_{3}a_{4}}+\delta_{a_{1}a_{3}}\delta_{a_{2}a_{4}}+\delta_{a_{1}a_{4}}\delta_{a_{2}a_{3}}\right) \quad \forall a_{1},a_{2},a_{3},a_{4},n\;,
\label{eq:DefGamma4m2posmixed1PIFRG0DONappendix}
\end{equation}
which are respectively the counterparts of~\eqref{eq:DefGamma2m2pos1PIFRG0DONappendix} and~\eqref{eq:DefGamma4m2pos1PIFRG0DONappendix} for the mixed theory, and:
\begin{equation}
\overline{\Gamma}_{\mathrm{mix},k,a_{1}\cdots a_{n}}^{(\mathrm{1PI})(n\phi,m\eta)} = 0 \mathrlap{\quad \forall a_{1},\cdots,a_{n},m, ~ \forall n ~ \mathrm{odd}\;.}
\label{eq:DefGammaoddm2posmixed1PIFRG0DONappendix}
\end{equation}
In particular, after combining~\eqref{eq:Pkmixed1PIFRG0DONappendix} with~\eqref{eq:mathcalRkmixed1PIFRG0DONAppendix},~\eqref{eq:DefGamma2m2posmixed1PIFRG0DONappendix} and~\eqref{eq:DefGammaoddm2posmixed1PIFRG0DONappendix}, we obtain:
\begin{equation}
\mathcal{P}_{k} = \begin{pmatrix}
\left(R_{k} + \overline{\Gamma}^{(\mathrm{1PI})(2\phi)}_{\mathrm{mix},k}\right) \mathbb{I}_{N} & \vec{0} \\
\vec{0}^{\mathrm{T}} & R_{k} + \overline{\Gamma}^{(\mathrm{1PI})(2\eta)}_{\mathrm{mix},k}
\end{pmatrix} \;,
\label{eq:PknoSSBmixed1PIFRG0DONappendix}
\end{equation}
and the inversion of $\mathcal{P}_{k}$ thus becomes trivial such that $\mathcal{P}_{k}^{-1}$ is now a diagonal matrix, i.e.:
\begin{equation}
\mathcal{P}_{k}^{-1} = \begin{pmatrix}
\left(R_{k} + \overline{\Gamma}^{(2\phi)}_{\mathrm{mix},k}\right)^{-1} \mathbb{I}_{N} & \vec{0} \\
\vec{0}^{\mathrm{T}} & \left(R_{k} + \overline{\Gamma}^{(2\eta)}_{\mathrm{mix},k}\right)^{-1}
\end{pmatrix} \equiv \begin{pmatrix}
\overline{G}^{(\phi)}_{k} \mathbb{I}_{N} & \vec{0} \\
\vec{0}^{\mathrm{T}} & \overline{G}^{(\eta)}_{k}
\end{pmatrix} \;.
\label{eq:Pkminus1noSSBmixed1PIFRG0DONappendix}
\end{equation}
This implies in particular that the number of terms generated by the matrix products in~\eqref{eq:expansionRHSmixed1PIFRGappendix} is significantly reduced as compared with the situation where SSB can occur (i.e. where~\eqref{eq:DefGammaoddm2posmixed1PIFRG0DONappendix} does not hold). Moreover, as $\vec{\overline{\phi}}_{k}=\vec{0}$ $\forall k$ in the unbroken-symmetry regime, the expressions of the components of $\mathcal{F}_{k}$ given by~\eqref{eq:Faamixed1PIFRGappendix} to~\eqref{eq:FNp1Np1mixed1PIFRGappendix} can be simplified as:
\begin{equation}
\mathcal{F}_{k,a_{1}a_{2}} = \sum^{\infty}_{n=1} \frac{1}{n!} \sum_{\underset{\lbrace \textcolor{red}{\text{m even}} \rbrace}{m=0}}^{n} \begin{pmatrix}
n \\
m
\end{pmatrix} \sum_{a_{3},\cdots,a_{m+2}=1}^{N} \overline{\Gamma}_{\mathrm{mix},k,a_{1} \cdots a_{m+2}}^{(\mathrm{1PI})((m+2)\phi,(n-m)\eta)} \phi_{a_{3}} \cdots \phi_{a_{m+2}} \left(\eta-\overline{\eta}_{k}\right)^{n-m} \;,
\label{eq:FaaBismixed1PIFRGappendix}
\end{equation}
\begin{equation}
\scalebox{0.94}{${\displaystyle \mathcal{F}_{k,a_{1} \hspace{0.04cm} N+1} = \mathcal{F}_{k,N+1 \hspace{0.04cm} a_{1}} = \sum^{\infty}_{n=1} \frac{1}{n!} \sum_{\underset{\lbrace \textcolor{red}{\text{m odd}} \rbrace}{m=0}}^{n} \begin{pmatrix}
n \\
m
\end{pmatrix} \sum_{a_{2},\cdots,a_{m+1}=1}^{N} \overline{\Gamma}_{\mathrm{mix},k,a_{1} \cdots a_{m+1}}^{(\mathrm{1PI})((m+1)\phi,(n-m+1)\eta)} \phi_{a_{2}} \cdots \phi_{a_{m+1}} \left(\eta-\overline{\eta}_{k}\right)^{n-m} \;,}$}
\label{eq:FaNp1Bismixed1PIFRGappendix}
\end{equation}
\begin{equation}
\mathcal{F}_{k,N+1 \hspace{0.04cm} N+1} = \sum^{\infty}_{n=1} \frac{1}{n!} \sum_{\underset{\lbrace \textcolor{red}{\text{m even}} \rbrace}{m=0}}^{n} \begin{pmatrix}
n \\
m
\end{pmatrix} \sum_{a_{1},\cdots,a_{m}=1}^{N} \overline{\Gamma}_{\mathrm{mix},k,a_{1} \cdots a_{m}}^{(\mathrm{1PI})(m\phi,(n-m+2)\eta)} \phi_{a_{1}} \cdots \phi_{a_{m}} \left(\eta-\overline{\eta}_{k}\right)^{n-m} \;,
\label{eq:FNp1Np1Bismixed1PIFRGappendix}
\end{equation}
on the one hand and, on the other hand, the LHS of~\eqref{eq:WetterichEqmixed1PIFRG0DONappendix} given by~\eqref{eq:TaylorExpBismixed1PIFRG0DONappendix} reduces to:
\begin{equation}
\begin{split}
\scalebox{0.85}{${\displaystyle \dot{\Gamma}^{(\mathrm{1PI})}_{\mathrm{mix},k}\Big(\vec{\phi},\eta\Big) = }$} & \ \scalebox{0.85}{${\displaystyle \dot{\overline{\Gamma}}^{(\mathrm{1PI})}_{\mathrm{mix},k} - \dot{\overline{\eta}}_{k} \overline{\Gamma}_{\mathrm{mix},k}^{(\mathrm{1PI})(2\eta)} \left(\eta-\overline{\eta}_{k}\right) }$} \\
& \scalebox{0.85}{${\displaystyle + \sum^{\infty}_{n=2} \frac{1}{n!} \sum_{\underset{\lbrace \text{\textcolor{red}{m even}} \rbrace}{m=0}}^{n} \begin{pmatrix}
n \\
m
\end{pmatrix} \sum_{a_{1},\cdots,a_{m}=1}^{N} \left(\dot{\overline{\Gamma}}_{\mathrm{mix},k,a_{1} \cdots a_{m}}^{(\mathrm{1PI})(m\phi,(n-m)\eta)} - \dot{\overline{\eta}}_{k} \overline{\Gamma}_{\mathrm{mix},k,a_{1} \cdots a_{m}}^{(\mathrm{1PI})(m\phi,(n-m+1)\eta)} \right) \phi_{a_{1}} \cdots \phi_{a_{m}} \left(\eta-\overline{\eta}_{k}\right)^{n-m} \;,}$}
\end{split}
\label{eq:TaylorExpnoSSBmixed1PIFRG0DONappendix}
\end{equation}
where the restrictions ``$\text{\textcolor{red}{m even}}$'' and ``$\text{\textcolor{red}{m odd}}$'' for the sums are a direct consequence of~\eqref{eq:DefGammaoddm2posmixed1PIFRG0DONappendix}. Finally, we carry out the matrix products $(\mathcal{P}_{k}^{-1}\mathcal{F}_{k})^{n}$ involved in~\eqref{eq:expansionRHSmixed1PIFRGappendix} (where the expansion is performed up to a finite truncation order $n=N_{\mathrm{max}}$ in practice) with $\mathcal{P}_{k}^{-1}$ given by~\eqref{eq:Pkminus1noSSBmixed1PIFRG0DONappendix} and $\mathcal{F}_{k}$ specified by~\eqref{eq:FaaBismixed1PIFRGappendix} to~\eqref{eq:FNp1Np1Bismixed1PIFRGappendix}. In this way, we have expanded the RHS of the Wetterich equation given by~\eqref{eq:WetterichEqmixed1PIFRG0DONappendix}. By identifying the terms of the relation thus derived with those of~\eqref{eq:TaylorExpnoSSBmixed1PIFRG0DONappendix} (which coincides with the LHS of the Wetterich equation in the form of~\eqref{eq:WetterichEqmixed1PIFRG0DONappendix}) involving the same powers of the fields $\vec{\phi}$ and $\eta-\overline{\eta}_{k}$, we obtain the following tower of differential equations for $N=1$:
\begin{equation}
\scriptstyle \dot{\overline{\Gamma}}^{(\mathrm{1PI})}_{\mathrm{mix},k} = \ \frac{1}{2}\left(\overline{G}^{(\phi)}_{k}-\overline{G}^{(\phi)(0)}_{k}\right) + \frac{1}{2}\left(\overline{G}^{(\eta)}_{k}-\overline{G}^{(\eta)(0)}_{k}\right) \;,
\label{eq:DiffEqGam00N1mixed1PIFRG0DONappendix}
\end{equation}
\begin{equation}
\scriptstyle \dot{\overline{\eta}}_{k} = \frac{\dot{R}_{k}}{2\overline{\Gamma}^{(\mathrm{1PI})(2\eta)}_{\mathrm{mix},k}} \left(\overline{\Gamma}^{(\mathrm{1PI})(3\eta)}_{\mathrm{mix},k} \left(\overline{G}^{(\eta)}_{k}\right)^2 + \overline{\Gamma}^{(\mathrm{1PI})(2\phi,1\eta)}_{\mathrm{mix},k} \left(\overline{G}^{(\phi)}_{k}\right)^2\right) \;,
\label{eq:DiffEqetaN1mixed1PIFRG0DONappendix}
\end{equation}
\begin{equation}
\scriptstyle \dot{\overline{\Gamma}}^{(\mathrm{1PI})(2\phi)}_{\mathrm{mix},k} = \ \dot{\overline{\eta}}_{k}\overline{\Gamma}_{\mathrm{mix},k}^{(\mathrm{1PI})(2\phi,1\eta)} - \frac{1}{2} \dot{R}_{k} \bigg( \overline{\Gamma}_{\mathrm{mix},k}^{(\mathrm{1PI})(2\phi,2\eta)} \left(\overline{G}^{(\eta)}_{k}\right)^2 + \overline{G}^{(\phi)}_{k} \left(\overline{\Gamma}_{\mathrm{mix},k}^{(\mathrm{1PI})(4\phi)} \overline{G}^{(\phi)}_{k} - 2 \left(\overline{\Gamma}_{\mathrm{mix},k}^{(\mathrm{1PI})(2\phi,1\eta)}\right)^2 \overline{G}^{(\eta)}_{k} \left(\overline{G}^{(\eta)}_{k} + \overline{G}^{(\phi)}_{k}\right)\right)\bigg) \;,
\label{eq:DiffEqGam20N1mixed1PIFRG0DONappendix}
\end{equation}
\begin{equation}
\scriptstyle \dot{\overline{\Gamma}}^{(\mathrm{1PI})(2\eta)}_{\mathrm{mix},k} = \ \dot{\overline{\eta}}_{k}\overline{\Gamma}_{\mathrm{mix},k}^{(\mathrm{1PI})(3\eta)} -\frac{1}{2} \dot{R}_{k} \bigg( \overline{\Gamma}_{\mathrm{mix},k}^{(\mathrm{1PI})(4\eta)} \left(\overline{G}^{(\eta)}_{k}\right)^2 - 2 \left(\overline{\Gamma}_{\mathrm{mix},k}^{(\mathrm{1PI})(3\eta)}\right)^2 \left(\overline{G}^{(\eta)}_{k}\right)^3 + \left(\overline{G}^{(\phi)}_{k}\right)^2 \left(\overline{\Gamma}_{\mathrm{mix},k}^{(\mathrm{1PI})(2\phi,2\eta)} - 2 \left(\overline{\Gamma}_{\mathrm{mix},k}^{(\mathrm{1PI})(2\phi,1\eta)}\right)^2 \overline{G}^{(\phi)}_{k}\right) \bigg) \;,
\label{eq:DiffEqGam02N1mixed1PIFRG0DONappendix}
\end{equation}
\begin{equation}
\begin{split}
\scriptstyle \dot{\overline{\Gamma}}^{(\mathrm{1PI})(2\phi,1\eta)}_{\mathrm{mix},k} = & \ \scriptstyle \dot{\overline{\eta}}_{k} \overline{\Gamma}_{\mathrm{mix},k}^{(\mathrm{1PI})(2\phi,2\eta)} \\
& \scriptstyle + \dot{R}_{k} \bigg( \overline{\Gamma}_{\mathrm{mix},k}^{(\mathrm{1PI})(3\eta)} \left(\overline{G}^{(\eta)}_{k}\right)^2 \left( \overline{\Gamma}_{\mathrm{mix},k}^{(\mathrm{1PI})(2\phi,2\eta)} \overline{G}^{(\eta)}_{k} - \left(\overline{\Gamma}_{\mathrm{mix},k}^{(\mathrm{1PI})(2\phi,1\eta)}\right)^2 \overline{G}^{(\phi)}_{k} \left(2 \overline{G}^{(\eta)}_{k} + \overline{G}^{(\phi)}_{k}\right)\right) \\
& \scriptstyle \hspace{0.75cm} + \overline{\Gamma}_{\mathrm{mix},k}^{(\mathrm{1PI})(2\phi,1\eta)} \overline{G}^{(\phi)}_{k} \Big(2 \overline{\Gamma}_{\mathrm{mix},k}^{(\mathrm{1PI})(2\phi,2\eta)} \overline{G}^{(\eta)}_{k} \left(\overline{G}^{(\eta)}_{k} + \overline{G}^{(\phi)}_{k}\right) + \overline{G}^{(\phi)}_{k} \Big(\overline{\Gamma}_{\mathrm{mix},k}^{(\mathrm{1PI})(4\phi)} \overline{G}^{(\phi)}_{k} \\
& \scriptstyle \hspace{0.75cm} - \left(\overline{\Gamma}_{\mathrm{mix},k}^{(\mathrm{1PI})(2\phi,1\eta)}\right)^2 \overline{G}^{(\eta)}_{k} \left(\overline{G}^{(\eta)}_{k} + 2 \overline{G}^{(\phi)}_{k}\right)\Big)\Big)\bigg) \\
& \scriptstyle + \mathcal{F}_{1}^{(N=1)}\big(\overline{\Gamma}^{(\mathrm{1PI})(n\phi,m\eta)}_{\mathrm{mix},k}; \ n+m \leq 5\big) \;,
\label{eq:DiffEqGam21N1mixed1PIFRG0DONappendix}
\end{split}
\end{equation}
\begin{equation}
\begin{split}
\scriptstyle \dot{\overline{\Gamma}}^{(\mathrm{1PI})(3\eta)}_{\mathrm{mix},k} = & \ \scriptstyle \dot{\overline{\eta}}_{k}\overline{\Gamma}_{\mathrm{mix},k}^{(\mathrm{1PI})(4\eta)} \\
& \scriptstyle -3 \dot{R}_{k} \bigg(-\overline{\Gamma}_{\mathrm{mix},k}^{(\mathrm{1PI})(3\eta)} \overline{\Gamma}_{\mathrm{mix},k}^{(\mathrm{1PI})(4\eta)} \left(\overline{G}^{(\eta)}_{k}\right)^3 + \left(\overline{\Gamma}_{\mathrm{mix},k}^{(\mathrm{1PI})(3\eta)}\right)^3 \left(\overline{G}^{(\eta)}_{k}\right)^4 + \overline{\Gamma}_{\mathrm{mix},k}^{(\mathrm{1PI})(2\phi,1\eta)} \left(\overline{G}^{(\phi)}_{k}\right)^3 \Big(-\overline{\Gamma}_{\mathrm{mix},k}^{(\mathrm{1PI})(2\phi,2\eta)} \\
& \scriptstyle \hspace{0.9cm} + \left(\overline{\Gamma}_{\mathrm{mix},k}^{(\mathrm{1PI})(2\phi,1\eta)}\right)^2 \overline{G}^{(\phi)}_{k}\Big)\bigg) \\
& \scriptstyle + \mathcal{F}_{2}^{(N=1)}\big(\overline{\Gamma}^{(\mathrm{1PI})(n\phi,m\eta)}_{\mathrm{mix},k}; \ n+m \leq 5\big) \;,
\end{split}
\label{eq:DiffEqGam03N1mixed1PIFRG0DONappendix}
\end{equation}
\begin{equation}
\begin{split}
\scriptstyle \dot{\overline{\Gamma}}^{(\mathrm{1PI})(4\phi)}_{\mathrm{mix},k} = & \ \scriptstyle 3 \dot{R}_k \bigg(\left(\overline{\Gamma}_{\mathrm{mix},k}^{(\mathrm{1PI})(2\phi,2\eta)}\right)^2 \left(\overline{G}^{(\eta)}_{k}\right)^3 - 2 \left(\overline{\Gamma}_{\mathrm{mix},k}^{(\mathrm{1PI})(2\phi,1\eta)}\right)^2 \overline{\Gamma}_{\mathrm{mix},k}^{(\mathrm{1PI})(2\phi,2\eta)} \left(\overline{G}^{(\eta)}_{k}\right)^2 \overline{G}^{(\phi)}_{k} \left(2 \overline{G}^{(\eta)}_{k} + \overline{G}^{(\phi)}_{k}\right) \\
& \scriptstyle \hspace{0.75cm} + \left(\overline{\Gamma}_{\mathrm{mix},k}^{(\mathrm{1PI})(4\phi)} - 2 \left(\overline{\Gamma}_{\mathrm{mix},k}^{(\mathrm{1PI})(2\phi,1\eta)}\right)^2 \overline{G}^{(\eta)}_{k}\right) \left(\overline{G}^{(\phi)}_{k}\right)^2 \left( \overline{\Gamma}_{\mathrm{mix},k}^{(\mathrm{1PI})(4\phi)} \overline{G}^{(\phi)}_{k} - 2 \left(\overline{\Gamma}_{\mathrm{mix},k}^{(\mathrm{1PI})(2\phi,1\eta)}\right)^2 \overline{G}^{(\eta)}_{k} \left( \overline{G}^{(\eta)}_{k} + \overline{G}^{(\phi)}_{k}\right)\right)\bigg) \\
& \scriptstyle + \mathcal{F}_{3}^{(N=1)}\big(\overline{\Gamma}^{(\mathrm{1PI})(n\phi,m\eta)}_{\mathrm{mix},k}; \ n+m \leq 6\big) \;,
\end{split}
\label{eq:DiffEqGam40N1mixed1PIFRG0DONappendix}
\end{equation}
\begin{equation}
\begin{split}
\scriptstyle \dot{\overline{\Gamma}}^{(\mathrm{1PI})(4\eta)}_{\mathrm{mix},k} = & \ \scriptstyle 3 \dot{R}_{k} \bigg(\left(\overline{\Gamma}_{\mathrm{mix},k}^{(\mathrm{1PI})(4\eta)}\right)^2 \left(\overline{G}^{(\eta)}_{k}\right)^3 - 6 \left(\overline{\Gamma}_{\mathrm{mix},k}^{(\mathrm{1PI})(3\eta)}\right)^2 \overline{\Gamma}_{\mathrm{mix},k}^{(\mathrm{1PI})(4\eta)} \left(\overline{G}^{(\eta)}_{k}\right)^4 + 4 \left(\overline{\Gamma}_{\mathrm{mix},k}^{(\mathrm{1PI})(3\eta)}\right)^4 \left(\overline{G}^{(\eta)}_{k}\right)^5 \\
& \scriptstyle \hspace{0.75cm} + \left(\overline{G}^{(\phi)}_{k}\right)^3 \left(\left(\overline{\Gamma}_{\mathrm{mix},k}^{(\mathrm{1PI})(2\phi,2\eta)}\right)^2 - 6 \left(\overline{\Gamma}_{\mathrm{mix},k}^{(\mathrm{1PI})(2\phi,1\eta)}\right)^2 \overline{\Gamma}_{\mathrm{mix},k}^{(\mathrm{1PI})(2\phi,2\eta)} \overline{G}^{(\phi)}_{k} + 4 \left(\overline{\Gamma}_{\mathrm{mix},k}^{(\mathrm{1PI})(2\phi,1\eta)}\right)^4 \left(\overline{G}^{(\phi)}_{k}\right)^2\right)\bigg) \\
& \scriptstyle + \mathcal{F}_{4}^{(N=1)}\big(\overline{\Gamma}^{(\mathrm{1PI})(n\phi,m\eta)}_{\mathrm{mix},k}; \ n+m \leq 6\big) \;,
\end{split}
\label{eq:DiffEqGam04N1mixed1PIFRG0DONappendix}
\end{equation}
\begin{equation}
\begin{split}
\scriptstyle \dot{\overline{\Gamma}}^{(\mathrm{1PI})(2\phi,2\eta)}_{\mathrm{mix},k} = & \ \scriptstyle \dot{R}_{k} \bigg(4 \overline{\Gamma}_{\mathrm{mix},k}^{(\mathrm{1PI})(3\eta)} \overline{\Gamma}_{\mathrm{mix},k}^{(\mathrm{1PI})(2\phi,1\eta)} \left(\overline{G}^{(\eta)}_{k}\right)^2 \overline{G}^{(\phi)}_{k} \left(\left(\overline{\Gamma}_{\mathrm{mix},k}^{(\mathrm{1PI})(2\phi,1\eta)}\right)^2 \overline{G}^{(\phi)}_{k} \left(\overline{G}^{(\eta)}_{k} + \overline{G}^{(\phi)}_{k}\right) - \overline{\Gamma}_{\mathrm{mix},k}^{(\mathrm{1PI})(2\phi,2\eta)} \left(2 \overline{G}^{(\eta)}_{k} + \overline{G}^{(\phi)}_{k}\right)\right) \\
& \scriptstyle \hspace{0.75cm} + \overline{\Gamma}_{\mathrm{mix},k}^{(\mathrm{1PI})(4\eta)} \left(\overline{G}^{(\eta)}_{k}\right)^2 \left(\overline{\Gamma}_{\mathrm{mix},k}^{(\mathrm{1PI})(2\phi,2\eta)} \overline{G}^{(\eta)}_{k} - \left(\overline{\Gamma}_{\mathrm{mix},k}^{(\mathrm{1PI})(2\phi,1\eta)}\right)^2 \overline{G}^{(\phi)}_{k} \left(2 \overline{G}^{(\eta)}_{k} + \overline{G}^{(\phi)}_{k}\right)\right) \\
& \scriptstyle \hspace{0.75cm} + \left(\overline{\Gamma}_{\mathrm{mix},k}^{(\mathrm{1PI})(3\eta)}\right)^2 \left(\overline{G}^{(\eta)}_{k}\right)^3 \left(-3 \overline{\Gamma}_{\mathrm{mix},k}^{(\mathrm{1PI})(2\phi,2\eta)} \overline{G}^{(\eta)}_{k} + 2 \left(\overline{\Gamma}_{\mathrm{mix},k}^{(\mathrm{1PI})(2\phi,1\eta)}\right)^2 \overline{G}^{(\phi)}_{k} \left(3 \overline{G}^{(\eta)}_{k} + \overline{G}^{(\phi)}_{k}\right)\right) \\
& \scriptstyle \hspace{0.75cm} + \overline{G}^{(\phi)}_{k} \Big(2 \left(\overline{\Gamma}_{\mathrm{mix},k}^{(\mathrm{1PI})(2\phi,2\eta)}\right)^2 \overline{G}^{(\eta)}_{k} \left(\overline{G}^{(\eta)}_{k} + \overline{G}^{(\phi)}_{k}\right) + \overline{\Gamma}_{\mathrm{mix},k}^{(\mathrm{1PI})(2\phi,2\eta)} \overline{G}^{(\phi)}_{k} \Big(\overline{\Gamma}_{\mathrm{mix},k}^{(\mathrm{1PI})(4\phi)} \overline{G}^{(\phi)}_{k} \\
& \scriptstyle \hspace{0.75cm} - 5 \left(\overline{\Gamma}_{\mathrm{mix},k}^{(\mathrm{1PI})(2\phi,1\eta)}\right)^2 \overline{G}^{(\eta)}_{k} \left(\overline{G}^{(\eta)}_{k} + 2 \overline{G}^{(\phi)}_{k}\right)\Big) + \left(\overline{\Gamma}_{\mathrm{mix},k}^{(\mathrm{1PI})(2\phi,1\eta)}\right)^2 \left(\overline{G}^{(\phi)}_{k}\right)^2 \Big(-3 \overline{\Gamma}_{\mathrm{mix},k}^{(\mathrm{1PI})(4\phi)} \overline{G}^{(\phi)}_{k} \\
& \scriptstyle \hspace{0.75cm} + 2 \left(\overline{\Gamma}_{\mathrm{mix},k}^{(\mathrm{1PI})(2\phi,1\eta)}\right)^2 \overline{G}^{(\eta)}_{k} \left(\overline{G}^{(\eta)}_{k} + 3 \overline{G}^{(\phi)}_{k}\right)\Big)\Big)\bigg) \\
& \scriptstyle + \mathcal{F}_{5}^{(N=1)}\big(\overline{\Gamma}^{(\mathrm{1PI})(n\phi,m\eta)}_{\mathrm{mix},k}; \ n+m \leq 6\big) \;,
\end{split}
\label{eq:DiffEqGam22N1mixed1PIFRG0DONappendix}
\end{equation}
and for $N=2$:
\begin{equation}
\scriptstyle \dot{\overline{\Gamma}}^{(\mathrm{1PI})}_{\mathrm{mix},k} = \ \left(\overline{G}^{(\phi)}_{k}-\overline{G}^{(\phi)(0)}_{k}\right) + \frac{1}{2}\left(\overline{G}^{(\eta)}_{k}-\overline{G}^{(\eta)(0)}_{k}\right) \;,
\label{eq:DiffEqGam00N2mixed1PIFRG0DONappendix}
\end{equation}
\begin{equation}
\scriptstyle \dot{\overline{\eta}}_{k} = \frac{\dot{R}_{k}}{2\overline{\Gamma}^{(\mathrm{1PI})(2\eta)}_{\mathrm{mix},k}} \left(\overline{\Gamma}^{(\mathrm{1PI})(3\eta)}_{\mathrm{mix},k} \left(\overline{G}^{(\eta)}_{k}\right)^2 + 2 \overline{\Gamma}^{(\mathrm{1PI})(2\phi,1\eta)}_{\mathrm{mix},k} \left(\overline{G}^{(\phi)}_{k}\right)^2\right) \;,
\label{eq:DiffEqetaN2mixed1PIFRG0DONappendix}
\end{equation}
\begin{equation}
\scriptstyle \dot{\overline{\Gamma}}^{(\mathrm{1PI})(2\phi)}_{\mathrm{mix},k} = \ \dot{\overline{\eta}}_{k}\overline{\Gamma}_{\mathrm{mix},k}^{(\mathrm{1PI})(2\phi,1\eta)} -\frac{1}{6} \dot{R}_{k} \bigg(3 \overline{\Gamma}^{(\mathrm{1PI})(2\phi,2\eta)}_{\mathrm{mix},k} \left(\overline{G}^{(\eta)}_{k}\right)^2 + 4 \overline{\Gamma}_{\mathrm{mix},k}^{(\mathrm{1PI})(4\phi)} \left(\overline{G}^{(\phi)}_{k}\right)^2 - 6 \left(\overline{\Gamma}_{\mathrm{mix},k}^{(\mathrm{1PI})(2\phi,1\eta)}\right)^2 \overline{G}^{(\eta)}_{k} \overline{G}^{(\phi)}_{k} \left(\overline{G}^{(\eta)}_{k} + \overline{G}^{(\phi)}_{k}\right)\bigg) \;,
\label{eq:DiffEqGam20N2mixed1PIFRG0DONappendix}
\end{equation}
\begin{equation}
\scriptstyle \dot{\overline{\Gamma}}^{(\mathrm{1PI})(2\eta)}_{\mathrm{mix},k} = \ \dot{\overline{\eta}}_{k}\overline{\Gamma}_{\mathrm{mix},k}^{(\mathrm{1PI})(3\eta)} + \dot{R}_{k} \left(-\frac{1}{2} \overline{\Gamma}^{(\mathrm{1PI})(4\eta)}_{\mathrm{mix},k} \left(\overline{G}^{(\eta)}_{k}\right)^2 + \left(\overline{\Gamma}^{(\mathrm{1PI})(3\eta)}_{\mathrm{mix},k}\right)^2 \left(\overline{G}^{(\eta)}_{k}\right)^3 - \overline{\Gamma}^{(\mathrm{1PI})(2\phi,2\eta)}_{\mathrm{mix},k} \left(\overline{G}^{(\phi)}_{k}\right)^2 + 2 \left(\overline{\Gamma}^{(\mathrm{1PI})(2\phi,1\eta)}_{\mathrm{mix},k}\right)^2 \left(\overline{G}^{(\phi)}_{k}\right)^3\right) \;,
\label{eq:DiffEqGam02N2mixed1PIFRG0DONappendix}
\end{equation}
\begin{equation}
\begin{split}
\scriptstyle \dot{\overline{\Gamma}}^{(\mathrm{1PI})(2\phi,1\eta)}_{\mathrm{mix},k} = \ & \scriptstyle \dot{\overline{\eta}}_{k}\overline{\Gamma}_{\mathrm{mix},k}^{(\mathrm{1PI})(2\phi,2\eta)} + \dot{R}_{k} \overline{\Gamma}^{(\mathrm{1PI})(3\eta)}_{\mathrm{mix},k} \left(\overline{G}^{(\eta)}_{k}\right)^2 \bigg( \overline{\Gamma}^{(\mathrm{1PI})(2\phi,2\eta)}_{\mathrm{mix},k} \overline{G}^{(\eta)}_{k} - \left(\overline{\Gamma}^{(\mathrm{1PI})(2\phi,1\eta)}_{\mathrm{mix},k}\right)^2 \overline{G}^{(\phi)}_{k} \left(2 \overline{G}^{(\eta)}_{k} + \overline{G}^{(\phi)}_{k}\right)\bigg) \\
& \scriptstyle + \frac{1}{3} \dot{R}_{k} \overline{\Gamma}^{(\mathrm{1PI})(2\phi,1\eta)}_{\mathrm{mix},k} \overline{G}^{(\phi)}_{k} \bigg(6 \overline{\Gamma}^{(\mathrm{1PI})(2\phi,2\eta)}_{\mathrm{mix},k} \overline{G}^{(\eta)}_{k} \left( \overline{G}^{(\eta)}_{k} + \overline{G}^{(\phi)}_{k}\right) + \overline{G}^{(\phi)}_{k} \Big(4 \overline{\Gamma}^{(\mathrm{1PI})(4\phi)}_{\mathrm{mix},k} \overline{G}^{(\phi)}_{k} \\
& \scriptstyle \hspace{3.2cm} - 3 \left(\overline{\Gamma}^{(\mathrm{1PI})(2\phi,1\eta)}_{\mathrm{mix},k}\right)^2 \overline{G}^{(\eta)}_{k} \left(\overline{G}^{(\eta)}_{k} + 2 \overline{G}^{(\phi)}_{k}\right)\Big)\bigg) \\
& \scriptstyle + \mathcal{F}_{1}^{(N=2)}\big(\overline{\Gamma}^{(\mathrm{1PI})(n\phi,m\eta)}_{\mathrm{mix},k}; \ n+m \leq 5\big) \;,
\end{split}
\label{eq:DiffEqGam21N2mixed1PIFRG0DONappendix}
\end{equation}
\begin{equation}
\begin{split}
\scriptstyle \dot{\overline{\Gamma}}^{(\mathrm{1PI})(3\eta)}_{\mathrm{mix},k} = & \ \scriptstyle \dot{\overline{\eta}}_{k}\overline{\Gamma}_{\mathrm{mix},k}^{(\mathrm{1PI})(4\eta)} \\
& \scriptstyle + \dot{R}_{k} \bigg(3 \overline{\Gamma}^{(\mathrm{1PI})(3\eta)}_{\mathrm{mix},k} \overline{\Gamma}^{(\mathrm{1PI})(4\eta)}_{\mathrm{mix},k} \left(\overline{G}^{(\eta)}_{k}\right)^3 - 3 \left(\overline{\Gamma}^{(\mathrm{1PI})(3\eta)}_{\mathrm{mix},k}\right)^3 \left(\overline{G}^{(\eta)}_{k}\right)^4 + 6 \overline{\Gamma}^{(\mathrm{1PI})(2\phi,1\eta)}_{\mathrm{mix},k} \left(\overline{G}^{(\phi)}_{k}\right)^3 \Big(\overline{\Gamma}^{(\mathrm{1PI})(2\phi,2\eta)}_{\mathrm{mix},k} \\
& \scriptstyle \hspace{0.75cm} - \left(\overline{\Gamma}^{(\mathrm{1PI})(2\phi,1\eta)}_{\mathrm{mix},k}\right)^2 \overline{G}^{(\phi)}_{k}\Big)\bigg) \\
& \scriptstyle + \mathcal{F}_{2}^{(N=2)}\big(\overline{\Gamma}^{(\mathrm{1PI})(n\phi,m\eta)}_{\mathrm{mix},k}; \ n+m \leq 5\big) \;,
\end{split}
\label{eq:DiffEqGam03N2mixed1PIFRG0DONappendix}
\end{equation}
\begin{equation}
\begin{split}
\scriptstyle \dot{\overline{\Gamma}}^{(\mathrm{1PI})(4\phi)}_{\mathrm{mix},k} = & \ \scriptstyle \frac{1}{3} \dot{R}_{k} \bigg(9 \left(\overline{\Gamma}^{(\mathrm{1PI})(2\phi,2\eta)}_{\mathrm{mix},k}\right)^2 \left(\overline{G}^{(\eta)}_{k}\right)^3 - 18 \left(\overline{\Gamma}^{(\mathrm{1PI})(2\phi,1\eta)}_{\mathrm{mix},k}\right)^2 \overline{\Gamma}^{(\mathrm{1PI})(2\phi,2\eta)}_{\mathrm{mix},k} \left(\overline{G}^{(\eta)}_{k}\right)^2 \overline{G}^{(\phi)}_{k} \left(2 \overline{G}^{(\eta)}_{k} + \overline{G}^{(\phi)}_{k}\right) \\
& \scriptstyle \hspace{0.9cm} + 2 \left(\overline{G}^{(\phi)}_{k}\right)^2 \Big(5 \left(\overline{\Gamma}^{(\mathrm{1PI})(4\phi)}_{\mathrm{mix},k}\right)^2 \overline{G}^{(\phi)}_{k} + 18 \left(\overline{\Gamma}^{(\mathrm{1PI})(2\phi,1\eta)}_{\mathrm{mix},k}\right)^4 \left(\overline{G}^{(\eta)}_{k}\right)^2 \left(\overline{G}^{(\eta)}_{k} + \overline{G}^{(\phi)}_{k}\right) \\
& \scriptstyle \hspace{0.9cm} - 9 \left(\overline{\Gamma}^{(\mathrm{1PI})(2\phi,1\eta)}_{\mathrm{mix},k}\right)^2 \overline{\Gamma}^{(\mathrm{1PI})(4\phi)}_{\mathrm{mix},k} \overline{G}^{(\eta)}_{k} \left(\overline{G}^{(\eta)}_{k} + 2 \overline{G}^{(\phi)}_{k}\right)\Big)\bigg) \\
& \scriptstyle + \mathcal{F}_{3}^{(N=2)}\big(\overline{\Gamma}^{(\mathrm{1PI})(n\phi,m\eta)}_{\mathrm{mix},k}; \ n+m \leq 6\big) \;,
\end{split}
\label{eq:DiffEqGam40N2mixed1PIFRG0DONappendix}
\end{equation}
\begin{equation}
\begin{split}
\scriptstyle \dot{\overline{\Gamma}}^{(\mathrm{1PI})(4\eta)}_{\mathrm{mix},k} = & \ \scriptstyle 3 \dot{R}_{k} \bigg(\left(\overline{\Gamma}^{(\mathrm{1PI})(4\eta)}_{\mathrm{mix},k}\right)^2 \left(\overline{G}^{(\eta)}_{k}\right)^3 - 6 \left(\overline{\Gamma}^{(\mathrm{1PI})(3\eta)}_{\mathrm{mix},k}\right)^2 \overline{\Gamma}^{(\mathrm{1PI})(4\eta)}_{\mathrm{mix},k} \left(\overline{G}^{(\eta)}_{k}\right)^4 + 4 \left(\overline{\Gamma}^{(\mathrm{1PI})(3\eta)}_{\mathrm{mix},k}\right)^4 \left(\overline{G}^{(\eta)}_{k}\right)^5 + 2 \left(\overline{G}^{(\phi)}_{k}\right)^3 \Big(\left(\overline{\Gamma}^{(\mathrm{1PI})(2\phi,2\eta)}_{\mathrm{mix},k}\right)^2 \\
& \scriptstyle \hspace{0.9cm} - 6 \left(\overline{\Gamma}^{(\mathrm{1PI})(2\phi,1\eta)}_{\mathrm{mix},k}\right)^2 \overline{\Gamma}^{(\mathrm{1PI})(2\phi,2\eta)}_{\mathrm{mix},k} \overline{G}^{(\phi)}_{k} + 4 \left(\overline{\Gamma}^{(\mathrm{1PI})(2\phi,1\eta)}_{\mathrm{mix},k}\right)^4 \left(\overline{G}^{(\phi)}_{k}\right)^2\Big)\bigg) \\
& \scriptstyle + \mathcal{F}_{4}^{(N=2)}\big(\overline{\Gamma}^{(\mathrm{1PI})(n\phi,m\eta)}_{\mathrm{mix},k}; \ n+m \leq 6\big) \;,
\end{split}
\label{eq:DiffEqGam04N2mixed1PIFRG0DONappendix}
\end{equation}
\begin{equation}
\begin{split}
\scriptstyle \dot{\overline{\Gamma}}^{(\mathrm{1PI})(2\phi,2\eta)}_{\mathrm{mix},k} = & \ \scriptstyle \frac{1}{3} \dot{R}_{k} \bigg(12 \overline{\Gamma}^{(\mathrm{1PI})(3\eta)}_{\mathrm{mix},k} \overline{\Gamma}^{(\mathrm{1PI})(2\phi,1\eta)}_{\mathrm{mix},k} \left(\overline{G}^{(\eta)}_{k}\right)^2 \overline{G}^{(\phi)}_{k} \left(\left(\overline{\Gamma}^{(\mathrm{1PI})(2\phi,1\eta)}_{\mathrm{mix},k}\right)^2 \overline{G}^{(\phi)}_{k} \left(\overline{G}^{(\eta)}_{k} + \overline{G}^{(\phi)}_{k}\right) - \overline{\Gamma}^{(\mathrm{1PI})(2\phi,2\eta)}_{\mathrm{mix},k} \left(2 \overline{G}^{(\eta)}_{k} + \overline{G}^{(\phi)}_{k}\right)\right) \\
& \scriptstyle \hspace{0.9cm} + 3 \overline{\Gamma}^{(\mathrm{1PI})(4\eta)}_{\mathrm{mix},k} \left(\overline{G}^{(\eta)}_{k}\right)^2 \left(\overline{\Gamma}^{(\mathrm{1PI})(2\phi,2\eta)}_{\mathrm{mix},k} \overline{G}^{(\eta)}_{k} - \left(\overline{\Gamma}^{(\mathrm{1PI})(2\phi,1\eta)}_{\mathrm{mix},k}\right)^2 \overline{G}^{(\phi)}_{k} \left(2 \overline{G}^{(\eta)}_{k} + \overline{G}^{(\phi)}_{k}\right)\right) \\
& \scriptstyle \hspace{0.9cm} + 3 \left(\overline{\Gamma}^{(\mathrm{1PI})(3\eta)}_{\mathrm{mix},k}\right)^2 \left(\overline{G}^{(\eta)}_{k}\right)^3 \left(-3 \overline{\Gamma}^{(\mathrm{1PI})(2\phi,2\eta)}_{\mathrm{mix},k} \overline{G}^{(\eta)}_{k} + 2 \left(\overline{\Gamma}^{(\mathrm{1PI})(2\phi,1\eta)}_{\mathrm{mix},k}\right)^2 \overline{G}^{(\phi)}_{k} \left(3 \overline{G}^{(\eta)}_{k} + \overline{G}^{(\phi)}_{k}\right)\right) \\
& \scriptstyle \hspace{0.9cm} + \overline{G}^{(\phi)}_{k} \Big(6 \left(\overline{\Gamma}^{(\mathrm{1PI})(2\phi,2\eta)}_{\mathrm{mix},k}\right)^2 \overline{G}^{(\eta)}_{k} \left(\overline{G}^{(\eta)}_{k} + \overline{G}^{(\phi)}_{k}\right) + \overline{\Gamma}^{(\mathrm{1PI})(2\phi,2\eta)}_{\mathrm{mix},k} \overline{G}^{(\phi)}_{k} \Big(4 \overline{\Gamma}^{(\mathrm{1PI})(4\phi)}_{\mathrm{mix},k} \overline{G}^{(\phi)}_{k} \\
& \scriptstyle \hspace{0.9cm} - 15 \left(\overline{\Gamma}^{(\mathrm{1PI})(2\phi,1\eta)}_{\mathrm{mix},k}\right)^2 \overline{G}^{(\eta)}_{k} \left( \overline{G}^{(\eta)}_{k} + 2 \overline{G}^{(\phi)}_{k}\right)\Big) + 6 \left(\overline{\Gamma}^{(\mathrm{1PI})(2\phi,1\eta)}_{\mathrm{mix},k}\right)^2 \left(\overline{G}^{(\phi)}_{k}\right)^2 \Big(-2 \overline{\Gamma}^{(\mathrm{1PI})(4\phi)}_{\mathrm{mix},k} \overline{G}^{(\phi)}_{k} \\
& \scriptstyle \hspace{0.9cm} + \left(\overline{\Gamma}^{(\mathrm{1PI})(2\phi,1\eta)}_{\mathrm{mix},k}\right)^2 \overline{G}^{(\eta)}_{k} \left( \overline{G}^{(\eta)}_{k} + 3 \overline{G}^{(\phi)}_{k}\right)\Big)\Big)\bigg) \\
& \scriptstyle + \mathcal{F}_{5}^{(N=2)}\big(\overline{\Gamma}^{(\mathrm{1PI})(n\phi,m\eta)}_{\mathrm{mix},k}; \ n+m \leq 6\big) \;,
\end{split}
\label{eq:DiffEqGam22N2mixed1PIFRG0DONappendix}
\end{equation}
where the propagators $\overline{G}^{(\phi)(0)}_{k}$ and $\overline{G}^{(\eta)(0)}_{k}$ are introduced for the same reason as for $\overline{G}^{(0)}_{k}$ in the framework of the original theory. The functions $\mathcal{F}_{n}^{(N)}$, involved in~\eqref{eq:DiffEqGam21N1mixed1PIFRG0DONappendix} to~\eqref{eq:DiffEqGam22N1mixed1PIFRG0DONappendix} as well as in~\eqref{eq:DiffEqGam21N2mixed1PIFRG0DONappendix} to~\eqref{eq:DiffEqGam22N2mixed1PIFRG0DONappendix}, depend on vertex functions of order 5 (and 6 for $n \geq 3$) so that they do not contribute to the above equations for $N_{\mathrm{max}} \leq 4$, according to the initial conditions set by~\eqref{eq:CIGammaOrder4ormoremix1PIFRG0DON} in section~\ref{sec:1PIFRG0DON}.

%% file: 7Appendix/Derivations2PIFRG.tex
\section{Bosonic index formalism}
\label{ann:BosonicIndices2PI}

The formalism related to bosonic indices involves additional subtleties as compared to that based on the fermionic ones. This stems from the symmetry properties (given in~\eqref{eq:SymmetryW2PIFRG} and~\eqref{eq:SymmetryG2PIFRG} notably) of objects like $G_{\alpha\alpha'}$ or $K_{\alpha\alpha'}$. In particular, the identity matrix is constructed so as to exhibit such symmetries:
\begin{equation}
\mathcal{I}_{\gamma_{1}\gamma_{2}}=\mathcal{I}_{\gamma_{2}\gamma_{1}}\equiv\frac{\delta G_{\gamma_{1}}}{\delta G_{\gamma_{2}}}=\frac{\delta K_{\gamma_{1}}}{\delta K_{\gamma_{2}}}=\delta_{\alpha_{1}\alpha_{2}}\delta_{\alpha'_{1}\alpha'_{2}}+\zeta\delta_{\alpha_{1}\alpha'_{2}}\delta_{\alpha'_{1}\alpha_{2}}\;.
\label{eq:2PIfrgInverseBosonicIndices}
\end{equation}
An expression for the functional derivative $\frac{\delta W[K]}{\delta K_{\gamma}}$ in terms of the propagator $G_{\gamma}$ directly follows from definition~\eqref{eq:2PIfrgInverseBosonicIndices} and from the generating functional expressed by \eqref{eq:2PIFRGgeneratingFunc}:
\begin{equation}
\begin{split}
\frac{\delta W[K]}{\delta K_{\alpha_{1}\alpha'_{1}}} = & \ \frac{1}{Z[K]}\int\mathcal{D}\widetilde{\psi} \left[\frac{1}{2}\int_{\alpha_{2},\alpha'_{2}} \widetilde{\psi}_{\alpha_{2}}\frac{\delta K_{\alpha_{2}\alpha'_{2}}}{\delta K_{\alpha_{1}\alpha'_{1}}}\widetilde{\psi}_{\alpha'_{2}}\right] e^{-S\big[\widetilde{\psi}\big] + \frac{1}{2}\int_{\alpha_{2},\alpha'_{2}}\widetilde{\psi}_{\alpha_{2}}K_{\alpha_{2}\alpha'_{2}}\widetilde{\psi}_{\alpha'_{2}}} \\
= & \ \frac{1}{Z[K]}\int\mathcal{D}\widetilde{\psi} \left[\frac{1}{2} \int_{\alpha_{2},\alpha'_{2}} \widetilde{\psi}_{\alpha_{2}}\left(\delta_{\alpha_{2}\alpha_{1}}\delta_{\alpha'_{2}\alpha'_{1}}+\zeta\delta_{\alpha_{2}\alpha'_{1}}\delta_{\alpha'_{2}\alpha_{1}}\right)\widetilde{\psi}_{\alpha'_{2}}\right] e^{-S\big[\widetilde{\psi}\big] + \frac{1}{2}\int_{\alpha_{2},\alpha'_{2}}\widetilde{\psi}_{\alpha_{2}}K_{\alpha_{2}\alpha'_{2}}\widetilde{\psi}_{\alpha'_{2}}} \\
= & \ \frac{1}{Z[K]}\int\mathcal{D}\widetilde{\psi} \ \frac{1}{2} \Big(\widetilde{\psi}_{\alpha_{1}}\widetilde{\psi}_{\alpha'_{1}}+\zeta\underbrace{\widetilde{\psi}_{\alpha'_{1}}\widetilde{\psi}_{\alpha_{1}}}_{\zeta\widetilde{\psi}_{\alpha_{1}}\widetilde{\psi}_{\alpha'_{1}}}\Big) \ e^{-S\big[\widetilde{\psi}\big] + \frac{1}{2}\int_{\alpha_{2},\alpha'_{2}}\widetilde{\psi}_{\alpha_{2}} K_{\alpha_{2}\alpha'_{2}} \widetilde{\psi}_{\alpha'_{2}}} \\
= & \ \frac{1}{Z[K]}\int\mathcal{D}\widetilde{\psi} \ \widetilde{\psi}_{\alpha_{1}}\widetilde{\psi}_{\alpha'_{1}} \ e^{-S\big[\widetilde{\psi}\big] + \frac{1}{2}\int_{\alpha_{2},\alpha'_{2}}\widetilde{\psi}_{\alpha_{2}}K_{\alpha_{2}\alpha'_{2}}\widetilde{\psi}_{\alpha'_{2}}} \\
= & \ \left\langle\widetilde{\psi}_{\alpha_{1}}\widetilde{\psi}_{\alpha'_{1}}\right\rangle_{K} \\
= & \ G_{\alpha_{1}\alpha'_{1}} \;,
\end{split}
\label{eq:2PIfrgPropagatorBosonicIndicesWithDetails}
\end{equation}
or, in terms of bosonic indices,
\begin{equation}
\frac{\delta W[K]}{\delta K_{\gamma}} = G_{\gamma}\;.
\label{eq:2PIfrgPropagatorBosonicIndices}
\end{equation}

\pagebreak

Another important point is the matrix multiplication with respect to bosonic indices. For two bosonic matrices $M$ and $N$, this gives us:
\begin{equation}
\left(MN\right)_{\gamma_{1}\gamma_{2}} = \frac{1}{2}\int_{\gamma_{3}} M_{\gamma_{1}\gamma_{3}} N_{\gamma_{3}\gamma_{2}} = \frac{1}{2} \int_{\alpha_{3},\alpha'_{3}} M_{\gamma_{1}(\alpha_{3},\alpha'_{3})} N_{(\alpha_{3},\alpha'_{3})\gamma_{2}} = M_{\gamma_{1}\hat{\gamma}} N_{\hat{\gamma}\gamma_{2}} \;,
\label{eq:MatrixMultBosonicIndices}
\end{equation}
\begin{equation}
\mathrm{Tr}_{\gamma}(M) = \frac{1}{2} \int_{\gamma} M_{\gamma\gamma}\;.
\end{equation}
Hence, it involves an extra $1/2$ factor as compared to the standard matrix multiplication with respect to $\alpha$-indices, which is convenient considering the symmetry properties discussed above~\eqref{eq:2PIfrgInverseBosonicIndices}. For instance, the bosonic identity matrix involving two terms in its definition~\eqref{eq:2PIfrgInverseBosonicIndices} so as to satisfy those symmetries, such a $1/2$ factor is usually canceled out as follows:
\begin{equation}
\begin{split}
\frac{\delta}{\delta G_{\gamma_{1}}} \frac{1}{2}\int_{\gamma_{3}} G_{\gamma_{3}} M_{\gamma_{3}\gamma_{2}} = & \ \frac{1}{2}\int_{\gamma_{3}} \underbrace{\frac{\delta G_{\gamma_{3}}}{\delta G_{\gamma_{1}}}}_{\mathcal{I}_{\gamma_{1}\gamma_{3}}} M_{\gamma_{3}\gamma_{2}} \\
= & \ \frac{1}{2}\int_{\gamma_{3}} \left( \delta_{\alpha_{1}\alpha_{3}}\delta_{\alpha'_{1}\alpha'_{3}} + \zeta \delta_{\alpha_{1}\alpha'_{3}}\delta_{\alpha'_{1}\alpha_{3}} \right) M_{\gamma_{3}\gamma_{2}} \\
= & \ \frac{1}{2} \big( M_{\gamma_{1}\gamma_{2}} + \zeta \underbrace{M_{(\alpha'_{1},\alpha_{1})\gamma_{2}}}_{\zeta M_{(\alpha_{1},\alpha'_{1})\gamma_{2}}} \big) \\
= & \ M_{\gamma_{1} \gamma_{2}} \;,
\end{split}
\end{equation}
with $M$ being an arbitrary bosonic matrix (independent of $G$) satisfying the symmetry properties of~\eqref{eq:SymmetryW2PIFRG}. Note in addition that a specific realization of~\eqref{eq:MatrixMultBosonicIndices} is the chain rule:
\begin{equation}
\frac{\delta W[K]}{\delta G_{\gamma_{1}}}=\frac{1}{2}\int_{\gamma_{2}} \frac{\delta K_{\gamma_{2}}}{\delta G_{\gamma_{1}}} \frac{\delta W[K]}{\delta K_{\gamma_{2}}}\;.
\label{eq:2PIfrgchainRuleBosonicIndices}
\end{equation}
For $n$ bosonic matrices $M_{1}$,..., $M_{n}$,~\eqref{eq:MatrixMultBosonicIndices} can be generalized to:
\begin{equation}
\begin{split}
\left(M_{1} \cdots M_{n}\right)_{\gamma_{1}\gamma_{2}} = & \ \frac{1}{2^{n-1}} \int_{\gamma_{3},...,\gamma_{n+1}} M_{1,\gamma_{1}\gamma_{3}} \cdots M_{n,\gamma_{n+1}\gamma_{2}} \\
= & \ \frac{1}{2^{n-1}}\int_{\alpha_{3},\alpha'_{3},...,\alpha_{n+1},\alpha'_{n+1}} M_{1,\gamma_{1}(\alpha_{3},\alpha'_{3})} \cdots M_{n,(\alpha_{n+1},\alpha'_{n+1})\gamma_{2}} \\
= & \ M_{1,\gamma_{1}\hat{\gamma}_{1}} \cdots M_{n,\hat{\gamma}_{n-1}\gamma_{2}} \;.
\end{split}
\end{equation}
We will often exploit the commutativity of the matrix product based on bosonic indices, i.e. $\left(M_{1} M_{2}\right)_{\gamma_{1}\gamma_{2}}=\left(M_{2} M_{1}\right)_{\gamma_{1}\gamma_{2}}$. This property is verified assuming that the matrices involved in the product are symmetric, i.e. $M_{i,\gamma_{1}\gamma_{2}}=M_{i,\gamma_{2}\gamma_{1}}$ with $i=1~\mathrm{or}~2$ for our example. The latter property is indeed often exhibited by entities manipulated in this work, as shown by~\eqref{eq:SymmetryW2PIFRG} and~\eqref{eq:2PIFRGUflowSymmetryPairPropagator}. The inverse with respect to bosonic indices can be defined from~\eqref{eq:2PIfrgInverseBosonicIndices} and~\eqref{eq:MatrixMultBosonicIndices}:
\begin{equation}
\mathcal{I}_{\gamma_{1}\gamma_{2}} = \left(M M^{\mathrm{inv}}\right)_{\gamma_{1}\gamma_{2}} = \frac{1}{2}\int_{\gamma_{3}} M_{\gamma_{1}\gamma_{3}} M^{\mathrm{inv}}_{\gamma_{3}\gamma_{2}}\;.
\label{eq:2PIfrgBosonicIndicesInverse}
\end{equation}

\vspace{0.5cm}

We will also evaluate the derivative $\frac{\delta\Gamma^{(\mathrm{2PI})}[G]}{\delta G_{\gamma}}$ by considering the definition of the 2PI EA given by~\eqref{eq:LegendreTransform2PIeffActionWithTrace} and recalled below:
\begin{equation}
\Gamma^{(\mathrm{2PI})}[G] = -W[K] + \mathrm{Tr}_{\gamma}(KG)\;.
\label{eq:LegendreTransform2PIeffActionWithTraceAppendix}
\end{equation}
Using the chain rule expressed by~\eqref{eq:2PIfrgchainRuleBosonicIndices}, we then differentiate both sides of~\eqref{eq:LegendreTransform2PIeffActionWithTraceAppendix} with respect to $G$:
\begin{equation}
\begin{split}
\scalebox{0.98}{${\displaystyle\frac{\delta\Gamma^{(\mathrm{2PI})}[G]}{\delta G_{\alpha_{1}\alpha'_{1}}} =}$} & \scalebox{0.98}{${\displaystyle-\frac{\delta W[K]}{\delta G_{\alpha_{1}\alpha'_{1}}} + \frac{1}{2}\int_{\gamma_{2}} \frac{\delta K_{\gamma_{2}}}{\delta G_{\alpha_{1}\alpha'_{1}}} G_{\gamma_{2}} + \frac{1}{2}\int_{\alpha_{2},\alpha'_{2}} K_{\alpha_{2}\alpha'_{2}} \frac{\delta G_{\alpha_{2}\alpha'_{2}}}{\delta G_{\alpha_{1}\alpha'_{1}}} }$} \\
\scalebox{0.98}{${\displaystyle = }$} & \scalebox{0.98}{${\displaystyle -\frac{1}{2}\int_{\gamma_{2}} \frac{\delta K_{\gamma_{2}}}{\delta G_{\alpha_{1}\alpha'_{1}}} \underbrace{\frac{\delta W[K]}{\delta K_{\gamma_{2}}}}_{G_{\gamma_{2}}} + \frac{1}{2}\int_{\gamma_{2}} \frac{\delta K_{\gamma_{2}}}{\delta G_{\alpha_{1}\alpha'_{1}}} G_{\gamma_{2}} + \frac{1}{2}\int_{\alpha_{2},\alpha'_{2}} K_{\alpha_{2}\alpha'_{2}} \left(\delta_{\alpha_{2}\alpha_{1}}\delta_{\alpha'_{2}\alpha'_{1}}+\zeta\delta_{\alpha_{2}\alpha'_{1}}\delta_{\alpha'_{2}\alpha_{1}}\right) }$} \\
\scalebox{0.98}{${\displaystyle = }$} & \ \scalebox{0.98}{${\displaystyle \frac{1}{2}\big(K_{\alpha_{1}\alpha'_{1}} + \zeta \underbrace{K_{\alpha'_{1}\alpha_{1}}}_{\zeta K_{\alpha_{1}\alpha'_{1}}}\big) }$} \\
\scalebox{0.98}{${\displaystyle = }$} & \ \scalebox{0.98}{${\displaystyle K_{\alpha_{1}\alpha'_{1}} \;, }$}
\end{split}
\end{equation}
and, in terms of bosonic indices,
\begin{equation}
\frac{\delta\Gamma^{(\mathrm{2PI})}[G]}{\delta G_{\gamma}} = K_{\gamma}\;.
\label{eq:2PIfrgDGammaDG}
\end{equation}

\section{Dyson equation}
\label{ann:DysonEq}

As a next step, we derive an expression for the free 2PI EA $\Gamma^{(\mathrm{2PI})}_{0}[G]$ starting from the generating functional expressed by~\eqref{eq:2PIFRGgeneratingFunc}:
\begin{equation}
Z[K]=e^{W[K]}=\int\mathcal{D}\widetilde{\psi} \ e^{-S\big[\widetilde{\psi}\big] + \frac{1}{2}\int_{\alpha,\alpha'}\widetilde{\psi}_{\alpha}K_{\alpha\alpha'}\widetilde{\psi}_{\alpha'}}\;.
\end{equation}
More specifically, the free part $S_{0}$ of action $S$ can be written explicitly in terms of the free propagator $C$ as formulated by~\eqref{eq:2PIFRGmostgeneralactionS}:
\begin{equation}
S\Big[\widetilde{\psi}\Big]=S_{0}\Big[\widetilde{\psi}\Big]+S_{\mathrm{int}}\Big[\widetilde{\psi}\Big]=\frac{1}{2}\int_{\alpha,\alpha'}\widetilde{\psi}_{\alpha} C^{-1}_{\alpha\alpha'}\widetilde{\psi}_{\alpha'} + S_{\mathrm{int}}\Big[\widetilde{\psi}\Big]\;.
\end{equation}
From this, we infer that the corresponding free generating functional is given by:
\begin{equation}
\begin{split}
Z_{0}[K]=e^{W_{0}[K]}= & \int\mathcal{D}\widetilde{\psi} \ e^{-S_{0}\big[\widetilde{\psi}\big] + \frac{1}{2}\int_{\alpha,\alpha'}\widetilde{\psi}_{\alpha}K_{\alpha\alpha'}\widetilde{\psi}_{\alpha'}} \\
= & \int\mathcal{D}\widetilde{\psi} \ e^{-\frac{1}{2}\int_{\alpha,\alpha'}\widetilde{\psi}_{\alpha} \left(C^{-1}-K\right)_{\alpha\alpha'}\widetilde{\psi}_{\alpha'}} \\
= & \left[\mathrm{Det}\big(C^{-1}-K\big)\right]^{-\zeta/2} \;,
\end{split}
\end{equation}
where we have used \eqref{eq:GaussianArbitraryComplexfield} to derive the last line. For later purposes, it will be more convenient to replace the functional determinant so as to obtain:
\begin{equation}
W_{0}[K]=-\frac{\zeta}{2} \mathrm{Tr}_{\alpha} \left[ \mathrm{ln}\big(C^{-1}-K\big) \right] \;.
\label{eq:2PIfreegeneratingFuncTrLog}
\end{equation}
The propagator $G_{0}$ can be deduced from the free generating functional $W_{0}[K]$ as follows:
\begin{equation}
\begin{split}
G_{0,\alpha_{1}\alpha'_{1}} = \ \frac{\delta W_{0}[K]}{\delta K_{\alpha_{1}\alpha'_{1}}} = & \ \frac{\zeta}{2} \int_{\alpha_{2},\alpha'_{2}} \left(C^{-1}-K\right)^{-1}_{\alpha_{2}\alpha'_{2}} \frac{\delta K_{\alpha'_{2}\alpha_{2}}}{\delta K_{\alpha_{1}\alpha'_{1}}} \\
= & \ \frac{\zeta}{2} \int_{\alpha_{2},\alpha'_{2}} \left(C^{-1}-K\right)^{-1}_{\alpha_{2}\alpha'_{2}} \left(\delta_{\alpha'_{2}\alpha_{1}}\delta_{\alpha_{2}\alpha'_{1}}+\zeta\delta_{\alpha'_{2}\alpha'_{1}}\delta_{\alpha_{2}\alpha_{1}}\right) \\
= & \ \frac{\zeta}{2} \left[\left(C^{-1}-K\right)^{-1}_{\alpha'_{1}\alpha_{1}}+\zeta\left(C^{-1}-K\right)^{-1}_{\alpha_{1}\alpha'_{1}}\right] \\
= & \ \left(C^{-1}-K\right)^{-1}_{\alpha_{1}\alpha'_{1}} \;.
\end{split}
\label{eq:2PIfreegeneratingFuncTrLogNext}
\end{equation}
From \eqref{eq:2PIfreegeneratingFuncTrLog} and~\eqref{eq:2PIfreegeneratingFuncTrLogNext}, we infer the two useful relations:
\begin{subequations}\label{eq:free2PIeffectiveActionIntermediaryResults}
\begin{empheq}[left=\empheqlbrace]{align}
& W_{0}[K]=\frac{\zeta}{2} \mathrm{Tr}_{\alpha} \left[ \mathrm{ln}(G_{0}) \right] \;. \label{eq:free2PIeffectiveActionIntermediaryResultsUp}\\
\nonumber \\
& K=C^{-1}-G_{0}^{-1} \;. \label{eq:free2PIeffectiveActionIntermediaryResultsDown}
\end{empheq}
\end{subequations}
We then consider the definition of the 2PI EA given by~\eqref{eq:LegendreTransform2PIeffActionWithTrace} which reads in the free case:
\begin{equation}
\Gamma_{0}^{(\mathrm{2PI})}[G_{0}]=-W_{0}[K]+\mathrm{Tr}_{\gamma}(KG_{0})\;.
\end{equation}
With the help of~\eqref{eq:free2PIeffectiveActionIntermediaryResultsUp} and~\eqref{eq:free2PIeffectiveActionIntermediaryResultsDown} as well as the conventions related to bosonic indices (see appendix~\ref{ann:BosonicIndices2PI}), this relation can be rewritten as follows:
\begin{equation}
\begin{split}
\Gamma_{0}^{(\mathrm{2PI})}[G_{0}]= & -W_{0}[K] + G_{0,\hat{\gamma}} K_{\hat{\gamma}} \\
= & -\frac{\zeta}{2} \mathrm{Tr}_{\alpha} \left[ \mathrm{ln}(G_{0}) \right] + \frac{1}{2} \int_{\alpha,\alpha'} G_{0,\alpha\alpha'} \underbrace{K_{\alpha\alpha'}}_{\zeta K_{\alpha'\alpha}} \\
= & -\frac{\zeta}{2} \mathrm{Tr}_{\alpha} \left[ \mathrm{ln}(G_{0}) \right] + \frac{\zeta}{2} \mathrm{Tr}_{\alpha}(G_{0}K) \\
= & -\frac{\zeta}{2} \mathrm{Tr}_{\alpha} \left[ \mathrm{ln}(G_{0}) \right] + \frac{\zeta}{2} \mathrm{Tr}_{\alpha}\left[G_{0}\left(C^{-1}-G_{0}^{-1}\right)\right]\;. \\
\end{split}
\end{equation}
The free part of the full 2PI EA can thus be defined as:
\begin{equation}
\Gamma_{0}^{(\mathrm{2PI})}[G]=-\frac{\zeta}{2} \mathrm{Tr}_{\alpha} \left[ \mathrm{ln}(G) \right] + \frac{\zeta}{2} \mathrm{Tr}_{\alpha}\left[GC^{-1}-\mathbb{I}\right]\;,
\label{eq:ExpressionFree2PIeffectiveAction}
\end{equation}
which could also have been deduced from the previous result~\eqref{eq:2PIEAzerovevfinalexpression} derived via the IM. Finally, in order to deduce Dyson equation from~\eqref{eq:ExpressionFree2PIeffectiveAction}, we combine the latter equality with the definition of the Luttinger-Ward functional established by~\eqref{eq:DefLWfunctional}:
\begin{equation}
\Phi[G]\equiv\Gamma^{(\mathrm{2PI})}[G]-\Gamma_{0}^{(\mathrm{2PI})}[G]=\Gamma^{(\mathrm{2PI})}[G] + \frac{\zeta}{2} \mathrm{Tr}_{\alpha} \left[ \mathrm{ln}(G) \right] - \frac{\zeta}{2} \mathrm{Tr}_{\alpha}\left[GC^{-1}-\mathbb{I}\right]\;.
\end{equation}
By differentiating this relation with respect to $G$, we obtain:
\begin{equation}
\begin{split}
\Sigma_{\gamma_{1}}[G] \equiv -\frac{\delta\Phi[G]}{\delta G_{\gamma_{1}}} = & - \underbrace{\frac{\delta \Gamma^{(\mathrm{2PI})}[G]}{\delta G_{\gamma_{1}}}}_{K_{\gamma_{1}}} - \frac{\zeta}{2} \int_{\alpha_{2},\alpha'_{2}} G^{-1}_{\alpha_{2}\alpha'_{2}} \frac{\delta G_{\alpha'_{2}\alpha_{2}}}{\delta G_{\alpha_{1}\alpha'_{1}}} + \frac{\zeta}{2} \int_{\alpha_{2},\alpha'_{2}} \frac{\delta G_{\alpha_{2}\alpha'_{2}}}{\delta G_{\alpha_{1}\alpha'_{1}}} C^{-1}_{\alpha'_{2}\alpha_{2}} \\
= & - K_{\gamma_{1}} - \frac{\zeta}{2} \int_{\alpha_{2},\alpha'_{2}} G^{-1}_{\alpha_{2}\alpha'_{2}} \left(\delta_{\alpha'_{2}\alpha_{1}}\delta_{\alpha_{2}\alpha'_{1}}+\zeta\delta_{\alpha'_{2}\alpha'_{1}}\delta_{\alpha_{2}\alpha_{1}}\right) \\
& + \frac{\zeta}{2} \int_{\alpha_{2},\alpha'_{2}} \left(\delta_{\alpha_{2}\alpha_{1}}\delta_{\alpha'_{2}\alpha'_{1}}+\zeta\delta_{\alpha_{2}\alpha'_{1}}\delta_{\alpha'_{2}\alpha_{1}}\right) C^{-1}_{\alpha'_{2}\alpha_{2}} \\
= & - K_{\gamma_{1}} - \frac{\zeta}{2} \Big(\underbrace{G^{-1}_{\alpha'_{1}\alpha_{1}}}_{\zeta G^{-1}_{\alpha_{1}\alpha'_{1}}}+\zeta G^{-1}_{\alpha_{1}\alpha'_{1}}\Big) + \frac{\zeta}{2} \big(\underbrace{C^{-1}_{\alpha'_{1}\alpha_{1}}}_{\zeta C^{-1}_{\alpha_{1}\alpha'_{1}}} + \zeta C^{-1}_{\alpha_{1}\alpha'_{1}}\big) \\
= & - K_{\gamma_{1}} - G^{-1}_{\alpha_{1}\alpha'_{1}} + C^{-1}_{\alpha_{1}\alpha'_{1}}\;,
\end{split}
\end{equation}
which is equivalent to Dyson equation in the form:
\begin{equation}
G^{-1}_{\gamma} = C^{-1}_{\gamma} - \Sigma_{\gamma}[G] - K_{\gamma}\;.
\label{eq:DysonEqAppendix}
\end{equation}
Note also that, as can be deduced from~\eqref{eq:2PIfreegeneratingFuncTrLogNext} and~\eqref{eq:DysonEqAppendix}, we can define the free propagator $G_{0}$ as $G_{0} \equiv G[\Sigma=0]$.

\section{Bethe-Salpeter equation}
\label{ann:BetheSalpeterEq}

We first consider the definition of the free propagator $G_{0}$ used in~\eqref{eq:2PIfreegeneratingFuncTrLogNext}:
\begin{equation}
G_{0,\gamma} = W^{(1)}_{0,\gamma}[K] = \frac{\delta W_{0}[K]}{\delta K_{\gamma}}\;.
\end{equation}
We then determine an expression for $W^{(2)}_{0}[K]$ by differentiating the latter equality with respect to $K$:
\begin{equation}
\begin{split}
W^{(2)}_{0,\gamma_{1}\gamma_{2}}[K(G_{0})] = & \ \frac{\delta^{2} W_{0}[K(G_{0})]}{\delta K_{\gamma_{1}} \delta K_{\gamma_{2}}} \\
= & \ \frac{\delta G_{0,\gamma_{2}}}{\delta K_{\gamma_{1}}} \\
= & \int_{\alpha_{3},\alpha'_{3}} G_{0,\alpha_{2} \alpha_{3}} \frac{\delta K_{\alpha_{3} \alpha'_{3}}}{\delta K_{\alpha_{1} \alpha'_{1}}} G_{0,\alpha'_{3} \alpha'_{2}} \\
= & \int_{\alpha_{3},\alpha'_{3}} G_{0,\alpha_{2} \alpha_{3}} \left(\delta_{\alpha_{3} \alpha_{1}}\delta_{\alpha'_{3} \alpha'_{1}}+\zeta\delta_{\alpha_{3} \alpha'_{1}}\delta_{\alpha'_{3} \alpha_{1}}\right) G_{0,\alpha'_{3} \alpha'_{2}} \\
= & \ \underbrace{G_{0,\alpha_{2} \alpha_{1}}}_{\zeta G_{0,\alpha_{1} \alpha_{2}}} G_{0,\alpha'_{1} \alpha'_{2}} + \zeta \underbrace{G_{0,\alpha_{2} \alpha'_{1}}}_{\zeta G_{0,\alpha'_{1} \alpha_{2}}} G_{0,\alpha_{1} \alpha'_{2}} \\
= & \ G_{0,\alpha_{1} \alpha'_{2}} G_{0,\alpha'_{1} \alpha_{2}} + \zeta G_{0,\alpha_{1} \alpha_{2}} G_{0,\alpha'_{1} \alpha'_{2}}\;.
\end{split}
\label{eq:2PIfrgDerivationPairPropagator}
\end{equation}
We have evaluated the derivative of $G_{0}$ with respect to $K$ in the above calculation by making use of the relation $G_{0}=\big(C^{-1}-K\big)^{-1}$ derived in~\eqref{eq:2PIfreegeneratingFuncTrLogNext}. We recall that the pair propagator $\Pi[G]$ is defined as:
\begin{equation}
\Pi[G] \equiv W^{(2)}_{0}[K(G)]\;.
\end{equation}
According to~\eqref{eq:2PIfrgDerivationPairPropagator}, this is equivalent to:
\begin{equation}
\Pi_{\gamma_{1} \gamma_{2}}[G] = G_{\alpha_{1} \alpha'_{2}} G_{\alpha'_{1} \alpha_{2}} + \zeta G_{\alpha_{1} \alpha_{2}} G_{\alpha'_{1} \alpha'_{2}} \;.
\label{eq:2PIfrgExpressionPairPropagator}
\end{equation}
From~\eqref{eq:2PIfrgExpressionPairPropagator} together with the definition of the inverse of a bosonic matrix set by~\eqref{eq:2PIfrgBosonicIndicesInverse}, we infer that the inverse pair propagator reads:
\begin{equation}
\Pi^{\mathrm{inv}}_{\gamma_{1} \gamma_{2}}[G] = G^{-1}_{\alpha_{1} \alpha'_{2}} G^{-1}_{\alpha'_{1} \alpha_{2}} + \zeta G^{-1}_{\alpha_{1} \alpha_{2}} G^{-1}_{\alpha'_{1} \alpha'_{2}} \;.
\end{equation}
This can be checked through the following calculation:
\begin{equation}
\begin{split}
\left(\Pi[G] \Pi^{\mathrm{inv}}[G]\right)_{\gamma_{1} \gamma_{2}} = & \ \Pi_{\gamma_{1} \hat{\gamma}_{3}}[G] \Pi^{\mathrm{inv}}_{\hat{\gamma}_{3} \gamma_{2}}[G] \\
= & \ \frac{1}{2} \int_{\gamma_{3}} \Pi_{\gamma_{1} \gamma_{3}}[G] \Pi^{\mathrm{inv}}_{\gamma_{3} \gamma_{2}}[G] \\
= & \ \frac{1}{2} \int_{\alpha_{3},\alpha'_{3}} \left(G_{\alpha_{1} \alpha'_{3}} G_{\alpha'_{1} \alpha_{3}} + \zeta G_{\alpha_{1} \alpha_{3}} G_{\alpha'_{1} \alpha'_{3}}\right) \left(G^{-1}_{\alpha_{3} \alpha'_{2}} G^{-1}_{\alpha'_{3} \alpha_{2}} + \zeta G^{-1}_{\alpha_{3} \alpha_{2}} G^{-1}_{\alpha'_{3} \alpha'_{2}}\right) \\
= & \ \frac{1}{2} \underbrace{\int_{\alpha_{3},\alpha'_{3}} G_{\alpha_{1} \alpha'_{3}} G_{\alpha'_{1} \alpha_{3}} G^{-1}_{\alpha_{3} \alpha'_{2}} G^{-1}_{\alpha'_{3} \alpha_{2}}}_{\delta_{\alpha_{1} \alpha_{2}}\delta_{\alpha'_{1} \alpha'_{2}}} + \frac{\zeta}{2} \underbrace{\int_{\alpha_{3},\alpha'_{3}} G_{\alpha_{1} \alpha'_{3}} G_{\alpha'_{1} \alpha_{3}} G^{-1}_{\alpha_{3} \alpha_{2}} G^{-1}_{\alpha'_{3} \alpha'_{2}}}_{\delta_{\alpha_{1} \alpha'_{2}}\delta_{\alpha'_{1} \alpha_{2}}} \\
& + \frac{\zeta}{2} \underbrace{\int_{\alpha_{3},\alpha'_{3}} G_{\alpha_{1} \alpha_{3}} G_{\alpha'_{1} \alpha'_{3}} G^{-1}_{\alpha_{3} \alpha'_{2}} G^{-1}_{\alpha'_{3} \alpha_{2}}}_{\delta_{\alpha_{1} \alpha'_{2}}\delta_{\alpha'_{1} \alpha_{2}}} + \frac{1}{2} \underbrace{\int_{\alpha_{3},\alpha'_{3}} G_{\alpha_{1} \alpha_{3}} G_{\alpha'_{1} \alpha'_{3}} G^{-1}_{\alpha_{3} \alpha_{2}} G^{-1}_{\alpha'_{3} \alpha'_{2}}}_{\delta_{\alpha_{1} \alpha_{2}}\delta_{\alpha'_{1} \alpha'_{2}}} \\
= & \ \delta_{\alpha_{1} \alpha_{2}}\delta_{\alpha'_{1} \alpha'_{2}} + \zeta \delta_{\alpha_{1} \alpha'_{2}}\delta_{\alpha'_{1} \alpha_{2}} \\
= & \ \mathcal{I}_{\gamma_{1}\gamma_{2}}\;.
\end{split}
\end{equation}
An important relation for the 2PI-FRG formalism is that linking $W^{(2)}[K]$ with the second-order derivative of the 2PI EA, i.e. $\Gamma^{(\mathrm{2PI})(2)}[G]$. In the bosonic index formalism, it can be derived from~\eqref{eq:2PIfrgInverseBosonicIndices},~\eqref{eq:2PIfrgPropagatorBosonicIndices},~\eqref{eq:2PIfrgBosonicIndicesInverse} and~\eqref{eq:2PIfrgDGammaDG} as follows:
\begin{equation}
\begin{split}
\mathcal{I}_{\gamma_{1}\gamma_{2}} \equiv \frac{\delta G_{\gamma_{1}}}{\delta G_{\gamma_{2}}} = \frac{\delta^{2} W[K]}{\delta G_{\gamma_{2}} \delta K_{\gamma_{1}}} = \frac{1}{2} \int_{\gamma_{3}} \frac{\delta K_{\gamma_{3}}}{\delta G_{\gamma_{2}}} \frac{\delta^{2} W[K]}{\delta K_{\gamma_{3}} \delta K_{\gamma_{1}}} = \frac{1}{2} \int_{\gamma_{3}} \frac{\delta^{2} \Gamma^{(\mathrm{2PI})}[G]}{\delta G_{\gamma_{2}} \delta G_{\gamma_{3}}} \frac{\delta^{2} W[K]}{\delta K_{\gamma_{3}} \delta K_{\gamma_{1}}}\;,
\end{split}
\end{equation}
which yields, in a more compact notation,
\begin{equation}
\mathcal{I}_{\gamma_{1}\gamma_{2}} = \left(\Gamma^{(\mathrm{2PI})(2)}[G] W^{(2)}[K] \right)_{\gamma_{1}\gamma_{2}}\;,
\end{equation}
or, equivalently,
\begin{equation}
W^{(2)}[K] = \left(\Gamma^{(\mathrm{2PI})(2)}[G]\right)^{\mathrm{inv}}\;.
\label{eq:2PIfrgW2Gamma2Interacting}
\end{equation}
In the free case, this reduces to:
\begin{equation}
W_{0}^{(2)}[K(G)] \equiv \Pi[G] = \left(\Gamma_{0}^{(\mathrm{2PI})(2)}[G]\right)^{\mathrm{inv}} \;.
\label{eq:2PIfrgW2Gamma2Free}
\end{equation}
Furthermore, we also know from the definition of the Luttinger-Ward functional given by~\eqref{eq:DefLWfunctional} that:
\begin{equation}
\Gamma^{(\mathrm{2PI})(2)}[G] = \Gamma_{0}^{(\mathrm{2PI})(2)}[G] + \Phi^{(2)}[G] \;.
\label{eq:2PIfrgGamma2LWfunc}
\end{equation}
Combining~\eqref{eq:2PIfrgW2Gamma2Interacting},~\eqref{eq:2PIfrgW2Gamma2Free} and~\eqref{eq:2PIfrgGamma2LWfunc} allows for writing:
\begin{equation}
\begin{split}
W^{(2)}[K] = & \ \left(\Gamma^{(\mathrm{2PI})(2)}[G]\right)^{\mathrm{inv}} \\
= & \ \left(\Gamma_{0}^{(\mathrm{2PI})(2)}[G] + \Phi^{(2)}[G]\right)^{\mathrm{inv}} \\
= & \ \left(\Pi^{\mathrm{inv}}[G] + \Phi^{(2)}[G]\right)^{\mathrm{inv}} \\
= & \ \Pi[G] \left(\mathcal{I} + \Pi[G] \Phi^{(2)}[G]\right)^{\mathrm{inv}} \;,
\end{split}
\label{eq:2PIfrgW2Gamma2Pi}
\end{equation}
where matrix multiplication (with respect to bosonic indices) is left implicit in the last line. We then consider the Taylor expansion:
\begin{equation}
\left(\mathcal{I}+M\right)^{\mathrm{inv}}= \sum_{n=0}^{\infty}(-1)^{n} M^{n}\;,
\label{eq:BosonicMatrixTaylorExpAppendix}
\end{equation}
where $M^{0}=\mathcal{I}$ and $M_{\gamma_{1} \gamma_{2}}^{n} = \frac{1}{2^{n-1}}\int_{\gamma_{3}...\gamma_{n+1}} M_{\gamma_{1} \gamma_{3}} \cdots M_{\gamma_{n+1} \gamma_{2}} = M_{\gamma_{1} \hat{\gamma}_{1}} \cdots M_{\hat{\gamma}_{n-1} \gamma_{2}}$. Sticking to the matrix notation used in the last line of~\eqref{eq:2PIfrgW2Gamma2Pi},~\eqref{eq:BosonicMatrixTaylorExpAppendix} enables us to rewrite~\eqref{eq:2PIfrgW2Gamma2Pi} as:
\begin{equation}
\begin{split}
W^{(2)}[K] = & \ \Pi[G] \sum_{n=0}^{\infty} (-1)^{n} \left(\Pi[G] \Phi^{(2)}[G]\right)^{n} \\
= & \ \Pi[G] + \Pi[G]\sum_{n=1}^{\infty} (-1)^{n} \left(\Pi[G] \Phi^{(2)}[G]\right)^{n} \\
= & \ \Pi[G] + \Pi[G] \underbrace{\Pi[G] \Phi^{(2)}[G]}_{\Phi^{(2)}[G] \Pi[G]} \sum_{n=1}^{\infty} (-1)^{n} \left(\Pi[G] \Phi^{(2)}[G]\right)^{n-1} \\
= & \ \Pi[G] - \Pi[G] \Phi^{(2)}[G] \underbrace{\Pi[G] \sum_{n=0}^{\infty} (-1)^{n} \left(\Pi[G] \Phi^{(2)}[G]\right)^{n}}_{W^{(2)}[K]} \\
= & \ \Pi[G] - \Pi[G] \Phi^{(2)}[G] W^{(2)}[K]\;,
\end{split}
\label{eq:DerivationBetheSalpeterEquationImplicit}
\end{equation}
where we have used the commutative property of the matrix product based on bosonic indices (see appendix~\ref{ann:BosonicIndices2PI}). Hence, we have just derived the Bethe-Salpeter equation:
\begin{equation}
W^{(2)}[K] = \Pi[G] - \Pi[G] \Phi^{(2)}[G] W^{(2)}[K]\;,
\label{eq:BetheSalpeterEqAppendix}
\end{equation}
which is equivalent to~\eqref{eq:2PIfrgW2Gamma2Interacting}, as we have just proven. Therefore, $W^{(2)}[K]$ can either be determined in a self-consistent manner from the Bethe-Salpeter equation in the form of~\eqref{eq:BetheSalpeterEqAppendix} or via the inversion of a bosonic matrix from~\eqref{eq:2PIfrgW2Gamma2Interacting}.

\section{Tower of flow equations}
\label{ann:2PIfrgFlowEquation}
\subsection{C-flow}
\label{ann:2PIfrgFlowEquationCflow}

The starting point of the derivation of the flow equations for the C-flow version of the 2PI-FRG is the generating functional $Z[K]$ in the form of~\eqref{eq:2PIFRGgeneratingFunc} combined with the classical action $S\big[\widetilde{\psi}\big]$ given by~\eqref{eq:2PIFRGmostgeneralactionS}:
\begin{equation}
Z[K]=e^{W[K]}=\int\mathcal{D}\widetilde{\psi} \ e^{-\frac{1}{2}\int_{\alpha,\alpha'}\widetilde{\psi}_{\alpha}C^{-1}_{\alpha \alpha'}\widetilde{\psi}_{\alpha'} - S_{\mathrm{int}}\big[\widetilde{\psi}\big] + \frac{1}{2}\int_{\alpha,\alpha'}\widetilde{\psi}_{\alpha}K_{\alpha \alpha'}\widetilde{\psi}_{\alpha'}}\;.
\label{eq:2PIFRGgeneratingFuncCflow}
\end{equation}
After performing the substitution $C^{-1} \rightarrow C^{-1}_{\mathfrak{s}} \equiv C^{-1} + R_{\mathfrak{s}}$ (or, equivalently, $C^{-1}_{\mathfrak{s}} \equiv R_{\mathfrak{s}} C^{-1}$) and differentiating both sides of~\eqref{eq:2PIFRGgeneratingFuncCflow} with respect to $\mathfrak{s}$ at a fixed configuration of the source $K$, we obtain:
\begin{equation}
\begin{split}
\left.\dot{W}_{\mathfrak{s}}[K]\right|_{K}= & -\frac{1}{2 Z_{\mathfrak{s}}[K]} \int_{\alpha,\alpha'}\dot{C}^{-1}_{\mathfrak{s},\alpha \alpha'} \int\mathcal{D}\widetilde{\psi} \ \widetilde{\psi}_{\alpha}\widetilde{\psi}_{\alpha'} \ e^{-\frac{1}{2}\int_{\alpha,\alpha'}\widetilde{\psi}_{\alpha}C^{-1}_{\mathfrak{s},\alpha \alpha'}\widetilde{\psi}_{\alpha'} - S_{\mathrm{int}}\big[\widetilde{\psi}\big] + \frac{1}{2}\int_{\alpha,\alpha'}\widetilde{\psi}_{\alpha}K_{\alpha\alpha'}\widetilde{\psi}_{\alpha'}} \\
= & -\frac{1}{2} \int_{\alpha,\alpha'}\dot{C}^{-1}_{\mathfrak{s},\alpha \alpha'} \underbrace{\left( \frac{1}{Z_{\mathfrak{s}}[K]} \int\mathcal{D}\widetilde{\psi} \ \widetilde{\psi}_{\alpha}\widetilde{\psi}_{\alpha'} \ e^{- S_{\mathfrak{s}}\big[\widetilde{\psi}\big] + \frac{1}{2}\int_{\alpha,\alpha'}\widetilde{\psi}_{\alpha}K_{\alpha\alpha'}\widetilde{\psi}_{\alpha'}} \right)}_{G_{\alpha\alpha'}} \\
= & - \dot{C}^{-1}_{\mathfrak{s},\hat{\gamma}} G_{\hat{\gamma}}\;,
\end{split}
\label{eq:2PIfrgCflowIntermediaryResult}
\end{equation}
where $G$ was introduced with the help of~\eqref{eq:GEqualGk2PIFRG}. Then, the procedure to deduce a flow equation for the 2PI EA from~\eqref{eq:2PIfrgCflowIntermediaryResult} is very similar to that used in appendix~\ref{sec:DerivMasterEq1PIFRG} to derive the Wetterich equation. Let us start by considering the chain rule relating derivatives with respect to the flow parameter at fixed source $K$ and that at fixed propagator $G$:
\begin{equation}
\left.\frac{\partial}{\partial \mathfrak{s}}\right|_{K} = \left.\frac{\partial}{\partial \mathfrak{s}}\right|_{G} + \frac{1}{2} \int_{\gamma} \left.\dot{G}_{\gamma}\right|_{K} \frac{\delta}{\delta G_{\gamma}}\;,
\label{eq:chainRule2PIFRGCflowAppendix}
\end{equation}
which is the counterpart of~\eqref{eq:ChainRule1PIFRG} for the 2PI EA. Applying this operator to the 2PI EA yields:
\begin{equation}
\begin{split}
\left.\dot{\Gamma}_{\mathfrak{s}}^{(\mathrm{2PI})}[G]\right|_{K} = & \ \left.\dot{\Gamma}_{\mathfrak{s}}^{(\mathrm{2PI})}[G]\right|_{G} + \frac{1}{2} \int_{\gamma} \left.\dot{G}_{\gamma}\right|_{K} \underbrace{\frac{\delta\Gamma_{\mathfrak{s}}^{(\mathrm{2PI})}[G]}{\delta G_{\gamma}}}_{K_{\gamma}} \\
= & \ \left.\dot{\Gamma}_{\mathfrak{s}}^{(\mathrm{2PI})}[G]\right|_{G} + \frac{1}{2} \int_{\gamma} \left.\dot{G}_{\gamma}\right|_{K} K_{\gamma} \;,
\end{split}
\label{eq:Gamma2PIdotKconstCflowAppendix}
\end{equation}
where we have used~\eqref{eq:2PIfrgDGammaDG} to introduce the source $K$. As a next step, we consider once again~\eqref{eq:LegendreTransform2PIeffActionWithTrace}:
\begin{equation}
\Gamma_{\mathfrak{s}}^{(\mathrm{2PI})}[G] = -W_{\mathfrak{s}}[K] + \mathrm{Tr}_{\gamma}(K G)\;,
\label{eq:LegendreTransform2PIeffActionCflow}
\end{equation}
that we rewrite via differentiation as follows:
\begin{equation}
\left.\dot{\Gamma}_{\mathfrak{s}}^{(\mathrm{2PI})}[G]\right|_{K} = -\left. \dot{W}_{\mathfrak{s}}[K]\right|_{K} + \frac{1}{2} \int_{\gamma} K_{\gamma} \left. \dot{G}_{\gamma} \right|_{K}\;.
\end{equation}
The latter relation is equivalent to:
\begin{equation}
\underbrace{\left.\dot{\Gamma}_{\mathfrak{s}}^{(\mathrm{2PI})}[G]\right|_{K} - \frac{1}{2} \int_{\gamma} K_{\gamma} \left. \dot{G}_{\gamma} \right|_{K}}_{\left.\dot{\Gamma}_{\mathfrak{s}}^{(\mathrm{2PI})}[G]\right|_{G}} = -\left. \dot{W}_{\mathfrak{s}}[K]\right|_{K}\;,
\end{equation}
where we are now in a position to exploit~\eqref{eq:Gamma2PIdotKconstCflowAppendix}, as indicated by the underbrace. Combining the result thus obtained with~\eqref{eq:2PIfrgCflowIntermediaryResult}, we can deduce the master equation of the 2PI EA underlying the C-flow:
\begin{equation}
\begin{split}
\dot{\Gamma}_{\mathfrak{s}}^{(\mathrm{2PI})}[G] \equiv \left. \dot{\Gamma}_{\mathfrak{s}}^{(\mathrm{2PI})}[G] \right|_{G} = & -\left.\dot{W}_{\mathfrak{s}}[K]\right|_{K} \\
= & \ \dot{C}^{-1}_{\mathfrak{s},\hat{\gamma}} G_{\hat{\gamma}} \;,
\end{split}
\label{eq:2PIfrgCflowDGammaDk}
\end{equation}
which resembles noticeably the Wetterich equation for the 1PI EA, as can be seen after comparison with~\eqref{eq:WetterichEqGammaDotAppendix}. In order to derive the corresponding master equation for the Luttinger-Ward functional, we will exploit expression~\eqref{eq:ExpressionFree2PIeffectiveAction} of the free 2PI EA after the substitution $C^{-1} \rightarrow C^{-1}_{\mathfrak{s}}$:
\begin{equation}
\Gamma_{0,\mathfrak{s}}^{(\mathrm{2PI})}[G]=-\frac{\zeta}{2} \mathrm{Tr}_{\alpha} \left[ \mathrm{ln}(G) \right] + \frac{\zeta}{2} \mathrm{Tr}_{\alpha}\left[GC_{\mathfrak{s}}^{-1}-\mathbb{I}\right]\;.
\label{eq:2PIfrgExpressionGamma0bis}
\end{equation}
The differentiation with respect to $\mathfrak{s}$ at fixed $G$ leads to:
\begin{equation}
\begin{split}
\dot{\Gamma}_{0,\mathfrak{s}}^{(\mathrm{2PI})}[G] \equiv \left.\dot{\Gamma}_{0,\mathfrak{s}}^{(\mathrm{2PI})}[G]\right|_{G} = & \ \frac{\zeta}{2} \int_{\alpha,\alpha'} G_{\alpha\alpha'}\underbrace{\dot{C}^{-1}_{\mathfrak{s},\alpha'\alpha}}_{\zeta \dot{C}^{-1}_{\mathfrak{s},\alpha\alpha'}} \\
= & \ \frac{1}{2} \int_{\alpha,\alpha'} G_{\alpha\alpha'} \dot{C}^{-1}_{\mathfrak{s},\alpha \alpha'} \\
= & \ G_{\hat{\gamma}} \dot{C}^{-1}_{\mathfrak{s},\hat{\gamma}}\;.
\end{split}
\label{eq:2PIfrgCflowDGamma0Dk}
\end{equation}
From \eqref{eq:2PIfrgCflowDGammaDk} and~\eqref{eq:2PIfrgCflowDGamma0Dk}, we infer that:
\begin{equation}
\dot{\Gamma}_{\mathfrak{s}}^{(\mathrm{2PI})}[G] = \dot{\Gamma}_{0,\mathfrak{s}}^{(\mathrm{2PI})}[G] \;,
\end{equation}
which proves that the Luttinger-Ward functional is an invariant of the flow in the present case:
\begin{equation}
\dot{\Phi}_{\mathfrak{s}}[G] \equiv \left.\dot{\Phi}_{\mathfrak{s}}[G]\right|_{G} = \dot{\Gamma}_{\mathfrak{s}}^{(\mathrm{2PI})}[G] - \dot{\Gamma}_{0,\mathfrak{s}}^{(\mathrm{2PI})}[G] = 0 \;.
\label{eq:2PIfrgCflowPhiInvariant}
\end{equation}
Furthermore, at vanishing external source, the 2PI EA becomes:
\begin{equation}
\overline{\Gamma}_{\mathfrak{s}}^{(\mathrm{2PI})} = \overline{\Gamma}_{0,\mathfrak{s}}^{(\mathrm{2PI})} + \overline{\Phi}_{\mathfrak{s}} = -\frac{\zeta}{2} \mathrm{Tr}_{\alpha} \left[ \mathrm{ln}\big(\overline{G}_{\mathfrak{s}}\big) \right] + \frac{\zeta}{2} \mathrm{Tr}_{\alpha}\left[\overline{G}_{\mathfrak{s}}C_{\mathfrak{s}}^{-1}-\mathbb{I}\right]+\overline{\Phi}_{\mathfrak{s}}\;,
\label{eq:2PIfrgCflowExpressionGammas2PIbarGam0PhiGCPhi}
\end{equation}
where we have used \eqref{eq:2PIfrgExpressionGamma0bis}. We then differentiate~\eqref{eq:2PIfrgCflowExpressionGammas2PIbarGam0PhiGCPhi} with respect to the flow parameter:
\begin{equation}
\begin{split}
\dot{\overline{\Gamma}}_{\mathfrak{s}}^{(\mathrm{2PI})} = & -\frac{\zeta}{2} \int_{\alpha,\alpha'} \dot{\overline{G}}_{\mathfrak{s},\alpha\alpha'} \overline{G}^{\hspace{0.01cm} -1}_{\mathfrak{s},\alpha'\alpha} + \frac{\zeta}{2} \int_{\alpha,\alpha'} \dot{\overline{G}}_{\mathfrak{s},\alpha\alpha'} C_{\mathfrak{s},\alpha'\alpha}^{-1} + \frac{\zeta}{2} \int_{\alpha,\alpha'} \overline{G}_{\mathfrak{s},\alpha\alpha'}\dot{C}_{\mathfrak{s},\alpha'\alpha}^{-1} + \dot{\overline{\Phi}}_{\mathfrak{s}} \\
= & \ \frac{\zeta}{2} \int_{\alpha,\alpha'} \dot{\overline{G}}_{\mathfrak{s},\alpha\alpha'} \underbrace{\left(-\overline{G}^{\hspace{0.01cm} -1}_{\mathfrak{s}}+C_{\mathfrak{s}}^{-1}\right)_{\alpha'\alpha}}_{\overline{\Sigma}_{\mathfrak{s},\alpha'\alpha}} + \frac{\zeta}{2} \int_{\alpha,\alpha'} \overline{G}_{\mathfrak{s},\alpha\alpha'}\dot{C}_{\mathfrak{s},\alpha'\alpha}^{-1} + \dot{\overline{\Phi}}_{\mathfrak{s}} \\
= & \ \frac{\zeta}{2} \int_{\alpha,\alpha'} \dot{\overline{G}}_{\mathfrak{s},\alpha\alpha'} \underbrace{\overline{\Sigma}_{\mathfrak{s},\alpha'\alpha}}_{\zeta\overline{\Sigma}_{\mathfrak{s},\alpha\alpha'}} + \frac{\zeta}{2} \int_{\alpha,\alpha'} \overline{G}_{\mathfrak{s},\alpha\alpha'} \underbrace{\dot{C}_{\mathfrak{s},\alpha'\alpha}^{-1}}_{\zeta\dot{C}_{\mathfrak{s},\alpha\alpha'}^{-1}} + \dot{\overline{\Phi}}_{\mathfrak{s}} \\
= & \ \dot{\overline{G}}_{\mathfrak{s},\hat{\gamma}} \overline{\Sigma}_{\mathfrak{s},\hat{\gamma}} + \overline{G}_{\mathfrak{s},\hat{\gamma}} \dot{C}_{\mathfrak{s},\hat{\gamma}}^{-1} + \dot{\overline{\Phi}}_{\mathfrak{s}} \;,
\end{split}
\label{eq:2PIfrgCflowGammaBarDot}
\end{equation}
where the self-energy was introduced by making use of Dyson equation in the form of~\eqref{eq:DysonEqAppendix}. The derivative $\dot{\overline{\Phi}}_{\mathfrak{s}}$ can be rewritten via the chain rule combined with~\eqref{eq:2PIfrgCflowPhiInvariant}:
\begin{equation}
\begin{split}
\dot{\overline{\Phi}}_{\mathfrak{s}} = & \underbrace{\overline{\dot{\Phi}}_{\mathfrak{s}}}_{0} + \dot{\overline{G}}_{\mathfrak{s},\hat{\gamma}} \underbrace{\overline{\Phi}_{\mathfrak{s},\hat{\gamma}}^{(1)}}_{-\overline{\Sigma}_{\mathfrak{s},\hat{\gamma}}} \\
= & -\dot{\overline{G}}_{\mathfrak{s},\hat{\gamma}} \overline{\Sigma}_{\mathfrak{s},\hat{\gamma}} \;.
\end{split}
\label{eq:2PIfrgCflowPhiBarDot}
\end{equation}
According to~\eqref{eq:2PIfrgCflowGammaBarDot} and~\eqref{eq:2PIfrgCflowPhiBarDot}, we have:
\begin{equation}
\dot{\overline{\Gamma}}_{\mathfrak{s}}^{(\mathrm{2PI})} = \overline{G}_{\mathfrak{s},\hat{\gamma}} \dot{C}_{\mathfrak{s},\hat{\gamma}}^{-1} \;.
\label{eq:2PIfrgGammakBar}
\end{equation}
We also consider $\Delta \overline{\Omega}_{\mathfrak{s}}$ defined by~\eqref{eq:2PIfrgCflowDefDeltaOmega}:
\begin{equation}
\Delta \overline{\Omega}_{\mathfrak{s}} \equiv \frac{1}{\beta}\left(\overline{\Gamma}_{\mathfrak{s}}^{(\mathrm{2PI})} - \Gamma_{0,\mathfrak{s}}^{(\mathrm{2PI})}[C_{\mathfrak{s}}] \right) = \frac{1}{\beta}\left(\overline{\Gamma}_{\mathfrak{s}}^{(\mathrm{2PI})} + \frac{\zeta}{2} \mathrm{Tr}_{\alpha} \left[ \mathrm{ln}(C_{\mathfrak{s}}) \right] \right) = \frac{1}{\beta}\left(\overline{\Gamma}_{\mathfrak{s}}^{(\mathrm{2PI})} - \frac{\zeta}{2} \mathrm{Tr}_{\alpha} \left[ \mathrm{ln}\big(C^{-1}_{\mathfrak{s}}\big)\right] \right) \;.
\label{eq:2PIfrgCflowDeltaOmegakBar}
\end{equation}
An expression for $\Delta \dot{\overline{\Omega}}_{\mathfrak{s}}$ can then be found from \eqref{eq:2PIfrgGammakBar} and~\eqref{eq:2PIfrgCflowDeltaOmegakBar}:
\begin{equation}
\begin{split}
\Delta \dot{\overline{\Omega}}_{\mathfrak{s}} = & \ \frac{1}{\beta}\bigg(\dot{\overline{\Gamma}}^{(\mathrm{2PI})}_{\mathfrak{s}} - \frac{\zeta}{2} \int_{\alpha,\alpha'} \dot{C}^{-1}_{\mathfrak{s},\alpha \alpha'} \underbrace{C_{\mathfrak{s},\alpha'\alpha}}_{\zeta C_{\mathfrak{s},\alpha\alpha'}} \bigg) \\
= & \ \frac{1}{\beta}\left(\overline{G}_{\mathfrak{s},\hat{\gamma}} \dot{C}_{\mathfrak{s},\hat{\gamma}}^{-1} - \dot{C}^{-1}_{\mathfrak{s},\hat{\gamma}} C_{\mathfrak{s},\hat{\gamma}}\right) \\
= & \ \frac{1}{\beta} \dot{C}_{\mathfrak{s},\hat{\gamma}}^{-1} \left(\overline{G}_{\mathfrak{s}}-C_{\mathfrak{s}}\right)_{\hat{\gamma}} \;.
\end{split}
\label{eq:2PIfrgCflowDerivDeltaOmega}
\end{equation}
The latter equation adds up to \eqref{eq:2PIfrgFlowEqGbarviaDysonEq} and the set of \eqref{eq:2PIfrgFlowEqCflowVersion2} so that the tower of flow equations for the C-flow is given by:
\begin{equation}
\dot{\overline{G}}_{\mathfrak{s},\alpha_{1} \alpha'_{1}}=-\int_{\alpha_{2},\alpha'_{2}}\overline{G}_{\mathfrak{s}, \alpha_{1} \alpha_{2}}\left(\dot{C}_{\mathfrak{s}}^{-1}-\dot{\overline{\Sigma}}_{\mathfrak{s}}\right)_{\alpha_{2} \alpha'_{2}} \overline{G}_{\mathfrak{s}, \alpha'_{2} \alpha'_{1}} \mathrlap{\;,}
\end{equation}
\begin{equation}
\Delta \dot{\overline{\Omega}}_{\mathfrak{s}} = \frac{1}{\beta} \dot{C}_{\mathfrak{s},\hat{\gamma}}^{-1} \left(\overline{G}_{\mathfrak{s}}-C_{\mathfrak{s}}\right)_{\hat{\gamma}} \mathrlap{\;,}
\end{equation}
\begin{equation}
\dot{\overline{\Phi}}_{\mathfrak{s}} = - \dot{\overline{G}}_{\mathfrak{s},\hat{\gamma}} \overline{\Sigma}_{\mathfrak{s},\hat{\gamma}} \mathrlap{\;,}
\end{equation}
\begin{equation}
\dot{\overline{\Sigma}}_{\mathfrak{s},\gamma} = - \dot{\overline{G}}_{\mathfrak{s},\hat{\gamma}} \overline{\Phi}_{\mathfrak{s},\hat{\gamma}\gamma}^{(2)} \mathrlap{\;,}
\end{equation}
\begin{equation}
\dot{\overline{\Phi}}_{\mathfrak{s},\gamma_{1}\cdots\gamma_{n}}^{(n)} = \dot{\overline{G}}_{\mathfrak{s},\hat{\gamma}} \overline{\Phi}_{\mathfrak{s},\hat{\gamma}\gamma_{1}\cdots\gamma_{n}}^{(n+1)} \mathrlap{\quad \forall n \geq 2 \;.}
\end{equation}

\subsection{U-flow}
\label{ann:2PIfrgFlowEquationUflow}

We start once again from~\eqref{eq:2PIFRGgeneratingFunc} expressing the generating functional $Z[K]$ together with the classical action $S\big[\widetilde{\psi}\big]$ given by~\eqref{eq:2PIFRGmostgeneralactionS}:
\begin{equation}
Z[K]=e^{W[K]}=\int\mathcal{D}\widetilde{\psi} \ e^{-S_{0}\big[\widetilde{\psi}\big]-\frac{1}{4!}\int_{\alpha_{1},\alpha'_{1},\alpha_{2},\alpha'_{2}} U_{\alpha_{1} \alpha'_{1} \alpha_{2} \alpha'_{2}} \widetilde{\psi}_{\alpha_{1}}\widetilde{\psi}_{\alpha'_{1}}\widetilde{\psi}_{\alpha_{2}}\widetilde{\psi}_{\alpha'_{2}} + \frac{1}{2}\int_{\alpha_{1},\alpha'_{1}}\widetilde{\psi}_{\alpha_{1}}K_{\alpha_{1}\alpha'_{1}}\widetilde{\psi}_{\alpha'_{1}}}\;.
\label{eq:2PIFRGgeneratingFuncUflow}
\end{equation}
Replacing the two-body interaction via $U \rightarrow U_{\mathfrak{s}}=U+R_{\mathfrak{s}}$ (or, equivalently, $U_{\mathfrak{s}}=R_{\mathfrak{s}}U$) and differentiating with respect to $\mathfrak{s}$ at fixed $K$ leads to:
\begin{equation}
\begin{split}
\scalebox{0.99}{${\displaystyle \left.\dot{W}_{\mathfrak{s}}[K]\right|_{K} = }$} & \scalebox{0.99}{${\displaystyle -\frac{1}{4!} \int_{\alpha_{1},\alpha'_{1},\alpha_{2},\alpha'_{2}} \dot{U}_{\mathfrak{s},\alpha_{1} \alpha'_{1} \alpha_{2} \alpha'_{2}} \underbrace{\left(\frac{1}{Z_{\mathfrak{s}}[K]} \int\mathcal{D}\widetilde{\psi} \ \widetilde{\psi}_{\alpha_{1}}\widetilde{\psi}_{\alpha'_{1}}\widetilde{\psi}_{\alpha_{2}}\widetilde{\psi}_{\alpha'_{2}} \ e^{-S_{\mathfrak{s}}\big[\widetilde{\psi}\big] + \frac{1}{2}\int_{\alpha_{1},\alpha'_{1}}\widetilde{\psi}_{\alpha_{1}}K_{\alpha_{1}\alpha'_{1}}\widetilde{\psi}_{\alpha'_{1}}}\right)}_{\left\langle \widetilde{\psi}_{\alpha_{1}}\widetilde{\psi}_{\alpha'_{1}}\widetilde{\psi}_{\alpha_{2}}\widetilde{\psi}_{\alpha'_{2}} \right\rangle_{K,\mathfrak{s}}} }$} \\
\scalebox{0.99}{${\displaystyle = }$} & \scalebox{0.99}{${\displaystyle-\frac{1}{4!} \int_{\alpha_{1},\alpha'_{1},\alpha_{2},\alpha'_{2}} \dot{U}_{\mathfrak{s},\alpha_{1} \alpha'_{1} \alpha_{2} \alpha'_{2}} \left\langle \widetilde{\psi}_{\alpha_{1}}\widetilde{\psi}_{\alpha'_{1}}\widetilde{\psi}_{\alpha_{2}}\widetilde{\psi}_{\alpha'_{2}} \right\rangle_{K,\mathfrak{s}} \;, }$}
\end{split}
\label{eq:IntermediaryResult2PIFRGUflowAppendix}
\end{equation}
where we have exploited the flow-dependent expectation value:
\begin{equation}
\big\langle \cdots \big\rangle_{K,\mathfrak{s}} = \frac{1}{Z_{\mathfrak{s}}[K]} \int\mathcal{D}\widetilde{\psi} \ \cdots \ e^{-S_{\mathfrak{s}}\big[\widetilde{\psi}\big] + \frac{1}{2}\int_{\alpha_{1},\alpha'_{1}}\widetilde{\psi}_{\alpha_{1}}K_{\alpha_{1}\alpha'_{1}}\widetilde{\psi}_{\alpha'_{1}}} \;.
\label{eq:FlowdependentExpectationValue2PIFRG}
\end{equation}
Using~\eqref{eq:GEqualGk2PIFRG} once more, we then express the connected correlation function $W_{\mathfrak{s}}^{(2)}[K]$ as follows:
\begin{equation}
\begin{split}
W_{\mathfrak{s},\gamma_{2} \gamma_{1}}^{(2)}[K] \equiv \frac{\delta^{2} W_{\mathfrak{s}}[K]}{\delta K_{\gamma_{2}} \delta K_{\gamma_{1}}} = & \ \frac{\delta}{\delta K_{\alpha_{2}\alpha'_{2}}} \left\langle \widetilde{\psi}_{\alpha_{1}} \widetilde{\psi}_{\alpha'_{1}} \right\rangle_{K,\mathfrak{s}} \\
= & \ \frac{1}{2} \int_{\alpha_{3},\alpha'_{3}} \left\langle \widetilde{\psi}_{\alpha_{1}} \widetilde{\psi}_{\alpha'_{1}} \widetilde{\psi}_{\alpha_{3}} \frac{\delta K_{\alpha_{3}\alpha'_{3}}}{\delta K_{\alpha_{2}\alpha'_{2}}} \widetilde{\psi}_{\alpha'_{3}} \right\rangle_{K,\mathfrak{s}} - \underbrace{\left\langle \widetilde{\psi}_{\alpha_{2}} \widetilde{\psi}_{\alpha'_{2}} \right\rangle_{K,\mathfrak{s}}}_{G_{\gamma_{2}}} \underbrace{\left\langle \widetilde{\psi}_{\alpha_{1}} \widetilde{\psi}_{\alpha'_{1}} \right\rangle_{K,\mathfrak{s}}}_{G_{\gamma_{1}}} \\
= & \ \frac{1}{2} \int_{\alpha_{3},\alpha'_{3}} \left\langle \widetilde{\psi}_{\alpha_{1}} \widetilde{\psi}_{\alpha'_{1}} \widetilde{\psi}_{\alpha_{3}} \left(\delta_{\alpha_{3} \alpha_{2}}\delta_{\alpha'_{3} \alpha'_{2}} + \zeta \delta_{\alpha_{3} \alpha'_{2}}\delta_{\alpha'_{3} \alpha_{2}}\right) \widetilde{\psi}_{\alpha'_{3}} \right\rangle_{K,\mathfrak{s}} - G_{\gamma_{2}} G_{\gamma_{1}} \\
= & \ \frac{1}{2} \left(\left\langle \widetilde{\psi}_{\alpha_{1}} \widetilde{\psi}_{\alpha'_{1}} \widetilde{\psi}_{\alpha_{2}} \widetilde{\psi}_{\alpha'_{2}} \right\rangle_{K,\mathfrak{s}} + \zeta \left\langle \widetilde{\psi}_{\alpha_{1}} \widetilde{\psi}_{\alpha'_{1}} \widetilde{\psi}_{\alpha'_{2}} \widetilde{\psi}_{\alpha_{2}} \right\rangle_{K,\mathfrak{s}}\right) - G_{\gamma_{2}} G_{\gamma_{1}} \\
= & \ \left\langle \widetilde{\psi}_{\alpha_{1}} \widetilde{\psi}_{\alpha'_{1}} \widetilde{\psi}_{\alpha_{2}} \widetilde{\psi}_{\alpha'_{2}} \right\rangle_{K,\mathfrak{s}} - G_{\gamma_{2}} G_{\gamma_{1}} \;,
\end{split}
\label{eq:RewriteW2UflowAppendix}
\end{equation}
which is equivalent to:
\begin{equation}
\left\langle \widetilde{\psi}_{\alpha_{1}} \widetilde{\psi}_{\alpha'_{1}} \widetilde{\psi}_{\alpha_{2}} \widetilde{\psi}_{\alpha'_{2}} \right\rangle_{K,\mathfrak{s}} = W_{\mathfrak{s},\gamma_{2} \gamma_{1}}^{(2)}[K] + G_{\gamma_{2}} G_{\gamma_{1}}\;.
\end{equation}
From this,~\eqref{eq:IntermediaryResult2PIFRGUflowAppendix} can be rewritten as:
\begin{equation}
\left.\dot{W}_{\mathfrak{s}}[K]\right|_{K} = -\frac{1}{6} \dot{U}_{\mathfrak{s},\hat{\gamma}_{1} \hat{\gamma}_{2}} \left(W_{\mathfrak{s},\hat{\gamma}_{2} \hat{\gamma}_{1}}^{(2)}[K] + G_{\hat{\gamma}_{2}} G_{\hat{\gamma}_{1}}\right)\;.
\label{eq:2PIfrgUflowExpressionWdot}
\end{equation}
We rewrite specifically the following term by renaming dummy indices:
\begin{equation}
\begin{split}
\dot{U}_{\mathfrak{s},\hat{\gamma}_{1} \hat{\gamma}_{2}} G_{\hat{\gamma}_{2}} G_{\hat{\gamma}_{1}} = & \ \frac{1}{4} \int_{\gamma_{1},\gamma_{2}} \dot{U}_{\mathfrak{s},\gamma_{1} \gamma_{2}} G_{\gamma_{2}} G_{\gamma_{1}} \\
= & \ \frac{1}{8} \int_{\alpha_{1},\alpha'_{1},\alpha_{2},\alpha'_{2}} \dot{U}_{\mathfrak{s},\alpha_{1} \alpha'_{1} \alpha_{2} \alpha'_{2}} G_{\alpha_{2} \alpha'_{2}} G_{\alpha_{1} \alpha'_{1}} + \frac{1}{8} \int_{\alpha_{1},\alpha'_{1},\alpha_{2},\alpha'_{2}} \dot{U}_{\mathfrak{s},\alpha_{1} \alpha'_{1} \alpha_{2} \alpha'_{2}} G_{\alpha_{2} \alpha'_{2}} G_{\alpha_{1} \alpha'_{1}} \\
= & \ \frac{1}{8} \int_{\alpha_{1},\alpha'_{1},\alpha_{2},\alpha'_{2}} \underbrace{\dot{U}_{\mathfrak{s},\alpha_{2} \alpha'_{1} \alpha'_{2} \alpha_{1}}}_{\dot{U}_{\mathfrak{s},\gamma_{1} \gamma_{2}}} G_{\alpha'_{2} \alpha_{1}} G_{\alpha_{2} \alpha'_{1}} + \frac{1}{8} \int_{\alpha_{1},\alpha'_{1},\alpha_{2},\alpha'_{2}} \underbrace{\dot{U}_{\mathfrak{s},\alpha'_{2} \alpha'_{1} \alpha_{2} \alpha_{1}}}_{\zeta \dot{U}_{\mathfrak{s},\gamma_{1} \gamma_{2}}} G_{\alpha_{2} \alpha_{1}} G_{\alpha'_{2} \alpha'_{1}} \\
= & \ \frac{1}{8} \int_{\alpha_{1},\alpha'_{1},\alpha_{2},\alpha'_{2}} \dot{U}_{\mathfrak{s},\gamma_{1} \gamma_{2}} \underbrace{\left(G_{\alpha'_{2} \alpha_{1}} G_{\alpha_{2} \alpha'_{1}} + \zeta G_{\alpha_{2} \alpha_{1}} G_{\alpha'_{2} \alpha'_{1}}\right)}_{\Pi_{\gamma_{2}\gamma_{1}}[G]} \\
= & \ \frac{1}{2} \dot{U}_{\mathfrak{s},\hat{\gamma}_{1} \hat{\gamma}_{2}} \Pi_{\hat{\gamma}_{2} \hat{\gamma}_{1}}[G] \;,
\end{split}
\label{eq:TraceTrickmUflowAppendix}
\end{equation}
where we have used the property $U_{\alpha_{1}\alpha_{2}\alpha_{3}\alpha_{4}}=\zeta^{N(P)}U_{\alpha_{P(1)}\alpha_{P(2)}\alpha_{P(3)}\alpha_{P(4)}}$. According to this,~\eqref{eq:2PIfrgUflowExpressionWdot} is equivalent to:
\begin{equation}
\left.\dot{W}_{\mathfrak{s}}[K]\right|_{K} = -\frac{1}{6} \dot{U}_{\mathfrak{s},\hat{\gamma}_{1} \hat{\gamma}_{2}} \left(W_{\mathfrak{s}}^{(2)}[K] + \frac{1}{2}\Pi[G]\right)_{\hat{\gamma}_{2}\hat{\gamma}_{1}}\;.
\label{eq:WdotUflowAppendix}
\end{equation}
The reasoning leading to the first line of~\eqref{eq:2PIfrgCflowDGammaDk} still holds in the framework of the U-flow. Let us then recall the relation thus obtained:
\begin{equation}
\dot{\Gamma}_{\mathfrak{s}}^{(\mathrm{2PI})}[G] \equiv \left.\dot{\Gamma}_{\mathfrak{s}}^{(\mathrm{2PI})}[G]\right|_{G} = -\left.\dot{W}_{\mathfrak{s}}[K]\right|_{K}\;,
\end{equation}
which, after inserting~\eqref{eq:WdotUflowAppendix}, becomes:
\begin{equation}
\dot{\Gamma}_{\mathfrak{s}}^{(\mathrm{2PI})}[G] = \frac{1}{6} \dot{U}_{\mathfrak{s},\hat{\gamma}_{1} \hat{\gamma}_{2}} \left(W_{\mathfrak{s}}^{(2)}[K] + \frac{1}{2}\Pi[G]\right)_{\hat{\gamma}_{2}\hat{\gamma}_{1}}\;.
\label{eq:2PIfrgGammasdot}
\end{equation}
By definition, the free 2PI EA is independent of the two-body interaction $U_{\mathfrak{s}}$ (and therefore independent of the flow parameter $\mathfrak{s}$ in the present case) so that:
\begin{equation}
\dot{\Gamma}_{0,\mathfrak{s}}^{(\mathrm{2PI})}[G] \equiv \left.\dot{\Gamma}_{0,\mathfrak{s}}^{(\mathrm{2PI})}[G]\right|_{G} = 0\;.
\end{equation}
According to~\eqref{eq:DefLWfunctional}, this gives us:
\begin{equation}
\dot{\Phi}_{\mathfrak{s}}[G] \equiv \left.\dot{\Phi}_{\mathfrak{s}}[G]\right|_{G} = \dot{\Gamma}_{\mathfrak{s}}^{(\mathrm{2PI})}[G] - \dot{\Gamma}_{0,\mathfrak{s}}^{(\mathrm{2PI})}[G] = \dot{\Gamma}_{\mathfrak{s}}^{(\mathrm{2PI})}[G] \;.
\label{eq:2PIfrgUflowPhiGammadot}
\end{equation}
Combining~\eqref{eq:2PIfrgUflowPhiGammadot} with~\eqref{eq:2PIfrgGammasdot} yields:
\begin{equation}
\dot{\Phi}_{\mathfrak{s}}[G] = \frac{1}{6} \dot{U}_{\mathfrak{s},\hat{\gamma}_{1} \hat{\gamma}_{2}} \left(W_{\mathfrak{s}}^{(2)}[K] + \frac{1}{2}\Pi[G]\right)_{\hat{\gamma}_{2}\hat{\gamma}_{1}} \;.
\label{eq:2PIfrgPhisdot}
\end{equation}
Most equations constituting the tower of flow equations for the U-flow scheme can be determined from~\eqref{eq:2PIfrgPhisdot} in particular:
\begin{itemize}
\item Expressions of $\dot{\overline{G}}_{\mathfrak{s}}$:\\
Just like the free 2PI EA, the free propagator $C$ does not depend on $U_{\mathfrak{s}}$, which translates into:
\begin{equation}
\dot{C}_{\gamma} = 0 \mathrlap{\quad \forall \gamma\;,}
\end{equation}
in the framework of the U-flow. In this situation,~\eqref{eq:2PIfrgFlowEqGbarviaDysonEq} reduces to:
\begin{equation}
\dot{\overline{G}}_{\mathfrak{s},\alpha_{1}\alpha'_{1}}=\int_{\alpha_{2},\alpha'_{2}}\overline{G}_{\mathfrak{s},\alpha_{1} \alpha_{2}} \dot{\overline{\Sigma}}_{\mathfrak{s},\alpha_{2} \alpha'_{2}} \overline{G}_{\mathfrak{s},\alpha'_{2} \alpha'_{1}} \;.
\label{eq:2PIfrgUFlowGbarviaDysonEq}
\end{equation}
The flow of $\overline{G}_{\mathfrak{s}}$ can actually also be determined via the Bethe-Salpeter equation instead of the Dyson one, in the U-flow as well as in the C-flow and CU-flow schemes. Even though we do not investigate this alternative in this work, we determine below the associated flow equation, which will be useful in further derivations. To that end, we recall that, in the calculations of~\eqref{eq:2PIfrgW2Gamma2Pi}, it was shown that:
\begin{equation}
W_{\mathfrak{s}}^{(2)}[K] = \left(\Pi^{\mathrm{inv}}[G] + \Phi_{\mathfrak{s}}^{(2)}[G]\right)^{\mathrm{inv}} \;.
\label{eq:2PIfrgW2PiInvPhi2}
\end{equation}
Furthermore, the 2PI EA under consideration is extremal at $K_{\gamma}=0$ $\forall\gamma$, i.e.:
\begin{equation}
\overline{\Gamma}_{\mathfrak{s},\gamma}^{(\mathrm{2PI})(1)} = 0 \mathrlap{\quad \forall \gamma, \mathfrak{s} \;,}
\label{eq:Gamma2PIextremalAppendix}
\end{equation}
with
\begin{equation}
\Gamma_{\mathfrak{s},\gamma_{1}\cdots\gamma_{n}}^{(\mathrm{2PI})(n)}[G] \equiv \frac{\delta^{n}\Gamma_{\mathfrak{s}}^{(\mathrm{2PI})}[G]}{\delta G_{\gamma_{1}}\cdots\delta G_{\gamma_{n}}} \;.
\end{equation}
Differentiating~\eqref{eq:Gamma2PIextremalAppendix} with respect to $\mathfrak{s}$ and making use of~\eqref{eq:2PIfrgW2Gamma2Interacting} and~\eqref{eq:2PIfrgUflowPhiGammadot} leads to:
\begin{equation}
\dot{\overline{\Gamma}}_{\mathfrak{s},\gamma}^{(\mathrm{2PI})(1)}=\underbrace{\overline{\dot{\Gamma}}_{\mathfrak{s},\gamma}^{(\mathrm{2PI})(1)}}_{\overline{\dot{\Phi}}_{\mathfrak{s},\gamma}^{(1)}} + \dot{\overline{G}}_{\mathfrak{s},\hat{\gamma}} \hspace{-0.08cm} \underbrace{\overline{\Gamma}_{\mathfrak{s},\hat{\gamma}\gamma}^{(\mathrm{2PI})(2)}}_{\left(\overline{W}_{\mathfrak{s}}^{(2)}\right)_{\hat{\gamma}\gamma}^{\mathrm{inv}}} \hspace{-0.08cm} = 0 \;,
\end{equation}
which, according to the definition $\Sigma_{\mathfrak{s}}[G]\equiv-\Phi^{(1)}_{\mathfrak{s}}[G]$, gives us:
\begin{equation}
\dot{\overline{G}}_{\mathfrak{s},\gamma} = \overline{W}_{\mathfrak{s},\gamma \hat{\gamma}}^{(2)} \overline{\dot{\Sigma}}_{\mathfrak{s},\hat{\gamma}} \;.
\label{eq:2PIfrgGdotW2Phi1BetheSalpeter}
\end{equation}
After inserting~\eqref{eq:2PIfrgW2PiInvPhi2} into~\eqref{eq:2PIfrgGdotW2Phi1BetheSalpeter}, we obtain:
\begin{equation}
\dot{\overline{G}}_{\mathfrak{s},\gamma} = \left(\overline{\Pi}_{\mathfrak{s}}^{\mathrm{inv}} + \overline{\Phi}_{\mathfrak{s}}^{(2)}\right)_{\gamma\hat{\gamma}}^{\mathrm{inv}} \overline{\dot{\Sigma}}_{\mathfrak{s},\hat{\gamma}}\;.
\label{eq:2PIfrgGdotBetheSalpeter}
\end{equation}
Hence, the flow of $\overline{G}_{\mathfrak{s}}$ can either be determined from~\eqref{eq:2PIfrgUFlowGbarviaDysonEq} or~\eqref{eq:2PIfrgGdotBetheSalpeter}.

\item Expression of $\dot{\overline{\Omega}}_{\mathfrak{s}}$:\\
The 2PI EA under consideration satisfies the extremization condition~\eqref{eq:Gamma2PIextremalAppendix}, which enables us to simplify the following implementation of the chain rule:
\begin{equation}
\dot{\overline{\Gamma}}_{\mathfrak{s}}^{(\mathrm{2PI})} = \overline{\dot{\Gamma}}_{\mathfrak{s}}^{(\mathrm{2PI})} + \dot{\overline{G}}_{\mathfrak{s},\hat{\gamma}} \underbrace{\overline{\Gamma}^{(\mathrm{2PI})(1)}_{\mathfrak{s},\hat{\gamma}}}_{0} = \overline{\dot{\Gamma}}_{\mathfrak{s}}^{(\mathrm{2PI})} \;.
\label{eq:2PIfrgGammaBarDotVsGammaDotBar}
\end{equation}
From~\eqref{eq:2PIfrgUflowPhiGammadot} and~\eqref{eq:2PIfrgGammaBarDotVsGammaDotBar}, it directly follows that:
\begin{equation}
\dot{\overline{\Gamma}}^{(\mathrm{2PI})}_{\mathfrak{s}} = \overline{\dot{\Phi}}_{\mathfrak{s}}\;.
\label{eq:2PIfrgdotGammaDotPhiUflowAppendix}
\end{equation}
According to the definition $\Omega_{\mathfrak{s}}\equiv\frac{1}{\beta}\Gamma^{(\mathrm{2PI})}_{\mathfrak{s}}$ together with~\eqref{eq:2PIfrgdotGammaDotPhiUflowAppendix}, we infer that:
\begin{equation}
\dot{\overline{\Omega}}_{\mathfrak{s}} =\frac{1}{\beta} \overline{\dot{\Phi}}_{\mathfrak{s}} \;.
\label{eq:2PIfrgOmegadotUflowAppendix}
\end{equation}
With the help of~\eqref{eq:2PIfrgPhisdot} and~\eqref{eq:2PIfrgOmegadotUflowAppendix}, we finally obtain:
\begin{equation}
\dot{\overline{\Omega}}_{\mathfrak{s}} = \frac{1}{6\beta} \dot{U}_{\mathfrak{s},\hat{\gamma}_{1} \hat{\gamma}_{2}} \left(\overline{W}_{\mathfrak{s}}^{(2)} + \frac{1}{2}\overline{\Pi}_{\mathfrak{s}}\right)_{\hat{\gamma}_{2}\hat{\gamma}_{1}}\;.
\label{eq:2PIfrgUflowExpressionOmegaDot}
\end{equation}

\item Expression of $\dot{\overline{\Phi}}_{\mathfrak{s}}$:\\
From~\eqref{eq:2PIfrgGeneralFlowEqPhi} and~\eqref{eq:2PIfrgGdotW2Phi1BetheSalpeter}, it follows that:
\begin{equation}
\dot{\overline{\Phi}}_{\mathfrak{s}} = \overline{\dot{\Phi}}_{\mathfrak{s}} + \dot{\overline{G}}_{\mathfrak{s},\hat{\gamma}} \overline{\Phi}^{(1)}_{\mathfrak{s},\hat{\gamma}} = \overline{\dot{\Phi}}_{\mathfrak{s}} - \dot{\overline{G}}_{\mathfrak{s},\hat{\gamma}} \overline{\Sigma}_{\mathfrak{s},\hat{\gamma}} = \overline{\dot{\Phi}}_{\mathfrak{s}} - \overline{\Sigma}_{\mathfrak{s},\hat{\gamma}_{1}} \overline{W}_{\mathfrak{s},\hat{\gamma}_{1} \hat{\gamma}_{2}}^{(2)} \overline{\dot{\Sigma}}_{\mathfrak{s},\hat{\gamma}_{2}} \;.
\label{eq:2PIfrgUflowdotPhiBarFirstStep}
\end{equation}
As a next step, we derive an expression for $\dot{\Sigma}_{\mathfrak{s}}$ from~\eqref{eq:2PIfrgPhisdot}:
\begin{equation}
\dot{\Sigma}_{\mathfrak{s},\gamma}[G] \equiv -\dot{\Phi}^{(1)}_{\mathfrak{s},\gamma}[G] = -\frac{1}{6} \dot{U}_{\mathfrak{s},\hat{\gamma}_{1} \hat{\gamma}_{2}} \left(\frac{\delta W_{\mathfrak{s},\hat{\gamma}_{2} \hat{\gamma}_{1}}^{(2)}[K]}{\delta G_{\gamma}} + \frac{1}{2} \frac{\delta \Pi_{\hat{\gamma}_{2} \hat{\gamma}_{1}}[G]}{\delta G_{\gamma}}\right)\;.
\label{eq:2PIfrgUflowPhi1IntermediaryResult}
\end{equation}
We then evaluate the derivative:
\begin{equation}
\begin{split}
\scalebox{0.99}{${\displaystyle\frac{\delta W_{\mathfrak{s},\gamma_{2} \gamma_{3}}^{(2)}[K]}{\delta G_{\gamma_{1}}} =}$} & \scalebox{0.99}{${\displaystyle\ \frac{\delta}{\delta G_{\gamma_{1}}} \left(\Pi^{\mathrm{inv}}[G] + \Phi_{\mathfrak{s}}^{(2)}[G]\right)_{\gamma_{2} \gamma_{3}}^{\mathrm{inv}}}$} \\
\scalebox{0.99}{${\displaystyle = }$} & \scalebox{0.99}{${\displaystyle - \underbrace{\left(\Pi^{\mathrm{inv}}[G] + \Phi_{\mathfrak{s}}^{(2)}[G]\right)_{\gamma_{2} \hat{\gamma}_{1}}^{\mathrm{inv}}}_{W_{\mathfrak{s},\gamma_{2} \hat{\gamma}_{1}}^{(2)}[K]} \left(\frac{\delta \Pi^{\mathrm{inv}}[G]}{\delta G_{\gamma_{1}}} + \frac{\delta\Phi_{\mathfrak{s}}^{(2)}[G]}{\delta G_{\gamma_{1}}}\right)_{\hat{\gamma}_{1} \hat{\gamma}_{2}} \underbrace{\left(\Pi^{\mathrm{inv}}[G] + \Phi_{\mathfrak{s}}^{(2)}[G]\right)_{\hat{\gamma}_{2} \gamma_{3}}^{\mathrm{inv}}}_{W_{\mathfrak{s},\hat{\gamma}_{2} \gamma_{3}}^{(2)}[K]}}$} \\
\scalebox{0.99}{${\displaystyle = }$} & \scalebox{0.99}{${\displaystyle \ W_{\mathfrak{s},\gamma_{2} \hat{\gamma}_{1}}^{(2)}[K] \left(\Pi_{\hat{\gamma}_{1}\hat{\gamma}_{2}}^{\mathrm{inv}}[G] \frac{\delta \Pi_{\hat{\gamma}_{2} \hat{\gamma}_{3}}[G]}{\delta G_{\gamma_{1}}} \Pi_{\hat{\gamma}_{3} \hat{\gamma}_{4}}^{\mathrm{inv}}[G] - \Phi_{\mathfrak{s},\gamma_{1} \hat{\gamma}_{1} \hat{\gamma}_{4}}^{(3)}[G]\right) W_{\mathfrak{s},\hat{\gamma}_{4} \gamma_{3}}^{(2)}[K]\;,}$}
\end{split}
\label{eq:2PIfrgUflowDW2DG}
\end{equation}
where we have replaced $W_{\mathfrak{s}}^{(2)}[K]$ in the first line using~\eqref{eq:2PIfrgW2PiInvPhi2}. Inserting the latter expression of $\frac{\delta W_{\mathfrak{s},\gamma_{2} \gamma_{3}}^{(2)}[K]}{\delta G_{\gamma_{1}}}$ into~\eqref{eq:2PIfrgUflowPhi1IntermediaryResult} and setting $K_{\gamma}=0$ $\forall\gamma$ yields:
\begin{equation}
\overline{\dot{\Sigma}}_{\mathfrak{s},\gamma} = -\frac{1}{6} \dot{U}_{\mathfrak{s},\hat{\gamma}_{1} \hat{\gamma}_{2}} \left[\overline{W}_{\mathfrak{s},\hat{\gamma}_{2} \hat{\gamma}_{3}}^{(2)} \left(\overline{\Pi}_{\mathfrak{s},\hat{\gamma}_{3}\hat{\gamma}_{4}}^{\mathrm{inv}} \frac{\delta \overline{\Pi}_{\mathfrak{s},\hat{\gamma}_{4}\hat{\gamma}_{5}}}{\delta \overline{G}_{\mathfrak{s},\gamma}} \overline{\Pi}_{\mathfrak{s},\hat{\gamma}_{5}\hat{\gamma}_{6}}^{\mathrm{inv}} - \overline{\Phi}_{\mathfrak{s},\gamma \hat{\gamma}_{3} \hat{\gamma}_{6}}^{(3)} \right) \overline{W}_{\mathfrak{s},\hat{\gamma}_{6} \hat{\gamma}_{1}}^{(2)} + \frac{1}{2}\frac{\delta \overline{\Pi}_{\mathfrak{s},\hat{\gamma}_{2}\hat{\gamma}_{1}}}{\delta \overline{G}_{\mathfrak{s},\gamma}} \right] \;,
\label{eq:2PIfrgUflowPhi1}
\end{equation}
using the shorthand notation~\eqref{eq:ShorthandNotation2PIFRGUflow}. Hence,~\eqref{eq:2PIfrgUflowdotPhiBarFirstStep} can be rewritten by replacing $\overline{\dot{\Phi}}_{\mathfrak{s}}$ and $\overline{\dot{\Sigma}}_{\mathfrak{s}}$ using respectively~\eqref{eq:2PIfrgPhisdot} and~\eqref{eq:2PIfrgUflowPhi1}:
\begin{equation}
\begin{split}
\scalebox{0.95}{${\displaystyle\dot{\overline{\Phi}}_{\mathfrak{s}} =}$} & \scalebox{0.95}{${\displaystyle \ \frac{1}{6} \dot{U}_{\mathfrak{s},\hat{\gamma}_{1} \hat{\gamma}_{2}} \left(\overline{W}_{\mathfrak{s}}^{(2)} + \frac{1}{2}\overline{\Pi}_{\mathfrak{s}}\right)_{\hat{\gamma}_{2}\hat{\gamma}_{1}}}$} \\
& \scalebox{0.95}{${\displaystyle +\frac{1}{6} \overline{\Sigma}_{\mathfrak{s},\hat{\gamma}_{1}} \overline{W}_{\mathfrak{s},\hat{\gamma}_{1} \hat{\gamma}_{2}}^{(2)} \dot{U}_{\mathfrak{s},\hat{\gamma}_{3} \hat{\gamma}_{4}} \left[ \overline{W}_{\mathfrak{s},\hat{\gamma}_{4} \hat{\gamma}_{5}}^{(2)} \left(\overline{\Pi}_{\mathfrak{s},\hat{\gamma}_{5}\hat{\gamma}_{6}}^{\mathrm{inv}} \frac{\delta \overline{\Pi}_{\mathfrak{s},\hat{\gamma}_{6}\hat{\gamma}_{7}}}{\delta \overline{G}_{\mathfrak{s},\hat{\gamma}_{2}}} \overline{\Pi}_{\mathfrak{s},\hat{\gamma}_{7}\hat{\gamma}_{8}}^{\mathrm{inv}} - \overline{\Phi}_{\mathfrak{s},\hat{\gamma}_{2} \hat{\gamma}_{5} \hat{\gamma}_{8}}^{(3)} \right) \overline{W}_{\mathfrak{s},\hat{\gamma}_{8} \hat{\gamma}_{3}}^{(2)} + \frac{1}{2} \frac{\delta \overline{\Pi}_{\mathfrak{s},\hat{\gamma}_{4} \hat{\gamma}_{3}}}{\delta \overline{G}_{\mathfrak{s},\hat{\gamma}_{2}}} \right] \;.}$}
\end{split}
\label{eq:2PIfrgUflowExpressionPhiDot}
\end{equation}

\item Expression of $\dot{\overline{\Sigma}}_{\mathfrak{s}}$:\\
From~\eqref{eq:2PIfrgGeneralFlowEqSigma} and~\eqref{eq:2PIfrgGdotW2Phi1BetheSalpeter}, we also obtain:
\begin{equation}
\dot{\overline{\Sigma}}_{\mathfrak{s},\gamma} = \overline{\dot{\Sigma}}_{\mathfrak{s},\gamma} - \dot{\overline{G}}_{\mathfrak{s},\hat{\gamma}} \overline{\Phi}^{(2)}_{\mathfrak{s},\hat{\gamma}\gamma} = \overline{\dot{\Sigma}}_{\mathfrak{s},\gamma} - \overline{\Phi}^{(2)}_{\mathfrak{s},\gamma\hat{\gamma}_{1}} \overline{W}_{\mathfrak{s},\hat{\gamma}_{1} \hat{\gamma}_{2}}^{(2)} \overline{\dot{\Sigma}}_{\mathfrak{s},\hat{\gamma}_{2}}\;.
\label{eq:2PIfrgUflowdotSigmaBarFirstStep}
\end{equation}
Leaving once again summations underlying multiplications between bosonic matrices implicit and following a reasoning similar to that of~\eqref{eq:DerivationBetheSalpeterEquationImplicit} in the derivation of the Bethe-Salpeter equation, we have:
\begin{equation}
\begin{split}
\dot{\overline{\Sigma}}_{\mathfrak{s}} = & \ \overline{\dot{\Sigma}}_{\mathfrak{s}} - \overline{\Phi}^{(2)}_{\mathfrak{s}} \overline{W}_{\mathfrak{s}}^{(2)} \overline{\dot{\Sigma}}_{\mathfrak{s}} \\
= & \left(\mathcal{I}-\overline{\Phi}^{(2)}_{\mathfrak{s}} \overline{W}_{\mathfrak{s}}^{(2)}\right) \overline{\dot{\Sigma}}_{\mathfrak{s}} \\
= & \left[\mathcal{I}-\overline{\Phi}^{(2)}_{\mathfrak{s}} \overline{\Pi}_{\mathfrak{s}} \left(\mathcal{I} + \overline{\Pi}_{\mathfrak{s}} \overline{\Phi}_{\mathfrak{s}}^{(2)}\right)^{\mathrm{inv}} \right] \overline{\dot{\Sigma}}_{\mathfrak{s}} \\
= & \left[\mathcal{I} - \overline{\Pi}_{\mathfrak{s}} \overline{\Phi}^{(2)}_{\mathfrak{s}} \sum_{n=0}^{\infty}(-1)^{n} \left(\overline{\Pi}_{\mathfrak{s}} \overline{\Phi}_{\mathfrak{s}}^{(2)}\right)^{n} \right] \overline{\dot{\Sigma}}_{\mathfrak{s}} \\
= & \left[\sum_{n=0}^{\infty}(-1)^{n} \left(\overline{\Pi}_{\mathfrak{s}} \overline{\Phi}_{\mathfrak{s}}^{(2)}\right)^{n}\right] \overline{\dot{\Sigma}}_{\mathfrak{s}} \\
= & \left(\mathcal{I} + \overline{\Pi}_{\mathfrak{s}} \overline{\Phi}_{\mathfrak{s}}^{(2)}\right)^{\mathrm{inv}} \overline{\dot{\Sigma}}_{\mathfrak{s}}\;,
\end{split}
\label{eq:2PIFRGSigmaSBetheSalpeter}
\end{equation}
where~\eqref{eq:2PIfrgW2Gamma2Pi} was exploited to replace $\overline{W}_{\mathfrak{s}}^{(2)}$ in the third line. Combining~\eqref{eq:2PIFRGSigmaSBetheSalpeter} with \eqref{eq:2PIfrgUflowPhi1} gives us the following expression for $\dot{\overline{\Sigma}}_{\mathfrak{s}}$:
\begin{equation}
\scalebox{0.89}{${\displaystyle\dot{\overline{\Sigma}}_{\mathfrak{s},\gamma} = -\frac{1}{6}\left(\mathcal{I} + \overline{\Pi}_{\mathfrak{s}} \overline{\Phi}_{\mathfrak{s}}^{(2)}\right)^{\mathrm{inv}}_{\gamma\hat{\gamma}_{1}} \dot{U}_{\mathfrak{s},\hat{\gamma}_{2} \hat{\gamma}_{3}} \Bigg[\overline{W}_{\mathfrak{s},\hat{\gamma}_{3} \hat{\gamma}_{4}}^{(2)} \left(\overline{\Pi}_{\mathfrak{s},\hat{\gamma}_{4}\hat{\gamma}_{5}}^{\mathrm{inv}} \frac{\delta \overline{\Pi}_{\mathfrak{s},\hat{\gamma}_{5}\hat{\gamma}_{6}}}{\delta \overline{G}_{\mathfrak{s},\hat{\gamma}_{1}}} \overline{\Pi}_{\mathfrak{s},\hat{\gamma}_{6}\hat{\gamma}_{7}}^{\mathrm{inv}} - \overline{\Phi}_{\mathfrak{s},\hat{\gamma}_{1} \hat{\gamma}_{4} \hat{\gamma}_{7}}^{(3)} \right) \overline{W}_{\mathfrak{s},\hat{\gamma}_{7} \hat{\gamma}_{2}}^{(2)} + \frac{1}{2} \frac{\delta \overline{\Pi}_{\mathfrak{s},\hat{\gamma}_{3}\hat{\gamma}_{2}}}{\delta \overline{G}_{\mathfrak{s},\hat{\gamma}_{1}}} \Bigg]\;.}$}
\label{eq:2PIfrgUflowSigmaDotCompact}
\end{equation}
As a next step, we evaluate the derivative of the pair propagator $\Pi[G]$ involved in the RHS of~\eqref{eq:2PIfrgUflowSigmaDotCompact}. Considering expression~\eqref{eq:2PIfrgExpressionPiandG} of $\Pi[G]$, we can show that:
\begin{equation}
\begin{split}
\frac{\delta \Pi_{\gamma_{2}\gamma_{3}}[G]}{\delta G_{\gamma_{1}}} = & \ \frac{\delta}{\delta G_{\gamma_{1}}}\left[G_{\alpha_{2} \alpha'_{3}} G_{\alpha'_{2} \alpha_{3}} + \zeta G_{\alpha_{2} \alpha_{3}} G_{\alpha'_{2} \alpha'_{3}}\right] \\
= & \ \frac{\delta G_{\alpha_{2} \alpha'_{3}}}{\delta G_{\gamma_{1}}} G_{\alpha'_{2} \alpha_{3}} + G_{\alpha_{2} \alpha'_{3}} \frac{\delta G_{\alpha'_{2} \alpha_{3}}}{\delta G_{\gamma_{1}}} + \zeta \frac{\delta G_{\alpha_{2} \alpha_{3}}}{\delta G_{\gamma_{1}}} G_{\alpha'_{2} \alpha'_{3}} + \zeta G_{\alpha_{2} \alpha_{3}} \frac{\delta G_{\alpha'_{2} \alpha'_{3}}}{\delta G_{\gamma_{1}}} \\
= & \ G_{\alpha'_{2} \alpha_{3}} \delta_{\alpha_{2} \alpha_{1}}\delta_{\alpha'_{3} \alpha'_{1}} + G_{\alpha_{2} \alpha'_{3}} \delta_{\alpha'_{2} \alpha_{1}} \delta_{\alpha_{3} \alpha'_{1}} + \zeta G_{\alpha'_{2} \alpha'_{3}} \delta_{\alpha_{2} \alpha_{1}}\delta_{\alpha_{3} \alpha'_{1}} + \zeta G_{\alpha_{2} \alpha_{3}} \delta_{\alpha'_{2} \alpha_{1}}\delta_{\alpha'_{3} \alpha'_{1}} \\
& + \zeta\left(\alpha_{1} \leftrightarrow \alpha'_{1}\right) \;.
\end{split}
\label{eq:2PIFRGuflowExpressionDerivPidG}
\end{equation}
From this, we calculate:
\begin{equation}
\begin{split}
& \scalebox{0.85}{${\displaystyle\left(\mathcal{I} + \overline{\Pi}_{\mathfrak{s}} \overline{\Phi}_{\mathfrak{s}}^{(2)}\right)^{\mathrm{inv}}_{\gamma\hat{\gamma}_{1}} \dot{U}_{\mathfrak{s},\hat{\gamma}_{2} \hat{\gamma}_{3}} \frac{\delta \overline{\Pi}_{\mathfrak{s},\hat{\gamma}_{3}\hat{\gamma}_{2}}}{\delta \overline{G}_{\mathfrak{s},\hat{\gamma}_{1}}} }$} \\
& \hspace{1.0cm} \scalebox{0.85}{${\displaystyle = \frac{1}{8} \int_{\gamma_{1},\gamma_{2},\gamma_{3}} \left(\mathcal{I} + \overline{\Pi}_{\mathfrak{s}} \overline{\Phi}_{\mathfrak{s}}^{(2)}\right)^{\mathrm{inv}}_{\gamma\gamma_{1}} \dot{U}_{\mathfrak{s},\gamma_{2} \gamma_{3}} \frac{\delta \overline{\Pi}_{\mathfrak{s},\gamma_{3}\gamma_{2}}}{\delta \overline{G}_{\mathfrak{s},\gamma_{1}}} }$} \\
& \hspace{1.0cm} \scalebox{0.85}{${\displaystyle = \frac{1}{8} \int_{\gamma_{1}} \Bigg[\int_{\alpha_{2},\alpha'_{3}} \left(\mathcal{I} + \overline{\Pi}_{\mathfrak{s}} \overline{\Phi}_{\mathfrak{s}}^{(2)}\right)^{\mathrm{inv}}_{\gamma \gamma_{1}} \underbrace{\dot{U}_{\mathfrak{s},\alpha_{2} \alpha'_{1} \alpha_{1} \alpha'_{3}}}_{\dot{U}_{\mathfrak{s},\alpha_{1} \alpha'_{3} \alpha_{2} \alpha'_{1}}} \overline{G}_{\mathfrak{s},\alpha'_{3} \alpha_{2}} + \int_{\alpha'_{2},\alpha_{3}} \left(\mathcal{I} + \overline{\Pi}_{\mathfrak{s}} \overline{\Phi}_{\mathfrak{s}}^{(2)}\right)^{\mathrm{inv}}_{\gamma \gamma_{1}} \underbrace{\dot{U}_{\mathfrak{s},\alpha'_{1} \alpha'_{2} \alpha_{3} \alpha_{1}}}_{\dot{U}_{\mathfrak{s},\alpha_{1} \alpha_{3} \alpha'_{2} \alpha'_{1}}} \overline{G}_{\mathfrak{s},\alpha_{3} \alpha'_{2}} }$} \\
& \hspace{2.3cm} \scalebox{0.85}{${\displaystyle + \zeta \int_{\alpha'_{2},\alpha'_{3}} \left(\mathcal{I} + \overline{\Pi}_{\mathfrak{s}} \overline{\Phi}_{\mathfrak{s}}^{(2)}\right)^{\mathrm{inv}}_{\gamma \gamma_{1}} \underbrace{\dot{U}_{\mathfrak{s},\alpha'_{1} \alpha'_{2} \alpha_{1} \alpha'_{3}}}_{\zeta \dot{U}_{\mathfrak{s},\alpha_{1} \alpha'_{3} \alpha'_{2} \alpha'_{1}}} \overline{G}_{\mathfrak{s},\alpha'_{3} \alpha'_{2}} + \zeta \int_{\alpha_{2},\alpha_{3}} \left(\mathcal{I} + \overline{\Pi}_{\mathfrak{s}} \overline{\Phi}_{\mathfrak{s}}^{(2)}\right)^{\mathrm{inv}}_{\gamma \gamma_{1}} \underbrace{\dot{U}_{\mathfrak{s},\alpha_{2} \alpha'_{1} \alpha_{3} \alpha_{1}}}_{\zeta \dot{U}_{\mathfrak{s},\alpha_{1} \alpha_{3} \alpha_{2} \alpha'_{1}}} \overline{G}_{\mathfrak{s},\alpha_{3} \alpha_{2}} \Bigg] }$} \\
& \hspace{1.25cm} \scalebox{0.85}{${\displaystyle + \zeta\left(\alpha_{1} \leftrightarrow \alpha'_{1}\right) }$} \\
& \hspace{1.0cm} \scalebox{0.85}{${\displaystyle = \int_{\gamma_{1},\gamma_{2}} \left(\mathcal{I} + \overline{\Pi}_{\mathfrak{s}} \overline{\Phi}_{\mathfrak{s}}^{(2)}\right)^{\mathrm{inv}}_{\gamma \gamma_{1}} \dot{U}_{\mathfrak{s},\alpha_{1} \alpha_{2} \alpha'_{2} \alpha'_{1}} \overline{G}_{\mathfrak{s},\gamma_{2}} }$} \\
& \hspace{1.0cm} \scalebox{0.85}{${\displaystyle = 4 \left(\mathcal{I} + \overline{\Pi}_{\mathfrak{s}} \overline{\Phi}_{\mathfrak{s}}^{(2)}\right)^{\mathrm{inv}}_{\gamma\hat{\gamma}_{1}} \dot{U}_{\mathfrak{s},\hat{\alpha}_{1} \hat{\alpha}_{2} \hat{\alpha}'_{2} \hat{\alpha}'_{1}} \overline{G}_{\mathfrak{s},\hat{\gamma}_{2}} \;. }$}
\end{split}
\label{eq:DetailedCalculation1Sigma2PIFRGUflowAppendix}
\end{equation}
In the same way, we obtain:
\begin{equation*}
\begin{split}
& \scalebox{0.87}{${\displaystyle \left(\mathcal{I} + \overline{\Pi}_{\mathfrak{s}} \overline{\Phi}_{\mathfrak{s}}^{(2)}\right)^{\mathrm{inv}}_{\gamma\hat{\gamma}_{1}} \dot{U}_{\mathfrak{s},\hat{\gamma}_{2} \hat{\gamma}_{3}} \overline{W}_{\mathfrak{s},\hat{\gamma}_{3} \hat{\gamma}_{4}}^{(2)} \overline{\Pi}_{\mathfrak{s},\hat{\gamma}_{4}\hat{\gamma}_{5}}^{\mathrm{inv}} \frac{\delta \overline{\Pi}_{\mathfrak{s},\hat{\gamma}_{5}\hat{\gamma}_{6}}}{\delta \overline{G}_{\mathfrak{s},\hat{\gamma}_{1}}} \overline{\Pi}_{\mathfrak{s},\hat{\gamma}_{6}\hat{\gamma}_{7}}^{\mathrm{inv}} \overline{W}_{\mathfrak{s},\hat{\gamma}_{7} \hat{\gamma}_{2}}^{(2)} }$} \\
& \hspace{1.0cm} \scalebox{0.87}{${\displaystyle = \frac{1}{128} \int_{\gamma_{1},\gamma_{2},\gamma_{3},\gamma_{4},\gamma_{5},\gamma_{6},\gamma_{7}} \left(\mathcal{I} + \overline{\Pi}_{\mathfrak{s}} \overline{\Phi}_{\mathfrak{s}}^{(2)}\right)^{\mathrm{inv}}_{\gamma\gamma_{1}} \dot{U}_{\mathfrak{s},\gamma_{2} \gamma_{3}} \overline{W}_{\mathfrak{s},\gamma_{3} \gamma_{4}}^{(2)} \overline{\Pi}_{\mathfrak{s},\gamma_{4} \gamma_{6}}^{\mathrm{inv}} \frac{\delta \overline{\Pi}_{\mathfrak{s},\gamma_{6} \gamma_{7}}}{\delta \overline{G}_{\mathfrak{s},\gamma_{1}}} \overline{\Pi}_{\mathfrak{s},\gamma_{7} \gamma_{5}}^{\mathrm{inv}} \overline{W}_{\mathfrak{s},\gamma_{5} \gamma_{2}}^{(2)} }$} \\
& \hspace{1.0cm} \scalebox{0.87}{${\displaystyle = \frac{1}{128}\int_{\gamma_{1},\gamma_{2},\gamma_{3},\gamma_{4},\gamma_{5}}\Bigg[\int_{\alpha'_{6},\alpha_{7}} \left(\mathcal{I} + \overline{\Pi}_{\mathfrak{s}} \overline{\Phi}_{\mathfrak{s}}^{(2)}\right)^{\mathrm{inv}}_{\gamma\gamma_{1}} \underbrace{\dot{U}_{\mathfrak{s},\gamma_{2} \gamma_{3}} \overline{W}_{\mathfrak{s},\gamma_{3} \gamma_{4}}^{(2)} \overline{\Pi}_{\mathfrak{s},\gamma_{4} (\alpha_{1},\alpha'_{6})}^{\mathrm{inv}}}_{\overline{\Pi}_{\mathfrak{s},(\alpha_{1},\alpha'_{6}) \gamma_{4}}^{\mathrm{inv}} \overline{W}_{\mathfrak{s},\gamma_{4} \gamma_{3}}^{(2)} \dot{U}_{\mathfrak{s},\gamma_{3} \gamma_{2}}} \overline{G}_{\mathfrak{s},\alpha'_{6} \alpha_{7}} \underbrace{\overline{\Pi}_{\mathfrak{s},(\alpha_{7},\alpha'_{1}) \gamma_{5}}^{\mathrm{inv}} \overline{W}_{\mathfrak{s},\gamma_{5} \gamma_{2}}^{(2)}}_{\overline{W}_{\mathfrak{s},\gamma_{2} \gamma_{5}}^{(2)} \overline{\Pi}_{\mathfrak{s},\gamma_{5} (\alpha_{7},\alpha'_{1})}^{\mathrm{inv}}} }$} \\
& \hspace{4.05cm} \scalebox{0.87}{${\displaystyle + \int_{\alpha_{6},\alpha'_{7}} \left(\mathcal{I} + \overline{\Pi}_{\mathfrak{s}} \overline{\Phi}_{\mathfrak{s}}^{(2)}\right)^{\mathrm{inv}}_{\gamma\gamma_{1}} \underbrace{\dot{U}_{\mathfrak{s},\gamma_{2} \gamma_{3}} \overline{W}_{\mathfrak{s},\gamma_{3} \gamma_{4}}^{(2)} \overline{\Pi}_{\mathfrak{s},\gamma_{4} (\alpha_{6},\alpha_{1})}^{\mathrm{inv}}}_{\zeta\overline{\Pi}_{\mathfrak{s},(\alpha_{1},\alpha_{6}) \gamma_{4}}^{\mathrm{inv}} \overline{W}_{\mathfrak{s},\gamma_{4} \gamma_{3}}^{(2)} \dot{U}_{\mathfrak{s},\gamma_{3} \gamma_{2}}} \overline{G}_{\mathfrak{s},\alpha_{6} \alpha'_{7}} \underbrace{\overline{\Pi}_{\mathfrak{s},(\alpha'_{1},\alpha'_{7}) \gamma_{5}}^{\mathrm{inv}} \overline{W}_{\mathfrak{s},\gamma_{5} \gamma_{2}}^{(2)}}_{\zeta\overline{W}_{\mathfrak{s},\gamma_{2} \gamma_{5}}^{(2)} \overline{\Pi}_{\mathfrak{s},\gamma_{5} (\alpha'_{7},\alpha'_{1})}^{\mathrm{inv}}} }$} \\
& \hspace{4.05cm} \scalebox{0.87}{${\displaystyle + \zeta\int_{\alpha'_{6},\alpha'_{7}} \left(\mathcal{I} + \overline{\Pi}_{\mathfrak{s}} \overline{\Phi}_{\mathfrak{s}}^{(2)}\right)^{\mathrm{inv}}_{\gamma\gamma_{1}} \underbrace{\dot{U}_{\mathfrak{s},\gamma_{2} \gamma_{3}} \overline{W}_{\mathfrak{s},\gamma_{3} \gamma_{4}}^{(2)} \overline{\Pi}_{\mathfrak{s},\gamma_{4} (\alpha_{1},\alpha'_{6})}^{\mathrm{inv}}}_{\overline{\Pi}_{\mathfrak{s},(\alpha_{1},\alpha'_{6}) \gamma_{4}}^{\mathrm{inv}} \overline{W}_{\mathfrak{s},\gamma_{4} \gamma_{3}}^{(2)} \dot{U}_{\mathfrak{s},\gamma_{3} \gamma_{2}}} \overline{G}_{\mathfrak{s},\alpha'_{6} \alpha'_{7}} \underbrace{\overline{\Pi}_{\mathfrak{s},(\alpha'_{1},\alpha'_{7}) \gamma_{5}}^{\mathrm{inv}} \overline{W}_{\mathfrak{s},\gamma_{5} \gamma_{2}}^{(2)}}_{\zeta\overline{W}_{\mathfrak{s},\gamma_{2} \gamma_{5}}^{(2)} \overline{\Pi}_{\mathfrak{s},\gamma_{5} (\alpha'_{7},\alpha'_{1})}^{\mathrm{inv}}} }$}
\end{split}
\end{equation*}
\begin{equation}
\begin{split}
& \hspace{4.05cm} \scalebox{0.87}{${\displaystyle + \zeta\int_{\alpha_{6},\alpha_{7}} \left(\mathcal{I} + \overline{\Pi}_{\mathfrak{s}} \overline{\Phi}_{\mathfrak{s}}^{(2)}\right)^{\mathrm{inv}}_{\gamma\gamma_{1}} \underbrace{\dot{U}_{\mathfrak{s},\gamma_{2} \gamma_{3}} \overline{W}_{\mathfrak{s},\gamma_{3} \gamma_{4}}^{(2)} \overline{\Pi}_{\mathfrak{s},\gamma_{4} (\alpha_{6},\alpha_{1})}^{\mathrm{inv}}}_{\zeta\overline{\Pi}_{\mathfrak{s},(\alpha_{1},\alpha_{6}) \gamma_{4}}^{\mathrm{inv}} \overline{W}_{\mathfrak{s},\gamma_{4} \gamma_{3}}^{(2)} \dot{U}_{\mathfrak{s},\gamma_{3} \gamma_{2}}} \overline{G}_{\mathfrak{s},\alpha_{6} \alpha_{7}} \underbrace{\overline{\Pi}_{\mathfrak{s},(\alpha_{7},\alpha'_{1}) \gamma_{5}}^{\mathrm{inv}} \overline{W}_{\mathfrak{s},\gamma_{5} \gamma_{2}}^{(2)}}_{\overline{W}_{\mathfrak{s},\gamma_{2} \gamma_{5}}^{(2)} \overline{\Pi}_{\mathfrak{s},\gamma_{5} (\alpha_{7},\alpha'_{1})}^{\mathrm{inv}}} \Bigg] }$} \\
& \hspace{1.25cm} \scalebox{0.87}{${\displaystyle + \zeta\left(\alpha_{1}\leftrightarrow \alpha'_{1}\right) }$} \\
& \hspace{1.0cm} \scalebox{0.87}{${\displaystyle = \frac{1}{16} \int_{\gamma_{1},\gamma_{2},\gamma_{3},\gamma_{4},\gamma_{5},\gamma_{6}} \left(\mathcal{I} + \overline{\Pi}_{\mathfrak{s}} \overline{\Phi}_{\mathfrak{s}}^{(2)}\right)^{\mathrm{inv}}_{\gamma\gamma_{1}} \overline{\Pi}_{\mathfrak{s},(\alpha_{1},\alpha_{6}) \gamma_{4}}^{\mathrm{inv}} \overline{W}_{\mathfrak{s},\gamma_{4} \gamma_{3}}^{(2)} \dot{U}_{\mathfrak{s},\gamma_{3} \gamma_{2}} \overline{W}_{\mathfrak{s},\gamma_{2} \gamma_{5}}^{(2)} \overline{\Pi}_{\mathfrak{s},\gamma_{5} (\alpha'_{6},\alpha'_{1})}^{\mathrm{inv}} \overline{G}_{\mathfrak{s},\gamma_{6}} }$} \\
& \hspace{1.0cm} \scalebox{0.87}{${\displaystyle = 4 \left(\mathcal{I} + \overline{\Pi}_{\mathfrak{s}} \overline{\Phi}_{\mathfrak{s}}^{(2)}\right)^{\mathrm{inv}}_{\gamma\hat{\gamma}_{1}} \left(\overline{\Pi}_{\mathfrak{s}}^{\mathrm{inv}} \overline{W}_{\mathfrak{s}}^{(2)} \dot{U}_{\mathfrak{s}} \overline{W}_{\mathfrak{s}}^{(2)} \overline{\Pi}_{\mathfrak{s}}^{\mathrm{inv}}\right)_{\hat{\alpha}_{1} \hat{\alpha}_{2} \hat{\alpha}'_{2} \hat{\alpha}'_{1}} \overline{G}_{\mathfrak{s},\hat{\gamma}_{2}} }$} \\
& \hspace{1.0cm} \scalebox{0.87}{${\displaystyle = 4 \left(\mathcal{I} + \overline{\Pi}_{\mathfrak{s}} \overline{\Phi}_{\mathfrak{s}}^{(2)}\right)^{\mathrm{inv}}_{\gamma\hat{\gamma}_{1}} \left[ \left(\mathcal{I} + \overline{\Pi}_{\mathfrak{s}} \overline{\Phi}_{\mathfrak{s}}^{(2)}\right)^{\mathrm{inv}} \dot{U}_{\mathfrak{s}} \left(\mathcal{I} + \overline{\Pi}_{\mathfrak{s}} \overline{\Phi}_{\mathfrak{s}}^{(2)}\right)^{\mathrm{inv}} \right]_{\hat{\alpha}_{1} \hat{\alpha}_{2} \hat{\alpha}'_{2} \hat{\alpha}'_{1}} \overline{G}_{\mathfrak{s},\hat{\gamma}_{2}} \;, }$}
\end{split}
\label{eq:DetailedCalculation2Sigma2PIFRGUflowAppendix}
\end{equation}
where the last line was derived using the identity:
\begin{equation}
\overline{\Pi}_{\mathfrak{s},\gamma_{1}\hat{\gamma}}^{\mathrm{inv}} \overline{W}_{\mathfrak{s},\hat{\gamma}\gamma_{2}}^{(2)} = \left(\mathcal{I} + \overline{\Pi}_{\mathfrak{s}} \overline{\Phi}_{\mathfrak{s}}^{(2)}\right)_{\gamma_{1}\gamma_{2}}^{\mathrm{inv}} \;,
\label{eq:2PIfrgW2PiInvPhi2tUflow}
\end{equation}
which follows from~\eqref{eq:2PIfrgW2PiInvPhi2} at $K_{\gamma}=0$ $\forall\gamma$. From both~\eqref{eq:DetailedCalculation1Sigma2PIFRGUflowAppendix} and~\eqref{eq:DetailedCalculation2Sigma2PIFRGUflowAppendix}, we prove that the flow equation of $\overline{\Sigma}_{\mathfrak{s}}$ given by~\eqref{eq:2PIfrgUflowSigmaDotCompact} is equivalent to\footnote{Note that the present derivation of the flow equation expressing $\dot{\overline{\Sigma}}_{\mathfrak{s},\gamma}$ reproduces the main results of section~6.1. of ref.~\cite{ren15}. In particular,~\eqref{eq:2PIfrgmUflowSelfEnergyNpert1} is to be compared with the equations numbered (81) and (83) in the latter reference.}:
\begin{equation}
\begin{split}
\dot{\overline{\Sigma}}_{\mathfrak{s},\gamma} = & -\frac{1}{3}\left(\mathcal{I} + \overline{\Pi}_{\mathfrak{s}} \overline{\Phi}_{\mathfrak{s}}^{(2)}\right)^{\mathrm{inv}}_{\gamma\hat{\gamma}_{1}} \left[ 2 \left(\mathcal{I} + \overline{\Pi}_{\mathfrak{s}} \overline{\Phi}_{\mathfrak{s}}^{(2)}\right)^{\mathrm{inv}} \dot{U}_{\mathfrak{s}} \left(\mathcal{I} + \overline{\Pi}_{\mathfrak{s}} \overline{\Phi}_{\mathfrak{s}}^{(2)}\right)^{\mathrm{inv}} + \dot{U}_{\mathfrak{s}} \right]_{\hat{\alpha}_{1} \hat{\alpha}_{2} \hat{\alpha}'_{2} \hat{\alpha}'_{1}} \overline{G}_{\mathfrak{s},\hat{\gamma}_{2}} \\
& +\frac{1}{6} \left(\mathcal{I} + \overline{\Pi}_{\mathfrak{s}} \overline{\Phi}_{\mathfrak{s}}^{(2)}\right)^{\mathrm{inv}}_{\gamma\hat{\gamma}_{1}} \dot{U}_{\mathfrak{s},\hat{\gamma}_{2} \hat{\gamma}_{3}} \overline{W}_{\mathfrak{s},\hat{\gamma}_{3} \hat{\gamma}_{4}}^{(2)} \overline{\Phi}_{\mathfrak{s},\hat{\gamma}_{1} \hat{\gamma}_{4} \hat{\gamma}_{5}}^{(3)} \overline{W}_{\mathfrak{s},\hat{\gamma}_{5} \hat{\gamma}_{2}}^{(2)} \;.
\end{split}
\label{eq:2PIfrgmUflowSelfEnergyNpert1}
\end{equation}

\item Expression of $\dot{\overline{\Phi}}_{\mathfrak{s}}^{(2)}$:\\
From~\eqref{eq:2PIfrgUflowPhi1}, we infer the following expression for $\overline{\dot{\Phi}}_{\mathfrak{s}}^{(2)}$:
\begin{equation*}
\begin{split}
\overline{\dot{\Phi}}_{\mathfrak{s},\gamma_{1}\gamma_{2}}^{(2)} = \frac{1}{6} \dot{U}_{\mathfrak{s},\hat{\gamma}_{1} \hat{\gamma}_{2}} & \Bigg[ 2\frac{\delta \overline{W}_{\mathfrak{s},\hat{\gamma}_{2} \hat{\gamma}_{3}}^{(2)}}{\delta \overline{G}_{\mathfrak{s},\gamma_{1}}} \left(\overline{\Pi}_{\mathfrak{s},\hat{\gamma}_{3}\hat{\gamma}_{4}}^{\mathrm{inv}} \frac{\delta \overline{\Pi}_{\mathfrak{s},\hat{\gamma}_{4}\hat{\gamma}_{5}}}{\delta \overline{G}_{\mathfrak{s},\gamma_{2}}} \overline{\Pi}_{\mathfrak{s},\hat{\gamma}_{5}\hat{\gamma}_{6}}^{\mathrm{inv}} - \overline{\Phi}_{\mathfrak{s},\gamma_{2} \hat{\gamma}_{3} \hat{\gamma}_{6}}^{(3)} \right) \overline{W}_{\mathfrak{s},\hat{\gamma}_{6} \hat{\gamma}_{1}}^{(2)} \\
& + \overline{W}_{\mathfrak{s},\hat{\gamma}_{2} \hat{\gamma}_{3}}^{(2)} \Bigg(\frac{\delta \overline{\Pi}_{\mathfrak{s},\hat{\gamma}_{3}\hat{\gamma}_{4}}^{\mathrm{inv}}}{\delta \overline{G}_{\mathfrak{s},\gamma_{1}}} \frac{\delta \overline{\Pi}_{\mathfrak{s},\hat{\gamma}_{4}\hat{\gamma}_{5}}}{\delta \overline{G}_{\mathfrak{s},\gamma_{2}}} \overline{\Pi}_{\mathfrak{s},\hat{\gamma}_{5}\hat{\gamma}_{6}}^{\mathrm{inv}} + \overline{\Pi}_{\mathfrak{s},\hat{\gamma}_{3}\hat{\gamma}_{4}}^{\mathrm{inv}} \frac{\delta^{2} \overline{\Pi}_{\mathfrak{s},\hat{\gamma}_{4}\hat{\gamma}_{5}}}{\delta \overline{G}_{\mathfrak{s},\gamma_{1}} \delta \overline{G}_{\mathfrak{s},\gamma_{2}}} \overline{\Pi}_{\mathfrak{s},\hat{\gamma}_{5} \hat{\gamma}_{6}}^{\mathrm{inv}} \\
& + \overline{\Pi}_{\mathfrak{s},\hat{\gamma}_{3}\hat{\gamma}_{4}}^{\mathrm{inv}} \frac{\delta \overline{\Pi}_{\mathfrak{s},\hat{\gamma}_{4} \hat{\gamma}_{5}}}{\delta \overline{G}_{\mathfrak{s},\gamma_{2}}} \frac{\delta \overline{\Pi}_{\mathfrak{s},\hat{\gamma}_{5}\hat{\gamma}_{6}}^{\mathrm{inv}}}{\delta \overline{G}_{\mathfrak{s},\gamma_{1}}} - \frac{\delta \overline{\Phi}_{\mathfrak{s},\gamma_{2} \hat{\gamma}_{3} \hat{\gamma}_{6}}^{(3)}}{\delta \overline{G}_{\mathfrak{s},\gamma_{1}}} \Bigg) \overline{W}_{\mathfrak{s},\hat{\gamma}_{6} \hat{\gamma}_{1}}^{(2)} \\
& + \frac{1}{2} \frac{\delta^{2} \overline{\Pi}_{\mathfrak{s},\hat{\gamma}_{2} \hat{\gamma}_{1}}}{\delta \overline{G}_{\mathfrak{s},\gamma_{1}} \delta \overline{G}_{\mathfrak{s},\gamma_{2}}} \Bigg] \\
= \frac{1}{3} \dot{U}_{\mathfrak{s},\hat{\gamma}_{1}\hat{\gamma}_{2}} & \Bigg[ \overline{W}_{\mathfrak{s},\hat{\gamma}_{2} \hat{\gamma}_{3}}^{(2)} \left(\overline{\Pi}_{\mathfrak{s},\hat{\gamma}_{3} \hat{\gamma}_{4}}^{\mathrm{inv}} \frac{\delta \overline{\Pi}_{\mathfrak{s},\hat{\gamma}_{4} \hat{\gamma}_{5}}}{\delta \overline{G}_{\mathfrak{s},\gamma_{1}}} \overline{\Pi}_{\mathfrak{s},\hat{\gamma}_{5} \hat{\gamma}_{6}}^{\mathrm{inv}} - \overline{\Phi}_{\mathfrak{s},\gamma_{1}\hat{\gamma}_{3}\hat{\gamma}_{6}}^{(3)}\right)\overline{W}_{\mathfrak{s},\hat{\gamma}_{6} \hat{\gamma}_{7}}^{(2)} \\
& \hspace{0.2cm} \times \left(\overline{\Pi}_{\mathfrak{s},\hat{\gamma}_{7} \hat{\gamma}_{8}}^{\mathrm{inv}} \frac{\delta \overline{\Pi}_{\mathfrak{s},\hat{\gamma}_{8} \hat{\gamma}_{9}}}{\delta \overline{G}_{\mathfrak{s},\gamma_{2}}} \overline{\Pi}_{\mathfrak{s},\hat{\gamma}_{9} \hat{\gamma}_{10}}^{\mathrm{inv}} - \overline{\Phi}_{\mathfrak{s},\gamma_{2} \hat{\gamma}_{7} \hat{\gamma}_{10}}^{(3)}\right) \overline{W}_{\mathfrak{s},\hat{\gamma}_{10} \hat{\gamma}_{1}}^{(2)} \\
& - \overline{W}_{\mathfrak{s},\hat{\gamma}_{2} \hat{\gamma}_{3}}^{(2)} \overline{\Pi}_{\mathfrak{s},\hat{\gamma}_{3} \hat{\gamma}_{4}}^{\mathrm{inv}} \frac{\delta \overline{\Pi}_{\mathfrak{s},\hat{\gamma}_{4} \hat{\gamma}_{5}}}{\delta \overline{G}_{\mathfrak{s},\gamma_{1}}} \overline{\Pi}_{\mathfrak{s},\hat{\gamma}_{5} \hat{\gamma}_{6}}^{\mathrm{inv}} \frac{\delta \overline{\Pi}_{\mathfrak{s},\hat{\gamma}_{6} \hat{\gamma}_{7}}}{\delta \overline{G}_{\mathfrak{s},\gamma_{2}}} \overline{\Pi}_{\mathfrak{s},\hat{\gamma}_{7} \hat{\gamma}_{8}}^{\mathrm{inv}} \overline{W}_{\mathfrak{s},\hat{\gamma}_{8} \hat{\gamma}_{1}}^{(2)}
\end{split}
\end{equation*}
\begin{equation}
\begin{split}
\hspace{3.4cm} & + \frac{1}{2} \overline{W}_{\mathfrak{s},\hat{\gamma}_{2} \hat{\gamma}_{3}}^{(2)} \left(\overline{\Pi}_{\mathfrak{s},\hat{\gamma}_{3} \hat{\gamma}_{4}}^{\mathrm{inv}} \frac{\delta^{2} \overline{\Pi}_{\mathfrak{s},\hat{\gamma}_{4} \hat{\gamma}_{5}}}{\delta \overline{G}_{\mathfrak{s},\gamma_{1}} \delta \overline{G}_{\mathfrak{s},\gamma_{2}}} \overline{\Pi}_{\mathfrak{s},\hat{\gamma}_{5} \hat{\gamma}_{6}}^{\mathrm{inv}} - \overline{\Phi}_{\mathfrak{s},\gamma_{1}\gamma_{2}\hat{\gamma}_{3}\hat{\gamma}_{6}}^{(4)}\right) \overline{W}_{\mathfrak{s},\hat{\gamma}_{6} \hat{\gamma}_{1}}^{(2)} \\
& + \frac{1}{4} \frac{\delta^{2}\overline{\Pi}_{\mathfrak{s},\hat{\gamma}_{2}\hat{\gamma}_{1}}}{\delta \overline{G}_{\mathfrak{s},\gamma_{1}} \delta \overline{G}_{\mathfrak{s},\gamma_{2}}}\Bigg] \;,
\end{split}
\label{eq:2PIfrgUflowDPhi2Dot}
\end{equation}
where the derivative of $W_{\mathfrak{s}}^{(2)}[K]$ with respect to $G$ was replaced with the help of~\eqref{eq:2PIfrgUflowDW2DG} in the latter equality. Let us then recall the chain rule from~\eqref{eq:2PIfrgGeneralFlowEqPhin}:
\begin{equation}
\dot{\overline{\Phi}}_{\mathfrak{s},\gamma_{1}\gamma_{2}}^{(2)} = \overline{\dot{\Phi}}_{\mathfrak{s},\gamma_{1}\gamma_{2}}^{(2)} + \dot{\overline{G}}_{\mathfrak{s},\hat{\gamma}} \overline{\Phi}_{\mathfrak{s},\hat{\gamma}\gamma_{1}\gamma_{2}}^{(3)} \;.
\label{eq:2PIFRGphi2SBetheSalpeter}
\end{equation}
Inserting~\eqref{eq:2PIfrgUflowDPhi2Dot} into~\eqref{eq:2PIFRGphi2SBetheSalpeter} gives us the differential equation for $\overline{\Phi}^{(2)}_{\mathfrak{s}}$ that we seek:
\begin{equation}
\begin{split}
\dot{\overline{\Phi}}_{\mathfrak{s},\gamma_{1}\gamma_{2}}^{(2)} = \frac{1}{3} \dot{U}_{\mathfrak{s},\hat{\gamma}_{1}\hat{\gamma}_{2}} & \Bigg[\overline{W}_{\mathfrak{s},\hat{\gamma}_{2} \hat{\gamma}_{3}}^{(2)} \left(\overline{\Pi}_{\mathfrak{s},\hat{\gamma}_{3} \hat{\gamma}_{4}}^{\mathrm{inv}} \frac{\delta \overline{\Pi}_{\mathfrak{s},\hat{\gamma}_{4} \hat{\gamma}_{5}}}{\delta \overline{G}_{\mathfrak{s},\gamma_{1}}} \overline{\Pi}_{\mathfrak{s},\hat{\gamma}_{5} \hat{\gamma}_{6}}^{\mathrm{inv}} - \overline{\Phi}_{\mathfrak{s},\gamma_{1}\hat{\gamma}_{3}\hat{\gamma}_{6}}^{(3)}\right)\overline{W}_{\mathfrak{s},\hat{\gamma}_{6} \hat{\gamma}_{7}}^{(2)} \\
& \hspace{0.2cm} \times \left(\overline{\Pi}_{\mathfrak{s},\hat{\gamma}_{7} \hat{\gamma}_{8}}^{\mathrm{inv}} \frac{\delta \overline{\Pi}_{\mathfrak{s},\hat{\gamma}_{8} \hat{\gamma}_{9}}}{\delta \overline{G}_{\mathfrak{s},\gamma_{2}}} \overline{\Pi}_{\mathfrak{s},\hat{\gamma}_{9} \hat{\gamma}_{10}}^{\mathrm{inv}} - \overline{\Phi}_{\mathfrak{s},\gamma_{2} \hat{\gamma}_{7} \hat{\gamma}_{10}}^{(3)}\right) \overline{W}_{\mathfrak{s},\hat{\gamma}_{10} \hat{\gamma}_{1}}^{(2)} \\
& - \overline{W}_{\mathfrak{s},\hat{\gamma}_{2} \hat{\gamma}_{3}}^{(2)} \overline{\Pi}_{\mathfrak{s},\hat{\gamma}_{3} \hat{\gamma}_{4}}^{\mathrm{inv}} \frac{\delta \overline{\Pi}_{\mathfrak{s},\hat{\gamma}_{4} \hat{\gamma}_{5}}}{\delta \overline{G}_{\mathfrak{s},\gamma_{1}}} \overline{\Pi}_{\mathfrak{s},\hat{\gamma}_{5} \hat{\gamma}_{6}}^{\mathrm{inv}} \frac{\delta \overline{\Pi}_{\mathfrak{s},\hat{\gamma}_{6} \hat{\gamma}_{7}}}{\delta \overline{G}_{\mathfrak{s},\gamma_{2}}} \overline{\Pi}_{\mathfrak{s},\hat{\gamma}_{7} \hat{\gamma}_{8}}^{\mathrm{inv}} \overline{W}_{\mathfrak{s},\hat{\gamma}_{8} \hat{\gamma}_{1}}^{(2)} \\
& + \frac{1}{2} \overline{W}_{\mathfrak{s},\hat{\gamma}_{2} \hat{\gamma}_{3}}^{(2)} \left(\overline{\Pi}_{\mathfrak{s},\hat{\gamma}_{3} \hat{\gamma}_{4}}^{\mathrm{inv}} \frac{\delta^{2} \overline{\Pi}_{\mathfrak{s},\hat{\gamma}_{4} \hat{\gamma}_{5}}}{\delta \overline{G}_{\mathfrak{s},\gamma_{1}} \delta \overline{G}_{\mathfrak{s},\gamma_{2}}} \overline{\Pi}_{\mathfrak{s},\hat{\gamma}_{5} \hat{\gamma}_{6}}^{\mathrm{inv}} - \overline{\Phi}_{\mathfrak{s},\gamma_{1}\gamma_{2}\hat{\gamma}_{3}\hat{\gamma}_{6}}^{(4)}\right) \overline{W}_{\mathfrak{s},\hat{\gamma}_{6} \hat{\gamma}_{1}}^{(2)} \\
& + \frac{1}{4} \frac{\delta^{2}\overline{\Pi}_{\mathfrak{s},\hat{\gamma}_{2}\hat{\gamma}_{1}}}{\delta \overline{G}_{\mathfrak{s},\gamma_{1}} \delta \overline{G}_{\mathfrak{s},\gamma_{2}}}\Bigg] + \dot{\overline{G}}_{\mathfrak{s},\hat{\gamma}} \overline{\Phi}_{\mathfrak{s},\hat{\gamma}\gamma_{1}\gamma_{2}}^{(3)} \;.
\end{split}
\label{eq:2PIfrgUflowPhi2Dot}
\end{equation}

\item Expression of $\dot{\overline{\Phi}}_{\mathfrak{s}}^{(3)}$:\\
We start from the chain rule expressed by~\eqref{eq:2PIfrgGeneralFlowEqPhin} in the form:
\begin{equation}
\dot{\overline{\Phi}}_{\mathfrak{s},\gamma_{1}\gamma_{2}\gamma_{3}}^{(3)} = \overline{\dot{\Phi}}_{\mathfrak{s},\gamma_{1}\gamma_{2}\gamma_{3}}^{(3)} + \dot{\overline{G}}_{\mathfrak{s},\hat{\gamma}} \overline{\Phi}_{\mathfrak{s},\hat{\gamma}\gamma_{1}\gamma_{2}\gamma_{3}}^{(4)} \;.
\label{eq:2PIFRGUflowChainRulePhi3}
\end{equation}
From~\eqref{eq:2PIfrgUflowDPhi2Dot}, we deduce an expression for $\overline{\dot{\Phi}}_{\mathfrak{s}}^{(3)}$ which can be combined with~\eqref{eq:2PIFRGUflowChainRulePhi3} to obtain:
\begin{equation*}
\begin{split}
\dot{\overline{\Phi}}^{(3)}_{\mathfrak{s},\gamma_{1}\gamma_{2}\gamma_{3}} = & \ \frac{1}{3} \dot{U}_{\mathfrak{s},\hat{\gamma}_{1}\hat{\gamma}_{2}} \Bigg[ 3 \overline{W}^{(2)}_{\mathfrak{s},\hat{\gamma}_{2}\hat{\gamma}_{3}}\left(\overline{\Pi}_{\mathfrak{s},\hat{\gamma}_{3}\hat{\gamma}_{4}}^{\mathrm{inv}}\frac{\delta \overline{\Pi}_{\mathfrak{s},\hat{\gamma}_{4}\hat{\gamma}_{5}}}{\delta\overline{G}_{\mathfrak{s},\gamma_{1}}}\overline{\Pi}_{\mathfrak{s},\hat{\gamma}_{5}\hat{\gamma}_{6}}^{\mathrm{inv}} - \overline{\Phi}^{(3)}_{\mathfrak{s},\gamma_{1}\hat{\gamma}_{3}\hat{\gamma}_{6}}\right)\overline{W}^{(2)}_{\mathfrak{s},\hat{\gamma}_{6}\hat{\gamma}_{7}} \\
& \hspace{1.75cm} \times \left(\overline{\Pi}_{\mathfrak{s},\hat{\gamma}_{7}\hat{\gamma}_{8}}^{\mathrm{inv}}\frac{\delta \overline{\Pi}_{\mathfrak{s},\hat{\gamma}_{8}\hat{\gamma}_{9}}}{\delta\overline{G}_{\mathfrak{s},\gamma_{2}}}\overline{\Pi}_{\mathfrak{s},\hat{\gamma}_{9}\hat{\gamma}_{10}}^{\mathrm{inv}} - \overline{\Phi}^{(3)}_{\mathfrak{s},\gamma_{2}\hat{\gamma}_{7}\hat{\gamma}_{10}}\right) \overline{W}^{(2)}_{\mathfrak{s},\hat{\gamma}_{10}\hat{\gamma}_{11}} \\
& \hspace{1.75cm} \times \left(\overline{\Pi}_{\mathfrak{s},\hat{\gamma}_{11}\hat{\gamma}_{12}}^{\mathrm{inv}}\frac{\delta \overline{\Pi}_{\mathfrak{s},\hat{\gamma}_{12}\hat{\gamma}_{13}}}{\delta\overline{G}_{\mathfrak{s},\gamma_{3}}}\overline{\Pi}_{\mathfrak{s},\hat{\gamma}_{13}\hat{\gamma}_{14}}^{\mathrm{inv}} - \overline{\Phi}^{(3)}_{\mathfrak{s},\gamma_{3}\hat{\gamma}_{11}\hat{\gamma}_{14}}\right) \overline{W}^{(2)}_{\mathfrak{s},\hat{\gamma}_{14}\hat{\gamma}_{1}} \\
& \hspace{1.52cm} + \Bigg( \overline{W}^{(2)}_{\mathfrak{s},\hat{\gamma}_{2}\hat{\gamma}_{3}} \left(\overline{\Pi}_{\mathfrak{s},\hat{\gamma}_{3}\hat{\gamma}_{4}}^{\mathrm{inv}}\frac{\delta \overline{\Pi}_{\mathfrak{s},\hat{\gamma}_{4}\hat{\gamma}_{5}}}{\delta\overline{G}_{\mathfrak{s},\gamma_{1}}}\overline{\Pi}_{\mathfrak{s},\hat{\gamma}_{5}\hat{\gamma}_{6}}^{\mathrm{inv}} - \overline{\Phi}^{(3)}_{\mathfrak{s},\gamma_{1}\hat{\gamma}_{3}\hat{\gamma}_{6}}\right) \overline{W}^{(2)}_{\mathfrak{s},\hat{\gamma}_{6}\hat{\gamma}_{7}} \\
& \hspace{2.35cm} \times \left( \overline{\Pi}_{\mathfrak{s},\hat{\gamma}_{7}\hat{\gamma}_{8}}^{\mathrm{inv}}\frac{\delta^{2} \overline{\Pi}_{\mathfrak{s},\hat{\gamma}_{8}\hat{\gamma}_{9}}}{\delta\overline{G}_{\mathfrak{s},\gamma_{2}}\delta\overline{G}_{\mathfrak{s},\gamma_{3}}}\overline{\Pi}_{\mathfrak{s},\hat{\gamma}_{9}\hat{\gamma}_{10}}^{\mathrm{inv}} - \overline{\Phi}_{\mathfrak{s},\gamma_{2}\gamma_{3}\hat{\gamma}_{7}\hat{\gamma}_{10}}^{(4)} \right) \overline{W}^{(2)}_{\mathfrak{s},\hat{\gamma}_{10}\hat{\gamma}_{1}} \\
& \hspace{2.18cm} + (\gamma_{2},\gamma_{1},\gamma_{3}) + (\gamma_{3},\gamma_{1},\gamma_{2}) \Bigg)
\end{split}
\end{equation*}
\begin{equation}
\begin{split}
\hspace{2.1cm} & \hspace{1.52cm} -2 \Bigg( \overline{W}^{(2)}_{\mathfrak{s},\hat{\gamma}_{2}\hat{\gamma}_{3}} \left(\overline{\Pi}_{\mathfrak{s},\hat{\gamma}_{3}\hat{\gamma}_{4}}^{\mathrm{inv}}\frac{\delta \overline{\Pi}_{\mathfrak{s},\hat{\gamma}_{4}\hat{\gamma}_{5}}}{\delta\overline{G}_{\mathfrak{s},\gamma_{1}}}\overline{\Pi}_{\mathfrak{s},\hat{\gamma}_{5}\hat{\gamma}_{6}}^{\mathrm{inv}} - \overline{\Phi}^{(3)}_{\mathfrak{s},\gamma_{1}\hat{\gamma}_{3}\hat{\gamma}_{6}}\right) \overline{W}^{(2)}_{\mathfrak{s},\hat{\gamma}_{6}\hat{\gamma}_{7}} \\
& \hspace{2.53cm} \times \overline{\Pi}_{\mathfrak{s},\hat{\gamma}_{7}\hat{\gamma}_{8}}^{\mathrm{inv}} \frac{\delta\overline{\Pi}_{\mathfrak{s},\hat{\gamma}_{8}\hat{\gamma}_{9}}}{\delta\overline{G}_{\mathfrak{s},\gamma_{2}}} \overline{\Pi}_{\mathfrak{s},\hat{\gamma}_{9}\hat{\gamma}_{10}}^{\mathrm{inv}} \frac{\delta\overline{\Pi}_{\mathfrak{s},\hat{\gamma}_{10}\hat{\gamma}_{11}}}{\delta\overline{G}_{\mathfrak{s},\gamma_{3}}} \overline{\Pi}_{\mathfrak{s},\hat{\gamma}_{11}\hat{\gamma}_{12}}^{\mathrm{inv}} \overline{W}^{(2)}_{\mathfrak{s},\hat{\gamma}_{12}\hat{\gamma}_{1}} \\
& \hspace{2.36cm} + (\gamma_{2},\gamma_{1},\gamma_{3}) + (\gamma_{3},\gamma_{1},\gamma_{2}) \Bigg) \\
& \hspace{1.52cm} + 3 \overline{W}^{(2)}_{\mathfrak{s},\hat{\gamma}_{2}\hat{\gamma}_{3}} \overline{\Pi}_{\mathfrak{s},\hat{\gamma}_{3}\hat{\gamma}_{4}}^{\mathrm{inv}} \frac{\delta\overline{\Pi}_{\mathfrak{s},\hat{\gamma}_{4}\hat{\gamma}_{5}}}{\delta\overline{G}_{\mathfrak{s},\gamma_{1}}} \overline{\Pi}_{\mathfrak{s},\hat{\gamma}_{5}\hat{\gamma}_{6}}^{\mathrm{inv}} \frac{\delta\overline{\Pi}_{\mathfrak{s},\hat{\gamma}_{6}\hat{\gamma}_{7}}}{\delta\overline{G}_{\mathfrak{s},\gamma_{2}}} \overline{\Pi}_{\mathfrak{s},\hat{\gamma}_{7}\hat{\gamma}_{8}}^{\mathrm{inv}} \frac{\delta\overline{\Pi}_{\mathfrak{s},\hat{\gamma}_{8}\hat{\gamma}_{9}}}{\delta\overline{G}_{\mathfrak{s},\gamma_{3}}} \overline{\Pi}_{\mathfrak{s},\hat{\gamma}_{9}\hat{\gamma}_{10}}^{\mathrm{inv}} \overline{W}^{(2)}_{\mathfrak{s},\hat{\gamma}_{10}\hat{\gamma}_{1}} \\
& \hspace{1.52cm} - \Bigg( \overline{W}^{(2)}_{\mathfrak{s},\hat{\gamma}_{2}\hat{\gamma}_{3}} \overline{\Pi}_{\mathfrak{s},\hat{\gamma}_{3}\hat{\gamma}_{4}}^{\mathrm{inv}} \frac{\delta\overline{\Pi}_{\mathfrak{s},\hat{\gamma}_{4}\hat{\gamma}_{5}}}{\delta\overline{G}_{\mathfrak{s},\gamma_{1}}} \overline{\Pi}_{\mathfrak{s},\hat{\gamma}_{5}\hat{\gamma}_{6}}^{\mathrm{inv}} \frac{\delta^{2}\overline{\Pi}_{\mathfrak{s},\hat{\gamma}_{6}\hat{\gamma}_{7}}}{\delta \overline{G}_{\mathfrak{s},\gamma_{2}} \delta \overline{G}_{\mathfrak{s},\gamma_{3}}} \overline{\Pi}_{\mathfrak{s},\hat{\gamma}_{7}\hat{\gamma}_{8}}^{\mathrm{inv}} \overline{W}^{(2)}_{\mathfrak{s},\hat{\gamma}_{8}\hat{\gamma}_{1}} \\
& \hspace{2.18cm} + (\gamma_{2},\gamma_{1},\gamma_{3}) + (\gamma_{3},\gamma_{1},\gamma_{2}) \Bigg) \\
& \hspace{1.52cm} -\frac{1}{2} \overline{W}^{(2)}_{\mathfrak{s},\hat{\gamma}_{2}\hat{\gamma}_{3}} \overline{\Phi}^{(5)}_{\mathfrak{s},\gamma_{1}\gamma_{2}\gamma_{3}\hat{\gamma}_{3}\hat{\gamma}_{4}} \overline{W}^{(2)}_{\mathfrak{s},\hat{\gamma}_{4}\hat{\gamma}_{1}} \Bigg] + \dot{\overline{G}}_{\mathfrak{s},\hat{\gamma}} \overline{\Phi}^{(4)}_{\mathfrak{s},\hat{\gamma}\gamma_{1}\gamma_{2}\gamma_{3}}\;,
\end{split}
\label{eq:2PIfrgUflowPhi3Dot}
\end{equation}
where we have used $\frac{\delta^{3}\Pi_{\mathfrak{s}}[G]}{\delta G_{\gamma_{1}}\delta G_{\gamma_{2}}\delta G_{\gamma_{3}}}=0$ $\forall\gamma_{1},\gamma_{2},\gamma_{3}$ (which follows directly from \eqref{eq:2PIfrgExpressionPiandG}), \eqref{eq:2PIfrgUflowDW2DG} (to evaluate the derivatives of $W_{\mathfrak{s}}^{(2)}[K]$ with respect to $G$) as well as the notation:
\begin{equation}
\mathcal{F}_{\gamma_{1}\gamma_{2}\gamma_{3}} + (\gamma_{2},\gamma_{1},\gamma_{3}) + (\gamma_{3},\gamma_{1},\gamma_{2}) = \mathcal{F}_{\gamma_{1}\gamma_{2}\gamma_{3}} + \mathcal{F}_{\gamma_{2}\gamma_{1}\gamma_{3}} + \mathcal{F}_{\gamma_{3}\gamma_{1}\gamma_{2}}\;,
\label{eq:Notation2PIFRGFgamma123}
\end{equation}
valid for any functional $\mathcal{F}$.

\end{itemize}

\paragraph{Additional calculations for the tU-flow:}

We have so far derived (from~\eqref{eq:2PIFRGgeneratingFuncUflow} to~\eqref{eq:Notation2PIFRGFgamma123}) the tower of differential equations for the U-flow, in a form suited to implement the pU-flow. In the following calculations, we rewrite these flow equations at $N_{\mathrm{max}}=2$ using the approximation of~\eqref{eq:2PIfrgExampleNmax2} and dropping terms of order $\mathcal{O}\left(\dot{U}_{\mathfrak{s}}\overline{\Phi}_{\mathfrak{s}}^{(2)}\right)$ in the differential equation expressing $\dot{\overline{\Phi}}_{\mathfrak{s},\gamma_{1}\gamma_{2}}^{(2)}$. This amounts to deriving the tower of flow equations underlying the tU-flow at $N_{\mathrm{max}}=2$. Introducing the identity matrix $\mathcal{I}$ via~\eqref{eq:2PIfrgUflowUsefulIdentitytUflow}, we proceed as follows:
\begin{itemize}
\item Expression of $\dot{\overline{\Phi}}_{\mathfrak{s}}^{(2)}$:\\
We first set all components of $\overline{\Phi}_{\mathfrak{s}}^{(3)}$ and $\overline{\Phi}_{\mathfrak{s}}^{(4)}$ equal to zero in~\eqref{eq:2PIfrgUflowDPhi2Dot} in order to implement the truncation order $N_{\mathrm{max}}=2$. Using~\eqref{eq:2PIfrgUflowUsefulIdentitytUflow}, the relation thus obtained can be further simplified as shown below:
\begin{equation}
\begin{split}
\overline{\dot{\Phi}}_{\mathfrak{s},\gamma_{1}\gamma_{2}}^{(2)} = & \ \frac{1}{3} \dot{U}_{\mathfrak{s},\hat{\gamma}_{1}\hat{\gamma}_{2}}\Bigg[\underbrace{\overline{W}_{\mathfrak{s},\hat{\gamma}_{2} \hat{\gamma}_{3}}^{(2)} \overline{\Pi}_{\mathfrak{s},\hat{\gamma}_{3} \hat{\gamma}_{4}}^{\mathrm{inv}}}_{\mathcal{I}_{\hat{\gamma}_{2}\hat{\gamma}_{4}}} \frac{\delta \overline{\Pi}_{\mathfrak{s},\hat{\gamma}_{4} \hat{\gamma}_{5}}}{\delta \overline{G}_{\mathfrak{s},\gamma_{1}}} \underbrace{\overline{\Pi}_{\mathfrak{s},\hat{\gamma}_{5} \hat{\gamma}_{6}}^{\mathrm{inv}}\overline{W}_{\mathfrak{s},\hat{\gamma}_{6} \hat{\gamma}_{7}}^{(2)}}_{\mathcal{I}_{\hat{\gamma}_{5}\hat{\gamma}_{7}}}\overline{\Pi}_{\mathfrak{s},\hat{\gamma}_{7} \hat{\gamma}_{8}}^{\mathrm{inv}} \frac{\delta \overline{\Pi}_{\mathfrak{s},\hat{\gamma}_{8} \hat{\gamma}_{9}}}{\delta \overline{G}_{\mathfrak{s},\gamma_{2}}} \underbrace{\overline{\Pi}_{\mathfrak{s},\hat{\gamma}_{9} \hat{\gamma}_{10}}^{\mathrm{inv}} \overline{W}_{\mathfrak{s},\hat{\gamma}_{10} \hat{\gamma}_{1}}^{(2)}}_{\mathcal{I}_{\hat{\gamma}_{9}\hat{\gamma}_{1}}} \\
& \hspace{1.52cm} - \underbrace{\overline{W}_{\mathfrak{s},\hat{\gamma}_{2} \hat{\gamma}_{3}}^{(2)} \overline{\Pi}_{\mathfrak{s},\hat{\gamma}_{3} \hat{\gamma}_{4}}^{\mathrm{inv}}}_{\mathcal{I}_{\hat{\gamma}_{2}\hat{\gamma}_{4}}} \frac{\delta \overline{\Pi}_{\mathfrak{s},\hat{\gamma}_{4} \hat{\gamma}_{5}}}{\delta \overline{G}_{\mathfrak{s},\gamma_{1}}} \overline{\Pi}_{\mathfrak{s},\hat{\gamma}_{5} \hat{\gamma}_{6}}^{\mathrm{inv}} \frac{\delta \overline{\Pi}_{\mathfrak{s},\hat{\gamma}_{6} \hat{\gamma}_{7}}}{\delta \overline{G}_{\mathfrak{s},\gamma_{2}}} \underbrace{\overline{\Pi}_{\mathfrak{s},\hat{\gamma}_{7} \hat{\gamma}_{8}}^{\mathrm{inv}} \overline{W}_{\mathfrak{s},\hat{\gamma}_{8} \hat{\gamma}_{1}}^{(2)}}_{\mathcal{I}_{\hat{\gamma}_{7}\hat{\gamma}_{1}}} \\
& \hspace{1.52cm} + \frac{1}{2} \underbrace{\overline{W}_{\mathfrak{s},\hat{\gamma}_{2} \hat{\gamma}_{3}}^{(2)} \overline{\Pi}_{\mathfrak{s},\hat{\gamma}_{3} \hat{\gamma}_{4}}^{\mathrm{inv}}}_{\mathcal{I}_{\hat{\gamma}_{2}\hat{\gamma}_{4}}} \frac{\delta^{2} \overline{\Pi}_{\mathfrak{s},\hat{\gamma}_{4} \hat{\gamma}_{5}}}{\delta \overline{G}_{\mathfrak{s},\gamma_{1}} \delta \overline{G}_{\mathfrak{s},\gamma_{2}}} \underbrace{\overline{\Pi}_{\mathfrak{s},\hat{\gamma}_{5} \hat{\gamma}_{6}}^{\mathrm{inv}} \overline{W}_{\mathfrak{s},\hat{\gamma}_{6} \hat{\gamma}_{1}}^{(2)}}_{\mathcal{I}_{\hat{\gamma}_{5}\hat{\gamma}_{1}}} + \frac{1}{4} \frac{\delta^{2}\overline{\Pi}_{\mathfrak{s},\hat{\gamma}_{2}\hat{\gamma}_{1}}}{\delta \overline{G}_{\mathfrak{s},\gamma_{1}} \delta \overline{G}_{\mathfrak{s},\gamma_{2}}}\Bigg] \\
= & \ \frac{1}{4} \dot{U}_{\mathfrak{s},\hat{\gamma}_{1}\hat{\gamma}_{2}}\frac{\delta^{2}\overline{\Pi}_{\mathfrak{s},\hat{\gamma}_{2}\hat{\gamma}_{1}}}{\delta \overline{G}_{\mathfrak{s},\gamma_{1}} \delta \overline{G}_{\mathfrak{s},\gamma_{2}}} \;.
\end{split}
\label{eq:2PIfrgUflowPhi2DottUflow}
\end{equation}
Since $\overline{\Phi}_{\mathfrak{s},\gamma_{1}\gamma_{2}\gamma_{3}}^{(3)}=0$ $\forall \gamma_{1},\gamma_{2},\gamma_{3}$ in the present case, we have in addition:
\begin{equation}
\dot{\overline{\Phi}}_{\mathfrak{s},\gamma_{1}\gamma_{2}}^{(2)} = \overline{\dot{\Phi}}_{\mathfrak{s},\gamma_{1}\gamma_{2}}^{(2)} + \dot{\overline{G}}_{\mathfrak{s},\hat{\gamma}} \overline{\Phi}_{\mathfrak{s},\hat{\gamma}\gamma_{1}\gamma_{2}}^{(3)} = \overline{\dot{\Phi}}_{\mathfrak{s},\gamma_{1}\gamma_{2}}^{(2)} \;.
\label{eq:2PIfrgDotPhiBar2WithPhiBar3null}
\end{equation}
Therefore,~\eqref{eq:2PIfrgUflowPhi2DottUflow} is equivalent to:
\begin{equation}
\dot{\overline{\Phi}}_{\mathfrak{s},\gamma_{1}\gamma_{2}}^{(2)} = \frac{1}{4} \dot{U}_{\mathfrak{s},\hat{\gamma}_{1}\hat{\gamma}_{2}}\frac{\delta^{2}\overline{\Pi}_{\mathfrak{s},\hat{\gamma}_{2}\hat{\gamma}_{1}}}{\delta \overline{G}_{\mathfrak{s},\gamma_{1}} \delta \overline{G}_{\mathfrak{s},\gamma_{2}}} \;.
\label{eq:2PIfrgtUflowEqPhi2dot}
\end{equation}
With the help of~\eqref{eq:2PIfrgExpressionPiandG}, we then evaluate the second-order derivative of $\Pi[G]$ with respect to $G$ involved in the RHS of~\eqref{eq:2PIfrgtUflowEqPhi2dot}:
\begin{equation}
\begin{split}
\scalebox{0.85}{${\displaystyle \frac{\delta^{2} \Pi_{\gamma_{3}\gamma_{4}}[G]}{\delta G_{\gamma_{1}} \delta G_{\gamma_{2}}} = }$} & \scalebox{0.85}{${\displaystyle \ \frac{\delta^{2}}{\delta G_{\gamma_{1}} \delta G_{\gamma_{2}}} \left(G_{\alpha_{3} \alpha'_{4}} G_{\alpha'_{3} \alpha_{4}} + \zeta G_{\alpha_{3} \alpha_{4}} G_{\alpha'_{3} \alpha'_{4}}\right) }$} \\
= & \scalebox{0.85}{${\displaystyle \ \frac{\delta}{\delta G_{\gamma_{1}}} \Bigg(\frac{\delta G_{\alpha_{3} \alpha'_{4}}}{\delta G_{\alpha_{2} \alpha'_{2}}} G_{\alpha'_{3} \alpha_{4}} +  G_{\alpha_{3} \alpha'_{4}} \frac{\delta G_{\alpha'_{3} \alpha_{4}}}{\delta G_{\alpha_{2} \alpha'_{2}}} + \zeta \frac{\delta G_{\alpha_{3} \alpha_{4}}}{\delta G_{\alpha_{2} \alpha'_{2}}} G_{\alpha'_{3} \alpha'_{4}} + \zeta G_{\alpha_{3} \alpha_{4}} \frac{\delta G_{\alpha'_{3} \alpha'_{4}}}{\delta G_{\alpha_{2} \alpha'_{2}}}\Bigg) }$} \\
= & \scalebox{0.85}{${\displaystyle \ \frac{\delta}{\delta G_{\gamma_{1}}} \Big[G_{\alpha'_{3} \alpha_{4}} \delta_{\alpha_{3} \alpha_{2}}\delta_{\alpha'_{4} \alpha'_{2}} + G_{\alpha_{3} \alpha'_{4}} \delta_{\alpha'_{3} \alpha_{2}}\delta_{\alpha_{4} \alpha'_{2}} + \zeta G_{\alpha'_{3} \alpha'_{4}} \delta_{\alpha_{3} \alpha_{2}}\delta_{\alpha_{4} \alpha'_{2}} + \zeta G_{\alpha_{3} \alpha_{4}} \delta_{\alpha'_{3} \alpha_{2}}\delta_{\alpha'_{4} \alpha'_{2}} }$} \\
& \hspace{1.0cm} \scalebox{0.85}{${\displaystyle + \zeta\left(\alpha_{2} \leftrightarrow \alpha'_{2} \right)\Big] }$} \\
= & \scalebox{0.85}{${\displaystyle \ \frac{\delta G_{\alpha'_{3} \alpha_{4}}}{\delta G_{\alpha_{1} \alpha'_{1}}} \delta_{\alpha_{3} \alpha_{2}}\delta_{\alpha'_{4} \alpha'_{2}} + \frac{\delta G_{\alpha_{3} \alpha'_{4}}}{\delta G_{\alpha_{1} \alpha'_{1}}} \delta_{\alpha'_{3} \alpha_{2}}\delta_{\alpha_{4} \alpha'_{2}} + \zeta \frac{\delta G_{\alpha'_{3} \alpha'_{4}}}{\delta G_{\alpha_{1} \alpha'_{1}}} \delta_{\alpha_{3} \alpha_{2}}\delta_{\alpha_{4} \alpha'_{2}} + \zeta \frac{\delta G_{\alpha_{3} \alpha_{4}}}{\delta G_{\alpha_{1} \alpha'_{1}}} \delta_{\alpha'_{3} \alpha_{2}}\delta_{\alpha'_{4} \alpha'_{2}} }$} \\
& \scalebox{0.85}{${\displaystyle + \zeta\left(\alpha_{2} \leftrightarrow \alpha'_{2} \right) }$} \\
\scalebox{0.85}{${\displaystyle = }$} & \scalebox{0.85}{${\displaystyle \ \Big[ \delta_{\alpha'_{3} \alpha_{1}} \delta_{\alpha_{4} \alpha'_{1}} \delta_{\alpha_{3} \alpha_{2}}\delta_{\alpha'_{4} \alpha'_{2}} + \delta_{\alpha_{3} \alpha_{1}} \delta_{\alpha'_{4} \alpha'_{1}} \delta_{\alpha'_{3} \alpha_{2}}\delta_{\alpha_{4} \alpha'_{2}} + \zeta \delta_{\alpha'_{3} \alpha_{1}} \delta_{\alpha'_{4} \alpha'_{1}} \delta_{\alpha_{3} \alpha_{2}}\delta_{\alpha_{4} \alpha'_{2}} + \zeta \delta_{\alpha_{3} \alpha_{1}} \delta_{\alpha_{4} \alpha'_{1}} \delta_{\alpha'_{3} \alpha_{2}}\delta_{\alpha'_{4} \alpha'_{2}} }$} \\
& \hspace{0.2cm} \scalebox{0.85}{${\displaystyle + \zeta\left(\alpha_{2} \leftrightarrow \alpha'_{2} \right) \Big] + \zeta\left(\alpha_{1} \leftrightarrow \alpha'_{1} \right) \;. }$}
\end{split}
\end{equation}
The latter result enables us to drastically simplify~\eqref{eq:2PIfrgtUflowEqPhi2dot} as follows:
\begin{equation}
\begin{split}
\dot{\overline{\Phi}}_{\mathfrak{s},\gamma_{1}\gamma_{2}}^{(2)} = & \ \frac{1}{16} \int_{\alpha_{3},\alpha'_{3},\alpha_{4},\alpha'_{4}} \dot{U}_{\mathfrak{s},\alpha_{3} \alpha'_{3} \alpha_{4} \alpha'_{4}} \Bigg[ \delta_{\alpha'_{4} \alpha_{1}} \delta_{\alpha_{3} \alpha'_{1}} \delta_{\alpha_{4} \alpha_{2}}\delta_{\alpha'_{3} \alpha'_{2}} + \delta_{\alpha_{4} \alpha_{1}} \delta_{\alpha'_{3} \alpha'_{1}} \delta_{\alpha'_{4} \alpha_{2}}\delta_{\alpha_{3} \alpha'_{2}} \\
& \hspace{4.5cm} + \zeta \delta_{\alpha'_{4} \alpha_{1}} \delta_{\alpha'_{3} \alpha'_{1}} \delta_{\alpha_{4} \alpha_{2}}\delta_{\alpha_{3} \alpha'_{2}} + \zeta \delta_{\alpha_{4} \alpha_{1}} \delta_{\alpha_{3} \alpha'_{1}} \delta_{\alpha'_{4} \alpha_{2}}\delta_{\alpha'_{3} \alpha'_{2}} \\
& \hspace{4.5cm} + \zeta\left(\alpha_{2} \leftrightarrow \alpha'_{2} \right) \Bigg] + \zeta\left(\alpha_{1} \leftrightarrow \alpha'_{1} \right) \\
= & \ \frac{1}{16} \Bigg[ \underbrace{\dot{U}_{\mathfrak{s},\alpha'_{1} \alpha'_{2} \alpha_{2} \alpha_{1}}}_{\dot{U}_{\mathfrak{s},\gamma_{1}\gamma_{2}}} + \underbrace{\dot{U}_{\mathfrak{s},\alpha'_{2} \alpha'_{1} \alpha_{1} \alpha_{2}}}_{\dot{U}_{\mathfrak{s},\gamma_{1}\gamma_{2}}} + \zeta \underbrace{\dot{U}_{\mathfrak{s},\alpha'_{2} \alpha'_{1} \alpha_{2} \alpha_{1}}}_{\zeta \dot{U}_{\mathfrak{s},\gamma_{1}\gamma_{2}}} + \zeta \underbrace{\dot{U}_{\mathfrak{s},\alpha'_{1} \alpha'_{2} \alpha_{1} \alpha_{2}}}_{\zeta \dot{U}_{\mathfrak{s},\gamma_{1}\gamma_{2}}} + \zeta\left(\alpha_{2} \leftrightarrow \alpha'_{2} \right) \Bigg] \\
& + \zeta\left(\alpha_{1} \leftrightarrow \alpha'_{1} \right) \\
= & \ \dot{U}_{\mathfrak{s},\gamma_{1}\gamma_{2}} \;.
\end{split}
\end{equation}
The Luttinger-Ward functional characterizes by definition the interaction part of the 2PI EA, which implies in particular that $\overline{\Phi}_{\mathfrak{s},\gamma_{1}\gamma_{2}}^{(2)}=0$ $\forall \gamma_{1},\gamma_{2}$ if $U_{\mathfrak{s},\gamma_{1}\gamma_{2}}=0$ $\forall \gamma_{1},\gamma_{2}$. This provides a boundary condition to integrate the latter differential equation, thus yielding:
\begin{equation}
\overline{\Phi}_{\mathfrak{s},\gamma_{1}\gamma_{2}}^{(2)} = U_{\mathfrak{s},\gamma_{1}\gamma_{2}} \;.
\label{eq:2PIfrgUflowExpressionPhi2}
\end{equation}

\item Expression of $\dot{\overline{\Sigma}}_{\mathfrak{s}}$:\\
Using~\eqref{eq:2PIfrgW2PiInvPhi2tUflow} and setting $\overline{\Phi}_{\mathfrak{s},\gamma_{1}\gamma_{2}\gamma_{3}}^{(3)}=0$ $\forall \gamma_{1},\gamma_{2},\gamma_{3}$,~\eqref{eq:2PIfrgUflowSigmaDotCompact} can be rewritten as:
\begin{equation}
\begin{split}
\scalebox{0.95}{${\displaystyle\dot{\overline{\Sigma}}_{\mathfrak{s},\gamma} =}$} & \ \scalebox{0.95}{${\displaystyle -\frac{1}{6}\left(\mathcal{I} + \overline{\Pi}_{\mathfrak{s}} \overline{\Phi}_{\mathfrak{s}}^{(2)}\right)^{\mathrm{inv}}_{\gamma\hat{\gamma}_{1}} \dot{U}_{\mathfrak{s},\hat{\gamma}_{2} \hat{\gamma}_{3}} \Bigg[\underbrace{\overline{W}_{\mathfrak{s},\hat{\gamma}_{3} \hat{\gamma}_{4}}^{(2)} \overline{\Pi}_{\mathfrak{s},\hat{\gamma}_{4}\hat{\gamma}_{5}}^{\mathrm{inv}}}_{\left(\mathcal{I} + \overline{\Pi}_{\mathfrak{s}} \overline{\Phi}_{\mathfrak{s}}^{(2)}\right)_{\hat{\gamma}_{3}\hat{\gamma}_{5}}^{\mathrm{inv}}} \frac{\delta \overline{\Pi}_{\mathfrak{s},\hat{\gamma}_{5}\hat{\gamma}_{6}}}{\delta \overline{G}_{\mathfrak{s},\hat{\gamma}_{1}}} \underbrace{\overline{\Pi}_{\mathfrak{s},\hat{\gamma}_{6}\hat{\gamma}_{7}}^{\mathrm{inv}} \overline{W}_{\mathfrak{s},\hat{\gamma}_{7} \hat{\gamma}_{2}}^{(2)}}_{\left(\mathcal{I} + \overline{\Pi}_{\mathfrak{s}} \overline{\Phi}_{\mathfrak{s}}^{(2)}\right)_{\hat{\gamma}_{6}\hat{\gamma}_{2}}^{\mathrm{inv}}} + \frac{1}{2} \frac{\delta \overline{\Pi}_{\mathfrak{s},\hat{\gamma}_{3}\hat{\gamma}_{2}}}{\delta \overline{G}_{\mathfrak{s},\hat{\gamma}_{1}}} \Bigg] }$} \\
\scalebox{0.95}{${\displaystyle = }$} & \ \scalebox{0.95}{${\displaystyle -\frac{1}{6}\left(\mathcal{I} + \overline{\Pi}_{\mathfrak{s}} \overline{\Phi}_{\mathfrak{s}}^{(2)}\right)^{\mathrm{inv}}_{\gamma\hat{\gamma}_{1}} \dot{U}_{\mathfrak{s},\hat{\gamma}_{2} \hat{\gamma}_{3}} \Bigg[\left(\mathcal{I} + \overline{\Pi}_{\mathfrak{s}} \overline{\Phi}_{\mathfrak{s}}^{(2)}\right)_{\hat{\gamma}_{3}\hat{\gamma}_{4}}^{\mathrm{inv}} \frac{\delta \overline{\Pi}_{\mathfrak{s},\hat{\gamma}_{4}\hat{\gamma}_{5}}}{\delta \overline{G}_{\mathfrak{s},\hat{\gamma}_{1}}} \left(\mathcal{I} + \overline{\Pi}_{\mathfrak{s}} \overline{\Phi}_{\mathfrak{s}}^{(2)}\right)_{\hat{\gamma}_{5}\hat{\gamma}_{2}}^{\mathrm{inv}} + \frac{1}{2} \frac{\delta \overline{\Pi}_{\mathfrak{s},\hat{\gamma}_{3}\hat{\gamma}_{2}}}{\delta \overline{G}_{\mathfrak{s},\hat{\gamma}_{1}}} \Bigg] \;.}$}
\end{split}
\end{equation}
The truncation inherent to the tU-flow can then be introduced by replacing $\overline{\Phi}_{\mathfrak{s}}^{(2)}$ via \eqref{eq:2PIfrgUflowExpressionPhi2}:
\begin{equation}
\dot{\overline{\Sigma}}_{\mathfrak{s},\gamma} = -\frac{1}{6}\left(\mathcal{I} + \overline{\Pi}_{\mathfrak{s}} U_{\mathfrak{s}}\right)^{\mathrm{inv}}_{\gamma\hat{\gamma}_{1}} \dot{U}_{\mathfrak{s},\hat{\gamma}_{2} \hat{\gamma}_{3}} \Bigg[\left(\mathcal{I} + \overline{\Pi}_{\mathfrak{s}} U_{\mathfrak{s}}\right)_{\hat{\gamma}_{3}\hat{\gamma}_{4}}^{\mathrm{inv}} \frac{\delta \overline{\Pi}_{\mathfrak{s},\hat{\gamma}_{4}\hat{\gamma}_{5}}}{\delta \overline{G}_{\mathfrak{s},\hat{\gamma}_{1}}} \left(\mathcal{I} + \overline{\Pi}_{\mathfrak{s}} U_{\mathfrak{s}}\right)_{\hat{\gamma}_{5}\hat{\gamma}_{2}}^{\mathrm{inv}} + \frac{1}{2} \frac{\delta \overline{\Pi}_{\mathfrak{s},\hat{\gamma}_{3}\hat{\gamma}_{2}}}{\delta \overline{G}_{\mathfrak{s},\hat{\gamma}_{1}}} \Bigg] \;.
\end{equation}

\item Expression of $\dot{\overline{\Phi}}_{\mathfrak{s}}$:\\
After replacing $\overline{W}_{\mathfrak{s}}^{(2)}$ through \eqref{eq:2PIfrgW2PiInvPhi2} or~\eqref{eq:2PIfrgW2PiInvPhi2tUflow} and setting once again $\overline{\Phi}_{\mathfrak{s},\gamma_{1}\gamma_{2}\gamma_{3}}^{(3)}=0$ $\forall \gamma_{1},\gamma_{2},\gamma_{3}$,~\eqref{eq:2PIfrgUflowExpressionPhiDot} becomes:
\begin{equation}
\begin{split}
\scalebox{0.92}{${\displaystyle\dot{\overline{\Phi}}_{\mathfrak{s}} =}$} & \ \scalebox{0.92}{${\displaystyle \frac{1}{6} \dot{U}_{\mathfrak{s},\hat{\gamma}_{1}\hat{\gamma}_{2}} \left[\left(\overline{\Pi}^{\mathrm{inv}}_{\mathfrak{s}}+\overline{\Phi}_{\mathfrak{s}}^{(2)}\right)^{\mathrm{inv}} + \frac{1}{2}\overline{\Pi}_{\mathfrak{s}}\right]_{\hat{\gamma}_{2}\hat{\gamma}_{1}} }$} \\
& \scalebox{0.92}{${\displaystyle +\frac{1}{6} \overline{\Sigma}_{\mathfrak{s},\hat{\gamma}_{1}} \left(\overline{\Pi}^{\mathrm{inv}}_{\mathfrak{s}}+\overline{\Phi}_{\mathfrak{s}}^{(2)}\right)^{\mathrm{inv}}_{\hat{\gamma}_{1} \hat{\gamma}_{2}} \dot{U}_{\mathfrak{s},\hat{\gamma}_{3} \hat{\gamma}_{4}} \Bigg[ \underbrace{\overline{W}_{\mathfrak{s},\hat{\gamma}_{4} \hat{\gamma}_{5}}^{(2)} \overline{\Pi}_{\mathfrak{s},\hat{\gamma}_{5}\hat{\gamma}_{6}}^{\mathrm{inv}}}_{\left(\mathcal{I} + \overline{\Pi}_{\mathfrak{s}} \overline{\Phi}_{\mathfrak{s}}^{(2)}\right)^{\mathrm{inv}}_{\hat{\gamma}_{4} \hat{\gamma}_{6}}} \frac{\delta \overline{\Pi}_{\mathfrak{s},\hat{\gamma}_{6}\hat{\gamma}_{7}}}{\delta \overline{G}_{\mathfrak{s},\hat{\gamma}_{2}}} \underbrace{\overline{\Pi}_{\mathfrak{s},\hat{\gamma}_{7}\hat{\gamma}_{8}}^{\mathrm{inv}} \overline{W}_{\mathfrak{s},\hat{\gamma}_{8} \hat{\gamma}_{3}}^{(2)}}_{\left(\mathcal{I} + \overline{\Pi}_{\mathfrak{s}} \overline{\Phi}_{\mathfrak{s}}^{(2)}\right)^{\mathrm{inv}}_{\hat{\gamma}_{7} \hat{\gamma}_{3}}} + \frac{1}{2} \frac{\delta \overline{\Pi}_{\mathfrak{s},\hat{\gamma}_{4} \hat{\gamma}_{3}}}{\delta \overline{G}_{\mathfrak{s},\hat{\gamma}_{2}}} \Bigg] }$} \\
\scalebox{0.92}{${\displaystyle = }$} & \ \scalebox{0.92}{${\displaystyle \frac{1}{6} \dot{U}_{\mathfrak{s},\hat{\gamma}_{1}\hat{\gamma}_{2}} \left[\left(\overline{\Pi}^{\mathrm{inv}}_{\mathfrak{s}}+\overline{\Phi}_{\mathfrak{s}}^{(2)}\right)^{\mathrm{inv}} + \frac{1}{2}\overline{\Pi}_{\mathfrak{s}}\right]_{\hat{\gamma}_{2}\hat{\gamma}_{1}} }$} \\
& \scalebox{0.92}{${\displaystyle +\frac{1}{6} \overline{\Sigma}_{\mathfrak{s},\hat{\gamma}_{1}} \left(\overline{\Pi}^{\mathrm{inv}}_{\mathfrak{s}}+\overline{\Phi}_{\mathfrak{s}}^{(2)}\right)^{\mathrm{inv}}_{\hat{\gamma}_{1} \hat{\gamma}_{2}} \dot{U}_{\mathfrak{s},\hat{\gamma}_{3} \hat{\gamma}_{4}} \Bigg[\left(\mathcal{I} + \overline{\Pi}_{\mathfrak{s}} \overline{\Phi}_{\mathfrak{s}}^{(2)}\right)^{\mathrm{inv}}_{\hat{\gamma}_{4} \hat{\gamma}_{5}} \frac{\delta \overline{\Pi}_{\mathfrak{s},\hat{\gamma}_{5}\hat{\gamma}_{6}}}{\delta \overline{G}_{\mathfrak{s},\hat{\gamma}_{2}}} \left(\mathcal{I} + \overline{\Pi}_{\mathfrak{s}} \overline{\Phi}_{\mathfrak{s}}^{(2)}\right)^{\mathrm{inv}}_{\hat{\gamma}_{6} \hat{\gamma}_{3}} + \frac{1}{2} \frac{\delta \overline{\Pi}_{\mathfrak{s},\hat{\gamma}_{4} \hat{\gamma}_{3}}}{\delta \overline{G}_{\mathfrak{s},\hat{\gamma}_{2}}} \Bigg] \;.}$}
\end{split}
\label{eq:ExpressionPhis2PIFRGtUflowAppendix}
\end{equation}
We then approximate~\eqref{eq:ExpressionPhis2PIFRGtUflowAppendix} by introducing $U_{\mathfrak{s}}$ at the expense of $\overline{\Phi}_{\mathfrak{s}}^{(2)}$ using~\eqref{eq:2PIfrgUflowExpressionPhi2}:
\begin{equation}
\begin{split}
\scalebox{0.98}{${\displaystyle \dot{\overline{\Phi}}_{\mathfrak{s}} = }$} & \ \scalebox{0.98}{${\displaystyle \frac{1}{6} \dot{U}_{\mathfrak{s},\hat{\gamma}_{1}\hat{\gamma}_{2}} \left[\left(\overline{\Pi}^{\mathrm{inv}}_{\mathfrak{s}} + U_{\mathfrak{s}}\right)^{\mathrm{inv}} + \frac{1}{2}\overline{\Pi}_{\mathfrak{s}}\right]_{\hat{\gamma}_{2}\hat{\gamma}_{1}} }$} \\
& \scalebox{0.98}{${\displaystyle +\frac{1}{6} \overline{\Sigma}_{\mathfrak{s},\hat{\gamma}_{1}} \left(\overline{\Pi}^{\mathrm{inv}}_{\mathfrak{s}} + U_{\mathfrak{s}}\right)^{\mathrm{inv}}_{\hat{\gamma}_{1} \hat{\gamma}_{2}} \dot{U}_{\mathfrak{s},\hat{\gamma}_{3} \hat{\gamma}_{4}} \Bigg[\left(\mathcal{I} + \overline{\Pi}_{\mathfrak{s}} U_{\mathfrak{s}}\right)^{\mathrm{inv}}_{\hat{\gamma}_{4} \hat{\gamma}_{5}} \frac{\delta \overline{\Pi}_{\mathfrak{s},\hat{\gamma}_{5}\hat{\gamma}_{6}}}{\delta \overline{G}_{\mathfrak{s},\hat{\gamma}_{2}}} \left(\mathcal{I} + \overline{\Pi}_{\mathfrak{s}} U_{\mathfrak{s}}\right)^{\mathrm{inv}}_{\hat{\gamma}_{6} \hat{\gamma}_{3}} + \frac{1}{2} \frac{\delta \overline{\Pi}_{\mathfrak{s},\hat{\gamma}_{4} \hat{\gamma}_{3}}}{\delta \overline{G}_{\mathfrak{s},\hat{\gamma}_{2}}} \Bigg] \;. }$}
\end{split}
\end{equation}

\item Expression of $\dot{\overline{\Omega}}_{\mathfrak{s}}$:\\
Similarly to the previous derivations, $\overline{W}_{\mathfrak{s}}^{(2)}$ is replaced in~\eqref{eq:2PIfrgUflowExpressionOmegaDot} using~\eqref{eq:2PIfrgW2PiInvPhi2}. The tU-flow is then implemented via~\eqref{eq:2PIfrgUflowExpressionPhi2}, thus yielding:
\begin{equation}
\dot{\overline{\Omega}}_{\mathfrak{s}} = \frac{1}{6\beta} \dot{U}_{\mathfrak{s},\hat{\gamma}_{1}\hat{\gamma}_{2}} \left[\left(\overline{\Pi}^{\mathrm{inv}}_{\mathfrak{s}} + U_{\mathfrak{s}}\right)^{\mathrm{inv}} + \frac{1}{2}\overline{\Pi}_{\mathfrak{s}}\right]_{\hat{\gamma}_{2}\hat{\gamma}_{1}} \;.
\end{equation}

\end{itemize}

\paragraph{Additional calculations for the mU-flow:}

We now aim at deriving the tower of differential equations underlying the mU-flow with a Hartree-Fock starting point, i.e. with $N_{\mathrm{SCPT}}=1$. To that end, let us first recall the transformations outlined in section~\ref{sec:Uflow2PIFRG} defining the flowing quantities in the framework of the mU-flow with $N_{\mathrm{SCPT}}=1$:
\begin{subequations}\label{eq:2PIfrgmUflowNpert1Appendix}
\begin{empheq}[left=\empheqlbrace]{align}
& \boldsymbol{\Omega}_{\mathfrak{s}}[G] = \Omega_{\mathfrak{s}}[G] + \frac{1}{2\beta}\left(U-U_{\mathfrak{s}}\right)_{\hat{\gamma}_{1}\hat{\gamma}_{2}}G_{\hat{\gamma}_{1}}G_{\hat{\gamma}_{2}} \;. \label{eq:2PIfrgmUflowNpert1AppendixNb1} \\
\nonumber \\
& \boldsymbol{\Phi}_{\mathfrak{s}}[G] = \Phi_{\mathfrak{s}}[G] + \frac{1}{2}\left(U-U_{\mathfrak{s}}\right)_{\hat{\gamma}_{1}\hat{\gamma}_{2}}G_{\hat{\gamma}_{1}}G_{\hat{\gamma}_{2}} \;. \label{eq:2PIfrgmUflowNpert1AppendixNb2}\\
\nonumber \\
& \boldsymbol{\Sigma}_{\mathfrak{s},\gamma}[G] = \Sigma_{\mathfrak{s},\gamma}[G] - \left(U-U_{\mathfrak{s}}\right)_{\gamma\hat{\gamma}}G_{\hat{\gamma}} \;. \label{eq:2PIfrgmUflowNpert1AppendixNb3} \\
\nonumber \\
& \boldsymbol{\Phi}^{(2)}_{\mathfrak{s},\gamma_{1}\gamma_{2}}[G] = \Phi^{(2)}_{\mathfrak{s},\gamma_{1}\gamma_{2}}[G] + U_{\gamma_{1}\gamma_{2}} - U_{\mathfrak{s},\gamma_{1}\gamma_{2}} \;. \label{eq:2PIfrgmUflowNpert1AppendixNb4} \\
\nonumber \\
& \boldsymbol{\Phi}^{(n)}_{\mathfrak{s},\gamma_{1}\cdots\gamma_{n}}[G] = \Phi^{(n)}_{\mathfrak{s},\gamma_{1}\cdots\gamma_{n}}[G] \quad \forall n \geq 3 \;. \label{eq:2PIfrgmUflowNpert1AppendixNb5}
\end{empheq}
\end{subequations}
The flow equations corresponding to the mU-flow with $N_{\mathrm{SCPT}}=1$ are basically determined by following the same reasoning as that leading to the pU-flow equations derived beforehand, except that $\Omega_{\mathfrak{s}}$, $\Phi_{\mathfrak{s}}$ and the 2PI vertices $\Phi_{\mathfrak{s}}^{(n)}$ must be replaced by their bold counterparts via~\eqref{eq:2PIfrgmUflowNpert1AppendixNb1} to~\eqref{eq:2PIfrgmUflowNpert1AppendixNb5} before setting $K_{\gamma}=0$ $\forall\gamma$ to deduce $\overline{\dot{\boldsymbol{\Omega}}}_{\mathfrak{s}}$, $\overline{\dot{\boldsymbol{\Sigma}}}_{\mathfrak{s}}$ and $\overline{\dot{\boldsymbol{\Phi}}}^{(n)}_{\mathfrak{s}}$ from $\dot{\boldsymbol{\Omega}}_{\mathfrak{s}}$, $\dot{\boldsymbol{\Sigma}}_{\mathfrak{s}}$ and $\dot{\boldsymbol{\Phi}}^{(n)}_{\mathfrak{s}}$, respectively.

\vspace{0.5cm}

We will now explain this procedure in further details. To begin with, we derive the differential equation expressing $\dot{\overline{\boldsymbol{\Omega}}}_{\mathfrak{s}}$ in two different manners for the sake of clarity. Firstly, in order to replace $\Omega_{\mathfrak{s}}$ by $\boldsymbol{\Omega}_{\mathfrak{s}}$ in the corresponding pU-flow equation (i.e.~\eqref{eq:2PIfrgUflowExpressionOmegaDot}), one can apply the substitution:
\begin{equation}
\Omega_{\mathfrak{s}}[G] \rightarrow \boldsymbol{\Omega}_{\mathfrak{s}}[G] - \frac{1}{2\beta}\left(U-U_{\mathfrak{s}}\right)_{\hat{\gamma}_{1}\hat{\gamma}_{2}}G_{\hat{\gamma}_{1}}G_{\hat{\gamma}_{2}} \;,
\label{eq:2PIfrgmUflowReplacementOmegaNpert1Appendix}
\end{equation}
in the form:
\begin{equation}
\dot{\Omega}_{\mathfrak{s}}[G] \rightarrow \dot{\boldsymbol{\Omega}}_{\mathfrak{s}}[G] + \frac{1}{2\beta} \dot{U}_{\mathfrak{s},\hat{\gamma}_{1}\hat{\gamma}_{2}}G_{\hat{\gamma}_{1}}G_{\hat{\gamma}_{2}} = \dot{\boldsymbol{\Omega}}_{\mathfrak{s}}[G] + \frac{1}{4\beta}\dot{U}_{\mathfrak{s},\hat{\gamma}_{1}\hat{\gamma}_{2}}\Pi_{\hat{\gamma}_{2}\hat{\gamma}_{1}}[G] \;,
\label{eq:2PIfrgmUflowReplacementOmegaDotNpert1Appendix}
\end{equation}
which follows from~\eqref{eq:2PIfrgmUflowNpert1AppendixNb1} as well as~\eqref{eq:TraceTrickmUflowAppendix}. In this fashion,~\eqref{eq:2PIfrgUflowExpressionOmegaDot} is turned into:
\begin{equation}
\dot{\overline{\boldsymbol{\Omega}}}_{\mathfrak{s}} = \frac{1}{6\beta} \dot{U}_{\mathfrak{s},\hat{\gamma}_{1} \hat{\gamma}_{2}} \left(\overline{W}_{\mathfrak{s}}^{(2)} - \overline{\Pi}_{\mathfrak{s}}\right)_{\hat{\gamma}_{2}\hat{\gamma}_{1}} \;.
\label{eq:2PIfrgUflowExpressionOmegaBoldDotAppendixV1}
\end{equation}
Secondly,~\eqref{eq:2PIfrgUflowExpressionOmegaBoldDotAppendixV1} can also be derived by noticing that the configuration $\overline{\boldsymbol{G}}_{\mathfrak{s}}$ of the bold propagator extremizes by definition the flowing bold 2PI EA $\boldsymbol{\Gamma}_{\mathfrak{s}}^{(\mathrm{2PI})}[G]$. This enables us to follow the lines set out by~\eqref{eq:2PIfrgGammaBarDotVsGammaDotBar} by showing:
\begin{equation}
\dot{\overline{\boldsymbol{\Gamma}}}_{\mathfrak{s}}^{(\mathrm{2PI})} = \overline{\dot{\boldsymbol{\Gamma}}}_{\mathfrak{s}}^{(\mathrm{2PI})} + \dot{\overline{\boldsymbol{G}}}_{\mathfrak{s},\hat{\gamma}} \underbrace{\overline{\boldsymbol{\Gamma}}^{(\mathrm{2PI})(1)}_{\mathfrak{s},\hat{\gamma}}}_{0} = \overline{\dot{\boldsymbol{\Gamma}}}_{\mathfrak{s}}^{(\mathrm{2PI})} \;,
\label{eq:2PIfrgGammaBarDotVsGammaDotBarmUflow}
\end{equation}
so that, according to the definition $\boldsymbol{\Omega}_{\mathfrak{s}}\equiv\frac{1}{\beta}\boldsymbol{\Gamma}^{(\mathrm{2PI})}_{\mathfrak{s}}$, we have:
\begin{equation}
\dot{\overline{\boldsymbol{\Omega}}}_{\mathfrak{s}} = \overline{\dot{\boldsymbol{\Omega}}}_{\mathfrak{s}} \;,
\label{eq:mUflowdotBarOmegaAppendix}
\end{equation}
where the RHS can be directly transformed via the substitution:
\begin{equation}
\dot{\boldsymbol{\Omega}}_{\mathfrak{s}}[G] \rightarrow \dot{\Omega}_{\mathfrak{s}}[G] - \frac{1}{4\beta}\dot{U}_{\mathfrak{s},\hat{\gamma}_{1}\hat{\gamma}_{2}}\Pi_{\hat{\gamma}_{2}\hat{\gamma}_{1}}[G] \;,
\end{equation}
resulting from~\eqref{eq:2PIfrgmUflowNpert1AppendixNb1} combined with~\eqref{eq:TraceTrickmUflowAppendix}, just like~\eqref{eq:2PIfrgmUflowReplacementOmegaDotNpert1Appendix}. Hence, we replace in this way $\boldsymbol{\Omega}_{\mathfrak{s}}$ by $\Omega_{\mathfrak{s}}$ (and not the reverse) in~\eqref{eq:mUflowdotBarOmegaAppendix}. This leads to:
\begin{equation}
\begin{split}
\dot{\overline{\boldsymbol{\Omega}}}_{\mathfrak{s}} = & \ \hspace{-0.15cm} \underbrace{\overline{\dot{\Omega}}_{\mathfrak{s}}}_{\frac{1}{\beta}\overline{\dot{\Gamma}}^{(\mathrm{2PI})}_{\mathfrak{s}}} \hspace{-0.15cm} - \frac{1}{4\beta}\dot{U}_{\mathfrak{s},\hat{\gamma}_{1}\hat{\gamma}_{2}}\overline{\Pi}_{\mathfrak{s},\hat{\gamma}_{2}\hat{\gamma}_{1}} \\
= & \ \frac{1}{\beta}\overline{\dot{\Phi}}_{\mathfrak{s}} - \frac{1}{4\beta}\dot{U}_{\mathfrak{s},\hat{\gamma}_{1}\hat{\gamma}_{2}}\overline{\Pi}_{\mathfrak{s},\hat{\gamma}_{2}\hat{\gamma}_{1}} \;,
\end{split}
\end{equation}
where~\eqref{eq:2PIfrgUflowPhiGammadot} was used to obtain the last line. After insertion of~\eqref{eq:2PIfrgPhisdot}, this becomes:
\begin{equation}
\begin{split}
\dot{\overline{\boldsymbol{\Omega}}}_{\mathfrak{s}} = & \ \frac{1}{6\beta} \dot{U}_{\mathfrak{s},\hat{\gamma}_{1} \hat{\gamma}_{2}} \left(\overline{W}_{\mathfrak{s}}^{(2)} + \frac{1}{2}\overline{\Pi}_{\mathfrak{s}}\right)_{\hat{\gamma}_{2}\hat{\gamma}_{1}} - \frac{1}{4\beta}\dot{U}_{\mathfrak{s},\hat{\gamma}_{1}\hat{\gamma}_{2}}\overline{\Pi}_{\mathfrak{s},\hat{\gamma}_{2}\hat{\gamma}_{1}} \\
= & \ \frac{1}{6\beta} \dot{U}_{\mathfrak{s},\hat{\gamma}_{1} \hat{\gamma}_{2}} \left(\overline{W}_{\mathfrak{s}}^{(2)} - \overline{\Pi}_{\mathfrak{s}}\right)_{\hat{\gamma}_{2}\hat{\gamma}_{1}} \;.
\end{split}
\label{eq:2PIfrgUflowExpressionOmegaBoldDotAppendix}
\end{equation}
The latter equality coincides with~\eqref{eq:2PIfrgUflowExpressionOmegaBoldDotAppendixV1}, as expected, and it is the mU-flow counterpart of~\eqref{eq:2PIfrgUflowExpressionOmegaDot} for $N_{\mathrm{SCPT}}=1$.

\vspace{0.5cm}

As a next step for the derivation of the mU-flow equations, we apply, in the expressions of $\dot{\Sigma}_{\mathfrak{s}}$ and $\dot{\Phi}^{(n)}_{\mathfrak{s}}$ (with $n \geq 2$) deduced from~\eqref{eq:2PIfrgPhisdot}, the following transformations set by~\eqref{eq:2PIfrgmUflowNpert1AppendixNb2} to~\eqref{eq:2PIfrgmUflowNpert1AppendixNb5}:
\begin{subequations}\label{eq:2PIfrgmUflowReplacementNpert1Appendix}
\begin{empheq}[left=\empheqlbrace]{align}
& \Phi_{\mathfrak{s}}[G] \rightarrow \boldsymbol{\Phi}_{\mathfrak{s}}[G] - \frac{1}{2}\left(U-U_{\mathfrak{s}}\right)_{\hat{\gamma}_{1}\hat{\gamma}_{2}}G_{\hat{\gamma}_{1}}G_{\hat{\gamma}_{2}} \;, \label{eq:2PIfrgmUflowReplacementPhiNpert1Appendix}\\
\nonumber \\
& \Sigma_{\mathfrak{s},\gamma}[G] \rightarrow \boldsymbol{\Sigma}_{\mathfrak{s},\gamma}[G] + \left(U-U_{\mathfrak{s}}\right)_{\gamma\hat{\gamma}}G_{\hat{\gamma}} \;, \label{eq:2PIfrgmUflowReplacementSigmaNpert1Appendix} \\
\nonumber \\
& \Phi^{(2)}_{\mathfrak{s},\gamma_{1}\gamma_{2}}[G] \rightarrow \boldsymbol{\Phi}^{(2)}_{\mathfrak{s},\gamma_{1}\gamma_{2}}[G] - U_{\gamma_{1}\gamma_{2}} + U_{\mathfrak{s},\gamma_{1}\gamma_{2}} \;, \label{eq:2PIfrgmUflowReplacementPhi2Npert1Appendix} \\
\nonumber \\
& \Phi^{(n)}_{\mathfrak{s},\gamma_{1}\cdots\gamma_{n}}[G] \rightarrow \boldsymbol{\Phi}^{(n)}_{\mathfrak{s},\gamma_{1}\cdots\gamma_{n}}[G] \quad \forall n \geq 3 \;, \label{eq:2PIfrgmUflowReplacementPhinNpert1Appendix}
\end{empheq}
\end{subequations}
from which follow notably:
\begin{subequations}\label{eq:2PIfrgmUflowReplacementDotNpert1Appendix}
\begin{empheq}[left=\empheqlbrace]{align}
& \dot{\Phi}_{\mathfrak{s}}[G] \rightarrow \dot{\boldsymbol{\Phi}}_{\mathfrak{s}}[G] + \frac{1}{2} \dot{U}_{\mathfrak{s},\hat{\gamma}_{1}\hat{\gamma}_{2}}G_{\hat{\gamma}_{1}}G_{\hat{\gamma}_{2}} = \dot{\boldsymbol{\Phi}}_{\mathfrak{s}}[G] + \frac{1}{4}\dot{U}_{\mathfrak{s},\hat{\gamma}_{1}\hat{\gamma}_{2}}\Pi_{\hat{\gamma}_{2}\hat{\gamma}_{1}}[G] \;, \label{eq:2PIfrgmUflowReplacementPhiDotNpert1Appendix} \\
\nonumber \\
& \dot{\Sigma}_{\mathfrak{s},\gamma}[G] \rightarrow \dot{\boldsymbol{\Sigma}}_{\mathfrak{s},\gamma}[G] - \dot{U}_{\mathfrak{s},\gamma\hat{\gamma}}G_{\hat{\gamma}} \;, \label{eq:2PIfrgmUflowReplacementSigmaDotNpert1Appendix} \\
\nonumber \\
& \dot{\Phi}^{(2)}_{\mathfrak{s},\gamma_{1}\gamma_{2}}[G] \rightarrow \dot{\boldsymbol{\Phi}}^{(2)}_{\mathfrak{s},\gamma_{1}\gamma_{2}}[G] + \dot{U}_{\mathfrak{s},\gamma_{1}\gamma_{2}} \;, \label{eq:2PIfrgmUflowReplacementPhi2DotNpert1Appendix} \\
\nonumber \\
& \dot{\Phi}^{(n)}_{\mathfrak{s},\gamma_{1}\cdots\gamma_{n}}[G] \rightarrow \dot{\boldsymbol{\Phi}}^{(n)}_{\mathfrak{s},\gamma_{1}\cdots\gamma_{n}}[G] \quad \forall n \geq 3 \;, \label{eq:2PIfrgmUflowReplacementPhinDotNpert1Appendix}
\end{empheq}
\end{subequations}
where transformation~\eqref{eq:2PIfrgmUflowReplacementPhiDotNpert1Appendix} was derived using~\eqref{eq:TraceTrickmUflowAppendix}. Regarding the flow equation expressing the derivative $\dot{\overline{\boldsymbol{\Sigma}}}_{\mathfrak{s}}$, we can still exploit~\eqref{eq:2PIFRGSigmaSBetheSalpeter} in the form\footnote{We recall that, in the framework of the mU-flow, the upper bars always indicate an evaluation at $G=\overline{\boldsymbol{G}}_{\mathfrak{s}}$ instead of $G=\overline{G}_{\mathfrak{s}}$, e.g. $\overline{\Pi}_{\mathfrak{s}} \equiv \Pi[G=\boldsymbol{G}_{\mathfrak{s}}]$.}:
\begin{equation}
\dot{\overline{\boldsymbol{\Sigma}}}_{\mathfrak{s}} = \left(\mathcal{I} + \overline{\Pi}_{\mathfrak{s}} \boldsymbol{\overline{\Phi}}_{\mathfrak{s}}^{(2)}\right)^{\mathrm{inv}} \overline{\dot{\boldsymbol{\Sigma}}}_{\mathfrak{s}} \;,
\label{eq:2PIFRGSigmaBoldSBetheSalpeter}
\end{equation}
with $\overline{\dot{\boldsymbol{\Sigma}}}_{\mathfrak{s}}$ determined via the recipe outlined above (involving transformation~\eqref{eq:2PIfrgmUflowReplacementSigmaDotNpert1Appendix} in particular). Furthermore, the derivatives $\dot{\boldsymbol{\Phi}}^{(n)}_{\mathfrak{s}}[G]$ resulting from this very recipe based on~\eqref{eq:2PIfrgmUflowReplacementNpert1Appendix} and~\eqref{eq:2PIfrgmUflowReplacementDotNpert1Appendix} are finally fed to the chain rule:
\begin{equation}
\dot{\overline{\boldsymbol{\Phi}}}^{(n)}_{\mathfrak{s},\gamma_{1}\cdots\gamma_{n}} = \overline{\dot{\boldsymbol{\Phi}}}^{(n)}_{\mathfrak{s},\gamma_{1}\cdots\gamma_{n}} + \dot{\overline{\boldsymbol{G}}}_{\mathfrak{s},\hat{\gamma}} \overline{\boldsymbol{\Phi}}^{(n+1)}_{\mathfrak{s}, \hat{\gamma} \gamma_{1} \cdots \gamma_{n}} \;.
\end{equation}
Let us finally summarize the tower of differential equations derived in this manner for the mU-flow with $N_{\mathrm{SCPT}}=1$:
\begin{equation}
\dot{\overline{\boldsymbol{G}}}_{\mathfrak{s},\alpha_{1}\alpha'_{1}}=\int_{\alpha_{2},\alpha'_{2}}\overline{\boldsymbol{G}}_{\mathfrak{s},\alpha_{1} \alpha_{2}} \dot{\overline{\boldsymbol{\Sigma}}}_{\mathfrak{s},\alpha_{2} \alpha'_{2}} \overline{\boldsymbol{G}}_{\mathfrak{s},\alpha'_{2} \alpha'_{1}} \;,
\label{eq:2PIfrgmUflowExpressionBoldGDot}
\end{equation}
\begin{equation}
\dot{\overline{\boldsymbol{\Omega}}}_{\mathfrak{s}} = \frac{1}{6\beta} \dot{U}_{\mathfrak{s},\hat{\gamma}_{1} \hat{\gamma}_{2}} \left(\overline{W}_{\mathfrak{s}}^{(2)} - \overline{\Pi}_{\mathfrak{s}}\right)_{\hat{\gamma}_{2}\hat{\gamma}_{1}} \;,
\label{eq:2PIfrgmUflowExpressionBoldOmegaDot}
\end{equation}
\begin{equation}
\begin{split}
\dot{\overline{\boldsymbol{\Phi}}}_{\mathfrak{s}} = & \ \frac{1}{6} \dot{U}_{\mathfrak{s},\hat{\gamma}_{1} \hat{\gamma}_{2}} \left(\overline{W}_{\mathfrak{s}}^{(2)} - \overline{\Pi}_{\mathfrak{s}}\right)_{\hat{\gamma}_{2}\hat{\gamma}_{1}} \\
& +\frac{1}{6} \left[\overline{\boldsymbol{\Sigma}}_{\mathfrak{s},\gamma_{1}} + \left(U-U_{\mathfrak{s}}\right)_{\gamma_{1}\hat{\gamma}_{2}}\overline{\boldsymbol{G}}_{\mathfrak{s},\hat{\gamma}_{2}}\right] \\
& \hspace{0.8cm} \times \overline{W}_{\mathfrak{s},\hat{\gamma}_{1} \hat{\gamma}_{3}}^{(2)} \dot{U}_{\mathfrak{s},\hat{\gamma}_{4} \hat{\gamma}_{5}} \left[ \overline{W}_{\mathfrak{s},\hat{\gamma}_{5} \hat{\gamma}_{6}}^{(2)} \left(\overline{\Pi}_{\mathfrak{s},\hat{\gamma}_{6}\hat{\gamma}_{7}}^{\mathrm{inv}} \frac{\delta \overline{\Pi}_{\mathfrak{s},\hat{\gamma}_{7}\hat{\gamma}_{8}}}{\delta \overline{\boldsymbol{G}}_{\mathfrak{s},\hat{\gamma}_{3}}} \overline{\Pi}_{\mathfrak{s},\hat{\gamma}_{8}\hat{\gamma}_{9}}^{\mathrm{inv}} - \overline{\boldsymbol{\Phi}}_{\mathfrak{s},\hat{\gamma}_{3} \hat{\gamma}_{6} \hat{\gamma}_{9}}^{(3)} \right) \overline{W}_{\mathfrak{s},\hat{\gamma}_{9} \hat{\gamma}_{4}}^{(2)} + \frac{1}{2} \frac{\delta \overline{\Pi}_{\mathfrak{s},\hat{\gamma}_{5} \hat{\gamma}_{4}}}{\delta \overline{\boldsymbol{G}}_{\mathfrak{s},\hat{\gamma}_{3}}} \right] \;,
\end{split}
\label{eq:2PIfrgmUflowExpressionBoldPhiDot}
\end{equation}
\begin{equation}
\begin{split}
\dot{\overline{\boldsymbol{\Sigma}}}_{\mathfrak{s},\gamma} = & - \frac{1}{3} \left(\mathcal{I} + \overline{\Pi}_{\mathfrak{s}} \overline{\boldsymbol{\Phi}}_{\mathfrak{s}}^{(2)}\right)^{\mathrm{inv}}_{\gamma\hat{\gamma}_{1}} \\
& \times \Bigg( \left[ 2 \left(\mathcal{I} + \overline{\Pi}_{\mathfrak{s}} \left(\overline{\boldsymbol{\Phi}}_{\mathfrak{s}}^{(2)} - U + U_{\mathfrak{s}} \right)\right)^{\mathrm{inv}} \dot{U}_{\mathfrak{s}} \left(\mathcal{I} + \overline{\Pi}_{\mathfrak{s}} \left(\overline{\boldsymbol{\Phi}}_{\mathfrak{s}}^{(2)} - U + U_{\mathfrak{s}} \right)\right)^{\mathrm{inv}} + \dot{U}_{\mathfrak{s}} \right]_{\hat{\alpha}_{1} \hat{\alpha}_{2} \hat{\alpha}'_{2} \hat{\alpha}'_{1}} \hspace{-0.15cm} \overline{\boldsymbol{G}}_{\mathfrak{s},\hat{\gamma}_{2}} \\
& \hspace{0.6cm} - \frac{1}{2} \dot{U}_{\mathfrak{s},\hat{\gamma}_{2} \hat{\gamma}_{3}} \overline{W}_{\mathfrak{s},\hat{\gamma}_{3} \hat{\gamma}_{4}}^{(2)} \overline{\boldsymbol{\Phi}}_{\mathfrak{s},\hat{\gamma}_{1} \hat{\gamma}_{4} \hat{\gamma}_{5}}^{(3)} \overline{W}_{\mathfrak{s},\hat{\gamma}_{5} \hat{\gamma}_{2}}^{(2)} - 3 \dot{U}_{\mathfrak{s},\hat{\gamma}_{1}\hat{\gamma}_{2}} \overline{\boldsymbol{G}}_{\mathfrak{s},\hat{\gamma}_{2}} \Bigg) \;,
\end{split}
\label{eq:2PIfrgmUflowExpressionBoldSigmaDot}
\end{equation}
\begin{equation}
\begin{split}
\dot{\overline{\boldsymbol{\Phi}}}_{\mathfrak{s},\gamma_{1}\gamma_{2}}^{(2)} = \frac{1}{3} \dot{U}_{\mathfrak{s},\hat{\gamma}_{1}\hat{\gamma}_{2}} & \Bigg[\overline{W}_{\mathfrak{s},\hat{\gamma}_{2} \hat{\gamma}_{3}}^{(2)} \left(\overline{\Pi}_{\mathfrak{s},\hat{\gamma}_{3} \hat{\gamma}_{4}}^{\mathrm{inv}} \frac{\delta \overline{\Pi}_{\mathfrak{s},\hat{\gamma}_{4} \hat{\gamma}_{5}}}{\delta \overline{\boldsymbol{G}}_{\mathfrak{s},\gamma_{1}}} \overline{\Pi}_{\mathfrak{s},\hat{\gamma}_{5} \hat{\gamma}_{6}}^{\mathrm{inv}} - \overline{\boldsymbol{\Phi}}_{\mathfrak{s},\gamma_{1}\hat{\gamma}_{3}\hat{\gamma}_{6}}^{(3)}\right)\overline{W}_{\mathfrak{s},\hat{\gamma}_{6} \hat{\gamma}_{7}}^{(2)} \\
& \hspace{0.2cm} \times \left(\overline{\Pi}_{\mathfrak{s},\hat{\gamma}_{7} \hat{\gamma}_{8}}^{\mathrm{inv}} \frac{\delta \overline{\Pi}_{\mathfrak{s},\hat{\gamma}_{8} \hat{\gamma}_{9}}}{\delta \overline{\boldsymbol{G}}_{\mathfrak{s},\gamma_{2}}} \overline{\Pi}_{\mathfrak{s},\hat{\gamma}_{9} \hat{\gamma}_{10}}^{\mathrm{inv}} - \overline{\boldsymbol{\Phi}}_{\mathfrak{s},\gamma_{2} \hat{\gamma}_{7} \hat{\gamma}_{10}}^{(3)}\right) \overline{W}_{\mathfrak{s},\hat{\gamma}_{10} \hat{\gamma}_{1}}^{(2)} \\
& - \overline{W}_{\mathfrak{s},\hat{\gamma}_{2} \hat{\gamma}_{3}}^{(2)} \overline{\Pi}_{\mathfrak{s},\hat{\gamma}_{3} \hat{\gamma}_{4}}^{\mathrm{inv}} \frac{\delta \overline{\Pi}_{\mathfrak{s},\hat{\gamma}_{4} \hat{\gamma}_{5}}}{\delta \overline{\boldsymbol{G}}_{\mathfrak{s},\gamma_{1}}} \overline{\Pi}_{\mathfrak{s},\hat{\gamma}_{5} \hat{\gamma}_{6}}^{\mathrm{inv}} \frac{\delta \overline{\Pi}_{\mathfrak{s},\hat{\gamma}_{6} \hat{\gamma}_{7}}}{\delta \overline{\boldsymbol{G}}_{\mathfrak{s},\gamma_{2}}} \overline{\Pi}_{\mathfrak{s},\hat{\gamma}_{7} \hat{\gamma}_{8}}^{\mathrm{inv}} \overline{W}_{\mathfrak{s},\hat{\gamma}_{8} \hat{\gamma}_{1}}^{(2)} \\
& + \frac{1}{2} \overline{W}_{\mathfrak{s},\hat{\gamma}_{2} \hat{\gamma}_{3}}^{(2)} \left(\overline{\Pi}_{\mathfrak{s},\hat{\gamma}_{3} \hat{\gamma}_{4}}^{\mathrm{inv}} \frac{\delta^{2} \overline{\Pi}_{\mathfrak{s},\hat{\gamma}_{4} \hat{\gamma}_{5}}}{\delta \overline{\boldsymbol{G}}_{\mathfrak{s},\gamma_{1}} \delta \overline{\boldsymbol{G}}_{\mathfrak{s},\gamma_{2}}} \overline{\Pi}_{\mathfrak{s},\hat{\gamma}_{5} \hat{\gamma}_{6}}^{\mathrm{inv}} - \overline{\boldsymbol{\Phi}}_{\mathfrak{s},\gamma_{1}\gamma_{2}\hat{\gamma}_{3}\hat{\gamma}_{6}}^{(4)}\right) \overline{W}_{\mathfrak{s},\hat{\gamma}_{6} \hat{\gamma}_{1}}^{(2)} \\
& + \frac{1}{4} \frac{\delta^{2}\overline{\Pi}_{\mathfrak{s},\hat{\gamma}_{2}\hat{\gamma}_{1}}}{\delta \overline{\boldsymbol{G}}_{\mathfrak{s},\gamma_{1}} \delta \overline{\boldsymbol{G}}_{\mathfrak{s},\gamma_{2}}}\Bigg] - \dot{U}_{\mathfrak{s},\gamma_{1}\gamma_{2}} + \dot{\overline{\boldsymbol{G}}}_{\mathfrak{s},\hat{\gamma}} \overline{\boldsymbol{\Phi}}_{\mathfrak{s},\hat{\gamma}\gamma_{1}\gamma_{2}}^{(3)} \;,
\end{split}
\label{eq:2PIfrgmUflowExpressionBoldPhi2Dot}
\end{equation}
where $\overline{W}^{(2)}_{\mathfrak{s}}$ can be expressed by applying substitution~\eqref{eq:2PIfrgmUflowReplacementPhi2Npert1Appendix} to~\eqref{eq:2PIfrgW2PiInvPhi2}:
\begin{equation}
\overline{W}^{(2)}_{\mathfrak{s}} = \left(\overline{\Pi}_{\mathfrak{s}}^{\mathrm{inv}} + \overline{\boldsymbol{\Phi}}^{(2)}_{\mathfrak{s}} - U + U_{\mathfrak{s}} \right)^{\mathrm{inv}} \;.
\end{equation}
Results~\eqref{eq:2PIfrgmUflowExpressionBoldGDot} to~\eqref{eq:2PIfrgmUflowExpressionBoldPhi2Dot} are the mU-flow counterparts of the pU-flow equations \eqref{eq:2PIfrgUFlowGbarviaDysonEq}, \eqref{eq:2PIfrgUflowExpressionOmegaDot}, \eqref{eq:2PIfrgUflowExpressionPhiDot}, \eqref{eq:2PIfrgmUflowSelfEnergyNpert1} and \eqref{eq:2PIfrgUflowPhi2Dot}, respectively. Note that, among~\eqref{eq:2PIfrgmUflowExpressionBoldGDot} to~\eqref{eq:2PIfrgmUflowExpressionBoldPhi2Dot},~\eqref{eq:2PIfrgmUflowExpressionBoldGDot} is the only equation which is not obtained from pU-flow equations via substitutions~\eqref{eq:2PIfrgmUflowReplacementOmegaNpert1Appendix}, \eqref{eq:2PIfrgmUflowReplacementOmegaDotNpert1Appendix}, \eqref{eq:2PIfrgmUflowReplacementNpert1Appendix} or \eqref{eq:2PIfrgmUflowReplacementDotNpert1Appendix}. It simply follows from the fact that $\overline{\boldsymbol{G}}_{\mathfrak{s}}$ satisfies a Dyson equation with $\overline{\boldsymbol{\Sigma}}_{\mathfrak{s}}$ as self-energy.


\subsection{CU-flow}
\label{ann:2PIfrgFlowEquationCUflow}

As pointed out in section~\ref{sec:CUflow2PIFRG}, the differential equations implementing the CU-flow are composed of C-flow and U-flow contributions. We can therefore directly deduce these from previous results:
\begin{itemize}
\item Expression of $\dot{\overline{G}}_{\mathfrak{s}}$:\\
As for the C-flow and the U-flow, the differential equation expressing $\dot{\overline{G}}_{\mathfrak{s}}$ is obtained by differentiating the corresponding Dyson equation with respect to the flow parameter:
\begin{equation}
\dot{\overline{G}}_{\mathfrak{s},\alpha_{1} \alpha'_{1}}=-\int_{\alpha_{2},\alpha'_{2}}\overline{G}_{\mathfrak{s},\alpha_{1}\alpha_{2}}\left(\dot{C}_{\mathfrak{s}}^{-1}-\dot{\overline{\Sigma}}_{\mathfrak{s}}\right)_{\alpha_{2} \alpha'_{2}} \overline{G}_{\mathfrak{s},\alpha'_{2} \alpha'_{1}} \;.
\end{equation}

\item Expression of $\dot{\overline{\Omega}}_{\mathfrak{s}}$:\\
Combining~\eqref{eq:2PIfrgCflowDGammaDk} for the C-flow and~\eqref{eq:2PIfrgGammasdot} for the U-flow, we obtain the relation:
\begin{equation}
\dot{\Gamma}^{(\mathrm{2PI})}_{\mathfrak{s}}[G] = \dot{C}^{-1}_{\mathfrak{s},\hat{\gamma}} G_{\hat{\gamma}} + \frac{1}{6} \dot{U}_{\mathfrak{s},\hat{\gamma}_{1} \hat{\gamma}_{2}} \left(W_{\mathfrak{s}}^{(2)}[K] + \frac{1}{2}\Pi[G]\right)_{\hat{\gamma}_{2}\hat{\gamma}_{1}} \;,
\label{eq:2PIfrgCUflowGammaDot}
\end{equation}
which, according to~\eqref{eq:2PIfrgGammaBarDotVsGammaDotBar}, reduces at $K_{\gamma}=0$ $\forall\gamma$ to:
\begin{equation}
\dot{\overline{\Gamma}}^{(\mathrm{2PI})}_{\mathfrak{s}} = \dot{C}^{-1}_{\mathfrak{s},\hat{\gamma}} \overline{G}_{\mathfrak{s},\hat{\gamma}} + \frac{1}{6} \dot{U}_{\mathfrak{s},\hat{\gamma}_{1} \hat{\gamma}_{2}} \left(\overline{W}_{\mathfrak{s}}^{(2)} + \frac{1}{2}\overline{\Pi}_{\mathfrak{s}}\right)_{\hat{\gamma}_{2}\hat{\gamma}_{1}} \;.
\label{eq:2PIfrgCUflowDotGammakBar}
\end{equation}
Similarly to the previous C-flow derivations,~\eqref{eq:2PIfrgCUflowDotGammakBar} is translated into a flow equation for:
\begin{equation}
\scalebox{0.94}{${\displaystyle \Delta \overline{\Omega}_{\mathfrak{s}} \equiv \frac{1}{\beta}\left(\overline{\Gamma}_{\mathfrak{s}}^{(\mathrm{2PI})} - \Gamma_{0,\mathfrak{s}}^{(\mathrm{2PI})}[C_{\mathfrak{s}}] \right) = \frac{1}{\beta}\left(\overline{\Gamma}_{\mathfrak{s}}^{(\mathrm{2PI})} + \frac{\zeta}{2} \mathrm{Tr}_{\alpha} \left[ \mathrm{ln}(C_{\mathfrak{s}}) \right]\right) = \frac{1}{\beta}\left(\overline{\Gamma}_{\mathfrak{s}}^{(\mathrm{2PI})} - \frac{\zeta}{2} \mathrm{Tr}_{\alpha} \left[ \mathrm{ln}\big(C^{-1}_{\mathfrak{s}}\big) \right]\right) \;. }$}
\end{equation}
The derivative of $\Delta \overline{\Omega}_{\mathfrak{s}}$ with respect to $\mathfrak{s}$ is expressed as in~\eqref{eq:2PIfrgCflowDerivDeltaOmega}:
\begin{equation}
\begin{split}
\Delta \dot{\overline{\Omega}}_{\mathfrak{s}} = & \ \frac{1}{\beta}\bigg(\dot{\overline{\Gamma}}^{(\mathrm{2PI})}_{\mathfrak{s}} - \frac{\zeta}{2} \int_{\alpha,\alpha'} \dot{C}^{-1}_{\mathfrak{s},\alpha \alpha'} \underbrace{C_{\mathfrak{s},\alpha'\alpha}}_{\zeta C_{\mathfrak{s},\alpha\alpha'}} \bigg) \\
= & \ \frac{1}{\beta}\left(\dot{\overline{\Gamma}}^{(\mathrm{2PI})}_{\mathfrak{s}} - \dot{C}^{-1}_{\mathfrak{s},\hat{\gamma}} C_{\mathfrak{s},\hat{\gamma}}\right) \;,
\end{split}
\end{equation}
and we finally replace $\dot{\overline{\Gamma}}^{(\mathrm{2PI})}_{\mathfrak{s}}$ using~\eqref{eq:2PIfrgCUflowDotGammakBar}:
\begin{equation}
\Delta \dot{\overline{\Omega}}_{\mathfrak{s}} = \frac{1}{\beta} \dot{C}_{\mathfrak{s},\hat{\gamma}}^{-1} \left(\overline{G}_{\mathfrak{s}}-C_{\mathfrak{s}}\right)_{\hat{\gamma}} + \frac{1}{6\beta} \dot{U}_{\mathfrak{s},\hat{\gamma}_{1} \hat{\gamma}_{2}} \left(\overline{W}_{\mathfrak{s}}^{(2)} + \frac{1}{2}\overline{\Pi}_{\mathfrak{s}}\right)_{\hat{\gamma}_{2}\hat{\gamma}_{1}} \;.
\end{equation}

\item Expression of $\dot{\overline{\Phi}}_{\mathfrak{s}}$:\\
From the definition of the Luttinger-Ward functional, we have:
\begin{equation}
\dot{\Phi}_{\mathfrak{s}}[G]=\dot{\Gamma}^{(\mathrm{2PI})}_{\mathfrak{s}}[G]-\dot{\Gamma}^{(\mathrm{2PI})}_{0,\mathfrak{s}}[G] \;,
\label{eq:2PIfrgCUflowDotPhi}
\end{equation}
where every derivative with respect to $\mathfrak{s}$ is implicitly carried out at fixed propagator $G$, as in~\eqref{eq:2PIfrgCflowPhiInvariant} and~\eqref{eq:2PIfrgUflowPhiGammadot}. Since the free EA $\Gamma^{(\mathrm{2PI})}_{0,\mathfrak{s}}[G]$ is independent from the interaction $U$, $\dot{\Gamma}^{(\mathrm{2PI})}_{0,\mathfrak{s}}$ only contains a contribution from the C-flow. In other words, $\dot{\Gamma}_{0,\mathfrak{s}}[G]$ is now given by~\eqref{eq:2PIfrgCflowDGamma0Dk}. By exploiting~\eqref{eq:2PIfrgCUflowGammaDot} as well,~\eqref{eq:2PIfrgCUflowDotPhi} becomes:
\begin{equation}
\begin{split}
\dot{\Phi}_{\mathfrak{s}}[G] = & \ \cancel{\dot{C}^{-1}_{\mathfrak{s},\hat{\gamma}} G_{\hat{\gamma}}} + \frac{1}{6} \dot{U}_{\mathfrak{s},\hat{\gamma}_{1} \hat{\gamma}_{2}} \left(W_{\mathfrak{s}}^{(2)}[K] + \frac{1}{2}\Pi[G]\right)_{\hat{\gamma}_{2}\hat{\gamma}_{1}}-\cancel{\dot{C}^{-1}_{\mathfrak{s},\hat{\gamma}} G_{\hat{\gamma}}} \\
= & \ \frac{1}{6} \dot{U}_{\mathfrak{s},\hat{\gamma}_{1} \hat{\gamma}_{2}} \left(W_{\mathfrak{s}}^{(2)}[K] + \frac{1}{2}\Pi[G]\right)_{\hat{\gamma}_{2}\hat{\gamma}_{1}} \;.
\end{split}
\label{eq:2PIfrgCUflowDotPhiBar}
\end{equation}
Therefore, the latter expression of $\dot{\Phi}_{\mathfrak{s}}$ only contains a contribution from the U-flow, which is consistent with~\eqref{eq:2PIfrgCflowPhiInvariant} showing that the Luttinger-Ward functional is an invariant of the C-flow. This remark can be extended to the derivatives of the 2PI vertices $\dot{\Phi}^{(n)}_{\mathfrak{s}}$. The flow equation expressing $\dot{\overline{\Phi}}_{\mathfrak{s}}$ is therefore identical to that of $\overline{\Phi}_{\mathfrak{s}}$ given by~\eqref{eq:2PIfrgUflowExpressionPhiDot} in the framework of the U-flow, i.e.:
\begin{equation}
\begin{split}
\scalebox{0.95}{${\displaystyle\dot{\overline{\Phi}}_{\mathfrak{s}} =}$} & \scalebox{0.95}{${\displaystyle \ \frac{1}{6} \dot{U}_{\mathfrak{s},\hat{\gamma}_{1} \hat{\gamma}_{2}} \left(\overline{W}_{\mathfrak{s}}^{(2)} + \frac{1}{2}\overline{\Pi}_{\mathfrak{s}}\right)_{\hat{\gamma}_{2}\hat{\gamma}_{1}}}$} \\
& \scalebox{0.95}{${\displaystyle +\frac{1}{6} \overline{\Sigma}_{\mathfrak{s},\hat{\gamma}_{1}} \overline{W}_{\mathfrak{s},\hat{\gamma}_{1} \hat{\gamma}_{2}}^{(2)} \dot{U}_{\mathfrak{s},\hat{\gamma}_{3} \hat{\gamma}_{4}} \left[ \overline{W}_{\mathfrak{s},\hat{\gamma}_{4} \hat{\gamma}_{5}}^{(2)} \left(\overline{\Pi}_{\mathfrak{s},\hat{\gamma}_{5}\hat{\gamma}_{6}}^{\mathrm{inv}} \frac{\delta \overline{\Pi}_{\mathfrak{s},\hat{\gamma}_{6}\hat{\gamma}_{7}}}{\delta \overline{G}_{\mathfrak{s},\hat{\gamma}_{2}}} \overline{\Pi}_{\mathfrak{s},\hat{\gamma}_{7}\hat{\gamma}_{8}}^{\mathrm{inv}} - \overline{\Phi}_{\mathfrak{s},\hat{\gamma}_{2} \hat{\gamma}_{5} \hat{\gamma}_{8}}^{(3)} \right) \overline{W}_{\mathfrak{s},\hat{\gamma}_{8} \hat{\gamma}_{3}}^{(2)} + \frac{1}{2} \frac{\delta \overline{\Pi}_{\mathfrak{s},\hat{\gamma}_{4} \hat{\gamma}_{3}}}{\delta \overline{G}_{\mathfrak{s},\hat{\gamma}_{2}}} \right] \;.}$}
\end{split}
\label{eq:2PIFRGCUflowPhidotSameasUflow}
\end{equation}

\item Expression of $\dot{\overline{\Sigma}}_{\mathfrak{s}}$:\\
There is however a subtlety regarding the flow equation for the self-energy because~\eqref{eq:2PIFRGSigmaSBetheSalpeter} and therefore~\eqref{eq:2PIfrgmUflowSelfEnergyNpert1} (which is derived from~\eqref{eq:2PIFRGSigmaSBetheSalpeter}) are not valid in the framework of the CU-flow. This stems from the fact that the derivation of~\eqref{eq:2PIFRGSigmaSBetheSalpeter} itself relies on~\eqref{eq:2PIfrgGdotW2Phi1BetheSalpeter} which is satisfied if and only if $\dot{\Gamma}_{\mathfrak{s}}^{(\mathrm{2PI})(1)}=\dot{\Phi}_{\mathfrak{s}}^{(1)}$ (or, equivalently, $\dot{\Gamma}_{\mathfrak{s}}^{(\mathrm{2PI})(1)}=-\dot{\Sigma}_{\mathfrak{s}}$), the latter equality requiring that $\dot{C}_{\mathfrak{s},\gamma}=0$ $\forall\gamma$, which is not valid for all $\mathfrak{s}$ in the framework of the CU-flow (by definition of the CU-flow). Nonetheless, we can directly exploit the chain rule in the form~\eqref{eq:2PIfrgGeneralFlowEqSigma} combined with expression~\eqref{eq:2PIfrgUflowPhi1} of $\overline{\dot{\Sigma}}_{\mathfrak{s},\gamma}$, thus leading to:
\begin{equation}
\begin{split}
\dot{\overline{\Sigma}}_{\mathfrak{s},\gamma} = & -\frac{1}{6} \dot{U}_{\mathfrak{s},\hat{\gamma}_{1} \hat{\gamma}_{2}} \left[\overline{W}_{\mathfrak{s},\hat{\gamma}_{2} \hat{\gamma}_{3}}^{(2)} \left(\overline{\Pi}_{\mathfrak{s},\hat{\gamma}_{3}\hat{\gamma}_{4}}^{\mathrm{inv}} \frac{\delta \overline{\Pi}_{\mathfrak{s},\hat{\gamma}_{4}\hat{\gamma}_{5}}}{\delta \overline{G}_{\mathfrak{s},\gamma}} \overline{\Pi}_{\mathfrak{s},\hat{\gamma}_{5}\hat{\gamma}_{6}}^{\mathrm{inv}} - \overline{\Phi}_{\mathfrak{s},\gamma \hat{\gamma}_{3} \hat{\gamma}_{6}}^{(3)} \right) \overline{W}_{\mathfrak{s},\hat{\gamma}_{6} \hat{\gamma}_{1}}^{(2)} + \frac{1}{2}\frac{\delta \overline{\Pi}_{\mathfrak{s},\hat{\gamma}_{2}\hat{\gamma}_{1}}}{\delta \overline{G}_{\mathfrak{s},\gamma}} \right] \\
& - \dot{\overline{G}}_{\mathfrak{s},\hat{\gamma}} \overline{\Phi}^{(2)}_{\mathfrak{s},\hat{\gamma}\gamma} \;.
\end{split}
\label{eq:DotOverlienSigma2PIfrgCUflowAppendix}
\end{equation}
As in~\eqref{eq:DetailedCalculation1Sigma2PIFRGUflowAppendix} and~\eqref{eq:DetailedCalculation2Sigma2PIFRGUflowAppendix}, we can show that:
\begin{equation}
\dot{U}_{\mathfrak{s},\hat{\gamma}_{1} \hat{\gamma}_{2}} \frac{\delta \overline{\Pi}_{\mathfrak{s},\hat{\gamma}_{2}\hat{\gamma}_{1}}}{\delta \overline{G}_{\mathfrak{s},\gamma}} = 4 \dot{U}_{\mathfrak{s},\alpha \hat{\alpha} \hat{\alpha}' \alpha'} \overline{G}_{\mathfrak{s},\hat{\gamma}} \;,
\label{eq:trick12PIfrgCUflowAppendix}
\end{equation}
\begin{equation}
\scalebox{0.93}{${\displaystyle\dot{U}_{\mathfrak{s},\hat{\gamma}_{1} \hat{\gamma}_{2}} \overline{W}_{\mathfrak{s},\hat{\gamma}_{2} \hat{\gamma}_{3}}^{(2)} \overline{\Pi}_{\mathfrak{s},\hat{\gamma}_{3}\hat{\gamma}_{4}}^{\mathrm{inv}} \frac{\delta \overline{\Pi}_{\mathfrak{s},\hat{\gamma}_{4}\hat{\gamma}_{5}}}{\delta \overline{G}_{\mathfrak{s},\gamma}} \overline{\Pi}_{\mathfrak{s},\hat{\gamma}_{5}\hat{\gamma}_{6}}^{\mathrm{inv}} \overline{W}_{\mathfrak{s},\hat{\gamma}_{6} \hat{\gamma}_{1}}^{(2)} = 4 \left[ \left(\mathcal{I} + \overline{\Pi}_{\mathfrak{s}} \overline{\Phi}_{\mathfrak{s}}^{(2)}\right)^{\mathrm{inv}} \dot{U}_{\mathfrak{s}} \left(\mathcal{I} + \overline{\Pi}_{\mathfrak{s}} \overline{\Phi}_{\mathfrak{s}}^{(2)}\right)^{\mathrm{inv}} \right]_{\alpha \hat{\alpha} \hat{\alpha}' \alpha'} \overline{G}_{\mathfrak{s},\hat{\gamma}} \;. }$}
\label{eq:trick22PIfrgCUflowAppendix}
\end{equation}
According to~\eqref{eq:trick12PIfrgCUflowAppendix} and~\eqref{eq:trick22PIfrgCUflowAppendix},~\eqref{eq:DotOverlienSigma2PIfrgCUflowAppendix} is equivalent to:
\begin{equation}
\begin{split}
\dot{\overline{\Sigma}}_{\mathfrak{s},\gamma} = & -\frac{1}{3} \left[ 2 \left(\mathcal{I} + \overline{\Pi}_{\mathfrak{s}} \overline{\Phi}_{\mathfrak{s}}^{(2)}\right)^{\mathrm{inv}} \dot{U}_{\mathfrak{s}} \left(\mathcal{I} + \overline{\Pi}_{\mathfrak{s}} \overline{\Phi}_{\mathfrak{s}}^{(2)}\right)^{\mathrm{inv}} + \dot{U}_{\mathfrak{s}} \right]_{\alpha \hat{\alpha} \hat{\alpha}' \alpha'} \overline{G}_{\mathfrak{s},\hat{\gamma}} \\
& +\frac{1}{6} \dot{U}_{\mathfrak{s},\hat{\gamma}_{1} \hat{\gamma}_{2}} \overline{W}_{\mathfrak{s},\hat{\gamma}_{2} \hat{\gamma}_{3}}^{(2)} \overline{\Phi}_{\mathfrak{s},\gamma \hat{\gamma}_{3} \hat{\gamma}_{4}}^{(3)} \overline{W}_{\mathfrak{s},\hat{\gamma}_{4} \hat{\gamma}_{1}}^{(2)} - \dot{\overline{G}}_{\mathfrak{s},\hat{\gamma}} \overline{\Phi}^{(2)}_{\mathfrak{s},\hat{\gamma}\gamma} \;.
\end{split}
\end{equation}

\item Expressions of $\dot{\overline{\Phi}}^{(n)}_{\mathfrak{s}}$ (with $n \geq 2$):\\
Since there are no C-flow contributions to the derivatives $\dot{\Phi}_{\mathfrak{s}}^{(n)}$ as explained above~\eqref{eq:2PIFRGCUflowPhidotSameasUflow}, the differential equations expressing $\dot{\overline{\Phi}}_{\mathfrak{s}}^{(n)}$ for the CU-flow are identical to those derived previously for the pU-flow. In particular, the CU-flow equations for $\overline{\Phi}^{(2)}_{\mathfrak{s}}$ and $\overline{\Phi}^{(3)}_{\mathfrak{s}}$ are already given by~\eqref{eq:2PIfrgUflowPhi2Dot} and~\eqref{eq:2PIfrgUflowPhi3Dot}, respectively.

\end{itemize}

\section{Initial conditions for the C-flow}
\label{ann:2PIfrgCflowInitialConditions}
\subsection{General case}
\label{ann:2PIfrgCflowInitialConditionsGeneral}

\begin{itemize}
\item For $N_{\mathrm{SCPT}}=1$:\\
As exposed by~\eqref{eq:2PIFRGperturbativeExpLWfunc}, the expression of the Luttinger-Ward functional resulting from self-consistent PT and truncated at the Hartree-Fock level (i.e. at $N_{\mathrm{SCPT}}=1$) is:
\begin{equation}
\Phi_{\mathrm{SCPT},N_{\mathrm{SCPT}}=1}[U,G]\equiv\frac{1}{8}\int_{\gamma_{1},\gamma_{2}}U_{\gamma_{1}\gamma_{2}}G_{\gamma_{1}}G_{\gamma_{2}}\;.
\label{eq:Phi1SCPTCI2PIfrgcflowAppendix}
\end{equation}
From~\eqref{eq:Phi1SCPTCI2PIfrgcflowAppendix}, we infer that:
\begin{equation}
\begin{split}
\Phi^{(2)}_{\mathrm{SCPT},N_{\mathrm{SCPT}}=1,\gamma_{1}\gamma_{2}}[U,G] \equiv & \ \frac{\delta^{2}\Phi_{\mathrm{SCPT},N_{\mathrm{SCPT}}=1}[U,G]}{\delta G_{\gamma_{1}} \delta G_{\gamma_{2}}} \\
= & \ \frac{1}{4} \frac{\delta}{\delta G_{\gamma_{1}}}\int_{\gamma_{3},\gamma_{4}} U_{\gamma_{3}\gamma_{4}} G_{\gamma_{3}} \frac{\delta G_{\gamma_{4}}}{\delta G_{\gamma_{2}}} \\
= & \ \frac{1}{4} \frac{\delta}{\delta G_{\gamma_{1}}}\int_{\gamma_{3},\gamma_{4}} U_{\gamma_{3}\gamma_{4}} G_{\gamma_{3}} \left(\delta_{\alpha_{4} \alpha_{2}}\delta_{\alpha'_{4} \alpha'_{2}} + \zeta\delta_{\alpha_{4} \alpha'_{2}}\delta_{\alpha'_{4} \alpha_{2}}\right) \\
= & \ \frac{1}{4} \frac{\delta}{\delta G_{\gamma_{1}}}\left(\int_{\gamma_{3}} U_{\gamma_{3}\gamma_{2}} G_{\gamma_{3}}+\zeta\int_{\gamma_{3}} U_{\gamma_{3} (\alpha'_{2},\alpha_{2})} G_{\gamma_{3}}\right) \\
= & \ \frac{1}{4} \int_{\gamma_{3}} U_{\gamma_{3}\gamma_{2}} \frac{\delta G_{\gamma_{3}}}{\delta G_{\gamma_{1}}} + \frac{\zeta}{4}\int_{\gamma_{3}} \underbrace{U_{\gamma_{3} (\alpha'_{2},\alpha_{2})}}_{\zeta U_{\gamma_{3} \gamma_{2}}} \frac{\delta G_{\gamma_{3}}}{\delta G_{\gamma_{1}}} \\
= & \ \frac{1}{2} \int_{\gamma_{3}} U_{\gamma_{3} \gamma_{2}} \frac{\delta G_{\gamma_{3}}}{\delta G_{\gamma_{1}}} \\
= & \ \frac{1}{2} \int_{\gamma_{3}} U_{\gamma_{3}\gamma_{2}} \left(\delta_{\alpha_{3}\alpha_{1}}\delta_{\alpha'_{3}\alpha'_{1}} + \zeta\delta_{\alpha_{3}\alpha'_{1}}\delta_{\alpha'_{3}\alpha_{1}} \right) \\
= & \ \frac{1}{2} U_{\gamma_{1}\gamma_{2}} + \frac{\zeta}{2} \underbrace{U_{(\alpha'_{1},\alpha_{1}) \gamma_{2}}}_{\zeta U_{\gamma_{1} \gamma_{2}}} \\
= & \ U_{\gamma_{1}\gamma_{2}} \;,
\end{split}
\end{equation}
where we have made use of the property $U_{\alpha_{1}\alpha_{2}\alpha_{3}\alpha_{4}}=\zeta^{N(P)}U_{\alpha_{P(1)}\alpha_{P(2)}\alpha_{P(3)}\alpha_{P(4)}}$. We are more specifically interested in the situation where the external source $K$ vanishes:
\begin{equation}
\overline{\Phi}^{(2)}_{\mathrm{SCPT},N_{\mathrm{SCPT}}=1,\mathfrak{s},\gamma_{1}\gamma_{2}} = U_{\gamma_{1}\gamma_{2}} \;.
\end{equation}

\item For $N_{\mathrm{SCPT}}=2$:\\
We consider once again the expression of the Luttinger-Ward functional resulting from self-consistent PT, i.e.~\eqref{eq:2PIFRGperturbativeExpLWfunc}, in order to evaluate $\Phi_{\mathrm{SCPT},N_{\mathrm{SCPT}}=2}^{(4)}[U,G]$. At $N_{\mathrm{SCPT}}=2$, it reduces to:
\begin{equation}
\scalebox{0.96}{${\displaystyle \Phi_{\mathrm{SCPT},N_{\mathrm{SCPT}}=2}[U,G]\equiv\frac{1}{8}\int_{\gamma_{1},\gamma_{2}}U_{\gamma_{1}\gamma_{2}}G_{\gamma_{1}}G_{\gamma_{2}} - \frac{1}{48} \int_{\gamma_{1},\gamma_{2},\gamma_{3},\gamma_{4}} U_{\alpha_{1}\alpha_{2}\alpha_{3}\alpha_{4}} U_{\alpha'_{1}\alpha'_{2}\alpha'_{3}\alpha'_{4}} G_{\gamma_{1}}G_{\gamma_{2}}G_{\gamma_{3}}G_{\gamma_{4}} \;. }$}
\label{eq:Phi2SCPTCI2PIfrgcflowAppendix}
\end{equation}
Differentiating both sides of~\eqref{eq:Phi2SCPTCI2PIfrgcflowAppendix} with respect to $G$ leads to:
\begin{equation}
\begin{split}
\scalebox{0.73}{${\displaystyle \Phi_{\mathrm{SCPT},N_{\mathrm{SCPT}}=2,\gamma_{1}\gamma_{2}\gamma_{3}\gamma_{4}}^{(4)} [U,G] \equiv }$} & \ \scalebox{0.73}{${\displaystyle \frac{\delta^{4}\Phi_{\mathrm{SCPT},N_{\mathrm{SCPT}}=2}[U,G]}{\delta G_{\gamma_{1}} \delta G_{\gamma_{2}} \delta G_{\gamma_{3}} \delta G_{\gamma_{4}}} }$} \\
\scalebox{0.73}{${\displaystyle = }$} & \scalebox{0.73}{${\displaystyle -\frac{1}{48} \frac{\delta^{4}}{\delta G_{\gamma_{1}} \delta G_{\gamma_{2}} \delta G_{\gamma_{3}} \delta G_{\gamma_{4}}} \int_{\gamma_{5},\gamma_{6},\gamma_{7},\gamma_{8}} U_{\alpha_{5} \alpha_{6} \alpha_{7} \alpha_{8}} U_{\alpha'_{5} \alpha'_{6} \alpha'_{7} \alpha'_{8}} G_{\gamma_{5}} G_{\gamma_{6}} G_{\gamma_{7}} G_{\gamma_{8}} }$} \\
\scalebox{0.73}{${\displaystyle = }$} & \scalebox{0.73}{${\displaystyle -\frac{1}{12} \frac{\delta^{3}}{\delta G_{\gamma_{1}} \delta G_{\gamma_{2}} \delta G_{\gamma_{3}}} \int_{\gamma_{5},\gamma_{6},\gamma_{7},\gamma_{8}} U_{\alpha_{5} \alpha_{6} \alpha_{7} \alpha_{8}} U_{\alpha'_{5} \alpha'_{6} \alpha'_{7} \alpha'_{8}} G_{\gamma_{5}} G_{\gamma_{6}} G_{\gamma_{7}} \frac{\delta G_{\gamma_{8}}}{\delta G_{\gamma_{4}}} }$} \\
\scalebox{0.73}{${\displaystyle = }$} & \scalebox{0.73}{${\displaystyle -\frac{1}{12} \frac{\delta^{3}}{\delta G_{\gamma_{1}} \delta G_{\gamma_{2}} \delta G_{\gamma_{3}}} \left[\int_{\gamma_{5},\gamma_{6},\gamma_{7}} U_{\alpha_{5} \alpha_{6} \alpha_{7} \alpha_{4}} U_{\alpha'_{5} \alpha'_{6} \alpha'_{7} \alpha'_{4}} G_{\gamma_{5}} G_{\gamma_{6}} G_{\gamma_{7}} + \zeta \left(\alpha_{4} \leftrightarrow \alpha'_{4} \right) \right] }$} \\
\scalebox{0.73}{${\displaystyle = }$} & \scalebox{0.73}{${\displaystyle -\frac{1}{4} \frac{\delta^{2}}{\delta G_{\gamma_{1}} \delta G_{\gamma_{2}}} \left[\int_{\gamma_{5},\gamma_{6},\gamma_{7}} U_{\alpha_{5} \alpha_{6} \alpha_{7} \alpha_{4}} U_{\alpha'_{5} \alpha'_{6} \alpha'_{7} \alpha'_{4}} G_{\gamma_{5}} G_{\gamma_{6}} \frac{\delta G_{\gamma_{7}}}{\delta G_{\gamma_{3}}} + \zeta \left(\alpha_{4} \leftrightarrow \alpha'_{4} \right) \right] }$} \\
\scalebox{0.73}{${\displaystyle = }$} & \scalebox{0.73}{${\displaystyle -\frac{1}{4} \frac{\delta^{2}}{\delta G_{\gamma_{1}} \delta G_{\gamma_{2}}} \left\lbrace\left[\int_{\gamma_{5},\gamma_{6}} U_{\alpha_{5} \alpha_{6} \alpha_{3} \alpha_{4}} U_{\alpha'_{5} \alpha'_{6} \alpha'_{3} \alpha'_{4}} G_{\gamma_{5}} G_{\gamma_{6}} + \zeta \left(\alpha_{3} \leftrightarrow \alpha'_{3}\right) \right] + \zeta \left(\alpha_{4} \leftrightarrow \alpha'_{4} \right) \right\rbrace }$} \\
\scalebox{0.73}{${\displaystyle = }$} & \scalebox{0.73}{${\displaystyle -\frac{1}{2} \frac{\delta}{\delta G_{\gamma_{1}}} \left\lbrace\left[\int_{\gamma_{5},\gamma_{6}} U_{\alpha_{5} \alpha_{6} \alpha_{3} \alpha_{4}} U_{\alpha'_{5} \alpha'_{6} \alpha'_{3} \alpha'_{4}} G_{\gamma_{5}} \frac{\delta G_{\gamma_{6}}}{\delta G_{\gamma_{2}}} + \zeta \left(\alpha_{3} \leftrightarrow \alpha'_{3}\right) \right] + \zeta \left(\alpha_{4} \leftrightarrow \alpha'_{4} \right) \right\rbrace }$} \\
\scalebox{0.73}{${\displaystyle = }$} & \scalebox{0.73}{${\displaystyle -\frac{1}{2} \frac{\delta}{\delta G_{\gamma_{1}}} \left(\left\lbrace\left[ \int_{\gamma_{5}} U_{\alpha_{5} \alpha_{2} \alpha_{3} \alpha_{4}} U_{\alpha'_{5} \alpha'_{2} \alpha'_{3} \alpha'_{4}} G_{\gamma_{5}} + \zeta \left(\alpha_{2} \leftrightarrow \alpha'_{2}\right) \right] + \zeta \left(\alpha_{3} \leftrightarrow \alpha'_{3}\right) \right\rbrace + \zeta \left(\alpha_{4} \leftrightarrow \alpha'_{4} \right) \right) }$} \\
\scalebox{0.73}{${\displaystyle = }$} & \scalebox{0.73}{${\displaystyle -\frac{1}{2} \left(\left\lbrace\left[ \int_{\gamma_{5}} U_{\alpha_{5} \alpha_{2} \alpha_{3} \alpha_{4}} U_{\alpha'_{5} \alpha'_{2} \alpha'_{3} \alpha'_{4}} \frac{\delta G_{\gamma_{5}}}{\delta G_{\gamma_{1}}} + \zeta \left(\alpha_{2} \leftrightarrow \alpha'_{2}\right) \right] + \zeta \left(\alpha_{3} \leftrightarrow \alpha'_{3}\right) \right\rbrace + \zeta \left(\alpha_{4} \leftrightarrow \alpha'_{4} \right) \right) }$} \\
\scalebox{0.73}{${\displaystyle = }$} & \scalebox{0.73}{${\displaystyle -\frac{1}{2} \left[\left(\left\lbrace\left[ U_{\alpha_{1} \alpha_{2} \alpha_{3} \alpha_{4}} U_{\alpha'_{1} \alpha'_{2} \alpha'_{3} \alpha'_{4}} + \zeta \left(\alpha_{1} \leftrightarrow \alpha'_{1}\right) \right] + \zeta \left(\alpha_{2} \leftrightarrow \alpha'_{2}\right) \right\rbrace + \zeta \left(\alpha_{3} \leftrightarrow \alpha'_{3}\right) \right) + \zeta \left(\alpha_{4} \leftrightarrow \alpha'_{4} \right) \right] \;, }$}
\end{split}
\end{equation}
where the equality $U_{\alpha_{1}\alpha_{2}\alpha_{3}\alpha_{4}}=\zeta^{N(P)}U_{\alpha_{P(1)}\alpha_{P(2)}\alpha_{P(3)}\alpha_{P(4)}}$ was used once more. After setting $K_{\gamma}=0$ $\forall\gamma$, we obtain:
\begin{equation}
\scalebox{0.76}{${\displaystyle \overline{\Phi}_{\mathrm{SCPT},N_{\mathrm{SCPT}}=2,\mathfrak{s},\gamma_{1}\gamma_{2}\gamma_{3}\gamma_{4}}^{(4)} = -\frac{1}{2} \left[\left(\left\lbrace\left[ U_{\alpha_{1} \alpha_{2} \alpha_{3} \alpha_{4}} U_{\alpha'_{1} \alpha'_{2} \alpha'_{3} \alpha'_{4}} + \zeta \left(\alpha_{1} \leftrightarrow \alpha'_{1}\right) \right] + \zeta \left(\alpha_{2} \leftrightarrow \alpha'_{2}\right) \right\rbrace + \zeta \left(\alpha_{3} \leftrightarrow \alpha'_{3}\right) \right) + \zeta \left(\alpha_{4} \leftrightarrow \alpha'_{4} \right) \right] \;. }$}
\end{equation}

\end{itemize}

\vspace{0.3cm}

Let us then recall that, according to~\eqref{eq:2PIfrgInitialConditionsGki}, the propagator satisfies:
\begin{equation}
\overline{G}_{\mathfrak{s} = \mathfrak{s}_{\mathrm{i}},\gamma}=0 \mathrlap{\quad \forall \gamma \;,}
\end{equation}
at the starting point of the C-flow. Therefore, the latter expressions of $\overline{\Phi}_{\mathrm{SCPT},N_{\mathrm{SCPT}}=1,\mathfrak{s}}^{(2)}$ and $\overline{\Phi}_{\mathrm{SCPT},N_{\mathrm{SCPT}}=2,\mathfrak{s}}^{(4)}$ can be generalized to all orders of self-consistent PT in this situation, i.e. $\overline{\Phi}_{\mathfrak{s}=\mathfrak{s}_{\mathrm{i}}}^{(2)}=\overline{\Phi}_{\mathrm{SCPT},N_{\mathrm{SCPT}}=1,\mathfrak{s}=\mathfrak{s}_{\mathrm{i}}}^{(2)}$ and $\overline{\Phi}_{\mathfrak{s}=\mathfrak{s}_{\mathrm{i}}}^{(4)}=\overline{\Phi}_{\mathrm{SCPT},N_{\mathrm{SCPT}}=2,\mathfrak{s}=\mathfrak{s}_{\mathrm{i}}}^{(4)}$ for the C-flow. We can thus deduce in this way the initial conditions for 2PI vertices in the framework of the C-flow, i.e. for $\overline{\Phi}_{\mathfrak{s}}^{(2)}$ and $\overline{\Phi}_{\mathfrak{s}}^{(4)}$:
\begin{equation}
\overline{\Phi}_{\mathfrak{s} = \mathfrak{s}_{\mathrm{i}},\gamma_{1}\gamma_{2}}^{(2)} = U_{\gamma_{1}\gamma_{2}} \;,
\label{eq:2PIFRGCflowInCondPhi2Appendix}
\end{equation}
\begin{equation}
\scalebox{0.89}{${\displaystyle \overline{\Phi}_{\mathfrak{s} = \mathfrak{s}_{\mathrm{i}},\gamma_{1}\gamma_{2}\gamma_{3}\gamma_{4}}^{(4)} = -\frac{1}{2} \left[\left(\left\lbrace\left[ U_{\alpha_{1} \alpha_{2} \alpha_{3} \alpha_{4}} U_{\alpha'_{1} \alpha'_{2} \alpha'_{3} \alpha'_{4}} + \zeta \left(\alpha_{1} \leftrightarrow \alpha'_{1}\right) \right] + \zeta \left(\alpha_{2} \leftrightarrow \alpha'_{2}\right) \right\rbrace + \zeta \left(\alpha_{3} \leftrightarrow \alpha'_{3}\right) \right) + \zeta \left(\alpha_{4} \leftrightarrow \alpha'_{4} \right) \right] \;. }$}
\label{eq:2PIFRGCflowInCondPhi4Appendix}
\end{equation}

\subsection{Application to the (0+0)-D $O(N)$-symmetric $\varphi^4$-theory}
\subsubsection{Original representation}
\label{ann:2PIFRGCflowExpressionPhinsi0DONOrigRepr}

We then determine the components of $\overline{\Phi}^{(2)}_{\mathfrak{s}=\mathfrak{s}_{\mathrm{i}}}$ and $\overline{\Phi}^{(4)}_{\mathfrak{s}=\mathfrak{s}_{\mathrm{i}}}$ for the original version of the studied $O(N)$ model. Such results can be directly deduced respectively from~\eqref{eq:2PIFRGCflowInCondPhi2Appendix} and~\eqref{eq:2PIFRGCflowInCondPhi4Appendix} after replacing $\alpha$-indices by color ones (i.e. after replacing $\alpha_{i}$ by $a_{i}$ for all integers $i$). Another possibility is to differentiate the perturbative expression of the Luttinger-Ward functional that can be inferred from~\eqref{eq:2PIEAzerovevfinalexpression}. This perturbative expression is:
\begin{equation}
\Phi_{\mathrm{SCPT}}[G] = \frac{1}{24} \hspace{0.08cm} \begin{gathered}
\begin{fmffile}{DiagramsFRG/2PIEAzerovev_Hartree}
\begin{fmfgraph}(30,20)
\fmfleft{i}
\fmfright{o}
\fmf{phantom,tension=10}{i,i1}
\fmf{phantom,tension=10}{o,o1}
\fmf{plain,left,tension=0.5,foreground=(1,,0,,0)}{i1,v1,i1}
\fmf{plain,right,tension=0.5,foreground=(1,,0,,0)}{o1,v2,o1}
\fmf{zigzag,foreground=(0,,0,,1)}{v1,v2}
\end{fmfgraph}
\end{fmffile}
\end{gathered}
+\frac{1}{12}\begin{gathered}
\begin{fmffile}{DiagramsFRG/2PIEAzerovev_Fock}
\begin{fmfgraph}(15,15)
\fmfleft{i}
\fmfright{o}
\fmf{phantom,tension=11}{i,v1}
\fmf{phantom,tension=11}{v2,o}
\fmf{plain,left,tension=0.4,foreground=(1,,0,,0)}{v1,v2,v1}
\fmf{zigzag,foreground=(0,,0,,1)}{v1,v2}
\end{fmfgraph}
\end{fmffile}
\end{gathered} -\frac{1}{72} \hspace{0.38cm} \begin{gathered}
\begin{fmffile}{DiagramsFRG/2PIEAzerovev_Diag1}
\begin{fmfgraph}(12,12)
\fmfleft{i0,i1}
\fmfright{o0,o1}
\fmftop{v1,vUp,v2}
\fmfbottom{v3,vDown,v4}
\fmf{phantom,tension=20}{i0,v1}
\fmf{phantom,tension=20}{i1,v3}
\fmf{phantom,tension=20}{o0,v2}
\fmf{phantom,tension=20}{o1,v4}
\fmf{plain,left=0.4,tension=0.5,foreground=(1,,0,,0)}{v3,v1}
\fmf{phantom,left=0.1,tension=0.5}{v1,vUp}
\fmf{phantom,left=0.1,tension=0.5}{vUp,v2}
\fmf{plain,left=0.4,tension=0.0,foreground=(1,,0,,0)}{v1,v2}
\fmf{plain,left=0.4,tension=0.5,foreground=(1,,0,,0)}{v2,v4}
\fmf{phantom,left=0.1,tension=0.5}{v4,vDown}
\fmf{phantom,left=0.1,tension=0.5}{vDown,v3}
\fmf{plain,left=0.4,tension=0.0,foreground=(1,,0,,0)}{v4,v3}
\fmf{zigzag,tension=0.5,foreground=(0,,0,,1)}{v1,v4}
\fmf{zigzag,tension=0.5,foreground=(0,,0,,1)}{v2,v3}
\end{fmfgraph}
\end{fmffile}
\end{gathered} \hspace{0.3cm} - \frac{1}{144} \hspace{0.38cm} \begin{gathered}
\begin{fmffile}{DiagramsFRG/2PIEAzerovev_Diag2}
\begin{fmfgraph}(12,12)
\fmfleft{i0,i1}
\fmfright{o0,o1}
\fmftop{v1,vUp,v2}
\fmfbottom{v3,vDown,v4}
\fmf{phantom,tension=20}{i0,v1}
\fmf{phantom,tension=20}{i1,v3}
\fmf{phantom,tension=20}{o0,v2}
\fmf{phantom,tension=20}{o1,v4}
\fmf{plain,left=0.4,tension=0.5,foreground=(1,,0,,0)}{v3,v1}
\fmf{phantom,left=0.1,tension=0.5}{v1,vUp}
\fmf{phantom,left=0.1,tension=0.5}{vUp,v2}
\fmf{zigzag,left=0.4,tension=0.0,foreground=(0,,0,,1)}{v1,v2}
\fmf{plain,left=0.4,tension=0.5,foreground=(1,,0,,0)}{v2,v4}
\fmf{phantom,left=0.1,tension=0.5}{v4,vDown}
\fmf{phantom,left=0.1,tension=0.5}{vDown,v3}
\fmf{zigzag,left=0.4,tension=0.0,foreground=(0,,0,,1)}{v4,v3}
\fmf{plain,left=0.4,tension=0.5,foreground=(1,,0,,0)}{v1,v3}
\fmf{plain,right=0.4,tension=0.5,foreground=(1,,0,,0)}{v2,v4}
\end{fmfgraph}
\end{fmffile}
\end{gathered} \hspace{0.38cm} + \mathcal{O}\big(\lambda^{3}\big) \;,
\end{equation}
and its differentiation leads to:
\begin{equation}
\begin{split}
\overline{\Phi}_{\mathfrak{s}=\mathfrak{s}_{\mathrm{i}},(a_{1},a'_{1})(a_{2},a'_{2})}^{(2)} = & \ \frac{\partial^{2}}{\partial G_{a_{1} a'_{1}}\partial G_{a_{2} a'_{2}}} \left(\rule{0cm}{1.2cm}\right. \frac{1}{24} \hspace{0.08cm} \begin{gathered}
\begin{fmffile}{DiagramsFRG/2PIEAzerovev_Hartree}
\begin{fmfgraph}(30,20)
\fmfleft{i}
\fmfright{o}
\fmf{phantom,tension=10}{i,i1}
\fmf{phantom,tension=10}{o,o1}
\fmf{plain,left,tension=0.5,foreground=(1,,0,,0)}{i1,v1,i1}
\fmf{plain,right,tension=0.5,foreground=(1,,0,,0)}{o1,v2,o1}
\fmf{zigzag,foreground=(0,,0,,1)}{v1,v2}
\end{fmfgraph}
\end{fmffile}
\end{gathered}
+\frac{1}{12}\begin{gathered}
\begin{fmffile}{DiagramsFRG/2PIEAzerovev_Fock}
\begin{fmfgraph}(15,15)
\fmfleft{i}
\fmfright{o}
\fmf{phantom,tension=11}{i,v1}
\fmf{phantom,tension=11}{v2,o}
\fmf{plain,left,tension=0.4,foreground=(1,,0,,0)}{v1,v2,v1}
\fmf{zigzag,foreground=(0,,0,,1)}{v1,v2}
\end{fmfgraph}
\end{fmffile}
\end{gathered} \left.\rule{0cm}{1.2cm}\right) \left.\rule{0cm}{1.2cm}\right|_{G=\overline{G}_{\mathfrak{s}=\mathfrak{s}_{\mathrm{i}}}=0} \\
= & \ \frac{\partial^{2}}{\partial G_{a_{1} a'_{1}}\partial G_{a_{2} a'_{2}}} \left(\rule{0cm}{1.2cm}\right. \frac{\lambda}{24} \left(\sum_{a_{3}=1}^{N} G_{a_{3}a_{3}}\right)^{2} + \frac{\lambda}{12} \sum_{a_{3},a_{4}=1}^{N} G_{a_{3}a_{4}}^{2} \left.\rule{0cm}{1.2cm}\right) \left.\rule{0cm}{1.2cm}\right|_{G=\overline{G}_{\mathfrak{s}=\mathfrak{s}_{\mathrm{i}}}=0} \;,
\end{split}
\end{equation}
\begin{equation}
\begin{split}
\scalebox{0.9}{${\displaystyle\overline{\Phi}_{\mathfrak{s}=\mathfrak{s}_{\mathrm{i}},(a_{1},a'_{1})(a_{2},a'_{2})(a_{3},a'_{3})(a_{4},a'_{4})}^{(4)} =}$} & \scalebox{0.9}{${\displaystyle - \frac{\partial^{4}}{\partial G_{a_{1} a'_{1}}\partial G_{a_{2} a'_{2}}\partial G_{a_{3} a'_{3}}\partial G_{a_{4} a'_{4}}} \left(\rule{0cm}{1.2cm}\right. \frac{1}{72} \hspace{0.38cm} \begin{gathered}
\begin{fmffile}{DiagramsFRG/2PIEAzerovev_Diag1}
\begin{fmfgraph}(12,12)
\fmfleft{i0,i1}
\fmfright{o0,o1}
\fmftop{v1,vUp,v2}
\fmfbottom{v3,vDown,v4}
\fmf{phantom,tension=20}{i0,v1}
\fmf{phantom,tension=20}{i1,v3}
\fmf{phantom,tension=20}{o0,v2}
\fmf{phantom,tension=20}{o1,v4}
\fmf{plain,left=0.4,tension=0.5,foreground=(1,,0,,0)}{v3,v1}
\fmf{phantom,left=0.1,tension=0.5}{v1,vUp}
\fmf{phantom,left=0.1,tension=0.5}{vUp,v2}
\fmf{plain,left=0.4,tension=0.0,foreground=(1,,0,,0)}{v1,v2}
\fmf{plain,left=0.4,tension=0.5,foreground=(1,,0,,0)}{v2,v4}
\fmf{phantom,left=0.1,tension=0.5}{v4,vDown}
\fmf{phantom,left=0.1,tension=0.5}{vDown,v3}
\fmf{plain,left=0.4,tension=0.0,foreground=(1,,0,,0)}{v4,v3}
\fmf{zigzag,tension=0.5,foreground=(0,,0,,1)}{v1,v4}
\fmf{zigzag,tension=0.5,foreground=(0,,0,,1)}{v2,v3}
\end{fmfgraph}
\end{fmffile}
\end{gathered} \hspace{0.3cm} + \frac{1}{144} \hspace{0.38cm} \begin{gathered}
\begin{fmffile}{DiagramsFRG/2PIEAzerovev_Diag2}
\begin{fmfgraph}(12,12)
\fmfleft{i0,i1}
\fmfright{o0,o1}
\fmftop{v1,vUp,v2}
\fmfbottom{v3,vDown,v4}
\fmf{phantom,tension=20}{i0,v1}
\fmf{phantom,tension=20}{i1,v3}
\fmf{phantom,tension=20}{o0,v2}
\fmf{phantom,tension=20}{o1,v4}
\fmf{plain,left=0.4,tension=0.5,foreground=(1,,0,,0)}{v3,v1}
\fmf{phantom,left=0.1,tension=0.5}{v1,vUp}
\fmf{phantom,left=0.1,tension=0.5}{vUp,v2}
\fmf{zigzag,left=0.4,tension=0.0,foreground=(0,,0,,1)}{v1,v2}
\fmf{plain,left=0.4,tension=0.5,foreground=(1,,0,,0)}{v2,v4}
\fmf{phantom,left=0.1,tension=0.5}{v4,vDown}
\fmf{phantom,left=0.1,tension=0.5}{vDown,v3}
\fmf{zigzag,left=0.4,tension=0.0,foreground=(0,,0,,1)}{v4,v3}
\fmf{plain,left=0.4,tension=0.5,foreground=(1,,0,,0)}{v1,v3}
\fmf{plain,right=0.4,tension=0.5,foreground=(1,,0,,0)}{v2,v4}
\end{fmfgraph}
\end{fmffile}
\end{gathered} \hspace{0.35cm} \left.\rule{0cm}{1.2cm}\right) \left.\rule{0cm}{1.2cm}\right|_{G=\overline{G}_{\mathfrak{s}=\mathfrak{s}_{\mathrm{i}}}=0} }$} \\
\scalebox{0.9}{${\displaystyle = }$} & \scalebox{0.9}{${\displaystyle - \frac{\partial^{4}}{\partial G_{a_{1} a'_{1}}\partial G_{a_{2} a'_{2}}\partial G_{a_{3} a'_{3}}\partial G_{a_{4} a'_{4}}} \left(\rule{0cm}{1.2cm}\right. \frac{\lambda^{2}}{72} \sum_{a_{5},a_{6},a_{7},a_{8}=1}^{N} G_{a_{5}a_{6}}G_{a_{6}a_{7}}G_{a_{7}a_{8}}G_{a_{8}a_{5}} }$} \\
& \hspace{5.0cm} \scalebox{0.9}{${\displaystyle + \frac{\lambda^{2}}{144} \left( \sum_{a_{5},a_{6}=1}^{N} G_{a_{5}a_{6}}^{2} \right)^{2} \left.\rule{0cm}{1.2cm}\right) \left.\rule{0cm}{1.2cm}\right|_{G=\overline{G}_{\mathfrak{s}=\mathfrak{s}_{\mathrm{i}}}=0} \;. }$}
\end{split}
\end{equation}
In order to recover results that are consistent with~\eqref{eq:2PIFRGCflowInCondPhi2Appendix} and~\eqref{eq:2PIFRGCflowInCondPhi4Appendix}, the differentiation of the propagator $G$ must be carried out according to the relation:
\begin{equation}
\frac{\partial G_{a_{1}a'_{1}}}{\partial G_{a_{2}a'_{2}}} = \mathcal{I}_{(a_{1},a'_{1})(a_{2},a'_{2})} = \delta_{a_{1}a_{2}}\delta_{a'_{1}a'_{2}} + \delta_{a_{1}a'_{2}}\delta_{a'_{1}a_{2}} \;,
\label{eq:BosonicIdentityMatrix2PIFRG0DON}
\end{equation}
which is the counterpart of definition~\eqref{eq:2PIfrgInverseBosonicIndices} in the framework of the toy model under consideration. This procedure leads to:
\begin{equation}
\overline{\Phi}_{\mathfrak{s} = \mathfrak{s}_{\mathrm{i}},(a_{1},a'_{1})(a_{2},a'_{2})}^{(2)} = \frac{\lambda}{3}\left(\delta_{a_{1}a'_{1}}\delta_{a_{2}a'_{2}}+\delta_{a_{1}a_{2}}\delta_{a'_{1}a'_{2}}+\delta_{a_{1}a'_{2}}\delta_{a'_{1}a_{2}}\right) = U_{(a_{1},a'_{1})(a_{2},a'_{2})} \;,
\label{eq:2PIfrgCflowICPhi20DONAppendix}
\end{equation}
\begin{equation}
\begin{split}
\scalebox{0.89}{${\displaystyle \overline{\Phi}_{\mathfrak{s} = \mathfrak{s}_{\mathrm{i}},(a_{1},a'_{1})(a_{2},a'_{2})(a_{3},a'_{3})(a_{4},a'_{4})}^{(4)} = }$} & \scalebox{0.89}{${\displaystyle -\frac{1}{9} \lambda^2 \Big(\left(\delta_{a_{1} a'_{4}} \delta_{a'_{1} a'_{3}} + \delta_{a_{1} a'_{3}} \delta_{a'_{1} a'_{4}}\right) \delta_{a_{2} a_{4}} \delta_{a'_{2} a_{3}} + \left(\delta_{a_{1} a_{4}} \delta_{a'_{1} a'_{3}} + \delta_{a_{1} a'_{3}} \delta_{a'_{1} a_{4}}\right) \delta_{a_{2} a'_{4}} \delta_{a'_{2} a_{3}} }$}\\
& \hspace{1.0cm} \scalebox{0.89}{${\displaystyle + \left(\delta_{a_{1} a'_{4}} \delta_{a'_{1} a_{3}} + \delta_{a_{1} a_{3}} \delta_{a'_{1} a'_{4}}\right) \delta_{a_{2} a_{4}} \delta_{a'_{2} a'_{3}} + \left(\delta_{a_{1} a_{4}} \delta_{a'_{1} a_{3}} + \delta_{a_{1} a_{3}} \delta_{a'_{1} a_{4}}\right) \delta_{a_{2} a'_{4}} \delta_{a'_{2} a'_{3}} }$} \\
& \hspace{1.0cm} \scalebox{0.89}{${\displaystyle + \left(\delta_{a_{1} a'_{4}} \delta_{a'_{1} a'_{3}} + \delta_{a_{1} a'_{3}} \delta_{a'_{1} a'_{4}}\right) \delta_{a_{2} a_{3}} \delta_{a'_{2} a_{4}} + \left(\delta_{a_{1} a'_{4}} \delta_{a'_{1} a_{3}} + \delta_{a_{1} a_{3}} \delta_{a'_{1} a'_{4}}\right) \delta_{a_{2} a'_{3}} \delta_{a'_{2} a_{4}} }$} \\
& \hspace{1.0cm} \scalebox{0.89}{${\displaystyle + \left(\delta_{a_{1} a_{4}} \delta_{a'_{1} a'_{3}} + \delta_{a_{1} a'_{3}} \delta_{a'_{1} a_{4}}\right) \delta_{a_{2} a_{3}} \delta_{a'_{2} a'_{4}} + \left(\delta_{a_{1} a_{4}} \delta_{a'_{1} a_{3}} + \delta_{a_{1} a_{3}} \delta_{a'_{1} a_{4}}\right) \delta_{a_{2} a'_{3}} \delta_{a'_{2} a'_{4}} }$} \\
& \hspace{1.0cm} \scalebox{0.89}{${\displaystyle + \left(\delta_{a_{1} a'_{4}} \delta_{a'_{1} a'_{2}} + \delta_{a_{1} a'_{2}} \delta_{a'_{1} a'_{4}}\right) \delta_{a_{2} a'_{3}} \delta_{a_{3} a_{4}} + \left(\delta_{a_{1} a'_{3}} \delta_{a'_{1} a'_{2}} + \delta_{a_{1} a'_{2}} \delta_{a'_{1} a'_{3}}\right) \delta_{a_{2} a'_{4}} \delta_{a_{3} a_{4}} }$} \\
& \hspace{1.0cm} \scalebox{0.89}{${\displaystyle + \left(\delta_{a_{1} a'_{4}} \delta_{a'_{1} a_{2}} + \delta_{a_{1} a_{2}} \delta_{a'_{1} a'_{4}}\right) \delta_{a'_{2} a'_{3}} \delta_{a_{3} a_{4}} + \left(\delta_{a_{1} a'_{3}} \delta_{a'_{1} a_{2}} + \delta_{a_{1} a_{2}} \delta_{a'_{1} a'_{3}}\right) \delta_{a'_{2} a'_{4}} \delta_{a_{3} a_{4}} }$} \\
& \hspace{1.0cm} \scalebox{0.89}{${\displaystyle + \left(\delta_{a_{1} a_{4}} \delta_{a'_{1} a'_{2}} + \delta_{a_{1} a'_{2}} \delta_{a'_{1} a_{4}}\right) \delta_{a_{2} a'_{3}} \delta_{a_{3} a'_{4}} + \left(\delta_{a_{1} a'_{3}} \delta_{a'_{1} a'_{2}} + \delta_{a_{1} a'_{2}} \delta_{a'_{1} a'_{3}}\right) \delta_{a_{2} a_{4}} \delta_{a_{3} a'_{4}} }$} \\
& \hspace{1.0cm} \scalebox{0.89}{${\displaystyle + \left(\delta_{a_{1} a_{4}} \delta_{a'_{1} a_{2}} + \delta_{a_{1} a_{2}} \delta_{a'_{1} a_{4}}\right) \delta_{a'_{2} a'_{3}} \delta_{a_{3} a'_{4}} + \left(\delta_{a_{1} a'_{3}} \delta_{a'_{1} a_{2}} + \delta_{a_{1} a_{2}} \delta_{a'_{1} a'_{3}}\right) \delta_{a'_{2} a_{4}} \delta_{a_{3} a'_{4}} }$} \\
& \hspace{1.0cm} \scalebox{0.89}{${\displaystyle + \left(\delta_{a_{1} a'_{4}} \delta_{a'_{1} a'_{2}} + \delta_{a_{1} a'_{2}} \delta_{a'_{1} a'_{4}}\right) \delta_{a_{2} a_{3}} \delta_{a'_{3} a_{4}} + \left(\delta_{a_{1} a_{3}} \delta_{a'_{1} a'_{2}} + \delta_{a_{1} a'_{2}} \delta_{a'_{1} a_{3}}\right) \delta_{a_{2} a'_{4}} \delta_{a'_{3} a_{4}} }$} \\
& \hspace{1.0cm} \scalebox{0.89}{${\displaystyle + \left(\delta_{a_{1} a'_{4}} \delta_{a'_{1} a_{2}} + \delta_{a_{1} a_{2}} \delta_{a'_{1} a'_{4}}\right) \delta_{a'_{2} a_{3}} \delta_{a'_{3} a_{4}} + \left(\delta_{a_{1} a_{3}} \delta_{a'_{1} a_{2}} + \delta_{a_{1} a_{2}} \delta_{a'_{1} a_{3}}\right) \delta_{a'_{2} a'_{4}} \delta_{a'_{3} a_{4}} }$} \\
& \hspace{1.0cm} \scalebox{0.89}{${\displaystyle + \left(\delta_{a_{1} a_{4}} \delta_{a'_{1} a'_{2}} + \delta_{a_{1} a'_{2}} \delta_{a'_{1} a_{4}}\right) \delta_{a_{2} a_{3}} \delta_{a'_{3} a'_{4}} + \left(\delta_{a_{1} a_{3}} \delta_{a'_{1} a'_{2}} + \delta_{a_{1} a'_{2}} \delta_{a'_{1} a_{3}}\right) \delta_{a_{2} a_{4}} \delta_{a'_{3} a'_{4}} }$} \\
& \hspace{1.0cm} \scalebox{0.89}{${\displaystyle + \left(\delta_{a_{1} a_{4}} \delta_{a'_{1} a_{2}} + \delta_{a_{1} a_{2}} \delta_{a'_{1} a_{4}}\right) \delta_{a'_{2} a_{3}} \delta_{a'_{3} a'_{4}} + \left(\delta_{a_{1} a_{3}} \delta_{a'_{1} a_{2}} + \delta_{a_{1} a_{2}} \delta_{a'_{1} a_{3}}\right) \delta_{a'_{2} a_{4}} \delta_{a'_{3} a'_{4}}\Big) }$} \\
& \scalebox{0.89}{${\displaystyle -\frac{2}{9} \lambda^2 \Big(\left(\delta_{a_{1} a'_{4}} \delta_{a'_{1} a_{4}} + \delta_{a_{1} a_{4}} \delta_{a'_{1} a'_{4}}\right) \left(\delta_{a_{2} a'_{3}} \delta_{a'_{2} a_{3}} + \delta_{a_{2} a_{3}} \delta_{a'_{2} a'_{3}}\right) }$} \\
& \hspace{1.0cm} \scalebox{0.89}{${\displaystyle + \left(\delta_{a_{1} a'_{3}} \delta_{a'_{1} a_{3}} + \delta_{a_{1} a_{3}} \delta_{a'_{1} a'_{3}}\right) \left(\delta_{a_{2} a'_{4}} \delta_{a'_{2} a_{4}} + \delta_{a_{2} a_{4}} \delta_{a'_{2} a'_{4}}\right) }$} \\
& \hspace{1.0cm} \scalebox{0.89}{${\displaystyle + \left(\delta_{a_{1} a'_{2}} \delta_{a'_{1} a_{2}} + \delta_{a_{1} a_{2}} \delta_{a'_{1} a'_{2}}\right) \left(\delta_{a_{3} a'_{4}} \delta_{a'_{3} a_{4}} + \delta_{a_{3} a_{4}} \delta_{a'_{3} a'_{4}}\right)\Big) \;,}$}
\end{split}
\label{eq:2PIfrgCflowICPhi40DONAppendix}
\end{equation}
where $U_{(a_{1},a'_{1})(a_{2},a'_{2})}$ was introduced in~\eqref{eq:2PIfrgCflowICPhi20DONAppendix} with the help of~\eqref{eq:2bodyinteraction2PIFRG0DON}. We finally deduce from~\eqref{eq:2PIfrgCflowICPhi20DONAppendix} and~\eqref{eq:2PIfrgCflowICPhi40DONAppendix} the initial conditions for the following components of 2PI vertices involved in the equation system made of~\eqref{eq:2PIfrgFlowEquationsCflowG0DONN2} to~\eqref{eq:2PIfrgFlowEquationsCflowPhi3s1111220DONN2}:
\begin{equation}
\overline{\Phi}_{\mathfrak{s} = \mathfrak{s}_{\mathrm{i}},(1,1)(1,1)}^{(2)} = \lambda \;,
\end{equation}
\begin{equation}
\overline{\Phi}_{\mathfrak{s} = \mathfrak{s}_{\mathrm{i}},(1,1)(2,2)}^{(2)} = \frac{\lambda}{3} \;,
\end{equation}
\begin{equation}
\overline{\Phi}_{\mathfrak{s} = \mathfrak{s}_{\mathrm{i}},(1,1)(1,1)(1,1)(1,1)}^{(4)} = -8\lambda^{2} \;,
\end{equation}
\begin{equation}
\overline{\Phi}_{\mathfrak{s} = \mathfrak{s}_{\mathrm{i}},(1,1)(1,1)(1,1)(2,2)}^{(4)} = 0 \;,
\end{equation}
\begin{equation}
\overline{\Phi}_{\mathfrak{s} = \mathfrak{s}_{\mathrm{i}},(1,1)(1,1)(2,2)(2,2)}^{(4)} = -\frac{8\lambda^{2}}{9} \;.
\end{equation}

\subsubsection{Mixed representation}
\label{ann:2PIFRGCflowExpressionPhinsi0DONMixedRepr}

We derive in the rest of section~\ref{ann:2PIfrgCflowInitialConditions} the expressions of the components of $\overline{\Phi}_{\mathrm{mix},\mathfrak{s} = \mathfrak{s}_{\mathrm{i}}}^{(2G,1D)}$ which are required to implement the C-flow in the framework of the mixed representation of the studied $O(N)$ model. To that end, we recall definitions~\eqref{eq:Cflow2PIFRGDefPhimixSCPThbarExp0DON} and~\eqref{eq:Cflow2PIFRGDefPhimixSCPTlambdaExp0DON}:
\begin{equation}
\Phi_{\mathrm{mix},\mathrm{SCPT},\hbar\text{-exp}}(G,D) = \frac{\hbar^{2}}{12}\begin{gathered}
\begin{fmffile}{DiagramsFRG/2PIFRGmixedCflow_PhiSCPT_Fock}
\begin{fmfgraph}(15,15)
\fmfleft{i}
\fmfright{o}
\fmfv{decor.shape=circle,decor.size=2.0thick,foreground=(0,,0,,1)}{v1}
\fmfv{decor.shape=circle,decor.size=2.0thick,foreground=(0,,0,,1)}{v2}
\fmf{phantom,tension=11}{i,v1}
\fmf{phantom,tension=11}{v2,o}
\fmf{plain,left,tension=0.4,foreground=(1,,0,,0)}{v1,v2,v1}
\fmf{wiggly,foreground=(1,,0,,0)}{v1,v2}
\end{fmfgraph}
\end{fmffile}
\end{gathered} + \mathcal{O}\big(\hbar^3\big) \;,
\label{eq:Cflow2PIFRGDefPhimixSCPThbarExp0DONAppendix}
\end{equation}
\begin{equation}
\Phi_{\mathrm{mix},\mathrm{SCPT},\lambda\text{-exp}}(G,D) = \frac{1}{24}\begin{gathered}
\begin{fmffile}{DiagramsFRG/2PIFRGmixedCflow_PhiSCPT_Hartree}
\begin{fmfgraph}(30,20)
\fmfleft{i}
\fmfright{o}
\fmfv{decor.shape=circle,decor.size=2.0thick,foreground=(0,,0,,1)}{v1}
\fmfv{decor.shape=circle,decor.size=2.0thick,foreground=(0,,0,,1)}{v2}
\fmf{phantom,tension=10}{i,i1}
\fmf{phantom,tension=10}{o,o1}
\fmf{plain,left,tension=0.5,foreground=(1,,0,,0)}{i1,v1,i1}
\fmf{plain,right,tension=0.5,foreground=(1,,0,,0)}{o1,v2,o1}
\fmf{wiggly,foreground=(1,,0,,0)}{v1,v2}
\end{fmfgraph}
\end{fmffile}
\end{gathered}
+\frac{1}{12}\begin{gathered}
\begin{fmffile}{DiagramsFRG/2PIFRGmixedCflow_PhiSCPT_Fock}
\begin{fmfgraph}(15,15)
\fmfleft{i}
\fmfright{o}
\fmfv{decor.shape=circle,decor.size=2.0thick,foreground=(0,,0,,1)}{v1}
\fmfv{decor.shape=circle,decor.size=2.0thick,foreground=(0,,0,,1)}{v2}
\fmf{phantom,tension=11}{i,v1}
\fmf{phantom,tension=11}{v2,o}
\fmf{plain,left,tension=0.4,foreground=(1,,0,,0)}{v1,v2,v1}
\fmf{wiggly,foreground=(1,,0,,0)}{v1,v2}
\end{fmfgraph}
\end{fmffile}
\end{gathered} + \mathcal{O}\big(\lambda^2\big) \;.
\label{eq:Cflow2PIFRGDefPhimixSCPTlambdaExp0DONAppendix}
\end{equation}
We then use once again identity~\eqref{eq:BosonicIdentityMatrix2PIFRG0DON} along with $\frac{\partial D}{\partial D} = \mathcal{I} = 2$ in order to calculate:
\begin{itemize}
\item From the $\hbar$-expansion with $\hbar=1$ (i.e. from~\eqref{eq:Cflow2PIFRGDefPhimixSCPThbarExp0DONAppendix}):
\begin{equation}
\begin{split}
\overline{\Phi}_{\mathrm{mix},\mathfrak{s} = \mathfrak{s}_{\mathrm{i}},(a_{1},a'_{1})(a_{2},a'_{2})}^{(2G,1D)} = & \ \frac{\partial^{3}}{\partial G_{a_{1} a'_{1}} \partial G_{a_{2} a'_{2}} \partial D} \left(\rule{0cm}{1.0cm}\right. \frac{1}{12}\begin{gathered}
\begin{fmffile}{DiagramsFRG/2PIFRGmixedCflow_PhiSCPT_Fock}
\begin{fmfgraph}(15,15)
\fmfleft{i}
\fmfright{o}
\fmfv{decor.shape=circle,decor.size=2.0thick,foreground=(0,,0,,1)}{v1}
\fmfv{decor.shape=circle,decor.size=2.0thick,foreground=(0,,0,,1)}{v2}
\fmf{phantom,tension=11}{i,v1}
\fmf{phantom,tension=11}{v2,o}
\fmf{plain,left,tension=0.4,foreground=(1,,0,,0)}{v1,v2,v1}
\fmf{wiggly,foreground=(1,,0,,0)}{v1,v2}
\end{fmfgraph}
\end{fmffile}
\end{gathered} \left.\rule{0cm}{1.0cm}\right) \left.\rule{0cm}{1.0cm}\right|_{G=\overline{G}_{\mathfrak{s}=\mathfrak{s}_{\mathrm{i}}}=0 \atop D=\overline{D}_{\mathfrak{s}=\mathfrak{s}_{\mathrm{i}}}=0} \\
= & \ \frac{\partial^{3}}{\partial G_{a_{1} a'_{1}} \partial G_{a_{2} a'_{2}} \partial D} \left.\left( \frac{1}{12} \lambda D \sum_{a_{3}, a_{4}= 1}^{N} G_{a_{3} a_{4}}^{2} \right) \right|_{G=\overline{G}_{\mathfrak{s}=\mathfrak{s}_{\mathrm{i}}}=0 \atop D=\overline{D}_{\mathfrak{s}=\mathfrak{s}_{\mathrm{i}}}=0} \\
= & \ \frac{2\lambda}{3}\left(\delta_{a_{1}a_{2}}\delta_{a'_{1}a'_{2}}+\delta_{a_{1}a'_{2}}\delta_{a'_{1}a_{2}}\right) \;.
\end{split}
\end{equation}

\item From the $\lambda$-expansion (i.e. from~\eqref{eq:Cflow2PIFRGDefPhimixSCPTlambdaExp0DONAppendix}):
\begin{equation}
\begin{split}
\scalebox{0.93}{${\displaystyle \overline{\Phi}_{\mathrm{mix},\mathfrak{s} = \mathfrak{s}_{\mathrm{i}},(a_{1},a'_{1})(a_{2},a'_{2})}^{(2G,1D)} = }$} & \ \scalebox{0.93}{${\displaystyle \frac{\partial^{3}}{\partial G_{a_{1} a'_{1}} \partial G_{a_{2} a'_{2}} \partial D} \left(\rule{0cm}{1.0cm}\right. \frac{1}{24}\begin{gathered}
\begin{fmffile}{DiagramsFRG/2PIFRGmixedCflow_PhiSCPT_Hartree}
\begin{fmfgraph}(30,20)
\fmfleft{i}
\fmfright{o}
\fmfv{decor.shape=circle,decor.size=2.0thick,foreground=(0,,0,,1)}{v1}
\fmfv{decor.shape=circle,decor.size=2.0thick,foreground=(0,,0,,1)}{v2}
\fmf{phantom,tension=10}{i,i1}
\fmf{phantom,tension=10}{o,o1}
\fmf{plain,left,tension=0.5,foreground=(1,,0,,0)}{i1,v1,i1}
\fmf{plain,right,tension=0.5,foreground=(1,,0,,0)}{o1,v2,o1}
\fmf{wiggly,foreground=(1,,0,,0)}{v1,v2}
\end{fmfgraph}
\end{fmffile}
\end{gathered}
+\frac{1}{12}\begin{gathered}
\begin{fmffile}{DiagramsFRG/2PIFRGmixedCflow_PhiSCPT_Fock}
\begin{fmfgraph}(15,15)
\fmfleft{i}
\fmfright{o}
\fmfv{decor.shape=circle,decor.size=2.0thick,foreground=(0,,0,,1)}{v1}
\fmfv{decor.shape=circle,decor.size=2.0thick,foreground=(0,,0,,1)}{v2}
\fmf{phantom,tension=11}{i,v1}
\fmf{phantom,tension=11}{v2,o}
\fmf{plain,left,tension=0.4,foreground=(1,,0,,0)}{v1,v2,v1}
\fmf{wiggly,foreground=(1,,0,,0)}{v1,v2}
\end{fmfgraph}
\end{fmffile}
\end{gathered} \left.\rule{0cm}{1.0cm}\right) \left.\rule{0cm}{1.0cm}\right|_{G=\overline{G}_{\mathfrak{s}=\mathfrak{s}_{\mathrm{i}}}=0 \atop D=\overline{D}_{\mathfrak{s}=\mathfrak{s}_{\mathrm{i}}}=0} }$} \\
\scalebox{0.93}{${\displaystyle = }$} & \ \scalebox{0.93}{${\displaystyle \frac{\partial^{3}}{\partial G_{a_{1} a'_{1}} \partial G_{a_{2} a'_{2}} \partial D} \left.\left( \frac{1}{24} \lambda D \left(\sum_{a_{3}=1}^{N} G_{a_{3}a_{3}} \right)^{2} + \frac{1}{12} \lambda D \sum_{a_{3}, a_{4} = 1}^{N} G_{a_{3} a_{4}}^{2} \right) \right|_{G=\overline{G}_{\mathfrak{s}=\mathfrak{s}_{\mathrm{i}}}=0 \atop D=\overline{D}_{\mathfrak{s}=\mathfrak{s}_{\mathrm{i}}}=0} }$} \\
= & \ \frac{2\lambda}{3}\left(\delta_{a_{1}a'_{1}}\delta_{a_{2}a'_{2}}+\delta_{a_{1}a_{2}}\delta_{a'_{1}a'_{2}}+\delta_{a_{1}a'_{2}}\delta_{a'_{1}a_{2}}\right) \;.
\end{split}
\end{equation}

\end{itemize}

\vspace{0.3cm}

The components $\overline{\Phi}_{\mathrm{mix},\mathfrak{s},(1,1)(1,1)}^{(2G,1D)}$ and $\overline{\Phi}_{\mathrm{mix},\mathfrak{s},(1,1)(2,2)}^{(2G,1D)}$ involved in the tower of differential equations made of~\eqref{eq:2PIfrgFlowEquationsmixedCflowG0DONN2} to~\eqref{eq:2PIfrgFlowEquationsmixedCflowPhi1G1Ds110DONN2} must therefore satisfy:
\begin{equation}
\overline{\Phi}_{\mathrm{mix},\mathfrak{s} = \mathfrak{s}_{\mathrm{i}},(1,1)(1,1)}^{(2G,1D)} = \left\{
\begin{array}{lll}
        \displaystyle{\frac{4\lambda}{3} \quad \text{from the $\hbar$-expansion} \;,} \\
        \\
        \displaystyle{2\lambda \quad \text{from the $\lambda$-expansion} \;,}
    \end{array}
\right.
\end{equation}
\begin{equation}
\overline{\Phi}_{\mathrm{mix},\mathfrak{s} = \mathfrak{s}_{\mathrm{i}},(1,1)(2,2)}^{(2G,1D)} = \left\{
\begin{array}{lll}
        \displaystyle{0 \quad \text{from the $\hbar$-expansion} \;.} \\
        \\
        \displaystyle{\frac{2\lambda}{3} \quad \text{from the $\lambda$-expansion} \;.}
    \end{array}
\right.
\end{equation}

\section{(0+0)-D limit of the bosonic index formalism at $N=1$}
\label{ann:ZeroDimLimitBosonicIndexFormalism}

As pointed out below~\eqref{eq:OneEqualTwoN12PIFRG0DON}, there are essentially two different manners to derive the 2PI-FRG flow equations for the studied (0+0)-D $O(N)$ model with $N=1$: either using standard derivation rules or by taking the (0+0)-D limit of their more general versions involving summations over bosonic indices. In the latter case, the identity reduces to $2$ according to:
\begin{equation}
\frac{\partial G}{\partial G} = \mathcal{I} = 2 \;,
\label{eq:OneEqualTwoN12PIFRG0DONAppendixV1}
\end{equation}
as a result of the definition of the bosonic identity matrix (given by~\eqref{eq:2PIfrgInverseBosonicIndices}) and the chain rule~\eqref{eq:2PIfrgchainRuleBosonicIndices} still introduces a $1/2$ factor:
\begin{equation}
\frac{\partial W(K)}{\partial G} = \frac{1}{2} \frac{\partial K}{\partial G} \frac{\partial W(K)}{\partial K}\;,
\label{eq:2PIfrgchainRuleBosonicIndicesAppendixV1}
\end{equation}
whereas, in the former case, standard derivation rules simply give us:
\begin{equation}
\frac{\partial G}{\partial G} = 1 \;,
\label{eq:OneEqualTwoN12PIFRG0DONAppendixV2}
\end{equation}
and
\begin{equation}
\frac{\partial W(K)}{\partial G} = \frac{\partial K}{\partial G} \frac{\partial W(K)}{\partial K}\;.
\label{eq:2PIfrgchainRuleBosonicIndicesAppendixV2}
\end{equation}
For the sake of clarity, we will compare in this appendix the equation systems determined from these two procedures for the tC-flow at $N_{\mathrm{max}}=2$. These equation systems are obtained by following the recipe outlined in section~\ref{sec:2PIFRGstateofplay} for the C-flow together with that of appendix~\ref{ann:2PIfrgFlowEquationCflow}, either by exploiting~\eqref{eq:OneEqualTwoN12PIFRG0DONAppendixV1} and~\eqref{eq:2PIfrgchainRuleBosonicIndicesAppendixV1} or~\eqref{eq:OneEqualTwoN12PIFRG0DONAppendixV2} and~\eqref{eq:2PIfrgchainRuleBosonicIndicesAppendixV2} as derivation rules. This leads to:
\begin{itemize}
\item From the (0+0)-D limit of the bosonic index formalism at $N=1$:
\begin{itemize}
\item Flow equations:
\begin{equation}
\dot{\overline{G}}_{\mathfrak{s}} =- \overline{G}_{\mathfrak{s}}^{2} \left(\dot{C}_{\mathfrak{s}}^{-1}-\dot{\overline{\Sigma}}_{\mathfrak{s}}\right) \;,
\label{eq:2PIfrgFlowEquationsCflowG0DONN1AppendixV1}
\end{equation}
\begin{equation}
\Delta \dot{\overline{\Omega}}_{\mathfrak{s}} = \frac{1}{2} \dot{C}_{\mathfrak{s}}^{-1} \left(\overline{G}_{\mathfrak{s}}-C_{\mathfrak{s}}\right)\;,
\label{eq:2PIfrgFlowEquationsCflowDOmega0DONN1AppendixV1}
\end{equation}
\begin{equation}
\dot{\overline{\Sigma}}_{\mathfrak{s}} = - \textcolor{red}{\frac{1}{2}} \dot{\overline{G}}_{\mathfrak{s}} \overline{\Phi}_{\mathfrak{s}}^{(2)}\;,
\label{eq:2PIfrgFlowEquationsCflowSigma0DONN1AppendixV1}
\end{equation}
\begin{equation}
\dot{\overline{\Phi}}_{\mathfrak{s}}^{(2)} = \textcolor{red}{\frac{1}{2}} \dot{\overline{G}}_{\mathfrak{s}} \overline{\Phi}_{\mathfrak{s}}^{(3)} \;.
\label{eq:2PIfrgFlowEquationsCflowPhi20DONN1AppendixV1}
\end{equation}

\item Initial conditions:
\begin{equation}
\overline{G}_{\mathfrak{s}=\mathfrak{s}_{\mathrm{i}}}=0 \;,
\label{eq:2PIfrgInitialConditionsGki0DONAppendixV1}
\end{equation}
\begin{equation}
\Delta \overline{\Omega}_{\mathfrak{s}=\mathfrak{s}_{\mathrm{i}}} = 0 \;,
\label{eq:2PIfrgCflowICDeltaOmega0DONAppendixV1}
\end{equation}
\begin{equation}
\overline{\Sigma}_{\mathfrak{s}=\mathfrak{s}_{\mathrm{i}}} = 0 \;,
\label{eq:2PIfrgCflowICSigma0DONAppendixV1}
\end{equation}
\begin{equation}
\overline{\Phi}_{\mathfrak{s} = \mathfrak{s}_{\mathrm{i}}}^{(2)} = \lambda \;,
\label{eq:2PIfrgCflowICPhi20DONAppendixV1}
\end{equation}
\begin{equation}
\overline{\Phi}_{\mathfrak{s} = \mathfrak{s}_{\mathrm{i}}}^{(3)} = 0 \;.
\label{eq:2PIfrgCflowICPhiodd0DONAppendixV1}
\end{equation}

\item Truncation:
\begin{equation}
\overline{\Phi}_{\mathfrak{s}}^{(3)} = \overline{\Phi}_{\mathfrak{s}=\mathfrak{s}_{\mathrm{i}}}^{(3)} \mathrlap{\quad \forall \mathfrak{s} \;.}
\label{eq:2PIfrgtruncationtCflowPhi30DONAppendixV1}
\end{equation}

\end{itemize}

\item From standard derivation rules:
\begin{itemize}
\item Flow equations:
\begin{equation}
\dot{\overline{G}}_{\mathfrak{s}} =- \overline{G}_{\mathfrak{s}}^{2} \left(\dot{C}_{\mathfrak{s}}^{-1}-\dot{\overline{\Sigma}}_{\mathfrak{s}}\right) \;,
\label{eq:2PIfrgFlowEquationsCflowG0DONN1AppendixV2}
\end{equation}
\begin{equation}
\Delta \dot{\overline{\Omega}}_{\mathfrak{s}} = \frac{1}{2} \dot{C}_{\mathfrak{s}}^{-1} \left(\overline{G}_{\mathfrak{s}}-C_{\mathfrak{s}}\right)\;,
\label{eq:2PIfrgFlowEquationsCflowDOmega0DONN1AppendixV2}
\end{equation}
\begin{equation}
\dot{\overline{\Sigma}}_{\mathfrak{s}} = - \textcolor{red}{2} \dot{\overline{G}}_{\mathfrak{s}} \overline{\Phi}_{\mathfrak{s}}^{(2)}\;,
\label{eq:2PIfrgFlowEquationsCflowSigma0DONN1AppendixV2}
\end{equation}
\begin{equation}
\dot{\overline{\Phi}}_{\mathfrak{s}}^{(2)} = \dot{\overline{G}}_{\mathfrak{s}} \overline{\Phi}_{\mathfrak{s}}^{(3)} \;.
\label{eq:2PIfrgFlowEquationsCflowPhi20DONN1AppendixV2}
\end{equation}

\item Initial conditions:
\begin{equation}
\overline{G}_{\mathfrak{s}=\mathfrak{s}_{\mathrm{i}}}=0 \;,
\label{eq:2PIfrgInitialConditionsGki0DONAppendixV2}
\end{equation}
\begin{equation}
\Delta \overline{\Omega}_{\mathfrak{s}=\mathfrak{s}_{\mathrm{i}}} = 0 \;,
\label{eq:2PIfrgCflowICDeltaOmega0DONAppendixV2}
\end{equation}
\begin{equation}
\overline{\Sigma}_{\mathfrak{s}=\mathfrak{s}_{\mathrm{i}}} = 0 \;,
\label{eq:2PIfrgCflowICSigma0DONAppendixV2}
\end{equation}
\begin{equation}
\overline{\Phi}_{\mathfrak{s} = \mathfrak{s}_{\mathrm{i}}}^{(2)} = \textcolor{red}{\frac{1}{4}} \lambda \;,
\label{eq:2PIfrgCflowICPhi20DONAppendixV2}
\end{equation}
\begin{equation}
\overline{\Phi}_{\mathfrak{s} = \mathfrak{s}_{\mathrm{i}}}^{(3)} = 0 \;.
\label{eq:2PIfrgCflowICPhiodd0DONAppendixV2}
\end{equation}

\item Truncation:
\begin{equation}
\overline{\Phi}_{\mathfrak{s}}^{(3)} = \overline{\Phi}_{\mathfrak{s}=\mathfrak{s}_{\mathrm{i}}}^{(3)} \mathrlap{\quad \forall \mathfrak{s} \;.}
\label{eq:2PIfrgtruncationtCflowPhi30DONAppendixV2}
\end{equation}

\end{itemize}

\end{itemize}

\vspace{0.3cm}

The equation system made of~\eqref{eq:2PIfrgFlowEquationsCflowG0DONN1AppendixV1} to~\eqref{eq:2PIfrgFlowEquationsCflowPhi20DONN1AppendixV1} corresponds to the differential equations~\eqref{eq:2PIfrgFlowEquationsCflowG0DONN1} to~\eqref{eq:2PIfrgFlowEquationsCflowPhin0DONN1}. Moreover, the initial conditions~\eqref{eq:2PIfrgInitialConditionsGki0DONAppendixV1} to~\eqref{eq:2PIfrgCflowICPhiodd0DONAppendixV1} are identical to~\eqref{eq:2PIfrgCflowICGki0DON} to~\eqref{eq:2PIfrgCflowICPhiodd0DON} at $N=1$.

\vspace{0.5cm}

The expressions of $\overline{\Phi}_{\mathfrak{s}=\mathfrak{s}_{\mathrm{i}}}^{(2)}$ given by~\eqref{eq:2PIfrgCflowICPhi20DONAppendixV1} and~\eqref{eq:2PIfrgCflowICPhi20DONAppendixV2} are both obtained from~\eqref{eq:PertExpressionPhiN12PIFRGtCflow0DON} recalled below:
\begin{equation}
\Phi_{\mathrm{SCPT}}(G) = \frac{1}{8} \lambda G^{2} + \mathcal{O}\big(\lambda^{2}\big) \;.
\label{eq:PertExpressionPhiN12PIFRGtCflow0DONAppendix}
\end{equation}
To obtain~\eqref{eq:2PIfrgCflowICPhi20DONAppendixV1}, we have differentiated~\eqref{eq:PertExpressionPhiN12PIFRGtCflow0DONAppendix} as follows:
\begin{equation}
\begin{split}
\overline{\Phi}_{\mathfrak{s}=\mathfrak{s}_{\mathrm{i}}}^{(2)} = & \ \frac{\partial^{2}}{\partial G^{2}} \left.\left( \frac{1}{8} \lambda G^{2} \right)\right|_{G=\overline{G}_{\mathfrak{s}=\mathfrak{s}_{\mathrm{i}}}=0} \\
= & \ \frac{\partial}{\partial G} \bigg.\bigg( \frac{1}{4} \lambda G \underbrace{\frac{\partial G}{\partial G}}_{2} \bigg)\bigg|_{G=\overline{G}_{\mathfrak{s}=\mathfrak{s}_{\mathrm{i}}}=0} \\
= & \ \bigg.\bigg( \frac{1}{2} \lambda \underbrace{\frac{\partial G}{\partial G}}_{2} \bigg)\bigg|_{G=\overline{G}_{\mathfrak{s}=\mathfrak{s}_{\mathrm{i}}}=0} \\
= & \ \lambda \;,
\end{split}
\end{equation}
whereas, in the derivation of~\eqref{eq:2PIfrgCflowICPhi20DONAppendixV2}, we have calculated:
\begin{equation}
\begin{split}
\overline{\Phi}_{\mathfrak{s}=\mathfrak{s}_{\mathrm{i}}}^{(2)} = & \ \frac{\partial^{2}}{\partial G^{2}} \left.\left( \frac{1}{8} \lambda G^{2} \right)\right|_{G=\overline{G}_{\mathfrak{s}=\mathfrak{s}_{\mathrm{i}}}=0} \\
= & \ \frac{\partial}{\partial G} \bigg.\bigg( \frac{1}{4} \lambda G \underbrace{\frac{\partial G}{\partial G}}_{1} \bigg)\bigg|_{G=\overline{G}_{\mathfrak{s}=\mathfrak{s}_{\mathrm{i}}}=0} \\
= & \ \bigg.\bigg( \frac{1}{4} \lambda \underbrace{\frac{\partial G}{\partial G}}_{1} \bigg)\bigg|_{G=\overline{G}_{\mathfrak{s}=\mathfrak{s}_{\mathrm{i}}}=0} \\
= & \ \frac{1}{4} \lambda \;.
\end{split}
\end{equation}
Furthermore, the difference between~\eqref{eq:2PIfrgFlowEquationsCflowPhi20DONN1AppendixV1} and~\eqref{eq:2PIfrgFlowEquationsCflowPhi20DONN1AppendixV2} is due to the presence of the $1/2$ factor in the functional chain rule of the bosonic index formalism, as explained with~\eqref{eq:2PIfrgchainRuleBosonicIndicesAppendixV1} and~\eqref{eq:2PIfrgchainRuleBosonicIndicesAppendixV2}. For this reason, one might expect a factor $1$ instead of $\textcolor{red}{2}$ in~\eqref{eq:2PIfrgFlowEquationsCflowSigma0DONN1AppendixV2} but there is an additional factor $2$ entering the arena in the definition of the self-energy:
\begin{equation}
\Sigma_{\mathfrak{s}}(G) = -2 \Phi_{\mathfrak{s}}^{(1)}(G) \;,
\label{eq:DefSelfEnergy2PIFRGN1StandardDerivAppendix}
\end{equation}
whereas we have:
\begin{equation}
\Sigma_{\mathfrak{s}}(G) = -\Phi_{\mathfrak{s}}^{(1)}(G) \;,
\label{eq:DefSelfEnergy2PIFRGN1ZeroDimLimitBosIndexAppendix}
\end{equation}
in the (0+0)-D limit of the bosonic index formalism at $N=1$. In order to justify~\eqref{eq:DefSelfEnergy2PIFRGN1StandardDerivAppendix}, we briefly go through some derivations of appendices~\ref{ann:BosonicIndices2PI} and~\ref{ann:DysonEq} with standard derivation rules, starting with:
\begin{equation}
Z(K) = e^{W(K)} = \int_{-\infty}^{\infty} d\widetilde{\varphi} \ e^{-S(\widetilde{\varphi}) + \frac{1}{2} K \widetilde{\varphi}^{2}} \;,
\label{eq:ZkN1standardDerivAppendixV2}
\end{equation}
from which we infer:
\begin{equation}
\begin{split}
\frac{\partial W(K)}{\partial K} = & \ \frac{1}{2} \frac{1}{Z(K)} \int_{-\infty}^{\infty} d\widetilde{\varphi} \ \widetilde{\varphi}^{2} \ e^{-S[\widetilde{\varphi}] + \frac{1}{2} K \widetilde{\varphi}^{2}} \\
= & \ \frac{1}{2} \left\langle \widetilde{\varphi}^{2} \right\rangle_{K} \\
= & \ \frac{1}{2} G \;,
\end{split}
\label{eq:GN1standardDerivAppendixV2}
\end{equation}
and the corresponding 2PI EA $\Gamma^{(\mathrm{2PI})}(G)$ satisfies:
\begin{equation}
\begin{split}
\Gamma^{(\mathrm{2PI})}(G) = & -W(K) + K \frac{\partial W(K)}{\partial K} \\
= & -W(K) + \frac{1}{2} K G \;.
\end{split}
\label{eq:Gamma2PIN1standardDerivAppendixV2}
\end{equation}
The free version of~\eqref{eq:ZkN1standardDerivAppendixV2} is:
\begin{equation}
\begin{split}
Z_{0}(K) = e^{W_{0}(K)} = & \ \int_{-\infty}^{\infty} d\widetilde{\varphi} \ e^{-S_{0}(\widetilde{\varphi}) + \frac{1}{2} K \widetilde{\varphi}^{2}} \\
= & \ \int_{-\infty}^{\infty} d\widetilde{\varphi} \ e^{-\frac{1}{2}(C^{-1} - K) \widetilde{\varphi}^{2}} \\
= & \ \sqrt{\frac{2\pi}{C^{-1}-K}} \;,
\end{split}
\label{eq:Z0kN1standardDerivAppendixV2}
\end{equation}
which leads to:
\begin{equation}
W_{0}(K) = \frac{1}{2}\left(\ln(2\pi) - \ln\big(C^{-1}-K\big)\right) \;.
\label{eq:W0kN1standardDerivAppendixV2}
\end{equation}
According to~\eqref{eq:GN1standardDerivAppendixV2} and~\eqref{eq:W0kN1standardDerivAppendixV2}, we have:
\begin{equation}
G_{0} = 2 \frac{\partial W_{0}(K)}{\partial K} = \frac{1}{C^{-1}-K} \;,
\end{equation}
or, equivalently,
\begin{equation}
K = C^{-1} - G_{0}^{-1} \;.
\label{eq:KN1standardDerivAppendixV2}
\end{equation}
We then deduce from~\eqref{eq:W0kN1standardDerivAppendixV2} to~\eqref{eq:KN1standardDerivAppendixV2} an expression for the free part $\Gamma_{0}^{(\mathrm{2PI})}(G)$ of $\Gamma^{(\mathrm{2PI})}(G)$ by calculating:
\begin{equation}
\Gamma_{0}^{(\mathrm{2PI})}(G_{0}) = -W_{0}[K] + \frac{1}{2} K G_{0} = -\frac{1}{2}\ln(2 \pi G_{0}) + \frac{1}{2}\left( C^{-1}G_{0} - 1 \right) \;,
\end{equation}
which implies that $\Gamma_{0}^{(\mathrm{2PI})}(G)$ reads:
\begin{equation}
\Gamma_{0}^{(\mathrm{2PI})}(G) = -\frac{1}{2}\ln(2 \pi G) + \frac{1}{2}\left( C^{-1}G - 1 \right) \;.
\label{eq:DefinitionGamma02PIFRGAppendixV2}
\end{equation}
Then, we define the Luttinger-Ward functional from~\eqref{eq:DefinitionGamma02PIFRGAppendixV2} as:
\begin{equation}
\Phi(G) \equiv \Gamma^{(\mathrm{2PI})}(G) - \Gamma_{0}^{(\mathrm{2PI})}(G) = \Gamma^{(\mathrm{2PI})}(G) + \frac{1}{2}\ln(2 \pi G) - \frac{1}{2}\left( C^{-1}G - 1 \right) \;.
\label{eq:DefinitionPhi2PIFRGAppendixV2}
\end{equation}
By differentiating~\eqref{eq:DefinitionPhi2PIFRGAppendixV2} with respect to $G$, we obtain:
\begin{equation}
\Phi^{(1)}(G) = \underbrace{\frac{\partial \Gamma^{(\mathrm{2PI})}(G)}{\partial G}}_{\frac{1}{2}K} + \frac{1}{2}G^{-1} - \frac{1}{2} C^{-1} = \frac{1}{2} \left( K + G^{-1} - C^{-1} \right) \;,
\label{eq:DifferentiatePhi2PIFRGAppendixV2}
\end{equation}
where the derivative $\frac{\partial \Gamma^{(\mathrm{2PI})}(G)}{\partial G}$ was replaced with the help of~\eqref{eq:Gamma2PIN1standardDerivAppendixV2}. We can thus see from \eqref{eq:DifferentiatePhi2PIFRGAppendixV2} that the self-energy $\Sigma$ must indeed be defined by~\eqref{eq:DefSelfEnergy2PIFRGN1StandardDerivAppendix} to recover Dyson equation in the form:
\begin{equation}
G^{-1} = C^{-1} - \Sigma(G) - K \;,
\end{equation}
in accordance with~\eqref{eq:DysonEqAppendix}. This implies the C-flow equation:
\begin{equation}
\dot{\overline{\Sigma}}_{\mathfrak{s}} = -2\dot{\overline{\Phi}}_{\mathfrak{s}}^{(1)} = - 2 \dot{\overline{G}}_{\mathfrak{s}} \overline{\Phi}_{\mathfrak{s}}^{(2)} \;,
\end{equation}
which coincides with~\eqref{eq:2PIfrgFlowEquationsCflowSigma0DONN1AppendixV2}.

\vspace{0.5cm}

In conclusion, solving the system of equations set either by~\eqref{eq:2PIfrgFlowEquationsCflowG0DONN1AppendixV1} to~\eqref{eq:2PIfrgtruncationtCflowPhi30DONAppendixV1} or by~\eqref{eq:2PIfrgFlowEquationsCflowG0DONN1AppendixV2} to~\eqref{eq:2PIfrgtruncationtCflowPhi30DONAppendixV2} leads to identical results (and notably to the green curves labeled ``2PI-FRG tC-flow $N_{\mathrm{max}}= 1 ~ \mathrm{or} ~ 2$'' in fig.~\ref{fig:original2PIFRGCflowlambdaN1}). This equivalence ensures the consistency of the definitions underlying the bosonic index formalism presented in appendix~\ref{ann:BosonicIndices2PI} and holds for any other 2PI-FRG approach applied to the (0+0)-D $O(N)$ model under consideration with $N=1$.

\section{Mixed C-flow for the (0+0)-D $O(N)$-symmetric $\varphi^4$-theory}
\label{ann:2PIfrgmCflowTruncationConditions0DON}

We give in this appendix the truncation conditions underlying the mC-flow for the mixed representation of the (0+0)-D $O(N)$-symmetric $\varphi^4$-theory at the truncation order $N_{\mathrm{max}}=1$ and at $N=1$. These relations are obtained by differentiating the following expressions of $\Phi_{\mathrm{mix},\mathrm{SCPT},\hbar\text{-exp}}(G,D)$ and $\Phi_{\mathrm{mix},\mathrm{SCPT},\lambda\text{-exp}}(G,D)$ (which are already given by~\eqref{eq:Cflow2PIFRGDefPhimixSCPThbarExp0DONN1} and~\eqref{eq:Cflow2PIFRGDefPhimixSCPTlambdaExp0DONN1}) with respect to the propagators $G$ and $D$:
\begin{equation}
\Phi_{\mathrm{mix},\mathrm{SCPT},\hbar\text{-exp}}(G,D) = \frac{1}{12} \hbar^{2} \lambda D G^{2} - \frac{1}{72} \hbar^{3} \lambda^{2} D^{2} G^{4} + \frac{5}{324} \hbar^{4} \lambda^{3} D^{3} G^{6} + \mathcal{O}\big(\hbar^{5}\big) \;,
\label{eq:Cflow2PIFRGDefPhimixSCPThbarExp0DONN1Appendix}
\end{equation}
\begin{equation}
\Phi_{\mathrm{mix},\mathrm{SCPT},\lambda\text{-exp}}(G,D) = \frac{1}{8} \lambda D G^{2} - \frac{1}{192} \lambda^{2} D^{2} G^{4} + \frac{1}{64} \lambda^{3} D^{3} G^{6} + \mathcal{O}\big(\lambda^{4}\big) \;.
\label{eq:Cflow2PIFRGDefPhimixSCPTlambdaExp0DONN1Appendix}
\end{equation}
This differentiation is carried out using the identity:
\begin{equation}
\frac{\partial G}{\partial G} = \frac{\partial D}{\partial D} = \mathcal{I} = 2 \;.
\end{equation}
This leads to:
\begin{itemize}
\item From the $\hbar$-expansion at $\hbar=1$ (i.e. from~\eqref{eq:Cflow2PIFRGDefPhimixSCPThbarExp0DONN1Appendix}):
\begin{itemize}
\item At $N_{\mathrm{SCPT}}=1$:
\begin{equation}
\overline{\Phi}^{(2G)}_{\mathfrak{s}} = \left. \overline{\Phi}^{(2G)}_{\mathrm{mix},\mathrm{SCPT},\hbar\text{-exp},N_{\mathrm{SCPT}}=1,\mathfrak{s}} \right|_{\lambda \rightarrow \frac{3}{4}\overline{\Phi}^{(2G,1D)}_{\mathfrak{s}}} = \frac{1}{2} \overline{\Phi}^{(2G,1D)}_{\mathfrak{s}} \overline{D}_{\mathfrak{s}} \;,
\label{eq:2PIFRGmixedmCflowPhi2GhbarNSCPT20DONAppendix}
\end{equation}
\begin{equation}
\overline{\Phi}^{(2D)}_{\mathfrak{s}} = \left. \overline{\Phi}^{(2D)}_{\mathrm{mix},\mathrm{SCPT},\hbar\text{-exp},N_{\mathrm{SCPT}}=1,\mathfrak{s}} \right|_{\lambda \rightarrow \frac{3}{4}\overline{\Phi}^{(2G,1D)}_{\mathfrak{s}}} = 0 \;,
\label{eq:2PIFRGmixedmCflowPhi2DhbarNSCPT20DONAppendix}
\end{equation}
\begin{equation}
\overline{\Phi}^{(1G,1D)}_{\mathfrak{s}} = \left. \overline{\Phi}^{(1G,1D)}_{\mathrm{mix},\mathrm{SCPT},\hbar\text{-exp},N_{\mathrm{SCPT}}=1,\mathfrak{s}} \right|_{\lambda \rightarrow \frac{3}{4}\overline{\Phi}^{(2G,1D)}_{\mathfrak{s}}} = \frac{1}{2} \overline{\Phi}^{(2G,1D)}_{\mathfrak{s}} \overline{G}_{\mathfrak{s}} \;.
\label{eq:2PIFRGmixedmCflowPhi1G1DhbarNSCPT20DONAppendix}
\end{equation}

\item At $N_{\mathrm{SCPT}}=2$:
\begin{equation}
\overline{\Phi}^{(2G)}_{\mathfrak{s}} = \left.\overline{\Phi}^{(2G)}_{\mathrm{mix},\mathrm{SCPT},\hbar\text{-exp},N_{\mathrm{SCPT}}=2,\mathfrak{s}}\right|_{\lambda \rightarrow \frac{3}{4}\overline{\Phi}^{(2G,1D)}_{\mathfrak{s}}} = \frac{1}{2} \overline{\Phi}^{(2G,1D)}_{\mathfrak{s}} \overline{D}_{\mathfrak{s}} - \frac{3}{8} \left(\overline{\Phi}^{(2G,1D)}_{\mathfrak{s}}\right)^{2} \overline{D}^{2}_{\mathfrak{s}} \overline{G}^{2}_{\mathfrak{s}} \;,
\label{eq:2PIFRGmixedmCflowPhi2GhbarNSCPT30DONAppendix}
\end{equation}
\begin{equation}
\overline{\Phi}^{(2D)}_{\mathfrak{s}} = \left.\overline{\Phi}^{(2D)}_{\mathrm{mix},\mathrm{SCPT},\hbar\text{-exp},N_{\mathrm{SCPT}}=2,\mathfrak{s}}\right|_{\lambda \rightarrow \frac{3}{4}\overline{\Phi}^{(2G,1D)}_{\mathfrak{s}}} = - \frac{1}{16} \left(\overline{\Phi}^{(2G,1D)}_{\mathfrak{s}}\right)^{2} \overline{G}^{4}_{\mathfrak{s}} \;,
\label{eq:2PIFRGmixedmCflowPhi2DhbarNSCPT30DONAppendix}
\end{equation}
\begin{equation}
\overline{\Phi}^{(1G,1D)}_{\mathfrak{s}} = \left.\overline{\Phi}^{(1G,1D)}_{\mathrm{mix},\mathrm{SCPT},\hbar\text{-exp},N_{\mathrm{SCPT}}=2,\mathfrak{s}}\right|_{\lambda \rightarrow \frac{3}{4}\overline{\Phi}^{(2G,1D)}_{\mathfrak{s}}} = \frac{1}{2} \overline{\Phi}^{(2G,1D)}_{\mathfrak{s}} \overline{G}_{\mathfrak{s}} - \frac{1}{4} \left(\overline{\Phi}^{(2G,1D)}_{\mathfrak{s}}\right)^{2} \overline{D}_{\mathfrak{s}} \overline{G}^{3}_{\mathfrak{s}} \;.
\label{eq:2PIFRGmixedmCflowPhi1G1DhbarNSCPT30DONAppendix}
\end{equation}

\item At $N_{\mathrm{SCPT}}=3$:
\begin{equation}
\begin{split}
\overline{\Phi}^{(2G)}_{\mathfrak{s}} = & \ \left.\overline{\Phi}^{(2G)}_{\mathrm{mix},\mathrm{SCPT},\hbar\text{-exp},N_{\mathrm{SCPT}}=3,\mathfrak{s}}\right|_{\lambda \rightarrow \frac{3}{4}\overline{\Phi}^{(2G,1D)}_{\mathfrak{s}}} \\
= & \ \frac{1}{2} \overline{\Phi}^{(2G,1D)}_{\mathfrak{s}} \overline{D}_{\mathfrak{s}} - \frac{3}{8} \left(\overline{\Phi}^{(2G,1D)}_{\mathfrak{s}}\right)^{2} \overline{D}^{2}_{\mathfrak{s}} \overline{G}^{2}_{\mathfrak{s}} + \frac{25}{32} \left(\overline{\Phi}^{(2G,1D)}_{\mathfrak{s}}\right)^{3} \overline{D}^{3}_{\mathfrak{s}} \overline{G}^{4}_{\mathfrak{s}} \;,
\label{eq:2PIFRGmixedmCflowPhi2GhbarNSCPT40DONAppendix}
\end{split}
\end{equation}
\begin{equation}
\begin{split}
\overline{\Phi}^{(2D)}_{\mathfrak{s}} = & \ \left.\overline{\Phi}^{(2D)}_{\mathrm{mix},\mathrm{SCPT},\hbar\text{-exp},N_{\mathrm{SCPT}}=3,\mathfrak{s}}\right|_{\lambda \rightarrow \frac{3}{4}\overline{\Phi}^{(2G,1D)}_{\mathfrak{s}}} \\
= & - \frac{1}{16} \left(\overline{\Phi}^{(2G,1D)}_{\mathfrak{s}}\right)^{2} \overline{G}^{4}_{\mathfrak{s}} + \frac{5}{32} \left(\overline{\Phi}^{(2G,1D)}_{\mathfrak{s}}\right)^{3} \overline{D}_{\mathfrak{s}} \overline{G}^{6}_{\mathfrak{s}} \;,
\end{split}
\label{eq:2PIFRGmixedmCflowPhi2DhbarNSCPT40DONAppendix}
\end{equation}
\begin{equation}
\begin{split}
\overline{\Phi}^{(1G,1D)}_{\mathfrak{s}} = & \ \left.\overline{\Phi}^{(1G,1D)}_{\mathrm{mix},\mathrm{SCPT},\hbar\text{-exp},N_{\mathrm{SCPT}}=3,\mathfrak{s}}\right|_{\lambda \rightarrow \frac{3}{4}\overline{\Phi}^{(2G,1D)}_{\mathfrak{s}}} \\
= & \ \frac{1}{2} \overline{\Phi}^{(2G,1D)}_{\mathfrak{s}} \overline{G}_{\mathfrak{s}} - \frac{1}{4} \left(\overline{\Phi}^{(2G,1D)}_{\mathfrak{s}}\right)^{2} \overline{D}_{\mathfrak{s}} \overline{G}^{3}_{\mathfrak{s}} + \frac{15}{32} \left(\overline{\Phi}^{(2G,1D)}_{\mathfrak{s}}\right)^{3} \overline{D}^{2}_{\mathfrak{s}} \overline{G}^{5}_{\mathfrak{s}} \;.
\end{split}
\label{eq:2PIFRGmixedmCflowPhi1G1DhbarNSCPT40DONAppendix}
\end{equation}

\end{itemize}

\item From the $\lambda$-expansion (i.e. from~\eqref{eq:Cflow2PIFRGDefPhimixSCPTlambdaExp0DONN1Appendix}):
\begin{itemize}
\item At $N_{\mathrm{SCPT}}=1$:
\begin{equation}
\overline{\Phi}^{(2G)}_{\mathfrak{s}} = \left. \overline{\Phi}^{(2G)}_{\mathrm{mix},\mathrm{SCPT},\lambda\text{-exp},N_{\mathrm{SCPT}}=1,\mathfrak{s}} \right|_{\lambda \rightarrow \frac{1}{2}\overline{\Phi}^{(2G,1D)}_{\mathfrak{s}}} = \frac{1}{2} \overline{\Phi}^{(2G,1D)}_{\mathfrak{s}} \overline{D}_{\mathfrak{s}} \;,
\label{eq:2PIFRGmixedmCflowPhi2GlambdaNSCPT20DONAppendix}
\end{equation}
\begin{equation}
\overline{\Phi}^{(2D)}_{\mathfrak{s}} = \left. \overline{\Phi}^{(2D)}_{\mathrm{mix},\mathrm{SCPT},\lambda\text{-exp},N_{\mathrm{SCPT}}=1,\mathfrak{s}} \right|_{\lambda \rightarrow \frac{1}{2}\overline{\Phi}^{(2G,1D)}_{\mathfrak{s}}} = 0 \;,
\label{eq:2PIFRGmixedmCflowPhi2DlambdaNSCPT20DONAppendix}
\end{equation}
\begin{equation}
\overline{\Phi}^{(1G,1D)}_{\mathfrak{s}} = \left. \overline{\Phi}^{(1G,1D)}_{\mathrm{mix},\mathrm{SCPT},\lambda\text{-exp},N_{\mathrm{SCPT}}=1,\mathfrak{s}} \right|_{\lambda \rightarrow \frac{1}{2}\overline{\Phi}^{(2G,1D)}_{\mathfrak{s}}} = \frac{1}{2} \overline{\Phi}^{(2G,1D)}_{\mathfrak{s}} \overline{G}_{\mathfrak{s}} \;.
\label{eq:2PIFRGmixedmCflowPhi1G1DlambdaNSCPT20DONAppendix}
\end{equation}

\item At $N_{\mathrm{SCPT}}=2$:
\begin{equation}
\overline{\Phi}^{(2G)}_{\mathfrak{s}} = \left. \overline{\Phi}^{(2G)}_{\mathrm{mix},\mathrm{SCPT},\lambda\text{-exp},N_{\mathrm{SCPT}}=2,\mathfrak{s}} \right|_{\lambda \rightarrow \frac{1}{2}\overline{\Phi}^{(2G,1D)}_{\mathfrak{s}}} = \frac{1}{2} \overline{\Phi}^{(2G,1D)}_{\mathfrak{s}} \overline{D}_{\mathfrak{s}} - \frac{1}{16} \left(\overline{\Phi}^{(2G,1D)}_{\mathfrak{s}}\right)^{2} \overline{D}^{2}_{\mathfrak{s}} \overline{G}^{2}_{\mathfrak{s}} \;,
\label{eq:2PIFRGmixedmCflowPhi2GlambdaNSCPT30DONAppendix}
\end{equation}
\begin{equation}
\overline{\Phi}^{(2D)}_{\mathfrak{s}} = \left. \overline{\Phi}^{(2D)}_{\mathrm{mix},\mathrm{SCPT},\lambda\text{-exp},N_{\mathrm{SCPT}}=2,\mathfrak{s}} \right|_{\lambda \rightarrow \frac{1}{2}\overline{\Phi}^{(2G,1D)}_{\mathfrak{s}}} = - \frac{1}{96} \left(\overline{\Phi}^{(2G,1D)}_{\mathfrak{s}}\right)^{2} \overline{G}^{4}_{\mathfrak{s}} \;,
\label{eq:2PIFRGmixedmCflowPhi2DlambdaNSCPT30DONAppendix}
\end{equation}
\begin{equation}
\overline{\Phi}^{(1G,1D)}_{\mathfrak{s}} = \left. \overline{\Phi}^{(1G,1D)}_{\mathrm{mix},\mathrm{SCPT},\lambda\text{-exp},N_{\mathrm{SCPT}}=2,\mathfrak{s}} \right|_{\lambda \rightarrow \frac{1}{2}\overline{\Phi}^{(2G,1D)}_{\mathfrak{s}}} = \frac{1}{2} \overline{\Phi}^{(2G,1D)}_{\mathfrak{s}} \overline{G}_{\mathfrak{s}} - \frac{1}{24} \left(\overline{\Phi}^{(2G,1D)}_{\mathfrak{s}}\right)^{2} \overline{D}_{\mathfrak{s}} \overline{G}^{3}_{\mathfrak{s}} \;.
\label{eq:2PIFRGmixedmCflowPhi1G1DlambdaNSCPT30DONAppendix}
\end{equation}

\item At $N_{\mathrm{SCPT}}=3$:
\begin{equation}
\begin{split}
\overline{\Phi}^{(2G)}_{\mathfrak{s}} = & \ \left. \overline{\Phi}^{(2G)}_{\mathrm{mix},\mathrm{SCPT},\lambda\text{-exp},N_{\mathrm{SCPT}}=3,\mathfrak{s}} \right|_{\lambda \rightarrow \frac{1}{2}\overline{\Phi}^{(2G,1D)}_{\mathfrak{s}}} \\
= & \ \frac{1}{2} \overline{\Phi}^{(2G,1D)}_{\mathfrak{s}} \overline{D}_{\mathfrak{s}} - \frac{1}{16} \left(\overline{\Phi}^{(2G,1D)}_{\mathfrak{s}}\right)^{2} \overline{D}^{2}_{\mathfrak{s}} \overline{G}^{2}_{\mathfrak{s}} + \frac{15}{64} \left(\overline{\Phi}^{(2G,1D)}_{\mathfrak{s}}\right)^{3} \overline{D}^{3}_{\mathfrak{s}} \overline{G}^{4}_{\mathfrak{s}} \;,
\end{split}
\label{eq:2PIFRGmixedmCflowPhi2GlambdaNSCPT40DONAppendix}
\end{equation}
\begin{equation}
\begin{split}
\overline{\Phi}^{(2D)}_{\mathfrak{s}} = & \ \left. \overline{\Phi}^{(2D)}_{\mathrm{mix},\mathrm{SCPT},\lambda\text{-exp},N_{\mathrm{SCPT}}=3,\mathfrak{s}} \right|_{\lambda \rightarrow \frac{1}{2}\overline{\Phi}^{(2G,1D)}_{\mathfrak{s}}} \\
= & - \frac{1}{96} \left(\overline{\Phi}^{(2G,1D)}_{\mathfrak{s}}\right)^{2} \overline{G}^{4}_{\mathfrak{s}} + \frac{3}{64} \left(\overline{\Phi}^{(2G,1D)}_{\mathfrak{s}}\right)^{3} \overline{D}_{\mathfrak{s}} \overline{G}^{6}_{\mathfrak{s}} \;,
\end{split}
\label{eq:2PIFRGmixedmCflowPhi2DlambdaNSCPT40DONAppendix}
\end{equation}
\begin{equation}
\begin{split}
\overline{\Phi}^{(1G,1D)}_{\mathfrak{s}} = & \ \left. \overline{\Phi}^{(1G,1D)}_{\mathrm{mix},\mathrm{SCPT},\lambda\text{-exp},N_{\mathrm{SCPT}}=3,\mathfrak{s}} \right|_{\lambda \rightarrow \frac{1}{2}\overline{\Phi}^{(2G,1D)}_{\mathfrak{s}}} \\
= & \ \frac{1}{2} \overline{\Phi}^{(2G,1D)}_{\mathfrak{s}} \overline{G}_{\mathfrak{s}} - \frac{1}{24} \left(\overline{\Phi}^{(2G,1D)}_{\mathfrak{s}}\right)^{2} \overline{D}_{\mathfrak{s}} \overline{G}^{3}_{\mathfrak{s}} + \frac{9}{64} \left(\overline{\Phi}^{(2G,1D)}_{\mathfrak{s}}\right)^{3} \overline{D}^{2}_{\mathfrak{s}} \overline{G}^{5}_{\mathfrak{s}} \;.
\end{split}
\label{eq:2PIFRGmixedmCflowPhi1G1DlambdaNSCPT40DONAppendix}
\end{equation}

\end{itemize}

\end{itemize}

\section{U-flow and CU-flow for the (0+0)-D $O(N)$-symmetric $\varphi^4$-theory}
\label{ann:2PIfrgUflowEquations0DON}
\subsection{pU-flow, mU-flow and CU-flow equations at $N=1$}
\label{ann:2PIfrgpUandmUflow0DON1}

We give below the differential equations required to implement the pU-flow, mU-flow and CU-flow versions of the 2PI-FRG up to $N_{\mathrm{max}}=3$ for the studied $O(N)$ model at $N=1$:
\begin{itemize}
\item For the pU-flow:
\begin{equation}
\dot{\overline{G}}_{\mathfrak{s}} = \overline{G}_{\mathfrak{s}}^{2} \dot{\overline{\Sigma}}_{\mathfrak{s}} \;,
\label{eq:2PIFRGpuflowFlowEqG0DON1Appendix}
\end{equation}
\begin{equation}
\dot{\overline{\Omega}}_{\mathfrak{s}} = \frac{\lambda}{24} \left( 4\left( 2\overline{G}_{\mathfrak{s}}^{-2} + \overline{\Phi}_{\mathfrak{s}}^{(2)} \right)^{-1} + \overline{G}_{\mathfrak{s}}^{2} \right) \;,
\label{eq:2PIFRGpuflowFlowEqOmega0DON1Appendix}
\end{equation}
\begin{equation}
\dot{\overline{\Sigma}}_{\mathfrak{s}} = -\frac{\lambda}{3} \left( 2 + \overline{G}_{\mathfrak{s}}^{2} \overline{\Phi}_{\mathfrak{s}}^{(2)} \right)^{-1} \left( \left(2\overline{G}_{\mathfrak{s}}^{-2} + \overline{\Phi}_{\mathfrak{s}}^{(2)}\right)^{-2} \left(8\overline{G}_{\mathfrak{s}}^{-3} - \overline{\Phi}_{\mathfrak{s}}^{(3)}\right) + \overline{G}_{\mathfrak{s}} \right) \;,
\label{eq:2PIFRGpuflowFlowEqSigma0DON1Appendix}
\end{equation}
\begin{equation}
\begin{split}
\dot{\overline{\Phi}}_{\mathfrak{s}}^{(2)} = & \ \frac{\lambda}{6} \bigg( 2\left( 2\overline{G}_{\mathfrak{s}}^{-2} + \overline{\Phi}_{\mathfrak{s}}^{(2)} \right)^{-3} \left( 8 \overline{G}_{\mathfrak{s}}^{-3} - \overline{\Phi}_{\mathfrak{s}}^{(3)} \right)^{2} - 64 \left( 2\overline{G}_{\mathfrak{s}}^{-2} + \overline{\Phi}_{\mathfrak{s}}^{(2)} \right)^{-2} \overline{G}_{\mathfrak{s}}^{-4} \\
& \hspace{0.5cm} + \left( 2\overline{G}_{\mathfrak{s}}^{-2} + \overline{\Phi}_{\mathfrak{s}}^{(2)} \right)^{-2} \left(16\overline{G}_{\mathfrak{s}}^{-4} - \overline{\Phi}_{\mathfrak{s}}^{(4)} \right) + 2 \bigg) + \frac{1}{2} \dot{\overline{G}}_{\mathfrak{s}} \overline{\Phi}_{\mathfrak{s}}^{(3)} \;,
\end{split}
\label{eq:2PIFRGpuflowFlowEqPhi20DON1Appendix}
\end{equation}
\begin{equation}
\begin{split}
\dot{\overline{\Phi}}_{\mathfrak{s}}^{(3)} = & \ \frac{\lambda}{6} \left( 2\overline{G}_{\mathfrak{s}}^{-2} + \overline{\Phi}_{\mathfrak{s}}^{(2)} \right)^{-2} \bigg( 6 \left( 2\overline{G}_{\mathfrak{s}}^{-2} + \overline{\Phi}_{\mathfrak{s}}^{(2)} \right)^{-2} \left( 8 \overline{G}_{\mathfrak{s}}^{-3} - \overline{\Phi}_{\mathfrak{s}}^{(3)} \right)^{3} \\
& \hspace{3.7cm} + 6 \left( 2\overline{G}_{\mathfrak{s}}^{-2} + \overline{\Phi}_{\mathfrak{s}}^{(2)} \right)^{-1} \left( 8 \overline{G}_{\mathfrak{s}}^{-3} - \overline{\Phi}_{\mathfrak{s}}^{(3)} \right) \left( 16 \overline{G}_{\mathfrak{s}}^{-4} - \overline{\Phi}_{\mathfrak{s}}^{(4)} \right) \\
& \hspace{3.7cm} - 384 \overline{G}_{\mathfrak{s}}^{-4} \left( \left( 2 \overline{G}_{\mathfrak{s}}^{-2} + \overline{\Phi}_{\mathfrak{s}}^{(2)} \right)^{-1}\left(8\overline{G}_{\mathfrak{s}}^{-3}-\overline{\Phi}_{\mathfrak{s}}^{(3)}\right) - \overline{G}_{\mathfrak{s}}^{-1} \right) - \overline{\Phi}_{\mathfrak{s}}^{(5)} \bigg) \\
& + \frac{1}{2} \dot{\overline{G}}_{\mathfrak{s}} \overline{\Phi}_{\mathfrak{s}}^{(4)} \;.
\end{split}
\label{eq:2PIFRGpuflowFlowEqPhi30DON1Appendix}
\end{equation}

\item For the mU-flow at $N_{\mathrm{SCPT}}=1$:
\begin{equation}
\dot{\overline{\boldsymbol{G}}}_{\mathfrak{s}} = \overline{\boldsymbol{G}}_{\mathfrak{s}}^{2} \dot{\overline{\boldsymbol{\Sigma}}}_{\mathfrak{s}} \;,
\label{eq:2PIFRGmuflowNSCPT1FlowEqG0DON1Appendix}
\end{equation}
\begin{equation}
\dot{\overline{\boldsymbol{\Omega}}}_{\mathfrak{s}} = \frac{\lambda}{24} \left( 4\left( 2\overline{\boldsymbol{G}}_{\mathfrak{s}}^{-2} + \overline{\boldsymbol{\Phi}}_{\mathfrak{s}}^{(2)} - \lambda\left(1-\mathfrak{s}\right) \right)^{-1} + \overline{\boldsymbol{G}}_{\mathfrak{s}}^{2} \right) - \frac{1}{8} \lambda \overline{\boldsymbol{G}}_{\mathfrak{s}}^{2} \;,
\label{eq:2PIFRGmuflowNSCPT1FlowEqOmega0DON1Appendix}
\end{equation}
\begin{equation}
\dot{\overline{\boldsymbol{\Sigma}}}_{\mathfrak{s}} = -\frac{\lambda}{3} \left( 2 + \overline{\boldsymbol{G}}_{\mathfrak{s}}^{2} \overline{\boldsymbol{\Phi}}_{\mathfrak{s}}^{(2)} \right)^{-1} \bigg( \left(2\overline{\boldsymbol{G}}_{\mathfrak{s}}^{-2} + \overline{\boldsymbol{\Phi}}_{\mathfrak{s}}^{(2)} - \lambda\left(1-\mathfrak{s}\right) \right)^{-2} \left(8\overline{\boldsymbol{G}}_{\mathfrak{s}}^{-3} - \overline{\boldsymbol{\Phi}}_{\mathfrak{s}}^{(3)} \right) - 2\overline{\boldsymbol{G}}_{\mathfrak{s}} \bigg) \;,
\label{eq:2PIFRGmuflowNSCPT1FlowEqSigma0DON1Appendix}
\end{equation}
\begin{equation}
\begin{split}
\dot{\overline{\boldsymbol{\Phi}}}^{(2)}_{\mathfrak{s}} = & \ \frac{\lambda}{6} \bigg( 2 \left(2\overline{\boldsymbol{G}}_{\mathfrak{s}}^{-2} + \overline{\boldsymbol{\Phi}}_{\mathfrak{s}}^{(2)} - \lambda\left(1-\mathfrak{s}\right) \right)^{-3} \left(8\overline{\boldsymbol{G}}_{\mathfrak{s}}^{-3} - \overline{\boldsymbol{\Phi}}_{\mathfrak{s}}^{(3)} \right)^{2} \\
& \hspace{0.55cm} -64 \overline{\boldsymbol{G}}_{\mathfrak{s}}^{-4} \left(2\overline{\boldsymbol{G}}_{\mathfrak{s}}^{-2} + \overline{\boldsymbol{\Phi}}_{\mathfrak{s}}^{(2)} - \lambda\left(1-\mathfrak{s}\right) \right)^{-2} \\
& \hspace{0.55cm} + \left(2\overline{\boldsymbol{G}}_{\mathfrak{s}}^{-2} + \overline{\boldsymbol{\Phi}}_{\mathfrak{s}}^{(2)} - \lambda\left(1-\mathfrak{s}\right) \right)^{-2} \left( 16\overline{\boldsymbol{G}}_{\mathfrak{s}}^{-4} - \overline{\boldsymbol{\Phi}}_{\mathfrak{s}}^{(4)} \right) + 2 \bigg) \\
& - \lambda + \frac{1}{2} \dot{\overline{\boldsymbol{G}}}_{\mathfrak{s}} \overline{\boldsymbol{\Phi}}_{\mathfrak{s}}^{(3)} \;,
\end{split}
\label{eq:2PIFRGmuflowNSCPT1FlowEqPhi20DON1Appendix}
\end{equation}
\begin{equation}
\begin{split}
\dot{\overline{\boldsymbol{\Phi}}}^{(3)}_{\mathfrak{s}} = & \ \frac{\lambda}{6} \left(2\overline{\boldsymbol{G}}_{\mathfrak{s}}^{-2} + \overline{\boldsymbol{\Phi}}_{\mathfrak{s}}^{(2)} - \lambda\left(1-\mathfrak{s}\right) \right)^{-2} \\
& \times \bigg( 6 \left(2\overline{\boldsymbol{G}}_{\mathfrak{s}}^{-2} + \overline{\boldsymbol{\Phi}}_{\mathfrak{s}}^{(2)} - \lambda\left(1-\mathfrak{s}\right) \right)^{-2} \left(8\overline{\boldsymbol{G}}_{\mathfrak{s}}^{-3} - \overline{\boldsymbol{\Phi}}_{\mathfrak{s}}^{(3)} \right)^{3} \\
& \hspace{0.55cm} + 6 \left(2\overline{\boldsymbol{G}}_{\mathfrak{s}}^{-2} + \overline{\boldsymbol{\Phi}}_{\mathfrak{s}}^{(2)} - \lambda\left(1-\mathfrak{s}\right) \right)^{-1} \left(8\overline{\boldsymbol{G}}_{\mathfrak{s}}^{-3} - \overline{\boldsymbol{\Phi}}_{\mathfrak{s}}^{(3)} \right) \left( -48\overline{\boldsymbol{G}}_{\mathfrak{s}}^{-4} - \overline{\boldsymbol{\Phi}}_{\mathfrak{s}}^{(4)} \right) \\
& \hspace{0.55cm} + 384 \overline{\boldsymbol{G}}_{\mathfrak{s}}^{-5} - \overline{\boldsymbol{\Phi}}_{\mathfrak{s}}^{(5)} \bigg) + \frac{1}{2} \dot{\overline{\boldsymbol{G}}}_{\mathfrak{s}} \overline{\boldsymbol{\Phi}}_{\mathfrak{s}}^{(4)} \;.
\end{split}
\label{eq:2PIFRGmuflowNSCPT1FlowEqPhi30DON1Appendix}
\end{equation}

\item For the mU-flow at $N_{\mathrm{SCPT}}=2$:
\begin{equation}
\dot{\overline{\boldsymbol{G}}}_{\mathfrak{s}} = \overline{\boldsymbol{G}}_{\mathfrak{s}}^{2} \dot{\overline{\boldsymbol{\Sigma}}}_{\mathfrak{s}} \;,
\label{eq:2PIFRGmuflowNSCPT2FlowEqG0DON1Appendix}
\end{equation}
\begin{equation}
\dot{\overline{\boldsymbol{\Omega}}}_{\mathfrak{s}} = \frac{\lambda}{24} \left( 4\left( 2\overline{\boldsymbol{G}}_{\mathfrak{s}}^{-2} + \overline{\boldsymbol{\Phi}}_{\mathfrak{s}}^{(2)} - \lambda\left(1-\mathfrak{s}\right) + \lambda^{2}\overline{\boldsymbol{G}}_{\mathfrak{s}}^{2}\left(1-\mathfrak{s}^{2}\right) \right)^{-1} + \overline{\boldsymbol{G}}_{\mathfrak{s}}^{2} \right) - \frac{1}{8} \lambda \overline{\boldsymbol{G}}_{\mathfrak{s}}^{2} + \frac{1}{24} \mathfrak{s} \lambda^{2} \overline{\boldsymbol{G}}_{\mathfrak{s}}^{4} \;,
\label{eq:2PIFRGmuflowNSCPT2FlowEqOmega0DON1Appendix}
\end{equation}
\begin{equation}
\begin{split}
\dot{\overline{\boldsymbol{\Sigma}}}_{\mathfrak{s}} = & -\frac{\lambda}{3} \left( 2 + \overline{\boldsymbol{G}}_{\mathfrak{s}}^{2} \overline{\boldsymbol{\Phi}}_{\mathfrak{s}}^{(2)} \right)^{-1} \bigg( \left(2\overline{\boldsymbol{G}}_{\mathfrak{s}}^{-2} + \overline{\boldsymbol{\Phi}}_{\mathfrak{s}}^{(2)} - \lambda\left(1-\mathfrak{s}\right) + \lambda^{2}\overline{\boldsymbol{G}}_{\mathfrak{s}}^{2}\left(1-\mathfrak{s}^{2}\right) \right)^{-2} \\
& \hspace{4.25cm} \times\left(8\overline{\boldsymbol{G}}_{\mathfrak{s}}^{-3} - \overline{\boldsymbol{\Phi}}_{\mathfrak{s}}^{(3)} - 4 \lambda^{2} \overline{\boldsymbol{G}}_{\mathfrak{s}} \left(1-\mathfrak{s}^{2}\right) \right) - 2\overline{\boldsymbol{G}}_{\mathfrak{s}} + 2\mathfrak{s}\lambda\overline{\boldsymbol{G}}_{\mathfrak{s}}^{3} \bigg) \;,
\end{split}
\label{eq:2PIFRGmuflowNSCPT2FlowEqSigma0DON1Appendix}
\end{equation}
\begin{equation}
\begin{split}
\scalebox{0.935}{${\displaystyle \dot{\overline{\boldsymbol{\Phi}}}^{(2)}_{\mathfrak{s}} = }$} & \ \scalebox{0.935}{${\displaystyle \frac{\lambda}{6} \bigg( 2 \left(2\overline{\boldsymbol{G}}_{\mathfrak{s}}^{-2} + \overline{\boldsymbol{\Phi}}_{\mathfrak{s}}^{(2)} - \lambda\left(1-\mathfrak{s}\right) + \lambda^{2}\overline{\boldsymbol{G}}_{\mathfrak{s}}^{2}\left(1-\mathfrak{s}^{2}\right) \right)^{-3} \left(8\overline{\boldsymbol{G}}_{\mathfrak{s}}^{-3} - \overline{\boldsymbol{\Phi}}_{\mathfrak{s}}^{(3)} - 4 \lambda^{2} \overline{\boldsymbol{G}}_{\mathfrak{s}} \left(1-\mathfrak{s}^{2}\right) \right)^{2} }$} \\
& \hspace{0.6cm} \scalebox{0.935}{${\displaystyle -64 \overline{\boldsymbol{G}}_{\mathfrak{s}}^{-4} \left(2\overline{\boldsymbol{G}}_{\mathfrak{s}}^{-2} + \overline{\boldsymbol{\Phi}}_{\mathfrak{s}}^{(2)} - \lambda\left(1-\mathfrak{s}\right) + \lambda^{2}\overline{\boldsymbol{G}}_{\mathfrak{s}}^{2}\left(1-\mathfrak{s}^{2}\right) \right)^{-2} }$} \\
& \hspace{0.6cm} \scalebox{0.935}{${\displaystyle + \left(2\overline{\boldsymbol{G}}_{\mathfrak{s}}^{-2} + \overline{\boldsymbol{\Phi}}_{\mathfrak{s}}^{(2)} - \lambda\left(1-\mathfrak{s}\right) + \lambda^{2}\overline{\boldsymbol{G}}_{\mathfrak{s}}^{2}\left(1-\mathfrak{s}^{2}\right) \right)^{-2} \left( 16\overline{\boldsymbol{G}}_{\mathfrak{s}}^{-4} - \overline{\boldsymbol{\Phi}}_{\mathfrak{s}}^{(4)} - 8\lambda^{2}\left(1-\mathfrak{s}^{2}\right) \right) + 2 \bigg) }$} \\
& \scalebox{0.935}{${\displaystyle - \lambda + 2 \mathfrak{s} \lambda^{2} \overline{\boldsymbol{G}}_{\mathfrak{s}}^{2} + \frac{1}{2} \dot{\overline{\boldsymbol{G}}}_{\mathfrak{s}} \overline{\boldsymbol{\Phi}}_{\mathfrak{s}}^{(3)} \;, }$}
\end{split}
\label{eq:2PIFRGmuflowNSCPT2FlowEqPhi20DON1Appendix}
\end{equation}
\begin{equation}
\begin{split}
\scalebox{0.97}{${\displaystyle \dot{\overline{\boldsymbol{\Phi}}}^{(3)}_{\mathfrak{s}} = }$} & \ \scalebox{0.97}{${\displaystyle \frac{\lambda}{6} \left(2\overline{\boldsymbol{G}}_{\mathfrak{s}}^{-2} + \overline{\boldsymbol{\Phi}}_{\mathfrak{s}}^{(2)} - \lambda\left(1-\mathfrak{s}\right) + \lambda^{2}\overline{\boldsymbol{G}}_{\mathfrak{s}}^{2}\left(1-\mathfrak{s}^{2}\right) \right)^{-2} }$} \\
& \scalebox{0.97}{${\displaystyle \times \bigg( 6 \left(2\overline{\boldsymbol{G}}_{\mathfrak{s}}^{-2} + \overline{\boldsymbol{\Phi}}_{\mathfrak{s}}^{(2)} - \lambda\left(1-\mathfrak{s}\right) + \lambda^{2}\overline{\boldsymbol{G}}_{\mathfrak{s}}^{2}\left(1-\mathfrak{s}^{2}\right) \right)^{-2} \left(8\overline{\boldsymbol{G}}_{\mathfrak{s}}^{-3} - \overline{\boldsymbol{\Phi}}_{\mathfrak{s}}^{(3)} - 4 \lambda^{2} \overline{\boldsymbol{G}}_{\mathfrak{s}} \left(1-\mathfrak{s}^{2}\right) \right)^{3} }$} \\
& \hspace{0.45cm} \scalebox{0.97}{${\displaystyle + 6 \left(2\overline{\boldsymbol{G}}_{\mathfrak{s}}^{-2} + \overline{\boldsymbol{\Phi}}_{\mathfrak{s}}^{(2)} - \lambda\left(1-\mathfrak{s}\right) + \lambda^{2}\overline{\boldsymbol{G}}_{\mathfrak{s}}^{2}\left(1-\mathfrak{s}^{2}\right) \right)^{-1} \left(8\overline{\boldsymbol{G}}_{\mathfrak{s}}^{-3} - \overline{\boldsymbol{\Phi}}_{\mathfrak{s}}^{(3)} - 4 \lambda^{2} \overline{\boldsymbol{G}}_{\mathfrak{s}} \left(1-\mathfrak{s}^{2}\right) \right) }$} \\
& \hspace{1.1cm} \scalebox{0.97}{${\displaystyle \times \left( -48\overline{\boldsymbol{G}}_{\mathfrak{s}}^{-4} - \overline{\boldsymbol{\Phi}}_{\mathfrak{s}}^{(4)} - 8\lambda^{2}\left(1-\mathfrak{s}^{2}\right) \right) }$} \\
& \hspace{0.45cm} \scalebox{0.97}{${\displaystyle + 384 \overline{\boldsymbol{G}}_{\mathfrak{s}}^{-5} - \overline{\boldsymbol{\Phi}}_{\mathfrak{s}}^{(5)} \bigg) + 8 \mathfrak{s} \lambda^{2} \overline{\boldsymbol{G}}_{\mathfrak{s}} + \frac{1}{2} \dot{\overline{\boldsymbol{G}}}_{\mathfrak{s}} \overline{\boldsymbol{\Phi}}_{\mathfrak{s}}^{(4)} \;. }$}
\end{split}
\label{eq:2PIFRGmuflowNSCPT2FlowEqPhi30DON1Appendix}
\end{equation}

\item For the mU-flow at $N_{\mathrm{SCPT}}=3$:
\begin{equation}
\dot{\overline{\boldsymbol{G}}}_{\mathfrak{s}} = \overline{\boldsymbol{G}}_{\mathfrak{s}}^{2} \dot{\overline{\boldsymbol{\Sigma}}}_{\mathfrak{s}} \;,
\label{eq:2PIFRGmuflowNSCPT3FlowEqG0DON1Appendix}
\end{equation}
\begin{equation}
\begin{split}
\dot{\overline{\boldsymbol{\Omega}}}_{\mathfrak{s}} = & \ \frac{\lambda}{24} \left( 4\left( 2\overline{\boldsymbol{G}}_{\mathfrak{s}}^{-2} + \overline{\boldsymbol{\Phi}}_{\mathfrak{s}}^{(2)} - \lambda\left(1-\mathfrak{s}\right) + \lambda^{2}\overline{\boldsymbol{G}}_{\mathfrak{s}}^{2}\left(1-\mathfrak{s}^{2}\right) - \frac{5}{2}\lambda^{3}\overline{\boldsymbol{G}}_{\mathfrak{s}}^{4}\left(1-\mathfrak{s}^{3}\right) \right)^{-1} + \overline{\boldsymbol{G}}_{\mathfrak{s}}^{2} \right) \\
& - \frac{1}{8} \lambda \overline{\boldsymbol{G}}_{\mathfrak{s}}^{2} + \frac{1}{24} \mathfrak{s} \lambda^{2} \overline{\boldsymbol{G}}_{\mathfrak{s}}^{4} - \frac{1}{16} \mathfrak{s}^{2} \lambda^{3} \overline{\boldsymbol{G}}_{\mathfrak{s}}^{6} \;,
\end{split}
\label{eq:2PIFRGmuflowNSCPT3FlowEqOmega0DON1Appendix}
\end{equation}
\begin{equation}
\begin{split}
\scalebox{0.96}{${\displaystyle \dot{\overline{\boldsymbol{\Sigma}}}_{\mathfrak{s}} = }$} & \scalebox{0.96}{${\displaystyle -\frac{\lambda}{3} \left( 2 + \overline{\boldsymbol{G}}_{\mathfrak{s}}^{2} \overline{\boldsymbol{\Phi}}_{\mathfrak{s}}^{(2)} \right)^{-1} \bigg( \left(2\overline{\boldsymbol{G}}_{\mathfrak{s}}^{-2} + \overline{\boldsymbol{\Phi}}_{\mathfrak{s}}^{(2)} - \lambda\left(1-\mathfrak{s}\right) + \lambda^{2}\overline{\boldsymbol{G}}_{\mathfrak{s}}^{2}\left(1-\mathfrak{s}^{2}\right) - \frac{5}{2}\lambda^{3}\overline{\boldsymbol{G}}_{\mathfrak{s}}^{4}\left(1-\mathfrak{s}^{3}\right) \right)^{-2} }$} \\
& \hspace{4.0cm} \scalebox{0.96}{${\displaystyle \times\left(8\overline{\boldsymbol{G}}_{\mathfrak{s}}^{-3} - \overline{\boldsymbol{\Phi}}_{\mathfrak{s}}^{(3)} - 4 \lambda^{2} \overline{\boldsymbol{G}}_{\mathfrak{s}} \left(1-\mathfrak{s}^{2}\right) + 20 \lambda^{3} \overline{\boldsymbol{G}}_{\mathfrak{s}}^{3} \left(1-\mathfrak{s}^{3}\right) \right) }$} \\
& \hspace{3.7cm} \scalebox{0.96}{${\displaystyle - 2\overline{\boldsymbol{G}}_{\mathfrak{s}} + 2\mathfrak{s}\lambda\overline{\boldsymbol{G}}_{\mathfrak{s}}^{3} - \frac{9}{2}\mathfrak{s}^{2}\lambda^{2}\overline{\boldsymbol{G}}_{\mathfrak{s}}^{5} \bigg) \;,}$}
\end{split}
\label{eq:2PIFRGmuflowNSCPT3FlowEqSigma0DON1Appendix}
\end{equation}
\begin{equation}
\begin{split}
\dot{\overline{\boldsymbol{\Phi}}}^{(2)}_{\mathfrak{s}} = & \ \frac{\lambda}{6} \bigg( 2 \left(2\overline{\boldsymbol{G}}_{\mathfrak{s}}^{-2} + \overline{\boldsymbol{\Phi}}_{\mathfrak{s}}^{(2)} - \lambda\left(1-\mathfrak{s}\right) + \lambda^{2}\overline{\boldsymbol{G}}_{\mathfrak{s}}^{2}\left(1-\mathfrak{s}^{2}\right) - \frac{5}{2}\lambda^{3}\overline{\boldsymbol{G}}_{\mathfrak{s}}^{4}\left(1-\mathfrak{s}^{3}\right) \right)^{-3} \\
& \hspace{0.985cm} \times\left(8\overline{\boldsymbol{G}}_{\mathfrak{s}}^{-3} - \overline{\boldsymbol{\Phi}}_{\mathfrak{s}}^{(3)} - 4 \lambda^{2} \overline{\boldsymbol{G}}_{\mathfrak{s}} \left(1-\mathfrak{s}^{2}\right) + 20 \lambda^{3} \overline{\boldsymbol{G}}_{\mathfrak{s}}^{3} \left(1-\mathfrak{s}^{3}\right) \right)^{2} \\
& \hspace{0.45cm} -64 \overline{\boldsymbol{G}}_{\mathfrak{s}}^{-4} \left(2\overline{\boldsymbol{G}}_{\mathfrak{s}}^{-2} + \overline{\boldsymbol{\Phi}}_{\mathfrak{s}}^{(2)} - \lambda\left(1-\mathfrak{s}\right) + \lambda^{2}\overline{\boldsymbol{G}}_{\mathfrak{s}}^{2}\left(1-\mathfrak{s}^{2}\right) - \frac{5}{2}\lambda^{3}\overline{\boldsymbol{G}}_{\mathfrak{s}}^{4}\left(1-\mathfrak{s}^{3}\right) \right)^{-2} \\
& \hspace{0.45cm} + \left(2\overline{\boldsymbol{G}}_{\mathfrak{s}}^{-2} + \overline{\boldsymbol{\Phi}}_{\mathfrak{s}}^{(2)} - \lambda\left(1-\mathfrak{s}\right) + \lambda^{2}\overline{\boldsymbol{G}}_{\mathfrak{s}}^{2}\left(1-\mathfrak{s}^{2}\right) - \frac{5}{2}\lambda^{3}\overline{\boldsymbol{G}}_{\mathfrak{s}}^{4}\left(1-\mathfrak{s}^{3}\right) \right)^{-2} \\
& \hspace{0.892cm} \times \left( 16\overline{\boldsymbol{G}}_{\mathfrak{s}}^{-4} - \overline{\boldsymbol{\Phi}}_{\mathfrak{s}}^{(4)} - 8\lambda^{2}\left(1-\mathfrak{s}^{2}\right) + 120 \lambda^{3} \overline{\boldsymbol{G}}_{\mathfrak{s}}^{2} \left(1-\mathfrak{s}^{3}\right) \right) + 2 \bigg) \\
& - \lambda + 2 \mathfrak{s} \lambda^{2} \overline{\boldsymbol{G}}_{\mathfrak{s}}^{2} - \frac{15}{2} \mathfrak{s}^{2} \lambda^{3} \overline{\boldsymbol{G}}_{\mathfrak{s}}^{4} + \frac{1}{2} \dot{\overline{\boldsymbol{G}}}_{\mathfrak{s}} \overline{\boldsymbol{\Phi}}_{\mathfrak{s}}^{(3)} \;,
\end{split}
\label{eq:2PIFRGmuflowNSCPT3FlowEqPhi20DON1Appendix}
\end{equation}
\begin{equation}
\begin{split}
\dot{\overline{\boldsymbol{\Phi}}}^{(3)}_{\mathfrak{s}} = & \ \frac{\lambda}{6} \left(2\overline{\boldsymbol{G}}_{\mathfrak{s}}^{-2} + \overline{\boldsymbol{\Phi}}_{\mathfrak{s}}^{(2)} - \lambda\left(1-\mathfrak{s}\right) + \lambda^{2}\overline{\boldsymbol{G}}_{\mathfrak{s}}^{2}\left(1-\mathfrak{s}^{2}\right) - \frac{5}{2}\lambda^{3}\overline{\boldsymbol{G}}_{\mathfrak{s}}^{4}\left(1-\mathfrak{s}^{3}\right) \right)^{-2} \\
& \times \bigg( 6 \left(2\overline{\boldsymbol{G}}_{\mathfrak{s}}^{-2} + \overline{\boldsymbol{\Phi}}_{\mathfrak{s}}^{(2)} - \lambda\left(1-\mathfrak{s}\right) + \lambda^{2}\overline{\boldsymbol{G}}_{\mathfrak{s}}^{2}\left(1-\mathfrak{s}^{2}\right) - \frac{5}{2}\lambda^{3}\overline{\boldsymbol{G}}_{\mathfrak{s}}^{4}\left(1-\mathfrak{s}^{3}\right) \right)^{-2} \\
& \hspace{1.1cm} \times \left(8\overline{\boldsymbol{G}}_{\mathfrak{s}}^{-3} - \overline{\boldsymbol{\Phi}}_{\mathfrak{s}}^{(3)} - 4 \lambda^{2} \overline{\boldsymbol{G}}_{\mathfrak{s}} \left(1-\mathfrak{s}^{2}\right) + 20 \lambda^{3} \overline{\boldsymbol{G}}_{\mathfrak{s}}^{3} \left(1-\mathfrak{s}^{3}\right) \right)^{3} \\
& \hspace{0.55cm} + 6 \left(2\overline{\boldsymbol{G}}_{\mathfrak{s}}^{-2} + \overline{\boldsymbol{\Phi}}_{\mathfrak{s}}^{(2)} - \lambda\left(1-\mathfrak{s}\right) + \lambda^{2}\overline{\boldsymbol{G}}_{\mathfrak{s}}^{2}\left(1-\mathfrak{s}^{2}\right) - \frac{5}{2}\lambda^{3}\overline{\boldsymbol{G}}_{\mathfrak{s}}^{4}\left(1-\mathfrak{s}^{3}\right) \right)^{-1} \\
& \hspace{1.3cm} \times \left(8\overline{\boldsymbol{G}}_{\mathfrak{s}}^{-3} - \overline{\boldsymbol{\Phi}}_{\mathfrak{s}}^{(3)} - 4 \lambda^{2} \overline{\boldsymbol{G}}_{\mathfrak{s}} \left(1-\mathfrak{s}^{2}\right) + 20 \lambda^{3} \overline{\boldsymbol{G}}_{\mathfrak{s}}^{3} \left(1-\mathfrak{s}^{3}\right) \right) \\
& \hspace{1.3cm} \times \left( -48\overline{\boldsymbol{G}}_{\mathfrak{s}}^{-4} - \overline{\boldsymbol{\Phi}}_{\mathfrak{s}}^{(4)} - 8\lambda^{2}\left(1-\mathfrak{s}^{2}\right) + 120 \lambda^{3} \overline{\boldsymbol{G}}_{\mathfrak{s}}^{2} \left(1-\mathfrak{s}^{3}\right) \right) \\
& \hspace{0.55cm} + 384 \overline{\boldsymbol{G}}_{\mathfrak{s}}^{-5} - \overline{\boldsymbol{\Phi}}_{\mathfrak{s}}^{(5)} + 480 \lambda^{3} \overline{\boldsymbol{G}}_{\mathfrak{s}} \left(1-\mathfrak{s}^{3}\right) \bigg) \\
& + 8 \mathfrak{s} \lambda^{2} \overline{\boldsymbol{G}}_{\mathfrak{s}} - 60 \mathfrak{s}^{2} \lambda^{3} \overline{\boldsymbol{G}}_{\mathfrak{s}}^{3} + \frac{1}{2} \dot{\overline{\boldsymbol{G}}}_{\mathfrak{s}} \overline{\boldsymbol{\Phi}}_{\mathfrak{s}}^{(4)} \;.
\end{split}
\label{eq:2PIFRGmuflowNSCPT3FlowEqPhi30DON1Appendix}
\end{equation}

\item For the CU-flow:
\begin{equation}
\dot{\overline{G}}_{\mathfrak{s}} = -\overline{G}_{\mathfrak{s}}^{2} \left( \dot{C}_{\mathfrak{s}}^{-1} - \dot{\overline{\Sigma}}_{\mathfrak{s}} \right) \;,
\label{eq:2PIFRGCUflowExpressionGdot0DON1Appendix}
\end{equation}
\begin{equation}
\Delta\dot{\overline{\Omega}}_{\mathfrak{s}} = \frac{1}{2} \dot{C}_{\mathfrak{s}}^{-1} \left( \overline{G}_{\mathfrak{s}} - C_{\mathfrak{s}} \right) + \frac{\lambda}{24} \left( 4\left(2\overline{G}_{\mathfrak{s}}^{-2} + \overline{\Phi}_{\mathfrak{s}}^{(2)} \right)^{-1} + \overline{G}_{\mathfrak{s}}^{2} \right) \;,
\label{eq:2PIFRGCUflowExpressionDeltaOmegadot0DON1Appendix}
\end{equation}
\begin{equation}
\dot{\overline{\Sigma}}_{\mathfrak{s}} = \frac{\lambda}{3} \overline{G}_{\mathfrak{s}} \left( 2 + \overline{G}_{\mathfrak{s}}^{2} \overline{\Phi}_{\mathfrak{s}}^{(2)} \right)^{-2} \left( \frac{1}{2} \overline{G}_{\mathfrak{s}}^{3} \overline{\Phi}_{\mathfrak{s}}^{(3)} - 4 \right) - \frac{\lambda}{6} \overline{G}_{\mathfrak{s}} - \frac{1}{2} \dot{\overline{G}}_{\mathfrak{s}} \overline{\Phi}_{\mathfrak{s}}^{(2)} \;,
\label{eq:2PIFRGCUflowExpressionSigmadot0DON1Appendix}
\end{equation}
\begin{equation}
\begin{split}
\dot{\overline{\Phi}}_{\mathfrak{s}}^{(2)} = & \ \frac{\lambda}{6} \bigg( 2\left( 2\overline{G}_{\mathfrak{s}}^{-2} + \overline{\Phi}_{\mathfrak{s}}^{(2)} \right)^{-3} \left( 8 \overline{G}_{\mathfrak{s}}^{-3} - \overline{\Phi}_{\mathfrak{s}}^{(3)} \right)^{2} - 64 \left( 2\overline{G}_{\mathfrak{s}}^{-2} + \overline{\Phi}_{\mathfrak{s}}^{(2)} \right)^{-2} \overline{G}_{\mathfrak{s}}^{-4} \\
& \hspace{0.5cm} + \left( 2\overline{G}_{\mathfrak{s}}^{-2} + \overline{\Phi}_{\mathfrak{s}}^{(2)} \right)^{-2} \left(16\overline{G}_{\mathfrak{s}}^{-4} - \overline{\Phi}_{\mathfrak{s}}^{(4)} \right) + 2 \bigg) + \frac{1}{2} \dot{\overline{G}}_{\mathfrak{s}} \overline{\Phi}_{\mathfrak{s}}^{(3)} \;,
\end{split}
\label{eq:2PIFRGCUflowExpressionPhi2dot0DON1Appendix}
\end{equation}
\begin{equation}
\begin{split}
\dot{\overline{\Phi}}_{\mathfrak{s}}^{(3)} = & \ \frac{\lambda}{6} \left( 2\overline{G}_{\mathfrak{s}}^{-2} + \overline{\Phi}_{\mathfrak{s}}^{(2)} \right)^{-2} \bigg( 6 \left( 2\overline{G}_{\mathfrak{s}}^{-2} + \overline{\Phi}_{\mathfrak{s}}^{(2)} \right)^{-2} \left( 8 \overline{G}_{\mathfrak{s}}^{-3} - \overline{\Phi}_{\mathfrak{s}}^{(3)} \right)^{3} \\
& \hspace{3.7cm} + 6 \left( 2\overline{G}_{\mathfrak{s}}^{-2} + \overline{\Phi}_{\mathfrak{s}}^{(2)} \right)^{-1} \left( 8 \overline{G}_{\mathfrak{s}}^{-3} - \overline{\Phi}_{\mathfrak{s}}^{(3)} \right) \left( 16 \overline{G}_{\mathfrak{s}}^{-4} - \overline{\Phi}_{\mathfrak{s}}^{(4)} \right) \\
& \hspace{3.7cm} - 384 \overline{G}_{\mathfrak{s}}^{-4} \left( \left( 2 \overline{G}_{\mathfrak{s}}^{-2} + \overline{\Phi}_{\mathfrak{s}}^{(2)} \right)^{-1}\left(8\overline{G}_{\mathfrak{s}}^{-3}-\overline{\Phi}_{\mathfrak{s}}^{(3)}\right) - \overline{G}_{\mathfrak{s}}^{-1} \right) - \overline{\Phi}_{\mathfrak{s}}^{(5)} \bigg) \\
& + \frac{1}{2} \dot{\overline{G}}_{\mathfrak{s}} \overline{\Phi}_{\mathfrak{s}}^{(4)} \;.
\end{split}
\label{eq:2PIFRGCUflowExpressionPhi3dot0DON1Appendix}
\end{equation}

\end{itemize}

\subsection{mU-flow equations at $N_{\mathrm{SCPT}}=1$ and for all $N$}
\label{ann:2PIfrgmUflow0DON}

As we have seen with the derivations of~\eqref{eq:DetailedCalculation1Sigma2PIFRGUflowAppendix} and~\eqref{eq:DetailedCalculation2Sigma2PIFRGUflowAppendix} to rewrite the U-flow equation~\eqref{eq:2PIfrgUflowSigmaDotCompact} expressing $\dot{\overline{\Sigma}}_{\mathfrak{s}}$, rewriting the 2PI-FRG flow equations by evaluating the derivatives $\frac{\delta\overline{\Pi}_{\mathfrak{s}}}{\delta\overline{G}_{\mathfrak{s},\gamma}}$ and/or $\frac{\delta^{2}\overline{\Pi}_{\mathfrak{s}}}{\delta\overline{G}_{\mathfrak{s},\gamma_{1}}\delta\overline{G}_{\mathfrak{s},\gamma_{2}}}$ is a cumbersome procedure\footnote{Recall that, for $n\geq 3$, we have $\frac{\delta^{n}\overline{\Pi}_{\mathfrak{s}}}{\delta\overline{G}_{\mathfrak{s},\gamma_{1}}\cdots\delta\overline{G}_{\mathfrak{s},\gamma_{n}}}= 0$ $\forall\gamma_{1},\cdots,\gamma_{n},\mathfrak{s}$ according to expression~\eqref{eq:2PIfrgExpressionPiandG} of $\Pi[G]$.}. Instead, we could of course just replace these derivatives by their expressions in terms of $\overline{G}_{\mathfrak{s}}$, Kronecker deltas and Dirac delta functions, such as~\eqref{eq:2PIFRGuflowExpressionDerivPidG} for $\frac{\delta\overline{\Pi}_{\mathfrak{s}}}{\delta\overline{G}_{\mathfrak{s},\gamma}}$, and treat numerically the (quite lengthy) differential equations thus obtained. However, we can also exploit a trick for the studied (0+0)-D model based on the fact that its 2PI and 2PPI EAs coincide in the absence of SSB, i.e. $\Gamma^{(\mathrm{2PI})}(G)=\Gamma^{(\mathrm{2PPI})}(\rho)$ (with $\Gamma^{(\mathrm{2PPI})}(\rho)$ defined from~\eqref{eq:LegendreTransform2PPIeffActionFor2PPIFRG} in arbitrary dimensions). We can therefore formulate a mU-flow from the master equation of the 2PPI-FRG (see appendix~\ref{sec:DerivMasterEq2PPIFRG}) which is treated as a next step with a vertex expansion up to $N_{\mathrm{max}}=3$. Such a procedure would therefore be equivalent to carrying out the mU-flow version of the 2PI-FRG up to $N_{\mathrm{max}}=3$ as well, owing to the connection between the 2PI-FRG and the vertex expansion highlighted via~\eqref{eq:VertexExpansion2PIFRG}. We will thus follow this recipe here to construct a mU-flow for the 2PPI-FRG with a Hartree-Fock starting point, i.e. with $N_{\mathrm{SCPT}}=1$. In order to achieve this, we start from the master equation of the 2PPI-FRG written for the (0+0)-D $O(N)$-symmetric $\varphi^{4}$-theory, i.e.:
\begin{equation}
\dot{\Gamma}^{(\mathrm{2PPI})}_{\mathfrak{s}}(\rho) = \frac{\lambda}{24} \left( \sum_{a_{1},a_{2}=1}^{N} \left(\Gamma_{\mathfrak{s}}^{(\mathrm{2PPI})(2)}(\rho)\right)_{a_{1}a_{2}}^{-1} + \sum_{a_{1},a_{2}=1}^{N} \rho_{a_{1}} \rho_{a_{2}} \right) \;,
\label{eq:2PPIFRGmasterEquation0DONAppendix}
\end{equation}
as was already given by~\eqref{eq:2PPIFRGmasterEquation0DON} and~\eqref{eq:standard2PPIFRGvertExpPropagator}. We then split the 2PPI EA $\Gamma^{(\mathrm{2PPI})}_{\mathfrak{s}}(\rho)$ according to:
\begin{equation}
\Gamma^{(\mathrm{2PPI})}_{\mathfrak{s}}(\rho) = \Gamma^{(\mathrm{2PPI})}_{0,\mathfrak{s}}(\rho) + \Phi_{\mathfrak{s}}(\rho) \;,
\label{eq:2PPIFRGmUflowsplitting2PPIEAAppendix}
\end{equation}
with $\Gamma^{(\mathrm{2PPI})}_{0,\mathfrak{s}}(\rho)$ and $\Phi_{\mathfrak{s}}(\rho)$ being respectively the free and interacting parts of $\Gamma^{(\mathrm{2PPI})}_{\mathfrak{s}}(\rho)$. Since $\Gamma^{(\mathrm{2PI})}(G)=\Gamma^{(\mathrm{2PPI})}(\rho)$, $\Phi_{\mathfrak{s}}(\rho)$ coincides with the Luttinger-Ward functional in the present case. As for the 2PI-FRG, the mU-flow is implemented by introducing the modified Luttinger-Ward functional defined by~\eqref{eq:DefinitionModifiedLWFunctionalNSCPT10DON} and~\eqref{eq:DefinitionmUflowPhiboldNscpt10DON} recalled below:
\begin{equation}
\boldsymbol{\Phi}_{\mathfrak{s}}(G) \equiv \Phi_{\mathfrak{s}}(G) + \Phi_{\mathrm{SCPT},N_{\mathrm{SCPT}}=1}(U,G) - \Phi_{\mathrm{SCPT},N_{\mathrm{SCPT}}=1}(U_{\mathfrak{s}},G) \;,
\label{eq:DefinitionModifiedLWFunctionalNSCPT10DONAppendix}
\end{equation}
with
\begin{equation}
\Phi_{\mathrm{SCPT},N_{\mathrm{SCPT}}=1}(U,G) = \lambda \left( \frac{1}{24}\left(\sum_{a_{1}=1}^{N} G_{a_{1}a_{1}} \right)^{2} + \frac{1}{12} \sum_{a_{1},a_{2}=1}^{N} G_{a_{1}a_{2}}^{2} \right) \;,
\label{eq:2PPIFRGmUflowDefPhiUAppendix}
\end{equation}
and
\begin{equation}
\Phi_{\mathrm{SCPT},N_{\mathrm{SCPT}}=1}(U_{\mathfrak{s}},G) = \mathfrak{s} \lambda \left( \frac{1}{24}\left(\sum_{a_{1}=1}^{N} G_{a_{1}a_{1}} \right)^{2} + \frac{1}{12} \sum_{a_{1},a_{2}=1}^{N} G_{a_{1}a_{2}}^{2} \right) \;.
\label{eq:2PPIFRGmUflowDefPhiUsAppendix}
\end{equation}
Due to the absence of SSB, we know that the propagator $G$ is a scalar in color space, i.e. $G_{a_{1} a_{2}} = G \ \delta_{a_{1} a_{2}}$ $\forall a_{1}, a_{2}$, which implies that the less constraining relation:
\begin{equation}
G_{a_{1} a_{2}} = \rho_{a_{1}} \delta_{a_{1} a_{2}} \mathrlap{\quad \forall a_{1}, a_{2} \;,}
\label{eq:2PPIFRGmUflowLink2PI2PPIEAAppendix}
\end{equation}
must be satisfied as well. Combining~\eqref{eq:2PPIFRGmUflowLink2PI2PPIEAAppendix} with~\eqref{eq:DefinitionModifiedLWFunctionalNSCPT10DONAppendix} to~\eqref{eq:2PPIFRGmUflowDefPhiUsAppendix} leads to:
\begin{equation}
\boldsymbol{\Phi}_{\mathfrak{s}}(\rho) \equiv \Phi_{\mathfrak{s}}(\rho) + \Phi_{\mathrm{SCPT},N_{\mathrm{SCPT}}=1}(U,\rho) - \Phi_{\mathrm{SCPT},N_{\mathrm{SCPT}}=1}(U_{\mathfrak{s}},\rho) \;,
\label{eq:DefinitionModifiedLWFunctional2PPIEANSCPT10DONAppendix}
\end{equation}
with
\begin{equation}
\Phi_{\mathrm{SCPT},N_{\mathrm{SCPT}}=1}(U,\rho) = \lambda \left( \frac{1}{24}\left(\sum_{a=1}^{N} \rho_{a} \right)^{2} + \frac{1}{12} \sum_{a=1}^{N} \rho_{a}^{2} \right) \;,
\label{eq:2PPIFRGmUflow2PPIEADefPhiUAppendix}
\end{equation}
and
\begin{equation}
\Phi_{\mathrm{SCPT},N_{\mathrm{SCPT}}=1}(U_{\mathfrak{s}},\rho) = \mathfrak{s} \lambda \left( \frac{1}{24}\left(\sum_{a=1}^{N} \rho_{a} \right)^{2} + \frac{1}{12} \sum_{a=1}^{N} \rho_{a}^{2} \right) \;.
\label{eq:2PPIFRGmUflow2PPIEADefPhiUsAppendix}
\end{equation}
According to~\eqref{eq:DefinitionModifiedLWFunctional2PPIEANSCPT10DONAppendix}, the splitting~\eqref{eq:2PPIFRGmUflowsplitting2PPIEAAppendix} is equivalent to:
\begin{equation}
\begin{split}
\Gamma^{(\mathrm{2PPI})}_{\mathfrak{s}}(\rho) = & \ \underbrace{\Gamma^{(\mathrm{2PPI})}_{0,\mathfrak{s}}(\rho) + \boldsymbol{\Phi}_{\mathfrak{s}}(\rho)}_{\boldsymbol{\Gamma}^{(\mathrm{2PPI})}_{\mathfrak{s}}(\rho)} - \Phi_{\mathrm{SCPT},N_{\mathrm{SCPT}}=1}(U,\rho) + \Phi_{\mathrm{SCPT},N_{\mathrm{SCPT}}=1}(U_{\mathfrak{s}},\rho) \\
= & \ \boldsymbol{\Gamma}^{(\mathrm{2PPI})}_{\mathfrak{s}}(\rho) - \Phi_{\mathrm{SCPT},N_{\mathrm{SCPT}}=1}(U,\rho) + \Phi_{\mathrm{SCPT},N_{\mathrm{SCPT}}=1}(U_{\mathfrak{s}},\rho) \;,
\end{split}
\label{eq:2PPIFRGmUflowsplittingbis2PPIEAAppendix}
\end{equation}
which enables us to rewrite the master equation~\eqref{eq:2PPIFRGmasterEquation0DONAppendix} as:
\begin{equation}
\dot{\boldsymbol{\Gamma}}^{(\mathrm{2PPI})}_{\mathfrak{s}}(\rho) = \frac{\lambda}{24} \left( \sum_{a_{1},a_{2}=1}^{N} \mathfrak{G}_{\mathfrak{s},a_{1}a_{2}}(\rho) + \sum_{a_{1},a_{2}=1}^{N} \rho_{a_{1}} \rho_{a_{2}} \right) - \dot{\Phi}_{\mathrm{SCPT},N_{\mathrm{SCPT}}=1}(U_{\mathfrak{s}},\rho) \;,
\label{eq:2PPIFRGmasterEquation0DONbisAppendix}
\end{equation}
where
\begin{equation}
\begin{split}
\mathfrak{G}^{-1}_{\mathfrak{s},a_{1}a_{2}}(\rho) \equiv & \ \Gamma_{\mathfrak{s},a_{1}a_{2}}^{(\mathrm{2PPI})(2)}(\rho) \\
= & \ \boldsymbol{\Gamma}_{\mathfrak{s},a_{1}a_{2}}^{(\mathrm{2PPI})(2)}(\rho) - \frac{\partial^{2}\Phi_{\mathrm{SCPT},N_{\mathrm{SCPT}}=1}(U,\rho)}{\partial\rho_{a_{1}}\partial\rho_{a_{2}}} + \frac{\partial^{2}\Phi_{\mathrm{SCPT},N_{\mathrm{SCPT}}=1}(U_{\mathfrak{s}},\rho)}{\partial\rho_{a_{1}}\partial\rho_{a_{2}}} \;.
\end{split}
\label{eq:2PPIFRGmuflowpropagatorAppendix}
\end{equation}
We then extract a tower of differential equations from~\eqref{eq:2PPIFRGmasterEquation0DONbisAppendix} by performing a vertex expansion of $\boldsymbol{\Gamma}_{\mathfrak{s}}^{(\mathrm{2PPI})}(\rho)$ around its flowing extremum at $\rho=\overline{\boldsymbol{\rho}}_{\mathfrak{s}}$, i.e.:
\begin{equation}
\boldsymbol{\Gamma}_{\mathfrak{s}}^{(\mathrm{2PPI})}(\rho) = \overline{\boldsymbol{\Gamma}}_{\mathfrak{s}}^{(\mathrm{2PPI})} + \sum_{n=2}^{\infty} \frac{1}{n!} \sum_{a_{1},\cdots,a_{n}=1}^{N} \overline{\boldsymbol{\Gamma}}_{\mathfrak{s},a_{1},\cdots,a_{n}}^{(\mathrm{2PPI})(n)} \left(\rho-\overline{\boldsymbol{\rho}}_{\mathfrak{s}}\right)_{a_{1}} \cdots \left(\rho-\overline{\boldsymbol{\rho}}_{\mathfrak{s}}\right)_{a_{n}} \;,
\label{eq:2PPIFRGmuflowVertexExpansionAppendix}
\end{equation}
with
\begin{equation}
\overline{\boldsymbol{\Gamma}}^{(\mathrm{2PPI})}_{\mathfrak{s}} \equiv \boldsymbol{\Gamma}^{(\mathrm{2PPI})}_{\mathfrak{s}}\big(\rho=\overline{\boldsymbol{\rho}}_{\mathfrak{s}}\big) \mathrlap{\quad \forall \mathfrak{s} \;,}
\label{eq:2PPIFRGmUflowDefGammabar1PIFRG0DONAppendix}
\end{equation}
\begin{equation}
\overline{\boldsymbol{\Gamma}}^{(\mathrm{2PPI})(n)}_{\mathfrak{s},a_{1} \cdots a_{n}} \equiv \left. \frac{\partial^{n} \boldsymbol{\Gamma}^{(\mathrm{2PPI})}_{\mathfrak{s}}(\rho)}{\partial \rho_{a_{1}} \cdots \partial \rho_{a_{n}}} \right|_{\rho=\overline{\boldsymbol{\rho}}_{\mathfrak{s}}} \mathrlap{\quad \forall a_{1},\cdots,a_{n},\mathfrak{s} \;,}
\label{eq:2PPIFRGmUflowDefGammanbar1PIFRG0DONAppendix}
\end{equation}
and, in particular,
\begin{equation}
\overline{\boldsymbol{\Gamma}}_{\mathfrak{s},a}^{(\mathrm{2PPI})(1)} = 0 \mathrlap{\quad \forall a, \mathfrak{s} \;.}
\label{eq:2PPIFRGmUflowDefGamma1bar1PIFRG0DONAppendix}
\end{equation}
Differentiating~\eqref{eq:2PPIFRGmuflowVertexExpansionAppendix} with respect to $\mathfrak{s}$ yields:
\begin{equation}
\begin{split}
\scalebox{0.99}{${\displaystyle \dot{\boldsymbol{\Gamma}}_{\mathfrak{s}}^{(\mathrm{2PPI})}(\rho) = }$} & \ \scalebox{0.99}{${\displaystyle \dot{\overline{\boldsymbol{\Gamma}}}_{\mathfrak{s}}^{(\mathrm{2PPI})} - \sum_{a_{1},a_{2}=1}^{N} \dot{\overline{\boldsymbol{\rho}}}_{a_{2}} \overline{\boldsymbol{\Gamma}}^{(\mathrm{2PPI})(2)}_{\mathfrak{s},a_{2}a_{1}} \left(\rho-\overline{\boldsymbol{\rho}}_{\mathfrak{s}}\right)_{a_{1}} }$} \\
& \scalebox{0.99}{${\displaystyle + \sum_{n=2}^{N} \frac{1}{n!} \sum_{a_{1},\cdots,a_{n}=1}^{N} \left( \dot{\overline{\boldsymbol{\Phi}}}_{\mathfrak{s},a_{1} \cdots a_{n}}^{(n)} - \sum_{a_{n+1}=1}^{N} \dot{\overline{\boldsymbol{\rho}}}_{\mathfrak{s},a_{n+1}} \overline{\boldsymbol{\Phi}}_{\mathfrak{s},a_{n+1} a_{1} \cdots a_{n}}^{(n+1)} \right) \left(\rho-\overline{\boldsymbol{\rho}}_{\mathfrak{s}}\right)_{a_{1}} \cdots \left(\rho-\overline{\boldsymbol{\rho}}_{\mathfrak{s}}\right)_{a_{n}} \;, }$}
\end{split}
\label{eq:2PPIFRGmUflowTaylorExpandGamma2PPIdotAppendix}
\end{equation}
where we have used the relation:
\begin{equation}
\dot{\overline{\boldsymbol{\Gamma}}}^{(\mathrm{2PPI})(n)}_{\mathfrak{s},a_{1} \cdots a_{n}} - \sum_{a_{n+1}=1}^{N} \dot{\overline{\boldsymbol{\rho}}}_{\mathfrak{s},a_{n+1}} \overline{\boldsymbol{\Gamma}}^{(\mathrm{2PPI})(n+1)}_{\mathfrak{s}, a_{n+1} a_{1} \cdots a_{n}} = \dot{\overline{\boldsymbol{\Phi}}}^{(\mathrm{2PPI})(n)}_{\mathfrak{s},a_{1} \cdots a_{n}} - \sum_{a_{n+1}=1}^{N} \dot{\overline{\boldsymbol{\rho}}}_{\mathfrak{s},a_{n+1}} \overline{\boldsymbol{\Phi}}^{(n+1)}_{\mathfrak{s}, a_{n+1} a_{1} \cdots a_{n}} \;,
\end{equation}
which results from the following chain rule (derived from the equality $\dot{\boldsymbol{\Gamma}}_{\mathfrak{s}}^{(\mathrm{2PPI})(n)}=\dot{\boldsymbol{\Phi}}_{\mathfrak{s}}^{(n)}$ $\forall n,\mathfrak{s}$):
\begin{equation}
\begin{split}
\dot{\overline{\boldsymbol{\Gamma}}}^{(\mathrm{2PPI})(n)}_{\mathfrak{s},a_{1} \cdots a_{n}} = & \ \overline{\dot{\boldsymbol{\Gamma}}}^{(\mathrm{2PPI})(n)}_{\mathfrak{s},a_{1} \cdots a_{n}} + \sum_{a_{n+1}=1}^{N} \dot{\overline{\boldsymbol{\rho}}}_{\mathfrak{s},a_{n+1}} \overline{\boldsymbol{\Gamma}}^{(\mathrm{2PPI})(n+1)}_{\mathfrak{s}, a_{n+1} a_{1} \cdots a_{n}} \\
= & \ \overline{\dot{\boldsymbol{\Phi}}}^{(n)}_{\mathfrak{s},a_{1} \cdots a_{n}} + \sum_{a_{n+1}=1}^{N} \dot{\overline{\boldsymbol{\rho}}}_{\mathfrak{s},a_{n+1}} \overline{\boldsymbol{\Gamma}}^{(\mathrm{2PPI})(n+1)}_{\mathfrak{s}, a_{n+1} a_{1} \cdots a_{n}} \\
= & \ \dot{\overline{\boldsymbol{\Phi}}}^{(n)}_{\mathfrak{s},a_{1} \cdots a_{n}} - \sum_{a_{n+1}=1}^{N} \dot{\overline{\boldsymbol{\rho}}}_{\mathfrak{s},a_{n+1}} \overline{\boldsymbol{\Phi}}^{(n+1)}_{\mathfrak{s}, a_{n+1} a_{1} \cdots a_{n}} + \sum_{a_{n+1}=1}^{N} \dot{\overline{\boldsymbol{\rho}}}_{\mathfrak{s},a_{n+1}} \overline{\boldsymbol{\Gamma}}^{(\mathrm{2PPI})(n+1)}_{\mathfrak{s}, a_{n+1} a_{1} \cdots a_{n}} \;.
\end{split}
\end{equation}
Furthermore, we calculate the derivatives of $\dot{\Phi}_{\mathrm{SCPT},N_{\mathrm{SCPT}}=1}(U_{\mathfrak{s}},\rho)=\Phi_{\mathrm{SCPT},N_{\mathrm{SCPT}}=1}(U,\rho)$ with respect to $\rho$ from~\eqref{eq:2PPIFRGmUflow2PPIEADefPhiUAppendix} or~\eqref{eq:2PPIFRGmUflow2PPIEADefPhiUsAppendix}:
\begin{equation}
\frac{\partial\dot{\Phi}_{\mathrm{SCPT},N_{\mathrm{SCPT}}=1}(U_{\mathfrak{s}},\rho)}{\partial\rho_{a_{1}}} = \frac{\partial\Phi_{\mathrm{SCPT},N_{\mathrm{SCPT}}=1}(U,\rho)}{\partial\rho_{a_{1}}} = \frac{\lambda}{12} \sum_{a_{2}=1}^{N} \rho_{a_{2}} + \frac{\lambda}{6} \rho_{a_{1}} \mathrlap{\;,}
\label{eq:2PPIFRGmUflowDerivPhiSCPT1Appendix}
\end{equation}
\begin{equation}
\frac{\partial^{2}\dot{\Phi}_{\mathrm{SCPT},N_{\mathrm{SCPT}}=1}(U_{\mathfrak{s}},\rho)}{\partial\rho_{a_{1}}\partial\rho_{a_{2}}} = \frac{\partial^{2}\Phi_{\mathrm{SCPT},N_{\mathrm{SCPT}}=1}(U,\rho)}{\partial\rho_{a_{1}}\partial\rho_{a_{2}}} = \frac{\lambda}{12} \left( 1 + 2 \delta_{a_{1} a_{2}} \right) \mathrlap{\;,}
\label{eq:2PPIFRGmUflowDerivPhiSCPT2Appendix}
\end{equation}
\begin{equation}
\frac{\partial^{n}\dot{\Phi}_{\mathrm{SCPT},N_{\mathrm{SCPT}}=1}(U_{\mathfrak{s}},\rho)}{\partial\rho_{a_{1}} \cdots \partial\rho_{a_{n}}} = \frac{\partial^{n}\Phi_{\mathrm{SCPT},N_{\mathrm{SCPT}}=1}(U,\rho)}{\partial\rho_{a_{1}} \cdots \partial\rho_{a_{n}}} = 0 \mathrlap{\quad \forall n \geq 3 \;.}
\label{eq:2PPIFRGmUflowDerivPhiSCPT3Appendix}
\end{equation}
From this, we infer the Taylor expansion of $\dot{\Phi}_{\mathrm{SCPT},N_{\mathrm{SCPT}}=1}(U_{\mathfrak{s}},\rho)$ around $\rho=\overline{\boldsymbol{\rho}}_{\mathfrak{s}}$:
\begin{equation}
\begin{split}
\dot{\Phi}_{\mathrm{SCPT},N_{\mathrm{SCPT}}=1}(U_{\mathfrak{s}},\rho) = & \ \dot{\Phi}_{\mathrm{SCPT},N_{\mathrm{SCPT}}=1}\big(U_{\mathfrak{s}},\rho=\overline{\boldsymbol{\rho}}_{\mathfrak{s}}\big) \\
& + \sum_{n=1}^{\infty} \frac{1}{n!} \sum_{a_{1},\cdots,a_{n}=1}^{N} \left.\frac{\partial^{n}\dot{\Phi}_{\mathrm{SCPT},N_{\mathrm{SCPT}}=1}(U_{\mathfrak{s}},\rho)}{\partial\rho_{a_{1}}\cdots\partial\rho_{a_{n}}}\right|_{\rho=\overline{\boldsymbol{\rho}}_{\mathfrak{s}}} \left(\rho-\overline{\boldsymbol{\rho}}_{\mathfrak{s}}\right)_{a_{1}} \cdots \left(\rho-\overline{\boldsymbol{\rho}}_{\mathfrak{s}}\right)_{a_{n}} \\
= & \ \lambda \left( \frac{1}{24}\left(\sum_{a_{1}=1}^{N} \overline{\boldsymbol{\rho}}_{\mathfrak{s},a_{1}} \right)^{2} + \frac{1}{12} \sum_{a_{1}=1}^{N} \overline{\boldsymbol{\rho}}_{\mathfrak{s},a_{1}}^{2} \right) \\
& + \frac{\lambda}{12} \sum_{a_{1}=1}^{N} \left( \sum_{a_{2}=1}^{N} \overline{\boldsymbol{\rho}}_{\mathfrak{s},a_{2}} + 2 \overline{\boldsymbol{\rho}}_{\mathfrak{s},a_{1}} \right) \left(\rho-\overline{\boldsymbol{\rho}}_{\mathfrak{s}}\right)_{a_{1}} \\
& + \frac{\lambda}{24} \sum_{a_{1},a_{2}=1}^{N} \left(1+2\delta_{a_{1}a_{2}}\right) \left(\rho-\overline{\boldsymbol{\rho}}_{\mathfrak{s}}\right)_{a_{1}} \left(\rho-\overline{\boldsymbol{\rho}}_{\mathfrak{s}}\right)_{a_{2}} \;.
\end{split}
\label{eq:2PPIFRGmUflowTaylorExpandPhidotSCPTAppendix}
\end{equation}
In the same way, the middle term in the RHS of~\eqref{eq:2PPIFRGmasterEquation0DONbisAppendix} can also be rewritten exactly in terms of $\rho-\overline{\boldsymbol{\rho}}_{\mathfrak{s}}$:
\begin{equation}
\begin{split}
\frac{\lambda}{24} \sum_{a_{1},a_{2}=1}^{N} \rho_{a_{1}} \rho_{a_{2}} = & \ \frac{\lambda}{24} \sum_{a_{1},a_{2}=1}^{N} \overline{\boldsymbol{\rho}}_{\mathfrak{s},a_{1}} \overline{\boldsymbol{\rho}}_{\mathfrak{s},a_{2}} \\
& + \frac{\lambda}{12} \left(\sum_{a_{2}=1}^{N}\overline{\boldsymbol{\rho}}_{\mathfrak{s},a_{2}}\right) \sum_{a_{1}=1}^{N}\left(\rho-\overline{\boldsymbol{\rho}}_{\mathfrak{s}}\right)_{a_{1}} \\
& + \frac{\lambda}{24} \sum_{a_{1},a_{2}=1}^{N} \left(\rho-\overline{\boldsymbol{\rho}}_{\mathfrak{s}}\right)_{a_{1}} \left(\rho-\overline{\boldsymbol{\rho}}_{\mathfrak{s}}\right)_{a_{2}} \;.
\end{split}
\label{eq:2PPIFRGmUflowTaylorExpandSumRho2Appendix}
\end{equation}
The only term left to expand in the master equation~\eqref{eq:2PPIFRGmasterEquation0DONbisAppendix} is:
\begin{equation}
\frac{\lambda}{24} \sum_{a_{1},a_{2}=1}^{N} \mathfrak{G}_{\mathfrak{s},a_{1}a_{2}}(\rho) \;.
\end{equation}
One can first expand the propagator $\mathfrak{G}_{\mathfrak{s}}(\rho)$ very conveniently in a diagrammatic fashion to derive such an expansion. The recipe to achieve this was already described in detail in section~\ref{sec:VertexExpansionApp1PIFRG} on the vertex expansion in the framework of the 1PI-FRG and we will therefore not repeat it here. We just point out instead that the expansion thus obtained is then inserted into the flow equation~\eqref{eq:2PPIFRGmasterEquation0DONbisAppendix}, alongside with~\eqref{eq:2PPIFRGmUflowTaylorExpandGamma2PPIdotAppendix},~\eqref{eq:2PPIFRGmUflowTaylorExpandPhidotSCPTAppendix} and~\eqref{eq:2PPIFRGmUflowTaylorExpandSumRho2Appendix}. The LHS and RHS of~\eqref{eq:2PPIFRGmasterEquation0DONbisAppendix} are both expanded in terms of $\rho-\overline{\boldsymbol{\rho}}_{\mathfrak{s}}$ in this way and identifying the terms with identical powers of $\rho-\overline{\boldsymbol{\rho}}_{\mathfrak{s}}$ gives us the tower of differential equations for the present mU-flow approach. Up to the truncation order $N_{\mathrm{max}}=3$ (i.e. up to order $\mathcal{O}((\rho-\overline{\boldsymbol{\rho}}_{\mathfrak{s}})^{3})$ in the expansion~\eqref{eq:2PPIFRGmuflowVertexExpansionAppendix}), we obtain:
\begin{equation}
\dot{\overline{\boldsymbol{\Gamma}}}_{\mathfrak{s}}^{(\mathrm{2PPI})} = \frac{\lambda}{24} \left( \sum_{a_{1},a_{2}=1}^{N} \overline{\mathfrak{G}}_{\mathfrak{s},a_{1}a_{2}} - 2 \sum_{a_{1}=1}^{N} \overline{\boldsymbol{\rho}}_{\mathfrak{s},a_{1}}^{2} \right) \;,
\label{eq:2PPIFRGmuflowEqExpressionGammadotAppendix}
\end{equation}
\begin{equation}
\scalebox{0.95}{${\displaystyle \dot{\overline{\boldsymbol{\rho}}}_{\mathfrak{s},a_{1}} = \frac{\lambda}{24} \sum_{a_{2}=1}^{N} \left(\overline{\Gamma}_{0,\mathfrak{s}}^{(\mathrm{2PPI})(2)}+\overline{\boldsymbol{\Phi}}_{\mathfrak{s}}^{(2)}\right)^{-1}_{a_{1}a_{2}} \left( \sum_{a_{3},a_{4},a_{5},a_{6}=1}^{N} \overline{\mathfrak{G}}_{\mathfrak{s},a_{3}a_{5}} \left(\overline{\Gamma}_{0,\mathfrak{s}}^{(\mathrm{2PPI})(3)}+\overline{\boldsymbol{\Phi}}_{\mathfrak{s}}^{(3)}\right)_{a_{2}a_{5}a_{6}} \overline{\mathfrak{G}}_{\mathfrak{s},a_{6}a_{4}} + 4 \overline{\boldsymbol{\rho}}_{\mathfrak{s},a_{2}} \right) \;, }$}
\label{eq:2PPIFRGmuflowEqExpressionrhodotAppendix}
\end{equation}
\begin{equation}
\begin{split}
\scalebox{0.94}{${\displaystyle \dot{\overline{\boldsymbol{\Phi}}}_{\mathfrak{s},a_{1}a_{2}}^{(2)} = }$} & \ \scalebox{0.94}{${\displaystyle \sum_{a_{3}=1}^{N} \dot{\overline{\boldsymbol{\rho}}}_{\mathfrak{s},a_{3}} \overline{\boldsymbol{\Phi}}^{(3)}_{\mathfrak{s},a_{3}a_{1}a_{2}} }$} \\
& \scalebox{0.94}{${\displaystyle + \frac{\lambda}{24} \Bigg( 2 \sum_{a_{3},a_{4},a_{5},a_{6},a_{7},a_{8}=1}^{N} \overline{\mathfrak{G}}_{\mathfrak{s},a_{3}a_{5}} \left(\overline{\Gamma}_{0,\mathfrak{s}}^{(\mathrm{2PPI})(3)} + \overline{\boldsymbol{\Phi}}^{(3)}_{\mathfrak{s}}\right)_{a_{1}a_{5}a_{6}} \overline{\mathfrak{G}}_{\mathfrak{s},a_{6}a_{7}} \left(\overline{\Gamma}_{0,\mathfrak{s}}^{(\mathrm{2PPI})(3)} + \overline{\boldsymbol{\Phi}}^{(3)}_{\mathfrak{s}}\right)_{a_{2}a_{7}a_{8}} \overline{\mathfrak{G}}_{\mathfrak{s},a_{8}a_{4}} }$} \\
& \ \hspace{0.85cm} \scalebox{0.94}{${\displaystyle - \sum_{a_{3},a_{4},a_{5},a_{6}=1}^{N} \overline{\mathfrak{G}}_{\mathfrak{s},a_{3}a_{5}} \left(\overline{\Gamma}_{0,\mathfrak{s}}^{(\mathrm{2PPI})(4)} + \overline{\boldsymbol{\Phi}}^{(4)}_{\mathfrak{s}}\right)_{a_{1}a_{2}a_{5}a_{6}} \overline{\mathfrak{G}}_{\mathfrak{s},a_{6}a_{4}} - 4 \delta_{a_{1}a_{2}} \Bigg) \;, }$}
\end{split}
\label{eq:2PPIFRGmuflowEqExpressionPhi2dotAppendix}
\end{equation}
\begin{equation}
\begin{split}
\scalebox{0.89}{${\displaystyle \dot{\overline{\boldsymbol{\Phi}}}_{\mathfrak{s},a_{1}a_{2}a_{3}}^{(3)} = }$} & \ \scalebox{0.89}{${\displaystyle\sum_{a_{4}=1}^{N} \dot{\overline{\boldsymbol{\rho}}}_{\mathfrak{s},a_{4}} \overline{\boldsymbol{\Phi}}^{(4)}_{\mathfrak{s},a_{4}a_{1}a_{2}a_{3}} }$} \\
& \scalebox{0.89}{${\displaystyle + \frac{\lambda}{24} \Bigg( -2 \sum_{a_{4},a_{5},a_{6},a_{7},a_{8},a_{9},a_{10},a_{11}=1}^{N} \overline{\mathfrak{G}}_{\mathfrak{s},a_{4}a_{6}} \left(\overline{\Gamma}_{0,\mathfrak{s}}^{(\mathrm{2PPI})(3)} + \overline{\boldsymbol{\Phi}}^{(3)}_{\mathfrak{s}}\right)_{a_{1}a_{6}a_{7}} \overline{\mathfrak{G}}_{\mathfrak{s},a_{7}a_{8}} \left(\overline{\Gamma}_{0,\mathfrak{s}}^{(\mathrm{2PPI})(3)} + \overline{\boldsymbol{\Phi}}^{(3)}_{\mathfrak{s}}\right)_{a_{2}a_{8}a_{9}} }$} \\
& \hspace{4.9cm} \scalebox{0.89}{${\displaystyle \times \overline{\mathfrak{G}}_{\mathfrak{s},a_{9}a_{10}} \left(\overline{\Gamma}_{0,\mathfrak{s}}^{(\mathrm{2PPI})(3)} + \overline{\boldsymbol{\Phi}}^{(3)}_{\mathfrak{s}}\right)_{a_{3}a_{10}a_{11}} \overline{\mathfrak{G}}_{\mathfrak{s},a_{11}a_{5}} }$} \\
& \scalebox{0.89}{${\displaystyle \hspace{1.25cm} -2 \sum_{a_{4},a_{5},a_{6},a_{7},a_{8},a_{9},a_{10},a_{11}=1}^{N} \overline{\mathfrak{G}}_{\mathfrak{s},a_{4}a_{6}} \left(\overline{\Gamma}_{0,\mathfrak{s}}^{(\mathrm{2PPI})(3)} + \overline{\boldsymbol{\Phi}}^{(3)}_{\mathfrak{s}}\right)_{a_{2}a_{6}a_{7}} \overline{\mathfrak{G}}_{\mathfrak{s},a_{7}a_{8}} \left(\overline{\Gamma}_{0,\mathfrak{s}}^{(\mathrm{2PPI})(3)} + \overline{\boldsymbol{\Phi}}^{(3)}_{\mathfrak{s}}\right)_{a_{1}a_{8}a_{9}} }$} \\
& \hspace{4.9cm} \scalebox{0.89}{${\displaystyle \times \overline{\mathfrak{G}}_{\mathfrak{s},a_{9}a_{10}} \left(\overline{\Gamma}_{0,\mathfrak{s}}^{(\mathrm{2PPI})(3)} + \overline{\boldsymbol{\Phi}}^{(3)}_{\mathfrak{s}}\right)_{a_{3}a_{10}a_{11}} \overline{\mathfrak{G}}_{\mathfrak{s},a_{11}a_{5}} }$} \\
& \scalebox{0.89}{${\displaystyle \hspace{1.25cm} -2 \sum_{a_{4},a_{5},a_{6},a_{7},a_{8},a_{9},a_{10},a_{11}=1}^{N} \overline{\mathfrak{G}}_{\mathfrak{s},a_{4}a_{6}} \left(\overline{\Gamma}_{0,\mathfrak{s}}^{(\mathrm{2PPI})(3)} + \overline{\boldsymbol{\Phi}}^{(3)}_{\mathfrak{s}}\right)_{a_{1}a_{6}a_{7}} \overline{\mathfrak{G}}_{\mathfrak{s},a_{7}a_{8}} \left(\overline{\Gamma}_{0,\mathfrak{s}}^{(\mathrm{2PPI})(3)} + \overline{\boldsymbol{\Phi}}^{(3)}_{\mathfrak{s}}\right)_{a_{3}a_{8}a_{9}} }$} \\
& \hspace{4.9cm} \scalebox{0.89}{${\displaystyle \times \overline{\mathfrak{G}}_{\mathfrak{s},a_{9}a_{10}} \left(\overline{\Gamma}_{0,\mathfrak{s}}^{(\mathrm{2PPI})(3)} + \overline{\boldsymbol{\Phi}}^{(3)}_{\mathfrak{s}}\right)_{a_{2}a_{10}a_{11}} \overline{\mathfrak{G}}_{\mathfrak{s},a_{11}a_{5}} }$} \\
& \scalebox{0.89}{${\displaystyle \hspace{1.25cm} +2 \sum_{a_{4},a_{5},a_{6},a_{7},a_{8},a_{9}=1}^{N} \overline{\mathfrak{G}}_{\mathfrak{s},a_{4}a_{6}} \left(\overline{\Gamma}_{0,\mathfrak{s}}^{(\mathrm{2PPI})(4)} + \overline{\boldsymbol{\Phi}}^{(4)}_{\mathfrak{s}}\right)_{a_{1}a_{2}a_{6}a_{7}} \overline{\mathfrak{G}}_{\mathfrak{s},a_{7}a_{8}} \left(\overline{\Gamma}_{0,\mathfrak{s}}^{(\mathrm{2PPI})(3)} + \overline{\boldsymbol{\Phi}}^{(3)}_{\mathfrak{s}}\right)_{a_{3}a_{8}a_{9}} \overline{\mathfrak{G}}_{\mathfrak{s},a_{9}a_{5}} }$} \\
& \scalebox{0.89}{${\displaystyle \hspace{1.25cm} +2 \sum_{a_{4},a_{5},a_{6},a_{7},a_{8},a_{9}=1}^{N} \overline{\mathfrak{G}}_{\mathfrak{s},a_{4}a_{6}} \left(\overline{\Gamma}_{0,\mathfrak{s}}^{(\mathrm{2PPI})(4)} + \overline{\boldsymbol{\Phi}}^{(4)}_{\mathfrak{s}}\right)_{a_{1}a_{3}a_{6}a_{7}} \overline{\mathfrak{G}}_{\mathfrak{s},a_{7}a_{8}} \left(\overline{\Gamma}_{0,\mathfrak{s}}^{(\mathrm{2PPI})(3)} + \overline{\boldsymbol{\Phi}}^{(3)}_{\mathfrak{s}}\right)_{a_{2}a_{8}a_{9}} \overline{\mathfrak{G}}_{\mathfrak{s},a_{9}a_{5}} }$} \\
& \scalebox{0.89}{${\displaystyle \hspace{1.25cm} +2 \sum_{a_{4},a_{5},a_{6},a_{7},a_{8},a_{9}=1}^{N} \overline{\mathfrak{G}}_{\mathfrak{s},a_{4}a_{6}} \left(\overline{\Gamma}_{0,\mathfrak{s}}^{(\mathrm{2PPI})(4)} + \overline{\boldsymbol{\Phi}}^{(4)}_{\mathfrak{s}}\right)_{a_{2}a_{3}a_{6}a_{7}} \overline{\mathfrak{G}}_{\mathfrak{s},a_{7}a_{8}} \left(\overline{\Gamma}_{0,\mathfrak{s}}^{(\mathrm{2PPI})(3)} + \overline{\boldsymbol{\Phi}}^{(3)}_{\mathfrak{s}}\right)_{a_{1}a_{8}a_{9}} \overline{\mathfrak{G}}_{\mathfrak{s},a_{9}a_{5}} }$} \\
& \scalebox{0.89}{${\displaystyle \hspace{1.25cm} - \sum_{a_{4},a_{5},a_{6},a_{7}=1}^{N} \overline{\mathfrak{G}}_{\mathfrak{s},a_{4}a_{6}} \left(\overline{\Gamma}_{0,\mathfrak{s}}^{(\mathrm{2PPI})(5)} + \overline{\boldsymbol{\Phi}}^{(5)}_{\mathfrak{s}}\right)_{a_{1}a_{2}a_{3}a_{6}a_{7}} \overline{\mathfrak{G}}_{\mathfrak{s},a_{7}a_{5}} \Bigg) \;, }$}
\end{split}
\label{eq:2PPIFRGmuflowEqExpressionPhi3dotAppendix}
\end{equation}
where all upper bars label quantities evaluated at $\rho=\overline{\boldsymbol{\rho}}_{\mathfrak{s}}$, like:
\begin{equation}
\overline{\mathfrak{G}}^{-1}_{\mathfrak{s}} \equiv \mathfrak{G}^{-1}_{\mathfrak{s}}(\rho=\overline{\boldsymbol{\rho}}_{\mathfrak{s}}) \;.
\label{eq:2PPIFRGmUflowDefGbarAppendix}
\end{equation}
According to~\eqref{eq:2PPIFRGmuflowpropagatorAppendix},~\eqref{eq:2PPIFRGmUflowDerivPhiSCPT2Appendix} and~\eqref{eq:2PPIFRGmUflowDefGbarAppendix}, we have:
\begin{equation}
\overline{\mathfrak{G}}^{-1}_{\mathfrak{s},a_{1}a_{2}} = \overline{\Gamma}^{(2)}_{0,\mathfrak{s},a_{1}a_{2}} + \overline{\boldsymbol{\Phi}}^{(2)}_{\mathfrak{s},a_{1}a_{2}} - \frac{\lambda}{12} \left(1-\mathfrak{s}\right)\left(1 + 2\delta_{a_{1}a_{2}}\right) \mathrlap{\quad \forall a_{1},a_{2} \;.}
\end{equation}

\vspace{0.5cm}

To clarify the link with the mU-flow formulated from the 2PI-FRG, we point out that, for example, the set of three flow equations including~\eqref{eq:2PIFRGmUflowNSCPT1EquationG0DON},~\eqref{eq:2PIFRGmUflowNSCPT1EquationOmega0DON} and~\eqref{eq:2PIFRGmUflowNSCPT1EquationSigma0DON} (obtained from a 2PI-FRG formulation) are altogether equivalent to the equation system made of~\eqref{eq:2PPIFRGmuflowEqExpressionGammadotAppendix} and~\eqref{eq:2PPIFRGmuflowEqExpressionrhodotAppendix} (obtained from a 2PPI-FRG formulation). This can be seen after inserting~\eqref{eq:2PIFRGmUflowNSCPT1EquationSigma0DON} (that expresses $\dot{\overline{\Sigma}}_{\mathfrak{s}}$) into~\eqref{eq:2PIFRGmUflowNSCPT1EquationG0DON} (that expresses $\dot{\overline{G}}_{\mathfrak{s}}$), which yields~\eqref{eq:2PPIFRGmuflowEqExpressionrhodotAppendix}. We illustrate in this way that, for the (0+0)-D model under consideration and at equal values of $N_{\mathrm{SCPT}}$ and $N_{\mathrm{max}}$, the present mU-flow version of the 2PPI-FRG is equivalent to its 2PI-FRG counterpart outlined in section~\ref{sec:2PIFRGuflow0DON}.

\vspace{0.5cm}

The expressions of the derivatives of the free 2PPI EA involved in~\eqref{eq:2PPIFRGmuflowEqExpressionGammadotAppendix} to~\eqref{eq:2PPIFRGmuflowEqExpressionPhi3dotAppendix} are found by differentiating the corresponding free 2PI EA $\Gamma_{0}^{(\mathrm{2PI})}(G)$ with respect to $G$, where $\Gamma_{0}^{(\mathrm{2PI})}(G)$ reads (according to~\eqref{eq:ExpressionFree2PIeffectiveAction}):
\begin{equation}
\Gamma_{0}^{(\mathrm{2PI})}(G) = -\frac{1}{2}\mathrm{Tr}_{a}\big[\ln(2\pi G)\big] + \frac{1}{2}\mathrm{Tr}_{a}\big(C^{-1}G-\mathbb{I}_{N}\big) \;,
\end{equation}
and then replacing $G_{a_{1}a_{2}}$ by $\rho_{a_{1}} \delta_{a_{1}a_{2}}$ with~\eqref{eq:2PPIFRGmUflowLink2PI2PPIEAAppendix}. In this way, we obtain:
\begin{equation}
\overline{\Gamma}_{0,\mathfrak{s},a_{1}a_{2}}^{(\mathrm{2PPI})(2)} = \frac{1}{2} \overline{\boldsymbol{\rho}}_{\mathfrak{s},a_{1}}^{-2} \delta_{a_{1}a_{2}} \mathrlap{\quad \forall a_{1},a_{2} \;,}
\label{eq:2PPIFRGmUflowExpressionGamma020DONAppendix}
\end{equation}
\begin{equation}
\overline{\Gamma}_{0,\mathfrak{s},a_{1}a_{2}a_{3}}^{(\mathrm{2PPI})(3)} = - \overline{\boldsymbol{\rho}}_{\mathfrak{s},a_{1}}^{-3} \delta_{a_{1}a_{2}} \delta_{a_{1}a_{3}} \mathrlap{\quad \forall a_{1},a_{2},a_{3} \;,}
\label{eq:2PPIFRGmUflowExpressionGamma030DONAppendix}
\end{equation}
\begin{equation}
\overline{\Gamma}_{0,\mathfrak{s},a_{1}a_{2}a_{3}a_{4}}^{(\mathrm{2PPI})(4)} = 3 \overline{\boldsymbol{\rho}}_{\mathfrak{s},a_{1}}^{-4} \delta_{a_{1}a_{2}} \delta_{a_{1}a_{3}} \delta_{a_{1}a_{4}} \mathrlap{\quad \forall a_{1},a_{2},a_{3},a_{4} \;,}
\label{eq:2PPIFRGmUflowExpressionGamma040DONAppendix}
\end{equation}
\begin{equation}
\hspace{4.1cm} \overline{\Gamma}_{0,\mathfrak{s},a_{1}a_{2}a_{3}a_{4}a_{5}}^{(\mathrm{2PPI})(5)} = - 12 \overline{\boldsymbol{\rho}}_{\mathfrak{s},a_{1}}^{-5} \delta_{a_{1}a_{2}} \delta_{a_{1}a_{3}} \delta_{a_{1}a_{4}} \delta_{a_{1}a_{5}} \quad \forall a_{1},a_{2},a_{3},a_{4},a_{5} \;.
\label{eq:2PPIFRGmUflowExpressionGamma050DONAppendix}
\end{equation}
Note also that the flow parameter $\mathfrak{s}$ runs from $\mathfrak{s}_{\mathrm{i}}=0$ to $\mathfrak{s}_{\mathrm{f}}=1$ as usual. The initial conditions for the derivatives $\overline{\boldsymbol{\Phi}}^{(n)}_{\mathfrak{s}}$ (with $n \geq 2$) are directly inferred from~\eqref{eq:2PPIFRGmUflowDerivPhiSCPT2Appendix} and~\eqref{eq:2PPIFRGmUflowDerivPhiSCPT3Appendix} at $\mathfrak{s}=\mathfrak{s}_{\mathrm{i}}$:
\begin{equation}
\overline{\boldsymbol{\Phi}}_{\mathfrak{s}=\mathfrak{s}_{\mathrm{i}},a_{1}a_{2}}^{(2)} = \left.\frac{\partial^{2}\Phi_{\mathrm{SCPT},N_{\mathrm{SCPT}}=1}(U,\rho)}{\partial\rho_{a_{1}}\partial\rho_{a_{2}}}\right|_{\rho=\overline{\boldsymbol{\rho}}_{\mathfrak{s}}} = \frac{\lambda}{12} \left( 1 + 2 \delta_{a_{1} a_{2}} \right) \mathrlap{\quad \forall a_{1},a_{2} \;,}
\label{eq:2PPIFRGmUflowICDerivPhiSCPT2Appendix}
\end{equation}
\begin{equation}
\hspace{4.25cm} \overline{\boldsymbol{\Phi}}_{\mathfrak{s}=\mathfrak{s}_{\mathrm{i}},a_{1} \cdots a_{n}}^{(n)} = \left. \frac{\partial^{n}\Phi_{\mathrm{SCPT},N_{\mathrm{SCPT}}=1}(U,\rho)}{\partial\rho_{a_{1}} \cdots \partial\rho_{a_{n}}} \right|_{\rho=\overline{\boldsymbol{\rho}}_{\mathfrak{s}}} = 0 \quad \forall a_{1}, \cdots, a_{n}, ~\forall n \geq 3 \;,
\label{eq:2PPIFRGmUflowICDerivPhiSCPT3Appendix}
\end{equation}
whereas $\overline{\boldsymbol{\Gamma}}_{\mathfrak{s}=\mathfrak{s}_{\mathrm{i}}}$ and $\overline{\boldsymbol{\rho}}_{\mathfrak{s}=\mathfrak{s}_{\mathrm{i}}}$ coincide respectively with the gs energy and density calculated with self-consistent PT (from $\Gamma^{(\mathrm{2PPI})}(\rho)$ or, equivalently, from $\Gamma^{(\mathrm{2PI})}(G)$) at $N_{\mathrm{SCPT}}=1$, i.e.:
\begin{equation}
\overline{\boldsymbol{\Gamma}}_{\mathfrak{s}=\mathfrak{s}_{\mathrm{i}}} = E_{\mathrm{gs},\mathrm{SCPT},N_{\mathrm{SCPT}}=1} \;,
\end{equation}
\begin{equation}
\overline{\boldsymbol{\rho}}_{\mathfrak{s}=\mathfrak{s}_{\mathrm{i}}} = \rho_{\mathrm{gs},\mathrm{SCPT},N_{\mathrm{SCPT}}=1} \;.
\end{equation}
Finally, the truncation of the infinite tower of differential equations containing~\eqref{eq:2PPIFRGmuflowEqExpressionGammadotAppendix} to \eqref{eq:2PPIFRGmuflowEqExpressionPhi3dotAppendix} is implemented via the condition:
\begin{equation}
\overline{\boldsymbol{\Phi}}_{\mathfrak{s}}^{(n)}=\overline{\boldsymbol{\Phi}}_{\mathfrak{s}=\mathfrak{s}_{\mathrm{i}}}^{(n)} \mathrlap{\quad \forall \mathfrak{s}, ~ \forall n > N_{\mathrm{max}} \;.}
\label{eq:2PPIfrgModifiedUflowTruncation0DONAppendix}
\end{equation}
We conclude the mU-flow derivations of this section with two important caveats:
\begin{itemize}
\item We can not bypass the 2PI-FRG in this way for more realistic models since the relation $\Gamma^{(\mathrm{2PI})}(G)=\Gamma^{(\mathrm{2PPI})}(\rho)$ does not hold in finite dimensions.
\item The mU-flow thus designed for the 2PPI-FRG is not directly applicable to finite-dimensional problems since $\Phi_{\mathrm{SCPT},N_{\mathrm{SCPT}}=1}(U,\rho)$ is not written explicitly in terms of $\rho$ in the framework of self-consistent PT for the 2PPI EA in such situations, which implies that we can not simply differentiate it with respect to $\rho$ to determine the initial conditions $\dot{\overline{\boldsymbol{\Phi}}}_{\mathfrak{s}=\mathfrak{s}_{\mathrm{i}}}^{(n)}$ (with $n \geq 2$) given here by~\eqref{eq:2PPIFRGmUflowICDerivPhiSCPT2Appendix} and~\eqref{eq:2PPIFRGmUflowICDerivPhiSCPT3Appendix}. However, following the ideas of the IM outlined in section~\ref{sec:2PPIEA}, one might attempt to work out some formulae expressing the initial conditions $\dot{\overline{\boldsymbol{\Phi}}}_{\mathfrak{s}=\mathfrak{s}_{\mathrm{i}}}^{(n)}$ at any dimensions but we leave this as an outlook for the present study.
\end{itemize}

%% file: 7Appendix/Derivations2PPIFRG.tex
\section{\label{sec:DerivMasterEq2PPIFRG}Master equation}

Following the lines set out by ref.~\cite{pol02}, we derive the master equation of the 2PPI-FRG in this appendix. The corresponding reasoning is very close to that leading to the CU-flow equations for the 2PI-FRG (see appendix~\ref{ann:2PIfrgFlowEquationCUflow}). To that end, we start again from an expression of the relevant generating functional, i.e.:
\begin{equation}
Z[K] = \int \mathcal{D}\widetilde{\psi}^{\dagger}\mathcal{D}\widetilde{\psi} \ e^{-S\big[\widetilde{\psi}^{\dagger},\widetilde{\psi}\big] + \int_{\alpha} K_{\alpha}\widetilde{\psi}_{\alpha}^{\dagger}\widetilde{\psi}_{\alpha}} \;,
\label{eq:GeneratingFunctional2PPIFRGAppendix}
\end{equation}
assuming the following analytical form for the classical action:
\begin{equation}
S\Big[\widetilde{\psi}^{\dagger},\widetilde{\psi}\Big] = \int_{\alpha} \widetilde{\psi}_{\alpha}^{\dagger}\left(\hat{O}_{\mathrm{kin},\alpha} + V_{\alpha} - \mu\right)\widetilde{\psi}_{\alpha} + \frac{1}{2}\int_{\alpha_{1},\alpha_{2}} \widetilde{\psi}_{\alpha_{1}}^{\dagger}\widetilde{\psi}_{\alpha_{2}}^{\dagger} U_{\alpha_{1}\alpha_{2}} \widetilde{\psi}_{\alpha_{2}}\widetilde{\psi}_{\alpha_{1}} \;.
\end{equation}
The source $K$ is now local in~\eqref{eq:GeneratingFunctional2PPIFRGAppendix}, as opposed to the source considered in the 2PI-FRG framework. The dependence with respect to the flow parameter $\mathfrak{s}$ is introduced via the substitutions $V \rightarrow V_{\mathfrak{s}}$ and $U \rightarrow U_{\mathfrak{s}}$, thus clarifying the link with the CU-flow version of the 2PI-FRG. In other words, the one-body potential (and therefore the free propagator $C$ as introduced in our presentation of the 2PI-FRG) and the two-body interaction are both \textit{a priori} flow-dependent in the framework of the 2PPI-FRG. The next step consists in differentiating the corresponding Schwinger functional with respect to $\mathfrak{s}$ while keeping the source $K$ constant, thus yielding:
\begin{equation}
\begin{split}
\left.\dot{W}_{\mathfrak{s}}[K]\right|_{K} = & - \int_{\alpha} \dot{V}_{\mathfrak{s},\alpha} \underbrace{\left(\frac{1}{Z_{\mathfrak{s}}[K]}\int \mathcal{D}\widetilde{\psi}^{\dagger}\mathcal{D}\widetilde{\psi} \ \widetilde{\psi}_{\alpha}^{\dagger} \widetilde{\psi}_{\alpha} \ e^{-S_{\mathfrak{s}}\big[\widetilde{\psi}^{\dagger},\widetilde{\psi}\big] + \int_{\alpha} K_{\alpha}\widetilde{\psi}_{\alpha}^{\dagger}\widetilde{\psi}_{\alpha}}\right)}_{\left\langle \widetilde{\psi}_{\alpha}^{\dagger}\widetilde{\psi}_{\alpha} \right\rangle_{K,\mathfrak{s}}} \\
& - \frac{1}{2} \int_{\alpha_{1},\alpha_{2}} \dot{U}_{\mathfrak{s},\alpha_{1}\alpha_{2}} \underbrace{\left(\frac{1}{Z_{\mathfrak{s}}[K]}\int \mathcal{D}\widetilde{\psi}^{\dagger}\mathcal{D}\widetilde{\psi} \ \widetilde{\psi}_{\alpha_{1}}^{\dagger}\widetilde{\psi}_{\alpha_{2}}^{\dagger}\widetilde{\psi}_{\alpha_{2}}\widetilde{\psi}_{\alpha_{1}} \ e^{-S_{\mathfrak{s}}\big[\widetilde{\psi}^{\dagger},\widetilde{\psi}\big] + \int_{\alpha} K_{\alpha}\widetilde{\psi}_{\alpha}^{\dagger}\widetilde{\psi}_{\alpha}}\right)}_{\left\langle \widetilde{\psi}_{\alpha_{1}}^{\dagger}\widetilde{\psi}_{\alpha_{2}}^{\dagger}\widetilde{\psi}_{\alpha_{2}}\widetilde{\psi}_{\alpha_{1}} \right\rangle_{K,\mathfrak{s}}} \\
= & - \int_{\alpha} \dot{V}_{\mathfrak{s},\alpha} W^{(1)}_{\mathfrak{s},\alpha}[K] - \frac{1}{2} \int_{\alpha_{1},\alpha_{2}} \dot{U}_{\mathfrak{s},\alpha_{1}\alpha_{2}} \left\langle \widetilde{\psi}_{\alpha_{1}}^{\dagger}\widetilde{\psi}_{\alpha_{2}}^{\dagger}\widetilde{\psi}_{\alpha_{2}}\widetilde{\psi}_{\alpha_{1}} \right\rangle_{K,\mathfrak{s}} \;,
\end{split}
\label{eq:ExpressionWdot2PPIFRGAppendix}
\end{equation}
where we have made use of the flow-dependent expectation value:
\begin{equation}
\big\langle \cdots \big\rangle_{K,\mathfrak{s}} \equiv \frac{1}{Z_{\mathfrak{s}}[K]}\int \mathcal{D}\widetilde{\psi}^{\dagger}\mathcal{D}\widetilde{\psi} \ \cdots \ e^{-S_{\mathfrak{s}}\big[\widetilde{\psi}^{\dagger},\widetilde{\psi}\big] + \int_{\alpha} K_{\alpha}\widetilde{\psi}_{\alpha}^{\dagger}\widetilde{\psi}_{\alpha}} \;,
\end{equation}
not to be confused with definition~\eqref{eq:FlowdependentExpectationValue2PIFRG} used for the 2PI-FRG approaches. Besides, similarly to~\eqref{eq:RewriteW2UflowAppendix} for the U-flow implementation of the 2PI-FRG, we calculate:
\begin{equation}
\begin{split}
W_{\mathfrak{s},\alpha_{1} \alpha_{2}}^{(2)}[K] \equiv \frac{\delta^{2} W_{\mathfrak{s}}[K]}{\delta K_{\alpha_{1}} \delta K_{\alpha_{2}}} = & \ \frac{\delta}{\delta K_{\alpha_{1}}} \left\langle \widetilde{\psi}^{\dagger}_{\alpha_{2}} \widetilde{\psi}_{\alpha_{2}} \right\rangle_{K,\mathfrak{s}} \\
= & \ \left\langle \widetilde{\psi}^{\dagger}_{\alpha_{1}} \widetilde{\psi}_{\alpha_{1}} \widetilde{\psi}^{\dagger}_{\alpha_{2}} \widetilde{\psi}_{\alpha_{2}} \right\rangle_{K,\mathfrak{s}} - \left\langle \widetilde{\psi}^{\dagger}_{\alpha_{1}} \widetilde{\psi}_{\alpha_{1}} \right\rangle_{K,\mathfrak{s}} \left\langle \widetilde{\psi}^{\dagger}_{\alpha_{2}} \widetilde{\psi}_{\alpha_{2}} \right\rangle_{K,\mathfrak{s}} \\
= & \ \left\langle \widetilde{\psi}^{\dagger}_{\alpha_{1}} \widetilde{\psi}^{\dagger}_{\alpha_{2}} \widetilde{\psi}_{\alpha_{2}} \widetilde{\psi}_{\alpha_{1}} \right\rangle_{K,\mathfrak{s}} - W^{(1)}_{\mathfrak{s},\alpha_{1}}[K] W^{(1)}_{\mathfrak{s},\alpha_{2}}[K] \;,
\end{split}
\label{eq:RewriteW2for2PPIFRGAppendix}
\end{equation}
which gives us:
\begin{equation}
\left\langle \widetilde{\psi}^{\dagger}_{\alpha_{1}} \widetilde{\psi}^{\dagger}_{\alpha_{2}} \widetilde{\psi}_{\alpha_{2}} \widetilde{\psi}_{\alpha_{1}} \right\rangle_{K,\mathfrak{s}} = W_{\mathfrak{s},\alpha_{1} \alpha_{2}}^{(2)}[K] + W^{(1)}_{\mathfrak{s},\alpha_{1}}[K] W^{(1)}_{\mathfrak{s},\alpha_{2}}[K] \;.
\label{eq:RewriteW2for2PPIFRGAppendixBis}
\end{equation}
According to~\eqref{eq:RewriteW2for2PPIFRGAppendixBis},~\eqref{eq:ExpressionWdot2PPIFRGAppendix} is equivalent to:
\begin{equation}
\begin{split}
\left.\dot{W}_{\mathfrak{s}}[K]\right|_{K} = & -\int_{\alpha} \dot{V}_{\mathfrak{s},\alpha} W^{(1)}_{\mathfrak{s},\alpha}[K] - \frac{1}{2} \int_{\alpha_{1},\alpha_{2}} \dot{U}_{\mathfrak{s},\alpha_{1}\alpha_{2}} W_{\mathfrak{s},\alpha_{1} \alpha_{2}}^{(2)}[K] \\
& - \frac{1}{2} \int_{\alpha_{1},\alpha_{2}} \dot{U}_{\mathfrak{s},\alpha_{1}\alpha_{2}} W^{(1)}_{\mathfrak{s},\alpha_{1}}[K] W^{(1)}_{\mathfrak{s},\alpha_{2}}[K] \;.
\end{split}
\label{eq:FlowequationWs2PPIFRGApendix}
\end{equation}
Hence, we have just derived a flow equation for the Schwinger functional $W_{\mathfrak{s}}[K]$. Let us then turn this into a flow equation for the corresponding 2PPI EA $\Gamma_{\mathfrak{s}}^{(\mathrm{2PPI})}[\rho]$. In that respect, we first rewrite the derivative $\left.\dot{\Gamma}^{(\mathrm{2PPI})}_{\mathfrak{s}}[\rho]\right|_{\rho}$ in terms of $\left.\dot{W}_{\mathfrak{s}}[K]\right|_{K}$. As was done earlier for the 1PI and 2PI EAs via~\eqref{eq:ChainRule1PIFRG} and~\eqref{eq:chainRule2PIFRGCflowAppendix} respectively, this can be achieved from the chain rule:
\begin{equation}
\left.\frac{\partial}{\partial\mathfrak{s}}\right|_{K} = \left.\frac{\partial}{\partial\mathfrak{s}}\right|_{\rho} + \int_{\alpha} \left.\dot{\rho}_{\alpha}\right|_{K} \frac{\delta}{\delta\rho_{\alpha}} \;.
\end{equation}
From this, we just need to repeat the reasoning outlined between~\eqref{eq:chainRule2PIFRGCflowAppendix} and~\eqref{eq:2PIfrgCflowDGammaDk} for the 2PI EA $\Gamma^{(\mathrm{2PI})}[G]$ in order to obtain:
\begin{equation}
\dot{\Gamma}^{(\mathrm{2PPI})}_{\mathfrak{s}}[\rho] \equiv \left.\dot{\Gamma}^{(\mathrm{2PPI})}_{\mathfrak{s}}[\rho]\right|_{\rho} = -\left.\dot{W}_{\mathfrak{s}}[K]\right|_{K} \;.
\label{eq:DerivativeGammavsW2PPIFRGApendix}
\end{equation}
As a next step, we will rewrite the derivatives of the Schwinger functional involved in the RHS of~\eqref{eq:FlowequationWs2PPIFRGApendix}. For the first-order derivative, we simply use~\eqref{eq:rhoEqualrhok2PPIFRG} recalled below:
\begin{equation}
W^{(1)}_{\mathfrak{s},\alpha}[K] = \rho_{\alpha}\;.
\label{eq:IntroduceRho2PPIFRGAppendix}
\end{equation}
For the second-order derivative, we follow the lines set out in our previous derivation of the Bethe-Salpeter equation (see appendix~\ref{ann:BetheSalpeterEq}):
\begin{equation}
\delta_{\alpha_{1}\alpha_{2}} = \frac{\delta\rho_{\alpha_{1}}}{\delta\rho_{\alpha_{2}}} = \int_{\alpha_{3}} \frac{\delta \rho_{\alpha_{1}}}{\delta K_{\alpha_{3}}} \frac{\delta K_{\alpha_{3}}}{\delta \rho_{\alpha_{2}}} = \int_{\alpha_{3}} \underbrace{\frac{\delta \rho_{\alpha_{1}}}{\delta K_{\alpha_{3}}}}_{W^{(2)}_{\mathfrak{s},\alpha_{1}\alpha_{3}}} \frac{\delta K_{\alpha_{3}}}{\delta \rho_{\alpha_{2}}} = \int_{\alpha_{3}} W^{(2)}_{\mathfrak{s},\alpha_{1}\alpha_{3}}[K] \frac{\delta K_{\alpha_{3}}}{\delta \rho_{\alpha_{2}}} \;,
\label{eq:RewriteW2step12PPIFRGAppendix}
\end{equation}
as a result of~\eqref{eq:IntroduceRho2PPIFRGAppendix} notably. Since $K_{\alpha}=\frac{\delta\Gamma^{(\mathrm{2PPI})}[\rho]}{\delta\rho_{\alpha}}$ according to~\eqref{eq:LegendreTransform2PPIeffActionFor2PPIFRG}, we can introduce $\Gamma_{\mathfrak{s},\alpha_{1}\alpha_{2}}^{(\mathrm{2PPI})(2)}[\rho]\equiv\frac{\delta^{2}\Gamma_{\mathfrak{s}}^{(\mathrm{2PPI})}[\rho]}{\delta\rho_{\alpha_{1}}\delta\rho_{\alpha_{2}}}$ into~\eqref{eq:RewriteW2step12PPIFRGAppendix} as:
\begin{equation}
\delta_{\alpha_{1}\alpha_{2}} = \int_{\alpha_{3}} W^{(2)}_{\mathfrak{s},\alpha_{1}\alpha_{3}}[K] \Gamma_{\mathfrak{s},\alpha_{3}\alpha_{2}}^{(\mathrm{2PPI})(2)}[\rho] \;,
\label{eq:RewriteW2step22PPIFRGAppendix}
\end{equation}
or, equivalently,
\begin{equation}
W^{(2)}_{\mathfrak{s},\alpha_{1}\alpha_{2}}[K] = \left(\Gamma_{\mathfrak{s}}^{(\mathrm{2PPI})(2)}[\rho]\right)_{\alpha_{1}\alpha_{2}}^{-1} \;.
\label{eq:RewriteW2step32PPIFRGAppendix}
\end{equation}
Finally, we just need to combine~\eqref{eq:FlowequationWs2PPIFRGApendix} with~\eqref{eq:DerivativeGammavsW2PPIFRGApendix},~\eqref{eq:IntroduceRho2PPIFRGAppendix} and~\eqref{eq:RewriteW2step32PPIFRGAppendix} to obtain the master equation of the 2PPI-FRG:
\begin{equation}
\dot{\Gamma}^{(\mathrm{2PPI})}_{\mathfrak{s}}[\rho] = \int_{\alpha} \dot{V}_{\mathfrak{s},\alpha} \rho_{\alpha} + \frac{1}{2} \mathrm{STr}\left[\dot{U}_{\mathfrak{s}} \left(\Gamma_{\mathfrak{s}}^{(\mathrm{2PPI})(2)}[\rho]\right)^{-1} \right] + \frac{1}{2} \int_{\alpha_{1},\alpha_{2}} \dot{U}_{\mathfrak{s},\alpha_{1}\alpha_{2}} \rho_{\alpha_{1}} \rho_{\alpha_{2}} \;.
\end{equation}

\section{Vertex expansion}
\label{sec:2PPIFRGVertexExpansionAppendix}
\subsection{Standard 2PPI functional renormalization group}
\label{sec:standard2PPIFRGVertexExpansionAppendix}

We give in this appendix the coupled differential equations extracted from the master equation of the standard 2PPI-FRG in the framework of the (0+0)-D $O(N)$-symmetric $\varphi^{4}$-theory. This master equation was already given by~\eqref{eq:2PPIFRGmasterEquation0DON} recalled below:
\begin{equation}
\dot{\Gamma}^{(\mathrm{2PPI})}_{\mathfrak{s}}(\rho) = \frac{\lambda}{24} \left( \sum_{a_{1},a_{2}=1}^{N} \boldsymbol{G}_{\mathfrak{s},a_{1}a_{2}}(\rho) + \sum_{a_{1},a_{2}=1}^{N} \rho_{a_{1}} \rho_{a_{2}} \right) \;,
\label{eq:2PPIFRGmasterEquationVertExp0DONAppendix}
\end{equation}
with the propagator $\boldsymbol{G}_{\mathfrak{s}}(\rho)$ defined by:
\begin{equation}
\boldsymbol{G}^{-1}_{\mathfrak{s},a_{1}a_{2}}(\rho) \equiv \Gamma_{\mathfrak{s},a_{1}a_{2}}^{(\mathrm{2PPI})(2)}(\rho) \;.
\label{eq:standard2PPIFRGvertExpPropagatorAppendix}
\end{equation}
The vertex expansion is then carried out by expanding both sides of~\eqref{eq:2PPIFRGmasterEquationVertExp0DONAppendix} around $\rho=\overline{\rho}_{\mathfrak{s}}$, where $\overline{\rho}_{\mathfrak{s}}$ is a configuration of $\rho$ that extremizes the 2PPI EA $\Gamma_{\mathfrak{s}}^{(\mathrm{2PPI})}(\rho)$ (i.e. $\overline{\Gamma}_{\mathfrak{s},a}^{(\mathrm{2PPI})(1)}=0$ $\forall a,\mathfrak{s}$). The LHS of~\eqref{eq:2PPIFRGmasterEquationVertExp0DONAppendix} is expanded using:
\begin{equation}
\Gamma_{\mathfrak{s}}^{(\mathrm{2PPI})}(\rho) = \overline{\Gamma}_{\mathfrak{s}}^{(\mathrm{2PPI})} + \sum_{n=2}^{\infty} \frac{1}{n!} \sum_{a_{1},\cdots,a_{n}=1}^{N} \overline{\Gamma}_{\mathfrak{s},a_{1} \cdots a_{n}}^{(\mathrm{2PPI})(n)} \left(\rho-\overline{\rho}_{\mathfrak{s}}\right)_{a_{1}} \cdots \left(\rho-\overline{\rho}_{\mathfrak{s}}\right)_{a_{n}} \;.
\end{equation}
The rest of the recipe of the vertex expansion is presented in detail in section~\ref{sec:VertexExpansionApp1PIFRG} for the 1PI-FRG, which was notably based on a diagrammatic representation of the expansion of the propagator $\boldsymbol{G}_{k}\big(\vec{\phi}\big)$ that is straightforwardly adapted to $\boldsymbol{G}_{\mathfrak{s}}(\rho)$ in the present situation. Therefore, we will not detail again the vertex expansion procedure but we refer instead the reader to section~\ref{sec:VertexExpansionApp1PIFRG} for the corresponding derivations. We thus directly give the differential equations that we have inferred from~\eqref{eq:2PPIFRGmasterEquationVertExp0DONAppendix}:
\begin{equation}
\dot{\overline{\Gamma}}_{\mathfrak{s}}^{(\mathrm{2PPI})} = \frac{\lambda}{24} \sum_{a_{1},a_{2}=1}^{N} \left( \overline{\boldsymbol{G}}_{\mathfrak{s},a_{1}a_{2}} + \overline{\rho}_{\mathfrak{s},a_{1}} \overline{\rho}_{\mathfrak{s},a_{2}} \right) \;,
\label{eq:standard2PPIFRGflowEqExpressionGammadotAppendix}
\end{equation}
\begin{equation}
\dot{\overline{\rho}}_{\mathfrak{s},a_{1}} = \frac{\lambda}{24} \sum_{a_{2}=1}^{N} \overline{\boldsymbol{G}}_{\mathfrak{s},a_{1}a_{2}} \left( \sum_{a_{3},a_{4},a_{5},a_{6}=1}^{N} \overline{\boldsymbol{G}}_{\mathfrak{s},a_{3}a_{5}} \overline{\Gamma}_{\mathfrak{s},a_{2}a_{5}a_{6}}^{(\mathrm{2PPI})(3)} \overline{\boldsymbol{G}}_{\mathfrak{s},a_{6}a_{4}} - 2 \sum_{a_{3}=1}^{N} \overline{\rho}_{\mathfrak{s},a_{3}} \right) \;,
\label{eq:standard2PPIFRGflowEqExpressionrhodotAppendix}
\end{equation}
\begin{equation}
\begin{split}
\dot{\overline{\Gamma}}_{\mathfrak{s},a_{1}a_{2}}^{(\mathrm{2PPI})(2)} = & \sum_{a_{3}=1}^{N} \dot{\overline{\rho}}_{\mathfrak{s},a_{3}} \overline{\Gamma}^{(\mathrm{2PPI})(3)}_{\mathfrak{s},a_{3}a_{1}a_{2}} \\
& + \frac{\lambda}{24} \Bigg( 2 + 2 \sum_{a_{3},a_{4},a_{5},a_{6},a_{7},a_{8}=1}^{N} \overline{\boldsymbol{G}}_{\mathfrak{s},a_{3}a_{5}} \overline{\Gamma}_{\mathfrak{s},a_{1}a_{5}a_{6}}^{(\mathrm{2PPI})(3)} \overline{\boldsymbol{G}}_{\mathfrak{s},a_{6}a_{7}} \overline{\Gamma}_{\mathfrak{s},a_{2}a_{7}a_{8}}^{(\mathrm{2PPI})(3)} \overline{\boldsymbol{G}}_{\mathfrak{s},a_{8}a_{4}} \\
& \ \hspace{1.05cm} - \sum_{a_{3},a_{4},a_{5},a_{6}=1}^{N} \overline{\boldsymbol{G}}_{\mathfrak{s},a_{3}a_{5}} \overline{\Gamma}_{\mathfrak{s},a_{1}a_{2}a_{5}a_{6}}^{(\mathrm{2PPI})(4)} \overline{\boldsymbol{G}}_{\mathfrak{s},a_{6}a_{4}} \Bigg) \;,
\end{split}
\label{eq:standard2PPIFRGflowEqExpressionGamma2dotAppendix}
\end{equation}
\begin{equation}
\begin{split}
\scalebox{0.97}{${\displaystyle \dot{\overline{\Gamma}}_{\mathfrak{s},a_{1}a_{2}a_{3}}^{(\mathrm{2PPI})(3)} = }$} & \ \scalebox{0.97}{${\displaystyle \sum_{a_{4}=1}^{N} \dot{\overline{\rho}}_{\mathfrak{s},a_{4}} \overline{\Gamma}^{(\mathrm{2PPI})(4)}_{\mathfrak{s},a_{4}a_{1}a_{2}a_{3}} }$} \\
& \scalebox{0.97}{${\displaystyle + \frac{\lambda}{24} \Bigg( -2 \sum_{a_{4},a_{5},a_{6},a_{7},a_{8},a_{9},a_{10},a_{11}=1}^{N} \overline{\boldsymbol{G}}_{\mathfrak{s},a_{4}a_{6}} \overline{\Gamma}_{\mathfrak{s},a_{1}a_{6}a_{7}}^{(\mathrm{2PPI})(3)} \overline{\boldsymbol{G}}_{\mathfrak{s},a_{7}a_{8}} \overline{\Gamma}_{\mathfrak{s},a_{2}a_{8}a_{9}}^{(\mathrm{2PPI})(3)} \overline{\boldsymbol{G}}_{\mathfrak{s},a_{9}a_{10}} \overline{\Gamma}_{\mathfrak{s},a_{3}a_{10}a_{11}}^{(\mathrm{2PPI})(3)} \overline{\boldsymbol{G}}_{\mathfrak{s},a_{11}a_{5}} }$} \\
& \hspace{1.21cm} \scalebox{0.97}{${\displaystyle -2 \sum_{a_{4},a_{5},a_{6},a_{7},a_{8},a_{9},a_{10},a_{11}=1}^{N} \overline{\boldsymbol{G}}_{\mathfrak{s},a_{4}a_{6}} \overline{\Gamma}_{\mathfrak{s},a_{2}a_{6}a_{7}}^{(\mathrm{2PPI})(3)} \overline{\boldsymbol{G}}_{\mathfrak{s},a_{7}a_{8}} \overline{\Gamma}_{\mathfrak{s},a_{1}a_{8}a_{9}}^{(\mathrm{2PPI})(3)} \overline{\boldsymbol{G}}_{\mathfrak{s},a_{9}a_{10}} \overline{\Gamma}_{\mathfrak{s},a_{3}a_{10}a_{11}}^{(\mathrm{2PPI})(3)} \overline{\boldsymbol{G}}_{\mathfrak{s},a_{11}a_{5}} }$} \\
& \hspace{1.21cm} \scalebox{0.97}{${\displaystyle -2 \sum_{a_{4},a_{5},a_{6},a_{7},a_{8},a_{9},a_{10},a_{11}=1}^{N} \overline{\boldsymbol{G}}_{\mathfrak{s},a_{4}a_{6}} \overline{\Gamma}_{\mathfrak{s},a_{1}a_{6}a_{7}}^{(\mathrm{2PPI})(3)} \overline{\boldsymbol{G}}_{\mathfrak{s},a_{7}a_{8}} \overline{\Gamma}_{\mathfrak{s},a_{3}a_{8}a_{9}}^{(\mathrm{2PPI})(3)} \overline{\boldsymbol{G}}_{\mathfrak{s},a_{9}a_{10}} \overline{\Gamma}_{\mathfrak{s},a_{2}a_{10}a_{11}}^{(\mathrm{2PPI})(3)} \overline{\boldsymbol{G}}_{\mathfrak{s},a_{11}a_{5}} }$} \\
& \hspace{1.21cm} \scalebox{0.97}{${\displaystyle +2 \sum_{a_{4},a_{5},a_{6},a_{7},a_{8},a_{9}=1}^{N} \overline{\boldsymbol{G}}_{\mathfrak{s},a_{4}a_{6}} \overline{\Gamma}_{\mathfrak{s},a_{1}a_{2}a_{6}a_{7}}^{(\mathrm{2PPI})(4)} \overline{\boldsymbol{G}}_{\mathfrak{s},a_{7}a_{8}} \overline{\Gamma}_{\mathfrak{s},a_{3}a_{8}a_{9}}^{(\mathrm{2PPI})(3)} \overline{\boldsymbol{G}}_{\mathfrak{s},a_{9}a_{5}} }$} \\
& \hspace{1.21cm} \scalebox{0.97}{${\displaystyle +2 \sum_{a_{4},a_{5},a_{6},a_{7},a_{8},a_{9}=1}^{N} \overline{\boldsymbol{G}}_{\mathfrak{s},a_{4}a_{6}} \overline{\Gamma}_{\mathfrak{s},a_{1}a_{3}a_{6}a_{7}}^{(\mathrm{2PPI})(4)} \overline{\boldsymbol{G}}_{\mathfrak{s},a_{7}a_{8}} \overline{\Gamma}_{\mathfrak{s},a_{2}a_{8}a_{9}}^{(\mathrm{2PPI})(3)} \overline{\boldsymbol{G}}_{\mathfrak{s},a_{9}a_{5}} }$} \\
& \hspace{1.21cm} \scalebox{0.97}{${\displaystyle +2 \sum_{a_{4},a_{5},a_{6},a_{7},a_{8},a_{9}=1}^{N} \overline{\boldsymbol{G}}_{\mathfrak{s},a_{4}a_{6}} \overline{\Gamma}_{\mathfrak{s},a_{2}a_{3}a_{6}a_{7}}^{(\mathrm{2PPI})(4)} \overline{\boldsymbol{G}}_{\mathfrak{s},a_{7}a_{8}} \overline{\Gamma}_{\mathfrak{s},a_{1}a_{8}a_{9}}^{(\mathrm{2PPI})(3)} \overline{\boldsymbol{G}}_{\mathfrak{s},a_{9}a_{5}} }$} \\
& \hspace{1.21cm} \scalebox{0.97}{${\displaystyle - \sum_{a_{4},a_{5},a_{6},a_{7}=1}^{N} \overline{\boldsymbol{G}}_{\mathfrak{s},a_{4}a_{6}} \overline{\Gamma}_{\mathfrak{s},a_{1}a_{2}a_{3}a_{6}a_{7}}^{(\mathrm{2PPI})(5)} \overline{\boldsymbol{G}}_{\mathfrak{s},a_{7}a_{5}} \Bigg) \;, }$}
\end{split}
\label{eq:standard2PPIFRGflowEqExpressionGamma3dotAppendix}
\end{equation}
\begin{equation}
\begin{split}
\dot{\overline{\Gamma}}_{\mathfrak{s}}^{(\mathrm{2PPI})(4)} = & \ \dot{\overline{\rho}}_{\mathfrak{s}} \overline{\Gamma}^{(\mathrm{2PPI})(5)}_{\mathfrak{s}} \\
& + \frac{\lambda}{24} \Big( 8 \overline{\boldsymbol{G}}_{\mathfrak{s}}^{3} \overline{\Gamma}_{\mathfrak{s}}^{(\mathrm{2PPI})(3)} \overline{\Gamma}_{\mathfrak{s}}^{(\mathrm{2PPI})(5)} + 6 \overline{\boldsymbol{G}}_{\mathfrak{s}}^{3} \left(\overline{\Gamma}_{\mathfrak{s}}^{(\mathrm{2PPI})(4)}\right)^{2} - 36 \overline{\boldsymbol{G}}_{\mathfrak{s}}^{4} \left(\overline{\Gamma}_{\mathfrak{s}}^{(\mathrm{2PPI})(3)}\right)^{2} \overline{\Gamma}_{\mathfrak{s}}^{(\mathrm{2PPI})(4)} \\
& \ \hspace{0.95cm} + 24 \overline{\boldsymbol{G}}_{\mathfrak{s}}^{5} \left(\overline{\Gamma}_{\mathfrak{s}}^{(\mathrm{2PPI})(3)}\right)^{4} - \overline{\boldsymbol{G}}_{\mathfrak{s}}^{2} \overline{\Gamma}_{\mathfrak{s}}^{(\mathrm{2PPI})(6)} \Big) \;,
\end{split}
\label{eq:standard2PPIFRGflowEqExpressionGamma4N1dotAppendix}
\end{equation}
where we have used the shorthand notations $\overline{\rho}_{\mathfrak{s}}\equiv\overline{\rho}_{\mathfrak{s},1}$ and $\overline{\Gamma}_{\mathfrak{s}}^{(\mathrm{2PPI})(n)}\equiv\overline{\Gamma}_{\mathfrak{s},1 \cdots 1}^{(\mathrm{2PPI})(n)}$ $\forall n$ in~\eqref{eq:standard2PPIFRGflowEqExpressionGamma4N1dotAppendix} which is only valid at $N=1$.

\subsection{Kohn-Sham functional renormalization group}
\label{sec:KSFRGVertexExpansionAppendix}

Remaining in the framework of the (0+0)-D $O(N)$ model under consideration, we then present the corresponding equation system for the KS-FRG, which is obtained by treating the master equation of the KS-FRG expressing the derivative $\dot{\gamma}_{\mathfrak{s}}(\rho)$ in the same footing. This master equation was given by~\eqref{eq:KSFRGmasterEquation0DON}, i.e.:
\begin{equation}
\dot{\gamma}_{\mathfrak{s}}(\rho) = \frac{1}{2}\dot{V}_{\mathfrak{s}} \sum_{a_{1}=1}^{N} \rho_{a_{1}} + \sum_{a_{1},a_{2}=1}^{N} \rho_{a_{1}} \overline{\Gamma}^{(2)}_{\mathrm{KS},\mathfrak{s},a_{1}a_{2}} \dot{\overline{\rho}}_{\mathfrak{s},a_{2}} + \frac{\lambda}{24} \left( \sum_{a_{1},a_{2}=1}^{N} \boldsymbol{G}_{\mathfrak{s},a_{1}a_{2}}(\rho) + \sum_{a_{1},a_{2}=1}^{N} \rho_{a_{1}} \rho_{a_{2}} \right) \;,
\label{eq:KSFRGmasterEquation0DONAppendix}
\end{equation}
where the propagator $\boldsymbol{G}_{\mathfrak{s}}(\rho)$ is given by~\eqref{eq:standard2PPIFRGvertExpPropagatorAppendix} which is equivalent to:
\begin{equation}
\boldsymbol{G}^{-1}_{\mathfrak{s},a_{1}a_{2}}(\rho) \equiv \Gamma_{\mathfrak{s},a_{1}a_{2}}^{(\mathrm{2PPI})(2)}(\rho) = \Gamma_{\mathrm{KS},\mathfrak{s},a_{1}a_{2}}^{(2)}(\rho) + \gamma^{(2)}_{\mathfrak{s},a_{1}a_{2}}(\rho) \;.
\label{eq:KSFRGvertExpPropagatorAppendix}
\end{equation}
The vertex expansion is now applied by expanding both sides of~\eqref{eq:KSFRGmasterEquation0DONAppendix} around the configuration $\rho=\overline{\rho}_{\mathfrak{s}}$ that extremizes $\gamma_{\mathfrak{s}}(\rho)$ (i.e. $\overline{\gamma}^{(1)}_{\mathfrak{s},a}=0$ $\forall a,\mathfrak{s}$). Notably, the expansion of the LHS of~\eqref{eq:KSFRGmasterEquation0DONAppendix} directly follows from:
\begin{equation}
\gamma_{\mathfrak{s}}(\rho) = \overline{\gamma}_{\mathfrak{s}} + \sum_{n=2}^{\infty} \frac{1}{n!} \sum_{a_{1},\cdots,a_{n}=1}^{N} \overline{\gamma}^{(n)}_{\mathfrak{s},a_{1} \cdots a_{n}} \left(\rho-\overline{\rho}_{\mathfrak{s}}\right)_{a_{1}} \cdots \left(\rho-\overline{\rho}_{\mathfrak{s}}\right)_{a_{n}} \;.
\end{equation}
We refer once again to section~\ref{sec:VertexExpansionApp1PIFRG} for the details of the derivations underlying the vertex expansion and directly give instead the differential equations that we obtained by applying this procedure to~\eqref{eq:KSFRGmasterEquation0DONAppendix}:
\begin{equation}
\dot{\overline{\gamma}}_{\mathfrak{s}} = \frac{1}{2}\dot{V}_{\mathfrak{s}} \sum_{a_{1}=1}^{N} \overline{\rho}_{\mathfrak{s},a_{1}} + \frac{1}{2} \sum_{a_{1}=1}^{N} \overline{\rho}^{-1}_{\mathfrak{s},a_{1}} \dot{\overline{\rho}}_{\mathfrak{s},a_{1}} + \frac{\lambda}{24} \sum_{a_{1},a_{2}=1}^{N} \left( \overline{\boldsymbol{G}}_{\mathfrak{s},a_{1}a_{2}} + \overline{\rho}_{\mathfrak{s},a_{1}} \overline{\rho}_{\mathfrak{s},a_{2}} \right) \;,
\label{eq:KSFRGEqExpressionGammadotAppendix}
\end{equation}
\begin{equation}
\begin{split}
\dot{\overline{\rho}}_{\mathfrak{s},a_{1}} = & - \frac{1}{2}\dot{V}_{\mathfrak{s}} \sum_{a_{2}=1}^{N} \overline{\boldsymbol{G}}_{\mathfrak{s},a_{1}a_{2}} \\
& + \frac{\lambda}{24} \sum_{a_{2}=1}^{N} \overline{\boldsymbol{G}}_{\mathfrak{s},a_{1}a_{2}} \left( \sum_{a_{3},a_{4},a_{5},a_{6}=1}^{N} \overline{\boldsymbol{G}}_{\mathfrak{s},a_{3}a_{5}} \left(\overline{\Gamma}_{\mathrm{KS},\mathfrak{s}}^{(3)}+\overline{\gamma}_{\mathfrak{s}}^{(3)}\right)_{a_{2}a_{5}a_{6}} \overline{\boldsymbol{G}}_{\mathfrak{s},a_{6}a_{4}} - 2 \sum_{a_{3}=1}^{N} \overline{\rho}_{\mathfrak{s},a_{3}} \right) \;,
\end{split}
\label{eq:KSFRGEqExpressionrhodotAppendix}
\end{equation}
\begin{equation}
\begin{split}
\scalebox{0.99}{${\displaystyle \dot{\overline{\gamma}}_{\mathfrak{s},a_{1}a_{2}}^{(2)} = }$} & \ \scalebox{0.99}{${\displaystyle \sum_{a_{3}=1}^{N} \dot{\overline{\rho}}_{\mathfrak{s},a_{3}} \overline{\gamma}^{(3)}_{\mathfrak{s},a_{3}a_{1}a_{2}} }$} \\
& \scalebox{0.99}{${\displaystyle + \frac{\lambda}{24} \Bigg( 2 + 2 \sum_{a_{3},a_{4},a_{5},a_{6},a_{7},a_{8}=1}^{N} \overline{\boldsymbol{G}}_{\mathfrak{s},a_{3}a_{5}} \left(\overline{\Gamma}_{\mathrm{KS},\mathfrak{s}}^{(3)} + \overline{\gamma}^{(3)}_{\mathfrak{s}}\right)_{a_{1}a_{5}a_{6}} \overline{\boldsymbol{G}}_{\mathfrak{s},a_{6}a_{7}} \left(\overline{\Gamma}_{\mathrm{KS},\mathfrak{s}}^{(3)} + \overline{\gamma}^{(3)}_{\mathfrak{s}}\right)_{a_{2}a_{7}a_{8}} \overline{\boldsymbol{G}}_{\mathfrak{s},a_{8}a_{4}} }$} \\
& \ \hspace{0.85cm} \scalebox{0.99}{${\displaystyle - \sum_{a_{3},a_{4},a_{5},a_{6}=1}^{N} \overline{\boldsymbol{G}}_{\mathfrak{s},a_{3}a_{5}} \left(\overline{\Gamma}_{\mathrm{KS},\mathfrak{s}}^{(4)} + \overline{\gamma}^{(4)}_{\mathfrak{s}}\right)_{a_{1}a_{2}a_{5}a_{6}} \overline{\boldsymbol{G}}_{\mathfrak{s},a_{6}a_{4}} \Bigg) \;, }$}
\end{split}
\label{eq:KSFRGEqExpressiongamma2dotAppendix}
\end{equation}
\begin{equation*}
\begin{split}
\scalebox{0.97}{${\displaystyle \dot{\overline{\gamma}}_{\mathfrak{s},a_{1}a_{2}a_{3}}^{(3)} = }$} & \ \scalebox{0.97}{${\displaystyle\sum_{a_{4}=1}^{N} \dot{\overline{\rho}}_{\mathfrak{s},a_{4}} \overline{\gamma}^{(4)}_{\mathfrak{s},a_{4}a_{1}a_{2}a_{3}} }$} \\
& \scalebox{0.97}{${\displaystyle + \frac{\lambda}{24} \Bigg( -2 \sum_{a_{4},a_{5},a_{6},a_{7},a_{8},a_{9},a_{10},a_{11}=1}^{N} \overline{\boldsymbol{G}}_{\mathfrak{s},a_{4}a_{6}} \left(\overline{\Gamma}_{\mathrm{KS},\mathfrak{s}}^{(3)} + \overline{\gamma}^{(3)}_{\mathfrak{s}}\right)_{a_{1}a_{6}a_{7}} \overline{\boldsymbol{G}}_{\mathfrak{s},a_{7}a_{8}} \left(\overline{\Gamma}_{\mathrm{KS},\mathfrak{s}}^{(3)} + \overline{\gamma}^{(3)}_{\mathfrak{s}}\right)_{a_{2}a_{8}a_{9}} }$} \\
& \hspace{5.3cm} \scalebox{0.97}{${\displaystyle \times \overline{\boldsymbol{G}}_{\mathfrak{s},a_{9}a_{10}} \left(\overline{\Gamma}_{\mathrm{KS},\mathfrak{s}}^{(3)} + \overline{\gamma}^{(3)}_{\mathfrak{s}}\right)_{a_{3}a_{10}a_{11}} \overline{\boldsymbol{G}}_{\mathfrak{s},a_{11}a_{5}} }$} \\
& \hspace{1.21cm} \scalebox{0.97}{${\displaystyle -2 \sum_{a_{4},a_{5},a_{6},a_{7},a_{8},a_{9},a_{10},a_{11}=1}^{N} \overline{\boldsymbol{G}}_{\mathfrak{s},a_{4}a_{6}} \left(\overline{\Gamma}_{\mathrm{KS},\mathfrak{s}}^{(3)} + \overline{\gamma}^{(3)}_{\mathfrak{s}}\right)_{a_{2}a_{6}a_{7}} \overline{\boldsymbol{G}}_{\mathfrak{s},a_{7}a_{8}} \left(\overline{\Gamma}_{\mathrm{KS},\mathfrak{s}}^{(3)} + \overline{\gamma}^{(3)}_{\mathfrak{s}}\right)_{a_{1}a_{8}a_{9}} }$} \\
& \hspace{5.3cm} \scalebox{0.97}{${\displaystyle \times \overline{\boldsymbol{G}}_{\mathfrak{s},a_{9}a_{10}} \left(\overline{\Gamma}_{\mathrm{KS},\mathfrak{s}}^{(3)} + \overline{\gamma}^{(3)}_{\mathfrak{s}}\right)_{a_{3}a_{10}a_{11}} \overline{\boldsymbol{G}}_{\mathfrak{s},a_{11}a_{5}} }$} \\
& \hspace{1.21cm} \scalebox{0.97}{${\displaystyle -2 \sum_{a_{4},a_{5},a_{6},a_{7},a_{8},a_{9},a_{10},a_{11}=1}^{N} \overline{\boldsymbol{G}}_{\mathfrak{s},a_{4}a_{6}} \left(\overline{\Gamma}_{\mathrm{KS},\mathfrak{s}}^{(3)} + \overline{\gamma}^{(3)}_{\mathfrak{s}}\right)_{a_{1}a_{6}a_{7}} \overline{\boldsymbol{G}}_{\mathfrak{s},a_{7}a_{8}} \left(\overline{\Gamma}_{\mathrm{KS},\mathfrak{s}}^{(3)} + \overline{\gamma}^{(3)}_{\mathfrak{s}}\right)_{a_{3}a_{8}a_{9}} }$} \\
& \hspace{5.3cm} \scalebox{0.97}{${\displaystyle \times \overline{\boldsymbol{G}}_{\mathfrak{s},a_{9}a_{10}} \left(\overline{\Gamma}_{\mathrm{KS},\mathfrak{s}}^{(3)} + \overline{\gamma}^{(3)}_{\mathfrak{s}}\right)_{a_{2}a_{10}a_{11}} \overline{\boldsymbol{G}}_{\mathfrak{s},a_{11}a_{5}} }$} \\
& \hspace{1.21cm} \scalebox{0.97}{${\displaystyle +2 \sum_{a_{4},a_{5},a_{6},a_{7},a_{8},a_{9}=1}^{N} \overline{\boldsymbol{G}}_{\mathfrak{s},a_{4}a_{6}} \left(\overline{\Gamma}_{\mathrm{KS},\mathfrak{s}}^{(4)} + \overline{\gamma}^{(4)}_{\mathfrak{s}}\right)_{a_{1}a_{2}a_{6}a_{7}} \overline{\boldsymbol{G}}_{\mathfrak{s},a_{7}a_{8}} \left(\overline{\Gamma}_{\mathrm{KS},\mathfrak{s}}^{(3)} + \overline{\gamma}^{(3)}_{\mathfrak{s}}\right)_{a_{3}a_{8}a_{9}} \overline{\boldsymbol{G}}_{\mathfrak{s},a_{9}a_{5}} }$} \\
& \hspace{1.21cm} \scalebox{0.97}{${\displaystyle +2 \sum_{a_{4},a_{5},a_{6},a_{7},a_{8},a_{9}=1}^{N} \overline{\boldsymbol{G}}_{\mathfrak{s},a_{4}a_{6}} \left(\overline{\Gamma}_{\mathrm{KS},\mathfrak{s}}^{(4)} + \overline{\gamma}^{(4)}_{\mathfrak{s}}\right)_{a_{1}a_{3}a_{6}a_{7}} \overline{\boldsymbol{G}}_{\mathfrak{s},a_{7}a_{8}} \left(\overline{\Gamma}_{\mathrm{KS},\mathfrak{s}}^{(3)} + \overline{\gamma}^{(3)}_{\mathfrak{s}}\right)_{a_{2}a_{8}a_{9}} \overline{\boldsymbol{G}}_{\mathfrak{s},a_{9}a_{5}} }$}
\end{split}
\end{equation*}
\begin{equation}
\begin{split}
\hspace{1.7cm} & \hspace{1.21cm} \scalebox{0.97}{${\displaystyle +2 \sum_{a_{4},a_{5},a_{6},a_{7},a_{8},a_{9}=1}^{N} \overline{\boldsymbol{G}}_{\mathfrak{s},a_{4}a_{6}} \left(\overline{\Gamma}_{\mathrm{KS},\mathfrak{s}}^{(4)} + \overline{\gamma}^{(4)}_{\mathfrak{s}}\right)_{a_{2}a_{3}a_{6}a_{7}} \overline{\boldsymbol{G}}_{\mathfrak{s},a_{7}a_{8}} \left(\overline{\Gamma}_{\mathrm{KS},\mathfrak{s}}^{(3)} + \overline{\gamma}^{(3)}_{\mathfrak{s}}\right)_{a_{1}a_{8}a_{9}} \overline{\boldsymbol{G}}_{\mathfrak{s},a_{9}a_{5}} }$} \\
& \hspace{1.21cm} \scalebox{0.97}{${\displaystyle - \sum_{a_{4},a_{5},a_{6},a_{7}=1}^{N} \overline{\boldsymbol{G}}_{\mathfrak{s},a_{4}a_{6}} \left(\overline{\Gamma}_{\mathrm{KS},\mathfrak{s}}^{(\mathrm{2PPI})(5)} + \overline{\gamma}^{(5)}_{\mathfrak{s}}\right)_{a_{1}a_{2}a_{3}a_{6}a_{7}} \overline{\boldsymbol{G}}_{\mathfrak{s},a_{7}a_{5}} \Bigg) \;, }$}
\end{split}
\label{eq:2PPIFRGEqExpressiongamma3dotAppendix}
\end{equation}
\begin{equation}
\begin{split}
\scalebox{0.98}{${\displaystyle\dot{\overline{\gamma}}_{\mathfrak{s}}^{(4)} = }$} & \ \scalebox{0.98}{${\displaystyle \dot{\overline{\rho}}_{\mathfrak{s}} \overline{\gamma}^{(5)}_{\mathfrak{s}} }$} \\
& \scalebox{0.98}{${\displaystyle + \frac{\lambda}{24} \bigg( 8 \overline{\boldsymbol{G}}_{\mathfrak{s}}^{3} \left(\overline{\Gamma}_{\mathrm{KS},\mathfrak{s}}^{(3)}+\overline{\gamma}_{\mathfrak{s}}^{(3)}\right) \left(\overline{\Gamma}_{\mathrm{KS},\mathfrak{s}}^{(5)}+\overline{\gamma}_{\mathfrak{s}}^{(5)}\right) + 6 \overline{\boldsymbol{G}}_{\mathfrak{s}}^{3} \left(\overline{\Gamma}_{\mathrm{KS},\mathfrak{s}}^{(4)}+\overline{\gamma}_{\mathfrak{s}}^{(4)}\right)^{2} }$} \\
& \hspace{0.95cm} \scalebox{0.98}{${\displaystyle - 36 \overline{\boldsymbol{G}}_{\mathfrak{s}}^{4} \left(\overline{\Gamma}_{\mathrm{KS},\mathfrak{s}}^{(3)}+\overline{\gamma}_{\mathfrak{s}}^{(3)}\right)^{2} \left(\overline{\Gamma}_{\mathrm{KS},\mathfrak{s}}^{(4)}+\overline{\gamma}_{\mathfrak{s}}^{(4)}\right) + 24 \overline{\boldsymbol{G}}_{\mathfrak{s}}^{5} \left(\overline{\Gamma}_{\mathrm{KS},\mathfrak{s}}^{(3)}+\overline{\gamma}_{\mathfrak{s}}^{(3)}\right)^{4} - \overline{\boldsymbol{G}}_{\mathfrak{s}}^{2} \left(\overline{\Gamma}_{\mathrm{KS},\mathfrak{s}}^{(6)}+\overline{\gamma}_{\mathfrak{s}}^{(6)}\right) \bigg) \;. }$}
\end{split}
\label{eq:KSFRGflowEqExpressionGamma4N1dotAppendix}
\end{equation}
Note that~\eqref{eq:KSFRGflowEqExpressionGamma4N1dotAppendix}, which is only valid at $N=1$, involves the shorthand notations $\overline{\rho}_{\mathfrak{s}}\equiv\overline{\rho}_{\mathfrak{s},1}$, $\overline{\Gamma}_{\mathrm{KS},\mathfrak{s}}^{(n)}\equiv\overline{\Gamma}_{\mathrm{KS},\mathfrak{s},1 \cdots 1}^{(n)}$ $\forall n$ and $\overline{\gamma}_{\mathfrak{s}}^{(n)}\equiv\overline{\gamma}_{\mathfrak{s},1 \cdots 1}^{(n)}$ $\forall n$.

%% file: 7Appendix/SummaryFrench.tex
Les approches de type intégrale fonctionnelle résultent d'une formulation particulière de la théorie quantique des champs, produit de la fusion entre la théorie des champs et la formulation de la mécanique quantique basée sur les intégrales de chemin~\cite{fey10,fri20}. Celles-ci sont donc souvent aussi appelées approches de type intégrale de chemin et se sont avérées extrêmement pertinentes dans des domaines de recherche très variés, incluant notamment plusieurs domaines de la physique qui traitent du problème à $N$ corps quantique (physique de la matière condensée, physique nucléaire, physique des particules, ...). Ce travail de thèse vise à expliquer comment différentes techniques de ce type, et surtout celles de l'état de l'art, permettent de décrire les systèmes quantiques à $N$ corps. Il paie une attention particulière aux systèmes fortement corrélés de taille finie, ce qui est le cas des noyaux atomiques notamment. Les travaux visant à décrire des systèmes nucléaires à l'aide d'approches de type intégrale de chemin demeurent peu nombreux, préliminaires mais prometteurs~\cite{pug03,bha05,fur07,kem16,kem16bis,yok19,yok19bis,yok19bis3}, et cette thèse s'inscrit ainsi dans cette lignée, tout en s'adressant à l'ensemble des communautés intéressées par le problème à $N$ corps quantique.

\vspace{0.5cm}

Le \textbf{chapitre}~\ref{chap:Intro1} de cette thèse présente un état des lieux des approches utilisées à l'heure actuelle pour traiter le problème à $N$ corps nucléaire. Pour une telle discussion, il faut tout d'abord avoir en tête les caractéristiques principales des noyaux atomiques~: i) Les noyaux sont constitués de fermions, à savoir les nucléons (i.e. les protons et les neutrons), qui sont fortement corrélés~; ii) Les nucléons sont eux-mêmes des particules composites puisqu'ils sont constitués de quarks et de gluons, ce qui implique que l'interaction nucléon-nucléon (NN) dérive de la dynamique des quarks et des gluons~; iii) Les noyaux sont des systèmes auto-confinés, i.e. les nucléons d'un noyau sont confinés dans un potentiel qu'ils génèrent eux-mêmes~; iv) Les noyaux sont des systèmes mésoscopiques (i.e. de taille finie) et ne peuvent donc pas manifester de brisure de symétrie spontanée (SSB).

\vspace{0.5cm}

Concernant tout d'abord la construction formelle de l'interaction NN, une approche fiable permettrait selon le point ii) de connecter directement l'interaction ainsi construite à celle gouvernant la physique des quarks et des gluons, décrite notamment par la chromodynamique quantique (QCD). Une approche de l'état de l'art consiste à développer des interactions NN à travers des théories effectives de basse énergie de la QCD, appelées théories effectives des champs (EFTs) chirales~\cite{epe09,mac20,ham19}. De cette manière, on construit une interaction NN nue, c'est-à-dire qui ne prend pas en compte les effets du milieu. Ces derniers sont alors capturés par l'étape du traitement du problème à $N$ corps, i.e. l'étape de traitement de cette interaction dans le but d'en extraire des observables. Ce processus en deux étapes (construction de l'interaction + traitement du problème à $N$ corps) caractérise toute approche microscopique visant à décrire un système nucléaire. Une telle philosophie basée sur une interaction NN nue définit ainsi les approches dites \textit{ab initio}, dont les implémentations les plus modernes sont souvent basées sur des EFTs chirales, et donc explicitement reliées à la QCD. Il est également important de mentionner que ces techniques \textit{ab initio} sont systématiquement améliorables dans le sens où l'on peut y définir un certain ordre ou niveau d'approximation, que l'on peut éventuellement améliorer en calculant un ordre suivant. Autrement dit, les approches \textit{ab initio} sont particulièrement adaptées pour contrôler la physique qui est prise en compte dans nos modèles. Ces avantages ont bien entendu un prix, plus précisément sur le plan numérique, si bien que les techniques \textit{ab initio} ne peuvent pas encore être appliquées aux noyaux les plus lourds (pas à la majorité des noyaux composés de plus de 100 nucléons notamment), même si leur portée d'application ne cesse de s'accroître à l'heure actuelle.

\vspace{0.5cm}

Il y a également l'approche énergie fonctionnelle de la densité (EDF)~\cite{ben03,sch19} qui constitue une alternative aux techniques \textit{ab initio}. Il s'agit aujourd'hui de la seule approche microscopique qui peut être appliquée à tous les systèmes nucléaires (i.e. à tous les noyaux atomiques mais aussi e.g. aux étoiles à neutrons). Contrairement aux techniques \textit{ab initio}, l'approche EDF n'est pas systématiquement améliorable et repose sur une interaction NN effective (c'est-à-dire une interaction NN incorporant déjà au moins partiellement les effets du milieu nucléaire). Alors qu'une interaction résultant d'une EFT dans le cadre d'une méthode \textit{ab initio} prend la forme d'un lagrangien ou d'un hamiltonien, celle-ci est encodée dans une fonctionnelle de la matrice densité à un corps (aussi appelée fonctionnelle de la densité ou EDF) dérivant d'un pseudo hamiltonien pour l'approche EDF. \`{A} cet égard, cette dernière ressemble ainsi à la théorie fonctionnelle de la densité (DFT)~\cite{hoh64,koh65,koh65bis}. Il y a cependant deux différences essentielles entre EDF et DFT~: i) L'approche EDF est phénoménologique dans le sens où la construction de la fonctionnelle de la densité sous-jacente repose sur une ansatz et non sur des théorèmes d'existence comme les théorèmes de Hohenberg-Kohn sur lesquels se base la DFT. De plus, les théorèmes de Hohenberg-Kohn~\cite{hoh64} sont définis pour des systèmes soumis à un potentiel externe~: leur formulation dans le cadre de systèmes auto-confinés comme les noyaux atomiques est toujours sujette à débats~; ii) Contrairement à l'EDF, la DFT n'est pas adaptée à décrire les systèmes à couches ouvertes (et qui exhibent donc des comportements collectifs), ce qui se traduit techniquement par le fait que les théorèmes de Hohenberg-Kohn et le schéma de Kohn-Sham ne sont formulés que pour des systèmes ne manifestant aucune brisure de symétrie. Le caractère phénoménologique discuté dans le point i) implique que l'approche EDF n'est pas explicitement connectée à la QCD, contrairement aux techniques \textit{ab initio}. En contrepartie, il induit également, en combinaison avec les méthodes de traitement des EDFs, que le coût numérique sous-jacent demeure suffisamment faible pour étudier tous les systèmes nucléaires.

\vspace{0.5cm}

Concernant justement les méthodes de traitement des EDFs, à savoir le traitement du problème à $N$ corps, nous pouvons à présent expliciter quelques points importants liés au point ii). Le traitement des EDFs se scinde généralement en deux étapes~\cite{sch19}. La première, appelée \textit{single-reference} (SR) EDF, consiste à appliquer un principe variationnel à la fonctionnelle utilisée~: les relations résultantes à résoudre sont en général des équations Hartree-Fock-Bogoliubov (HFB)~\cite{bog58,val58,deg66,rin80}, aussi appelées équations Bogoliubov-de Gennes. Cette étape est ainsi basée sur une approximation de champ moyen, ce dernier prenant en général la forme d'un déterminant de Slater brisant certaines symétries dont la nature dépend du type de comportement collectif que l'on cherche à décrire. En physique nucléaire, ces SSBs concernent typiquement le groupe des translations spatiales (les noyaux sont des systèmes mésoscopiques et donc localisés dans l'espace), le groupe $U(1)$ (la plupart des noyaux sont superfluides) et le groupe $SU(2)$ (les noyaux peuvent également être non-sphériques, ce qui est qualifié de déformation nucléaire). Comme dit précédemment, les noyaux atomiques sont des systèmes mésoscopiques et ne peuvent donc pas manifester de SSB. Par conséquent, les symétries brisées au niveau SR EDF doivent être restaurées via une seconde étape appelée \textit{multi-reference} (MR) EDF. Ainsi, dans le cadre de l'approche EDF, ce mécanisme de brisures et de restaurations de symétrie permet de capturer des corrélations associées à divers phénomènes collectifs de manière particulièrement efficace pour des systèmes nucléaires, d'où le point ii) évoqué ci-dessus. Plus concrètement, différentes réalisations de l'étape MR EDF sont mentionnées dans le chapitre~\ref{chap:Intro1}, et en particulier la \textit{projected generator coordinate method} (PGCM) qui repose sur un autre principe variationnel optimisant les poids d'une superposition d'états restaurant ainsi la ou les symétrie(s) spontanément brisée(s) au niveau SR EDF. Ces superpositions impliquent une somme sur les différentes valeurs du ou des paramètre(s) d'ordre associé(s) à cette ou ces symétrie(s) brisée(s). Ce paramètre est un degré de liberté (bosonique) collectif, de même que la densité qui intervient dans la fonctionnelle utilisée. Cela souligne l'importance de tels degrés de liberté pour traiter efficacement le problème à $N$ corps nucléaire.

\vspace{0.5cm}

Il serait bien sûr extrêmement intéressant de construire des modèles possédant à la fois la fiabilité des méthodes \textit{ab initio} (qui découle de leur caractère systématiquement améliorable et de leur lien explicite avec la QCD) et l'efficacité de l'approche EDF (liée à son coût numérique gérable pour les outils actuels, même pour les systèmes nucléaires les plus complexes). Notre but général est de construire des modèles nucléaires connectés à la QCD et dont l'implémentation reste abordable d'un point de vue numérique. Il y a bien entendu plusieurs directions possibles à suivre dans cette optique. En particulier, la formulation de la théorie quantique des champs basée sur les intégrales de chemin est une option pertinente pour diverses raisons. Parmi ces dernières, on peut notamment mentionner que les approches de ce formalisme sont systématiquement améliorables et qu'elles sont adaptées pour traiter des lagrangiens ou des hamiltoniens (qui résulteraient par exemple d'EFTs de la QCD). De plus, puisque l'importance des degrés de liberté collectifs a été soulignée ci-dessus, il convient également d'insister sur le fait qu'il existe différentes techniques inhérentes à ce formalisme pour introduire de tels objets, comme e.g. les transformations de Hubbard-Stratonovich (HSTs)~\cite{str57,hub59} ou encore les méthodes utilisant des actions effectives (EAs) \textit{n-particle-(point-)irreducible} ($n$P(P)I). Comme mentionné précédemment, il se trouve que les approches basées sur des intégrales de chemin ont été peu exploitées en physique nucléaire et les applications effectuées demeurent préliminaires mais prometteuses~\cite{pug03,bha05,fur07,kem16,kem16bis,yok19,yok19bis,yok19bis3}. Ceci dit, le formalisme des intégrales de chemin pourrait permettre de répondre à des questions ouvertes soulevées par les modèles nucléaires actuels. Par exemple, les EDFs dérivent de pseudo hamiltoniens dépendant de la densité et cette dépendance est source de spuriosités mais permet de construire une interaction sous la forme d'une EDF qui prend en compte les effets du milieu nucléaire~: l'origine de cette dépendance, qui est au cœur de l'approche EDF, est toujours sujette à débats à l'heure actuelle. Il y a donc de nombreux indicateurs appuyant la pertinence des approches de type intégrale de chemin pour construire des modèles nucléaires fiables. Ce travail de thèse est une étude exploratoire de telles approches dans le cadre d'un \textit{toy model}, à savoir la théorie $\varphi^4$ (0+0)-D et $O(N)$-symétrique ou plus simplement modèle $O(N)$ (0+0)-D. Considérant l'importance des modèles $O(N)$~\cite{sta68,zin02,mos03} et des techniques basées sur des intégrales de chemin~\cite{neg98,zin02,kle06} en physique théorique, la pertinence du travail ainsi réalisé dépasse clairement les frontières de la physique nucléaire. Néanmoins, le lien avec le problème à $N$ corps nucléaire sera également mis en exergue régulièrement à travers cette étude. Le chapitre~\ref{chap:Intro1} se termine par une annonce de plan indiquant que les méthodes testées sont séparées en deux catégories~: les approches de type groupe de renormalisation fonctionnel (FRG) et les autres qualifiées de techniques diagrammatiques. Celles-ci font respectivement l'objet des chapitres~\ref{chap:FRG} et~\ref{chap:DiagTechniques}. Le chapitre~\ref{chap:Intro2} s'attache à définir techniquement ces deux catégories d'approches ainsi que le \textit{toy model} choisi et le chapitre~\ref{chap:Conclusion} contient la conclusion de cette thèse.

\vspace{0.5cm}

Le \textbf{chapitre}~\ref{chap:Intro2} de cette thèse définit tout d'abord les principales fonctionnelles manipulées dans le cadre des approches de type intégrale de chemin, à savoir la fonction de partition, la fonctionnelle de Schwinger et les EAs $n$PI~\cite{jon64,ded64,cor74,vas72,vas73,vas73bis,vas74,pis74,vas74bis}. Les EAs $n$PPI sont abordées dans la présente étude mais discutées en détail dans le chapitre~\ref{chap:DiagTechniques} uniquement, où l'on explicite notamment le lien entre EA 2PPI et DFT~\cite{fuk94,fuk95,val97,sch04,fer08}. Le début du chapitre~\ref{chap:Intro2} fournit également les équations permettant d'obtenir à partir de ces fonctionnelles les grandeurs physiques d'intérêt pour ce travail de thèse, à savoir l'énergie et la densité à l'état fondamental, notées respectivement $E_{\mathrm{gs}}$ et $\rho_{\mathrm{gs}}$, ainsi que le potentiel effectif $V_{\mathrm{eff}}$. Ensuite, une section clarifie ce que l'on qualifie d'approches diagrammatiques dans le cadre de cette étude~: les diagrammes en question sont générés par le théorème de Wick~\cite{wic50} à travers une relation que l'on démontre (on fait ainsi référence à la notion d'intégration gaussienne traitée en détail dans l'annexe~\ref{ann:GaussianInt}). Des diagrammes peuvent également être utilisés dans le cadre d'approches de type FRG mais (en général) pas dans le but de représenter les contractions associées au théorème de Wick. Concernant le FRG justement, quelques caractéristiques de la philosophie sous-jacente sont aussi mises en avant, à savoir notamment l'implémentation dans le cadre du FRG basé sur l'équation de Wetterich~\cite{wet93} (que l'on appelle également 1PI-FRG) de l'intégration par couches d'impulsion proposée par Wilson~\cite{wil71,wil71bis,wil72} et particulièrement efficace pour décrire les phénomènes critiques. Le FRG basé sur l'équation de Wetterich est aujourd'hui clairement l'approche FRG la plus répandue. Dans le cadre d'applications à des systèmes fermioniques, cette méthode est souvent combinée avec une bosonification dépendante de l'échelle (appelée \textit{scale-dependent bosonization} ou \textit{flowing bosonization}), la bosonification étant mise en œuvre via une HST~\cite{flo09,flo10,gie02,gie02bis,gie04,paw07,fu20}. Une telle approche est en général utilisée pour~: i) gérer l'ambiguïté de Fierz~\cite{jae03}~; ii) changer de degrés de liberté au cours du \textit{flow} (e.g. passer des quarks aux mésons)~; iii) étudier les SSBs et les transitions de phase. L'ambiguïté de Fierz se traduit par une dépendance non-physique (par rapport à un paramètre de champ moyen) des résultats obtenus par la théorie du champ moyen (MFT) combinée avec le 1PI-FRG, ce qui découle du caractère fermionique du problème traité. Cependant, ce problème est absent du modèle $O(N)$ étudié dans le cadre de la présente thèse puisque celui-ci n'implique aucun champ fermionique. On définit ainsi ce \textit{toy model} dans la dernière section du chapitre~\ref{chap:Intro2}. Celui-ci est basé sur l'action classique:
\begin{equation}
S\Big(\vec{\widetilde{\varphi}}\Big)=\frac{m^{2}}{2}\vec{\widetilde{\varphi}}^{2}+\frac{\lambda}{4!}\left(\vec{\widetilde{\varphi}}^{2}\right)^{2} \;,
\label{eq:S0DSummary}
\end{equation}
où $m$ est la masse associée au champ $\widetilde{\varphi}_{a}$ ($a=1,\cdots,N$), $\lambda$ une constante de couplage et le champ réel fluctuant $\widetilde{\varphi}_a$ est défini par l'application $\widetilde{\varphi}_a: \lbrace\bullet\rbrace\rightarrow\mathbb{R}$, où $\lbrace\bullet\rbrace$ indique un espace-temps ponctuel (i.e. sans dimension d'espace ni de temps), résultant du caractère (0+0)-D du modèle. Un champ $\widetilde{\varphi}_{a}$ correspond à une composante du vecteur $\vec{\widetilde{\varphi}}$, soit:
\begin{equation}
\vec{\widetilde{\varphi}} \equiv \begin{pmatrix}
\widetilde{\varphi}_{1} \\
\vdots \\
\widetilde{\varphi}_{N}
\end{pmatrix} \;.
\label{eq:originaldofSummary}
\end{equation}
L'action classique~\eqref{eq:S0DSummary} est également invariante sous les transformations du groupe $O(N)$ que l'on définit mathématiquement dans le chapitre~\ref{chap:Intro2}. La simplicité d'un tel modèle peut être considérée comme un avantage important pour la présente étude puisque l'implémentation numérique des méthodes testées peut être réalisée à moindre coût. En d'autres termes, le \textit{toy model} choisi est un terrain propice pour effectuer une étude comparative exhaustive des techniques considérées. Il s'agit également d'un modèle exactement soluble, ce qui nous permettra de juger la performance des méthodes testées en comparant les résultats obtenus avec les solutions exactes correspondantes. En particulier, la solution exacte pour le potentiel effectif $V_{\mathrm{eff}}$ nous permet de montrer que, même si la symétrie $O(N)$ peut être spontanément brisée par la théorie classique (e.g. lorsque $m^{2}<0$, $\mathcal{R}e(\lambda)>0$ et $\mathcal{I}m(\lambda)=0$), aucune SSB n'est présente au niveau de la solution exacte. Cette dernière jouant ici le rôle du système quantique, cette caractéristique permet ainsi de faire le lien avec l'étude de systèmes mésoscopiques qui ne manifestent eux non plus aucune SSB. Enfin, on précise que les calculs réalisés pour l'ensemble de ce travail de thèse seront effectués dans deux régimes, l'un brisant la symétrie $O(N)$ spontanément au niveau classique ($m^{2}<0$), l'autre ne brisant aucune symétrie ($m^{2}>0$), toujours avec les conditions $\mathcal{R}e(\lambda)>0$ et $\mathcal{I}m(\lambda)=0$. De plus, la majorité des applications numériques sera réalisée pour $N=1$ ou $N=2$. Il faut noter que la symétrie $O(N)$ n'est continue que pour $N \geq 2$~: il s'agit de la situation dans laquelle des modes de Goldstone sont générés si la symétrie $O(N)$ est brisée spontanément (ce qui ne peut donc se produire qu'au niveau classique pour le \textit{toy model} étudié). On s'intéressera également particulièrement au régime non-perturbatif ($\lambda/4! \gtrsim 1$) du modèle choisi, pour faire notamment le lien avec le problème à $N$ corps nucléaire.

\vspace{0.5cm}

Le \textbf{chapitre}~\ref{chap:DiagTechniques} se concentre donc sur les techniques diagrammatiques. Dans le cadre de ce chapitre, le formalisme est d'abord développé pour un modèle $O(N)$ en dimensions arbitraires, puis le cas (0+0)-D est considéré pour les applications numériques. Afin de préparer les études des chapitres~\ref{chap:DiagTechniques} et~\ref{chap:FRG}, on commence par utiliser une HST pour définir différentes représentations du modèle $O(N)$ étudié. Sans HST, il y a la représentation originale de ce modèle basée sur $N$ degrés de liberté (comme ceux impliqués dans~\eqref{eq:originaldofSummary} en (0+0)-D) interagissant via une interaction quartique $O(N)$-symétrique (comme celle impliquée dans~\eqref{eq:S0DSummary} en (0+0)-D). En appliquant une HST à la fonction de partition de la représentation originale, on peut introduire un champ collectif $\widetilde{\sigma}$ en plus des $N$ degrés de liberté originaux. L'interaction quartique originale disparaît dans ce processus de telle sorte que ces $N+1$ degrés de liberté interagissent désormais via une interaction de Yukawa (elle aussi $O(N)$-symétrique) et la représentation ainsi obtenue est qualifiée de mixte dans la présente étude. Enfin, il se trouve que, dans la fonction de partition de la représentation mixte, l'intégrale (fonctionnelle) sur les $N$ degrés de liberté originaux est gaussienne et peut donc être effectuée exactement, de telle sorte que la fonction de partition résultante n'implique plus que $\widetilde{\sigma}$ comme champ fluctuant. Cela mène donc à la troisième et dernière représentation considérée dans cette thèse, à savoir la représentation collective. Les représentations originale, mixte et collective sont toutes des versions mathématiquement équivalentes du modèle étudié puisque les HSTs et l'intégration gaussienne sont toutes basées sur des relations mathématiques exactes. En d'autres termes, les fonctions de partition de ces trois représentations contiennent exactement la même information car aucune approximation n'est réalisée pour obtenir la représentation mixte ou collective à partir de l'originale. Cependant, l'application d'une technique donnée (avec un ordre de troncation donné) pourra mener à des performances différentes selon la représentation choisie, ce qui est analysé en détail par la suite.

\vspace{0.5cm}

On passe ensuite à la première technique diagrammatique traitée dans le chapitre~\ref{chap:DiagTechniques}, à savoir la \textit{loop expansion} (LE). Toutes les techniques diagrammatiques reposent sur une séparation (\textit{splitting} en anglais) de l'action classique (avec possiblement des termes de source) en une partie résiduelle et une partie non-perturbée autour de laquelle le développement en série diagrammatique de la fonctionnelle de Schwinger (ou de la fonction de partition) du modèle est réalisé selon le théorème de Wick (la détermination des diagrammes résultant de la LE est détaillée dans l'annexe~\ref{ann:Diag}). Le paramètre organisant les contributions à ce développement en série est la constante de Planck réduite $\hbar$ et la série résultante est typiquement asymptotique. Dans le cadre de la LE, cette séparation est réalisée de la manière suivante~: l'action classique de la représentation considérée (ajoutée à un ou plusieurs terme(s) de source permettant d'avoir accès à toutes les fonctions de corrélation du modèle) est développée autour d'un de ces extrema et les termes qui sont au maximum quadratiques par rapport à (aux) champ(s) fluctuant(s) constituent la partie non-perturbée. On montre que la LE est équivalente pour les représentations originale et mixte. En particulier, les LEs originale et mixte ne donnent aucun résultat fini lorsque $m^{2}<0$ et $N \geq 2$, ce qui peut être attribué à une manifestation du théorème de Goldstone~\cite{col74} en (0+0)-D. De plus, toujours pour les représentations originale et mixte, la LE produit des séries (e.g. pour la fonctionnelle de Schwinger, $E_{\mathrm{gs}}$ ou $\rho_{\mathrm{gs}}$) qui sont toutes organisées par rapport à la constante de couplage $\lambda$ du modèle $O(N)$ étudié (en plus de l'être par rapport à $\hbar$). Ce n'est en revanche pas le cas pour la LE formulée dans la représentation collective, qui est ainsi intrinsèquement non-perturbative. On montre alors que, pour un ordre de troncation donné par rapport à $\hbar$, la LE collective surpasse clairement ses homologues original et mixte. Cela est illustré jusqu'au troisième ordre non-trivial de la LE (i.e. jusqu'à l'ordre $\mathcal{O}(\hbar^{4})$ pour $E_{\mathrm{gs}}$ ou $\rho_{\mathrm{gs}}$) et, à notre connaissance, la LE collective n'avait jamais été appliquée à un ordre aussi élevé précédemment. On prouve également que la symétrie $O(N)$ est automatiquement restaurée par la LE collective alors que les LEs originale et mixte mènent toujours à une brisure spontanée non-physique (i.e. pas au niveau classique) de cette symétrie. Les résultats montrent également que les performances de la LE collective s'améliorent lorsque $N$ augmente. Cela est cohérent avec le fait que le développement en $1/N$, qui est une méthode répandue (et traitée explicitement dans l'annexe~\ref{ann:LargeN} reprenant notamment les travaux de Keitel et Bartosch~\cite{kei12} ainsi que ceux de Rosa et collaborateurs~\cite{ros16})~\cite{wil73,mos03}, et la LE collective coïncident au niveau de leur premier ordre non-trivial, ce qui est montré par les résultats de la présente étude (cependant, cette équivalence n'est plus valable à des ordres de troncation plus élevés).

\vspace{0.5cm}

Il a été mentionné précédemment que la LE est réalisée autour d'un extremum de l'action classique mais que se passe-t-il lorsque cette action a plusieurs extrema ? Dans la formulation standard de la LE, il faut effectivement en choisir un mais celle-ci peut être généralisée à l'aide de la théorie de Picard-Lefschetz~\cite{wit10,tan15} pour prendre en compte tous les extrema de l'action classique. Pour ce faire, on détaille comment décomposer le contour d'intégration de la fonction de partition du \textit{toy model} étudié (uniquement à $N=1$) pour obtenir une représentation en transsérie (et non en série) de cette fonction de partition, et pour en déduire ensuite les quantités calculées dans le cadre de cette étude, à savoir $E_{\mathrm{gs}}$ et $\rho_{\mathrm{gs}}$. La représentation en transséries correspond à une combinaison de séries obtenues via la LE standard évaluée à chacun des extrema de l'action classique et ces séries sont toutes pondérées par un terme exponentiel dont l'argument implique l'action classique évaluée au même extremum. Pour obtenir une information fiable (et donc une approche systématiquement améliorable) à partir de la LE combinée ou non avec la théorie de Picard-Lefschetz (et à partir de toute autre technique basée sur des séries asymptotiques), on peut appliquer diverses méthodes de resommation aux séries asymptotiques déterminées. Trois de ces méthodes ont été exploitées dans ce travail de thèse~: i) La resommation de Padé-Borel~\cite{pad1892,bor28,ell96,ben99,kle01} est une technique très connue et répandue, qui utilise des approximants (appelés approximants de Padé) sous forme de fractions rationnelles~; ii) Le \textit{conformal mapping}~\cite{leg80} repose sur une application du plan de Borel dans un cercle de rayon unité~: c'est par cette application que l'on peut redistribuer l'information sur le comportement de la série originale (asymptotique) aux ordres élevés à tous les coefficients d'une nouvelle série à partir de laquelle on peut donc construire de meilleures approximations~; iii) La resommation Meijer-G ou resommation Borel-hypergéométrique~\cite{mer15,mer16,ped16a,ped16b,san17,mer18,ant19} est quant à elle basée sur d'autres approximants (qui peuvent être exprimés à l'aide de fonctions G de Meijer~\cite{mar83,bea13}) qui manifestent des coupures de branchement, ce qui les différencie des pôles des approximants de Padé. A notre connaissance, la présente étude est la première à appliquer la resommation Borel-hypergéométrique à un modèle $O(N)$ (mis à part le cas (0+0)-D à $N=1$) ainsi qu'à combiner la LE collective avec une méthode de resommation. Concernant les résultats obtenus, l'apport de la représentation en transséries est clairement illustré dans un régime pertinent (à savoir pour $m^{2}<0$ à $N=1$). Cependant, même si la resommation Borel-hypergéométrique se démarque, il s'avère que, pour les LEs originale et mixte, la resommation ne permet d'obtenir une description satisfaisante du régime non-perturbatif du \textit{toy model} choisi qu'à partir du troisième ordre non-trivial, ce qui peut être extrêmement difficile à atteindre dans le cadre d'une théorie réaliste.

\vspace{0.5cm}

On passe ensuite à la deuxième technique diagrammatique traitée dans le chapitre~\ref{chap:DiagTechniques}, à savoir la théorie des perturbations optimisée (OPT)~\cite{ste81,kil81,oko87,oko87bis,oko88,dun88,oko89,dun89,jon89,jon90,coo91,kli91,buc92,buc92bis,sis91,sis92,sis92bis,kor93,kor93bis, sis93,sis94,sis94bis,kle98,kle11,kle01,kle06,kle11bis,jon90bis,sch98,sch08,yuk99,sez79,zin10}. Dans le cadre de cette approche, on ajoute et on retranche un terme quadratique à l'action classique, de telle sorte à habiller les parties libre (i.e. quadratique) et interaction de cette action. Les versions habillées de ces parties libre et interaction constituent respectivement les parties non-perturbée et résiduelle à partir desquelles le développement diagrammatique de l'OPT est réalisé (la détermination des diagrammes résultant de l'OPT est détaillée dans l'annexe~\ref{ann:Diag}). La partie quadratique ainsi ajoutée implique un champ classique $\sigma$ que l'on peut ensuite ajuster pour optimiser ce développement, ce qui explique le nom de la méthode OPT. Ainsi, l'OPT peut être considérée comme une alternative aux HSTs qui introduisent un nouveau champ quantique et non classique dans la théorie considérée. Différents critères existent pour ajuster le champ classique de l'OPT et trois d'entre eux ont été testés dans ce travail de thèse~: i) Le principe de moindre sensibilité (PMS) consiste à extrémiser la série diagrammatique tronquée par rapport au champ classique $\sigma$~; ii) La méthode du \textit{turning point} (TP) revient à chercher les points d'inflexion (et non les extrema comme le PMS) de la série tronquée par rapport à $\sigma$~; iii) La condition auto-cohérente (SCC) impose que certaines quantités choisies soient reproduites exactement à l'ordre zéro ainsi qu'à l'ordre de troncation du développement de l'OPT (si la quantité choisie est la densité, on a ainsi une connexion claire avec le schéma de Kohn-Sham de la DFT). Il se trouve que le PMS et la méthode du TP peuvent être soit appliqués directement à la série décrivant la quantité que l'on cherche à calculer (e.g. $E_{\mathrm{gs}}$), soit à la série exprimant la fonction de partition (à partir de laquelle la grandeur physique d'intérêt est donc déterminée \textit{a posteriori}). Les résultats obtenus (notamment avec le PMS pour déterminer $E_{\mathrm{gs}}$) montrent que la première option est meilleure que la seconde. On note que le PMS et la méthode du TP ont d'ailleurs déjà été appliqués au modèle $O(N)$ de cette étude mais uniquement dans le régime où $m^{2}>0$~\cite{ros16}. Dans cette thèse, on étend cette étude au régime avec $m^{2}<0$ (où l'on rappelle que la symétrie $O(N)$ est spontanément brisée au niveau classique) et compare les résultats obtenus avec la SCC. Bien que le critère du PMS s'avère être clairement le plus performant à tous les ordres testés de l'OPT, il ne mène pas aux équations les plus simples à résoudre. Pour cette raison, la présente étude met en avant la SCC comme une alternative viable pour construire des modèles abordables d'un point de vue numérique. On peut aussi noter que, contrairement à la LE, la symétrie $O(N)$ est automatiquement conservée dans le cadre de l'OPT qui ne peut donc pas mener à une SSB non-physique pour le \textit{toy model} traité. Enfin, il convient de préciser que l'ensemble des calculs réalisés pour l'OPT ont été effectués dans la représentation originale du modèle $O(N)$. Le \textit{toy model} traité dans cette thèse ne possède qu'un seul canal mais, dans le cadre d'applications à des théories plus réalistes reposant sur plusieurs canaux, l'OPT peut en principe être généralisée de telle sorte à traiter tous ces canaux de manière équitable en introduisant un champ classique, i.e. en ajoutant et en soustrayant un terme quadratique à l'action classique, pour chacun de ces canaux. Une telle généralisation est aussi possible pour les approches formulées dans le cadre de la représentation mixte ou collective en utilisant des HSTs à plusieurs canaux~\cite{lan15,kug18}.

\vspace{0.5cm}

Ni la LE ni l'OPT n'implique de fonctionnelle reliant l'énergie à la densité et donc aucune méthode exploitée à ce stade ne permet d'établir un lien clair avec l'approche EDF nucléaire. Ce n'est pas le cas des théories des perturbations auto-cohérentes, qui sont basées sur des EAs, vers lesquelles on se tourne ensuite. Une EA est définie comme transformée de Legendre de la fonctionnelle de Schwinger correspondante~: différents types d'EAs (EAs 1PI, 2PI, 3PI ... ou 2PPI, 3PPI, ...) sont obtenues selon les termes de source introduits dans cette fonctionnelle de Schwinger. Lorsqu'une expression diagrammatique de cette dernière est établie (e.g. via la LE), on peut toujours en principe en déduire une expression pour l'EA correspondante par e.g. la méthode de l'inversion (qui est appliquée dans l'annexe~\ref{ann:InversionMethod} aux cas des EAs 1PI, 2PI et 4PPI)~\cite{fuk94,fuk95}. Les quantités $E_{\mathrm{gs}}$ et $\rho_{\mathrm{gs}}$ (ou d'autres observables d'intérêt) sont déterminées en extrémisant l'EA considérée (tronquée à un ordre donné dans son développement) par rapport à ses arguments~: les équations ainsi obtenues sont appelées équations de \textit{gap}. Lorsque le paramètre choisi pour organiser le développement est $\hbar$, une EA $n$PI peut être directement exprimée en sélectionnant tous les diagrammes $n$PI dans l'expression de la fonctionnelle de Schwinger associée et résultant de la LE. Deux paramètres différents ont été considérés pour les approches diagrammatiques basées sur des EAs, à savoir $\hbar$ et $\lambda$. Pour l'EA 1PI, on considère les représentations originale et collective (la représentation mixte n'est pas exploitée pour des raisons techniques expliquées dans le chapitre~\ref{chap:DiagTechniques}). Pour l'EA 1PI originale avec $\hbar$ ou $\lambda$ comme paramètre de développement, on constate que la symétrie $O(N)$ est toujours conservée, ce qui implique que la solution des équations de \textit{gap} pour la fonction de corrélation à 1-point (associée au champ original $\vec{\widetilde{\varphi}}$) est toujours nulle. Par conséquent, l'EA 1PI originale n'a aucun moyen de capturer des corrélations efficacement pour décrire le régime non-perturbatif du modèle $O(N)$ étudié de manière satisfaisante, ce qui transparaît clairement dans les résultats obtenus. En revanche, l'EA 1PI collective fait intervenir la fonction de corrélation à 1-point du champ auxiliaire $\widetilde{\sigma}$, qui n'est quant à elle pas contrainte par la symétrie $O(N)$. On montre ainsi que cette approche basée sur l'EA 1PI collective, connue notamment sous le nom d'\textit{auxiliary field loop expansion} (LOAF)~\cite{coo10,coo11,mih11bis,mih11,coo12,daw12,chi14,coo14,coo14bis,coo16}, mène à de bien meilleurs résultats que son homologue utilisant l'EA 1PI originale. Cependant, l'EA 1PI collective possède aussi certains défauts, à savoir e.g. que les développements diagrammatiques du modèle $O(N)$ sont bien plus difficiles à déterminer pour la représentation collective que pour l'originale ou la mixte. On étudie donc ensuite le cas de l'EA 2PI dans le cadre de ces deux dernières représentations. Il convient de rappeler que des degrés de liberté collectifs sont automatiquement introduits lorsque l'on manipule des EAs $n$P(P)I avec $n \geq 2$. Les EAs 2PI~\cite{lee60,lut60,ded62,ded64,ded64bis,bay62,cor74} et 2PPI~\cite{he90,ver92} peuvent être considérées respectivement comme des fonctionnelles de fonctions de Green et des fonctionnelles de la densité, d'où le lien avec l'approche EDF nucléaire. Dans le cadre du développement par rapport à $\hbar$ et par rapport à $\lambda$, les EAs 2PI originale et mixte sont toutes les deux poussées jusqu'à leur troisième ordre non-trivial, ce qui est une première pour le cas mixte~\cite{col74,ben77,cho99,mih01,bla01,coo05,aar02} à notre connaissance (la détermination des diagrammes contribuant à l'EA 2PI mixte est aussi détaillée dans l'annexe~\ref{ann:Diag}). Pour la représentation originale (et contrairement à la situation mixte), les développements en $\hbar$ et en $\lambda$ coïncident lorsque l'on impose que les fonctions de corrélation à 1-point s'annulent. On peut aussi noter que les premiers ordres non-triviaux de l'OPT via le PMS ou la SCC et de la théorie des perturbations auto-cohérente pour l'EA 2PI originale sont identiques pour $E_{\mathrm{gs}}$~: il s'agit du résultat de la théorie Hartree-Fock. De plus, il se trouve que, dans le cadre des théories des perturbations auto-cohérentes, les EAs sont elles aussi typiquement représentées par des séries asymptotiques. Une resommation de Padé-Borel est développée pour l'EA 2PI originale (en imposant que la fonction de corrélation à 1-point du champ $\vec{\widetilde{\varphi}}$ s'annule) ainsi que pour l'EA 2PI mixte avec $\hbar$ comme paramètre de développement, ce qui constitue aussi une première pour l'EA 2PI mixte. Cette dernière approche s'avère être la plus performante parmi toutes celles testées dans le chapitre~\ref{chap:DiagTechniques}, avec la LE collective et l'OPT basée sur le PMS. L'étude réalisée montre notamment que les solutions retenues des équations de \textit{gap} de l'EA 2PI mixte (qui mènent donc à d'excellents résultats) ne manifestent aucune brisure non-physique de la symétrie $O(N)$, et se distinguent donc de la LE originale ou mixte. Enfin, pour conclure le chapitre~\ref{chap:DiagTechniques}, une section clarifie le lien entre EA 2PPI et DFT (à la Kohn-Sham~\cite{val97})~\cite{fuk94,fuk95}. Cependant, il s'avère que, pour le \textit{toy model} étudié, les EAs 2PI et 2PPI coïncident en l'absence de brisure spontanée de la symétrie $O(N)$. Enfin, une dernière section traite l'EA 4P(P)I~\cite{vas74bis,oko97,car04,car13,car13bis} dans la représentation originale, avec un développement en $\hbar$ et en imposant que les fonctions de corrélation à 1-point s'annulent. Cependant, on montre que l'EA 4P(P)I coïncide avec l'EA 2P(P)I (qui est pourtant bien moins complexe à traiter, que ce soit sur le plan formel ou numérique) pour les deux premiers ordres non-triviaux de la théorie des perturbations auto-cohérente dans le cas étudié.

\vspace{0.5cm}

Le \textbf{chapitre}~\ref{chap:FRG} traite ensuite les approches de type FRG. L'idée sous-jacente du FRG est aussi d'ajouter un ou plusieurs terme(s) (contenant une fonction de \textit{cutoff}) à l'action classique du modèle considéré, ce(s) terme(s) impliquant une dépendance par rapport à un paramètre d'échelle ou de \textit{flow} qui peut par exemple correspondre à une échelle d'impulsion. De cette façon, toutes les fonctionnelles du formalisme (e.g. la fonctionnelle de Schwinger ou les EAs) deviennent dépendantes de ce paramètre. On dérive ainsi une équation différentielle exacte (d'ordre 1 par rapport au paramètre de \textit{flow}) pour l'une de ces fonctionnelles, appelée équation de \textit{flow}, qui est formulée la plupart du temps pour une EA. L'approche $n$P(P)I-FRG est donc basée sur une équation de \textit{flow} pour une EA $n$P(P)I. Cette équation différentielle est ensuite traitée pour en tirer un système infini d'équations intégro-différentielles (qui doit donc être tronqué). Il convient alors de pointer du doigt un aspect clé des approches FRG, à savoir le point de départ utilisé pour résoudre ces équations différentielles couplées. Ce point de départ doit bien entendu correspondre à un système que l'on sait traiter exactement (e.g. système libre, théorie Hartree-Fock) mais on peut s'attendre à ce que plus le point de départ utilisé incorpore des corrélations, plus l'approche FRG sous-jacente soit performante. La qualité du point de départ des approches FRG testées joue ainsi un rôle prépondérant dans l'étude du chapitre~\ref{chap:FRG}.

\vspace{0.5cm}

L'approche FRG la plus répandue est traitée dans un premier temps~: il s'agit du 1PI-FRG qui est basé sur l'équation de Wetterich~\cite{wet93} (cette équation est démontrée dans l'annexe~\ref{ann:Derivations1PIFRG}). Pour cette approche, la fonction de \textit{cutoff} est introduite dans la partie libre de l'action classique, de manière à habiller le propagateur libre de la théorie sous-jacente. Le début du chapitre~\ref{chap:FRG} explique pourquoi les conditions initiales de l'approche de Wetterich coïncident avec la théorie classique, ce qui résulte notamment d'une modification de la transformée de Legendre de l'EA 1PI à l'aide de cette fonction de \textit{cutoff}. L'extraction d'un système d'équations intégro-différentielles couplées à partir de l'équation de Wetterich peut être effectuée par diverses méthodes, notamment par la \textit{vertex expansion}~\cite{mor94,kop10bis}, la \textit{derivative expansion} (DE)~\cite{mor94bis,ber96,can03,can03bis,kop10bis2,jak14} et l'approximation Blaizot-M\'{e}ndez-Galain-Wschebor (BMW)~\cite{bla05,bla06,bla06bis,bla06bis2,bla07}. Cependant, pour des raisons techniques que l'on explique, la DE et l'approximation BMW ne peuvent pas être testées dans le cadre du \textit{toy model} (0+0)-D traité dans cette thèse. Seule la \textit{vertex expansion} sera donc utilisée dans la présente étude. Le 1PI-FRG sera ainsi appliqué au modèle $O(N)$ (0+0)-D dans le cadre des représentations originale, mixte et collective (les démonstrations des équations différentielles couplées résultant de la \textit{vertex expansion} appliquée à l'équation de Wetterich de ce \textit{toy model} sont détaillées dans l'annexe~\ref{ann:Derivations1PIFRG}). Cette application a déjà été réalisée pour la représentation originale dans le régime où $m^{2}>0$~\cite{kei12}. On teste donc ici une nouvelle fois la pertinence de l'introduction de degrés de liberté collectifs à travers une HST. En particulier, l'implémentation de la MFT via le 1PI-FRG mixte est discutée et illustrée par des applications numériques. Des problèmes de raideur des équations différentielles à traiter se manifestent dans le régime où $m^{2}<0$ et dans le cadre des représentations originale et mixte. Le 1PI-FRG collectif traite en revanche le modèle étudié avec la même efficacité quelque soit le signe de $m^{2}$. Cependant, dans le régime où $m^{2}>0$, le 1PI-FRG reste le plus performant dans la représentation mixte.

\vspace{0.5cm}

On se concentre ensuite sur le 2PI-FRG~\cite{dup05,dup14} et le 2PPI-FRG~\cite{pol02} qui sont tous deux définis dans un cadre théorique imposant que toutes les fonctions de corrélation à 1-point s'annulent. Cela implique donc que la symétrie $O(N)$ ne peut pas être brisée spontanément par ces approches. A notre connaissance, ces dernières n'ont jamais été appliquées à un modèle $O(N)$. L'impact de la symétrie $O(N)$ sur les formalismes du 2PI-FRG et du 2PPI-FRG est donc discuté en détail dans le chapitre~\ref{chap:FRG}. Le 2PI-FRG est traité avant le 2PPI-FRG. Différents avantages d'une formulation basée sur une EA 2PI (qui ne dépend donc pas de variables de Grassmann puisque toutes les fonctions de corrélation à 1-point sont nulles) sont tout d'abord mis en avant, comme notamment la possibilité de construire des schémas de troncation non-perturbatifs possiblement inaccessibles dans le cadre d'une formulation 1PI. Trois formulations du 2PI-FRG sont traitées~: i) le \textit{C-flow}~\cite{dup05} pour lequel la fonction de \textit{cutoff} est introduite en habillant le propagateur libre de la théorie considérée, comme pour le 1PI-FRG formulé par Wetterich~; ii) le \textit{U-flow}~\cite{dup14} pour lequel la fonction de \textit{cutoff} est introduite de manière à habiller la partie interaction (et non la partie libre) de l'action classique du modèle traité~; iii) le \textit{CU-flow}~\cite{dup14} pour lequel la fonction de \textit{cutoff} habille à la fois la partie libre et la partie interaction de l'action classique et qui est donc une combinaison du \textit{C-flow} et du \textit{U-flow}. Les équations de \textit{flow} et les systèmes d'équations différentielles résultant de leur \textit{vertex expansion} sont démontrés en détail en dimensions arbitraires dans le chapitre~\ref{chap:FRG} (ainsi que dans l'annexe~\ref{ann:Derivations2PIFRG}). On considère tout d'abord le \textit{C-flow} à partir duquel des applications numériques ont été réalisées dans le cadre des représentations originale et mixte. Deux versions du \textit{C-flow} sont testées~: i) le \textit{truncated C-flow} (\textit{tC-flow}) qui est l'implémentation standard du \textit{C-flow} et connue pour être équivalente à la théorie des perturbations auto-cohérente~\cite{ren15}, ce qui est vérifié à diverses reprises dans la présente étude~; ii) le \textit{modified C-flow} (\textit{mC-flow}) qui repose sur une ansatz des vertex 2PI pour tronquer le système infini d'équations différentielles sous-tendant le \textit{C-flow}. Pour le \textit{tC-flow} et le \textit{mC-flow}, quelques problèmes de raideur sont une nouvelle fois mis en exergue et discutés notamment en exploitant le lien entre le \textit{tC-flow} et la théorie des perturbations auto-cohérente. Quelque soit la représentation choisie, les conclusions pour le \textit{C-flow} sont identiques~: i) Bien que le \textit{tC-flow} permet finalement d'appliquer la théorie des perturbations auto-cohérente en résolvant des équations différentielles couplées d'ordre 1 (au lieu des équations de \textit{gap} traitées au chapitre~\ref{chap:DiagTechniques}), les résultats associés se dégradent plus la troncation du système d'équations différentielles couplées sous-jacent est raffinée, conformément aux propriétés des séries divergentes mises en avant dans le chapitre~\ref{chap:DiagTechniques}. Le \textit{tC-flow} n'est donc pas une approche systématiquement améliorable en soit~; ii) L'ansatz sous-tendant le \textit{mC-flow} ne s'avère pas suffisamment fiable pour construire des approches améliorables de manière systématique. Pour pallier la limitation soulevée par le point i), une combinaison du \textit{tC-flow} avec des méthodes de resommation pourrait être testée mais n'est pas traitée par la présente étude. Il est également expliqué que le \textit{C-flow}, dans sa version exploitée dans cette thèse~\cite{dup05}, n'est pas adapté pour décrire le régime du modèle $O(N)$ étudié avec $m^{2}<0$. Pour cette raison, une piste de généralisation du \textit{C-flow} incluant des fonctions de corrélation à 1-point non-triviales est suggérée mais présente plusieurs désavantages conséquents mis en avant. On passe ensuite au \textit{U-flow} dont trois implémentations sont traitées uniquement dans le cadre de la représentation originale (pour des raisons techniques expliquées en détail)~: i) le \textit{truncated U-flow} (\textit{tU-flow}) dont le point de départ associé est la théorie libre et qui est basé sur une approximation particulièrement drastique permettant d'éviter de résoudre l'équation de Bethe-Salpeter pour traiter les équations différentielles correspondantes~; ii) le \textit{plain U-flow} (\textit{pU-flow}) qui est l'implémentation standard du \textit{U-flow} dont le point de départ associé est aussi la théorie libre~; iii) le \textit{modified U-flow} (\textit{mU-flow}) qui constitue une modification ou une généralisation du \textit{pU-flow} telle que le point de départ de l'approche coïncide désormais avec les résultats de la théorie des perturbations auto-cohérente (tronquée à l'ordre choisi). Le \textit{pU-flow} et le \textit{mU-flow} peuvent s'avérer particulièrement coûteux numériquement car le traitement des équations différentielles associées nécessite de résoudre l'équation de Bethe-Salpeter. Pour cette raison, le \textit{tU-flow} pourrait être une alternative intéressante mais la présente étude montre que cette approche ne permet pas de décrire le régime non-perturbatif du \textit{toy model} étudié de manière satisfaisante. De plus, le \textit{mU-flow} se montre bien plus performant que le \textit{tU-flow} et le \textit{pU-flow}, ce qui pouvait être anticipé de par la qualité supérieure de son point de départ. Le \textit{mU-flow} est implémenté jusqu'au troisième ordre non-trivial de la procédure FRG, pour les trois premiers ordres non-triviaux de la théorie des perturbations auto-cohérente constituant son point de départ, et pour les deux signes de $m^{2}$. Le point de départ Hartree-Fock (i.e. le premier ordre non-trivial de la théorie des perturbations auto-cohérente pour l'EA 2PI originale) s'avère être le meilleur choix et le \textit{mU-flow} constitue d'ailleurs dans ce cas l'approche FRG la plus performante du chapitre~\ref{chap:FRG}, quelque soit le signe de $m^{2}$. L'extension du \textit{mU-flow} à la représentation mixte est aussi discutée~: il s'agit d'une direction intéressante puisqu'elle permettrait d'utiliser les excellents résultats obtenus avec l'EA 2PI mixte (via la théorie des perturbations auto-cohérente) au chapitre~\ref{chap:DiagTechniques} comme point de départ de la procédure FRG. Cependant, ce travail requiert encore des développements conséquents sur le plan du formalisme, ce qui est laissé à de futures investigations. Enfin, le \textit{CU-flow} est aussi traité et appliqué numériquement. Une seule implémentation du \textit{CU-flow} est testée (à savoir son implémentation standard qui peut être considérée comme une fusion entre le \textit{tC-flow} et le \textit{pU-flow})~: ses performances sont satisfaisantes et comparables à celle du \textit{mU-flow} avec point de départ Hartree-Fock dans le régime où $m^{2}>0$. Cependant, le \textit{CU-flow} ne permet pas de traiter le cas avec $m^{2}<0$ pour les mêmes raisons que le \textit{C-flow}.

\vspace{0.5cm}

Le chapitre~\ref{chap:FRG} se termine par une section sur le 2PPI-FRG~\cite{pol02}. L'idée d'une implémentation de type \textit{CU-flow} a déjà été mise en avant pour le 2PPI-FRG~\cite{sch04}, auquel cas le point de départ de la procédure FRG pourrait coïncider avec le système de Kohn-Sham. Cependant, pour les applications numériques, on se concentre sur la formulation de type \textit{U-flow} qui est la plus utilisée dans les travaux actuels basés sur le 2PPI-FRG~\cite{kem16,kem16bis,lia18,yok19,yok19bis,yok19bis2,yok19bis3,yok21,yok21bis}. En parallèle des implémentations de type \textit{U-flow} ou \textit{CU-flow} (qui conditionnent uniquement la manière dont la fonction de \textit{cutoff} est introduite dans la fonction de partition ou la fonctionnelle de Schwinger), deux versions du 2PPI-FRG existent~: i) le 2PPI-FRG standard qui correspond à la formulation originelle du 2PPI-FRG~\cite{pol02}~; ii) le Kohn-Sham FRG (KS-FRG) qui est une modification du 2PPI-FRG standard avec une meilleure convergence (par rapport aux troncations canoniques utilisées pour le 2PPI-FRG standard, à savoir le \textit{standard U-flow} (\textit{sU-flow}) et le \textit{plain U-flow} (\textit{pU-flow}) que l'on définit) obtenue en résolvant l'équation de Kohn-Sham~\cite{koh65,koh65bis} en plus des équations différentielles sous-jacentes. Les formalismes de ces deux versions sont développés en détail en dimensions arbitraires dans le chapitre~\ref{chap:FRG} (ainsi que dans l'annexe~\ref{ann:Derivations2PPIFRG}). Les résultats numériques obtenus pour le modèle $O(N)$ (0+0)-D confirment que le KS-FRG surpasse les versions canoniques du 2PPI-FRG standard (i.e. le 2PPI-FRG standard avec le \textit{sU-flow} et le \textit{pU-flow}). On montre également qu'un schéma de troncation plus raffiné (que le \textit{sU-flow} et le \textit{pU-flow}), appelé \textit{improved U-flow} (\textit{iU-flow}), permet de retrouver les résultats du KS-FRG avec le 2PPI-FRG standard~: cela permet de mettre en avant la puissance du formalisme du KS-FRG qui est plus facilement généralisable à des modèles plus réalistes (par rapport au \textit{toy model} traité dans cette thèse) que le 2PPI-FRG standard basé sur l'\textit{iU-flow}. Ainsi, le KS-FRG est la meilleure implémentation du 2PPI-FRG testée dans la présente étude comparative. On montre notamment qu'avec un choix de fonction de \textit{cutoff} judicieux, cette technique peut être utilisée pour décrire le régime du modèle $O(N)$ (0+0)-D avec $m^{2}<0$, et ce avec la même efficacité que pour le cas où $m^{2}>0$. Il est aussi illustré que, pour un problème (0+0)-D, les résultats du KS-FRG et du \textit{pU-flow} du 2PI-FRG coïncident.

\pagebreak

Le \textbf{chapitre}~\ref{chap:Conclusion} conclut ce travail de thèse en résumant les résultats des chapitres~\ref{chap:DiagTechniques} et~\ref{chap:FRG} ainsi qu'en donnant différentes directions possibles à suivre pour de futurs travaux. Une comparaison est faite entre les meilleures approches testées dans cette thèse, à savoir la LE collective, l'OPT basée sur le PMS, la théorie des perturbations auto-cohérente basée sur l'EA 2PI mixte et l'implémentation \textit{mU-flow} du 2PI-FRG. Le succès des approches basées sur une EA 2P(P)I et/ou sur une HST (on rappelle que les représentations mixte et collective sont toutes les deux basées sur une HST) témoigne de la pertinence des degrés de liberté collectifs. Dans ce chapitre de conclusion, on insiste également sur les limitations du \textit{toy model} considéré pour l'étude comparative réalisée, à savoir~: i) son caractère (0+0)-D~; ii) le fait qu'il ne possède qu'un seul canal. Le point i) est à la fois un avantage et un désavantage puisqu'il permet de tester un grand nombre de techniques différentes avec des temps de calculs raisonnables mais certaines approches ne peuvent pas être testées de manière pertinente pour un modèle (0+0)-D, comme le 1PI-FRG basé sur la DE qui a prouvé son efficacité~\cite{mor94bis,ber96,can03,can03bis,kop10bis2,jak14} mais se réduit à une méthode exacte en (0+0)-D dès son premier ordre. Un modèle à dimension finie serait aussi bien plus pertinent pour analyser le poids numérique des différentes méthodes testées, comme le 2PI-FRG. \`{A} cet égard, on peut mentionner les récents travaux de Katanin qui ont montré des résultats encourageants du 2PI-FRG dans le cadre d'un modèle (2+1)-D~\cite{kat19}. Concernant le point ii), on a notamment expliqué comment l'OPT ou encore les approches basées sur des HSTs peuvent être généralisées pour traiter des modèles à plusieurs canaux. Il est à noter que, dans le cadre de tels modèles, une partie des techniques utilisées dans cette thèse peut en principe souffrir de l'ambiguïté de Fierz et une analyse detaillée des effets associés pourrait s'avérer nécessaire. Ainsi, une poursuite naturelle de ce travail de thèse serait par exemple une étude comparative des méthodes testées ici dans le but de traiter un système fermionique à dimension finie avec différents canaux d'importance comparable. Concernant l'approche EDF utilisée en physique nucléaire, un lien entre l'étape MR EDF et la formulation \textit{mU-flow} du 2PI-FRG pourrait également être clarifié.

%% file: 8LastPage/LastPage.tex
\thispagestyle{empty}
\newgeometry{top=0.5cm, bottom=0.5cm, left=2.0cm, right=2.0cm}
\lhead{}
\rhead{}
\rfoot{}
\cfoot{}
\lfoot{}

\noindent 

\begin{flushleft}
\includegraphics[height=2.5cm]{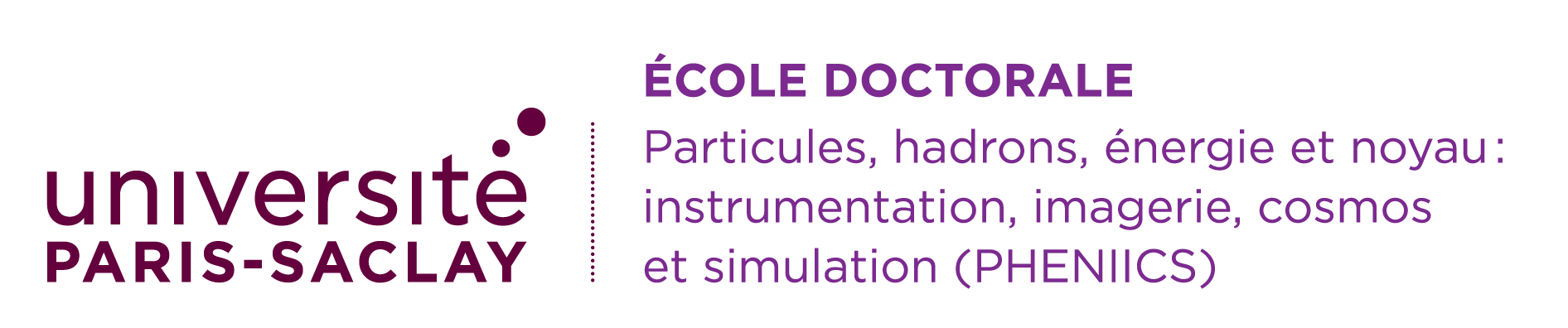}
\end{flushleft}


\vspace{-0.1cm}

\begin{mdframed}[linecolor=Prune,linewidth=1]
\vspace{-.4cm}
\paragraph*{Titre :} Approches de type intégrale de chemin pour l'étude de systèmes quantiques à $N$ corps fortement corrélés

\begin{small}
\vspace{-.4cm}
\paragraph*{Mots clés :} {\footnotesize Groupe de renormalisation fonctionnel, Diagrammes de Feynman, Action effective, Théorie de la resommation, Brisure de symétrie spontanée, Modèle $O(N)$}

\vspace{-.55cm}
\begin{multicols}{2}
\paragraph*{Résumé :} {\footnotesize Le cœur de ce travail de thèse est la formulation de la théorie quantique des champs basée sur les intégrales de chemin et sa capacité à décrire les systèmes quantiques à $N$ corps fortement corrélés de taille finie. Les phénomènes collectifs gouvernant la phénoménologie de tels systèmes peuvent être efficacement décrits par l'implémentation de brisures de symétrie spontanées (SSB) dans le cadre d'approches de type champ moyen. Cependant, la limite thermodynamique n'étant pas pertinente pour des systèmes de taille finie, ces derniers ne peuvent manifester de SSB et les symétries brisées au niveau du champ moyen doivent donc être restaurées. L'efficacité d'approches théoriques à traiter les systèmes quantiques de taille finie peut donc être étudiée à travers leur capacité à restaurer les symétries brisées spontanément. Dans ce travail de thèse, nous prenons pour cadre théorique un modèle $O(N)$ à zéro dimension pour réaliser une telle étude avec différentes techniques de type intégrale de chemin~: théorie des perturbations combinée avec diverses méthodes de resommation (Padé-Borel, Borel-hypergéométrique, \textit{conformal mapping}), versions améliorées de la théorie des perturbations (transséries déterminées via le formalisme des Lefschetz \textit{thimbles}, théorie des perturbations optimisée), théorie des perturbations auto-cohérente basée sur des actions effectives (\textit{auxiliary field loop expansion} (LOAF), formalisme Cornwall-Jackiw-Tomboulis (CJT), action effective 4PPI, ...), techniques de type groupe de renormalisation fonctionnel (FRG) (FRG basé sur l'équation de Wetterich, DFT-FRG, 2PI-FRG). Des connexions entre ces différentes méthodes sont aussi mises en exergue. De plus, le formalisme des intégrales de chemin nous offre la possibilité d'introduire des degrés de liberté collectifs de manière exacte à l'aide de transformations de Hubbard-Stratonovich~: l'effet de telles transformations sur les méthodes susmentionnées est également étudié en détail.}
\end{multicols}
\end{small}
\end{mdframed}

\begin{mdframed}[linecolor=Prune,linewidth=1]
\vspace{-.4cm}
\paragraph*{Title:} Path-integral approaches to strongly-coupled quantum many-body systems

\begin{small}
\vspace{-.4cm}
\paragraph*{Keywords:} {\footnotesize Functional renormalization group, Feynman diagrams, Effective action, Resummation theory, Spontaneous symmetry breaking, $O(N)$ model}

\vspace{-.55cm}
\begin{multicols}{2}
\paragraph*{Abstract:} {\footnotesize The core of this thesis is the path-integral formulation of quantum field theory and its ability to describe strongly-coupled quantum many-body systems of finite size. Collective behaviors can be efficiently described in such systems through the implementation of spontaneous symmetry breaking (SSB) in mean-field approaches. However, as the thermodynamic limit does not make sense in finite-size systems, the latter can not exhibit any SSB and the symmetries which are broken down at the mean-field level must therefore be restored. The efficiency of theoretical approaches in the treatment of finite-size quantum systems can therefore be studied via their ability to restore spontaneously broken symmetries. In this thesis, a zero-dimensional $O(N)$ model is taken as a theoretical laboratory to perform such an investigation with many state-of-the-art path-integral techniques: perturbation theory combined with various resummation methods (Padé-Borel, Borel-hypergeometric, conformal mapping), enhanced versions of perturbation theory (transseries derived via Lefschetz thimbles, optimized perturbation theory), self-consistent perturbation theory based on effective actions (auxiliary field loop expansion (LOAF), Cornwall-Jackiw-Tomboulis (CJT) formalism, 4PPI effective action,~...), functional renormalization group (FRG) techniques (FRG based on the Wetterich equation, DFT-FRG, 2PI-FRG). Connections between these different techniques are also emphasized. In addition, the path-integral formalism provides us with the possibility to introduce collective degrees of freedom in an exact fashion via Hubbard-Stratonovich transformations: the effect of such transformations on the aforementioned methods is also examined in detail.}
\end{multicols}
\end{small}
\end{mdframed}